# Understanding Quantum Technologies

**Fifth edition**

2022

**Olivier Ezratty**

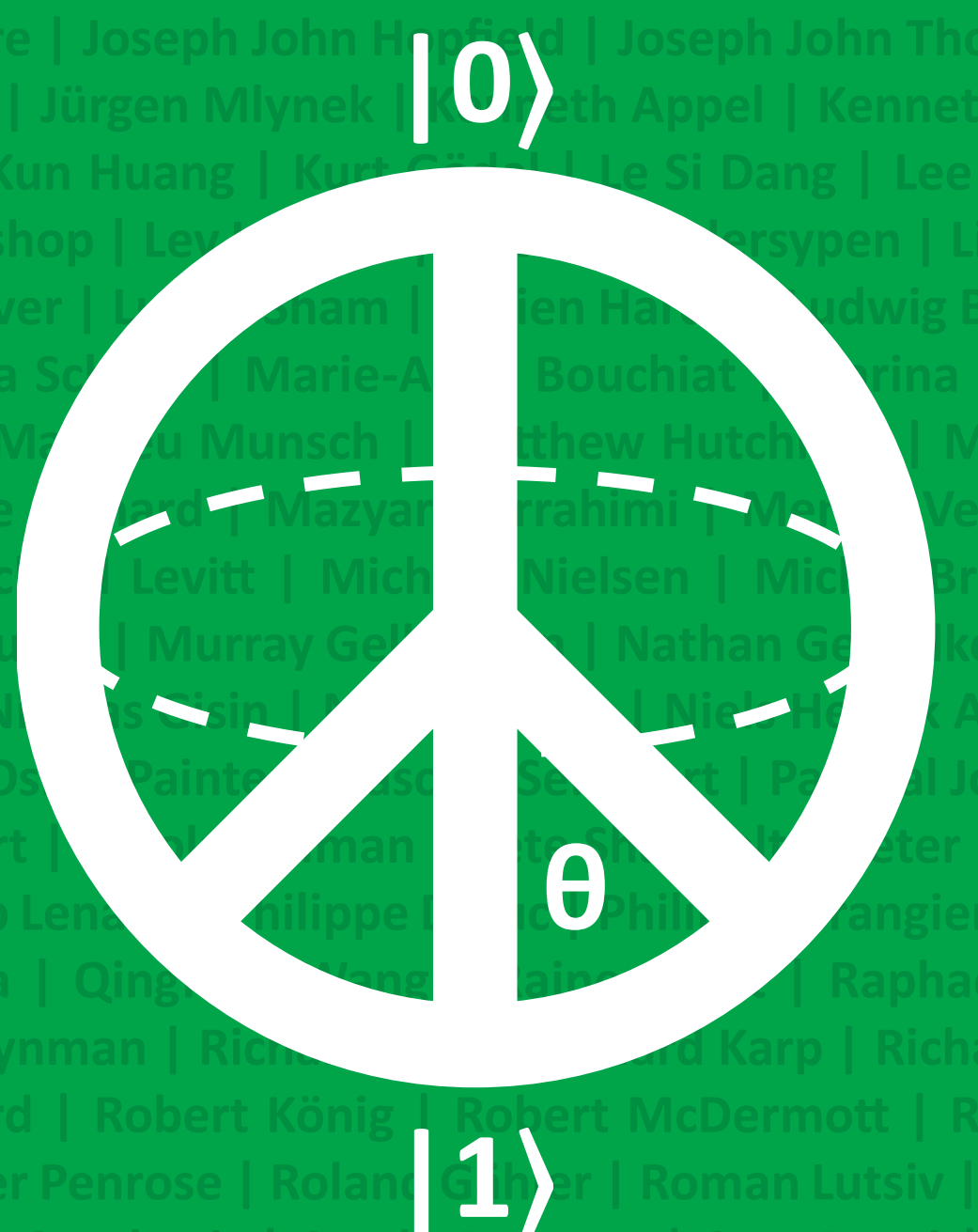

le lab quantique

cover back page

the quantum energy initiative

discover the QEI in page 251

# Understanding Quantum Technologies

**Fifth edition**

2022

**Olivier Ezratty**

## About the author

**Olivier Ezratty**
consultant and author
0000-0003-3944-2896
olivier (at) oezratty.net, www.oezratty.net, @olivez
+33 6 67 37 92 41

Olivier Ezratty advises and trains businesses and public services in the development of their innovation strategies in the quantum technologies realm. He brings them a 360° understanding of these: scientific, technological, marketing as well as the knowledge of the quantum ecosystems.

He has covered many other topics since 2005, with among others digital television, Internet of things and artificial intelligence. As such, he carried out various strategic advisory missions of conferences or training in different verticals and domains such as the **media and telecoms** (Orange, Bouygues Telecom, TDF, Médiamétrie, BVA, Astra), **finance and insurance** (BPCE group, Caisse des Dépôts, Société Générale, Swiss Life, Crédit Agricole, Crédit Mutuel-CIC, Generali, MAIF), **industry and services** (Schneider, Camfil, Vinci, NTN-STR, Econocom, ADP, Air France, Airbus) and the **public sector** (CEA, Météo France, Bpifrance, Business France).

In the quantum realm:

- He is a keynote speaker in a large number of quantum technology events since 2018.
- He published the reference book **Understanding Quantum Technologies** (September 2021 and 2022) following three previous editions in French in 2018, 2019 and 2020. The 2021 and 2022 editions are also available in paperback version on Amazon.
- He runs **two series of podcasts** on quantum technologies with Fanny Bouton (in French): a monthly « Quantum » on tech news (since September 2019) and Decode Quantum, with entrepreneurs and researchers since March 2020, with a total of over 80 episodes.
- He is a trainer on quantum technologies for **Capgemini Institut** and for **CEA INSTN**. In September 2021, he took in charge an elective curriculum on quantum technologies for **EPITA**, an IT engineering school in France.
- He is the cofounder of the **Quantum Energy Initiative** with Alexia Auffèves (CNRS MajuLab Singapore) and Robert Whitney (CNRS LPMMC).
- He is advising **Bpifrance** on quantum projects evaluations, a member of the strategic committee for **France 2030**, the French government innovation strategy plan and also a lecturer at IHEDN.

He also lectures in various universities such as CentraleSupelec, Ecole des Mines de Paris, Télécom Paristech, EPITA, Les Gobelins, HEC, Neoma Rouen and SciencePo, on artificial intelligence, quantum technologies as well as entrepreneurship and product management, in French and English as needed. He is also the author of many open source ebooks in French on entrepreneurship (2006-2019), the CES of Las Vegas yearly report (2006-2020) and on artificial intelligence (2016-2021).

Before all that, Olivier Ezratty started in 1985 at **Sogitec,** a subsidiary of the Dassault group, where he was successively Software Engineer, then Head of the Research Department in the Communication Division. He initialized developments under Windows 1.0 in the field of editorial computing as well as on SGML, the ancestor of HTML and XML. Joining **Microsoft France** in 1990, he gained experience in many areas of the marketing mix: products, channels, markets and communication. He launched the first version of Visual Basic in 1991 and Windows NT in 1993. In 1998, he became Marketing and Communication Director of Microsoft France and in 2001, of the Developer Division, which he created in France to launch the .NET platform and promote it to developers, higher education and research, as well as to startups.

Olivier Ezratty is a software engineer from **Centrale Paris** (1985), which became CentraleSupelec in 2015.

## Credits

Cover illustration: personal creation associating a Bloch sphere describing a qubit and the symbol of peace (my creation, first published in 2018) above a long list of over 400 scientists and entrepreneurs who are mentioned in the ebook.

This document contains over 1600 illustrations. I have managed to give credits to their creators as much as possible. Most sources are credited in footnotes or in the text. Only scientists' portraits are not credited since it's quite hard to track it. I have added my own credit in most of the illustrations I have created. In some cases, I have redrawn some third-party illustrations to create clean vector versions or used existing third-party illustrations and added my own text comments. The originals are still credited in that case.

# Table of contents

# *Foreword*

Quantum technologies hold the promise of major disruptions in computing, communications and sensing. But scientific and technological challenges to their large-scale deployment are still important, and it is quite difficult for public decision makers, users, investors, professionals, and the public at large to anticipate when these will happen. This is of paramount importance for companies to stay competitive, for governments to position their country in this technology race, or for students to make decisions about their career. While some quantum devices are already in use with practical impact, e.g. sophisticated microscopes taking benefit of the exquisite sensitivity of the spin of point defects in diamonds, other technologies will take years if not decades to reach the markets.

But the situation is changing fast. When I co-founded the Quantonation investment fund in 2018, most of the fund's presentation was about the promises of quantum, and about the science. Today, with 21 seed investments made in startups in Europe and North America, the situation has already radically changed since, for the most mature, we are talking about products and customers, and, at least, proofs of concepts. Consulting firms are busy assessing future markets, their size keeps increasing and the horizon is getting closer with significant practical achievements not much further down the road. I'm often asked whether there is not too much "hype" in the field. I don't think so, particularly when I am comparing quantum technologies with other sectors. This is the beginning of market recognition, for a sector which impact is slowly being assessed properly.

But to do that, make proper assessments and keep control of the quantum narrative, we need deep experts who have a proper understanding of all the facets of the technology, from the fundamentals of the science to its applications, including questions about their deployment, their funding, how to teach them, and more. It is necessary to be able to mobilize academic experts to provide an opinion on the science at the base of the innovation, on the ability to make robust products, but we must also be able to imagine their use cases, and scientists alone are not equipped to do so. There is a need for a multidisciplinary collaboration involving scientists, engineers and users capable of taking a forward-looking posture. And here enters my friend Olivier Ezratty, the author of this most wonderful book "Understanding Quantum Technologies", who embodies multidisciplinarity. He has the unique ability to listen, question, gather facts, and synthesize his learnings in a book that stands out as unique in the whole world, as far as I know.

I first met Olivier when I started Quantonation back in 2018. From the start I was impressed by his extremely methodic approach that he had applied with success on an earlier publication on artificial intelligence, and his very unique ambition. The book was first published in French, later in English, and it grew with the field he was "decoding" to use the title of Olivier's famous podcast with Fanny Bouton on quantum technologies. The book has gone only better with time, with thorough updates and new chapters about exciting topics *e.g.* "Quantum Matter" in this new edition. Olivier has also been among the very first supporters of the not-for profit that I co-founded and chaired, Le Lab Quantique. Le Lab Quantique is proud to promote "Understanding Quantum Technologies", an instrument that will benefit its ecosystem building mission.

I am convinced that this book will become a primer for professionals, from scientists to engineers, technicians, investors, and also for teachers, students, and the public at large. We're all extremely lucky to see the second quantum revolution happening before our eyes, science and technology are progressing at an amazing pace and it is essential to invent a new model of knowledge sharing, of collaboration. Olivier Ezratty's book is an indispensable instrument to read this revolution.

**Christophe Jurczak**, Partner at Quantonation, Paris and co-founder, Le Lab Quantique

# *Why*

This book is the 5th edition of a book originally compiling a series of 18 articles that I published in French between June and September 2018. After two enriched editions in French in 2019 and 2020, I switched to English in the fourth, in September 2021 and here we are with an even larger sequel.

This book is a kaleidoscope for quantum technologies with a 360° perspective encompassing historical, scientific, technological, engineering, entrepreneurial, geopolitical, philosophical, and societal dimensions. It is not a quantum for dummies, babies, or your mother-in-law book. It mainly targets three audiences: information technologies (IT) specialists and engineers who want to understand what quantum physics and technologies are all about and decipher its ambient buzz, all participants to the quantum ecosystem from researchers to industry vendors and policy makers, and at last scientific students who would like to investigate quantum technologies as an exploratory field. For them, this book is also the largest review paper they could imagine with over 3,500 bibliographical references.

"Understanding Quantum Technologies" bears a lot of specificities compared to the existing quantum literature. While being rather technical in many parts, it tries to explain things and translate the complex quantum lingua in other tech's lingua, particularly for IT and computer science professionals. It looks at the history of science and ideas and pays tribute to key people, from the past and the present. It investigates rarely covered aspects of quantum technologies and quantum engineering like various enabling technologies (cryogenics, cryo-electronics, new materials design, semiconductors, cabling and lasers), their thermodynamic and energetic dimension and what raw materials are used and where they come from. I cover quantum matter and describe how quantum circuits and manufactured. I even explain how research works in general and in the quantum realm and its codes.

It also extensively covers quantum sensing, telecommunications and cryptography. I also crafted a lot of precisely documented custom illustrations. Another differentiation is in the tone, relaxed when possible and calling out the bs and nonsense when necessary. It is abundant, particularly with when media, analysts and consultants are fueling the quantum hype. I'm always puzzled by how they sometimes cover vendor news without having a real clue about what they are writing about. It motivated me in the first place back in 2015 to start investigating this field. It is always true as quantum technologies are more commonplace but are still largely misunderstood by general audiences as well as by many IT professionals. One striking example shows up when some folks explain that thanks to quantum cryptography, quantum computers will help make cryptography more secure!

Large vendors and the quantum startups funding craze have elevated quantum technologies to the rank of strategic sectors for developed countries. Most governments have launched their national quantum plans, starting with Singapore, the UK, China, USA, Germany, Japan, Australia, France, Russia, Israel, Taiwan, India and The Netherlands. The worldwide quantum technologies race is on. Countries are embattled to acquire or preserve their technological sovereignty, like if it was the last chance to achieve it, particularly for those countries who felt they lost the digital battle against the USA and Asia (mostly China, South Korea and Taiwan). Also, like many deep techs, quantum technologies are dual-use ones, with both civilian and military use cases, increasing the strategic stakes.

While it has not yet reached the volume and funding of other sectors such as artificial intelligence or the digital cloud, the quantum startups and small business ecosystem continues to expand worldwide. In this book, I mention about 550 such companies in many different categories (hardware, software, telecommunications, cryptography, sensing, enabling technologies, services). In most cases, hardware are in the deep techs realm if not in hard tech territory, with many still at an applied research stage with a rather low technology readiness level. Being still very uncertain, this market remains quite open to opportunities for scientists and creative innovators, while in other markets like with semiconductors and large consumer Internet players, the game looks like it is less open.

Quantum technologies are also surrounded by a fair share of hype. A few scientists, their laboratory's communication department, startups and large vendors frequently exaggerate the impact of their work. Many companies also integrate "quantum" into their positioning if not branding in many fancy ways. Either in a totally artificial way, or based on using technologies from the first quantum revolution.

Transistors, lasers and image sensors are quantum, so most digital technologies can claim to be quantum. As a consequence, we must learn to distinguish the old (first quantum revolution related) from the new (second quantum revolution related). However, even stronger bs shows up elsewhere, with false science-based quantum medicine and other charlatanism. I showcase it in a unique section dedicated to quantum hoaxes and scams, starting page 1015.

This book has another flavor. It is the result of an unprecedented human adventure at the heart of the quantum ecosystem. I started the journey back in 2016. I had then decided to select the theme of quantum computing for my usual techno-screening activities, ranging from preparing conferences and training to writing educational ebooks for professionals. I was joined by my friend **Fanny Bouton** to run a popularization conference on quantum computing in Nantes. She brought and still brings a different perspective, including some science fiction derived inspirations. This led to the conference **Le quantique, c'est fantastique** on June 14th, 2018 (video) and to numerous subsequent presentations. On top of that, we launched two series of podcasts (in French) covering quantum tech news and with interviews with researchers, entrepreneurs and also users. We also worked on gender balance and contributed as early as possible to this sector feminization and attract new talents[1]. Fanny took an interesting turn in 2020, starting to work on **OVHcloud**'s startup program. She plays a key role to embark this European cloud vendor in the quantum adventure and now leads this effort. We both went from a role of observer to a very different one.

In this journey that is still going on, we've had the opportunity to meet with top researchers and entrepreneurs, first in France, and then internationally. It started with **Alain Aspect** (IOGS), **Philippe Grangier** (IOGS), **Daniel Esteve** (CEA), **Patrice Bertet** (CEA), **Maud Vinet** (CEA), **Tristan Meunier** (CNRS Institut Néel), **Eleni Diamanti** (CNRS LIP6), **Iordanis Kerenidis** (CNRS IRIF), **Pascale Senellart** (CNRS & UPS C2N and Quandela), **Elham Kashefi** (CNRS LIP6 and VeriQloud), **Alexia Auffèves** (CNRS Institut Néel in Grenoble and now MajuLab in Singapore), **Philippe Duluc** and **Cyril Allouche** (Atos), **Xavier Waintal** (CEA), **Robert Whitney** (CNRS LPMMC), **Théau Peronnin** (Alice&Bob), **Georges-Olivier Reymond** and **Antoine Browaeys** (Pasqal) and many others afterwards. We also toured almost all quantum **startups** in France. And of course, **Christophe Jurczak** from Quantonation and Le Lab Quantique, who kindly wrote this book foreword.

Our outreach then expanded internationally, particularly in Canada, the USA, the UK, Austria and The Netherlands. I had the opportunity to discuss with **Artur Ekert**, **Peter Knight**, **Tommaso Calarco** and many startup founders, from **PsiQuantum**, **IQM**, **ParityQC**, **ProteinQure**, **Qilimanjaro**, **Qblox**, **Jay Gambetta** from IBM and **Rainer Blatt** from AQT. It is not enough. I want more!

In short, during these years, we have been "embedded" in the scientific and entrepreneurial ecosystem. We also applied one of Heisenberg's principles derivatives, namely that a measurement device may influence the measured quantity. It was and remains a beautiful adventure with real people, passions, convictions, ups and downs, and in the end, a nice result with French and European research and entrepreneurship in quantum technologies that are more dynamic and better positioned than a few years ago. And the adventure is just beginning!

[1] With a one-day training session with Roland Berger and Axelle Lemaire in April 2019, with high school students at Magic Makers in September 2019, with young people and parents at the Startup4Teens event in February 2020, and a debate in early March 2020 with Alexia Auffèves, Elham Kashefi and Pascale Senellart hosted by Fanny Bouton and organized at Talan, another event with a dominant female audience of all ages in the Tech4All event organized by Ecole 42 and Digital Ladies in March 2020, each time in partnership with the association *Quelques Femmes du Numérique ! (Some Digital Women).*

You may wonder why this book is free and what is its business model. I have published all my books like this since 2006 and fared well so far (on entrepreneurship, artificial intelligence and other technology and science related topics).

I favor distribution breadth over revenue. It makes knowledge easily accessible to broad audiences, particularly with students. Also, being distributed in digital format, books are easy to correct and update. It is quite practical when you mention hundreds of people and organizations, and deal with complicated scientific matters. Afterwards, I sell my time in a rather traditional way with speaking, training and consulting missions. The business model is simple: the (very) long version is free and the (too) short versions are charged. Since the people who don't have time usually have money and the other way around, it works quite well even if it may be counterintuitive in the first place.

## A complex domain in search of pedagogy

After having swept through many areas of science and deep techs, I can definitively position quantum physics and quantum computing at the complexity scale apex. Quantum physics is difficult to apprehend since relying on counter-intuitive phenomena like wave-particle duality and entanglement, and on a mathematical formalism that is not obvious to most people, including IT specialists and developers, one of the key audiences for this book. It is still an open challenge to translate this scientific field lingua into natura language for most people, even with a strong engineering background.

There's the rehashed famous quote from **Richard Feynman** who pointed out that when you study quantum physics, if you think you understood everything, you are making a fool of yourself. **Alain Aspect** confirms this, always expressing doubts about his own understanding of the quantum entanglement phenomenon that he experimented with photons in his famous 1982 experiment.

Explaining quantum computing is thus a new and difficult art. When reading quantum physics books, you discover a mathematical formalism and many terms like observables, degeneracy, gentle measurement, Hermitian operators and the likes and wonder how they relate to the physical world. Sometimes, it takes quite a while before being able to make this connection! On the other hand, you hear simplistic descriptions of quantum physics, noticeably on superposition and entanglement, and quantum computing, some coming from quantum computing vendors themselves[2].

Once you think you understand it after having created a mental view of how it works, your explanations become quickly inaccessible for the profane. How do you avoid this side effect? Probably with finding analogies and use more visual tools to explain things than too much mathematics. I try this in many sections of this book, but, still, mathematics are useful in many parts. Also, to make sure it does not lose its scientific soundness in the process, many parts of this book have been fact-checked and proof-read by quantum scientists. I'd say, not enough. You'll be the judge.

This book frequently responds to questions like what, why, where and how? Particularly with linking theory, maths and the real world. Has Moore's empirical law really stalled? What being "quantum" means for a product or technology? Why are we using this convoluted mathematical formalism? Do we really have objects sitting simultaneously at two different locations? Why parallel opposite vectors in the Bloch sphere are mathematically orthogonal? Why and where density matrices are useful? What are pure and mixed states describing in the physical world? Why superposition and entanglement are the two sides of the same coin? Why do we need to cool many qubit types? How are cryostats working? What is the energy consumption of a quantum computer? How much data sits in quantum registers? How is data loaded in a quantum program? What data is generated by quantum algorithms and how is it decoded? Are quantum computers made for big data applications? How can you compare such and such quantum computer technology? Is Shor algorithm a serious threat for cybersecurity? When will we have a "real" quantum computer? Have we really achieved quantum supremacy?

[2] See the interesting point in [What Makes Quantum Computing So Hard to Explain?] by Scott Aaronson, June 2021.

And on and on… What is the real speedup of quantum algorithms? Are the case studies from D-Wave and the like real production grade applications? Will a quantum Internet replace the existing Internet? Why do many physicists dislike D-Wave and say it is not quantum? Can quantum telecommunications enable either faster than light communications or high-throughput data links? How are classical computing technologies competing with quantum computers? Why are quantum random number generators not that random? Are the Chinese going to kill us (metaphorically) with their (not so) huge R&D investments in quantum technologies? Can Europe take its fair share in this new market? Oh, and if I'm in an organization... what should I do? Am I late in the game by doing nothing?

To properly address this broad laundry list of questions, this book is positioned above the average media coverage of quantum computing, as well as analyst reports, and below classical scientific publications that are generally largely inaccessible to non-specialists, or to specialists from other domains.

# A new technology wave

Quantum computing stays on top of the various applications of the second quantum revolution. Quantum sensing is more exotic and fragmented, and quantum telecommunications and cryptography are less fascinating. Why is quantum computing becoming an important topic? Firstly, because large IT companies such as IBM, Google, Intel and Microsoft are making headlines with impressive announcements that we must, however, take with a grain of salt, with a lot of hindsight, and decipher calmly. There's also the obvious impact of Peter Shor's factoring algorithm. It drives fuzzy fears on the future of Internet security and for your own digital privacy.

Above all, it is linked to the broad impact that quantum technologies could have on many scientific fields and digital markets. It may theoretically make it possible to solve problems belonging to classes of complexity that even the largest giant supercomputers will never be able to tackle with. Then, the hype builds exaggerated stories on how quantum computing will for sure fix climate change, predict the weather, cure cancer, and other miracles.

The other reason for this sudden interest is that we are still at the beginning of the story. New leaders will show up. A new ecosystem is being built. This in a field where there are still enormous scientific and technology challenges to overcome. It is a land of opportunities for science, technology and innovation. To resume quantum physics, we are in a highly indeterministic world.

It is quite difficult to evaluate the feasibility of large-scale quantum computing. For most scientists, we are still many decades away from it. Some believe it will never show up. Others are more optimistic. The main enemy is quantum decoherence and qubits errors happening during computing, and which are difficult to avoid and correct. The plan is to fix that with quantum error corrections and logical qubits made of physical qubits. It then becomes, at least, a physical scalability issue with a bunch of complex engineering issues related to cooling, cryo-electronics, cabling, classical computing, miniaturization, as well as fundamental thermodynamic and energetic dimensions.

It is a very interesting living case study of how mankind builds upon scientific progress and addresses the most difficult challenges around. For this respect, it is on par with controlling nuclear fusion.

# Reading guide

Here is a tentative to prioritize which parts of this book you could read according to your business and scientific level.

Physicists can find a state-of-the-art tour covering all dimensions of quantum technologies beyond the field they already master.

Computer scientists, engineers and students in various scientific fields are the core target audience for this book, as it presents, popularizes and contextualizes the various scientific, mathematical and engineering concepts used in quantum technologies.

The required mathematical and computer basics level is at the bachelor's degree level for most parts. Afterwards, it can also depend on your age since many of these concepts were not in current programs a couple decades ago unless you were already specialized in quantum physics. Non-technical and decision-makers can still read the sections dealing with usages as well as with how countries are faring and societal issues.

| Book sections | Quantum physicists | Computer scientists and developers | Students in sciences (STEM) | Non technical audiences | Business audiences |
|---|---|---|---|---|---|
| Why | | | | | |
| History and scientists | | | | | |
| Quantum Physics 101 | known | optional | | | |
| Gate-based Quantum Computing | | | | | |
| Quantum Computing Engineering | | | | | |
| Quantum Enabling Technologies | | optional | | | |
| Quantum Computing Hardware | | | | | |
| Quantum Algorithms | | | | | |
| Quantum Software Development tools | | | | | |
| Quantum Computing Business applications | | | | | |
| Unconventional computing | | | | | |
| Quantum Telecommunications and Cryptography | | | | | |
| Quantum Sensing | | | | | |
| Quantum Technologies around the world | | | | | |
| Corporate Adoption | | | | | |
| Quantum technologies in society | | | | | |
| Quantum Fake Sciences | | | | | |

*Figure 1: Understanding Quantum Technologies parts and audiences relevance. (cc) Olivier Ezratty 2021-2022.*

Here's another view of the table of contents showcasing the overall logic between the lower « physics » layers and the upper hardware, software and solutions layers.

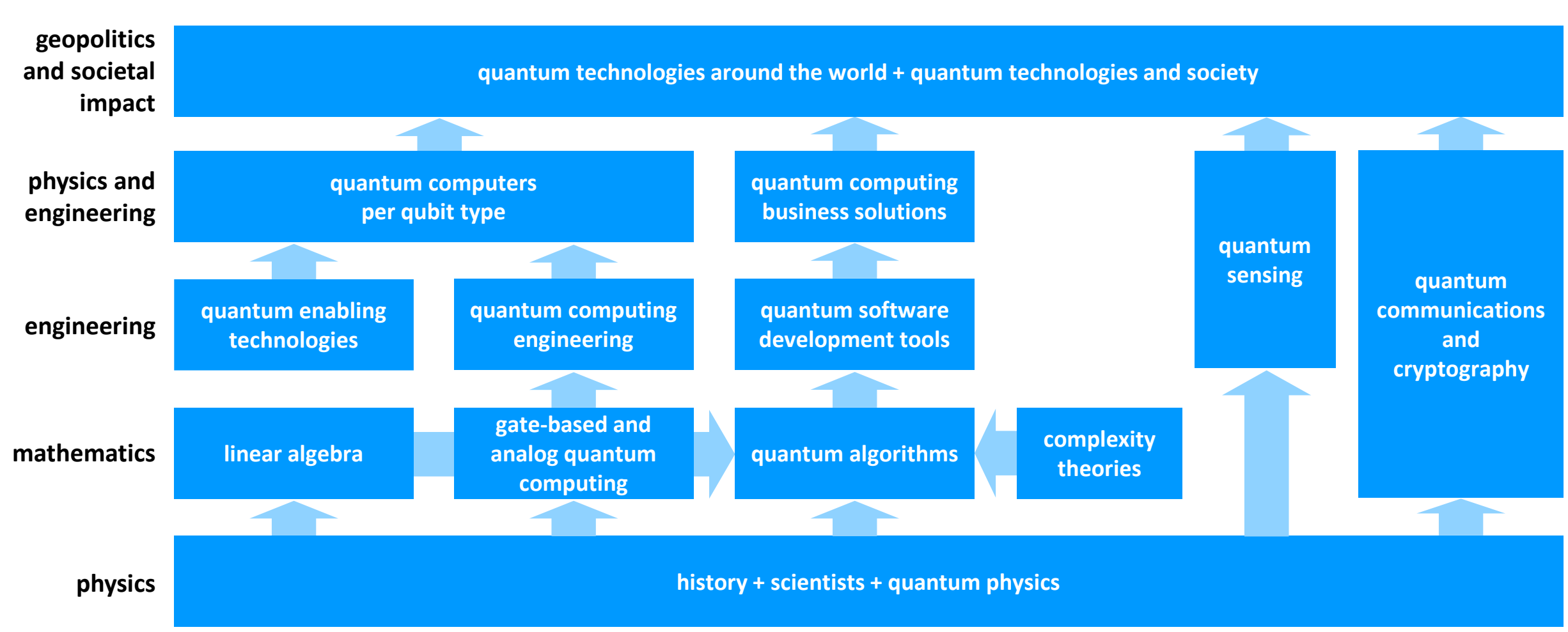

*Figure 2: how the topics covered in Understanding Quantum Technologies are related with each other. (cc) Olivier Ezratty.*

At last, let's mention one of the reasons why a curious mind may like quantum technologies: they encourage you to explore many scientific disciplines, even human and social sciences, like a scientific Pandora's box.

**physics**
electromagnetism
quantum physics
quantum matter
thermodynamics
fluids mechanics
photonics

**engineering**
materials design
electronics engineering
cryogenics

**mathematics**
linear algebra
groups theory
analysis
complexity theories

**computer science**
information theory
algorithms design
programming
classical computing
telecommunications

**human sciences**
philosophy
epistemology
sociology
technology ethics
economics of innovation
R&D policy making
geopolitics
startups ecosystem

*Figure 3: the many scientific domains to explore when being interested in quantum technologies. That's why you'll love this book if you are a curious person. (cc) Olivier Ezratty, 2021-2022.*

If you have some scientific background, you'll play in familiar territory but if you've had your degree a couple decades ago, this overview will provide you with some interesting intellectual upgrades. On top of that, learning quantum science is probably more efficient than Sudoku or crosswords to train your brain muscle as it ages!

## First and second quantum revolutions applications

Quantum physics has been implemented since the post-war period in almost all products and technologies in electronics, computing and telecommunications.

This corresponds to the **first quantum revolution**. It includes transistors, invented in 1947, which use the tunnel effect and are the basis of all our existing digital world, photovoltaic cells which rely on the photoelectric effect, and lasers which also exploit the interaction of light and matter and are used in a very large number of applications, particularly in telecommunications.

**first quantum revolution**
manipulating
**groups of quantum particles**
photons, electrons and atoms interactions

**second quantum revolution**
manipulating
**superposition** and **entanglement**
and/or **individual particles**

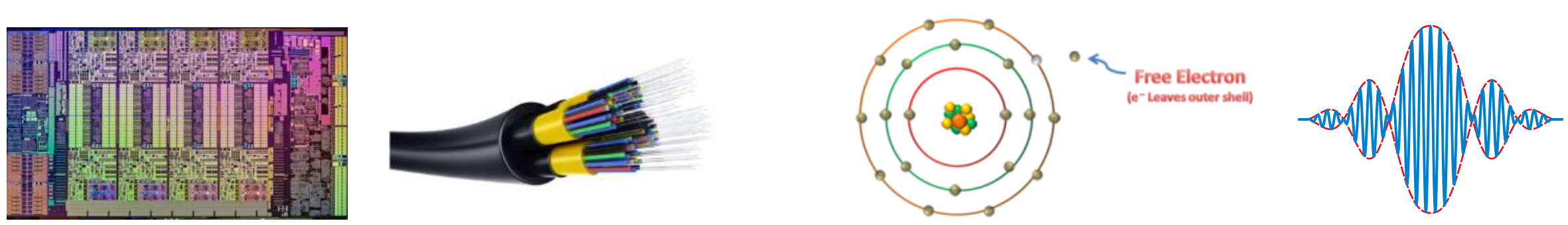

**transistors, lasers, GPS**
**photovoltaic cells, atom clocks**
**medical imaging, digital photography and video**
**LEDs, LCD TV quantum dots**
1947-*

**quantum computing**
**quantum telecommunications**
**quantum cryptography**
**quantum sensing**
1982-*

*Figure 4: first and second quantum revolution definition and related use cases. (cc) Olivier Ezratty, 2020-2022.*

Many medical imaging solutions rely on various quantum effects, including nuclear magnetic resonance imaging (MRI). LEDs are also based on quantum effects. The GPS is relying on atomic clocks synchronization. Quantum dots used in high-end LCD displays and Smart TVs also use variations of the photoelectric effect. The list is long, and we will not detail all these use cases!

The **second quantum revolution** covers the technologies combining all or part of the ability to control individual quantum objects (atoms, electrons, photons), use quantum superposition and/or entanglement. We owe the names of the first and second quantum revolutions to Alain Aspect, Jonathan Dowling and Gerard Milburn in 2003[3]. The first and the two following ones created it simultaneously and independently. In the United States, the paternity is attributed to the latter, while in France, it is attributed to the former! Who knows why?

The scope of the second quantum revolution covers various recent applications of quantum physics that integrate quantum computing, quantum telecommunications, quantum cryptography and quantum sensing. Said simply, it's about improving our digital world performance and security, and to increase the precision of all sorts of sensors.

- **Quantum computing** is the broad domain of using quantum physics to find solutions to various computing problems. It includes various computing paradigms like gate-based computing, quantum annealing and quantum simulations. Hundred pages are covering this topic in this book.
- **Quantum cryptography** is a mean of communicating inviolable public cryptography keys thanks to quantum physics phenomena and rules, like photon entanglement and the no-cloning theorem. It relies either on fiber optic communications or on space links with satellites as China has tested with its Micius satellite since 2017. **Post-quantum cryptography** is a different field which is intended to replace current classical cryptographic solutions with new solutions that are supposed to be resistant to attacks carried out by future quantum computers. It is not belonging to the second quantum revolution per se but is rather a consequence of it.
- **Quantum telecommunications** enables distributed computing, connecting quantum computers enabling qubit to qubit distant entanglement, and, potentially, quantum sensors, which can be implemented to improve their accuracy. This field still in the making could become the base for a very secure quantum Internet and quantum cloud infrastructures. We cannot exploit it to transmit classic information faster than today[4]. However, it can be used to distribute quantum processing on several quantum processors. It could provide a mean to "scale-out" quantum computers, when it's becoming difficult to "scale-in". This requires a lot of engineering, particularly to convert solid qubits into photon qubits and share entanglement resources.

[3] See Speakable and unspeakable in quantum mechanics by John S. Bell, June 2004 edition (289 pages) which contains a preface by Alain Aspect on the second quantum revolution, dated February 2003, pages 18 to 40. We find the expression in Quantum technology: the second quantum revolution by Jonathan P. Dowling and Gerard J. Milburn, June 2003 (20 pages) as well as in Quantum Technology Second Quantum Revolution by Jonathan Dowling, 2011 (60 pages). Dowling's writings make a very large inventory of various quantum technologies embedded in this second quantum revolution. The Second Quantum Revolution: From Entanglement to Quantum Computing and Other Super-Technologies by Lars Jaeger, 2018 (331 pages) is a broader overview of the different sides of the second quantum revolution.

[4] But..." *Entangled states cannot be used to communicate from one point to another in space-time faster than light. Indeed, the states of these two particles are only coordinated and do not allow to transmit any information: the result of the measurement relative to the first particle is always random. This is valid in the case of entangled states as well as in the case of non-entangled states. The modification of the state of the other particle, however instantaneous it may be, leads to a result that is just as random. Correlations between the two measurements can only be detected once the results have been compared, which necessarily implies a classical exchange of information, respectful of relativity. Quantum mechanics thus respects the principle of causality*". Source: https://fr.wikipedia.org/wiki/Intrication_quantique.

- **Quantum sensing** makes it possible to measure most physical dimensions with several orders of magnitude better precision than existing classical sensing technologies, even existing atomic clocks. It is a vast scientific field that is the subject of numerous research projects and industrial solutions. It includes ultra-precise atomic clocks[5], cold atom accelerometers and gyroscopes that use atomic interferometry, SQUIDs (superconducting based) and NV center magnetometers.

  Microgravimeters measure gravity with extreme precision, enabling discoveries of underground anomalies like holes, water and various materials. This domain also includes various advanced medical imaging systems with higher precision and non-destructive imaging and measurement tools[6]. A dedicated section of this book is covering quantum sensing, starting page 565. The diversity of quantum sensing solutions or prospect solutions is staggering.

**Table 1. Quantum Metrology and Sensing Technologies**

| Technology | Technological Readiness[a] | Potential Market |
|---|---|---|
| Measurement | | |
| Atomic clocks | Commercial | \$50–\$500 million |
| Meters for voltage, current, and resistance | Commercial | — |
| Sensors | | |
| Gravimeters and other atomic interferometers | Commercial | < \$50 million |
| Quantum inertial motion units | Medium-term | \$50–\$500 million |
| Atomic magnetometers | Commercial | \$50–\$500 million |
| Magnetoencephalography | Commercial | \$50–\$500 million |
| Quantum electron microscopes | Medium-term | \$50–\$500 million |
| Quantum-assisted nuclear spin imaging | Long-term | < \$50 million |
| Signal measurement | Medium-term | — |

*Sources*: European Commission (2017) United States Air Force Scientific Advisory Board 2015; interviews.

*Figure 5: various quantum sensing use cases. Source: EU and US Air Force, 2015.*

## Why quantum computing?

The main goal for using quantum computing is to solve complex problems that are and will stay inaccessible to classical computers. This happens when these problems solutions scale exponentially in computing time on classical machines. Problems that scale polynomially on classical hardware are not very interesting for quantum computing. The promise of quantum computing is to address this need. But a big warning and legal disclaimer: it is still a *promise*! We are still far off from delivery.

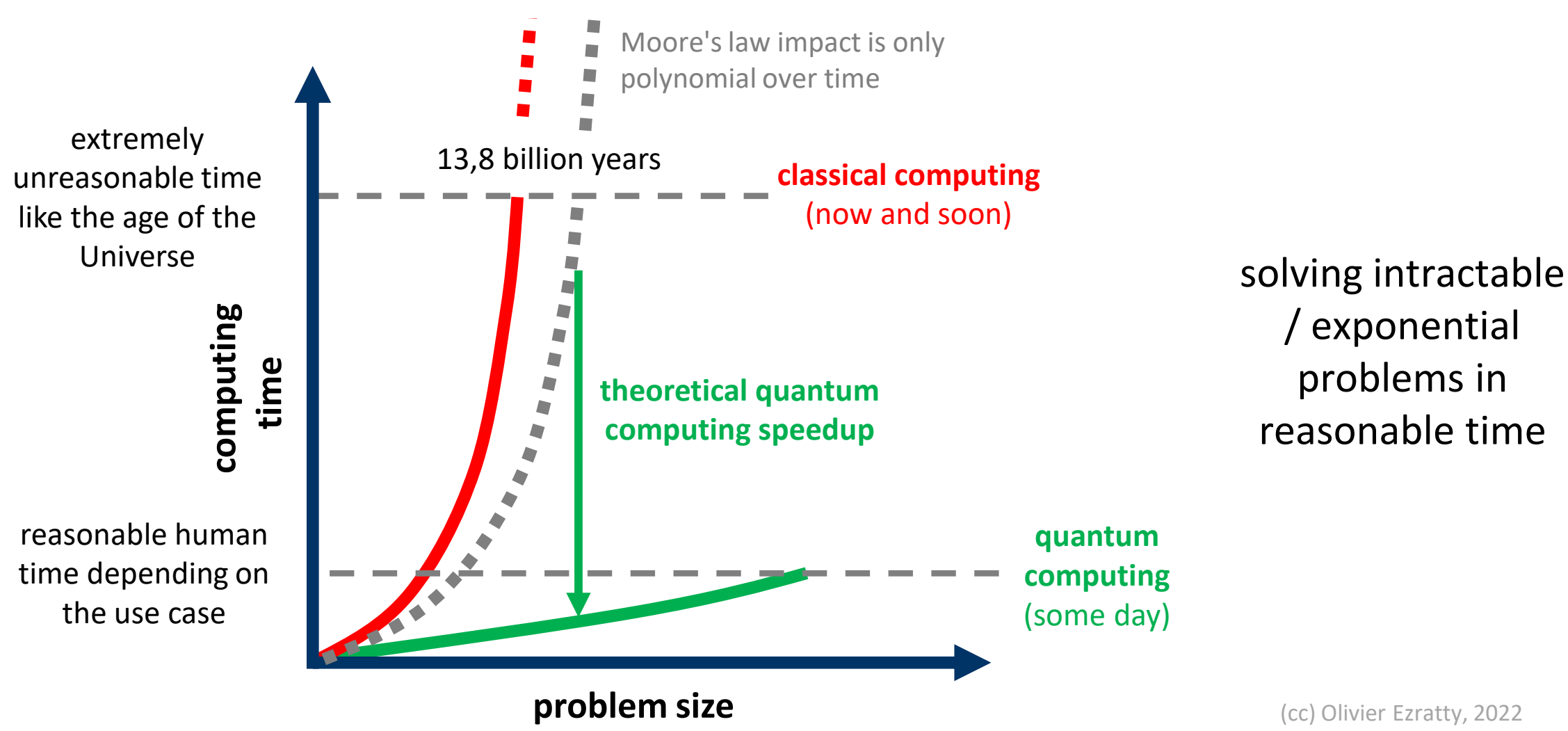

*Figure 6: simplified view of the quantum computing theoretical promise. Before delivering this promise, quantum computers may bring other benefits like producing better and more accurate results and/or doing this with a smaller energy footprint. (cc) Olivier Ezratty, 2022.*

---

[5] See for example this NIST work on an atomic clock based on rubidium, the element most frequently used in atomic clocks. NIST Team Demonstrates Heart Of Next-Generation  Chip-Scale Atomic Clock, May 2019.

[6] See Quantum camera snaps objects it cannot 'see', by Belle Dume, May 2018. This is a variant of Diffraction Free Light Source for Ghost Imaging of Objects Viewed Through Obscuring Media by Ronald Meyers, 2010 (22 pages). Yanhua Shih (University of Maryland) US Army Research Laboratory, has been working on the subject since 2005. Quantum Imaging by Yanhua Shih, 2007 (25 pages). Also, see Quantum Imaging - UMBC (47 slides).

## Quantum computing promise

Typical exponential problems are combinatorial optimization searches and chemical simulations. Their size is usually expressed in a number of items like a number of steps for solving a travelling salesperson problem. Exponential problems are said to be "intractable" because their computation time evolves in crazy proportions with their size.

It starts with various optimization problems such as the above-mentioned traveling salesperson problem, with its contemporary equivalents applied to product delivery or autonomous vehicles routing. Today, you optimize your route with Google Maps or Waze, based on traffic conditions. Traffic conditions are variable and your actual journey time is not always what was planned nor optimal.

With fully autonomous fleets, it may theoretically be possible to optimize the individual path of each and every vehicle based on their departure and destination locations. Conventional algorithms could work with a limited number of vehicles, but beyond a few hundred vehicles and trips, traditional computing capacities would be largely saturated. Quantum computing may then come to the rescue!

Secondly, we have physics and molecular simulations, themselves governed by quantum mechanics equations. It usually boils down to finding the minimum energy configuration of a system, in order to simulate the interaction of atoms in molecules, complex crystal structures or even how magnetism works in various materials. This deals with both classical chemical engineering and biochemistry. Rest assured, this will not go so as far as to simulate an entire living being or even a cell. It will already be a fantastic feat when we are able to simulate some simple de-novo protein folding in a better way than what AlphaFold 3 from DeepMind is doing today, the next step being protein interactions simulations[7].

A third area for quantum computing is the training and inferences of machine learning models and neural networks. It is now within the reach of conventional computers equipped with GPGPUs (general purpose GPUs) such as Nvidia's V100, A100 and H100 and their tensor processing specialized units, optimizing matrices-based operations. Quantum advantage is less obvious in this field, particularly since machine learning must usually be trained with a lot of data. Nowadays, however, using quantum computing for machine learning happens to potentially bring another benefit: creating better solutions instead of creating it faster.

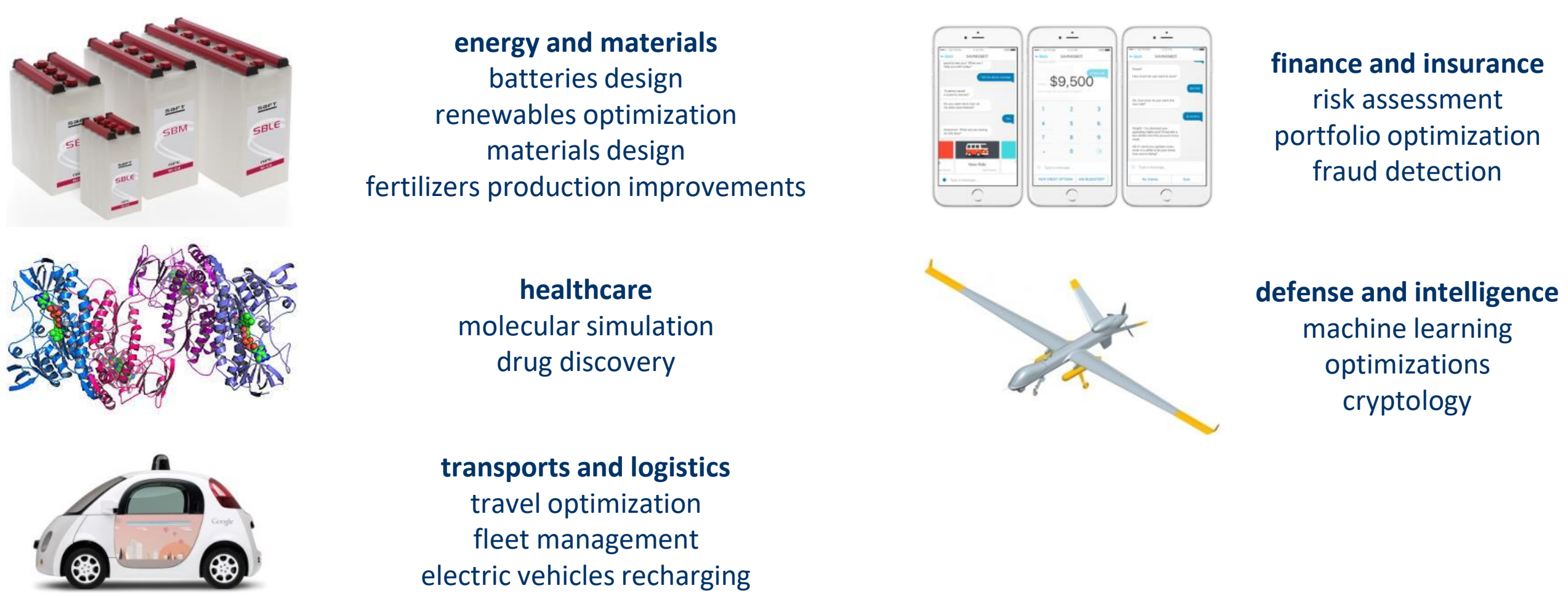

*Figure 7: typical quantum computing use cases where a quantum speedup brings clear benefits. These are still "promises" since the capable hardware to implement many of these solutions with a quantum speedup remains to be created and it may take a while up to several decades! (cc) Olivier Ezratty, 2020.*

[7] The competition from classical machine learning is still significant and growing. See Scientists are using AI to dream up revolutionary new proteins by Ewen Callaway, Nature, September 2022.

Finally, you can't avoid integer factorization, which is of particular interest to the NSA and their peers to break RSA-type public-key encryption security. We'll dig into this in details starting page 787.

> *"Building a quantum computer is a race between humans and nature, not between countries"*
> Lu Chaoyang, China
> December 2020.

Other applications are investigated for different markets such as finance, insurance and even marketing. Many businesses have complex optimization problems to solve. Like with most technology-driven disruptions, businesses will progressively discover quantum computing use case as its market and related skills grow.

In extreme cases, computing times on conventional computers for exponential problems, even with the most powerful supercomputers of the moment, would exceed the age of the Universe, i.e. 13.85 billion years.

Most of these promises are dependent on the ability to create large scale and fault-tolerant quantum computers, which are years if not decades away. In the interim, we may end-up having quantum systems able to deliver other benefits like producing better and more accurate results and/or doing this with a smaller energy footprint, but not with a real exponential speedup.

**Moore's law limitations**

Moore's empirical law application, or "More than Moore" as its successor is now labelled, would have a marginal impact, dotted in the graph. First of all, it has been slowing down since 2006, and even if it did not slow down, it would not bring the capacity to solve exponential problems. Computation times for exponential problems would remain exponential despite the supposed doubling of machine power every 18 months to two years. The addition of a single qubit theoretically doubles quantum computers power, both in terms of internal memory space and computing parallelism capacity[8].

In comparison, quantum computers could theoretically, one of these days, solve these same problems within a reasonable time span on the scale of a human life, in hours, days, weeks or months. Reasonableness obviously depends on the nature of the problem to be solved.

The main benefit of quantum computation is to modify the time scales for solving a problem and turn problems whose classical solution requires some exponential time into quantum solutions requiring at most some polynomial time. It can become useful when the size of the problem is large, sometimes with only about fifty items in a combinatorial optimization search! Quantum computation also makes it possible to gain space, particularly memory, to perform these calculations.

However, the scientific and technological barriers to overcome to make this real are still immense. Some of these use case promises may even be frequently oversold.

Meanwhile, quantum computing is not a "jack of all trades" solution. It is not a replacement tool but more a complement to current High-Performance Computers (HPC). Many, if not most of today's classical computing problems and software are not at all relevant use cases for quantum computing.

From an economy historical perspective, the consequence is that quantum computing won't probably be a Schumpeterian innovation. It will not entirely replace classical legacy technologies. It will complement it. It's an incremental instead of being a replacement technology. You probably won't have a quantum desktop, laptop or smartphone to run your usual digital tasks although quantum technologies can be embedded in these devices like quantum sensors and quantum random number generators.

---

[8] One could though argue that adding a single functional qubit to a quantum computer appears to be exponentially difficult with the number of qubits.

Quantum computers will be hidden from users and sit in cloud data centers, like Nvidia GPGPUs racks. This will be even amplified by the progress we can anticipate with wireless telecoms.

When quantum computers will scale after 2030, we'll probably use 6G or 7G networks with even better latency and bandwidth! Of course, it's still hard to anticipate the usages brought by quantum computers when they will scale.

Let's still boil in the fact that, as we'll see later, quantum computers are not excellent to handle big data not for real-time computing. This makes it less relevant to use a local quantum processor, as it makes sense today to have local neural networks capacities to handle your in-camera image recognition processing and voice recognition in smartphones. Less data means more relevance for distant quantum computation done in the cloud.

### Classical computing technology developments

How are we currently making progress with conventional computing? We rely on a few known techniques, some of which have not yet been fully explored.

Multi-core architectures enable parallel processing but with limits formalized by **Amdahl's law,** which describes the upper limits of parallel computing systems acceleration.

We have the ongoing sluggish increase of transistors density in processors coupled with so-called **Domain Specific Architectures** using ad-hoc circuits like tensors (matrix multipliers) used to run specialized algorithms like neural networks. One key technology development is to make sure memory is as close as possible to processing units.

**Neuromorphic processors** mimic biological neurons features with integrated memory and processing using memristors[9]. They can be implemented with spintronics electronics, that imitate how brain cells work with their own memory[10].

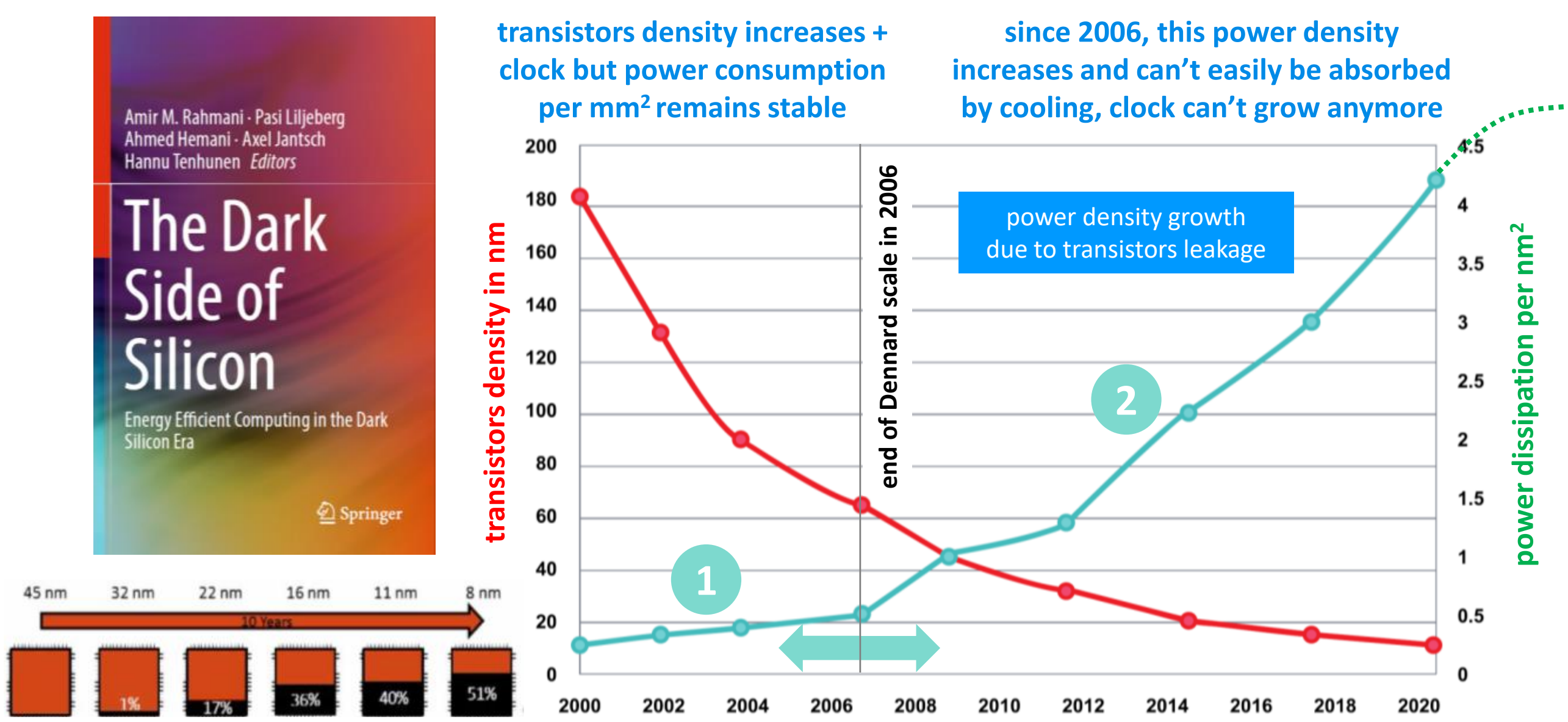

*Figure 8: Dennard's scale which explains the dark silicon phenomenon where all CMOS chipsets components cannot be used simultaneously. Compilation (cc) Olivier Ezratty.*

[9] One famous work with neuromorphic processor is the Loihi project from Intel. See Intel's Neuromorphic Chip Gets A Major Upgrade Loihi 2 packs 1 million neurons in a chip half the size of its predecessor by Samuel K. Moore, IEEE Spectrum, October 2021.

[10] See the review paper Quantum materials for energy-efficient neuromorphic computing: Opportunities and challenges by Axel Hoffmann, Julie Grollier et al, April 2022 (24 pages).

The heat barrier limits our capacity to increase processor clock speed beyond 5 GHz. It can reach 6 GHz with liquid cooling[11].

This is due to the end, in 2006, of **Robert Dennard**'s (1932, American) scale established in 1974.

According to this scale or rule, as the transistors density increased, the power consumed per unit area of the chipsets was stable. This happened since the transistors voltage and current could decrease with their density, while increasing the clock frequency. Starting with 65 nm integration, this rule was broken. That's why, among other phenomena, your laptop computer is also heating your legs when you use it in public transportation or in your coach.

The transistors current leaks started to grow and power consumption soared. This is what prevents the growth of processors clock. At the beginning of the 2000s, Intel planned in its roadmaps to raise their CPU clock frequency up to 20 GHz.

Intel then stopped playing this game and instead entered the multicore realm. However, in June 2021, Intel released a new microprocessor for high-end laptops running at a 2.9 GHz base clock but with a 5 GHz turbo mode for a single core, the 4-core i7-1195G7, etched in 10 nm, and with a 28W TDP[12].

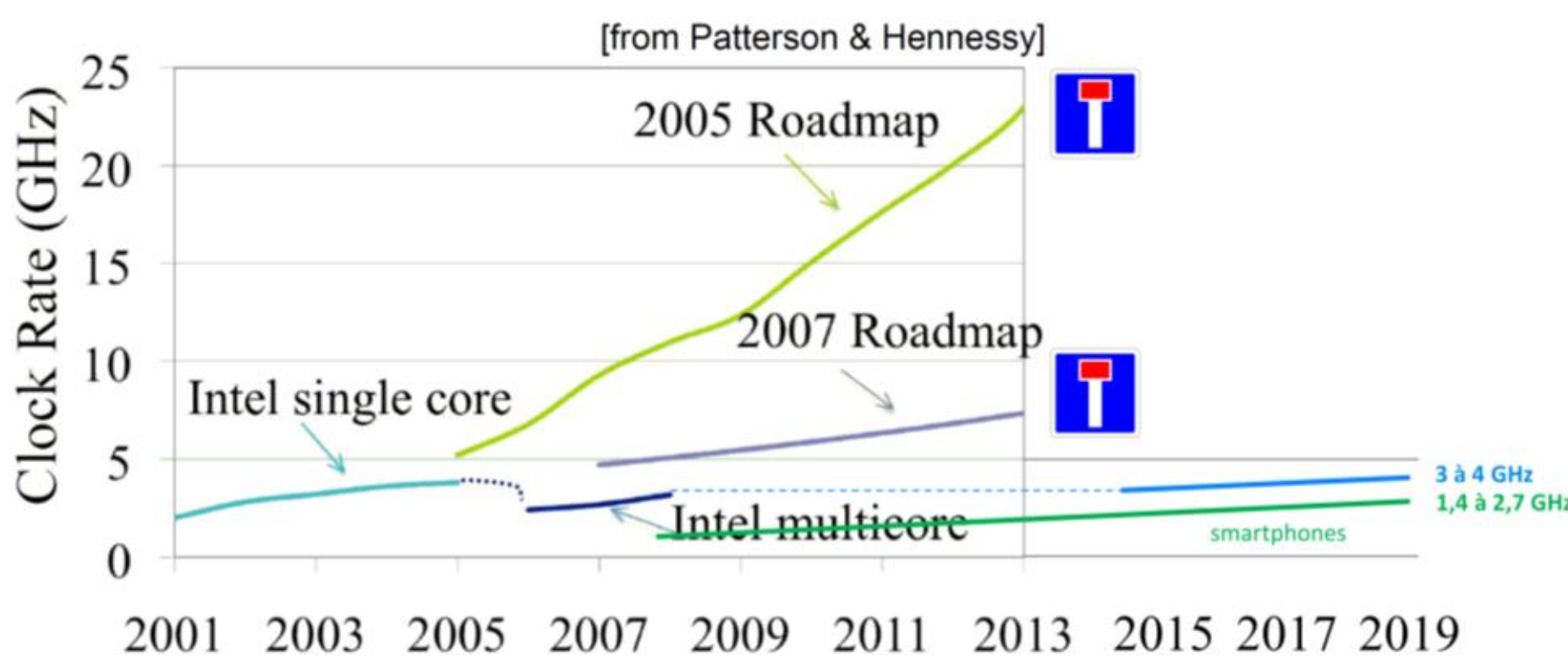

*Figure 9: how CMOS chipsets clock was supposed to increase… and didn't. Source: High Performance Computing - The Multicore Revolution by Andrea Marongiu (41 slides), 2019. Additions: Olivier Ezratty.*

The semiconductor demand switched in 2007 towards low-power multi-functions chipsets for smartphones. This opened a boulevard for Arm core-based processors and growth for corporations like **Qualcomm**.

The available computing power per consumed kW increased steadily, doubling every 1.57 years between 1946 and 2009, according to **Jonathan Koomey**'s empirical law enacted in 2010. However, this doubling slowed down to 2.6 years after 2000, due to the end of Dennard's scale.

There are many techniques used to optimize classical computing footprint, particularly around memory management, with making sure memory is as close as possible to computing, including in-memory processing[13]. After 2006, transistors density still continued to increase.

However, the end of Dennard's scale led to the rarely mentioned **dark silicon** phenomenon. As the chipsets get too hot, it becomes difficult to use it entirely. Various methods are then combined to circumvent this inconvenience: on-demand cores or functions deactivations according to usage needs, a shutdown of certain portions or cores, a voltage drop, or a selective clock frequency adjustment. This is what is used in the Arm core-based processors of smartphone chipsets, whose cores do not use the same clock rates, in the so-called big.LITTLE architectures created in 2011, and replaced with the more flexible DynamIQ architecture in 2017 [14].

[11] See on this subject Minimum Energy of Computing, Fundamental Considerations by Victor Zhirnov, Ralph Cavin and Luca Gammaitoni, 2014 (40 pages) which compares the energy efficiency of living things and electronics.

[12] Thermal dissipation power.

[13] See Energy Efficient Computing Systems: Architectures, Abstractions and Modeling to Techniques and Standards by Rajeev Muralidhar et al, July 2020 (35 pages) which makes a good inventory of the various ways to save energy with classical computing. And Processing-in-memory: A workload-driven perspective by S. Ghose et al, IBM Research, 2019 (19 pages).

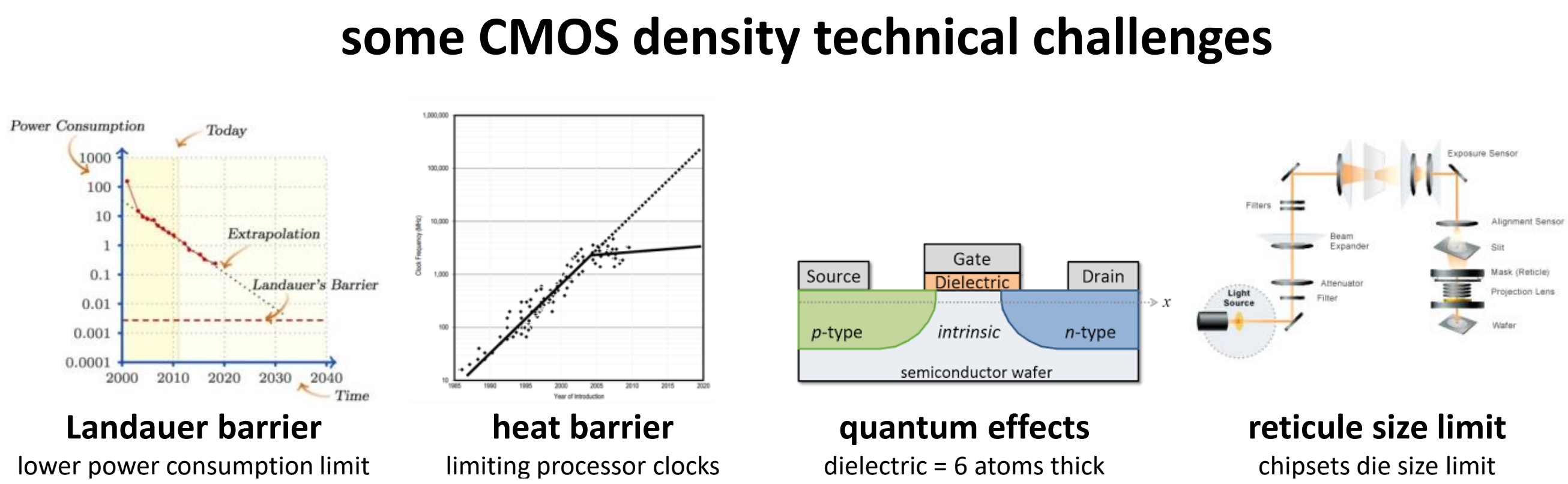

*Figure 10: some of the key CMOS density technical challenges to overcome by the semiconductor industry. One source: [Reversible Circuits: Recent Accomplishments and Future Challenges for an Emerging Technology](#) by Rolf Drechsler and Robert Wille, 2012 (8 pages).*

To lower transistors density below 10 nm, etching systems using extreme ultraviolet are required, coming from **ASML**. Etching resolution depends on the wavelength of the light used to project a mask on a photoresist.

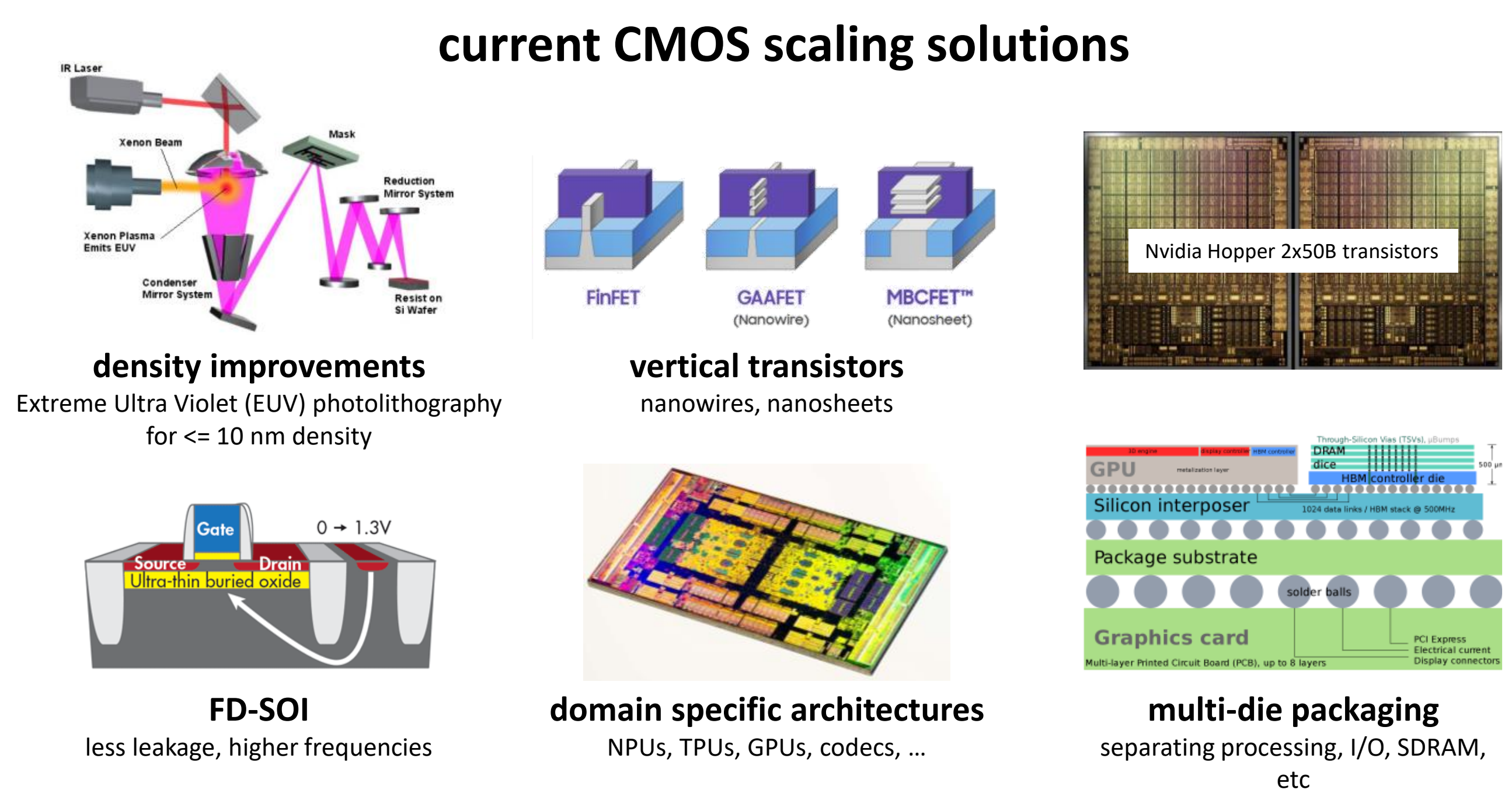

*Figure 11: current CMOS scaling solutions adopted by the semiconductor industry. (cc) Olivier Ezratty with uncredited image sources.*

Lowering the transistors size requires increasing this frequency to decrease the wavelength, and thus go from the current deep ultra-violet to extreme ultra-violet. It took more than 10 years to develop these EUV lithography systems. It is in production since 2019 in TSMC and Samsung 5 nm nodes fabs. One of key benefits of EUV etching is to reduce the usage of the costly multiple patterning process to improve lithography resolution.

[14] There are many other techniques to improve classical processors energy efficiency. See for example [Energy Efficient Computing Systems: Architectures, Abstractions and Modeling to Techniques and Standards](#) by Rajeev Muralidhar et al, AWS and Melbourne University, July 2020 (35 pages).

ASLM's latest EUV lithography generation is dubbed High-NA (for high numerical aperture). A bit like in photography, High-NA optics will convey more light onto masks and silicon targets and will be required for nodes under 3 nm. It requires both new UV optics but also new light sources. And the EUV machines are much bigger and costly. These machines will be deployed around 2024. The generation after High-NA would be Hyper-NA but even ASML is doubting it will be economically viable[15].

For a while, scientists warned about undesirable quantum effects appearing below 10 nm nodes. But it didn't prevent going down to 5 nm and then below dimensions. TSMC started producing 2 nm chipsets in 2022, combining EUV etching with the traditional FinFET technology that has been in use for more than 10 years. They are expecting to mass-product 2 nm chipsets by 2025, thanks to nanowires and nanosheets techniques[16]. In July 2021, Intel even announced a new density scale using angstrom sized transistors, with 20Å and 18Å by 2025 (meaning... about 2 nm, given 1 Å = 0,1 nm).

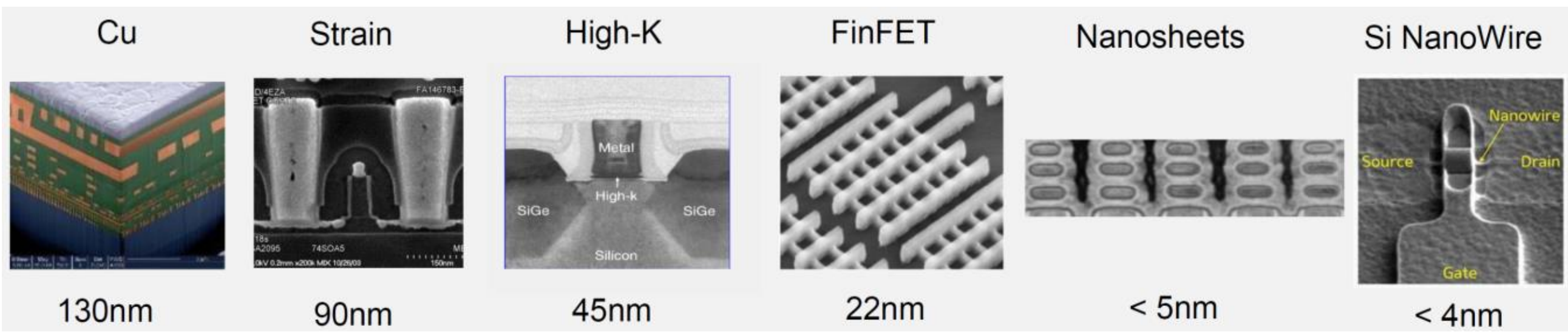

*Figure 12: the various CMOS transistor technologies used as density increased.*

In May 2021, IBM announced it had prototyped 2 nm nanosheet-based chipsets, manufactured by Samsung, and also using EUV lithography[17].

As far as integration is concerned, two other limits must be taken care of, such as **Rolf Landauer'**s (1927-1999, researcher at IBM in 1961) principle which defines the minimum energy required to erase a bit of information. It is a very low theoretical barrier contested by some physicists. And it can be circumvented as we will see with the technique of adiabatic and reversible computing that is covered page 766.

[15] See Hyper-NA after high-NA? ASML CTO Van den Brink isn't convinced, Bit Chips, September 2022.

[16] See Beyond CMOS, Superconductors, Spintronics, and More than Moore Enablers by Jamil Kawa, Synopsys, March 2019 (43 slides), a good presentation describing the various ways to improve the power of components including cold CMOS, semiconductors operating at liquid nitrogen temperature levels (-70°C) and superconducting Josephson effect based transistors.

[17] See IBM Introduces the World's First 2-nm Node Chip by Dexter Johnson, IEEE Journal, May 2021.

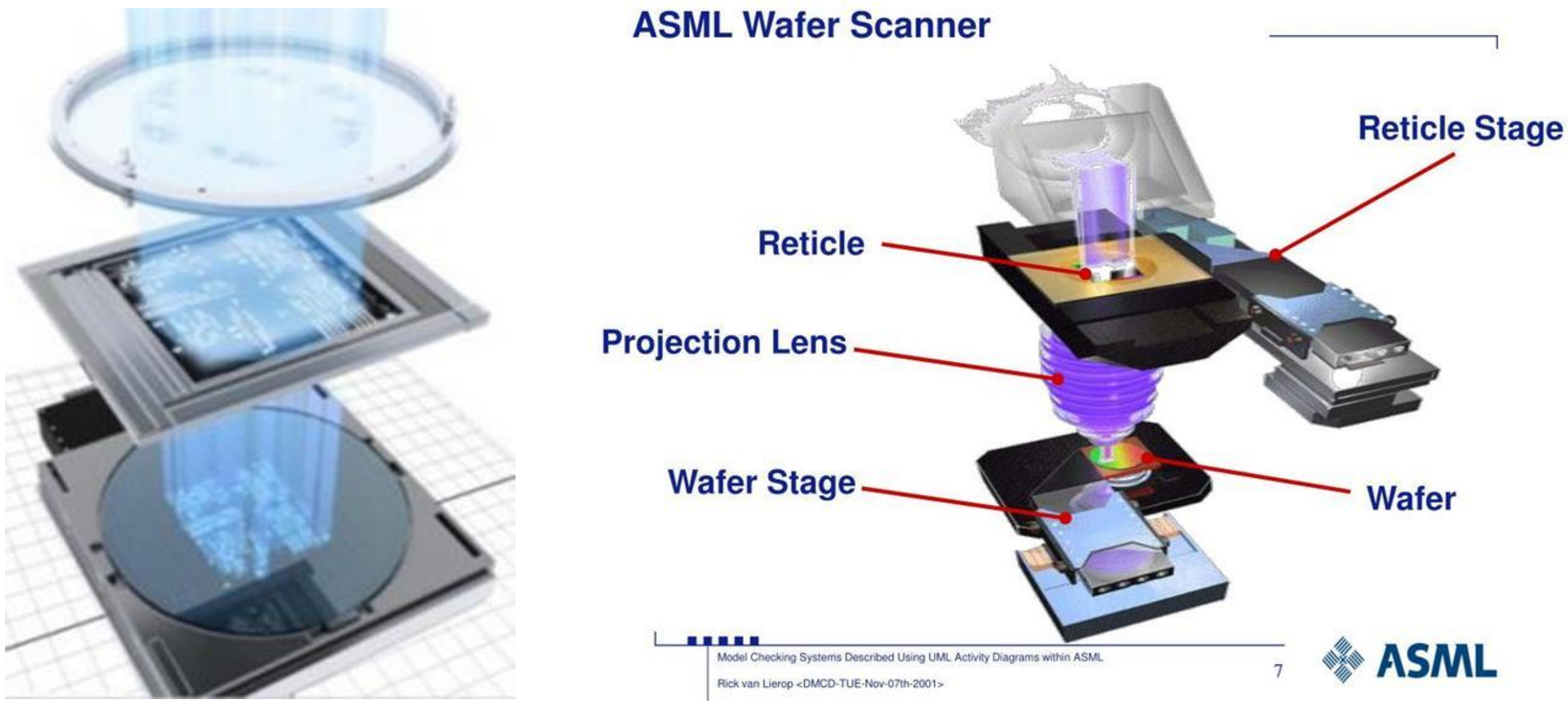

*Figure 13: reticle used in photolithography and its related optics, explaining the size limitation of dies in semiconductor manufacturing.*

Finally, there is a limit coming from the reticles size, these optical systems used in lithography whose size is physically limited, especially optically. It's explained in above illustration in Figure 13, coming from **ASML,** the world leader in semiconductor lithography. This limit has been reached with the largest recent processors.

The largest single-die processors of 2020 were the **Nvidia** A100 with its 54.4 billion transistors etched in 7 nm, superseded closely in size by the **Graphcore** GC200 with its 59.4 billion transistors and 1,472 cores, launched in July 2020 and the Nvidia H100 launched in 2022 with 80 billion transistors, consolidating two adjacent 4 nm chipsets in a single package.

**Cerebras** (USA) nevertheless launched in 2019 an amazingly large 21.5 cm x 21.5 cm square processor, fitting in an entire 300 mm wafer, which circumvents the reticle size limit by being etched in several runs, for its 84 main processing units connected by metal layers. The second version of this chipset launched in 2021 contains 2,6 trillion transistors and 40 GB of cache SRAM memory and has a memory bandwidth of 20 PB/s, allowing it to significantly accelerate neural networks training.

This massive Cerebras chipset, shown in Figure 14, burns about 15 kW/h which are evacuated by a specific water-cooling system. Manufacturing techniques generate defects and more than a couple percent of the 850,000 processing units are defective and are short-circuited during software execution[18]. In September 2022, Cerebras announced its own Wafer-Scale Cluster computer using up to 192 15U rack CS-2 systems. It is competing aggressively against Intel/Nvidia and AMD-based supercomputers that are currently dominating the HPC landscape.

Quantum computing may make it possible to overcome the various limitations of current CMOS processors for certain tasks. However, it will not replace them at all for tasks currently performed by today's computers and mobile devices.

[18] With its D1 chipset presented in July 2021, Tesla chose another approach. Engraved in 7 nm, it has a computing capacity of 22.6 TFLOPS FP32, with 50 billion transistors and a 400W TDP. It contains 354 computing units with 1,25 MB SRAM per unit. They assemble these D1 in 25-chipsets tiles, consuming 15 kW, exactly like a Cerebras chipset.

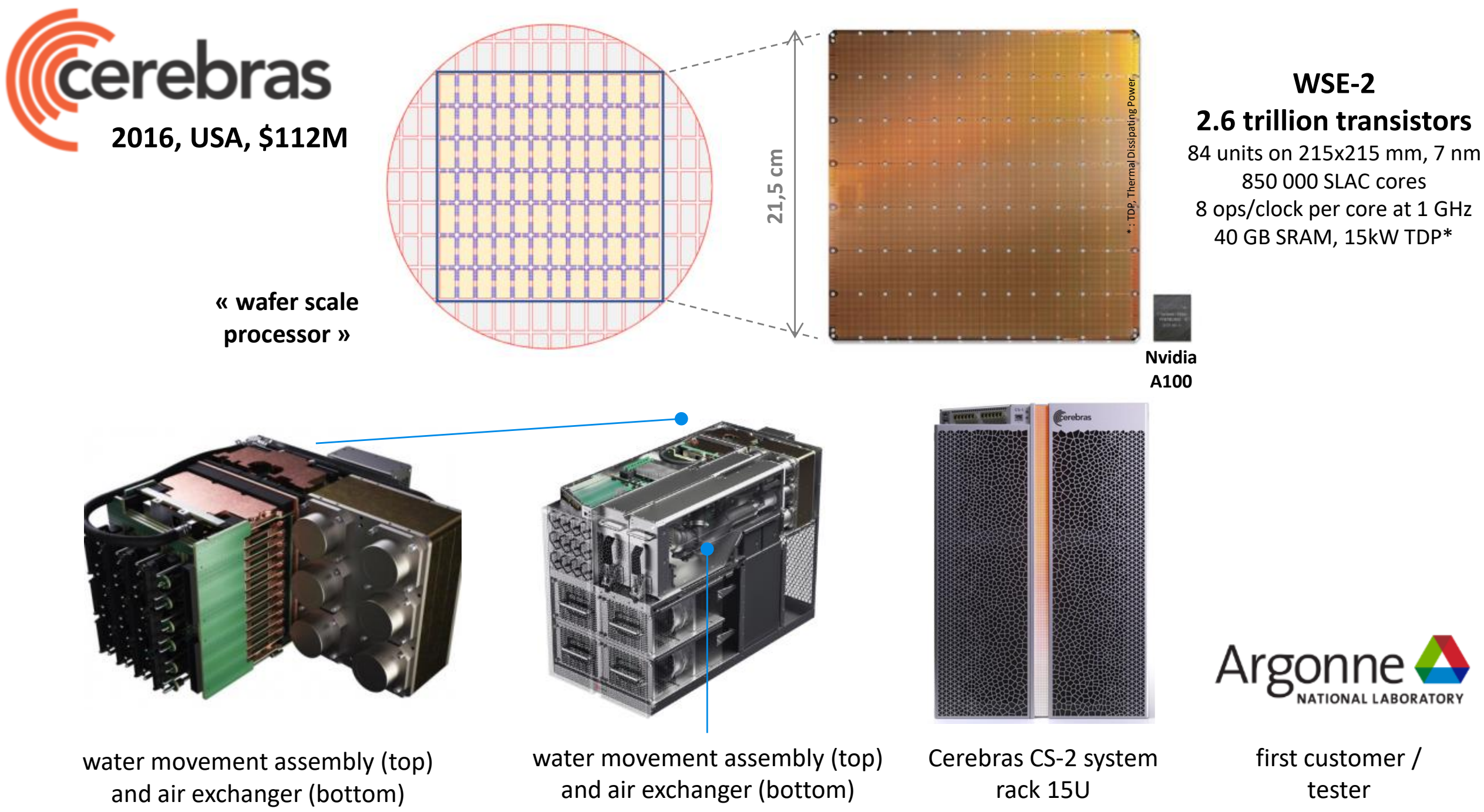

*Figure 14: the impressive Cerebras wafer-scale chipset. Source: Cerebras.*

Typically, video and audio compression and decompression are not relevant tasks for quantum computing. They are usually carried out in specialized chipset processing units, known as DSPs (for digital signals processing). Similarly, applications handling very large volumes of data are not suitable for quantum computing for a whole host of reasons that we will study, mainly because data loading speed into qubits is quite low, whatever the qubit type.

As its use cases will be different, it is hard to anticipate the IT landscape that will emerge with powerful quantum computers when they show up. Even with the advent of quantum computers, Ray Kurzweil's singularity predictions, which rely on the ad vitam extension of Moore's empirical law, will need to be adjusted!

### Unconventional computing

In a dedicated part starting page 754, we will evaluate some the other avenues considered to overcome the current limitations of classical computing, which may provide some power or efficiency gains positioned between classical and quantum computing. These belong to the broad category of "unconventional computing".

This includes **superconducting computing** operating at low temperatures (investigated in the USA and Japan), **digital annealing** computing (proposed by **Fujitsu**), **reversible** and/or **adiabatic computing** that could reduce energy consumption and circumvents Dennard's scale end, **probabilistic computing** as well as different breeds of **optical computing**.

I also delve into some of the inner workings of supercomputers and specialized processors to better understand their strengths and weaknesses. When comparing quantum computers to classical computers, we are better off with knowing both sides of the equation, not just the loud new kid in town!

These are sort of backup solutions, should science fail to create scalable quantum computers. It will also complement quantum computing used in the context of hybrid computing. Interestingly, some unconventional computing avenues, such as superconducting electronics, are potential enabling technologies for scaling certain types of quantum computers.

However, at this point, none of these solutions seem positioned to solve intractable problems although some of these are claiming they have this capacity, which is quite hard to fact-check at large scales.

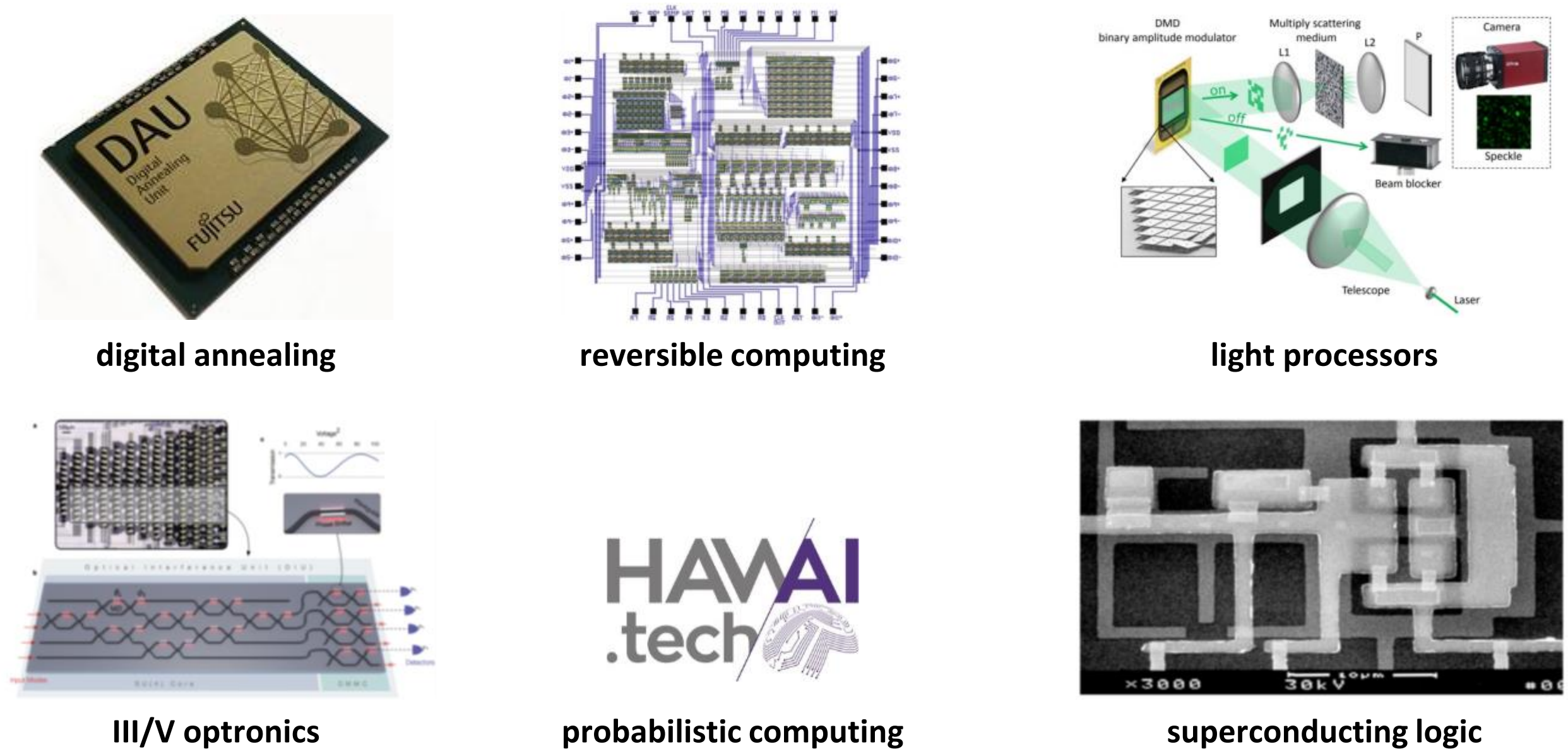

*Figure 15: various unconventional computing approaches besides quantum computing. (cc) Olivier Ezratty with uncredited images.*

The history of technology is about exploring multiple branches. Some do not succeed. Some help each other. Also, some can suddenly wake up after being frozen for decades. The game is open!

### Why... key takeaways

- All existing digital technologies are already quantum and belong to the first quantum revolution including transistors, lasers and the likes, leveraging our control of light-matter interactions with large ensembles of quantum objects (electrons, atoms, photons). The second quantum revolution is about using a variable mix of superposition, entanglement and individual quantum objects. It usually contains quantum computing, quantum telecommunications, quantum cryptography and quantum sensing.
- Quantum technologies are at the crossroads of many scientific domains encompassing physics, mathematics, computing, social sciences and the likes. It creates new educational and pedagogy challenges that must be addressed in innovative ways and customized according to various audiences. This book targets broad audiences with some technical background, including computer science engineers.
- Quantum computing promise is to solve so-called intractable problems whose computing complexity grows exponentially with their size. These can't be solved with classical computing, whatever happens with Moore's law. But we're not there yet since there are many challenges to scale quantum computers beyond what can be done today. In the interim, some marginal improvements will come with noisy intermediate scale computers, including better and more precise solutions in various domains.
- Other new technologies may compete with quantum computing, belonging to the broad "unconventional computing" category. Only a very few of these could also bring some exponential computing capacity. Most others bring other benefits compared with classical computing like in the energy consumption domain. Some of these technologies like superconducting electronics and adiabatic/reversible computing could also be helpful as enablers of quantum computing scalability.
- This book is unique in its shape and form. It covers quantum technologies with a 360° approach. It's more scientific than most publications, outside research review papers. It's a good appetizer for those who want to investigate the matter whatever the angle.

# History and scientists

After having set the stage, we'll make an history detour to discover the origins of quantum physics. As any scientific and technological endeavor, it's above all a great human story. I pay tribute here to the many scientists who, step by step, made all this possible and are still working on it for those who are still in this world.

**Nanoscopic physics**. Quantum physics deals with atomic and sub-atomic level particles and with the interactions between electromagnetic waves and matter. It differs from classical Newtonian physics, which predictably governs the dynamics of macrophysical objects, beyond a few microns and up to the size of planets and stars. Classical physics is governed by Newton's laws for matter, by Maxwell's laws for electromagnetic fields and associated forces and by statistical physics which describes continuous media such as gases and fluids and from which the principles of thermodynamics are derived.

When the speed of objects becomes close to the speed of light or when we reach large object's mass, the theory of relativity comes in, explaining the curvature of space-time and modelling the impact of gravity. It helps describing extreme phenomena such as black holes or neutron stars. It allows us to interpret the History of the Universe, but not entirely. But relativistic electrons are also hidden in our body's atoms and in many elements on earth as we'll quickly discover with the weird field of relativistic quantum chemistry.

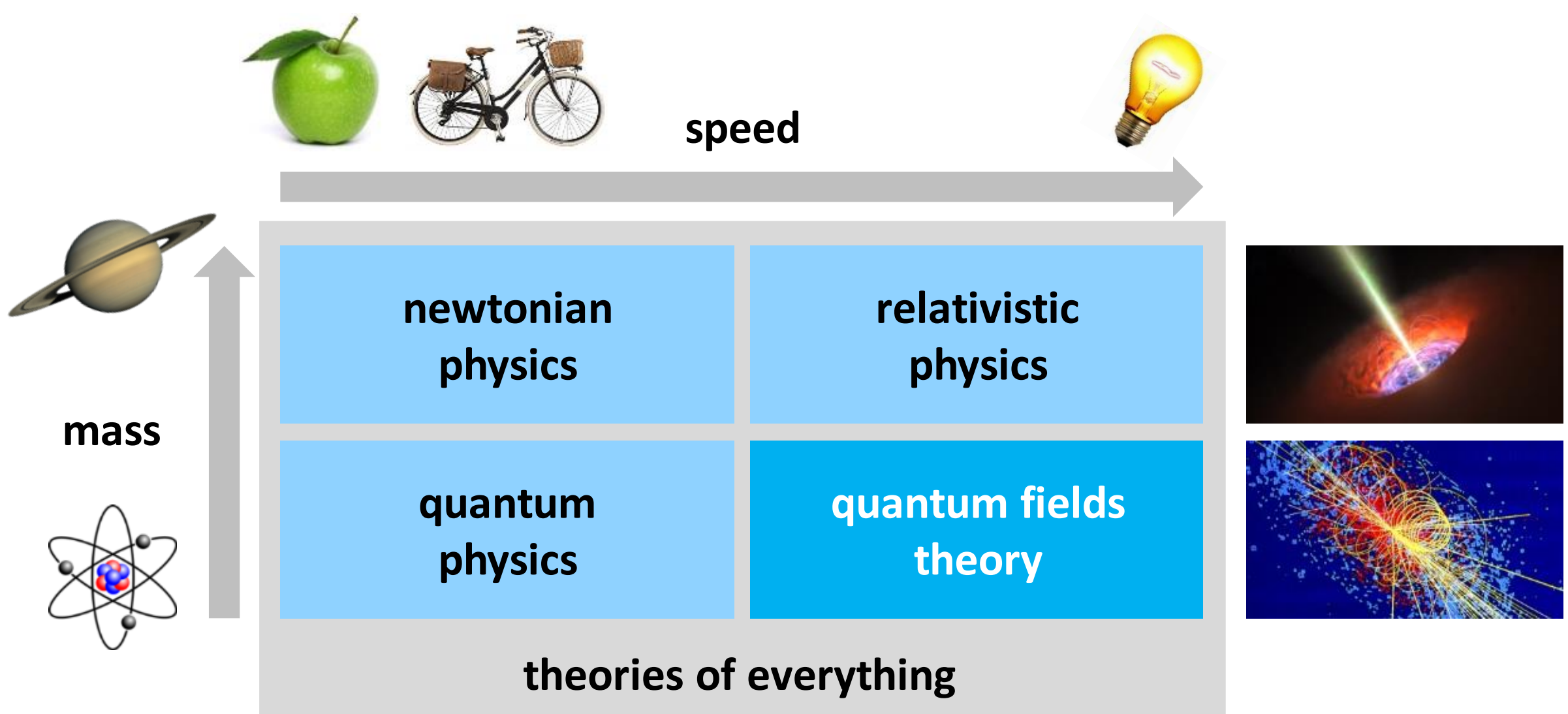

*Figure 16: high-level classification of the branches of physics. (cc) Olivier Ezratty, 2020.*

The fourth domain of physics in this quadrant is the quantum fields theory. It describes the physics of high-speed elementary particles, such as those observed in particle accelerators like quarks and the famous Higgs boson. Richard Feynman is one of the founders of quantum electrodynamics, a subset of quantum field theory.

In a way, quantum physics was a mean to unify classical matter physics and electromagnetic waves physics. It helps describe how matter was organized at the atomic and electrons levels and how these interacted with quantized electromagnetic waves, aka photons, including visible light.

**Unification still in the making**. Physics is still not yet complete nor unified. Some observable physical phenomena still resist it. We do not know how to explain the origins of gravitation and we are still looking for the dark matter and energy that would explain the cohesion of galaxies and the Universe current expansion. Scientists would like to explain everything, but some knowledge may never be accessible such as the shape and form of the Universe before the Big Bang.

The so-called theory of everything (ToE) or unification theory sought after by some physicists would be a formalism unifying all the theories of physics and in particular relativity, gravity and quantum physics. This very serious field of physics is still in the making[19]. Numerous proposals emerge and sorting it out is not easy[20].

**Connecting the dots**. This part will help you memorize who's who in the History of quantum physics and quantum computing. It will also cover some important science basics such as the Maxwell and Schrödinger equations that I'll try to explain in layman's terms, at least for readers having basic sciences knowledge. Explaining quantum computing inevitably starts with some quantum physics 101 explanations. Some of its basics, although sometimes quite abstract, must be understood. I still always try to connect the dots between quantum physics and quantum computing from a practical basis. It's a vast puzzle. I'll add its pieces one by one and even though the puzzle may not be fully completed, you'll get a picture enabling you to become fairly well educated on quantum computing.

**Experiments and theories**. Quantum physics took shape in 1900. Like almost all sciences, it is the result of the incremental work of many scientists with interactions between experimentation, theories and mathematical creativity. Sometimes, quantum physics is better explained with its underlying mathematical models than with incomplete physical interpretations. Representation models such as the broad field of linear algebra plays a key role to describe quantum states and their evolution in space and time. Linear algebra is also an essential tool to understand how quantum computer qubits are manipulated and measured. Even if we can trace the beginning of quantum physics to Max Planck's 1900 quanta discovery, it was based on earlier work from many other scientists who devised about the particle or wave nature of light, on the discovery of electromagnetism and atoms. Quantum physics is a human adventure that brought together immense talents who confronted each other and evolved step by step their understanding of the nanoscopic world. New generations of scientists have always questioned the state of the art built by their predecessors[21]. Physicists conducted numerous experiments, build theories and then verified it experimentally, sometimes with several decades of latency. They also had to pour philosophy into their work to interpret the deep significance of their discoveries, and quantum physics was not an exception. Despite its constant enrichment, quantum physics has shown an astonishing robustness to stand the test of time and with extreme precision.

**Misrepresentations**. Many quantum physics scientists are famous even for general audiences, even though their work has been overly simplified. Schrödinger's famous cat and Heisenberg's indeterminacy principle are commonplace... even when their underlying details are quite different from their related clichés. Schrödinger's key work is his non-relativistic particles wave equation, not the 10 lines he wrote in 1935 on his eponymous cat thought experiment that is usually grossly misinterpreted!

---

[19] The American-Japanese physicist Michio Kaku estimates that some theory of everything will be finalized by 2100. See Michio Kaku thinks we'll prove the theory of everything by 2100, April 2019. Michio is at the origin of string theory. He defines very well the connection between the different branches of physics and this theory of everything in A theory of everything?. But for many reasons too long to explain here, he happens to be very optimistic in his prediction!

[20] This is the case of the Wolfram Physics Project launched in April 2020 by Stephen Wolfram, a prolific Anglo-American physicist, mathematician and computer scientist. Building on his 2002 book "A new kind of science", the author's idea is to explain everything, the world, physics, the universe, whatever, with cellular automata, graphs and fractals. The world would be discrete on a small scale, including time. His Physics Project focuses on the unification of physics with the same set of tools. See the hundred pages presentation of the project, the white paper which contains a section on quantum physics. Physicists' views on this theory are more than circumspect. The paper does not develop a theory that would be verifiable with an experimental approach as was the case for quantum physics (superposition, wave function, wave function collapse, atomic transition spectral lines, ...). Wolfram's theory was critically analyzed in 2002 by Scott Aaronson in a 14-page review, particularly about his Bell's inequalities interpretation, and in A New Kind of Science by Cosma Rohilla Shalizi of Carnegie Mellon University, who does not mince his words. The same "hammer/nail explains everything" approach was created by a team of scientists who describe the Universe physics laws self-learning capabilities with a giant neural network approach, in The Autodidactic Universe by Stephon Alexander, Jaron Lanier, Lee Smolin et al, 2021 (79 pages).

[21] Max Planck's cynically explained in 1950 the evolution of science with the death of old generation of scientists: "*A new scientific truth does not triumph by convincing its opponents and making them see the light, but rather because its opponents eventually die and a new generation grows up that is familiar with it*".

Like life in general, science is a great relay race, with many players. Hundreds of other less-known contributors have also grown the field and must be recognized. Sometimes, genius scientists were so prolific than we forget their contributions. This is the case of John Von Neumann who is better-known for his "Von Neumann model" that is the cornerstone of classical computing and for his contribution to the development of EDVAC in 1949, the first stored program-based computer, rather than for his huge contribution to quantum physics mathematical formalism with density matrices and quantum measurement. It depends on the field you are working in, classical computing or quantum physics.

You won't find here inventors or entrepreneurs *a la* Steve Jobs or Elon Musk, even though the founders of startups like D-Wave, IonQ, Rigetti and PsiQuantum are among the entrepreneurial pioneers of this burgeoning industry, all being high-level scientists with a PhD!

**Hall of fame**. The History of 20$^{th}$ century quantum physics is embodied in the mythical Fifth **Solvay Conference in 1927,** held at the Institute of Physiology in Brussels. It brought together the greatest mathematicians and physicists of the time including almost all the historical founders of quantum physics with Max Planck, Albert Einstein, Niels Bohr, Louis de Broglie, Erwin Schrödinger, Max Born, Werner Heisenberg and Paul Dirac[22]. All this happened as the foundations of 20$^{th}$ quantum physics theories were fairly well laid out. 17 of its 29 participants got a Nobel Prize, 6 of which before the congress (names underlined in green) and the others afterwards (in blue). It was probably one of the largest concentrations and density of scientific brains per square meter in the history of mankind!

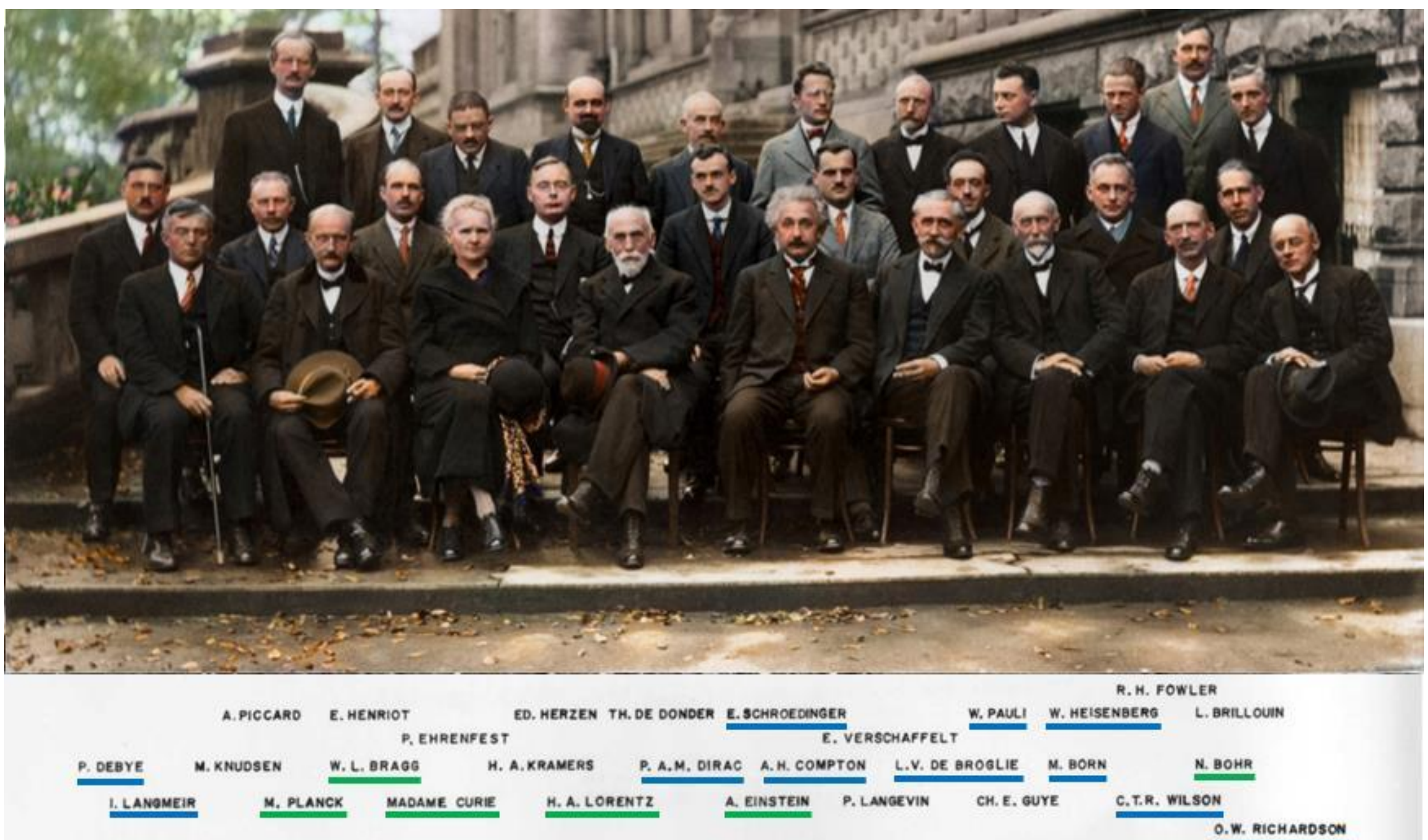

*Figure 17: the famous Solvay 1927 conference photo with its 17 Nobel prizes (6 back then, and 11 after the conference). Photo credit: Benjamin Couprie, Institut International de Physique de Solvay.*

Solvay conferences on physics are held every 3 to 4 years since their creation in 1911 by the entrepreneur and chemist **Ernest Solvay**. The 1927 congress's topic was electrons and photons, which are at the heart of quantum physics. Half of these conferences are dedicated to quantum physics, the other on different branches of physics. The 28$^{th}$ edition was held in May 2022 and gathered a contemporary hall of fame of quantum scientists from quantum physics to quantum information science.

[22] Only fathers and no mother! Marie Curie was present but was not specialized in quantum physics. She worked on radioactivity.

The major contributions of early scientists in quantum physics are generally arranged in chronological order, with some indication of who influenced whom.

## Precursors

We begin with the classical physicists and mathematicians of the 18th and 19th centuries who laid the scientific groundwork that allowed their 20th century successors to formalize the foundations of quantum physics[23].

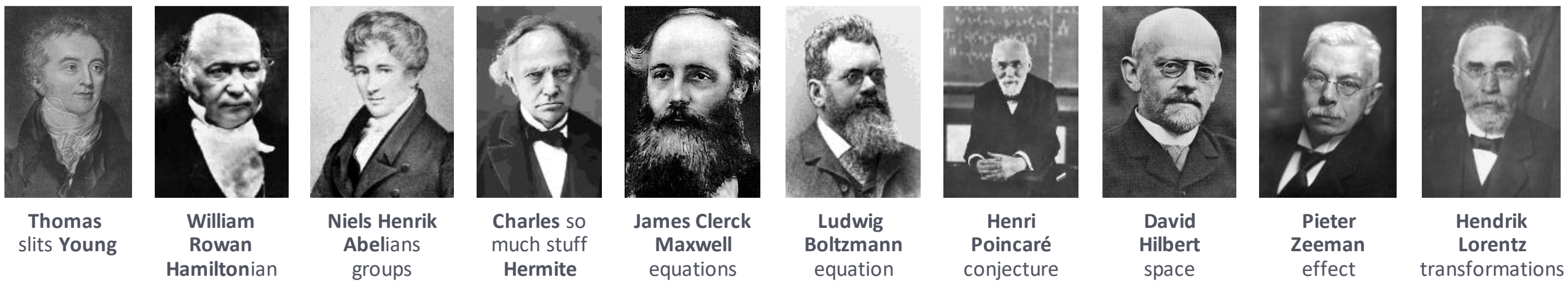

*Figure 18: precursor scientists who laid the ground particularly in the electromagnetic fields and mathematics domains.*

It's roughly organized in scientific contributions chronological order.

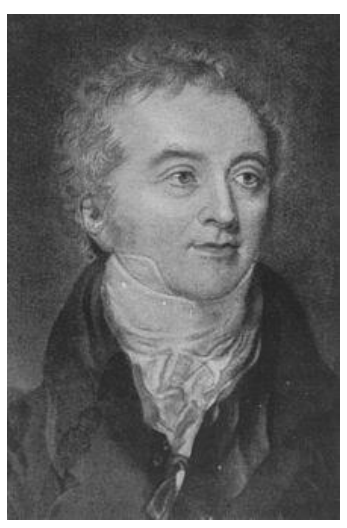

**Thomas Young** (1773-1829, English) was one of the great sciences and arts polymaths of his time, working in optics, medicine, linguistics, Egyptology and music. He determined that light behaved like a wave, which he proved with the double-slit experiment around 1806, illustrated in Figure 19, that now bears his name. When reducing the size of both slits, it generates interference fringes creating alternating light and dark zones related to the wave nature of light. We had to wait till Albert Einstein's work in 1905 to determine that light was also made of particles.

His experiment used red filtered sunlight going through a first slit. Contemporary experiments use coherent laser light sources. This experiment is one of the cornerstones leading much later to the creation of the electromagnetism theory by James Maxwell.

The slit experiment was implemented with electrons in 1961, with a similar result, illustrating the electron wave-particle duality, devised first by Louis de Broglie in 1924. It was then also done with atoms in 1991 and with various molecules starting in 2002.

Thomas Young also worked on the principles of refraction and human trichromatic vision as well as in fluid mechanics, including on the notion of capillarity and surface tension.

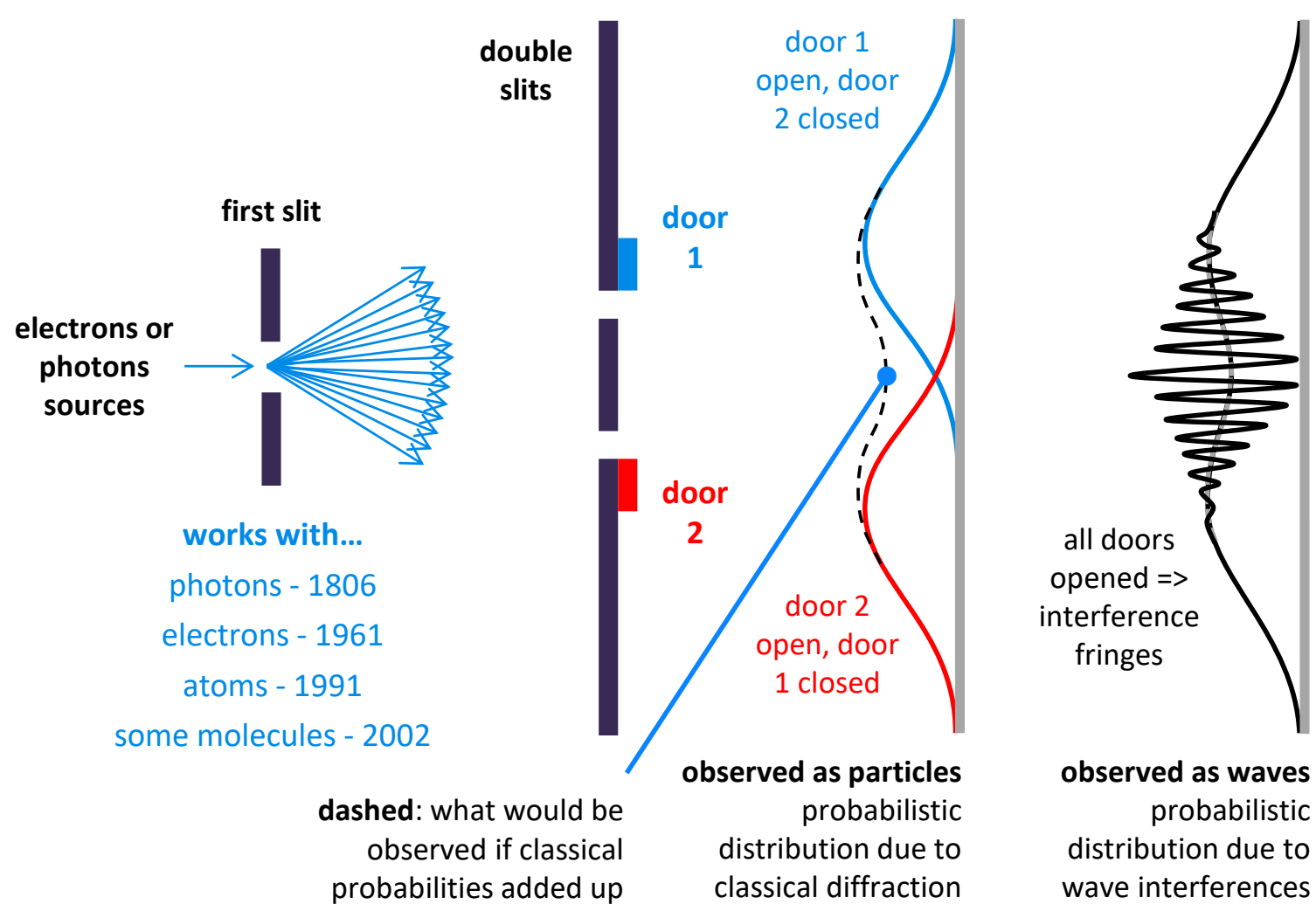

*Figure 19: the double-slit experiment principle (cc) Olivier Ezratty, sources compilation.*

As an Egyptologist, Thomas Young contributed to the study of the hieroglyphs of the famous Rosetta Stone, which was later used by **Jean-François Champollion** to decipher the whole stone texts. Champollion was then sponsored and helped by a certain **Joseph Fourier**. Yes, the mathematician!

[23] I do not always indicate the source of the diagrams used in this text. These are part of common scientific knowledge that are now in the public domain.

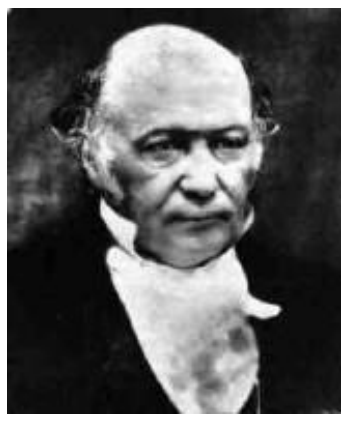

**William Rowan Hamilton** (1805-1865, Irish) was a mathematician and astronomer. He invented around 1827 a set of new mathematical formulations of the laws of physics incorporating electromagnetism. In quantum mechanics, we often speak of Hamiltonians or Hamiltonian functions. These are mathematical operators used to evaluate the total energy of a system of elementary particles including their kinetic and potential energies. This energy is evaluated over time.

Schrödinger's 1926 wave equation describes the evolution of a system's Hamiltonian over time. Among other domains, this concept is used in analog quantum computing with quantum simulators and quantum annealers, like with D-Wave's systems. We'll have the opportunity to cover this in detail in this book, starting page 278.

Hamilton is also behind the creation of quaternions in 1843 which generalize complex numbers, with using i, j and k as imaginary numbers with $i^2 = j^2 = k^2 = ijk = -1$. It can be used to compute three-dimensions rotations and have some applications in quantum computing like for the representation of two-qubit entanglement and of single qubit gates from the Pauli group, in topological quantum computing. This is an exotic domain that we won't cover in this book.

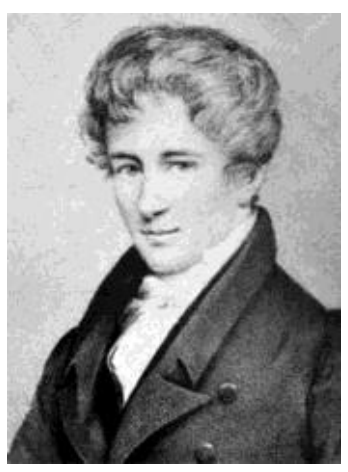

**Niels Henrik Abel** (1802-1829, Norwegian) is a prolific mathematician at the origin of the so-called Abelian groups. His work focused on the semi-convergence of numerical series, sequences and series of functions, the convergence criteria of generalized integrals, the notion of elliptic functions and integrals (used in cryptography) and the resolution of algebraic equations including his proof of the impossibility of solving general quintic equations.

He died way too early at the age of 26 from tuberculosis while visiting Paris and meeting his fiancée! Along with William Rowan Hamilton, Charles Hermite and Emmy Noether, he is one of the main 'suppliers' of the mathematical foundations used in quantum mechanics.

The adjectives "Abelian" and "non-Abelian" are associated with anyons, the quasiparticles that are the basis of topological quantum computing.

Why do these concepts of quantum mechanics invented long before his death refer to this mathematician? Mainly because the distinction between Abelian and non-Abelian is linked to their commutative mathematical representation. A system with A and B is Abelian when A*B = B*A or non-commutative and non-Abelian when A*B is not equal to B*A. The most common non-commutative operations are non-square matrices multiplications. The multiplication of a matrix (p x q) * (q x p) will give a matrix (p x p) whereas in the other direction, (q x p) * (p x q) will generate a matrix (q x q), q and p being here numbers of rows and/or columns.

Non-commutativity is frequently found in quantum physics and particularly with quantum measurement. The order in which quantum objects properties are measured may influence the results because the used measurement operators are non-commutative. In some cases, though, operators are commutative, like with the Measurement-Based Quantum Computing (MBQC) technique that we will have the opportunity to describe later when dealing with photon-based quantum systems.

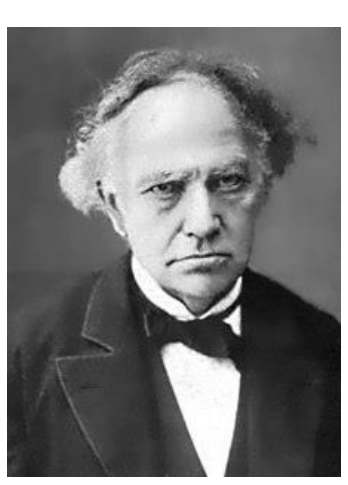

**Charles Hermite** (1822-1901, French) was another prolific $19^{th}$ century mathematician. He worked on numbers theory, quadratic forms, the theory of invariants, orthogonal polynomials, elliptic functions and algebra. His main works were concentrated on the 1848-1860 period. We owe him the notion of Hermitian functions and matrices, which are widely used in quantum physics and quantum computing. A Hermitian matrix is composed of real numbers in the diagonal and can be complex in the rest, and is equal to its transconjugate.

Namely, their transpose matrix whose value of complex numbers has been inverted (i becomes -i and vice-versa).

Quantum measurement operations in quantum physics and computers are defined by Hermitian matrices.

$$A = \begin{bmatrix} 2 & i & -2i \\ -i & 1 & 3 \\ 2i & 3 & -1 \end{bmatrix}^{\dagger} \qquad \bar{A} = \begin{bmatrix} 2 & -i & 2i \\ i & 1 & 3 \\ -2i & 3 & -1 \end{bmatrix} \qquad A = \overline{(A)}^{*}$$

transposed matrix

**Hermitian matrix**
matrix equals its transconjugate

*Figure 20: how a Hermitian matrix is constructed.*

**Achille Marie Gaston Floquet** (1847-1920, French) was a mathematician who developed mathematical analysis in the theory of differential equations. His name is used in Floquet codes (quantum error correction codes) and we also find him in the physics of quantum matter.

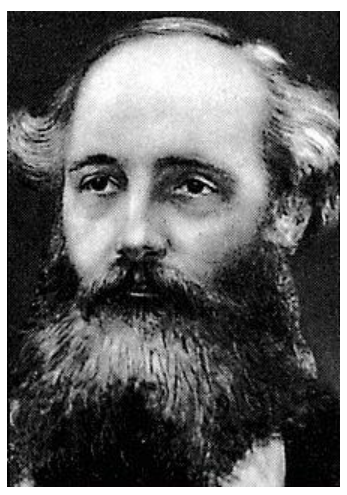

**James Clerk Maxwell** (1831-1879, Scottish) created in 1865 the theory of electromagnetic fields, combining an electric field and a magnetic field orthogonal to the direction of wave propagation as in the diagram *below*, and moving at the speed of light. This theory explains light-light interactions such as reflection, diffraction, refraction and interferences. Maxwell's work built on and improved the formalism from Faraday, Gauss, and Ampère.

Maxwell's equations illustrate that when they are constant, electric, and magnetic fields are independent, and in variable regime (with a wavelength λ), it becomes interdependent ($\vec{E}$ and $\vec{B}$), one generating the other and vice-versa, hence the notion of electromagnetic waves and fields.

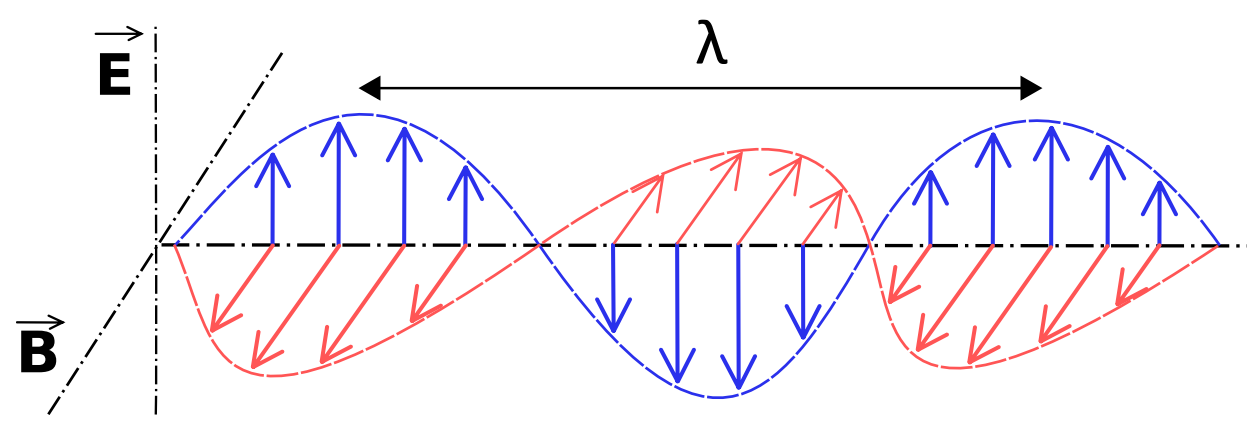

*Figure 21: electromagnetic wave electric and magnetic fields components.*

In Maxwell's equations, the electromagnetic field is represented by an electromagnetic tensor, a 4x4 matrix whose diagonal is zero and whose half of the components describe the electric field and the other half the magnetic field. These four dimensions correspond to space (3) and time (1).

In fact, there are four main Maxwell equations[24]:

- The **Maxwell-Gauss** equation describes how an electric field is generated by electric charges. At each point in space, the electric field is directed from positive to negative charges in directions depending on the charges space position.

$$\operatorname{div}\left(\vec{E}\right) = \frac{\rho}{\epsilon_0}$$

electric field

divergent, measures field variation orientation

charges distribution

vacuum dielectric permittivity

*Figure 22: Maxwell-Gauss equation describing the electric field created by electric charges.*

- The **Maxwell-flux** equation states that a magnetic field is always generated by a dipole with positive and negative charges that are connected and inseparable. Mathematically, this translates into the fact that the divergence of the magnetic field is zero and that there is no magnetic monopole.

$$\operatorname{div}\left(\vec{B}\right) = 0$$

magnetic field

$$\oiint_{(\boldsymbol{\Sigma})} \vec{\boldsymbol{B}} \cdot \mathbf{d}\vec{\boldsymbol{S}} = \mathbf{0}$$

surface integral

closed surface

surface vector **Σ** derivative

*Figure 23: Maxwell-flux equation.*

[24] See these well done and visual explanations of Maxwell's equations: A plain explanation of Maxwell's equations.

Namely, that there is no magnetic field line that escapes to infinity as we have with an electric field.

- The **Maxwell-Faraday** equation describes how the variation of a magnetic field creates an electric field. This is the principle used in electric alternators. The rotational operator using a nabla sign ∇ corresponds to a differential vector operation. It is equal to the first derivative of the magnetic field over time.
- The **Maxwell-Ampere** equation states that magnetic fields are generated by electric currents or by the variation of an electric field. This interdependence between magnetic fields and varying electric fields explains the circulation of self-sustaining electromagnetic waves. On the left of the equation is the rotational magnetic field.

rotational → ; electric field ; magnetic field

$$\vec{\nabla} \times \vec{E} = -\frac{\partial \vec{B}}{\partial t}$$

*Figure 24: Maxwell-Faraday equation connecting the magnetic and electric fields.*

rotational ; current density vector ; vacuum magnetic permeability ; vacuum dielectric permittivity ; electric field ; magnetic field

$$\vec{\nabla} \times \vec{B} = \mu_0 \vec{J} + \mu_0 \epsilon_0 \frac{\partial \vec{E}}{\partial t}$$

*Figure 25: Maxwell-Ampere equation connecting magnetic field to electric field*

second derivative over time of electric field ; electric field vector ; second derivative over space of electric field ; speed of light

$$\frac{1}{c_0^2}\frac{\partial^2 E}{\partial t^2} - \nabla^2 \mathbf{E} = 0$$

second derivative over time of magnetic field ; magnetic field pseudo-vector ; second derivative over space of magnetic field

$$\frac{1}{c_0^2}\frac{\partial^2 \boldsymbol{B}}{\partial t^2} - \nabla^2 \mathbf{B} = 0$$

*Figure 26: Maxwell's equations in vacuum.*

As with Schrödinger's equation, Maxwell's equations have several variations, which may be confusing. Maxwell first published twenty equations with twenty unknown variables in 1865. In 1873, he reduced them to eight equations. In 1884, **Oliver Heaviside** (1850-1925, English) and **Willard Gibbs** (1839-1903, American) downsized the whole stuff to the four partial differential vector equations mentioned above. These four vector equations are reduced to two tensor equations for electromagnetic waves propagated in vacuum.

The non-interaction with other elements explains the independence in this equation between electric and magnetic fields.

Maxwell predicted that electromagnetic waves were travelling at the speed of light.

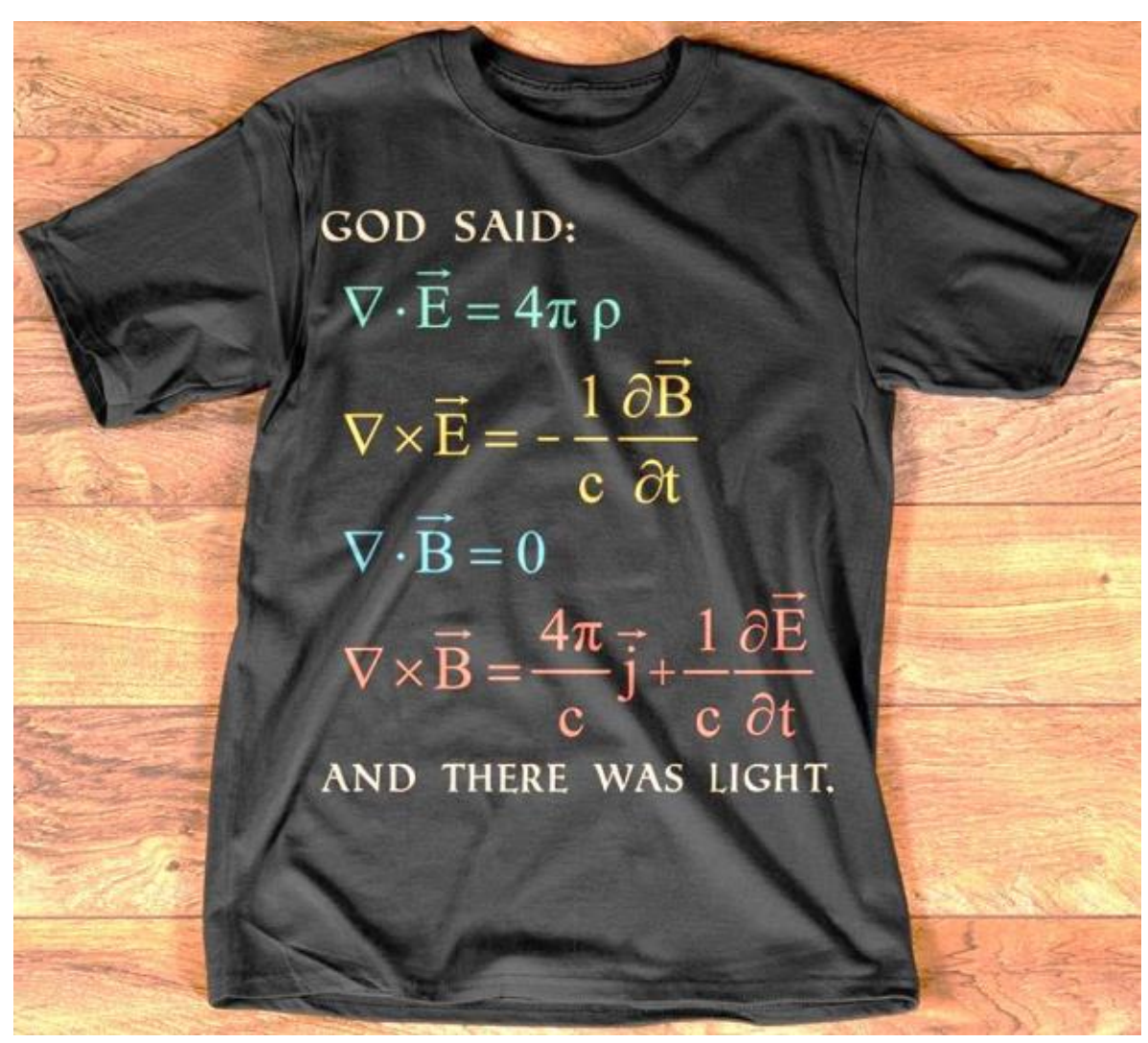

Electromagnetic waves were only experimentally discovered after Maxwell's death, by **Heinrich Hertz** (1857-1894) between 1886 and 1888. Hertz also discovered the photoelectric effect in 1887. Maxwell's description of electromagnetic waves had a phenomenal impact in electromagnetic telecommunications and optronics. It also served as a foundation for the first quantum physics laws developed by Max Planck in 1900 which led progressively to the quantization of the electromagnetic waves.

Maxwell is also at the origin of the **Maxwell-Boltzmann** statistical law of gas distribution. It models the particle velocity distribution of a perfect gas. It does not take into account the interactions between particles and is not applicable to extreme conditions, such as very low temperatures.

In particular, it is replaced by the **Bose-Einstein condensate** statistic for bosons (integer spin particles such as helium 4, which can be gathered in the same quantum state and energy level) and by the **Fermi-Dirac** statistic for fermions (particles with half-integer spins such as electrons or helium-3, which cannot cohabit in the same quantum and energy state).

Maxwell is the designer in 1867 of the so-called **Maxwell's demon** thought experiment which would make possible the reversibility of thermodynamic exchange processes and invalidate the second law of thermodynamics.

It rests on two boxes containing two different gases where a gas at two different temperatures is separated by a hole and a closure controlled by a "demon". When the door is opened, the gases mix.

Once mixed (see Figure 27), the demon would control which molecules could go from one box to another, taking advantage of the natural kinetic energy of the gases. This would allow in theory and after a certain time to return to the previous equilibrium in a non-equilibrium situation (on the right in Figure 27).

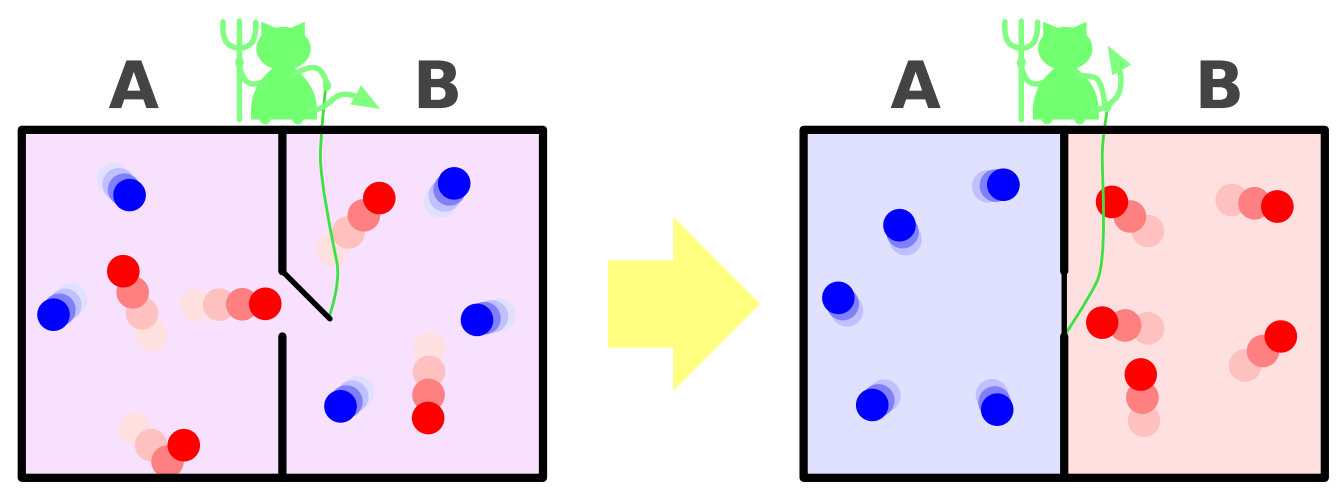

*Figure 27: Maxwell's demon principle. Source: Wikipedia.*

It took several decades to find the fault, notably via Léo Szilard in 1929 and Léon Brillouin in 1948. Initially, the explanation was that the demon needs to consume some energy to obtain information about the state of the gas molecules to sort them out. Therefore, energy is consumed to modify the stable equilibrium obtained to mix the gases.

The "up to date" explanation is somewhat different. The energy cost comes from resetting the demon's memory, which ultimately consists of a single bit of information[25].

All this had repercussions on the notion of the energy value of information and led, much later, to the creation of the field of information thermodynamics, i.e., the study of the energetic and entropic footprints of information, particularly in quantum computing.

This field was then investigated by **Rolf Landauer**, known for his study of irreversible information management circuits heat generation, and by **Charles Bennett** and **Gilles Brassard**, the co-inventors of the QKD based BB84 protocol, which we will discuss later, and then by **Paul Benioff**, who was at the origin of the idea of gate-based quantum computing.

We finally owe Maxwell the creation of color photography in 1855, that was implemented in 1861, based on the three primary colors of human vision.

Maxwell's electromagnetic field equations has very well survived the test of time. It's still the basis of classical optics and quantum optics. Even when studying quantized light, researchers and students still rely on Maxwell's equations and their subsequent derivations created since then.

---

[25] Here is the detailed explanation by Alexia Auffèves (CNRS Institut Néel / MajuLab): we can understand the operation of resetting a bit of memory by considering an ultimate Carnot engine, consisting of a single particle that can be located to the left or right of a certain thermostated volume. Left = 0, Right = 1 There are two possible operations. The first one is compression. The particle is initially to the left or to the right of the volume that contains it (we don't know) and we compress the said volume so that at the end it is necessarily on the left. It is an initialization operation where the bit is reset to state 0. As with any compression, you have to pay, here in this ultimate case, the work to be expended is kT log 2. This is Landauer's famous work, which sets an energy bound to all logically irreversible operations. The second operation is relaxation. In the beginning, we know whether the particle is on the left or on the right. We position a wall, a pulley with a mass at the end and let the trigger operate while extracting an elementary work equivalent to kT log 2. This is a Szilard machine. These two manipulations were performed experimentally in 2011 at ENS Lyon. It shows the energy footprint of information and are the ultimate solution to Maxwell's demon paradox. See Information and thermodynamics: Experimental verification of Landauer's erasure principle by Antoine Bérut, Artyom Petrosyan and Sergio Ciliberto, Université de Lyon and ENS Lyon, 2015 (26 pages).

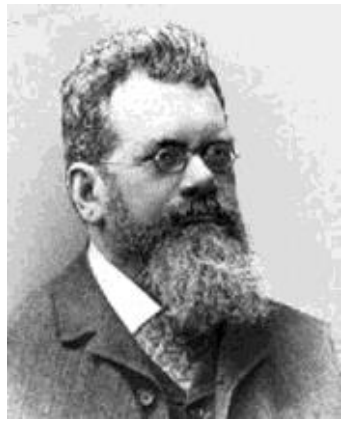

**Ludwig Boltzmann** (1844-1906, Austrian) was a physicist, the father of statistical physics, defender of the existence of atoms, facing a strong opposition from scientists until the beginning of the 20th century, and creator of equations describing fluid and gas dynamics in 1872. He is also at the origin of the probabilistic interpretation of the second law of thermodynamics, which establishes the irreversibility of physical phenomena, particularly during thermal exchanges.

Irreversibility is associated with the creation of entropy. Boltzmann tried his hand at philosophy while defending the existence of atoms. Depressed, he died by committing suicide.

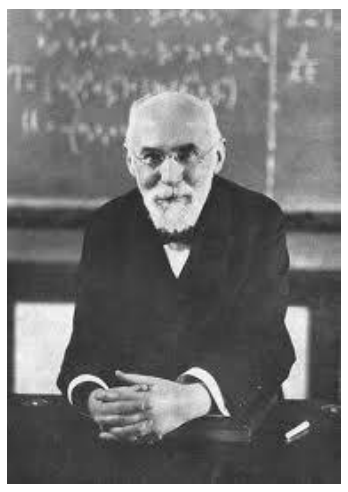

**Henri Poincaré** (1854-1912, French) was a mathematician and physicist, precursor of the theory of relativity and gravitational waves. We owe him a probabilistic function that bears his name, and which is the optical equivalent of the Bloch representation that we will see later, which mathematically describes the state of qubits. He is also the author in 1904 of the mathematical conjecture that bears his name and that was demonstrated in 2003 by the Russian mathematician Grigori Perlman. It is relative to hypersphere bounding the unit ball in a 4-dimensional space.

He was a first cousin of Raymond Poincaré (1860-1934), president of France during the First World War, a lesser-known figure than Georges Clémenceau who was then Prime Minister and drove the war efforts against Germany and with allies from the UK and the USA.

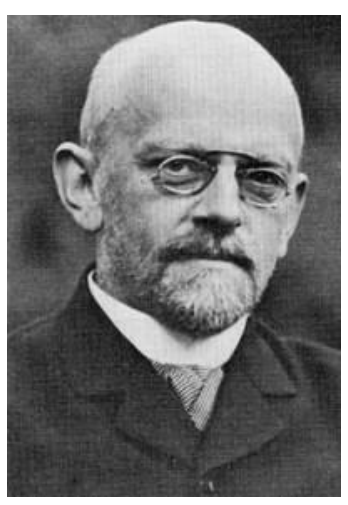

**David Hilbert** (1862-1943, German) is yet another prolific mathematician who, at the end of the $19^{th}$ century, was the creator of the mathematical foundations widely used in quantum physics, in particular his so-called Hilbert spaces using vectors to measure lengths, angles and define orthogonality. They are used to represent the state of quantum objects and qubits with vectors and complex numbers with an inner product, distances and an orthonormal basis (see Figure 28). Still, his work had nothing to do with quantum physics, which was not yet formulated at the time.

His work was used by Paul Dirac in 1930 and John Von Neumann in 1932 to lay the groundworks of quantum physics mathematical foundations like the Dirac Bra-Ket notation and the Von Neumann quantum measurement formalism.

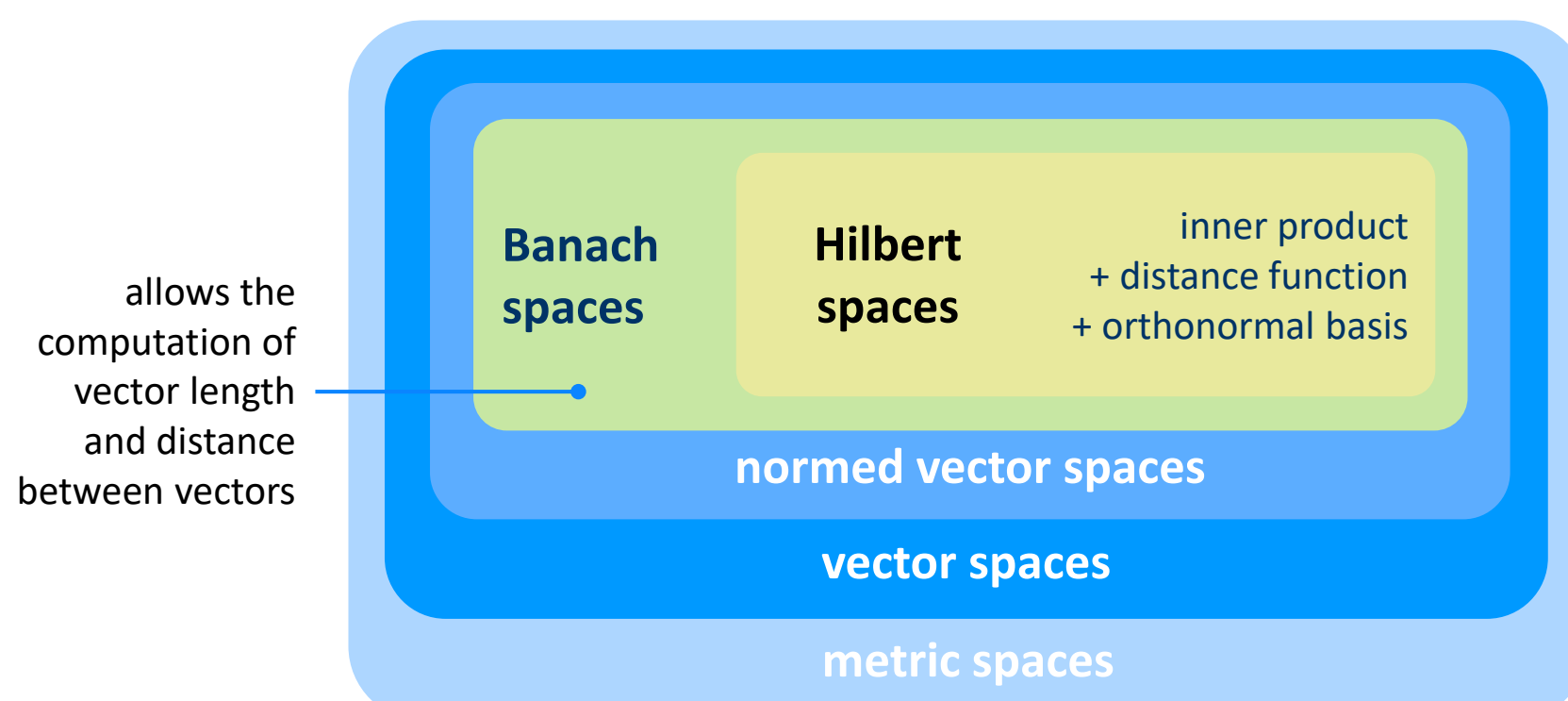

*Figure 28: a Hilbert space is a vector space with an inner product. It enables the measurement of vector distances, angles and lengths. Source: compilation Olivier Ezratty, 2022.*

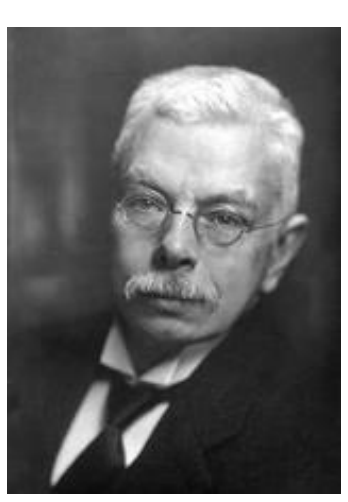

**Pieter Zeeman** (1865-1943, Dutch) was a physicist, Nobel Prize in Physics in 1902 with Hendrik Lorentz, for the discovery of the effect that bears his name between 1896 and 1897. The Zeeman effect occurs when excited atoms are exposed to a magnetic field. This affects their emission or absorption spectrum, that displays many discrete spectral lines. The effect is observed with spectroscopy, which breaks down light rays of different wavelengths with a prism.

In his experiment, spectral lines are broken down into an odd number of lines (normal Zeeman effect, as shown in Figure 29 for cadmium atoms) or an even number of lines (abnormal Zeeman effect). The decomposition depends on the intensity of the magnetic field passing through the analyzed atoms. There is also a nuclear Zeeman effect explained by the spin of atom nucleus.

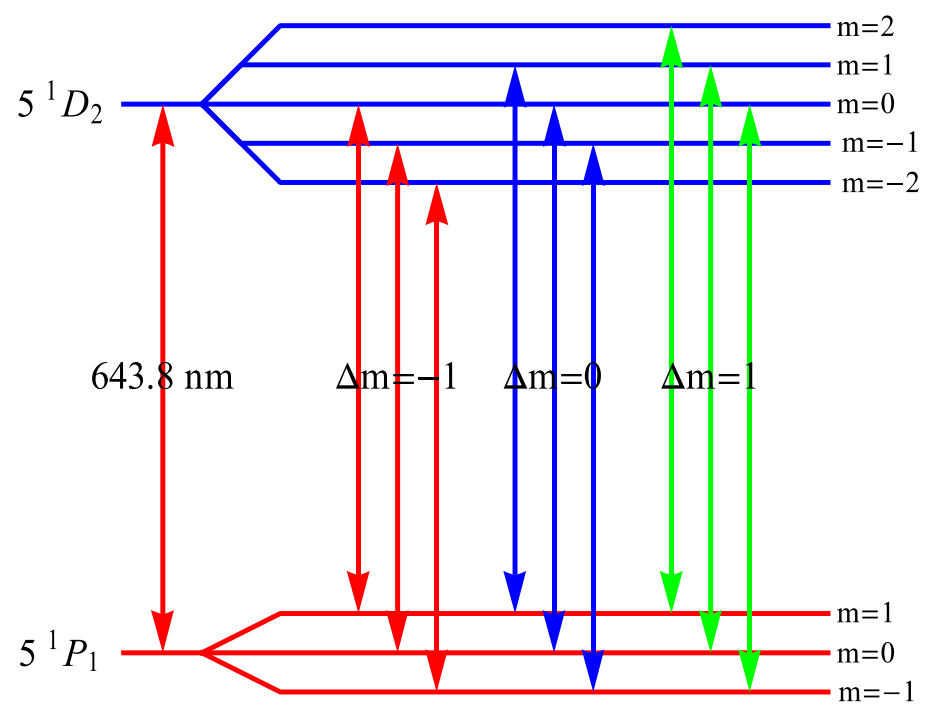

*Figure 29: normal Zeeman's effect energy transitions. Source: [Lecture Note on Zeeman effect in Na, Cd, and Hg](#) by Masatsugu Sei Suzuki and Itsuko S. Suzuki, 2011.*

It is matched by a polarization of the generated light whose nature and intensity depends on the orientation of the magnetic field relative to the light beam as shown here. The Zeeman effect can be explained by Pauli's exclusion principle, elaborated in 1925, and by the transitions in the energy level of the electrons in the same atom layer and having different orbital angular momentum (normal) and spin (abnormal). In astronomy, the Zeeman effect measurement is used to evaluate the intense magnetic fields in stars as well as within the Milky Way. The nuclear Zeeman effect is used in magnetic resonance spectroscopy in MRI scanners.

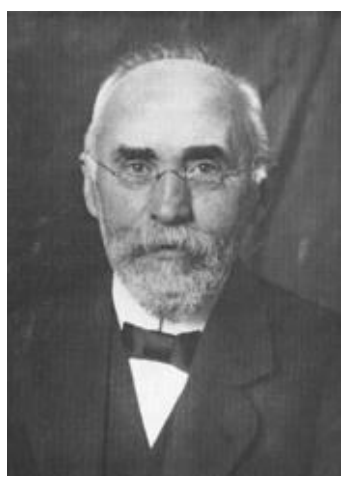

**Hendrik Antoon Lorentz** (1853-1928, Dutch) was a physicist who worked on the nature of light and the constitution of matter and made the link between light and Maxwell's electromagnetism equations. We owe him the Lorentz transformations that explain the results of Michelson-Morley's experiments between 1881 and 1887 which showed that the speed of light is stable, whatever the reference frame. With Henri Poincaré and George Francis FitzGerald (1851-1901, Irish), he was a key contributor to the theory of relativity formalized later by Albert Einstein between 1905 and 1915.

Let's also add **Joseph Larmor** (1857-1942, Irish/British) who, among other various contributions, was one of the first to associate electric charges with electron particles in 1894. We also own him the notion of Larmor precession, the rotation of the magnetic moment of an object when it is exposed to an external magnetic field, discovered with protons in 1919 and later extended to electrons.

## Founders

The foundations of quantum physics started with Max Planck's black-body explanation with energy quanta and, then took shape over three and a half decades, roughly until 1935. It involved the successive contributions from Einstein, Bohr, De Broglie, Schrödinger, Born, Heisenberg and Dirac to mention only the best-known contributors who were all theoreticians and not experimentalists. In the timeline from Figure 30, the gold coins represent a Nobel prize.

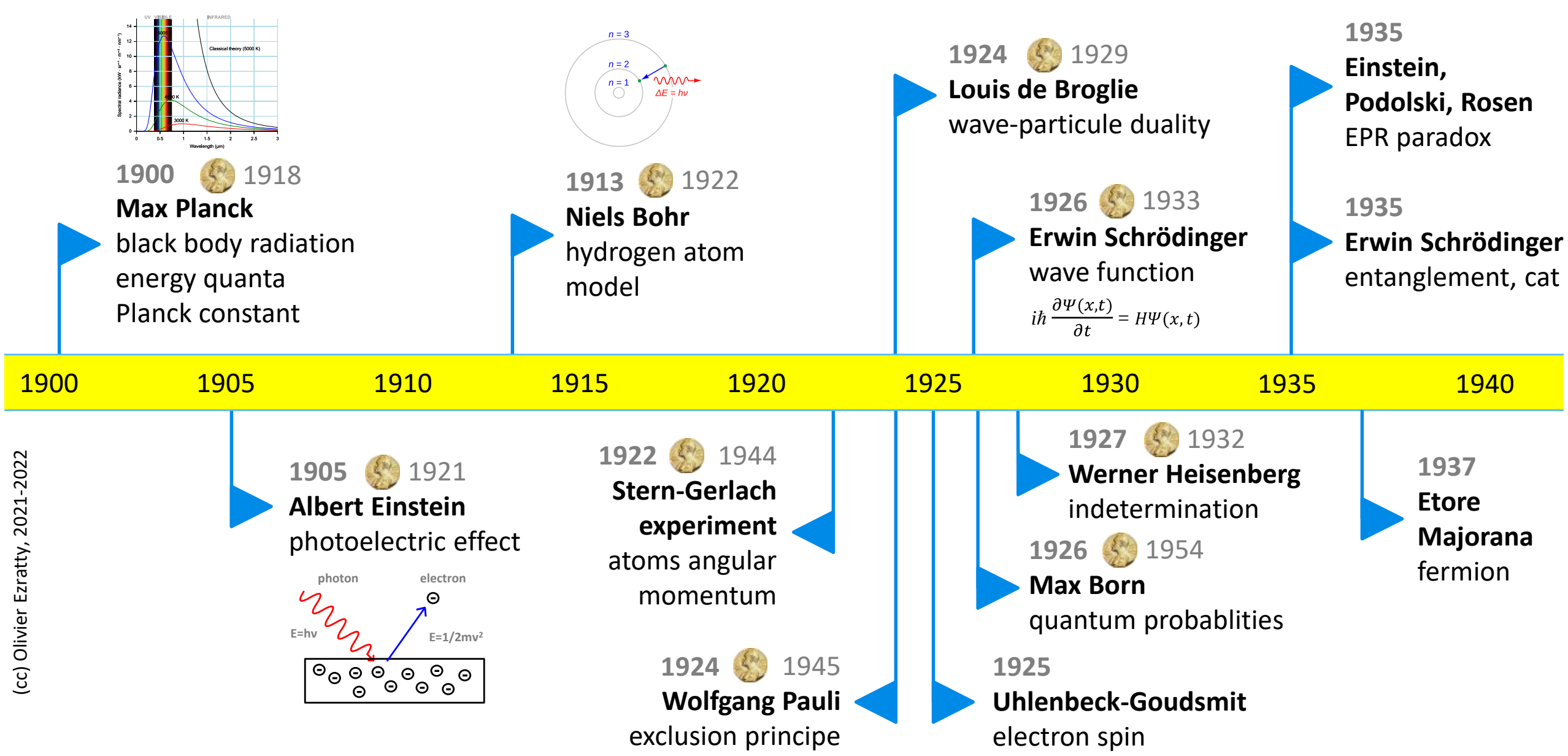

*Figure 30: quantum physics foundational years timeline. (cc) Olivier Ezratty, 2021-2022.*

Things were relatively quiet during World War II as lots of scientists were focused on creating the atomic bomb in the USA under the umbrella of the then very secret Manhattan project while Europe was not the best place in the world for travel and international scientific collaborations. German scientists who initially led quantum physics became isolated or emigrated to the USA or the UK because they were Jews, like Albert Einstein or Max Born.

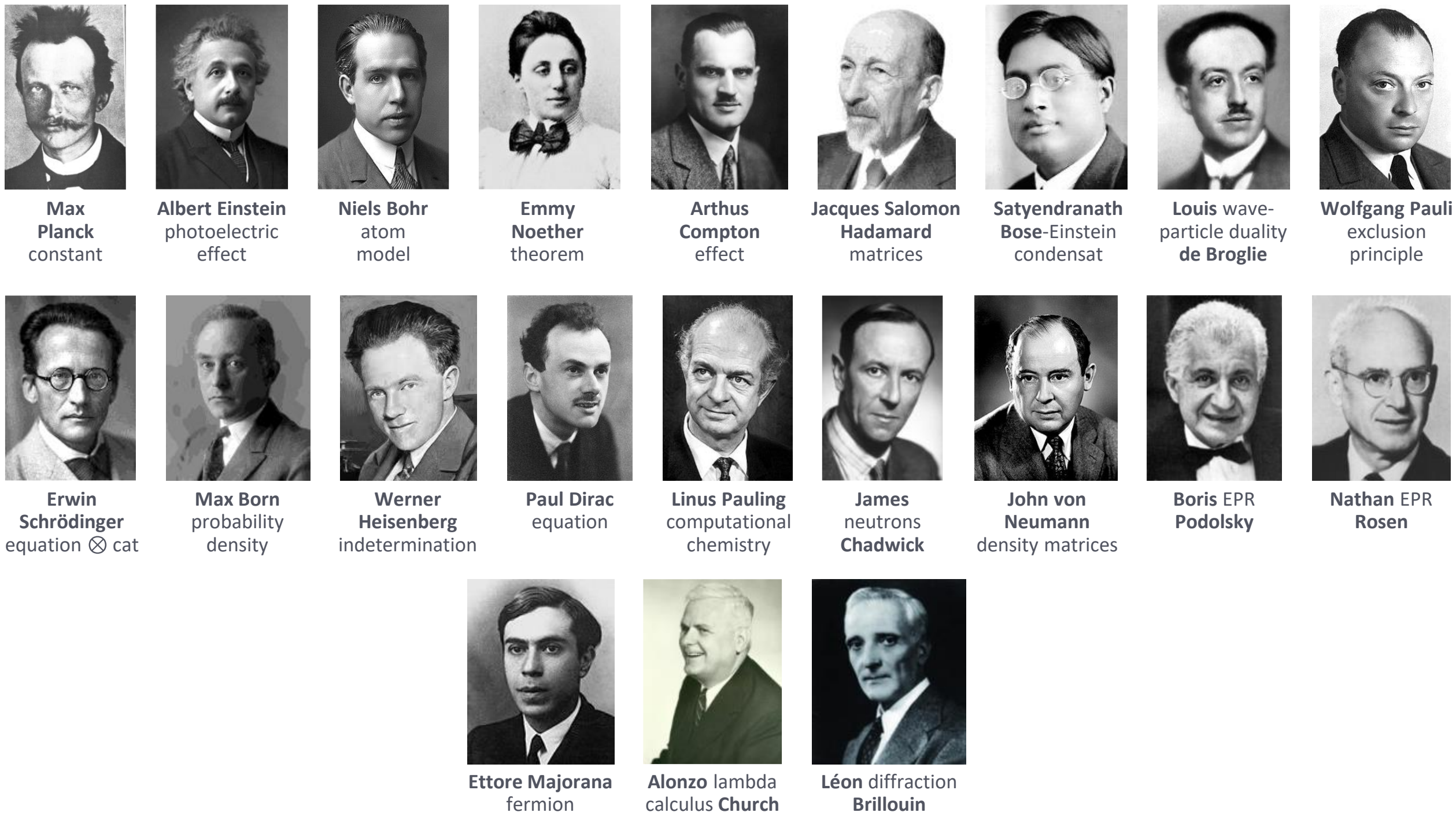

*Figure 31: the key founders of quantum physics in the first part of the 20th century. (cc) Olivier Ezratty, 2020.*

Here is a broader tour of the great physicists and mathematicians who laid the foundations of quantum physics. They are all Europeans who, some of whom emigrated from Europe to the United States before World War II.

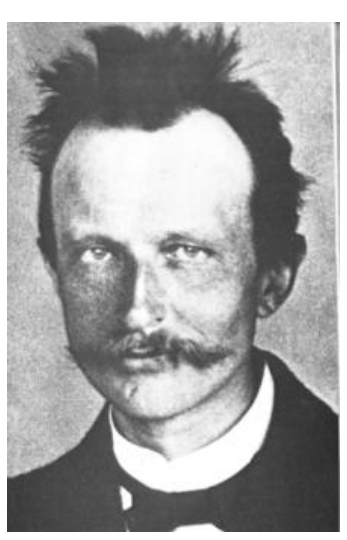

**Max Planck** (1858-1947, German) was a physicist, initially specialized in thermodynamics. In 1900, he developed the first basis of quantum physics, hypothesizing that energy exchanges between light and matter are made by discrete quanta. This radiation is not continuous but varies by thresholds, in steps of a certain amount of energy, hence the term "quantum" and "quantum physics" or “quantum mechanics”. His theory allowed him to roughly explain for the first time the enigmatic radiation of black bodies, that absorbs all incident magnetic radiation.

Examples of black bodies are a closed cavity like an oven, a heated metal that becomes red, orange, then white depending on the temperature, or a star like our own Sun. The spectrum of electromagnetic waves emitted by a black body depends only on its temperature and not at all on its material. The higher the temperature is, the more the electromagnetic spectrum emitted by the black body slides towards higher frequencies on the left, therefore towards purple and ultraviolet. The theory solved the ultraviolet catastrophe.

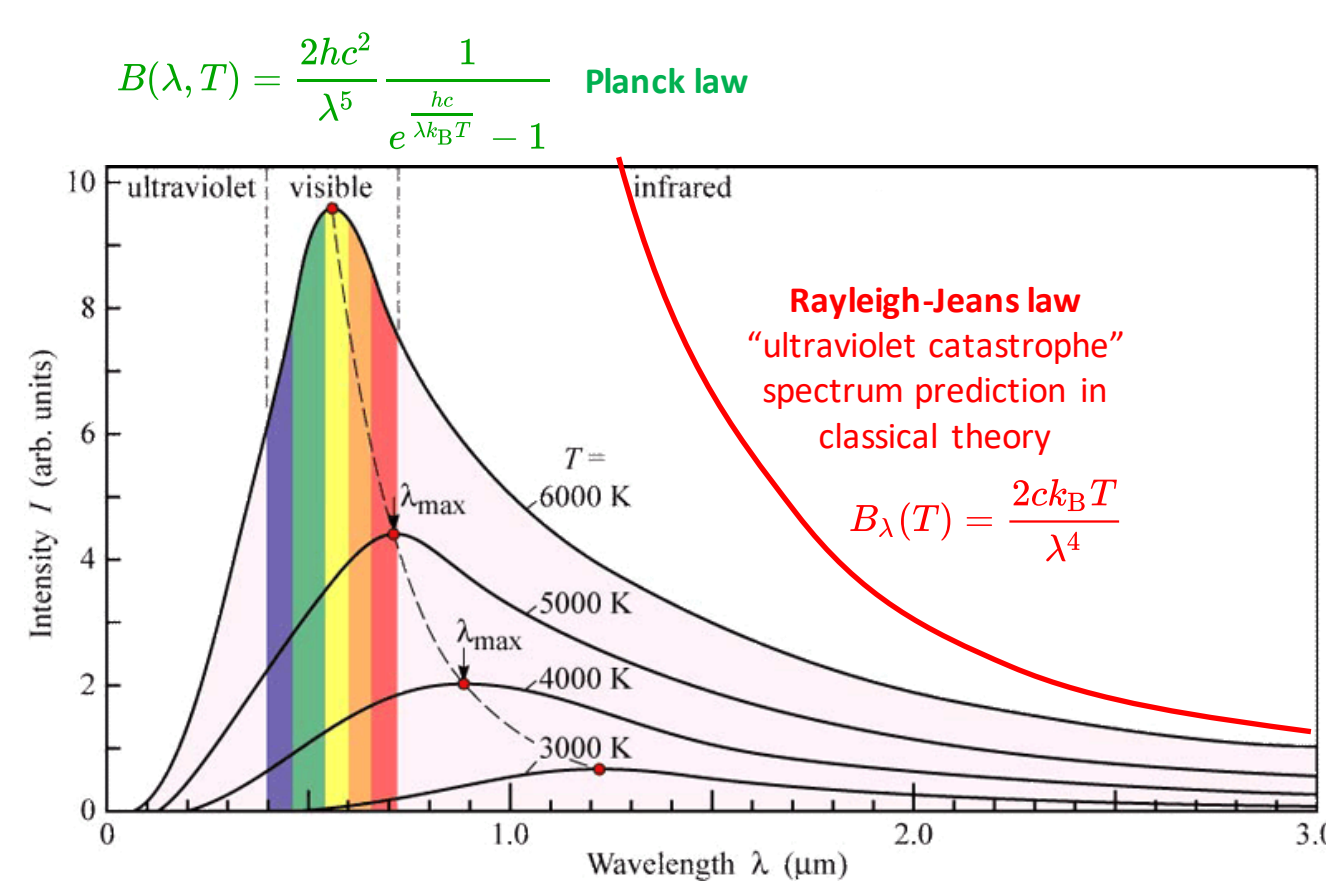

*Figure 32: black-body spectrum and the ultra-violet catastrophe.*

This so-called ultraviolet catastrophe, an expression **Paul Ehrenfest** (1880-1933, Austrian) created later in 1911, happened with the Rayleigh-Jeans law also proposed in 1900, which was trying to predict the shape of the light spectrum with the black body temperature. It was diverging to infinite values as the temperature was growing. Planck's law solved the problem and avoided the ultraviolet catastrophe. He found his spectrum equation empirically and only then, a related explanation based on harmonic oscillators and energy quanta exchanged between the radiation and the black body "wall". For this work, Max Planck was awarded the Nobel Prize in Physics in 1918.

We also owe him the constant which bears his name ($h$) and which is used in his blackbody radiation equation. The Planck constant ($6.626 \times 10^{-34}$ Js) was then used in the equation according to which atomic state energy shifts equals to the radiation frequency multiplied by Planck's constant. The constant appears in most quantum physics equations (De Broglie, Schrödinger, Dirac, etc.).

When an electron changes its orbit in a hydrogen atom, it emits or absorbs an electromagnetic wave whose energy is equal to Planck's constant multiplied by the emitted light frequency. More generally, a system can evolve only with multiples of Planck's constant. Despite the numerous experimental validations carried out a few years later, Max Planck expressed until his death a lot of doubts about the very principles of quantum physics!

Planck is also at the origin of several infinitesimal constants as shown in Figure 33: Planck time, which is $t_P=10^{-44}$s and Planck distance which is $l_P=1.616255*10^{-35}$m. Planck's time is the time it would take for a photon to travel the Planck distance.

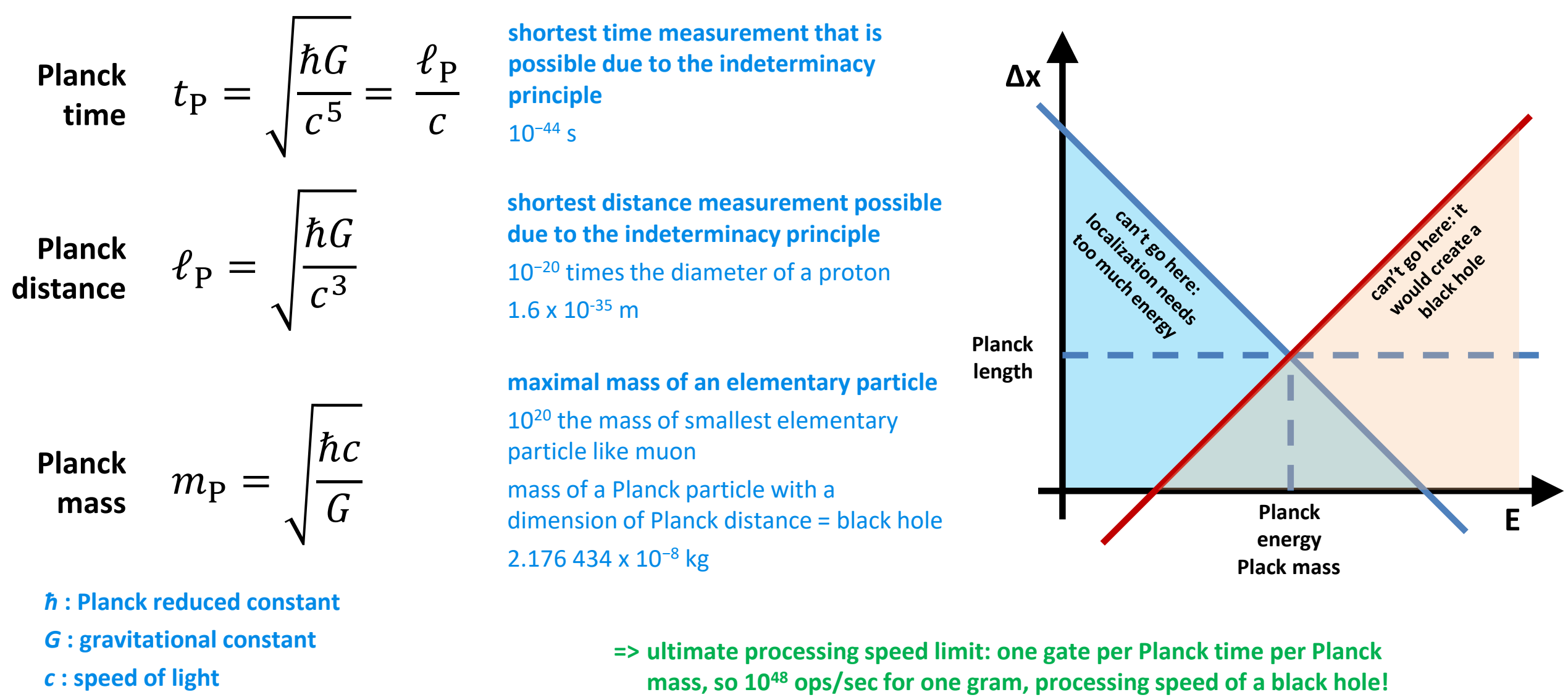

*Figure 33: Planck time, distance and mass constants (cc) Olivier Ezratty, 2021.*

These are the dimensions of the infinitely small below which any observation would be impossible. The length of Planck $l_P$ is so small that a photon used to observe it would have such a high frequency and energy that it would generate a black hole around it and would therefore become unobservable!

At last, Planck mass is the maximum mass of an elementary particle. A particle with this mass and the size of Planck's distance would be a black hole. These are quite extreme physics. In today's classical cosmology, Planck's wall corresponds in the history of the expansion of the Universe to the moment when $10^{-43}$s after the big bang, its size would have been $10^{-35}$m, which is respectively the Planck time and Planck distance. Needless to say that the experimental conditions of the big bang are difficult to reproduce. It doesn't prevent some physicists to try to simulate it digitally[26].

[26] See A new algorithm that simulates the intergalactic medium of the Universe in seconds is developed by the Instituto de Astrofisica de Canarias, May 2022.

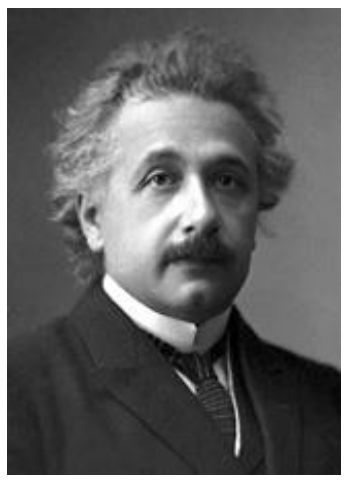

**Albert Einstein** (1879-1955, German then American) got his Nobel Prize in 1921 for his interpretation of the photoelectric effect in 1905, which became one of the foundations of quantum mechanics after Planck and before De Broglie, Heisenberg and Schrödinger. Einstein determined that Planck's quanta are elementary grains of energy $E = h\upsilon$ (Planck constant times frequency) with a momentum of $p = H\upsilon/c$ [27]. These were named "photons" in 1926 by **Gilbert Lewis** (1875-1946, American). Symbolically, 1905 is also the year of Jules Verne's death.

Symmetrically to what Louis De Broglie would later do with electrons, he hypothesized that a photon behaves both as a wave and as a particle.

This was coming out of just one out of his four 1905 "annus mirabilis" papers sent between March and June to Annalen der Physik, the others being on special relativity, Brownian motion and mass-energy equivalence, published when he was just 26. This was on top of his own 24 pages PhD thesis on a theoretical method to calculate molecular sizes using fluid mechanics and hydrodynamics.

With the photoelectric paper, he reconciled the corpuscular theories of **René Descartes** (1596-1650, French, in 1633) and **Isaac Newton** (1642-1726, English, in 1704) with the wave-based theories of **Christiaan Huygens (**1629-1695, Dutch, in 1678) to describe light.

This was followed by the works from **Augustin-Jean Fresnel** (1788-1827, French), **Léon Foucault** (1819-1868, French, who measured first the speed of light), **Hippolyte Fizeau** (1819-1896, French, who co-discovered the Doppler effect) and of course **James Clerk Maxwell**.

The photoelectric effect corresponds to the capacity of a photon to dislodge an electron from a generally inner orbit of an atom and to create some electric current[28].

It is exploited in the cells of silicon-based photovoltaic solar panels. It also explains photosynthesis in plants, which is the metabolic starting point of glucose production.

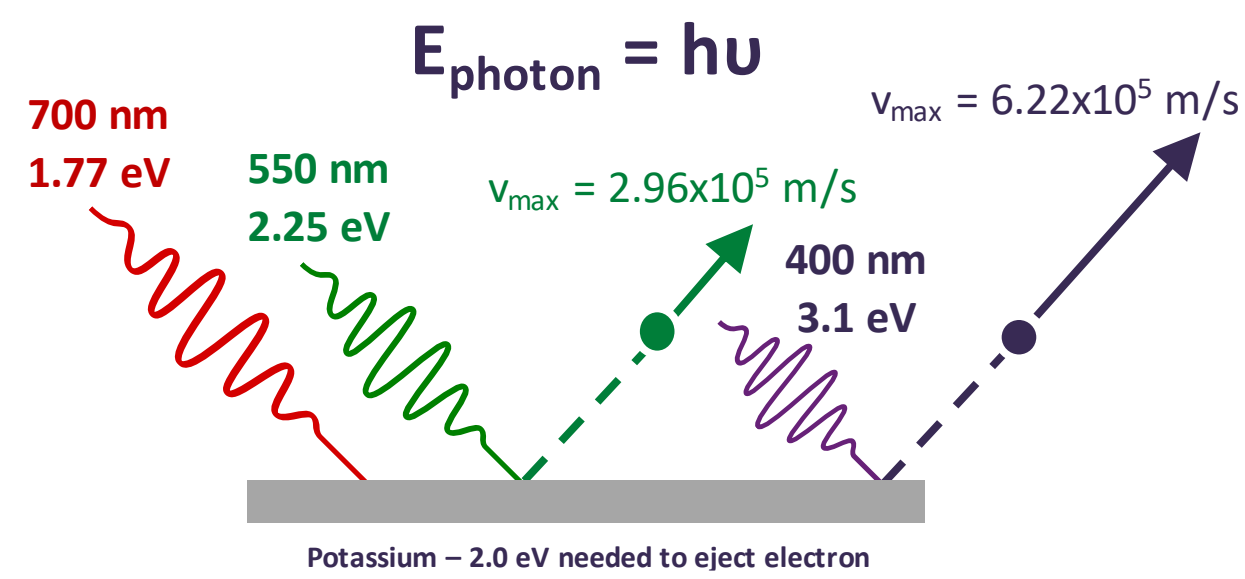

*Figure 34: the photoelectric effect.*

In addition to Max Planck's work on black body radiation, Einstein's interpretation was based on the earlier work of **Heinrich Hertz** (1857-1894, German) who discovered in 1887 that light can extract an electron out of metal, and **Philipp Lenard** (1862-1947, German) who, in 1902, studied the photoelectric effect and determined that it is only triggered at a certain frequency for the projected light. The latter was awarded the Nobel Prize in Physics in 1905. Becoming a fervent Nazi and opposed to Einstein by scientific rivalry and then by explicit anti-Semitism, he mostly disappeared from quantum physics hall of fame.

Einstein's photoelectric effect equations were then verified by the experiments of **Robert Andrews Millikan** (1868-1953, American) between 1909 and 1914. It enabled him to measure the electric charge of a single electron, which earned him the Nobel Prize in Physics in 1923.

---

[27] In On a Heuristic Viewpoint Concerning the Production and Transformation of Light, 1905.

[28] The electron layers of the atoms are numbered from 1 to N, their quantum number. One also starts the numbering by K (first layer close to the nucleus with a maximum of 2 electrons) then L (8 electrons maximum), M (with a maximum of 18 electrons but in practice 8), etc. The photoelectric effect mainly concerns the layers K and L. The ejected electron is then replaced by an electron of external orbit, which generates a new photon, in X-rays or in fluorescence, according to the energy of the incident photon. This then emits an X-ray photon due to the energy differential between electronic layers or an electron called "Auger" from the name of Pierre Auger. This phenomenon was discovered around 1923 by the latter and by Lise Meitner. Another variant of the photoelectric effect is the Compton effect, when the high energy of an incident photon in gamma rays will release an electron from the valence layer and generate another photon. Finally, when the energy of the incident photon is even higher, the interaction takes place at the nucleus of the target atom and generates an electron and a positron.

Of course, Einstein is also at the origin of the special and general theories of relativity. He didn't obtain a Nobel Prize for his work on relativity despite its considerable impact on science.

This is due, among other things, to his theories being based on earlier work from **Hendrick Anton Lorentz** and **Henri Poincaré** as well as the contribution of his former professor **Hermann Minkowski** (1864-1909, German) who created the four-dimensional space-time notion in 1908.

On top of many other contributions in quantum physics, Einstein predicted the photons stimulated emission effect in 1917, that would later lead to the creation of lasers. He also predicted in 1925 a particular behavior of matter, the Bose-Einstein condensate, which occurs when gases are cooled to very low temperatures. Atoms are then in a minimum energy quantum state showing particular physical properties.

This is the case of superfluid helium, discovered in 1938, which is superfluid at very low temperatures, i.e., it can move without dissipating energy. Bose is the name of **Satyendra Nath Bose** (1894-1974, India) with whom Einstein had worked during the 1920s and to whom we owe the "bosons", which verify the characteristics of Bose-Einstein's condensates.

Bosons include elementary particles without mass such as photons and gluons but also certain atoms such as deuterium or Helium 4 as well as certain quasi-particles such as the superconducting electron pairs that are Cooper's pairs. We will see a little later that it is a question of the spin sum of these particles that determines the fact that they are bosons as opposed to fermions.

Albert Einstein also contributed to the philosophical-scientific debates on quantum physics realism, confronting Niels Bohr. He focused on the fact that quantum physics did not seem to completely describe the physical world with its probabilistic bias. Einstein wanted to find a realistic interpretation of quantum physics. He could not be satisfied with a probabilistic description of the state of electrons and other quantum objects. He could not find sufficient the interpretation of quantum physics according to which the observer and the measurement "make" the real world. He thought that the real world exists independently of measurements and observers.

The debate between Albert Einstein and Niels Bohr revolved around various thought experiments on determinism discussed during the 1927 Solvay Congress.

MAY 15, 1935 PHYSICAL REVIEW VOLUME 47

**Can Quantum-Mechanical Description of Physical Reality Be Considered Complete?**

A. Einstein, B. Podolsky and N. Rosen, *Institute for Advanced Study, Princeton, New Jersey*

(Received March 25, 1935)

In a complete theory there is an element corresponding to each element of reality. A sufficient condition for the reality of a physical quantity is the possibility of predicting it with certainty, without disturbing the system. In quantum mechanics in the case of two physical quantities described by non-commuting operators, the knowledge of one precludes the knowledge of the other. Then either (1) the description of reality given by the wave function in quantum mechanics is not complete or (2) these two quantities cannot have simultaneous reality. Consideration of the problem of making predictions concerning a system on the basis of measurements made on another system that had previously interacted with it leads to the result that if (1) is false then (2) is also false. One is thus led to conclude that the description of reality as given by a wave function is not complete.

*Figure 35: the famous EPR paper from Albert Einstein, Boris Podolsky and Nathan Rosen published in 1935.*

It culminated later, in 1935, with the famous **EPR paradox** paper, named after its authors Albert Einstein, Boris Podolsky and Nathan Rosen. The paper raised the question of the incompleteness of quantum mechanics at the time[29].

---

[29] See Can Quantum-Mechanical Description of Physical Reality Be Considered Complete?, by Albert, Einstein, Boris Podolsky and Nathan Rosen 1935 (4 pages).

It sought to explain the non-locality of the correlated quantum state measurement results of entangled particles which was a consequence of Schrödinger's wave function. It was not yet physically observed as of 1935[30]. For the EPR paper authors, the quantum theory based on Schrödinger's wave function was either incomplete or two quanta could not be instantaneously synchronized at a distance at measurement time. Their measurement outcome being random and correlated, entangled quantum objects had to convey with them a sort of “information switch” indicating where the random measurement should land. A physical theory is complete if each component of reality has a counterpart in the theory that makes it possible to predict its behavior, such as some tuning happening at the source when entangled quanta are created, and transmitted to each one with some hidden variables that would determine the outcome of their measurement. This underlies the notion of determinism, a principle that is absent in Schrödinger's wave function which is entirely probabilistic in nature.

Einstein thought that quantum physics was an incomplete theory that didn’t describe reality precisely enough. Einstein was then often credited with the idea that there were hidden variables. It seems, however, that he never mentioned them in his writings despite what John Bell later said. The EPR paper ends with indicating that it should be possible to build a complete theory of quantum mechanics[31]. Hidden variables are a consequence rather than a hypothesis in the EPR paradox paper.

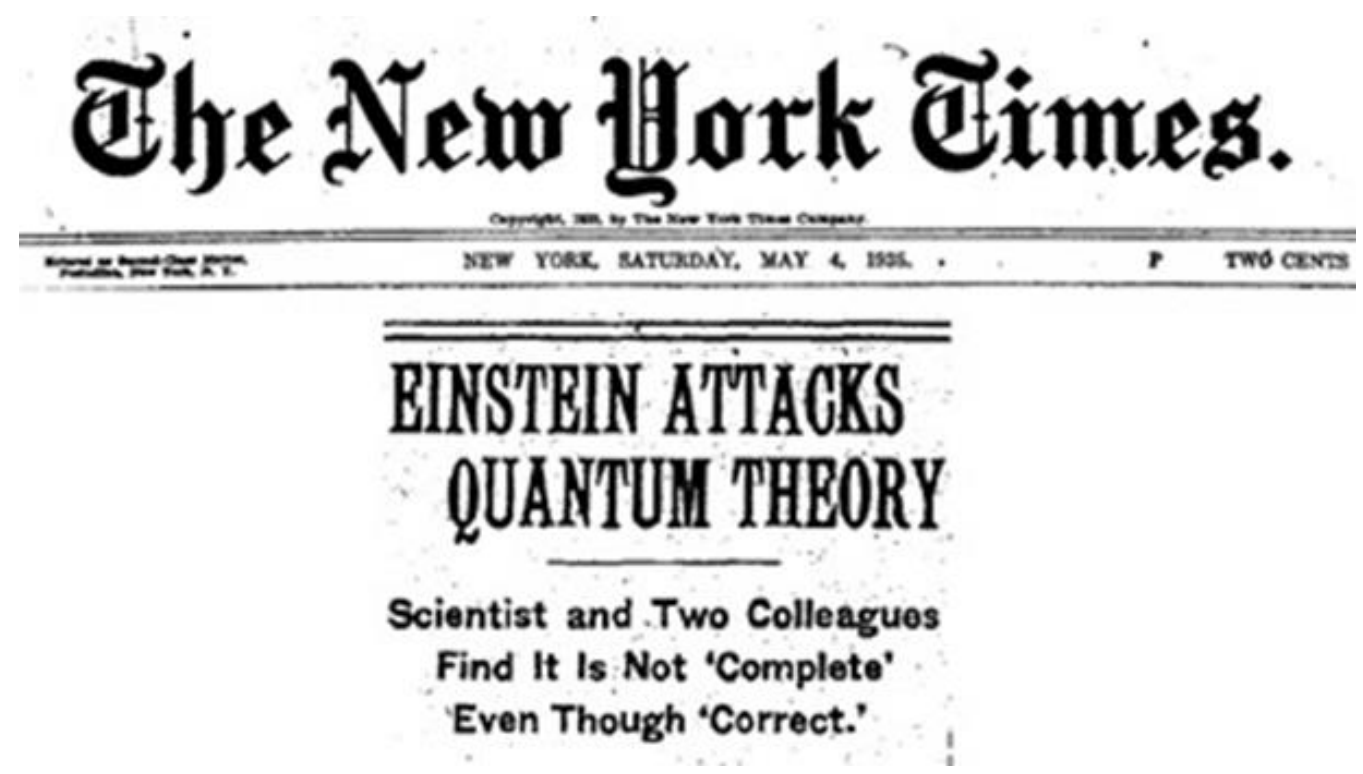
The New York Times.

NEW YORK, SATURDAY, MAY 4, 1935. TWO CENTS

EINSTEIN ATTACKS QUANTUM THEORY

Scientist and Two Colleagues Find It Is Not 'Complete' Even Though 'Correct.'

*Figure 36: New York Times coverage of the EPR paper.*

The explanation of entanglement by "hidden variables" comes rather from Louis de Broglie with his pilot wave hypothesis elaborated in 1927, an idea later pursued by David Bohm in the 1950s[32]. With his "inequalities", John Stewart Bell demonstrated in 1964 that the existence of such hidden local variables was incompatible with the principles of quantum mechanics. Alain Aspect's 1982 experiment on photon entanglement confirmed this hypothesis. In the end, Einstein could not finish his work on his theory of general relativity which was, for him, as incomplete as quantum mechanics. In particular, he wanted to reconcile quantum mechanics and gravity.

Be careful with the simplistic views that Einstein was “against” quantum mechanics, had it all wrong or did not believe in it[33]. He first questioned the principle of indeterminacy in 1927 and 1930, then estimated that the theory was incomplete to explain entanglement, with the EPR paradox paper published in 1935, and finally, he opposed the lack of realism of quantum theory. This incompleteness is still being discussed more than 80 years later. The origins of entanglement are still not physically explained under certain conditions, particularly with long distance. It is only observed physically and described mathematically[34]. This remains an open debate as scientists continue to ponder the different possible interpretations of quantum physics. This is part of the field of quantum foundations and quantum physics philosophy that we cover later in this book, page 987.

---

[30] Einstein's landmark was classical and relativistic physics that acted locally. Gravity is local and is transmitted at the speed of light. All physical theories before quantum physics were local or EPR-local. Remote actions all involve a delay, usually coupled with attenuation with distance as it’s the case for gravity.

[31] The 1935 New York Times article was published thanks to a "leak" provoked by Boris Podolsky, the youngest of the EPR 3 gang.

[32] See Albert Einstein, David Bohm and Louis de Broglie on the hidden variables of quantum mechanics by Michel Paty, 2007 (29 pages) which sets the record straight on Albert Einstein's position on the subject of hidden variables. The author, born in 1938, is a physicist and a philosopher of science.

[33] This story is well told in Einstein and the Quantum - The Quest of the Valiant Swabian by A. Douglas Stone, 2013 (349 pages).

[34] See the abundant Einstein Bohr debates and Interpretations of quantum mechanics pages on Wikipedia, from which the table on the next page is taken.

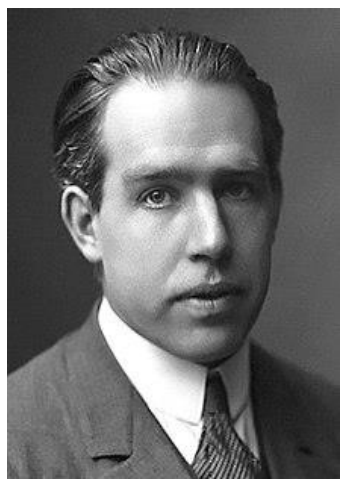

**Niels Bohr** (1885-1962, Danish) was a physicist, Nobel Prize in Physics in 1922, who created in 1913, aged 28, a descriptive model of the hydrogen atom with its nucleus made of a proton and an electron rotating around the nucleus on precise orbits corresponding to levels of kinetic energy, multiple of h/2π, h being Planck's constant and n = 1, 2, 3 and so on. This model explained hydrogen spectral lines observed in the experiments of **Johann Balmer** (1825-1898) in 1885, **Theodore Lyman (**1874-1954) in 1906 and **Friedrich Paschen (**1865-1947) in 1908.

It also explained why electrons didn't crash on atom nucleus! Niels Bohr followed the work of **Ernest Rutherford** (1871-1937) who discovered in 1911 the structure of atoms with their positively charged nucleus, thanks to its protons, and their electrons revolving around the nucleus. The latter, with whom Niels Bohr was doing his post-doc in 1911, relied himself on **Hantaro Nagaoka** (1865-1950, Japan) who predicted in 1903 the structure of atoms with a positively charged nucleus and negatively charged electrons revolving around it, called the "Saturnian model".

Electrons had been discovered by **Joseph John Thomson** (1856-1940, English) in 1897 by analyzing the rays emitted by a cathode in a cathode ray tube (CRT), deflected by an electric field as well as by a magnetic field, and detected by a layer of phosphorus. He was awarded the Nobel Prize in Physics in 1906.

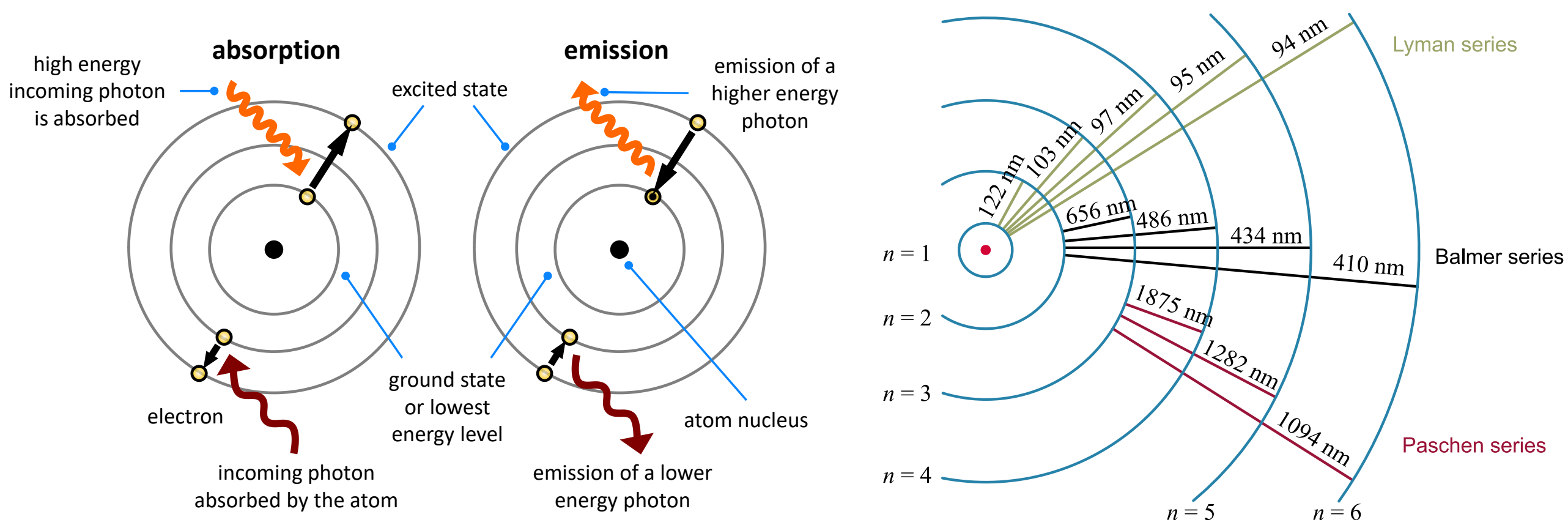

*Figure 37: the Bohr atomic model. Source: Wikipedia and other open sources.*

Ernest Rutherford had also imagined the existence of neutrons, which was not verified experimentally until 1932 by **James Chadwick** (1891-1974, English). **Marie Curie** (1867-1934, Polish and French) had discovered polonium and radium in 1898 and some effects of radioactivity but not the existence of neutrons.

According to Niels Bohr, electrons emit or absorb a photon when they change orbit. Subsequently, Louis de Broglie's work on wave-particle duality interpreted that the orbits of the electrons were an integer multiple of their associated wavelength.

Together with Werner Heisenberg, Pascual Jordan and Max Born, Niels Bohr is at the origin of the so-called **Copenhagen** interpretation of quantum physics which is based on three key principles[35] :

- The description of a wave-particle is realized by its wave function, and no other "hidden" local information or variable can be used to describe its state. We must accept the wave function probabilistic used to describe a quantum state.

[35] See also Richard Webb's Seven ways to skin Schrödinger's cat, 2016 which describes the different schools of thought in quantum physics. See also other interpretations of quantum physics in Ethan Siegel's The Biggest Myth In Quantum Physics Starts With A Bang in Forbes, 2018.

- When a quantum state measurement is performed, its composite wave function of several states is reduced to the wave function of one of the possible states of the quantum with a probability define by Born's rule (we'll see that later). This is the collapse of the wave function.
- When two properties are linked by an uncertainty relationship, the two properties cannot be measured with a greater precision than that allowed by the uncertainty relationship (Heisenberg principle of indeterminacy). Moreover, when we measure the position of a particle, we affect its motion, and vice versa. It comes from the bare fact that speed and position do not have any meaning before measurement in quantum physics. Variables linked through an indetermination link are conjugate with regards to actions which can change only by quantum leaps.

This is the main interpretation of quantum mechanics. There are many other interpretations available, listed below in the table from Figure 38. We will have the opportunity to detail the interpretation of Copenhagen towards the end of the book in the part dedicated to the philosophy of quantum physics, page 987.

| Interpretation | Year published | Author(s) | Deterministic? | Ontologically real wavefunction? | Unique history? | Hidden variables? | Collapsing wavefunctions? | Observer role? | Local dynamics? | Counterfactually definite? | Extant universal wavefunction? |
|---|---|---|---|---|---|---|---|---|---|---|---|
| Ensemble interpretation | 1926 | Max Born | Agnostic | No | Yes | Agnostic | No | No | No | No | No |
| Copenhagen interpretation | 1927 | Niels Bohr, Werner Heisenberg | No | No[1] | Yes | No | Yes[2] | Causal | Yes | No | No |
| de Broglie–Bohm theory | 1927-1952 | Louis de Broglie, David Bohm | Yes | Yes[3] | Yes[4] | Yes | Phenomenological | No | No | Yes | Yes |
| Quantum logic | 1936 | Garrett Birkhoff | Agnostic | Agnostic | Yes[5] | No | No | Interpretational[6] | Agnostic | No | No |
| Time-symmetric theories | 1955 | Satosi Watanabe | Yes | No | Yes | Yes | No | No | No[55] | No | Yes |
| Many-worlds interpretation | 1957 | Hugh Everett | Yes | Yes | No | No | No | No | Yes | Ill-posed | Yes |
| Consciousness causes collapse | 1961-1993 | John von Neumann, Eugene Wigner, Henry Stapp | No | Yes | Yes | No | Yes | Causal | No | No | Yes |
| Stochastic interpretation | 1966 | Edward Nelson | No | No | Yes | Yes[14] | No | No | No | Yes[14] | No |
| Many-minds interpretation | 1970 | H. Dieter Zeh | Yes | Yes | No | No | No | Interpretational[7] | Yes | Ill-posed | Yes |
| Consistent histories | 1984 | Robert B. Griffiths | No | No | No | No | No | No | Yes | No | Yes |
| Transactional interpretation | 1986 | John G. Cramer | No | Yes | Yes | No | Yes[8] | No | No[12] | Yes | No |
| Objective collapse theories | 1986-1989 | Ghirardi–Rimini–Weber, Penrose interpretation | No | Yes | Yes | No | Yes | No | No | No | No |
| Relational interpretation | 1994 | Carlo Rovelli | No[56] | No | Agnostic[9] | No | Yes[10] | Intrinsic[11] | Yes[57] | No | No |
| QBism | 2010 | Christopher Fuchs, Ruediger Schack | No | No[16] | Agnostic[17] | No | Yes[18] | Intrinsic[19] | Yes | No | No |

*Figure 38: the various interpretation of quantum physics. Source: Wikipedia.*

Note that Niels Bohr's son, **Aage Niels Bohr** (1922-2009, Danish), was awarded the Nobel Prize in Physics in 1975 for his work on the structure of atom nucleus[36]!

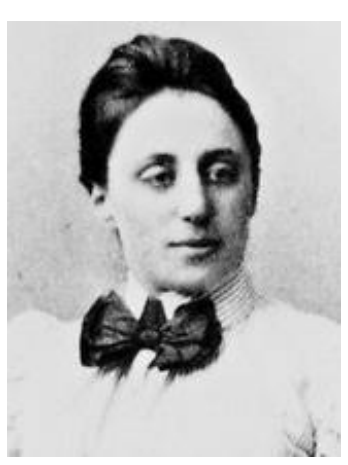

**Emmy Noether** (1882-1935, German) is the creator of the theorem that bears her name in 1915 at the University of Göttingen in Germany and which says that if a system has a continuous symmetry property, then there are corresponding quantities whose values are conserved in time[37]. At the origin of the field of abstract algebra, it is a foundation to Lagrangian mechanics, precursor of Hamilton's formalism. At that time, she could not teach at the University because this role was forbidden to women. Her theorem was only published in 1918 and she could not officially teach until 1919.

She did not receive a salary from the University until 1923. Her theorem links conservation principles and symmetries. It is one of the foundations of particle physics. Her work helped Albert Einstein to refine the foundations of the theory of general relativity he developed in 1915[38]. She died relatively young, at 53.

$$\frac{d}{dt}\left(\sum_{a}\frac{\delta L}{\delta\frac{dq_a}{dt}}\delta q_a\right)=0$$

*Figure 39: Emmy Noether's main equation.*

[36] See Quantum Model of the Atom by Helen Klus, 2017.

[37] See In her short life, mathematician Emmy Noether changed the face of physics Noether linked two important concepts in physics: conservation laws and symmetries by Emily Conover, 2018. She created a second important and more general theorem that is the basis of gauge fields theories in quantum fields theory.

[38] See Women in Science: How Emmy Noether rescued relativity, by Robert Lea, February 2019.

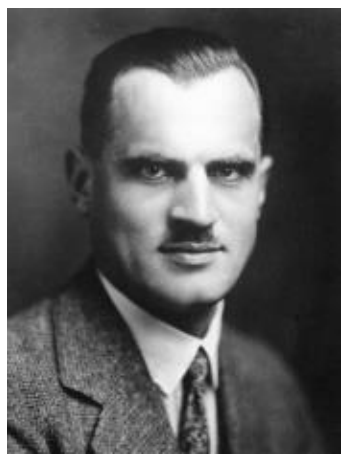

**Arthur Holly Compton** (1892-1962, American) was a physicist who got the 1927 Nobel Prize in physics for the discovery in 1922/1923 of the effect which demonstrates that photons can have momentum and behave as particles. His experiment makes a photon interact with a free electron around an atom, validating the photoelectric effect theories of Planck and Einstein. The Compton effect is a variant of this effect, applied to X and gamma rays which are high energy photons.

Compton scattering deals with the reception of an X or gamma photon which has an energy higher than that of the ejected electron. The X ray photon is slowed down and deflected with a lower energy and becomes a scattered photon. This is also called an elastic shock. The effect is used in X-ray radios. X-rays are emitted during electronic transitions between the atomic layers K, L and M (the first around the nucleus of the atom). The emission angles of the ejected electron and the re-emitted photon depend on the energy level of the incident photon.

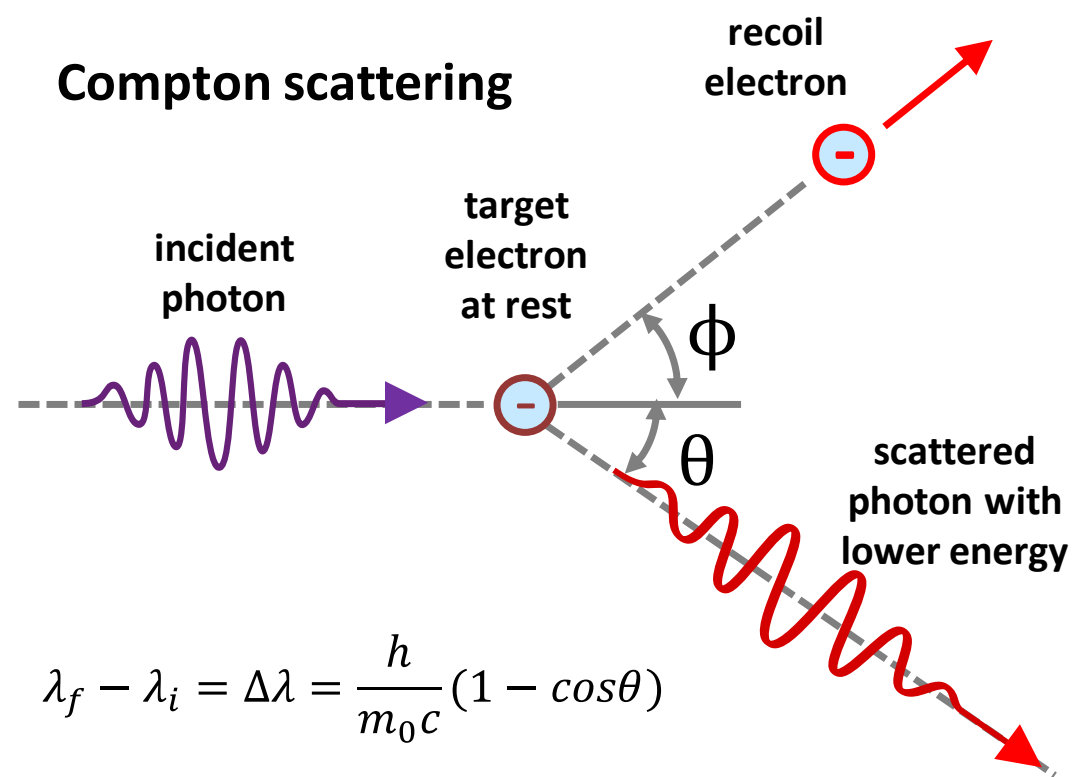

*Figure 40: Compton scattering phenomenon.*

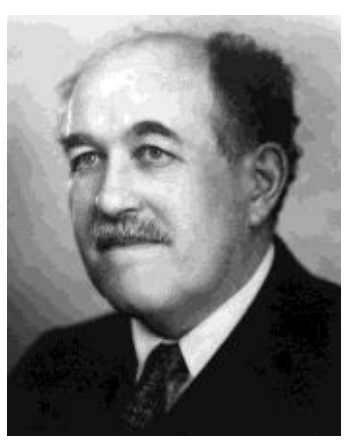

**Otto Stern** (1888-1969, German-American) and **Walther Gerlach** (1889-1979, German) respectively conceived in 1921 and together realized in 1922 in Frankfurt the famous Stern-Gerlach experiment which discovered the intrinsic angular momentum quantization in a magnetic field using a beam of electrically neutral silver atoms as shown in Figure 41 [39]. In the experiment, this momentum came from the 47th electron spin from heated silver atoms.

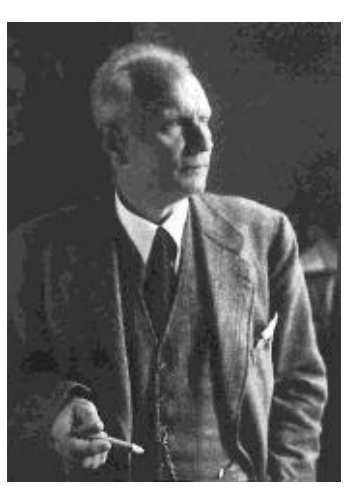

It did show that these atoms have a quantized angular dipole that deflects the beam in a given direction upward or downward. It later became known as particle spins. The experiment also did show that spin measurement along a given direction was incompatible with being done in another direction, corresponding to the notion of observables complementarity.

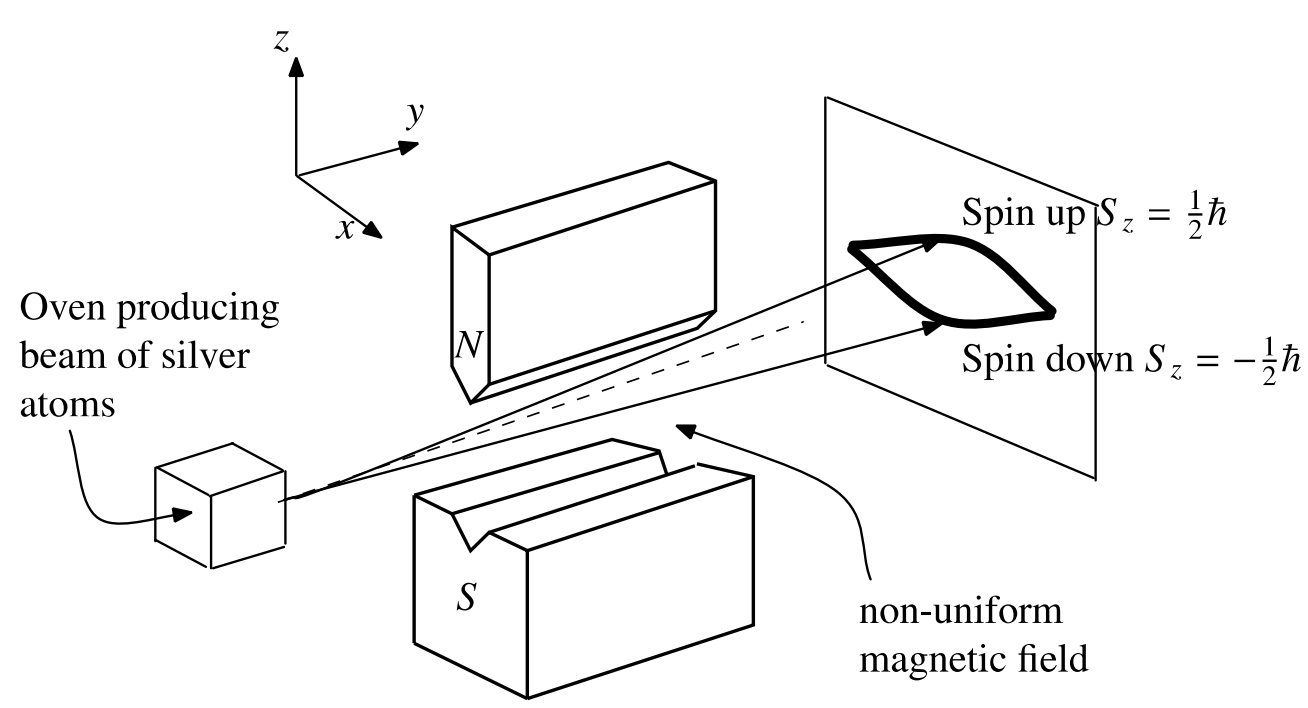

*Figure 41: the Stern-Gerlach experiment where an atomic stream of silver is deviated in two discrete directions by a magnetic field.*

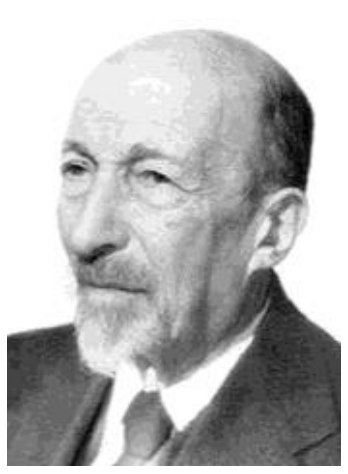

**Jacques Salomon Hadamard** (1865-1963, French) was a mathematician who gave his name to the Hadamard gate used in quantum computers and quantum algorithms. He had worked on complex numbers, differential geometry and partial differential equations, particularly during the 1920s. He also became interested in the creative process of mathematicians with studying the creative process of hundreds of colleagues.

---

[39] Illustration coming from: [Chapter 6, Particle Spin and the Stern-Gerlach Experiment](). See [Stern and Gerlach, how a bad cigar helped reorient atomic physics]() by Bretislav Friedrich and Dudley Herschbach, Physics Today, December 2003 (7 pages). The X, Y and Z components of the electron spin measured in the Stern-Gerlach experiment are complementary variables. Measuring one of the three variables prevents from doing so with the two others.

We owe him in particular the Hadamard transforms, square matrix operations with $2^n$ complex or integer values on each side. The quantum gate named after Hadamard is used in quantum computation to create a superposition of the states |0⟩ and |1⟩ with a transform of Hadamard of type H1 as described in Figure 42.

This superposition enables computing parallelism in quantum computing, in addition to the principle of entanglement which links the qubits together conditionally and is the real source of quantum exponential acceleration. Superposition is only responsible for a potential polynomial acceleration.

$$H_0=1$$

$$H_1=\frac{1}{\sqrt{2}}\begin{bmatrix}0 & 1\\ 1 & 0\end{bmatrix}$$

$$H_2=\frac{1}{2}\begin{bmatrix}1 & 0 & 0 & 0\\ 0 & 1 & 0 & 0\\ 0 & 0 & 0 & 1\\ 0 & 0 & 1 & 0\end{bmatrix}$$

$$H_3=\frac{1}{2^{3/2}}\begin{bmatrix}1&0&0&0&0&0&0&0\\0&1&0&0&0&0&0&0\\0&0&1&0&0&0&0&0\\0&0&0&1&0&0&0&0\\0&0&0&0&1&0&0&0\\0&0&0&0&0&1&0&0\\0&0&0&0&0&0&0&1\\0&0&0&0&0&0&1&0\end{bmatrix}$$

*Figure 42: Hadamard matrices of various dimensions.*

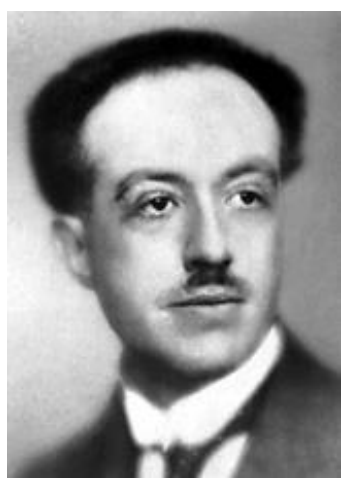

**Louis de Broglie** (1892-1987, French) was a mathematician and physicist who, in 1923 and 1924, extended the particle-waves duality, then only applied to photons, to massive particles, mainly electrons, and also atoms, protons and neutrons[40]. According to this principle, elementary particles behave like particles (with a position, a trajectory and possibly a mass) and like waves (potentially delocalized and scattering in all directions and generating interference) depending on the circumstances.

This is the case of electrons which have a mass and can interfere with each other. Louis de Broglie turned this duality into an equation: λp=h, where λ is a wavelength, p is a quantity of motion and h is Planck's constant.

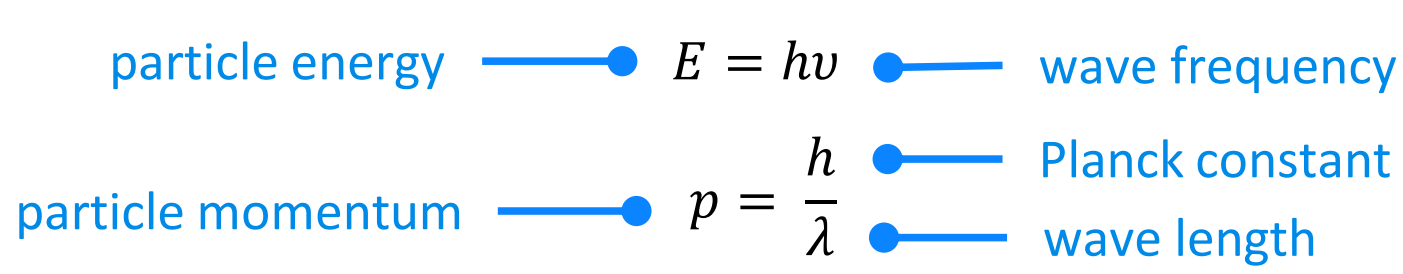

*Figure 43: De Broglie wave-particle equation with electrons.*

This earned him the Nobel Prize in Physics in 1929. He is the main French contributor to quantum physics during the inter-war period. The wave-particle duality of electrons was confirmed in 1927.

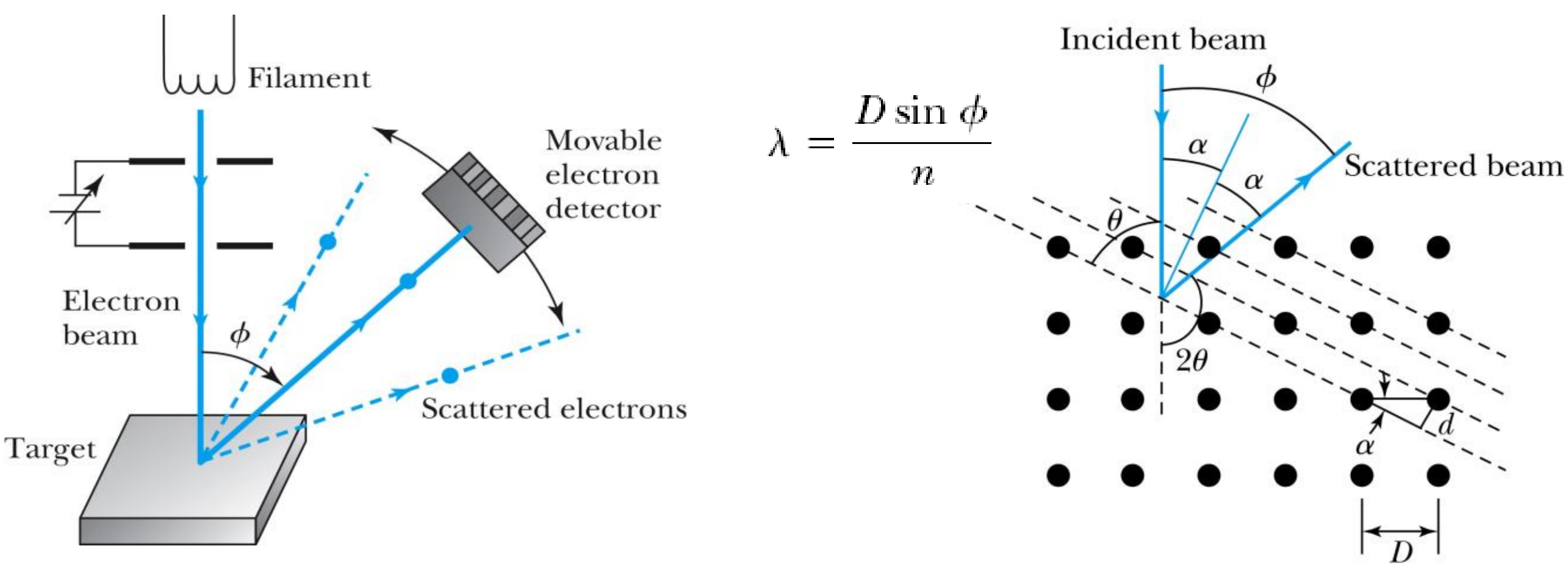

*Figure 44: electron wave-particle diffraction experiment. Source: Wave Properties of Matter and Quantum Mechanics I (48 slides).*

It was done as shown above in Figure 44 with a nickel crystal based diffraction experiment by **Clinton Davisson** (1881-1958) and **Lester Germer** (1896-1971) from the Bell Labs in the USA, who shared a Nobel Prize in Physics in 1937.

[40] Louis de Broglie's brother, Maurice de Broglie (1875-1960), was also a physicist. He had studied X-rays and spectrography. Both brothers were members of the Academy of Sciences in France.

**George Paget Thomson** (1892-1975) from the University of Aberdeen in Scotland did a similar experiment also in 1927. However, the Young double-slit experiment done with electrons was realized much later, in 1961, by **Claus Jönsson** (1939, German).

The confirmation of the wave-particle duality was then verified for neutrons much later in 1988 by **Roland Gähler** and **Anton Zeilinger**[41] and for atoms in 1991 by **Olivier Carnal** and **Jürgen Mlynek**, using double-slit diffraction and by **Mark Kasevich** and **Steven Chu**, who created the first cold atom interferometer using a light-beam splitter based on Raman transitions. It became the basis of atom interferometry used in quantum absolute gravimeters, using an equivalent of a Mach-Zehnder interferometer replacing light with so-called matter-wave made of atoms[42]. It is even verifiable with molecules of several atoms.

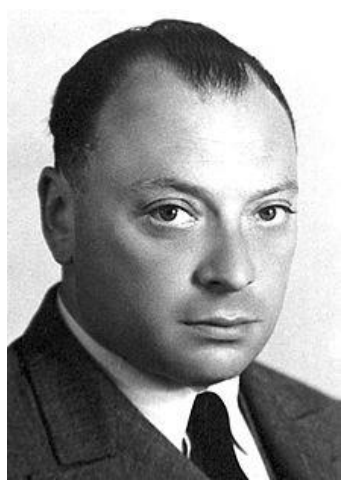

**Wolfgang Pauli** (1900-1958, Austrian/American) is at the origin of the principle of exclusion which bears his name elaborated in 1925 and according to which two electrons cannot have the same quantum state in an atom. He had an early role in the discovery of electron spin between 1925 and 1927, as well as the neutrino in 1930, the existence of which was only experimentally proven in 1956, and on works on quantum electrodynamics. He was awarded the Nobel Prize in Physics in 1945. The history of his discoveries is more complex than it seems.

He first discovered in 1924 the atom nucleus spin, used to explain the hyperfine structure of atomic spectra, i.e., the existence of very close spectral lines observed during their excitation. It cannot be explained by the quanta and energy levels of the electron layers in the atoms. He then introduced in 1925 a new degree of freedom for electrons that he did not qualify at first.

It adds to the first three parameters describing the state of an electron in an atom, aka quantum numbers. The first is the energy level of the electron in the atom (the layer where it is located), the second is the azimuthal quantum number (which defines the electron sub-layer) and the third is the magnetic quantum number (which makes it possible to distinguish the orbitals of the electron in the atom)[43]. This fourth degree of freedom was identified by **George Uhlenbeck** (1900-1988, Netherlands/USA) and **Samuel Goudsmit (**1902-1978, Netherlands/USA) as an electron spin[44] .

In 1925, Wolfgang Pauli also formulated the exclusion principle according to which electrons in the same system (an atom) cannot be simultaneously in the same quantum state, a principle that was later extended to all fermions, i.e. half-integer spin particles (electrons have a spin ½ but fermion atoms can have 3/2, 5/2, 7/2 and even 9/2 spins, like $^{40}K$).

The quantum state of an electron is defined with the four quantum numbers, or degrees of freedom, that we have just mentioned. An electron spin is described as a direction of magnetic polarization or as an angular rotation of the electron in one direction or the other, but it is only an image and not a physical representation[45]. Electron spins are used in silicon qubits that we cover later, starting page 292.

---

[41] See Single- and double-slit diffraction of neutrons by Anton Zeilinger et al, Review of Modern Physics, 1988 (7 pages).

[42] In this setup, the Mach-Zehnder beamsplitter is replaced by a series of three lasers pulses creating a superposition of two atomic energy states driving a diffraction effect, then a mirror effect and at last for a recombination of split wavepackets.

[43] The second and third electron quantum numbers were introduced by Arnold Sommerfeld (1868-1951, German). Among others, Wolfgang Pauli and Werner Heisenberg were his PhD students. The alpha constant or fine structure constant is also called the Sommerfeld constant per his work from 1916! See Electron spin and its history by Eugene D. Commins, May 2012 (28 pages).

[44] Georges Uhlenbeck and Samuel Goudsmit were students of Paul Ehrenfest (1880-1933, Austria/Netherlands). His laboratory had welcomed some illustrious future physicists such as Enrico Fermi, Robert Oppenheimer, Werner Heisenberg and Paul Dirac. Ehrenfest was a specialist in statistical physics. In particular, he contributed to the understanding of phase changes in matter.

[45] See How Electrons Spin by Charles T. Sebens, California Institute of Technology, July 2019 (27 pages) which provides a good background on electron spin's physical interpretations, particularly with regards to electron's size. Pauli did demonstrate in 1924 that if the electron spin corresponded to an angular momentum, the electron's rotation would exceed the speed of light.

137 is a number that played a weird role in Pauli's life. It turns out that 1/137 is a value that roughly corresponds to the fine-structure constant, a ratio that is found in many places in quantum physics and compares data of the same dimension[46]. It is for example the ratio between the velocity of an electron in the lower layer of a hydrogen atom and the speed of light or the probability of emission of the absorption of a photon for an electron (complete list). "137" is a sort of "42" of quantum physics. Wolfgang Pauli died after some pancreatic cancer surgery, while his hospital room number was 137!

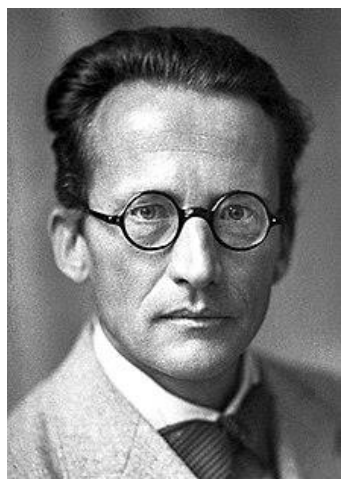

**Erwin Schrödinger** (1887-1961, Austrian) is a physicist who was awarded the Nobel Prize in 1933 for the creation of his famous wave function in 1926, *aka* Schrödinger equation, which describes the evolution in time and space of the quantum state of a massive quantum particle and the probabilities of finding the quantum at a given place and time. Schrödinger's equation is a variant of the Newtonian mechanics equations that define the total energy of an object as the sum of its kinetic energy and its potential energy. We describe this equation in detail in a dedicated section page 97.

Erwin Schrödinger also created his famous alive and dead cat in a box thought experiment[47]. In the scenario, an opaque box contains a vial of poison, the opening of which is caused by the disintegration of a radioactive radium atom generating alpha particles ("alpha decay"), made of two protons and two neutrons, that are detected by a Geiger counter. Since radium has a 50/50 chance of disintegrating at its mid-life, the cat has a 50/50 chance of being alive and dead, at deadline. When opened, it is either alive or dead. As long as the door is not opened, the cat is said to be superposed in the alive and dead states and entangled with the radium atom state. This story has been repeated ad-nauseam since 1935. But his thought experiment was created to show the absurdity of the measurement postulate, the wave function collapse and Born's rule. Unfortunately, the contrary has been memorized.

The caveat is that a cat can't be superposed in two states because it is a macroscopic object of a size well beyond the quantum/classical limit. It's either alive or dead, never both. These are exclusive states. On top of that, the radium atom disintegration as well as the cat's death are both irreversible processes. They can't be implemented as linear superpositions of waves. When the cat is dead, he's not in a superposition. He's just plain dead.

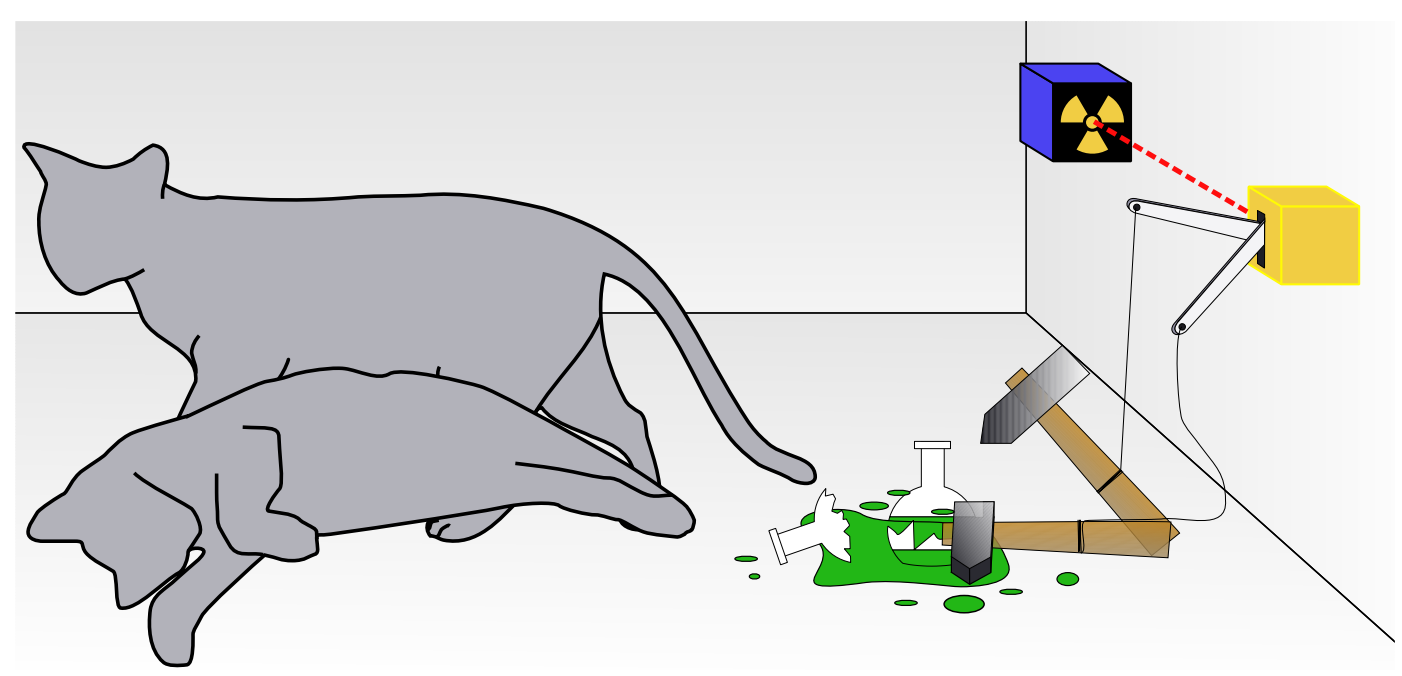

*Figure 45: the infamous Schrodinger's cat thought experiment.*

We can consider that the cat's death is provoked by a not yet read measurement when the box is closed, corresponding to a non-selective measurement as described page 190. The cat state uncertainty is a classical one, not a quantum one. The cat is in a maximally "mixed state" where the uncertainty of its death is classical, not in a "pure state" where it would be quantum (we define these notions starting page 150). If you used a webcam inside the box and made sure it didn't influence the radium half-life period, you could track the cat state all along, from alive to dead or alive to alive, which are the only two possible paths and observe the absence of superposition.

[46] The fine-structure constant was measured with a precision of $2.0\times10^{-10}$ in 2020 using cold atoms interferometry. See Determination of the fine-structure constant with an accuracy of 81 parts per trillion by Léo Morel, 2020 (36 pages).

[47] The Cat Thought Experiment was published in a series of three papers in 1935, shortly after the publication of the EPR paradox paper by Einstein, Podolsky and Rosen. See The Present Status of Quantum Mechanics by Erwin Schrödinger, 1935 (26 pages). The history of the cat occupies only nine lines in this long document which deals with superposition, measurement and entanglement. That's even where Schrödinger coined the term entanglement in the first chapter "*The Lifting of Entanglement. The Result Depends on the Will of the Experimenter*". Schrödinger translated himself the German word Verschränkung into entanglement. The cat that appears only three times in all and for all is therefore anecdotal but that is what everyone has remembered. Which is quite normal: the rest is much less easy to apprehend!

This thought experiment was intended to highlight two things. First, that superposition and entanglement only applied to the infinitely small and not to macroscopic objects. History retained the principle of superposition and not this difference between the microscopic and macroscopic worlds. Second, that there was and still is an uncertain limit between the quantum and classical worlds. Schrodinger's thought experiment also dealt with the entanglement between the radium atom and the cat. Could this entanglement work with a macro-object[48]? The paper containing this thought experiment was about entanglement and that was forgotten. Also, this paper's publication was the one generating the publication of the EPR paradox piece by Einstein et al. We should retain Schrödinger's wave function and the notion of states superposition which only makes sense on a microscopic scale. Let's forget that poor cat!

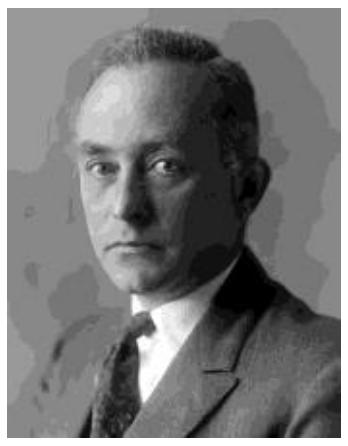

**Max Born** (1892-1970, German) is a physicist and mathematician who developed the mathematical representation of quantum in a matrix form. We owe him in 1926 the statistical explanation of the probability of finding an electron in a given energy state from its wave function, elaborated by Schrödinger the same year. This principle is applied to qubits, where the sum of the square of the probabilities of the two states of the qubit is equal to 1, given the probabilities are complex numbers.

In 1925, he created the non-commutativity relation of two conjugate quantities, one being the Fourier transform of the other (the commutator $[X,P]=XP-PX=i\hbar I$, where X is a position and P a momentum and I, the identity). It led to the indeterminacy principle creation. He also created the first version of the adiabatic theorem with Vladimir Fock in 1928. He got the Nobel prize in physics in 1954. Fun fact, the British singer Olivia Newton-John is his grand-daughter[49].

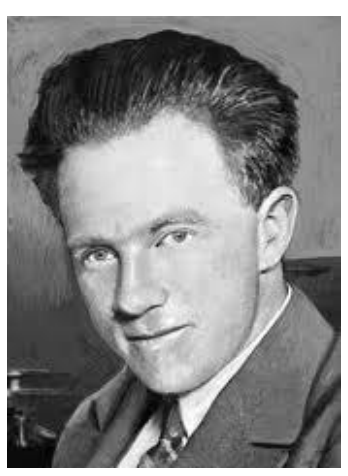

**Werner Heisenberg** (1901-1976, German) is a physicist, Nobel Prize in Physics in 1932, to whom we owe in 1927 the creation of the famous principle of uncertainty, or rather indeterminacy, according to which one cannot accurately measure both the position and the velocity of an elementary particle, or, more generally, two arbitrary unrelated quantities. He is at the origin, with Max Born and Pascual Jordan in 1925, of the quantum matrix formalism describing physical quantities.

The indeterminacy principle is a consequence of this formalism. It was described mathematically in a simplified manner in 1928 by **Earle Hesse Kennard** (1885-1968, American) in the famous equation in Figure 46, where the product of the standard deviation of position and velocity is greater than half the Dirac constant.

$$\Delta x \, \Delta p \geqslant \frac{\hbar}{2}$$

*Figure 46: Heisenberg inequality, created by Earle Hesse Kennard.*

This principle can be used to improve the accuracy of a measurement of any quantity by lowering the accuracy of another quantity characterizing a quantum[50] . These quantities can be for example an energy level, a position, a wavelength, or a speed.

One consequence of Heisenberg's principle is that all particles in the Universe are in permanent motion. If they were stable, we would know their position (fixed) and their velocity (zero), violating the indeterminacy principle.

---

[48] You can apply this thought experiment to the baking of the half-cooked chocolate. As long as you don't take it out of the oven after the mandatory baking of 9 minutes, but with an oven with an unknown power, you don't know if it is well done or not, and run it through the middle before you take it out. It is in a state of superposition between undercooked, well done and overcooked. On the other hand, if it is overcooked, it will be difficult to go back, like Schrödinger's half-dead cat in case he died. Overcooking as well as the death of the cat are irreversible. It is therefore not a true superposition of quantum states. But here, I have no clue about how the oven and the half-baked chocolate are entangled. It's about statistical physics and thermodynamics, not quantum physics. Even through the oven is a black body! Cheers!

[49] See Olivia Newton-John's grandfather Max Born was friend of Albert Einstein by Matthew Alice, 1995.

[50] This measurement technique is used in "quantum squeezing" which is integrated in the latest version of LIGO for the measurement of gravitational waves: NIST Team Supersizes 'Quantum Squeezing' to Measure Ultra Small Motion, 2019.

Another consequence is that a perfect vacuum could not exist because the value and evolution of the magnetic and gravitational fields that pass through it would be stable, violating once again Heisenberg's indeterminacy. This explains the astonishing vacuum quantum fluctuations we discover a little further starting in page 134. The no-cloning theorem of a qubit state also derives from the principle of indeterminacy.

For some, this indeterminacy principle is a simplified interpretation of the corpuscular nature of matter. It leads to the question of the position and velocity of an electron, when it has no precise position. According to the Copenhagen interpretation of quantum mechanics, we shouldn't try to determine where the electron is located. Try to apply the concepts of classical mechanics to electrons is vain.

In practice, quantum particles are not classical physical particles and therefore their velocity and position cannot be measured. They can only be described by their (Schrödinger) wave function and position probabilities. More generally, in the infinitely small, the measurement device influences the measured quantity. One example illustrates this phenomenon at the macroscopic level: if you illuminate an insect with sunlight and a magnifying glass to better observe it, you may burn it! The same happens with a photon that is used to detect an electron, in the Heisenberg microscope thought experiment, as shown in Figure 47. It will change the speed and position of the electron.

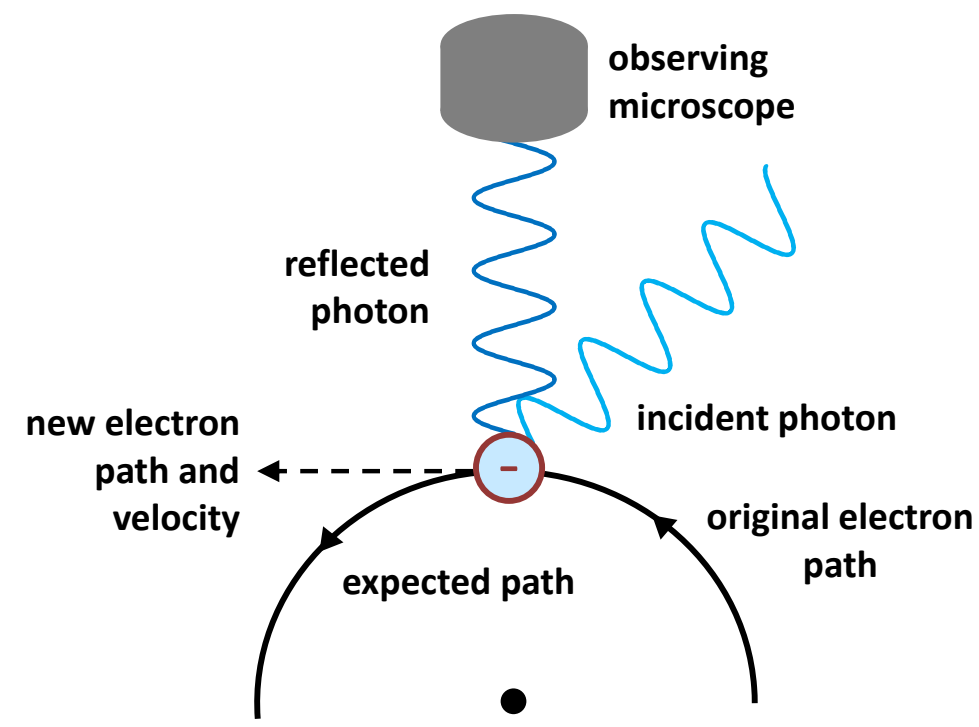

*Figure 47: Heisenberg microscope thought experiment. Source.*

Finally, like many of the colleagues of his time, Werner Heisenberg was interested in the links between science, quantum mechanics and philosophy, and as early as 1919. He was assistant to Niels Bohr between 1924 and 1927, before leaving for the University of Leipzig. He also had Max Born as a professor!

During World War II, he was asked with other German scientists to work on the Reich's atomic bomb project. Later revelations did show that he was not very active on this project and did not believe it was an achievable goal.

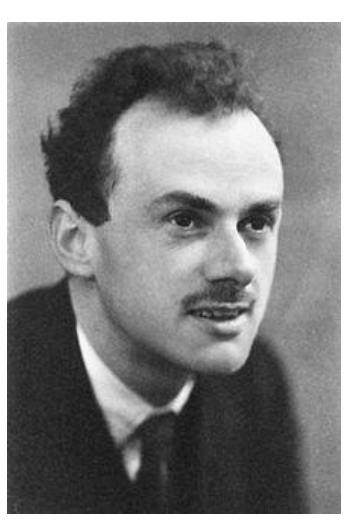

**Paul Dirac** (1902-1984, English) is a mathematician and physicist among the founders of 20th century quantum physics. He is credited with the 1928 electron spin equation, which is one of the foundations of relativistic quantum physics (*below*). His equation is a kind of variant of Schrödinger's equation for free relativistic particles, fermions (electrons, protons, neutrons, quarks, neutrinos) which are half-integer spin particles. Relativistic particles are those moving at a speed close to the speed of light, which contains electrons if lower shells of heavy atoms.

In Dirac's equation, the wave function of the electron ψ includes four components of complex numbers that integrate time and space. Dirac's equation enabled him to predict the existence of a particle that was later be called the positron, an opposite of the electron with a positive charge[51].

$$\left(\beta mc^2 + c\sum_{n=1}^{3}\alpha_n p_n\right)\Psi(x,t) = i\hbar\frac{\partial\Psi(x,t)}{\partial x}$$

*Figure 48: Dirac's relativistic wave-function equation.*

[51] Positrons were discovered experimentally by Carl Anderson in 1932. He was awarded the Nobel Prize in Physics in 1936.

Dirac formalized the quantization of the free electromagnetic field in 1927. He also introduced in 1939 the bra-ket notation, known as Dirac's notation, which simplified the notation and manipulation of quantum states and operators in linear algebra (example: $\langle \phi | \psi \rangle$). The Dirac constant also named reduced Planck constant is the Planck constant $h$ divided by $2\pi$, also called "h-bar" for its italicized strikethrough h symbol: $\hbar$. This Dirac constant is used in the Schrödinger wave function.

Paul Dirac was awarded the Nobel Prize in Physics in 1933, at the age of 31. The Nobel Prizes of the early 20th century were frequently awarded to young scientists, which seems to be out of fashion since then! The youngest Nobel Prize in Physics was awarded to Lawrence Bragg, who won it at the age of 25 in 1915 for his discovery of X-ray refraction at the age of 22[52]. In which case do we have to deal with relativistic particles, in particular with electrons? It is generally considered that an electron becomes relativistic when the total of its mass and kinetic energy is at least twice the rest mass. This ratio corresponds to the Lorentz factor. It represents a speed of at least 86% of the speed of light. But relativistic phenomena may occur before that speed is reached. In Newtonian equivalent, the speed of an electron around the nucleus of a hydrogen atom is about c/137. With electrons from heavy atoms inner shells, this velocity can exceed c/2.

This is the case for electrons of the first layer of the gold atom, which move at 85% of the speed of light. This affects the position of relativistic electrons in the low orbits of heavy atoms such as lanthanides, which belong to the rare earths. The Bohr radius that defines the average orbital of an electron decreases inversely proportional to the apparent mass of the electron. Because the electron's apparent mass increases, this Bohr radius is smaller for relativistic electrons. This modifies the structure of the electron orbitals of heavy atoms and the transition energy levels between orbitals that absorb or emit photons.

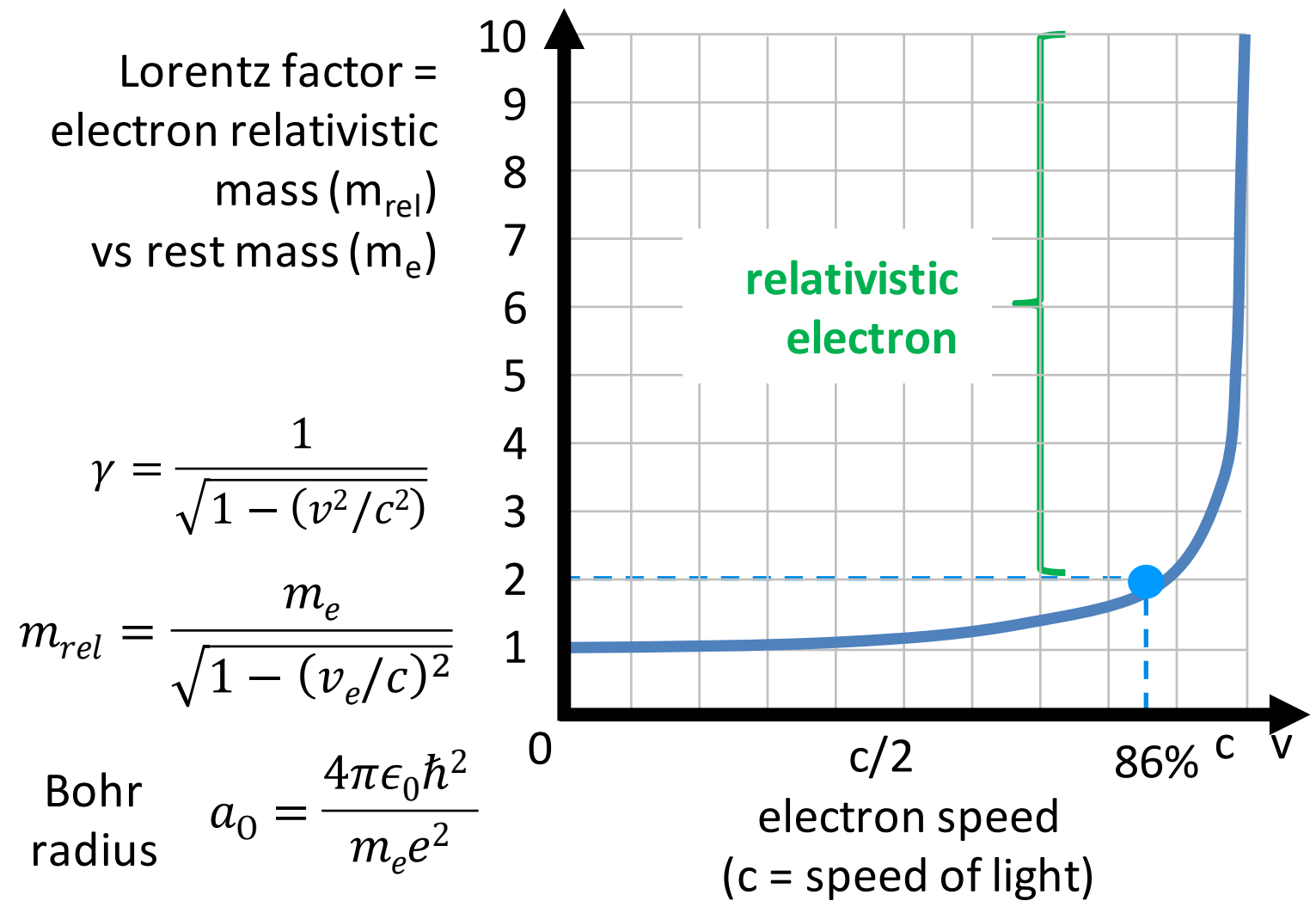

*Figure 49: relativistic electrons and Lorentz factor.*

This explains the color of gold and silver, due to relativistic modification of orbits of electron layers between which transitions occur due to the absorption of photons. Blue is absorbed in the case of gold, explaining its yellow color. Without the relativistic effect, gold would be white. This has a lot of implications in the chemistry of these materials and with their crystal organization[53]. This quantum relativistic effect also explains why mercury is liquid at room temperature[54]. All this gives rise to a field of chemistry called relativistic quantum chemistry[55].

---

[52] Paul Dirac was distinguished by his shyness and parsimonious oral expression in meetings or during meals. So much so that his Cambridge colleagues had defined the "dirac" unit as the most concise way to express himself in a meeting, namely, at the rate of a single word per hour. His behavior was equivalent at the Solvay Congresses he attended, notably that of 1927. However, he must have broken a record in his speech accepting his Nobel Prize at the end of 1933. It is still six pages long! Half, however, of the 12 pages of the speech of Erwin Schrödinger, also winner of the Nobel Prize in Physics that year. Another anecdote: Dirac was married to one of the sisters of Eugene Wigner, Nobel prize in physics in 1963 and famous for his function and also his “friend” paradox.

[53] See more examples in Relativistic Effects in Chemistry More CommonThan You Thought by Pekka Pyykko, 2012 (24 pages).

[54] See Why is mercury liquid?Or, why do relativistic effects not get into chemistry textbooks? by Lars J. Norrby, 2018 (4 pages).

[55] See Relativistic quantum chemistry by Trond Saue, 2019 (110 slides) and An introduction to Relativistic Quantum Chemistry by Lucas Visscher (107 slides). The mathematical formalism of relativistic quantum chemistry is well documented in the voluminous Introduction to Relativistic Quantum Chemistry by Kenneth Dyall and Knut Faegri, 2007 (545 pages).

It also explains why the size of atoms is not proportional to their number of protons and electrons[56] .

Particles also become relativistic in **particle accelerators** such as the CERN LHC near Geneva (the largest in the world), the ESRF in Grenoble (European Synchrotron Radiation Facility, specialized in the generation of "hard", very high-frequency X-rays) or the SOLEIL light synchrotron located in Saint-Aubin near Saclay just next to the CEA, also in France, or its equivalent from PSI in Switzerland.

The SOLEIL synchrotron uses electrons accelerated to a relativistic speed and inverters that generate beams of light 10,000 times denser than sunlight[57] . Equivalent instruments exist such as the Advanced Photon Source at the Argonne National Laboratory from the US Department of Energy near Chicago.

**Free Electron Lasers** (**FEL**) exploit relativistic electron sources. These are lasers generating coherent light (spatially and temporally, the emitted photons have the same frequency, phase and in that case, also polarization) and exploit relativistic electron sources from synchrotrons. The interaction between these electrons and a strong alternating magnetic field makes it possible to generate coherent light in electromagnetic frequency ranges from infrared to X-rays, through visible light and ultraviolet. The FEL are used to explore all sorts of matter, particularly in biomedical research like with X-rays crystallography.

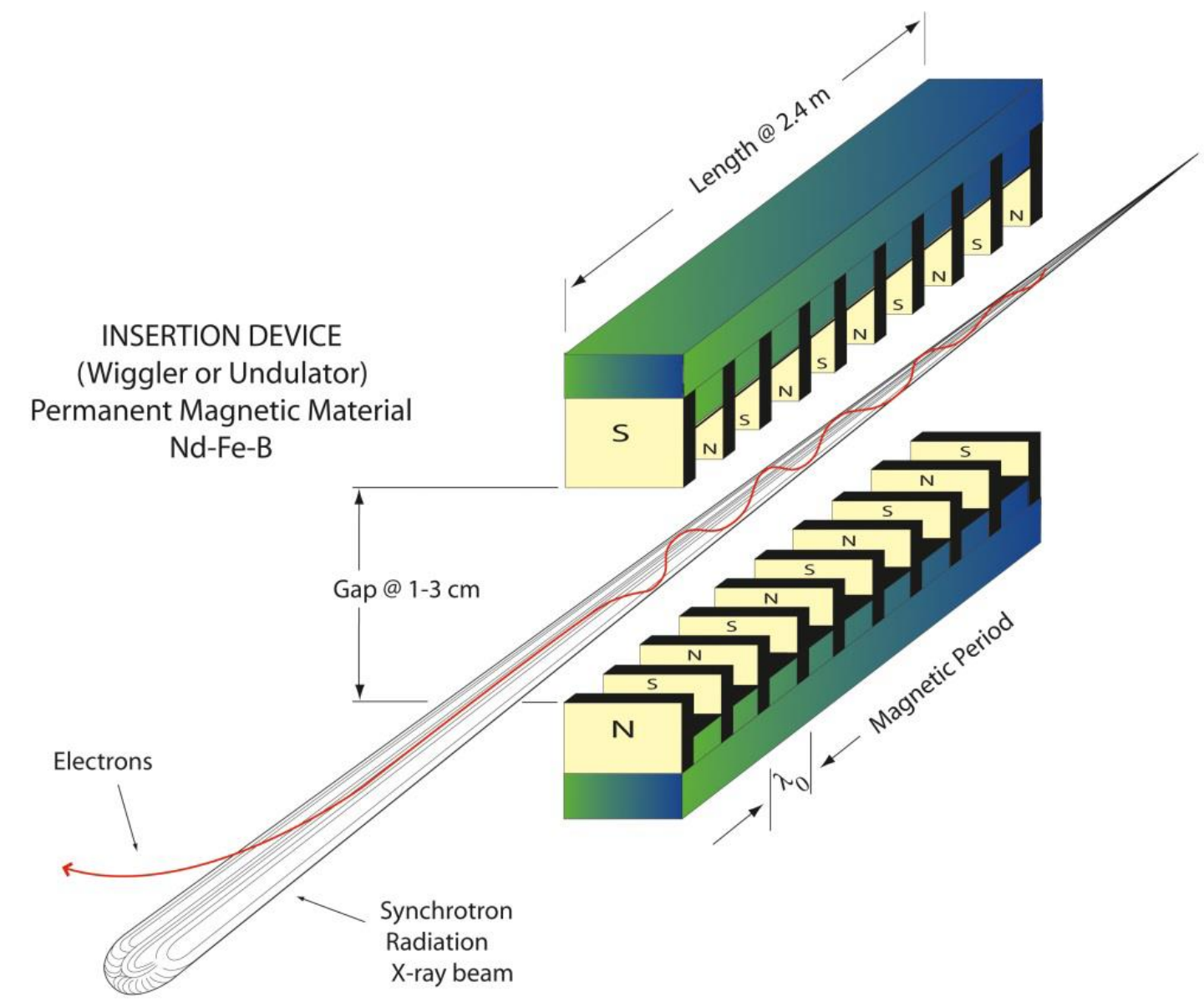

*Figure 50: free-electron laser. Source: X-ray diffraction: the basics by Alan Goldman (31 slides).*

Finally, relativistic particles can be found in **astrophysics** and, for example, in cosmic ray sources as well as in relativistic plasma jets produced at the center of galaxies and quasars[58].

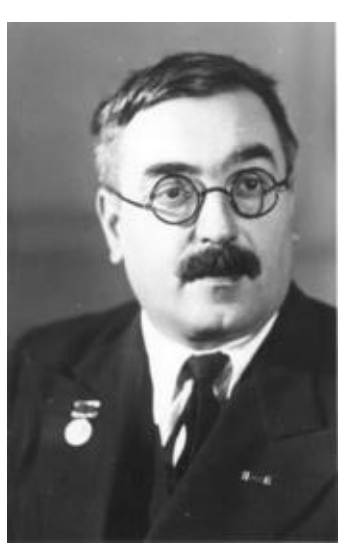

**Vladimir Fock** (1898-1974, Russian) was a theoretician physicist who worked on quantum physics, the theory of gravitation and theoretical optics. We own him the Fock space, representation and state, used in quantum photonics to represent the state of bosons many-body systems having the same quantum state. He co-created the Klein-Gordon equation in 1926, the relativist version of Schrödinger's equation for zero spin massive particles, the adiabatic theorem with Max Born in 1928 and the Hartree–Fock quantum simulation method in 1930. He also worked on quantum electrodynamics and quantum foundations.

[56] See this periodic table of elements with an indication of the sizes of the atoms.

[57] See the conference Electrons relativists as light sources by Marie-Emmanuelle Couprie, Synchrotron Soleil, 2011 (1h25). Electrons circulate in the synchrotron at a speed close to that of light. SOLEIL powers more than 25 analytical instruments covering the spectrum from infrared to X-rays, with numerous applications in precision microscopy, including a microscopy using very well collimated and polarized white light. These instruments can be used to analyze the three-dimensional structure of organic molecules such as complex proteins, such as the glycoproteins that surround viruses. This even allows one to study how these proteins combine with those of the attacked cells, or ribosomes, which are used to produce the proteins in the cells, are also analyzed.

[58] Dirac's equation is linked to the **Klein-Gordon equation (**1926) which applies to bosons such as elementary gluon particles and pions, particles having integer or zero spin. Relativistic quantum mechanics is a broad field of physics, used in particular in elementary particles physics. I have not yet found any use cases of this branch of physics in current quantum technologies. See the main foundations of relativistic quantum mechanics in Relativistic Quantum Mechanics by David J. Miller, University of Glasgow, 2008 (116 slides).

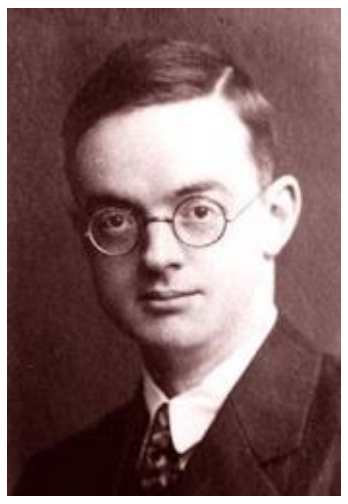

**Pascual Jordan** (1902-1980, German) was a physicist who collaborated with Max Born and Werner Heisenberg and contributed to laying the mathematical foundations of quantum mechanics, especially in matrix computation. Like Philipp Lenard, he was somewhat forgotten because of his membership in the Nazi Party during the 1930s, although he was rehabilitated after the Second World War thanks to the help of Wolfgang Pauli. He became interested in the philosophical notion of free will.

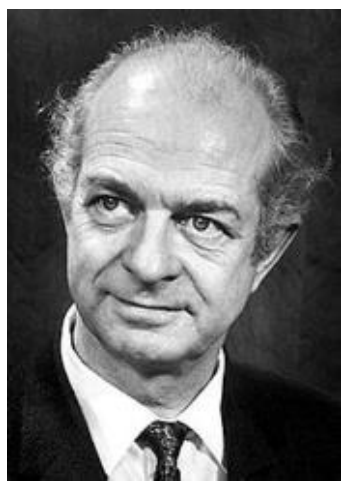

**Linus Pauling** (1901-1994, American) was a biochemist known to have co-founded the scientific fields of quantum chemistry and molecular biology. He had the opportunity to meet in Europe the founders of quantum physics like Erwin Schrödinger and Niels Bohr in 1926-1927. He described chemical bonds over a period between 1928 and 1932 and in particular the hybridization of orbitals which explains the geometry of molecules. He published "The Nature of the Chemical Bond" in 1939.

He was awarded the Nobel Prize in Chemistry in 1954 and the Nobel Peace Prize in 1962 for his political activism in favor of nuclear disarmament. He is considered to be at the origin of computational chemistry, which makes it possible to numerically simulate the structure of molecules and which we discuss in the section on quantum applications in health on page 693.

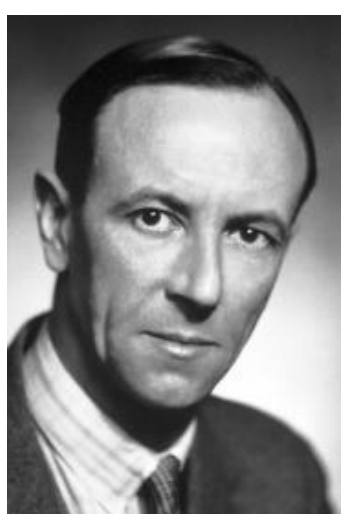

**James Chadwick** (1891-1974) is an English physicist who was responsible for the discovery of neutrons in 1932, which earned him the Nobel Prize in Physics in 1935. This discovery was late compared to quantum physics and the discovery of electrons. Nuclear physics has indeed progressed in parallel with quantum physics, which was mainly concerned with the interactions between electrons and photons. Before the discovery of neutrons, scientists thought that the nucleus of atoms contained protons and electrons.

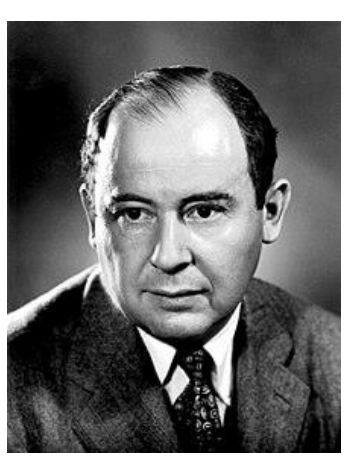

**John Von Neumann** (1903-1957, Hungarian, then American) was a polymath and an extremely prolific mathematician. He participated in the creation of the mathematical foundations of quantum mechanics, notably in the "Mathematical Foundations of Quantum Mechanics" published in 1932. He transposed the main principles of quantum mechanics into models and equations of linear algebra. He devised the key mathematical principles behind quantum measurement models.

This deals, for example, with the representation of quantum states as a position in a Hilbert space, the observables which are projections into Hilbert spaces and the indeterminacy principle which can be explained by the non-commutativity of measurement operators. These principles are also named Birkhoff-von Neumann *quantum logic*, in connection with their seminal paper published in 1936[59].

Von Neumann also affirmed that the introduction of hidden variables to incorporate determinism was a lost cause because it would contradict other (verified) predictions of quantum physics. Three years before Einstein/Podolsky/Rosen's EPR paper!

We owe him the creation of the notion of entropy (by Von Neumann), in 1932, which is associated with the notions of operators and density matrices that he created in 1927 and which describe the state of a multi-partite quantum system. He participated in the Manhattan Project in the USA.

[59] See The Logic of Quantum Mechanics by Garrett Birkhoff and John Von Neumann, 1936 (22 pages).

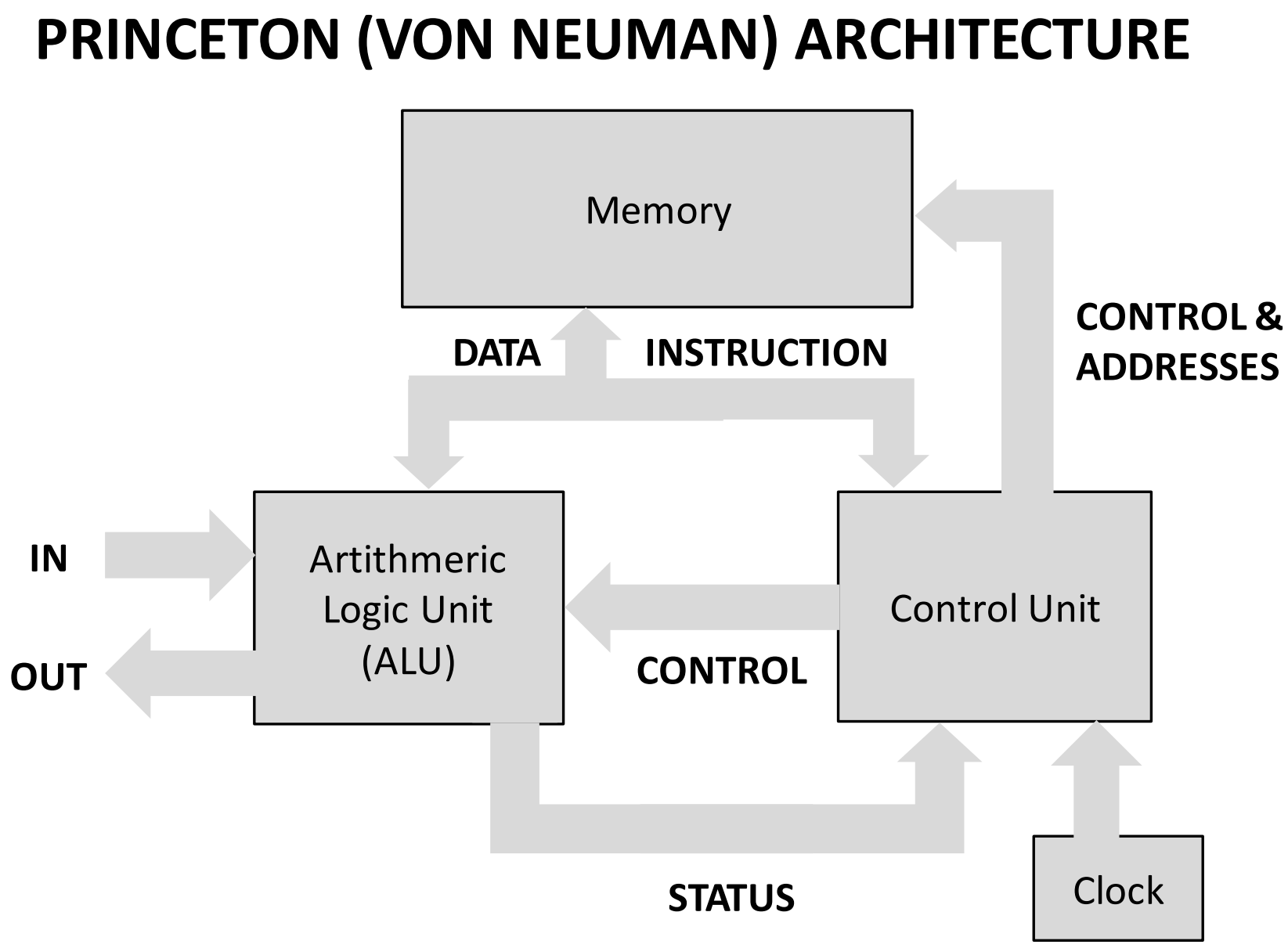

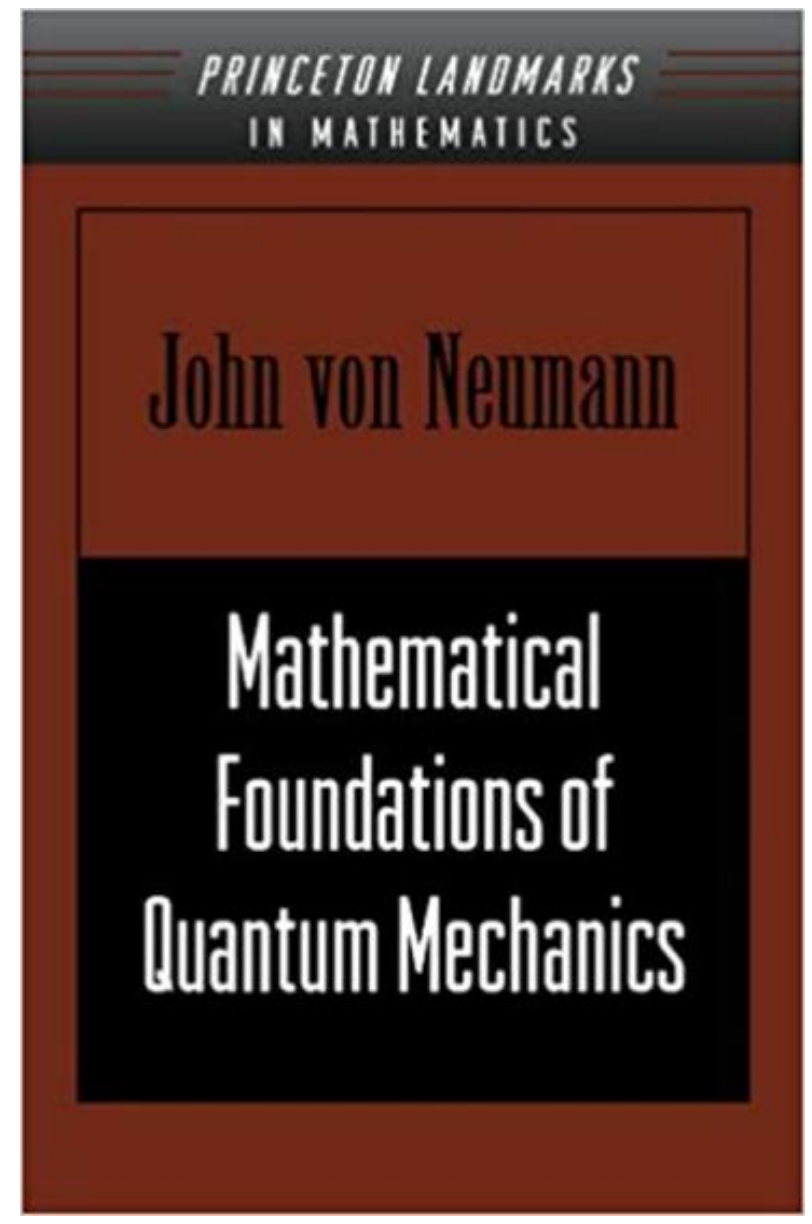

*Figure 51: the Von Neuman Princeton architecture which still defines classical computing.*

He modelled explosions and lenses for compressing plutonium in A-bombs. He is also responsible for the basic concepts in game theory and classical computers that are still in use. Thus, almost all computers use a Von Neumann architecture with memory, registers, control unit, computing unit, inputs and outputs. What a contribution!

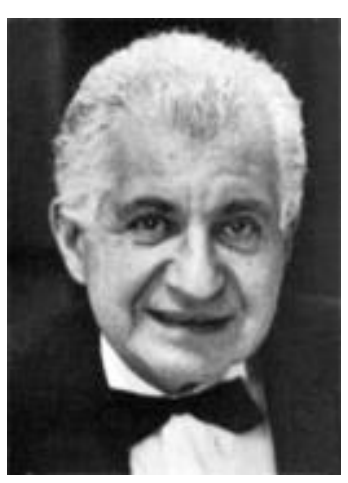

**Boris Podolsky** (1896-1966, Russian then American) wrote the EPR paradox paper with Albert Einstein and Nathan Rosen in 1935 on quantum entanglement and questions of non-locality of the properties of entangled quanta. He was a specialist in electrodynamics which deals with the analysis of electric and electromagnetic fields. He emigrated to the USA and, according to Russian archives, was a post-war KGB spy and informer of the USSR on the American atomic program between 1942-1943. His code name was... " Quantum".

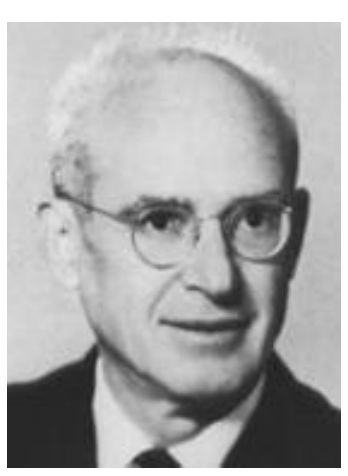

**Nathan Rosen** (1909-1995, American then Israeli) is the third EPR paradox author when working as an assistant to Albert Einstein in Princeton. After moving to Israel in 1953, he created the Institute of Physics at Technion University in Haifa. He was mainly working on astrophysics and relativity theory. He devised the concept of wormholes, a theoretical link between different points in space and time. He also thought neutrons were built out of a proton coupled to an electron.

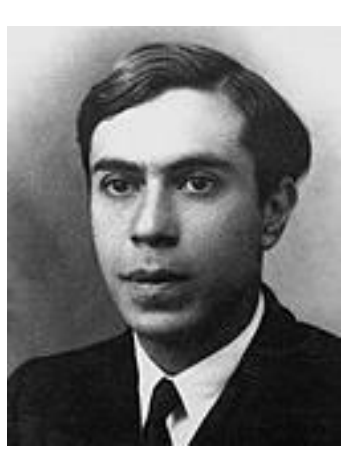

**Ettore Majorana** (1906-circa 1938, Italian) imagined the existence of a fermion in 1937 based on Dirac's equations, an elementary particle that would be its own anti-particle. The Majorana fermion naming is also abusively applied in condensed matter physics to quasi-particles having similar properties. Their existence was discovered in 2012 and verified in 2016 and then in 2018, even if it is still disputed by many physicists and two related 2018 papers had to be retracted in 2021.

These Majorana quasi-particles (or “Majorana Zero Modes”) could make it possible to design universal quantum computers called topological computers that can handle very efficient error correction codes requiring a small number of physical qubits. This is the exploration path chosen by Microsoft after the work of Michael Freedman and Alexei Kitaev in the late 1990s. Ettore Majorana is said to have committed suicide after a depression, because he could hardly stand the pressure of his genius! But his disappearance remains enigmatic because his body has never been found!

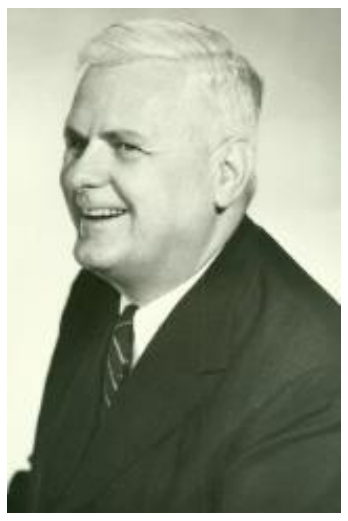

**Alonzo Church** (1903-1995, American) was a mathematician who was a key contributor to the foundations of theoretical computer science and on the notion of computability. Among other things, he created the lambda calculus in 1936, a universal abstract programming language which inspired the creation of LISP. He also created the so-called Church-Turing thesis. For this last one, any automatic calculation can be carried out with a Turing machine. Church and Turing also proved an equivalence between being λ-computable and Turing computable.

Many variations of the Church-Turing thesis were elaborated after them to extend the broad field of complexity theories. For example, the extended Church-Turing thesis states that the computation time of a problem is equivalent at worst to a polynomial depending on the size of the problem. It is not demonstrable.

What about the others, known, unknown or less famous from the 1927 Solvay Congress? Two participants deserve to be mentioned who had some connections with quantum physics.

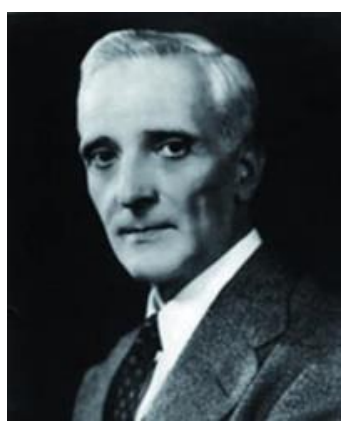

**Léon Brillouin** (1889-1969, Franco-American) who is less known in France because of his expatriation to the USA during World War II contributed to advances in quantum physics between the two World Wars. In particular, he brought quantum mechanics closer to crystallography. He especially discovered the phenomena of diffraction of waves traversing crystals, called Brillouin scattering.

And then, finally, **Hendrik Anthony Kramers** (1894-1952, Dutch) who assisted Niels Bohr in the creation of quantum theory. Many of the participants were not quantum physics scientists. They were invited because the Belgium organizers tried to have a stable proportion of Belgians, French, Germans and English participants. Were there, for example, **Émile Henriot** and **Marie Curie** who were focused on radioactivity, **Paul Langevin** (with whom Marie Curie had had an affair in 1910, after the accidental death of her husband Pierre Curie in 1906), as well as a good number of chemists.

What was striking during this prolific period were the way the social network of physicists worked, without smartphones and the Internet. They had many encounters, cross-University tenures, meetings, letter exchanges and conferences. It was slow according to today's references, but the results were still astounding.

At last, here's a simple chart reminding us how young the founders of quantum physics were when they published their seminal work in the key years from 1900 to 1935. Back then, scientific research didn't work the same way. They also were frequently awarded Nobel prizes at less than 40! Nowadays, most of the times, you have to wait until you are at least 50 if not 70.

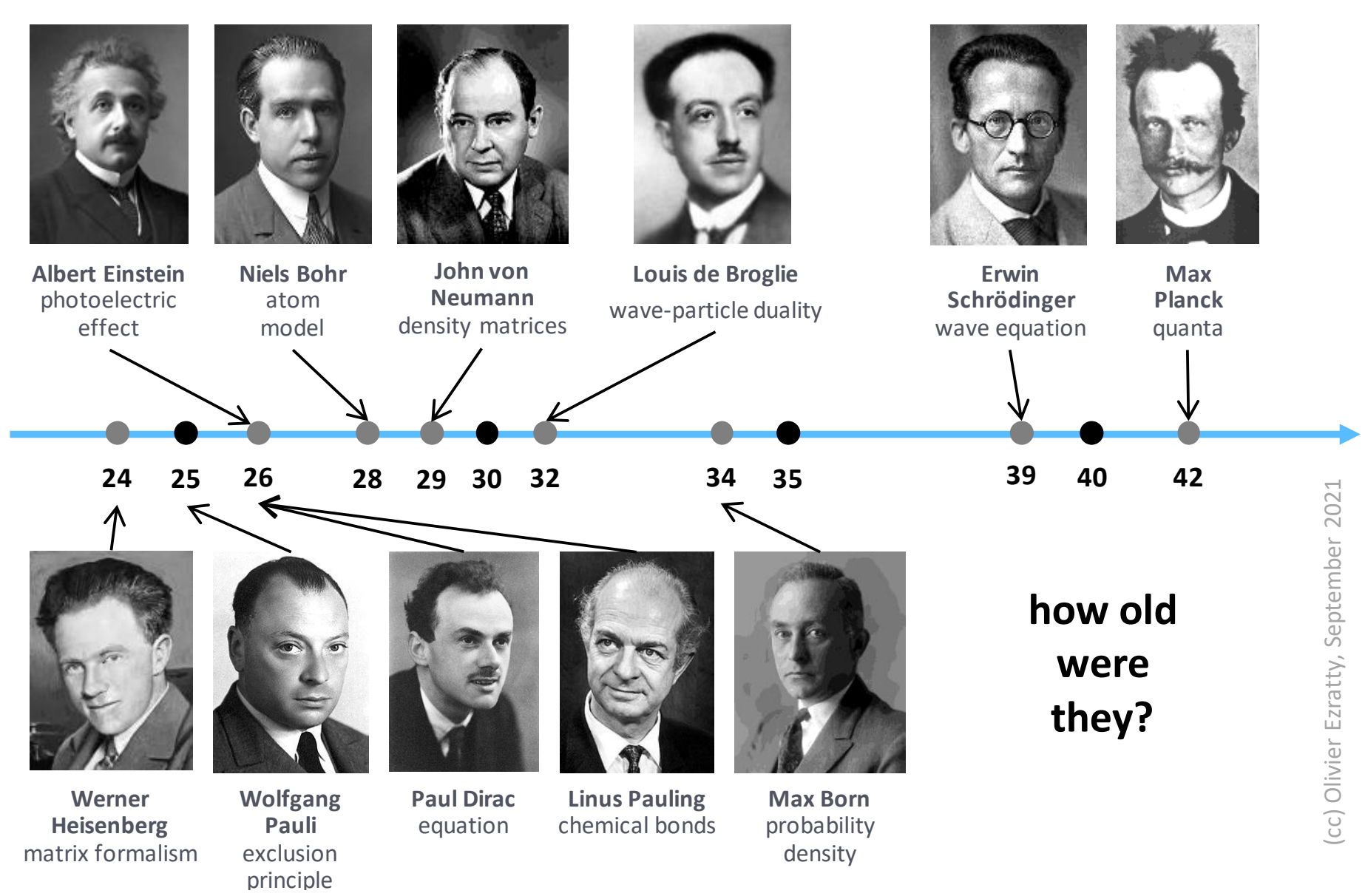

*Figure 52: how old were quantum scientists when they were awarded the Nobel prize in physics? (cc) Olivier Ezratty, 2021.*

## Post-war

As mentioned before, quantum physics developments seemed to slow down between 1935 and 1960. Physicists were then busy with nuclear physics. The Manhattan project mobilized an amazingly large number of physicists like John Von Neumann and **Enrico Fermi** (1901-1954, Italian American, Nobel prize in physics in 1938) whose contributions were centered in nuclear physics and statistical physics, leading to the Fermi-Dirac ideal gas statistics.

Quantum physics still led, after World War II, to an incredible wealth of technologies that revolutionized the world. We can mention three important branches resulting from the applications of the first quantum revolution: **transistors**, invented in 1947 by William Shockley, John Bardeen and Walter Brattain from the Bell Labs[60], **masers** and **lasers** invented between 1953 and 1960 by Gordon Gould, Theodore Maiman, Nikolay Basov, Alexander Prokhorov, Charles Hard Townes and Arthur Leonard Schawlow, only a few of whom received the Nobel Prize associated with these discoveries, **photovoltaic cells** that convert light into electricity, and the **GPS.** Transistors and lasers are the basis of much of today's digital technology. All our digital devices are already quantum! The field of quantum optics started in the early 1960s with the laser invention and Roy J. Glauber's work, with his seminal work in 1963 on light classification where he formalized the coherent states generated by lasers, *aka* Glauber states.

The post-war period was also dominated in quantum physics by advances made on superconductivity with the BCS theory in 1957 and the Josephson junction in 1962, and by the theoretical work of John Stewart Bell in 1964.

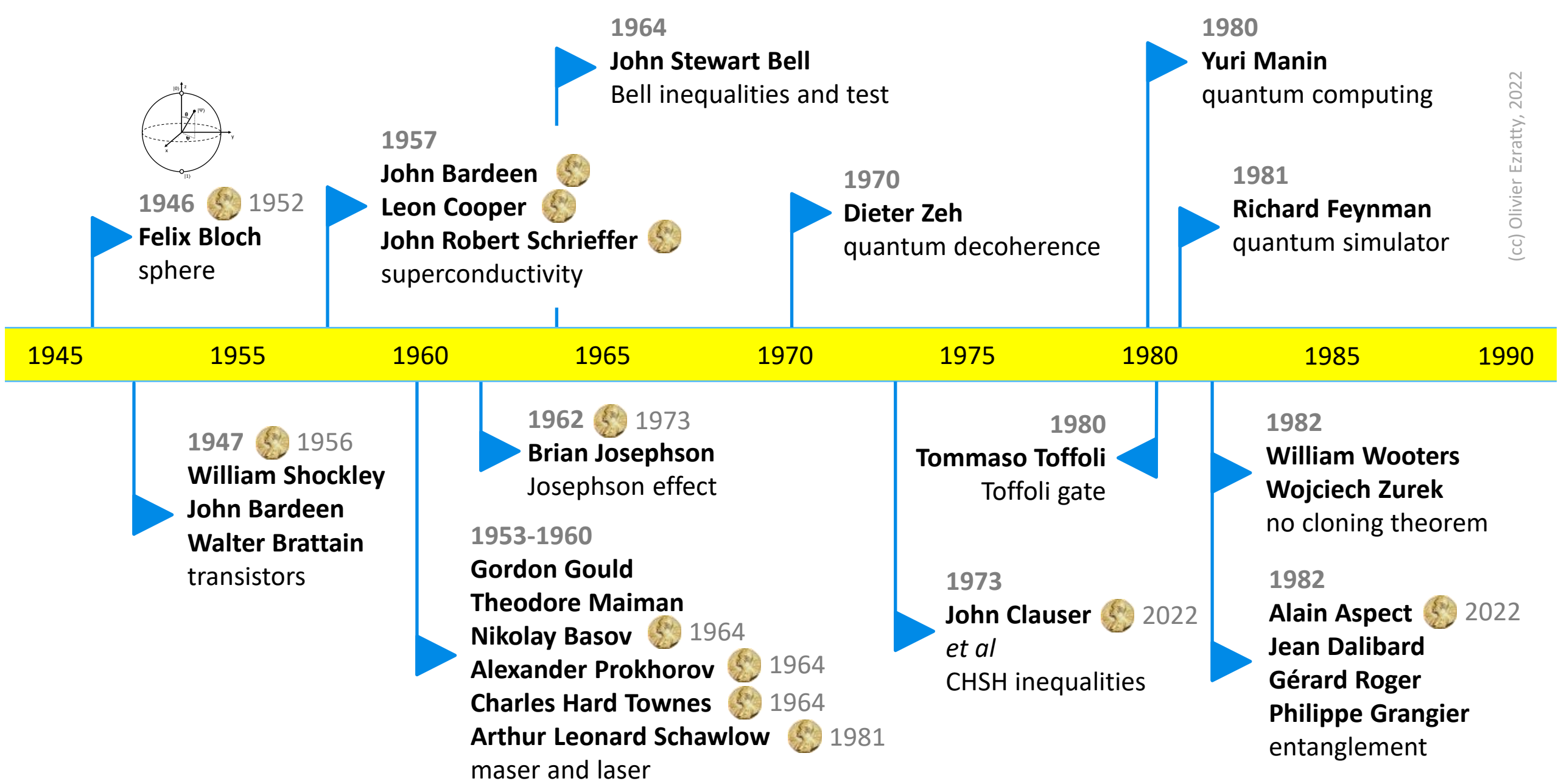

*Figure 53: timeline of key events in quantum physics after World-War II. (cc) Olivier Ezratty, 2022.*

---

[60] Transistors are based on many quantum phenomena, particularly the electronic structure of atoms in semiconductors crystals that was discovered during the 1930s and creates forbidden energy levels named band gaps (found by Sir Alan Herries Wilson, UK, in 1931), the impact of defects in crystals leading to doping and the tunneling effect due to the wave-particle duality of electrons. It also uses the field effect, which modulates the electrical conductivity of a material by the application of an external electric field. It was invented by Julius Edgar Lilienfeld (1882-1963, Austro-Hungarian and American) who got a related patent granted in 1926 using copper-sulfide semiconductor materials. It corresponds to what we today call a "Field Effect Transistor" (FET). The first transistor invented in 1947 was made of germanium, not silicon. See [The Transistor, an Emerging Invention: Bell Labs as a Systems Integrator Rather Than a 'House of Magic'] by Florian Metzler, October 2020 (57 pages) which shows the flow of discoveries that led to the creation of the first transistor by the Bell labs in 1947. This first computers using transistors was the TRADIC Phase One computer that was built in 1954.

We then have the verification of entanglement by Alain Aspect's experiment in 1982. 1980 and 1981 are other key dates which mark the symbolic beginnings of quantum computing, imagined by Yuri Manin (gate-based quantum computing) and Richard Feynman (quantum simulation).

The term **second quantum revolution** covers advances from the 1990s and later, when the quantum properties of individual particles could be controlled at the level of photons (polarization, ...), electrons (spin) and atoms or ions, as well as superposition and entanglement. This led to the emergence of quantum cryptography and quantum telecommunications, in addition to the premises of quantum computing. The original definition of this second quantum revolution is however not as precise as that [61].

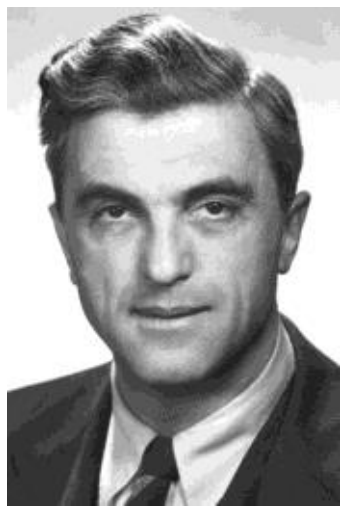

**Felix Bloch** (1905-1983, Swiss then American) is a physicist who created the geometrical representation of a qubit state in a sphere, Bloch's sphere was elaborated in 1946 in a paper on nuclear magnetism, his main specialty. Like other physicists of his time, he contributed to the Manhattan Project, although quite shortly. He was awarded the Nobel Prize in Physics in 1952 for his work on nuclear magnetic resonance and magnons conceptualization. He was also the first director of the international particle physics laboratory CERN in 1954.

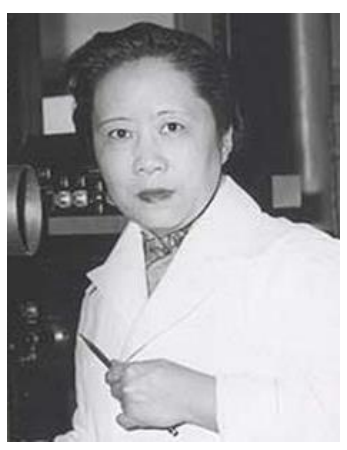

**Chien-Shiung Wu** (1912-1997, Chinese then American) was a scientist who contributed to the development of nuclear physics and to the Manhattan Project, with her gaseous diffusion process used for separating uranium 238 from uranium 235. She also contributed to the development of quantum physics by conducting the first experiment related to the synchronization of photon pairs and entanglement in 1949, before Alain Aspect's experiment in 1982[62].

This experiment was different and was based on the measurement of the angular correlation of gamma ray photons (with very high-frequency and high-energy) generated by the encounter of electrons and positrons.

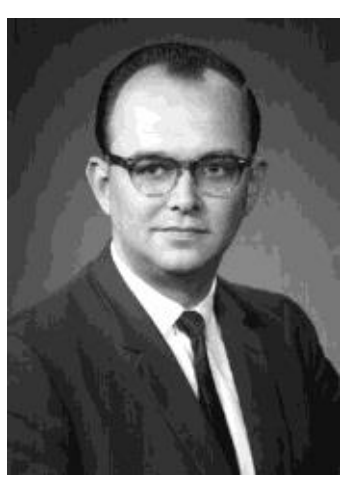

**Hugh Everett** (1930-1982, American) is a physicist who created the formulation of relative states and a global wave function of the Universe integrating observations, observers and tools for observing quantum phenomena. He met Niels Bohr with other physicists in Copenhagen in 1959 to present his theory. He was politely listened to, but his interlocutors said that he understood nothing about quantum physics.

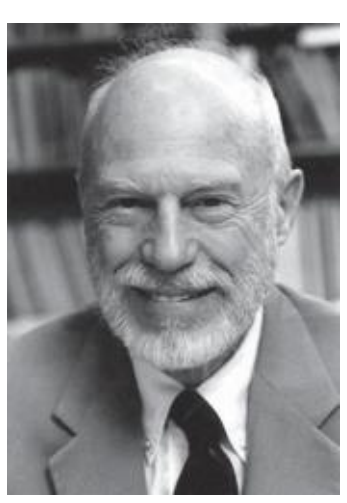

Everett was also a contributor to the connections between the theory of relativity and quantum physics, especially around quantum gravitation. He is credited with the hypothesis of multiple or multiverse worlds, or many-worlds interpretation, explaining quantum entanglement and non-locality. It is in fact coming from **Bryce DeWitt** (1922-2004, American) who interpreted his work in 1970. DeWitt also worked on the formulation of quantum gravity theories.

[61] The second quantum revolution expression was created simultaneously and independently in 2003 by Alain Aspect and by Jonathan Dowling and Gerard Milburn. The latter is also known to be one of the three protagonists of the KLM model of photon-based quantum computing, created in 2001 jointly with Emanuel Knill and Raymond Laflamme.

[62] See The Angular Correlation of Scattered Annihilation Radiation, Wu and Shaknov, 1949.

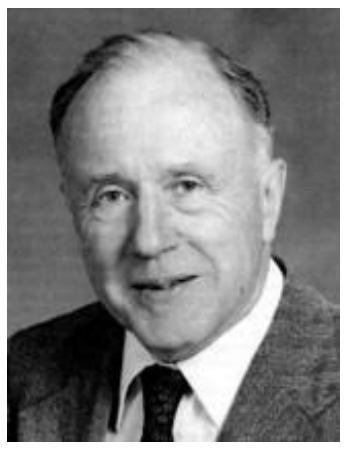

**John Wheeler** (1911-2008, American) supervised Hugh Everett's thesis. He was a specialist in quantum gravitation. He worked in the field of nuclear physics, notably in the Manhattan project, on the first American H-bombs and on very high-density nuclear matter found in neutron stars. He popularized the term black hole in 1967. He imagined a delayed-choice experiment to decide when a quantum object decides to travel as a wave or as a particle.

He collaborated with Niels Bohr and among his PhD students were Richard Feynman and Wojciech Zurek!

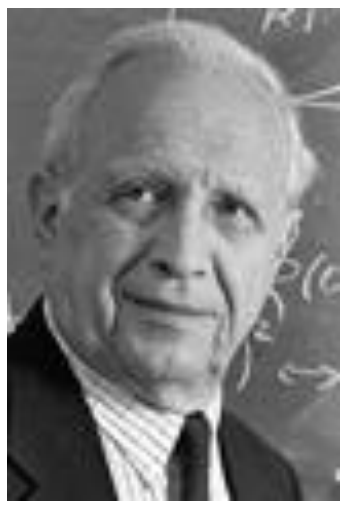

**Roy J. Glauber** (1925-2018, USA) was a theoretical physicist, teaching at Harvard and at the University of Arizona. He got the Nobel Prize in Physics in 2005 for his foundational work on the quantum theory of optical coherence. He is considered to a father of non-classical light description and of the quantum optics field, with his work in 1963, describing the various types of light (coherent, not coherent, ...). He also worked in the field of high-energy particle physics, which we don't cover in this book since out of scope of the "second quantum revolution".

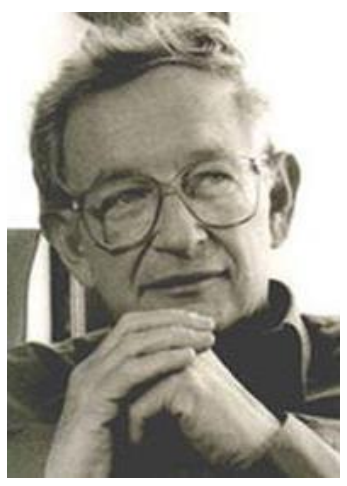

**Philip W. Anderson** (1923-2020, USA) was a theoretical physicist who contributed to the theories of localization (*aka* "Anderson localization" according to which extended states can be localized by the presence of disorder in a system), antiferromagnetism and quantum spin liquid, symmetry breaking leading to the creation of the Standard Model, superconductivity (at high-temperature, pseudospin approach to the BCS theory, Anderson's theorem on impurity scattering in superconductors).

He created the "condensed matter physics" naming. He got the Nobel prize in physics in 1977 for his work on the electronic structure of magnetic and disordered systems. He worked at the Bell Labs and was also a teacher at Cambridge University, UK.

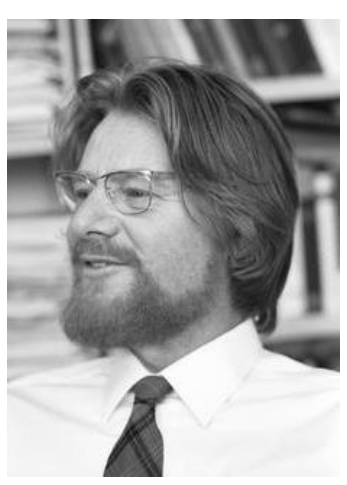

**John Stewart Bell** (1928-1990, Irish) relaunched research in quantum mechanics in the 1960s on the notion of entanglement. We owe him the Bell inequalities that highlight the paradoxes raised by quantum entanglement. Bell's 1964 theorem indicates that no theory of local hidden variables - imagined by Einstein in 1935 - can reproduce the phenomena of quantum mechanics [63]. He was rather pro-Einsteinian in his approach and favorable to a realistic interpretation of quantum physics [64].

His Bell inequalities define the means to verify or invalidate the hypothesis of the existence of hidden variables explaining quantum entanglement. Bell's inequalities were violated by the experiments of **Alain Aspect** in 1982, demonstrating the inexistence of these local hidden variables.

Prior to this experiment, Bell's inequalities had been formulated for pairs of entangled photons by **John Clauser** (1942, American, 2022 Nobel prize in physics), **Michael Horne** (1943-2019, American), **Abner Shimony** (1928-2015, American) and **Richard Holt** in 1969 with their so-called CHSH inequalities with some experimental settings proposals[65].

---

[63] See this explanation of Bell's theorem in a paper by Tim Maudlin on the occasion of the 50th anniversary of the theorem: What Bell Did, 2014 (28 pages). And Bell's original document: On the Einstein-Podolsky-Rosen paradox, John S. Bell, 1964 (6 pages). In 1964, Bell worked at the University of Wisconsin.

[64] See What Bell Did by Tim Maudlin, 2014 (28 pages) which describes the EPR paradox and Bell's contribution.

[65] See Proposed experiment to test local hidden-variable theories, 1969 (5 pages).

John Bell's work was completed in 2003 by **Anthony Leggett** (1938, Anglo-American, Nobel Prize in Physics in 2003 for his work on superfluid helium) with his inequalities applicable to hypothetical non-local hidden variables[66]. Anthony Leggett was also an initial key contributor to what led to the creation of superconducting qubits.

**Anton Zeilinger** (1945, Austrian) managed to experimentally violate these inequalities in 2007. According to Alain Aspect, however, this did not call into question the non-local hidden variable model proposed by David Bohm.

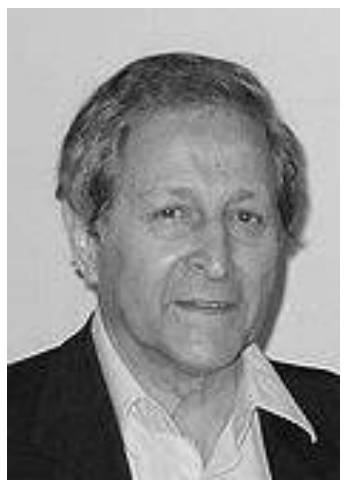

**Claude Cohen-Tannoudji** (1933, French) is a former student of Ecole Normale Supérieure (ENS Paris) where he followed the teachings of mathematicians Henri Cartan and Laurent Schwartz and physicist Alfred Kastler. He was awarded the Nobel Prize in Physics in 1997 at the same time as Steven Chu, who was later Secretary of Energy during Barack Obama's first term. This Department (DoE, Department of Energy) is one of the federal agencies most invested in quantum technologies, notably because they operate the largest supercomputers in the country.

Claude Cohen-Tannoudji owes his Nobel Prize to his work on atoms laser cooling which made it possible to reach extremely low temperatures, below the milli-Kelvin[67]. Alain Aspect once worked in his team. Alain Aspect says that he discovered quantum physics with reading the reference book on quantum physics by Claude Cohen-Tannoudji, Bernard Diu and Franck Laloë published in 1973[68]. It totals over 2300 pages. So, this book is quite small in comparison. And also, more accessible!

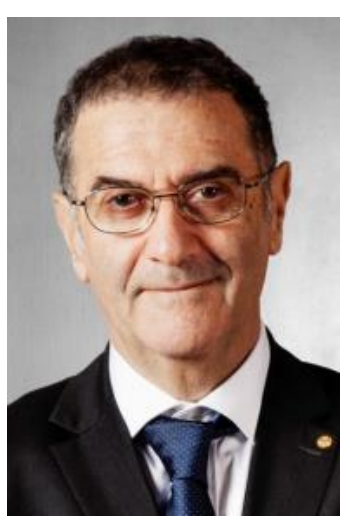

**Serge Haroche** (1944, French), Nobel Prize in Physics in 2012, is a founder of Cavity Electrodynamics (CQED) which describes the interaction between photons and atoms in cavities. He used it to create cold atom based qubits. **Jean-Michel Raimond**[69] and **Michel Brune** were among his key collaborators. Serge Haroche was the first to measure the phenomenon of quantum decoherence (loss of superposition) in an experiment in 1996. This experiment was conducted at the ENS with rubidium atoms. Serge Haroche is also a member of Atos Scientific Council.

CQED was later applied in the field of superconducting qubits with Circuit Electrodynamics (cQED), where atoms are replaced by an artificial atom made with a Josephson junction and the cavity by a planar microwave resonator. Serge Haroche is one of the most circumspect scientists on the future of quantum computing, at least for universal gate computing. He believes more in the advent of quantum simulation[70].

Other scientists brought key contributions in atoms science. **Daniel Kleppner** (1932, American) was the first to create a Bose-Einstein condensate with Rubidium atoms in 1995, and then in 1998 with hydrogen. **Herbert Walther** (1935-2006, German) did pioneering work in cavity quantum electrodynamics and also with trapped ions. He created the Max Planck Institute of Quantum Optics in 1981. **Gerhard Rempe** (1956, German) developed cavity quantum electrodynamics with the control of neutral atoms using microwaves, in connection with **Jeff Kimble** (1949, American, Caltech).

---

66 See Nonlocal Hidden-Variable Theories and Quantum Mechanics: An Incompatibility Theorem by Anthony Leggett, 2003 (25 pages).

67 See his Nobel lecture.

68 This book is published in three tomes that were last revised in 2019. The first one is Quantum Mechanics, Volume 1: Basic Concepts, Tools, and Applications. The second deals with Angular Momentum, Spin, and Approximation Methods and the third one with Fermions, Bosons, Photons, Correlations, and Entanglement. These are classical quantum physics student textbooks.

69 See his interesting conference Quantum Computing or how to use the strangeness of the microscopic world, Jean-Michel Raimond, 2015 (1h36mn). See also his presentation material (56 slides).

70 See Quantum Computing: Dream or Nightmare? by Serge Haroche and Jean-Michel Raimond, Physics Today, 1995 (2 pages) who expressed their skepticism about quantum computing. Serge Haroche continues to convey this skepticism.

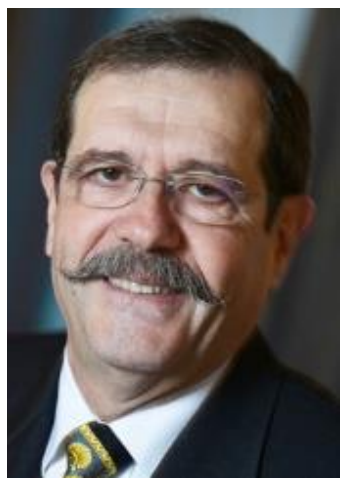

**Alain Aspect** (1947, French, 2022 Nobel prize in physics) observed violations of Bell's inequalities with a series of experiments conducted between 1980 and 1982 at the Institut d'Optique (Orsay University in the southern suburb of Paris with Jean Dalibard, Philippe Grangier and Gérard Roger. Taking the principles of quantum physics for granted, it validated the non-locality of quantum properties and "spooky action at a distance"[71]. One other option is you need to reject these principles and use a local variable model to explain the phenomenon. But it is not the only one[72].

The experiment avoided any potential synchronization between the polarizers, using a 50 MHz random optical switch on both sides, feeding two orthogonal polarizers and photon detectors. From 1988 to 2015, other experiments were conducted elsewhere and implemented loophole-free Bell tests, first closing individual loopholes and then, in 2015, closing them altogether. It confirmed then that there were no local variables explaining entanglement and validated the non-locality condition: long distance between analyzers to avoid any interactions made possible by special relativity.

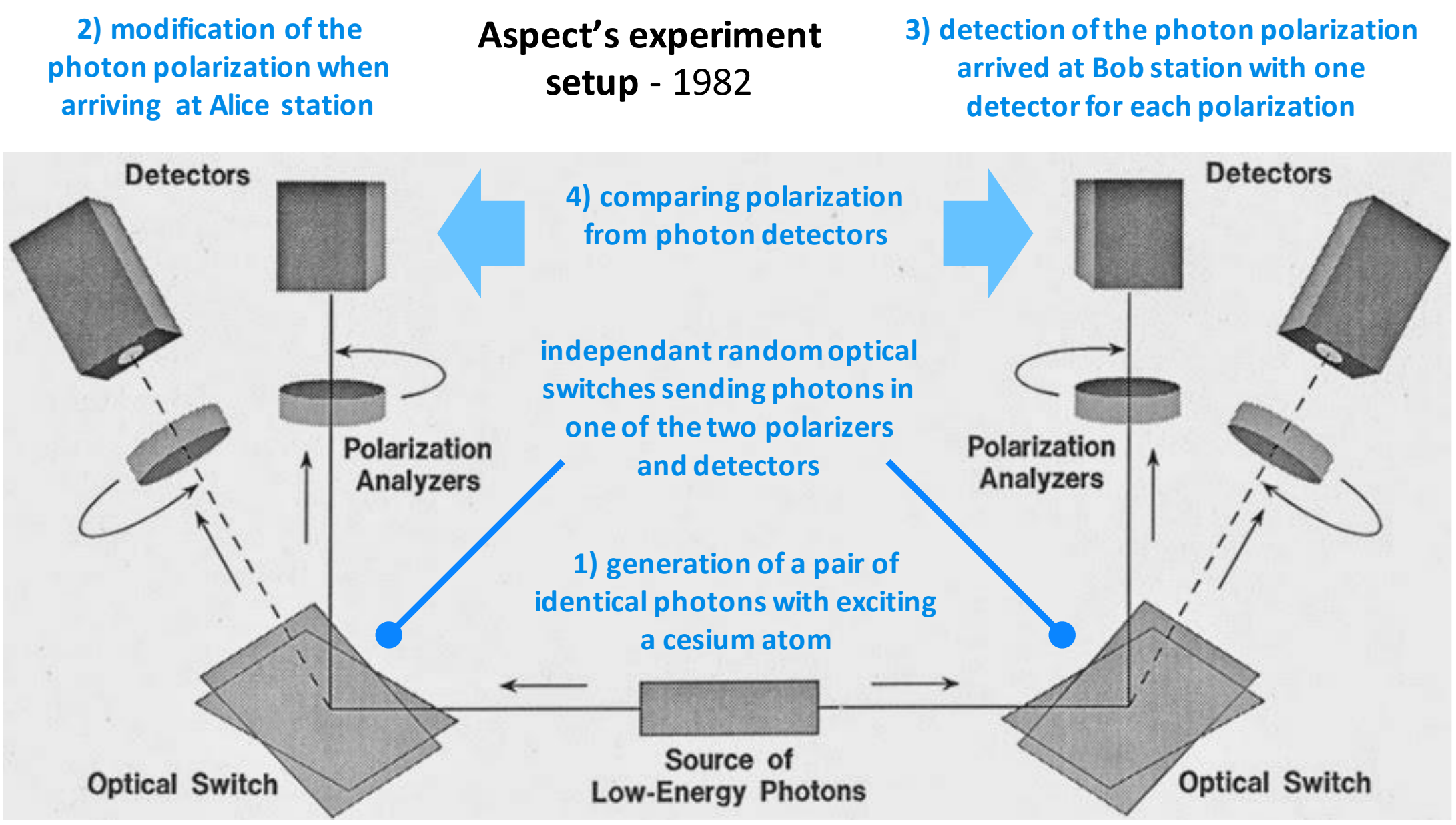

*Figure 54: Alain Aspect et al 1982 Bell inequality test experiment setup.*

It avoided detection loopholes with high-efficiency photon detectors on top of escaping 'memory loopholes', which was already obtained by Alain Aspect et al in their seminal 1982 experiment[73]. After his work on photon entanglement, Alain Aspect shifted gear on cold atoms control with lasers, starting with helium.

---

[71] Alain Aspect's experiments were using calcium atoms as source of photons, using some laser excitement and an atomic cascade generating pairs of entangled photons in the visible spectrum at 551 nm and 423 nm. There were actually several experiments: in 1981 with Philippe Grangier and Gérard Roger with one way polarizers, 1982 also with Grangier and Roger with two-channels polarizers and also 1982, with Jean Dalibard and Gérard Roger, using variable polarizers based on acousto-optical 10 ns switches. These could act faster than light propagation between the polarizers (40 ns) and even than the photons time of flight between the source and each switch (20 ns). See Experimental Test of Bell's Inequalities Using Time-Varying Analyzers by Alain Aspect, Gérard Roger and Jean Dalibard, PRL, December 1982 (4 pages).

[72] You have superdeterminism-based theories promoted by Carl H. Brans, Sabine Hossenfelder and Tim Palmer that are based on the hypothesis of superdeterministic hidden variables theory and could still violate Bell's inequalities, but also the CSM ontology which pertains that the Psi function is lacking information on the measurement context, like described in Why ψ is incomplete indeed: a simple illustration by Philippe Grangier, October 2022 (2 pages).

[73] See Experimental loophole-free violation of a Bell inequality using entangled electron spins separated by 1.3 km by B. Hensen et al, ICFO and ICREA in Spain and Oxford, UK, August 2015 (8 pages) and also A strong loophole-free test of local realism by Lynden K. Shalm et al, September 2016 (9 pages).

This led to the creation of a promising field of quantum computing in France, using cold atoms, embodied by the startup **Pasqal**, whose scientific director is Antoine Browaeys, a former PhD student of Philippe Grangier, himself the first PhD student of Alain Aspect.

Along with other scientists, Alain Aspect is also a member of Atos Scientific Council and in the scientific board of **Quandela**. He teaches quantum physics, notably in MOOCs created for Ecole Polytechnique and distributed by Coursera.

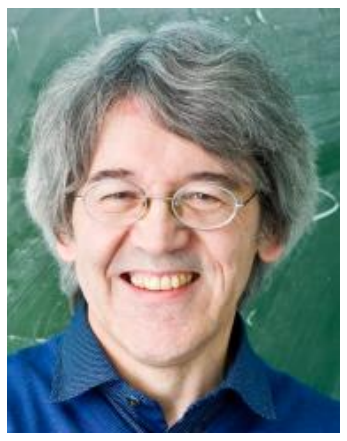

**Philippe Grangier** (1957, French) was a PhD student of Alain Aspect with whom he worked on the 1982 experiment with Gérard Roger and Jean Dalibard. He is one of the world's leading specialists in quantum cryptography, especially on CV-QKD. He was involved in the creation of the associated startup, Sequrnet, in 2008 and closed in 2017, probably created a little too early in relation to the needs of the market. He is also invested in cold atoms control with lasers at IOGS (Institut d'Optique).

At last, he cocreated the CSM ontology of quantum foundations with Alexia Auffèves and Nayla Farouki, starting in 2013 and with a series of 7 foundational papers published between 2015 and 2019. CSM ontology is quickly covered in the Quantum Foundations section.

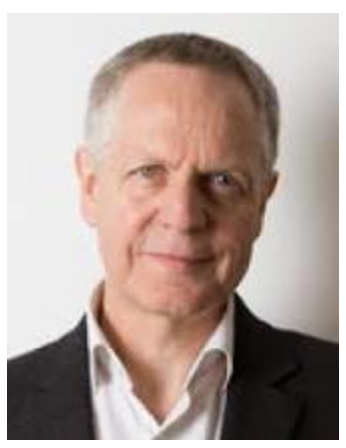

**Jean Dalibard** (1958, French) is a research physicist at the ENS and teacher at the Polytechnique and the Collège de France. He is a specialist in quantum optics and interactions between photons and matter[74]. He participated with Philippe Grangier in the set-up of Alain Aspect's experiment in 1982 when he was a contingent scientist at the Institut d'Optique. He created the magneto-optical trap (MOT) system in 1987 that is used to cool neutral atoms using a mix of variable magnetic fields and lasers.

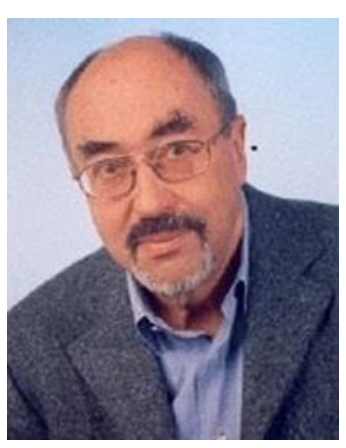

**Dieter Zeh** (1932-2018, German) is the discoverer of the quantum decoherence phenomenon in 1970. It marks the progressive end of the phenomenon of superposition of quantum states, when particles are disturbed by their environment and their amplitude and phase is modified. The notion of decoherence is key in the design of quantum computers. The objective is to delay it as much as possible resulting from the interaction between quanta and their environment[75].

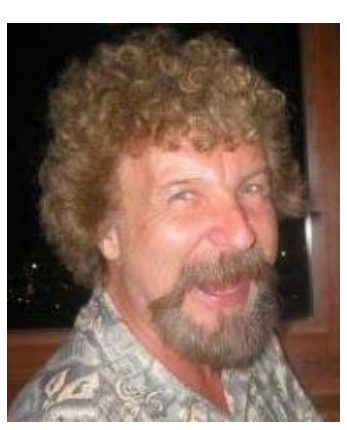

**Wojciech Zurek** (1951, Polish) is a quantum decoherence physicist who contributed to the foundations of quantum physics applied to quantum computers. We owe him the no-cloning theorem, which states that it is impossible to clone a qubit identically without the resulting qubits then being entangled. He is also at the origin of the concept of quantum Darwinism which would explain the link between the quantum world and the macrophysical world.

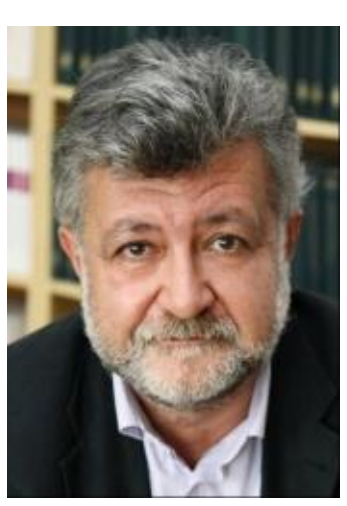

**Maciej Lewenstein** (1955, Polish) is a theoretical physicist, specialized in quantum optics of dielectric media and cavity quantum electrodynamics, teaching at ICFO in Spain. He worked with many leading worldwide scientists including Roy J. Glauber (Nobel in Physics in 2005) at Harvard, Thomas W. Mossberg, Andrzej Nowak, Bibb Latané, Anne L'Huillier (CEA, France), Peter Zoller and Eric Allin Cornell (Nobel in Physics in 2001 for his work on Bose-Einstein condensates in 1995), in the USA, France, Spain, Poland and Germany.

[74] See in particular his lesson on cold atoms at the Collège de France which describes well how atoms are cooled at very low temperatures with lasers.

[75] Dieter Zeh is notably the author of On the Interpretation of Measurement in Quantum Theory in 1970 (8 pages).

His contributions span an incredible number of fields like the physics of ultra-cold gases, quantum information, quantum optical systems, quantum communications, quantum cryptography, quantum computers, mathematical foundations of quantum physics, tensor networks and entanglement theory, laser-matter interactions atto-second physics, quantum optics (cQED), atoms cooling and trapping, non-classical states of light and matter and quantum physics foundations.

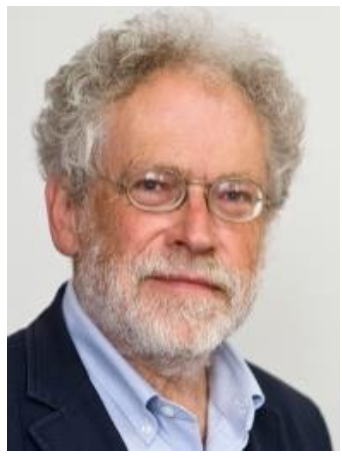

**Anton Zeilinger** (1945, Austrian, 2022 Nobel prize in physics) is a physicist who advanced the field of quantum teleportation in the 2000s. He also proved in 1991 the wave-particle duality of neutrons. He was also the first to experiment a qubit teleportation in 2009. He is a specialist in quantum entanglement, having proved that it is possible to entangle more than two quantum objects or qubits. He created theoretical and experimental foundations for quantum cryptography.

With two colleagues, he also developed the GHZ (Greenberger-Horne-Zeilinger) entangled state, which enables yet another demonstration of the inexistence of hidden variables which would explain quantum entanglement of at least three particles and with a finite number of measurements. The concept was created in 1989 and was validated experimentally in 1999. Anton Zeilinger also supervised the thesis of **Jian-Wei Pan**, who became later the quantum research czar in China with the development of many advances, particularly in quantum communications and photonics.

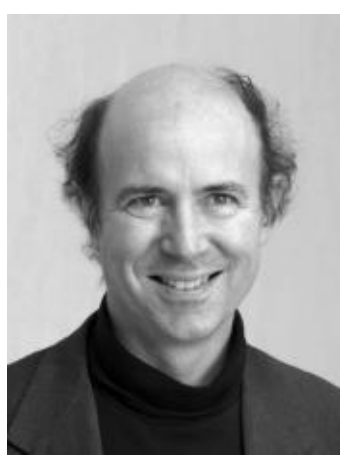

**Frank Wilczek** (1951, American) is a professor of physics at MIT and the chief scientist at the Wilczek Quantum Center in Shanghai. He was awarded in 2004 the Nobel Prize for Physics shared with David Gross and H. David Politzer, for his work on the theory of strong interaction and quantum chromodynamics. He is known for his work on quasi-particles and anyons in 1982 and he also predicted the existence of time crystals in 2012.

## Quantum technologies physicists

This story now provides an overview of key contributors to the physics of quantum computing. They are often specialized in condensed matter, such as for superconducting qubits, and in photonics.

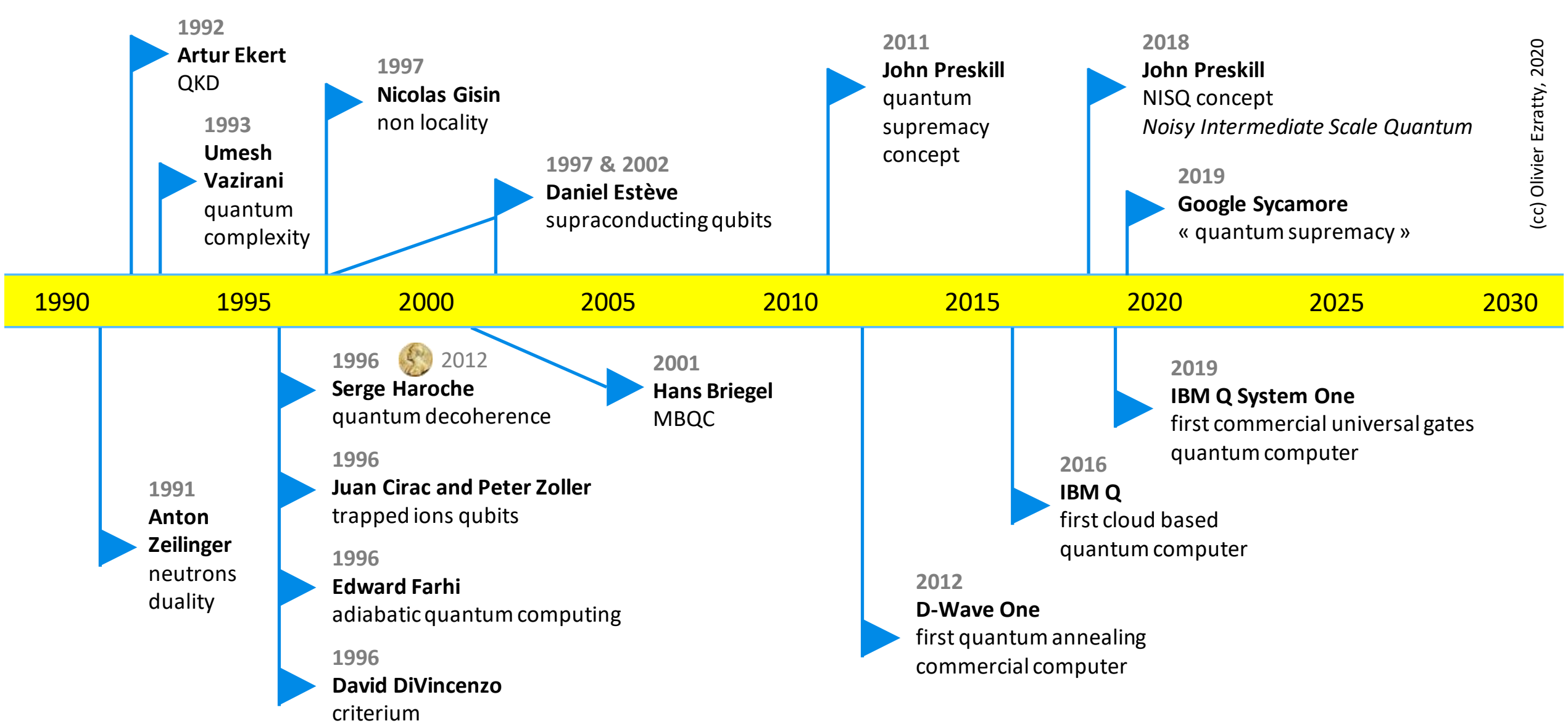

*Figure 55: quantum computing key events timeline from 1990 to 2020. (cc) Olivier Ezratty, 2020.*

I highlight many European and French physicists, particularly those I have had the opportunity to meet for the last three years in my journey in the quantum ecosystem. This inventory is both objective and subjective.

Objective because it includes a broad and worldwide hall of fame in the field. Subjective because I have added a good dose of physicists I know. It creates a measurement bias which is easy to understand in social science as well as in quantum physics.

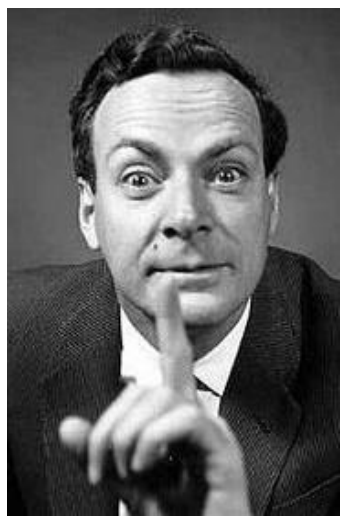

**Richard Feynman** (1918-1988, American) is one of the fathers of quantum electrodynamics, which earned him the Nobel Prize in Physics in 1965. He is at the origin of the quantum explanation of helium superfluidity at very low temperature in a series of papers published between 1953 and 1958. He theorized in 1981 the possibility of creating quantum simulators, capable of simulating quantum phenomena, which would be useful to design new materials and molecules in various fields like chemistry and biotechs[76] . He was also known for his great presentation skills.

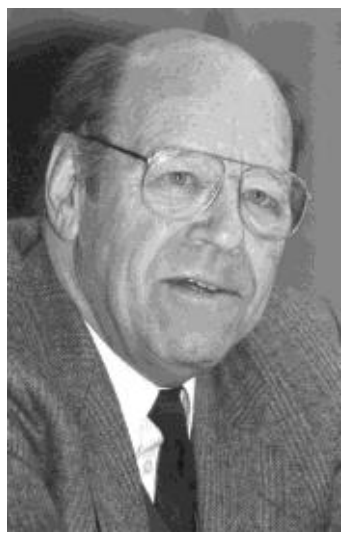

**Wolfgang Paul** (1913-1993, Germany), not to be confused with Wolfgang Pauli, is a physicist who conceptualized trapped ions in the 1950s. He got the Nobel Prize in Physics in 1989. We owe him the traps that bear his name and are used to control trapped ions. He shared his Nobel prize with **Hans Georg Dehmelt** (1922-2017, Germany) who codeveloped these traps with him. The physicists **Juan Ignacio Cirac** (1965, Spanish) and **Peter Zoller** (1952, Austria) theorized, designed and tested the first trapped ion qubits in 1996, based on the work of Wolfgang Paul.

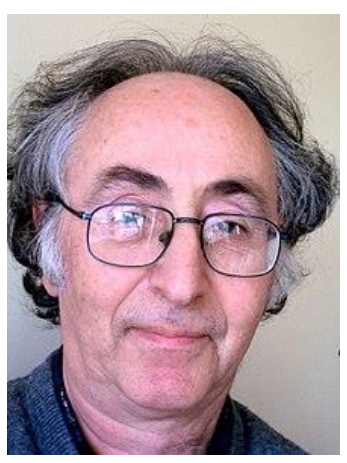

**Brian Josephson** (1940, English) is a physicist from the University of Cambridge. He was awarded the Nobel Prize in Physics in 1973 at the age of 33[77], for his prediction in 1962 of the effect that bears his name when he was only 22 years old and a PhD student at the University of Cambridge. The Josephson effect describes the passage of current in a superconducting circuit through a thin insulating barrier a few nanometers thick, using tunneling effect, and the associated threshold effects.

Below a certain voltage, the current starts to oscillate. It is generated by electrons with opposite spins organized in Cooper pairs named after Leon Cooper who discovered it in 1952. These pairs behave as bosons.

These electrons pairs have opposite spins (magnetic polarity). The system behaves as a resistance associated with a loop inductance, the oscillation being controllable by a magnetic field and having two distinct energy states. Superconductivity was discovered in 1911 by **Heike Kamerlingh Onnes** (1853-1926, Netherlands). This is the basis of superconducting qubits and their quantum gates!

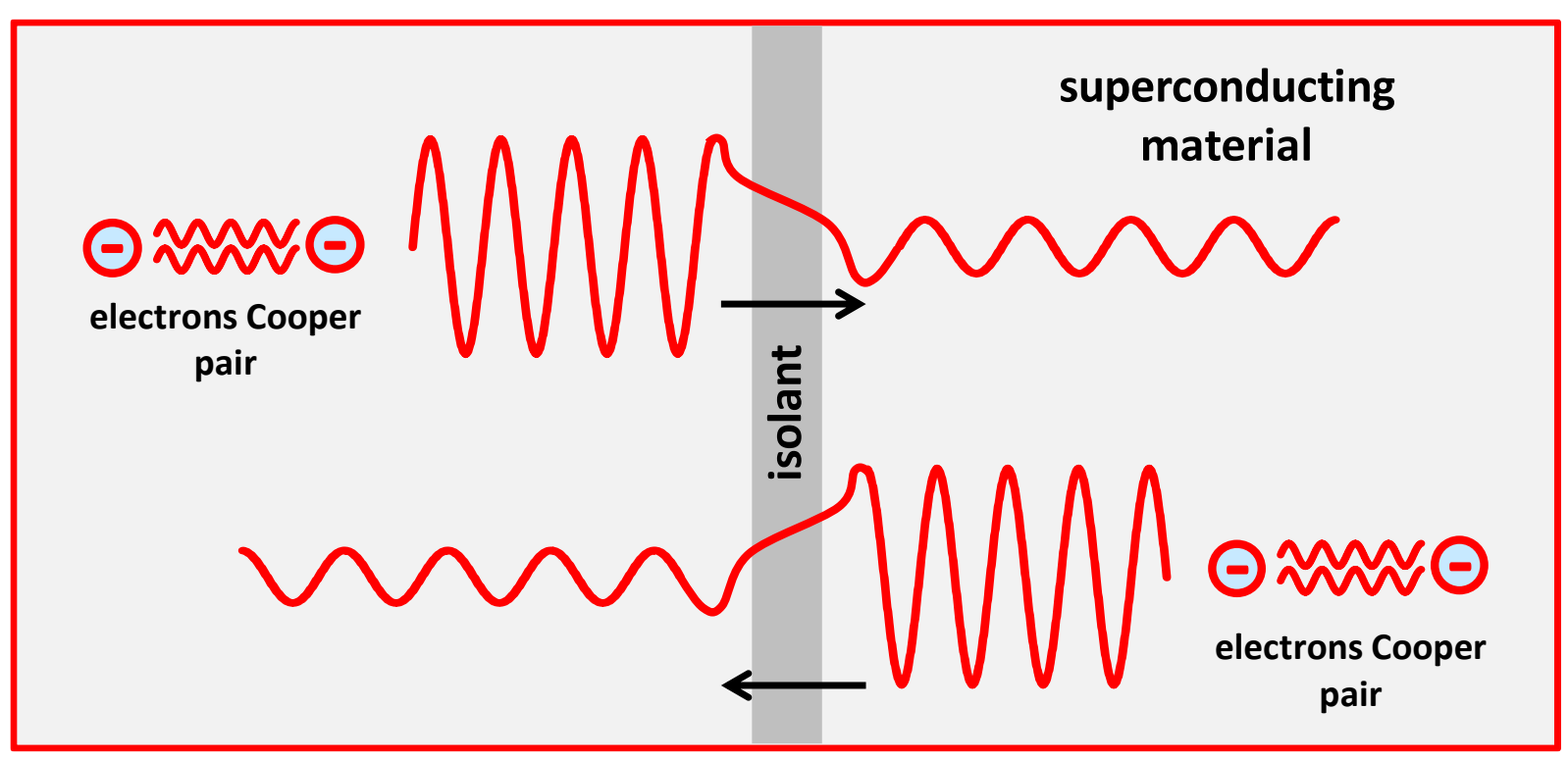

*Figure 56: Josephson effect and Cooper pairs of opposite spin electrons.*

[76] See Simulating Physics with Computers submitted in May 1981 to the International Journal of Theoretical Physics and published in June1982 and Quantum Mechanical Computers also by Richard Feynman, published in 1985 (10 pages). He describes how a quantum computer could perform mathematical operations similar to those of traditional computers. He concludes by saying that it should be possible to create computers where a bit would fit into a single atom!

[77] Brian Josephson shared the 1973 Nobel Prize in Physics with two scientists who had worked before him in the same field: Leo Esaki (1925, Japan, still alive in early 2020) for his discovery of the tunnel effect in semiconductors in 1958 and Ivar Giaever (1929, Norway, also still alive) who found that this effect could occur in superconducting materials in 1960.

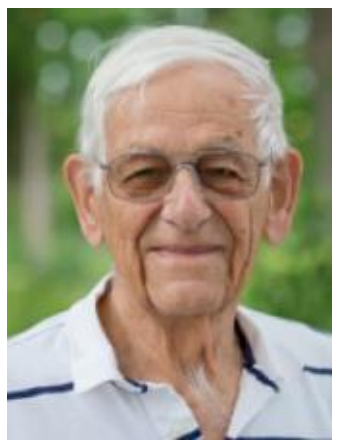

**Paul Benioff** (1930-2022, American) proposed in 1979/1980 the concept of a reversible and non-dissipative quantum Turing machine using 2D lattices of spins ½, based on earlier work from Rolf Landauer on the thermodynamics of computing and Charles Bennett on reversible computing [78]. It was a semi-classical machine concept that didn't yet exploit entanglement and interferences. His work was extended by the "universal quantum computer" concept from David Deutsch in 1985.

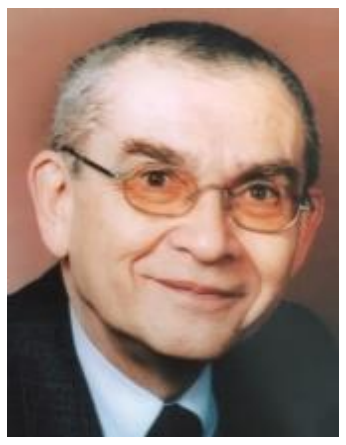

**Yuri Manin** (1937, Russian and German) is a mathematician who proposed the idea of creating gate-based quantum computers, in his 1980 book "Computable and Uncomputable", then in the USSR.

Then, **Richard Feynman** devised in 1981 the idea of a quantum simulator. Feynman and Benioff were participants of the famous "Physics & Computation" conference in 1981 that was co-organized by IBM and the MIT at the MIT Endicott House[79].

It brought together a number of well-known scientists in quantum information technology such as Tommaso Toffoli and Edward Fredkin.

Rolf Landauer was also among them. It was for this conference that Richard Feynman published his famous paper "Simulating Physics with Computers" which created the concept of quantum simulation[80].

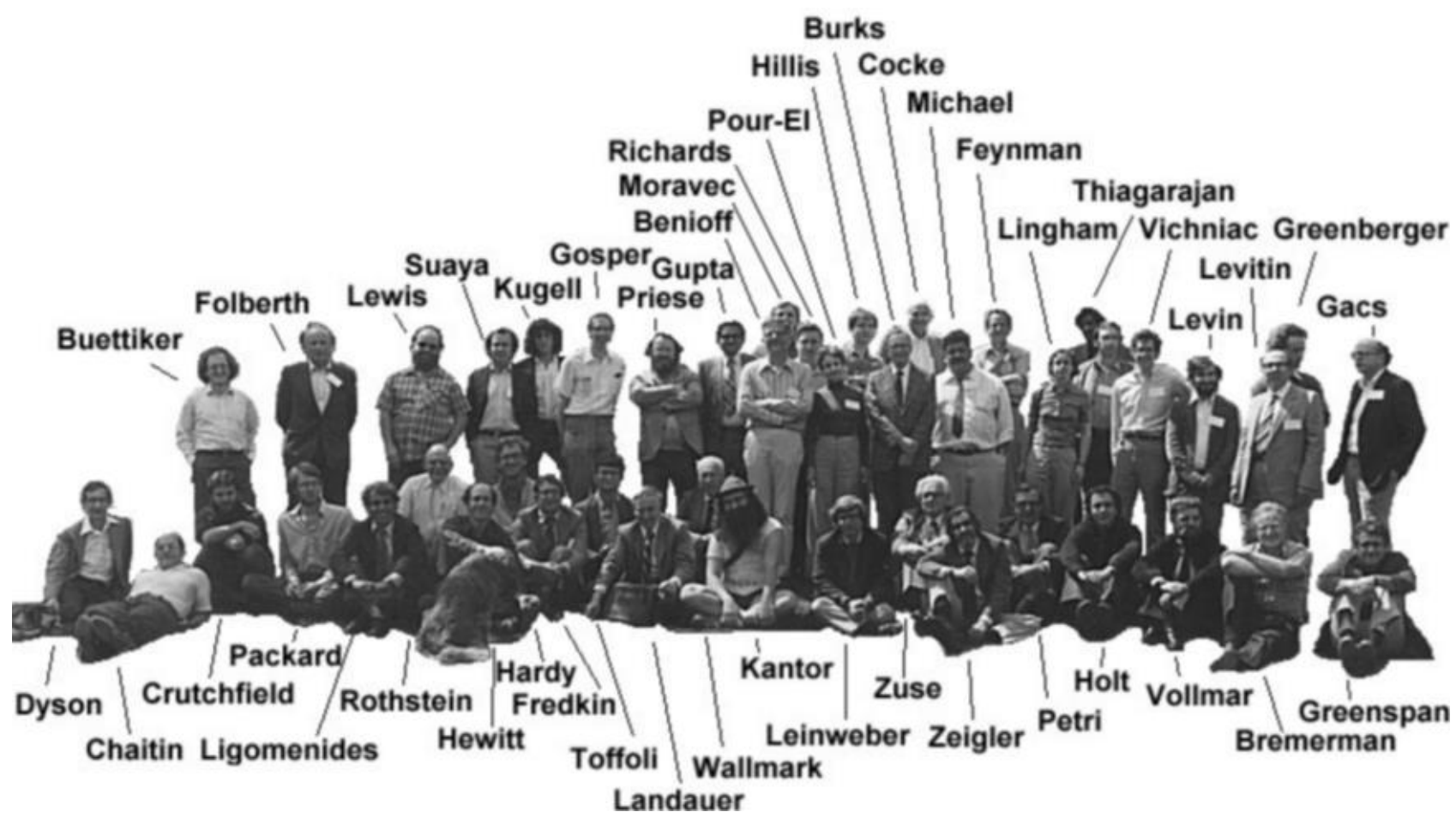

*Figure 57: participants of the first quantum computing conference in 1981. Source: Simulating Physics with Computers by Pinchas Birnbaum and Eran Tromer (28 slides).*

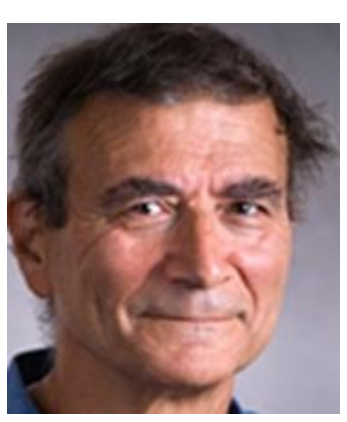

**Tommaso Toffoli** (1943, Italian then American) is an engineer known for the creation, at the beginning of the 1980s, of the quantum gate bearing his name, a conditional gate with three inputs that is widely used in quantum programming. After working at MIT, he became a Boston University professor, where he has served since 1995. Like Stephen Wolfram, his interests include cellular automata and artificial life.

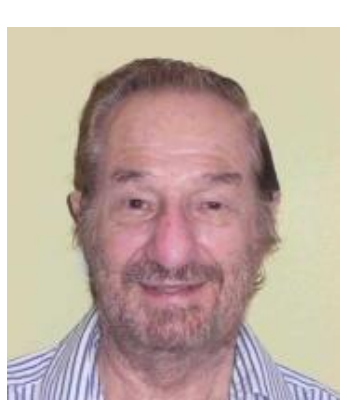

**Edward Fredkin** (1934, American) is a professor at Carnegie Mellon University. He is the author of the two-way conditional swap quantum gate (SWAP). He is also the designer of the concept of reversible classical computer with Tommaso Toffoli at MIT. He is also a prolific inventor far beyond quantum computing and is the originator of vehicle identification transponders and automotive geonavigation.

---

[78] See The computer as a physical system: A microscopic quantum mechanical Hamiltonian model of computers as represented by Turing machines by Paul Benioff, Journal of Statistical Physics, June 1979, published in May 1980 (30 pages). Paul Benioff was then in a visiting stay at the Centre de Recherche Théorique from CNRS in Marseille, France while being affiliated with the DoE Argonne National Laboratory in the USA. The paper was followed by Quantum Mechanical Hamiltonian Models of Turing Machines by Paul Benioff, October 1981 and June 1982, also in the Journal of Statistical Physics (32 pages). This theoretical system was based on using a two-dimensional lattice of spin ½ systems (today, it would be electron spins based qubits). Back in the 1980s, the very notion of qubits was not yet in the radar. It appeared much later, in 1995. In Benioff's model, a quantum gate was a Hamiltonian transformation of individual spins that was driven by the Turing quantum machine.

[79] See How a 1981 conference kickstarted today's quantum computing era by Harry McCracken, FastCompany, May 2021.

[80] See Simulating Physics with Computers by Richard Feynman, 1981 (103 pages).

He is also a promoter of the notion of "digital philosophy" which reduces the world and its functioning to a giant quantum program, a theory he shares with Seth Lloyd, an idea that has been revived by Elon Musk who believes that the Universe is a gigantic program and that we live in a simulation. Is the "automatic" respect of elementary physical laws a "program"? A thorny philosophical and semantic question!

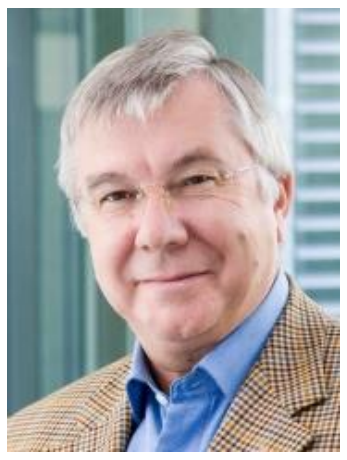

**Rainer Blatt** (1952, Austrian and German) from the University of Innsbruck is an experimental physicist specialized, among other things, in trapped ions qubits. He was the first to entangle the quantum states of two trapped ions in 2004 and then with eight ions in 2006. He co-founded Alpine Quantum Technologies (AQT), whose ambition is to create and commercialize a trapped ions based quantum computer. He also works at TUM in Munich, Germany and is the coordinator of the Munich Quantum Valley since 2021.

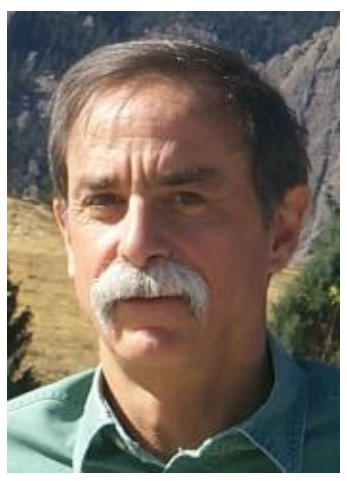

**David Wineland** (1944, American) is a Boulder-based NIST physicist known for his advances in trapped ions and their laser-based cooling in 1978. He also created in 1995 the first single quantum gate operating on a single atom. He was awarded the Nobel Prize in Physics in 2012 jointly with Serge Haroche for his advances in atoms and ions laser cooling, a technique he first experimented in 1978, followed by the first quantum gate applied to a trapped ion in 1995 and the entanglement between four trapped ions in 2000.

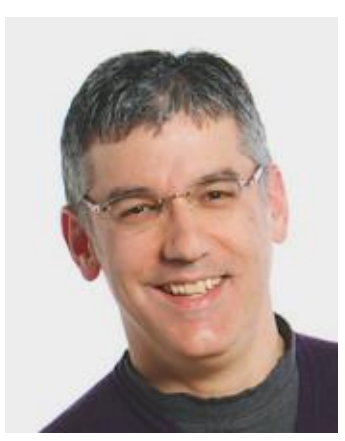

**Christopher Monroe** (1965, American) is an American physicist known for his work on trapped ions and for co-founding IonQ in 2015, one of the two best funded quantum startups worldwide with PsiQuantum. He worked on trapped ions with David Wineland at the NIST Maryland laboratory. He demonstrated the ability to entrap ions, create ions-based quantum memory and create analog quantum simulators. He also ran a laboratory at the University of Michigan in the early 2000s.

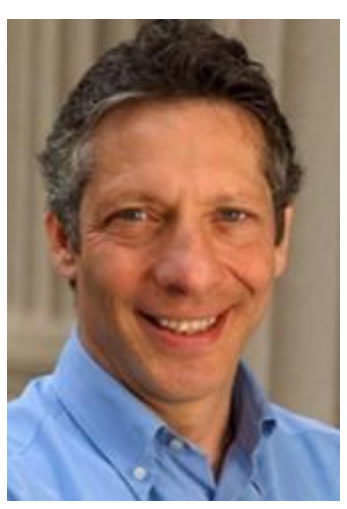

**Edward Farhi** (1952, American) is a theoretical physicist who has worked in many fields, including high-energy particle physics, particularly at the CERN LHC in Geneva and then at MIT. He worked with Leonard Susskind on unified theories with electro-weak dynamical symmetry breaking. He and Larry Abbott proposed a model in which quarks, leptons, and massive gauge bosons are composite. He is the creator of adiabatic quantum algorithms and quantum walks. He also introduced with Peter Shor the concept of quantum money in 2010.

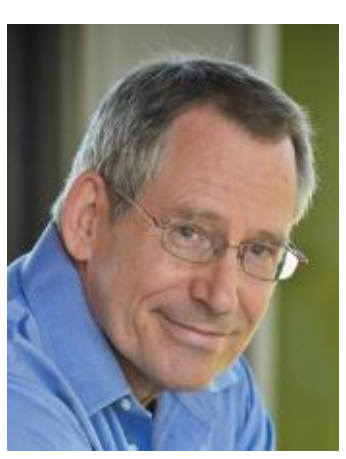

**John Preskill** (1953, American) is a professor at Caltech. Among many other contributions, he is the creator of quantum supremacy notion in 2011 and of NISQ in 2018, the Noisy Intermediate-Scale Quantum, qualifying current and future noisy quantum computers. He is a regular speaker at conferences where he reviews the state of the art of quantum computing[81]. He's now involved with Amazon and their cat-qubits superconducting project revealed in December 2020.

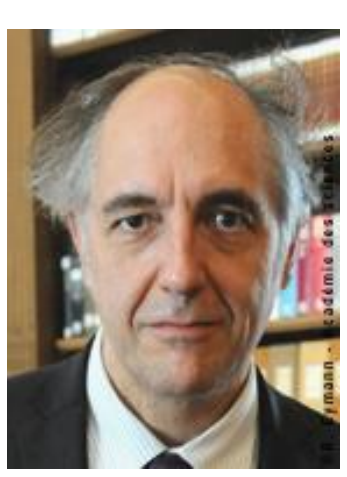

**Daniel Esteve** (1954, French) is a physicist in charge of the CEA's Quantronics laboratory in Saclay, France, launched in 1984 with Michel Devoret and Cristian Urbina, and part of the IRAMIS laboratory. He contributed to the development of transmon superconducting qubits. He created a first operational qubit in 1997, the quantronium, followed by another controllable prototype in 2002, with Vincent Bouchiat. He continues to work on improving the quality of superconducting qubits.

---

[81] See his presentation that provides an overview of the state of the art of quantum computing Quantum Computing for Business, John Preskill, December 2017 (41 slides).

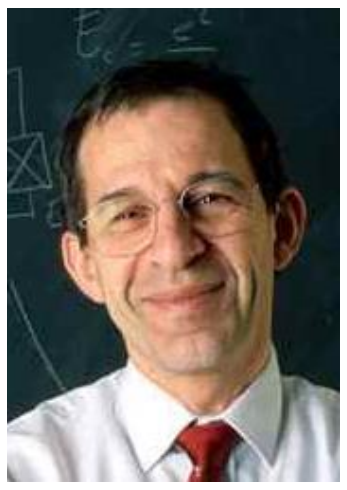

**Michel Devoret** (1953, French) is a telecom engineer turned physicist, co-founder of the Quantronics laboratory with Daniel Esteve at the CEA in Saclay between 1985 and 1995, which is one of the world pioneers of superconducting qubits. He is a professor at Yale University since 2002. He was a co-founder of the American startup QCI with his Yale colleague Rob Schoelkopf (1964, USA), which he left in 2019/2020. He preferred to be entirely dedicated to research.

He worked several times with John Martinis, when John was a PhD student in UCSB, then when he was a post-doc at CEA in Saclay in the early 2000s, and at last at the University of Santa Barbara (UCSB), where they wrote together a review paper in 2004 on superconducting qubits[82].

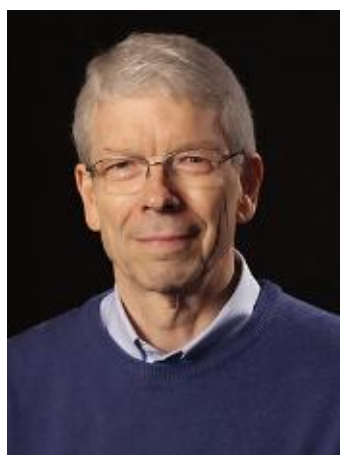

**Steven Girvin** (1950, USA) is a professor of physics at Yale University, specialized in condensed matter physics, and Director of the Co-design center for Quantum Advantage, at Brookhaven University since 2020. He is a key contributor to works on circuit quantum electrodynamics (cQED) and superconducting qubits. At Yale, he works with Robert Schoelkopf and Michel Devoret on the various engineering problems associated with superconducting qubits.

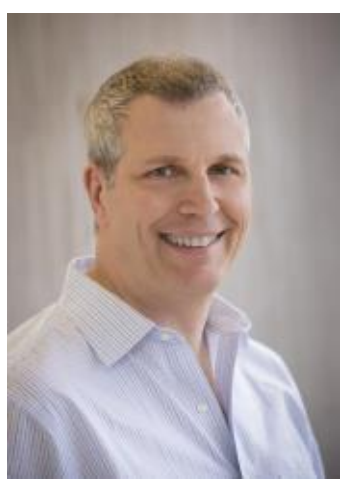

**Rob Schoelkopf** (1964, USA) a physicist and director of the Yale Quantum Institute. Along with Steve Girvin and Michel Devoret, he made key advances in superconducting qubits. He particularly worked on single-electron devices, being the inventor of the Radio-Frequency Single-Electron Transistor. He also created the field of circuit quantum electrodynamics (cQED) with Andreas Wallraff and Alexandre Blais who were respectively Yale post-doc and PhD student around 2002-2004.

In 2007, with Steven Girvin, he engineered a superconducting communication bus to store and transfer information between distant qubits on a chip. In 2009, their team, also including Alexandre Blais and Jay Gambetta, demonstrated the quantum processor running some quantum computation, with two qubits[83].

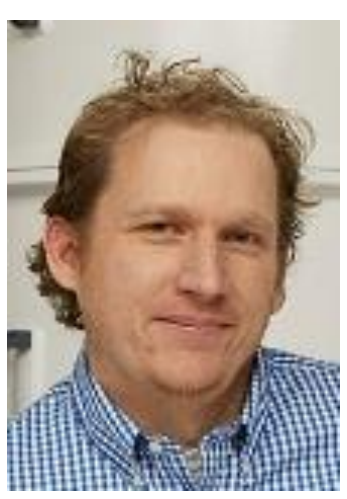

**Jay Gambetta** (1979, USA) is the scientist leading as a VP since 2019 IBM's research team working on superconducting qubits quantum computers after running the IBM team that created and launched IBM Quantum Experience, Qiskit and the IBM Quantum System One in 2019. He joined IBM in 2011. After a thesis in quantum foundations and non-Markovian open quantum systems done in Australia in 2004, he focused on developing superconducting qubits, first in a post-doc tenure at Yale University and then at the Institute for Quantum Computing in Waterloo. He also worked on quantum validation techniques, quantum codes and applications.

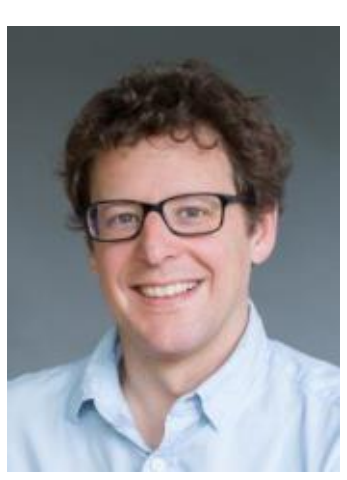

**Alexandre Blais** (Canada) is a Professor in the Department of Physics and Director of the Université de Sherbrooke's Institut Quantique. He is one of the key contributors to the development of circuit quantum electrodynamics (cQED) that enable the creation of superconducting qubits. He is also a cofounder of Nord Quantique, a Quebec startup developing bosonic code qubits. Like Jay Gambetta, he did a post-doc at Yale, the US epicenter of the early developments of superconducting qubits.

---

[82] In Implementing Qubits with Superconducting Integrated Circuits by Michel Devoret and John Martinis, 2004 (41 pages).

[83] See Demonstration of Two-Qubit Algorithms with a Superconducting Quantum Processor by L. DiCarlo, Rob Schoelkopf et al, 2009 (9 pages).

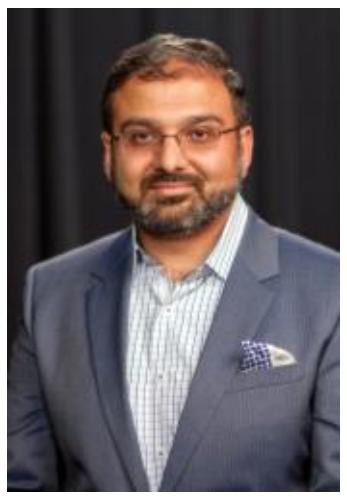

**Irfan Siddiqi** (1976, American-Pakistani) is one key contributor to advancements in superconducting qubits. He did his PhD and post-doc at Yale, working initially in aluminum hot-electron bolometers for microwave astronomy and then, high frequency measurement techniques for superconducting qubits. He developed the Josephson Bifurcation Amplifier that uses the non-dissipative and nonlinear nature of the Josephson junction to create high gain and minimal back action readout of qubits.

This led to the creation of superconducting parametric amplifiers and Josephson traveling wave parametric amplifiers. He then moved at Berkeley University and the DoE Lawrence Berkeley National Laboratory. He works on quantum electrodynamics, quantum error correction, multi-partite entanglement generation and single photon detection. He runs there the Advanced Quantum Testbed, an integrated research platform on superconducting qubits and enabling technologies.

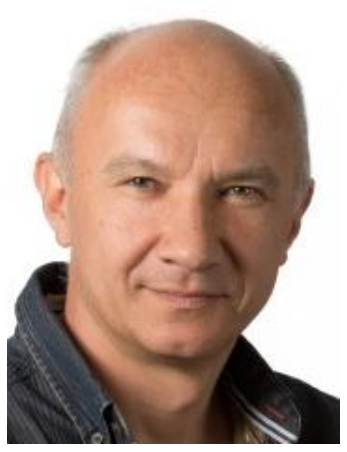

**Artur Ekert** (1961, Polish and English) is a quantum physicist known to be one of the founders of quantum cryptography. He had met Alain Aspect in 1992 to talk to him about this inspiration after discovering the latter's experiments. This is a fine example of step-by-step inventions, one researcher inspiring another! He was the director of the Singapore Center for Quantum Technology from 2007 to 2020. He is also a teacher at Oxford University and a member of Atos's Scientific Council.

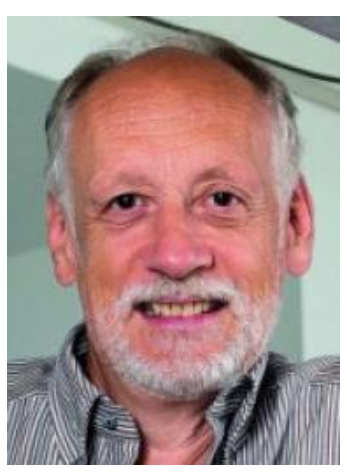

**Nicolas Gisin** (1952, Switzerland) is a physicist specialized in quantum communication. He demonstrated quantum non-locality with an experiment in 1997 over a 10 km distance, extending the performance achieved in the laboratory by Alain Aspect in 1982. He co-founded IDQ in 2001, a Swiss startup initially specialized in quantum random number generators using photons passing through a dichroic mirror. It was acquired by SK Telecom in 2018.

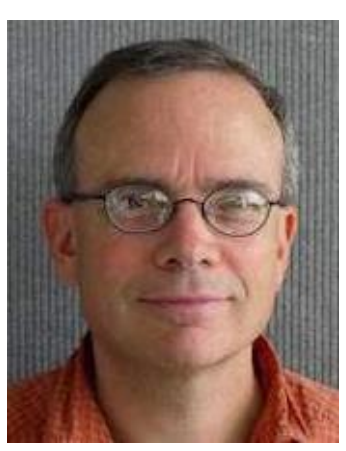

**David DiVincenzo** (1959, American) was a researcher at IBM and the creator of the criteria that define the minimum requirements for a quantum computer with universal gates. He is now a researcher and professor at the University of Aachen in Germany. He is a member of the Atos Scientific Council, along with Alain Aspect, Serge Haroche, Artur Ekert and Daniel Esteve, among others.

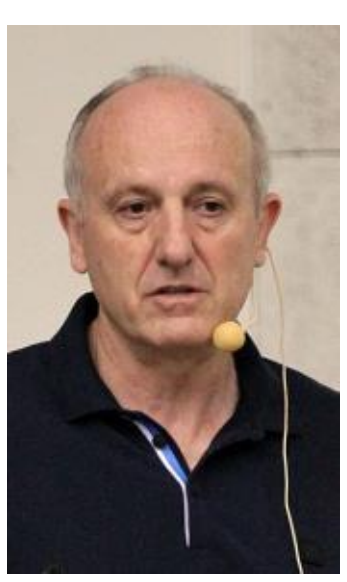

**Daniel Loss** (1958, Swiss) proposed in 1998 with David DiVincenzo to use electron spins in quantum dots to create a quantum computer. He currently is the Co-Director and founding member of National Center on Spin Qubits (NCCR SPIN) that gathers the University of Basel, EPFL and IBM Zurich, an initiative from the Swiss Nanoscale Center SNI. He is the Director of the Center for Quantum Computing at the University of Basel. After a PhD in theoretical physics at the University of Zurich in 1985 he was a post-doc in the group of Anthony J. Leggett in the USA and at IBM Research. After a stint in Vancouver, he went back to Switzerland.

He works on condensed matter physics and spin-dependent and phase-coherent phenomena in semiconducting nanostructures and molecular magnets with applications in quantum computing.

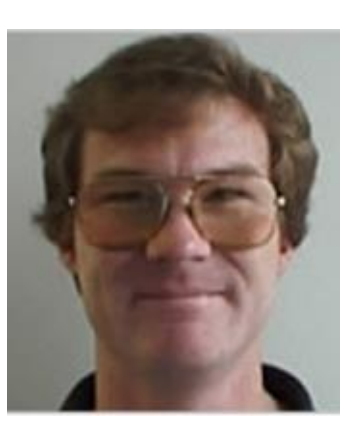

**Bruce Kane** (c. 1958, American) is a researcher at the Joint Quantum Institute from the University of Maryland (a JV with NIST). While he was doing research at UNSW, he presented in 1998 the “donors spin” model, a spin-based qubit concept based on using individual phosphorous atoms in pure silicon lattice structures. This is the principle on which Michelle Simmons works at both UNSW and her startup SQC.

The jury's still out to demonstrate that this technology can scale among the various spin qubits proposals.

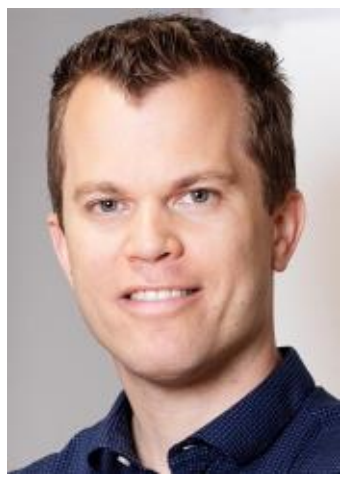

**Menno Veldhorst** (1984, Dutch) is a group leader at QuTech. He got his PhD in 2012 on superconducting and topological hybrids at the University of Twente. He then worked on silicon quantum dots at UNSW where he demonstrated in 2015 the first two qubit operations in silicon. At QuTech, he works on silicon and silicon/germanium (SiGe) qubits to build scalable quantum computers. His team is currently pioneering work on SiGe/Ge qubits with qubits manipulation in arrays up to 16 quantum dots. He proposed a crossbar array architecture to create logical qubits.

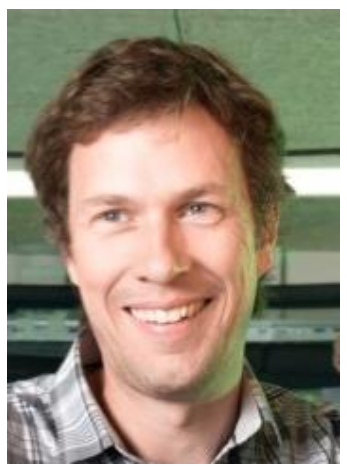

**Lieven Vandersypen** (1972, Belgian) started as a mechanical engineer and a PhD at Stanford, then went to IBM in Almaden, California, where he became interested in MEMS. He demonstrated the use of Shor's algorithm for factoring the number 15 with NMR qubits, and then became a researcher at TU Delft University in the Netherlands and in its QuTech spin-off, which he currently runs. He is a pioneer of electron spin qubits. In this capacity, he works notably with Intel, and is testing their FinFET-based qubit chipsets at QuTech with Intel, which invested $50M in QuTech in 2015.

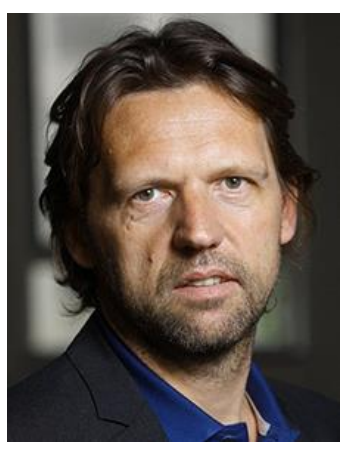

**Leo Kouwenhoven** (1963, Dutch) is a quantum physicist who got his PhD at TU Delft in 1992 and became a professor there in 1999. He led experimental results on the potential "signatures" of Majorana fermion quasiparticles in 2012 and later on their "definitive" existence in 2018. The related Nature paper had to be retracted in 2021 due to experimental data mismanagement and reporting. From 2016 till 2022, he was a researcher at Microsoft Research. He left Microsoft in 2022 and has returned to his home based at QuTech and the Kavli Institute of Nanoscience from TU Delft.

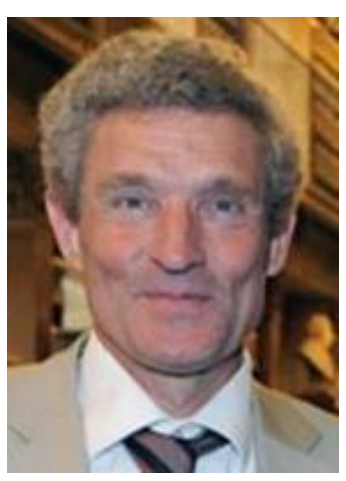

**Christophe Salomon** (1953, French) is a physicist specialized in photonics and cold atoms, research director at the LKB (Normale Sup in Paris). He is particularly interested in quantum gases superfluidity (Bose-Einstein condensates) and in time measurement with cesium atomic clocks. He did a thesis in laser spectroscopy and then did a post-doc at the joint JILA laboratory between NIST and the University of Colorado. He is also a member of the Academy of Sciences since 2017.

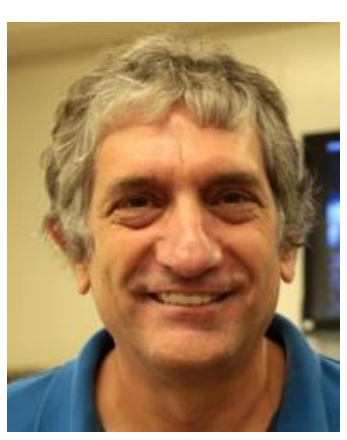

**John Martinis** (1958, American), is a physicist from UCSB who famously worked at Google between 2014 and 2020 where he led the hardware team in charge of superconducting qubits up to creating the Sycamore processor and its related "quantum supremacy experiment", published in Nature in October 2019. After his thesis at Berkeley on superconducting qubits, he did a post-doc in Daniel Esteve's Quantronics laboratory at the CEA in Saclay.

In September 2020, he started to work with Michelle Simmons at SQC in Australia. He also created Quantala in 2020, a quantum computing company selling IP and protecting his own patents.

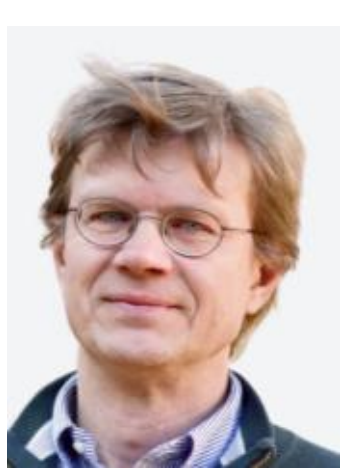

**Mikhail Lukin** (USA) is a Russian born quantum physics professor at Harvard. He's a prolific scientist with a skyrocketing h-index of 163, working on quantum optics, quantum control of atomic and nanoscale solid-state systems, quantum sensing, nanophotonics and quantum information science. He's behind many feats in cold atoms physics as well as in the NV centers field, being the inventor of NV centers based magnetometry.

He cofounded QuEra (USA) that develops a cold atoms gate-based quantum computer, reaching 256 qubits as of 2021. He is also a cofounder and scientific advisor of QDTI (USA).

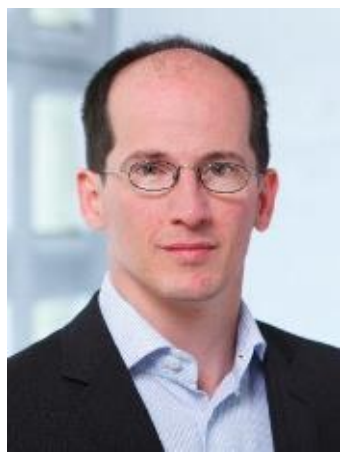

**Andreas Wallraff** (German) is a Professor for Solid State Physics at ETH Zurich after having obtained degrees in physics from the London Imperial College and RWTH Aachen in Germany and worked at the Jülich Research Center also in Germany, Yale University in the USA and the LKB in France. He is specialized in the coherent interaction of single photons with quantum electronic circuits and quantum effects as well as on hybrid quantum systems combining microwave control, superconducting circuits and semiconductor quantum dots.

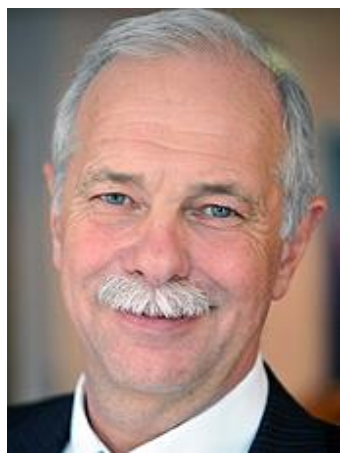

**Jürgen Mlynek** (1951, German) is a physicist specialized in optronics and interferometry. He was the coordinator of the strategic advisory board behind the launch of the European Flagship project on quantum in 2018. We owe him, as mentioned in connection with Louis De Broglie, the experiment validating the wave-particle duality of atoms carried out using helium in 1990 with Olivier Carnal at the University of Konstanz.

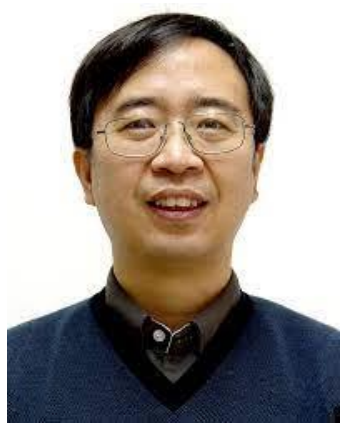

**Jian-Wei Pan** (1970, China) is the leading quantum physics scientist in China. He is a professor and Executive VP at USTC (University of Science and Technology of China) and a member of CAS (China Academy of Science). He did his PhD in Vienna under the supervision of Anton Zeilinger. He and his team are famous for premiere experiments on photons quantum entanglement in 2004, quantum key distribution over a satellite (2017), with boson sampling (2019) and superconducting qubits (2021).

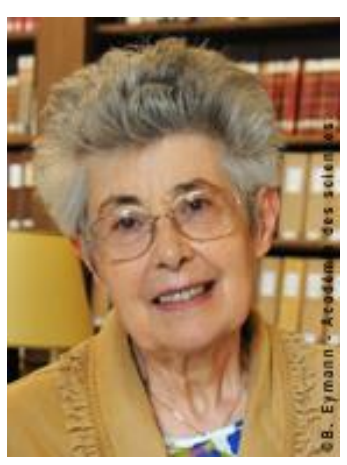

**Marie-Anne Bouchiat** (1934, French) is a specialist in rubidium atoms physics and their control by optical pumping. This is the basis for the creation of quantum computers based on cold atoms. Her daughter **Hélène Bouchiat** (1958, French) is also a physicist, specialized in condensed matter at the LPS laboratory of the University Paris-Saclay and member of the Académie des Sciences since 2010, like her mother who has been there since 1988.

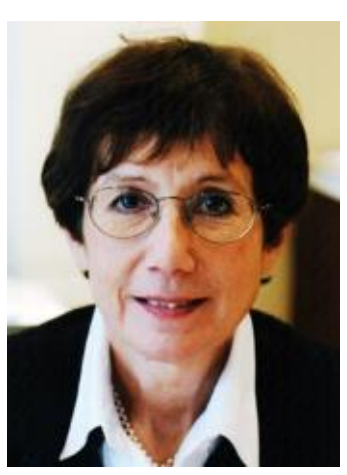

**Elisabeth Giacobino** (1946, French) is a specialist in laser physics, nonlinear optics, quantum optics and superfluidity, particularly in relation to the control of cold atoms. She worked at the CNRS in the ENS LKB (Laboratoire Kastler-Brossel). She is a member of the scientific selection committee of the European Quantum Flagship and also for the ANR (Agence Nationale de la Recherche).

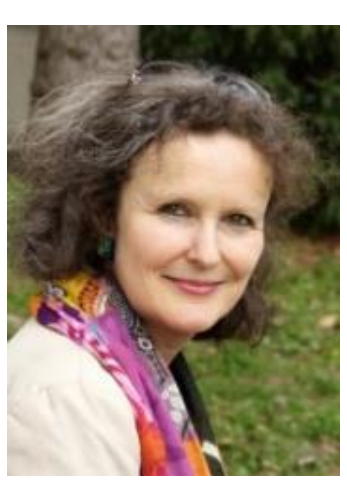

**Jacqueline Bloch** (1967, French) is a research director at CNRS (PI) in the Centre de Nanosciences et de Nanotechnologies (C2N) lab from CNRS and Université Paris-Saclay, working on polaritons, quasi-particles coupling light and semiconductor matter, mainly built in gallium arsenide (GaAs). These have potential applications in the creation of quantum simulators based on polariton arrays as well as for quantum metrology.

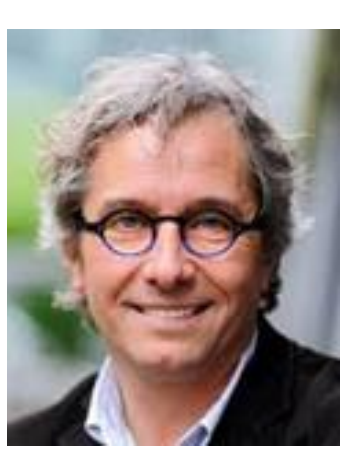

**Jean-Michel Gérard** (1962, French) is a physicist from the CEA IRIG laboratory in Grenoble and director of the joint PHELIQS laboratory (PHotonics, ELectronics and Quantum Engineering) from UGA (University of Grenoble) and CEA. He works in particular on the creation of single photon sources based on quantum dots as well as single photon detectors based on superconducting nanowires and OPO laser diodes.

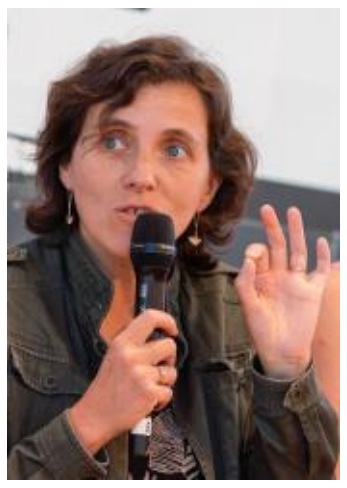

**Pascale Senellart** (1972, French) is a physicist, CNRS research director at the C2N laboratory. She designed and invented a process for manufacturing sources of unique and indistinguishable photons used in quantum telecommunications and computing. These are GaAsAl semiconductor quantum dot trapped in a multi-layered 3D structure, powered by a laser and directly feeding an optical fiber. She co-founded the startup Quandela in 2017 with **Valérian Giesz** (CEO) and **Niccolo Somaschi** (CTO and Chairman) who were a PhD student and a post-doc in her team.

Quandela is selling these photon sources and is creating photon qubit-based quantum computers. She is their scientific advisor. Pascale Senellart also launched the Quantum hub of the University Paris-Saclay in November 2019, which brings together public and private research laboratories as well as higher education institutions. She was awarded the CNRS Silver Medal in 2014.

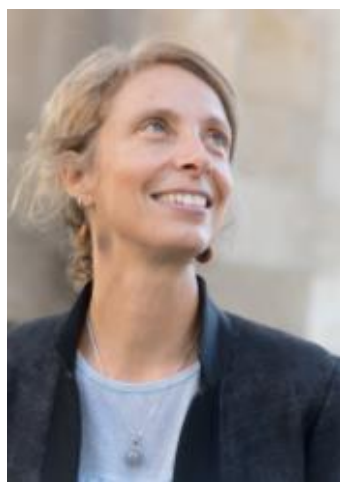

**Maud Vinet** (1975, French) started as physics engineer and was granted a PhD in physics from Grenoble University. She then spent 20 years working in silicon technologies development and transfer for the semiconducting industry. She led the silicon qubit project at CEA-Leti in Grenoble. Since 2016, CEA-Leti was focused on silicon spin qubits leveraging the strong relationships between fundamental science and technology in Grenoble ecosystem. Then, there's a startup being created, which we'll discuss about later in time in an intermediate revision of this book.

The silicon qubit ecosystem in Grenoble involves several laboratories in addition to CEA-Leti: IRIG (also from CEA), CNRS's Institut Néel, LPMMC, and various entities of UGA (Université Grenoble Alpes). Maud is also driving QLSI, the European Quantum Flagship research project on silicon spins qubits, awarded in March 2020, after obtaining with **Tristan Meunier** (1977, French, at CNRS Institut Néel) and **Silvano de Franceschi** (1970, Italian, at CEA IRIG) an ERC funding of €14M in 2018 for the QuCube silicon qubit project. Before her journey in quantum computing, she had previously contributed to the industrialization of the FD-SOI technology with CEA and STMicroelectronics[84], Globalfoundries and IBM.

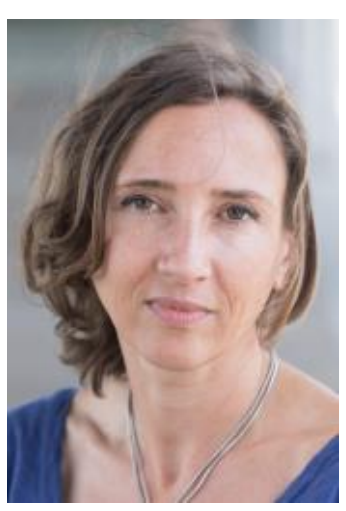

**Alexia Auffèves** (1976, French) is a CNRS research director at Singapore's CNRS MajuLab international laboratory since October 2022 after having conducted her research for over 15 years in Grenoble at CNRS Institut Néel. She is specialized in quantum thermodynamics and collaborates with various teams in France (C2N, ENS Lyon) and around the world (Center for Quantum Technologies in Singapore, Chapman University and Saint-Louis University in the USA, Oxford and Exeter Universities in the UK, Madrid University in Spain and Luxembourg University).

Alexia Auffèves started as an experimentalist, doing he PhD thesis at the ENS LKB in Paris, with Serge Haroche. She then became a theoretician although with quite abroad perspective. She developed the CSM ontology of quantum mechanics (Contexts, Systems and Modalities) with Philippe Grangier and the philosopher Nayla Farouki that we cover later in this book, when discussing quantum foundations, page 987[85]. She launched and coordinated QuEnG (Quantum Engineering Grenoble), the Grenoble quantum ecosystem, which became the QuantAlps federation in January 2022. Her recent work focuses on the energetic aspects of quantum technologies, both from fundamental and full-stack perspectives, which explains why she cofounded the Quantum Energy Initiative in the summer of 2022 with **Robert Whitney** (a physicist from CNRS LPMMC in Grenoble) and Olivier Ezratty. Yes, that's me, the writer of this book.

[84] FD-SOI = Fully-Depleted Silicon on Insulator. The technology uses on the one hand a layer of silicon oxide insulator and on the other hand, channels of undoped silicon between the drain and the source, limiting leakage between the latter two.

[85] See Contexts, Systems and Modalities: a new ontology for quantum mechanics by Alexia Auffèves and Philippe Grangier, 2015 (9 pages). See also the associated Wikipedia page. This work has been articulated on a total of seven papers released between 2015 and 2019.

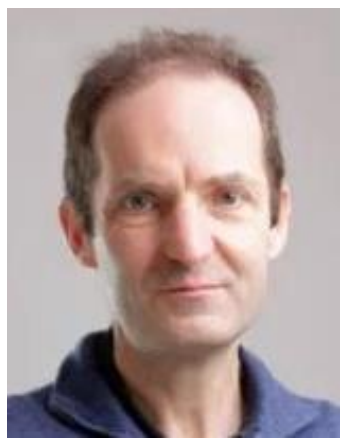

**Antoine Browaeys** (c. 1970, French) is a CNRS research director leading the quantum optics-atom team in the Charles Fabry Laboratory at Institut d'Optique specialized in the control of cold atoms. He is also a cofounder and the scientific director of Pasqal, a startup designing a cold atoms computer that will be first used as a quantum simulator, and then, as a universal gates quantum computer. He was awarded the CNRS silver medal in 2021.

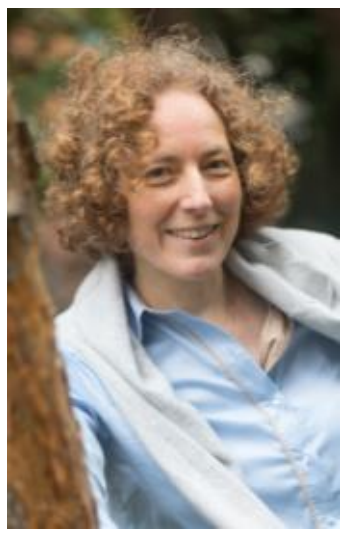

**Hélène Perrin** (c. 1975, French) is CNRS research director working at the Laboratoire de Physique des Lasers (LPL) from Université Sorbonne Paris Nord, working on Bose-Einstein condensates and cold atoms control. Together with Pascal Simon, she drives the Quantum Simulation SIM project, a cold atom-based quantum simulator. She also gives lessons on quantum computing. She did her PhD thesis with Christophe Salomon at the ENS LKB in Claude Cohen-Tannoudji's group. At CEA-Saclay, she also worked on fractional quantum Hall effect. Since 2022, she is the director of QuanTIP, the Paris region quantum ecosystem network.

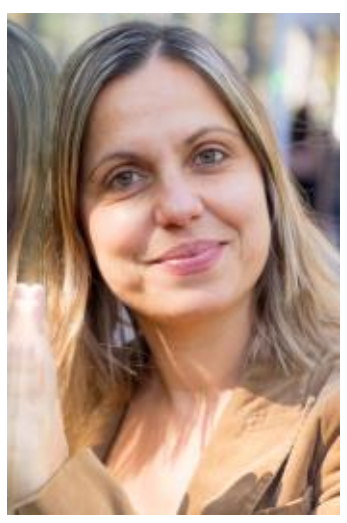

**Eleni Diamanti** (1977, Franco-Greek) is a leading specialist and experimenter in the development of photonic resources for quantum cryptography, also working on quantum communication complexity. She's a CNRS research Director and faculty at LIP6 laboratory from Paris-Sorbonne University. She is the vice-director of the Paris Centre for Quantum Computing since April 2020. She is also involved in many European projects around quantum key distribution, like the Quantum Internet Alliance and OpenQKD. She is a recipient of a European Research Council Starting Grant.

At last, she's a cofounder and a scientific advisor with Julien Laurat for the startup WeLinQ, created in 2022, which creates cold atom based quantum memories for quantum computer interconnects and quantum repeaters.

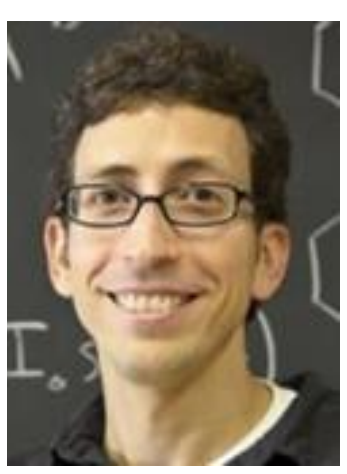

**Jason Alicea** (American) is a Professor of Theoretical Physics at Caltech University's IQIM (Institute for Quantum Information and Matter). He is specialized in condensed matter physics and topological phase of matter which could lead on creating non-Abelian anyons and Majorana fermions, a qubit type mainly explored by Microsoft.

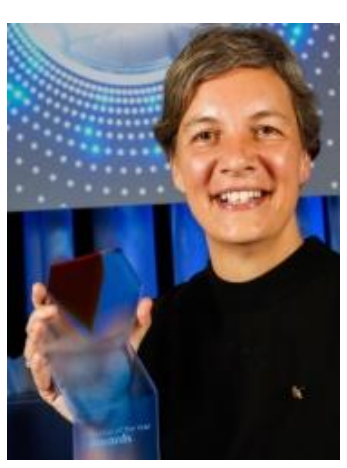

**Michelle Simmons** (1967, British-Australian) is a physicist from the University of New Wales in Australia (UNSW), working on silicon spin qubits. She is the director of CQC2T (Centre of Excellence for Quantum Computation and Communication Technology) from UNSW. She is also the co-founder of SQC (Silicon Quantum Computing), the leading quantum computing Australian startup ($66M), a spin-off from her university and from QQC2T.

In 2019, her team built the first two-qubit gate between phosphorous atom qubits in silicon, operating in only 0.8 ns. It became a full-fledged 10 qubit processor in 2022. She is using STM (scanning tunneling microscopes) to position phosphorus dopants in the silicon substrate.

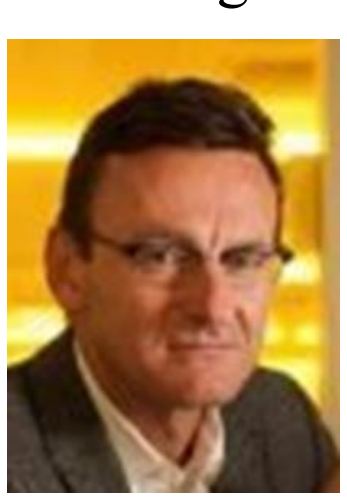

**Andrew S. Dzurak** (Australian) is the Director of the Nanotechnology Fabrication Unit at UNSW's Australian National Fabrication Facility from the CQC2T research center. This facility's white room is used to manufacture silicon qubits chipsets. Andrew Dzurak is a pioneer of silicon qubits since 1998. He is leading research at CQC2T on silicon qubit control and reading. He created the first phosphorus-based silicon double qubits in 2015. He was a lead scientist for SQC, founded by Michelle Simmons, but seemingly left the company in 2021.

He created Diraq in 2022, a startup dedicated to the creation of scalable quantum computers using quantum dot silicon spin qubits.

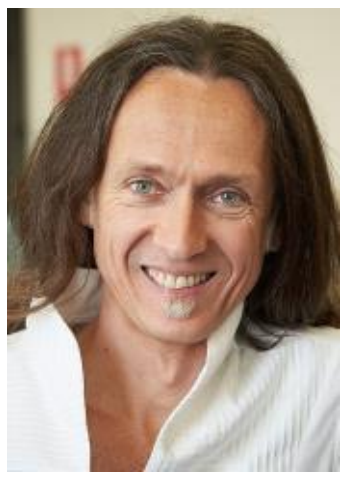

**Andrea Morello** (1972, Italian) is one of the star researchers at UNSW in Australia. He is Program Manager of the ARC Centre of Excellence at CQC2T and leads the Fundamental Quantum Technologies Laboratory at UNSW. During his studies, he attended the Laboratoire National des Champs Magnétiques Intenses of the CNRS in Grenoble. Today he is one of the specialists in silicon-based qubits. He is also a quantum engineering teacher at UNSW.

His team was the first to demonstrate coherent control and readout of an individual phosphorus atom electron and nuclear spin in silicon and held for many years the record for the longest quantum memory time of 35.6 s in a single solid-state qubit.

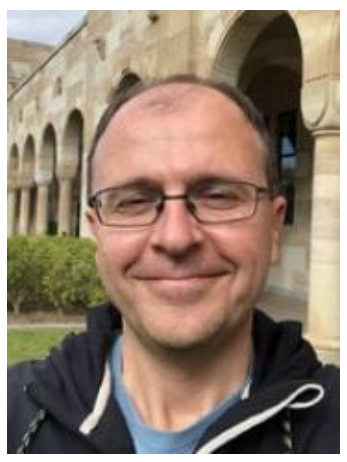

**Andrew G. White** (c. 1970, Australian) is a leading Australian quantum scientist who is the Director of the University of Queensland Quantum Technology Laboratory. He is most known for his work in quantum photonics, including a first demonstration of an optical CNOT entangling gate realized in 2004 and based on the Knill, Laflamme and Milburn (KLM) protocol and linear optics. He is also very eclectic, having also worked on nuclear physics and marine biology. He's a scientific advisor for Quandela.

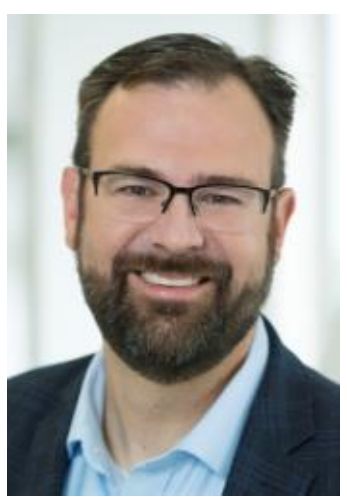

**James Clarke** (c. 1971, American) launched Intel's quantum computing research efforts and the Director of Quantum Hardware at Intel since 2015. He's also behind Intel's partnership with QuTech in The Netherlands. He is currently focused with his team of about 100 researchers and engineer on creating scalable quantum computers with silicon and SiGe qubits. He started working at Intel as a process engineer in 2001 after having studied and worked on organic chemistry (PhD in Harvard and post-doc at ETH Zurich).

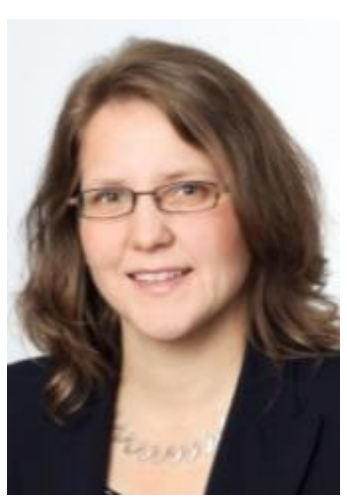

**Christine Silberhorn** (1974, German) is a researcher and professor working on photon-based quantum computing at the University of Paderborn located between Dortmund and Hanover. She leads there the Integrated Quantum Optics group. Her laboratory designs and manufactures integrated optronics components, entangled photon sources and quantum array systems. Her team designed a system to convert photon qubits between infrared and visible wavelengths. She also works on optical quantum memories. She was awarded the Leibnitz prize in 2011.

She cofounded It'sQ in 2022, a quantum photonic computing startup and is one of the very few lead researchers in Germany who created a quantum computing hardware company.

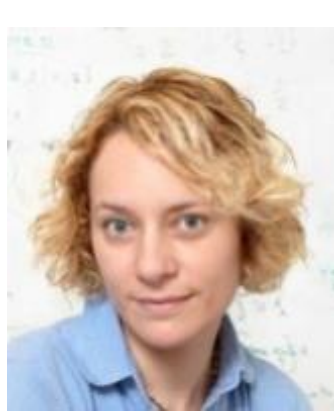

**Stephanie Wehner** (1977, German) is a physicist working on quantum communication protocols, based at the University of Delft in the Netherlands. She coordinates the "Quantum Internet Alliance", one of the projects of the European Quantum Flagship, which plans to deploy a quantum key distribution (QKD) Internet network running in mesh mode. She started her professional life in cybersecurity, detecting system flaws. She is also producing many quantum tech MOOCs.

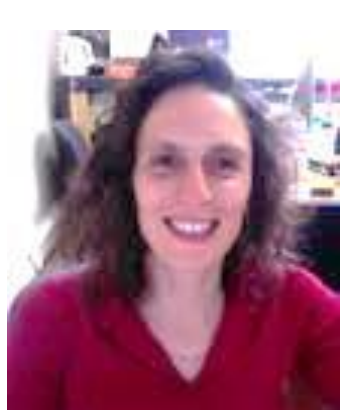

**Perola Milman** (c. 1975, French) is a specialist in the theory of quantum computing and in particular with trapped photons and ions. In particular, she has demonstrated the entanglement capacity of molecules. She is a lecturer-researcher at the Laboratory of Quantum Materials and Phenomena of the University Paris Diderot. She is a professor of quantum theory of light and on quantum entanglement.

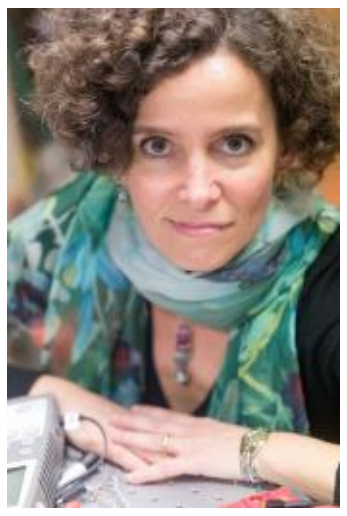

**Sara Ducci** (1971, French) is another teacher-researcher at the same Laboratoire Matériaux et Phénomènes Quantiques (MPQ) where she co-founded in 2002 a team in charge of nonlinear optical devices. She is working on producing pairs of entangled photons sources based on III-V semiconductors. She is also interested in the characterization (state measurement...) and manipulation of photons. At last, she teaches quantum physics at Ecole Polytechnique.

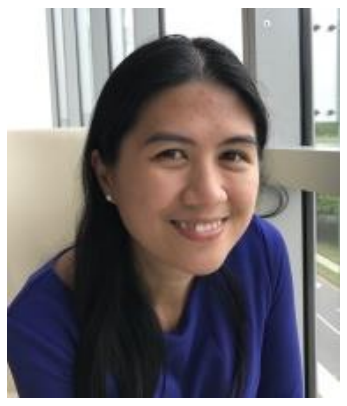

**Jacquiline Romero** (c. 1985, Philippines) is a quantum optics physicist doing research in Australia at the University of Queensland, after completing her PhD in Glasgow, UK. She is working on optical neuromorphic architectures and on dense encoding of information in photons using several of their characteristics in addition to the usual polarization.

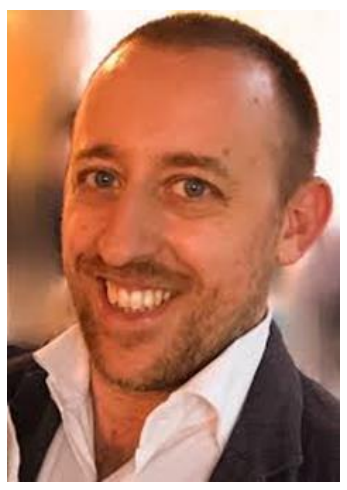

**Fabio Sciarrino** (1978, French Italian) is the director of the Quantum Information Lab at the Sapienza University of Rome and specialized in photonics. His team is at the origin of many advances in the field, notably in boson sampling, a key experiment in the path of photon-based quantum computers. He collaborates with Quandela's team and the C2N of Palaiseau (Pascale Senellart).

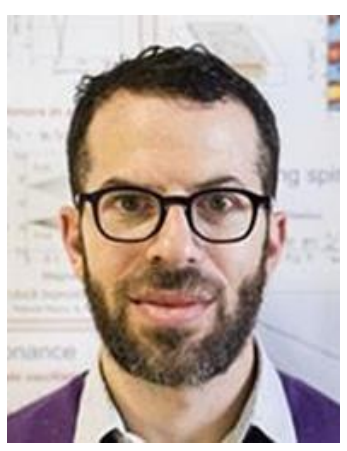

**Patrice Bertet** (c. 1976, France) is part of Daniel Esteve's team at CEA-SPEC. He did his thesis at Serge Haroche on Rydberg atoms and then went to Delft University. He participated in the early days of superconducting qubits (quantronium at CEA and TU Delft). He then worked on QED (quantum electrodynamics) circuits based on cavities and then on transmon qubits. He is working on the association of superconducting qubits and the measurement of their state with electron spins, notably based on NV centers, which can also be used for quantum memories.

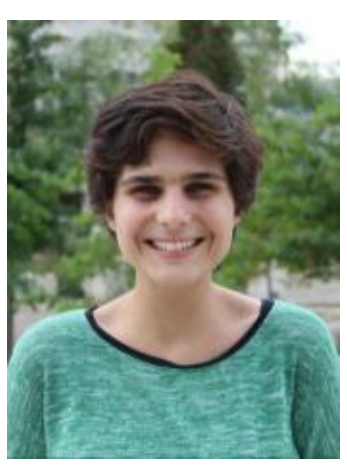

**Audrey Bienfait** (c. 1990, France) is a former PhD student of Patrice Bertet at CEA-SPEC who is now doing her research at ENS Lyon in the team of **Benjamin Huard** (1979, French). She was awarded the Bruker Prize 2018 for her thesis on electron paramagnetic resonance or "ESR - Electron Spin Resonance" in quantum regime and the Michelson Postdoctoral Prize 2019 in March 2020 for her work on the entanglement of superconducting qubits via phonons.

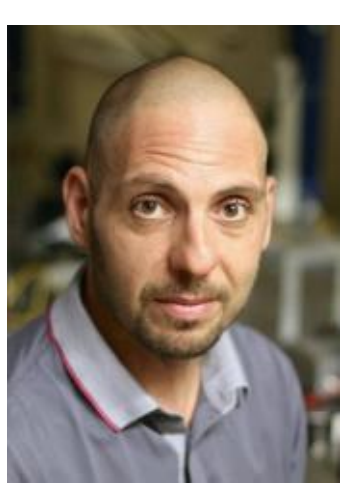

**Sébastien Tanzilli** (France) is the director of the InPhyNi physics laboratory in Nice and also the CNRS national quantum program director. He works on quantum cryptography with continuous or discrete keys (CV-QKD and DV-QKD), in fundamental quantum optics as well as in hybrid quantum systems for the study and realization of quantum communication networks. He was also the president of the GDR-IQFA, a community of quantum physics researchers in France (IQFA = Information Quantique, Fondements & Applications) from its creation in 2011 until 2021.

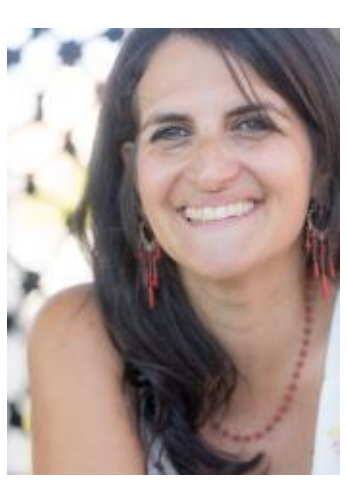

**Virginia D'Auria** (Italy) is a researcher working on quantum optics transmission systems using continuous and discrete variables and DV/QV hybridization. Having worked at the ENS LKB in Paris, she also worked on photon detectors. Since 2010, she is part of the photonics group of InPhyNi and works on discrete and continuous variable quantum communications compatible with optical fibers of telecom operators.

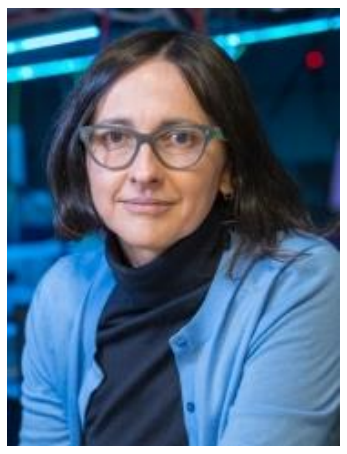

**Jelena Vucokic** (c. 1975, Serbian) is a research professor at Stanford, working in quantum photonics. She directs the Nanoscale and Quantum Photonics Lab and the Q-FARM (Quantum Fundamentals, ARchitecture and Machines initiative), an interdisciplinary quantum laboratory. She contributes to developments in photonics for the development of optical quantum computers. She did her PhD at Caltech in 2002.

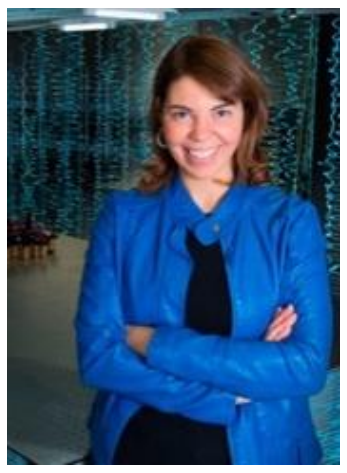

**Francesca Ferlaino** (1977, Italian) is a typically European researcher, having worked in many laboratories from different countries. She is research director at the IQOQI in Innsbruck, Austria, where she leads the Dipolar Quantum Gases laboratory. She is a specialist in cold atoms and erbium-based Bose-Einstein condensates.

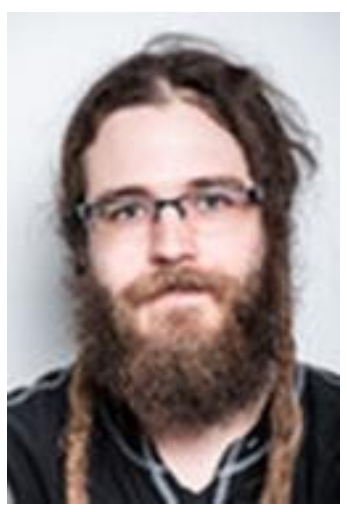

**Marcus Huber** (Austria) is a research group leader at the IQOQI in Vienna, working on quantum entanglement, qubit state measurement and quantum thermodynamics in general. In addition to the IQOQI, he has also worked at the Universities of Bristol, Geneva and Barcelona. He is a great advocate of the open publication of research work, being at the origin of the Quantum-Journal.org website, a kind of arXiv for quantum science.

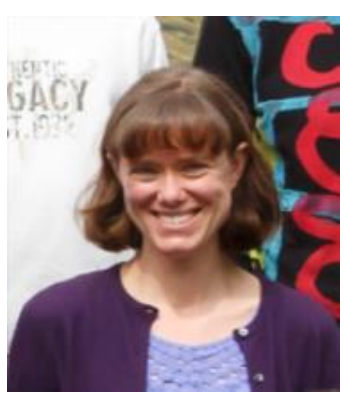

**Tracy Northup** (c. 1975, Austria) is a researcher working on trapped ions and optical cavities, one of the major branches of quantum computing. She leads the Quantum Interfaces Group laboratory at the University of Innsbruck, which is one of the most active in the field of trapped ions, a major Austrian specialty.

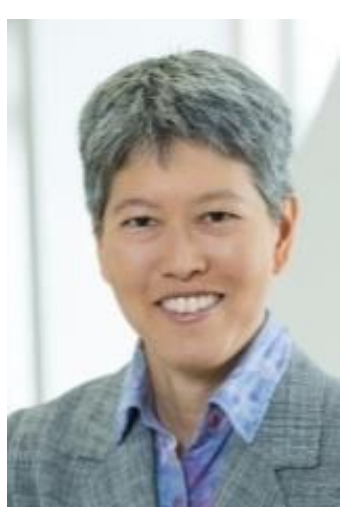

**Anne Matsuura** (c. 1970, Japanese-American) is a physicist who is leading the Quantum & Molecular Technologies team from the Intel Quantum Research Laboratory since 2014. She leads the American's efforts in the creation of superconducting and silicon qubits quantum computers, with an overall vision of the hardware architecture. Her impressive career starts with a thesis at Stanford in synchrotrons, then in US Air Force labs and In-Q-Tel (the CIA investment fund). She also directed the European Theoretical Spectroscopy Facility in Belgium.

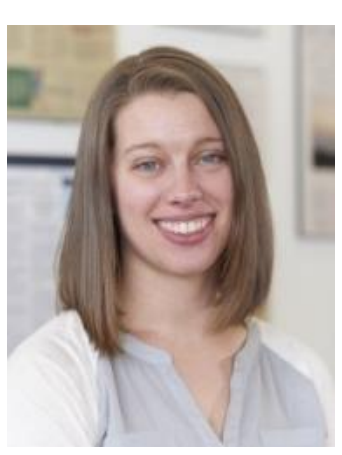

**Sarah Sheldon** (c. 1986, American) has been a member of IBM's quantum computing teams based at the Thomas J. Watson Research Center in Yorktown, New York, since 2013. She is particularly active in improving the quality of superconducting qubits, their quantum gates and error correction codes. She obtained her PhD at MIT in 2013 before doing a post-doc with IBM.

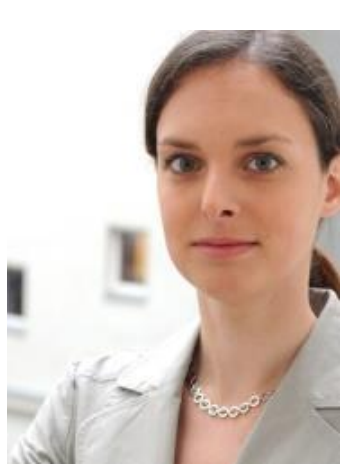

**Stefanie Barz** (c. 1980, German) is a quantum optics professor and researcher at the University of Stuttgart. Her interests include quantum cryptography and quantum telecommunications. She worked in particular on blind computing with Elham Kashefi and Anne Broadbent. She leads the SiSiQ project funded by the German Ministry of Research with €3.6M of European funding, which aims to create quantum communication infrastructure with silicon photonics.

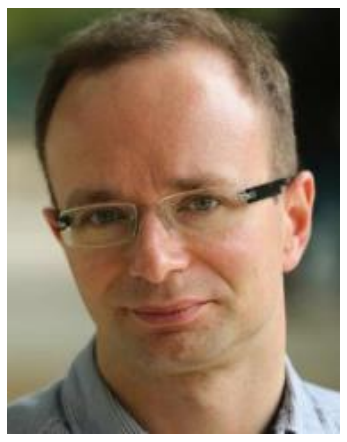

**Alexei Grinbaum** (1978, Franco-Russian) is a researcher at CEA-Saclay in Etienne Klein's LARSIM laboratory. He works on the quantum foundations and quantum physics philosophy[86]. He is notably the author of the book "Les robots et le mal" (Robots and evil) published in 2018. He is particularly interested in the ethics of science, its acceptance by society and responsible innovation.

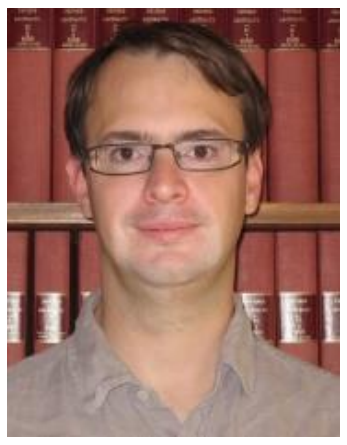

**Frédéric Grosshans** (1976, French) is a CNRS researcher at LIP6 from Université Paris-Sorbonne, specialized in QKD, repeaters and quantum networks. He was the creator with Philippe Grangier of the continuous variable QKD. He is also the co-director with Nicolas Treps (from LKB) of the Quantum Information Center Sorbonne of the Alliance Paris-Sorbonne launched in September 2020, which federates quantum research and training of several Parisian quantum groups.

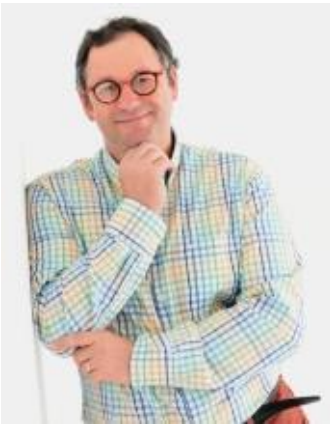

**Jean-François Roch** (1964, French) is a quantum physics professor at ENS Paris Saclay. He is a pioneer of the usage of NV centers in many applications, particularly in quantum sensing, including for studying matter and magnetism at very high-pressure, which could be helpful for the discovery of high-temperature superconducting materials. He conducts these researches in partnership with Thales and with the CEA. He also led the founding Wheeler delayed choice experiment in 2006.

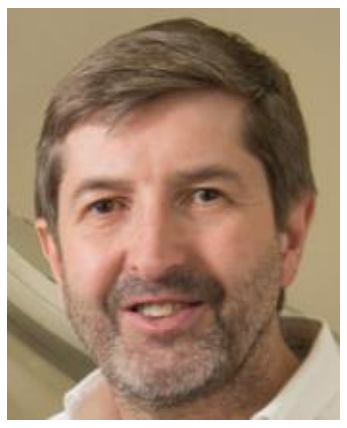

**Ronald Walsworth** (c. 1972, American) is a pioneer in the usage of NV centers for quantum sensing in various fields, from life science to physics and astrophysics like for the detection of dark matter. He leads the Walsworth group at the University of Maryland and is the founding director of the UMD Quantum Technology Center. Several startups emerged from his lab like qdm.io, Hyperfine.io (MRI) and QDTI (which he both cofounded).

He also launched the Quantum Catalyzer quantum startups accelerator (Q-CAT) that creates quantum startups from scratch. He got a PhD in physics from Harvard in 1991.

## Quantum information science and algorithms creators

Let's end this long "hall of fame" with some of the main contributors to the creation of quantum information science and algorithms. It is a relatively new discipline that emerged in the early 1990s.

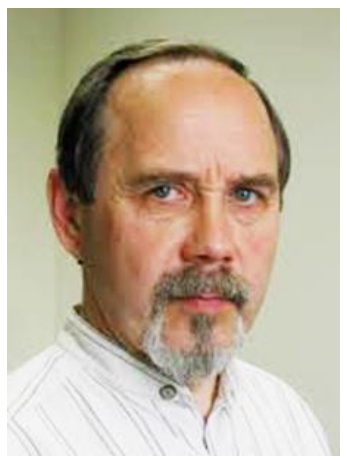

**Alexander Holevo** (1943, Russian) is a mathematician working in quantum information science and who devised the 1973 Holevo theorem according to which we cannot retrieve more than N bits of useful information from a register of N qubits[87]. This is the consequence of the wave packet reduction that reduces the qubit state to its basis states $|0\rangle$ and $|1\rangle$ after measurement. He also developed the mathematical basis of quantum communications.

[86] See Narratives of Quantum Theory in the Age of Quantum Technologies by Alexei Grinbaum, 2019 (20 pages).

[87] This theorem indirectly validates the fact that it is difficult to do "big data" with a quantum computer in the sense of storing and analyzing large volumes of information. On the other hand, Grover's algorithm makes it possible to quickly find a needle in a haystack, as we will see later.

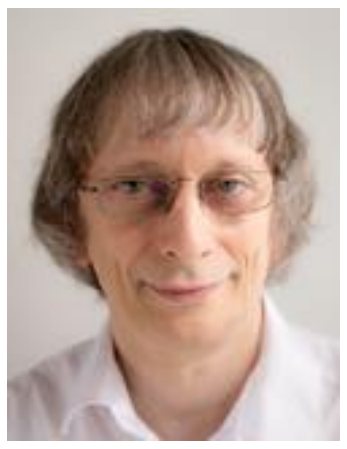

**David Deutsch** (1953, Israeli and English) is a physicist from the Quantum Computing Laboratory at Oxford University in the UK. He devised in 1985 the idea of creating a universal quantum computer using a quantum Turing machine which led him to create in 1989 the gate-based circuits programming model, completing Yuri Manin's and Paul Benioff's 1980 ideas[88]. He is also the author of a search algorithm, with two variants, a first one from 1985 and a second one in 1992 that he co-created with Richard Jozsa.

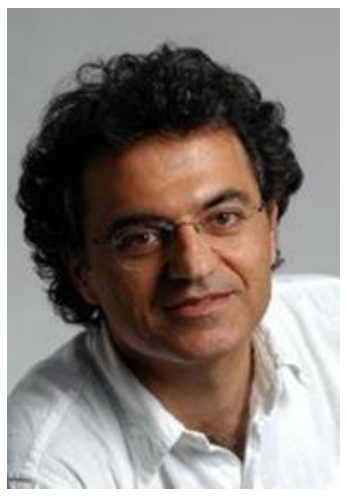

**Umesh Vazirani** (1945, Indian-American) is a professor at the University of Berkeley. He is one of the founders of quantum computing, with his paper co-authored in 1993 with his student Ethan Bernstein, Quantum Complexity Theory. He is also the creator of the Quantum Fourier Transform (QFT) algorithm, which was used less than a year later by Peter Shor to create his famous integer factoring algorithm that served as a spur to funding research in quantum computing in the USA. The QFT is a founding algorithm used in many other quantum algorithms.

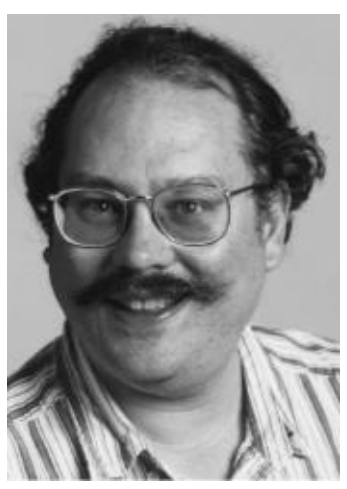

**Peter Shor** (1959, American) is a mathematician who became the father of the algorithm of the same name in 1994 which allows the factorization of integers into prime numbers, based on quantum Fourier transforms (QFT). Before that, he created the first quantum discrete-log algorithm (dlog) and, later, the famous nine-qubit flip error and phase error correction algorithm for quantum computers called the "Shor code"[89]. We indirectly owe to him the whole movement of post-quantum cryptography (PQC).

PQC is about creating cryptography codes resisting to public keys breaking using the Shor algorithm and other quantum algorithms... with quantum computers that do not yet exist. Peter Shor created his famous factorization algorithm while working at Bell Labs. He has been teaching applied mathematics at MIT since 2003.

**Daniel R. Simon** (American) is the creator of another search algorithm in 1994, bearing his name. Precisely, his quantum algorithm solves the hidden subgroup problem (HSP) using an oracle based model, providing an exponential acceleration compared to classical computing[90]. Daniel Simon worked at Microsoft Research when he created his famous algorithm. He later worked on cybersecurity research until his retirement, always with Microsoft Research.

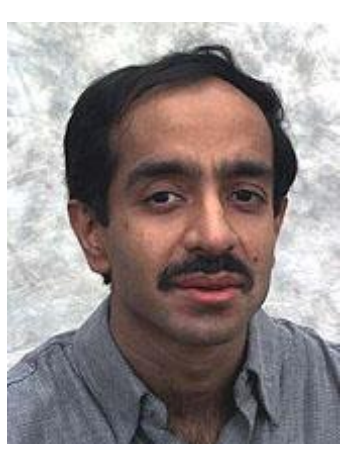

**Lov Grover** (1961, Indian-American) is a computer scientist who created the seminal quantum algorithm in 1996 that is said to be a search algorithm in a database but has many more use cases as we'll see in the quantum algorithms part of this book (page **Error! Bookmark not defined.**). He currently works in the Department of Mathematics of the Guru Nanak Dev University, in Punjab, India. His full name is Lovleen Kumar Grover.

---

[88] See Quantum theory, the Church-Turing principle and the universal quantum computer by David Deutsch, 1985 (21 pages). This is a foundational paper describing a lot of concepts, including the unitaries used in single qubit gates, the notion of quantum computing complexity, etc. It was also followed by Quantum computational networks by David Deutsch, September 1989 where networks correspond to series of gate operations. Back then, the very name of qubit didn't exist yet, and was created only in 1995.

[89] See the excellent The Early Days of Quantum Computation by Peter Shor, August 2022 (10 pages) where Peter Shor recount the history of the early years of quantum computing and how he discovered his various algorithms with try and error.

[90] See On the power of quantum computation by Daniel Simon, 1994 (11 pages) also updated in 1997.

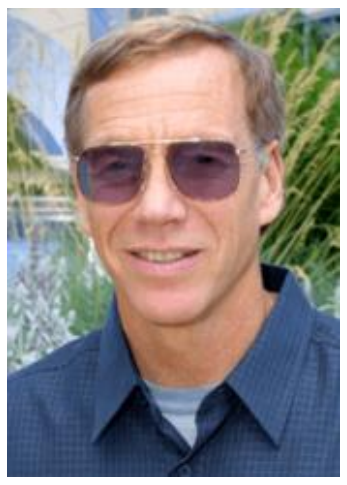

**Michael Freedman** (1951, American) is a mathematician who founded and runs the Microsoft Station Q laboratory in Santa Barbara, California. He is one of the fathers of topological quantum computing along with Alexei Kitaev. He was also awarded the Fields Medal in 1986 for his work on the Poincaré conjecture, later demonstrated in 2006 by Grigori Perelman.

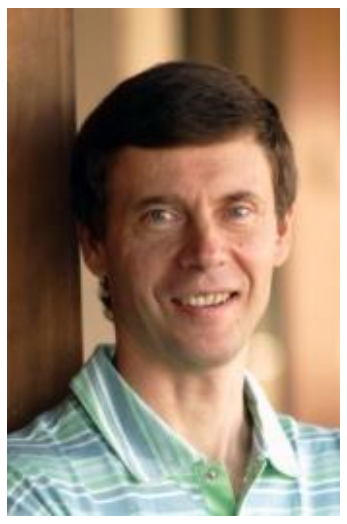

**Alexei Kitaev** (1963, Russian and American) is with Michael Freedman one of the fathers of the topological quantum computer concept in 1997, investigated by Microsoft. He was a researcher at Microsoft Research in the early 2000s and is now working at Caltech University and with Google. He has also done a lot of work on error correction codes, including the creation of toric codes, surface codes and magic states distillation (with **Sergey Bravyi**) and the Quantum Phase Estimate algorithm, used in Shor's integer factorization algorithm.

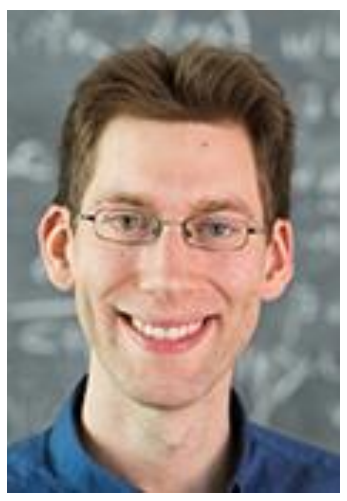

**Aram Harrow** (American) is a prolific specialist in quantum algorithms. He teaches both quantum physics and quantum computing at MIT. At MIT, he is surrounded by Peter Shor and Charles Bennett. He is the co-author of the HHL quantum algorithm used to solve linear equations which he created jointly with Avinatan Hasidim and Seth Lloyd[91] . He is also interested in the creation of hybrid classical/quantum algorithms.

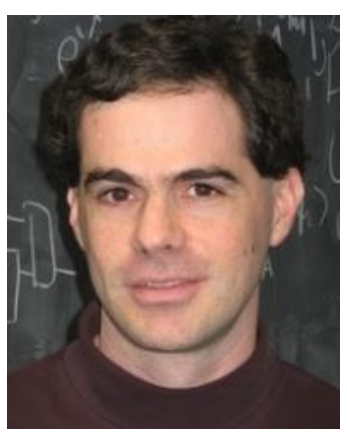

**Daniel Gottesman** (1970, American) is a physicist from the Perimeter Institute in Waterloo, Canada. He did his PhD thesis at Caltech under the supervision of John Preskill. He is known for his work on quantum error correction codes (QEC) and is co-author of the famous Gottesman-Knill's theorem according to which a quantum algorithm using only Clifford gates can be efficiently simulated (meaning, polynomially) on a classical computer.

Clifford group quantum gates are based on half and quarter-turn rotations (of the qubit in the Bloch sphere), Hadamard gate and the C-NOT conditional gate. This theorem thus indirectly proves that a basic gate set is insufficient to generate an exponential quantum advantage. We need to add a T gate to make it possible to approximate any arbitrary unitary transformation, meaning, any move within the Bloch sphere for single qubit operations. This is particularly important for the Shor algorithm.

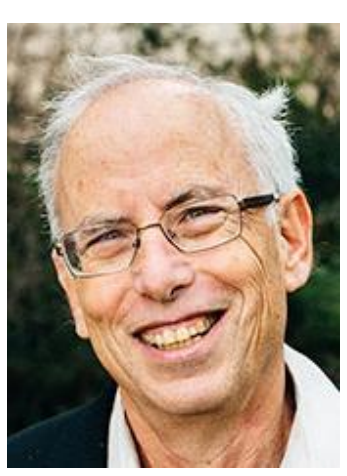

**Gil Kalai** (1955, Israeli) is a professor of mathematics at the Hebrew University of Jerusalem and at Yale University. His main ambition is to demonstrate mathematically that it will be impossible to create real universal quantum computers, due to their error rate, even with error correction codes and the notion of logical qubits that assemble physical qubits. He also questioned the reality of the October 2019 Google supremacy performance in several of his writings and conference talks.

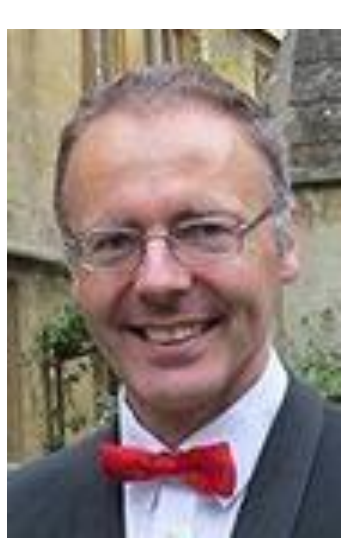

**Andrew Steane** (1965, English) is a Professor of Physics at Oxford University. He created the so-called Steane quantum error correction code in 1996. This code corrects flip and phase errors on a single qubit. Looking at how it works provides good insights on the inner workings of quantum error correction codes, although this particular code will probably not be used when we'll have scalable quantum computers. Other more sophisticated QEC codes are investigated like color codes, surface codes and Floquet codes.

---

[91] See Quantum algorithm for linear systems of equations, 2009 (24 pages).

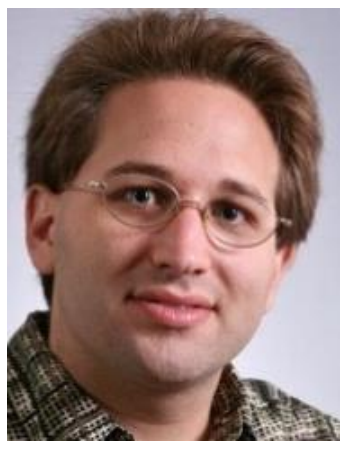

**Scott Aaronson** (1981, American) teaches information science at the University of Austin in Texas. He is a leading expert in quantum algorithms and complexity theories. He is notably at the origin of a quantum algorithm used for boson sampling, a way to demonstrate some quantum advantage for photonic based experiments. Bosons are integer spin particles such as photons, while particles such as electrons, neutrons and protons are fermions, with a spin 1/2.

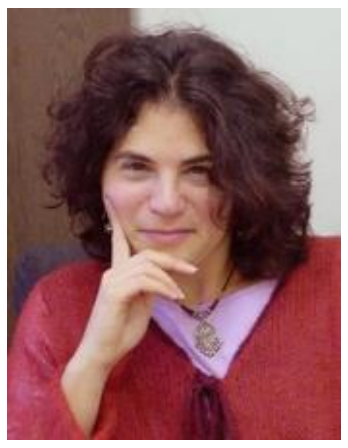

**Dorit Aharonov** (1970, Israeli) is a quantum algorithms researcher. She received her PhD in Computer Science in 1999 at the Hebrew University of Jerusalem on "Noisy Quantum Computation" and then did a post-doc at Princeton and Berkeley. She is credited with the "quantum threshold theorem" co-demonstrated with Michael Ben-Or which states that below a certain error rate threshold, error correction codes can be recursively applied to obtain an arbitrarily low error rate of logical qubits.

This is a very theoretical mathematical approach that doesn't take into account the way noise is also scaling as we increase the number of qubits. Dorit Aharonov's uncle is **Yakir Aharonov** (1932, Israeli), a physicist who had worked with David Bohm, among others.

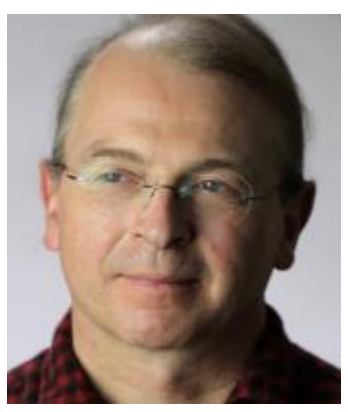

**Seth Lloyd** (1960, American) is a professor at MIT who is a prolific contributor to quantum information and quantum algorithms. He is the initiator of Quantum Machine Learning, of the concept of qRAM (quantum random access memory), of continuous variables gates-based quantum computing (1999), of quantum radars (2008). He's also the L in the famous HHL quantum linear equation solving algorithm and worked on quantum error correction codes and quantum biology.

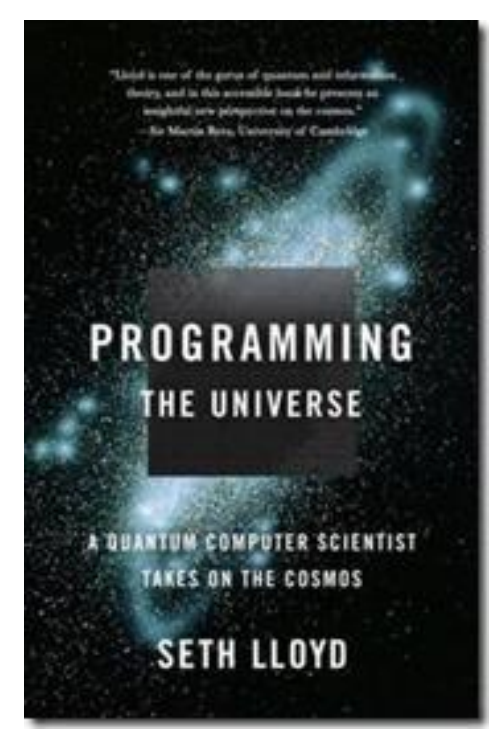

In his 2006 book, Programming the Universe, Lloyd contends that the universe itself is one big quantum computer producing what we see around us, and ourselves, as it runs a cosmic program. According to Lloyd, once we understand the laws of physics completely, we will be able to use small-scale quantum computing to understand the universe completely as well. In about 600 years.

Seth Lloyd was laid off from MIT in 2019 then put on leave, then on disciplinary actions for a period of five years starting in 2020 because he had not informed his management of some Jeffrey Epstein originated funding.

In 2016, he created Turing (2016, USA) with Michele Reilly, a software company working on hybrid classical-NISQ software solutions using AI and quantum machine learning techniques.

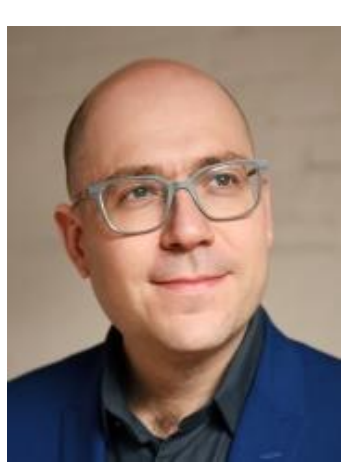

**Alán Aspuru-Guzik** (circa-1978, American) is a research director at the University of Toronto, formerly at Harvard, who, among other things, created various quantum chemistry algorithms, a topic we will cover in the section dedicated on quantum algorithms. He is also the co-founder of the Zapata Computing, a startup developing quantum computing software frameworks, particularly in chemical simulation.

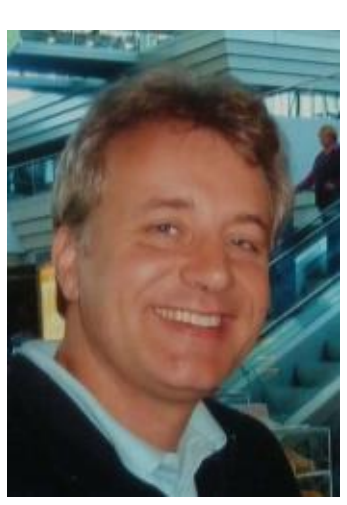

**Robert Raussendorf** (c. 1975, German) is well known for having invented one-way quantum computing and measurement-based quantum computing (MBQC) along with **Hans Briegel** (1962, German) in the early 2000's. He is an Associate Professor at the Department of Physics and Astronomy of the University of British Columbia. He did his thesis at the Ludwig Maximilians University in Munich, Germany in 2003 on MBQC.

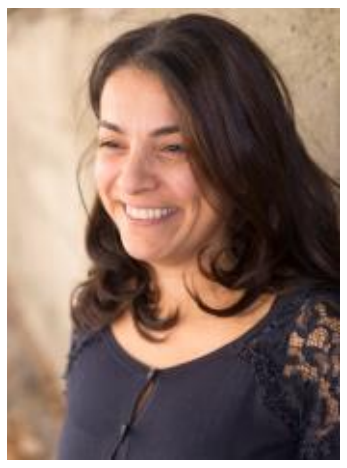

**Elham Kashefi** (1973, British Iranian) is a research director at CNRS in France, in the LIP6 laboratory from Sorbonne University. She is also the co-founder with Marc Kaplan of VeriQloud, a secure quantum telecommunications startup, and teaching quantum information science at the University of Edinburgh. Originally a mathematician and computer scientist, she became a specialist in quantum communication protocols and quantum algorithms, around topics like code verification and blind quantum computing.

She did her PhD thesis "Complexity Analysis and Semantics for Quantum Computation" at the Imperial College of London in 2003 under the co-supervision of Peter Knight. She created the BFK blind computing protocol in 2009 with Anne Broadbent and Joe Fitzsimons (who created Horizon Quantum Computing in Singapore).

With her team at LIP6, she is at the origin of the creation of a site on the zoo of quantum communication protocols[92]. And as this was not enough, she is also versed in Quantum Physical Unclonable Functions (QPUF), physical identifiers of quantum and tiltable objects, a topic we briefly cover in this book in page 844.

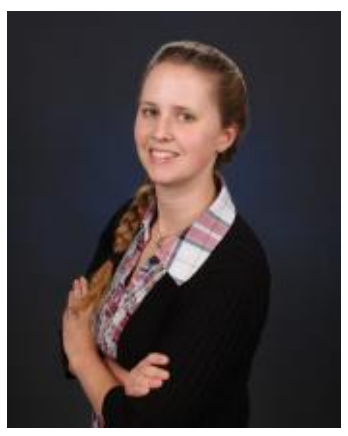

**Anne Broadbent** (Canadian) is a mathematician from the University of Ottawa specialized in quantum computing, quantum cryptography and quantum information. She was a student of Alain Tapp and Gilles Brassard at the Université de Montréal. She created the BFK blind computing protocol in 2009 along with Elham Kashefi and Joe Fitzsimons.

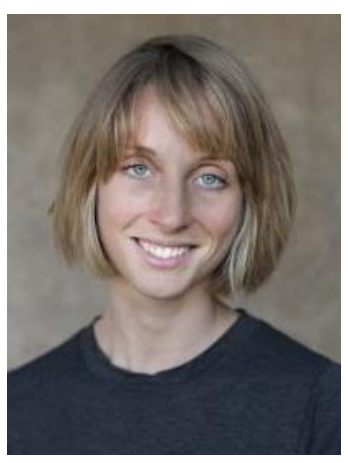

**Maria Schuld** (c. 1989, German) is a senior researcher and software developer at Xanadu since 2017, based in South Africa at the University of KwaZulu-Natal in Durban where she got her PhD in quantum machine learning and was then a post-doc after a short internship at Microsoft Research in the USA. She is a key contributor to the development of quantum machine learning algorithms, particularly in the field of pattern recognition.

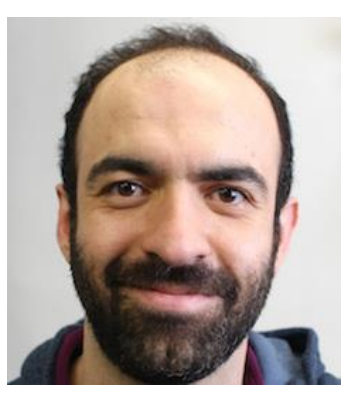

**Mazyar Mirrahimi** (circa 1980, Iranian) is a mathematician who moved to quantum physics. He is currently the director of Inria's Quantic laboratory, which specializes in error correction codes and quantum algorithms, among other topics. He did his post-doc with Michel Devoret at Yale University. Back in 2013, he published a seminal paper on cat-qubits.

These are physical qubits using a cavity and a superconducting qubit that self-corrects some errors, starting with flip errors. These cat-qubits are used by the startup Alice&Bob as well as by Amazon, as announced in December 2020.

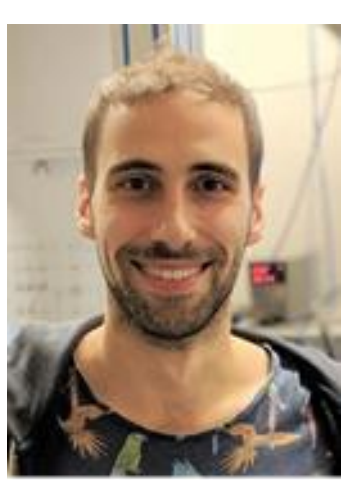

**Zaki Leghtas** (Morocco/France) is a researcher based in France in Mazyar Mirrahimi's team and is also specialized in error correction codes and systems. He is notably one of the creators of cat-qubits mentioned above. These are supposed to enable the creation of logical qubits with fewer than 100 physical qubits. He worked in Michel Devoret's laboratory at Yale University before joining Inria's Quantic team in 2015. He is also affiliated with ENS and Mines ParisTech.

[92] See the Protocol Library wiki.

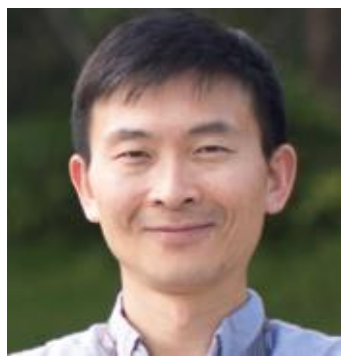

**Shi Yaoyun** (1976, Chinese) is a professor at the University of Michigan and also leading the Alibaba Quantum Laboratory which develops fluxionium superconducting qubit computers. He created various records of quantum simulation on server clusters that we will describe in this book. He earned a computer science PhD from Stanford. He also worked on quantum cryptography and certifiable randomness.

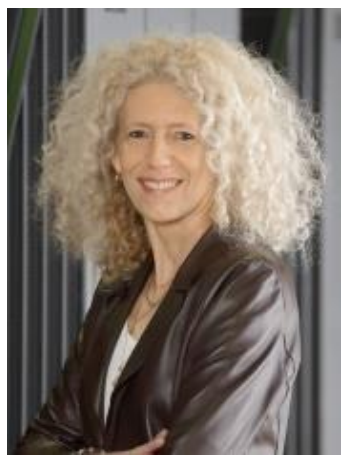

**Kristel Michielsen** (circa-1969, Belgian) is a physicist working at the University of Aachen in Germany and at the Jülich Supercomputing Centre (JSC) where she leads the Quantum Information Processing (QIP) research group. She has contributed to numerous works in quantum computing both in physics and algorithms. She created the QTRL scale, for Quantum Technology Readiness Level, that is used to evaluate the level of maturity of quantum technologies and which we will discuss in the section dedicated to practices in research.

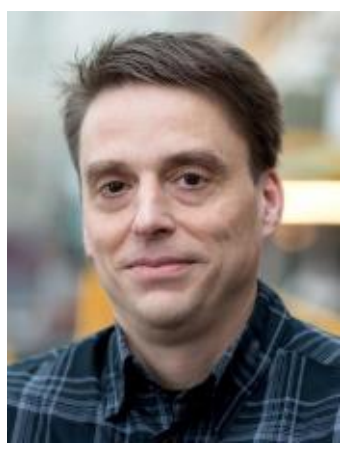

**John Watrous** (Canadian) is a researcher working at the University of Waterloo, Canada, specialized in quantum algorithms and complexity theory. He demonstrated some complexity classes equivalencies like QIP is in EXP and QIP=PSPACE. He also worked on cellular automata. He had previously collaborated with Scott Aaronson. He is the author of the voluminous The Theory of Quantum Information, 2018 (598 pages).

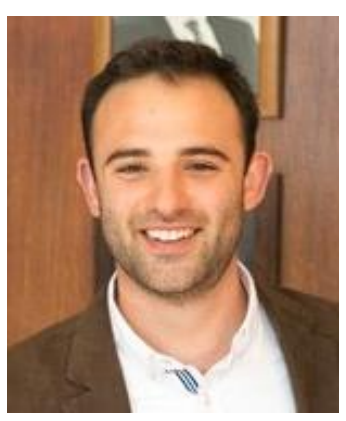

**Ryan Babbush** (circa-1989, American) is a Google researcher working on quantum simulation algorithms. His goal is to create commercial quantum chemistry solutions. In a February 2020 presentation, he did show that chemical simulation with Google's Sycamore 53 qubits processor could not use more than 12 qubits because of its high error rate.

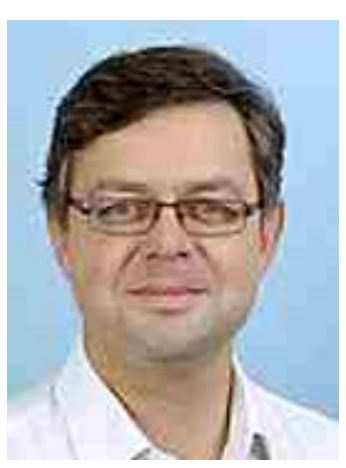

**Matthias Troyer** (1968, Austrian) is Professor of Computational Physics at ETH Zurich. He joined Microsoft Research in Redmond at the beginning of 2017. He is one of the creators of the Q# language for quantum programming and of the open source framework ProjectQ launched in 2016 by ETH Zurich. He is particularly interested in chemical simulation with quantum computers. He received his PhD from ETH Zurich in 1994.

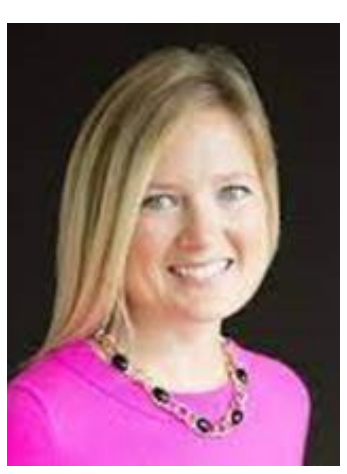

**Krysta Svore** (c.1978, American) is currently the general manager of quantum software at Microsoft. She has a Ph.D. in Computer Science from Columbia University. Her contribution in quantum information science covers a broad range of topics: MBQC, quantum machine learning, contributing to the creation of the LIQUi|> quantum programming language, surface codes, fault-tolerance quantum computing.

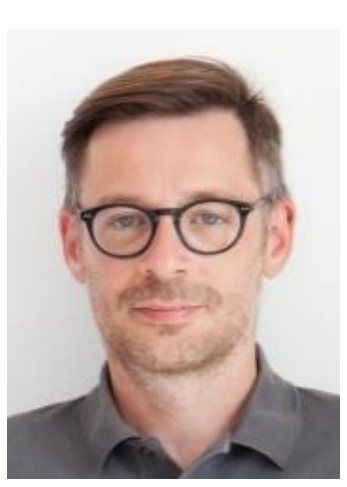

**Iordanis Kerenidis** (c. 1980, Greek) is a director of research from CNRS at IRIF (Institut de Recherche en Informatique Fondamentale), working on cryptography, quantum communication, quantum complexity theories and quantum machine learning, his latest specialty. He did his thesis at MIT under the supervision of Peter Shor and worked in the same office as Scott Aaronson and also worked at Berkeley with Umesh Vazirani. He is part of the founding team of QC Ware.

There he leads the R&D in quantum algorithms. He also co-leads the Paris Quantum Ecosystem (PCQC) with Eleni Diamanti. He was one of the members of the parliamentary mission on quantum technologies led by MP Paula Forteza between April 2019 and January 2020.

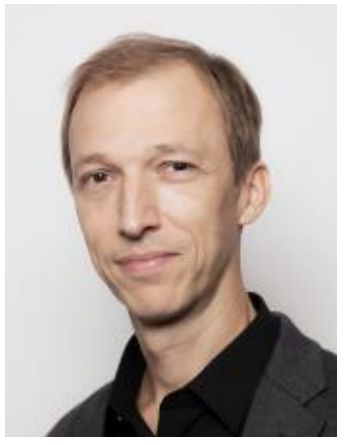

**Frédéric Magniez** (French) is the Director of the CNRS IRIF laboratory mentioned above. He also did run a Chair at Collège de France in Spring 2021. His research focuses on the design and analysis of randomized algorithms for processing large datasets, as well as the development of quantum computing, particularly algorithms, cryptography and its interactions with physics. In 2006, he founded and led the national working group for quantum computing, bringing together 20 research groups.

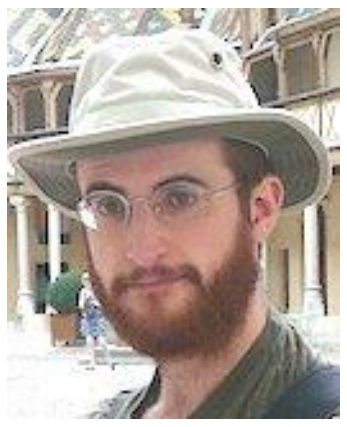

**Benoît Valiron** (1980**,** France) is a researcher at the CNRS LIR laboratory from Université Paris-Saclay and teaching quantum programming and algorithms, including at CentraleSupelec. This quantum programming specialist is the co-author of the open source quantum programming language Quipper, which he contributed to create while being at the University of Pennsylvania.

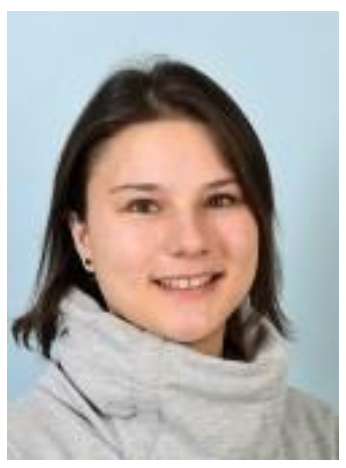

**Bettina Heim** (c. 1980) is a Microsoft developer specializing in quantum software. She is responsible for the development of the quantum programming language Q# compiler, promoted by Microsoft since 2017 and which is part of their Quantum Development Kit, currently running on quantum emulators on traditional processors and now supported on third party hardware proposed on the cloud, including IonQ and Honeywell trapped ion based quantum processors.

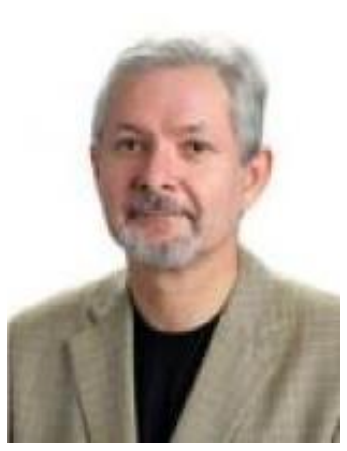

**Cristian Calude** (1952, Romanian/New Zealander) and **Elena Calude** (Romanian/New Zealander) are researchers from the Institute of Information Sciences, University of Albany in Auckland, New Zealand. They work on quantum algorithms, hybrid quantum algorithms and complexity theories.

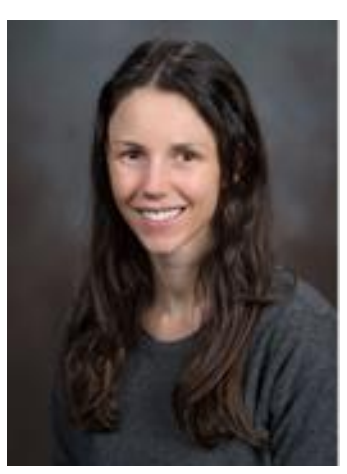

**Sophia Economou** (c. 1980, Greek-American) is an Associate Professor in the Department of Physics at Virginia Tech College of Science. She previously worked at the US Naval Research Laboratory. She is a physicist specialized in the control of quantum dot semiconductor spins and their spin-photon interfaces. She is also a creator of advanced molecular simulation algorithms on quantum computers.

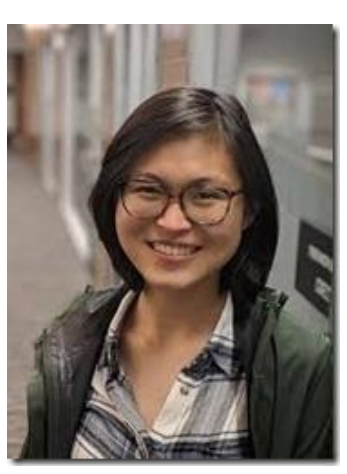

**Ewin Tang** (2000, American) published in July 2018 a paper demonstrating a classical recommendation algorithm as efficient as an algorithm designed for D-Wave quantum computers by Iordanis Kerenidis and Anupam Prakash in 2016[93]. They responded by finding a flaw in the reasoning. On close inspection, the quantum algorithm would scale better in some extreme conditions. She was 18 years old at the time. Ewin Tang is now a computer scientist at the University of Washington.

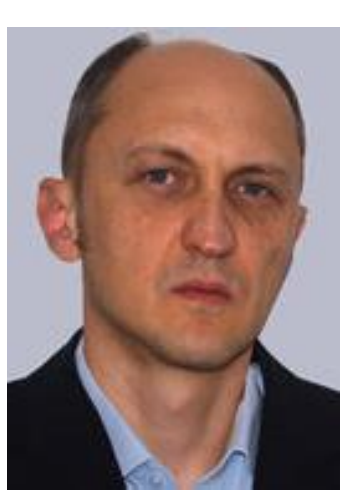

**Cyril Allouche** (French) has been leading Atos R&D efforts in Quantum Computing since its beginning in 2015. Cyril Allouche are the "implementers" of the quantum vision of Thierry Breton, CEO of Atos until 2019. His work encompasses developing the aQASM (Atos Quantum Assembly Language) quantum programming language and the myQLM quantum programming emulator running on regular personal computers and servers.

[93] See A quantum-inspired classical algorithm for recommendation systems, Ewin Tang, July 2018 (32 pages) and Major Quantum Computing Advance Made Obsolete by Teenager by Kevin Harnett, July 2018.

Here we are. We've covered a whole lot of people and probably missed many who should be in this hall of fame list! I'll update it whenever required. We will encounter many of these scientists in this book.

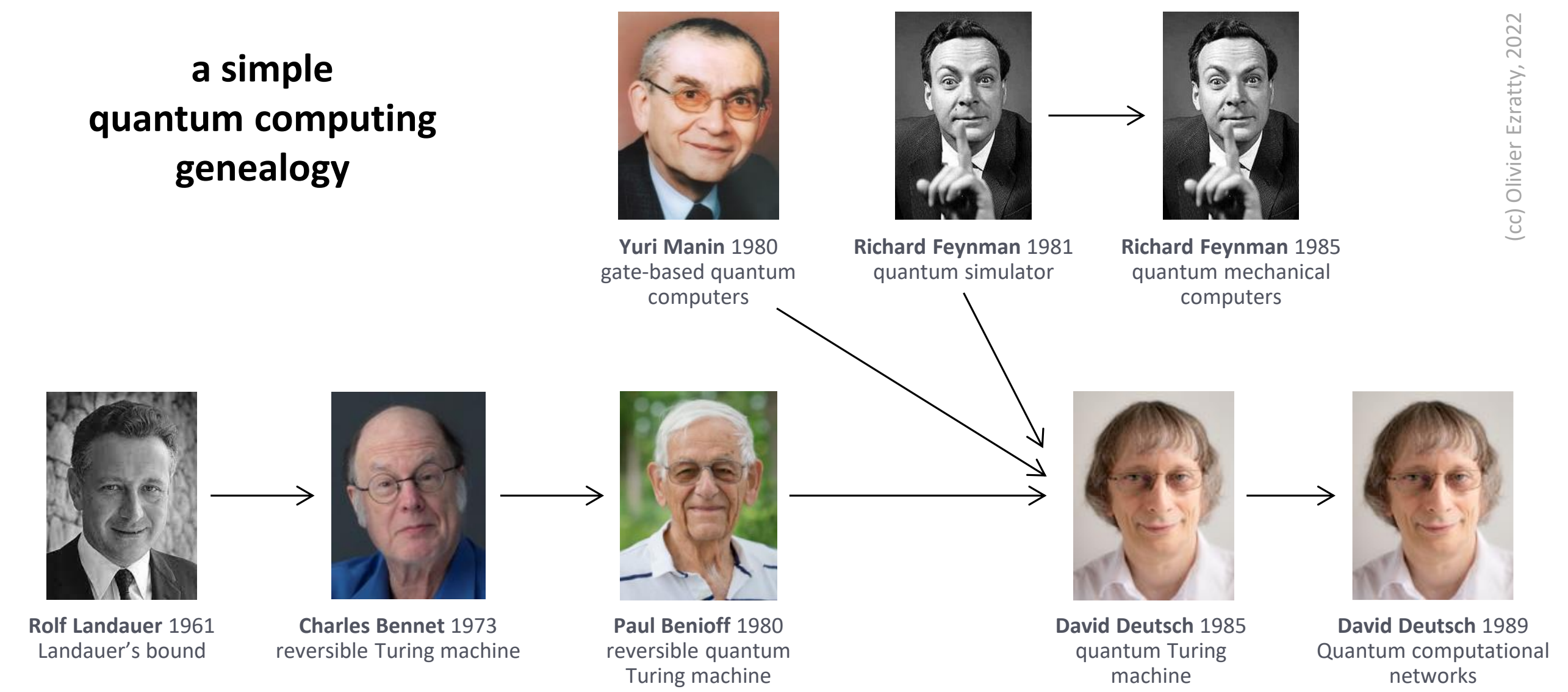

*Figure 58: quantum computing genealogy to remind us that other scientists than Richard Feynman have to be remembered for their contribution. (cc) compilation Olivier Ezratty, 2022.*

# Research for dummies

As I investigated the broad quantum science and technology landscape, I learned more on how fundamental and applied research was operating.

I did not know much about it before this adventure. Working in the 'digital world', as a developer, marketer and in the entrepreneurial ecosystem doesn't necessarily make you look deeply into the inner workings of research. I discovered many aspects that I am detailing here, particularly with regards to practices, lingua-franca, careers and evaluations.

If you're a researcher, this is very basic stuff that you already know fairly well. For others, it will clarify some of vague understanding you might have on how research works.

### Long-term

The first key point is the long-term approach in quantum technologies. It can also be found in other branches of physics and so-called deep-tech related sciences. Time scales are measured in decades. It starts with intuitions, creativity, passion, rigor and hard work. These ideas are not always broadly adopted right away. There's always some resistance with the current scientific establishment.

This long-term history can be observed in condensed matter physics. Brian Josephson devised the Josephson junction in 1962. IBM tried to use it unsuccessfully to build superconducting computers. Anthony Leggett made significant discoveries in the early 1980s which led to the creation of the first superconducting qubits in the early 2000s and to Google and IBM's superconducting machines between 2016 and 2020. And we're not done there since this technology's scalability has not yet been proven.

Alain Aspect's work, which started in the late 1970s and culminated with his 1982 experiment had no immediate industrial application. Fortunately, he was well supported by many laboratories, particularly to build the necessary instrumentation. His work led to the creation of many of the branches of quantum technology. For example, Artur Ekert was inspired by Alain Aspect's work to advance the field of quantum cryptography in the early 1990s.

All of this cannot be meticulously planned in advance. Research serendipity must prevail. Commercialization comes later, through meetings between specialists from different and complementary disciplines. Innovators are either the researchers themselves, or more generally others, engineers and entrepreneurs, who know how to detect research work having some business potential. Hence the importance of bringing them together in innovation ecosystems. However, in its current shape, the quantum startup ecosystem is mostly made of researchers turned into entrepreneurs.

This generates its share of misunderstandings with public authorities. They are tempted to over-evaluate and measure the performance of basic research, if not to fund it, using only criteria from the business world. On the other hand, and this is particularly true for quantum technologies, research work requires peer reviews. This may give the impression that researchers are both judge and jury. To prevent this from driving decision-makers and people suspicious, research work must honestly be translated in layman's terms. This should encourage researchers to communicate with broader audiences than their peers. It requires leadership. Scientists must be more involved there, particularly in those times where people are more and more skeptic on science and innovation.

**Publications**

This book contains many references to scientific publications. I do this almost systematically and always look for the original scientific publication whatever the news.

Research is now frequently published first in open access in the famous **arXiv** site managed by Cornell University. These are articles pre-prints that have not yet gone through peer reviewing and be published in peer-reviewed journals. These articles must sometimes be taken with a grain of salt. However, they allow authors to collect comments from informed readers. Their quantity and quality depend on the author's fame, the topic and the number of researchers who master it[94].

Between 9 and 18 months later, a paper publication in a peer-reviewed journal may follow. If the delay is too short, it may mean the journal is a predatory one. It is usually published mostly as is, includes some revisions suggested by the "referees" of the review committees, or even with a change of title. In these cases, the version published on arXiv is not necessarily the most recent. It is sometimes updated. The benefits are openness and free access.

As a general rule, when I discover the existence of an article, I search for it on Google Search with the name followed by "filetype:PDF" and I find it free of charge in more than 90% of the cases on arXiv or on the ResearchGate site, the researchers' reference social network.

Quantum technologies peer-reviewed[95] journals include **Nature** and its various thematic variations like **Nature Communications**, **Science**, **Physical Review X**, **Physical Review Research, Physical Review Letters**, **Quantum Science and Technology**, **Journal of Applied & Computational Mathematics, International Journal of Quantum Information, Quantum Engineering, Advanced Quantum Technologies, Quantum Journal, Quantum Information Processing, IEEE Journal of Quantum Electronics, and** IEEE **Transactions on Quantum Engineering.** Fortunately, in this field, there are only a few predatory journals that do not have peer-review process and charge researchers for their work publication.

---

[94] See Comment bien lire et comprendre une étude scientifique par Gary Dagorn, Mathilde Damgé et Bessma Sikouk, May 2021. It provides a lot of insights on how to read a scientific paper. You can translate this article in French in your browser. Also look at Ten simple rules for reading a scientific paper by Maureen A. Carey, Kevin L. Steiner and William A. Petri Jr, July 2020.

[95] In peer-reviews journals, the reviewers are unknown to the paper authors. They provide some feedback on the paper and expect a paper update. The authors provide an updated version and comments that are either accepted or rejected by the reviewer. It can lead authors to modify their claims and even their paper title. When everything's finalized, the paper can be published. Nowadays, the initial paper published on arXiv is also updated to reflect these changes. There is also a special double-blind review process where the authors are unknown from the reviewers to avoid any reviewer bias. I have bumped only once on such a case in quantum technologies, on a QML algorithm: On the universal approximability and complexity bounds of deep learning in hybrid quantum-classical computing, 2021 (15 pages).

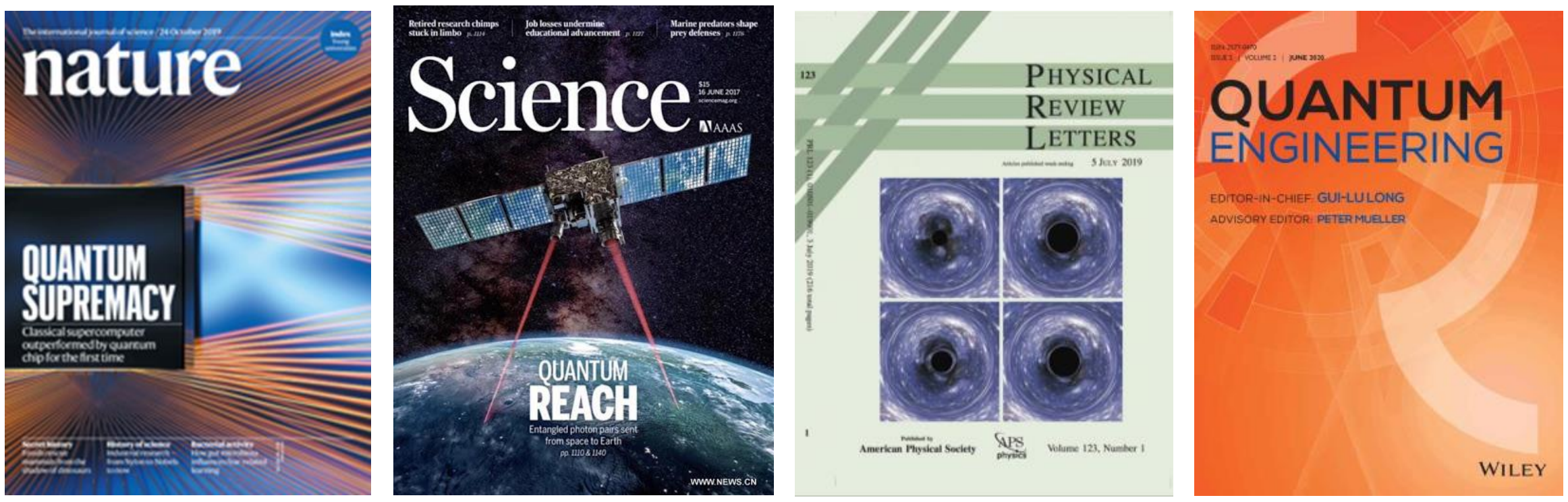

*Figure 59: some key quantum physics peer-review publications.*

In most scientific fields, including quantum science, there are many publications but not enough skilled reviewers. This job is sometimes done by PhD students. Sometimes, innovative papers are locked by reviewers, particularly when they are cross-discipline, which is frequently the case with quantum science and is a problem when publications are over-segmented.

arXiv is unlocking this situation and is now common practice. It enables fast turnaround for debates between scientists[96]. It also makes it easier for students and others to create their bibliography and review papers. It however doesn't seem that quantum research is prone to significant paper-milling or even to papers being retracted[97]. On the **RetractionWatch** database, you can find only a few retracted papers in quantum physics, mostly coming from China and India (102 items with "quantum" in the title). It includes the famous retracted papers from The Netherlands and Denmark on Majorana fermions.

There are other sites for pre-prints like arXiv, with for example **engrXiv** on engineering (with some papers related to quantum technologies. And **viXra** is an arXiv for the preprints that will never be published in peer-reviews publication and are too fringe to be accepted on arXiv (vixra.org/quant).

PhD theses are easier to retrieve and are generally published freely. These are usually good sources of bibliographical information. Beyond the main thesis goal that is to advance science in a usually narrow domain, it generally starts with making an inventory of the state of the art, like in review papers. Review papers present a state of the art of a field. Their bibliography is generally impressive, sometimes as long as the paper itself. They are a good starting point to study a subject, especially if the paper is not too old. I provide links to many such review papers, particularly on specific qubit types. If the author's pedagogy is good, it can be very useful for learning on your own. A bibliography generally allows you to go deeper into the subject by discovering the need-to-know fundamental texts.

Several authors are usually mentioned in scientific papers, up to a very large number. In general, beyond three authors, the first is the one who was the owner and done the bulk of the work. It's usually a PhD student or a post-doc. He/she has processed the experience and written a large part of the document, but this may depend on countries, laboratories and thesis supervisors. The last one is the thesis or research laboratory supervisor[98] . In the latter case, the penultimate author is the thesis director who supervised the work. In between are the other contributors, experimenters or simple reviewers.

[96] Like with Reply to arXiv:2203.14555 by Margaret Hawton, May 2022 (1 page) that is a reply to A Comment on the "Photon position operator with commuting components" by Margaret Hawton and A. Jadczyk, March 2022 (4 pages). See also Is the Moon there if nobody looks: A reply to Gill and Lambare by Marian Kupczynski, September 2022 (8 pages) which is typical of the debates going on with quantum foundation topics and on the nature of reality.

[97] See The fight against fake-paper factories that churn out sham science, Nature, March 2021.

[98] This is the case of these hundreds of publications with the famous Didier Raoult who is cited as the last contributor, as laboratory director but not necessarily thesis director.

Some papers have a very large number of authors. It is typical from the papers published by Google AI which can have upwards of 80 coauthors, which means about half of their whole team. They probably all contributed to the published work but certainly not equally[99].

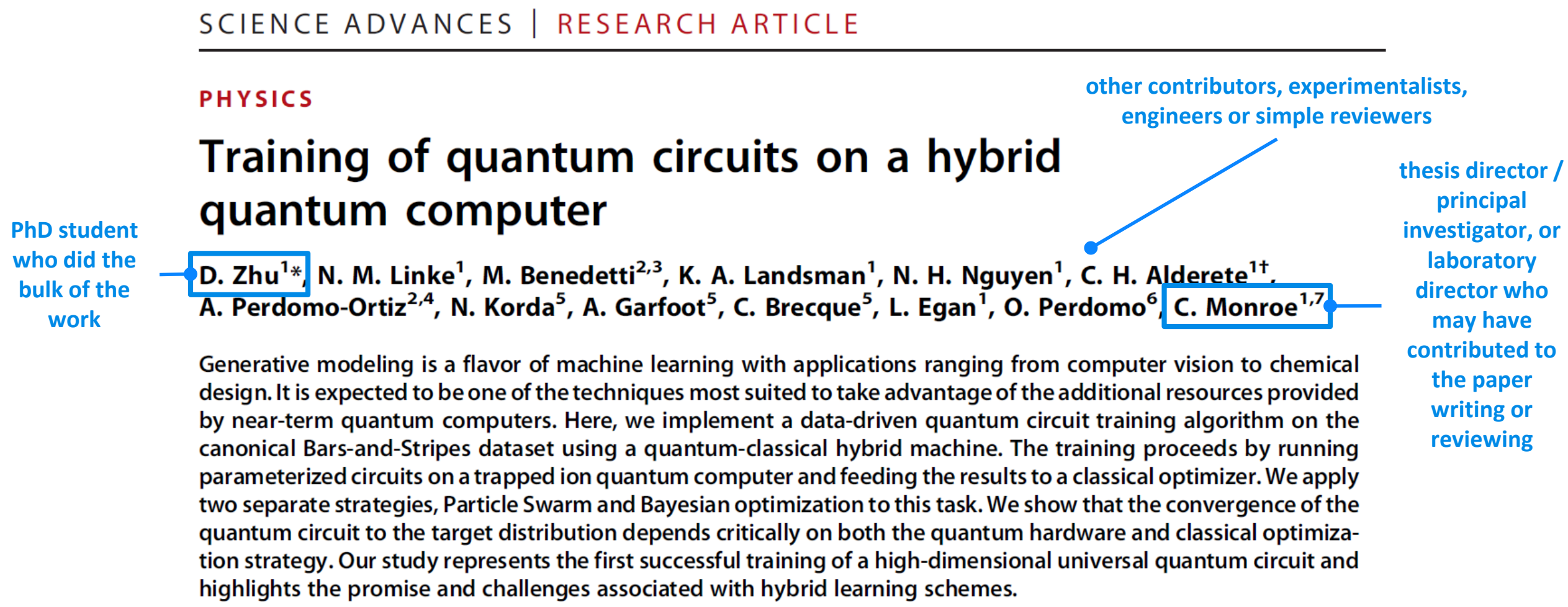
SCIENCE ADVANCES | RESEARCH ARTICLE

PHYSICS

Training of quantum circuits on a hybrid quantum computer

D. Zhu[1*], N. M. Linke[1], M. Benedetti[2,3], K. A. Landsman[1], N. H. Nguyen[1], C. H. Alderete[1†], A. Perdomo-Ortiz[2,4], N. Korda[5], A. Garfoot[5], C. Brecque[5], L. Egan[1], O. Perdomo[6], C. Monroe[1,7]

Generative modeling is a flavor of machine learning with applications ranging from computer vision to chemical design. It is expected to be one of the techniques most suited to take advantage of the additional resources provided by near-term quantum computers. Here, we implement a data-driven quantum circuit training algorithm on the canonical Bars-and-Stripes dataset using a quantum-classical hybrid machine. The training proceeds by running parameterized circuits on a trapped ion quantum computer and feeding the results to a classical optimizer. We apply two separate strategies, Particle Swarm and Bayesian optimization to this task. We show that the convergence of the quantum circuit to the target distribution depends critically on both the quantum hardware and classical optimization strategy. Our study represents the first successful training of a high-dimensional universal quantum circuit and highlights the promise and challenges associated with hybrid learning schemes.

*Figure 60: typical presentation of scientific paper's co-authorship. Source: [Training of quantum circuits on a hybrid quantum computer](#) by D. Zhu, Christopher Monroe et al, 2019 (7 pages).*

Well crafter papers don't forget to mention the respective contribution of all the authors, like in the example below. It also mentions reviewers (not those from a peer-review publication), research funding source, any potential competing interest, how the research data can be accessed and the availability of any supplemental material, that is now usually placed at the end of papers in their pre-print format. These supplemental materials can contain technical details and can be very interesting, like for example, to describe the experimental setup and its hardware and/or software engineering.

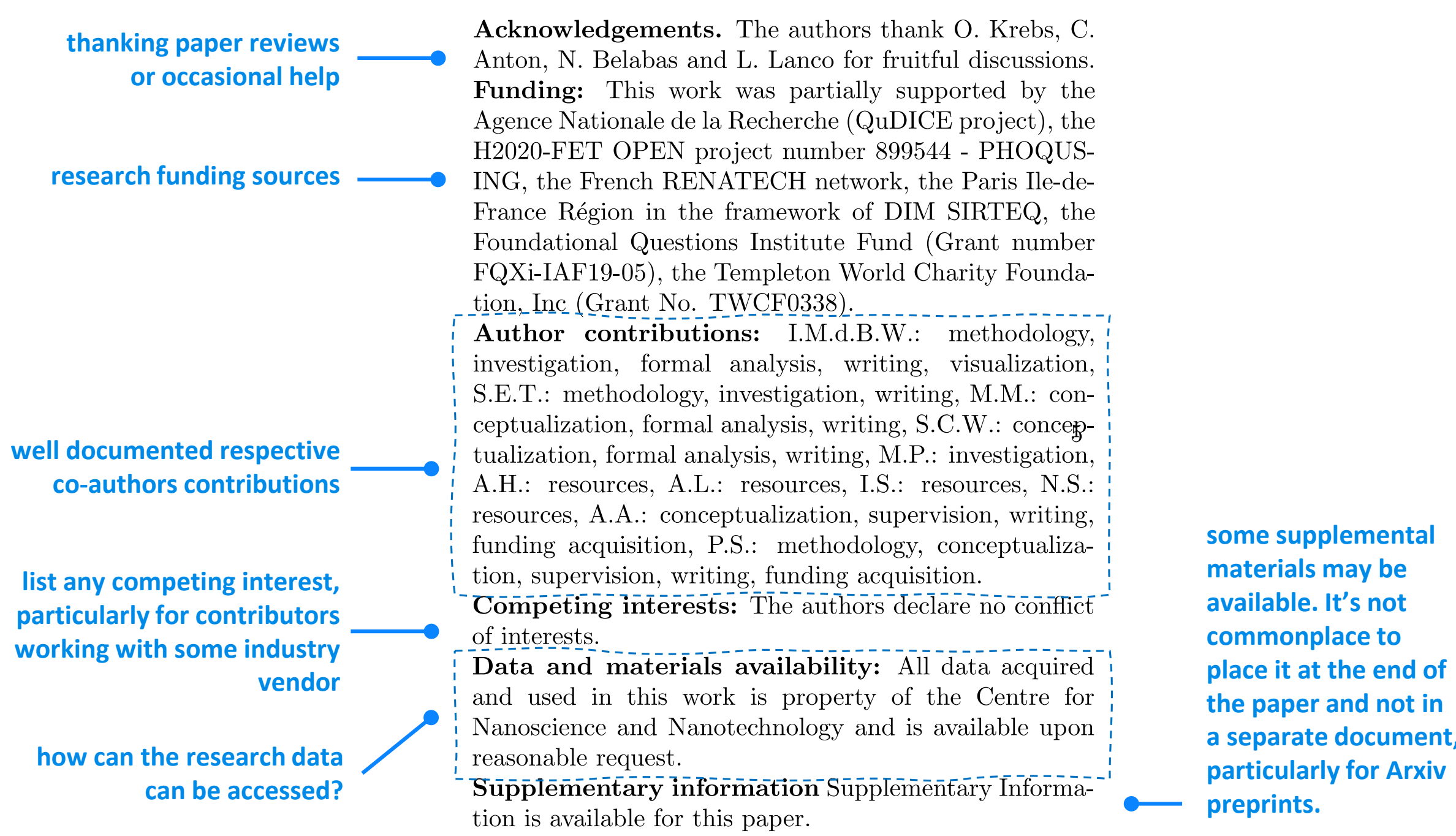
**Acknowledgements.** The authors thank O. Krebs, C. Anton, N. Belabas and L. Lanco for fruitful discussions. **Funding:** This work was partially supported by the Agence Nationale de la Recherche (QuDICE project), the H2020-FET OPEN project number 899544 - PHOQUSING, the French RENATECH network, the Paris Ile-de-France Région in the framework of DIM SIRTEQ, the Foundational Questions Institute Fund (Grant number FQXi-IAF19-05), the Templeton World Charity Foundation, Inc (Grant No. TWCF0338). **Author contributions:** I.M.d.B.W.: methodology, investigation, formal analysis, writing, visualization, S.E.T.: methodology, investigation, writing, M.M.: conceptualization, formal analysis, writing, S.C.W.: conceptualization, formal analysis, writing, M.P.: investigation, A.H.: resources, A.L.: resources, I.S.: resources, N.S.: resources, A.A.: conceptualization, supervision, writing, funding acquisition, P.S.: methodology, conceptualization, supervision, writing, funding acquisition. **Competing interests:** The authors declare no conflict of interests. **Data and materials availability:** All data acquired and used in this work is property of the Centre for Nanoscience and Nanotechnology and is available upon reasonable request. **Supplementary information** Supplementary Information is available for this paper.

*Figure 61: typical credits at the end of a scientific paper. Source: [Coherence-powered work exchanges between a solid-state qubit and light fields](#) by Ilse Maillette De Buy Wenniger, Maria Maffei, Niccolo Somaschi, Alexia Auffèves, Pascale Senellart et al, April 2022 (17 pages). This is the typical requirement for some peer-reviewed publications like Nature.*

---

[99] I found out this extreme case in [Search for a massless dark photon in $\Lambda_c^+ \to p\gamma'$ decay](#) by BESIII Collaboration, August 2022 (8 pages) with 573 authors from 75 research organizations, in China. For just 8 pages!

The other extreme case is a paper having only a single author. It means first that it is probably not a PhD student, otherwise his PhD supervisor would be a coauthor, or the author is a PhD but he lost the support of his/her supervisor for whatever reason, which is bad omen and very rare. Second, you can look at whether he/she works in a research institution and his CV. At last, you can assess the author's network if he/she mentions and thanks reviewers or contributors. The author may be already famous like say a John Preskill, Peter Shor, Seth Lloyd or Scott Aaronson, so no problem. Other cases with no attached institution, record or network may mean that the author may be working on some fringe theories in a very isolated fashion, particularly if the there's no mention of any help or thanks to anybody for the research.

In many countries, such as the USA, it is common practice to mention authors with the initials of their first and middle names initials. It does not make it easy to search them online, especially for Chinese authors. This is particularly the case when there are many contributors. I try to quote authors with their first name when they are easy to be found.

In the thousand footnotes in this book, I otherwise take the liberty of not using the cryptic description convention that is used in the abundant bibliographies of scientific publications, sometimes using authors, publication references but not the paper title!

[17] K. O'Brien, C. Macklin, I. Siddiqi, and X. Zhang, Phys. Rev. Lett. **113**, 157001 (2014).
[18] T. C. White, J. Y. Mutus, I.-C. Hoi, R. Barends, B. Campbell, Y. Chen, Z. Chen, B. Chiaro, A. Dunsworth, E. Jeffrey *et al.*, Appl. Phys. Lett. **106**, 242601 (2015).
[19] C. Macklin, K. O'Brien, D. Hover, M. E. Schwartz, V. Bolkhovsky, X. Zhang, W. D. Oliver, and I. Siddiqi, Science **350**, 307 (2015).

*Figure 62 : why (t.h.) these long bibliographies do not contain any title?*

I use a clear title convention followed by first author/authors, sometimes their research laboratories or companies, publication date and then number of pages or slides, which helps you identify at a glance the volume and depth of the referenced documents. And footnotes may be cumbersome, but they prevent you from looking at bibliographical references at the end of the document, which is never very practical whether you read a paperback or electronic version of the document. When I don't mention all a paper's contributors, I use the expression "*et al*" which is an abbreviation of the Latin "*et alia*", meaning "*and the others*". I'm usually selecting the first and last authors, then in the middle those I happen to know some way or the other, as described in Figure 63.

[386] See Correlated charge noise and relaxation errors in superconducting qubits by C.D. Wilne, Roger McDermott et al, Nature, December 2020 on Arxiv and June 2021 in Nature (19 pages) which describes the correlated errors appearing in superconducting qubits and how it could impact the architecture of quantum error correction codes and A potential hangup for quantum computing: Cosmic rays - For quantum chips, the problems they cause are too big for error correction by John Timmer, ArsTechnica, December 2021, referring to Resolving catastrophic error bursts from cosmic rays in large arrays of superconducting qubits by Matt McEwen, Rami Barends et al, Google AI, Nature Physics, December 2021 (13 pages) who developed a test protocol to assess the impact of radiations on 26 qubits in its Sycamore processor.

[387] See Impact of ionizing radiation on superconducting qubit coherence by Antti P. Vepsäläinen, William D Oliver et al, August 2020 (24 pages), the source of the illustration.

*Figure 63: bibliographical references as presented in this book. I find it more practical although it doesn't seem to be orthodoxal.*

## Paper communication

These scientific publications can be discovered by following the RSS feeds of arXiv, reference specialized papers, in addition, from scientific news feeds of online media or popular scientific press. I also discover new interesting papers with scanning scientific conferences presentations[100].

In the case of quantum technologies, the "tech" media often broadcasts scientific news dressed-up with sensationalism and exaggerations. This often stems from the propensity of laboratory communicators or sometimes researchers themselves to make shortcuts between their work and its potential usage that may be very long-term[101]. It is even stronger when the communication comes from a large company such as Google or when the article was written by the laboratory's communication branch.

[100] Here is an example with a list of many IEEE superconducting technologies presentations.

The job of the technology screener consists in sorting this out. When your local non-English speaking media broadcasts such information, it is often necessary to start by identifying the original paper which is possibly quoted at the end of the article. Sometimes, you discover blatant translation error that entirely twists the scope of the covered scientific advance.

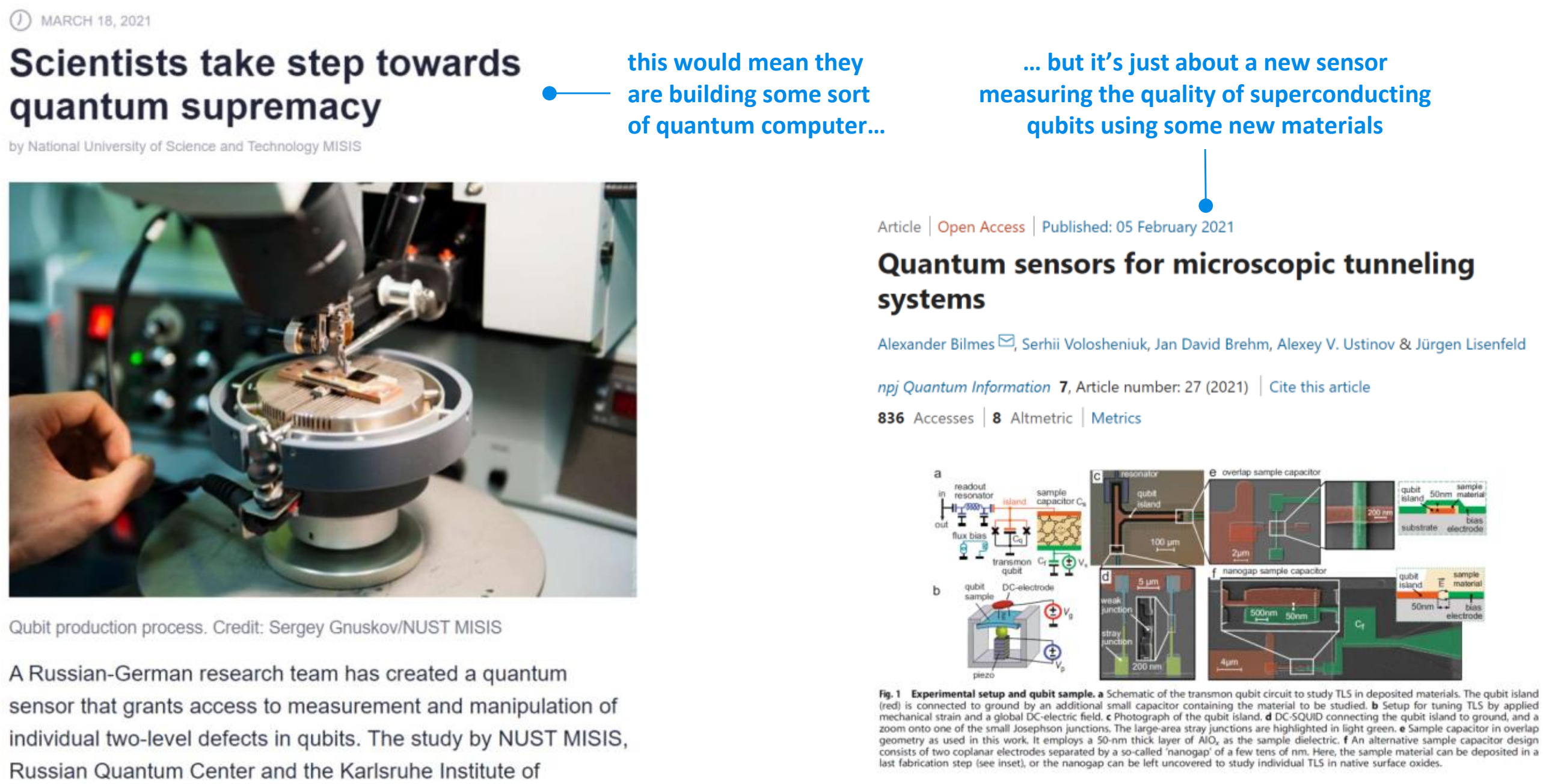

*Figure 64: example of a scientific paper presented with outrageous claims by its lab communication department. Sources: Scientists Take Step Towards Quantum Supremacy, MISIS, March 2021 and Quantum sensors for microscopic tunneling systems by Alexander Bilmes et al, February 2021 (6 pages).*

## Analysis and classification

Next, one must find the original scientific article with the methods described above. Once all this has been done, the bulk of the work consists in classifying the information: what is it about and how does it fit into the web of quantum technologies? As far as I know, no artificial intelligence can automatize this process[102]. This classification task is a tedious one and you can be easily misled with reading a paper title or press release too quickly. Here is one interesting example with a post from James Dargan in The Quantum Insider which wrongly described the European LSQuanT project as an initiative to provide quantum computing solutions to the transportation industry[103]. Wrong! It is a project related to fundamental quantum physics and digital simulation of quantum transport, a condensed matter phenomenon!

What is the actual progress made with regards to the state of the art? You can rely on classical recommendations: read the introduction and not just the abstract, identify the problem that the writers are trying to solve and how they are advancing the state of the art, look at the data and identify any missing data, and read the conclusion. If you can't decipher the paper content, make a search of other more generalists web sites mentioning it.

---

[101] The example below comes from Scientists take step towards quantum supremacy by National University of Science and Technology MISIS, March 2021. The supremacy from the article title is very far away considering the paper is about some sensing technology to measure the efficiency of some superconducting qubit.

[102] Various tools attempt to automate this sorting work, such as In Layman's Terms: Semi-Open Relation Extraction from Scientific Texts by Ruben Kruiper et al, May 2020 (13 pages). It is currently applied to the field of biology.

[103] See LSQuant: Novel Initiative Created To Improve Quantum Transport Methodologies, May 2021.

In general, a paper presenting a breakthrough that will allow the quantum computer to be realized at room temperature or ahead of all others becomes a simple very one-time breakthrough in the development of a particular type of qubit. It looks like your tiny hairy dog after the shower! In many cases, quantum science-related papers are inaccessible, requiring solid mathematical and/or physics background. Even quantum science specialists have a hard time interpreting many papers.

You frequently come across a set of Russian dolls concepts with unknown concepts referring to other unknown concepts, and so on. This is some sort of involuntary humor of scientific complexity[104]. However, hopefully, some papers do not use too much jargon and manage to deal with a big fundamental question by making it understandable to many specialists in their discipline and well beyond. This is often the case with publications in Nature.

How can I check the whole thing, particularly given the specialists in my own network have not yet had the time to do so? You either need to be patient, do it on your own, or look for someone who has done the job. For big news related to quantum computing, one can wait for the next post from Scott Aaronson or a laconic tweet from John Preskill.

Finally, I use arXiv as soon as I come across a startup that defines in too broad terms what it does without any technology specifics. It's so commonplace now! A search starts with finding the startup scientific founder, then with identifying their research work that they are probably willing to package in their freshly created startup. In their bibliography building work, researchers also look at **Google Scholar** and also on **SciRate**, where discussions take place around pre-print papers published on arXiv.

We must recognize our limits and understand that we're not protected from believing scientific hoaxes like the famous one created by **Alan D. Sokal** in 1996. It merged social sciences and quantum gravity and was published in a social science publication, not a quantum physics one [105].

Hopefully, quantum scientific publications are way more serious than most of the quantum hype that is conveyed by general news with their amazing amplification capabilities. You'll read time and again that quantum computing will drive autonomous cars, create quantum intelligent robots, reduce $CO_2$ emissions, cure cancers, help Tesla (but not others) build top-notch batteries or that quantum communications will teleport your data faster than light around the Earth. Most of these assertions will flourish when the IBMs and Googles of this world make fancy announcement or after your government launches its own "billion dollars" national quantum plan. But they are at least unproven if not entirely false. Who's going to reveal it to you?

### Roles

In most countries and in all disciplines, several roles can be distinguished in research organizations.

**Doctoral students** are students who are undertaking a doctoral thesis (PhD, for Philosophy Doctorate, for any science). It lasts from three to five years depending on the country. This thesis completes a higher education program in the University.

---

[104] Here are a couple interesting examples of papers whose title refers to mostly unknown concepts: The Franke-Gorini-Kossakowski-Lindblad-Sudarshan (FGKLS) Equation for Two-Dimensional Systems by Alexander A. Andrianov et al, April 2022 (27 pages), Floquet integrability and long-range entanglement generation in the one-dimensional quantum Potts model by A.I. Lotkov et al, October 2021-April 2022 (24 pages), Probing Lorentz-Invariance-Violation Induced Nonthermal Unruh Effect in Quasi-Two-Dimensional Dipolar Condensates by Zehua Tian et al, May 2022 (12 pages), Emergent quantum mechanics of the event-universe, quantization of events via Denrographic Hologram Theory by Oded Shor et al, August 2022 (12 pages) and Emergent Sasaki-Einstein geometry and AdS/CFT by Robert J. Berman et al, Nature Communications, January 2022 (8 pages) which I found has some connections with Exploring uberholography by Dmitry S. Ageev, August-September 2022 (14 pages) which deals with some quantum error correction code. To some extent, this complexity can be fun. See also Variational quantum algorithm for measurement extraction from the Navier-Stokes, Einstein, Maxwell, Boussniesq-type, Lin-Tsien, Camassa-Holm, Drinfeld-Sokolov-Wilson, and Hunter-Saxton equations by Pete Rigas, September 2022 (144 pages) which requires a significant mathematical background.

[105] See Transgressing the Boundaries: Towards a Transformative Hermeneutics of Quantum Gravity by Alan D. Sokal, 1996 (39 pages).

**Post-docs** or post-doctoral researchers are researchers who, after having obtained their PhD, conduct research in a laboratory under a fixed-term contract. They sometimes do several post-docs in different locations, frequently out of their originating country. It is the anteroom of a full-time research position.

**Researchers** have a full-time tenure in a research organization whether in the industry or with government funded research organizations. In many countries, they are also civil servant researchers recruited through some open competitions process .

**Habilitation to Direct Research** (HDR in France) allows a tenured researcher to direct the thesis of one or more doctoral students as a thesis director and to obtain a university professorship. The rules vary from country to country, such as having completed two doctoral theses and having published internationally recognized work in one's field[106].

**Research Directors** are researchers with the possibility to autonomously determine the field of their research work. They supervise several doctoral students and post-docs when they are successful with finding the related public and/or private funding. They are also selected by competition in research institutions. Depending on the country and research organization, there are several grades in the function, linked to advancement over time and merit.

**Principal Investigators** are lead researchers who are in charge of the preparation, conduct, resources allocation and administration of a research grant for which they are the project lead researcher and main holder. Sometimes, a PI is synonym of laboratory director or research group leader.

In addition to these roles, let's not forget the **laboratory technicians** who set up the experiments and about whom less is said and the **engineers** who can play a role in the creation of many scientific instruments.

### h-index

The h-index, named after its creator Jorge Hirsch in 2005, is an index that quantifies a researcher's productivity and scientific impact. It is based on the level of citations of his scientific publications in peer-reviewed journals. It is a bit like a PageRank for a website, but a simpler one. It is an integer corresponding to the number of papers h that have each obtained more h citations in other papers.

The level of h-index can be used as a quantitative data for obtaining a position as a resident researcher (10-12), professor (>18) or member of an academy of science (>45). As with any composite index[107], it generates side effects: a race to "publish or perish" papers of little incremental value, cross-referencing between researchers, self-citation, an abundance of co-authors[108], etc.

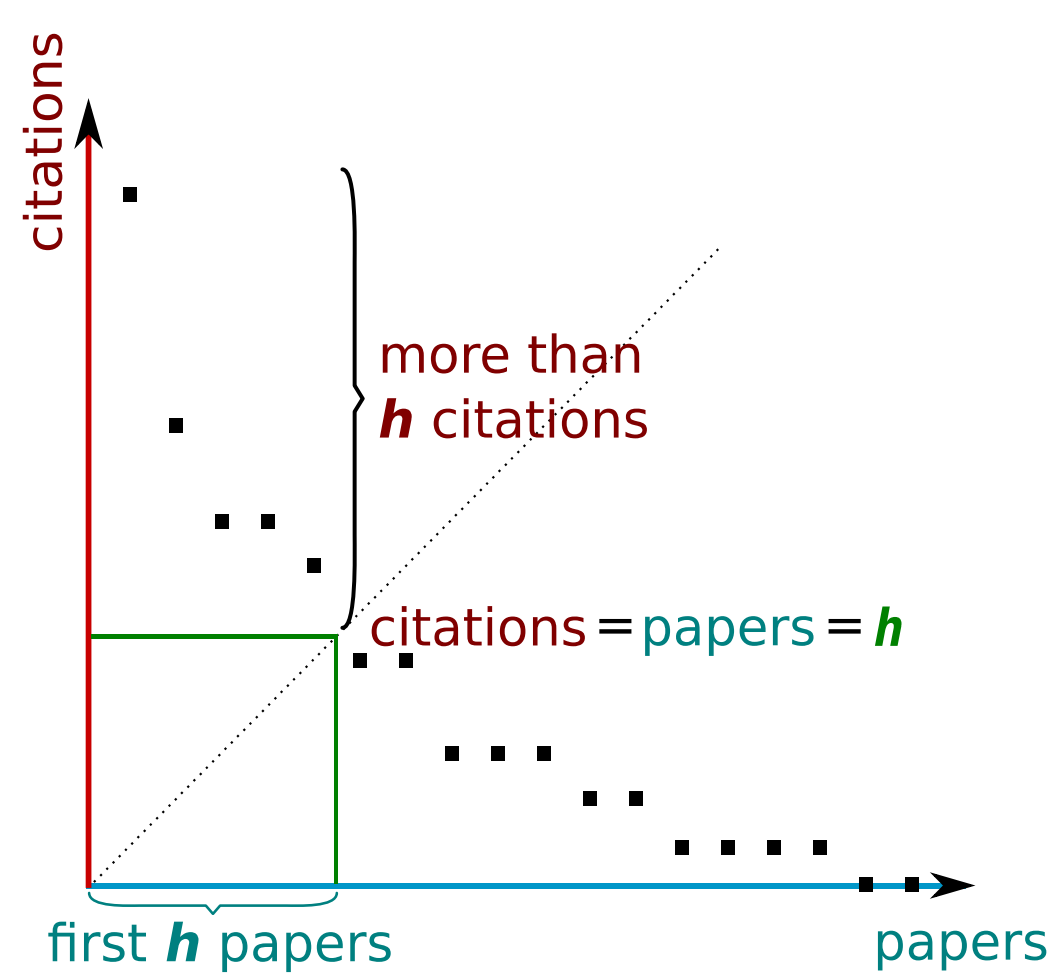

*Figure 65: h-index explained graphically.*

The discrepancy of h-index is quite high with researchers with a Nobel prize in physics with low index like with John Clauser (29, Nobel in 2022) and Brian Josephson (22, Nobel in 1973) and very high index like Anton Zeilinger (139, Nobel in 2022) or David Wineland (122, Nobel in 2012).

---

[106] This habilitation replaced the Doctorat d'Etat in 1984 in France. The HDR is considered to be a diploma. It is awarded on free application by the research commission of the Universities which deliberates in the form of a jury.

[107] The Shanghai ranking list of universities comes to mind.

[108] In this paper from Google, we have no less than 85 co-authors: [Implementing a quantum approximate optimization algorithm on a 53-qubit NISQ device]() by Bob Yirka, February 2021 (19 pages). It's a bit too much and we can wonder about their all contributions!

Some alternatives indexes have been proposed like the recent h-frac, but not yet adopted[109]. It remains, however, an interesting indicator of the influence of researchers and their production volume.

On average, the h-index of a researcher in physics is close to the length of his career since his PhD. It obviously evolves over time. It is full of flaws like all quantitative indicators. For example, the basic h-index does not distinguish between main author and co-author. Hence the abundance of authors cited in many papers, some of them having made only marginal contributions.

The index is usually calculated from **Google Scholar** data, but it is sometimes found calculated only on the SemanticScholar website. The most serious index is provided by the Website of **Science** because its database is the cleanest.

### Open research

On top of being published in open source as pre-prints, research results and datasets can be published in various platforms like **Zenodo**, which was developed under the European OpenAIRE program and is operated by CERN. The deposits can contain research papers, the experiments data sets, research software, detailed reports and any other digital artefacts. Using this sort of service is becoming common practice, to make sure experiments are reproducible. Other services like **OSF** (Open Science Framework) also promote open research practices.

### Fake news

Science is not exempt of fake news. In all scientific fields, some researchers may publish questionable results for their experiments, aggregate and compile tinkered data, or simply avoid taking into account embarrassing data, generating a survivor bias. This can happen in quantum technologies, particularly when evaluating the quality of experimental qubits or, for instance, finding Majorana zero modes, *aka* fermions. In general, you need to be an expert in the field to identify this kind of abuse. They however seem rare in quantum technologies.

With a generalist technological knowledge in the domain, one can start to detect tricks of the trade or exaggerations. This is easier to do with commercial vendors like with IBM and their quantum volume, Honeywell and their "*most powerful quantum computer in the world*" or with the Google and Chinese quantum supremacy experiments.

### Poster sessions

In a scientific conference, a "poster session" is usually a part of the conference dedicated to the presentation of researchers' projects during a break, in a dedicated area.

Researchers display a poster describing their research work and talk with conference participants as they stroll through the conference exhibition area during dedicated breaks. It is an exercise in humility reminding what Jehovah's witnesses are doing in the streets.

### Figures of merit

This common expression broadly describes a set of specifications and the success metrics to be achieved to bring a given technology to fruition. DiVincenzo's qubit technology criteria can be considered a figure of merit for success for quantum computing. It usually provides a roadmap and set of goals for researchers and technology vendors.

---

[109] See The h-index is no longer an effective correlate of scientific reputation by Vladlen Koltun and, David Hafner, Intel Labs, February 2021 (26 pages). Among other things, the authors found out that the correlation between h-index and scientific awards in physics is declining. They propose an alternative index named h-frac, for h-fractional, that improves the correlation between the index and other scientometric measures like scientific awards. It allocates citations fractionally and evenly among all coauthors of scanned papers to avoid the phenomenon of low-contribution hyperauthors.

### International

Nowadays, all modern countries have crafted their "quantum national plan" with a certain willingness to better control their sovereignty. It's like being the first with the atomic bomb during World War II.

But let's remember that international collaboration between researchers is intense. Most of those I met in French laboratories collaborate with colleagues either in Europe within the framework of Europe 2020 projects, the European Flagship or for some ERCs.

They also collaborate with researchers outside the European Union, particularly in Asia (Japan, Singapore), as well as in the USA, UK, Switzerland and Australia[110].

Quantum science knowledge is quite open and is rather well shared on a global scale. This is encouraged by many international scientific conferences where knowledge is being built, researchers get to know each other, and joint projects are being launched. This is one of the reasons why I don't believe in the existence of a supposed quantum computer whose capabilities would defy understanding and which would be hidden in the basement of a secret NSA datacenter to break all the RSA keys of the Internet.

Scientific nationalism in quantum technologies finally comes into play further downstream of research, when it comes to transforming it into industrial advantage. Technologies often have their "magic sauce", as in semiconductor manufacturing processes. This has always been the case in digital technologies.

### Technology Readiness Level

This technology readiness level notion is commonly used in deep techs. It describes the level of maturity of a technology with a scale from 1 to 9. It follows a relatively standardized classification initially created by NASA in 1975[111], then used by the European Union and various other organizations. It was initially mainly used in the aerospace, defense and energy industries.

This scale can have several use cases. It is used to assess the level of risk and maturity for an investor in a startup. Very advanced deep techs are also the playground of TRL and quantum technologies are no exception.

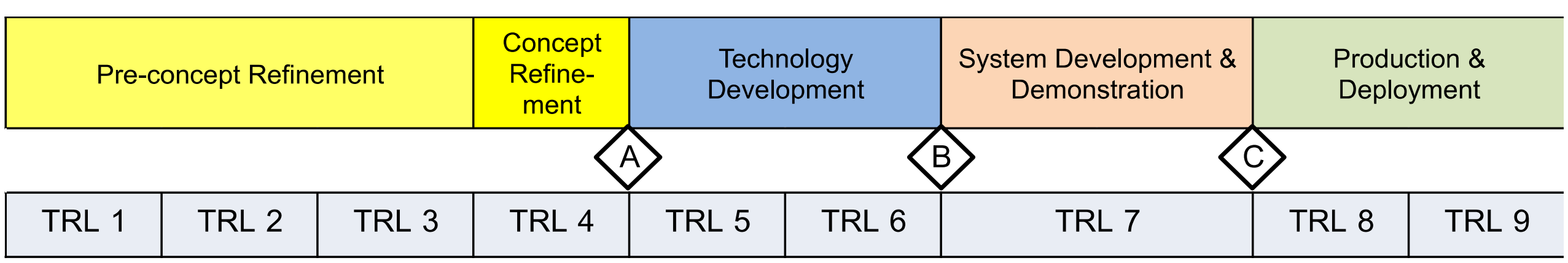

*Figure 66: the scale of technology readiness level. Source: Some explanations on the TRL (Technology readiness level) scale, DGA, 2009 (15 pages).*

The TRL scale has 9 levels[112]:

- **TRL 1**: basic principles are described or observed, at the theoretical or experimental stage.
- **TRL 2**: technological concepts are formulated and not yet necessarily tested.
- **TRL 3**: proof of concept is carried out in a laboratory, at the level of the technical process.
- **TRL 4**: the technology is validated in the laboratory as a whole.

---

110 This can also take the form of CNRS International Mixed Units such as those established in Japan and Singapore.

111 See Technology Readiness Levels at 40: A Study of State-of-the-Art Use, Challenges, and Opportunities by Alison Olechowski et al, 2015 (11 pages) which is the source of the diagram.

112 See Technology Development Stages and Market Readiness by Surya Raghu, June 2017 (35 slides).

- **TRL 5**: a technology model in a production grade environment is created.
- **TRL 6**: a technology prototype is demonstrated in an environment representative of the intended use case.
- **TRL 7**: a prototype is evaluated in an operational environment.
- **TRL 8**: a complete system has been evaluated and qualified.
- **TRL 9**: a complete system is operational and qualified in production.

The relevance of the solution to market needs is missing at this scale, but it is a marketing rather than a technical consideration[113]. Most of the time, it more or less coincides with TRL levels 7 to 9 since reaching this scale requires funding and finding customers willing to test the solution.

**Kristel Michielsen** has proposed a scale suitable for quantum computing, the QTRL, for the Quantum Technology Readiness Level in Figure 67. Her assessment of some technologies can be argued. For example, she positions D-Wave's quantum-annealed computers in TRL 8 and 9. This is commercially correct since these computers are well marketed. This being said, if they are well available physically, it is not proven that they are of much use at the moment.

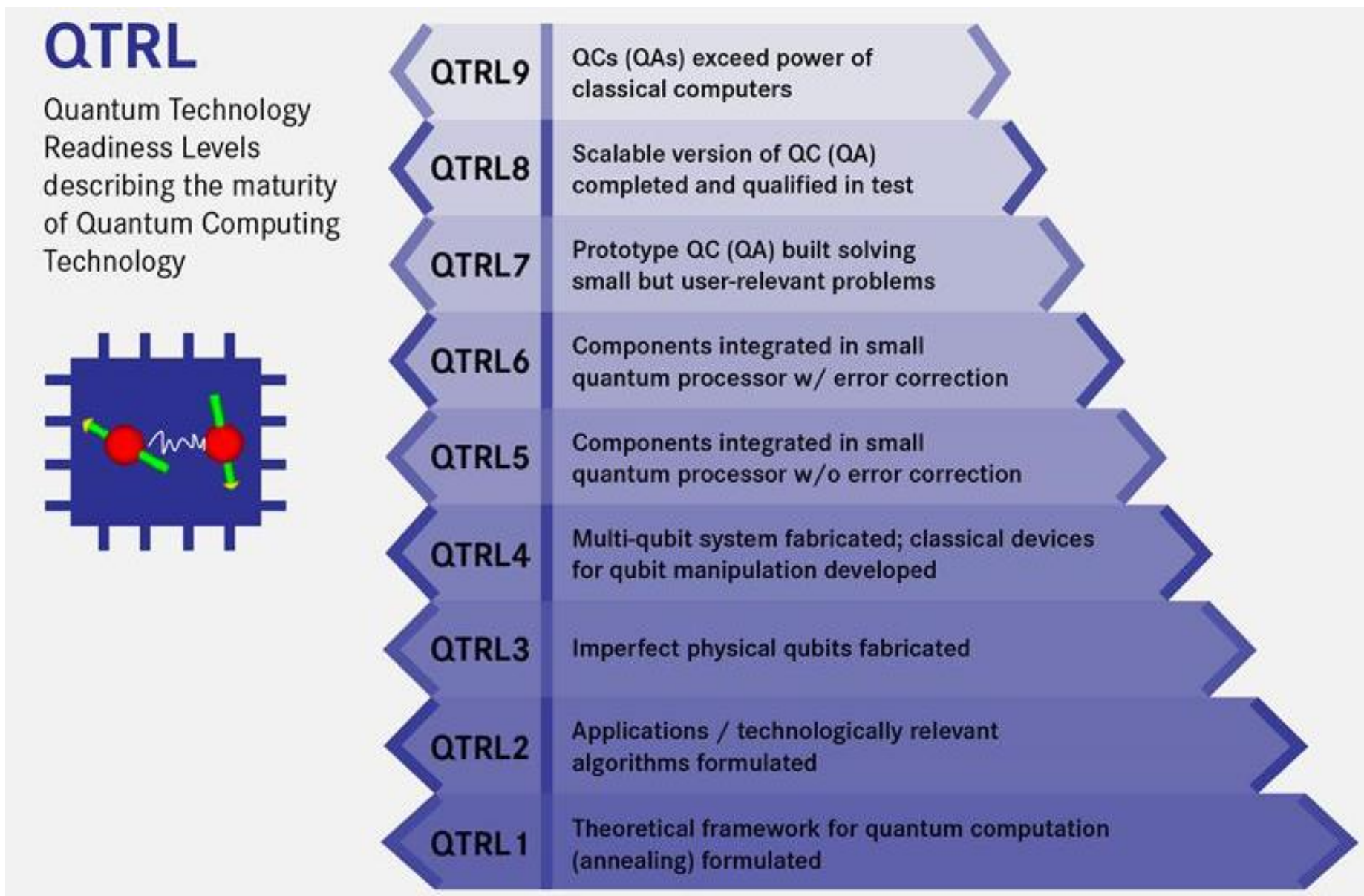

*Figure 67: the quantum TRL scale, created by Kristel Michielsen. Source: Simulation on/of various types of quantum computers by Kristel Michielsen, March 2018 (40 slides).*

The specificity of quantum technologies is that many hardware startups are created with very low TRLs. This is particularly true for those who are starting to design qubits using technologies that have not yet been proven, even in the labs. In quantum technologies, the notions of "MVP" (minimum viable product) are very different from the classical digital world. It's based on scientific rather than functional metrics. We have many such startups around in quantum technologies because of the famous FOMO (fear of missing out) syndrome with investors.

This shows up with investors who fear of missing the future golden goose or unicorn. They are ready to overinvest in companies they perceived will be the future market champion. This explains for example the level of funding for startups like **Rigetti** and **PsiQuantum** or the new SPAC funding mechanism (special purpose acquisition company) implemented by **IonQ**, **Rigetti** and **D-Wave** and the recent quantum business spin-off from **Honeywell** and its merger with **CQC** (becoming **Quantinuum** in December 2021).

[113] See TRL, MRL, POC, WTF? by Massis Sirapian of the Defense Innovation Agency, April 2019.

## Quantum physics history and scientists key takeaways

- A first wave of 19th century scientists laid the groundwork that helped create quantum physics afterwards (Young, Maxwell, Boltzmann, mathematicians). The photoelectric effect, black body spectrum and atoms emission or absorption spectrum were not explained with the current theoretical frameworks.
- Starting with Max Planck, a second wave of scientists (Einstein, De Broglie, Schrodinger, Heisenberg, Dirac, Born, Von Neumann) created quantum physics to describe light/matter interactions, energy quantification and wave-particle duality. It solved most of the $19^{th}$ century unexplained physics experiments.
- These scientists were theoreticians while many lesser-known researchers were experimentalists with landmark discoveries (superconductivity, electron interferences, Stern-Gerlach experiment, ...).
- After World War II all digital technologies (transistors, lasers, telecommunications) were based and are still based on quantum physics, as part of what is now called the first quantum revolution.
- Since the 1980s and thanks to advances in individual quantum objects control and the usage of quantum superposition and entanglement, new breeds of technologies were created, most of them belonging to the "quantum information science" field and being part of the second quantum revolution.
- Many of these research programs were funded by governments after Peter Shor's integer factoring algorithm was created.
- While the first quantum revolution was driven by research coming mostly out of Europe, the last wave comes out of all developed countries across several continents (North America, Europe, Asia/Pacific).

# Quantum physics 101

After a historical review of quantum physics and computing with its most important contributors, let's look at the fundamentals of quantum physics in a more structured way. Whatever the ups and downs, this field has gone through the test of time for nearly a whole century. Thousands of experiments have validated the theory and mathematical formalism behind it even though we still can't explain what's happening at a physical level, particularly with quantum entanglement or, even, with the wave-particle duality phenomenon with electrons and photons.

Several years of undergraduate and graduate studies are usually necessary to master quantum physics notwithstanding its rich mathematical foundations. This part will save you some of this time and provide some scientific background knowledge that will help you better understand the various quantum information systems exposed in the remainder of this book.

As seen before, quantum physics appeared at the beginning of the 20th century to explain the dynamics of elementary particles, particularly to study how **photons**, **electrons** and **atoms** behave and interact[114]. Quantum physics also deals with elementary particles from the standard model like quarks and neutrinos, but it's out of scope in the second quantum revolution and quantum information science[115]. In some cases, we still care about atom nucleus spins, which relate to proton spins, itself linked to its quark constituents. Nucleus spin plays a role in NV centers-based technologies. We also care about it with electron spin-based qubits since nucleus spin can have a detrimental impact on electron spins handling qubits information. It relates to the kinds of isotopes of carbon and silicon that are used in carbon nanotubes and silicon wafers used to create electron spin qubits.

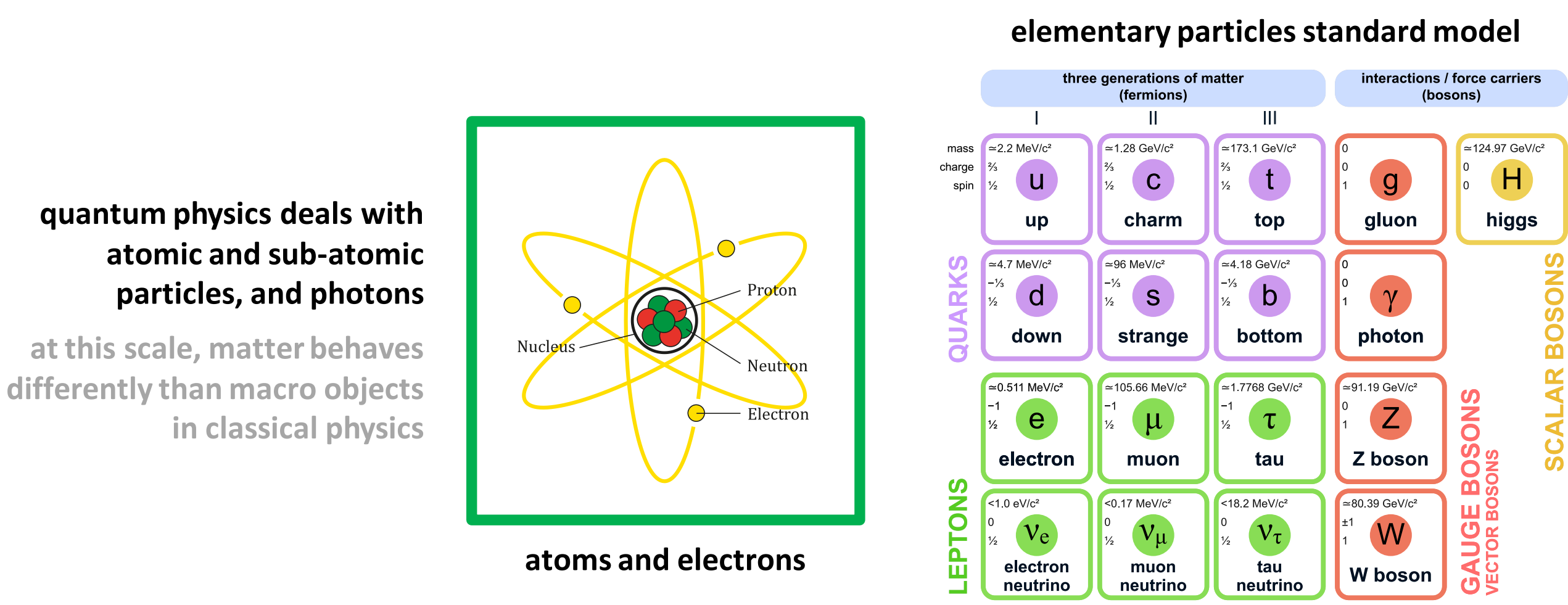

*Figure 68: what particles are we dealing with quantum physics? All of them, but in the second quantum revolution, we mainly use electrons, photons and atoms. Source: Wikipedia.*

Quantum physics first helped explain various observations such as the **black-body radiation** (solved by Max Planck in 1900), the **photoelectric effect** (solved by Albert Einstein in 1905) and the **sharp spectral lines** observed with excited atoms like hydrogen (solved by Niels Bohr and its atom model in 1913).

---

114 As a reminder, here are the dimensions of elementary particles: $10^{-10}$m for an atom, $10^{-15}$m for the diameter of a hydrogen atom nucleus, thus of a single proton, and $10^{-18}$m for that of an electron.

115 See Neutrinos as Qubits and Qutrits by Abhishek Kumar Jha et al, March 2022 (30 pages) which makes a proposal to use neutrinos for quantum computing, without taking care of the related engineering problems. It's very hard to contain and control neutrinos!

Later on, in the mid 1920's, quantum physics was built upon a **mathematical formalism** using multi-dimensional Hilbert spaces and vectors. It centered around the **Schrödinger wave equation** which describes how a massive particle like the electron behaves over space and time, using complex number probability amplitudes and differential equations over time and space.

These provide a probabilistic insight on the outcome of the measurement of a particle's energy, momentum, and many other physical properties.

Quantum mechanics differs from classical physics with demonstrating how and why particles energy, momentum, angular momentum and other metrics are restricted to discrete values (**quantization**), objects can behave as particles or waves depending on the context (**wave-particle duality**), and there are limits to how accurately the value of a physical quantity can be predicted prior to its measurement, given a complete set of initial conditions (**indeterminacy principle**).

It also refers to **state superposition** which is at the basis of qubit operations and one of the sources of the quantum computers processing parallelism, **entanglement** which is a direct consequence of superposition applied to several quantum objects and is used with multi-qubits quantum gates and is also related to quantum communications and cryptography. Quantum objects **no-cloning** is a particular aspect of quantum physics that limits what we can do with qubits and how memory is managed. At last, **quantum tunnelling effect** has some impact in quantum technologies, like with the Josephson junctions used in superconducting qubits and with D-Wave quantum annealers.

Quantum physics explains other physical phenomena belonging to the broad **quantum matter** category, like **superconductivity** which plays a key role in superconducting qubits, **superfluidity,** used with liquid helium in dilution refrigerators and **quantum vacuum fluctuation** and its role in quantum decoherence. It also enabled the creation of **lasers**, used in many places like for controlling cold atom and trapped ion qubits and for all photonic based quantum computing and telecommunications. At last, **polaritons** are sets of interactions between light and semiconductors which could become useful in quantum sensing and quantum simulation. The quantum objects bestiary also includes **skyrmions** and **magnons**!

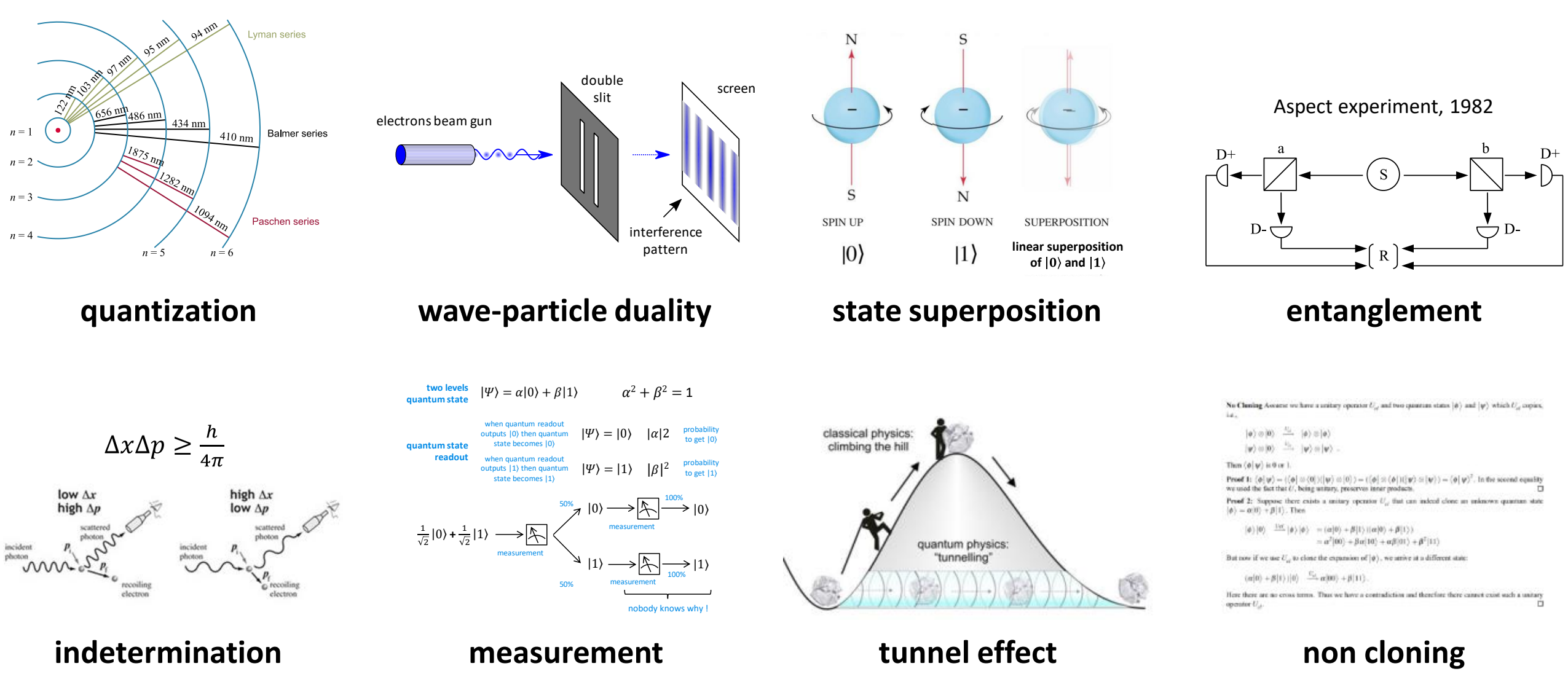

*Figure 69: eight key dimensions of quantum physics that we are dealing with. (cc) compilation Olivier Ezratty, 2021.*

## Postulates

Quantum physics formalism is based on a set of postulates that follows[116]. Why are these postulates and not laws? Mainly because they describe a mathematical formalism that cannot be proved per se.

One of the other reasons is that quantum physics does not rely on an ontology describing the physical objects it's based upon. I'll try whenever possible to connect these postulates with some physical meaning. If all of this seems gibberish for you, skip it!

**Postulate I - Quantum state**: the state of an isolated physical system is represented, at a given time t, by a state vector $|\psi\rangle$ (psi) belonging to a Hilbert space $H$ called the state space with vectors of length 1, using complex numbers. This is the canonical definition of a quantum state. The $|\psi\rangle$ vector contains the knowledge we can have of a quantum system, represented by the values taken by its measurable and compatible properties. A broader definition of a quantum state is the ensemble of values taken by compatible physical properties of a system made of one or several quantum objects. These compatible properties must be measurable simultaneously or in any order. The $|\psi\rangle$ vector is a mathematical object that helps determine and predict over time the probabilistic distribution of the various values of the quantum object compatible properties. The immediate consequence of this first postulate is the notion of superposition where a linear combination of several $|\psi\rangle$ vectors can form another valid quantum state. For a generic qubit, its quantum state defines its amplitude and phase as we'll see later in the Bloch sphere description. $|\psi\rangle$ is then a vector in a two-dimensional Hilbert vector space combining the $|0\rangle$ and $|1\rangle$ basis states with their related complex amplitudes.

**Postulate II - Physical quantities**: are related in quantum physics with observables that are mathematical operators $\hat{A}$ acting on the $|\psi\rangle$ vector as $\hat{A}|\psi\rangle$. With the quantum matrix formalism, $\hat{A}$ is a Hermitian (linear) matrix operator acting on the state vector $|\psi\rangle$ to evaluate quantized or continuous physical properties of quantum objects. This operator is a self-adjoint matrix, with the implication that several consecutive measurements generate the same (vector) result. A projector operator like a Pauli matrix $\sigma_x$, $\sigma_y$ or $\sigma_z$ used to measure a qubit state is a specific case of an observable operator.

By the way, let's clearly define properties and their variations:

**Properties** correspond to a quantum system's various observables. For a photon, it can be, for example its phase, polarization, and wavelength. In quantum physics, it is not possible to evaluate the values of all properties of quantum systems to describe it, due to Bohr's complementarity principle. Properties can also be continuous like a quantum object momentum or position.

**Exclusive property values** are the possible results of a quantum measurement of a quantized property. The classical examples are vertical and horizontal polarization for a photon or spin up or down for an electron spin along a projection axis. These are mutually exclusive since it corresponds to two results of a physical measurement. Mathematically speaking, two properties are exclusive if their projector operators (*aka* observables...) are orthogonal. Otherwise, these are non-exclusive properties.

**Compatible properties** of a quantum system can be measured in any order or simultaneously[117]. In that case, their observable operators A and B commute (AB=BA), or their commutator is equal to zero ([A,B]=AB-BA=0)[118].

[116] Source: Wikipedia.

[117] The notion of properties compatibility must not be confused with complementarity. There is complementarity between incompatible properties, like position and momentum! Incompatible observables are related to conjugate variables, defined by one being a Fourier transform of the other and Heisenberg's indeterminacy principle being consequently applied to both these variables measurement. See Bohr's Complementarity and Kant's Epistemology by Michel Bitbol and Stefano Osnaghi, 2013 (22 pages) which lay out well these different concepts.

[118] Compatible properties are well explained in Mathematical Foundations of Quantum Mechanics: An Advanced Short Course by Valter Moretti, 2016 (103 pages).

Compatible properties have commuting observables. Measuring a complete set of commuting observables (CSCO) constitutes the most complete measurement of a quantum system.

**Incompatible properties** aka conjugate variables cannot be measured simultaneously and their observable operators A and B do not commute (AB <> BA or $[A,B] \neq 0$). This is a particularity of quantum mechanics.

However, revealing one property value with a measurement doesn't exclude revealing another property afterwards. But it is not possible to obtain exact knowledge of both properties at the same time (in the probabilistic sense and following Born's rule). At least one will be totally probabilistic. For a single particle, one example of incompatible properties or observables are two different spin components (X and Y or X and Z). After measuring the X spin component, a Z measurement will yield a random result. Also, the energy and position of an electron are incompatible properties.

**Postulate III - Measurement**: is the result of a physical quantity measurement with an observable operator A. The measurement result is one of the observable operator eigenvalues. We define eigenvalues later starting page 146 and cover the related mathematical formalism in the measurement section of this book starting page 184. This postulate is sometimes embedded or associated with the previous one. The observable operator doesn't generate a measurement result per se. It helps create a probabilistic distribution of the possible measurement outcomes of a property given what is mathematically known of the quantum object state vector. When applied to a quantum object vector, it creates another state vector along the eigenvectors of the observable operator. It can then serve to create a series of real numbers describing the probabilities of the various exclusive values a given property can take. The **expectation value,** or **predicted mean value**, is the average value of repeated measurements that would be obtained with the physical implementation of the observable. We'll come back to this later starting page 152. The measurement postulate is also named the Von Neumann measurement postulate.

**Postulate IV - Born rule**: when the physical quantity A is measured on a system in a normalized state $|\psi\rangle$, the probability of obtaining an eigenvalue $\alpha_n$ for discrete values or $\alpha$ for continuous values of the corresponding observable A is given by its squared amplitude of the related wave function. It is a projection on the corresponding eigenvector. This is related to Max Born's probability rule. A quantum state can be generally represented by a density operator, which is a square matrix, nonnegative self-adjoint operator ρ normalized to be of trace 1. The average expected value of A in the state $\rho$ is $tr(A\rho)$, the trace (sum of diagonal matrix values) of the observable operator applied to the density matrix[119]. This postulate is sometimes merged with the measurement postulate. This postulate is associated with the principle of spectral decomposition. For a single qubit, the Born rule is simple to describe with $\alpha^2$ being the probability of getting a $|0\rangle$ and $\beta^2$ of getting a $|1\rangle$ when the qubit state is described as $|\psi\rangle = \alpha|0\rangle + \beta|1\rangle$ with $\alpha$ and $\beta$ being complex numbers. And due to probabilities normalization, $\alpha^2 + \beta^2 = 1$.

**Postulate V - State collapse**: only one result is obtained after a quantum measurement. Two sequential measurements based on the same observable operator will always output the same value. For a qubit, after we measure its state, whatever it is, we get a $|0\rangle$ or a $|1\rangle$ and this becomes the new qubit state after measurement.

[119] There are variations of this postulate for various quantum spectrum (discrete and nondegenerate, discrete and degenerate, continuous and non-degenerate). Degenerate spectrum is defined in the glossary.

**Postulate VI - Time evolution**: the time evolution of the state vector $|\psi(t)\rangle$ is governed by the Schrödinger wave equation[120]. We don't directly deal much with time evolutions to understand quantum computing with qubits and gates, but it still plays a key role in quantum annealing and quantum simulation and, behind the scenes, in gate-based computing, with qubits decoherence, quantum noise, quantum error corrections mechanisms and measurement.

There is also a **Composition** postulate, which defines the notion of tensor product applied to separable composite quantum systems. *Aka* "Composite Systems" with John Preskill's axioms. We'll talk about it abundantly when covering linear algebra starting page 144 and qubit registers starting page 169.

There are indeed many variations of these postulates in shape, form, name and number, which ranges from 4 to 9 depending on the source[121]. Quantum State can become State Space and Physical Quantities become Unitary Dynamics[122]. John Preskill lists five 'axioms', considering that postulates are axioms since they are not contradicted experimentally[123]. There is not really a single "bible" of quantum postulates even when reading quantum physics founders writings (Bohr, Heisenberg and others) who didn't agree on all of it. I have consolidated below a table with some of these variations of postulates. Imagine if there were various versions of the Bible with 5, 7, 9, 10 and 12 commandments!

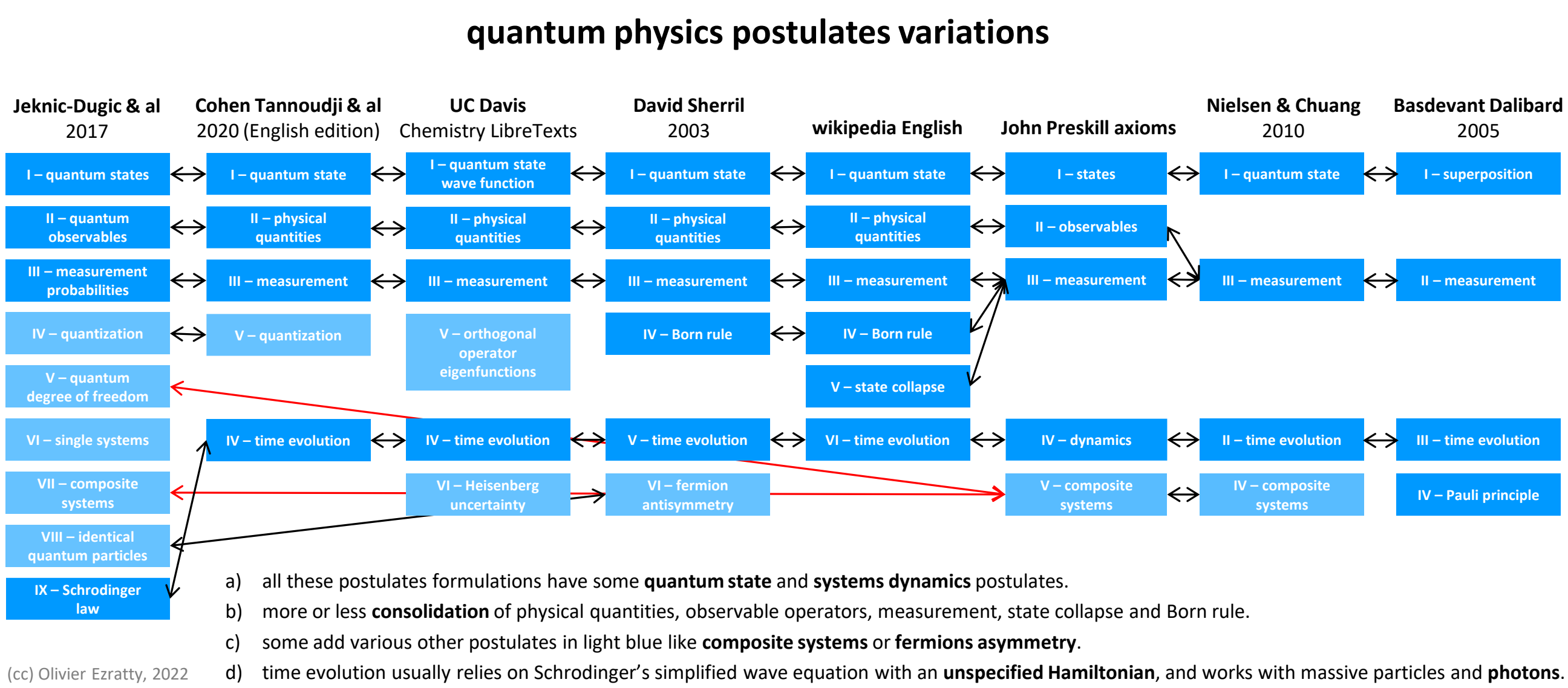

*Figure 70: a compilation of various inconsistent lists of quantum postulates and axioms. (cc) Olivier Ezratty, 2022.*

Mostly covered in linear algebra section starting page 144, the main related quantum physics mathematical tools are:

- **Linear algebra**: complex numbers, eigenvectors, eigenvalues and eigenstates.
- **Functional analysis**: Hilbert spaces, Hermitian matrices, linear operators, spectral theory.
- **Differential equations**: partial differential equations, separation of variables, ordinary differential equations, Sturm–Liouville problems, eigenfunctions.
- **Harmonic analysis**: Fourier transforms and series.

---

[120] As a result, the postulates are applicable for massive non-relativistic particles. Relativistic massive particles time evolution is described by the Dirac and Klein-Gordon equations while photons are covered by Maxwell's equations and their various derivations.

[121] 9 postulates are listed in Axiomatic quantum mechanics: Necessity and benefits for the physics studies by J. Jeknic-Dugic et al, 2017 (23 pages).

[122] In Quantum mechanics distilled by Andy Matuschak and Michael Nielsen on the Quantum Country site.

[123] See Lecture Notes for Ph219/CS219: Quantum Information Chapter 2 by John Preskill, California Institute of Technology, July 2015 (53 pages).

# Quantization

In quantum physics, material or immaterial quantum objects have some physical properties that are discontinuous and not continuous like distances in classical physics. This frequently corresponds to the orbits of electrons around atomic nuclei which are defined in a discrete way, to atom energy levels, but also deals with photons various properties, electrons and atom nucleons and nucleus spins, and other properties of matter and waves as well as other particles from the standard model (quarks, gluons, neutrinos, …) that are studies in the physics of high energy particles (HEP).

## Principle

There is a correspondence between the discontinuous energetic transitions of electrons in orbit around atoms and the related absorbed or emitted photons. Quantization shows up in other various places like in crystals. Atoms also form harmonic oscillators and vibrate at quantified amplitudes in crystalline structures, according to a model Einstein developed in 1907.

You'll actually find many quantum oscillators all over the place, like with superconducting qubits.

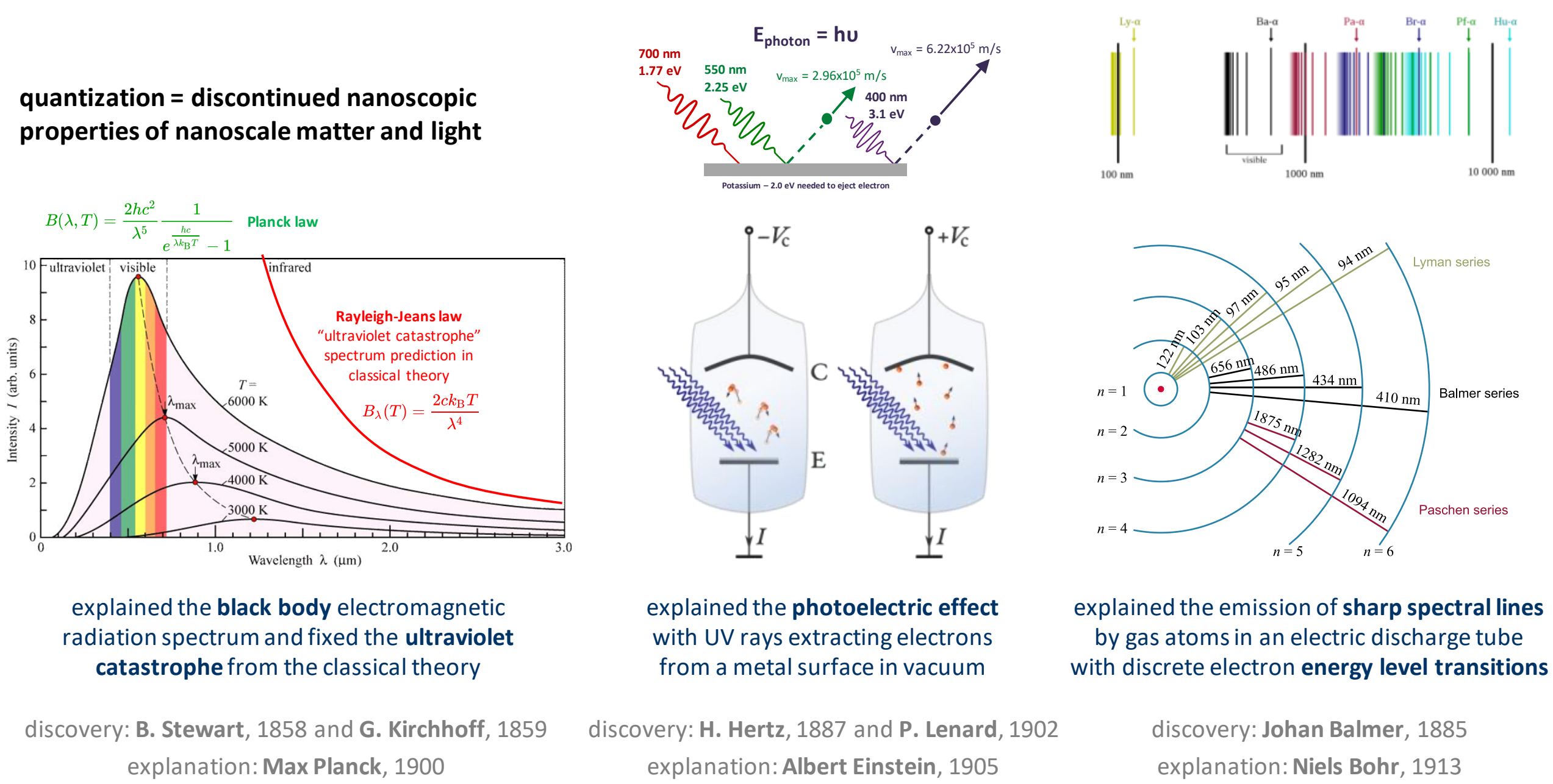

*Figure 71: the three fundamental 19th century electro-magnetic waves experimental results which were later explained by quantum physics, all explained by quantization of the electro-magnetic wave field. (cc) Olivier Ezratty compilation. Various schema sources.*

Quantization was a way to progressively explain experiments done beforehand, the first being the blackbody radiation spectrum. This one marked the beginnings of quantum physics.

Before explaining black body spectrum, let's recall the three kinds of spectrum that can be usually found experimentally and are pictures in Figure 72.

- A **continuous spectrum** comes from a hot and dense body like the sun, a heated solid or a perfect such body *aka* black body. It contains light in all visible frequencies that come from the random excitement of atoms in the examined body.
- An **absorption spectrum** is usually made of a continuous source of light traversing an absorbing medium like a cold gas. The resulting spectrum will be a continuous one with black lines corresponding to the frequencies absorbed by the medium.
- An **emission spectrum** is created by some rarified hot gas. It shows discrete spectrum lines corresponding to photons emitted by the excited gas atoms at specific frequencies.

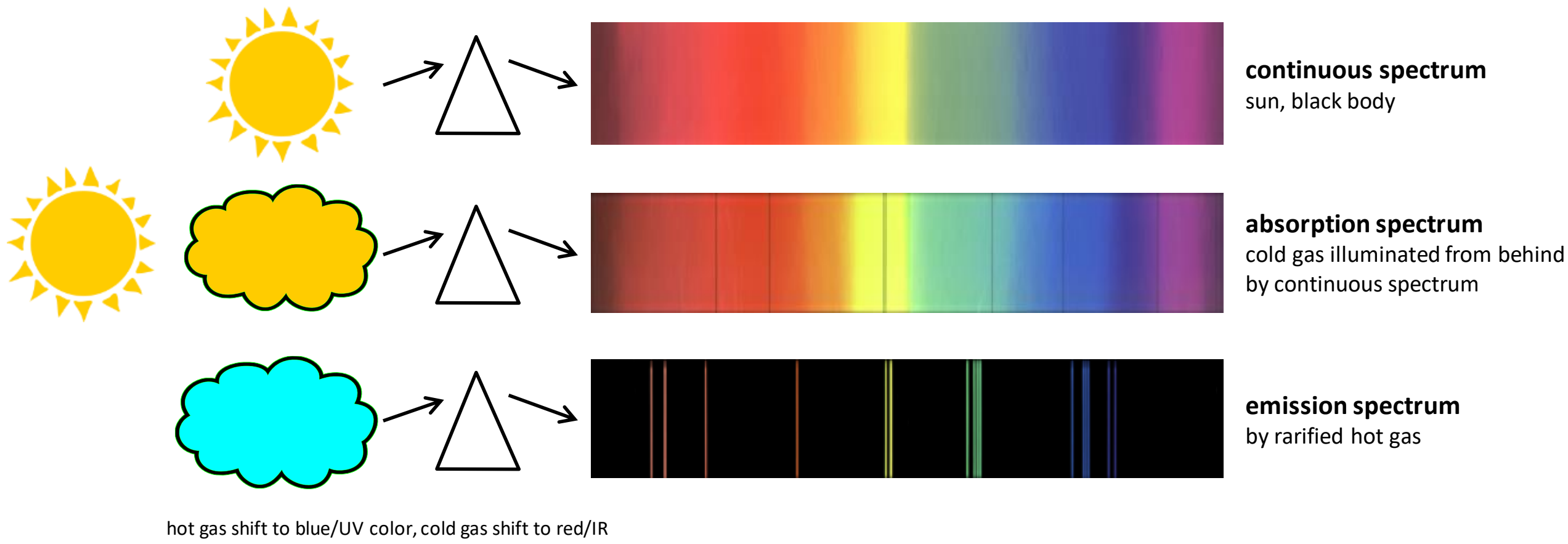

*Figure 72: differences between continuous spectrum, absorption spectrum and emission spectrum.*

Black bodies were theorized by **Gustav Kirchhoff** in 1859. These are ideal physical bodies in thermal equilibrium that absorb all incident electromagnetic waves radiations and reflects or transmits none. Since it absorbs all wavelengths, it's supposed to be black, although stars like the Sun are good approximations of black bodies and are not black at all. In usual experiments, a black body has a little hole that emits radiations which are analyzed by a spectrograph. The challenge which took four decades to be resolved was to evaluate the spectrum of the cavity radiation.

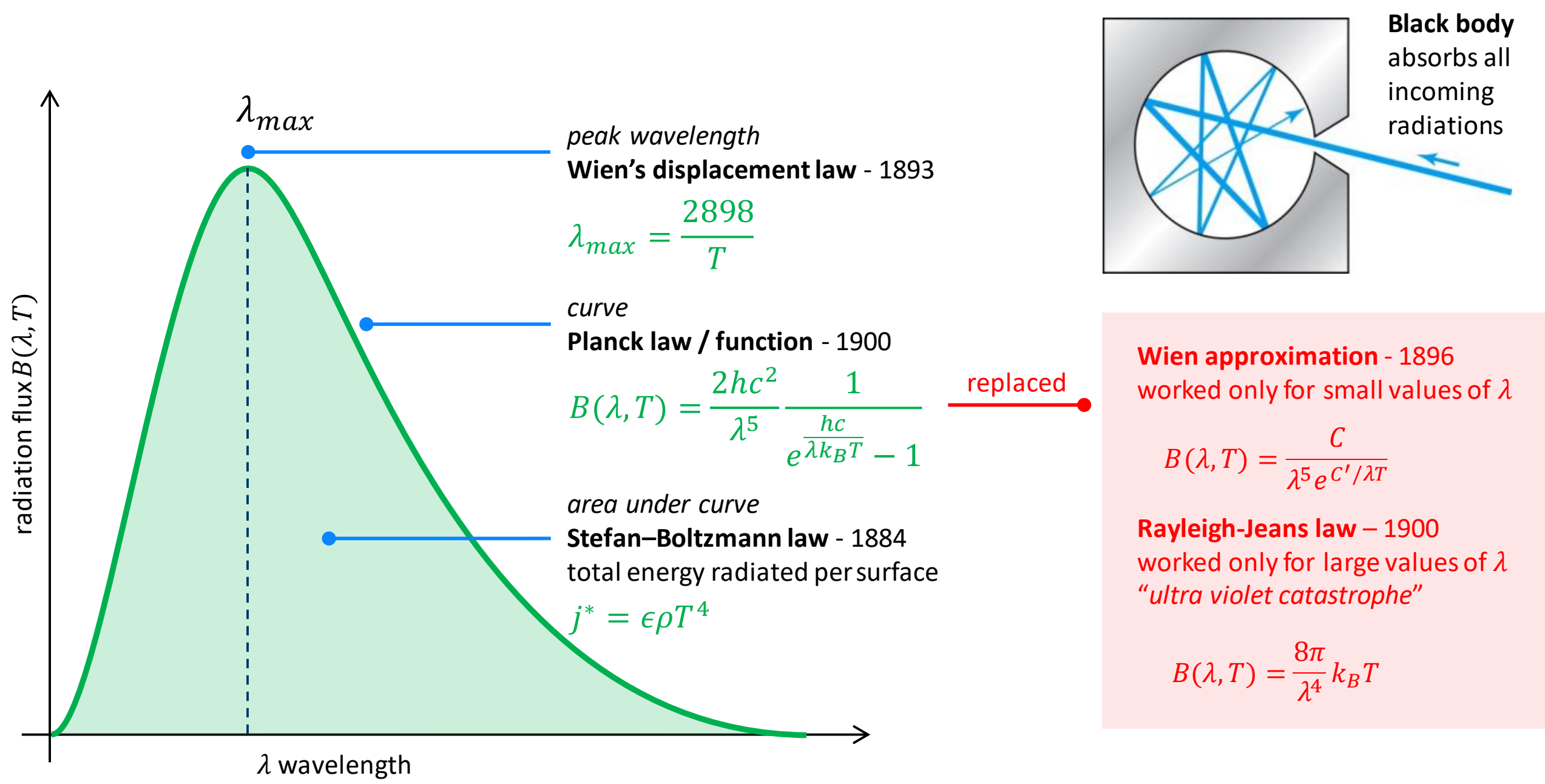

*Figure 73: blackbody spectrum explanations over time. Compilation (cc) Olivier Ezratty, 2021.*

It was first discovered that the spectrum didn't depend on the body radiation and only on its temperature T and wavelength λ (lambda). It also proved that thermal radiation was an electromagnetic one. Hot objects like lightbulbs and heated metals are close to black bodies.

As the temperature increases, the black body color, corresponding to the spectrum peak shown in Figure 73, shifts from red to blue. There were various attempts to explain the blackbody radiation with thermodynamics and oscillators and to predict the spectrum curve.

Before Planck's work, Stefan-Boltzmann's law (1884) described the relation between temperature and total energy radiated per surface area ($\epsilon \rho T^4$) and Wien's displacement law (1893) described the relationship between peak wavelength and temperature. These two laws worked well. **Wilhelm Wien** (1864-1926, Germany) even won the 1911 Nobel Prize in physics for this discovery.

Predicting the spectrum curve didn't work so well. First, Wien devised another law in 1896, Wien's approximation or radiation law that didn't work well with large wavelengths. The Rayleigh-Jeans formula created in 1900 didn't work for small wavelengths, leading to the so-called ultra-violet catastrophe. It was based on Boltzmann's statistical methods.

To make a better curve prediction, Max Planck guessed that the energy of the oscillators in the cavity was quantized and was a multiple of some quantity with the formula E = nhν, n being an integer, h being Planck's constant and ν the wave frequency. With this discretization, oscillators couldn't afford having many energy quanta for high energy levels. Thus, their number decreased as the frequency increased instead of growing exponentially as in Rayleigh-Jeans law. See the details in Figure 73.

But, at this point in time, there was no clear explanation on the origin of these quanta. The second step was Albert Einstein's work on the photoelectric effect in 1905, explaining how light and electrons interacted in quantized form. He guessed that the energy from an electromagnetic field is not spread over a spherical wavefront but is localized in individual directional quanta, which were later described as wave packets with a speed (of light) and length. But light quantization can show up in many other photon's characteristics: their polarization, their frequency, their phase and other various characteristics.

**Electrons quantum numbers**

At last, the Niels Bohr's atomic model in 1913 helped describe the electron energy transitions within atoms that explained the various hydrogen emission spectrums discovered by **Johan Balmer** in 1885, **Theodore Lyman** in 1906 and **Friedrich Paschen** in 1908, corresponding to transitions starting from the second, first and third atom electron layers. These are known as Balmer series, Lyman series and Paschen series. But other energy transitions like those from the **Zeeman** effect could only be explained by the existence of other quantum numbers.

During the 1920s, a better understanding of the quantum nature of electrons was achieved. It was progressively discovered that electrons had actually four quantum numbers:

- $n =$ **principal quantum number** corresponding to their energy level or electron shell in the atom electron shells, numbered from 1=K, 2=L, 3=M to n, n being very high for so-called Rydberg (high-energy) states close to atom ionization[124]. This number may correspond to some energy levels used in cold atoms and trapped ions qubits. It corresponds to the rows shown in Figure 74.
- $\ell =$ **orbital angular momentum** numbered from 0 to n-1 or letters (s, p, d, f, g, h, i, etc.) also named azimuthal or orbital quantum number, describes the electron subshell. It corresponds to different types of elliptic orbitals around the atom and to the columns shown in Figure 74.
- $m_l =$ **magnetic quantum number** describing the electron energy level within its subshell. Its value is an integer between $-\ell$ and $\ell$ which describes the number of different orbitals in the subshell and their orientation.
- $m_s =$ **spin projection quantum number** being either +1/2 or -1/2, in a given spatial direction (usually x, y or z in an orthonormal basis), called spin component, also named intrinsic angular momentum. This is the property used in so-called electron spin or silicon qubits. But… what is the unit of the spin? It is rarely mentioned but the spin unit is the Dirac constant $\hbar$, which equals the Planck constant h divided by 2π. What physical property is it describing? Nobody really knows. It's an intrinsic property which doesn't depend on the situation like temperature. It doesn't describe a rotation of the electron around an axis. Spin is the only quantum number that has no physical meaning equivalent in the macroscopic world.

---

[124] The principal quantum number is limited to 7 for non-excited atoms and is theoretically illimited with excited atoms. A record of n=766 was observed with hydrogen atoms in interstellar medium.

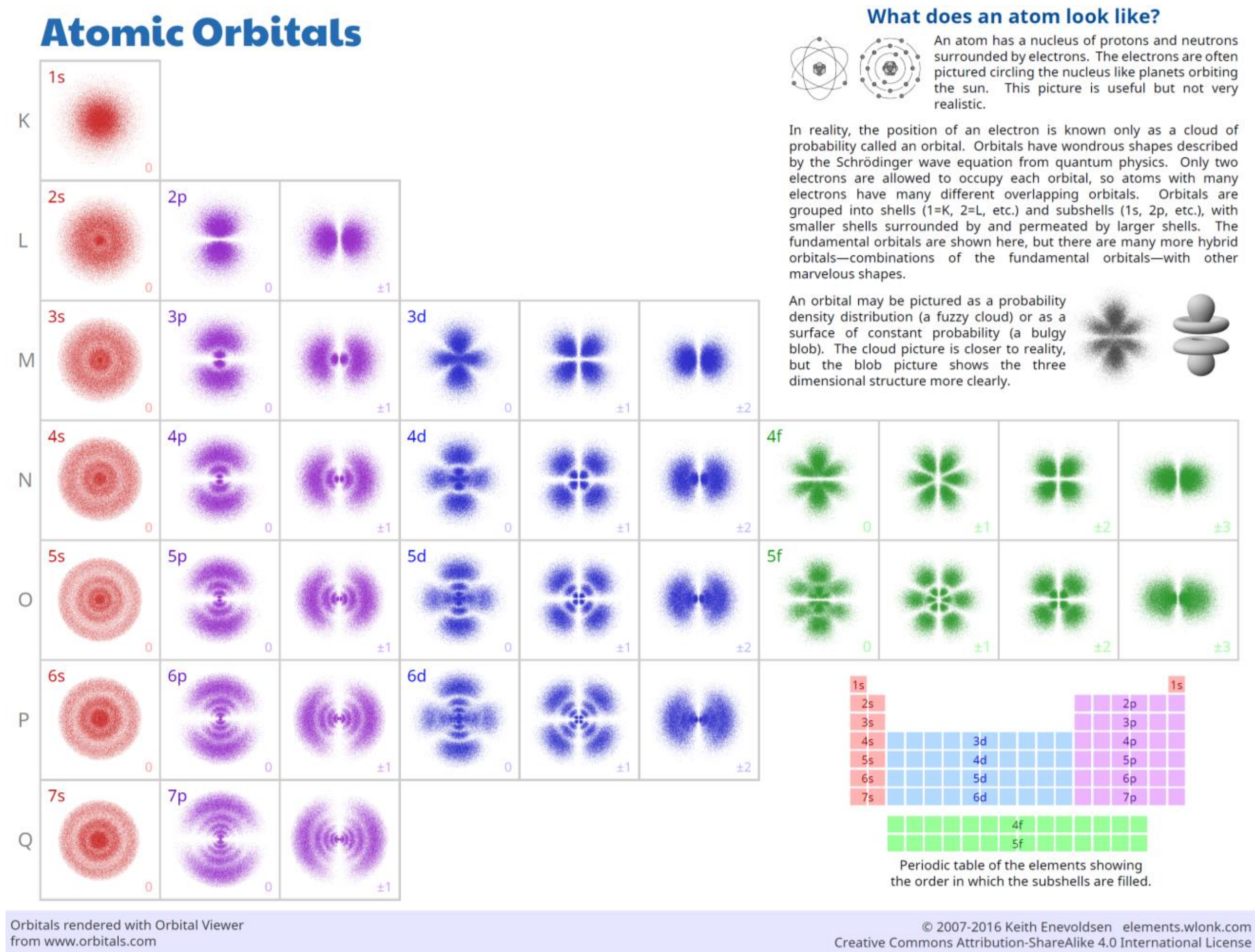

*Figure 74: electron atomic orbitals corresponding to their angular momentum quantum number. Source: Keith Enevoldsen.*

It's key to understand the effect of these various electron quantum numbers in many fields like with NV centers and silicon spin qubits, quantum dot photon sources and many others.

### Nucleons and nucleus quantum numbers

Not only electrons have a spin but also atom nucleus and their nucleons constituents that are neutrons and protons. An atom nucleus has Z protons corresponding to the element atomic number and N neutrons which add up to a total of A=Z+N nucleons. Protons and neutron have a spin of $1/2$. The nucleus has a half-integer spin ($1/2$, $3/2$, $5/2$, …) when the number of neutrons plus the number of protons is odd, an integer spin (1, 2, 3, …) when the number of neutrons and protons are both odd and no spin at all when its number of neutrons and protons are both even.

Nucleus spins are either something we need to avoid like in silicon qubits produced with $^{28}$Si, the silicon isotope with a null spin, or that we use to store qubit information like in NV centers and SiC cavities and also electron donor qubits based on atoms like phosphorus where there is a coupling between some atom nucleus spins and some free electrons.

But how is it possible to have a zero spin when you add-up the spins of protons and neutrons which are positive? Let's take a pause and provide some answer. This is due to way a nuclear spin is calculated.

Nucleons have quantum numbers that are similar to electrons quantum numbers, but with different possible values bounds and meanings:

- $n$ = **nucleon shell number** or layer with integer values ranging from 0 to 6. It is bounded and there is no equivalent of Rydberg states in nucleus with large principal quantum number. The nuclear shell model is the equivalent of the atomic Bohr model related to electron shells.
- $\ell$ = **orbital angular momentum** quantum number which is also quantized with an integer value starting at 0, the angular momentum itself being $L^* = \hbar\sqrt{\ell(\ell+1)}$.
- $m_l$ = **magnetic quantum number** with integer values ranging from $-\ell$ to $\ell$. In each nucleon shell, nucleons of the same type have a tendency to regroup by pairs with opposite magnetic quantum number.
- $m_s$ = **spin quantum number**, being either +1/2 or -1/2, in each spatial direction, also named intrinsic angular momentum. A nucleon spin $s$ equal to 1/2 is the size of the vector $\vec{s}$. The spin angular momentum is $S^* = \hbar\sqrt{s(s+1)}$ with s=1/2.

A nucleon total angular momentum is a vector $\vec{j} = \vec{\ell} + \vec{s}$ and $j = \ell + s$ in scalar representation with $\ell$ being the nucleon orbital angular momentum and $s = 1/2$ its intrinsic angular momentum or spin. The scalar representation is a good approximation of the vector representation since nucleons move in an average magnetic field orienting them in a similar direction. In the end, the atom's **nuclear spin** is the sum of its nucleon's total angular momentum $j$.

As atom nucleus size grows, nucleus shells are filled progressively. Filled layers have a number of neutrons or protons called "magic numbers" (2, 8, 20, 28, 50, 82, 126) as shown below in Figure 75. Atoms with entirely filled layers of either neutrons or protons are more stable. Nucleons pair in orbits with projections $\pm m_\ell$ such that their momenta cancel. The notion of layer magic number applies separately for protons and for neutrons. For example, $^{116}$Sn (selenium) has a magic number of 50 protons and $^{54}$Fe (iron) has a magic number of 28 neutrons. Filled shells have a total angular momentum of zero since made of pairs of neutrons or protons with opposite projections of total angular momentum. That's why an even number of protons and neutrons lead us to have a zero nuclear spin. And when you have both a magic number of neutrons and protons, your nucleus is doubly magic like with $^{40}$Ca and $^{208}$Pb, and has an exceptional stability. All this refers to atoms and nucleus in their ground state.

On top of these numbers, nucleus have a parity that is $\lambda = (-1)^\ell$ where $\ell$ is the total orbital angular momentum of the nucleus. Its value corresponds to the symmetrical or asymmetrical structure of the nucleus.

Analyzing some material nuclear spin is the basis of **NMR spectroscopy**. It exposes it to a strong homogeneous magnetic field, usually generated with cooled superconducting magnets which create it thanks to their support of large electric currents with no resistance. A sample is then irradiated by a radiofrequency (TF) field around the Larmor precession frequency of the searched elements, in the hundred MHz range. This is the frequency of the nucleus spin vector rotation in a cone around the axis of the ambient magnetic field. A receiver captures the transmitted RF signal, amplifies it and pass it through a quadrature demixer fed by a reference frequency tone. It down converts the signal to a lower frequency and decomposes it into its in-phase and quadrature which is then converted into digital format through with ADCs (analog-to-digital converters) before being analyzed digitally.

Two other notions are related to atom's nucleus and are frequently mentioned elsewhere in this book:

**Spin–orbit coupling** or spin–orbit interaction is a relativistic interaction of a particle's spin with its motion inside a potential. One example if the shifts in an electron's energy levels that due to electromagnetic interaction between the electron's magnetic dipole, its orbital motion and the electrostatic field of the atom nucleus.

**Hyperfine structure** are small differences in otherwise degenerate (equivalent, equal) energy levels in atoms, molecules and ions that are explained by the electromagnetic multipole interaction between the nucleus and electron clouds. In atoms, hyperfine structure come from the energy of the nuclear magnetic dipole moment interacting with the magnetic field generated by the atom electrons and the energy of the nuclear electric quadrupole moment in the electric field gradient due to the distribution of charge within the atom.

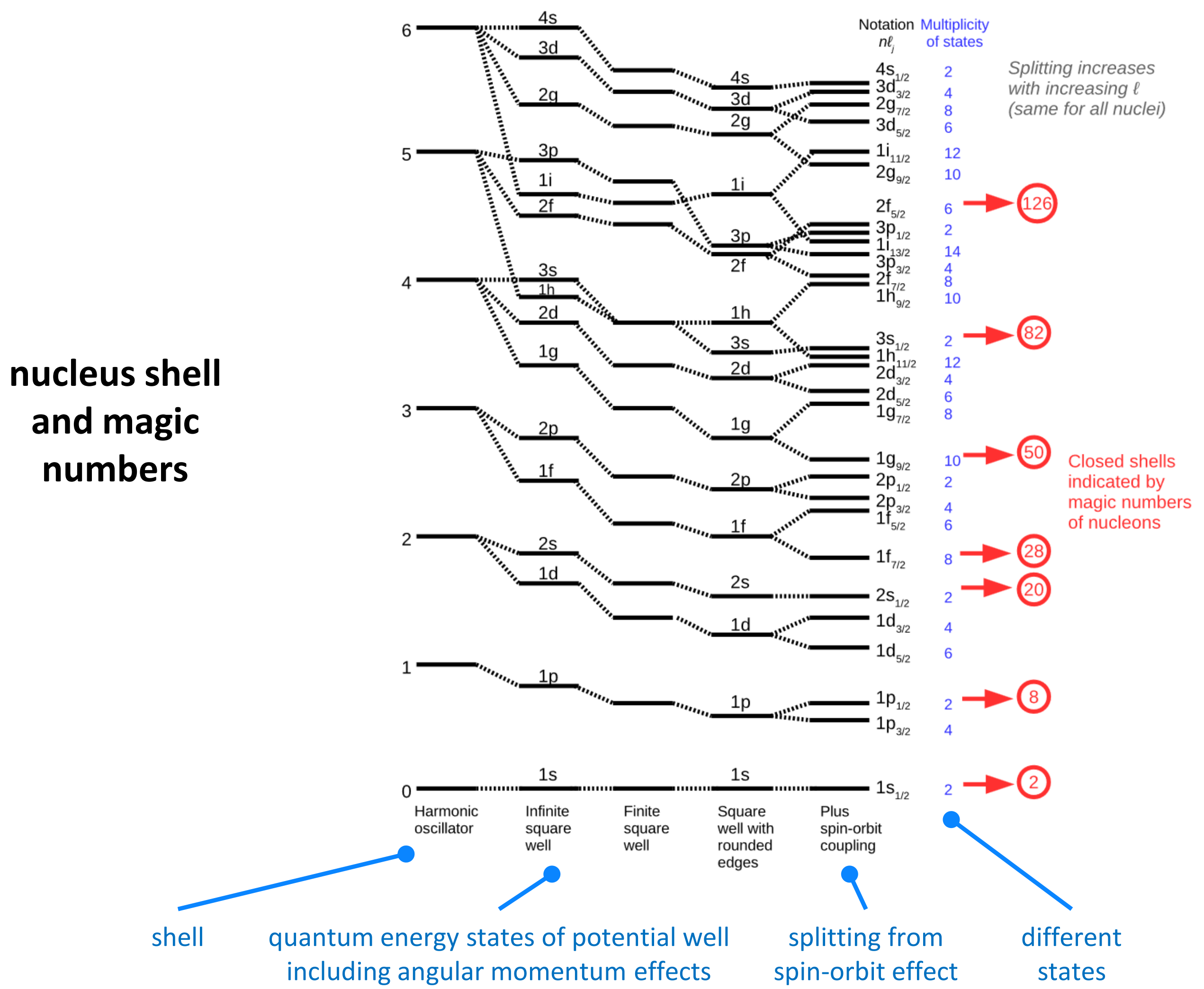

*Figure 75 : nucleus shells and magic numbers. Source: Particle and Nuclear Physics Handout #3 by Tina Potter, 2022 (124 slides).*

The cohesion of atom's nucleus comes from the strength of the nuclear force that binds nucleons together. It is countered by Coulomb's force that creates a repulsion between same charge particles like protons. The relative value of the nuclear force and the Coulomb repulsion force explain nuclear fusion for small atoms and fission for large atoms, iron being in the neutral zone in the elements table.

### Photon quantum numbers

Photons also have their quantum numbers but they are different than with electrons and nucleons. We describe it in the section dedicated to photon qubits, starting page 427.

In quantum information systems, we use quantum objects which can usually have two different separable states that can be initialized, modified and measured. Even superconducting loops in superconducting qubits rely on two systems levels clearly distinct for the oscillating current flowing through their Josephson effect insulator.

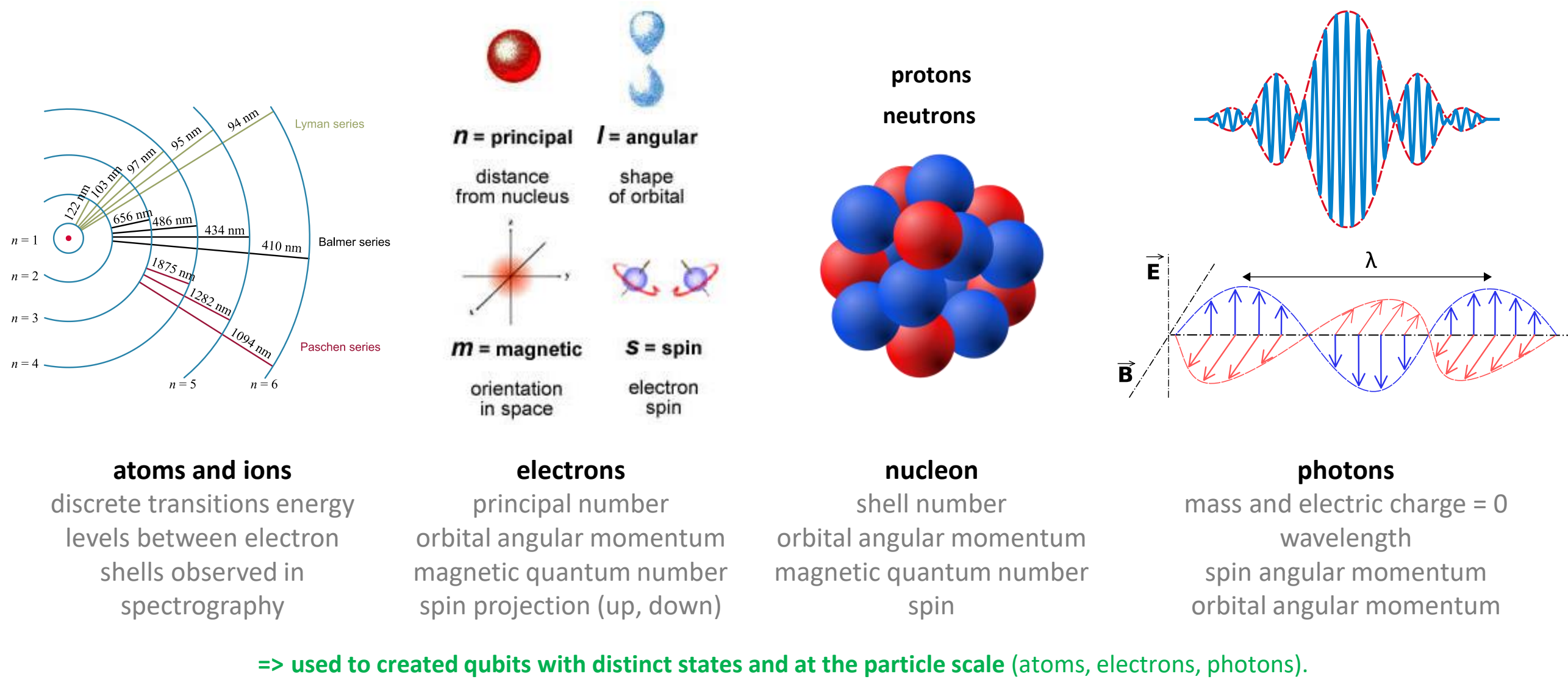

*Figure 76: quantization applied to atoms, ions, electrons and photons. (cc) Olivier Ezratty, 2022, with Wikipedia images source.*

# Wave-particle duality

We often read and hear that quantum objects like photons and electrons are both waves and particles.

The right manner to describe this would be to say that they behave differently depending on the way they are observed. In some experiments, these quantum objects behave like classical waves, are not localized in space and generate interferences when added together, a bit like colors can mix (photons) and sounds can mix (acoustic waves). In other experiments, they behave as classical particles and can be localized in space and have a kinetic momentum and mass[125]. One simpler way to interpret things is to say that quantum objects act as a particle when observed and as waves when not observed.

Various experiments such as Young's double-slit experiment show that both photons and electrons behave both as particles and as waves depending on the context and measurement system, generating interference fringes when observed as waves. You can observe the path of a quantum object or the interferences it creates, but not both simultaneously.

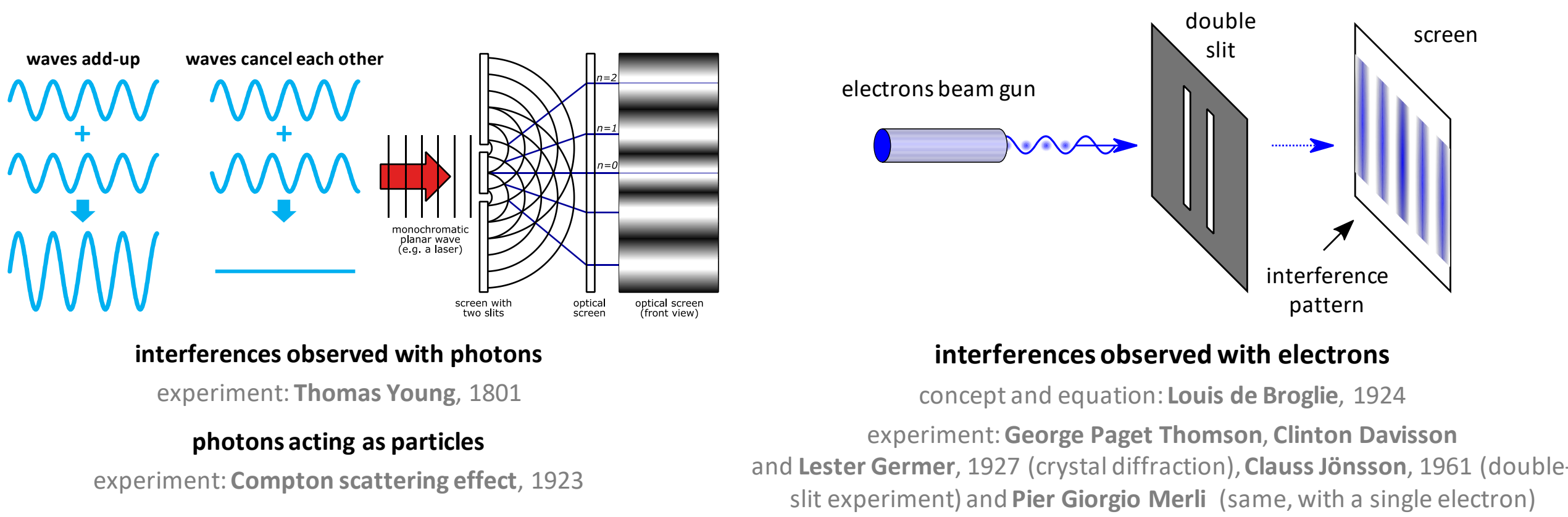

*Figure 77: experiments showing wave-particle duality with photons and electrons.*

This is the Bohr's principle of complementarity according to which it is not possible to apply observables simultaneously in terms of particles and waves. It shows up in the Young experiment: if we let the quantum object traverse both slits, it behaves like a wave and creates interferences.

[125] Usually, it is impossible to observe these two behaviors simultaneously although there are some exceptions.

If we detect the quantum object in each of the slits, practically done with closing one of the slits, it creates a measurement-based decoherence and the quantum object behaves and is observed as a particle. And the classical probabilities of particle observation don't add up to make for the interferences observed with the wave observation. This wave-particle duality is linked to the quantum physics mathematical formalism that relies on vectors that can add up linearly like waves.

It led to a still unsolved mystery, the "which-way" question. When interference fringes appear on the screen, by superposition of paths coming from the two slits, which path did the single photon or electron take?

The wave-particle duality is used in many quantum computers to make physical qubits such as trapped ions interact with energy in the form of photons emitted by lasers. Qubits can also interfere with each other thanks to interferences.

### Delayed choice experiment

John Wheeler proposed various thought experiments between 1978 and 1984 to determine if light choses its path with sensing the experimental devices. The Wheeler's delayed-choice or which-way experiment asked the question: when does a quantum object decide to travel as a wave or as a particle?

It led to various experiments like the 1999 quantum eraser but the most decisive experiment was conducted by a team of French researchers in 2006 as shown in Figure 78[126].

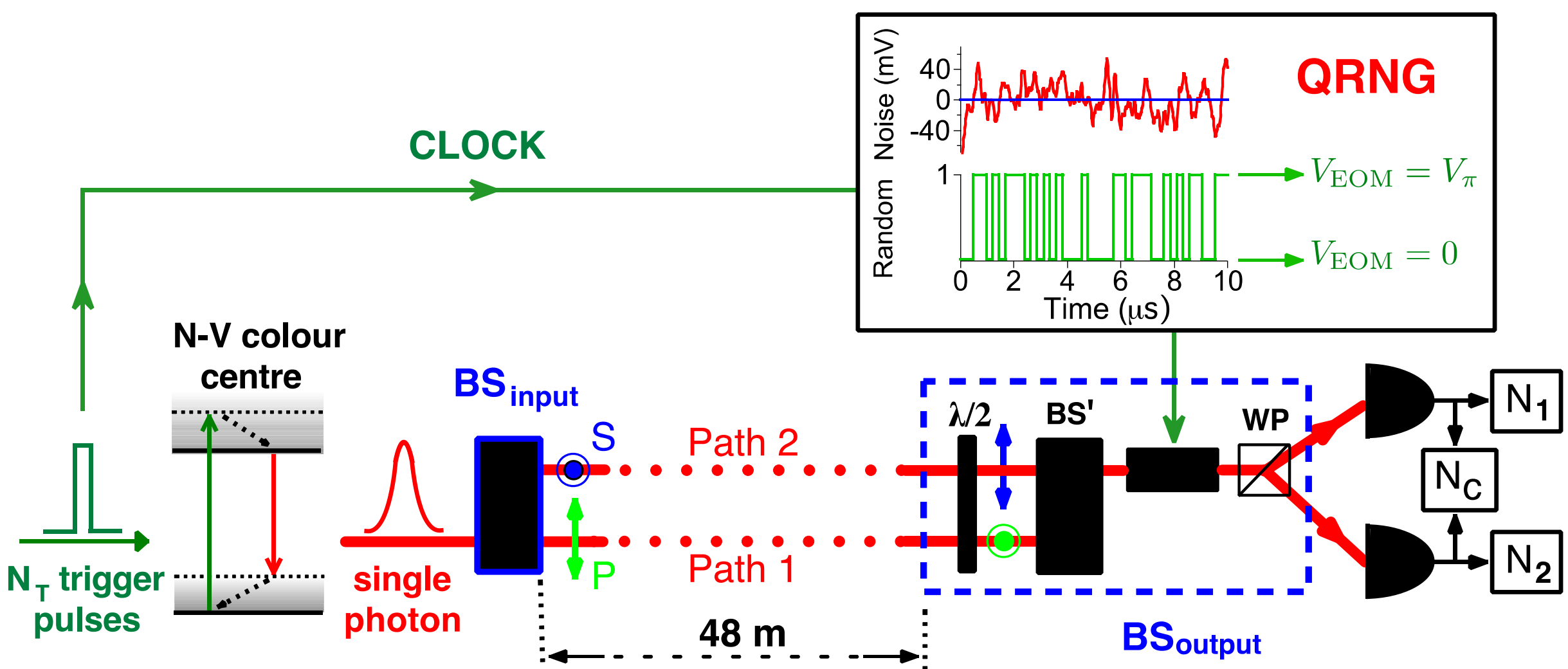

*Figure 78: delayed choice experiment and its quantum eraser. Source: Experimental realization of Wheeler's delayed-choice GedankenExperiment by Vincent Jacques, Frédéric Grosshans, Philippe Grangier, Alain Aspect, Jean-François Roch et al, 2006 (9 pages).*

They generated pulses of single photons with an NV centers source created by Jean-François Roch, a pioneer in this domain, that were sent through a first beam splitter ($BS_{input}$) and a delay line of 48 meters. Then, the two beams traversed a dynamic-controlled beam-splitter by electro-optical modulator driven ($BS_{output}$) by a quantum random number generator (QRNG).

At last, two photon detectors ($N_1$ and $N_2$) could determine if the photon behaved as a particle (no interference due to the inactive beamsplitter) or as a wave (with interferences due to the activated beamsplitter).

---

126 See Experimental realization of Wheeler's delayed-choice GedankenExperiment by Vincent Jacques, Frédéric Grosshans, Philippe Grangier, Alain Aspect, Jean-François Roch et al, 2006 (9 pages). The experiment used a single photon source using NV centers. The experiment has been reproduced many times since then with many variations. See for example A generalized multipath delayed-choice experiment on a large-scale quantum nanophotonic chip by Xiaojiong Chen et al, 2021 (10 pages) which is based on a nanophotonic component.

The QRNG clock was near the photon source, but the QRNG was positioned close to the dynamic beamsplitter. The experiment determined that the wave/particle behavior of the photons in the interferometer was dependent on the choice of the measured observable at the end of the photon journey, not the beginning.

And even when that choice was made at a position and a time sufficiently separated from the entrance of the photons in the interferometer. Although it's still debated, it does not require a backward in time effect explanation.

Other more delayed-choice sophisticated experiments are regularly done. A Chinese team demonstrated a generalized multipath wave-particle duality implemented by a large-scale silicon-integrated multipath interferometers[127].

### Schrödinger's wave equation

The wave-particle duality led Schrödinger to create his famous wave equation which describes a massive non-relativistic quantum object with a wave function with the probabilities of finding the particle at a particular position in space at a given time. Ladies and gentlemen, here is this wave equation and its constituents, unleashed below in Figure 79.

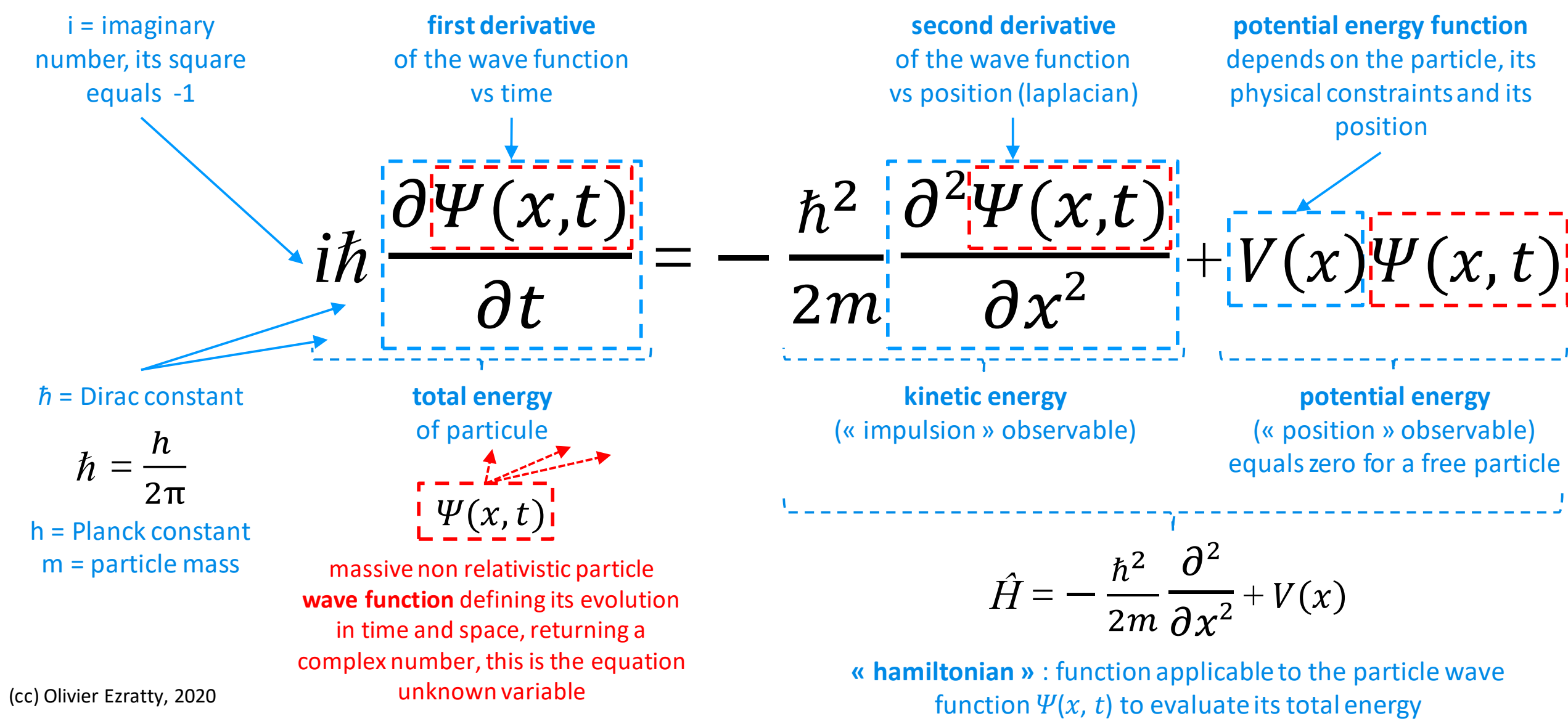

*Figure 79: the famous Schrodinger's wave equation explained in detail (cc) Olivier Ezratty, 2021.*

Here's how to understand the components of this equation and their implications:

- Its **unknown** is the wave function of the particle ψ(x, t) that describes its probabilistic behavior in space and time. x indicates the position of the particle in space, with one, two or three dimensions depending on its constraints, and t is the time. This function returns a complex number that encodes the wave amplitude and phase.
- The full Schrodinger wave equation illustrates the principle of **energy conservation**. The item to the left of the equation describes the total energy of the particle at a given time and place. The element on the right includes the kinetic energy of the particle and its potential energy. Like said about the quantum physics postulates, starting page 86, the Schrodinger's Hamiltonian, which is a time-dependent unitary matrix operator, is expressed differently with photons and with relativistic massive particles.

---

[127] See A generalized multipath delayed-choice experiment on a large-scale quantum nanophotonic chip by Xiaojiong Chen et al, 2021 (10 pages).

- The **wave function square** is equal to the probability of finding the particle at location x at time t. For an electron, which is the most commonly analyzed particle with this equation, it is an indication of the probability of finding it at a given distance from the nucleus of the atom around which it orbits. Logically, as a result, the sum of the probabilities of finding the particle somewhere is equal to 1.

  This is called a normalization constraint. One of its derivatives is the Max Born function that we will see later. The modulus of a complex number is the size of its vector. If z = a + ib, the modulus $|z|$ of z is thus the square root of the sum of the squares of a and b, see *below*.

- It is a **partial differential** equation, i.e. it connects its components via derivative functions, in this case of first degree (a slope on a curve) and second degree (an acceleration). The particle wave function appears three times in the equation: to the left of the equation with a first derivative on the time of the wave function, to the right with a second derivative on its position and with a simple multiplication with the function V(x).

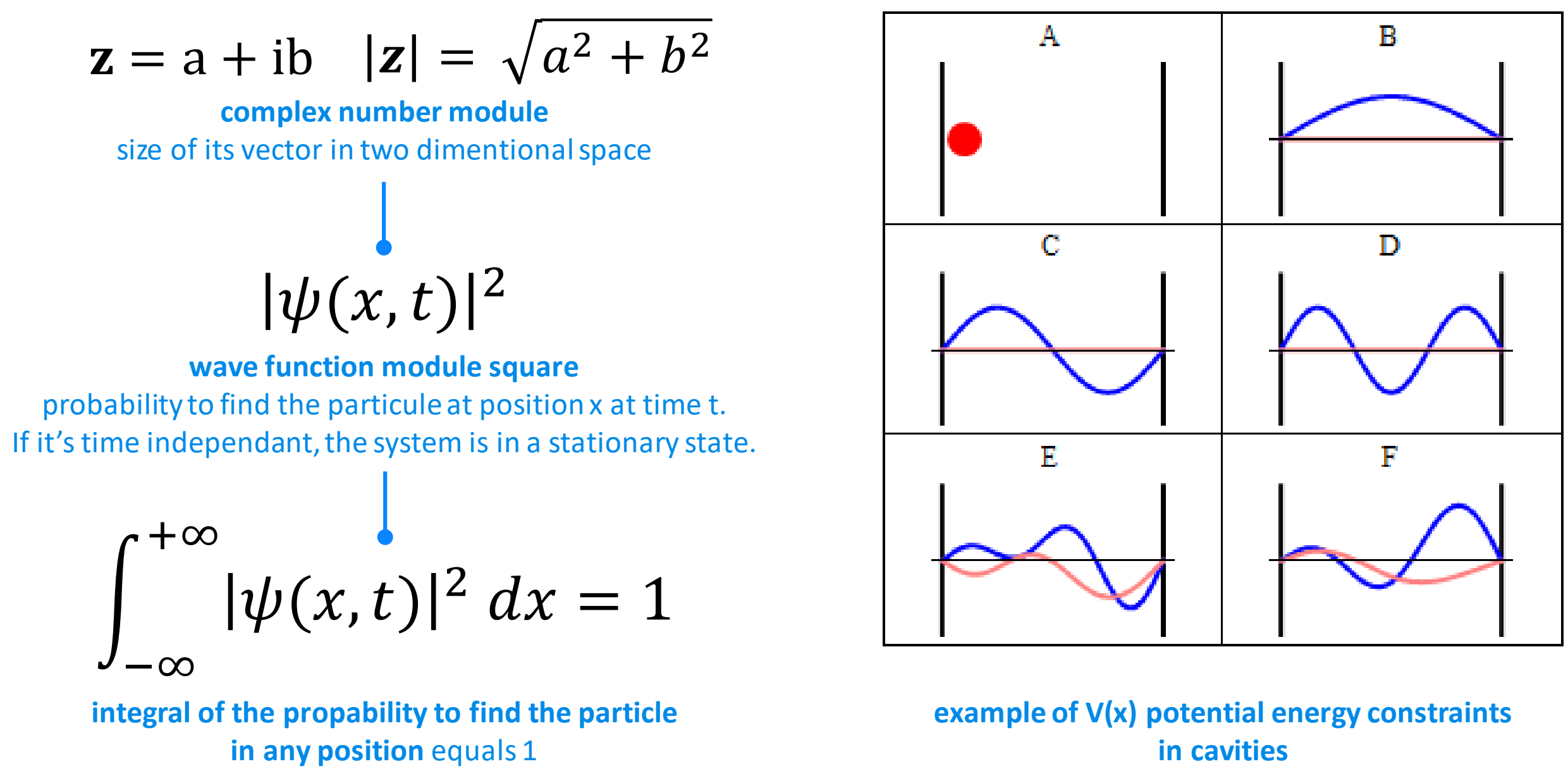

*Figure 80: constraints of the Schrodinger's equation (cc) Olivier Ezratty, 2021.*

- The **potential energy of the particle** is defined by the function V(x) which depends only on the particle position in space and its physical constraints, in particular electromagnetic ones. When a particle is free and moves without constraints, this function returns zero. This function V(x) is the main variable of Schrödinger's equation.

- Schrödinger's equation is **analytically solved** in a limited number of cases such as for the electron of a hydrogen atom, a free particle, a particle in a potential well or box or a quantum harmonic oscillator. In the most complex cases, the resolution of the equation requires non-analytical methods, raw calculation and simulation. It is one of the fields of application of quantum simulators to solve the Schrödinger equation in cases where analytical methods are not available. Any micro or macro-object has a Schrödinger wave function, all the way to the entire Universe. But the equation only makes practical sense for nanoscopic objects.

- The equation is **linear** over time. This means, among other things, that any combination of solutions of the equation becomes a new solution of the equation. This makes it possible to decompose a wave function into several elementary wave functions that are called the "eigenstates" of the quantum object. They correspond to the different energy levels of the particle that are discrete when the particle is constrained in space, like the electrons in an atom.

One can indeed in this case derive the notion of quantification of the particle states from the Schrödinger equation (demonstration). The linearity of this equation has a lot of consequences like superpositions, entanglement as well as the no-cloning theorem.

- The operator who acts on the right side and accumulates the second derivative and the potential energy function is called a **Hamiltonian,** which describes the total energy of the system. We find this expression in the quantum annealing calculation with D-Wave and with quantum simulators.
- This equation is a **general postulate** that has been experimentally validated in a large number of cases. Its interpretation has given rise to much debate, namely, is it a simple probabilistic model or does it describe reality? We deal with this in the chapter on the philosophy of quantum physics.
- The generic Schrödinger equation presented so far is said to be **time dependent**. This equation is presented in various ways depending on the needs and annotations. The second derivative of the wave function on the position of the particle is sometimes presented with the nabla sign squared ($\nabla^2$).

  A nabla operates a derivative on a scalar or vector function. The $\nabla^2$ operates a second derivative, also called Laplacian. The most concise form of Schrödinger's equation is on the bottom right, with a Hamiltonian operator on the left ($\hat{H}$) and the energy operator on the right ($\hat{E}$), both of which apply to the particle wave function.

$$\left[-\frac{\hbar^2}{2m}\nabla^2 + V\right]\Psi = i\hbar\frac{\partial}{\partial t}\Psi$$

$$\hat{H}\psi(x,t) = \hat{E}\psi(x,t)$$

*Figure 81: concise versions of Schrodinger's wave equation.*

- There is a **time-independent** form of Schrödinger's equation that applies to particles in a stationary state[128]. In this version of Schrödinger's equation, the energy operator E is a simple constant, a real number.

$$\left[-\frac{\hbar^2}{2m}\nabla^2 + V(\mathrm{r})\right]\Psi(\mathrm{r}) = \hat{E}\Psi(\mathrm{r})$$

*Figure 82: time-dependent version of the Schrodinger's equation.*

- The Schrödinger equation is **symmetric** or **antisymmetric** depending on the particle type. When applied to two quantum objects $r_1$ and $r_2$, $\psi(r_1, r_2) = \psi(r_2, r_1)$ when the equation is symmetric (meaning, the wave equation is not differentiated by the given particles order) and $\psi(r_1, r_2) = -\psi(r_2, r_1)$ when it's antisymmetric. The first case corresponds to bosons which can be indistinguishable and "live" together and have a zero or integer spin and the second, to fermions, which can't cohabit with the same quantum state at the same location and have half-integer spins. All this is a consequence of Pauli's exclusion principle.
- The $\psi(x,t)$ function must be a **continuous function** and "filled" everywhere in space. Its value is bounded by 0 and 1, with no infinite value anywhere. It also has a single value, even in the case of superposition. In that case, the $\psi(x,t)$ is a linear superposition of two Psi functions and is itself a psi function. A quantum superposition is just another wave function.

For a system with several quantum objects, the wave function describes the quantum system state, or quantum state. According to the Copenhagen interpretation of quantum physics, the wave function from the Schrödinger equation contains the best description possible of a quantum system.

---

[128] According to Wikipedia: "*A standing wave is the phenomenon resulting from the simultaneous propagation in opposite directions of several waves of the same frequency and amplitude, in the same physical medium, which forms a figure, some elements of which are fixed in time. Instead of seeing a wave propagating, we see a standing vibration but of different intensity at each observed point. The characteristic fixed points are called pressure nodes.* ».

If electrons and photons both can behave as waves, they have not the same wavelengths. Indeed, a photon with an energy of 1 eV (electron-volt) has a wavelength λ of 1240 nm (in the infrared spectrum) while an electron with the same energy has a much shorter wavelength of 1.23 nm (in the X-ray spectrum). This short wavelength explains why we use electron microscopes to probe matter with a much better resolution than light-based microscopes.

Relativistic particles obey to Dirac and Klein-Gordon wave equations while photons are described with Maxwell's equations combined with a formalism coming from the so-called second quantization which regroups superposed photons, use photon numbers, and creation/annihilation operators.

### Large objects wave behavior

The wave-particle duality was verified with atoms in 1991 in interferometry experiments involving lasers and classical optics. A Young double-slit experiment was also carried out in Austria in 2002 with fullerene molecules ($C_{60}$, formed of 60 carbon atoms as in Figure 83 [129], but also with a 70 atoms variant) and in 2012 with molecules containing 58 and 114 atoms, the latter named $F_{24}PcH_2$ being made of fluorine, carbon, oxygen, hydrogen and nitrogen[130]. Figure 84 shows the shape of the molecule. In 2019, the same kind of experiment was done with a slightly more complex molecule, a polypeptide of 15 amino acids which serves as an antibiotic, gramicidin A1[131] .

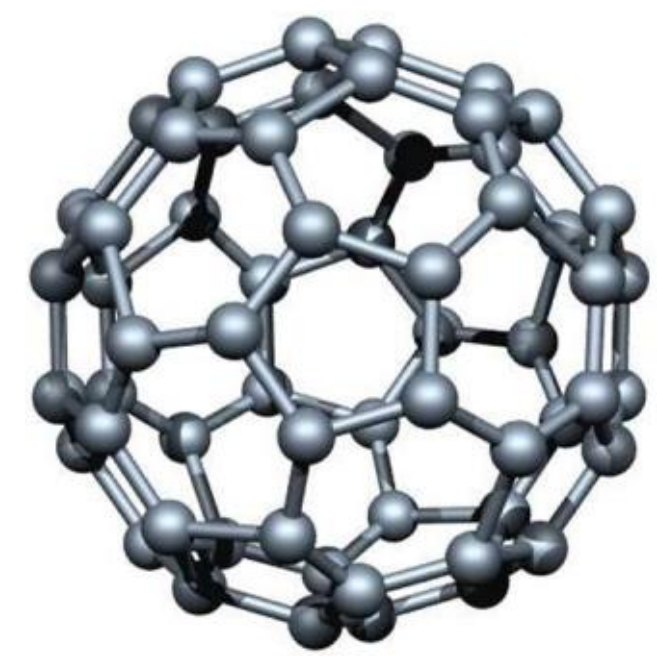

*Figure 83: $C_{60}$ fullerene molecule.*

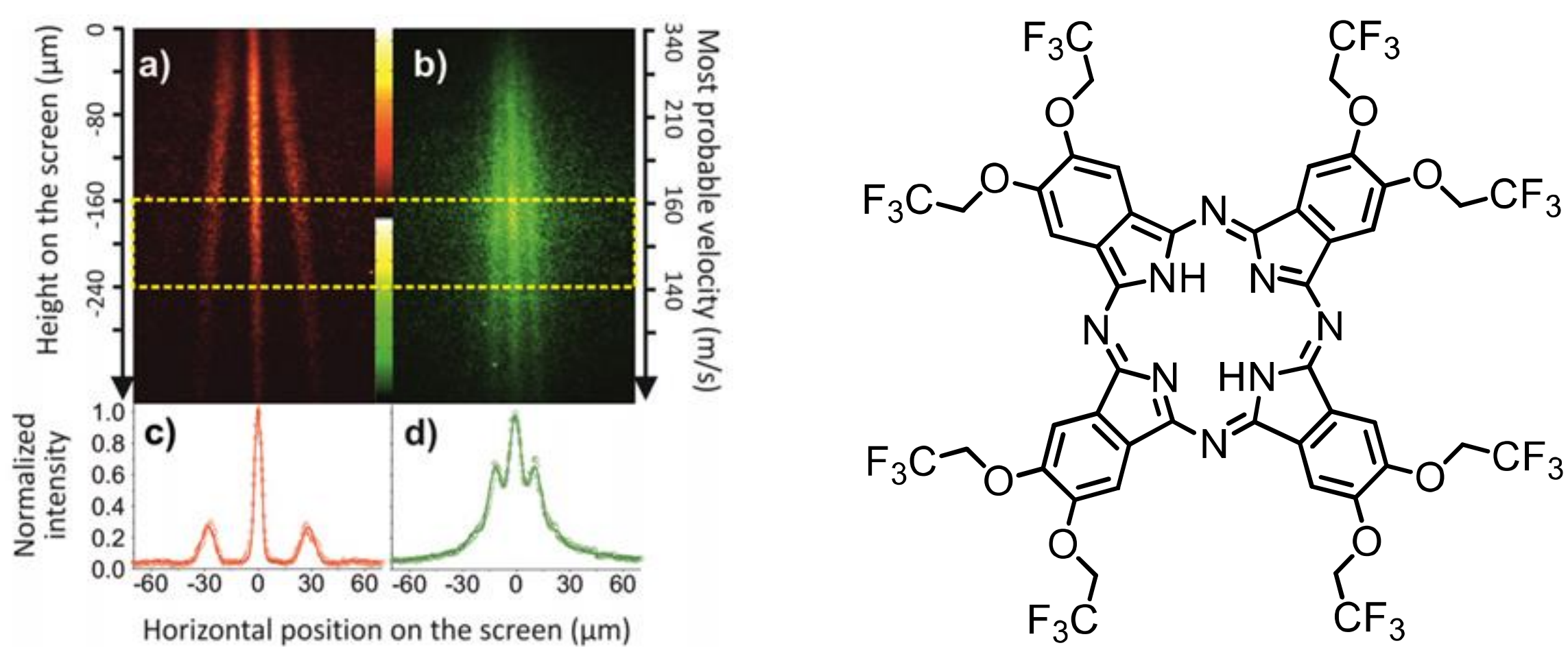

*Figure 84: $F_{24}PcH_2$ made of fluorine, carbon, oxygen, hydrogen and nitrogen. Sources: Real-time single-molecule imaging of quantum interference by Thomas Juffmann et al, 2012 (16 pages) and Highly Fluorinated Model Compounds for Matter-Wave Interferometry by Jens Tüxen, 2012 (242 pages).*

In 2021, other experiment led to the creation of larger quantum objects, sized 100 and 140 nm, and cooled at ultra-low temperatures[132].

---

[129] See Quantum interference experiments with large molecules by Olaf Nairz, Markus Arndt and Anton Zeilinger, 2002 (8 pages).

[130] See Real-time single-molecule imaging of quantum interference by Thomas Juffmann et al, 2012 (16 pages). See also the video of the experiment. Highly Fluorinated Model Compounds for Matter-Wave Interferometry by Jens Tüxen, 2012 (242 pages) describes the experimental device for the verification of the wave-matter duality of large molecules.

[131] See A natural biomolecule has been measured acting like a quantum wave for the first time, November 2019, which refers to Matter-wave interference of a native polypeptide by Armin Shayeghi et al, October 2019 (10 pages).

[132] See How Big Can the Quantum World Be? Physicists Probe the Limits by Philip Ball, Quanta Magazine, July 2021, Real-time optimal quantum control of mechanical motion at room temperature by Lorenzo Magrini et al, July 2021 (36 pages) and Quantum control of a nanoparticle optically levitated in cryogenic free space by Felix Tebbenjohanns et al, Nature, July 2021 (26 pages).

### Photon's wave-particle duality

On the other hand, photons can behave under certain conditions like particles. When they reach an atom, they can transmit it some kinetic motion. This is what makes it possible to generate the somewhat counter-intuitive physical phenomenon of atoms laser cooling using lasers and a Doppler effect. Temperature is related to the movement of atoms in their gaseous, liquid or solid medium. Lowering the temperature means slowing down the movement of atoms. A Doppler effect is used to do this. The moving atoms are illuminated with a laser whose frequency is tuned just below the energy absorption level of the atoms.

The atoms moving towards the light will absorb the photons because these have an apparent frequency that is higher than the absorption level. This reduces the kinetic energy of the atoms receiving the photon.

The photons moving in the other direction will not absorb them because the apparent frequency of the incident photon is below the absorption level, so it's unable to change the energy state of the atoms.

Thanks to the random movement of the atoms in all directions, after a certain time, the overall temperature drops. This phenomenon slows down once the velocity of the atoms falls below a certain threshold, which explains the Doppler effect attenuation ("Doppler shift").

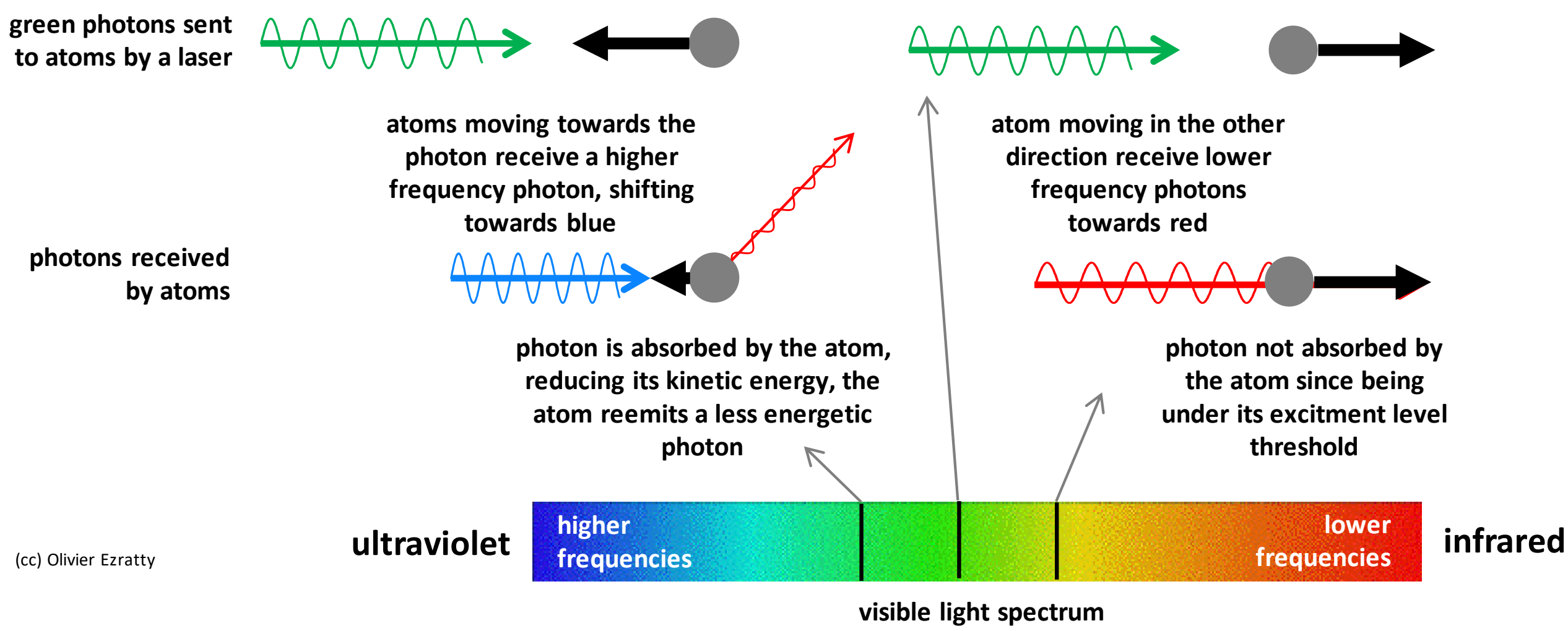

*Figure 85: explanation of Doppler effect with photons, (cc) Olivier Ezratty, 2021.*

These techniques are used to cool atoms to temperatures close to absolute zero. It is used to prepare cold atoms and trapped ions used in certain types of quantum computers, often in combination with magnetic and/or electronic traps to control the atoms position[133].

The record low temperature was reached in 2019 with 50 nK, achieved by researchers from JILA, the joint laboratory of NIST and the University of Colorado[134].

## Superposition and entanglement

Superposition and entanglement are directly related to the wave nature of quantum objects and to the linearity of the underlying mathematical models expressed in quantum physics postulates.

---

[133] Doppler measurement is also used to evaluate the speed at which stars and galaxies move away from each other and to evaluate the rate of expansion of the Universe. Other atoms laser-based cooling methods crafted to reach lower temperatures include Sisyphus cooling first proposed by Claude Cohen-Tannoudji in 1989 and using two counter-propagating lasers using orthogonal polarization, evaporative cooling using magneto-optical traps (MOT) and optical molasses with 3D Doppler effect.

[134] See JILA Researchers Make Coldest Quantum Gas of Molecules, February 2019. The 50 nK record was obtained with laser cooling of a gas containing 25,000 potassium-rubidium molecules.

## Superposition

The strawman's version of superposition is that quantum objects can be simultaneously in several states or locations, such as the direction of electron spin, upward or downward, the linear polarization of photons, horizontal or vertical, or the frequency, phase or energy of an oscillating current in a superconducting loop crossing the barrier of a Josephson junction. It is not correct according to canonical interpretations of quantum mechanics. It is more related to quantum objects behaving as waves when not being measured. Superposition is also a mathematical consequence of quantum postulates and wave-particle duality. It results from the fact that a linear combination of solutions to the Schrödinger equation is also a solution to this equation.

A quantum states of a given quantum object can be added together or superposed. Superposition explains the interferences obtained with electrons in the 1961 double-slit experiment.

A quantum object is not per se in a superposition of various states. It has a single and predictable quantum state described by a probability distribution of given observables. Measuring this property can provide different values according to the probability distribution. That's all.

According to the Copenhagen interpretation of quantum physics, one shouldn't try to give a physical meaning to superposition before any measurement. In a classical physics interpretation, superposition could be explained by a very high frequency of quantum state changes. It is considered to be totally inaccurate for specialists, but it is still an intuitive way to figure out how superposition looks like in the physical world.

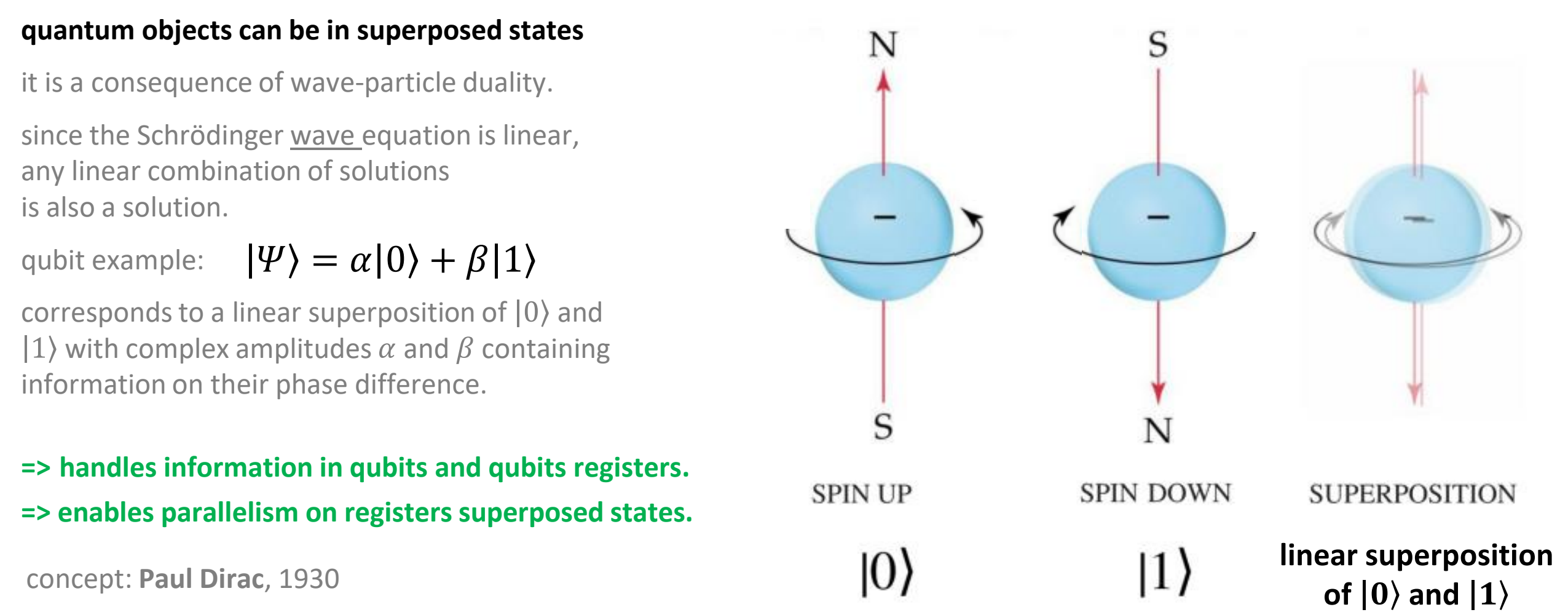

*Figure 86: electron spin superposition. (cc) Compilation Olivier Ezratty, 2021.*

Superposition can happen with various weird situations, as we'll later see. For example, you can create superposition between several photon Fock-states, meaning, superposing 0 photon, 1 photon and 2 photons, or even photon frequencies. You can even superpose temperatures[135] and thermodynamic evolutions with opposite time arrows[136] which can challenge your willingness to visualize what's it all about!

In quantum computing, superposition shows up with qubits, allowing which have an internal "value" linearly combining their basis states $|0\rangle$ and $|1\rangle$ with complex amplitudes instead of having one of the two values as with classical bits. This mathematical view is expanded to N qubits whose internal state is characterized by $2^N$ complex amplitude values. This contributes to the massive parallelism enabled by quantum computers. It looks like it should enable some exponential computing capacity but it's not the case. As we'll investigate later in this book, superposition alone is not sufficient. We also need entanglement and some specific quantum gates to really bring some exponential acceleration.

[135] See Quantum Superposition of Two Temperatures by Arun Kumar Pati and Avijit Misra, December 2021 (7 pages).

[136] See Quantum superposition of thermodynamic evolutions with opposing time's arrows by Giulia Rubino, Gonzalo Manzano and Časlav Brukner, November 2021 (10 pages).

In the case of a single quantum object, superposition is a combination of states corresponding to several exclusive states of an observable. Coherence is another name describing a superposition. And decoherence is a phenomenon that destroys superposition, particularly with quantum measurement.

**Entanglement**

The simplest way to describe an entangled state of two quantum object is to say these have a correlated state, whatever the distance between them. They form sort of a single object. You touch one, it will affect the other. You measure one object, then the other, and the related results will be correlated. This can be checked with repeated tests on a system repeatedly prepared in a similar way, using a so-called Bell test.

Entanglement can also be described with a mathematical viewpoint based on superposition. The mathematical representation of a quantum system $AB$ made of two subsystems $A$ and $B$ is the tensor product of the two subsystems, meaning, a large vector or matrix: $H_{AB} = H_A \otimes H_B$. We'll described the form of the matrices representing quantum systems a little later. In that case, the system $AB$ can be described by or decomposed with its individual parts $A$ and $B$. There are however situations where you can linearly combine several of these composite quantum states, which becomes a new quantum state. In many cases, such a composite state can't anymore be decomposed as the tensor product of two states. The composite quantum state becomes inseparable. That's where entanglement shows up! Entanglement is a direct consequence of superposition applied to multi-object systems.

Entangled quantum objects cannot be considered as separated objects. With a pair of entangled quantum objects, a measurement made on one quantum will instantly have an effect on the other quantum, without waiting for a delay in the transmission of information at the speed of light between the two quanta. This is the principle of the "non-locality" of quantum properties that disturbed Einstein in 1935 and spurred his famous EPR paper with Rosen and Podolsky.

Using qubit's representation that we'll describe later, classical entangled two-qubit states are Bell pairs, like $\frac{|00\rangle+|11\rangle}{\sqrt{2}}$ or $\frac{|01\rangle+|10\rangle}{\sqrt{2}}$. You see that they are a simple linear combination of separable states ($|00\rangle$ and $|11\rangle$ or $|01\rangle$ and $|10\rangle$. If you measure the first qubit in both cases, you have an even 50%/50% chance to get a $|0\rangle$ or a $|1\rangle$. When you measure the second qubit, you then have a 100% chance to get respectively the same value of the opposite values $|1\rangle$ or $|0\rangle$. But you can't decide in advance what is the first measurement outcome (on Alice's side). So, you observe some synchronicity between two measurements but no determinism on the first readout value. It's all about having two simultaneous synchronized random values. It is described as the "no-signaling principle": there is no statistical difference between a "first" or "second" measurement of entangled pairs, meaning Bob doing the measurement before or after Alice and Alice didn't send any actual pre-determined information to Bob when doing the measurement on her side.

But that is a mathematical representation of entanglement. You can wonder how these composite objects are created in the real world. Of course, some physical interaction must be created to entangle electrons, atoms and/or photons. Photons can be prepared to be entangled with being generated by some excitement of atoms like cesium which generates photon couples of different wavelengths but with some correlated properties like their polarization. Neutral atoms can be entangled with exciting them with lasers, raising their energy levels to a so-called Rydberg state. Electron spins are entangled with lowering a potential energy barrier between them. Quantum objects of different types can also be entangled, like photons with atoms or electrons with photons[137].

These entangled particles are not linked by chance. They usually had a common past or some past interactions. For example, two entangled photons can be produced with a birefringent mirror and separated by dichroic mirrors, creating two photons of orthogonal polarizations. The action on one of the two photons has an impact on the other photon as demonstrated by Alain Aspect in his famous 1982 experiment. But the values that are generated are completely random!

It is not defined at one end and transmitted to the other end. It is a random value that can be uncovered at two different places with some quantum measurement.

A 2019 experiment conducted at the University of Glasgow has even allowed to photograph a representation of the state of entangled photons[138]. Nevertheless, we are still able to entangle particles that do not necessarily have a common past[139]. Bell's inequalities were first validated with photons in the visible spectrum. It's been extended to other parts of the spectrum, of course in the infrared bands that are used in fiber optics and free space quantum communications and even in the X-ray band[140]. It has also been done with all sorts of qubits (superconducting, silicon spin, trapped ions, neutral atoms).

Despite its randomness, entanglement is a very powerful resource. It helps generate random secret keys for two parties with the QKD (quantum key distribution) protocols. It powers quantum computing with creating interdependencies between qubits. Multi-qubits quantum gates conditionally link them together. Once entangled, qubits have inseparable quantum states. Without it, no useful quantum algorithm could work. But quantum entanglement does not mean we can transmit some useful information faster than light since the entangled objects properties are random.

How are we checking that a quantum system is entangled? This is done with correlation statistic tests. With two quantum objects, it's a **Bell inequality test** or Bell experiment (see glossary, page 1054) that looks at the statistical correlation between the states of two quantum objects, with an experiment done a large number of times with the same settings. This test was extended with the **Mermin inequalities test** created by David Mermin in 1990 to extend Bell's inequalities test to the entanglement of a higher number of quantum object like a GHZ state with three or more qubits[141]. These tests are very costly as you increase the number of correlated quantum objects. Another variation is to conduct a state tomography for a set of qubits as described page 188. Again, its cost grows exponentially with the number of qubits, which explains why most qubit tomographies are not done beyond 6 qubits.

---

137 In 2017, researchers in Warsaw were able to entangle a photon with billions of rubidium atoms. See Quantum entanglement between a single photon and a trillion of atoms, 2017.

138 See Scientists unveil the first-ever image of quantum entanglement by Paul-Antoine Moreau, July 2019.

139 See Qubits that never interact could exhibit past-future entanglement by Lisa Zyga, July 2012.

140 See Entangled X-ray Photon Pair Generation by Free Electron Lasers by Linfeng Zhang et al, August 2022 (13 pages).

141 See Extreme quantum entanglement in a superposition of macroscopically distinct states by David Mermin. PRL, 1990 (no free access).

In science at the frontier of science fiction, some imagine exploiting quantum entanglement to analyze a quantum state inside a black hole[142]! This is beyond the scope of this book[143]!

## Indetermination

Heisenberg's principle of indeterminacy or indetermination states that one cannot accurately measure both the position and velocity of a particle or two complementary quantities describing a quantum object state. It is mathematically described as an inequality showing that the multiplication of both precisions can't be lower than the Planck constant divided by 4π. Surprisingly, this inequality was not created by Werner Heisenberg but devised by **Earle Hesse Kennard** in 1927 as he was doing a sabbatical at the University of Göttingen. It is even named the Kennard inequality.

The indeterminacy principle has another consequence: one cannot observe at the same time a quantum object in its particle state and in its wave state, per the principle of complementarity enacted by Niels Bohr around 1928, that we already mentioned in the wave-particle duality section. It also explains vacuum quantum fluctuations that we cover later in page 134.

For purists, the notions of particle speed and position are even meaningless for electrons. Its characterization is based on its wave nature and its probabilistic description via Schrödinger's wave function. Don't even try to understand where it is at a given place and time.

When it deals with velocity and position or waves, Heisenberg's indeterminacy principle is closely related to a characteristic of Fourier transforms: a nonzero function and its Fourier transform cannot both be sharply concentrated, so, precisely measurable. The more concentrated a signal is in the time domain, the more spread out it is in the frequency domain and vice-versa. We have here a mathematical balance between a pulse length precision and its spectral analysis precision.

Since complementary (or incompatible) properties can't both be measured with an arbitrary precision, we can improve one property measurement precision with decreasing the measurement precision of the complementary property. It's being implemented with the so-called photons squeezing technique. This technique is implemented in the latest LIGO (USA) and VIRGO (Italy) huge interferometers that are used to measure gravitational waves generated by huge astrophysical phenomena like dual black hole collapses. These instruments increase the precision of photons time arrival in the interferometer at the price of a greater imprecision in the number of photons[144].

## Measurement

Measuring quantum object properties follows a very different path than with classical physics due to the back action induced by quantum measurement on the measured system and to its probabilistic dimension.

With classical mechanics, you can usually predict over time the results of the measurement of macro-objects properties (dimension, speed, position) based on their dynamics. In quantum mechanics, given the knowledge of the position of the measured object, one cannot measure precisely its momentum. More generally, the knowledge we have about two non-commuting observables is bounded such that we can never assign them a well define value simultaneously, due to the Heisenberg uncertainty principle.

---

142 See Can entangled qubits be used to probe black holes? by Robert Sanders, 2019.

143 Superposition also happens within benzene $C_6H_6$ with two carbon-carbon links with their neighbors, using one or two electrons.

144 See Squeezing More from Gravitational-Wave Detectors, December 2019. Kip Thorne (1940, USA), Rainer Weiss and Barry C. Barish got the Nobel Prize in Physics in 2017 for their contributions to the creation of the LIGO detector and the observation of gravitational waves.

Moreover, a quantum measurement readout requires some interaction with a macroscopic object that selects automatically one specific outcome. In strawman language, quantum measurement is in the eye of the beholder! Measuring the same initial state several times can lead to different outcomes. However, even if each measurement yields a probabilistic result, when repeated several times, their statistical distribution is not probabilistic. It corresponds to the knowledge that can be obtained from the evaluated quantum state created experimentally in a similar way several times.

Before measurement, a single isolated quantum object is said to be in a pure state, represented by a vector in a Hilbert space, or its “Psi” ($\psi$) vector. It is a superposition, or linear combination of basis states or one of the object basis state, like “ground state” or “excited”. When a quantum object is measured against one observable, the state of the quantum object become one of the observable basis states, like a spin direction up or down or a discrete energy level. The quantum object collapses in a probabilistic way into one of the available basis states. If we conduct another measurement, we’ll always get the same result being the basis state that was obtained beforehand in the first measurement. This is also called "*Schrödinger wave function* collapse" or “*wave packet collapse*” which however works only with so-called projective measurement, as defined by John Von Neumann.

With a photon of intermediate polarization between horizontal or vertical linear polarization, it will become a horizontally or vertically polarized photon after its polarization measurement.

In quantum computing, this principle of reduction is implemented when measuring the state of a qubit. It modifies its value by collapsing it to the basis states $|0\rangle$ or $|1\rangle$.

The outcome is probabilistic with a chance of retrieving a $|0\rangle$ or a $|1\rangle$ depending on the qubit state. However, when the quantum state is a basis state, say $|0\rangle$ or $|1\rangle$ for a qubit, its measurement should return this basis state in 100% of the cases and is therefore not probabilistic but deterministic. This works however only in a perfect world without any quantum noise. Even when a qubit is in a basis state, its measurement doesn’t return a perfect basis state 100% all the time. You get a % that is inferior to 100% and corresponds to the readout qubit fidelity. It turns a basis state measurement into a probabilistic one.

The subtle information contained in a qubit that is represented by a complex number or a two-dimensional vector is reduced to $|0\rangle$ or $|1\rangle$ at the time of its measurement. It becomes a classical bit. A single measurement is then making us lose all the wealth of information contained in the qubit. We turn the equivalent of two floating point numbers to a single bit! However, this measurement is supposed to happen only at the end of quantum algorithms. During computing, the whole wealth of qubit internal information is leveraged, particularly with the creation of interferences between qubits.

All this is illustrated in the diagram below in Figure 87. We will come back to the meaning of α and β complex numbers in the next section on qubits.

This reduction should occur theoretically only at the end of computing. During computing, qubits are modified by quantum gates preserving the richness of their information, the combinatorial nature of their values based on superposition and entanglement. However, quantum measurement is to be implemented during computing with systems implementing quantum error corrections.

The subject of quantum measurement is quite broad. In a forthcoming more detailed section page 184, we will cover several additional concepts such as projective (Von Neumann) measurement, non-selective measurement, weak measurement, gentle measurement and non-destructive measurement.

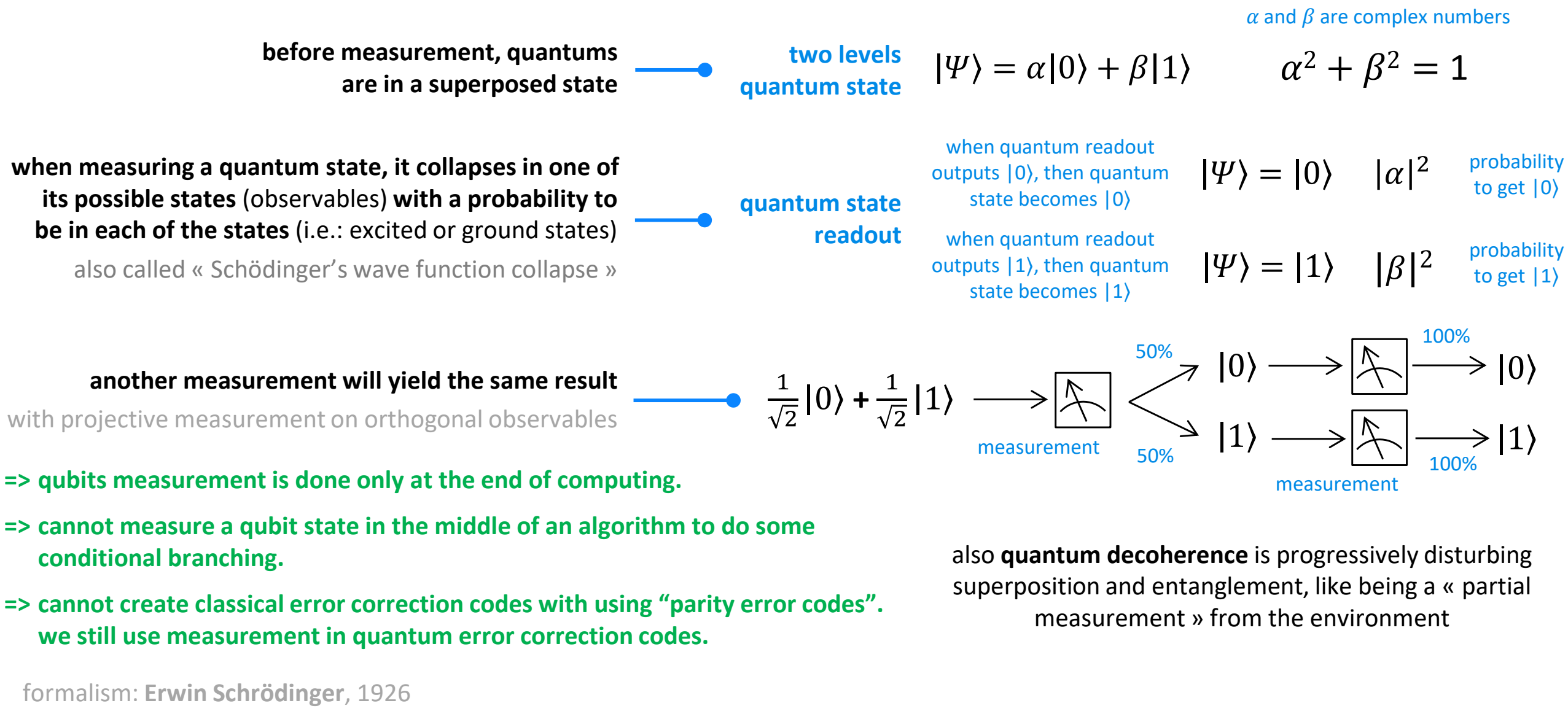

*Figure 87: quantum measurement explained with qubits, (cc) Olivier Ezratty, 2021.*

## No-cloning

The no-cloning theorem prohibits the identical copy of the state of a quantum object onto another quantum object. The theorem is mathematically demonstrated in six lines in Figure 88. It is also described page 148.

The theorem was demonstrated in 1982 by William Wootters, Wojciech Zurek and Dennis Dieks[145]. The article is still not available in open source on a site such as arXiv, self-applying the no-cloning principle[146]!

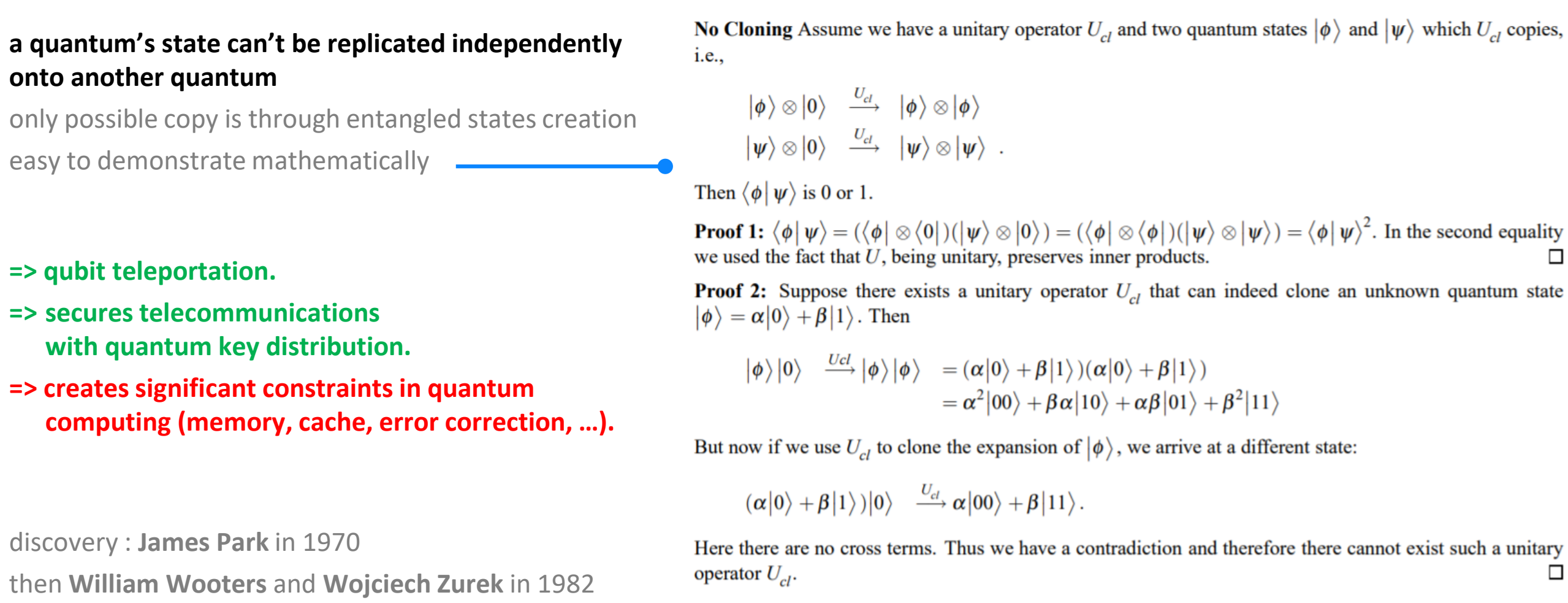

*Figure 88: no-cloning theorem demonstration, source: Wikipedia.*

As a consequence, it is impossible to copy the state of a qubit to exploit it independently of its original, contrarily to a classical bit that can be copied from/to memory and from/to storage. It also prevents quantum computers to implement the Von Neumann computing model with separate processing and memory.

[145] See A single quantum cannot be cloned by William Wootters and Wojciech Zurek, Nature, 1982.

[146] A summarized version if available in The no-cloning theorem by William K. Wooters and Wojciech H. Zurek, Physics Today 2009 (2 pages).

In quantum computers, qubits can be duplicated via quantum gates and entanglement, but the resulting qubits are entangled and therefore somehow synchronized, inseparable and... random. Reading the copy destroys the original by projecting the state of the two qubits to the 0 or 1 closest to their initial state and in a probabilistic way.

This has a direct impact on the design of quantum algorithms and in particular on the error correction codes of quantum computers. These error-correction codes use the trick of projective measurement on a different computational basis as we'll see later.

A derivative of no-cloning is non-deleting. In the case of a qubit, it means it's impossible to reset a qubit from an entangled set of two qubits $|\psi\rangle$, meaning to transform $|\psi\rangle|\psi\rangle$ into $|\psi\rangle|0\rangle$.

## Tunnel effect

The wave-particle nature of matter allows it to cross physical barriers also known as energy walls in some circumstances, depending on the wall thickness and quantum object wavelength. The transmitted wave is usually attenuated after crossing the barrier and its strength depends on the wavelength with regards to the barrier length and composition.

This phenomenon was first accidentally unveiled by **Henri Becquerel** in 1896 when he discovered radioactivity. It did show up with uranium salts decaying, producing alpha rays comprised of two neutrons and two protons. This phenomenon was explained later thanks to quantum physics and wave-particle duality by **George Gamow** (1904-1968, Ukrainian-Russian-American) in 1928. Just before in 1927, the German physicist **Friedrich Hund** (1896-1997, German) created the formalism explaining electron based tunneling effect.

The tunnel effect is used in superconducting Josephson junctions and exploited in D-Wave quantum annealers where it is used to converge a system of spin qubits ("Hamiltonian", with a given total energy level) towards an energy minimum corresponding to the resolution of a complex combinatorial problem or a search for energy minimum as in chemistry or molecular biology.

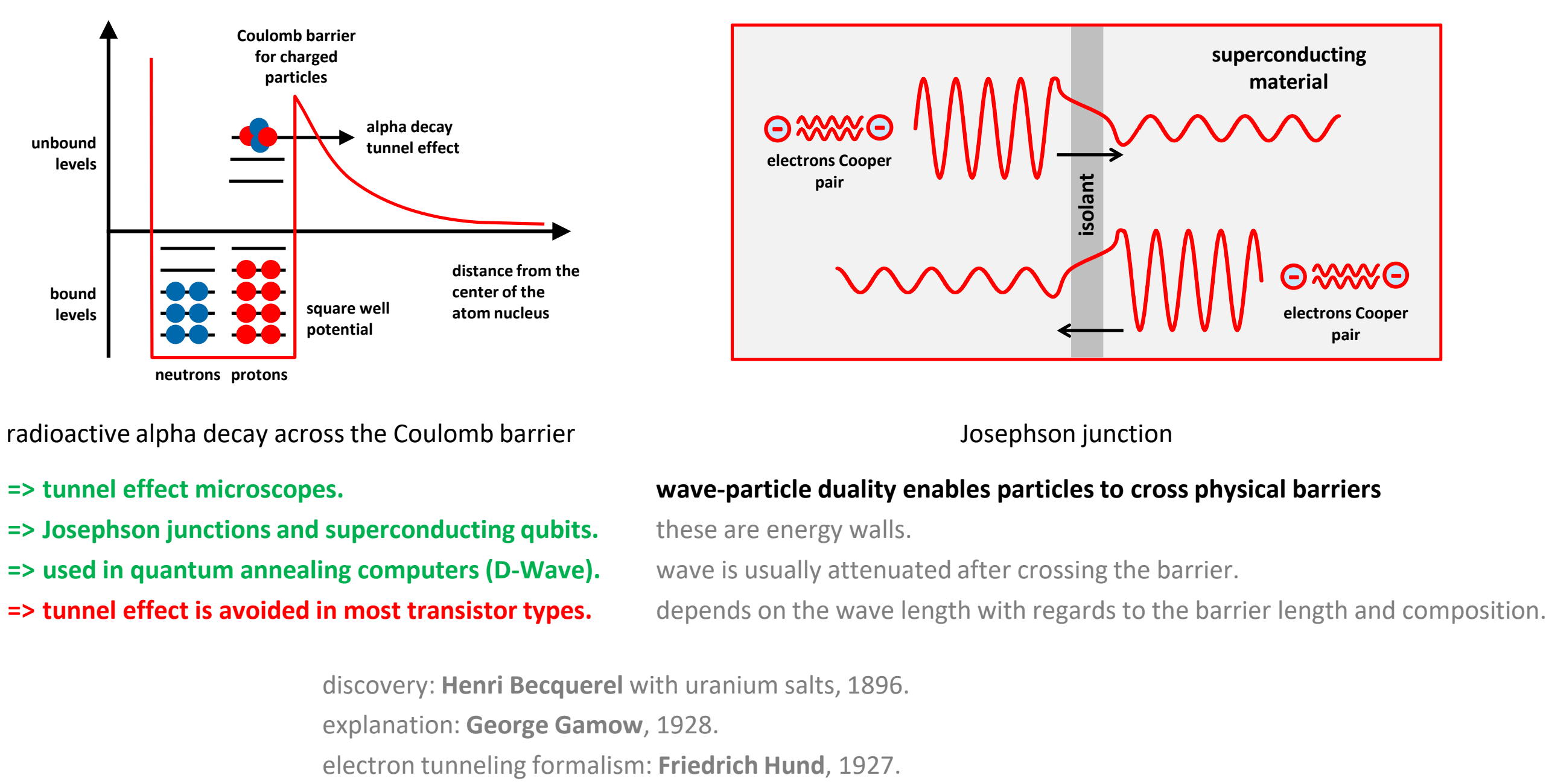

*Figure 89: overview of the tunnel effect and its use cases, (cc) Olivier Ezratty, 2021.*

But contrarily to what I wrote in the previous editions of this book, the tunnel effect is not exploited in transistors. Most transistors make use of the field effect which was patented by Julius Edgar Lilienfeld in 1926. It is implemented in MOSFETs (metal-oxide-semiconductor field effect transistor) and in CMOS (complementary-metal-oxide-semiconductor, that use variants of MOSFETs). These transistors use a metal gate deposited on a silicon-dioxide ($SiO_2$) and now a "high-K dielectric" as the gate dimension is decreasing with higher densities, to reduce the tunnel effect. The gate voltage determines the transistor conductivity.

# Quantum matter

Quantum matter refers to materials or assemblies of few atoms which, for specific conditions, physical observables such as magnetism, electronic state or optical properties are only described by advanced quantum physics. They are at the crossroads of statistical physics. Iconic quantum materials are superconducting materials in which, below a certain temperature, electrons behave collectively as a sort of fluid dubbed "quantum fluid". Other known quantum fluids in physics are superfluid helium, Bose-Einstein condensates, polariton condensates and ultra-cold neutral atoms. They all exhibit quantum mechanical effects at a macroscopic collective level. These phenomena are usually reported at very low temperatures, close to -273°C, and sometimes high-pressures but some of them start to emerge in less drastic conditions.

### Definitions

Given all standard matter such as metal, semiconductor or insulator rely on quantum description, starting with electrons quantum numbers and the atomic structure, how are quantum materials and quantum matter being accurately defined? Where is the frontier? Well, there's no real consensus on this, a bit like how postulates are formalized in quantum physics.

One of the reasons is that quantum materials range from yet untested theoretical concepts, to lab-based experiments, up to industry applications like with graphene. It's an entire new research field that is still in the making with a lot of fundamental research.

It is also a field that is really hard to dig into, even more than many other fields that are covered in that book, like quantum error correction. So, forgive some of the vagueness of this part. I have not really understood *all* the sentences I wrote here!

The simplest definition I found is "*materials where electrons do interesting things*". Then, I opened quantum matter's Pandora's box and found many other definitions.

The US Department of Energy created its own definition in 2016 with "*solids with exotic physical properties that arise from the interactions of their electrons, beginning at atomic and subatomic scales where the extraordinary effects of quantum mechanics cause unique and unexpected behaviors*"[147].

A more precise definition was proposed by Philip Ball in 2017 which is based on materials where electrons are operating collectively as quasi-particles and are frequently confined in some 2D geometries like graphene sheets, with derivatives in 3D assemblies of graphene sheets with small angle rotations called **magic angle**, creating the new field of **twistronics**[148].

---

[147] Seen in Basic Research Needs for Quantum Materials, DoE, 2016 (4 pages), with a slightly simpler one "*solids with exotic physical properties, arising from the quantum mechanical properties of their constituent electrons; such materials have great scientific and/or technological potential*" seen in Quantum Materials for Energy Relevant Technology by the DoE Office of Science, 2016 (170 pages).

[148] In Quantum materials: Where many paths meet by Philip Ball, MRS Bulletin, 2017 (8 pages), Magic angle, a new twist on by Pablo Jarillo-Herrero and Senthil Todadri, MIT, January 2021 (12 pages) and Magic-Angle Multilayer Graphene: A Robust Family of Moiré Superconductors by Jeong Min Park, Pablo Jarillo-Herrero, December 2021 (15 pages). This could lead to interesting superconducting effects.

Quantum materials are also grouped as **strongly correlated materials** where magnetism is important and their behavior is "*dictated by quantum mechanical correlations between electrons*", and **topological materials** where some symmetry of the material lattice provides protected electronic states on the surface or in the bulk of the crystal.

And I didn't try to find any semantic nuance between quantum matter and quantum materials!

In another source[149], quantum matter deals with "*novel phases of matter at zero temperature with exotic properties*". It adds:

"*The main ways of characterizing and manipulating quantum matter are with entanglement, symmetry, and topology:*

***Entanglement** is the quantum property of correlated physical attributes among particles (position, momentum, spin, polarization).*

***Symmetry** refers to features of particles and spacetime that are unchanged under some transformation, seen as the property of a system looking the same from different points of view (a face, a cube, or the laws of physics) and its partner, symmetry breaking (in phase transitions)*[150].

***Topology** is the property of geometric form being preserved under deformation (bending, stretching, twisting, and crumpling, but not cutting or gluing). Physical systems may have global symmetric and topological properties that remain invariant across system scales*". Usually, obtaining some topological order requires cooling at very low temperatures like with superconducting materials.

As stated before, quantum matter is characterized with being based on collective excitations. These excitations are composite entities that are analogous in their behavior to a single particle[151] named quasiparticles

It can be quasiparticles that are assemblies of several fermions, mostly electrons and holes, like two electrons in Cooper pairs explaining superconductivity, polaritons, excitons and vortex magnetic phenomena like skyrmions, etc... It can also be collective excitations of bosons like phonons in crystal lattices. There are over 30 identified quasiparticles classes including some that are very exotic and less talked about like the Bogoliubon (a quasiparticle found in superconductors) and the wrinklon[152].

Philip Ball proposes a classification of these quasiparticles in seven categories[153]:

- **Cooper pairs** of electrons in classical superconductivity (high-temperature superconductivity with cuprates requires a more complicated explanation). We cover their various use cases with superconducting qubits and sensors.

[149] See Quantum Matter Overview by Melanie Swan, Renato P. dos Santos and Frank Witte, April 2022 (23 pages). See also The 2021 Quantum Materials Roadmap by Feliciano Giustino et al, February 2021 (93 pages) and Introduction to Quantum Materials by Leon Balents, KITP, 2018 (51 slides).

[150] Classical matter phases transitions are traditionally described with Lev Landau's symmetry breaking model elaborated in 1937. It describes in a simplified way what happens at phase transitions (like gas↔liquid and liquid↔solid) with the evolution of a symmetry-breaking order parameter (OP) named η (eta). It also describes various types of ordering phenomena like ferromagnetic, ferroelectric, ferroelastic or other types of electronic orders like Mott or insulator-metal transitions systems. In most cases, quantum matter is described by a "topological order" that can't be explained by Landau's model. Some examples include topological insulators, topological semimetals, fractional quantum Hall states, quantum spin liquids and Fermi liquids. A Mott transition is a particular type of topological phase transition. Mott insulators are materials that are expected to conduct electricity but are insulators, particularly at low temperatures, and under certain conditions which can be controlled, leading to so-called Mott transitions.

[151] Source: Webster. Quasiparticle were first defined by Lev Landau in the 1930s.

[152] Source : https://en.wikipedia.org/wiki/List_of_quasiparticles.

[153] See Quantum materials: Where many paths meet by Philip Ball, 2017 (8 pages)

- **Relativistic Dirac fermions** such as many-electron excitations in Dirac semimetals and in graphene[154]. Graphene has many applications in sensing and electronics. There was even a European Union Graphene Flagship program launched in 2013 with 1B€.
- **Weyl fermions** are massless fermions related to Dirac fermions whose existence was predicted by Herman Weyl in 1929 and discovered in 2015 at Princeton[155]. These fermions are massless, have a high degree of mobility and are quasiparticle excitations in Weyl semimetals. Topological semimetals could be used in low-consumption spintronic and magnetic memory devices and ultrafast photodetectors[156].
- **Laughlin quasiparticles** proposed by Robert Laughlin in 1983 and who received the Nobel in Physics in 1998 for his theoretical explanation of the fractional quantum Hall effect, together with Horst Störmer and Daniel Tsui, who discovered the effect experimentally. They relate to the "fractional quantum Hall effect" (FQHE, discovered in 1982) in a 2D "electron gas" placed in a magnetic field. It involves electron quasiparticles behaving like if they had a fractional charge, such as 1/3, 2/5 or 3/7, 1 being the charge of a single electron. One use case is to create an electron interferometer[157].
- **Majorana fermions** are hypothetical particles proposed in 1937 by Ettore Majorana, which are their own antiparticles. Their existence is still questioned. They could lead to creating topological qubits quantum computers with a better resistance to quantum noise and errors.
- **Anyons** are hypothetical particles proposed by Frank Wilczek in 1982. Anyons have quantum statistics positioned in a continuum between fermions (1/2 spin) and bosons (integer spin). They could show up in quantum spin liquids[158]. These quantum spin liquids which can show up in magnetic materials where electron spins are not orderly aligned but are entangled. The first spin liquids were experimentally detected in 2020[159]. It could help to create innovative electronic memories. This state of matter was envisioned in 1973 by Philip W. Anderson[160].
- **Skyrmions** take the form of vortex-like topological quasiparticle excitations of spins in some magnetic materials. They were envisioned in 1962.

We could still add here various classes and subclasses of quantum materials:

- **Spin glasses** where electron spins freeze in a disordered fashion at some non-zero temperature. It leads to the notion of **quantum glasses**[161].
- **Plasmons** which are collective oscillations of electrons on the surface of a conductor that can interact with photons. It could also help creating energy savings and faster data storage solutions.
- **Topological insulators** are materials whose bulk part is insulating and whose surface (2D or 3D) presents counterpropagating spin channels with no charge current. It could help create a new breed of energy-saving and fast-switching transistors.

---

[154] See the thesis Relativistic Phases in Condensed Matter by Thibaud Louvet, 2018 (165 pages).

[155] See After 85-year search, massless particle with promise for next-generation electronics discovered by Morgan Kelly, 2015.

[156] See Topological Semimetals by Andreas P. Schnyder, 2020 (32 pages).

[157] See Realization of a Laughlin quasiparticle interferometer: Observation of fractional statistics by F. E. Camino, Wei Zhou and V. J. Goldman, 2005 (25 pages).

[158] See A Field Guide to Spin Liquids by Johannes Knolle and Roderich Moessner, 2018 (17 pages).

[159] See Scale-invariant magnetic anisotropy in $RuCl_3$ at high magnetic fields by K. A. Modic et al, October 2020 (32 pages).

[160] See Quantum Spin Liquids by C. Broholm et al, May 2019 (21 pages).

[161] See the review paper Quantum Glasses by Leticia F. Cugliandolo and Markus Müller, Sorbonne Université CNRS LPTHE and Paul Scherrer Institute, August-September 2022 (23 pages).

- **Quantum wires** are conducting wires with quantum confinement effects modifying the transport properties, mostly when the wires have a diameter of a few nanometers, event down to a single atom[162]. They are usually called nanowires. Carbon nanotubes are a class of quantum wires.
- **Spin-torque materials** that are already used in some low-power non-volatile magnetic memories (STT-RAM or STT-MRAM).

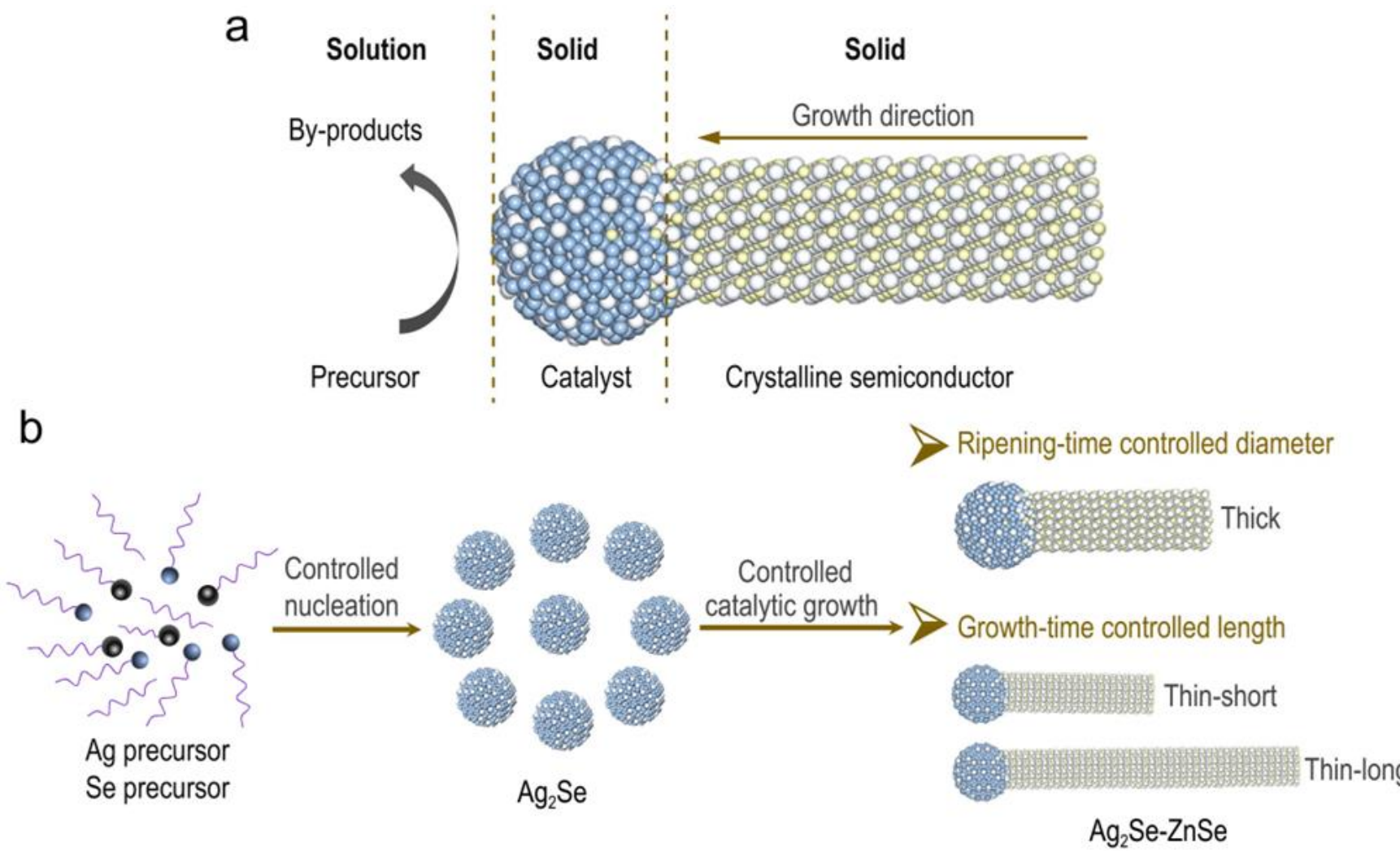

*Figure 90: quantum wires. Source: [On demand defining high-quality, blue-light-active ZnSe colloidal quantum wires](#) from Yi Li et al, National Review Science, April 2022 (29 pages).*

- **Time crystals** which we'll cover later, and it is the source of a lot of headaches.
- **Wigner crystals** are another very weird phenomenon. Predicted by Eugene Wigner in 1934 (the same Wigner of the Wigner function used in quantum photonics), it consists in crystals made of electrons, of course also at very low temperatures

  They were experimentally observed in 2018 by an Israeli-US-Hungarian team in one dimension at 10 mK using carbon nanotubes for their measurement[163] and 2020 in 2D at 80 mK by a team from ETH Zurich (as shown here on the right in Figure 91) [164].

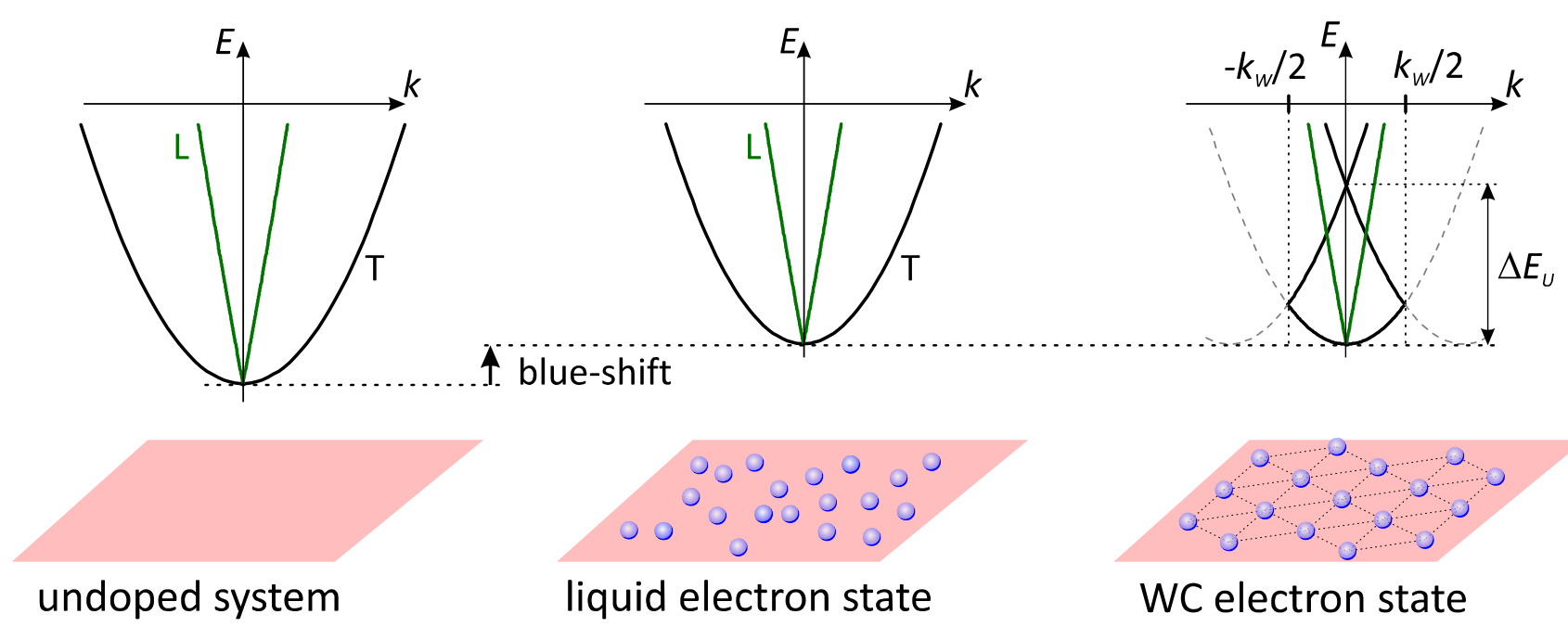

*Figure 91: Wigner crystals. Source: [Observation of Wigner crystal of electrons in a monolayer semiconductor](#) by Tomasz Smoleńsk et al, 2020 (26 pages).*

- **Quantum batteries** are still theoretical devices that would be more efficient than traditional batteries with a shorter recharging cycle.

[162] See one recent example of quantum nanowire in On demand defining high-quality, blue-light-active ZnSe colloidal quantum wires from Yi Li et al, National Review Science, April 2022 (29 pages).

[163] See Imaging the electronic Wigner crystal in one dimension by I. Shapir et al, Science, 2019 (38 pages).

[164] See Observation of Wigner crystal of electrons in a monolayer semiconductor by Tomasz Smoleńsk et al, 2020 (26 pages).

- **Quantum dots** that are used in LCD screens and are not considered as being quantum materials since their behavior is explained by single electrons and classical quantum light/matter exchanges. They are made of powder with tiny compound grains of different sizes between 2 and 6 nm which are used to down-convert the blue light coming from LEDs into red and green light, creating a better balanced coverage of primary colors, as shown in Figure 92 [165]. The main problem is to replace cadmium that is a pollutant. These LCD screens quantum dots must not be confused with the quantum dots used in silicon qubits to trap single electrons and control their spin as well as the quantum dots used in unique photon sources like the ones from Quandela.

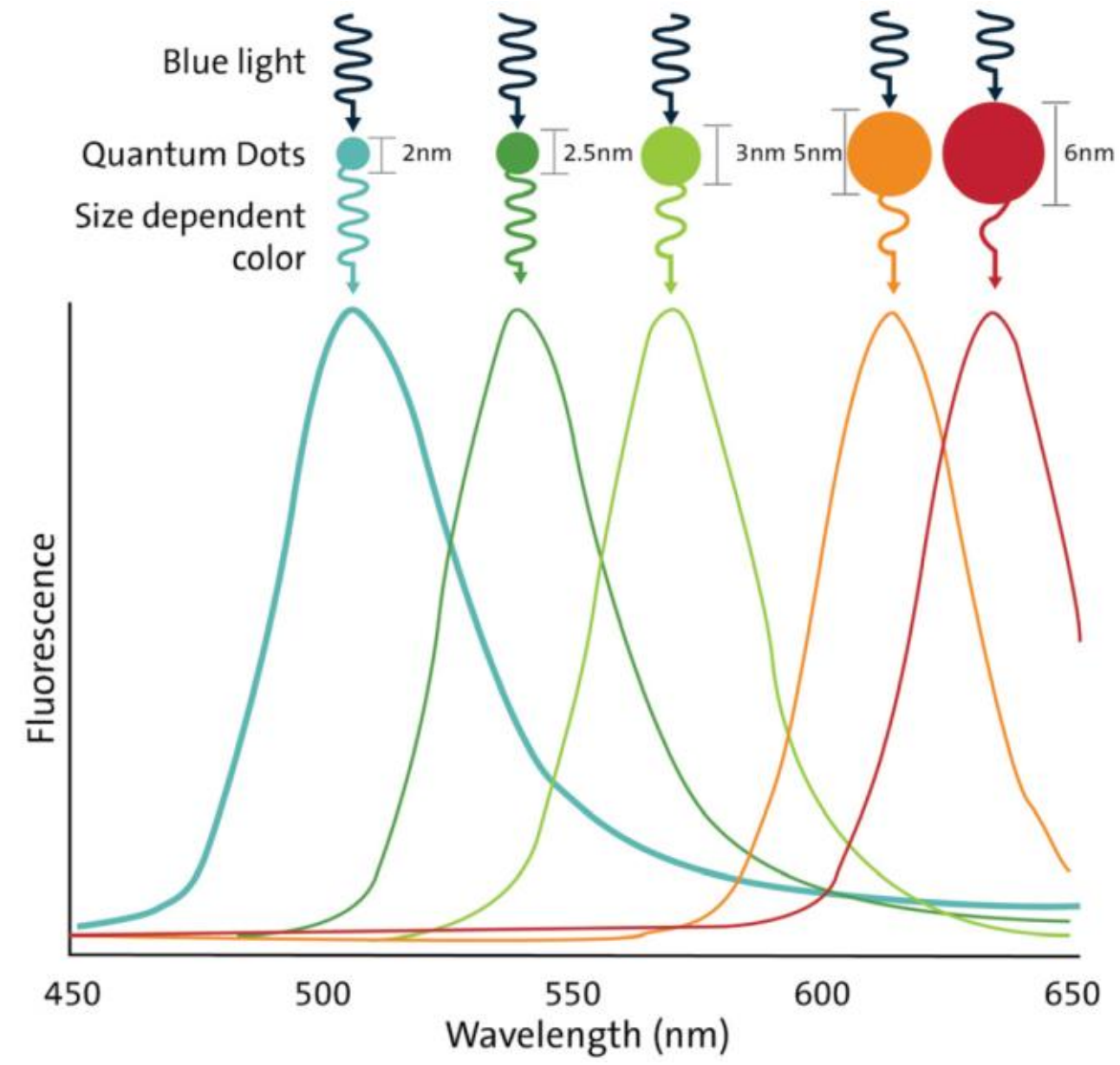

*Figure 92: quantum dots used in LCD screen and lighting. Source: Nanomatériaux et nanotechnologies : quel nanomonde pour le futur? by Pierre Rabu, 2018.*

Some other concepts related to quantum matter mandate some explanations:

Many of the quantum matter species happen in crystals. And there are a lot of types of crystals classified by their crystallographic order! There are 230 crystallographic space groups organized in 7 crystal systems named triclinic, monoclinic, orthorhombic, tetragonal, trigonal, hexagonal and cubic and subclasses with primitive centering, centered on a single face, body centered and face centered[166].

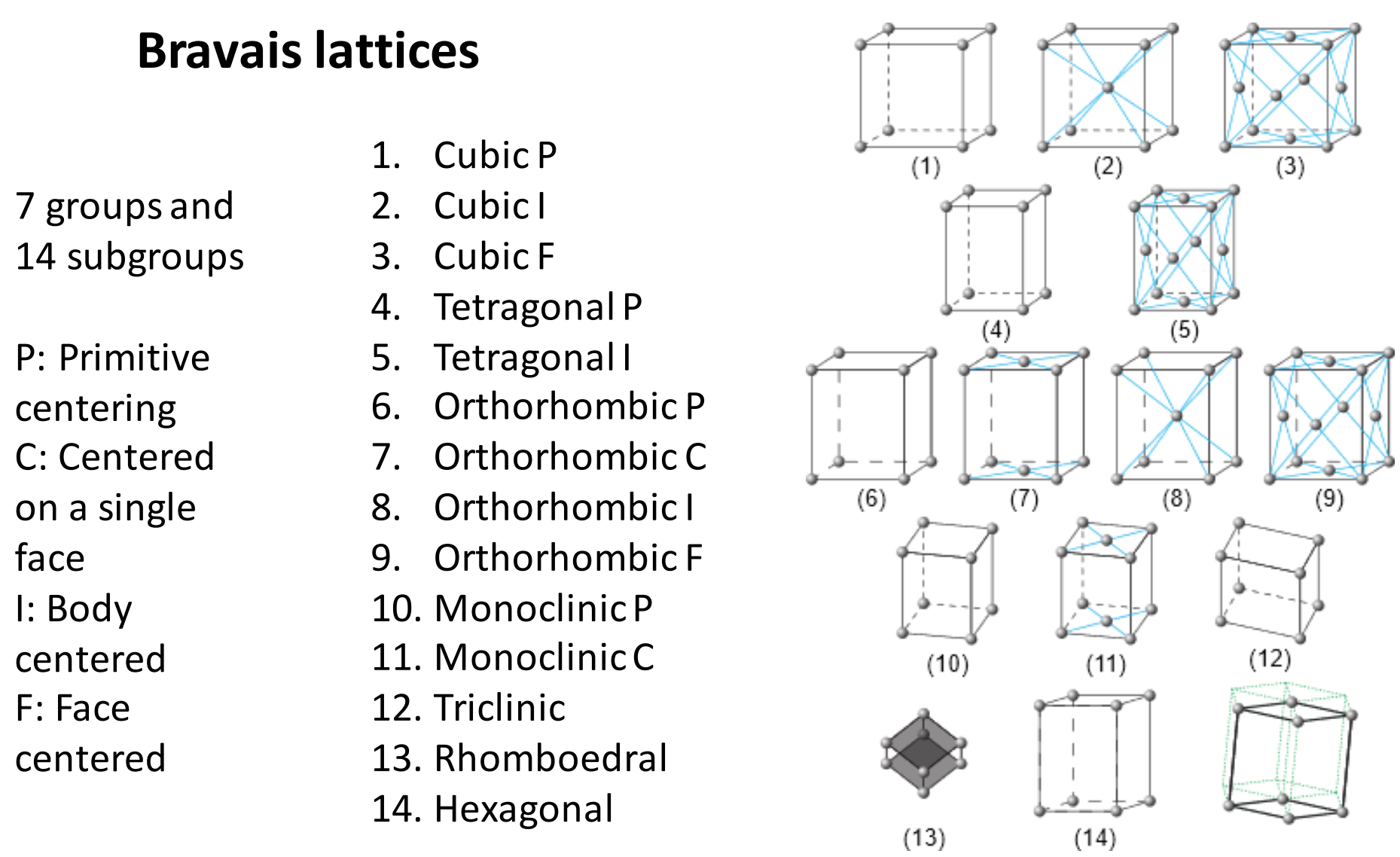

*Figure 93: Bravais lattices and crystal structure classification. Source: Wikipedia.*

---

[165] It was first discovered at the end of the 1970s by Alexei Ekimov in Russia and explained in 1982 by Alexander Efros, also from Russia. From The Quantum Dots Discovery. See Advances in Quantum-Dot-Based Displays by Yu-Ming Huang et al, 2020 (29 pages), schema from Quantum dots and their potential impact on lighting and display applications by Paul W. Brazis, 2019 (18 pages).

[166] See Crystal Systems and Space Groups by Paul D. Boyle, University of Western Ontario (44 slides) and Cristallographie et techniques expérimentales associées (in English) by Béatrice Grenier, 2014 (67 slides).

One key notion in crystallography is chirality which describes how crystal structures break spatial symmetry and are not identical to their mirrored structure[167]. There are also 1651 magnetic space groups which describe magnetism configurations at the atom level in crystal lattices[168].

Another key notion in quantum matter is time reversal symmetry. A time reversal symmetry means that the material looks the same when looking at a time scale backwards and forward.

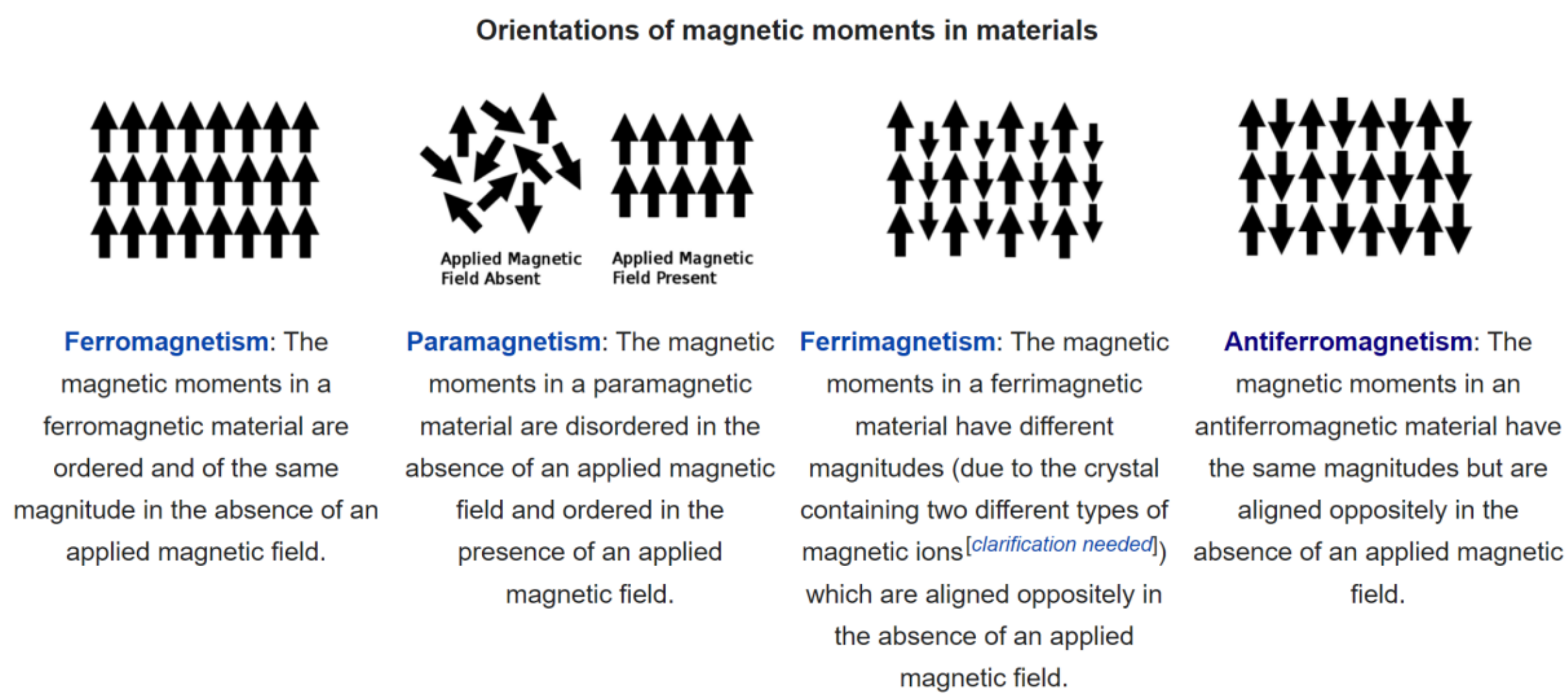

*Figure 94: ferromagnetism, paramagnetism, ferrimagnetism and antiferromagnetism explained. Source: Wikipedia.*

Reversing time means looking backwards in time only from a mathematical standpoint, not physically reversing time. There's no way you can change the arrow of time backwards. Time reversal is not a time machine!

Figure 95 presents an inventory of some physical properties that change or do not change with time reversal.

**time reversal symmetry**

| do not change with time reversal | changes with time reversal |
|---|---|
| position of particle in space | time of events |
| particle acceleration in space | particle velocity |
| force on the particle | particle linear momentum |
| particle energy | electric vector potential |
| electric potential and field | magnetic field |
| density of electric charge | electric current density |
| energy density of the EM field | power / rate of work done |

*Figure 95: time reversal symmetry explained.*

## Superconductivity

Superconductivity occurs when under a low-level temperature, some conducting materials no longer oppose resistance to electric current. With usual electric current, electrons move from atom to atom and transform part of their kinetic energy into heat related to the movement of the atoms hit by electrons, also known as the Joule effect.

With superconductivity, electrons arrange themselves in pairs, called Cooper's pairs, circulating between atoms without friction. The structure of the atoms of the conductive metal is also modified. Waves of atoms occur that follow and accompany the movement of Cooper's pairs. These are specific breeds of phonons.

---

[167] See A Chirality-Based Quantum Leap by Clarice D. Aiello and many al, November 2021 (93 pages) described in Chirality and the next revolution in quantum devices by César Tomé López, Mapping Ignorance, May 2022. See also Topology and Chirality by Claudia Felser and Johannes Gooth, May 2022 (27 pages) which makes a good classification including chiral and topological matter.

[168] See Magnetic Group Table, Part 2 Tables of Magnetic Groups, by Daniel B. Livin, 2014 (11976 pages). I hope the author found some way to automatize the production of all these pages! See also Exhaustive constructions of effective models in 1651 magnetic space groups by Feng Tang et al, March 2021 (25 pages) and Structure and Topology of Band Structures in the 1651 Magnetic Space Groups by Haruki Watanabe et al, August 2018 (43 pages).

Cooper's pairs are electrons of opposite spins forming composite bosons (ensemble with zero spin), allowing them to have the same quantum state[169].

Superconductivity was discovered experimentally in 1911 by **Heike Kamerlingh Onnes** (1853-1926), Cornelis Dorsman, Gerrit Jan Flim and Gilles Holst at the University of Leiden in the Netherlands, with solid mercury at 4.2K. Kamerlingh Onnes also discovered that a magnetic field whose intensity depends on temperature could make the superconducting effect disappear[170]. Its interpretation was formulated much later, in 1957.

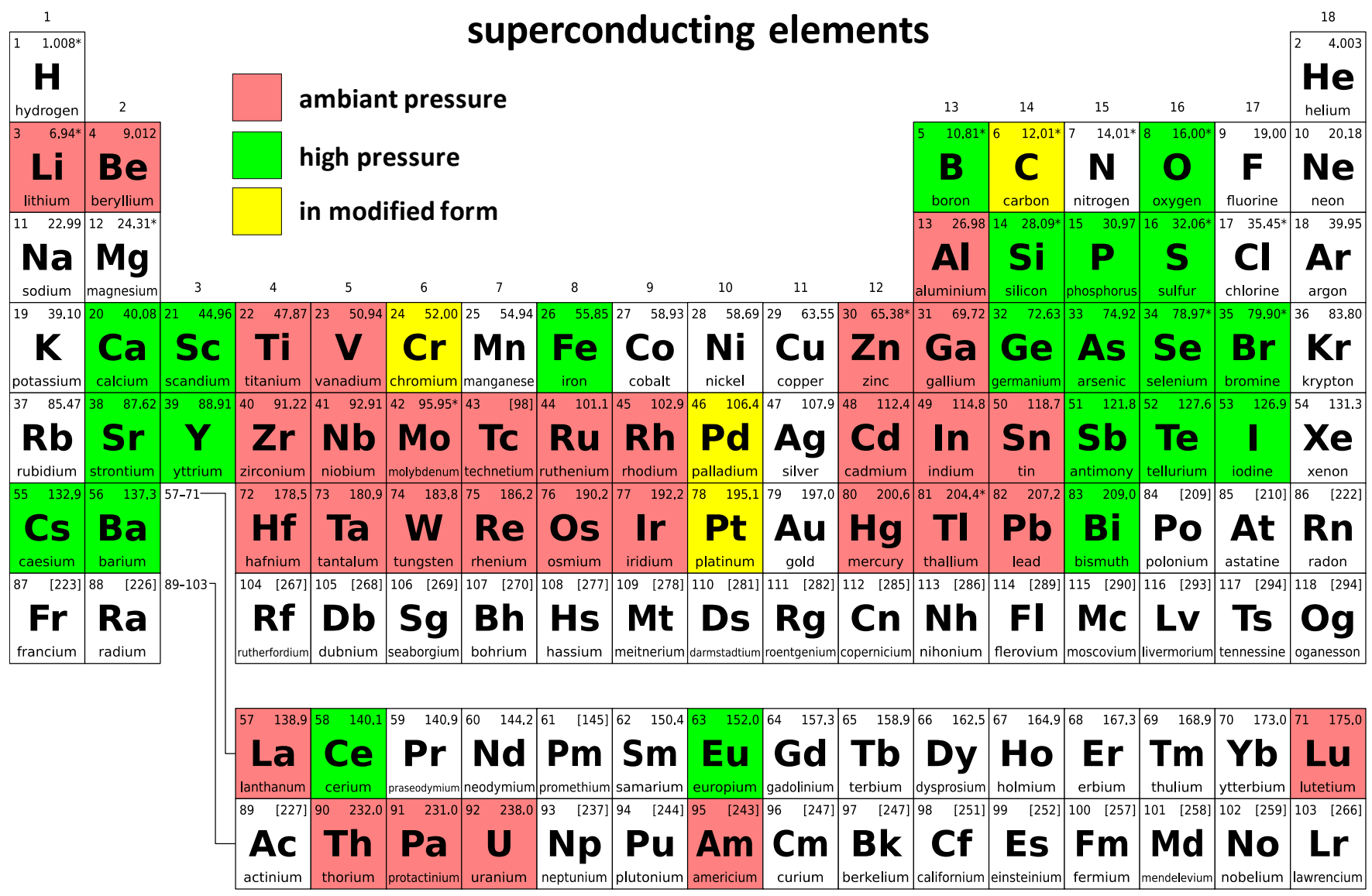

*Figure 96: table of chemical elements with those which are superconducting. Source: Wikipedia and various other sources.*

It was achieved by **John Bardeen**[171], **Leon Neil Cooper** and **John Robert Schrieffer** of the University of Illinois. They built the so-called **BCS theory**[172]. Later experiments and extrapolations on the persistence of circulating currents injected into macroscopic superconducting rings found that the lower bound of these permanent currents was around $10^5$ years.

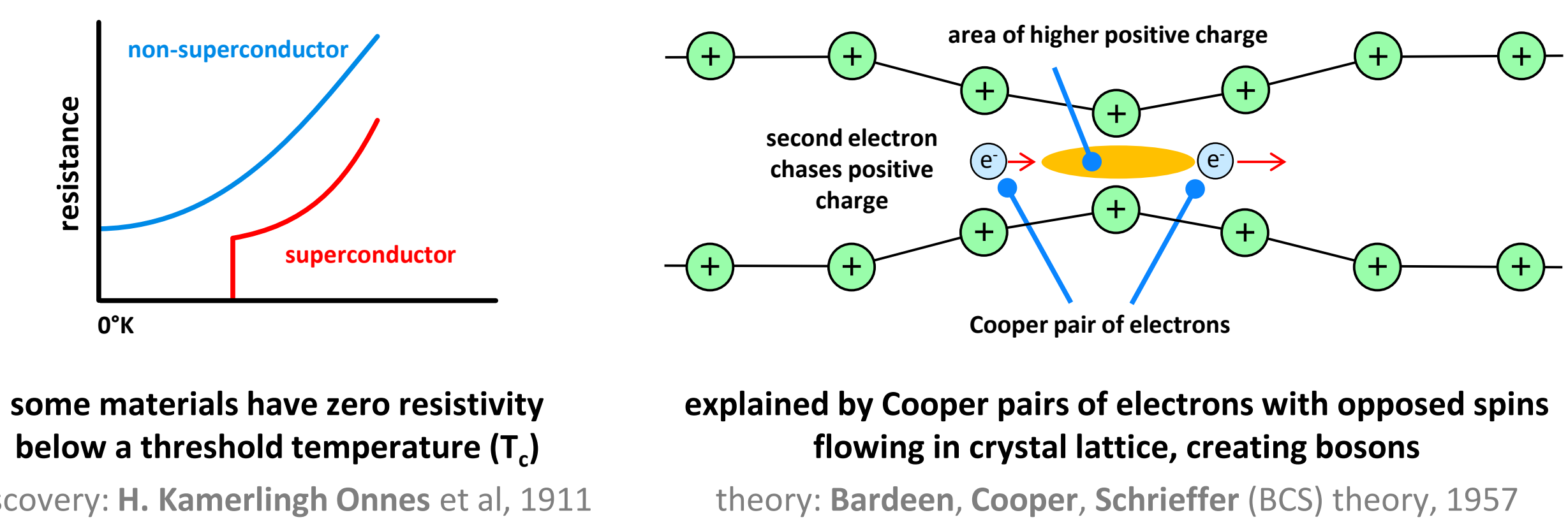

*Figure 97: superconductivity explained.*

---

169 Cooper's pairs can also be formed with atoms as with helium 3, a fermion, in its superfluid state named a fermionic condensate.

170 See this detailed presentation: Superconductivity and Electronic Structure by Alexander Kordyuk, 2018 (145 slides).

171 John Bardeen holds two Nobel prizes in physics, one in 1956 for the invention of the transistor with William Shockley and Walter Brattain and the other for the interpretation of superconductivity in 1972 with Leon Neil Cooper and John Robert Schrieffer. Cooper co-created the BCS theory at the age of 27 and won the corresponding Nobel Prize at the age of 42. Born in 1930, he is still with us today.

172 An accurate timeline of the discovery of the principle of superconductivity is provided in the presentation 50 Years of BCS Theory "A Family Tree" Ancestors BCS Descendants, by Douglas James Scalapino, John Rowell and Gordon Baym, 2007 (52 slides). See also the excellent book The rise of superconductors by P.J. Ford and G.A. Saunders 2005 (224 pages) which tells the story of the discovery and then interpretation of superconductivity. Before the BCS theory, many physicists had broken their teeth on the explanation of superconductivity: Albert Einstein, Niels Bohr, Lev Landau, Max Born, Felix Bloch, Léon Brillouin, John Bardeen (co-inventor of the transistor), Werner Heisenberg and Richard Feynman.

About 50 chemical elements are superconducting at low temperature but the superconductivity temperature and pressure thresholds are very variable. In general, metals that are superconductors are poor conductors in their normal state and most good conductors like copper, gold and silver are not superconductors.

Superconductivity is possible with composite alloys such as germanium, titanium and niobium alloys or copper-based materials (as cuprates). This is particularly the case with aluminum and mercury. The most common superconducting materials are aluminum and a niobium and titanium alloy, used in superconducting wires in MRI imaging systems and superconducting qubit cryostats[173].

The superconducting effect is maximum for atoms that have a large number of valence electrons, i.e., in the last orbital layer, with the highest quantum number. Superconductivity explains unexpected phenomena such as the levitation of magnets above superconductors immersed in liquid nitrogen. Superconducting ceramics, discovered since 1986, can be used in this striking experiment.

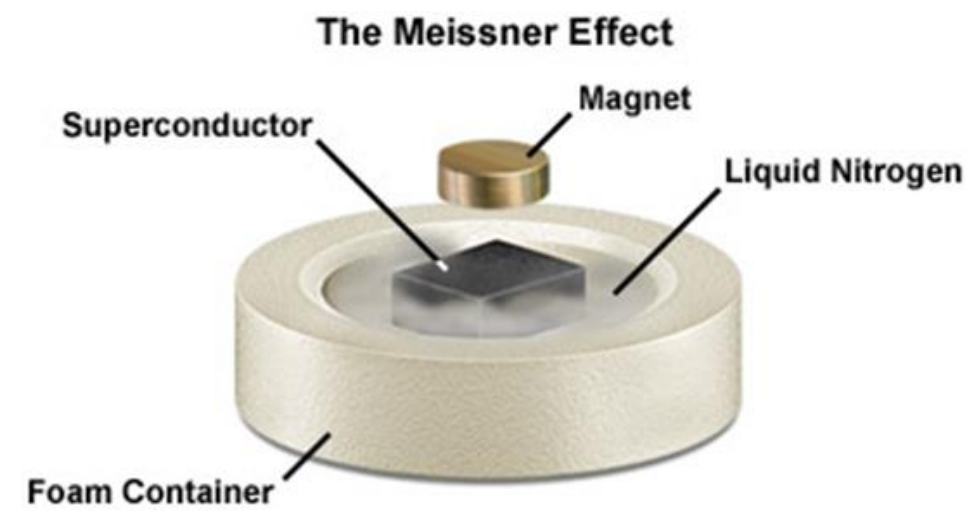

*Figure 98: the Messner effect.*

The magnetic field is then expelled from inside the superconducting material. This is the Meissner effect, discovered in 1933 by **Walther Meissner** (1882-1974, German), which only applies to certain so-called type I superconductors. It explains the repulsion demonstrated in numerous experiments. Type II which does not generate this phenomenon includes niobium titanium alloys which are frequently used with a 1:1 ratio of each in the alloy.

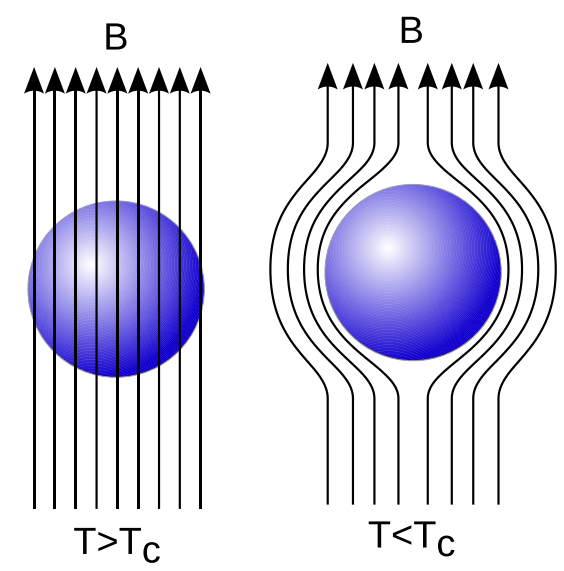

*Figure 99: Meissner effect explanation.*

In type II superconductors, an intermediate phase between the classical metallic phase and the superconducting phase allows the magnetic field to pass partially. The Holy Grail of superconductivity would be to obtain it at room temperature, allowing, for example, to reduce transmission losses in grid electric power lines[174]. Out of the various metals used in quantum technologies, titanium becomes superconducting at 390 mK, aluminum at 1.2K, indium at 3.4K and niobium at 9.26K.

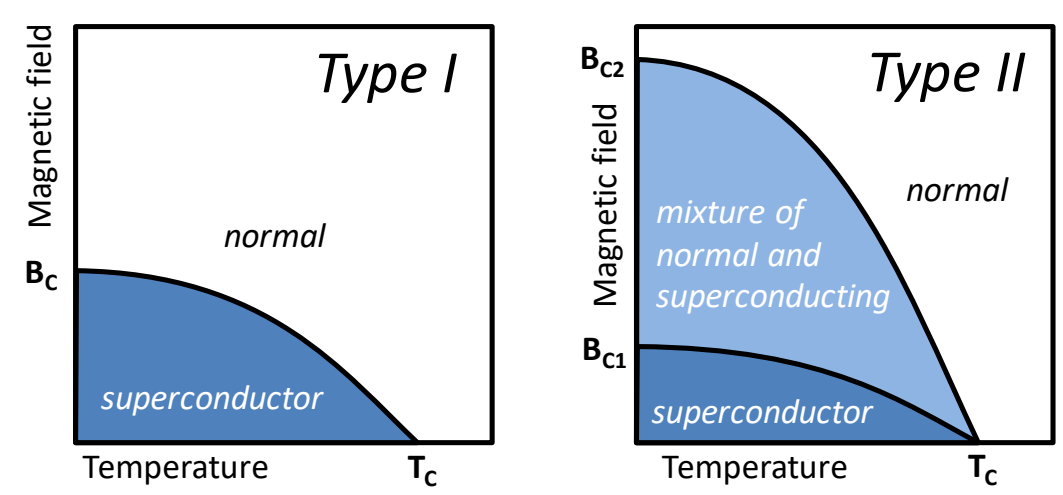

*Figure 100: type I and II superconductors characteristics. Source: [Critical Magnetic Field](), undated.*

Scientists from IBM began discovering superconducting metal alloys above 77K (-196°C) in the late 1980s, the temperature of liquid nitrogen.

[173] See Superconductivity 101. The superconducting properties of the niobium-titanium alloy were discovered in 1962. It is widely used in the cooling of MRI scanners but also in many scientific instruments, notably in the ITER experimental nuclear fusion reactor at Caradache. The Periodic Table of Elements comes from Wikipedia.

[174] Type I and II superconductors are mathematically and quantumly explained by the **Ginzburg–Landau** theory created in 1950. See Theory of Superconductivity by Carsten Timm, TU Dresden, February 2022 (150 pages).

Most of them are cuprates alloys (copper-based). A record was achieved in 2019 with a molecule combining lanthanum and hydrogen ($LaH_{10}$, illustrated in Figure 101) and at -23°C, thus a near-ambient temperature. In the latter case, however, it works at a huge pressure of 218 GPa, representing more than 2 million times the atmospheric pressure, which is 101,325 Pa [175]. Other records were broken with metallic hydrogen in 2020 by CEA, operating at 17°C and at an even greater pressure of 400 GPa[176]. Another record of 15°C with 270 GPa was achieved in the USA also in 2020, using a carbonaceous sulfur hydride[177]. A less impressive 2022 record was created in China with clathrate calcium hydride ($CaH_6$) being superconducting at 215K and 172 GPa[178].

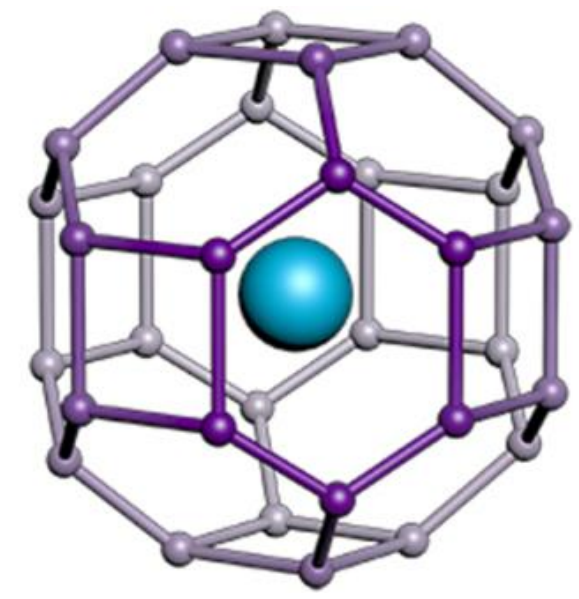

Figure 101: $LaH_{10}$ high superconducting temperature molecule.

You always see this trade-off between superconducting temperature and pressure. At this very high pressure, practical use cases are not easy to implement! But at lower temperatures, interesting used cases arise like with single photons detectors[179].

Hence the willingness to use quantum simulators or computers to run superconductivity quantum equations and identify materials that would be superconducting at room or near-room temperature[180].

By the way, we may wonder why scientists are not using high-temperature superconducting materials to build superconducting qubits? The main reason is that their low temperature of about 15 mK is related to the controlled noise environment linked to using driving micro-waves in the 5-10 GHz range (~0.040 meV). These microwaves have the benefit of being photons adapted to the anharmonic excitement levels of Josephson gates and to be transportable on coaxial cables which are themselves made of superconducting materials like niobium-titanium. The superconducting qubits cooling temperature of 15 mK creates an ambient thermal noise that is one order of magnitude lower than the temperature corresponding to these controlling microwaves (a few kT, with k=Boltzmann constant and T being the temperature = ~0.004 meV).

Superconductivity is commonly used in **MRI scanners**[181], using large superconducting magnets that are cooled with liquid helium. Scanners are encased in a protective coating to constrain the magnetic field inside the scanner. The niobium-titanium coil wiring is enveloped in a copper matrix.

This combination is also used in large physics instruments like the **CERN LHC** in Geneva with 1200 tons of cables including 470 tons of NbTi (niobium-titanium), the rest being copper, in cables totaling 21 km. Superconductivity creates a current of 11,850 A generating a powerful magnetic field of 8.33 tesla creating a centripetal force holding the accelerated particles. These magnets are cooled by 10,000 tons of superfluid helium-4 at 1.9K. Their cables are made of niobium-titanium filaments surrounded by copper. The whole unit power is 40MW with an electricity consumption estimated at 750 GWh per year according to CERN. It is the largest and most powerful refrigerator in the world!

---

[175] See Quantum Crystal Structure in the 250K Superconducting Lanthanum Hydride by Ion Errea, July 2019 (20 pages).

[176] See Here comes metallic hydrogen - at last! by Jean-Baptiste Veyrieras, May 2020. Another record was broken in 2019 with $YH_6$ (yttrium hybrid) at a pressure of 110 GPa. See Anomalous High-Temperature Superconductivity in YH6 by Ivan A. Troyan et al, 2019 (36 pages).

[177] See Room-temperature superconductivity in a carbonaceous sulfur hydride by Elliot Snider et al, Nature, October 2020 (14 pages).

[178] See High-Temperature Superconducting Phase in Clathrate Calcium Hydride $CaH_6$ up to 215 K at a Pressure of 172 GPa by Liang Ma et al, PRL, April 2022 (not open access).

[179] See Single-photon detection using high-temperature superconductors by I. Charaev et al, August 2022 (8 pages).

[180] Another fancy solution consists in lowering the room temperature as described in Novel approach to Room Temperature Superconductivity problem by Ivan Timokhin and Artem Mishchenko, April 1st, 2020 (4 pages).

[181] Nuclear magnetic resonance imaging.

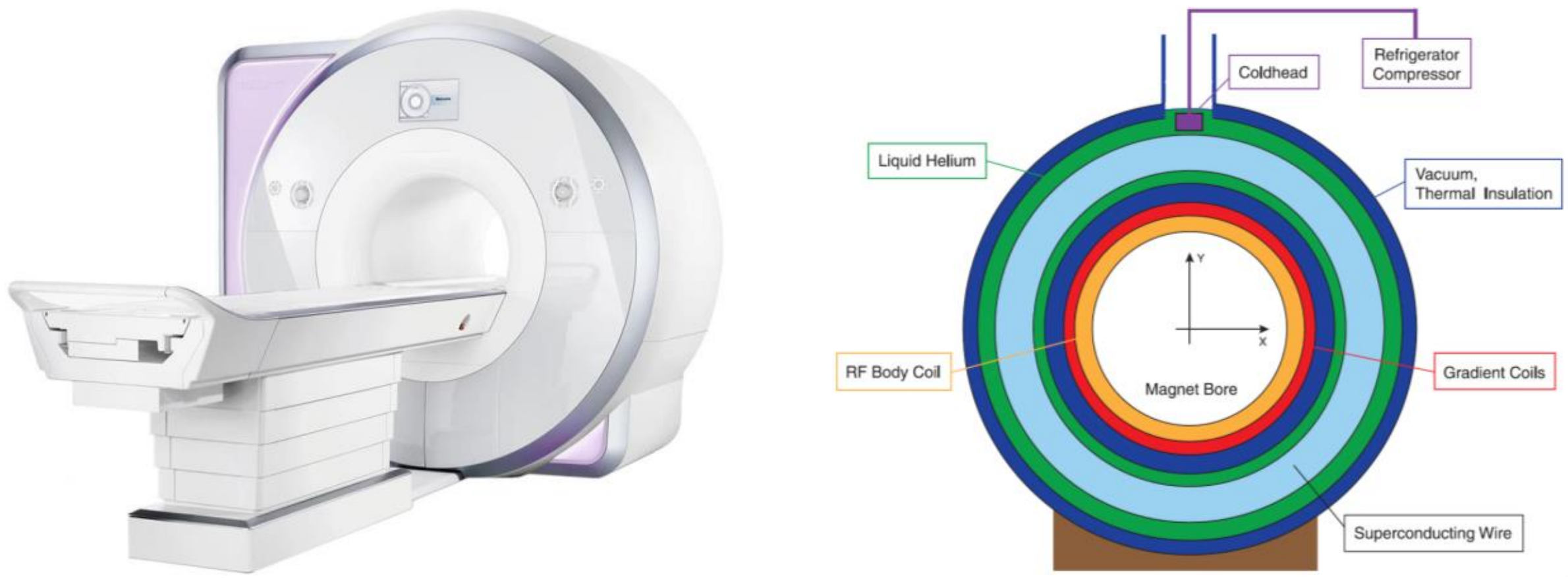

*Figure 102: MRI principle. Source of illustration on the right: [Helium Reclaiming Magnetic Resonance Imagers](#) by Dan Hazen, MKS Instruments (5 pages).*

Superconductivity is operated in the **Chuo Shinkansen** Maglev high-speed train in Japan, which has been undergoing trials since 2013 and is expected to reach a commercial speed of 505 km/hr. It uses a superconductive based magnetic suspension with a rather expensive infrastructure. Power consumption per passenger/kilometer is three times that of traditional Shinkansen, but it is still competitive with airplanes. A 286 km Tokyo-Nagoya line is planned for commercial service in 2027.

Superconductivity has also been studied to improve the efficiency of electric motors and generators with HTS Synchronous Motors (High-Temperature Superconducting). It allows a reduction of motors size and efficiency improvements. It is based on superconducting materials that only require liquid nitrogen cooling, but some systems still use helium-based cooling.

Studies began in the 1980s and these engines and generators are beginning to be deployed in the military navy and in wind power generation, notably at **ASMC**, **Sumitomo Electric**[182] and with the European **EcoSwing** project, which involves Sumitomo's cryostat division.

Superconducting cables have also been introduced to transmit electricity without power loss and greater capacity to meet the ever-increasing demand. They are offered by the French cable manufacturer **Nexans**, which installed one in Long Island. Their 600 m underground cable has been in operation since 2008. It can supply electricity to 300,000 homes[183]. But it is rather complex to implement and was not seemingly replicated in many places. The project cost was $46.9M.

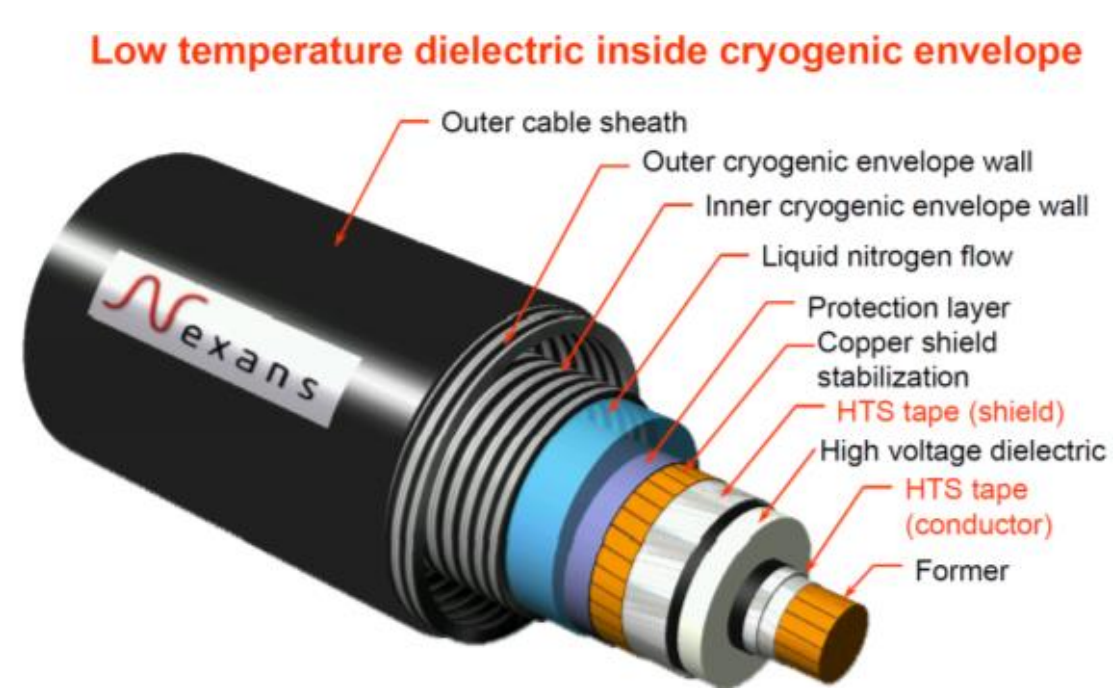

*Figure 103: Nexans superconducting cable.*

As far as quantum computers are concerned, superconductivity is used in particular in superconducting qubits that exploits the Josephson effect that we have already described in another section. This technology is also used in variations of SQUIDs (superconducting quantum interference device) in quantum sensing. Josephson junctions have a relationship between voltage and frequency which enables the creation of various sensors. It can convert a voltage to frequency as well as a frequency to voltage (with the inverse AC Josephson effect using a microwave impulse). We also find it in the type II niobium-titanium based superconducting cables used for reading the state of superconducting and electron spin qubits.

---

[182] See [Design of MW-Class Ship Propulsion Motors for US Navy by AMSC](#) by Swarn S. Kalsi, 2019 (50 slides).

[183] Information Source: [Long Island HTS Power Cable](#), Department of Energy, 2008 (2 pages). In addition to Nexans, the cryogenic system was supplied by Air Liquide.

Superconductivity could also be used to create processors operating at low temperatures and capable of operating up to 700 GHz, much faster than current server processors running at a peak 4 to 5 GHz[184]. An MIT team announced in July 2019 a proposal for a technique to create spiking neurons with superconducting Josephson effect circuits using nanowires[185].

This is still a research field with very few industry applications at this point. We'll investigate this field in a specific section on unconventional computing. Superconducting electronics could be very useful to create and analyze the microwaves used in superconducting and electron spin qubits.

**Superfluidity**

Superfluidity is yet another quantum physics phenomenon to cover here. It occurs only with superfluid helium which, at ambient pressure, never freezes, no matter how low the temperature can be.

Superfluid liquid has zero viscosity and flows without any loss of kinetic energy. When poured into a recipient, it tends to rise up by capillary action on its rim and flow out of it. It can even pass through very fine capillaries.

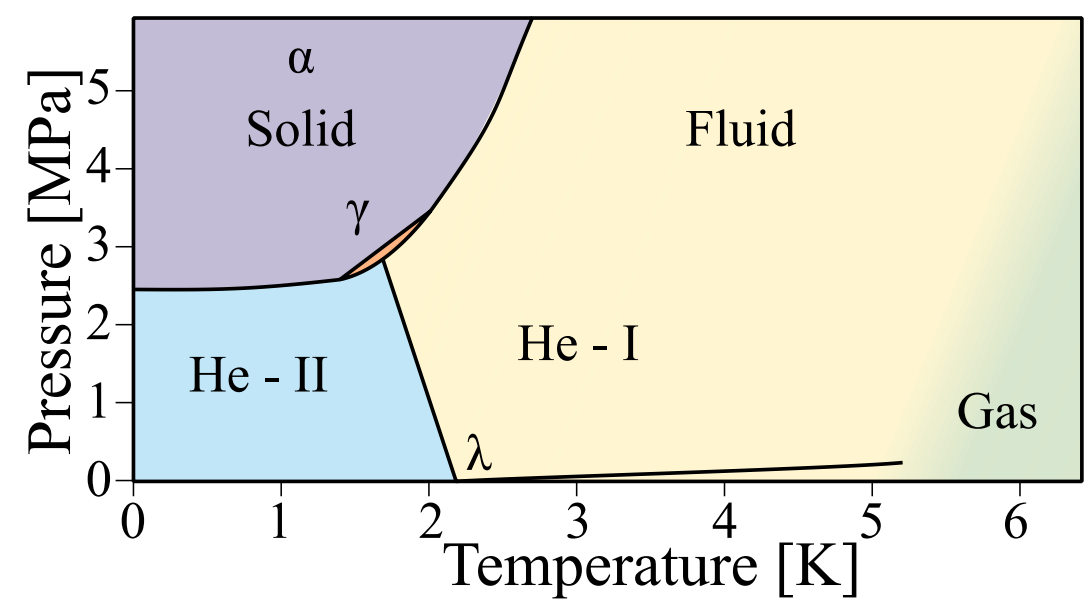

*Figure 104: superfluidity. Source: Wikipedia.*

Helium was first liquefied in 1908 at 4.2K by Heike Kamerlingh Onnes, the discoverer of superconductivity in 1911. Its superfluidity was highlighted independently in 1938 by **Pyotr Kapitsa** (1894-1984, USSR), **John Frank Allen** (1908-2001, USA) and **Don Misener** (1911-1996, USA)[186].

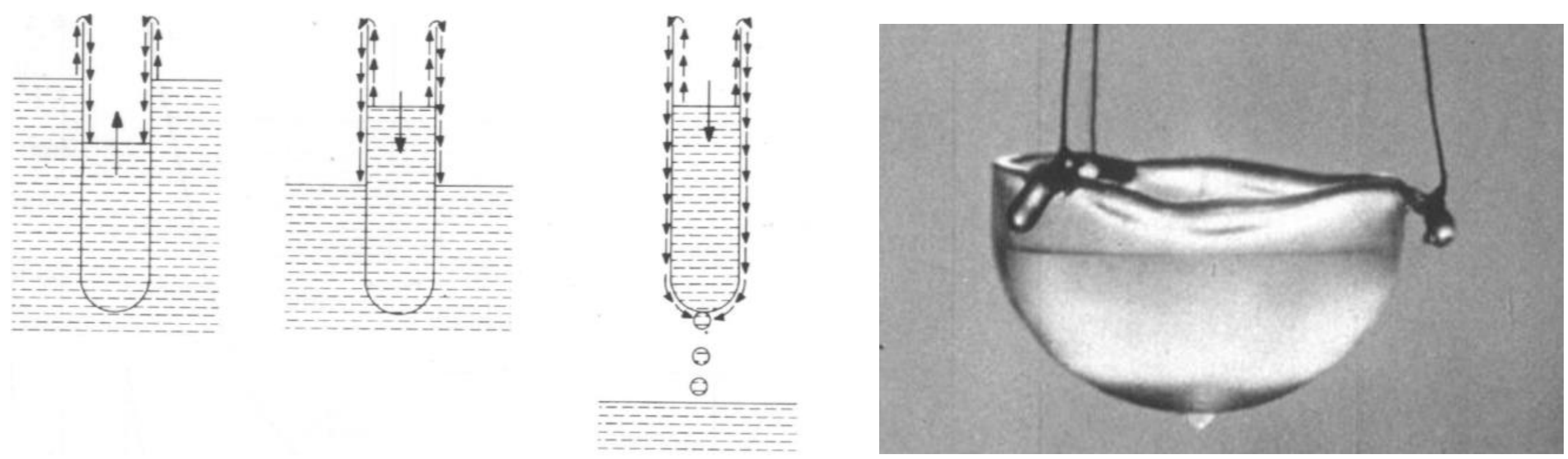

*Figure 105: visualization of the superfluidity phenomenon. Source: Helium 4 (14 slides).*

There are two isotopes of helium: $^3$He with a single neutron, which is the least abundant in nature, and $^4$He, with two neutrons, the most common. The latter is a boson, with an integer spin, giving it different properties from helium 3, which is a fermion with a half-integer spin. At low temperature, $^4$He behaves like Bose-Einstein condensates since being bosons. $^3$He behaves differently, being fermions, and assemble in pairs similar to electron Cooper pairs. It becomes superfluid at lower temperatures than $^4$He, at around 1 mK in the absence of a magnetic field (see the phase diagram in Figure 104), vs. 2.17K for $^4$He.

[184] See Superconductor ICs: the 100-GHz second generation by Darren Brock, Elie Track and John Rowell of Hypres, 2000 (7 pages).

[185] See A Power Efficient Artificial Neuron Using Superconducting Nanowires by Emily Toomey, Ken Segall et Karl Berggren, 2019 (17 pages).

[186] See Viscosity of Liquid Helium below theλ-Point, Pyotr Kapitsa, Nature (1938) and Flow of liquid helium II, Joan F. Allen, Don Misener, 1938 (1 page). Pyotr Kapitsa was awarded the Nobel Prize in 1978 for his work in the field of low temperatures.

Its superfluidity was only discovered in 1973[187]. The different properties of $^3$He and $^4$He are used to operate the dilution cryogenics systems that equip many quantum computers whose operating temperature is between 10mK and 1K. We will study this in detail in this book, starting page 465.

Industrial demand for helium is spread across many industries: medical imaging for MRI systems magnets cooling, then microelectronics industries.

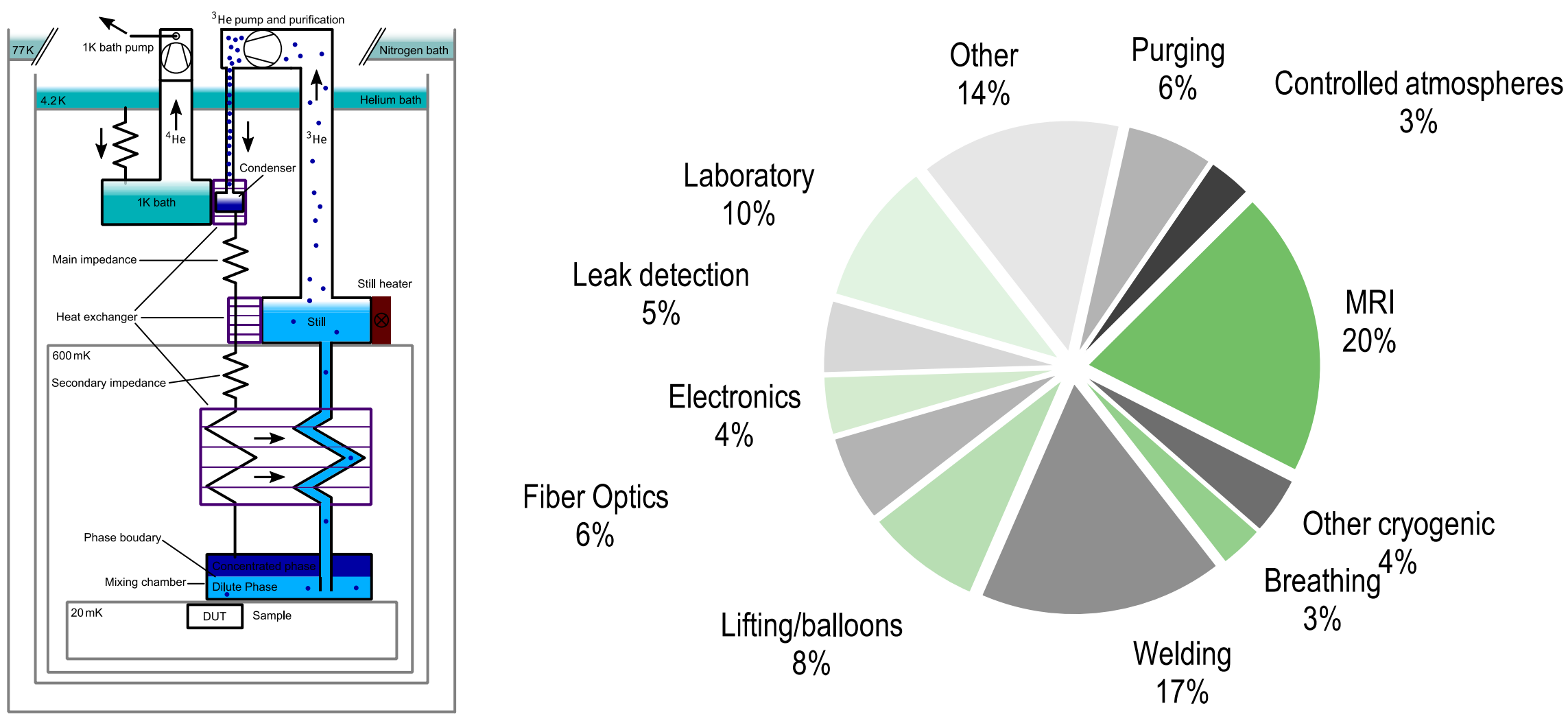

*Figure 106: Sources: left diagram: Wikimedia, right diagram: Edison Investment Research, February 2019, referring to Kornbluth Helium Consulting.*

## Bose-Einstein Condensates

Bose-Einstein condensates are extremely low-density gases of bosons cooled down to very low temperatures, at the lowest energy level we can set matter in, below solid state. $^4$He is the most famous element that was experimented in this matter state.

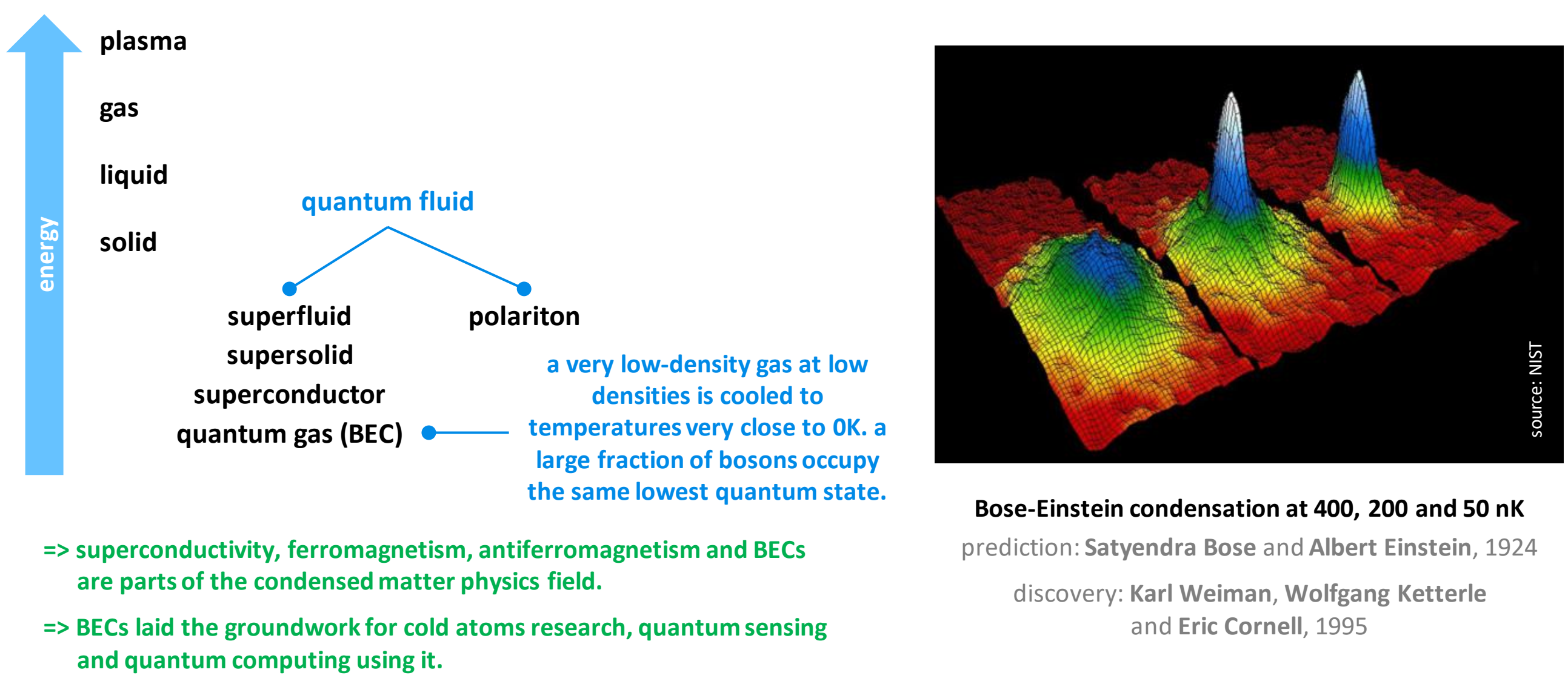

*Figure 107: Bose-Einstein condensates positioned within the various states of matter.*

It took a while between the work of Bose and Einstein in 1924 and the experimental discovery of BECs in 1995 by **Carl Wieman**, **Wolfgang Ketterle** and **Eric Cornell** with rubidium 87 at 170 nK. It was cooled with laser-based Doppler effect and magnetic evaporating technique.

---

187 David Morris Lee (1931), Douglas Dean Osheroff (1945) and Robert Coleman Richardson (1937-2013) were awarded the Nobel Prize in Physics in 1996 for their discovery of helium-3 superfluidity.

BECs play an important role in quantum technologies. They led to the control of individual atoms that are used in quantum simulators and in quantum gravimeters. Together with superfluids and supersolids, BECs belong to the field of quantum hydrodynamics.

### Supersolids

Supersolidity is another weird quantum state of matter showing up at ultracold temperatures, when atoms behave as a crystal and as a superfluid at the same time. This is made possible with crystal lattice with holes (like in an NV center).

The vacancies behave quantumly as bosons and can switch position in a quantum manner like a Bose Einstein Condensate. It's a vacancies quantum tunnelling phenomenon.

This state of matter was predicted in 1969[188] and it was first demonstrated, although debated for a long time, in 2004 with $^4He$ at a pressure of about 60 bar and below 170 mK[189]. The related fundamental research is going on in various places in the world like in the USA, Innsbruck[190], Pisa[191], Stuttgart, Warsaw, Geneva, and Paris. It is now possible to create supersolids with ultracold dipolar quantum gases of highly magnetic lanthanide atoms like erbium and dysprosium. The supersolidity effect can be controlled by a magnetic field.

There are no known practical applications of this phenomenon to date although it could lead to new forms of quantum simulation systems like the ones using cold atoms.

### Polaritons

Polaritons is a field of quantum physics that is rarely mentioned in the context of quantum technologies. It mostly belongs to fundamental research but could be of interest in various fields such as quantum computing and quantum sensing.

Polaritons are quantum quasi-particles in the domain of strong coupling between light and matter. They result from the coupling between photons and an electrical polarization wave.

These waves occur in particular in plasmons (oscillations of free electrons in metals), phonons (oscillations of atoms, especially in crystal structures) and excitons (pairs of electron holes generated by photons in semiconductors[192]). The materials can be atoms gas, massive classical semiconductors, thin films inserted in optical cavities or superconducting Josephson junctions.

Excitation photons have a wavelength corresponding to the resonance frequency of the associated medium, often in the visible light or infrared ranges. Polaritons have mixed properties of photons dressed by electronic excitations.

They behave like bosons (having an integer spin) that can occupy the same quantum state and operate in groups, such as superconducting currents forms with paired electrons named Cooper pairs or Bose-Einstein condensates (BEC).

---

188 By David J. Thouless (1934-2019, British, 2016 Nobel prize in physics) and, independently, by Alexander Andreev (1939, Russian) and Ilya Mikhailovich Lifshitz (1917-1982, Russian). See The flow of a dense superfluid by David J. Thouless, 1969 (25 pages) and Quantum theory of defects in crystals by Alexander Andreev and Ilya Mikhailovich Lifshitz, 1969 (7 pages).

189 See Probable observation of a supersolid helium phase by E Kim and M H W Chan, 2004, The enigma of supersolidity by Sébastien Balibar, Nature, 2010 (7 pages) and the review paper Saga of Superfluid Solids by Vyacheslav I. Yukalov, 2020 (26 pages).

190 Research in Austria is led by Francesca Ferlaino from the University of Innsbruck, IQOQI.

191 See The supersolid phase of matter by Giovanni Modugno, 2020 (37 slides).

192 The name of polariton was created by Joseph John Hopfield (1933, American) in 1958 and at that time concerned polariton excitons. See Theory of the Contribution of Excitons to the Complex Dielectric Constant of Crystals by Joseph John Hopfield, 1958 (14 pages). Hopfield is also known in the field of neural networks in AI with his "Hopfield networks".

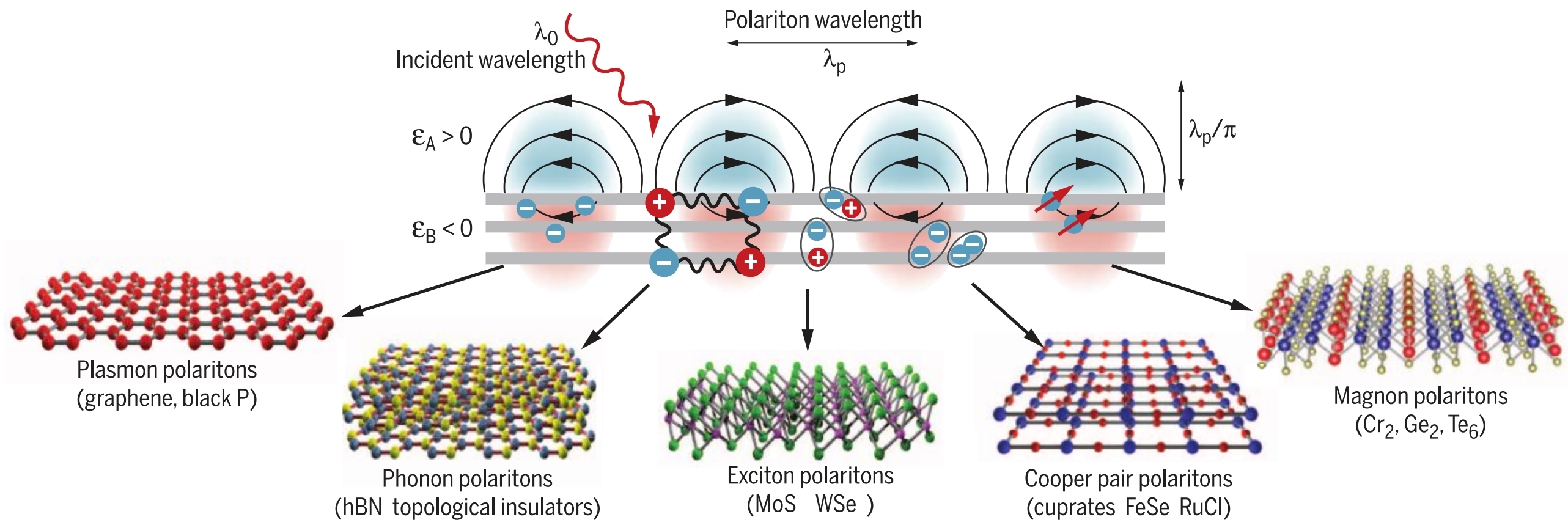

*Figure 108: various forms of polaritons. Source: Polaritons in van der Waals materials by D. N. Basov et al, 2016 (9 pages) which makes a good inventory of different types of polaritons and their fields of application.*

Depending on the interaction scale, polaritons operate in a semiclassical or quantum regime. In the first case, the electromagnetic field interacts with a macroscopic polarization field. The polariton field then has the properties of a classical field but its elementary quantum is the result of a dipole-photon "wrapping" that can only be described by quantum mechanics. In the second case, the electromagnetic field interacts with a single polarization field quantum that has been isolated in one way or another, such as a superconducting qubit or an exciton in a quantum box. We are then in the quantum regime of strong coupling, known as the "Jaynes-Cummings Hamiltonian", where the energy levels are discrete and each level correlate to a given number of excitation quanta in the system. Cavity-excited polaritons are generally in the first regime.

In polaritons, semiconductor matter receives photons that excite it. It then emits photons to get out of its excited state, all of this in a very fast iterative cycle, the photons circulating in a closed circuit in the cavity. In practice, electromagnetic and polarization fields co-propagate in the medium in an identical way, notably in polarization and frequency, and with a fixed phase relation (without phase shift or with a 180° phase shift, i.e., π). Polaritons are particularly interesting for generating strong nonlinearities which are searched in photonics[193].

Thanks to the degenerate states in which polaritons can be prepared and to the fact that they interact with each other, polaritons constitute an out-of-equilibrium quantum fluid called "light quantum fluid", often abusively referred to as "liquid light". Polaritons can thus generate surface waves and propagation phenomena typical of quantum fluids such as superfluids. Polaritons also interact with each other, which is not the case for photons in vacuum[194] . We can experimentally control the spatial distribution of the density, phase and velocity of these fluids of light[195].

[193] See also this very dense review paper Quantum Fluids of Light by Iacopo Carusotto and Cristiano Ciuti, 2013 (68 pages).

[194] See the pedagogical presentation Swimming in a sea of light: the adventure of photon hydrodynamics by Iacopo Carusotto, 2010 (28 slides). Presentation realized with the help of, among others, Elisabeth Giacobino and Alberto Bramati from CNRS. See also the very well-illustrated presentation Quantum fluids of light by Jacqueline Bloch, February 2020 (58 slides).

[195] Source: description of the ANR project: Quantum Light Fluids - QFL launched in 2016.

There are many variants of polaritons which depend on the nature of the electronic excitation of the matter:

- **Phonon-polaritons** resulting from the coupling between an infrared photon and an optical phonon caused by the mechanical oscillation of two adjacent ions of opposite charge in a crystalline structure. This oscillation produces an oscillating electric dipole moment. This phenomenon was discovered by **Kirill Tolpygo** (1916-1994, Russian) in 1950 and, independently, by **Kun Huang** (1919-2005, Chinese) in 1951. One application of phonon polaritons are thermal emitted and imagers[196].

- **Exciton-polaritons** result from the coupling of a photon with an exciton in a semiconductor cavity. An exciton is a quasi-particle consisting of an electron-hole pair connected by Coulomb forces, generated by excitation photons. The notion of exciton was created by **Yakov Frenkel** (1894-1952, Russian) in 1931. Like all types of polaritons, these have two energy bands: the high and low polariton. It is a general property of the strong coupling regime between electric dipole and electromagnetic field. Here, the level is high when the photon and the semiconductor are excited and in phase, and low when they are in opposite phase[197].

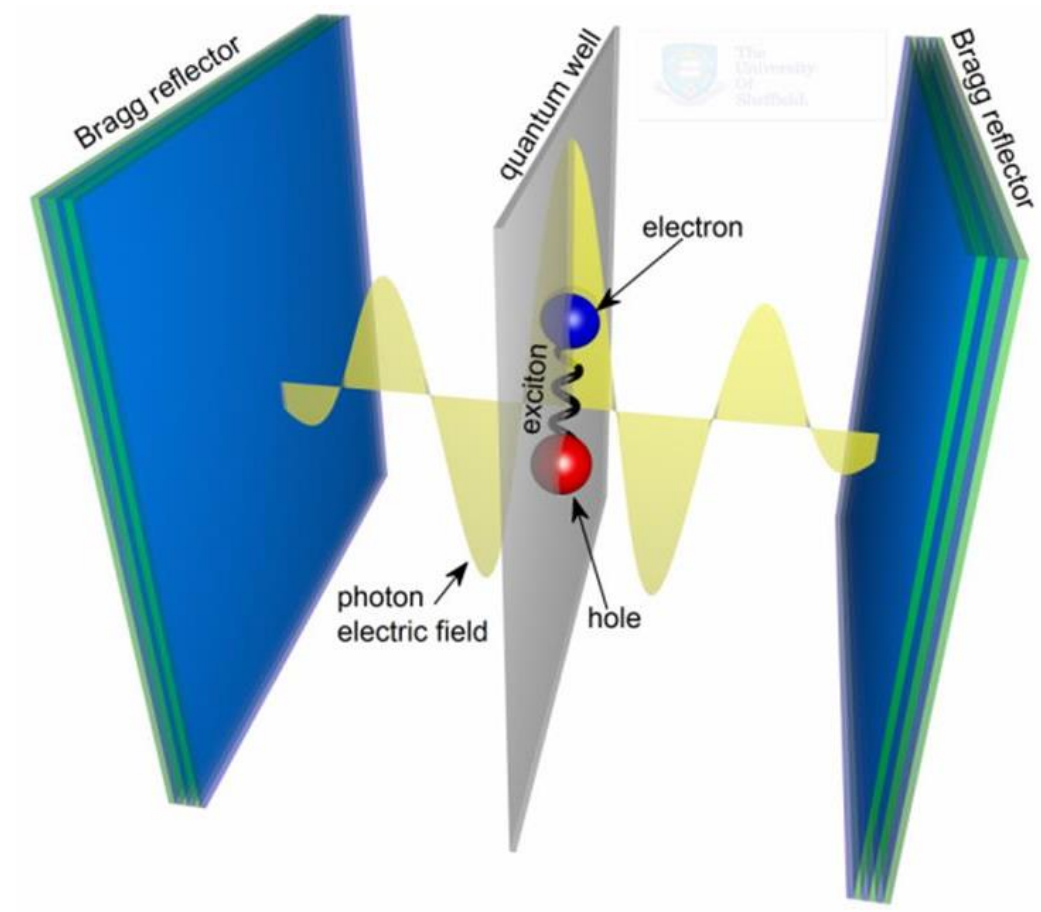

*Figure 109: exciton-polariton. Source: Polariton: The Krizhanovskii Group. University of Sheffield.*

  Researchers are trying to create transistors using polariton-exciton as well as on single quantum control[198].

- **Surface plasmon polaritons** (SPP) result from coupling surface plasmons and photons. A plasmon is a quantized oscillation of high-density electron gases. A surface plasmon is a coherent electron oscillation occurring at the interface between two different materials, often a metal and a dielectric or between metal and air. A surface plasmon polariton is an oscillation caused by an incident photon.

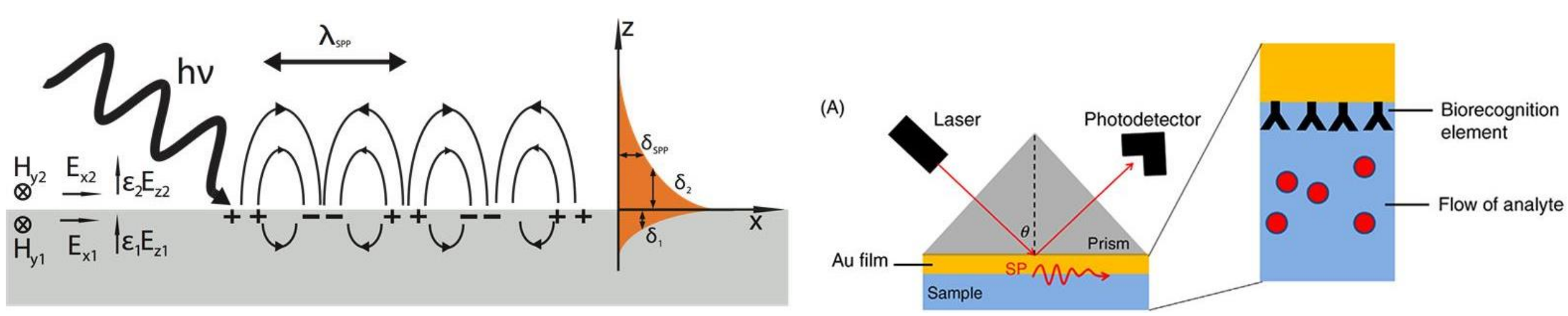

*Figure 110: surface-plasmon polariton phenomenon. Source: Wikipedia.*

---

[196] See Surface phonon polaritons for infrared optoelectronics by Christopher R. Gubbin et al, January 2022 (23 pages).

[197] Source of illustration: Low Dimensional Structures & Devices Group. University of Sheffield, mentioned here.

[198] The "polariton blockade" mechanism allows in principle to manipulate excitonic cavity polaritons at the single quantum scale. See Towards polariton blockade of confined exciton-polaritons by Aymeric Delteil, 2019 (4 pages).

SPPs are used in optical quantum sensors for temperature and for the detection of the concentration of different components by refractivity and then spectroscopy, especially in medtechs (detection of various organic molecules and of interactions between proteins), biological analyses (toxins, drugs, additives) or for the detection of gases[199].

SPRs (Surface Plasmon Resonance Plasma) can be much more powerful than near-infrared spectroscopy sensors such as those from Scio[200] . They measure the polarized light reflected from a laser diode in terms of intensity, angle, wavelength, phase and polarization.

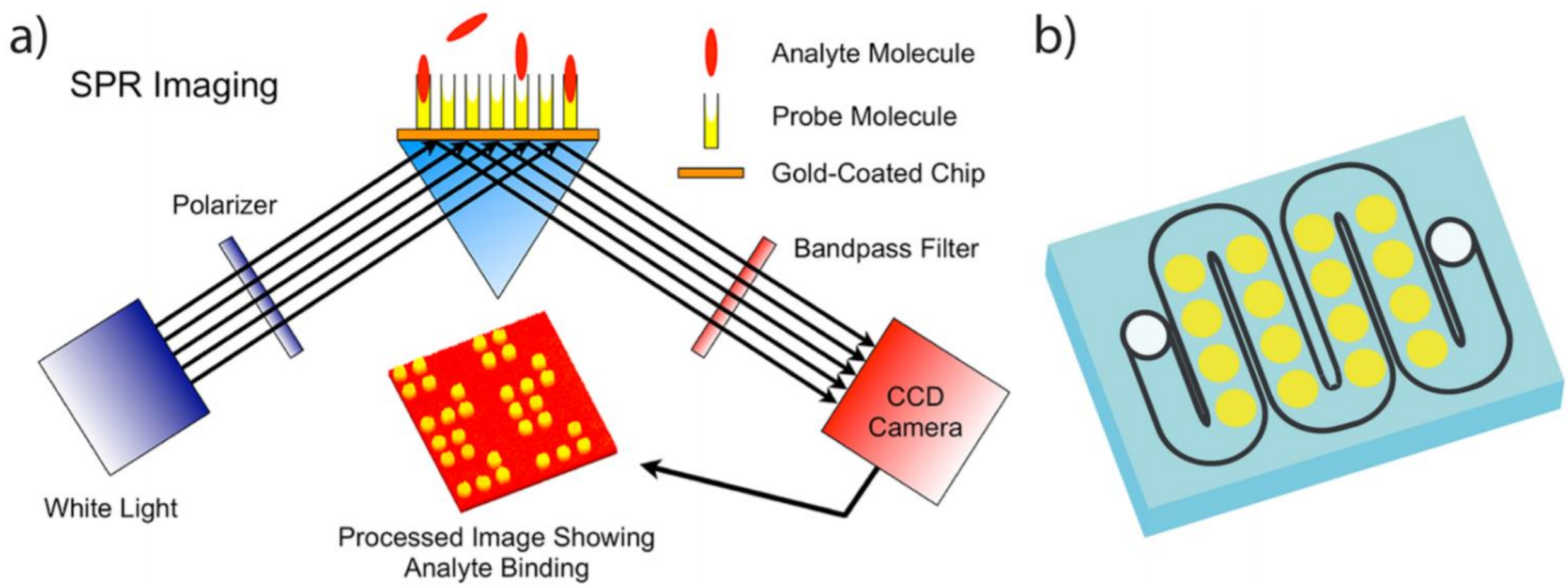

*Figure 111: surface plasmon resonance plasma. Source: Surface Plasmon Resonance (SPR) by Lifeasible.*

As in many biological analysis systems, it is possible to create 2D matrices (microarrays) integrating many detection molecules and to detect a lot of components in the sample to be analyzed[201].

SPRs are commonly marketed by companies such as **Cytiva** (USA), **Carterra** (USA), **Horiba** (Japan)[202], **IBIS Technologies** (Netherlands), **Lifeasible** (USA), **Polaritons Technologies** (Switzerland) and **XanTec** (Germany).

- **Cavity polaritons** are a variant of the polariton excitons where the photon is trapped in a microcavity, and the exciton is confined in a quantum well. They are made of III-V semiconductors like indium, arsenic and gallium.

  Photons trapping is often performed using two Bragg mirrors facing each other to create an optical cavity using layers of dielectrics to reflect light very efficiently and of all wavelengths. These mirrors are fabricated from molecular beam epitaxy allowing coherent crystal growth on a gallium arsenide (GaAs) crystal substrate. The result is monocrystalline and can contain more than a hundred layers of different alloys, with thicknesses ranging from 5 nm to 50 nm, controlled to the nearest atomic monolayer[203]. These microcavity polaritons were discovered in 1992 by Claude Weisbuch (France)[204].

[199] The general principle of this instrument is to use a laser diode to illuminate a gold surface at an angle (via a mechanically controllable angle) and to capture the reflected beam with a detector. The gold surface is coated with a specific molecule ("biorecognition element" in the diagram) that tends to associate itself with a molecule that we want to detect (in the liquid phase "flow of analyte"). The molecules detected can be peptides, polypeptides, proteins, enzymes, vitamins, DNA or RNA sequences, or antibodies (in particular for cancers diagnosis). The association modifies the reflectivity of gold and allows the detection of the target molecule.

[200] See Recent advances in Surface Plasmon Resonance for bio sensing applications and future prospects by Biplob Mondal and Shuwen Zeng, August 2020 (31 pages). The second author is from the Limoges XLIM laboratory in France.

[201] See Surface Enzyme Chemistries for Ultra sensitive Microarray Biosensing with SPR Imaging by Jennifer B. Fasoli et al, 2015 (10 pages) where the associated illustration comes from.

[202] Horiba's European research center is located in Palaiseau next to the C2N of the CNRS, Télécom Paris, Thales and the Institut d'Optique. Horiba is specialized in spectrometers and various other optical instruments like near-IR photoluminescence characterization of InGaAs/GaAs quantum dots. They acquired Yvon Jobin, a French optical instruments manufacturer in 1997.

- **Intersubband-polaritons** result from the coupling of an infrared or terahertz photon with an intersubband excitation. They can be used to create infrared detectors.
- And then **Bragg-polaritons** (Braggoritons), **plexcitons** (plasmons + excitons), **magnon polaritons** (magnon, spin waves in ferromagnetic materials + photons) and **similaritons** (amplified photons in optical fibers).

In short, all these "*-ons" are the result of the interaction between photons and different forms of matter, noticeably electrons. What does this have to do with quantum computing? Polaritons are used in various optical devices related to photon qubits, including photon transport and single photon detectors.

They could eventually allow the creation of photon qubits that can interact with each other. This is what emerged from an MIT and Harvard publication by Vladan Vuletić and Mikhail Lukin in 2018 which demonstrated the interaction of three photons in an atom placed in a Rydberg state, constituting a "Rydberg polariton"[205]. Another research project in Singapore uses polariton excitons to create photon qubits with the particularity of being able to operate at room temperature, using single-qubit gates and $\sqrt{\text{SWAP}}$ two-qubits gates[206].

Microcavities polaritons can be used to create quantum simulators[207]. They are implanted in III-V semiconductor structures as 2D arrays. One field of application is the simulation of gravitational structures such as a Hawking radiation on the horizon of a black hole. And why not, to simulate the operation of a dilution refrigerator associating helium 3 and 4 at very low temperature.

Polaritons are also the field of topological behaviors of matter and are perhaps an alternative way to the Majorana fermions to create error corrected qubits. These are longer term pathways than the qubit technologies studied in this book, but worthy of interest.

Other applications, already mentioned, target the very diverse field of quantum sensing, including optomechanical systems[208].

In France, polaritons are the specialty of Cristiano Ciuti (UPC MPQ), **Elisabeth Giacobino** (CNRS LKB), **Jacqueline Bloch** (CNRS C2N[209]), **Alberto Bramati** (ENS LKB), **Alberto Amo** (Phlam-CNRS Lille), **Le Si Dang** and **Maxime Richard** (CNRS Institut Néel Grenoble).

---

[203] See Cavity polaritons for new photonic devices by Esther Wertz, Jacqueline Bloch, Pascale Senellart et al, 2010 (12 pages).

[204] See Observation of the coupled exciton-photon mode splitting in a semiconductor quantum microcavity by Claude Weisbuch et al, 1992 (4 pages).

[205] See Physicists create new form of light by Jennifer Chu, 2018 referencing Observation of three-photon bound states in a quantum non linear medium by Qi-Yu Liang et al, 2018 (5 pages).

[206] We will define this type of quantum gate in a dedicated section of this book. See Quantum computing with exciton- polariton condensates by Sanjib Ghosh and Timothy C. H. Liew, October 2019 (6 pages). Tim Liew is a researcher at the joint MajuLab laboratory between CNRS and the National University of Singapore.

[207] See Microcavity Polaritons for Quantum simulation by Thomas Boulier, Alberto Bramati, Elisabeth Giacobino, Jacqueline Bloch et al, May 2020 (21 pages) as well as Polaritonic XY-Ising machine by Kirill P. Kalinin, Alberto Amo, Jacqueline Bloch and Natalia G. Berloff, 2020 (12 pages).

[208] See Enhanced Cavity Optomechanics with Quantum-well Exciton Polaritons by Nicola Carlon Zambon, Zakari Denis, Romain De Oliveira, Sylvain Ravets, Cristiano Ciuti, Ivan Favero and Jacqueline Bloch, February-September 2022 (22 pages).

[209] The clean room of the C2N in Palaiseau, France, allows the prototyping of a whole bunch of nanostructures. The semiconductors used to manage polaritons are moreover manufactured with techniques similar to the single photon sources of Pascale Senellart's team, also from the C2N, and the associated startup, Quandela.

## Magnons

Quantum matter also includes **magnons**, a category of quasi-particles that take the form of quantized spin waves in magnetic materials, usually crystalline lattices. Magnons were conceptualized by **Felix Bloch** in 1930 and experimentally detected in 1957 by **Bertram Brockhouse** (1918-2003, Canadian). These objects which behave as bosons could be used in quantum information systems.

Current physics experiments are done at the control low-level like with controlling these magnons with microwaves[210] or measured with superconducting qubits[211]. Magnons can also be used at low temperature to create some topological materials[212] and even for some species of SiC-based spin qubit control[213].

## Skyrmions

Order is not restricted to the periodic atomic array of a crystal and can also be associated with magnetic order in a solid where spins align parallel to each other in ferromagnets and antiparallel in antiferromagnets. More complex magnetic nanostructures are skyrmions that form mesoscopic magnetic vortex with particle-like properties[214].

Then, how do you distinguish between magnons and skyrmions which are both magnetic quasiparticles? Magnons are quantized dynamic magnetic excitations that travel through magnetic materials while skyrmions are static.

The skyrmion naming comes from **Tony Hilton Royle Skyrme** (1922-1987) who in 1961 formulated a nonlinear field theory of massless pions in which particles can be represented by topological solitons. Skyrmions existence in magnetic materials was predicted in 1989 by Bogdanov et al[215]. In 2008, **Sebastian Mühlbauer** discovered skyrmions in MnSi crystals at the Munich reactor using neutrons[216].

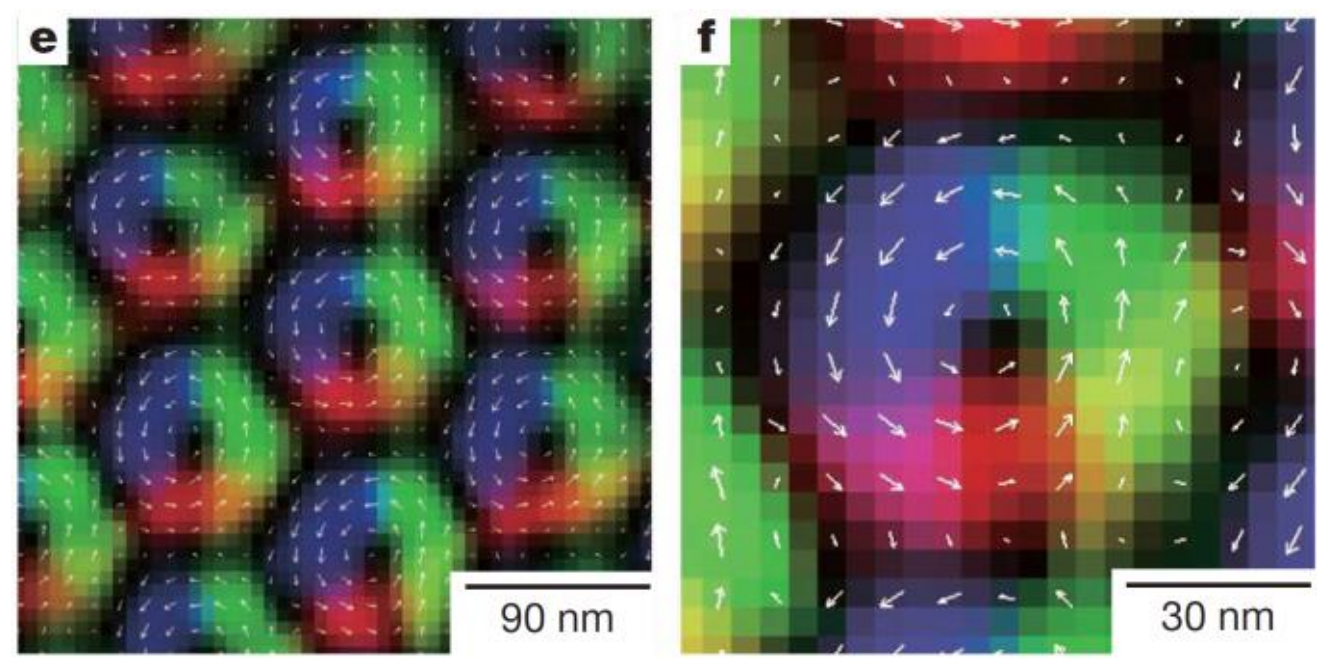

*Figure 112:visualizing a skyrmion. Source: Real-space observation of a two-dimensional skyrmion crystal by X. Z. Yu et al, 2010, Nature (5 pages).*

Then, Japanese and Korean researchers implement real-space imaging of a two-dimensional hexagonally arranged skyrmion lattices spaced by 90 nm in a thin film of $Fe_{0.5}Co_{0.5}Si$ and exposed to a magnetic field of 50–70mT, using Lorentz transmission electron microscopy[217]. This helicoidal structure can also be 3D and create superposition of various magnetic skyrmion states.

---

210 See Floquet Cavity Electromagnonics by Jing Xu et al, Argonne Lab and University of Chicago, October 2020 (9 pages).

211 See Dissipation-Based Quantum Sensing of Magnons with a Superconducting Qubit by S. P. Wolsk et al, University of Tokyo, September 2020 (6 pages).

212 See Topological Magnons: A Review by Paul McClarty, 2021 (21 pages).

213 See Nonlinear magnon control of atomic spin defects in scalable quantum devices by Mauricio Bejarano et al, August 2022 (17 pages).

214 I found these insights on skyrmions in the presentation Introduction to Contemporary Quantum Matter Physics Lecture 11: Skyrmions I by Marc Janoschek and Johan Chang, 2021 (26 slides) and Part II (24 slides). See also the review paper The 2020 skyrmionics roadmap by C Back et al, 2020 (38 pages).

215 See Thermodynamically stable "vortices" in magnetically ordered crystals. The mixed state of magnets by A. N. Bogdanov and D. A. Yablonskii, 1989 (3 pages).

216 See Skyrmion Lattice in a Chiral Magnet by S. Mühlbauer et al, Science, 2009 (44 pages) which also mentions hedgehogs or instantons, composed of two merons. An endless story. These skyrmions are observed at a critical temperature of 29.5K. And Instantons: thick-wall approximation by V. F. Mukhanov and A.S. Sorin, June 2022 (12 pages).

217 See Real-space observation of a two-dimensional skyrmion crystal by X. Z. Yu et al, 2010, Nature (5 pages).

This could lead to the creation of new ultra-high-density memories[218] particularly with the room-temperature Néel skyrmions that can be made with thin-film systems[219], to in-memory processing architectures[220], to create QRNGs[221], in low-power spintronic applications[222] and in a new breed of qubits with skyrmions in magnetic nano disks bounded by electrical contacts, where static electric and magnetic fields control the skyrmions quantized energy levels corresponding to their helicity. You may probably then need to find a way to entangle them[223]!

### Topological matter

The very concept of topological quantum states leading to topological matter was discovered with a specific insulating phenomenon that can be explained by the **quantum Hall effect**, with electrons moving through a strong magnetic field and accumulating in some parts of the material depending on its shape. This electron conductivity is quantized, as discovered in 1980 by **Klaus von Klitzing** (Germany) who was awarded the Nobel Prize in Physics in 1985. This "integer" quantum Hall effect was later completed by the discovery of the fractional quantum Hall effect by Tsui et al. in 1982 in two-dimensional electron systems in semiconductor devices, followed by the theoretical discovery of the entangled gapped quantum spin-liquid state of integer-spin "quantum spin chains" by **Frederick Duncan** and **Michael Haldane** in 1981, who was awarded the Nobel prize in physics in 2016 along with **David J. Thouless** and **J. Michael Kosterlitz**[224].

In 2005, **Eugene Mele** and **Charles Kane** predicted that topological insulation could happen in graphene sheet submitted to strong spin-orbit coupling creating the quantum Hall effect without any applied magnetic field[225]. This phenomenon is named the "quantum spin Hall effect" and relates to the Kane-Mele invariant[226]. It was demonstrated to occur in wafers of mercury telluride. It was experimented by **Shou-Cheng Zhang** et al from Stanford University in 2007[227]. The same year, the first 3D topological insulator was discovered by **Zahid Hasan** from Princeton[228].

[218] See for example Skyrmion-Electronics: Writing, Deleting, Reading and Processing Magnetic Skyrmions Toward Spintronic Applications by Xichao Zhang et al, 2019 (80 pages).

[219] See Mobile Néel skyrmions at room temperature: status and future by Wanjun Jiang et al, 2016 (15 pages) and Observation of Robust Néel Skyrmions in Metallic PtMnGa by Abhay K. Srivastava et al, Advanced Materials, December 2019 (5 pages).

[220] See Skyrmion Logic-In-Memory Architecture for Maximum/Minimum Search by Luca Gnoli et al, January 2021 (15 pages) and Robust and programmable logic-in-memory devices exploiting skyrmion confinement and channeling using local energy barriers by Naveen Sisodia et al, May 2022, UGA, CNRS and CEA (11 pages).

[221] See Single skyrmion true random number generator using local dynamics and interaction between skyrmions by Kang Wang et al, Nature Communications, 2022 (8 pages).

[222] See The skyrmion switch: turning magnetic skyrmion bubbles on and off with an electric field by Marine Schott et al, CNRS Institut Néel, UGA and CEA IRIG, 2016 (31 pages).

[223] See Skyrmion qubits: A new class of quantum logic elements based on nanoscale magnetization by Christina Psaroudaki and Christos Panagopoulos, Caltech and NTU Singapore, PRL, August 2021 (11 pages) and also Universal quantum computation based on nanoscale skyrmion helicity qubits in frustrated magnets by Jing Xia et al, April 2022 (7 pages).

[224] See Topological Quantum Matter by F. Duncan M. Haldane, Nobel Lecture, December 2016 (23 pages).

[225] See Quantum spin Hall effect in graphene by Charles Kane and Eugene Mele, University of Pennsylvania, 2005 (4 pages).

[226] See Topological Insulators and the Kane-Mele Invariant: Obstruction and Localisation Theory by Severin Bunk and Richard J. Szabo, 2019 (81 pages) and Quantum spin Hall effect: a brief introduction (34 slides).

[227] See Quantum Spin Hall Insulator State in HgTe Quantum Wells by Markus Koenig, Shou-Cheng Zhang et al, October 2007 (16 pages).

[228] See A topological Dirac insulator in a quantum spin Hall phase (experimental realization of a 3D Topological Insulator) by D. Hsieh, Zahid Hasan et al, Princeton University, 2009 (12 pages).

Since then, over 20 topological insulators materials were discovered and there are probably hundreds of them[229]. A French American research team devised in 2020 a machine learning model to detect such topological insulators out of an initial database of 4009 candidates[230]. Again, spintronics are a potential use case of topological insulators to create power-saving electronics where the on/off of a bit would be an electron spin instead of the on/off path of an electron stream.

In topology, an invariant can be described by a single winding number which describes the type of structure with its domain walls, vortices and vector order. It related to the **Chern number**. This number changes over quantum phase transitions. These are other various physics concepts to consider, way beyond what I can do at this point in my quantum journey[231].

It is interesting to note that some materials can showcase 3D topological behavior at ambient temperature, like bismuth-selenide ($Bi_2Se_3$). It is a semiconductor and a thermoelectric material that has a topological insulator ground-state. It could be used in targeted cancer treatments and X-ray to mammography[232]. You can also potentially build **magnetic monopoles** quasiparticles, breaking the convention that magnetism always shows up with dipoles[233].

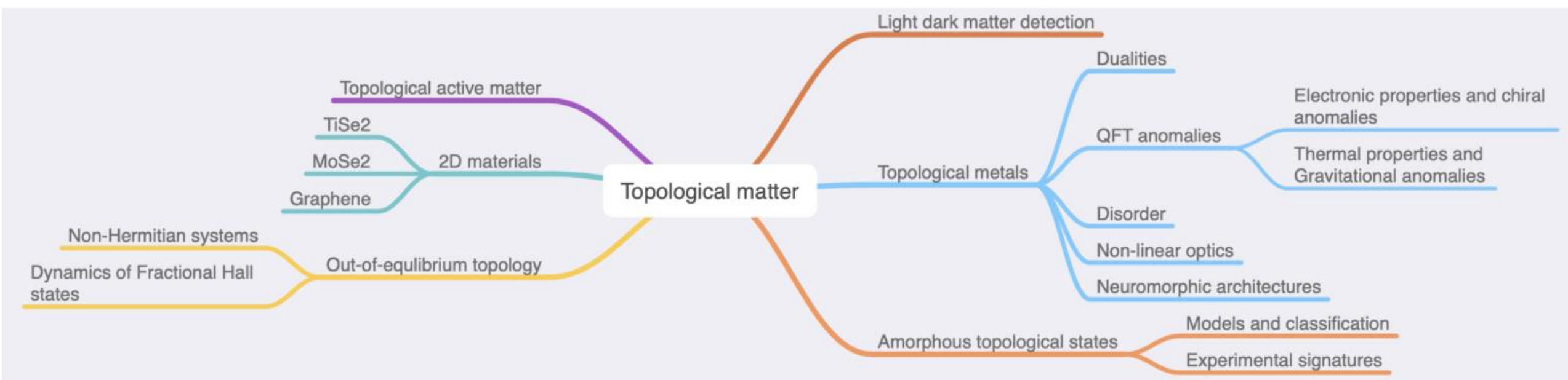

*Figure 113: a classification of topological matter. Source: Research Lines - Theory of Topological Matter by Adolfo Grushin, CNRS.*

Like me, you're certainly willing to "visualize" the different types of topological materials identified. I found this nice and highly detailed table showing their great diversity in a review paper, below in Figure 114.

Topological matter can have several applications related to light-matter interactions in the Terahertz regime. It can help create waveguides, optical isolators and diodes who are more resistant to their environment perturbations in the recent field of **topological photonics** which is related to polaritons[234].

---

[229] See Topological phases of amorphous matter by Adolfo G. Grushin, January 2021 (45 pages) which describes the physics of topological phases and Introduction to topological Phases in Condensed Matter by Adolfo G. Grushin (28 pages) which provides some background information on the way to classify topological matter.

[230] See Detection of Topological Materials with Machine Learning by Nikolas Claussen et al, ENS Paris, Princeton, June 2020 (15 pages).

[231] See Topological Materials : Some Basic Concepts by Ion Garate, 2016 (35 slides), Core Concept: Topological insulators promise computing advances, insights into matter itself by Stephen Ornes, 2016 and Topological phases by Nicholas Read, Physics Today, 2012 (6 pages).

[232] See Topological insulator bismuth selenide as a theranostic platform for simultaneous cancer imaging and therapy by Juan Li and al, 2013 (7 pages).

[233] See Emergent magnetic monopoles isolated using quantum-annealing computer by Los Alamos National Laboratory, Physorg, July 2021, which refers to Qubit spin ice by Andrew D. King, Science, July 2021 (18 pages) which simulates a new topological material with a D-Wave quantum annealer.

[234] See Roadmap on Topological Photonics by Hannah Price et al, Journal of Physics, 2022 (63 pages), the well illustrated presentation Introduction to Topological Photonics by Mikael C. Rechtsman, Penn State, AMOLF Nanophotonics Summer School, June 2019 (42 slides), Topological photonic crystals: a review by Hongfei Wang et al, 2020 (23 pages) and Topological photonic crystals: physics, designs and applications by Guo-Jing Tang et al, January 2022 (60 pages).

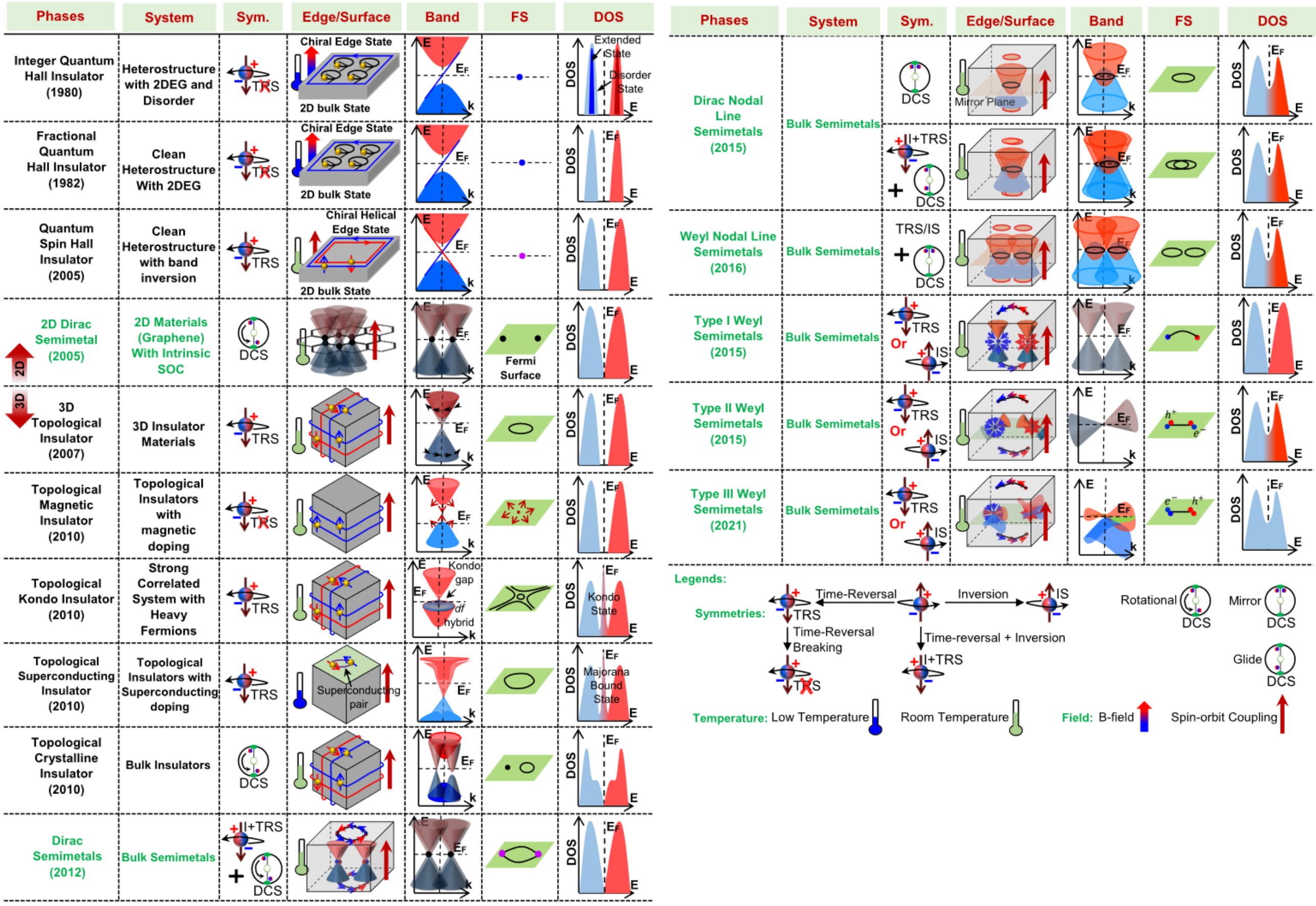

*Figure 114: a table with a classification of various topological materials in 2D and 3D, and indicating time reversal and operating temperature. Source: Topological Quantum Matter to Topological Phase Conversion: Fundamentals, Materials, Physical Systems for Phase Conversions, and Device Applications by Md Mobarak Hossain Polash et al, February 2021 (83 pages).*

We even have **topological lasers**[235], which can for example consolidate multiple sources in a coherent way, leading to even more powerful lasers, using a topological insulator vertical-cavity surface-emitting array (VCSEL)[236].

Then of course, one key application of topological matter is topological qubits, often associated with Majorana fermions sought after by Microsoft. But topological qubits are way more diverse with many competing definitions and architectures. For example, you also can count with Fibonacci anyons[237].

### Time crystals

Time crystals is a beast we hear a lot about since mid-2021, when Google announced it had created such thing in its Sycamore processor[238]. It shed some light on this weird phenomenon that was devised in a 2012 paper by Frank Wilczek from the MIT (and 2004 Nobel prize in physics) and by another paper by him and Alfred Shapere from the University of Kentucky[239].

---

[235] See Topological lasing, PhLAM Laboratory, Lille France.

[236] See Topological-cavity surface-emitting laser by Lechen Yang et al, Nature Photonics, 2021 (6 pages) and Topological insulator vertical-cavity laser array by Alex Dikopoltsev et al, Science, 2021 (5 pages).

[237] See Fibonacci Anyons Versus Majorana Fermions: A Monte Carlo Approach to the Compilation of Braid Circuits in $SU(2)_k$ Anyon Models by Emil Génetay Johansen and Tapio Simula, 2021 (23 pages).

[238] See Eternal Change for No Energy: A Time Crystal Finally Made Real by Natalie Wolchover, July 2021 referring to Observation of Time-Crystalline Eigenstate Order on a Quantum Processor by Xiao Mi et al, Google, July 2021 (24 pages) and Realizing topologically ordered states on a quantum processor by K. J. Satzinger et al, Google AI, April 2021 (6 pages).

[239] See Quantum Time Crystals by Frank Wilczek, MIT, 2012 (6 pages) and Classical Time Crystals by Alfred Shapere and Frank Wilczek, PRL, 2012 (5 pages).

This thing is somewhat linked to the history of the search for a perpetuum mobility, an isolated object supposed to keep in motion indefinitely. It was dismissed by the French Academy of Science in 1775 due to the limits of friction and, later, to the second law of thermodynamics[240].

In classical crystals, the atoms are periodically arranged in space structured according to one of the 230 structured already described. In time crystals, these atoms are periodically arranged in both space and time. It simply means that their structure is in a permanent oscillating mode with a given period, for so-called discrete time crystals[241].

But the scientific description of the phenomenon is the less explicit "spontaneous time symmetry breaking". Then, you quickly lose grounds with common wisdom[242].

Time crystals do not lose energy to the environment. They are the stage of motion without energy. It is a type or phase of non-equilibrium matter. But they are still initially driven, sometimes even out of their equilibrium level. Some real time crystals were first observed in lab experiments, starting in 2017 with some constantly rotating ring of charged ions spin (which by the way, shows some signal damping, in Figure 115)[243]. It can also happen with some continuous change of spin for some particles, when the change periods is up to 100 times longer than the system drive period. It was tested in 2021 by a QuTech team in The Netherlands using controllable $^{13}C$ nuclear spins in diamond structures[244].

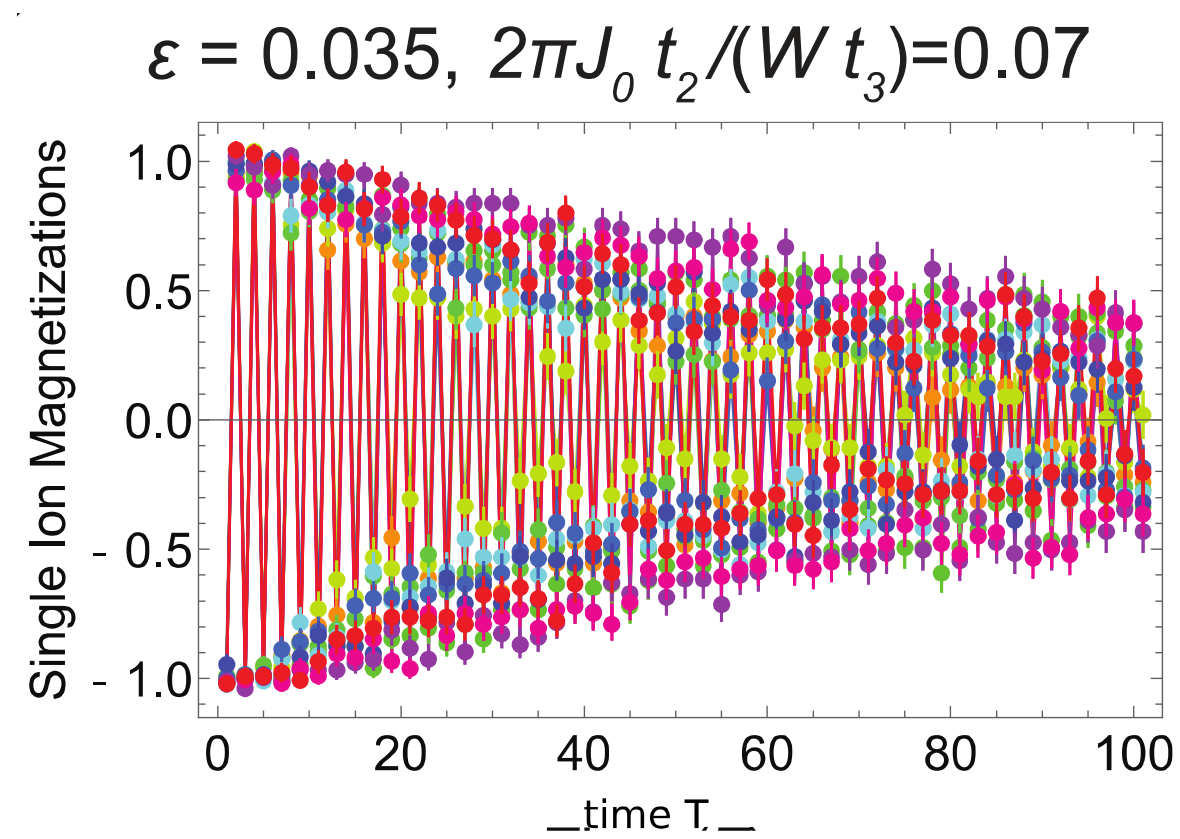

*Figure 115: time crystal oscillations over time. Source: Observation of a Discrete Time Crystal by J. Zhang, Christopher Monroe et al, September 2016 (9 pages).*

So why all this fuss around time crystals and how could they become useful? Some think they may be useful to create some form of quantum memory.

Things get complicated when you learn that time crystals have also been experimented with superconducting qubits like with the Google 2021 experiments and other subsequent ones with a continuous line of 57 qubits in a 65 qubits IBM QPU[245]. How could a series of connected superconducting qubits become a "crystal" per se?

They may behave as a continuously oscillating system but are not a single crystal since they are a complex assembly of Josephson junctions, capacitances, resonators and microwave drives mixing various elements (aluminum, aluminum-oxide, niobium, titanium…).

### Quantum batteries

Quantum matter research is leading some labs to investigate the possibility of creating innovative batteries for energy storage relying on some quantum phenomenon including entanglement.

---

240 See A Decade of Time Crystals: Quo Vadis? by Peter Hannaford and Krzysztof Sacha, April 2022 (8 pages) and A Brief History of Time Crystals by Vedika Khemani et al, Harvard, October 2019 (79 pages).

241 There are also continuous time crystals that were observed first in 2022 in Germany. See Observation of a continuous time crystal by Phatthamon Kongkhambut et al, February-August 2022 (13 pages).

242 There's even an acronym for this, TTSB which means time translation symmetry breaking.

243 See Observation of a Discrete Time Crystal by J. Zhang, Christopher Monroe et al, September 2016 (9 pages).

244 See Many-body–localized discrete time crystal with a programmable spin-based quantum simulator by J. Randall et al, Qutech, Science, November 2021 (7 pages).

245 See Realization of a discrete time crystal on 57 qubits of a quantum computer by Philipp Frey and Stephan Rachel, January 2022 (12 pages).

Work in this field started around 2012 with some fundamental research by Robert Alicki and Mark Fannes from Poland and Belgium on how much work could be stored and extracted from quantum batteries[246]. Quantum batteries could store energy in high energy states of quantum objects and extracted efficiently. Some of these batteries rely on various quantum principles, some of them being not far from classical quantum photonics. This is a different field than classical batteries whose design could be improved with using quantum computers, as covered page 703 in this book.

All the papers I've found in that field are very theoretical and quite far from practical batteries. The main benefit of these quantum batteries seems to be fast charging, with the caveat of fast discharging, which is quite inconvenient[247]. I have not found yet any quantum battery that would improve energy density in a real documented manner with a full-stack product packaging, one of the main showstoppers for various use cases like for long distance electric vehicles or aerial vehicles. So, you're far from buying your next Tesla equipped with a 1000-mile range quantum battery[248].

So, what do we have in-store here? Mainly scientific work with very low TRLs.

Scientists from Australia and Italy are working on an **organic battery** with fast charging using a process called superextensive scaling of absorption, meaning that the larger the system is, the faster it absorbs energy[249]. It's based on of a thin active layer of a low-mass molecular semiconductor named LFO (Lumogen F-orange) that is dispersed into a polymer matrix that is sandwiched between two dielectrics made of 8 and 10 pairs of Brag mirrors, creating a microcavity. The battery cell is then controlled by a laser in the 500 nm red-light range, a noncollinear optical parametric amplifier, beam splitters and delay lines and a detector. In a word, we could say it's a "light" battery, absorbing energy as light, and rendering it as light, in a different wavelength. Like in many other papers of this kind, it's quite difficult to infer the practicality of these quantum batteries.

If researchers are not overselling it, the news media are doing it, touting "*batteries with one million miles autonomy*"[250].

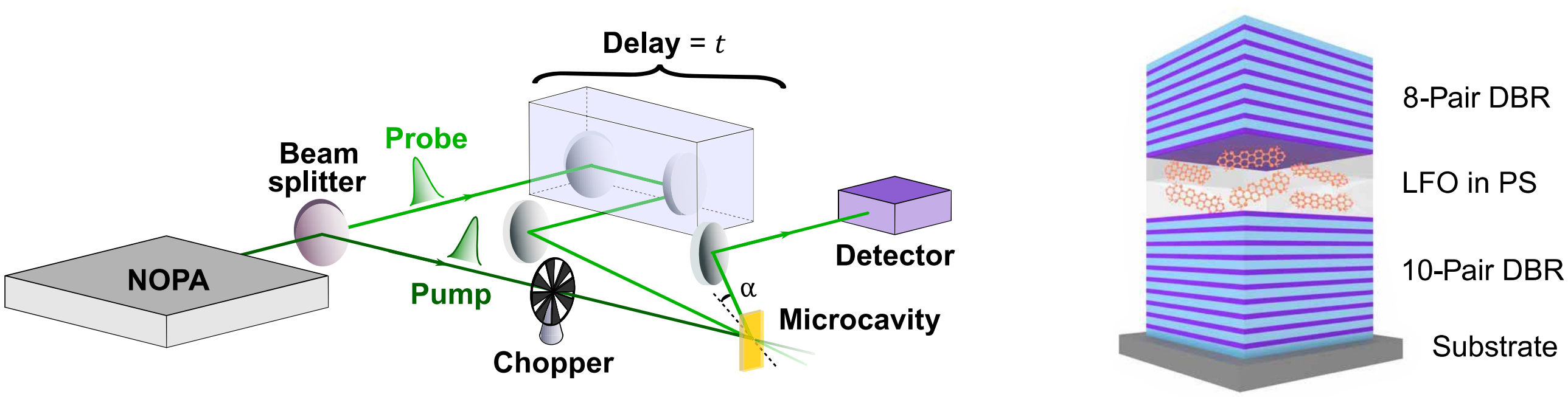

*Figure 116: source: Superabsorption in an organic microcavity: Toward a quantum battery by James Q. Quach et al, Heriot-Watt University, 2022 (9 pages).*

[246] See Extractable work from ensembles of quantum batteries. Entanglement helps by Robert Alicki and Mark Fannes, Physical Review E, November 2012 (4 pages).

[247] See Sizing Up the Potential of Quantum Batteries by Sourav Bhattacharjee, Indian Institute of Technology, April 2022.

[248] Despite what you can read in Quantum technology could make charging electric cars as fast as pumping gas by Institute for Basic Science, March 2022 that is linked to Quantum charging advantage cannot be extensive without global operations by J.-Y. Gyhm et al, PRL, April 2022 (13 pages).

[249] See Superabsorption in an organic microcavity: Toward a quantum battery by James Q. Quach et al, Heriot-Watt University, 2022 (9 pages).

[250] See How quantum batteries could lead to EVs that go a million miles between charges, The Next Web, June 2022.

This comes from another paper, authored by Canadian scientists and an engineer from Tesla which proposes an improved Li-Ion battery that could last 1.5 million miles over its lifespan but, of course, not with a single recharge[251]. And it's even not a quantum battery.

In another approach, other scientists from Australia are looking at ways to store energy in light-induced spin state trapping in spin crossover materials[252]. And a team from Italy and Korea wants to use micromasers to store energy[253].

Another paper from a Korean American German Singaporean team describes quantum batteries as isolated quantum systems undergoing unitary charging protocols (unitary in the mathematical sense)[254]. With ensembles of such batteries, some collective effects enhance work extraction or boost the charging power thanks to entanglement between the component quantum batteries. The described system is based on an Otto engine which can serve as an engine and as a refrigerator.

In another work from US and Japanese researchers, we are closer to classical battery designs. It's about using lithium-dopped samarium nickelate, a quantum crystalline material with strongly correlated electron systems[255].

Lithium ions are usually the main compound of batteries electrolytes.

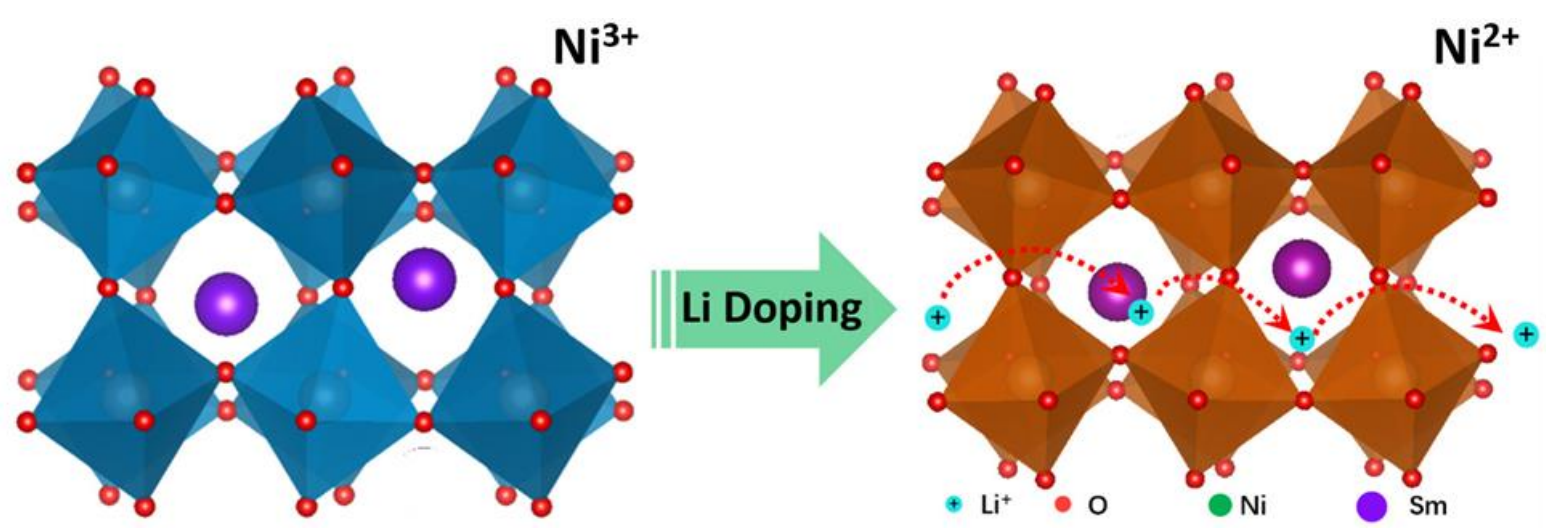

*Figure 117: lithium-dopped samarium nickelate quantum battery. Source: Strongly correlated perovskite lithium ion shuttles by Yifei Sun et al, 2018 (6 pages).*

The quantum crystal structure improves the conduction of these ions that could also be sodium ions. It could enable better electrolytes but another effect of the structure where additional electron modifies the material conductivity could be used in neuromorphic synapses for storing neural networks connections weights.

Other research work deal with microscopic batteries which don't seem to be useful for energy storage[256]. They can help better understand the thermodynamics of qubits manipulation and provide innovative insights on how to fight decoherence and noise[257].

Higher TRLs can be found with rather classical batteries that would use topological semi-metallic porous carbon materials as potential more efficient anodes for Li-Ion, sodium-ion and potassium-ion batteries. Other topological materials could be useful for supercapacitors.

---

[251] See A Wide Range of Testing Results on an Excellent Lithium-Ion Cell Chemistry to be used as Benchmarks for New Battery Technologies by Jessie E. Harlow, J.R. Dahn et al, 2019 (15 pages).

[252] See UQ discovery paves the way for faster computers, longer-lasting batteries, June 2022 referring to Toward High-Temperature Light-Induced Spin-State Trapping in Spin-Crossover Materials: The Interplay of Collective and Molecular Effects by M. Nadeem, Jace Cruddas, Gian Ruzzi and Benjamin J. Powell, May 2022 (55 pages). A similar spin-based approach is described in Quantum advantage in charging cavity and spin batteries by repeated interactions by Raffaele Salvia et al, April 2022 (14 pages).

[253] See Micromasers as Quantum Batteries by Vahid Shaghaghi et al, April 2022 (6 pages).

[254] See Charging Quantum Batteries via Otto machines: The influence of monitoring by Jeongrak Son et al, May 2022 (16 pages). Hard to understand what are the characteristics of this kind of battery and how it performs compared to classical Li-ion batteries!

[255] See Quantum material is promising 'ion conductor' for research, new technologies by Emil Venere, Physorg, 2018. Pointing to Strongly correlated perovskite lithium ion shuttles by Yifei Sun et al, 2018 (6 pages).

[256] Like with IBM Quantum Platforms: A Quantum Battery Perspective by Giulia Gemme et al, April 2022 (13 pages) which is using an IBM superconducting processor to store energy in qubits. It's actually using the Armonk processor which has exactly one qubit. A similar experiment done in China is described in Optimal charging of a superconducting quantum battery by Chang-Kang Hu et al, August 2021 (4 pages).

[257] Like with Coherence-powered work exchanges between a solid-state qubit and light fields by Ilse Maillette De Buy Wenniger, Maria Maffei, Niccolo Somaschi, Alexia Auffèves, Pascale Senellart et al, April 2022 (17 pages).

Topological materials could also be useful to create more efficient catalyzers for water electrolysis, with the production of hydrogen in sight coming from renewable originated electricity[258].

# Extreme quantum

Beyond the basics of quantum physics, many other branches of quantum physics deserve to be examined in this book. They can have various impacts on quantum technologies, noticeably on quantum sensing. They are also used in cosmology. Finally, they are unfortunately used by many false sciences and scams that we will discuss in the section dedicated to quantum hoaxes, starting page 1015.

### Quantum field theory

Quantum Field Theory (QFT[259]) is a branch of quantum physics that deals with the physics of elementary particles in the relativistic realm, including their creation or disappearance during various interactions, such as electron and positron pairs. These phenomena are generally reproduced in particle accelerators[260].

QFT also covers the mechanisms of condensed matter such as Bose-Einstein condensates or superfluid helium and more generally, the behavior of quasiparticles, complex collective behaviors such as Cooper's (electron) pairs in superconducting materials.

QFT combines elements of quantum mechanics, special relativity, and classical notions of electromagnetic fields. It is based on a mathematical formalism that is even more difficult to assimilate than the one of non-relativistic quantum physics.

It exploits the notion of Lagrangian and Lagrangian integrals over time describing the evolution of fields and the interactions between the fields of several particles.

QFT is used to explain or modelize the fine structure of the hydrogen atom (corresponding to close spectral lines not explainable by classical quantum energy jumps), the existence of particle spin (which explains these spectral lines), the spontaneous emission of photons by atoms during their return to their fundamental state and the mechanisms of radioactivity.

The foundations of QFT were created by many scientists starting in 1928: **Paul Dirac**, **Wolfgang Pauli**, **Vladimir Fock** (1898-1974, Russian), **Shin'ichirō Tomonaga (**1906-1979, Japanese), **Julian Schwinger (**1918-1994, American), **Richard Feynman** and **Freeman John Dyson (**1923-2020, American[261]). Shin'ichirō Tomonaga, Julian Schwinger and Richard Feynman received the 1965 Nobel Prize in Physics for their work on quantum electrodynamics which is part of QFT.

In the early 1950s, they solved the problem of infinite energy values generated by the initial QFT models by using an adjustment technique called **renormalization**.

Physicists are still struggling to integrate the theory of general relativity into the QFT, preventing it from becoming a "theory of the whole" or unified theory explaining all known physical phenomena in the Universe.

QFT operates in three main areas:

- In the physics of **high-energy particles** explored in particle accelerators such as the CERN LHC. It has been supplemented on this point by the standard model that we will see below.

[258] See Topological quantum materials for energy conversion and storage by Huixia Luo, Peifeng Yu, Guowei Li and Kai Yan, Nature Review Physics, July 2022 (14 pages).

[259] Later on, we'll use the QFT acronym with another meaning, Quantum Fourier Transform!

[260] See The History of QFT, a Stanford site, which summarizes the history of QFT.

[261] It also gave rise to the notion of the Dyson sphere, which dimensions the level of technological control of energy sources by extraterrestrial civilizations, with a sphere capturing the totality of a star's energy.

- In the **physics of condensed matter** with superconductivity, superfluidity and the quantum Hall effect. This is the framework of **QED** (quantum electrodynamics), launched by Paul Dirac in 1928, which studies in particular the production of positrons and positron/electron interactions (attraction, annihilation, pair creation, Compton effect). The **CQED** (cavity QED) sub-branch studies the relations between matter and photons in optical cavities. It is used by condensed matter physicists working on superconducting qubits.
- In **cosmology** to contribute to modeling the origin and evolution of the Universe as well as certain mechanisms of interaction between black holes and quantum fields.

### Quantum vacuum fluctuation

One of the consequences of QFT is the notion of quantum vacuum fluctuation, also called vacuum energy. Based on Heisenberg's principle of indeterminacy that quantum fields are in perpetual fluctuation, QFT models zero-point fluctuations or vacuum energy, which is the minimum energy level of quantum systems.

In this framework, Heisenberg's principle can be considered as a generalized predicate. According to these models, total vacuum cannot exist. Elementary fluctuations lead to spontaneous electromagnetic waves creation, given all fields are fluctuating.

One scenario devised by Paul Dirac is the creation of pairs of virtual electron and positron particles, which rapidly annihilate each other, generating photons in the process. But this is not the only solution to his equations. It can come from electromagnetic fields moving at the speed of light.

Under the influence of a surrounding electromagnetic field, this leads to a polarization of the vacuum. The latter even leads to make the vacuum birefringent, its refractive index depending on the polarization of the light that gets through it. The phenomenon is however potentially observable only with some very intense electromagnetic field.

Theoretical models initially indicated that this vacuum energy would be infinite on the scale of the Universe. They were then corrected using the renormalization method, already mentioned above. These elementary vacuum fluctuations would explain the spontaneous emission of radiation by the electrons in the atoms as well as the spontaneous radioactivity[262].

The concept of vacuum energy originated with **Max Planck** in 1911 when he published an article containing an energy equation for a medium containing a fixed constant, a kind of energy floor for this medium, without being able to interpret it. It was not until 1916 that the chemist **Walther Nernst** (1864-1941, German[263]) interpreted this constant as the energy level of the vacuum in the absence of any radiation. It happens when you cool down a black body to a very low temperature, below a couple millikelvins (mK).

According to the QFT, the Universe is a vast soup containing constantly fluctuating fields, both fermions (leptons and quarks) and bosons (force fields like gluons mediate the strong force that stick together the quarks that are the elementary constituents of protons and protons, and photons, and the cohesion between nucleons is coming from a residual force from strong interactions). This notion of minimum energy level is a modern version of the notion of ether - a not completely empty void - which dominated 19th century physics, notably for James Clerk Maxwell. The electromagnetic bath in which the vacuum is immersed, supplemented by the energy of the vacuum, would give vacuum some viscosity properties.

---

[262] In addition to these elementary fluctuations, vacuum is constantly traversed, even in the remotest regions of space, by electromagnetic waves, not to mention the effects of gravitation. The Universe is thus filled with radiations including the cosmological background noise which is a remnant of the big bang, having a temperature of 2.7K. It is the same in a vacuum-packed box because all matter emits radiation.

[263] Walther Nernst played a key role in launching the Solvay Congresses from 1911 onwards.

Still, these theories are less complete than classical quantum mechanics. One of the solutions is to assume that fermions have a negative vacuum energy and bosons have a positive vacuum energy, both balancing each other. But this has not been demonstrated experimentally, particularly with non-relativistic energy particles.

Some link could be found between vacuum energy and the dark energy of the Universe as well as gravity[264]. This is very speculative. It could help explain the 73% of the energy contained in the Universe, sometimes called dark energy. Its density is very low, at $10^{-13}$ Joules/cm$^2$.

There are different ways to verify the existence of quantum vacuum fluctuations. The best-known is related to the Casimir effect that we will study in the next part. Recently, French and German scientists have also managed to interact with this quantum vacuum fluctuation in a semiconductor[265].

### Casimir effect

The physicist **Hendrik Casimir** (1909-2000, Dutch) predicted in 1948 the existence of an attractive force between two parallel electrically conductive and uncharged plates[266]. He obtained his PhD in 1931 at the University of Leiden in the Netherlands. He also visited Niels Bohr in Copenhagen and was a research assistant to Wolfgang Pauli in 1938. The Casimir effect is interpreted as being related to the existence of quantum vacuum energy. The experiment imagined by Casimir uses parallel mirrored metal surfaces that are as perfectly flat as possible. They create a Fabry-Perot cavity similar to the one that used in lasers.

The Casimir effect is commonly attributed to quantum fluctuations in vacuum. Temporary changes in the energy level at points in the space between the two mirrors would spontaneously generate pairs of very short-lived particles and antiparticles and photons associated with their annihilation. These vacuum fluctuations take place in and out of the volume of the cavity.

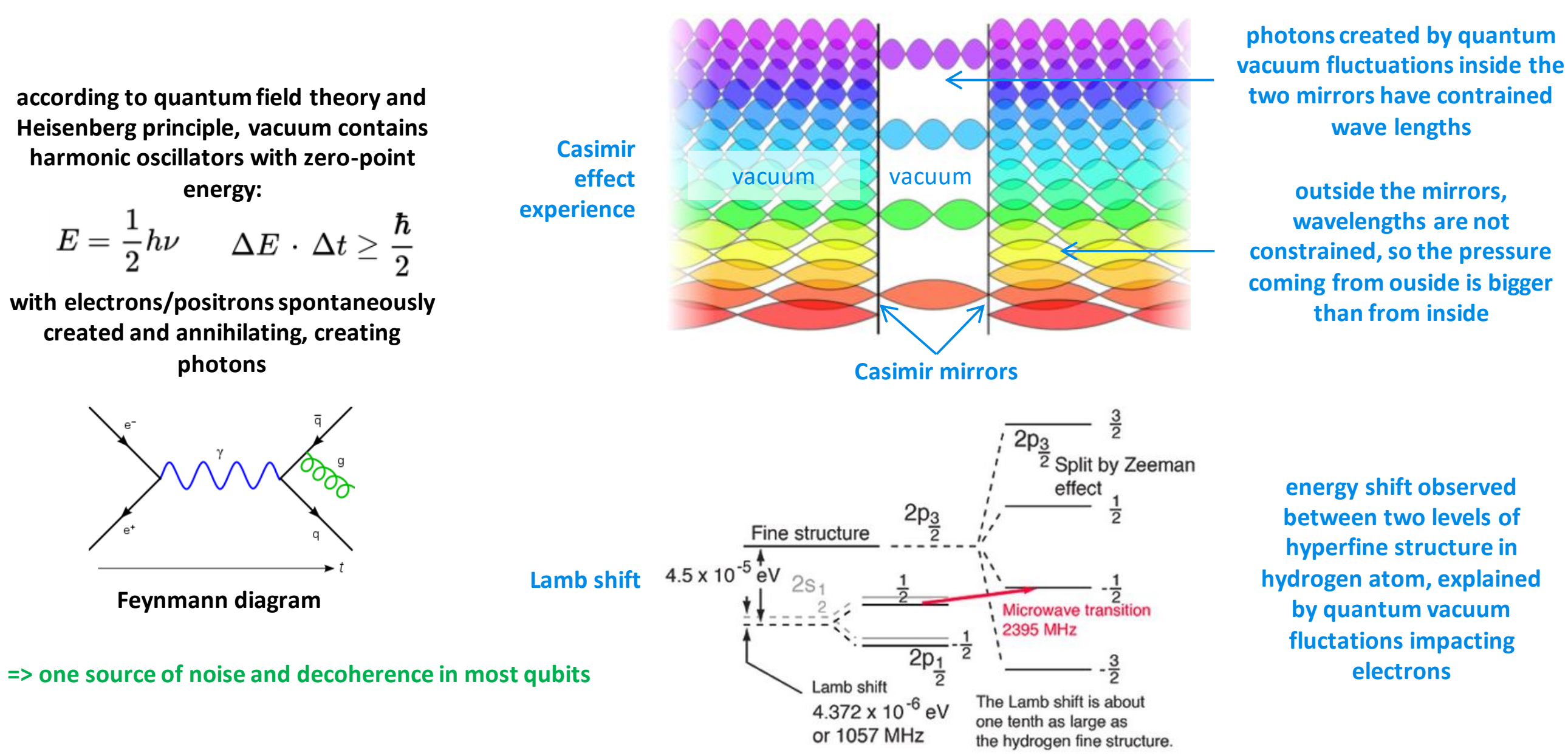

*Figure 118: vacuum fluctuations measurement. Sources: The Lamb Shift and The Casimir Effect by Kyle Kingsbury, 2014 (82 slides).*

---

[264] See Casimir cosmology by Ulf Leonhardt, February 2022 (41 pages).

[265] See Understanding vacuum fluctuations in space, August 2020 and Electric field correlation measurements on the electromagnetic vacuum state by Ileana-Cristina Benea-Chelmus, Jérôme Faist et al, 2018/2020.

[266] See On the attraction between two perfectly conducting plates by Hendrik Casimir, 1948 (3 pages) and Electromagnetic vacuum fluctuations, Casimir and Van der Waals forces by Cyriaque Genet, Astrid Lambrecht et al, 2004 (18 pages).

Because of the interference effect induced by the cavity, fluctuations at certain frequencies are reduced. The density of electromagnetic energy in the cavity is thus lower than the density of energy outside the cavity as shown above in Figure 118 [267]. These are spontaneous quantum fluctuations.

The effect cannot be explained by the simple pressure that is higher on the outside than the pressure between the two plates. In detail, the wavelengths of the photons generated by the vacuum outside the plates can be of any size and especially long while inside the plates, these wavelengths are constrained by the distance between the plates and can only be 1/n of this distance.

The spontaneous electromagnetic spectrum of the vacuum is therefore wider outside the plates than inside, creating a stronger pressure inside than inside, which therefore tends to make the plates move closer together, but very slightly.

For two parallel mirrors of surface A and a distance L between the two mirrors, the force of attraction between the two mirrors follows the formula on the right. In practice, L is between 0.2 µm and 5 µm and is usually 1 µm. This is a "macroscopic" scale.

$$F_{C_{as}} = \frac{\hbar c \pi^2 A}{240 L^4}$$

According to Heisenberg's principle, which is used to explain the effect, energy and time can be linked by the formula on the right. It shows indirectly that during a very short time, a small amount of energy can be created.

$$\Delta E . \Delta t \geq \frac{\hbar}{2}$$

The macroscopic accumulation of these operations is annihilated, making it possible to avoid a violation of the energy conservation principle. So, be uber-skeptic when hearing anyone claiming they can harvest energy from vacuum to produce free electricity.

The experiments are not necessarily 100% conclusive and the data generated do not fit perfectly with the models unlike many classical quantum mechanics experiments. The reason for this is that it is difficult to obtain perfect surfaces.

The first experiments validating the Casimir effect were carried out almost 50 years after the definition of this effect[268]. The first one is that of **Steve Lamoreaux** (American) in 1996, using parallel plates.

His measurement gave a result that was 5% off the predictions. The precision instruments used then detected a force of one billionth of a Newton. The model was improved in other experiments carried out in 1998 and again in 2012 using an electrode geometry combining a plane and a polystyrene sphere with a diameter of 200 µm and covered with gold (diagrams *below*)[269]. The differences between the models and the measurements decreased to 1%, which remains significant in physics.

The Casimir effect could explain several other commonly observed physical phenomena such as the electron's abnormal magnetic moment and the Lamb shift. The first phenomenon describes a drift of this magnetic moment with respect to Dirac's equations.

The second comes from **Willis Eugene Lamb** (1913-2008, American), Nobel Prize in Physics in 1955, who had done his thesis under the supervision of Robert Oppenheimer. Lamb shift is an energy gap observed between two levels of fine structure of the hydrogen atom, two very close energy levels.

---

[267] See a good panorama of the Casimir effect with The Casimir effect and the physical vacuum by G. Takács, 2014 (111 slides). See also The Casimir Effect by Kyle Kingsbury, 2014 (82 slides) which describes well the experimental devices for the evaluation of the Casimir effect and evokes some cases of use in MEMS. And then Zero-Point Energy and Casimir Effect by Gerold Gründler, 2013 (47 pages), which casts the history of the Casimir effect, going back to Planck's work in 1911.

[268] The experimental difficulty consists in cancelling out all the other forces between the two plates and they are all much larger than the Casimir effect, particularly electrostatic and van der Waals forces.

[269] See Physicists solve Casimir conundrum by Hamish Johnston, 2012 which refers to Casimir Force and In Situ Surface Potential Measurements on Nanomembranes by Steve Lamoreaux et al, 2012 (6 pages).

The effect is explained with the perturbations coming from vacuum fluctuations and affecting the electron in these two neighboring energy levels, creating the spontaneous generation of photons that are rapidly absorbed by the electron.

The effect was discovered in 1947 by Willis Eugene Lamb and interpreted the same year by **Hans Bethe** (1906-2005, German) for the hydrogen spectrum using the idea of mass renormalization. It was used in the development of post-war quantum electrodynamics.

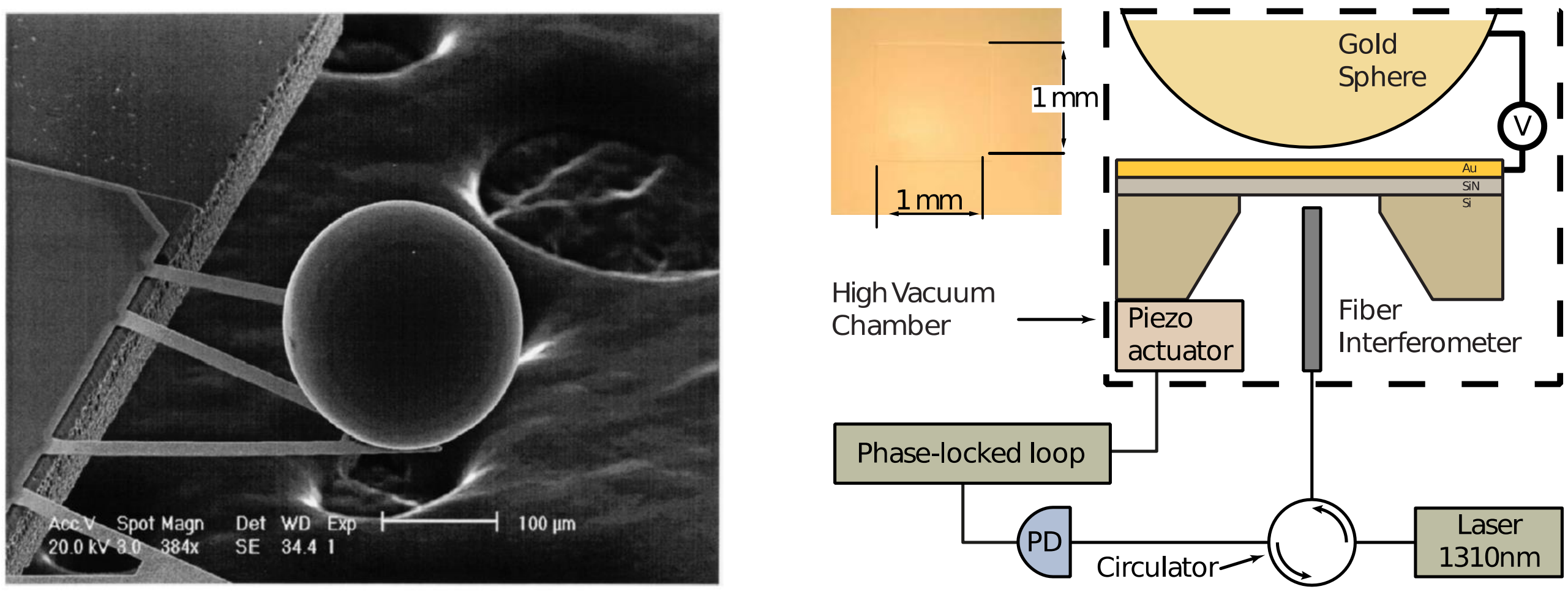

*Figure 119: vacuum source measurement with a dynamic Casimir effect. Sources: The Casimir Effect by Kyle Kingsbury, 2014 (82 slides) and Casimir Force and In Situ Surface Potential Measurements on Nanomembranes by Steve Lamoreaux et al, 2012 (6 pages).*

The polarization of vacuum explains part of this shift at 27 MHz for a total of 1057 MHz[270]. The calculation uses the fine-structure constant α (about 1/137) which describes the contribution of vacuum energy to the electron's anomalous magnetic moment. The α constant is also used to quantify the strength of the electromagnetic interaction between elementary charged particles.

There is also a **Dynamic Casimir Effect** (DCE), discovered by **Gerald Moore** in 1969. It generates pairs of particles by the movement of the mirrors used in the Casimir experiment[271].

As with the Casimir Effect, the energy observed is infinitesimal. For the energy to be significant, the mirrors would have to move at relativistic velocities, which is not very practical. And there is no problem with energy conservation, the necessary energy being provided by the mirror movement. The vacuum simply serves as a nonlinear medium!

The interpretation of the Casimir effect is still debated. Some physicists explain it by other mechanisms than vacuum energy.

They rely on the **van der Waals** (1837-1923, another Dutch) forces, where atoms attract or repel each other depending on their distances[272]. However, this infinitesimal force works at a microscopic scale, where the Casimir effect operates at a macroscopic scale.

---

[270] This phenomenon of vacuum polarization in the Lamb effect is described in The Vacuum Polarisation Contribution to the Lamb Shift Using Non-Relativistic Quantum Electrodynamics by Jonas Frafjord, 2016 (61 pages).

[271] See Electro-mechanical Casimir effect by Mikel Sanz, Enrique Solano et al, 2018 (10 pages).

[272] See The origin of Casimir effect: Vacuum energy or van der Waals force? by Hrvoje Nikolic, 2018 (41 slides) and the even more skeptic The Casimir-Effect: No Manifestation of Zero-Point Energy by Gerold Gründler, 2013 (15 pages) and All wrong with the Casimir effect by Astrid Karnassnigg, 2014 (3 pages). Then, The Casimir effect: a force from nothing by Astrid Lambrecht, 2007 (5 pages).

French physicists are quite active in the field, and, in particular **Astrid Lambrecht,** formerly director of the INP of the CNRS, the Institute of Physics which oversees the physics laboratories of the CNRS[273].

The Casimir effect could be of interest in quantum metrology to create sensors and in particular NEMS/MEMS.

These theories on quantum vacuum fluctuation and the Casimir effect are also fraudulently exploited by the creators of so-called machines capable of capturing vacuum energy, which collect nothing at all in practice. The fluctuation-dissipation theorem ensures that quantum vacuum fluctuations does not violate the second principle of thermodynamics. No energy can be recovered thanks to these fluctuations! Forget it.

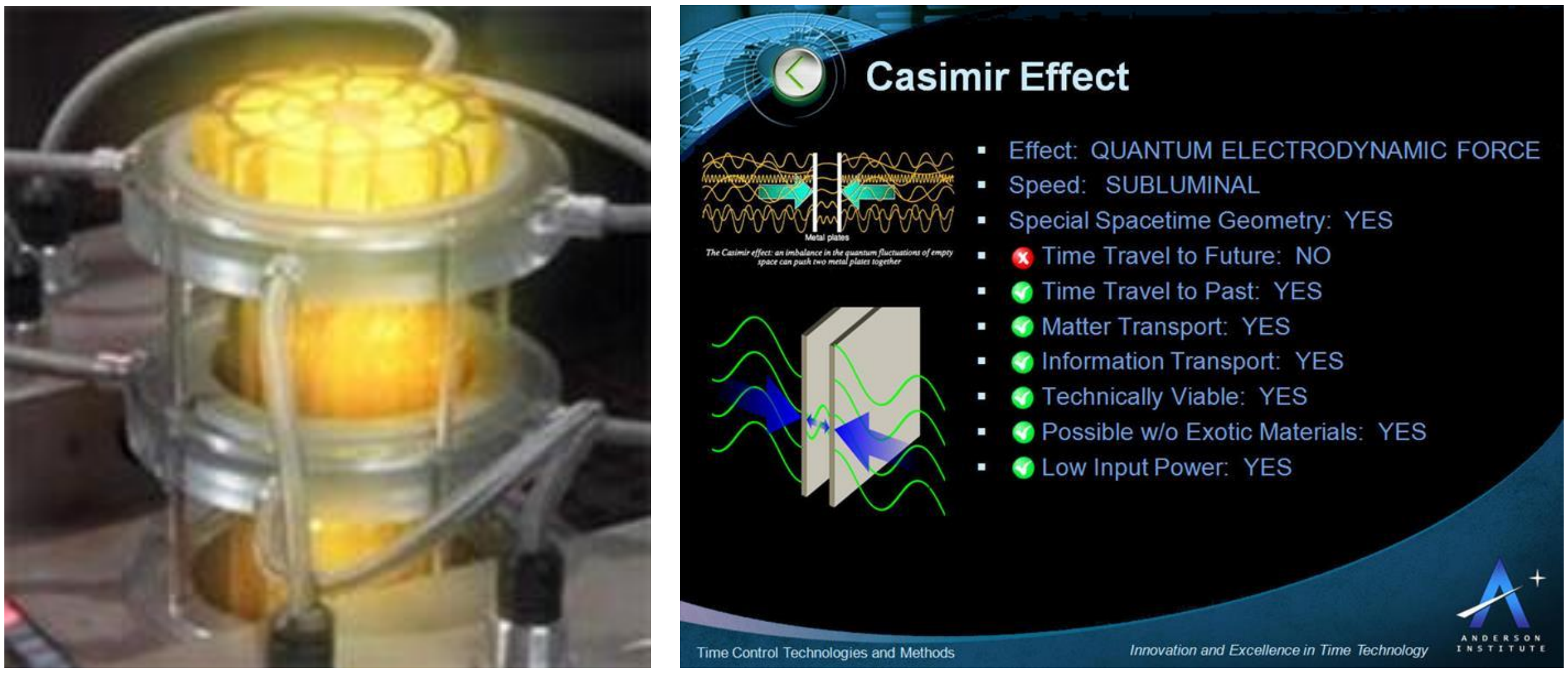

*Figure 120: Anderson Institute claims about using the Casimir effect.*

For example, you have a certain **David Lewis Anderson**, who started the **Anderson Institute** in 1990, who claims to be able to use the Casimir effect to travel back in time and create a "free" electricity generator[274].

In other cases, the Casimir effect is exploited in a scientific but borderline way to imagine science fiction scenarios like ways to cross wormholes[275].

The **NASA** even explored the idea to use sails and vacuum fluctuation to propel a space vessel between 1996 and 2002, to no avail. It was one of the ideas explored as part of the fancy Breakthrough Propulsion Physics Program, which was awarded a tiny budget of $1.2M and later cancelled.

---

[273] See The Casimir effect theories and experiments by Romain Guérout, Astrid Lambrecht and Serge Reynaud, LKB, 2010 (28 slides) and Casimir effect and short-range gravity tests, LKB, 2013 (15 slides). Astrid Lambrecht chaired the Casimir RNP group, which brought together researchers from around the world working on the Casimir effect. The group was active between 2009 and 2014.

[274] Its website seems to be inactive since 2012. See this radio interview from 2019 with the guy who defies the laws of bullshit in his talk. It shows how an interviewer lacking some scientific background can be fooled by a good talker. In See Is Time Travel Real? 2019 and the Anderson Institute website.

[275] See One Theory Beyond the Standard Model Could Allow Wormholes that You Could Actually Fly Through - Universe Today by Matt Williams, August 2020, mentioning Humanly traversable wormholes by Juan Maldacena and Alexey Milekhin, August 2020.

### Unifying theories

The quest of a **unified theory** has occupied many physicists for nearly a century. Its goal would be to consolidate all the physics theories and in particular, quantum physics, relativity and gravity into a single formalism. In addition to the QFT, a very large number of explanatory and unifying theories of physics have been developed.

No such theory is considered today as being complete. Here's a rough map showing how these different theories are related.

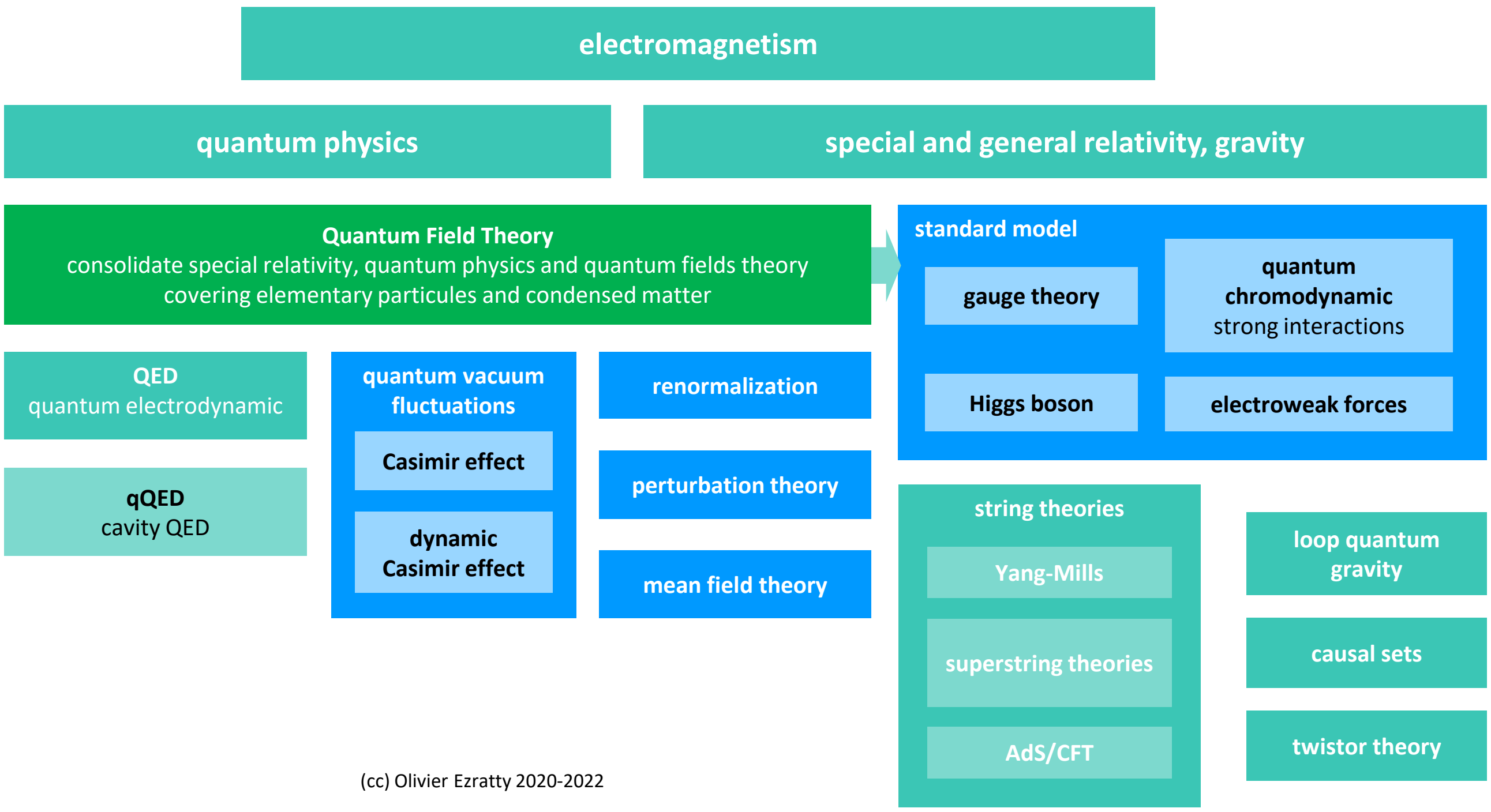

*Figure 121: vague classification of quantum physics theories and unification theories. (cc) Olivier Ezratty, 2020.*

**Quantum chromodynamics** provides a description of the strong interactions binding quarks together via gluons to form particles called hadrons, namely, protons and neutrons. Murray Gell-Mann (1929-2019, American, Nobel Prize in Physics in 1969) and Georges Zweig (1935, Russian then American, former PhD student of Richard Feynman) each proposed the existence of quarks in 1963. Quantum chromodynamics is an extension of the quantum field theory developed in 1972 by Murray Gell-Mann and Harald Fritzsch.

**Standard model** describes the architecture of known elementary particles and their interactions. It models the fundamental weak and strong electromagnetic forces. It only lacks gravity to be complete. This model predicted the existence of quarks, these massive particles forming neutrons and protons, in addition to other elementary particles such as the famous Higgs boson whose existence was proven at CERN's LHC in 2012. The expression "standard model" was created in 1975. It relies on a gauge theory because of its mathematical symmetries.

It is not the first of its kind because Maxwell's electromagnetism is also a gauge theory, between magnetic and electric fields. The standard model particles do not cover the famous dark matter whose nature is not yet known.

**String theory** combines general relativity and quantum physics to propose a quantum explanation of gravity, using a new massless particle, the graviton. According to this theory, elementary particles are tiny strings, open or closed, with vibration types defining the nature of the particle. Their size is of the order of magnitude of $10^{-35}$ m, the Planck length. According to this theory, the Universe would be a set of vibrating strings.

The graviton would join the three other forces of nature intermediated by particles without mass: electromagnetic waves mediated by photons, strong interactions mediated by gluons that link quarks together in protons and neutrons and weak interactions mediated by W and Z bosons that govern atomic nuclei and in particular radioactivity[276] . String theory essentially covers bosons of all kinds.

**Superstring theory** is an extension of the string theory that adds fermions to the code theory model that focused on bosons. It tries to consolidate the description of all forces in a single unified theory. It quantifies gravity and ties it to other forces. It is based on the notion of supersymmetry which extends the standard model by making each type of boson correspond to a type of fermion. The theory took shape in 1943 with Werner Heisenberg in the form of the S-matrix theory, and then was reborn in 1984. It uses 10 dimensions to describe physics, far beyond the four classical dimensions (three for position and one for time). It also uses the notion of "branes" which describes point particles in these multidimensional spaces. However, this theory is not unique since there are five variants, which some people try to unify in the **M-theory**, which is based on 11 dimensions. A never-ending story!

**Loop quantum gravity theory** is another tentative to explain gravity with a quantum model. It discretizes the effects of gravity by presenting space as a meshed structure with quantized areas and volumes of space, and gravitational field quanta connected to each other by links characterized by a spin (that has nothing to do with usual particles spin)[277]. For this theory created in the 1980s, the Universe would be a gigantic spin foam. Its main promoters are **Carlo Rovelli** (Center for Theoretical Physics in Marseille) and **Lee Smolin** (Perimeter Institute for Theoretical Physics in Waterloo[278]). The seeds of the theory date back to 1952, with many intermediate stages as described in Figure 122. It is, above all, a mathematical and topological model. It does not seem to formulate an experimental validation method even though it is used to model that the big bang was coming after a big bounce in a cyclical phenomenon with contractions and expansions. It may be possible to detect some fossil signature of these phenomenon.

A brief history of quantum gravity:

1952 Flat space quantization (Rosenfeld, Pauli, Fierz, Gupta,...)
1959 Canonical structure of general relativity (Dirac, Bergmann, Arnowit, Deser, Misner)
1964 Penrose introduces the idea of spin networks
1967 Wheeler-DeWitt equation
1974 Hawking radiation and black hole entropy
1984 String theory
1986 New variables for general relativity (Ashtekar, Sen)
1988 Loop representation and solutions to the Wheeler-DeWitt equation (Jacobson, Smolin)
1989 Extra dimensions from string theory
1995 Hilbert space of loop quantum gravity, geometric operators
2000' Spin foam models, group field theory, loop quantum cosmology,...

*Figure 122: history of quantum gravity. Source: The philosophy behind loop quantum gravity by Marc Geiller, 2001 (65 slides).*

[276] A proton has two up quarks and one down quark. A neutron has two down quarks and one up quark. An up quark can desintegrate in a down quark, a positron and a neutrino via a W boson and a down quark can disintegrate in an up quark, one electron, one antineutrino and a W boson. A quark has a size close to that of an electron, about $10^{-16}$ cm. Radioactivity emits alpha rays via strong forces, particles comprising two protons and two neutrons (helium 4 atom without electron), beta rays generated by weak forces which are electrons or positrons and finally gamma rays which are photons of very high energy level.

[277] It is reminiscent of the recent theory of the whole built by Stephen Wolfram and published in 2020.

[278] See Lee Smolin Public Lecture Special: Einstein's Unfinished Revolution, 2019 (1h13mn) where he describes the shortcomings of quantum mechanics.

These are only a few of the many theories being devised. Some amateurs also try to create their own theory of the whole, without usually obtaining any feedback from the scientific community[279].

### Quantum physics 101 key takeaways

- Quantum physics is based on a set of postulates and a strong linear algebra mathematical formalism. Surprisingly, there are many variations of these postulates. There is not a single bible or reference for these, illustrating the diversity of pedagogies and opinions in quantum physics. But although deemed incomplete, the theory has been validated by an incredible number of experiments.
- Quantum physics describe the behavior of matter and light at nanoscopic levels. It deals not only with atoms, electrons and photons which are used in quantum information technologies but also with all elementary particles from the standard model (quarks, ...).
- Quantumness comes from the quantification of many properties of light and matter that can take only discrete values, from the wave-particle duality of massive (atoms, electrons) and non-massive (photons) particles, and from wave-particle duality and its consequences like superposition and entanglement. By the way, a cat can't be both alive and dead since it's not a nanoscopic quantum object. Forget the cat and instead, learn Schrodinger's equation!
- Indetermination principle states it's impossible to measure with an infinite precision quantum objects properties that are complementary like speed and position. You can use this principle to improve measurement precision in one dimension at the expense of the other. It is used in photons squeezing, itself applied in the LIGO giant gravitational waves interferometer.
- Quantum matter and fluids are showing up with composite elements associating light and matter, or with superfluidity and superconductivity where boson quantum objects can behave like a single quantum object. You find there a wealth of strange phenomenon such as skyrmions, magnons, topological insulators and quantum batteries. They could lead to a new chapter in the second quantum revolution.
- Quantum physics also explains weird effects like vacuum quantum fluctuation, although it doesn't violate the second principle of thermodynamics, nor can it lead to the creation of some free energy sources.
- Most of quantum physics phenomena as described in this section have or will have some use cases in quantum information science and technologies.

[279] See, for example, the Unified Theory Research Team website, which announced the publication in September 2020 of a theory model of the whole called MME for Model of Material and Energy. The site claims that its model, which is presented as an algorithmic approach, can explain everything, from the functioning of all particles to the bricks of life. The team behind this project includes two Pierre and Frédéric Lepeltier from France. The first has been the CEO of the Unified Theory Research Team for 32 years.

# Gate-based quantum computing

As a computer scientist, you may have skipped all the previous parts to get here right away. One can indeed understand how quantum computers operate without delving too deeply into quantum physics beyond grasping its basic mechanisms. Some mathematical knowledge is however required on trigonometry and linear algebra, including vectors, matrices and complex numbers[280].

The first basic element of a quantum computer is its inevitable qubit. You've probably already heard about this mysterious object having "simultaneously" the values 0 and 1. As a result, you've been told that a set of N qubits create an exponential $2^N$ superposed state that explains the power of quantum computing. Unfortunately, most explanations usually stop there and you then end up wondering how it actually works to make some calculation. What comes in and out of a quantum computer? How is it programmed? How do you feed it with data and code? Where is it useful? This book is there to provide you with some educated answers to all these critical questions.

We will cover here the logical and mathematical aspects of qubits, qubit registers, quantum gates and measurement[281]. Each and every time, when possible, we'll draw parallels with traditional computing. In the following part, we'll look at quantum computer engineering and hardware and even describe the complete architecture of a superconducting qubits quantum computer.

## In a nutshell

Before digging into qubits, qubit registers and the likes, here's a tentative to summarize the key elements of gate-based quantum computing that we'll cover in detail afterwards. It shows how physics and mathematics are intertwined.

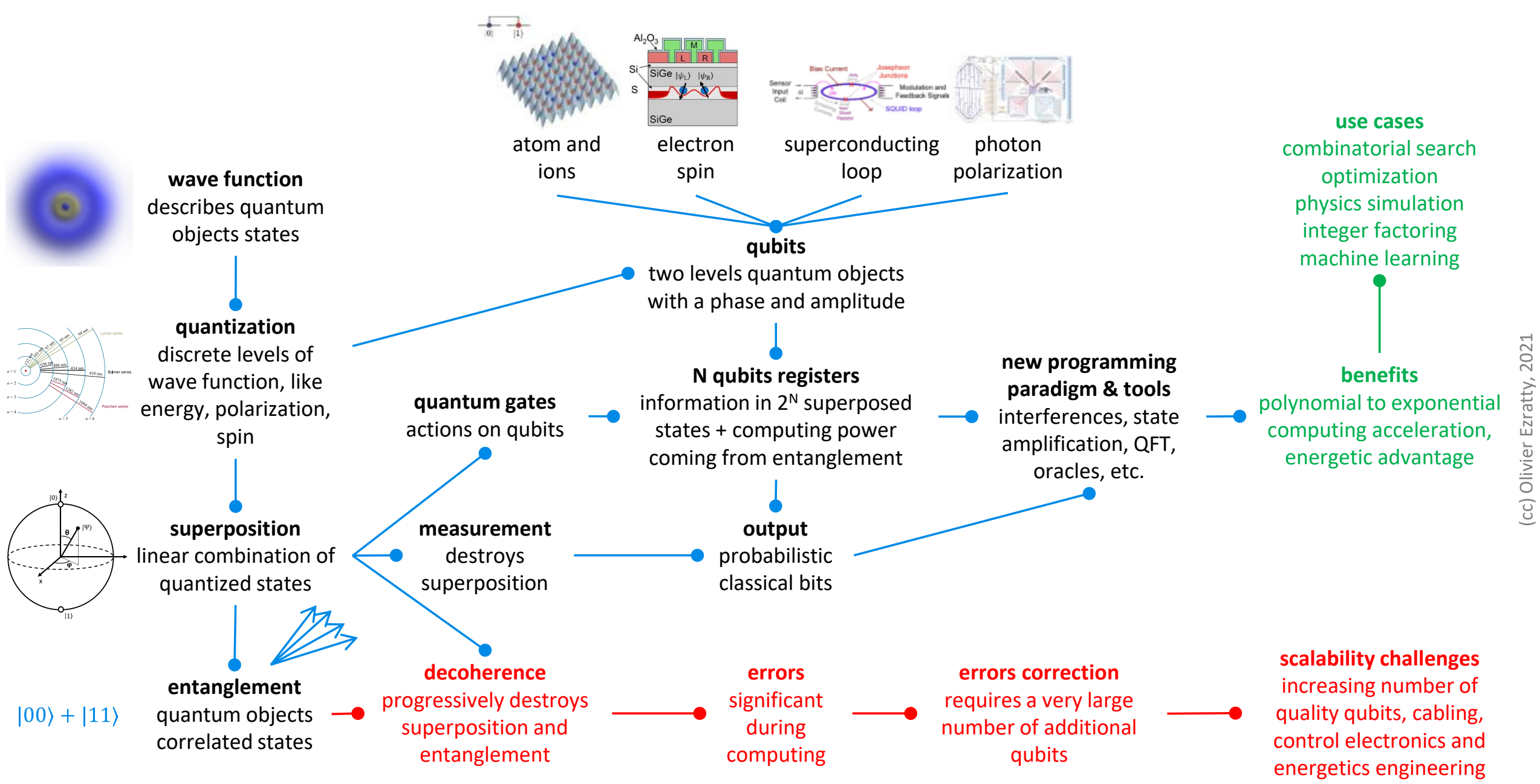

*Figure 123: a single schematic to describe quantum physics and quantum computing. (cc) Olivier Ezratty, 2021.*

[280] Complex numbers were created by the polymath Girolamo Cardano (1501-1576, Italian) and the Algerian mathematician Raffaele Bombelli (1526-1572, Italian) between 1545 and 1569. They were used to solve polynomial equations associating cubes and squares that kept Italian mathematicians busy since the end of the fifteenth century. See A Short History of Complex Numbers by Orlando Merino, 2006 (5 pages).

[281] The name qubit, for 'quantum' and 'bit', appeared in 1995 in Quantum coding by Benjamin Schumacher, April 1995 (34 pages).

**Wave function**
mother equation of quantum physics, created by Erwin Schrödinger. It describes particles properties probabilities in space and time with a complex number. This equation is specific to non-relativistic massive particles like electrons. We also use photons in quantum computing, whose properties are defined by Maxwell's electromagnetic equations and the second quantization equations (Glauber states, Wigner function, Fock states, etc.).

**Quantization**
properties of quantum objects, having discrete, not continuous and exclusive values. It enables the creation of qubit physical and logical objects having two levels.

**Superposition**
qubits are quantized quantum objects having two basis computational states |0⟩ and |1⟩. These can be combined linearly, thanks to the linearity over space of Schrödinger's wave equation. Solutions of this equation can be linearly combined with complex numbers. Thus, a wave adding two solution waves is still a solution. This doesn't mean the qubit is really simultaneously in two states.

**Entanglement**
often presented as a situation where several quantum objects have properties that are correlated. Actually, entanglement is the consequence of superposition of multiple qubit states. This is the phenomenon that provides both a real theoretical exponential acceleration to quantum computing but also enables conditional relations between qubits. Without it, qubits would be independent and no useful computing could be done.

**Qubits**
mathematical objects with two levels 0 and 1. It's described by two complex number amplitudes. But due to normalization and getting rid of their global phase (we'll explain all of that), they are described by two real numbers for their amplitude and phase. Physical qubits are based on massive (electron, controlled atoms, superconducting currents) or non massive quantum objects (photons) and one of their quantum properties or observables (spin, energy level, current direction of phase, polarity).

**Registers**
physical and logical assemblies of several qubits. With N qubits, they can handle computing on a space $2^N$ computational basis states together represented by complex number amplitudes. Each basis state is one of the possible combinations of N 0s and 1s. Computing power comes from entanglement.

**Quantum gates**
logical operations exerted on qubits. We have single qubit gates which are changing single qubit states and several qubit gates conditionally changing one or two qubits based on the state of a control qubit, and leveraging entanglement. Gates are the only mechanism used to feed a quantum register with data and instructions. These are not separated as in classical computing based on a Von Neumann / Turing machine model.

**Programming paradigms**
quantum programming is based on very different paradigms than classical programming. In a nutshell, it's analog-based. We play with interferences, states amplification, quantum Fourier transforms and the concept of oracles.

**Measurement**
the way to extract information from qubits. Unfortunately, you can't read the two real numbers describing the qubit state nor the combination of qubit registers computational basis states. You get just classical 0s and 1s for each qubit. Quantum algorithms toy with the wealth of superposition and entanglement during computing to recover a simple result at the end. Measurement is also used during quantum error corrections. Since qubit measurement output is probabilistic, you generate a deterministic output with running your algorithm several times (up to several thousand times) and computing an average of the obtained results.

**Output**
for a register of N qubits, you get N 0s and 1s. But these are probabilistic results. You usually need to run your algorithm several times and compute an average of the results to get a deterministic result. Noise and decoherence are additional reasons why you need to do this several times.

**Benefit**
an acceleration of computing time com-pared to the best classical computers. Accelerations can be from polynomial to exponential. The benefit can also be economic like with the energetic cost of quantum computing that many expect to be fairly low compared to classical computing.

**Use cases**
quantum computing will not replace most use cases of classical computing. It brings value for complex combinatorial problems, optimization problems, quantum physics simulation, some machine learning problems and at last, fast integer factoring.

**Decoherence**
the enemy with quantum computing. This is when qubit states is degraded, both for superposition and entanglement. It results from the interactions between the qubits and their environment despite of all the care implemented to isolate it.

**Errors**
result of decoherence and other perturbations affecting the qubits. Other sources of errors are the imprecision of the control electronics driving qubit gates. Qubit phase and amplitude is degraded over time. Existing error rates are many order of magnitude higher that with classical computing. These are the reasons why we don't have yet quantum computers with a very high number of functional qubits.

**Error corrections**
set of techniques used to correct these errors. It requires assembling so-called logical qubits made of a great number of physical qubits. The needed ratio at this point is ranging from 30 to 10,000 physical qubits to create a logical qubit. The ratio depends on the qubit quality and technology but also on the target logical qubit fidelity (from $10^{-8}$ to $10^{-15}$ error rates).

**Scalability challenges**
assembling these huge logical qubit is the mother of the challenges with quantum computing. It's not easy to assemble that many qubits and keeping them stable, limit their decoherence and the likes. On top of that, assembling a great number of qubits creates huge engineering challenges with cryogenics cooling power, thermal dissipation, cabling and control electronics. These are the reason why quantum computers don't scale yet to bring their expected benefits.

*Figure 124: the key concepts behind gate-based quantum computing in one page. (cc) Olivier Ezratty, 2021-2022.*

# Linear algebra

Quantum physics and computing require some understanding of a whole bunch of concepts of linear algebra that we will quickly scan here. They are associated with a mathematical formalism describing quantum phenomena. This mathematical formalism is also the cornerstone of quantum physics postulates, already covered in an earlier section, page 86. It is also essential to create quantum algorithms.

I will try to explain some of these concepts and mathematical conventions that are used with quantum computing. This will mainly allow you to find your way through some of the scientific publications I mention in this book.

### Linearity

Linear algebra is the branch of mathematics using vector spaces, matrices and linear transformations. In the case of quantum physics and computing, it also deals with complex numbers.

A phenomenon is linear if its effects are proportional to its causes. This translates into the verification of two simple equations pertaining to homogeneity and additivity as shown in Figure 125.

homogeneity $f(\lambda x) = \lambda f(x)\ for\ all\ x \in \mathbb{R}$

additivity $f(x+y) = f(x) + f(y)\ for\ all\ x, y \in \mathbb{R}$

*Figure 125: homogeneity and additivity in linear algebra.*

ℝ being a vector space, λ a real number, x being a vector of the vector space ℝ and f(x) a function applying to this vector. In a one-dimensional space, a classic example of a linear function is f(x) = ax. A polynomial function of the type $f(x) = ax^2 + b$ is obviously not linear because it evolves non-proportionally to x. Even $f(x) = ax + b$ is not linear, and for the same reason.

As already defined, an observable is a mathematical operator, a Hermitian matrix, used to measure (mathematically) a property of a physical system. It's frequently assimilated to the measured property. For a qubit, it corresponds to some measurable value by a sensor on a quantum object outputting a classical 0 or 1. The measurement causes the qubit quantum object wave function to collapse on one of the basis states. If the state of a quantum or qubit is measured twice, the measurement will yield the same result. With qubits, observables are usually based on projections on a two-level properties system, mathematically materialized by a $|0\rangle$ or $|1\rangle$, aka qubit computational basis states. But, if the physics permits it, other computational basis can be used. It's the case with photons and polarization measurement where their angle can be easily made different in different parts of an experiment.

### Hilbert spaces and orthonormal basis

A quantum state of a single or several quantum objects can be described by a vector in a Hilbert space. A qubit state is represented in a two-dimensional orthonormal space formed with the basis states vectors $|0\rangle$ and $|1\rangle$. It is a vector of complex numbers in a two-dimensional Hilbert space allowing lengths and angles measurements. A complex number is defined as a+ib where a and b are real and $i^2$=-1.

Complex numbers are very useful in quantum physics. It relates to the wave-particle duality of all quantum objects and to the need to handle their amplitude (complex number norm, vector length or modulus) and phase (the complex number angle when using polar coordinates).

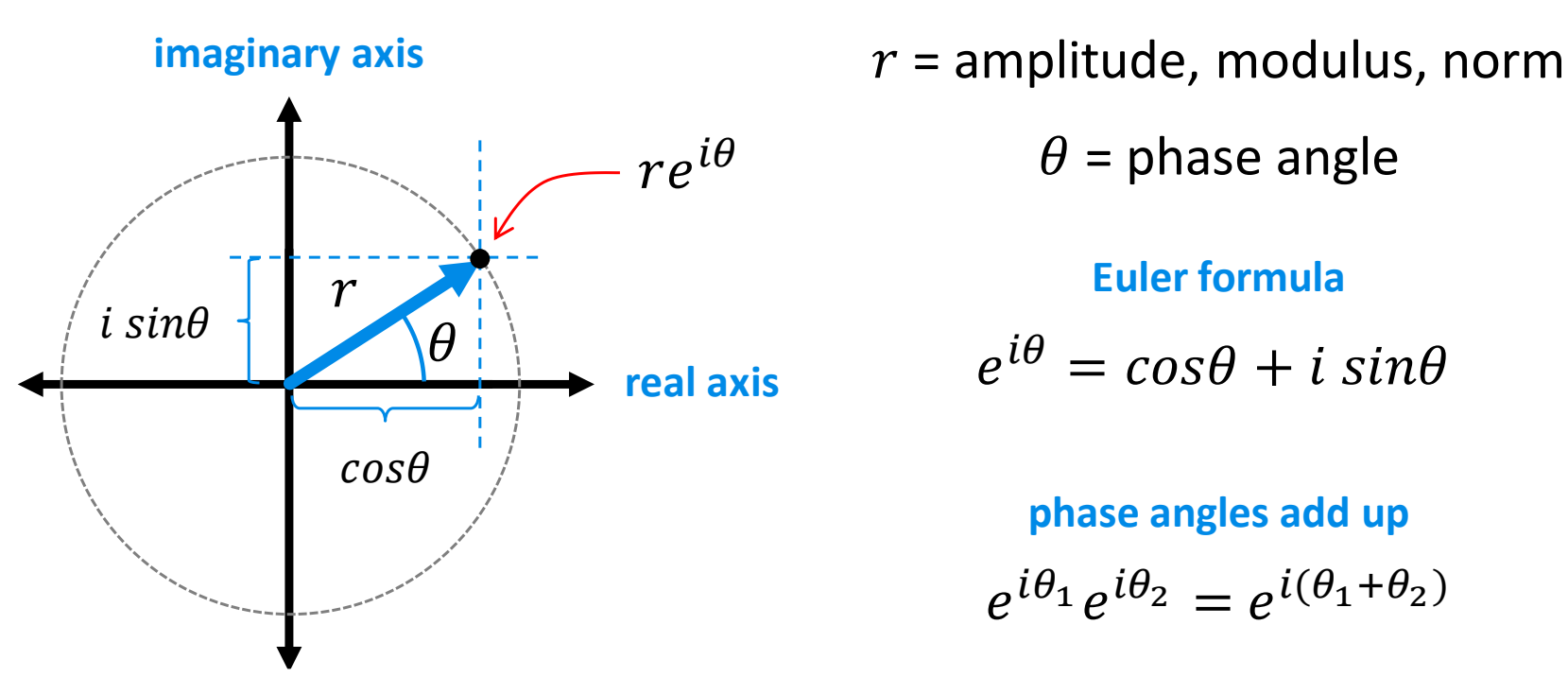

*Figure 126: complex number explained by geometry and trigonometry.*

With qubits, it is represented with the complex numbers α and β associated with the states $|0\rangle$ and $|1\rangle$ and whose sum of squares makes 1. This linear combination of the states $|0\rangle$ and $|1\rangle$ describes the phenomenon of superposition within a qubit.

This two-dimensional space replaces the infinite-dimensional space that characterizes a Schrödinger wave function $f(x)$, where x can take any value in space. It is thus a simplified representation of the quantum state of a qubit. By manipulating these symbols, the vectors and matrices, we forget a little the wave-like nature of the manipulated quanta, even though it is still present in the phase information embedded in the imaginary part of α and β for one qubit. It also can deal with photons which do not obey to Schrödinger's equation but to Maxwell's electromagnetic equations.

An orthonormal basis of a vector space consists of base vectors which are all mathematically orthogonal with each other and whose length is 1. In the representation of a qubit state, the most common orthonormal basis is made of the states $|0\rangle$ and $|1\rangle$.

Other orthonormal reference basis can be used for measurement, particularly with photons, and polarization references different from the starting reference (0°/90° then 45°/135°, obtained with rotating a simple polarizer).

Another example of an orthonormal basis is the states located on the Bloch sphere on the x-axis and represented with $|+\rangle$ and $|-\rangle$. These are often called Schrodinger cats.

$$|+\rangle = \frac{|\mathbf{0}\rangle + |\mathbf{1}\rangle}{\sqrt{\mathbf{2}}} \qquad |-\rangle = \frac{|\mathbf{0}\rangle - |\mathbf{1}\rangle}{\sqrt{\mathbf{2}}}$$

Figure 127: another orthonormal basis.

### Dirac Notation

In Dirac notation, a quantum object state is represented by $|\Psi\rangle$, the **ket** of quantum state Ψ. The **bra** of the same state vector, represented by $\langle\Psi|$ is the conjugate (or transconjugate, or adjoint) transpose of the "ket". It is the "horizontal" vector $[\bar{\alpha}, \bar{\beta}]$ where $\bar{\alpha}$ and $\bar{\beta}$ are the conjugates of α and β, inverting the sign of the imaginary part of the number (-i instead of +i, or the opposite).

vectors **Dirac notation**

$$|\Psi\rangle = \begin{bmatrix}\alpha\\ \beta\end{bmatrix} \qquad \bar{\alpha} = \alpha^* \quad \langle\Psi| = [\bar{\alpha}, \bar{\beta}]$$

**Ψ ket** **Ψ bra**

$$(1+i)^* = 1 - i$$

complex number **conjugate**

Figure 128: introduction to Dirac vector notation.

The **scalar product** of two qubits $\langle\Psi_1|\Psi_2\rangle$ is the mathematical projection of the state vector $\Psi_2$ onto the vector $\Psi_1$. This yields a complex number. When the vectors are orthogonal, the scalar product is equal to 0. When the two vectors are identical, $\langle\Psi|\Psi\rangle$ is Ψ's norm and is always equal to 1. A scalar product is also named an inner product.

$$\langle\Psi_1|\Psi_2\rangle = [\overline{\alpha_1}, \overline{\beta_1}] \times \begin{bmatrix}\alpha_2\\ \beta_2\end{bmatrix} = \overline{\alpha_1}\alpha_2 + \overline{\beta_1}\beta_2$$

inner **scalar** product: vector similarity

$$\langle\Psi|\Psi\rangle = [\bar{\alpha}, \bar{\beta}] \times \begin{bmatrix}\alpha\\ \beta\end{bmatrix} = \alpha^2 + \beta^2 = 1$$

Figure 129: inner scalar product.

An **inner product** is a generalization of a dot vector product applied to complex number vectors, according to the sigma in Figure 130.

**complex vectors dot product**

$$A.B = \sum_i a_i \bar{b_i}$$

Figure 130: dot product.

The **outer product** of two vectors representing a qubit, one in bra and the other in ket, gives an operator or density matrix which is a 2x2 matrix.

When the bra corresponds to the transconjugate of the ket, it is a density operator of a pure state. This notion of density operator will then be extended to a combination of qubits.

$$|\Psi\rangle\langle\Psi| = \begin{bmatrix}\alpha\\ \beta\end{bmatrix} \times [\bar{\alpha}, \bar{\beta}] = \begin{bmatrix}\alpha\bar{\alpha} & \alpha\bar{\beta}\\ \beta\bar{\alpha} & \beta\bar{\beta}\end{bmatrix}$$

Figure 131: outer product.

What are the use cases of this Dirac notation? It is particularly helpful for manipulating quantum states, to simplify tensor products representations and with measurement, which we'll cover later starting page 184.

### Eigenstuff

We also need to define the notions of **eigenvector**, **eigenvalue**, **eigenstate** and **eigenspace** which are often used in quantum mechanics and quantum computing as well as in machine learning, particularly in dimension reduction algorithms such as PCA (Principal Components Analysis). These notions allow to define the structure of certain square matrices[282].

For a square matrix A, an eigenvector x or eigenvector of A is a vector that verifies the equation Ax = λx, λ being a complex number called eigenvalue.

These eigenvectors have the particularity of not changing direction once multiplied by the matrix A. For an eigenvalue λ, the associated eigenspace, or eigenspace, is the set of vectors x that satisfy Ax = λx. These eigenvalues are evaluated by calculating the determinant of the matrix A - λI, where I is the identity matrix (1 in the diagonal boxes and 0 elsewhere). We then find the values of which solves 0 = A - λI. It is a polynomial equation having a degree less than or equal to the size of the square matrix[283].

The reference eigenvectors of a matrix A allow to reconstitute an orthonormal space linked to the matrix. For example, a projection matrix in a 3D plane will have as main eigenvectors two orthogonal vectors located in the plane and one vector orthogonal to the plane. This multiplication gives λx with λ being non-zero if the eigenvector is in the plane in question and 0 if the vector is orthogonal to the plane [284]. A matrix A can be that of a quantum gate. An eigenvector of a quantum gate is therefore a ket whose value is not modified by the quantum gate.

This is easy to imagine for the S gate, phase change, which we will see later. The $|0\rangle$ and $|1\rangle$ kets being in the rotation axis, they are not modified by it.

They are thus eigenvectors of the S gate and the corresponding eigenvalues are 1 and -1. This is always the case for quantum gate matrices since the vectors representing the quantum states, the kets, always have a length of 1. These eigenvalues are the only ones enabling this!

The search for the eigenvectors and eigenvalues of a matrix A is like diagonalizing it. For this it must be diagonalizable ("non-defective"). Hermitian and unitary matrices commonly used in quantum physics are all non-defective and diagonalizable. The diagonalization of a square matrix consists in finding the matrix which will multiply it to transform it into a matrix filled only in its diagonal. A matrix A is diagonalizable if we can find a matrix P and a diagonal matrix D such that $P^{-1}AP = D$ ($P^{-1}$ being the inverse matrix of P, such that $P^{-1}P= PP^{-1}=I$, I being the matrix identity with 1's in the diagonal and 0's elsewhere). A square matrix of dimension n is diagonalizable if it has n mutually independent eigenvectors. The diagonalized matrix diagonal contains the eigenvalues $\lambda_i$ of the origin matrix, with i=1 to N being the size of the matrix.

A diagonalized quantum state of a quantum object can look like $A = \sum_i \lambda_i |i\rangle\langle i|$. This decomposition of a pure state vector in a Hilbert space in eigenstates $|i\rangle$ and eigenvalues $\lambda_i$ is also named a **spectral decomposition**. It's linked to the wave-duality aspect of all quantum objects.

---

282 See a good quick review of linear algebra in Linear Algebra Review and Reference by Zico Kolter and Chuong Don 2015 (26 pages).

283 See this nice visual explanation of eigenvectors and eigenvalues: Eigenvectors and eigenvalues | Chapter 14, Essence of linear algebra, 2016 (17 minutes).

284 This is well explained in Gilbert Strang's lecture at MIT, 2011 (51 minutes).

A quantum object is indeed decomposed into a coherent superposition of elementary waves. In the case of photons, it's easy to grasp with several photons of different frequencies being superposed and forming a gaussian wave packet. It constitutes a coherent superposition of the electromagnetic field. These wave packets are commonly generated by femtosecond pulse lasers[285].

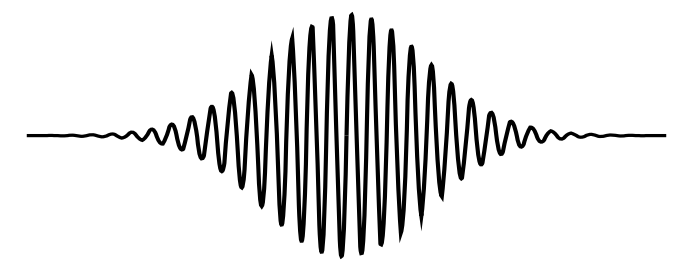
*Figure 132: a photon gaussian wave packet.*

And the eigenstates? This is another name given to eigenvectors, but by physicists!

### Tensor products

The tensor product of two vectors of dimension m and n gives a vector of dimension m*n while the tensor product of a matrix of dimension m*n by a matrix of dimension k*l will give a matrix of dimension mk*nl. Tensor products use the sign ⊗.

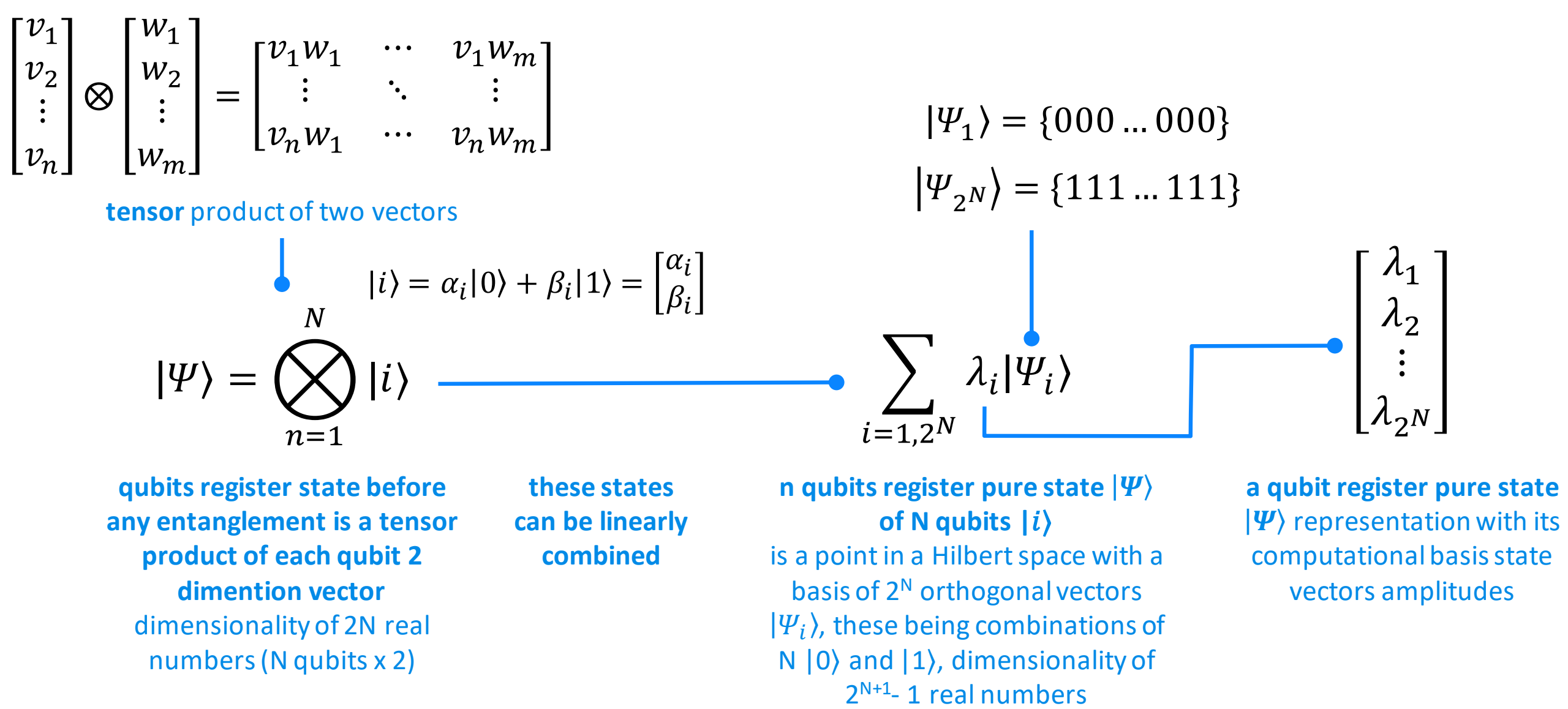

*Figure 133: tensor products construction. (cc) Olivier Ezratty, 2020.*

Tensor products are used to compute "manually" the state of quantum registers containing several unentangled qubits. The state of a register of N non-entangled qubits is the tensor product of these N qubits represented by their vertical ket vector.

This gives a ket, a vertical vector that has $2^N$ different values, each representing the complex number weight of different combinations of 0s and 1s. A quantum register is a superposition of these $2^N$ different states complex amplitudes. The sum of these squared amplitudes gives 1 per the Born rule. By the way, the tensor product of qubits is represented by a vector, after vectorization of the tensor product matrix of $2^N$ dimensions.

### Entanglement

Quantum states are separable when they are mathematically the result of the tensor product of each of the pure states that compose it. But these values can be assembled linearly to create another quantum state, modulo a normalization rule. This combines several vectors resulting from tensor products. These combinations can become inseparable.

[285] And when the carrier frequency is growing or decreasing through the pulse, it's named a chirp pulse.

That's when entanglement comes into play. An entangled state of two or more qubits occurs when it cannot be factorized as the tensor product of two pure states. In other words, it cannot be the combination of independent qubits. The qubits become dependent.

This is demonstrated mathematically for the states $|00\rangle$ and $|11\rangle$ of a register of two qubits. In these pairs, the measurement of the value of one of the qubits determines that of the other, here identical. The creation of such entangled pairs of qubits requires preparation operations like using a combination of Hadamard and CNOT gates.

Two qubits placed side by side are not magically entangled! The pair used in the example can be generated by two quantum gates, an H gate (Hadamard) and a CNOT gate, as shown just below.

**an entangled EPR pair can't be a tensor product of two qubits $|\Psi_1\rangle$ and $|\Psi_2\rangle$**

$|\Psi_1\rangle = \alpha_1|0\rangle + \beta_1|1\rangle$ $\quad$ $|\Psi_2\rangle = \alpha_2|0\rangle + \beta_2|1\rangle$

$|\Psi_1\rangle \otimes |\Psi_2\rangle = (\alpha_1|0\rangle + \beta_1|1\rangle)(\alpha_2|0\rangle + \beta_2|1\rangle)$

$\frac{1}{\sqrt{2}}(|00\rangle + |11\rangle) = \alpha_1\alpha_2|00\rangle + \alpha_1\beta_2|01\rangle + \beta_1\alpha_2|10\rangle + \beta_1\beta_2|11\rangle)$

$\alpha_1\beta_2 = 0$ and $\beta_1\alpha_2 = 0$
are incompatibles with $\alpha_1\alpha_2 = \frac{1}{\sqrt{2}}$ and $\beta_1\beta_2 = \frac{1}{\sqrt{2}}$
if $\alpha_1 = 0$ then $\alpha_1\alpha_2 = 0$
if $\beta_2 = 0$ then $\beta_1\beta_2 = 0$

implications: the density matrix mathematical representation of qubits registers

*Figure 134: non separability of two entangled qubits.*

We will define this CNOT gate later on, after page 171. This is described as both qubits having correlated values. But these values are... random since being a perfect superposition of 0 and 1!

Only multi-qubit quantum gates generate entangled qubits in a qubit register, besides the SWAP gate which doesn't. Here with an example of creating a Bell pair associating the states $|00\rangle$ and $|11\rangle$ with a mix of Hadamard and CNOT gates.

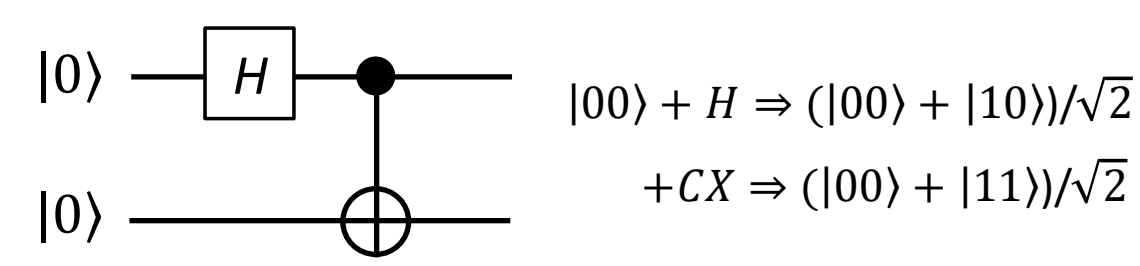

*Figure 135: a Bell pair.*

A so-called **GHZ** state (for Greenberger-Horne-Zeilinger, distinguishable from GHz frequencies with a capital Z) with three entangled qubits is superposing the states $|000\rangle$ and $|111\rangle$. It is a generalization of the 2-qubit Bell state $(|00\rangle + |11\rangle)/\sqrt{2}$. A GHZ is usually prepared with a Hadamard gate and two consecutive CNOTs.

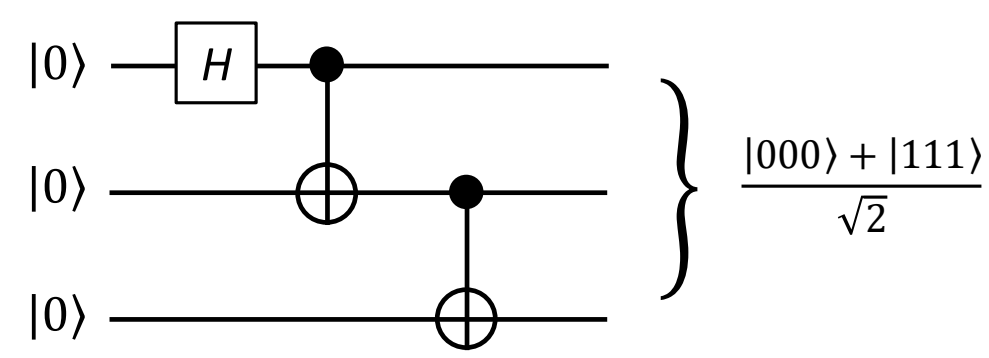

*Figure 136: a GHZ state.*

These pairs of Bell and GHZ states are used in error correction codes as well as in telecommunications, among other things.

Another typical entangled state is the **W state**, created in 2000, that has the property of being maximally entangled and robust against particle loss. It is a generalized version of another of the four possible Bell states, $(|01\rangle + |10\rangle)/\sqrt{2}$ [286]:

$$|W\rangle = \frac{1}{\sqrt{3}}(|001\rangle + |010\rangle + |100\rangle)$$

*Figure 137: a W state.*

At last, the level of entanglement of a qubit register depends on the Hamming distance between the basis states involved in the linear superposition of basis states.The far apart they are, with the greater number of non-identical 0s and 1s, the greatest the entanglement is.

[286] See Three qubits can be entangled in two inequivalent ways by Wolfgang Dür (which explains the W in W states), G. Vidal, and J. Ignacio Cirac, 2000 (12 pages) and the thesis Symmetry and Classification of Multipartite Entangled States by Adam Burchardt, September 2021 (126 pages).

## Matrices

Various matrix transformations must be understood here:

- **Matrix conjugate** when all complex number see their complex part negated, or $a_{ij} = a_{ij}^*$.
- **Matrix transpose** when all matrix $a_{ij}$ values are transformed into $a_{ji}$value, with i=line and j=column indices of matrix “cells”.
- **Matrix transconjugate** which is a conjugate of the transpose or vice-versa, also named adjoint. It’s notated as $A^\dagger$, for A « dagger ».
- **Matrix traces** are the sum of their diagonal values, usually normalized to 1, like with density matrices. It is also the sum of their eigenvalues.

We also have three important classes of matrices:

- **Hermitian matrices** are equal to their transconjugate, meaning that $a_{ij} = a_{ji}^*$ .
- **Projectors** are matrix operators using a Hermitian matrix that is equal to its square. A diagonalized projector contains only zeros and a single 1. A projector is a non-unitary operation. It relates with the irreversibility of quantum measurement.

  If $|\psi\rangle$ is a unit vector, the outer product $|\psi\rangle\langle\psi|$ is a projector that can project any vector $|\phi\rangle$ on $|\psi\rangle$.

| Notation | Description |
|---|---|
| $z^*$ | Complex conjugate of the complex number $z$. $(1+i)^* = 1-i$ |
| $\lvert\psi\rangle$ | Vector. Also known as a *ket*. |
| $\langle\psi\rvert$ | Vector dual to $\lvert\psi\rangle$. Also known as a *bra*. |
| $\langle\varphi\lvert\psi\rangle$ | Inner product between the vectors $\lvert\varphi\rangle$ and $\lvert\psi\rangle$. |
| $\lvert\varphi\rangle \otimes \lvert\psi\rangle$ | Tensor product of $\lvert\varphi\rangle$ and $\lvert\psi\rangle$. |
| $\lvert\varphi\rangle\lvert\psi\rangle$ | Abbreviated notation for tensor product of $\lvert\varphi\rangle$ and $\lvert\psi\rangle$. |
| $A^*$ | Complex conjugate of the $A$ matrix. |
| $A^T$ | Transpose of the $A$ matrix. |
| $A^\dagger$ | Hermitian conjugate or adjoint of the $A$ matrix, $A^\dagger = (A^T)^*$. $\begin{bmatrix} a & b \\ c & d \end{bmatrix}^\dagger = \begin{bmatrix} a^* & c^* \\ b^* & d^* \end{bmatrix}$. |
| $\langle\varphi\lvert A\rvert\psi\rangle$ | Inner product between $\lvert\varphi\rangle$ and $A\lvert\psi\rangle$. Equivalently, inner product between $A^\dagger\lvert\varphi\rangle$ and $\lvert\psi\rangle$. |

*Figure 138: linear algebra key rules. Source: [Quantum Computation and Quantum Information](#) by Nielsen and Chuang, 2010 (10th edition, 704 pages).*

Indeed, $(|\psi\rangle\langle\psi|)|\phi\rangle = |\psi\rangle(\langle\psi||\phi\rangle) = (\langle\psi|\phi\rangle)|\psi\rangle$, given $\langle\psi|\phi\rangle$ is a real number being the inner product of both vectors. Some of these elements are summarized in Figure 138.

- **Unitary matrices** are square matrices whose inverse equals their transconjugate $(A^\dagger = A)$. A unitary matrix has several properties, one of which is to have orthogonal eigenvectors and to be diagonalizable. Unitary matrices define the reversible gates applied to qubits or sets of qubits.

$$A = \begin{bmatrix} 2 & i & -2i \\ -i & 1 & 3 \\ 2i & 3 & -1 \end{bmatrix}^\dagger \quad \bar{A} = \begin{bmatrix} 2 & -i & 2i \\ i & 1 & 3 \\ -2i & 3 & -1 \end{bmatrix} \quad A^\dagger = \overline{(A)}^* = \begin{bmatrix} 2 & i & -2i \\ -i & 1 & 3 \\ 2i & 3 & -1 \end{bmatrix} \quad \begin{matrix} U|x\rangle = |y\rangle \\ |x\rangle = U^\dagger|y\rangle \end{matrix}$$

transposed matrix — hermitian matrix, transconjugate = identity — unitary reversibility

*Figure 139: unitary matrices. (cc) Olivier Ezratty, 2021.*

A unitary operation is the application of a unitary matrix to a computational state vector that we’ll later see. Quantum computing reversibility comes from this unitary property. A unitary matrix $U$ can also be expressed as $U = e^{iH}$, with H being a Hermitian matrix, but finding $H$ given $U$ is a complicated calculation problem.

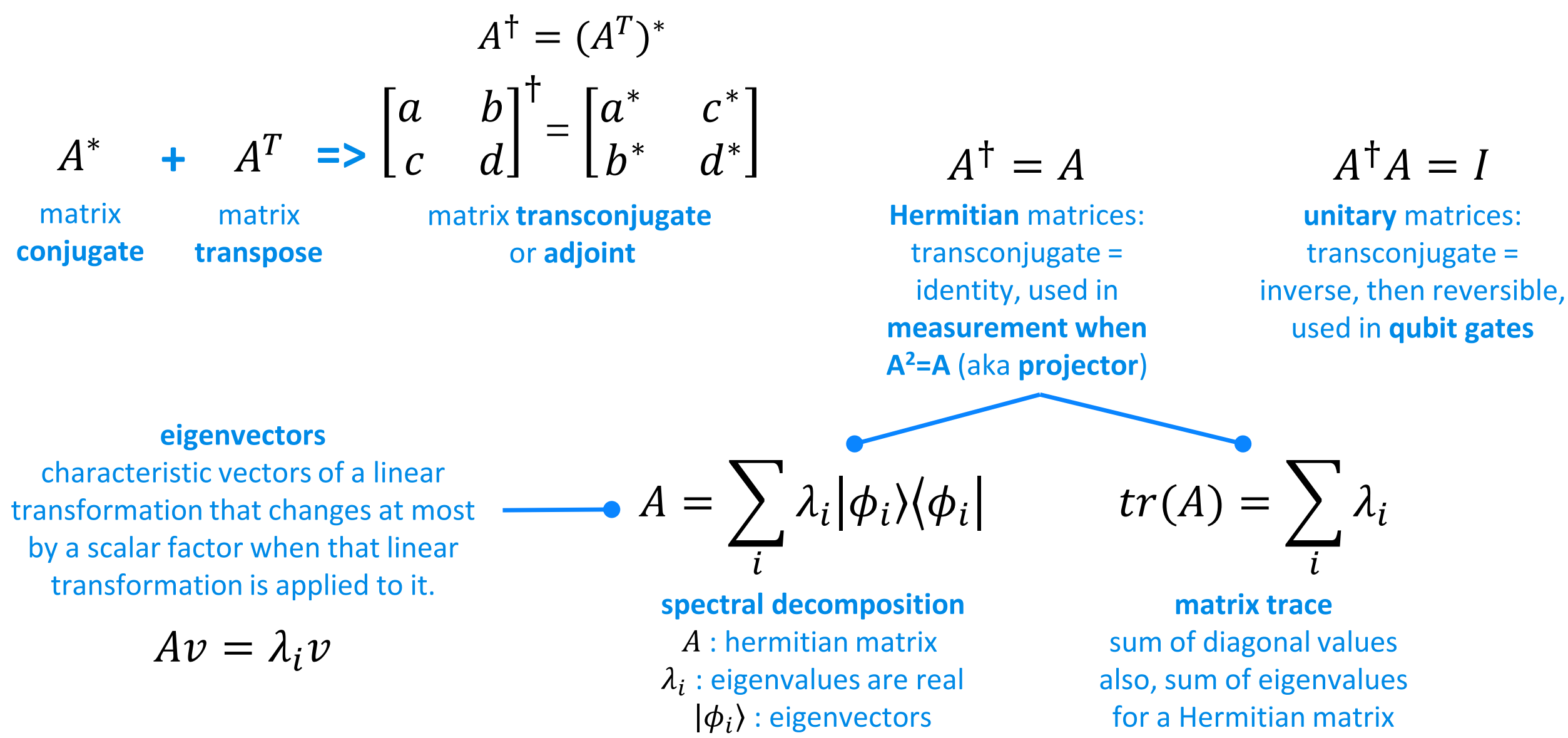

*Figure 140: difference between unitary matrices and Hermitian matrices. (cc) Olivier Ezratty, 2021.*

## Pure and mixed states

Let's now explain what the three main states of quantum objects are, basis, pure and mixed. We'll apply it to the case of qubits, given these notions are valid with any quantum system. We are dealing with mathematical models that describe quantum objects states[287].

| | basis states | pure states | mixed states |
|---|---|---|---|
| definitions | aka computational basis states, are N dimensions vectors combining 0s and 1s, with $2^N$ different such vectors for a N qubits register. | vectors in a Hilbert space of norm 1, specified by a single ket describing coherent superpositions of basis states with complex numbers. | or statistical mixture of pure states, are classical statistical ensemble of combination $p_i$ of pure states $\Psi_i$. $\Psi_i$ can be any combination of pure states but is usually a set of computational basis states. |
| randomness origin | no randomness with perfect qubits | quantum | quantum and classical |
| with a single qubit | $\|0\rangle$ and $\|1\rangle$ | $\|\Psi\rangle = \alpha\|0\rangle + \beta\|1\rangle$<br>$\|\alpha\|^2 + \|\beta\|^2 = 1$ | $p_1\|\psi_1\rangle, p_2\|\psi_2\rangle$<br>we don't add them, it's just a statistical ensemble, statistical mixture or convex sum of several systems. |
| with a N qubits register<br>$i$ = 1 to $2^N$ | $\|i\rangle$ $\|01101011\rangle$ for N=8<br>all $\|i\rangle$ form the computational basis states of the N qubits register, contains N combinations of 0 and 1, all basis states are mathematically orthogonal. | $\|\Psi\rangle = \sum_i \alpha_i\|i\rangle$<br>$\sum_i \alpha_i^2 = 1$ $\alpha_i$ = complex number<br>a pure state is a linear superposition of computational basis states. | $\{(p_i\|\Psi_i\rangle\}$ ensemble notation<br>$\sum_i p_i = 1$<br>$p_i$ = positive real number probability to find $\Psi_i$ in the mixed state given all $p_i$ are 0 or a 1 in a pure state. |

(cc) Olivier Ezratty, September 2021

*Figure 141: differences between basis states, pure states and mixed states. (cc) Olivier Ezratty, 2021.*

**Basis states** correspond to given combinations of 0 and 1 values in a qubit register. For a single qubit, these are the states $|0\rangle$ and $|1\rangle$. For a register of N qubits, it is one of the $2^N$ different basis states combinations of 0s and 1s, or a tensor product of N single qubit basis states. It constitutes the computational basis in a complex numbers Hilbert space of dimension $2^N$.

[287] See The Many Inconsistencies of the Purity-Mixture Distinction in Standard Quantum Mechanics by Christian de Ronde and César Massri, August 2022 (19 pages) that provides an interesting historical perspective on the pure and mixed states nuances and shortcomings.

The vectors of this basis are all mathematically orthogonal. A basis state is also named a computational basis state. When measuring individual qubits in these states, you get a deterministic result, at least with theoretically perfect qubits.

**Pure states** describe the state of an isolated quantum system of one or several objects as a linear superposition of the states from its computational basis. It's a vector in a Hilbert space. That's when superposition and entanglement come in. With massive particles, basis and pure states are solutions to Schrödinger's equation. It's applicable to one or several quantum objects or qubits. During computation, a qubit register is theoretically in a pure state, but quantum decoherence will gradually turn it into a mixed state. A pure state is also presented as a quantum state where we have exact information about the quantum system. This information corresponds to the famous $\psi$ vector in the Hilbert space. When preparing a quantum state, we indeed know the parameters of the vector $\psi$ even though actual property measurements will generate random results if the quantum state is not measured along with one of its eigenstates. The information we have about measurement potential results is their probabilistic distribution.

**Mixed states** are weird beasts. Literally, these are "*statistical ensembles of classical probabilistic combinations of pure states*", these being usually computational basis states, but they can also be expressed as real number linear combinations of any pure states. Basis states and pure states describe the information available for a single quantum object or qubit, or a group of such objects. A mixed state describes a large number of such systems, prepared in a similar manner, and the states they could be in when repeating an experiment followed by some measurement.

However, a pure state measurement generating random results most of the time, we still also experimentally prepare and measure it on a repeated basis to have an idea of its state probability distribution. In the end, both pure states and mixed states describe the information we can extract from a system after doing repeated experiments and measurements. Their difference lies with the origin of measurement randomness. Its origin is entirely quantum for pure states and both quantum and classical (or "non-quantum") for mixed states. Got it? If not, we have a couple practical examples below to figure out what it looks like in the real world!

Typically, mixed states provide the available information describing two sorts of systems:

**Random quantum objects** like photons coming from an unpolarized photons source, or, when photons with different polarities are merged like in the below illustration on the right. The photon polarization at this point is a statistical mixture of horizontal and vertical polarization photons. Let's say this is the case where quantum objects that are prepared differently and are then mixed together. The two sources are not "coherently" prepared. In the example in the left, a 45° polarizing beam splitter applied to horizontalized prepared photons produces superposed H and V photons in a pure state. On the right, the polarizing beam splitter creates 50% vertically and 50% horizontally polarized photons that can be merged by a 45° non-polarizing beam splitter. They are statistically merged, but not superposed, thus creating a mixed state.

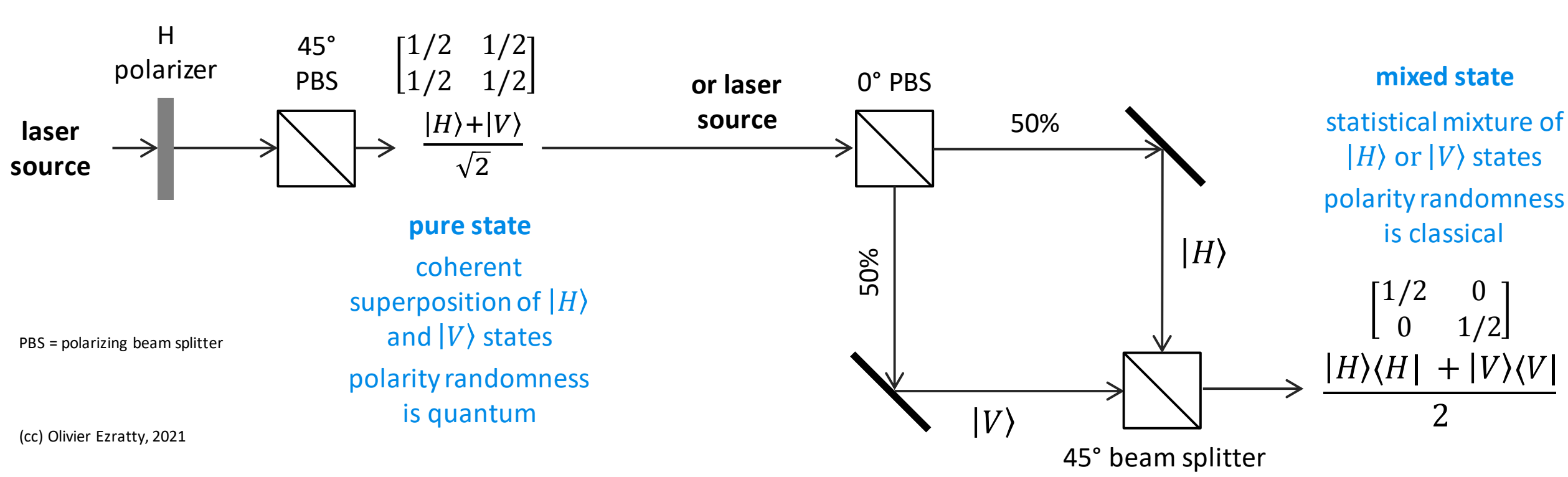

*Figure 142: how to generate mixed states with photons. (cc) Olivier Ezratty, 2021.*

In this other example, two lasers are preparing coherent light that is polarized respectively horizontally and vertically and then merged by a beam combiner. The resulting photons represent a totally mixed state with uncorrelated and incoherent photons. Their statistical distribution is entirely classical with a density matrix void of any off-diagonal values.

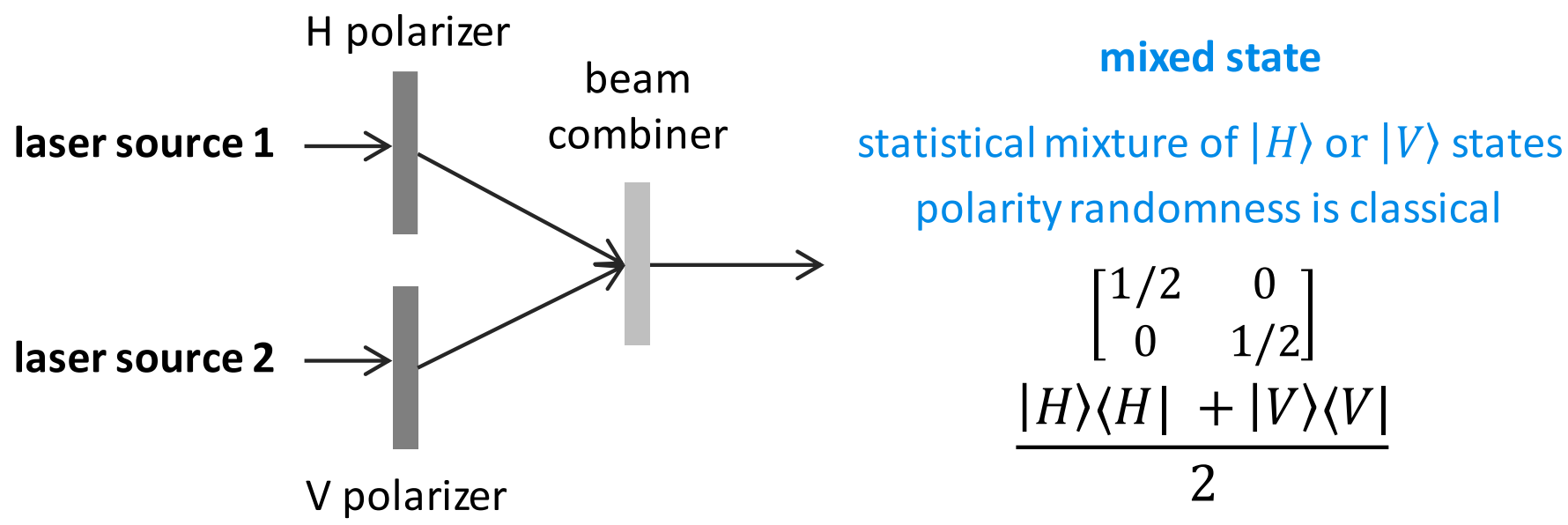

*Figure 143: another method to generate a mixed state with photons. (cc) Olivier Ezratty, 2021.*

**Subsystems** of an inseparable entangled system of several quantum objects. It helps understand what we are measuring at the end of computing when the resulting qubits are still entangled. One particular case is a set of qubits affected by decoherence coming from interactions with the environment. It helps understand the effect of decoherence on the state of a qubits register during computing and how error correction codes are mitigating it. Decoherence comes from the entanglement between a system and its environment, thus, the observed system is not yet isolated and becomes a subsystem of a larger entangled system. Thus, it becomes a mixed state. Want to grasp it clearly? You need to toy with density matrices representations of these pure and mixed states.

Note that these concepts are applicable to both a single qubit and a register of N qubits.

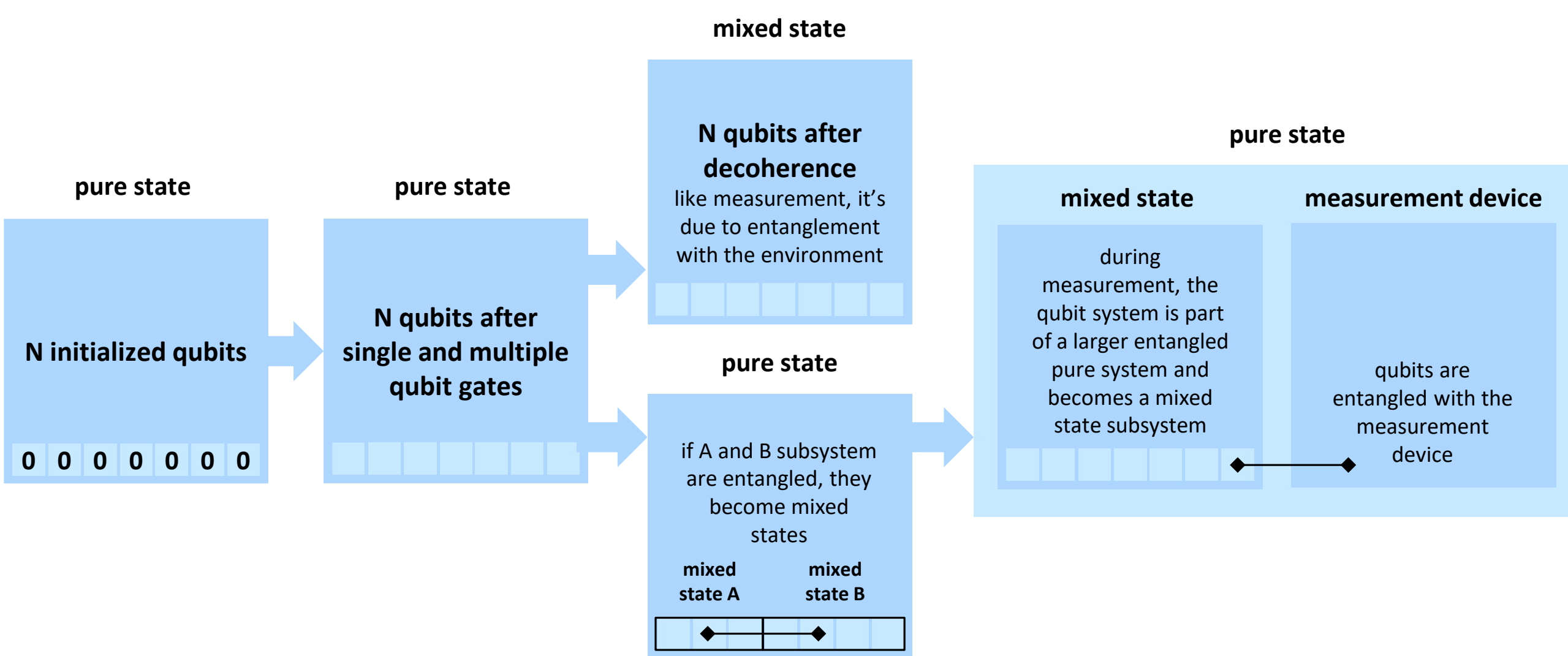

*Figure 144: mixed states and pure states when using qubits. (cc) Olivier Ezratty, 2021.*

## Density matrices

Density matrices, also named density operators, were introduced in 1927 by **John von Neumann** and **Lev Landau** and later expanded by **Felix Bloch**. Von Neumann created this formalism to develop his theory of quantum measurements.

A density matrix is a mathematical tool used to describe quantum systems in pure or mixed states. Compared to the state vector that we saw earlier, a density matrix is the only way to mathematically describe a mixed state. It consolidates all the physically significant information that could be retrieved from a set of quantum objects given what we know about them. Quantum and classical probabilities are boiled in the density matrix.

Usually represented by the sign ρ (rho), a density matrix is a square matrix of complex numbers used to describe a quantum system, like a register of several qubits. Its size is $2^N$x$2^N$ where N is the number of qubits in the register.

The density matrix of a quantum register in **pure state** is the outer product of its computational basis state vector |Ψ⟩⟨Ψ| as described below, with an example using a Bell pair of two qubits. There is no more information in the density matrix than in the basis state vector at this stage.

A density matrix for a **mixed state** adds several pure states matrices with real probability coefficients $p_i$. The $|\Psi_i\rangle$ pure states that are combined to form a mixed state can be themselves states from the computational basis (combination of 0s and 1s) but not necessarily. They can be any vector in the $2^N$ Hilbert space and made of (normalized) linear superpositions of these basis states. Mathematically speaking, a pure state density matrix is a special case of mixed state density matrix where only one $p_i$ is not zero.

We'll repeat here what was said with pure and mixed states: a mixed state density matrix consolidates both **quantum uncertainties** (that persists even when the system state if well known) and **classical uncertainties** (due to a lack of knowledge of individual quantum sources and preparation conditions) when a pure state density matrix contains only information pertaining to **quantum uncertainties**.

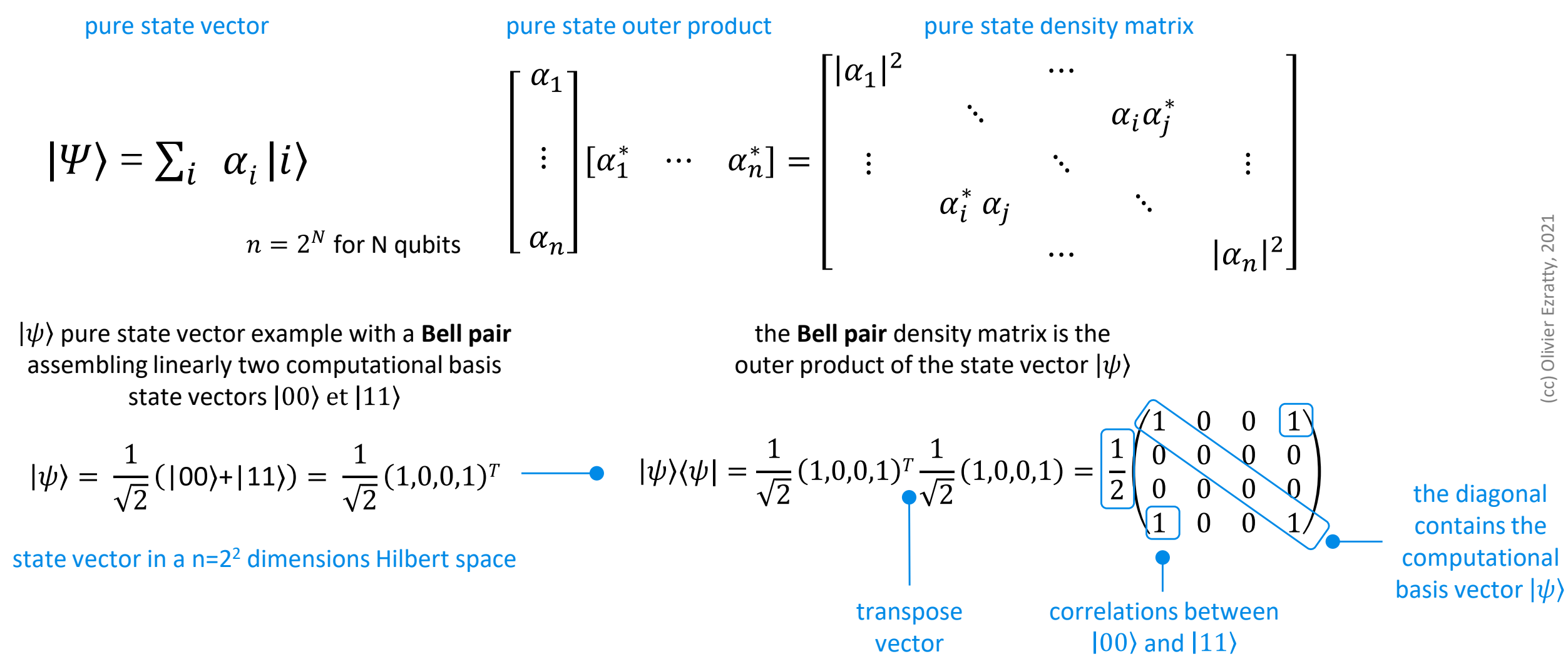

*Figure 145: how a pure state matrix is constructed. (cc) Olivier Ezratty, 2021.*

A density matrix has several mathematical properties as described *below* and detailed afterwards with some differences between pure and mixed states density matrices.

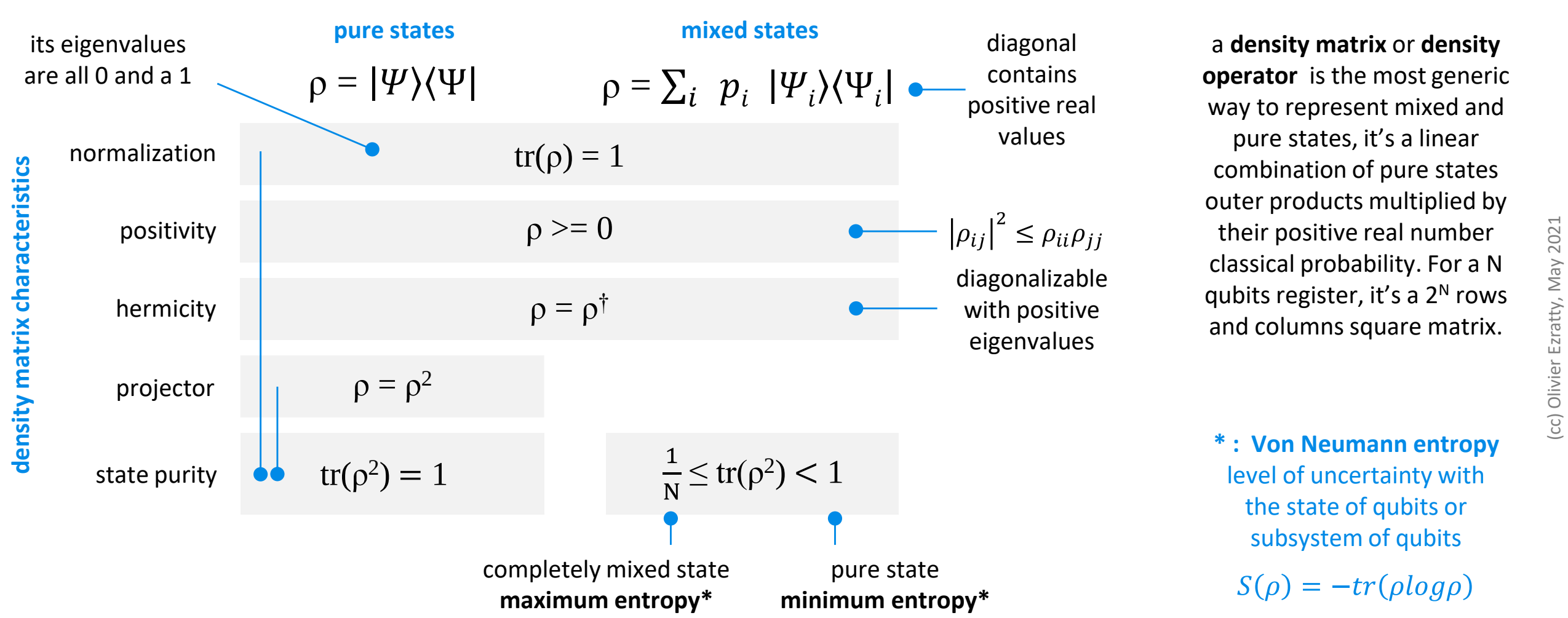

*Figure 146: the various mathematical properties of pure and mixed states density matrices.*

**Hermicity**. A density matrix is Hermitian, meaning that it's equal to its transconjugate matrix. As a consequence, the density matrix can be diagonalized in a different basis, with positive real number eigenvalues. Hermicity comes from the density matrix construction: it's real number linear sum of Hermitian matrices resulting from the Hermitian inner product of pure states vectors. One consequence is that it removes any global phase from the quantum system it describes. You can easily understand it by evaluating on your own a density matrix of a given qubit and its global phase.

**Positivity**. A density matrix M is positive semi-definite, meaning that $\langle x|M|x\rangle \geq 0$ for all x vectors. It's also defined as a symmetric matrix with non-negative eigenvalues (meaning... positive or zero). These eigenvalues being the values in the diagonal after matrix diagonalization. But even before diagonalization, all density matrices diagonal values are positive due to hermicity and the way they are constructed as positive probabilities combinations of outer products of pure states whose diagonal are always containing positive values.

**Normalization**. A density matrix trace equals 1 for both pure and mixed states. A density operator is said to be "*normalized to unit trace*". That's the sum of its diagonal values which are all positive real numbers. It comes from two rules: Born's rule applied to a pure state ($\sum_i \alpha_i^2 = 1$) and classical probabilities rules applied to the mixed state ($\sum_i p_i = 1$). As a result, a density matrix diagonal value at position j = $\sum_i p_i \alpha_{ij}^2$, $\alpha_{ij}$ being the weight $\alpha_j$ from the pure state i composing the mixed state. The diagonal is also referred to as a statistical mixture or as a population.

There are some differences between pure and mixed states density matrices.

**Projector**. A pure state density matrix is a projector, i.e. equal to its square and the trace of its square density matrix $\rho^2$ is equal to 1. Being a projector means that its eigenvalues are all zeros except a single one that is 1. The eigenvector associated with the eigenvalue one is the state vector of the system. Being a projector means the density matrix can be used as the way to measure a quantum state using this vector as a basis reference. In a single qubit system and the Bloch sphere, it would be any vector in the sphere and the related measurement observable, a geometrical projection of the evaluated qubit on this vector. In the case of a mixed state, the density matrix trace is inferior to 1 and its minimum is 1/N, when the state is maximally mixed with equal probabilities for all basis values. The average value obtained with applying an observable A to a pure state quantum system state vector $\psi$ is evaluated with the formula $\langle\psi|A|\psi\rangle$, also named an expectation value. In other words, it's the dot vector product of $\psi$ and the vector obtained by applying matrix A to vector $\psi$. The expectation value of a mixed state represented by a density matrix $\rho$ is $tr(\rho A)$, a trace of the density matrix multiplied by the observable A matrix.

**Off-diagonal elements** can have a time-dependent phase that will describe the evolution of coherent superpositions. These elements are also named "coherences". As decoherence starts due to interactions with the environment, any pure state will progressively turn into a mixed state and the off-diagonal values will be affected. This evolution follows the Liouville–von Neumann equation.

$$\frac{|0\rangle\langle 0| + |1\rangle\langle 1|}{2} = \frac{|+\rangle\langle +| + |-\rangle\langle -|}{2} = \begin{bmatrix} 1/2 & 0 \\ 0 & 1/2 \end{bmatrix} = \frac{1}{2}\mathbb{I}$$

**Mixedness** defines how much "mixed" is a quantum state defined by its density matrix. It's computed with $tr(\rho^2)$ and is equal to 1 for a pure state and 1/N for a mixed state with N quantum objects. As a result, any time-dependent unitary transformation U applied to this quantum state won't affect the mixedness. Indeed, the density matrix over time is $\rho(t) = U(t,t_0)\rho(t_0)U^\dagger(t,t_0)$. Its mixeness is $tr(\rho^2(t)) = tr(U(t,t_0)\rho(t_0)U^\dagger(t,t_0)U(t,t_0)\rho(t_0)U^\dagger(t,t_0))$ which equals $tr(\rho^2(t_0))$.

**Combinations**. A mixed state can be the result of an infinite number of combinations of pure states, the most common example being, for two qubits, the half-identity mixed state being an equally mixed state of both $|0\rangle$ and $|1\rangle$ or $|+\rangle$ and $|-\rangle$. Given a density matrix, you can't compute the pure states that were combined to create it. Said otherwise, quantum states with the same density matrix can't be distinguished operationally (i.e. by a set of measurements). Also, when a unitary operation $U$ (defined later, sorry) is applied to a mixed state defined by its density matrix ρ, the resulting state density matrix is $U\rho U^\dagger$.

For the fun of a better understanding, I've added below in Figure 147 a graphical segmentation of all the various matrix types we've been mentioning in the previous pages and how they are related with each other.

We forgot to define a **non-defective matrix**, which is a diagonalizable matrix. And a normal matrix A verifies $AA^\dagger = A^\dagger A$. A **trivial** matrix is both Hermitian and unitary and have orthonormal eigenvectors with eigenvalues being +1 or -1.

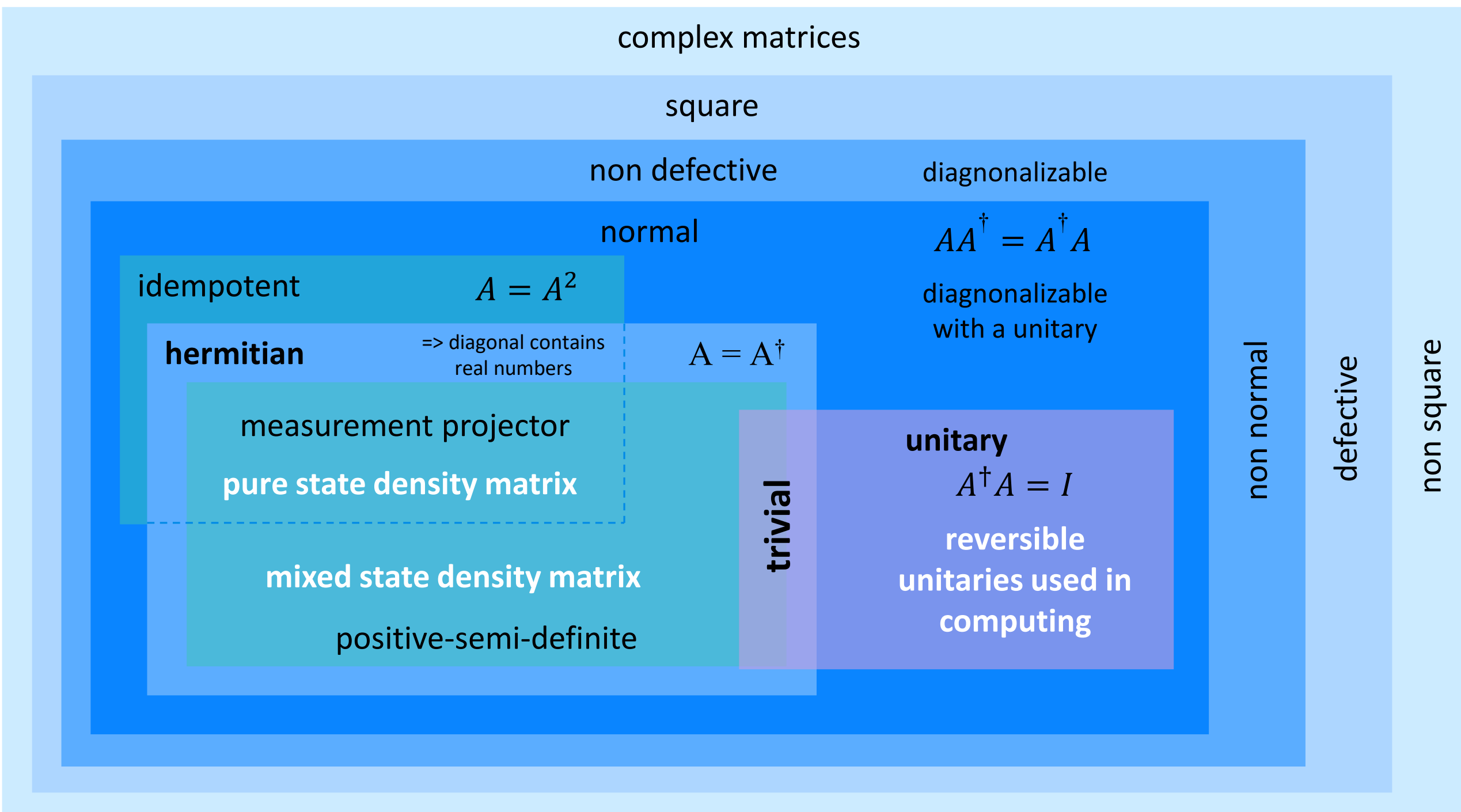

*Figure 147: a Russian dolls map of matrices. (cc) Olivier Ezratty, 2021.*

**Single qubit mixed states** can be represented by a point inside the Bloch sphere as shown below in a "Death Star" representation, with a statistical mixture of two pure qubit states. The mixed state is a convex sum of pure states inner products, 'convex' meaning it's a sum using positive real coefficients that sum up to 1. The geometric representation is a good way to figure out why a given mixed state can result from an infinite number of combinations of two pure states. We can combine more than two pure states to create a mixed state. By the way, the Bloch sphere becomes a Bloch ball.

**Density matrix dimensionality**. Although it contains $2^{2N}$ complex values, due to normalization, the dimensionality of a density matrix is $2^{2N}$-1 real numbers. The explanation is reconstructed below. For a starter, we have $2^{2N}$ complex values which is the square or $2^N$, the number of lines and columns in the density matrix. We separate the matrix diagonal from the off-diagonal values. The diagonal values are real numbers because they are the positive probability sums of the diagonal values of pure states density matrices, themselves being positive as $|\alpha_i|^2$.

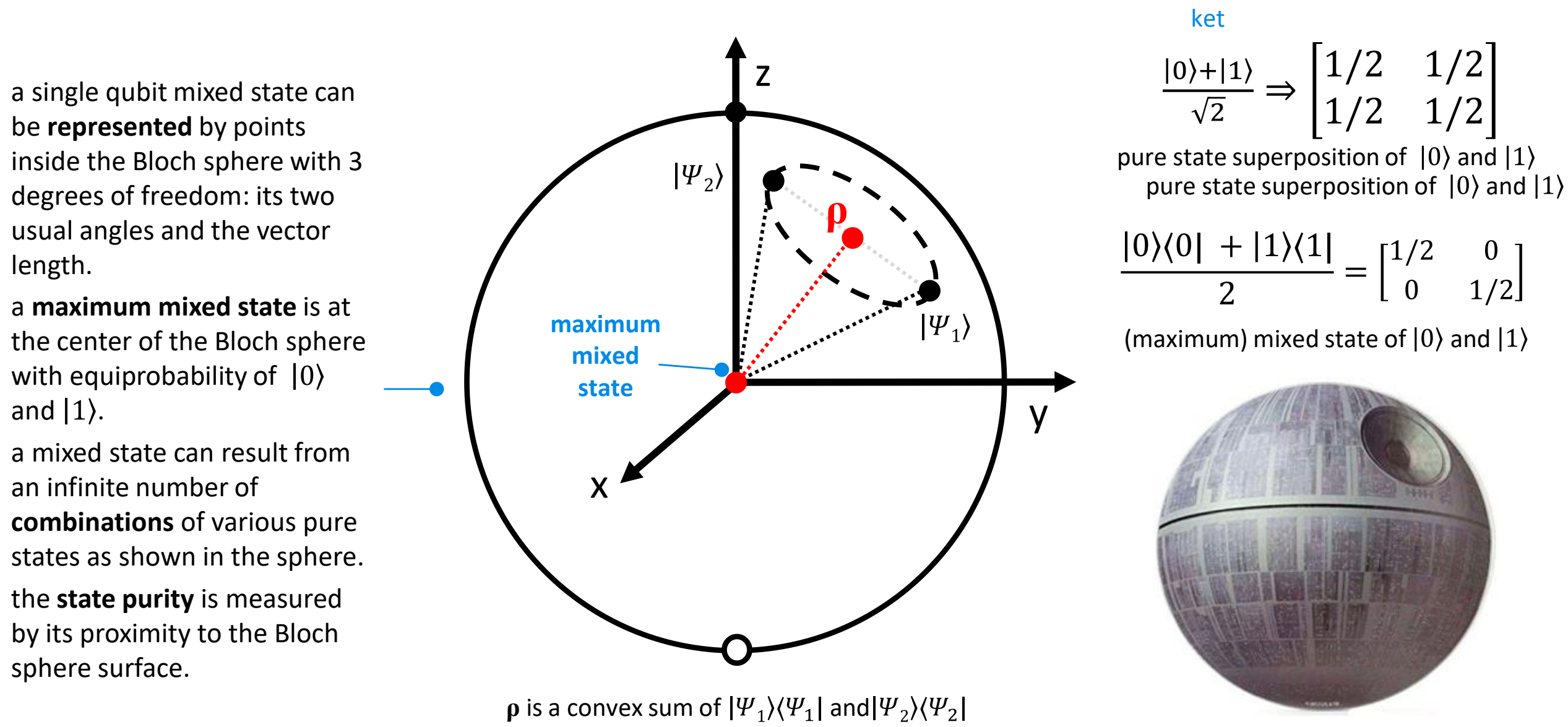

*Figure 148: representation of a single qubit mixed state in the Bloch sphere. (cc) Olivier Ezratty, 2021.*

The matrix trace equals 1, removing another useful dimension. The off-diagonal values are redundant since the matrix is equal to its transadjoint. So, we divide by two their dimensionality. Since these are complex numbers, we multiply it by two to get a number of real numbers. When summing this up, we find $2^{2N}$ -1 different real numbers. This dimensionality is usually presented as $2^{2N-1}$ complex numbers or $2^{2N}$ real numbers, avoiding the minus 1 which is quickly negligible as N grows.

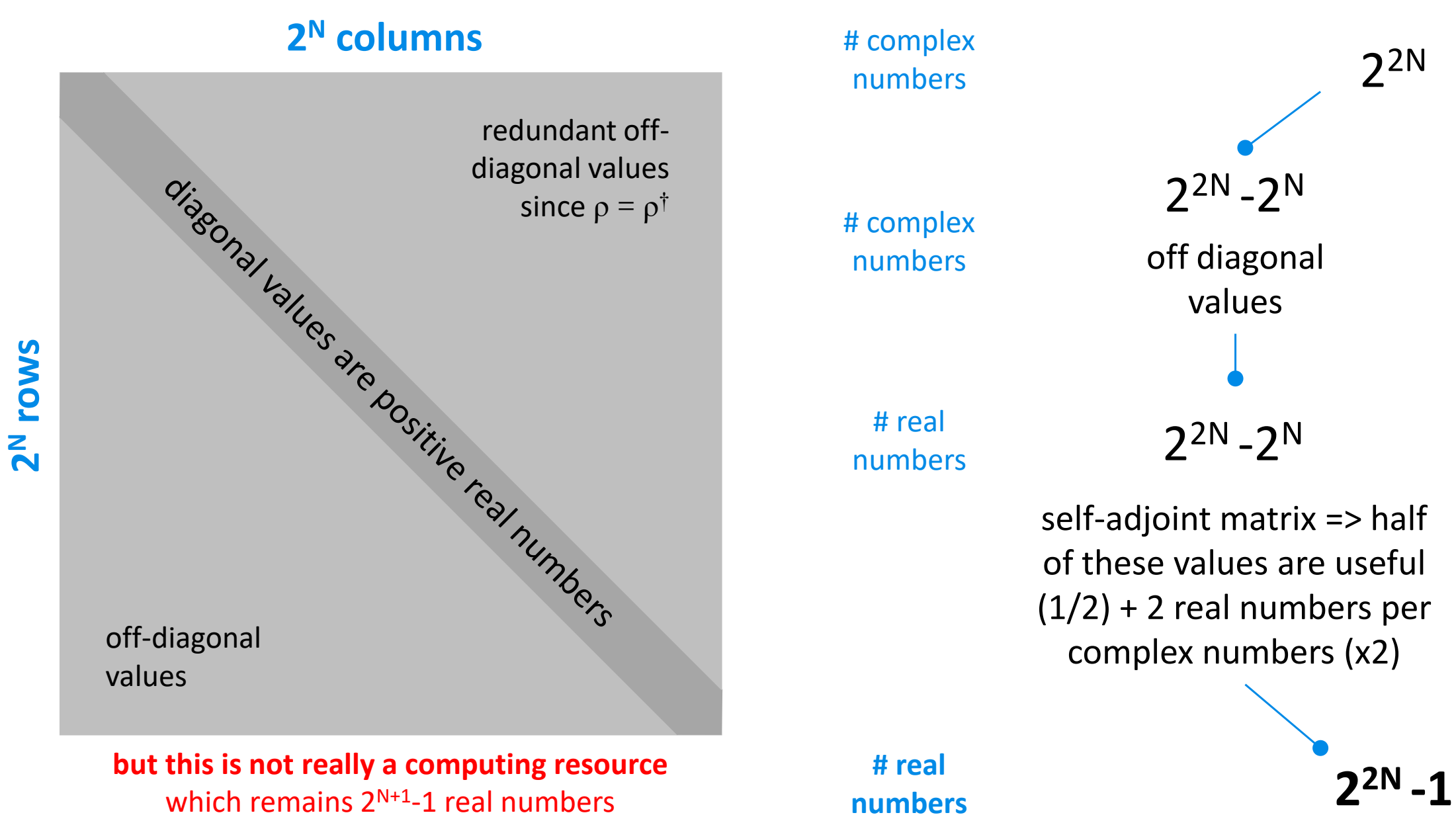

*Figure 149: computing the dimensionality of a density matrix. (cc) Olivier Ezratty, 2021.*

However, this dimensionality does not correspond to some useful computing resource in standard gate-based programming models although some work has been done to exploit it, but with no additional computing acceleration[288].

[288] See Quantum Circuits with Mixed States by Dorit Aharonov, Alexis Kitaev and Noam Nissam, 1998 (20 pages). It describes a model using not only unitary matrix operator-based quantum gates. It enables the usage of subroutines in programming. But this programming model doesn't seem adopted so far except for quantum error correction codes which implement measurement during computing. Mixed states based programming is implemented in the qGCL extension of the language pGCL as described in Quantum programming with mixed states by Paolo Zuliani, 2005 (14 pages).

A theoretical perfect gate-based quantum computer is using qubits registers that are in a pure state until measurement, representing thus a dimensionality of $2^{N+1}$-1 real numbers, the -1 standing for the normalization constraint of the computational basis vector[289]. So why do we care about these density matrices for mixed states? These are mostly used to understand the effects of decoherence and measurement and with qubits registers tomography which helps determine their fidelities.

The sequence of quantum gates in a quantum circuit can also be represented by a large unitary matrix of dimension $2^N*2^N=2^{2N}$ complex numbers. So, with a dimensionality close to a density matrix. But this is not an actual computing resource. It deals more with the extensive computing resources required to emulate in-memory an entire unitary algorithm in a classical computer instead of just executing gates one by one on the computational state vector.

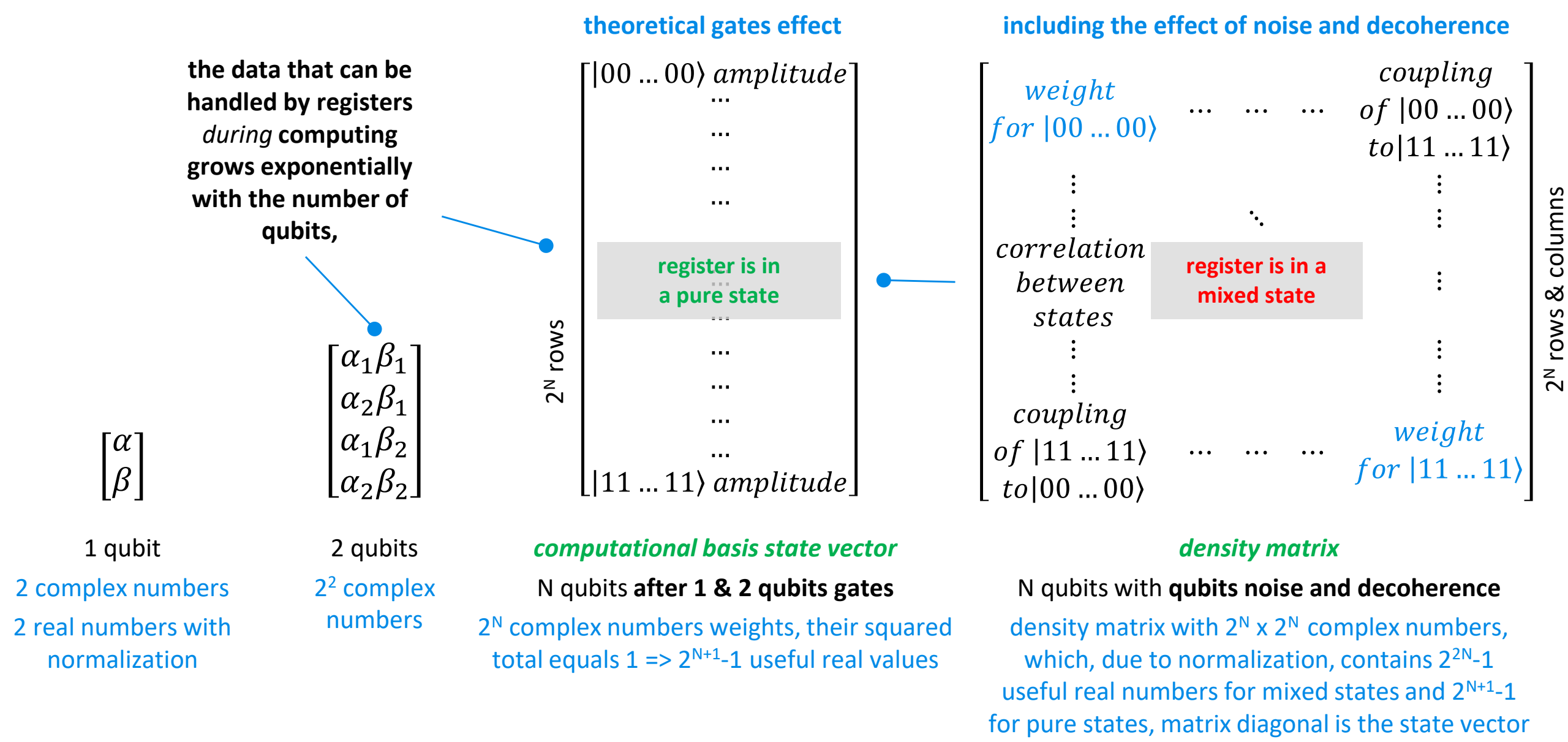

*Figure 150: dimensionality of a qubit register. (cc) Olivier Ezratty, 2022.*

There are many other subtleties with density matrices that we can't detail in the book. For example:

**Diagonalization** is possible for any mixed state density matrix. It will decompose the state into classical probabilistic combination of pure states eigenvectors forming an orthonormal basis.

**Reduced density matrices** are the density matrices of subsystems of composite systems. The reduced density matrix for an entangled pure state is a mixed state or mixed ensemble.

**Mixed state purification** consists, inversely, in integrating a mixed state in a larger system to create or reconstruct a pure state. It is used in some error-correcting codes.

**Bipartite pure states** are tensor products of two systems that are not entangled. A pure state system is entangled if and only if some of its reduced states are mixed rather than pure. If all were pure, it would mean that the pure state density matrix ρ would be separable into several pure states, one for each qubit in the case of a qubits register.

**Schmidt decompositions** are used to decompose bipartite systems and evaluate their level of entanglement. This level of entanglement can be determined with the Schmidt coefficients coming from the Schmidt decomposition.

---

[289] Thus, wrong is the statement that "*A calculation using n number of qubits on a quantum computer would need $2^n$ classical bits on a standard computer*" as seen in [Simulating subatomic physics on a quantum computer] by Sarah Charley, October 2020. Why? Because one of the $2^N$ quantum amplitudes in a N qubit register cannot be stored or emulated on a single bit!

**Matrix rank**. A matrix rank is the number of non-zero values in its diagonalized version. The rank of a density matrix gives an indication of the purity of the state it represents. A pure state density matrix has a rank 1, since it can be diagonalized into a matrix where only one value in the diagonal is non-zero. A maximally mixed state has a rank of $2^N$, i.e. the number or lines and columns in the density matrix representing N qubits.

**Schmidt rank** is an indication of the level of entanglement in a density matrix. Not to be confused with the matrix rank which deals with its purity level.

**Quantum Channels** are transformations of a quantum state resulting from any kind of interaction with a quantum environment. They are modelized with an operator, called a superoperator, transforming a density matrix into another density matrix. Technically speaking, a superoperator is a completely positive (we've defined that already) and trace-preserving operator (self-explainable), or CPTP. Its form is a linear map from one Hilbert space to another Hilbert space. Its dimension is a square matrix with $2^{2N}$ columns and as many rows, so with $2^{4N}$ (or $16^N$) complex numbers, before normalization, N being the number of qubits. It is useful to modelize quantum subsystems (which are in mixed state), decoherence, quantum error correction and qubits noise[290]. It is even possible to build a tomography with a superoperator, aka a quantum process tomography (QPT). One for example can build a QPT of a quantum gate to detect its imperfections. A QPT can also be done for a more complex operation, or unitary applied to a set of qubits, like a Quantum Fourier Transform[291].

### Grad, curls and divs

In the equations of Maxwell, Schrödinger, Dirac and others that we have seen are used notations good to remember here around the symbol nabla: $\boldsymbol{\nabla}$, sometimes used with an arrow $\vec{\boldsymbol{\nabla}}$.

Nabla generally designates the gradient of a scalar or vector function, i.e. its first derivative. A scalar function applies to a vector, often of three dimensions x, y and z of a Euclidean space. It returns a number. A vector function returns a vector! This leads to the notions of **gradient** and **Laplacian** which apply to a scalar function and correspond to first and second derivatives in space, and to divergence and rotational (or curl) which apply to a vector function. A Laplacian can also be applied to a vector function. We won't go far in this book with respect to these functions.

$$\boldsymbol{\nabla} = \left(\frac{\partial}{\partial x}, \frac{\partial}{\partial y}, \frac{\partial}{\partial z}\right)$$

**del** or **nabla** operator, first space derivative of a vector

$$\boldsymbol{\nabla} \boldsymbol{f} = \left(\frac{\partial f}{\partial x}, \frac{\partial f}{\partial y}, \frac{\partial f}{\partial z}\right)$$

scalar function **gradient**, scalar field vector of space variations

$$\boldsymbol{\nabla}.\vec{\mathrm{G}} = \left(\frac{\partial G_x}{\partial x}, \frac{\partial Gy}{\partial y}, \frac{\partial Gz}{\partial z}\right)$$

vector function **divergence**, showing its local evolution

$$\boldsymbol{\nabla} \times \vec{\mathrm{G}} = \left(\frac{\partial G_z}{\partial y} - \frac{\partial G_y}{\partial z}, \frac{\partial G_x}{\partial z} - \frac{\partial G_z}{\partial x}, \frac{\partial G_y}{\partial x} - \frac{\partial G_x}{\partial y}\right)$$

**rotational** or **curl** of a vector function G transforming a vector field in a vector field describing the field variation in space

$$\boldsymbol{\nabla}^2 \boldsymbol{f} = \boldsymbol{\nabla}.\boldsymbol{\nabla} \boldsymbol{f} = \left(\frac{\partial^2 f}{\partial x^2}, \frac{\partial^2 f}{\partial y^2}, \frac{\partial^2 f}{\partial z^2}\right)$$

scalar function **Laplacian**

$$\boldsymbol{\nabla}^2 \vec{\mathrm{G}} = \left(\frac{\partial^2 G_x}{\partial x^2} + \frac{\partial^2 G_x}{\partial y^2} + \frac{\partial^2 G_x}{\partial z^2}, \frac{\partial^2 G_y}{\partial x^2} + \frac{\partial^2 G_y}{\partial y^2} + \frac{\partial^2 G_y}{\partial z^2}, \frac{\partial^2 G_z}{\partial x^2} + \frac{\partial^2 G_z}{\partial y^2} + \frac{\partial^2 G_z}{\partial z^2}\right)$$

vector function **Laplacian**

*Figure 151: del, nabla, gradient, di vergence, rotational, curl, Laplacian. You won't need them in the rest of this book, sort of. This is just informative.*

[290] See Quantum Channels by Stéphane Attal (65 pages).

[291] See Quantum Process Tomography of the Quantum Fourier Transform by Yaakov S. Weinstein, Seth Lloyd et al, 2004 (45 pages).

### Permanent and determinant

This inventory would not be complete without describing an even stranger mathematical object: the **permanent** of a square **matrix** n*n, invented by Louis Cauchy in 1812. The formula in Figure 152 describes its content.

$$\mathrm{per}(A) = \sum_{\sigma \in \mathfrak{S}} \prod_{i=1}^{n} a_{i,\sigma(i)}$$

Figure 152: a permanent.

The Π denotes a multiplication of values from the index matrix i and σ(i). σ is a permutation function of integers between 1 and n, the dimension of the matrix (number of columns and rows). The sigma relates to the set of σ functions of the permutation group $S_n$ (also called symmetrical group) which has a size of n! (factorial of n). The values $a_{i,\sigma(i)}$ are the cells of the coordinate matrix i and σ(i).

Here is what it gives with n=2 and n=3 knowing that beyond that, it becomes less readable:

$$\mathrm{perm}\begin{pmatrix} a & b \\ c & d \end{pmatrix} = ad + bc \qquad \mathrm{perm}\begin{pmatrix} a & b & c \\ d & e & f \\ g & h & i \end{pmatrix} = aei + bfg + cdh + ceg + bdi + afh$$

Figure 153: computing the permanent of 2x2 and 3x3 matrices.

The permanent is therefore a real number resulting from n! (factorial of n) additions of multiplications of n values of the matrix. The permanents are notably used to evaluate matrices that represent graphs.

They are also used in the classical numerical simulation of boson sampling that we will describe in the section dedicated to photon qubits, page 445[292]. Contrary to the calculation of the determinant, in Figure 154*,* which can be simplified, that of the permanent remains a classical intractable problem.

The **determinant of a matrix** is a variant of its permanent. sgn(σ) is the sign of permutations, which is +1 if the number of permutations needed to create the permutation is even and -1 if it is odd. Olé!

$$\det(A) = \sum_{\sigma \in Sn} \left( \mathrm{sgn}(\sigma) \prod_{i=1}^{n} a_{i,\sigma(i)} \right)$$

Figure 154: a determinant.

And this is what it gives for n=3. Note that the group of permutations includes the permutation that does not change the order of the elements.

$$\det\begin{pmatrix} a & b & c \\ d & e & f \\ g & h & i \end{pmatrix} = aei + bfg + cdh - ceg - bdi - afh$$

Figure 155: computing the determinant of a 3x3 matrix.

Determinants have particular properties such as det(AB)=det(A).det(B)=det(B).det(A)=det(BA) which can facilitate the calculation of the determinant of a matrix if it can be factorized into several matrices. Also, the determinant of a matrix is the product of its eigenvalues.

So much for the definition of the basics of the linear algebra of quantum computing. I've skipped a lot of other definitions and rules of computation. It was a question of clarifying certain notions that are frequently used in the scientific literature on quantum computing and in many of the reference works cited in this book. What we have just seen may be useful for you to compare some of the scientific literature on quantum computing.

If you like maths, linear algebra and complexity, you can have some fun exploring type III factors algebra that describes the observables in relativistic quantum fields theory[293]! Classical quantum physics and computing is based on a simplistic type I factors algebra. Simpler, but still complicated.

---

[292] The calculation time of a permanent increases faster than an exponential of a fixed value (Mn) as soon as n becomes very large compared to M. So, for example, with M=2, 2n is much smaller than n! as soon as n is greater than 4. As the numerical simulation of the boson requires a determinant that depends on the size of the simulation, it is even more cumbersome to compute than an exponential problem.

[293] See The Role of Type III Factors in Quantum Field Theory by Jakob Yngvason, 2004 (15 pages).

### Fourier transforms

Since quantum physics deals a lot with wave-particle duality and particularly with waves, waves signals decomposition is a key mathematical tool. That's the role of a Fourier transform that we mentioned already when dealing with Heisenberg's indeterminacy principle. It's about maths but not linear algebra.

The Fourier Transform implements a mathematical decomposition of a function $f(x)$ into a function $\hat{f}(\xi)$ returning a complex number containing an amplitude and phase for single frequencies $\xi$. It's a more generic version of Fourier series which work with periodic signals. Fourier transform are Fourier series where the signal period can approach infinite.

It can be used for example to decompose a wave packet pulse signal that is concentrated in time. A Fourier transform usually operates in the time domain with $x$ being a time in second and $\xi$ a frequency in Hertz.

$$\hat{f}(\xi) = \int_{-\infty}^{\infty} f(x) e^{-2\pi i x \xi} dx$$

*Figure 156: a Fourier transform in the time domain.*

It can be decomposed using Euler's formula in its real and complex parts separating the amplitude and phase of the Fourier transformed signal:

$$\hat{f}(\xi) = \int_{-\infty}^{\infty} f(x) \cos(2\pi i x \xi)\, dx - i \int_{-\infty}^{\infty} f(x)\ \sin(2\pi i x \xi)\, dx$$

*Figure 157: Fourier transform decomposed in real and complex part.*

The inverse Fourier transforms that frequency decomposition function $\hat{f}(\xi)$ back into its original compound time domain signal $f(x)$.

$$f(x) = \int_{-\infty}^{\infty} \hat{f}(\xi) e^{2\pi i x \xi} d\xi$$

*Figure 158: inverse Fourier transform.*

All of this is easier to understand with examples like in the schema below decomposing a time domain signal into five frequencies constituents with their respective magnitude and (equal) phases.

Computing Fourier series and transforms is done in many ways:

**Discrete-time Fourier Transform** (DTFT) is a form of Fourier analysis that is applicable to a sequence of values. It is often used to analyze samples of a continuous function. The term discrete-time refers to the fact that the transform operates on discrete data, often samples whose interval has some units of time.

**Discrete Fourier Transform** (DFT) converts a finite sequence of equally-spaced samples of the function into a same-length sequence of equally-spaced samples of the Discrete-Time Fourier transform (DTFT). The samples are complex numbers coming from a DTFT.

**Fast Fourier Transform** (FFT) computes the discrete Fourier transform (DFT) of a sequence, or its inverse (IDFT). It's an efficient variation of the DFT.

**Quantum Fourier Transform** (QFT) is a linear transformation applied on qubits. It is the quantum analogue of the DFT and reverse DFT. A QFT is a Discrete Fourier Transform applied to the data stored in the $2^n$ computational basis states of a n qubits register. The Quantum Fourier Transform, implements a DFT on the complex amplitudes of a quantum state. We cover it later page 581.

Fourier series were created by **Joseph Fourier** (1768-1830, French) as part of his work in the book "The Analytical Theory of Heat" published in 1822. Beforehand, he accompanied Napoleon Bonaparte in his 1798-1801 Egyptian expedition as a scientific advisor. He then became a Prefect for the Isère department, based in Grenoble. Afterwards, he also drove the young Jean-François Champollion to get interested in deciphering the Rosetta Stone.

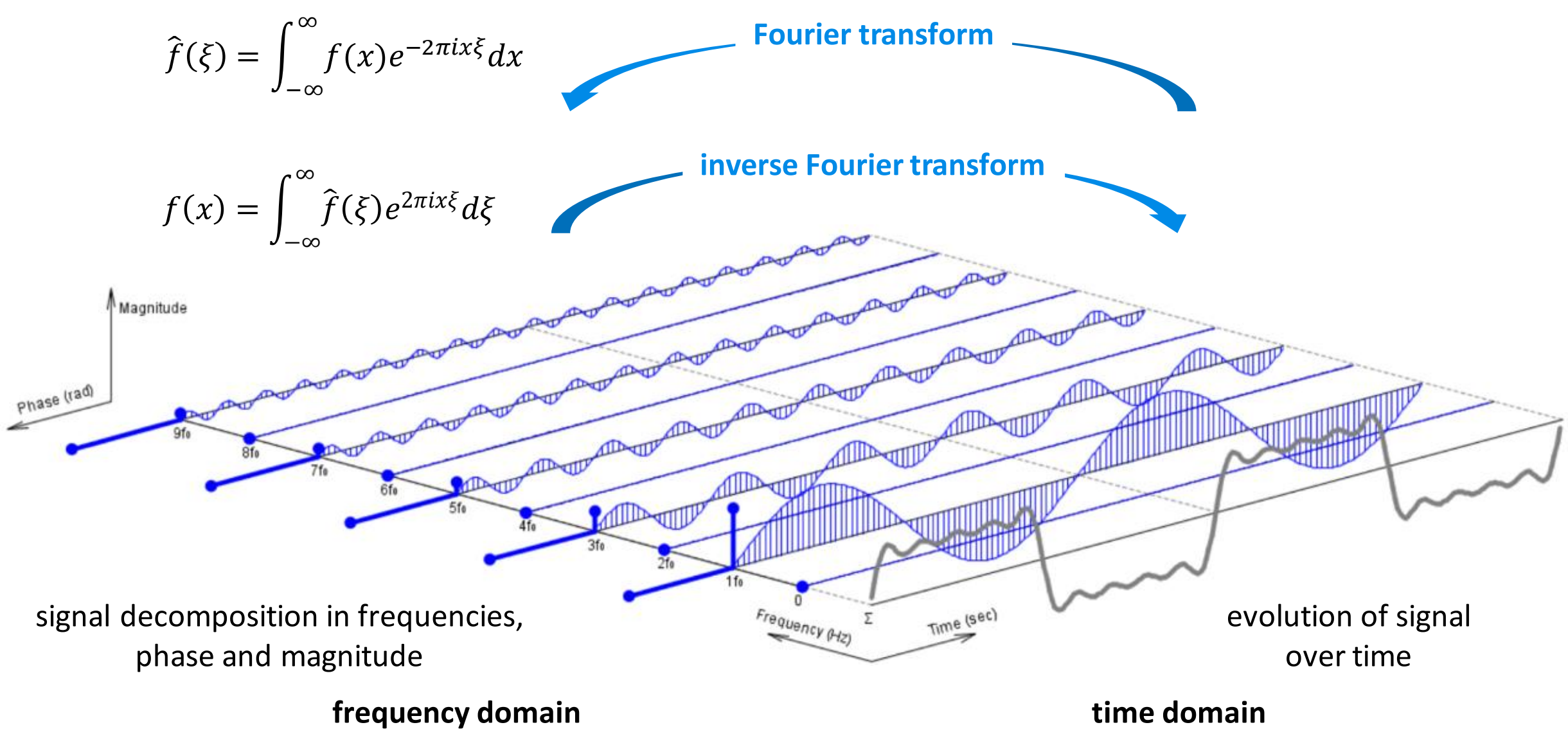

*Figure 159: Fourier transform and inverse Fourier transform and signal decomposition. Source: https://www.tomasboril.cz/files/myprograms/screenshots/fourierseries3d.png, comments (cc) by Olivier Ezratty, 2021.*

## Nonlinearities

We often hear about nonlinearities with quantum physics, particularly with the difficulty to implement it with qubits. It's also used in neural networks activation functions in classical computing. But their meaning is not the same in these different scenarios.

Superconducting qubits exploit the Josephson effect and an anharmonic oscillator to prevent the energy states of the superconducting loop oscillating current from being separated by the same energy level. This is a nonlinear effect linked to the way harmonic oscillators work when dampened in a certain way. It enables microwaves controls for changing qubits state between $|0\rangle$ and $|1\rangle$ with a larger frequency than the one that would allow a switch from the $|1\rangle$ state to the $|2\rangle$ state, which is what we are trying to avoid.

Nonlinearities are also sought after in photonics, especially to create quality two-photon quantum gates. Nonlinearities occur when solid media modify the characteristics of photons such as their polarization P and in a nonlinear way with respect to the electric field applied to the solid. The dominant chi $\chi^{(i)}$ of a nonlinear medium defines its order. A $\chi^{(3)}$ is a third order nonlinear medium.

$$P = \epsilon_0(\chi^{(1)}E + \chi^{(2)}E^2 + \chi^{(3)}E^3 + \cdots) \text{ with } \epsilon_0 \text{ being the vacuum permittivity.}$$

This phenomenon happens in the Kerr effect which sees some materials refractive index changing in a nonlinear (quadratic, second order $\chi^{(2)}$ medium) manner as a function of the electric field applied to them. Conversely, the Pockels effect used in optical modulators sees the refraction changed in a linear manner as a function of the electric field applied. This nonlinearity in optics also occurs in many devices such as power lasers.

Finally, nonlinearities are classically used in neural networks activation functions. These are, for example, sigmoid based on exponential fractions.

So how can such activation functions be performed in quantum computation that relies only on linear algebra? One of the first imagined solutions consists in using a nonlinear, non-reversible and dissipative quantum gate called D [294]. Others consists in handling the nonlinearity part of algorithms in their classical parts before feeding a quantum algorithm. That's what can be done in algorithms solving Navier-Stokes fluid mechanics equations.

---

[294] Method proposed by Sanjay Gupta in Quantum Neural Networks, 2001 (30 pages) and Quantum Algorithms for Deep Convolutional Neural Network, by Iordanis Kerenidis et al, 2020 (36 pages).

# Qubits

Qubits are the basic elements of data manipulation in quantum computers. They are the quantum equivalents of classical computing bits. With them, we move from a deterministic to a probabilistic world but with the capability to handle more information during computing.

In conventional computing, bits used in processing units like microprocessors correspond to circulating electrical charges that reflect the passage or absence of an electrical current. A classical bit has a value of 1 if the current is flowing or 0 if the current is not flowing. The logic is transistors based. A bit readout gives 1 or 0 and deterministically, i.e. if the read operation is repeated several times, or the read operation is repeated after a re-edition of the calculation, it will yield the same result. This is true for data storage of information, for its transport and processing. This is valid modulo the errors that can occur during this journey. These most often occur in storage and memory and are corrected via error correction systems using some data redundancy, usually with some parity bits for each stored byte, so with a rather low data overhead. In data storage, complicated redundancy systems are used like RAID disks organization mixing and matching several disks and parity error codes to consider the physical errors coming from storage.

In a qubit, everything is different! While qubits are usually initialized at $|0\rangle$, operations on them called quantum gates create a mathematical linear superposition between states $|0\rangle$ and $|1\rangle$. These two states correspond to two different discrete possible values of a physical property of a quantum object like an electron spin (up or down in a given direction), a photon polarization or an atom energy level. Qubits are represented mathematically by a vector in a two-dimension Hilbert space which describes its amplitude and phase, reminding us of the "wave" nature of quantum objects.

We'll see later how we use the Bloch sphere geometrical representation to understand how amplitude and phase are visualized. And it gets more complicated when we conditionally connect qubits together using multi-qubits quantum gates implementing quantum entanglement.

At the end of computing, we read the value of a qubit. Like all quantum object measurement, it results in a wave packet collapse onto one of the two qubit basis states. So, we get a $|0\rangle$ or a $|1\rangle$ and the result is probabistic, not deterministic. The wealth of information handled by a qubit during computing is lost at the end of calculation.

| | | bits: 0 or 1 | | qubits: 0 and 1 |
|---|---|---|---|---|
| states | mathematical bit | 2 possible exclusive states, 0 or 1 | mathematical qubit | linear combination of $|0\rangle$ and $|1\rangle$ |
| initialization | | 0 or 1 | | $|0\rangle$ basis state |
| dimensionality | | 1 binary digit => two possible values | | 2 real numbers => one point on Bloch sphere |
| modifications | | logic decision tables, irreversible | | reversible unitary transformations |
| readout | | 0 ou 1, deterministic | | 0 or 1, probabilistic |
| errors | | no error, perfect mathematical object | | no error, perfect mathematical object |
| physical implementation | physical bit | current/no current in a logic electronic circuit and two states memory device bit | physical qubit | quantum object with two exclusive states for one property |
| computing vs memory | | separate devices and parts of processors | | all done in qubits and with quantum gates |
| computing operations | | transistors based logic | | amplitude and phase change + entanglement |
| errors sources | | cosmic rays, transistors leakage | | decoherence, thermal, electromagnetic, radioactivity… |
| error levels | | $2.5 \times 10^{-11}$ bit per hour error rate | | usually $\gg$ 0,1% with qubit gates and readout |
| information redundancy | logical bit | parity bits | logical qubit | large number of physical per logical qubits, $10^2$ to $10^6$ |
| error correction | | ECC (memory), MCA (CPU), RAID (storage) | | quantum error correction codes |
| error level after correction | | negligible | | under an acceptable threshold for fault tolerance |

*Figure 160: detailed comparison between classical bits and qubits with separating the mathematical logic, the physical implementation and error correction techniques. (cc) Olivier Ezratty, 2021.*

The role of a quantum algorithm is to leverage this wealth of information during computing so that a simple result is generated at the end. We turn this probabilistic outcome in a deterministic one with executing the algorithm a great number of times, up to thousand times, and averaging the obtained results. It's also dependent on the structure of quantum algorithms which are designed to generate a result with qubits being as close as possible to their so-called "computational basis states", namely, $|0\rangle$ and $|1\rangle$.

To sort things out, it's still useful to differentiate three levels of 'qubit objects' used in computing as described in Figure 160:

**Mathematically**. Bits and qubits are idealized mathematical objects that implement a pure mathematical formalism with no errors. What is named a "qubit" is above all a mathematical object. Its dimensionality is different than with a bit. It's represented by two complex numbers, the amplitudes $\alpha$ and $\beta$ from the qubit quantum state description $\alpha|0\rangle+\beta|1\rangle$. Due to normalization ($\alpha^2+\beta^2=1$) and getting rid of the qubit global phase, its dimensionality becomes two real numbers, usually represented by two angles in the Bloch sphere. Bits and qubit measurement are both mathematical and physical operations. With qubits, it's mathematically based on a projective measurement on the computational basis comprised of $|0\rangle$ and $|1\rangle$, using a Hermitian matrix. Physically, it's using a measurement apparatus operating on the qubit quantum object.

**Physically**. Bits and qubits are implemented with different sorts of physical devices. With bits, we use to say they correspond to currents circulating or not circulating in transistor-based devices. While this is true with processing, this is different with memory and storage[295]. Qubits are implemented with quantum systems comprised of a single quantum object (atom, electron, photon) or several quantum objects (particularly with superconducting qubits and topological matter qubits like Majorana fermions). The $|0\rangle$ or a $|1\rangle$ states correspond to two exclusive states for one given property of a quantum object or system, that is clearly separable at measurement, like a photon polarization that is detected with a polarizer and a photon detector or an electron spin that can be detected with some magnetic sensor and a technique called electron spin resonance (ESR). These are also called two-level systems (TLS).

Physical qubits processing is using physical operations: **amplitude and phase changes** implemented by single-qubit gates and provoking **superposition** and **entanglement** which conditionally connects qubits together with two or more qubits gates, **interferences** resulting from the previous operations and are at the core of most quantum algorithms, and **quantum measurement** yielding $|0\rangle$ or $|1\rangle$ for each qubit when computing has ended or when executing quantum error correction codes. Both bits and qubit physical objects are prone to physical errors. While error rates are very small with classical bits, it's currently quite high with qubits. One simple operation like a two qubits quantum gate can generate over 0,4% error rates, which is unacceptable for most algorithms.

Qubits errors, namely decoherence, come from the various interactions between the qubit quantum objects and their environment like thermal noise, electro-magnetic noise, cosmic rays and gravity[296]. These errors require quantum error correction codes, which, as we'll later see, require a significant overhead of physical qubits.

---

[295] These rely on electronic systems storing information like some magnetic encoding in hard disks drives or with two states transistor-based objects in SRAM (used in processors), DRAM (used around processors) or Flash memory (used in SSD and your usual USB memory key).

[296] It explains why many qubit types requires some sort of isolation: vacuum and low temperature to avoid thermal and electro-magnetic noise and multi-layered shielding to avoid other sources of electromagnetic noise. But we'll see later that for superconducting and electron spin qubits, the required low temperature is also linked to the microwaves used to control qubits.

**Logically**. Error correction is thus required to create usable computing devices. In classical computing and telecommunications, "bits" are corrected with different techniques including using parity bits[297]. Bits are processed, stored and transmitted with a very low-level of errors.

Qubits must be assembled in groups called logical qubits, which are physical assemblies of a much great number of physical qubits, up to 10 000's[298]. Redundancy overhead becomes much bigger than with parity bits used in classical computing. In logical qubits, physical qubits are processed with quantum error correcting codes. The number of physical qubits assembled into logical qubits depends on their physical error rate and on the logical qubit error rate that is expected to enable practical quantum computing. For example, the famous integer factoring Shor algorithm is very demanding since using very precise small angles phase rotation gates.

While qubits are everywhere in quantum computing, these are not the only quantum objects available to manage quantum information.

Quantum computers can also theoretically be built with **qutrits** (with three possible quantum states), **ququarts** (with four possible states) and more generically, with **qudits** (d being the number of possible quantum states of the qubit underlying quantum system) [299].

It can deliver some computing power with a smaller number of quantum objects than with qubits. These are still mostly research labs tools. For example, researchers at Berkeley are investigating superconducting qudits with more than two levels[300].

The most common qudits are implemented with photons by managing several of their properties.

Using qudits would have an impact on quantum algorithms design and programming. Most of quantum algorithms are designed for quantum computers using qubit-based gates. However, compilers could probably automatically transform classical quantum gates into qudits-based gates.

The record so far is about creating quvigints, qudits with 20 different exclusive values for photons, that are efficiently measured with state tomography[301].

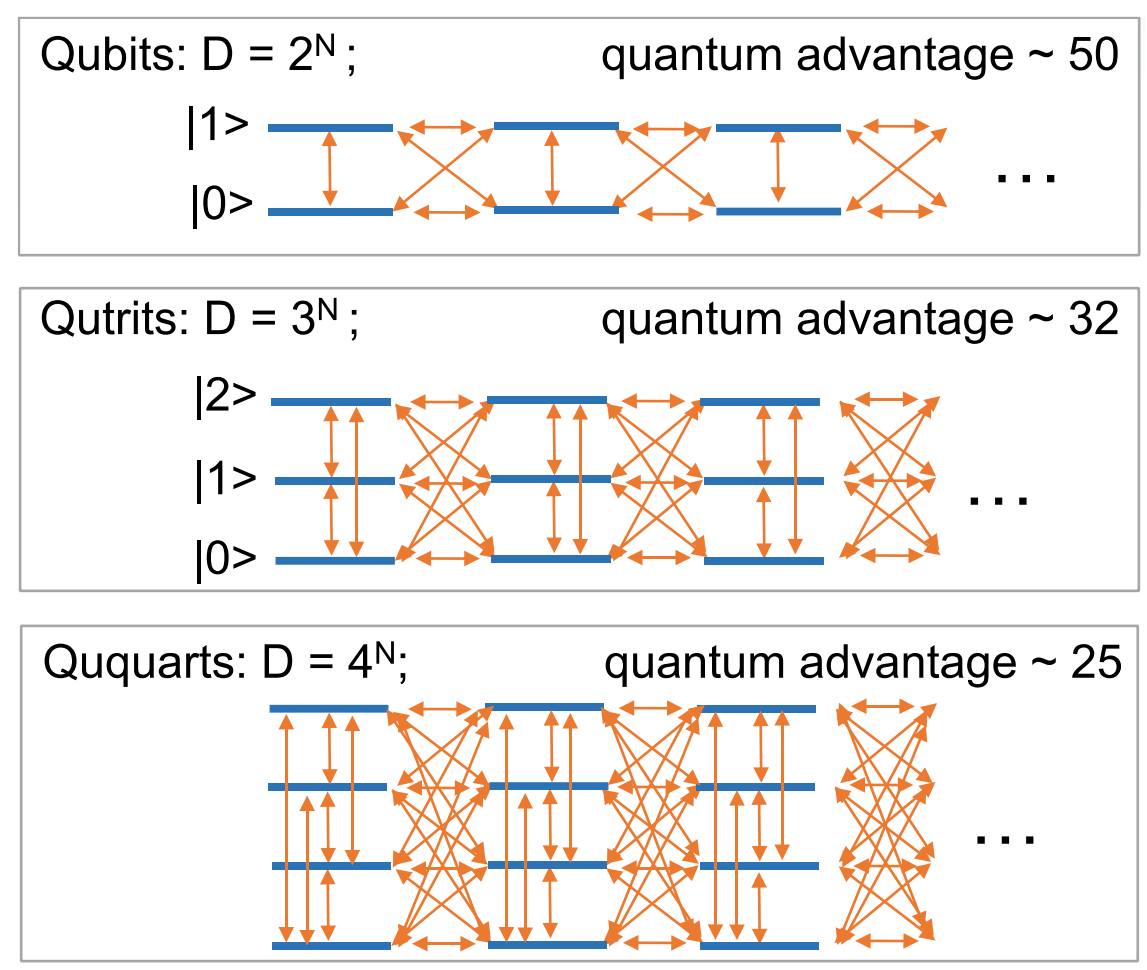

*Figure 161: qubits, qutrits and ququarts. Source: Quantum Simulations with Superconducting Qubits by Irfan Siddiqi, 2019 (66 slides).*

[297] ECC (error correcting codes) are used in memories. Some systems are used in processors like the Intel MCA (Machine Check Architecture) which detects and reports errors in microprocessor. Other systems correct errors in storage like RAID redundancy for hard-disk drives and SSDs. We also have error correction codes used in classical telecoms.

[298] As of 2021, there are no commercial computers using real logical qubits. The reason is simple: the number of available physical qubits in gate-based processing units, topping at 127 with IBM's last generation of superconducting qubits, is still *under* the number of physical qubits required to build just one logical qubit!

[299] See for example Ultracold polar molecules as qudits by JM Hutson et al, 2020 which deals with qudits using fluorine-calcium and rubidium-cesium diatomic molecules allowing four quantum levels per molecule. This reduces the number of necessary qubits of $\log_2(d)$, d being the number of state levels of the qubits.

[300] See Quantum Simulations with Superconducting Qubits by Irfan Siddiqi, 2019 (66 slides).

[301] See Finding quvigints in a quantum treasure map by University of Queensland, March 2021 and Robust and Efficient High-Dimensional Quantum State Tomography by Markus Rambach et al, March 2021 (6 pages).

## Bloch sphere

Let's first dig into the mathematical models of qubit representation. These models do not depend on the qubits underlying quantum object types. Physical qubit types have an impact on their error level and types as well as on the low-level quantum gates operations available to control qubits.

In a classical probabilistic model, a probabilistic pbit would have a probability p of having the value 0 and 1-p of having the value 1[302]. It would be a linear probabilistic model. We cover the niche market of probabilistic computers in a dedicated section, page 776.

Well, with qubits, these probabilistic laws are quite different!

A qubit vector state is defined by two complex numbers $\alpha$and β according to the formula describing the qubit quantum object state |Ψ⟩ as α|0⟩+β|1⟩. Quantumly speaking, |Ψ⟩ is a linear superposition of basis states |0⟩ and |1⟩ with coefficients α and β, aka amplitudes. α is a complex number whose square describes the probability of having the state |0⟩ and β is a complex number whose square describes the probability of having the state |1⟩.

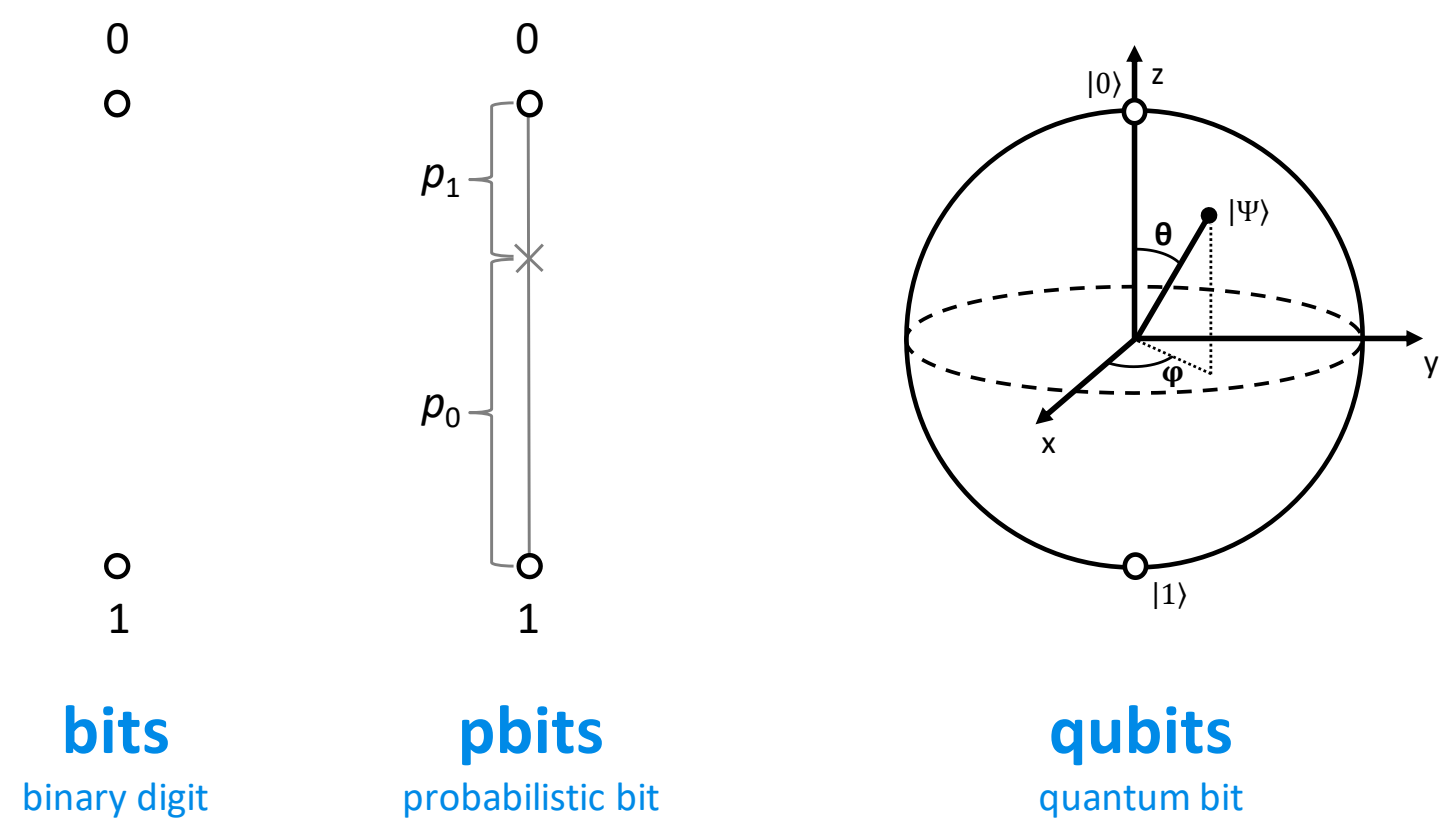

*Figure 162: bits, probabilistic bits and qubits.*

The sum of the probabilities of the two basis states must give 1. It is indeed not α+β but $\alpha^2+\beta^2$ that give 1. It comes from the generic probabilistic model developed by **Max Born** in 1926 and from one of the postulates of quantum physics. It gives to the square of the modulus of the wave function of a quantum the meaning of a probability density of the presence of an elementary particle in space (mostly, for electrons).

The mathematical representation model of the state of a qubit is based on complex numbers and on the geometrical metaphor of the famous **Bloch sphere**. This model is linked to the representation of the state of a qubit or any two-state quantum by a two-dimensional vector whose length, called "norm", is always 1.

**Angles**. The qubit state |0⟩ is a length 1 vector going from the center of the sphere to the North pole of the sphere and the state |1⟩ is a vector going from the center of the sphere to its South pole. An arbitrary qubit state $|\Psi\rangle$ is represented by a vector with an angle θ (0 to $\pi$, latitude) with respect to the vertical z-axis and an angle φ (0 to $2\pi$, longitude) with respect to the x-axis located from the center of the sphere to its equator and around the z-axis. θ corresponds to the qubit amplitude and φ to its phase.

**Orthogonality**. The basis states |0⟩ and |1⟩ are opposite in the Bloch sphere and are mathematically orthogonal. This is highly counter-intuitive and linked to the angle θ that is divided by two in the formulae. When θ equals $\pi$, corresponding to a half turn in the sphere, moving from |0⟩ to |1⟩, cos(θ/2) = cos(90°) = 0 illustrating the fact that |0⟩ and |1⟩ are indeed mathematically orthogonal states. This is true for any opposing states within the sphere as with the $|\Psi\rangle$ and $|\Psi'\rangle$ examples below in Figure 163.

[302] Linear probabilistic models are used in the probabilistic processors discussed in a small dedicated chapter of this book.

These opposite states are antiparallel or antipodal, meaning parallel but in opposite directions. It explains why angle θ is halved in the equations describing a quantum state in Bloch sphere in the sine and cosine calculations of the formulas giving α and β [303]!

So, we divide θ by 2 to link the geometric representation in the sphere with the mathematical representation of the qubit state, and to allow a spreading of all the states of a qubit over the whole sphere. The whole sphere occupation of qubits representations makes it easier to describe how single qubit gates work as we'll show later in a graphical way.

By the way, sin(θ) is a marker of the qubit coherence or level of superposition. It's easy to grasp since the sinus will be equal to zero when the qubit is in the $|0\rangle$ and $|1\rangle$ states. It will be maximal, at 1, when the qubit vector will sit on the equator in the Bloch sphere with an even superposition of $|0\rangle$ and $|1\rangle$.

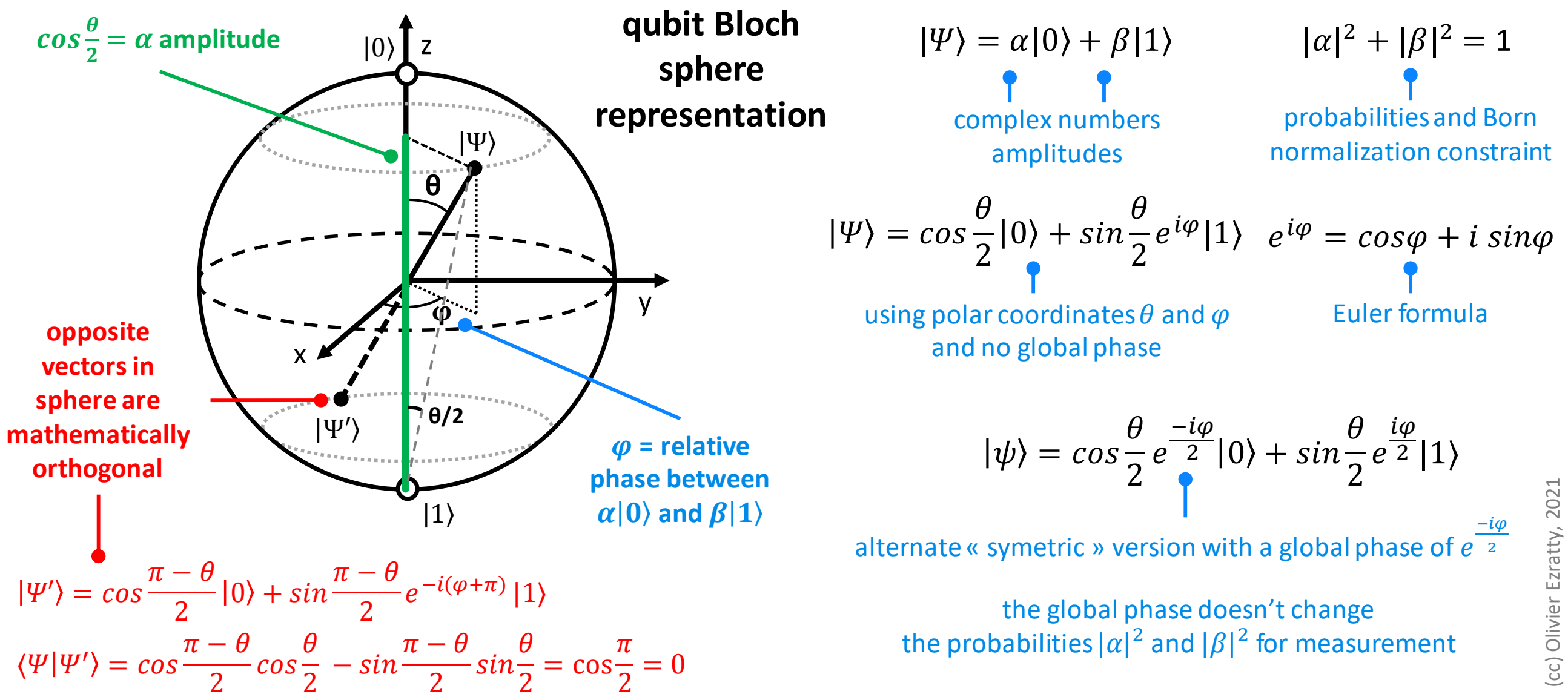

*Figure 163: a thorough explanation of the Bloch sphere representation of qubits. (cc) Olivier Ezratty, 2021.*

**Global phase.** A qubit representation is usually independent of its global phase. It can be removed from the equation to turn α into a real number. Still, a qubit is sometimes represented with a global phase of $\frac{-i\varphi}{2}$ as shown in Figure 163. When removing the global phase from α, the complex part of β integrates the phase difference between the amplitudes α and β. In that case, β is a complex number when the qubit is not in the plane crossing the x-axis (θ = 0) and the z-axis (φ = 0) of the Bloch sphere, meaning it has a non-zero phase. This complex number associates a real part for the direction z and a complex part for the dimensions x and y which are orthogonal to z. Applying a rotation around the z-axis will generally reintroduce a complex number in the α of the transformed qubit, which we do not necessarily factorize to remove the global phase of the qubit when doing hand calculations.

**Information**. The paradox to be understood is the following: since there is an infinite number of positions in Bloch's sphere, a single qubit could theoretically store a large amount of information, at least much more than a bit. Let's say it could be two floating point numbers, like the two angles θ and φ in the Bloch sphere.

[303] This is deciphered in Ian Glendinning's The Bloch Sphere, 2005 (33 slides) which explains this by the mathematical orthogonality of the two states $|0\rangle$ and $|1\rangle$ which are nevertheless opposed in the Bloch sphere. It is even better explained in Why is theta/2 used for a Bloch sphere instead of theta? which definitely clears up this mystery.

Unfortunately, we can only obtain a classical 0 or 1 after measurement because of that damn Holevo theorem[304]! We could theoretically retrieve some floating-point number with averaging the results of a large number of runs of the algorithm. Their precision will depend on several factors: the number or runs or "shots", the qubit error rates and the efficiency of quantum error correction codes. Given the overhead of all of this, forget about using qubits as a high-precision floating-point number storage device!

When the state vector of the qubit is horizontal in the Bloch sphere, i.e. it sits in its equator, and we have an even superposition of $|0\rangle$ and $|1\rangle$, but with a variable relative phase between the $|0\rangle$ and $|1\rangle$ amplitudes which is related to the horizontal angle of the vector φ with respect to the z axis as in the diagram on the right. Two usually superposed states are $|+\rangle$ and $|-\rangle$.

These are orthogonal states. These equatorial states share the same α component of $1/\sqrt{2}$ but opposite β values. This qubit-rich information is then modified by phase rotation quantum gates. If all qubits in the equator share the same 50%/50% amplitude probabilities, they have a different phase.

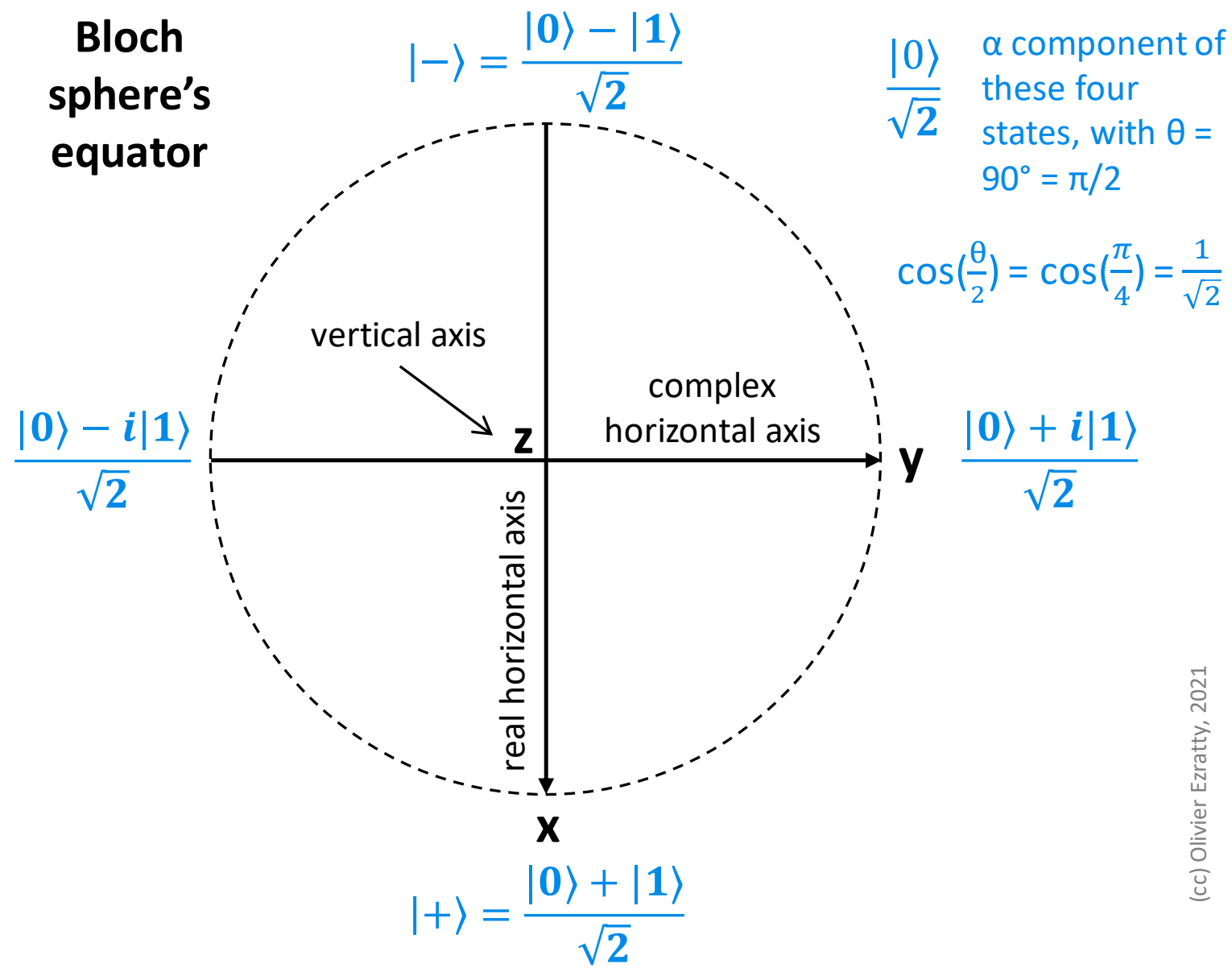

*Figure 164: Bloch sphere equator and superposed states (cc) Olivier Ezratty, 2021.*

A significant part of the quantum computing power comes with playing with the qubit phase that generates interferences between qubits. We'll see that later with algorithms such as phase amplitude and phase kickback.

As a general rule, most quantum gates do not generate all vector positions in the Bloch sphere. They are often half or quarter turns. The points of the sphere most often used are the cardinal points: the $|0\rangle$, $|1\rangle$, then the four points corresponding to the superposition of $|0\rangle$ and $|1\rangle$ on the equator.

To obtain all the quantum computing power, we need to make smaller turns than quarter turns, with the variable-phase R gates, usually composed with T gates, which we will see later and is outside the so-called Clifford gates group. Only these gates are supposed to enable some exponential speedup with gate-based quantum computing. Another way to look at this is that quantum advantage comes from using the full power of "analog" qubits.

**Origins**. We owe this Bloch sphere to three scientists: **Erwin Schrödinger** for his wave function of 1926, **Max Born** for his associated probabilistic model, created the same year, and to **Felix Bloch** (1903-1983, Switzerland) who represented the state of a two-level quantum on the sphere in 1946.

[304] To learn more and with a better scientific accuracy, you can consult the Wikipedia sheet of the wave function and amplitude probability. Other explanations can be found in the example of the electron orbit levels in the hydrogen atom in Quantum Mechanics and the hydrogen atom (19 slides). The physical interpretation of Max Born's statistical rule remains in any case open, as explained in Arkady Bolotin's June 2018 paper, Quantum probabilities and the Born rule in the intuitionistic interpretation of quantum mechanics (14 pages).

Bloch's sphere is frequently assimilated to **Poincaré's sphere**, named after **Henri Poincaré** (1854-1912, France) and created in 1892[305]. It is used to describe the polarization of light. The sphere polar coordinates represent the various types of light polarization: linear polarization (on the sphere equator), left elliptical polarization (upper hemisphere), right elliptical polarization (lower hemisphere) then left and right circular polarization (North and South poles).

The vertical axis (circular polarization) and one of the horizontal axis (linear polarization) represent two observables for a photon. All other states can be described as linear superpositions of these couples of basis states. And contrarily to massive particle-based quantum objects whose quantum probabilities are described by Schrödinger's equation, light equations used here are just Maxwell's electromagnetic waves equations.

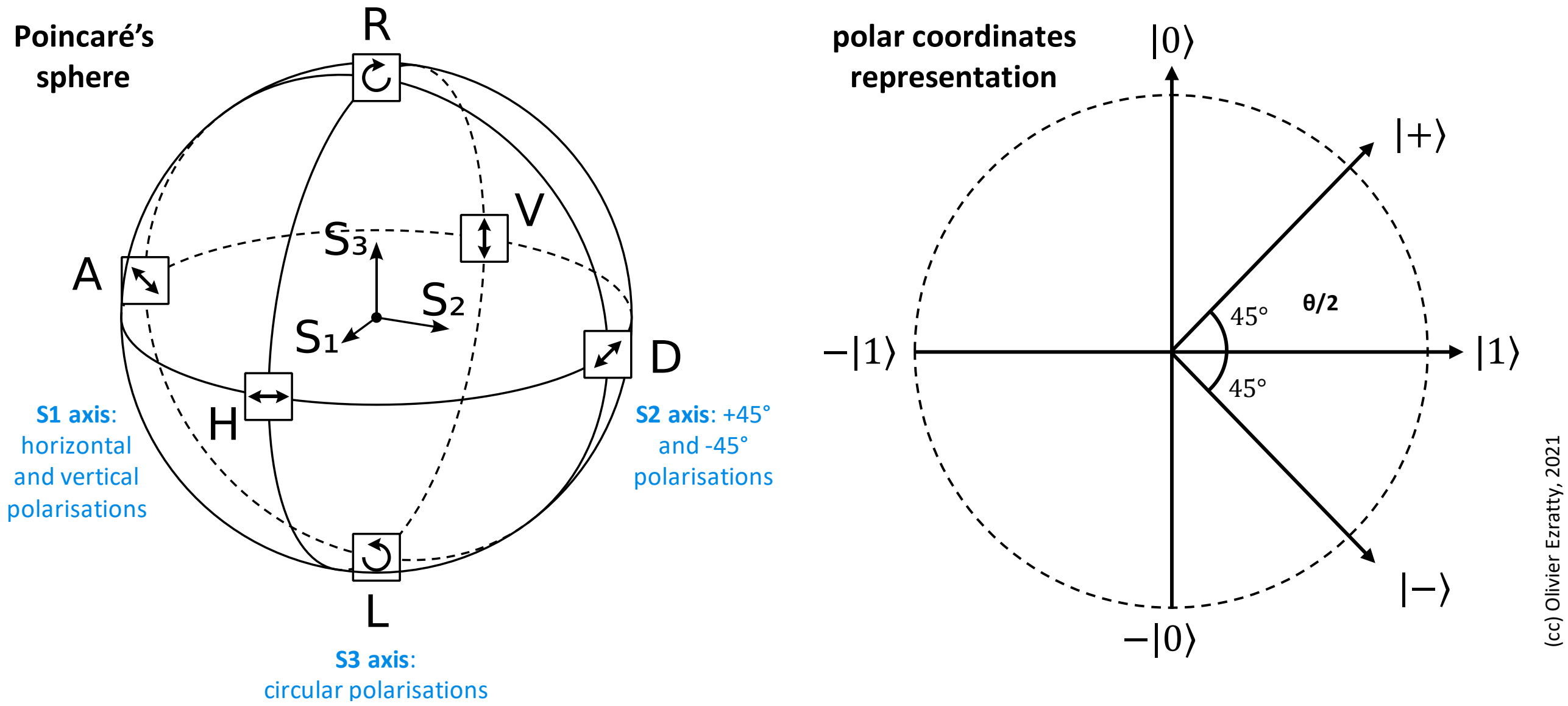

*Figure 165: the Poincaré photon sphere which inspired the Bloch sphere creation and another, Euclidian, representation of a qubit.*

The Bloch sphere representation is also used for representing an electron spin measured along three orthogonal axis (X, Y, Z), showing how superposition works with spins.

Sometimes, a system of polar coordinates is used on one circle, positioning the computational basis states of $|0\rangle$ and $|1\rangle$ as geometrically orthogonal vectors. It somewhat duplicates values since of $-|0\rangle$ and $-|1\rangle$ are similar to $|0\rangle$ and $|1\rangle$, with just a different global phase. Only the right half of the circle is useful.

Many other fancy qubits representations have been created with projection of the Bloch sphere onto a plane, representations of several qubit states with many Bloch spheres, even some representation of quantum entanglement with three Bloch spheres for two qubits[306] or with tetrahedrons[307]. None of these have been standardized and have a practical value for most quantum developers.

[305] Here are some sources of information associated with this section: Lectures on Quantum Computing by Dan C. Marinescu and Gabriela M. Marinescu, 2003 (274 pages), The Bloch Sphere by Ian Glendinning, 2005 (33 slides), The statistical interpretation of quantum mechanics, Max Born's 1954 Nobel Prize acceptance speech in physics (12 pages) and the excellent book The mathematics of quantum mechanics by Martin Laforest, 2015 (111 pages), which describes the mathematical basics of quantum computing with complex numbers, vectors, matrices and everything.

[306] See Two-Qubit Bloch Sphere by Chu-Ryang Wie, 2020 (14 pages).

[307] See Geometry of Qubits - A picture book by Yosi Avron and Oded Kenneth, 2018 (20 slides).

## Registers

In a quantum computer, qubits are organized in registers: a bit like the 32- or 64-bit registers of today's classical processors. One key difference is for now, a quantum computer has only one register and not many as with current classical microprocessors.

The main difference between an n-qubit register and a traditional n-bit register is the amount of information that can be manipulated simultaneously. In conventional computers, 32- or 64-bit registers store integers or floating-point numbers on which elementary mathematical operations are performed.

A register of n qubits is a vector in a $2^n$ dimensional space of complex numbers. Its dimensionality is exponentially larger than a n-bits register. Let's take for instance a register of 3 bits and 3 qubits. The first one will store one value at a time as 101 (5 in base 2) while the register of three qubits will contain complex numbers attached to each of the possible values of this register, 2 to the power of 3, i.e. 8, *aka* computational state basis. These complex numbers are the amplitude of each computational state. The total of their squares equals 1 since these are probabilities.

| n bits register | n qubits register |
|---|---|
| $2^n$ possible states **once at a time** | $2^n$ possible states **linearly superposed** |
| **evaluable** | **partially evaluable** |
| **independent copies** | **no copy** |
| **individually erasable** | **non individually erasable** |
| **non destructive readout** | **value changed after readout** |
| **deterministic** | **probabilistic** |

101 — n=3 example — 000 001 010 011 100 101 110 111

*aka* **register pure states**

*Figure 166: key differences between a classical bit register and a qubit register. (cc) Olivier Ezratty, 2021.*

However, these $2^n$ states amplitudes do not really constitute some information storage capacity. Quantum algorithm's main goal is to amplify the computational basis state amplitude that is the sought result, while reducing all the other amplitudes to near zero. Logically, it is like testing many hypotheses in parallel to bring out the best one.

The output information is a set of n classical bits. The $2^n$ amplitudes handled during computation are not some useful information that we exploit outside the register. We'll always end with one computational state and its related classical bits. So, in the end, you don't really process "big data" with quantum computing or at least, you don't output any big data. You may still use some sort of big data to prepare the state of the register before or during calculation[308].

But it's not to the advantage of quantum computing since feeding a quantum register with classical data is quite slow[309].

---

[308] However, exceptions are beginning to appear with hybrid methods for accelerating database access combining traditional computer-based and quantum algorithms. See Quantum computers tackle big data with machine learning by Sarah Olson, Purdue University, October 2018.

[309] It's well explained in the excellent overview Quantum Computing: Progress and Prospects from the US Academy of Sciences, 2019 (272 pages) : "*Large data inputs cannot be loaded into a QC efficiently. While a quantum computer can use a small number of qubits to represent an exponentially larger amount of data, there is not currently a method to rapidly convert a large amount of classical data to a quantum state (this does not apply if the data can be generated algorithmically). For problems that require large inputs, the amount of time needed to create the input quantum state would typically dominate the computation time, and greatly reduce the quantum advantage.*".

The graphic representation below in Figure 167 was built using the Quirk open source simulator. It is a sample of a quantum Fourier transform algorithm run on 4 qubits. The column numbers vector is showing the computational base probabilities. In the beginning we have a 100% |0000⟩.

After applying an X gate on the first qubit, we get a 100% amplitude for a |1000⟩**.** After applying Hadamard gates to all qubits, we get even amplitudes of 6,3% for all computational basis states. Then the QFT finds out the result, |1001⟩ which shows up on the last column[310].

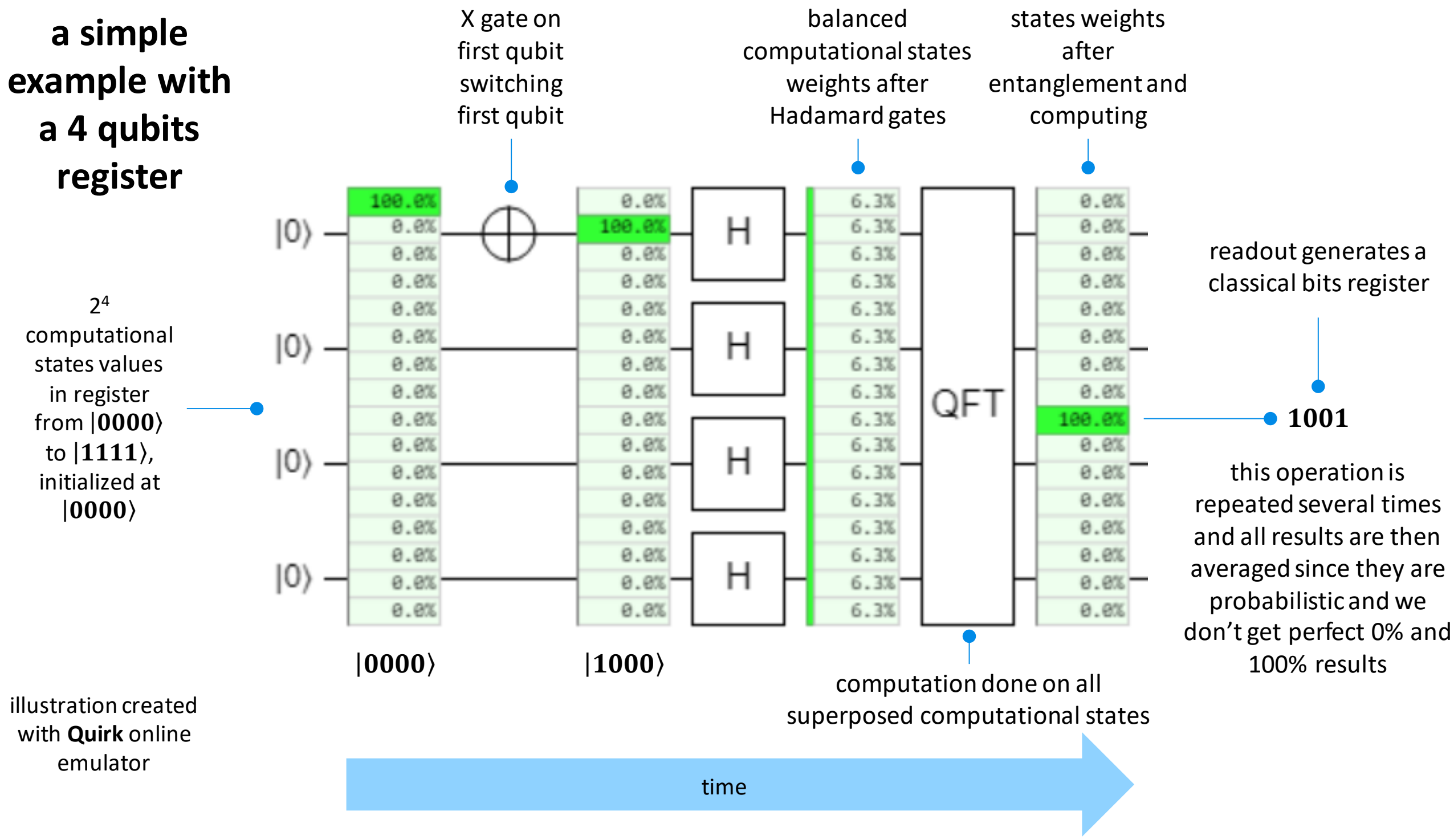

*Figure 167: manipulating a 4-qubit register vector state with Quirk. (cc) Olivier Ezratty, 2021.*

Another way of presenting things is a little simpler and more graphical: all the register states are on the left, the calculation generates interference between these states to make one of the states on the right come out which is the answer to the problem posed[311]. The example is based on the use of only two qubits that give four different "binary" states of the qubits.

So, we do not recover $2^n$ values in practice, but n bits. The operation can be repeated several times to obtain an average in the form of floating numbers. But it depends on the algorithms. For the majority of them, a binary output is sufficient, as for Peter Shor's integer factorization algorithm.

We are anyway constrained by **Holevo's theorem** of 1973 which proves that with n qubits, we cannot recover more than n bits of information after a quantum calculation!

At the current stage of qubit development, one and two-qubit gates error rate is between 0,1% and 0.5% and ideally it should be less than 0.0001%. This error rate can be evaluated for each isolated qubit.

[310] In A quantum computer only needs one universe by Andrew Steane, 2003 (10 pages), the latter insists on the key role of entanglement. He considers that entanglement does not so much explain the gain in quantum computing power.

[311] See Introduction to Quantum Computing by William Oliver from MIT, December 2019 (21 slides).

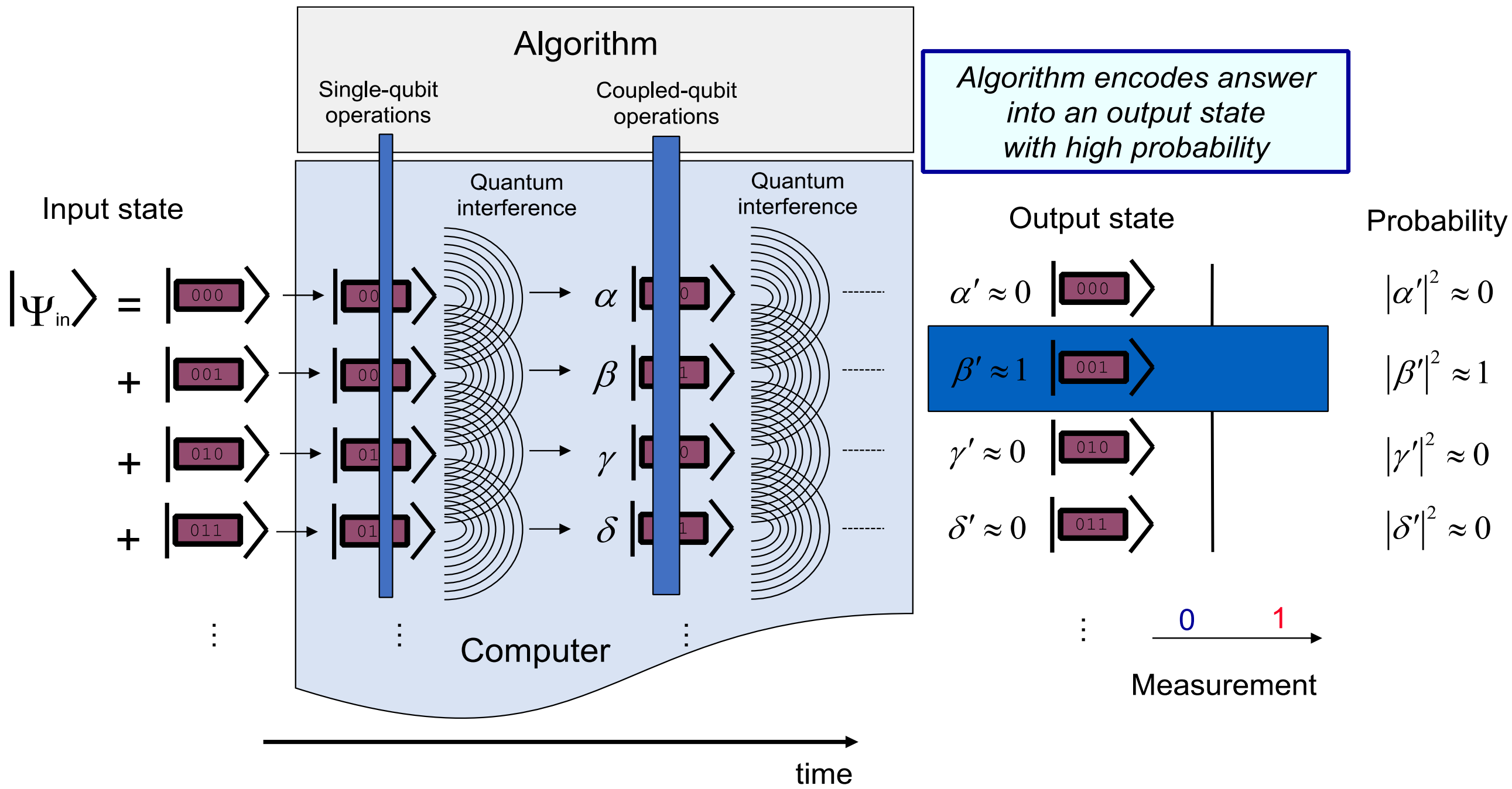

*Figure 168: representing qubits manipulations with interferences. Source: [Introduction to Quantum Computing](#) by William Oliver from MIT, December 2019 (21 slides).*

By the way, don't believe the nonsense that is the comparison of the exponential size of the qubit registers computational basis state with the number of particles in the Universe. These are not equivalent dimensions. A number of objects combination is not homothetic with a number of objects! With a given number of objects, combinations of these objects will always represent a number that is much higher than the number of objects taken as a reference. And… exponentially!

On the other hand, besides this exponential combination sizing, qubits have a lot of drawbacks in total opposition with classical bits. One can neither copy classically nor erase the value of qubits individually. Their measurement modifies their values. These are probabilistic objects that are difficult to manipulate.

**Ancilla qubits**. Universal gate quantum computing uses ancilla qubits or control qubits that can be combined with the computing qubits. The value of these qubits is not read at the end of the processing. It is a kind of trash can of qubits used during computations. They are used in various algorithms as well as to implement the error correction codes (QEC) explained later. We still always use a single qubit register. It can be just logically partitioned between computation qubits and ancilla qubits, these last playing more or less the role of classical registers in a microprocessor. Their content may be scrapped at the end of some parts of computing. It's sometimes done using the "uncompute trick" which reverses part of the processing affecting these ancilla without erasing the other qubits containing the intermediate computing result.

# Gates

In classical computing, logic gates execute Boolean algebra using bit-dependent decision tables as an input. Several types of logic gates with one and two inputs are used, including the NAND gate which is interesting because it is universal and uses only two transistors. The other one- and two-bit Boolean gates can theoretically be created with NAND gates. In general, however, logic gates are mixed in the circuits.

An Intel Core i5/7 processor with over 10 billion transistors contains several billion logic gates. A processor is obviously very complex, with gates managing access to a cache memory and registers, and instruction pipeline executing the code defining the gates to be used in calculations. These operations are generated at the processor's clock frequency, most often expressed in GHz.

The classic two-bit logic gates (NAND, NOR, XOR, AND, OR) are irreversible because they destroy information during their execution.

Qubits undergo operations via quantum gates that can be applied to one or more qubits.

**Single-qubit gates** apply a 2x2 unitary matrices of complex numbers to the qubit state vector containing the famous α and β complex amplitudes. These always generate some rotation of the qubit vector in the Bloch sphere. The norm of the vector remains stable at 1 at least, before any decoherence happens. And quantum gates modify qubits information without reading it. A single qubit gate on a register of N qubits is a unitary operator, a large square matrix of $2^N$ lines and columns which results from the tensor product of the gate matrix applied to a qubit and the identity operator acting on all the other qubits, in the qubits order.

**Two qubit gates** apply 4x4 unitary matrices to the computational basis state vector containing 4 entries ($2^2$).

**Three qubit gates** apply a 8x8 matrix to a state vector containing $8=2^3$ entries.

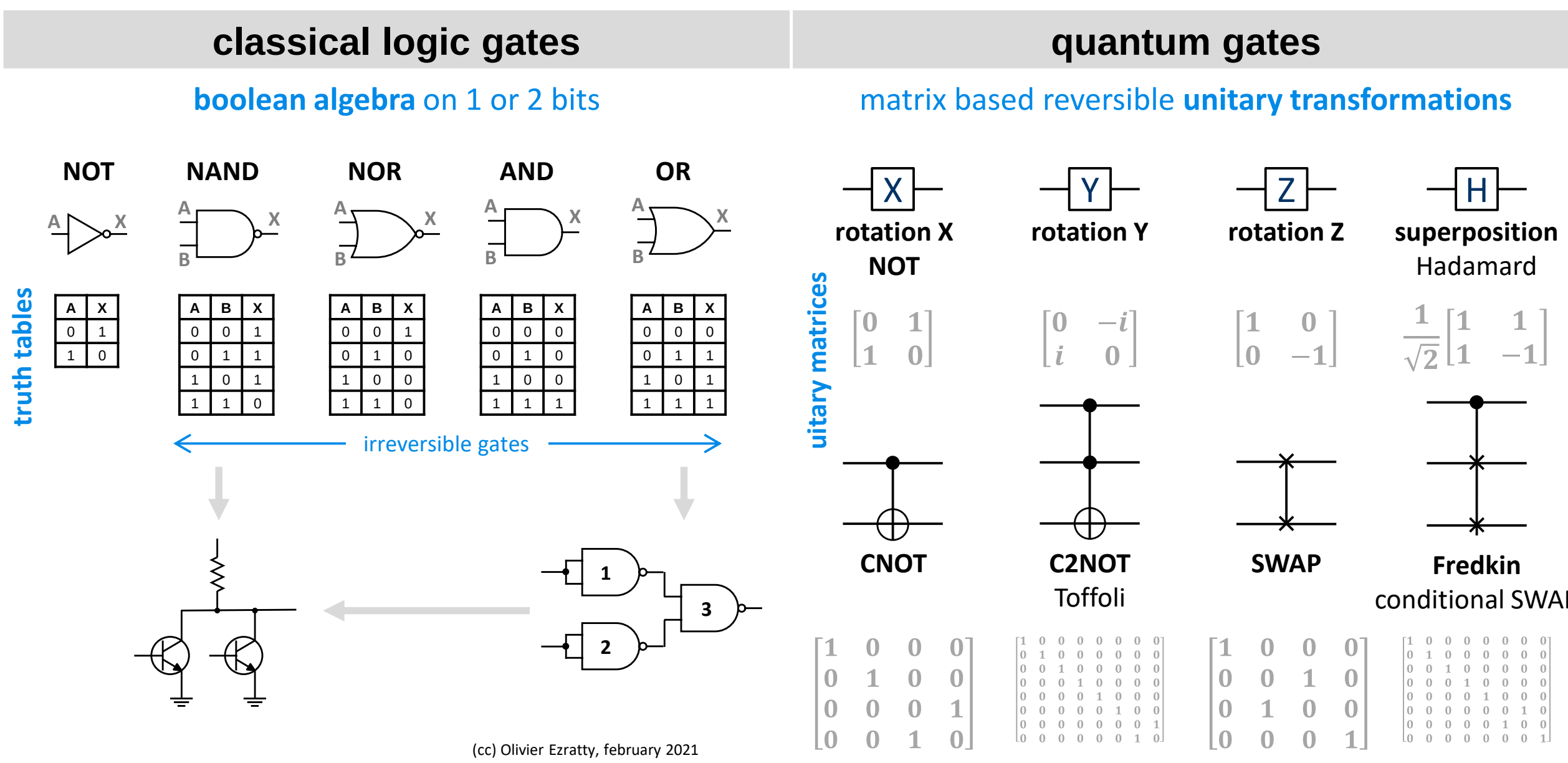

*Figure 169: comparison between classical logic gates and qubit gates. (cc) Olivier Ezratty, 2021.*

We'll now look at the various quantum gates made available to quantum developers[312]. The variations come from the rotation axis in the Bloch sphere (usually, X, Y or Z) and the angle of the rotation (1/2 turn, 1/4 turn, 1/8 turn or arbitrary rotation angle)[313].

- **X** gate (or NOT) performs an inversion or bit flip. A $|0\rangle$ becomes $|1\rangle$ and vice-versa. Mathematically, it inverts the α and the β of the two-component vector that represents qubit state. It generates a 180° rotation in the Bloch sphere around the X axis.

  This gate is often used to initialize to $|1\rangle$ the state of a qubit at the beginning of a process which is by default initialized at $|0\rangle$.

  $$X = \begin{bmatrix} 0 & 1 \\ 1 & 0 \end{bmatrix}$$

- **Y** gate performs a 180° rotation around the Y-axis in the Bloch sphere. It also turns a $|0\rangle$ into $|1\rangle$.

  $$Y = \begin{bmatrix} 0 & -i \\ i & 0 \end{bmatrix}$$

[312] Single qubit gates can be classified in XY and Z gates. XY gates are rotations around an axis in Bloch's sphere equator and can be viewed as amplitude change gates while Z gates are rotations around the Z axis and can be described as phase change gates.

[313] The formalism and classification of quantum gates is more sophisticated, as very well explained in the excellent lecture notes Gates, States, and Circuits - Notes on the circuit model of quantum computation by Gavin E. Crooks, January 2022 (79 pages).

- **Z** gate applies a sign change to the β component of the qubit vector (phase flip), i.e. a phase inversion and a 180° rotation with respect to the Z axis. The X, Y and Z gates complemented by the identify I are the **Pauli gates**. $Z = \begin{bmatrix} 1 & 0 \\ 0 & -1 \end{bmatrix}$

  They have several characteristics like ZX=iY and $X^2=Y^2=Z^2=I$. Their unitary matrices are noted $\sigma_x$, $\sigma_y$ and $\sigma_z$. Any single qubit unitary transformation can be written as a linear combination of Pauli gates with real number coefficients, plus the identity I.

- **S** gate generates a phase change, or a quarter turn rotation around the Z-axis (vertical). This is the equivalent of a half Z-gate. It is also called a "phase gate". $S = \begin{bmatrix} 1 & 0 \\ 0 & i \end{bmatrix}$

- **T** gate equivalent to a half S, which generates a phase change of one eighth of a turn. With two of these gates, an S gate is generated. This gate that is not part of Clifford's group (defined ... later) has the particularity of allowing by approximation the creation of any rotation in Bloch's sphere. $T = \begin{bmatrix} 1 & 0 \\ 0 & e^{\frac{i\pi}{4}} \end{bmatrix}$

  It's the key to universal gate-based quantum computing. It is indispensable to run a quantum Fourier transform and all derived algorithms like Shor integer factoring, HHL (linear algebra) and most quantum machine learning algorithms.

- **R** phase shift gates are variations of Pauli gates, with an arbitrary rotation angle in the Bloch sphere. The $R_z$ gate rotates around the z axis, $R_x$ around the x axis and $R_y$ around the y axis[314]. A $R_z$(angle) gate is also called a $P_{angle}$ gate (P for phase). $R_m = \begin{bmatrix} 1 & 0 \\ 0 & e^{\frac{2i\pi}{2^m}} \end{bmatrix}$

  When the x, y and z axes are not specified, it is z, the vertical axis of the Bloch sphere, as in the matrix *above*. When x, y and z are specified without an angle or m, it is 90° or $\pi/2$. The rotation is carried out on a complete round divided by m. The $R_z$ gates modify the phase of a qubit and not its amplitude. Thus, the measurement of its state $|0\rangle$ or $|1\rangle$ is not affected by this gate. It will return both $|0\rangle$ or $|1\rangle$ with the same proportions, before and after the use of an $R_z$ gate. Only two points of a sphere do not move during a rotation around an axis connecting them.

- **H** gate aka Hadamard-Walsh: puts a qubit at $|0\rangle$ or $|1\rangle$ in a superposed state "$|0\rangle$ and $|1\rangle$". It is fundamental to generate this superposition in the registers. $H = \frac{1}{\sqrt{2}}\begin{bmatrix} 1 & 1 \\ 1 & -1 \end{bmatrix}$

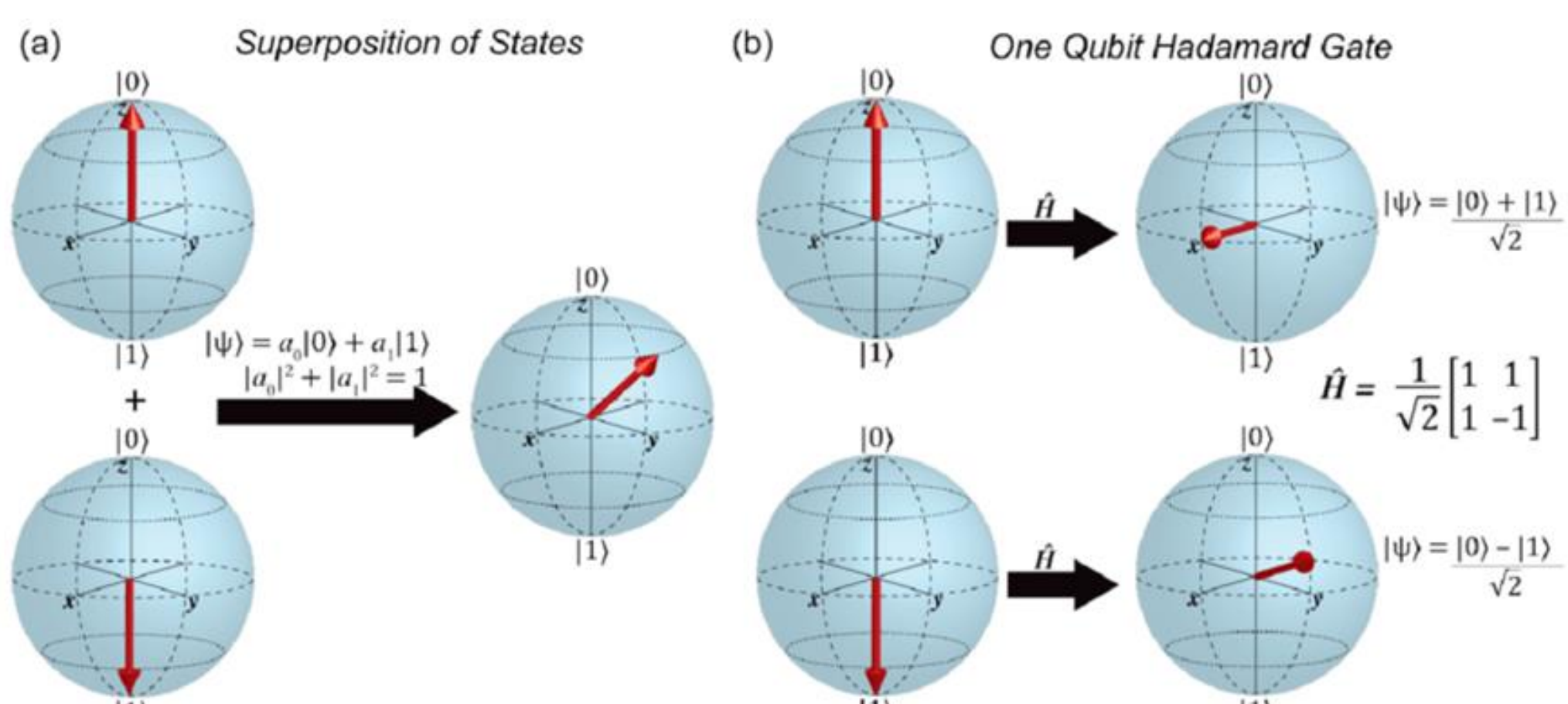

*Figure 170: example of application of an Hadamard gate on |0⟩ or |1⟩ qubits. Source: Molecular spin qudits for quantum algorithms by Eufemio Moreno-Pineda, Clément Godfrin, Franck Balestro, Wolfgang Wernsdorfer and Mario Ruben, 2017 (13 pages).*

---

[314] This is well explained in The Prelude, Microsoft, 2017.

It is often used to initialize a quantum register before executing an oracle-based algorithm like Grover or Simon algorithms. Here is a representation of the effect of this gate on a qubit initialized at $|0\rangle$ or $|1\rangle$. If we apply two Hadamard gates to a qubit, we return to the starting point. In other words: HH = I (I = identity operator)[315].

- **I** gate is the identity gate. It may be used as a pause. In the real physical world, a real I gate is not an exact identity due to decoherence! If you "run" 20 identity gates on a $|1\rangle$ qubit, you'll end up having some amplitude flipping error transforming progressively the qubit into a $|0\rangle$.

$$I = \begin{bmatrix} 1 & 0 \\ 0 & 1 \end{bmatrix}$$

- $|\mathbf{0}\rangle$ reset gate is sometimes indicated at the beginning of an algorithm to indicate that we start with initialized qubits. It is obviously irreversible.

$$|\mathbf{0}\rangle = \begin{bmatrix} 1 & 0 \\ 0 & 0 \end{bmatrix}$$

The mathematical formalism applied to a single qubit simply illustrates this. But this works only in theory, only if the gate error rate is zero. Since it is not zero, you don't ever a perfect $|0\rangle$ or $|1\rangle$.

A qubit reset operation may also be used to clean up ancilla qubits after their usage, when we are not using the uncompute trick, which is a way to cleanly reset ancilla qubits and remove potential entanglements with other qubits.

Below are representations of the effect of these single qubit gates, also labelled unary gates, on qubits initialized in $|0\rangle$ for the gates H, X, Y, $R_x$ and $R_y$ and with $|+\rangle$ for the phase change gates S, T, Z and $R_z$. Indeed, phase shift gates have no effect on $|0\rangle$ as well as on $|1\rangle$. For $|1\rangle$, it may just change the qubit global phase, and not its relative phase between the qubit amplitudes $\alpha$ and $\beta$, with no material impact on most algorithms. In the examples, the R gates use an angle of 90° or $\pi/2$.

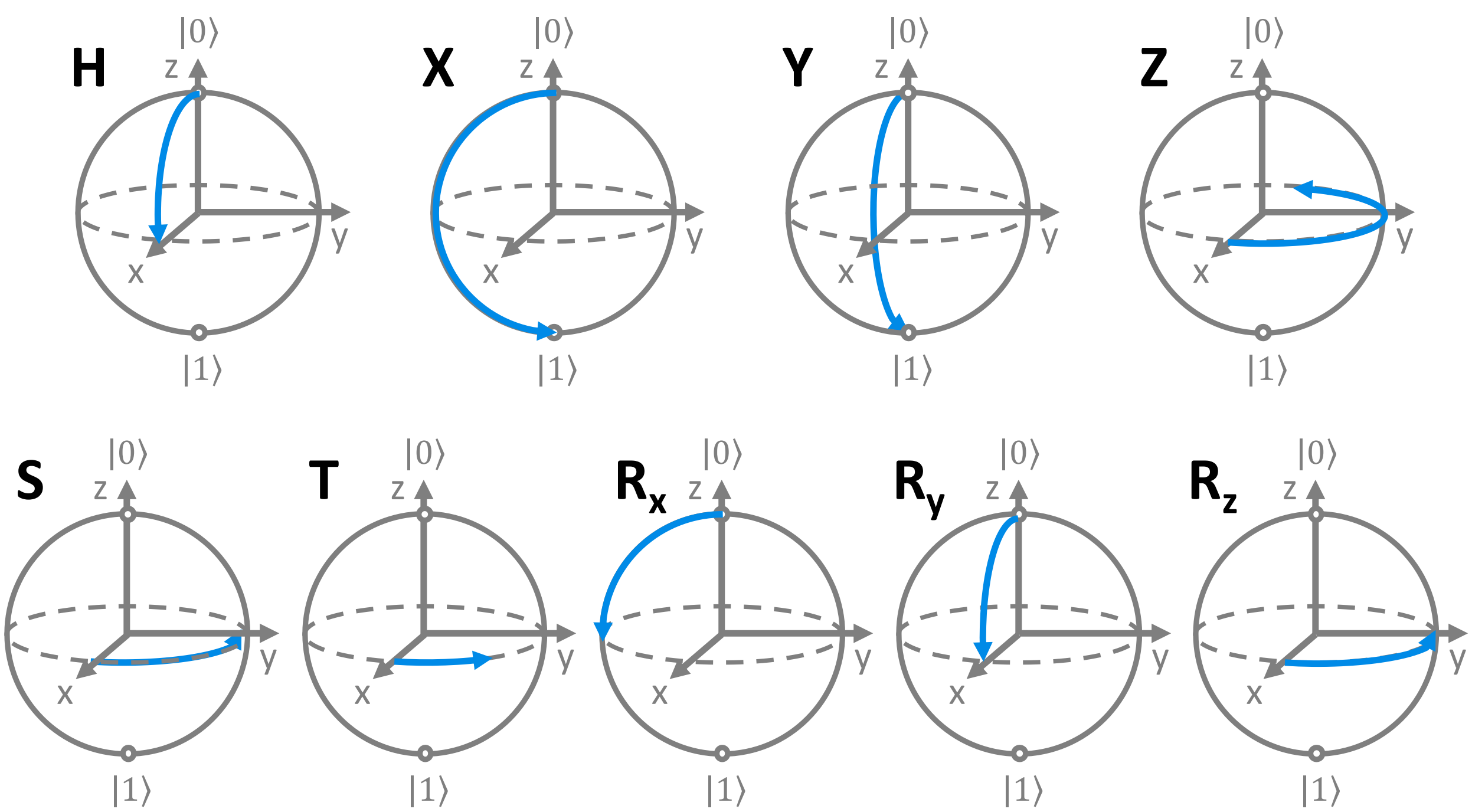

*Figure 171: Bloch sphere representation of various single-qubit gates. (cc) Olivier Ezratty, 2021.*

[315] This is also valid with X, Y and Z gates. In the usual notation, an H gate applied to $|0\rangle$ gives a state $|+\rangle$ and an H gate applied to $|1\rangle$ gives a state $|-\rangle$.

We now cover two and three qubit gates. Apart from the SWAP gate, all of these gates are conditional gates that apply a transformation of the state of one or two target qubits according to the state of one control qubit. These conditional gates create entanglement between the qubits that are in play. The entanglement between the involved qubits is persistent after executing these gates.

- **CNOT gate** is an inversion of the value of a qubit conditioned by the $|1\rangle$ value of another qubit. It is a quantum equivalent of the XOR gate in classical computing. Formerly called Feynman gate (C).
- **C2NOT** or **Toffoli gate** is an inversion of the value of a qubit conditioned by the $|1\rangle$ value of two other qubits.
- **CZ gate**, or Control-Z, is a conditional phase change Z gate.
- **CS gate**, or Control-S, allows a phase change of a qubit controlled by the state of a qubit.
- **SWAP gate** inverts the quantum values of two qubits. It can be generated from the chaining of three consecutive CNOT gates. The SWAP gate is the only two-qubit gate that is not creating a new entanglement between the two qubits. If they were separable before the gate, they will still be separable afterwards.

$$\text{SWAP} = \begin{bmatrix} 1 & 0 & 0 & 0 \\ 0 & 0 & 1 & 0 \\ 0 & 1 & 0 & 0 \\ 0 & 0 & 0 & 1 \end{bmatrix}$$

*Figure 172: the two-qubit SWAP gate unitary matrix.*

The key role of SWAP gates is to connect qubits that are physically distant in the register physical layout. A SWAP gate may also displace some entanglement. For example, if qubits A and B are entangled, but C is not entangled with A and B, a SWAP between B and C will displace entanglement to A and C and leave B unentangled with A and C. SWAP is usually a costly gate. It is not used a lot when the qubit topology enables all to all qubits direct connections like with some trapped ions qubits. As a consequence, most SWAP gates are created by compilers.

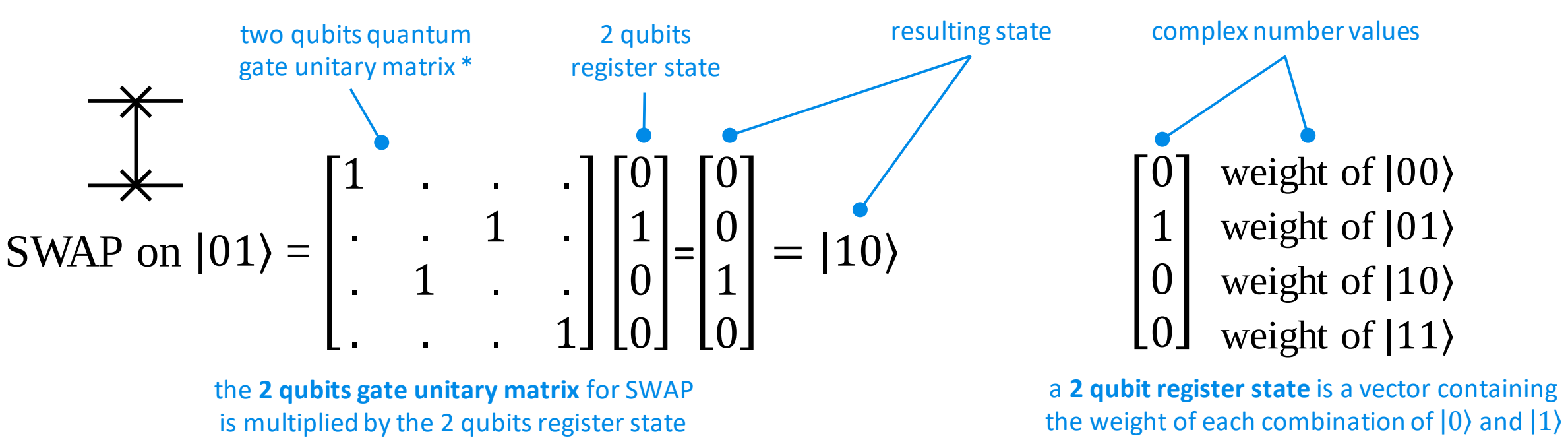

*Figure 173: example of SWAP gate operation. (cc) Olivier Ezratty, 2021.*

- **Fredkin gate** is a SWAP gate between two qubits that is conditioned by the state of a third qubit. So it has three inputs.
- **Generic Control-U gate** is a two qubits gate applying a generic one qubit unitary to a qubit based on the state of a control qubit.

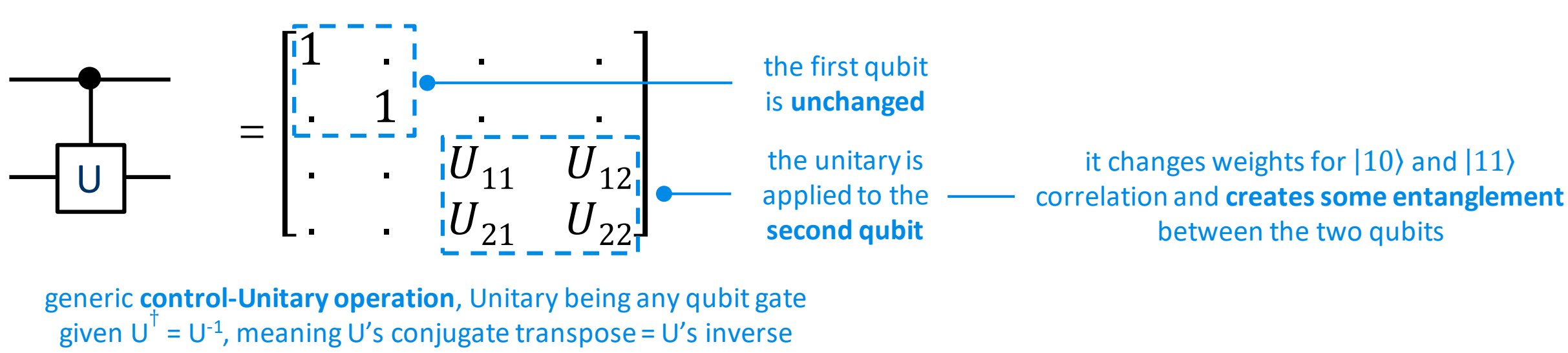

*Figure 174: control-U two-qubit gate unitary matrix. (cc) Olivier Ezratty, 2021.*

- **Phase-controlled R** gates are the equivalent of single-qubit phase-change R gates, conditioned by the state of a control qubit. If the algorithm, like a quantum Fourier transform, requires m to be large, it is not easy to ensure the reliability of the gate because the required precision becomes very large compared to the phase errors generated by the quantum system. However, phase errors are difficult to correct!

  A precision record of such a gate seems to have been reached by Honeywell with its ion trapped qubits presented in 2020 which have a rotation precision of 1/500 turn. This reminds us that during operations, quantum computing is analog. It is digital only at the level of commands and measured results, which become classical bits again[316].

  There are some reasons to get confused with S, T and R phase gates angles. For example, a S gate is sometimes branded as a $\pi/2$ and sometimes as a $\pi/4$. The same is applied to a T gate that is sometimes a $\pi/4$ and sometimes a $\pi/8$. The explanation is in the chart below and is related to the way a global phase is applied to the gate unitary operator. We can split hairs with using a "rotation" for the large one and a "round" for the small one.

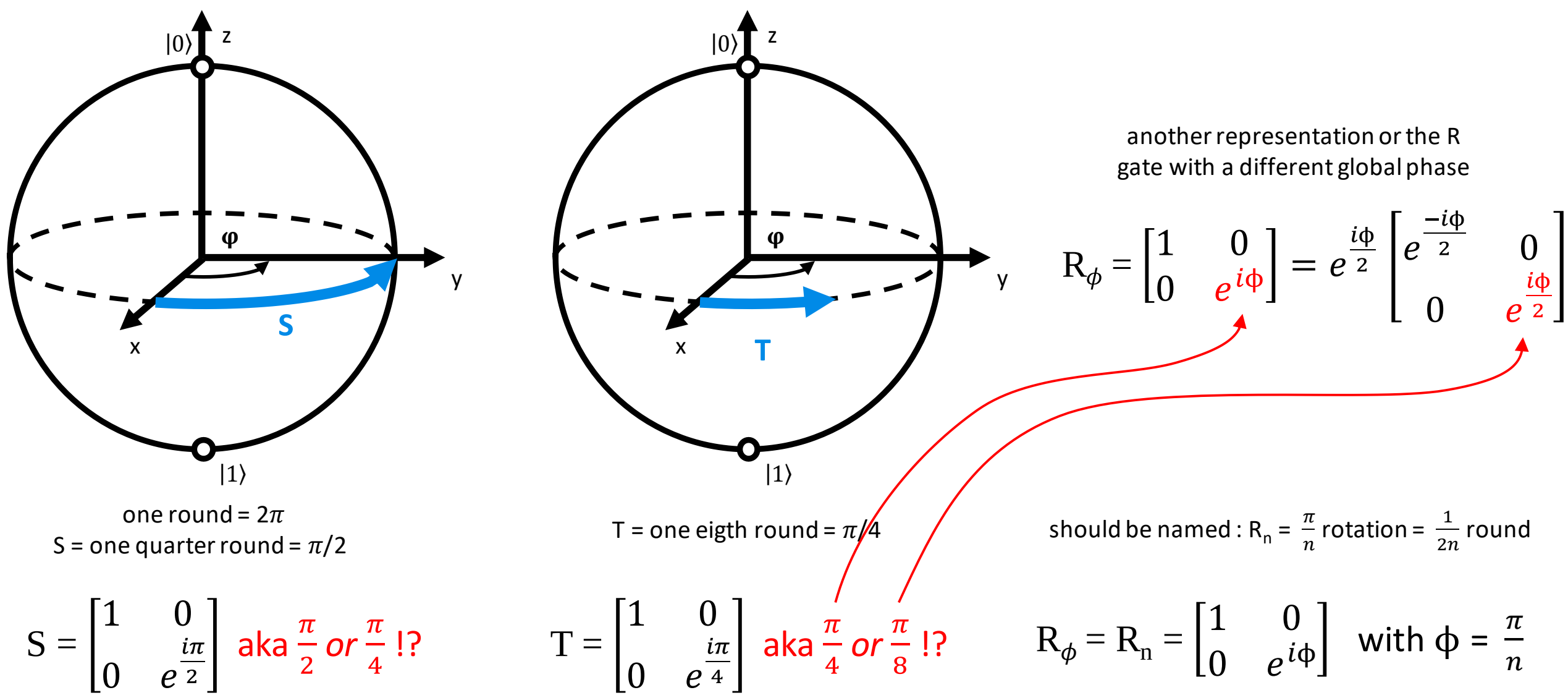

*Figure 175: solving the ambiguity of phase gates labelling. (cc) Olivier Ezratty, 2021.*

The effect of two-qubit gates is mostly always presented with using $|0\rangle$s and $|1\rangle$s as starting points in the control qubit, like with "*a CNOT inverts the state of a target qubit when the control qubit is* $|1\rangle$ ». But the CNOT will always have an effect on the target qubit when the control qubit is not exactly in the $|0\rangle$ state.

You just need to have a non-null $\beta$ complex amplitude component in the first qubit. So, the only case a CNOT will do nothing on the target qubit is when the control qubit is exactly a $|0\rangle$.

To fully understand the effect of these gates on any qubit state and computational basis vectors for several qubits, you have to look at the unitary matrices implementing these gates and their linear effects on the qubits and/or register computational basis vectors.

---

316 Here are a few sources of information on the subject of quantum gates: Gates, States, and Circuits by Gavin E. Crooks, July 2021 (82 pages), Universality of Quantum Gates by Markus Schmassmann, 2007 (22 slides), An introduction to Quantum Algorithms by Emma Strubell, 2011 (35 pages), Equivalent Quantum Circuits by Juan Carlos Garcia-Escartin and Pedro Chamorro-Posada, 2011 (12 pages), The Future of Computing Depends on Making It Reversible by Michael P. Frank, 2017.

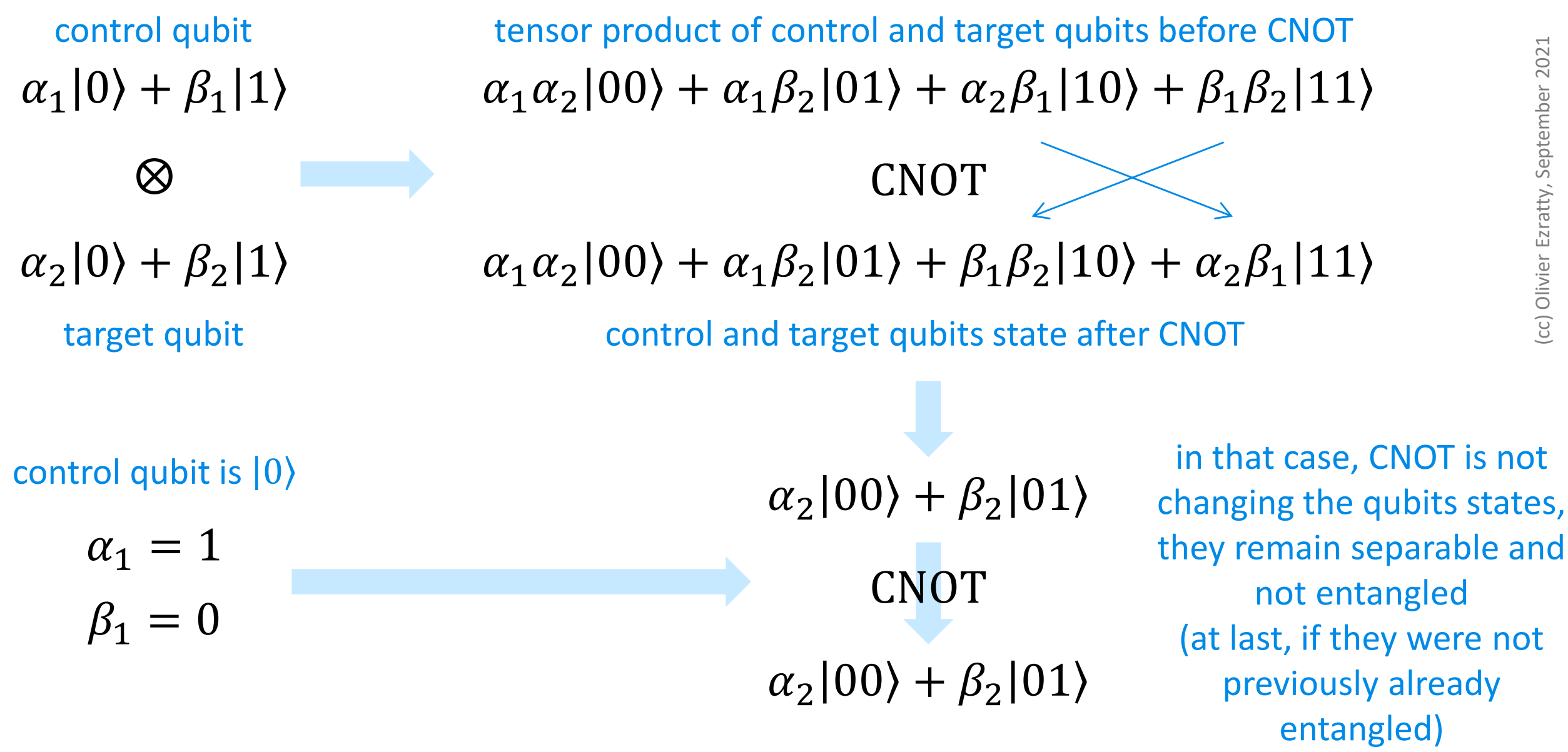

*Figure 176: visualization of a CNOT two-qubit gate effect, generically and with a control qubit at $|0\rangle$, the only case when it won't generate any qubit entanglement. (cc) Olivier Ezratty, 2021.*

In other words, and as demonstrated in Figure 176, unless the control qubit is $|0\rangle$, a CNOT gate will create some new entanglement between the control and target qubit. But one could argue two things: first, after a couple of operations, we never have a perfect $|0\rangle$ and are rapidly off-bounds, creating tiny entanglements with CNOT gates in that case, and second, most CNOT gates are run after a Hadamard gate was applied on the control qubit, getting off the $|0\rangle$ state!

Let's add a number of two-qubit gates which play a particular role. These are physical gates implemented at the lowest control level depending on the qubit type. They are not necessarily directly useful for developers but are the basis of some specific universal gates sets with some qubit types.

- **$\sqrt{\textbf{SWAP}}$ gate**, or square root SWAP, stops halfway through a SWAP. It is a physical level gate used to entangle electron spin qubits.
- **iSWAP gate** is a two-qubit gate that is implemented in superconducting qubits like those from IBM.

$$\sqrt{\text{SWAP}} = \begin{bmatrix} 1 & 0 & 0 & 0 \\ 0 & \frac{1}{2}(1+i) & \frac{1}{2}(1-i) & 0 \\ 0 & \frac{1}{2}(1-i) & \frac{1}{2}(1+i) & 0 \\ 0 & 0 & 0 & 1 \end{bmatrix}$$

*Figure 177: a $\sqrt{SWAP}$ unitary matrix.*

- **XY gate** is a generic two-qubits gate implementing a rotation by some angles $\beta$ and $\theta$ between the states $|01\rangle$ and $|10\rangle$ and iSWAP=XY$(0,\pi)$. It's a physical gate proposed by Rigetti that can be implemented on superconducting qubits to reduce the number of two-qubits gates required to run many algorithms[317].

$$\text{XY}(\beta,\theta) = \begin{pmatrix} 1 & 0 & 0 & 0 \\ 0 & \cos(\frac{\theta}{2}) & i\sin(\frac{\theta}{2})e^{i\beta} & 0 \\ 0 & i\sin(\frac{\theta}{2})e^{-i\beta} & \cos(\frac{\theta}{2}) & 0 \\ 0 & 0 & 0 & 1 \end{pmatrix}$$

*Figure 178: an $XY(\theta,\pi)$ two-qubit gate unitary matrix.*

- **ZZ gate** that is implement with qubits coupling is a technique that can be used with qubit couplers to connect two superconducting qubits and implement as a CZ gate[318].

[317] See Implementation of the XY interaction family with calibration of a single pulse by Deanna M. Abrams et al, 2019 (13 pages).

[318] See Implementation of Conditional Phase Gates Based on Tunable ZZ Interactions by Michele C. Collodo, Andreas Wallraff et al, PRL, May 2020 (10 pages).

- **$R_{xx}$**, **$R_{yy}$** and **$R_{zz}$** are two-qubit gates that are implemented natively in some trapped ion quantum computers as a Mølmer-Sørensen gate. They are called Ising coupling gates. The gate $R_{xx}$ is implemented natively in IonQ systems. These gates were also created for the first NMR quantum computing systems.
- **Mølmer-Sørensen gate**, **Cirac-Zoller gate** (C-NOT), **AC Stark shift gate** and **Bermudez gate** are various two-qubit gates implemented at the physical level with trapped ions qubits. The Mølmer-Sørensen gate is a "mixed-species" entangling gate that can couple different breeds of ions. It is also less sensitive to motion temperature. It's the main entangling gate for IonQ trapped ion computers.

| | atoms | | electrons & spins | | | | photons |
|---|---|---|---|---|---|---|---|
| **qubit type** | **trapped ions** | **cold atoms** | **super-conducting** | **silicium** | **NV centers** | **Majorana fermions** | **photons** |
| single qubit gates | rotations | $U_{xyz}(\theta,\psi,\mu)$ | $R_x(\pm\pi/2)$, $R_z(\lambda)$ (IBM, Rigetti) | $R_x$, $R_y$ | $R_x$, $R_y$ | T, H | X, Z, H<br>X, Z, R, CZ (Xanadu) |
| two qubit gates | XX Mölmer-Sorensen | C-Z<br>$C\text{-}U_{xy}(\theta,\psi)$ | CNOT, iSWAP (IBM)<br>C-Z (Rigetti) | $\sqrt{SWAP}$ | CNOT | CNOT | CNOT |

(cc) Olivier May, July 2021

*Figure 179: examples of physical qubit gates implement by specific qubit types. Consolidation (cc) Olivier Ezratty, 2021.*

**Logical reversibility**. Quantum gates have the particularity of being logically reversible. It can easily be visualized for a single qubit gate, which is a simple rotation in the Bloch's sphere and therefore, reversible with the inverse rotation. A multi-qubit gate is a rotation in a wider dimensional space, with $2^N$ dimensions, N being the number of qubits. Likewise, it's logically reversible with an inverse rotation, but harder to visualize.

We can rewind some parts of algorithms by applying in reverse order the quantum gates that have just been applied to a set of qubits[319]. One benefit of this process is the so-called uncompute trick used in some oracle-based algorithms. It enables resetting the ancilla qubits used in computation without doing any reading. It avoids damaging the useful qubits that we need to use for the rest of the algorithm.

That being said, qubits can undergo other operations. They could be stored, meaning transferred, in or from quantum memory. They can also be used to encode two bits instead of one, in what is called "superdense coding", which is mainly used in quantum telecommunications[320].

**Gates classes**. The science of quantum gates has led to the creation of many concepts, theorems about groups of quantum gates. They are associated with the notion of **universal gate sets**, capable of generating all other quantum gates.

Figure 180 contains a custom diagram summarizing these classes of quantum gates. In short, $SU(2^n)$ is the space of unitary transformations applicable on n qubits. It covers all the quantum computations that can be performed on n qubits. SU(2) includes all the unitary transformations that can be performed on one qubit (with n=1!). Clifford's group includes gates with one and discrete qubits quarter-turn rotation plus conditional gates. T (eighth turn) and R as Control-R gates with different angles from $\pi$ and $\pi/2$ are not in Clifford's group. They are needed to cover SU(2) and $SU(2^n)$ well. In practice, the addition of the T gate is enough to create a universal gate set with using approximations.

319 See Synthesis and Optimization of Reversible Circuits - A Survey by Mehdi Saeedi and Igor Markov, 2011 (34 pages), which reviews the algorithmic impact of reversibility in both classical and quantum computing.

320 See From Classical to Quantum Shannon Theory, 2019 (774 pages) which describes the application of Shannon's information theory to quantum computing. As well as On superdense coding, August 2018, by Fred Bellaiche, an Econocom engineer who publishes very interesting and popularized scientific articles on quantum.

The classification of the gates begins with the **Pauli gates** that apply half-turn rotations around the X, Y and Z axes of the Bloch sphere of representation of the qubits.

**Pauli group** includes the gates resulting from the combination of these three Pauli gates and the sign inversion operations on the α or the β of the qubits (±1 and ±i). On one qubit, the Pauli group includes the gates ±I, ±iI, ±X, ±iX, ±Y, ±iY, ±Z, and ±iZ (where I is the identity).

**Clifford group** includes single and multiple qubit gates that standardize the Pauli group applicable to n qubits, i.e., the U gates of this group combined with the Pauli group gates σ with UσU* generate Pauli group gates. A Clifford gate is a quantum gate that can be decomposed into Clifford group gates. These include Pauli gates (X, Y, Z) and H, S (90° rotation) and CNOT (also called CX for *control-X*) gates. The Clifford group is very large as soon as n>1. Its size is respectively 24, 11,520 and 92,897.280 elements for n=1, 2 and 3[321]. It is usually said that Clifford group gates are digital quantum gates while non-Clifford gates are analog.

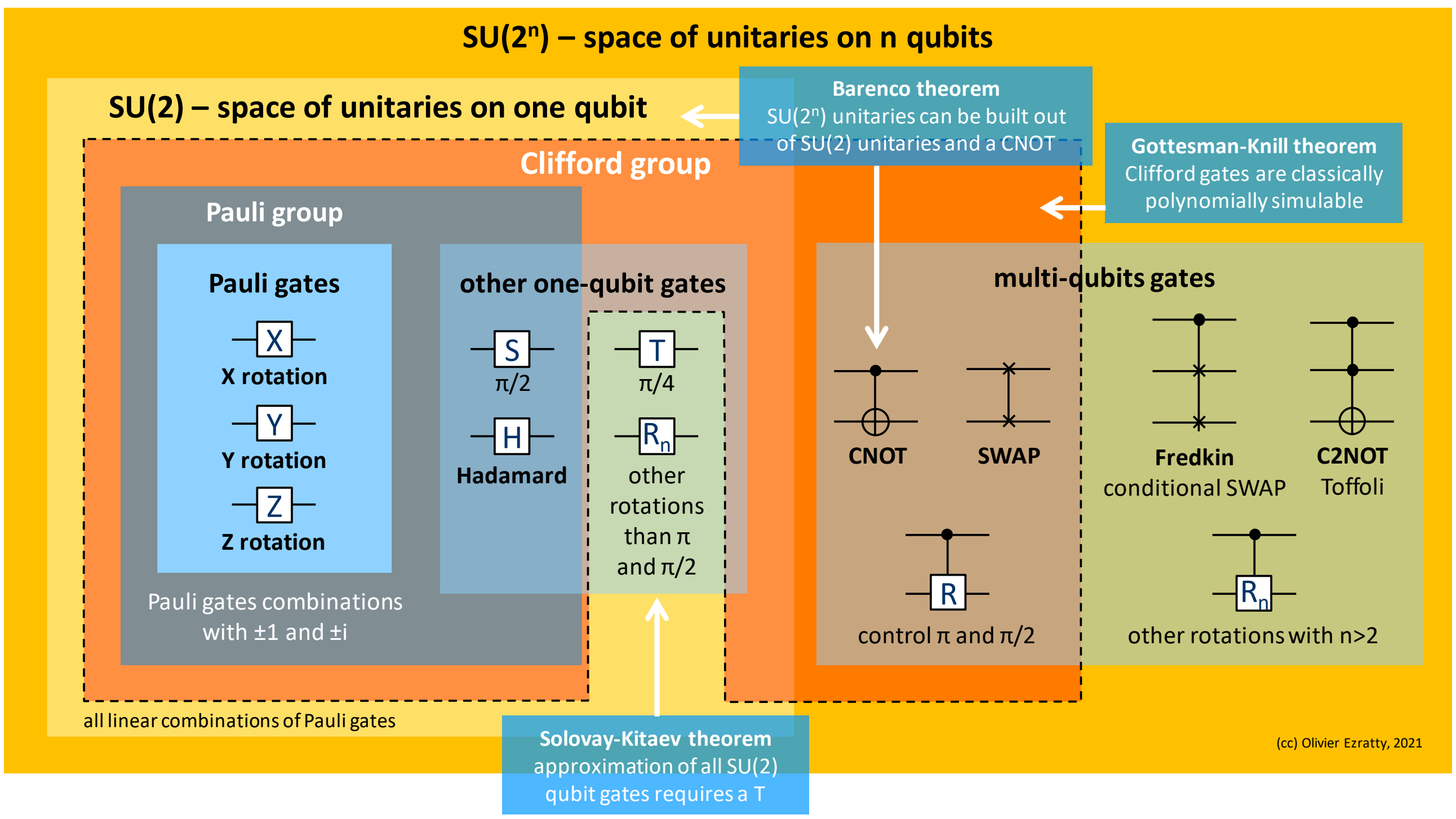

*Figure 180: a visual taxonomy of qubit gates explaining the Pauli gates, the Pauli group, the Clifford group and the role of T and R gates to create a universal gate set. (cc) Olivier Ezratty, 2021.*

**Gottesman-Knill's theorem** demonstrates that algorithms using gates in the Clifford group can be simulated in polynomial time on classical computers. It means that they are insufficient to provide an exponential speedup compared to classical computing[322]. Another variant of this theorem from Leslie G. Valiant defines conditions for a quantum algorithm to be classically simulable in polynomial time on a classical computer[323].

---

[321] See Clifford group by Maris Ozols, 2008 (4 pages). Clifford is the name of an English mathematician, William Kingdon Clifford (1845-1879) who is not related to the group that bears his name.

[322] See Positive Wigner Functions Render Classical Simulation of Quantum Computation Efficient by A. Mari and J. Eisert, December 2021 (7 pages) that generalizes the Gottesman-Knill theorem to quantum systems that preserve the positivity of the Wigner function (aka, do not use non-Gaussian photon states). It creates additional constraints on how to obtain exponential speedups with photon based quantum computers. It is also discussed in Quantum computational advantage implies contextuality by Farid Shahandeh, December 2021 (6 pages).

[323] See Quantum Computers that can be Simulated Classically in Polynomial Time by Leslie G. Valiant, 2002 (10 pages).

So, how can we obtain an exponential acceleration? It is necessary to use gates with more than two qubits implementing entanglement to obtain this acceleration like the Toffoli gate[324]. This can also be achieved with using phase-controlled R gates that are not part of Clifford's group, which can be approximated with adding a T gate. These non-Clifford gates have a particularity: they are difficult to correct with quantum error correction codes and to be implemented in a fault-tolerant manner. We'll see that later in page 235. On top of that, a maximally entangled state is required among the used qubits which makes sense since separable subsets of the qubit register wouldn't provide a large Hilbert space for computation. To create a universal gate set, you need to use two gates that don't commute or anticommute. T and H gates don't commute whereas all Pauli gates anticommute (XY=-YX, ZX=-XZ, YZ=-ZY). Geometrically, commuting and anticommuting happens when the related gates rotation axis are respectively parallel (like S and T) and orthogonal in the Bloch sphere (X and Y, or X and Z). With T and H, they are neither parallel or orthogonal but separated by a 45° turn [325].

**Continuous gates** make it possible to generate rotations of any angle in the Bloch sphere. The latter allow to generate all the phase-controlled R gates we have just seen and which are indispensable for QFT (Quantum Fourier Transform) based algorithms. Only a few qubits technologies can generate these gates at the hardware level, and usually with a poor precision.

**Discrete gates** are sets of (Hadamard, Z, S, CNOT) that make at best only half and quarter turns in the Bloch sphere.

**Universal gate set** is a group of gates that has the property of allowing the creation of all unitary operations on a set of qubits. From a practical point of view, also it allows to create all known quantum gates for one, two and three qubits. Such a gate-set must be able to create superpositions, entanglement and it must have at least one gate with no-real parameters (i.e. complex numbers instead of real numbers).

Here are some known sets of universal gates:

- CNOT + all single qubit unitaries can enable the creation of any unitary transformation on any number of qubits. This is demonstrated in the **Barenco theorem** according to which $SU(2^n)$ unitaries can be built out of SU(2) unitaries and a CNOT two qubit gate[326]. It also demonstrates that any unitary transformation $SU(2^n)$ on n qubits can be built with a maximum of $4^n$ elementary quantum gates.
- CNOT + T (eighth of turn) + Hadamard, using approximations, linked to the **Solovay-Kitaev's theorem**. It proves that a dense and finite set of quantum gates in SU(2) space allows can be used to reconstruct any gate in this space with a maximum error rate ε.

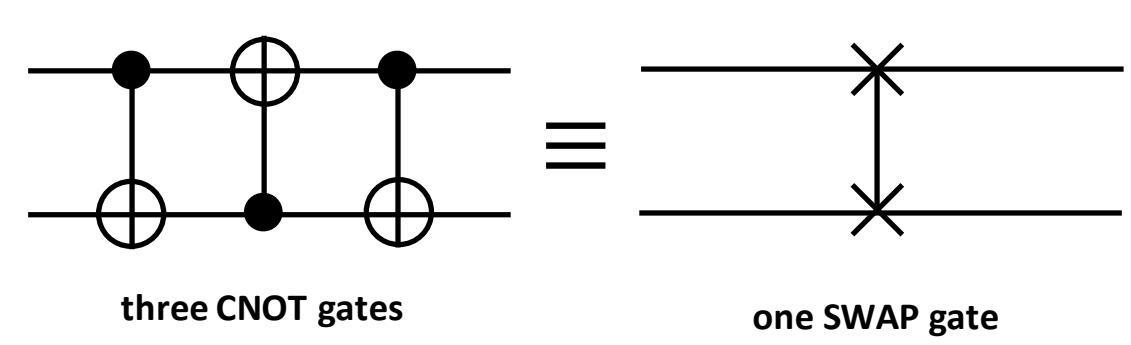

*Figure 181: how to create a SWAP gate with three CNOT gates.*

  The number of gates to be chained is a polynomial order of magnitude of log(1/ε). The SU(2) space is the Special Unitary group of dimension two.

  It includes unit matrices (from determinant 1) with complex coefficients and dimension 2.

$$SU(2) = \left\{ \begin{pmatrix} \alpha & -\overline{\beta} \\ \beta & \overline{\alpha} \end{pmatrix} : \ \alpha, \beta \in \mathbb{C}, |\alpha|^2 + |\beta|^2 = 1 \right\}$$

[324] See On the role of entanglement in quantum computational speed-up by Richard Jozsa et Noah Linden, 2002 (22 pages).

[325] See Quantum computing 40 years later by John Preskill, June 2021 (49 pages).

[326] See Elementary gates for quantum computation by Adriano Barenco, Charles Bennett, David DiVincenzo, Peter Shor and al, 1995 (31 pages).

This search for a set of discrete quantum gates allowing by approximation to generate a set of continuous gates of arbitrary rotations is important for some algorithms that we will see later, notably the discrete Fourier transform that is exploited in Shor's algorithm. You can see below in Figure 182 the effect of the sequence of T and H gates which, according to the combinations, allow to cover the different positions of Bloch's sphere, validating **Solovay-Kitaev**'s theorem[327]. Transpilers are the parts of quantum code compilers that convert any quantum gate in the underlying universal gate set implemented by the quantum processor and handle related optimizations.

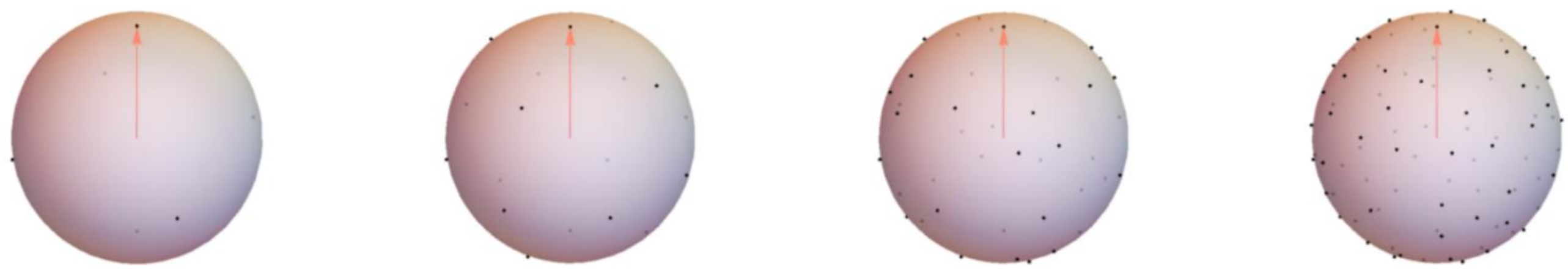

*Figure 182: a visual description of Solovay-Kitaev's theorem. Source: TBD.*

## Inputs and outputs

Traditional microprocessors are composed of fixed logic gates, etched into the silicon, and 'moving' bits, which are electrical pulses that propagate through the circuit through the various gates. All this at a certain frequency, often in GHz, set by a quartz clock.

In a quantum computer, the first stage of processing consists of resetting the quantum register into an initial state. This is called "preparing the system". The various registers are first physically configured in the $|0\rangle$ state. The following initialization consists in using different operators such as the Hadamard transformation to create $|0\rangle+|1\rangle$ superposition or the X gate to change this value $|0\rangle$ to $|1\rangle$. Sometimes, more preparation is required to prepare a denser register state, like with quantum machine learning algorithms. Once this initialization is done, computing gates operations are sequentially applied to the qubits according to the algorithm to be executed.

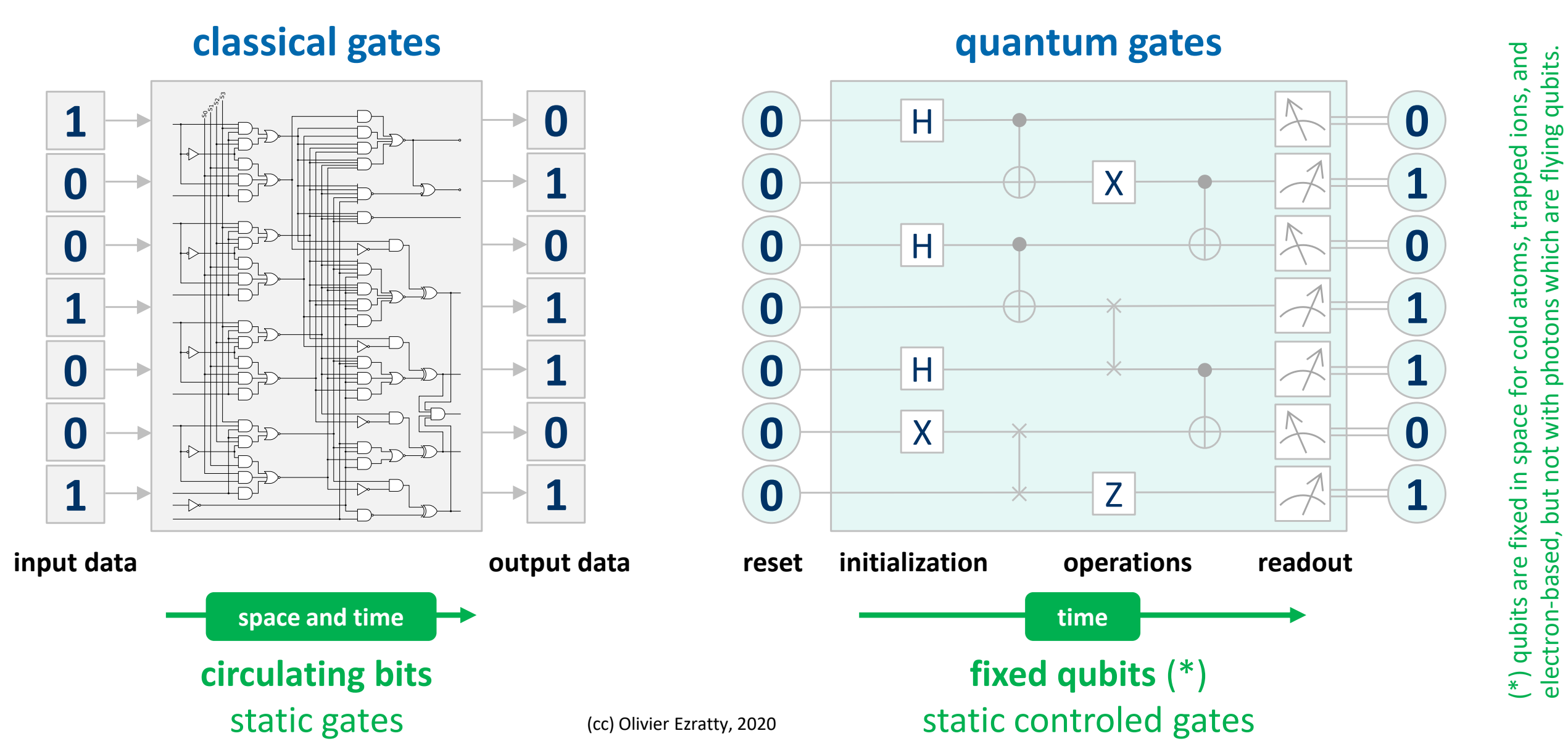

*Figure 183: time and space differences with classical logic and quantum gates. (cc) Olivier Ezratty, 2021.*

327 See Shorter quantum circuits by Vadym Kliuchnikov et al, Microsoft, Facebook and the Universities of Birmingham, Oxford, Bristol and Brussels, March 2022 (83 pages) which proposes an efficient method to generate any unitary with fewer gates, and T-count and T-depth of any multi-qubit unitary by Vlad Gheorghiu, Michele Mosca, Priyanka Mukhopadhyay, October 2021-October 2022 (28 pages) which defines lower bounds for these T gate usage.

They always include some multi-qubits gates implementing entanglement between qubits. Finally, qubits are measured at the end of the processing, which has the effect of modifying their quantum state.

Quantum algorithms diagrams for universal gates computers (*below* right) are most often time diagrams, whereas for classical logic gates it is also a physical diagram. In the right part describing a quantum algorithm, there are no physical wires connecting the qubits between an input and an output, the gates being in their path. It is a time-based schema!

A quantum algorithm is the description of a quantum circuit made of a series of sequenced timely quantum gates operating on 1, 2 and sometimes 3 qubits. It's the way to create a large unitary transformation on the initialized qubits.

Now, let's toy a little bit with qubits and gates with Quirk, particularly to identify pure and mixed states with single or two qubits. It also shows the role of off-diagonal values in density matrices.

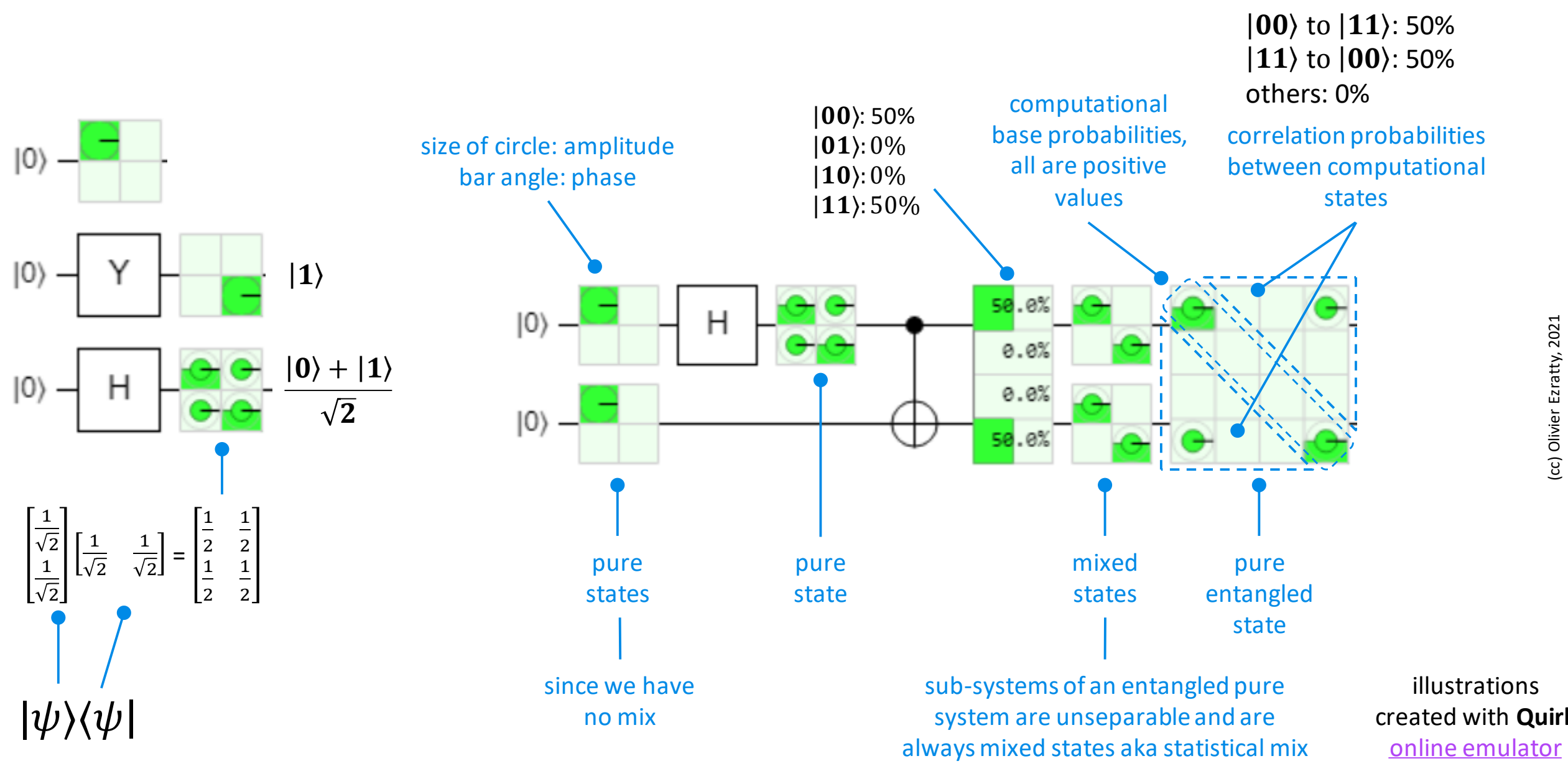

*Figure 184: on examples of toying with Quirk to see how pure and mixed states look with two qubits. (cc) Olivier Ezratty, 2021.*

Here, we describe a mixed state generated on two qubits after one of them is entangled with a third qubit.

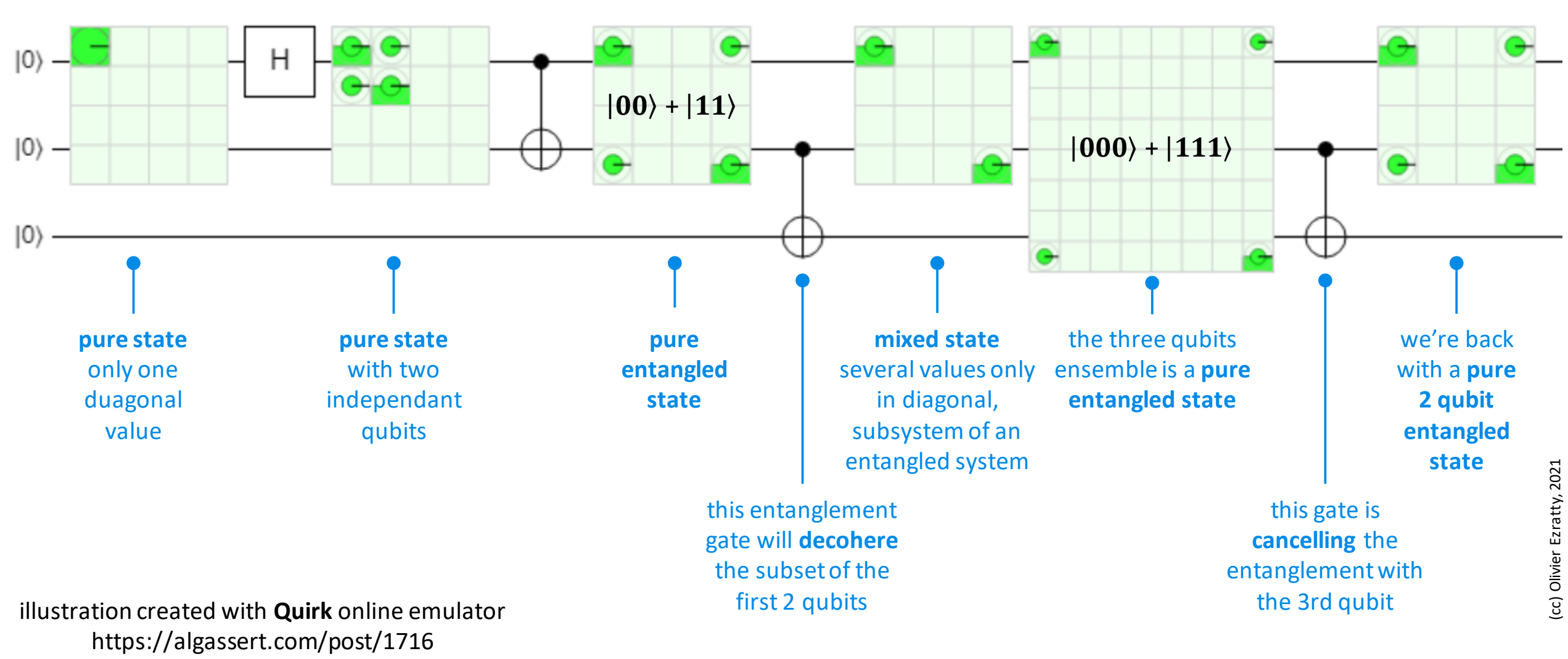

*Figure 185: three examples of toying with Quirk to see how pure and mixed states look with three qubits. (cc) Olivier Ezratty, 2021.*

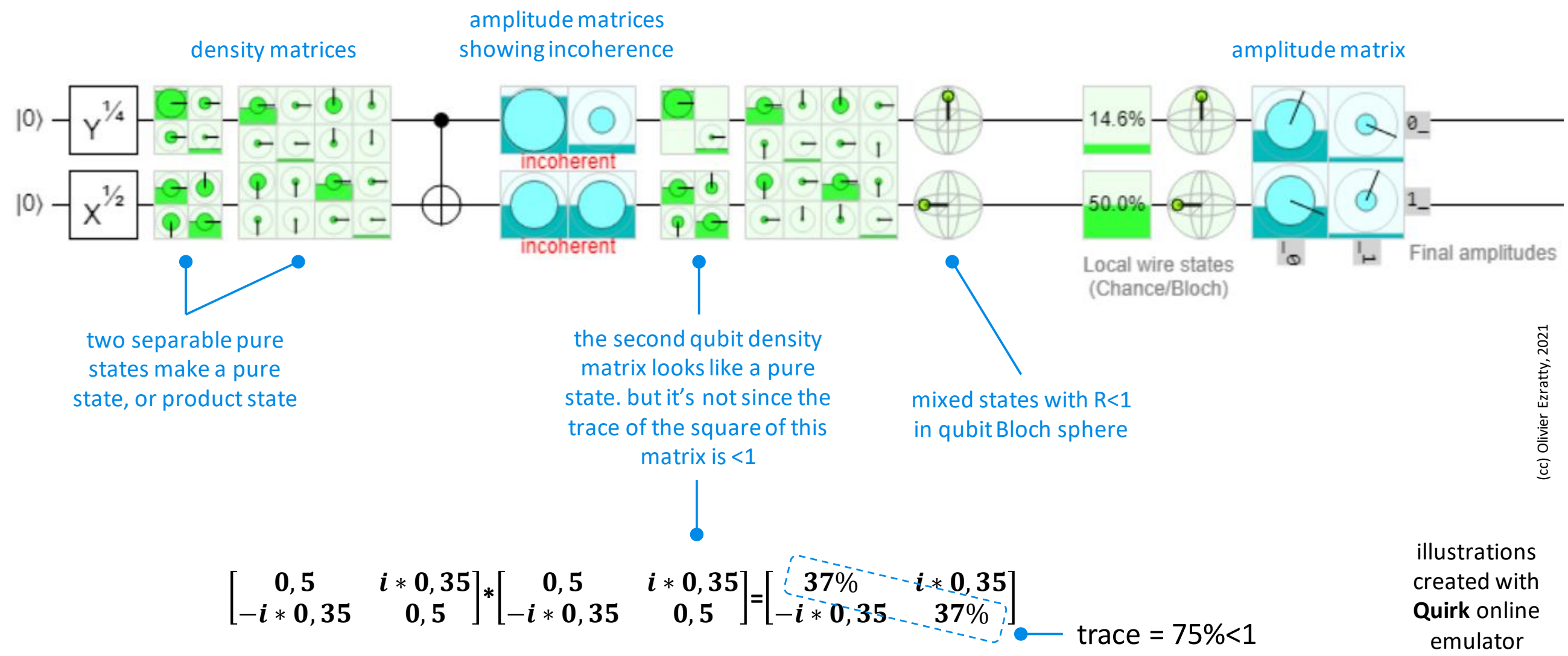

*Figure 186: three examples of toying with Quirk to see how pure and mixed states look with two qubits. (cc) Olivier Ezratty, 2021.*

# Qubit lifecycle

One way to understand how a universal gates quantum computer works is to track the life of a qubit during processing:

**Initialization**. A qubit is always initialized at |0⟩, corresponding to the base state, usually at rest, of the qubit. This initialization consumes some energy with all known types of qubits.

**Preparation**. It is then programmatically prepared with quantum gates to adjust its values that are vectors in the Bloch sphere. The Hadamard gate is one of the most common one and creates a superposed state of |0⟩ and |1⟩ Single qubit gates apply a rotation of the qubit vector in the Bloch sphere. These rotations are based on unitaries, 2x2 complex number matrix operations applied to the qubit vector [α, β]. These unitaries have a trace of 1, maintaining the vector length of 1. For most quantum algorithms, qubit preparation is usually simple with a set of X gates to set them and H gates to create superposed states. In some cases, like with quantum machine learning, qubit states preparation can be more complex, requiring a lot of gates.

**Multiple-qubit gates** then conditionally link qubits together. Without these quantum gates, little could be done with qubits.

**Data manipulation**. The qubits information that is manipulated during computing is "rich" with a dimension of two real numbers, the angles θ and ϕ, or the vector [α, β] for each qubit. But a set of N qubits holds $2^N$ complex number values, representing the proportion of each of the computational basis states made of the various combinations of N 0s and 1s. It creates a dimensionality of $2^{N+1}$-1 real numbers, to take into account the normalization constraint for the computational basis states amplitudes. As these gates are operated on the qubits, quantum computing works in an analog way[328].

**Measurement**. When we measure the value of a qubit, we obtain a classical binary 0 or 1 with a probabilistic return depending on the qubit state. So, for each qubit, we have a 0 as input, a 0 or a 1 as output, and an infinite number of states in between during calculations.

---

[328] This is the position stated in Harnessing the Power of the Second Quantum Revolution by Ivan H. Deutsch, November 2020 (13 pages). Or more precisely, the author states that gate-based quantum computers are both digital and analog.

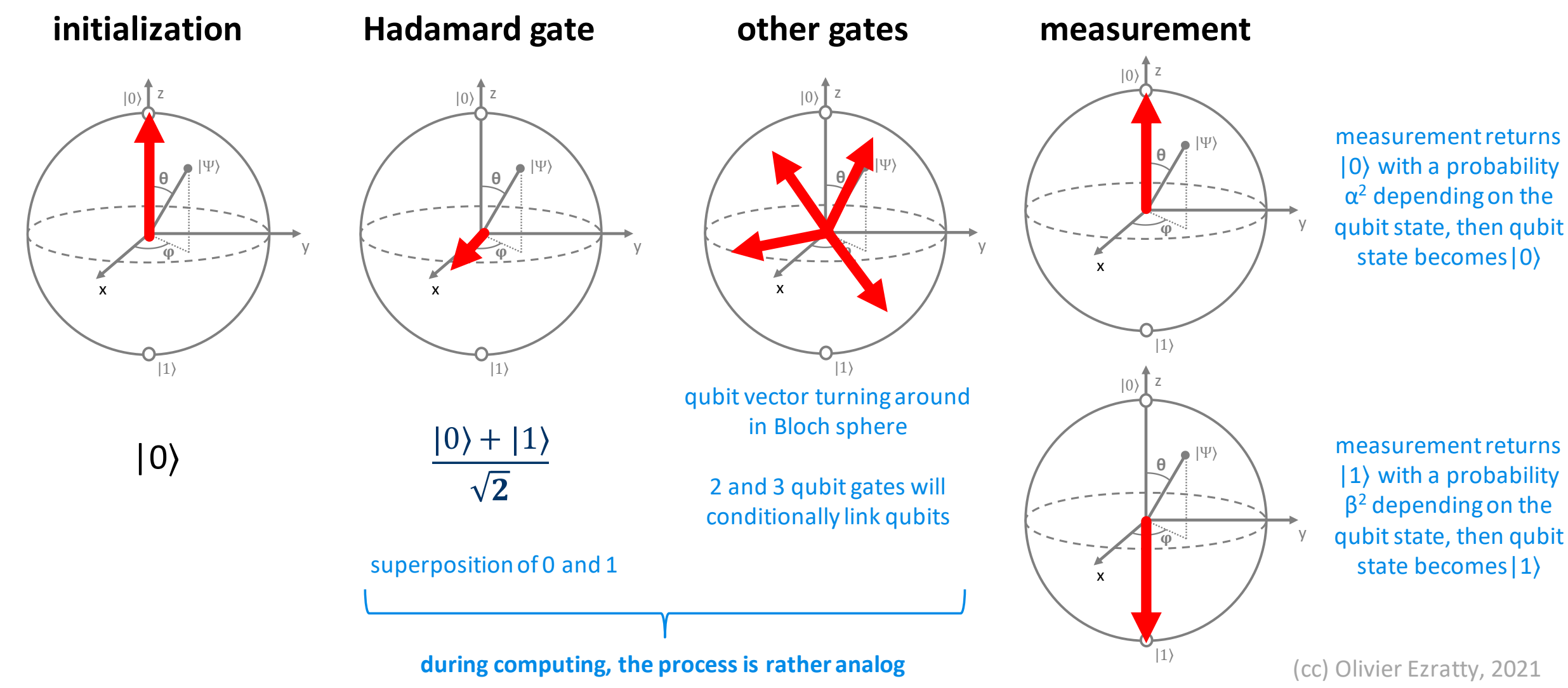

*Figure 187: the effect of measurement on a single qubit. (cc) Olivier Ezratty, 2021.*

All this to say that the mathematical richness of qubit-based quantum computing happens only during processing. This is the life cycle of the qubit illustrated in the above diagram in Figure 187.

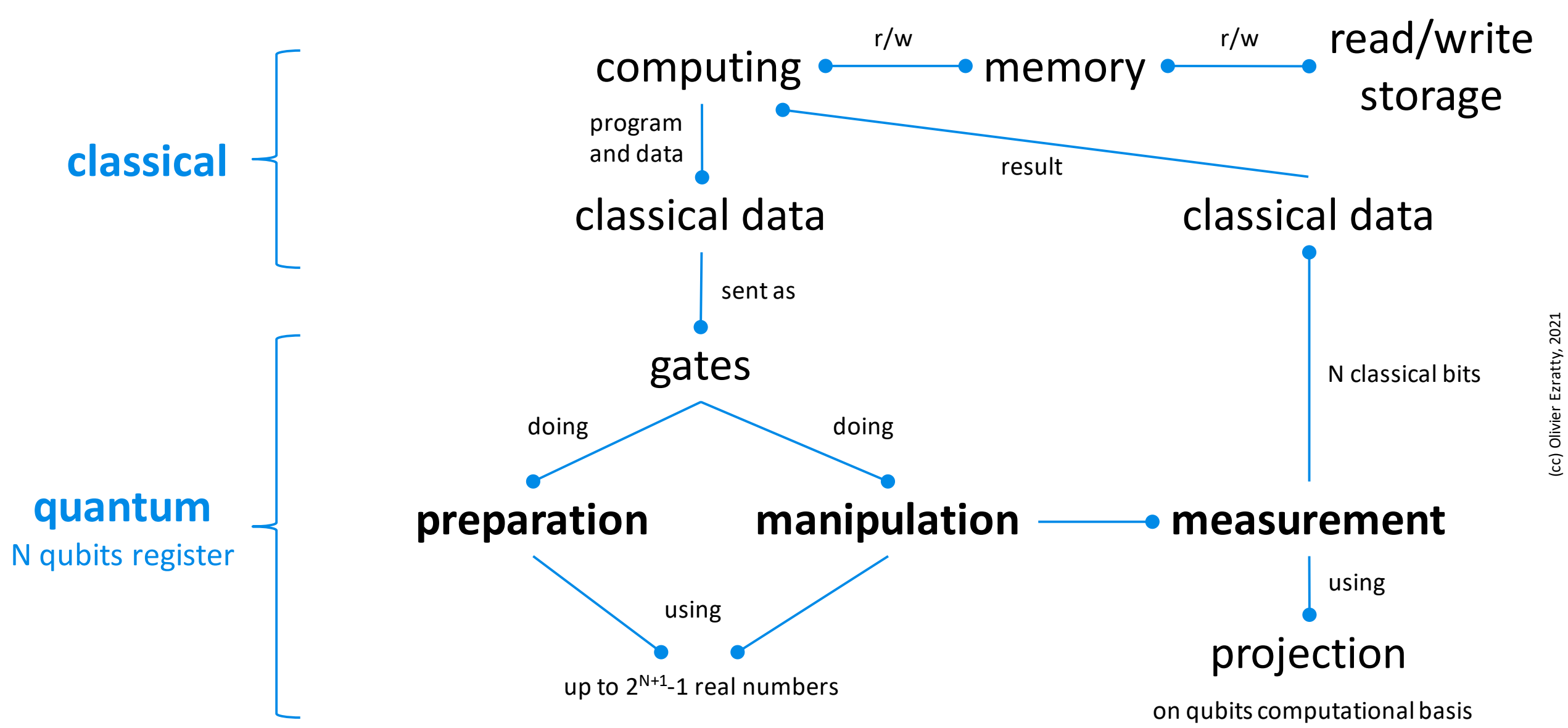

*Figure 188: classical and quantum data flow in gate-based quantum computing. (cc) Olivier Ezratty, 2021.*

Another schematic view of how classical and quantum computing are intertwined and the format of data that is handled in provided in Figure 188. What is specific to quantum computing is that the same instructions handle data and computing, i.e. quantum gates. The wealth of data in registers exists only during computing but not at the end, after measurement, where it is back in classical mode, turning the computational basis state vector of dimension $2^{N+1}$-1 real numbers to a meager N classical bits.

**Quantum switch** is a curious artefact worth mentioning here. It consists in creating a series of qubit transformations that can be implemented simultaneously in different orders. Like say, A then B and B then A, on a given register state. It defies logic and understanding of time flow, creating an indefinite causal order[329].

[329] See Comparing the quantum switch and its simulations with energetically-constrained operations by Marco Fellous-Asiani, Raphaël Mothe, Léa Bresque, Hippolyte Dourdent, Patrice A. Camati, Alastair Abbott, Alexia Auffèves and Cyril Branciard, August 2022 (20 pages).

It can even be a useful resource for improving reliability of quantum communications[330].

## Measurement

We'll now look into quantum measurement, a much broader topic than you may think. We have already explained that quantum measurement is assimilated to a wave function collapse onto basis states, in the case of qubits, $|0\rangle$ or $|1\rangle$. We've also seen that quantum computing is highly probabilistic, requiring executing several times your calculation and making an average of the obtained results.

But quantum measurement is way more subtle than that. We'll see here what can be measured in qubits and when, what is a projective measurement, what is a POVM, a CPTP map, what are gentle and weak measurements, non-selective and selective measurement, state tomography and the likes. Some of these techniques are related to quantum computing, including error corrections and some hardware benchmarking tasks and others with quantum telecommunications.

### Projective measurement

A projective measurement is the most generic form of measurement used in quantum computing. We'll first describe it geometrically and then with some mathematical formalism. Projective measurement is also named a von Neumann measurement since **John Von Neumann** elaborated its formalism in 1932.

It's easy to intuitively understand what it looks like with using the Bloch sphere for a qubit. A projective measurement consists in doing a geometrical vector projection of your qubit pure state on any axis in the Bloch sphere.

The simplest case of all is a projection on the z axis containing the $|0\rangle$ and $|1\rangle$ orthogonal vectors. It's about doing a measurement in the qubit computational basis. It could also be, theoretically, a projection on any other axis, like the $|+\rangle$ and $|-\rangle$ states that sit on the Bloch sphere equator along the x axis. We'll see later how to achieve this feat.

While quantum gates are reversible operations based on unitary operators, reading the state of the qubits is an irreversible operation. It is not a rotation in Bloch's sphere but a projection on an axis, which will yield a binary result with a probability depending on the qubit state. The projection is using a self-adjoint matrix operator, meaning that if executed several times, you'll always get the same result. Of course, the measurement of the qubit modifies its state unless it's already a perfect $|0\rangle$ or $|1\rangle$.

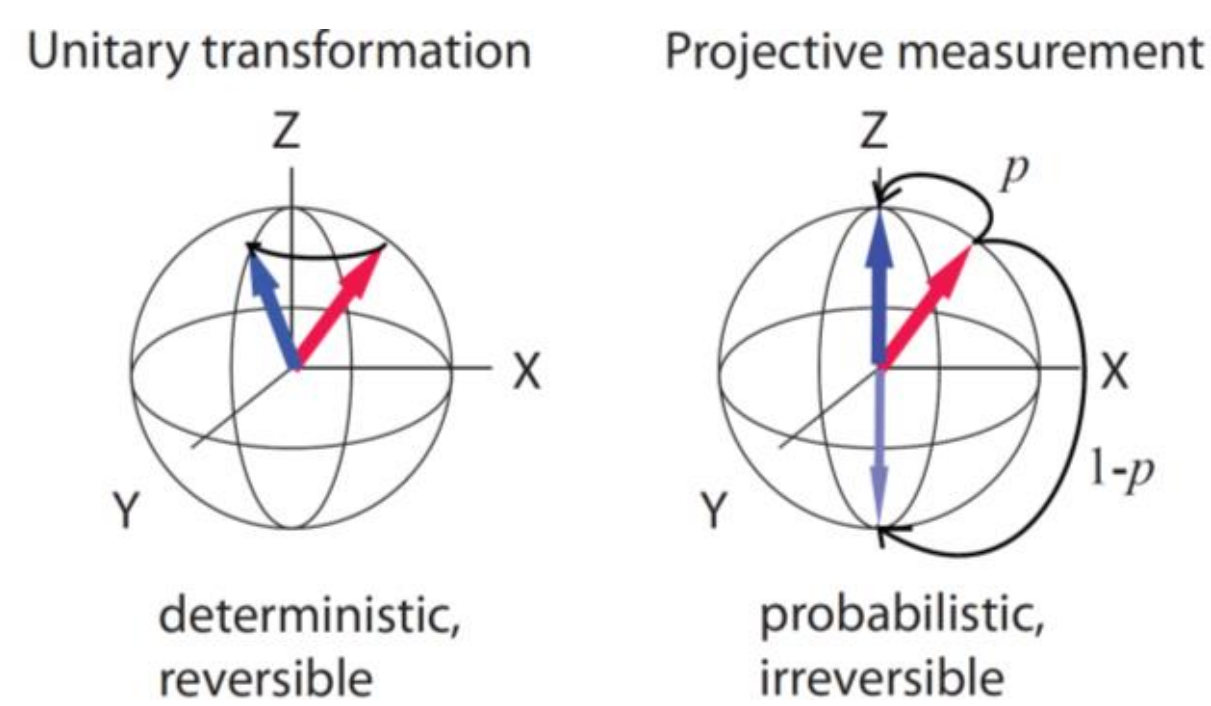

*Figure 189: visual difference between a unitary transformation (gate) and a projective measurement. Source: A computationally universal phase of quantum matter by Robert Raussendorf (41 slides).*

After a projective measurement on the Z axis, the qubit will irreversibly collapse in the states $|0\rangle$ or $|1\rangle$. Qubits measurement is reversible only in the case when they are already perfectly in the computational basis states $|0\rangle$ or $|1\rangle$. In that case, the measurement along the Z axis is not changing the qubit value and is therefore reversible since it's an identity operation.

Mathematically, a projective measurement is using Projection-Valued Measures (PVMs) on a closed system. On a given qubit, it uses two orthogonal measurement operators, in the form of 2x2 self-adjoined (Hermitian) matrices.

[330] See Improvement in quantum communication using quantum switch by Arindam Mitra, Himanshu Badhani and Sibasish Ghosh, September 2022 (14 pages).

When measuring a qubit along the Z axis, also named the observable Z with eigenvalues +1 and −1 and eigenvectors $|0\rangle$ and $|1\rangle$ (the observable Z is the matrix representation of a Z single qubit quantum gate!), these PVMs operators are respectively:

$$M_0 = |0\rangle\langle 0| = \begin{bmatrix}1\\0\end{bmatrix}\begin{bmatrix}1 & 0\end{bmatrix} = \begin{bmatrix}1 & 0\\0 & 0\end{bmatrix} \quad \text{and} \quad M_1 = |1\rangle\langle 1| = \begin{bmatrix}0\\1\end{bmatrix}\begin{bmatrix}0 & 1\end{bmatrix} = \begin{bmatrix}0 & 0\\0 & 1\end{bmatrix}$$

Given the Z observable operator is Z=$M_0 - M_1$, which returns $+1$ for $|0\rangle$ and $-1$ for $|1\rangle$.

On a general basis, with a quantum object with several distinct states, a measurement operator is a matrix $M_m$ and the probability to get the outcome m (with m=0 and 1 in the case of a qubit, or m=0 to N-1 in the case of a N states quantum object) is $p(m) = \langle\psi|M_m^\dagger M_m|\psi\rangle$ with the completeness constraint $\sum_m M_m^\dagger M_m = I$ ($I$ being the identity matrix).

For m=0, it reads as $p(0) = \begin{bmatrix}\alpha & \beta\end{bmatrix}\begin{bmatrix}1 & 0\\0 & 0\end{bmatrix}\begin{bmatrix}1 & 0\\0 & 0\end{bmatrix}\begin{bmatrix}\alpha\\\beta\end{bmatrix} = \begin{bmatrix}\alpha & \beta\end{bmatrix}\begin{bmatrix}\alpha\\0\end{bmatrix} = \alpha^2$! Since $\beta^2 = 1 - \alpha^2$ due to the Born normalization rule, only one measurement is required to get both $\alpha^2$ and $\beta^2$, these being not individual measurement results but their respective probabilities.

Any global phase added to $|\psi\rangle$ will disappear during measurement. If we define $|\psi'\rangle = e^{i\theta}\,|\psi\rangle$ and apply a measurement operator $M_m$ on $|\psi'\rangle$:

$$p'(m) = \langle\psi'|M_m^\dagger M_m|\psi'\rangle = \langle\psi'|e^{-i\theta} M_m^\dagger M_m e^{i\theta}|\psi'\rangle = \langle\psi'|M_m^\dagger M_m|\psi'\rangle = p(m)$$

After the measurement with the operator $M_m$, the system state $|\psi\rangle$ becomes the projection of $|\psi\rangle$ on $M_m$ divided by the probability of getting state m:

$$\frac{M_m|\psi\rangle}{\sqrt{\langle\psi|M_m^\dagger M_m|\psi\rangle}} \quad \text{also often written} \quad \frac{M_m|\psi\rangle}{\sqrt{\langle\psi|M_m|\psi\rangle}}$$

since $M_m^\dagger = M_m$ (self-adjoint matrix) and $M_m M_m = M_m$ (projector matrix)

All these measurement equations are part of the measurement postulate (usually the third) from quantum mechanics postulates.

In Figure 190, let's make a pause to understand the $\langle A|B|C\rangle$ Dirac notation. You usually read it from the right. The ket on the right is a vertical vector that is multiplied by the middle object that is a square matrix. It creates a similar vertical vector. Then, you multiply it with the bra on the left which is a horizontal vector. It is a dot product of an inner scalar product. The result is a complex number and it is a real number when Ψ = ϕ.

Now, let's be a bit practical.

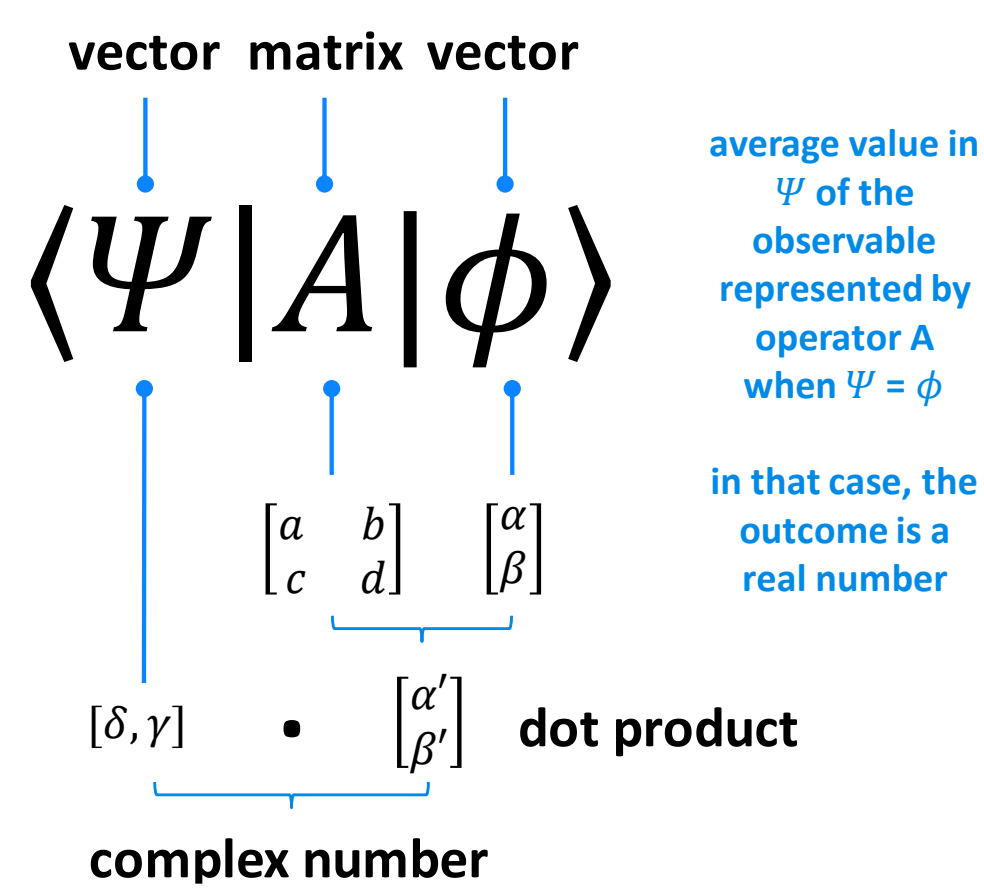

Figure 190: understanding the $\langle A|B|C\rangle$ Dirac notation.

How can we change the measurement basis with qubits, for implementing a measurement along another axis than Z? At least two options are available:

- It may be possible to *physically* implement a measurement on a different basis than the computational basis. This is, for example, the case with polarization-based photon qubits where the polarizer angle can be dynamically and programmatically modified with some electrically controlled optical settings. It looks more difficult to implement for other types of qubits.

- When the *only* supported measurement is a projective measurement in the computational basis $|0\rangle$ and $|1\rangle$, any another projective measurement can be implemented with first applying a unitary transformation to the qubit that creates a rotation in the Bloch sphere equivalent to moving the measurement axis to the Z axis ($|0\rangle$ and $|1\rangle$). When we say we do an "X" or "Y measurement", it means that we first apply a H or $HS^{\dagger}$ single gate rotation (H = Hadamard gate and S = half a Z gate or quarter phase turn) to handle this axis rotation and then, apply a (computational basis) Z-axis measurement. This is what is regularly done with quantum error correction codes as well as with MBQC (measurement-based quantum computing).

  With QECs (quantum error correction codes), this sort of projective measurement is applied to ancilla qubits, these additional qubits that detect errors in entangled computing qubits. So, when physicists say they are doing a measurement on a basis of two orthogonal vectors, they mean they are applying first a unitary transformation and then a measurement on the computational basis.

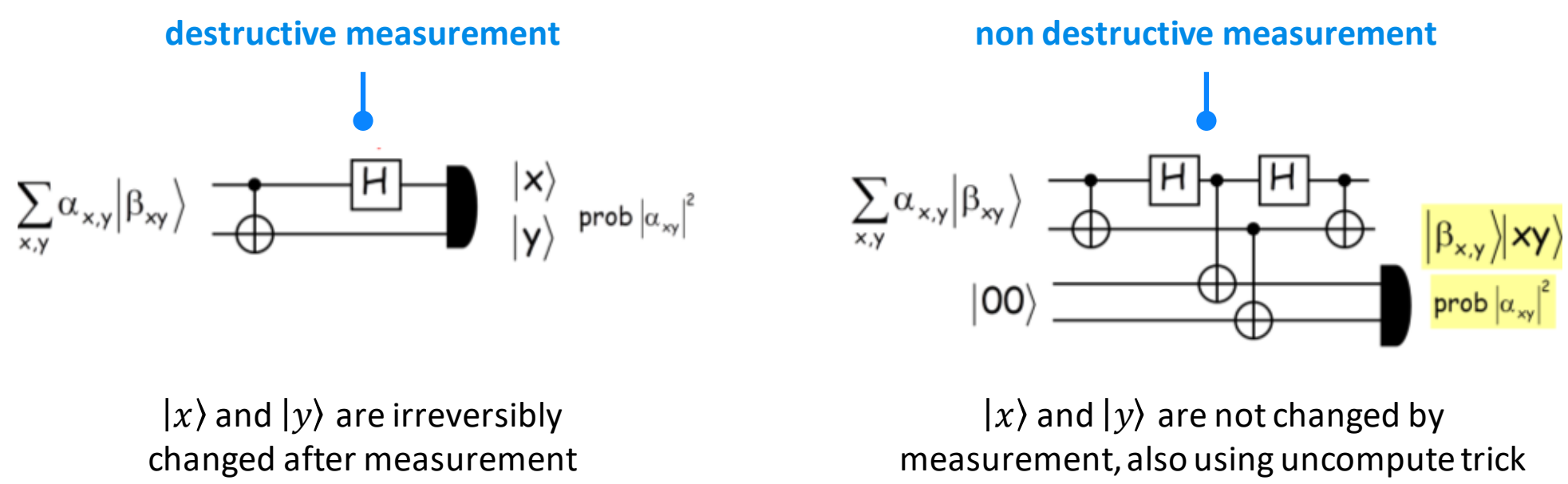

*Figure 191: how a projective measurement in a different basis can implement non-destructive measurement. Which is actually different from the notion of QND (quantum-non-destructive measurement) that we'll define later.*

### Qubits register measurement

So far, we've just elaborated on measurement mathematical underlying tools and dealt with only one qubit. How about measuring a whole qubit register?

A N qubit register has $2^N$ possible computational basis states, from $|00\ldots00\rangle$ to $|11\cdots11\rangle$. When measuring once a qubit register, you get one of these states, being a combination of N 0s and 1s.

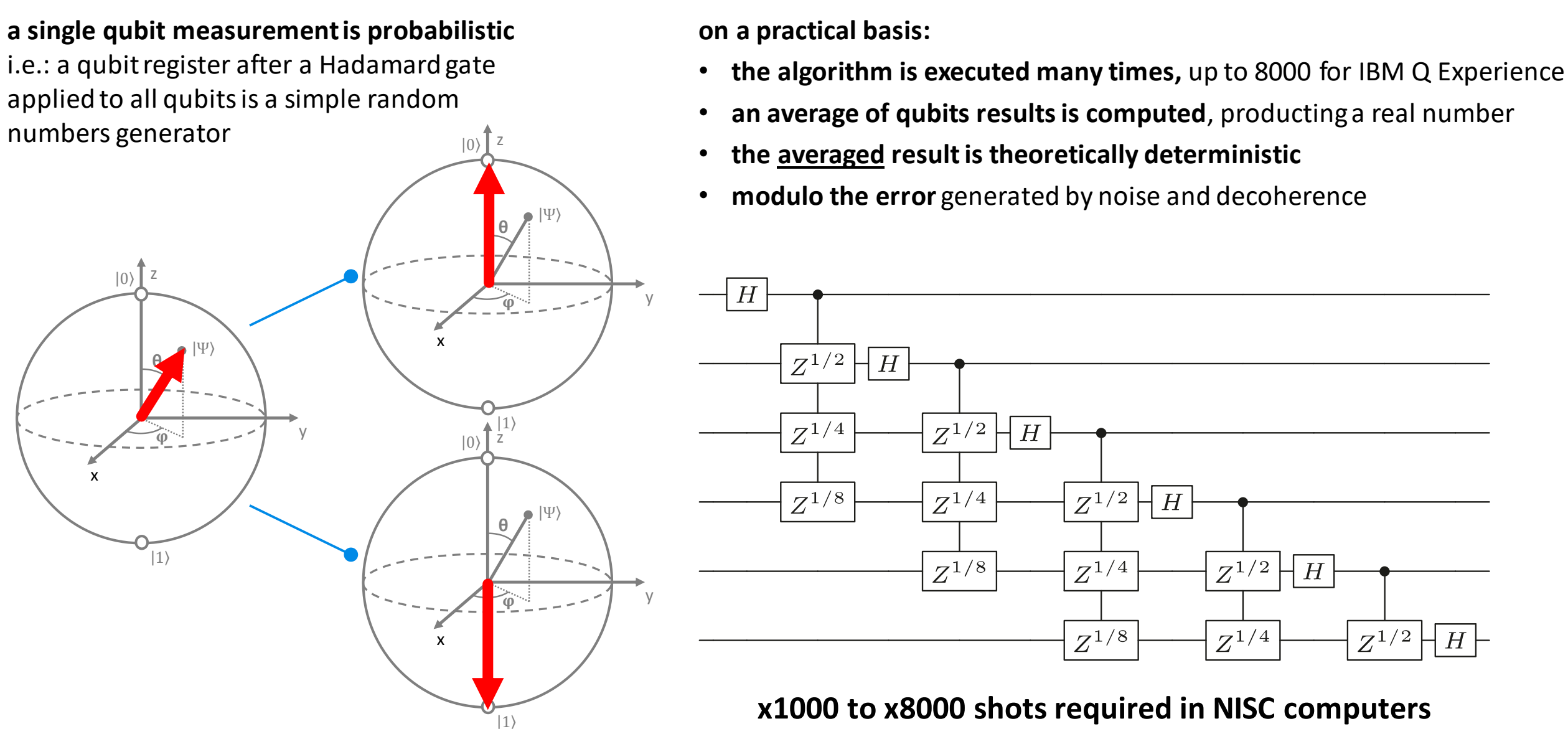

*Figure 192: a qubit probabilistic measurement and the notion of computing shots. (cc) Olivier Ezratty, 2021.*

You could stop there and think, that's my result, fine, I'm done! Well, no! Since the measurement outcome is probabilistic and prone with errors, you need to run your algorithm a certain number of times and count the number of times you'll get each computational basis state. If doing so a great number of times, you'll end up recovering a probability distribution for each computational basis state and reconstruct a full state vector. But to do that, you'll need to execute your algorithm an exponential number of times with regards to the number of qubits, losing any gain coming from quantum computing.

The process you'll implement will depend on what data you want to extract from your prepared qubits register and the run algorithm. Usually, a quantum algorithm is supposed to generate a simple computational basis state (one given combination of 0s and 1s) and not a combination of several states and their respective probabilities.

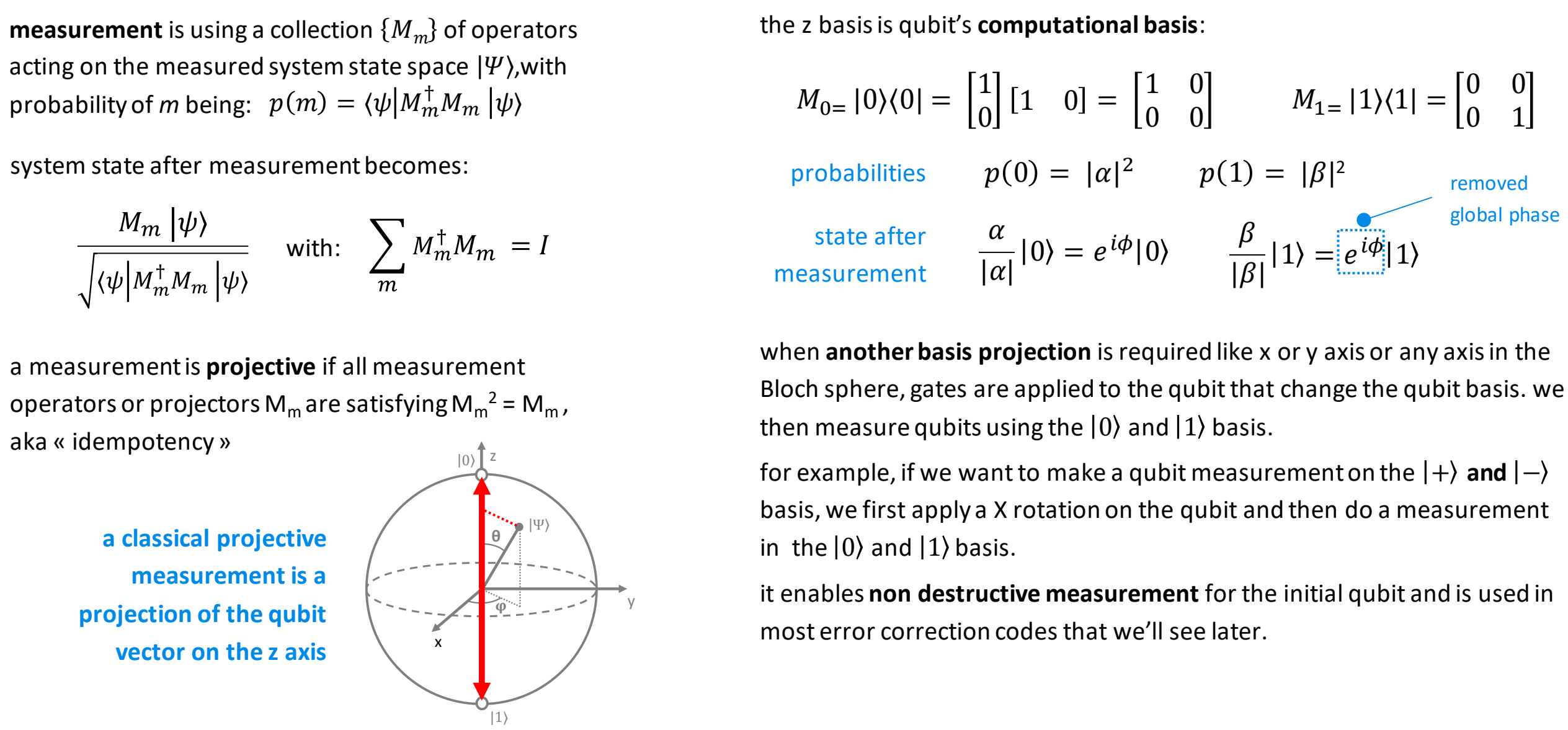

*Figure 193: another explanation of projective measurement on a different basis and its usage in non-destructive measurement techniques like with error correction codes. (cc) Olivier Ezratty, 2021.*

You can then run several times your algorithm and compute the average values of each qubit, giving a % of 0/1 for each then round up to the nearest 0 and 1. And there you are. What is "several"? It depends. IBM proposes to run your algorithm a couple thousand times on its cloud Q Experience platform with 5 to 65 qubits and states that this number will grow with the number of qubits, we hope linearly. I have not yet found the rule of thumbs used to define the number of runs, or "shots".

All in all, you must remember that one run of an algorithm is **probabilistic** and with many runs, you'll converge progressively to a **deterministic** solution being the average of all runs results.

### From computational vector state to full state tomography

What are we measuring? A single computational state, a statistical weight of 0 and 1 or a full vector state? It depends on the algorithm and also on the actual technical need of the undertaken measurement. For most algorithms, a series of runs and qubit measurement and their average will output after roundup the found computational basis state.

For algorithms debugging with a reasonable number of qubits and for characterizing the quality of a small group of qubits, it may be useful to compute either a histogram of the whole computational state vector or even, a so-called quantum state tomography which will reconstitute the density matrix of the quantum register.

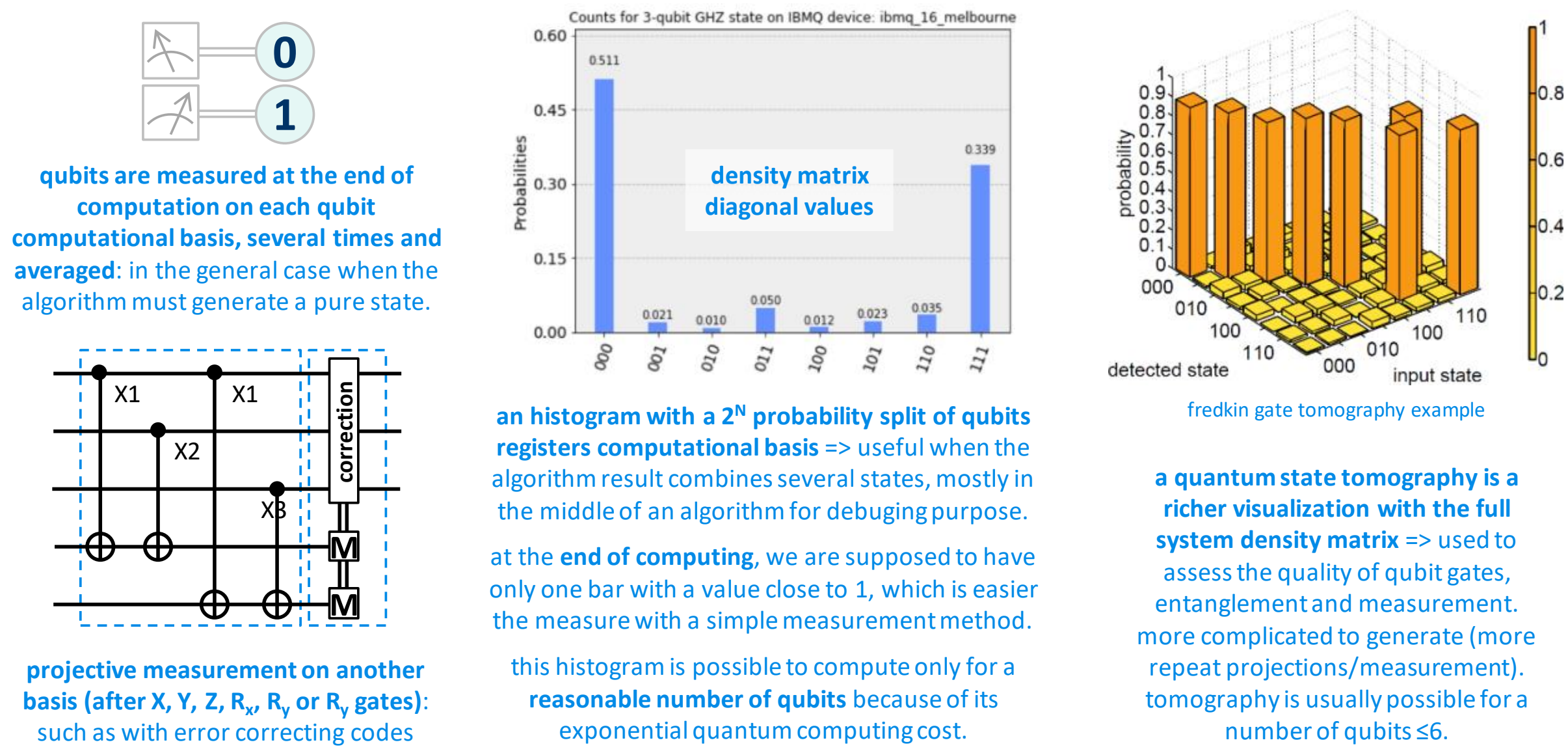

*Figure 194: from a vector state to a full density matrix, the various ways to measure the state of a qubit register. Compilation (cc) Olivier Ezratty, 2021.*

The computational state vector is assembled with a lot of repeat runs and measurements with a number growing exponentially with the number of qubits. It will eventually provide the statistical distribution of each and every computational basis states. Since the number of runs grows exponentially, you understand quickly why it won't make sense to use this technique when we'll exploit a large number of qubits.

Development tools like IBM Quantum Experience dumps the vector state of your qubits only for helping you learn about how their system work and also understand the impact of noise and decoherence.

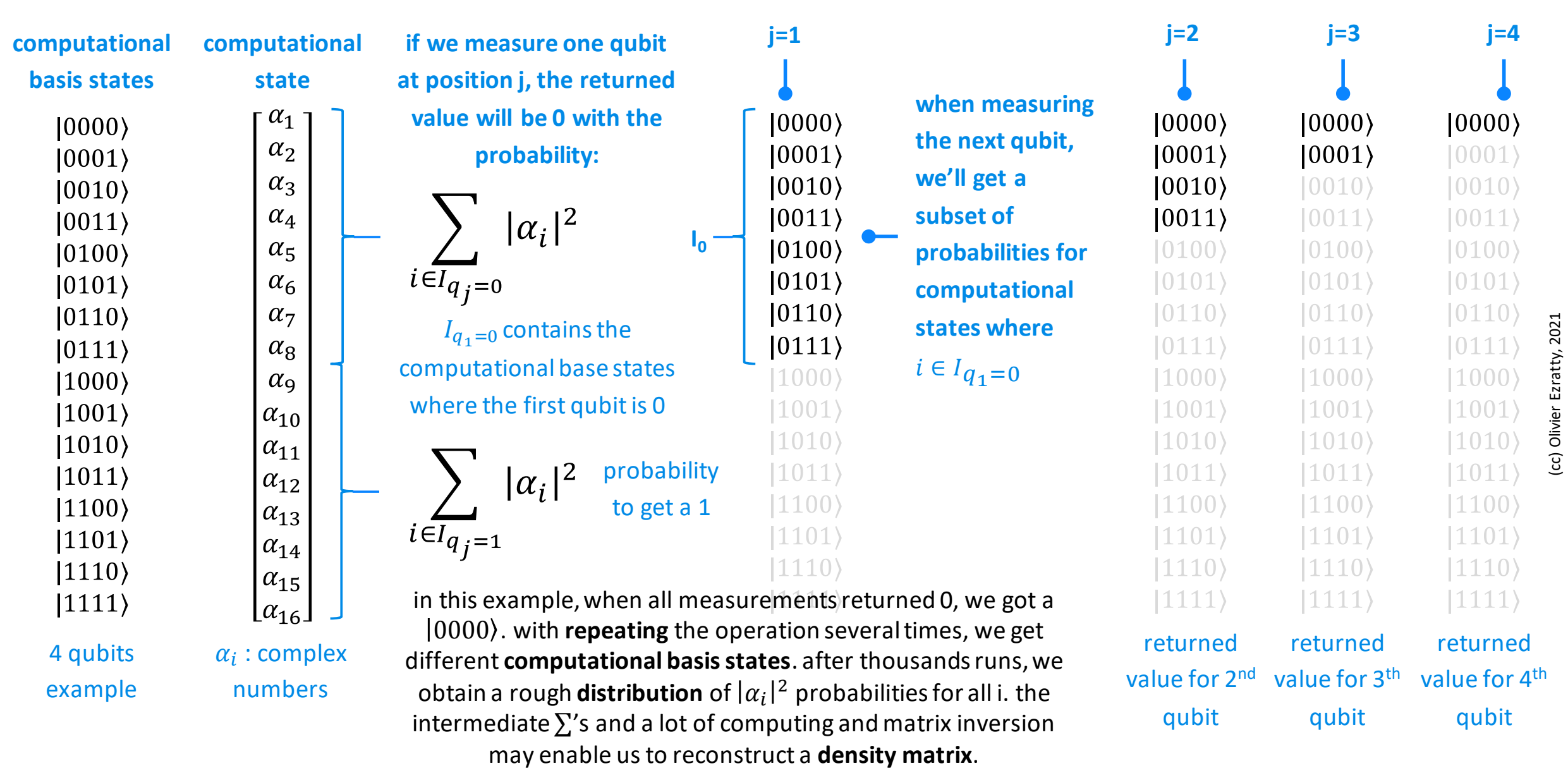

*Figure 195: what happens to your qubits when you progressively measure them. (cc) Olivier Ezratty, 2021.*

Reconstituting the whole system density matrix is a more tedious process. In the most basic technique used, we are keeping track of all intermediate measurements leading to getting the computational state vector and some matrix inversion is required to create it in the end. The process requires even more quantum and classical computation than for reconstituting the computational state vector. And it scales with $2^{3N}$, N being the number of qubits!

This is usually applied with up to 6 qubits, and particularly with 2 qubits to characterize the quality of two qubit gates. A record state tomography of 8 qubits was achieved in 2005 with trapped ions by Rainer Blatt's group in Innsbruck[331].

The graphical representation of these density matrices is often used to evaluate the fidelity of 2 or 3-qubit gates in research publications. The example in Figure 196 illustrates this with comparing the theoretical state of a density matrix for 2 and 4 qubits and measurement results. It also helps qualify the quality of qubits entanglement. Various techniques are proposed to speed-up quantum state tomographies and achieve it with a better precision.

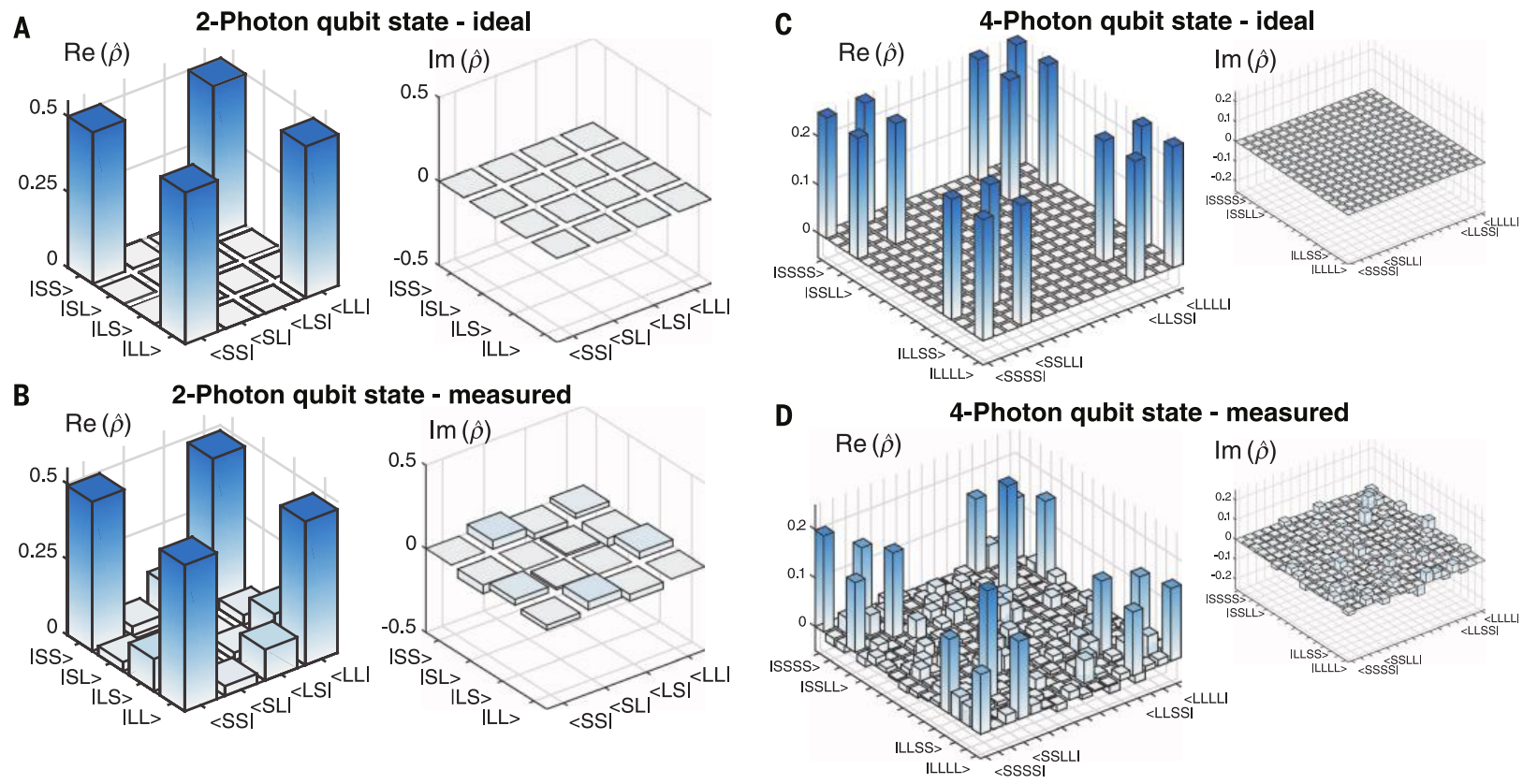

*Figure 196: the difference between an ideal 2 and 4-photon density matrices and as measured in experiments. Source: [Generation of multiphoton entangled quantum states by means of integrated frequency combs](#) by Christian Reimer et al, Science, 2016 (7 pages).*

However, this is a tool for researchers and hardware designers, not for quantum software developers[332]. The next step is a Quantum Process Tomography which qualifies the quantum channel of a given process, like a series of gates, one gate, or quantum noise and decoherence. It creates an even richer matrix with $2^{2N}$ columns and rows, representing a linear operator on the system density matrix, *aka* a superoperator.

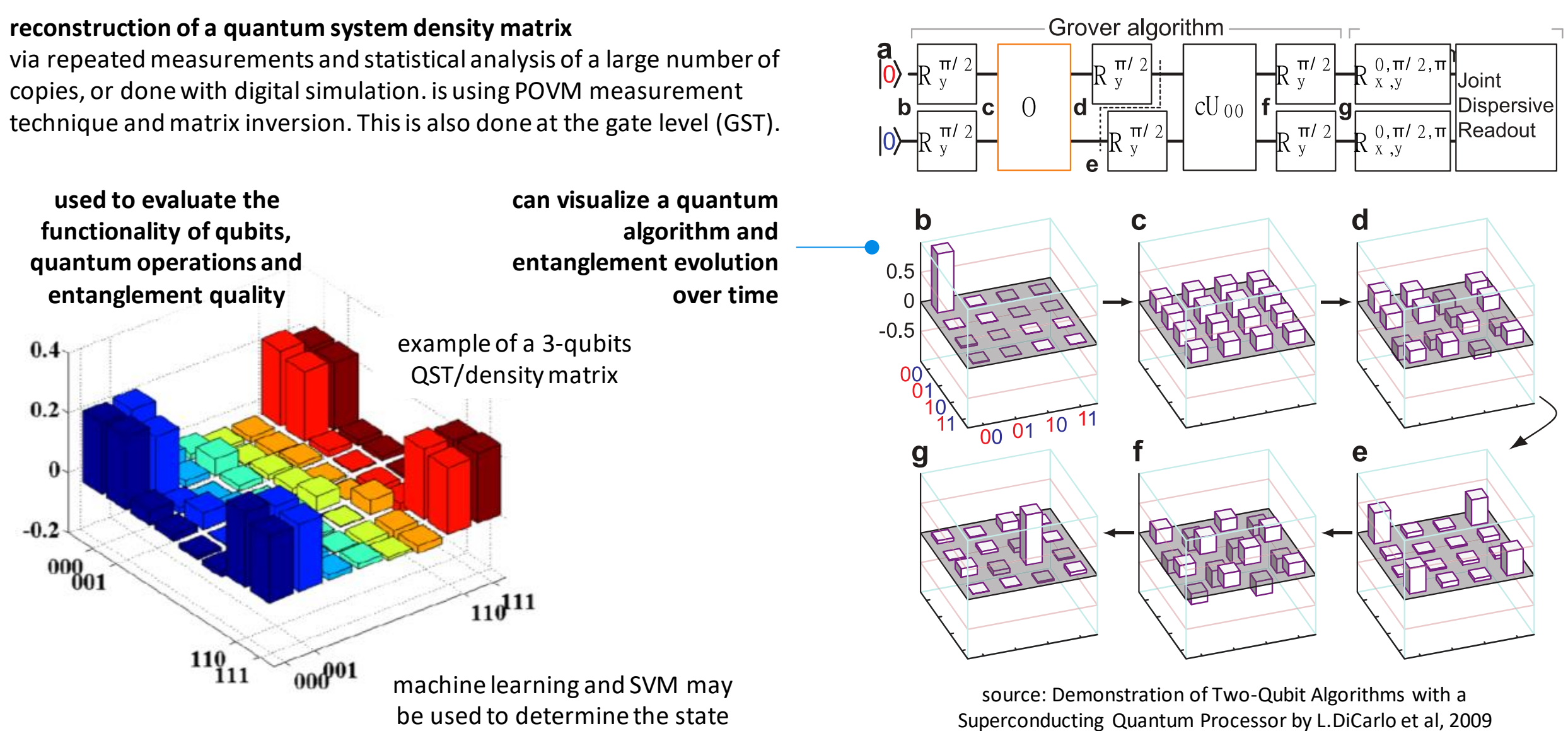

*Figure 197: how do you reconstruct a quantum system density matrix.*

## Non-selective and selective measurements

A non-selective measurement is a measurement that is physically done but not yet read. For any reason, its outcome is not available either because it wasn't yet used or because it's really inaccessible when measurement is done by the environment. How is it different from a real measurement? It deals with the information available about the quantum states we are evaluating. This is explained in the example below using photons polarization and relates with pure states and mixed states.

[331] See [Scalable multi-particle entanglement of trapped ions](#) by H. Haffner, Rainer Blatt et al, 2006 (17 pages).

[332] See for example [Quantum process tomography via completely positive and trace-preserving projection](#) by George C. Knee et al, UK, 2020 (13 pages). But it requires some background knowledge!

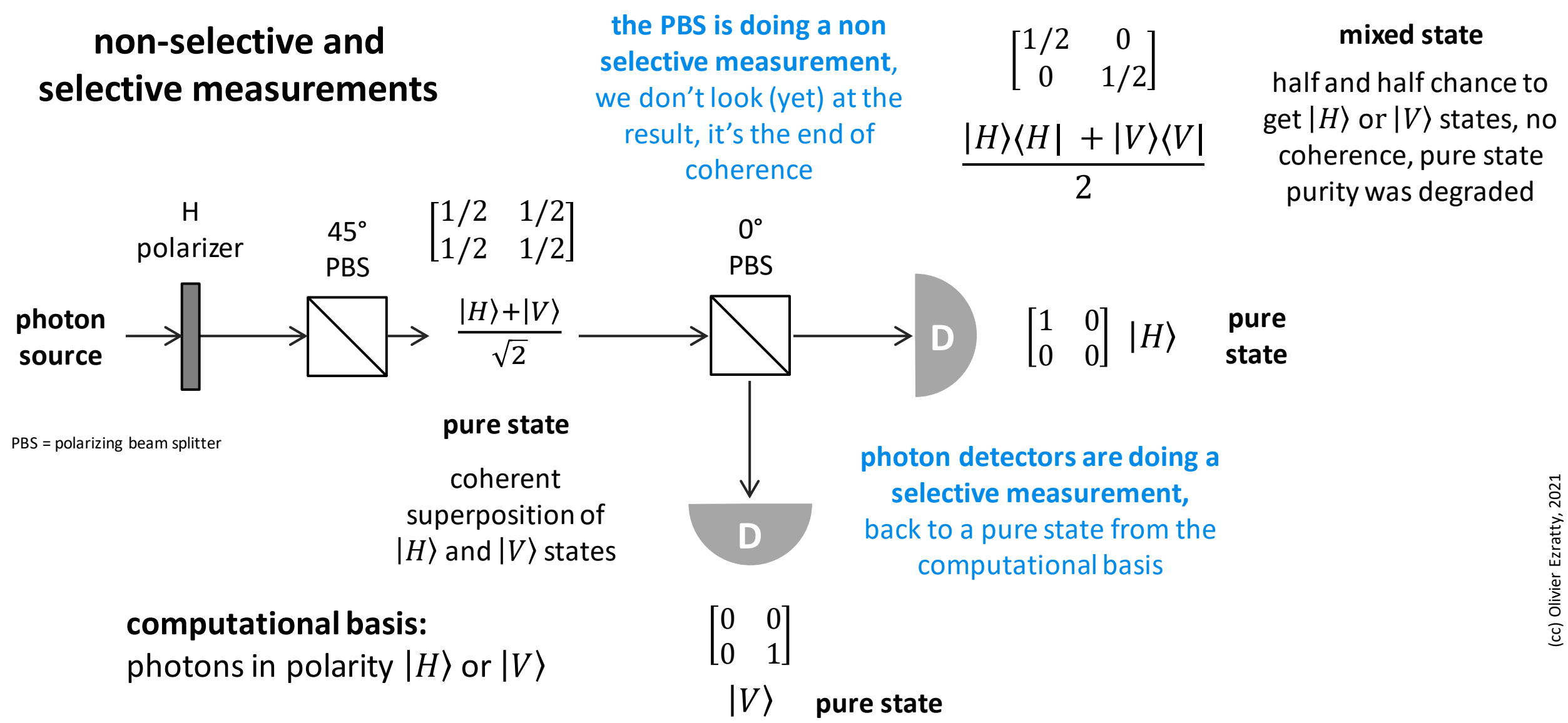

*Figure 198: non-selective and selective measurements. (cc) Olivier Ezratty, 2021.*

A single photons source generates photons that traverse first a horizontal polarizing filter and then a 45° polarizing beam splitter (PBS). The PBS create a pure state coherent superposition of $|H\rangle$ and $|V\rangle$ states (horizontally and vertically polarized photons). Then, this coherent superposition traverses a 0° PBS. The outcome can be measured in the two PBS exits with single photon detectors. Before being measured, this output is a mixed state of $|H\rangle$ and $|V\rangle$.

There is no more coherent superposition (exit the pure state) and we don't know yet what both detectors will read. But we know that there's a 50% chance that the detector on the PBS horizontally polarized exit will detect a photon and 50% for the other detector. After detection, we'll end up with finding a single photon on one of the detectors, giving a related pure state. And nothing for the other.

This means that after measurement of a qubit in a given basis, the coherences in its density matrix in the measurement basis are erased. There's no more any coherence and superposition. This happens before looking at any measurement outcomes. In other words, a non-selective measurement of a pure state degrades its purity by turning it into a totally mixed state.

This could be used in a new updated Schrodinger' cat thought experiment, replacing the disintegrating radium atom by a simple qubit in a superposed state (after a H gate). A measurement at time T would trigger the poison release if the result is $|1\rangle$. All this in a closed box. Keeping the box closed at time T+whatever would be an equivalent of a non-selective measurement, then opening the box at time T+after whatever, would become a classical measurement of an already totally mixed state.

### Positive Operator-Valued Measurement (POVM)

A Positive Operator-Valued Measure (POVM) is a quantum measure generalizing Projection-Valued Measures (PVMs) which is useful when the measurement basis is not made of orthogonal states in their Hilbert space. It is of particularly interest when measuring a photon qubit in a telecommunication link with two non-orthogonal polarization basis (0° and 45° like in the BB84 protocol). Like in PVMs, the measurement operators of a POVM add up to identity matrix. POVMs are also interesting when measuring a subsystem of an open system.

POVMs that are not PVMs are called non-projective measurements. They have many use cases like enhancing quantum states tomography, help detect entanglement and allow unambiguous state discrimination of non-orthogonal states, with applications in quantum cryptography and randomness generation[333].

**Other measurements concepts**

I'll cover here other measurement-related tools and concepts I have encountered in various courses and scientific papers. You probably don't need to understand this if you are just a quantum software developer. It may be interesting, however, if you are involved in designing quantum systems, error correction systems, measurement systems, quantum firmware and the likes.

**Gentle or Weak Measurement**. It is one type of quantum measurement that retrieves little information of the measured system in average with the benefit of only slightly disturbing it. In a weak measurement, the correlations in the off-diagonal values of the system density matrix are only slightly altered. The system purity and entanglement remain mostly unaltered.

**Postselected Measurement**. It is a measurement where the result is chosen by the user, usually after a weak measurement. Surprising! As all measurements, it also turns a pure state into a mixed state. It refers to the process of conditioning on the outcome of a measurement on some other qubit values. The process consists in throwing away any outcome which does not allow you to do what you want to do. If the outcome you are trying to select has probability 0<p<1, you will have to try an expected number 1/p times before you manage to obtain the outcome you are trying to select. If p=1/2n for some large integer n, you may be waiting a very long time.

This weird technique is noticeably used to better understand quantum physics and phenomenon like measurement non-commutativity[334].

**CPTP map**. A Completely Positive and Trace Preserving map also referenced as a quantum channel is used to describe non-selective measurements, conditional expectations and quantum filters, as well as feedback networks in quantum control theory. It corresponds to the most generic operation that can be applied to a quantum system. The state of the target system is associated to a trace-one, positive semidefinite density operator and, under the assumption that no initial correlations are present with the environment, its evolution over some specified time interval is described by a completely positive, trace-preserving (CPTP) linear map.

For open quantum systems, however, the interaction between the system and environment leads to non-unitary evolution of the system (e.g., dissipation), which requires CPTP maps for full characterization[335].

In other words, a CPTP map is the mathematical operation that transforms the density matrix $\rho$ of a quantum system during a measurement on the basis $\langle m_k|$ into the density matrix $\rho'$ as described in Figure 199. A CPTP map is a superoperator of dimension $2^{4N}$ complex numbers.

$$\rho' = \sum_k M_k \, \rho M_k = \sum_k p_k \, M_k$$

$$\text{with } p_k = \langle m_k|\rho|m_k\rangle$$

*Figure 199: defining a CPTP map.*

---

[333] See Understanding the basics of measurements in Quantum Computation by Nimish Mishra, 2019. But what is $\delta_{mm'}$ in these formulas? It is the Kronecker Delta function which is equal to 0 when m≠m' and equal to 1 when m=m'. Meaning that inner product of all measurement operators is equal to 0 when they are different. This is the definition of orthonormality between a set of operators.

[334] See for example Quantum advantage in postselected metrology by David R. M. Arvidsson-Shukur, Seth Lloyd et al, Nature Communications, 2020 (9 pages).

[335] Source: Quantum and classical resources for unitary design of open-system evolutions by Francesco Ticozzi and Lorenza Viola, 2017 (27 pages).

**Quantum Non-Demolition measurement**. It is a type of measurement in which the uncertainty of the measured observable does not increase from its measured value during the subsequent normal evolution of the system. For a qubit measurement, it means that after its measurement, its value won't change anymore in subsequent measurements. QND measurements are the least disturbing type of measurement in quantum mechanics. QND measurements are extremely difficult to implement. Note that the term "non-demolition" does not imply that the wave function fails to collapse[336]. It can be implemented with photons, particularly to measure a photon number (number of photons in a superposed states of similar photon, or a single-mode Fock state), using a secondary probe field interfering with the signal field[337]. It has also been experimented to measure an electron spin with an additional ancilla quantum dot next to an operational quantum dot[338]. It also currently works well with superconducting qubits. What would be a "demolition measurement"? It would be one that, after retrieving the result, would create so significant a back-action on the measured quantum that it would either destroy it (like a classical photon counting device that absorbs the counted photons) or turn it into a state outside the computational basis (such as a different energy level than ground/excited levels for a qubit).

**Quantum Steering** is a quantum measurement phenomenon when one subsystem can influence the wave function of another subsystem by performing specific measurements. It is a variation of non-local correlations intermediate between Bell nonlocality and quantum entanglement[339].

**Quantum Measurement Thermodynamics**. We have already mentioned the theoretical reversible aspect of gates-based quantum computing which relates to the unitary transformations applied with quantum gates. But most of the time, particularly with solid qubits, there is always some energy exchanges between qubits and their control as well as measurement devices. Fundamental research is undertaken to better understand the evolution of the thermodynamic equilibrium of qubit operations particularly during entanglement and also, measurement and error correction. Since measurement is done on a repeated basis due to the implementation of quantum error correction codes, it makes sense to wonder whether this could be optimized. Depending on the qubit state (ground level or excited level, and also in intermediate states), measurement can absorb or release some energy that is quantum and microscopic in nature and it's also powered by entanglement[340].

---

[336] QND was initially introduced in 1975 by VB Braginsky and YI Vorontsov in USSR. Source: Quantum Nondemolition Measurement,Wikipedia. See also Quantum Non-Demolition Measurement of Photons by Keyu Xia, March 2018. It was demonstrated with the detection of a single photon as described in Seeing a single photon without destroying it by G. Nogues et al, 1999 (4 pages).

[337] See Detecting an Itinerant Optical Photon Twice without Destroying It by Emanuele Distante et al, Max Planck Institute, June 2021 (6 pages) which deals with detecting twice a photon with some non demolition quantum measurement. The detectors use a single atom coupled to an optical cavity. Other methods consist in using the cross-Kerr effect where a measured photon traverses an optical medium and changes its refraction index. It provokes a phase shift for a probe photon traversing the same media, its phase being measured with a Mach-Zehnder interferometer. See a description of this old technique in Quantum non-demolition measurements in optics by Philippe Grangier, Juan Ariel Levenson and Jean-Philippe Poizat, 1998 (7 pages).

[338] See Quantum non-demolition readout of an electron spin in silicon by J. Yoneda et al, Nature, 2020 (7 pages).

[339] See Quantum Steering by Roope Uola et al, 2020 (43 pages) and Quantum steering on IBM quantum processors by Lennart Maximilian Seifert et al, PRA, April 2022 (11 pages) which shows poor entanglement with 15 qubits.

[340] See also Probing nonclassical light fields with energetic witnesses in waveguide quantum electrodynamics by Maria Maffei, Patrice Camati and Alexia Auffèves, September 2021 (6 pages) which studies the thermodynamics of a qubit coupled to a waveguide, which relates well to superconducting qubit gates and readout operations but also other qubit operations (photons, cold atoms). They demonstrate that the work performed by a coherent pulse on the qubit is always larger than the work that can later be extracted from the qubit, *aka* its ergotropy. But this classical ergotropy bound is violated if the input field is a resonant single-photon pulse. This opens the door to some energy recovery at the end of computing.

This research field could lead to a better understanding of the whereabouts of the energetic footprints of quantum measurement and entanglement and how it can impact the energy cost of quantum computing, particularly as it scales up[341].

**Quantum Reservoir Engineering** is a set of qubits management techniques using a quantum bath in order to reduce its energetic footprint, its measurement readout times and enable quantum non-demolition measurement[342]. It's about tightly controlling the qubit coupling with its environment. It is connected to quantum error correction techniques. The approach was initially imagined for NMR qubits, leveraging the Nuclear Overhauser effect. Then it was tested with trapped ions, using some coupling between the qubit harmonic oscillator and a reservoir of oscillator with laser radiations[343]. The technique is also branded "quantum bath", "engineered dissipation", "autonomous feedback" and "coherent feedback". It has since been tested with superconducting qubits and is the basis of the cat-qubits from Inria, Alice&Bob and Amazon[344].

**Algorithmic Cooling** is a related technique also named heat-bath algorithmic cooling, which consists in balancing the entropy transfers between qubits and with ancilla qubits as part of error correction codes[345]. It is used to improve the purity of a target subset of qubits quantum states in a qubit register.

---

[341] The thermodynamics of quantum measurement is involving a few groups worldwide including the team of Alexia Auffèves from Institut Néel in Grenoble, France, IQOQI and the University of Innsbruck in Austria and Andrew Jordan's team at the University of Rochester, USA. See A two-qubit engine powered by entanglement and local measurements by Ingrid Fadelli, April 2021 which refers to Two-Qubit Engine Fueled by Entanglement and Local Measurements by Léa Bresque, Andrew Jordan, Alexia Auffèves et al, March 2021, PRL (5 pages), Alternative experimental ways to access entropy production by Zheng Tan, Alexia Auffèves, Igor Dotsenko et al, May 2021 (15 pages) and the colloquium A short story of quantum and information thermodynamics by Alexia Auffèves, March 2021 (14 pages). See also Stochastic Thermodynamic Cycles of a Mesoscopic Thermoelectric Engine by R David Mayrhofer, Cyril Elouard, Janine Splettstoesser and Andrew Jordan, October 2020 (18 pages) and Thermodynamics of quantum measurements by Noam Erez, 2018 (3 pages).

[342] Quantum Reservoir Engineering must not be confused with Quantum Reservoir Computing which is an entirely different beast. Introduced by Keisuke Fujii and Kohei Nakajima in 2017, it is the quantum equivalent of a similar technique used in classical deep learning where a low-dimensional data input is projected onto a higher-dimensional dynamical system, the reservoir, generating transient dynamics that facilitates the separation of input states. It is particularly useful to analyze time series of complex data structures. See Quantum reservoir computing: a reservoir approach toward quantum machine learning on near-term quantum devices by Keisuke Fujii and Kohei Nakajima, November 2020 (13 pages).

[343] See Quantum Reservoir Engineering by J.F. Poyatos, J.I. Cirac and Peter Zoller, 1996 (14 pages) and the associated presentation Quantum Reservoir Engineering by Peter Zoller, 2013 (86 slides).

[344] See Measurement, Dissipation, and Quantum Control with Superconducting Circuits by Patrick Michael Harrington, 2020 (154 pages), Reservoir engineering using quantum optimal control for qubit reset by Daniel Basilewitsch et al, 2019 (13 pages), Reservoir (dissipation) engineering and autonomous stabilization of quantum systems, Quantic team, Inria, 2018 and Quantum reservoir engineering and single qubit cooling by Mazyar Mirrahimi, Zaki Leghtas and Uri Vool, 2013 (6 pages).

[345] See Novel Technique for Robust Optimal Algorithmic Cooling by Sadegh Raeisi, Mária Kieferová and Michele Mosca, June 2019 (10 pages).

## Gate-based quantum computing key takeaways

- Gate-based quantum computing is the main quantum computing paradigm. It relies on qubits and finite series of quantum gates acting on individual qubits or two and three qubits. The main other paradigms belong to analog quantum computing and include quantum simulators and quantum annealers.
- To understand the effect of qubits and quantum gates, you need to learn a bit of linear algebra. It deals with Hilbert vector spaces made of vectors in highly multidimensional spaces, matrices and complex numbers. The Dirac Bra-Ket notation helps manipulate vectors and matrices in that formalism.
- A qubit is usually represented in a Bloch sphere, reminding us of the wave nature of quantum objects during computation. This wave nature is exploited with qubits phase control and entanglement which provokes interferences between qubits. Qubits entanglement is created by conditional multi-qubit gates like the CNOT.
- A qubit register of N qubits can store a linear superposition of $2^N$ basis states corresponding to the qubit computational basis, each associated with a complex number. But surprisingly, this exponential growth in size is not enough to create an exponential speedup for quantum computing.
- While the computational space grows exponentially with the number of qubits, a qubit register measurement at the end of quantum algorithms yields only N classical bits. You have to deal with it when designing quantum algorithms.
- Computation must usually be done a great number of times and its results averaged due to the probabilistic nature of qubits measurement.
- Qubits measurement can be done in various ways, the main one being a classical projective measurement, if possible a non-demolition one (QND) that will maintain the qubit in its collapsed state after measurement. Other techniques are used that are useful for qubits quality characterization and for quantum error corrections.

# Quantum computing engineering

After reviewing the basic principles of quantum physics and the logical dimension of gate-based quantum computing, let's look at the operational and physical operations of a quantum computer[346].

Quantum computer architectures depend closely on the characteristics of their qubits. In this section, we will rely on the most common universal quantum gate computer architecture, that of superconducting qubits based on the Josephson effect. It is notably used by IBM, Google, Intel, Rigetti and IQM. However, many of the architectural principles mentioned here are applicable to quantum computers using other types of qubits.

First and as a reminder, here are the main components of a classical computer that you also find in various shapes and forms in smartphones, tablets, personal computers, game consoles and servers. Its key component is its microprocessor. It retrieves data and programs from a storage system and copies them to memory (RAM) entirely or on the fly as needed. The microprocessor then reads the program's instructions from memory in its cache to execute it one after the other and use conditional branching.

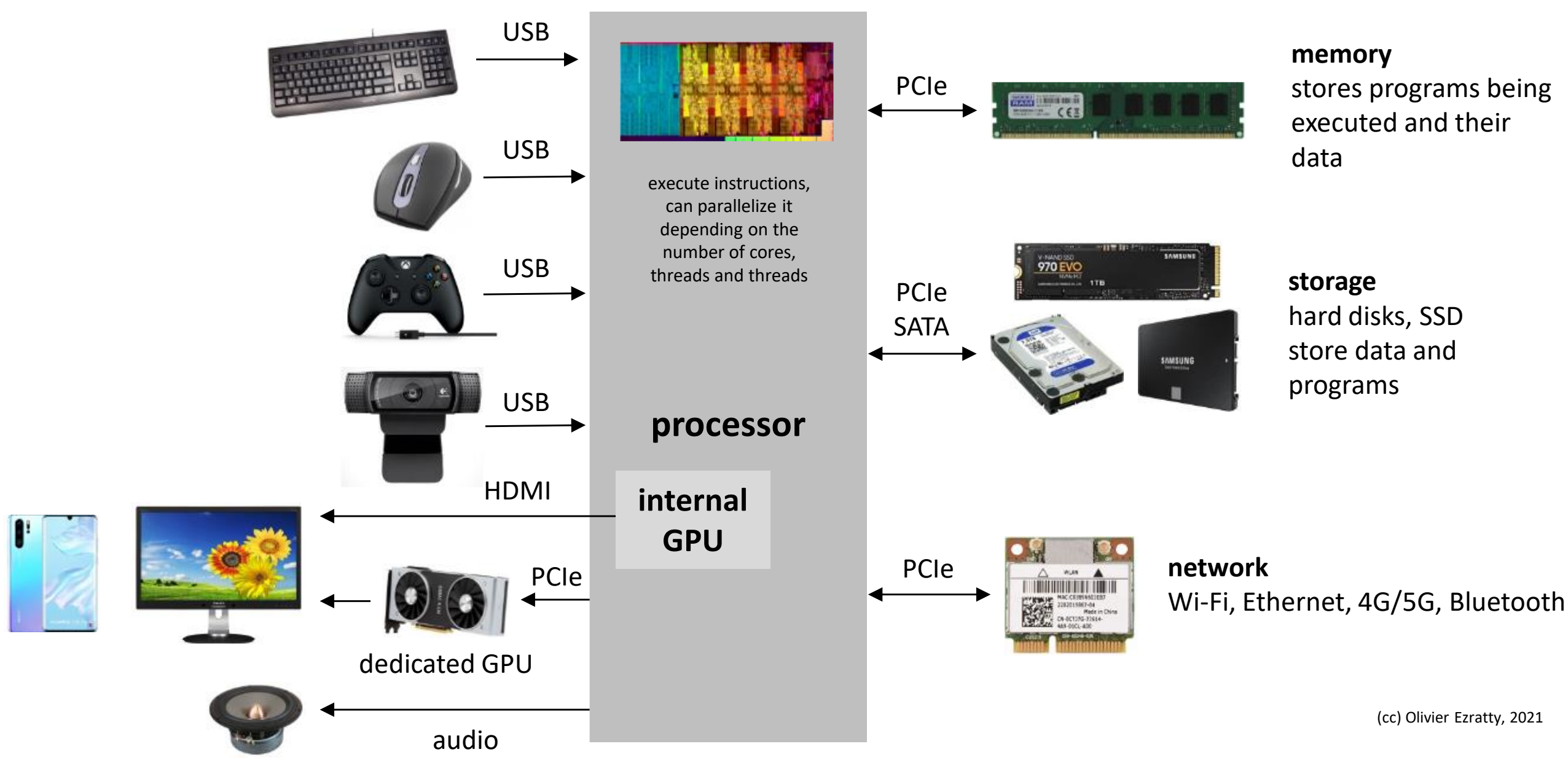

*Figure 200: a classical personal computer hardware architecture. (cc) Olivier Ezratty, 2021.*

Data and programs can be retrieved remotely over a network and from remote servers on the Internet. The whole system is controlled by physical interfaces at input (keyboard, mouse, touchpad, joystick, webcam, microphones, scanners) and generates output (displays, audio, printers, other peripherals). The processor can be complemented by a graphics processor (GPU). It is either external to the microprocessor, for demanding requirements such as in CAD and video games or integrated into the microprocessor as is the case for all most laptops and most desktops processors.

Depending on the configuration, the processor is surrounded by a variable number of external components that are soldered in the motherboard.

---

[346] I consulted a very large number of information sources to carry out this part, both on the research side and on the supplier side, such as IBM or D-Wave. Note Quantum Computing Gentle Introduction from MIT, published in 2011 (386 pages) which describes precisely some mechanisms of quantum computers such as qubit state reading methods. It also describes quite well the mathematical foundations used in quantum computers. You can also enjoy an 8-minute video from Dominic Walliman, who explains the basics of the quantum computer!

This is the case of the Intel chipsets like the Z390, which complements the core processors and manages a large part of the computer's inputs/outputs. Wi-Fi and cellular modems are associated with antennas. Of course, an internal and external power supply and a battery for mobile devices must be added.

On the energy side, it is the processor and GPU that heat up the most and require passive or active cooling depending on their power drain. In embedded systems such as smartphones, this is done with heat conducts and air. In PCs, it is supplemented by one or more fans. In the most extreme cases, liquid cooling uses a water circuit to improve heat dissipation. One of the reasons why heat is generated by classical processing is the non-reversibility of classical computing.

# Key parameters

Let's look at the definition of the key performance indicators of gate-based quantum computers. The best-known set of indicators was created by **David DiVincenzo** in 2000 when he was an IBM researcher. He is now a research professor at the University of Aachen in Germany[347].

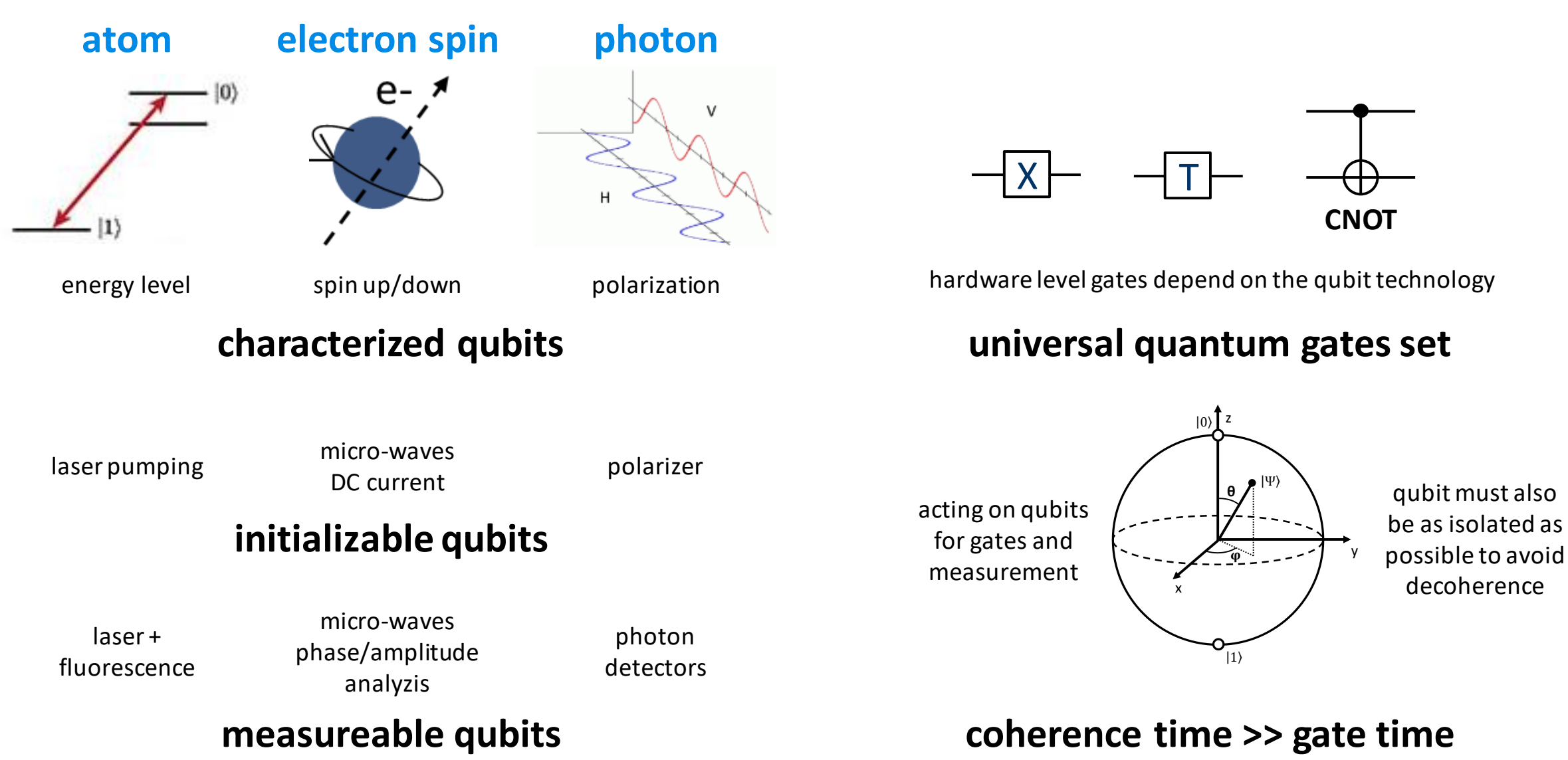

*Figure 201: DiVincenzo gate-based quantum computing criteria. (cc) Olivier Ezratty, 2021, inspired by Pascale Senellart.*

While individual qubits barely existed, he defined the basic technical characteristics of a universal gate-based quantum computer as follows:

**Well-characterized qubits**. Quantum computers use qubits that exploit quantum objects that can have two distinct and measurable states. Their physical characteristics are well known. The architecture is scalable in the sense that it can exploit many physical qubits and then, logical qubits relying on these physical qubits and quantum error correction codes.

**Initializable qubits**. In general, to the value $|0\rangle$ often called "ground state" for the associated quantum objects, corresponding, for example, to the lowest energy level of an elementary particle or an artificial atom as for superconducting qubits.

**Coherence times**. It must be greater than quantum gates activation times. The time during which the qubits are in a coherent state must be greater than the quantum gates activation time in order to be able to execute an algorithm containing a sufficiently long sequence of quantum gates. Error correction codes using a large number of physical qubits have the benefit of extending this usable computing time.

[347] See The Physical Implementation of Quantum Computation by David DiVincenzo, 2000 (9 pages).

**Universal quantum gates set**. The quantum hardware must allow the creation of a universal gate set. It depends on the qubit technology. It requires a minimum set of single-qubit gates allowing the creation of any rotation in the Bloch sphere, completed by a CNOT two-qubits gate.

**Measurement**. With the ability to measure qubits state at the end of computing, which seems obvious. This measurement should not influence the state of other qubits in the system. Ideally, the measurement error rate should be well below 0.1%.

David DiVincenzo added two other optional criteria that are used instead for quantum communications:

**Flying qubits conversion**. The ability to convert static qubits into flying qubits, who are usually photons, and sometimes electrons.

**Transport these moving qubits**. from one point to another reliably and remotely. This will allow to manage quantum telecommunications, distributed architectures of quantum computers and to set up *blind computing* architectures allowing to distribute treatments while protecting their confidentiality. The technology will quickly become essential to enable the distribution of quantum computations over several quantum processors, a bit like we do with multi-core chipsets or with processing distribution architectures over several CPUs and several servers. Some vendors like IonQ have announced that they will rely on this architecture. This will be useful for qubit architectures that will be limited in the number of qubits, which may only be able to consolidate a few hundred at most. It will thus be necessary to be able to link remote processor qubits and keep them entangled. Different quantum interconnection techniques are possible. The most generic is optical and is not much constrained by distance. At rather short distances, microwave links are possible, particularly to couple superconducting qubits, as well as shuttling electrons[348].

DiVincenzo's criteria are quite basic. From a practical and operational point of view, quantum computers can also be characterized by another set of parameters as follows:

**Number of qubits**. It will condition the available computing power. As this power theoretically increases exponentially with the number of qubits, it is a key parameter. As of late 2021, the commercial record was 127 qubits with the largest IBM Quantum System available in the cloud. The number of qubits should be evaluated in its capacity to scale. Some technologies are easier to miniaturize and scale than others. It is necessary to integrate in this miniaturization both the quantum qubit chipsets and the elements that control them. On top of that, we must ensure that decoherence and noise does not increase as the number of qubits is growing. Today, trapped ions qubits have an excellent fidelity but don't scale well. Superconducting qubits seem to scale-up better but their fidelity is not stable as the number of qubits grows with existing industry vendors hardware although it could change in the future. Cold atom qubits scale a little better but with some practical limits in the number of controllable atoms. Electron spins qubits could scale best in theory.

**Qubits connectivity**. It will condition the quantum algorithms execution speed. The greater this physical connectivity, the faster the code execution will be. With a low connectivity, the compiler of the quantum code will have to add a lot more operations to link the qubits together, particularly relying on SWAP gates. This connectivity varies greatly from one technology to another. In 2D technologies, as with superconducting and silicon qubits, it is limited to neighboring qubits. It seems better with some types of trapped ion qubits.

**Qubit parallel operation**. How qubit gates can be parallelized over different qubit zones without disruption will also condition the speed of execution of quantum algorithms.

---

[348] Princeton University and Konstanz University in Germany are working on optical interconnection between CMOS quantum processors. This is documented in [Quantum Computing Advances With Demo of Spin-Photon Interface in Silicon](), 2018. The magic consists in transferring the quantum state of an electron spin to a photon at its phase level.

**Qubits fidelities.** When executing quantum gates and reading their state, qubit fidelity conditions the ability to execute long algorithms. It has a direct impact on the supported algorithm depth. It also impacts the capacity to run quantum error correction codes and create logical qubits with an arbitrary fidelity level.

**Execution time**. For both quantum gates and qubit state measurement. The first is obviously important to make the algorithms run as fast as possible. But the second is equally important because it is involved in error correction codes and therefore conditions the execution time of all algorithms.

**Operating temperature**. For the processor and their equipment which is very dependent on the type of qubits. The Holy Grail is of course to operate at room temperature. The currently operational quantum computers based on superconductors operate at a very low temperature of 15 mK (1 mK = 1 milli-kelvin, 0 kelvin = -273.15°C), but some types of qubits still in the research stage are supposed to operate at room temperature, such as those based on photons and NV centers (cavities in nitrogen-doped diamond structures like with Quantum Brilliance). However, this is not necessarily the case for associated equipment such as photon generators and detectors for photon qubits. Operating at very low temperature is a way to preserve the coherence of the qubits. But the lower the temperature, the smaller the energy that can be radiated by the qubits and their control electronics. Operating qubits at 100 mK or 1K, like with electron spin qubits, creates a much larger available cooling budget to control the qubits than operation at 15 mK.

**Total energy consumption**. We will investigate this and study it in a global manner with incorporating all quantum computer components: the processor itself, all its control electronics as well as the involved cryogenic systems, starting page 259. As of 2022, quantum computers had a power drain sitting between 2 kW and 35 kW depending on the qubit type and number of qubits.

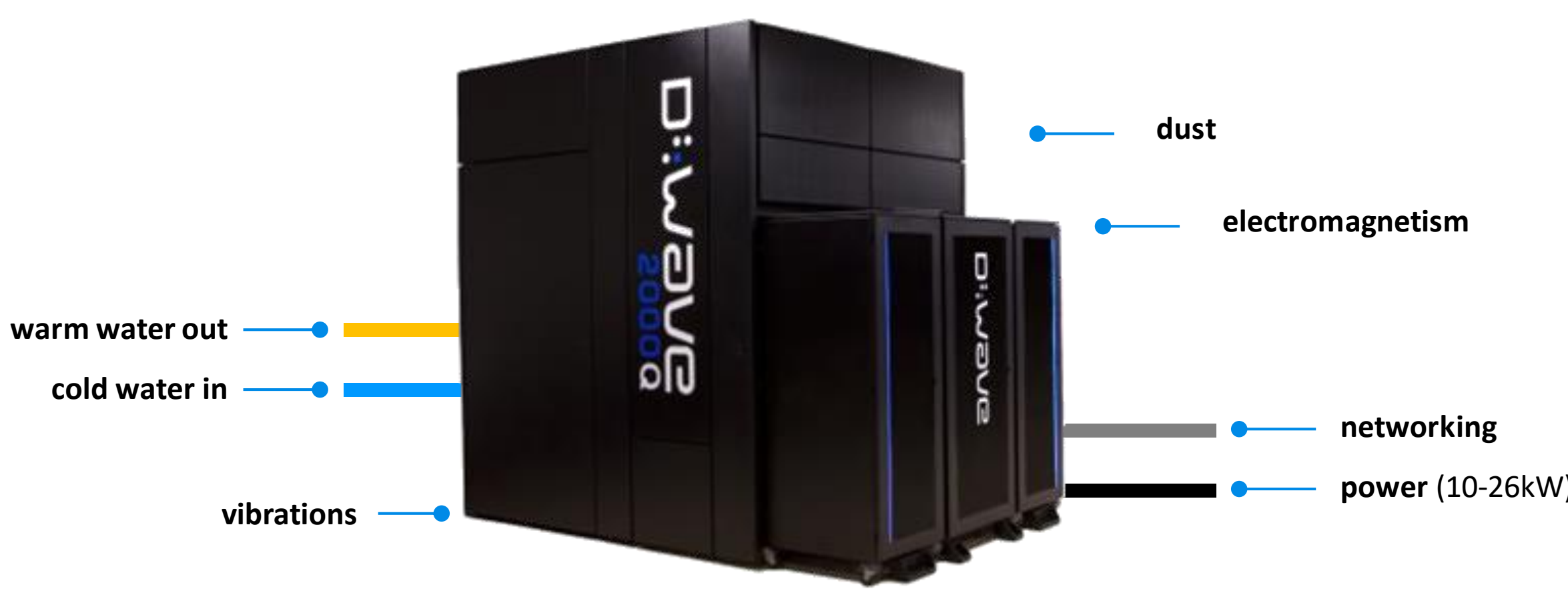

*Figure 202: datacenters integration topics quantum for quantum computers. (cc) Olivier Ezratty, 2021.*

**System rackability.** How will quantum computers be deployed in data centers? Does it fit in standard rack systems? It is notably planned by the startup Pasqal, as well as for Quandela's photon generators and LightOn's optical processors, as well as micro-wave external electronics from companies like Zurich Instruments and Qblox. Alpine Quantum Technologies from Austria also announced in 2021 it would fit its trapped ion computing in two standard 19-inch racks. It is associated with issues of weight, space, cooling and power supply. What kind of fluids must be used for cooling, usually cold water, connected to the first stage compressor of cryostats, whatever their size? Quantum computers must also withstand the usual data centers conditions like vibrations, dust and electromagnetic environment, or be separated in special isolated facilities. They could site in the modular building blocks used in the most recent data centers.

These last three operational parameters play a role when deploying computers or quantum accelerator in data centers. It plays a critical role since, for most applications, quantum computers will be offered through cloud services.

All these considerations to gauge the capabilities of a quantum computer involve the discipline of quantum computers benchmarking! As Kristel Michielsen points out, benchmarks can be used when the number of qubits is below 50 when comparing the rendering of algorithms between quantum computers and their emulation on supercomputers[349]. Beyond that, it will be more difficult.

Benchmarked quantum computers will generally have dissimilar characteristics: different universal quantum gates requiring compilers to assemble different quantum gates to execute the same algorithm, and different error correction codes, adapted to the error rate of the qubits and quantum gates of the compared computers. The dissimilarities will be much greater than between two Intel and AMD processors or two smartphone chipsets!

## Quantum computers segmentation

There is not just one category of quantum computers, but many. We must at least distinguish gate-based quantum computers and analog computers, including quantum annealing computers such as the ones from D-Wave.

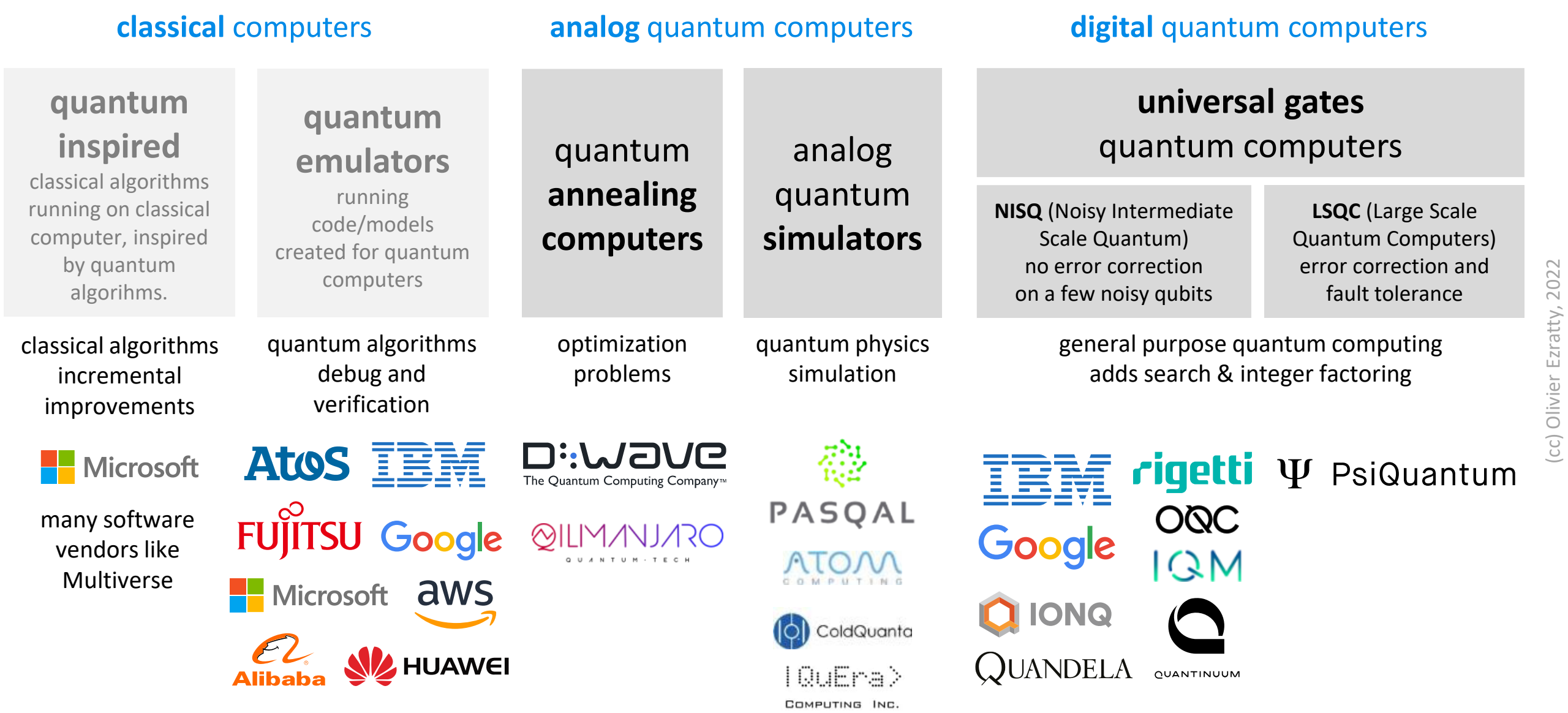

*Figure 203: the different computing paradigms with quantum systems, hybrid systems and classical systems. (cc) Olivier Ezratty, 2022.*

But there are at least six categories of quantum computers:

**Quantum emulators**. These are used to execute quantum algorithms on traditional computers ranging from simple laptops to supercomputers, depending on the number of qubits to be emulated. They execute these algorithms, quantum gates and qubits with the processing capabilities of traditional computers, using large vectors and matrices. It is used to test quantum algorithms without quantum computers. Quantum emulators are sometimes called quantum simulators, but this name should be avoided to prevent the confusion with... quantum simulators. These are analog quantum computers simulating quantum physics phenomena, for example magnetism or the tridimensional structure of molecules. Quantum emulators may however also simulate qubit noise model like Atos QLM emulator[350]. They can also reproduce the (quantum) physical characteristics of various qubits and in that case, they also implement some form of digital quantum simulation. To date, supercomputers can fully emulate up to the equivalent of 40 to 50 qubits. We detail this starting page 638.

[349] In Benchmarking gate-based quantum computers, 2017 (33 pages).

[350] We can make a distinction between an exact digital simulation and approximate digital simulation, emulating a digital error rate that is equal or below NISQ hardware. This can help simulate a greater number of qubits.

Records have however been broken with more than 100 qubits, with a low number of quantum gates and using various techniques like tensor compression. Emulating quantum computers requires a lot of power both on the memory side, to store $2^N$ quantum register states for N qubits, if not the full $2^{2N}$ real numbers of the density matrix, and for the associated processing that relies on floating-point matrix multiplications. Still, records in this field are regularly broken.

**Quantum inspired** computing is about using classical algorithms running on classical hardware that are inspired by quantum algorithms and bring some new efficiencies. They are not about emulating quantum code on a classical computer. Typical quantum inspired algorithms use tensor network libraries.

**Quantum annealing computers** use the adiabatic theorem which consists in using a slow and controlled evolution of a set of qubits linked together according to a particular topology ("Chimera", "Pegasus" or for the new generation "Zephyr" in the case of D-Wave). The process is first initialized in the ground state of the Hamiltonian and the adiabatic theorem guarantees the convergence of the system towards a low energy state, ideally the ground state. This technique is used to search for an energy minimum in a complex problem, such as the simulation of atomic interactions in molecules or the optimization of the duration of a path. The coefficients of the Hamiltonian are the couplings (weights of the interactions between qubits) and the self-couplings (weights of the qubits) and the variables of an instance are the spins of each qubit. Many optimization and quantum simulation algorithms can be translated into quantum annealing algorithms. Until now, D-Wave seems to bring interesting gains in computation time but this is strongly disputed by some specialists.

**Quantum simulators** serve as simulators of quantum phenomena without using gates-based qubit systems. They work in an analog and not digital way, i.e., the parameters linking the qubits together are continuous. For the moment, they are mainly laboratory tools. The most commonly used technique are neutral atoms cooled and controlled by lasers, like the ones from Pasqal, ColdQuanta and Atom Computing. Trapped ions, superconducting qubits, spin qubits[351] and other qubit types can also be used in simulators but no commercial vendor is promoting it when they can also implement gate-based quantum computing which is supposed to be more generic[352].

**Universal quantum computers** use qubits with quantum gates capable of executing all quantum algorithms. At a later time, it will provide a speedup compared to the best classical algorithms running on supercomputers as well as vs quantum annealers. They are currently limited to 127 qubits. The quantum noise levels of qubits is detrimental to computing and requires the usage of logical qubits made of many physical qubits and quantum error correction codes (QEC). While waiting for these fault-tolerant quantum computers to ramp up with logical qubits, we are using non corrected qubits in the so-called NISQ for "Noisy Intermediate-Scale Quantum", an expression from John Preskill[353]. It describes existing and future general-purpose quantum computers supporting 50 to a few hundred physical qubits. These can run algorithms with a limited circuit depth due to the qubit error rates like variational quantum circuits and quantum machine learning algorithms driven by classical algorithms in so-called hybrid algorithms. They are supposed at some point to exceed supercomputers computing capacities for specific algorithms. Then, much later, we'll have fault tolerant (FTQC) and large scale quantum computers (LSQC), with a very large number of physical qubits and over 100 logical qubits compatible with quantum software requirements.

There are now several other variations of universal quantum computers that deserve some description:

---

[351] See Analog Quantum Simulation of the Dynamics of Open Quantum Systems with Quantum Dots and Microelectronic Circuits by Chang Woo Kim, John M. Nichol, Andrew N. Jordan and Ignacio Franco, March-October 2022 (20 pages).

[352] Amazon is investigating it, in A scalable superconducting quantum simulator with long-range connectivity based on a photonic bandgap metamaterial by Xueyue Zhang, Oskar Painter et al, August 2022 (34 pages).

[353] In Quantum Computing in the NISQ era and beyond in 2018.

**Continuous variables quantum computers**, or analog quantum computers with universal gates. They use qubits that store variable quantities between 0 and 1 and can be manipulated with quantum gates, also named 'qunats'[354].

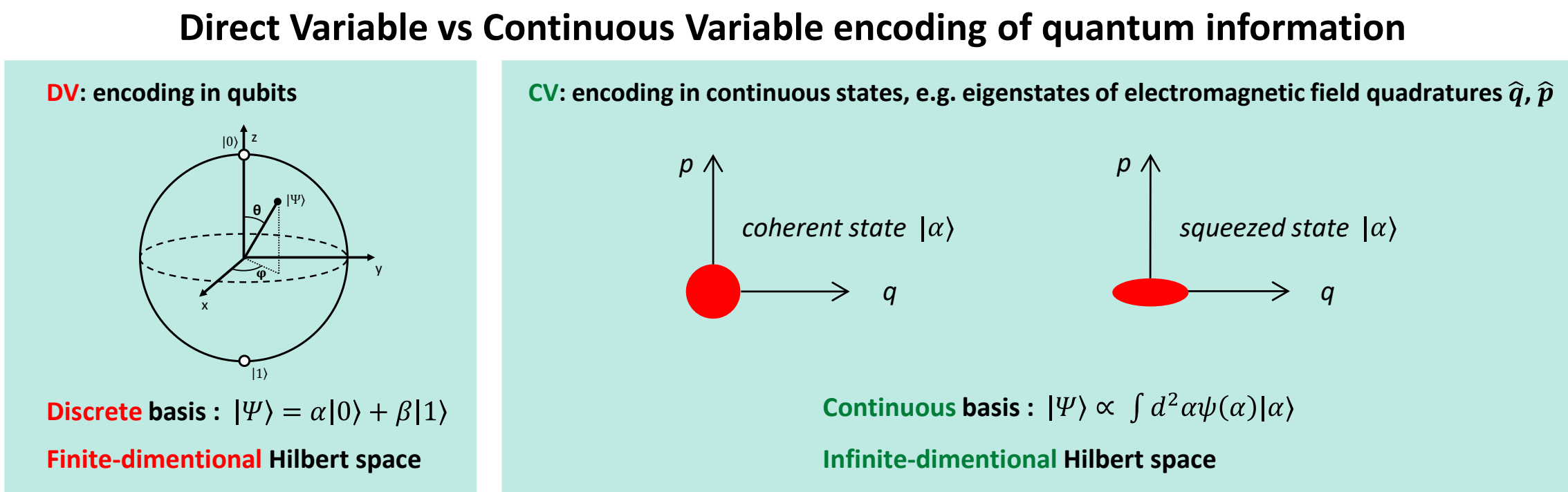

*Figure 204: direct variable and continuous variable encoding of quantum information. inspired from Sub-Universal Models of Quantum Computation in Continuous Variables by Giulia Ferrini, Chalmers University of Technology, Genova, June 2018. (35 slides).*

This category of quantum computing was proposed in 1999 by Seth Lloyd and Samuel L. Braunstein[355]. They are usually based on continuous variable photons but other qubit types like trapped are used.

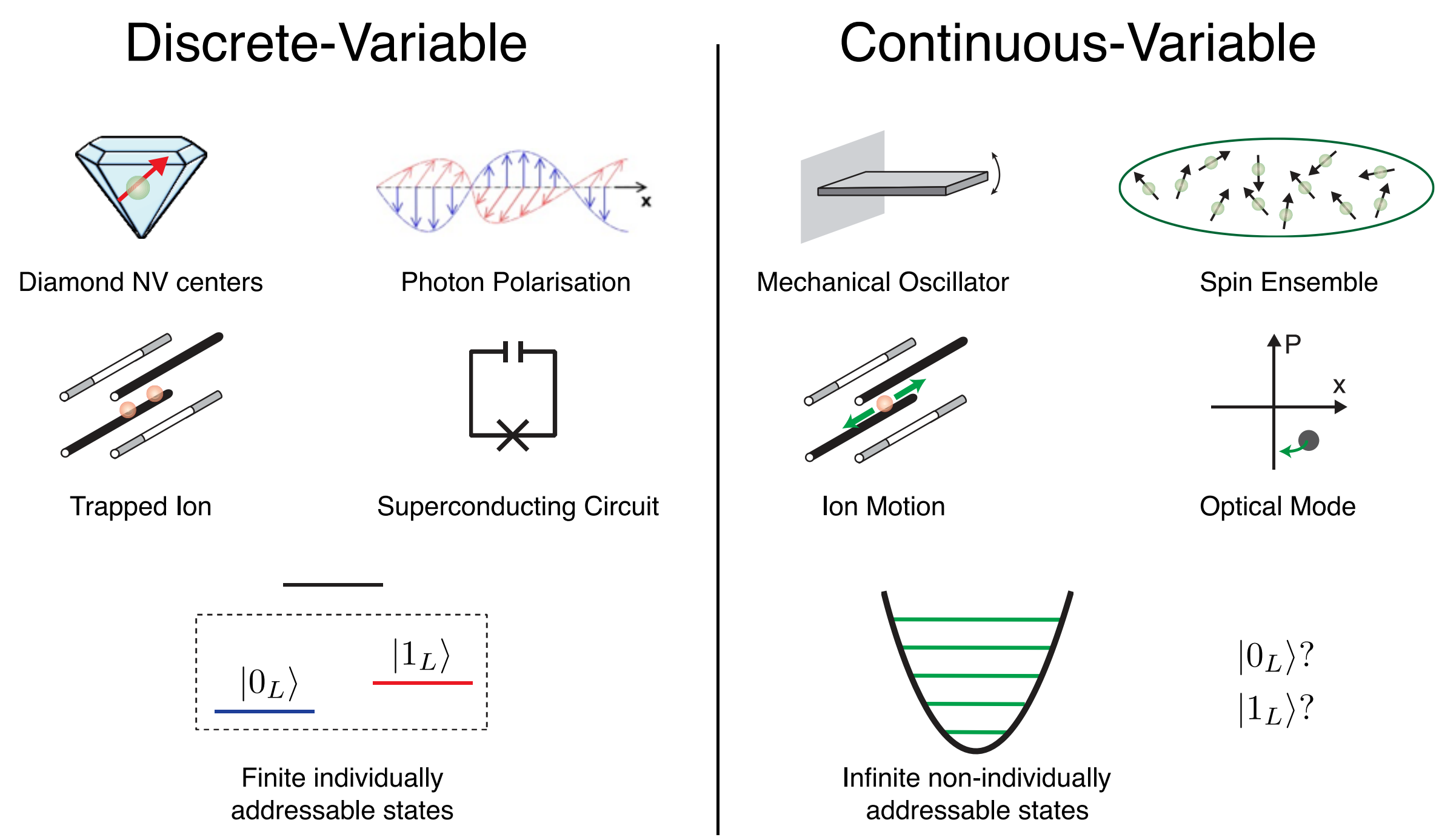

*Figure 205: various implementations of discrete-variable and continuous-variable quantum computing. Source: TBD.*

**MBQCs**, or Measurement Based Quantum Computers, is an architecture variant of NISQ/LSQ adapted to flying qubits and particularly to photon qubits which can't easily be entangled with two qubits gates. The process consists in entangling all qubits at the beginning of computing. It's followed by qubits readouts in an ordered way, enabling the implementation of traditional gates. MBQC also implements some massive parallelism, adapted to the limited and finite processing depth of flying qubits. The startup PsiQuantum plans to use a variant of this technique named FBQC.

---

354 See Universal Quantum Computing with Arbitrary Continuous-Variable Encoding by Hoi-Kwan Lau and Martin B. Plenio, 2016 (5 pages) as well as Continuous-variable quantum computing in the quantum optical frequency comb by Olivier Pfister, 2019 (16 pages).

355 See Quantum Computation over Continuous Variables by Seth Lloyd and Samuel L. Braunstein, February 1999 (9 pages).

**Topological quantum computing** is based on specific anyon qubits that are self-corrected. The low-level programming model of these qubits is much different from universal quantum computers. This is the path chosen by Microsoft, together with QuTech. Its development seems to be quite sluggish.

Here's another segmentation of these models with two dimensions: discrete or continuous data encoding and discrete of continuous variables computing given the vendor position is rough, some being positioned in various slots (Pasqal also wants to do gate-based computing)[356]:

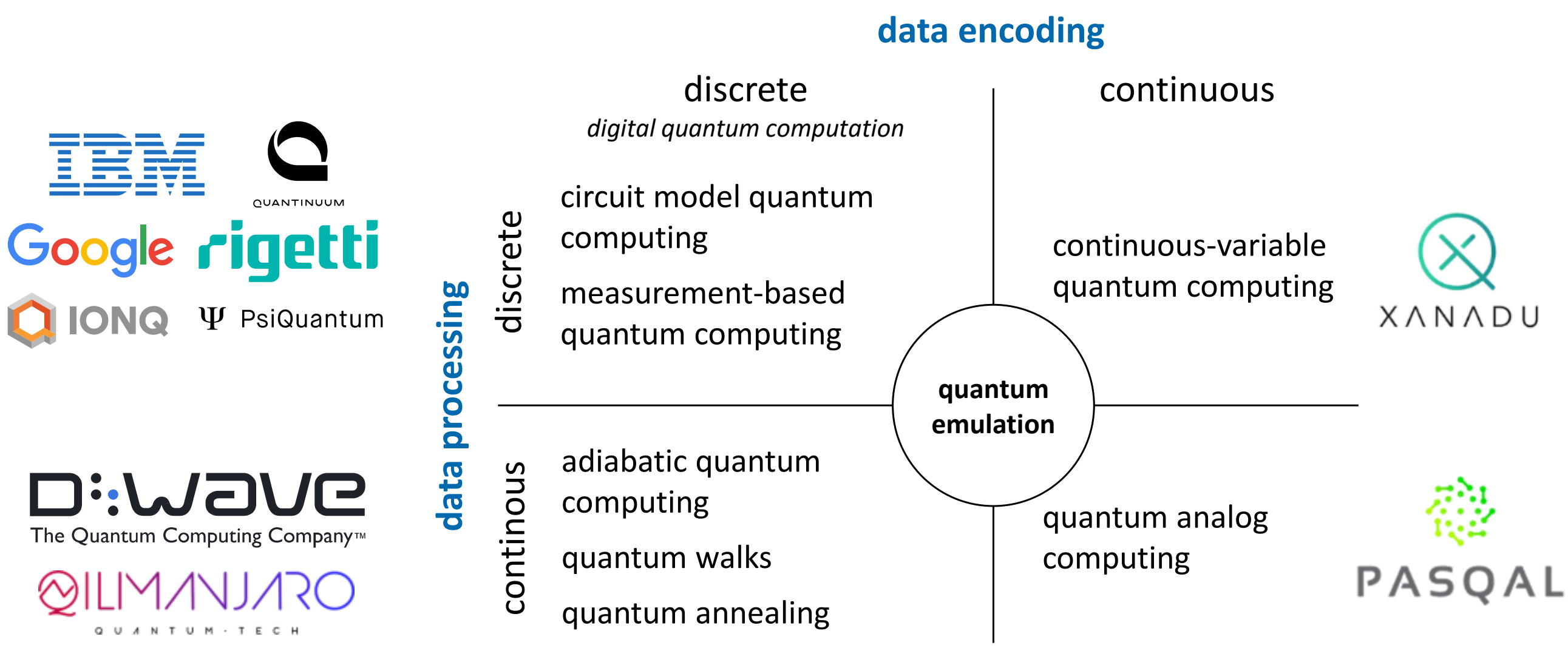

*Figure 206: discrete vs continuous data encoding vs data processing. Source: Quantum computing using continuous-time evolution by Viv Kendon, 2020 (19 pages).*

**Quantum Accelerator**. It is a quantum computer used as a complement to a supercomputer or HPC, usually to run hybrid algorithms like VQE (Variational Quantum Eigensolvers) combining a classical part that prepares the data structure that feeds a quantum accelerator[357]. The QPU serve as an accelerator for the HPC which can be a node or the whole HPC, using CPU and/or GPUs/TPUs. GPUs/TPUs are themselves also accelerators for the CPUs. There are some design issues requiring tight integration between the HPC and the QPU, particularly with regards to batch loading and to the way the quantum algorithm is executed multiple times. A QPU contains itself a classical computer. It converts digital signals (gates) into analog signals (the micro-waves or lasers controlling the qubits and handling their readout). This QPU computer will need to be as close as possible to the HPC computing capacities to improve the turnaround. It may lead to create custom designs integrating an HPC and one or several quantum accelerators[358].

Other quantum accelerator designs contain more or less generic upper software layers with connectors driving various quantum and classical architectures (annealers, gate-based, emulators)[359].

This inventory is only an appetizer. We will have the opportunity to detail these architectures.

And we are always in for many surprise and new programming paradigms that nearly nobody in the ecosystem is evaluating like the "dark path holonomic qudit computation" coming from Sweden[360].

---

[356] See Quantum computing using continuous-time evolution by Viv Kendon, 2020 (19 pages).

[357] See Quantum Accelerators for High-performance Computing Systems by Keith A. Britt et al, 2017 (7 pages).

[358] See Quantum Accelerator Stack: A Research Roadmap by K. Bertels et al, 2021 (39 pages) which proposes a detailed architecture for a quantum accelerator and See QPU-System Co-Design for Quantum HPC Accelerators by Karen Wintersperger, Hila Safi and Wolfgang Mauerer, Siemens AG and Technical University of Applied Sciences Regensburg, September 2022 (15 pages).

[359] See for example the proposals in Quantum Computer Architecture: Towards Full-Stack Quantum Accelerators by Koen Bertels et al, 2019 (20 pages).

[360] See Dark path holonomic qudit computation by Tomas André and Erik Sjoqvist, August 2022 (6 pages).

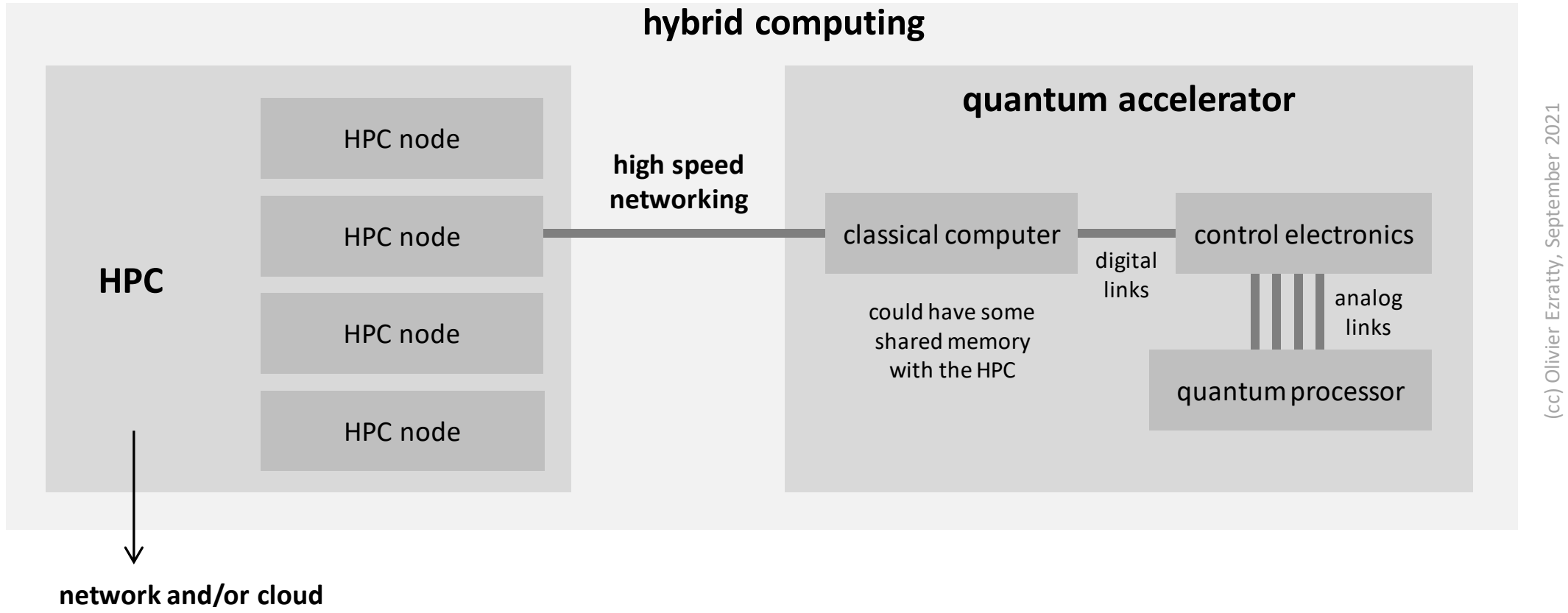

*Figure 207: basics of a hybrid classical/quantum computing hardware architecture. (cc) Olivier Ezratty, 2021.*

# Qubit types

Quantum computers physical qubits are devices that handle particles or quasiparticles with one physical property or observable that can have two possible mutually exclusive states, that can be initialized, modified with quantum gates and then measured.

They are sometimes individual quantum objects, as with atoms (trapped ions and cold atoms), electrons (quantum dots silicon qubits) or photons! And only one at a time! In the case of superconducting qubits or Majorana fermions, the quantum state is based on a large number of electrons arranged in Cooper pairs that share the same quantum state, the pairs of electrons that are created at superconducting temperature. With NV centers and some exotic qubits, qubits are constructed with ensembles of quantum objects or with heterogeneous quantum objects like mixing electron spins and atom nuclear spins.

Qubits can also be classified in two meta-breeds: stationary or moving (flying). Those based on trapped ions, cold atoms, electrons spin, NV centers and superconducting loops are stationary. Flying qubits are based on photons that physically circulate from quantum gate to quantum gate as well as on flying electrons. They move around from a source, through physical devices implementing quantum gates and land on detectors. In all cases, the quantum gates are dynamically activated by electronic circuits or lasers and operate on the qubits where they are (stationary qubits) or in the path of their transit (flying qubits).

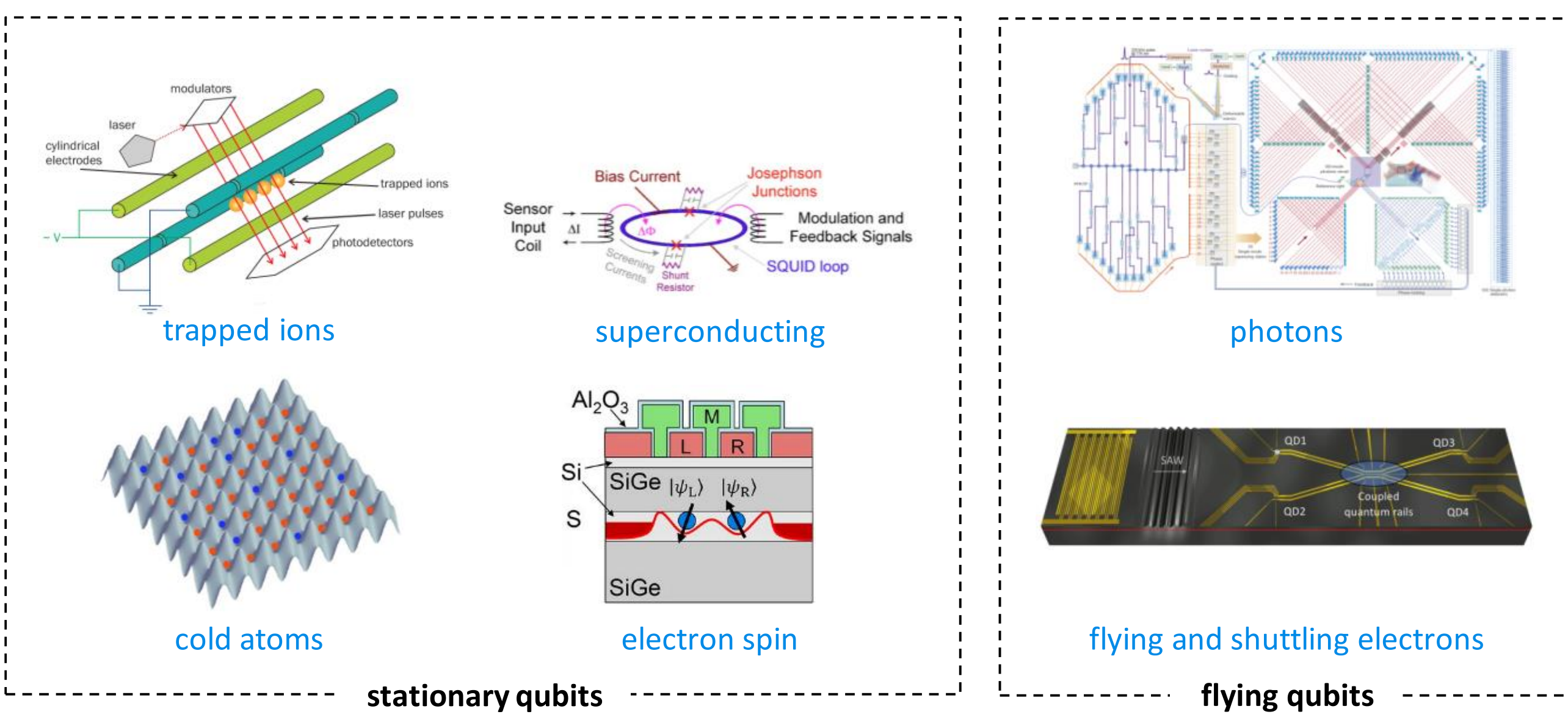

*Figure 208: separating stationary and flying qubits. (cc) Olivier Ezratty, 2021.*

Here are the main types of qubits that are currently being studied, tested and sometimes commercialized[361]. We'll detail later starting in page 273 and with much details all these different options. Looking at these technologies reminds me of the Wacky Races movie and cartoons vehicles as well as the Tatooine podracers in Star Wars I, with an amazing technology diversity and true believers in their fate. The only difference is we may end up with no single winner but several winners if not some forms of technology hybridization. Can we compare it to the Manhattan project from 1940-1045? It had only two main uranium and plutonium combustible options and some variations with the explosives. Here, with quantum computing, the world is looking at many more options.

### Controlled atoms

This is one of the oldest types of qubits. It consists in controlling atoms in vacuum with lasers, one qubit per atom. Cold atoms are neutral while trapped ions are ionized atoms. One key difference is how these atoms are controlled in space. Ions can be positioned with electrodes and magnetic fields while non-ionized atoms are only controlled by lasers. They both share a similar measurement technique using laser excitation, fluorescence and visual readout with some CCD or CMOS imaging sensor.

**Trapped ions** are atom ions that are kept in a vacuum and suspended by electrostatic suspension. Their initialization is done with laser optical pumping.

Lasers are used to cool and stabilize the ions, exploiting the Doppler effect, with different energy transitions than those used to modify the state of the qubits. The most frequently used ions are calcium and strontium. Single-qubit quantum gates are activated by microwaves, lasers or magnetic dipoles. Lasers or electrodes are used for two-qubit quantum gates. While trapped ions are best-in-class for qubits fidelity and connectivity, it seems currently difficult to scale it beyond a hundred ions. And it's very slow.

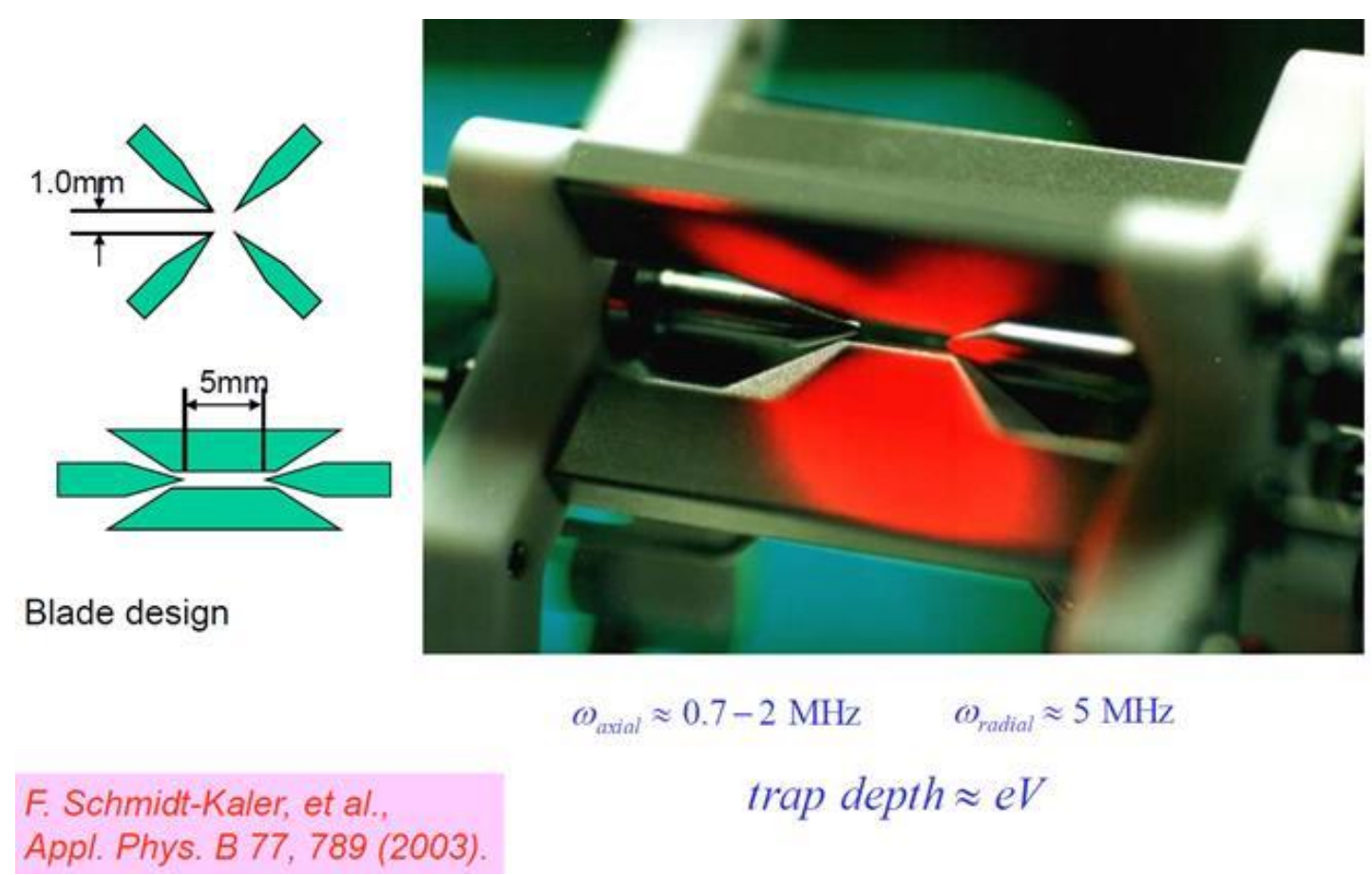

*Figure 209: a typical Paul trap for trapped ions, created in 2003.*

**Cold atoms** are cooled at very low temperatures, also using the Doppler effect and other laser-based techniques. The used atoms are neutral atoms and quite often rubidium, an alkaline metal. The quantum state of these cold atoms is their energy level, which can use their Rydberg high-excitation states on some occasions. Cold atoms are used to create both gate-based quantum computer and analog quantum simulators.

**Nuclear magnetic resonance** (NMR) qubits were tested over 20 years ago and are nearly completely abandoned. Most of the time, they are based on using ensemble of atoms or molecules. They do not scale at all. It is a good demonstration that qubits research must remain open and cannot be settled too early around one or two technologies. Even now, it's too early to tell which qubit type will really scale to create useful quantum computers.

### Controlled electrons

This other category of qubits is about electrons that are controlled most of the time in solid-state circuits instead of vacuum like with cold atoms and trapped ions.

[361] See Roadmap on quantum nanotechnologies by Arne Laucht et al, 2021 (49 pages) which reviews some of these qubit types.

**Superconducting qubits** are based on the state of a superconducting current that crosses a very thin barrier in a loop, usually a metal oxide such as aluminum, using the Josephson effect[362] . There are several types of superconducting qubits: flux, phase and charge.

The most common one is the transmon, a variation of charge superconducting qubits. In all cases, qubit observables are two very distinct states of a high-frequency oscillating current flowing through the Josephson junction.

The oscillation is made possible by the fact that the loop integrates the equivalent of an inductance and a capacitance. The current oscillation is activated by microwaves pulses using frequencies between 4 and 8 GHz and transmitted by coaxial wires. In transmon qubits, the qubit observable is measured with a resonator integrated in the circuit which receives a microwave and sends it back. The readout electronic system splits out the amplitude and phase of the readout microwave to detect the qubit value. On the right is a schematic of two superconducting Google Sycamore qubits, themselves connected by a third qubit - in green - which acts as a dynamic coupler between two qubits, to create controlled entanglement.

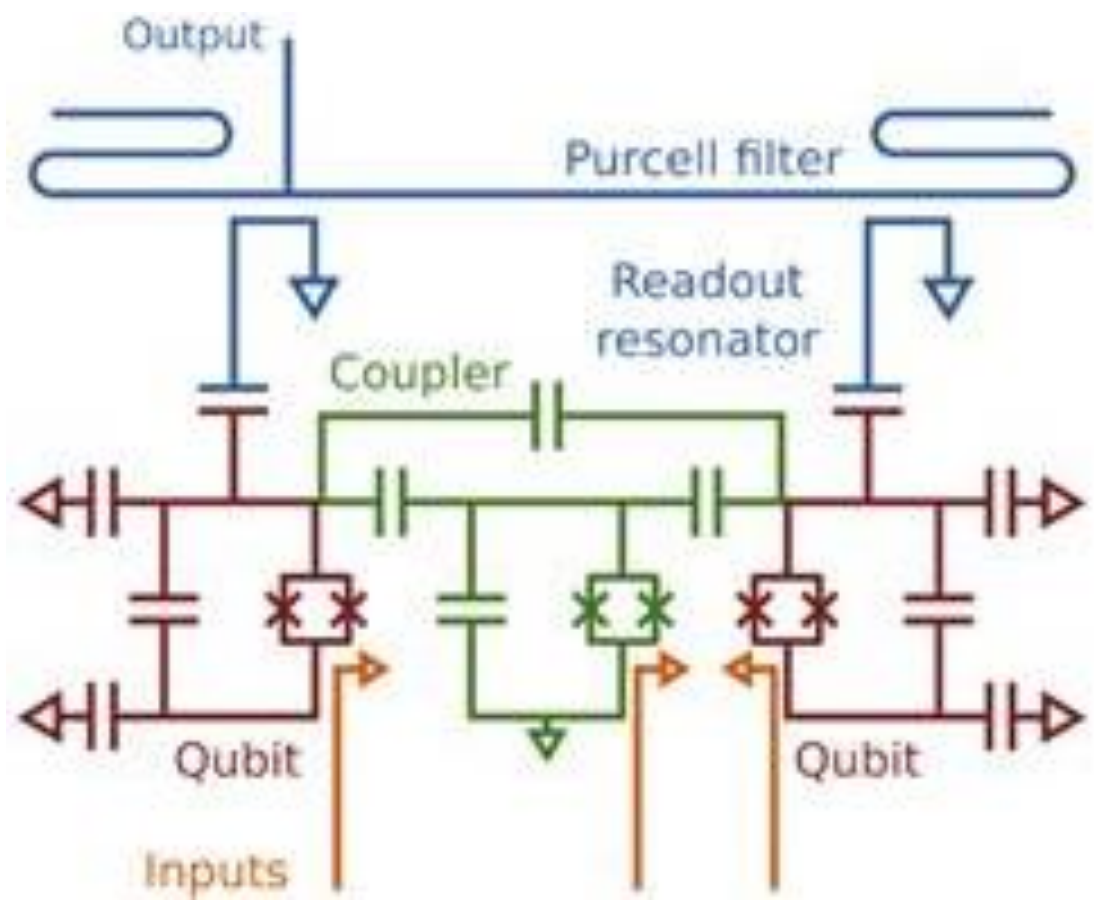

*Figure 210: Google Sycamore superconducting electronic architecture. Source: Google.*

In some transmons, individual qubits activation frequency is tuned by a direct current flux bias line.

Superconducting qubits are relatively easy to manufacture because they are based on semiconductor circuit creation techniques even if some of the materials are different, such as niobium and aluminum[363]. They are built on a dielectric substrate, usually with silicon or sapphire. These qubits are operating at 15 mK, requiring a dilution refrigerator. This temperature is required for a chain of reasons: qubits are driven by microwaves in the 4-8 GHz range and the current thermal noise is constrained by order of magnitude below the temperature corresponding to these microwaves' energy. The 4-8GHz corresponds to off-the-shelf microwave generation equipment and to the size of the capacitor and resonator used in the vicinity of the qubit Josephson junctions.

Superconducting qubits have many challenges dealing with scalability. The microwave RF generators are usually located outside the cryogenic enclosure of the quantum processor, which create a lot of wiring with about 3 to 4 cables per qubit. Qubits control frequencies must be different and tuned for adjacent qubits. Their fidelity is not best-in-class and seems to decrease as we grow the number of qubits.

**Quantum dots electron spin** qubits are developed with scalability in sight. Most of them use two electrons trapped in a quantum well, one containing the qubit and the other one used to measure it. These qubits are usually manufactured using silicon-based CMOS circuits. Silicon is often supplemented with various dopants. They benefit from the reuse of CMOS manufacturing processes that are already well mastered. These qubits are easy to miniaturize down to below 100 nm. They work at temperatures between 100 mK and 1K, higher than superconducting qubits, allowing the use of more electronics around the chipset, to generate the microwaves and other electric signals required to create qubit gates and handle qubit readout. This promising technology is however less mature than superconducting qubits. No lab or company has really exceeded 15 functional qubits as of 2022.

---

[362] See Digital readout and control of a superconducting qubit by Caleb Jordan Howington, 2019 (127 pages).

[363] See Practical realization of Quantum Computation Superconducting Qubits (36 slides).

**NV centers** (Nitrogen Vacancy) are artificial diamond structures in which a carbon atom has been replaced by a nitrogen atom near a carbon atom gap. Qubit states and control rely on a combination of electron, nitrogen and carbon $^{13}C$ nucleus spins. Qubit gates are implemented with microwaves, a magnetic field and an electric field. Entanglement is handled with photons, magnetic coupling or with controlling the core spin of neighboring $^{13}C$ carbon atoms via the use of microwaves to create a CNOT gate. Qubit readout is using a laser and fluorescence detection.

**Majorana fermions** are anyons or quasi-particles which are particular states of Cooper's pairs in condensed matter at very low temperature.

These qubits use braiding, a special topology that makes it possible to implement error correction at the qubit level. The promise is to enable the creation of scalable fault-tolerant quantum computers. These must also be cooled to a temperature close to absolute zero, around 10mK. This is the path chosen by Microsoft. The existence of the fermions of Majorana is not yet proven. It is one of the most hazardous paths to quantum computing. Majorana fermions are often discussed but they belong to a broader category named "topological matter" and "many-body systems".

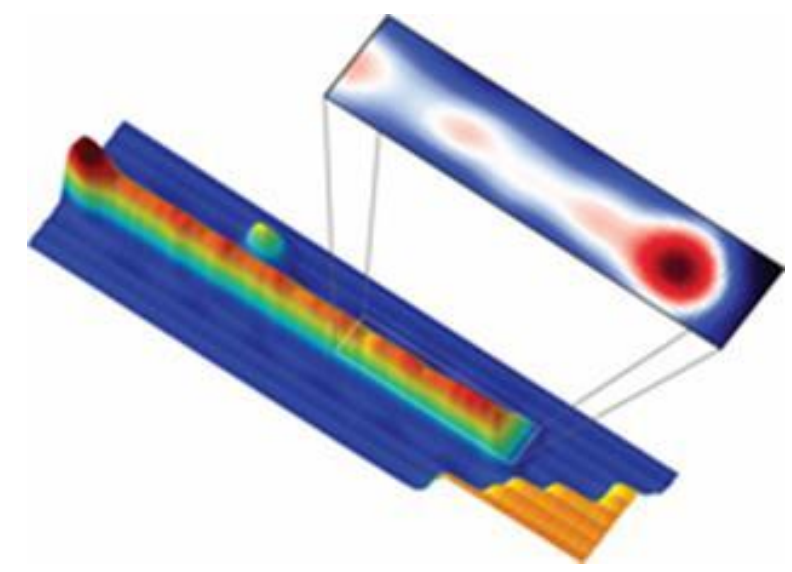

*Figure 211: researchers may have seen Majorana fermions, but that's not really sure.*

The main problem is… we are not sure these anyons and Majorana fermions really exist. It's still work in progress with ups and downs.

### Flying qubits

Flying qubits are special because they travel from the place where they originated, traverse physical devices acting on them and terminate their journey on a sensor measuring one observable. They have a limited time available to run any computing, including a finite and small number of quantum gates.

**Photons** are the most common flying qubit and there are many varieties of implementations. One type is based on a horizontal/vertical polarization observable. Others use continuous variables qubits. Boson sampling systems use multi-modes photons. It is quite difficult to implement two qubit gates with these photon qubits, thus the alternative of the MBQC architecture that is an interesting workaround. Also, photon generation follows a probabilistic pattern which makes things complicated when the number of qubits grows. Most of these qubits operate at room temperature, but the photon sources and their detectors must however usually be cooled to temperatures between 4K and 10K, which is much less demanding than the 15 mK of superconducting qubits or the 1K of silicon qubits.

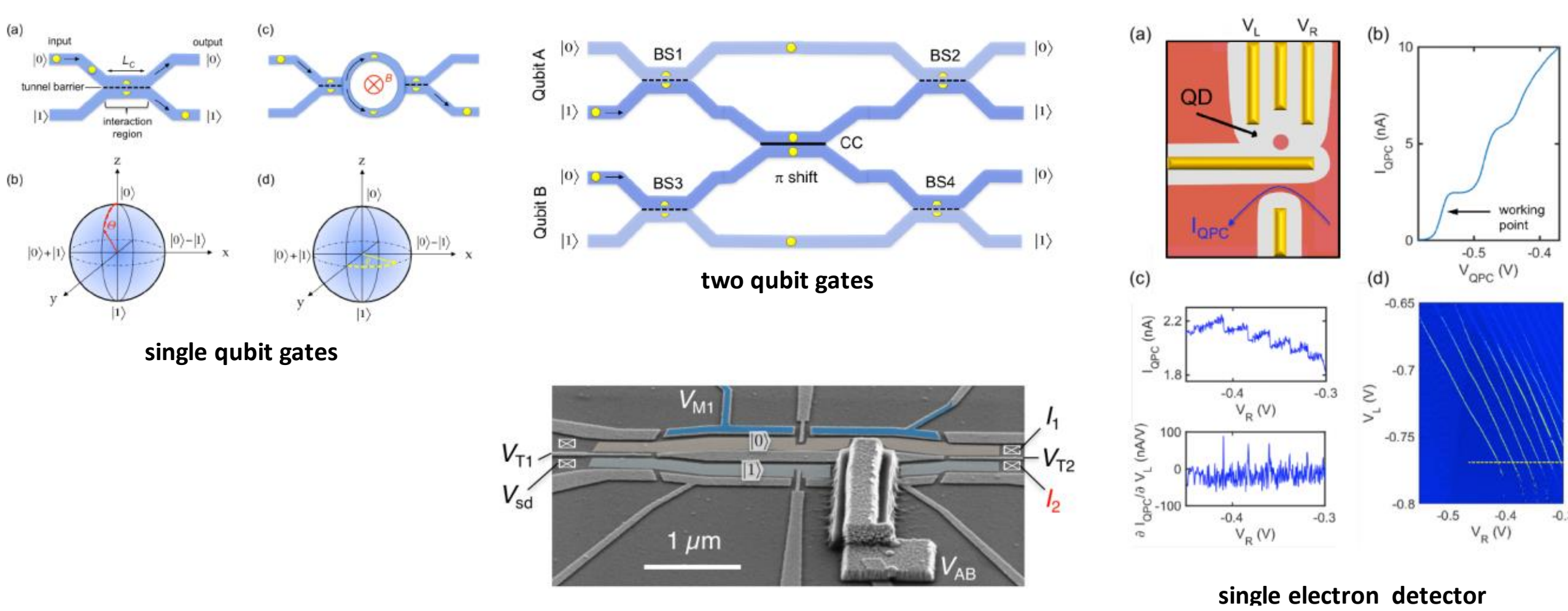

*Figure 212: flying electrons in their waveguides. Their circuit architecture has some commonalities with photon circuits. Source: [Coherent control of single electrons: a review of current progress](#) by Christopher Bäuerle, Xavier Waintal et al, 2018 (35 pages).*

**Flying electrons** are at a pure research stage qubit technology using traveling electrons[364]. It is based on using single-electron transport circulating on wave guide nanostructures build on semiconductors circuits, mostly GaAs, leveraging Coulomb coupling, quantum charge Hall effect and surface acoustic waves. Single- and two-qubit quantum gates can be realized on such circuits. Electrons can fly on distance of 6 to 250 microns. Electrons are created with producing THz photons which are converted into electrons. One qubit uses two-electron paths for states $|0\rangle$ or $|1\rangle$[365]. At the end of processing, these flying electrons are detected by a quantum dot.

This technique could also be used to create shuttling electrons qubits connecting static quantum dots-based qubits together. A few labs in the world are working on this including NPL in the UK, Ruhr-Universität Bochum, ERATO-JST and AIST in Japan, CEA-Leti and Institut Néel in Grenoble, France.

### Exotic qubits

Many research labs are working on using exotic qubits of various kinds. Most of the time, these qubits are at the fundamental research stage and far away from industrialization or even, sometimes, are now yet materialized with a real single functional qubit.

In the **atom realm**, we can count with rare-earth ions in an insulating solid-state matrix[366], molecular ions[367], cold atom ensembles[368], 2D organic molecule networks[369], LCD base nematic qubits[370] and chemical compounds that have photon-controlled state transitions[371].

Molecular magnets are being explored, one variant being made with terbium and with four possible spin related quantum levels, creating qudits, with d=4. The small name of these magnets is SMM for Single-Molecule Magnets. The molecule used is TbPc2 also called bis (phthalocyaninato) terbium(III). Their state is measured with a phase-measuring interferometer. Their advantage is their stability. But they are relatively difficult to control[372].

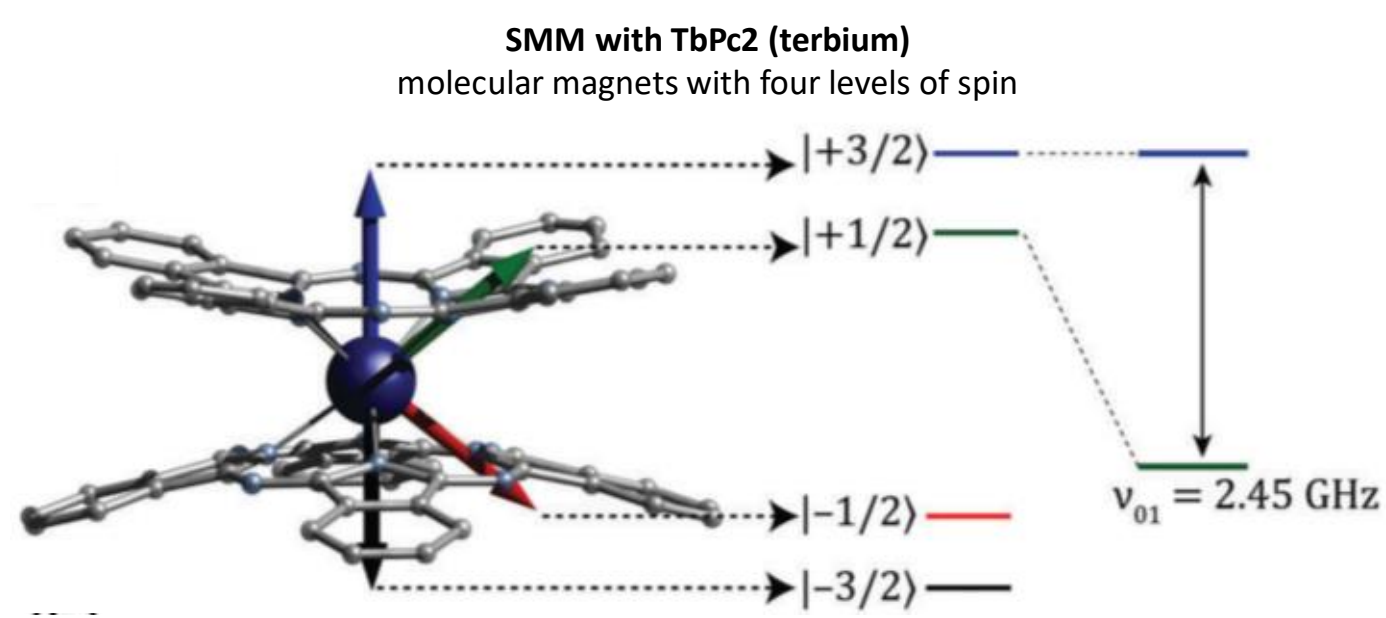

*Figure 213: TbPc2 is a molecular magnet molecule used in prototype quantum processors. Source: Molecular spin qudits for quantum algorithms by Eufemio Moreno-Pineda, Clément Godfrin, Franck Balestro, Wolfgang Wernsdorfer and Mario Ruben, 2017 (13 pages).*

---

[364] See the review paper Semiconductor-based electron flying qubits: Review on recent progress accelerated by numerical modelling by Hermann Edlbauer, Xavier Waintal, et al, July 2022 (44 pages).

[365] See Electrical control of a solid-state flying qubit by Michihisa Yamamoto, Christopher Bäuerle et al, 2017 (17 pages), Coherent control of single electrons: a review of current progress by Christopher Bäuerle, Xavier Waintal et al, 2018 (35 pages) and Macroscopic Electron Quantum Coherence in a Solid-State Circuit by H. Duprez et al, 2019 (10 pages).

[366] See Universal Quantum Computing Using Electronuclear Wavefunctions of Rare-Earth Ions by Manuel Grimm et al, 2021 (19 pages).

[367] See the review paper Molecular-ion quantum technologies by Mudit Sinhal and Stefan Willitsch, University of Basel, April 2022 (15 pages). Difficult to cool molecules with lasers. Destructive measurement.

[368] See Quantum supremacy with spin squeezed atomic ensembles by Yueheng Shi et al, April 2022 (12 pages)

[369] See Blueprint of optically addressable molecular network for quantum circuit architecture by Jiawei Chang et al, September 2022 (11 pages).

[370] See Nematic bits and universal logic gates by Ziga Kos and Jörn Dunkel, August 2022 (10 pages).

[371] See Functionalizing aromatic compounds with optical cycling centres by Guo-Zhu Zhu et al, UCLA, Nature Chemistry, July 2022 (6 pages).

[372] See Molecular spin qudits for quantum algorithms by Eufemio Moreno-Pineda, Clément Godfrin, Franck Balestro, Wolfgang Wernsdorfer and Mario Ruben, 2017 (13 pages). This work was carried out in partnership with the Karlsruhe Institute of Technology in Germany. And also the thesis Quantum information processing using a molecular magnet single nuclear spin qudit by Clement Godfrin, 2017 (191 pages).

In the **electrons realm**, various topological materials[373], various forms of graphene based qubits[374], carbon nanotubes-based mechanical oscillators[375], electron spin in magnetic materials in Van der Waals crystals made of chromium[376], electrons on solid neon[377], quantum neural networks using variations of superconducting qubits[378], quantum memristors[379], toponomic quantum computing which is a variant of topological computing[380] and various qubit hybridization techniques to couple fast operating qubits and long coherence time qubits for implementing some sort of quantum memories, like with associating superconducting qubits with NV centers, or superconducting qubits with yttrium iron garnet magnons[381].

Figuring out the TRL of these proposals is usually easy: it is very low! Particularly when you don't have any published one and two qubit fidelities data.

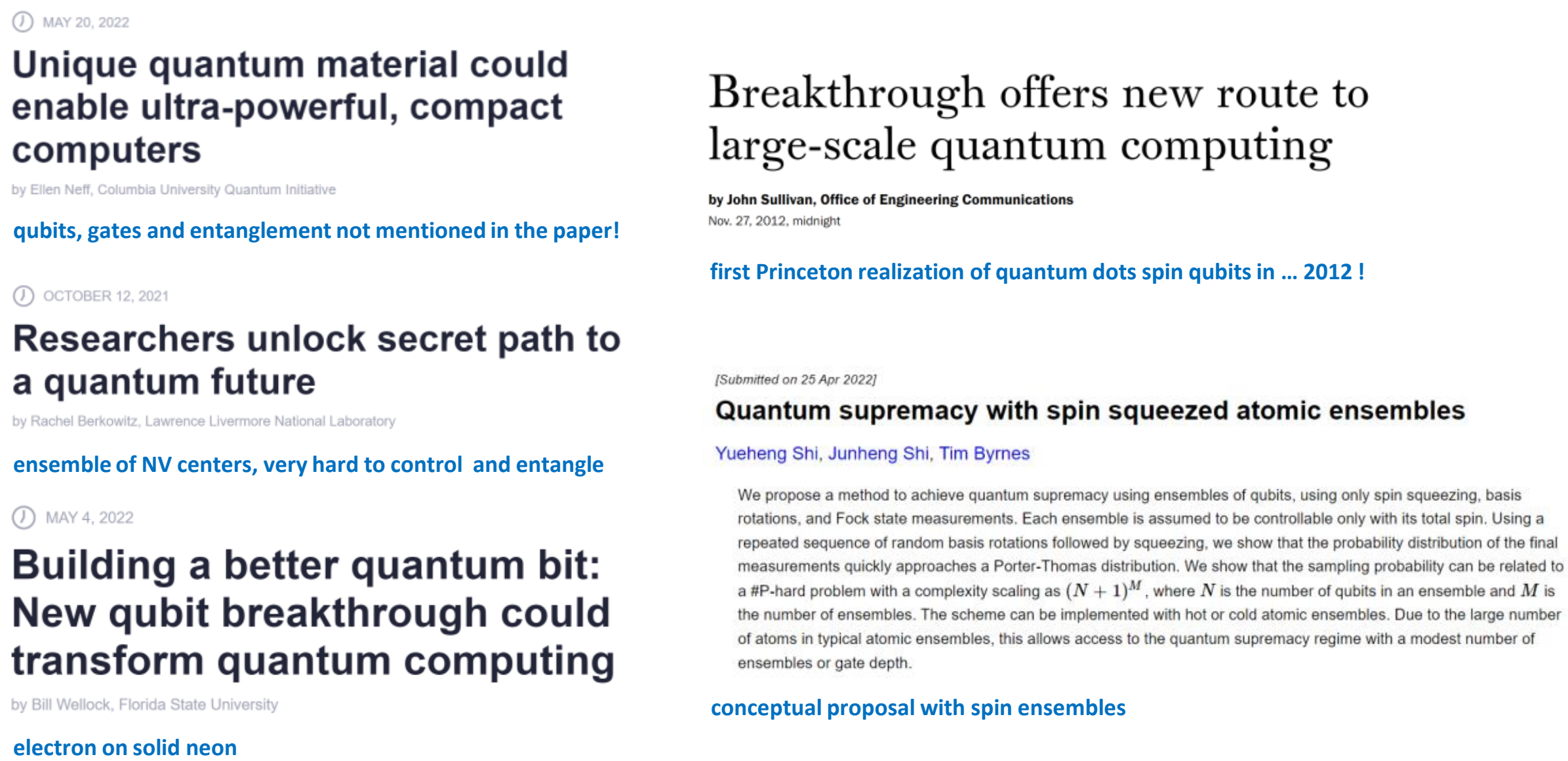

*Figure 214: examples of research laboratories communication on new exotic qubits with very low TRL!*

---

[373] See Anomalous normal fluid response in a chiral superconductor $UTe_2$ by Seokjin Bae et al, July 2021 (5 pages) and Multicomponent superconducting order parameter in UTe2 by I. M. Hayes, July 2021.

[374] See Visualization and Manipulation of Bilayer Graphene Quantum Dots with Broken Rotational Symmetry and Non trivial Topology by Zhehao Ge et al, 2021 (19 pages).

[375] See Proposal for a nanomechanical qubit by F. Pistolesi, Andrew Cleland, A. Bachtold, August 2021 (19 pages).

[376] See Unique quantum material could enable ultra-powerful, compact computers by Ellen Neff, Columbia University Quantum Initiative, May 2022, referring to Coupling between magnetic order and charge transport in a two-dimensional magnetic semiconductor by Evan J. Telford et al, Nature Materials, May 2022 (15 pages). The initial title is of course quite overselling. One simple indication in the scientific paper: the words qubits, gates and entanglement are not even mentioned. So, a powerful quantum computer is very far in this roadmap even though it could operate at 132 K which is considered to be "hot" in quantum computing (ambient temperature is 300K)!

[377] See Building a better quantum bit: New qubit breakthrough could transform quantum computing by Bill Wellock, Florida State University, May 2022, referring to Single electrons on solid neon as a solid-state qubit platform by Xianjing Zhou, David I. Schuster et al, Nature, May 2022 (16 pages). The team created its qubit by freezing neon gas into a solid at very low temperatures, spraying electrons from a light bulb onto the solid and trapping a single electron there.

[378] See Coherently coupled quantum oscillators for quantum reservoir computing by Julien Dudas, Julie Grollier and Danijela Marković, April 2022 (4 pages), a quantum reservoir neural network implementation on a Josephson parametric converter.

[379] See Quantum Memristors with Quantum Computers by Y.-M. Guo, F. Albarrán-Arriagada, H. Alaeian, E. Solano, G. Alvarado Barrios, PRA, December 2021 -August 2022(7 pages) and Entangled quantum memristors by Shubham Kumar, Enrique Solano et al, arXiv and PRA, July & December 2021 (9 pages).

[380] See Toponomic Quantum Computation by C. Chryssomalakos et al, February 2022 (5 pages).

[381] See Analog quantum control of magnonic cat states on-a-chip by a superconducting qubit by Marios Kounalakis et al, PRL, TU Delft, Tohoku University and CAS in China, July 2022 (14 pages).

These are most of the time interesting physics experiments but no functional qubit, a fortiori, entangled qubits and related fidelities. Sometimes, there are real use cases but not in quantum computing and more in quantum sensing. It won't of course present research laboratories communications departments to fuel the hype with their stack of overpromises.

**Figures of merits**

Let's now inventory the various figures of merit of these qubit architectures:

**Qubits stability** which is evaluated with their coherence time (the $T_1$ we'll describe later when discussing error correction page 216). Associated with quantum gate times and error rate, it conditions the number of quantum gates that can be chained in an algorithm. The most stable qubits so far are trapped ions based but as far as you don't have too many of them.

**Qubits fidelity** is related to the errors level that is evaluated with single and double gates as well as with measurement. Again, the best-in-class are trapped ions. We cover that starting 220.

**Qubits connectivity** is the way they are linked together, which will condition many parameters such as the execution speed and the depth of the algorithms that can be exploited. Best-in-class qubits for this respect are again trapped ions in 1D structures, although it does not scale well.

**Large scale entanglement** if possible, without being limited to the immediately neighboring qubits. So far, nobody does it really well.

**Operating temperature** and for the accompanying electronics. The best are NV centers which are supposed to work at ambient temperature, and the worst are superconducting qubits, requiring 15 mK.

**Qubits density** and their control electronics which impacts scalability. This rather favors quantum dots electron spin qubits.

**Manufacturing process** which depends on many parameters. In the case of cold atoms, for example, it is not necessary to create specialized circuits, whereas it is necessary for all other technologies.

**Scalability potential** which depends on many systems parameters, both at the fundamental level with the qubit stability and fidelities at large scale but also with the various enabling technologies. Unfortunately for your forecasting, scalability potentials do not align with qubits technologies present maturities!

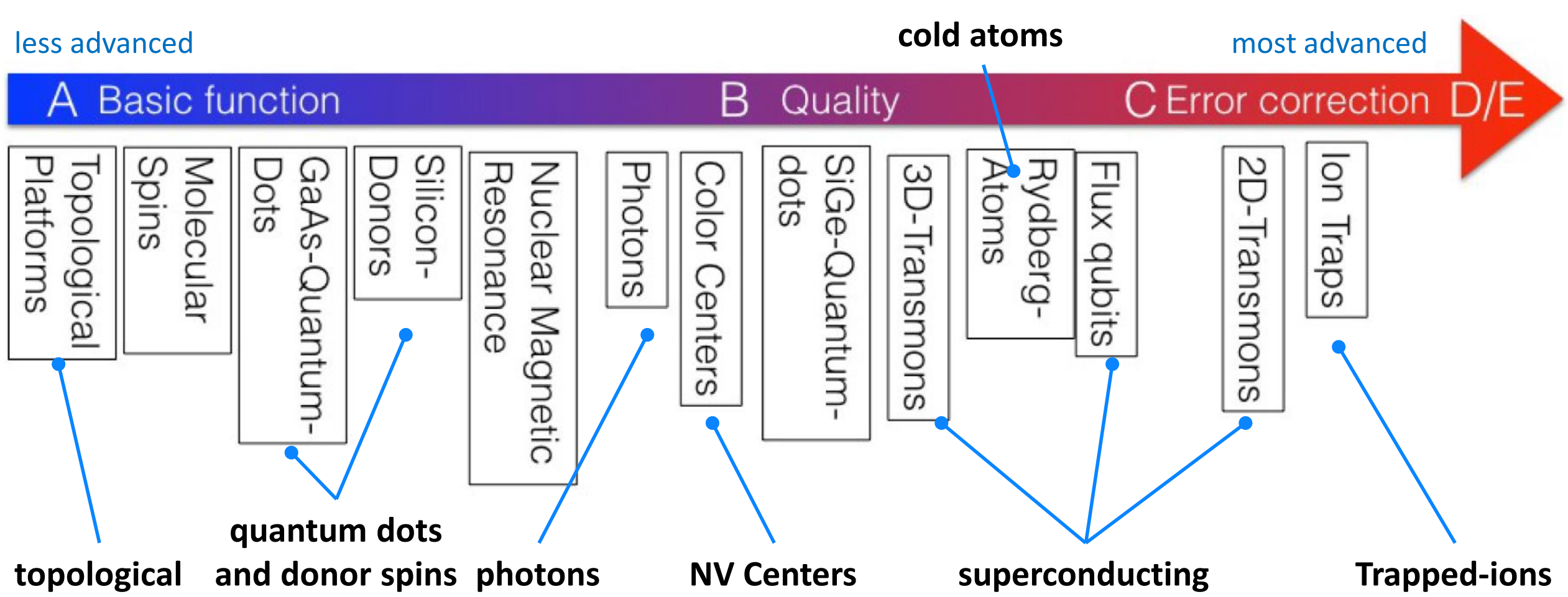

*Figure 215: degree of maturity of various qubit technologies. Entwicklungsstand Quantencomputer (State of the art of quantum computing, in English, June 2020 (266 pages).*

The level of qubits is evolving rapidly. It is described in this excellent document from the German cybersecurity agency[382]. It mentions other technologies not listed in this inventory.

Here's another way to put it[383]. It segments the types of qubits according to three dimensions: the clock frequency of the quantum gates (roughly, the gates number that can be executed per second), the number of operations before errors occur, and the quantum gates fidelity (separating the one- and two-qubit gates). These last two axis are roughly homothetic because the number of operations before errors are generated depends on the error rate.

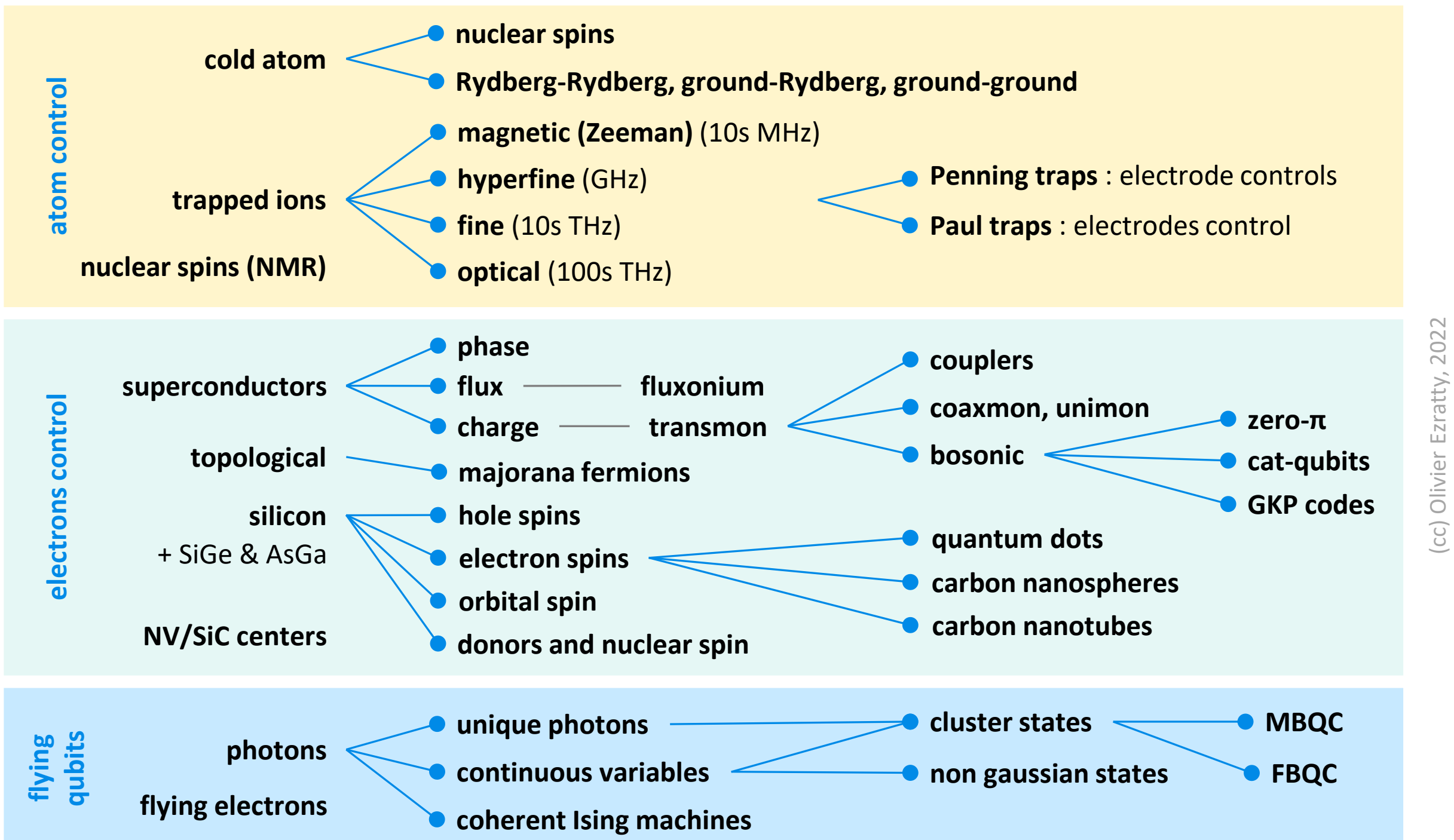

*Figure 216: rough zoology of qubits classes and sub-classes. (cc) Olivier Ezratty, 2022.*

Trapped ions have better gates than superconducting qubits but are quite slow. Silicon qubits are for the moment quite fast (at least, as fast as superconducting qubits). Cold atoms are slower. A last axis is missing: the number of qubits as of today and technology scalability. The chart was made in 2019 and may be outdated for some qubit types. In the section devoted to quantum computing commercial vendors, we cover with more detail the science and technology behind each of these types of qubits, starting page 292.

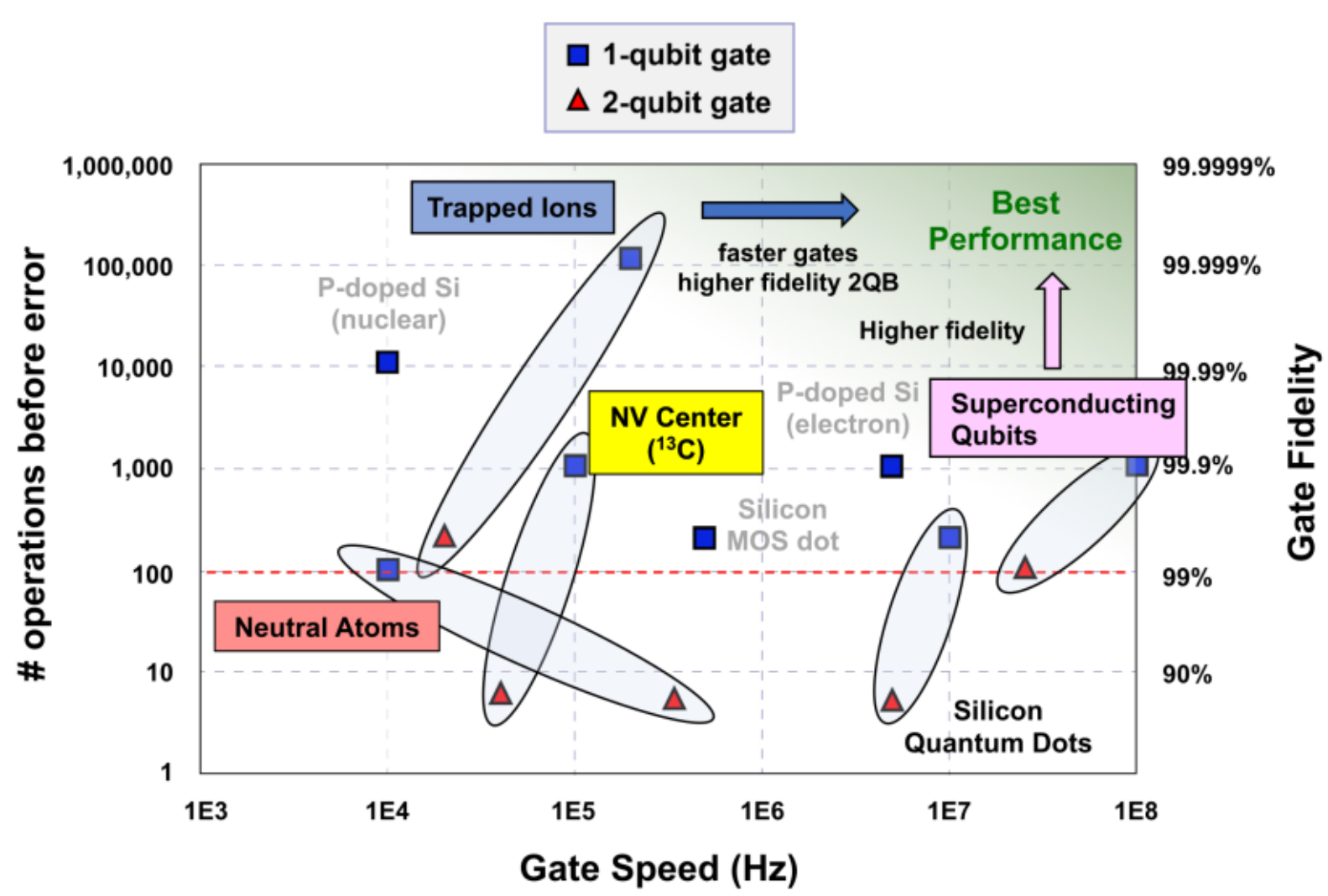

*Figure 217: comparison of qubit computing depth and gate speed. Source: [Engineering Quantum Computers](#) by William D. Oliver, December 2018 (15 slides).*

---

[382] See Entwicklungsstand *Quantencomputer* (*State of the art of quantum computing,* in English, June 2020 (266 pages).

[383] See Introduction to Quantum Computing by William Oliver from MIT, December 2019 (21 slides) and Engineering Quantum Computers by William D. Oliver, December 2018 (15 slides).

# Architecture overview

We will provide here an overview of the general architecture of a quantum computer, using the example of a superconducting qubit accelerator.

First of all, a bit like some external GPUs, quantum computers are implemented as co-processors or accelerators of classical computers that power and control them. A quantum computer is always driven by a classical computer, as can be a GPU for video games or for training neural networks in deep learning. These conventional computers are used to run the programs that are driving the quantum processor with physical operations to be performed on the qubits and are interpreting qubits readout results.

The classical computer closely controls the operation of the quantum computer by triggering at a precise rate the operations on the qubits that are performed by various electronic devices creating various electronics and photonic signals controlling quantum gates and quantum readout. It takes into account quantum gates execution time and the known qubits coherence time, i.e., the time during which the qubits remain in a state of superposition and entanglement. You will always need a classical computer to drive all of these tasks, unless you are Donald Trump, who can certainly do this with his thoughts.

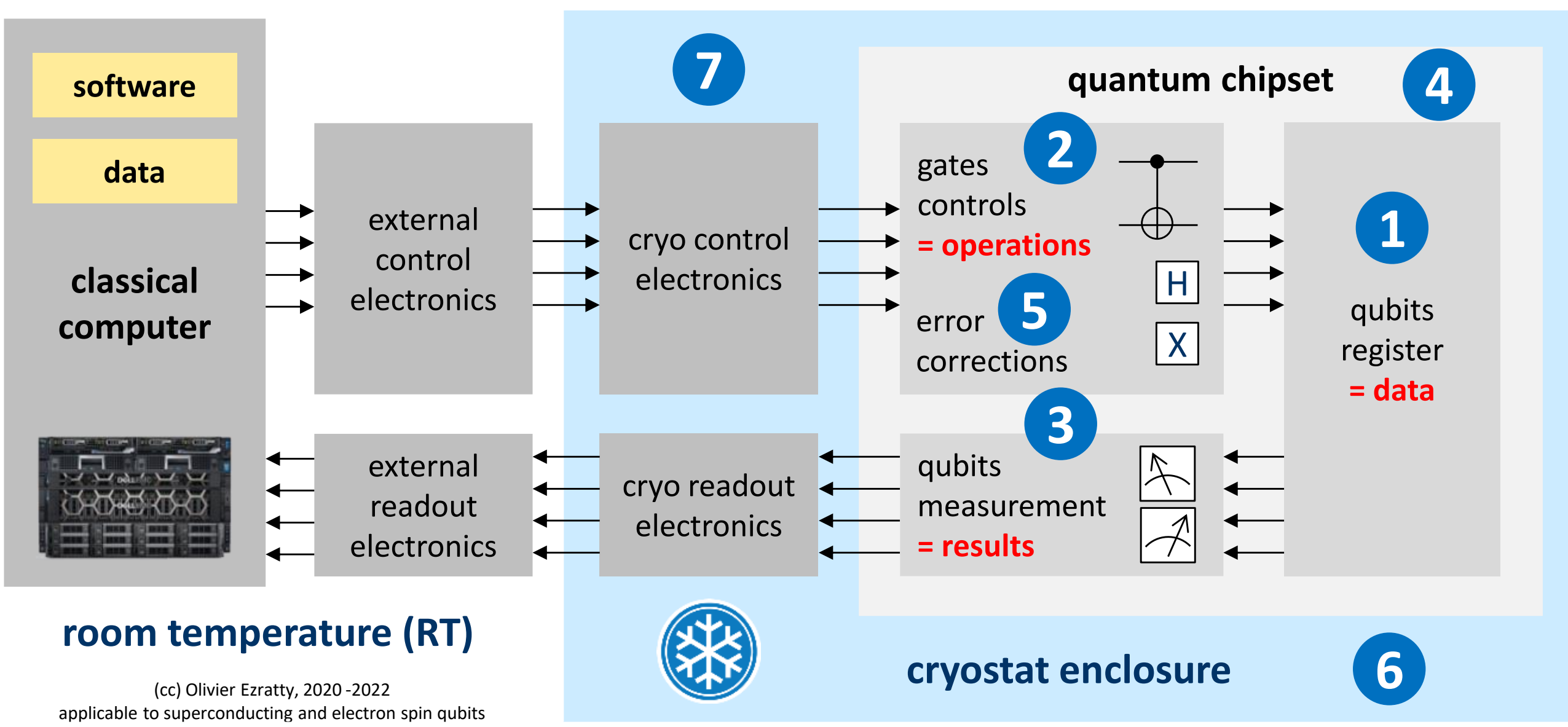

*Figure 218: typical high-level architecture of a gate-based quantum computer. (cc) Olivier Ezratty, 2020.*

In addition to its classical control computer, our quantum computer includes at least the components labeled from 1 to 7 that we will describe one by one, first with an overview below, then later, with a more detailed view. The other types of quantum computers have similarities and differences that we will mention whenever relevant.

❶ **Quantum registers** are collections of qubits. In 2021, the benchmarked record was 65 superconducting qubits with IBM. Quantum registers store the information manipulated in the computer and exploit the principle of superposition and entanglement allowing simultaneous operations on a large number of register values. To make a parallel with classical computing, this is memory. But processing is done directly in memory.

❷ **Quantum gate** controllers are physical devices that act on the quantum register qubits, both to initialize them and to perform quantum gates on them. These gates are applied iteratively, according to the algorithms to be executed. They can also be used to manage error correction codes. Quantum gates feed registers with both data and instruction. These are not separated operations like with classical microprocessors.

❸ **Measurement** qubit states is used to obtain the result at the end of the sequential execution of an algorithm's quantum gates and to evaluate error syndromes during quantum error correction. This cycle of initialization, calculation and measurement is usually applied several times to evaluate an algorithm result. The result is then averaged to a value between 0 and 1 for each qubit in the quantum computer's registers. The signals coming from qubit readout are then converted into digital values and transmitted to the conventional computer which controls the whole and implements results interpretation. In common cases, such as with D-Wave and IBM, computing is repeated at least a couple thousand times. The reading devices are connected to their control electronics via superconducting wires in the case of superconducting computer qubits.

❹ **Quantum chipset** usually includes quantum registers, quantum gates controls and measuring devices when it comes to superconducting or electron spin qubits. These are fed by microwaves coming from outside the chipset. Devices are more heterogeneous for other types of qubits, such as those that use lasers for initialization, quantum gates and qubit measurement like with trapped ions and cold atoms. Current chipsets are not very large. They have the size of a full-frame or dual-format photo sensor for the largest of them. Each qubit is relatively large, their size being measured in microns for superconducting qubits or down to 100 nm for electron spin qubits whereas modern CMOS processor transistors now have transistor sizes around 5 nanometers. The chipset for superconducting and electron spin qubits is a chip of a few square centimeters. It is usually integrated in an OFHC (Oxygen-Free High thermal Conductivity) copper packaging which is purified and freed from oxygen, limiting thermal conductivity. This package is fitted with coaxial connectors so that it can be fed by the microwaves controlling qubit gates. In the latest superconducting processors from IBM and Google with 53 qubits, more than 160 of these connectors are required. The chipset package is integrated in two small concentric aluminum and Cryoperm (from **MµShield**) magnetically insulated enclosures.

❺ **Error correction** is implemented with special code operating on a large number of consolidated qubits named logical qubits. They can be physically organized to optimize error correction, such as with surface codes and color codes. It's one of the biggest challenges ahead for creating scalable quantum computers. As of 2021, no quantum computer is large enough to accommodate logical qubits, given the number of physical per logical qubits exceeds the maximum number of qubits currently available.

❻ **Cryogeny** usually keeps the qubit chipset and its surrounding control electronics at a temperature close to absolute zero. It contains part of the control electronics and the quantum chip(s) to avoid generating disturbances that prevent the qubits from working. The Holy Grail would be to operate qubits at room temperature but the corresponding architectures such as in NV centers are not yet operational and there are still practical performance reasons to operate it at low-temperatures. The cryostat uses a mix of helium 3 and 4 to cool the components inside the chandelier while its compressor is itself cooled with cold water coming from another compressor, similar to the compressors used in classical air conditioning. Other types of qubits use cryostats in different places: with cooling photon sources or detectors in photon qubits systems, or for cooling ultra-vacuum pumps with cold atoms.

❼ **Control electronics** in the cryostat enclosure. The qubit control electronics drive the physical devices used to initialize, modify, and read the qubit status. In superconducting qubits, quantum gates are activated with microwave generators of frequencies between 4 and 8 GHz generally located outside the cryostat. These microwaves are transmitted on coaxial electrical wires between their source and the quantum processor, with superconducting cables below 4K. Their generators still take up a lot of space. They are not very miniaturized at this stage. Interesting work aims at integrating these microwave generators and readers inside the cryostat enclosure, if only to limit the wiring. These are frequently based on cryo-CMOS technology, CMOS components that are tailored to work at low temperature, 4K for many and as low as 20 mK for some. Figure 219 provides a rough representation of an entire superconducting qubits based quantum computer. The blue equipment corresponds to the microwave generators and analyzers from Zurich Instruments.

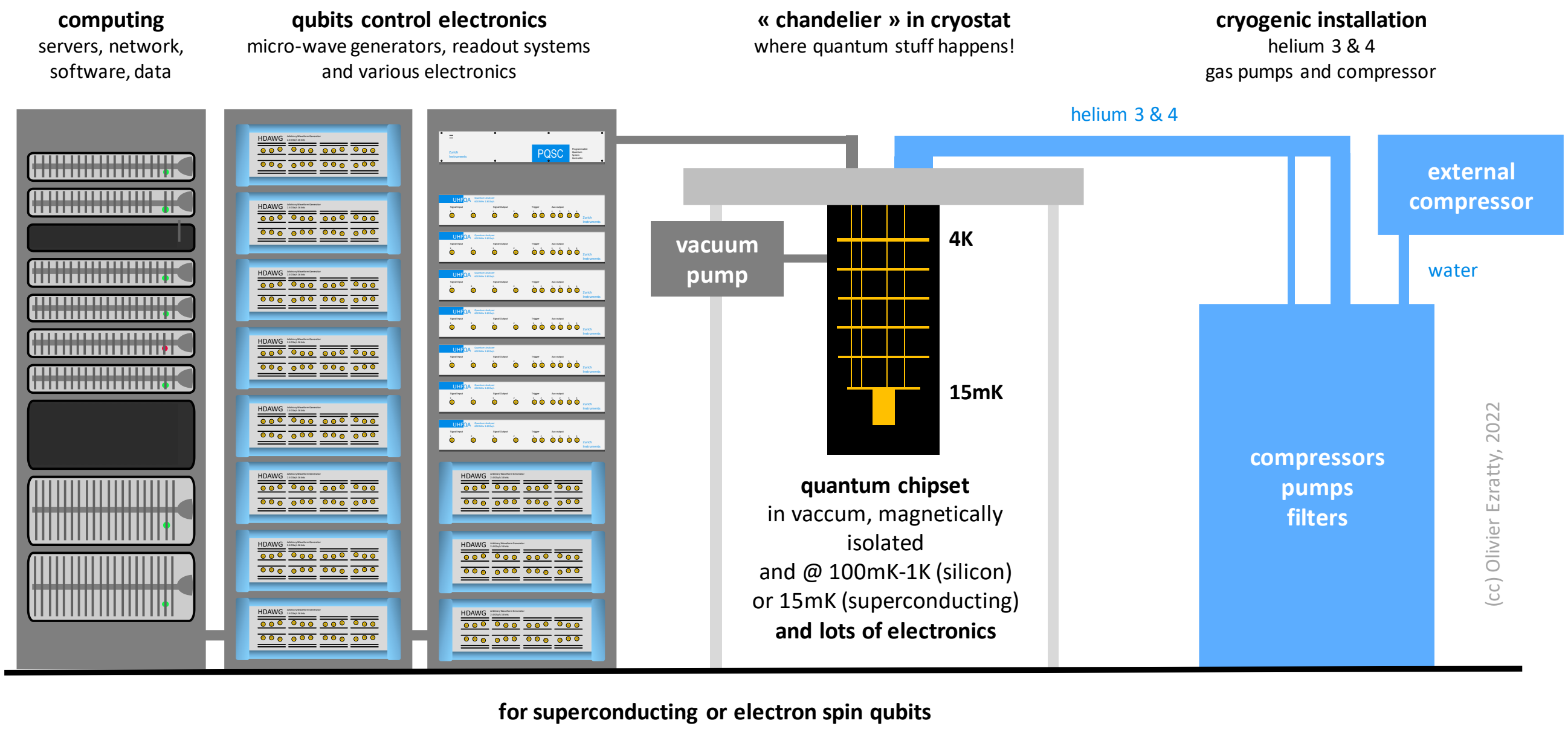

*Figure 219: typical physical components of a superconducting qubit quantum computer. It contains a classical computer that drives the whole system. (cc) Olivier Ezratty, 2020-2022.*

## Processor layout

To better understand the previous explanation, here is a chipset layout with 8 superconducting qubits, from ETH Zurich. Although it's already a few years old, the underlying concepts are generic.

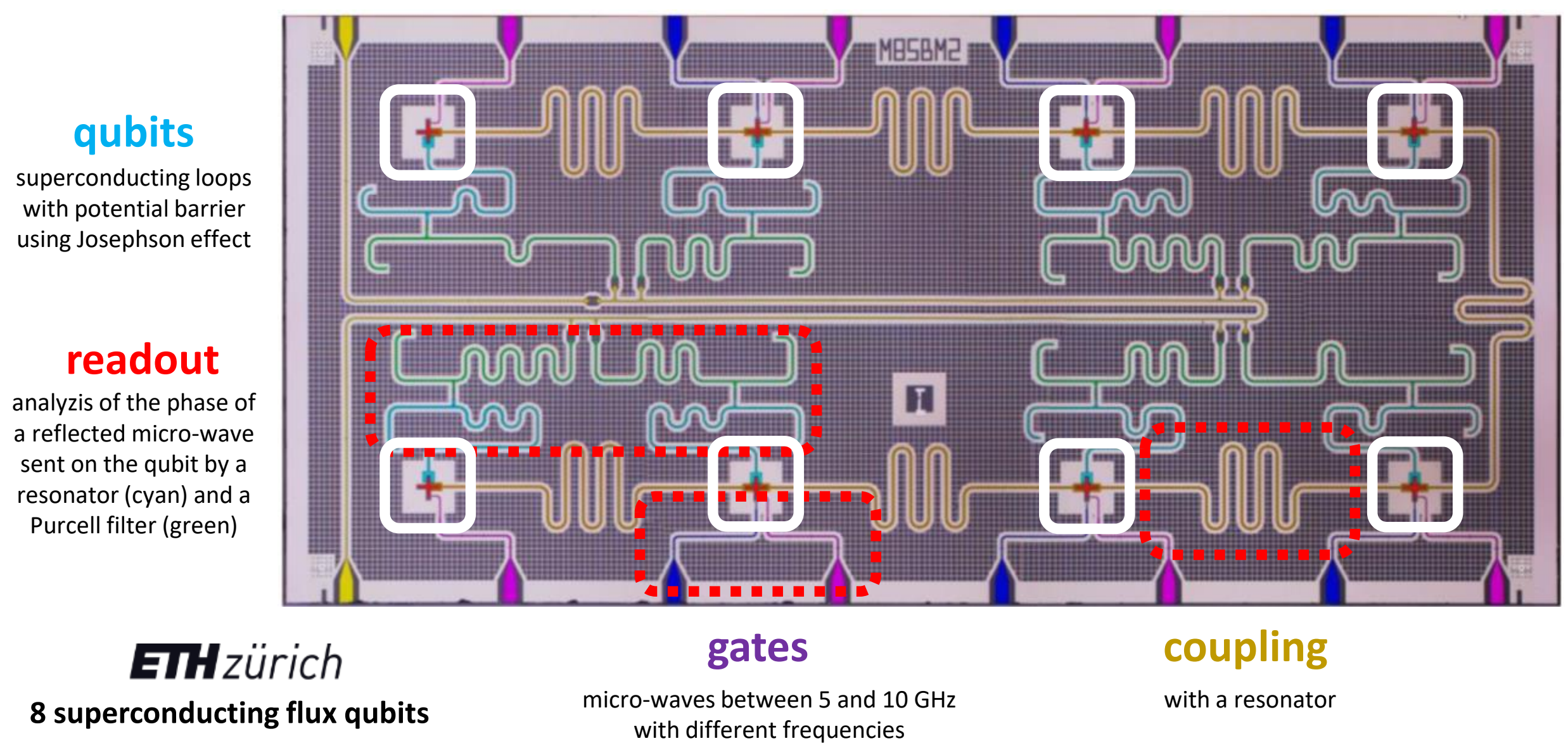

*Figure 220: a small 8-qubit superconducting processor from ETH Zurich showing its various components controlling the qubits. source: [The European Quantum Technologies Roadmap](), 2017 (30 pages) and the thesis [Digital quantum computation with superconducting qubits]() by Johannes Heinsoo, ETH Zurich, 2019 (271 pages).*

- **Qubits** are located in the white rectangles. These are tiny Josephson effect superconducting circuit loop.
- **Coupling circuits** link them together. It's used to control entanglement between pairs of qubits.
- **Single-qubit gates** use the blue and purple contacts. It sends microwaves to the qubits. These pins are powered via cables by very high frequency current sources, sending microwaves photons, between 4 and 8 GHz. These frequencies must be different between adjacent qubits of the same circuit to avoid crosstalk. It is the combination of these frequencies that will trigger different types of quantum gates and entanglements between adjacent qubits.

* **Measurement** takes place with other circuits, also fixed in the component. In superconducting qubits, these are magnetometers which are then connected to the outside of the vacuum chamber and cooled by superconducting cables. These are driven by microwaves.

Qubits must interact with each other but as little as possible with their environment until measurement. This is one of the reasons why they are usually cooled to a temperature close to absolute zero and magnetically isolated from the outside. The choice of materials for the chipsets also plays a role in minimizing the noise that could affect the qubits and bring them out of their coherent state.

In the diagram *below* is the how and why of the relationship between qubit gates time and coherence time during which the qubits remain stable. The orders of magnitude of these times for a typical quantum computer, particularly with superconducting qubits, give at best a ratio of 1 to 500 between gate times and coherence time. This means that the number of quantum gates that can be used in an algorithm is limited on NISQ systems. In the first generations of IBM quantum computers, the X, Hadamard and CNOT gates lasted 130 ns, 130 ns and 650 ns respectively.

These indications provide an upper limit on the number of gates that can be chained in a quantum algorithm. Note that these times are longer for quantum computers with ion traps, but the gate times are also longer. In CMOS qubits, coherence times are longer and gate times are low.

However, the available computing time is more limited by the quantum gates error rate. It constrains what is called the computation depth, i.e. the number of quantum gates that can be chained together without the error rate of the gates mitigating the results. Algorithms must therefore optimize the number of gate cycles to be executed, which is furthermore constrained by the physical connectivity between qubits.

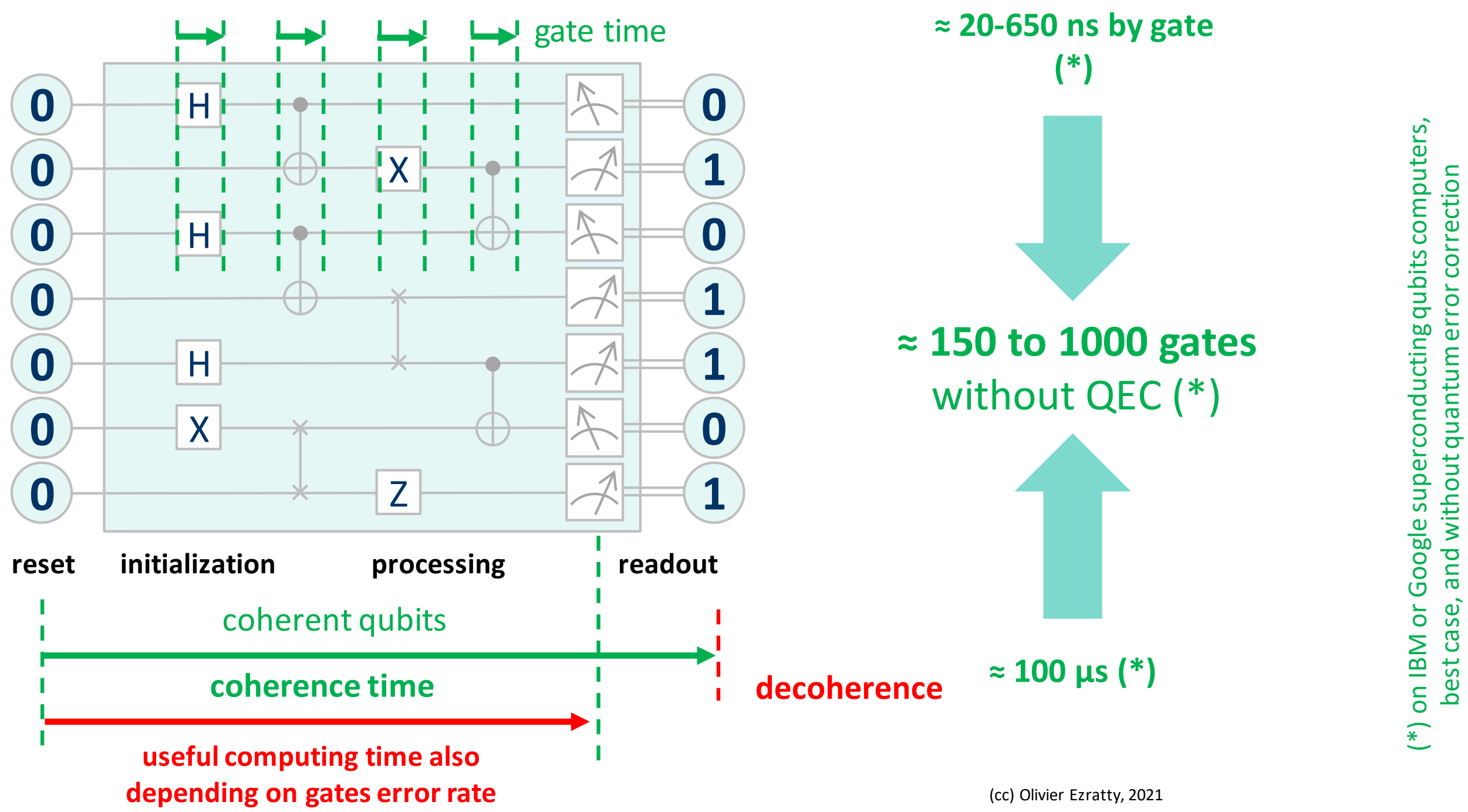

*Figure 221: a timeline showing the relation between useful computing time and gate coherence time and fidelities. (cc) Olivier Ezratty, 2021.*

In diagrams describing quantum algorithms, such as the one in Figure 221, the double bar after measuring the state of a qubit conventionally indicates that a normal bit has been recovered, at 0 or 1. By the way, all this reminds us that there are as many output qubits as input qubits in a quantum computation since they are physically the same!

# Error correction

One of the pitfalls of existing qubits is their significant error rates generated with quantum gates and measurement and coming mostly from the fateful quantum decoherence.

Decoherence is mostly generated by the interactions between the qubits and their environment. It progressively destroys the quantum information sitting in the qubits, and particularly the entanglement between these. It leads to an inevitable failure in computation after a short time. Error rates for each operation and readouts are commonly between 0.1% and several %, depending on the qubit type and quality. But even 0.1% is an intolerable level for most calculations.

In conventional computing, errors are way less frequent but must still be corrected. While some errors may be detected and fixed during computing in microprocessors, most errors are happening in memory, storage and telecommunications. These errors are discrete, corresponding to some unwanted bit flip. In quantum technologies, errors happen first and foremost in computing and within qubits and they are continuous and analog by nature.

So, we correct them with various techniques that we'll cover here, without necessary going into their details. This field is quite broad and very technical. We'll describe the various Quantum Error Correction (QEC) techniques that will be related to Fault-Tolerant Quantum Computing and Quantum Error Mitigation techniques that will be applicable in the near-term NISQ generation of quantum computers.

### Errors sources and typologies

A qubit register is coherent when its qubits superposition and entanglement is preserved over some period of time. At the individual qubit level, qubit stability is measured against its amplitude and phase stability, related to flip and phase error rates. Qubits coherence time is an indication of how long register qubits remain coherent, with stable superposition. Qubits amplitude stability is evaluated with a $T_1$ time while phase stability is measured with a $T_2$ and a $T_2^*$, these being sometimes confused with each other in the literature[384].

Real single qubit errors can be decomposed as linear combinations of these flip and phase errors[385]. Quantum error corrections codes are designed to separately correct flip and phase errors, the integration of which corrects any linear combination of both error types.

- **Flip errors** as shown in the Bloch sphere in Figure 222, are amplitude errors that tend to push the amplitude back to $|0\rangle$. These errors correspond mathematically to a decay of the diagonal part of the qubit density matrix in its eigenstate basis. It is related to the $T_1$, which is linked to a loss of amplitude ("energy relaxation"). It is also called "longitudinal coherence time", "spontaneous emission time", "population lifetime" or "amplitude damping" and corresponds to a loss of energy in qubits. It is a dissipative process that releases some energy[386].

[384] I found a good explanations in Dancing with qubits by Robert Suttor, pages 415 to 421, 2019 (516 pages).

[385] More precisely, single qubits errors can be decomposed in quantum channels: depolarizing channel (with a bit flip error, a phase flip error or a combination of both, in which case, the qubit remains in a pure state, and the qubit moves with some rotation in the Bloch sphere), a dephasing or phase damping channel (vanishing off-diagonal values in the qubit density matrix, in which case the qubit moves in a mixed state and inside the Bloch sphere) and an amplitude-damping channel. Source: Lecture Notes for Ph219/CS219: Quantum Information Chapter 3 by John Preskill, Caltech, October 2018 (65 pages).

[386] In superconducting qubit circuits, $T_1$ is proportional to the circuit quality factor $Q = \omega_q T_1$, which itself is proportional to the ratio between the energy stored in the qubit resonator and its rate of energy loss. $T_1$ comes from different phenomena: spontaneous emission, quasiparticle tunneling, flux coupling and dielectric losses in the Josephson junction. The Purcell effect is a spontaneous emission through the readout cavity. The Purcell decay rate is related to the speed of this phenomenon. It is reduced with using a Purcell filter which suppresses signal propagation at the qubit transition frequency. The filter is a pass-through with the readout cavity frequency that protects the qubit from decoherence channels while enabling its readout

The flip error is measured with a simple experiment with using an X gate and measuring the result n times at different t times. $T_1$ corresponds to the time when the probability of obtaining a $|1\rangle$ reaches 1/e. Such an error can damage the entanglement of the qubits related to the one affected by this error.

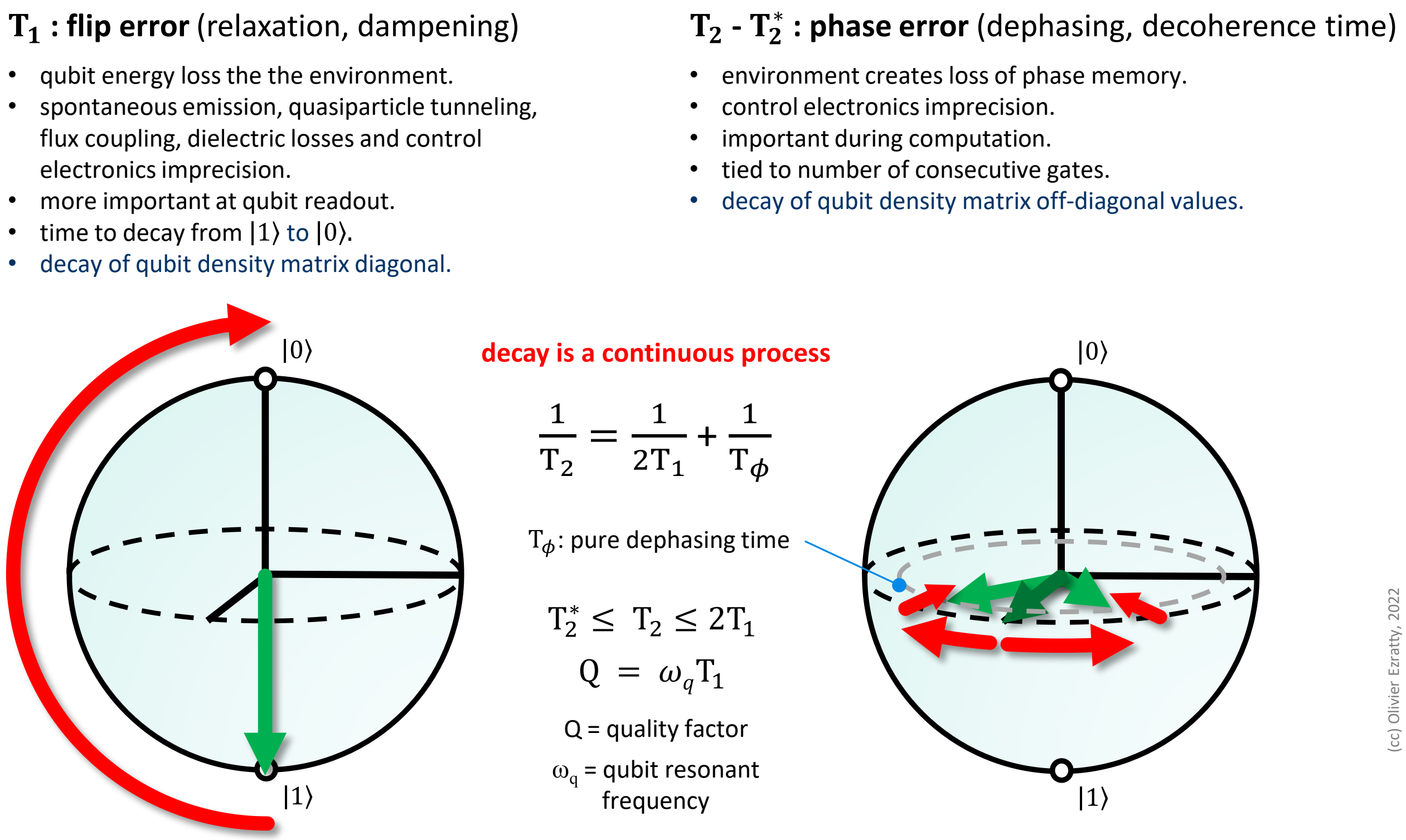

*Figure 222: flip error and phase errors and their effect in the qubit Bloch sphere. (cc) Olivier Ezratty, 2022.*

- **Phase errors** are rotations around the z-axis in the Bloch sphere (coherent noise, mostly due to control electronics imprecision) coupled with a move of the qubit vector within the sphere (decoherence noise, pure dephasing). These errors are not dissipative, meaning, they are thermodynamically neutral. Pure dephasing is related to the decay of the non-diagonal part of the qubit density matrix, creating a mixed state explaining why the qubit state moves inside the Bloch sphere. Coherent phase errors are measured in two manners, with $T_2^*$ and $T_2$ and are also called the "transverse coherence time", "transverse relaxation", "phase coherence time" or "phase damping". $T_2^*$ is evaluated with a Ramsey experiment, applying one Hadamard gate, wait time t, then apply another Hadamard gate, and measure the output. Without phase errors, the probability should look like a sinusoidal curve. With it, the curve slowly converges around a probability 0.5. $T_2^*$ is obtained when the probability reaches 1/e. On the other hand, $T_2$ is obtained with a Hahn echo experiment where an X gate is added at t/2, which removes some error sources not pertaining to qubit defects.
- **Leakage errors** is a third error type that sees a qubit drifting and stabilizing in another energy state than the basic $|0\rangle$ or $|1\rangle$. This can occur in the $|2\rangle$ level of a superconducting qubit, which we are trying to avoid, or with variations in the hyperfine energy levels of trapped ion qubits. This type of error can be corrected with specific reset protocols[387]. You don't observe such leakage errors with photon qubits using polarization as information encoding.

[387] See an example in [Removing leakage-induced correlated errors in superconducting quantum error correction]() by M. McEwen et al, Google AI, February 2021 (12 pages).

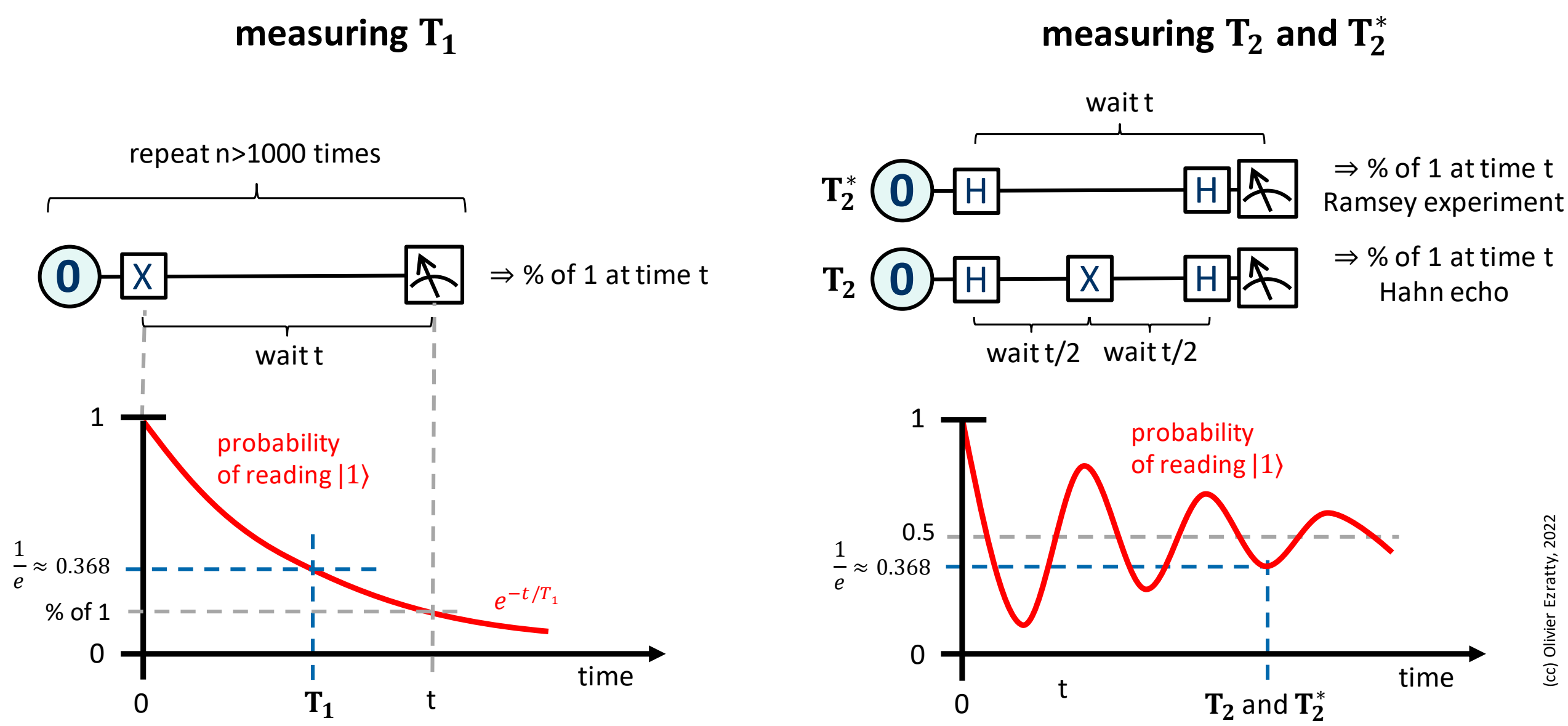

*Figure 223: how are measured T1, $T_2$ and $T_2^*$. (cc) Olivier Ezratty, 2022.*

The goal of having long qubit energy relaxation times is in competition with that of achieving high-fidelity qubit control and measurement. One key concern is to be able to apply error corrections as fast as possible after they are detected and before qubits decoherence takes effect or gets amplified. This is the reason why readout gates and phase detection electronics (for superconducting/quantum dots spin qubits) must be as fast as possible. But fast gates and readouts can drive leakage errors! They also require high-bandwidth microwave pulses, which reduce the capacity to frequency multiplex it in readout microwave circuits.

Error sources are multiple[388], leading notably to the progressive decoherence of qubits which affects qubits superposition and entanglement. They are linked to the various interactions between qubits and their immediate environment[389]. These include:

- **Control electronics** imperfections like clock, phase and amplitude jitter. It can be triggered by calibration errors of quantum gates that occur in particular in the calibration of superconducting qubits. They can notably trigger leakage errors. These small errors can create imprecisions with qubit operations. They are mitigated with improving the precision of control electronics. With superconducting and quantum dots spin qubits, it's mainly located at the local oscillators, mixers, AWG (arbitrary wave generation) and DAC levels (digital to analog conversion). These are coherent errors while other errors drive qubits decoherence.
- **Thermal noise** from components around the qubits. This is the reason for the existence of attenuators around superconducting qubits. It comes among other things from shocks between atoms.
- **Electrical and magnetic noise** which can have many origins depending on the qubits. It explains why D-Wave isolates its quantum computer with 16 metal layers to limit the impact of terrestrial magnetism on its qubits. Most solid state quantum processors are packaged in tight metal shielding.
- **Material defects** which are commonplace with solid state qubits (superconducting, quantum dots spins, NV centers). One common defect comes from dielectric losses in the Josephson junction of superconducting qubits.

---

[388] Here is a small inventory of noise sources for superconducting qubits: Sources of decoherence, ETH Zurich, 2005 (23 slides).

[389] Any operation will generate an error. An error can be generated at the time of correction, at the time of detection or at the time of application of a gate. Doing nothing on a qubit can also generate errors because of its finite coherence time.

- **Vacuum quantum fluctuations** originating errors which we studied quickly in a previous section, page 134 [390]. It's an endogamous source of errors within qubits while all others are exogamous.
- **Radioactivity**, particularly coming from cosmic rays**.** Radiations can be X-rays, gamma rays (whose electromagnetic nature was discovered in 1914), beta particles and their ionizing effects. The phenomenon is now well characterized. It creates phonons quasi-particles in the chipset substrate that can propagate to many surrounding qubits, endangering the efficiency of quantum error correction codes[391]. Radioactivity is one of the many sources of long range correlated noise. Envisioned solutions are to shield the qubit processing unit with lead[392] or copper backside metallization[393], to trap phonons using high impedance resonators made of granular aluminum[394] and to implement distributed error corrections schemes[395].

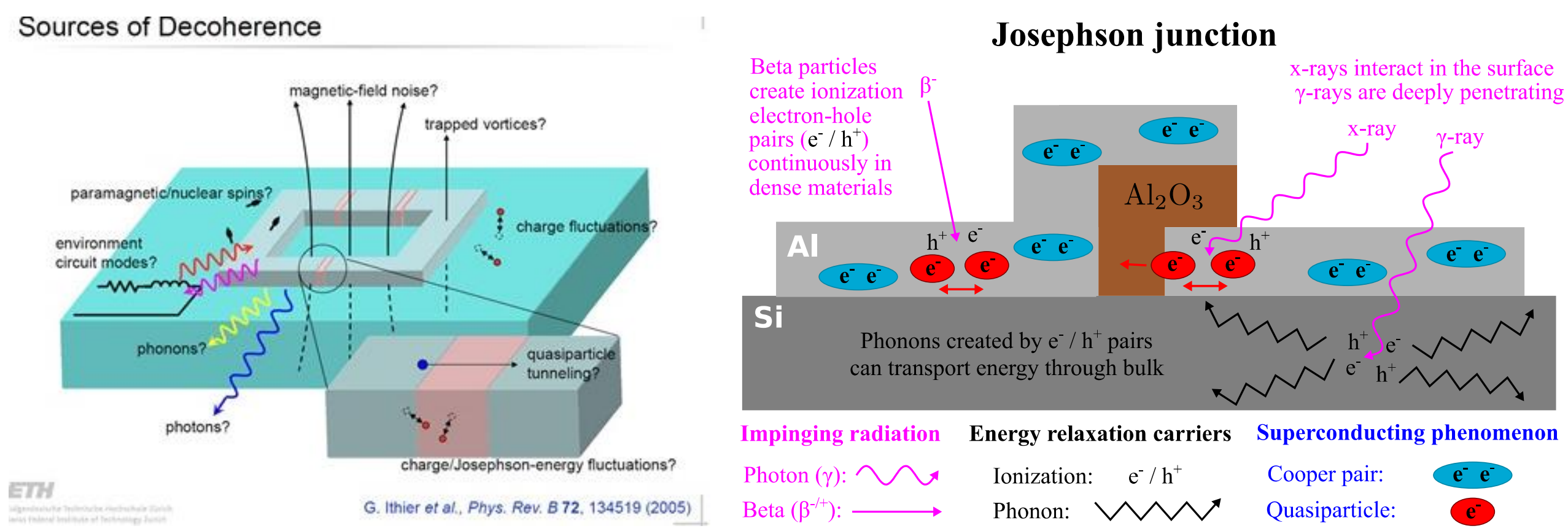

*Figure 224: sources of decoherence and cosmic radiations affecting superconducting qubits. Sources: Sources of decoherence, ETH Zurich, 2005 (23 slides) and Impact of ionizing radiation on superconducting qubit coherence by Antti P. Vepsäläinen, William D Oliver et al, August 2020 (24 pages).*

- **Gravity**, given this type of error and vacuum fluctuation ones appear to be minor compared to the previous ones[396].

Generally speaking, errors are generated by various interactions, electromagnetic or mechanical, between qubits and their immediate environment and are associated with the phenomenon of quantum decoherence. The first objective of physicists is obviously to reduce these physical sources of error. They are progressing steadily but are barely managing to gain one or two orders of magnitude in error rates, whereas in an ideal world, we would need ten orders of magnitude improvements.

---

[390] See Observation of quantum many-body effects due to zero point fluctuations in superconducting circuits by Sébastien Léger, Nicolas Roch et al, Institut Néel, 2019 (8 pages) which describes the phenomenon on superconducting qubits.

[391] See Correlated charge noise and relaxation errors in superconducting qubits by C.D. Wilne, Roger McDermott et al, Nature, December 2020-June 2021 (19 pages) which describes the correlated errors appearing in superconducting qubits and how it could impact the architecture of quantum error correction codes, Resolving catastrophic error bursts from cosmic rays in large arrays of superconducting qubits by Matt McEwen, Rami Barends et al, Google AI, Nature Physics, December 2021 (13 pages) who developed a test protocol to assess the impact of radiations on 26 qubits in its Sycamore processor and TLS Dynamics in a Superconducting Qubit Due to Background Ionizing Radiation by Ted Thorbeck et al, IBM, October 2022 (14 pages) which identifies the impact of ionizing radiations on qubit lifetimes.

[392] See Impact of ionizing radiation on superconducting qubit coherence by Antti P. Vepsäläinen, William D Oliver et al, August 2020 (24 pages).

[393] See Phonon downconversion to suppress correlated errors in superconducting qubits by V. Iaia, Robert McDemott et a, Wisconsin and Syracuse Universities, March 2022 (21 pages).

[394] See Phonon traps reduce the quasiparticle density in superconducting circuits by Fabio Henriques et al, Applied Physics Letters, 2019 (14 pages).

[395] See Distributed quantum error correction for chip-level catastrophic errors by Qian Xu, Lian Jiang et al, March 2022 (11 pages).

[396] See about gravity: A model of quantum collapse induced by gravity by Franck Laloë, 2020 (14 pages) and Gravitational Decoherence, 2017 (78 pages).

Some of these effects are avoided by cooling the qubits to a temperature close to absolute zero, but this is not enough. Researchers are therefore working hard to ensure that the noise affecting the qubits is as low as possible so that qubits coherence time can be as long as possible.

We have to manage this contradiction: qubits remain coherent, in a state of superposition and entanglement, if we do not disturb them, but we spend our time disturbing them with quantum gates operations! There are three ways to address these issues: improving gates fidelity, implementing quantum error correction codes and at last, reduce the number of gates needed to run algorithms.

### Qubits fidelities

In a gate-based quantum computer, three types of errors (or related fidelities) are usually evaluated: errors in single-qubit quantum gates, errors in two-qubit gates, and errors with qubits readout. These error rates are currently sitting between 0.1% and several 1%, which is much higher than the current error rates of traditional computing, which are negligible[397]. Qubits "fidelity" for any of these three dimensions is 100% minus the related error rate. In typical quantum parlance, when a "three-nines" 2 qubit-fidelity is mentioned, it means that it is better than 99,9%.

The chart below in Figure 225 consolidates a comparison of some fidelity levels of superconducting, trapped ion and cold atom quantum computer qubits, this information being provided by their vendors[398]. Qubit fidelities encompasses these three fidelities/errors dimensions (1Q, 2Q, readout).

Two-qubit gates and readout error rates are generally higher than one-qubit gates error rates[399]. We must therefore always pay attention to two-qubit gates, particularly given these gates are the source of much of the quantum computing power.

But these fidelities are not always measured in the same conditions. Some are measured with only a couple interacting qubits while others are done with all the register's qubits being active. It can create significant differences favoring the first kind of measurements, particularly due to the significant crosstalk between qubits.

The best fidelity achieved as of 2022 was 99.989% for Honeywell's 4 ion trapped single-qubit gate and 99.91% for IBM's Falcon R10 (27 qubit) dual-qubit gates [400]. Google's Sycamore single qubit gates fidelity is 99.84% with 53 qubits. The two-qubit CZ gates fidelity of China's Zuchongzhi 2.1 66 qubits superconducting processor is 99.4%[401].

Another observation relates to IBM's most recent fidelities with their best-in-class 27, 65 and 127 qubit systems as of November 2021. It did show 2-gates and readout fidelities that are lower as the number of qubits grows. Still, for a given a qubits number, IBM improves its processor fidelities over time.

---

[397] In classical calculation, errors are very rare. We talk about single particle perturbations (PPI) and single event upset (SEU) which trigger "soft errors" or logical errors. The SER (Soft Error Rate) combines the SDC (Silent Data Corruption, not detected) and the DUE (Detected and Unrecoverable Error, detected but not correctable). The unit of error measurement is the FIT (Failure in Time), which corresponds to one error per billion hours of use. The MTBF of electronic equipment (Mean Time Between Failure) is generally measured in years or decades. Errors are generally caused by isolated particles (ions, electrons, photons), particularly from cosmic rays like high-energy gammy rays. This affects in particular the electronics used in aerospace, which must be hardened to withstand them, as well as those used on Earth but at altitude. Memory is often more affected than processors. Hence error correction systems that use for example a parity bit and cyclic redundancy check used in telecommunications.

[398] Source for qubit reliability data mainly comes from Qubit Quality on Quantum Computing Report website, 2020. Plus some additional data coming from vendor sites.

[399] See An introduction to quantum error correction by Mazyar Mirrahimi, 2018 (31 slides) as well as Introduction to quantum computing by Anthony Leverrier and Mazyar Mirrahimi, March 2020 (69 slides) which completes it well.

[400] See the NASA and Google paper describing Google's performance: Quantum Supremacy Using a Programmable Superconducting Processor by Eleanor G. Rieffel et al, August 2019 (12 pages).

[401] Data source: Superconducting Quantum Computing by Xiaobo Zhu, June 2019 (53 slides).

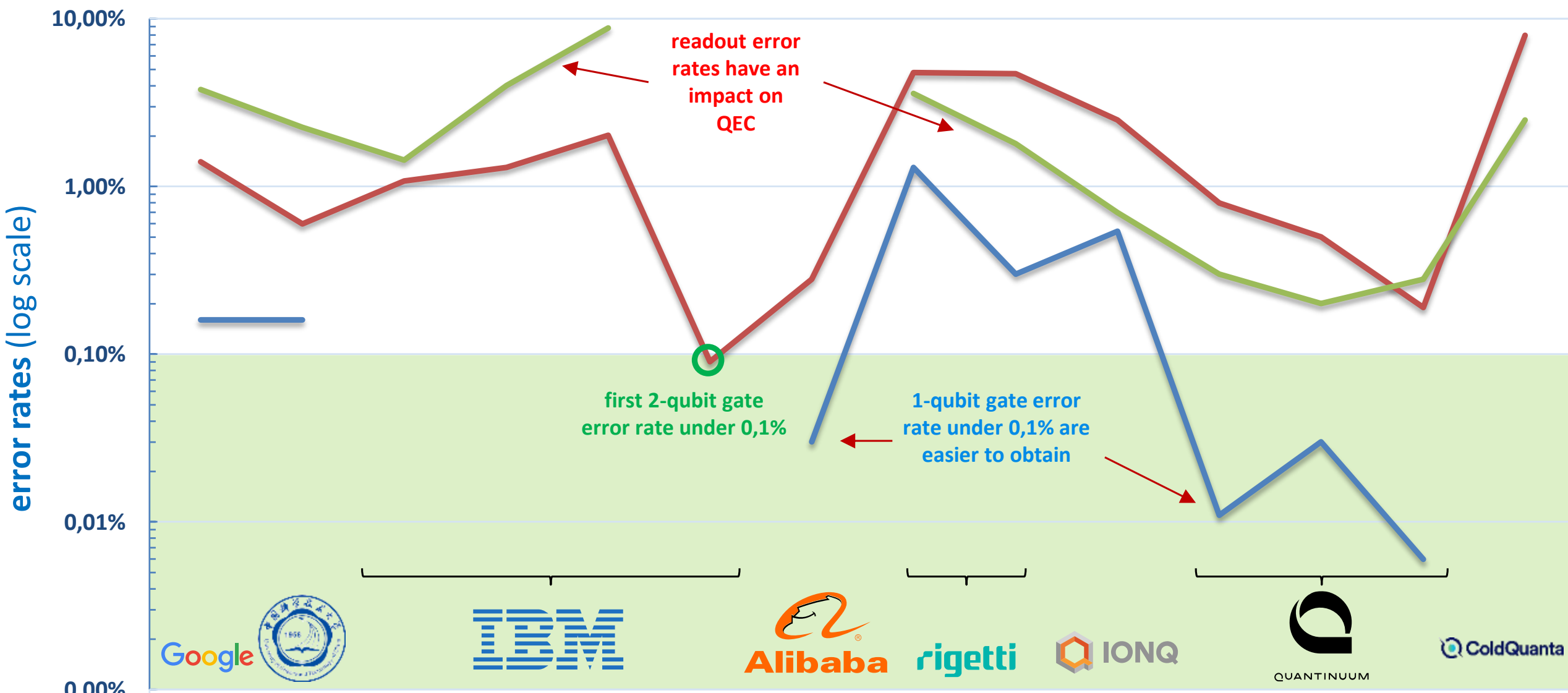

| | Google 53 qubits Sycamore 2019 | China Zuchongzhi 2.1 66 qubits | IBM Q27 Kolkata | IBM Q65 Brooklyn | IBM Q127 Washington | IBM Q27 Falcon R10 | Alibaba fluxonium | Rigetti 32Q Aspen | Rigetti 80Q Aspen-M-1 | IonQ 11 Qubit | Quantinuum H0 4 Qubits | Quantinuum H1 10 Qubits | Quantinuum H1-2 12 qubits | ColdQuanta 49 atoms |
|---|---|---|---|---|---|---|---|---|---|---|---|---|---|---|
| 1 qubit gate | 0,16% | 0,16% | | | | | 0,03% | 1,30% | 0,30% | 0,54% | 0,01% | 0,03% | 0,01% | |
| 2 qubit gate | 1,40% | 0,60% | 1,08% | 1,30% | 2,02% | 0,09% | 0,28% | 4,80% | 4,70% | 2,50% | 0,80% | 0,50% | 0,19% | 8,00% |
| readout | 3,80% | 2,26% | 1,44% | 4,00% | 8,82% | | | 3,60% | 1,80% | 0,70% | 0,30% | 0,20% | 0,28% | 2,50% |

superconducting | ions | cold atoms

*Figure 225: comparison of some qubit error rates with recent quantum processors. The most important rate is the two-qubit error rate. At this point, only IBM has a 2QB error rate below 0,1% with an experimental Falcon R10 27-qubit QPU. Compilation (cc) Olivier Ezratty, 2022.*

These error rates are currently prohibitive when executing many quantum gates in a row. With each operation, error rates add up and the reliability rate decreases. Imagine chaining a few dozen two-qubit gates! At this rate, the error rate can very fast exceed 50% at the end of a rather simple algorithm and, generally, well before the fateful qubit coherence time limit.

Hence the fact that the power of a quantum computer is always evaluated not simply by the number of available qubits but by the number of operations that can be done with a reasonable error rate at the end of the calculation. To avoid this quantitative constraint, we should have qubits with quantum gate error rates of $10^{-10}$ or even $10^{-15}$.

Figure 226 illustrates this discrepancy between today's physical qubits and the need to perform reliable calculations (without error correction).

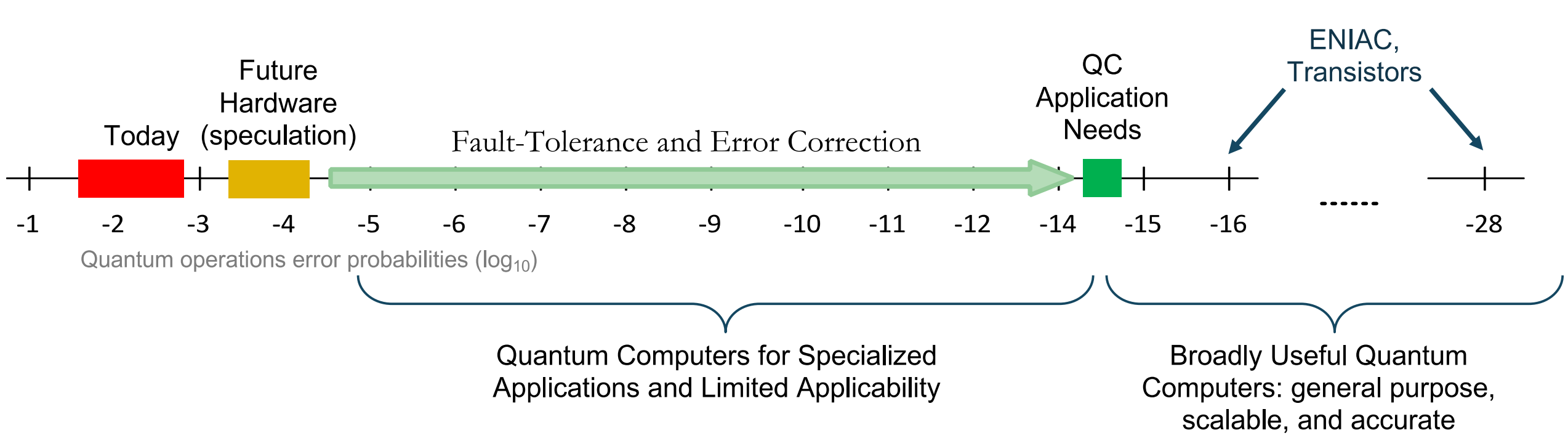

*Figure 226: comparison of error levels between existing quantum hardware and what is required, with error correction codes. Source: How about quantum computing? by Bert de Jong, DoE Berkeley Labs, June 2019 (47 slides).*

A formula is used to evaluate the dependency between quantum gates error rates (e), the number of qubits (n) and the number of usable gates (d), called "circuit depth": $nd < 1/e$. As the error rate decreases, the usable circuit depth increases, and the range of usable algorithms expands.

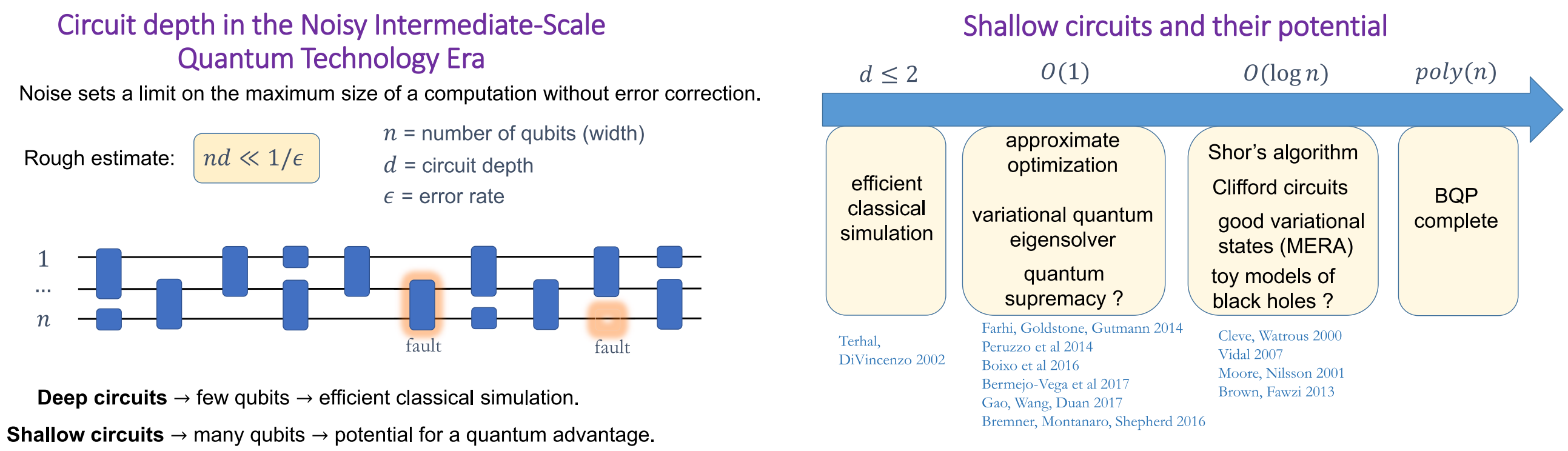

*Figure 227: relationship between circuit depth and their use case. Source: [Quantum advantage with shallow circuits](#) by Robert König, 2018 (97 slides).*

It is presented in another way in this **Quantum Benchmark** diagram, with the number of qubits on the abscissa and the depth of the circuits on the ordinate (number of gates that can be linked in a quantum calculation), conditioned by the skewed dotted lines that correspond to the error rates of the quantum gates.

The white zone is the *quantum* supremacy zone, also known as the *quantum discovery regime*[402].

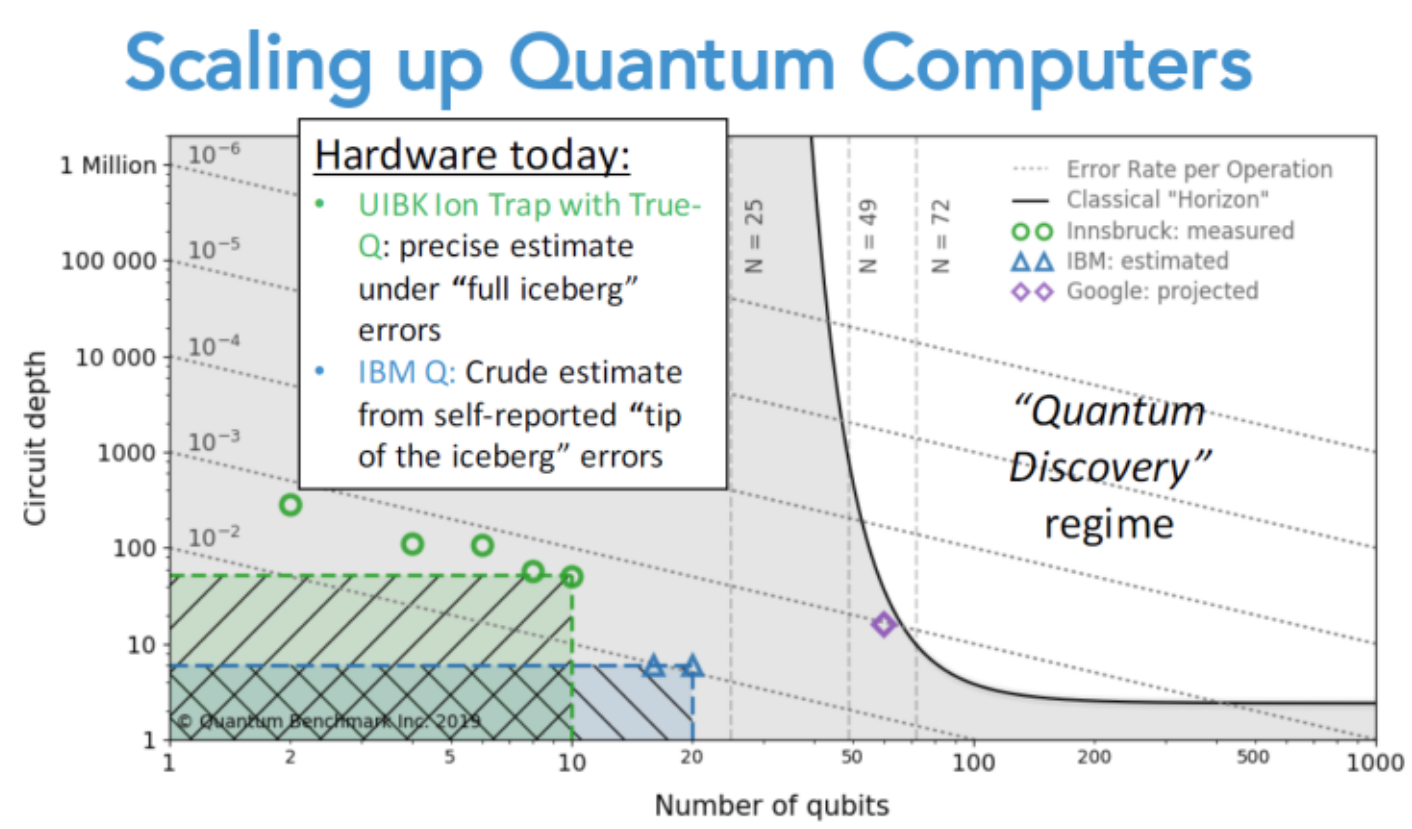

*Figure 228: circuit depth vs number of qubits. Source: Joseph Emerson, Quantum Benchmark. 2019.*

For a quantum computer to be useful and scalable, you need a lot of qubits, a low error rate for quantum gates and qubits readout, and a long qubit coherence time to be able to execute algorithms without much time constraints although quantum error correction codes are indispensable workarounds for this last constraint.

How are quantum gates and measurement error rates evaluated? We've seen previously how individual qubits flip and phase error rates are usually measured. Other methods are required to have an idea of the fidelities of registers with entangled qubits.

One method is the **Randomized Benchmarking** (RBM) process which consists in chaining a random sequence of quantum gates whose result is known in advance and with comparing the results obtained with the right responses. Usually, a random sequence of Clifford gates is launched and then executed backwards. The error rate increases with the number of chained quantum gates and depends on their type. We can evaluate the error rate of a given gate with the **Interleaved RBM** which injects the gate periodically into the random gate set used. We then measure the difference in error rate between the sequence with and without these added quantum gates[403].

[402] Slides presented by Joseph Emerson of Quantum Benchmark at the Quantum Computing Business conference organized in Paris on June 20, 2019 by Bpifrance. They position Google very close to the area of interest with their 72 qubits, but public benchmarks of these qubits have not yet been published after their announcement in March 2018. The 53 qubits of the Sycamore generation announced in October 2019 are however at about the same place (purple diamond).

[403] See Efficient measurement of quantum gate error by interleaved randomized benchmarking by Easwar Magesan, Jay Gambetta et al, March 2012 (5 pages). And Quantum Computing: Progress and Prospects, 2018 (206 pages), page 2-20. The process of benchmarking quantum gates is detailed in Randomized benchmarking for individual quantum gates by Emilio Onorati et al, 2018 (16 pages). The origin of the method is Scalable noise estimation with random unitary operators by Joseph Emerson et al, 2005 (8 pages).

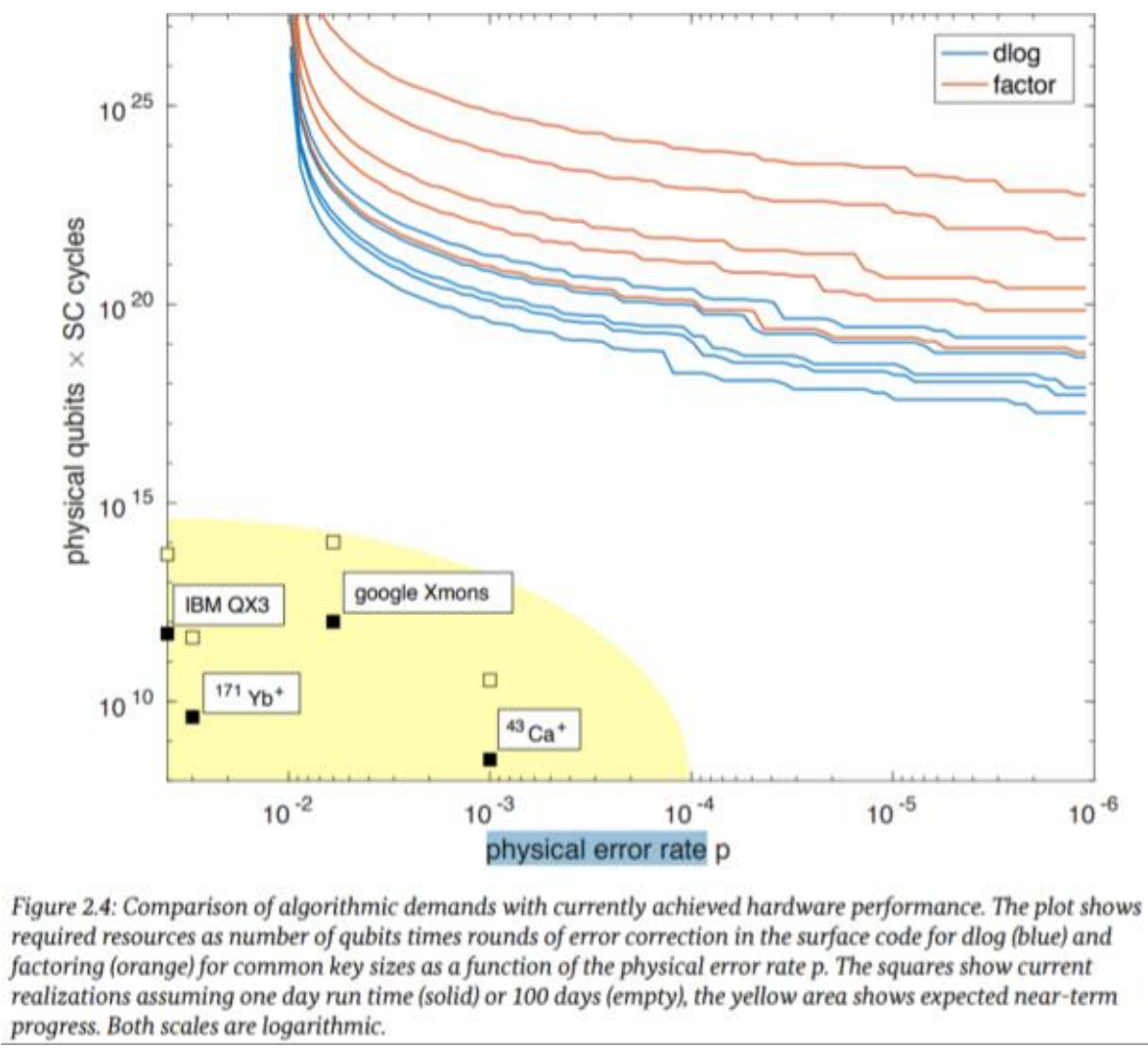

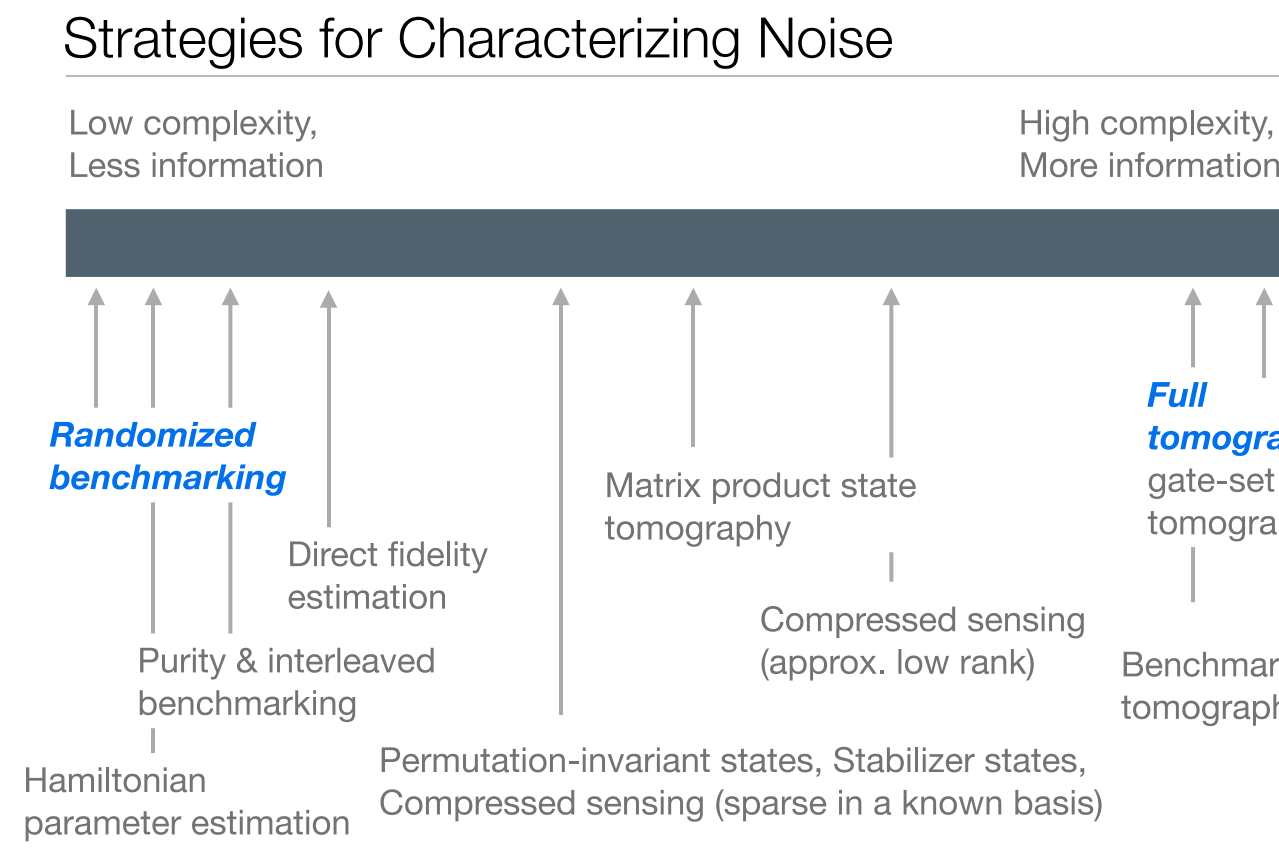

*Figure 229: source: Entwicklungsstand Quantencomputer. 2018.*

*Figure 230: comparing the various strategies to characterize qubit noise. Source: Characterization of quantum devices by Steve Flammia, University of Sydney, 2017 (118 slides).*

You'll have to look elsewhere to find out more data[404]. The RBM method has some drawbacks for clean noise quantification. It is apparently not suitable for the detection of any noise patterns[405].

Several other methods exist, such as **quantum state tomography** (QST) that we already covered in the section dedicated to measurement, page 184, which is based on a repeated measurement of qubit states that allows the reconstruction of a mean density matrix and the associated errors, for one or two qubits after a calculation.

Another method is used to assess the reset/gate/readout cycle, **SPAM** (state preparation and measurement). SPAM measures the cumulative fidelities of a single qubit state preparation and readout. It's used and advertised by Quantinuum. It however doesn't provide any indication on multiple qubit gates and qubits entanglement quality and scaling[406].

Yet another method exists that is based on some mathematical tools identifying a match between the noise rate of one and two-qubit gates of an algorithm and the total noise rate of the complete algorithm. In short, it links macro noise (algorithm) to micro noise (quantum gates).

### Error correction codes zoo

Quantum error correction can't work the same as classical error correction. Qubits cannot be independently replicated with some measurement that would be performed on one replicated qubit. On top of that, we are correcting analog errors in multiple dimensions, not just a 0/1 error flip that could be labelled as a simple "digital error"[407].

[404] As in the aforementioned German report Entwicklungsstand Quantencomputer (*State of the art of quantum computing*), which dates from 2018 and highlights the huge gap between the performance of qubits, particularly at IBM and Google, and the need for integer factorization to break common RSA keys. See also Efficient learning of quantum noise by Robin Harper et al, Nature Communications, 2019 (15 pages) and Characterization, certification and validation of quantum systems by Martin Kliesch, April 2020 (87 pages).

[405] See Characterization of quantum devices by Steve Flammia, University of Sydney, 2017 (118 slides) which provides an excellent overview of the various qubits benchmarking techniques.

[406] See 99.9904% SPAM Fidelity with barium-137 sets the standard and creates a further step towards solving some of the world's most intractable problems by Kortny Rolston-Duce, Quantinuum, March 2022.

[407] The stakes of QEC are very well explained in Approaches to Quantum Error Correction by Julia Kempe, 2005 (29 pages). See also the review paper Quantum Error Correction for Quantum Memories by Barbara M. Terhal, April 2015 (47 pages) and the excellent Introduction to Quantum Error Correction and Fault Tolerance by Steven M. Girvin, August 2022 (99 pages) which is a transcript from a lecture at Les Houches Summer School in 2019 and (brilliantly) covers both classical and quantum error correction techniques.

The techniques explored for more than two decades consists in implementing quantum error correction codes called **QEC** for Quantum Error Correction or rather **QECC** for QEC Codes [408]. Most of these QEC schemes correct errors that are small and independent, meaning, not correlated between several close of distant qubits.

Error correction codes apply to both universal gate quantum computing and quantum telecommunications. In the first case, they are integrated into the concept of fault-tolerance quantum computing (**FTQC)**. Error correction is a mean to slow down qubits decoherence and extend the available computation time.

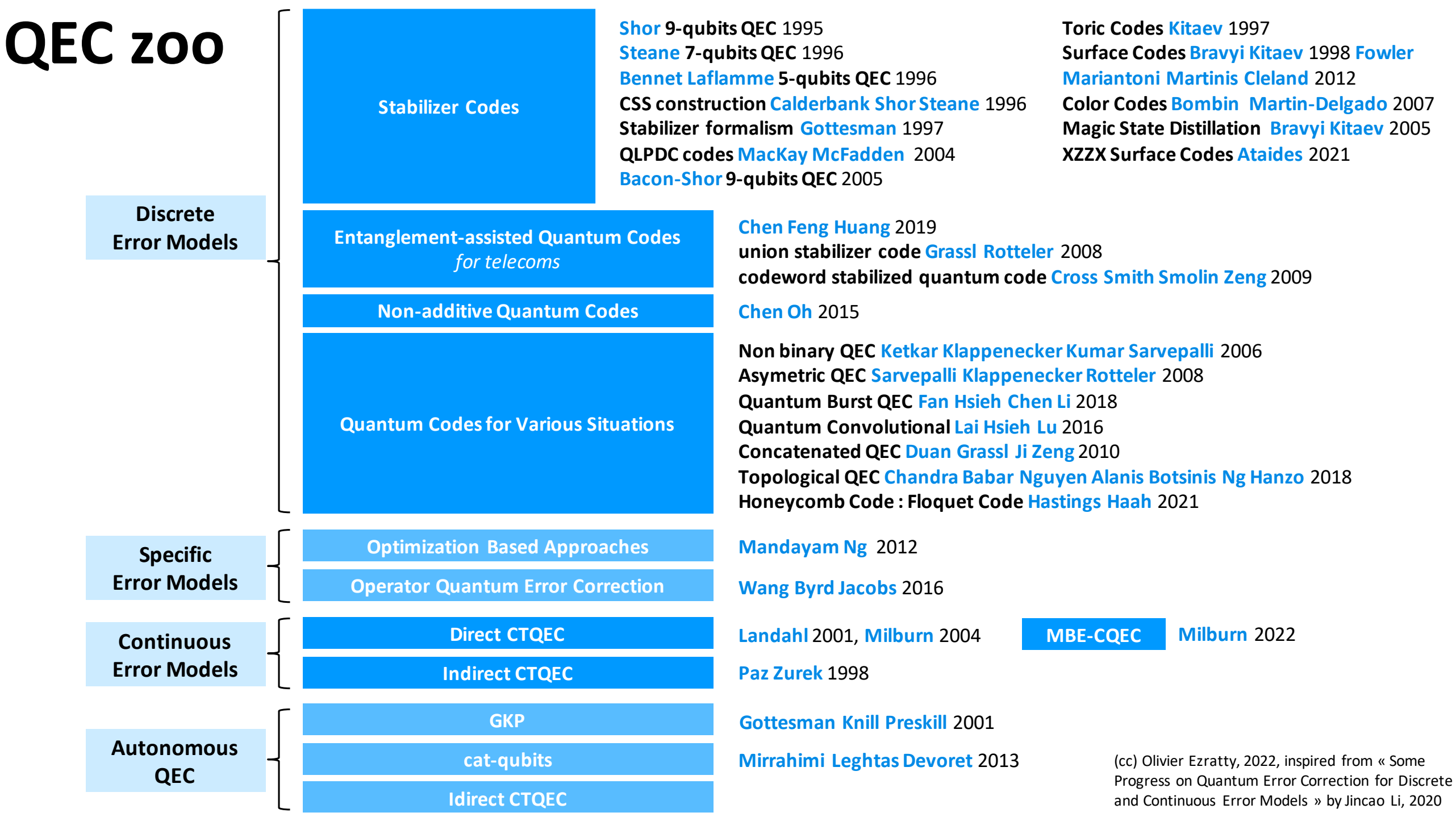

*Figure 231: inventory of key quantum error correction codes. (cc) Olivier Ezratty, 2022, inspired from Some Progress on Quantum Error Correction for Discrete and Continuous Error Models by Jincao Li, 2020 (16 pages).*

The chart in Figure 231 makes an inventory of the main quantum error correction codes with their origin and date of creation[409]. This error correction zoo is very dense[410].

It is a very rich scientific field of quantum technologies and has been growing regularly since 1995. It includes several families of error correction codes.

The most important QEC codes family are the **stabilizer codes**. The first ones correct flip and/or phase errors with three, five (Laflamme), seven (Steane) or nine qubits (Shor). These codes replicate qubits several times with entanglement. They follow the same processing in parallel. Then, the code compares the results at the output of algorithms to keep the statistically dominant results. All this is done without reading the value of the qubits which would make the whole system collapse. This is implemented with ancilla qubits that are used to detect error syndromes without affecting the qubits used in calculation.

---

[408] This theme has, like many quantum specialties, its own conference. See International Conference on Quantum Error Correction and the videos with all the presentations of the 2019 edition.

[409] Illustration inspired by a scheme discovered in Some Progress on Quantum Error Correction for Discrete and Continuous Error Models by Jincao Li, 2020 (15 pages).

[410] See the Error Correction zoo and its section on quantum error correction codes. And Quantum Error Correction for Beginners by Simon J. Devitt, William J. Munro, and Kae Nemoto, 2013 (41 pages).

The trick consists in duplicating the information on several qubits and do qubits rotations in the Bloch sphere then some projective measurement. This will not deteriorate the information contained in the qubits. This measurement enables the detection of error syndromes. We then use single-qubit gates to correct the qubits for which an error was detected. It goes through some classical processing that must be as fast as possible.

Stabilizers codes have many variants including:

**Topological codes** including **surface codes** and **color codes**[411] themselves derived from **toric codes**[412], and many other specimens such as the **DFS** (Decoherence Free Subspaces) protocol encode quantum information in a subspace that is unaffected by physical errors or so-called holographic codes[413] and also **Fractal Surface Codes**[414]. Color codes have much less overhead than surface codes and render possible the implementation of FTQC (Fault-Tolerant Quantum Computing). It is due to one key feature of these correction codes: they can be implemented with the transversal gates described in the section on FTQC. It could be used with superconducting and quantum dots spin quits. However, what is "colored" in these colored codes and how does it work? I have no clear idea[415].

**Magic state distillation**. A Shor and Steane code can correct any Pauli error, including Y gate, which is equal to iZX. It can correct any linear combination of I, X, Y and Z gates with complex numbers. This comes from the fact that any unit operation on a qubit can be expressed as a combination of IXYZ with complex factors: U = aI + bX + cY + dZ. This means, indirectly, that these QECs should be able to correct analog and continuous errors such as slight variations in amplitude or phase, i.e. rotations of a few degrees in the Bloch sphere. To correct these errors corresponding to gates outside the Clifford group such as a T gate (eighth of rotation in the Bloch sphere, these gates that provide an exponential speedup for gate-based computing), however, magic states are also used which feed circuits made with gates from the Clifford group. These states are prepared by a process called magic state distillation[416]. It has an enormous overhead with the number of required physical qubits to create a single logical qubit, of about two orders of magnitude (x100). Magic state distillation is implemented in surface codes. This overhead could be avoided or reduced with using 3D correction codes, which are difficult to implement with actual qubits at this time.

**ZXXZ surface code** that would reduce the number of required physical qubits to create a logical qubit thanks to a lower error threshold. In April 2021, University of Sydney science undergraduate Pablo Bonilla Ataides ZXXZ paper published in Nature Communications brought the attention of Amazon researchers[417].

---

411 This is however not the only solution to the magic state distillation physical qubits cost. See Fault-tolerant magic state preparation with flag qubits by Christopher Chamberland and Andrew Cross, IBM, May 2019 (26 pages) which describes an alternative using more ancilla qubits ("flag qubits").

412 See Fault-tolerant quantum computation by anyons by Alexei Kitaev, 1997 and 2008 (27 pages).

413 Color codes are variations of stabilizing codes. See some explanations in The Steep Road Towards Robust and Universal Quantum Computation by Earl T. Campbell, Barbara M. Terhal and Christophe Vuillot, 2016 (10 pages).

414 See Topological Order, Quantum Codes, and Quantum Computation on Fractal Geometries by Guanyu Zhu, Tomas Jochym-O'Connor, and Arpit Dua, IBM, PRX, Quantum, September 2022 (55 pages).

415 See for example The ABCs of the color code by Aleksander Marek Kubica, 2018 (205 pages), a rich thesis done under the supervision of John Preskill at Caltech with the help from Jason Alicea, Fernando Brandão and Alexei Kitaev. And The cost of universality: A comparative study of the overhead of state distillation and code switching with color codes, by Michael E. Beverland, Aleksander Kubica and Krysta M. Svore, 2021 (69 pages).

416 See Universal quantum computation with ideal Clifford gates and noisy ancillas by Sergey Bravyi and Alexei Kitaev, 2004 (15 pages). There are other solutions such as A fault-tolerant non-Clifford gate for the surface code in two dimensions by Benjamin J. Brown, May 2020, which applies to surface codes.

417 See Student's physics homework picked up by Amazon quantum researchers by Marcus Strom, University of Sydney, April 2021, Sydney student helps solve quantum computing problem with simple modification by James Carmody April 2021 and The XZZX surface code by J. Pablo Bonilla Ataides et al, April 2021, Nature Communications (12 pages).

This surface code could be used by Amazon who made a choice to use a relatively low number of photons per cat qubit (8 to 10, compared to about 30 for Alice&Bob), still requiring some first level bit-flip error correction on top of phase-flip correction. That's where a ZXXZ surface code QEC could come into play. ZXXZ QEC codes are indeed mentioned as an option QEC technique in Amazon's technical paper from December 2020. A team from Amazon, Caltech and the University of Chicago improved these codes in 2022 to work with biased noise (which has more phase than flip noise)[418].

**Planar Honeycomb Code**, *aka* **Floquet Code**, was created by Matthew B. Hastings and Jeongwan Haah from Microsoft in 2021 to simplify toric codes with fewer qubits and stabilizers[419]. It is adapted to qubits architectures implementing pair-wise qubit measurements (XX, YY, ZZ) like with Majorana fermions. The technique was improved by Google researchers[420], by Christophe Vuillot from Inria Nancy[421] and with the $XYZ^2$ variant in 2022[422].

**Quantum LDPC codes** that are inspired by classical LDPC (low-density parity check) codes used in telecommunications (Wi-Fi, 5G, DVB) and seem to scale better than surface codes[423]. They are also adapted to Majorana fermions but for a longer term. But they require an increased connectivity between qubits that is not possible on a 2D layout and which will require 3D circuitry to create distance connections between qubits.

**Autonomous QEC**. Here, we must distinguish "logical error correction codes" and "physical codes" that are directly managed at the qubit hardware level. Also branded "autonomous QEC", these contain **bosonic codes**, including GKP error codes[424], binomial codes, 0-π and cat-codes[425]. The latter implement in a cavity a "Schrödinger cat" that allows to manage a projection space used for error correction, as in the error correction algorithms based on stabilizing codes that we will see later. It usually corrects flip errors autonomously, but new variations are able to correct both flip and phase errors[426].

**Continuous error correction**. In opposition to discrete quantum error correction codes (DQEC) that we've covered, there are alternative QEC using continuous measurement and correction. Juan Pablo Paz and Wojciech Zurek proposed in 1998 a continuously operating error correction code, the CTQEC, for "continuous-time QEC", or CQEC, based on differential equations and acting at reduced time intervals.

---

[418] See Tailored XZZX codes for biased noise by Qian Xu, Lian Jiang et al, March 2022 (16 pages). Biased noises is well explained in the thesis Quantum Error Correction with Biased Noise by Peter Brooks, Caltech, 2013 (198 pages).

[419] See Dynamically Generated Logical Qubits by Matthew B. Hastings and Jeongwan Haah, Microsoft, October 2021 (18 pages).

[420] See Benchmarking the Planar Honeycomb Code by Craig Gidney and Michael Newman, Google, February-September 2022 (16 pages).

[421] See Planar Floquet Codes by Christophe Vuillot, Inria, December 2021 (16 pages) and A Pair Measurement Surface Code on Pentagons by Craig Gidney, June 2022 (16 pages).

[422] See The $XYZ^2$ hexagonal stabilizer code by Basudha Srivastava et al, Chalmers, April 2022 (15 pages).

[423] See Quantum Low-Density Parity-Check Codes by Nikolas P. Breuckmann and Jens Niklas Eberhardt, October 2021 (17 pages), Qubits Can Be as Safe as Bits, Researchers Show by Mordechai Rorvig, January 2022 referring to See Asymptotically Good Quantum and Locally Testable Classical LDPC Codes by Pavel Panteleev and Gleb Kalachev, 2022 (51 pages).

[424] See Encoding a qubit in an oscillator by Daniel Gottesman, Alexei Kitaev and John Preskill, PRA, 2001 (22 pages) and the perspective Quantum error correction with the Gottesman-Kitaev-Preskill code by Arne Grimsmo and Shruti Puri, PRX Quantum, June 2021 (20 pages).

[425] Cat-codes are used by the startup Alice&Bob. Knowing that their creation goes back to the work of Mazyar Mirrahimi and Zaki Leghtas in 2013, with whom the founders of Alice&Bob worked. Error correction codes are constantly being updated. Thus, a proposal recently emerged from QEC that goes further than cat-code and does not depend on hardware architecture. See Novel error-correction scheme developed for quantum computers, March 2020 which refers to Quantum computing with rotation-symmetric bosonic codes by Arne L. Grimsmo, Joshua Combes and Ben Q. Baragiola, September 2019.

[426] See Quantum error correction using squeezed Schrödinger cat states by David S. Schlegel et al, January 2022 (20 pages) which provides protection for both flip and phase errors.

There are two methods for acting directly on the information (direct CTQEC) or via auxiliary qubits (indirect CTQEC). CQEC avoids using ancilla qubits to measure the stabilizer operators by weakly measuring the physical qubits. It also enables faster measurements and error detection, reducing undetected errors.

These methods were later improved by various contributors including Andrew J. Landahl and Gerard J. Milburn[427]. The later recently proposed to use some real-time measurement-based estimator (MBE) of the real logical qubit to be protected to accurately track the actual errors occurring within the real qubits in real-time. This leads to the **MBE-CQEC** scheme that protects the logical qubit to a high degree and allows the error correction to be applied either immediately or at a later time.

**Quantum Error Mitigation**. At last, one other solution being considered deals with using NISQ, for Noisy Intermediate Scale Quantum computers, those current quantum computers that use noisy and uncorrected qubits. This is done with algorithms, often hybrid classical/quantum algorithms, which are supposed to be errors resilient and with using some Quantum Error Mitigation techniques (QEM). We detail it later in this section.

*"Whatever comes out of these gates, we have a better chance to survive if we work together. You understand? We stay together, we survive."*
General Maximus Decimus Meridius (Russell Crowe) in Gladiator, 2000.

### Error correction principles

The general principle of a classical quantum error correction code is illustrated in the diagram in Figure 232 with a six-step correction[428]:

1. **Encoding**: the qubit to be corrected will first be replicated a certain number of times via CNOT gates on several auxiliary qubits (here 2). The resulting qubits are entangled. In the example, we get the state $\alpha|000\rangle+\beta|111\rangle$ for an input state $|\psi\rangle = \alpha|0\rangle+\beta|1\rangle$.
2. **Processing**: it will potentially generate an error coming from various sources of noise. This can be a calculation as well as some telecom transmission of a qubit.
3. **Detection**: one or more error syndromes are detected via quantum gates that associate qubits with other ancilla qubits. In the example below, it detects pure flip errors.
4. **Measurement**: the state of these ancilla qubits is measured to become classical bits in the syndrome extraction process. It helps create the index of the qubit to be corrected in the replicated entangled qubits. This is some non-destructive measurement for the corrected entangled qubits since it's done in a different basis. These measurements are labelled "mid-circuit measurements" since they occur before the end of your circuit execution, only on a subset of the register qubits and without changing the quantum state of other qubits in the register.
5. **Correction**: the address obtained with syndrome measurement is used to correct the faulty qubits with an X gate (for a phase error, we'd use a Z gate). There are alternative forms of QEC that do not involve the measurement of the syndrome by qubit reading but by its direct use with quantum gates that correct the defective qubit without going through conventional bits.
6. **Consolidation**: finally, the corrected qubits are disentangled to recreate an isolated corrected qubit $|\psi\rangle$. This consolidation seems to be used with error correction for quantum telecommunications. When applied to quantum computing, the corrected entangled qubits can be kept to move on to the next step, i.e. another computing operation to be corrected.

---

[427] See Continuous quantum error correction via quantum feedback control by Charlene Ahn, Andrew C. Doherty and Andrew J. Landahl, PRA, March 2002 (12 pages), Practical scheme for error control using feedback by Mohan Sarovar, Charlene Ahn, Kurt Jacobs and Gerard Milburn, PRA, May 2004 (12 pages) and Measurement based estimator scheme for continuous quantum error correction by Sangkha Borah, Gerard Milburn et al, March 2022 (9 pages).

[428] Based on A Tutorial on Quantum Error Correction by Andrew M. Steane, 2006 (24 pages). See also An introduction to quantum error correction by Mazyar Mirrahimi, 2018 (31 slides).

7. **Reuse**: the correct qubit or qubits can now be used for subsequent operations that will also be corrected with the same process.

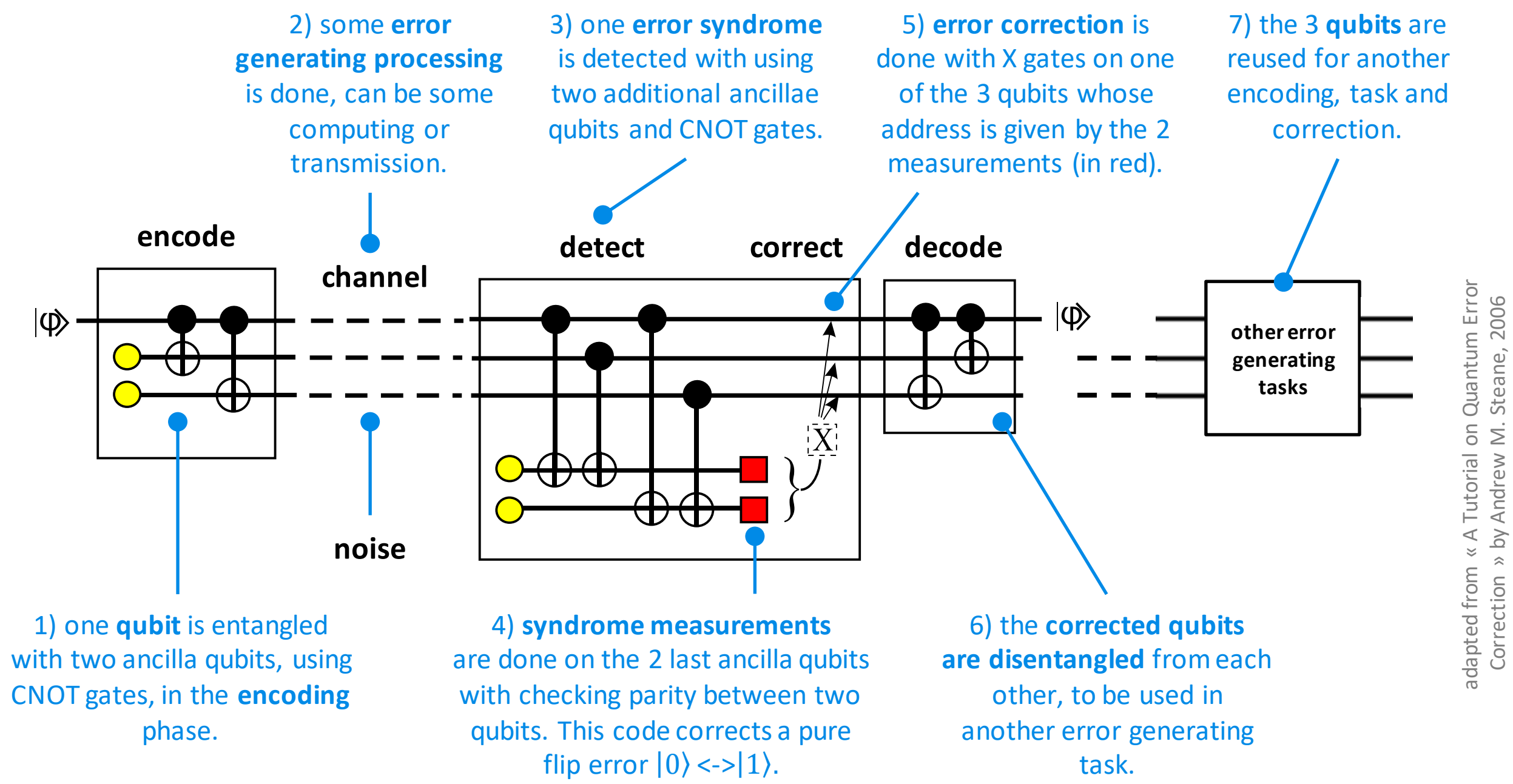

*Figure 232: a simple error correction code, adapted from A Tutorial on Quantum Error Correction by Andrew M. Steane, 2006 (24 pages).*

Error correction codes charts such as Shor's on Wikipedia are usually not complete. They usually lack the measure/correction steps. They can also rely on direct error correction[429] .

They also do not specify where to place the error correction codes in a quantum algorithm. It seems that this is required at each stage of some quantum computation. Error correction codes will be repeated several times that is roughly proportional to the computational depth of the quantum algorithm. It will be the role of the compilers to position QEC in the code executed on the quantum processor. It may depend on their knowledge of the fidelity rates of the quantum gates used in the hardware. In the end, the QEC will increase computation time by one to several orders of magnitude depending on the ratio of physical qubits per logical qubits and the qubit life-extension obtained with the code. It has to be taken into account when evaluating the time-based quantum computing advantage brought by a given algorithm.

Looking at the genealogy of these error correction codes, we must start with the simplest ones that correct qubit sign errors with three qubits like the example in Figure 232 from Andrew Steane. A similar QEC corrects qubit phase errors by exploiting Hadamard gates.

The 1995 **Shor's 9-qubit error correction code** consolidates these two methods, with the corrected qubit being replicated 8 times. This code corrects both flip and phase errors[430]. It uses a total of 15 qubits. What such a complete code looks like is shown in Figure 233. In the first phase the corrected qubit is replicated two times and each resulting qubit is again replicated two times with CNOT gates. The first three blocks of 3 qubits implement a flip error correction. It outputs 3 qubits which then implement a phase error correction.

---

[429] See Quantum Error Correction An Introductory Guide by Joschka Roffe, 2019 (29 pages) which explains the generic operation of error correction codes and Quantum Error Correction for Beginners by Simon Devitt, William Munro and Kae Nemoto, 2013 (41 pages). These are the two main sources of information that allowed me to write these pages on QEC. See also a description of various error correction codes in Software for Quantum Computation, a thesis by Daniel Matthias Herr from ETH Zurich, 2019 (164 pages).

[430] The details of the process are well documented in the Wikipedia sheet of quantum error correction.

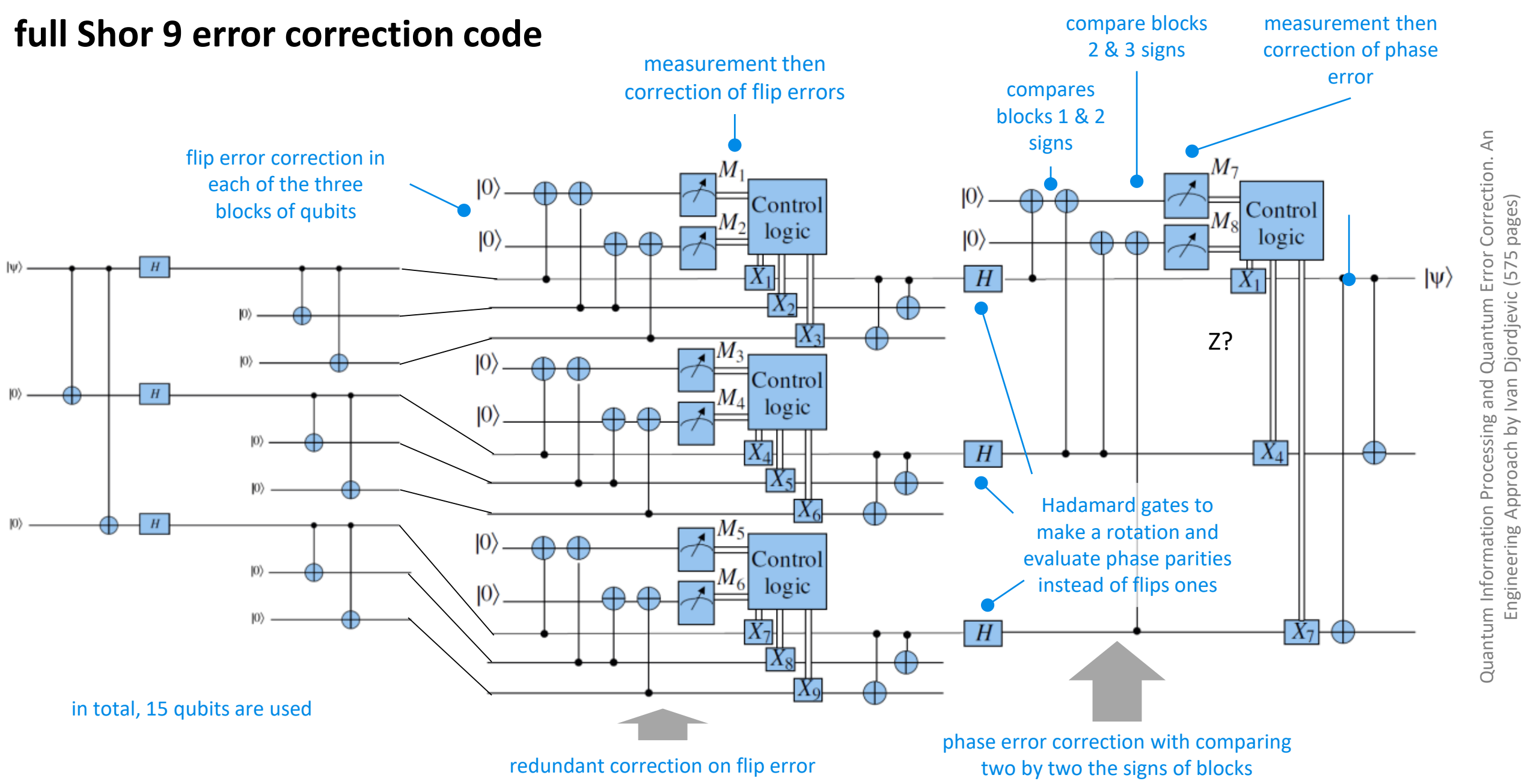

*Figure 233: a full Shor 9 error correction code correcting both flip and phase errors. Source: Quantum Information Processing and Quantum Error Correction. An Engineering Approach by Ivan Djordjevic (575 pages).*

**Raymond Laflamme** (1960, Canada) demonstrated in 1996 that at least five physical qubits are needed to create a "logical qubit" integrating flip and phase error correction. With **Emanuel Knill**, he also demonstrated that any single qubit error was a linear combination of flip and phase errors, leading to factoring error correction to flip and phase errors corrections[431].

In practice, the 7-qubit **Steane** code is the most referenced because it is not redundant like the Shor code. These 3-, 5-, 7- and 9-qubit codes are part of a generic group called **stabilizer codes** formalized by **Daniel Gottesman** in 1997. We are now going to dig a little deeper into how they operate.

We will better understand how an error correction works without reading the state of the qubit to be corrected. Let's take the case of a simple flip error correction code with three qubits.

These three entangled qubits can have an error $X_1$, $X_2$ or $X_3$ or no error (I=identity). X is an amplitude inversion Pauli gate. It creates an amplitude inversion of the corresponding entangled qubit as shown in the equations in Figure 234. These new states correspond to three errors and the absence of errors.

$$|\psi_L\rangle = \alpha|000\rangle + \beta|111\rangle \xrightarrow{I} \alpha|000\rangle + \beta|111\rangle,$$
$$|\psi_L\rangle = \alpha|000\rangle + \beta|111\rangle \xrightarrow{X_1} \alpha|100\rangle + \beta|011\rangle,$$
$$|\psi_L\rangle = \alpha|000\rangle + \beta|111\rangle \xrightarrow{X_2} \alpha|010\rangle + \beta|101\rangle,$$
$$|\psi_L\rangle = \alpha|000\rangle + \beta|111\rangle \xrightarrow{X_3} \alpha|001\rangle + \beta|110\rangle.$$

*Figure 234: amplitude inversions.*

These four states have the interest of being mathematically orthogonal for all the values of the α and β defining the state of the qubit to be corrected. The trick is to perform a measurement of these values in the vector space corresponding to these four values and not in the original qubit computational base. This will not deteriorate the superposition of the original qubit. The syndrome extraction is called a "Stabilizer code" or “stabilization code”, which will feed the ancilla qubits. The process is the same to evaluate and correct a phase error but with Z gates instead of X gates.

---

[431] This is demonstrated in A Theory of Quantum Error-Correcting Codes by Emanuel Knill and Raymond Laflamme, 1996 (34 pages). But also independently in Mixed State Entanglement and Quantum Error Correction by Charles Bennett, David DiVincenzo, John A. Smolin and William K. Wootters, 1996 (82 pages). See also Magic States by Nathan Babcock (28 slides).

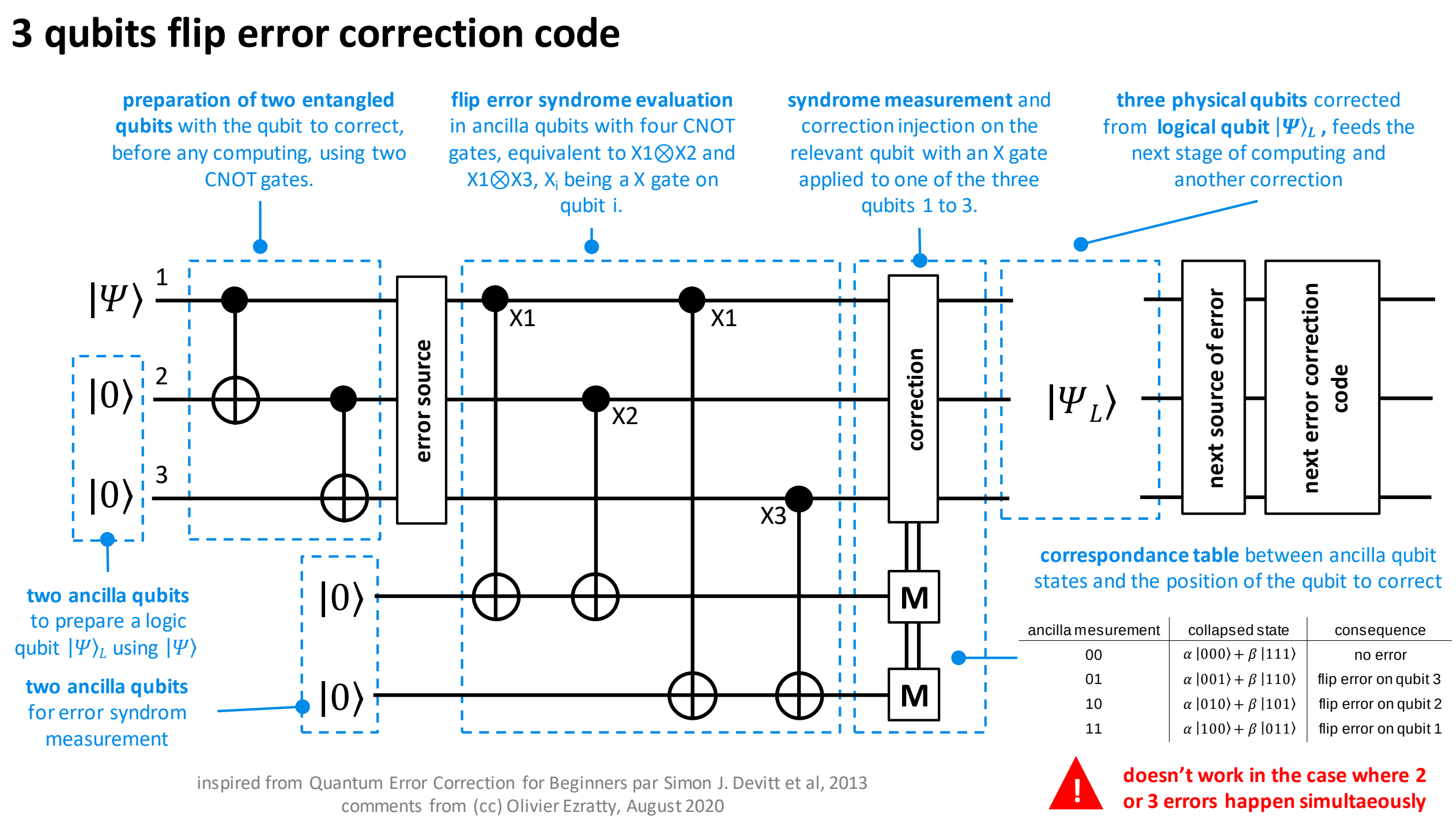

*Figure 235: 3 qubits flip error correction code explained. (cc) Olivier Ezratty, 2021.*

The disadvantage of the solution is that it cannot detect errors that would occur at the same time on two or three of the entangled qubits. No error-correcting code can correct all errors!

The stabilizer codes formalism generically describes the error correction codes we have just studied with three parameters: **[[n, k, d]]** with:

- **k** = number of logical qubits, usually 1 which is the qubit that needs to be corrected.
- **n** = number of physical qubits used in the code, with n > k. The n-k qubits store the redundant information thanks to entanglement.
- **d** = smallest number of simultaneous qubit errors that can transform one valid codeword into another, *aka* the **code distance**. The complete definition is actually more complicated.

More precisely, an error correction code with a code distance d can correct errors for up to (d − 1)/2 (replicated) qubits, or say differently d ≥ 2m+1, m being the number of redundant qubits that can be corrected. You need d to be at least 3 to correct both flip and phase errors.

In this notation, Shor's 9 qubit code is a [[9, 1, 3]], Steane's is a [[7, 1, 3]] and Laflamme's is a [[5, 1, 3]]. They all have a code distance of 3. A simple 3-qubit flip or phase correction code is a [[3, 1, 1]] stabilizer code with a code distance of 1. There are larger cases like with the [[512, 174, 8]] CSS code[432].

The stabilizer codes use a syndrome table that provides a match between the errors on each qubit and the detected syndrome. The number of ancilla qubits used to create this table must therefore be sufficient to identify the qubits to be corrected in the logical qubit. In the example in Figure 236 with a logical qubit with 7 physical qubits, the 3 ancilla qubits allow the identification of eight scenarios, sufficient to determine which of the 7 physical qubits must be corrected. The 8$^{th}$ scenario is the absence of error, therefore needing no correction.

[432] See Classical product code constructions for quantum Calderbank-Shor-Steane codes by Dimiter Ostrev et al, 2022 (19 pages).

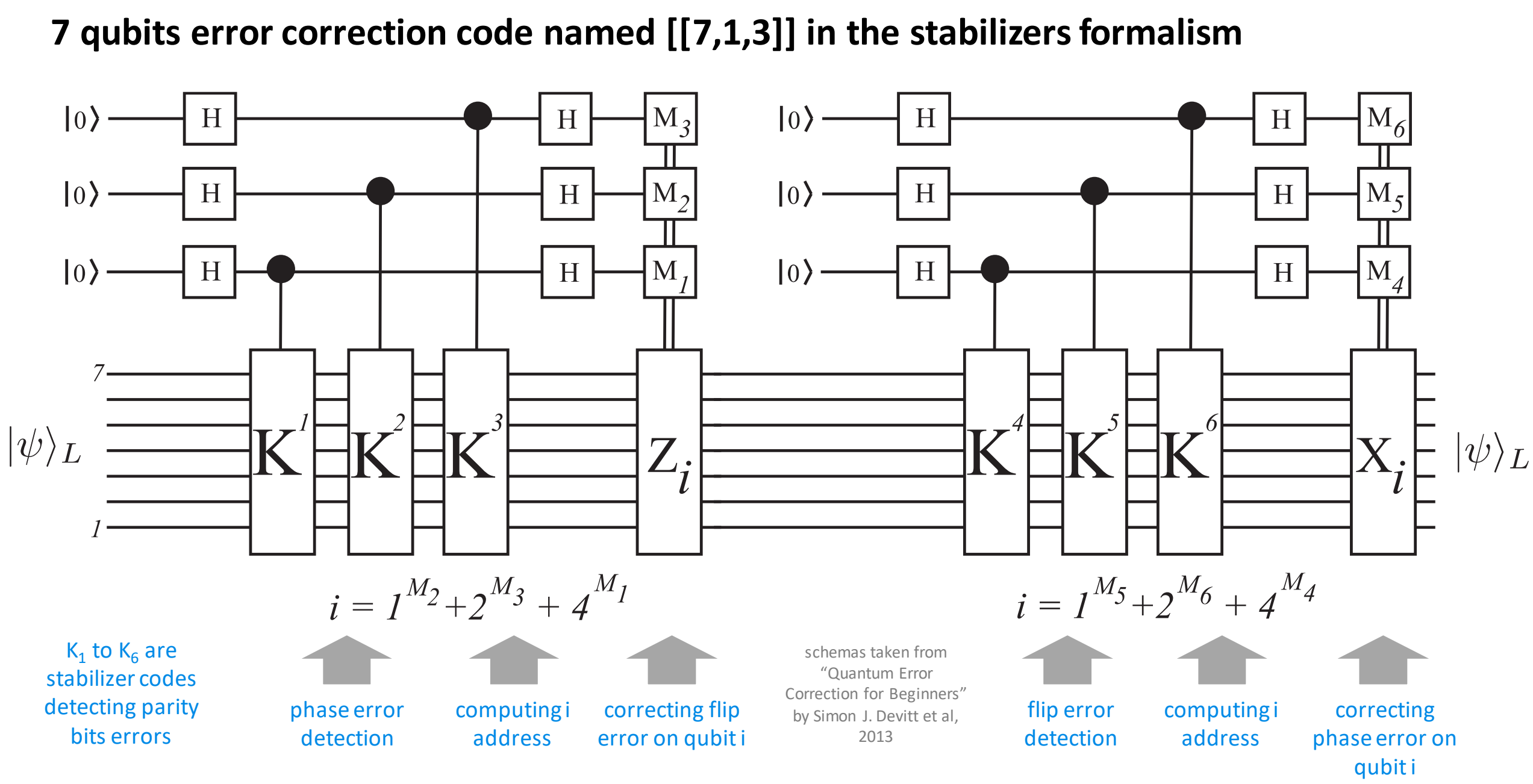

*Figure 236: 7 qubits correction code with a code distance 3. Source: [Quantum error corrections for beginners](#) by Simon J. Devitt et al, 2013.*

It seems that the qubits correction can be applied in two manners: the one presented so far with a measurement of ancilla parity qubits generating classical bits allowing to determine on which qubits to apply a quantum error correction gate, and other methods allowing this without the measurement and to apply the correction directly with quantum gates[433]. The first solution seems to be more commonly used. So why is the second solution less used or even recommended?

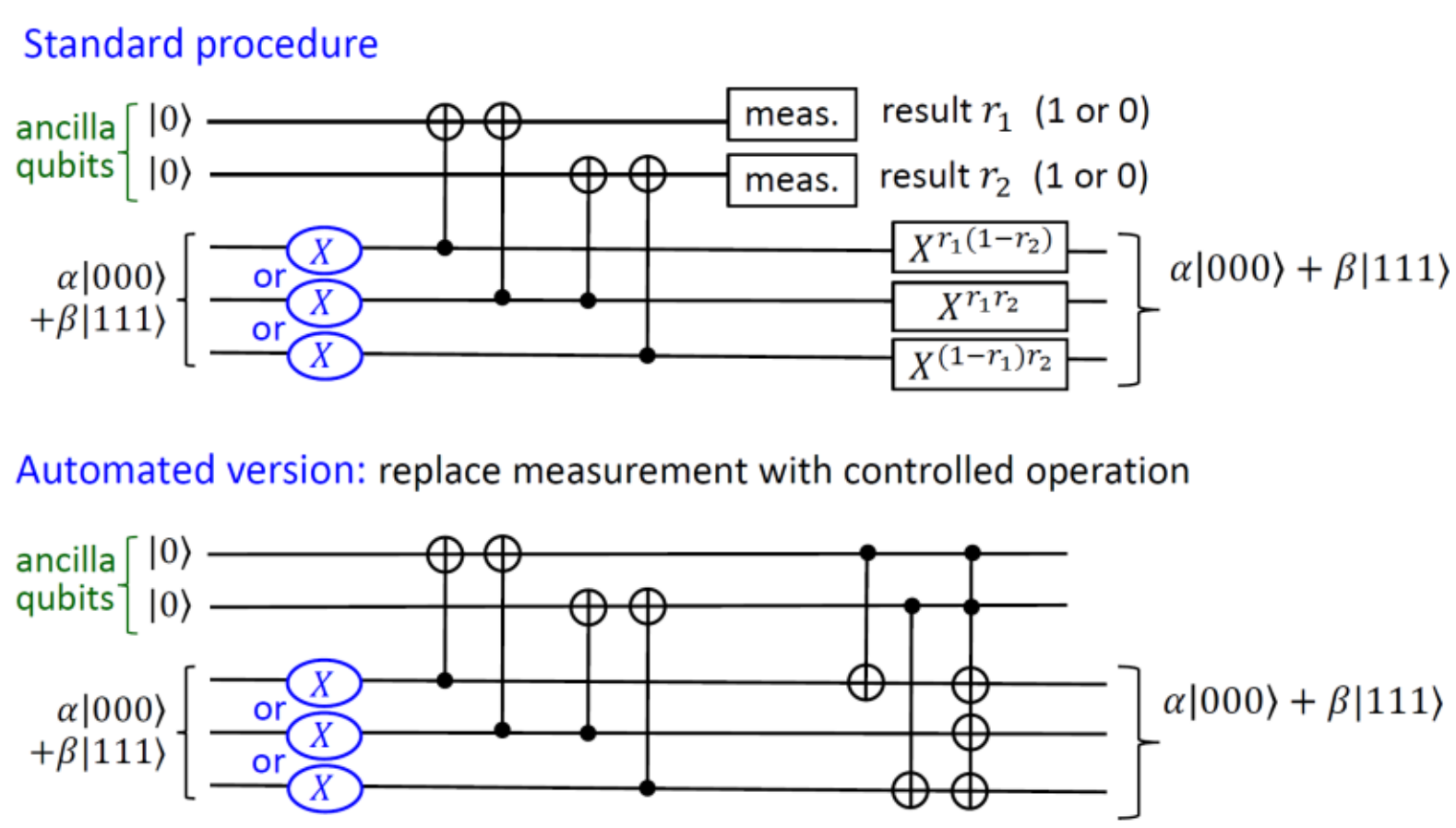

*Figure 237: error correction replacing measurement with a controlled operation. Source: [Quantum error correction (QEC)](#) by Alexander Korotkov, 2017 (39 slides).*

The autonomous method also branded **aQEC** (autonomous quantum error correction) would be more energy and time saving (comparison in Figure 237[434]). It is also a way to possibly run the error correction autonomously within a quantum processor, without going through the classical part, should qubits control be performed very close to the qubits. But in that case, the ancillas can't be reused. Some energy dissipation must be handled, using a technology called reservoir engineering, which is actually implemented in cat-qubits[435]. Otherwise, whatever, the ancilla qubits used in QEC must be reset to $|0\rangle$ and this reset is a dissipative process. The thermal bath is just elsewhere!

---

433 Like the "No ancilla error detection" code (NAED), an error detection scheme that does not employ ancilla qubits or mid-circuit measurements and encodes qubits in pairs of qubits with $|0\rangle_L = |01\rangle$ and $|1\rangle_L = |10\rangle$. See Quantum Error Detection Without Using Ancilla Qubits by Nicolas J. Guerrero and David E. Weeks, US Air Force Institute of Technology, April 2022 (8 pages).

434 Seen in Quantum error correction (QEC) by Alexander Korotkov, 2017 (39 slides). List of all courses on quantum computing.

435 See Protecting a Bosonic Qubit with Autonomous Quantum Error Correction by Jeffrey M. Gertler et al, University of Massachusetts-Amherst and Northwestern University, October 2020 (23 pages). This study investigates autonomous QEC on bosonic codes qubits using reservoir engineering. See also Autonomous quantum error correction and quantum computation by Jose Lebreuilly et al, Yale University, Amazon and University of Chicago, March 2021 (18 pages) and Autonomous quantum error correction with superconducting qubits by Joachim Cohen, ENS Paris, 2017 (164 pages).

## Logical Qubits

With quantum computers available online such as those from IBM having up to a few dozen qubits, it is the role of software to implement dynamic error correction codes and more precisely, compilers that will transform the developer's code into executable machine code at the physical level of the qubits and integrating QEC code. Given that we have at this point just enough qubits to test small QEC like Steane's 7-qubits codes or small sized surface codes.

Conceptually, a logical qubit sits between a physical qubit (with a small lifetime and prone to significant error rates) and a mathematically perfect qubit (with infinite computing time and zero error). It lasts longer than a physical qubit and should have an error rate in the range $10^{-8}$ to $10^{-15}$ that is compatible with the constraints of your algorithm. This error rate is more or less the inverse of the number of quantum gates in your circuit.

At one point in time, logical qubits will maybe be implemented entirely in the hardware architecture, exposing logical qubits to the classical computer driving the quantum accelerator. This will simplify the connection between the classical control computer and the quantum processor.

A QEC (Quantum Error Correction) could be performed at the hardware level by creating qubit assemblies that generate ready-to-use physical logical qubits. Here is an old example with seven superconducting physical qubits to create one simple logical qubit.

The number of physical qubits to be assembled to create a logical qubit depends on the error rate of the qubits.

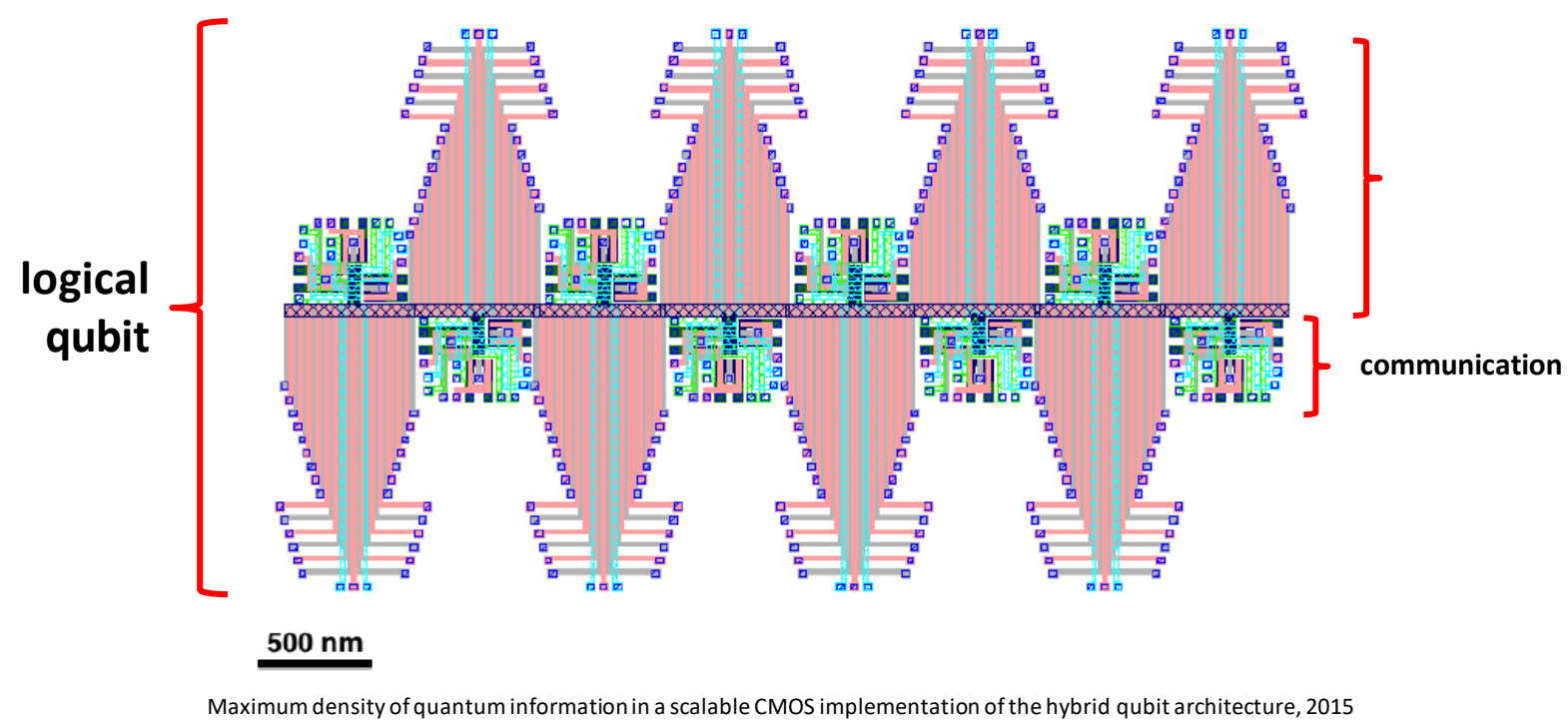

*Figure 238: a concept of logical qubit implemented at the physical level. Source: Maximum density of quantum information in a scalable CMOS implementation of the hybrid qubit architecture, 2015 (17 pages).*

The higher the qubit error rate, the more qubits must be assembled. This number can reach several thousand qubit physical qubits[436]. But QEC seems to be bound to be implemented mainly with software and on generic QPU architectures. The number of physical qubits in a logical qubit depend on many factors such as the quantum error correction code used, the underlying physical qubits fidelities, their connectivity and their target error rate.

We're still a long way from that! Current estimates are around 1,000 physical qubits to create a logical qubit. This corresponds to the plans published by **IBM**, **Google** and **PsiQuantum** with 100 logical qubits created out of one million physical qubits. On the physical architecture side, **topological qubits** are an analog version of surface codes that should allow to reduce this ratio of logical/physical qubits, just like cat-qubits, which are forecasted to require fewer than 100 physical qubits to create one logical qubit.

Trapped ions can use **lattice surgery** to connect and entangle these topologically corrected physical qubits[437].

[436] See What determines the ultimate precision of a quantum computer? by Xavier Waintal, 2019 (6 pages) which describes the limits of error correction codes. Other useful contents on error correction include: Error mitigation in quantum simulation, Xiao Yuan, IBM Research, 2017 (42 minutes), Code Used To Reduce Quantum Error In Logic Gates For First Time, 2019, Scientists find a way to enhance the performance of quantum computers by the University of Southern California, 2018 and Cramming More Power Into a Quantum Device by Jay Gambetta and Sarah Sheldon, March 2019 about the error level of the IBM Q System One announced in January 2019.

[437] See Error protected quantum bits entangled, University of Innsbruck, January 2021 referring to Entangling logical qubits with lattice surgery by Alexander Erhard et al, Nature, 2020 (15 pages).

**IonQ** is planning to create logical qubits corrected with a Bacon-Shor QEC (a variation of Shor's code with 13 qubits[438]) thanks to their much better fidelities[439].

For qubits that can be physically well connected with their immediate neighbors, the most often considered error correction is the **surface code**, created between 1998 and 2001.

As shown in the diagram in Figure 241, it uses matrices of processing qubits (in white) connected to measuring qubits (in black) via **Pauli X (**amplitude flip) and **Pauli Z** (phase flip) gates operating on these data qubits as shown in yellow and green. This gives two ancilla qubits for two physical qubits organized to detect and correct flip and phase errors over 4 replicated qubits. This constitutes a stabilizer code of type [[5, 1, 2]] using four blocks with four cycles.

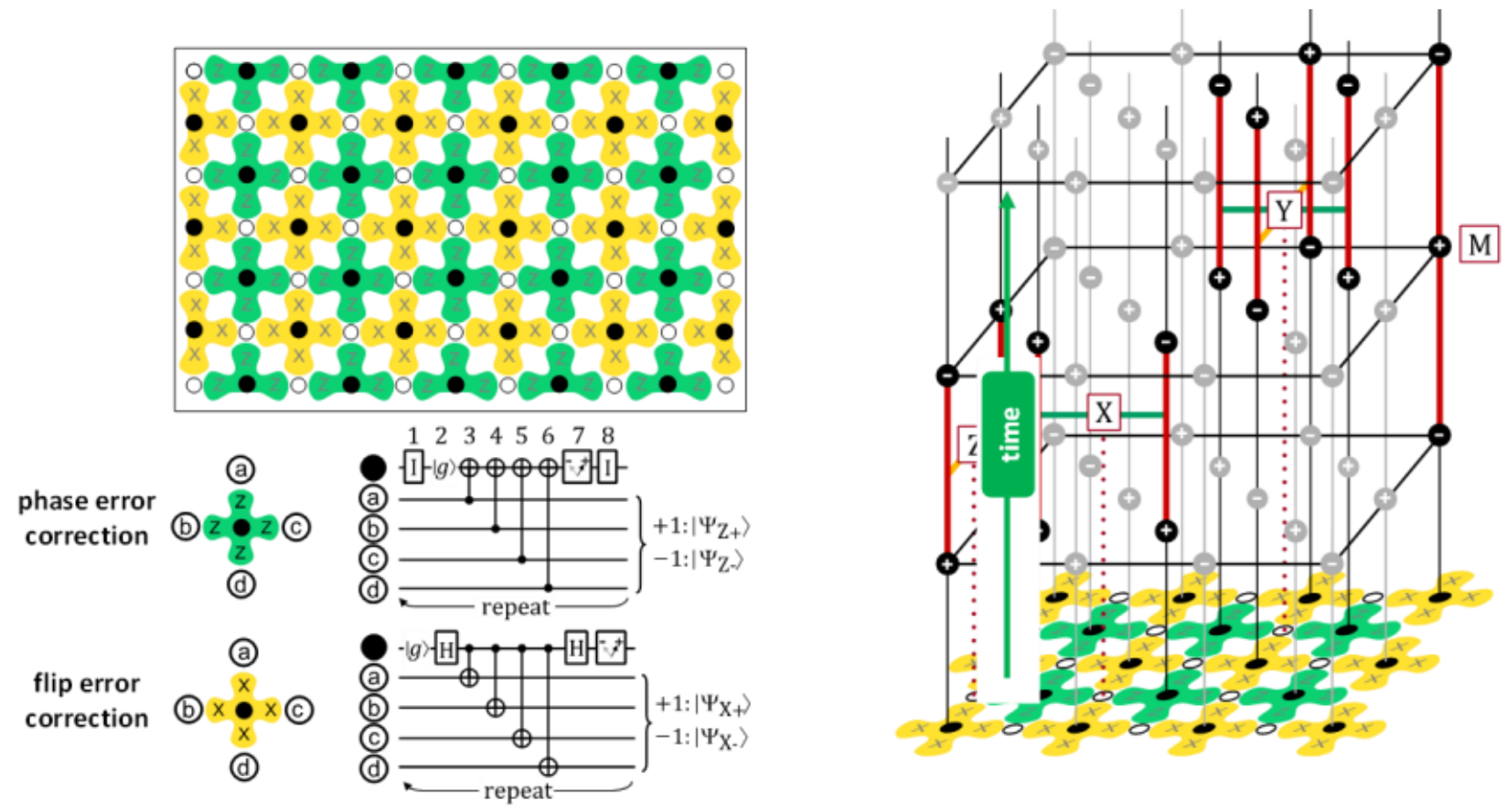

*Figure 239: surface code physical layout and process. Source: Surface codes: Towards practical large-scale quantum computation by Austin G. Fowler, Matteo Marianton, John M. Martinis and Andrew Cleland, 2012 (54 pages).*

A surface code with distance d requires $d^2$ replicated qubits and $d^2 - 1$ measurement qubits, so a total of $2d^2 - 1$ physical qubits to correct a single qubit. $d$ being usual odd, you then have an even number of measurement qubits divided in two equal parts with flip (Z) measurement) and phase (X) measurement qubits as shown below with two example of distance 3 and 5 surface codes. A surface code logical qubit error rate is $P_L \approx 0.03(p/p_{th})^{d_e}$ with $p$ being the physical qubit error rate, $p_{th}$ being the threshold physical error rate below which logical errors falls with $d$, and $d_e$ being linked to the surface code distance $d$ with $d_e = d/2$ when d is even, and $= (d+1)/2$ when $d$ is odd. Evaluating how many physical qubits are required to create a logical qubit of a given fidelity is quite complicated with many logarithms using qubit fidelities ratios with fidelities threshold and surface code distance. How are logical qubits delimited and logical gates implemented is another (complicated) story.

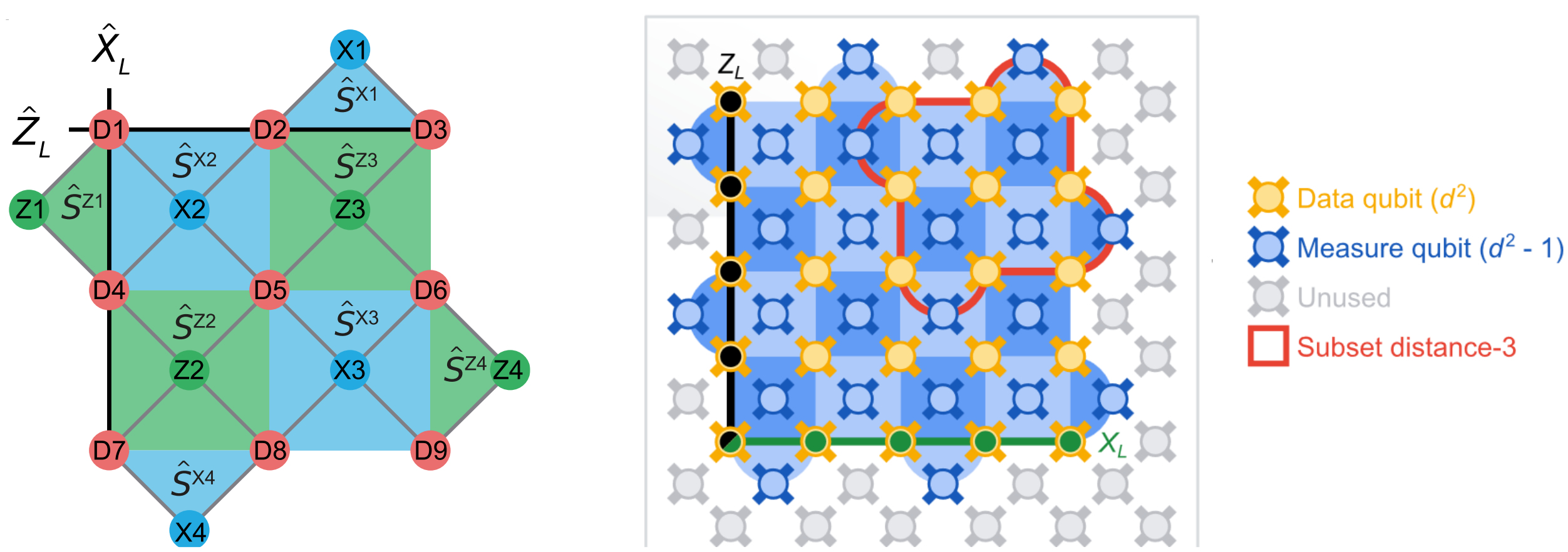

*Figure 240: two examples of surface codes, with a distance 3 using 17 qubits (left) and 5 using 49 qubits (right). On the left, the replicated qubits are in red and the measurement qubits are in green (Z, for flip error correction) and blue (X, for phase error correction). Sources: Realizing Repeated Quantum Error Correction in a Distance-Three Surface Code by Sebastian Krinner, Alexandre Blais, Andreas Wallraff et al, December 2021 (28 pages) and Suppressing quantum errors by scaling a surface code logical qubit by Rajeev Acharya et al, Google AI, July 2022 (44 pages).*

---

[438] Bacon-Shor code is documented in Operator Quantum Error Correcting Subsystems for Self-Correcting Quantum Memories by Dave Bacon, 2006 (17 pages).

[439] And Fault-Tolerant Operation of a Quantum Error-Correction Code by Laird Egan, Christopher Monroe et al, 2020 (23 pages).

A quantum error correcting code has a certain **threshold** level that defines the higher bound of physical qubit error rates when a logical qubit will have an error rate inferior to that of the physical qubits. It depends on the code itself and on the qubit type.

Surface codes have a higher threshold in the 1% range and are thus tolerant to higher qubit error rates. But they require a larger number of physical qubits per logical qubits. Also, physical qubits must be connected to their immediate neighbors in a 2D structure or with honeycomb variations [440].

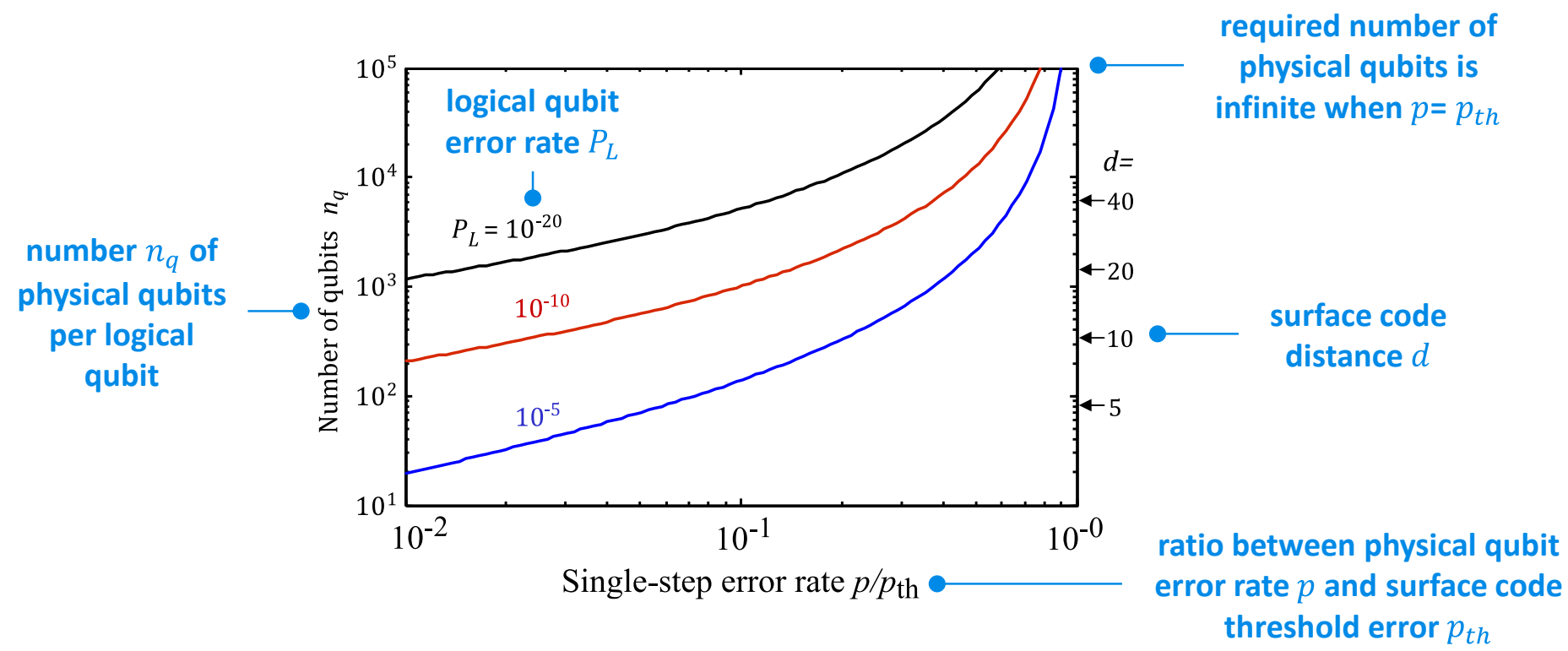

*Figure 241: relationship between physical and logical qubit error rate with the number of physical qubits in a logical qubit and the surface code distance. Source: the excellent review paper An introduction to the surface code by Andrew Cleland, University of Chicago, 2022 (68 pages).*

How about real implementations of logical qubits? They are now plentiful but are not yet creating logical qubits with higher fidelities than their underlying physical qubits.

- A team from **Maryland** led by Christopher Monroe implemented in January 2021 a logical qubit using a Bacon-Shor 13 code with a chain of 15 trapped ytterbium ions that was correcting single qubit errors[441]. They then used two such logical qubits in a configuration of 32 qubits to implement fault-tolerant 2-qubit gates.
- **Google** announced in July 2021 the creation of their first logical qubits with 5 and 21 physical qubits, showing a x100 improvement in the error rate between the 5 and 21 version[442].
- **Quantinuum** created a single Steane color-code also using 10 trapped ion qubits. They defined the pseudothreshold as a crossover point where the logical qubit error rates is below the physical level error rates. They created a logical qubit with better fidelities than their underlying physical qubits in August 2022[443].
- A **China** research team implemented a distance 3 surface code using 17 physical qubits ($=3^2+(3^2-1)$) on the 66 qubits Zuchongzhi 2.1 superconducting qubits QPU. It implements repeated error corrections and post-processing error corrections[444].

[440] Surface codes are well documented in Surface codes: Towards practical large-scale quantum computation, by Austin G. Fowler, Matteo Mariantoni, John M. Martinis and Andrew Cleland, 2012 (54 pages) but their source of inspiration is older and comes from Quantum codes on a lattice with boundary by Sergey Bravyi and Alexei Kitaev, 1998 (6 pages). In practice, the structure of surface codes is quite complex and involves activated and deactivated substructures in the qubit matrix.

[441] See Fault-tolerant control of an error-corrected qubit by Laird Egan, Christopher Monroe et al, January 2021 (9 pages). The 15 used qubits contain the 9 for the Bacon-Shor correction code, 4 for the stabilizers ancilla and two unused ions at the edges of the 1D set of ions.

[442] For Google's logical qubit, see Exponential suppression of bit or phase errors with cyclic error correction by Zijun Chen et al, February 2021 in arXiv and in Nature in July 2021 (6 pages) and supplemental materials (30 pages).

[443] See Realization of Real-Time Fault-Tolerant Quantum Error Correction by C. Ryan-Anderson et al, PRX, December 2021 (29 pages), a follow-up from the previous paper. It uses a 10 qubit trapped ion quantum computer to encode a single logical qubit using the Steane [[7, 1, 3]] color code. They later implemented a [[7, 1, 3]] color code with 20 qubits with a transversal CNOT gate on two logical qubits in Implementing Fault-tolerant Entangling Gates on the Five-qubit Code and the Color Code by C. Ryan-Anderson et al, August 2022 (17 pages) and could obtain a logical qubit with 99.94% fidelity, compared to 99.68% for the underlying physical qubits.

[444] See Realization of an Error-Correcting Surface Code with Superconducting Qubits by Youwei Zhao et al, PRL, December 2021 (10 pages). "*Future work will concentrate on realizing larger-scale surface codes, to achieve the important goal of suppressing the logical error rate as the code distance increases. This necessitates further improvements to the quantum computing system's performance, such as the number and quality of qubits, the fidelity of quantum gate operations, and rapid feedback of digital electronics*".

- A team led by **Andreas Wallraff** from ETH Zurich did a similar experiment a little later[445].
- Another team in **Austria** and **Germany** developed a proof of concept of two logical qubits made with trapped ions and using a T gate and magic state distillation[446].
- A joint **QuTech-Fujitsu-Element Six** team demonstrated in 2022 a fault-tolerant operation of a NV centers based QPU with logical qubits made of 5 physical spin qubits and two additional measurement qubits in a 29-qubit QPU running at 10K[447].
- In 2022, **Google AI** created a distance 5 surface code logical qubit with 49 qubits (=$5^2$+($5^2$-1)) that improves logical qubit error rates as physical qubits per logical qubits grows. But they have not yet reached the QEC efficiency threshold where logical qubit errors would be lower than physical qubit errors[448]. Google researchers indicate that their logical qubits would be better than their underlying physical qubit starting with a distance 7 surface code, requiring about 100 physical qubits, which they are not far from obtaining.

What are the figures of merit of a quantum error correction code and a logical qubit architecture? A key one is the logical qubit fidelities. You can't claim the creation of a logical qubit without adding its target fidelity. The end goal is to reach between $10^{-9}$ and $10^{-15}$ error rates. These rates differ according to the target algorithms. $10^{-9}$ may be enough for condensed matter and Hubbard models simulations, $10^{-15}$ is required for Shor's integer factoring for 2048-bit RSA keys while even better error rates are required for complex chemical simulations. The inverse of these error rates corresponds approximatively to the number of T gates to execute in these algorithms, which are the costlier to correct in surface codes. Then comes the overhead with the number of physical qubits per logical qubit but also code time cost, meaning, how will it slow down quantum computing. A logical qubit must also be able to implement non-Clifford quantum gates in a fault-tolerant manner, an aspect we'll describe in the next part.

### Fault-tolerant quantum computing

**FTQC** (fault-tolerance quantum computing) was defined by Peter Shor in 1997[449] and is based on a few general principles related to implementing a practically useful QEC scheme with logical qubits: error-tolerant state preparation, error-tolerant quantum gates, error-tolerant measurement and error-tolerant error correction. Error correction codes can themselves introduce errors since they use quantum gates and state measurements which themselves generate errors. Moreover, error correction codes do not correct all possible errors. They just increase the apparent fidelity rate of the corrected qubits.

Also, QEC codes used repeatedly during long calculations must not introduce more errors than are corrected and should not spread errors in an uncontrollable way to various qubits in the computing register. As Peter Shor recounts: "*To be able to build a quantum computer, it's not enough to be able to correct errors with noiseless gates; you need to be able to correct errors using noisy gates. This means you have to correct the errors faster than you introduce new ones*"[450]. This is where you understand why qubit gates, qubit readout time and the classical processing of readout data have all to be as fast as possible.

---

[445] See Realizing Repeated Quantum Error Correction in a Distance-Three Surface Code by Sebastian Krinner, Alexandre Blais, Andreas Wallraff et al, Nature, December 2021-May 2022 (28 pages).

[446] See Demonstration of fault-tolerant universal quantum gate operations by Lukas Postler, Rainer Blatt et al, Nature, December 2021 (14 pages).

[447] See QuTech and Fujitsu realise the fault-tolerant operation of a qubit by Qutech, May 2022.

[448] See Suppressing quantum errors by scaling a surface code logical qubit by Rajeev Acharya et al, Google AI, July 2022 (44 pages).

[449] See Fault-tolerant quantum computation by Peter Shor, March 1997 (11 pages).

[450] In The Early Days of Quantum Computation by Peter Shor, August 2022 (10 pages).

FTQC theoretically allows the execution of algorithms of arbitrary length, whereas without it, we are limited to a few series of gates. The challenge is to ensure that the calculation and QEC prevents errors from cascading. We must avoid linking one qubit with too many qubits with multi qubit gates in QECs. For this respect, a 7-qubits Steane code is appropriate.

And let's not forget that a CNOT gate propagates flip errors from the control qubit to the target qubit and phase errors from the target to the control. From an operational standpoint, FTQCs creation involves minimizing the number of ancilla qubits and optimizing the choice of QECs according to the type of errors generated by each type of qubit and quantum gates[451].

**Transversal gates** are implemented with FTQC to avoid propagating errors beyond the corrected qubits. It is an arrangement of links between logical qubits linked together by two-qubit gates.

The diagram on the right illustrates these links between two logic qubits using a 7 qubit Steane code via CNOT gates. Each of the physical qubits of the logical qubits is connected one by one between the two logical qubits. This is still very theoretical, besides trapped ions, no qubit topology enables this kind of connectivity.

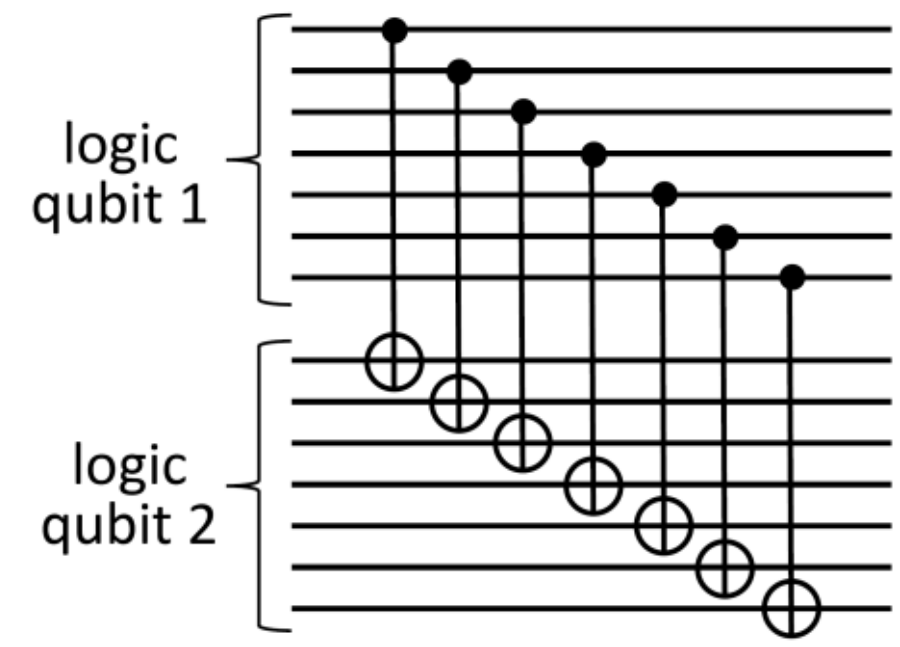

*Figure 242: how transversality connects two logical qubits.*

However, transversal gates can only be implemented within the Clifford group. According to the **Eastin-Knill** no-go theorem, no QEC code can transversally implement a universal gate set. That's why we usually need a costly QEC named **magic state distillation** to implement FTQC with T and Toffoli gates which lie outside the Clifford group. It has a huge cost of two orders of magnitude for physical per logical gates, explaining why it's often estimated said that logical qubits require overs 10K physical qubits (on top of the effect of code concatenations or surface code)[452].

One of the problems is that error correction generates an overhead that grows faster than the exponential gain of the quantum computer ($2^{4n}$ vs $2^n$ according to Quantum Benchmark).

We can get some comfort from the **threshold theorem** demonstrated by Dorit Aharonov and Michael Ben-Or in 1999 according to which it is possible to perform error correction up to an arbitrary desired apparent error rate if the error rate of the single-qubit gates is below a given threshold which is dependent on the error correction code used and the characteristics of the qubits[453].

**Concatenation of codes $C_1$ (size $n_1$) and C2 (size $n_2$)**

We construct a code of size $n_1n_2$, where each qubit of $C_2$ is replaced by a block of $n_1$ qubits encoded in $C_1$.

**Higher order QEC by concatenation**

| Level of concatenation | Error probability |
|---|---|
| Physical qubits | $\epsilon_0 = p$ |
| 1st encoded level | $\epsilon_1 = cp^2 = c^{-1}(cp)^2$ (*) |
| 2nd encoded level | $\epsilon_2 = c(cp^2)^2 = c^{-1}(cp)^{2^2}$ |
| ⋮ | ⋮ |
| r'th encoded level | $\epsilon_r = c(\epsilon_{r-1})^2 = c^{-1}(cp)^{2^r}$ |

(*) For the Steane code $c \approx 10^4$

*Figure 243: how concatenated codes are reducing the error rate. Source: Introduction to quantum computing by Anthony Leverrier and Mazyar Mirrahimi, March 2020 (69 slides).*

[451] See A Comparative Code Study for Quantum Fault Tolerance by David DiVincenzo, Barbara Terhal and Andrew Cross, 2009 (34 pages).

[452] See Roads towards fault-tolerant universal quantum computation by Earl T. Campbell et al, 2018 (9 pages).

[453] See Fault-Tolerant Quantum Computation With Constant Error Rate by Dorit Aharonov and Michael Ben-Or, 1999 (63 pages).

This rate would be between 0.1% and 1% but is subject to change. The consequence of this theorem is to allow the application of error correction codes recursively until reaching the desirable error rate to execute a given algorithm. This is however based on the assumption that qubit fidelities are stable as their number is growing, a feat that is not yet achievable!

It also doesn't take into account various sources of errors like isotropic errors affecting simultaneously several qubits[454]. On the other hand, a variation of the threshold theorem was recently demonstrated that takes into account a stable percentage of defects in planar arrays of qubits and includes a QEC protocol for large arrays of defective qubits[455].

The standard specification in vendor roadmaps for FTQC QPUs is one million physical qubits with 99.9% two-qubit gate fidelity running a surface code QEC. It might enable 100 to 1000 logical qubits, taking into account the significant overhead of T-gates or Toffoli gates synthesis and correction.

How about the number of logical qubits of a FTQC QPU? It should provide a space-related quantum advantage vs classical computing so we need at least 50 to 55 data qubits. Most algorithms requiring an equivalent of about 50 additional qubits (ancilla, transit, …), we end up with needing about 100 logical qubits and a number of physical qubits that depends on the architecture, ranging from 30 to 10.000 physical qubits per logical qubits. The sheer number of qubits required to build a FTQC awarded it another nickname: Fairy-Tale Quantum Computing!

**QEC concatenation** is exploiting this recursivity of error correction codes. A QEC generates logical qubits which can then be used as virtual physical qubits for a new QEC, and so on. With each recursion, the apparent error rate decreases. We stop concatenating QEC codes when we reach an error rate compatible with the expected usage of the qubits. Concatenation can be optimized by using different types of QEC at each level of recursivity[456]. This theorem was demonstrated only for a 7-qubit Steane error correction code and for error rates that are not growing with the number of physical qubits. This is unfortunately not what is currently observed with the majority of qubit types! Surface codes and their various derivatives are not concatenated but rather expanded in 2D with a growing number of qubits. But their relative noise and number of qubit scaling are different.

With concatenated codes, the noise is reduced by the exponential law $\epsilon^{2^k}$ with a number of qubits in $X^k$, $\epsilon$ being the error rate, k the number of concatenations and X the number of qubits within a single concatenation, which can reach about 91 depending on the implementation and on the way ancilla qubits are managed and recycled, using Steane's method (not to be confused with Steane's code)[457]. But concatenated codes threshold is quite low, in the range of $10^{-6}$, which is currently inaccessible for all breeds of qubits.

[454] See Quantum codes do not increase fidelity against isotropic errors by J. Lacalle et al, January 2022 (18 pages).

[455] See Quantum computing is scalable on a planar array of qubits with fabrication defects by Armands Strikis, Simon C. Benjamin and Benjamin J. Brown, November 2021 (16 pages).

[456] See Dynamic Concatenation of Quantum Error Correction in Integrated Quantum Computing Architecture by Ilkwon Sohn et al, 2019 (7 pages).

[457] 91 is based on using a Steane 7-qubit [[7; 1; 3]] code, including the ancilla factory and 4x7=28 qubits ancilla factory times 3 because the preparation, verification and measurement of ancillas is three times longer than the data qubit operations (9 versus 3 time-steps). Hence, while ancillas are in used for a given gate, ancillas must be being prepared for the next two gates. Thus each level of error correction replaces one qubit by 91 qubits (7 data qubits and 3x28 ancilla qubits). Source: Optimizing resource efficiencies for scalable full-stack quantum computers by Marco Fellous-Asiani, Jing Hao Chai, Yvain Thonnart, Hui Khoon Ng, Robert S. Whitney and Alexia Auffèves, arXiv, September 2022 (39 pages). Flag qubits could reduce this significant overhead and reduce X. But this doesn't take into account T gates magic state distillation that adds a minimum of 15 qubits! See Fault-tolerant quantum error correction on near-term quantum processors using flag and bridge qubits by Lao Lingling et al, 2020 (12 pages) and Fault-tolerant quantum error correction for Steane's seven-qubit color code with few or no extra qubits by Ben W. Reichardt, April 2018 (11 pages). See Overhead analysis of universal concatenated quantum codes by Christopher Chamberland, Raymond Laflamme et al, 2017 (25 pages) which describes a fault-tolerant QEC of 105 qubits.

With surface codes, the noise is reduced according to $\epsilon^{\frac{d}{2}}$ and the number of required physical qubits grows by $d^2$, d being the distance of the surface code, more or less corresponding to the edge of the surface code squares as shown in Figure 240 for distances 3 and 5.

All in all, concatenated error correction codes have a better impact on noise, but at the expense of a large number of physical qubits while surface codes scale slower in error reduction and physical qubits requirements. It seems that surface codes are more appropriate for more noisy physical qubits while concatenated codes will be better, for less noisy qubits.

**Qubits lifetime extension**. A nagging question may arise: if we need to accumulate error correction codes, don't we risk running into the wall of qubit decoherence, particularly with superconducting qubits? Well, no. As said before, error correction codes have the direct effect of artificially extending the coherence time of the qubit registers by several orders of magnitude[458]. Each correction is equivalent to a reset of the qubits decoherence times $T_1$ (flip) and $T_2$ (phase). This explains how Google could publish an optimized version of the Shor integer factoring algorithm with 20 million qubits and requiring 8 hours of run-time, which is many orders of magnitude longer than their qubits coherence time that sits way under a tiny 100 µs.

**Instruction bandwidth bottleneck** is yet another engineering challenge for FTQC and error correction. Thousands of physical qubits must be driven by software-based quantum error correction. It creates a digital workload from the classical control computer down to the physical qubits and their many ancilla qubits, in a range exceeding several tens of TB/s just for factoring a 1024 bits integer with Shor's algorithm! Specific architectures can be designed to handle QEC as close as possible to the physical qubits, ideally in cryo-electronics components and with some microcode sitting at the lowest possible stage in the cryostat (for solid-state qubits), starting at 4K[459].

### Approximate QEC

One QEC group named Approximate QEC or Quasi-Exact fault-tolerant Quantum (QEQ) computation sits in-between NISQ and FTQC[460]. It is an intermediate solution implementing some error correction, but not to a point of creating perfect logical qubits. It still uses some variations of stabilizers and surface codes. One of these methods named NISQ+ combines aQEC and SFQ driving circuits[461]. It can help boost the "simple" quantum volume of NISQ QPUs. The simple quantum volume is computed with multiplying the number of useful qubits and a number of doable quantum gates under a certain error threshold.

The related paper raises an interesting point: slow QEC decoders make applications take exponential time to complete, which is kind of problematic! It's explained by the ratio between QEC data generation and QEC data processing is around 2 for syndrome data processing ratio using classical controls. With superconducting SFQ circuits, the ratio is of 0.125 thanks to a very low latency. The proposed SFQ circuit uses a circuit map similar to the qubit circuits topology. It implements an "Approximate SFQ decoder" stabilizer-based algorithm using a union-find algorithm, resets (stopping signal propagation once pairs are found), boundaries (match signals to boundaries) and tie-breaking (chooses single paths among an equal set).

---

[458] See Extending the lifetime of a quantum bit with error correction in superconducting circuits by Nissim Ofek, Zaki Leghtas, Mazyar Mirrahimi, Michel Devoret et al, 2016 (5 pages) which shows that thanks to a cat-code-based QEC, the lifetime of superconducting qubits can be extended by a factor of 20!

[459] See the QuEST architecture proposal in Taming the Instruction Bandwidth of Quantum Computers via Hardware Managed Error Correction by Swamit Tannu et al, GeorgiaTech, Stanford and Microsoft, 2017 (13 pages slides).

[460] See Theory of quasi-exact fault-tolerant quantum computing and valence-bond-solid codes by Dong-Sheng Wang, Raymond Laflamme et al, May 2021 (22 pages).

[461] See NISQ+: Boosting quantum computing power by approximating quantum error correction by Adam Holmes et al, Intel, University of Chicago and USC, April 2020 (13 pages) and explained in this video (21 mn).

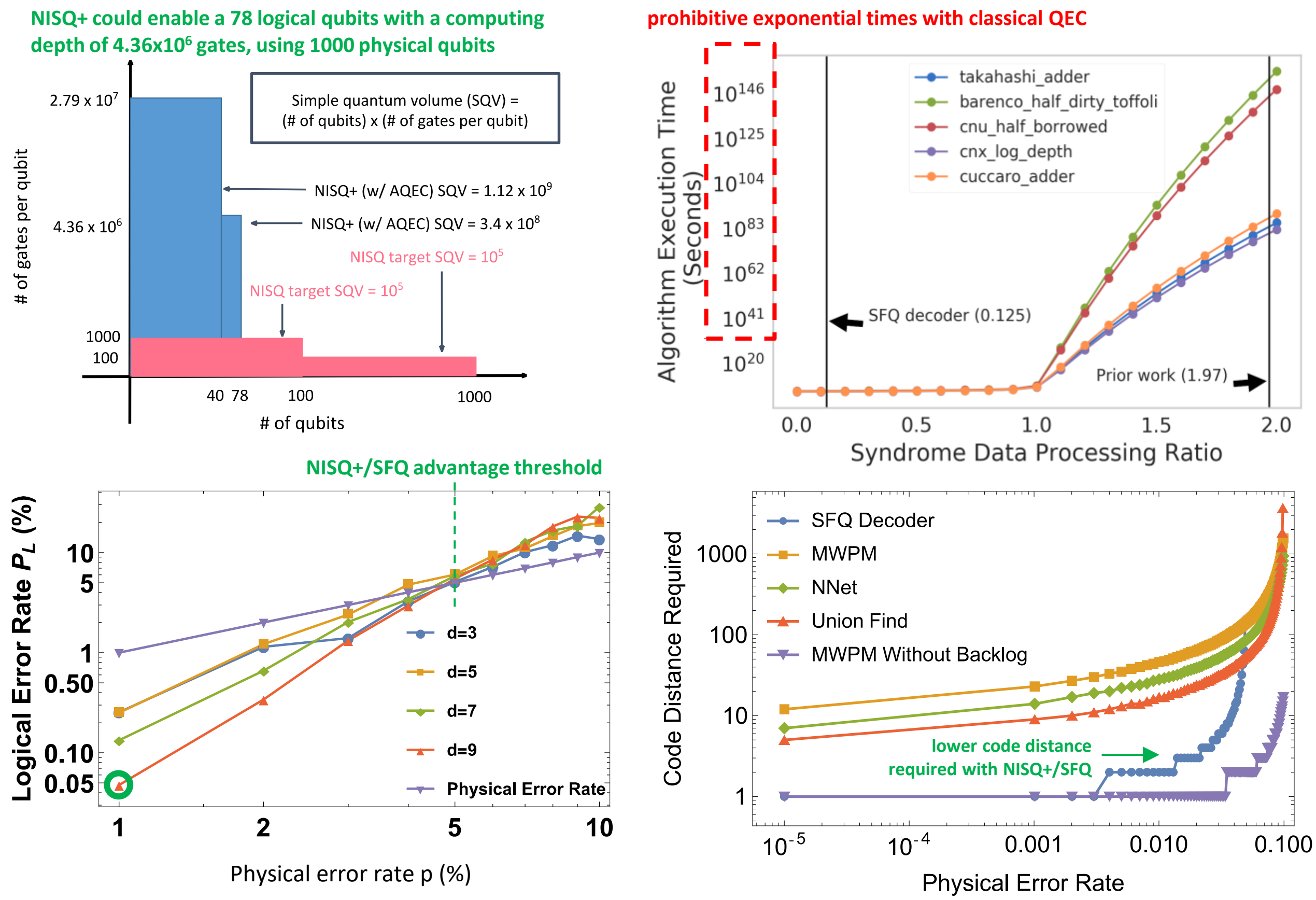

*Figure 244: the NISQ+ architecture and benefits. Source: [NISQ+: Boosting quantum computing power by approximating quantum error correction](#) by Adam Holmes et al, Intel, University of Chicago and USC, April 2020 (13 pages).*

The accuracy threshold of this SFQ circuit is at 5% of physical error rate and significantly interesting at 1%, yielding then a logical qubit error rate of 0,05% with a code distance of d=9. It is a much lower required code distance compared to other correction techniques like those using neural networks.

The SFQ circuit power consumption is 13 µW for a full circuit with a logical depth of 6, has a real-estate of 1.3 $mm^2$ and a latency of 20 ns for QEC. It seems made to run at 4K. It could enable the creation of a 78 logical qubits system using 1000 physical qubits and a computing depth of $4{,}36 \times 10^6$ gates.

In another work, a team led by Microsoft created a concept architecture to implement a FTQC with a scalable decoder running the QEC, but without details on the required hardware (room temperature or cryoelectronics, CMOS or SFQ)[462].

### Quantum error mitigation

Quantum error mitigation (QEM) is about reducing quantum algorithms errors with combining classical post-processing techniques with some potential circuits modifications on top of running the algorithm several times and averaging its results (*aka* the "*expectation values of an observable*", i.e., the combination of 0s and 1s). QEM has a much lower overhead in qubits and running time vs QEC but its scalability is still questioned. It is a NISQ-era solution aiming at creating a quantum computing advantage before FTQC shows up in the longer term. QEM reduces the influence of quantum errors using multiple runs and subsequent measurements coupled to some classical processing as opposed to QEC-based active qubits measurement and fast feedback-based corrections impacting the results of individual runs.

---

[462] See [A Scalable Decoder Micro-architecture for Fault-Tolerant Quantum Computing](#) by Poulami Das, Krysta Svore, Nicolas Delfosse et al, Microsoft, GeorgiaTech and Caltech, January 2020 (19 pages).

QEM proposals started to pop-up around 2016[463]. Most of them consist in learning the effects of noise on qubit evolutions and create predictive (if not linear) noise models that can be applied to tune the results of quantum computations. It is adapted to rather shallow circuits[464] and we're not really sure yet it brings a quantum advantage on useful problems. Most QEM methods do not increase the required qubits count for a given algorithm. You'll notice that many contributions in the QEM space come from IBM Research.

Here are some identified QEM techniques:

**Zero noise extrapolation** (ZNE) builds error models based on solving linear equations. It supposes the noise is stable. It cancels noise perturbations by an application of Richardson's deferred approach to the limit and works on short-depth (or shallow) circuits[465]. A similar protocol was designed for (hybrid) variational quantum simulators used to simulate the dynamics of quantum systems[466].

**Probabilistic error cancellation** is about detecting circuit bias with finding noise quantum channels, represented as density matrices for quantum gates, using quasi-probability decomposition. There's a sampling overhead in the process. It then inverts a well-characterized noise channel to produce noise-free estimates of the algorithm observables (the 0s and 1s they're supposed to generate)[467]. It's also called Bayesian error mitigation and Bayesian read-out error mitigation (BREM).

**Learning Based Methods** QEM are based on machine learning techniques using training data to learn the effect of quantum noise in various situations. It's proposed by companies like QuantrolOx (UK), by the University of Erlangen in Germany[468] and by Quantum Machines (Israel). One of these is Clifford circuit learning or Clifford Data Regression is a variation of the previous technique that learns the effect of noise from Clifford gates using data comparing quantum emulation on classical hardware and runs on quantum processors. It then uses rather simple linear regression techniques to correct errors in post-processing[469]. It can also be applicable to fault-tolerant T gates[470]. You have also QuantumNAS, a noise adaptative search method[471] as well as some specific deep reinforcement learning techniques to improve qubit control precision[472].

---

463 See the review papers Hybrid Quantum-Classical Algorithms and Quantum Error Mitigation by Suguru Endo, Zhenyu Cai, Simon C. Benjamin and Xiao Yuan, 2020 (39 pages), Quantum Error Mitigation by Zhenyu Cai, Ryan Babbush, Simon C. Benjamin et al, October 2022 (40 pages) and Testing platform-independent quantum error mitigation on noisy quantum computers by Vincent Russo, Andrea Mari, Nathan Shammah, Ryan LaRose and William J. Zeng, October 2022 (17 pages).

464 See Fundamental limits of quantum error mitigation by Ryuji Takagi et al, npj, September 2022 (11 pages).

465 See the beginning of Error mitigation for short-depth quantum circuits by Kristan Temme, Sergey Bravyi and Jay M. Gambetta, 2016 (15 pages) and Scalable error mitigation for noisy quantum circuits produces competitive expectation values by Youngseok Kim, Jay M. Gambetta, Kristan Temme et al, August 2021 (7 pages).

466 See Efficient Variational Quantum Simulator Incorporating Active Error Minimization by Ying Li and Simon C. Benjamin, PRX, 2017 (14 pages).

467 See the second part of Error mitigation for short-depth quantum circuits by Kristan Temme, Sergey Bravyi and Jay M. Gambetta, 2016 (15 pages), Probabilistic error cancellation with sparse Pauli-Lindblad models on noisy quantum processors by Ewout van den Berg, Zlatko K. Minev, Abhinav Kandala and Kristan Temme, January 2022 (30 pages), Probabilistic error cancellation with sparse Pauli-Lindblad models on noisy quantum processors by Ewout van den Berg et al, IBM, January 2022 (30 pages) and Unfolding Quantum Computer Readout Noise by Benjamin Nachman et al, October 2019-May 2020 (13 pages).

468 Seer Neural networks enable learning of error correction strategies for quantum computers, October 2018 and Reinforcement Learning with Neural Networks for Quantum Feedback, Thomas Fösel et al, 2018 (7 pages).

469 See Error mitigation with Clifford quantum-circuit data by Piotr Czarnik et al, May 2020 (16 pages) and Improving the efficiency of learning-based error mitigation by Piotr Czarnik, Michael McKerns, Andrew T. Sornborger and Lukasz Cincio, April 2022 (13 pages).

470 See Error mitigation for universal gates on encoded qubits by Christophe Piveteau, David Sutter, Sergey Bravyi, Jay M. Gambetta and Kristan Temme, IBM Research, March 2021 (11 pages).

471 See QuantumNAS: Noise-Adaptive Search for Robust Quantum Circuits by Hanrui Wang et al, January 2022 (19 pages).

472 See Deep Reinforcement Learning for Quantum State Preparation with Weak Nonlinear Measurements by Riccardo Porotti, Antoine Essig, Benjamin Huard and Florian Marquardt, June 2021 (15 pages).

**Error suppression by derangement** (ESD) which provides an exponential error suppression with increasing the qubit count by n≥2 but is still adapted to NISQ architecture and shallow circuits[473].

**Dynamic Decoupling** involves decoupling idle qubits from other qubits under certain conditions. The technique is proposed by IBM[474]. It seems that under certain circumstances, it can generate a good quantum speedup for oracle-based algorithms[475].

**Other methods** include symmetry constraints verification, distillation using randomized benchmarking[476], randomized compiling[477], applying gates simulating the reverse effect of errors[478], depolarizing noise[479], quantum verification and post-selection[480], virtual distillation with derangement operators[481], using matrix product operators[482], read-out noise mitigation and also mixing various QEM and QEC techniques[483]. Detailing and comparing these various methods is way above my quantum computing pay grade!

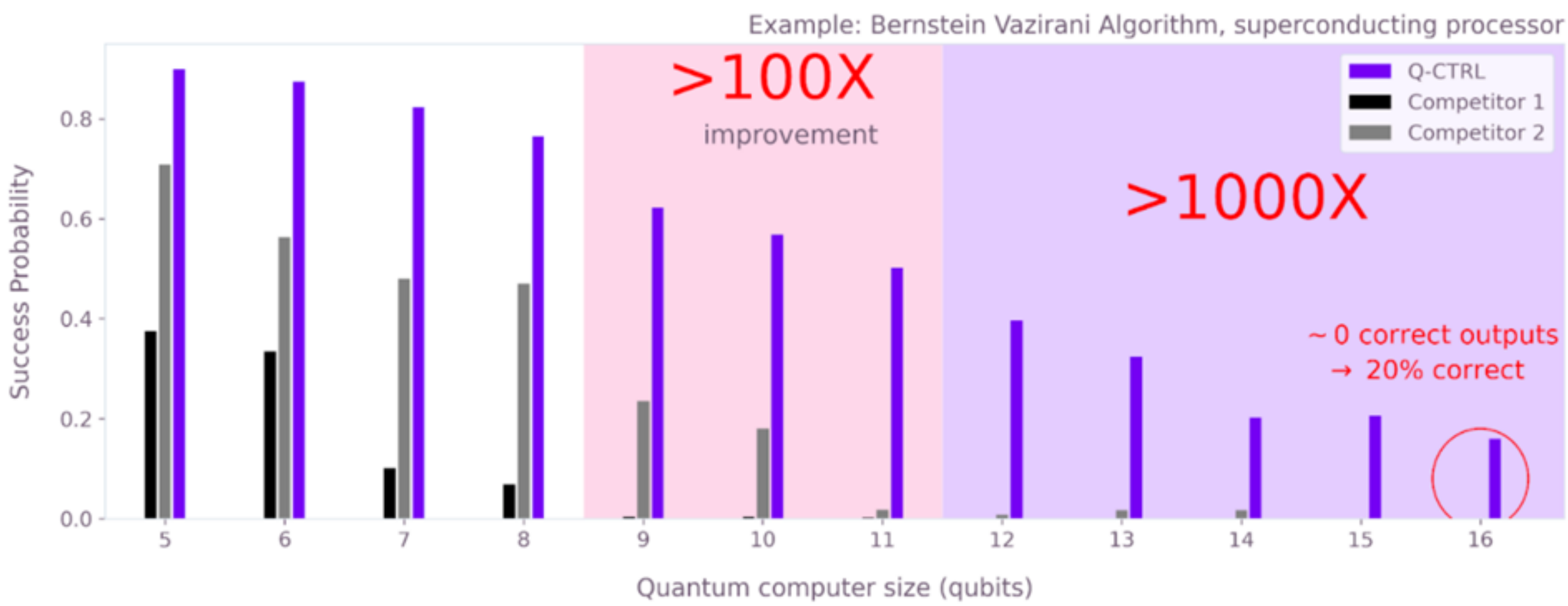

*Figure 245: charting the Q-CTRL improvements. Firing up quantum algorithms - boosting performance up to 9,000x with autonomous error suppression by Michael J. Biercuk, March 2022 and Experimental benchmarking of an automated deterministic error suppression workflow for quantum algorithms by Pranav S. Mundada, Michael J. Biercuk, Yuval Baum et al, September 2022 (16 pages).*

473 See Exponential error suppression for near-term quantum devices by Balint Koczor, PRX, 2021 (34 pages).

474 See Pulse-level Noise Mitigation on Quantum Applications by Siyuan Niu and Aida Todri-Sanial, LIRMM Montpellier France, April 2022 (11 pages) and Analyzing Strategies for Dynamical Decoupling Insertion on IBM Quantum Computer by Siyuan Niu and Aida Todri-Sainal, LIRMM France, April 2022 (11 pages).

475 See Demonstration of algorithmic quantum speedup by Bibek Pokharel and Daniel A. Lidar, July 2022 (12 pages).

476 See Shadow Distillation: Quantum Error Mitigation with Classical Shadows for Near-Term Quantum Processors by Alireza Seif, Liang Jiang, March 2022 (16 pages) and Virtual Distillation for Quantum Error Mitigation by William J. Huggins et al, Google AI, PRX, 2021 (25 pages).

477 See Crucial leap in error mitigation for quantum computers by Monica Hernandez and William Schulz, Lawrence Berkeley National Laboratory, December 2021, referring to Randomized Compiling for Scalable Quantum Computing on a Noisy Superconducting Quantum Processor by Akel Hashim, Irfan Siddiqi et al, 2021 (12 pages).

478 See Quantum Error Mitigation via Quantum-Noise-Effect Circuit Groups by Yusuke Hama et al, May 2022 (22 pages).

479 See Mitigating Depolarizing Noise on Quantum Computers with Noise-Estimation Circuits by Miroslav Urbanek, Benjamin Nachman, Vincent R. Pascuzzi, Andre He, Christian W. Bauer, and Wibe A. de Jong, PRA, December 2021 (7 pages).

480 See Mitigating errors by quantum verification and post-selection by Rawad Mezher, James Mills and Elham Kashefi, September 2021 and May 2022 (15 pages).

481 See Virtual Distillation for Quantum Error Mitigation by William J. Huggins, Ryan Babbush et al, August 2021 (26 pages).

482 See Quantum error mitigation via matrix product operators by Yuchen Guo et al, January-October 2022 (13 pages) which accounts for correlated errors between different gates.

483 Like in Quantum error mitigation as a universal error-minimization technique: applications from NISQ to FTQC eras by Yasunari Suzuki, October 2021 (33 pages).

In the commercial world, QEM can lead some vendors to display some outlandish claims like when Q-CTRL announces that its error correction scheme boosts algorithms performance by up to 9000x thanks to some autonomous error correction[484]. Why not, but 9000x of what? Looking at the details, this is achieved with back-end and front-end optimization compilation and some error mitigation techniques. When you read their chart, shown in Figure 245, you find that the x9000 ratio pertains to the success rate of running a Bernstein-Vazirani algorithm on a superconducting qubits processor, in the case of 16 qubits. But the related success factor is below 20% and is an extreme case.

You must remind yourself that 16 qubits can be easily emulated on your own laptop and is way below any quantum computing advantage. If you were to extend their chart beyond 30 qubits, you'd be hundreds of thousands better than their competitors but with a very small success probability.

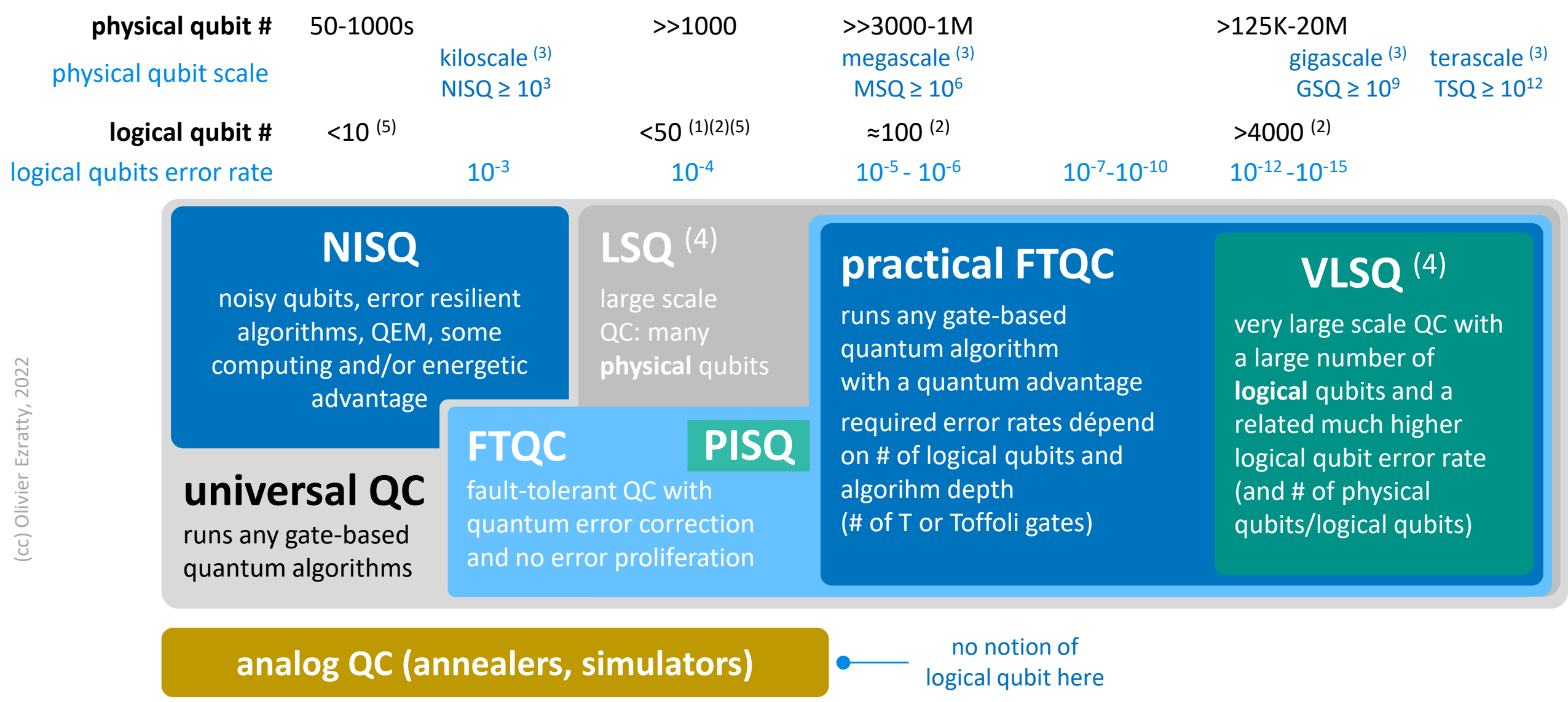

(1) 100 logical qubits # is based on a mathematical advantage starting with about 55 qubits and for a similar number of ancilla qubits for most algorithms (QFT, HHL, QML).
(2) the # of logical qubits depends on autonomous error correction and the number of physical qubits required per logical qubits, the best case being with cat-qubits.
(3) this scale was presented in March 2022 by Dave Bacon from Google. It positions NISQ above the 1000 physical qubits threshold.
(4) VLSQ is an equivalent to VLSI in classical semiconductors computing.
(5) a few <50 logical qubits are not sufficient to run a quantum gate-based algorithm with some quantum computing advantage. Thus, the practical FTQC threshold at 100 logical qubits.

*Figure 246: positioning all the concepts: NISQ, PISQ, LSQ, FTQC, Universal quantum computing and the related error correction codes. (cc) Olivier Ezratty, 2022.*

I summarized these various gate-based computing classes in the above schema in Figure 246.

It requires a lot of comments and annotations and is still work in progress:

- **Universal quantum computing** is the quantum computing paradigm in which all quantum algorithms can be implemented from a mathematical standpoint. It must support non-Clifford quantum gates. This feature is implemented at a narrow and noisy scale with NISQ systems and with FTQC.
- **NISQ** definition is not really agreed upon. Is it starting today, or will we need more physical qubits and generate some proven generic quantum computing advantage? I added in blue a scale proposal presented by Dave Bacon at Google in March 2022[485], which deals with a simple scale of number of physical qubits with NISQ being in the thousand, MSQ in the million, GSQ in the billion and TSQ in the tera number of qubits. NISQ is powered by quantum error mitigation in the near term and with approximated QEC in the mid-term. It will extend the usability of quantum computers with a larger number of qubits and circuit depth.

[484] See Firing up quantum algorithms - boosting performance up to 9,000x with autonomous error suppression by Michael J. Biercuk, March 2022.

[485] See QIP 2022 | Software of QIP, by QIP, and for QIP by Dave Bacon from Google, March 2022 (1 hour).

- **LSQ** stands for large scale quantum computer and is about having a QPU with a large number of qubits. But are these physical or logical qubits and how does it relate to FTQC? The dust has not yet settled for its definition. What we know is a large scale quantum system without error correction would not be very usable. On average, the depth of gate-based computing is limited by qubit error rates and many quantum algorithms have a breadth (number of qubits) that is in line with their depth (number of gate cycles). So, we have a sort of gap between the upper stages of NISQ and early stages of practical FTQC with logical qubits.
- **FTQC** can start with a few logical qubits of average error correction with target error rates of about $10^{-3}$ to $10^{-4}$. We'll maybe have a continuum in the FTQC progress with error rates growing progressively until it reaches $10^{-15}$ in the long term as the number of logical qubits will grow. These error rates will have to shrink at a faster rate than the increase of logical qubit numbers.
- **Practical FTQC** is about FTQC providing a generic quantum advantage. It would require at least 100 logical qubits given half of them are used for data with a space exceeding the memory capabilities of equivalent classical systems, and the other half providing the ancilla qubits required for many algorithms like the QFT and its derivatives. The number of physical qubits corresponding to the logical qubit thresholds depend on autonomous error correction and the number of physical qubits required per logical qubits, the best case being with cat-qubits[486]. The target error rate of $10^{-5}$ to $10^{-6}$ is a rough estimate, below the inverse square of the number of logical qubits.
- **PISQ** for Perfect Intermediate Scale Quantum is a proposal from Qutech scientists that corresponds to the arrival of 50+ FTQC "perfect" qubits QPUs. They advocate to get ready for it in parallel with all the efforts related to NISQ systems[487].
- **VLSQ**: this is large scale FTQC with several orders of magnitude larger number of logical qubits used to run chemical simulations, Shor integer factoring and large scale optimization and industry scale algorithms. So, in the above chart in Figure 246, I position these various definitions for universal quantum computing, FTQC, LSQ and VLSQ, with one scale for physical qubits and one for logical qubits as well as with logical qubit error rates.
- **FTDQC** is a new term, meaning « fault tolerant distributed quantum computation », which could potentially be implement with long distance quantum communication, even with satellites[488].

### QEC impact on computing time

There are only a few studies and research done to evaluate how long it would take to execute specific quantum algorithms in an "end-to-end" fashion. We know that, theoretically, with a FTQC of 20 million qubits, we could factorize an RSA 2048 bits key in 8 hours with superconducting qubits. Gate time is quite variable from 12 ns for superconducting qubits to 100 µs for trapped ions qubits.

You can get an idea of the timing overhead coming from three mechanisms:

- **Non-Clifford gates** creation overhead like R/Control-R gates with arbitrary phases, based on the Solovay-Kitaev theorem. It creates a x127 to x235 gates overhead!
- **Quantum error correction** (QEC) overhead in the case of FTQC. It creates a x10 to x20 gates overhead, minimum! It may be much bigger for large surface codes and concatenated codes. With surface code QEC, this runtime overhead scales with the code distance.

---

[486] I saved you the EFTQC variation, for early FTQC that is used in On proving the robustness of algorithms for early fault-tolerant quantum computers by Rutuja Kshirsagar et al, September 2022 (27 pages) which deals with an interesting question: what is the error rate of logical qubits in the FTQC realm that would be required for some key algorithms?

[487] See Quantum Computing -- from NISQ to PISQ by Koen Bertels et al, April 2022 (11 pages).

[488] See Upper Bounds for the Clock Speeds of Fault-Tolerant Distributed Quantum Computation using Satellites to Supply Entangled Photon Pairs by Hudson Leone et al, University of Technology Sidney, August 2022 (9 pages).

* **Number of runs** or shots required to average probabilistic results. IBM advises using 4000 runs but this number may grow with the number of used qubits. So, a x4000 overhead! We can anticipate that this number will remain high with logical (error corrected) qubits given quantum computing will always have a probabilistic dimension, even with error correction.

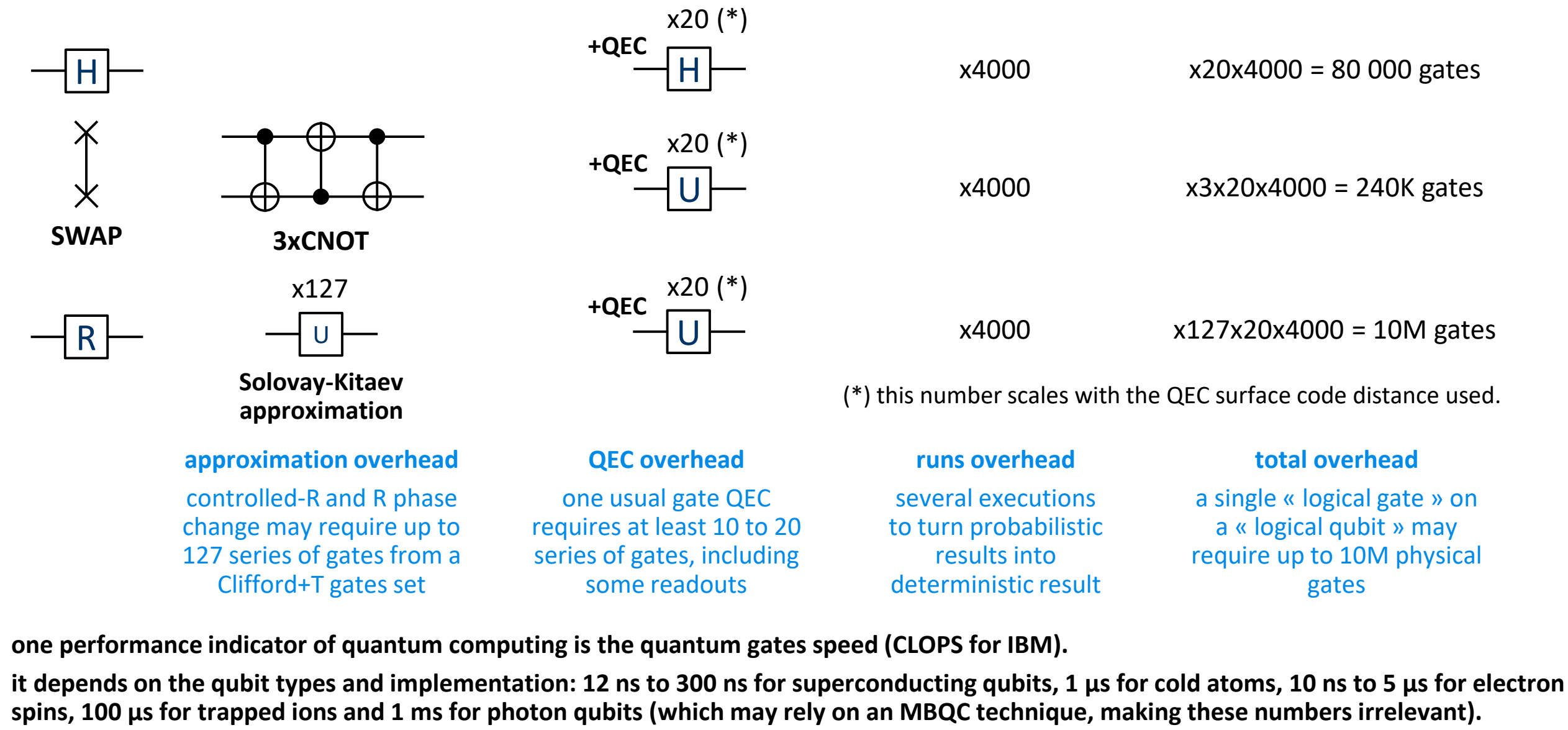

*Figure 247: assessing the overhead of quantum error correction on a practical basis. (cc) Olivier Ezratty, 2021.*

I tentatively added these three mechanisms for three scenarios: an H gate, a SWAP gate assembled with three CNOT gates and an arbitrary R gate created with a Clifford gate set plus a T gate using the Solovay-Kitaev approximation theorem. Adding all these timing overheads, you obtain between 80K and 10M gates to run to execute a single physical gate. That's quite significant!

Interestingly, the longer the gates, like with trapped ions qubits, the better fidelity they have, creating a balancing effect between the QEC overhead and the gates times. All this should be taken into account when dealing with so-called quantum algorithms speedups, particularly with non-exponential speedups.

But these estimates are very raw and deserve scrutiny. It depends on the QEC technique that is being used, on the qubit type, on their fidelities, and so on.

# Quantum memory

We would guess that quantum memory is some memory capable of storing the quantum state of qubits and then using them to feed quantum computer registers[489]. It should be able to store superposed and entangled qubits and deliver it to whatever computing is needed. But it is part of a broader category defined as "quantum RAM" or qRAM, which is able to store either classical or quantum data, the data being queried with superposed quantum addresses.

Quantum memory is also required in quantum key distribution repeaters[490] and can be useful in various situations like with quantum sensing and for creating deterministic sources of photons[491].

[489] See Architectures for a quantum random access memory, by the Italians Vittorio Giovannetti and Lorenzo Maccone and the American Seth Lloyd, 2008 (12 pages).

[490] Here's one example with One-hour coherent optical storage in an atomic frequency comb memory by Yu Ma et al, April 2021 (6 pages) and another one with Space-borne quantum memories for global quantum communication by Mustafa Gündoğan et al, 2020 (11 pages).

[491] See Quantum memories - A review based on the European integrated project "Qubit Applications (QAP)" by C. Simon et al, 2010 (22 pages).

However, we focus here on the first category of quantum memory, aimed at quantum computing. It is a very diverse one with different logical and physical architectures. We'll look at quantum memories for repeaters in the section dedicated to quantum telecommunications hardware.

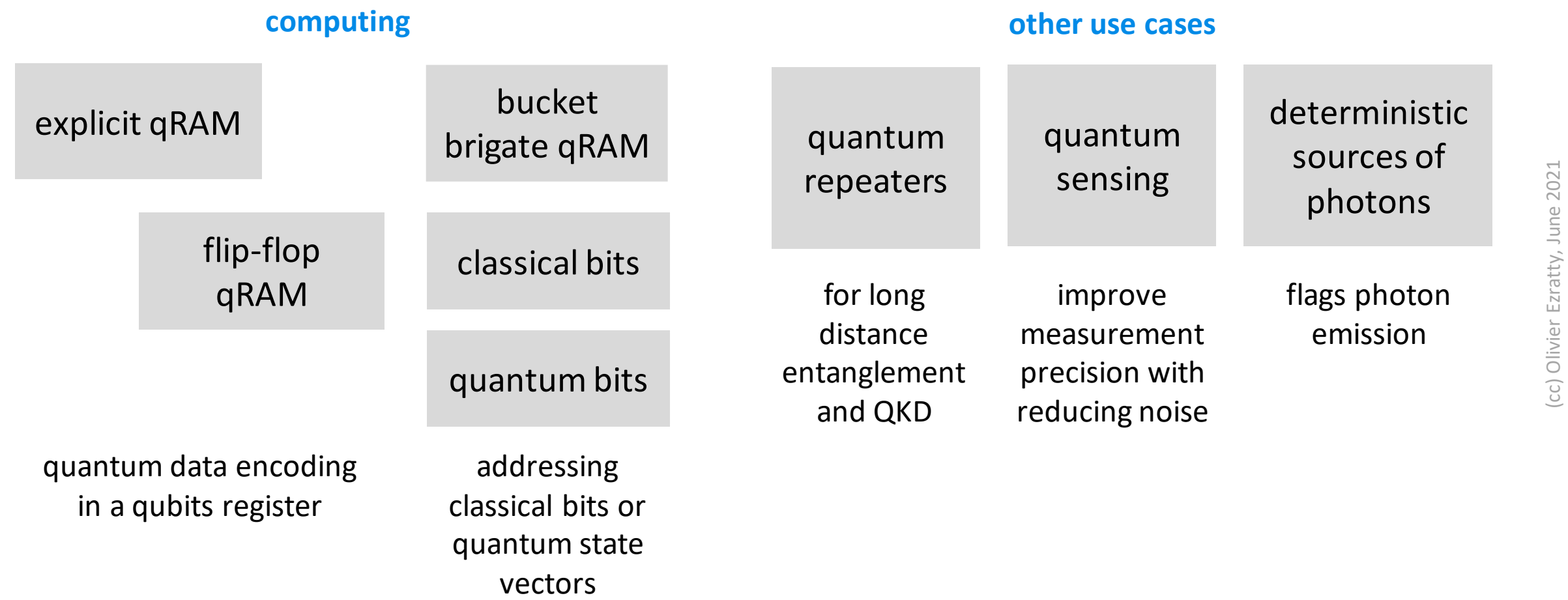

*Figure 248: various classes of quantum memories and use cases. (cc) Olivier Ezratty, 2021.*

## Quantum algorithms requirements

One anticipated usage of quantum memory is to temporarily store the state of a qubit register during a data preparation process, a usual lengthy process, before transferring it to a faster quantum processing unit. With N qubits, this memory would be able to store in theory $2^N$ different computational vector states amplitude values.

According to the no-cloning theorem, the content of this memory cannot be the copy of the state of other quantum registers. In computing, quantum memory is used to store data into some quantum memory to be later used in quantum processing. Data preparation and encoding depends on the algorithm. It is necessary for certain types of quantum algorithms such as Grover's search and quantum machine learning algorithms that we will describe later on[492].

The most demanding encoding is when you encode a vector of $2^N$ values (well, minus 1 for normalization constraints) in the whole computational state vector[493]. This creates a superposition with all or some of the basis states from the computational basis. Namely, we encode a vector $\boldsymbol{x}$ containing $2^N$ real (or even complex) number values from $\boldsymbol{x_0}$ to $\boldsymbol{x}_{2^N-1}$ with the normalization constraint that the square of these values is equal to 1. It ends up creating the state vector on the right with $2^N$ amplitudes $\boldsymbol{x_i}$ associated with the vectors $|i\rangle$ from the computational basis. This is called amplitude encoding.

$$\sum_{i=1,\,2^N} x_i^2 = 1$$

normalization constraint

$$\sum_{i=1,\,2^N} x_i|i\rangle = \begin{bmatrix} x_0 \\ x_1 \\ \vdots \\ x_{2^N-1} \end{bmatrix}$$

encoded state vector

Since this data encoding grows exponentially with the number of qubits, it may erase any computing speedup we would gain later. So, this is efficient only if we find a way to make this fast. One solution is to encode sparse vectors where only a few values are nonzero.

[492] See Quantum Machine Learning and qRAM by Behnam Kia, 2018 (59 slides) as well as Quantum Algorithms for Linear Algebra and Machine Learning by Anupam Prakash, 2014 (91 pages).

[493] See Quantum 101: Do I need a quantum RAM? by Olivia Di Matteo, May 2020 (58 slides).

### Quantum memory types

There are several types of qRAM and other quantum memory types:

- **Explicit qRAM** encodes data in physical qubits and then, use quantum circuits to extract the encoded data[494]. There is no specific addressing system to selectively access parts of this memory. This is the scenario depicted above. Also named QAQM for Quantum Access Quantum Memory and Quantum Access Memory[495].
- **Flip-flop qRAM** is a variant of explicit qRAM based on qubits circuit algorithms used to efficiently load classical data in a qubit register[496].
- **Implicit qRAM** was proposed by Seth Lloyd et al in 2008 with the **bucket brigade** addressing system, based on a qutrits tree (three states quantum objects) containing wait/left/right flags[497], sort of decision trees to reach the right memory cell. Also named QACM for Quantum Access Classical Memory.

  This quantum addressing system can be used for accessing both *classical* bits and *coherent states* in qubits. The first case may be useful when building some oracles for algorithms like a Grover search. In the full quantum case, the coherent superposition of these addresses enables a readout of a superposition of many state amplitudes in the computational basis. Namely, we can query a given amplitude $\alpha_i$ of the computational basis vector at the i address, encoded in binary with N classical bits or several of these, encoded in superposition[498].

$$\sum_j \alpha_j |j\rangle|0\rangle$$

$\alpha_j$ weighted superposition of adresses corresponding to computational basis states $|j\rangle$

$$\sum_j \alpha_j \; |j\rangle|b_j\rangle$$

result of query, weights are applied to $|b_j\rangle$ j-th memory location

  In classical RAM, the memory array of N bits ($2^n$) is usually organized in a 2-dimensional lattice which requires $O(\sqrt{N})$ switches, precisely, usually a fixed number of address data to address lines and columns in memory chipsets.

  In bucket brigade qRAM, this can decrease to $O(logN)$ to adress a particular computational basis vector amplitude. But this has to take into account the burden of any quantum error correction[499].

---

[494] See Optimal QRAM and improved unitary synthesis by quantum circuits with any number of ancillary qubits by Pei Yuan and Shengyu Zhang, Tencent Quantum Laboratory, February 2022 (19 pages) proposes an optimized method for feeding QRAM with amplitude QSP (quantum state preparation).

[495] See Quantum Associative Memory by Dan Ventura and Tony Martinez, 1998 (31 pages) and this implementation proposal that sees quite optimistic despite the support of prestigious folks like John Preskill, in Quantum Data Center: Theories and Applications by Junyu Liu et al, University of Chicago, Caltech and AWS, July 2022 (24 pages).

[496] See Circuit-based quantum random access memory for classical data with continuous amplitudes by Tiago M. L. de Veras et al, 2020 (11 pages) referring to Circuit-based quantum random access memory for classical data with continuous amplitudes by Daniel K. Park et al, 2019 (9 pages).

[497] See Quantum random access memory by Vittorio Giovannetti, Seth Lloyd et al, 2008 (4 pages) and Architectures for a quantum random access memory by Vittorio Giovannetti, Seth Lloyd and Lorenzo Maccone, 2008 (12 pages).

[498] See Circuit-Based Quantum Random Access Memory for Classical Data by Daniel K. Park et al, Nature, 2019 (8 pages) which proposes an optimized implementation.

[499] The QEC burden may be significant. also On the Robustness of Bucket Brigade Quantum RAM by Srinivasan Arunachalam et al, 2015 (19 pages) which shows that the timing advantage of qRAM bucket brigade addressing may be quickly lost due to QEC overhead. See also Quantum Random Access Memory by Aaron Green and Emily Kaplitz, 2019 (12 pages) and Methods for parallel quantum circuit synthesis, fault-tolerant quantum RAM, and quantum state tomography by Olivia Di Matteo, 2019 (111 pages) and Fault tolerant resource estimation of quantum random-access memories by Olivia Di Matteo et al, 2020 (14 pages).

Various implementations of the bucket brigade solution have been proposed so far, including one using quantum walks, with the benefit of being more robust to decoherence and easier to parallelize[500].

Before any qRAM data transfer to computation qubits can be done, an uncompute processing must be implemented that remove the selected computational basis vectors addresses from the related data.

- There are also proposals for creating **Quantum Read Only Memories** (QROM) which allows only retrieval of stored quantum information; the stored information cannot be updated[501].
- **Probabilistic Quantum Memory** (PQM) stores and simultaneously analyzes r patterns while using only n qubits. A quantum computer therefore would need O(n) qubits as opposed to O(rn) bits of associative memory on a classical computer[502].

In the end, when quantum data is transferred from quantum memory to computing qubits, it is achieved with teleporting the memory qubits to the computing one by one, usually with using entangled photons and, in many cases, some conversion from solid qubits to photon qubits (spin or charge to photons and the other way around). This teleportation is supposed to preserve the superposition and entanglement between the memory qubits during this transfer. Given there must be some errors generated during the transfer, which will require their own error correction codes.

### Quantum memory physical implementations

None of the different quantum memory architectures studied over the last two decades is working yet. However, research is making progress, with targeted use cases that are more related to secure telecommunications and for quantum optical repeaters. At this stage, the advent of qRAM for quantum computing is more difficult to predict than scalable quantum computing!

The most promising quantum memory technologies are coupling cold atoms and photon polarization[503]:

- **Cold atoms and light polarization**. Chinese scientists used in 2019 the storage of the circular polarization state of a single photon trapped in a laser-cooled rubidium structure in a magneto-optical trap and thus made transparent[504].

  Rubidium atoms are cooled with lasers to 200 µK. The same year, another team in China created a 105 qubits memory using 210 memory cells and dual-rail representation of a photon-based qubits with fidelities of 90% but these qubits seem not entangled and thus, not able to store a full state vector with $2^N$ values, but only a N or 2N values using basis encoding in each individual qubit[505]. Other techniques are based on cesium with fidelities reaching 99%[506].

---

[500] See Quantum random access memory via quantum walk by Ryo Asaka et al, 2021 (13 pages).

[501] See Optimization of Quantum Read-Only Memory Circuits by Koustubh Phalak et al, PennState and IBM, April 2022 (6 pages). It uses amplitude encoding with qubits for address and qubits for memory.

[502] See Probabilistic Quantum Memories by Carlo A. Trugenberger, PRL, 2000 (4 pages) and a recent implementation improvement proposal in EP-PQM: Efficient Parametric Probabilistic Quantum Memory with Fewer Qubits and Gates by Mushahid Khan et al, University of Toronto, January 2022 (27 pages).

[503] As in Highly-efficient quantum memory for polarization qubits in a spatially-multiplexed cold atomic ensemble, 2017 (13 pages), a paper to which Julien Laurat from CNRS contributed.

[504] As reported in HKUST Physicist Contributes To New Record Of Quantum Memory Efficiency, 2019, which refers to Efficient quantum memory for single-photon polarization qubits by Yunfei Wang et al, 2019 (8 pages).

[505] See Experimental realization of 105-qubit random access quantum memory by N. Jiang et al, 2019 (6 pages).

[506] See Highly-efficient quantum memory for polarization qubits in a spatially-multiplexed cold atomic ensemble by Pierre Vernaz-Gris, Julien Laurat et al, Nature Communications, January 2018 (6 pages) and Efficient reversible entanglement transfer between light and quantum memories by M. Cao, Julien Laurat et al, LKB France, April 2021 (6 pages).

This is also the technique developed by Julien Laurat at ENS LKB in Paris and implemented by WeLinQ.

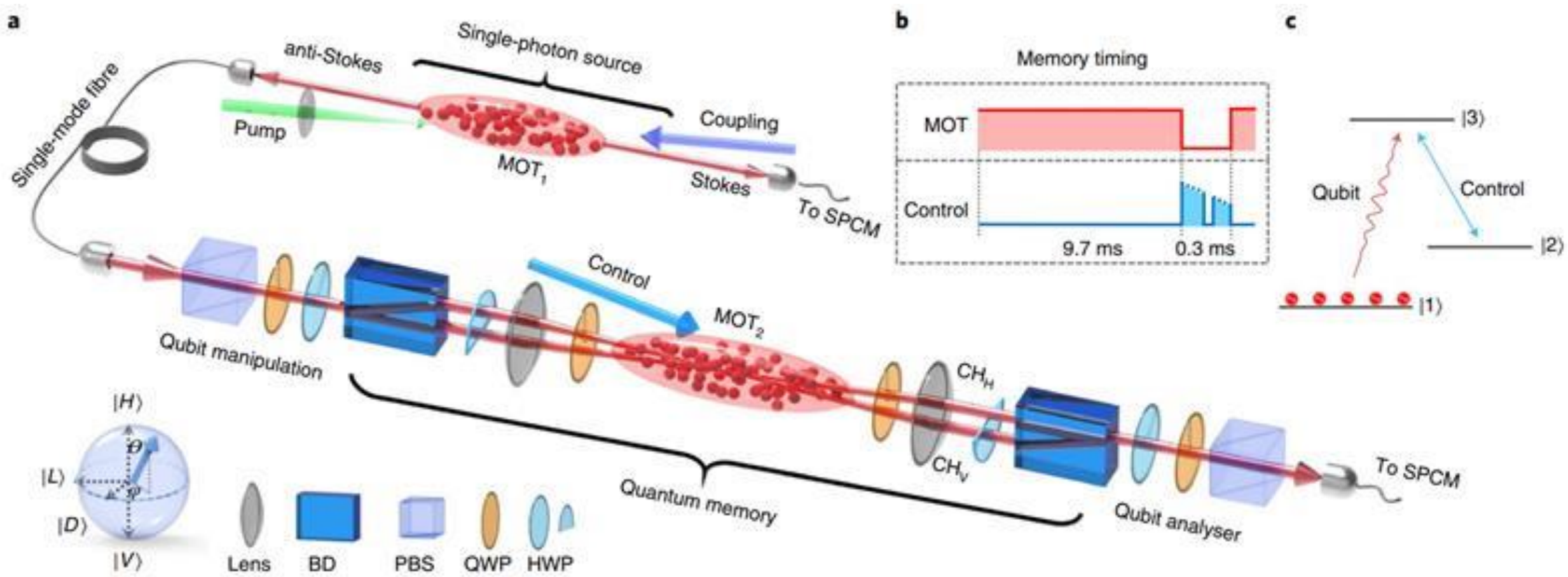

**Fig. 1 | Experimental set-up and energy level scheme of the single-photon quantum memory. a**, Schematic of the experimental optical set-up. The cold atoms in the first magneto-optical trap ($MOT_1$) serve as a nonlinear optical medium for producing time-frequency entangled photon pairs, while the cold atoms in the second magneto-optical trap ($MOT_2$) are the medium for the quantum memory. The anti-Stokes photon is coded with an arbitrary polarization state through the QMU consisting of a QWP and HWP. After the QMU, the two orthogonal linear polarizations are separated into two beams by a polarization beam displacer (BD) that are coupled into the two balanced spatial channels $CH_H$ and $CH_V$ of the quantum memory. The memory read-outs are recombined at the second BD and the polarization state is measured by the qubit analyser. **b**, The memory operation timing shows the MOT sequence and the optimized control laser intensity time-varying profile in each experimental cycle. **c**, The atomic energy level scheme of the quantum memory based on EIT.

*Figure 249: a cold atom base single qubit memory. Source: Efficient quantum memory for single-photon polarization qubits by Yunfei Wang et al, 2019 (8 pages).*

- A related work in **Canada** is dynamically controlling rubidium's transparency to trap single photons[507]. In practice, photons are stored for a thousandth of a second, but this would be sufficient for optical telecommunication repeaters. Another work in France from the Pasqal team used cold atoms to store quantum information[508].
- **Optical memories** are also tested with ytterbium[509], a rare earth that can be controlled at high frequency. The process is similar to the previous one and consists in preserving the polarization of a single photon in a magnetic trap, rather for optical repeater applications in long-distance secure communication lines[510].
- The storage of quantum states is also possible in **electron spins**[511] and **donors spins**[512].

---

507 See Physicists create new, simpler-than-ever quantum 'hard drive for light', by Kate Willis, University of Alberta, 2018, which refers to Coherent storage and manipulation of broadband photons via dynamically controlled Autler-Townes splitting, October 2017 (17 pages).

508 See Storage and Release of Subradiant Excitations in a Dense Atomic Cloud by Giovanni Ferioli, Antoine Glicenstein, Loic Henriet, Igor Ferrier-Barbut and Antoine Browaeys, PRX, May 2021 (12 pages).

509 See Nuclear spin-wave quantum register for a solid-state qubit by Andrei Ruskuc et al, Caltech, Nature, February 2022 (32 pages). It uses ytterbium nuclear spin in yttrium orthovanadate crystal ($YVO_4$, V for vanadium) arranged in nanophotonic cavity. It stores polarization information in spin ensembles. Bell states are created with ytterbium and vanadium. Control is made with 675 and 991 MHz microwaves and optical readout at 984 nm. It operated at 460 mK.

510 See Simultaneous coherence enhancement of optical and microwave transitions in solid-state electronic spins, December 2017 (10 pages). This is a joint work between the University of Geneva, notably Nicolas Gisin, and the CNRS in France.

511 See Researchers achieve on-demand storage in integrated solid-state quantum memory by Liu Jia, Chinese Academy of Sciences, January 2021.

512 See Random-access quantum memory using chirped pulse phase encoding by James O'Sullivan, March 2021-June 2022 (27 pages) which deals with using ensembles of bismuth donors spin in natural silicon, coupled to a planar superconducting niobium resonator, all operating at 100 mK with a resonant frequency of 7.093 GHz. Pulses are made of 1,200 photons. It seems to be used with individual qubits memory, not entangled qubits and amplitude encoding.

- **NV centers**[513] and other crystal defects are also tested, storing qubits in **nuclear spins**[514].
- Passively **corrected quantum memory** can be implemented with cat-qubits[515].
- **Photons** trapped in cavities[516].
- **Trapped ions** as experimented in 2022 in China with 218 ions in a 1D trap with over 300 ms stability [517].
- Memory with **error correction** using the honeycomb technique or Floquet codes[518].

Many new quantum memory proposals pop up from time to time. An interesting one from the University of Cambridge stores some quantum bit information in an electron spin hidden in haystack of 100,000 atom nuclei. The electron spin and the whole haystack are controlled by a laser. But the nuclei surrounding the electron make it difficult to entangle several qubits. End of story[519]!

# Quantum technologies energetics

The main motivation for creating quantum computers is their computing capacity, which theoretically increases exponentially with their number of high fidelity qubits. This should make it possible to perform calculations that will someday be inaccessible to conventional supercomputers. In some other cases, like with some NISQ architecture using quantum error mitigation, it will only be "just" faster or sometimes, provide better results, like in quantum machine learning or quantum physics simulations.

How does this computing capacity translate in terms of energy consumption is a key question. At first glance, it looked like the energetic cost of quantum computing was several orders of magnitude lower than classical computers. That was a naïve interpretation of Google Sycamore's 2019 quantum supremacy demonstration which did show a ratio of about one to one million in energy consumption compared to the IBM Summit supercomputer that was used as a comparison, and even when using the optimized algorithm and configuration proposed afterwards by IBM.

But the benchmark was comparing apples and oranges with a randomized benchmark with no input data nor any useful output data. It was later shown by Waintal et al that, with accounting for its high error rate and noise and using tensor networks, Sycamore's performance could be emulated on a rather simpler classical server cluster[520].

---

[513] See Storing quantum information in spins and high-sensitivity ESR, by two researchers including Patrice Bertet of the Quantronics group at CEA/CNRS, September 2017 (13 pages). See also A Ten-Qubit Solid-State Spin Register with Quantum Memory up to One Minute by C. E. Bradley et al, QuTech and TU Delft, 2019 (12 pages) and Multiplexed control of spin quantum memories in a photonic circuit by D. Andrew Golter et al, MITRE Corporation, Sandia Labs, University of Arizona, September 2022 (18 pages).

[514] See Nuclear Spin Quantum Memory in Silicon Carbide by Benedikt Tissot et al, April-August 2022 (12 pages). It uses an all-optical O-band (in the 1260 nm-1360 nm range, adapted to long distance communication) to control vanadium defect spins in SiC

[515] See Candidate for a self-correcting quantum memory in two dimensions by Simon Lieu et al, May 2022 (11 pages).

[516] See Toward a Quantum Memory in a Fiber Cavity Controlled by Intracavity Frequency Translation by Philip J. Bustard et al, March 2022 (7 pages). Here, the memory traps photons in a low-loss cavity.

[517] See Experimental realization of a 218-ion multi-qubit quantum memory by R. Yao et al, September 2022 (6 pages).

[518] See A Fault-Tolerant Honeycomb Memory by Craig Gidney et al, August 2021 (17 pages).

[519] See Light used to detect quantum information stored in 100,000 nuclear quantum bits by University of Cambridge, February 2021 and A different type of cloud computing: Quantum breakthrough uses lasers to find data in a giant cloud of atomic nuclei by Daphne Leprince-Ringuet, February 2021. And Quantum sensing of a coherent single spin excitation in a nuclear ensemble by D. M. Jackson et al, Nature Physics, 2021 (21 pages).

[520] See What limits the simulation of quantum computers? by Yiqing Zhou, Edwin Miles Stoudenmire and Xavier Waintal, PRX, November 2020 (14 pages) and A density-matrix renormalization group algorithm for simulating quantum circuits with a finite fidelity by Thomas Ayral, Thibaud Louvet, Yiqing Zhou, Cyprien Lambert, E. Miles Stoudenmire and Xavier Waintal, August 2022 (25 pages).

On the other hand, another commonplace view is that the sheer power of about 15kW that is required for cooling superconducting qubits processors is a showstopper. It gives the impression that quantum computers will be high-power consuming devices. This may not be true and forgets that a rack of Nvidia GPGPUs used for machine learning tasks has a power consumption above 30kW.

Real comparisons should be made in the future, with large-scale quantum computers that will bring a quantum computing advantage to classical supercomputers. These will require a large number of physical qubits to implement error correction. Controlling these qubits uses energy-consuming conventional electronics. The question remains open: will quantum computers provide some energy advantage on top of a computing advantage, or do they risk to turn into energy hogs[521]?

The same questions should be asked for other quantum technologies that could potentially be deployed at a large scale like quantum telecommunications and cryptography as well as quantum sensors.

**Digital energy footprint**

Quantum computers are usually compared in performance and energy footprint with supercomputers. So, what are we dealing with? The world's largest supercomputers consume several MW (megawatts) like the recent Frontier from the DoE Oak Ridge Laboratory in Tennessee and its 21 MW for 1.1 exaflops and 700 petabytes of storage, 9,400 AMD CPUs and 37,000 AMD GPUs.

It followed the IBM Summit in 2019 and its 13 MW of peak power for 200 petaflops, including 3.9 MW just for cooling. These MW came from the thousands of Power9s CPU chipsets and general purpose Nvidia GPUs requiring a complex water-cooling system that uses two tons of water per minute. IBM Summit occupies 500 $m^2$ and weighs 349 tons, compared to about 2 tons for a superconducting quantum computer that fits into a room of about 20 $m^2$, the device being a square cube of about 2.75m, which also gives a "mass advantage" and a "surface advantage" in its current state, provided we also obtain a computing advantage, which has yet to be proven.

New supercomputers are launched each and every year, but their scale doesn't change fast. These supercomputers won't be replaced by quantum computers. Many of the scientific applications they are used for are not suitable for quantum computing, like any digital simulation requiring large sets of data such as in weather forecasts or using the finite elements method and other methods to solve differential equations. We will always need them. On the other hand, when quantum computers scale up, they will be able to perform computations inaccessible to conventional supercomputers, like molecular simulations and, probably with a smaller energy footprint.

The energy efficiency of classical systems is the ratio of their performance to their energy consumption. A classical server efficiency can be expressed in FLOPS/W, where FLOPS is the number of floating-point operations per second. Since the birth of computing, this efficiency has doubled about every 18 months.

This is Koomey's law, with a current record for supercomputers of 52 GFLOPS/W for the DoE's full size Frontier HPC launched in 2022[522].

However, this sort of amazing progress has not prevented an explosion in global energy consumption to power digital technologies. Digital technologies now consume 11% of the world's electricity, with computer datacenters accounting for a quarter of this energy footprint[523].

---

[521] Like in this evaluation of Shor's energetic cost seen in Energy Cost of Quantum Circuit Optimisation: Predicting That Optimising Shor's Algorithm Circuit Uses 1 GWh by Alexandru Paler et al, ACM Transactions on Quantum Computing, March 2022.

[522] See https://www.top500.org/ and the June 2022 Top 500 charts. See also Compute and energy consumption trends in deep learning inference by R. Desislavov, F. Martinez-Plumed, and J. Hernandez-Orallo, 2021 (26 pages).

[523] See Spintronic devices for energy-efficient data storage and energy harvesting by Jorge Puebla et al, Communication Materials, 2020 (9 pages).

It is increasing as usage grows. These phenomena are simply a new manifestation of the rebound effect, formalized by William Stanley Jevons in 1865.

Efficiency gains automatically lead to a decrease in the cost of resources. Without regulation of markets and uses, they lead to an increase in global consumption (see Figure 250). However, this does not mean that improving the energy efficiency of computers is inherently wrong. On the contrary, it is the only solution to maintain performance with limited energy and material resources.

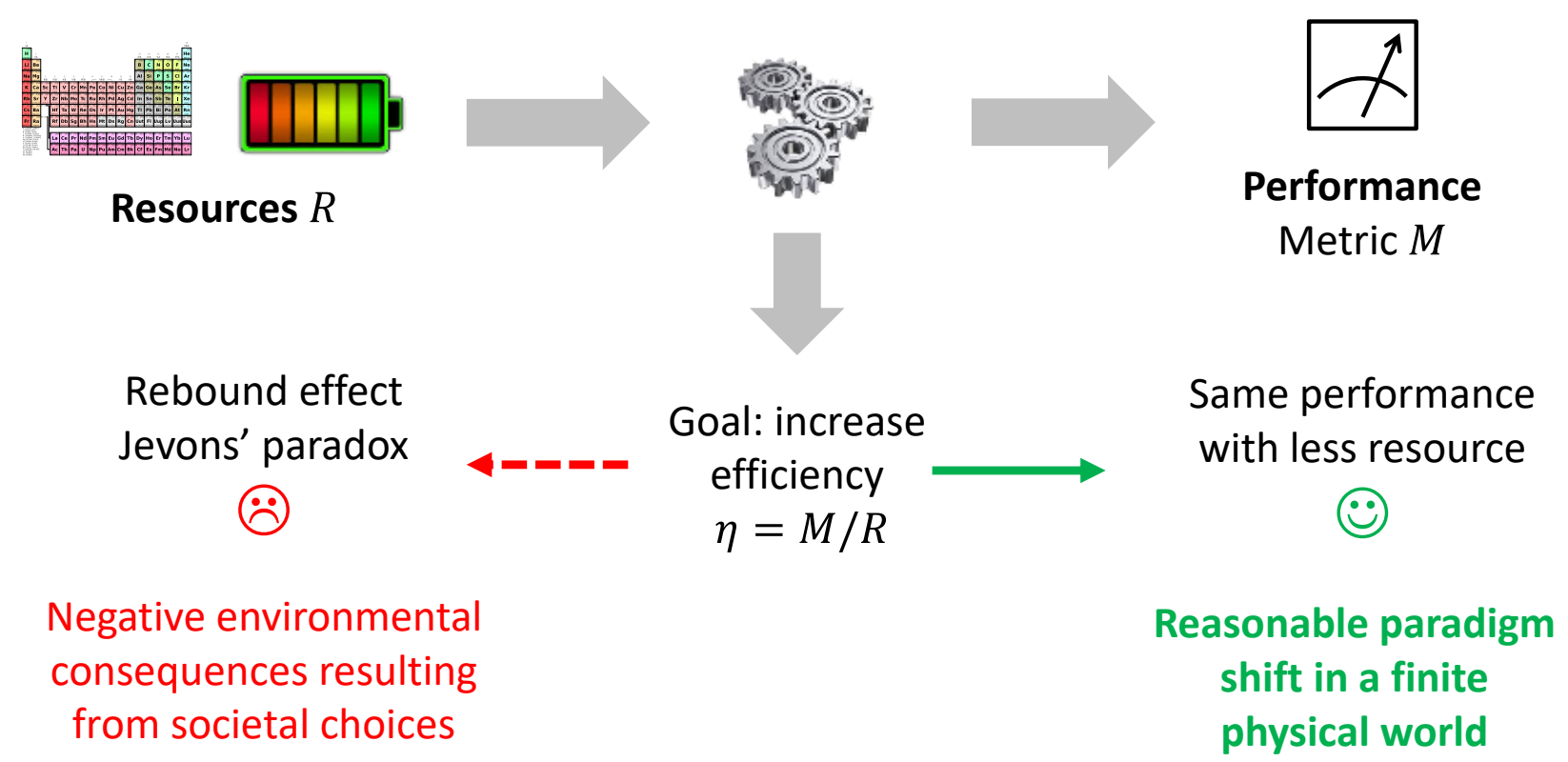

*Figure 250 Energy efficiency and the rebound effect. A machine consumes material and energy resources to perform a task with a performance M. Its efficiency is defined by the ratio $\eta = M/R$. Source: Alexia Auffèves, France Quantum June 2022 presentation.*

The debate is also raging about the potentially large energetic footprint of cryptocurrencies, with various more or less questionable comparison methodologies[524]. And let's forget about the metaverse which may itself be yet another digital energetic hog.

## Quantum Energy Initiative

Scaling quantum computing is one of the most challenging scientific and technology endeavors ever launched by mankind on top of space exploration, nuclear fusion and DNA sequencing and genome based therapies creations.

It should be undertaken with behaving responsibly as early as possible. One way is to embed in research and systems design an approach integrating the environmental footprint of quantum technologies. This footprint is of course energetic but also encompasses raw materials, manufacturing processes and product lifecycle handling. Addressing these questions are both scientific, technology and societal challenges.

We can learn a couple lessons from what happened with artificial intelligence and deep learning. It became trendy starting in 2012 with a peak around 2020 when deep learning use cases became mainstream and embedded from smartphones to cloud datacenters. Suddenly, it was discovered that AI had a significant energetic cost, both for training large deep learning models and to run them whether on end-user devices or on servers[525]. The "frugal AI" topic then emerged. Solutions were proposed to reduce the energetic footprint on AI mainly with less data-hungry machine learning models[526], so-called data "quantization" (using 8-bit and even 1-bit numbers instead of 16-32-64 floating-point numbers) and also with optimizing the power consumption of dedicated hardware like GPGPUs and embedded systems chipsets (in smartphones, connected objects and also cars). What if the environmental footprint or AI had been taken care of earlier?

---

[524] On Bitcoin's Energy consumption: a quantitative approach on a subjective question by Rachek Rybarczyk, Galaxy Digital Mining, May 2021 (13 pages) and Fact sheet: Climate and Energy Implications of Crypto-Assets in the United States, White House, September 2022. In the Blockchain realm, Ethereum switched in September 2022 from proof-of-work to proof-of-stake for mining, with a significant energy saving of several orders of magnitude. See Ethereum energy consumption, Ethereum, September 2022 which provides a lot of energy consumption related data for various Internet services.

[525] See Compute and Energy Consumption Trends in Deep Learning Inference by Radosvet Desislavov, 2021 (26 pages) which describes how GLOPS/W have recently evolved depending on the type of AI problem (CNN for convolutional networks, NLP for natural language processing).

[526] See Frugal Machine Learning by Mikhail Evchenko, Joaquin Vanschoren, Holger H. Hoos, Marc Schoenauer and Michèle Sebag, November 2021 (31 pages).

The same question deserves to be asked for quantum technologies. Why not take care right now of their environmental footprint? One could argue that the first challenge is scientific before being environmental. Some are advocating to first obtain high-fidelity qubits and useful fault-tolerant quantum computers and later address the environmental problem. Looking at how research labs and industry vendors were working until now on addressing the scalability challenges of quantum computers demonstrate that despite its relative technology immaturity and high scientific uncertainty, it is time to take environmental concerns into account right now. In a world of doubts on the role of science and technology, it's also a way to demonstrate that in emerging technologies, it's possible to implement responsible innovation practices from the start and not as afterthoughts and under pressure.

#QEI
the quantum energy initiative

All of this is the reasoning behind the creation of the **Quantum Energy Initiative** (QEI) in 2022[527]. The idea came out from a research team in France led by Alexia Auffèves (MajuLab in Singapore) with Robert Whitney (CNRS LPMMC in Grenoble) and Olivier Ezratty (myself).

It quickly got the support of researchers (in quantum computing and also quantum telecommunications) and industry vendors (in quantum computing and enabling technologies) throughout the world. It's about bringing a scientific and technology community together to address these issues in a concerted way and foster a cross-disciplinary research-industry collaboration.

The QEI is first about answering several key questions related to quantum computing:

- Is there a **quantum energy advantage** as the processors scale up and how different is it from the quantum computational advantage?
- What is the fundamental **minimal energetic cost** of quantum computing?
- How to **avoid energetic dead-ends** on the road to large scale quantum computing? Can we create optimization tools and models for qubit technology, enabling technologies and software engineering?

The seed of Quantum Energy Initiative is described in a thorough perspective paper from Alexia Auffèves published in PRX Quantum in June 2022[528]. It lays the ground for a transversal initiative, connecting quantum thermodynamics, quantum information science, quantum physics and engineering. It makes the connection between classical and quantum thermodynamics, qubit architectures, qubit noise models, room temperature control electronics and cryo-electronics, quantum error correction codes, algorithms and compiler designs. It proposes a methodology to assess the energetic performance of quantum technologies, dubbed MNR[529].

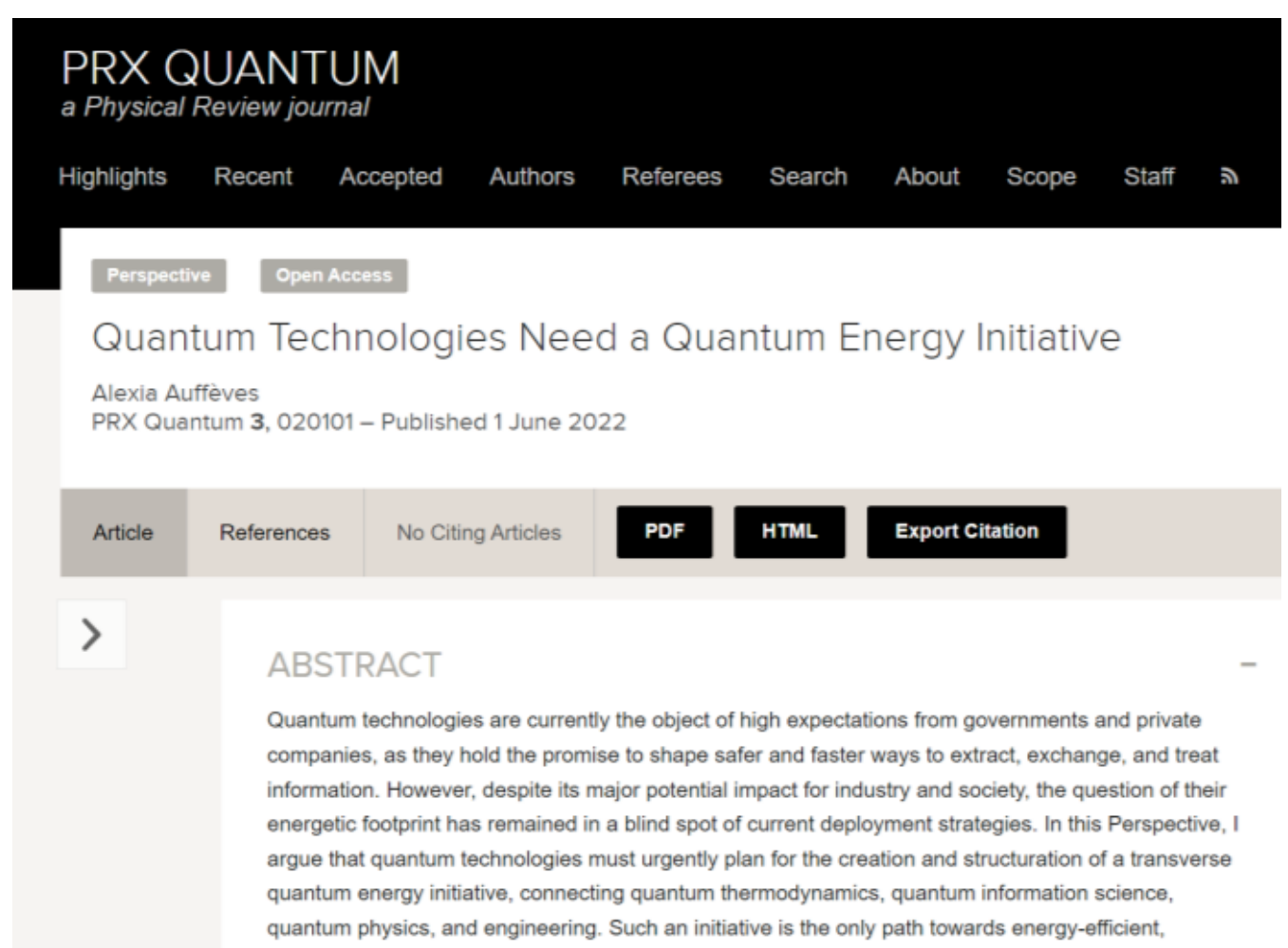
PRX QUANTUM
a Physical Review journal
Highlights Recent Accepted Authors Referees Search About Scope Staff

Perspective Open Access

Quantum Technologies Need a Quantum Energy Initiative

Alexia Auffèves
PRX Quantum **3**, 020101 – Published 1 June 2022

Article | References | No Citing Articles | PDF | HTML | Export Citation

ABSTRACT

Quantum technologies are currently the object of high expectations from governments and private companies, as they hold the promise to shape safer and faster ways to extract, exchange, and treat information. However, despite its major potential impact for industry and society, the question of their energetic footprint has remained in a blind spot of current deployment strategies. In this Perspective, I argue that quantum technologies must urgently plan for the creation and structuration of a transverse quantum energy initiative, connecting quantum thermodynamics, quantum information science, quantum physics, and engineering. Such an initiative is the only path towards energy-efficient, sustainable quantum technologies, and to possibly bring out an energetic quantum advantage.

*Figure 251: the QEI position paper. Quantum technologies need a Quantum Energy Initiative by Alexia Auffèves, PRX Quantum, June 2022 (11 pages).*

[527] See the QEI website: https://quantum-energy-initiative.org/. It contains a poll ("Join us") to build the QEI community and a Manifesto that quantum professionals can sign to support the initiative. As of the publishing of this book late September 2022, about 200 signatures had been collected coming from 32 countries.

[528] See Quantum technologies need a Quantum Energy Initiative by Alexia Auffèves, PRX Quantum, June 2022 (11 pages).

[529] See also the thesis The resource cost of large scale quantum computing by Marco Fellous-Asiani, November 2021 (215 pages).

After having first modeled the energy consumption of a scalable fault-tolerant superconducting qubits quantum computer and learned some lessons on the conditions of a related energetic advantage, the QEI aims to apply this methodology to all types of qubits developed by research laboratories and companies around the world. This includes silicon spin qubits, trapped ion qubits, neutral atom qubits and photon qubits. The three main paradigms of quantum computing will also need to be evaluated, namely programmable quantum gate computing, quantum annealing and quantum simulation. This will allow the energy dimension to be exploited for comparison and scaling. These efforts also involve the entire quantum computing software chain, in particular error correction codes, algorithms and compilers.

This work could lead to the implementation of a "Q-Green 500" type benchmarking system to compare the best quantum computers in terms of their energy efficiency. They will also provide a basis for the creation of tools and models to dimension quantum computer architectures from an energy point of view by integrating all their hardware components - quantum and classical - and software. They will provide roadmap elements and specifications for companies in the enabling technology sector, such as benchmarks for the energy consumption of control electronics systems.

The QEI self-mandate is not limited to quantum computing. It goes beyond and is intended to expand to all quantum technologies, namely quantum telecommunications[530] and quantum sensing.

### Modeling a quantum computing energetic advantage

Thanks to quantum coherence, superposition and entanglement, quantum computers could showcase an exponential computing speedup compared to their classical counterparts, depending on the size and nature of the problems to be solved and on the used quantum algorithm. This computational advantage is usually predicted for ideal, error-free processors. In reality, quantum processors are noisy, with error rates currently exceeding 0.1% per operation, a prohibitive level for most algorithms and many quantum error correction codes.

In the short term, algorithms are created that can run on such noisy processors in the quantum computing paradigm called NISQ (Noisy Intermediate Scale Quantum) and with using quantum error mitigation techniques.

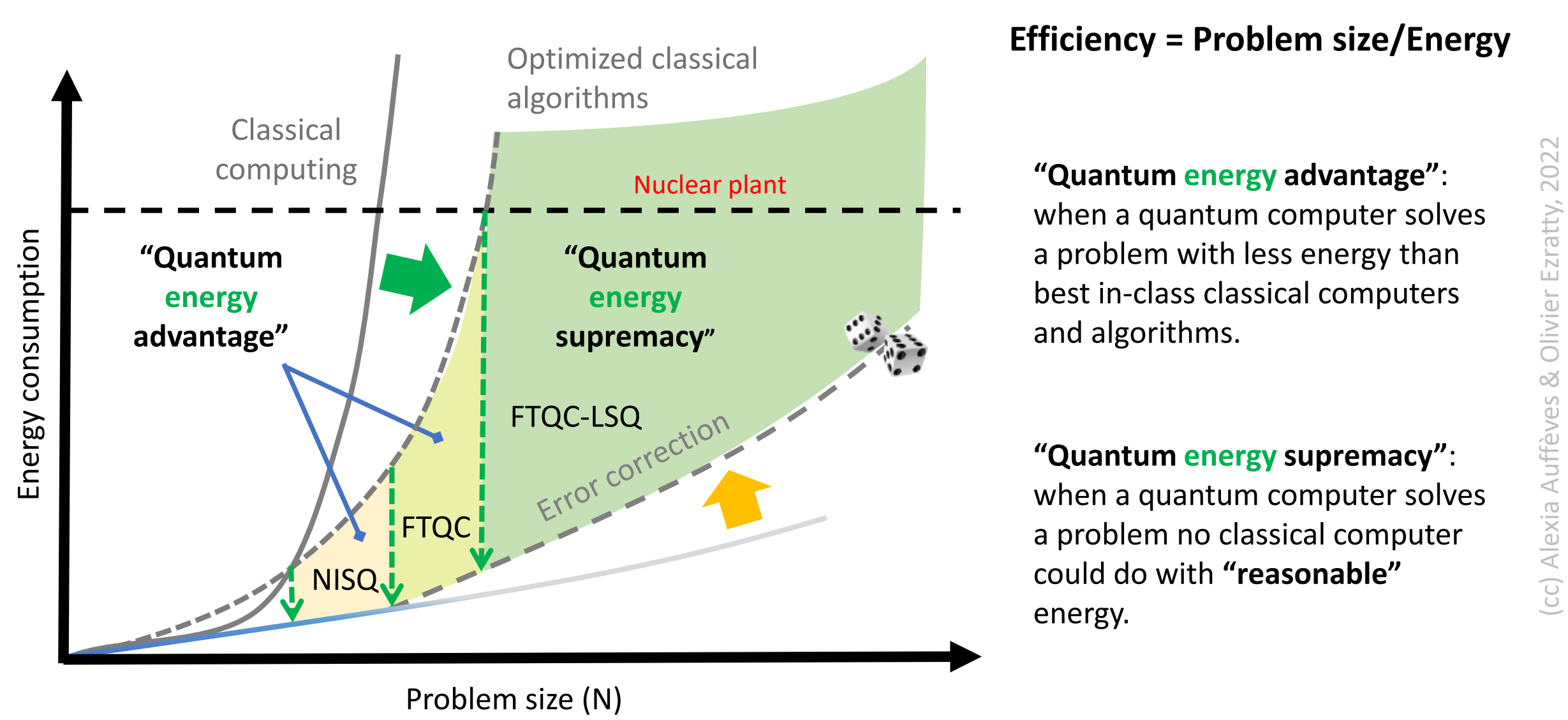

*Figure 252 Different regimes of quantum energy advantage. Source: Alexia Auffèves and Olivier Ezratty, 2022.*

---

[530] Work has already started there. See for example Reducing energy consumption of fiber networks via quantum communication technology by Janis Notzel and Matteo Rosati, February 2022 (25 pages). With some proposal of a quantum receiver that would reduce the power consumption of classical fiber optic lines amplifiers.

In the longer term, we'll rely on quantum error correction using a large number of so-called physical qubits assembled as logical qubits, that will enable longer calculations and at an acceptable error rate. This number varies from 30 to 10,000 depending on the qubit technology. The associated error correction mechanisms also considerably increase the number of calculation steps.

In both cases, demonstrating a computational advantage for real quantum processors is an open question. In some cases, quantum computers could be less energy-intensive than conventional computers to solve the same problem as illustrated in Figure 252.

With larger scale FTQC could emerge a sort of quantum energy supremacy when a quantum computer solves a problem that no classical computer could process with a reasonable energy footprint like what comes out of a nuclear plant reactor (1 GW)[531].

Modeling and optimizing the energetic efficiency of quantum computers must take into account the resources used for control and error correction. As a ratio of a performance to a resource, this energy efficiency is a hybrid quantity:

- Computational performance emerges at the fundamental quantum level, and results from the ability to control the noisy quantum processor to perform an algorithm with a certain accuracy. Understanding and optimizing these mechanisms is a matter of quantum control, quantum thermodynamics, quantum error correction, algorithms and compilers.
- Establishing satisfactory control at the quantum level requires the provision of resources at the macroscopic level, which determines the energy consumption necessary to carry out the calculation. This is the domain of enabling technologies including cryogenics, control electronics, cabling, lasers, amplifiers, detectors, whose mix depends on the qubit type.

It is essential to set up a full-stack quantum computer model coupling these different levels, as well as common language and concepts[532]. On this basis, the methodology proposed in the QEI is simple.

It sets a target performance at the microscopic level defining an implicit relationship between the different parameters of the model and a macroscopic energy consumption that is then minimized under this constraint. It was applied on a superconducting qubits model and considered typical algorithms used for optimization, physical simulations, quantum machine learning, and cryptanalysis for integer factorization.

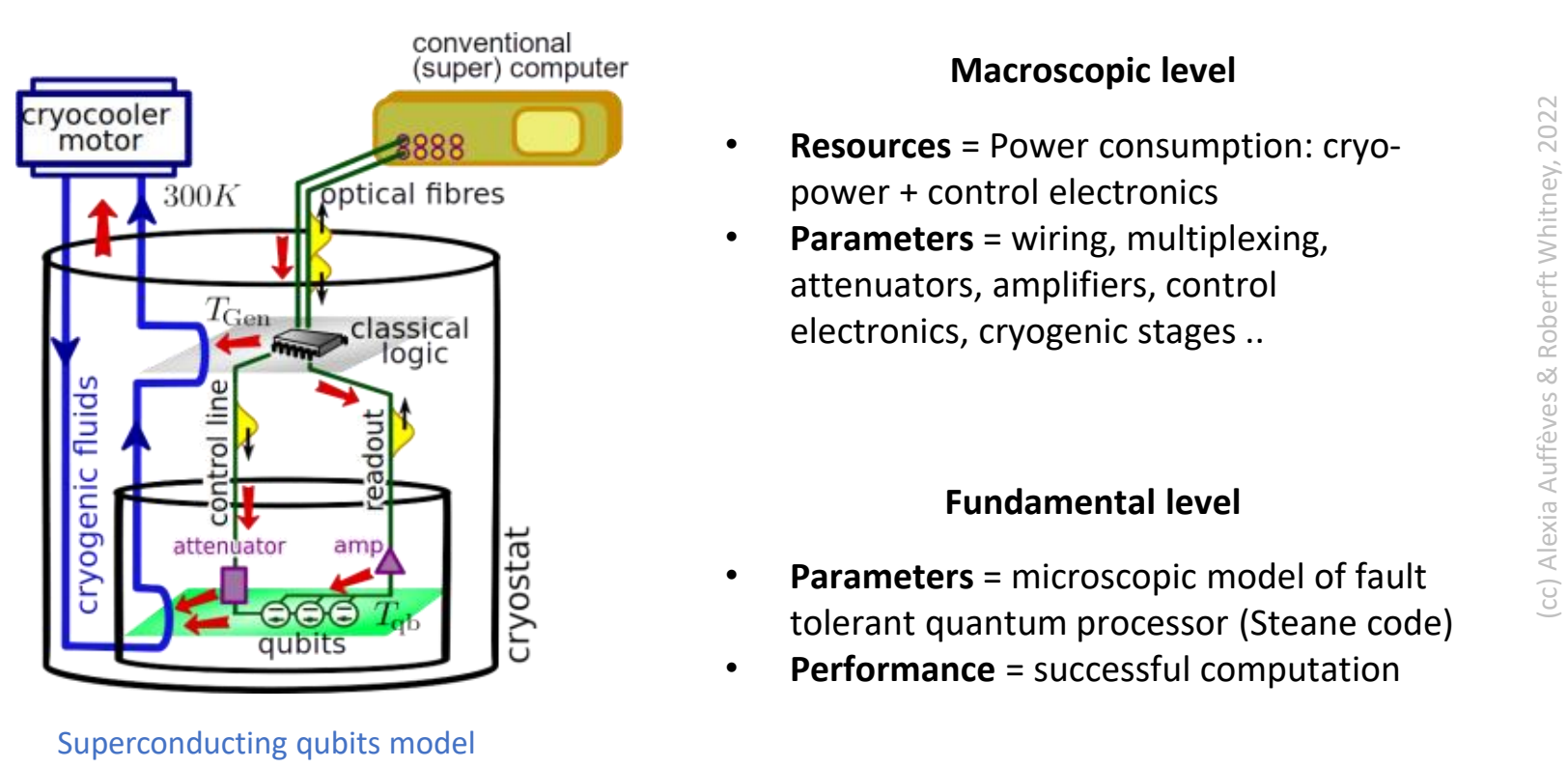

*Figure 253 Full-stack model of a superconducting quantum computer coupling a quantum level and a macroscopic level of description. Source: Alexia Auffèves and Robert Whitney.*

[531] Interestingly, one paper shows how Shor algorithm is reaching this exact threshold, in Energy Cost of Quantum Circuit Optimisation: Predicting That Optimising Shor's Algorithm Circuit Uses 1 GWh by Alexandru Paler and Robert Basmadjian, ACM Transactions on Quantum Computing, March 2022 (no free access).

[532] See Energy use in quantum data centers: Scaling the impact of computer architecture, qubit performance, size, and thermal parameters by Michael James Martin et al, NREL, 2021 (18 pages) that proposes a modelling of quantum processors energy consumption but not in a full-stack manner. It doesn't take into account the characteristics of the algorithm and is very generic with regards to all enabling technologies where many technology choices can impact the total system power consumption.

A full-stack modeling integrates the sources of quantum noise affecting qubits, the conventional qubit control resources such as electronics that generate microwave pulses and voltages, filters and attenuators, cryogenics, cables and amplifiers used for reading the state of the qubits, then the sources of heat dissipation involved in the whole material chain and in particular in the cryostat (see Figure 253). At last, it takes into account the size of the error correction code, initially a concatenated Steane code and then a surface code.

The model establishes a relationship between microscopic processor parameters such as qubit fidelity, with macroscopic qubit control parameters. It allows to minimize the energy consumption of the whole computer, under the constraint of reaching a targeted computational performance[533].

Naturally, the results depend strongly on the qubits fidelity. A gain of a factor of 10 could lead to an energy gain of a factor of 100. The model can help find out the optimal temperature for control electronics. For CMOS type technologies and even with a highly optimistic assumed power consumption of 2mW per qubit[534], room temperature is preferable to run the electronics. It would be similar with higher power drains electronics. The technological constraint then lies in the wiring, which must be simplified, essentially by using advanced (and future) multiplexing techniques. Another option is to use superconducting electronics running at the processor level or at the 4K stage.

With this model, the possibility of an energy-based quantum advantage was investigated. It computes the minimum energy consumption of a fault-tolerant quantum computer to factor an N-bit integer and compare it to the classical record[535].

A classical record was obtained in 2021 by an Inria team on a Germany-based supercomputer, for factoring a 829-bit key[536] with a consumption of 965 GJ, or 1.3MW of power over 8.6 days.

The models show that a quantum computer operating with qubits 2000 times more faithful than Google Sycamore, combined with Steane's code would require 2.7GJ = 2.9MW for 16 minutes, which is the amount of energy contained in about 75 liters of fuel oil. That would be 350 times less energy than used on the supercomputer. Breaking a 2048-bit RSA key is beyond the reach of a conventional supercomputer. On a quantum computer of the same type as above, the energy consumption would be 38 GJ =7 MW for 1.5 hours. Using surface codes error correction would lighten the constraint of qubit fidelities.

Estimations were made for different key sizes in the classical and quantum cases (see Figure 254), giving access to an energy efficiency in each case. An energetic quantum advantage is clear with N=848.

---

[533] All of this modeling comes out of Optimizing resource efficiencies for scalable full-stack quantum computers by Marco Fellous-Asiani, Jing Hao Chai, Yvain Thonnart, Hui Khoon Ng, Robert S. Whitney and Alexia Auffèves, arXiv, September 2022 (39 pages). See also the thesis The resource cost of large scale quantum computing by Marco Fellous-Asiani, November 2021 (215 pages).

[534] A Scalable Cryo-CMOS 2-to-20GHz Digitally Intensive Controller for 4×32 Frequency Multiplexed Spin Qubits/Transmons in 22nm FinFET Technology for Quantum Computers by Bishnu Patra et al, 2020 (4 pages). This consumption model should still be full stack, up to analyzing readout microwaves after traversing parametric amplifiers, HEMTs and ADCs. It is not sure 2 mW are enough to do all of this. One key question to ask is what is the theorical lower bound of microwave packets generation energetic costs?

[535] The method is different from the one proposed in Is quantum computing green? An estimate for an energy-efficiency quantum advantage by Daniel Jaschke et al, May 2022 (12 pages) which compares NISQ systems and their classical emulation equivalent, but not best in-class classical algorithms equivalents. This makes the energetic reasoning incomplete. They also remind us that a quantum advantage comes from maximally entangled states, the overarching question of quantum computing scalability.

[536] See The State of the Art in Integer Factoring and Breaking Public-Key Cryptography by Fabrice Boudot, Pierrick Gaudry, Aurore Guillevic, Nadia Heninger, Emmanuel Thomé and Paul Zimmermann, June 2022 (9 pages).

This energy advantage is different in nature from the computational advantage, which considers only the computation time[537].

Both advantages are thus achieved for different key sizes. Let us recall that the proposed corrector code is resource-intensive and that the result would be much lower with, for example, a surface code.

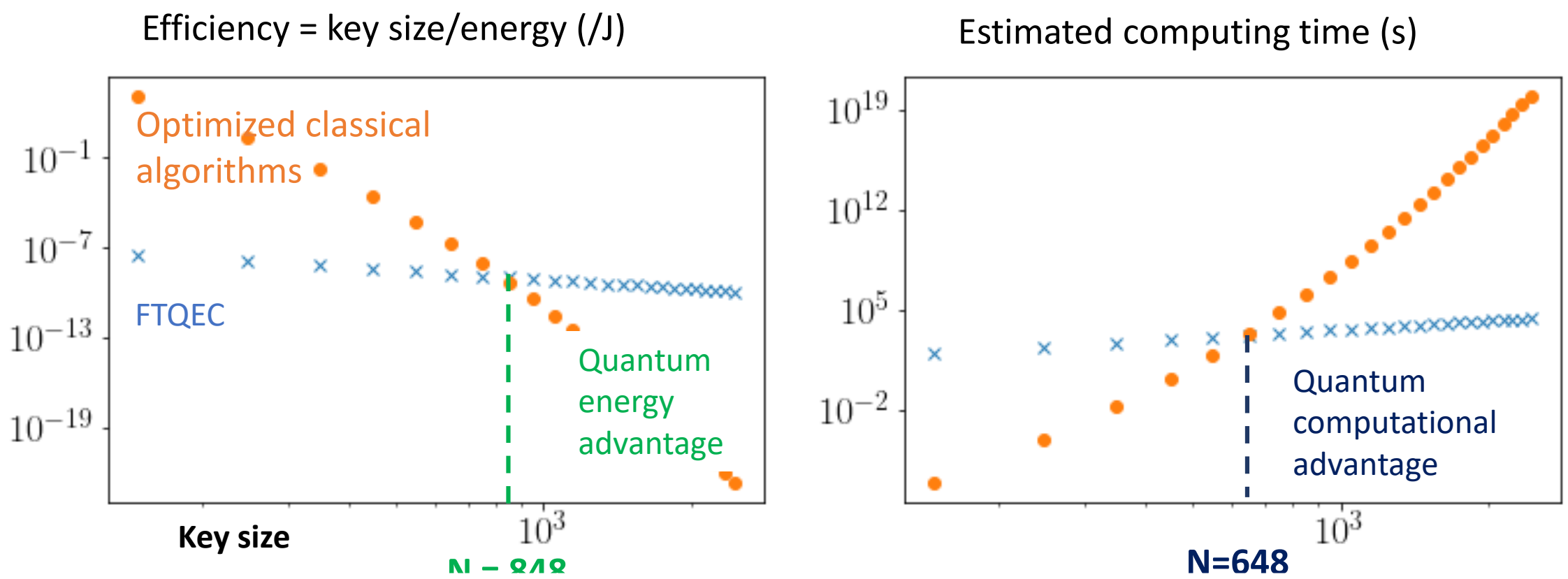

*Figure 254 First results for estimating the quantum energy advantage. Source: Marco Fellous-Asiani, Alexia Auffèves, Robert Whitney.*

This is a theoretical model with many optimistic technology assumptions and complex interdependencies that have to be discussed with many stakeholders, particularly in the enabling technologies vendor space. The beauty of the model is to highlight these interdependencies, which can help make sound choices in quantum computer design.

### Microscopic energetics of quantum technologies

The fundamental quantum level mentioned before is already a rich field of research[538]. The energy and entropy at stake when dealing with quantum systems are the kingdom of quantum thermodynamicians[539]. Some of them even investigate the interconnection and optimization of qubit technologies and the energetic cost of quantum computing and other quantum technologies like quantum communications.

As Kater Murch & al write in a review paper[540] "*Quantum information processing relies on precise control of non-classical states in the presence of many uncontrolled environmental degrees of freedom—requiring careful orchestration of how the relevant degrees of freedom interact with that environment. These interactions are often viewed as detrimental, as they dissipate energy and decohere quantum states. Nonetheless, when controlled, dissipation is an essential tool for manipulating quantum information: Dissipation engineering enables quantum measurement, quantum state preparation, and quantum state stabilization*".

---

[537] This is the topic of The impact of hardware specifications on reaching quantum advantage in the fault tolerant regime by Mark Webber et al, September 2021 (16 pages) which shows that the number of qubits to achieve a given task that is inaccessible to a classical computer depends on the target precision and computing time. 13 to 317 million qubits would be necessary to break Bitcoin signatures (which has no real use case…) and about the same order of magnitude to simulate FeMoCo (with potential use cases in reducing the energetic footprint of fertilizers production). See also Nitrogen, Bitcoin, and Qubits The Shape of Transmons to Come by The Observer, September 2021, and From FeMoco to Bitcoin: Universal Quantum answers two major quantum advantage questions by Universal Quantum, January 2022, which advertises the benefits of trapped ions qubits, and points to Blueprint for a microwave trapped ion quantum computer by Bjoern Lekitsch et al, 2017 (12 pages).

[538] See the colloquium A short story of quantum and information thermodynamics by Alexia Auffèves, March 2021 (14 pages).

[539] See for example Third law of thermodynamics and the scaling of quantum computers by Lorenzo Buffoni et al, March-October 2022 (9 pages) which looks a fundamental issue related to the limits of the preparation of a qubit ground state.

[540] See the review papers Engineered Dissipation for Quantum Information Science by Patrick M. Harrington, Erich Mueller and Kater Murch, February 2022 (28 pages) and Energy dynamics, heat production and heat-work conversion with qubits: towards the development of quantum machines by Liliana Arrachea, May 2022 (63 pages).

There are many quantum thermodynamics concepts in play in qubits inner working and with other quantum technologies. Each qubit technology comes with its own assets and challenges with regard to their energy consumption.

**Superconducting qubits** are a well investigated area where qubits microwaves spontaneous emission is a dissipative process engendering errors and decoherence, with energy exchanges between the qubit and its controlling microwave during a qubit gate operation[541], error mitigation which can make use (among various other techniques) engineering dissipation with energy baths whether it is handled with bosonic qubits like cat-qubits or with programmed error correction, engineering dissipation which can also help efficiently prepare (entangled) Bell states with two qubits, the Zeno effect with measurement backactions, how to optimize the measurement operation with various types of microwave light (single photon, coherent light, thermal light)[542], the connection between quantum measurement and error correction[543] and ways to purify the state of a single qubit with a quantum thermodynamic method[544]. There are also specific cooling mechanisms for superconducting qubits and even some connections between qubit thermodynamics and the way to optimize computing at the compiler level. At last, in the internal debates between the types of superconducting qubits, let's note that fluxonium qubits have a lower energy consumption since being driven by lower frequency microwaves and need less cooling, but at the price of various constraints.

**Silicon spin** energetics are also studied. Their operating parameter and controls are a bit different than with superconducting qubits, with a richer mix of microwave pulses and direct current and operations at a potentially higher temperature. Some quantum energetic advantage can even be found at the scale of one-qubit full added implemented with a three quantum dots silicon spin qubits with a three orders of magnitude gain[545] and with entanglement generation between electron spin and photons in devices mixing static and flying qubits like quantum memories, repeaters and interconnections between quantum computing units[546].

**Trapped-ions** qubits operations can also be optimized with regards to the energetics of their gates[547].

**Photon qubits** are different beasts with regards to the thermodynamics of the whole food chain between single photon generations, entanglement preparation, computing (usually, using MBQC) and photon readout. A first seed of optical computing energetics was launched in Pascale Senellart's C2N team[548]. Also, the H2020 OPTOlogic project aims to create light-induced and controlled topology for energy-efficient logic operations in quantum photonic computing systems.

---

[541] See Energetics of a Single Qubit Gate by J. Stevens, Andrew Jordan, Audrey Bienfait, Alexia Auffèves, Benjamin Huard et al, PRL, September 2021-September 2022 (19 pages).

[542] See Energetic cost of measurements using quantum, coherent, and thermal light by Xiayu Linpeng, Léa Bresque, Maria Maffei, Andrew N. Jordan, Alexia Auffèves and Kater W. Murch, PRL, June 2022 (13 pages).

[543] There are also debates about what are heat and work in quantum thermodynamics and qubits.

[544] See Quantum thermodynamic method to purify a qubit on a quantum processing unit by Andrea Solfanelli, Alessandro Santini and Michele Campisi, March 2022 (5 pages).

[545] See Quantum dynamics for energetic advantage in a charge-based classical full-adder by João P. Moutinho, Silvano De Franceschi et al, July 2022 (18 pages).

[546] See Energy-efficient entanglement generation and readout in a spin-photon interface by Maria Maffei, Andrew Jordan, Alexia Auffèves et al, May 2022 (6 pages).

[547] See Classical Half-Adder using Trapped-ion Quantum Bits: Towards Energy-efficient Computation by Sagar Silva Pratapsi et al, October 2022 (7 pages).

[548] See Coherence-Powered Charge and Discharge of a Quantum Battery by Ilse Maillette de Buy Wenniger, M. Maffei, N. Somaschi, A. Auffèves, P. Senellart et al, February 2022 (19 pages).

And in general, there is a direct link between quantum thermodynamics and physics with the speed of the quantum gates a quantum computer could execute, with fundamental Quantum Speed Limits (QSL)[549].

**Quantum sensing** is also an interesting field of research with regards to quantum thermodynamics and energetics, particularly to find theoretical lower bounds of energy consumption in quantum sensors[550]. Quantum thermodynamics can also help optimize quantum sensors precision.

### About the reversibility of classical and quantum calculations

Here we study the impact of theoretical reversibility of gate-based quantum computing on its energetic cost. We first need to define the notion of logical reversibility of computation and its thermodynamic impact.

**Logical reversibility** of a calculation is linked to the ability to reverse it after one or more operations and recover input data from output data. This can be done at the scale of a classical logic gate or an elementary quantum gate and then up to a complete calculation. If logical reversibility is possible at the level of any gate used, then it becomes ipso-facto doable for a complete calculation[551]. Today's classical computers are logically irreversible. They rely on two-bit logic gates that destroy information since they generate one bit with two bits and we don't keep the information from the two initial qubits. One bit is thrown away every time. You can't reverse a simple NAND, OR or AND logic operation. We could use reversible logic gates that do not destroy information and generate as many output bits as input bits. This would lead to a logically reversible calculation. All of this was theorized by Charles Bennett in 1973 and Tommaso Toffoli in 1980. Classical computing is a big energy spender because logic gates are not logically reversible. The lower bound of energy consumption of current classical computing comes from Landauer's famous limit of kT ln(2) energy dissipated per irreversible bit operation, which can be the erasure of a bit or the merging of two computation paths. Even though we are far off this limit with current classical computing technologies, this lower bound could be avoided with logical reversible computing.

The implementation of this logical reversibility by rewinding calculations would reduce the energetic cost of classical computing, the energy spent in the forward calculation being potentially recovered in the reverse calculation. It was not a chosen path for various reasons. First was the steadiness of Moore's empirical law for many decades. Second is reversible classical architecture have significant overhead in the number of transistors used.

**Thermodynamic reversibility** is another matter and can be obtained when the system is continuously balanced with its thermal bath. It requires handling operations in a quasi-static way, namely, slowly and with logical gates requiring a minimum energy spending. This is the field of adiabatic computing.

Gate-based quantum computing is logically reversible because it uses unitary operations which are all mathematically reversible. Qubits readout is the only logically irreversible operation when it collapses qubit states to a basis state[552].

---

[549] See From quantum speed limits to energy-efficient quantum gates by Maxwell Aifer and Sebastian Deffner, February 2022 (19 pages). It mentions that Amazon Web Services (AWS) classical computing is charged with about $4x10^{-13}$ cents per classical floating point operation when a single quantum circuit evaluation currently costs 1 cent on an AWS-owned Rigetti QPU (pricing source).

[550] See Thermodynamic principle for quantum metrology by Yaoming Chu and Jianming Cai, Huazhong University of Science and Technology, March 2022 (19 pages) and Notes on Thermodynamic Principle for Quantum Metrology by Yaoming Chu and Jianming Cai, Huazhong University of Science and Technology, August 2022 (6 pages).

[551] See these detailed explanations on the reversibility of classical calculus: Synthesis of Reversible Logic Circuits by Vivek Shende et al, 2002 (30 pages).

[552] Measurement Based Quantum Computing, which relies mainly on measurement during the entire calculation, is irreversible by construction. This is why it is also called 1WQC for one way quantum computing.

Qubits readout is reversible only when the qubit states are perfectly aligned with the basis qubit states |0⟩ and |1⟩, i.e., when the readout doesn't change the qubit quantum state.

However, quantum computing is not really thermodynamically reversible. It would be reversible in the absence of noise and if measurements were not changing qubit's internal states. Achieving physical irreversibility would also mandate that all non-quantum qubit control electronics rely on physically and thermodynamically reversible processes or at least be energy-saving operations.

One way to achieve this would be to use adiabatic and reversible electronic components working from within the cryostat, but it's not really possible, particularly at the DAC/ADC levels, given these analog/digital pulse signals conversions are not reversible processes.

Another explored avenue is ABQC for **Asynchronous Ballistic Quantum Computing**, promoted by Michael P. Frank's team at the DoE Sandia Labs in the USA. They plan to implement it with Josephson junction circuits[553].

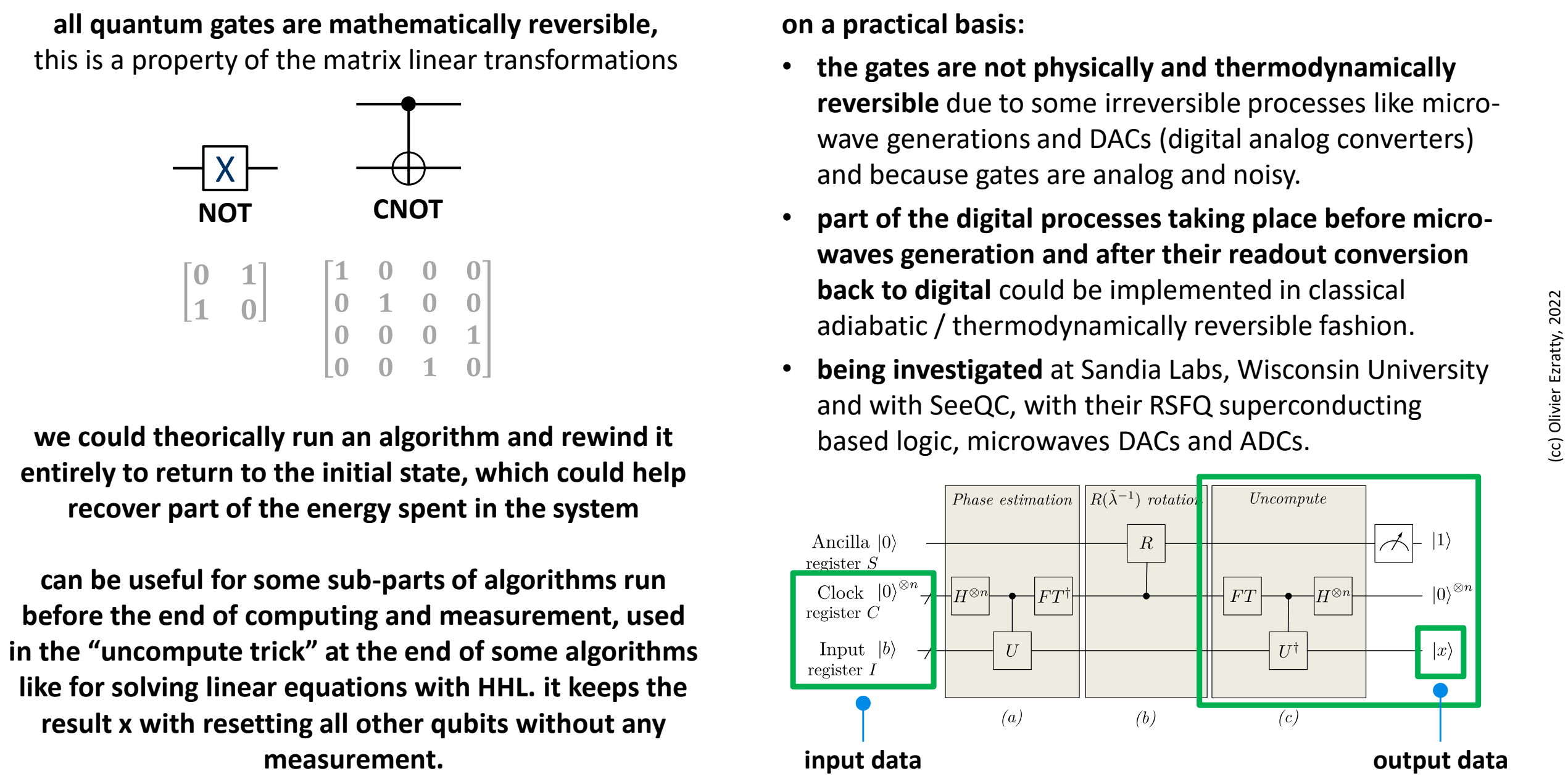

*Figure 255:reversibility in quantum computing. Source: Olivier Ezratty, 2021.*

Quantum reversible computing can also be used in quantum memory and with the uncompute trick of results that are no longer necessary, such as those sitting in ancilla qubits[554]. However, quantum reversibility is not the key to reducing the energy consumption of quantum computing.

### Macroscopic energetics of quantum computing

We'll look here into more details about the "classical" and "macroscopic" power consumption of a quantum computer, taking first the example of a superconducting qubits QPU.

To date, the energy consumption of a quantum computer is relatively reasonable. A current quantum computer with superconducting qubits consumes about 25 kW, of which 16 kW comes from cryogenics. Cold atoms or photons quantum computers consume even less energy, in particular because they do not require cryogenic cooling to 15 mK.

---

553 See Pathfinding Thermodynamically Reversible Quantum Computation by Karpur Shukla and Michael P. Frank, January 2020 (28 slides) and Asynchronous Ballistic Reversible Computing using Superconducting Elements by Michael P. Frank et al, April 2020 (27 slides).

554 See Putting Qubits to Work - Quantum Memory Management by Yongshan Ding and Fred Chong, July 2020.

Photon qubits only require photon sources and detectors cooled at between 3K and 10K which is less energy hungry in face-value provided it scales well with the number of used qubits.

When thousands of qubits will fit in these machines, their power consumption will increase due to the energetic cost of qubit control for initialization, quantum gates, error correction and qubit readout[555]. Most of qubits energetic costs come from the signals used for gate activations and readout. These signals are microwaves pulses, direct current pulses and laser beams. The related spent power seems to increase linearly with the number of qubits. But error correction requires a large number of physical qubits per logical qubits, adding another power consumption multiplying factor. It will depend on the fidelity of the physical qubits and the ratio of physical qubits per logical qubits. The higher the fidelity, the lower this ratio will be. On top of that, the cryogenic cost of the qubits grows very fast as the temperature is lower in proportion of the mass to cool.

Let's breakdown the power consumption of a typical quantum computer:

**Control electronics** power consumption varies greatly from one technology to another and depends on the number of physical qubits managed, which will be counted in millions with large scale quantum computers (LSQC). It is currently high for the control of superconducting qubits based on microwaves produced outside the cryostat with electronics coming from Zurich Instruments, Qblox, Quantum Machines and the likes. Microwave readouts is costly in bandwidth, requiring Gbits/s of data streams per qubit. Microwave production with cryo-CMOS components sitting in the cryostat looks promising and is studied at Google, Intel, Microsoft, CEA-LIST and elsewhere. It can significantly reduce microwave generation related power consumption and may make sense for silicon spin qubits who run at higher temperatures than superconducting qubits and have a higher cooling budget of 500 to 1500µW at 100 mK. Trapped ion-based qubits control is performed with lasers and conventionally generated microwave pulses. With cold atoms, qubits control exploits a couple lasers and an SLM matrix that potentially supports a thousand qubits with modest power consumption. With photon qubits, the power drain seems more important for photon detection (about 7.5W per qubit) than for photon generation (about 1mW per qubit, source: Quandela). Superconducting based photon detectors are more demanding with cooling.

**Cryogenics** consumes up to 16 kW[556] for superconducting and silicon qubits and a little less for other types of qubits due to higher temperatures, such as the 3K to 10K of photon generators and detectors used with photon qubits. Cryogenics will be required for cold atoms at the ultra-vacuum pump level, but will not significantly scale with the number of injected atoms. These are cooled with laser beams and tweezers and under ultra-high vacuum.

The cryogenics consumption is usually continuous, without variations between thermalization and production. Thermalization lasts about 24 hours for dilution refrigerator systems used with superconducting and electron spins qubits.

**Vacuum** is generated with superconducting and silicon spin qubits while cold atoms and trapped ions qubits use ultra-high vacuum. Photons do not need it. With superconducting and silicon qubits, vacuum comes from pumps and dilution refrigeration cooling. Cold atoms require only 100W for the ultra-vacuum pump plus about 300W for its cooling at 4K. Trapped ions systems use heating strips covering the vacuum chamber with a process that can take weeks. This is a fixed cost because when vacuum is in place, heating is stopped, and vacuum remains stable during computations.

---

[555] This is the thesis of Joni Ikonen, Juha Salmilehto and Mikko Mottonen in Energy-Efficient Quantum Computing 2016 (12 pages).

[556] This cooling power usually doesn't take into account the cost of cooling the water circulating in the cryostat compressor.

**Computer control** is used with all types of qubits. They all require one to three control servers that drive the qubit gates and readout devices by exploiting compiled quantum software, that transforms qubit gates into low-level instructions for qubits initialization, control and readout. These servers are networked, either on premise or in the cloud and via conventional network switches. They represent a limited fixed cost with an estimated consumption of between 300W and 1 kW. Part of the control computing could be moved into the cryostat for superconducting and electron spin qubits, in order to implement autonomous error correction codes. The control computer would then only drive logical qubits and not the physical qubits of the configuration.

**Error correction** is an important parameter that will condition the power consumption of a quantum computer. The key parameter is the ratio between the number of physical qubits and logical qubits, which depends on physical qubit fidelities. The higher this one is, the lower the ratio of physical/logical qubits. It also depends on the algorithm size and its target performance. To perform LSQC (large scale quantum computing), the number of qubits to control will be multiplied and generate high energy consumption. However, error correction codes may run up against another wall: the scale dependence of qubit noise. Namely, qubit gates and readout fidelities usually decrease with the number of qubits.

This may have the consequence of reversing the effect of increasing the number of qubits used in error correction. The error rate of logical qubits gates then increases, instead of decreasing[557]. The same problem arises with surface codes although their overhead seems lower than with concatenated codes. For error correction codes to be effective, the error rate of qubits should be at least ten times lower than their current level. In addition, we must also consider the fact that quantum algorithms are executed thousands of times and their result are then averaged. This increases the power drain of a quantum computation because it extends its duration by three orders of magnitude, at least, for the time being. In their work modelling a full-stack energetic cost of a superconducting qubits computer, Fellous-Asiani, Whitney, Auffèves et al found out that, from the energetic footprint standpoint, it is way more efficient to use room temperature electronics than cryo-CMOS due to the overhead cost of their cooling. This result moves the scalability burden cost on the wiring and its multiplexing. On top of that, control electronics have an energetic bill that is much bigger than the cryogeny used for the electronic components sitting in the cryostat (cables, filters, attenuators, qubit chipsets, circulators, amplifiers).

Many of these quantum computer components have a variable energy cost depending on the number of qubits, including the cryogenic side. Indeed, the electronics embedded in cryostats release heat in approximate proportion to the number of physical qubits used. This heat must be evacuated within the cryostat. The consumption of the control electronics also generally depends on the number of qubits. It seems that, up to a thousand qubits, this control electronics is a fixed cost for cold atoms. Only vacuum creation and the control computer seem to be fixed costs.

In the entrepreneurial scene, it's interesting to observe that the total power consumption of a quantum computer recently starting to become a selling point, although not yet being perceived as being an important one. For example, AQT (trapped ions) explain that their 20-qubit system can be attached to a regular 220V/110W plug with their <2kW power drain, similar to a kitchen oven. Pasqal is using a similar selling point for its Fresnel quantum simulator.

Creating a scalable quantum computer is clearly an optimization problem taking into account many energetic constraints. Qubits systems that operate at cryogenic temperature are constrained by the cryostat cooling power and by the heat released within the cryostat.

---

[557] This is what comes out of Limitations in quantum computing from resource constraints by Marco Fellous-Asiani, Jing Hao Chai, Robert S. Whitney, Alexia Auffèves and Hui Khoon Ng, PRX Quantum, November 2021 (8 pages).

Superconducting and silicon spins qubits are the most challenging for that respect. Heat is generated by the inbound cable microwave attenuation filters and in the qubit readout related microwave amplifiers. In addition, the part of microwave generation and readout systems that is integrated in the cryostat have their own thermal footprint. All this must fit into the current thermal budget of the cryostats that is currently limited to 1W at the 4K stage and to about 25 µW at the lower 15 mK stage. We will probably be able to create even more powerful cryostats with more pulse heads and as dilutions with a gain of an order of magnitude for the available cooling power. Other optimizations can be implemented to increase the available cooling power at very low temperature with getting closer to the theoretical Carnot efficiency. which seems possible with large cryostats. These are still constraints for the scalability in number of qubits, particularly for LSQC (large scale quantum computers) which will require millions of physical qubits!

Several options are investigated to reduce the power consumption of the qubits-related classical electronics. An interesting one is superconducting components such as those from SeeQC. D-Wave has integrated its own superconducting controls in its own quantum processor. With cryo-CMOS control electronics, the heat dissipation is greater.

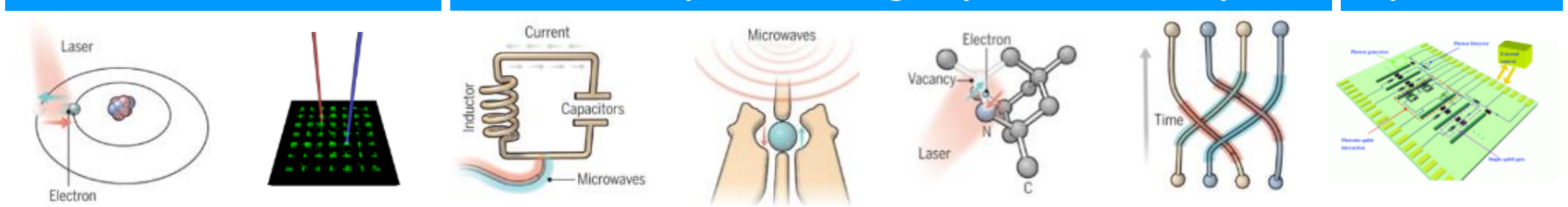

| | atoms | | electron superconducting loops & controlled spin | | | | photons |
|---|---|---|---|---|---|---|---|
| **qubit type** | **trapped ions** | **cold atoms** | **super-conducting** | **silicium** | **NV centers** | **Majorana fermions** | **photons** |
| cryogeny | <300W | N/A | 16 KW | 12 KW | < 1 KW | 16 KW | 3KW |
| vacuum pumps[1] | Vacuum | ultra-vacuum 100W | vacuum | vacuum | vacuum | vacuum | vacuum |
| qubits gate controls | <1.4KW ions heating, lasers, micro-aves generation, CMOS readout electronics | 5,8KW atoms heater, lasers, control (SLM, etc) and readout image sensor + electronics | 1 to 5 KW depends on architectures with micro-wave generation outsideor inside the cryostat | | N/A | N/A | 300 W for photons sources and detectors, qubit gates controls |
| computing | 300W | 1 KW | 1 KW | 1 KW | <1 KW | 1 KW | 700 W |
| # qubits used | 24 | 100-1000 | 65 | 4 | N/A | N/A | 20 |
| **total** | **2 KW** (4) | **7 KW** (1) | **25 KW** (2) | **21 KW** | **N/A** | **N/A** | **4 KW** (3) |

[1] : fixed energetic cost, for preping stage

typical configurations for Pasqal (1), Google (2), Quandela/QuiX (3), AQT (4) rough estimates for others

*Figure 256: current quantum computers total power and decomposition. (cc) Olivier Ezratty, 2022.*

The other way to be less constrained is to run the qubits at higher temperatures. This is what is possible with silicon spin qubits, which only require a temperature between 100mK and 1K instead of 15 mK for superconductors. This increases the thermal budget for the control electronics at the qubit stage.

Some significant engineering is required to optimize a multi-parameter system, at least with superconducting and silicon spin qubits:

1. Physical scalability requires putting as much as possible **qubits control electronics inside the cryostat**… but it is not efficiency energy wise unless control electronics are of the superconducting breed (SFQ).
2. These electronics thermal dissipation is **constrained** by the available cryostat cooling power.
3. Two paths must be investigated simultaneously: increase the available **cryostat cooling power** and reduce these **electronics thermal footprint** as low as possible.

4. Find an efficient way to **handle digital communication** between the inside and outside of the cryostat. Fiber optics, wireless, multiplexing, up/down signal conversions, whatever!
5. Look at various ways to **reduce qubits power drain**, with optimizing their own quantum thermodynamics, particularly when implementing error correction codes. It can also come from algorithm and compiler designs. Reducing the number of physical qubits per logical qubits is an option pursued for example with bosonic codes and cat-qubits who are self-correcting flip errors.
6. With **scale-out solutions** involving connecting several quantum computing processing units with some microwaves and photonic links, look at the energetic footprint of this connectivity, on top of its probable impact on qubit links fidelity.

**Energetic cost of distributed architectures**

The temptation is great to create ever larger quantum computers, with giant cryostats in the case of superconducting qubits, like we'll see with IBM and Google's roadmaps. Another approach would be to create distributed architectures of quantum computers linked together by quantum connection based on entangled photons, a choice made by IonQ with their trapped ions qubits, noticeably because it is difficult to scale these qubits beyond a couple dozens.

In theory, this would make it possible to create computing clusters that, seen from the outside, would create a single computer, a bit like a large classical server cluster.

This will be conditioned by the capability of converting qubits states to photons qubits states and by the resulting qubit connectivity between the various quantum processors units of this quantum cluster. But this is not just about "connecting" qubits. Interconnect architectures are about creating fragile entangled states between qubits using the intermediary of photons. These may create some statistical overhead on their own, which has to be boiled in for both assessing the real obtained quantum computing scalability and the related energetic footprint.

**Use cases energetic assessment**

Another longer-term question deserves to be asked: does the potential energetic advantage of quantum computing depend on algorithms and applications? What will happen if and when quantum computing becomes widespread? Are we finally going to create a new source of energy consumption that will be added to existing sources, which are already growing fast in the digital world? What will be its impact? How can it be limited?

At this stage, it is too early to have a clear idea. Answers will largely come from the emergence or not of quantum solutions for volume applications, such as autonomous vehicle routing or personalized health solutions.

Without volume-oriented applications, quantum computers will be dedicated to niche applications equivalent to those of current supercomputers, which are mainly used in fundamental and applied research or for public services like weather forecasts.

On their end, volume applications will only be achievable once the quantum computing scalability will work and millions of low-noise qubits can be operated. This scalability will probably come from fixing some of energetic consumption issues of quantum computing. And we'll close the loop!

Then, we'll have to look at the externalities of these applications and potential Jevon's effects. Namely, some new solutions will have a given quantum computing energetic cost but may help reduce the environmental footprint in other domains like in transportation. If it's well balanced, that's fine. If, on the other hand, the externalities are not positive, like, say in finance portfolio optimization tasks, you will have to think about it.

# Economics

Given we are at the very early stage of the quantum computing era, it's still difficult to assess the economics of this industry. It's too small to generate economies of scale giving some indications on the cost and price of a regular quantum computer. Still, we can make some projections based on a couple assumptions.

The only "priced" quantum computers on the market today are coming from D-Wave. Their units are priced at about $14M. They have sold only a few of these. Most D-Wave customers are using D-Wave computers sitting on the cloud either with D-Wave itself or with Amazon. Some customers pay in excess of $200K per year to benefit from a premium access to these machines. As far as we know, the other "volume" manufacturer of quantum computers is IBM, but they haven't sold any unit so far, at least publicly. They installed a couple ones in Germany, South Korea, Japan and Canada in their own facilities, to serve these markets through various local research, university and industry partners. The rest is provided through their own cloud services.

One can economically make a distinction between **cost** (of R&D, goods and manufacturing), **price** (how much is it sold or rented) and **value** (what value is it bringing to customers, particularly, compared with existing classical computing solutions). Right now, the equation is simple: costs are high, prices are high as well when computers are sold (particularly superconducting qubits ones) and value is low at this point, and is positioned in the educational and proof-of-concept realms.

A quantum computer cost and price depend on several parameters including its underlying R&D, bill of materials of off-the-shelf and custom-designed components, manufacturing and integration costs, economies of scale, marketing and sales costs, the cost of maintenance and consumables if any, and finally, the manufacturer's profit. The higher the sales volume, the greater the economies of scale. Volumes are currently very low given most quantum computers are just prototypes that are not yet useful for production-grade applications.

At some point, when and if we reach some quantum advantage threshold, useful applications will emerge. It will first target niche b2b and government markets[558]. Then, when applications and innovation ramp-up, we may have a larger number of corporate users. It will justify scaling manufacturing capacities. R&D fixed costs will then be easier to amortize with volume. Cost of goods may also decrease, particularly if technology progress can help get rid of the complicated wirings and electronics that we have today in some of these devices.

Let's look one by one at the major hardware components of a quantum computer looking at how it will benefit from economies of scale:

- **Control computer(s)**: these are standard rack-mounted servers as well as the associated networking connection. These are the most generic parts of a quantum computer.
- **Chipset**: quantum registers chipsets are the cornerstone of electron-based quantum computers, such as with superconducting and electron spin qubits. Even if they are manufactured in CMOS or similar technologies, their manufacturing volume is very low. Economies of scale are therefore almost non-existent. You don't need such components with cold atoms and trapped ions qubits. It is replaced by specialized optical components to direct the laser beams controlling the qubit atoms. With NV centers, chipsets can be cheap to manufacture if done in volume.

[558] Some economists think that quantum computers may offer an economic advantage compared to classical computing even without reaching a computing advantage, thanks to asymmetries in cost structures. This still is conjecture based since these economists didn't really analyze the real possibility of pre-quantum-advantage NISQ computers to bring any usefulness. The proposed model is only based on economies of scale and the effects of competition. See Quantum Economic Advantage by Francesco Bova, Avi Goldfarb and Roger G. Melko, National Bureau of Economic Research, February 2022 (28 pages).

- **Electronic components**: these are used to create, process, transmit and send the quantum gate signals to the qubits. Their technology depends on the type of qubit. These signals are microwaves for superconducting and electron spins qubits, laser-based photons for cold atoms and trapped ions, and some other varieties of electro-magnetic signals otherwise. Standard and expensive laboratory equipment are used for microwave generations such as those from Rohde&Schwarz. More integrated equipment is sold by companies like Zurich Instruments, Qblox, Quantum Machines and Keysight. It's using customized FPGA and rather standard electronic components. When these tools are miniaturized as ASICs running at room temperature or cryo-CMOS running at temperatures below 4K, their small economies of scale make it rather expensive.
- **Cabling**: niobium-titanium superconducting cabling used to feed superconducting and electron spin qubits with microwaves are very expensive, costing about $3K each. And we need about three such cables for each and every qubit. This drives high-costs for manufacturing superconducting and electron-spin based qubits systems. The main companies providing these cables are Coax&Co and Delft Circuits.
- **Cryogenics**: these are standard systems but marketed in low volumes. They can cost up to $1M for superconducting and silicon qubits. Their cost is one to two orders of magnitude lower for the cryogenics of components such as photon qubits. Large cryostats use an enclosed cooled system with many cylindrical layers of protection, a GHS (gas handling system), a compressor (such as those coming from CryoMech and Sumitomo) and another compressor used to cool the water feeding the primary compressor.
- **Consumables**: in quantum computers operating at very low temperatures, there is at least some liquid nitrogen and gaseous helium 3 and 4. The latter two are not consumables and operate in a closed circuit in dry dilution systems. It's still expensive.
- **Casing**: this is about steel, glass and design with some specifics linked to vibrations isolation.

As quantum technologies mature, some cost structures will increase and others will decrease. Economies of scale will do the rest. We can therefore forecast that the $15M D-Wave computer price tag will remain for some time in the top range of quantum computers prices, at least at superconducting qubits. Some computers will be less expensive than $1M. In practice, many quantum computers will be usable as resources in the cloud and at a more moderate cost. This is what IBM, Rigetti, D-Wave, Microsoft and Amazon (with third-party machines for the latter two) are already offering. Microsoft and Amazon quantum cloud pricing is already quite high. Then, one can wonder about its added value.

## Quantum uncertainty

Estimating if and when scalable and useful quantum computers will be available is a difficult art and science. The opinion spread between optimists and pessimists is quite large. Some entrepreneurs expect to achieve miracles in less than one decade while some scientists, on the other hand, think that this will never happen. In between, other scientists are moderately optimists and expect the wait to last at least a couple decades. Let's look at these various opinions.

### Optimism

Google said it achieved quantum supremacy in October 2019. It was not a true supremacy since their Sycamore setting was doing no programmable computing solving a specific problem. It was found later that, due to the qubit noise in their system, it was relatively easy to emulate it on a classical server cluster, and discussed here in this book, starting page 680. So much for any quantum supremacy or advantage! It was the same with the so-called boson sampling experiments quantum advantages coming from China in 2019 and 2020. These were unprogrammable random photon mixers. Later boson sampling experiments in 2021 and 2022, like withy Xanadu, were programmable, but did not show a practical computing advantage.

As published in their 2020 roadmaps, Google, IBM and Amazon expect to achieve true quantum supremacy relatively quickly and create a quantum computer with 100 logical qubits in less than a decade.

**Kenneth Regan** thought in 2017 that an industry vendor - probably Google - would claim to have reached quantum supremacy in 2018 and that it would quickly be contradicted by the scientific community[559]. This happened in 2019 and the contradictions came fast. That was quite a good prediction! For the specialists who can dissect their scientific publications, the view is obviously more nuanced, especially concerning the reliability of the qubits they generate. They communicate a lot about their efforts to reduce the noise of qubits to make them more reliable[560].

**Alain Aspect** does not see any strong scientific obstacle preventing the creation of reliable quantum computers. He believes that the uncertainty is mostly a technological and engineering one, but that it will take a few decades to create one reliable advantage-grade quantum computer. However, there is nothing to prevent this process from being accelerated, if it is fueled by good talent and public/private investments. **John Preskill** has the same opinion: it will work, but it will take several decades. **Nicolas Gisin** estimates that the pace to quantum usability is accelerating [561]. **Jian-Wei Pan** is even more optimistic, forecasting some regular quantum advantage before 2027 [562].

Optimists also include the many hardware quantum computing startups, all with solutions that are expected to work on a large scale within the next five years. They are found in all types of qubits: superconductors (IQM, QCI), electron spins (Quantum Motion, SQC), cold atoms (Pasqal, Atom Computing, ColdQuanta, QuEra), trapped ions (IonQ, Quantinuum, Universal Quantum) and photons (PsiQuantum which predicts one million qubits in less than five to ten years, Orca Computing, Quandela, Xanadu).

At last, you have ultra-optimists like **Michio Kaku**, a Japanese physicists turned into a futurist and best-selling author and who cocreated the string field theory seems affected by a variant of the Nobel disease. His upcoming book “Quantum Supremacy”, to be published in 2023, predicts that quantum computing will soon “*solve some of humanity's biggest problems, like global warming, world hunger, and incurable disease*”. That’s a little overpromising!

### Pessimism

Pessimism comes from a few researchers, who are not necessarily specialized in the field in which they express themselves. Above all, they are pessimistic about the ability to fix the noise that affects qubits, whatever their type[563].

The first and best-known of these pessimists is the Israeli researcher **Gil Kalai** who believes that we will never be able to create quantum computers with a low error rate[564]. According to him, we cannot create stable quantum computers because of the noise that affects the qubits. This is illustrated in the scale below in Figure 257, which sets the lowest reasonably achievable noise level well above the level required to create a scalable quantum computer.

---

[559] In Predictions we didn't make, January 2018.

[560] See The Era of quantum computing is here.Outlook: cloudy by Philipp Ball, in Science, April 2018.

[561] See Quantum computing at the quantum advantage threshold: a down-to-business review by A.K. Fedorov, Nicolas Gisin, Sergei Beloussov et al, March 2022 (55 pages).

[562] See Jian-Wei Pan Sees Routine Quantum Advantage Within Five Years by Matt Swayne, The Quantum Insider, February 2022.

[563] See The different forms of quantum computing skepticism by Boaz Barak, 2017.

[564] See Why Quantum Computers Cannot Work, 2013 (60 slides) illustrating How Quantum Computers Fail: Quantum Codes, Correlations in Physical Systems, and Noise Accumulation, 2011 (16 pages) and The Argument Against Quantum Computers by Katia Moskwitch, February 2017. Gil Kalai declares: "*My expectation is that the ultimate triumph of quantum information theory will be in explaining why quantum computers cannot be built*".

He is working on the creation of some mathematical model that would prove the impossibility of overriding these errors, even with quantum error correction codes. In 2022, he published another paper to prove his point, with a philosophical approach related to the notion of free will[565].

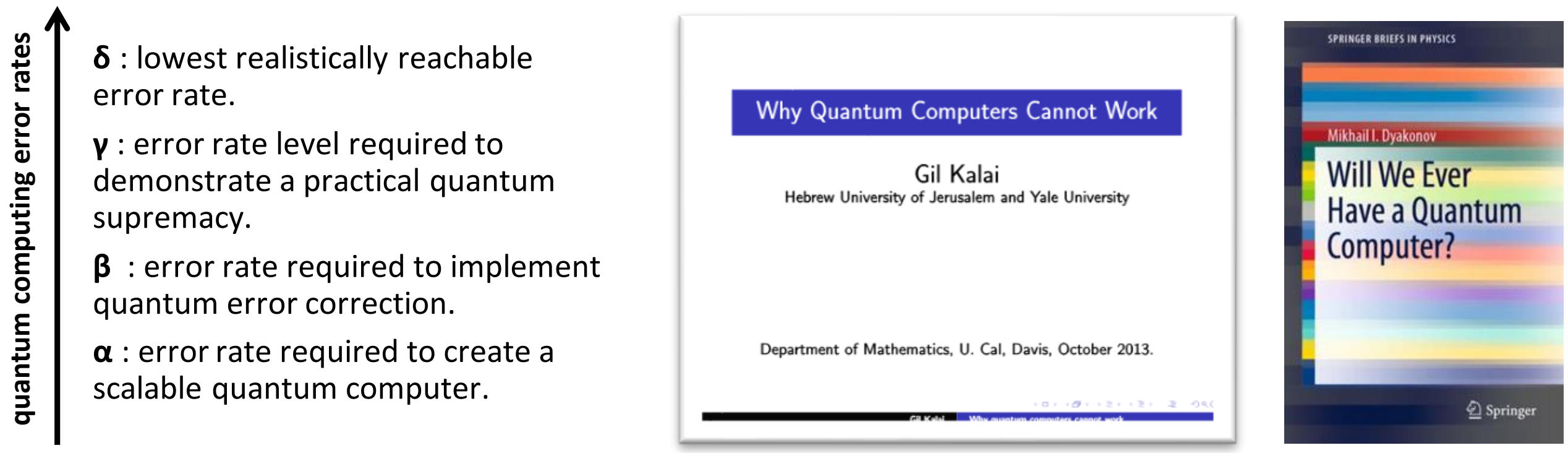

*Figure 257: Gil Kalai's quantum computing errors complexity scale.*

Another skeptic of quantum computing is the **Mikhail Dyakonov** (born in 1940 in the USSR) who works in the Charles Coulomb Laboratory (L2C) of the CNRS and the University of Montpellier in France. He detailed his views in an article at the end of 2018, which he later turned into a book[566]. His argument is more intuitive but less documented than the work of Gil Kalai[567].

We also have **Serge Haroche** for whom universal quantum computing is an unreachable dream, also because of that damned noise. On the other hand, he thinks that the path of quantum simulation, especially based on cold atoms, is reasonable and realistic.

**Xavier Waintal** (CEA-IRIG in Grenoble, France) also has serious reservations about the possibility of creating large-scale quantum computers. Here again, the culprit is noise. His reasoning is based on physical explanations different from those of Gil Kalai. Qubit operations are relying on very complex n-body quantum problems and error correction codes generally deal with only two types of errors (flip, phase) but not with all sources of error. He recommends exploiting the mean-fields theory which allows to model the complex interactions between qubits and their environment[568]. These are very fundamental questions to address. I'd say his arguments are both the most documented I've seen but which, well used, may fuel interesting research to find solutions.

**Cristian Calude** and **Alastair Abbott** point out that the advantage of the main quantum algorithms usable in practice would generate a modest quadratic acceleration (square root of classical computing time) that could be achieved on classical computers with heuristic approaches[569].

Quantum skepticism is also evident in **Ed Sperling's** November 2017 review of the field, which included a reminder of all the obstacles to be overcome[570].

---

565 See Quantum Computers, Predictability, and Free Will by Gil Kalai, April 2022 (33 pages).

566 See The Case Against Quantum Computing, 2018. He even made a short book about it, Will We Ever Have a Quantum Computer?, 2020 (54 pages, free download). As well as a debate on the subject launched by Scott Aaronson in Happy New Year! My response to M. I.Dyakonov. See also Skepticism of Computing by Scott Aaronson who dissects 11 objections on quantum computing capabilities. See also Noise stability, noise sensitivity and the quantum computer puzzle by Gil Kalai, 2018 (1h04mn).

567 See a response to this argument in The Case Against 'The Case Against Quantum Computing' by Ben Crige, January 2019 and a highly documented response from Scott Aaronson in Happy New Year! My response to M. I. Dyakonov, 2013.

568 See What determines the ultimate precision of a quantum computer? by Xavier Waintal, 2017 (6 pages) that we have already mentioned. Xavier Waintal has notably developed classical algorithms for the simulation of N-body problems. They are used by various teams of researchers in condensed matter physics, notably those working on topological matter and Majorana fermions. They run on laptops and supercomputers.

569 In The development of a scientific field by Alastair Abbott and Cristian Calude, June 2016.

570 In Quantum Madness by Ed Sterling, November 2017.

**Leonid Levin** (a Russian scientist who defined the NP complete complexity class in 1973) and **Oded Goldreich** (an Israeli professor in computer science from the Weizmann Institute) are other quantum computing skeptics mentioned by Scott Aaronson in [Lecture 14: Skepticism of Quantum Computing](#).

Another argument against scalable quantum computing deals with the computational state vector amplitudes values becoming tiny as the number of qubits grows. After just applying a set of H gates on N qubits, this amplitude becomes $1/2^N$ for each computational basis state in the qubits register. It becomes quite small as N grows beyond 50. Are these values corresponding to some physical observable that would have a value way below the physical error rate in the system? Or even below some physical Planck constant? Well, this is good food for thought. At least, the computational state vector always has a norm of 1. And the physical observables in the system remain based on the individual qubits basis states $|0\rangle$ and $|1\rangle$.

> *"For me, the most important application of a quantum computer is disproving the people who said it's impossible. The rest is just icing on the cake."*
> Scott Aaronson
> 2019.

### Managing uncertainty

One key challenge is to make a distinction between scientific unfeasibility, scientific uncertainty and technological uncertainty. This set of uncertainties raises existential questions about how to manage such a long innovation cycle. When should we invest? When are market positions being settled? Is fundamental research decoupled from industrialization? Is quantum computing a simple engineering matter? Or, on the other hand, will it be impossible to control very large swaths of maximally entangled physical qubits?

Note that the pessimists quoted above are not Americans and most of the optimists are. Is there a cultural bias here? These variations in innovation and economic cultures have an impact on industry approaches. Major IT companies such as IBM, Google, Intel, Amazon and Microsoft can fund quantum computing R&D investments with a very long-term vision. They have the profitability, the cash and the ability to attract skills to do so. You may still note that, at this point in time, these large IT vendors have not yet engaged in a startup acquisition spree like they did in the fields of artificial intelligence and other emerging technologies.

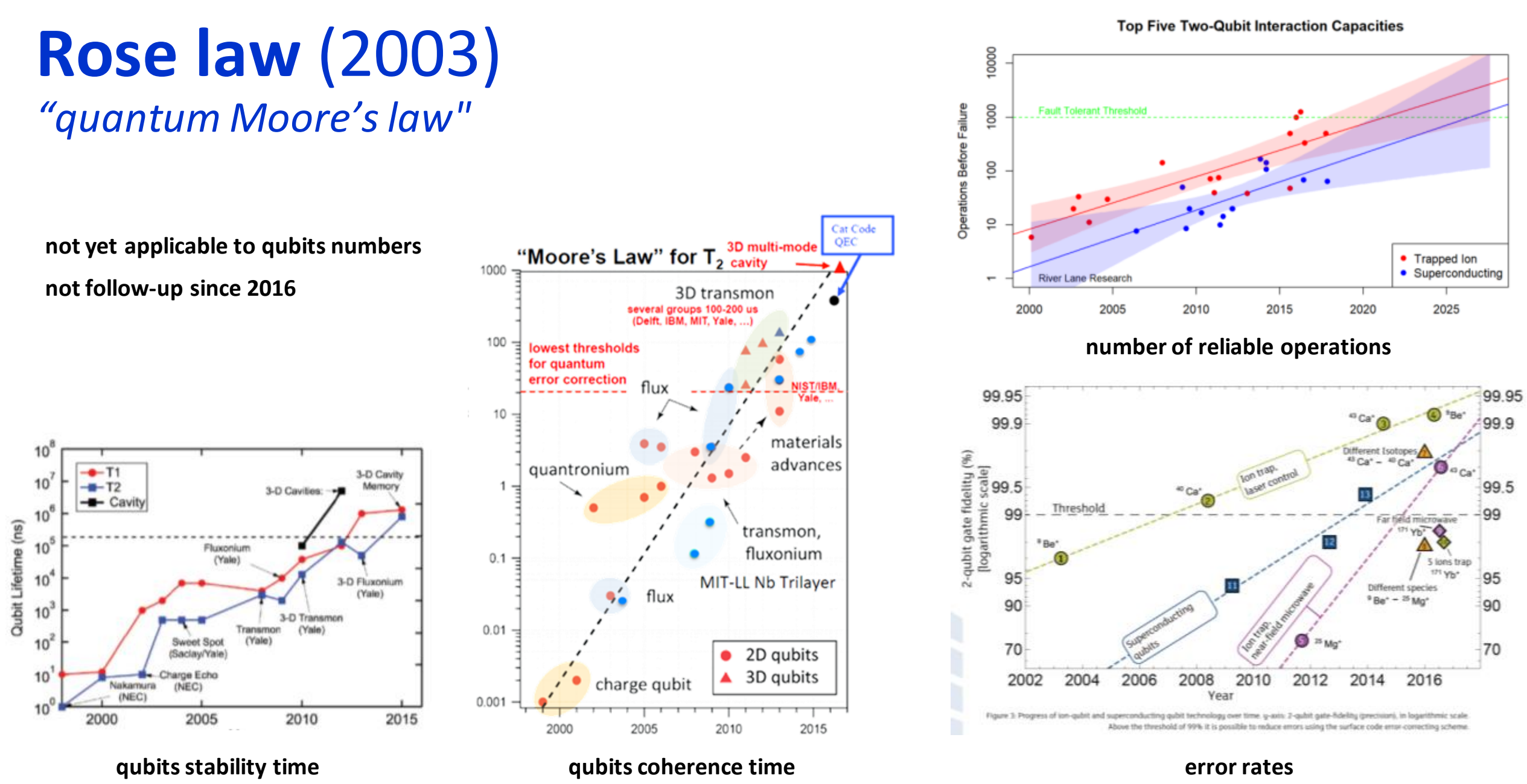

*Figure 258: a compilation of the putative equivalents of Moore's law in quantum computing. They all need updates! Otherwise, you can't prove there's a real ongoing acceleration of progress in quantum computing science and technology.*

Well-funded startups such as D-Wave, Rigetti, IonQ, PsiQuantum, OQC and IQM can also adopt a fairly long-term view, even if it still depends on their ability to commercialize quantum computer prototypes and to attract long-term oriented investors. The corresponding amounts are not necessarily crazy.

The engineering problems to be solved deal with qubits materials and manufacturing techniques, quantum error correction, control electronics, large-scale cryogenics and of course algorithmic and software advances. The required approach is multidisciplinary with mathematics, fundamental quantum physics, thermodynamics and chemistry, and finally, code, including machine learning which is notably used for qubits calibration.

We can try to extrapolate the evolutions of the last ten years in quantum computing. When he was the co-founder of D-Wave, Geordie Rose enacted in 2003 his own equivalent of Moore's empirical law, Rose's Law, predicting a doubling every year of the number of qubits in a quantum computer, as show in Figure 258. Since 2007, D-Wave delivered well on this promise but the progress has been sluggish for gate-based quantum computers.

Most of these charts have not been updated and some even include numbers corresponding to nonoperational systems like Google's 2018 72-qubit Bristlecone or IonQ's 129 qubits which never saw the day of light. You then understand why you must be cautious when interpreting these "exponential charts" with looking at a similar chart created in 2015 that positioned NMR qubits as best-in-class fort their scalability potential[571]. In reality, NMR qubits didn't really scale well.

Some exponential law can however be observed in the evolution of other operating parameters of quantum computers such as the stability time of qubits, their error rate and the number of consecutive operations performed reliably[572]. Recently, **Rob Schoelkopf** from Yale University created his own law showcasing the progress with superconducting qubits coherence times and gates fidelities and times.

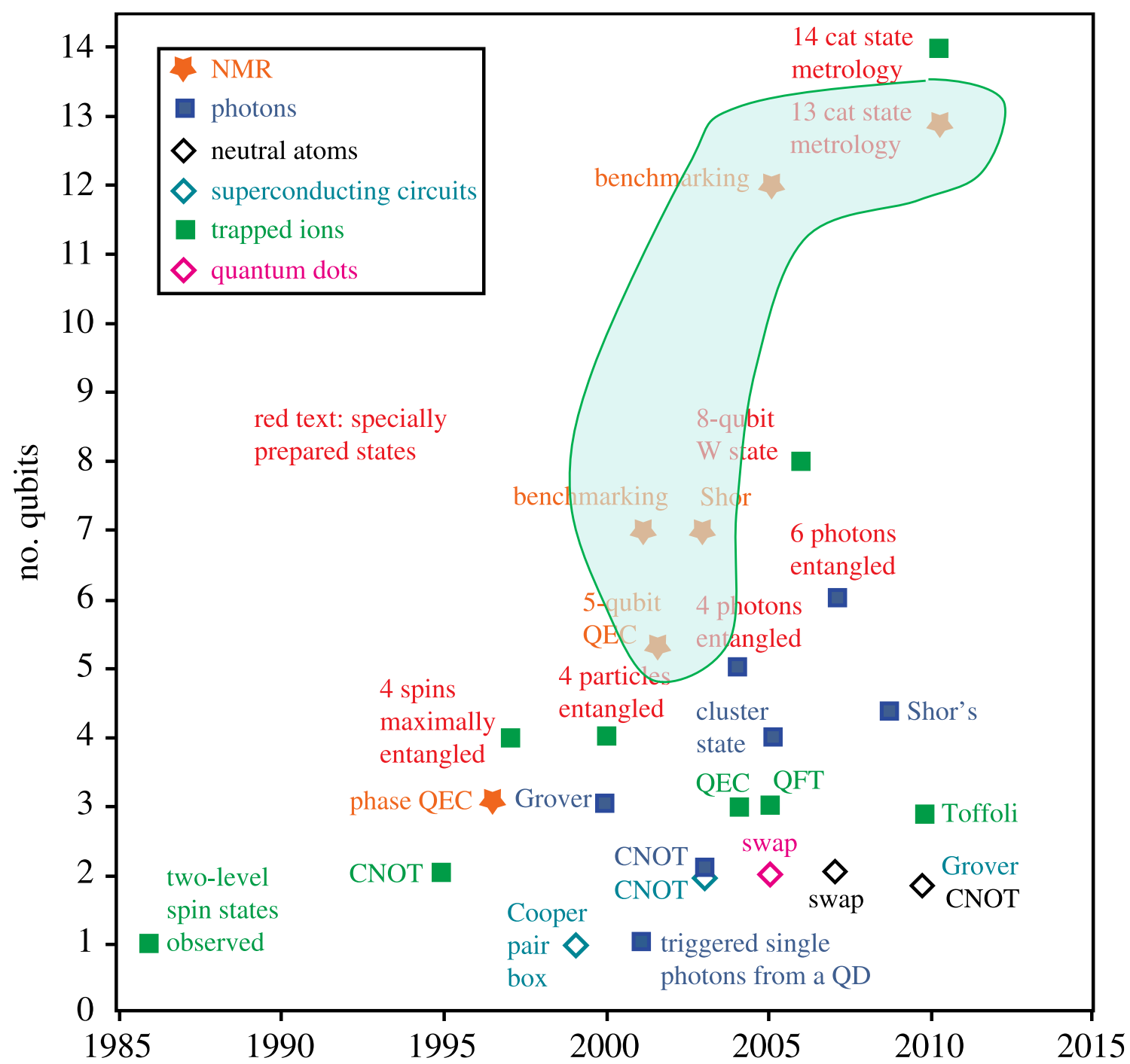

*Figure 259: a chart with number of qubits per technology and year, as of 2015. It gave the impression, back then, that NMR qubits were the most scalable. They are not! Source: Recent advances in nuclear magnetic resonance quantum information processing by Ben Criger, Gina Passante, Daniel Park and Raymond Laflamme, The Royal Society Publishing, 2015 (16 pages).*

I tried to understand why the predictions of creating viable quantum computers were always quite long-term, between 5 and 50 years. One of the answers comes from the length of cycles in the associated research and manufacturing processes.

[571] See Recent advances in nuclear magnetic resonance quantum information processing by Ben Criger, Gina Passante, Daniel Park and Raymond Laflamme, The Royal Society Publishing, 2015 (16 pages).

[572] Some of the diagrams below come from the Technical Roadmap for Fault-Tolerant Quantum Computing, a UK report published in October 2016 and from this other source.

For example, creating a prototype silicon-based qubit chipset takes up to nine months of manufacturing with up to 160 manufacturing steps. After this manufacturing process, component characterization and packaging stage add more time.

Components characterization qualitatively tests and selects the manufactured components. This can last up to a month and in the best case down to a week. Then, to carry out the tests, the thermalization of the computer takes about 24 hours and the change of the chipset to be tested takes at least 3 to 7 hours as we'll discovered in the section on cryogenics, page 473.

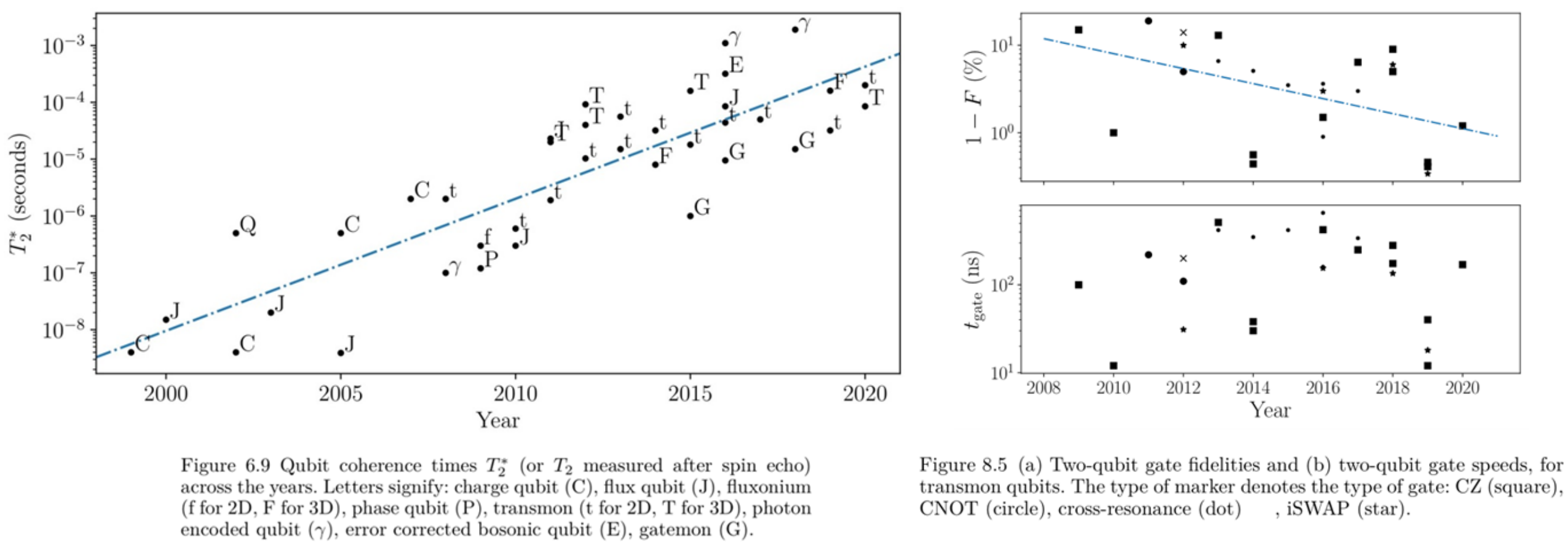

*Figure 260: Rob Schoelkopf charts on progress with superconducting qubits coherence times and gates fidelities and times, Twitter source (didn't find a better one), August 2020.*

The design to manufacturing whole cycle lasts about 2,5 years. Finally, the *test & learn* cycles are often very long, much longer than with software! This long cycle may be shorter with other solid-state qubits like supercomputing qubits, and also with photon qubits for which semiconducting photonic circuits are also required.

## Challenges ahead

Whether you are an optimist or a pessimist with regards to the advent of scalable quantum computers, you need to adopt an educated view of the challenges ahead. Over time, as my understanding of these challenges was growing, I tended to shift from "optimism" to "neutralism" or, at least, to being a "documented optimist". Some of the challenges ahead are indeed enormous.

Xavier Waintal uses the scale shown in Figure 261, with 5 difficulty levels, to build a quantum computing machine. It positions where we are right now and the challenges ahead. It goes beyond large scale computing given some quantum memory would be mandatory for some key algorithms like QML and HHL.

| difficulty scale | technology | use cases | examples |
|---|---|---|---|
| 1 | quantum simulator (analog-no gates) | quantum simulations | D-Wave, Pasqal |
| 2 | gates-based analog systems, low fidelity | system validation NISQ algorithms | Google, IBM, Rigetti, … |
| 3 | gates-based analog systems, low fidelity | variational calculations in quantum chemistry | Possibly PsiQuantum |
| 4 | ideal quasi-deterministic gates-based systems (FTQC/LSQ) | factoring large numbers, exact quantum chemistry | TBD |
| 5 | 4 + quantum memory | quantum machine learning, linear algebra (HHL) | TBD |

*Figure 261: Source: Xavier Waintal.*

My chart in Figure 262 goes into some details with laying out some of these challenges, most of which being covered extensively in various parts of this document.

Two things come to mind: one is that quantum computers scalability is the most challenging issue to tackle with. If quantum computing capacity is known to theoretically scale exponentially with the number of qubits, you may wonder whether the scale challenge itself is also an exponential one.

One way to grasp it is to look at IBM and Google's progress with their superconducting qubits. It's been sluggish since 2019 with 72/127 qubits, given that most benchmarks show that only fewer than 20 qubits are practically usable[573]. There's still some hope with bosonic codes and cat-qubits which could limit the logical/physical scaling overhead. Also, scale-out options with qubits interconnect options using microwaves or photons are interesting but have their own scalability challenges. Other qubit types like electron spins and photons also look promising.

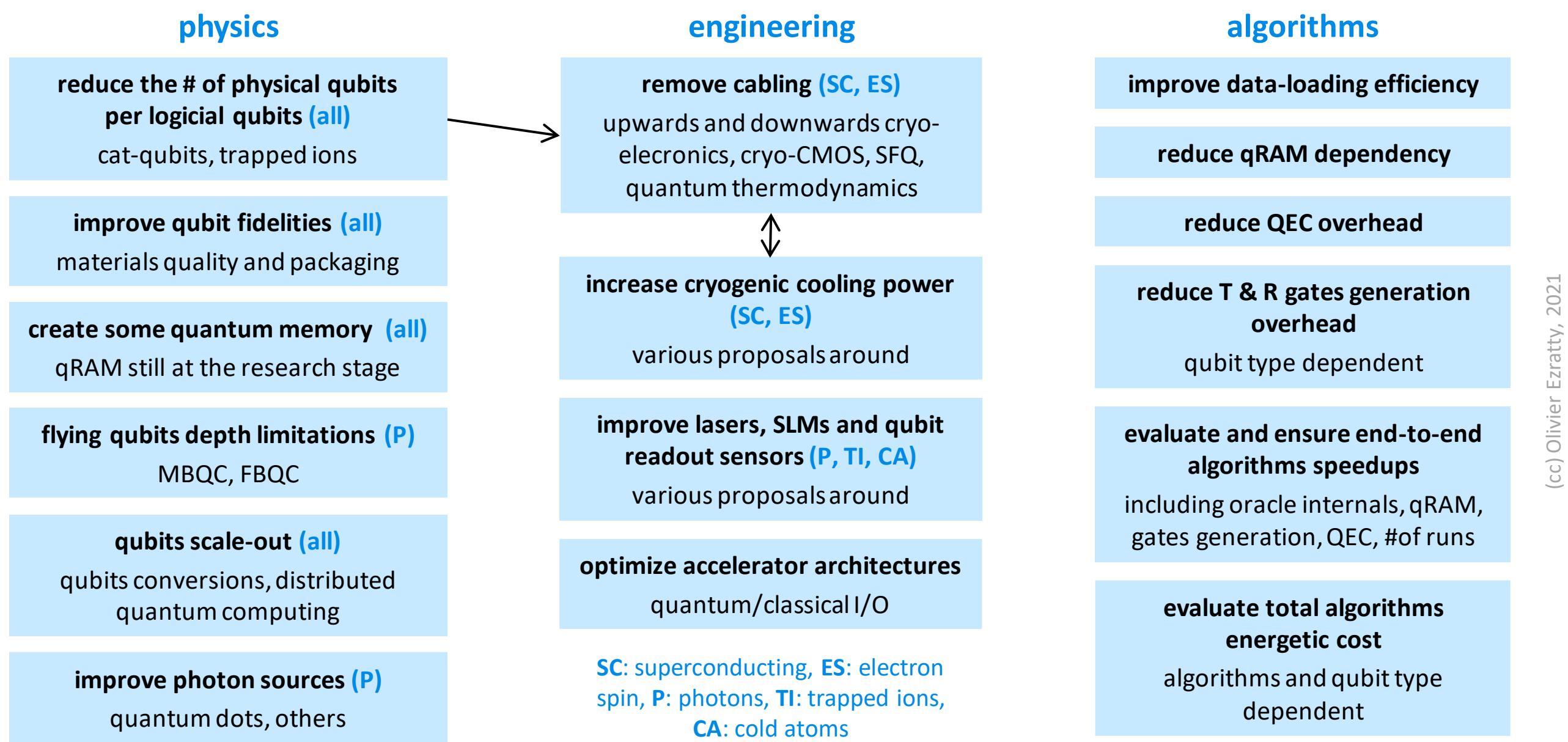

*Figure 262: a map of the various challenges to make quantum computing a practical reality. (cc) Olivier Ezratty, 2021.*

The second challenge deals with real algorithms speedups. Not all algorithms showcase an exponential theoretical speedup. Grover's algorithm speedup is only polynomial. All non-exponential speedups may end-up being useless due to their implementation cost. The trick of the trade is that all speedups are theoretical but not yet practical. Another way to look at this is to envision, even with moderate algorithms speedups, an energetic advantage for quantum computing, as discussed in the related part that we just saw, starting on page 249.

These speedups are never documented with taking into account all the quantum computing food chain: data preparation, oracle operations, quantum memory access when it is required, quantum error correction, non-Clifford group quantum gates generation (particularly for all algorithms using a quantum Fourier transform, and there are many) and the number of shots/runs required (with or without quantum error corrections).

I wish somebody did produce such evaluations with actual and projected data on these different aspects of gate-based quantum computing, even if it brings bad news! When bad news travel fast, fixes arrive faster, if there are any! And this is valid for the dominant wave of NISQ hybrid algorithms.

One broader challenge for the industry is to spur developer creativity for the design of even more algorithms and ways to assemble many quantum algorithms to create innovative solutions.

In the end, it looks like quantum simulators may be one very viable short-term option but we also still lack data to prove it. Some algorithms are being evaluated to run on these quantum simulators, like the ones from Pasqal and ColdQuanta but they have only been tested so far on classical computers emulators. The quantum software ecosystem will have to look at this!

573 See one example here from Google with experiments on Sycamore stopping at 20 qubits: Efficient and Noise Resilient Measurements for Quantum Chemistry on Near-Term Quantum Computers by William J. Huggins et al, Google AI, June 2020 (17 pages).

Quantum annealing and photonic coherent Ising machines could also bring their share of hope. The debate is still out to assess what is the quantum advantage of D-Wave with its latest annealer generation, relying on their 5000 qubits Pegasus chipset.

I still count on two things to reach quantum computing scalability and practicality. One is the great diversity of paths chosen by scientists and entrepreneurs. I have never seen so many variations in hardware technologies than in the quantum space. This creates a sort of fault-tolerance for innovation. The second, more generally, is I still believe and bet on scientists and engineers' creativity to solve these highly complicated problems. It is still a very generic commonplace reasoning that has not much value per se.

### Quantum computing engineering key takeaways

- A quantum computer is based on physical qubits of different nature, the main ones being superconducting qubits, electron spin qubits, NV centers, cold atoms, trapped ions and photons. They all have pros and cons and no one is perfect at this stage. Future systems may combine several of these technologies.
- Many key parameters are required to create a functional quantum computer. It must rely on two-states quantum objects (qubits). These must be initializable and manipulable with a set of universal gates enabling the implementation of any linear transformation of qubit states. Qubits must be measurable at the end of algorithms. Their coherence time must allow the execution of a sufficient number of quantum gates. Decoherence and errors must be as low as possible.
- Most quantum computers are composed of several parts: the qubit circuit (for solid-state qubits), vacuum enclosures (particularly with cold atoms and trapped ions) or waveguides (photons), usually housed in a cryogenic vacuum chamber (with the exceptions of photons and NV centers), some electronics sending laser rays or coaxial cables guided microwaves or direct currents onto qubits and a classical computer driving these electronic components.
- Since qubits are noisy, scientists have devised quantum error correcting schemes. These rely on creating logical "corrected" qubits composed of a lot of physical qubits, up to 10,000. This creates huge scalability challenges, many of them with classical enabling technologies like cabling, electronics and cryogeny. The science of quantum error correction, quantum error mitigation (a new section in this edition) and fault-tolerant quantum computing is a realm in itself.
- Many quantum algorithms also require some form of quantum memory, either for data preparation and loading (such as with quantum machine learning) or to access efficiently classical data (such as with oracle based algorithms like a Grover search). These quantum memories don't exist yet.
- The energetic cost of quantum computing is both a potential benefit but also an immense challenge, particularly when a large number of physical qubits are required to create large scale fault-tolerant computers. All components must be carefully designed to take into account the cryogenic cooling power as well as the available space to house cabling and cryo-electronics. This explains the creation of the Quantum Energy Initiative by Alexia Auffèves, Olivier Ezratty and Robert Whitney, which creates a community of researchers and industry vendors and organizations working collectively on this topic.
- The economics of quantum computers are still uncertain due to their immaturity and the current low manufacturing volume. Uncertainty is also strong with regards to the feasibility of scalable quantum computers. The scalability challenges ahead are enormous. One of them is to benefit from actual algorithm speedups when including all end-to-end computing operations.

# Quantum computing hardware

In a bottom-up approach, we've covered successively the basics of quantum physics, the mathematical aspects of gate-based quantum computing, then quantum computing engineering and enabling technologies. Let's now move to the last stack, quantum computers, with focusing on their specifics depending on the types of physical qubits they are using. We are dealing here with all sorts of players: public research laboratories, large established companies as well as startups.

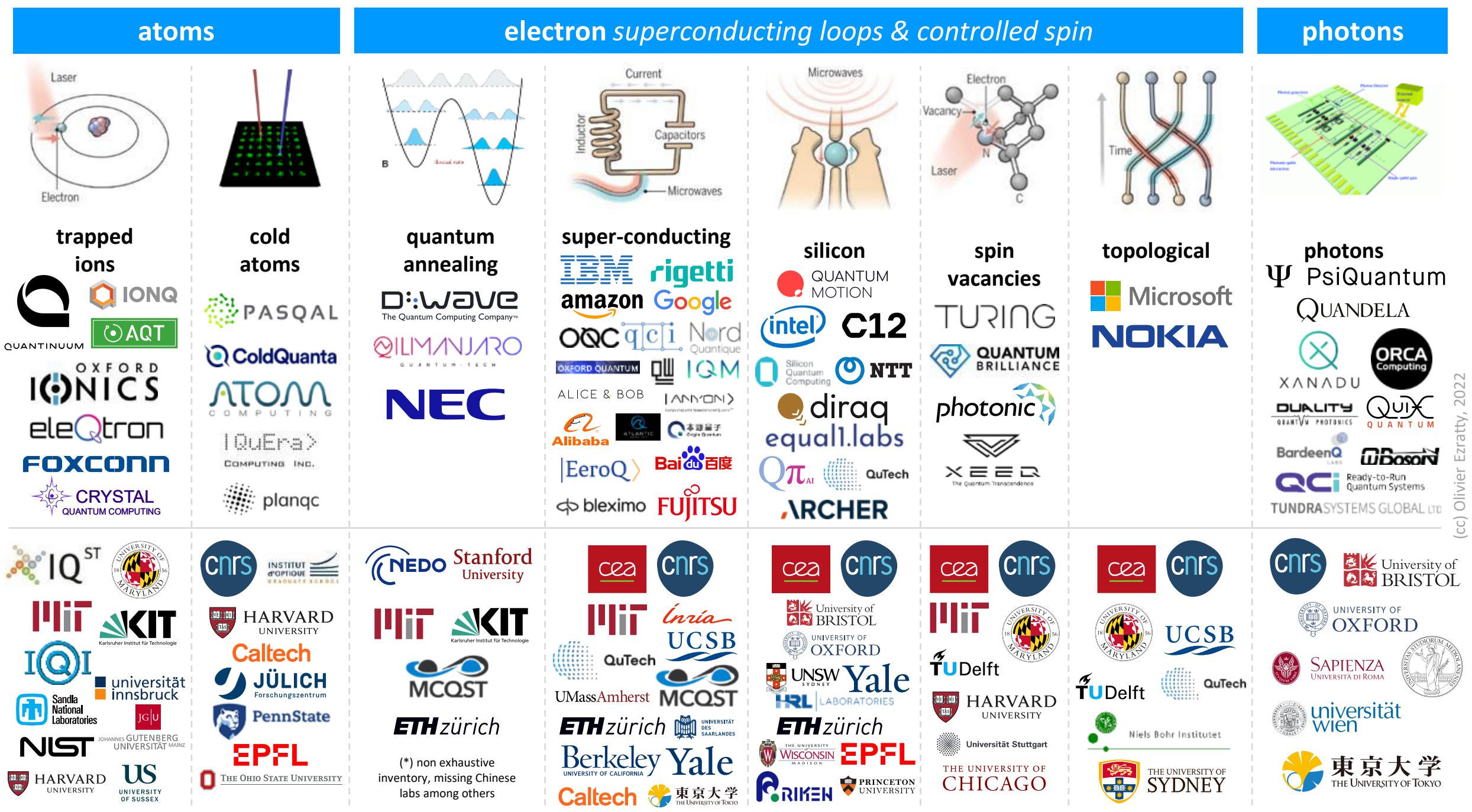

*Figure 263: a map of key research lab and industry vendors in quantum computing hardware per qubit type. (cc) Olivier Ezratty, 2022. Qubits drawing source: [Scientists are close to building a quantum computer that can beat a conventional one](#) by Gabriel Popkin in Science Mag, December 2016. I consolidated the logos lists since 2018. It's incomplete for the research labs at the bottom but rather exhaustive for the vendors at the top.*

There number startups in this inventory keeps growing, in Figure 263. They do not shy away from Google and IBM. There are not many Chinese startups yet. For the moment, the country's investments in quantum computing are concentrated in well-funded public research like with Jian Wei-Pan's giant lab in Hefei and with large cloud companies like Baidu, Alibaba and Alibaba. Notice that Chinese labs are missing in the chart. There are eight main categories of quantum computers grouped into three categories:

**Atoms**:

- **Trapped ions** found in particular at IonQ, a spin-off from the University of Maryland, as well as at Honeywell and the Austrian startup Alpine Quantum Technologies.
- **Cold atoms** like rubidium, cesium and strontium are used by Pasqal, QuEra, ColdQuanta and Atom Computing to create both analog quantum computers and gate-based quantum computers.
- **Nuclear Magnetic Resonance**, which is nearly abandoned as a path for quantum computing, NISQ or beyond despite one China company selling a desktop version for educational use cases.

**Electrons**:

- **NV centers** with only a few industrial players like Quantum Brilliance. Most NV centers applications are in quantum sensing.

- **Superconducting** effect qubits are used in IBM's, Google, Rigetti, OQC and IQM quantum computers as well as in D-Wave quantum annealers. This is a broad category with Josephson qubits (transmon, fluxonium, coaxmon, unimon, ...) and photon cavities-based qubits using superconducting control qubits (cat-qubits, GKP) with Alice&Bob, Amazon, Nord Quantique and QCI.
- **Quantum dots spins qubits** pushed notably by Intel, Quantum Motion, SQC, C12, Archer and the consortium between CEA-Leti CNRS Institut Néel and CEA-IRIG in France. There are many variations there as well.
- **Topological qubits** with the hypothetical Majorana fermions from Microsoft whose existence is yet to be proven. Other topological qubits avenues are investigated in research laboratories at the fundamental research level.

**Flying qubits**:

- **Photon qubits**, with a lot of variations like with the use of MBQC architectures to circumvents the difficulty to handle two-qubit gates and the limited computing depth of flying qubits, boson sampling and coherent Ising machines. The current main photon qubit vendors are PsiQuantum, Xanadu, Orca Computing and Quandela.
- **Flying electrons**, a separate track of qubits, with no commercial vendor yet involved in it. It's a fundamental research field.

Solid-states qubits usually refer to the electron qubits category with superconducting and electron spin qubits. However, trapped ions and photon qubit also rely on semiconductor circuits for their operations. But the "ions" are flying above their control circuits and the photons are circulating in nanophotonic circuits.

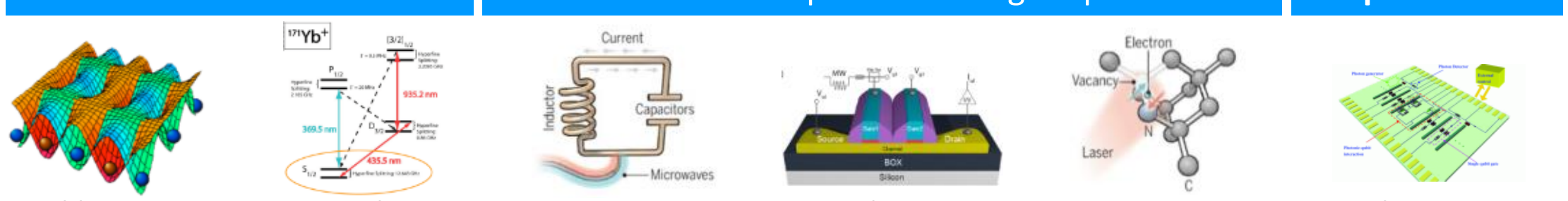

| | atoms | | electrons superconducting & spins | | | photons |
|---|---|---|---|---|---|---|
| | cold atoms | trapped ions | superconducting | silicon | NV centers | photons |
| qubit size | about 1 µm space between atoms | about 1 µm space between atoms | $(100\mu)^2$ | $(100nm)^2$ | $<(100nm)^2$ | nanophotonics waveguides lengths, MZI, PBS, etc |
| best two qubits gates fidelities | 99.4% | 99.9% | 99.97% (IBM Falcon R10) | >99% (SiGe) | 99.2% | 98% |
| best readout fidelity | 99.1% | 99.9% | 99.4% | 99% (SiGe) | 98% | 50% |
| best gate time | 1 ns | 100 µs | 20 ns - 300 ns | ≈5 µs | 10-700 ns | <1 ns |
| best $T_1$ | > 1 s | 0,2s-10mn | 100-400µs | 20-120µs | 2.4 ms | ∞ & time of flight |
| qubits temperature | < 1mK<br>4K for vacuum pump | <1mK<br>4K cryostat | 15mK<br>dilution cryostat | 100mK-1K<br>dilution cryostat | 4K to RT | RT<br>4K-10K cryostats for photons gen. & det. |
| operational qubits | 324 (Pasqal) | 32 (IonQ)<br>20 (AQT) | 127 (IBM)<br>56-66 (China) | 15 (Delft) in SiGe | 5 (Quantum Brilliance)-10 | 216 modes GBS (Xanadu) |
| scalability | up to 10,000 | <50 | 1000s | millions | 100s | 100s-1M |

(cc) Olivier Ezratty, 2022. RT = room temperature.

*Figure 264: figures of merit per qubit type. Best gate time covers only the electronics drive part but not the whole classical drive computing time. Best $T_1$ is the best qubit relaxation time. (cc) Olivier Ezratty, 2022. Data sources: cold atoms (ColdQuanta summer 2022, In situ equalization of single-atom loading in large-scale optical tweezer arrays by Kai-Niklas Schymik, Florence Nogrette, Antoine Browaeys, Thierry Lahaye et al, PRA, August 2022 and Ultrafast energy exchange between two single Rydberg atoms on a nanosecond timescale by Y. Chew et al, Nature Photonics, 2022 (7 pages)), trapped ions (Trapped Ion Quantum Computing: Progress and Challenges, 2019, Materials Challenges for Trapped Ion Quantum Computers, 2020, Infineon, IonQ and Quantinuum), silicon (Roadmap on quantum nanotechnologies, 2020), superconducting (IBM papers), NV centers (Quantum computer based on color centers in diamond, 2021). I list only the most demanding two qubit gates and readouts fidelities. Cold atoms systems are usually simulators, but data pertains to gate-based implementations.*

Many of the commercial companies in this panorama are associated with American or European research laboratories. Google collaborates with the University of Santa Barbara in California, IBM and Microsoft with the University of Delft in the Netherlands, and IBM with the University of Zurich, among other publicly funded research organizations.

These categories of technologies have different levels of maturity. Superconducting qubits are the most proven to date. Trapped ions are best-in-class with regards to fidelity and connectivity but do not scale well. Neutral atoms are starting to scale better. Linear optics and NV centers have also some difficulties to scale. Electron spin-based systems could scale but are less mature. Finally, Majorana fermions are still in limbo. But other qubit types are looming around and may become promising (other topological materials, Silicon Carbide, etc.). Creating assessment on the maturity of these technology pertains more to weather forecast than climate change predictions. Meaning, while you can forge some ideas on the relative maturity of these technologies with a short term view, it's much harder to make sound predictions in the longer term. For example, scaling these various technologies face very different challenges. As such, the table in Figure 264 is a sketchy and probably highly questionable comparison between these different qubits, particularly with cold atoms which are so far, used in quantum simulation mode and not gate-based architectures.

Another way of comparing qubit classes is to look at where the industry bucks are going. I created the chart below in Figure 265 with doing some guesswork on how much was invested by the large IT companies (IBM, Google, Microsoft, Amazon, Intel). But again, investors are not necessary in position to guess which technology will really scale.

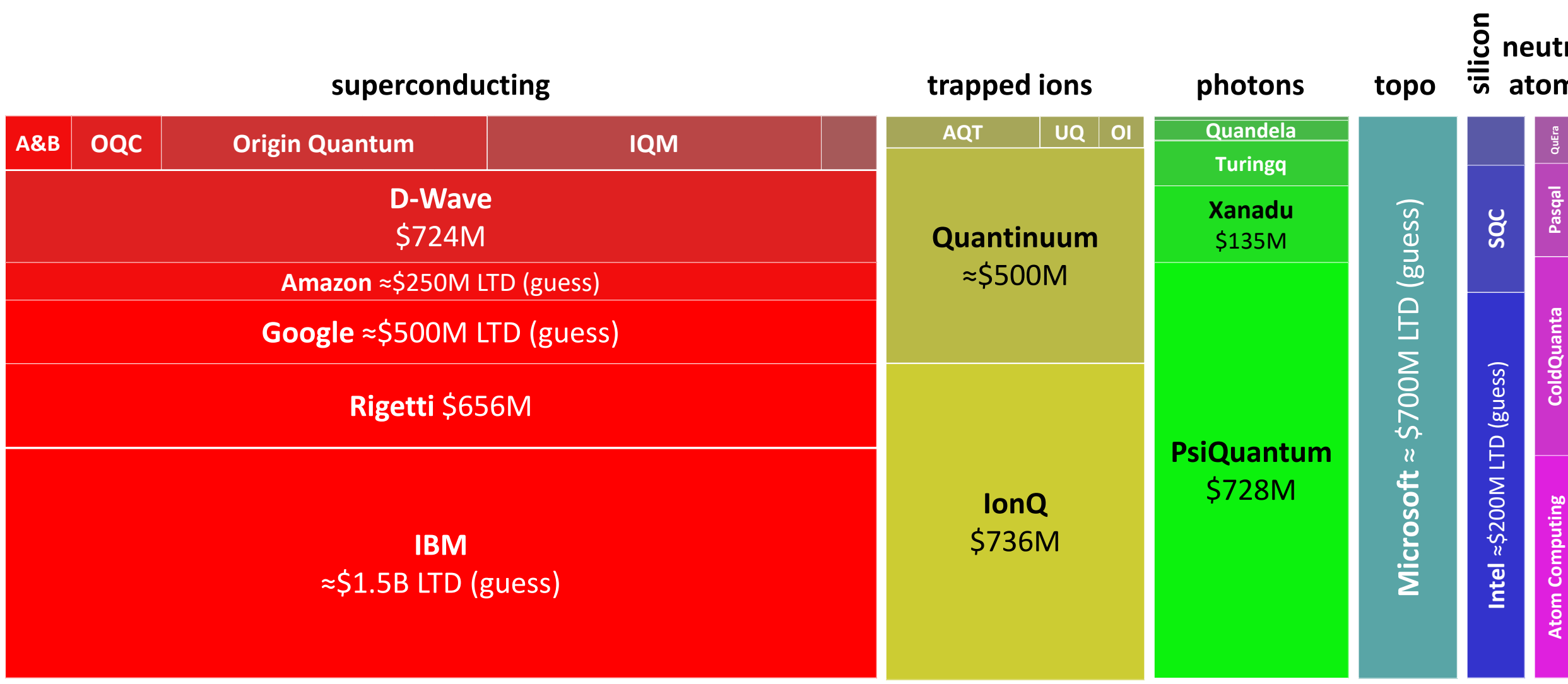

*Figure 265: fuzzy logic assessment of industry investments per qubit type mixing capital investment for startups and internal investments for legacy companies. LTD=life to date. UQ = Universal Quantum. OI = Oxford Ionics. cc Olivier Ezratty, 2022.*

These numbers are approximate and if you have some information to correct or update it, I'll take it! For startups, it's easier and consolidate the amount publicly raised either in venture capital or through a SPAC (for D-Wave, IonQ and Rigetti). What do you learn from this chart? That superconducting qubits are kings, followed by trapped ions and photon qubits. Therefore, silicon and cold atoms qubits seem under invested. Both because it seems Intel has not yet invested much in silicon qubits that have a rather low TRL and since not large company invested in cold atoms. I would seriously advise not to determine future winners based on these amounts.

There is so much scientific and technology uncertainty! Some solutions even not in this chart could show up in the future, either in the topological qubits space, beyond Microsoft's sole endeavor, or with SiC and other variations of spin cavities (beyond NV centers which don't show up in the chart due to the very small amount raised by the couple startups from this field).

Another summary view, below, shows how various qubit types have use cases beyond quantum computing and even, within the three paradigm of quantum computing with all qubit types being usable for quantum simulations although it seems it is only available with neutral atom qubits with commercial vendors. Photons, cavity spins and cold atoms have the broadest use cases given they have many applications in quantum sensing. Silicon qubits seem to have a narrower usage scope but don't discount them too soon. They may showcase one of the best scalability potential for quantum computing.

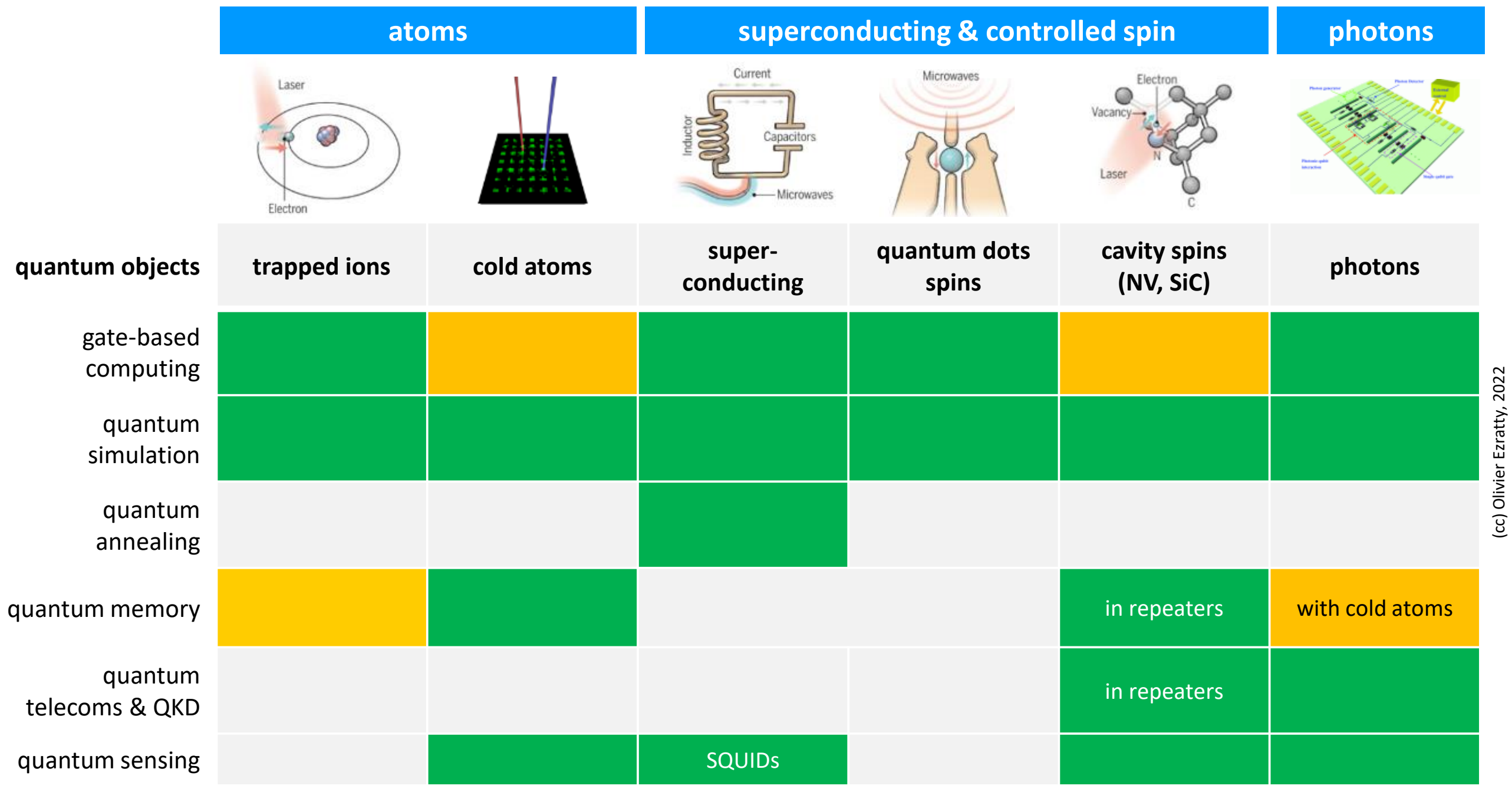

*Figure 266: qubit types and their use cases in all quantum computing paradigms and on telecommunications and sensing. Orange means: less commonplace and/or harder to implement. (cc) Olivier Ezratty, 2022.*

At last, here is a summary of the scalability challenges faced by the main qubit types around (I left aside topological and NV centers qubits). All these topics are detailed in this part[574].

| | superconducting | neutral atoms | trapped ions | silicon spins | photons |
|---|---|---|---|---|---|
| **challenges** | • noise and crosstalk ↗ with # of qubits.<br>• electronics energetic cost.<br>• cabling.<br>• scaling cryostats. | • atom controls beyond 10,000 qubits.<br>• harder to implement gate-based QC.<br>• SLM resolution. | • entanglement beyond 30 qubits.<br>• overall scaling beyond 40 qubits. | • controlled electrostatic potential.<br>• error correction.<br>• qubits entanglement. | • photon sources power.<br>• photon statistics.<br>• creating large cluster states of entangled photons. |
| **solutions** | • materials improvement.<br>• 3D chipset stacking.<br>• cryo-CMOS or SFQ.<br>• microwave signals multiplexing.<br>• scale-out with photons.<br>• more powerful cryostats. | • scale-out with atoms/photon conversion.<br>• higher resolutions SLMs.<br>• various atoms controls (microwaves, lasers). | • ions shuttling.<br>• switched to baryum (IonQ).<br>• Rydberg states ions (Crystal Quantum Computing).<br>• QPU photonic interconnect. | • material and interfaces improvement.<br>• integrated cryoelectronics.<br>• more powerful cryostats. | • MBQC.<br>• bright and deterministic photon sources (Quandela).<br>• deterministic sources of cluster states.<br>• integrated nanophotonics. |
| **caveats** | • photonic interconnect overhead and statistics.<br>• energetic cost of microwave multiplexing. | • precision of gate control.<br>• potential applicability limited to mid-scale simulations. | • photonic interconnect viability.<br>• photonic interconnect statistics and impact on speedups. | • long learning & chipset prod. cycles.<br>• scalability potential is capital intensive.<br>• fidelities still improving. | • photon statistics.<br>• small cluster states so far. |

(cc) Olivier Ezratty, 2022

(cc) Olivier Ezratty, 2022

*Figure 267: the various pathways to scalability in quantum computing per qubit type. (cc) Olivier Ezratty, 2022.*

[574] See 2022 Roadmap for Materials for Quantum Technologies by Christoph Becher et al, February 2022 (38 pages) which provides an overview of many qubit types and their various challenges.

## Quantum annealing

Quantum annealing is a particular quantum computing paradigm and technology that is also based on quantum physics and qubits but works entirely differently compared to gate-based quantum computing as we'll detail here. It has characteristics and performance levels that are intermediate between traditional supercomputers and general-purpose (fault-tolerant) gate-based quantum computers. D-Wave is the main commercial player in this category. Some research laboratories are also involved in quantum annealing but not as many as those involved in the different types of gate-based quantum computers.

In the 2000s, the interest in Shor's factoring algorithm created more traction for gates-based quantum computing, at the expense of quantum annealing. Ironically, the largest integer factoring (of 6 digits: 376,289) by a quantum processor was achieved in 2018 on a D-Wave quantum annealer and not with Shor's algorithm on a gate-based system and it is still the record to date[575].

### History

The quantum annealing paradigm (QA) is an optimization process for finding the global minimum of a given objective function by a process using quantum fluctuations and the tunnel effect[576]. It is mostly used for solving problems like combinatorial optimization problems where the search space is discrete, with many local minima. It is still usable for chemical simulations and non-discrete problems.

The idea to implement quantum annealing using quantum tunnelling effect came first in 1988 and 1989 in Italy and Germany[577]. It was then perfected in Japan by **Tadashi Kadowaki** and **Hidetoshi Nishimori** in 1998 with the "*introduction of quantum fluctuations into the simulated annealing process of optimization problems, aiming at faster convergence to the optimal state. Quantum fluctuations cause transitions between states and thus play the same role as thermal fluctuations in the conventional approach. The idea is tested by the transverse Ising model, in which the transverse field is a function of time similar to the temperature in the conventional method. The goal is to find the ground state of the diagonal part of the Hamiltonian with high accuracy as quickly as possible*"[578].

As described in Wikipedia: "*Quantum annealing starts from a quantum-mechanical superposition of all possible states (candidate states) with equal weights. Then, the system evolves following the time-dependent Schrödinger equation […]. The amplitudes of all candidate states keep changing, realizing a quantum parallelism, according to the time-dependent strength of the transverse field, which causes quantum tunneling between states. If the rate of change of the transverse field is slow enough, the system stays close to the ground state of the instantaneous Hamiltonian. The transverse field is finally switched off, and the system is expected to have reached the ground state of the classical Ising model that corresponds to the solution to the original optimization problem*". In here, what is a "transverse field"? It is a transverse magnetic field that is applied in an homogenous way on qubits and then slowly decreased to control the evolution of the Ising model[579].

A year after Tadashi Kadowaki and Hidetoshi Nishimori's paper, **D-Wave** was created in Canada and it produced its first commercial annealer 13 years later, in 2012. But Japan did not surrender to the idea to create annealing industry solutions.

---

[575] See Quantum Annealing for Prime Factorization by Shuxian Jiang et al, 2018 (9 pages).

[576] Source: https://en.wikipedia.org/wiki/Quantum_annealing

[577] See A numerical implementation of quantum annealing by S. Albeviero et al, University of Milan, July 1988 (10 pages) which refers to Quantum stochastic optimization by Bruno Apolloni et al, 1989 (12 pages).

[578] See Quantum annealing in the transverse Ising model by Tadashi Kadowaki and Hidetoshi Nishimori, 1998 (9 pages).

[579] An Ising model is a statistical physics model created to solve problems of simulation of ferromagnetic and para-magnetic materials associating particles having two state levels (a magnetic moment +1 or -1) which are linked together.

It took the form of a digital annealer created by **Fujitsu** and plans by **NEC** to create a quantum annealer. Also, many Japanese software startups are dedicated to quantum annealing.

At some point, John Martinis was working on creating a quantum annealer at **UCSB** but the idea was abandoned in favor of gate-based superconducting qubits when he started working at Google in 2004[580]. In Europe, **Qilimanjaro** (Spain) is in the process of creating a quantum annealer that would have greater capacities than the ones from D-Wave.

In 2000, **Edward Farhi** et al from the MIT created an algorithm to solve a SAT problem using adiabatic evolution which is considered an algorithmic cornerstone of quantum annealing[581].

In 2013, Google and NASA set-up a joint quantum computing lab, QUAIL, and toyed with D-Wave systems. It drove some awareness on the first supposed and oversold "quantum advantage" in 2015. Quantum annealing and D-Wave drove relatively bad press in the quantum community. They oversold their capacity and didn't explain well how it worked. Things fared better starting in 2017 and 2020 with their latest 2000 and 5000 qubit releases.

### Science

We must start here with explaining the following schema which connects the (quantum) adiabatic theorem and the various forms of digital simulated and quantum annealing:

- The **adiabatic theorem** states if you are in the ground state of a slowly varying quantum system, you stay in the ground state. It was created by Max Born and Vladimir Fock in 1928.
- The **diabatic theorem** is related to quick Hamiltonian evolutions and can be implemented in gate-based quantum computing although a recent proposal emerged to implement it with a quantum annealer using a locally managed transverse field[582].
- It can be implemented to solve various optimization and simulation problems in three manners:
  - Classically digitally with a **simulated annealer** (including Fujitsu's digital annealers),
  - With a **quantum annealer** (D-Wave).
  - With a **gate-based quantum computer**, using a time discretization of the system Hamiltonian evolution.

  Quantum annealing is faster than digital annealing as found in a 2022 benchmark[583]. It was also found that digitized simulated annealing can be used for more efficient integer factoring on NISQ QPUs than with Shor's algorithm[584].
- **Quantum annealing** is implemented with converting an optimization problem into a generic **QUBO** optimization problem (Quadratic unconstrained binary optimization) which itself can be turned into solving an **Ising model** on a quantum annealer.
- **Reverse annealing** uses classical methods such as simulated annealing to find a trivial solution and find better solutions using quantum annealing. This has been recently implemented with D-Wave annealers and is more efficient than classical quantum annealing.

---

[580] See the thesis Superconducting flux qubits for high-connectivity quantum annealing without lossy dielectrics by Christopher M. Quintana, 2017 (413 pages), directed by John Martinis who was then at Google.

[581] See Quantum Computation by Adiabatic Evolution by Edward Farhi, Jeffrey Goldstone, Sam Gutmann and Michael Sipser, MIT and Northwestern University, 2000 (24 pages).

[582] See Locally Suppressed Transverse-Field Protocol for Diabatic Quantum Annealing by Louis Fry-Bouriaux et al, UCL October 2021 (18 pages).

[583] See Benchmarking Quantum(-inspired) Annealing Hardware on Practical Use Cases by Tian Huang et al, March 2022 (35 pages).

[584] See Digitized Adiabatic Quantum Factorization by Narendra N. Hegade, Enrique Solano et al, November 2021 (10 pages).

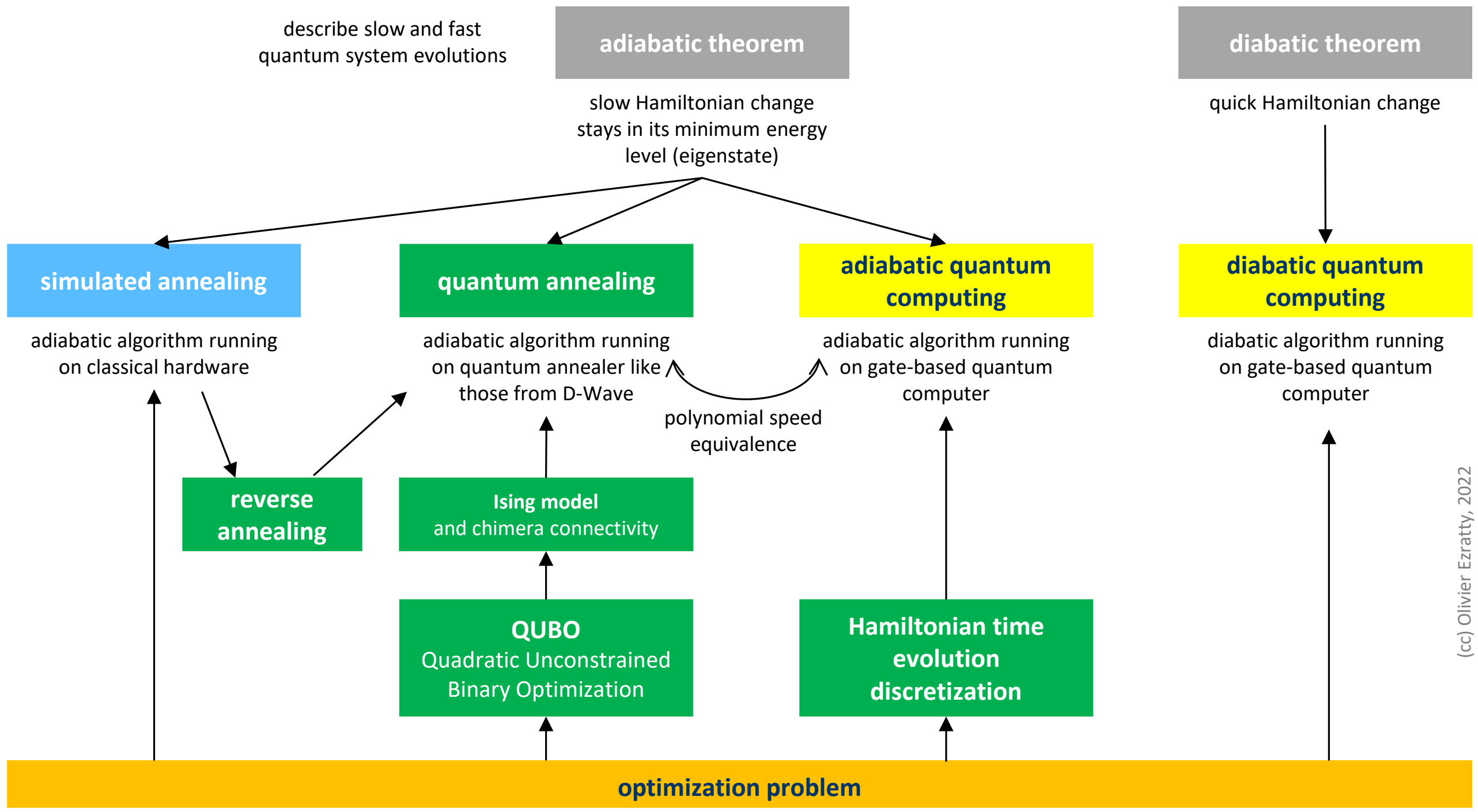

*Figure 268: from the adiabatic theorem to quantum annealing. (cc) Olivier Ezratty, 2022. And some references found in [Adiabatic quantum computation](#) by Tameem Albash and Daniel A. Lidar, 2018 (71 pages).*

Quantum annealing starts with preparing a system of interconnected qubits with establishing links between them using weights that are defined by couplers, a bit like in neural networks used in artificial intelligence. It sets the relative qubits connections energy in the longitudinal field (z) and the absolute qubits energy as a linear coefficient or bias on each qubit. These values $h_i$ and $J_{ij}$ are discretized depending on the capabilities of the DACs integrated in the chipset. It was using a 4 bit encoding in the first commercial D-Wave systems with values ranging from -1 to +1 by steps of 1/8. The precision of these DACs has since then improved with each new generation of D-Wave annealer.

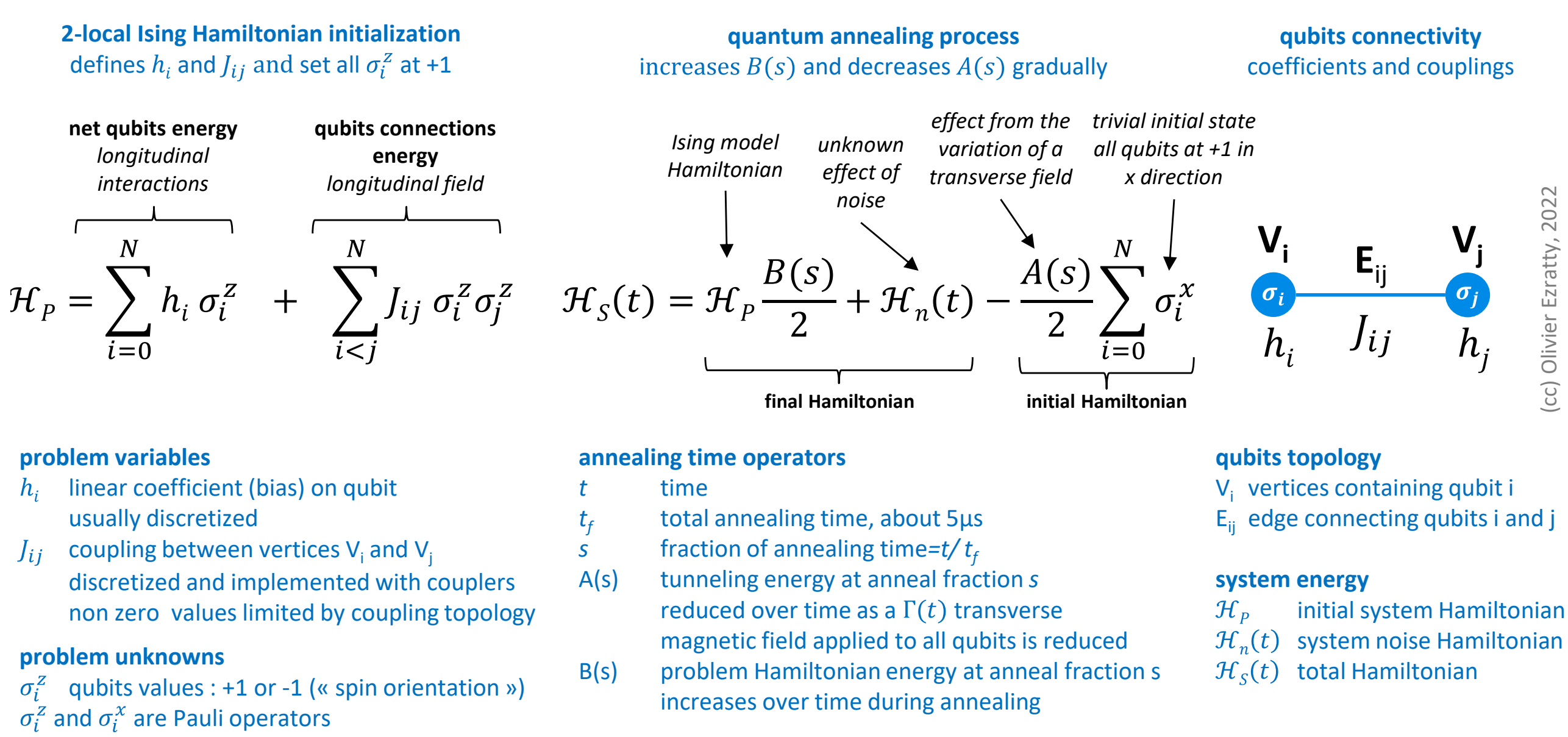

*Figure 269: uncovering a quantum annealing Hamiltonian. (cc) Olivier Ezratty, 2022.*

The system is then initialized with setting the qubits at $|+\rangle$, a perfect superposition between $|0\rangle$ and $|1\rangle$, corresponding to the lowest-energy state of the system, also called the tunneling Hamiltonian.

The annealing process then takes place with progressively reducing this magnetic transverse field down to zero, which, thanks to the quantum tunnel effect, will drive the system to an equilibrium state that corresponds to a minimum energy level. In the equations in Figure 269, it means reducing the value of scalar $A$(s) and increasing the value of $B$(s) accordingly, usually not linearly. This leads to automatically modifying the values of the qubits (spin up or down in the z direction) towards a result that corresponds to the solution of the submitted problem. This annealing process is a magnetic equivalent of a classical thermal annealing.

From the mathematical standpoint, the system Hamiltonians $\mathcal{H}_p$ $\mathcal{H}_S$ and $\mathcal{H}_n$ described in Figure 269 are square matrix operators of dimension $2^N$, N being the number of used qubits. The $\sigma_j^z$ notation in the Ising and annealing Hamiltonian are somewhat confusing. It is actually the tensor product of the identity operator for all qubits non equal to j and the Pauli z operator for the given qubit j, as follows. A Pauli operator is a 2x2 matrix equivalent to the X, Y and Z single-qubit gates from gate-based quantum computing. Thus, $\sigma_j^z$ is a simplified notation of the whole tensor product of dimension $2^N$, the full version being the following, $I_k$ being identity matrices of dimention 2x2 for all k between 1 and N at the exception of j:

$$\sigma_j^z = I_1 \otimes \cdots \otimes I_{j-1} \otimes \sigma_j^z \otimes I_{j+1} \otimes \cdots \otimes I_N$$

The same can be said of the product $\sigma_i^z \sigma_j^z$ which is also a tensor product of dimension $2^N$ using the following equation with $I_k$ for all k between 1 and N at the exception of i and j:

$$\sigma_i^z \sigma_j^z = I_1 \otimes \cdots \otimes I_{i-1} \otimes \sigma_i^z \otimes I_{i+1} \otimes \cdots \otimes I_{j-1} \otimes \sigma_j^z \otimes I_{j+1} \otimes \cdots I_N$$

System Hamiltonian matrix eigenvalues are the possible energy levels of the system. The annealing process has the effect of maintaining the system in its minimum energy level. It doesn't compute these eigenvalues but finds the qubit spin values ($\sigma_i^z$) that are minimizing the Hamiltonian. Another notation of the effect of quantum annealing is to find the combination of N spins noted $\hat{\sigma}$ that minimizes the Hamiltonian. The searched space is $\Omega_N$ that contains all N combinations of -1 and +1: $\{-1,+1\}^N$.

$$\hat{\sigma} = \underset{\sigma \subset \Omega_N}{\arg\min}\left(\sum_{i=0}^{N} hi\, \sigma_i^z \quad + \quad \sum_{i<j}^{N} J_{ij}\, \sigma_i^z \sigma_j^z\right)$$

At the end of the annealing, the qubits state is read and generates a +1 or a -1 for each of them depending on the direction of the magnetic flux of the superconducting loop (and, by the way, we don't care about it being a QND – aka non demolition - measurement). As a result, the solved problem search space is discrete and finite. The process is however iterative with several annealing passes and results being averaged. Like with gate-based quantum computing, the process is also probabilistic, and not just because noise gets involved with an unknown time-evolving Hamiltonian $\mathcal{H}_n(t)$.

**quantum annealers**

- mature **development tools** offering.
- large number of **software startups**, particularly in Japan and Canada.
- quantum annealers are available in the **cloud** by D-Wave and Amazon Web Services.
- the greatest number of well documented **case studies** in many industries although still at the proof of concept stage.
- most universal qubits gates algorithms can be have an equivalent on quantum annealing.

- only **one operational commercial vendor**, D-Wave.
- computing **high error rate.**
- no **operational proof** of quantum advantage.
- **most commercial applications** are still at the pilot stage and not production-scale grade but this is also the case for all gate-based quantum computers.
- **all algorithms are hybrid**, requiring some preparation on classical computers.

*Figure 270: quantum annealers pros and cons. (cc) Olivier Ezratty, 2022.*

There are variations in this model's implementation with regards to the qubits coupling mechanism. It can be made on one degree (z for D-Wave) or two and three degrees of freedom (x, y and z, in a so-called Heisenberg model) like what **Qilimanjaro** (Spain) is planning to implement.

### Qubit operations

Using a quantum annealer works as described in Figure 271. Algorithms are prepared classically with converting the given problem into a QUBO problem that is then translated in an Ising model with setting up the links between qubits in the initialization process and the qubits "weights".

The annealing process then takes place with controlled evolutions of A(s) and B(s) as described in the previous chart, with tuning the magnetic transverse field affecting the qubits chipset. When s=1, the system proceeds with reading the qubits values +1 or -1.

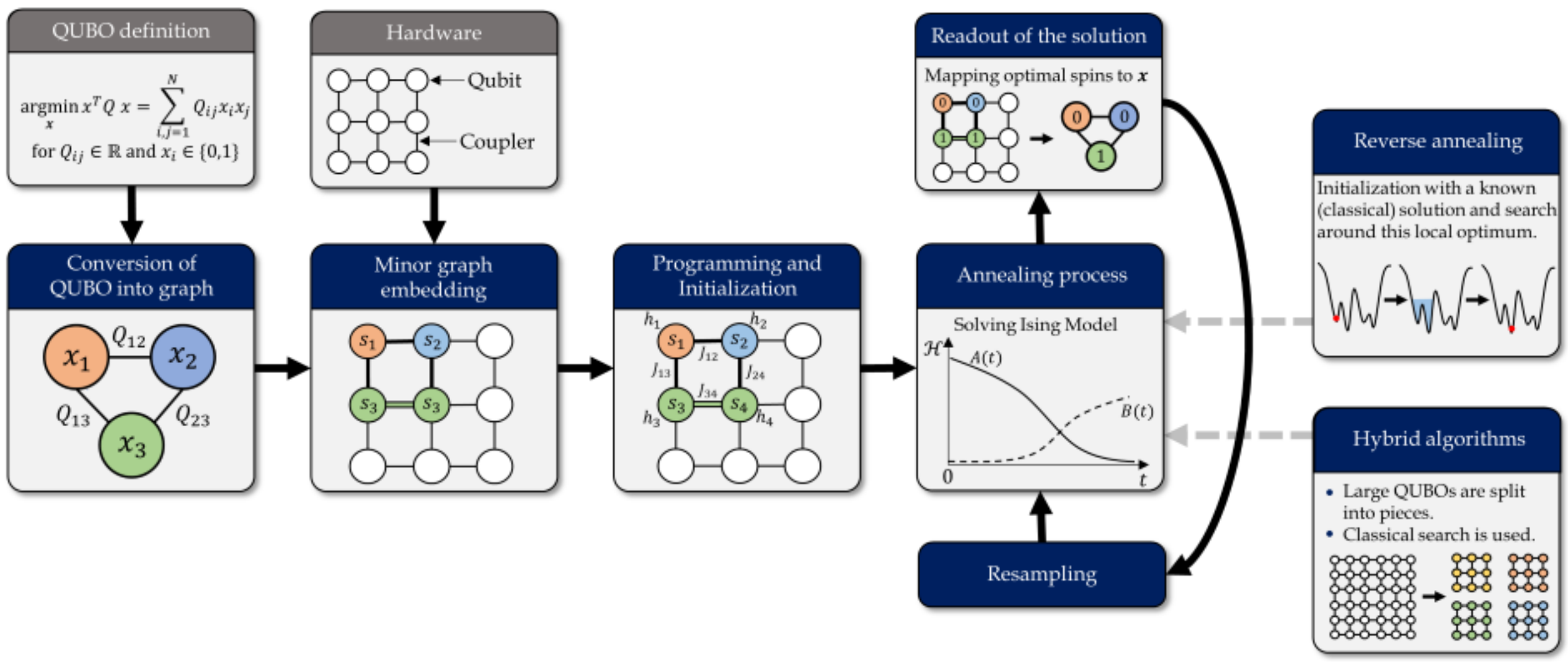

*Figure 271: a quantum annealing computing process. Source: Quantum Annealing for Industry Applications: Introduction and Review by Sheir Yarkoni et al, Leiden University and Honda Research, December 2021 (43 pages).*

D-Wave qubits are niobium-based rf-SQUID exploiting superconducting current loops interrupted by two Josephson effect barriers that are controlled by variable magnetic fluxes.

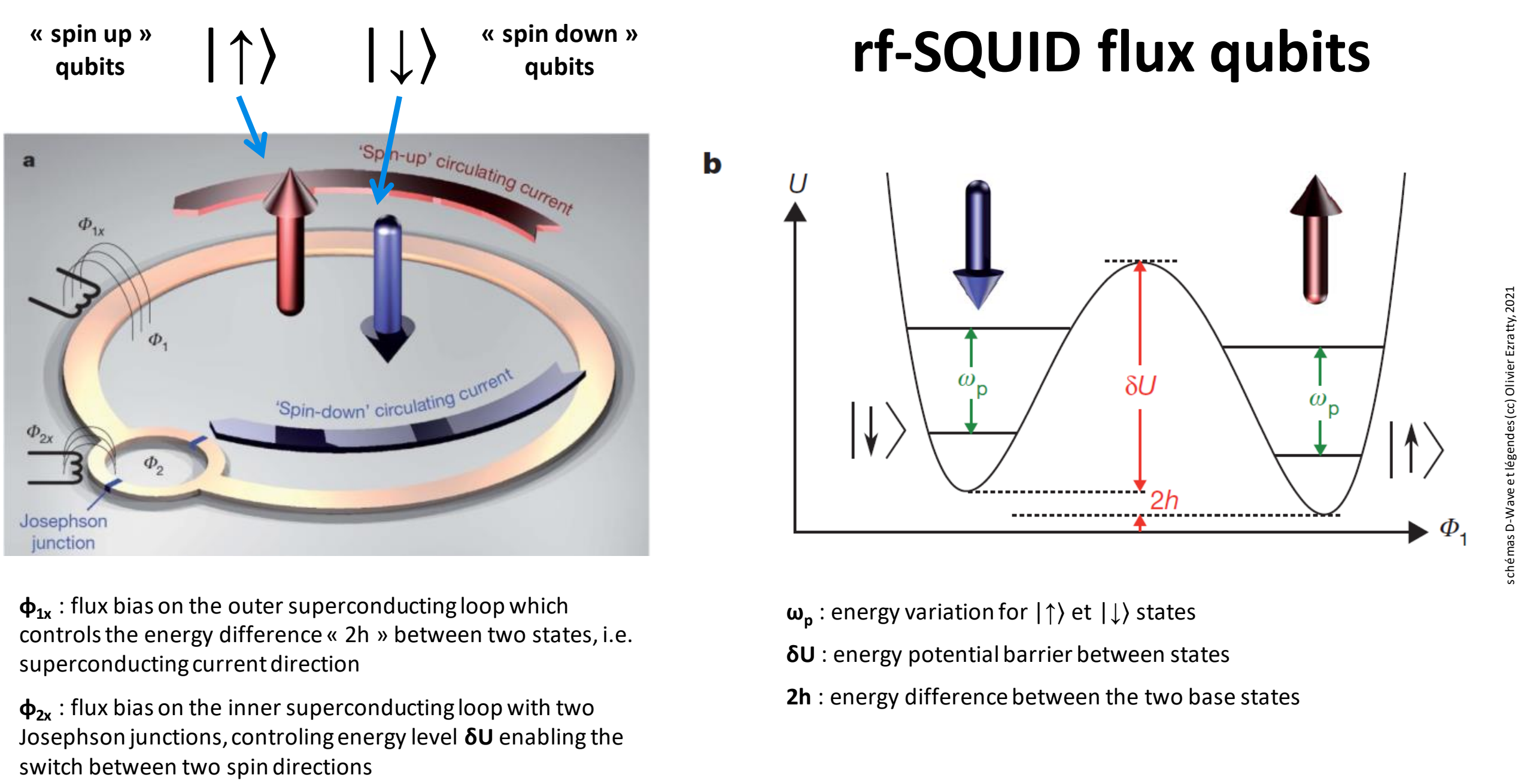

*Figure 272: rf-SQUIDs used in a D-Wave quantum annealer. Source: D-Wave.*

Taking always the example of D-Wave annealers, we can dig into how the qubits are controlled to prepare and handle an annealing process[585]. The process needs to control several parameters:

$J_{ij}$ the coupling between vertices $V_i$ and $V_j$ is controlled by DACs (digital to analog converters). These couplers are set with a static DC flux current applied to their compound-junction.

$h_i$ the energy between two states of qubits i is also controlled via a DAC. But other DACs are there to correct the drift created by the annealing process.

A(s) is a transverse magnetic field applied to all qubits and controlled in the chipset by analog lines via the CCJJ DACs.

B(s) increases over time during annealing, also controlled by CCJJ analog lines in the chipset.

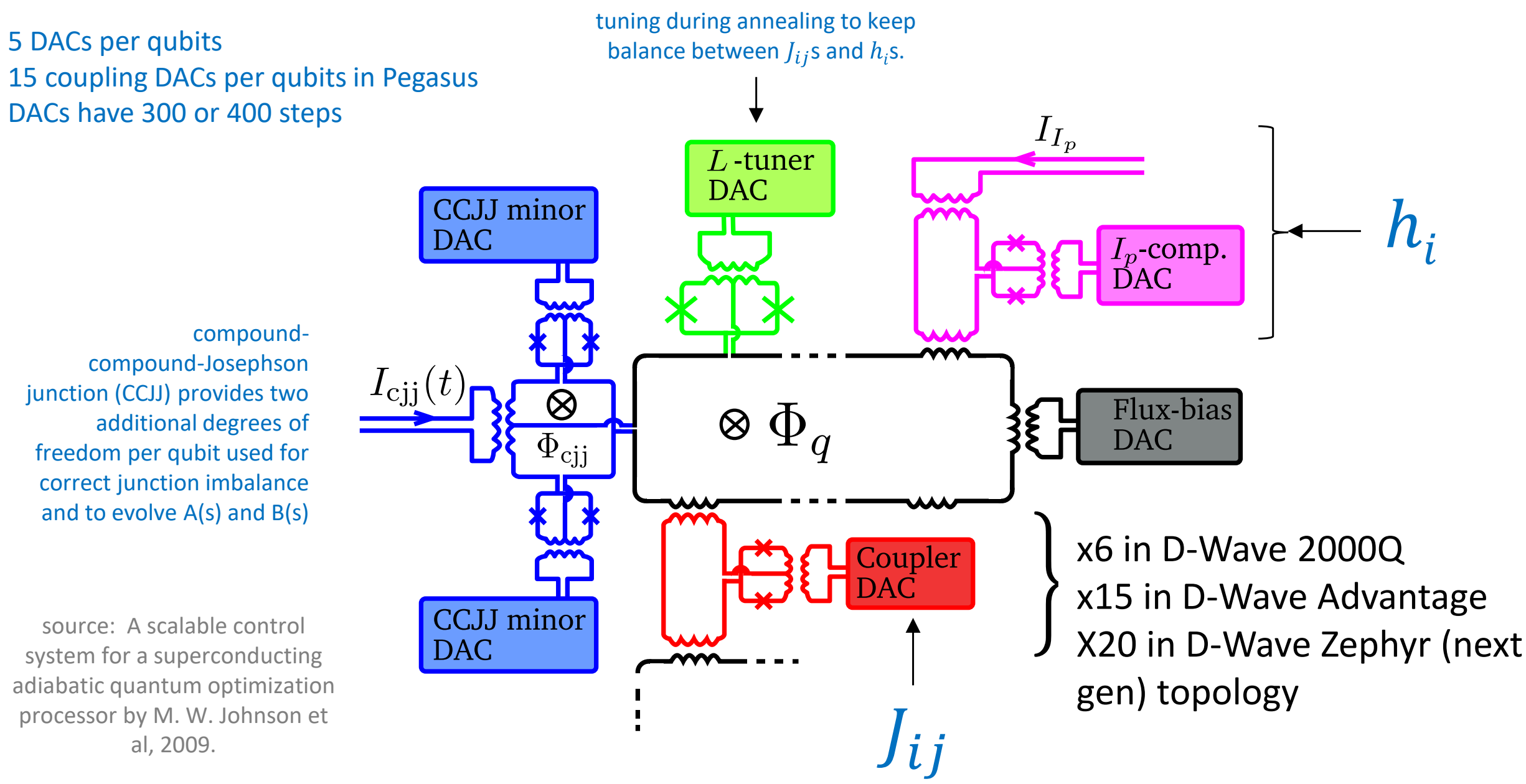

*Figure 273: source: how D-Wave qubits are controlled at the physical level. Source: A scalable control system for a superconducting adiabatic quantum optimization processor by M. W. Johnson et al, 2009.*

These are optimization problems where the variables $J_{ij}$ can only take two values (-1 or +1 for solving an Ising model or 0 and 1 for solving a QUBO problem) and where they are linked together by different fixed parameters which are defined as floating numbers (FP32) with the boundary constraints $-1 \leq J_{ij} \leq 1$ and $-2 \leq h_i \leq 2$. However, the related DACs introduce significant sampling noise due to their sampling rate with a couple hundred different steps. In the end, the precision of the data of the problem to be solved is much lower, probably below one single byte. We're far from high-precision floating point scientific computation. It should be mentioned that D-Wave systems require frequent recalibration.

The initialization of a D-Wave 2000Q takes 25 ms, the annealing itself usually lasts 20 µs but can be extended to 2 ms, and readout time is 260 µs. If we repeat the annealing process a thousand times, we end-up running the whole thing in less than a second. Between each run, some time was required to enable heat dissipation, with the chipset temperature rising to 500 mK with early generations of D-Wave annealers.

[585] The electronics architecture of superconducting qubits control is described in A scalable control system for a superconducting adiabatic quantum optimization processor by M. W. Johnson et al, 2009 (14 pages) and Architectural considerations in the design of a superconducting quantum annealing processor by P. I. Bunyk et al, 2014 (9 pages).

The initialization signals of the Hamiltonian are multiplexed and sent in digital format from the outside to the chip. A chipset requires an order of magnitude of $O(\sqrt[3]{N})$ external control lines for N qubits, using signals multiplexing.

This greatly simplifies the cabling of the cryogenic enclosure of the computer compared to the superconducting IBM and Google computers, as shown in the *adjacent* illustration of a 2000Q. Most of what can be seen in the intermediate stages in the cryostat corresponds to the dilution cryogenics system.

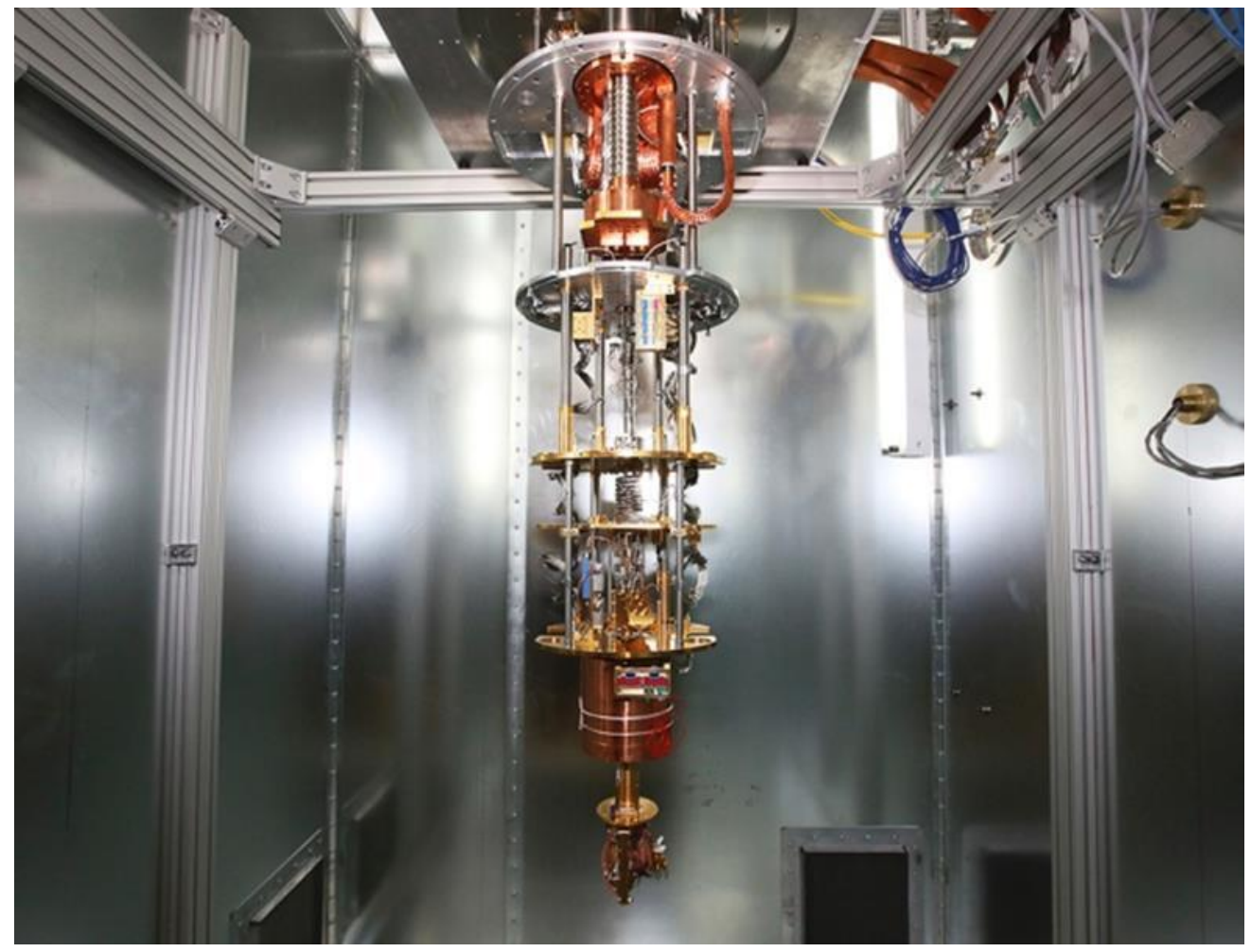

Figure 274: Inside a D-Wave system, with the cryostat open. Source: D-Wave.

Annealers don't need to send microwave pulses to qubits and thus, avoid the related coaxial cables used in gate-based superconducting qubits. Some of the magic comes from the integrated DC ramp pulses generation circuits that are sitting in the quantum chip. These circuits use SFQ components implementing DACs, basically superconducting transistors using Josephson junctions close to those of the qubits. Still, these components are noisy and may contribute to the noise affecting the qubits in this architecture[586].

**Research**

Let's mention some research work beyond what D-Wave is doing.

Quantum annealing was explored in 2016 by the IARPA agency in its **Quantum-Enhanced Optimization** (QEO) project, which aimed to create an adiabatic computer void of some of the limitations from D-Wave, particularly in terms of connectivity and quality of qubits. Appropriately, in view of IARPA's mission, the goal was to accelerate the production of quantum computers capable of executing Shor's integer factoring algorithm to break the public keys coming from intercepted communications. This project was folded into DARPA's **QAFS** project (Quantum Annealing Feasibility Study) in February 2020 which produced a 25 coherent annealer system.

**Stanford University** is also working on quantum annealing. In 2016, they created a prototype photonic based annealer with 100 qubits having an all-to-all connectivity (so... 10,000 connections)[587]. This connectivity is what makes such a system "coherent". This research was still going on in 2021 and involves **NTT** in Japan.

At last, the H2020 European project **AVaQus** (Annealing-based VAriational QUantum processorS) launched in October 2020 brings together five research laboratories (Institut de Física d'Altes Energies of Barcelona, Karlsruhe Institut für Technologie (KIT) of Karlsruhe, CNRS Institut Néel in France, the University of Glasgow and the Consejo Superior de Investigaciones Científicas in Madrid), associated with three startups **Delft Circuits (**Netherlands), **Qilimanjaro** (Spain) and **HQS** (Germany). The project is scheduled to end in 2023 and got a funding of €3M, independently of the Quantum Flagship program.

586 See Analog errors in quantum annealing: doom and hope by Adam Pearson, 2019 (9 pages).

587 See A fully-programmable 100-spin coherent Ising machine with all-to-all connections by Peter L. McMahon, Yoshihisa Yamamoto et al, 2016 (9 pages).

**Vendors**

We'll of course start with D-Wave and will follow-on with Qilimanjaro and NEC.

Located in Vancouver, **D-Wave** (1999, Canada, $724M + SPAC) was for a very long time the only supplier of commercial quantum processors.

D-Wave was created by Geordie Rose (their first CTO and for some time also their CEO[588]), Haig Farris, Bob Wiens and Alexandre Zagoskin, formerly in charge of research. Geordie Rose received his PhD in Materials Physics in the mid-1990s from the University of British Columbia. The creation of D-Wave is a direct result of this work. He met Haig Farris during his studies while the latter was teaching economics.

It took D-Wave eight years to prototype its first chip containing four qubits and a total of 13 years to sell their first quantum computer, the D-Wave One. During these years, they raised $31M, then $1.2M in 2012 from InQTel, the CIA's investment fund. In 2011, D-Wave signed a partnership with Lockheed Martin, which does some work for the NSA. All in all, the startup went through 13 rounds of funding, ending with a SPAC finalized in 2022!

In February 2022, D-Wave announced its own SPACification via the dedicated investment fund **DPCM Capital** created by Eric Schmidt (former Google, now SandboxAQ chairman), Peter Diamandis (Singularity University), Shervin Pishenar (Hyperloop One founder)[589]. They were to raise $340M, including $40M coming from a Canadian pension fund, PSP Investments. Were also involved Goldman Sachs and NEC Corporation. This valued the company at $1.6B which was IPOed in August 2022. They have another funding source with $150M coming from Lincoln Park Capital. Public financial information uncovered that in 2021, D-Wave revenue was $11M with losses of $59M. The company expects a fast growth starting in 2025, generating a $551M turnover in 2026. As of early 2022, they had a staff of 180 including 36 PhDs and a 200 patents portfolio.

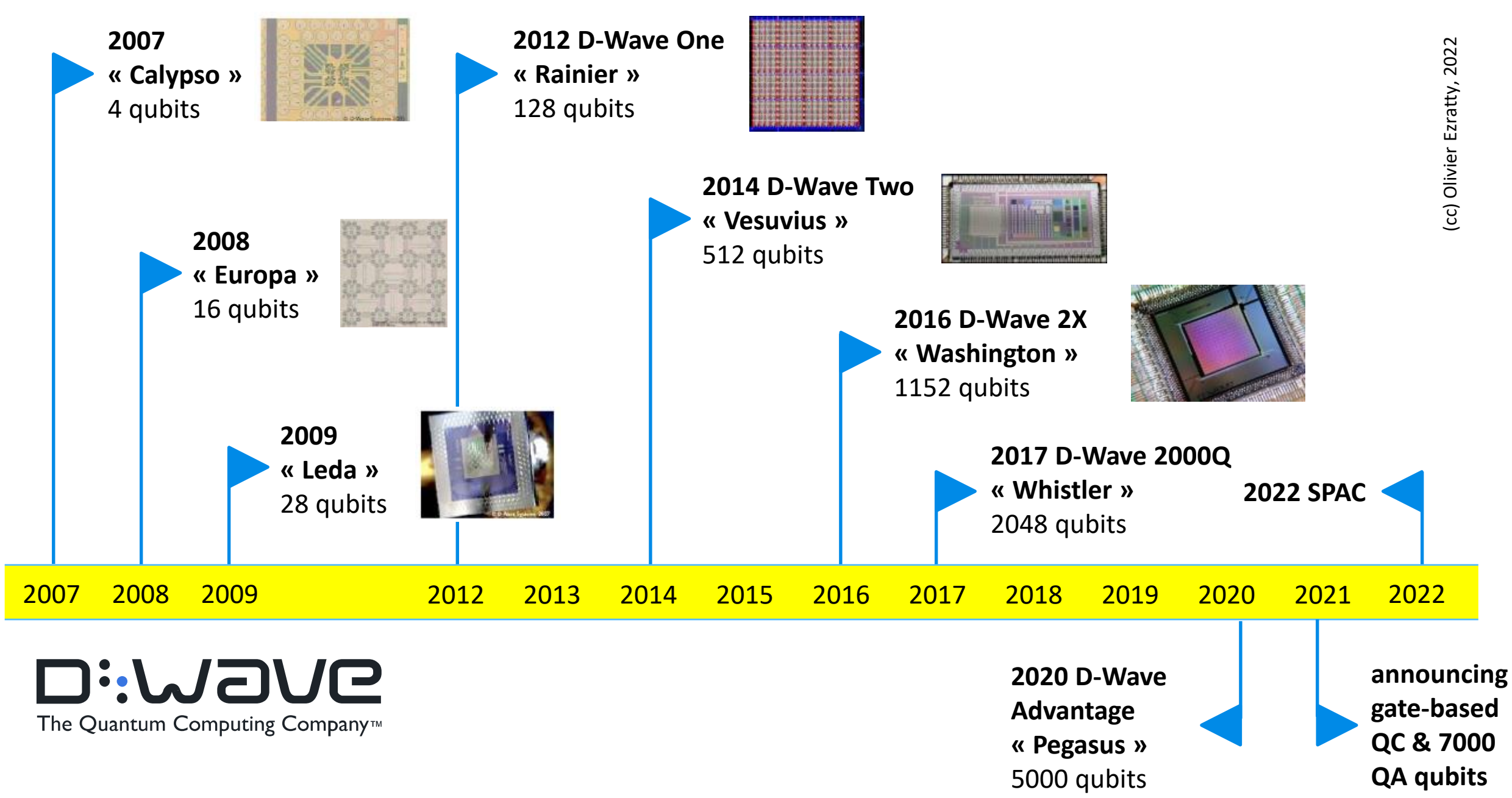

*Figure 275: timeline of D-Wave's history. (cc) Olivier Ezratty, 2022.*

588 Co-founder Geordie Rose then created **Kindred.ai,** a startup that aims to integrate General Intelligence (GIA) into robots. He leaved Kindred.ai in 2018 to create Sanctuary, a spin-off of Kindred, dedicated to AGI, the quest for the Holy Grail of general artificial intelligence.

589 See their February 2022 investor presentation (25 slides).

D-Wave's 2021 management team is quite different. Only one of the co-founders is still part of it, Eric Ladizinsky, who is their Chief Scientist. The CEO from 2009 to 2020 was Vern Brownell. Their CTO Alan Baratz joined the company in 2017 and became CEO in 2020.

Although quantum annealing accelerators have technical limitations compared to general-purpose quantum computers, they have the advantage of being available and are surrounded by a strong software ecosystem. However, most of the 250+ case studies solutions published by D-Wave and their customers and partners seem to be proofs of concept. Few seem to have been deployed, be production-grade, or at least provide a proven quantum advantage over classical computing. D-Wave has developed its end-to-end quantum annealing computer solution. It is the first full-featured quantum computer in history with a design that allows it to be easily integrated into a clean room.

Their roadmap has progressed steadily with the first three generations of prototypes created between 2007 and 2009 and then, starting from 2012, five generations of commercial computers, starting with the D-Wave One in 2012 with 128 qubits, the D-Wave 2000Q in 2017 with 2048 qubits and 128,000 Josephson junctions on a $(5.5\ mm)^2$ chipset and a list price of $15M, up to the D-Wave Advantage with the Pegasus chipset launched in September 2020 with 5,640 qubits and one million Josephson junctions[590], each qubit being coupled to 15 other qubits compared to 6 in the 2000Q[591]. It allows more complex problems to be solved with an equivalent number of qubits. It is a performance given gate-based superconducting qubits have a 1 to 3 (IBM) to 1 to 4 (Google) connectivity at best. The Pegasus chipset is larger, being a square of 8,5 mm. It is manufactured in the USA in Skywater's cleanrooms (formerly a Cypress Semiconductor fab) located in Bloomington, Minnesota. The embedding graph or qubits connectivity is branded a chimera their D-Wave 2000Q annealers and a Pegasus graph for their D-Wave Advantage annealers[592].

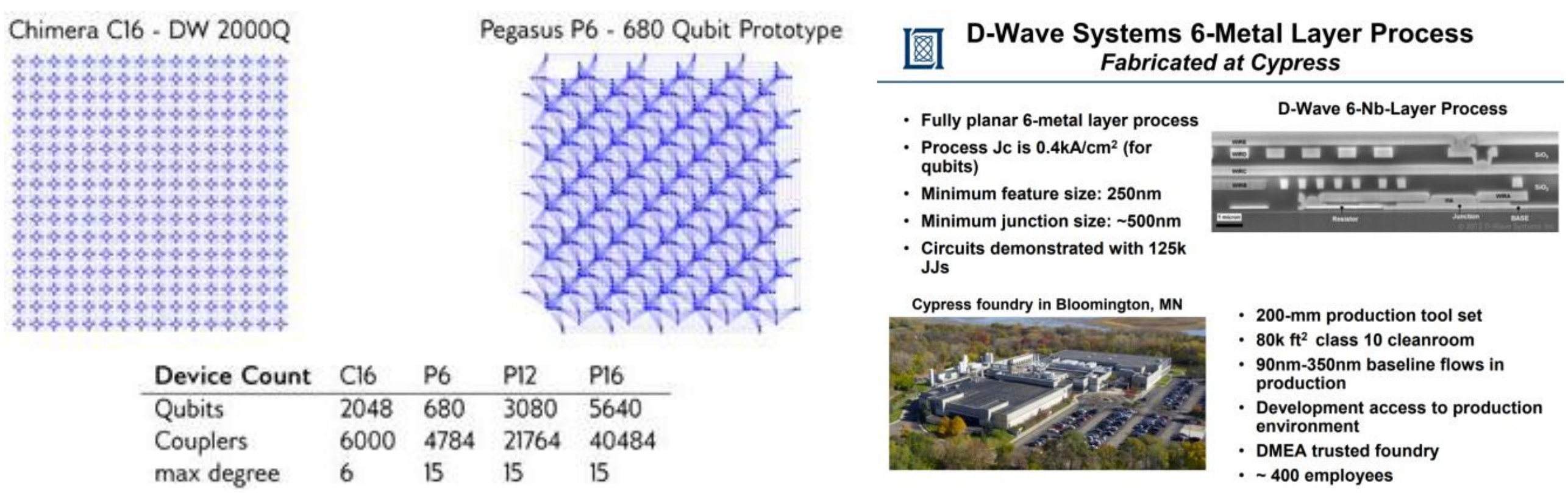

*Figure 276: evolution of D-Wave's qubit connectivity. And their chipset manufacturing process. Source: D-wave.*

[590] See Quantum annealing with manufactured spins by Mark Johnson et al, 2011 (6 pages) which outlines the D-Wave process. As well as Technical Description of the D-Wave Quantum Processing Unit by D-Wave, 2020 (56 pages) and related supplemental information (19 pages). The Pegasus architecture from the D-Wave advantage is described in Next Generation Quantum Annealing System by Mark Johnson, March 2019 (27 slides) and in Next-Generation Topology of D-Wave Quantum Processors by Kelly Boothby et al, 2019 (24 pages). See D-Wave Announces General Availability of First Quantum Computer Built for Business by D-Wave, September 2020.

[591] The chimera uses cells with 8 qubits with internal and external couplings. It has 4 internal couplings within cells and 2 external couplings in pre-Pegasus chipsets and 12 internal and 3 external couplings in Pegasus chipsets.

[592] D-Wave's chimera matrix requires a conversion process of its qubit mesh problem. This process is so far mostly exploited for problems that fit well with this qubit organization. For an arbitrary optimization problem, the conversion gives a result that is not convincing in terms of efficiency and acceleration. This is what emerges from the work of Daniel Vert, then PhD student at CEA LIST, in On the limitations of the chimera graph topology in using analog quantum computers by Daniel Vert, Renaud Sidney and Stéphane Louise, CEA LIST, 2019 (5 pages) and in Revisiting old combinatorial beasts in the quantum age: quantum annealing versus maximal matching by the same authors, October 2019 (36 pages). D-Wave's chimera structure limits the way a QUBO or other optimization problem can be converted into an Ising problem solvable with D-Wave's chimera structure.

In October 2021, D-Wave made significant announcements as part of their Clarity roadmap with an upcoming Advantage 2 generation codename Zephyr with 7000 qubits to be released by 2024 (with a first 500-qubit prototype which was delivered in June 2022), a 20-way qubits connectivity in a new graph architecture[593]. They announced also the future release of a gate-based QPU that will be implemented in a separate (flux-based) multi-layer superconducting processor architecture, starting with 60 and then 1000 qubits to implement error correction. They plan to use surface code QECs and to use some combination of RSFQ and other cryo-electronics to control these qubits. They could be highly differentiated for that respect and only challenged by other superconducting qubit vendors partnering with SeeQC.

How about error rates? They are not computed the same way as gate-based single and two qubit gate fidelities. The error rate is measured as a precision with the Ising model parameters, which is about 2% given the DACs precision[594]. The high-error rate of D-Wave annealing systems can be mitigated with some quantum error correction technique, created in 2019[595] and more recently in 2022 for post-processing error correction[596] and machine learning aided error correction[597].

Other researchers found that the thermal noise involved in the annealing process could be used as a resource to enable faster and more reliable computations, involving the curious process or reverse annealing[598]. It could help finding a better solution than an existing solution already computed with a regular annealing process.

The qubits operate at 10 to 15 mK like all gate-based superconducting qubits. The cryostat is thus the same, using a dry dilution system. Cryogeny consumes about 16kW out of a total of 25kW. The remaining 9kW is related to traditional computer control systems that are outside the cryostat. The cryogenic part includes an enclosure with five layers of magnetic isolation.

So, what is quantum in D-Wave? Beyond the many Josephson junctions used in their chipset, it comes from the tunnelling effect that allows the system to quickly search for a global energy minimum of an N-body system[599]. It is coupled with superposition of the qubit states. According to D-Wave, the system also uses entanglement, which is poor and probably circumvented to nearest neighbor qubits. This has been questioned by some scientists[600].

Algorithms designed for classical gate-based quantum computers can theoretically be converted into algorithms executable on D-Wave and vice versa at a maximum polynomial time overhead cost, which can be substantial[601]. However, a similar problem will require many more qubits with D-Wave than with a universal quantum computer.

---

593 See Zephyr Topology of D-Wave Quantum Processors by Kelly Boothby, Andrew D. King and Jack Raymond, D-Wave, September 2021 (18 pages).

594 The question remains open as to whether this architecture is scalable and provides a real quantum advantage. This is questioned in Fundamental Limitations to the Scalability of Quantum Annealing Optimizers by Tameen Albash et al, 2019. The reasons: issues of noise and thermodynamics.

595 See Analog errors in quantum annealing: doom and hope by Adam Pearson et al, 2019 (16 pages).

596 See Post-Error Correction for Quantum Annealing Processor using Reinforcement Learning by Tomasz Śmierzchalski et al, March 2022 (14 pages).

597 See Boosting the Performance of Quantum Annealers using Machine Learning by Jure Brence et al, March 2022 (14 pages).

598 See Thermodynamic study of D-Wave processor could lead to better quantum calculations by Hamish Johnston, June 2020.

599 See the review paper Quantum Annealing: An Overview by Atanu Rajak et al, India, July 2022 (36 pages).

600 Jonathan Dowling thought in the previous reference that the only quantum effects of D-Waves were tunneling and superposition, but without quantum entanglement.

601 This is documented in Adiabatic quantum computation is equivalent to standard quantum computation, 2005 (30 pages) and in How Powerful is Adiabatic Computation? by Wim van Dam, Michele Mosca and Umesh Vazirani, 2001 (12 pages).

On a D-Wave, the number of qubits would need to be up to 32 times the number of quantum gates of the classical quantum algorithm but it depends on the problem[602].

According to D-Wave, their annealers can solve NP-complete problems, a category of combinatorial problems theoretically solved in polynomial time on D-Wave but which are solved in exponential time on a classical computer[603], like routing problems, the traveling salesperson (TSP) problem and the likes. D-Wave annealers can also be used to solve statistical problems[604].

In 2017, **John Preskill** estimated that there is no convincing theoretical basis for the advantage of quantum annealing, which is one form of adiabatic quantum computing[605]. He thinks this architecture is not theoretically as scalable as general-purpose quantum computers. The arguments about D-Wave's annealers quantumness revolve around the low-scale coherence between qubits which may prevent an efficient implementation of quantum annealing[606]. It is also related to the limited connectivity between qubits[607]. Others think that D-Wave systems can generate at best some quadratic acceleration and not an exponential one, compared to traditional computing[608]. **Daniel Lidar** from the University of Southern California is investigating variations of quantum annealing algorithms that could solve intractable problems on classical computers[609].

D-Wave's software development environment is Ocean, a suite of open source Python tools and libraries accessible via the Ocean SDK on both the D-Wave GitHub repository and in their Leap quantum cloud service (that is also accessible on Amazon Braket).

It contains a large set of libraries to solve various optimization and constraint satisfaction problems as described on the right. We cover it with more details later in this book in the section dedicated to hardware vendor software development tools.

In October 2021, D-Wave released its **Constrained Quadratic Model** solver (CQM), working with both discrete and continuous variables. It expanded the optimization problems D-Wave's annealers can solve, with up to 100.000 variable constraints[610] on top of the binary quadratic model (BQM) solver problems defined with binary values (0,1) and the discrete quadratic model (DQM) solver for problems on nonbinary (multiple choices) values.

One of the oldest and famous D-Wave publicized case study came from Google and NASA using a 2012 D-Wave annealer to solve an optimization and combinatorial problem in a graph whose algorithm was designed in 1994.

---

[602] From "Automatically Translating Quantum Programs from a Subset of Common Gates to an Adiabatic Representation" by Malcolm Regan et al, seen in Reversible Computation, conference proceedings, 11th International Conference, RC 2019, Lausanne, Switzerland, June 2019 (246 pages).

[603] See Practical Annealing-Based Quantum Computing by Catherine McGeoch et al of D-Wave, June 2019 (16 pages) which makes an inventory of the benefits of quantum annealing computing, especially in terms of the size of the problems to be solved, which should be neither too small because they are trivial, nor too large because they must then be broken down into sub-problems that are manageable with the capacity of current D-Wave processors. It seems that the problems to be solved must have global minimums and local minimums, the first being difficult to find with classical methods.

[604] See Applications of Quantum Annealing in Statistics by Robert C. Foster, 2019 (30 pages).

[605] In Quantum Computing for Business, 2017 (41 slides).

[606] See How "Quantum" is the D-Wave Machine? by Seung Woo Shin, Umesh Vazirani et al, 2014 (8 pages).

[607] See the example of Phase-coded radar waveform AI-based augmented engineering and optimal design by Quantum Annealing by Timothé Presles et al, Thales, August 2021 (9 pages). In this use case, no quantum advantage can be seen with D-Wave due to limited qubits connectivity.

[608] This was the opinion of Jonathan P. Dowling in Schrödinger's Killer App - Race to Build the World's First Quantum Computer by Jonathan P. Dowling, 2013 (445 pages), pages 208 to 216.

[609] See his Adiabatic quantum computing page on USC Quantum Computation and Open Quantum Systems web site.

[610] See Hybrid Solver for Constrained Quadratic Models, 2021 (8 pages).

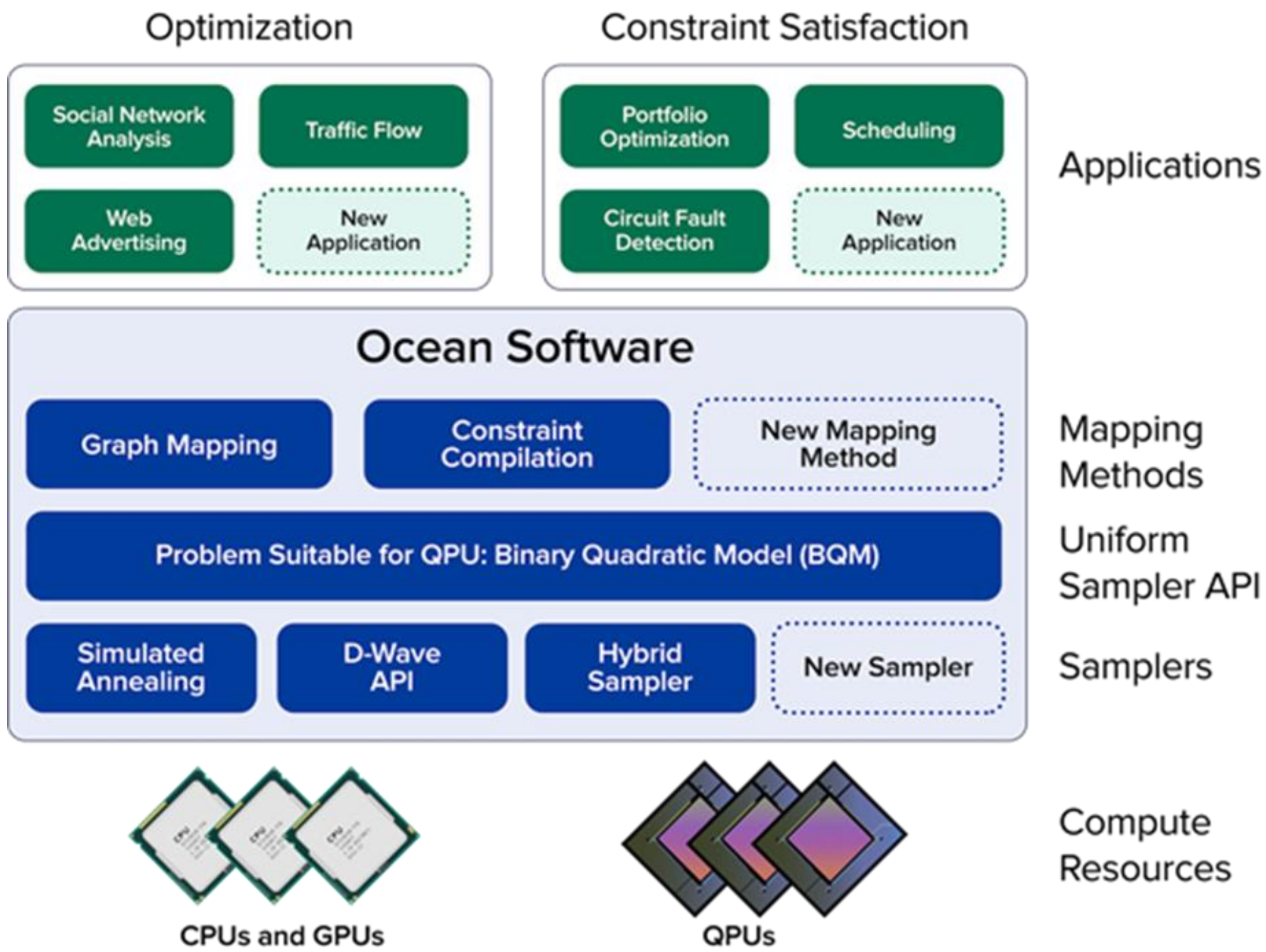

*Figure 277: D-Wave Ocean software platform. Source: D-Wave.*

Google announced in 2015 that it had achieved a performance 100 million times better than that of traditional computers, a single core of an Intel Xeon server processors[611]. Like many such claims, they were questionable since based on a single optimized algorithm, here, a Quantum Monte Carlo simulating quantum tunneling on a classical computer. Critics abounded about this performance[612]. It was a first "hype" moment of quantum computing.

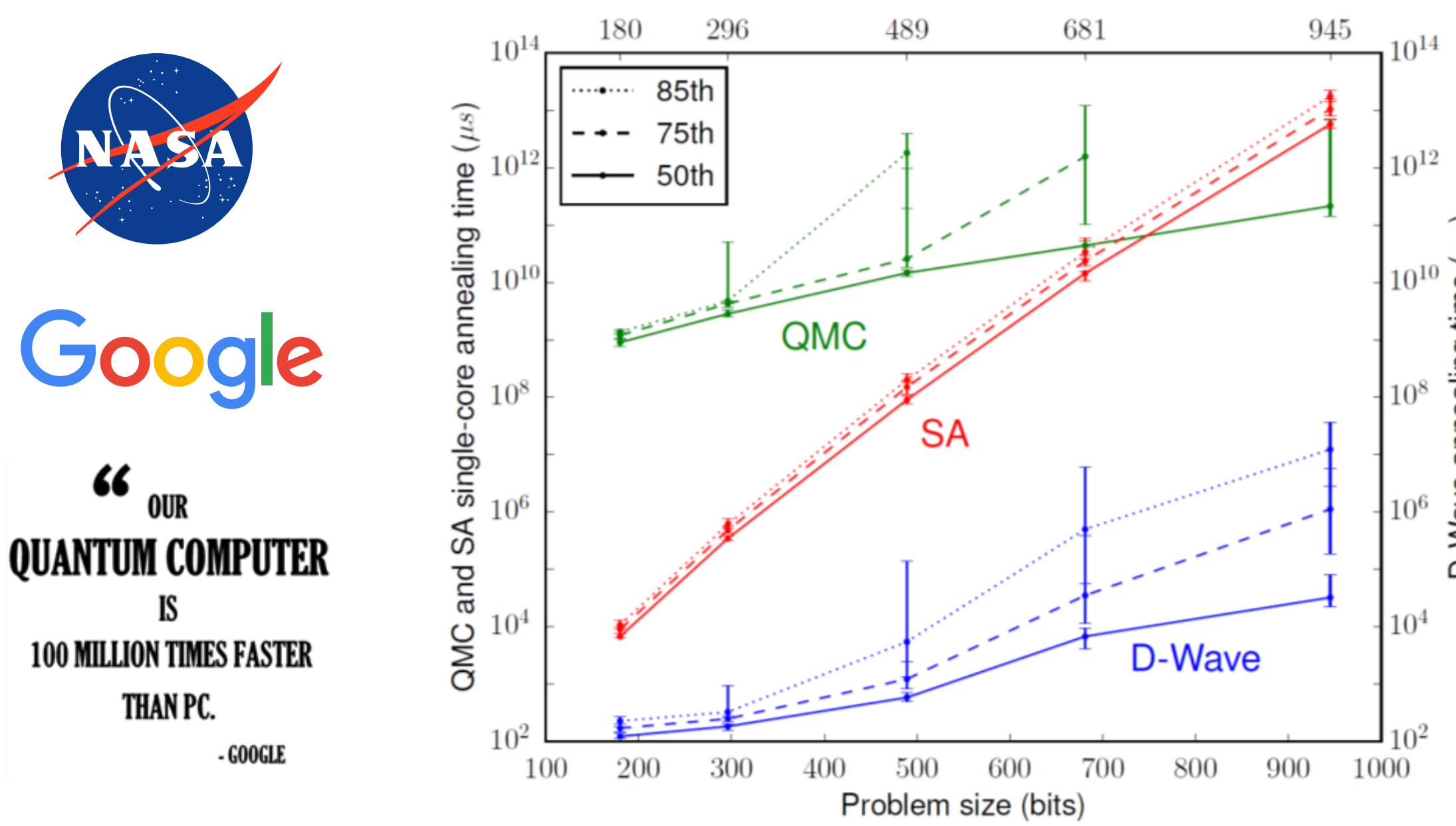

*Figure 278: how Google and NASA communicated in 2015 about the performance of a D-Wave annealer. Source: [What is the Computational Value of Finite Range Tunneling](#) by Vasil S. Denchev, John Martinis, Hartmut Neven et al, January 2016 (17 pages).*

---

[611] In Google's D-Wave 2X Quantum Computer 100 Million Times Faster Than Regular Computer Chip by Alyssa Navarro in Tech Times, November 2015 and documented in What is the Computational Value of Finite Range Tunneling by Vasil S. Denchev, Sergio Boixo, Sergei V. Isakov, Nan Ding, Ryan Babbush, Vadim Smelyanskiy, John Martinis and Hartmut Neven, January 2016 (17 pages).

[612] Including Temperature scaling law for quantum annealing optimizers, 2017 (13 pages), which points out the limitations of quantum annealing.

D-Wave communicates on many other of its pilot references [613]. Its web site references over 250 case studies, the largest number of any quantum computing vendor[614]. Here are a couple ones:

- **Denso**, a Japanese car equipment manufacturer presented at CES 2017 in Las Vegas a system for optimizing a fleet of Toyota delivery vehicles.
- **Biogen**, **1Qbit** and **Accenture** did prototype in 2012 a screening solution to identify molecules for drug retargeting, with a problem of map staining[615]. It is difficult to say what this has generated in practice. Their partner **Menten AI** performs protein analysis.
- **Lockheed-Martin** produced in 2014 some validation procedures for its embedded software in 6 weeks instead of 8 months with a D-Wave and its QVTRace tool[616].
- **Volkswagen** simulated the operations of a cab fleet in Beijing, and also a solution to develop new batteries[617]. The solution was used in November 2019 to optimize the shuttle route at Lisbon's WebSummit, in partnership with Here and Volkswagen's Data:Lab in Munich.
- **NASA** experimented D-Wave annealers in its joint QUAIL lab with Google in various fields, including the detection of exoplanets by analysis of telescopic observations using the transit method, as well as for various optimization and planning problems[618].
- **GE Research** experimented some hybrid maintenance resource allocation optimization application.
- **Ocado**, a British retailer, prototyped some optimization solution for its robots-based warehouse operations.
- **Los Alamos National Laboratory** with **Stanford University** prototyped the detection of the formation of terrorist networks in Syria with analyzing imbalances in social networks[619].
- **Volkswagen** experimented quantum annealing to optimize car paint-shop processing in order to minimize color switching, but with no clear quantum advantage[620].
- **KAIST** and **LG U+** in Korea used D-Wave annealers to optimize data-transfer routing using a fleet of low-earth orbit telecommunications satellites and a QUBO algorithm, with the help of **Qunova Computing**, a Korean quantum software startup[621].
- Researchers in Poland, Hungary, Brazil and the USA found a way to use D-Wave annealers to **optimize trains dispatching** on single-railways lines[622].

---

[613] I found this inventory in Quantum Applications by D-Wave, May 2019 (96 slides).

[614] See Quantum Annealing for Industry Applications: Introduction and Review by Sheir Yarkoni et al, Leiden University and Honda Research, December 2021 (43 pages) which provides an overview of how D-Wave annealers are "programmed" and what kinds of problems it can solve.

[615] Described in Programming with D-Wave Map Coloring Problem, 2013 (12 pages).

[616] See Quantum Computing Approach to V&V of Complex Systems Overview, 2014 (31 slides) and Experimental Evaluation of an Adiabatic Quantum System for Combinatorial Optimization, 2013 (11 pages).

[617] See Forget quantum supremacy: This quantum-computing milestone could be just as important by Steve Ranger, December 2019.

[618] See Quantum Computing at NASA: Current Status by Rupak Biswas, 2017 (21 slides) as well as Adiabatic Quantum Computers: Testing and Selecting Applications by Mark A. Novotny, 2016 (48 slides).

[619] See Using the D-Wave 2X Quantum Computer to Explore the Formation of Global Terrorist Networks by John Ambrosiano et al, 2017 (14 pages).

[620] See Multi-car paint shop optimization with quantum annealing by Sheir Yarkoni et al, September 2021 (7 pages).

[621] See KAIST & LG U+ Team Up for Quantum Computing Solution for Ultra-Space 6G Satellite Networking, KAIST press office, June 2022.

[622] See Quantum annealing in the NISQ era: railway conflict management by Krzysztof Domino et al, December 2021 (23 pages).

- In April 2020, D-Wave opened free access to its cloud computers to researchers looking for **solutions** to the **covid pandemic19**[623]. The solutions developed included solving optimization problems such as optimizing patient routing to hospitals in Japan, modeling the spread of the virus, managing nurses' schedules in hospitals, assessing the rate of virus mutation and screening molecules. It remains to be proven that D-Wave provides a real quantum advantage in solving these different problems.
- In February 2021, D-Wave and Google published a study showcasing a computational advantage of annealing with the **D-Wave Advantage** for simulating some condensed matter physics, 3 million times faster than with classical methods. It didn't exactly describe the classical hardware that is being used as a reference, but it looked like a traditional Intel-based server[624]. These comparisons with a single narrow algorithm are insufficient to draw any conclusions. Another similar work was published in 2022 with using 2000 qubits for accurate simulations of coherent quantum dynamics at large scales[625].

We can also mention some recent quantum annealing algorithms:

- **Sorting lists** and building search trees or heaps, which can be modeled as QUBO problems[626].
- **Neural network training** with using binary encoding the neural network free parameters, polynomial approximation of the activation function and reduction of binary higher-order polynomials into quadratic ones[627].
- **Parallelizing annealing** on several quantum annealings with decomposing a problem graph into several graphs with the DBK algorithm, used for the Maximum Clique problem with 120 nodes and 6395 edges[628]. These smaller disjointed graphs are executed in a D-Wave Advantage.
- **Physical simulations** of topological matter and phase change[629].

As of 2022, D-Wave had installed fewer than 10 quantum annealers at customer sites[630] and operates more than 30 of them in its own facilities, with more than half of them dedicated to their cloud access offering, some of them being available through the Amazon Braket cloud offering.

One D-Wave Advantage was ordered by the DoE Los Alamos National Laboratory in September 2019 and deployed since then[631]. In January 2022, a first D-Wave (Advantage) system was deployed in Europe, at the Forschungszentrum Jülich Supercomputing Center. It was also the first D-Wave coupled with a supercomputer as part of the Jülich UNified Infrastructure for Quantum computing (JUNIQ).

---

[623] See Can Quantum Computers Help Us Respond to the Coronavirus? by Mark Anderson, April 2020.

[624] See Scaling advantage over path-integral Monte Carlo in quantum simulation of geometrically frustrated magnets, February 2021 (6 pages).

[625] See Coherent quantum annealing in a programmable 2000-qubit Ising chain by Andrew D. King et al, Nature, February-September 2022 (24 pages).

[626] See QUBOs for Sorting Lists and Building Trees by Christian Bauckhage et al, March 2022 (6 pages).

[627] See Completely Quantum Neural Networks by Steve Abel et al, February 2022 (12 pages).

[628] See Solving Larger Optimization Problems Using Parallel Quantum Annealing by Elijah Pelofske et al, May 2022 (16 pages). See also Parallel Quantum Annealing by Elijah Pelofske et al, November 2021 (12 pages).

[629] See Observation of topological phenomena in a programmable lattice of 1,800 qubits, August 2018 (37 slides).

[630] Identified customers are the joint Google/NASA Quail research center, USRA (Universities Space Research Association), Lockheed Martin and the University of Southern California sharing one system, and Jülich Supercomputing Centre in Germany (since 2021). Other customers like in pharmaceutical companies are using D-Wave annealers through their Leap cloud offering.

[631] See Nuclear weapons lab buys D-Wave's next-gen quantum computer by Stephen Shankland, September 2019 and On the Emerging Potential of Quantum Annealing Hardware for Combinatorial Optimization by Byron Tassef et al, October 2022 (25 pages).

The Julich team has published several benchmarks of various algorithms since then[632].

In summary, quantum annealing may be a technique contested by many specialists, but it has the merit of existing and being testable in many use cases[633]. It will probably make some progress with newly published cases running on its Advantage generation. And we are looking forward at their capabilities to implement gate-based superconducting qubits computing.

**Qilimanjaro** (2019, Spain) is a startup based in Barcelona created by three physicists coming from different Spanish institutions (Barcelona Supercomputing Center, IFAE, University of Barcelona) and with a strong international experience and two experienced business managers!

The founding team assembles Jose Ignacio Latorre (Chief Science Officer, also now the Director of CQT in Singapore and Chief Research of CRO Quantum at TII in Abu Dhabi, went through MIT, CERN and Niels Bohr Institute), Pol Forn-Díaz (Chief Hardware Architect, TU Delft, MIT, Caltech and IQC Waterloo), Artur Garcia Sáez (Chief Software Architect, ICFO, Stony Brook), plus Víctor Canivell (Chief Business Officer) and Jordi Blasco (Chief Financial & Legal Officer).

They develop their own quantum annealer based on coherent flux qubits. Their differentiation lies with a better qubit coherence, qubits coupling designs and qubits connectivity.

These qubits will be first controlled by classical electronics working at room temperature. In a later stage, they plan to create cryogenic controls on a separate chipset. They rely on two fabs for their qubits designs, the one from IFAE and the one from the Institute of Microelectronics of Barcelona (IMB-CNM, which has similarities with the C2N in Palaiseau, France).

In a full-stack approach, they are also developing **QIBO**, a quantum software platform in the cloud. It is the cloud operating service to run software batches on the future Qilimanjaro quantum annealer, classical quantum emulators and gate-based quantum computers with a design pattern to create classical/quantum hybrid algorithms. But they also plan to sell their hardware to customers willing to use it on-premises.

They initially wanted to launch an ICO (initial coin offering) to fund the company when it was trendy but abandoned it. On top of benefiting from public grants, the company started to work for an unnamed French international company involved in logistics to develop quantum inspired optimization algorithms. They then established a partnership with Abu-Dhabi to help the Emirate create its Quantum Research Centre at the Technology Innovation Institute (QRC-TII). They provide Abu Dhabi with their know-how to build the QCR research lab and team, provide access to their technology with the goal to sell them a multi-qubit quantum processor before 2023, and let them then become self-sufficient. Jose-Ignacio Latorre became their TII's Chief Scientist after the deal was made. In 2020, he also became the director of CQT in Singapore after having been a visiting professor since 2013. CQT may play a role first in Qilimanjaro's software development efforts.

Qilimanjaro also benefits from European funding through the project **AVAQUS** already mentioned. This project coordinator is Pol Forn-Díaz, head of the IFAE Quantum Computing Technologies group on top of his role in Qilimanjaro. It involves the superconducting team from Nicolas Roch at Institut Néel in Grenoble who designs the microwave amplifiers used in flux qubits readouts.

---

[632] Like Improved variational quantum eigensolver via quasi-dynamical evolution by Manpreet Singh Jattana, Kristel Michielsen et al, February 2022 (19 pages) and Quantum annealing for hard 2-SAT problems : Distribution and scaling of minimum energy gap and success probability by Vrinda Mehta, Fengping Jin, Hans De Raedt and Kristel Michielsen, February 2022 (17 pages).

[633] To learn more about D-Wave, here are their explanations about the structure of their hardware, a video explaining the structure of D-Wave chipsets, a video from Linus, a blogger who gets into the bowels of a D-Wave 2000Q in quite a detailed way, the video of Colin Williams's presentation at USI in June 2018 in Paris (33 minutes) as well as Near-Term Applications of Quantum Annealing, 2016, an interesting Lockheed Martin presentation on the uses of a D-Wave computer (34 slides). And testimonials from their customers in Qubits 2017. See also Brief description on the state of the art of some local optimization methods: Quantum annealing by Alfonso de la Fuente Ruiz, 2014 (21 pages).

At last, let's mention that **NEC** (Japan) is also developing coherent quantum annealers using parametric oscillators and qubits as couplers.

They ambition to have an available system by 2023 with an "all-to-all" qubits connectivity (which is actually a nearest-neighbor one)[634]. They seem to reuse some research work done on superconducting qubits initially aimed at gate-based quantum computing. Meanwhile, they also work on some simulated annealing software running on their classical supercomputer, the SX-Aurora Tsubasa.

## Superconducting qubits

After describing superconducting-based quantum annealing, let's move on to gate-based superconducting qubits quantum computers. From a physical point of view, D-Wave's accelerators and superconducting qubits ones have in common to be based on Josephson junctions but their qubit physics, overall architectures, control signals and programming models are way different.

**superconducting qubits**

- **key technology** in public research and with commercial vendors (IBM, Google, Rigetti, Intel, Amazon, OQC, IQM, etc).
- **record of 127 programmable qubits** with IBM.
- constant progress in **noise reduction,** particularly with the cat-qubits variation which could enable a record low ratio of physical/logical qubits.
- many existing **enabling technologies**: cryostats, cabling, amplifiers, logic, sensors.
- **potentially scalable technology** and deployable in 2D geometries.

- **qubit coherence time usually** < 300 µs.
- high **qubits noise levels** with most vendors.
- **cryogeny constrained** technology at <15 mK.
- **heterogeneous qubits** requiring calibration and complex micro-wave frequency maps.
- **cabling complexity** and many passive and active electronic components to control qubits with micro-waves.
- **qubit coupling limited** to neighbor qubits in 2D structures (as compared with trapped ions).
- **qubits size** and uneasy miniaturization.

*Figure 279: superconducting qubits pros and cons. (cc) Olivier Ezratty, 2022.*

Superconducting qubits seem to be the kings in quantum computing town, being exploited or chosen by IBM, Google, Intel, Rigetti, Amazon, as well as many startups such as IQM (Finland) and OQC (UK). It is the currently best scalable architecture in the gate-based model, even if the results are still modest with a record of 127 operational qubits for with IBM and 66 in China as of mid-2022[635].

Like all existing gate-based quantum systems, superconducting qubits computers are in the pre-NISQ or NISQ realm, *aka* noisy intermediate-scale computers, with such a low qubit gates and readout fidelity that they are impractical for most industry use cases.

It is observable with the discrepancy between the number of available physical qubits (72, 80 and 127 with Google, Rigetti and IBM) and the number of qubits that are actually exploited with useful algorithms, that doesn't exceed 20 at this point in time (besides a correction error code using 49 qubits on a 72 qubit chipset by Google in July 2022). IBM's quantum volume is currently capped at 9 useful qubits with their best 27 qubits system using their Falcon R10 chipset.

The short term workaround of low fidelity are quantum error mitigation techniques and adapted quantum algorithms.

[634] See Quantum Computing Initiatives, NEC.

[635] See a general point on the issue in Superconducting Qubits: Current State of Play by Morten Kjaergaard et al, MIT & Chalmers, 2020 (30 pages).

Longer term workarounds are quantum error correction and qubits number scalability. It creates problems that are not yet solved with regards to fidelity stabilization with a growing number of qubits, solving the qubits cabling maze with cryogenic electronics or signals multiplexing and scaling cryostats cooling power.

Google and IBM's current approaches to scale their systems are way too optimistic as we'll see later but you can still bet on them solving both fundamental and engineering problems[636].

The Josephson effect is used in these qubits to control the flow of a circulating current through a thin nanometric insulating barrier between two superconducting metals, creating a tunnel junction. It creates a nonlinear and nondissipative physical system with a single degree of freedom, the number of Cooper pairs (electron pairs) traversing the tunnel junction conjugated to the superconducting phase difference across it. Superconducting qubits have the particularity of being the only mainstream ones that are macroscopic, in the sense that they are not linked to the control of a single particle as an atom, electron or photon, as in most other qubit technologies.

At superconducting temperature, the superconducting electrons in a Josephson loop look like a single particle, with billions of electron Cooper pairs behaving as bosons which can be condensed into the same quantum state. They form an artificial atom with precisely controllable energy levels according to their parameters comprising a Josephson barrier, some capacitances and inductances connected in series and/or in parallel and some non-destructive readout circuits using a nearby resonator[637]. One qubit is using about $10^{11}$ electrons (100 billion).

Superconducting qubits use non dissipative elements: capacitors, inductors and the Josephson junction which act as a nonlinear non-dissipative inductor. Capacitors store energy in the electric field while inductors store energy in the magnetic field. But at any non-zero frequency, superconductors still dissipate some power, through two channels: the transport by the Cooper pairs and by normal charge carriers (quasi-particles), that is proportional to the quasi-particle density, which diminishes exponentially at low temperatures.

### History

The history of superconducting qubits started in the mid-1980s but you need to fly back to 1957 and the elaboration of the **BCS theory** explaining (partially) how pairs of opposite spin electrons - *aka* Cooper pairs - behave at low temperatures, generating the superconducting effect. Then, 1962 marks the Josephson effect discovery by **Brian Josephson** completed by its experimental proof in 1963 by **Philip Anderson** and **John Rowell**. In 1980, **Antony Leggett** modelized the collective degrees of freedom of superconducting circuits. A bit like a Bose-Einstein condensate of cold atoms, Cooper-pairs of electrons in a superconducting material behave like a single quantum object with its own quantum wave[638].

In 1985, **John Clarke**, **Michel Devoret** (his post-doc) and **John Martinis** (his PhD), all at Berkeley, implemented the first spectroscopy of these artificial atom, using microwave radiations to excite it, creating the first phase qubit[639]. Back then, the JJ (the little nickname for Josephson junctions) was implemented with Nb-$NbO_x$-PbIn (niobium, lead, indium) and cooled with a $He_4$-based cryostat.

---

636 See for example the review paper A practical guide for building superconducting quantum devices by Yvonne Y. Gao et al, September 2021 (49 pages).

637 This artificial atom property was demonstrated in 1985. See Energy-Level Quantization in the Zero-Voltage State of a Current-Biased Josephson Junction by John Martinis, Michel Devoret and John Clarke, 1985 (2 pages).

638 See A Brief History of Superconducting Quantum Computing by Steven Girvin, August 2021 (39 mn).

639 See Energy-Level Quantization in the Zero-Voltage State of a Current-Biased Josephson Junction by John M. Martinis, Michel H. Devoret and John Clarke, PRL, 1985 (4 pages).

In 1998, **Vincent Bouchiat**, then a PhD in **Michel Devoret**, **Daniel Esteve** and **Cristian Urbina**'s Quantronics group at CEA-Saclay in France, implemented the first Cooper Pairs Box (CPB) in a loop and characterized its ground state.

The first demonstration of quantum coherent superposition with the first excited state was achieved in 1999 by **Yasunobu Nakamura** with **Yuri Pashkin** and **Jaw-Shen Tsai** at NEC Labs in Tsukuba, Japan[640]. It was the first "charge qubit" per se, with a tiny coherence time of 2 ns. They extended it in 2001, implementing the first measurement of Rabi oscillations associated with the transition between two Josephson levels in the Cooper pair box, using the configuration developed by Vincent Bouchiat and Michel Devoret in 1998. A first functional qubit version of the Cooper pair box, the quantronium, was demonstrated by the CEA-Saclay Quantronics team in 2002[641].

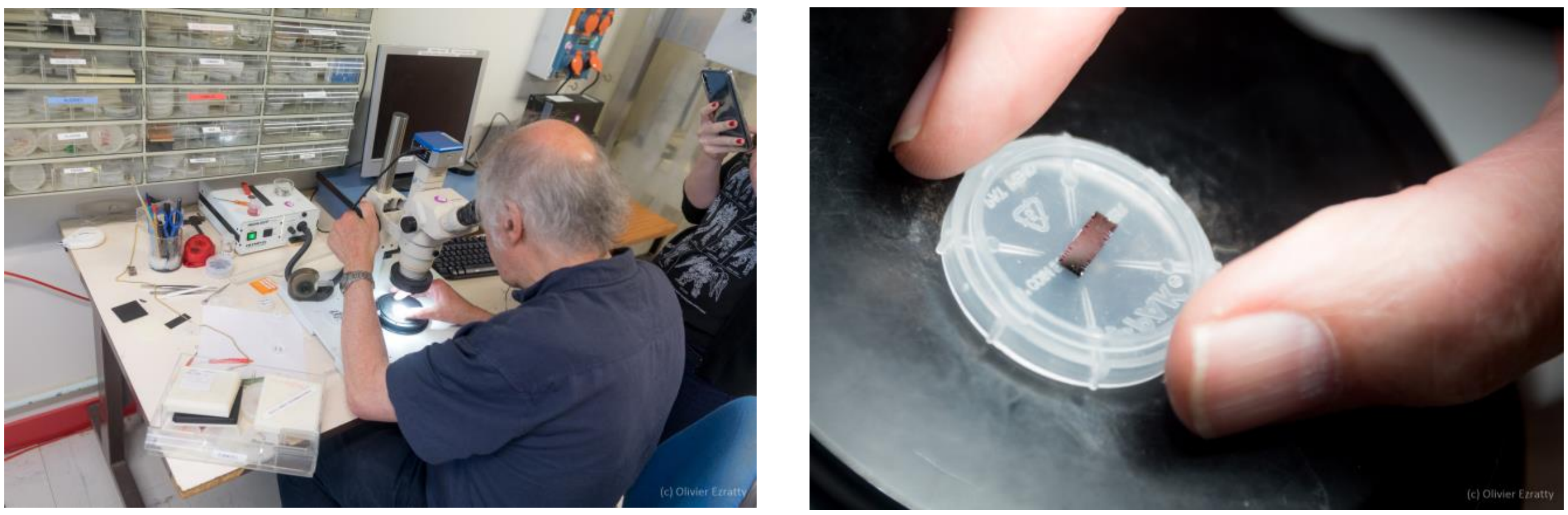

*Figure 280: Daniel Esteve (CEA Quantronics) showing to the author the first operational two-transmon processor in his laboratory, June 2018.*

The modern version of the CPB circuit, the transmon, was developed at Yale University in 2006. The Yale University research teams led by **Rob Schoelkopf**, **Michel Devoret** and **Steve Girvin** welcomed many talented theoreticians and experimentalists who were key contributors to the progress of transmon qubits. **Alexandre Blais** and **Andreas Wallraff** developed around 2003-2004 the key principles of circuit QED (cQED)[642]. It allowed quantum non-demolition readout of qubit state in the dispersive regime. A QND readout happens after measurement collapses the wave function onto $|0\rangle$ or $|1\rangle$ and a subsequent readout will yield the same $|0\rangle$ or $|1\rangle$[643]. Then, **David Schuster** and **Jay Gambetta** created between 2007 and 2011 2D and 3D cavity resonators designs[644]. **Jens Koch** created Cooper

[640] See Coherent control of macroscopic quantum states in a single-Cooper-pair box by Yasunobu Nakamura, Yuri Pashkin and Jaw-Shen Tsai, Nature, 1999 (4 pages).

[641] See Superconducting quantum bits by Hans Mooij, Physics World, December 2004 that provides more technical insights of what was achieved in Japan and France between 1999 and 2022. It took about 12 years to CEA's team to reach 4 qubits as described in the interesting thesis Design, fabrication and test of a four superconducting quantum-bit processor by Vivien Schmidt, 2015 (191 pages). Back then, IBM and Google teams were also at a similar stage.

[642] cQED was defined in Cavity quantum electrodynamics for superconducting electrical circuits: An architecture for quantum computation by Alexandre Blais, Ren-Shou Huang, Andreas Wallraff, Steve Girvin and Rob. Schoelkopf, PRA, 2004 (14 pages) and Strong coupling of a single photon to a superconducting qubit using circuit quantum electrodynamics by Andreas Wallraff, David Schuster, Alexandre Blais, Steve Girvin, Rob Schoelkopf et al, Nature, 2004 (7 pages). See also Superconducting Qubits: A Short Review by Michel H. Devoret, Andreas Wallraff and John M. Martinis, 2004 (41 pages) and Circuit QED and engineering charge based superconducting qubits by Steve Girvin, Michel Devoret and Rob Schoelkopf, 2009 (27 pages).

[643] QND was created by Vladimir Braginsky (1931-2016) in Russia in the early 1980s.

[644] See Circuit Quantum Electrodynamics by David Schuster, 2007 (255 pages), 3D microwave cavity with magnetic flux control and enhanced quality factor by Yarema Reshitnyk et al, 2016 (6 pages) and the foundational paper Observation of high coherence in Josephson junction qubits measured in a three-dimensional circuit QED architecture by Hanhee Paik, Michel Devoret et al, 2011 (5 pages).

pair boxes with a large shunting capacitance which created a modest reduction in anharmonicity and enabled strong coupling with microwave photons[645].

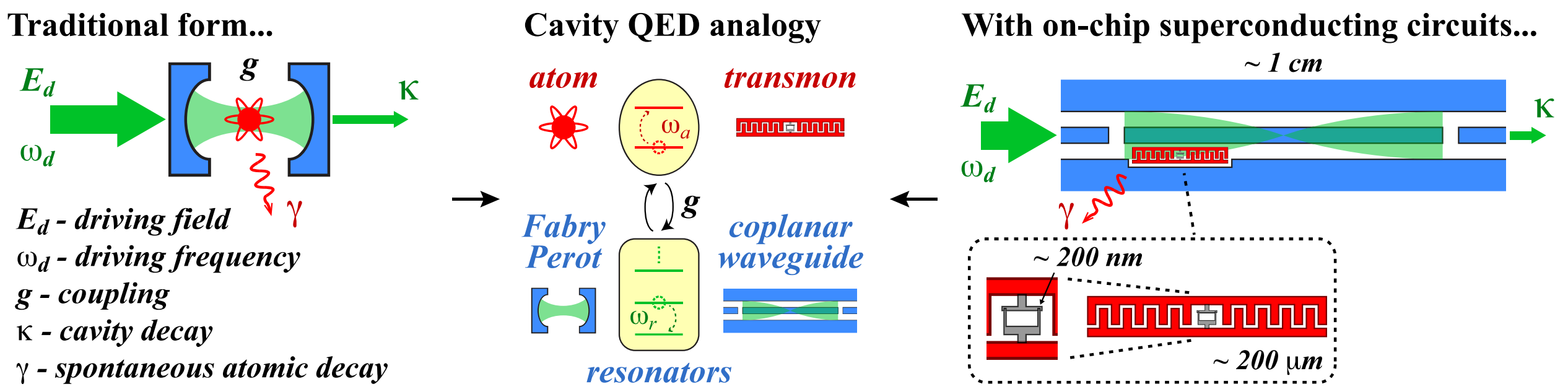

Figure 281: principles of circuit QED. Source: Circuit QED - Lecture Notes by Nathan K. Langford, 2013 (79 pages).

**Jerry Chow** was also a key contributor between 2005 and 2010 and has since then been at IBM, now leading their quantum hardware system developments in Jay Gambetta's team. In 2009. Devoret, Schoelkopf, **Leonardo Di Carlo** (now at TU Delft), **Jerry Chow** et al created the first programmable two-qubit processor and implemented a small Grover search on it. **Blake Robert Johnson** proposed in 2011 to use a Purcell filter to protect a qubit from spontaneous emission coming from the Purcell effect that is a relaxation through the readout resonator. It's a mix of low-pass and high-pass microwave filter[646]. **Matt Reagor** and **Hanhee Paik** improved in 2013 the stability of microwaves in 3D resonators used in superconducting qubits[647]. Hanhee Paik has been working as a researcher at IBM since 2014 after a two-year stint at Raytheon BBN. Nowadays, the Yale University team are working on variations of cat-qubits.

Other contributions worth mentioning are **Hans Mooij** (TU Delft) who created a flux-qubit with three Josephson junctions in 1999 with experiments done in 2000. **Andrew Hook** (Princeton) contributed to the development of the transmon qubit. In 2010, **Andrew Cleland**, **John Martinis** and their PhD **Arron O'Connell** were able to entangle three flux superconducting qubits and to control it with a mechanical resonator[648]. It led to the creation of the Xmon tunable qubit in 2013[649], which was later used by Martinis at Google after 2014. **Andrew Cleland** now runs his own lab at the University of Chicago. In 2017, **Peter Leek** then at Oxford created the coaxmon superconducting qubit, where the qubit and resonator are on opposing sides of a single chip, with control and readout wiring being provided by coaxial wiring running perpendicular to the chip plane[650].

It led the same year to the creation of OQC. At last, in 2022, **Mikko Möttönen** from IQM created the Unimon superconducting qubit with a simpler setting, better nonlinearity and fidelities[651].

---

645 See Charge insensitive qubit design derived from the Cooper pair box by Jens Koch, Terri M. Yu, Jay Gambetta, Andrew. Houck, David Schuster, J. Majer, Alexandre Blais, Michel Devoret, Steve Girvin and Rob Schoelkopf, 2007 (21 pages). That's quite a hall of fame for a paper!

646 See Controlling Photons in Superconducting Electrical Circuits by Blake Robert Johnson, a thesis under the direction of Rob Schoelkopf, 2011 (190 pages) which proposed the Purcell filter. See also Controlling the Spontaneous Emission of a Superconducting Transmon Qubit by Andrew Houck, Jay Gambetta, Michel Devoret, Rob Schoelkopf et al, 2008 (4 pages) and Quantum theory of a bandpass Purcell filter for qubit readout by Eyob A. Sete et al, 2015 (15 pages). The spontaneous emission rate (SER) is one key contributor that affects a superconducting qubit coherence time $T_1$.

647 See Reaching 10 ms single photon lifetimes for superconducting aluminum cavities by Matt Reagor, Hanhee Paik et al, 2013 (4 pages).

648 See Quantum ground state and single-phonon control of a mechanical resonator by Aaron O'Connell, John Martinis, Andrew Cleland et al, Nature, 2010 (7 pages).

649 See Coherent Josephson qubit suitable for scalable quantum integrated circuits by R. Barends, John Martinis and Andrew Cleland, April 2013 (10 pages).

650 See Double-sided coaxial circuit QED with out-of-plane wiring by J. Rahamim, Peter Leek et al, 2017 (4 pages).

651 See Unimon qubit by Eric Hyyppä, Mikko Möttönen et al, March 2022 (37 pages).

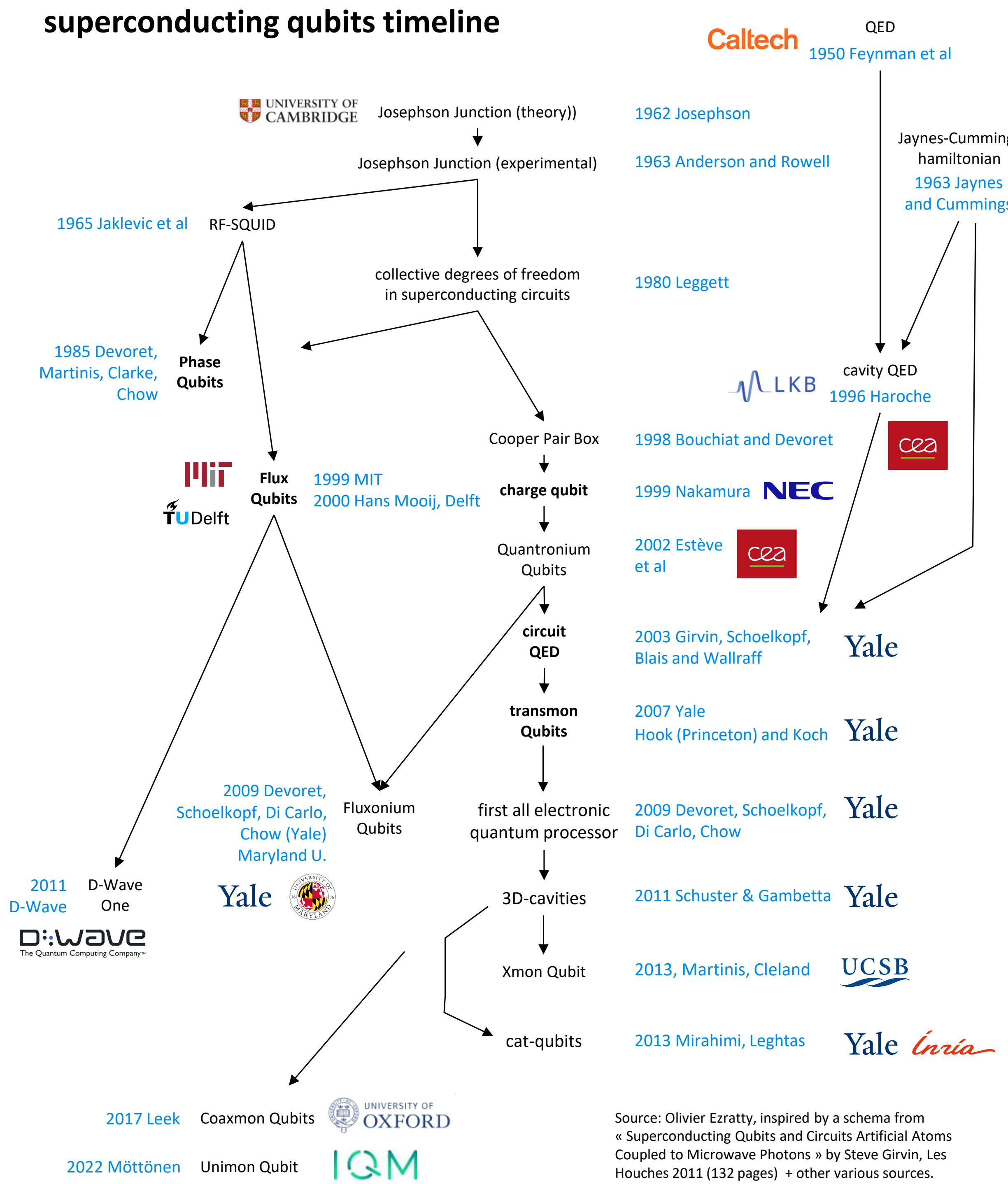

*Figure 282: a historical timeline of superconducting qubits. The contribution of scientists at Yale University seems dominant here, thus the nickname of the "Yale gang". (cc) Olivier Ezratty, 2022.*

We'll also talk later about the cat-qubits, created in 2013 by **Maryar Mirrahimi** and **Zaki Leghtas**.

Let's now circle back to the different types of superconducting qubits that differ in the way they encode quantum information in two distinct states[652]:

**Phase Qubits** use larger Josephson junctions than in charge qubits. Their state corresponds to two levels of current energy in a Josephson junction. This approach was tested by NIST in the USA among other places but no commercial vendor seems to use this type of superconducting qubit. John Martinis tested such qubits back in 2012 at UCSB in a 5-qubit system used to factorize the number 15 [653]. A German (Jülich, University of Munster) and Russian (Kotelnikov Institute) team proposed in early 2020 to use $YBa_2Cu_3O_7$-x nanotubes (also called YBCO, for yttrium, barium, copper and oxide, which is superconducting at 92K) to create phase qubits controllable by a single microwave photon[654].

[652] This is well explained in Practical realization of Quantum Computation, (36 slides).

**Flux Qubits**: their states correspond to the direction of flow of the superconducting current in its loop. It couples a capacitor, from one to three Josephson junctions and a superinductor and has high coherence and large anharmonicity which also enable the handling of qutrits instead of qubits like what Rigetti is experimenting. Measuring the state of such a qubit uses a SQUID (superconducting quantum interference device) with two Josephson junctions connected in parallel, a magnetometer that measures the current direction in the qubit, thus its basis state 0 or 1. This type of superconducting qubit is adopted by D-Wave (in annealing mode and soon, gate-based mode) and Qilimanjaro (in annealing mode), Rigetti, Alibaba[655], Bleximo and Atlantic Quantum in the vendors space. It is studied in research labs at the MIT, TU-Delft (until 2010), the Lawrence Berkeley National Laboratory (Irfan Siddiqi[656]), the University of Berkeley and Yale University (Shruti Puri), the University of Maryland (Vladimir Manucharyan[657]), in Russia[658].

In recent works, fluxonium qubits generated the best $T_1/T_2$ with $T_1$ exceeding 1 ms. They use control frequencies below 1 GHz which lowers down dielectric loss effects and leads to long relaxation time $T_1$. Single-qubit gates can have good speed in the range of 10 ns and errors levels around $10^{-4}$. In this architecture, both readout and control crosstalk are expected to be small[659]. Their main shortcomings are their lower control frequencies and bad protection from both relaxation ($T_1$) and dephasing ($T_2$).

**Charge Qubits**: their states correspond to current flow thresholds in the Josephson junction of the superconducting loop. Small Josephson junctions delimit a superconducting island with a well-defined electrical charge. The basis states of such charge qubits are the states of charge of the island in Cooper pairs. The most common variant is the **transmon**, for "transmission line shunted plasma oscillation qubit", which reduces the effect of charge noise but with a weaker anharmonicity[660]. With transmons, the Cooper pairs box is operated in the phase regime.

The nonlinear Josephson junction inductance makes the LC resonator slightly anharmonic, and its two lowest energy levels are the basis states of the qubit. Transmons are used by IBM, Google, IQM and others. To date, these are the qubits generating the lowest error rate in superconducting qubits but their low anharmonicity creates a toll on gate and readout speeds.

---

[653] See Computing prime factors with a Josephson phase qubit quantum processor by Erik Lucero, John Martinis et al, 2012 (5 pages).

[654] See Energy quantization in superconducting nanowires, February 2020, referring to Energy-level quantization and single-photon control of phase slips in YBa2Cu3O7-x nanowires by M. Lyatti, February 2020.

[655] See Fluxonium qubits for ultra-high-fidelity and scalable quantum processors by Chunqing Deng, (49 minutes) and Fluxonium: An Alternative Qubit Platform for High-Fidelity Operations by Feng Bao et al, 2022 (19 pages).

[656] See Scalable High-Performance Fluxonium Quantum Processor by Long B. Nguyen, Irfan Siddiqi Singh et al, January 2022 (29 pages).

[657] See Millisecond coherence in a superconducting qubit by Aaron Somoroff, Vladimir Manucharyan et al, University of Maryland, March 2021 (14 pages). Not only do they have a $T_2$ exceeding 1.35 ms but their single-qubit gate fidelity also exceeds 0.9999. See also The high-coherence fluxonium qubit by Long B. Nguyen, Vladimir E. Manucharyan et al, October 2018 (12 pages).

[658] See High fidelity two-qubit gates on fluxoniums using a tunable coupler by Ilya N. Moskalenko, Russia, March 2022 (18 pages) which proposes a fluxonium architecture with two qubits CZ gates fidelities of 99,23%.

[659] See Transmon and Fluxonium Qubits by Emanuel Hubenschmid, June 2020 (106 slides).

[660] See Charge insensitive qubit design derived from the Cooper pair box by Jens Koch, Jay Gambetta, Alexandre Blais, Michel Devoret, Rob Schoelkopf et al, 2007 (21 pages).

They are divided into at least two categories: qubits with a single Josephson junction (single junction transmon, used by IBM) or with two Josephson junctions connected in parallel (spit transmon, used by Google)[661].

Then, you have many variations with the **coaxmon** (OQC) and **unimon** (IQM) and the **mergemon** or merged element transmon where the Josephson junction is engineered to act as its own parallel shunt capacitor, reducing the size of the qubit[662].

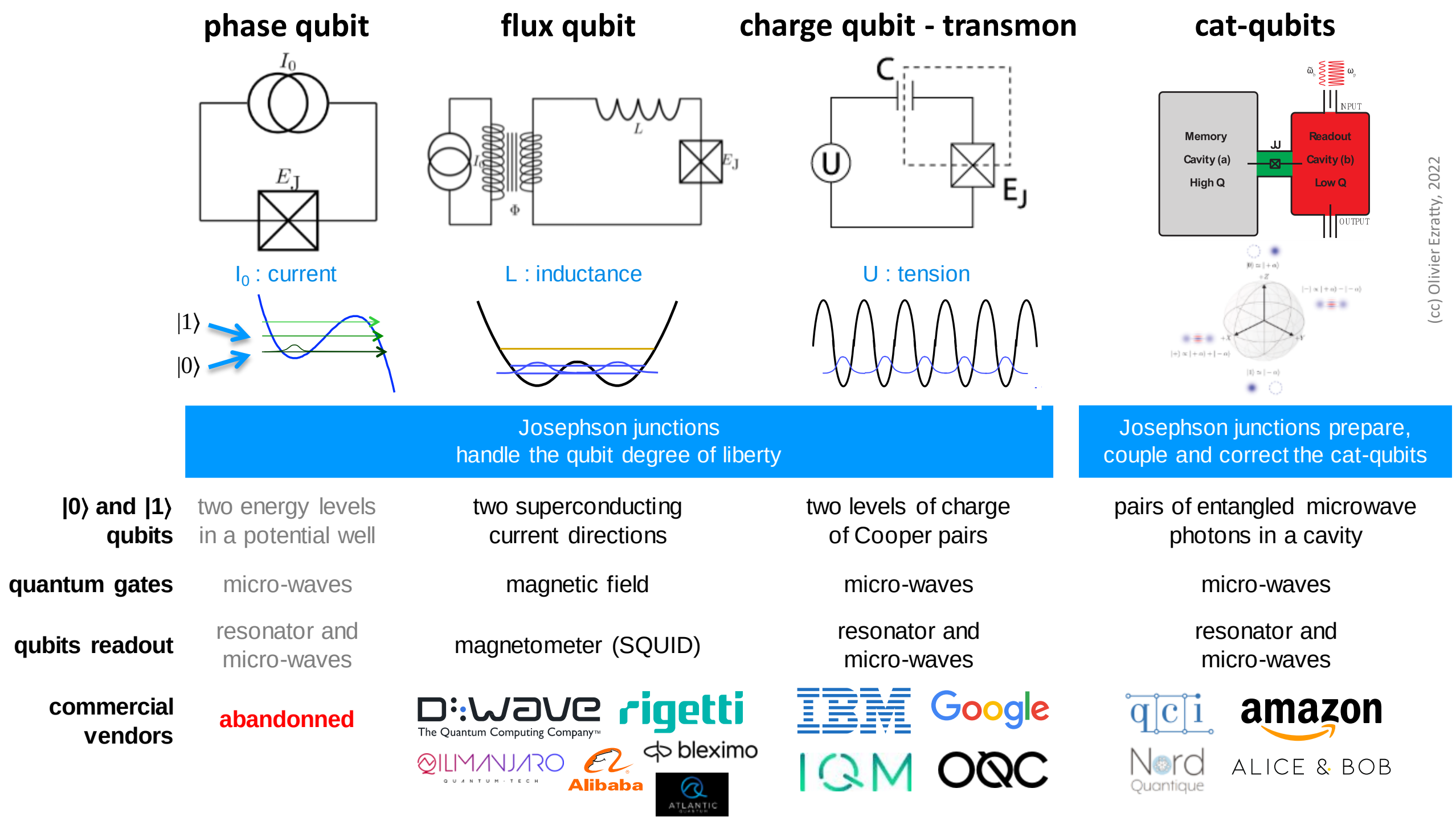

*Figure 283: the different types of superconducting qubits and the related industry vendors. inspired from Implementing Qubits with Superconducting Integrated Circuits by Michel Devoret, 2004 (41 pages) and Flux Noise in Superconducting Qubits, 2015 (44 slides).*

**Andreev Spin Qubits** (ASQ) is a research-level qubit that relies on a localized microscopic excitation of the BCS condensate that natively has only two levels and is based on a nanowire. It is not a collective excitation of the superconducting loop circuit. This qubit type was proposed at and is studied at Chalmers in Sweden[663] (funded as a H2020 program from 2019 to 2023[664]), at CEA Saclay in France[665], NBI in Denmark and also QuTech in The Netherlands, among other places.

Since it manipulates electron spins in relation to a superconducting resonator and makes use of circuit electrodynamics (cQED), it sits in between the categories of superconducting and silicon spin qubits.

---

661 Transmon is a diminutive of "Transmission line shunted plasmon oscillation circuit" created by Rob Schoelkopf, in other words, an oscillator circuit based on shunted Josephson junction. The shunt has become a capacitance that filters low frequencies. A plasmon is the collective behavior of free electrons of metals, here in the form of superconducting Cooper pairs.

662 See Merged-element transmon by R. Zhao et al, December 2020 (8 pages) and Merged-Element Transmons: Design and Qubit Performance by H. J. Mamin et al, IBM Research, August 2021 (8 pages).

663 See the initial proposal in Andreev Level Qubit by A. Zazunov et al, PRL, 2003 (4 pages), Dynamics and phonon-induced decoherence of Andreev level qubit by A. Zazunov et al, PRB, 2005 (22 pages) and the thesis Coherent manipulation of Andreev Bound States in an atomic contact by Camille Janvier, CEA Quantronic, 2016 (268 pages). And recent research in Coherent manipulation of an Andreev spin qubit by M. Hays, Michel Devoret et al, Science, 2021 (17 pages) and Direct manipulation of a superconducting spin qubit strongly coupled to a transmon qubit by Marta Pita-Vidal et al, August 2022 (24 pages).

664 See Andreev qubits for scalable quantum computation with an EU contribution of 3.5M€.

665 See Circuit-QED with phase-biased Josephson weak links by C. Metzger, Christian Urbina, Hugues Pothier et al, January 2021 (22 pages).

**cat-qubits** are cavity-based qubits connected to a transmon qubit used only for their preparation, readout and/or correction depending on the implementation. The cat-qubit technique was devised by Mazyar Mirrahimi and Zaki Leghtas around 2013, particularly during their work at Yale University with Michel Devoret. It was then adopted by Rob Schoelkopf's team at Yale.

**bosonic qubits** is a broad category of qubits that are resilient to noise or generating less noise and make it possible to assemble logical qubits with much fewer physical qubits, in the 10-100 range instead of 1,000-10,000 range[666]. It contains cat-qubits and **GKP codes**[667]. Other protected qubits include the **zero-π qubits** of Peter Brooks, Alexei Kitaev and John Preskill which use two Josephson junctions, the **bifluxon**[668] and other variants[669].

The cat-qubits approach is chosen by **Alice&Bob** (France), **Amazon** (USA) and **QCI** (USA) while **Nord Quantique** (Canada) seems to use another breed of bosonic code. Cat-qubits are also investigated in many other research labs like **RIKEN** in Japan[670]. The **QuCoS** QuantERA collaborative three-year European project is focused on demonstrating the scalability of cat-qubits. It combines the University of Innsbruck (Gerhard Kirchmair), ENS Lyon (Benjamin Huard), Mines ParisTech and ENS Paris (Zaki Leghtas), KIT (Ioan Pop), Inria (Mazyar Mirrahimi), the Romanian National Institute for Research and Development of Isotopic and Molecular Technologies (Luiza Buimaga-Iarinca) and Quantum Machines (Israel).

How about using superconducting qubits for implementing quantum simulations? It is not a common practice. One of the reasons is the lack of generic long-range connectivity that could enable some direct entanglement between all qubits. It would require a different physical arrangement of the qubits and to create specific long-range connections between the qubits. This is possible with using cross-resonance gates that create interactions between qubits with their respective resonance frequencies.

### Science

For what follows, we will focus on those transmon qubits that are the most common and exploited by IBM, Google and IQM. They are anharmonic and therefore nonlinear oscillators. Their nonlinearity comes from the Josephson junction which allows to better separate two energy states of the superconducting loop (*on the right in* Figure 284) than with a simple linear resonator coupling a capacitor and an inductor (*on the left in* Figure 284). In a harmonic oscillator, the energy levels are spaced equally and are multiples of the first energy level ($\hbar\omega_r$ in the diagram).

The capacitance has an electrical energy (kinetic) and the inductance has a magnetic energy (potential). With the transmon qubit, the Josephson tunnel junction has a nonlinear inductance which creates its anharmonicity. In both cases, the flowing current is quantized with discrete energy levels corresponding to the horizontal bars in the graph in Figure 284, with corresponding different current phases corresponding to the intersection between these bars and the parabolic (CPB) and cosinusoidal (JJ) curves.

---

666 See Quantum information processing with bosonic qubits in circuit QED by Atharv Joshi et al, 2021 (24 pages).

667 See Quantum Error Correction with the GKP Code and Concatenation with Stabilizer Codes by Yang Wang, July 2019 (59 pages).

668 See Moving beyond the transmon: Noise-protected superconducting quantum circuits by András Gyenis, Alexandre Blais, Andrew A. Hook, David I. Schuster et al, June 2021 (14 pages).

669 Like Superconducting circuit protected by two-Cooper-pair tunneling by W. C. Smith, A. Kou, X. Xiao, U. Vool and M. H. Devoret, 2020 (9 pages) which uses pairs of Cooper pairs to create a qubit that is insensitive to multiple relaxation and dephasing mechanisms. And also Encoding qubits in multimode grid states by Baptiste Royer, Shraddha Sing and Steven M. Girvin, January 2022 (38 pages) and Coherent control of a multi-qubit dark state in waveguide quantum electrodynamics by Maximilian Zanner et al, Nature Physics, March 2022 (8 pages).

670 See Fault-Tolerant Multi-Qubit Geometric Entangling Gates Using Photonic Cat Qubits by Ye-Hong Chen et al, RIKEN, 2021 (12 pages). About a realization of Mølmer-Sørensen multi-qubit cat-qubits gates.

These energy states are usually controlled by microwaves pulses. These interactions between superconducting qubits and microwave photons are part of a branch of quantum physics called **circuit quantum electrodynamics**, or cQED[671].

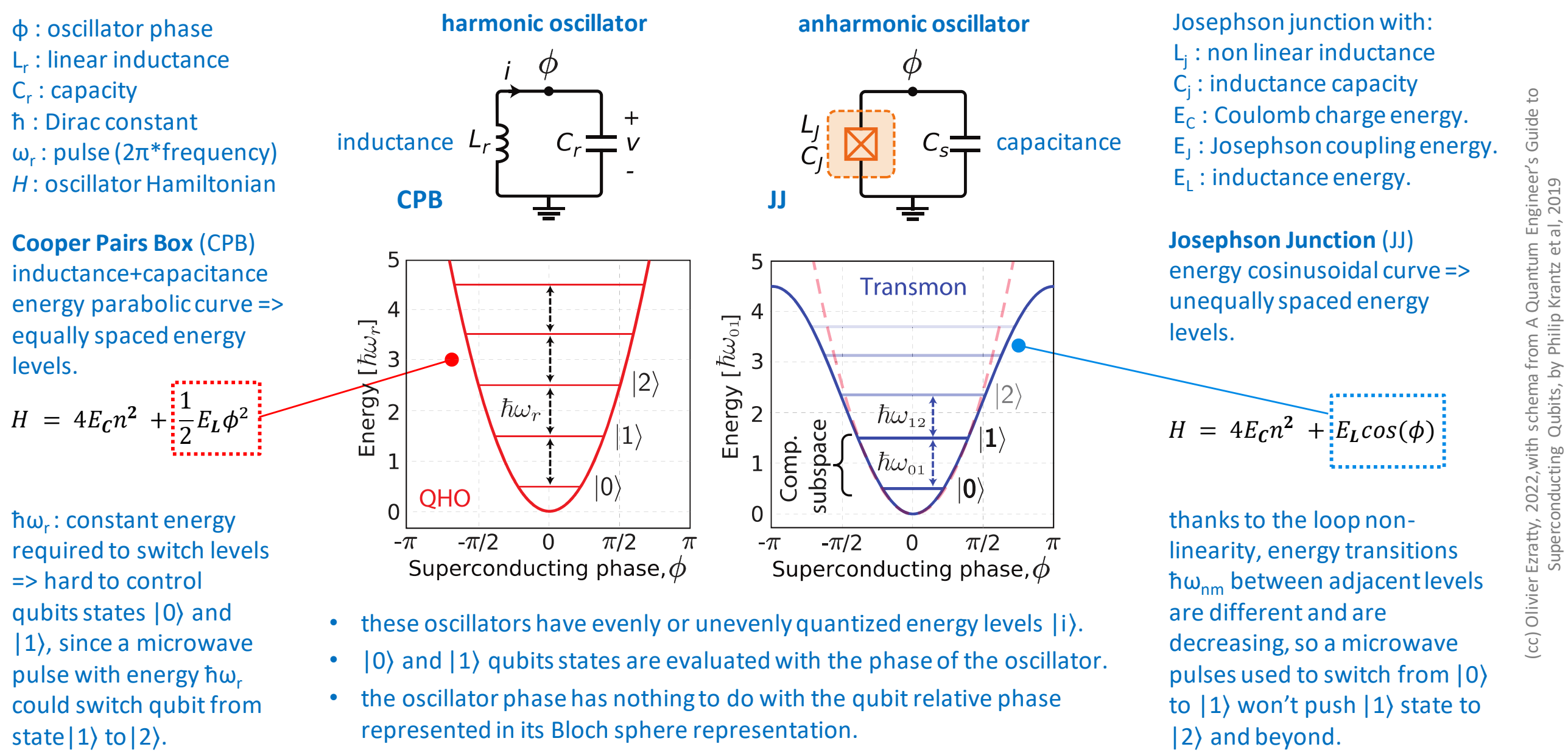

*Figure 284: why superconducting qubits use an anharmonic oscillator. (cc) Olivier Ezratty, 2022,with schema from "A Quantum Engineer's Guide to Superconducting Qubits" by Philip Krantz et al, 2019.*

Qubits use a linear superposition of the first two energy levels which have a different wave function relating the phase and current probabilities across the Josephson junction. The superposed states in the Bloch sphere equator like $(|0\rangle + |1\rangle)/\sqrt{2}$ and $(|0\rangle - |1\rangle)/\sqrt{2}$ correspond to an oscillating current that is dampened over time, as Rabi oscillations, in the 10 MHz range, shown in Figure 285. The $\hbar\omega_{01}$ energy level between the basis states |0⟩ and |1⟩ correspond to microwave frequencies in the 4 to 8 GHz band.

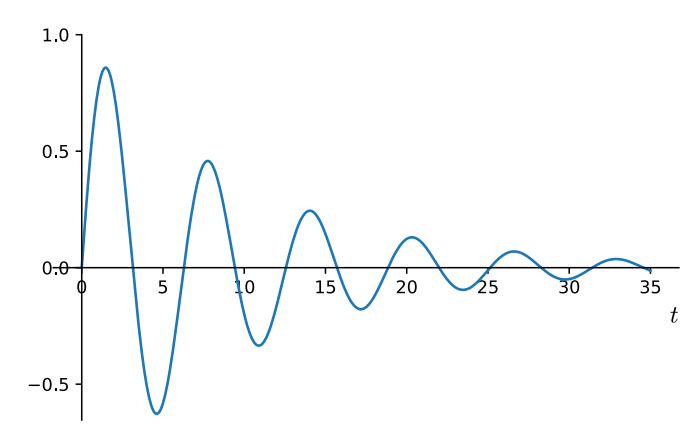

*Figure 285: a Rabi oscillation for superposed qubit states, at a frequency in the 10 MHz range.*

These frequencies must be well separable from the following ones. This separation is made possible because the (microwave photon) energy sent to move from one level to the other is different from one of these levels to other higher levels. Since the upper levels are less spaced, their related transition energy is lower. As the qubits are activated by microwaves, they are no longer likely to switch to a higher energy level. The anharmonic oscillator in the Josephson loop is provided by a nonlinear inductance $L_j$. The energy level between |0⟩ and |1⟩ of $\hbar\omega_{01}$ is higher than the energy levels needed to go to the upper levels $\hbar\omega_{12}$ and $\hbar\omega_{23}$. It is also compatible with the cooling temperature of the processor and the ambient noise.

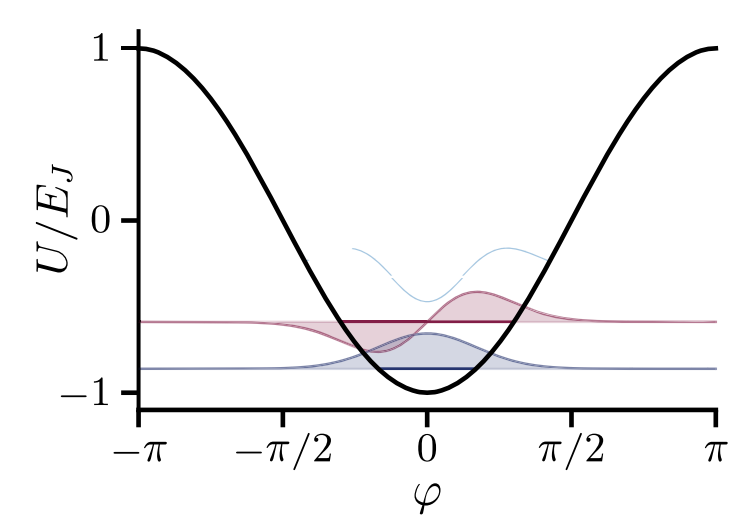

*Figure 286: |0⟩ and |1⟩ wave function giving the probability of phase φ in blue and green. Source: Superconducting circuit protected by two-Cooper-pair tunneling by W. C. Smith et al, 2020 (9 pages).*

[671] See Circuit-QED with phase-biased Josephson weak links by C. Metzger, Christian Urbina, Hugues Pothier et al, January 2021 (22 pages). Serge Haroche was awarded the Nobel Prize in Physics in 2012 for his work on the interaction between cold atoms and superconducting cavities. See on this subject the excellent Circuit Quantum Electrodynamics by Alexandre Blais, Andreas Wallraff et al, May 2020 (82 pages).

Those of the superconducting qubits control around 5 GHz have an energy level equivalent to a temperature of about 250 mK, much higher than the 15 mK temperature commonly used[672]. The microwaves for silicon qubit control are located between 8 and 26 GHz and enable qubit temperatures of 100 mK while some can even reach 1,5K.

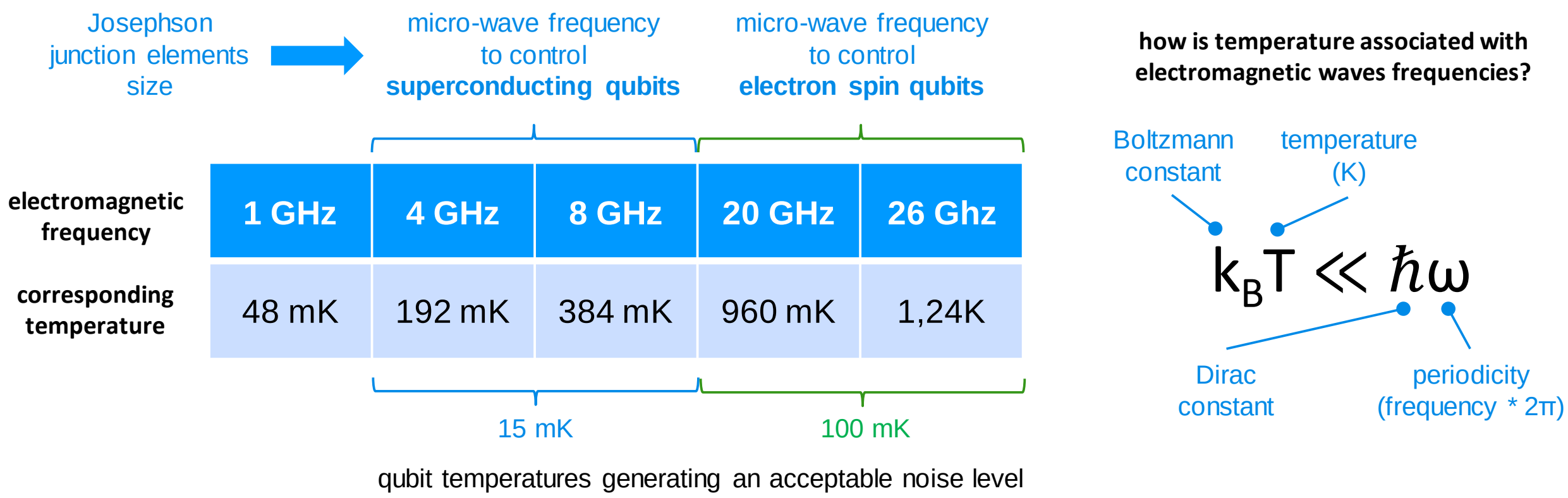

*Figure 287: the rationale behind the 15 mK operating temperature of superconducting qubits. (cc) Olivier Ezratty, 2021.*

There is another reason for running the qubit at around 15 mK. It takes a certain amount of energy, known as the energy gap, to break up the Cooper pairs running in a superconducting qubit. In aluminum that is the typical material used to create the Josephson junction and its surroundings, the energy gap corresponds to 90 GHz at 20 mK. It is an order of magnitude greater than the energy difference between the two levels in a qubit. It means that the qubit can be driven with lower energies (in the 4-8 GHz range) without breaking up the superconducting current Cooper pairs and altering the quantum coherence of the qubit[673].

What differentiates phase, charge (transmon) and flux qubits are the relative values of the charge energy ($E_C$, *aka* Coulomb charge energy), the Josephson coupling energy ($E_J$) and the qubit inductance energy ($E_L$)[674].

| $E_J/E_C$ \ $E_L/(E_J - E_L)$ | 0 | $\ll 1$ | $\sim 1$ | $\gg 1$ |
|---|---|---|---|---|
| $\ll 1$ | cooper-pair box | | | |
| $\sim 1$ | quantronium | fluxonium | | |
| $\gg 1$ | transmon | | | flux qubit |
| $\gg\gg 1$ | | | phase qubit | |

*Figure 288: periodic table of superconducting circuits. Source: Introduction to Quantum Electromagnetic Circuits by Uri Vool and Michel Devoret, 2017 (56 pages).*

Then, just with transmon qubits, you find other variations with:

- Fixed (IBM, MIT[675]) or tunable (Google) qubit frequencies.
- Tunable couplers (Google).
- Architectures mixing digital and analog superconducting computing[676].

[672] See Why 4-8 GHz? The rationale behind common qubit frequencies, The Observer, January 2022, explains the rationale for the microwaves frequencies being used with superconducting qubits. Above 8 GHz, electronics are too expensive and below 4 GHz, the ambient thermal noise is too important.

[673] Source: Superconducting quantum bits by Hans Mooij, Physics World, December 2004.

[674] This is well explained in Experiments on superconducting qubits coupled to resonators by Marcus Jerger aus Bühl, 2013 (140 pages).

[675] See Cancelling microwave crosstalk with fixed-frequency qubits by Wuerkaixi Nuerbolati et al, April 2022 (5 pages).

[676] See Superconducting Circuit Architecture for Digital-Analog Quantum Computing by J. Yu, Enrique Solano et al, March 2021 / May 2022 (23 pages).

- Controlled-phase gates with variable amplitude and frequency which could significantly reduce the depth of quantum circuits particularly for implementing a quantum Fourier transform required in many algorithms like Shor, HHL and QML (C-$R_\theta$)[677].
- New techniques to implement faster qubit readout[678].
- And techniques using qutrits instead of qubits (with Rigetti).

Of course, many researchers are looking for ways to improve qubits fidelities with better materials and designs[679].

In some cases, researchers invent useless things like with trying to entangle superconducting qubits with tardigrades[680].

But the physics of a superconducting qubit is much more complicated than that for the neophyte. The qubit itself is coupled to a cavity containing a resonator usually implemented as a **coplanar wave-guide** (CPW) resonator on a superconducting circuit. Its length usually corresponds to a quarter-wavelength or the resonator drive frequency. With a 6 GHz drive frequency, it turns into a 1.25 mm resonator that is usually squeezed in a serpentine layout.

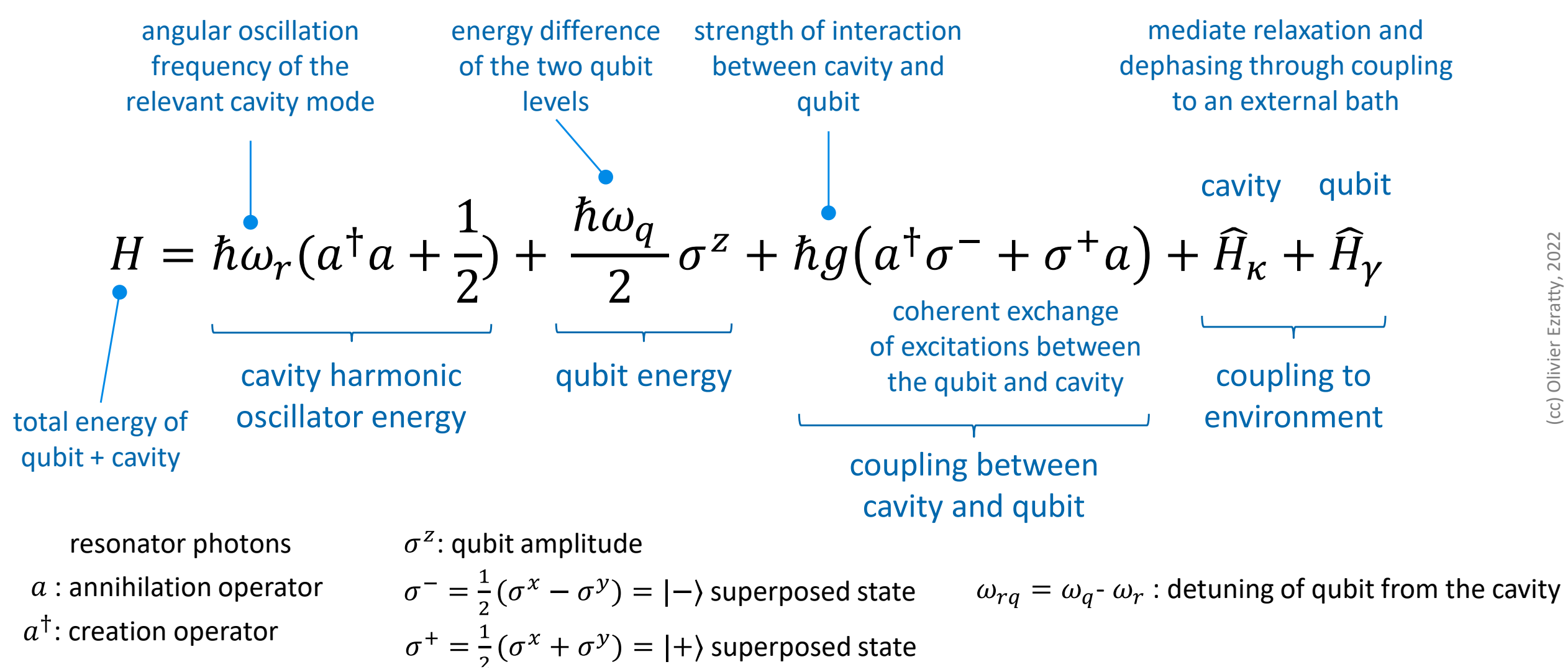

*Figure 289: Jaynes-Cumming cQED Hamiltonian, (cc) Olivier Ezratty, 2022.*

The energy of the ensemble is modelized by a **Jaynes-Cummings Hamiltonian** as shown in Figure 289 [681].

---

[677] See Extensible circuit-QED architecture via amplitude- and frequency-variable microwaves by Agustin Di Paolo, Alexandre Blais, William D. Oliver et al, MIT, April 2022 (29 pages).

[678] See Fast readout and reset of a superconducting qubit coupled to a resonator with an intrinsic Purcell filter by Yoshiki Sunada, Yasunobu Nakamura et al, February 2022 (12 pages) and Realization of fast all-microwave CZ gates with a tunable coupler by Shaowei Li, Jian-Wei Pan et al, February 2022 (12 pages).

[679] Like with Engineering superconducting qubits to reduce quasiparticles and charge noise by Xianchuang Pan et al, February 2022 (23 pages) which reduces quasiparticles generation coming from broken Cooper pairs.

[680] See Entanglement between superconducting qubits and a tardigrade by K. S. Lee et al, December (19 pages) and the pushback it generated in Peers dispute claim that tardigrades were entangled with qubits by Bob Yirka , Phys.org, December 2021, Schrödinger's Tardigrade Claim Incites Pushback At issue: Quantum-entangled water bears?! by Philippe Ross, IEEE Journal, December 2021 and Frequency Shifts Do Not Imply Quantum Entanglement by Ben Brubaker, January 2022.

[681] See The Jaynes-Cummings model and its descendants by Jonas Larson and Th. K. Mavrogordatos, February 2022 (237 pages).

This involves many notions like a Jaynes-Cummings spectrum, a resonant regime (the cavity-qubit are interoperating oscillators), dressed states (the different energy levels of the qubits) and a dispersive regime (enabling qubits readout with the resonator)[682].

Many parameters define a superconducting qubit's characteristics, like its **Q factor**, the ratio between the energy stored in an oscillator and the energy dissipated per oscillation cycle times $2\pi$. It characterizes the stability of a superconducting qubit and determines its $T_1$ or relaxation time. The greater the Q factor is, the longer $T_1$ will be[683] but it can be detrimental to noise sensitivity.

## Qubit operations

The general principle of superconducting qubits operations is as follows:

- **Qubit quantum state** in the generic case of a transmon is a two-level charge of Cooper pairs that correspond to a nonlinear oscillator containing at least a Josephson junction and a capacitance laid out in a current loop. A flux bias (direct current pulse) can be used to individually control each qubit resonant frequency if it is frequency tunable. It can help reduce control frequency crosstalk between qubits but at the cost of a lower lifetime ($T_1$). It is better to have fixed and different qubit frequencies.

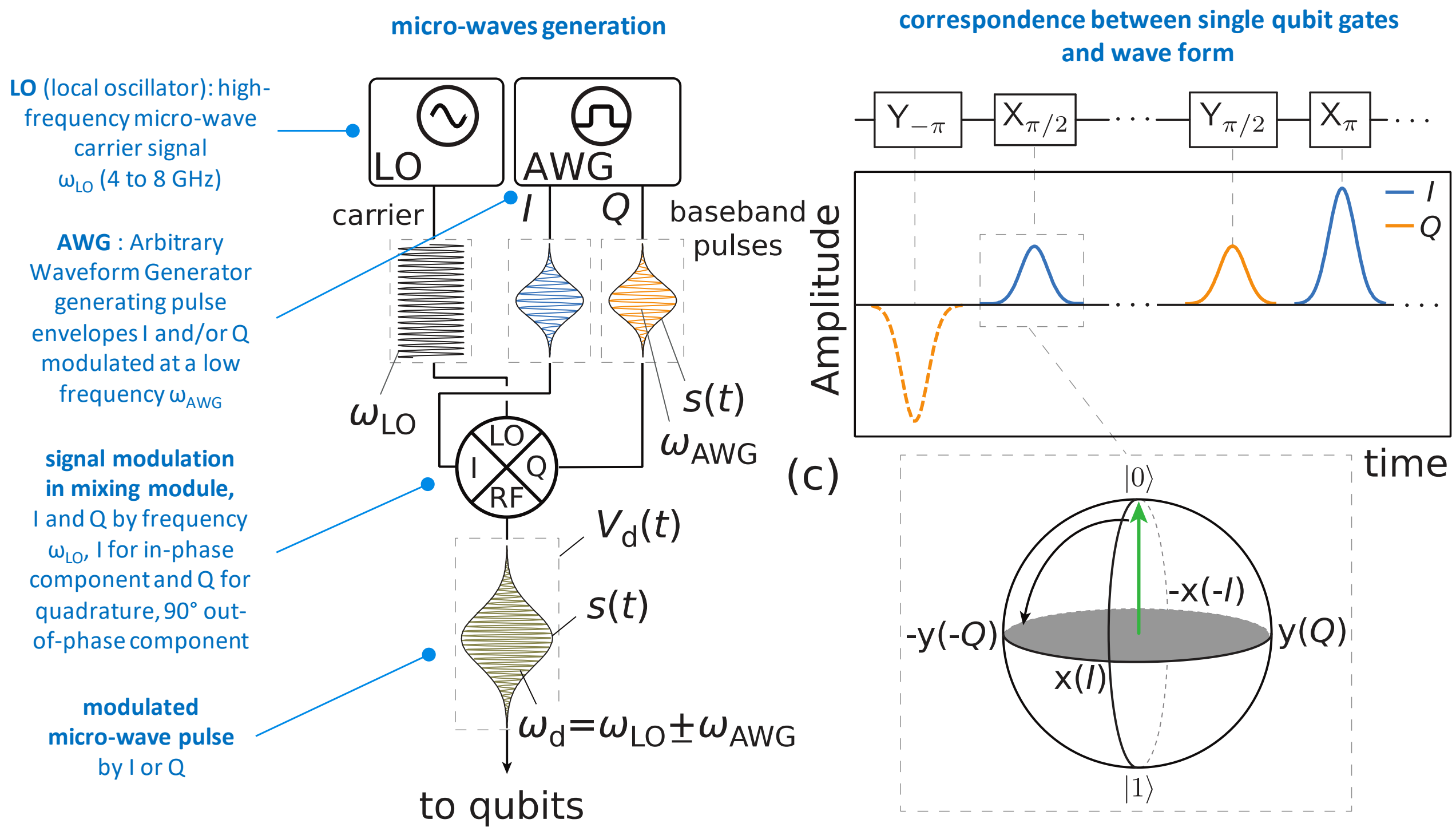

*Figure 290: qubit drive microwaves generation. Source: A Quantum Engineer's Guide to Superconducting Qubits, by Philip Krantz et al, 2019 (67 pages).*

[682] Some sources to learn cQED: the review paper Microwave photonics with superconducting quantum circuits by Xiu Gu et al, 2017 (170 pages) that describes well how superconducting qubits interact with microwaves. Superconducting Qubits and Circuits Artificial Atoms Coupled to Microwave Photons by Steve Girvin, Les Houches 2011 (132 pages), the review Practical Guide for Building Superconducting Quantum Devices by Yvonne Y. Gao, Adriaan Rol, Steven Touzard and Chen Wang, November 2021 (48 pages) and Superconducting qubit in a resonator test of the Leggett-Garg inequality and single-shot readout by Agustin Palacios-Laloy, 2010 (253 pages).

[683] See Decoherence benchmarking of superconducting qubits by Jonathan J. Burnett et al, Nature, 2019 (8 pages) and Extending Coherence in Superconducting Qubits: from microseconds to milliseconds by Adam Patrick Sears, a thesis under the supervision of Rob Schoelkopf, 2013 (178 pages).

- **Single-qubit quantum gates** are generated by microwave pulses sent via coaxial cables on the qubits. Their frequency is adjusted to the energy level $\hbar\omega_{01}$ mentioned above. This frequency is calibrated to be different on adjacent qubits to avoid crosstalk effects[684].The microwave pulse amplitude controls the rotation angle and its phase adjusts the axis of the gate rotation operation.

  This makes it possible to create T, S and R gates with a phase other than a quarter or half turn in the Bloch sphere[685]. In practice, two arbitrary waveform generators create a wave form for "in-phase" and "quadrature" (I and Q) signals which are two microwave pulses that have the same (local-oscillator originated) frequency and are 90° out of phase, as shown in Figure 290.

  The I signal is a cosine waveform and the Q signal is a sine waveform. They add-up in the mixer to create a pulse signal with an arbitrary phase depending on the relative amplitudes of the I and Q waveform signals[686]. The mixer then adds the local oscillator signal to the resulting signal[687]. At 5 GHz, an LO pulse lasts 0,2 ns. Most single qubit gate last at least 10 to 20 ns. It means in that case that a generated microwave packet contains about 50 to a hundred 5 GHz pulses shaped by its wave form. Microwave pulses generated at ambient temperature are progressively attenuated and filtered at every stage of the cryostat so that only a couple hundred microwave photons reach the qubit. The attenuators eliminate photons proportionally to the cooling budget available at each cold plate stage.

- **Two-qubit quantum gates** are realized with a coupling circuit positioned between the two qubits, which can be a simple capacitor or a dynamically controllable system. As we will see later, this coupling is managed with an intermediate qubit in Google's Sycamore processor and their Chinese equivalents. IBM is not using couplers but instead cross-resonance gates.

- **Qubits readout** depends on its type. With transmon qubits, a resonator is coupled to the qubit. It transmits a microwave pulse in a resonator that is coupled with the qubit using microwave reflectometry. The qubit state slightly affects the resonator frequency and phase. These readout microwaves are usually amplified in several stages. The method is called "dispersive readout" where for a fixed microwave drive frequency, the resonance frequency of the waveguide resonator shifts depending on the qubit measured state[688]. This measurement technique protects the qubit from all radiations except the readout microwave pulse at $\omega_r$, it amplifies the outgoing signal with the lowest added noise (near the quantum limit). The readout also creates a differentiated phase in the reflected microwave that is analyzed after demixing which generates the in-phase and quadrature signals (I/Q). Measuring the phase of the reflected microwave determines the state of the qubit after measurement without destroying it. It's a QND readout as already explained.

[684] A precise calibration of these frequencies is also necessary because of the variability of the behavior of Josephson loops, which are different from one another due to imprecise manufacturing techniques. This variability does not exist for qubits based on single particles such as trapped ions or cold atoms. See Control and mitigation of microwave crosstalk with frequency-tunable qubits by Ruixia Wang et al, Beijing Academy of Quantum Information Sciences, July 2022 (5 pages) which proposes a quantum error mitigation technique limiting the impact of crosstalk. Static ZZ crosstalk is the one many researchers are working to suppress like in Frequency adjustable Resonator as a Tunable Coupler for Xmon Qubits by Hui Wang et al, China, August 2022 (12 pages). ZZ crosstalk errors can be described as the amplitude of one qubit influencing the amplitude of other qubits. ZX or ZY errors are linked to the amplitude of a qubit influencing the phase of other qubits.

[685] These gates can be optimized by modulating the pulsation in an optimal way. See Implementing optimal control pulse shaping for improved single-qubit gates by J. M. Chow et al., May 2020 (4 pages) which anticipates the capacity to generate single-qubit gates in 1 ns, against a current minimum of around 20 ns.

[686] See Radio frequency mixing modules for superconducting qubit room temperature control systems by Yilun Xu, Irfan Siddiqi et al, Lawrence Berkeley National Laboratory, July 2021 (7 pages) that describes the role of a signal mixer.

[687] Another option consists in mixing the LO signal separately with the I and Q signals and then merge the resulting signal, removing unwanted spurious frequency components. See Frequency Up-Conversion Schemes for Controlling Superconducting Qubits by Johannes Herrmann, Andreas Wallraff et al, October 2022 (9 pages).

[688] See Dispersive Readout, Rabi- and Ramsey-Measurements for Superconducting Qubits by Can Knaut, 2018 (25 slides).

One first stage can use a low-noise superconducting Josephson Parametric Amplifier (JPA) or Traveling Wave Parametric Amplifier (TWPA) operating at the quantum limit, then with a high electron mobility transistor (HEMT) amplifier running at the 4K stage and, at last, with a Low Noise Amplifier (LNA) running at room temperature.

At last, the amplified microwave is converted in digital format with an ADC (analog to digital converter) and analyzed by a FPGA circuit to identify the qubit basis states |0⟩ or |1⟩ with a microwave phase analysis.

Frequency-based multiplexed readout can also be achieved to simplify the wiring exiting the qubit chipset. The readout microwave is modulated with a higher frequency than the quantum gates frequency, above 6 GHz[689].

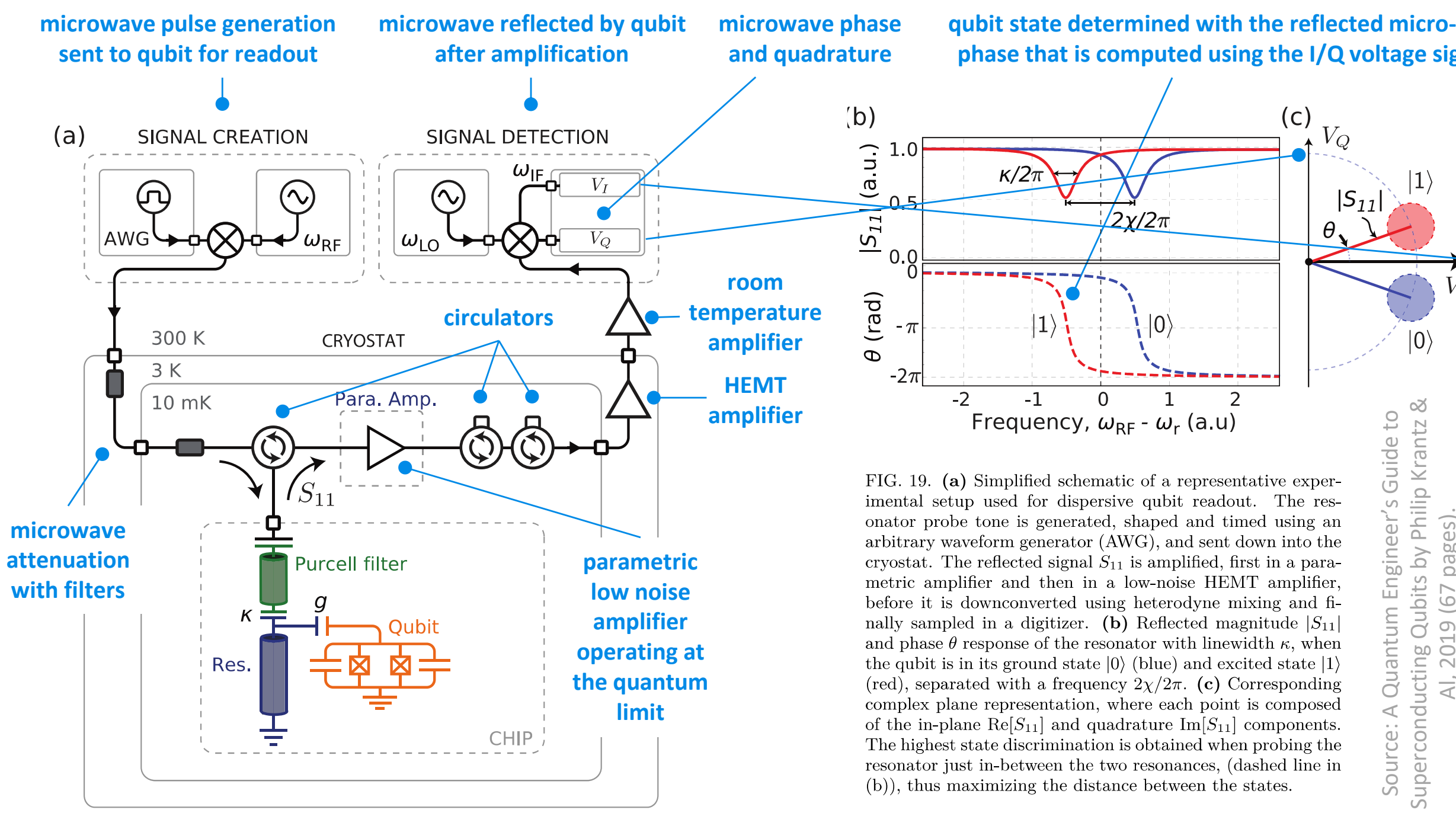

*Figure 291: superconducting qubit readout process. Source: A Quantum Engineer's Guide to Superconducting Qubits, by Philip Krantz et al, 2019 (67 pages).*

I have always wondered why measuring a simple phase of a microwave signal was so complicated. This is the most complex part of superconducting qubits engineering. Could it be possible to measure a reflected microwave pulse phase with a simpler analog setting running at the qubit temperature and then transmit a simple 0 or 1? Seemingly not.

- **Connectivity** is an important feature of a quantum processor. The more qubits are connected with each other, the fewer SWAP gates must be run to logically entangle them. With 2D structures, one of the problems to be solved lies in the internal connections in the chipset.

[689] Other techniques for measuring the state of superconducting qubits are being considered, such as the activation of qubit fluorescence. It is done by jumping from the |0⟩ to |2⟩ state of the qubit, the transition to the |1⟩ state not being possible with the fluorescence excitation photon. See the thesis Energy and Information in Fluorescence with Superconducting Circuits by Nathanaël Cottet, 2018 (227 pages).

3D architectures are used with one layer for qubit readout and another for qubit operations but the qubits topology connectivity is at best with 4 nearest neighbors like with Google's Sycamore. As shown in Figure 292, a Japanese team proposed in 2020 an original solution consisting in flattening the matrix and making it possible to connect the control elements in 2D. But at the price of overlapping part of the links between qubits[690] .

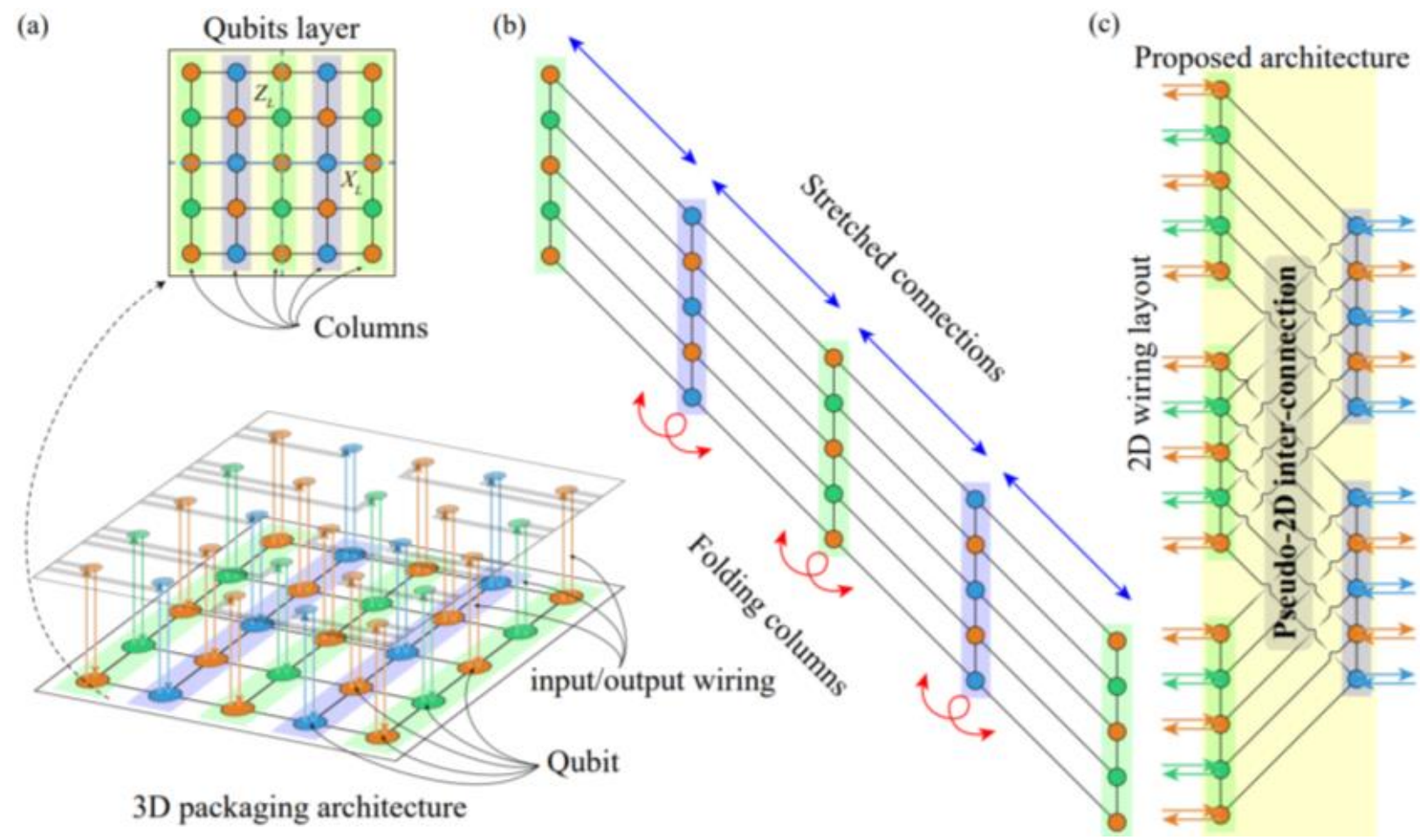

*Figure 292: a proposal to improve superconducting qubits connectivity. Source: Pseudo-2D superconducting quantum computing circuit for the surface code by H. Mukai, February 2019 (8 pages).*

- **Digital simulations**. Some software solutions are available to simulate the physical behavior of superconducting qubits at a low level. Among these, **CircuiQ** is a proposal from the MIT, ParityQC and the University of Innsbruck with the participation of Benoît Vermersch (LPMMC, Grenoble)[691]. It's a Python open source toolbox that can be used for analyzing superconducting circuits at the physical level, using their Hamiltonian. It can help estimate the qubits $T_1$ under various noise mechanisms.

  There are other similar software packages like **scqubits**, developed by researchers from Northwestern University and **SQcircuit** from Stanford[692].

### Setups

In the current state of the art, the cryostats housing these qubits are filled with many cables and microwave attenuators driving the qubits and with first stages amplifiers used in the qubits state readout[693]. Implementing quantum error correction will require 1,000 or 10,000 physical qubits per logical qubit[694]. It will create significant challenges for scaling up the architecture at least, with the existing cabling and external microwave generation and readout systems. Thus, the need for cryogenic electronics and miniaturized microwaves coaxial cabling that we will soon investigate, starting in page 485.

---

690 See Wiring the quantum computer of the future: A novel simple build with existing technology by Jaw-Shen Tsai (Japan), April 2020 which points to Pseudo-2D superconducting quantum computing circuit for the surface code by H. Mukai, February 2019 (8 pages).

691 See CircuitQ: An open source toolbox for superconducting circuits by Philipp Aumann, William D. Olivier et al, March 2022 (14 pages).

692 See Computer-aided quantization and numerical analysis of superconducting circuits by Sai Pavan Chitta, Jen Koch et al, Northwestern University, June 2022 (12 pages) and Analysis of arbitrary superconducting quantum circuits accompanied by a Python package: SQcircuit by Taha Rajabzadeh1, Zhaoyou Wang, Nathan Lee, Takuma Makihara, Yudan Guo, and Amir H. Safavi-Naeini, June 2022 (23 pages).

693 This is well explained in Superconducting Circuits Balancing Art and Architecture by Irfan Siddiqi of Berkeley Lab, 2019 (34 slides).

694 Source: Surface codes: Towards practical large-scale quantum computation by Austin G. Fowler et al, 2012 (54 pages).

Digital-to-analog converters, aka DACs and Analog-to-Digital converters convert microwaves at room temperature and handle a very large volume of outbound or inbound data of 8 to 14 Gbits/s as shown in the diagram in Figure 293 corresponding to Google's Sycamore.

This data is managed in real time. It does not however seem necessary to store them. It is not a big-data system!

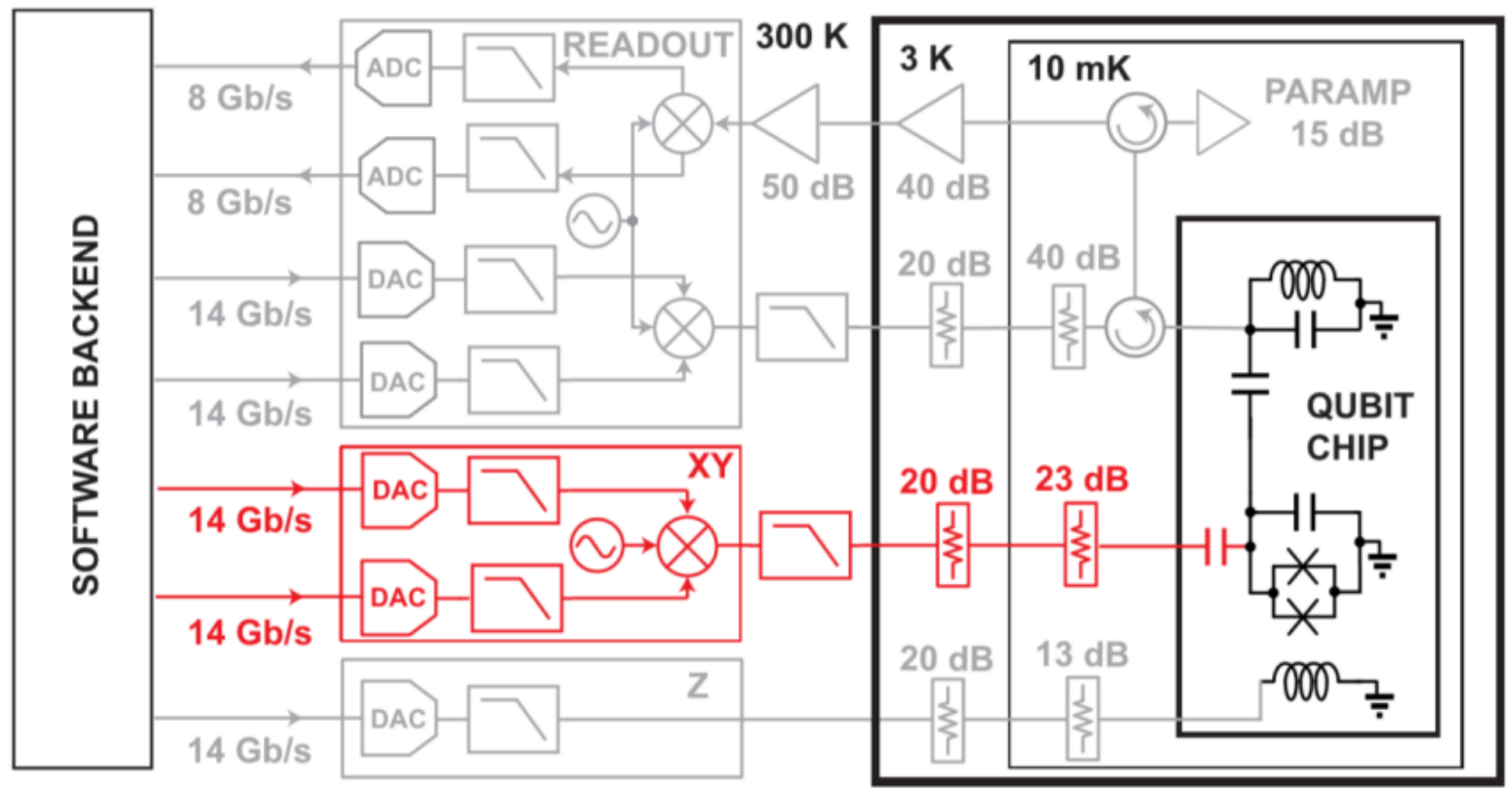

Figure 293: Sycamore's qubit control and readout architecture. Source: Google.

The electronics used in research laboratory equipment is illustrated with the example in Figure 294 of *a* configuration used to test a 5-qubit superconducting chipset in 2015.

Its uses classical off-the-shelf equipment from **Rohde & Schwarz** or **Tektronix**.

These external generators are appreciated for the quality of the microwave pulses they produce. For a larger number of qubits, multiple microwave generators are used from vendors like **Zurich Instruments**, **Qblox** and **Quantum Motion** that we cover in a dedicated section, page 485. Others, like **SeeQC**, are attempting to miniaturize all or part of these components with superconducting electronics.

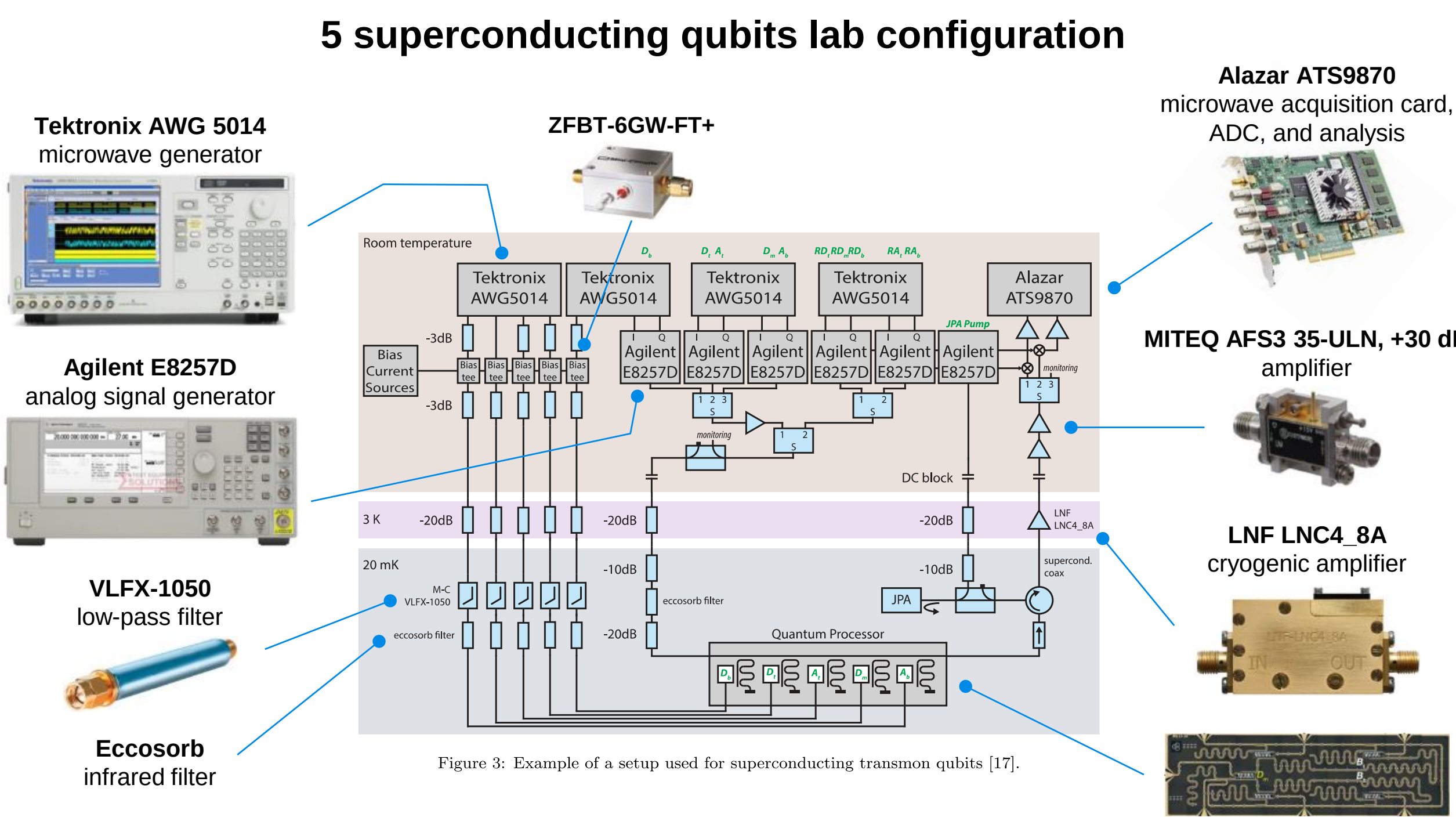

Figure 294: a superconducting qubits lab configuration. Source: The electronic interface for quantum processors by J.P.G. van Dijk et al, March 2019 (15 pages). I have added visuals of the electronic components used in the configuration.

Superconducting qubits fidelities are not best-in-class compared to trapped ions. It also decreases with the number of qubits. There is some progress being made to reduce qubit noise. It has several origins such as charge fluctuations, random electrons and materials impurities. Fidelity is currently not high enough to implement error correction codes. Some methods are proposed to improve readout fidelity.

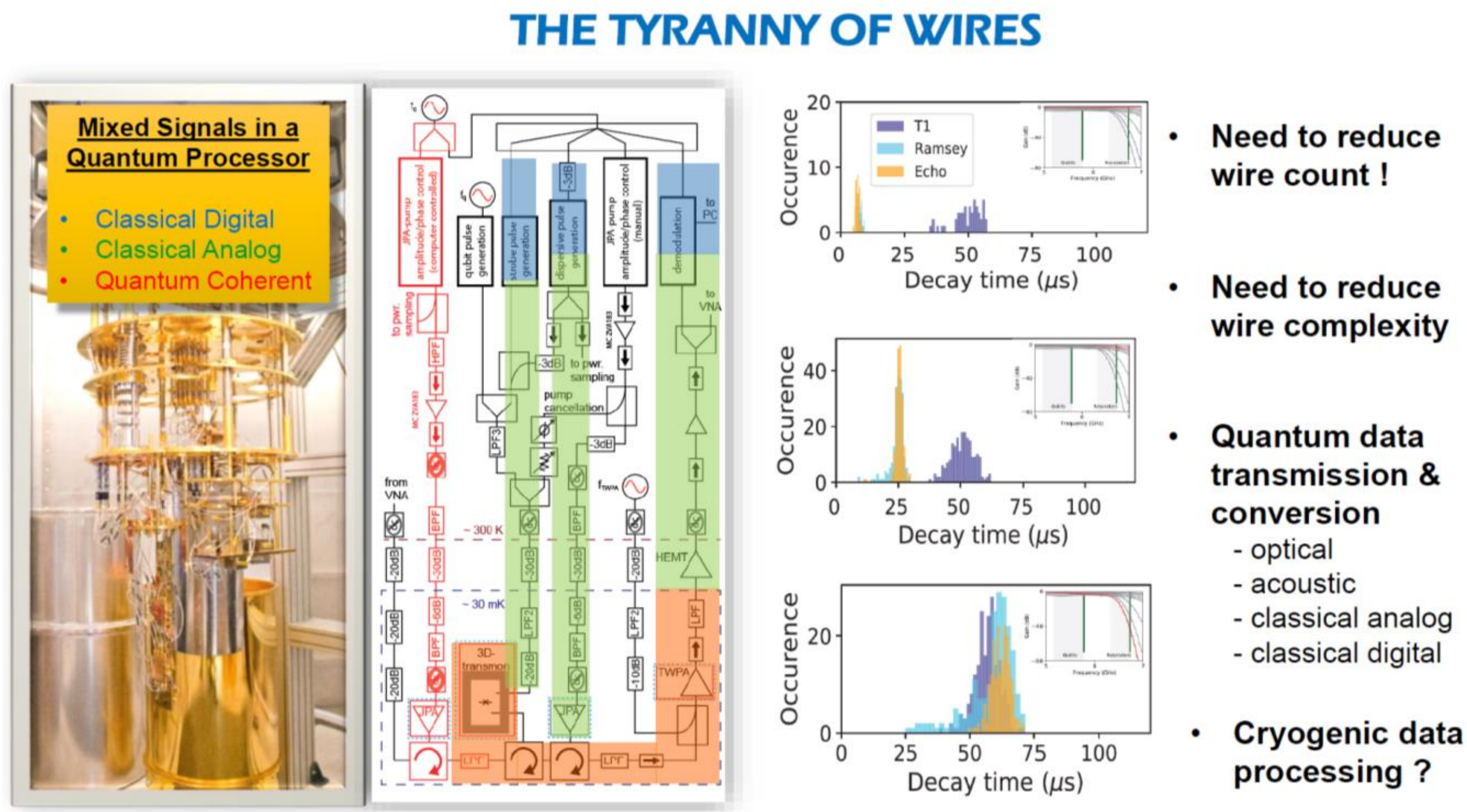

*Figure 295: the tyranny of wires in superconducting qubits. Source: Superconducting Circuits Balancing Art and Architecture by Irfan Siddiqi of Berkeley Lab, 2019 (34 slides).*

A team of Canadian and American researchers is proposing a miniaturizable optical measurement[695]. A variant was proposed in 2018 by Robert McDermott of the University of Wisconsin-Madison, with the objective of improving measurement fidelity to 99%[696].

The size of superconducting qubits is in the micron range, making it difficult to create large chips with millions of qubits. Miniaturization always seems possible but it is difficult to manage because the quality of the superconducting qubits seems to decrease with their size[697].

### Manufacturing

Superconducting qubits are electronic circuits built with techniques that are not that far from how classical analog circuits are being produced like in the radar and electronics markets, with some similarities with digital electronics, *aka* CMOS chipsets.

We'll describe later the specifics of the manufacturing of superconducting qubits. Most thesis coming out of superconducting labs contain a description of the manufacturing techniques being used although it changes over time[698].

---

695 See Heisenberg-limited qubit readout with two-mode squeezed light, 2015 (12 pages).

696 In Measurement of a Superconducting Qubit with a Microwave Photon Counter, March 2018 (11 pages).

697 See Investigating surface loss effects in superconducting transmon qubits by Jay Gambetta et al, 2016 (5 pages) and On-chip integrable planar NbN nano SQUID with broad temperature and magnetic-field operation range by Itamar Holzman and Yachin Ivry, Technion, April 2019 (7 pages) who prototyped miniaturized 45 nm x 165 nm SQUIDs.

698 See for example Design, fabrication and test of a four superconducting quantum-bit processor by Vivien Schmitt, 2015 (192 pages).

Materials used for manufacturing superconducting qubits include generally aluminum (for the Josephson junction, at least for the dielectric), niobium (for capacitors and resonators and sometimes the Josephson junction) and indium (for the chipset connectors), bore (in boron-nitride in Josephson junction dielectric[699]), titanium nitride (for capacitors, with a better quality factor) and occasionally selenium (associated with niobium and bore in capacitors), silicon or sapphire (for the wafer substrate) and tantalum ([700]).While most deposition techniques generate polycrystalline structures [701], some are starting to investigate epitaxial deposition to create monocrystalline structures.

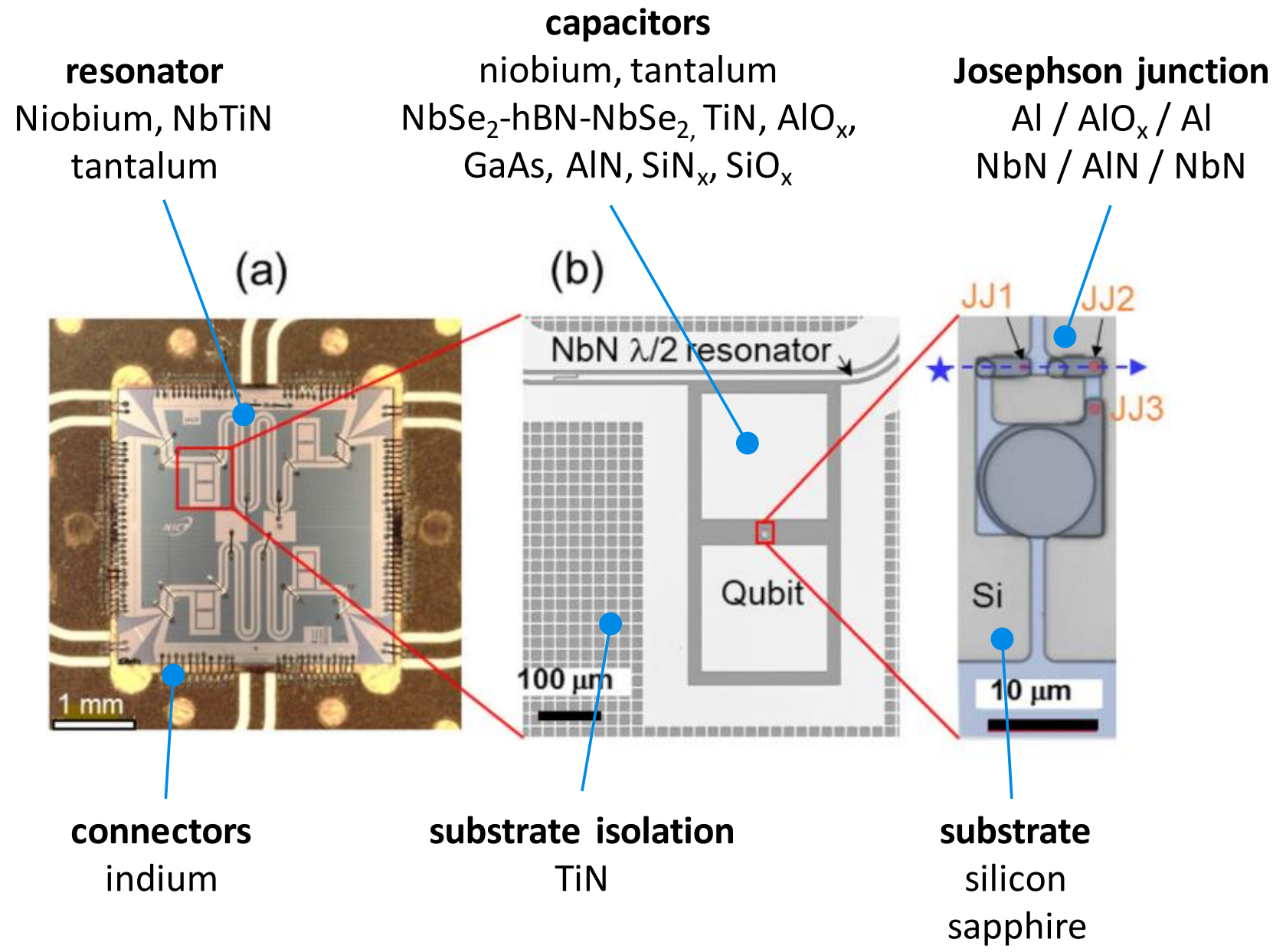

*Figure 296: the various components and materials used in a superconducting qubit. Source: Enhanced coherence of all-nitride superconducting qubits epitaxially grown on silicon substrate by Sunmi Kim et al, September 2021.*

IMEC already tests such processes avoiding lift-off and angled evaporation, all done with photolithography. They are part of the related EU project **Matqu** with CEA-Leti and others to build superconducting qubits on 300 mm wafers using existing CMOS fabs.

Superconducting qubits miniaturization is an interesting area of research given they are currently quite large, mainly due to the size of their resonators with a length of λ/4, λ corresponding to their control wavelength. It can exceed a surface of 1 $mm^2$. Resonators could be as small as 0.04 $mm^2$ using special fabrication techniques[702]. Similar efforts are undertaken to miniaturize capacitors with van der Waals materials ($NbSe_2$-hBN-$NbSe_2$)[703].

[699] See Hexagonal boron nitride as a low-loss dielectric for superconducting quantum circuits and qubits by Joel I-J. Wang, William D. Oliver et al, MIT, Nature Materials, January 2022 (30 pages) and Enhanced coherence of all-nitride superconducting qubits epitaxially grown on silicon substrate by Sunmi Kim et al, September 2021 (7 pages).

[700] See Towards practical quantum computers: transmon qubit with a lifetime approaching 0.5 milliseconds by Chenlu Wang et al, NPJ, January 2022 (6 pages).

[701] See Microscopic relaxation channels in materials for superconducting qubits by Anjali Premkumar, Andrew A. Houck et al, Nature Communications, July 2021 (9 pages).

[702] See Compact superconducting microwave resonators based on Al-AlOx-Al capacitor by Julia Zotova et al, March 2022 (10 pages) and See Tiny materials lead to a big advance in quantum computing by Adam Zewe, MIT News Office, January 2022.

[703] See Miniaturizing transmon qubits using van der Waals materials by Abhinandan Antony et al, Columbia University and Raytheon BBN. September 2021 (6 pages).

## Research

A significant number of research laboratories are working on superconducting qubits all over the world. In the USA, at **Yale University** and **MIT**[704], in Europe and in Germany, in Sweden at the **WACQT** of Chalmers University, in France at the **CEA**, in Switzerland at **ETH Zurich**[705], in Finland[706] and in **Japan**.

Other works aim at lengthening the coherence time of superconducting qubits, notably at Princeton in Andrew A. Houck's team[707]. Indeed, this coherence time of the order of one hundred micro-seconds (µs) is still quite limiting. It generates a constraint on the number of quantum gates that can be executed in a quantum software, even if the accumulated errors become prohibitive before this limit threshold. New records were broken in 2021 with 1.6 ms $T_1$ at Princeton and Sherbrooke with a 0–π circuit (but with a 25 µs dephasing time, aka $T_2$) and 210 µs with transmon qubits at Yale[708]. In May 2021, a China team obtained a 300 µs $T_1$ with a transmon qubit[709]. IBM reached the 1 ms $T_1$ barrier with one experimental planar transmon qubit in May 2021 as well (but the related paper is still pending). The best lab-level record was with a 1.48 ms $T_2$ coherence time on flux qubits at the University of Maryland in Vladimir Manucharyan's team[710]. These records are however not necessarily obtained with a great number of functional qubits... when more than 2 are used!

Superconducting qubits lifetime record is still way above this, with 3D SRF cavities (for superconducting radio frequency cavities). These are developed by the DoE Fermilab and have a very high Q-factor.

In 2020, they reached qubit lifetimes of about 2s with special materials design reducing the 2-level system losses. Fermilab researchers plan to implement qudits with these SRFs, packing between 63 and 128 effective qubits into 9 SRF cavities hosting qudits. These cavities are bulky, the size of the device being about one meter long in Figure 297[711].

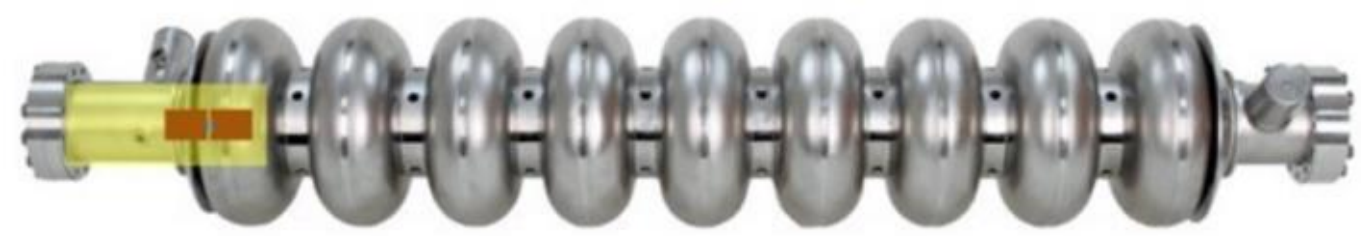

*Figure 297: the huge SRF superconducting qubits from the DoE Fermilab. Source: Superconducting Quantum Materials and Systems Center by Anna Grassellino, SQMS Center Director, Fermilab, June 2021 (40 slides),*

[704] See Quantum Computing @ MIT: The Past, Present, and Future of the Second Revolution in Computing by Francisca Vasconcelos, MIT, February 2020 (19 pages). They have developed a 16-qubit superconducting chipset, manufactured by Lincoln Labs at MIT.

[705] With Andreas Wallraff's QuSurf team working on superconducting qubits and their error correction codes. This project is funded by the American IARPA agency. In 2019, they were at 7 experimental qubits. It is also supported by the ScaleQIT project (Scalable Superconducting Processors for Entangled Quantum Information Technology) funded by the European Union and by the OpenSuperQ project of the European flagship.

[706] VTT's goal is to manage 50 to 100 superconducting qubits. VTT has its own circuit manufacturing unit with a 2600 $m^2$ clean room of a similar size to CNRS C2V clean room in Palaiseau, France. See Engineering cryogenic setups for 100-qubit scale superconducting circuit systems by S. Krinner et al, 2019 (29 pages).

[707] See New material platform for superconducting transmon qubits with coherence times exceeding 0.3 milliseconds by Alex P. M. Place, Andrew A. Houck et al, February 2020 (37 pages). Qubits are using tantalum instead of niobium and on a sapphire substrate. The paper describes starting page 8 the manufacturing process of these qubits.

[708] See Experimental Realization of a Protected Superconducting Circuit Derived from the 0–π Qubit by András Gyenis, Alexandre Blais et al, Sherbrooke, Princeton, U. Chicago and Northwestern University, March 2021 (31 pages) and Direct Dispersive Monitoring of Charge Parity in Offset-Charge-Sensitive Transmons by K Serniak, R Schoelkopf, Michel Devoret et al, Yale University, March 2019 (11 pages) with transmons at a $T_1$ of 210 µs.

[709] See Transmon qubit with relaxation time exceeding 0.5 milliseconds by Chenlu Wang et al, May 2021 (15 pages).

[710] See Millisecond coherence in a superconducting qubit by Aaron Somoroff, Vladimir E. Manucharyan et al, University of Maryland, 2021 (14 pages),

[711] See Superconducting Quantum Materials and Systems Center by Anna Grassellino, SQMS Center Director, Fermilab, June 2021 (40 slides), Materials and devices for fundamental quantum science and quantum technologies by Marco Polini et al, January 2022 (19 pages) and Three-Dimensional Superconducting Resonators at T<20 mK with Photon Lifetimes up to τ=2s by A. Romanenko, R. Pilipenko, S. Zorzetti, D. Frolov, M. Awida, S. Belomestnykh, S. Posen, and A. Grassellino, March 2020 (5 pages).

Other researchers work on using various qubits materials like titanium nitride and tantalum on sapphire substrates, at Princeton, ENS Lyon and Alice&Bob among other locations. These are used in complement to the Al/AlO/Al Josephson junctions, for various other parts of the qubit circuits (isolators, capacitances, resonators).

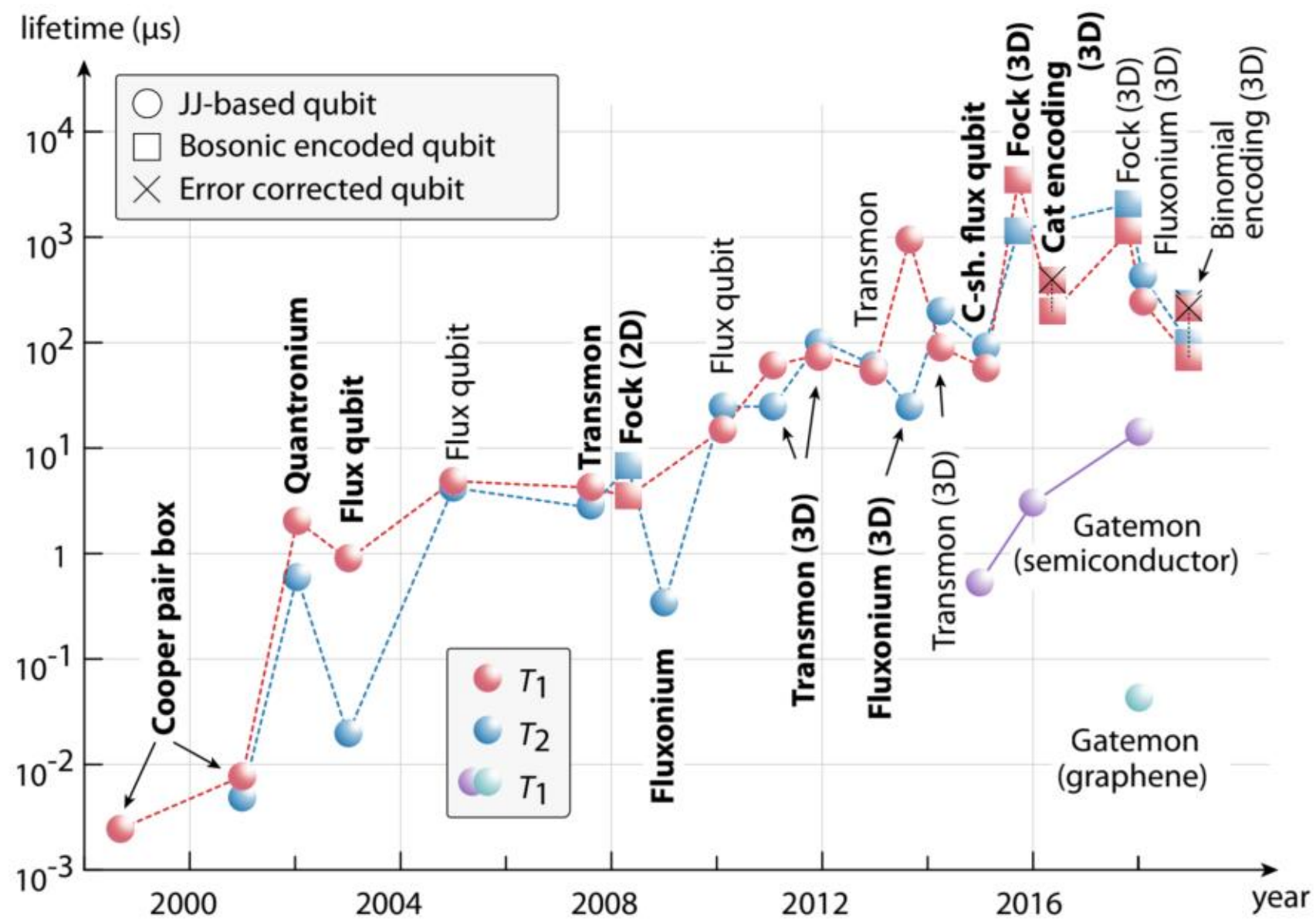

*Figure 298: logarithmic evolution of superconducting lifetime over time. Source: Superconducting Qubits Current State of Play by Morten Kjaergaard et al, 2020 (30 pages).*

The Quantronics team at CEA-Saclay uses transmons for its quantum circuits, but now explores another route based on high coherence impurity spins in insulators for making qubits, with superconducting quantum circuits for controlling them. The rationale is that the electro-nuclear spin levels of such systems may indeed provide more robust qubits for which quantum error correction could be more easily manageable than for transmon qubits.

Other research conducted at the CEA consists in associating superconducting qubits with NV centers, linked by microwaves, to be used as quantum memory as well as a means of more precise readout of superconducting qubits. NV centers spins can serve as quantum memory thanks to a spin coherence time that is 1000 times longer than that of superconducting qubits (100 milliseconds vs. 100 microseconds). Another field of research is the coupling of superconducting qubits with nuclear spins (instead of electron spins, on phosphorus or bismuth nuclei) via electron spins.

As with many solid-state qubits, one of the key research goals is to transform these microwave photons into photons in the visible/infrared band to allow their long-distance transport, in particular via fiber optic-based telecommunication, which would become the basis of distributed quantum computing[712].

There is another interesting field of research aimed at simplifying qubit readout that may avoid the burden of parametric microwave amplification and circulators at the 15 mK stage. One of these consists in using microwave photons counting and Josephson photomultipliers (JPM) that are embedded directly in the qubit chipset[713]. It has however some shortcomings to overcome like crosstalk and loss of qubit fidelity over time.

---

[712] See for example Microwave-to-optical conversion via four-wave mixing in a cold ytterbium together by Jacob P. Covey et al, July 2019 which discusses this conversion.

[713] See High-Fidelity Measurement of a Superconducting Qubit Using an On-Chip Microwave Photon Counter by A. Opremcak, Roger McDermott et al, PRX, February 2021 (15 pages) and the associated thesis Qubit State Measurement Using a Microwave Photon Counter by Alexander M. Opremcak, 2020 (159 pages).

In 2021, a **China** research team led by Jian-Wei Pan created a 66 superconducting qubits system and claimed having reached another quantum advantage. In this **Zuchongzhi** 2.1 system, they reproduced the Google supremacy experiment with a 2D array of qubits with 13 additional qubits, using the same coupling technology, with 110 couplers[714]. Their fidelities were not best-in-class with 99,86% for single qubit gates, 99,24% for two-qubit gates and 95,23% for qubits readout, on top of a rather low $T_1$ of 30.6µs. In their experiment, though, they did use only 56 of their 66 qubits, showing that qubits fidelities are probably not that good when all qubits are activated. In September 2021, they used 60 qubits on 24 cycles with an improved readout fidelity of 97.74%[715]. China researchers implemented some quantum neuronal sensing application of quantum many-body states[716].

**Vendors**

IBM is one of the few major players in the IT world that has been investing in fundamental research for a very long time in quantum computing[717]. It is one of the most advanced in universal quantum computing research, having focused on superconducting qubits for a while.

IBM's quantum activity is driven by Jay Gambetta with researchers in their Yorktown, Poughkeepsie, San Jose and Zurich labs, partnering with various American and other countries universities including ETH Zurich and EPFL in Switzerland. IBM has both the most significant full-stack physics, hardware, software tools and cloud R&D investment and strong market presence. Their in-house manufacturing capacity enable them to prototype and produce most of the components needed to build their machines, particularly for their qubit chipsets and control electronics. Their most visible outside vendor is Bluefors although they are also developing an in-house giant "super-fridge" cryostat codenamed Goldeneye. The most recent 2022 update added a scale-out strategy with their QPUs on top of their scale-in roadmap announced in 2020[718].

They are also very open and reliable, publishing roadmaps and roadmap updates, respecting their planned milestones and with a steady stream of open research publications showcasing their contribution to the complicated quantum computing field even though it sometimes looks like a puzzle that may be hard to reassemble.

IBM's choice technology is the fixed frequencies transmon superconducting qubits. It is using cross-resonance two-qubit gates which consists in applying a microwave drive to one qubit (the control) at the frequency of another qubit (the target), generating a ZX interaction that is mediated by a bus[719].

[714] See Strong quantum computational advantage using a superconducting quantum processor by Yulin Wu, Jian-Wei Pan et al, June 2021 (22 pages).

[715] See Quantum Computational Advantage via 60-Qubit 24-Cycle Random Circuit Sampling by Qingling Zhu, Jian-Wei Pan et al, September 2021 (15 pages).

[716] See Quantum Neuronal Sensing of Quantum Many-Body States on a 61-Qubit Programmable Superconducting Processor by Ming Gong et al, January 2022 (14 pages).

[717] Who does fundamental research? Mainly IBM, Microsoft, Google and large telecom companies. The Bell Labs coming from the dismantling of AT&T in 1982 are now part of Nokia after gone through Lucent and Alcatel-Lucent.

[718] See The Future of Quantum Computing with Superconducting Qubits by Sergey Bravyi, Oliver Dial, Jay M. Gambetta, Dario Gil and Zaira Nazario, IBM Quantum, September 2022 (20 pages), a well-crafted paper detailing their scientific roadmap including quantum error correction trade-off choices, chipset manufacturing and scale-in/scale-out architecture.

[719] See First-principles analysis of cross-resonance gate operation by Moein Malekakhlagh et al, IBM Research, May 2020 (30 pages) and Mitigating off-resonant error in the cross-resonance gate by Moein Malekakhlagh et al, August 2021 (20 pages).

Its number of qubits increased steadily from 5 in 2016 to 127 in November 2021. IBM's quantum systems have been running in the cloud since 2016. These are already used by thousands of researchers, students, startups and corporations around the world. After creating laboratory computers, IBM ventured into creating packaged ones when announcing the Q System One in January 2019 at the Las Vegas CES, initially a 20-qubit system[720].

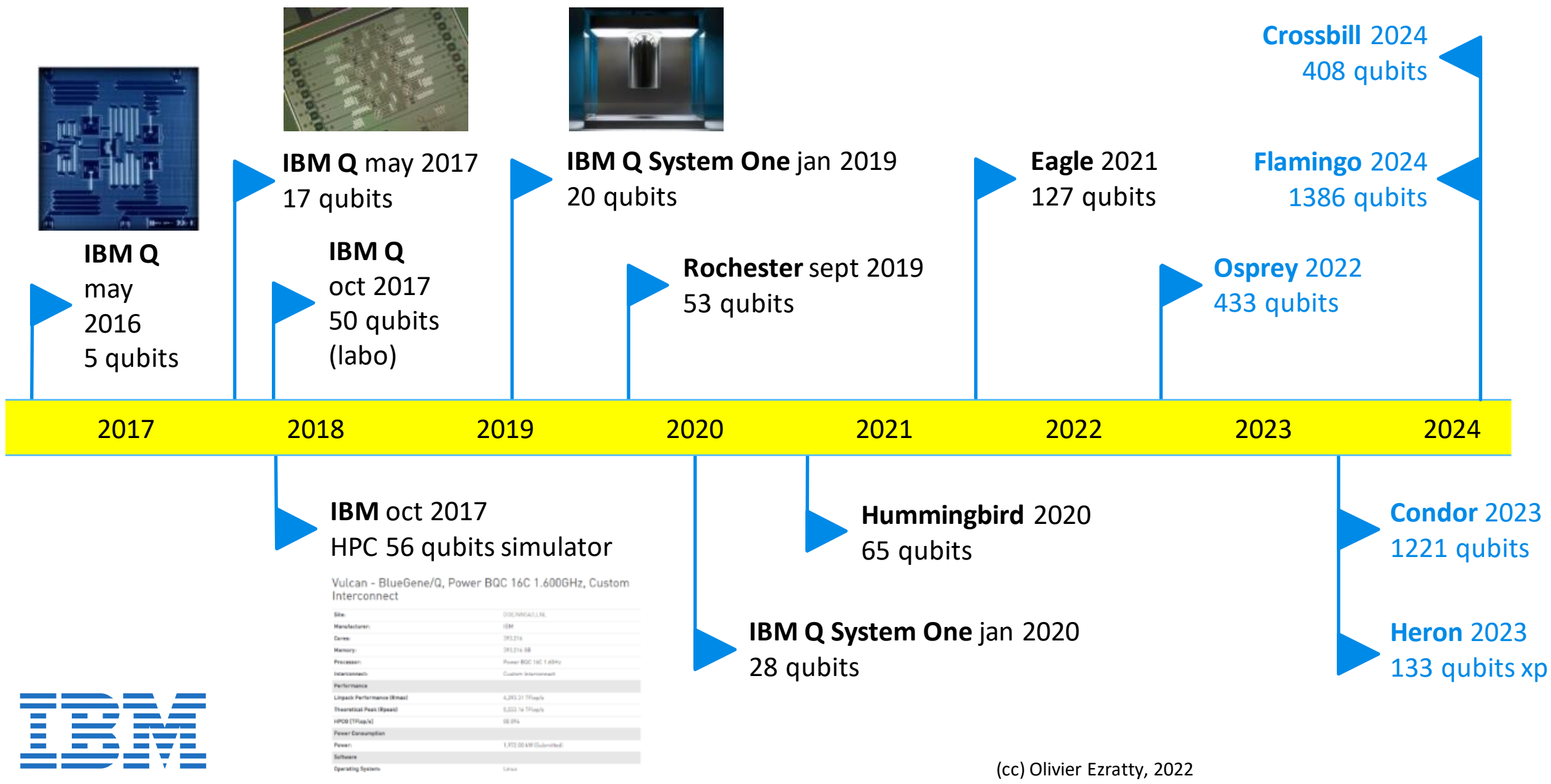

*Figure 299: IBM quantum computing timeline. (cc) Olivier Ezratty, 2022.*

The Quantum System One is 2.75 m wide, about the size of a D-Wave quantum annealer. In Figure 300, ci-dessous on the right, you can see the Quantum System assembly workshop. IBM is implementing a pre-industrial approach to the production of its quantum computers, despite their very limited capacities and low volume economics. The casing front contains the suspended cryostat while the back contains all the computing, electronics and cryostat compressor and pumps. The Quantum Systems are also self-calibrating. IBM is continuously improving these systems and updating the related qubits lifetime and gates fidelities data on their Quantum Experience web site.

Most of these units are sitting in IBM's own data center in Poughkeepsie, with some extra systems sitting in IBM sites in Germany (operated with the Fraunhofer Institute), Tokyo in Japan, in Korea at Yonsei University, in Canada near Sherbrooke Institut Quantique[721] and supposedly at Cleveland Clinic in the USA. In October 2022, Joe Biden visited the Poughkeepsie facility and was presented the Osprey 433 qubit processor that was to be unveiled later in 2022[722]. As of October 2022, IBM had already retired 26 quantum computers with 1, 5, 7, 15, 20, 27, 28, 53 and 65 qubits[723].

---

[720] Its design was created with the design studios Map Project Office and **Universal Design Studio** (UK) and **Goppion** (Italy), a manufacturer of high-end exhibition devices for museums, which notably designed the protective device for the Mona Lisa in the Louvre Museum and the Queen's jewels in the Tower of London.

[721] See IBM Research launches the first Discovery Accelerator in Canada, IBM, February 2022. This deployment represents in investment of CAN $65M by IBM matched by the Government of Québec. IBM will team up there with Alexandre Blais' team at Institut Quantique in Sherbrooke.

[722] See U.S. President Biden visits IBM's quantum data center — home of the world's largest fleet of quantum computers, IBM, October 2022.

[723] See Retired systems, IBM. Extracted on October 24th, 2022.

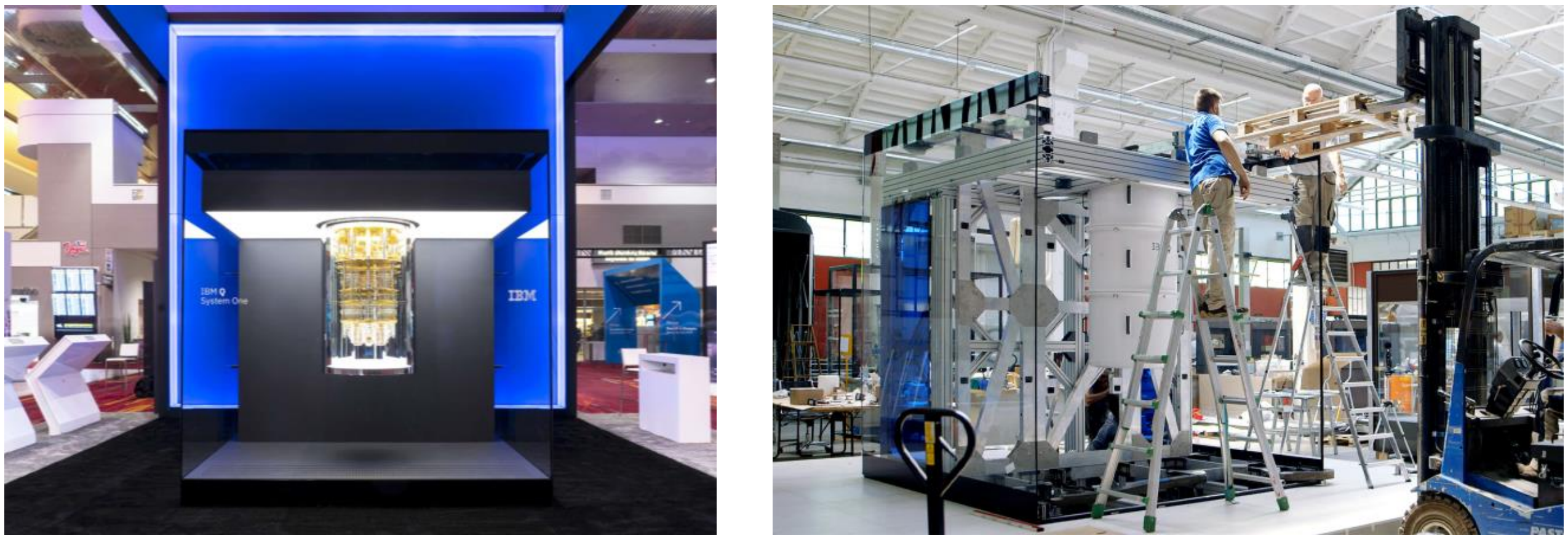

*Figure 300: IBM System Q packaging (left) and without packaging (right). Source: IBM.*

In September 2020, IBM announced their plan to “scale-in” the number of qubits of their quantum computers[724] with a 127 qubits version ("Eagle") introduced in November 2021[725], 433 qubits planned for late 2022 ("Osprey") and 1221 qubits in 2023 ("Condor").

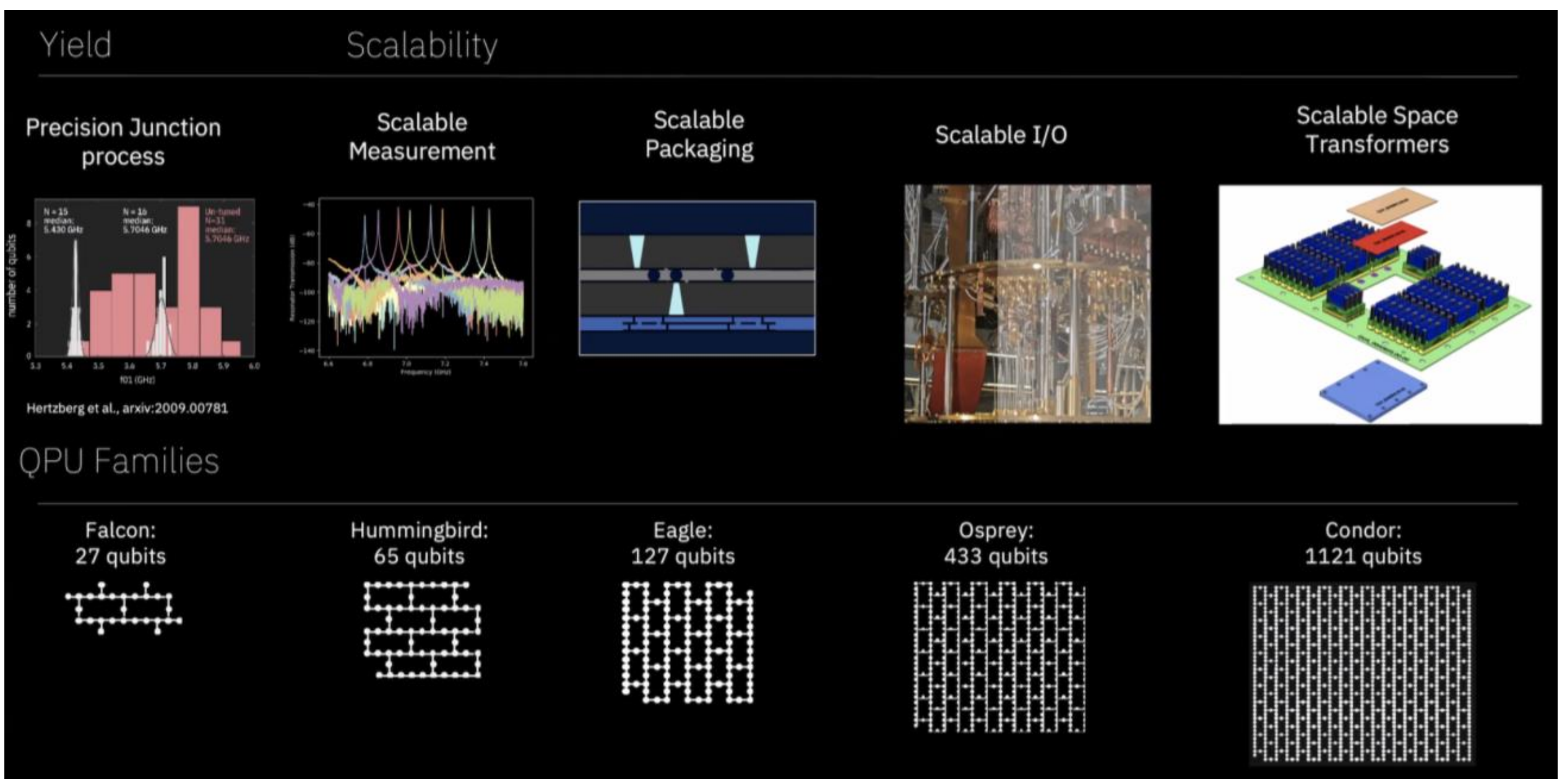

*Figure 301: IBM’s superconducting roadmap from 2020 to 2023. Source: IBM.*

This roadmap was updated in May 2022 with the addition of new processors[726].

This was a « scale-out » complementary approach, about how to assemble several quantum chipsets with three-steps: first, Heron (133 qubits, 2023) will be assembled in QPU with many units running the same algorithm in parallel, accelerating the thousands of runs that are necessary with NISQ systems algorithms.

---

[724] See IBM's Roadmap For Scaling Quantum Technology by Jay Gambetta, September 2020, completed by IBM publishes its quantum roadmap, says it will have a 1,000-qubit machine in 2023 by Frederic Lardinois in TechCrunch. See also IBM Envisions the Road to Quantum Computing Like an Apollo Mission by Dexter Johnson, September 2020.

[725] See Eagle’s quantum performance progress by Oliver Dial, IBM, March 2022.

[726] See Expanding the IBM Quantum roadmap to anticipate the future of quantum-centric supercomputing by Jay Gambetta, May 2022.

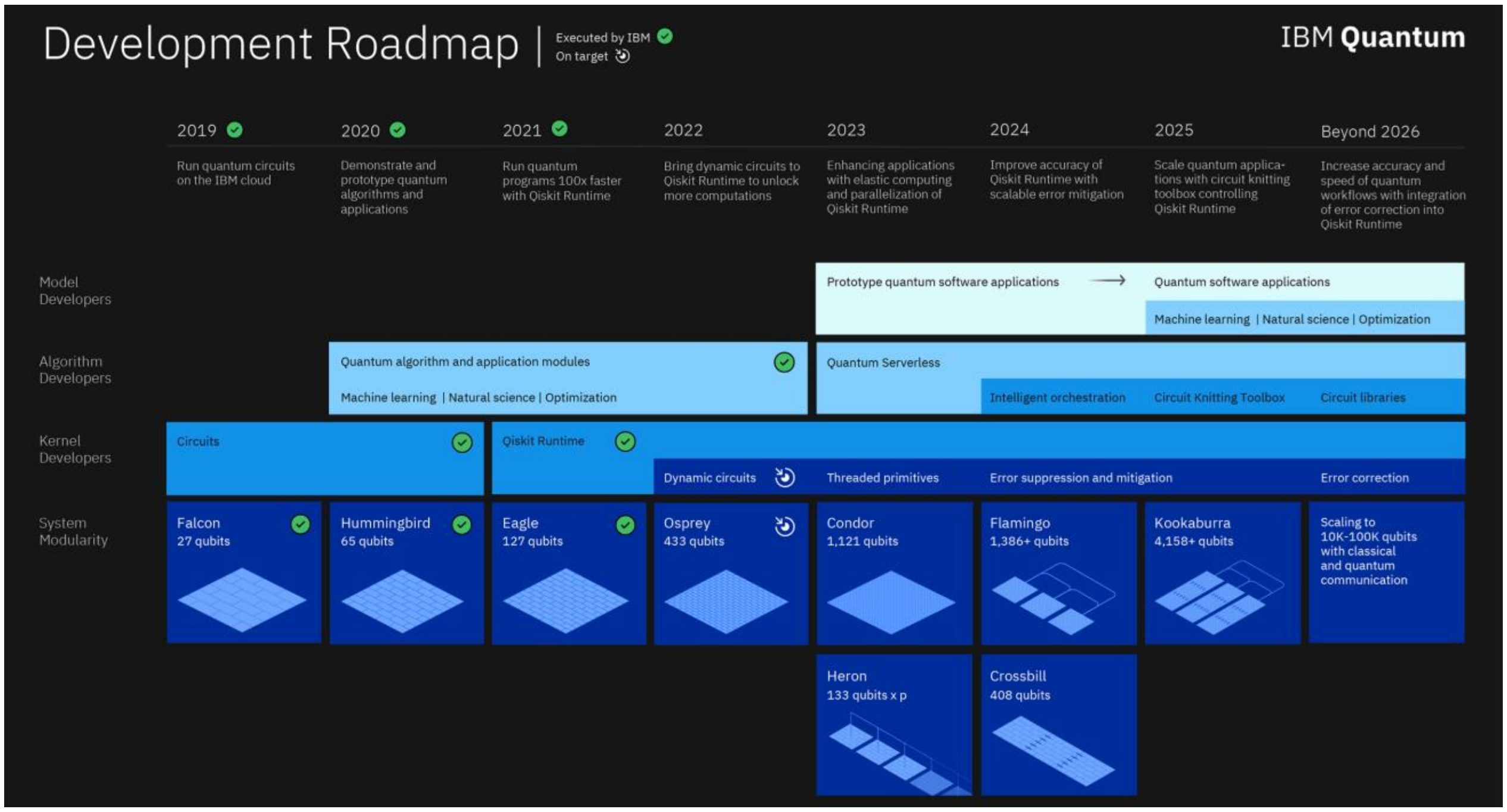

*Figure 302: IBM's scale-in and scale-out roadmap. Source: IBM.*

Second, Crossbill will assemble three chipsets similar to Heron and total 408 qubits in 2024 (3×133= 399, we can presume that the remaining 9 qubits are explained by some connectivity constraints). These chipsets will be tightly connected with chip-to-chip coupler qubit gates. Then, Flamingo will follow in 2024 with 1386 qubits (3×462) with three chipsets blocks interconnected through some short-range microwave-photonic link (less than one meter). Then, Kookaburra will reach 4158 qubits assembling three blocks of three chipsets associating chip-to-chip micro-wave coupler gates and longer range photonic-based interconnectivity.

We'll now look at the various technology improvements IBM is implementing or planning to implement in its various superconducting qubits systems. IBM is probably the most open industry vendor with regards to its scientific openness. Among others, their presentations at the Chicago APS March Meeting in 2022 was enlightening, particularly with the Industry Session talk from Hanhee Paik and another from Oliver Dial[727].

**Heavy-Hexagon layout**. In July 2021, IBM announced a generalization of their hexagon qubits topology. This heavy-hex lattice is the 4th version of IBM Quantum systems qubits topology and has been used upwards of its Falcon processor (27 qubits) as shown in the schema, where useful computing qubits are in yellow, black qubits in the red zones, the phase (Z) errors correction qubits and the white qubits in blue zones, the flip (X) errors corrections qubits [728]. It uses a hexagonal arrangement with an intermediate qubit on each side of hexagons. The topology is optimized for quantum errors correction, using custom hybrid surface and Bacon-Shor subsystem codes.

The IBM processors qubit numbers can be explained by this architecture and the lattice code distance for error correction (Falcon 27 qubits with a distance 3 code, Hummingbird 65 qubits and a distance 5 code and Eagle with 127 qubits and a distance 7 code). This topology is different from the square lattice chosen by Google in its Sycamore processors, which, however, uses coupling qubits, a solution that IBM was not relying on until it was announced in May 2022 that they would later implement it.

[727] See also this impressive list of over 800 papers from IBM research and its research partners published on arXiv (as of August 2022).

[728] See The IBM Quantum heavy hex lattice by Paul Nation et al, IBM Research, July 2021 and Topological and subsystem codes on low-degree graphs with flag qubits by Christopher Chamberland et al, IBM Research, December 2019 (20 pages). Heavy hex requires some tuning with algorithms like QAOA. See Scaling of the quantum approximate optimization algorithm on superconducting qubit based hardware by Johannes Weidenfeller et al, IBM, February 2022 (20 pages).

IBM was using fixed frequency qubits when Google uses tunable frequency ones. The hex lattice reduces the effect of frequency collision between qubits.

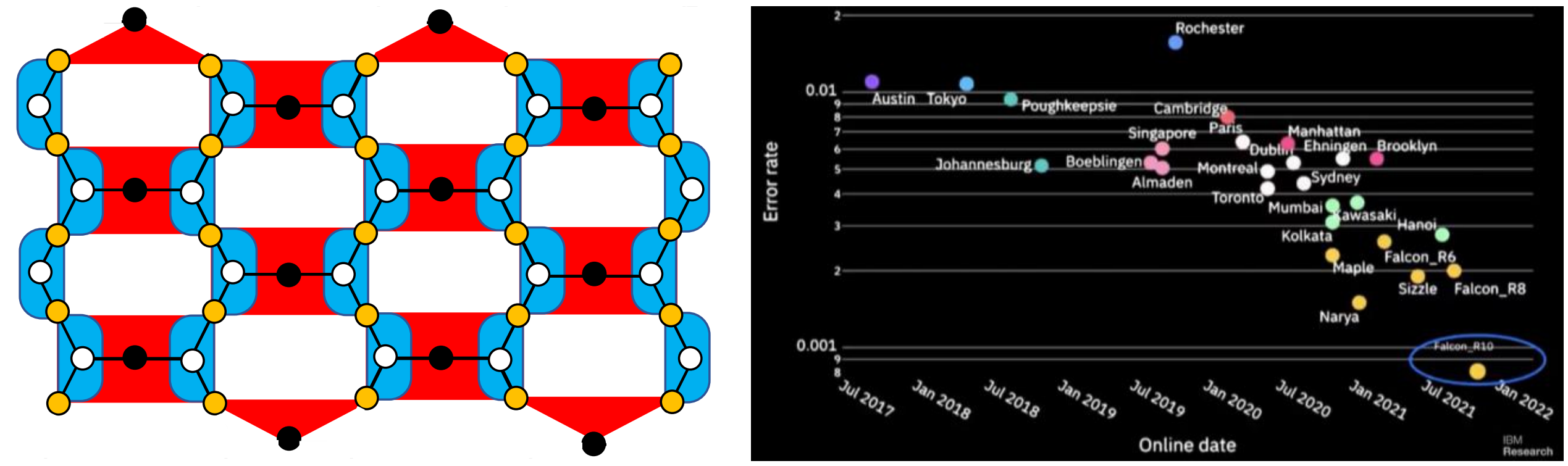

*Figure 303: Heavy-Hexagon layout (left) and evolution of IBM's superconducting qubits fidelities over time (right).*

**Qubits quality**. With improving qubit readouts fidelity using low noise amplifiers (QLA for quantum-limited amplifiers) [729].They reached fidelities records in November 2021 with Falcon R10 (27 qubits) with less than 0.001 errors (*aka* a "three nines", meaning 99,9% fidelities for two-qubit gates), but with no scientific paper describing Falcon R10 so far. It enabled them to obtain a quantum volume of 512 in May 2022 with this chipset (meaning: 9 operational qubits).

They also improve coherence times on a regular basis within a class of systems processors (like Falcon for 27 qubits)[730]. Their record is a $T_1$ of 300 µs with their Falcon R8 processor and 1 ms in research (source). They also continuously improve qubit physical properties[731]. Like with using laser annealing in their production[732], with new materials design improving their purity and avoiding contaminations and computer aided design[733], and with improving chipset vacuum isolation during its assembly[734]. Metallic superconducting nanowires could also be controlled with applying a moderate voltage to a nearby gate electrode. Switches using this effect would require very little power and could mitigate the well-known negative impact of phonons on the coherence of superconducting qubits[735].

IBM researchers are also quite prolific in finding ways to improve gates fidelities. Like with enabling efficient SWAP gates implementation with low crosstalk, using dispersively coupled fixed-frequency transmon qubits and simultaneous driving of the coupled qubits at the frequency of another qubit, with a fast two-qubit interaction equivalent to ZX + XZ entangling gates, implemented without strongly driving the qubits[736].

---

[729] See Rising above the noise: quantum-limited amplifiers empower the readout of IBM Quantum systems by Baleegh Abdo, January 2020.

[730] $T_1$ and $T_2$ reached about 260 µs in September 2021 with their Peekskill 27 qubit system. It was a 3-times improvement vs previous systems.

[731] See Materials challenges and opportunities for quantum computing hardware by Nathalie P. de Leon et al, Science, April 2021 (x pages).

[732] See High-fidelity superconducting quantum processors via laser-annealing of transmon qubits by Eric J. Zhang et al, December 2020 (9 pages) and Laser-annealing Josephson junctions for yielding scaled-up superconducting quantum processors by Jared B. Hertzberg et al, August 2021 (8 pages).

[733] See What if We Had a Computer-Aided Design Program for Quantum Computers?, IBM, October 2020 and Qiskit Metal: IBM Community Building a Computer-Aided Program for Quantum Device Design by Matt Swayne, October 2020.

[734] See Ultrahigh Vacuum Packaging and Surface Cleaning for Quantum Devices by M. Mergenthaler et al, 2020 (6 pages).

[735] See Vibrations could flip the switch on future superconducting devices by Markus F. Ritter, Andreas Fuhrer and Fabrizio Nichele, March 2022, pointing to Out-of-equilibrium phonons in gated superconducting switches by Markus F. Ritter et al, Nature Electronics, March 2022 (7 pages).

[736] See Cross-Cross Resonance Gate by Kentaro Heya and Naoki Kanazawa, IBM Research Japan, PRX, November 2021 (15 pages).

**3D circuits**. It started with stacked pairs of chipsets separating qubits from microwave controls, using TSV (through-silicon vias) with their 127 qubit systems in 2021[737].

This 3D chipset layout introduced with Eagle (127 qubits) adds multi-level wiring (MLW) on the backside of the interposer. Control and readout signals are routed as strip lines in the MLW and are well isolated from the interposer and qubit metal levels. Connections between the qubits and this MLW is done via superconducting TSVs (through-silicon vias). It reduces crosstalk when pairing qubits. They also improved readout multiplexing with using 9 lines instead of 5 in Eagle. However, this degrades their CNOT gates fidelity due to collisions between qubit readout frequencies.

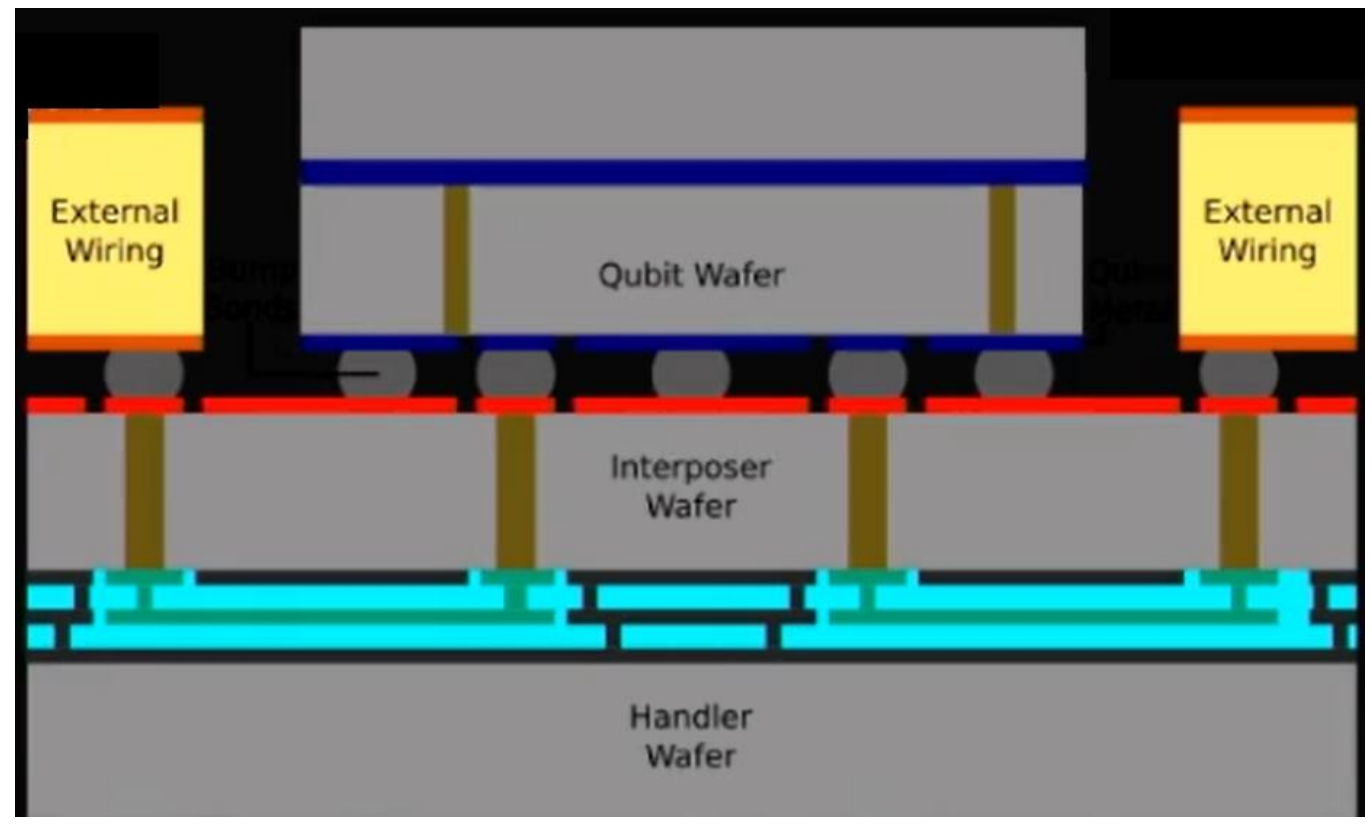

*Figure 304: the three stacked die chipset architecture used in Eagle's 127 qubit processor. Source: IBM.*

**Scalability**. With microwaves signals multiplexing and intelligent filtering for qubit states readouts, starting with Hummingbird 65 qubits system in 2020. They are also using microwave flexible cables to reduce the space used by microwave cabling in cryostats. They are also implementing tunable couplers to control qubits entanglement[738].

With Osprey (433 qubits), wires density will further improve using flexible cables and with control electronics of "generation 3", using custom-made FPGAs.

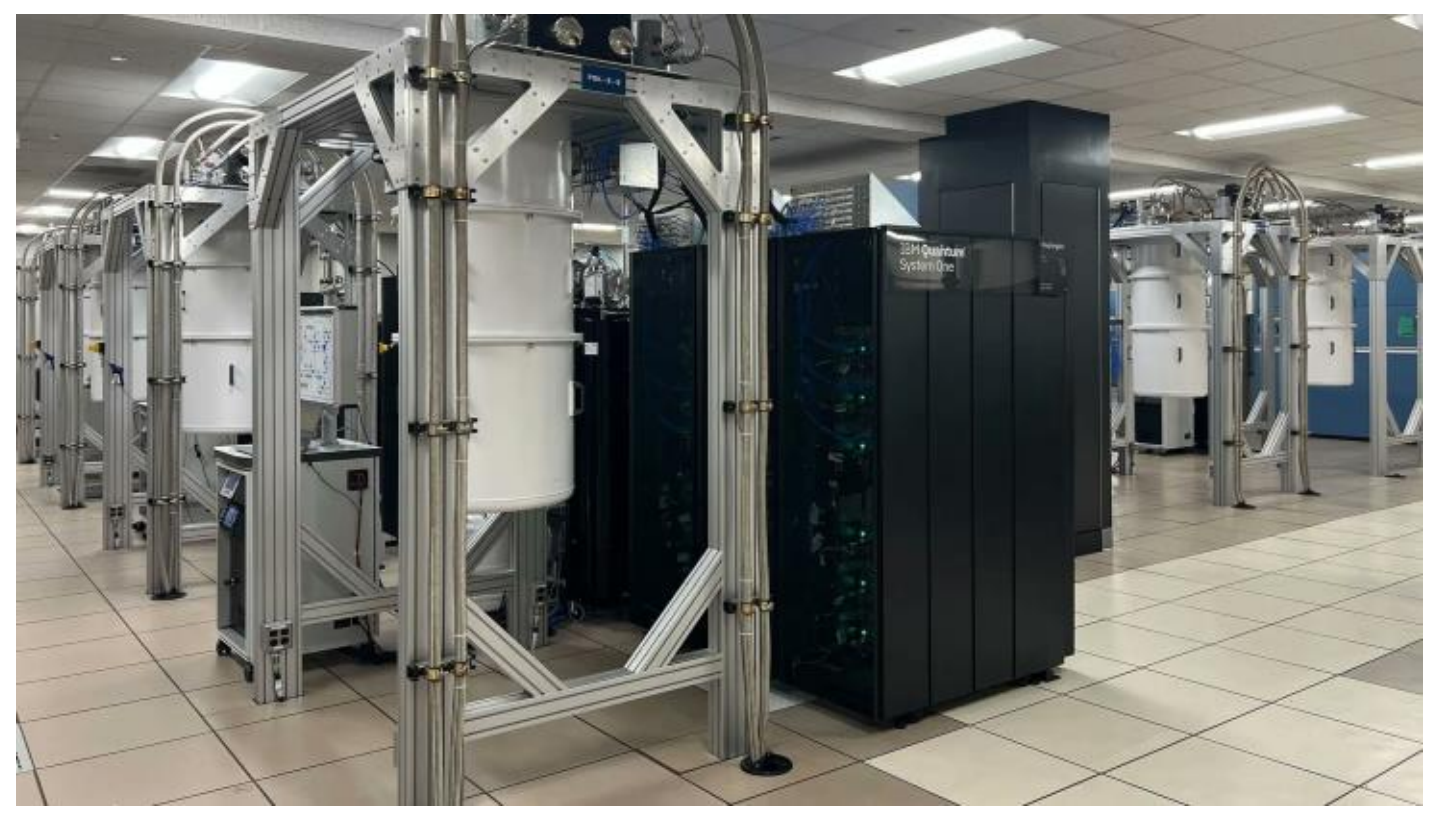

*Figure 305: IBM's quantum data center in Poughkeepsie, New York State. Source: IBM.*

Scalability will also come from qubits miniaturization coming from various paths like a simplification of the readout electronics replacing the usual circulator with a "microwave-controlled qubit readout multichip module" (QRMCM)[739].

**Error Mitigation**. They are optimizing their quantum error correction architecture[740], particularly to correct T gates errors while using classical QEC for Clifford gates as part of "Quantum readout-error mitigation" (QREM).

---

[737] See Merged-Element Transmons: Design and Qubit Performance by H. J. Mamin et al, March 2021 (7 pages).

[738] See Tunable Coupling Architecture for Fixed-frequency Transmons by J. Stehlik et al, IBM Research, February 2021 (7 pages) and With fault tolerance the ultimate goal, error mitigation is the path that gets quantum computing to usefulness by Kristan Temme, Ewout van den Berg, Abhinav Kandala and Jay Gambetta, July 2022.

[739] See High-Fidelity Qubit Readout Using Interferometric Directional Josephson Devices by Baleegh Abdo et al, December 2021 (34 pages). Avoiding the magnet-based circulator for qubit readout using a microwave-controlled qubit readout multichip module (QRMCM) that "*integrates interferometric directional Josephson devices consisting of an isolator and a reconfigurable isolator or amplifier device, and an off-chip low-pass filter*". Also, see Merged-Element Transmons: Design and Qubit Performance by H.J. Mamin et al, PRA, August 2021 (7 pages).

[740] Their views on QEC: Hardware-aware approach for fault-tolerant quantum computation by Guanyu Zhu, 2020.

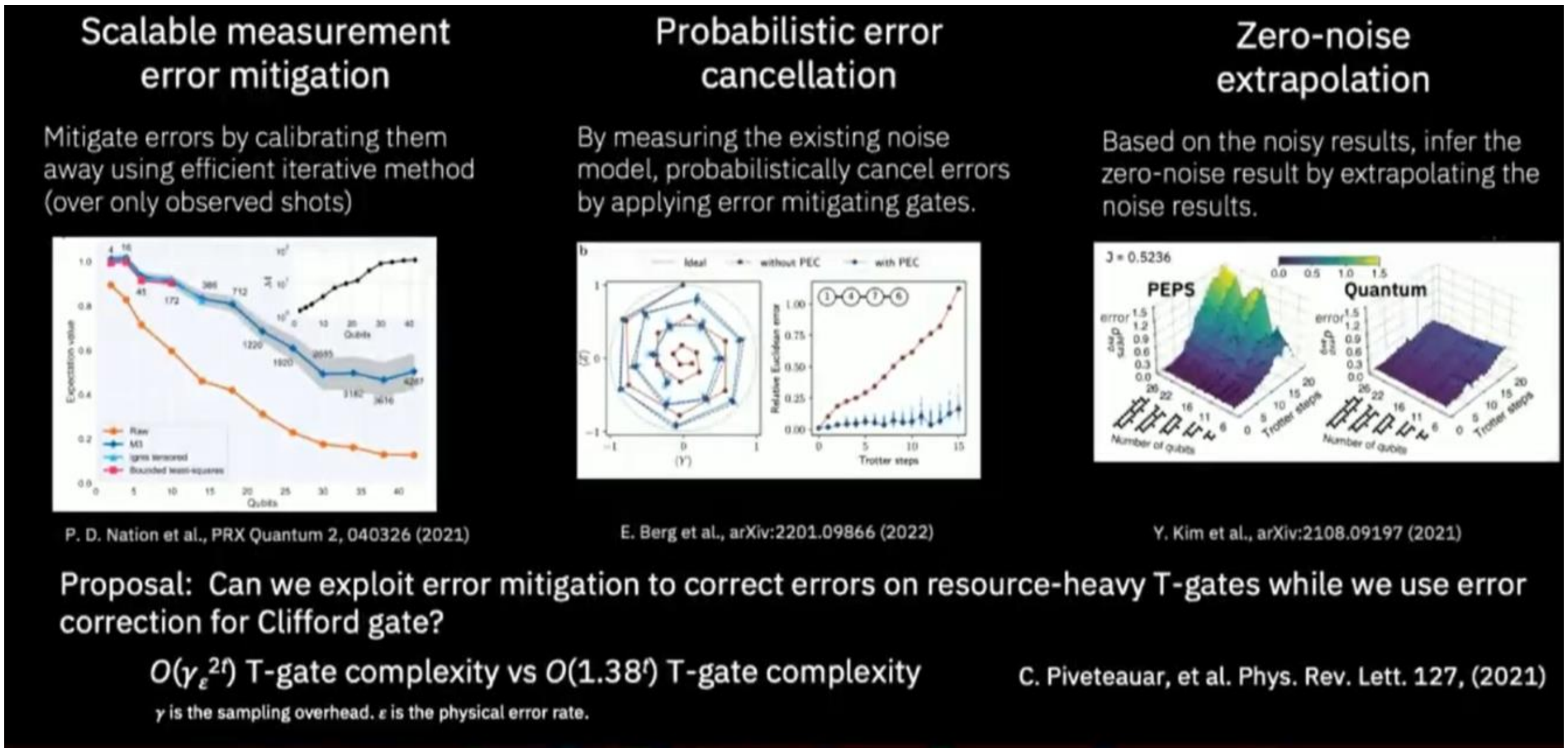

*Figure 306: the various quantum error mitigation techniques IBM is working on. Source: IBM.*

**Various algorithms optimizations**. IBM research teams are finding many ways, more or less efficient, to optimize algorithms run-time.

For example, they are allowing more computations with the Bernstein-Vazirani algorithm run on 12 qubits, where SWAP gates are replaced with resets, improving fidelity from 0.0007 to 0.80.

They also propose to use "circuit knitting", i.e., combining smaller circuits to simulate larger problems using entanglement forging[741]. However, this technique has a strong tendency to strongly attenuate algorithms quantum advantage and parallelism.

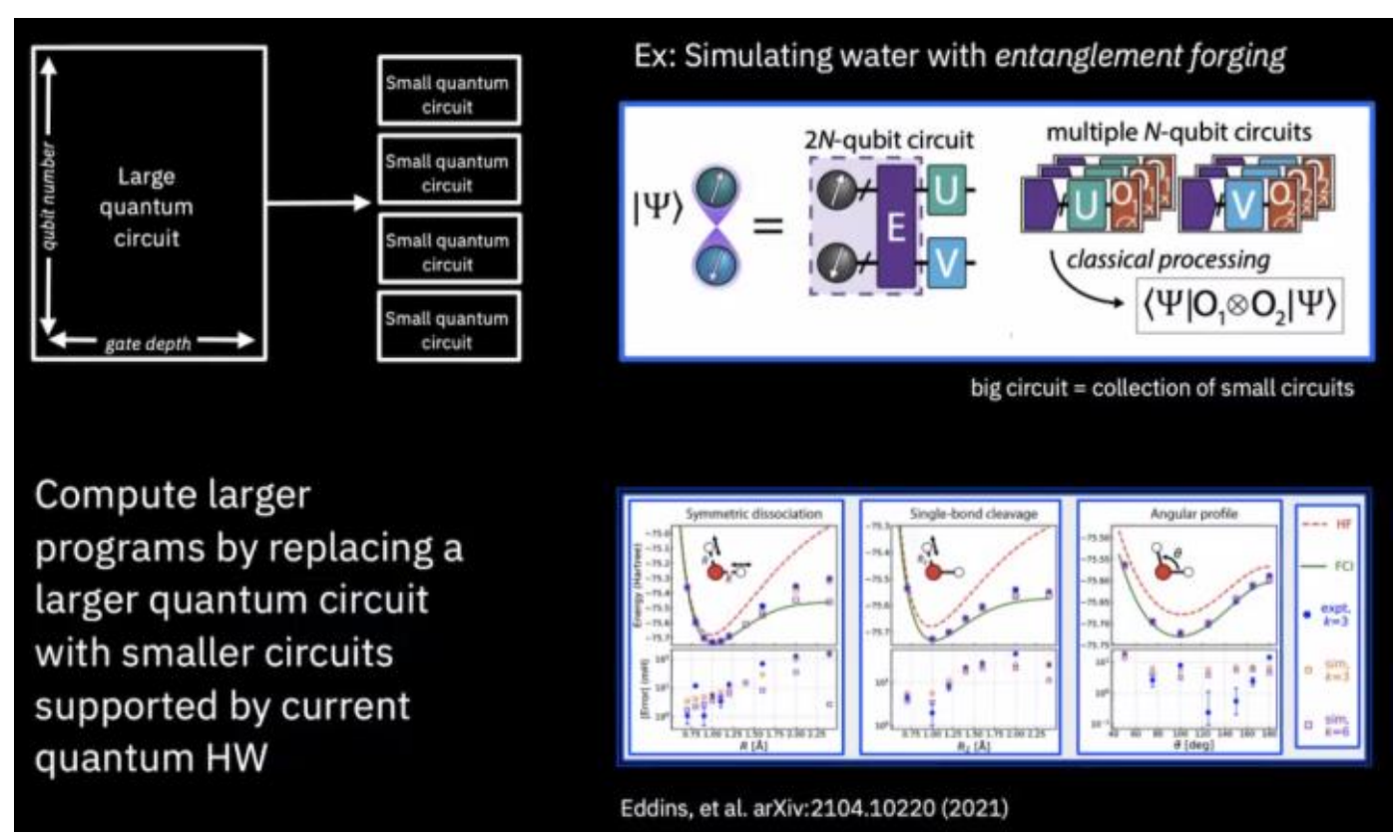

*Figure 307: entanglement forging technique. Source : IBM.*

**Cryogeny**. IBM announced in 2020 that it was working on a giant home-made cryostat called "Goldeneye" exceeding current market capacity, to host from a thousand to a million physical qubits[742], shown in Figure 308. It is due for 2023 and has already been tested as of May 2022 at 25 mK. It has about 12 pulse tubes and 6 dry dilutions with half of the dilution being inverted at the bottom.

[741] See Doubling the size of quantum simulators by entanglement forging by Andrew Eddins, Sergey Bravy, Sarah Sheldon et al, April 2021 (17 pages), Entanglement forging The 2x Gambit: IBM Tech Doubles Qubit Effectiveness by Charles Q. Cho, February 2022, Simulating Large Quantum Circuits on a Small Quantum Computer by Tianyi Peng, Aram Harrow et al, October 2020, PRL (20 pages), Quantum Divide and Compute: Exploring The Effect of Different Noise Sources by Thomas Ayral and F.M Le Régent (Atos) with Zain Saleem, Yuri Alexeev, Martin Suchara (DoE Argonne National Laboratory, February 2021 (21 pages) and Constructing a virtual two-qubit gate by sampling single-qubit operations by Kosuke Mitarai and Keisuke Fujii, Osaka University, JST PRESTO and RIKEN, January 2021 (13 pages).

[742] See IBM scientists cool down world's largest quantum-ready cryogenic concept system by Pat Gumann and Jerry Chow, September 2022. The device is 3m high and 2m wide with 1.7 $m^3$ of experimental volume and 10 internal plates. Its cooling power is of 10 mW at 100 mK and 24W at 4K.

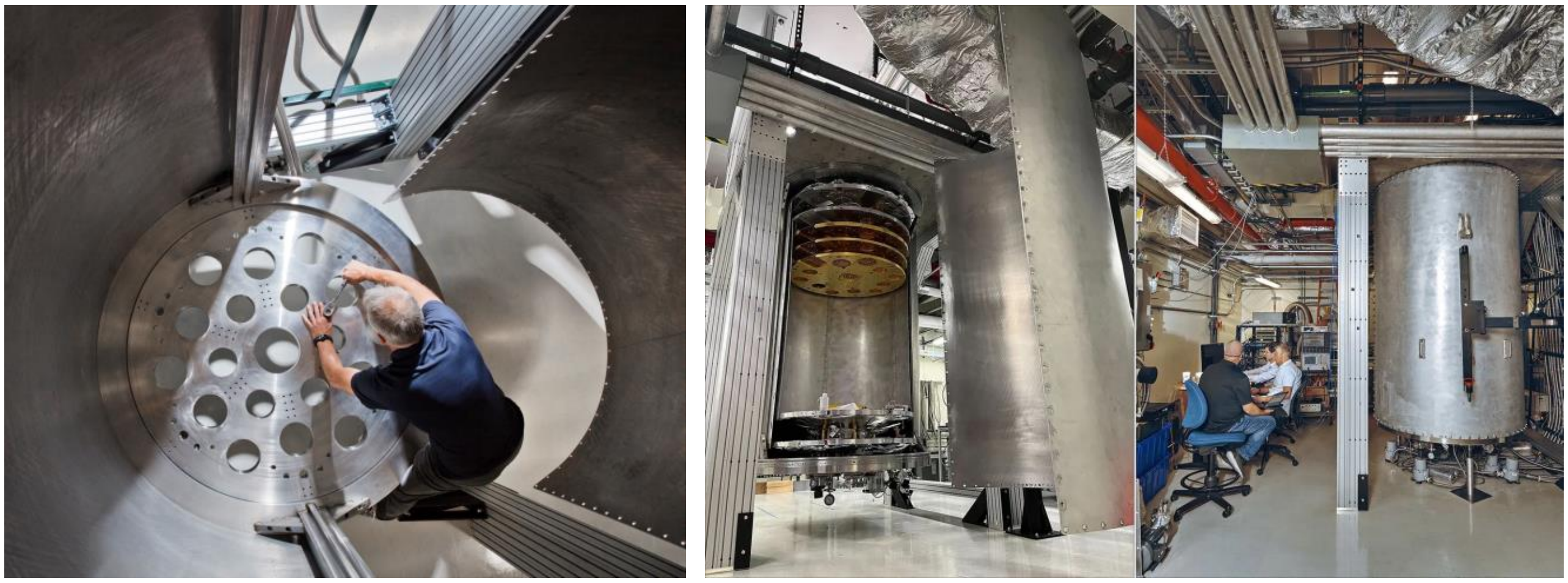

*Figure 308: IBM's giant Goldeneye dilution refrigerator. Source: IBM.*

**QPU interconnect**. Beyond 2023 and Condor, scale-out will involve interconnecting quantum processing unit microwave-optical transduction and optical fiber to connect QPUs, with optomechanical coupling or electro-optic coupling. One option is to use optical channels with SiGe/Si optical resonators[743]. These quantum units will be cooled with a new generation of cryostats, designed by Bluefors as part of their KIDE range using a hexagonal form factor, announced in November 2021. It will precede Goldeneye in their roadmap. IBM will start to implement this System Two modular architecture with their 1121 qubits systems in 2023.

**MBQC option**. Other longer-term plans consist in using constant depth circuits using entanglement and measurements ala "one way quantum computing" and MBQC that is also to be used with flying qubits like photons.

**Benchmarking**. IBM now uses three key metrics to benchmark its quantum systems. The first one is simply their number of qubits which define the **scale** of their system. The second is the quantum volume that was introduced in 2019 and described in details page 672. It defines their computing **quality**. IBM announced it would double every year.

In April 2022, they obtained a record quantum volume of 256, meaning 8 operational qubits and 8 depths of computing, which increased to 512 in May 2022[744]. There is some inconsistency with IBM's roadmap. They expect to more than double the number of qubits in their system while doubling the available quantum volume every year. This means adding one operational qubit per year! At last, in November 2021, IBM introduced CLOPS (circuit layers operations per seconds) which defines the speed of their processor.

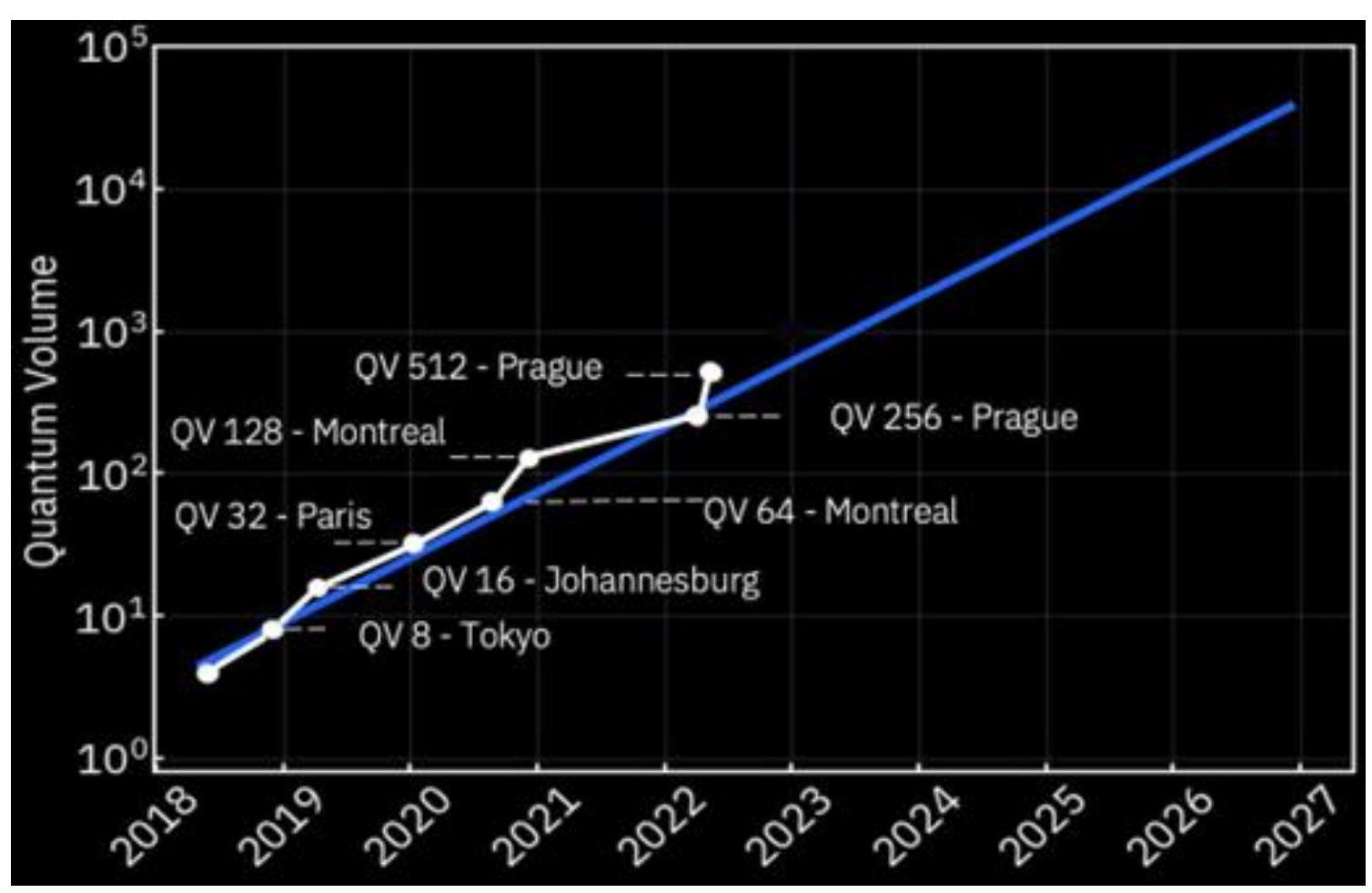

*Figure 309: IBM's quantum volume evolution over time. Source: IBM.*

[743] See Engineering electro-optics in SiGe/Si waveguides for quantum transduction by Jason Orcutt et al, Quantum Science Technology, 2020 and Ultrahigh-Q on-chip silicon–germanium microresonators by Ryan Schilling, Hanhee Paik et al, Optica, 2022 (4 pages).

[744] See Pushing quantum performance forward with our highest Quantum Volume yet by Petar Jurcevic et al, IBM, April 2022.

It currently sits between 800 and 2500 CLOPS[745].

It is also key to care about the level of large entanglements in these systems, *aka* genuine multipartite entanglement (GME). It was done in 2021 by a team of Australian researchers with fidelities of 54% for 27 qubits with using some quantum readout-error mitigation (QREM) [746]. It was also implemented by IBM Research with a 65-qubit QPU [747]. The schema in Figure 310 shows the history of experimentally prepared quantum states exhibiting N-qubit GME, where $N \geq 3$, with at least 95% confidence in gate-based quantum systems. It illustrates the challenges to create large high-fidelity entangled states.

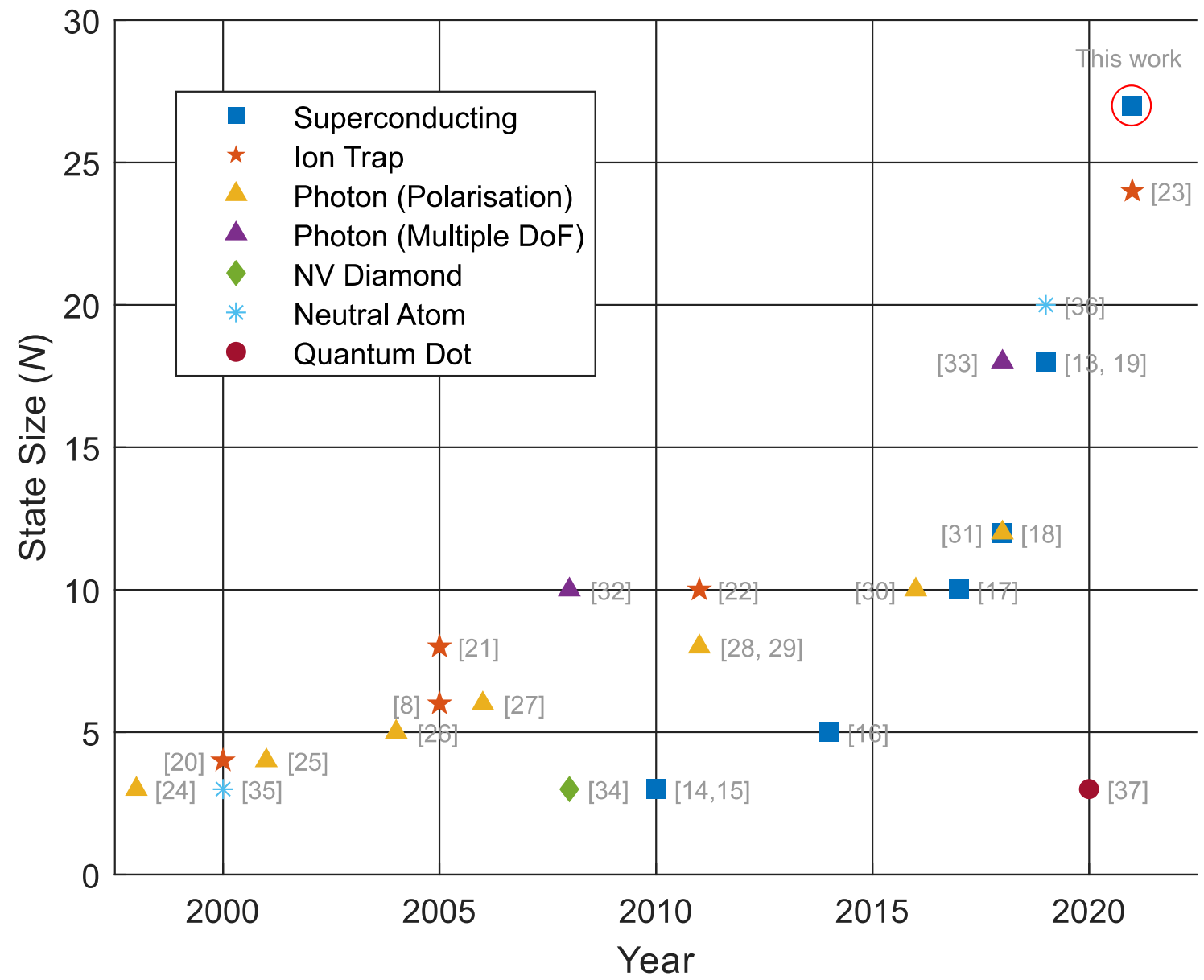

*Figure 310: largest multipartite entangled state over time. Source: Generation and verification of 27-quit Greenherger-Horne-Zellinger states in a superconducting quantum computer by Gary J. Mooney et al, August 2021 (16 pages).*

**Serverless**. As part of their May 2022 announcements, IBM explained it is adopted a "serverless" architecture, another name for a cloud service driving both classical and quantum computers with a pay-as-you-go pricing[748]. It involves three techniques: circuit knitting that leverages classical resources cut a quantum problem in smaller problems and circuits to run on NISQ QPUs and also use classical processing. This is interesting but reduces the quantum acceleration generated by the QPUs who will run quantum algorithms of smaller Hilbert space. Then, entanglement forging is a way to simplify knitting specific to solving chemistry problems. They then use "quantum embedding" to reframe a problem and split it between pieces running classically and others quantumly "*for only the classically difficult parts of the problem*". At last, error mitigation is based on classical post-processing to reduce the impact of some classes of quantum errors.

**Deployments and customer evangelism**. IBM has been investing a lot since 2016 to build a community of developers and users worldwide. They launched the IBM Quantum network in 2017. It brings together major Fortune 500 companies, research laboratories and startups interested in developing quantum solutions. This network offers access free access to quantum systems with one (crowded) 15 bits system, 8 5-bit systems and a 1-bit small-use system. Commercial systems have respectively 5, 27, 28, 53, 65 and 127 qubits. At last, a quantum emulator (branded a simulator) supports 32 qubits. As of August 2022, they had 27 quantum systems online.

IBM also launched a customer Quantum Computation Center in Poughkeepsie, New York, a quantum center in Montpellier, France, in 2018, and then a partnership in Germany with a Fraunhofer Institute in 2019 plus other quantum centers in Japan and Quebec, Canada.

---

[745] See Quality, Speed, and Scale: three key attributes to measure the performance of near-term quantum computers by Andrew Wack, Hanhee Paik, Jay Gambetta et al, 2021 (8 pages).

[746] See Generation and verification of 27-quit Greenherger-Horne-Zellinger states in a superconducting quantum computer by Gary J. Mooney et al, August 2021 (16 pages).

[747] See Whole-device entanglement in a 65-qubit superconducting quantum computer by Gary J. Mooney et al, October 2021 (15 pages).

[748] See Introducing Quantum Serverless, a new programming model for leveraging quantum and classical resources by Blake Johnson, Ismael Faro, Michael Behrendt and Jay Gambetta, May 2022

Their task is to evangelize developers and researchers to encourage them to develop software on their Qiskit platform and their quantum systems sitting in the cloud.

IBM publishes amazing data on their developer community activity with over 410,000 developers with a run-rate of 3.5 billion quantum circuits executed each and every day as of April 2022 which doesn't really mean anything.

At last, we should mention the quantum volume benchmark created by IBM in 2017 and updated in 2019. We cover it in detail in the section dedicated to benchmarks, page 670.

In April 2021, IBM finalized the deployment of a 28-qubit Quantum System in its own site near Stuttgart, Germany, in relationship with Fraunhofer as an intermediate to reach out the developer community[749]. It was even inaugurated remotely by Chancellor Angela Merkel on June 15th, 2021.

IBM also announced a partnership of 10 years with Cleveland Clinic in the USA, including the delivery of their 1,121 qubits system around 2024. Meanwhile, the customer will rely on the existing cloud-based Quantum Experience systems[750]. Then, in June 2021, IBM announced a five year $300M artificial intelligence and quantum computing research partnership with the UK. They plan to hire 60 scientists as part of the new Hartree National Centre for Digital Innovation (HNCDI)[751]. IBM has however not put all its eggs in the superconducting qubits basket. Their Zurich research center is also investigating electron spins and Majorana fermions qubits at a fundamental research level, working on this with ETH Zurich and EPFL. Also, in February 2022, IBM invested $25M in Quantinuum[752].

Google started to invest in quantum computing in the mid-2010s. In 2014/2015, it tested some algorithms on a **D-Wave** quantum annealing system installed in the joint QUAIL laboratory established with NASA and located at the Ames Research Center in Mountain View.

Google initially wanted to create its own quantum annealer *ala* D-Wave but quickly switched gears towards gate-based superconducting qubits quantum computing, under the direction of **Hartmut Neven** since 2006, who manages quantum hardware and software. In 2019, he put forward an empirical law called Dowling-Neven according to which the power of computers doubles exponentially. This was exaggerated when you look at their evaluation method[753]!

Hardware was developed by **John Martinis** et al between 2014 and 2020[754]. All this was done in connection with the University of Santa Barbara in California (UCSB), where he came from with part of his team.

---

[749] See Fraunhofer launches quantum computing research platform in Germany, April 2021.

[750] See Cleveland Clinic, IBM launch 10-year quantum computing partnership by Mike Miliard, March 2021.

[751] See UK STFC Hartree Centre and IBM Begin Five-Year, £210 Million Partnership to Accelerate Discovery and Innovation with AI and Quantum Computing, June 2021.

[752] See IBM invests in Quantinuum, and other quantum news updates by Dan O'Shea, Fierce Electronics, February 2022.

[753] The reasoning is as follows: the number of qubits would so far increase exponentially, and the power doubles with each addition of a single qubit. All this, every six months. Unfortunately, the available data on the actual power of today's quantum computers does not comply with this law. There is no doubling of the number of operational qubits every six months! There is even regression! Google announced 72 qubits in March 2018 and then 53 qubits in October 2019. At IBM, we are in the total confusion between the Q System One which went from 20 to 28 qubits between January 2019 and January 2020, which does not look like a doubling every six months. On the other hand, this doubling could eventually be achieved with other technologies such as Honeywell's trapped ions or Pasqal's cold atoms. In his presentation at Q2B in December 2019, John Preskill highlighted another exponential doubling: gate fidelity rates are steadily improving, which would increase quantum volume exponentially. At the same time, the cost of emulating quantum computing on conventional computers increases exponentially with quantum volume. Hence a doubly exponential evolution of computing power. The bug? Nothing says that the fidelity of quantum gates will continue to improve steadily. See A New Law to Describe Quantum Computing's Rise?, June 2019.

[754] John Martinis resigned from Google in April 2020 after being demoted to a scientific advisory role mid-2019. He explained this in an exit interview for Forbes: Google's Top Quantum Scientist Explains In Detail Why He Resigned by Paul Smith-Goodson, 2020. See also Google's Head of Quantum Computing Hardware Resigns by Tom Simonite, April 2020.

In 2017, Google stated its ambition to obtain some quantum supremacy as defined by John Preskill in 2011[755]. In April 2017 came a first 9 qubits chipset. In 2018, their Foxtail 22 qubits chipset was tested, but in quiet way. Then came in March 2018 the announcement of a record 72 qubits with their Bristlecone generation, promising a two-qubits gates fidelity of 99,56%. It seemed however a dead-end and was abandoned. Then came the famous October 2019 so-called quantum supremacy with their Sycamore processor using 53 qubits and a random algorithm similar to the boson sampling algorithm imagined by Scott Aaronson in 2012[756].

NASA and Google science papers were mistakenly posted on the Internet in September 2019 and then officially published in the journal Nature in October 2019[757], filing 70 pages with a level of detail never seen before[758]. Google compared their qubits with the most powerful supercomputer of the time, the IBM Summit installed at the Department of Energy's Oak Ridge National Laboratory in Tennessee[759]. Computing for 200 seconds on Sycamore would take 10,000 years once emulated on the IBM Summit. This comparison didn't make much sense as we discuss quantum supremacy and advantages in another part of this book, page 680.

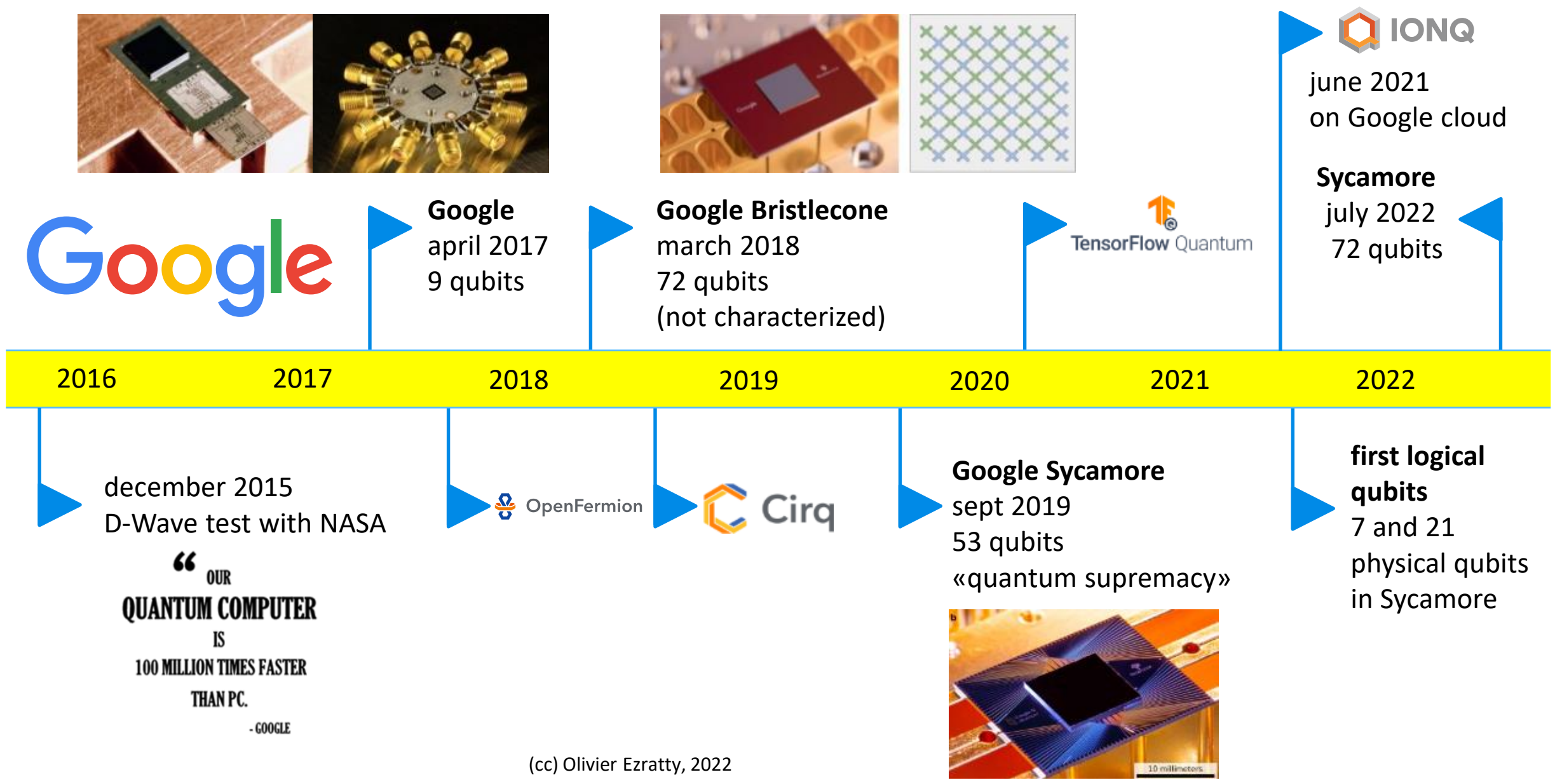

*Figure 311: Google's quantum computing timeline. (cc) Olivier Ezratty, 2022.*

---

[755] See Google says it is on track to definitively prove it has a quantum computer in a few months' time by Tom Simonite, April 2017. See also The Question of Quantum Supremacy, May 2018 which references two related papers : Characterizing Quantum Supremacy in Near-Term Devices, 2016 (23 pages) and A blueprint for demonstrating quantum supremacy with superconducting qubits, 2017 (22 pages).

[756] See Quantum Supremacy Using a Programmable Superconducting Processor by John Martinis, October 2019. The most detailed presentation on Google's hardware engineering with Sycamore is available on Google's quantum computer and pursuit of quantum supremacy by Ping Yeh, September 2019 (63 slides). See also Quantum Computer Datasheet, Google AI, May 2021 (6 pages) which provides detailed indications of Sycamore's qubit fidelities with the Weber version of the processor.

[757] See Hello quantum world! Google publishes landmark quantum supremacy claim by Elizabeth Gibney, October 2019.

[758] See Quantum supremacy using a programmable superconducting processor by Frank Arute, John Martinis et al, October 2019 (12 pages) and Supplementary information for "Quantum supremacy using a programmable superconducting processor" by Frank Arute, John Martinis et al, October 2019 (58 pages). See also Quantum supremacy using a programmable superconducting processor, a lecture by John Martinis at Caltech, November 2019 (one hour). And another version, played at QC Ware's Q2B conference in December 2019 (19 slides and 32-minute video). At last, here is this video promoting Google's supremacy: Demonstrating Quantum Supremacy, October 2019 (4'42").

[759] See Google researchers have reportedly achieved "quantum supremacy" by Martin Giles, in the MIT Technology Review, September 2019 and the source of the paper on the Internet, with illustrations. They use a type of algorithm that is of little use, but which clearly favors quantum computing and requires a limited number of quantum gates, which is good for noise-generating quantum processors. See also Why I Coined the Term 'Quantum Supremacy' by John Preskill, October 2019.

Let's have a look at Sycamore's engineering and its related supremacy benchmark:

**Cross-Entropy Benchmarking (XEB)**. The benchmark algorithm combined a set of random quantum gates with a homogeneous distribution. This last part scans all the possible values ($2^{53}$) of qubits superpositions[760]. In the supremacy regime, the so-called computation has a 0,2% chance to produce the right results. It is executed 3 million times to generate an average measurement mitigating this low fidelity[761]!

It uses superposition on all the qubits (53), allowing maximum performance. Usually, ancilla qubits are necessary to make some calculation. Ancilla qubits are used as buffer values. As a result, the exponential advantage decreases accordingly. Typical algorithms don't benefit from the superposition of $2^{53}$ states but, for example, a lesser $2^{30}$ or $2^{40}$ states. Any quantum advantage would then vanish. This explains why in most vendors roadmap, the end-goal is to create systems with 100 logical qubits and not just between 50 and 55 qubits. The benchmarks use a small 20 quantum gates computing depth. Namely, the algorithm tested at full load only chains 20 sequences of quantum gates executed simultaneously. This is related to the noise generated in the qubits which limits this depth. Many algorithms require a larger number of quantum gates, such as Shor's integer factorization.

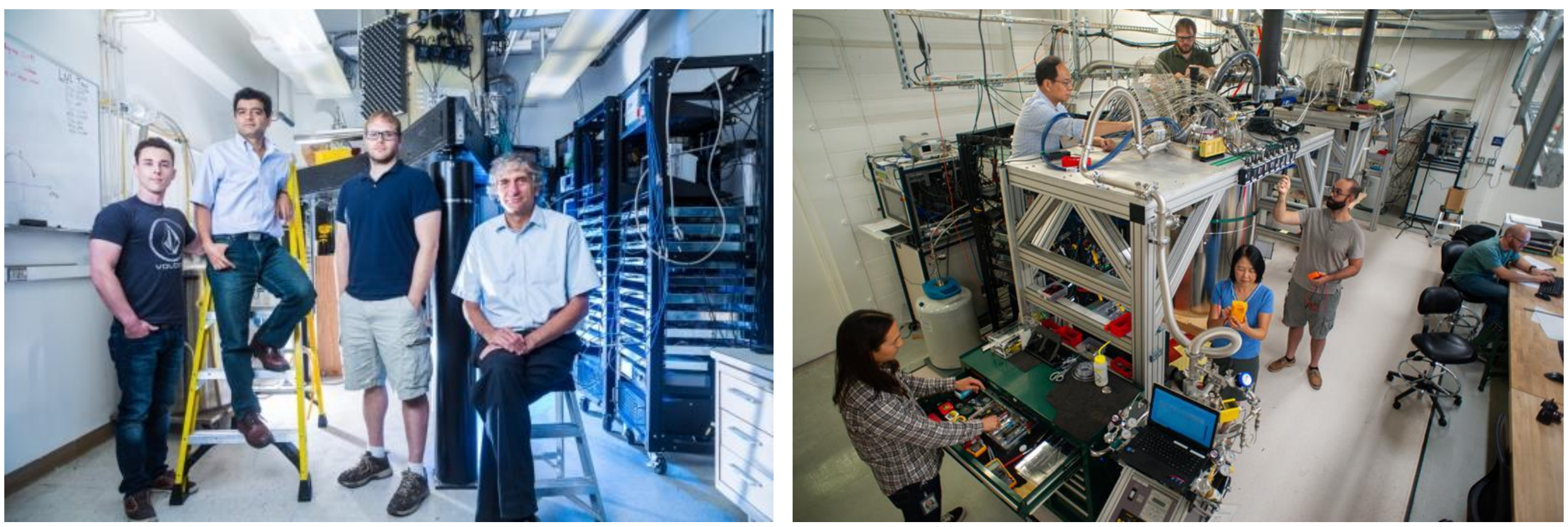

*Figure 312: John Martinis and his team when he was at Google and Google's Sycamore's assembly in their lab. Sources: Google.*

On October 21, 2019, IBM researchers published an article in which they questioned Google's performance, stating that they could run their algorithm in 2.5 days instead of 10,000 years on the IBM Summit supercomputer[762]. This would require adding 64 PB of SSD to the supercomputer, which they had not tested. That's about 7 racks full of SSDs at 2019 capacity. IBM wanted to contradict Google's claim of quantum supremacy, which they turned into some sort of quantum advantage[763].

---

[760] The following explanation can be found in Kevin Harnett's Quantum Supremacy Is Coming: Here's What You Should Know in Quanta Magazine, July 2019.

[761] See The Google Quantum Supremacy Demo and the Jerusalem HQCA debate by Gil Kalai, December 2019, where he questions the results of Google's quantum supremacy, particularly its evaluation of qubit noise at large scale.

[762] See On "Quantum Supremacy" | IBM Research Blog by Edwin Pednault, October 2019 and Leveraging Secondary Storage to Simulate Deep 54-qubit Sycamore Circuits by Edwin Pednault et al, October 2019 (30 pages).

[763] Google's quantum supremacy quibbles have gone a long way, including IBM's response. And then Has Google Finally Achieved Quantum Supremacy?, October 2019, which is quite well documented. Then Quantum supremacy: the gloves are off by Scott Aaronson, October 2019 where he discusses the fact that this case is the equivalent of Kasparov vs. Deep Blue, with IBM playing the role of Kasparov. Not to mention the debate on supremacy terminology that has once again generated a lot of fuss, as reported in Academics derided for claiming 'quantum supremacy' is a racist and colonialist term by Sarah Knapton, December 2019.

On top of that, randomized benchmarking used in Google's experiment is an approach that is not unanimously accepted to establish the superiority of quantum computing over classical computing[764].

| Metric | Value | Unit | Comments |
|---|---|---|---|
| Number of qubits | 53 | qubits | Computing qubits |
| Couplers | 86 | couplers | Qubits used for coupling computing qubits |
| Single qubits gates | 1,113 | gates | Number of single qubit gates executed in benchmark |
| Single qubits gates duration | 25 | nano-seconds | Duration of a single qubit gate |
| Single qubit error | 0,16% | percent | |
| Two qubits gates | 430 | gates | Number of two qubits gates executed in benchmark |
| Two qubits gates duration | 12 | nano-seconds | Duration of a two qubit gate |
| Two qubits gates error | 0,93% | percent | |
| Readout error | 3,80% | percent | |
| Gates depth | 20 | cycles | Number of series of quantum gates executed. |
| Gates per cycle | 55,65 | gates/cycle | Number of quantum gates executed per cycle |
| Measured fidelity | 0,20% | percent | Total fidelity of system in supremacy regime |
| Number of iterations | 3,000,000 | iterations | Number of full algorithm executions |
| Computing time | 6,000 | seconds | Total computing time |
| Quantum computing time | 30 | seconds | Total quantum computing time |
| Readout analog to digital convertors | 277 | number | Generating 8 bits at 1 GB/s |

*Figure 313: all the figures of merit of Sycamore processor in 2019. Sources: Quantum supremacy using a programmable superconducting processor by Frank Arute, John Martinis et al, October 2019 (12 pages) and Supplementary information for "Quantum supremacy using a programmable superconducting processor" by Frank Arute, John Martinis et al, October 2019 (58 pages).*

**Qubits couplers**. Sycamore uses controllable qubit couplers, a technique pioneered by William D. Oliver's research team at the MIT Lincoln Labs[765]. There are 86 of them in all, connecting the 53 qubits of the chipset. This makes a total of 139 qubits. These couplers are in fact qubits whose frequency is controlled by a direct current line (DC). It allows the implementation of fast two qubits quantum gates, acting in an average 12 ns. They implement CZ and CPHASE two-qubit gates. Sycamore's processor has qubit readout error ranging from 3% to 7% and two-qubit gates have an error rate ranging from 0.5% to 1.5% while single-qubit gates have error between 0.05% and 0.5%.

**Machine Learning based calibration**. These qubits and couplers are controlled with microwaves carried by coaxial cables, at frequencies between 5 and 7 GHz, adjusted by a DC flux line. Google developed a deep learning-based qubit calibration code, which has made it possible to refine the qubit microwaves activation frequencies to avoid crosstalk between neighboring qubits.

**Isolation**: the qubit chipset is protected by some Mu-metal shielding, another one in aluminum and a black coating to absorb infrared photons. The processor is made of aluminum and indium on silicon and includes two dies stacked on top of each other or next to the other.

[764] See Lecture 3: Boson sampling by Fabio Sciarrino (63 slides) and An introduction to boson-sampling by Bryan Gard, Jonathan P. Dowling et al, 2014 (13 pages). See the review Quantum computers: amazing progress (Google & IBM), and extraordinary but probably false supremacy claims (Google) by Gil Kalai, September 2019 as well as The Quest for Quantum Computational Supremacy by Scott Aaronson, September 2019, which was published three weeks before Google's announcement but was still valid (16 pages).

[765] See Tunable Coupling Scheme for Implementing High-Fidelity Two-Qubit Gates by Fei Yan, William D. Oliver et al, MIT Lincoln Labs, PRX, 2018 (10 pages).

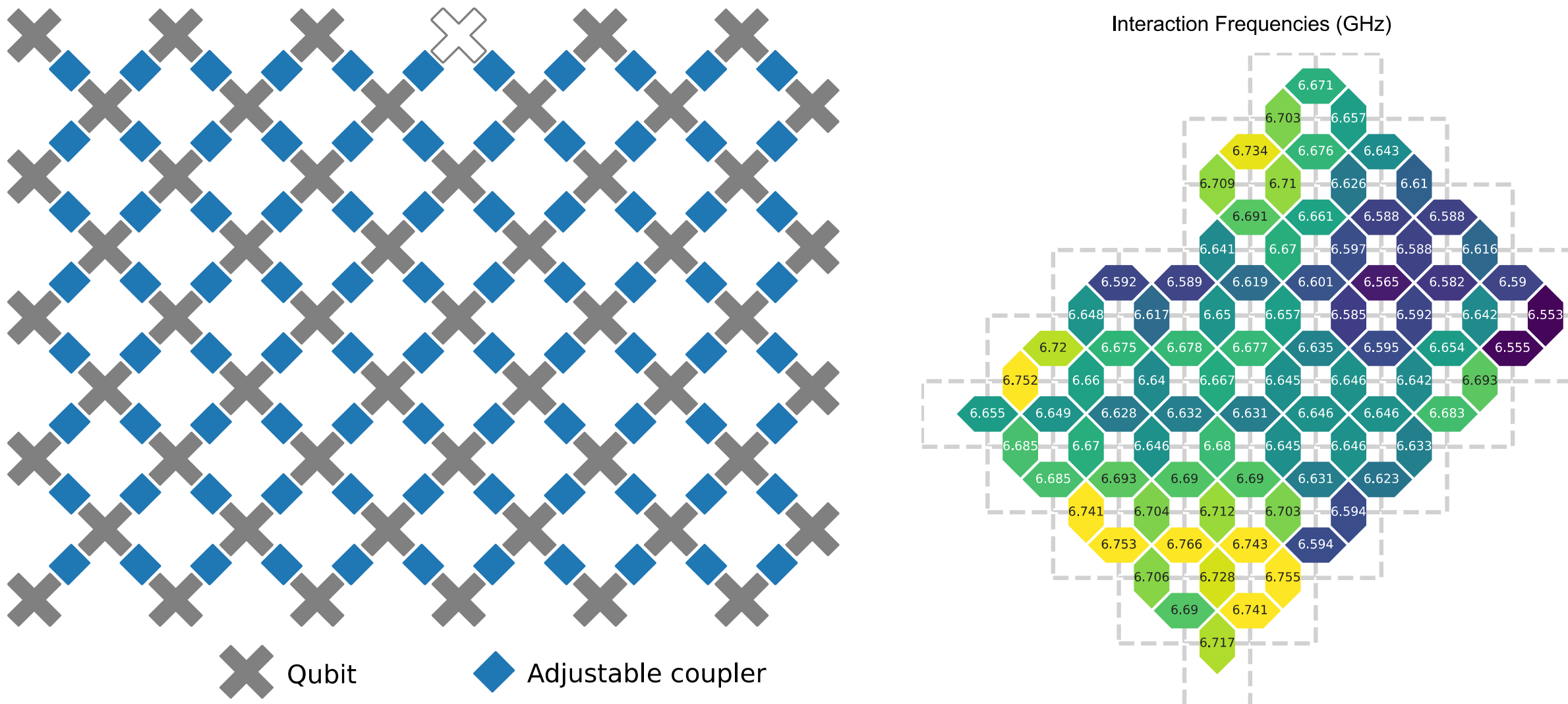

*Figure 314: Google's Sycamore qubits layout, with their data qubits and coupler qubits (in blue). On the right, the interaction frequencies with each qubit which were calibrated and optimized using a machine learning solution. Source: Sycamore's papers.*

**Microwave generation**. Below is a set of schematics of the control electronics inside and outside the cryostat. The system uses 54 external microwave signal generators for the single-qubit gates (X and Y), 54 for qubit frequency control and 86 for qubit control. This is completed by 9 readout microwave control signals, meaning they are frequency domain multiplexing readout by chunks of 6 qubits thanks to their use of a wideband parametric amplifier at the 15 mK stage, what they call an IMPA. The control electronics package includes 277 digital-to-analog converters that occupy 14 6U rack-mount cases. There is a similar number of coaxial cables ending in the cryostat.

**Z gates**: DC flux lines are also used to create Z gates, or phase gates. They are controlled with microwaves in IBM's superconducting qubits. Using DC flux lines is reducing the phase error observed with these gates.

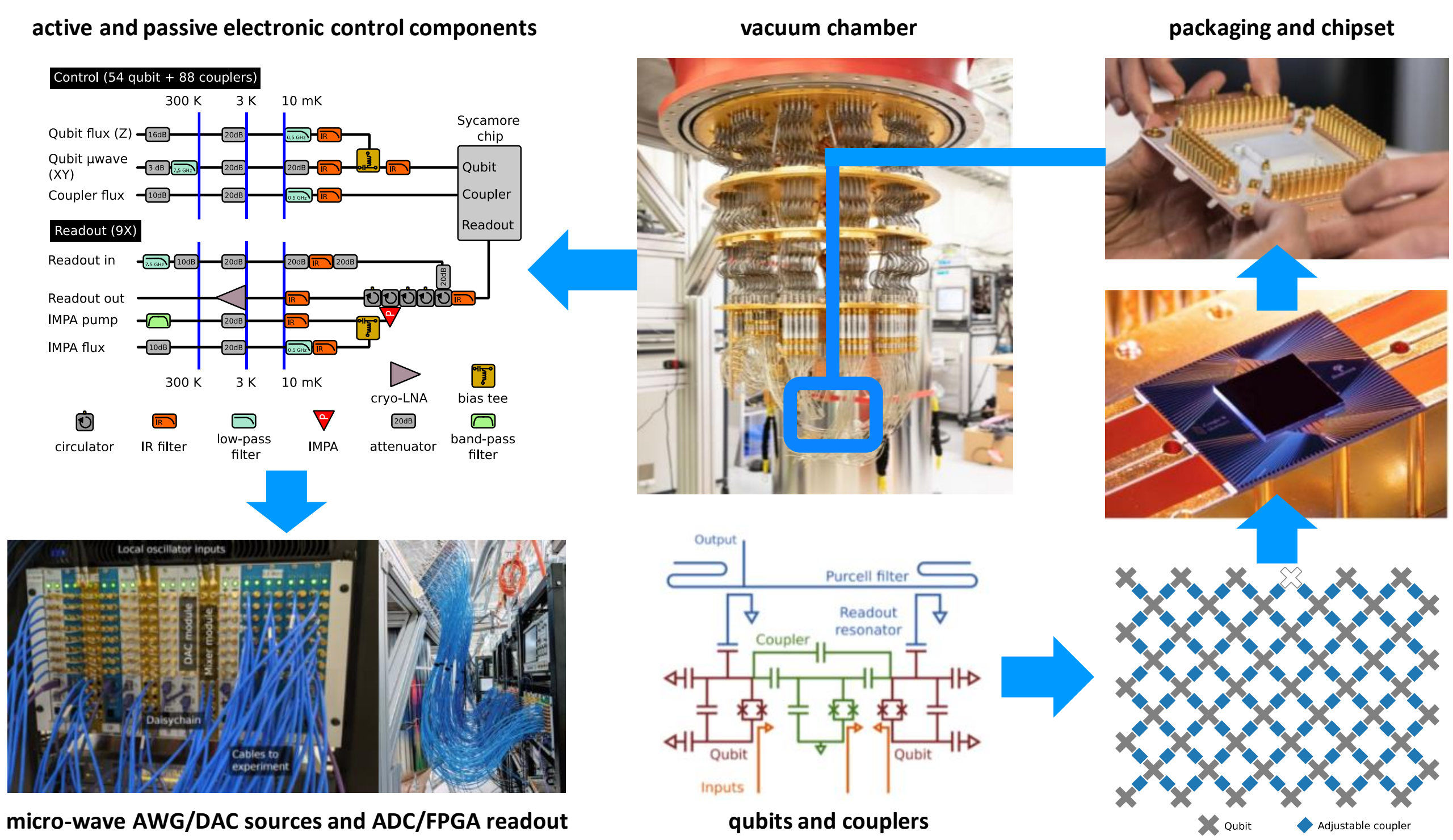

*Figure 315: a Russian doll description of Sycamore starting with the qubits and coupler, then with the chipset layout, its size, its packaging and connectors, where it is placed in the cryostat and the surrounding control electronics. Source: Google. Compilation (cc) Olivier Ezratty, 2020-2022 with sources from Google.*

**Qubit readouts** is done with only a few microwave photons sent to the qubits. The result is amplified by 100 dB in several steps, one at the 15 mK processor stage and another at the 3K stage. The resulting amplified microwaves are converted digitally by an ADC (analog-to-digital converter) and analyzed by a FPGA to detect their phase. The system multiplexes in the frequency domain the readout microwaves of 6 qubits groups conveyed by a single cable, between 5 and 7 GHz.

**Energetic advantage**: Sycamore showcases some energy consumption advantage, with a ratio of about one to a million. Its power consumption is about 25 kW and the ORNL IBM Summit is at 12 MW at full charge, and the computing time ratio is 2.5 minutes vs. 2.5 days (1/1440) in the most favorable IBM Summit case. But we are probably comparing apples and oranges given the supremacy doesn't relate to solving some useful problem with input data and parameters.

Between 2019 and mid-2022, Google had not yet released a new generation processor. Finally, they quietly disclosed that they had created a 72-qubit version of Sycamore in a paper related to surface codes, published in July 2022, as shown in Figure 316. The next step will be a 100+ qubits QPU to implement a distance-7 surface code logical qubit.

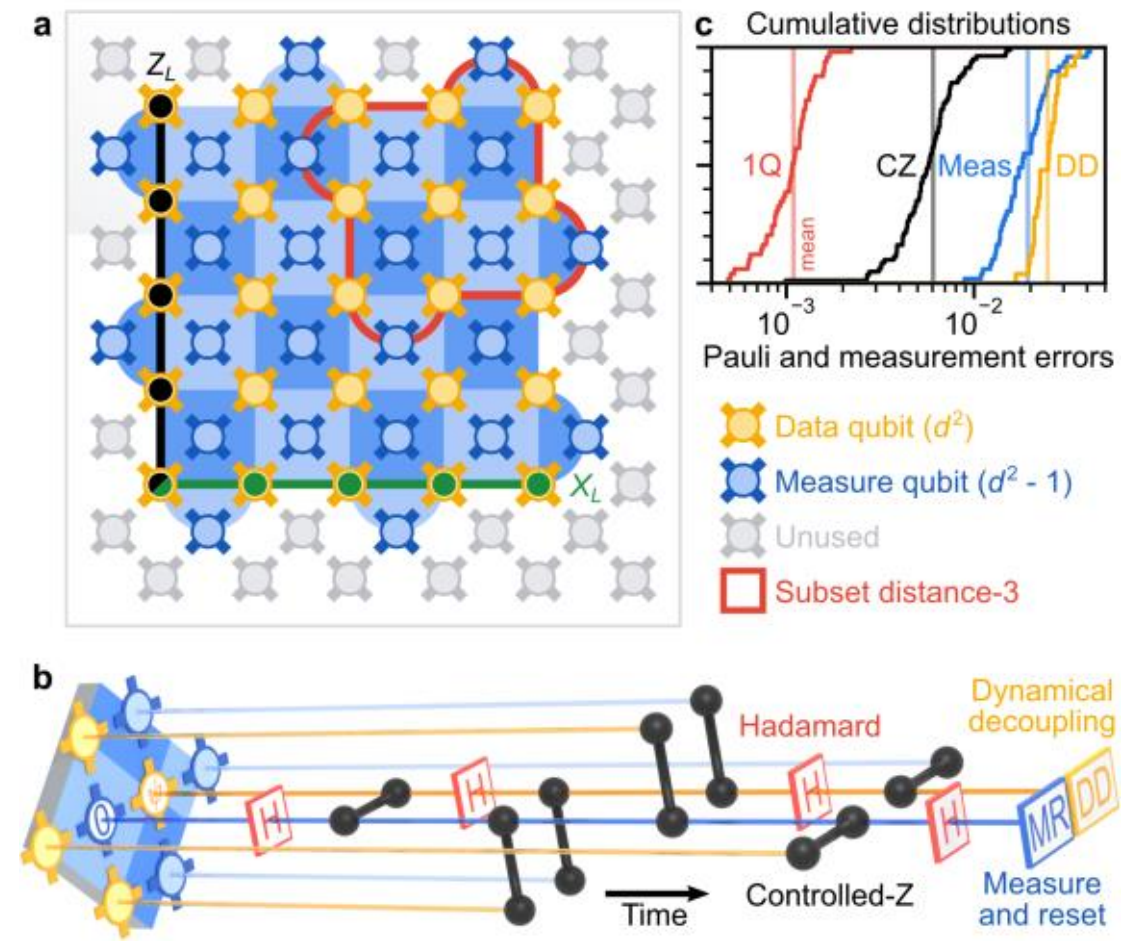

*Figure 316: Sycamore's 72 qubit version that implements a distance-5 surface code error correction for a single logical qubit, that is still insufficient to improve qubit fidelities. Source: [Suppressing quantum errors by scaling a surface code logical qubit](#) by Rajeev Acharya et al, Google AI, July 2022 (44 pages).*

It wants to reduce qubits error rates and scale up to a hundred logical qubits[766]. In a July 2020 conference, Hartmut Neven announced his 10-year plan to achieve this result, showing impressive mockups of a giant quantum computer containing 100 modules with 10 000 physical qubits each. It would be a giant installation, as shown on the impressive artist's rendering in Figure 317.

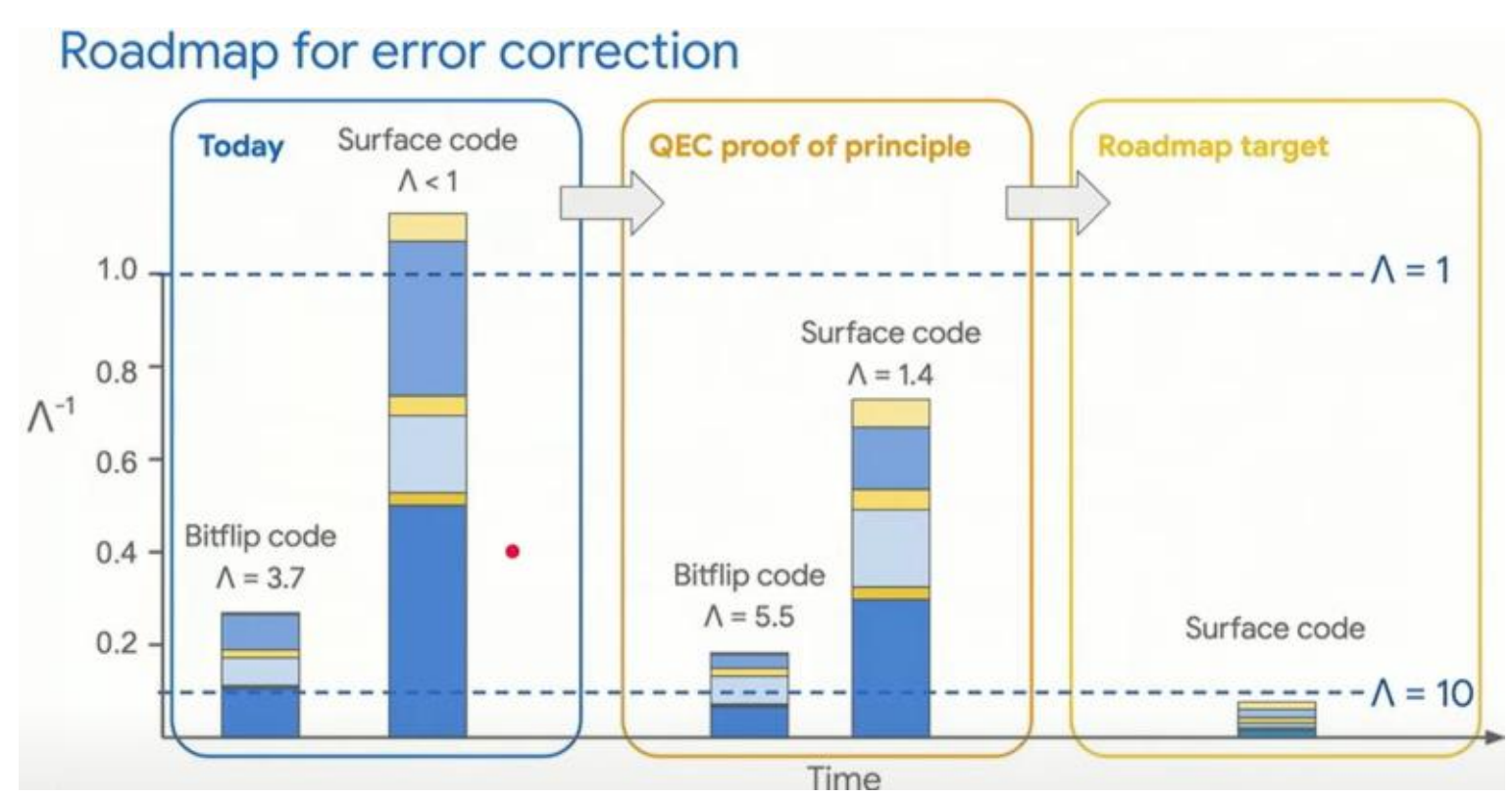

*Figure 317: Google's roadmap for error corrections. Source: Hartmut Neven, July 2020.*

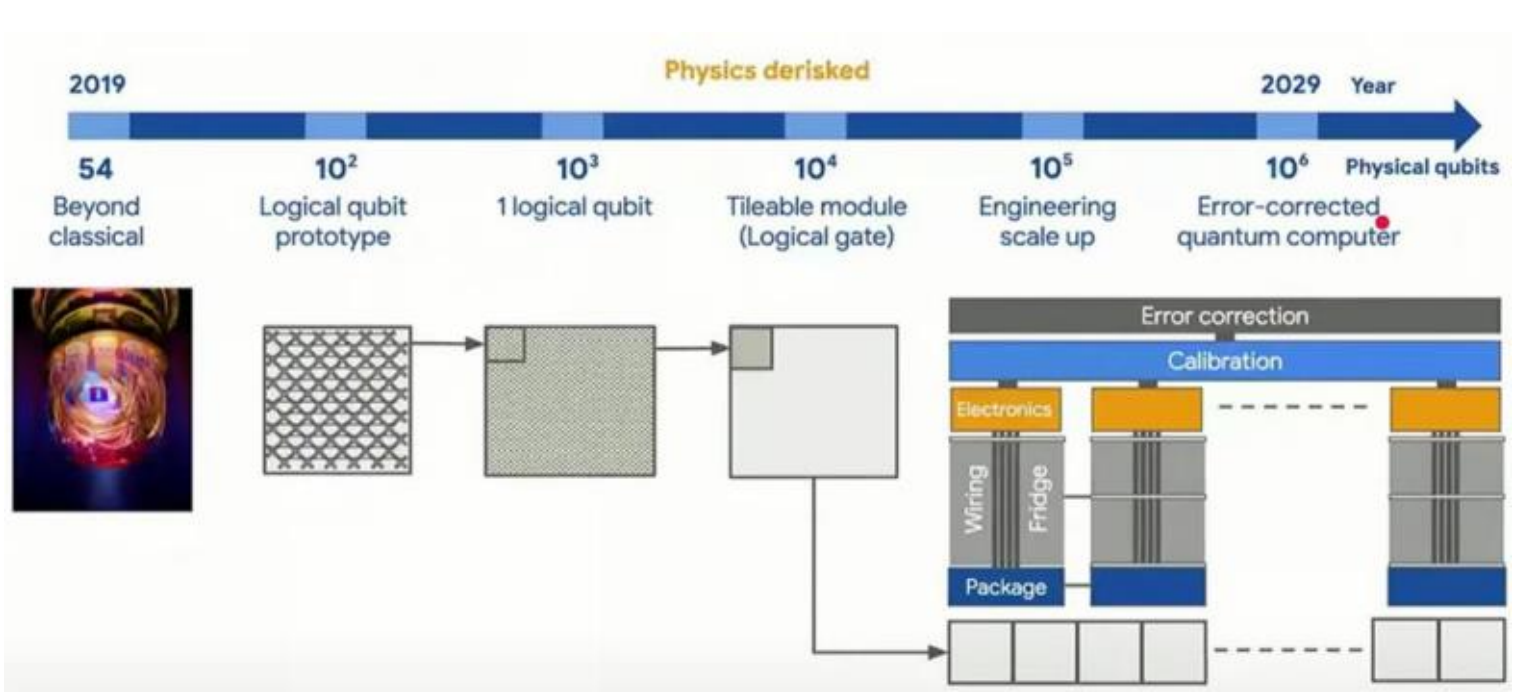

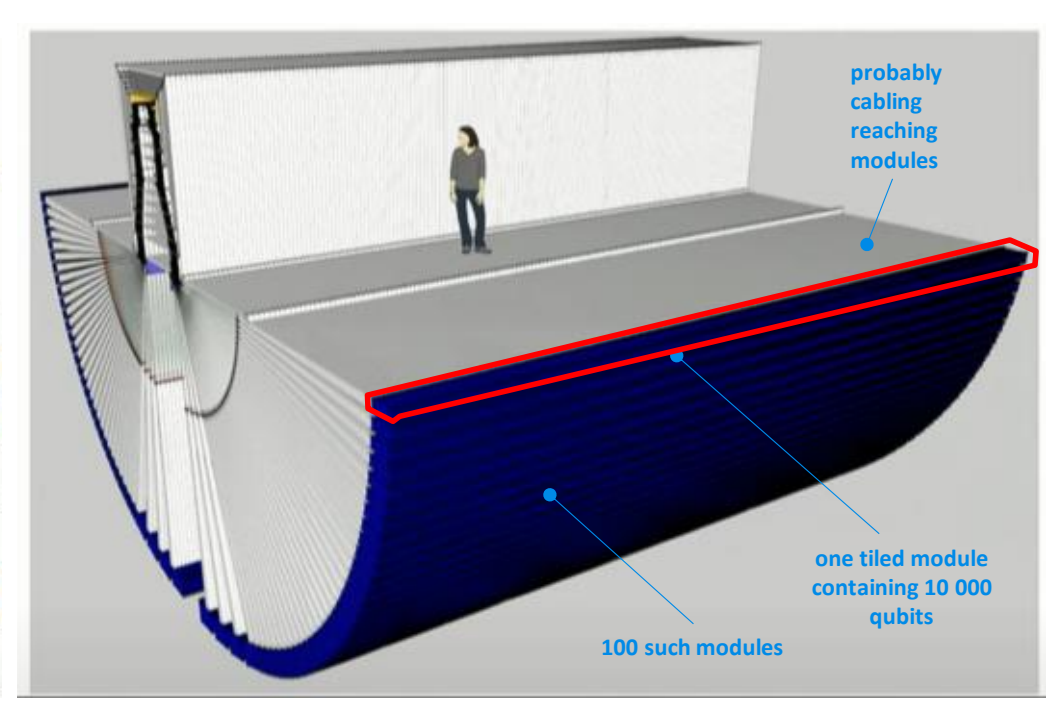

*Figure 318: Google's scalability roadmap with logical qubits made of 1000 physical qubits. And a giant system, as envisioned in 2020. Things may have changed since then. Source: Hartmut Neven, July 2020.*

[766] Source of the illustrations shown in Figure 317 and Figure 318: [Day 1 opening keynote by Hartmut Neven (Quantum Summer Symposium 2020)](#), July 2020 (30 mn) and the [whole symposium](#).

**Cryo-CMOS**. To scale up micro-waves generation and put it inside the cryostat, Google published some work on a cryo-CMOS chipset operating at 3K, using simple waveform generators consuming a minimum of energy to create only single qubit gates[767].

It has however not yet been deployed. It relies on cosinusoidal shape microwaves with the interest of creating spectral "holes" corresponding to the qubits frequencies harmonics of the state |1⟩ to the state |2⟩ transition, that must be avoided. It corresponds to the wavelength known as $\omega_{12}$ as seen in the illustration from page 299.

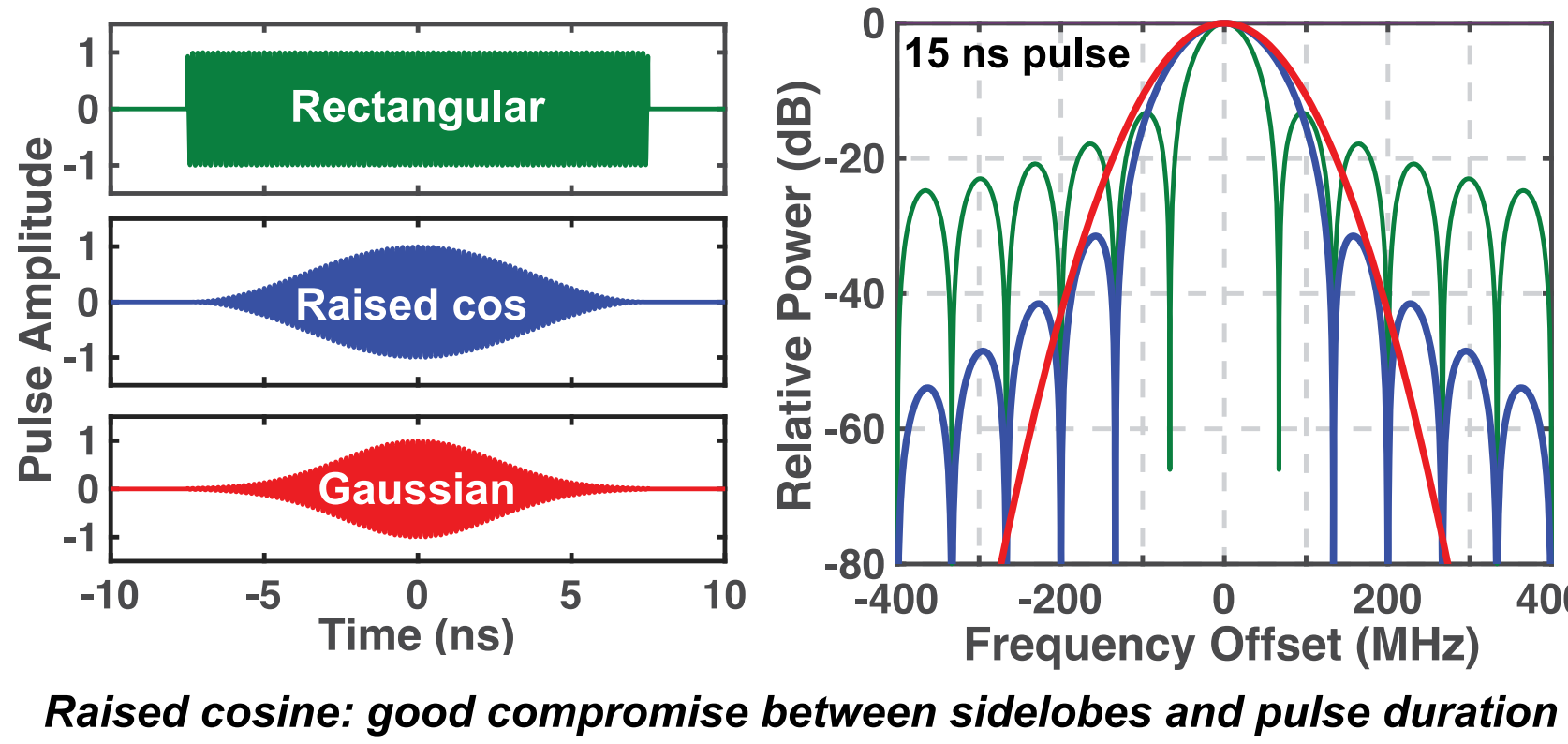

Figure 319: qubit control signals optimization with spectral holes matching qubit frequencies harmonics. Source: XY Controls of Transmon Qubits by Joseph Bardin, June 2019 (36 slides).

**Qubits improvements**. Several physical improvements are tested with their qubits. They reset qubits with high fidelity, allowing us to reuse qubits in quantum computations. They also test mid-circuit measurement that makes it possible to keep track of computation within quantum circuits. They also work on addressing cosmic radiation originated noise in circuits[768].

**Error correction codes**. At the APS March Meeting 2022 in Chicago, Kevin Satzinger provided an update on Google's ploy with quantum error correction using surface codes. Their mid-term goal is to create a "logical qubit" prototype with 100 physical qubits, then extend it to 1000 physical qubits[769]. The end goal is to build logical qubits with error rates around $10^{-12}$, a level that is required to execute many useful gate-based algorithms. With surface codes, these logical qubits are organized in squared arrays of about $d^2$ physical qubits (in blue, green ones are the qubit couplers) where d is the so-called code distance (5 in the example below).

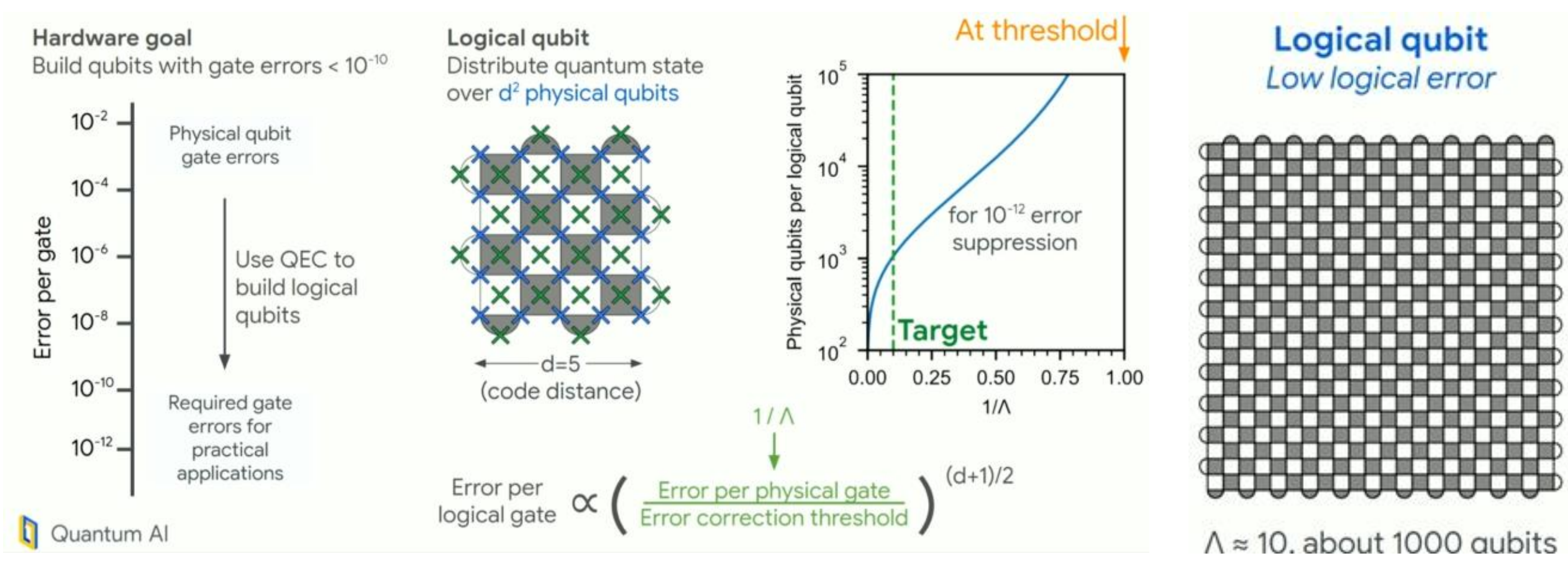

Figure 320: how Google plans to reach an error rate of $10^{-12}$ with its logical qubits. Source: APS March Meeting: Google, Intel and Others Highlight Quantum Progress Points by John Russell, HPCwire, March 2022.

767 See Control of transmon qubits using a cryogenic CMOS integrated circuit by Joseph Bardin, March 2020 (35 minutes) and A 28nm Bulk-CMOS 4-to-8GHz <2mW Cryogenic Pulse Modulator for Scalable Quantum Computing by Joseph Bardin, Craig Gidney, Charles Neil, Hartmut Neven, John Martinis et al, February 2019 (13 pages).

768 See Resolving catastrophic error bursts from cosmic rays in large arrays of superconducting qubits by Matt McEwen et al, April 2021 (13 pages).

769 See APS March Meeting: Google, Intel and Others Highlight Quantum Progress Points by John Russell, HPCwire, March 2022.

Kevin Satzinger explained how Google came out with the 1000 physical qubit per logical qubit number. It comes from the way logical gates error are evaluated, per the formula 1/Λ, Λ being the ratio between the threshold (theorem) error level and the physical gate error level. Λ must be at 10 to reach a $10^{-12}$ error rate with 1000 physical qubits per logical qubits (above, right). They first plan to implement d=3 and then d=5 surface codes and make sure the latter is better than the former with regards to logical gate errors[770].

In 2021, Google experimented two repetition code layouts on Sycamore with 21 qubits in a 1D chain correcting flip or phase errors and a distance-2 surface code of 7 qubits correcting both flip and phase errors. It did show that flip and phase errors could be exponentially suppressed with adding more physical qubits[771]. They did show a 100-factor improvement in error suppression as the size of their logical qubits was growing from 5 to 21 physical qubits. They could assess the impact of a good Λ value.

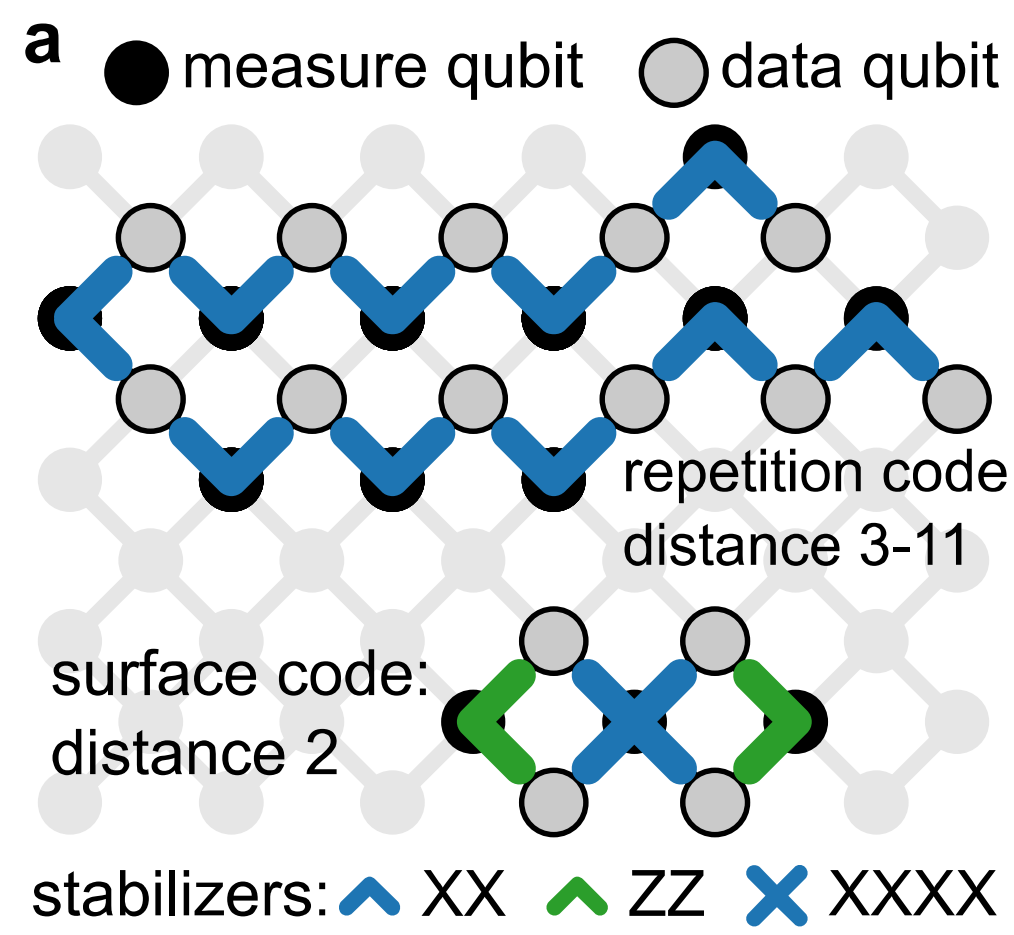

*Figure 321: the first logical qubits created on Sycamore in 2021. Source: Exponential suppression of bit or phase flip errors with repetitive error correction by Zijun Chen et al, Nature, Google AI, July 2021 (32 pages).*

The next step being to expand these tests on a yet-to-deliver successor of Sycamore and a larger number of qubits and the capacity to implement larger surface codes with d≥3. They also propose to use "pulse sequence" to correct unwanted crosstalk and dephasing that are disturbing surface codes[772].

They still have a huge challenge, like with IBM, to grow the number of physical qubits without degrading their fidelities.

**Sycamore at work**. Since 2020, Google tried to make some good use of Sycamore to test various algorithms. They "*simulated simple models of chemical bonding, high-temperature superconductivity, nanowires, and even exotic phases of matter such as time crystals*". They also solved some small combinatorial problems and managed a chemical simulation of a molecule of four atoms. In most cases, there was no more touted supremacy. This time, without mentioning the notion of supremacy[773]! Which makes sense given they didn't use more than 15 qubits to solve these small-scale problems. Other same scale algorithms were published in 2020[774]. Like with IBM's QPUs, the larger the number of qubits that are used in algorithms, the worse it is with their noise and computing depth. They end-up with using around 15 to 20 qubits for useful purposes. Of course, with that number of qubits, we're very far from any quantum advantage or supremacy.

---

[770] See also Progressing superconducting quantum computing at Google by Kevin Satzinger, April 2022 (47 mn).

[771] See Demonstrating the Fundamentals of Quantum Error Correction by Jimmy Chen et al, August 2021 and Exponential suppression of bit or phase flip errors with repetitive error correction by Zijun Chen et al, Nature, July 2021 (32 pages). Removing leakage-induced correlated errors in superconducting quantum error correction by M. McEwen et al, March 2021, Nature Communications (12 pages) deals with another error reduction technique named "multi-level reset" that consists in "pumping" the excess energy from superconducting qubits that leak to their higher energy levels |2⟩ or |3⟩.

[772] See Pulse sequence design for crosstalk mitigation by Murphy Yuezhen Niu.

[773] See Quantum Approximate Optimization of Non-Planar Graph Problems on a Planar Superconducting Processor by Google AI Quantum and Collaborators, April 2020 (17 pages) which deals with three families of combinatorial problems with the QAOA algorithm and Hartree-Fock on a superconducting qubit quantum computer by Google AI Quantum and Collaborators, April 2020 (27 pages) with a diimide $((NH)^2$ molecular simulation algorithm.

[774] See this theoretical paper on the use of quantum computing, not necessarily with Google qubits, to study black holes. See Google Scientists Are Using Computers to Study Wormholes by Ryan F. Mandelbaum, November 2019 which refers to Quantum Gravity in the Lab: Teleportation by Size and Traversable Wormholes by Adam R. Brown et al, November 2019 (20 pages).

In 2021[775], together with researchers from Columbia University, Google's teams created a chemical simulation classical/quantum hybrid algorithm using a Monte Carlo method[776]. It was used to compute the ground state of two carbon atoms in a diamond crystal, using 16 qubits. The method was however not more efficient than a full classical algorithm.

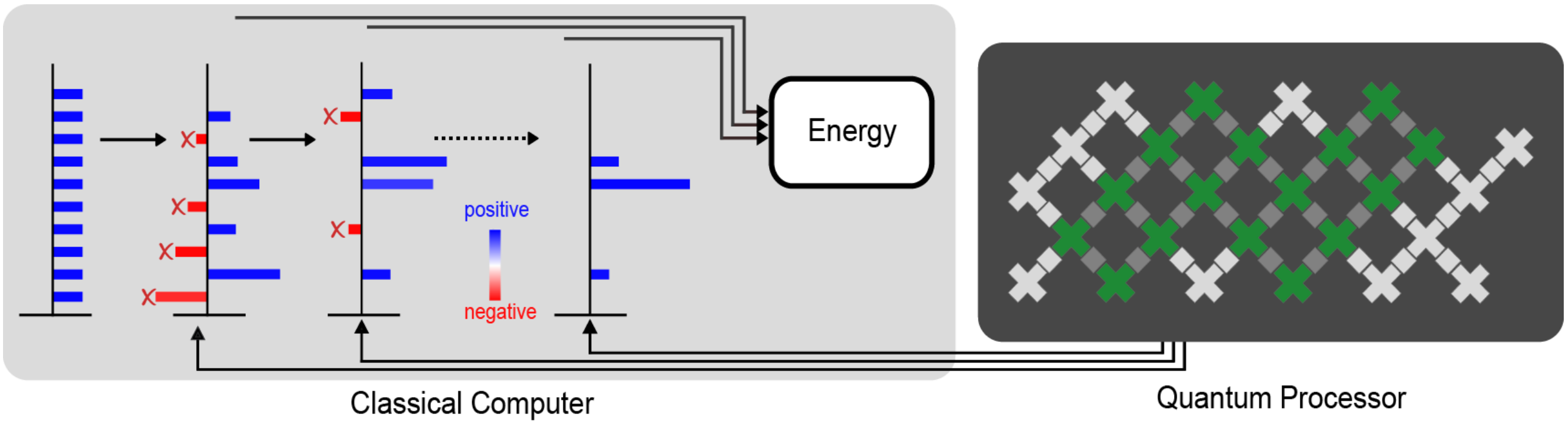

*Figure 322: simple schematic of a chemical simulation classical/quantum hybrid algorithm using a Monte Carlo method. Source: [Hybrid Quantum Algorithms for Quantum Monte Carlo](#) by William J. Huggins, March 2022.*

In 2022, another simulation of some molecules and materials as published, first with the iron-sulfur clusters of nitrogenase, including the FeMo-cofactor (aka FeMoCo), a component of the natural nitrogen cycle and second with the electronic structure of α-RuCl3, a candidate material for realizing spin liquid physics[777]. The experiments were using 5 to 11 qubits, far from any quantum advantage territory. All of this was running on Google's 53-qubit Weber processor, the second generation of Sycamore chipsets with slight performance improvements.

In 2021, Google also teamed up with **Caltech** to show some quantum superiority when quantum computers directly exploit quantum data coming from quantum sensors. Fewer experiments are required than if the communication between sensors and the quantum processor was classical. The experiment was done with 40 qubits and 1300 quantum operations[778].

**Google software tools**. Several Google teams are working on quantum software, including those working on Cirq, on TensorFlow Quantum and another Google X team working on applications, under the leadership of Jack Hidary[779]. They also released a **Fermionic Quantum Simulator** for quantum chemistry applications in collaboration with **QSimulate** (2018, USA) *aka* qsim. It can simulate noisy quantum circuits with Nvidia GPUs on Google Cloud. Google also published stim, an open source tool providing a 10000x speedup when simulating error correction circuits.

**Quantum cloud**. At last, it has a quantum cloud offering for quantum algorithm simulation and hosts an IonQ trapped ion system and, surprisingly, none of its Sycamore systems which are only accessible to a handful of academic partners, an outreach strategy that is very different from the broadscale one IBM adopted.

**Rigetti** (2013, USA, $656M) is another commercial superconductor vendor. With D-Wave, IonQ and PsiQuantum, it is the fourth best funded startup in the industry.

---

[775] See 2021 Year in Review: Google Quantum AI by Emily Mount, December 2021.

[776] See Hybrid Quantum Algorithms for Quantum Monte Carlo by William J. Huggins, March 2022 and Unbiasing Fermionic Quantum Monte Carlo with a Quantum Computer by William J. Huggins, Ryan Babbush, Joonho Lee et al, Nature, July 2021 (28 pages).

[777] See Simulating challenging correlated molecules and materials on the Sycamore quantum processor by Ruslan N. Tazhigulov et al, March 2022 (22 pages).

[778] See Quantum advantage in learning from experiments by Hsin-Yuan Huang et al, December 2021 (52 pages).

[779] See Alphabet Has a Second, Secretive Quantum Computing Team by Tom Simonite, January 2020. No secret anymore buddy!

It was launched by Chad Rigetti, who got a PhD and did a post-doc at Yale University on microwave driven two-superconducting qubit gates in 2009[780], and then worked as a researcher at IBM between 2010 and 2013.

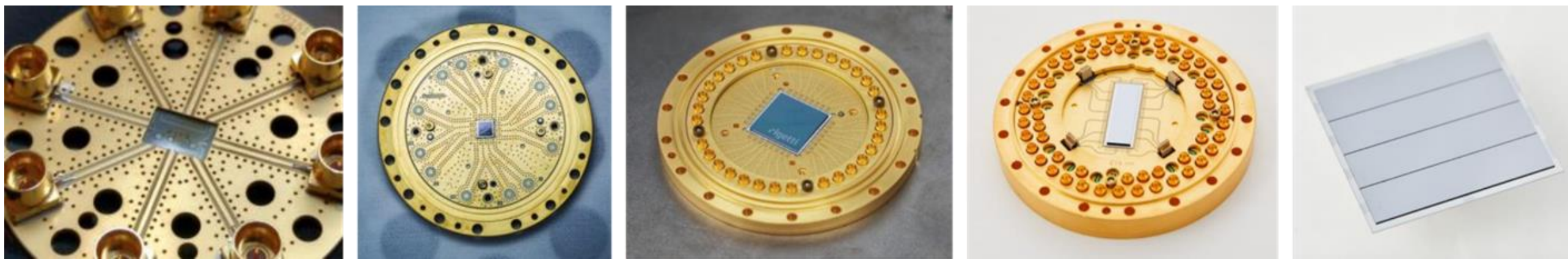

*Figure 323: evolution of Rigetti actual chipsets over time. Source: Rigetti investor presentation.*

Over about 8 years, their QPUs went from 3 to 80 qubits. They had a tendency to oversell their roadmap, having announced prematurely a 128 qubits test version in August 2018 that was never benchmarked or deployed. They started to deploy their system on Amazon Braket with its Aspen-9 processor of 31 qubits in 2020.

Their Aspen-M 80 qubits chipset was announced in December 2021 and commercially available in February 2022 on Rigetti Quantum Cloud Services and Amazon Braket, and subsequently on Azure Quantum, Strangeworks QC and Zapata Computing's Orquestra platform[781]. Their fidelities are not as good as with IBM and Google but it is continuously improving, even as they increase the number of qubits.

| metric | Aspen-9 | Aspen-11 | Aspen-M-1 |
|---|---|---|---|
| number of physical qubits | 31 | 40 | 80 |
| median $T_1$ | 27 µs | 25.7 µs | 30.7 µs |
| median $T_2$ | 19 µs | 14.9 µs | 23.0 µs |
| median simultaneous 1Q fidelity | 99,80% | 99.50% | 99.70% |
| median 2Q XY fidelity | 95,40 % | 93.70% | 95.30% |
| median 2Q CZ fidelity | 95,80 % | 90.20% | 93.10% |
| median RO fidelity | | 97.10% | 98.20% |
| median active reset fidelity | | 99.20% | 99.80% |

*Figure 324: Rigetti qubits figures of merit for their last generation chipset. These number are now fairly well detailed, but they show that it doesn't compete well with IBM at least on two qubit gates. Data source: Rigetti.*

In 2022, they demonstrated improved two-qubit gates fidelities of 99,5% but with a 9 qubits prototype[782]. Sideways, they are also experimenting the usage of qutrits with three level anharmonic oscillators in their superconducting loops. It was tested on a 5-transmon chipset with two qutrits entanglement[783].

So, what is special with Rigetti? Let's look at a couple aspects of their qubits and systems engineering.

**Coupling qubits**. Their flux qubits are entangled by dynamically configurable couplers, which reminds us of Google Sycamore. It brings more flexibility for managing two qubit gates[784]. These are adjustable transmon qubits using asymmetric SQUIDs (magnetometers).

---

[780] See Quantum Gates for Superconducting Qubits, 2009 (248 pages).

[781] See Rigetti Announces Commercial Availability of their 80 Qubit Aspen-M and a Teaming with NASDAQ to Explore Financial Applications of QC, February 2022.

[782] See Rigetti Computing Reports Fidelities as High as 99.5% on Next-Generation Chip Architecture, February 2022.

[783] See Beyond Qubits: Unlocking the Third State in Quantum Processors by Alex Hill, Rigetti, December 2021 and Quantum Information Scrambling on a Superconducting Qutrit Processor by M. S. Blok et al, April 2021 (21 pages).

[784] This is explained in Demonstration of Universal Parametric Entangling Gates on a Multi-Qubit Lattice by M. Reagor et al, 2018 (17 pages).

**Electronics optimization**. Rigetti made efforts to optimize the physical and electrical components of its accelerators. First, by integrating the control and measurement wiring of the qubits in compact sheets that they patented[785]. They developed their own microwave generation electronics. They also found a way to limit crosstalk between qubits[786]. They also work on merging microwave and DC flux lines into the same wires used for respectively XY and Z single qubit gates, between the 10 mK cold plate and the qubit chipset[787]. As a result, they tout performance improvements expressed in CLOPS (circuit layer operations per seconds) for their 40-qubit Aspen-11 and 80-qubit Aspen-M systems. The first had 844 CLOPS while the second reached 892. It can be compared to IBM's 1500 CLOPS on their 65 qubits system and 850 with 127 qubits (as of April 2022)[788].

**Modular chipsets**. Rigetti is now splitting qubits in multiple semiconductors dies connected with each other with indium-based flip-chip bonded on a single larger carrier die. This reduces qubits crosstalk between modules, at the expense of a smaller fidelity. They first tried this with 4 chips containing each 4 aluminum and niobium-based SQUIDs qubits and 4 tunable couplers, their fidelities are 99.1±0.5% and 98.3±0.3% for iSWAP and CZ gates[789]. They then expanded this to two 40 qubits dies in their Aspen M1 chipset released in December 2021.

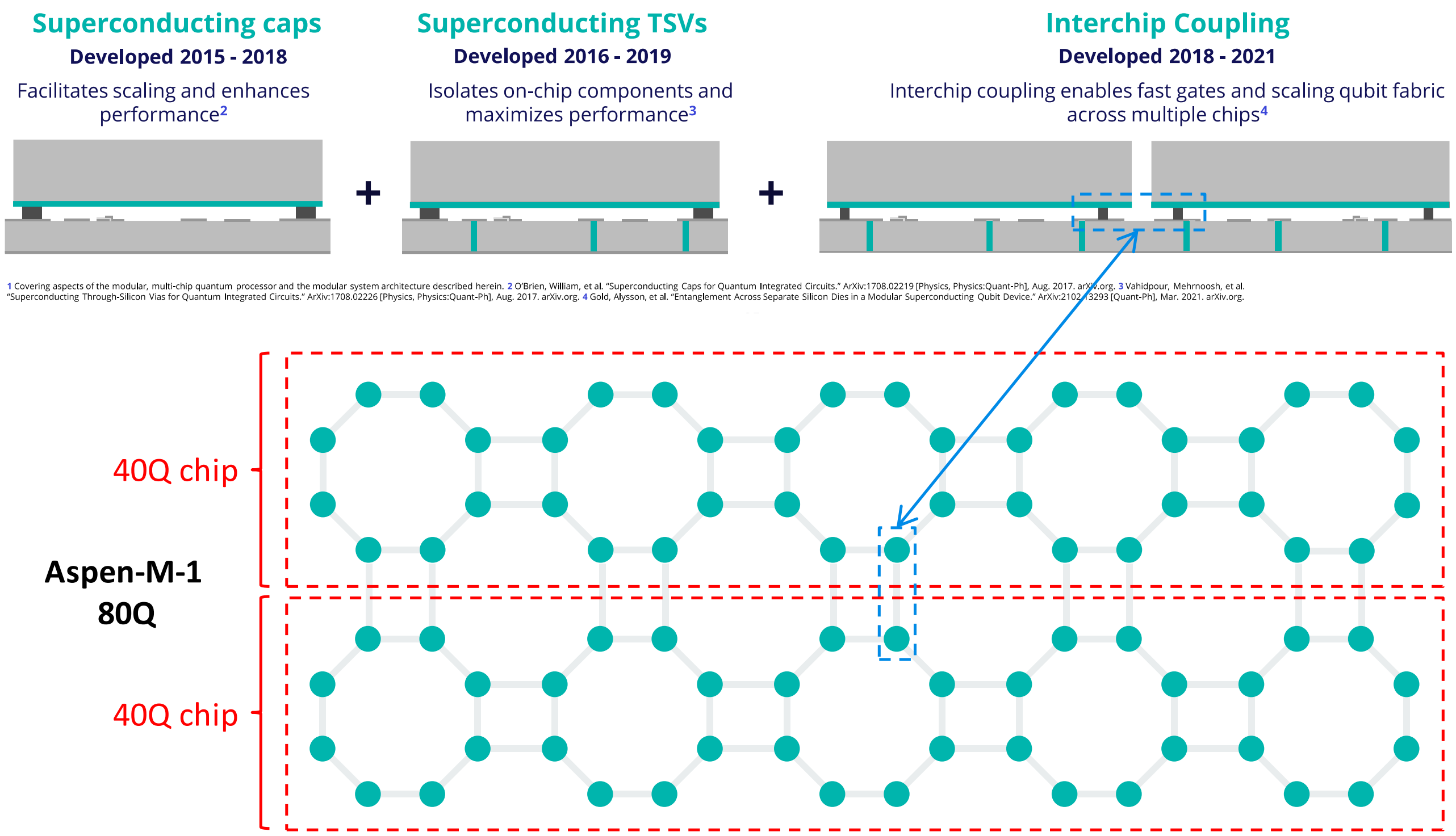

*Figure 325: interchip coupling implemented with their Aspen-M-1 80-qubit processor, assembling two dies of 40 qubits. Source: Rigetti.*

**Cleanroom**. Rigetti have their own small manufacturing unit producing their semiconducting chipsets, named Fab-1. This enables them to create new chipsets with a 5-15 month cycle. The required and initial investment of about $10M, which is reasonable even for a startup. The creation of superconducting qubit circuits is done with a very low-level of integration.

---

785 See Connecting Electrical Circuitry in a Quantum Computing System, USPTO 20190027800.

786 See Methods for Measuring Magnetic Flux Crosstalk Between Tunable Transmons by Deanna M. Abrams et al, August 2019 (12 pages).

787 See Full control of superconducting qubits with combined on-chip microwave and flux lines by Riccardo Manenti et al, July 2021 (8 pages).

788 See Optimizing full-stack throughput and fidelity with Rigetti's Aspen-M generation of quantum processors, Rigetti, February 2022.

789 See Entanglement Across Separate Silicon Dies in a Modular Superconducting Qubit Device by Alysson Gold, 2021 (9 pages).

We are far from the $20B 5 nm fabs from TSMC. In the case of silicon qubits, on the other hand, it is necessary to have an equipment of about $1B[790]!

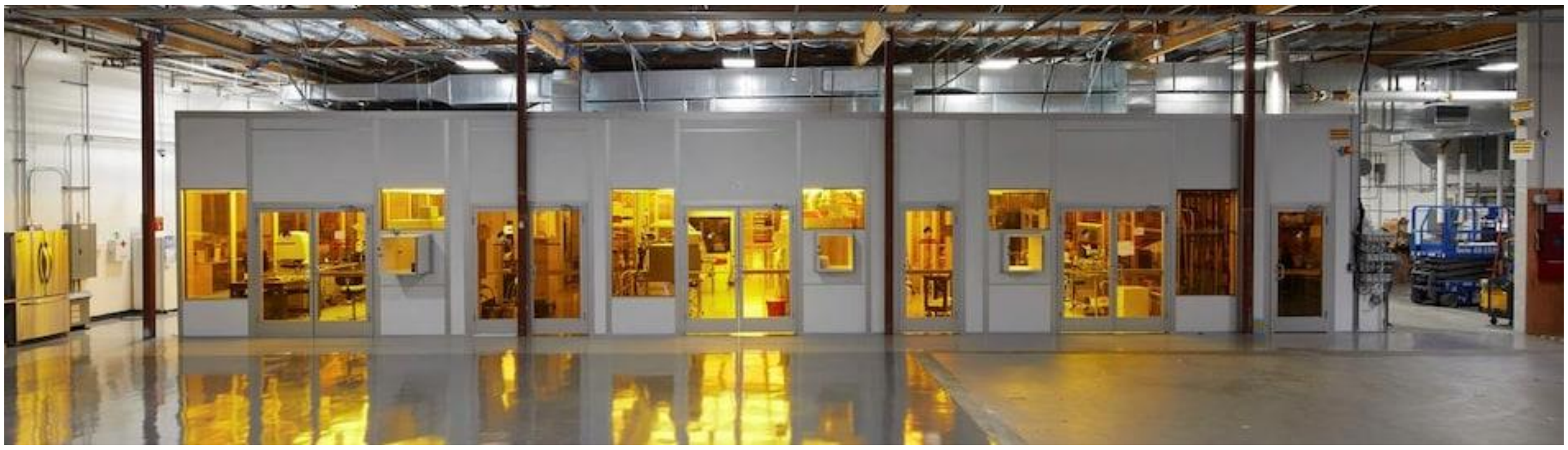

*Figure 326: Rigetti's superconducting cleanroom fab line in Fremont, California. Source: Rigetti.*

This multi-dies QPU approach is the main driving technology for them to scale their QPUs. The short term roadmap contains the 84-qbit Ankaa for 2023 and the 336-qubit Lyra for late 2023. Then, they plan to create 500 qubits chipsets by 2025 and assemble them in 1000, 4000 and larger QPUs. They have however not yet explained how they will interconnect these multi-dies chipsets[791].

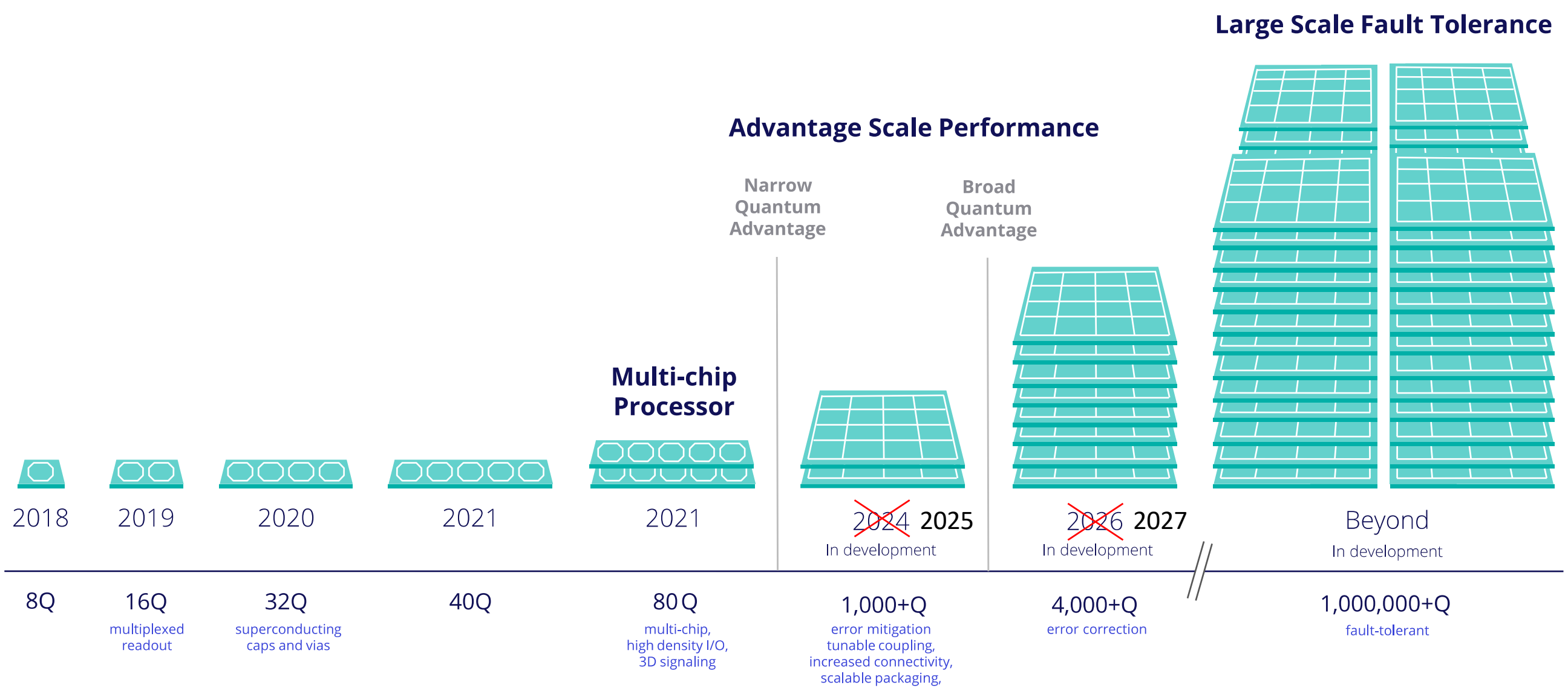

*Figure 327: Rigetti's scalability roadmap announced in October 2021. In May 2022, the company announced an additional one-year delay for their 1000 and 4000 qubits QPUs. They also expect to release 84 and 336 qubit chipsets in 2023. Source: Rigetti investor presentation, October 2021 and Rigetti Computing Reports First Quarter 2022 Financial Results and Provides Business Update, May 2022.*

**Full-stack software development**. It includes pyQuil for scripting and Quil for quantum gate management. These are both open source and published on Github. Quil allows to synchronize tasks between quantum and classical computing[792]. In 2018, they demonstrated the use of their quantum computer for a machine learning algorithm that does not require a hybrid algorithm[793].

---

790 See Quantum Cloud Computing Rigetti by Johannes Otterbach, 2018 (105 slides) and the corresponding video, and Manufacturing low dissipation superconducting quantum processors by Ani Nersisyan et al, Rigetti, 2019 (9 pages).

791 Source: Rigetti Investor Presentation, October 2021 (56 slides).

792 See A Practical Quantum Instruction Set Architecture by Robert S. Smith, Michael J. Curtis and William J. Zeng, Rigetti Computing, 2017 (15 pages).

793 In Quantum Kitchen Sinks: An algorithm for machine learning on near-term quantum computers, July 2018 (8 pages).

**Cloud**. Rigetti offers access to its quantum computers via the cloud, like IBM and D-Wave do with their Quantum Cloud Services. It started running in beta in January 2019. Since early 2020, they are also distributed in the cloud by Amazon in its Braket service.

**Acquisition**. Rigetti acquired QxBranch in July 2019 to complete its software offering. It was established in the USA, UK and especially in Australia. In September 2020, their UK-based subsidiary announced the launch of a collaborative project to accelerate the commercialization of quantum computers, funded with £10M private/public money. To do so, they will use a latest-generation Proteox cryostat from Oxford Instruments.

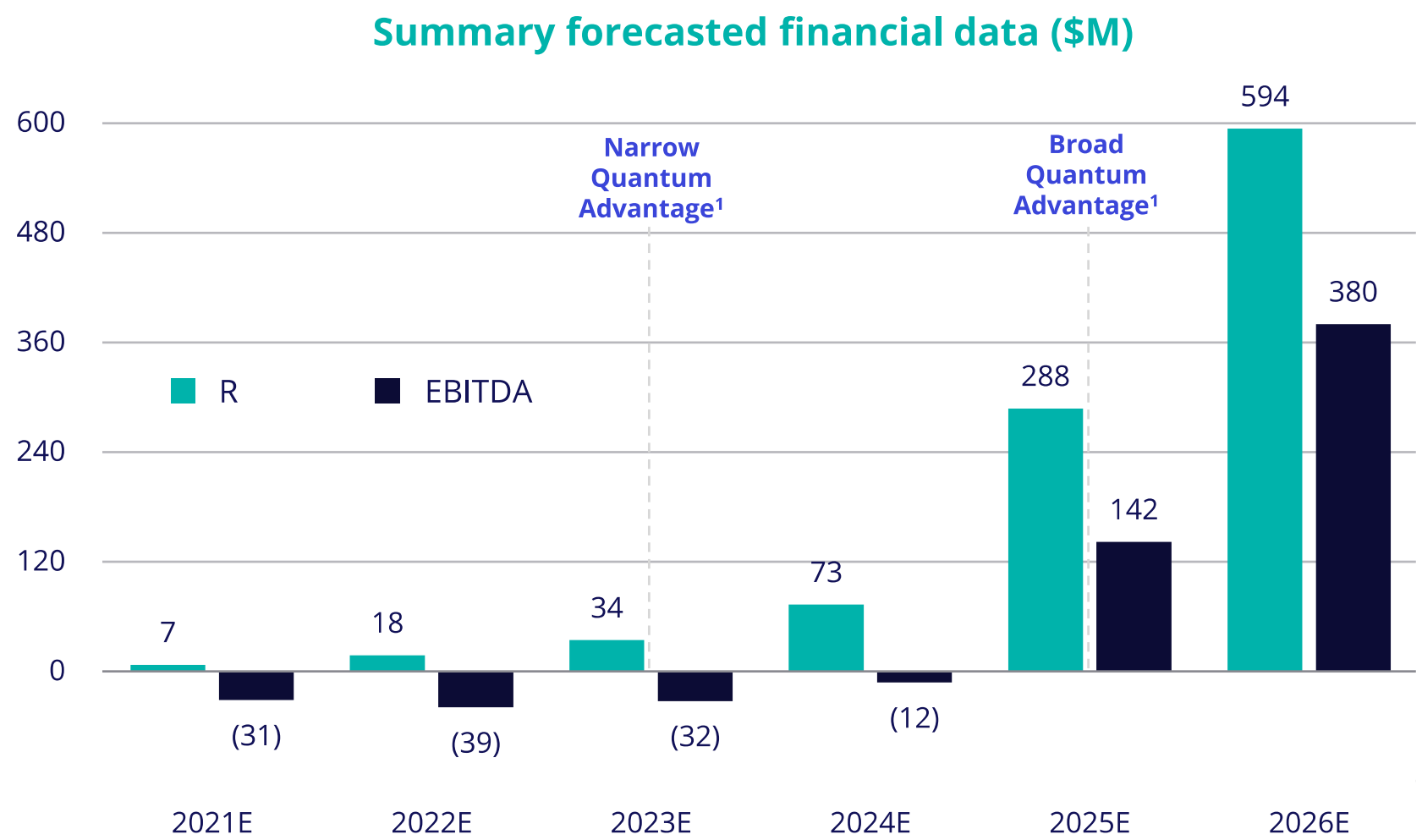

*Figure 328: Rigetti's revenue and EBITDA forecasts until 2026. In the first quarter of 2022, they made $2.1M. It seems their 2022 forecast was optimistic. Source: Q1 2022 quarterly report.*

**SPAC.** After having raised about $200M through classical VCs, they went into the stock market after being acquired by a SPAC company in March 2022, Supernova Partners Acquisition Company II. It brought on the table $345M of funding for a valuation of $1,152M. As of December 2021, the company had 140 people in the United States, UK and Australia. At their SPAC time, they were forecasting a revenue of $594M by 2026, based on the release of their 1000 qubits system in 2024.

In May 2022, they announced a one-year delay on their roadmap[794].

**Patents.** They have a portfolio of 100 patents and applications in interchip coupling and multi-die chipsets, cabling, processor design, cloud quantum computing and quantum software tools.

**Partnerships**. On top of **Amazon**, **Microsoft**, **Zapata Computing** and **Strangeworks** for cloud deployments, they announced in 2022 a new partnership with **Ampere Computing** (USA) to create hybrid quantum-classical computers designed to run machine learning applications, with **Keysight** for control electronics and **Bluefors** for cryogeny. Ampere is a fabless company designing 128-core arm chipsets for servers[795]. They have also various business and academic partnerships running in the UK, which led to the deployment in the UK of a 32-qubit Aspen system in June 2022 (why not the more recent 40 or 80 qubit Aspen?). With **Zapata**, they are building hybrid quantum-classical compilation tools with the support of the 80Q Aspen-M QPU and Rigetti cloud services. At last, they announced a partnership with **Riverlane** (UK) in June 2022 to work on error correction.

Rigetti is also working with DARPA in the USA, having been selected to provide hardware, software and benchmarks for phase two of the DARPA ONISQ program (Optimization with Noisy Intermediate-Scale Quantum). The aim is to create quantum computers able to solve complex optimization problems. This work is jointly done with Universities Space Research Association (USRA) and NASA's Quantum Artificial Intelligence Laboratory (QuAIL). Rigetti has also a strong R&D partnership with the DoE **Fermilab**, which conducts testing and material designs in its Superconducting Quantum Materials and Systems Center (SQMS) led by Anna Grassellino[796].

---

[794] See Rigetti Pushes Back Roadmap on Development of 1,000-Qubit, 4,000 Qubit Models by Matt Swayne, The Quantum Insider, May 2022.

[795] See Ampere Goes Quantum: Get Your Qubits in the Cloud by Ian Cutress, AnandTech, February 2022.

[796] See Superconducting Quantum Materials and Systems Center by Anna Grassellino, June 2021 (40 slides).

Customer wise, they have a couple early adopters like **Nasdaq** who plans to use their systems to detect fraud, optimize order matching and handle risk management.

**IQM** (2018, Finland, 167M€) is a spin-off from the Quantum Computing and Devices group of the Aalto University and from the VTT research center. Its funding is a mix of dilutive capital investment and debt financing through the European Investment Bank.

In June 2020, IQM received 15M€ capital funding from the European Commission's EIC Accelerator, supplemented by a 2.5M€ grant. A new funding round of 128M€ was announced in July. All-in-all, IQM raised 167M€ making it the best funded quantum startup in Europe.

They develop superconducting qubits QPUs after having initially created an on-chip refrigeration system technology for superconducting and silicon chipsets based on electron transfer using an electron tunnel-effect[797]. The company states that their qubits are operable with a faster clock speed than competing superconducting qubits thanks to optimizations applied to qubits reset, gates and readout. They use tunable couplers for qubits entanglement. They also developed a fast graphene-based bolometer for qubit readout able to detect a single microwave photon. Its benefits is some power saving compared to parametric amplifiers[798].

On the R&D stage, IQM participates to a research consortium including Aalto University and VTT which proposed in 2021 a qubit on-chip circuit to create microwaves pulses that could be used to drive superconducting qubits and working at 10 mK[799]. Its size is one mm and would remove the need to use cables to feed the processor with microwave pulses. So far, it only creates a sinus wave at 1 GHz, still far from what is needed to drive a superconducting qubit, i.e., a short duration pulse with a precise waveform added to some carrier frequency at around 5 GHz. It is closer to a local oscillator source! IQM also uses a TWPA from VTT in the first stage amplification of qubit readout microwaves[800]. This enables them to enable 5 to 10 qubits readout multiplexing with using different frequencies for qubits readouts with resonators of different lengths.

In March 2022, IQM introduced a new superconducting-qubit type nicknamed the unimon which has better fidelities, of about 99.9% with single-qubit gates. It is using a single Josephson junction in a resonator, combining high anharmonicity in the superconducting loop, different anharmonicities for each qubit, better insensitivity to low frequency charge noise and insensitivity to magnetic flux noise. As of 2022, they had tested three unimon qubits given they use tunable couplers to create two-qubit gates.

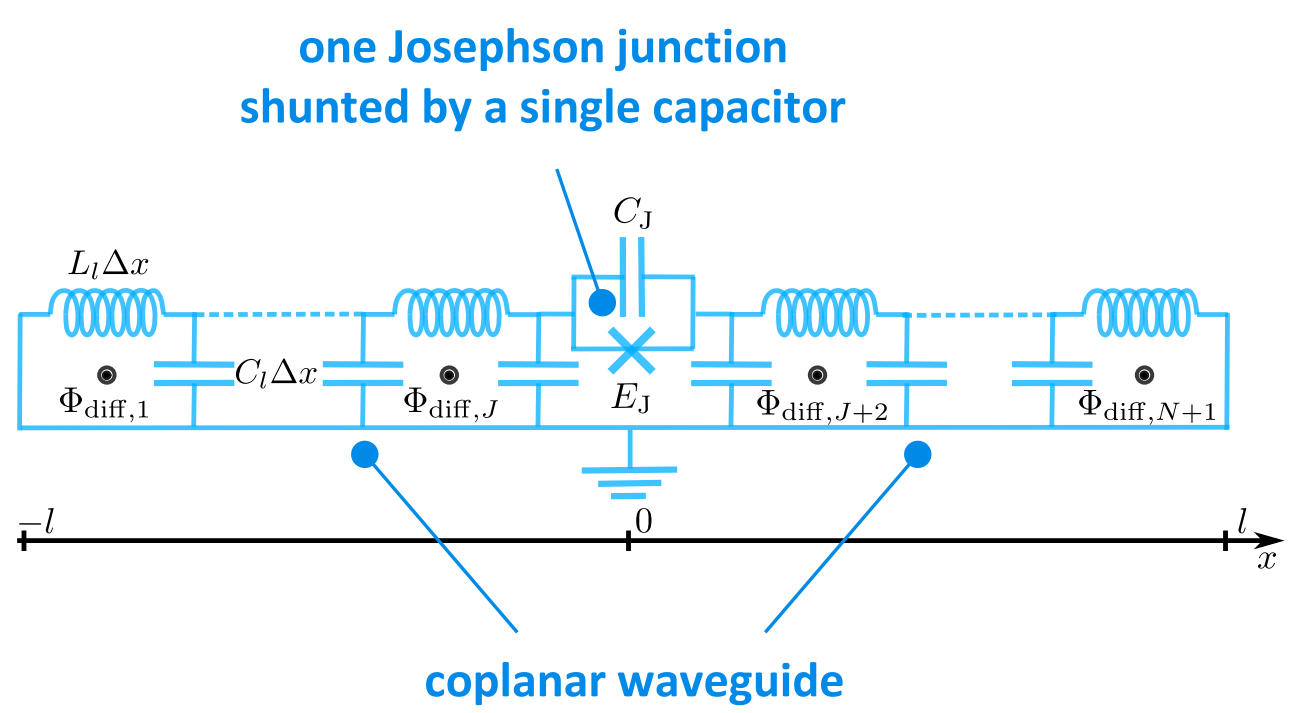

*Figure 329: IQM's unimon circuit layout. Source: Unimon qubit by Eric Hyyppä, Mikko Möttönen et al, IQM and VTT, April 2022 (72 pages).*

They obtained two-qubit CPHASE gates fidelity above 99% and fidelities of 99.9% for X and Y gates. All with fast 13 ns gate (good), readout probe pulses of 100 ns (fine) and a $T_1$ of 8,6 µs (not good).

[797] See Quantum-circuit refrigerator by Kuan Yen Tan et al, 2017 (8 pages) and video.

[798] See Bolometer operating at the threshold for circuit quantum electrodynamics by R. Kokkoniemi, Mikko Möttönen et al, Nature, September 2020 (19 pages).

[799] See A low-noise on-chip coherent microwave source by Chengyu Yan et al, Nature Electronics, December 2021 (14 pages) and A new super-cooled microwave source boosts the scale-up of quantum computers, December 2021 that is clearly overselling this technology development. It also requires pulse shaping with other techniques (AWG, DAC).

[800] See Broadband continuous variable entanglement generation using Kerr-free Josephson metamaterial by Michael Perelshtein, Pertti Hakonen et al, March 2022 (15 pages).

It now must be tested with a large number of qubits and with two qubit gates[801]. In August 2022, they also published an arXiv preprint explaining how they will implement long-distance reliable two-qubit gates[802].

After relying on VTT Micronova 2600 $m^2$ clean-room fab, they inaugurated their own Espoo 560 $m^2$ and 20M€ fab in November 2021 to manufacture their chipsets, a self-sufficiency strategy also seen with Rigetti. In 2020, the Finland government granted VTT with a 20,7M€ funding to acquire an IQM system. It should reach 50-qubit by 2024. They had 5 operational qubits as of November 2021.

The company had over 190 people as of September 2022. They opened a research lab in Germany in March 2020, one office in Spain in 2021 and one in Paris in 2022.

IQM's business model is based on selling quantum computing system to research and supercomputing centers as well as proposing customized hybrid analog/digital "Co-Design QC" quantum processors. The latter could be classified as "quantum ASICs", based on superconducting qubits[803]. These systems are adapted to the execution of hybrid algorithms such as VQE (Variational Quantum Eigensolvers) and QAOA (Quantum Approximate Optimization Algorithm). IQM will also implement a digital-analog quantum processor together with other partners like Infineon at the LRZ supercomputing center in Garching, near Munich in Germany[804].

Their co-design offering is based on **KQCcircuits**, an open sourced software tool based on KLayout for qubit design, using the OASIS format for masks that is lighter. It enables a graphic-based creation of circuits elements and contains a library with SQUIDs, complex waveguides, coplanar capacitors, qubits, flipchip connectors, indium bumps and other templates. They are partnering since August 2022 with **Multiverse Computing** for the software implementation of these co-designed QPUs, relying on Multiverse's Singularity SDK. They also announced a partnership with **Atos** together with the Finnish supercomputing center **CSC** which bought a classical QLM emulator for their services. This machine is used both to simulate the operation of IQM's quantum accelerator qubits and to drive it[805]. CSC will provide scientific computing resources to the country's researchers, much like GENCI/TGCC/IDRISS do in France and JSC in Germany. Atos has also announced its interest to distribute an IQM quantum accelerator, among other market solutions, including the Pasqal simulator.

OQC

**Oxford Quantum Circuits** (2017, UK, $45M) was launched by Peter Leek from Clarendon Laboratory Oxford. The startup is run by Ilana Wisby and had a team of 60 people as of mid-2022. The company wants to remove the identified barriers that prevent superconducting qubits from scaling.

[801] See Unimon qubit by Eric Hyyppä, Mikko Möttönen et al, IQM and VTT, April 2022 (72 pages) which provides a good scientific and technical documentation of unimon qubits.

[802] See Long-distance transmon coupler with CZ gate fidelity above $99.8% by Fabian Marxer, Mikko Möttönen, Johannes Heinsoo et al, IQM, QCD Lab and VTT, August 2022 (24 pages).

[803] Their method is described in Approximating the Quantum Approximate Optimization Algorithm by David Headley et al, February 2020 (14 pages) and Improving the Performance of Deep Quantum Optimization Algorithms with Continuous Gate Sets by Nathan Lacroix, Alexandre Blais, Andreas Wallraff et al, May 2020 (14 pages).

[804] See New EU Consortium shaping the future of Quantum Computing, IQM, February 2021. In November 2021, IQM was officially selected to provide its quantum computer to LRZ (Leibniz Supercomputing Centre) in association with an HPC to set-up an hybrid computing system as part of the Q-Exa project. It's part of a €45.3M consortium project funded by BMBF (German Federal Ministry of Education and Research) with €40.1M. Although it was not detailed in the announcement, we can suspect that the provided QPU will have 50 qubits as planned by IQM in 2025.

[805] See Atos, CSC and IQM join forces to accelerate the commercialization of European quantum technologies, June 2020.

OQC's technology is based on their "coaxmon" superconducting qubits that are composed of highly coherent planar qubits[806] and using a 3D structure connecting the qubit chipset with an interposer and using a layer for controlling the qubits on top of the chipset and another one below for qubit readouts[807].

It's based on various works from MIT and the University of Oxford, on an idea from Peter Leek[808].

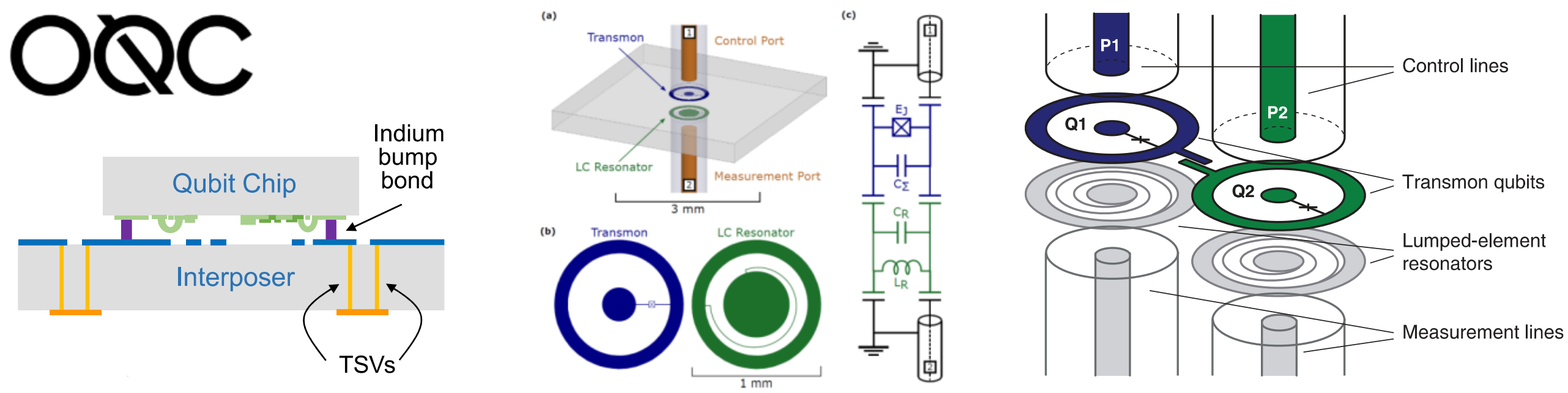

Solid-state qubits integrated with superconducting through-silicon vias by D. R. W. Yost et al, MIT, September 2020

Double-sided coaxial circuit QED with out-of-plane wiring by J. Rahamim et al, 2017

Calibration of a Cross-Resonance Two-Qubit Gate Between Directly Coupled Transmons by A.D. Patterson et al, 2019

*Figure 330: OQC coaxmon schematics showing how microwave controls are distributed vertically onto the qubits and their resonator. Source: OQC.*

They are partnering with Cambridge Quantum Computing (CQC) which is developing a quantum compiler dedicated to their qubits. In April 2020, OQC obtained collaborative project funding from the British government of £7M. As part of this project, they are associated with SeeQC UK, Oxford Instruments, Kelvin Nanotechnology, the University of Glasgow and the Royal Holloway University of London.

In July 2021, OQC announced that they were making their first system available only as a QCaaS solution, in private beta (quantum cloud as a service) without even saying how many qubits were deployed. They then announced in December 2021 that an 8-qubit version of their processor nick-named Lucy would be made available on Amazon Braket and revealed that their July 2021 system had a mere 4 qubits. The OQC system became live on Braket in February 2022.

**Intel** is another player in the superconducting qubits field. With no commercial solution so far as it's only a research field at this stage, completed by to the more natural avenue of electron spin silicon qubits they are also pursuing. At CES 2018, Intel's CEO proudly showcased a 49-qubit superconducting chipset during his keynote, stuck between a passenger drone demonstration and a broad talk on artificial intelligence.

Named Tangle Lake, the chipset was  tested at **Qutech** in the Netherlands. They were at 7 qubits at the end of 2016, 17 qubits at the end of 2017 and 49 (uncharacterized) qubits in January 2017. Since then, no news. It seems that Intel is now entirely focused on electron spin qubits, along with their partner Qutech in The Netherlands, where they invested $50M back in 2015.

ANYON
Computing with Manufactured Quanta™

**Anyon Systems** (2014, Canada) was created by Alireza Najafi-Yazdi. Their physics team is managed by Gabriel Éthier-Majcher and the startup has over 20 employees as of mid-2021.

---

806 See Surface acoustic wave resonators in the quantum regime, 2016 (40 slides).

807 The 3D layering and TSV structure is inspired from Solid-state qubits integrated with superconducting through-silicon vias by D. R. W. Yost et al, MIT, September 2020 (9 pages). This project was funded by IARPA.

808 See Double-sided coaxial circuit QED with out-of-plane wiring by J. Rahamim, Peter Leek et al, 2017 (4 pages), Calibration of a Cross-Resonance Two-Qubit Gate Between Directly Coupled Transmons by A.D. Patterson, Peter Leek et al, 2019 (8 pages) and Superconducting microwave circuits for quantum computing by Peter Leek, 2018 (42 slides).

They started with creating their Quantum Device Simulator (QDS), a software tool used in quantum computer design and simulation that can run on supercomputers. It was used by John Martinis' Google team in 2017 for their design of a superconducting 6- and then 20-qubit qubit processor[809]. Their software was mainly used to predict the level of adjacent qubits cross-talks. It's part of Snowflake, an open source library for creating quantum circuits running both on quantum emulators and quantum computers.

But their main goal then became to create gates-based superconducting qubits quantum computers in a full stack approach, creating their own control electronics and cryogenics systems. They are also implementing some sort of topological error correction codes. In December 2020, they announced that they would provide such a system to the Canadian Department of Defense using their Yukon processor. It was put online in 2021. They published qubit fidelities data in March 2022 with 99,7% for single qubit gates in parallel and 95,6% for two qubit gates[810]. Their $T_1$ is 10 µs and $T_2$ is 8 µs. Is that good? If it were with fewer than 20 qubits, it would be less than stellar, particularly compared with the recent IBM systems who have $T_1$ over 100 µs and 99,9% two-qubit gate fidelities with 27 qubits (Falcon R10 processor as of November 2021).

If it was achieved with over 50 qubits, that would be better! But there's no open way to have some clues on the number of their qubits. I'd guess it is fewer than 30 qubits. Otherwise, you'd hear about it. In June 2022, they announced the delivery of their quantum computer to Calcul Québec, a Canada's public supercomputing center hosting Narval, the 92th HPC in the world Top 500 and 28th in the Green 500 as of June 2022.

They also published an impressive 3D picture of their Qube computer which shows well that they use a cryostat, that is said to be homemade, at least for the dilution part, but with no more information on the of qubits that we thus suspect to be very low and under 10.

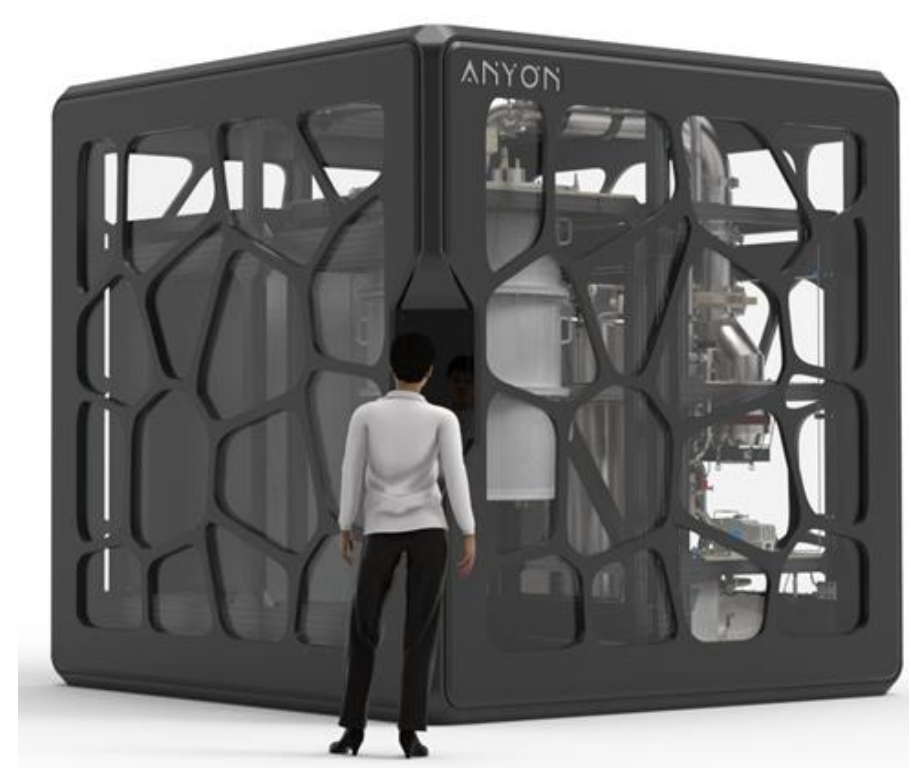

*Figure 331: artist rendering of Anyon's quantum computer, with all the traditional nuts and bolts of a superconducting quantum computer. Source: Anyon.*

**Bleximo** (2017, USA, $1.5M) was founded by Alexei Marchenkov and Richard Maydra, two former Rigetti employees.

If develops superconducting qubits processors tailored for specific needs and adapted to different markets including biotechs and financial services[811]. It seems similar to Finland's IQM strategy.

They focus on improving the classical control electronics driving their qubits and are partnering with Q-CTRL which develops error correction codes quantum software. As of 2022, they had created a 8-qubit processor which is less than stellar compared to IBM and Rigetti but in line with OQC from the UK[812]. Their qubits lifetime sits around 100 µs. They would need to assemble at least 40 to 50 functional physical qubits, if not 100, to enable real-life applications with their customers.

They also develop quantum software solutions (aka "QCO") for typical use cases like optimization, physics simulation and machine learning. In their team, Anastasia Marchenkova is a researcher producing a lot of educational video content.

---

809 See Google's 'supreme' 20-qubit quantum computer by Tushna Commissariat, 2017.

810 See Update on Performance Metrics, Anyon Systems, March 2022.

811 See Application-Specific Quantum Hardware is the Most Promising Approach for Early Practical Applications by Fabio Sanches, Chiara Pelletti, and Alexei Marchenkov, February 2022.

812 See Superconducting Quantum Processor Design at Bleximo by Chiara Pelletti and Fabio Sanches, March 2022 and Bleximo builds its competitive advantage with an application-specific approach, PhysicsWorld, June 2022.

Their customer based seems made of US research labs (Berkeley University, DoE Berkeley Lab[813], Syracuse University, John Hopkins Applied Physics Laboratory). So, we're not far from a contract research company. This is exemplified by their willingness to create application-specific hardware, like IQM, which doesn't make much sense from an economical and even practical standpoint[814]. It would make sense if it brought similar benefits like FPGA (slower operations, low fixed cost, higher variable costs) vs ASICs (fast operations, higher fixed costs, low variable costs). But the costs here are still high, and they are not yet providing any quantum advantage with their system.

**QuantWare** (2021, Netherlands, $1M) is a designer and manufacturer of superconducting qubits processors created by Matthijs Rijlaarsdam and Alessandro Bruno[815]. They offer their 25 qubits Contralto processor and a customizable connectivity that could for example help prototype specific quantum error corrections codes.

The chipsets have AirBridges and a proprietary TSV configuration (through-silicon via). It seems to be very classical transmon superconducting qubits. They propose custom processors and a product design-to-delivery cycle of 30 days, leveraging the Van Leeuwenhoek Lab cleanroom at TU Delft.

They don't build full-fledged quantum computers. Their first 5 -qubit Soprano QPU had a modest $T_1$ of 10 µs and a single-qubit gate fidelity of 99,99%. Contralto's 25-qubit processor (pictured *below*) could reach a $T_1$ of 60 µs. They don't provide data on the most important figures of merit: dual-qubits gates and readout fidelity. Who could use these QPUs? Seemingly, research labs and vendors developing enabling technologies for superconducting qubits, like their colleagues from Qblox and Delft Circuits. They plan to double the number of qubits in their QPUs each and every year. QuantWare has various partnerships in place, including with SeeQC (USA), with QuantrolOx (UK) and QphoX (The Netherlands).

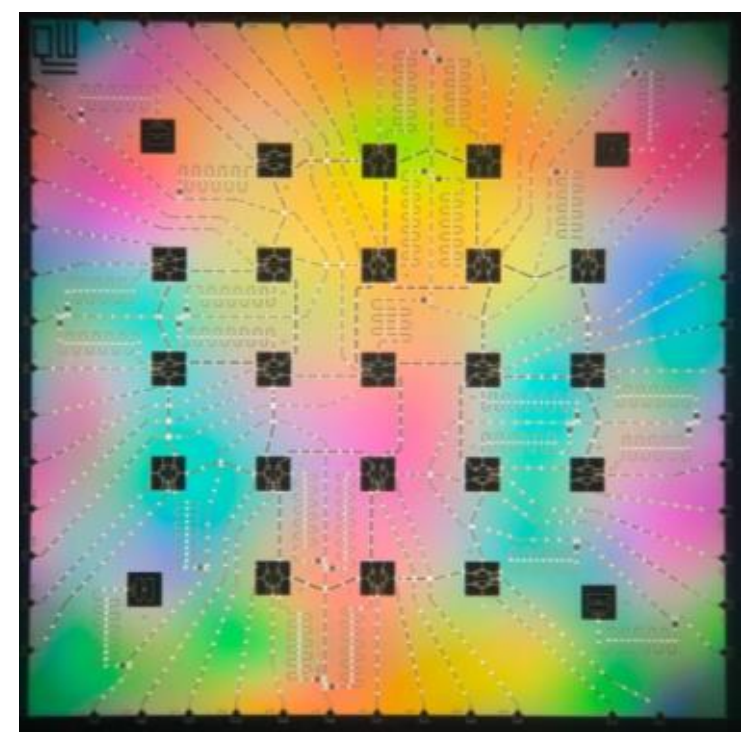

*Figure 332: QuantWare's 25 qubit processor.*

As we'll see later in the cryoelectronics section, QuantWare also designs Crescendo, a TWPA (qubits readout traveling waves parametric amplifiers). In September 2022, the company got a subsidy funding of 1.1M€ from Quantum Delta NL, the foundation running the Dutch quantum national plan, to develop superconducting qubits based on undefined novel materials.

**Atlantic Quantum** (2022, USA/Sweden, $9M) is a startup cofounded by Jonas Bylander from Chalmers University in Sweden, along with Bharath Kannan (CEO), Simon Gustavsson, Youngkyu Sung, William D. Oliver, Shereen Shermak and Tim Menke, from the MIT.

[813] See Raising the Bar in Error Characterization for Qutrit-Based Quantum Computing, Monica Hernandez, Lawrence Berkely National Laboratory in HPCwire, September 2021.

[814] See Application-Specific Quantum Hardware is the Most Promising Approach for Early Practical Applications by Fabio Sanches, Bleximo, February 2022.

[815] Among their scientific advisors are Charlie Marcus, formerly running the Microsoft Quantum Lab in Copenhagen, Denmark. He left Microsoft in November 2021.

The startup develops scalable fluxonium superconducting qubits-based quantum computers with experts covering all aspects of the quantum computing stack, from chip design and device fabrication to gate calibration and quantum algorithms. The cofounder's research group develops superconducting quantum electronic devices for quantum computing and simulation. Jonas Bylander recently published a paper on an efficient qubit readout solution using two microwave pulses and getting rid of the parametric amplifier, but it doesn't tell if it's in Atlantic Quantum's roadmap[816].

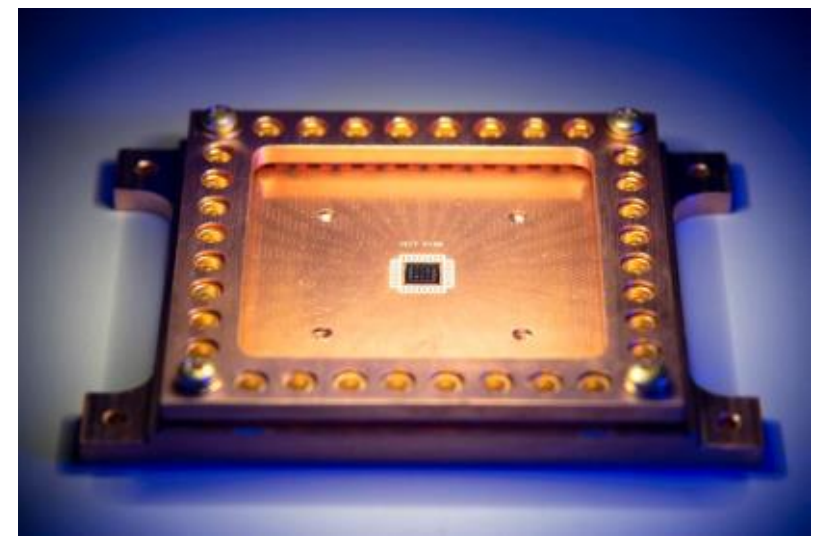

*Figure 333: Atlantic Quantum fluxonium superconducting chipset. Which is insufficient to have an idea of its qubit fidelities, that is not yet published. Source: Atlantic Quantum.*

In April 2021, the Japanese research center **RIKEN** and **Fujitsu** created the RIKEN RQC-Fujitsu Collaboration Center to do joint research and create a superconducting qubit computer, with a goal of reaching 1000 physical qubits and develop an associated software platform.

It will leverage RIKEN's existing work on superconducting qubits and Fujitsu's computing know-how. The research plan is quite classical: improving qubit manufacturing, reducing the size and noise of driving electronics components and wiring and improve error correcting codes.

Some plans roadmaps were uncovered in August 2022. Fujitsu plans to release a first 64 qubits QPU by spring 2023, to be expanded to 1000 qubits in 2026. Fujitsu has created a research center in Wako City, Japan, to work on these systems with Riken with about 20 researchers.

**Toshiba** has been conducting fundamental research in quantum computing since at least 2008, in quantum photonics and with superconducting qubits in its Frontier Research Laboratory.

Here, Hayato Goto is a prolific scientist working in many disciplines. In 2022, he created a double-transmon coupler that turns on/off coupling between two superconducting qubits in an efficient manner, enabling fast computing and two-qubit gate fidelities of 99.99% and time of 24 ns[817].

**Alibaba** is active in using the resources of its datacenters to simulate quantum algorithms exceeding 50 qubits. China's leading e-commerce company is also partnering with the **University of Science and Technology of China** (USTC) of the Chinese Academy of Sciences (CAS) to create superconducting quantum computers with superconducting qubits.

They offer cloud access to 11 qubits since early 2018, on a technology platform developed with USTC. They even announced in 2018 that they were creating a subsidiary, **Ping-Tou-Ge**, which develops NPUs (neuromorphic processors for AI) and, eventually, superconducting quantum chipsets[818]. They work on superconducting qubits, using the fluxonium variation, which could bring some coherence advantage. They announce qubit lifetimes $T_1$ and $T_2$ over 100 µs and a 99,5% iSWAP gate fidelity.

**Baidu** announced in August 2022 its Qian Shi QPU with 10 superconducting qubits, to be later expanded to 36 qubits using couplers to run two-qubit gates, reusing a concept first pioneered by Google in 2019.

---

[816] See Transmon qubit readout fidelity at the threshold for quantum error correction without a quantum-limited amplifier by Liangyu Chen, Jonas Bylander, Giovanna Tancredi et al, August 2022 (8 pages).

[817] See Double-Transmon Coupler: Fast Two-Qubit Gate with No Residual Coupling for Highly Detuned Superconducting Qubits by Hayato Goto, PRA, March-September 2022 (10 pages).

[818] See Alibaba Launches Chip Company "Ping-Tou-Ge"; Pledges Quantum Chip, September 2018.

These qubits are said to showcase high fidelities. Let's look at the numbers: $T_1$ = 31 µs, $T_2$ = 8.7 µs, single qubit gate at 99,8% and two-qubit gates at 96,4% (CX) and 96,8% (CZ). For 10 qubits, it's less than stellar and the numbers look like those of some average transmon qubits.

**Origin Quantum Computing** (2017, China, $163.4M) is a startup created in Hefei by Guo Guoping out of the CAS quantum lab in Hefei. It closed an amazing $148M Series B funding round in July 2022.

The company works on a full-stack quantum offering including a 24 superconducting qubits system, working on semiconductor quantum dots qubits as well, on qubits control electronics (Quantum AIO), cryogenic equipment, on quantum software with an operating system (Origin Pilot), a programming framework (QPanda), the EmuWare virtual machine, a quantum machine learning framework (VQNet), a quantum programming language (QRunes), an integrated development environment (Qurator), and some applications frameworks for quantum chemistry (ChemiQ), fluid dynamics (OriginQ QCFD) and financial optimization.

Initially, it worked mainly on developing quantum algorithms and quantum emulation software. They were behind one of the records for 64-qubit quantum algorithm emulation on a supercomputer[819]. They also created cloud based emulation appliances supporting 32 and 64 qubits. They then started to create their own quantum chipsets, including superconducting qubits chipsets with 6 qubits (KF C6-130) and 100 qubits (XW B2-100), using tunable couplers. A bit like IBM, they expect to reach 1024 qubits by 2025 with intermediate steps of 64 qubits in 2021 and 144 qubits in 2022.

Now, on to cat-qubit and other bosonic qubits vendors...

**Alice&Bob** (2020, France, $33M) was created by Théau Peronnin (ENS Lyon) and Raphaël Lescanne (ENS Paris). They are designing a fault-tolerant gate-based quantum computer associating superconducting technology and stabilized photon-based (in the microwave regime) cat-qubits. Their technology main benefit is its capability to implement a complete universal fault-tolerant quantum computer with a much lower ratio of physical per logical qubits than traditional transmon based superconducting qubits. It saves at least two orders of magnitude, moving from 1000 to 1 down to 30 to 1 (about $\sqrt{1000}$)!

Alice&Bob's technology is based on the PhD thesis from the startup founders and the associated work of the **Mazyar Mirrahimi**'s Quantic team from Inria where Raphaël Lescanne was a doctoral student and where **Zaki Leghtas** as well as **Jérémie Guillaud** also work or worked[820], the CNRS and ENS Lyon and ENS Paris. **Pierre Rouchon** from MinesParistech is also a key contributor[821]. We'll see how all these works were influential, up to inspire Amazon in its own cat-qubits engineering efforts, documented later.

Cat-qubits encode the state of a qubit with superposing opposite quantum states in micro-wave photon cavities, precisely, in the two-dimensional Hilbert space spanned by two coherent states of micro-waves of same amplitude and opposite phase[822].

---

819 See Researchers successfully simulate a 64-qubit circuit, June 2018.

820 Mazyar Mirrahimi did work in Michel Devoret's team at Yale University around 2012. See Dynamically protected cat-qubits: a new paradigm for universal quantum computation by Mazyar Mirrahimi, Zaki Leghtas and Michel Devoret, 2013 (28 pages). Jérémie Guillaud is now Chief of Theory at Alice&Bob.

821 See Quantum computation with cat qubits by Jérémie Guillaud, Joachim Cohen and Mazyar Mirrahimi, March 2022 (75 pages).

822 See Exponential suppression of bit-flips in a qubit encoded in an oscillator by Raphaël Lescanne et al, July 2019 (18 pages) and Repetition Cat Qubits for Fault-Tolerant Quantum Computation by Jérémie Guillaud and Mazyar Mirrahimi, July 2019 (23 pages).

These cat-qubits have a very low bit-flip error rate given it decreases exponentially with the average number of microwave photons used in the cat qubit cavity[823]. Phase-flip errors can be corrected with repetition error codes having a rather low overhead[824].

Cat-qubits can support a native implementation of 3-qubits Toffoli gates which, combined with Clifford gates, form a universal set of quantum gates. The implementation of such a universal gate set is a prerequisite to run quantum algorithms with a proven exponential speed-up. The Toffoli gate is an alternative to the usual (non-Clifford) T gate used in QFT-based algorithms. This gate can be corrected efficiently with avoiding magic state distillation, enabling fault-tolerance, and limiting error propagation between ancilla qubits[825].

These qubits are more complex to design and operate but it would only take about 30 of them to create a well-corrected logical qubit, which would make it possible to create a better scalable architecture whereas with the current technologies of IBM, Google and Rigetti, about 1,000 to 10,000 physical qubits are required to create a functional logical qubit given their expected fidelities. These corrected qubits could also play the role of associative quantum memory. Moreover, their system avoids microwave radiations leaks between adjacent qubits.

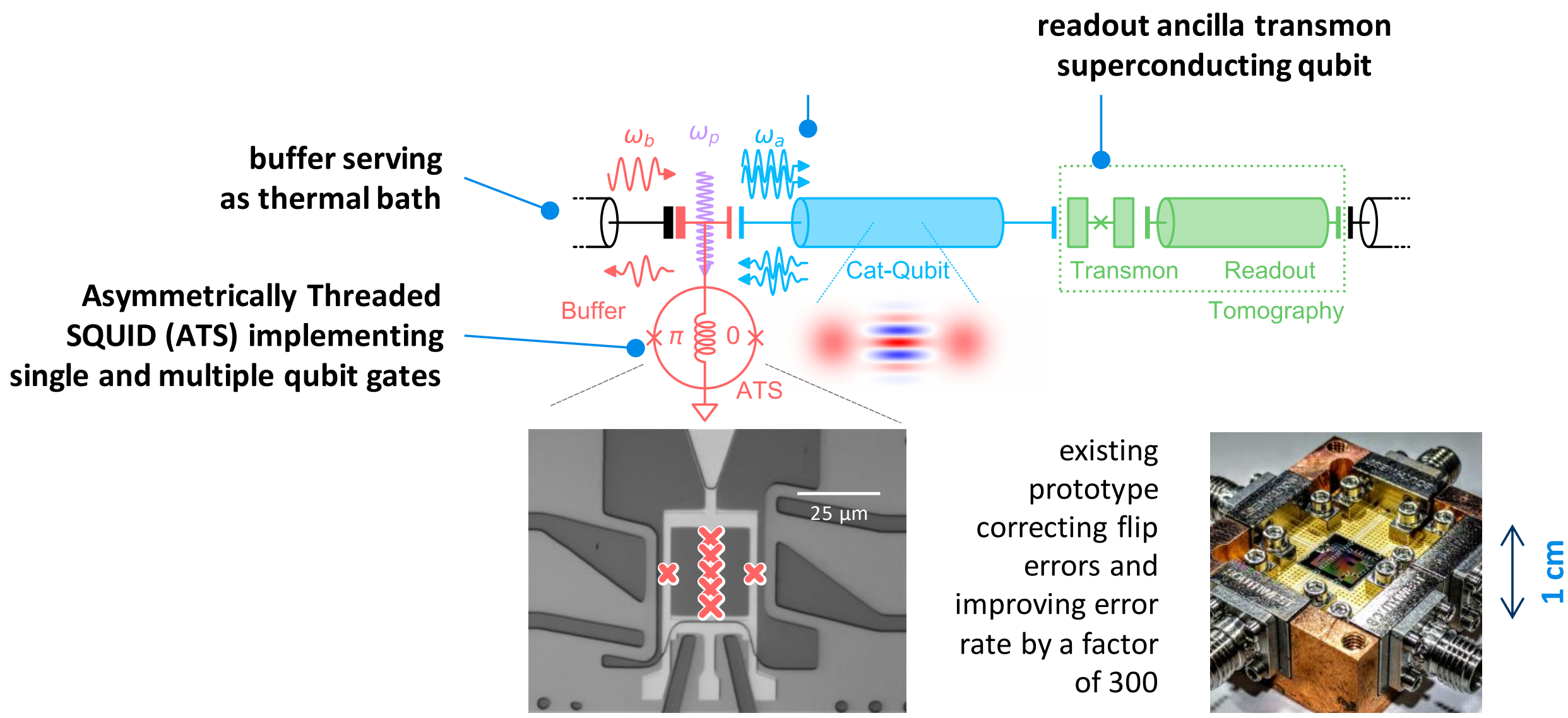

*Figure 334: Alice&Bob cat-qubit cavity and its coupling to a transmon qubit and the ATS (asymmetrically threaded SQUID) that implements single and multiple qubit gates. Source: Alice&Bob.*

The gates they implement on top of a Toffoli gate are a CNOT and a Hadamard gate. SWAP gates are built with three CNOTs in a classical fashion.

These are heavily used to circumvent the absence of many to many qubits connectivity in most 2D qubits layouts. All this will require a specific compiler, to be created later in the startup product lifecycle.

823 See One hundred second bit-flip time in a two-photon dissipative oscillator by C. Berdou, Zaki Leghtas, Maryar Mirrahimi, Pierre Rouchon, Raphael Lescanne, Théau Peronnin, Taki Kontons et al, April 2022 (20 pages).

824 See Error Rates and Resource Overheads of Repetition Cat Qubits by Jérémie Guillaud and Mazyar Mirrahimi, March 2021 (17 pages). Based on numerical simulation, it estimates that a fault-tolerant cat-qubits computer with a logical error probability of $10^{-10}$ can be realized using 140 physical cat-qubits for Clifford gates and an average number of 15 photons per mode. A Toffoli gate could be implemented with only 180 physical cat-qubits including all required ancilla qubits.

825 They propose a 'pieceable fault-tolerant' implementation of the Toffoli gate, following the method introduced in Universal Fault-Tolerant Gates on Concatenated Stabilizer Codes by Theodore J. Yoder, Ryuji Takagi and Isaac L. Chuang, September 2016 (23 pages). This is a substitute to the transversal gates technique.

On a practical way, these cat-qubits drive require some more corrections than regular transmon qubits. A continuous pulse tone drives the cat-qubit photons buffer. The qubit error corrections makes use of an ATS (Asymmetrically Threaded SQUID), a nonlinear element that is flux biased with an AC current in the 4-8 GHz range. A one qubit gate requires two pulses (cat drive and dissipation phase). A cat-qubit readout uses two pulses on the coupling transmon and a readout pulse. Readout uses 3 inbound pulses and one reflected pulse. And in total, cat-qubits require about 6 control lines.

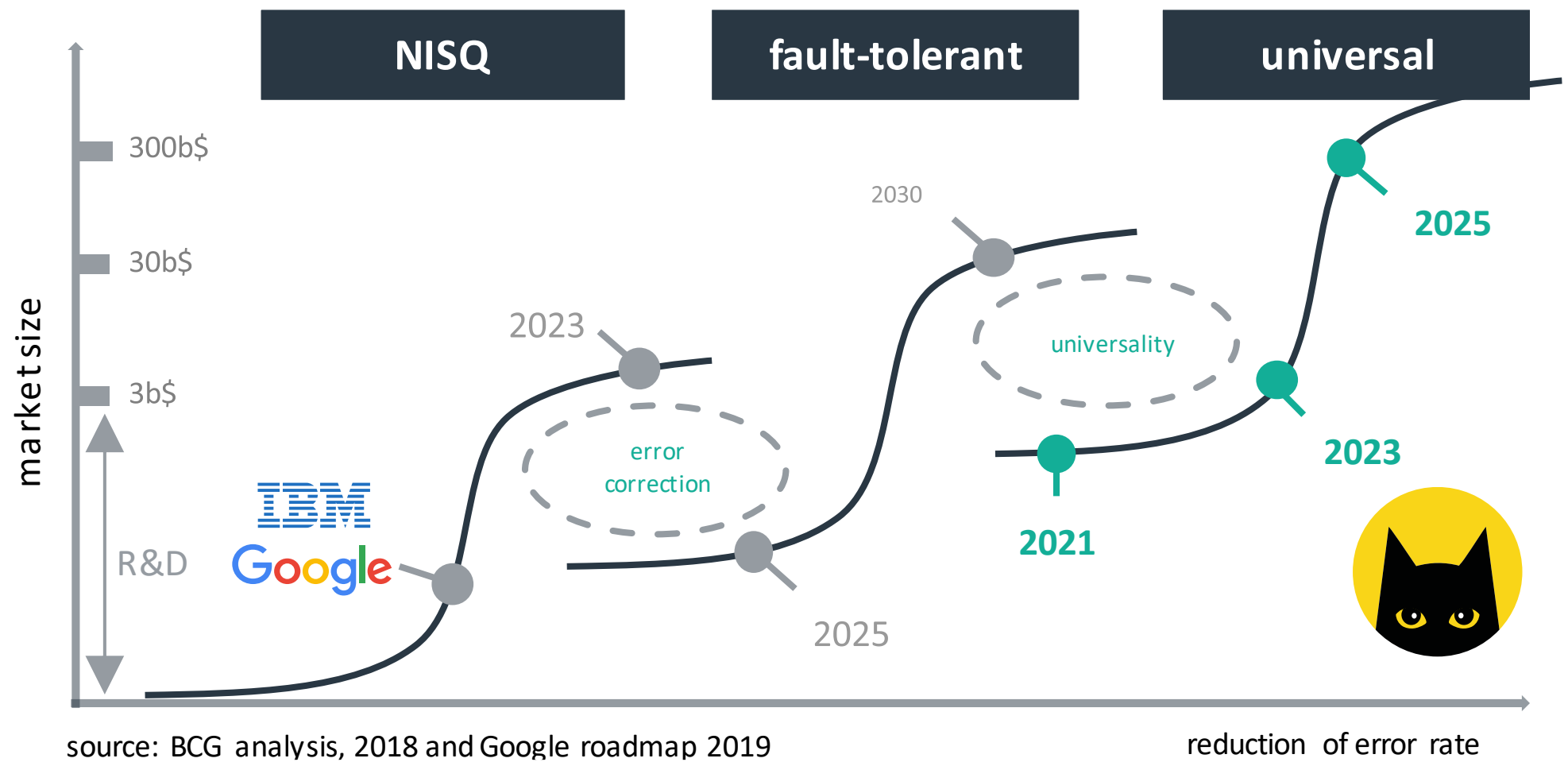

*Figure 335: Alice&Bob roadmap which ambitions to directly create a fault-tolerant QPU and then a universal quantum computer (although the definition of universal quantum computer is not agreed upon. Source: Alice&Bob.*

**Amazon** (USA) started first to announce late 2019 its Amazon Braket cloud offering, based on using third-party quantum computers from D-Wave, Rigetti and IonQ, covered in the cloud section of this book, page 663.

In December 2020, they went out of the woods with announcing their detailed plan to build their own quantum computers, using cat-qubits, in a thorough 118 pages paper[826].

This work is getting the help from Caltech, including John Preskill, in connection with Yale University where some Caltech students did their PhDs in the teams of Rob Schoelkopf and Michel Devoret. The Amazon effort is led by **Simone Severini** (Director of Quantum Computing at AWS), **Oskar Painter** (Head of Quantum Hardware at AWS), **Fernando G.S.L. Brandão** (Head of Quantum Algorithms at AWS and also researcher at Caltech) and **Richard Moulds** (GM Amazon Braket).

The bulk of Amazon's quantum team are based in the new 21,000-sqrt-ft AWS Center for Quantum Computing building next to Caltech in Pasadena, North of Los Angeles. It was inaugurated in October 2021.

The Amazon proposed architecture is largely inspired by what the French teams at Inria have investigated since 2013 with Mazyar Mirrahimi et al, including the founders of Alice&Bob[827]. They want to create a FTQC. As of 2020, Amazon was planning to use an electro-acoustic resonator to host the cat qubits while the circuit element, the Asymmetrically Threaded SQUID (ATS) invented by Raphaël Lescanne and Zaki Leghtas, used by Alice&Bob to stabilize the cat-qubit is superconducting. While Alice&Bob QEC is based on dissipating excess qubit energy to maintain it in low-energy states with encoding it in a linear oscillator driven by 10 GHz microwaves, Amazon chose a variant that uses linear harmonic oscillators-based cat-qubits using very compact piezoelectric nanostructures and phonons. Like with Alice&Bob, these cat-qubits self-corrects flip errors at the hardware level while phase errors are being handled by some QEC requiring, supposedly about 20 physical qubits.

[826] See Building a fault-tolerant quantum computer using concatenated cat codes by Christopher Chamberland, John Preskill, Oskar Painter, Fernando G.S.L. Brandão et al, 2020 (118 pages). It is summarized in Designing a fault-tolerant quantum computer based on Schrödinger-cat qubits by Patricio Arrangoiz-Arriola and Earl Campbell, April 2021. See also Fault-tolerant quantum computing with biased-noise hardware by Earl Campbell, November 2020 (40 mn).

[827] On top of France's founding work on cat-qubits, Amazon is also relying on many US Universities research like Caltech, Stanford, Chicago University and Yale University.

Cat-qubits encode information with microwaves put in coherent states with opposite phases, $|+\rangle$ and $|-\rangle$. The qubit computational basis states are defined as even and odds coherent states cats, meaning using positive and negative sign superpositions for these two cat-states. Like Alice&Bob, they will implement a universal gate set comprising X, Z, CNOT and Toffoli gates. They use two new ideas for implementing fault-tolerant Toffoli gates: an extremely small chip layout ("bottom-up Toffoli") and a technique to lower the bit-flip error rate ("top-down Toffoli"). They also avoid crosstalk between cat-qubits with using four cat-qubits connected to a single dissipating reservoir. This compact layout is compatible with a scalable architecture but may generate significant crosstalk errors, which could be mitigated with a well-chosen filter design cutting the frequencies to remove crosstalk errors.

They first plan to implement a 9-qubit QEC to obtain a logical error rate of $2.7 \times 10^{-8}$. As a result, they expect to use 2000 superconducting qubits to create a 100 logical qubits system. If this works, as with Alice&Bob, it will make a significant difference with IBM and Google who plan to obtain the same number of logical qubits with one million physical qubits. The scalability constraints are much different in both cases, whether it deals with cryogenics, microwave generations and readouts, or cabling.

In April 2021, University of Sydney science undergraduate Pablo Bonilla Ataides published in Nature Communications a paper on its ZXXZ surface code that would reduce the number of required physical qubits to create a logical qubit thanks to a lower error threshold. It brought the attention of Amazon researchers[828]. This surface code could be used by Amazon who made a choice to use a relatively low number of photons per cat qubit (8 to 10, compared to about 15 for Alice&Bob, but the optimum number of photons is still to be determined experimentally), still requiring some first level bit-flip error correction on top of phase-flip correction. That's where a ZXXZ surface code QEC could come into play. ZXXZ QEC codes are indeed mentioned as an option QEC technique in Amazon's technical paper from December 2020.

The AWS Center for Quantum Computing opened at Caltech in October 2021 and houses all Amazon teams working on quantum computing[829]. They even then showcased a picture of a prototype quantum processor, maybe the one with 9 qubits.

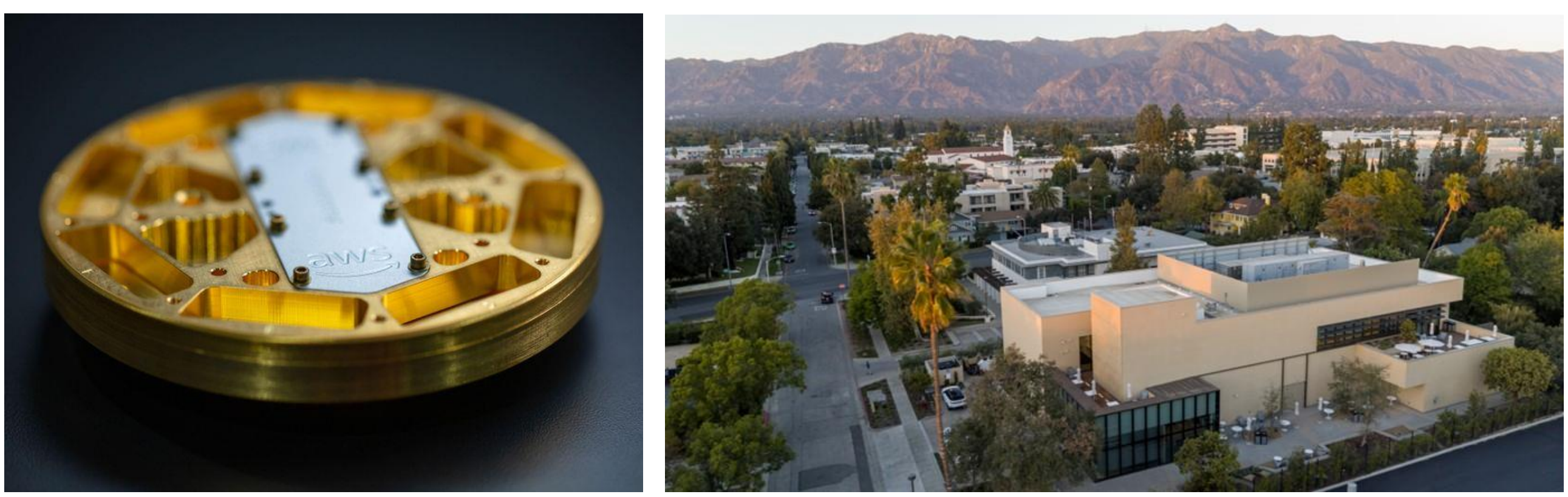

*Figure 336: the first prototype Amazon cat-qubit chipset of undisclosed characteristics, and their lab in Caltech opened in October 2021.*

[828] See Student's physics homework picked up by Amazon quantum researchers by Marcus Strom, University of Sydney, April 2021, Sydney student helps solve quantum computing problem with simple modification by James Carmody April 2021 and The XZZX surface code by J. Pablo Bonilla Ataides et al, April 2021, Nature Communications (12 pages).

[829] See Announcing the opening of the AWS Center for Quantum Computing by Nadia Carlsten, October 2021.

**QCI** (2015, USA, $18M) or Quantum Circuits Inc is a spin-off from Yale University co-founded by Rob Schoelkopf, Luigi Frunzio and Michel Devoret. Michel Devoret left the company in 2019, preferring to be a full-time researcher at Yale University.

Their technology is also based on cat-qubits that solve noise and coherence problems, using Rob Schoelkopf's teamwork at Yale. They have a long track-record in that space although they are not very talkative. They announced in 2019 that their system should be available some day on Microsoft Azure Quantum cloud. But so far, they have not delivered any functional QPU.

Technically, their planned cat-qubit are stabilized by discrete parity measurement using a transmon[830] and their two qubit gates are implemented using SNAP gates (Selective Number-dependent Arbitrary Phase (SNAP) using the same transmon[831]. They also planned to use micromachined 3D cavities with a good Q factor[832]. They plan to obtain a linear gain in qubit lifetime with the number of physical qubits. It will still require some surface codes to fully correct flip-errors (on top of phase errors).

They are also at the origin of the **qbsolv** framework that is part of their **Mukai** middleware and development platform launched in January 2020[833]. It supports D-Wave computers, Fujitsu digital-annealed computers and Rigetti superconducting qubits.

**Nord Quantique** (2019, Canada, $7.6M) is a startup from the Institut Quantique from the University of Sherbrooke that is working on creating a superconducting quantum computer using more efficient error correction, using another variation of bosonic codes that would have fast quantum gates, but with no further publicity on how they implement this and on their roadmap.

The company was created by Julien Camirand Lemyre and Philippe St-Jean with the scientific support from Alexandre Blais and their investors are BDC Capital (USA), Quantonation (France) and Real Ventures (Canada).

## Quantum dots spins qubits

Electron spins qubits are a new promising qubit technology with a lot of variations. Its related research started later than superconducting qubits. Its potential benefits are miniaturization and scalability. It could leverage existing manufacturing processes for standard CMOS semiconductors[834].

### History

Electron spin qubits quantum state is generally the spin orientation of an electron trapped in a potential well or of an electron hole, i.e. a missing electron and its virtual inverse impact on structural spin.

[830] See Demonstrating Quantum Error Correction that Extends the Lifetime of Quantum Information by Nissim Ofek, Zaki Leghtas, Steve Girvin, Liang Jiang, Mazyar Mirrahimi, Michel Devoret, Rob Schoelkopf et al, February 2016 (44 pages).

[831] See Cavity State Manipulation Using Photon-Number Selective Phase Gates by Reinier W. Heeres, Eric Holland (who now works at Keysight), Liang Jiang, Robert Schoelkopf et al, March 2015 (9 pages).

[832] See Multilayer microwave integrated quantum circuits for scalable quantum computing by Teresa Brecht, Michel Devoret, Rob Schoelkopf et al, Nature, 2015 (5 pages) and the related thesis Micromachined Quantum Circuits by Teresa Brecht, 2017 (271 pages).

[833] See QCI Qbsolv Delivers Strong Classical Performance for Quantum-Ready Formulation by Michael Booth et al, May 2020 (7 pages).

[834] CMOS ("Complementary Metal Oxide Semiconductor") is the dominant technology used to produce microprocessors, for CPUs (Intel, AMD), GPUs (Nvidia, AMD), chipsets for smartphones (Qualcomm, Samsung, MediaTek, HiSilicon, etc.) and in a whole host of specialized sectors (microcontrollers, radio components, etc.).

It is usually considered that Daniel Loss with David DiVincenzo and Bruce Kane are the first to have devised ideas to use electron spins to create a quantum computer, a couple years after Peter Shor created his famous integer factoring eponymous algorithm.

**Daniel Loss** (University of Basel) and **David DiVincenzo** (then at IBM Research) published a similar paper in 1996-1997 where they proposed the concept of quantum dots to create qubits with controlling the spin of electrons in a potential well[835]. Their design used a two-qubit gate (SWAP) using an electrical control of the tunneling barrier between neighboring quantum dots. A low gating voltage creates a coupling – *aka* Heisenberg coupling - between the neighbor qubits. The design also implemented single qubit gates. The Loss-Vincenzo concept was later extended with using pairs of quantum dots electron spins, one being the qubit itself, and the other, capacitively coupled with the first one and being used for qubit readout with a spin-to-charge conversion using conductance measurement, usually with some radio-frequency reflectometry using a microwave pulse, a bit like with superconducting qubit readout.

The first electron spin qubit was created in 2005 by a USA and Brazil team using GaAs on Si substrates[836]. But GaAs qubits suffers from a major limitation, a strong hyperfine coupling together with a nonzero nuclear spin that leads to dephasing. So physicists looked at Si and SiGe alternatives, which showcase lower hyperfine coupling. They could even provide quasi-noiseless environment for spins thanks to their nuclear-free isotopes $^{28}$Si and $^{72}$Ge isotopes. But they are much more challenging in terms of manufacturing, regardless of the supply of isotopically purified materials.

The first demonstration of silicon spin qubit was made in 2012 by UNSW teams. In 2016, silicon qubits were demonstrated using industry-grade manufacturing processes by a French team from CEA-Leti and IRIG in Grenoble[837].

This belongs to the Si-MOS category, the most generic and easier to manufacture. Qubits are derived from planar MOS bulk or FDSOI technologies as well as with Fin-FET that are inspired from the latest CMOS manufacturing technologies. These spin qubits have a size of about 100x100 nm, leading to potential high densities when it will scale[838].

**Bruce Kane** (UNSW) presented in 1998 another spin-based quantum computer concept based on placing individual phosphorous atoms ($^{31}$P) in a pure silicon lattice structure[839]. This approach is labelled "donors spin". It is a hybrid scheme using quantum dots and single atom nuclear magnetic resonance (NMR) since qubits associate phosphorus atoms nuclear spin and silicon donors electron spins. The qubits are controlled by electrical and magnetic fields[840]. The main benefit is the long coherence of nuclear spins, which can theoretically extend to several seconds[841].

---

835 See Quantum computation with quantum dots by Daniel Loss and David DiVincenzo, 1997 (20 pages).

836 See Coherent Manipulation of Coupled Electron Spins in Semiconductor Quantum Dots by Jason Petta et al, Science, 2005 (5 pages).

837 See A CMOS silicon spin qubit by Romain Maurand, Maud Vinet, Marc Sanquer, Silvano De Fransceschi et al, 2016 (12 pages).

838 See The path to scalable quantum computing with silicon spin qubits by Maud Vinet, Nature Nanotechnology, December 2021 and Scaling silicon-based quantum computing using CMOS technology by M. F. Gonzalez-Zalba, Silvano de Franceschi, E. Charbon, Tristan Meunier, Maud Vinet and Andrew S. Dzurak, Nature Electronics, December 2021 (16 pages).

839 See A silicon-based nuclear spin quantum computer by Bruce Kane, Nature, 1998 and Silicon-based Quantum Computation by Bruce E. Kane, 2000 (14 pages).

840 See The Race To Make Better Qubits by Katherine Derbyshire, Semiconductor Engineering, November 2021.

841 The Bruce Kane concept is well described in Toward a Silicon-Based Nuclear-Spin Quantum Computer by Robert G. Clark, P. Chris Hammel, Andrew Dzurak, Alexander Hamilton, Lloyd Hollenberg, David Jamieson, and Christopher Pakes, Los Alamos Science, 2022 (18 pages). It shows linear array of phosphorous donor atoms buried into a pure silicon wafer, operating in the presence of a large magnetic field and at sub-K temperatures. The donor atoms nuclear spins are be aligned either parallel or antiparallel with the magnetic field, corresponding to $|0\rangle$ and $|1\rangle$ qubit basis states. The metal gates are above an insulating barrier of $SiO_2$. The A-gates above the $^{31}$P atoms enable single qubits gates while the J-gates in between the donors regulate an electron-mediated coupling between adjacent nuclear spins, for two-qubit operations. At last, qubit readout is done with either a single electron transistor (SET) or with a magnetic-resonance force microscope (MRFM, not shown).

The challenges lie with the way to precisely position the phosphorus atoms in the silicon lattice and how to handle qubits entanglement and readout. This is the path chosen by Michelle Simmons at UNSW and in her startup SQC. The individual atoms are positioned in the silicon structure with lithography using a scanning tunneling microscope (STM).

The first processor fully implementing this architecture was announced by SQC in 2022 (we cover it later). Similar options are pursued like the use of antimony nucleus embedded in silicon lattice structures and controlled by microwaves by Andrea Morello from UNSW[842].

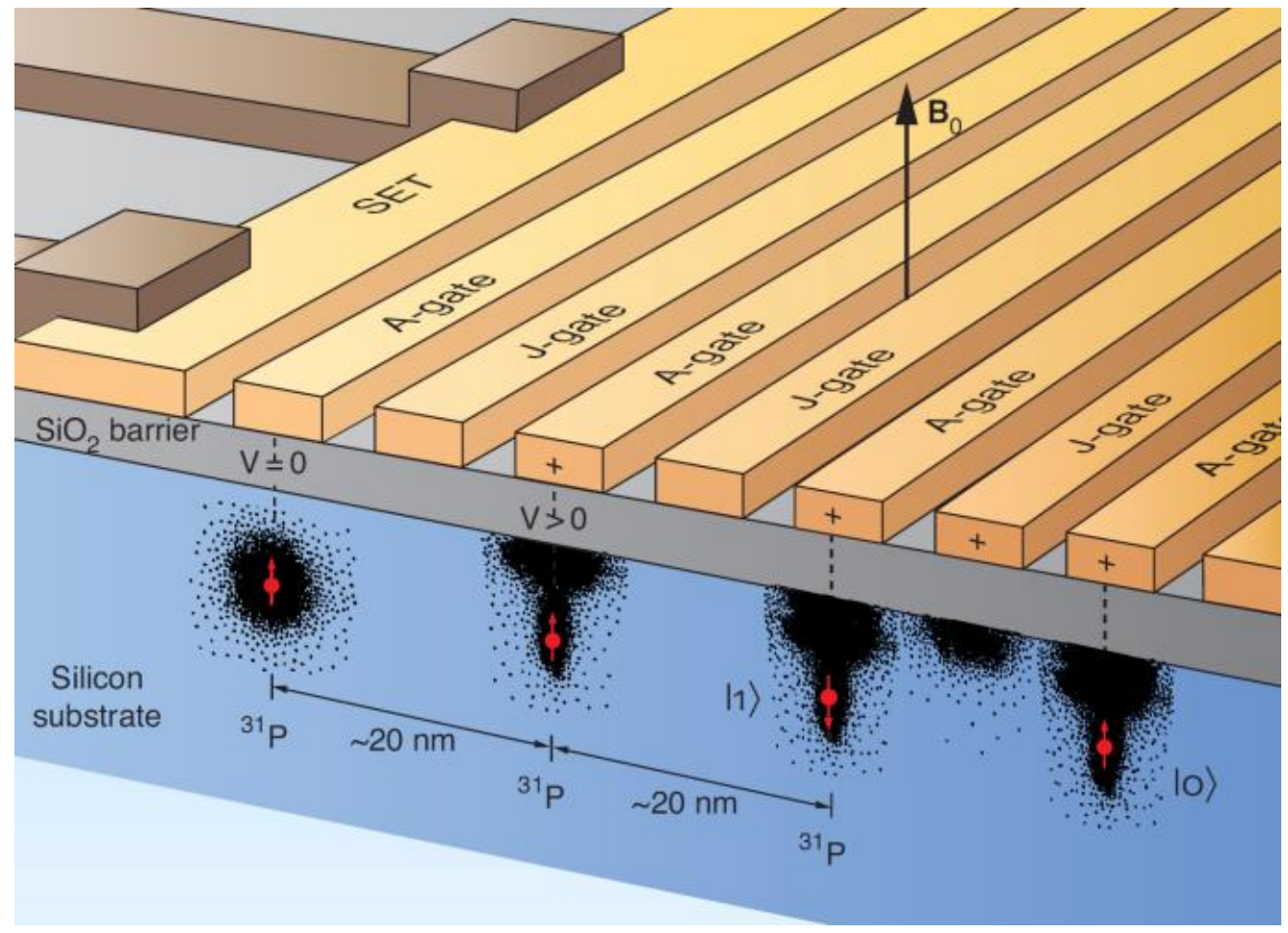

*Figure 337: the donor spin architecture with phosphorous atom implanted in a silicon substrate under a $SiO_2$ isolation layer. Source: Toward a Silicon-Based Nuclear-Spin Quantum Computer by Robert G. Clark, P. Chris Hammel, Andrew Dzurak, Alexander Hamilton, Lloyd Hollenberg, David Jamieson, and Christopher Pakes, Los Alamos Science, 2022 (18 pages).*

Later, Andrea Morello and Patrice Bertet designed another hybrid approach, coupling transmon superconducting qubits (for computing) and phosphorus in silicon nuclear spins with a donor electron (for creating a quantum memory)[843].

On top of the above two mainstream paths (Si-MOS/CMOS and donors), several other avenues are investigated in the quantum dots based spin qubit realm:

- **Silicon/silicon germanium** (Si/SiGe) heterostructures qubits (Qutech, CEA IRIG) where germanium is used for the stability of its spin holes, large band gaps, higher electron mobility, stronger spin-orbit coupling, its insensitiveness to exchange coupling oscillations and long coherence times. It is however more difficult to manufacture and scale, with gates that are far from the qubits[844]. A record breaking 4 entangled qubits was announced late 2020 by TU Delft, based on germanium. Germanium allows the creation of very fast quantum gates ranging from 0.5 to 5 ns[845].
- **Gallium-arsenide** (GaAs), first tested in 2005, but with very short coherence times due to spin interferences from gallium and arsenic atoms nuclei.
- **Electrons trapped** on solid (inert) neon[846] or on superfluid helium.
- **Electron spin trapped in carbon nanotubes** (C12 Quantum Electronics) or **carbon nanospheres** (Archer Materials). These structures better protect the spin of a trapped electron, at the expense of more complicated interfaces and controls

[842] See Coherent electrical control of a single high-spin nucleus in silicon by Serwan Asaad, Andrea Morello, Kohei M. Itoh, Andrew S. Dzurak et al, 2019 (56 pages).

[843] See Donor Spins in Silicon for Quantum Technologies by Andrea Morello, Patrice Bertet and Jarryd Pla, 2020 (17 pages).

[844] See this excellent germanium review paper: The germanium quantum information route by Giordano Scappucci, Silvano De Franceschi et al, 2020 (18 pages).

[845] See also Quantum control and process tomography of a semiconductor quantum dot hybrid qubit, 2014 (12 pages).

[846] See Single electrons on solid neon as a solid-state qubit platform by Xianjing Zhou, Kater W. Murch, David I. Schuster et al, Nature, May 2022 (16 pages). The trapped electrons are coupled to a superconducting resonator on top of solid neon at 10 mK. It should bring better coherence but their current $T_2$ is at 200 ns.

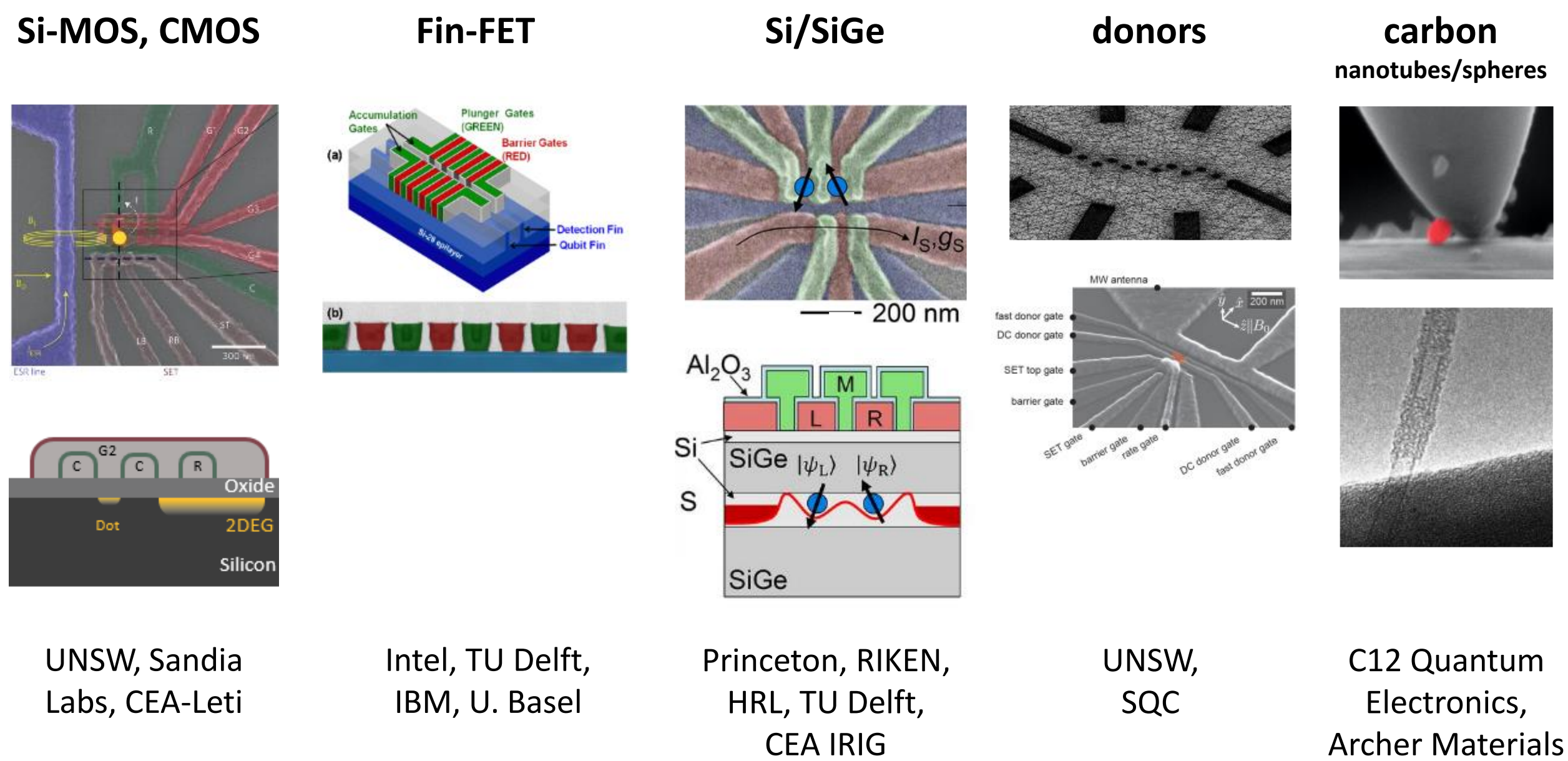

*Figure 338: various silicon spin qubits. (cc) Olivier Ezratty, 2022, inspired by a compilation by Maud Vinet, CEA Leti.*

These qubits small dimensions and the possibility to integrate control electronics in or around the qubit chipset make it an interesting candidate for large-scale quantum computing[847]. There are however many challenges to overcome, particularly with materials design[848].

| | specs | Pros | cons |
|---|---|---|---|
| **Si-MOS** | silicon quantum dot | mastered fabrication technique and scalability potential | qubit decoherence, poor entanglement, disordered potential in materials |
| **FinFET** | silicon quantum dot | mastered fabrication technique and scalability potential | disordered potential in materials |
| **SiGe** | holes or spin in Si Ge heterostructures | longer coherence time, entanglement, relatively high temperature (>1K), clean epitaxial barrier | valley degeneracy making it difficult to differentiate qubit states |
| **GaAs** | first Si- GaAs qubits in 2005 | stable dopants, single conduction band valley, easy to manufacture, test platform for many characteristics with knowledge transferable to Si-MOS/FinFET/SiGe qubits | nuclear spins effect and very small coherence time ($T_2$) |
| **atom donors** | phosphorus atom spin donor | long coherence time | complicated to manufacture, impact of impurities |
| **electrons on neon or helium** | trapped electrons on superfluid helium. | long cohérence, high connectivity, fast gates | not very vocal company (EeroQ) on progress made |
| **carbon nanotubes and nanospheres** | electron spin trapped in carbon nanotubes or nanospheres | long coherence time, reuse components from other qubit types for qubit control | more complex wiring and control, entanglement |

*Figure 339: specificities with pros and cons of each silicon spin qubit variety. (cc) Olivier Ezratty, 2022.*

[847] A good up-to-date overview of silicon qubits can be found in Scaling silicon-based quantum computing using CMOS technology: Challenges and Perspectives by Fernando Gonzalez-Zalba, Silvano de Franceschi, Tristan Meunier, Maud Vinet, Andrew Dzurak et al, 2020 (16 pages).

[848] See Democratizing Spin Qubits by Charles Tahan, November 2021 (19 pages) which describes the many challenges with quantum dots spin qubits and Quantum Technologies for Engineering: the materials challenge by Kuan Eng Johnson Goh, Leonid A Krivitsky and Dennis L Polla, IOP Publishing, March 2022 (14 pages).

## Science

Spin qubits may allow the integration of a large number of qubits in a circuit, with potentially up to billions of qubits on a single chipset. It seems to be the only technology that can achieve this level of integration. These qubits would have a rather long coherence time and an error rate at least as low as with superconducting qubits[849].

The control microwaves used have a higher energy level which explains why silicon qubits can theoretically operate around 1K instead of 15 mK for superconducting qubits. This level corresponds to microwaves with a frequency higher than 20 GHz, compared to the 4 to 8 GHz control microwaves of superconducting qubits. This higher temperature makes it possible to place denser control electronics around the qubits without heating up the circuit too much. The reference data are as follows: only one milliwatt of energy can be consumed at 100 mK[850].

This limits the control electronics to about 10,000 transistors in CMOS technology[851]. Once developed, silicon qubits will require the use of massive error correction codes, such as surface codes or color codes.

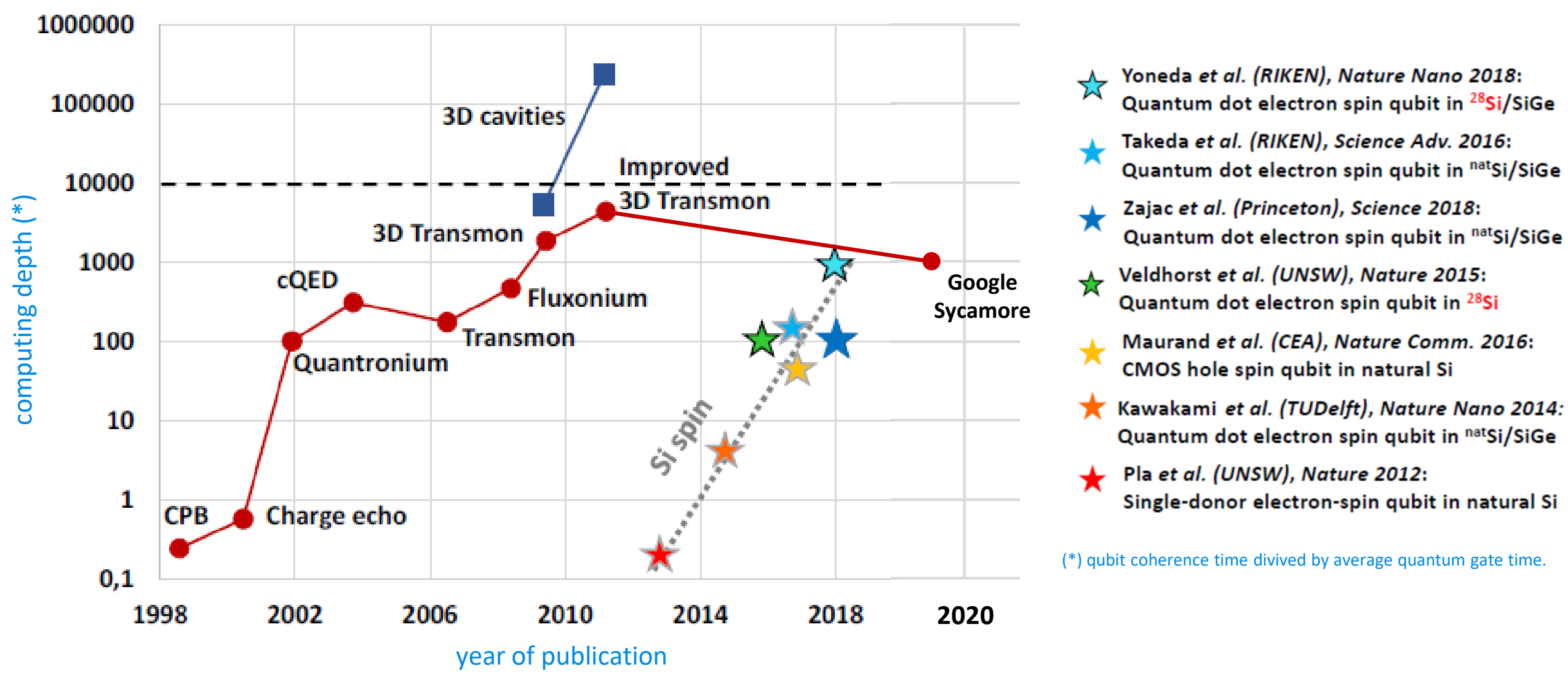

*Figure 340: a perspective chart showing how silicon qubit progressed in the last 10 years with respect of computing depth. It requires some updates. First adapted by Maud Vinet from Superconducting Circuits for Quantum Information: An Outlook by Michel Devoret and R. J. Schoelkopf, Science, 2013. Then updated by Olivier Ezratty in April 2020.*

Advances in spin qubits are more recent in a race against superconducting qubits. The diagram in Figure 340 illustrates this evolution over time between 2013 and 2020[852], and would require some updating.

---

[849] A record silicon qubit coherence time was broken in 2020 by a team from the University of Chicago, reaching 22 ms (T2). This is 10,000 times longer than the usual coherence times around 100µs found in superconducting qubits. These qubits use double gaps in silicon carbide structures. See Universal coherence protection in a solid-state spin qubit by Kevin C. Miao, David D. Awschalom et al, August 2020 (12 pages). University of Chicago.

[850] A milli-Watt of cooling power can be achieved with a double pulsed tube cryostat such as the BlueFors XLD1000 or the Oxford Instruments TritonXL.

[851] This is explained in 28nm Fully-Depleted SOI Technology Cryogenic Control Electronics for Quantum Computing, 2018 (2 pages), from CEA-Leti and STMicroelectronics. It discusses the good performance of CMOS components manufactured in FD-SOI technology and operating at 4K, where the available cooling budget is even higher than at 100 mK. A 4K, the cooling power is in the order of a quarter of a Watt to a Watt.

[852] This diagram is by Maud Vinet and is inspired by Superconducting Circuits for Quantum Information: An Outlook by Michel Devoret and Robert Schoelkopf, 2013 (7 pages). Being quite old, it does not indicate the progress made since then in superconducting qubits as well as on spin qubits. "*Operations per error*" is proportional to the ratio between the lifetime of the qubits and the speed of the quantum gates on these qubits. It does not take into account the impact of the qubit error rate, which generally occurs well before reaching the limit of the number of theoretically executable gates.

They use a single parameter of comparison, the number of quantum gates that can be executed before reaching qubit decoherence time $T_2$.

At the state-of-the-art level, the Australians, Dutch researchers from QuTech[853] and Jason Petta at Princeton have demonstrated two-qubit gates in different geometries. To get to the next step, the challenge is to control the electrostatic potential between the quantum wells where the electrons are stored - and thus their spin - with a number of grids that allow the qubits to be arranged not too far apart, typically on the order of a few tens of nanometers. Also, a lot has to be done in materials design and process manufacturing[854].

Note that these qubits can be associated with photonics for long range connectivity. The states of these qubits can be transmitted via photons, which would enable distributed quantum computing architectures[855].

### Qubit operations

The general principle of quantum dots spin qubits consists in confining a spin carrier (an electron or an electron hole) in an electrostatically defined quantum well that is surrounded by tunnel barriers. A static magnetic field enables the creation of a 2-level system with spin up and down. Single gate rotations are handled with submitting the spins/holes to a radio frequency magnetic field. The spin being affected only by this field protects it against electrical noise and other undesirable interactions like phonons (atomic vibrations). Two-qubit gates are created with lowering the tunnel barrier between adjacent dots. Spin state measurement is implemented with a spin-to-charge conversion.

Let's now look at the details[856]:

- **Qubit quantum state** is generally the spin of a trapped individual electron in a potential well. The electrons are confined in a two-dimensions layer semiconductor structure, such as at the interface between two different semiconductors as in GaAs/AlGaAs, in a small quantum well as in Si/SiGe, or between an insulator and semiconductor (MOS), with the two-dimensional conducting layer known as a two-dimensional electron gas (2DEG)[857]. In normal temperature, the energy levels of valence electrons with different spin is the same, or "degenerate". This spin degeneracy is lifted with using cryogeny at very low temperature (below a couple K) and with exposing the material under a magnetic field.
- **Single-qubit quantum gates** use the principle of electron spin resonance (ESR). As with superconducting qubits, these gates rely on the irradiation of the qubit by microwaves pulses, either using electromagnetic cavities, or with radio-frequency lines in which an alternating current creates a magnetic field, or finally, using micro-magnets. The related microwaves use frequencies between 8 and 20 GHz. These gates are usually $R_x$ and $R_y$ gates with the microwave pulse phase driving the gate rotation around axis X or Y and their amplitude and duration driving the rotation angle.

---

[853] See A Crossbar Network for Silicon Quantum Dot Qubits by R Li et al, 2017 (24 pages).

[854] See the review paper Materials for Silicon Quantum Dots and their Impact on Electron Spin Qubits by Andre Saraiva, Wee Han Lim, Chih Hwan Yang, Christopher C. Escott, Arne Laucht and Andrew S. Dzurak, December 2021 (22 pages).

[855] See Coherent shuttle of electron-spin states by Lieven Vandersypen et al, 2017 (21 pages).

[856] See Silicon Qubits by Thaddeus D. Ladd 2018 (19 pages) which describes various methods other than the one discussed here.

[857] See the review paper Quantum Dots / Spin Qubits by Shannon Harvey, April 2022 (20 pages).

- **Two-qubit quantum gates** are created by controlling a tunneling interaction between two neighboring qubits with a significant number of electrodes. These interact with each other by modifying the potential barrier that separates the two qubits. The manipulations, as in single-qubit gates, are performed by applying square pulse currents to qubit barrier and plunger gates.

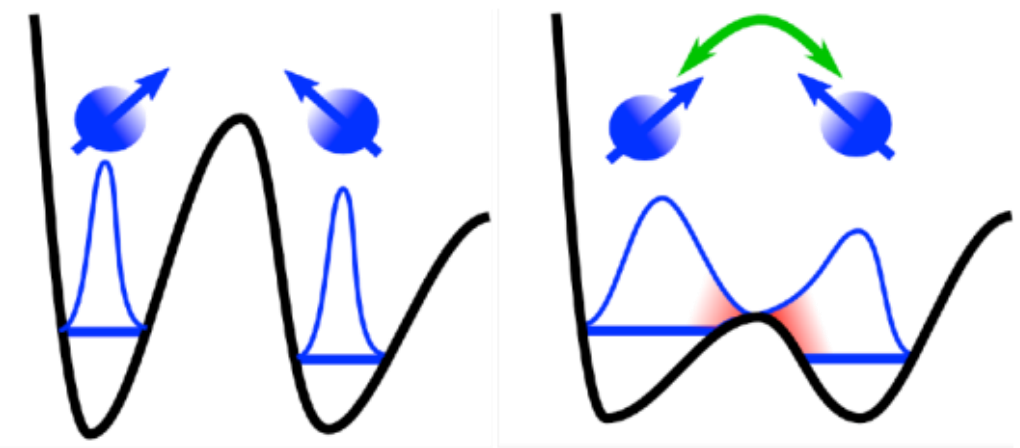

Figure 341: how a two-qubit gate is implemented with reducing the tunnel barrier between two spins. Source: Maud Vinet.

  Common low-level gates of this type are the square root of a SWAP gate and a phase controlled gate.

- **Qubit readout** uses the conversion of the electron spin into electrical charge ("spin to charge") which is then exploitable by traditional electronics. It's frequently using a second electron-spin positioned next to each and every computing qubit. It's based on a microwave pulse sent on the qubit and a reflected signal phase/amplitude analysis, *aka* gate reflectometry[858].

In typical circuits, qubit drive is usually implemented with several wirings: L (lead, providing the electron for the quantum dot), P (plunger, control the electron population), T (for tunnel coupling between quantum dots), S (source) and D (drain).

**Research**

Here are now the main research laboratories that are exploring the silicon spin path, very often in multi-laboratory and multi-country partnership ventures. We'll focus first on the Netherlands, Australia, France and the USA, and will then cover other countries.

In **The Netherlands**, TU Delft and QuTech are among the most active research organizations in Europe around quantum dots based qubits. This activity is centered in Lieven M. K. Vandersypen's lab at QuTech.

Two main technology paths are explored there with Si-MOS quantum dots in partnership with Intel since 2015, and on silicon-germanium qubits under the leadership of Menno Veldhorst and Giordano Scappucci. On top of his main role at Forschungszentrum Jülich in Germany, David DiVincenzo is also a professor at the EEMCS Department at the TU Delft and a contributing scientist at QuTech. TU Delft collaborates on germanium qubits with Purdue University in Indiana and Wisconsin-Madison University.

In 2020, Menno Veldhorst's team demonstrated the feasibility of germanium qubits with two entangled qubits with fidelities of respectively 99.9% and 98% for one and two qubit gates, using germanium electron holes and with two-qubit gate time of 75 ns over a coherence time of about 1µs[859]. These qubits are built on a SOI substrate[860].

---

[858] See an implementation with Gate-reflectometry dispersive readout and coherent control of a spin qubit in silicon by Alessandro Crippa, Silvano De Franceschi, Maud Vinet, Tristan Meunier et al, July 2019 (6 pages).

[859] See Fast two-qubit logic with holes in germanium by N.W. Hendrickx, Menno Veldhorst, Giordano Scappucci et al, January 2020 in Nature et on arXiv in April 2019 (6 pages) also described in Reliable and extremely fast quantum calculations with germanium transistors, Qutech, January 2020.

[860] The SOI for "silicon on insulator" is a technology from the French CEA-Leti and SOITEC. It adds a layer of silicon oxide insulator (SiO2 or "BOX" for "buried oxide") over the silicon wafers and on which are then etched transistors and other circuits conductors

Also in 2020, they implemented the same types of qubits in a 2x2 qubit array with a coherence time of 0.5 ms with 10 ns single qubit gate time[861] (the qubits are labelled $P_1$ to $P_4$ in Figure 342).

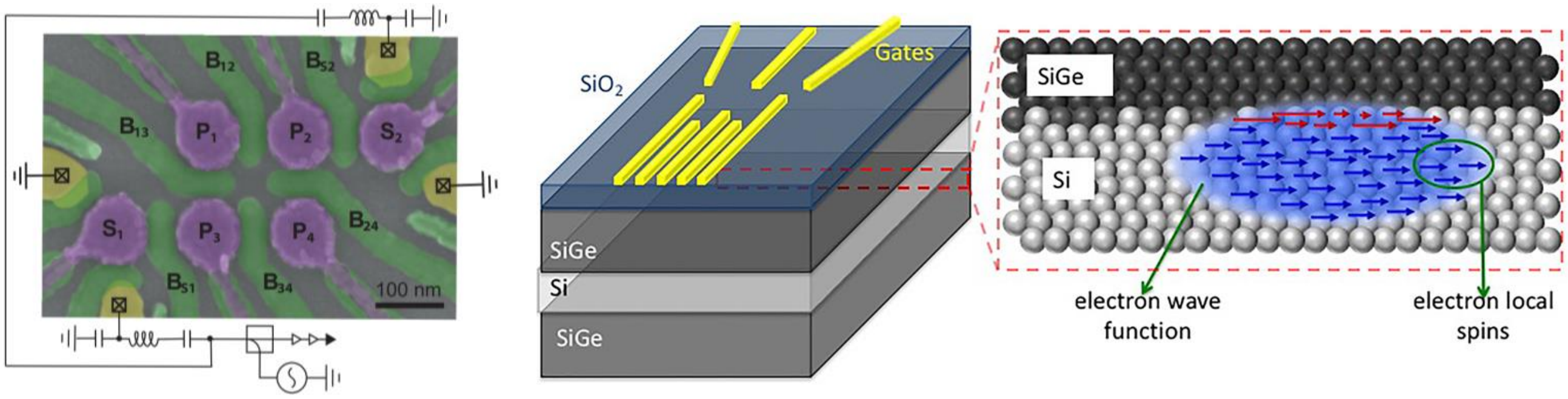

*Figure 342: silicon-germanium prototype qubits. Source: A two-dimensional array of single-hole quantum dots by F. van Riggelen, Giordano Scappucci, Menno Veldhorst et al, August 2020 (7 pages) and Silicon provides means to control quantum bits for faster algorithms by Kayla Wiles, Purdue University, June 2018.*

In 2022, they prototyped a Si-Ge spin qubits phase error correction proof of concept using the same 4 SiGe qubits[862]. They then extended the number of qubits to 6 while preserving good fidelities of 99.77% for single-qubit gates and reaching 71% to 84% for two-qubit gates, with a $T_2$ coherence time of several µs[863]. They also achieved interesting results based on two SiGe qubits, with two-qubit gate fidelities exceeding 99%, paving the road for fault-tolerance (although surface codes QEC would probably require much better fidelities in the 99,99% range and with way above 2 qubits)[864].

Sideways, the QuTech team in collaboration with research teams from ICREA and IC2N in Spain also made interesting inroads with interfacing their SiGe qubits with a germanosilicide superconducting material.

It could be useful in topological qubits design, to create gate-tunable superconducting qubits (gatemons), to create long-range coupling between spin qubits using superconductivity and microwave guides[865]. At last, they demonstrated also in 2022 a 36×36 gate electrode crossbar supporting 648 narrow-channel field effect transistors (FET) to drive SiGe qubits[866].

Qutech is also the testing arm of Intel with its HorseRidge system and Si-MOS double quantum dots qubits. In 2020, QuTech and Intel announced having developed "hot" silicon qubits operating at around 1K, more precisely at 1.1K[867].

---

[861] See A two-dimensional array of single-hole quantum dots by F. van Riggelen, Giordano Scappucci, Menno Veldhorst et al, August 2020 (7 pages) and A four-qubit germanium quantum processor by N.W. Hendrickx et al, September 2020 (8 pages).

[862] See Phase flip code with semiconductor spin qubits by F. van Riggelen, M. Veldhorst et al, QuTech, February 2022 (8 pages).

[863] See Universal control of a six-qubit quantum processor in silicon by Stephan G.J. Philips, Menno Veldhorst, Lieven M.K. Vandersypen et al, February 2022 (38 pages).

[864] See Quantum logic with spin qubits crossing the surface code threshold by Xiao Xue, Lieven M. K. Vandersypen et al, Nature, January 2022 (17 pages).

[865] See Hard superconducting gap in a high-mobility semiconductor by Alberto Tosato, Francesco Borsoi, Menno Veldhorst, Giordano Scappucci et al, June 2022 (20 pages).

[866] See A quantum dot crossbar with sublinear scaling of interconnects at cryogenic temperature by P. L. Bavdaz, James Clarke, Menno Veldhorst, G. Scappucci et al, Nature, 2022 (6 pages) and Shared control of a 16 semiconductor quantum dot crossbar array by Francesco Borsoi, Giordano Scappucci, Menno Veldhorst et al, September 2022 (33 pages).

[867] See Hot, dense and coherent: scalable quantum bits operate under practical conditions by QuTech, April 2020 which refers to Universal quantum logic in hot silicon qubits by L. Petit, Menno Veldhorst et al, April 2020 in Nature and October 2019 in pre-print (10 pages).

At the same time, UNSW researchers were testing similar qubits at 1.5K[868].

Australians are among the most active around silicon qubits, whether in the **CQC2T** teams at UNSW (University of New South Wales) or in other laboratories. Australian Universities are also teaming up with Microsoft Research[869].

UNSW's **CQC2T** (Center for Quantum Computing & Communication Technology) laboratory is led by Michelle Simmons. She also cofounded and runs SQC, a silicon qubit startup that spun-out of UNSW. Their specialty is donors spin qubits using phosphorous atoms implanted in silicon wafers, along the Bruce Kane model already described in the History part and created in 1998 at UNSW. This is a home run! In 2020, a team from the University of Melbourne showed how machine learning could help calibrate the placement of phosphorus atoms in a 2D structure of qubits on a silicon substrate[870]. We'll cover these phosphorus-based qubits in the part dedicated to SQC a bit later in the vendors part.

Two other key scientists at UNSW are Andrew S. Dzurak and Andrea Morello. Andrea Morello's team working with DoE's Sandia Labs published in 2022 results with a single-qubit gate fidelity of 99.95%, a C-Z two-qubit gate fidelity of 99.37% and 98.95% readout fidelity with two phosphorus nuclear spins coupled with a single silicon spin donor electron. They even produced a three-qubit entangled state (GHZ) with a fidelity 92.5%. This was interesting but was not a "scalable" demonstration[871].

Still, Morello and Dzurak usually work on more classical silicon quantum dots qubits. In 2018, they proved the feasibility of creating silicon qubits and developed protocols for reading the state of the spins of these qubits without the need for averaging via a process called "Pauli spin blockade", paving the way for error correction codes implementation and the creation of large-scale quantum computers[872]. They also obtained in 2019 a 2% error rate for two-qubit quantum gates and a 99.96% fidelity for one-qubit gates[873].

Andrew S. Dzurak found in 2021 a way to improve the scalability of spin qubits with removing some the microwave circuits within the qubit chipset and providing these microwaves to the qubits quantum dots with a dielectric microwave resonator (DR) made in potassium tantalate and activated by a discrete loop coupler, made of a simple wire[874]. It drives the ESR (Electron Spin Resonance) magnetic field that enables spin rotations and single qubit gates as well as spin state readout. All this saves at least two microwave circuits in the quantum dots chipset, reducing heating and simplifying the chipset design and, potentially, qubits topology.

---

[868] See Hot qubits made in Sydney break one of the biggest constraints to practical quantum computers by UNSW, April 2020 referring to Silicon quantum processor unit cell operation above one Kelvin by C. H. Yang, Andrea Morello, Andrew Dzurak et al, February 2019 (15 pages).

[869] UNSW also received in 2018 a funding of $53M from the telecom operator Telstra, the Commonwealth Bank and the governments of Australia and the New South Wales region.

[870] See Machine learning to scale up the quantum computer by Muhammad Usman and Lloyd Hollenberg, University of Melbourne, March 2020. Also seen in To Tune Up Your Quantum Computer, Better Call an AI Mechanic by NIST associated with UNSW, March 2020.

[871] See Precision tomography of a three-qubit donor quantum processor in silicon by Mateusz Madzik, Andrea Morello et al, January 2022 (51 pages).

[872] See Tests show integrated quantum chip operations possible, October 2018 and Integrated silicon qubit platform with single-spin addressability, exchange control and single-shot singlet-triplet readout by M. A. Fogarty, Andrea Morello, Andrew S. Dzurak et al, Nature Communications, October 2018 (7 pages).

[873] See Quantum World-First: Researchers Reveal Accuracy Of Two-Qubit Calculations In Silicon, May 2019.

[874] See Single-electron spin resonance in a nanoelectronic device using a global field by Ensar Vahapoglu, Andrew S. Dzurak et al, August 2021 (7 pages) and Supplemental Materials (12 pages).

The global magnetic field generated by this system comes from a dielectric microwave resonator of 0,7*0.55*0.3mm and the discrete loop coupler is even larger, while quantum spin qubits can scale down as low as 100 nm x100 nm. The team communicates on this technology as one that could enable scaling quantum dots to million qubits. So how are individual qubits controlled? Individual spin control and readout is activated by some classical direct current tension sent to each quantum dots in the qubit chipset, replacing the usual microwave signals sent and reflected in the chipset. The next step is to implement the qubit circuit on isotopically purified $^{28}Si$ and check qubits coherence. While the solution simplifies the qubit chipset wiring for some of the microwave lines, the prototype is based on using external microwave generators and readout systems, which doesn't scale at all. It circles back to a cryo-CMOS component that was developed by another Australian team and with Microsoft, which we describe in the cryo-CMOS section, page 498.

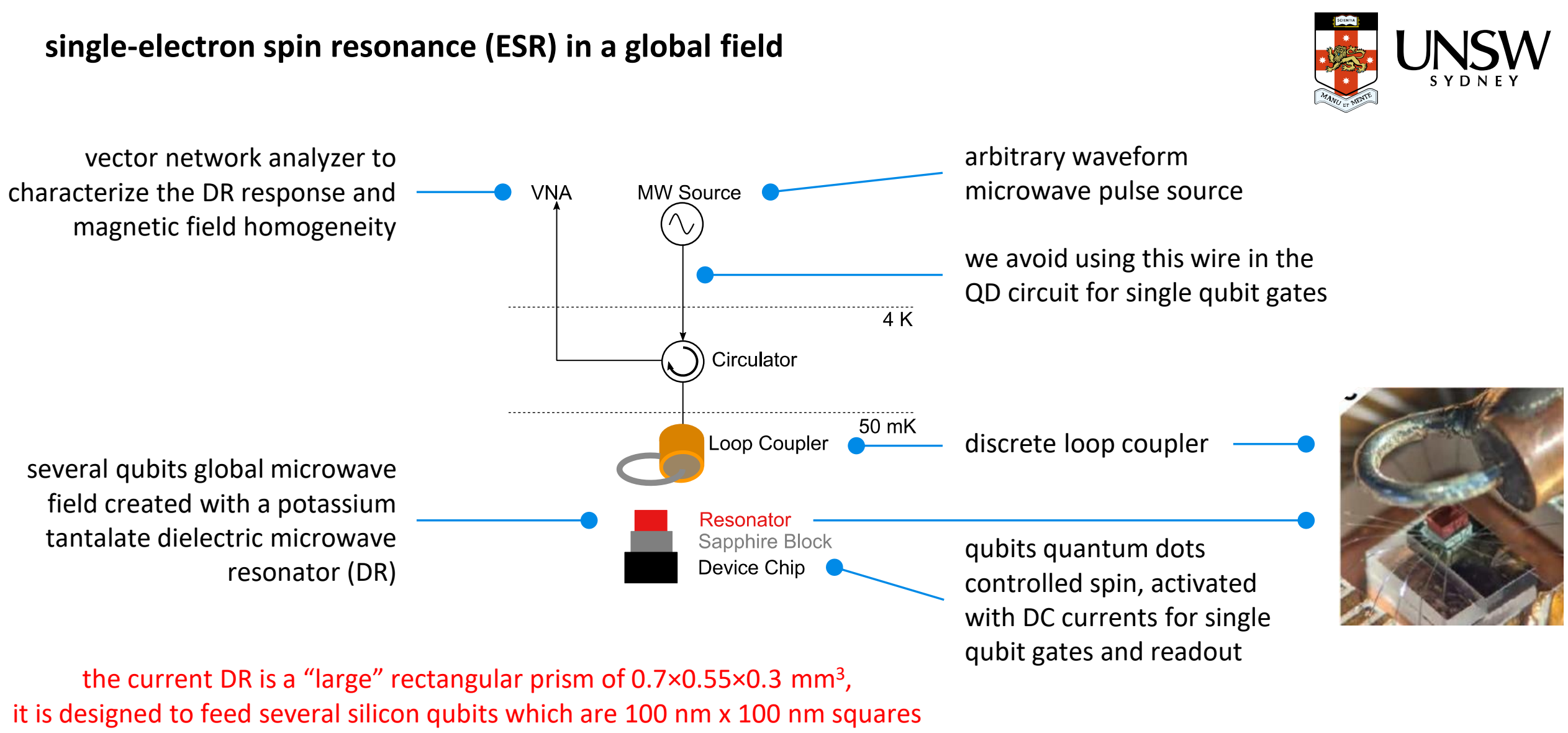

*Figure 343: a proposal to improve the scalability of spin qubits with removing some the microwave circuits within the qubit chipset and providing these microwaves to the qubits quantum dots with a dielectric microwave resonator. Source: Single-electron spin resonance in a nanoelectronic device using a global field by Ensar Vahapoglu, Andrew S. Dzurak et al, August 2021 (7 pages).*

Andrea Morello's team is also studying the remote coupling of electron spins via photons in the visible or in radio waves spectrum. His team created silicon qubits exploiting the control of electron spin of intermediate layers of silicon atoms, increasing stability and reducing flip (or charge) errors[875]. It also managed, by chance, to control the spin of antimony atomic nuclei with an oscillating electric field[876].

In France, the silicon qubit effort is driven out of a group of research labs in Grenoble with CEA-Leti, CEA-LIST, CEA-IRIG, CNRS Institut Néel and UGA (Université Grenoble Alpes).

[875] See UNSW use flat electron shells from artificial atoms as qubits by Chris Duckett, February 2020 and Engineers Just Built an Impressively Stable Quantum Silicon Chip From Artificial Atoms by Michelle Starr, February 2020 which refers to Coherent spin control of s-, p-, d- and f-electrons in a silicon quantum dot by Andrea Morello et al, 2020 (7 pages).

[876] See Engineers crack 58-year-old puzzle on way to quantum breakthrough by UNSW, March 2020 and Chance discovery brings quantum computing using standard microchips a step closer by Adrian Cho, March 2020.

In October 2018, the Grenoble-based team of Silvano De Franceschi (INAC, CEA), Tristan Meunier (Institut Néel, CNRS) and Maud Vinet (CEA-Leti) obtained a €14M **ERC Synergy Grant** for their QuQube project, spread over 6 years to produce a 100-qubit electron spin CMOS quantum processor[877]. Since March 2020, the Grenoble team is also coordinating the 4-year European Quantum Flagship project **QLSI** which was formally launched in February 2021[878].

It consolidates fundamental research in silicon qubits and brings together CEA, CNRS Institut Néel, Atos, SOITEC and STMicroelectronics for France, IMEC (Belgium), Quantum Motion and UCL (UK), Infineon, IHP, U Konstanz, Fraunhofer and RWTH Aachen (Germany), UCPH (Denmark), TU Delft, U Twente and TNO (Netherlands) and U Basel (Switzerland). With a budget of 15M€ to be shared between all these entities, the objective is to enable the manufacture and testing of 16 silicon qubits with gate fidelity of over 99%, and the preparation of a roadmap to be able to scale beyond a thousand qubits.

The Grenoble silicon qubit project is led by **Maud Vinet** (CEA-Leti) and **Tristan Meunier** (CNRS). This story started about 10 years earlier when CEA-Leti got interested in silicon qubits as a way to extend a then flailing Moore's law. CEA-Leti implemented their first silicon quantum dots qubits in 2016[879].

Here are some of the specifics of this endeavor.

**Different techniques**. The Grenoble team mainly bets on Si-MOS silicon spin qubits. But they are still investigating SiGe qubits (CEA-IRIG), well as GaAs (for testing and calibration purpose) as well as silicon spin holes qubits, which are easier to control with a tension on the quantum dot transistor gate[880].

**Manufacturing capacity** is a key asset from CEA-Leti, being one of the few public laboratories in the world with a CMOS component test production platform (IMEC is their counterpart in Europe, in Belgium). It includes all the tools required to produce 200 mm and 300 mm wafers.

It allows the production of all kinds of components in silicon, germanium and III-V materials (photonics, gallium arsenide, gallium nitride, etc.).

---

[877] See An ERC Synergy Grant for Grenoble research on quantum technologies, October 2018 (6 pages). A European Research Council Synergy Grand funds "moonshots" in European research involving at least two research laboratories. 14M€ is the maximum funding for such projects. 10M€ of core funding and 4M€ which can fund heavy investments or access to large infrastructures.

[878] See New EU Quantum Flagship consortium launches a project on silicon spin qubits as a platform for large-scale quantum computing, February 2021. QLSI was a follow-up project from the European collaborative project **Mos-quito**, a 3-year project from 2016 to 2019 with 4M€ funding for studying the performance of different types of individual silicon spin qubits to provide recommendations for their large-scale implementation.

[879] See A CMOS silicon spin qubit by Romain Maurand, Maud Vinet, Marc Sanquer, Silvano De Fransceschi et al, 2016 (12 pages) that was already mentioned.

[880] See Dispersively probed microwave spectroscopy of a silicon hole double quantum dot by Rami Ezzouch, Maud Vinet, Matias Urdampilleta, Tristan Meunier, Marc Sanquer, Silvano De Franceschi, Romain Maurand et al, 2020 (13 pages) and A single hole spin with enhanced coherence in natural silicon by Nicolas Piot, Boris Brun, Vivien Schmitt, Maud Vinet, Matias Urdampilleta, Tristan Meunier, Yann-Michel. Niquet, Silvano De Franceschi et al, Nature, September 2022 (15 pages).

The CEA-Leti clean rooms are spread over several buildings, with the main one being 185 m long on 8000 $m^2$ [881]. Here, CMOS qubits manufacturing process uses 300 mm SOI wafers on with an additional thin layer of 99.992% purified 28-isotope silicon[882]. Validated production is to be later transferred to volume production in commercial fabs such like those from STMicroelectronics in Crolles, France, or GlobalFoundries in the USA or Germany.

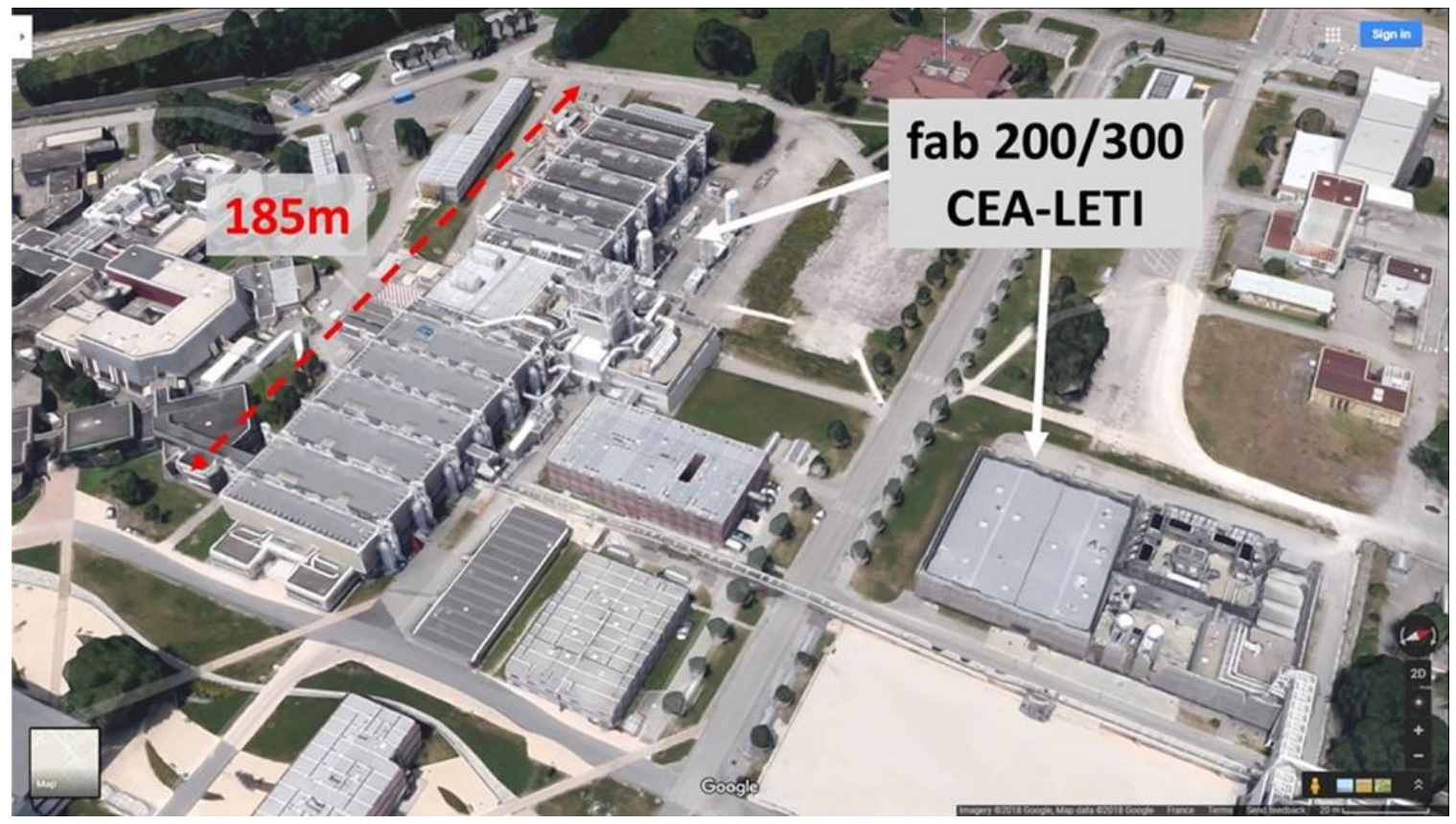

Figure 344: the CEA-Leti's pre-industrial 200/300 fabs in Grenoble.

In its early stages, the size of the quantum computer market will be modest. In a conventional batch of 25 wafers alone, you can produce several thousand quantum chips in a single run, enough to power a large base of quantum supercomputers. But industry-grade clean room ensure quality processes that are not necessarily found in pre-production clean rooms.

**Staging progress** with the Grenoble team expecting to progress in stages over a six-year period starting in 2019: demonstration of a two-qubit silicon-based gate, demonstration of quantum simulation in a 4x4 array based on III-V material, demonstration six qubits in silicon, development of error correction codes and adapted algorithms, and fabrication of 100 2D array qubits in silicon at the end of this journey.

**Control electronics** with the Grenoble team creating control electronics operating at cryogenic temperature. The 2D architecture of Leti's CMOS chipset contains several layers with silicon qubits and then the integrated electronics for control and qubit readout wiring. The qubits are distributed in 2D, but the integration of the components is also vertical within the components.

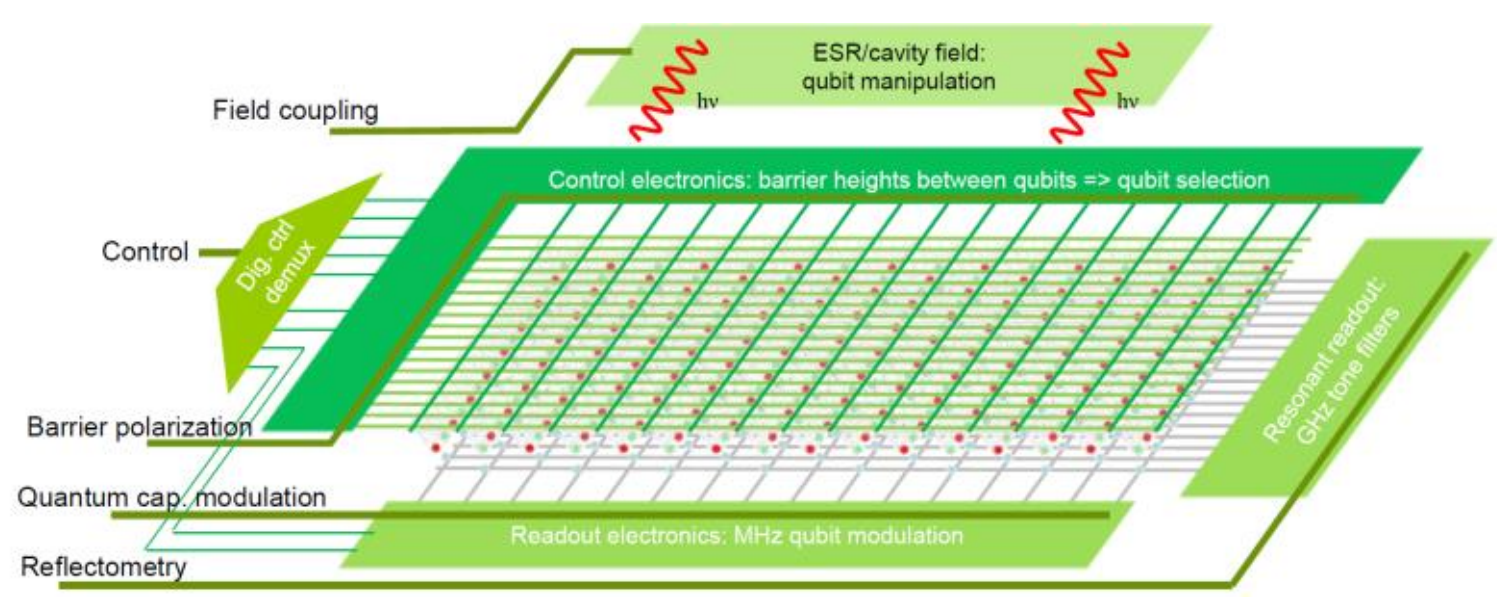

Figure 345: 2D wiring to access spin qubits with scalable wiring. Source: Silicon Based Quantum Computing, Maud Vinet 2018 (28 slides).

The measurement layer is located below the qubits while the layer for activating the qubits with quantum gates is above. For $N^2$ qubits, they would need 2N control lines (horizontal, vertical) instead of $2N^2$, which would generate an appreciable gain in connectivity. The technique would work to generate one- and two-qubit quantum gates[883].

[881] Other research clean rooms exist in France: the C2N clean room in Palaiseau, the IEF in Orsay, the Thales TRT clean room in Palaiseau, the IEMN clean room in Lille, Femto-ST clean room in Besançon and the Laas clean room in Toulouse. In production are mainly the fab 200 and 300 at STMicroelectronics in Crolles, near Grenoble. Some of these laboratories are associated in the National Network of Large Technology Plants (Renatech). They make their platforms available to companies in project and contract mode.

[882] The first test of spin control with isotopically purified silicon was achieved in 2011. See Electron spin coherence exceeding seconds in purified silicon by Alexei Tyryshskin, Kohei Itoh, John Morton et al, 2011 (18 pages).

[883] The technique is described in Towards scalable silicon quantum computing by Matias Urdampilleta, Maud Vinet, Tristan Meunier, Yvain Thonnart et al, 2020 (4 pages) as well as in the presentation Silicon Based Quantum Computing, Maud Vinet, 2018 (28 slides).

The great challenge of these architectures is their variability, i.e., the differences in behavior from one qubit to another and from one circuit to another. This leads to a need for precise calibration, qubit by qubit, of the microwaves controlling and reading the state of the qubits. As for superconducting qubits, this calibration can be done using dedicated machine learning software. They use superconducting materials for the metal layer of these circuits, based on titanium nitride. This provides low resistance and reduces the noise of qubit state measurement.

**3D stacking** is used to arrange chipsets components in 3D[884], which can help solve various scalability problems. CEA-Leti is using it CoolCube technology. The reference publications of these teams on CMOS qubits are numerous[885].

**Spin-photon coupling** could be used to create a communication link between remote qubits. At the Néel Institute, the aim is to move electron spins over long-distances ("Long-distance coherent spin shuttling"). Here, a long-distance means 5 µm! But it makes enough to link qubits together, so it's worth it[886].

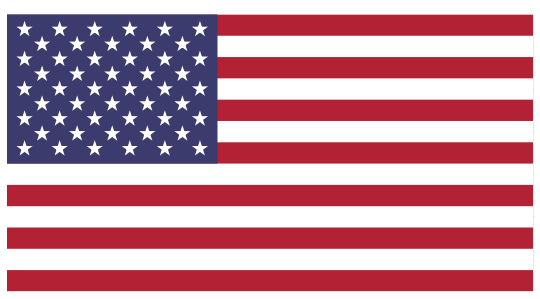

In the USA, on top of Intel, several research labs are working on electron spin qubits. Let's factor in the already mentioned **DoE Sandia Labs** with sites in New Mexico and California, and **Purdue** and **Wisconsin-Madison Universities.**

They work on the physics of silicon qubits and their error correction codes. They are targeting an operating temperature of 100 mK.

**Princeton University** and Jason Petta's team are working on the realization of a two-qubit silicon CNOT gate with a very high level of reliability and low operating time, respectively 200ns and 99%[887].

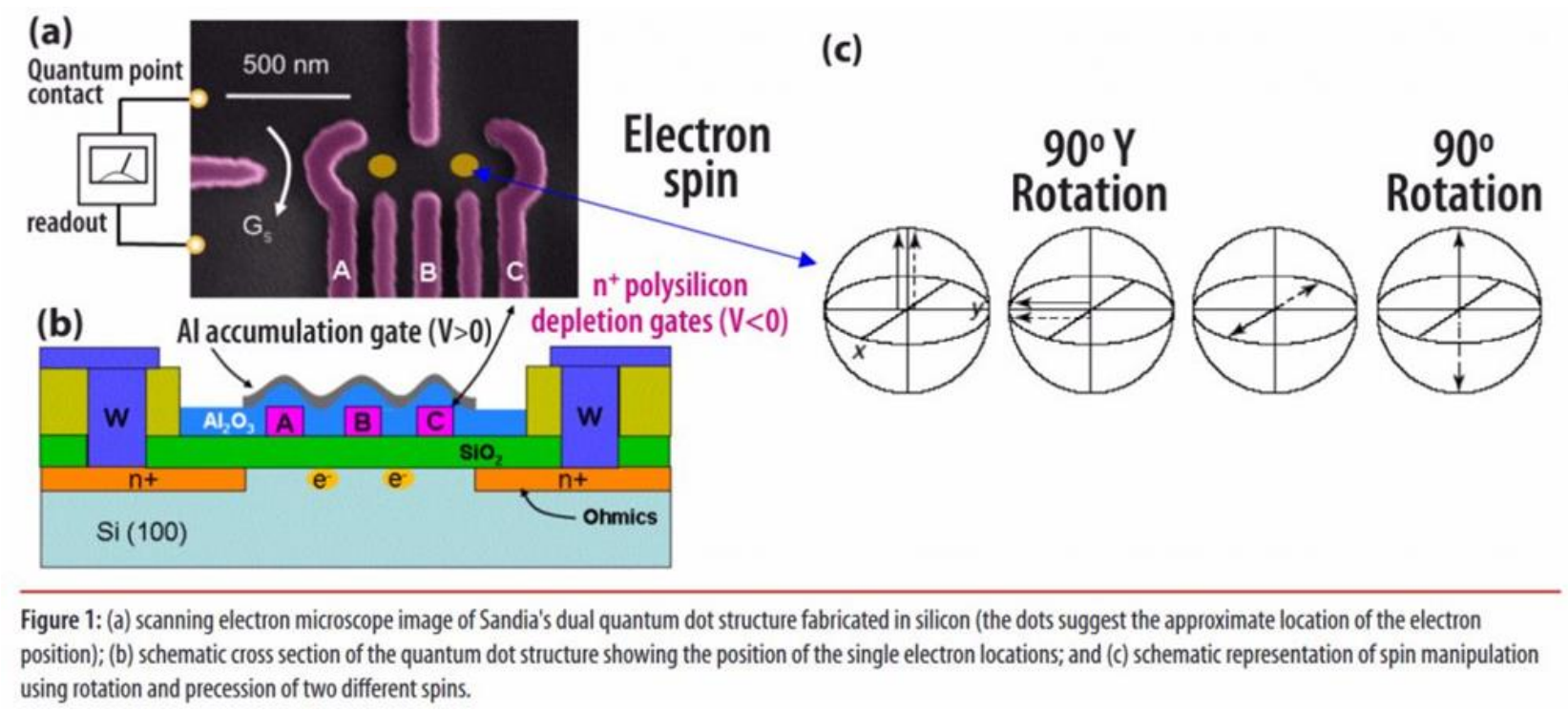

*Figure 346: a typical double quantum dots spin qubit. Source: Toward realization of a silicon-based qubit system for quantum computing by Malcolm Carroll, Sandia Labs, 2016.*

These are also double quantum dots qubits using silicon and germanium. In October 2018, this Princeton team had succeeded in monitoring the state of its CMOS qubits with light and exploiting a microwave field to exchange a quantum between an electron and a photon[888].

**UCLA** and **Virginia Commonwealth University** are working on nanomagnets to drive silicon qubits[889].

---

884 See CoolCube: A True 3DVLSI Alternative to Scaling Resource Library, Technologies Features by Jean-Eric Michallet, 2015.

885 These include A CMOS silicon spin qubit by Romain Maurand, Maud Vinet, Marc Sanquer, Silvano De Fransceschi et al, 2016 (12 pages) which defines the basis of double quantum dot CMOS qubit, SOI technology for quantum information processing, 2016 which completes this description as well as Conditional Dispersive Readout of a CMOS Single-Electron Memory Cell by Simon Schaal et al, 2019 (9 pages) which describes, in the framework of a partnership with the University of London, the work on reading the state of a CMOS quantum dot. And then Towards scalable silicon quantum computing by Maud Vinet et al, 2018 (4 pages).

886 See Coherent long-distance displacement of individual electron spins, 2017 (27 pages) and Quantum Silicon Grenoble, the project on which the Forteza report relies for a quantum computer made in France by Manuel Moragues, January 2020.

887 Seen in Quantum CNOT Gate for Spins in Silicon, 2017 (27 pages).

888 See How old-school silicon could bring quantum computers to the masses, October 2018 and In leap for quantum computing, silicon quantum bits establish a long-distance relationship by University of Princeton, December 2019.

889 See Quantum Control of Spin Qubits Using Nanomagnets by Mohamad Niknam et al, March 2022 (11 pages).

At last, **HRL Malibu**, a joint research subsidiary of Boeing and General Motors located in California is working on both GaAs and Si / SiGe spin qubits. In 2022, they achieved a CNOT gate fidelity of 96.3±0.7%, SWAP fidelity of 99.3±0.5% with 6 silicon qubits arranged in a 1D array[890]. They also work on triple-dot qubits with long lifetimes and better self-error correction[891].

Let's finish this list with a couple other countries:

**Switzerland**: the Swiss National Science Foundation launched **SPIN (**Spin Qubits in Silicon), an electron spin qubits project in December 2019 with a funding of $18M. The end goal is to create a scalable universal quantum computer with more than a thousand logical qubits. The project led by the University of Basel also gathers researchers from ETH Zurich, EPFL and IBM Research Zurich. It looks like a "Plan B" for IBM who is so far focused commercially on superconducting qubits[892]. We've also seen the key role of Daniel Loss in silicon qubits. He still works in Switzerland[893]! Another key researcher is EPFL's Andrea Ruffino, who is working with the Hitachi Cambridge Laboratory in the design of a mixed silicon qubit and cryo-CMOS readout control circuit[894].

**UK**: is another active country on silicon qubits, particularly in **Oxford University**, **Cambridge University** and **UCL**, and with their startup **Quantum Motion**.

**Germany**: at the **University of Aachen,** researchers created double quantum dots of silicon with graphene[895]. Also involved are Infineon, Leibnitz Institute IHP Microelectronics, Universität Konstanz and Fraunhofer IPMS with its own cleanroom.

**Denmark**: **Niels Bohr Institute** and **CEA-Leti** collaborated to build a silicon qubit 2x2 matrix using single electrons quantum dots. These were fabricated on a classical 300 mm SOI wafer coming out of the CEA-Leti fab in Grenoble. While not being operational qubits, these quantum dots electrons were controllable with voltage pulses bases gates. They also implemented electron swaps, that could be useful in optimizing SWAP gates in future systems[896]. In another work with Purdue University in the USA, NBI researchers implemented four GaAs qubits in a 2x2 array[897].

**China** also explores silicon qubits. Their work is difficult to evaluate and they don't publish much on this type of qubits compared to photon-based qubits, boson sampling and superconducting qubits[898].

**Japan** silicon qubits are investigated at **RIKEN**. Their Semiconductor Quantum Information Device Theory Research Team is led by… Daniel Loss, yes, the same one. They were able in 2020 to measure the state of silicon qubits without altering it.

---

[890] See Universal logic with encoded spin qubits in silicon by Aaron J. Weinstein et al, HRL, February 2022 (12 pages).

[891] See Full-permutation dynamical decoupling in triple-quantum-dot spin qubits by Bo Sun et al, HRL, August 2022 (12 pages).

[892] See for example A spin qubit in a fin field-effect transistor by Leon C. Camenzind et al, March 2021 (14 pages) which describes a FinFET hole spin qubit potentially operating at 4K.

[893] See Fully tunable longitudinal spin-photon interactions in Si and Ge quantum dots by Stefano Bosco, Daniel Loss et al, EPFL, March 2022 (18 pages) and A hot hole spin qubit in a silicon FinFET, IBM, March 2021.

[894] See A cryo-CMOS chip that integrates silicon quantum dots and multiplexed dispersive readout electronics by Andrea Ruffino, Edoardo Charbon et al, Nature Electronics, December 2021 (14 pages). The circuit was implemented for three qubits

[895] See Bilayer graphene double quantum dots tune in for single-electron control by Anna Demming, March 2020.

[896] See Single-electron operations in a foundry-fabricated array of quantum dots by Fabio Ansaloni, Benoit Bertrand, Louis Hutin, Maud Vinet et al, December 2020 (7 pages).

[897] See Simultaneous Operations in a Two-Dimensional Array of Singlet-Triplet Qubits by Federico Fedele et al, October 2021 (12 pages) and Roadmap for gallium arsenide spin qubits by Ferdinand Kuemmeth and Hendrik Bluhm, 2020 (4 pages).

[898] See Semiconductor quantum computation by Xin Zhang Hai-Ou Li et al, December 2018 (23 pages). The document provides an overview of CMOS quantum technology but does not specify the specific contribution of Chinese research laboratories.

This non-destructive measurement uses an Ising interaction model based on ferromagnetism that evaluates the spin of atoms neighboring the atom containing the qubit spin electron[899].

Teaming up with Qutech, they also work on shuttling electrons to connect distant silicon QPUs[900], on SiGe high fidelity qubits[901] and on quantum error correction[902].

**quantum dots spins qubits**

- **good scalability potential** to reach millions of qubits, thanks to their size of 100x100 nm.
- **works at around 100 mK - 1K** => larger coooling budget for control electronics.
- **average qubits fidelity** reaching 99% for two qubits gates in labs.
- adapted to **2D architectures** usable with surface codes or color codes QEC.
- can leverage existing semiconductor **fabs**.
- good quantum **gates speed**.

- **active research** in the field started later than with other qubit technologies and spread over severale technologies (full Si, SiGe, atom spin donors).
- **so far, only 4 to 15 entangled qubits** (QuTech, UNSW, Princeton, University of Tokyo).
- **qubits variability** to confirm.
- **scalability** remains to be demonstrated.
- **high fabs costs** and quality manufacturing constraints.
- **less funded** startup scene.

*Figure 347: quantum dots spin qubits pros and cons. (cc) Olivier Ezratty, 2022.*

### Vendors

Key silicon based qubits industry vendors and startups are **Intel** (USA), **SQC** (Australia), **Diraq** (Australia), **Quantum Motion** (UK), **C12 Quantum Electronics** (France), **Equal1.labs** (Ireland/USA) and **Archer Materials** (Australia).

Silicon Quantum Computing

**Silicon Quantum Computing** or **SQC** (2017, Australia, $66M) is a spin-off from UNSW and CQC2T launched by Michelle Simmons. In 2017, their goal was to reach 10 qubits by 2022[903] and they expect to reach 100 qubits by 2028.

The 40+ people company recruited John Martinis from Google at the end of September 2020 but he left the team less than a year later. SQC uses the spin donors technique, trapping phosphorus atoms on a silicon substrate. Their qubit is made with controlling the association of the phosphorus atom nucleus spin and a silicon donor electron spin. They create two-qubit gates with two phosphorus atoms that are a few nanometers apart, using quantum tunneling. They showcased single-gates fidelity of about 99.99% and two-qubit gates speed of less than one nanosecond[904].

---

899 See Scientists succeed in measuring electron spin qubit without demolishing it, RIKEN, March 2020, mentioning Quantum non-demolition readout of an electron spin in silicon by J. Yoneda et al, 2020 (7 pages).

900 See A shuttling-based two-qubit logic gate for linking distant silicon quantum processors by Akito Noiri et al, February 2022 (25 pages).

901 See Fast universal quantum control above the fault-tolerance threshold in silicon by Akito Noiri, Giordano Scappucci et al, 2022 (27 pages).

902 See Quantum error correction with silicon spin qubits by Kenta Takeda, Giordano Scappucci et al, Nature, January-August 2022 (23 pages).

903 This is documented in Silicon quantum processor with robust long-distance qubit couplings by Guilherme Tosi, Andrea Morello et al, 2017 (17 pages).

904 See Exploiting a Single-Crystal Environment to Minimize the Charge Noise on Qubits in Silicon by Ludwik Kranz, Michelle Simmons et al, 2020 and A two-qubit gate between phosphorus donor electrons in silicon by Y. He, Michelle Simmons et al, 2019.

In 2022, they touted having produced the "*first ever quantum circuit*"[905]. It was a way to fulfill their 2017 goal and showcase a 10-qubit processor, implemented in a 1D lattice. It was presented in an article published on Nature, implementing a particular physics simulation, the many-body Su–Schrieffer-Heeger (SSH) model. Of course, with just 10 qubits, it can't showcase any quantum advantage. Unfortunately, the paper doesn't provide qubit fidelities data[906]. The paper was published as SQC announced it was starting a new financial round with a goal to obtain $130M[907].

Mid-2021, Andrea Morello and Andrew Dzurak quit SQC where they were involved since the beginning. They preferred to focus on SiMOS qubit instead of phosphorus and donors qubits. It led to the creation of their startup Diraq, which we'll discuss later. As a result, SQC sold in May 2022 its patent portfolio and special equipment related to SiMOS to SiMOS Newco, a company created in December 2021, and probably related to Diraq, created a month later[908].

**Diraq** (2022, Australia) is a startup spun out of UNSW created by Andrew Dzurak that develops quantum dots electron spin qubits. The company set a goal to build a one billion qubits computer, the largest I've seen so far with commercial vendors.

Their first planned steps are to reach 9 and then 256 qubits. The team is already set up with a bunch of scientists and engineers like Arne Laucht, Henry Yang. Andrea Morello from UNSW is also a scientific advisor for the venture. Morello and Dzurak were previously working with Michelle Simmons in her company SQC and they parted away in 2021. The founding team has a good track record in the advancement of quantum dots based qubits with many "firsts" achieved since 2014, including many patented processes (SiMOS - Silicon-Metal-Oxide-Semiconductor qubits, resonators and qubit electrical control, etc).

**Quantum Motion Technologies** (2017, UK, $9.7M) is an Oxford University spin-off that wants to create high-density silicon quantum computers. They have received unspecified seed funding from the UK fund Parkwalk Advisors in 2017.

The startup co-founded by John Morton (UCL) and Simon Benjamin (Oxford University) wants to industrialize a process created by Joe O'Gorman's team at Oxford University, which consists of clearly separating silicon qubits and their measurement.

Measurement was supposed to be carried out with a magnetic probe mechanically moved on the surface and making "square" movements, guided by a MEMS (micro-electro-mechanical device). This probe system was designed to avoid the use of control electronics and allow a better separation between the qubits[909]. A data rate separation process with intermediate mediation rates was limiting leakage effects[910]. This process protected by one granted US patent and four patents pending. But it seems that this technology is finally not the one they will implement!

---

905 See how the media buy in such outrageous claim in A Huge Step Forward in Quantum Computing Was Just Announced: The First-Ever Quantum Circuit by Felicity Nelson, Science Alert, June 2022. Hopefully, The Quantum Insider is not parroting the fancy claim and titled Silicon Quantum Computing Announces its First Quantum Integrated Circuit by James Dargan, June 2022.

906 See Engineering topological states in atom-based semiconductor quantum dots by M. Kiczynski, Michelle Simmons et al, Nature, June 2022 (11 pages).

907 See Quantum star kicks off crucial $130m funding push by Paul Smith, Financial Review, June 2022.

908 See UNSW Sydney spin-out buys quantum computing hardware technology by Lauren Croft, Lawyers Weekly, May 2022.

909 See A silicon-based surface code quantum computer by Joe O'Gorman et al, 2015 (14 pages). The paper is co-authored by John Motin and Simon Benjamin who are two co-founders of the startup Quantum Motion Technologies. Their Si-MOS qubits are mixing planar and 3D SOI components and are laid out to enable surface code error correction.

910 See A Silicon Surface Code Architecture Resilient Against Leakage Errors by Zhenyu Cai (Quantum Motion Technologies) et al, April 2018 (19 pages).

In January 2021, Quantum Motion presented with Hitachi Cambridge, University of Cambridge and EPFL a 50 mK cryo-CMOS including quantum dots qubit arrays, row-column control electronics lines and analog LC resonators for multiplexed readout, using 6-8 GHz microwave resonators. This was a first step to implement time- and frequency-domain multiplexing scalable qubits readout[911].

In March 2021, Quantum Motion announced a record of stability of 9 seconds for an isolated silicon qubit. The chipsets were manufactured by CEA-Leti in Grenoble and the French team led by Maud Vinet coauthored the paper associated with this performance[912]. Quantum Motion and UCL are part of the Quantum Flagship QLSI on silicon qubit that is led by Maud Vinet. So, this explains that.

Their roadmap consists of producing 5 qubit "small cells" by 2022 in a structure that could then be reproduced in matrix patterns. They believe they can create a quantum computer with 100 logical qubits by 2029, a classical milestone for most quantum computer vendors. They provided an update on their architecture in an arXiv paper in August 2022[913].

In the software area, Quantum Motion developed QuEST, an open source, hybrid multithreaded and distributed, GPU accelerated simulator of quantum circuits. It works both on any laptop or on supercomputers. It supports pure (computational state vector) and mixed states (density matrices) to reproduce the effects of noise and decoherence[914].

**Intel** is a key contender in the race for silicon qubits. They started working on superconducting qubits but it seems it was a secondary route for them. They started with producing a wafer with 26 qubit chipsets in 2017 and made some progress since they, although it is rather hard to evaluate.

Intel's quantum work is managed under the direction of **Anne Matsuura**[915] and **James Clarke** for hardware.

In June 2018, Intel made another announcement with a highly integrated chip using SiMOS qubits, with 1500 qubits, fabricated in the D1D fab located in Portland, Oregon, with an etch density of 50 nm, six times greater than the early 2018 generation.

But this chipset, like many that did follow, were produced to test their manufacturing capacity and their material designs. They were not functional particularly with regards to two-qubit gates. In 2022, Intel did show again some interesting data related to their qubits manufacturing capacity, producing chipsets with 3 to 55 quantum dots on 300 mm wafers and a >95% production yield, using 193 nm Deep UV immersion photolithography instead of electron beam lithography[916]. These SiGe qubits have a relaxation time of >1s ($T_1$), coherence times of >3 ms ($T_2$) and single qubit gates of >99% (and no published data for two-qubit gates…). These qubits are to be characterized by the DoE Argonne lab in Chicago in its Q-NEXT research center[917].

[911] See Integrated multiplexed microwave readout of silicon quantum dots in a cryogenic CMOS chip by A. Ruffino, January 2021 (14 pages).

[912] See Spin Readout of a CMOS Quantum Dot by Gate Reflectometry and Spin-Dependent Tunneling, by Virginia N. Ciriano-Tejel, Maud Vinet, John Morton et al, 2021 (18 pages). This followed Remote capacitive sensing in two-dimensional quantum-dot arrays by Jingyu Duan, Michael A. Fogarty, James Williams, Louis Hutin, Maud Vinet and John J. L. Morton, 2020 (31 pages) which described the coupling technique using silicon nanowires (SiNW) to measure qubits spins with remote capacitive charge sensing.

[913] See Silicon edge-dot architecture for quantum computing with global control and integrated trimming by Michael A. Fogarty, August 2022 (13 pages).

[914] See QuEST and High Performance Simulation of Quantum Computers by Tyson Jones et al, December 2018 (8 pages).

[915] See Intel's quantum efforts tied to next-gen materials applications, January 2019 and Intel's spin on qubits and quantum manufacturability, both from Nicole Hemsoth, November 2018 and Leading the evolution of compute, Anne Matsuura, June 2018 (26 slides).

[916] See Qubits made by advanced semiconductor manufacturing by A.M.J. Zwerver, Menno Veldhorst, L.M.K. Vandersypen, James Clarke et al, 2021 (23 pages).

[917] See Intel to install quantum computing test bed for Q-NEXT, April 2022.

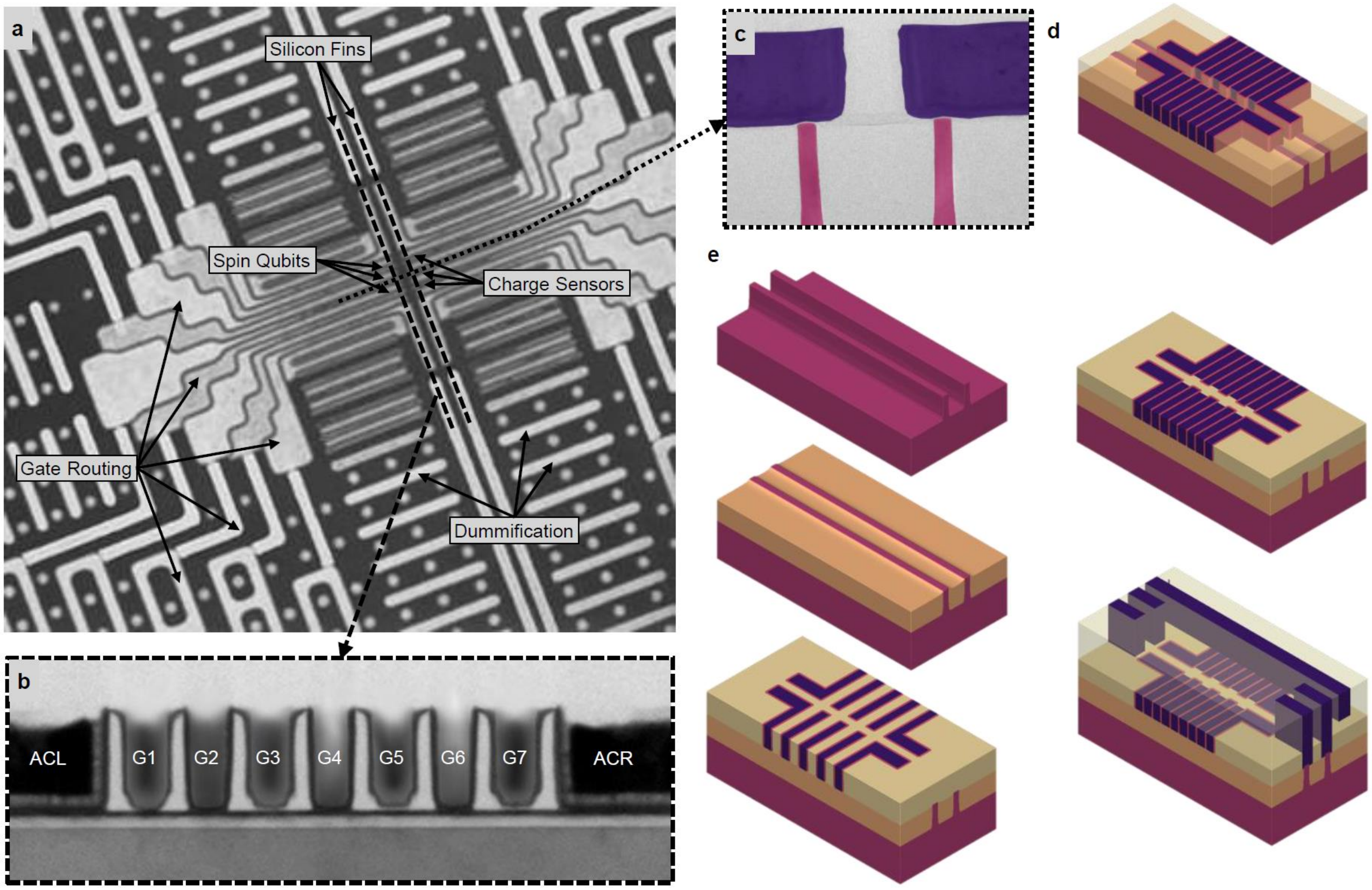

*Figure 348: Intel SiGe quantum dots circuit implementation and process quality. Source: [Qubits made by advanced semiconductor manufacturing](#) by A.M.J. Zwerver, Menno Veldhorst, L.M.K. Vandersypen, James Clarke et al, 2021 (23 pages).*

As part of their efforts in manufacturing, they are now using, like CEA-Leti, a cryo-wafer prober provided by Afore and Bluefors, that enables testing entire wafers at 1.6K, significantly accelerating the testing and characterization process (on the left, below).

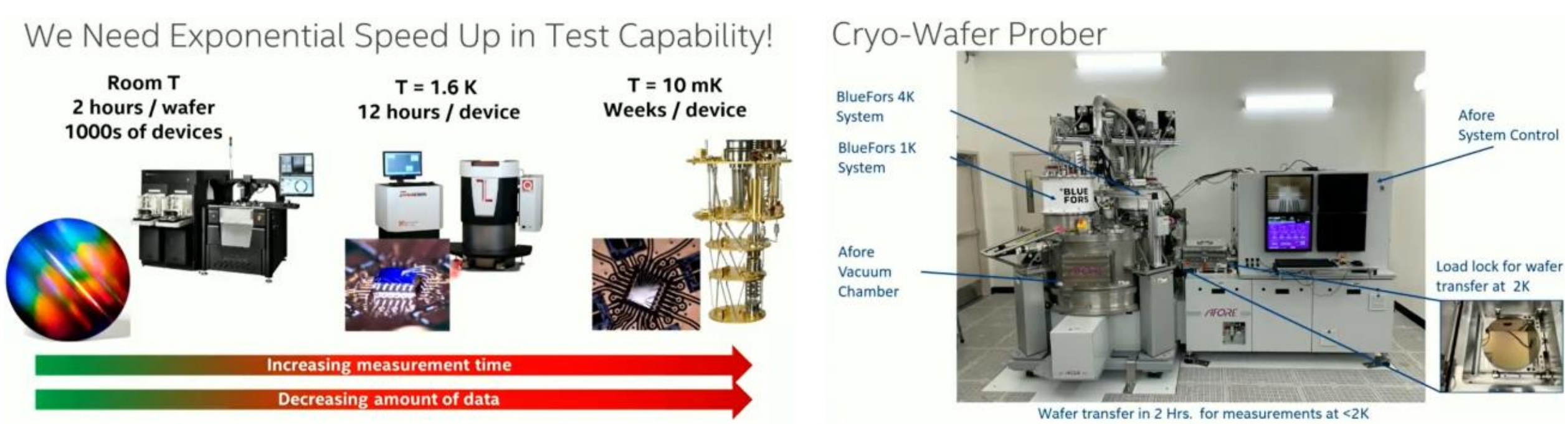

*Figure 349: how Intel is saving time with a Bluefors/a-Fore cryo-prober. Source: Intel.*

QuTech and Intel work well together on these qubits. QuTech got a $50M investment from Intel in 2015 to explore it. Intel announced in 2018 that it had succeeded in controlling a two-qubit SiMOS processor with single and two quantum gates running Deutsch-Jozsa and Grover algorithms on a very small scale. These silicon-germanium qubits manufactured by Intel were tested by the Vandersypen Laboratory at the University of Delft, part of QuTech[918]. Since 2018, Intel has kept a rather low profile on its silicon qubit advances[919].

[918] See [A programmable two-qubit quantum processor in silicon](#) by T F Watson et al, TU Delft, May 2018 (22 pages).

[919] See [What Intel Is Planning for The Future of Quantum Computing: Hot Qubits, Cold Control Chips, and Rapid Testing](#) by Samuel Moore, August 2020, which provides a rather pedagogical overview of Intel's approach to silicon qubits.

At the beginning of 2020, Intel announced that it had developed with QuTech the **HorseRidge** cryo-component. It is a CMOS component operating at 4K that is used to generate the microwaves used to drive both superconducting and silicon qubits. A second version was announced in 2021. We cover it in the section dedicated to cryo-CMOS electronics.

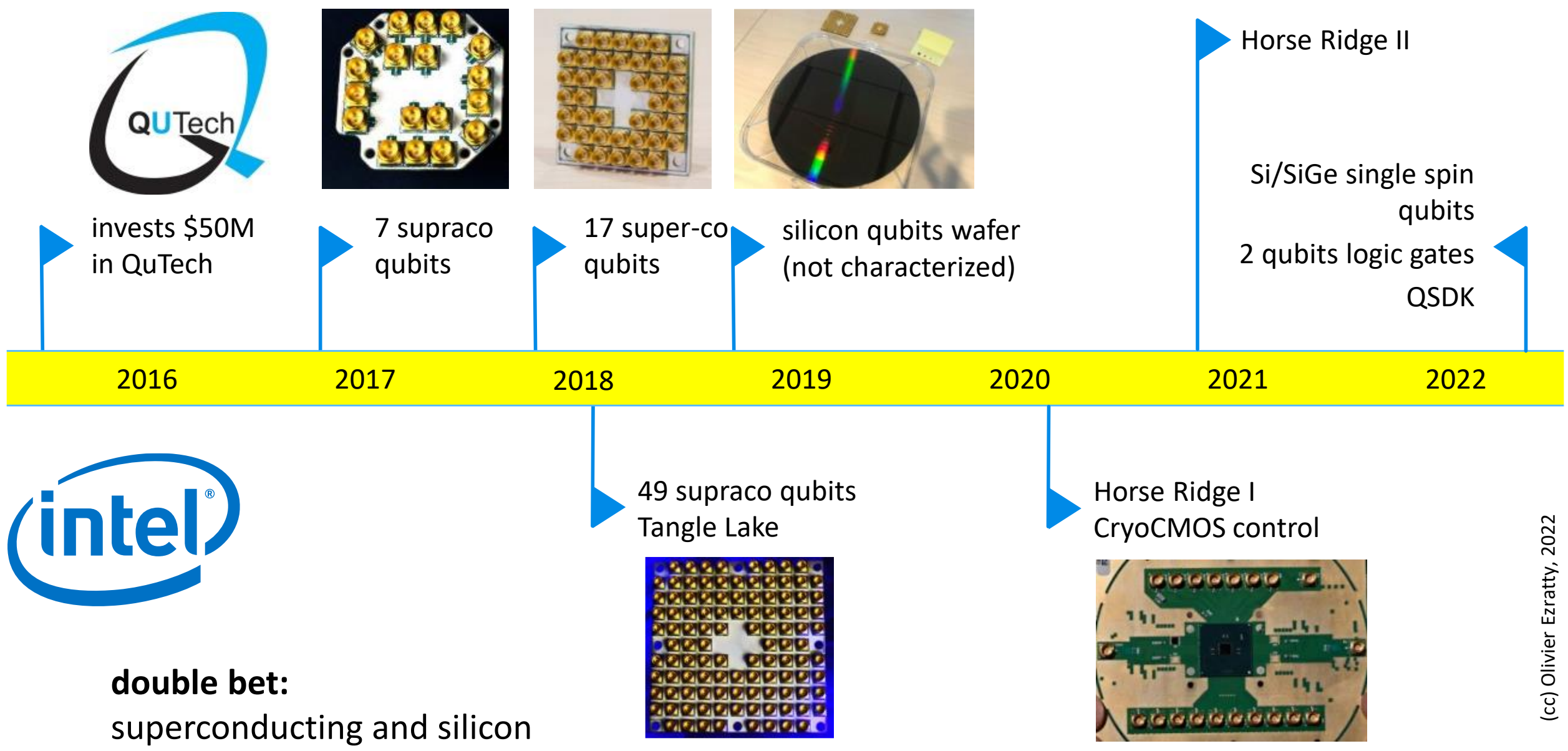

*Figure 350: Intel quantum computing timeline. (cc) Olivier Ezratty, 2022.*

equal1.labs

**equal1.labs** (2017, Ireland/USA, 6M€) is creating a charge electron spin qubits chipset manufactured in 22 nm FD-SOI technology at GlobalFoundries in Dresden, Germany.

They announced in 2021 a 422 qubits test chipset embedded in a full-rack system with its cryogeny, Alice mk1. At this stage, they are just able to inject single electrons in their quantum dots and simulate numerically some one- and two-qubit quantum gates, but not much more[920]. Their next generation Aquarius is to fit in a desktop packaging, planned for 2022, and is to house one million qubits. They position their systems to run quantum neural networks for imaging applications.

In May 2021, equal1 uncovered a prototype chipset operating at 3.7K and including 10 million transistors handling qubits controls and readout with arbitrary waves generation (AWGs), all coupled to an external FPGA, as well as some classical cryogenic memory. There's a caveat with their coherence time being only 150 ns. Equal1 also designs its own cryogenic system.

The company was created by Dirk Leipold, Mike Asker and Bogdan Staszewski from the University of Dublin. Elena Blokhina is their CTO and expect to raise $50M by 2022.

|EeroQ⟩

**EeroQ Quantum Hardware** (2016, USA, $7.5M) develops an exotic quantum processor using trapped (and more or less flying/moving) electrons on superfluid helium ("eHe"). The startup was created by Johannes Pollanen from the University of Michigan (CSO), Dave Ferguson, Nick Farina (CEO) and Faye Wattleton (EVP)[921].

---

[920] See A Single-Electron Injection Device for CMOS Charge Qubits Implemented in 22-nm FD-SOI by Imran Bashir, Elena Blokhina et al, 2020 (4 pages).

[921] See Helium surface fluctuations investigated with superconducting coplanar waveguide resonator by N.R. Beysengulov, Johannes. Pollanen et al, 2022 (10 pages). It deals with a superconducting resonator and not with a qubit.

In May 2021, they appointed Princeton University Professor Steve Lyon as CTO. It has benefited from US public (NSF) and private funding. Johannes Pollanen's bio indicates that he conducted research in superconducting and two-dimensional qubits (silicon, graphene) [922].

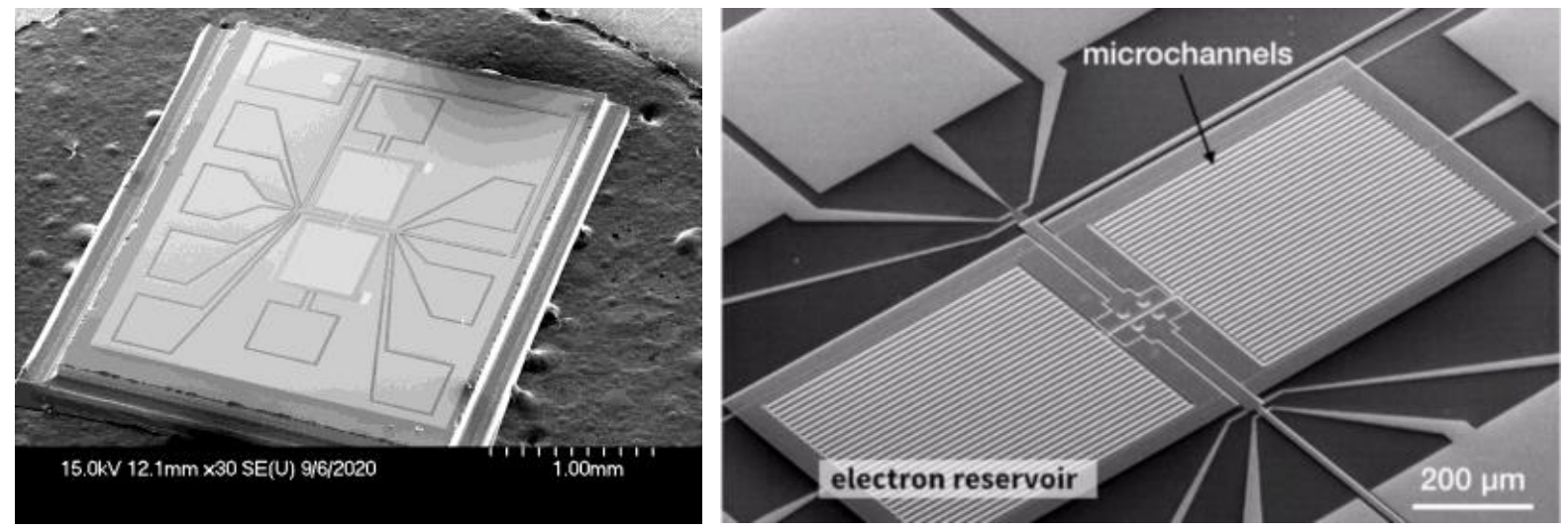

*Figure 351: EeroQ silicon qubit prototype processor. Source: EeroQ.*

They want to associate the long coherence and high connectivity of trapped ions and the fast gates of electron spin and superconducting qubits. The chipset is built using CMOS technology and electrodes driving electron spins. It is probably complicated to tune given we've not heard from them for a while, with any functional and programmable system. They still got an undisclosed investment from VCapital in 2022.

C12

**C12 Quantum Electronics** (2020, France, $10M) was launched by Matthieu Desjardins and his twin brother Pierre. It is a project originating from the LPENS at ENS Paris and 15 years of research from Takis Kontos in this lab, with contributions from Jérémie Vienot at Institut Néel Grenoble.

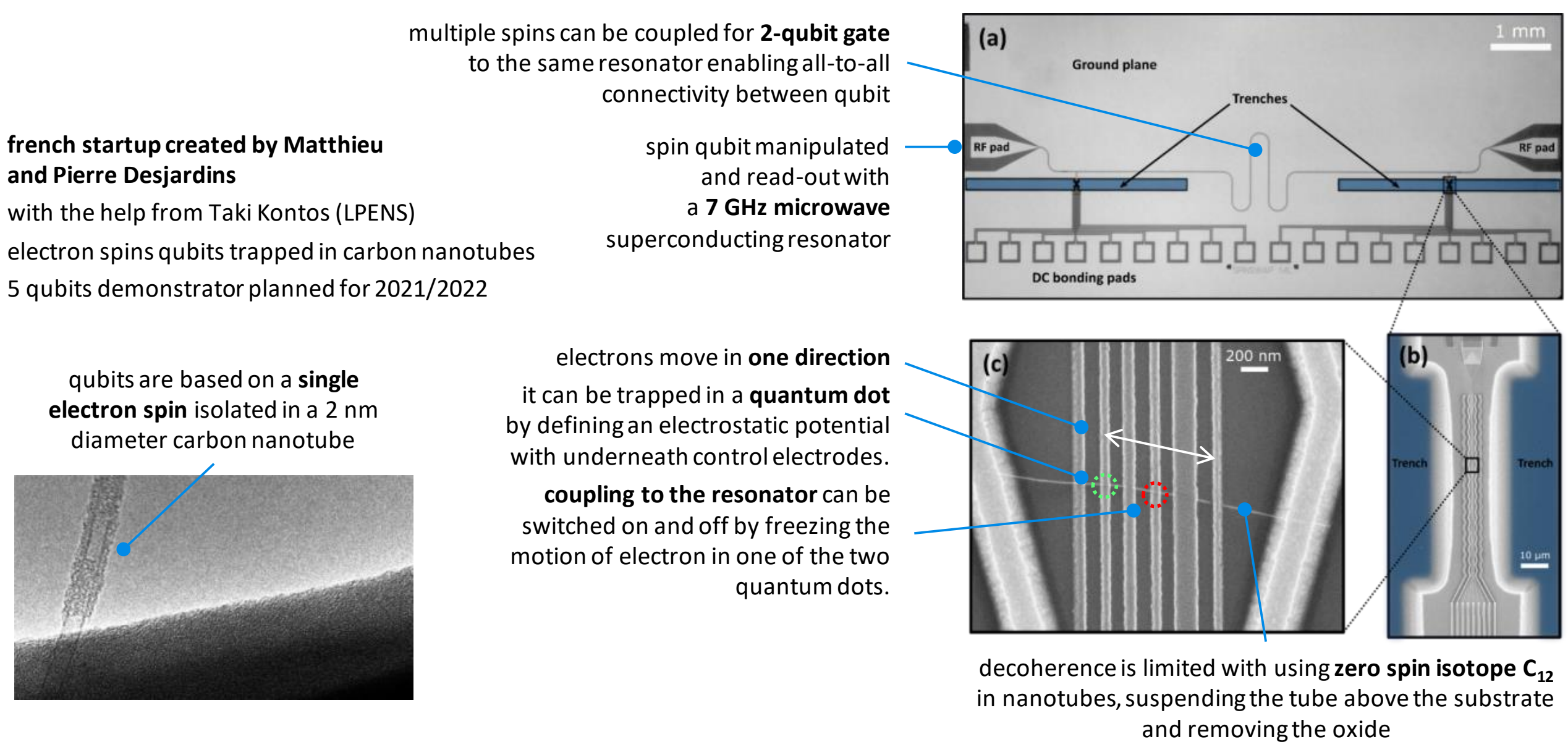

*Figure 352: C12 Quantum Electronics carbon nanotubes and how it is controlled. Source: C12.*

Their goal is to use carbon nanotubes to trap electrons used in electron spin qubits and build the surrounding control circuitry on silicon substrate. This technology can improve qubits isolation and coherence time by a factor of 100, up to one second, while keeping a strong coupling for fast qubit manipulation. The qubits are controlled by spin-photon coupling in the microwave regime, using frequency multiplexing to avoid crosstalk.

Qubit readout uses spin to charge coupling with a single charge coupling with 8 qubits[923]. The challenges sit in materials, design, control electronics, connectivity, topology and error correction codes.

[922] See Integrating superfluids with superconducting qubit systems by Johannes Pollanen et al, 2019 (11 pages).

[923] A related technique is described in Charge Detection in an Array of CMOS Quantum Dots by Emmanuel Chanrion, Pierre-André Mortemousque, Louis Hutin, Silvano de Franceschi, Franck Balestro, Maud Vinet, Tristan Meunier, Matias Urdampilleta et al, Grenoble CEA-LETI, CNRS Institut Néel and UGA, August 2020 (8 pages).

The nanotubes are mechanically integrated into the circuit at the end of the manufacturing process[924]. The carbon nanotubes are grown by C12 using a CVD process (chemical vapor deposition). The connection between two qubits is based on microwave cavities, exploiting QEDC (Quantum Electrodynamic Cavity). Of course, there are still many challenges to develop this kind of qubits but it is worth exploring. It could even have some use cases beyond computing, in quantum sensing. In March 2022, C12 announced it was starting a manufacturing partnership with CEA-Leti. These will provide their quantum chips on silicon on 200 mm wafers. These will contain the superconducting electronics controlling the state of their carbon nanotubes that will be positioned on the chipsets.

On the software engineering side, Atos works with C12 to develop its quantum compiler, to create digital simulation models of its qubits and for co-designing quantum gates.

ARCHER **Archer Materials** (2017, Australia) develops quantum computing and sensing technologies based on carbon nanospheres that operate at room temperature[925].

The company was cofounded by Mohammad Choucair who invented their $^{12}$CQ chipset design and by Martin Fuechsle who contributed to the development of the single electron transistor and worked at UNSW with Michelle Simmons[926]. The 12 is not a number of qubits but the isotopic weight of the zero spin carbon atoms used in these nanospheres trapping some electron and its spin. Electrons are delocalized around the nanosphere and not setting inside it, contrarily to the C12 Quantum Electronics electrons that are stored inside nanotubes.

They built their first three-nanospheres chipset in 2019, in red, in the picture, the 50 nm nanospheres being surrounded by driving electrodes, but without any visible coupling between these qubits. In February 2021, Archer announced that they had achieved "electronic transport" in a single qubit at room temperature in its $^{12}$CQ quantum computing qubit processor chip.

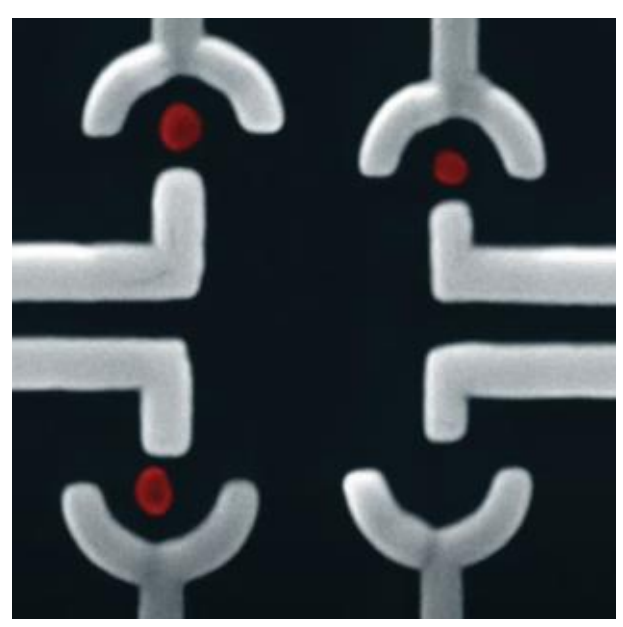

Figure 353: Archer qubits. Source: Archer.

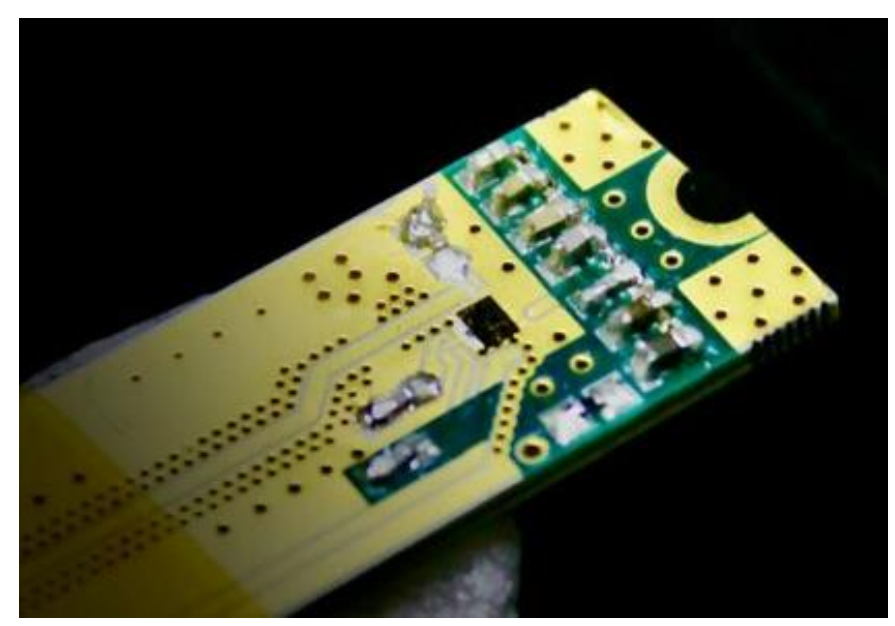

Figure 354: Archer-EPFL spin-resonance circuit. Source: Archer.

It however does not mean that it is a fully functional qubit that can be operated with quantum gates[927]. In July 2021, they announced that they were embedding some parts of their qubits control electronics in the qubit chipset, that records the Continuous Wave Electron Spin Resonance (cwESR) signals generated by a superconducting on-chip resonator.

In February 2022, through a collaboration with EPFL, they demonstrated the use of a single-chip integrated electron spin resonance (ESR, seemingly at 9.4 GHz according to their patent) detector based on high electron mobility transistor (HEMT) to detect and characterize their qubit at room temperature[928]. The resonator was manufactured by OMMIC (France), a fab company specialized in III-V wafer processes. This technology is as mature as superconducting qubits were in the mid 1990's when it became possible to create the first qubits.

---

924 It is described in Nanoassembly technique of carbon nanotubes for hybrid circuit-QED by Tino Cubaynes, Matthieu Desjardin, Audrey Cottet, Taki Kontos et al, September 2021 (6 pages).

925 See Archer Materials granted trading halt ahead of quantum computing chip agreement by Quantum Analyst, 2020 and Room temperature manipulation of long lifetime spins in metallic-like carbon nanospheres by Bálint Náfrádi, Mohammad Choucair et al, 2016 (32 pages) which describes in detail this technique of electron spin trapping in a 35 nm wide carbon nanosphere. In April 2021, Archer Materials announced it would sell off all its mineral traditional business to iTech Minerals, to focus on quantum technologies.

926 Their $^{12}$CQ processor is patented in the USA, Japan and Europe (since February 2022). See WO2017091870 - A QUANTUM ELECTRONIC DEVICE (55 pages). The patent dates from 2017 and is very sketchy with regards to single and multiple qubit gates operations. We can still learn they use $SiO^2$ isolation layers of 200 to 400 nm between control electrodes and nanospheres. Conductors can use unspecified graphene structures. The spin stability in their device is amazingly small, at around 115 ns, not sufficient to run several quantum gates.

927 Their electron spin T1/T2 is very low, at 175 ns at 27°C. They also said they tested microwave pulses from 4 GHz to 420 GHz. Seen in their video Archer's Quantum Computing Q&A Webinar, April 2020.

What would be needed are some tomographies for their qubit state, showing real superposition, and a full cycle of reset, flip and phase single qubit quantum gates and qubit readout, then implement this with a couple entangled qubits. One can wonder how they will drive their qubits with microwave pulses at room temperature given the ambient thermal noise will be larger than the pulse themselves. The road ahead is still quite long for them. Based on this 2022 milestone, the company CEO forecasted "*mobile quantum computing*" use-cases. That is quite an oversold proposal. It deserves scrutiny, if not strong skepticism.

In December 2020, Archer launched a partnership with **Max Kelsen**, another Australian company, specialized in QML software development. Max Kelsen and Archer will develop QML algorithms based on Qiskit, eying a future execution on Archer's processor. They also announce ghat GlobalFoundries would become the manufacturer of their 12Q qubit chipset.

The company is also developing graphene-based biosensor chip aka "lab on chip". This is a more short-term and credible product proposal than their room temperature spin qubits.

**Qpi** (2019, India) is a QML software and hardware development company, providing the QpiAI library. They are working on creating the ASGP (AI System Generating Processor), a hybrid classical-quantum compute chip.

Practically speaking, they planned first to introduce a qubit control chip in September 2021 operating at 4K and produced in a 22 nm TSMC CMOS process[929]. This chipset was to control the microwaves sent to both superconducting and electron spin qubits processors. They later announced a room temperature control chipset named QpiAISense. They are now developing a full-stack classical and silicon qubit solution in a single package. It is supposed to contain a classical chip for optimization (ASGP for "AI System Generating Processor"), their microwaves cryogenic control chip and a silicon spin-qubit based QPU (quantum processing unit). They signed a partnership with QuantrolOx in April 2022 to develop their QPU and set-up a lab in Finland for that purpose, where part of QuantrolOx team is installed. Their goal is to start with releasing a functional 25 qubits system. They plan later to create a one million electron spin quantum dots qubits processor. Overselling seems not to be an issue for them. QpiAI Tech is a subsidiary selling software services for quantum computing and AI to transportation, materials, manufacturing, finance, and pharmaceutical business.

## NV centers qubits

This qubit technology is based on the control of electron spins trapped in artificial defects of crystalline carbon structures in which one carbon atom is replaced by one nitrogen atom and another carbon atom is replaced by a void, gap or cavity. Practically speaking, it's a bit more complex since the qubits themselves are stored in nuclear spins of surrounding carbon and nitrogen atoms[930].

### History

Defects in diamonds have been studied from 1930 with the examination of infrared absorption. This made it possible to distinguish two categories of diamonds: type I with an absorption band of 8 μm in the infrared and type II without this band. The defects explain the color of diamond gems.

---

928 See Quantum information detected using mobile compatible chip technology, Archer Materials, February 2022.

929 Source: QpiAI in Partnership With IISc Launches Joint Certification for AI and Quantum Computing to Upskill Enterprises, Schools and Colleges, March 2021.

930 See the review paper Diamond Integrated Quantum Photonics: A Review by Prasoon K. Shandilya et al, July 2022 (31 pages) which provides a good 360° overview of NV centers, and not only for quantum computing.

It was not until 1959 that these impurities were found to be related to the presence of nitrogen, at 7.8 µm and that nitrogen atoms were well isolated in the diamond crystal. In 1975, it was discovered that some heat treatment could control the diffusion of nitrogen atoms in the diamond. These nitrogen centers explain the diamond color. It has four types: one nitrogen atom isolated in a gap, two nitrogen atoms, three nitrogen atoms surrounding a gap and four nitrogen atoms.

It is the first type that is interesting for both quantum computing and quantum sensing. We can visualize these defects with a confocal microscope (having a very shallow depth of field) by illuminating them with a green laser beam that will generate some red light.

These NV centers diamonds are slightly pink. These properties make it possible to generate single-photon sources thanks to the isolation of a NV center.

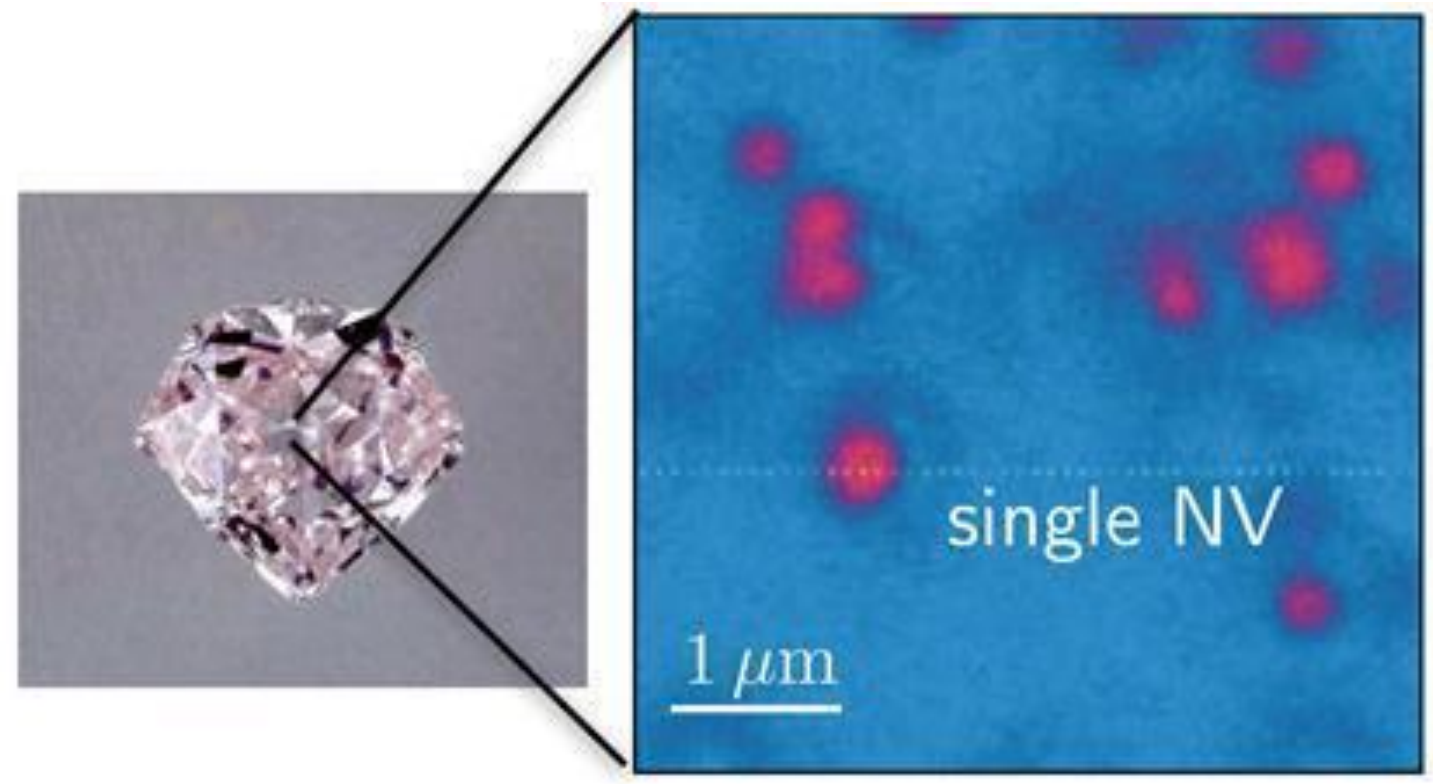

*Figure 355: how NV center cavities look in real diamonds. Source: TBD.*

Nitrogen-rich artificial diamonds are used to manufacture these NV centers. Gaps are generated with irradiation.

Vacuum annealing at about 800°C-900°C moves the vacancies next to the nitrogen atoms in the crystal structure[931].This is explained by nitrogen atoms being as large as carbon atoms. The gap creates a small bar of electrons that serve as a virtual magnet via their spin.

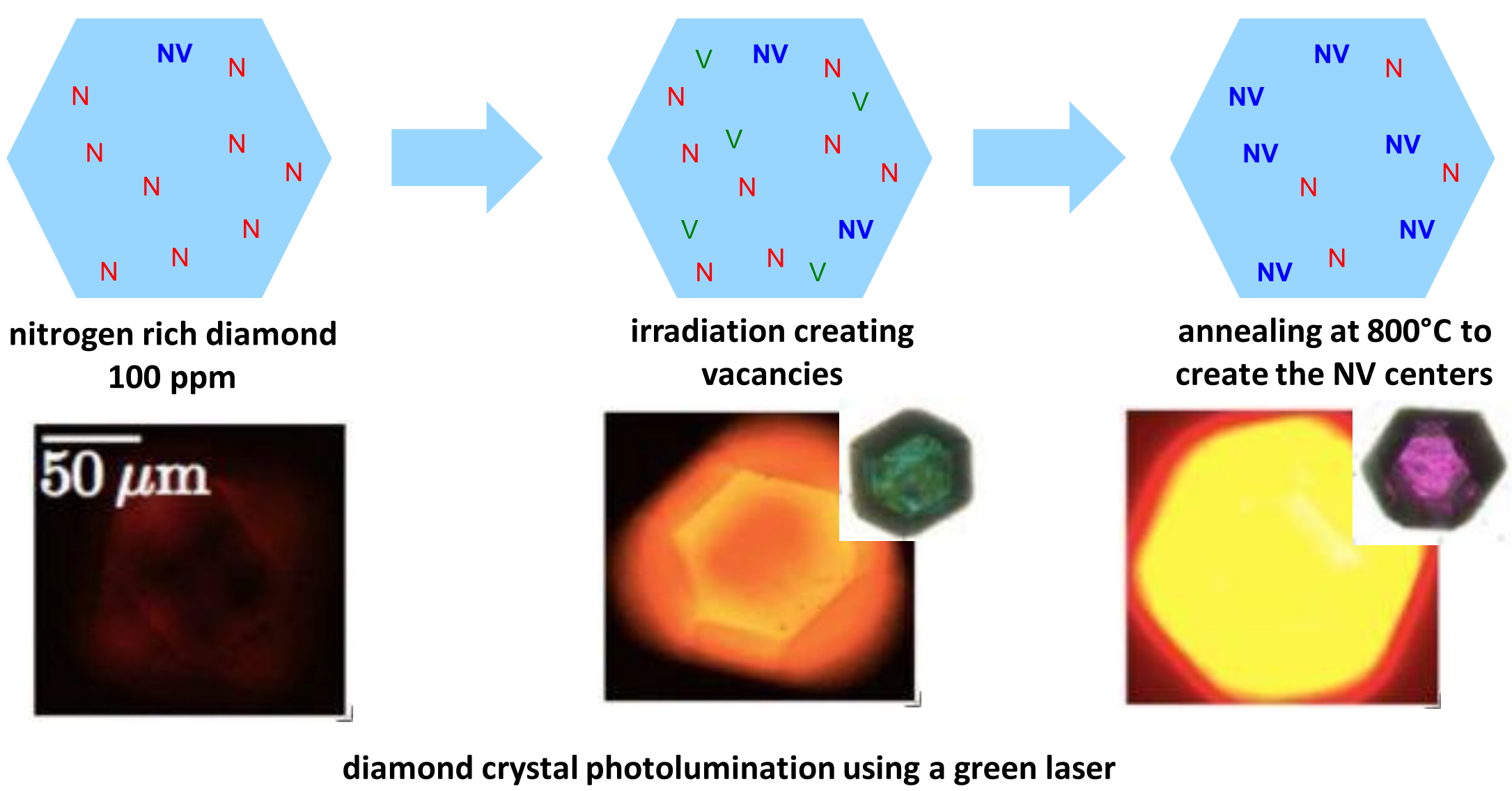

*Figure 356: how are nitrogen vacancies created. Source: [NV Diamond Centers from Material to Applications](#) by Jean-François Roch, 2015 (52 slides).*

Diamonds can also be produced at NV centers with vacuum deposition of hydrogen and methane (CVD, for Chemical Vapor Deposition) to create a perfect diamond crystal structure and then with ion implantation with nitrogen ion beams[932].

---

[931] See NV Diamond Centers from Material to Applications by Jean-François Roch, 2015 (52 slides) for an historical view of NV centers and a thesis that describes the different techniques of NV centers creation in Engineering of NV color centers in diamond for their applications in quantum information and magnetometry, Margarita Lesik, 2015 (139 pages).

[932] See a description of this manufacturing process in CVD diamond single crystals with NV centres: a review of material synthesis and technology for quantum sensing applications by Jocelyn Achard, Vincent Jacques and Alexandre Tallaire, 2019 (41 pages).

The carbon structure surrounding a NV center protects the cavity area well. The state of the gap is unstable and quantum. It is excited by lasers and microwaves. The reading of the qubit state is performed by a fluorescence brightness measurement.

### Science

The cavity contains a free electron that is generated by an electrical voltage applied to an n-p junction obtained by doping the diamond. The free electron is coupled to another one from the nitrogen atom near the cavity. The cavity includes two other pairs of electrons from the nitrogen atom in the cavity, with a total zero spin.

The process involves controlling the collective spin of these two free electrons as well as the spin of the nitrogen nucleus of the cavity and possibly of the neighboring $^{13}C$ carbon atoms[933]. The cumulative spin of the two electrons of the cavity is 0, 1 or -1 because it adds the spins of two electrons that are either ½ or -½. These electron spins are controlled by a combination of microwave and magnetic field.

Commonly used NV centers are called $NV^-$ because of the addition of an external electron into the cavity. The cavity has 6 electrons, three from the surrounding carbons, two from the nitrogen valence shell and one captured from the bulk. There are other variations like vacancies without this captured electrons ($NV^0$), or with missing electrons ($NV^{+1}$, $NV^{+2}$) that are not commonly used.

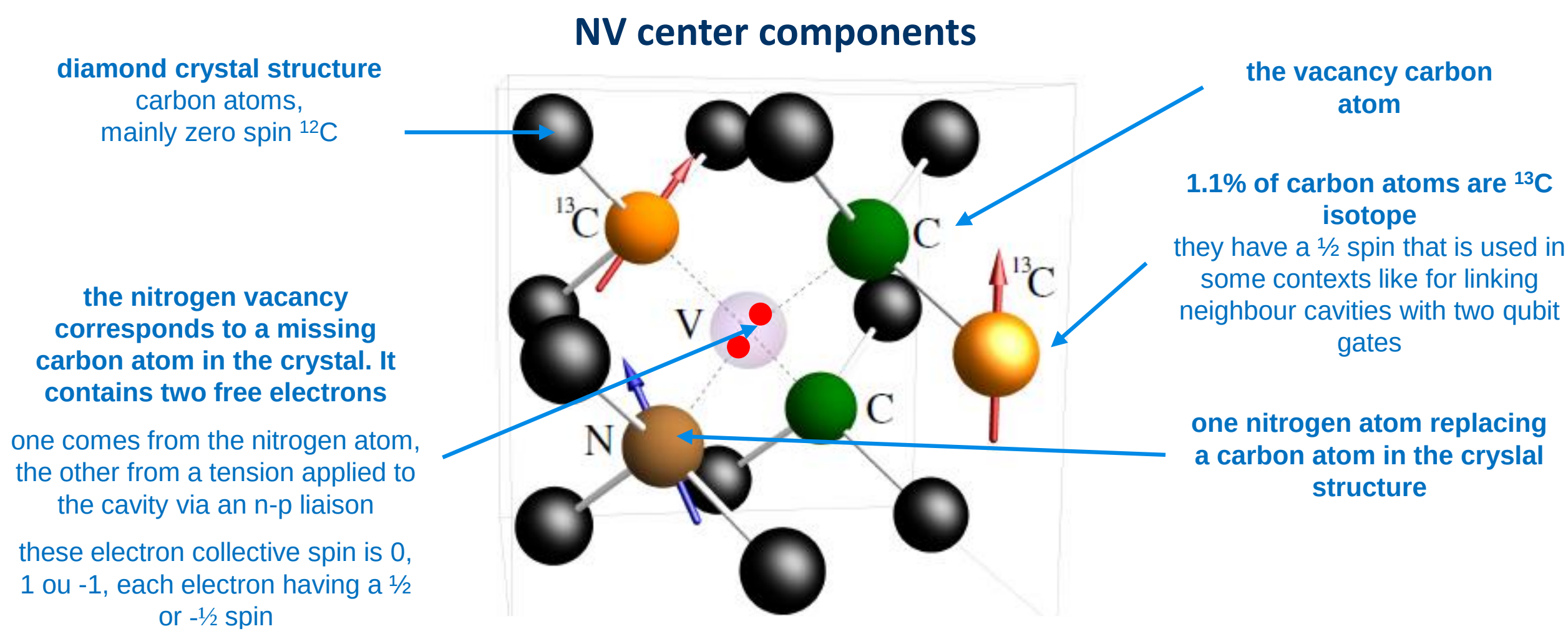

*Figure 357: the nitrogen vacancy contains two free electrons. Their spin is controlled as well as nuclear spins from surrounding $^{13}C$ and nitrogen atoms. (cc) Olivier Ezratty, 2021, image source TBD.*

Below is a diagram that describes what a NV center can look like in practice considering that there are many different implementations, knowing that NV centers are used not only for computing but, in a dominant manner, in quantum sensing. NV centers can be integrated in circuits fabricated on an SOI silicon wafer with a layer of $SiO_2$ insulator. It is covered with a matrix Fresnel lens, used to focus a control and readout laser[934]. It is frequently associated with other photonics components when the NV center spins are used to generate photons for quantum communications[935].

[933] Approximately 1.1% of the carbon atoms in diamond are of the $^{13C}$ isotope. The most common isotope is $^{12}C$. $^{14}C$ is present in trace amounts and is used to date carbonaceous objects due to its half-life of 5730 years. See Coherent control of an NV- center with one adjacent $^{13}C$ by Burkhard Scharfenberger et al, 2014 (24 pages).

[934] See Spin Readout Techniques of the Nitrogen-Vacancy Center in Diamond by David Hoper et al, 2018 (30 pages).

[935] See Hybrid Quantum Nanophotonics: Interfacing Color Center in Nanodiamonds with Si3N4-Photonics by Alexander Kubanek et al, July 2022 (55 pages) that describes such hybrid nanophotonic circuits.

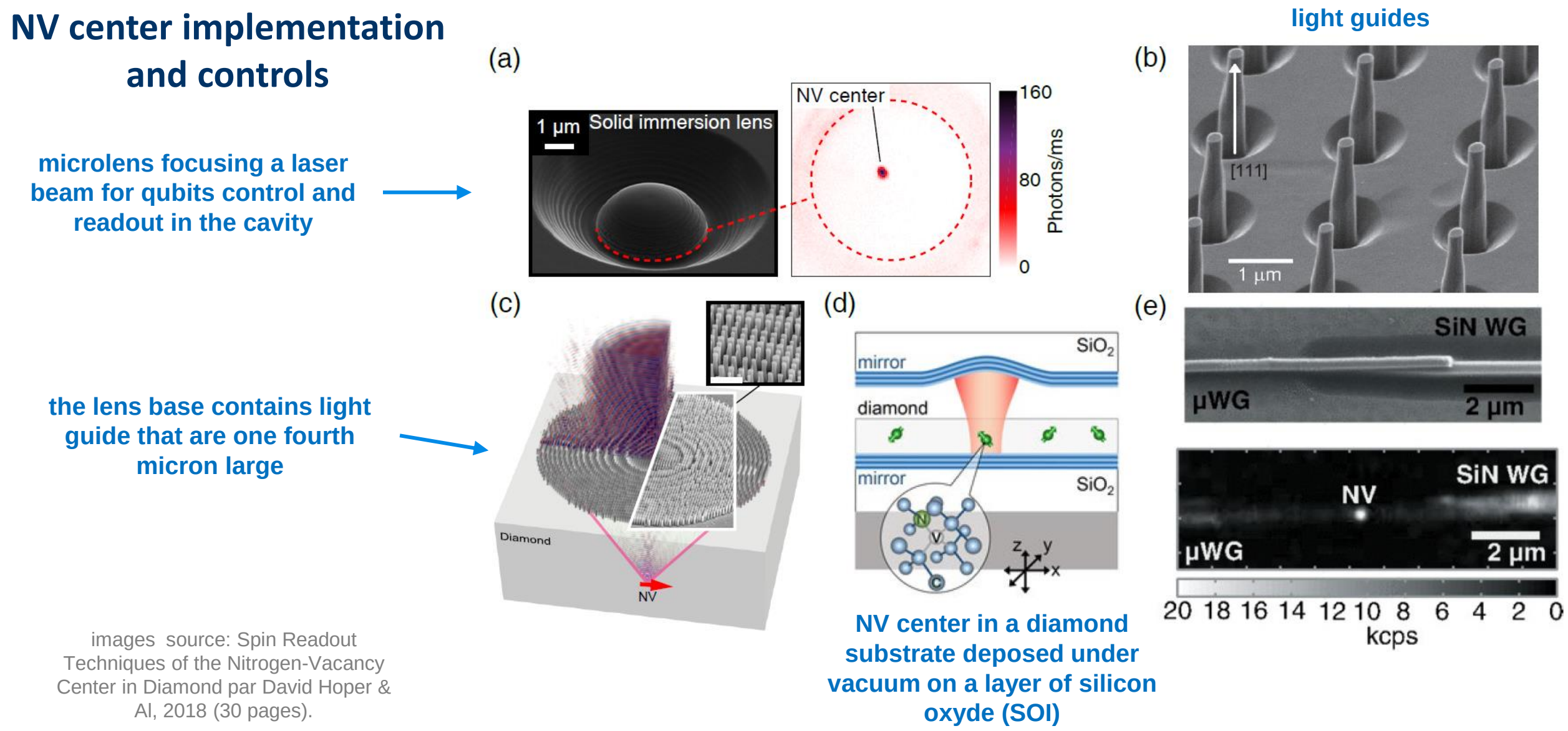

*Figure 358: examples of NV centers implementation and controls to guide laser light on the cavities. Source: [Spin Readout Techniques of the Nitrogen-Vacancy Center in Diamond](#) by David Hoper et al, 2018 (30 pages).*

Next is a diagram explaining how these rather complex qubits operate with the various energy levels and transitions of the cavity and its free electrons using microwaves and green photons. The vertical arrows represent useful energy transitions[936]. This is applicable to the $NV^-$ species.

The vacancy qubit state is controlled with 2.87 GHz microwaves that change the spin state of the vacancy electrons and switches the vacancy state between |0⟩ and |1⟩. The degeneracy of the spins 1 and -1 at the $^3A$ level (meaning: same energy level for different quantum properties) is removed with exposing the qubit to a static magnetic field. Other techniques are used to change the qubit state of the surrounding $^{13}C$ and $^{14}N$ nucleus spins.

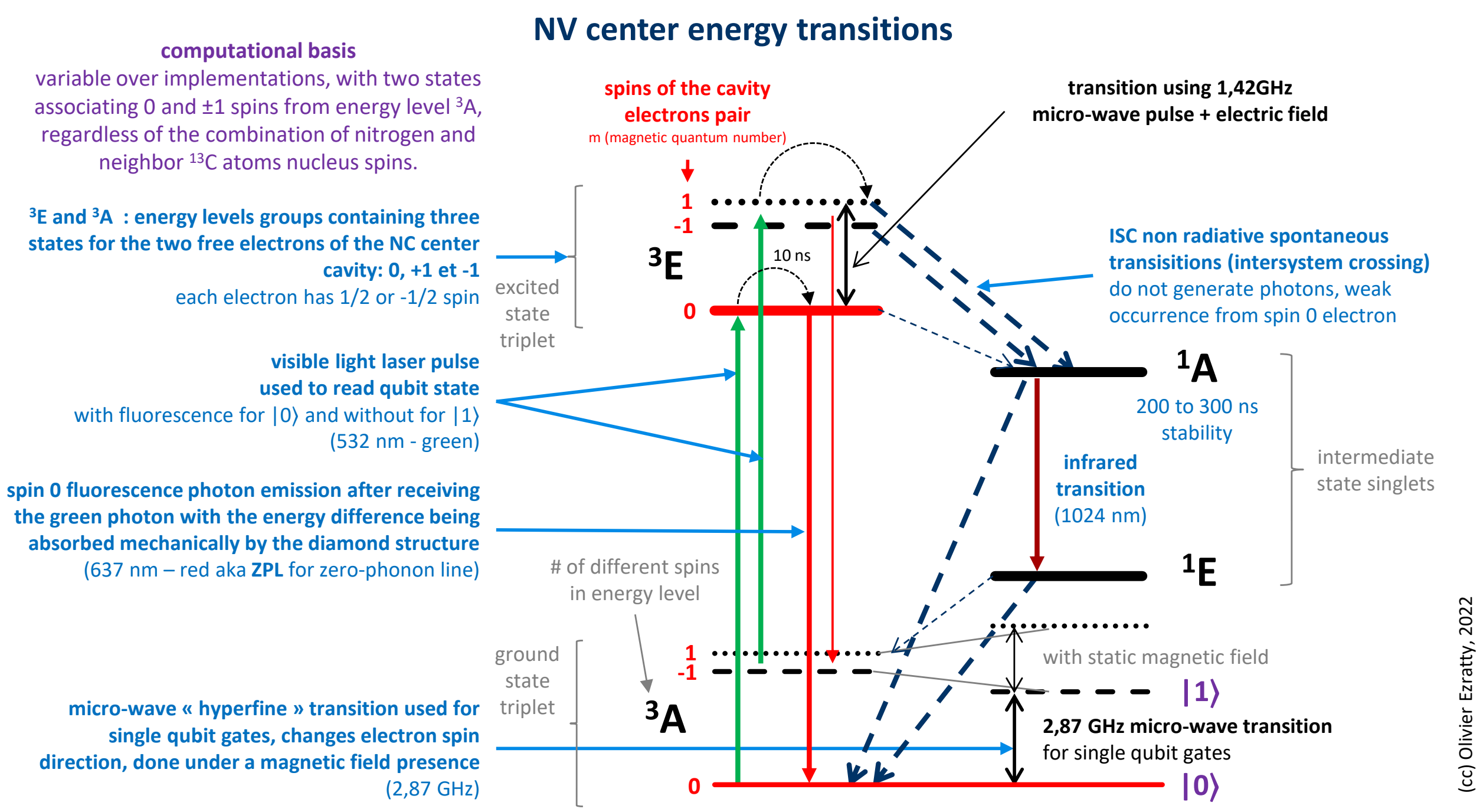

*Figure 359: energy transitions in an NV center. (cc) compilation by Olivier Ezratty, 2022.*

936 See the excellent review paper [Quantum computer based on color centers in diamond](#) by Sebastien Pezzagna and Jan Meijer, May 2020 (17 pages).

For qubit readout, an incoming green photon, usually at 532 nm, from a laser generates two possible outcomes:

- For |0⟩ at a zero spin, a change of state to $^3E$ which then degenerates via the $^1A$ state into a spontaneous non-radiative transition which does not emit a photon but transmits some mechanical energy to the crystal structure and returns to the basic state |0⟩. This happens when the related electron has a zero spin.
- For |1⟩ at a non-zero-spin, an emission of a fluorescence 637 nm red photon, part of the energy being also mechanically absorbed by the diamond structure. There is still a residual red photon emission, that creates some qubit readout noise.

NV centers qubits operate theoretically at room temperature[937]. Recent experiments have reached a 400 µs $T_1$ at ambient temperature[938].

NV centers have a too low DWF of 3%. This Debye-Waller factor which is measured by the ratio between the ZPL (zero-phonon lines) red emission and the total ZPL plus the phonon sideband emission (PSB). The ZPL is the sharp zero–phonon lines of luminescence of NV center in the visible spectrum that is of interest.

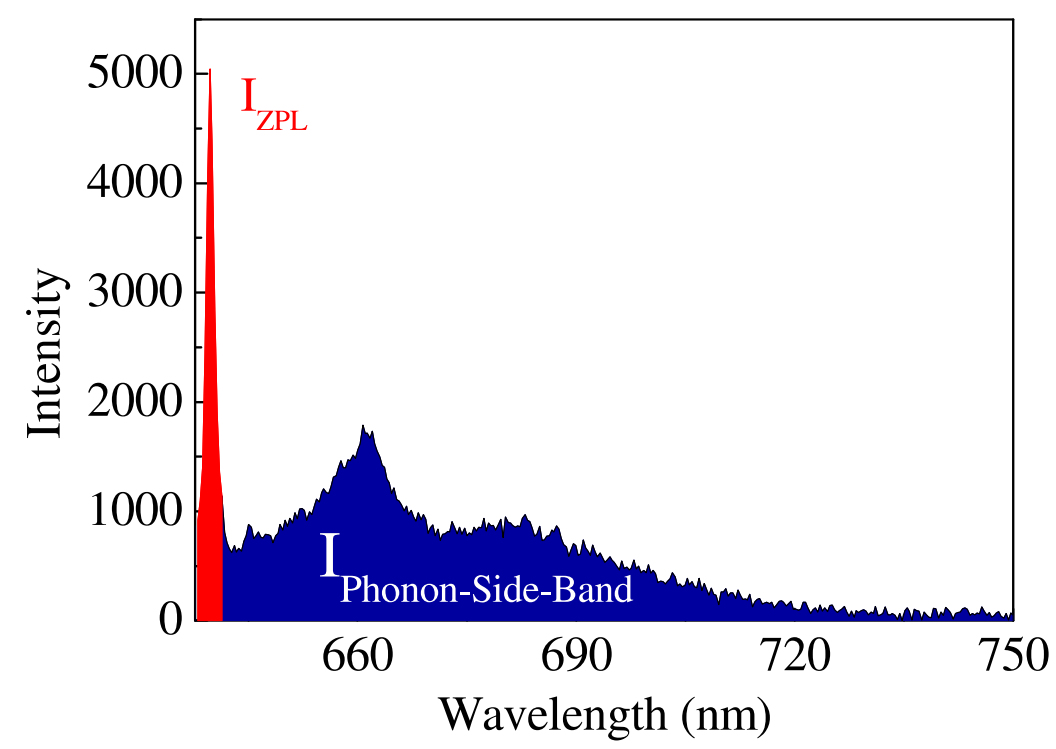

*Figure 360: visualizing a ZPL and phonon-side-band. Source: Suppression of fluorescence phonon sideband from nitrogen vacancy centers in diamond nanocrystals by substrate effect by Hong-Quan Zhao et al, Hokkaido and Osaka Universities, Japan, Optics Express, 2012 (8 pages).*

The phonon sideband is a thermal effect that is problematic[939]. Also, the DWF gets improved with low operating temperatures. This reduces qubit readout errors and explains why, on practice, a temperature of 4K is frequently used[940]! Another reason is that at low temperature, the spectral lines of the different energy states of the cavity are different, better spaced and easier to distinguish[941].

A joint QuTech-Fujitsu-Element Six team demonstrated in 2022 a fault-tolerant operation of a NV centers based QPU with logical qubits made of 5 physical spin qubits and two additional measurement qubits in a 29-qubit QPU running at 10K[942].

A shown in Figure 361 in page 370, the configuration was using different qubits: the NV center cavity was used as an auxiliary qubit with its two electrons, then the nuclear spin of the nearby nitrogen and five nuclear spins of $^{13}C$ were used for one flag and five data qubits.

---

937 See A programmable two-qubit solid-state quantum processor under ambient conditions by Yang Wu of Hefei's USTC in China, 2018 (5 pages). He describes an NV center managing two qubits at ambient temperature exploiting the cavity electron spin and the associated nitrogen atom nucleus spin. See also the review paper Quantum information processing with nitrogen–vacancy centers in diamond by Gang-Qin Liu et al, 2018 (15 pages).

938 See Success in mass production technology for ultra-high-purity 2-inch diamond wafer; expected to spur realization of quantum computing, August 2022 and Long spin coherence times of nitrogen vacancy centers in milled nanodiamonds by B. D. Wood et al, PRB, May 2022 (11 pages).

939 See Suppression of fluorescence phonon sideband from nitrogen vacancy centers in diamond nanocrystals by substrate effect by Hong-Quan Zhao et al, Hokkaido and Osaka Universities, Japan, Optics Express, 2012 (8 pages).

940 The technique is documented in Quantum information processing with nitrogen vacancy centers in diamond by Gang-Qin Liu and Xin-Yu Pan, 2018 (15 pages) and in Diamond NV centers for quantum computing and quantum networks by Lilian Childress and Ronald Hanson, 2017 (5 pages).

941 This interdependence between hyperfine spectral lines and temperature is not unique to diamond cavities. They are common in crystalline structures because temperature modifies many parameters such as the relative arrangement of the atoms in the crystals which leads to changes in electrical and magnetic gradients and therefore spins, etc.

942 See QuTech and Fujitsu realise the fault-tolerant operation of a qubit by Qutech, May 2022 and Fault-tolerant operation of a logical qubit in a diamond quantum processor by M. H. Abobeih et al, May 2022 (11 pages).

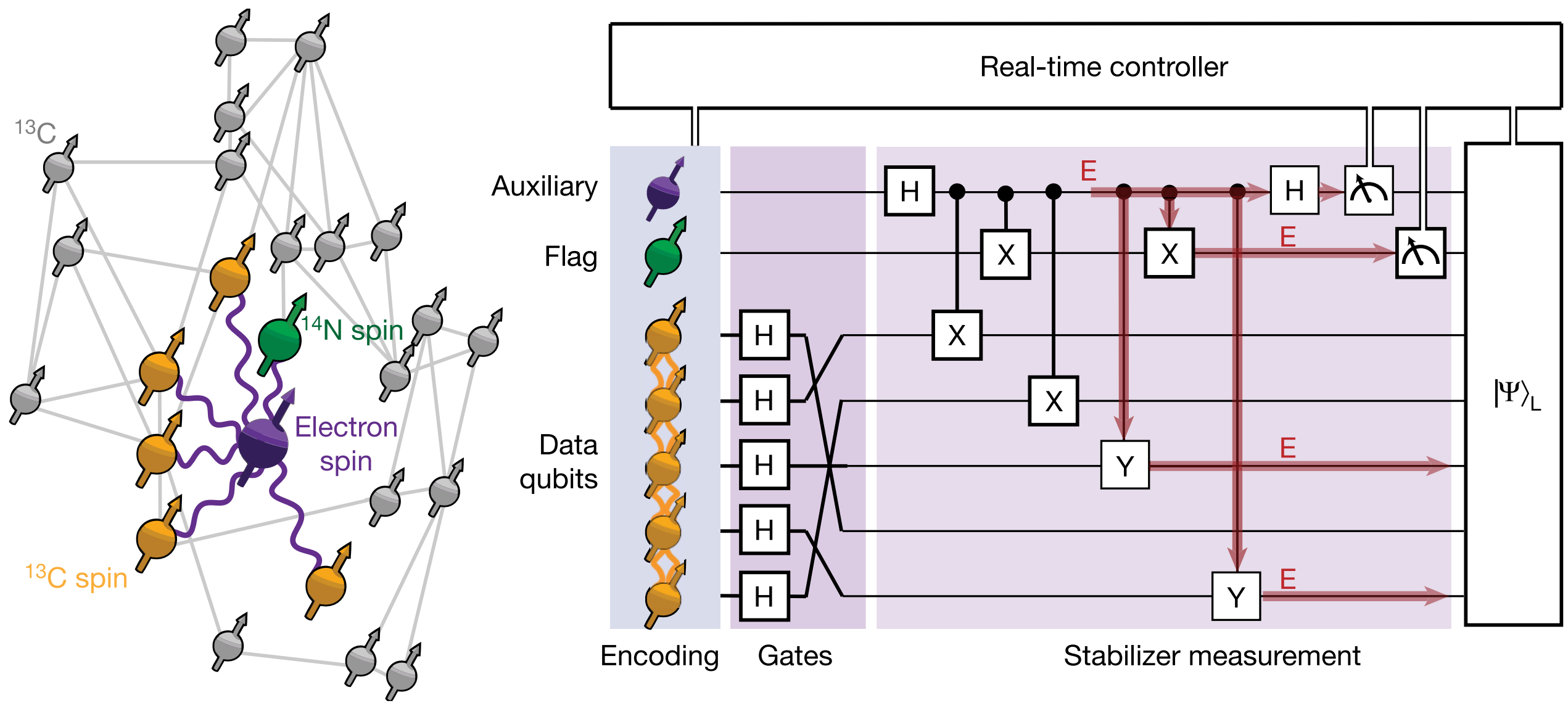

*Figure 361: an error correction code implemented with NV centers qubits. Source: Fault-tolerant operation of a logical qubit in a diamond quantum processor by M. H. Abobeih et al, May 2022 (11 pages).*

Another team, in Japan, implemented a similar error correction code with a Shor-3 codes using 6 qubits[943].

### Qubit operations

The general principle of operation for these qubits is as follows[944]:

- **Qubit quantum state** is based on a two-state computational basis, with $|0\rangle$ corresponding to the $^3$A energy zero spin base level and $|1\rangle$ to the same level but with a non-zero spin. The computational basis is sometimes $|+1\rangle$ and $|-1\rangle$ corresponding to the two non-zero spin levels of the $^3$A basis. Most techniques use the neighboring $^{13}$C and $^{14}$N atoms nuclear spins as qubits and are arranged in clusters. The NV center electrons spin is used as a mediator to control the neighbor atomic spins.
- **Single-qubit quantum gates** are microwave-activated and exploit hyperfine energy transitions at a frequency of 2.87 GHz[945]. These transitions work together with a magnetic field for zone A and an electric field for zone E.
- **Two-qubit quantum gates** use different methods: coupling NV centers with entangled photons which doesn't work well, magnetic coupling, or with coupling the NV center with the nucleus spin of neighboring $^{13}$C and $^{14}$N atoms with microwaves[946]. There are many variations of two-qubit gates like a CNOT[947], a CZ or even a weird $\exp(i\pi S_z \otimes I_z)$[948].

---

943 See Quantum error correction of spin quantum memories in diamond under a zero magnetic field by Takaya Nakazato et al, Nature Communications Physics, April 2022 (7 pages).

944 I was initially inspired by a diagram from lecture 3 of Hélène Perrin's course, February 2020. Then I integrated other sources of information. See in particular The nitrogen-vacancy color center in diamond by Marcus Doherty, Joerg Wrachtrup et al, 2013 (101 pages) which describes in particular the energy levels variations of NV centers as a function of their temperature.

945 As we have seen about trapped ions, hyperfine transitions are energetic transitions of low energy electrons, here in the microwave regime, which are generally related to the interaction between the magnetic polarities of the nucleus of the atoms with the magnetic field generated by the electrons. Knowing that here we are talking about electrons that do not rotate around the nucleus of an atom but in a cavity.

946 See some explanations in Entanglement in NV centers by Alexander Okupnik, Andrei Militaru and Ramon Gao, ETH Zurich, 2017 (34 slides).

947 See some detailed explanations in Colour centers in diamond by Joerg Wrachtrup, 2017 (36 slides).

948 See A programmable two-qubit solid-state quantum processor under ambient conditions by Yang Wu et al, NPJ, 2019 (5 pages).

- **Qubits readout** uses the capture of the fluorescence of the cavity activated by a laser and with an APD (avalanche photodiode) or a CCD sensor, similar to what is done with trapped ions and cold atoms. It consists in illuminating the cavity with a green (546 nm) laser. This excites level $^3A$ in $^3E$ but without changing the spin[949]. The non-zero spin state $^3E$ will generate a non-radiative transition passing through the $^1A$ state. The null spin state $^3E$ will generate the emission of a red photon (689 nm) which will be detected by the CCD sensor. This optical readout of single isolated qubits works only at low temperatures to avoid the creation of perturbation affecting neighbor qubits. The measurement of the cavity electron spin can exploit other techniques, each with their advantages and disadvantages: SCC (spin to charge conversion[950]), NMR (readout is assisted by the nucleus spin of neighboring atoms) and only by photonics means, knowing that lasers are used in all cases. Only a nuclear spin readout can be nondestructive (QND)[951].

The technology is not easy to industrialize on a large scale, whether it is the chipset itself or the control lasers. In Figure 362 is a schematic diagram of the control mechanism for these qubits[952].

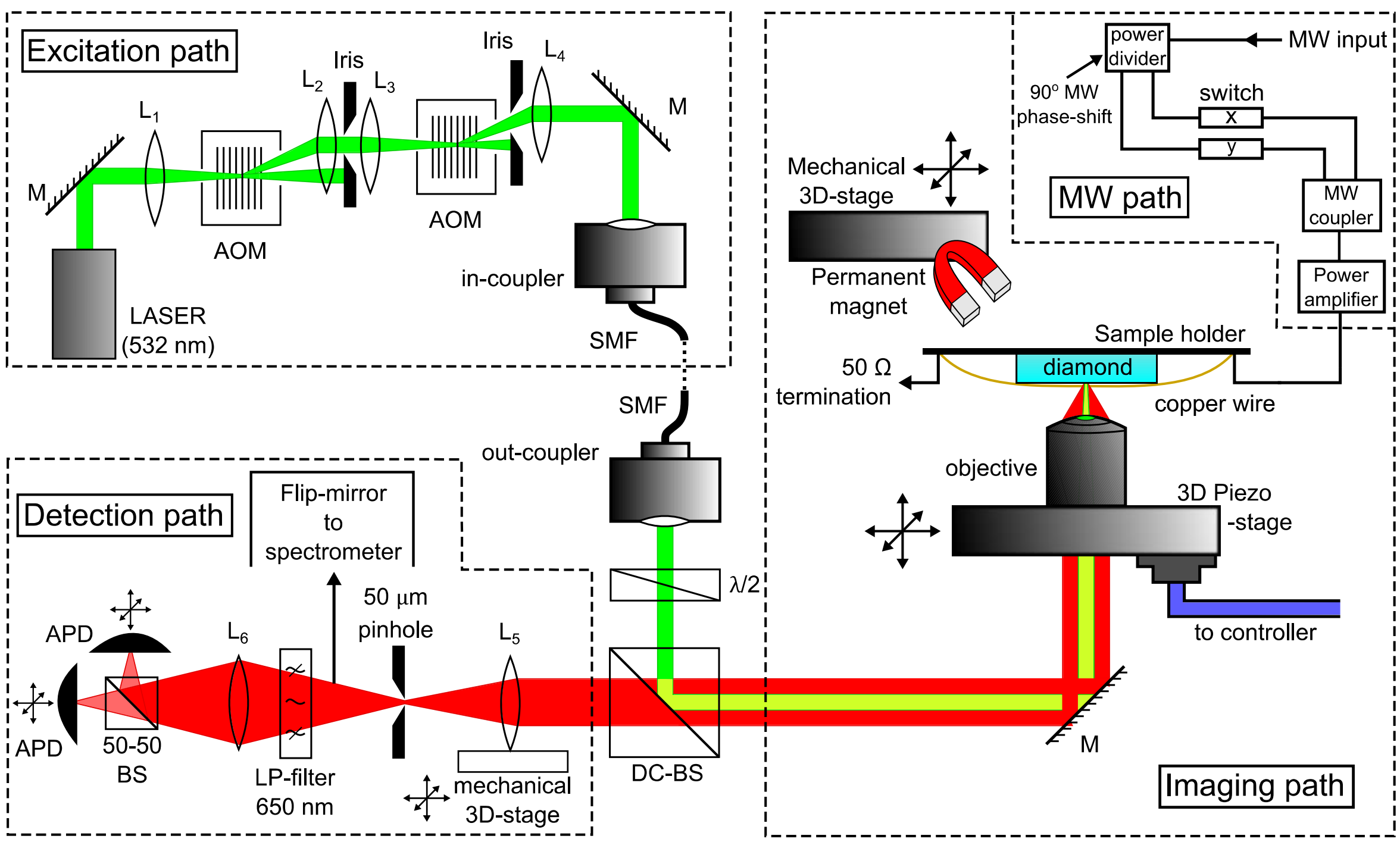

**Figure A.1.: Schematic representation of the utilized setup for the characterization of NV centers.** Experimental setup utilized for optical characterization and coherent spin manipulation of NV centers, comprising of a home-built confocal microscope, a scanning-stage for the imaging of diamond, and external magnet and microwave apparatus. The excitation wavelength is 532 nm. In the figure, mirrors are represented by M, lenses by $L_i$, single-mode optical fiber by SMF, beam-splitters by BS, and avalanche photo-diodes by APD.

*Figure 362: characterization of NV centers setup. Source: Forefront engineering of nitrogen-vacancy centers in diamond for quantum technologies by Felipe Favaro de Oliveira, 2017 (235 pages).*

[949] This technique is labelled ODMR for optically detected magnetic resonance.

[950] Explained in detail in Spin readout via spin-to-charge conversion in bulk diamond nitrogen-vacancy sets by Harishankar Jayakumar, September 2018 (5 pages).

[951] See Color Centers in Diamond by Andreas Wallraff, ETH Zurich, 2017 (34 slides).

[952] Seen in Forefront engineering of nitrogen-vacancy centers in diamond for quantum technologies by Felipe Favaro de Oliveira, 2017 (235 pages).

### Research

The main countries involved are China, the Netherlands (TU Delft and Qutech[953]), Australia (University of Melbourne, Quantum Brilliance), Germany (University of Ulm), Japan (NII and NTT), some laboratories in France (such as CEA SPEC) and of course in different labs in the USA.

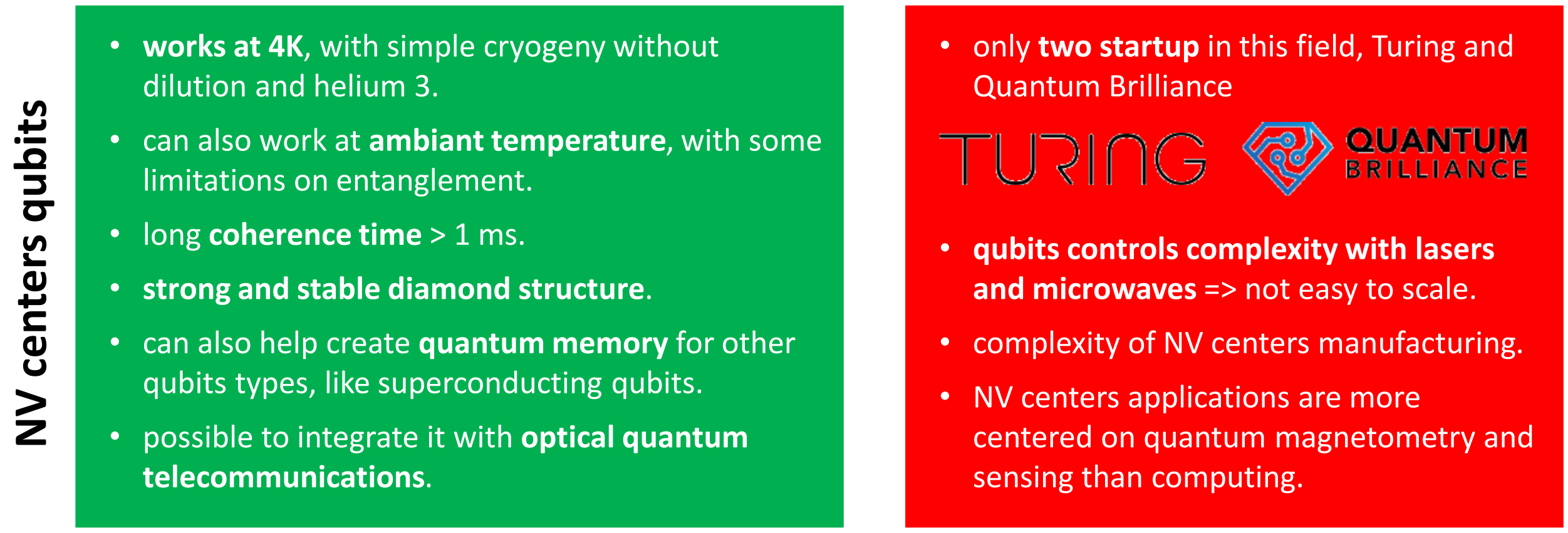

*Figure 363: pros and cons of NV centers qubits. (cc) Olivier Ezratty, 2022.*

One of the challenges of NV centers is their implantation in diamonds. One promising technique created by Berkeley Labs is using gold ions-based implantation that could scale to thousands of qubits. But this is just about fabrication and not functional qubits[954].

NV Centers based quantum computers have been very low key for a few years.

Indeed, it seems that NV centers have more promising uses in quantum sensing for the creation of precision magnetometers or for quantum memories interoperable with qubits realized with other technologies such as superconducting qubits in hybrid systems. This is a path recently explored by the University of Delft[955], in Japan[956] and by CEA-SPEC with Patrice Bertet as shown in Figure 364, with a superconducting qubit linked to a NV center memory qubit.

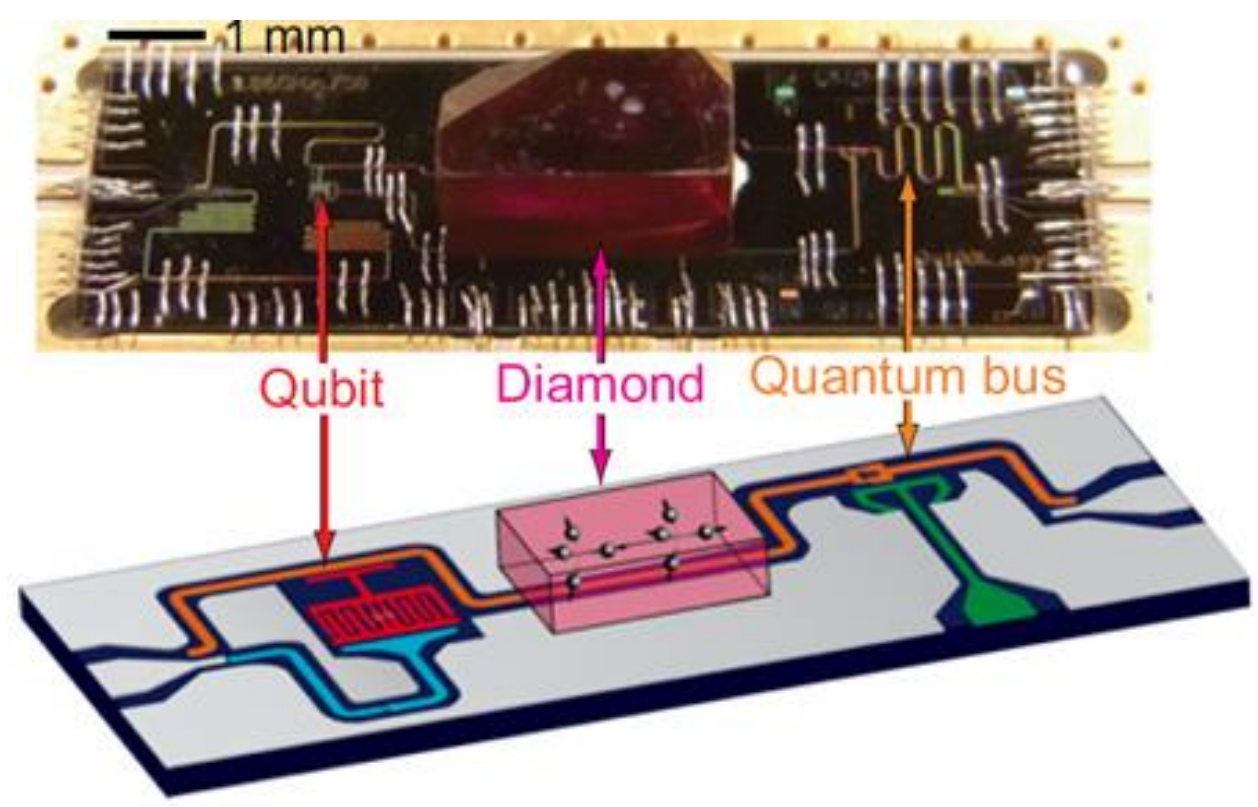

*Figure 364: NV center used as a quantum memory for a superconducting qubit, which could lead to create heterogeneous qubits. Source Quantum technologies with hybrid systems, Patrice Bertet et al, 2015 (8 pages).*

There are also variants of NV center techniques with defects introduced in phosphorus-doped silicon carbide.

It would have the advantage of creating qubits whose readout is more accurate since being based on the emission of a narrow frequency fluorescence[957].

---

[953] Qutech demonstrated a 10-qubit prototype with a coherence time of one minute and working at 3.7K in 2019. See Fully controllable and highly stable 10-qubit chip paves way for larger quantum processor, Qutech, 2019.

[954] See Ion-Trap Advance: Berkeley Lab Pioneers Way That Could Increase Scalability to Over 10,000 Qubits for Quantum Sensing, Quantum Computing by Matt Swayne, May which refers to Direct formation of nitrogen-vacancy centers in nitrogen doped diamond along the trajectories of swift heavy ions by Russell E. Lake et al, March 2021 (5 pages).

[955] See Diamond-based 10-qubit register with coherence more than one minute, November 2019.

[956] See Coherent Coupling between a Superconducting Qubit and a Spin Ensemble by Shiro Saito et al, 2012 (7 pages).

[957] See Study Takes Step Toward Mass-Producible Quantum Computers, 2017.

In a similar fashion, **MIT** prototyped in 2020 a NV centers chipset replacing nitrogen with silicon and germanium. They assembled 128 qubits but these are not operational [958]. China researchers are also experimenting NV center qubits but so far, their prototype has only 3 qubits [959]. In another work, Australian researchers were able to implement single Clifford group qubit gates with $NV^-$ centers with a fidelity of 99.3%[960].

## Manufacturing

Artificial diamonds are produced either with high-pressure high-temperature processes (HPHT) or with chemical vapor deposition (CVD)[961]. The latter is used for quantum use cases.

NV centers are manufactured with various methods of precise defects implantation. NV vacancies are usually produced by electrons, neutrons, protons or ions irradiation to create the vacancies followed by thermal annealing at temperatures above 650 °C to move the vacancies close to the defect atoms (here, nitrogen). A tiny share of nitrogen atoms impurities are deposed during the CVD (chemical vapor deposition) production of artificial diamond. One technique makes use of targeted ion depositions of ion beams with masking and ebeam lithography nanopatterning[962].

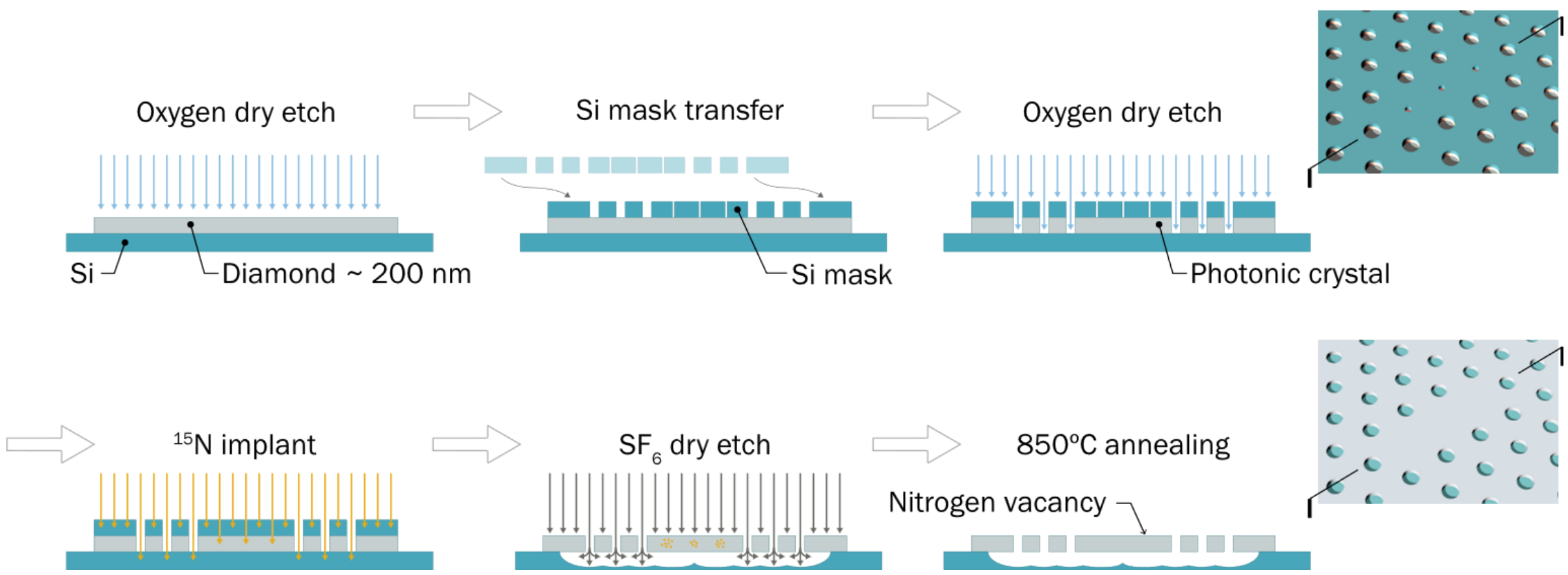

*Figure 365: example of NV center implantation technique using a mask. Source: Scalable fabrication of coupled NV center – photonic crystal cavity systems by self-aligned N ion implantation by T. Schroder and A. Stein, May 2017 (13 pages).*

## Vendors

**Quantum Brilliance** (2019, Australia), **Turing Inc** (2016, USA) and **XeedQ** (2021, Germany) are the few companies dedicated to creating NV center-based quantum computers.

**Quantum Brilliance** (2019, Australia/Germany, $11.4M) is developing a NV centers quantum processor that operates at room temperature, created by ANU (Australian National University) researchers, Andrew Horsley (CEO) and Marcus Doherty (CSO).

[958] See Large-scale integration of artificial atoms in hybrid photonic circuits by Noel H. Wan, Dirk Englund et al, MIT, UC Berkeley, Sandia Labs, Nature, 2020 (11 pages).

[959] See Quantum anomaly detection of audio samples with a spin processor in diamond by Zihua Chai et al, January 2022 (8 pages).

[960] See High Fidelity Control of a Nitrogen-Vacancy Spin Qubit at Room Temperature using the SMART Protocol by Hyma H. Vallabhapurapu et al, UNSW, August 2022 (7 pages). Clifford group gates are the simplest to implement and are not sufficient to create a universal gate set. It misses either a 3-qubit Toffoli gate or a T gate.

[961] See the thesis Engineering of NV color centers in diamond for their applications in quantum information and magnetometry by Margarita Lesik, 2015 (138 pages) and the review paper Chemical vapour deposition diamond single crystals with nitrogen-vacancy centres: a review of material synthesis and technology for quantum sensing applications by Jocelyn Achard, Vincent Jacques and Alexandre Tallaire, 2020 (30 pages).

[962] See Scalable fabrication of coupled NV center – photonic crystal cavity systems by self-aligned N ion implantation by T. Schroder and A. Stein, May 2017 (13 pages).

They estimate that their solution will be size/weight/performance/cost/power competitive and bring some quantum advantage earlier than competing systems from Google and IBM that they brand « quantum mainframes ». They want to create "quantum desktops" and why not pushing the envelope a bit too far with "smartphone quantum computers"[963].

They introduced in March 2021 a 5-qbits prototype fitting into a 2U classical 19-inch server form factor[964]. They expect to reach 50 qubits by 2026 and to then scale up this architecture with connecting several units together[965]. Probably with photons... and there, you'd probably need some cooling for photon sources and detectors!

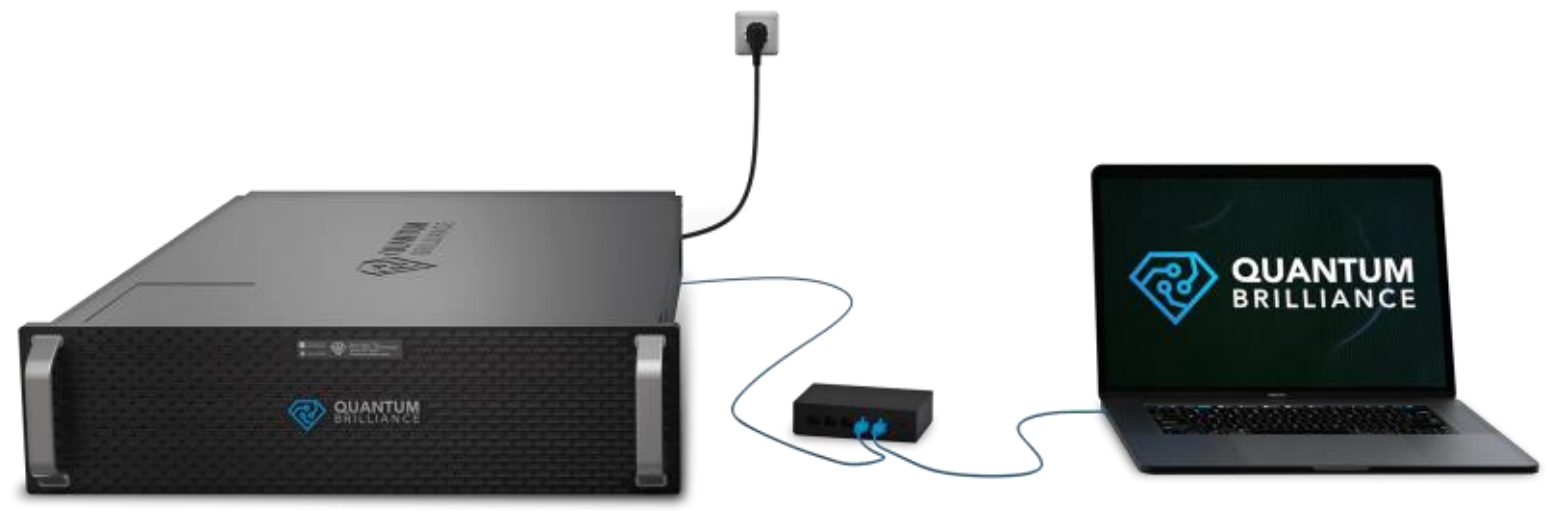

*Figure 366: a Quantum Brilliance computer fitting in a 19' rack and connected to a simple laptop. Source: Quantum Brilliance.*

One caveat is that all the NC centers specialists I've been talking to say that it's nearly impossible to operate NV centers qubits at room temperature for handling qubits entanglement properly. I wonder who should be believed here[966]! And the fact that they provide not a digit of data on their qubit fidelities doesn't drive more trust.

To investigate, you need to look at the scientific papers coauthored by Marcus Doherty[967]. You find that they have an error rate of $10^{-5}$ for single qubit gates, which is fine, and expect to have a similar level for two-qubit gates. But what is documented is a 99,2% two-qubit gate fidelity, which is not stellar nor sufficient to implement fault-tolerant quantum computing.

In April 2021, Quantum Brilliance also announced a partnership with **Quantum-South** (Uruguay) and to develop proof of concepts optimization quantum applications for air and maritime cargo companies. This is a bit early given their existing 5-qubits but why not exploring the path.

In January 2022, Quantum Brilliance announced DE-Brill, a joint research project with the Fraunhofer Institute for Applied Solid State Physics IAF and the University of Ulm funded by the German government quantum plan. The partnership is focused on the development of manufacturing (Fraunhofer) and control (Ulm) techniques of NV center qubits. The total investment of this 5-year project starting in December 2021 if €19.9M with 78.4% funded by BMBF, the German ministry of research. In April 2022, the company launched a joint R&D hub with La Trobe University and RMIT University around material design and manufacturing. This and other German universities working on NV centers[968] create a critical mass of skills in NV centers aimed at quantum computing.

At home in Australia, the **Pawsey Supercomputing Research Centre** announced in June 2022 the installation of a Gen1 Quantum Brilliance quantum accelerator.

---

[963] See Breakthrough: Quantum computers will soon fit in your phone by Maija Palme, Sifted, August 2021.

[964] In some sources, the number of available qubits is two and not five... !

[965] See Diamond-Based Quantum Accelerator Puts Qubits in a Server Rack by Charles Q. Choi, March 2021. The illustrative picture comes from Quantum Brilliance. See also some technical details in Quantum accelerators: a new trajectory of quantum computers by Marcus Doherty, Quantum Brilliance, March 2021.

[966] Their web site says "no need for absolute zero temperature". It may mean that OK, it's not 15 mK like with superconducting qubits, but it may still require some 4K cooling. Which is not, by any definition, "room temperature".

[967] See Optimisation of diamond quantum processors by YunHeng Chen, Marcus W. Doherty, et al, September 2020 (42 pages), Spin-to-Charge conversion with electrode confinement in diamond by Liam Hanlon, Marcus W. Doherty et al, August 2021 (16 pages) and Optical activation and detection of charge transport between individual color centers in room-temperature diamond by Artur Lozovoi, Marcus W. Doherty et al, October 2021 (15 pages).

[968] See Optically coherent nitrogen-vacancy defect centers in diamond nanostructures by Laura Orphal-Kobin et al, Humboldt-Universität and Ferdinand-Braun-Institut both in Berlin, 2022 (26 pages).

There was no precision whatsoever on the specifications of this system.

**Turing Inc** (2016, USA, $15.5M) is a startup willing to create quantum computing hardware and software, based on NV centers qubits and operating at 4K[969].

They also develop error correction systems that they market to other industry specialists. A way to avoid putting all your eggs in the same basket!

**XeedQ** (2021, Germany) is developing XQ1, an NV-center-based multi-qubit mobile quantum processor running at room-temperature. It fits in a desktop format and seemingly has 4+ qubits (what the + means is quite a mystery).

They plan to release a version with 256 qubits by 2026. You may wonder what are the exact quantum computing skills of the founding team when their web site touts that "*quantum systems offer encryption standards that are virtually impossible to breach. Welcome to future-proof quantum secure communications on a mobile footprint*". Someone should tell them that PQC doesn't need a quantum computer to run. The company created by Gopalakrishnan Balasubramanian, formerly at the Max Planck Institute for Biophysical Chemistry in Göttingen and is based in Leipzig. He published a wealth of papers on the physics of NV centers but not on NV-centers-based quantum computers nor how do you entangle such qubits at room temperatures.

### SiC and other spin cavities variants

Besides NV centers, another similar technique is investigated at the research stage that uses various vacancies in silicon carbide crystal structures (SiC) or silicon. Vacancies can be missing nearby couples of carbon and silicon atoms, called divacancies ($V_{Si}V_C^0$) or just a missing silicon atom ($V_{Si}^-$)[970]

Others use transition metal defects with chromium, vanadium, molybdenum, tungsten, erbium[971] and also hexagonal boron nitride (h-BN)[972]. In 2022, a research team from the DoE Argonne National Lab created a prediction model of the coherence time of vacancies depending on their characteristics, which can help investigate new materials[973].

As with NV centers, SiC qubits could theoretically work at ambient temperature[974]. While these vacancies could be used to create qubits and quantum processors like with NV centers, they are currently aimed mostly at quantum photonics applications. One of the reasons is that some SiC vacancies have fluorescence wavelengths corresponding to fiber optics telecom wavelengths in the near infrared band around 1,5 µm, in the so-called 4H-SiC hexagonal lattice version[975].

---

[969] See Turing Inc: Large Scale Universal Machines, 2017, which details this a little bit.

[970] See Single artificial atoms in silicon emitting at telecom wavelengths by W. Redjem et al, 2020 (4 pages).

[971] See Roadmap for Rare-earth Quantum Computing by Adam Kinos, Alexandre Tallaire et al, March 2021 (47 pages).

[972] See First-principles theory of extending the spin qubit coherence time in hexagonal boron nitride by Jaewook Lee, Huijin Park and Hosung Seo, npj, September 2022 (9 pages).

[973] See A mathematical shortcut for determining quantum information lifetimes by Leah Hesla, Argonne National Lab, April 2022 and Generalized scaling of spin qubit coherence in over 12,000 host materials by Shun Kanai, David D. Awschalom et al, PNAS, April 2022 (8 pages). It shows that the coherence time $T_2$ of vacancies is dependent on the cavity spin density ($n_i$) of nucleus i, the crystalline structure, the nuclear spin-factor ($g_i$), and the nuclear spin quantum number ($I_i$) according to the formula $T_{2,i} = 1.5 \times 10^{18} \left|g_i\right|^{-1.6} I_i^{-1.1} n_i^{-1.0}$ (s).

[974] The DoE Argonne National Laboratory together with researchers from Hungary, Sweden and Russia published in 2019 a work on SiC qubits operating at room temperature. See Scientists Find Yet Another Way to Get Qubits Working at Room Temperature by David Nield, March 2020 and Novel Qubit Design Could Lead to Quantum Computers That Work at Room Temperature by Matt Swayne, March 2020 which references Quantum well stabilized point defect spin qubits by Viktor Ivády et al, May 2019 (20 pages).

[975] See the excellent review paper Quantum Information Processing With Integrated Silicon Carbide Photonics by Sridhar Majety et al, March 2022 (50 pages).

SiC vacancies can indeed be used as interesting sources of single or entangled photons in quantum communications and cryptography. SiC photon sources are implanted on nanophotonic devices[976].

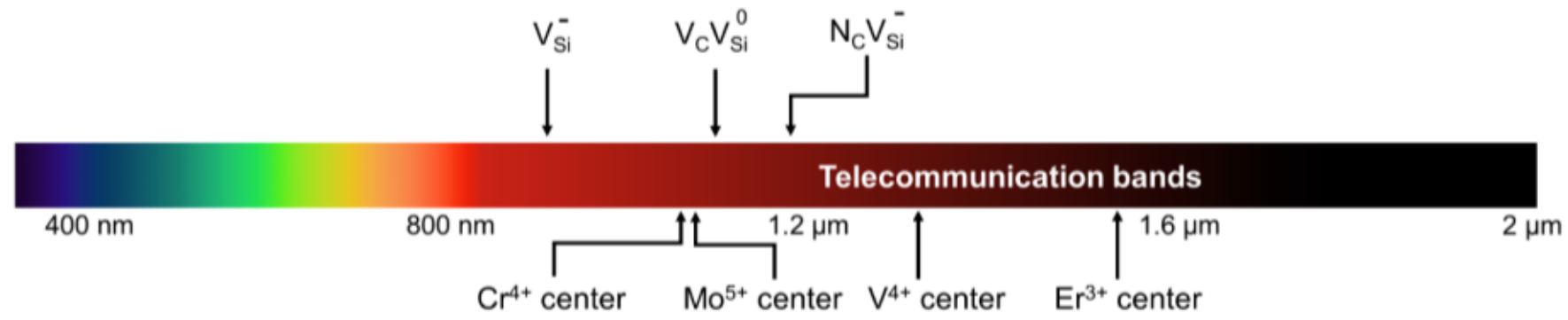

*Figure 367: other cavities are interesting due to their transition frequencies that sit in the telecommunication wavelengths. Source: Quantum Information Processing With Integrated Silicon Carbide Photonics by Sridhar Majety et al, March 2022 (50 pages).*

It could also be used in quantum repeaters thanks to relatively long spin coherence times above 50 ms. Also, SiC vacancies can show a much better DWF (Debye-Waller factor) than diamond NV centers. However, their readout contrast has to be improved[977].

In quantum information processing, SiC vacancies are investigated in various areas such as with quantum simulation and measurement-based quantum computing. There are some specific technology paths like using SiC spin qubits that are coupled by photons[978].

SiC vacancies have very long coherence times in the seconds range[979]. It leads to theoretically long computing time, although, it would work with solving many other problems like large entanglement capacities and high two-qubit gate fidelities. Thus, when you read in the media that SiC could achieve a hundred million operations, you may get skeptic and right to be so[980]. Indeed, the related paper simply compute this number of operations with dividing SiC qubit coherence time of 5 seconds by a single-qubit gate time. They have not implemented it yet, particularly given it's useless to do that on a single qubit. It would be nice to have a tomography of a 2-qubit gate...:)!

**Photonic** (2019, Canada) is a spin-off from the Silicon Development Lab at Simon Fraser University in Vancouver. They develop silicon based chipsets quantum computers with the aim of putting on million qubits on two square-cm.

They tout having reached 1000 electron spin qubits in 2020 and 100,000 in 2021. Their technology seems to be based on defects in silicon, using similar techniques as SiC and NV vacancies[981]. Too bad they are probably overselling their achievements.

## Topological qubits

In this category of qubits and quantum computing, we must create a distinction between the notion of "topological" which defines a type of qubit based on anyons and the "Majorana fermions" which are a variant of anyons to create topological qubits. Of all the types of qubits, they are the most mysterious and complex to understand[982]! It's part of the broad field of topological matter.

---

[976] See Integrated quantum photonics with silicon carbide: challenges and prospects by Daniil M. Lukin, Melissa A. Guidry and Jelena Vuckovic, October 2020 (20 pages) and Fabrication and nanophotonic waveguide integration of silicon carbide colour centres with preserved spin-optical coherence by Charles Babin, Florian Kaiser et al, November 2021 (18 pages).

[977] See Room temperature coherent manipulation of single-spin qubits in silicon carbide with a high readout contrast, by Qiang Li et al, July 2021 (10 pages).

[978] See Silicon photonic quantum computing with spin qubits by Xiruo Yan et al, 2021 (28 pages).

[979] See Five-second coherence of a single spin with single-shot readout in silicon carbide by Christopher P. Anderson, David D. Awschalom et al, 2021 (9 pages).

[980] See Quantum Computing: Researchers Achieve 100 Million Quantum Operations by Francisco Pires, Tom's Hardware, February 2022

[981] See Silicon-Integrated Telecommunications Photon-Spin Interface by L. Bergeron et al, 2020 (15 pages).

[982] See Topological Quantum Computing by Torri Yearwood, January 2020 and A Short Introduction to Topological Quantum Computing by Ville Lahtinen and Jiannis K. Pachos, May 2017 (44 pages).

## History

Ettore Majorana predicted in 1937 the existence of a new class of particles that are is its own anti-particles. It was labelled “Majorana fermions.

The notion of anyon, "quasi-particle", i.e. particle representation models that describe the state of electron clouds around atoms, in the superconducting regime and in 2D was first proposed by Jon Leinaas and Jan Myrheim from the University of Oslo in 1976 and then elaborated by Frank Wilczek in 1982[983]. Majorana fermions are a specific type of these quasiparticles organized along a small superconducting wire in 1D structures. They have collective electron behaviors in crystalline networks at very low temperature[984].

Alexei Kitaev then had the idea in 1997 to use anyons for quantum calculations when he was a researcher at Microsoft. He published two foundational papers in 2000 describing the category of fermionic quantum computers and a model of spinless fermions in a 1D superconducting nanowire, *aka* a Kitaev chain [985]. There are several various variations of fermionic quantum computing being investigated, including analog quantum computing with cold atoms[986].

In 2008, Liang Fu and C.L. Kane from the University of Pennsylvania predicted that Majorana bound states can appear at the interface between topological insulators and superconductors[987].

In 2012, Leo Kouwenhoven then at TU Delft announced the detection of Majorana Zero Modes quasiparticles at TU Delft and later on in 2018 when he was at Microsoft Research in Delft.

Some other advances came out in 2016 from the MIT and in 2018, from a group of three American universities UC Irvine, UCLA and Stanford, who said they discovered real Majorana fermions. In May/June 2019, German and Austrian researchers said they succeeded in creating two-dimensional topological phenomena like Majorana zero modes[988].

Princeton researchers also published in June 2019 the results of their work that led them to control the state of a quasi-particle[989].

---

[983] See On the theory of identical particles by J.M. Leinaas and J. Myrheim, 1976 (23 pages) and See Quantum Mechanics of Fractional-Spin Particles by Frank Wilczek, PRL, 1982. See also the essay Quanta of the Third Kind by Frank Wilczek, November 2021 that provides an excellent simplified view of what are anyons, braiding and quasi-particles. The author explains three specifics of quasi-particles: fractionalization (where quasiparticles have properties that are subparts of usual whole-number multiples like spins or unit of electron electric charge or angular momentum), flux tubes (fractional angular momentum made possible by particles orbiting around tubes of magnetic flux like in type II superconductors) and dimensional reduction (with point-like structures).

[984] This is the thesis of Hugo de Garis in Topological Quantum Computing The TQC Shock Wave and its Impact on University Computer Science Teaching, 2011 (29 pages).

[985] See Fermionic quantum computation by Serguei B. Bravyi and Alexei Y. Kitaev, 2000 (18 pages) and Unpaired Majorana fermions in quantum wires by Alexei Kitaev, 2000 (16 pages).

[986] See The Power of Noisy Fermionic Quantum Computation by Fernando de Melo et al, April 2013 (21 pages) and Quantum register of fermion pairs by Thomas Hartke et al, MIT, March 2021 (10 pages). Fermionic quantum computing was defined by Serguei B. Bravyi and Alexei Y. Kitaev in Fermionic quantum computation, 2000 (18 pages).

[987] See Superconducting proximity effect and Majorana fermions at the surface of a topological insulator by Liang Fu and C.L. Kane, 2008 (4 pages) and Josephson Current and Noise at a Superconductor-Quantum Spin Hall Insulator-Superconductor Junction by Liang Fu and C.L. Kane, 2008 (4 pages).

[988] See Computing Faster With Quasi-Particles, May 2019 referring to Topological superconductivity in a phase-controlled Josephson junction by Hechen Ren et al, Nature, April 2019.

[989] See Mysterious Majorana Quasiparticle Is Now Closer To Being Controlled For Quantum Computing, June 2019 mentioning Observation of a Majorana zero mode in a topologically protected edge channel by Ali Yazdani et al, Science, June 2019 (12 pages).

2021 marked the beginning of a crisis winter for Majorana fermion. It started with an expression of concern and a withdrawal of Leo Kouwenhoven's 2018 Nature paper[990]. In a March 2022 Twitter thread Sergey Frolov made an impressive inventory on many other unreliable Majorana research papers with other retracted papers[991] or that should be retracted.

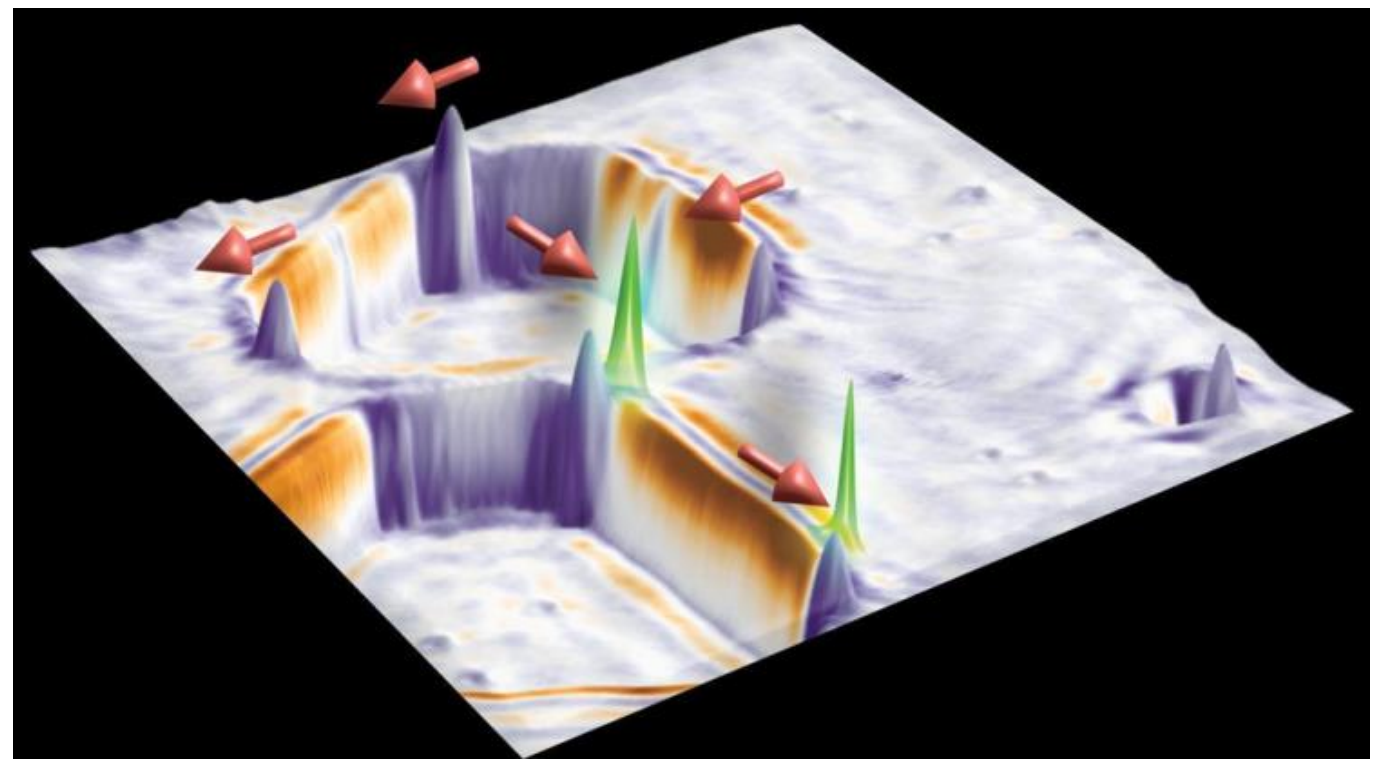

Figure 368: a Majorana Zero mode discovered at Princeton in 2019. Source: Mysterious Majorana Quasiparticle Is Now Closer To Being Controlled For Quantum Computing, June 2019.

He used to work at TU Delft with Leo Kouwenhoven on Majorana fermions and moved in 2012 to the University of Pittsburgh[992].

The withdrawal came after Sergey Frolov and his fellow Pittsburgh researcher Vincent Mourik tried to reproduce Kouwenhoven's experiment and couldn't reproduce its results[993]. Early in 2022, Leo Kouwenhoven left Microsoft in March 2022[994]. Charlie Marcus from Microsoft Research in Denmark also quit Microsoft late 2021 to return to academic research at the Niels Bohr Institute[995].

The same Sergei Frolov is himself making discoveries on Majorana bound states using the 4π Majorana-Josephson effect using a fluxonium superconducting qubit, the ensemble being branded a braidonium[996]. He also reported in 2022 that looking at research experimental data from 2012 did show the presence of Majorana modes in nanowire devices[997]. Other researchers in Germany succeeded in integrating a topological insulator into a superconducting qubit in 2022, following on a 2013 proposal from researchers from Caltech and Harvard[998]. Another 2020 proposal is to couple a Majorana qubit playing the role of a well-protected quantum memory and a superconducting qubit for computation and to implement a SWAP gate between these[999].

990 See Quantized Majorana conductance by Leo Kouwenhoven et al, 2017 (26 pages) which was followed by an "expression of concern" from the authors warning readers about the veracity of the published results, which were not reproducible due to a problem with the calibration of measuring instruments. The coverage on the paper withdrawal in 2021 was dense, starting with Data manipulation and omission in 'Quantized Majorana conductance', Zhang et al, Nature 2018 by Frolov et al, March 2021 (31 slides) which spurred Microsoft's Big Win in Quantum Computing Was an 'Error' After All, by Tom Simonite, Wired, February 2021. Another of their paper was later retracted. See Retraction Note: Epitaxy of advanced nanowire quantum devices by Sasa Gazibegovic, Leo Kouwenhoven et al, Nature, April 2022.

991 See Chiral Majorana fermion modes in a quantum anomalous Hall insulator–superconductor structure, Science, 2017 (7 pages) was the subject of an expression of concern in December 2021.

992 See Signatures of Majorana fermions in hybrid superconductor-semiconductor nanowire devices, Kavli Institute of Nanoscience, Delft University of Technology, 2012.

993 The story is well told in Major Quantum Computing Strategy Suffers Serious Setbacks by Philip Ball, Quanta Magazine, September 2021. The same Philip Ball that was mentioned in the part of this book related to quantum matter taxonomy, page 116.

994 See Kouwenhoven departs, Microsoft presents Majoranas, Delta GTU Delft, March 2022.

995 As shown in his LinkedIn profile: https://www.linkedin.com/in/charles-marcus-02984597.

996 See Braiding quantum circuit based on the 4π Josephson effect by John P. T. Stenger, Michael Hatridge, Sergey M. Frolov and David Pekker, University of Pittsburg, Physical Review B, 2019 (10 pages).

997 See We cannot believe we overlooked these Majorana discoveries by Sergey Frolov and Vincent Mourik, University of Pittsburg and FZ Jülich, March 2022 (9 pages).

998 See Integration of Topological Insulator Josephson Junctions in Superconducting Qubit Circuits by Tobias W. Schmitt et al, FZ Jülich, March 2022 (14 pages). And Proposal for Coherent Coupling of Majorana Zero Modes and Superconducting Qubits Using the 4π Josephson Effect by David Pekker, Chang-Yu Hou, Vladimir E. Manucharyan and Eugene Demler, PRL, 2013 (8 pages).

999 See SWAP gate between a Majorana qubit and a parity-protected superconducting qubit by Luca Chirolli et al, Berkeley, May 2022 (7 pages).

In August 2019, NIST physicists led by Nick Butch announced the discovery by chance of interesting properties of uranium ditelluride ($UTe_2$). It would be superconducting at 1.7K with the ability to do so via Cooper pairs with identical spins in addition to opposite spins, allowing three types of pairs. This would give it a rare ability to get a magnetic flux resistant superconductivity. This material would thus have topological properties in this framework allowing to create topological qubits that are more stable and less subject to decoherence[1000]. Related work was published by researchers from John Hopkins University in 2018 with superconducting topological qubits made of a bismuth-palladium alloy[1001].

And the story goes on and on around Majorana fermions that are discovered, believed to be discovered or rediscovered depending on the case.

They are found on gold[1002], on the surface of superconducting nanowires[1003], in crystals[1004], in 2D graphene[1005], not to mention other publications that are not obvious to analyze[1006], all this in 2020.

**Science**

The principle of topological quantum computing is based on the notion of anyon which are "quasi-particles" integrated in two-dimensional systems, given that there are Abelian and non-Abelian anyons! Anyons are asymmetrical and two-dimensional physical structures whose symmetry can be modified. This makes it possible to apply some topology principles with sets of successive permutations applied to pairs of anyons that are in close proximity in circuits.

The related algorithms are based on the concepts of topological braid or node organizations ("braids"). Their representation explains this, with a temporal evolution of the permutations of temporal anyons going from bottom to top, knowing that in other representations, it may go from top to bottom[1007].

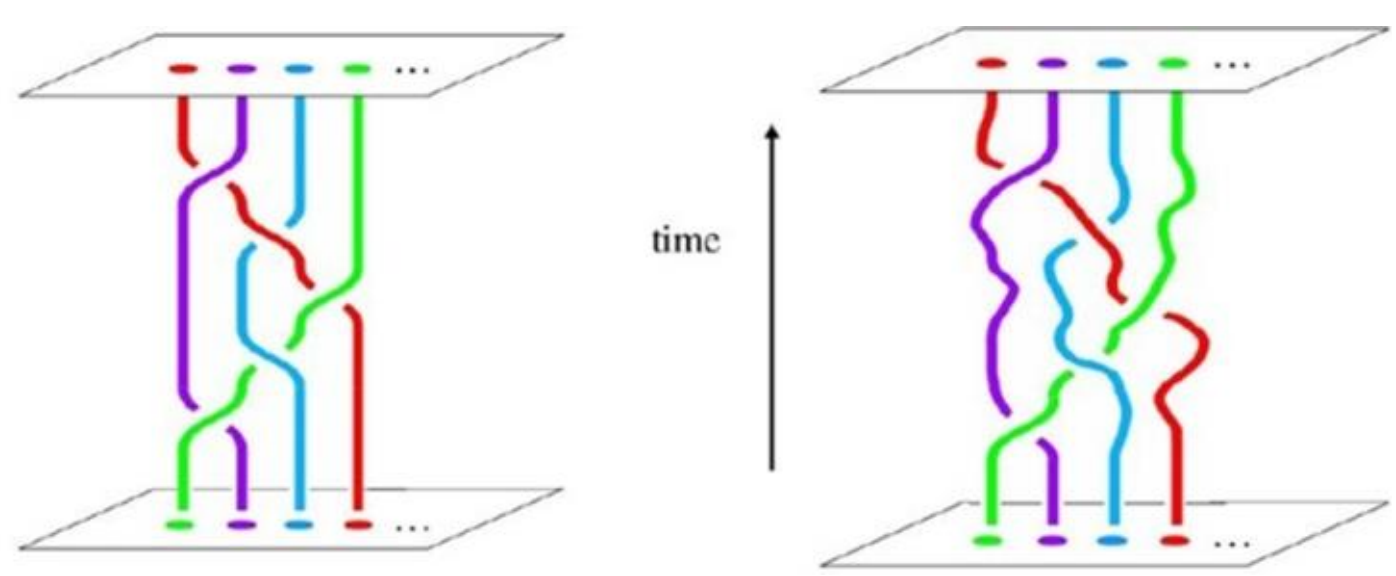

*Figure 369: anyon braiding explained topologically.*

The diagram in Figure 370 clarifies this a little. Topological quantum gates require a long sequence of anyonic permutations as with the CNOT gate shown at the bottom of the diagram. They are a sort of quantum error correction code.

---

1000 See Newfound Superconductor Material Could Be the 'Silicon of Quantum Computers' Possible "topological superconductor" could overcome industry's problem of quantum decoherence, August 2019, mentioning Nearly ferromagnetic spin-triplet superconductivity by Sheng Ran et al, 2019.

1001 See Observation of half-quantum flux in the unconventional superconductorβ-Bi2Pd by Yufan Li & Al, October 2018 (12 pages).

1002 See Quantum Computing Breakthrough: First Sighting of Mysterious Majorana Fermion on Gold by Jennifer Chu, MIT, Indian Institute of Technology, University of California & Hong Kong University, 2020. And Signature of a pair of Majorana zero modes in superconducting gold surface states by Sujit Manna et al, MIT, 2019 (35 pages).

1003 See Alternative route to topological superconductivity Hub, April 2020. University of Copenhagen in collaboration with Microsoft. Refers to Flux-induced topological superconductivity in full-shell nanowires by S. Vaitiekėnas et al, March 2020 (38 pages).

1004 See Building block for quantum computers more common than previously believed by Chanapa Tantibanchachai, Johns Hopkins University, April 2020.

1005 See Observation of Yu-Shiba-Rusinov states in superconducting graphene by E. Cortés-del Río, Pierre Mallet et al, 2020 (22 pages) and published in Advanced Materials in April 2021.

1006 See The observation of photon-assisted tunneling signatures in Majorana wires par Ingrid Fadelli, May 2020, Quantum computers do the (instantaneous) twist by Chris Cesare, August 2020 on a topological error correction system and Fractional statistics of anyons in a two-dimensional conductor, C2N, April 2020.

1007 Topological qubits could also be realized in photonics-based architecture. See New photonic chip promises more robust quantum computers, September 2018, involving researchers in Australia, Italy and Switzerland.

You need two Majorana fermions to create a single two-level quantum system, *aka* a qubit. With Majorana modes, you implement XX, YY and ZZ two-qubit measurements which are well protected and a CNOT is built with two consecutive such measurements. But a T gate is not natively protected making it hard to implement FTQC. Some researchers in China still found a workaround to create non-Clifford single-qubit phase gates[1008].

By the way, let's clarify a little bit the weird vocabulary from this field. Majorana fermions are non-Abelian excitations. They are also called Majorana Zero Modes (MZM) or Majorana bound states or even Majorinos. You have also Majorana Edge Modes (MEM) and Majorana Pi Modes (MPM)[1009]. Others non-Abelian anyons exist like Ising anyons, Fibonacci anyons[1010] and Jones-Kauffman anyons.

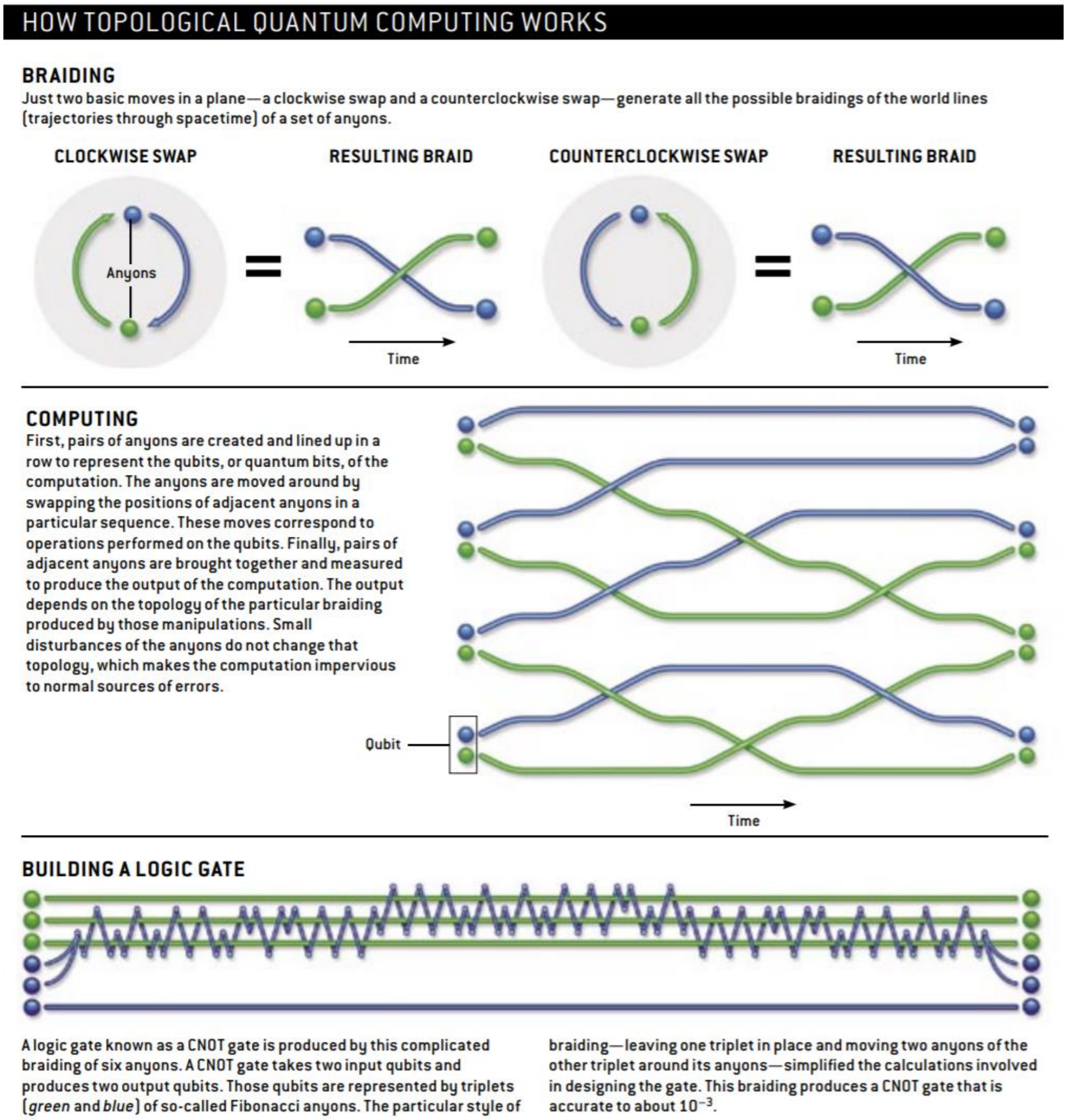

*Figure 370: how topological quantum computing is supposed to work. Source: Computing with Quantum Knots by Graham Collins, Scientific American, 2006 (8 pages).*

---

[1008] See Universal topological quantum computation with strongly correlated Majorana edge modes by Ye-Min Zhan et al, March 2022 (18 pages) which is about creating non-Clifford π/10 gates.

[1009] To sort things out, see the excellent thesis Quantum Field Theories, Topological Materials, and Topological Quantum Computing by Muhammad Ilyas, August 2022 (204 pages).

[1010] See A short introduction to Fibonacci anyon models by Simon Trebst, Matthias Troyer et al, 2009 (24 pages).

### Research

On top of the above-mentioned labs, different physics laboratories are working on topological qubits, notably in the USA, China, the Netherlands, Denmark, Finland[1011] and also in France.

**Maryland**[1012], **Caltech**[1013] and **Purdue**[1014] Universities also have teams working on topological quantum computing and/or Majorana fermions, the two later working with Microsoft Research.

It is also the case of the Quantum Science Center (QSC) from the DoE **ORNL**.

The KouBit Lab from the **University of Illinois** and led by Angela Kou also investigate topologically protected superconducting qubits.

Even IBM is working on Majorana fermions. In 2022, an **IBM Research** team published a paper showing how they could simulate Majorana Zero Modes (MZM), Majorana Pi Modes (MPM) and Majorana braiding on 27-qubits quantum computers[1015]. A large **Google AI** team did a similar experiment with 47 qubits from its Sycamore processor and simulating Majorana Edge Modes[1016].

Research on topological qubits still goes on at **TU Delft** with some recent work on the way to stabilize entanglement of topological qubits[1017]. Teams in **Finland** and **Russia** also work on topological qubits[1018].

In **France**, a team at the CEA-IRIG in Grenoble (Manuel Houzet, Julia Meyer and Xavier Waintal), Pierre Mallet of CNRS Institut Néel in Grenoble, Hugues Pothier at the CEA in Saclay and Pascal Simon at the LPS in Orsay[1019] do not work on Majorana fermions per se, but on topological matter at the fundamental level and particularly on Andreev's states, the linked states and the physics of weak links, different areas that remain to be explored in these lines. Some of these researchers are conducting joint projects with TU Delft.

Majorana zero modes (MZMs) were also found by researchers in **China** in 2022 with iron-based superconductors showcasing topological vertices[1020]. In February 2020, **John Preskill** (father of the notions of quantum supremacy and NISQ) predicted that by 2030, we will be able to demonstrate two entangled topological qubits, against **Jonathan Dowling (**photonicist) who did not believe it could

---

[1011] See Ultra-thin designer materials unlock quantum phenomena, Aalto University, December 2020 and Topological superconductivity in a van der Waals heterostructure by Shawulienu Kezilebieke et al, March 2021 (27 pages).

[1012] See On-demand large conductance in trivial zero-bias tunneling peaks in Majorana nanowires by Haining Pan and Sankar Das Sarma, University of Maryland, March 2022 (8 pages) and Euler-obstructed Cooper pairing: Nodal superconductivity and hinge Majorana zero modes by Jiabin Yu, Yu-An Chen and Sankar Das Sarma, University of Maryland, Physical Review B, March 2022 (47 pages).

[1013] See Dephasing and leakage dynamics of noisy Majorana-based qubits: Topological versus Andreev by Ryan V. Mishmash, Bela Bauer, Felix von Oppen and Jason Alicea, November 2019 (22 pages).

[1014] See Ternary Logic Design in Topological Quantum Computing by Muhammad Ilyas et al, Purdue and Portland Universities, April 2022 (53 pages).

[1015] See Observing and braiding topological Majorana modes on programmable quantum simulators by Nikhil Harle et al, Yale University, MIT & IBM Research, March 2022 (14 pages). Tested with 27-qubit superconducting systems.

[1016] See Noise-resilient Majorana Edge Modes on a Chain of Superconducting Qubits by Xiao Mi et many al, Google AI, April 2022 (24 pages).

[1017] See Topological Entanglement Stabilization in Superconducting Quantum Circuits by Guliuxin Jin and Eliska Greplova, TU Delft, May 2022 (10 pages).

[1018] See Half-quantum vortices and walls bounded by strings in the polar-distorted phases of topological superfluid $^3$He by J.T. Makinen, G.E. Volovik et al, 2018 (17 pages).

[1019] See Pascal Simon's presentation Majorana zero modes around skyrmionics textures, 2021 (75 slides).

[1020] See Ordered and tunable Majorana-zero-mode lattice in naturally strained LiFeAs by Meng Li et al, Nature, May 2022 (38 pages). See also Topologically protected quantum entanglement emitters by Jianwei Wang et al, February 2022 (10 pages) which deals with topologically protected matter although not formally a Majorana fermion.

be created! The object of this symbolic wager? A good beer and a pizza. Jonathan Dowling died in 2020 and will therefore not be able to see if he won or lost his bet in 2030.

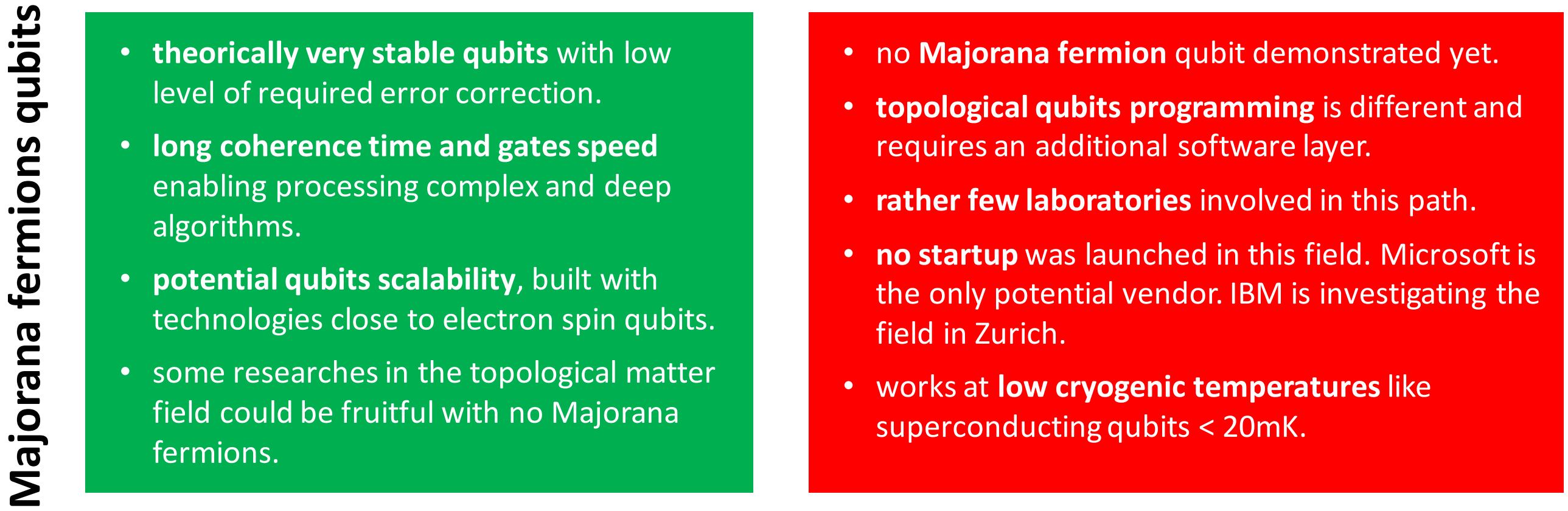

*Figure 371: pros and cons of Majorana fermions and topological qubits. (cc) Olivier Ezratty, 2022.*

**Vendors**

Microsoft Research has investigating topological quantum computing and Majorana fermions for quite a few years but has no prototype at this stage.

The company is making a bet there, up to say that if they fail, everybody will also fail in quantum computing. While being a little arrogant and a very risky bet, it would bring lots of strategic advantages if it worked!

Indeed, Majorana qubits would be much more reliable and generate fewer errors ($10^{-30}$), with the implication that we could avoid using some of the classical quantum error correction codes that are implemented with other types of qubits[1021].

A Fields Medal in 1986 for his work on the Poincaré conjecture, Fields Medal winner **Michael Freedman** joined Microsoft in 1997, coming from the University of Santa Barbara, the same where John Martinis came from when he joined Google in 2014. Freedman demonstrated with Alexei Kitaev the possibility of doing quantum computing with a hypothetical particle, the Majorana fermion, conceptualized in 1937 by the Italian Ettore Majorana from the resolution of mathematical equations of Dirac[1022]. This fermion is a strange particle, whose charge and energy are zero and which is its own antiparticle.

Freedman and Kitaev were recruited by Microsoft Research. Created by Michael Freedman, Microsoft Quantum Santa Barbara (Station Q) is located on the campus of the University of Santa Barbara. They were complemented by **Leo Kouwenhoven**'s team based in Microsoft's Delft Lab in the Netherlands and with **Charles Marcus** from the Niels Bohr Institute who also joined Microsoft Research who both left Microsoft in 2022 and 2021. Microsoft also collaborates with **Purdue University** in Indiana, where it has a dedicated research team **Microsoft Quantum Purdue**, working on III-V superconductors.

[1021] Here are a few leads to find out more: Microsoft Ready to Build a Quantum Computer by Juliette Raynal, 2016, A Software Design Architecture and Domain-Specific Language for Quantum Computing, 2014 (14 pages), Quantum Computing at Microsoft (56 slides) and Quantum Computing Research at Microsoft (59 slides) by Dave Wecker and A short introduction to topological quantum computation by Ville Lahtinen and Jiannis Pachos, 2017, (43 pages). And some videos: keynote of November 2017 with Leo Kouwenhoven (43 mn), Build conference of May 2018 on Q# (1h15mn) and Majorana qubits by Xiao Hu, in May 2017 (22 mn).

[1022] In Topological Quantum Computation published in 2002 and updated in 2008 (12 pages).

Majorana fermions are strange behaviors of electrons and their spin that are found at both ends of superconducting wires. They operate at very low temperatures, as for superconducting and silicon-based qubits, at about 15-20 mK[1023].

Seen up close, these qubits are sophisticated variants of superconducting qubits. These "topological" mesh associations provide protection against qubit decoherence because the shape of the braids does not matter as long as their topology is stable.

Microsoft announced at the Build conference in May 2018 that they would release their first fermion-based quantum computer from Majorana in 2023[1024]. After Leo Kouwenhoven's 2017 paper withdrawal in 2021, this planning seems somewhat challenging[1025]. But let's not count Microsoft out of the game too rapidly. All discoveries have their up and downs. If a failure meant *stop all research*, Thomas Edison would not have discovered the light bulb and many vaccines and cancer treatments wouldn't see the light of day!

In 2022, as Kouwenhoven was leaving Microsoft, the company published some new results related to Majorana fermions and scalable quantum computing. They seem not having learned from their past errors and were emphatically overselling their recent work[1026]. It dealt with the observation of a 30 μeV topological gap in indium arsenide-aluminum heterostructures. But it seemed that this topological qubit was only digitally simulated and not implemented practically.

Just before, in February 2022, another Microsoft team published a paper related to a quantum error correction code (planar Floquet code) that was suitable for topological qubits and with a very high threshold exceeding 1% (meaning, qubits with 1% error rates are sufficient to implement error correction which seems a very high error-rate compared to what would be needed with superconducting qubits and surface codes, that require errors way below 0.1%)[1027].

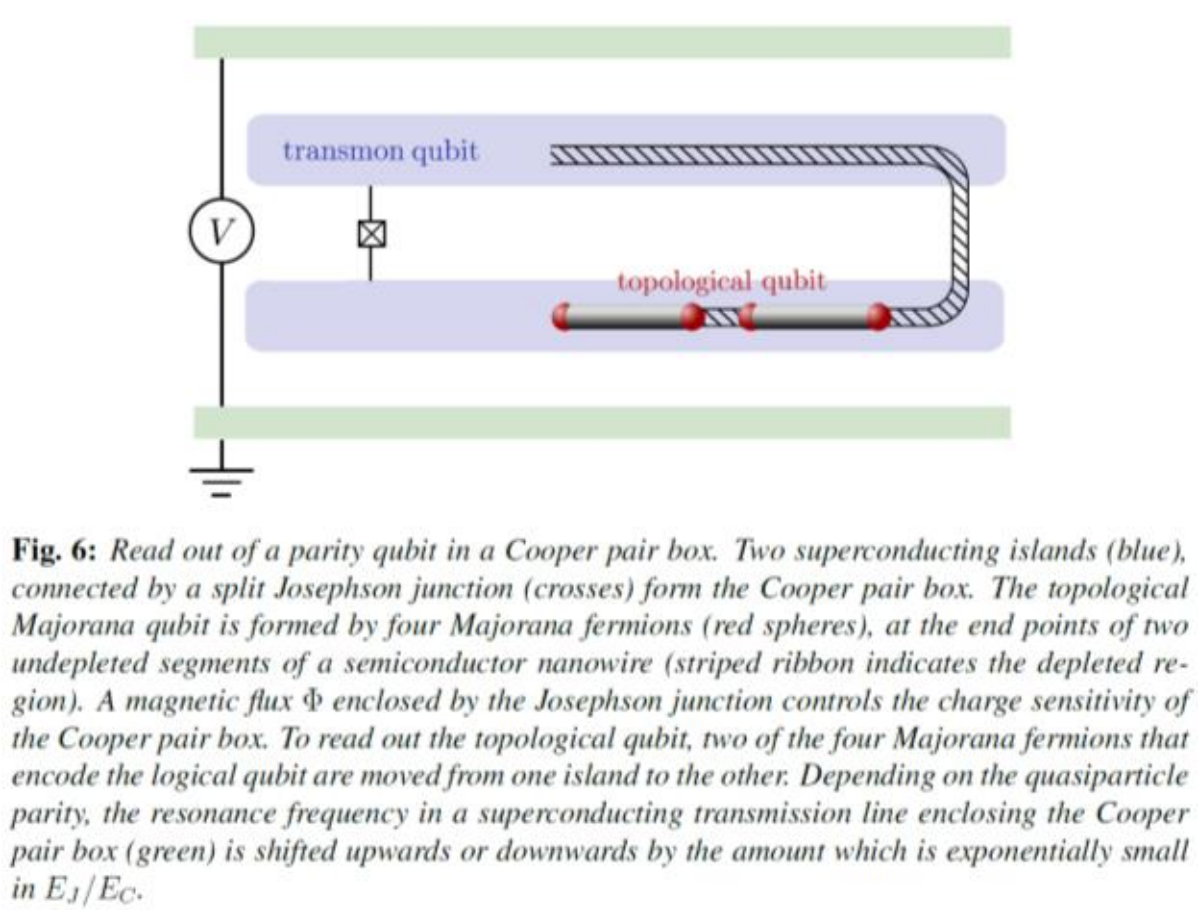

*Figure 372: typical combination of a topological and a superconducting qubit. Source: [Majorana Qubits](#) by Fabian Hassler, 2014 (21 pages).*

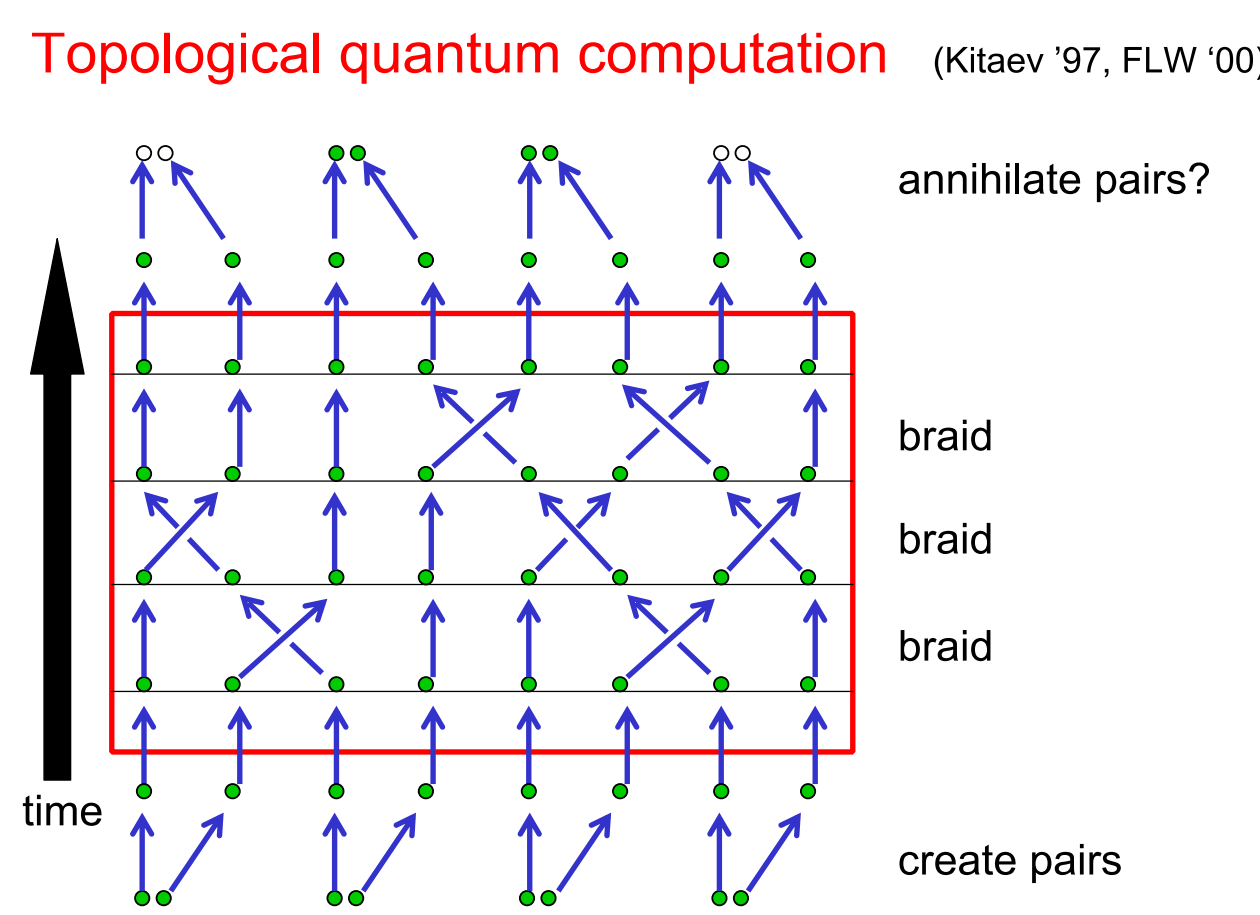

*Figure 373: how braiding is sequenced during topological computing. Source: [Topological quantum computing for beginners](#), by John Preskill (55 slides).*

---

1023 See [Majorana Qubits](#) by Fabian Hassler, 2014 (21 pages).

1024 See this video ad: [Introducing Quantum Impact (Ep. 0)](#), February 2020 (4 minutes).

1025 See [Topological quantum computing for beginners](#), by John Preskill (55 slides).

1026 See [In a historic milestone, Azure Quantum demonstrates formerly elusive physics needed to build scalable topological qubits](#) by Jennifer Langston, March 2022, [Microsoft has demonstrated the underlying physics required to create a new kind of qubit](#) by Chetan Nayak, March 2022, based on [Protocol to identify a topological superconducting phase in a three-terminal device](#) by Dmitry Pikulin et al, March 2021 (28 pages).

1027 See [Performance of planar Floquet codes with Majorana-based qubits](#) by Adam Paetznick, Nicolas Delfosse et al, February 2022 (20 pages).

Overall, these Microsoft's research results have a very low TRLs have not been successful so far. We have for example no idea about the number of physical or logical qubits that they could implement, and how these are driven from the control and readout standpoint.

Microsoft obviously also invested on software development, first with its Liquid platform, then with F# for scripting and with the Q# language used for quantum programming, launched at the end of 2017.

One of the contributors to these efforts is researcher **Krysta Svore** from Columbia University. In 2018, Microsoft recruited a certain **Helmut Katzgraber**, one of the apostles of D-Wave quantum annealing and MBQC (measurement-based quantum computers)[1028].

| | Realizations | Lifetimes | Gate Speed |
|---|---|---|---|
| | Topological (Majorana) | 1 minute | Nanoseconds |
| supraconducting | Flux Qubit | $/ 10^{10}$ | same |
| supraconducting | Charge Qubit | $/ 10^{10}$ | same |
| supraconducting | Transmon | $/ 10^{7}$ | same |
| | Ion Trap | $/ 10^{2}$ | $10^{3}$ slower |

better stability qubits
low decoherence noise
few errors
long coherence time
high gate speed

nothing demonstrated so far
no prototype
different algorithms

*Figure 374: timing benefits from Majorana fermions. Source: Microsoft, 2018.*

**Nokia's Bell Labs** in the USA, located in Murray Hill, New Jersey, work *or worked* on topological qubits[1029].

Nokia also supports Oxford University's Quopal initiative on the use of quantum in machine learning. Nokia likes to remind us that Grover and Shor's algorithms were discovered by their creators when they worked at the Bell Labs. Nokia is also working on quantum cryptography, at least at the level of its transport on optical fibers, as demonstrated by this partnership with SK Telecom of 2017.

There is even a private research company, **Quantum Gravity Research** (2009, USA) with an overarching goal to create a unified physics theory encompassing gravity and quantum physics that also undertakes research in topological computing[1030]. The organization created by Klee Irwin employs about 20 researchers like Fang Fang, Raymond Aschheim, Marcelo Amaral, Dugan Hammock, Richard Clawson and Michel Planat. They envision "*a specific substructure of spacetime at the smallest scale, so that in this view physical reality is like a mosaic tiling language of Planck scale, 3-dimensional, tetrahedron-shaped pixels*". Let's say it is quite difficult to fact-check these sorts of claims and their practicality!

## Trapped ions qubits

Trapped ions are positively ionized atoms that are trapped by electrodes and sometimes also magnetically in a confined space and placed next to each other. The atoms are generally alkaline metals from the second column of Mendeleev's table (called "Group IIA" in Mendeleev's notation or group 2 in the modern notation, with beryllium, magnesium, strontium, barium and calcium), then as ytterbium which is a rare earth in the lanthanide family or even mercury, and finally, quite rarely, metals of group IIB or 12 (zinc, cadmium, mercury) [1031].

[1028] See Quantum Driven Classical Optimizations, August 2018 (28 min video).

[1029] See Quantum computing using novel topological qubits at Nokia Bell Labs published in 2017, which describes their approach with topological qubits.

[1030] See Exploiting Anyonic Behavior of Quasicrystals for Topological Quantum Computing by Marcelo Amaral et al, Quantum Gravity Research, July 2022 (20 pages).

[1031] See this interesting perspective on trapped ions qubits in Introduction to Trapped Ion Quantum Computing by Gabriel Mintzer from MIT, February 2020.

**trapped ions qubits**

- **identical ions** => no calibration required like with superconducting/electron spin qubits.
- **good qubits stability** with best in class low error rate.
- **long coherence time** and high ratio between coherence time and gate time => supports deep algorihms in number of gates.
- **entanglement** possible between all qubits on 1D architecture. It speeds up computing.
- works at **4K to 10K** => simpler cryogeny than for superconducting/electron spins.
- **easy to entangle ions with photons** for long distance communications.

- **entanglement** doesn't seem to scale well with a large number of ions.
- **questionnable scalability options** beyond 50 qubits (ions shuttling, 2D architectures, photon interconnect).
- **relatively slow computing** due to slow quantum gates which may be problematic for deep algorithms like Shor integer factoring.

*Figure 375: pros and cons of trapped ions qubits. (cc) Olivier Ezratty, 2022.*

### History

Before the very notion of a qubit even existed, scientists tried to control ions in space. **Wolfgang Paul** created in 1953 a way to control ions with a mass spectrometer avoiding the use of a magnetic field, named the "Paul trap". He got the Nobel Prize in Physics in 1989. Later, in 1959, **Frans Penning** and **Hans Dehmelt** were able to control individual electrons with a magnetron trap that was later named the "Penning trap". Penning traps are still being studied, particularly in their 2D variant.

In the USA, **David Wineland** from NIST created the laser cooling technique starting in 1979, using Doppler effect with magnesium ions. He got the Nobel prize in physics in 2012 along with Serge Haroche. In 1989, he used the technique to cool ions at their zero-point energy of motion with the sideband cooling technique that goes farther than Doppler cooling[1032].

**Juan Ignacio Cirac** and **Peter Zoller** from the University of Innsbruck in Austria proposed in 1995 a blueprint to create a gate-based quantum processor with a linear trap of ions controlled by laser beams[1033]. They initiated a long-lasting experience in the field at Innsbruck, which led to the creation of the startup **AQT** in 2017.

### Science

Here Are some specifics of trapped ions qubits…

**Why ions?** The interest of exploiting ions is to allow to trap them magnetically or with electrodes. It is also possible to couple them at long distance, of the order of several tens of microns. It can also be hybridized with several ion types mixed together, like calcium and strontium, to get their related benefit such as fast gates for calcium and stability for strontium[1034]. The used elements have several common characteristics related to their electron layer configuration such as excitation levels from the ground state that are of short duration and allow their use for atoms cooling with laser and the Doppler effect. The basic energy state corresponding to the $|0\rangle$ and the excited energy level corresponding to $|1\rangle$ state are stable over time, which facilitates the implementation of quantum gate operations.

---

[1032] It is too complex to describe the Doppler cooling limit and how sideband cooling works. See Laser cooling of trapped ions by Jürgen Eschner, Giovanna Morigi, Ferdinand Schmidt-Kaler and Rainer Blatt, 2003 (13 pages) which describes various ions cooling techniques.

[1033] See Trapped Ion Quantum Computing: Progress and Challenges by Colin Bruzewicz et al from MIT, April 2019 (56 pages). This is a very well-documented state-of-the-art review of trapped ion technology. And the founding article Quantum Computations with Cold Trapped Ions by Juan Ignacio Cirac and Peter Zoller, 1995 (4 pages).

[1034] See Benchmarking a high-fidelity mixed-species entangling gate by A. C. Hughes et al, Oxford University, August 2020 (7 pages).

**Long coherence time**. Trapped ions have a rather long coherence time of up to several tens of seconds, but this is compensated by equally long gate times in proportion. The ratio between coherence time and gate time is however currently quite good at $10^6$, while it is $10^3$ for superconducting qubits and about 200 for cold atoms qubits.

**Qubit fidelities**. Trapped ions show a fairly low error rate with up to single-qubit 99.999% and two-qubit gates 99.9% fidelity. The table below illustrates these fidelities depending on the quantum gate implementation and used ions. This and long coherence time make it possible to theoretically execute "deep algorithms” with a large number of quantum gates and to obtain a good quantum volume, to use IBM’s terminology. However, this error rate seems to increase with the number of qubits, at least in 1D architectures like the one from IonQ.

| Type | Method | Fidelity | Time ($\mu$s) | Species | Ref. |
|---|---|---|---|---|---|
| 1-qubit | Optical | 0.99995 | 5 | $^{40}Ca^+$ | Bermudez 2017 |
| | Raman | 0.99993 | 7.5 | $^{43}Ca^+$ | Ballance 2016 |
| | Raman | 0.99996 | 2 | $^{9}Be^+$ | NIST 2016 |
| | Raman | 0.99 | 0.00005 | $^{171}Yb^+$ | Campbell 2010 |
| | Raman | 0.999 | 8 | $^{88}Sr^+$ | Keselman 2011 |
| | $\mu$wave | 0.999999 | 12 | $^{43}Ca^+$ | Harty 2014 |
| | $\mu$wave | | 0.0186 | $^{25}Mg^+$ | Ospelkaus 2011 |

Adapted from Bruzewicz et al.

source des schémas : lecture 1 de 77 slides du cours d'Hélène Perrin à l'Université Paris 13 en quatre parties, février 2020.

| Type | Method | Fidelity | Time ($\mu$s) | Species | Ref. |
|---|---|---|---|---|---|
| 2-qubit | Optical | 0.996 | – | $^{40}Ca^+$ | Erhard 2019 |
| (1 sp.) | Optical | 0.993 | 50 | $^{40}Ca^+$ | Benhelm 2008 |
| | Raman | 0.9991(6) | 30 | $^{9}Be^+$ | NIST 2016 |
| | Raman | 0.999 | 100 | $^{43}Ca^+$ | Ballance 2016 |
| | Raman | 0.998 | 1.6 | $^{43}Ca^+$ | Schafer 2018 |
| | Raman | 0.60 | 0.5 | $^{43}Ca^+$ | Schafer 2018 |
| | $\mu$wave | 0.997 | 3250 | $^{43}Ca^+$ | Harty 2016 |
| | $\mu$wave | 0.985 | 2700 | $^{171}Yb^+$ | Weidt 2017 |
| 2-qubit | Ram./Ram. | 0.998(6) | 27.4 | $^{40}Ca^+/^{43}Ca^+$ | Ballance 2015 |
| (2 sp.) | Ram./Ram. | 0.979(1) | 35 | $^{9}Be^+/^{25}Mg^+$ | Tan 2015 |

*Figure 376: some trapped ions fidelities obtained with different atoms. Source: lecture 1 on trapped ions, Hélène Perrin, February 2020 (77 slides).*

There’s a tendency with trapped ions vendors like IonQ and Quantinuum to measure qubit fidelities with the SPAM method which encompasses the whole process from state preparation to measurement. In March 2022, IonQ tested its new barium ions and improved fidelities with a record 99,96% SPAM fidelity. Quantinuum reached simultaneously a SPAM fidelity of 99,9904% also with barium ions[1035], which is clearly best-in-class in the qubit world.

**Connectivity**. Trapped ions qubits can all be entangled with each other with using phonons or microwaves, but it depends on how they are distributed in space[1036]. This simplifies the implementation of many algorithms, avoiding the usage of costly SWAP gates to connect distant qubits.

**No calibration**. Since these qubits are atoms, they are identical and do not require calibration adjustments like with superconducting qubits whose physical properties vary from one qubit to another depending on their materials and manufacturing.

**Ions variations**. There are five main variations of trapped ions being used, depending on the energy transitions applied to manage the two states of a qubit[1037]. Each of these modes correspond to different transition frequencies:

- **Zeeman qubits** use electromagnetic waves of a few MHz with magnetic field control. They are very sensitive to it but allow to have qubits with a very low error rate once this field is well controlled[1038]. They are rather used in quantum sensing since their control frequency is too low to allow a precision control of several qubits close to each other.

[1035] See High fidelity state preparation and measurement of ion hyperfine qubits with I>1/2 by Fangzhao Alex An et al, March 2022 (5 pages).

[1036] See Benchmarking an 11-qubit quantum computer by Christopher Monroe et al, March 2019 (8 pages).

[1037] See Ion traps you never knew existed by M. Malinowski, February 2022 which makes an interesting inventory of trapped ions settings.

[1038] See Comparing Zeeman qubits to hyperfine qubits in the context of the surface code: $^{174}Yb^+$ and $^{171}Yb^+$ by Natalie Brown, April 2018 (7 pages).

- **Hyperfine structure qubits** use microwaves of a few GHz and laser-based Raman transitions[1039]. This works with ions having a non-zero spin nucleus. The other cases concern ions with zero spin nuclei, i.e. those whose proton and neutron numbers are both even. This explains why some elements such as calcium are sometimes used in several of these categories, with different isotopes such as $^{40}Ca+$ in optical qubits and $^{43}Ca+$ in qubits of hyperfine structure. The number of neutrons in these ions changes the spin of the nucleus of atoms and its hyperfine energy states. In this category, IonQ and Honeywell are using hyperfine structure qubits driven by lasers and Oxford Ionics is using microwave gates.
- **Fine structure qubits** use submillimeter waves of a few THz.
- **Optical qubits** use photons of a few hundred THz. AQT is using this type of qubits.
- **Rydberg qubits** use so-called Rydberg energy states controlled by VUV ultraviolet rays (vacuum ultraviolet, not transmitted in air, needs vacuum), with wavelengths under 122 nm[1040].

Figure 377: contains a description of these variants based on ion energy levels and transitions. On the left, a generic structure of ion energy levels with the transitions allowing the change of qubit state and those used to prepare the qubit state or to read it. These charts showing atoms electronics energy transitions including fine and hyperfine transitions are called Grotrian diagrams. In the middle and on the right, the different energy transitions used to define the $|0\rangle$ and $|1\rangle$ of the qubit. The height between the two levels characterizes the energy level that separates these two states. The higher it is, the higher the frequency used to modify the qubit state, going from radio waves of a few MHz to extreme ultraviolet in the case of Rydberg qubits.

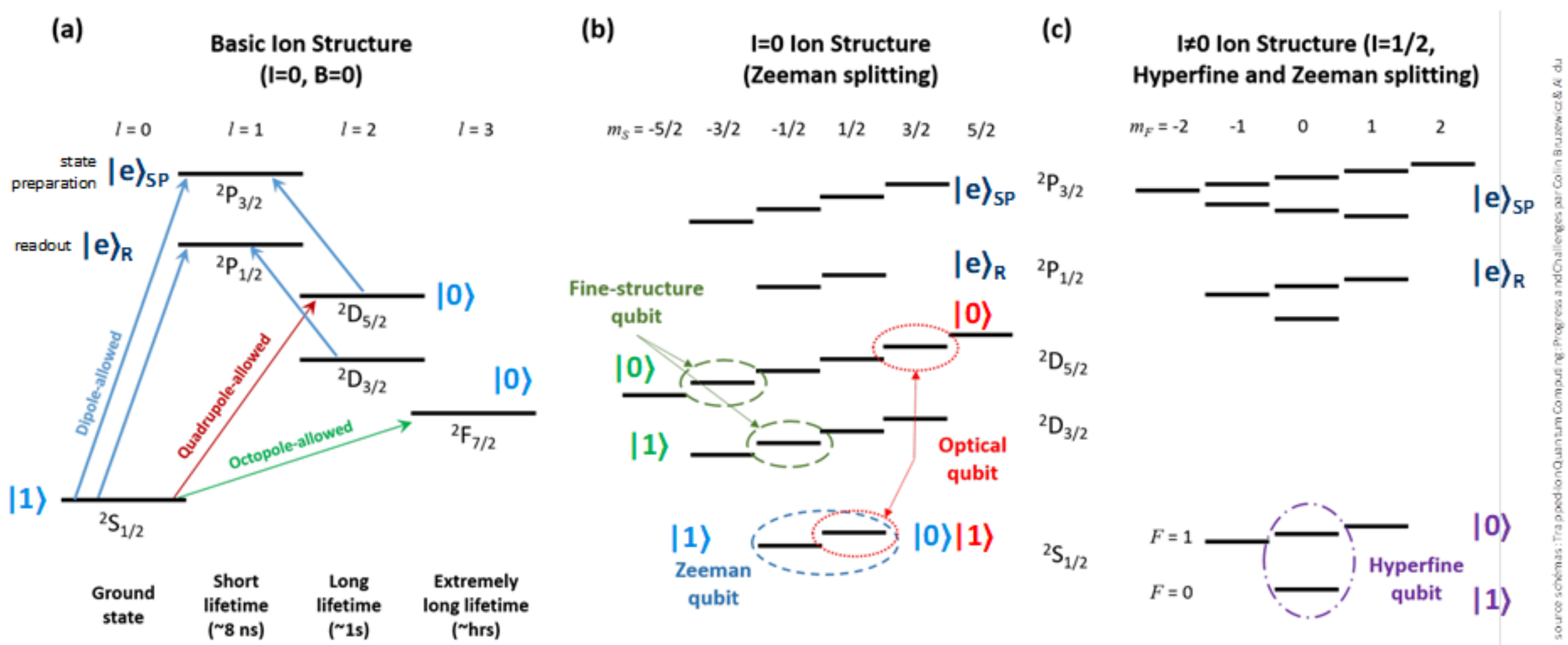

*Figure 377: various types of trapped ions and their respective energy transitions. Source: [Trapped Ion Quantum Computing: Progress and Challenges](#) by Colin Bruzewicz et al from MIT, April 2019 (56 pages).*

The spatial stabilization of trapped ions is achieved in two main ways with ion traps that allow individual control of their position[1041]:

---

[1039] See Controlling Qubits With Microwave Pulses Reduces Quantum Computer Error Rates, Increases Efficiency by Matt Swayne, 2020, which references Robust and resource-efficient microwave near-field entangling $^{9}Be+$ gate by G. Zarantonello, November 2019 (6 pages). See glossary for Raman transition.

[1040] See for example Speeding-up quantum computing using giant atomic ions by Stockholm University, April 2020.

[1041] See many details in Multi-wafer ion traps for scalable quantum information processing by Chiara Decaroli, 2021 (248 pages).

- With a **magnetic field** and an **electric quadrupole**: these are the Penning traps, invented in 1959. Among other places, they have been tested at the ETH Zurich in Jonathan P. Home's team and in a 2D version which has the advantage of being theoretically scalable[1042].

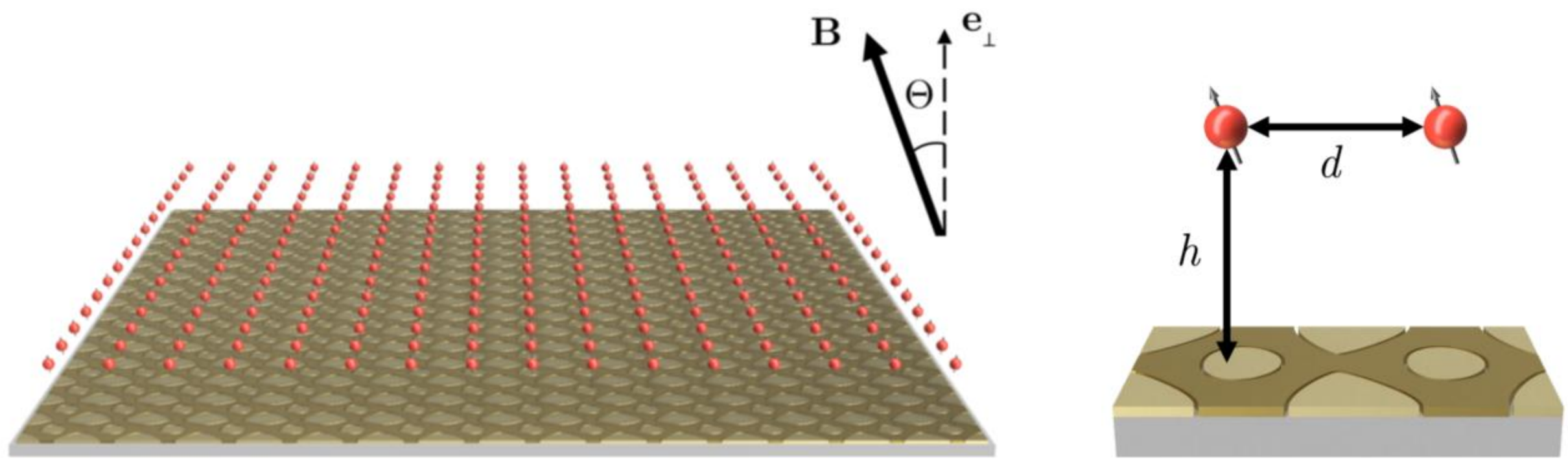

*Figure 378: proposal for an array of trapped ions. Source: [Scalable arrays of micro-Penning traps for quantum computing and simulation]() by S. Jain, Jonathan P. Home et al, April 2020 (21 pages).*

- With a **variable electric field**: these are the Paul traps named after Wolfgang Paul. These traps are either linear in 1D structure (in Figure 379 in (f)) or flattened to create 2D structures. They are the most often used. The flat version corresponds to the technique used by Quantinuum and IonQ.

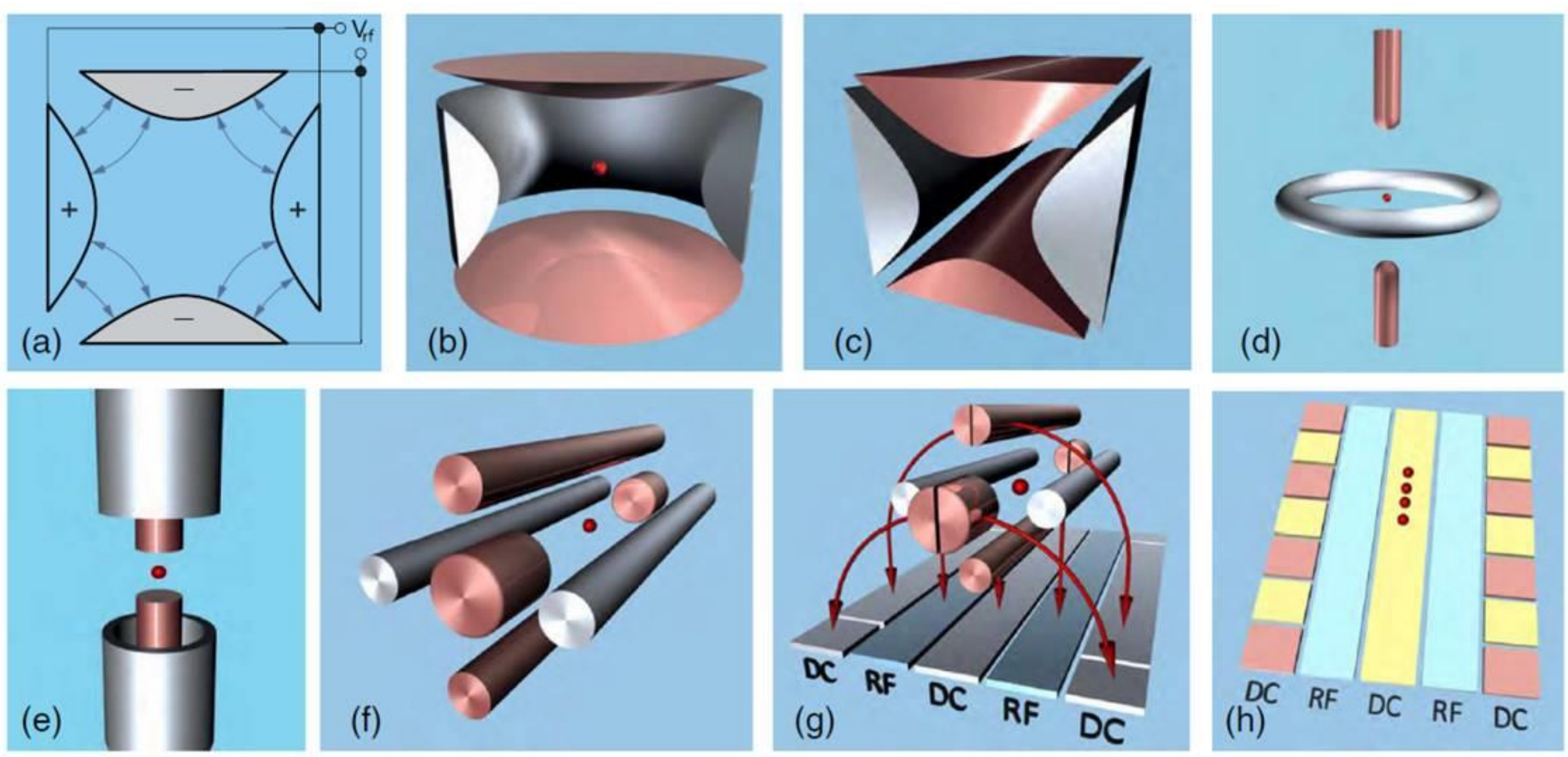

FIG. 2. (Reproduced from [68].) RF Paul trap geometries. (a) The basic concept of RF trapping, where quadrupolar fields oscillating at an RF frequency are produced using a set of (parabolic) electrodes. (b) The simplest cylindrically symmetric version of the basic RF trap. This is of the "ring and endcap" point-trap geometry. (c) The simplest translationally symmetric version of the basic RF trap. This will form a quadrupole mass filter and can be used to make a linear trap. (d,e) Topologically equivalent deformations of the geometry shown in (b). (f) Topologically equivalent deformations of the geometry shown in (c) with additional endcap electrodes added to form a four-rod, linear trap. (g) The four-rod trap in (f) may be deformed such that all electrodes reside in a single plane, forming a linear "surface-electrode trap." (h) A subset of the electrodes in a linear trap [a surface-electrode trap is depicted here, but segmentation may be applied to other linear trap geometries, such as that shown in (f)] may be segmented to allow trapping in multiple zones, along the axial direction.

*Figure 379: the various ways to trap ions. Source: [Trapped Ion Quantum Computing: Progress and Challenges]() by Colin Bruzewicz et al from MIT, April 2019 (56 pages).*

[1042] See [Scalable arrays of micro-Penning traps for quantum computing and simulation]() by S. Jain, Jonathan P. Home et al, April 2020 (21 pages).

These various traps are implemented on integrated circuits using variations of direct current and radio-frequency electrodes and/or laser wave guides. Lasers play several roles in trapped ions control: they are used to cool the ions with the Doppler effect and by sideband cooling to slow down phonons (these are inter-ions vibrations, kind of shock waves), to initialize the energy state of the qubits to its ground |0⟩ state, to create quantum gates and finally, for qubits state readout[1043].

The main disadvantage is that the solution will probably not scale well, particularly with laser light control that goes through a light splitter and some lenses to focus it on the controlled ions. The ions are aligned in rows and separated by about 2 to 5 µm.

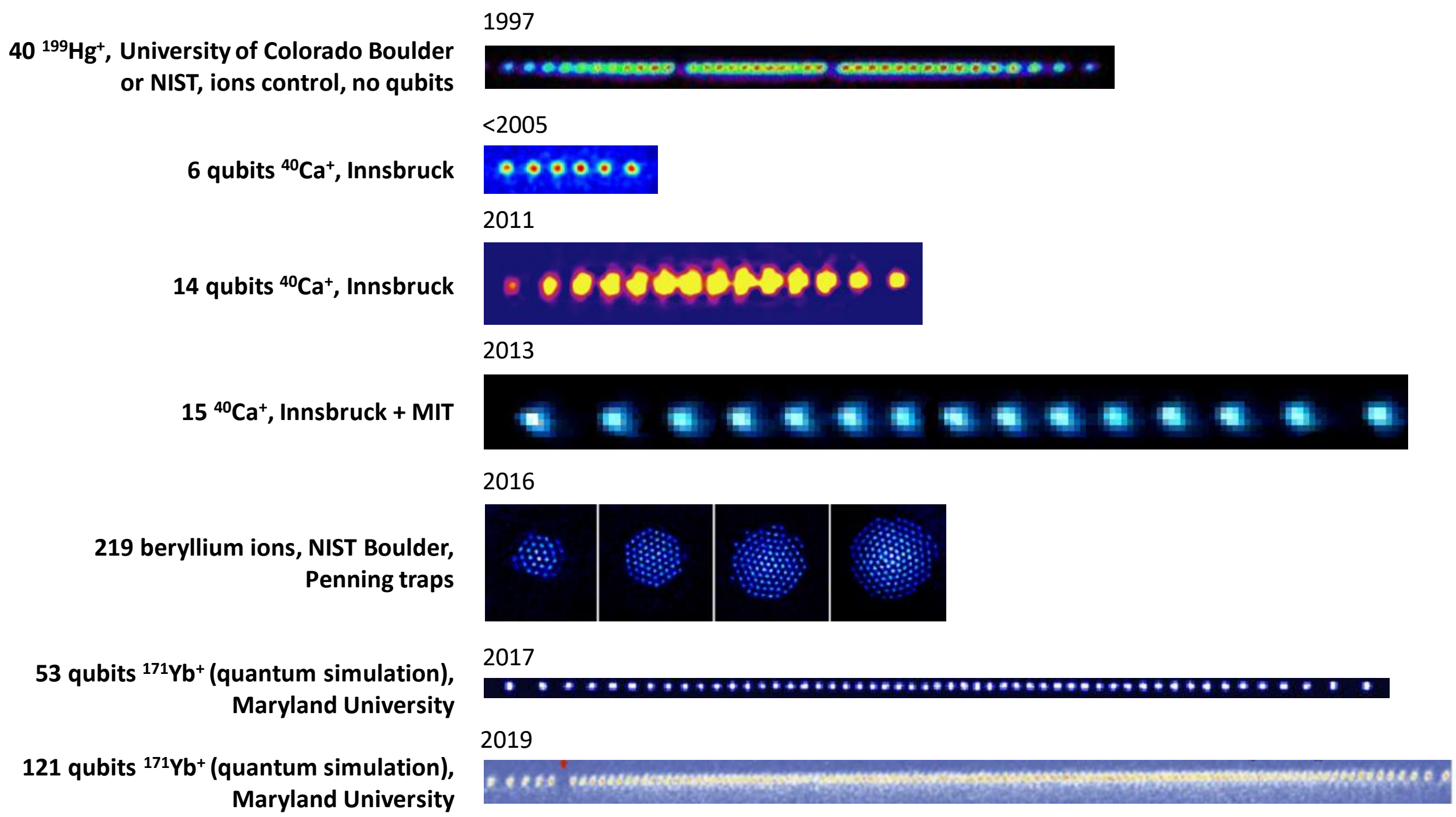

*Figure 380: different lines of trapped ions over time. Compilation: Olivier Ezratty. 2020.*

**Temperature**. Trapped ions are supposed to operate at room temperature. In practice, they generate an annoying heating effect, which is not fully explained at the moment. This requires some cooling between 4K and 10K[1044]. The interest of such a cooling is also to improve the quality of the ultra-high vacuum chamber. Lasers and ions readout imagers also need some cooling at reasonable temperatures (from 10K to -35°C depending on the case).

Researchers from the University of Innsbruck and from ETH Zurich are thinking about making the trapped ion technology "portable", forgetting the vacuum system and the cryostat, necessary for their operation[1045]. They are part of the EU-funded **PIEDMONS** project (E2020). It also involves Infineon Austria[1046].

---

[1043] See Quantum information processing with trapped ions by Christian Roos, 2012 (53 slides) on how trapped ion qubits are driven.

[1044] See Closed-cycle, low-vibration 4 K cryostat for ion traps and other applications by P. Micke et al, May 2019 (15 pages) which describes a cryostat for ion trapped processors using a pulsed head.

[1045] See Quantum computers to become portable, August 2019.

[1046] See 2D Linear Trap Array for Quantum Information Processing by Philip C. Holz, Rainer Blatt et al, September 2020 (20 pages).

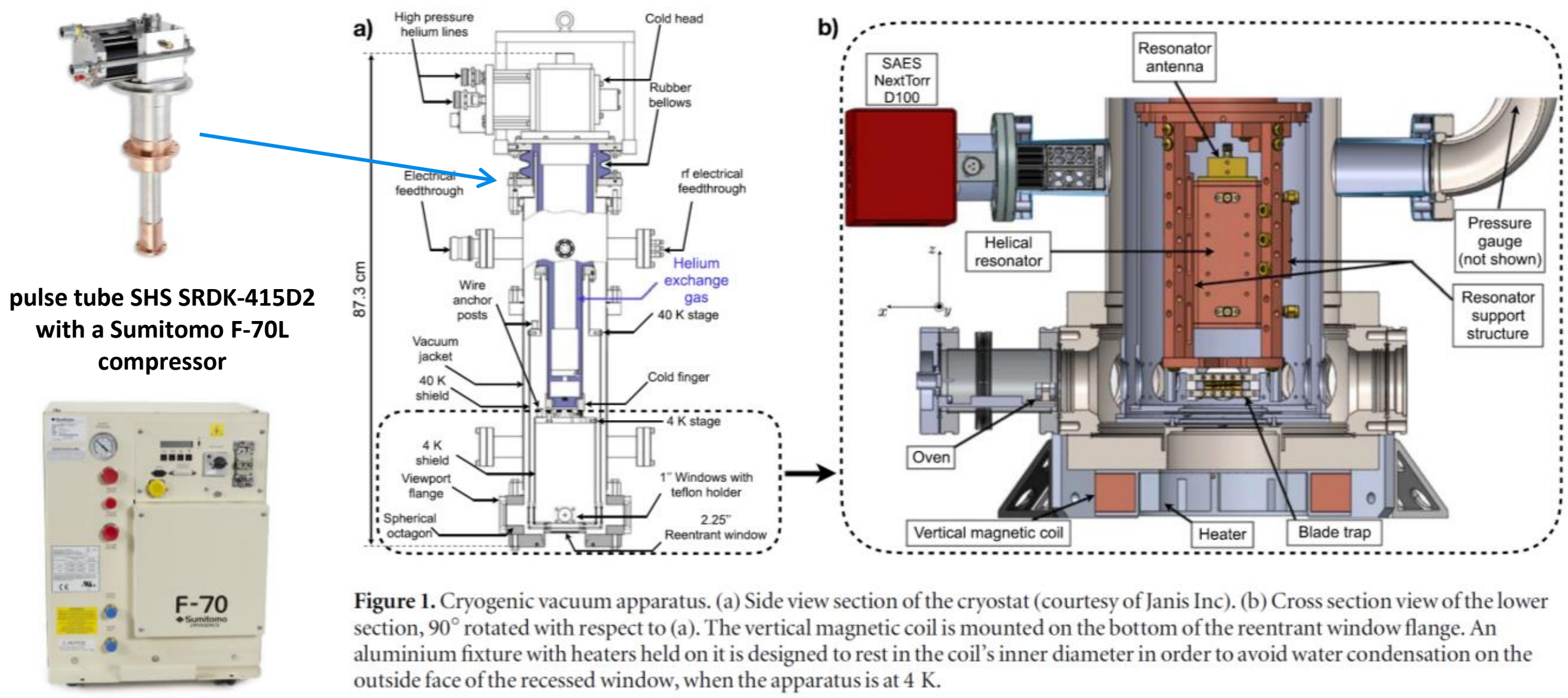

*Figure 381: the 4K cryostat used a while ago by Christopher Monroe's team at the University of Maryland to trap more than a hundred ytterbium ions. It operated a 4.2K SHS pulse tube and a Sumitomo compressor[1047]. Source : Cryogenic trapped ion system for large scale quantum simulation by Christopher Monroe et al, 2018 (17 pages).*

**Scale-out**. Trapped ions have at least two other use cases: quantum memories, and their integration in quantum repeaters for secure quantum telecommunications, including quantum keys distribution[1048]. This involves interactions between trapped ions and photons, using cavities. It is already possible to entangle trapped ions via a photonic link of several hundred meters, a feat done over a distance of 400 m at the University of Innsbruck. This would enable the creation of distributed quantum computing architectures, a plan devised by IonQ to circumvent the scalability limitations of their qubits[1049].

Other avenues are explored to scale-out trapped ions processors like with shuttling ions from computing units to other computing units[1050]. This is the basis of rather old research and is still pursued practically by Quantinuum in their future 2D architecture.

---

[1047] See also the thesis Towards Cryogenic Scalable Quantum Computing with Trapped Ions by Matthias Brandl, 2016 (138 pages) which documents very well the overall engineering of a quantum computer based on trapped ions.

[1048] See Single-qubit quantum memory exceeding 10-minute coherence time by Ye Wang (Chine), 2017 (6 pages).

[1049] See Large Scale Modular Quantum Computer Architecture with Atomic Memory and Photonic Interconnects by Christopher Monroe et al, 2014 (16 pages).

[1050] See Building a prototype for the world's first large-scale quantum computer, 2022, related to Blueprint for a microwave trapped ion quantum computer by Bjoern Lekitsch et al, ScienceAdvances, February 2017 (11 pages). It deals with microwave gates using a static magnetic field with high fidelities and low crosstalk, a technique proposed by Mintert and Wunderlich in 2001. It adds ions transport between trapped ions modules. Laser-based gate control doesn't scale well due to the way phonons work and also due to the complexity of laser beams alignment. They plan to use silicon circuits and X-junction plus on-chip control electronics. Trapping ions would require an area of 103.5×103.5 $m^2$ and 23x23 connected vacuum chambers corresponding to the guestimate in 2017 of what was needed to factorize an RSA key of 2048 bits. Now, it's 100 times fewer qubits. It still makes 10,3x10,3 $m^2$ with a power dissipation of 300W per module, so, with 4 modules now. It is surprisingly low and deserves some recalculation. See also A Shuttle-Efficient Qubit Mapper for Trapped Ion Quantum Computers by Suryansh Upadhyay et al, April 2022 (7 pages) which is also pursuing a ions shuttling approach.

## Qubit operations

The general principle of trapped ion qubits is as follows:

- **Preparation** with neutral atoms being first heated in a small heating oven and then ionized and cooled by laser beams . Ions are then confined in vacuum in different ways by a magnetic and/or electric fields with variants of Paul and Penning traps as seen earlier. Qubits initialization is relying on electric dipole or quadrupole transitions driven by lasers to set them up at the right energy level corresponding to the $|0\rangle$ ground state. All this happens in an ultra-vacuum chamber.
- **Qubit quantum state** corresponds to two relatively stable energy levels of the trapped ions that are controllable by optical or microwave transitions.
- **Single-qubit quantum gates** are activated by microwaves, lasers or magnetic dipoles electric fields.
- **Two-qubit quantum gates** often use lasers with entangled photons and exploit the phonon phenomenon that links atoms together by vibrations that propagate from one atom to another, which is valid for qubits aligned in linear Paul traps[1051]. It however doesn't scale well beyond a couple dozen ions. More scalable variants use microwave fields distributed through the ions supporting circuit[1052] or efficiently distribute laser beams on nanophotonic circuits[1053].
- **Qubits readout** uses the detection of the cavity fluorescence with either superconducting photon detectors[1054] or CCD image sensors after ions are excited by a laser[1055]. The excited ions corresponding to the $|1\rangle$ state are visible while the unexcited ions corresponding to the and $|0\rangle$ state are not.

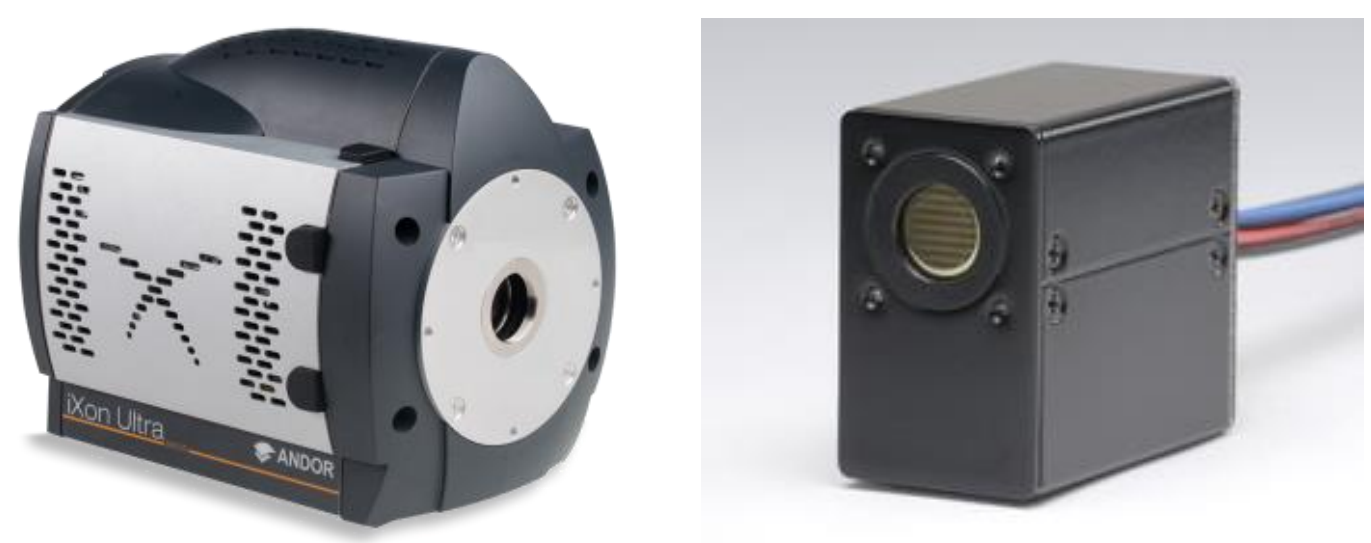

*Figure 382: examples of image sensors for trapped ions qubits readout with an **Oxford Instrument** Andor iXon Ultra 888 UVB (left) and a **Hamamatsu** H10682-210 PMT (right).*

## Setup

The typical trapped ions setup contains a vacuum chamber containing a chipset where the ions are "floating on". They are driven by microwave and laser pulses, thus the associated control electronics. Like with cold atoms, an imager sensor is involved in the qubit readout. Some cooling is frequently used, but at reasonable temperatures of about 4K. All of this can fit into two standard data-center racks.

---

[1051] Photon mediated entanglement was invented in 2004 by Christopher Monroe et al. See Scalable Trapped Ion Quantum Computation with a Probabilistic Ion-Photon Mapping by L.-M. Duan, B. B. Blinov, D. L. Moehring and Christopher Monroe, University of Michigan, 2004 (6 pages).

[1052] Trapped ions single and two qubit gates can be generated with only microwave magnetic fields and radiofrequency magnetic field gradients and no lasers. See High-fidelity laser-free universal control of two trapped ion qubits by R. Srinivas et al, February 2021 (40 pages).

[1053] See Integrated optical multi-ion quantum logic by Karan K. Mehta, Jonathan P. Home et al, ETH Zurich, Nature, October 2020 (12 pages).

[1054] See State Readout of a Trapped Ion Qubit Using a Trap-Integrated Superconducting Photon Detector by S. L. Todaro, David Wineland et al, NIST, University of Colorado Boulder, University of Oregon, PRL, 2020 (7 pages).

[1055] See Real-time capable CCD-based individual trapped ion qubit measurement by S. Halama et al, April 2022 (16 pages) which compares a fast CCD camera with PMT (photomultiplier tubes) and superconducting nanowire single-photon detectors (SNSPDs). It is an Andor iXon Ultra 888 UVB from Oxford Instruments of 1024x1024 pixels and operating at -35°C and 200 frames per seconds with reasonable noise. It works in UV light. The reference PMT comparison example is a Hamamatsu H10682-210.

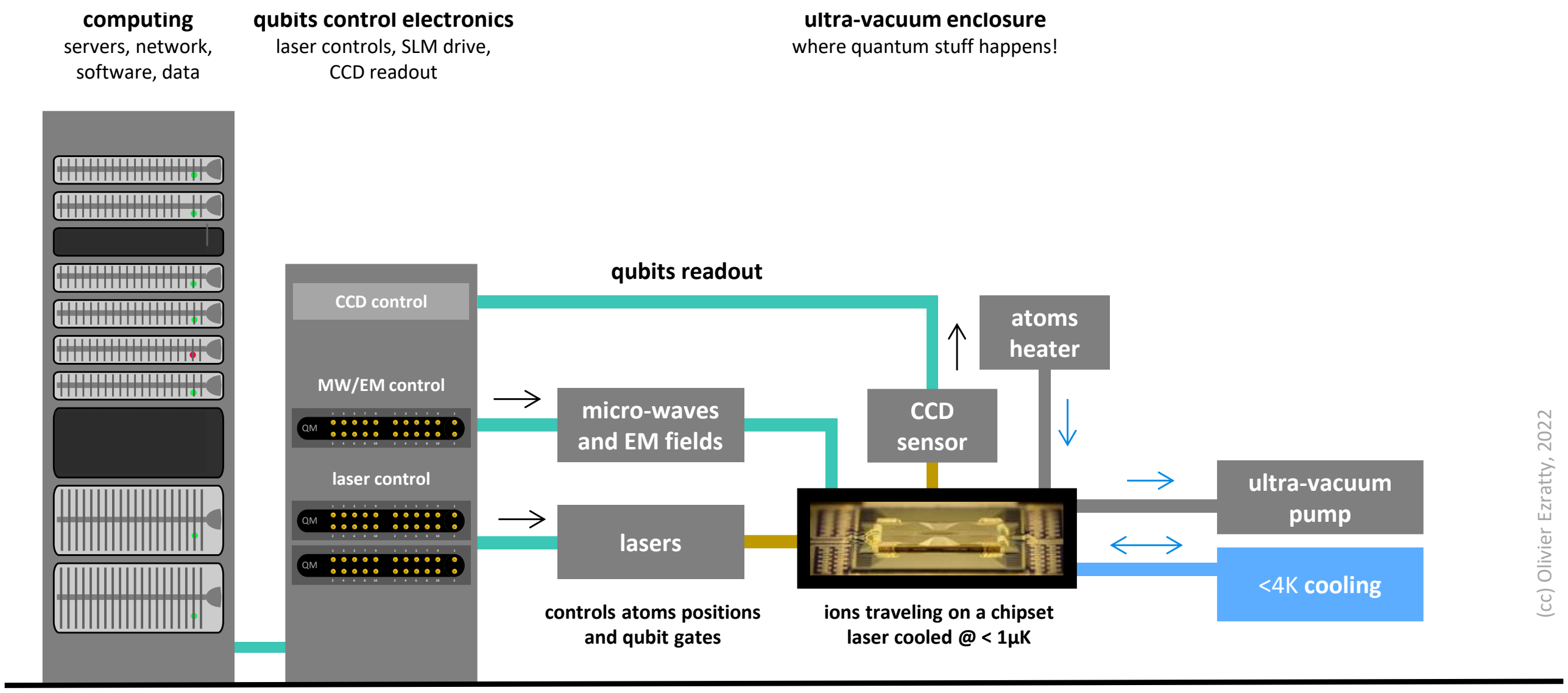

*Figure 383: generic architecture of a trapped ion quantum computer which fits into a 2-rack system. (cc) Olivier Ezratty, 2022.*

## Research

About a hundred research teams around the world are working on trapped ion qubits in almost every country working on quantum technologies (Australia, Austria, Canada, China, Denmark, Finland, France, Germany, India, Israel, Japan, Netherlands, Singapore, Switzerland, UK, USA)[1056].

**Rainer Blatt** from the University of Innsbruck is one of the pioneers in this field. He created a register of 14 addressable qubits in 2011 and increased it to 20 addressable and individually controllable qubits in 2018, using calcium ions. Rainer Blatt then cofounded **Alpine Quantum Technologies** (2017, Austria) where he characterized up to 10 high quality ion trapped qubits. In 2021, his team also demonstrated the use of trapped ions to create qudits with 3, 5 and 7 levels, potentially opening the path for more powerful trapped ion based quantum computing[1057].

Quantum simulation using trapped ions, and an Ising model as with the D-Wave, is also investigated by some laboratories such as at **ETH Zurich** in Jonathan P. Home's Trapped Ions Quantum Information Group (TIQG)**,** the **University of Maryland**, elsewhere in the USA[1058] and also in China[1059].

In May 2020, Wesley Campbell's **UCLA** team associated with UNSW announced that they had stabilized barium ions ($^{133}Ba^{+}$) to build quality qubits in a linear trap[1060]. The quality of these barium ions is compared to that of 2,014 qubits with a 10-fold improvement. This quality is evaluated only with the SPAM indicator which measures a fidelity on a qubit after preparation, some initialization single qubit gates and measurement (SPAM = "state preparation and measurement").

1056 There were 98 research laboratories in the world working on trapped ions in 2020. See this table listing them all in List of Ion Trapping Groups, February 2020.

1057 See A universal qudit quantum processor with trapped ions by Martin Ringbauer et al, September 2021 (14 pages). 8 levels for a calcium-based trapped ion qubit.

1058 See Digital Quantum Simulation with Trapped Ions by Kenny Choo and Tan Li Bing, 2016 (29 slides) and Programmable Quantum Simulations of Spin Systems with Trapped Ions by Christopher Monroe et al, 2019 (42 pages) and a follow-up with Programmable quantum simulations of bosonic systems with trapped ions by Or Katz and Christopher Monroe, July 2022 (7 pages).

1059 See Probing critical behavior of long-range transverse-field Ising model through quantum Kibble-Zurek mechanism by B.-W. Li et al, August 2022 (10 pages).

1060 See Physicists develop world's best quantum bits by Stuart Wolpert of UCLA, May 2020 which refers to High-fidelity manipulation of a qubit enabled by a manufactured nucleus by Justin Christensen et al, May 2020 (5 pages). First precaution of use: identify the author of the article. It happens to be a certain Stuart Wolpert from UCLA, in charge of media relations at the University where the published work comes from. So he does the PR for the laboratory and publishes his article on a site where it is possible (Physorg).

Let's also mention the **IQOQI** (Austria, see Rainer Blatt, one of their laboratories) and the **IQST** (Germany), and their calcium based 20 qubits prototype[1061] as well as the Ion Quantum Technology Group from the **University of Sussex** (UK) that is run by Winfried Hensinger and its 10 qubits prototype, proposing an architecture design to scale up to 1,000 qubits through a cluster of quantum processors[1062]. The group led to the creation of the startup **Universal Quantum** (2019, UK).

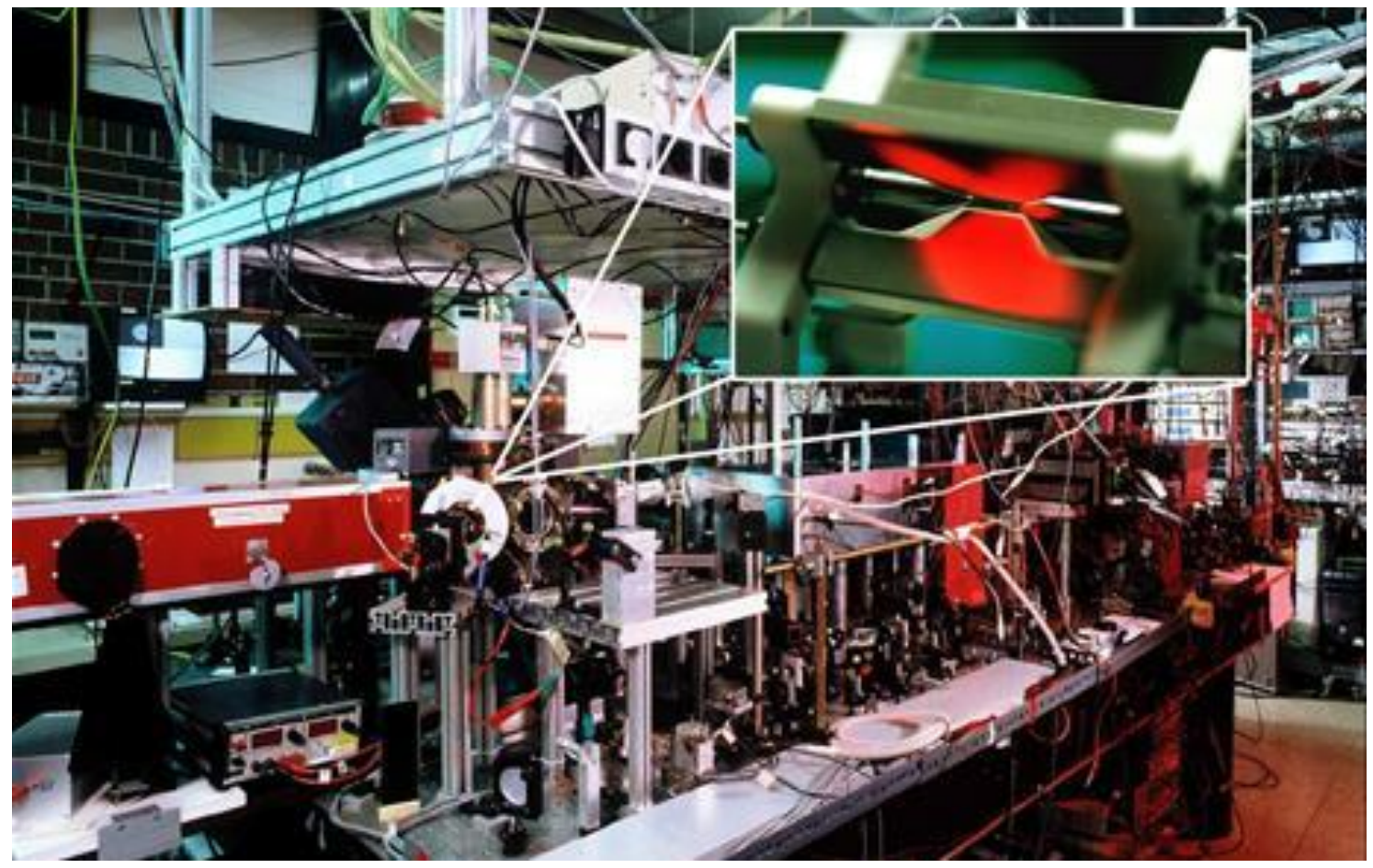

*Figure 384: Rainer Blatt's lab in Innsbruck.*

In March 2021, the DoE **Sandia Labs** launched the QSCOUT (Quantum Scientific Computing Open User Testbed), a cloud quantum computing resource available to selected researchers from universities and other government research agencies[1063]. It is an $^{171}$Yt based trapped ions system of 3 qubits used for benchmarking and for algorithms development, particularly in computational chemistry. It will later be expanded to a 10 and then 32 bits system, by 2023, on par with 2021's IonQ's capacity. At a low-level, this system is programmed with the in-house assembly language Jaqal ("Just Another Quantum Assembly Language").

Also, in the USA, a team from **Georgia Tech** led by Creston Herold and including NIST and DoE Oak Ridge's quantum lab and funded by DARPA is working on rare-earth trapped ion systems using the less used Penning traps, that controls the position of the trapped ions with magnetic and electric fields, using permanent magnets made of neodymium and samarium cobalt[1064]. At this point, they control 10 trapped ions.

The European Flagship includes the **AQTION** project, which is led by the University of Innsbruck and has a budget of €9.57M. The objective is to reach 50 operational qubits to prepare the next phase, beyond 100 qubits, by adopting a distributed architecture with photonic links. Alpine Quantum Technologies (AQT), the University of Oxford, ETH Zurich, Fraunhofer IOF and Atos are participating. Atos works on the solution software stacks and applications.

In **Israel**, a team of researchers from the Weissman Institute announced in 2022 the creation of the first local quantum computer using 5 strontium ions[1065]. It is less than stellar at this point. They plan to reach 64 qubits someday.

In **Russia**, the Russian Quantum Center, the P.N. Lebedev Physics Institute of the Russian Academy of Sciences and Rosatom presented in 2021 a prototype of a trapped ions computer, starting with 4 qubits. Looks like they are late in the catch-up game behind the USA, UK and Austria!

---

[1061] They coauthored Observation of Entangled States of a Fully Controlled 20-Qubit System, April 2018 (20 pages).

[1062] See Blueprint for a microwave trapped ion quantum computer by Winfried Hensinger et al, 2017 (12 pages) and their review paper Quantum control methods for robust entanglement of trapped ions by C H Valahu, Winfried Hensinger et al, Journal of Physics B: Atomic, Molecular and Optical Physics, 2022 (27 pages).

[1063] See Rare open-access quantum computer now operational, Sandia Labs, March 2021.

[1064] See DARPA Probing Quantum Computing Capabilities by Meredith Roaten, June 2022 and Universal Control of Ion Qubits in a Scalable Microfabricated Planar Trap by Creston D. Herold et al, February 2016 (17 pages). Back in 2016, their single qubit gate fidelities was 97% which is far from being stellar for trapped ions.

[1065] See Trapped Ion Quantum Computer with Robust Entangling Gates and Quantum Coherent Feedback by Tom Manovitz, Yotam Shapira, Lior Gazit, Nitzan Akerman and Roee Ozeri, March 2022 (12 pages).

## Vendors

**IonQ** (2016, USA, $736M[1066]) is a spin-off from the University of Maryland specialized in the design of universal quantum computers based on ytterbium trapped ions, and later, barium ions[1067].

Co-founded by Christopher Monroe, professor at the university who is also their Chief Scientist, the company CEO is Peter Chapman, a former e-commerce executive at Amazon. IonQ's cap table includes Google Ventures, Amazon, Samsung Ventures, Microsoft, Lockheed Martin, Bosch and HPE. In June 2020, they created an advisory board including David Wineland, Umesh Vazirani, Margaret Williams (ex Cray) and Kenneth Brown (Duke University). In March 2021, IonQ announced a new round of funding with a merger agreement through the Special Purpose Acquisition Company (SPAC) mechanism, with the fund dMY Technology Group III that will yield a $650 million investment. The funding was made of $350M coming from investors including Hyundai, Kia Corporation[1068] and Breakthrough Energy Ventures. The remaining $300M came from dMY and an IPO[1069]. The IPO was finalized in October 2021.

IonQ's architecture use 1D arrays of ions of variable length. They are controlled by lasers for both cooling, quantum gates and readout. Trapped ions enable all-to-all connectivity between ions, making it easier to run algorithms and avoiding the usage of costly SWAP gates.

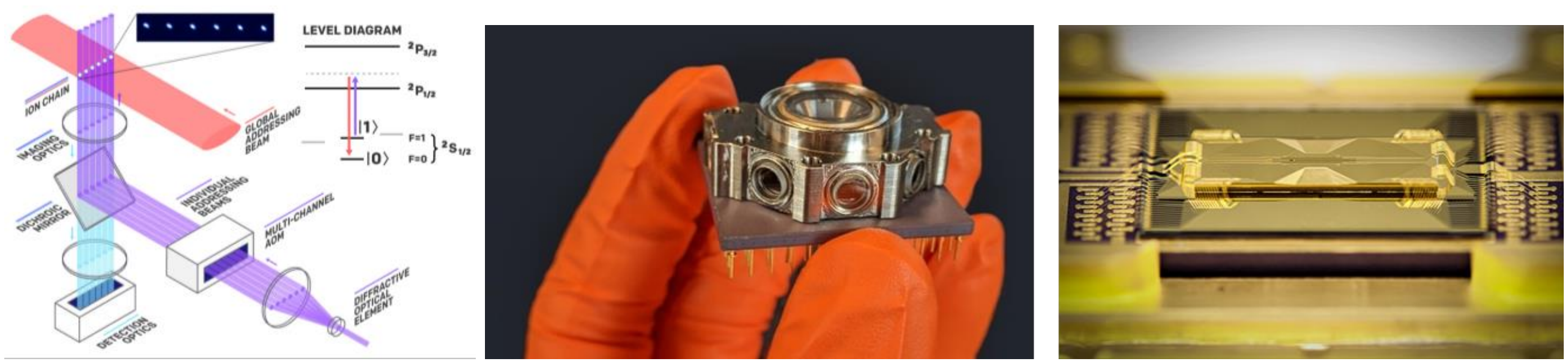

*Figure 385: IonQ trapped ion drive system, the small vacuum enclosure where the ions are located, and the chipset controlling the ions position. Sources: IonQ and Ground-state energy estimation of the water molecule on a trapped ion quantum by Yunseong Nam, Christopher Monroe et al, March 2019 (14 pages).*

This allows implementing a very good optimization of quantum algorithms to minimize the number of gates to be executed as shown in the example in Figure 386.

But, as with all trapped ions qubits QPUs, this is achieved at the expense of relatively slow gates in comparison with superconducting and silicon qubits QPUs. The scalability of trapped ions is being questioned and it shows-up well in the history of the company. At the beginning of 2018, they announced a record of 53 coherent and entangled qubits but these were used for quantum simulation and not with gate-based computing.

---

1066 This amount includes $84M from VCs and the 2021 SPAC. It excludes the total $165M grants the company and Christopher Monroe's lab in Maryland University got from the US government, per their 2021 investor presentation.

1067 See A Reconfigurable Quantum Computer by David Moehring, 2017 (20 slides).

1068 They seem to have closed links with South Korea. These investors add up with a partnership with Q Center. See IonQ and South Korea's Q Center Announce Three-Year Quantum Alliance, January 2021. To provide to the Q Center students to the IonQ computer online.

1069 See QC ethics and hype: the call is coming from inside the house by Scott Aaronson, October 2020, who found this IPO to be pushing the envelope of bullshit a bit too far.

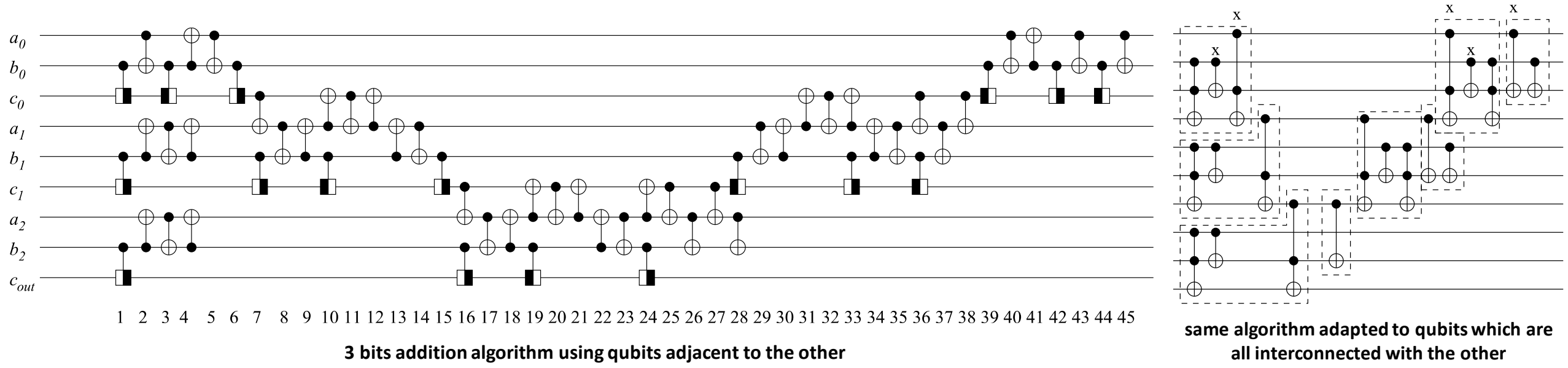

*Figure 386: how the good connectivity with trapped ions enables a good compression of the code. Source: [Fast Quantum Modular Exponentiation](#) by Rodney Van Meter and Kohei Itoh, 2005 (12 pages).*

At the end of 2018, they said they had reached 79 qubits associated with 160 storage qubits but with not fidelity numbers[1070]. In 2019, they had 11 characterized qubits[1071].

In October 2020, IonQ announced that it had created the world's most powerful quantum computer with 32 qubits and a quantum volume of 4,000,000 but it took over a year and a half for this system to become live and tested[1072] and made available on Amazon Braket, Microsoft Azure Quantum and Google cloud offerings. They claimed to handle error correction codes with 13 (and sometimes, 16) physical qubits per logical qubits. They also announced the creation of a Quantum Data Center sized to host 10 of their quantum computers.

In December 2020, IonQ unveiled its 5 years roadmap with plans to use rack-mounted modular quantum computers small enough to be networked together in a datacenter by 2023. IonQ adopted a new benchmark metric of their own: algorithmic qubits, using $\log_2$ of IBM's quantum volume and a different calculus mode that we cover in the benchmarking section of this book, using a set of QED-C benchmarks.

Their 32 qubits support 22 algorithmic qubits with plans to reach 29 algorithm qubits by 2023, 64 by 2025 with using a 16:1 error-correction encoding (meaning: 16 physical qubits per logical qubits). Later on, they will rely on a 32:1 ratio.

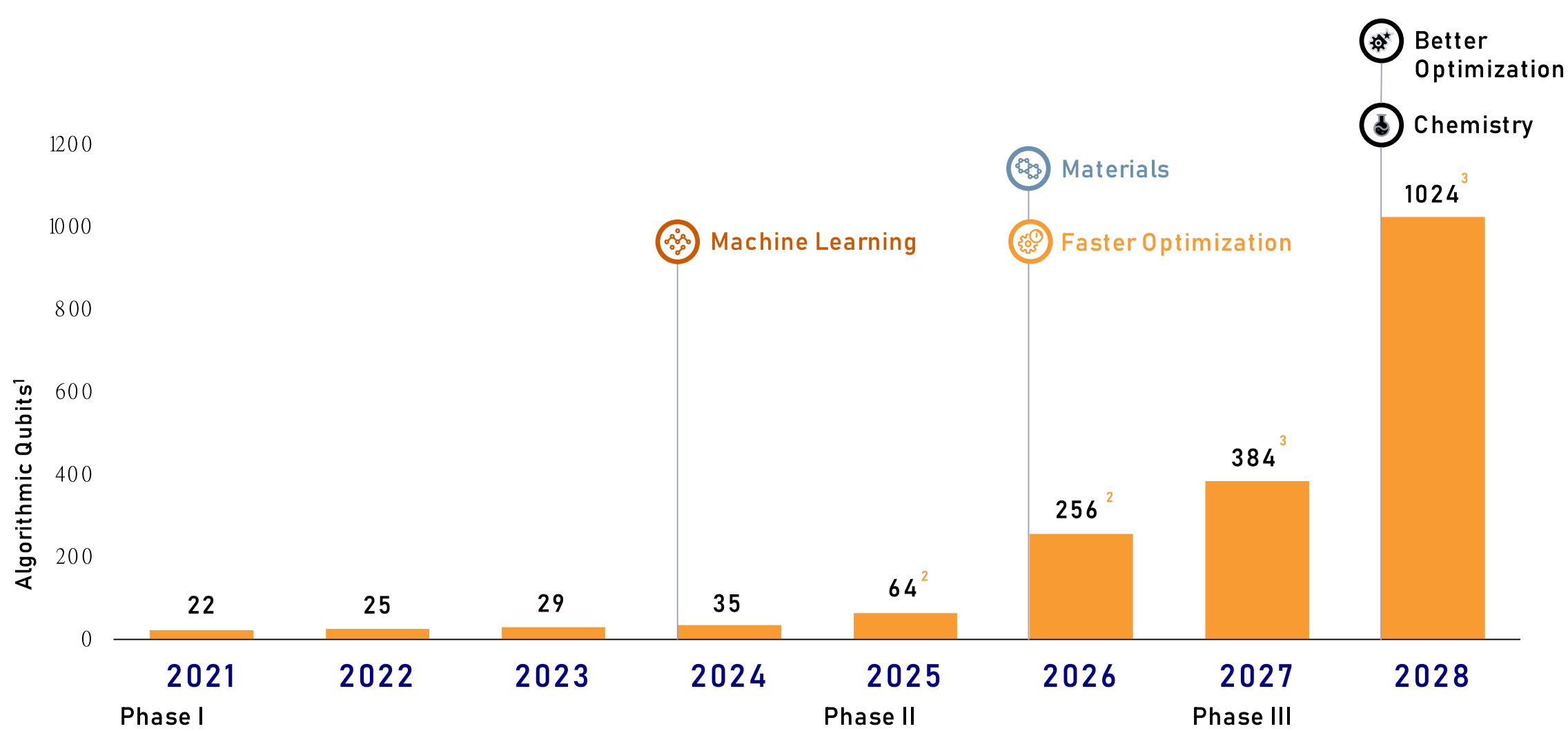

*Figure 387: IonQ's qubits roadmap as published in March 2021.*

[1070] See [IonQ Has the Most Powerful Quantum Computers With 79 Trapped Ion Qubits and 160 Stored Qubits](#) by Brian Wang, December 2018.

[1071] See [Benchmarking an 11-qubit quantum computer](#) by K. Wright et al, November 2019.

[1072] See [IonQ Unveils World's Most Powerful Quantum Computer](#), IonQ, October 2020.

Then, they expect to scale beyond 64 and reach a broad quantum advantage with 256 then 1024 algorithmic qubits by 2026 and 2028. The caveat is that this can be achieved only with scaling-out their quantum processors, assembling several units of 64 qubits through photonic links in a distributed computing manner. Something that has not been tested yet beyond one-to-one qubit connectivity[1073].

IonQ announced in August 2021 their Reconfigurable Multicore Quantum Architecture (RMQA) detailing how they would create 64 ions chipsets (video). It would assemble 4 chains or lines of 16 ions, 12 being usable as qubits and the 4 remaining for cooling, in a single chipset manufactured on a glass support (Evaporated Glass Traps) replacing their previous silicon-based platform built by Sandia Labs and Honeywell. These chunks of 16 ions can be moved around, paired and entangled, to create dynamic 32 ions units. IonQ stated that this architecture could scale-up and support even more blocks of 16 ions. Well, if that actually works in practice, why not!

A team associating IonQ, Duke University in Durham and ColdQuanta published an interesting paper describing the architecture of a trapped ions systems cryostat from Montana Instruments that is optimized to minimize the vibrations coming from the pulse tube. This seems to be one of the figures of merit to ensure the stability of the trapped ions qubits and their control devices likes lasers[1074]. The qubits are cooled at 5K while laser-based cooling using the Doppler effect cool it at an even lower temperature.

In February 2022, IonQ and Duke University presented a new way to create 3-qubit gates including a Toffoli gate using state squeezing[1075]. This sort of gate in interesting since it can be the basis for a universal gate-set enabling fault-tolerance. It can help speeding up many algorithms including Grover and variational quantum eigensolvers (VQEs).

In December 2021, after they had finalized their SPAC and IPO, IonQ announced they were switching from ytterbium to barium ions (precisely $^{133}Ba^{+}$[1076]). The reasons were well explained: it provides better gates and readout fidelities and the ions are primarily controlled with visible light rather than ultraviolet light, using standard silicon photonics technology, which can better enable QPU photonics interconnect[1077]. They also secured the provisioning of these atoms with a partnership with **DoE's PNNL** (Pacific Northwest National Laboratory) in February 2022.

---

[1073] The associated concepts were laid out in Scaling the ion trap quantum processor by Christopher Monroe and J. Kim, Science, 2013 (7 pages). It consists in associating one qubit of two ions QPUs with probabilistic entangled photons. See also Large-scale modular quantum-computer architecture with atomic memory and photonic interconnects by Christopher Monroe, Robert Raussendorf et al, PRA, 2013 (16 pages).

[1074] See High stability cryogenic system for quantum computing with compact packaged ion traps by Robert F. Spivey et al, August 2021 (12 pages). ColdQuanta seems involved here given a cold atoms system can reuse some of the experimental setting crafted for trapped ions. Interestingly, in its 2021 investor presentation, IonQ pretended that their system was operating at room temperature!

[1075] See $N$-body interactions between trapped ion qubits via spin-dependent squeezing by Or Katz, Marko Cetina and Christopher Monroe, February 2022 (7 pages).

[1076] See Ba-133: the Goldilocks qubit? by UCLA Hudson Lab.

[1077] See IonQ Announces New Barium Qubit Technology, Laying Foundation for Advanced Quantum Computing Architectures, IonQ, December 2021.

On May 3rd, 2022, IonQ was literally attacked by a short-seller financial company, **Scorpion Capital**, which published a scathing report on their business, presented as a scam[1078].

This 193-page report was very long, apparently detailed and based on many interviews. But it was misplaced.

**IONQ (NYSE: IONQ)**

**The "World's Most Powerful Quantum Computer" Is A Hoax With Staged Nikola-Style Photos – An Absurd VC Pump With A Recent Lock-Up Expiration Takes SPAC Abuses To New Extremes**

- A part-time side-hustle run by two academics who barely show up, dressed up as a "company"
- A useless toy that can't even add 1+1, as revealed by experiments we hired experts to run
- Fictitious "revenue" via sham transactions and related-party round-tripping
- A scam built on phony statements about nearly all key aspects of the technology and business
- CEO appears to be making up his MIT educational credentials

**$1.6B market cap | $8/share | ADV 6.4MM shares | Short interest 7mm shares** *5/2/22 per Capital IQ

*Figure 388: Scorpion Capital review cover page with extreme and misleading statements.*

It did criticize IonQ wrongly on many points like when explaining that quantum computers couldn't even do a 1+1 calculation. They pinpointed exaggerations that can be found in IonQ investor March 2021 presentation. They even said that their 32-qubit system was non-existent (which is not true at all). They also highlighted that their ions control chipset was produced by Sandia Labs, a DoE lab operated by Honeywell, but it was not a secret. The same with **Hyundai** being both one of their investors and also a customer, on far-fetched use-case plans related to battery designs. They could have been harsh on their aggressive roadmap, their scalability goals and their related QPU interconnect plans but lacked scientific background to do so [1079].

All of this was border line defamation as was shown later. IonQ was then defended by preeminent quantum computing analysts[1080]. A couple days later, IonQ announced the "select" availability of their 31 bits Forte system a couple months after having released its 23-qubits Aria system supporting 20 algorithmic qubits (as of August 2022[1081]). "Select" means, only for selected developers and customers, general availability being planned for 2023, a common and cautious practice with quantum hardware vendors. The system introduces an Acousto-Optic Deflector (AOD) which dynamically directs laser beams towards individual ions to drive qubit gates and supports up to 40 ions.

In November 2019, **Microsoft** announced the integration of IonQ's quantum accelerator support into its Azure Quantum cloud offering and its Q#, QDK and Visual Studio development tools. All this was made available to developers from late spring 2020. IonQ is also proposed by **Google** in its own cloud offering, on top of **Amazon** AWS Braket. IonQ became in 2021 the only quantum computer vendor available on Amazon, Google and Microsoft clouds (with 11 qubits, being extended to 32 qubits).

On the use cases side, IonQ works with a couple customers like **Goldman Sachs** on financial services on top of the above mentioned Hyundai. They also partner with **Accenture** to develop customer applications.

In September 2021, IonQ announced the creation of a joint laboratory with the University of Maryland (UMD), the **Q-Lab**, with $20M funding. Among other things, the lab is tasked with training UMD students on quantum computing. In 2022, they also established business development subsidiaries in Germany and Israel.

---

[1078] See The "World's Most Powerful Quantum Computer" Is A Hoax With Staged Nikola-Style Photos – An Absurd VC Pump With A Recent Lock-Up Expiration Takes SPAC Abuses To New Extremes by Scorpion Capital, May 2022 (183 slides).

[1079] I mention this in the paper Mitigating the quantum hype, January 2022 (26 pages) that is quoted in Scorpion's presentation on slide 14. They may have just read its title!

[1080] See A short report has placed a spotlight on IonQ, a quantum computing champion. This should not deflect long term interest in this or other quantum technologies by David Shaw, Doug Finke and André M. König, May 2022.

[1081] See IonQ Aria: Past and Future (Part Two) by IonQ, August 2022.

**Quantinuum** (USA/UK) is the result of the merger in June 2021 of Honeywell Quantum Systems (USA, a branch of Honeywell) and the software company Cambridge Quantum Computing (UK), with an investment of $300M for a stake of 55% for Honeywell in the resulting company[1082]. The Quantinuum renaming occurred in December 2021.

Quantinuum's CEO is Ilyas Khan who previously was the founder and CEO of CQC. The company now has more than 600 people overall. Honeywell started working in quantum computing in 2016 in "stealth" mode. Their team came in particular from the NIST Boulder lab and the University of Colorado with some alumni from the University of Maryland and Christopher Monroe's team (IonQ). In March 2020, they announced the development of a quantum computer that was bound to become be "*the most powerful in the world*", doubling the power of the previous record that was then held by IBM[1083]. The initial announcement dealt with a four-qubit trapped ions-based quantum processor[1084], its power being evaluated using IBM's quantum volume benchmark.

**Trapped ion QCCD** is the trapped ions technique they are using (for "quantum charge-coupled device"). It uses ytterbium-based ions coupled with barium ions to cool the device. This technique was developed in 2002 by Christopher Monroe, David Wineland and Dave Kielpinski[1085]. They are reusing many other works from various research laboratories spread out between 2008 and 2012.

Ions are generated from a jet of collimated atoms obtained by heating a solid ytterbium target. They are then "hit" by a laser, which removes an electron from the valence layer of the atom (the last one). Only one electron remains in this layer, giving rise to an ion with a positive charge, Yb+. The laser cooling of these ions is well-controlled thanks to their favorable energy level pattern. Thanks to their electrical charge, it is possible to trap and move these atoms using electrostatic and radiofrequency potentials. The ions quantum states correspond to two "hyperfine" energy states related to the interaction between the magnetic moment of the nucleus and that of the electrons of the ion. These hyperfine levels are also used in cesium atomic clocks. The transition frequency between the two hyperfine levels of ytterbium is 12.6 GHz[1086]. The hyperfine states of the ytterbium ion are well suited for quantum computation because they are very stable, which allows them to have a long coherence time.

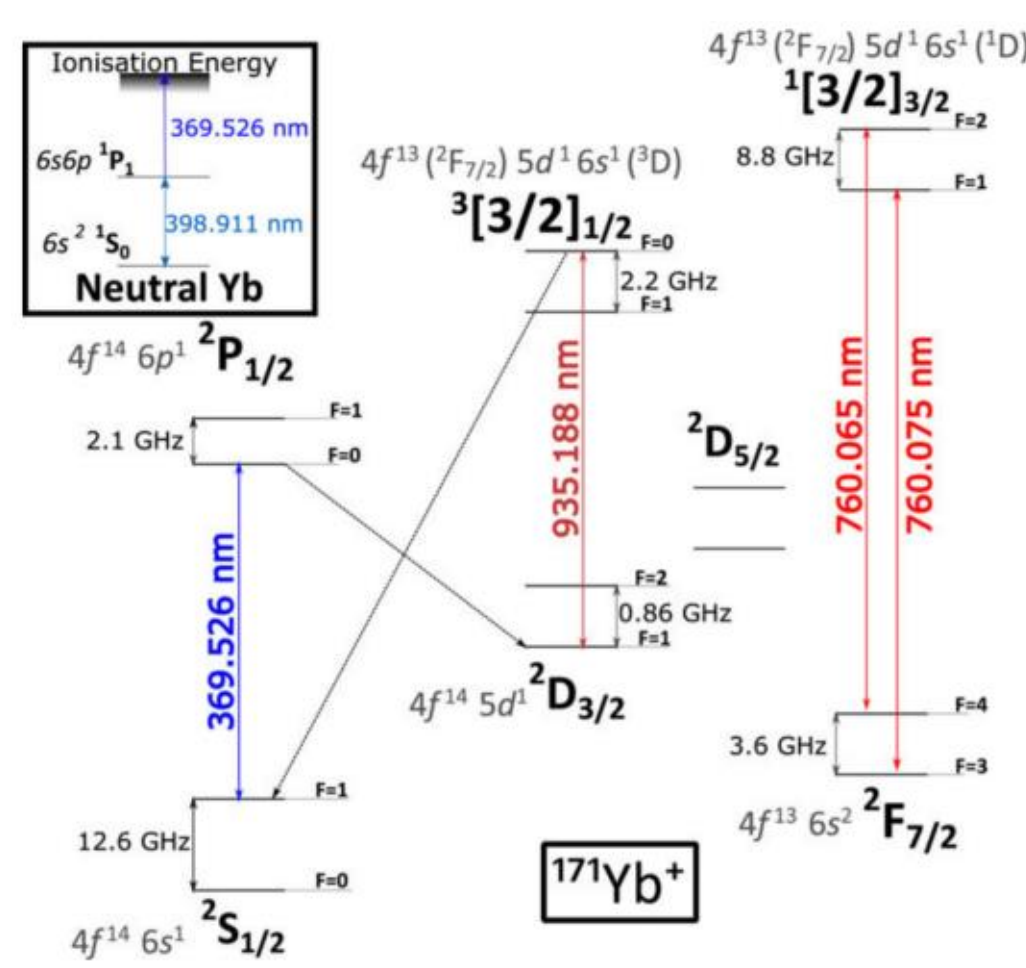

*Figure 389: ytterbium atomic transitions used by Quantinuum. Source: Laser-cooled ytterbium ion microwave frequency standard by S. Mulholland et al, 2019 (16 pages).*

[1082] See Honeywell Quantum Solutions And Cambridge Quantum Computing Merge With Go-Public In Mind by Paul Smith-Goodson, June 2021.

[1083] See Honeywell Achieves Breakthrough That Will Enable The World's Most Powerful Quantum Computer and How Honeywell Made the Leap into Quantum Computing by Honeywell, March 2020. In Honeywell has it created the world's most powerful quantum computer, March 2020, I analyze the ad in detail, with the text embedded in the book as a compacted version.

[1084] The performance is described in detail in: Demonstration of the QCCD trapped ion quantum computer architecture by J. M. Pino et al, 2020 (8 pages). This can be complemented by the presentation Shaping the future of quantum computing by Tony Uttley, the head of Honeywell's quantum team at the Q2B conference at QC Ware in California in December 2019 (slides).

[1085] It is described in Architecture for a large-scale ion-trap, 2002 (4 pages).

[1086] See Laser-cooled ytterbium ion microwave frequency standard by S. Mulholland et al, 2019 (16 pages).

**Shuttling ions** is a technique used to handle their connectivity. This is a rare case of shuttling qubits, the other one being shuttling electrons that still is in research labs. Usually, qubits based on electrons, cold atoms or ions don't move (too much) where they are installed. This idea was proposed in 2002 by Dave Wineland and co. This was the first working shuttling ions setup.

Their system prepares ytterbium atoms, ionizes them and sends them into a hole that feeds the chipset. It then uses about ten to twelve ions storage and sorting areas (in orange, yellow and blue in Figure 390).

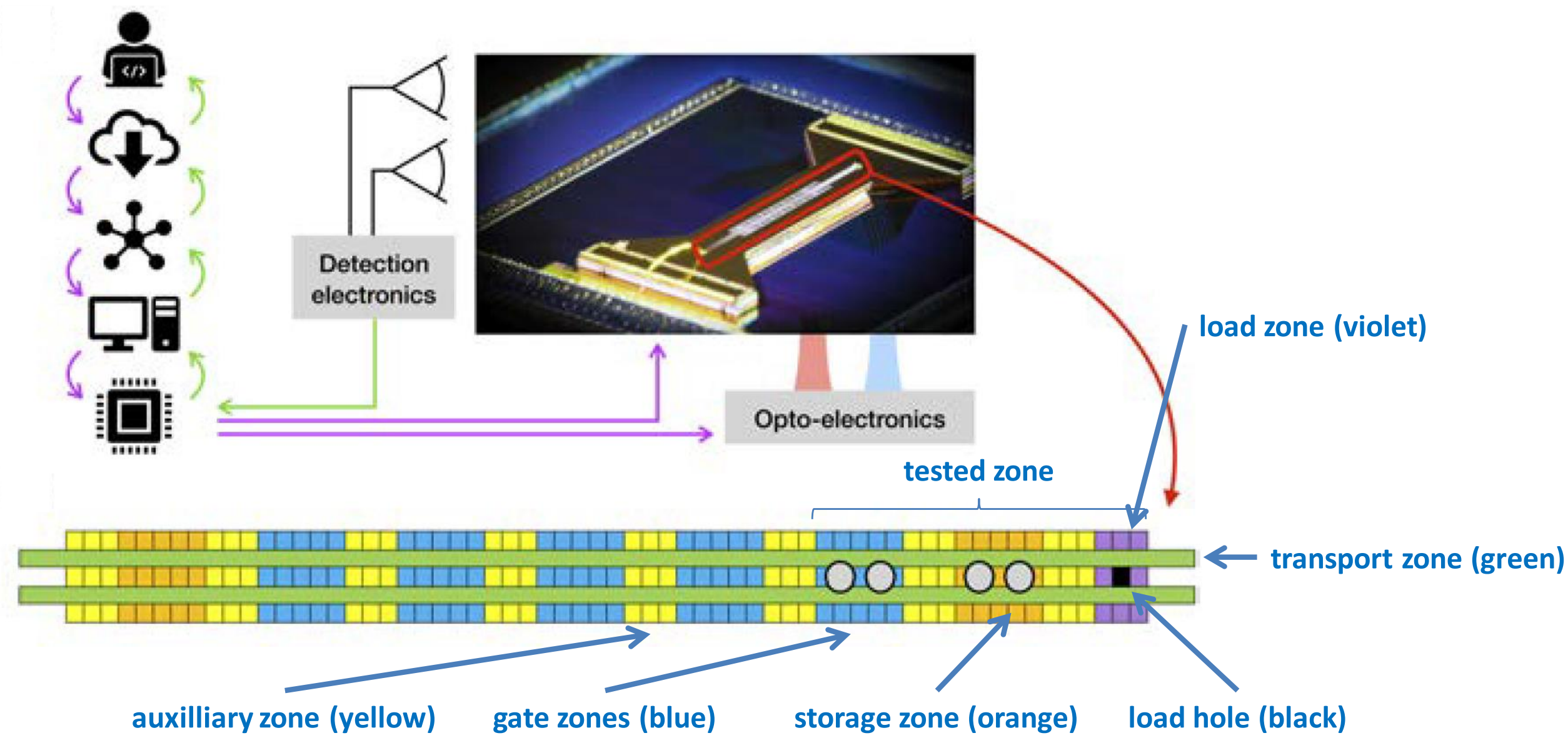

*Figure 390: overall control architecture in 1D versions of Quantinuum's trapped ions, as presented in 2020.*

The ytterbium ions are confined above a rail of three rows of electrodes whose variable voltage allows to control their position and to move them laterally. Their aim is to be able to demonstrate logical operations between several qubits while moving them at will between storage areas and interaction areas during operations.

Their initial system was using 198 direct current (DC) electrodes for controlling the displacement and positioning of ytterbium ions coupled with barium ions used for cooling. The chip uses cryogenic surface traps that dynamically rearrange the positioning of the ytterbium/barium ion pairs and implement quantum gates running in parallel on several areas of the circuit. Ions circulate above the green band, allowing arbitrary movements of the ions along the band. Once positioned, they are transferred to the middle band to get submitted to a single qubit quantum gate, or in the side bands for two-qubits quantum gates, as explained in the diagram *below*. One of these operations is a SWAP gate that allows the ions to be physically interchanged.

**Slow gates**. The disadvantage of the technique is its slow quantum gates. The time required to configure the ions to create a quantum gate is 3 to 5 ms, which is not negligible, especially for algorithms that require a large number of quantum gates.

**Cooling**. The system operates at a temperature of 12.6K and with a temperature stability of 2mK which avoids disturbing the ions and their superposed and entangled quantum states. Helium cooling is complemented by a so-called "sympathetic cooling" technique which combines the use of Doppler effect and Raman cooling on the barium ions next-door to the ytterbium ions. The Coulomb interaction between the barium ions cools the ytterbium ions next to the barium ions. A barium ion cooling operation takes place before each two-qubit gate execution. Ion laser cooling has been operating at room temperature for more than 30 years. Like many research groups, they cool the ion trap to 12.6K to minimize the effect of abnormal ions heating, which is a major problem that is not fully understood. This abnormal heating is greatly reduced when the trap is cooled.

**Qubit gates**. The system is built around four-qubit chunks and uses one- and two-qubit quantum gates that are activated by lasers, via the Raman effect that requires a pair of beams. The single-qubit gates are activated by a pair of 370.3 nm Raman beams in circular polarization. The system allows the generation of X, Y and Z gates for which quarter and half turns are performed around the three axes of the Bloch sphere. These rotations are done with very high precision according to Honeywell. This ensures a minimum error rate for single-qubit quantum gates.

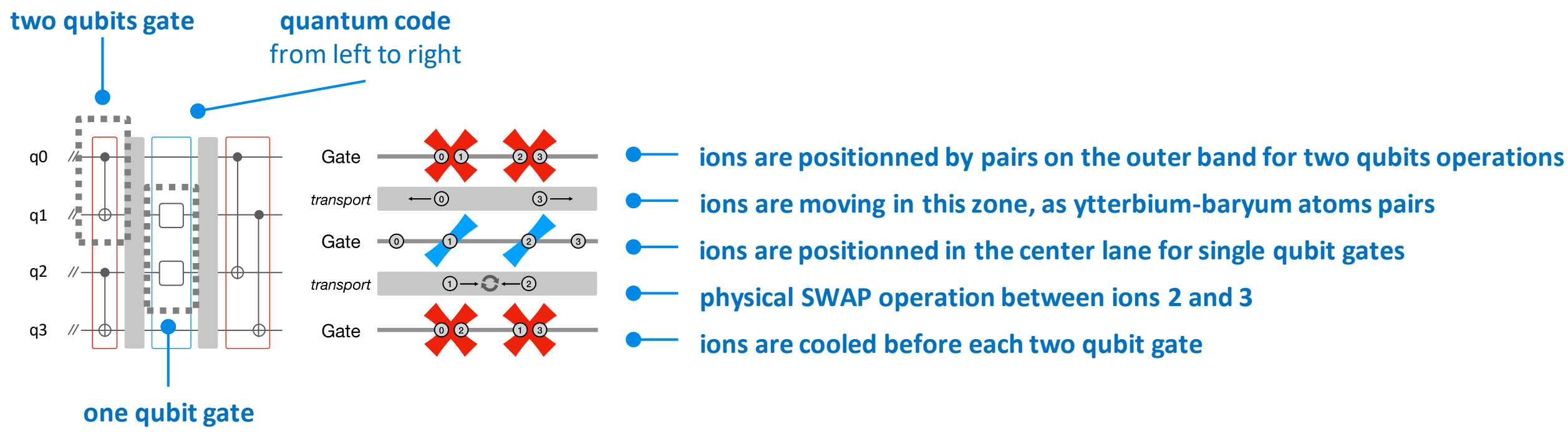

*Figure 391: how single and two-qubit gates are implemented in Quantinuum trapped ions systems. Source: Honeywell, 2020.*

Two-qubit gates use two additional pairs of laser beams that act on pairs of ytterbium atoms that have been brought closer together by the circuit's positioning control electrodes. Two ions are thus moved by the electrodes into the same potential well before being coupled by laser. The qubits can then be separated and moved elsewhere to interact with other qubits. I wrote all of this in 2020 and presume it has not changed since then.

**Qubits readout** is performed with a classical imager that detects the energetic state of the ions via their laser-activated fluorescence. This imager is a PMT array, i.e. a linear array of photomultipliers (Photo-Multiplier Tubes). Their architecture allows a qubit readout during processing, without disturbing the neighboring qubits. This would allow the implementation of conditional logic, with IF THEN ELSE like with classical programming. They are also using the mid-circuit measurement and qubit reuse technique (MCMR) which can be used to optimize the length of quantum algorithms.

The system includes an FPGA programmable electronic circuit for qubits controls, sitting outside the cryogenic enclosure.

**Qubit fidelities** seem very good. They launched their 6-qubit H0 system in June 2020, then their 10-qubits H1 system in October 2020 with an initial quantum volume of 128 (7 qubits x 7 gates depth). Their quantum volume reached 512 in March 2021 (9x9 qubits with 10 qubits). Single-qubit gate fidelity were above 99.991% and two-qubit gate fidelity above 99.76% while readout fidelity is at 99,75% with a measurement cross-talk at 0.2%, characterized as the decay of a qubit coherence in an equal superposition state, while repeatedly measuring the nearest qubit[1087]. In July 2021, HQS announced the creation of the first logical qubits using color codes with their 10 trapped ions qubits[1088].

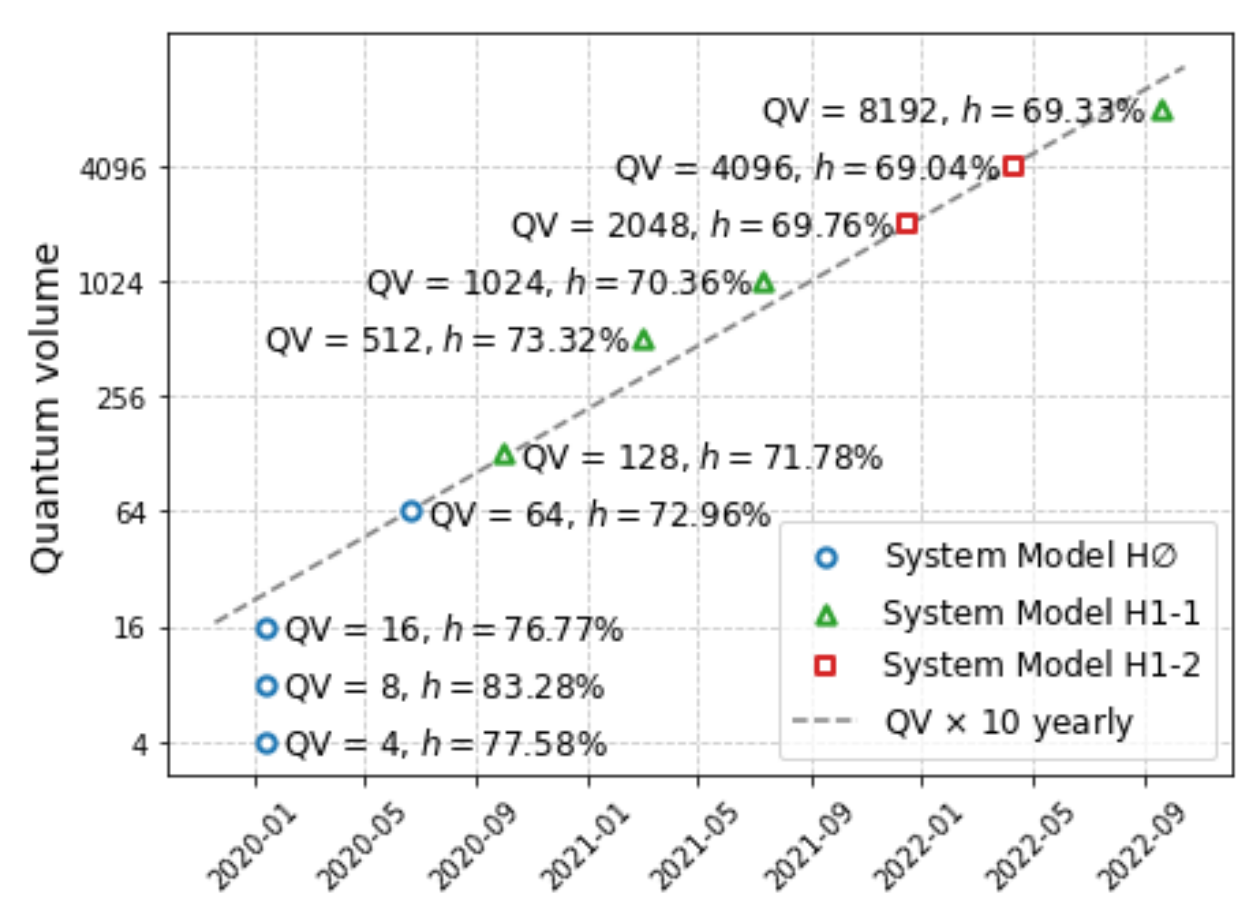

*Figure 392: evolution of Quantinuum systems quantum volume. Source: Quantinuum Sets New Record with Highest Ever Quantum Volume, Quantinuum, September 2022.*

[1087] See Get to Know Honeywell's Latest Quantum Computer System Model H1 by Honeywell, October 2020.

[1088] See Realization of real-time fault-tolerant quantum error correction by C. Ryan-Anderson et al, HQS, July 2021 (22 pages).

As of April 2022, they had 12 running qubits reaching a quantum volume of $2^{12}$ with their System Model H1-2. It is the only quantum processor where a quantum volume is reached with all its available qubits. Their related fidelities were 99.994% for single-qubit gates, 99.81% for two-qubit gates and 99.72% for qubits readout[1089]. Meanwhile, their H1-1 system launched in June 2022 has 20 qubits (despite a lower "version" number), which enabled them in September 2022 to reach a QV of 8,192 ($2^{13}$) [1090]. This system adds arbitrary angle two-qubit gates which helps shorten the length of many algorithms, particularly those relying on a QFT (quantum Fourier transform).

**From 1D to 2D**. For now, Quantinuum is using a 1D trapped ion bar. They plan to adopt a 2D bar layout that would allow them to move the ions in two directions, and accumulate more of them and connect them with their neighbors in two dimensions[1091].

**Partnerships**. Quantinuum initially touted several partnerships: with **Microsoft,** for the integration of its systems in Azure Quantum which became operational in July 2020, an investment in **Cambridge Quantum Computing** (2014, UK, which they later merged with) and **Zapata Computing** (2017, USA, $64M). In February 2022, **IBM** invested about $25M in Quantinuum, probably more interested by its software branch (CQC) with which they had been partnering for a while.

Their first customers include **DHL**, **Merck**, **Accenture** and **Samsung**, who works on new batteries designs and **JPMorgan Chase** to create quantum algorithms in the financial sector. All of this for pilot projects. 12 or 20 qubits are way too few to enable production grade applications. They also work with **JSR Corporation** (Japan) to improve semiconductor design and research with organic and inorganic materials. In May 2022, Quantinuum launched InQuanto, a quantum computational chemistry software platform that was developed with the support of **BMW**, Honeywell, JSR, **Mitsui & Co**, **Nippon Steel Corporation** and **TotalEnergies**. The platform makes it possible to associate various quantum algorithms coupled with chemistry-specific noise-mitigation techniques running on NISQ systems. It also breaks down larger problems into smaller subproblems that fits existing NISQ machines. InQuanto is based on Quantinuum's open source toolkit **TKET**, which had been downloaded 500,000 times as of September 2022. You can wonder whether there are that many quantum developers in the world!

Quantinuum is investing a lot in QNLP (quantum natural language). They released lambeq in March 2022, a Python library that "*converts any natural language sentence into a quantum circuit*" that contains Bobcat, a neural-based Combinatory Categorial Grammar (CCG) parser, Bobcat. It is used to pre-process natural language training data to be subsequently used for various NLP applications (classification, summaries, etc).

In July 2022, a team assembling researchers from **JP Morgan Chase Bank** and the **University of Maryland** published an amazing paper saying that a (Quantinuum) quantum computer may be better at summarizing long documents[1092]. That was not exactly true. First, it was a hybrid algorithm with a lot of classical data preparation. The classical part analyzed a dataset of 300,000 news articles from CNN and the Daily Mail and precomputed it with a BERT NLP classical deep learning model that handles sentences extraction and converts them into vectors. Second, the experiment worked to summarize text from respectively 20 to 8 and 14 to 8 sentences, corresponding exactly to the number of used qubits in Quantinuum QPUs versions H1-1 and H1-2. On the H1-1, the quantum computing part executed at most 765 two-qubit gates with a computing depth of 159 and 2000 shots.

---

1089 See Quantinuum Announces Quantum Volume 4096 Achievement by Kortny Rolston-Duce, Quantinuum, April 2022.

1090 See Quantinuum System Model H1 Product Data Sheet Version 5.00, June 14, 2022 (9 pages).

1091 See Transport of multispecies ion crystals through a junction in an RF Paul trap by William Cody Burton et al, June 2022 (6 pages) where they describe how they can transport ytterbium and barium in 2D structures.

1092 See Long Story Short: Researchers Say Quantum Computers May be Better at Summarizing Long Documents by Matt Swayne, The Quantum Insider, June 2022, referring to Constrained Quantum Optimization for Extractive Summarization on a Trapped ion Quantum Computer by Pradeep Niroula et al, June 2022 (16 pages).

The experiment was based on using three quantum optimization algorithms working under constraints: QAOA, L-VQE[1093] and XY-QAOA and the comparison was made vs a classical random guess, with XY-QAOA being the best. This was to date the best optimization under constraint problem ever solved by a quantum computer. But its capacity is obviously limited to simple and short texts. It could not summarize a 300 sentences document given there are not enough physical qubits available, and fidelities allowing very long depth computing accordingly. It couldn't of course summarize the scientific paper for you since it contains several hundreds of sentences that are way more complicated than short news from CNN and The Daily Mail. In the end, we always must find out if the thing scales well or not, and under which circumstances. That aspect wasn't addressed in the paper.

**Alpine Quantum Technologies** or **AQT** (2017, Austria, $34.8M) is a spin-off from the University of Innsbruck created by Rainer Blatt. Peter Zoller and Thomas Monz. Ignacio Cirac (MPI) and Jonathan Home (ETH Zurich) are among their scientific advisors.

AQT drives its microwave trapped ions without the use of lasers, which simplifies the device. They use only one laser for photoionization of their calcium ions, which creates the ions at start-up, and another for measuring the qubit state by fluorescence after calculations. The fidelity of their qubits is 99.6% for two qubits and drops to 86% for 10 qubits[1094].

Although small and less visible than IonQ and Quantinuum, AQT has the largest trapped ions systems available in-store, with 20 working qubits working out of two 19-inches datacenter rack[1095]. The associated research team had already entangled 14 ions back in 2011[1096]! Their PINE system uses a linear Paul trap that supports up to 50 ions, including from multiple species. It can be used beyond quantum computing for quantum clocks or spectroscopy experiments.

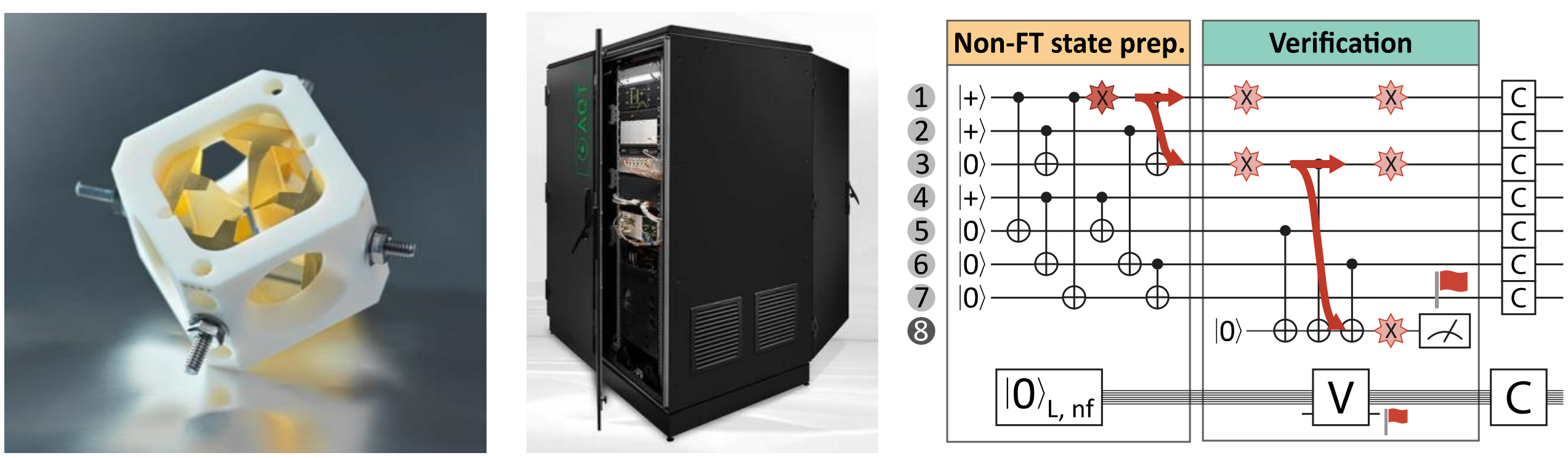

*Figure 393: AQT's pane system to trap their calcium ions, the 2-rack system, and how they implemented a fault-tolerant T gate with magic state preparation. Source: [Demonstration of fault-tolerant universal quantum gate operations](#) by Lukas Postler, Rainer Blatt, Thomas Monz et al, Nature, November 2021 and May 2022 (14 pages).*

AQT is also experimenting using qudits of dimension 5 with its ions[1097]. In May 2022, Thomas Monz's team announced the first realization of a fault-tolerant CNOT gate across two logical qubits made with 16 physical qubits and using 7-qubits color codes quantum error correction plus one

1093 See Layer VQE: A Variational Approach for Combinatorial Optimization on Noisy Quantum Computers by Xiaoyuan Liu et al, May 2022 (22 pages). So, we deal with a very recent algorithm!

1094 See Characterizing large-scale quantum computers via cycle benchmarking by Alexander Erhard et al, 2019 (13 pages).

1095 In EU Team Make Progress Toward European-Only Compact Quantum Computer That Could Run on Solar Power by Matt Swayne, The Quantum Insider, October 2021, we see them touting an energetic performance: a 24 qubits experimental system consumes only 1500W, like a kettle. Unfortunately, with 24 qubits can be emulated on a laptop that consumes less than 30W!

1096 See 14-Qubit Entanglement: Creation and Coherence by Thomas Monz et al, 2011 ( pages).

1097 See Native qudit entanglement in a trapped ion quantum processor by Pavel Hrmo, Rainer Blatt, Tomas Monz et al, June 2022 (9 pages).

measurement qubit (above, in Figure 393, *on the right*, so 16=2x(7+1)). They also separately implemented a fault-tolerant T gate using magic state preparation with flag qubits, fully using their 16 qubits[1098].

Their PINE system supports Qiskit, Cirq, PennyLane and Pytket. They team up with NTT on developing financial applications[1099].

**Oxford Ionics** (2019, UK, $3.6M) is a spin-off from the Department of Physics at Oxford University created by Chris Balance and Tom Harty which is developing a quantum computer based on trapped ions and low-noise control electronics.

They were originally called Nqie Limited. The company was founded by Thomas Harty and Christopher Ballance and also includes Jochen Wolf, all from Oxford University. They announced in July 2022 that they are teaming up with Infineon for the manufacturing of their trapped ions chipsets.

**Universal Quantum** (2018, UK, $14.6M) is a spin-off from the Ion Quantum Technology Group at the University of Sussex in the UK led by Winfried Hensinger. They are developing a trapped ion system that uses microwaves transmitted by electrical circuits, and magnetic fields to control them instead of lasers.

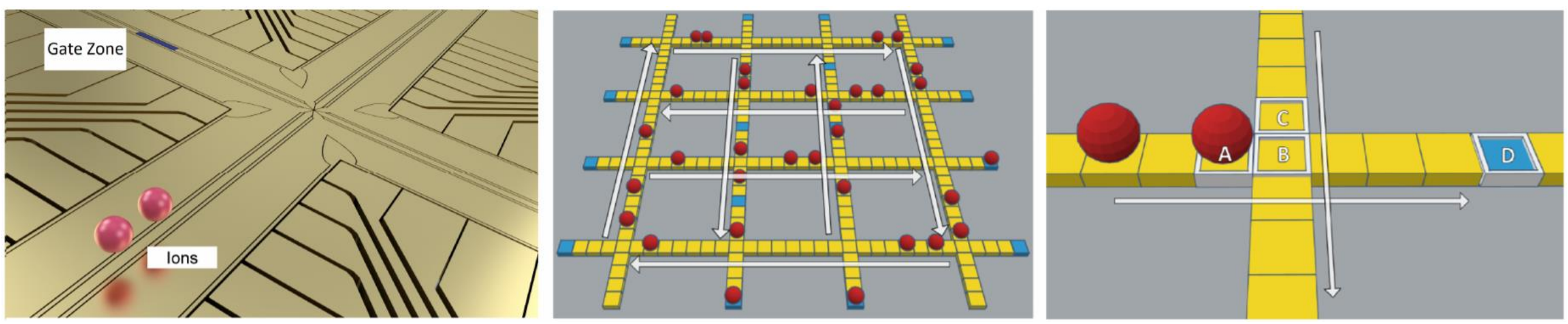

*Figure 394: Universal Quantum's shuttling ion architecture in their Penning traps. Source: Universal Quantum.*

They use Penning traps which are well known. The company presentation video gives the impression that they use a 2D process similar to Quantinuum's[1100]. The cooling required is around 70K, which is done with liquid nitrogen. They still need to use lasers at least for the Doppler based ions cooling during their preparation, then for the qubit state readout combining the usual laser excitation and fluorescence readout with a CMOS or CCD sensor[1101]. They use electrodes to drive qubit gates. In 2022, they announced their plan to reach one million qubits, some day[1102]. They plan to use electric fields to connect several modules on their silicon based wafer.

**Aquabits** (2021, Canada) is developing a trapped ions qubit processor using 'aquaporins', that trap ions inside artificial water channels. It is supposed to avoid using lasers and micro-nano fabrication techniques, making these qubits highly scalable. There's no public way to find out how all these qubits are controlled, entangled and measured.

---

[1098] See Demonstration of fault-tolerant universal quantum gate operations by Lukas Postler, Rainer Blatt, Thomas Monz et al, Nature, November 2021 and May 2022 (14 pages).

[1099] See Quantum computing in finance - Quantum readiness for commercial deployment and applications, NTT, February 2022 (17 pages).

[1100] The ion routing process is described in Efficient Qubit Routing for a Globally Connected Trapped Ion Quantum Computer by Winfried Hensinger et al, February 2020 (13 pages). This is the origin of the illustration used in these lines.

[1101] The ion control process with Penning Traps used by Universal Quantum seems to be described in Microfabricated Ion Traps by Winfried Hensinger et al, 2011 (28 pages).

[1102] See How Universal Quantum is rising to the million-qubit challenge, Universal Quantum, February 2022

**eleQtron Gmbh** (2020, Germany) develops a NISQ trapped ions quantum computer. They use their Magnetic Gradient Induced Coupling (MAGIC) to control the qubits.

The project involves the University of Siegen and Infineon. They are partnering with ParityQC (Austria) for software development within the ATIQ consortium with a total funding of 44.5M€ including 37M€ from the German government.

**Hon Hai** / **Foxconn** (Taiwan) announced in December 2021 it is starting the development of a trapped ions quantum computer in its quantum computing research center, part of its Research Institute[1103].

The lab is directed by Min-Hsiu Hsieh who was previously an associate professor at the Centre for Quantum Software and Information from University of Technology Sydney. He's more specialized in quantum machine learning than in trapped ions computing.

**Crystal Quantum Computing** (2021, France) was a stealth startup until 2022, created by Quentin Bodart (with a long-lasting experience with neutral atoms, including quantum microgravimeters) and Luca Guidon (from the CNRS MPQ laboratory in Paris).

Its goal is to create a trapped ion quantum computer using strontium 88 ions energized at Rydberg levels. Their ions would be easier to control than classical trapped ions, using UV lasers at 243 nm (generated with IR lasers and double frequency doubling) and THz microwave directive antennas, all being handled in a Penning trap on a chipset running at 30K.

## Neutral atoms qubits

Neutral atoms, *aka* cold atoms, are another atomic form of qubits in addition to trapped ions[1104]. They are both trapped, but not exactly in the same way. Since these atoms are not used in ionized form, they are not trapped with electrodes but with lasers. The atoms preparation is done in multiple steps.

An atom cloud is first trapped and cooled in a MOT (magneto-optical-trap)[1105]. Then other lasers using the method of "optical tweezing" or "optical traps" will precisely control the position of the atoms and arrange them in patterns like 2D matrices[1106].

Neutral atoms can be used to create qubits with their two states corresponding to different atomic energy levels, where transitions are controlled by a variable mix of laser beams and microwaves. These atoms have controllable high-energy so-called Rydberg states which can be used as the $|1\rangle$ qubit state and/or for coupling qubits with two-qubit gates or for setting-up a Hamiltonian for a quantum simulation run.

Indeed, neutral atoms qubits can be used in two ways: with gate-based computing using one and two-qubit gates and for quantum simulations, using a prepared state of interconnected qubits that are converging to a minimum energy level, helping to find a solution to chemical simulation and optimization problems.

---

1103 See Hon Hai to develop trapped ion quantum computers, Taipei Times, December 2021.

1104 See this excellent review paper: Quantum simulation and computing with Rydberg-interacting qubits by Manuel Agustin Morgado and Shannon Whitlock, Laboratory of Exotic Quantum Matter, University of Strasbourg, December 2020 (28 pages).

1105 It creates a variable magnetic field and associate three pairs of lasers to cool down the atoms below the Doppler limit, using the Zeeman variable shift effect. The frequency of the lasers used for atoms cooling would have to be changed as they are cooled. A workaround is to progressively change their resonance frequency with a varying magnetic field so that it's aligned with the cooling laser frequency.

1106 See Quantum information processing with individual neutral atoms in optical tweezers by Philippe Grangier, (47 slides).

Quantum simulation also simulate the "Hubbard model" (*aka* Fermi-Hubbard model) which modelizes strongly correlated electronic materials like condensed matter and high-temperature superconducting materials[1107].

However, neutral atoms are very versatile and have various other use cases on top of quantum computing and simulation, including quantum sensing, with applications in microgravity detection, electromagnetic spectrum analysis and atomic clocks and also quantum memories and repeaters[1108]. We cover these various use cases in other parts of this book.

### History

Neutral atom-based computing history starts a long time ago with fundamental physics research before quantum computing was even conceptualized. High energy atom states were formalized by Johannes Rydberg in Sweden in 1887 based on Johann Balmer's series. It was later explained by Niels Bohr in 1913 with his semiclassical model of the hydrogen atom with discrete energy levels. An extended understanding of the observed hydrogen spectrum was done by Wolfgang Pauli in 1926[1109].

In parallel with the many research on Bose-Einstein Condensate, which are not relevant for neutral atoms computing, mechanisms were developed in the 1980s to control individual atoms in vacuum using lasers[1110]. It was first demonstrated in 1985 at the Bell Laboratories by Steven Chu, creating what they called "optical molasses" due to the viscosity of the confined sodium atoms used in their experiment.

In 1985, Claude Cohen-Tannoudji, Alain Aspect and Jean Dalibard started to work on laser-based atoms cooling using Doppler effect. In 1987, David Wineland and Wayne Itano improved laser cooling, which led the way for various applications including ultrahigh resolution spectroscopy and atomic clocks. In 1988, scientists led by Claude Cohen-Tannoudji at ENS Paris and others from Stanford University developed new atoms cooling mechanisms based on laser optical pumping, light shifts and laser polarization gradients[1111]. They invented "Sisyphus cooling" in 1989, a cold atom cooling, *aka* polarization gradient cooling, reaching temperatures below the Doppler cooling limit. This led Claude Cohen-Tannoudji to be awarded the Nobel prize in physics in 1997, together with Steven Chu.

Laser cooling of atoms then reached very low temperatures, in the nK range. It contributed in 1995 to the discovery in the USA, of gaseous Bose-Einstein condensates that was devised in the mid-1920s by Bose and Einstein.

Ultra-cold atoms were used to create more precise atomic clocks than the cesium-based ones running at room temperature starting in 1998 in France.

In the 1980s, Serge Haroche, started to work with Rydberg atoms and their integration in superconducting cavities, pioneering cavity electrodynamics (CQED), light-atoms interactions, cold atoms control and the understanding of quantum decoherence.

So, how about neutral atoms and quantum computing? First, we have the raw idea of a quantum simulator by Richard Feynman in 1981. In 1996, Seth Llyod demonstrated that it "was possible" to implement such scheme, noticeably with controlled atoms[1112].

---

[1107] See Quantum simulation of the Hubbard model with ultracold fermions in optical lattices by Leticia Tarruell (ICFO, Spain) and Laurent Sanchez-Palencia (CPHT, France), January 2019 (38 pages).

[1108] See Highly-efficient quantum memory for polarization qubits in a spatially-multiplexed cold atomic ensemble by Julien Laurat et al, 2018 (6 pages) and Experimental realization of 105-qubit random access quantum memory by N. Jiang et al, 2019 (6 pages).

[1109] See Rydberg Physics by Nikola Šibalić and Charles S Adams, 2018 (28 pages).

[1110] See New Mechanisms for Laser Cooling by Claude Cohen-Tannoudji and William D. Phillips, 1990 (8 pages).

[1111] See Laser cooling and trapping of neutral atoms by Jean Dalibard and Claude Cohen-Tannoudji (20 pages).

[1112] See Universal Quantum Simulators by Seth Lloyd, 1996 (7 pages).

While quantum simulation can be theoretically implemented with trapped ions and superconducting qubits, the cold atom way is the only one that is seriously investigated and which has reached the commercial stage.

Then things started to get serious with two-qubit gates proposals from Dieter Jaksch, J. Ignacio Cirac and Peter Zoller[1113] and from Gavin K. Brennen et al in 1998[1114], with improvements from Jaksch, Zoller and Mikhail Lukin in 2000[1115]. In 2012, J. Ignacio Cirac and Peter Zoller proposed a set of criteria for quantum simulators similar to those from David DiVincenzo's 2000 for gate-based quantum computing (in Figure 395)[1116].

TABLE I. Criteria for quantum simulators and quantum computers

| Criteria | Quantum computers[54] | Quantum simulators[4] |
|---|---|---|
| Quantum system | A scalable physical system with well characterized qubits | A system of quantum particles (bosons, fermions, pseudo-spins) confined in space and collectively possessing a large number of degrees of freedom |
| Initialization | The ability to initialize the state of the qubits to a simple fiducial state, such as $\|000\ldots\rangle$ | The ability to prepare (approximately) a known quantum state (typically a pure state) |
| Coherence | Long relevant decoherence times, much longer than the gate operation time | |
| Interactions | A "universal" set of quantum gates | An adjustable set of interactions used to engineer Hamiltonians/quantum master equations including some that cannot be efficiently simulated classically |
| Measurement | A qubit-specific measurement capability | The ability to perform measurements on the system; either individual particles or collective properties |
| Verification | | A way to verify the results of the simulation are correct |

*Figure 395: comparisons of gate-based quantum computing (left) and quantum simulation (right). Source: Quantum simulation and computing with Rydberg-interacting qubits by Manuel Agustin Morgado and Shannon Whitlock, December 2020 (28 pages).*

The first two-qubits gates with pairs of Rydberg atoms were implemented in 2009 by Mark Saffman from the University of Wisconsin (with "gg-qubits", which we cover later) and at the Institut d'Optique in France ("gr-qubits").

Many progresses were made in the 2010's which encouraged many scientists to create their own companies. It started with the creation of ColdQuanta (2007), Muquans (2011), both using cold atoms for quantum sensing in micro-gravimetry, BraneCell (2015), Atom Computing (2018), Pasqal (2019) and QuEra (2020). Besides Muquans (now in France's ixBlue), all the others are positioned in the quantum computing market although ixBlue is also a technology provider for Pasqal for lasers.

## Science

Neutral atoms are simply non-ionized atoms with an equivalent number of protons in their nucleus and electrons in their shells. The neutral atoms that are used in cold atoms computing belong to the first column in the table of elements, having a single electron in the valence layer, such as hydrogen, sodium, lithium, cesium or rubidium, the last one being the most commonly used. This alkaline metal has interesting energy transitions that correspond to common lasers wavelengths as well as easily generated microwaves between 3 and 10 GHz. It is possible to manage with them so-called closed transitions which allow, with lasers, to make atoms transit between several states in a cyclic and controlled manner.

---

[1113] See Entanglement of atoms via cold controlled collisions by Dieter Jaksch, H.-J. Briegel, J. Ignacio Cirac, C. W. Gardiner and Peter Zoller, 1998 (4 pages).

[1114] See Quantum Logic Gates in Optical Lattices by Gavin K. Brennen, Carlton M. Caves, Poul S. Jessen, and Ivan H. Deutsch, PRL, 1998 (7 pages).

[1115] See Fast Quantum Gates for Neutral Atoms by D. Jaksch, J. Ignacio Cirac, Peter Zoller, Mikhail D. Lukin et al, PRL, 2000 (4 pages).

[1116] See Goals and opportunities in quantum simulation by J. Ignacio Cirac and Peter Zoller, Nature Physics, 2012 (3 pages).

On top of that, states are stable long enough to perform computations, i.e. about a hundred microseconds. Other elements are investigated like dysprosium and praseodymium who are lanthanide elements.

Cold atoms can be used in Rydberg states, which correspond to a very high level of energetic excitation, between 50 and 100 electron quantum number (layer position in atom against Bohr's model, labelled n or N). This creates very large electron orbits, scaling by $N^2$. These high energy states are used to create entanglement between atoms and thus to operate multi-qubit quantum gates or large Hamiltonians in quantum simulation modes. These excited states have a fairly good stability level of about 100 µs. They are several orders of magnitude longer than the classical excited states (hyperfine, which are used for cold atoms qubit states). This stability is somehow equivalent to the coherence time of superconducting qubits.

Cold atoms computing also exploits the Rydberg blockade effect, where a Rydberg atom excited with a high energy level (with n>50-70) prevents neighboring atoms from reaching that level.

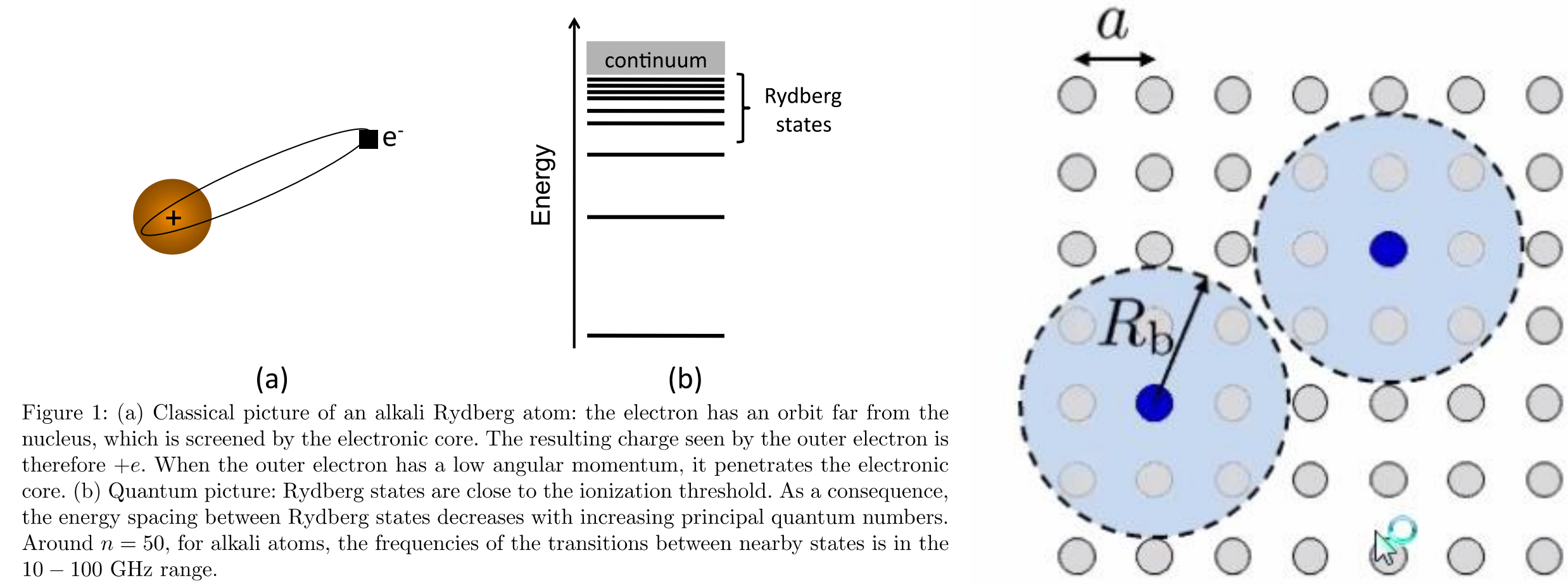

*Figure 396: Rydberg state are high-energy level of excited atoms that create a dipole in the atom. It enables entanglement with neighbor atoms. Source: Interacting Cold Rydberg Atoms: a Toy Many-Body System by Antoine Browaeys and Thierry Lahaye, 2013 (20 pages).*

When excited, these atoms behave like accentuated dipoles, the orbit of the electrons of the valence layer being very inclined as shown in Figure 396.

They also have a disproportionate size of up to one micron (µm) in diameter for n=100 with $^{87}$Ru. This is close to being in an ionized state[1117]. Their electromagnetic characteristics make the atoms react with their neighbors whose excitation they block within a perimeter of up to 20 µm, which is huge at the atomic scale.

Activated Rydberg atom can also be excited by lasers to generate well-isolated single photons that can be used in nonlinear optics[1118]. This provides yet another source of single photons, in addition to quantum dots. The Rydberg blockade phenomenon could also be implemented in quantum telecommunications, in spectroscopy and in atomic clocks[1119].

[1117] This presentation of 52 slides from 2014 describes well the history and geometry of the Rydberg atoms.

[1118] See Observation of coherent many-body Rabi oscillations by Yaroslav Dudin and Alex Kuzmich, GeorgiaTech, 2012 (5 pages) and Nonlinear quantum optics mediated by Rydberg interactions by Sebastian Hofferberth et al, 2016 (26 pages).

[1119] See Photon-Mediated Quantum Information Processing with Neutral Atoms in an Optical Cavity by Stephan Welte, 2019 (124 pages).

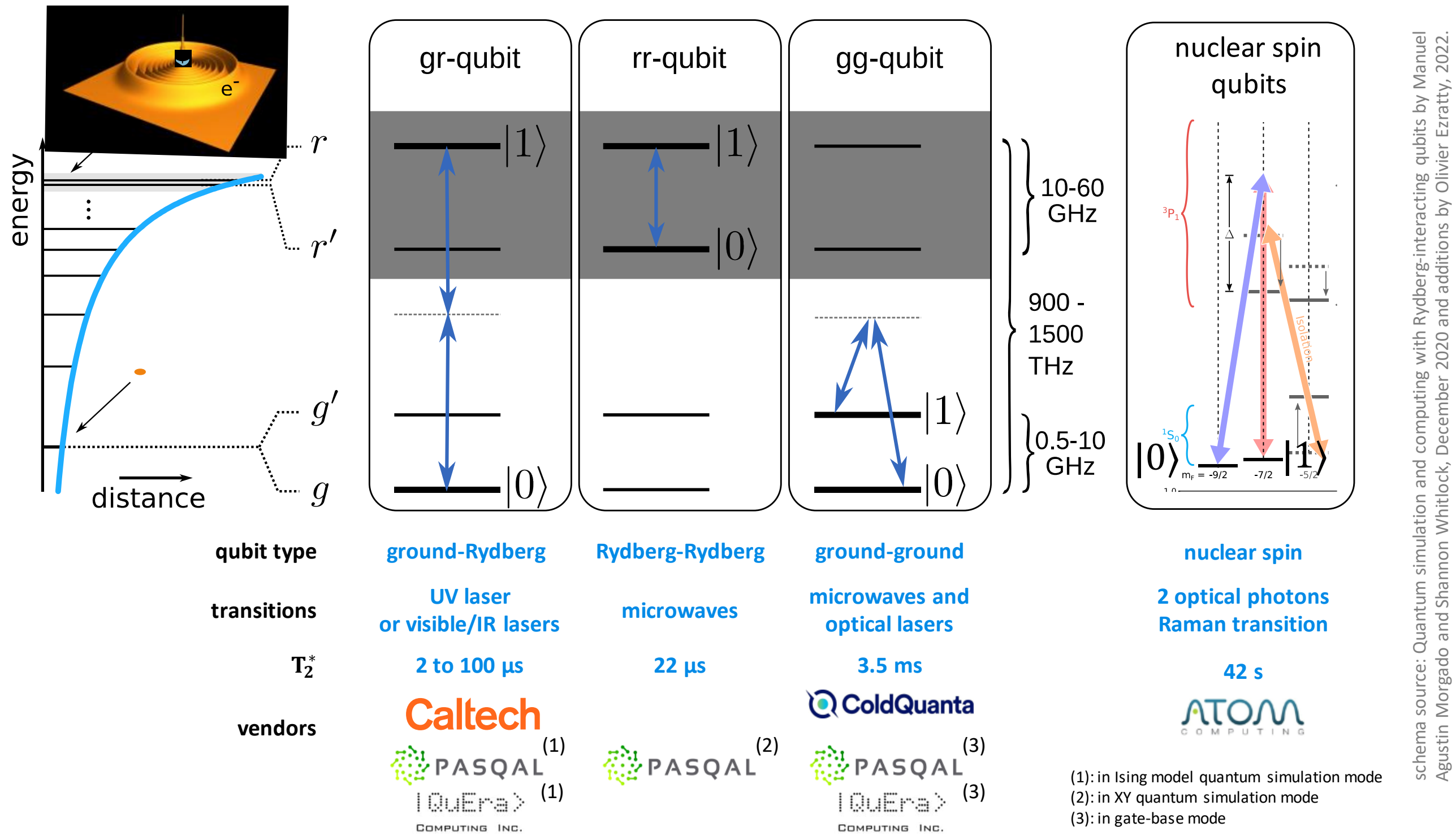

*Figure 397: the various ways to control cold atoms. Source: Quantum simulation and computing with Rydberg-interacting qubits by Manuel Agustin Morgado and Shannon Whitlock, December 2020 (28 pages) and additions by Olivier Ezratty, 2022.*

But as usual with all qubit types, there are many variations of cold atoms qubits. First, you have three breeds of qubits whose manifold is based on classical energy levels. With ground-Rydberg qubits which are controlled by UV, visible and infrared lasers, Rydberg-Rydberg qubits controlled by microwaves and lasers, ground-ground qubits controlled by microwaves and optical lasers, and at last, nuclear spin atoms controlled by optical lasers using Raman transitions. Some players like Pasqal have been investigating the first three types of qubits, seemingly favoring gr-qubits and rr-qubits for quantum simulation and gg-qubits for gate-based computing. Topological states allowing to create more reliable qubit-based computing systems are also studied[1120].

Cold atoms qubits are the most common ones that can be used in both gate-based quantum computing and in quantum simulation computing mode (*aka* analog quantum simulation). In the first case, qubits are individually controlled over time with single and two-qubit gates, to read qubit state at the end of processing[1121]. With quantum simulation[1122], a so-called Hamiltonian is prepared with specific atoms geometry and connectivity, usually using Rydberg states, which then converges itself into an energy minimum leading to qubits measurement. Individual qubits are controlled only at the initialization stage and for readout. There is no sequential programing.

---

[1120] See Topologically protected edge states in small Rydberg systems by Antoine Browaeys et al, 2018 (6 pages) and Observation of a symmetry protected topological phase of interacting bosons with Rydberg atoms by Antoine Browaeys, Thierry Lahaye et al, 2019 (20 pages). Quantum simulation using cold atoms is also a tool to simulate topological matter. See Scientists unveil first quantum simulation of 3-D topological matter with ultracold atoms by Hong Kong University of Science and Technology, July 2019.

[1121] See Versatile neutral atoms take on quantum circuits by Hannah J Williams, Nature, 2022 (2 pages) which describes two such methods, implemented by QuEra and ColdQuanta and mentioned later.

[1122] See Toward quantum simulation with Rydberg atoms by Thanh Long Nguyen, 2016 (182 pages), Quantum simulations with ultracold atoms in optical lattices by Christian Gross and Immanuel Bloch, 2017 (8 pages), Tunable two-dimensional arrays of single Rydberg atoms for realizing quantum Ising models by Thierry Lahaye and Antoine Browaeys, 2017 (13 pages), Quantum read-out for cold atomic quantum simulators, par J. Eisert et al, 2018 (20 pages), Quantum critical behaviour at the many-body localization transition by Markus Greiner et al, 2018 (10 pages), Quantum Kibble-Zurek mechanism and critical dynamics on a programmable Rydberg simulator by Alexander Keesling et al, 2019 (16 pages) and Many-body physics with individually controlled Rydberg atoms by Antoine Browaeys and Thierry Lahaye, 2020 (14 pages).

| cold atoms qubits | | |
| --- | --- | --- |
| | • long qubit **coherence time and fast gates**.<br>• **operational systems** with 100-300 atoms.<br>• **identical atoms**, that are controlled with the same laser and micro-wave frequencies (but dual-elements architectures are investigated).<br>• works in both **simulation** and **gate-based** paradigms, but still with difficulty for gate-based.<br>• **reuse trapped ions qubits tools** for qubits readout with fluorescence and CCD/CMOS detection.<br>• no need for specific **integrated circuits**.<br>• uses **standard apparatus**.<br>• low **energy consumption**. | • **acceptable quantum gates error rate** although not "best in class".<br>• **crosstalk** between qubits that can be mitigated with two-elements systems.<br>• **adapted to simulation** more than to universale gates computing.<br>• not yet operational QND (quantum non demolition) measurement that is required for QEC and FTQC.<br>• control **lasers and optical** not scaling well beyond one thousand qubits with the current state of the art. |

*Figure 398: pros and cons of cold atoms quantum computers and simulators. (cc) Olivier Ezratty, 2022.*

Most neutral atoms systems and qubit types can be used in both paradigms, but it seems that the gate-based model is the most demanding and complicated to handle. Thus, three situations in the market can be observed: startups like Pasqal are explicitly saying that they start first with the quantum simulation paradigm, others like ColdQuanta tout their positioning on gate-based quantum computing but actually start with quantum simulation and at last, others like Atom Computing start readily with a gate-based approach but to no real avail.

### Qubit operations

We'll look here at the way qubits lifecycle works, from initialization to readout, with quantum gates in-between (for gate-based systems). The general principle is as follows:

- **Quantum state** for the $|0\rangle$ and $|1\rangle$ qubit basis corresponds to a ground and excited state, which depends on the qubit type as seen previously with ground-Rydberg, Rydberg-Rydberg, ground-ground and nuclear spin atoms qubits. The most commonplace for gate-based computing seems the ground-ground case. The qubit $|0\rangle$ state is usually prepared with laser pumping or with some microwave pulse. Contrarily to superconducting and quantum dots spin qubits who are static in nature in their electronic circuits, atoms have to be first arranged in space before any computing can start. The qubits can be arranged in 1D, 2D[1123] or 3D matrices[1124].

  They are cooled, controlled, and positioned by several lasers organized in precision "optical tweezers". A qubit can be based on a single atom or on a group of atoms depending on the methods used. The atoms are prepared with a hot or cold source (some μK) which then feeds an ultra-vacuum chamber where laser control takes place.

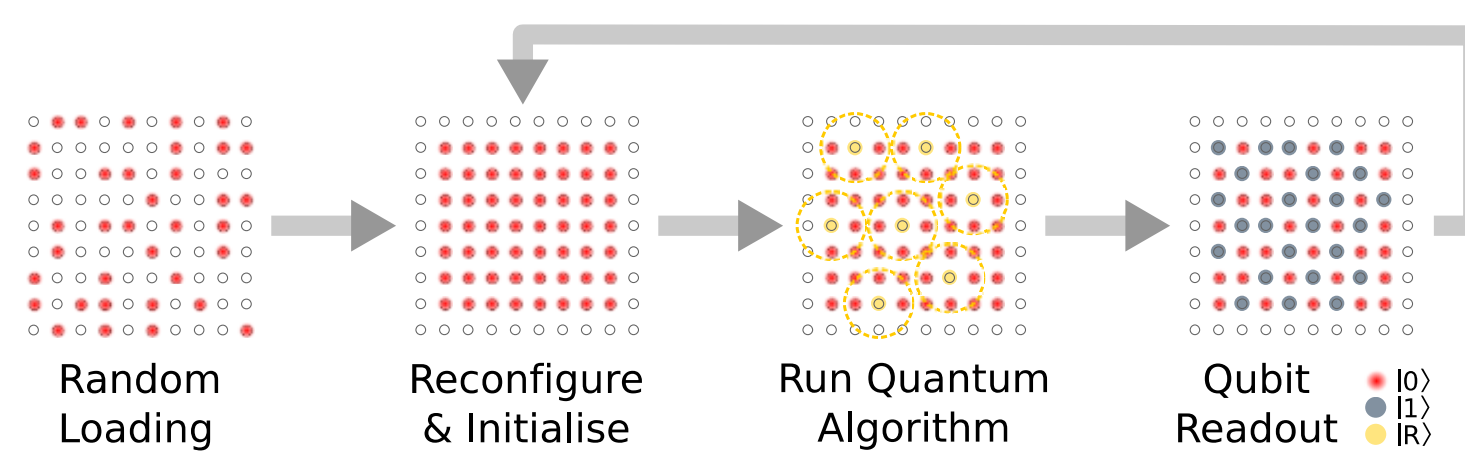

**Figure 2.** Schematic of a Rydberg array quantum computer. Atoms are initially loaded stochastically, followed by rearrangement to achieve a defect free qubit register. Coherent excitation to Rydberg states allows implementation of quantum algorithms exploiting long-range interactions to couple neighbouring qubits, followed by state-selective readout which is repeated many times to tomographically reconstruct the output state.

*Figure 399: how an array of cold atoms is being prepared. Source: [Rydberg atom quantum technologies](#) by James Shaffer, 2019 (24 pages).*

---

[1123] See the thesis Rydberg interactions in a defect-free array of single-atom quantum systems by Daniel Ohl de Mello, 2020 (147 pages) which describes the way to fill a 2D matrix of a hundred heavy atoms.

[1124] See Three-Dimensional Trapping of Individual Rydberg Atoms in Ponderomotive Bottle Beam Traps by Antoine Browaeys, Thierry Lahaye et al, 2019 (8 pages).

- **Single-qubit quantum gates** are activated by a mix of microwaves (a few GHz, compatible with hyperfine states in the case of Rydberg-Rydberg or ground-ground qubits) and laser pumping to change the energy state of the cold atom between its ground and excited state. These gates can also use Raman transitions driven by lasers on two frequencies or by a combination of the Stark effect of spectral line shifting under the effect of an electric field and microwaves. In most cases, cold atoms single qubit gates are $R_z(\theta)$ (arbitrary phase rotation). The best single-qubit gate fidelities are around 99.6% with a long-term objective of reaching an error rate of $10^{-4}$ [1125]. Fortunately, there are optical systems for multiplexing laser beams, which make it possible to avoid having more lasers than qubits.

- **Two-qubit quantum gates** also use a variable mix of microwaves and lasers most of the time, using Rydberg state, the related Rydberg blockage phenomenon and dipole-dipole interactions[1126]. They are applied to atoms in their ground or excited state, which projects its valence layer electrons into a high orbit. For rubidium, there is only one electron to manage in this layer. These quantum gates can in practice involve more than two qubits, which is useful to set up a Hamiltonian in quantum simulation mode. The fidelity of two qubit gates was quite low in 2016 with a maximum of 75% with rubidium and 81% in 2016 with cesium. It increased to a better level of 99,1% in 2020[1127]. The decoherence of cold atoms qubits has different origins: photoionization, spontaneous emission of photons, transitions induced by black body radiation, stability of control lasers and laser pulse timing and precision control of atoms in space[1128]. The two-qubit gate set is variable. It can for example contain a CPHASE, CZ (ColdQuanta, QuEra) and XY gate. These gates usually work in a nearest neighborhood fashion. With cold-atoms, two-qubit gates are usually faster to operate than single qubit gates[1129].

- **Qubit readout** uses a CCD or CMOS camera that detects the atoms fluorescence with a method similar to the one used with trapped ions and NV centers. In Figure 400 is a simplified description a cold atom qubits system with laser and microwaves-based control tools and qubit measurement using fluorescence and a camera. This method is destructive of the qubit state so it's not a QND measurement.

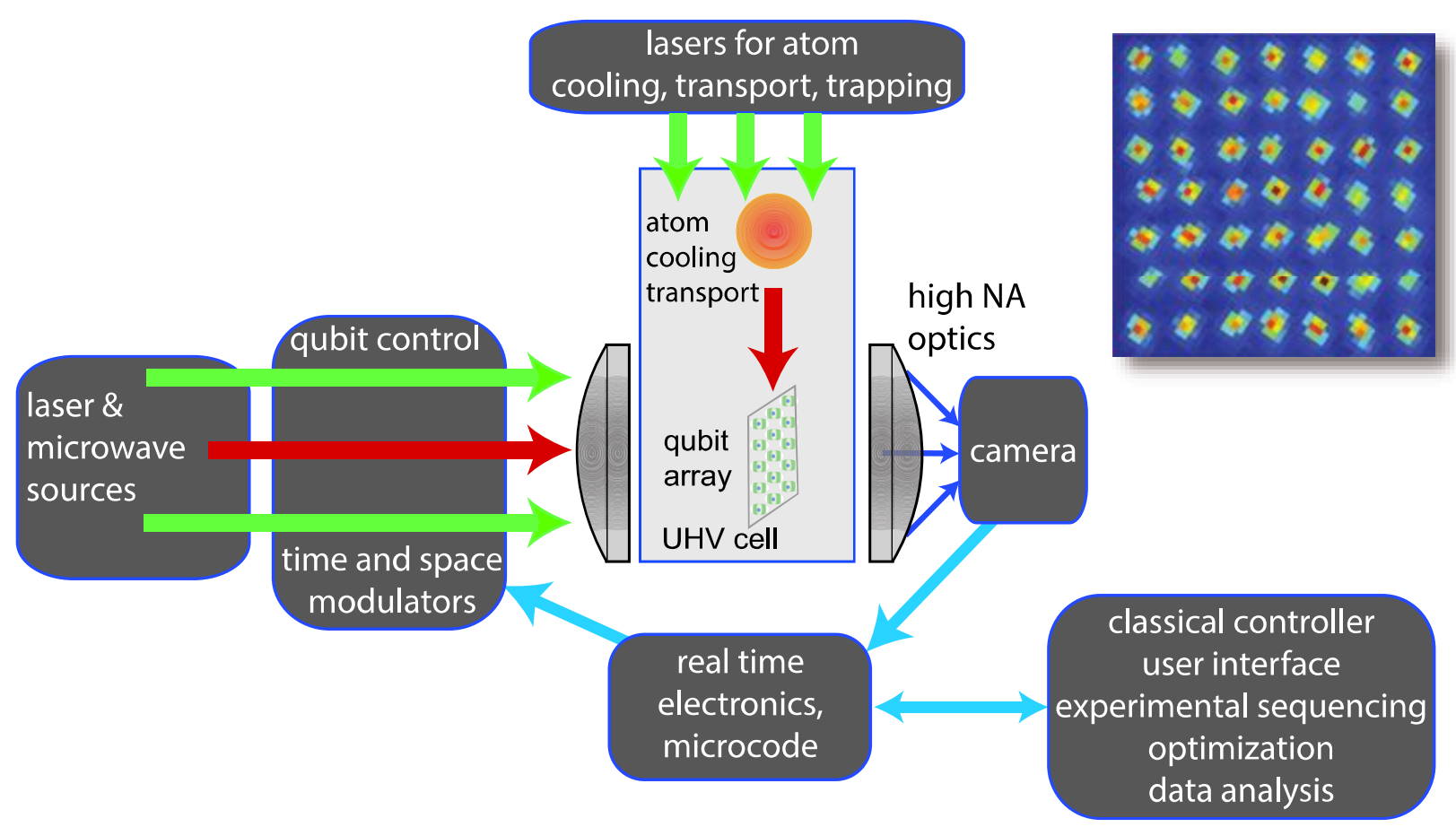

*Figure 400: typical devices arrangement to control cold atoms. Source: Quantum computing with atomic qubits and Rydberg interactions: Progress and challenges by Mark Saffman, 2016 (28 pages).*

---

[1125] See High-Fidelity Control, Detection, and Entanglement of Alkaline-Earth Rydberg Atoms by Ivaylo Madjarov, January 2020 (13 pages) which uses strontium.

[1126] In 2019, American researchers were able to create multi-qubit quantum gates with 95% fidelity based on cold atoms, in Parallel implementation of high-fidelity multi-qubit gates with neutral atoms by H. Levine et al, August 2019 (16 pages). Two-qubit gate bases in Direct Measurement of the van der Waals Interaction between Two Rydberg Atoms by Lucas Béguin, Antoine Browaeys et al, 2013 (5 pages). And Quantum information processing with individual neutral atoms optical tweezers by Philippe Grangier (47 slides).

[1127] See High-fidelity entanglement and detection of alkaline-earth Rydberg atoms by Ivaylo S. Madjarov et al, November 2020 (16 pages).

[1128] Source: Quantum Computing with Neutral Atoms, 2013 (42 slides).

[1129] It could even reach the nanosecond scale as experimented in Ultrafast energy exchange between two single Rydberg atoms on a nanosecond timescale by Y. Chew et al, Nature Photonics, 2022 (7 pages).

As a result, cold atoms are not the easiest candidate to implement measurement-based quantum error correction for creating FTQC systems. There are however various investigated solutions[1130].

## Setup

In general, cold atom-based systems operate at room temperatures but atoms are cooled at below 1 mK and in ultra-high vacuum. In practice, it is the ultra-high vacuum and the atoms laser cooling that ensures this thermalization. Preparing and controlling the qubits is based on a set of lasers, light structuring devices (SLM and AODs), polarizing beam splitters and microwaves devices. Microwaves are sent in a "one to many" mode and usually coupled with targeted photons to control individual gates. Neutral atom qubits vendors often argue that their system work at ambient temperature and do not require any refrigerant based cooling system. While they do not use cryostats like those that cool superconducting and silicon qubits, they still cool their qubits, using different methods combining ultra-vacuum pumps, magnetic traps and lasers. And now, some are even using a 4K cryostat to cool their ultra-vacuum pump to avoid the pollution of their tweezer-assembled 2D grid by spare atoms, at least, beyond 200 atoms.

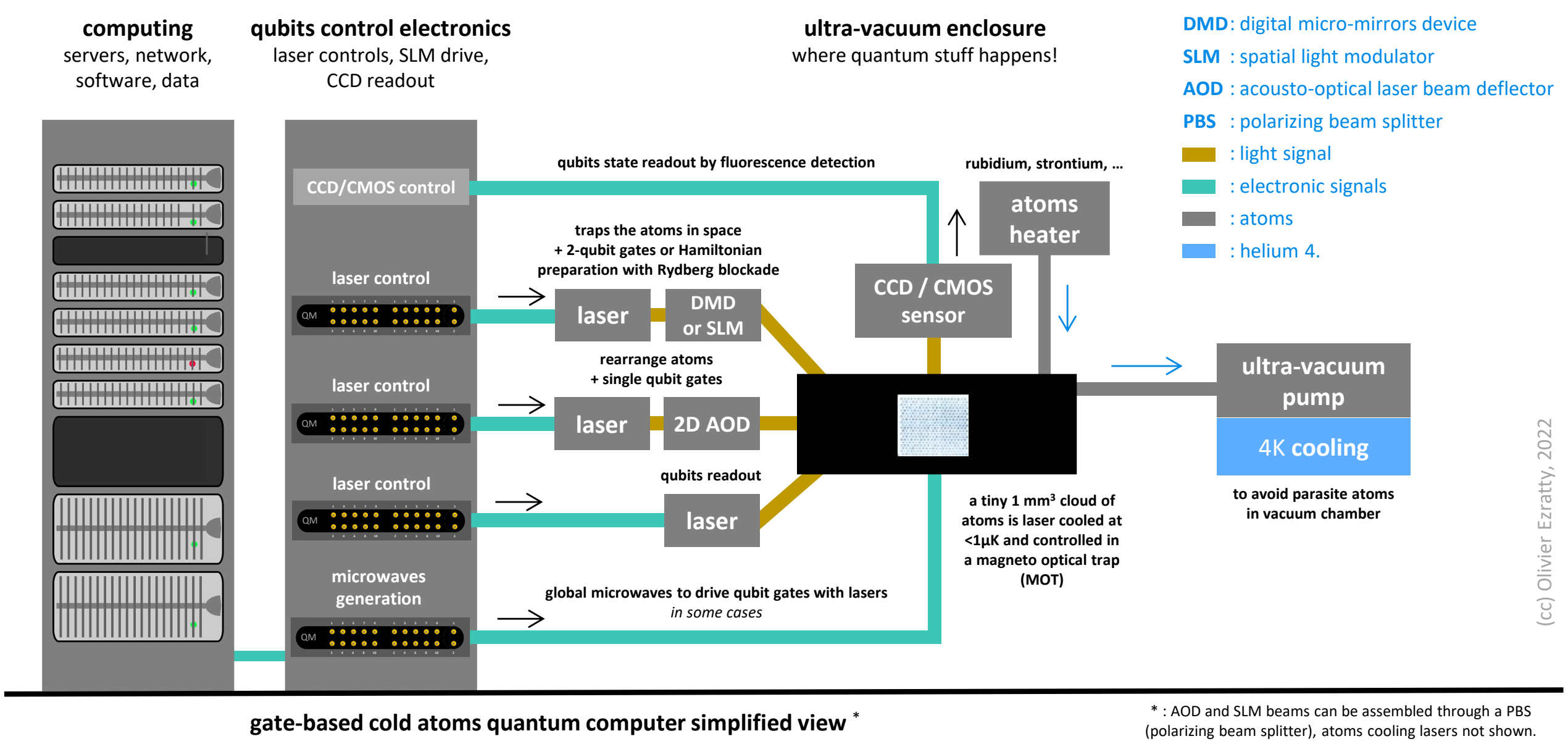

*Figure 401: overall architecture of a cold atoms based computer. (cc) Olivier Ezratty, 2022.*

I now owe you some explanations on the mentioned devices:

**MOT** (magneto-optical trap) uses laser cooling and a spatially-varying magnetic field to create a trap for our cold atoms where they are cooled with lasers and Doppler effect. The MOT contains a weak quadrupolar spatially-varying magnetic field generated by a coil of about 8 cm diameter and four to six glass doors letting through circularly-polarized red lasers beams which are slowing down the movement of the atoms at the center of the MOT. It enables the atoms to reach very low temperature under 1 mK. Next is the MOT chamber used by Pasqal. It's quite heavy, weighting in excess of 25 kg. The MOT chamber is pumped to be under ultra-vacuum.

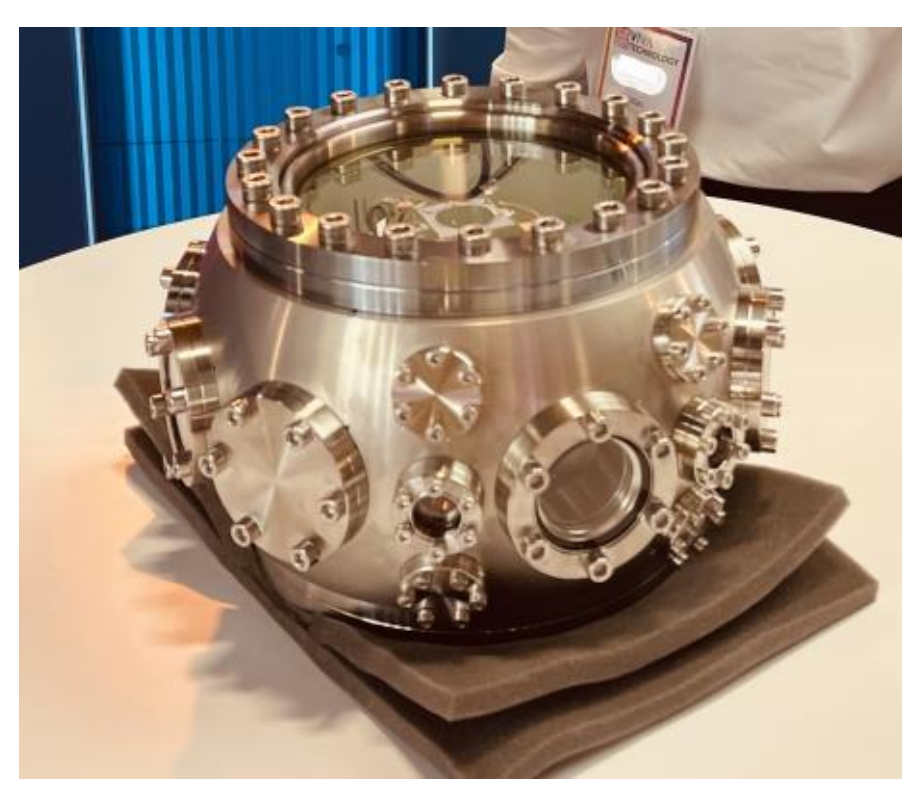

*Figure 402: a vacuum chamber from Pasqal, which contains a MOT.*

[1130] See for example Monitoring Quantum Simulators via Quantum Nondemolition Couplings to Atomic Clock Qubits by Denis V. Vasilyev, Andrey Grankin, Mikhail A. Baranov, Lukas M. Sieberer and Peter Zoller, October 2020 (22 pages).

**SLM** (spatial light modulator) are systems modulating light. Its main use is with video projectors, using LCD (liquid-crystal displays, transparent), LCOS (Liquid crystal on silicon, a reflective version of LCD chipsets) or DMD chipsets (using micro-mirrors reflecting light in various directions). The SLMs breeds used for cold atom tweezers modulate the light phase instead of just its intensity and creates sort of a hologram. Phase SLMs are usually based on LCOS chipsets thanks to their controllable birefringence. SLM main vendors are Hamamatsu, Holoeye and Thorlabs. SLM resolution can reach 4,160 x 2,160, equivalent to 4K in TV/PC formats. SLMs are used with neutral atoms to create optical tweezers that precisely control atoms position in vacuum and at the nanometer scale[1131].

**AOM** (acoustic-optical modulators) *aka* **AOD** (acousto-optic deflectors) uses an acousto-optic effect to diffract and shift the frequency of light using sound waves at radio frequencies. It contains a piezoelectric transducer that is attached to a material such as glass. An oscillating electric signal creates transducer vibrations, producing sound waves within the material, changing the refraction index[1132].

### Research

The most active research laboratories with cold atom-based qubits are in the USA (Harvard with **Mikail Lukin**, University of Wisconsin with **Mark Saffman**[1133], **Adam Kaufman** from JILA in Colorado, Caltech with **Vladan Vuletic** and **Manuel Endres**, Princeton with **Jeff Thompson**, GeorgiaTech), in the UK at the University of Cambridge, in Austria at the University of Innsbruck and the University of Vienna, Germany (Max-Planck Institute, Free University of Berlin, University of Stuttgart).

In France we have Institute d'Optique Graduate School with **Antoine Browaeys** and **Thierry Lahaye** who cofounded **Pasqal**, and the **Unistra** laboratory in Strasbourg run by **Shannon Whitlock**, who is behind the project **aQCess** and the European Flagship **EuRyQa** (with Pasqal).[1134]. Another European H2020 project, **AtomQT**, covers research with cold atoms, both qubits and sensing. In France, it involves the Bordeaux Optics Institute and the LPMMC in Grenoble.

In Germany, **Johannes Zeiher** from the Mack Planck Institute of Quantum Optics created SNACQ in 2022, a research group on cold atoms ("Scalable Neutral Atom Quantum Computing"). Its ambition it to create the first error-corrected logical qubit[1135]. It complements **Immanuel Bloch**'s team (Quantum Many-body Systems) and **Gerhard Rempe**'s group (Quantum Dynamics).

In 2017, Mikhail Lukin's team from **Harvard** University and a team from **MIT** assembled 51 rubidium atoms[1136]. It went up to 256 qubits in July 2021[1137]. Antoine Browaeys's team at **Institut d'Optique** reached 72 cold atoms in a 3D structure in 2018, 196 in 2020[1138]. and 500 in 2021. Four startups are positioned in the cold atoms computing market: **ColdQuanta** (2007, USA, working initially on quantum sensing), **Atom Computing** (2018, USA), **Pasqal** (2019, France) and **QuEra Computing (**2020, USA, linked to Lukin and Harvard).

---

[1131] See High-Precision Laser Beam Shaping and Image Projection by Jinyang Liang, 2012 (126 pages).

[1132] See alternative and more scalable approaches as proposed in An integrated photonic engine for programmable atomic control by Ian Christen et al, MIT and CSEM, August 2022 (16 pages).

[1133] See Quantum computing with atomic qubits and Rydberg interactions: Progress and challenges by Mark Saffman, 2016 (28 pages).

[1134] See Time-Optimal Two- and Three-Qubit Gates for Rydberg Atoms by Sven Jandura and Guido Pupillo, February 2022 (24 pages).

[1135] See Johannes Zeiher launches new research group, February 2022.

[1136] See Quantum simulator with 51 qubits is largest ever by Matt Reynolds, 2017 which refers to Probing many-body dynamics on a 51-atom quantum simulator by Hannes Bernien, Mikhail Lukin et al, 2017 (24 pages).

[1137] See Harvard-led physicists take big step in race to quantum computing, Harvard, July 2021. Their work is like Pasqal/IOGS based on the same technique with rubidium atoms and SLM tweezers.

[1138] See Synthetic three-dimensional atomic structures assembled atom by atom by Daniel Barredo, Antoine Browaeys et al, 2018 (4 pages).

You may obviously wonder whether China is also working on cold atom. You bet they are, like in all quantum fields. It includes the Key Laboratory of Quantum Optics in Shanghai and the Center for Cold Atom Physics in Wuhan, both from the **China Academy of Science**[1139]. These labs are mainly focused on building cold atom based quantum sensors and also repeaters. They also investigate cold atoms quantum computing.

**neutral atoms gate-based quantum computing scalability challenges**

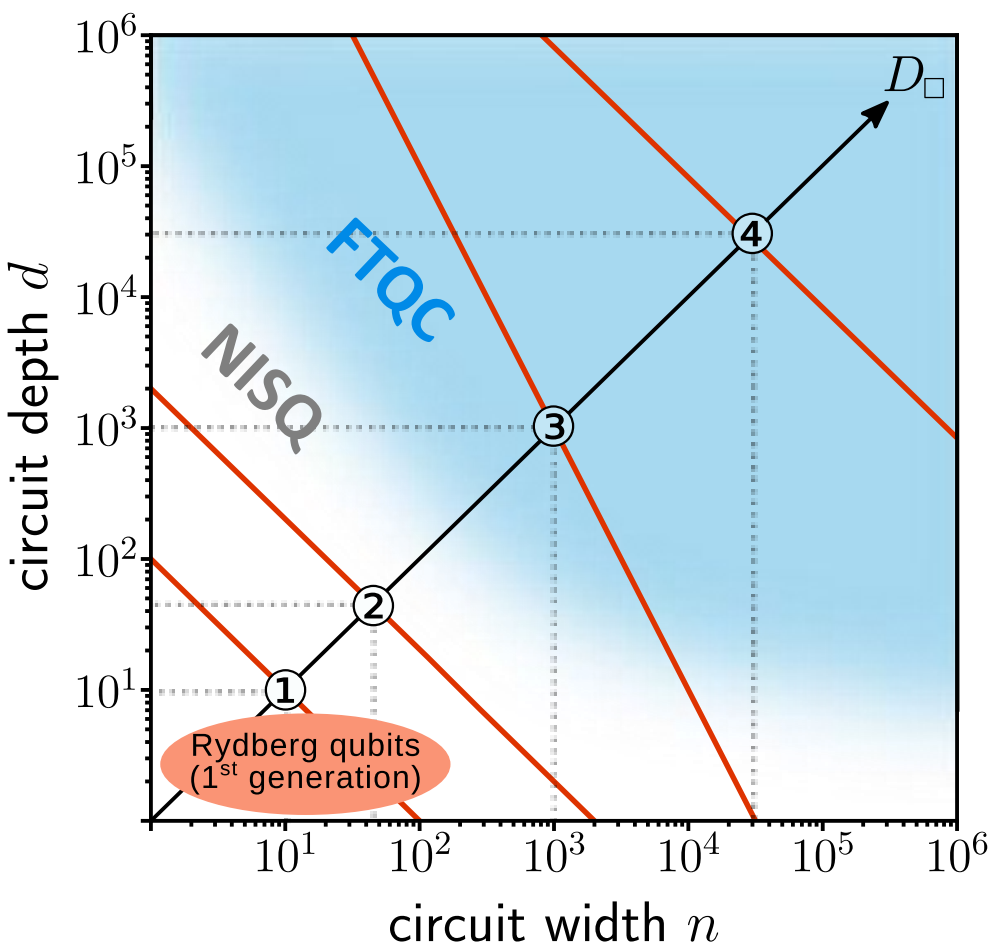

| | limitations | $T_1$ | d | $N_g$ | solution |
|---|---|---|---|---|---|
| ① | short dwell time of Rydberg excitations | 5 µs | 10 | 50 | blue detuned optical or magnetic traps that can simultaneously trap both ground and Rydberg states |
| ② | atoms interaction with thermal and vacuum electromagnetic fields | 100 µs | 44 | $10^3$ | embedding atoms inside cryogenic cavities to inhibit black-body transitions and spontaneous microwave emissions |
| ③ | trap loss due to background gas collisions | 100 s | $10^3$ | $5 \times 10^5$ | cooling vacuum chamber |
| ④ | | 100 s | $3 \times 10^4$ | $5 \times 10^8$ | apply gates in parallel |

red lines: successive barriers associated with the finite lifetime $T_1$ of Rydberg qubits with barriers assuming each gate has a duration of 50 ns.

$N_g = d*n/2$: multi-qubit gate count with n/2 two-qubit count per circuit layer.

schema source: Quantum simulation and computing with Rydberg-interacting qubits by Manuel Agustin Morgado and Shannon Whitlock, December 2020 and additions by Olivier Ezratty, 2022.

*Figure 403: sorting out the cold atoms computing challenges per generation. Source: Quantum simulation and computing with Rydberg-interacting qubits by Manuel Agustin Morgado and Shannon Whitlock, December 2020 (28 pages) and text formatting by Olivier Ezratty, 2022.*

The many challenges to overcome are, like with all other qubit types, about scalability and also, the implement non-destructive qubit measurement to enable QEC/FTQC.

Scalability can be improved mostly with increasing the lifetime of atom interactions as described in the above chart in Figure 403. It also requires the continuous improvement of gate fidelities, one goal being to reach three nines (99,9% for two-qubit gates)[1140]. One approach, also tested with trapped ions, consists in mixing two types of neutral atoms. It is evaluated in a configuration prototyped at the **University of Chicago**. Shown in Figure 404, it packs 512 atoms in an array using an equivalent proportion of cesium and rubidium atoms in alternating patterns. Since these atoms require a different laser wavelength for gate drive. The benefit is a reduction of qubit crosstalk[1141].

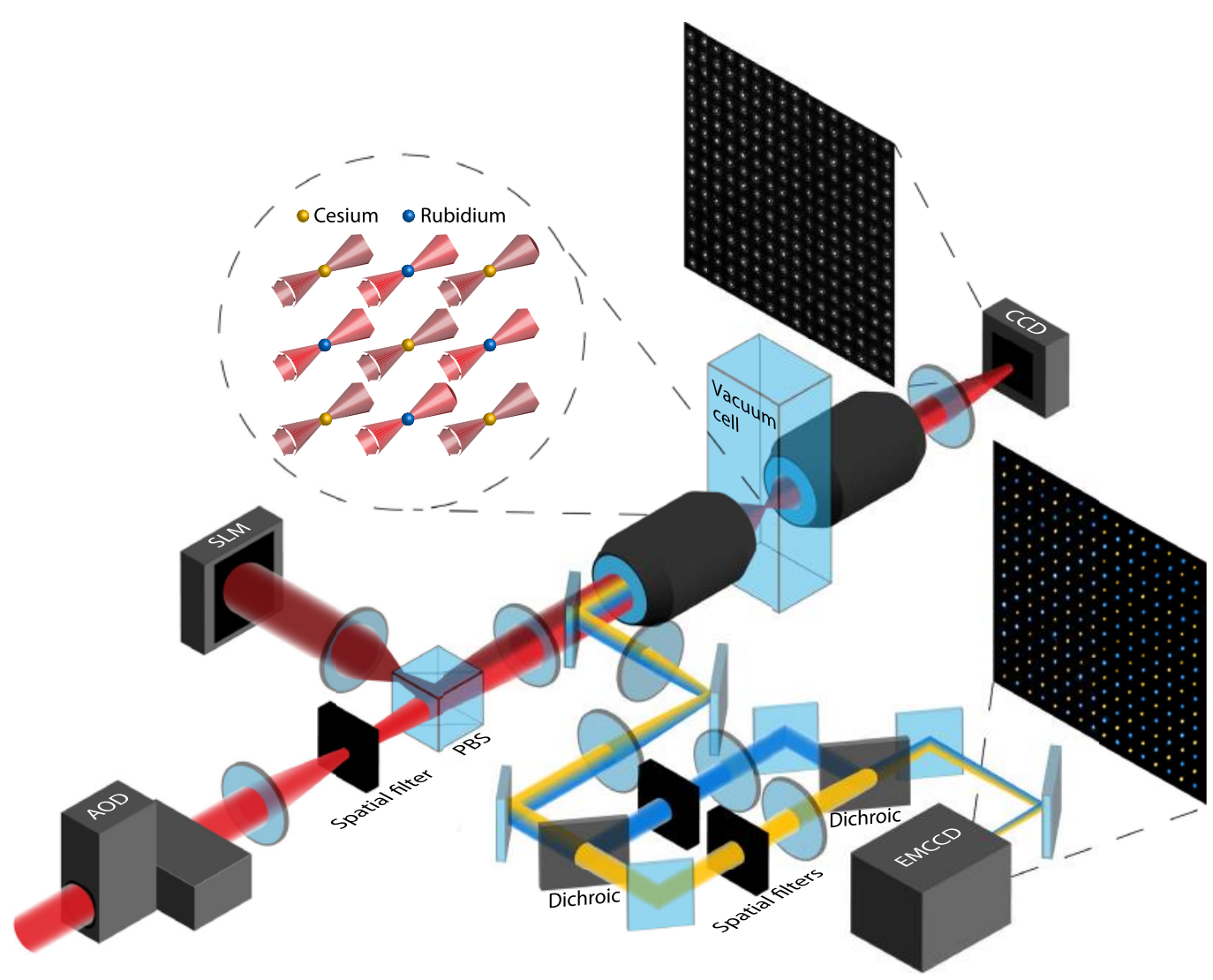

*Figure 404: mixing two types of atoms, cesium and rubidium. Source: Dual-Element, Two-Dimensional Atom Array with Continuous-Mode Operation by Kevin Singh et al, University of Chicago, February 2022 (11 pages).*

1139 See Two-qubit controlled-PHASE Rydberg blockade gate protocol for neutral atoms via off-resonant modulated driving within a single pulse by Yuan Sun et al, October 2019 (16 pages) which deals with the improvement of CPHASE two-qubit gates on gg-qubits, with gate time at 1 µs.

1140 See Two-qubit gate in neutral atoms using transitionless quantum driving by Archismita Dalal and Barry C. Sanders, University of Calgary, June 2022 (22 pages). These researchers could improve CZ two-qubit gate with cesium with fidelities of 99,85%.

1141 See Dual-Element, Two-Dimensional Atom Array with Continuous-Mode Operation by Kevin Singh et al, University of Chicago, February 2022 (11 pages).

## Vendors

Let's cover them by order of creation.

**ColdQuanta** (2007, USA, $71.6M) is a company created by Dana Anderson, now his CTO, which develops laser-based solutions for cooling cold atoms and also designs cold atoms-based computers.

It is located in Boulder, Colorado, not far from the NIST Quantum Laboratory. Mark Saffman from the University of Wisconsin is their Chief Scientist for Quantum Information while Fred Chong from the University of Chicago is their Chief Scientist for Quantum Software (he also works for QCI).

Their initial core technology was the Quantum Core (*left* in Figure 405), a light guide that converges laser beams to control cold atoms that are usually cooled to less than 50 μK. It is integrated in QuCAL, a complete Bose-Einstein condensate generator, and in the Physics Station, a complete optical device for the control of cold atoms that can be used for various purposes. Atom chips are chips that can be integrated in these systems including miniaturized cold atom control optics. The startup uses these generic technologies to create a wide variety of systems, and above all for quantum sensing, especially for geopositioning instead of GPS, microgravimetry or cesium quantum clocks. They also offer ultra-high vacuum pumps for the control of cold atoms, called RuBECi as well as a cryogenic trapped ions package and magneto-optical traps, magnetic coils and cold atoms sources.

Their approach to the market is truly diverse. They have equipped the ISS space station with measuring instruments for NASA and JPL.

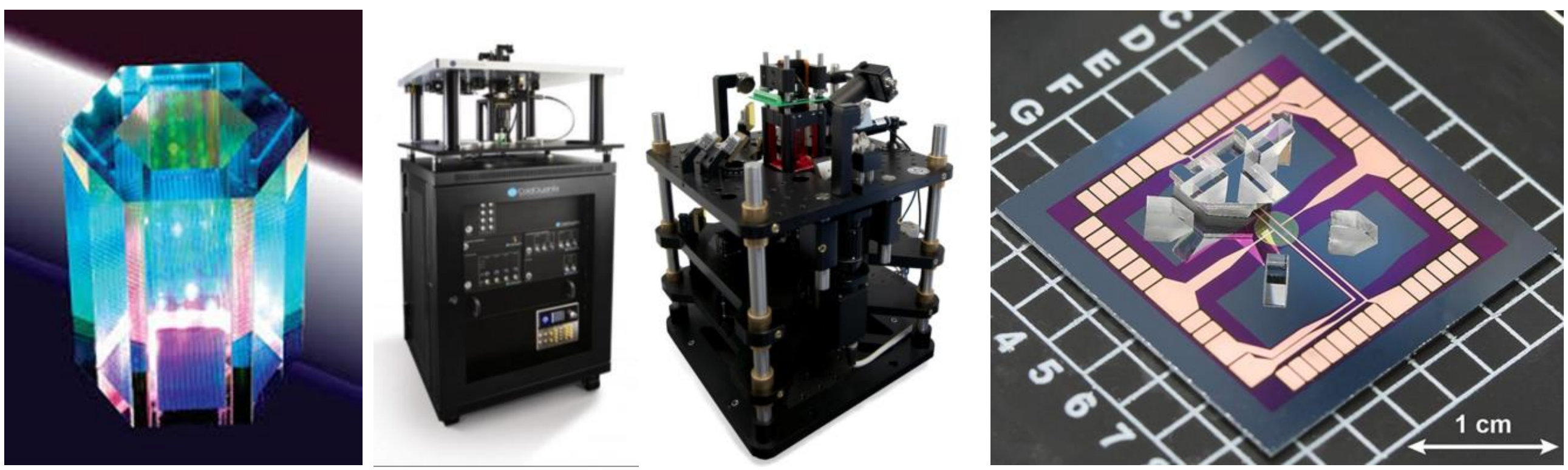

*Figure 405: ColdQuanta Quantum Core (left), Physics Station (middle) and the atoms control chipset (right). Source: ColdQuanta.*

Their overarching goal is to create and sell cold atoms-based quantum computers[1142]. They obtained some funding from DARPA in April 2020 under the ONISQ program with a $7.4M collaborative project involving numerous universities and Raytheon BBN. The DARPA asked them to develop a scalable (>1000 qubits) system that can demonstrate quantum advantage on real-world problems. ColdQuanta is focused on creating gate-based systems with 100 qubits[1143]. Their first 2021 100-qubit "Hilbert" cloud-based quantum system was supposed to work in July 2021 but was said to be commercially available only by May 2022. It is using cesium atoms.

---

1142 See ColdQuanta - Life in Quantum's Slow (and Cold) Lane Heats Up by John Russell, April 2020 and the webinar Powering the Quantum Information Age with Bo Ewald, April 2020 (53 minutes).

1143 See Demonstration of multi-qubit entanglement and algorithms on a programmable neutral atom quantum computer by T. M. Graham, M. Saffman et al, ColdQuanta, February 2022 (25 pages) and published as Multi-qubit entanglement and algorithms on a neutral-atom quantum computer in Nature in April 2022. It describes their state of the art: a preparation of entangled Greenberger-Horne-Zeilinger (GHZ) states with up to 6 qubits, implementation of quantum phase estimation for a chemistry problem, and Quantum Approximate Optimization Algorithm (QAOA) for the MaxCut graph problem. Their gates set is made of local $R_Z$ phase gates and two-qubit CZ gates, run on a 7x7 grid spaced by 3 μm. The system is using cesium atoms. See the related See ColdQuanta, Riverlane and University of Wisconsin–Madison Demonstrate Algorithms on a Programmable Neutral Atom Quantum Computer, April 2022.

During the summer of 2022, they released some interesting numbers on their gate-based system with fidelities of 99,4% for single qubit gates (lasting between 0.2 and 5 µs) and 96,5% for two qubit dates (CZ, lasting 0.75 µs), a measurement time of 1.5 ms, a $T_2$ of 1 second, all of this with 121 qubits.

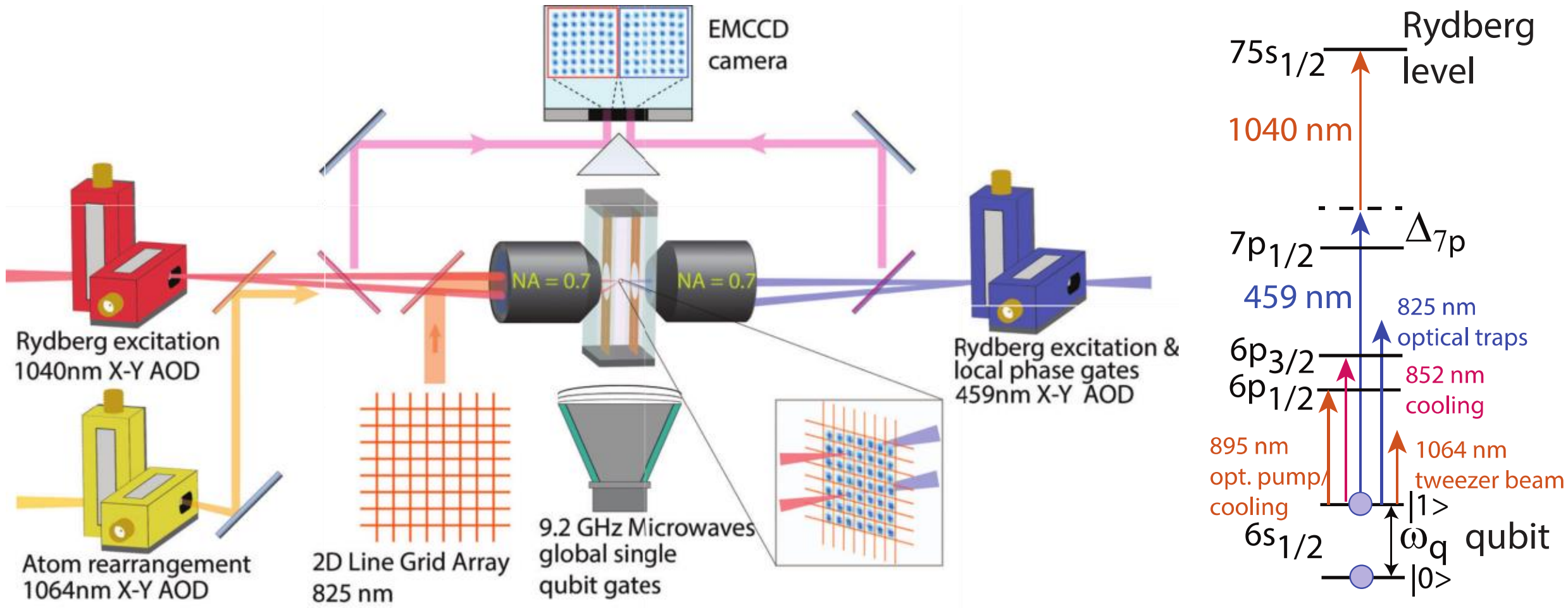

*Figure 406: ColdQuanta's gate-based system architecture. Source: Demonstration of multi-qubit entanglement and algorithms on a programmable neutral atom quantum computer by T. M. Graham, M. Saffman et al, ColdQuanta, February 2022 (25 pages).*

In May 2021, ColdQuanta joined a couple other quantum computers vendors like Pasqal with supporting IBM's Qiskit, formally joining the "IBM Quantum Network". The startup had about 170 people onboard as of mid-2022.

In May 2022, ColdQuanta acquired **Super.tech**, a software company providing pulse-level optimization, optimized transpilation and error mitigation techniques (SuperstaQ). It also provides benchmarking tools (SupermarQ). Fred Chong was their Chief Scientist from January 2021 until their acquisition in 2022. They are also teaming up with **Classiq** to handle code compilation[1144] and with **ParityQC** (Austria) for the development of quantum software targeting optimization problems.

**Atom Computing** (2018, USA, $81M) aims to create a quantum computer based on optically controlled neutral atoms with qubit states using atoms nuclear spins initially with cesium and later with strontium 87 [1145].

The company was created by Ben Bloom (CTO, coming from Rigetti) and Jonathan King (Chief Scientist, directly coming from Berkeley), joined in 2021 by Bob Hays (CEO). The company based in Berkeley, in California, established an R&D facility in Boulder, Colorado in 2022.

They demonstrated in October 2021 their strontium-based 100-qubit Phoenix system with a 40 second qubits coherence time. They control their qubits with individual microwave drives. The qubit manifold states are in the atom electronic ground state, avoiding spontaneous decay and with a coherence time of 42 seconds. They can run simultaneous gates on individual atoms using two photon Raman transitions[1146]. As pictured below, it was demonstrated on 21-qubit arrays of 3x7 qubits.

[1144] See Classiq and ColdQuanta partner to provide a complete solution to creating and executing 100-qubit quantum circuits and beyond, January 2022.

[1145] See Quantum computing with neutral atoms by David Weiss 2017 (7 pages) and Assembly and coherent control of a register of nuclear spin qubits by Katrina Barnes et al, August 2021 (10 pages).

[1146] See Assembly and coherent control of a register of nuclear spin qubits by Katrina Barnes et al, Atom Computing, August 2021 (11 pages).

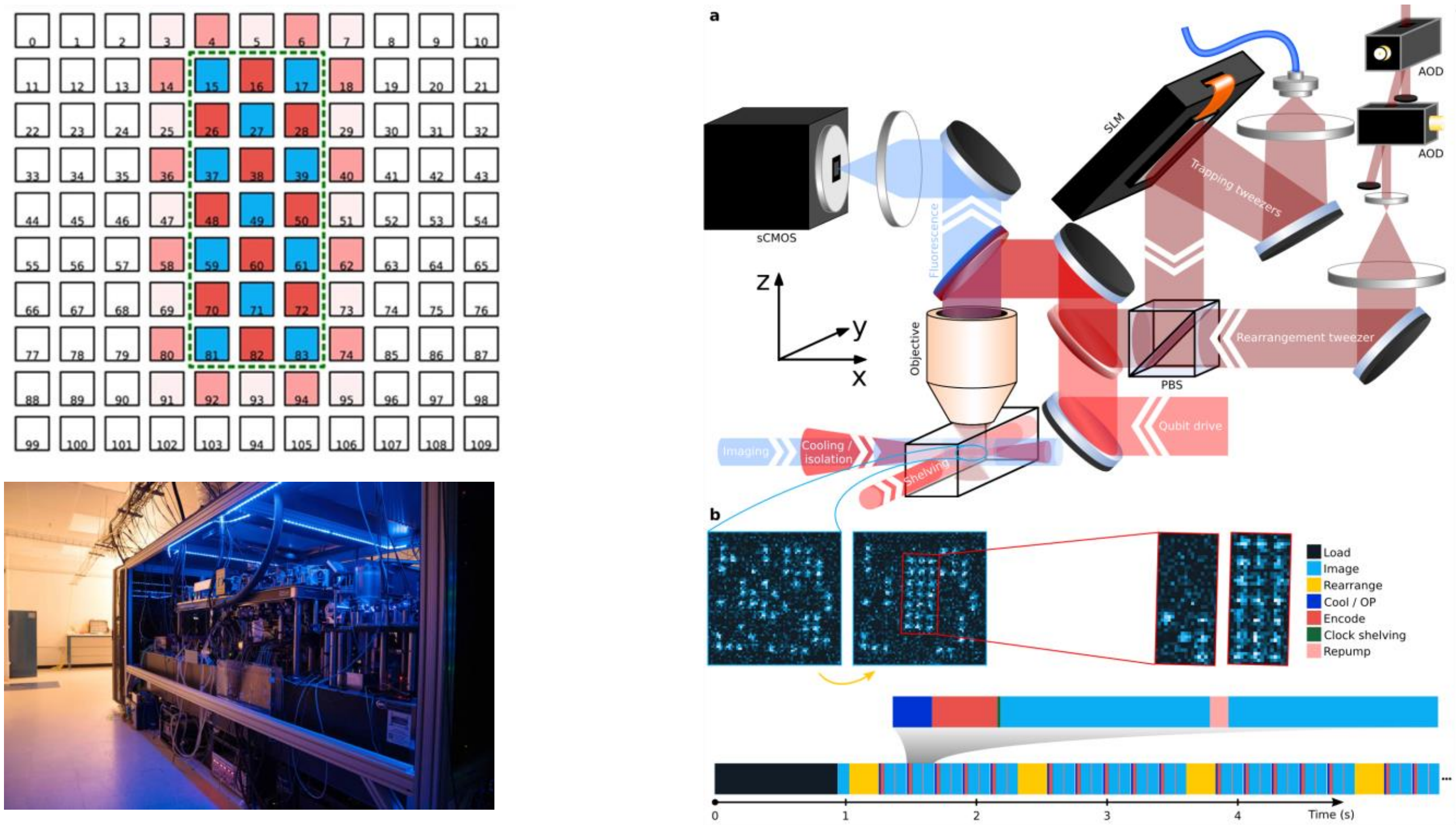

*Figure 407: Atom Computing architecture for over 100-qubit gate-based computing. Source: [Assembly and coherent control of a register of nuclear spin qubits](#) by Katrina Barnes et al, August 2021 (10 pages).*

They use the same apparatus as other vendors with SLMs for atoms trapping and rearrangement tweezers drive, AOD (acousto-optics deflector), EOM (electronic optical modulator) and a CMOS image sensor for qubits fluorescence readout. They plan to implement some form of QEC but don't provide many details on how they will handle non-demolition measurement (QND).

**Pasqal** (2019, France, 40M€) use magnetically confined rubidium atoms cooled by Doppler laser to reach mK and with a variant of the atomic Sisyphus effect to go down to 30 µK[1147]. The atoms are trapped in 2D arrays with a spacing of a few microns between each of them.

Qubit states are managed with various modes depending on the use case: two levels of Rydberg-level energy (for XY quantum simulation model), ground-Rydberg levels (for Ising quantum simulation models[1148]) or with ground-ground states (for gate-based model, in future versions). Quantum gates are laser-activated to control the atom energy state. Qubits entanglement comes from atoms excitement in the Rydberg state which allows them to interact with other at long distance[1149].

[1147] This method also uses lasers emitting orthogonally polarized photons. The method was invented by Claude Cohen-Tannoudji who was awarded the Nobel Prize in Physics in 1997.

[1148] See [Efficient protocol for solving combinatorial graph problems on neutral-atom quantum processors](#) by Wesley da Silva Coelho, Mauro D'Arcangelo and Louis-Paul Henry, Pasqal, July 2022 (16 pages) that describes how they use a variational analog quantum computing and machine learning to solve graph problems. In that case, the atoms are arranged in a graph with specific distances between atoms that match the problem to be solved. It is not a simple regular 2D array.

[1149] See [Quantum Computing with Arrays of Atoms](#) by Lucas Béguin and Adrien Signoles from Pasqal, April 2020, which details the functioning of the startup's quantum processors. And their white paper Quantum [Computing with Neutral Atoms](#), June 2020 (41 pages).

Pasqal plans first to implement quantum simulators *aka* PQS (Programmable Quantum Simulator, or analog quantum computers) and then, to move to NISQ gate-based quantum computing[1150], with some hybrid analog/digital approach [1151]. The technology currently works with 100 qubits in simulation mode with plans to reach a thousand qubits by 2023 [1152].

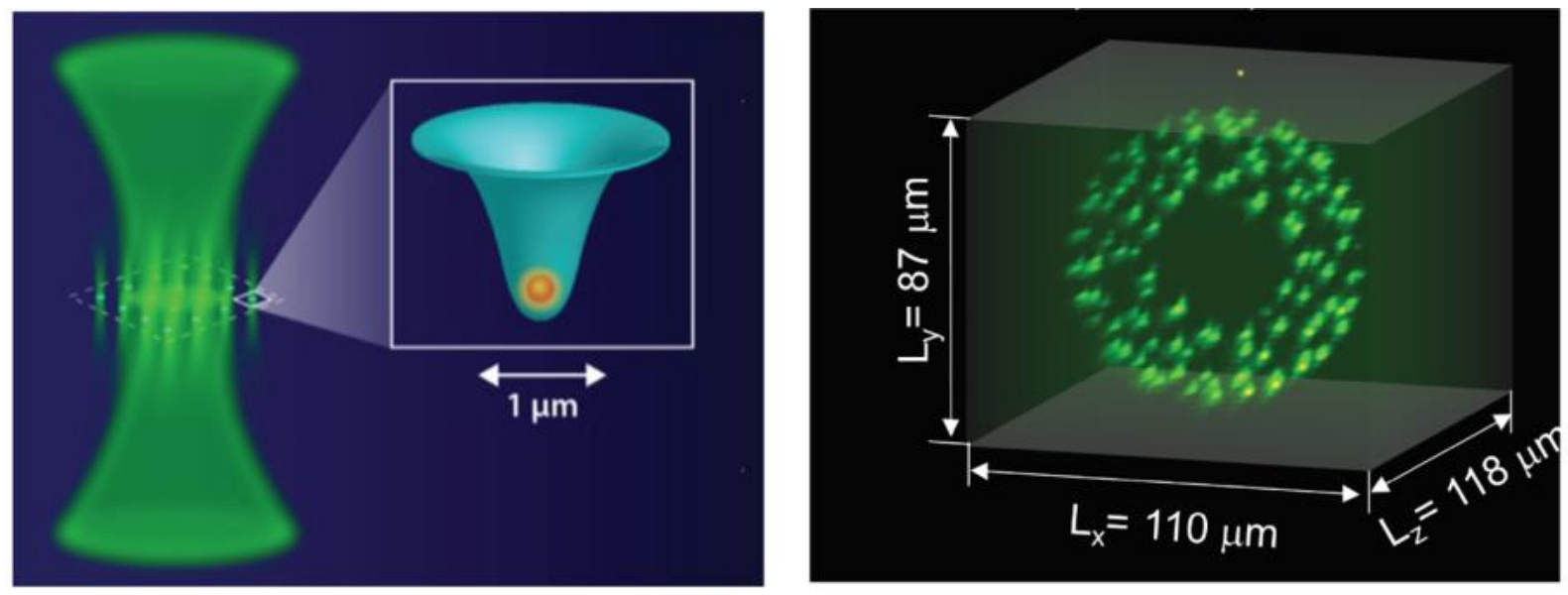

*Single atoms are trapped in an energy potential pattern created by lasers*

*Each green dot is a Rubidium atom arranged within a torus*

*Figure 408: how atoms can be arranged, even in 3D. Source: Pasqal.*

The computer will eventually fit into a 4-unit wide double-depth data center rack. It is based on rather standard components and does not require the creation of specific chipsets as it is the case for all other types of qubits.

In April 2020, startups **Pasqal** and **Muquans** announced a partnership that had been in preparation for a long time and on the use of a Muquans lasers system to control the cold atoms.

Milestone scientific papers and achievements happened between late-2020 and mid-2022, led by Pasqal and by the research team at IOGS behind Pasqal's cold atom-based system. The first one from December 2020 describes how a 2D matrix of 196 qubits was assembled using optical tweezers created with an SLM (a high-resolution Spatial Light Modulator) and an optimized arrangement technique using fewer than 200 steps[1153]. These SLMs enable phase grading and controlling phases spatial pattern. This atoms-positioning system will be extended to 3D arrays. It went on in 2022 with parallel gates execution[1154]. In July 2022, they announced having trapped up to 324 cold atoms in a 2D array, with using a 4K cryostat cooling the ultra-vacuum pump, which helps create a better vacuum and extend the atom qubit lifetime [1155].

[1150] See Why analog neutral atoms quantum computing is the most promising direction for early quantum advantage by Jean-Charles Calbelguen, June 2022.

[1151] See Microwave Engineering of Programmable XXZ Hamiltonians in Arrays of Rydberg Atoms by P. Scholl, Loic Henriet, Thierry Lahaye, Antoine Browaeys et al, PRX, April 2022 (10 pages) which presents an hybrid analog-digital architecture based on Hamiltonian evolutions and one-qubit gates. Qubits encoding will be different according to the use case. For gate-based computing, qubits are encoded with two hyperfine ground states. For Ising-like Hamiltonian, qubits use a ground state and a Rydberg state and for XY exchange Hamiltonian, they use two Rydberg states.

[1152] Rydberg atoms have unsuspected uses, such as managing random music. See Quantum music to my ears, June 2019. This is a change from music generated by deep learning!

[1153] See Enhanced atom-by-atom assembly of arbitrary tweezers arrays by Kai-Niklas Schymik, Antoine Browaeys, Thierry Lahaye et al, November 2020 (10 pages).

[1154] See Pulse-level Scheduling of Quantum Circuits for Neutral-Atom Devices by Richard Bing-Shiun Tsai, Loic Henriet et al, Pasqal, June 2022 (8 pages) and Pulser: An open source package for the design of pulse sequences in programmable neutral-atom arrays by Henrique Silvério, Nathan Shammah, Louis-Paul Henry, Loïc Henriet et al, January 2022 (21 pages).

[1155] See In-situ equalization of single-atom loading in large-scale optical tweezers arrays by Kai-Niklas Schymik, Antoine Browaeys, Thierry Lahaye et al, PRA, July 2022 (5 pages).

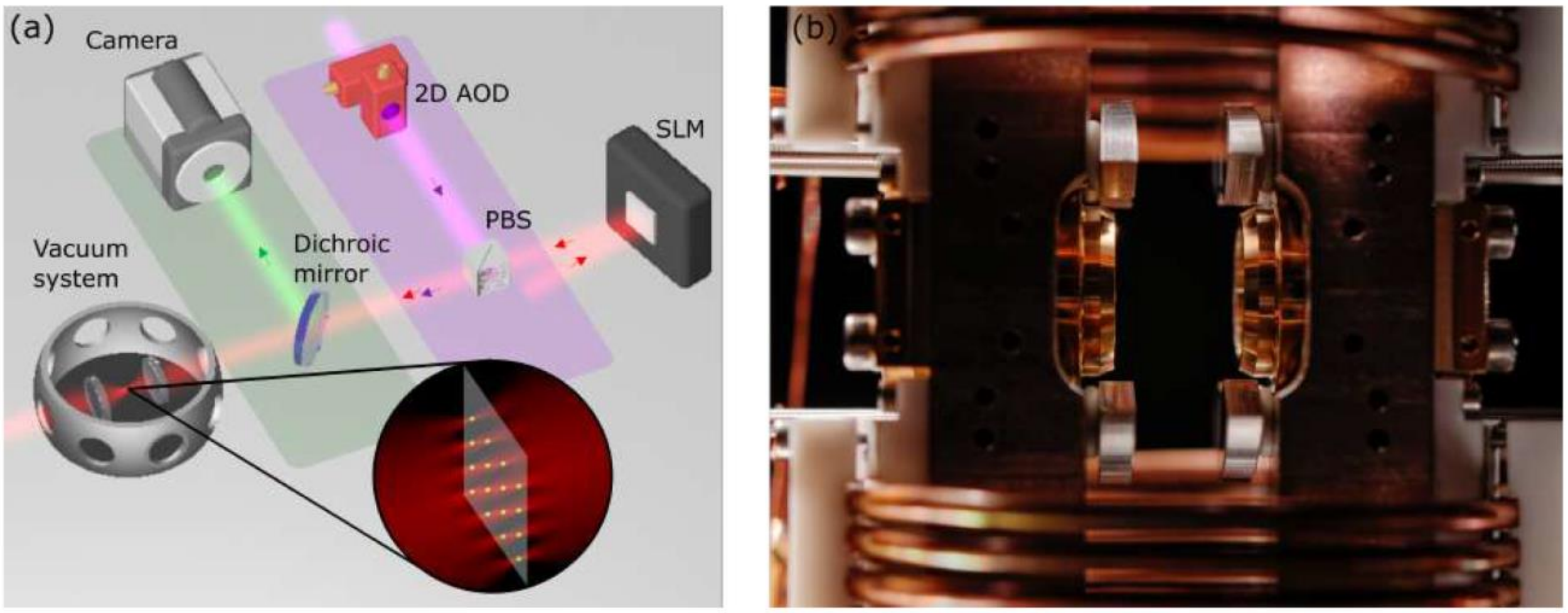

*Figure 409: Pasqal's cold atom-based qubit control system includes a spatial light modulator (Spatial Light Modulator, SLM, based on LCoS liquid crystals[1156]) that controls the phase of the transmitted light in a focal plane with optical micro-traps. Laser tweezers or traps/pinches for rearranging the atoms and preparing the Hamiltonian to solve are controlled by the AOD (Acousto-Optic laser beam Deflector) and added to the beam from the SLM by a PBS (Polarizing Separator Filter). The fluorescent light emitted by the atoms during qubit readout is filtered by a dichroic mirror and analyzed by a CCD camera. The controlled atoms are confined in a small space of 1 mm³.*

Applications wise, they published papers on quantum machine learning[1157], on a financial risks assessment case study with CA-CIB in France[1158] and for solving graph classification problems using the "quantum evolution kernel" method (QEK) with a first use case in 2021 in toxicity screening of chemical compounds and in the identification of optimal chemical reaction pathways. A 196 cold atoms setup was then used to program an Ising model simulating ferromagnetism and enabling quantum simulation[1159].

Then came the optimization of smart charging of electrical vehicles co-developed with EDF, using a QAOA hybrid algorithm and classical emulators of Pasqal systems and the QLM emulator from Atos[1160].

It was also funded through the PASQuanS flagship project. Atos researchers also evaluated the specifications of a cold atom simulator needed to reach some quantum advantage to solve an optimization task, the UD-MIS problem (Unit-Disk Maximum Independent Set problem)[1161].

---

[1156] See a description of an SLM in Spatial Light Modulators by Aurélie Jullien, 2020 (6 pages).

[1157] See Quantum evolution kernel: Machine learning on graphs with programmable arrays of qubits by Louis-Paul Henry, Slimane Thabet, Constantin Dalyac and Loïc Henriet, PRA, September 2021 (19 pages) that is better explained in Machine Learning – Pasqal's Quantum Computers can be used on concrete industrial problems, Pasqal, October 2021.

[1158] See Towards quantum advantage with efficient graph implementations, Pasqal, April 2022.

[1159] See Programmable quantum simulation of 2D antiferromagnets with hundreds of Rydberg atoms by Pascal Scholl, Thierry Lahaye, Antoine Browaeys et al, December 2020 (16 pages). Also published in Nature in July 2021. This was a result of a research funded through the European Flagship PASQuanS project in partnership with labs from Spain, Germany and Austria

[1160] See Qualifying quantum approaches for hard industrial optimization problems. A case study in the field of smart-charging of electric vehicles by Constantin Dalyac, Loïc Henriet, Emmanuel Jeandel, Wolfgang Lechner, Simon Perdrix, Marc Porcheron and Margarita Veshchezerova, 2021 (29 pages).

[1161] A MIS problem consists in determining the size of the largest possible independent set in a graph and returning an example of such a set. The Unit-Disk MIS (UD-MIS) problem is the MIS problem restricted to unit-disk graphs. A graph is a unit-disk graph if one can associate a position in the 2D plane to every vertex such that two vertices share an edge if and only if their distance is smaller than unity.

They found out that over 1,000 qubits were required with a time budget of 0.2 seconds, if and when the system coherence could be improved by a factor ten[1162].

Pasqal has created its own low-level programming environment that interfaces with high-level programming tools, including support for development platforms such as Google's **Cirq** that is supported in emulation mode, in digital gate-based programming mode[1163], **TensorFlow Quantum** and IBM's **Qiskit.**

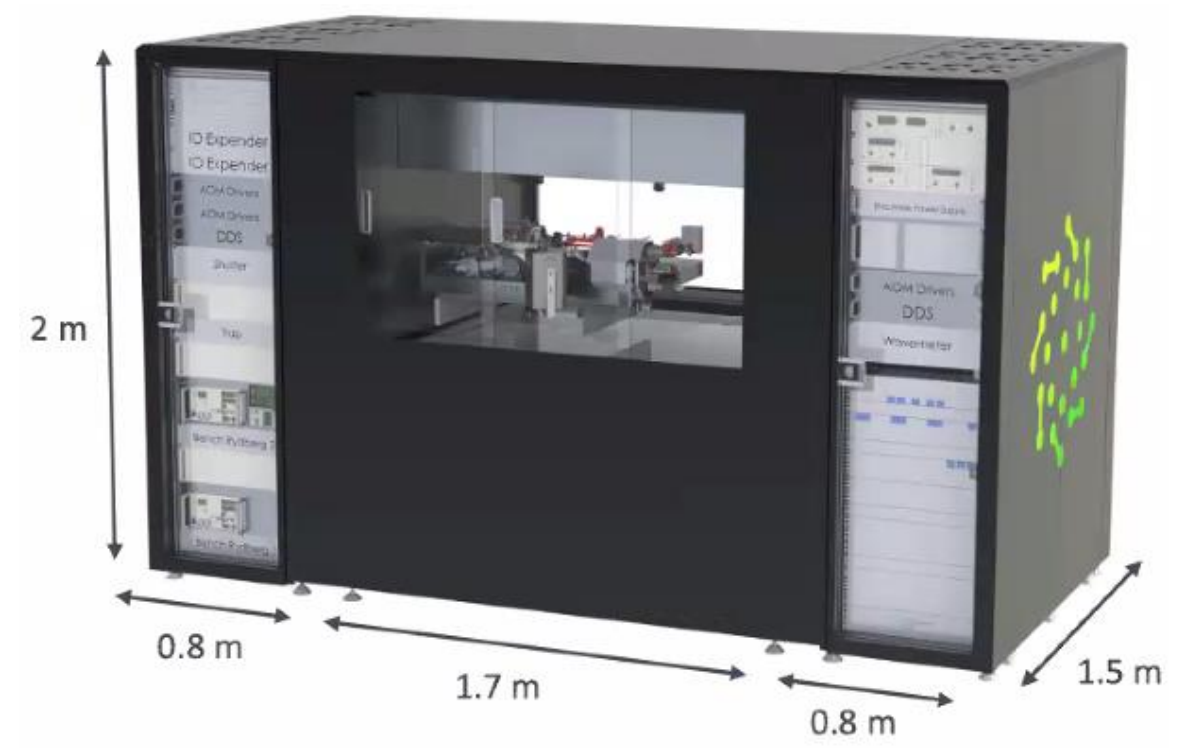

*Figure 410: Pasqal Fresnel packaging.*

In July 2020, **Cambridge Quantum Computing** (CQC) announced their support of Pasqal's qubits with their development tool t|ket⟩. Pasqal added a partnership with **ParityQC** with a 3-year collaboration to advance quantum optimization and parallelization[1164] and released the open source library **Pulser**, co-developed with the **Unitary Fund** enabling the control of their processor at the level of laser pulses[1165]. They also announced a partnership with **Atos** in November 2020 to integrate a Pasqal accelerator with Atos supercomputers. At last, they also work with **Rahko** as well as with **Multiverse Computing**, in association with Crédit Agricole CIB.

In January 2021, the Italian HPC consortium **CINECA** announced that will use Pasqal's Fresnel 100-qubits processor[1166]. The startup was also selected as part of the project HPC-QS from EuroHPC to provide two of their systems to HPC public datacenters, one in Germany (FZ Julich *aka* Jülich Research Centre) and one in France (CEA TGCC). Both Pasqal systems will be connected to an Atos QLM for emulation and quantum system drive.

Pasqal funding came through an initial round of 2.5M€ led by Quantonation and Christophe Jurczak who is their chairman, then a 4,5M€ grant from the European Union EIC Accelerator[1167], and finally a second round of funding of 25M€ announced in April 2021.

In January 2022, Pasqal and Qu&Co (The Netherlands) merged, creating an integrated hardware/software company.

2022 was a busy year with new partnership announced with **EDF**, **Thales**, **GENCI**, **Aramco** (for implementing QML algorithms in the energy business), **Siemens** (for computer-aided-design and engineering, simulation and testing), **BASF** (fluid dynamic problem solving algorithm), **BMW** (material science and material deformation algorithms) and the first sales of two 100-qubit quantum processor to **GENCI** and **FZJ** as part of the HPCQS EU consortium. They opened an office in Boston (USA) and in Sherbrooke (Canada) in June 2022. And their first 100-qubit quantum simulator web online with **OVHcloud** in May 2022 (in private beta) and will be online later with **Microsoft Azure Quantum**.

---

[1162] See Solving optimization problems with Rydberg analog quantum computers: Realistic requirements for quantum advantage using noisy simulation and classical benchmarks by Michel Fabrice Serret, Bertrand Marchand and Thomas Ayral, November 2020 (25 pages).

[1163] See Quantum circuits on Pasqal devices.

[1164] See ParityQC and Pasqal partner to build the first fully parallelizable quantum computer, Pasqal, October 2020. With other researchers in Austria, ParityQC is proposing a way to encode QAOA generic optimization problems into gate-based cold atoms computers using a specific 4-qubit gate as seen in Quantum computing on neutral atoms: the novel four-body Rydberg gate, ParityQC, 2022, referring to: Quantum Optimization via Four-Body Rydberg Gates by Clemens Dlaska, Kilian Ender, Glen Bigan Mbeng, Andreas Kruckenhauser, Wolfgang Lechner, and Rick van Bijnen, PRL, March 2022 (11 pages).

[1165] See Pulser: a control software at the pulse-level for Pasqal quantum processors by Pasqal, January 2021.

[1166] See CINECA-Pasqal agreement on quantum computing, January 2021.

[1167] See Europe is betting on quantum computing with neutral atoms, Pasqal, December 2020.

**QuEra Computing** (USA, 2018, $17M) develops a cold atom gate-based quantum computer. The startup was created by researchers from Harvard University and MIT.

With Nathan Gemelke, Alexei Bylinskii, Shengtao Wang and Mikhail D. Lukin, among others, who is one of their scientific advisors[1168]. They published a research paper on a 2D array 256 programmable cold atom system in 2021 and announced in November 2021 it would become a commercial product[1169]. What they brand "programmable quantum computers" are not gate-based systems and are still quantum simulators where they encode graph problems into a layout of cold atoms[1170]. A graph problem is expressed as a "maximum weight independent set" (MWIS) problem that is then turned into several unit-disk graph sets (UDG) which are assemblies of nearby neutral atoms using the Rydberg effect for local entanglement. These sets are connected to each other with entangled neutral atoms at their edge. Various problems like QUBO, Ising models and even integer factorization can be mapped onto MWIS problems.

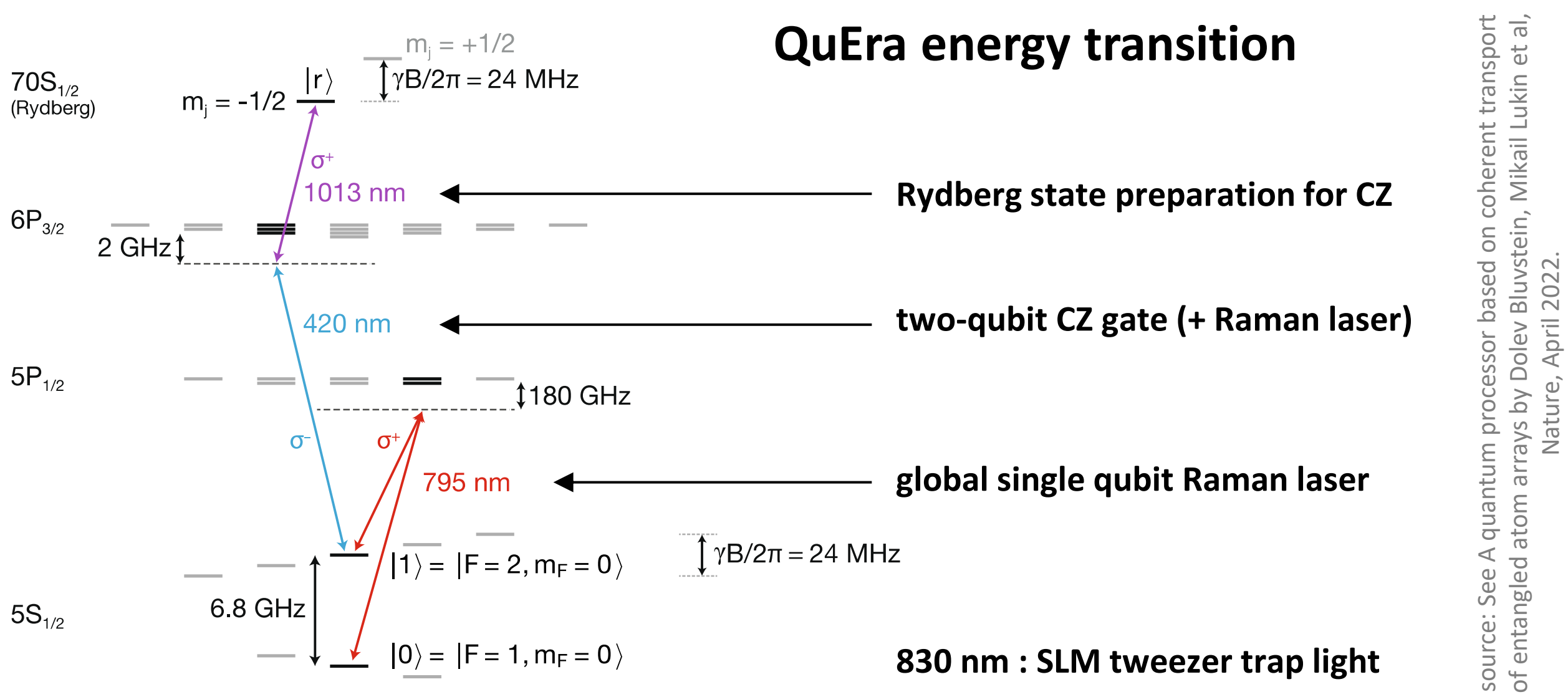

*Figure 411: QuEra atomic energy transitions used to control qubits and qubit gates. Source: A quantum processor based on coherent transport of entangled atom arrays by Dolev Bluvstein, Mikhail D. Lukin et al, Nature, April 2022 (21 pages).*

They proposed in 2022 a way to transport qubits to handle two-qubit gates using atoms shuttling with tweezers. They implemented it with a Steane-7 quantum error correction code graph (but not with a full QEC), topological surface code and toric code states using mobile ancilla and with hybrid analog-digital quantum circuits using dynamically reconfigurable array and CZ gates[1171].

They are also working on making non-demolition qubit measurement, a key feature to implement full QEC (quantum error correction) and FTQC (fault-tolerant quantum computing). This would be based on moving selected atoms "*into a readout zone where their qubit state can be rapidly detected via fast, resonant photon scattering on a cycling transition*".

---

[1168] See in particular Parallel Implementation of High-Fidelity Multiqubit Gates with Neutral Atoms by Harry Levine, Mikail D. Lukin et al, August 2019 (16 pages).

[1169] See Quantum Phases of Matter on a 256-Atom Programmable Quantum Simulator by Sepehr Ebadi, Mikhail D. Lukin et al, 2020 (20 pages).

[1170] See This new startup has built a record-breaking 256-qubit quantum computer by Siobhan Roberts, MIT Technology Review, November 2021 and the real stuff in Quantum Optimization of Maximum Independent Set using Rydberg Atom Arrays by Sepehr Ebadi, Mikhail Lukin et al, February 2022 (10 pages) and Quantum optimization with arbitrary connectivity using Rydberg atom arrays by Minh-Thi Nguyen, Mikhail D. Lukin et al, September 2022 (19 pages).

[1171] See A quantum processor based on coherent transport of entangled atom arrays by Dolev Bluvstein, Mikhail D. Lukin et al, Nature, April 2022 (21 pages).

They could also use arrays with two atoms species such as two isotopes of the same element or two different atom elements, with the data atoms being encoded in one atomic species and ancilla atoms encoded in another species that can be easily measured[1172].

In July 2022, QuEra launched Bloqade, their quantum emulation software package that is available on Github as a Julia package. It emulates on classical computers atom-based quantum simulations. This is akin to Pulser from Pasqal. They got help for this from Amazon and the Perimeter Institute.

**Planqc** (2022, Germany, 4.6M€) is a startup created in Garching near Munich by Alexander Glätzle (CEO, a researcher at the University of Oxford in the UK), Sebastian Blatt (CTO) who was a researcher at Ludwig-Maximilians-University Munich (and is a son of Rainer Blatt) and Johannes Zeiher (a researcher at MPQ).

Immanuel Bloch and Jose Ignacio Cirac from MPQ and Dieter Jaksch, Professor of Physics at the University of Oxford and the University of Hamburg are their scientific advisors. It's a neutral atom based qubits company that plans to scale its system to thousands of qubits. They are still semi-stealth and don't provide any details on their technology choices and roadmap. Looking at the research work from the founders, you may infer that they plan to implement MBQC on large cluster states of entangled cold atoms[1173].

# NMR qubits

### History and science

Nuclear Magnetic Resonance was an early technique investigated to create quantum computers. Qubit states are the spin of nuclei within large ensembles of up to $10^{15}$ molecules. The qubit states are readout using nuclear magnetic resonances, implementing a variant of nuclear magnetic resonance spectroscopy.

The NMR phenomenon was discovered in 1945 by **Edward Mills Purcell** (1912-1997, American) for which he was awarded the Nobel prize in physics in 1952, shared with Felix Bloch.

Using nuclear spins for quantum computing was first proposed in 1993 by **Seth Lloyd** from the MIT[1174] and refined in 1994 by **David DiVincenzo**, then at IBM Research, for the implementation of perfluorobutadienyl two-qubit gates[1175]. It was refined by various proposals coming from the MIT, UCSB and Stanford researchers in 1997 with a computing method using liquid state NMR (molecules in a liquid) and enabling the measurement of the expectation value of quantum observables, ensuring an exponential computing speedup[1176].

---

[1172] See Hardware-Efficient, Fault-Tolerant Quantum Computation with Rydberg Atoms by Iris Cong, Mikhail Lukin et al, May 2022 (31 pages) making references to A dual-element, two-dimensional atom array with continuous-mode operation by Kevin Singh et al, October 2021 (11 pages) on reduced cross-talk and QND with two mixed atom elements (cesium and rubidium), Fast Preparation and Detection of a Rydberg Qubit using Atomic Ensembles by Wenchao Xu et al, May 2021 (11 pages) and Interaction enhanced imaging of individual atoms embedded in dense atomic gases by G. Günter, Shannon Whitlock et al, June 2011 (6 pages) which also describes a QND measurement of cold atoms state.

[1173] See Realizing distance-selective interactions in a Rydberg-dressed atom array by Simon Hollerith, Immanuel Bloch, Johannes Zeiher et al, October 2021 (5 pages), See Quantum Information Processing in Optical Lattices and Magnetic Microtraps by Philipp Treutlein, Immanuel Bloch et al, Max-Planck Institute, June 2006 (15 pages). This is a variation of cold atoms qubits adapted to cluster states and MBQC. See also Quantum simulations with ultracold atoms in optical lattices by Christian Gross and Immanuel Bloch, 2017 (8 pages).

[1174] See A potentially realizable quantum computer by Seth Lloyd, Science, 1993 (4 pages).

[1175] See Two-Bit Gates are Universal for Quantum Computation by David DiVincenzo, July 1994 (21 pages) and Bulk Spin-Resonance Quantum Computation by Neil A. Gershenfeld and Isaac L. Chuang, 1997 (7 pages).

[1176] See Ensemble quantum computing by NMR spectroscopy by David G. Cory, Amr F. Fahmy and Timothy F. Havel, MIT, 1997 (6 pages).

An expectation value of an observable is the average value of the observable and not a random value obtained by a projective measurement, the type of qubit readout measurement usually implemented with other techniques. Using another explanation, this measures a statistical mixture of pure states.

In 2001, **IBM** researchers even implemented Shor's algorithm with the number 15 on a 7-qubit NMR quantum processor using a perfluorobutadienyl iron complex where the spins come from the nuclear spins of fluor and carbon atoms (in red)[1177]. It is still the record to date! In these systems, qubit states where the ensemble nuclear spins ½ in a magnetic field, quantum gates were operated by radiofrequency pulses and qubit readout was done with detecting spin states with a radiofrequency coil[1178].

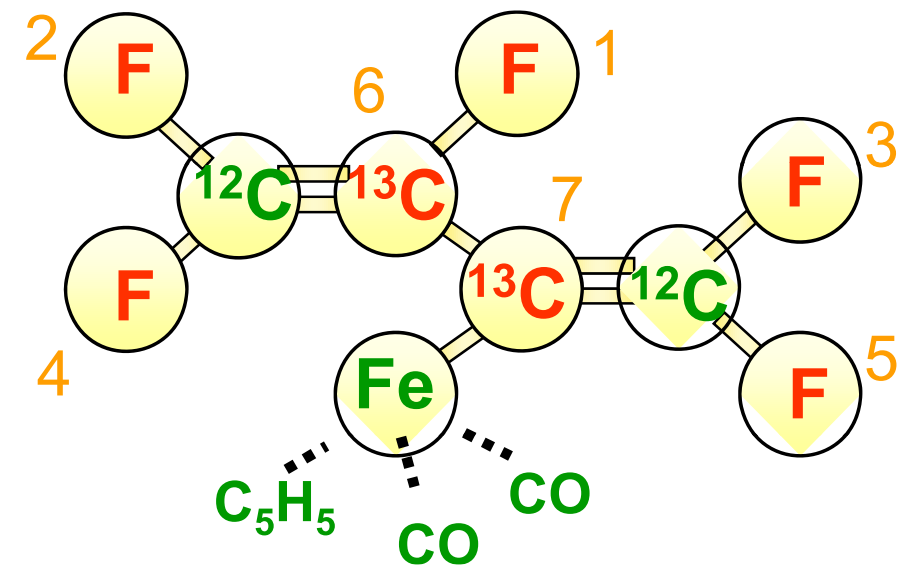

*Figure 412: NMR can rely on complex molecule like perfluorobutadienyl. Source: IBM.*

Afterwards, there were some developments with "Solid state NMR" (SSNMR) using nitrogen vacancies and interactions between carbon atomic spins and vacancies electron spins or nuclear spins of $^{29}Si$ in silicon structures[1179].

NMR was quite fashionable in quantum computing about 20 years ago. IBM was even planning for the existence of NMR-based tabletop quantum computers[1180]. 22 years ahead, their (superconducting qubits) quantum computers are 3-meter wide cubes consuming in excess of 25 kW.

Still, research is still ongoing on NMR quantum computing. It goes on in Europe in the Europe 2020 project **FATMOLS** run by a research consortium led by Spain[1181]. Its ambition is to create a molecular spin quantum processor using artificial magnetic molecules implementing spin qudits controlled, read-out and linked with some superconducting circuits.

Quantum features are implemented with nuclear spins, electronic spins and circuits. Programming models range from quantum simulations to gate-based FTQC. The FATMOLS project's goals is to create a proof-of-concept of one of the repetition unit cells of this platform with at least two molecules with multiple and fully addressable levels and related algorithms. The end-goal is to reach 100 to 1000 physical qubits[1182]. The project runs from March 2020 to August 2023 with a total cost of 3.2M€, entirely funded by the EU.

---

[1177] See Experimental realization of Shor's quantum factoring algorithm using nuclear magnetic resonance by Lieven M.K. Vandersypen, Isaac L. Chuang et al, December 2001 (18 pages).

[1178] See NMR Quantum Computing, 2012 (43 slides) and NMR Quantum Computation NMR Quantum Computation by Thaddeus Ladd, Stanford, 2003 (39 slides).

[1179] See Solid-State Silicon NMR Quantum Computer by E. Abe, K. M. Itoh, Y. Yamamoto et al, 2003 (5 pages).

[1180] See Toward a table-top quantum computer by Y. Maguire, E. Boyden and N. Gershenfeld, IBM, 2000 (17 pages).

[1181] With CSIC, Aragón Materials Science Institute (ICMA), The Centre of Astrobiology (CAB), University of Barcelona, Universidad de Valencia / Instituto de Ciencia Molecular (UVEG), the Keysight team in Barcelona, and outside Spain: University of Manchester Molecular Magnetism Group, University of Oxford Department of Condensed Matter Physics, University of Parma and the Department of Chemistry at the University of Florence, Universität Stuttgart / Institute for Functional Matter and Quantum Technologies, Wolfgang Pauli Institut, Technische Universität Wien and IBM Zurich (Ivano Tavernelli).

[1182] See A perspective on scaling up quantum computation with molecular spins by S. Carretta et al, Applied Physics Letters, May2021 (13 pages).

## Research

Various NMR variants are still investigated like the use of large molecular spins[1183], europium molecules with nuclear spins that can interact with luminescent photons carrying the nuclear spin information[1184], with molecular ensembles[1185] and optically addressable molecular spins[1186].

**DQC1** (Deterministic Quantum Computation with 1 quantum bit) is a curious quantum computing model created in1998 by Emanuel Knill and Raymond Laflamme. It was designed for NMR qubits which we were commonly developed in the late 1990s and were very noisy. It is described as a deterministic quantum computation using one qubit and a classical computer[1187].

It computes deterministic functions on one bit and a classical computer is performing probabilistic computing. But there's more than one qubit in the story: the input is indeed constrained to a single qubit in a pure state but a set of n other qubits is prepared in an identity state and subject to a random unitary transformation $U_n$ and in a completely mixed state. These systems use noise as a resource which is formalized with **quantum discord** mathematical tools[1188]. At the end of computing, a measurement takes place on the first qubit. It is not a universal quantum computing model, doesn't use massive entanglement and has a narrow set of use cases although it is supposed to bring some exponential speedups in some circumstances[1189].

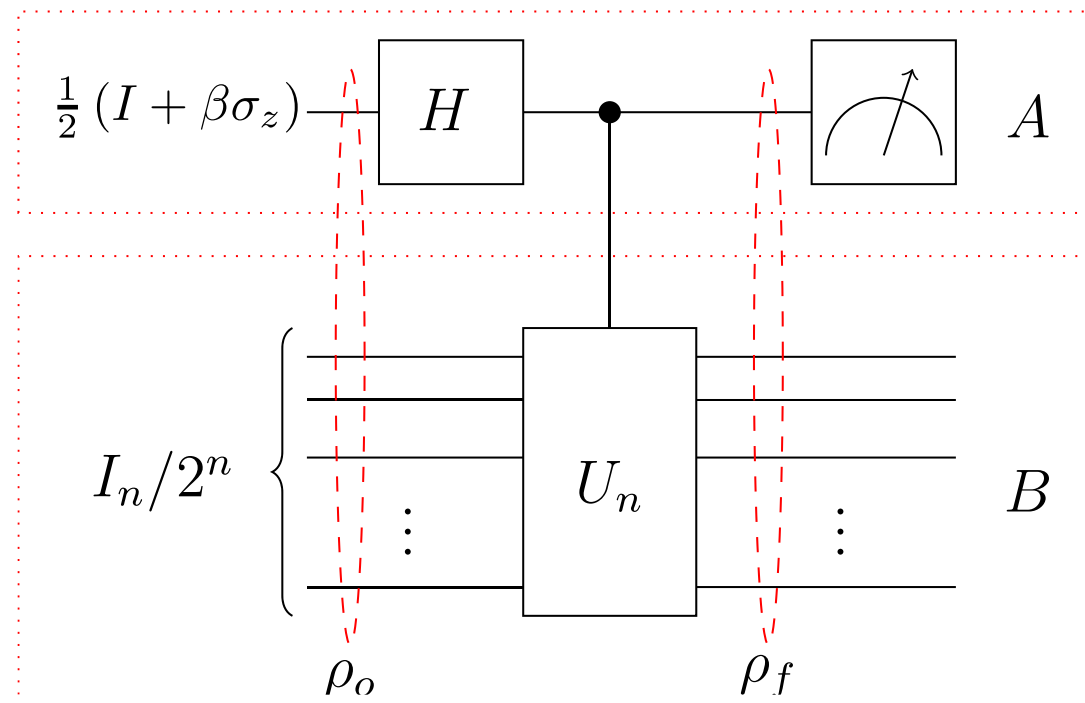

*Figure 413: description of the DQC1 model. A qubit at the top is the only input. It is prepared then subject to an Hadamard gate and the result controls the application of the $U_n$ unitary transformation to n other qubits. At the end of this processing, the first qubit is the only one measured, with the process being repeated several times. The output yields a trace of the unitary $U_n$. Source: Measurement-Based Quantum Correlations for Quantum Information Processing by Uman Khalid, Junaid ur Rehman and Hyundong Shin, Nature Research Scientific Reports, 2020 (9 pages).*

## Vendors

NMR quantum computers don't scale well due to noise affecting qubits and poor entanglement. It didn't prevent one company from China to manufacture and sell and NMR-based quantum computer. And fulfilling IBM's 2000 dream, it is a desktop product.

---

1183 See Optically addressable molecular spins for quantum information processing by S. L. Bayliss et al, April 2020 (9 pages) as well as Chemical tuning of spin clock transitions in molecular monomers based on nuclear spin-free Ni(ii) by Marcos Rubín-Osanz et al, 2021 (11 pages). It involves one lab in Spain and three in France (ICMM Orsay, LCPQ Toulouse and LNCMI Grenoble).

1184 See Ultra-narrow optical linewidths in rare-earth molecular crystals by Diana Serrano, Senthil Kumar Kuppusamy, Benoît Heinrich, Olaf Fuhr, David Hunger, Mario Ruben and Philippe Goldner, KIT, CNRS, University of Strasbourg and Chimie ParisTech, Nature, May 2021-March 2022 (19 pages).

1185 See A New Approach to Quantum Computing Using Magnetic Resonance Imaging by Zang-Hee Cho et al, June 2022 (9 pages).

1186 See Enhancing Spin Coherence in Optically Addressable Molecular Qubits through Host-Matrix Control by S. L. Bayliss, David Awschalom et al, April 2022 (13 pages).

1187 See On the Power of One Bit of Quantum Information by Emanuel Knill and Raymond. Laflamme, 1998 (5 pages).

1188 See Measuring geometric quantum discord using one bit of quantum information by G. Passante, O. Moussa and Raymond Laflamme, 2021 (5 pages), Introducing Quantum Discord by Harold Ollivier, October 2001 (5 pages) and Quantum Discord: A Measure of the Quantumness of Correlations by Harold Ollivier and Wojciech H. Zurek, PRL, December 2001 (5 pages).

1189 See Power of Quantum Computation with Few Clean Qubits by Keisuke Fujii et al, 2015 (45 pages).

**SpinQ Technology** (2018, China) started with announcing in January 2021 their Gemini $5K 2-qubit desktop quantum computer and their cloud quantum computing platform Taurus[1190]. It followed an initial version launched in 2020 and sold at $55K The computer weights 55 kg and works at ambient temperature. They plan to increase the number of qubits of this device in upcoming versions, up to a maximum 15 qubits. It would be nice since 2 qubits are totally useless even for quantum programming learning tasks. Meanwhile, you can test for free 7 real superconducting qubits on IBM Quantum Experience cloud systems.

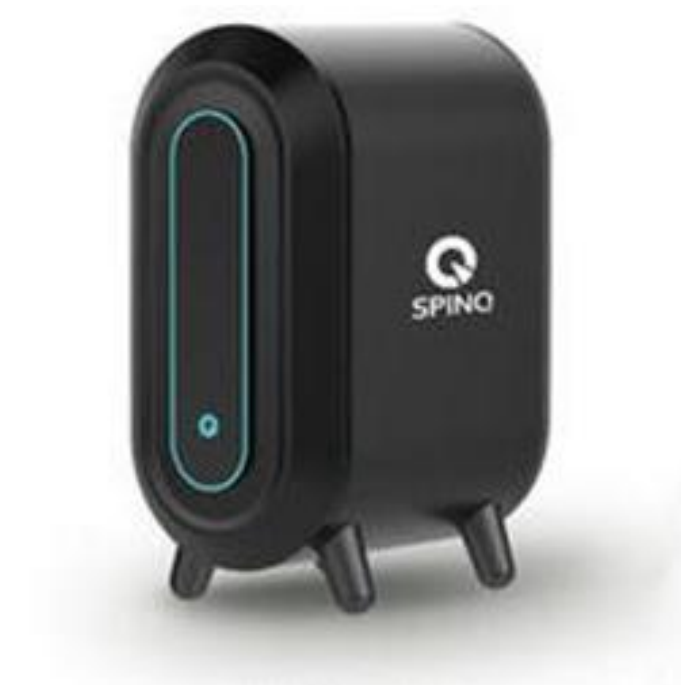

The SpinQ computers uses liquid dimethyl-phosphite molecules with two $OCH_3$ groups associated to a phosphorus atom plus one oxygen and one hydrogen atom. These molecules are controlled with permanent magnets and an RF pulse generation system. They followed-on with the Gemini mini version (also with two qubits) and the Triangulum (with a hefty three qubits).

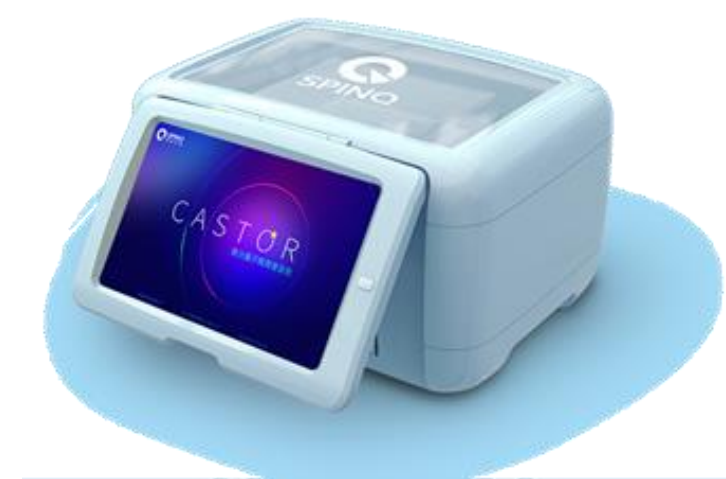

## Photons qubits

Contrarily to all the previous qubits, photons have no mass and move at about the speed of light, modulo the optical refractive index of the physical media they pass through. While photons were used everywhere in solid qubits in control and readout features with microwaves or laser beams, they can be used to create qubits exploiting polarization or other physical characteristics such as frequency, amplitude, phase, mode, path or photon number. This is the vast field of linear and nonlinear optics[1191]. It is found in both the generation of qubits for quantum computation or simulation and with their application in telecommunications and quantum cryptography which we study in another part of this document, starting page 805.

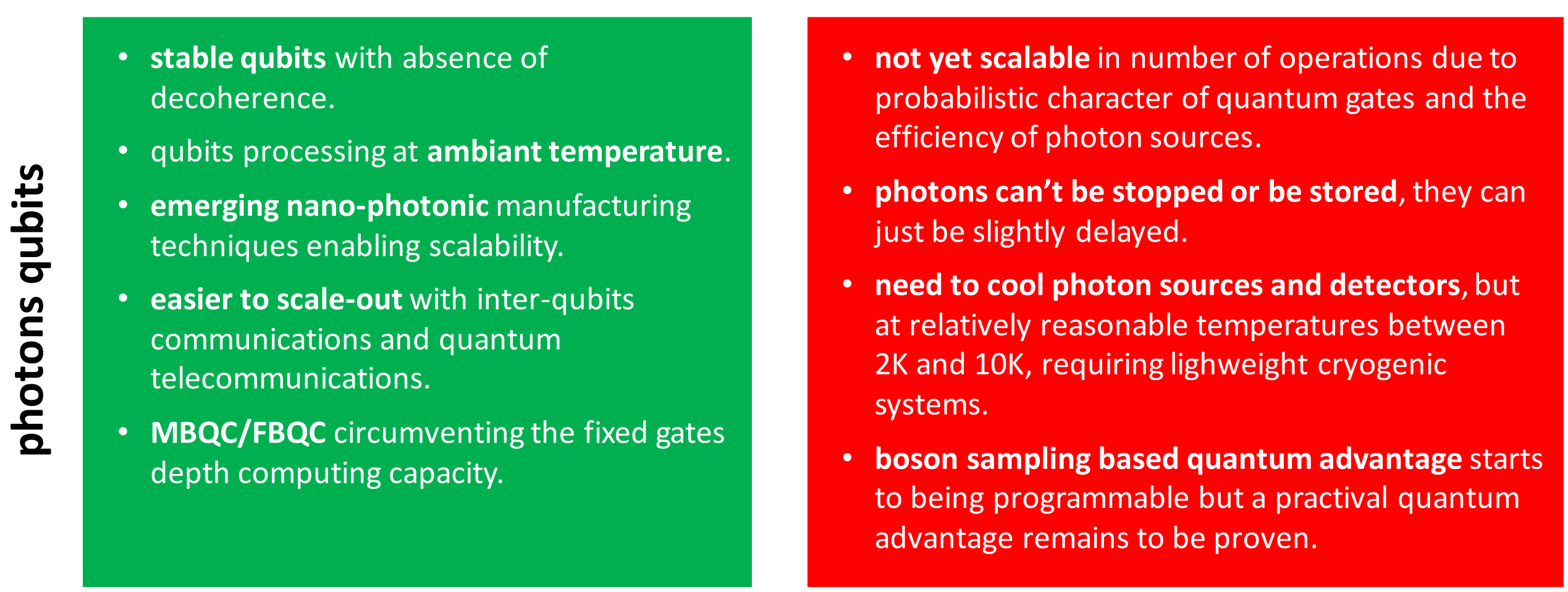

*Figure 414: pros and cons of photon qubits. (cc) Olivier Ezratty, 2022.*

[1190] See SpinQ Gemini: a desktop quantum computer for education and research by Shi-Yao Hou et al, 2021 (14 pages). It was updated with 3 qubits in September 2021 with their Triangulum version.

[1191] Nonlinear optics is well described in Nonlinear Optics by Franz X. Kärtner and Oliver D. Mücke, University of Hamburg, December 2016 (255 pages). See also the reference book Nonlinear Optics by Robert W. Boyd, 2007 (620 pages).

Photonics is both an interesting solution for creating qubits as well as a transversal technology that is indispensable to other types of qubits because it is the only one that allows long-distance communications between quantum sensors, quantum networks and quantum computers. Photons are also used directly in quantum sensing, particularly for precision time measurement and even for pressure measurement.

The advantages of photonics are that it allows to manage quite stable qubits with a very low error rate at the quantum gate level thanks to their weak coupling with the environment. The main source of decoherence is related to the optical losses happening with photons propagation. Photons also operate at any temperature[1192], do not require expensive nanoscopic manufacturing techniques and can be based on nanophotonic CMOS manufacturing processes[1193]. Their disadvantage lies in the fact that photons are even more probabilistic beasts than any of the other qubits. Scalability issues make it difficult to assemble more than a few dozens of qubits, at least for the moment. Photon sources must be more powerful to accommodate a larger number of entangled qubits.

Current technology developments are based on progresses made with more efficient single photon sources, better photon detectors, nonlinear optics, advanced quantum states preparation (multimode, spatial or spectral multiplexing, non-Gaussian states) with a larger computing space than traditional two-states qubits, using cluster-states measurement-based techniques (MBQC) to avoid the pitfalls of a physically limited quantum gates depth and quantum error corrections.

## History

The roots of quantum photonics date back from 1963 with the introduction of Glauber states by Roy J. Glauber which created the notion of coherent states of light explained by the quantization of the electromagnetic field. This corresponded to the beginnings of the laser era which led, among many things, to its broad industry impact, including fiber optics in the telecom realm.

While the first physical qubits were experimented in the mid-1990 (trapped ions, NMR) and early 2000s (superconducting), photon qubits used for computation saw the light much later. Starting in the mid-1980s, quantum photonics were envisioned for implementing quantum key distribution protocols. 2001 was a foundational year with the creation of the KLM theory by Emanuel Knill and Raymond LaFlamme (then at the Los Alamos National Lab) and Gerald Milburn (University of Queensland, Australia)[1194].

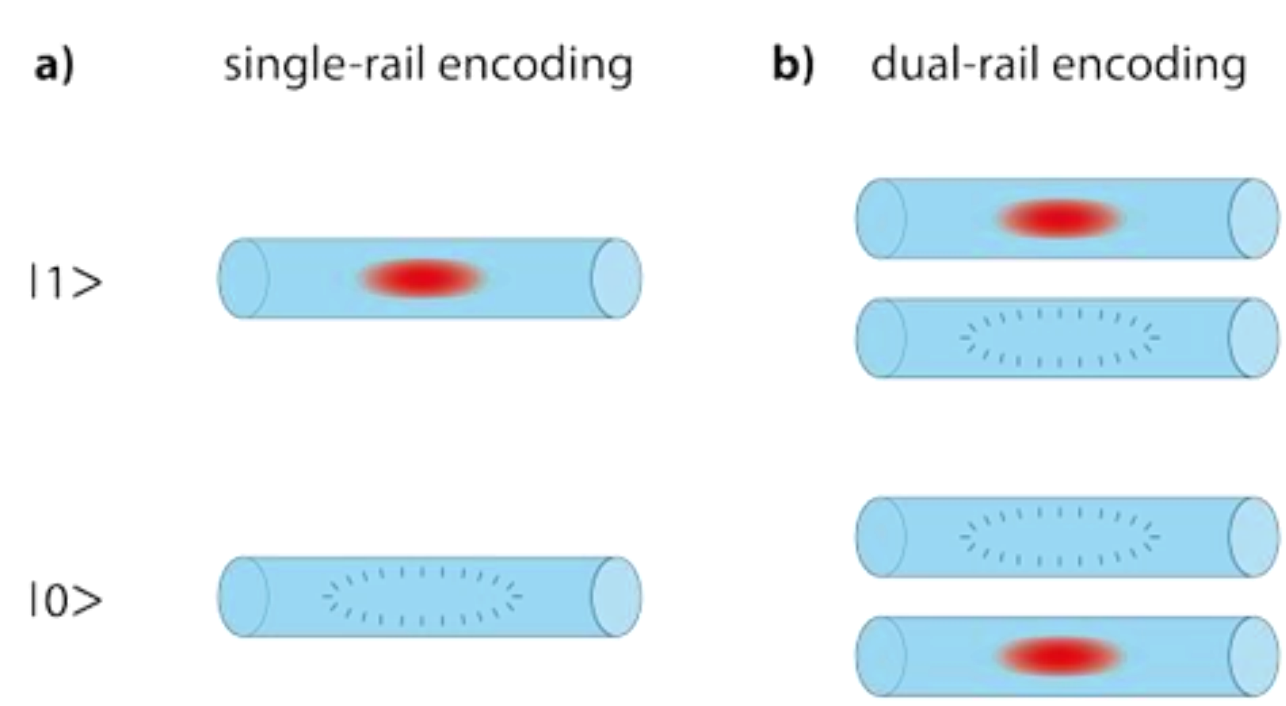

*Figure 415: how dual-rail encoding works. Source: No-go theorem for passive single-rail linear optical quantum computing by Lian-Ao Wu et al, Nature, 2013 (7 pages).*

[1192] In general, solid-state light source must be cooled to 10K and the photon detectors output to about 2K to 4K. At least, one avoids going below 1K, which allows the use of cryogenic systems that are satisfied with helium 4 and do not require helium 3. These cryogenic systems are miniaturizable and require much less energy than the dilution systems used for superconducting and silicon qubits.

[1193] See Photonic quantum information processing: A concise review by Sergei Slussarenko and Geoff Pryde of the Centre for Quantum Dynamics and the Centre for Quantum Computation and Communication Technology at Griffith University in Brisbane, Australia (20 pages) which describes the state of the art of photon qubits. This is the source of the diagram. See also the older Why I am optimistic about the silicon-photonic route to quantum computing by Terry Rudolph, a cofounder of PsiQuantum, published in 2016 (14 pages).

[1194] See A scheme for efficient quantum computation with linear optics by Emanuel Knill, Raymond Laflamme and Gerard Milburn, 2001 (7 pages).

This model could theoretically implement quantum computing without relying on some sort of non-linearity for creating entangling quantum gates[1195]. Implementing for example a CNOT gate with photons is not easy since photons do not easily interact with each other.

The KLM model was a breakthrough, making it possible to implement two qubit gates with using photon sources, beam splitters and photon detectors, using a dual-rail encoding, represented by the presence of a single photon in one of two spatial optical modes[1196]. It circumvents the need for non-linear interactions between photons with the use of post-selection with using ancilla photons, many executions and iterations being required. It creates significant overhead which makes the model impractical and not scaling well[1197].

There were then many "firsts" with the first universal optical quantum computer in 2015 using a chipset handling 6 photons with 15 Mach-Zehnder interferometers and 30 thermo-optic phase shifters and a 12-single-photon detector system[1198], a programmable photonic quantum computer created by Xanadu in 2021 with 8 photons[1199] and a first quantum computational advantage using gaussian boson sampling in China[1200]. Boson sampling was based on an idea from Scott Aaronson elaborated in 2011[1201]. It led in 2022 to a similar experiment achieved by Xanadu with a programmable gaussian boson sampler. These systems however are not yet providing a quantum computing advantage with a real useful problem defined by some entry data.

Many fundamental research advances were also achieved:

- The MBQC model was created in 2000 by Robert Raussendorf and Hans Briegel. It circumvented some limitations of photon qubits beyond the nonlinearity already discussed, like the finite number of gates that could be executed, the photons being "flying qubits". We'll describe this later. The key figure of merit of this architecture is the ability to create large cluster states of entangled photons.
- Quality photon sources, like single and deterministic photon sources, and better, entangled photon sources. Two entangled photons can be created with a photon source and a polarizing crystal. More entangled photons can be created with a single deterministic indistinguishable photon sources and delay lines.
- Silicon-based or III-V Quantum Photonic Integrated Circuits (QPICs) that implement quantum gates with waveguides and electronically controllable optical elements like beam splitters and polarizing filters as well as photon sources and photons detectors. They enable miniaturization of these components. Their figures of merit are miniaturization and low photon losses[1202].

---

1195 See the review paper The Category of Linear Optical Quantum Computing by Paul McCloud, March 2022 (34 pages).

1196 See No-go theorem for passive single-rail linear optical quantum computing by Lian-Ao Wu et al, Nature, 2013 (7 pages).

1197 This part is inspired from Twenty Years at Light Speed: The Future of Photonic Quantum Computing by David D. Nolte, December 2021. See also Photonic quantum technologies by Jeremy L. O'Brien, Akira Furusawa and Jelena Vučković, Nature Photonics, 2009 (11 pages).

1198 See Universal linear optics by Jacques Carolan, Jeremy O'Brien, Anthony Laing et al, Science, 2015 (6 pages).

1199 See Quantum circuits with many photons on a programmable nanophotonic chip by J. M. Arrazola et al, Nature, March 2021 (21 pages).

1200 See Quantum computational advantage using photons by Han-Sen Zhong et al, December 2020 (23 pages).

1201 See The computational Complexity of Linear Optics by Alex Arkhipov and Scott Aaronson, 2010 (94 pages).

1202 See the review paper Silicon photonic devices for scalable quantum information applications by Lantian Feng et al, August 2022 (20 pages).

Then, since about 2017, a wealth of startups have been created that are all pursuing the goal of creating photon qubits quantum computers, first in the NISQ and then the FTQC realms: **PsiQuantum**, **Xanadu**, **Quandela**, **Orca Computing**, **QuiX**, etc. They all adopt very diverse technological choices as we will see in the vendor section.

### Science

To understand photons in quantum information systems, one needs to get a bit deeper in quantum optics and statistical optics. This section constitutes a very rudimentary primer, enabling you to understand some of the vocabulary used by quantum information photonicians. It can also help us segment the various kinds of photonic qubits like discrete variables and continuous variables qubits.

So far, we've mainly mentioned photons as wave-particles interacting with matter, with the photoelectric effect and atoms energy transitions. But exactly, what are photons? How do we define and classify it?

A photon is a moving perturbation of the electromagnetic field with orthogonal magnetic and electric field variations themselves orthogonal to the photon propagation direction.

Photons are described with their quantum numbers which are:

- **Mass** and **electric charge** that are equal to zero.
- **Wavelength** $\lambda$ or frequency $\nu$ which define the photon energy and momentum. Several photons with same or different frequencies can be coherently superposed and create a "photon number" or a "wave packet". Wave packets are usually generated by femto-lasers pulses, mostly in the visible and infrared ranges, or by digital-to-analog microwave generators like those used to drive superconducting and electron spin qubits.
- **Spin angular momentum** (SAM), which corresponds to their angular momentum having quantized values $+\hbar$ or $-\hbar$ (spin = +1 or -1 since spins are expressed in $\hbar$ units) corresponding to circular right or left polarization. Any single monochromatic photon is a linear superposition of these two circular polarizations, including linearly horizontally or vertically polarized photons.
- **Orbital angular momentum** (OAM) where the electromagnetic field is rotating helically along its propagation axis or vector[1203]. Equals $\ell\hbar$ with $\ell$ being any integer.

Photons interact with other particles, mainly electrons either tied to atom nucleus, for photons absorption and/or emission, or free electrons like with the Compton effect.

They can be created, destroyed and modified by many of these interactions. Pairs of photons can also be generated by the collision between particles and their antiparticles. Their behavior is mainly described with Maxwell's equations and its derivatives.

**Photon directionality**. Is a photon directional? Textbooks usually make a distinction between spontaneous emission with photons going in any direction, from a lightbulb or the Sun, and stimulated emission, with directional light, coming from lasers. Radio-frequency antennas can also create spherical radiations going in many directions.

But whatever its source and wavelength, a single photon is mostly always directional and moving in space as a planar wave. A photon electromagnetic wave is represented by orthogonal electric and magnetic fields variations travelling along a vector orthogonal to them. A photon direction can change when it traverses various materials having different refraction indices.

[1203] See Quantum advantage using high-dimensional twisted photons as quantum finite automata by Stephen Z. D. Plachta et al, February 2022 (20 pages) which proposes to use qubits encoded in orbital angular momentum to implement a Quantum finite automata (QFA) to solve binary optimization problems.

Can we have non-planar photons? “Any direction” photons can come from a statistical view of random multidirectional photons emissions or from a coherent superposition of photons emitted in several directions.

With light bulbs, many photons are emitted in various directions by random thermal processes, with various photon wavelengths. Laser coherent light is made of photons with the same wavelength, phase and direction. The distinction between a wave and a point-like particle is as blurred with photons as it is with electrons as far as their exact physical nature and dimensional scope is concerned.

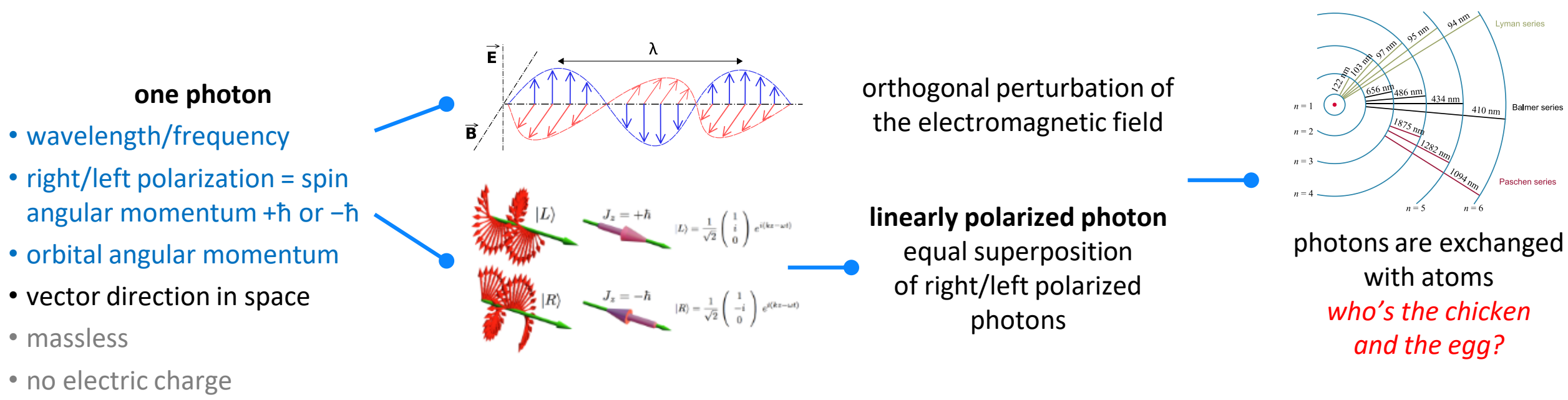

*Figure 416: photon characteristics, polarization.*

**Photon length** and size are thus notions that are rarely mentioned. According to the Hunter-Wadlinger electromagnetic theory of the photon established in 1985 and verified experimentally for some wavelengths, an optical photon has a shape similar to an elongated ellipsoid of length λ and diameter λ/π, λ being the photon’s wavelength.

What this means is the usual graphic representation in Figure 417 in green is not realistic! According to other literature, the longitudinal length of a single photon is half of its wavelength (λ/2)[1204].

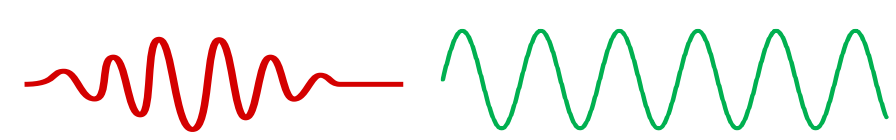

*Figure 417: a photon wave packet or pulse, and a single frequency photon… of undetermined length. The first has many harmonic frequencies shaped like a Gaussian curve in their Fourier transform while the single frequency photon Fourier transform is a single point.*

This length’s range is quite broad, with 1 nm for X-rays and several orders of magnitude smaller for gamma rays, to over one millimeter and up to several kilometers for radio waves. A classical representation in the above illustration with an EM field of 1,5 wavelengths doesn’t correspond to a single photon according to this interpretation, but to three or 1,5 consecutive single photons.

Nothing says that this can represent reality. On top of that, with a photon having half a wavelength, its Fourier transform won’t be decomposed with a single frequency, but with some harmonic frequencies. We’re safe since this can be explained with Heisenberg’s indeterminacy, related here to two complementary properties, the photon length, and its wavelength. In other words, if you try to describe with precision the length of the photon wave (time/space domain), you end up losing precision with its wavelength (energy domain). And vice versa!

**Photon modes** are not that easy to define and their simplified descriptions are diverse. These are defined by coherence and orthogonality properties of the EM field. These are orthogonal solutions of the EM wave equations. Different photon modes do not interfere. The energy of a linear superposition of modes equals the sum of the energy of the individual modes. Only photons with the same mode can be coherent and interfere.

[1204] See Electromagnetic fields, size, and copy of a single photon by Shan-Liang Liu, 2018 (4 pages) and The Size and Shape of a Single Photon by Zhenglong Xu, 2021 (22 pages).

There are two types of photon modes: spatial modes that are transverse to their direction of propagation and temporal modes in the direction of propagation (time and frequency)[1205].

We find these multimode photons in various quantum optics setups like with boson sampling experiment that we'll describe a bit later, as well as in quantum key distribution settings[1206].

**Photon number** is a way to describe groups of similar photons. Several photons with the same wavelength and polarization, can be at the same place and at the same time. They also share the same direction vector. These photons are indistinguishable. This is a property of bosons which are elementary or composite particles with the same quantum characteristics which can get together, following Bose-Einstein statistics, while fermions with the same quantum characteristics can't be together, following Fermi-Dirac statistics.

A group of similar photons form an electromagnetic wave whose energy $E = h\nu N = \hbar\omega N$, i.e. the energy of each photon multiplied by the number of added photons having that wavelength[1207].

A photon number is this number of "clustered" photons forming a higher energy EM field than a single photon[1208]. You can even create superpositions of multi-photons (or single-mode Fock state as we'll see later) with 0, 1, 2 and 3 photons. This can be used to create photon-number Bell states, namely entangled states of superposed photon numbers[1209]. This is head twisting and hard to visualize!

Quantum optics is heavily based on the model of the **quantum harmonic oscillator**, the quantum-mechanical analog of the classical harmonic oscillator, with quantized energy.

The energy of a quantum oscillator can be described with a simple equation, N being the photon number. When N=0, the oscillator energy corresponds to the vacuum state energy.

$$\text{energy } E \text{ and photon number } N: \quad E = \hbar\omega\left(N + \frac{1}{2}\right) \qquad \text{photon wavenumber } k = \frac{2\pi}{\lambda}.$$

**Photon wavenumber** is the spatial frequency of a wave, measured in radians per unit distance. It is defined as k with the above right formula using the photon wavelength.

**Creation and annihilation operators** or ladder operators are mathematical operators used with quantum harmonic oscillators and many-particles systems. An annihilation operator $\hat{a}$ reduces the number of particles in a given state by one and a creation operator $\hat{a}^\dagger$ increases this number by one. It is the adjoint operator of the annihilation operator. These operators act on states of various types of particles, and with photons, as adding or removing a quantum of energy to and oscillator system.

The use of these operators instead of wavefunctions is part of the second quantization formalism. It explains why the canonical quantum physics postulates that we described in an earlier part (page 86) are not entirely applicable to quantum optics, particularly the time evolution postulate related to Schrödinger's wave equation that is applicable only to non-relativistic massive particles and even the structure of the quantum state $\psi$.

---

[1205] See The concept of modes in optics and photonics by René Dändliker, 1999 (6 pages).

[1206] See the review paper Roadmap on multimode photonics by Ilaria Cristiani et al, 2022 (39 pages) which also covers classical use cases of multimode photonics.

[1207] Here, $\nu$ is the photon frequency, h in Planck's constant, $\hbar$ is Planck's reduced constant or Dirac's constant and $\omega$ is the photon angular frequency with $\omega = 2\pi\nu$, in radians per second, $2\pi$ radians corresponding to a 1 Hz frequency.

[1208] A powerful radio of digital TV emitter is creating these kinds of photons, in the radiowave range! Same for a radar.

[1209] See Generation of non-classical light in a photon-number superposition by J. C. Loredo, Pascale Senellart et al, November 2018 (13 pages), Generating superposition of up-to three photons for continuous variable quantum information processing by Mitsuyoshi Yukawa et al, 2013 (7 pages), and Generation of light in a photon-number quantum superposition, August 2019. And the entangled photon numbers in Photon-number entanglement generated by sequential excitation of a two-level atom by S. C. Wein, Pascale Senellart et al, June 2021 and in Nature Photonics, April 2022 (18 pages).

Mathematically, a photon occupation number operator is a Hermitian operator $\hat{N} = \hat{a}^\dagger \hat{a}$. And a photon number of n superposed photons created by the operator $\hat{a}^\dagger$ applied n times to the vacuum state $|0\rangle$ creates the state $|n\rangle = \frac{1}{\sqrt{n!}}(\hat{a}^\dagger)^n|0\rangle$.

**Second quantization** is the broad field of quantum physics that deals with many-body quantum systems. It was introduced by **Paul Dirac** in 1927 and developed afterwards by **Vladimir Fock** and **Pascual Jordan**. While the first quantization dealt with individual quantum objects and their description by the Schrödinger wave equation, the second quantization describes many-body systems which are represented mathematically by Fock states and Fock spaces.

Its formalism introduces creation and annihilation operators to construct and handle the Fock states, providing the mathematical tools to the study quantum many-body systems.

Instead of describing such a system as a tensor product of all its constituent quantum objects, it is simplified with chaining $|n_{k_i}\rangle$ describing the $n_i$ quantum objects that are in the same quantum state $k_i$ as described in the above equation related to the creation operator.

A many-body system is described as the tensor product of the Fock states $|n_{k_i}\rangle$ corresponding to each individual quantum states in the system: $|n_{k_0}\rangle \otimes \cdots \otimes |n_{k_n}\rangle$, given that the photon number $n_i$ for the Fock state $|n_{k_i}\rangle$ can be 0 or 1 for fermions and any positive number for a boson. When all occupation numbers are equal to zero, the Fock state corresponds to the vacuum state.

A Fock state with only one non-zero occupation number is a single-mode Fock state. Contrarily, a multi-mode Fock state has several non-zero occupation numbers.

**Quantum optics**. This field of quantum physics started quite late. In 1956, the Hanbury-Brown-Twiss (HBT) experiment was about observing the intensity correlations of the radiation of a mercury lamp and from some bright stars. After traversing a beamsplitter (with a mercury lamp) or at two spatially separated points (for stars), the intensities measured by two detectors were fluctuating, and these fluctuations were correlated.

It was then explained by the emission of photon bunches coming from thermal sources. But it could be explained without using photons and quantum physics.

Quantum optics really started when it became possible to create non classical light sources like pairs of photons and single photons, respectively in 1967 and 1977. Photon pairs were first created with using cascaded atom decay and parametric down conversion[1210].

**Semi-classical light**. It describes interactions between quantized matter such as atoms and electrons with classical light fields. Continuous laser light belongs to this category.

**Non-classical light**. Light and photons are always quantum, just because it comes from quantized energy exchanges with matter. Still, light is considered to show non-classical and quantum effects when the electromagnetic field is quantized and photons are handled individually. This happens in a couple situations: creation of entangled Bell states, antibunching, photon noise and negative probabilities with the Wigner function. We'll look at each of these phenomena.

**Bell states** where single photons behave probabilistically and in the general case have no *a priori* properties like polarization, wavelength, wavevector before being measured. These properties are revealed while being measured and show correlations between entangled photons whose measured properties will be random.

[1210] Source: Lecture 1. Basic concepts of statistical optics (7 pages).

**Anti-bunching** corresponds to a light field where photons are equally spaced in time, much better than with a coherent laser field. It is detected with a HBT (Hanbury Brown & Twiss) intensity autocorrelator... with no correlations. It refers to sub-Poissonian photon statistics, that is a narrow photon number distribution.

It can be generated by single photon sources as well as from pulse mode lasers. A coherent state from a laser has a Poissonian statistics generating random photon spacing and a thermal source light field has super-Poissonian statistics and yields bunched photon spacing. All these aspects belong to the field of statistical optics.

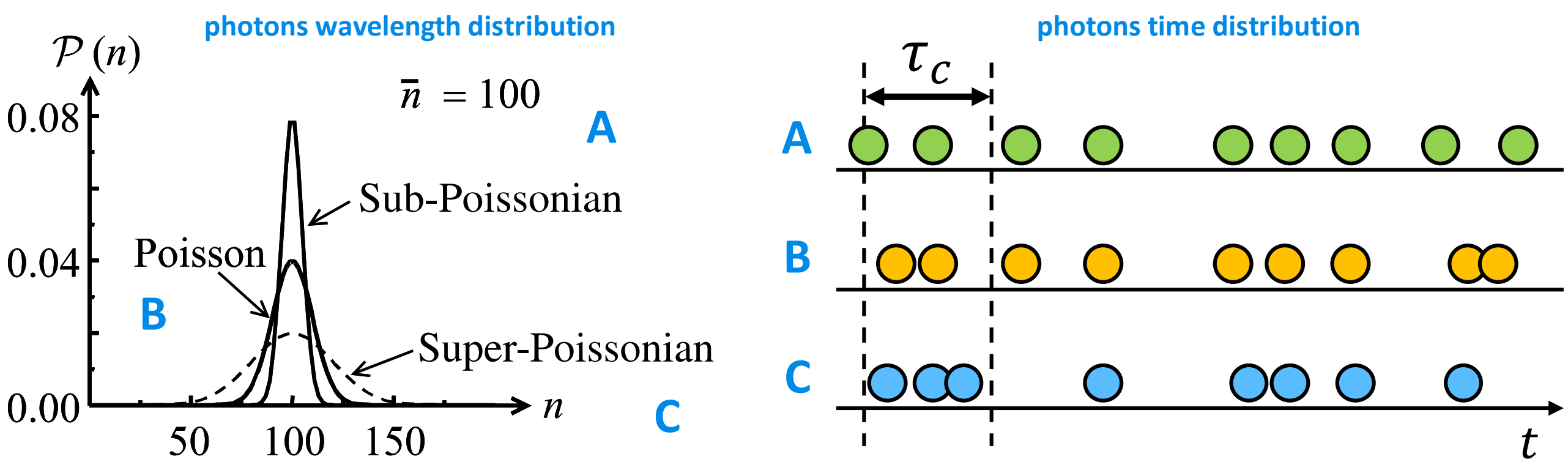

| | A | B | C |
|---|---|---|---|
| photon wavelength distribution | **sub-Poissonian distribution** narrow frequency gaussian | **Poissonian distribution** gaussian frequency | **super-Poissonian distribution** broad frequency gaussian |
| photon time distribution | **antibunched** even distribution | **random** random distribution | **bunched** random and regroupments |
| photon sources | **single photons sources squeezed light** | **laser light** | **thermal/chaotic light sun, lightbulb** |

*Figure 418: poissonian, sub-Poissonian and super-Poissonian photons wavelength distribution and photons time distribution. Compilation: Olivier Ezratty, 2021.*

The quality of single photons source is measured with the data from two experiments. The first uses a variant of a **Hanbury Brown and Twiss** (HBT) intensity autocorrelator that checks the photons are emitted in a very regular way, like a metronome. From a starting click on one of the two photon detectors, it analyzes the time distribution of the appearance of the following photons. This produces the plots in Figure 419. The ideal model would be that of a high peak on either side of the center. The low peaks represent the system noise[1211].

The second experiment called H.O.M. for **Hong-Ou & Mandel** and created in 1987 uses a Mach-Zehnder interferometer to validate the fact that the emitted photons are indeed identical and impossible to distinguish[1212].

---

[1211] This experiment, originally created to detect the size of stars, also allowed to validate the corpuscular nature of photons. The experiment can be easily interpreted in an intuitive way: photons pass through a one-way mirror, whether or not it crosses randomly. Behind this mirror are two photon counters, here with SPADs (avalanche diodes). The system detects when a photon is detected at the same time by both sensors. If the photons take the same way to reach both detectors, there will be no coincidence since the emitted photons are sent in well-ordered trains and can only be on one side or the other. By adding a delay line between the mirror and one of the sensors that is proportional to the period of emission of the photons, it creates many occurrences with photons arriving simultaneously in both sensors. This is what we see in the two curves, one of them being with a linear scale of coincidences (measured over a period of time sufficient to capture hundreds of them) and another logarithmic which allows to better characterize the noise of the system.

[1212] See High-performance semiconductor quantum-dot single-photon sources by Pascale Senellart, Glenn Solomon and Andrew G. White, 2017 (14 pages) which describes the various ways to characterize the quality of single photons sources.

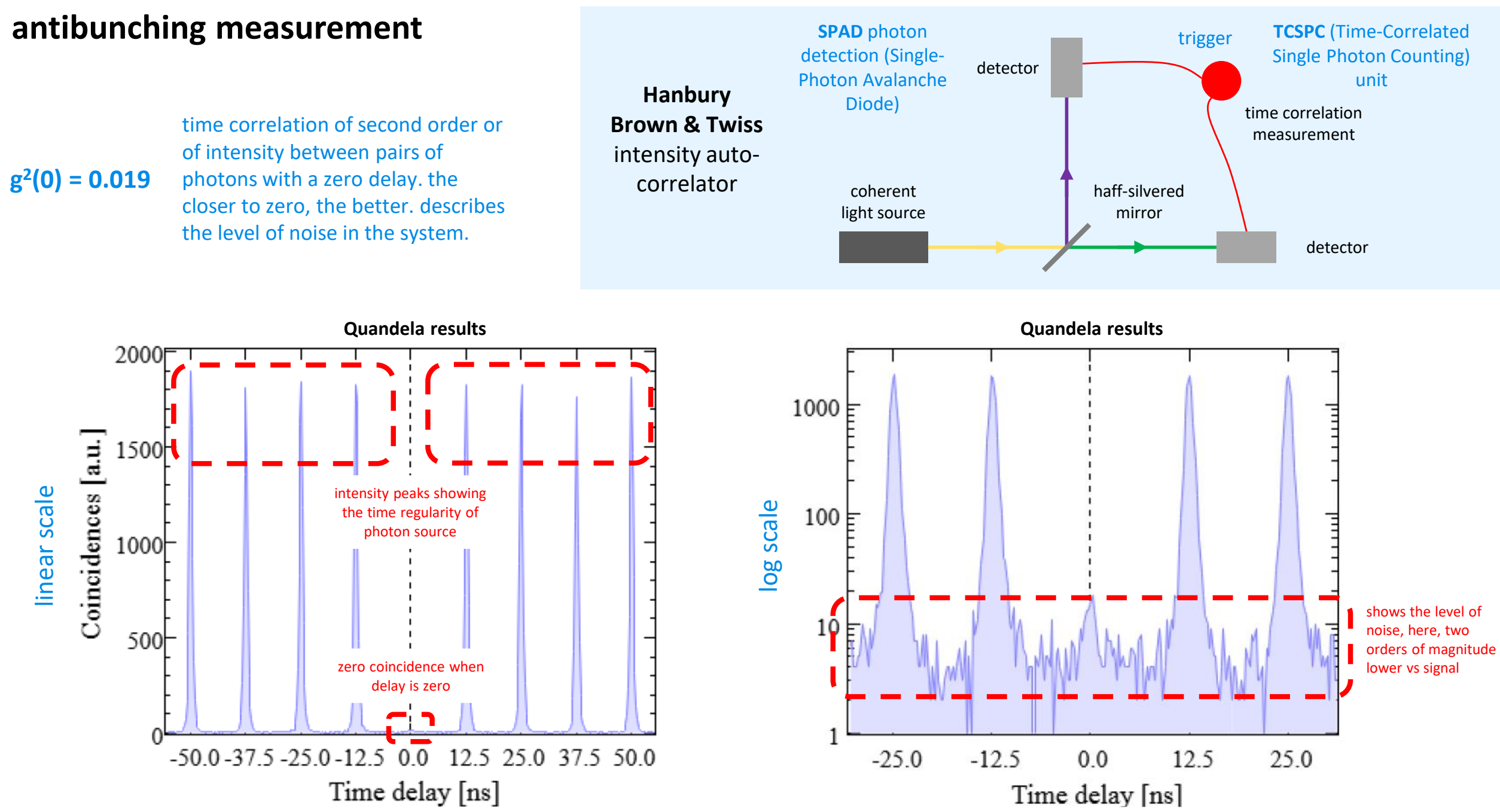

*Figure 419: how antibunching is measured. Sources: various.*

**Quadratures representation** is a way to describe the electromagnetic field and its related uncertainty. An EM wave is positioned in two axis X and Y or $X_1$ and $X_2$ corresponding to the rotation of the electric field in the EM field, thus the equations describing $X_1$ and $X_2$ below, with the cosine and sine of $\omega t$ (angular frequency × time). Said otherwise, a quadrature describes the real and imaginary parts of a complex amplitude. This EM field complex amplitude is rotating so what's interesting is not the grey circle position in the chart but its shape and size which represents the photon measurement uncertainty. It is represented by the variation of the length of the vector which is the photon number and of the width of the circle, orthogonally to the vector, which corresponds to the phase uncertainty. For unsqueezed coherent light, this uncertainty is the same as the vacuum state.

**Photon noise** aka shot noise is found in the detection of light and corresponds to quantum fluctuations in the electromagnetic field. This noise or imprecision can be squeezed in one dimension.

**Squeezed light** corresponds, in a quadrature or phasor diagram representation, to wave functions which have an uncertainty in one of the quadrature amplitudes (phase or photon number) smaller than for the ground-state corresponding to the vacuum state. It can be generated by different means like a parametric down conversion[1213]. Balanced homodyne detectors are used to detect squeezed light.

**Wigner function** is yet another representation of a quantum state, richer than the phase diagram above which is used to measure the level of quantumness of a light pulse. It's not far from a probability distribution of the electric field in the (Q, P) plane that can take negative values in some conditions, for so-called non-Gaussian states[1214]. With coherent states, W(Q, P) is a symmetric Gaussian function peaking at the average values of the sine and cosine components of the electric field with the peak width corresponding to the vacuum noise like in the quadrature representation.

[1213] See Generation of squeezed states by parametric down conversion by Ling-An Wu et al, University of Texas, Physical Review Letter, 1986 (4 pages).

[1214] See Conversion of Gaussian states to non-Gaussian states using photon number-resolving detectors by Daiqin Su et al, Xanadu, April 2019 (37 pages), the tutorial paper Non-Gaussian Quantum States and Where to Find Them by Mattia Walschaers, LKB - Collège de France, April 2021 (55 pages) and the review paper Production and applications of non-Gaussian quantum states of light by A. I. Lvovsky, Philippe Grangier et al, June 2020 (50 pages).

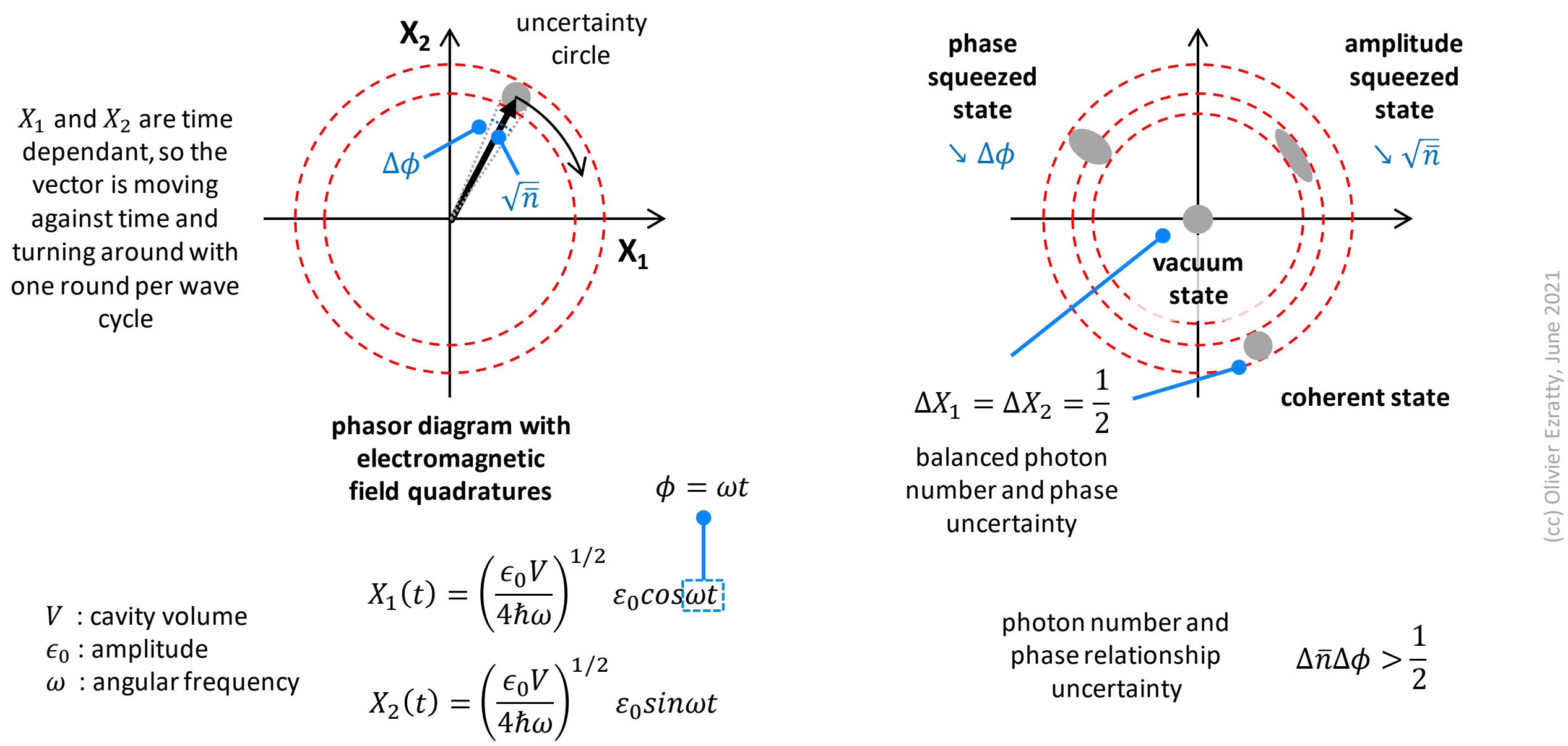

*Figure 420: what squeezed light looks like when using quadratures representation. Sources: various.*

The Wigner function equation looks like:

$$W(Q,P) = \frac{1}{\pi\hbar}\int_{-\infty}^{+\infty} \psi^*(Q+y)\psi(Q-y)e^{2iPy/\hbar}dy$$

*Figure 421: Wigner function which helps measure the quantumness of light.*

with $\psi$ being the quantum object wave function, Q and P the position and momentum and y, the variable used in the integral. It returns a real value that can be positive or negative. Q and P could be replaced by the sine and cosine components of the quantized electric field like in the phasor diagram.

In Figure 422 are presented a set of Wigner functions, ranging from the most classical to the most quantum fields. A is a coherent state [1215], B, a squeezed state, C a single-photon state and D a Schrödinger's-cat state. The projections or shadows of the Wigner function are the probability distributions of the quantum continuous variables Q or P. The Wigner function is a Gaussian function for A and B but takes negative values for the non-Gaussian strongly quantum states C and D[1216].

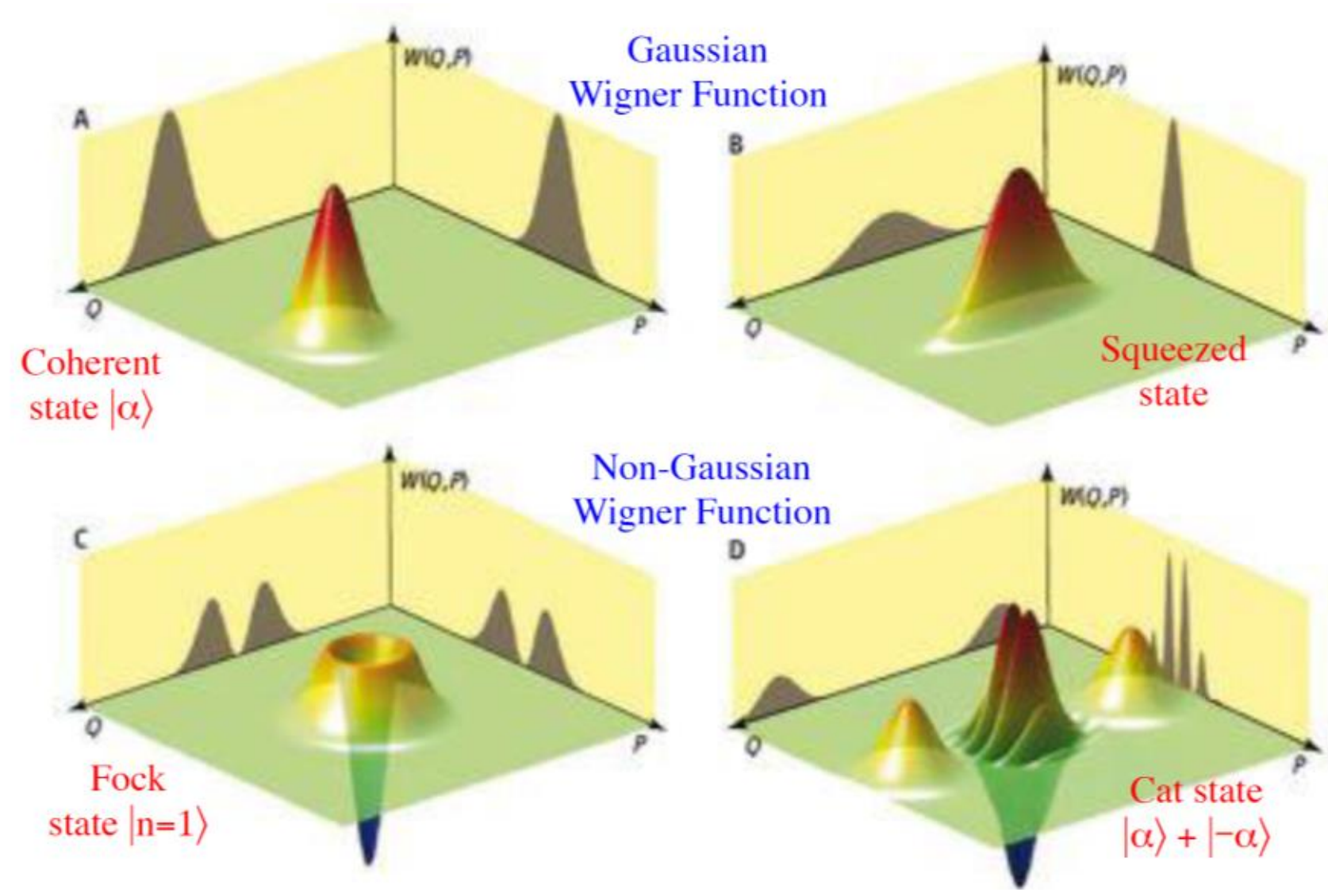

*Figure 422: various 3D representations of the Wigner function, for Gaussian and non-Gaussian light. Source: [Make it quantum and continuous](#) by Philippe Grangier, Science, 2011.*

---

[1215] The vacuum state has a similar Wigner function, but centered around P=0 and Q=0.

[1216] See Recent advances in Wigner function approaches by J. Weinbub and D. K. Ferry, 2018 (25 pages) which shows the various use cases of the Wigner function.

These negative values vanish very quickly with decoherence.

**Parametric down-conversion** is a nonlinear optical process converting one photon of high energy into a pair of photons of lower energy. It is used to generate pairs of entangled photons.

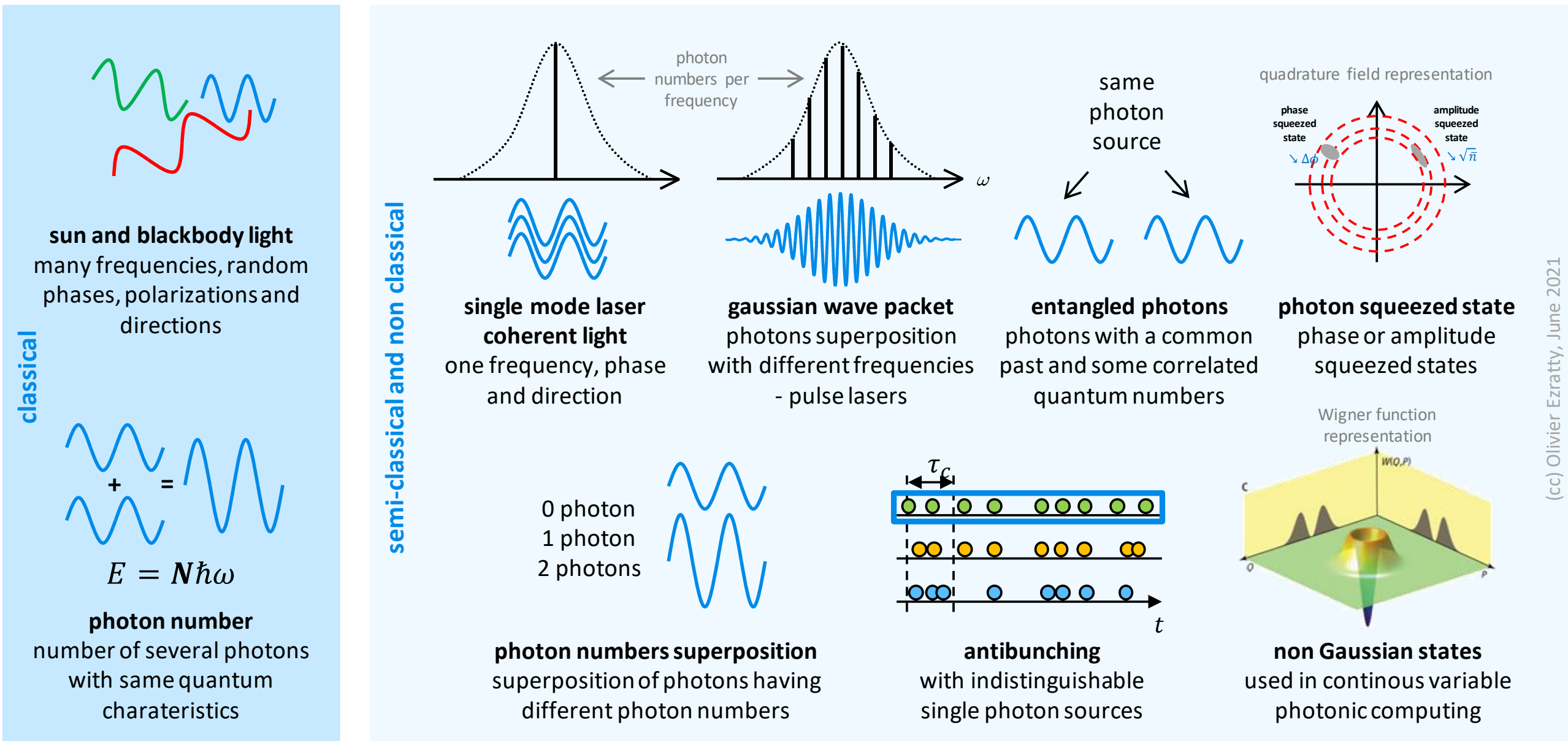

*Figure 423: a zoo of photons. (cc) Olivier Ezratty, 2021.*

**Photons zoo**. Figure 423 shows some various photon states as a summary of this section. Random photons in spontaneous light coming from the Sun or light bulb and “photon number” waves assembling several similar photons belong to classical light.

Other forms of light described here are semi-classical or non-classical: photon number superposition, squeezed states where the precision is improved in photon number, amplitude or phase at the expense of the others, single-mode coherent laser light, wave packets created by pulse lasers or microwave coming from waveform generators used with superconducting and electron spin qubits, entangled photons used in QKD and photon qubits, and non-gaussian states which are weird beasts too complicated to describe in a couple of words that are used to implement non-Clifford quantum gates with photon qubits.

### Qubit operations

Photons are "flying qubits". They are the only ones having this characteristic with flying electrons, which are investigated at the fundamental research level. There are two main classes of photon qubits: discrete variable and continuous variable qubits[1217].

**Discrete Variable** qubits use single photons and use a two-dimensional space like orthogonal polarizations or the absence and presence of single photons. DV systems can even be based on qudits using more than one degree of liberty. DV qubits rely on highly efficient, deterministic and indistinguishable single photon sources. They are using the “particle” side of photons. Their indistinguishability must exceed 95%, meaning this percentage of photons must be indistinguishable. The photon sources must also be efficiently connected to dynamically controllable photonic computing chipsets.

[1217] See this good review paper: Integrated photonic quantum technologies by Jianwei Wang et al, May 2020 (16 pages) and Hybrid entanglement of light for remote state preparation and quantum steering by Adrien Cavaillès et al, LKB (41 slides) which positions well the difference between DV and CV computing.

| | discrete variables | continous variables | boson sampling |
|---|---|---|---|
| quantum information | **discrete degree of freedom of a photon**<br>Fock states: \|0⟩, \|1⟩, \|2⟩ …<br>single or many photon properties | **quadrature of a light field**<br>coherent states, qumodes, spectral and time modes | **multimode photons** |
| photon sources | **single indistinguishable photon sources** | **entangled photons sources squeezed states, …** | **unique photons source** |
| representation | **density matrix** | **Wigner function** | **permanent** |
| gates | **KLM model, MZI (Mach-Zehnder Interferometer) gates** | **determinist gates**<br>modes measurement<br>gaussian and non gaussian gates | **MZI and interferometer** |
| photon detectors | **photon counters /detectors**<br>APD, SNSPD, VLPC, TES | **homodyne and heterodyne detectors** | **single photons detectors** |
| players | PsiQuantum<br>ORCA Computing<br>QUANDELA<br>DUALITY QUANTUM PHOTONICS | XANADU | |

*Figure 424: comparison of the main models of photon-based quantum computing. (cc) Olivier Ezratty, 2021.*

Efforts are also undertaken to create cluster states of entangled photons used in MBQC and to create deterministic multi-qubit gates using spin-photon interactions like in NV centers or other silicon spin defects[1218].

**Continuous Variable** qubits encode information in the fluctuations of the electromagnetic field, in their quadrature components, in qubits that are sometimes baptized qumodes[1219]. We are playing here with the wave nature of photons.

Photons readout can be done with a Gaussian measurement comprising homodyne detection for one of the two quadrature components and heterodyne detection on one of these[1220], and a non-Gaussian measurement implementing photon counting returning an integer. There, you hear about Wigner function amplitude, phase encoding, Gaussian states[1221], including squeezed states generated with nonlinear media and non-Gaussian gates to execute non-Clifford group gates bringing a real exponential speedup for quantum computing. Quantum gates can be deterministic, homodyne detectors are cheaper than single photons detectors, and quantum states are more robust. CV qubits are implementing larger cluster states for MBQC, using a large number of photon modes (in the thousands)[1222].

In the CV qubit domain, **PhotoQ** (Denmark) is a collaborative project implementing this architecture with using surface codes for fault-tolerance[1223]. The project is built around research from DTU with participating organizations being AMCS Group, Aarhus University, Kvantify and NKT Photonics (lasers) with a public funding of 3M€ from Innovation Fund Denmark. They are (wonder why) focused on solving logistics and pharmaceutical industries problems.

---

[1218] See Multidimensional cluster states using a single spin-photon interface coupled strongly to an intrinsic nuclear register by Cathryn P. Michaels et al, University of Cambridge, April 2021 (11 pages).

[1219] For an explanation of the difference between qumodes and qubits, see Introduction to quantum photonics from Xanadu.

[1220] On the measurement of CV qubits, see Optical hybrid architectures for quantum information processing by Kun Huang, LKB, 2017 (215 pages). This is not the same Kun Huang as the discoverer of phonon-polaritons in 1951.

[1221] Understanding how Gaussian states work is already quite a challenge. See Gaussian Quantum Information by Christian Weedbrook, Seth Lloyd et al, 2011 (51 pages).

[1222] See A fault-tolerant continuous-variable measurement-based quantum computation architecture by Mikkel V. Larsen, January 2021 (16 pages).

[1223] See A fault-tolerant continuous-variable measurement-based quantum computation architecture by Mikkel V. Larsen et al, August 2021 (19 pages) and Deterministic multi-mode gates on a scalable photonic quantum computing platform by Larsen, Mikkel V. et al, Nature Physics, August 2021 (32 pages).

There you will also find cat-qubits and GKP states[1224]. Hybrid DV/CV qubits approaches are also investigated[1225]. CV computing can be used with universal gates quantum computing as well as with quantum simulations.

**Quantum Walks based simulation** is another computing technique using photons. Similarly to the CV/DV computing segmentation, you have two classes of photon-based quantum walk systems: discrete-time quantum walks with discrete steps evolutions[1226] and continuous-time quantum walks with a continuous evolution of a Hamiltonian coupling different sites[1227]. A research team in China created a CV-quantum walk system handling a Hilbert space of dimension 400 as shown in Figure 425. It even takes the form of a seemingly packaged and designed product despite coming out of a public research lab and not a startup. There are even hybrid fermion/bosons approaches that are proposed but that are very theoretical and with very few hardware implementation details[1228].

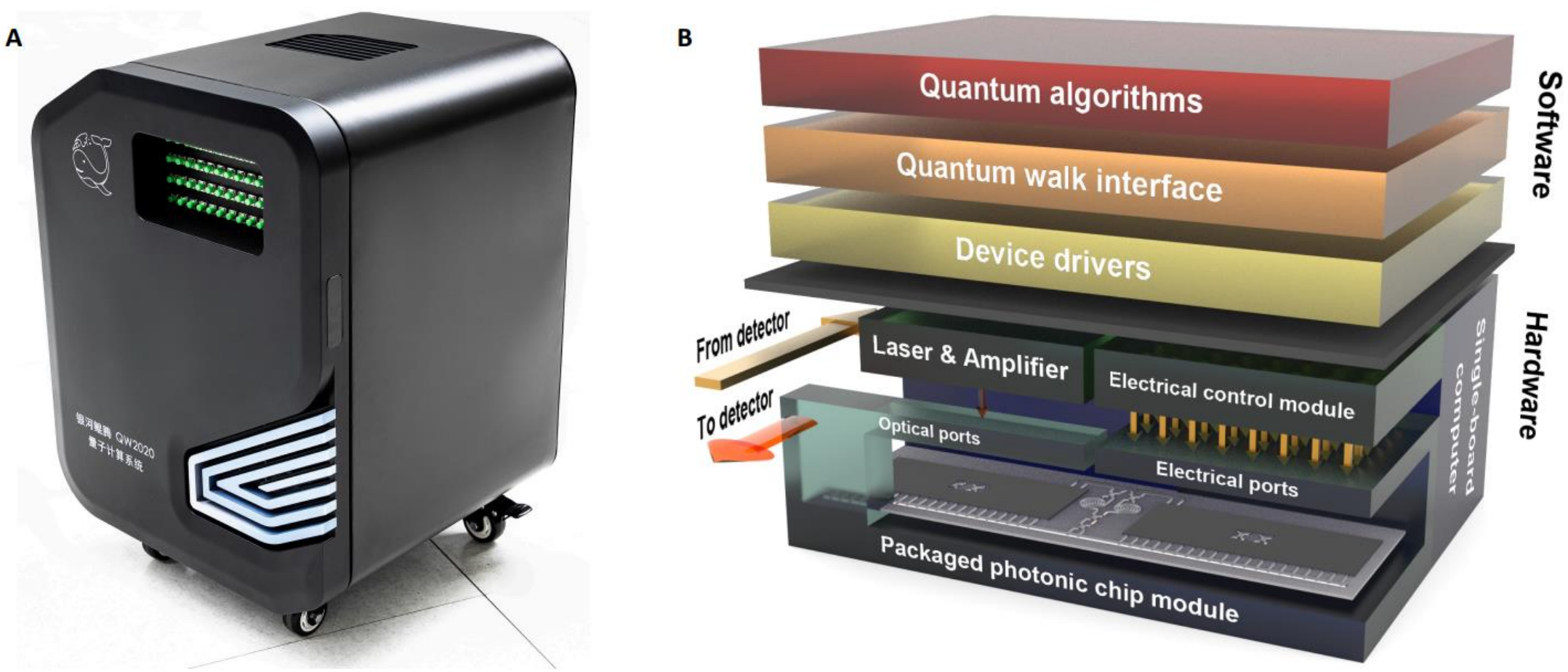

*Figure 425: the continuous-variable quantum walk system YH QUANTA QW2020 from China. Source: Large-scale full-programmable quantum walk and its applications by Yizhi Wang et al, August 2022 (73 pages).*

**Boson sampling** is a separate technique we'll cover later in a dedicated section, page 445. It's a research field that has not yet brought to life programmable computing.

**Coherent Ising Machines** is another technique based on using optical neural networks that can solve combinatorial optimization problems with mapping them onto NP-hard Ising problems [1229].

---

[1224] These light states require specific preparation techniques. See for example Robust Preparation of Wigner-Negative States with Optimized SNAP-Displacement Sequences by Marina Kudra, Jonas Bylander, Simone Gasparinetti et al, Chalmers University, PRX Quantum, September 2022 (12 pages) with interesting cavity-based preparation of various quantum light states usable in CV photon qubit systems.

[1225] See Hybrid Quantum Information Processing by Ulrik L. Andersen and al, 2014 (13 pages) and Hybrid discrete and continuous-variable quantum information by Ulrik L. Andersen et al, 2015 (11 pages), Visualization of correlations in hybrid discrete—continuous variable quantum systems by R P Rundle et al, February 2020 (15 pages) and Remote creation of hybrid entanglement between particle-like and wave-like optical qubits by Olivier Morin, Claude Fabre, Julien Laurat, LKB France, 2013 (7 pages).

[1226] See for example Quantum walks of two correlated photons in a 2D synthetic lattice by Chiara Esposito, Fabio Sciarrino et al, April 2022 (18 pages).

[1227] See Purdue University Scientists Say 'Quantum Rainbow' May Allow Room-Temperature Quantum Computing par Matt Swayne, 2021 referring to Probing quantum walks through coherent control of high-dimensionally entangled photons by Poolad Imany et al, July 2020 (9 pages).

[1228] See Two-level Quantum Walkers on Directed Graphs I: Universal Quantum Computing by Ryo Asaka et al, December 2021 (20 pages) and Two-level Quantum Walkers on Directed Graphs II: An Application to qRAM by Ryo Asaka et al, April 2022 (23 pages).

[1229] CIM is described in the presentation Coherent Ising Machines: non-von Neumann computing using networks of optical parametric oscillators by Peter McMahon, Cornell University, October 2020 (100 slides). It reminds us that HPE and Ray Beausoleil worked on a CIM before abandoning all quantum computing endeavors altogether. The source of the illustration was found on slide 55.

Practically speaking, Ising models can solve many problems: planning and scheduling, financial portfolio optimizations, graph problems and even material and molecular design. These problems are defined by couplings between a set of spins. The solution is the spin orientation that minimizes the energy function of the system. CIM systems use single-mode photon squeezing, oscillation at degenerate frequency, Optical Parametric Amplifiers in a Cavity (OPO) and a measurement feedback technique. Leveraging delay lines, time division multiplexing and measurement feedback, CIM can implement many-to-many connectivity. The largest CIM system was built in Japan in 2021 with 100,000-spins[1230]. It competed with quantum annealing (from D-Wave) and also classical CMOS-based annealing (from Fujitsu).

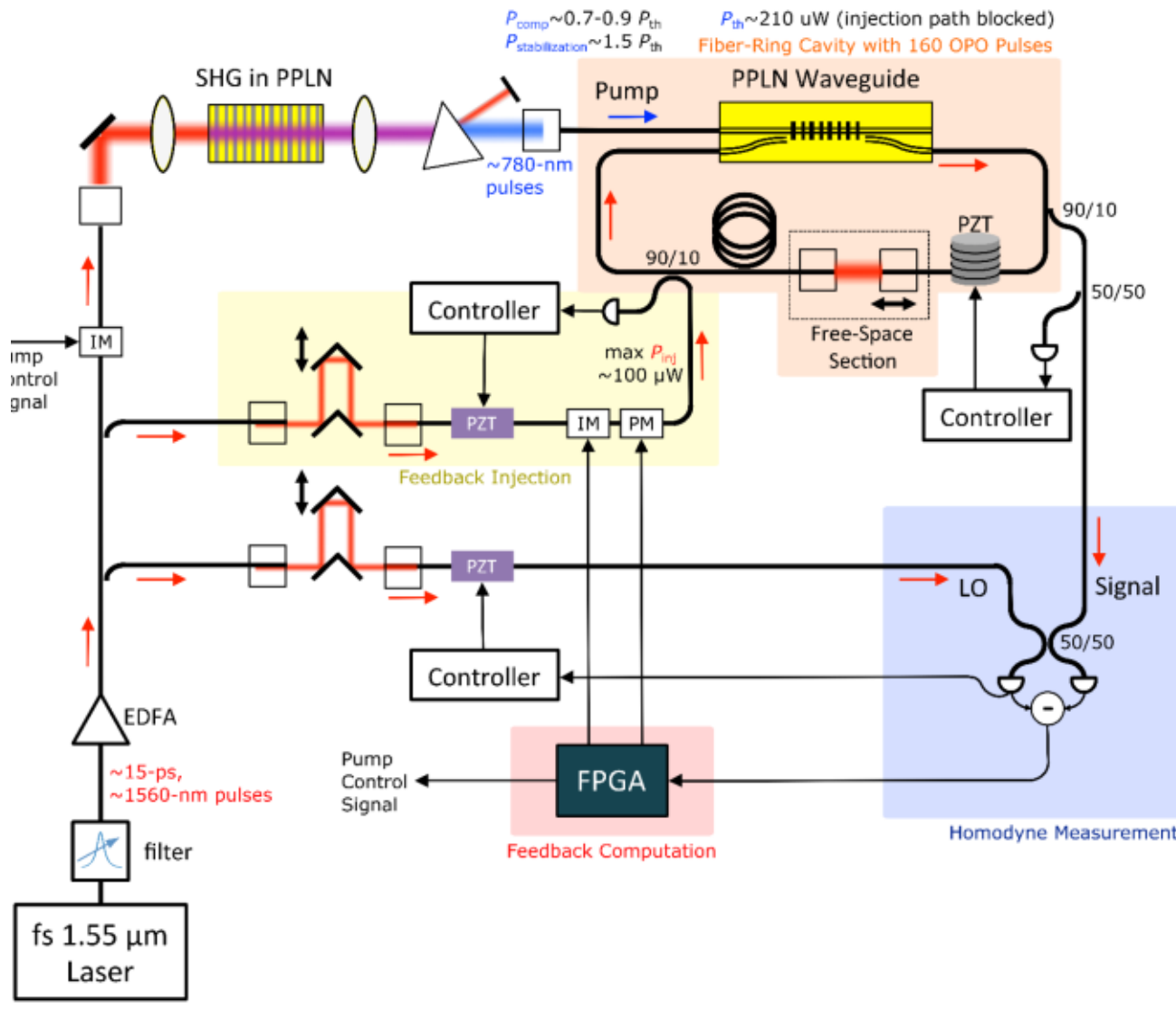

*Figure 426: example of realization of a coherent Ising Machine. Source: Coherent Ising Machines: non-von Neumann computing using networks of optical parametric oscillators by Peter McMahon, Cornell University, October 2020 (100 slides).*

A bit like the difference we discovered between (D-Wave) quantum annealing solving Ising "Z" problems[1231] and quantum simulation models implementing XY qubits connectivity, there are also photonic based coherent XY models that compete with CIM[1232].

**Hybrid atoms-photons qubits**. Stanford University researchers devised a new hybrid quantum photonic approach using a single atom that modifies photons states via quantum teleportation and implement quantum gates and qubit readout[1233].

It reduces the need for multiple photon emitters and greatly simplifies the hardware setting that makes use of a photons storage ring made of a fiber loop, optical switches in the loop, a beam splitter, a phase shifter, a photon scattering unit and a cavity containing a single atom controlled by a laser, the atom getting entangled with the photon.

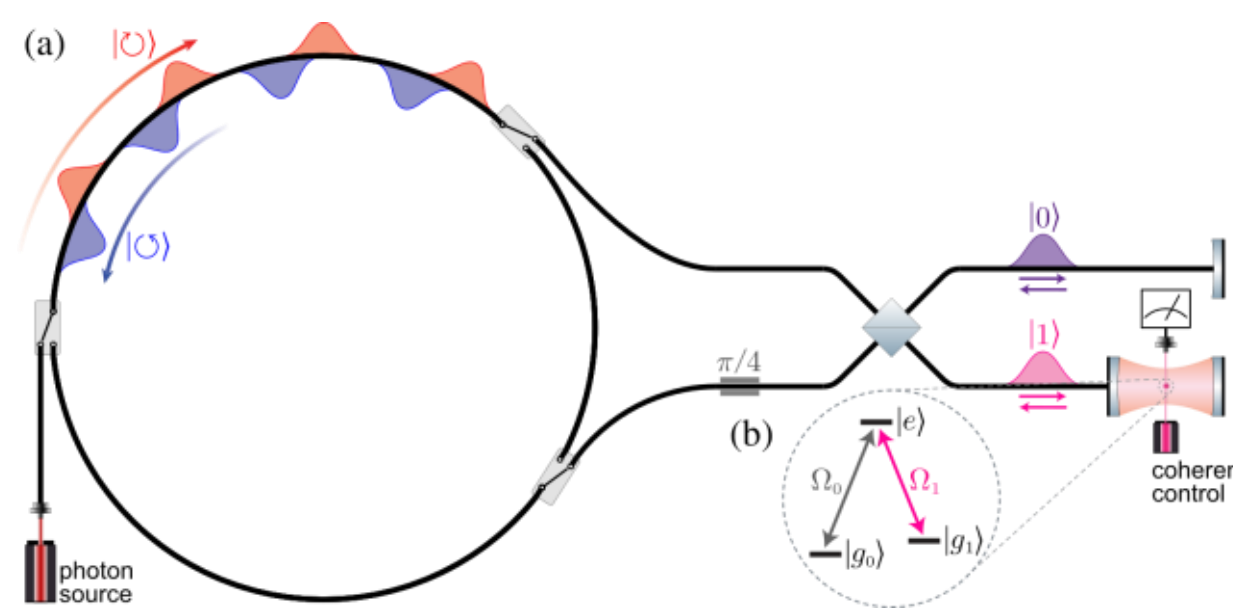

*Figure 427: an example of hybrid atoms-photons system. Source: Deterministic photonic quantum computation in a synthetic time dimension by Ben Bartlett, Avik Dutt and Shanhui Fan, Optica, November 2021 (9 pages).*

---

1230 See 100,000-spin coherent Ising machine by Toshimori Honjo et al, September 2021 (8 pages).

1231 CIM is supposed to work better than D-Wave that also implement Ising models according to Practical Application-Specific Advantage through Hybrid Quantum Computing by Michael Perelshtein et al, 2021 (14 pages) that is described in A poor man's coherent Ising machine based on opto-electronic feedback systems for solving optimization problems by Fabian Böhm et al, Nature Communications, 2019 (9 pages). See also Experimental investigation of performance differences between coherent ising machines and a quantum annealer by R. Hamerly et al, Sci. Adv., 2019 (26 pages) for the comparison with D-Wave 2000Q. The compared CIM machine is from NTT and Stanford. The Stanford CIM is described in A fully programmable 100-spin coherent Ising machine with all-to-all connections by Peter L. McMahon et al, 2016 (9 pages). The NTT CIM is described in A coherent Ising machine for 2000-node optimization problems by T. Inagaki et al, Science, 2016 (6 pages).

1232 Coherent Ising machines models and challenges are described in Coherent ising machines - optical neural networks operating at the quantum limit by Y. Yamamoto et al, npj Quantum, December 2017 (16 pages).

1233 See Deterministic photonic quantum computation in a synthetic time dimension by Ben Bartlett, Avik Dutt and Shanhui Fan, Optica, November 2021 (9 pages) also described in Researchers propose a simpler design for quantum computers by McKenzie Prillaman, Stanford University, Physorg, November 2021.

The photon source is a single photon train source. But this is not a perfect solution. It has some requirements on the cavity quality, on low fiber attenuation, on very low insertion losses optical switches and can't implement quantum gates in parallel.

The general principle of quantum computing systems using photon qubits is as follows:

- **Photon sources** are lasers, often coupled with single and indistinguishable photon generators [1234]. They are critical to generate simultaneously a large number of indistinguishable photons that will feed in parallel several qubits thanks to delay lines. These are well time-isolated unique and indistinguishable photons generated in well-spaced in time series. These single photons are individually detectable at the end of processing with single photon detectors. The key metrics of these photon sources are the system efficiency (probability that at least one photon is created per pulse), purity (probability of getting a maximum of one photon per pulse) and coherence (how generated photons are quantum mechanically identical or indistinguishable).

  The purity and high probability to get a photon per clock cycle are the enabler of quantum interferences based multiple qubit gates with discrete variable qubits. High-efficiency sources are qualified as "on-demand" or "deterministic" with the alternatives of heralded sources, where the emission time can be accurately measured.

  There are two main types of single photons sources (SPSs)[1235]:

  **Quantum dot single-photon source** are the best-in-class devices, able to generate photons with a 99.7% single-photon purity, and overs 65% extraction efficiency, which could potentially reach 80% (meaning, 4 photons generated out of 5 clock cycles). See below in Figure 428 how these efficiencies are improved. These sources also have an over 99% photon indistinguishability. In the second quantization formalism, they create a single Fock state with a photon number equal to one.

  The leaders in this market are **Quandela**[1236] and **Sparrow Quantum**[1237]. And many research labs are working on other varieties of quantum dots[1238]. These photon sources must be cooled at about 3K to 4K. In their latest Prometheus generation, Quandela directly couples the quantum dot to a fiber, avoiding the use of cumbersome confocal microscopes and significantly increasing the photon generation yield. It creates a path to reaching a combined source–detector efficiency closer to the 2/3 threshold that is mandatory for scalable discrete variable optical quantum computing.

  Others are trying to operate these quantum dots at ambient temperature. That's the case with the quantum dots single photon emitters developed at EPFL, with a mix of gallium nitride and aluminum nitride (GaN/AlN) on silicon substrate.

---

[1234] See Near-ideal spontaneous sources in silicon quantum photonics by S. Paesani et al, 2020 (6 pages) which describes a single photon source based on a photonics component. It is an Anglo-Italian research project.

[1235] See Integrated photonic quantum technologies by Jianwei Wang et al, May 2020 (16 pages).

[1236] See The race for the ideal single-photon source is on by Sarah Thomas and Pascale Senellart, Nature Nanotechnology, January 2021 (2 pages) which describes the various ways to improve the yields of single photon sources, Sequential generation of linear cluster states from a single photon emitter by D. Istrati, Niccolo Somaschi, Hélène Ollivier, Pascale Senellart et al, October 2020 and Reproducibility of high-performance quantum dot single-photon sources by Hélène Ollivier, Niccolo Somaschi, Pascale Senellart et al, October 2019 (10 pages) on benchmarking single photon sources.

[1237] See Scalable integrated single-photon source by Ravitej Uppu et al, December 2020 (7 pages) which describes the latest advancements of their technology.

[1238] See Planarized spatially-regular arrays of spectrally uniform single quantum dots as on-chip single photon sources for quantum optical circuits by Jiefei Zhang et al, University of Southern California and IBM, November 2020 (8 pages) describes an array with 32 quantum dots and a simultaneous purity of single-photon emission over 99.5%.

These are showcasing a single-photon purity of 95% at cryogenic temperatures (below 50K) and a purity of 83% at room temperature. The photon emission rates reaches 1 MHz with a single-photon purity exceeding 50%[1239].

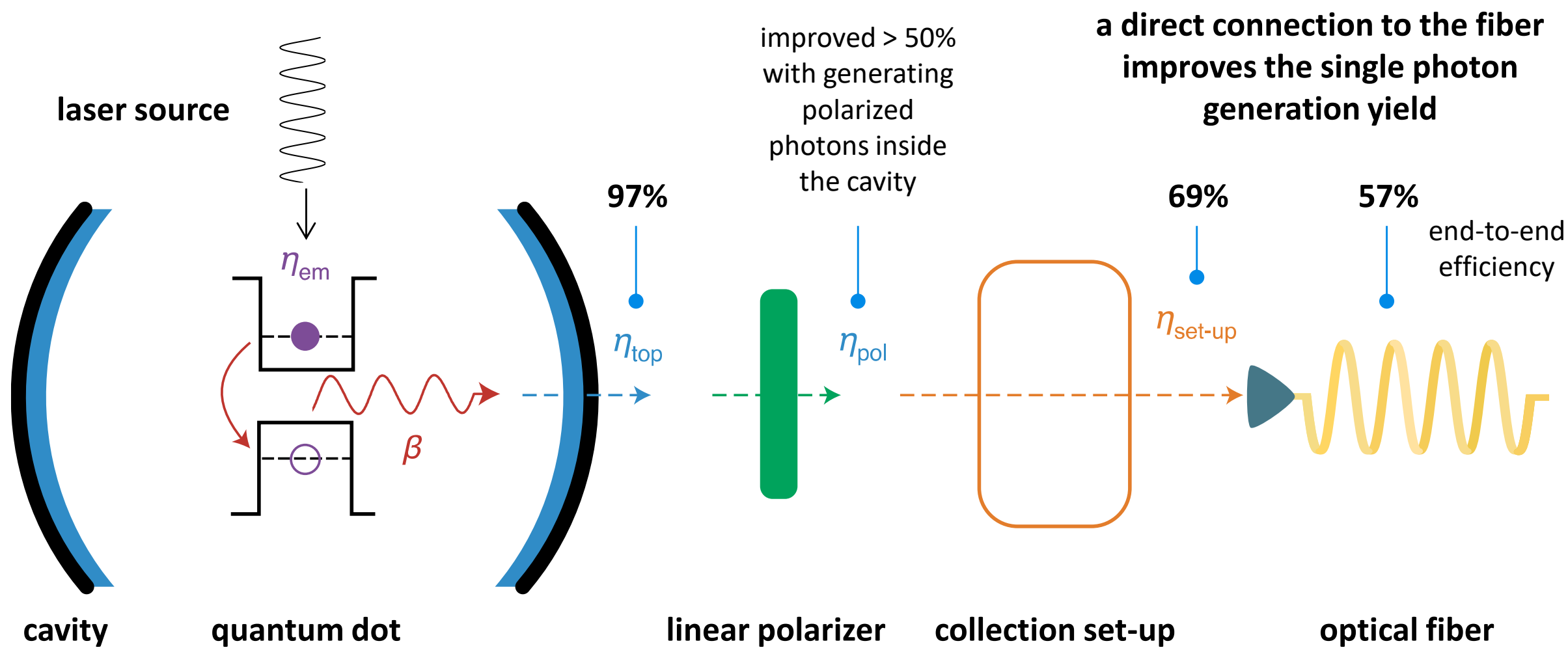

*Figure 428: what characterizes the efficiency of a quantum dots photon generator. Source: The race for the ideal single-photon source is on by Sarah Thomas and Pascale Senellart, Nature Nanotechnology, January 2021 (2 pages) and comments by Olivier Ezratty, 2021.*

**Parametric photon-pair sources** are laser pumping nonlinear optical waveguides or cavities that create photon-pairs. It can be integrated in nanophotonic circuits. They are using either spontaneous four-wave mixing (SFWM) or spontaneous parametric down-conversion (SPDC) processes in nonlinear crystals. The efficiency is lower than with quantum dots, reaching about 50% with a 95% photon indistinguishability from separated SPSs. Photons are created non-deterministically with a rather low 5% to 10% probability, which can be increased to above 60% with time and spatial domains multiplexing.

Such solutions are embedded in nanophotonic solutions like with **PsiQuantum**. SPDC sources work at room temperature but for efficient multiplexing (>95%), it is necessary to use SNSPD detectors running at low temperature. Progress is being made with nanophotonic-based single photons generation, although their performance still lags quantum-dots sources[1240]. Figure 429 shows a SPDC method to create pairs of entangled photons. The conversion creates pairs of orthogonally polarized photons in two light cones with entangled photons at their intersection.

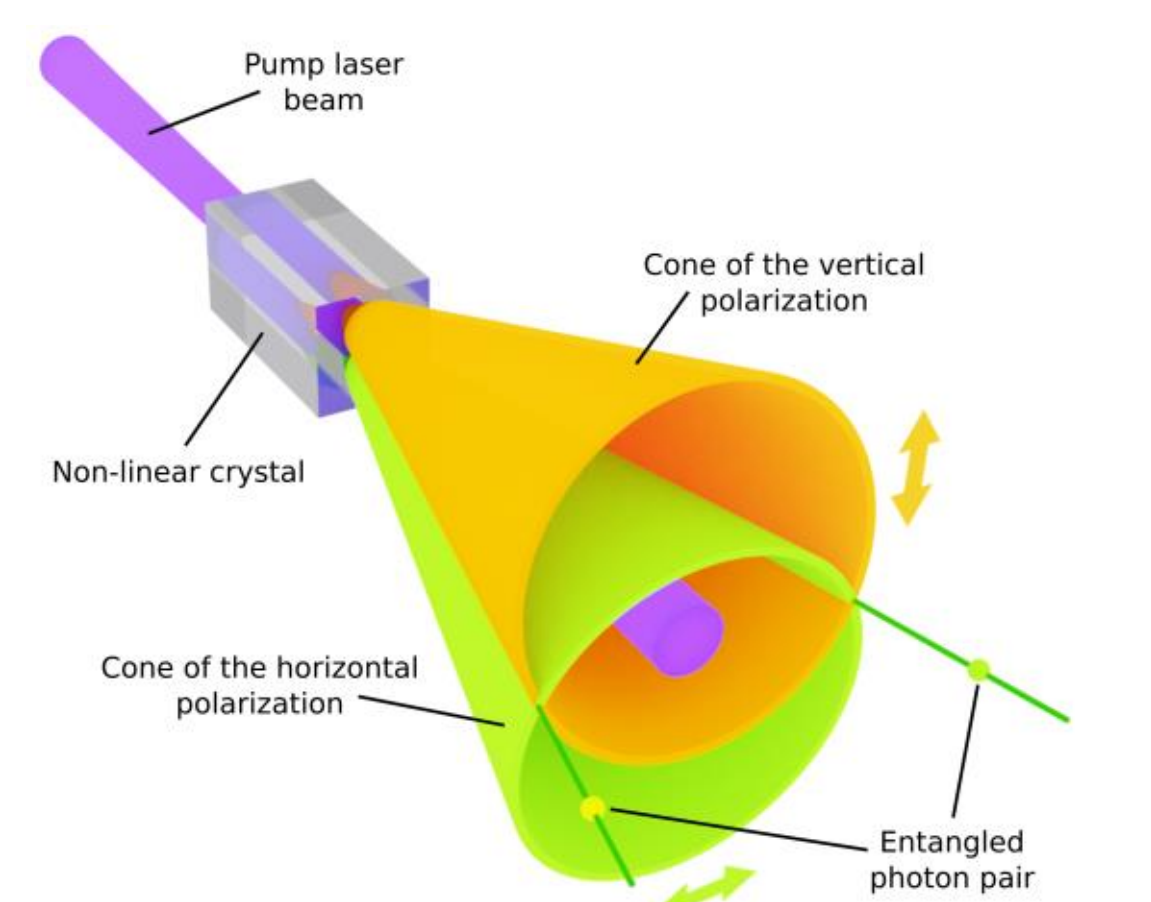

*Figure 429: how entangled photons are generated with SPDC method.*

[1239] See Toward Bright and Pure Single Photon Emitters at 300 K Based on GaN Quantum Dots on Silicon by Sebastian Tamariz, Nicolas Grandjean et al, January 2020 (19 pages).

[1240] See High-efficiency single-photon generation via large-scale active time multiplexing by F. Kaneda et al, October 2019 (7 pages), Researchers create entangled photons 100 times more efficiently than previously possible pointing to Ultra-bright Quantum Photon Sources on Chip by Zhaohui Ma et al, October 2020 (5 pages) and A bright and fast source of coherent single photons by Natasha Tomm et al, University of Basel and Ruhr-Universität Bochum, July 2020 (14 pages).

One key challenge with implementing MBQC, one dominant photonic quantum computing architecture that we'll cover later, is the ability to create large cluster-states of entangled photons. There are many options to implement it. Quantum dots source can be tailored for this need. Good indistinguishable and deterministic photon sources can be coupled with delay lines and mixers to create these cluster states. There are even solutions to create hundred pairs of entangled photons, as investigated by the University of Virginia, using frequency combs. Here, the photon source is a continuous laser emitting a single continuous wave, a small 3 mm Kerr microcavity creates a frequency comb generating pairs of entangled photons around the pump frequency as described below. This could lead to massive multimode photonic quantum computing[1241].

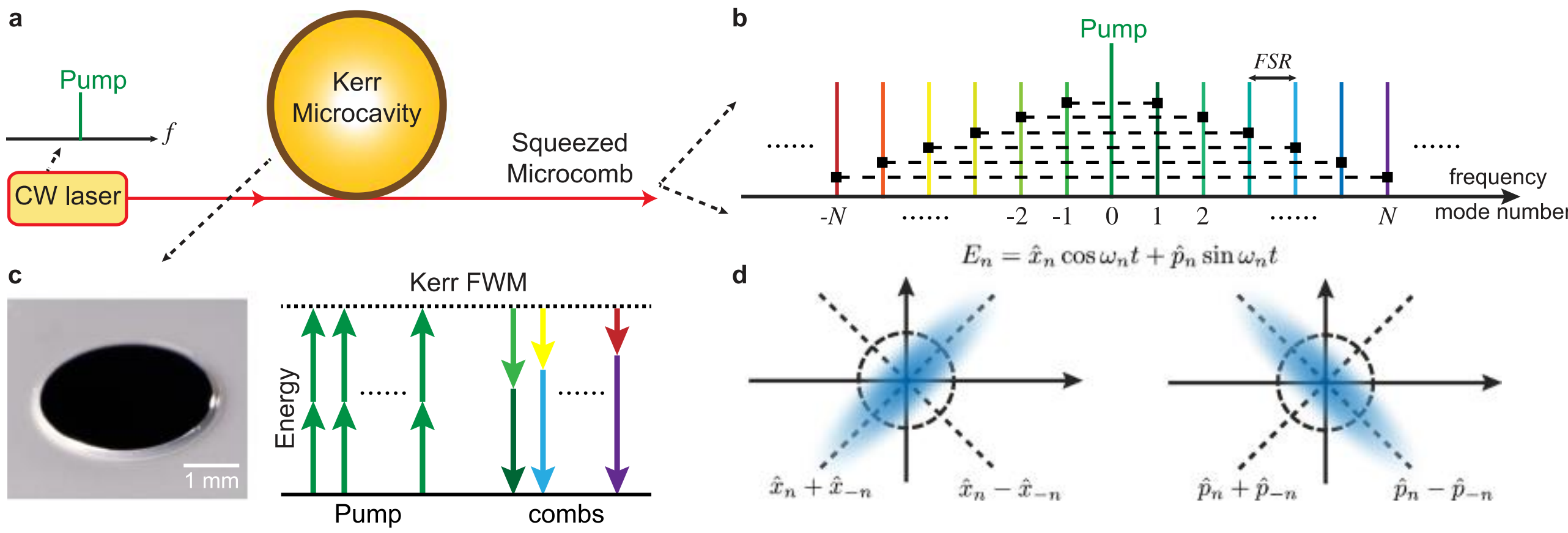

*Figure 430: a frequency comb method to generate a large cluster state of entangled photons. Source: A squeezed quantum microcomb on a chip by Zijiao Yang et al, Nature Communications, August 2021 (8 pages).*

- **Quantum state** is based on a single or several properties of the photons. The most common is their polarization with a computational basis based on horizontal and vertical polarization. Other parameters of photons are also explored to create qubits such as their phase, amplitude, frequency, path, photon number, spin orbital momentum and even orbital angular momentum[1242]. This potentially allows the creation of qutrits or qudits managing more than two exclusive values. Photons are "flying qubits" because they move in space and are not static or quasi-static at the macroscopic scale unlike most other types of qubits[1243].

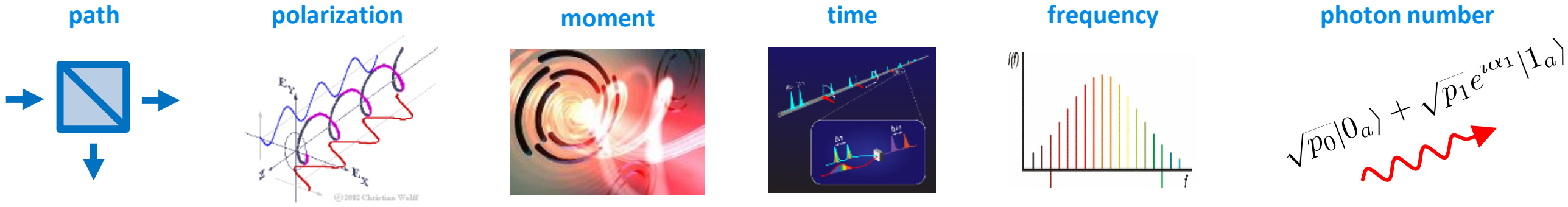

*Figure 431: the various properties or observables of photons that can be used to create a qubit. You have many more solutions than the old-fashioned polarization! Compilation (cc) Olivier Ezratty, 2021.*

---

[1241] See A squeezed quantum microcomb on a chip by Zijiao Yang et al, Nature Communications, August 2021 (8 pages) covered in Researchers open a path toward quantum computing in real-world conditions by Karen Walker, University of Virginia, August 2021.

[1242] This multiplicity of parameters also makes it possible to encode not only qubits but also qudits, with a greater number of states. But it is quite complex to manage and, if only to manage two-qubit quantum gates, we are happy with qubits instead of using qudits. See also Forget qubits -scientists just built a quantum gate with qudits by Kristin Houser, July 2019, which refers to High-dimensional optical quantum logic in large operational spaces by Poolad Imany et al, 2019 (10 pages). See the definition of orbital angular momentum in the glossary. It was discovered in 1992. See Orbital angular momentum of light and the transformation of Laguerre-Gaussian laser modes by Les Allen et al, 1992 (5 pages).

[1243] The other qubits are "non-flying": spin-controlled electrons trapped in a cavity, cold atoms (which are stabilized in space) and trapped ions (which can move, but in a limited space), NV centers (cavities do not move) and superconducting circuits (which are fixed in space even if they use pairs of circulating Cooper electrons).

One such implementation was achieved in 2022 by a China-international team using ququarts states (quantum objects with four dimensions instead of the two dimensions of qubits)[1244]. The experiment shown below in Figure 432 was implemented in a programmable silicon CMOS photonic integrated chipset of 15 x 1.5 mm implementing linear optics and enhancing parallelism. It was tested with a QFT, Deutsch-Jozsa and Bernstein-Vazirani, quaternary phase estimation and factorization algorithms. The system creates four-level entangled state in an array of four integrated identical SFWM (spontaneous four-wave mixing) sources. Then the photons traverse wavelength-division multiplexing filters (WDM), Mach-Zehnder interferometers (MZI). thermal-optic phase shifters (TOPSs), multimode interferometer beamsplitters (MMI) and qudit states measurement.

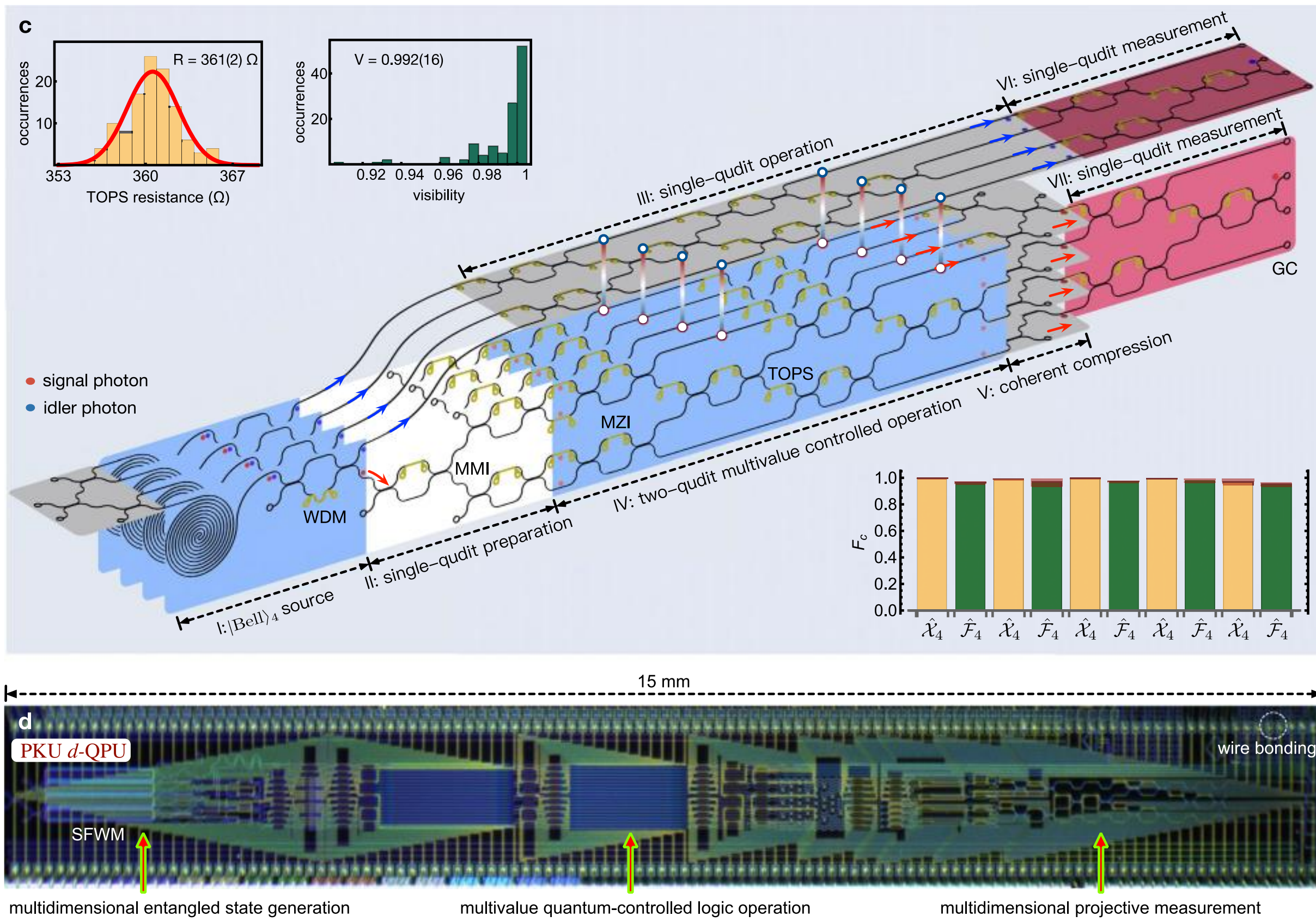

*Figure 432: a ququart photons processor created in China. A programmable qudit-based quantum processor by Yulin Chi, Jeremy O'Brien et al, Nature, March 2022 (10 pages).*

- **Single-qubit quantum gates** use simple optical circuitry, including beamsplitters, waveplates, mirrors and semi-reflective mirrors, and phase shifters[1245]. For example, a Hadamard gate (H) uses a beamsplitter or waveplate, a Pauli X gate (bit flip) combines a beamsplitter and a Hadamard gate, and a Pauli Z gate (phase flip) uses a phase shifter causing a 180° phase change (π).

[1244] See A programmable qudit-based quantum processor by Yulin Chi, Jeremy O'Brien et al, Nature, March 2022 (10 pages) covered in Scientists Make Advances in Programmable Qudit-based Quantum Processor by Matt Swayne, The Quantum Insider, March 2022.

[1245] This is based on the KLM scheme proposed in A scheme for efficient quantum computation with linear optics by Emanuel Knill, Raymond Laflamme and Gerard Milburn, 2001 (7 pages).

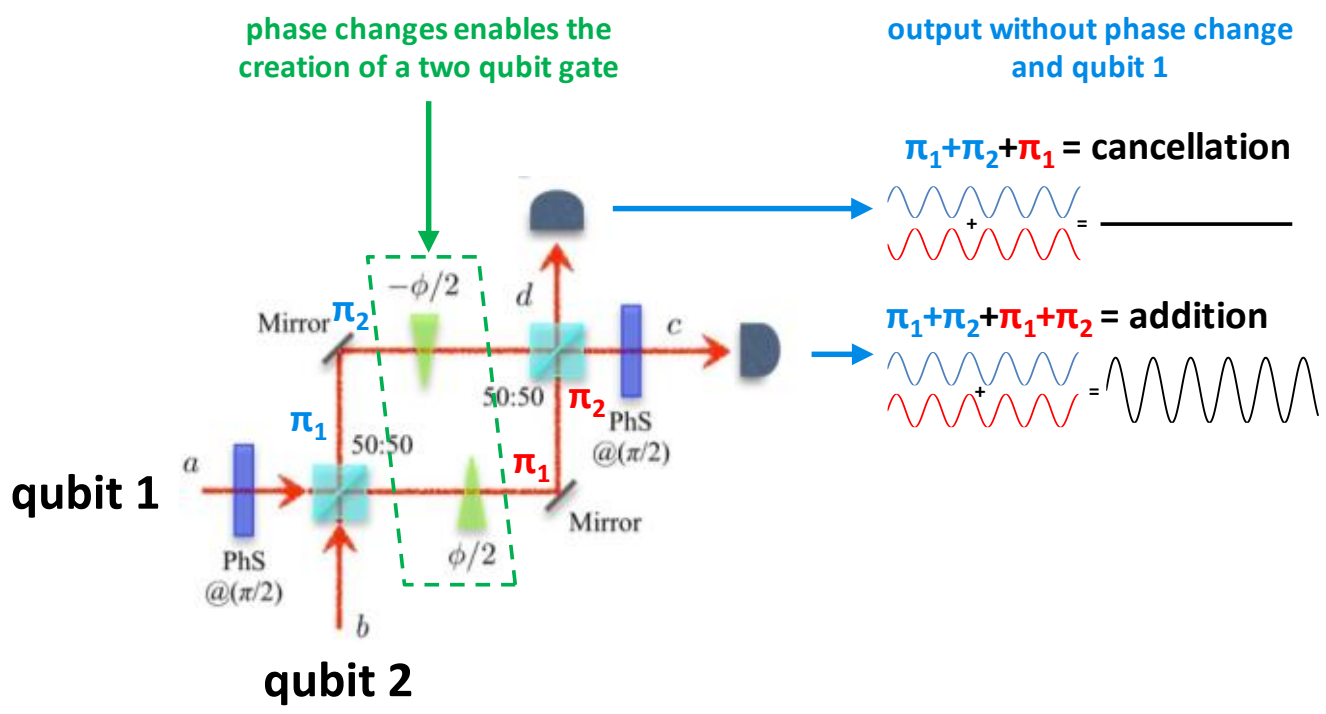

**Mach-Zehnder interferometer (MZI)**
creates an interference between two photons with introducing a dephasing linked to the number of mirrors reflections (the πs) and phase changers

The implementation of various quantum logic gates using optical MZI is tabulated in Table. 3.

| Quantum Logic Gate | Unitary Matrix | Relation for MZI implementation | Elements |
|---|---|---|---|
| Beam Splitter(B(θ)) | $\begin{bmatrix} cos\theta & -sin\theta \\ sin\theta & cos\theta \end{bmatrix}$ | | |
| 50-50 Beam Splitter(B) | $\frac{1}{\sqrt{2}}\begin{bmatrix} 1 & -1 \\ 1 & 1 \end{bmatrix}$ | | |
| Hadamard(H) | $\frac{1}{\sqrt{2}}\begin{bmatrix} 1 & 1 \\ 1 & -1 \end{bmatrix}$ | H=BZ | 50-50 Beam splitter |
| Phase flip gate (Z) | $\begin{bmatrix} 1 & 0 \\ 0 & -1 \end{bmatrix}$ | Z=HB | π Phase shifter |
| Bit Flip gate (X) | $\begin{bmatrix} 0 & 1 \\ 1 & 0 \end{bmatrix}$ | X=BH | Beam Splitter, Hadamard |
| T gate | $\begin{bmatrix} 1 & 0 \\ 0 & \exp(i\pi/4) \end{bmatrix}$ | | π/4 phase shifter |
| S gate | $\begin{bmatrix} 1 & 0 \\ 0 & i \end{bmatrix}$ | | Quarter wave plate |
| Pauli Y gate | $\begin{bmatrix} 0 & -i \\ i & 0 \end{bmatrix}$ | | |
| CNOT gate | $\begin{bmatrix} 1 & 0 & 0 & 0 \\ 0 & 1 & 0 & 0 \\ 0 & 0 & 0 & 1 \\ 0 & 0 & 1 & 0 \end{bmatrix}$ | (I⊗H)×K×(I⊗H) | Kerr Media(K) Hadamard(H) Identity(I) |

**MZI based quantum gates**
a one qubit gate can be created with introducting some dephasing in one or two of the circuits, and two qubit gate with using two entries and some dephasing. The table shows the correspondence between quantum gates and the used filters elements.

Quantum Logic Processor: A Mach Zehnder Interferometer based Approach, 2006 (19 pages).

*Figure 433: how a Mach-Zehnder Interferometer works. Source: Quantum Logic Processor: A Mach Zehnder Interferometer based Approach by Angik Sarkar and Ajay Patwardhan 2006 (19 pages).*

- **Two-qubit quantum gates** are difficult to realize because it is not easy to have photons interact with each other, particularly when they are not perfectly indistinguishable. They use optical circuits based on beamsplitters or Mach-Zehnder interferometers with two inputs integrating phase changes on the optical paths, based on the KLM method already quoted in footnote.

  This does not work well when the photons are uneven, such as those coming from lasers. Namely, in only a few % of the cases. With indistinguishable photons, gates are more than 95% efficient since photons can interfere with each other, and add or subtract. It facilitates Mach-Zehnder interferometry operations.

  These sources have the additional advantage of being very bright, which allows them to multiply the incoming photons and then to pass through many quantum gates. There are also solutions based on cavities. Research is also active on the creation of nonlinearities to improve the reliability of these quantum gates[1246]. Ideally, nonlinear separating cubes should be used[1247].

  How about photonic gates execution times? It must be fast since photons traverse optical devices at nearly the speed of light in vacuum. Traversing a one meter long series of optical instruments would last only 3 ns. If the circuit is nanophotonic based, we'll get into the cm realm and reach tens of pico-seconds. But that's not the right way to evaluate the speed of quantum computing here, particularly when dealing with non-deterministic photon sources where computing has to be repeated many times. Then, you may need to take into account the speed of the electronics that define the various quantum gates along the photon route and we may end-up adding milliseconds.

- **Qubit readout** uses single photon detectors that also capture their quantum state. This detection is still imperfect. Several single-photons detection technologies are competing: SPAD (avalanche photodiodes, which detect photon occurrences but not photon number)[1248], transition edge sensor

[1246] See Quantum Computing With Graphene Plasmons, May 2019 which refers to Quantum computing with graphene plasmons by Alonso Calafell et al, 2019. This is the creation of two-qubit quantum gates with graphene-based nonlinear structures. It comes from the University of Vienna in Austria and from Spanish and Serbian laboratories. As well as Researchers see path to quantum computing at room temperature by Army Research Laboratory, May 2020 which refers to Controlled-Phase Gate Using Dynamically Coupled Cavities and Optical Nonlinearities by Mikkel Heuck, Kurt Jacobs and Dirk R. Englund, 2020 (5 pages).

[1247] It is a function that can be realized with Quandela's single photon generation component, diverted from its original use. See also Researchers see path to quantum computing at room temperature, May 2020 which refers to Controlled-phase Gate using Dynamically Coupled Cavities and Optical Nonlinearities by Mikkel Heuck, September 2019 (5 pages) and discusses a nonlinear cavity optical quantum gate technique.

[1248] See recent progress with SPADs in Low-noise photon counting above 100 × 106 counts per second with a high-efficiency reach-through single-photon avalanche diode system by Michael A. Wayne et al, NIST, December 2020 (6 pages).

(TES, which can detect photon numbers) and Superconducting Nanowire SPDs (SNSPDs, which also detect photon numbers). Fully integrated SNSPDs are based on GaAs, Si and $Si_3N_4$ waveguides. In order to limit the dark count phenomenon coming from thermal effects, these SNSPDs are usually cooled between 800 mK and 3K which requires a dilution refrigeration system[1249]. NbTiN-based SNSPDs could work with higher-temperature cooling, between 2.5K to 7K[1250]. One goal is to integrate these photon detectors directly in photonic computing circuits. Other detectors are specialized for analyzing continuous variables qubits, like homodyne and heterodyne detectors. There are even concepts of single photon detectors that can also detect their frequency with high precision[1251]. In the continuous variable types of qubits, qubit readout uses homodyne detectors to detect the photon quadrature such as optical parametric amplifier (OPA)[1252]. Others are experimenting photon detectors operating at a relatively hot temperature of 20K, using cuprates who are known for their relatively high superconducting temperature[1253].

From a physical point of view, these items are classical photonic components: single and identical photon sources, light guides, optical delay lines (optical fibers or voltage-controlled Pockels cells), Mach-Zehnder interferometers, beam splitters (splitters, which divide an optical beam into two beams, generally identical), birefringent filters (which have two different refractive indices), phase shifters and single photon detectors[1254].

To conduct experiments, these discrete and very affordable components are installed on carefully calibrated optical tables of a few square meters with lots of instruments and photons that circulate largely in the free space of a darkened room.

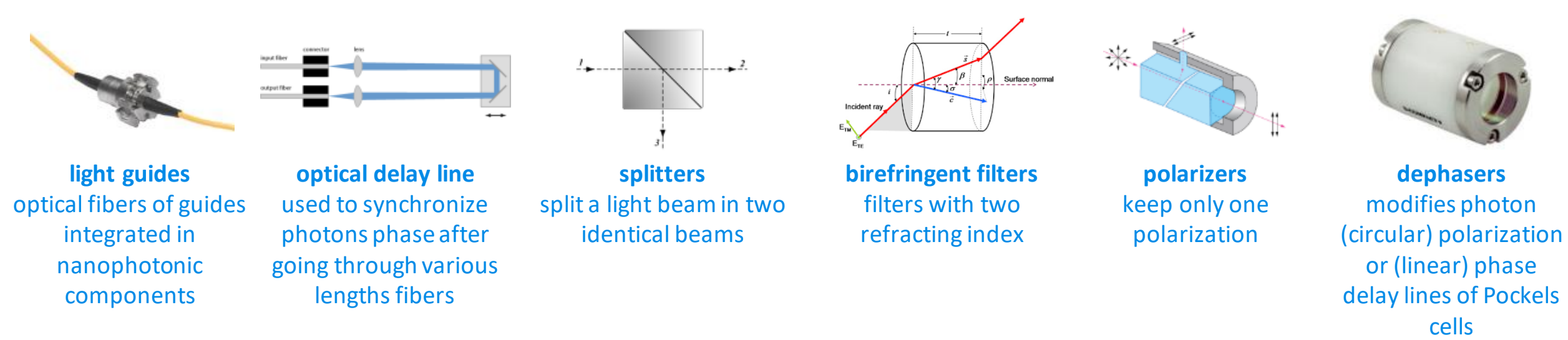

*Figure 434: the various optical tools to control light in a quantum processor. These are made for experiments and can be miniaturized in nanophotonic circuits. Compilation (cc) Olivier Ezratty, 2021.*

Fortunately, these optical components are miniaturizable on semiconductor integrated circuits. This is part of the vast field of nanophotonics. Nanophotonics components are etched with densities between 220 nm and 3 µm [1255].

---

1249 See The potential and challenges of time resolved single-photon detection based on current-carrying superconducting nanowires by Hengbin Zhang et al, October 2019 (19 pages) and Superconducting nanowire single-photon detectors for quantum information by Lixing You, June 2020 (20 pages). Dark counts are detected photons coming from the environment due to thermal or tunneling effects.

1250 See Superconducting nanowire single photon detectors operating at temperature from 4 to 7 K by Ronan Gourgues et al, Optics Express, 2019 (9 pages).

1251 See Nanoscale Architecture for Frequency-Resolving Single-Photon Detectors by Steve M. Young et al, Sandia Labs, May 2022 (20 pages).

1252 See Towards a multi-core ultra-fast optical quantum processor: 43-GHz bandwidth real-time amplitude measurement of 5-dB squeezed light using modularized optical parametric amplifier with 5G technology by Asuka Inoue et al, May 2022 (18 pages).

1253 See Two-dimensional cuprate nanodetector with single photon sensitivity at T = 20 K by Rafael Luque Merino, Dmitri K. Efetov et al, August 2022 (27 pages). I also found out Graphene-based Josephson junction single photon detector by Evan D. Walsh, Thomas A. Ohki, Dirk Englund et al, September 2017 (12 pages) but it seems not applicable for photonic qubit readout. It works at 25 mK.

1254 This is well explained in Silicon photonic quantum computing by Syrus Ziai, PsiQuantum, 2018 (72 slides) as well as in Large-scale quantum photonic circuits in silicon, by Nicholas C. Harris, Dirk Englund et al, Nanophotonics, 2016 (13 pages).

1255 See for example the work of InPhyNi discussed in High-quality photonic entanglement based on a silicon chip by Dorian Oser, Sébastien Tanzilli et al, 2020 (9 pages).

In nanophotonics, quantum gates are dynamically programmed by the conditional routing of photons in optical circuits and/or with their modes of generation (polarization, phase, frequency, …).

These circuits are often etched on CMOS (silicon) or III/V (especially germanium) components. These components could be assembled in a modular way as shown in the functional diagram in Figure 435[1256]. This enables a better management of processes heterogeneity used to create these different circuits.

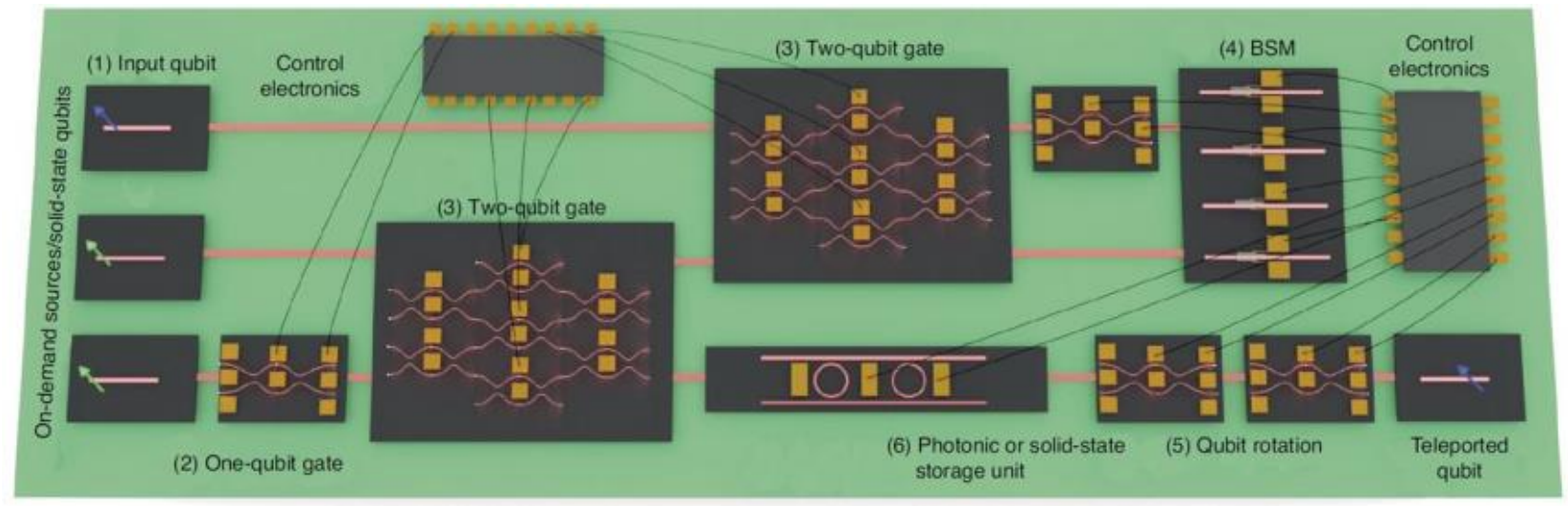

Figure 435: a nanophotonic circuit functional diagram. Source: Hybrid integrated quantum photonic circuits by Ali W. Elshaari et al, 2020 (14 pages).

Many semiconductor fabs in the world are helping photonicians design and prototype nanophotonic circuits to support photon qubits. We'll mention here only a few of them. Many fab technologies are investigated with classical silicon-based CMOS, hybrid CMOS with silicon nitride (SiN) and lithium niobate ($LiNbO_3$), III/V materials (GaAs[1257], InP, ...), etc[1258].

In France, **CEA-Leti** is also building an integrated silicon photonic qubits platform including single photons source, phase shifters and superconducting nanowire single-photon detectors (SNSPD) or CdHgTe avalanche photodiodes (APD), working at 2,5-4K that is compatible with single photon detectors. They are initially targeting secured QKD based telecommunications.

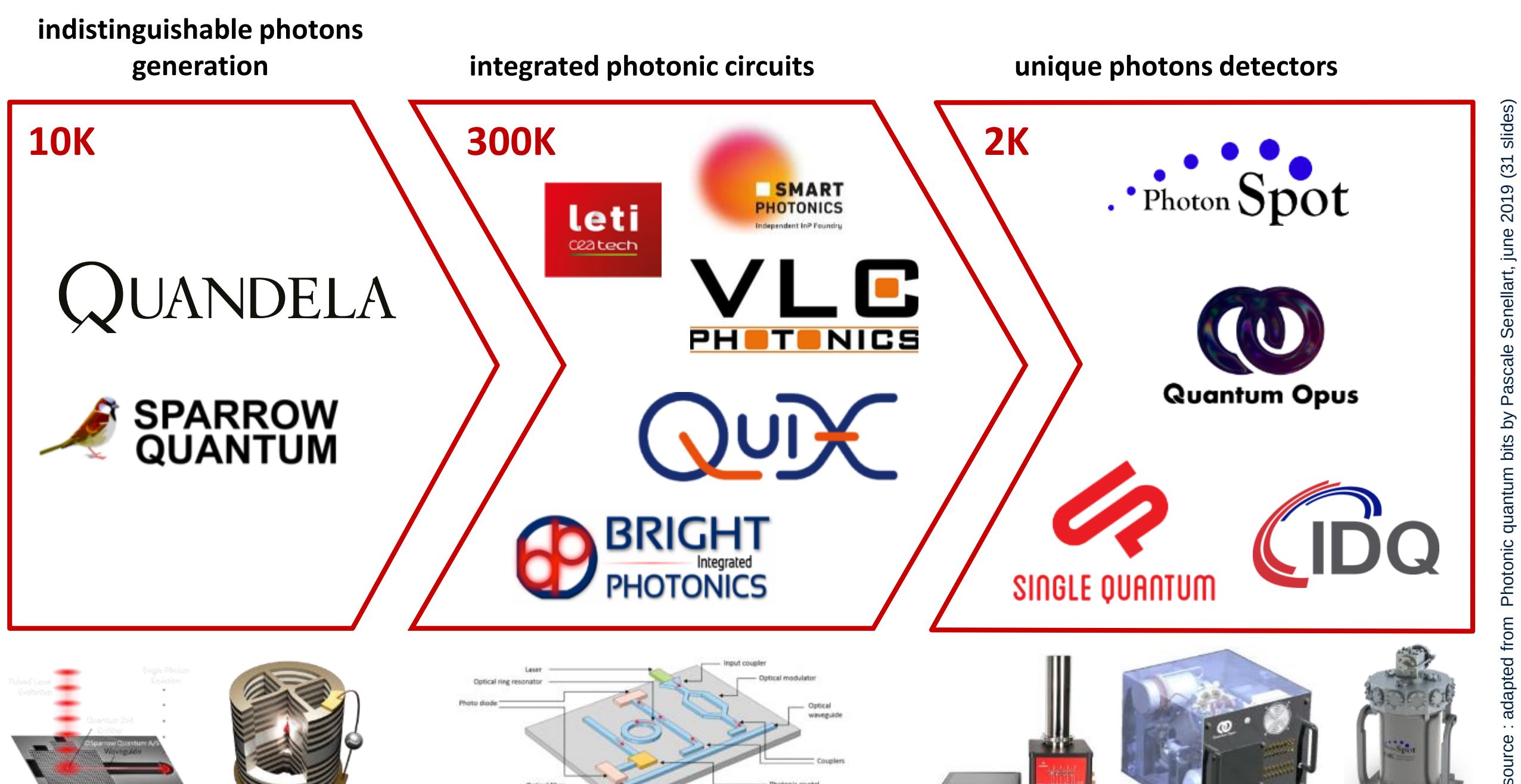

Figure 436: the key components of a photonic quantum computer: quality photon sources, preferably deterministic, nanophotonic circuits for processing, and photon detectors for readout. Source: adapted from Photonic quantum bits by Pascale Senellart, June 2019 (31 slides) in slide 11.

[1256] See Hybrid integrated quantum photonic circuits by Ali W. Elshaari et al, 2020 (14 pages).

[1257] See Expanding the Quantum Photonic Toolbox in AlGaAsOI by Joshua E. Castro et al, May 2022 (9 pages). They implement non linear elements, edge couplers, waveguide crossings, couplers, and MZIs in Aluminum gallium arsenide-on-insulator (AlGaAsOI).

[1258] See the review paper Roadmap on integrated quantum photonics by Galan Moody, Jacquiline Romero, Eleni Diamanti et al, August 2021 (108 pages).

Ultimately, a photon qubits quantum computer would consolidate three key components as shown in Figure 436: a single photon generator, integrated photonic circuits and single photon detectors. The first and last ones are integrated with a cryogenic system operating at about 10K and 2K-4K respectively. But it seems also possible to integrate photon sources and detectors in a single photonic chipset[1259].

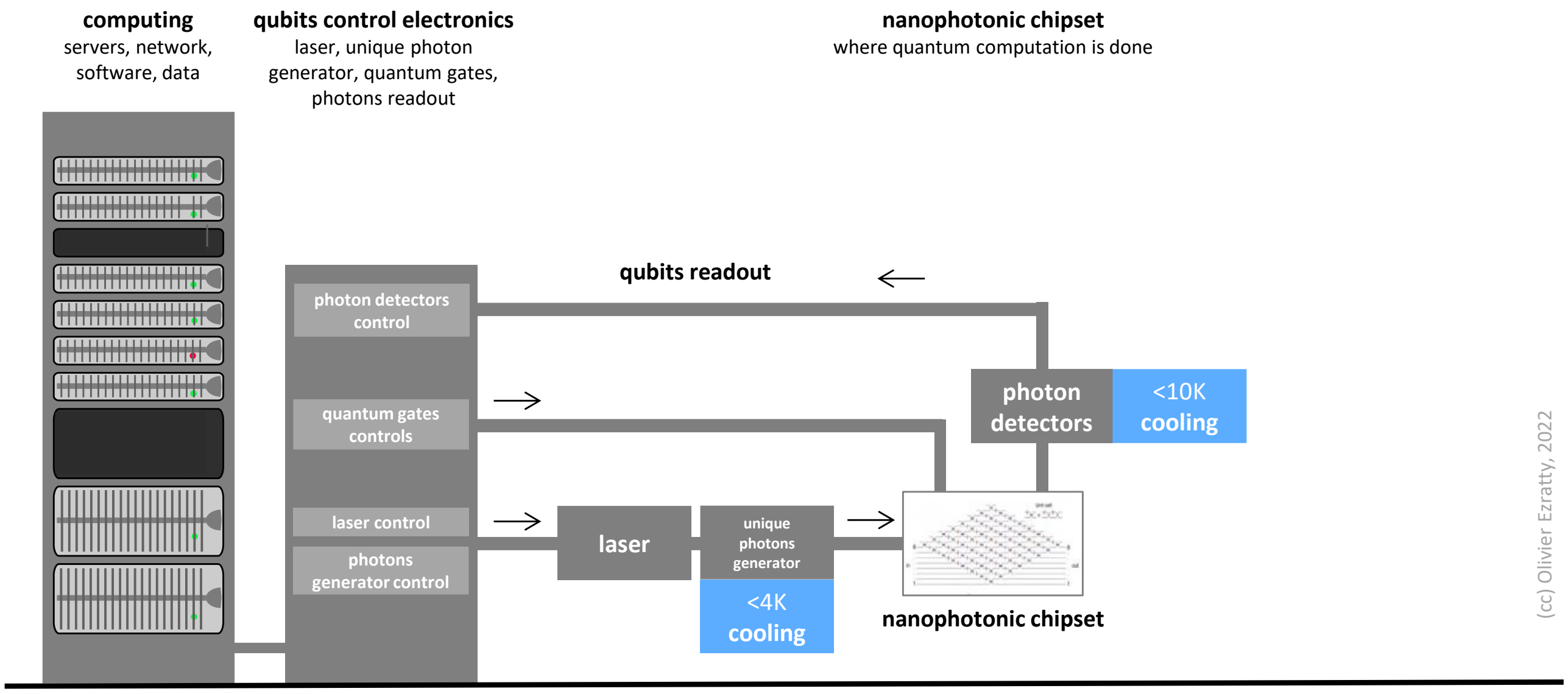

*Figure 437: typical architecture of a photon qubits quantum computer. (cc) Olivier Ezratty, 2022.*

The most active countries in the field seem to be China, the UK (particularly at the Universities of Oxford, Bristol, Cambridge and Southampton)[1260], France (C2N, LKB, ...), Italy[1261], Germany (Universities of Stuttgart and Paderborn), Austria, Australia, Japan and of course the USA.

Photon qubits are the specialty of some startups like **PsiQuantum**, **Orca Computing**, **Tundra Systems Global**, **QuiX, Quandela, Nu Quantum,** and **Xanadu**.

## Boson sampling

In photonics, the simulation of boson sampling is an experiment that is used to showcase the advancement of photon qubits. The idea of boson sampling came from **Scott Aaronson** and **Alex Arkhipov** from the MIT in a paper published in 2010[1262]. They devised a linear optics-based experiment that would be impossible to easily emulate on a classical supercomputer[1263].

Boson sampling is about solving a problem of sampling the probability distributions of identical and indistinguishable photons being mixed in an interferometer and reaching single photon detectors. This physical process is impossible to emulate above a certain threshold, which generates yet another so-called "quantum supremacy" or “advantage”.

[1259] See Integrated nanophotonics for the development of fully functional quantum circuits based on on-demand single-photon emitters by S. Rodt and S. Reitzenstein, APL Photonics, December 2020 (14 pages).

[1260] According to Quantum Age technological opportunities from the UK Government Office of Science in 2016 (64 pages).

[1261] Fabio Sciarrino of La Sapienza University in Rome, carried out in 2013 a sampling of bosons with a chip with 13 input ports and 13 output ports, with three photons. See Efficient experimental validation of photonic boson sampling against the uniform distribution, 2013 (7 pages).

[1262] See The computational Complexity of Linear Optics by Alex Arkhipov and Scott Aaronson, 2010 (94 pages).

[1263] In quantum computing, we rely on only one type of boson: the photon. The other bosons are elementary particles such as gluons or Higgs bosons that can only be observed in particle accelerators. There are also composite particles such as the Cooper pairs (double electron) which are at the origin of superconducting currents. But when we talk about boson sampling, we always mean "photon".

A classical emulation requires extremely heavy matrix computing: the evaluation of square matrix permanents[1264]. This sits in the "#P difficult" problem class of the complexity theories zoo[1265]. The verification of the obtained result can't even be carried out by a classical computer[1266].

Boson sampling is the quantum and photonic analogue of the famous **Galton** plate experiment where balls cross rows of nails in a random way and end up in columns, with a Gaussian distribution.

This experiment is based on various probability concepts: convergence of a binomial distribution law towards a normal or Gaussian distribution, Moivre-Laplace theorem, etc. In the photon-based experiment, photons are injected into a series of interferometers combining them with their neighbor in a random way. On the other hand, the distribution at the end does not follow a Gaussian curve.

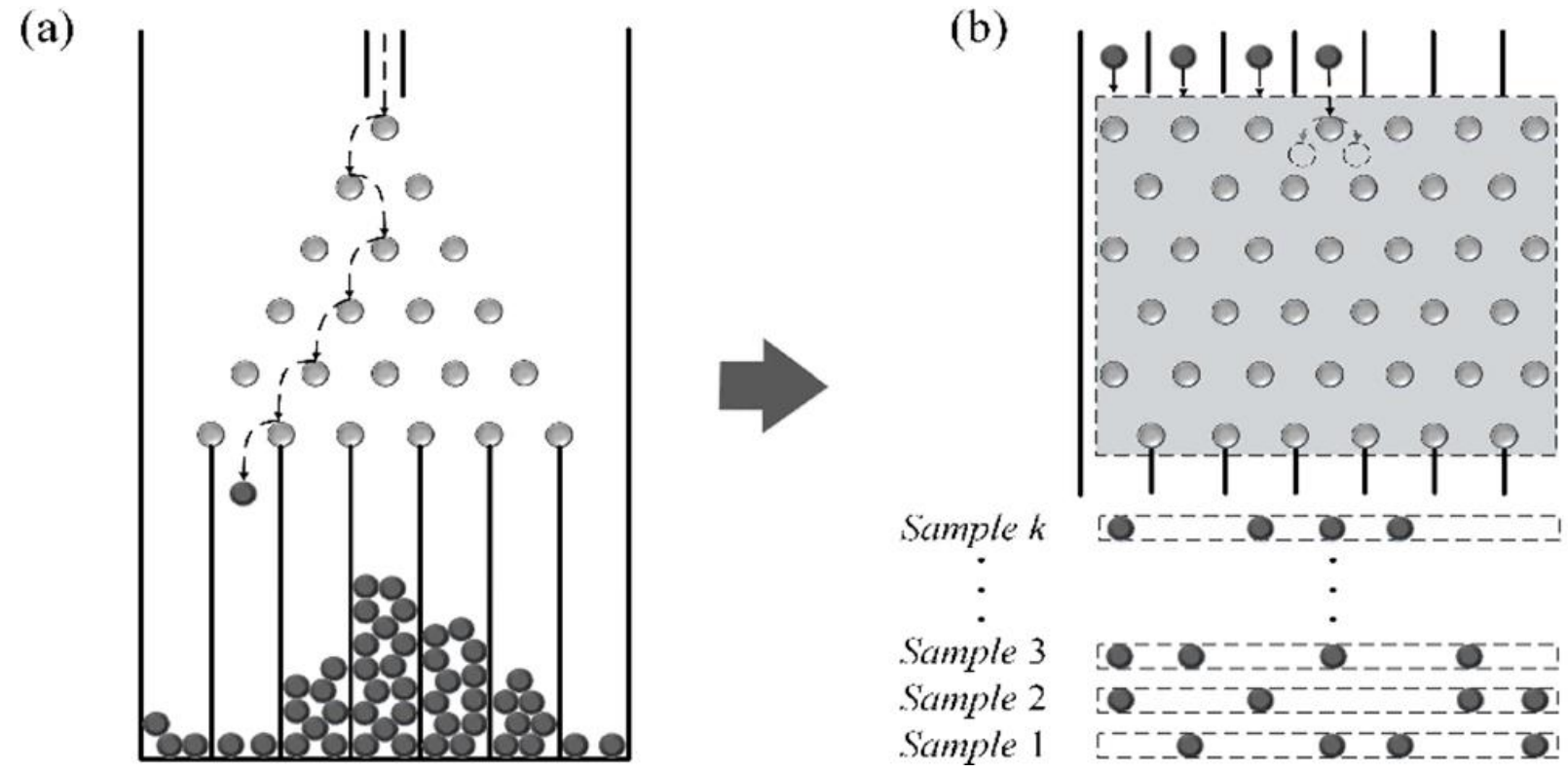

*Figure 438: the typical Galton plate experiment that inspires Boson sampling. Source: Quantum Boson-Sampling Machine by Yong Liu et al, 2015.*

It depends on the photons being sent upstream.

The appropriateness of the boson sampling style exercise is questionable. It implements a physical phenomenon with photons that is difficult to emulate in a classical way[1267]. However, it is not strictly a form of calculation with some problem input data. There is not even a real notion of qubits, quantum gates and programming, except in the choice of the photons we send into the system. The optical components used are all passive and static, except the photon generators and detectors[1268].

It is a physics experiment generating additive and subtractive interferences and superposition of quantum states[1269]. The difficulty of the experiment lies mainly in the complexity of the production of identical and indistinguishable photons. At this stage, nobody has managed to transform (or reduce) a useful algorithm into boson sampling. However, this could eventually lead to applications in homomorphic encryption and blind computing[1270].

---

1264 If you want to explore the question, see for example Lecture 3: Boson sampling by Fabio Sciarrino, University of Rome, (63 slides) and Experimental boson sampling with integrated photonics (33 slides) by the same author who describes laser-based techniques for etching integrated photonic components. As well as Permanents and boson sampling by Stefan Scheel, University of Rostock, 2018 (21 slides). As for the definition of the notion of permanent in Wikipedia, it uses notions and notations of linear algebra that are not even explained. The permanent of a matrix is a variant of its determinant. If the classical resolution of sampling requires the computation of matrix permanents, its resolution by linear optics system does not allow the computation of matrix permanents.

1265 #P is the class of function problems that counts the number of solutions of NP problems.

1266 In 2018, a Chinese team carried out a numerical simulation of 50 photon boson sampling using 320,000 processors from the Tianhe-2 supercomputer. See A Benchmark Test of Boson Sampling on Tianhe-2 Supercomputer, 2018 (24 pages). With the 20 photons and 60 modes of the Chinese experiment published in October 2019, a supercomputer is no longer able to follow.

1267 But this is a valid reality for the simulation of many complex physical phenomena, such as the folding of a protein or the functioning of a living cell, except that these remain in the realm of the living and are not simulated in a machine.

1268 See An introduction to boson-sampling by Jonathan Dowling et al, 2014 (13 pages) which describes well the issues involved in conducting boson sampling.

1269 See the animation Boson Sampling with Integrated Photonics, 2015 (3mn) which describes the path of photons in a boson sampling experiment as well as Photonic implementation of boson sampling: a review by Fabio Sciarrino, 2019 (14 pages) which describes in detail this kind of experiment.

1270 Seen in Introduction to boson-sampling by Peter Rohde, 2014 (34 minutes) which refers to A scheme for efficient quantum computation with linear optics by Emanuel Knill, Raymond Laflamme and Gerard Milburn, 2001 (7 pages) which theorized that quantum computation based on linear optics was plausible. We owe them the KLM scheme or protocol (their initials), a linear optics quantum computing (LOQC) programming model that has the disadvantage of being very heavy in terms of the number of hardware devices.

There are also some algorithms for simulating molecular vibrations based on boson sampling[1271]. In 2020, a Chinese team was conducting an experiment similar to boson sampling to play a variant of the Go game[1272].

Chinese researchers are particularly active in the field[1273]. In June 2019, the **Hefei** laboratory created a boson sampling using six photons with three degrees of freedom, therefore, based on qutrits (three-state qubits)[1274]. The states of the photons were their traveled path, polarization and orbital angular momentum. With a gate error rate of 29%. In October 2019, Chinese researchers upgraded the feat to 20 photons with an experiment presented as reaching quantum supremacy, at the same time as the announcement of Google Sycamore[1275]. In this experiment described in the diagram *below,* 20 indistinguishable photons were sent in a series of splitters and ended up in 60 photon detectors. The output Hilbert space was limited to 14 detectors, with a size of $3.7*10^{14}$ or $2^{48}$. With the 60 activated detectors, this space should be able to reach a size of $60^{20}$ or $2^{118}$.

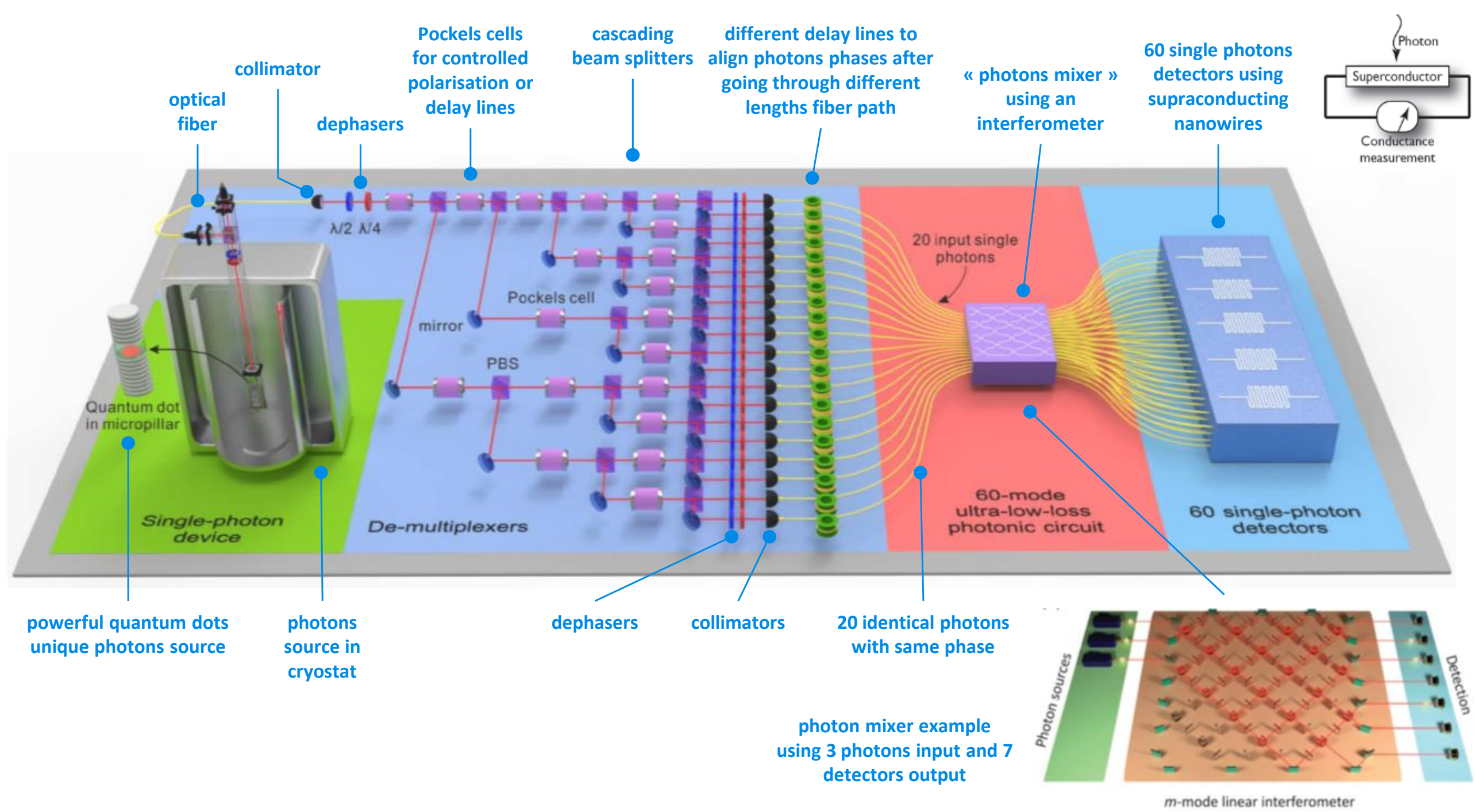

*Figure 439: one of the first Boson sampling experiment made in China, in 2019, with 20 photon modes. Source: Boson sampling with 20 input photons in 60-mode interferometers at $10^{14}$ state spaces by Hui Wang et al, October 2019 (23 pages).*

The size of Hilbert's space of such a device is evaluated with the size of the Fock space of M modes occupied by N photons. This would give a binomial space $\binom{M+N-1}{M}$ so $\binom{79}{60}$ which is equal in size to $\frac{79!}{60!*19!}$ (source).

[1271] See Boson sampling for molecular vibronic spectra by Joonsuk Huh, Alán Aspuru-Guzik et al, 2014 (7 pages) and Vibronic Boson Sampling: Generalized Gaussian Boson Sampling for Molecular Vibronic Spectra at Finite Temperature by Joonsuk Huh et al, 2017 (10 pages).

[1272] See Quantum Go Machine by Lu-Feng Qiao et al, July 2020 (16 pages).

[1273] See Chinese researchers on the road to the 'ultimate' quantum processor? by Bruno Cormier, September 2018 which points to Building Quantum Computers With Photons Silicon chip creates two-qubit processor by Neil Savage, September 2018 which discusses the creation of a two-qubit quantum processor. The original article is Large-scale silicon quantum photonics implementing arbitrary two-qubit processing, September 2018 (23 pages). The researchers involved were Chinese, English and Australian.

[1274] See 18-Qubit Entanglement with Six Photons Three Degrees of Freedom by Xi-Lin Wang et al, June 2019 (14 pages).

[1275] See Boson sampling with 20 input photons in 60-mode interferometers at $10^{14}$ state spaces by Hui Wang et al, October 2019 (23 pages).

The Chinese researchers indicated that they could use several hundred detectors in output and use a double encoding of photons (polarization and spatial encoding) to multiply the power of their system and thus make it able to create a NISQ (noisy intermediate scale quantum computer) system, except that the ability to program it does not seem obvious, nor its uses.

This represents the number of incoming photon detectors at the power of the incoming photon number. The previous record was 5 photons over 16 modes and the sampling was verifiable on a classical computer whereas with these 20 photons and 60 modes, it was no longer possible. The photon generator was realized with quantum dots in gallium and indium arsenide, placed in a 4K[1276] cryostat. The photon mixer used 396 beam splitters and 108 mirrors. For the experiment to work, one photon must arrive at the same time in all the inputs of the photon mixer.

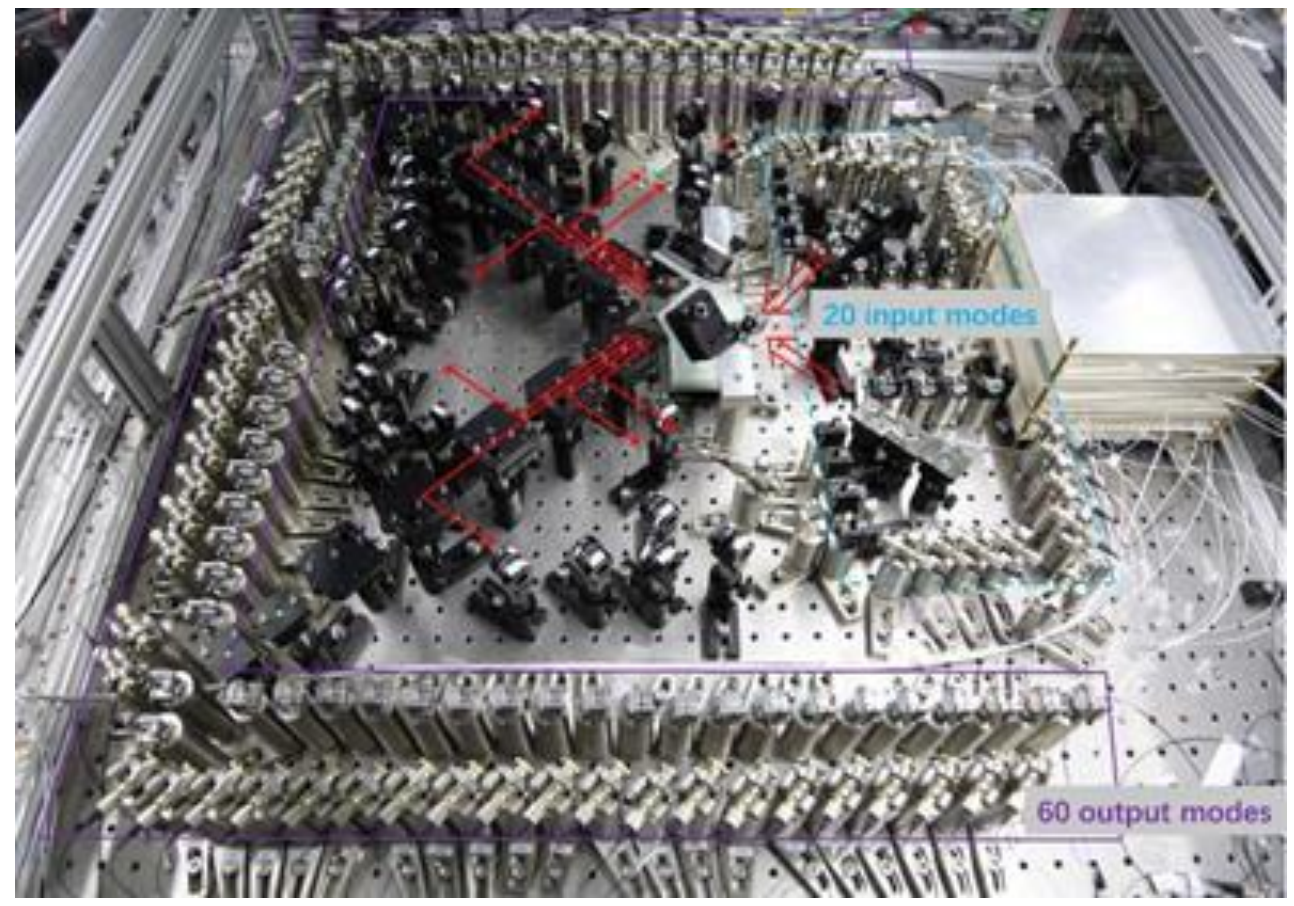

*Figure 440: optics table of the 20 photons/60 modes China experiment.*

The corresponding probability is very low. They use active demultiplexers with Pockels cells to demultiplex and direct the photons.

In 2020, other Chinese researchers used an optical quantum calculator to solve a useful problem, the subset-sum problem, which is complete NP. The system shown in Figure 441 uses a chipset much more miniaturized than the usual boson sampling experiments.

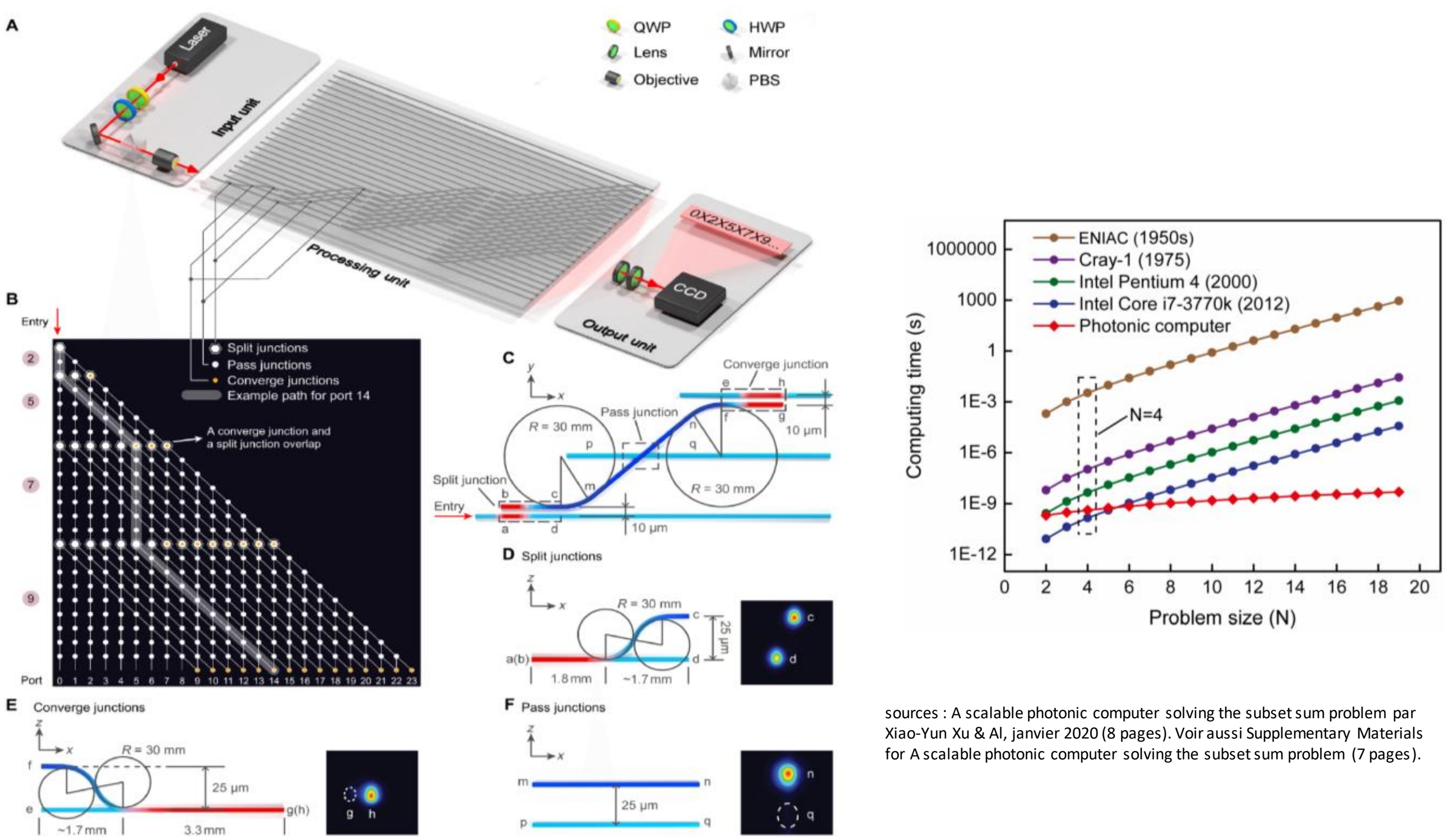

*Figure 441: a first optical calculator to solve a useful problem created in 2020. Source: [A scalable photonic computer solving the subset sum problem](#) by Xiao-Yun Xu et al, January 2020 (8 pages).*

[1276] The photon source would come from a German laboratory located in Würzburg, Bavaria. It is largely inspired by the reference work in the field of Pascale Senellart's team from the CNRS C2N.

The problem is to determine, apart from a set of signed integers, whether it is possible to add a subset of them together to obtain a given integer[1277]. The system uses a laser as a source of photons. The benchmark has been realized with N=4 integers. They indicate that by extrapolating, their system would beat all other known methods of solving this kind of problem.

One of the perspectives of photon-based qubits is to bypass their flaws with the use of MBQC and cluster states, which we have already defined on page 450. Indeed, these use the implementation of an entangled state between all qubits and then a measurement of the progressive state of the others. This avoids the complexity of optical quantum gates, which are difficult to implement, whereas we now know how to create a set of well-entangled photons.

In December 2020, the stakes went higher with a **gaussian bosons sampling** done with 70 photons modes[1278]. The experiment was even more impressive than the previous ones and the publicized quantum advantage reached new heights. But the system, shown in Figure 442, was not more programmable than the previous ones. So, any computing advantage claim was dubious.

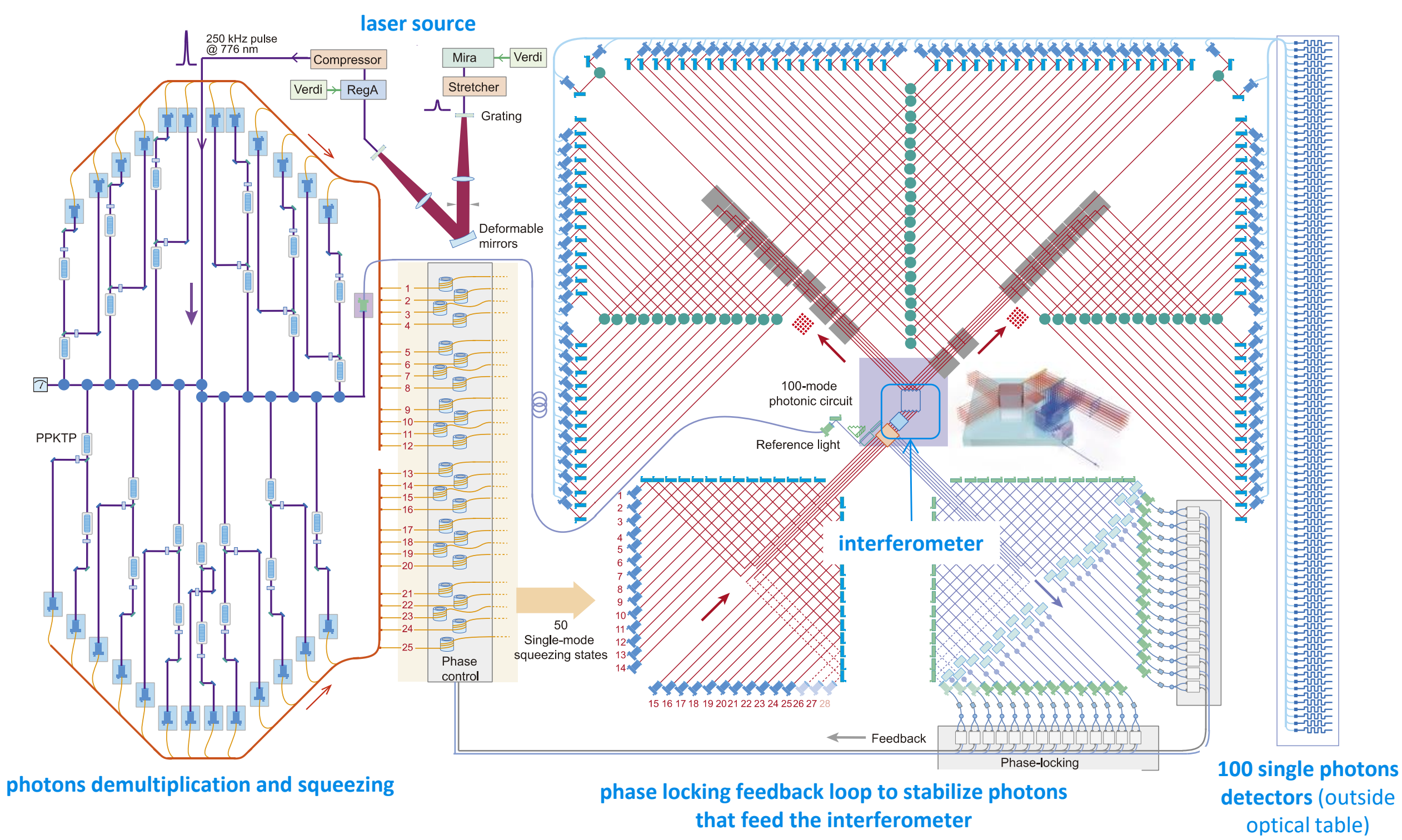

*Figure 442: a 2020 generation China boson sampling experiment with up to 70 simultaneous photon modes. Source: A scalable photonic computer solving the subset sum problem by Xiao-Yun Xu et al, January 2020 (8 pages).*

Late 2020, a competing Chinese team implemented another form of boson sampling using "membosonsampling" for which an emulation requires even more complicated Haar-random unitary matrix[1279]. But it was not programmable.

[1277] See Photonic computer solves the subset sum problem, February 2020 which points to A scalable photonic computer solving the subset sum problem by Xiao-Yun Xu et al, January 2020 (8 pages). See also Supplementary Materials for A scalable photonic computer solving the subset sum problem (7 pages).

[1278] See Chinese Scientists Begin Climb Toward Universal Quantum Computer by Matt Swayne, December 2020, Chinese scientists say they've achieved a quantum computing breakthrough by Shiyin Chen et al, December 2020 and Quantum computational advantage using photons by Han-Sen Zhong et al, December 2020 (23 pages) and the supplemental materials (64 pages). See Benchmarking 50-Photon Gaussian Boson Sampling on the Sunway TaihuLight by Yuxuan Li et al, 2020 (12 pages) for the classical emulation on classical supercomputers.

[1279] See Quantum Advantage with Timestamp Membosonsampling by Jun Gao, December 2020 (30 pages).

In June 2021, the China team from 2020's GBS upgraded their experiment and made it somewhat programmable, ramping it up to 113 detection events extracted from 144 photon modes circuit shown in Figure 443. The input squeezed photons are phase programmable before they enter the fixed part of the experiment in the interferometer. The experimenter still has to implement some real-world algorithms and benchmarks to demonstrate sort of quantum computing advantage[1280].

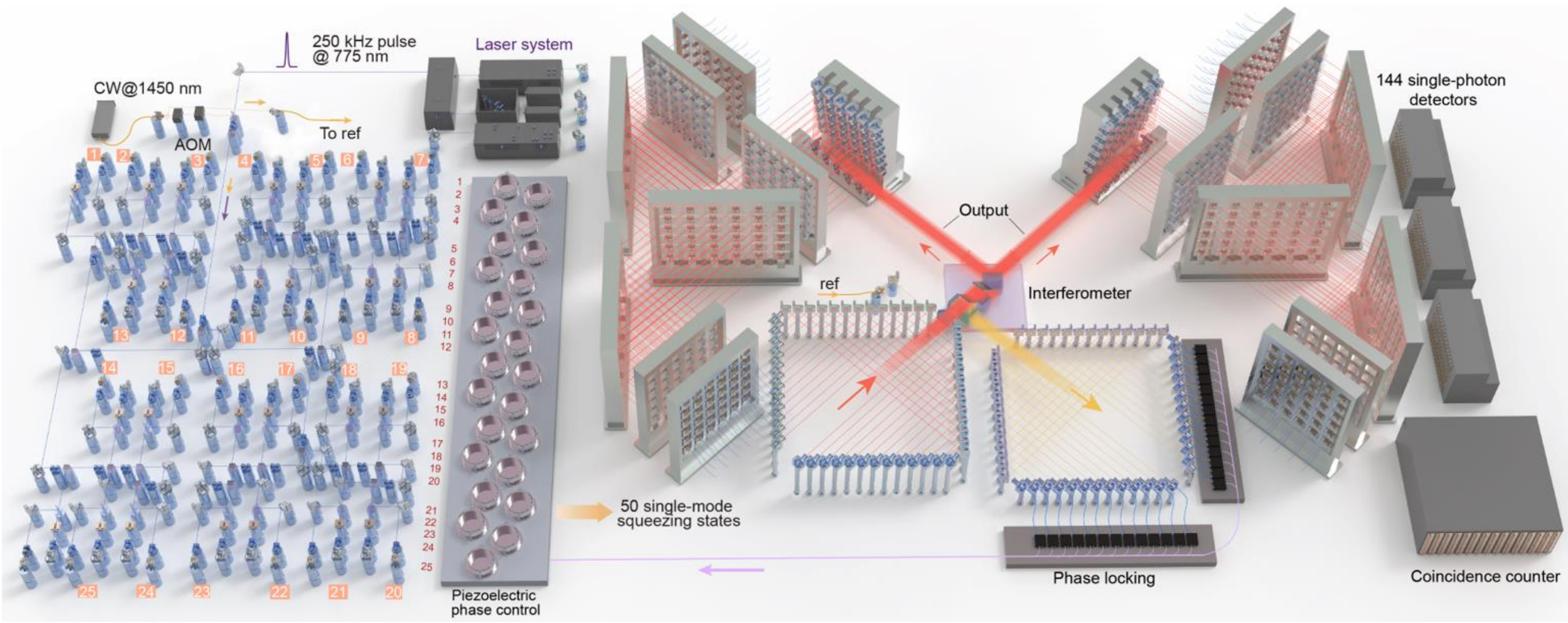

Figure 443: the latest Boson sampling experiment achieved in China in 2021 with 144 photon modes. Source: Phase-Programmable Gaussian Boson Sampling Using Stimulated Squeezed Light by Han-Sen Zhong, Chao-Yang Lu, Jian-Wei Pan et al, June 2021 (9 pages).

In 2022, Fabio Sciarrino did demonstrate that it was possible to create a programmable interferometer in a boson sampler[1281]. And in 2022, Xanadu did up the ante with a new GBS experiment, more powerful than the last one from China, but simpler thanks to some time-bin multiplexing. It was also programmable and even made available in the cloud but with caveats we'll describe in the photonic qubits vendor section.

### Measurement Based Quantum Computing

MBQC is a very particular approach to quantum computing. It consists in exploiting the initialization of entangled qubits and then performing step-by-step measurements on certain qubits to obtain a result on the last measured qubits at the end of the run. There are several variants, the *one-way quantum computing* (1WQC[1282]) which uses two-dimensional qubits matrices to create cluster states and the *measurement-only QC* which only measures qubits, without prior entanglement. We will focus here on the first method which seems to be the most commonplace.

---

[1280] See Phase-Programmable Gaussian Boson Sampling Using Stimulated Squeezed Light by Han-Sen Zhong, Chao-Yang Lu, Jian-Wei Pan et al, June 2021 (9 pages).

[1281] See Reconfigurable continuously-coupled 3D photonic circuit for Boson Sampling experiments by Francesco Hoch, Fabio Sciarrino et al, npj quantum information, May 2022 (7 pages).

[1282] MBQC was designed in 2000 by Robert Raussendorf and Hans Briegel. See A computationally universal phase of quantum matter by Robert Raussendorf, 2018 (41 slides), Measurement-based Quantum Computation by Elham Kashefi, University of Edinburgh (50 slides) and the extensive Introduction to measurement based quantum computation by Tzu-Chieh Wei from Stone Brook University, 2012- (88 slides) and a one pager: Universal measurement-based quantum computation with Mølmer-Sørensen interactions and just two measurement bases. Other information sources include Blind quantum computation by Charles Herder (10 pages), Cluster-state quantum computation by Michael Nielsen, 2005 (15 pages), Fault-tolerant quantum computation with cluster states by Michael Nielsen and Christopher Dawson, 2004 (26 pages), 2D cluster state (50 slides), Quantum Computing with Cluster States by Gelo Noel Tabia, 2011 (18 pages), Quantum picturalism for topological cluster-state Computing by Clare Horsman 2011 (18 pages) and Cluster State Quantum Computing by Dileep Reddy et al, 2018 (11 pages). See also Quantum computing with photons: introduction to the circuit model, the one-way quantum computer, and the fundamental principles of photonic experiments by Stephanie Barz, 2015 (26 pages). At last, see the review paper Realizations of Measurement Based Quantum Computing by Swapnil Nitin Shah, December 2021 (7 pages).

MBQC allows the execution of classical quantum algorithms with universal gates. Where is it relevant? It is particularly interesting in qubit-based quantum systems where it is difficult to create multi-qubit quantum gates exploiting entanglement and where the number of chained gates is limited for physical reasons.

The model was initially created for cold atoms qubits but it later made more sense with photon qubits for which two-qubit gates are difficult to create. Photons are also indicated because they allow to easily manage rotation angles in the Bloch sphere that are used in the single-qubit quantum gates of the process, via a phase control of the photon qubits. It can also be implemented with other types of qubits like silicon carbide defects[1283].

With MBQC, things are done a bit backwards with respect to classical quantum computing: we first apply single-qubit gates and measure them progressively, whereas in classical quantum computing with universal gates, we only involve qubits progressively and then make measurements at the end of computation.

An MBQC calculation is **logically irreversible**, unlike a quantum algorithm based on universal quantum gates. Indeed, the process of measuring qubit states cannot be logically reversed except when the state of the qubits read corresponds exactly to their basis states $|0\rangle$ and $|1\rangle$.

A quantum computation executed with universal gates is the equivalent of applying a unitary transformation embodied by a giant square matrix of dimension $2^N$ to a set of N qubits initialized in the state $|0\rangle$. This matrix can be inverted by scrolling backwards the quantum gates that were used to create it. With MBQC, this is not possible. This irreversibility of MBQC calculations explains why it is also called 1WQC for One Way Quantum Computing. There is no way going back.

This model is also **probabilistic**, due to the probabilistic nature of the state measures of qubits at each step of the calculation. The successive measurements provide information on the state of the qubits, which makes it possible to become determinist again in the rest of the computation by applying a kind of error correction on the fly. A bit like using 3-qubit error correction codes.

By definition, MBQC is a **hybrid algorithms** method since its implementation depends on interactions between quantum computing and the exploitation of qubits readout data by a classical computer controlling the system.

Qubits used in the cluster state-based MBQC are of four different classes: those that are prepared and measured (the ancilla qubits), those that are only measured during computing, those that are only prepared (but measured at the end of computing) and those that are neither prepared nor measured (and are used for the rest of computing).

The principle is based on the sequencing of so-called NEMC sequences with four steps[1284]:

- Using a set of **ancilla qubits** (step N), those of the first type which are measured with a Z projection.
- Creating **cluster-states of entangled qubits** (step E). With photons, there are many ways to generate these cluster states and it's one of the key scientific and technology challenges with MBQC. Theoretically, you could generate these cluster states with regular independent qubits and apply these a series of single and entangling quantum gates (H, CNOT, etc). Problem is, these entangling gates are difficult to create with photons and MBQC is a method that gets rid of these in the first place! So, scientists are looking for ways to generate these cluster states with other means.

[1283] In Quantum Information Processing With Integrated Silicon Carbide Photonics by Sridhar Majety et al, March 2022 (50 pages).

[1284] Information sources: Advanced Quantum Algorithms by Giulia Ferrini et al, 2019 (30 pages) and An introduction to Quantum Computing by Elham Kashefi, School of Informatics University of Edinburgh, 2020 (119 slides).

There are many figures of merits here: the source must be as deterministic as possible and avoid so-called heralding and post-selection methods that reduce the chance to get a full cluster state at a given moment. There are also 1D and 2D cluster states.

Photonic cluster states can be generated in many ways which have evolved over time with **SPDC** (spontaneous parametric down-conversion) using powerful laser single photons source heralding with a probabilistic outcome that is detected post-selectively and doesn't scale well beyond a dozen qubits[1285], atom based **cavity QED** generation[1286] that was later extended to **ensemble of Rydberg atoms**[1287], individual **neutral atoms**, **spin-photon entanglement** to deterministically generate linear cluster states aka the Lindner-Rudolph protocol [1288] with recent improvements[1289], with **quantum dots molecules**[1290], with the **entanglement of several single photon sources**[1291], **time-domain multiplexing** using indistinguishable photon sources which has the advantage to be theoretically unlimited[1292], **2D spin-photon** cluster states[1293], etc. You can also add **spectral domain multiplexing** on top of time-domain multiplexing as a complement to SPDC sources[1294].

---

[1285] See 12-Photon Entanglement and Scalable Scattershot Boson Sampling with Optimal Entangled-Photon Pairs from Parametric Down-Conversion by Han-Sen Zhong, Jian-Wei Pan et al, PRL, 2018 (17 pages) with a ~97% heralding efficiency and ~96% photons indistinguishability. They generated 12-photon entanglement with a state fidelity of $0.572 \pm 0.024$. It was use for early Boson sampling experiments.

[1286] See Sequential generation of matrix-product states in cavity QED C. Schön, K. Hammerer, M. M. Wolf, J. I. Cirac, and E. Solano, 2006 (11 pages) and Efficient generation of entangled multiphoton graph states from a single atom by Philip Thomas et al, Nature, August 2022 (12 pages) with a generation of 14 qubits GHZ states and linear cluster states of 12 photons.

[1287] See Sequential generation of multiphoton entanglement with a Rydberg superatom by Chao-Wei Yang, Jian-Wei Pan et al, December 2021 (11 pages). One disadvantage of this method is its slow emission rate.

[1288] See Proposal for Pulsed On-Demand Sources of Photonic Cluster State Strings by Netanel H. Lindner and Terry Rudolph, PRL, 2009, published initially as A photonic cluster state machine gun on arXiv (10 pages), a first demonstrations obtained with semiconductor quantum dots spins in Deterministic generation of a cluster state of entangled photons by I. Schwartz, D. Gershoni et al, Technion and University of Washington, Science, 2016 (28 pages) with series of 5 entangled photons and recent improvements in Probing the dynamics and coherence of a semiconductor hole spin via acoustic phonon-assisted excitation by Nathan Coste, Niccolo Somaschi, Loic Lanco, Pascale Senellart et al, C2N and Quandela, July 2022 (6 pages).

[1289] See the first results of high photon indistinguishability in A deterministic source of indistinguishable photons in a cluster state by Dan Cogan, David Gershoni et al, Technion, October 2021 (17 pages) where quantum dot emits indistinguishable polarization-entangled photons with a Gigahertz rate deterministic generation of >90% indistinguishable photons in a cluster state of over 10 photons and High-rate entanglement between a semiconductor spin and indistinguishable photons by Nathan Coste, Sophia Economou, Niccolo Somaschi, Alexia Auffèves, Loic Lanco, Pascale Senellart et al, July 2022 (17 pages).is about the efficient generation of three qubits cluster state with one semiconductor spin and two indistinguishable photons with 2 and 3 particle entanglement with fidelities of 80 % and 63 % respectively, with photon indistinguishability of 88%. The spin-photon and spin-photon-photon entanglement rates exceed by three and two orders of magnitude respectively the previous state of the art.

[1290] See Deterministic generation of entangled photonic cluster states from quantum dot molecules by Arian Vezvaee, Sophia Economou et al, June 2022 (5 pages).

[1291] See Multi-photon entanglement from distant single photon sources on demand by Almut Beige et al, 2006 (9 pages) and Protocol for generation of high-dimensional entanglement from an array of non-interacting photon emitters by Thomas J Bell et al, University of Bristol and NBI, New Journal of Physics, January 2022 (9 pages).

[1292] See Sequential generation of linear cluster states from a single photon emitter by D. Istrati, Pascale Senellart, H.S. Eisenberg et al, 2020 (8 pages), Deterministic generation of a two-dimensional cluster state by Mikkel Vilsbøll Larsen et al, Science, September 2019 (30 pages), using two OPOs (Optical Parametric Oscillator) and Generation of time-domain-multiplexed two-dimensional cluster state by Warit Asavanant et al, Science, 2019 (23 pages).

[1293] See Multidimensional cluster states using a single spin-photon interface coupled strongly to an intrinsic nuclear register by Cathryn P. Michaels et al, University of Cambridge, October 2021 (14 pages) and Deterministic multi-mode gates on a scalable photonic quantum computing platform by Mikkel V. Larsen et al, DTU, Nature Physics, July 2021 (30 pages) which deals with creating an universal gate set with cluster states, with CV qubits using telecommunication wavelengths (1550nm).

[1294] See Spectrally shaped and pulse-by-pulse multiplexed multimode squeezed states of light by Tiphaine Kouadou, Nicolas Treps, Valentina Parigi et al, September 2022 (9 pages) which is about generating continuous variables entangled field modes which could also be used for Gaussian boson sampling.

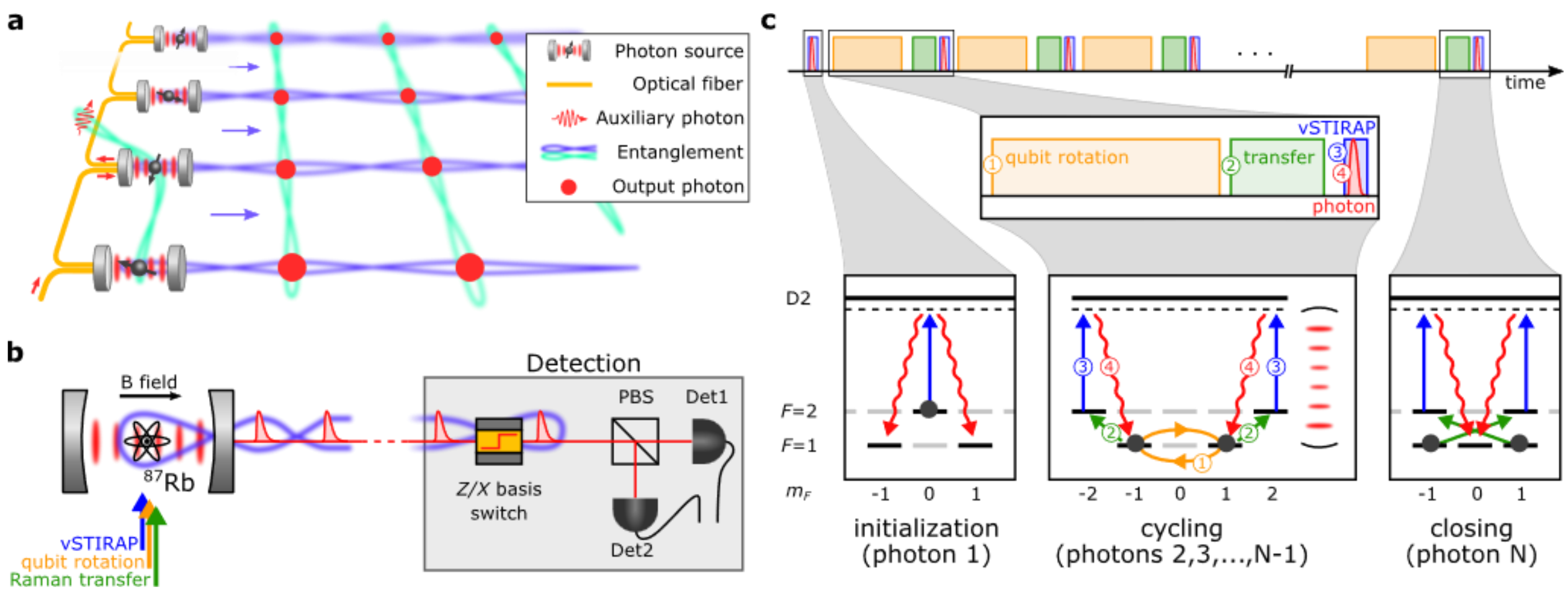

*Figure 444: one solution to generate a cluster state of entangled photons for MBQC. Source: Efficient generation of entangled multi-photon graph states from a single atom by Philip Thomas, Leonardo Ruscio, Olivier Morin and Gerhard Rempe, MPI, May 2022 (10 pages).*

- Measuring **state of intermediate qubits** during computing (M). It is carried out with a variation of projective measurement. It consists in first applying one or more X or Y gates to a qubit to create a rotation in their Bloch sphere and then to measure their state on the computational basis. It is a bit like rotating the Z (|0⟩/|1⟩) axis in the Bloch sphere to change the reference point.

$$|\pm\rangle = \frac{|0\rangle \pm |1\rangle}{\sqrt{2}}$$

$$|\pm_\alpha\rangle = \frac{|0\rangle \pm e^{i\alpha}|1\rangle}{\sqrt{2}}$$

  The projective measurement basis is in the form of states of the type $|\pm_\alpha\rangle$, $\alpha$ being generally a half or quarter turn in Bloch's sphere. A measured qubit is always an intermediate resource and is not an output resource. This helps obtaining an information that can be used to manipulate the qubits afterwards to propagate computation. Projective Z measurements have the effect of removing the measured qubits from the cluster.

- These successive **corrections** make computing deterministic (step C) with X and Z gates. They are applied according to the result of the projective measurements made in (M). No correction gate acts here on a qubit already measured. This model makes it possible to apply any gate to a qubit which is in fact a combination of Rz(γ)Rx(β)Rz(α), i.e. rotations around the three axis of the Bloch sphere of angles γ, β and α[1295].

What has just been described allows to interpret the lower right-hand part of the illustration in Figure 445 which explains how the MBQC equivalents of the CNOT (two-qubit), H or S quantum gate equivalents are realized in MBQC. Each *X* or *Y* circle is an X and Y projective measure that combines an X or Y gate followed by a qubit readout. The result conditions the type of projective measurement performed immediately afterwards in the order indicated (1 to 15 and 1 to 5).

Two forms of measurements affect the inner working of the qubit matrix: Z measurements separate the qubits by digging sort of grooves in the qubit matrix, a bit like pacmans, then classical measurements along the "wires" or on the "bridges" between these wires simulate single-qubit gates like Hadamard's and the two-qubit CNOT gates. The sequence of operations depends on the result of each measurement along the wires. The computation result is located in the last qubits whose state is not yet measured and which will be measured last[1296].

[1295] The decomposition of quantum gates into a computational method that can be used for MBQC has been patented by Krysta Svore of Microsoft, who leads the QuArC group there.

[1296] Illustrations sources: Basics of quantum computing and some recent results by Tomoyuki Morimae, 2018 (70 slides).

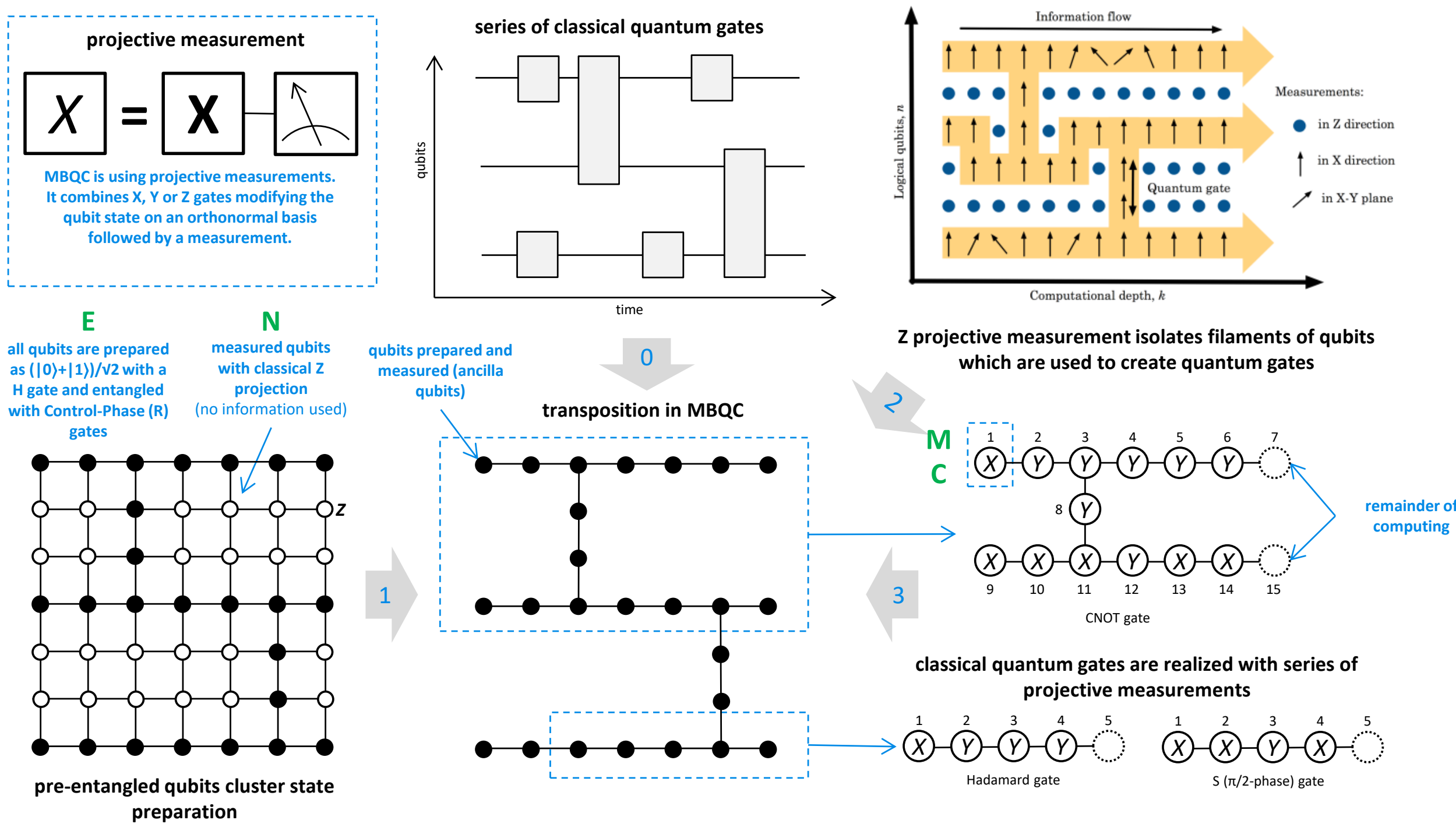

*Figure 445: a tentative summary of how MBQC works. Usually, learning it works like a Write Once Read Never (WORN) memory! (cc) compilation, Olivier Ezratty, 2021, dedicated to my friend Jean-Christophe Gougeon.*

This combination of NEMC sequences allows the reproduction of the operation of one- and two-qubit quantum gates. A complete quantum computation is a sequence of multiple NEMCs that ends with the measurement of the state of the remaining qubits!

The consequences of what we have just seen are multiple:

- MBQC requires way **more qubits** than in a conventional circuit-based model. We've seen that a single X or Y gate results from the combination of four X and Y gates and as many measurements. This in turn creates a "pressure" on the classical part of the calculation, linked to the measurement. But we'll catch up later with parallelism.
- MBQC still requires **error correction codes** such as those we have studied in a previous section, page 216. They too will multiply by several orders of magnitude the number of physical qubits necessary for computing any algorithm. It could be facilitated if we could organize the qubits in 3D matrices, the third dimension being used to align the qubits necessary for error correction, especially with surface codes. On the other hand, since MBQC models contains its own error correction mechanisms, it is less demanding in terms of additional qubits for error correction necessary for the creation of "fault tolerant" quantum computers[1297].
- The **temporal dimension** of computing is modified compared to classical gate-based quantum computing. As we can parallelize operations coupling gates and measurements, MBQC is a bit like Nutella on the breadcrumbs: we can spread it out! The depth of the available computation is no longer linked to the ability to chain quantum gates in time as in the middle-high diagram in the previous illustration, but to execute a large number of them in parallel over a very large number of qubits (modulo the required error correction). The sequences of measurements labeled 1, 2 ... n will be carried out simultaneously in groups 1, 2 ... n, n being limited to 15. Therefore, the required physical calculation depth if defined by the maximum number of physical gates to execute to create a CNOT. This is an argument in favor of photon qubits.

[1297] See one proposal of correction codes in Error-protected qubits in a silicon photonic chip by Caterina Vigliar et al, VTT, Nature Physics, September 2021 (31 pages).

The depth of an algorithm no longer depends on the ability to chain quantum gates with one and two qubits, but on the entanglement capacity of the qubits at startup in the model's cluster states. In short, sequential quantum computing is replaced by massively parallel quantum computing with a very shallow depth. This is the approach chosen by PsiQuantum.

- An MBQC model is easily exploitable to take advantage of teleportation and **distributed quantum computing** algorithms. Cluster states will be able to be linked together via remote optical links. It is also one of the tools of blind computing[1298].
- Finally, there is a direct link between the MBQC and the **ZX Calculus**. ZX Calculus is a graph model that help formalizing MBQC, its cluster states and the associated error corrections[1299].
- The **algorithms** are specific to this kind of architecture[1300]. It is not yet experimental because it requires a large number of qubits that are not yet practically available.

**Vendors**

**PsiQuantum** (2016, USA/Europe, $728M) is a startup created by Jeremy O'Brien, a former Stanford and Bristol University researcher, who wants to create a photon-based quantum processor in CMOS silicon technology.

Other cofounders are Pete Shadbolt (co-inventor of the VQE algorithm with Jeremy O'Brien and Alán Aspuru-Guzik), Mark Thompson and Terry Rudolph, who discovered when he finished his physics thesis that he was a grandson of Erwin Schrödinger, which may have helped with fundraising! The company already employs over 150 people, most of them in Palo Alto in the USA, but some of them work remotely all over the world, including a couple ones in Europe.

Early in 2021, the company started to be more open on its technology[1301]. It published a paper describing their qubit architecture, using an **FBQC** system, aka Fusion-based quantum computation, a variant of MBQC that we study a bit later. It uses micro-clusters states with groups of 4 qubits connected together and using Resource State Generators (RSGs). It's replacing measurement of entangled states by double measurement of non-connected adjacent qubits to create entanglements between them[1302]. Qubits are encoded in path, in what they call dual-rail encoding with lines for photon states $|0\rangle$ and $|1\rangle$.

Two qubit gates use XX nondeterministic and ZZ deterministic measurements (measuring two photons simultaneously with the same polarization basis), implemented with a beam splitter then combining fusions to create small cluster states.

[1298] See Measurement-based and Universal Blind Quantum Computation by Anne Broadbent, Joseph Fitzsimons and Elham Kashefi, 2016 (41 pages).

[1299] Seen in Universal MBQC with generalised parity-phase interactions and Pauli measurements by Aleks Kissinger and John van de Wetering, 2019 (21 pages).

[1300] See for example Changing the circuit-depth complexity of measurement-based quantum computation with hypergraph states, May 2019 (16 pages). The article describes an MBQC method based on the exploitation of Toffoli (CCZ) and Hadamard (H) gates. They allow to simulate topological quantum computation, reducing the error rate of quantum computation.

[1301] See Silicon Photonic Quantum Computing - PsiQuantum at 2021 APS March Meeting by Jeremy O'Brien, April 2021 (25 mn).

[1302] FBQC is fairly well explained in Quantum Computing at the Speed of Light by Terry Rudolph, November 2021 (1h13 video).

With that, the qubit computing depth is quite shallow, avoiding the pitfalls of qubits error rates. It's replaced by a large breadth of computing and commutative operations replacing "depth-computing" by "breadth-computing" [1303]. Their ambition is to produce a system with one million physical qubits generating the equivalent of 100 logical qubits. Their photonic chipsets manufacturing is handled at the 300 mm wafers **GlobalFoundries** Luther Forest Technology Campus in upstate New York.

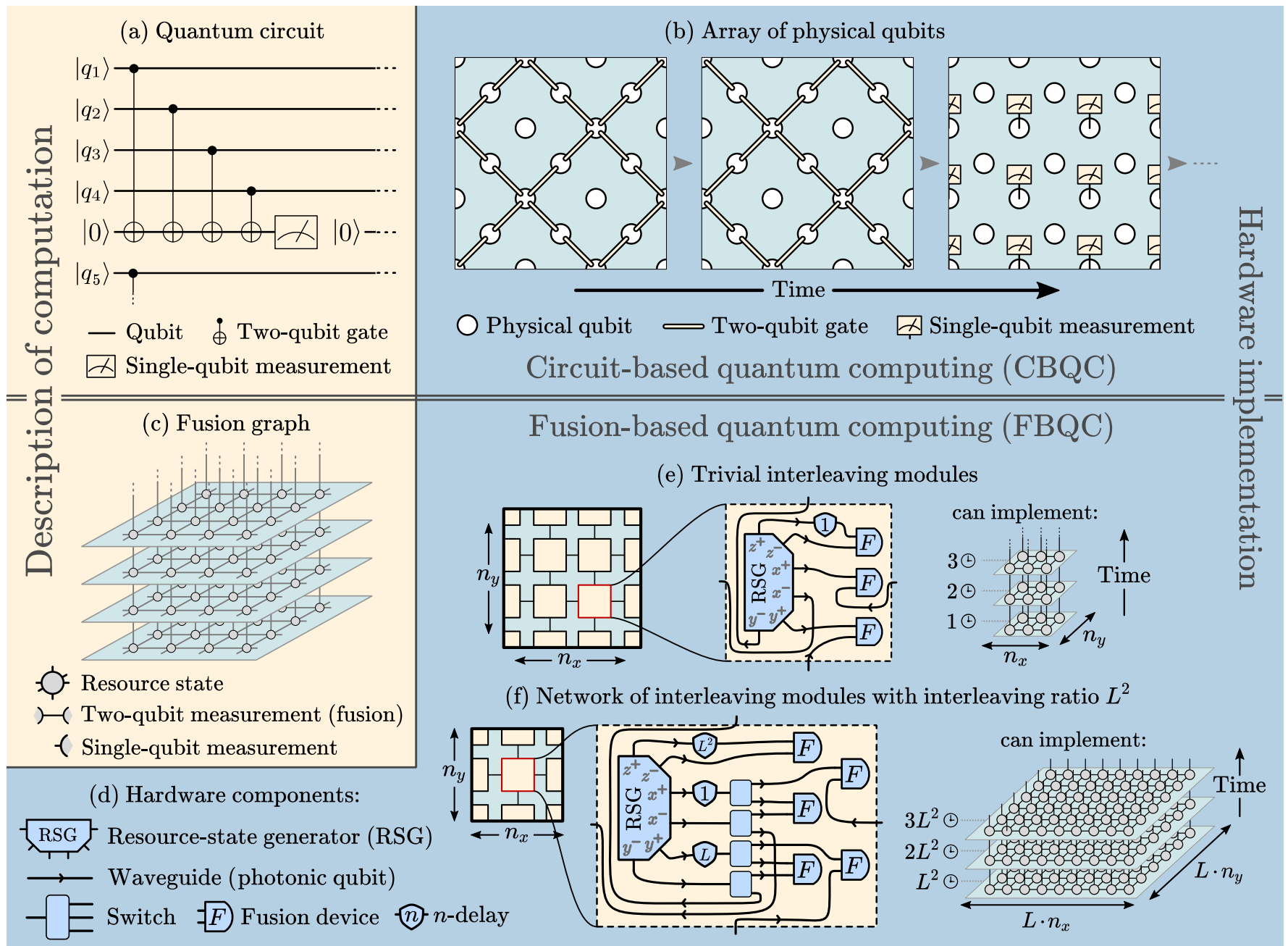

*Figure 446: a description of the FBQC method for the amateur photonicist. Source: [Interleaving: modular architecture for fault-tolerant photonic quantum computing] by Hector Bombin et al, 2021 (22 pages).*

They announced having produced a first q1 chipset sample in April 2021 integrating tens of thousands of single photon sources and detectors[1304]. Their physical architecture is using sandwiches assembling a 22 nm CMOS electronic chipset of 750M transistors using superconducting nanowires bonded with 100K connections to a photonic chipset containing thousands of photon sources, detectors and other optical devices. The photonic chipset has 200 optical fiber entries and exits that are used to interconnect similar photonic chipsets together in a distributed architecture manner.

The final PsiQuantum one million physical qubits computer will be made of thousands of computing chips connected together so we can presume each chip is implementing fewer than 1000 physical qubits. The whole system will run at a temperature of 4K, requiring only a pulse tube refrigeration system, that is much simpler than a dilution system for sub 100mK temperatures and with more cooling power. They are also using fiber delay lines as optical memory thanks to its low loss rate. It is mixed with topological fault tolerance codes. This is supposed to multiply by 5000x the number of usable qubits[1305].

To date, PsiQuantum is the best funded startup in the world in quantum computing, even ahead of D-Wave and Rigetti and on par with IonQ and its 2021 SPAC. Originally from the United Kingdom, it moved part of the team to the USA[1306]. They even have Microsoft as investors as well as Pascal Cagni's investment fund, C4 Ventures. Their last funding round of $450M in July 2021 cemented this funding lead. In October 2022, PsiQuantum also got a funding from the US Federal Government through an US Air Force Research Laboratory contract of $22.5M.

---

[1303] See Percolation thresholds for photonic quantum computing by Mihir Pant, 2017 (14 pages). The process is also documented in Towards practical linear optical quantum computing by Mercedes Gimeno-Segovia, 2015 (226 pages). This was the last publication on the PsiQuantum architecture until when they released Fusion-based quantum computation by Sara Bartolucci et al, January 2021 (25 pages). See also QIP2021 Tutorial: Architectures for fault tolerant quantum computing by Naomi Nickerson, January 2021 (3h).

[1304] See PsiQuantum partners with GLOBALFOUNDRIES to bring up Q1 quantum system by Mercedes Gimeno-Segovia, PsiQuantum, May 2021.

[1305] See Interleaving: modular architecture for fault-tolerant photonic quantum computing by Hector Bombin et al, 2021 (22 pages).

[1306] See the presentation Measurement-based fault tolerance beyond foliation by Naomi Nickerson of PsiQuantum in September 2019 and Quantum Computing With Particles Of Light: A $215 Million Gamble by Paul Smith-Goodson, April 2020.

**Xanadu** (2016, Canada, $135.6M) is a startup created by Christian Weedbrook, a prolific researcher having started at MIT and the University of Toronto, among others. The startup is developing a photon qubit quantum computer in the FTQC realm (fault tolerant quantum computer).

In September 2020, they launched a cloud-based testing platform of 8 and 12 qubits. Their qubits are qumodes based on squeezed states using continuous variables encoding[1307]. The 8-qubit silicon-nitride chipset is 4mm x 10 mm wide. It's fed by infrared laser pulses and generates "squeezed states" superposing multiple photons, then flowing through an interferometer made of beam splitters and phase shifters performing quantum gates, and exiting to superconducting photon detectors. They don't specify the detailed characteristics of these qubits, particularly in terms of fidelity[1308].

In 2021, a team of Xanadu and Canadian researchers published a blueprint with more details on the Xanadu FTQC architecture. It's based on MBQC and three-dimensional resource states comprising both GKP bosonic qubits and squeezed states of light. This hybridization enables the implementation of both Clifford and non-Clifford gates. All of this will be implemented on 2D photonic chipsets[1309]. In August 2021, Xanadu announced that their FTQC silicon-nitride chipsets would be manufactured by IMEC in Belgium. But in March 2022, Xanadu announced a partnership with GlobalFoundries for the manufacturing of their chipset on 300 mm silicon wafers, like PsiQuantum.

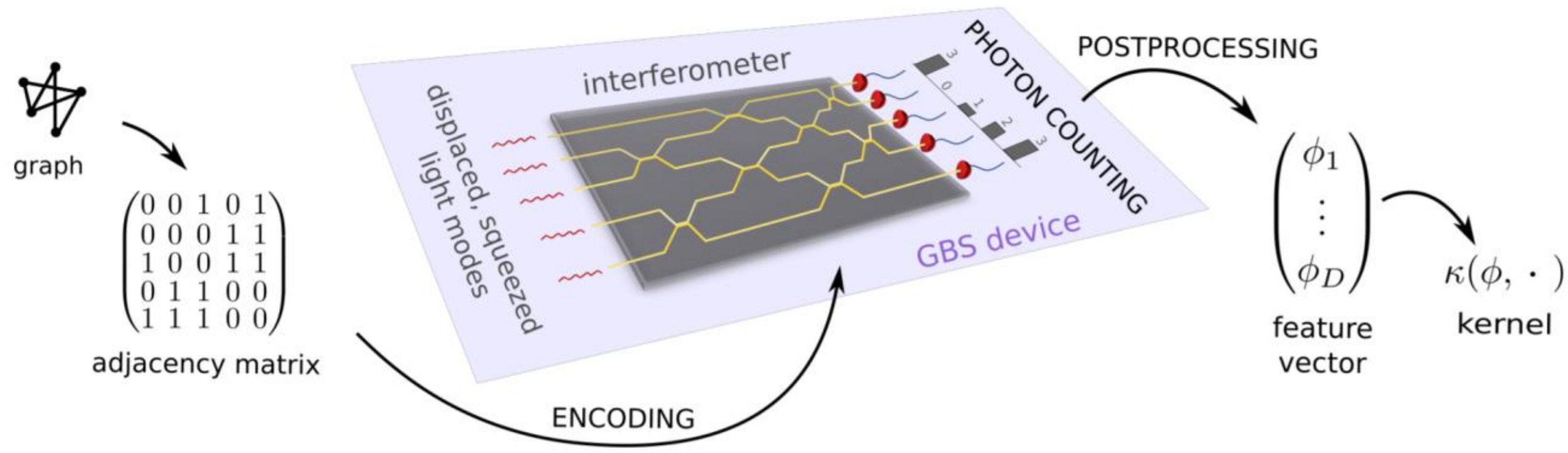

*Figure 447: Xanadu's architecture for their 2022 GBS. Source: Xanadu.*

In June 2022, Xanadu announced their own "quantum advantage" with their gaussian boson sampling architecture (GBS). Adding to China's 2021 similar performance, they increased the number of handled photon modes to 216 thanks to using frequency multiplexing and delay lines. Their system is programmable with parametrizable photon phases and it was put online on the cloud, seemingly with Amazon Bracket. However, Xanadu was cautious in saying that it didn't yet found use case with some useful quantum advantage[1310]. Since then, it seems that Xanadu decided not to pursue the path of parametrizable GBS in its photonic computer roadmap.

---

[1307] Their process is documented in The power of one qumode for quantum computation, 2016 (10 pages) with an example of implementation in Continuous-variable gate decomposition for the Bose-Hubbard model, 2018 (9 pages). See also Optical hybrid approaches to quantum information by Peter van Loock, 2010 (35 pages). See also Quantum computing with multidimensional continuous-variable cluster states in a scalable photonic platform by Bo-Han Wu et al, 2020 (22 pages) and the review paper Quantum computing overview: discrete vs. continuous variable models by Sophie Choe, June 2022 (12 pages).

[1308] See In the Race to Hundreds of Qubits, Photons May Have "Quantum Advantage" by Charles Q. Choi, March 2021.

[1309] See Programmable optical quantum computer arrives late, steals the show by Chris Lee, March 2021 referring to Blueprint for a Scalable Photonic Fault-Tolerant Quantum Computer by J. Eli Bourassa et al, February 2021 (38 pages).

[1310] See Quantum computational advantage with a programmable photonic processor by Lars S. Madsen et al, Xanadu, June 2022 (11 pages) and the earlier and more detailed Quantum Computational Advantage via High-Dimensional Gaussian Boson Sampling by Abhinav Deshpande et al, February 2021 and January 2022 (24 pages).

Xanadu develops the software platform **Strawberry Fields** and **PennyLane** in Python[1311] (wondering about the inspiration...). It includes the Blackbird language and targets chemistry use cases, graph theory problems and quantum machine learning. All this is distributed in open source.

Their main application is the analysis of similarities between graphs to identify those that are similar and/or separate them into several classes of similarity. Classical methods for solving this kind of problem are similar to finding a matrix determinant[1312].

**Quandela** (2017, France, €35M) decided in 2020 to expand its historical single photon source activity to create photon qubits computers as part of their project ROQC (Reconfigurable Optical Quantum Computer).

Their first quantum computer is MosaiQ, which handles 12 photon modes and consumes 1kW. They first target use cases are certified QRNGs using Bell states with the Entropy solution, hybrid quantum machine learning algorithms and chemical simulations. The Quandela team planned to deploy a platform that will handle a few digital qubits in the cloud in 2022. They announced in October 2022 that their future photonic chipsets will be designed and manufactured by CEA-Leti in Grenoble.

Quandela is always teaming up with Pascale Senellart's C2N research lab. With Fabio Sciarrino's team from Sapienza University in Rome, Italy, they qualified the ability of Quandela's photon source to create entangled states that are used in MBQC computation. They developed an interferometer to assess the indistinguishability of 4 entangled photons generated by their quantum dots source[1313].

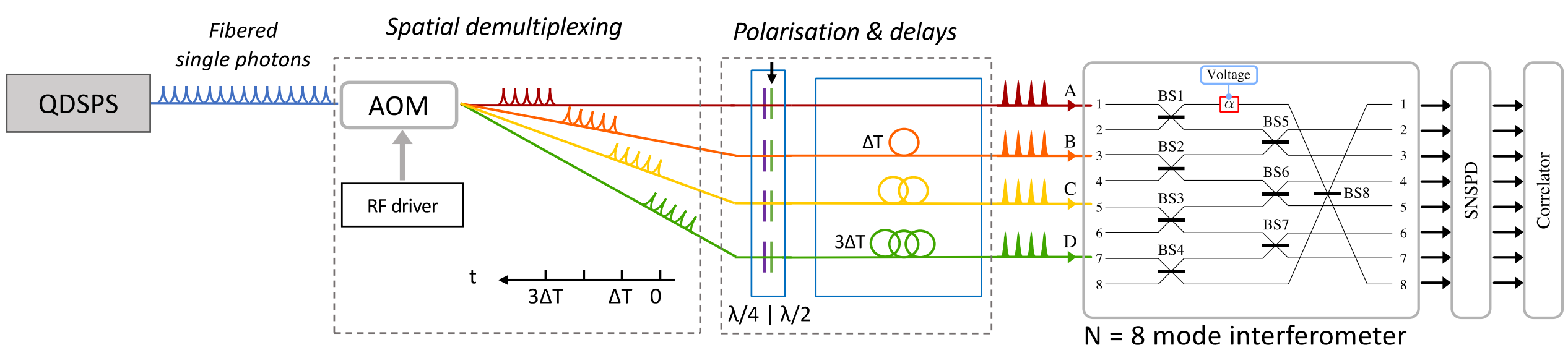

*Figure 448: an interferometer used to validate the indistinguishability of a set of generated photons paving the way for the creation of cluster states of entangled photons. Source: Quantifying n-photon indistinguishability with a cyclic integrated interferometer by Mathias Pont, Fabio Sciarrino, Pascale Senellart, Andrea Crespi et al, PRX, January-September 2022 (21 pages).*

In February 2022, Rawad Mezher and Shane Mansfield from Quandela proposed a single-number benchmark metric, the **Photonic Quality Factor** (PQF), defined as the largest number of input photons for which the output statistics pass all tests. It covers photons quantum computing using single photon sources, multi-mode linear optics and photon detectors, including boson sampling experiments[1314].

In April 2022, Quandela released **Perceval**, their photon qubits physical classical simulation software. It enables the simulation at low level of photonic linear circuits (PBS, …), help understand how photon qubits work and create adapted algorithms like Grover, Shor, GBS, VQE and QML[1315]. They later announced that Perceval was proposed in the cloud in partnership with **OVHcloud** and connector with popular programming frameworks like Qiskit.

---

[1311] This is documented in Strawberry Fields: A Software Platform for Photonic Quantum Computing, 2018 (25 pages).

[1312] See Measuring the similarity of graphs with a Gaussian Boson Sampler by Maria Schuld et al, 2019 (11 pages).

[1313] See Quantifying n-photon indistinguishability with a cyclic integrated interferometer by Mathias Pont, Fabio Sciarrino, Pascale Senellart, Andrea Crespi et al, PRX, January-September 2022 (21 pages).

[1314] See Assessing the quality of near-term photonic quantum devices by Rawad Mezher and Shane Mansfield, Quandela, February 2022 (30 pages).

[1315] See Perceval: A Software Platform for Discrete Variable Photonic Quantum Computing by Nicolas Heurtel et al, April 2022 (31 pages).

**ORCA Computing** (2019, UK, $18.7M including some UK public funding) is developing a quantum computing platform based on photons and a proprietary photonic memory[1316]. It's also relying on MBQC, a quantum computation management method which consists in starting by creating a cluster state of entangled qubits (in this case GHZ states) and progressively reading the qubit states to carry out its computations progressively. They also use optical frequency combs and continuous variables. Their chipset is manufactured by **Ligentec** in Switzerland.

The startup was cofounded by Richard Murray (CEO, former head of the UK quantum program), Josh Nunn (CTO, former Oxford University, and also working with VeriQloud) and Cristina Escoda (COO), an entrepreneur with a background in finance and deep tech[1317]. Their roadmap consists in creating 3 qubits by 2024 and hundreds of qubits by 2026. Quantonation is among their investors.

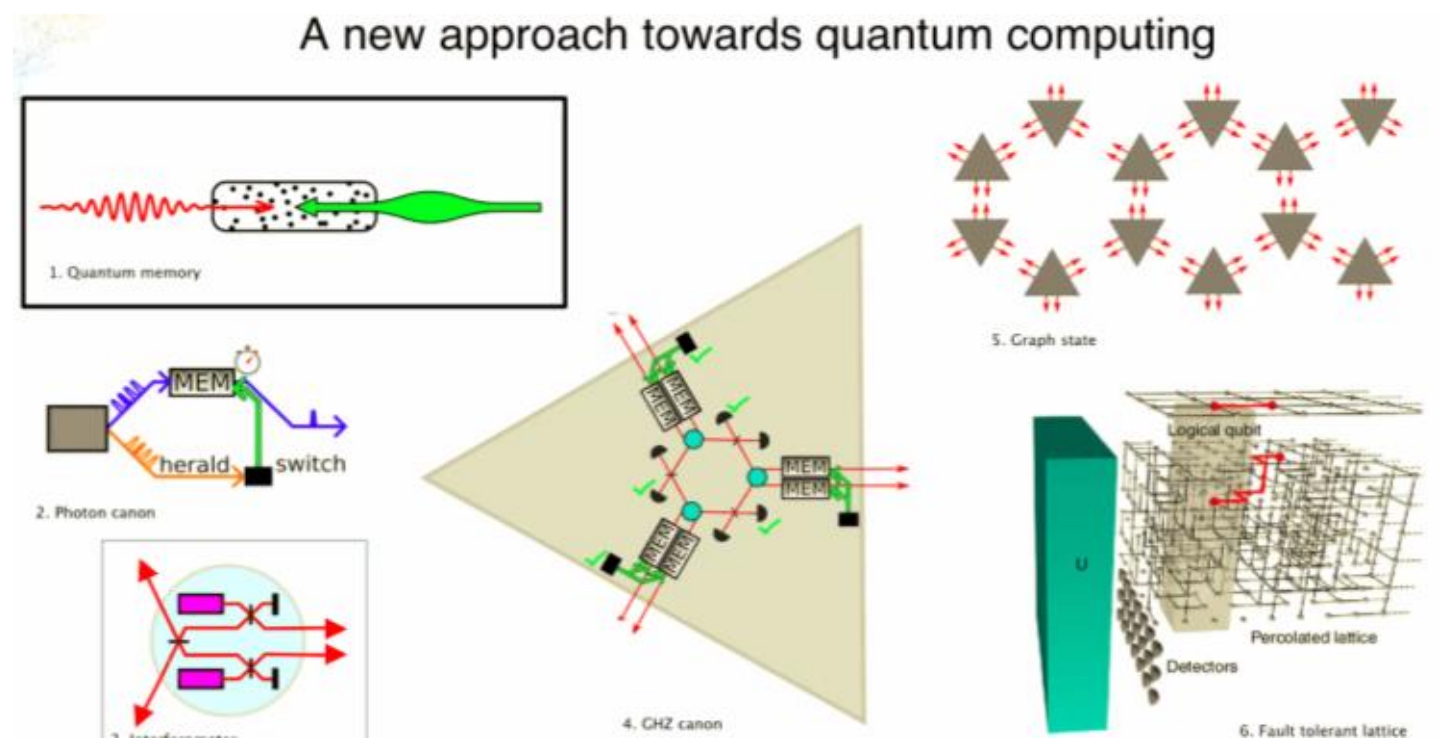

*Figure 449: Orca's view of quantum computing. Source: Orca Computing.*

The startup leverages research done by Ian Walmsley's Ultra-fast and Nonlinear Quantum Optics Group from the University of Oxford. In June 2022, the UK Minister of Defense announced the procurement of Orca's PT-1 quantum computer. It is supposed to run machine learning and QUBO algorithms with their Python software library, seemingly with between 30 and 70 photon modes[1318].

Orca also sold a PT-1 QPU to Israel's Quantum Computing Centre managed by Quantum Machines in July 2022.

**TundraSystems** (2014, UK) is developing a linear optics quantum processor operating supposedly at room temperature. They seem to create a photonic microprocessor and not necessarily, a quantum computer with qubits using linear optics.

Their Advisory Board includes two Chinese scientists, Xinliang Zhang and Pochi Yeh who are specialized in optronics (site).

**QuiX Quantum** (2019, Netherlands, 5.5M€) is developing a photonic quantum processor using silicon nitrides ($Si_3N_4$) waveguides. It came out of a project from the University of Twente and the AMOLF laboratory in Amsterdam. The company is a subsidiary of the fab Lionix.

Their fab also provides photonic components to other industry vendors, like Quandela. They presented in 2021 a record 12x12 programmable photonic processor. It uses thermo-optic phase shifters and tunable beam splitters. The circuit is labelled a 12x12 because it has 12 input photons and a depth of 12 quantum gates[1319].

1316 See One-Way Quantum Computing in the Optical Frequency Comb by Nicolas C. Menicucci, Steven T. Flammia and Olivier Pfister, April 2018 (4 pages) and High-speed noise-free optical quantum memory by K. T. Kaczmarek et al, April 2018 (12 pages).

1317 See some details on their approach in Photonic quantum processors, Orca Computing, April 2020 (27 slides).

1318 See Certain properties and applications of shallow bosonic circuits by Kamil Bradler and Hugo Wallner, December 2021 (34 pages).

1319 See A 12-mode Universal Photonic Processor for Quantum Information Processing by Caterina Taballione et al, 2020 (11 pages).

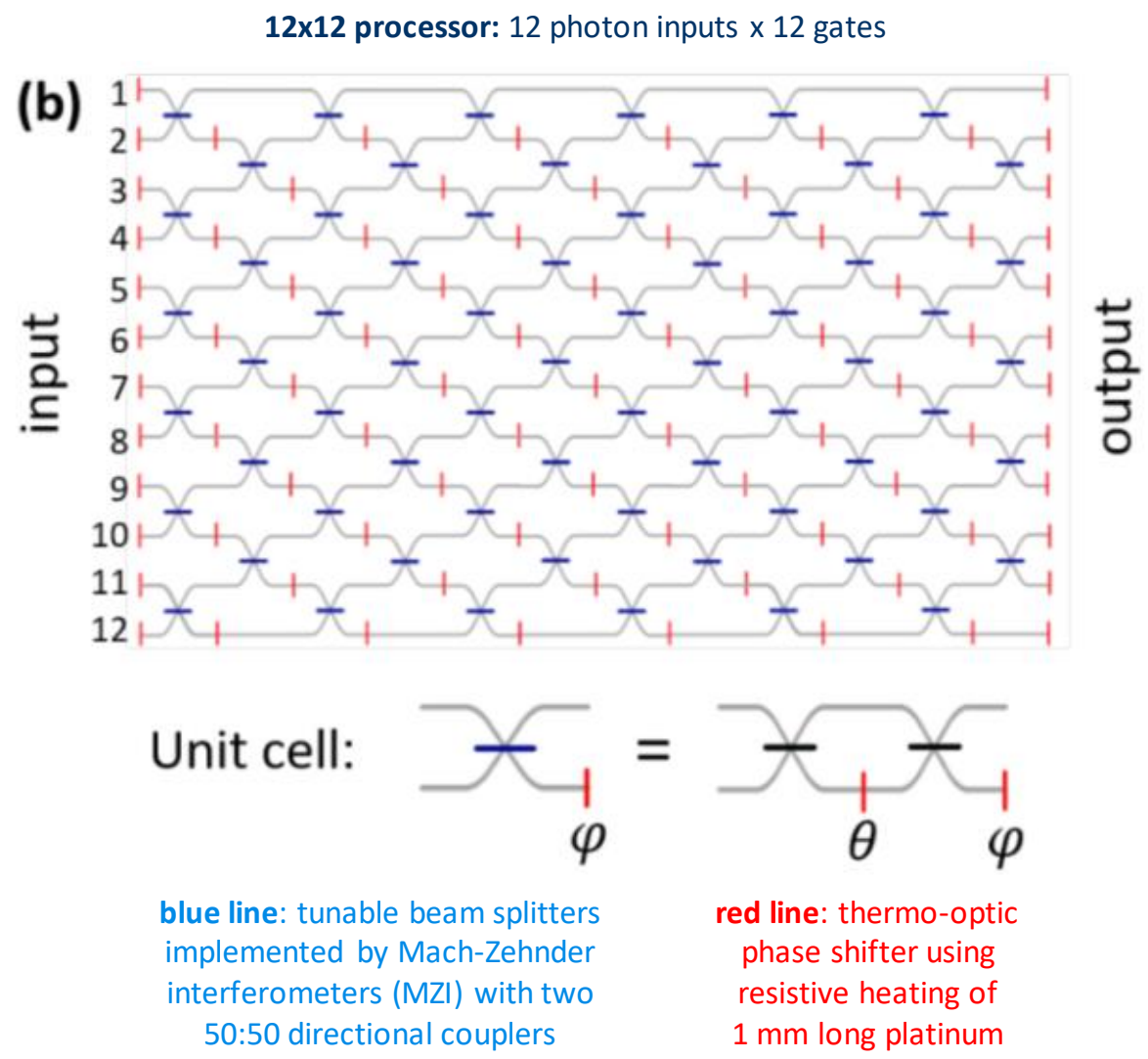

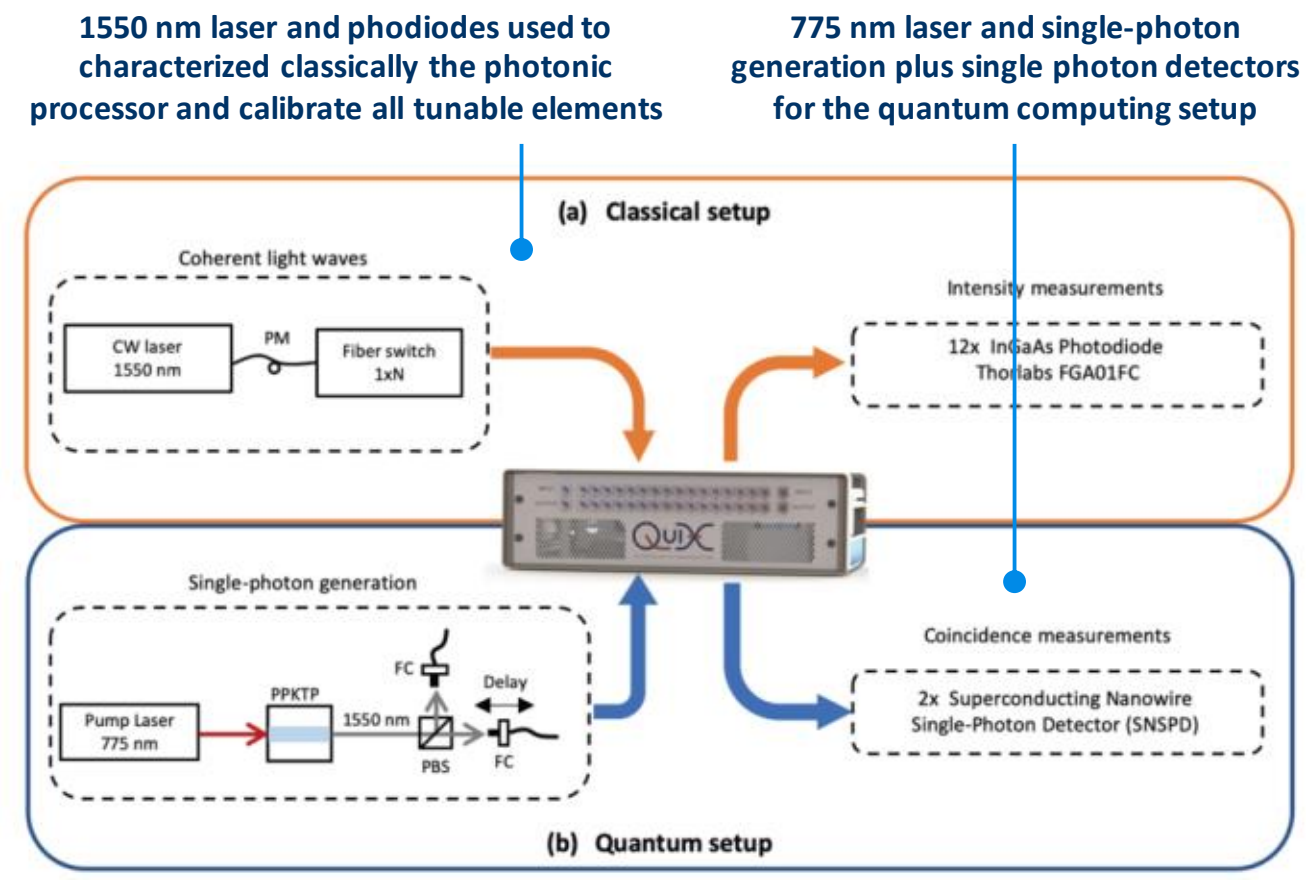

| | # | $\mathcal{F}_{ave} \pm \sigma$ |
|---|---|---|
| Random Perm. | 12 | $0.930 \pm 0.013$ |
| Pauli-$X^{n=0,\ldots,12}$ | 12 | $0.945 \pm 0.007$ |
| Switching | 12 | $0.985 \pm 0.006$ |
| Haar Random | 1000 | $0.904 \pm 0.024$ |
| Logo | 2 | $0.926 \pm 0.004$ |

Table 1 Summary of measured fidelities.

qubit gates fidelities are good for photon based systems

*Figure 450: a QuiX circuit handling 12x12 photons (12 photons and 12 quantum gate depth using MZIs). Source: A 12-mode Universal Photonic Processor for Quantum Information Processing by Caterina Taballione et al, 2020 (11 pages).*

In March 2022, QuiX announced the "*world's largest photonic quantum processor*" expanding the previous 2021 performance from 12 to 20 "qumodes"[1320]. It contains 380 thermo-optic tunable elements and a photon source made of a Ti:Sapphire laser pumping a crystal. QuiX is now teaming up with **PHIX** (The Netherlands), an assembly subcontractor, to create an even larger quantum photonic processor with 50 qumodes, using over a hundred optical fiber connections and about 5000 electrical connections. In September 2022, the company signed a 14M€ contract with DLR, the German Aerospace Center, to build a 64-qubit quantum computer.

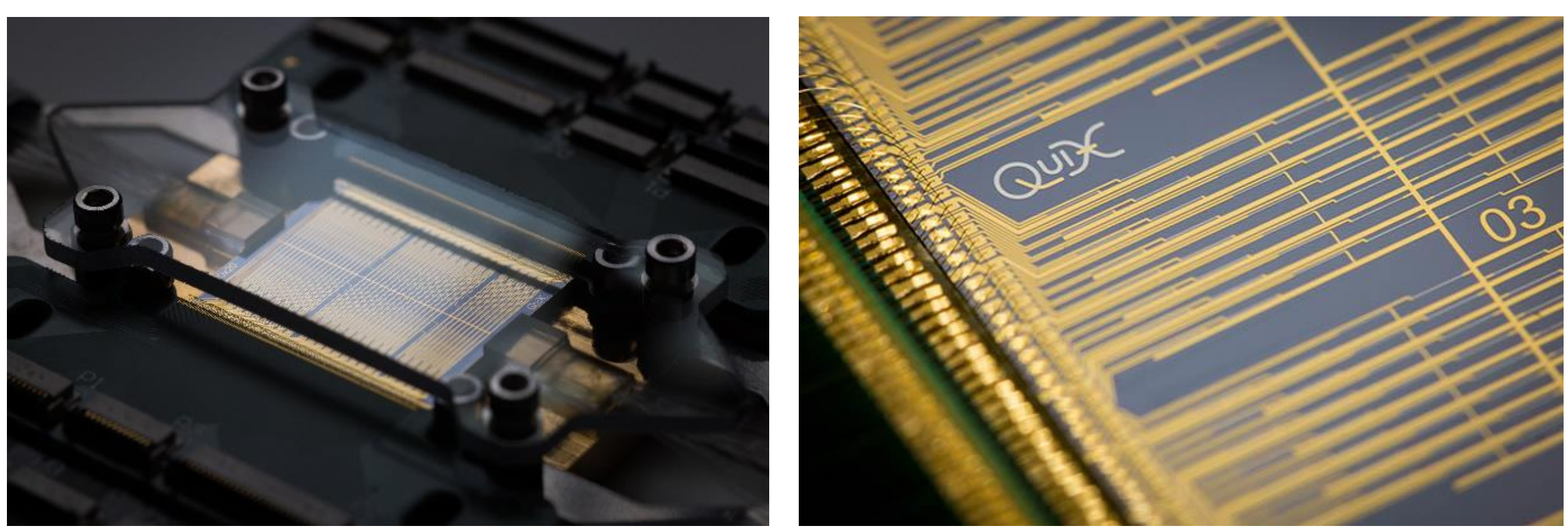

*Figure 451: QuiX photonic processor.*

**QBoson** (2020, China, $3M) aka "Bose Quantum" was founded by Wen Kai in Chaoyang (200 km North-East of Beijing), who studied at Tsinghua University and later got a PhD from Stanford in quantum computing.

He also worked at Google AI in the USA. The company is creating photon-based quantum computers with, in sights, a hybrid AI applications approach.

[1320] See 20-Mode Universal Quantum Photonic Processor by Caterina Taballione, June 2022 (9 pages).

They claim to have completed the construction of a laboratory and of a 1,000 photon-based qubit quantum computer with a plan to reach 1 million-qubits in 3 to 4 years. The first part is probably a little oversold even if the second part is not far from PsiQuantum promises[1321]. Another promise is that this computer works at ambient temperature, which is a highly dubious claim since you generally need some form of cooling for your light sources and photon detectors. On the left, the only visual of the laboratory that was inaugurated in July 2021 (source)!

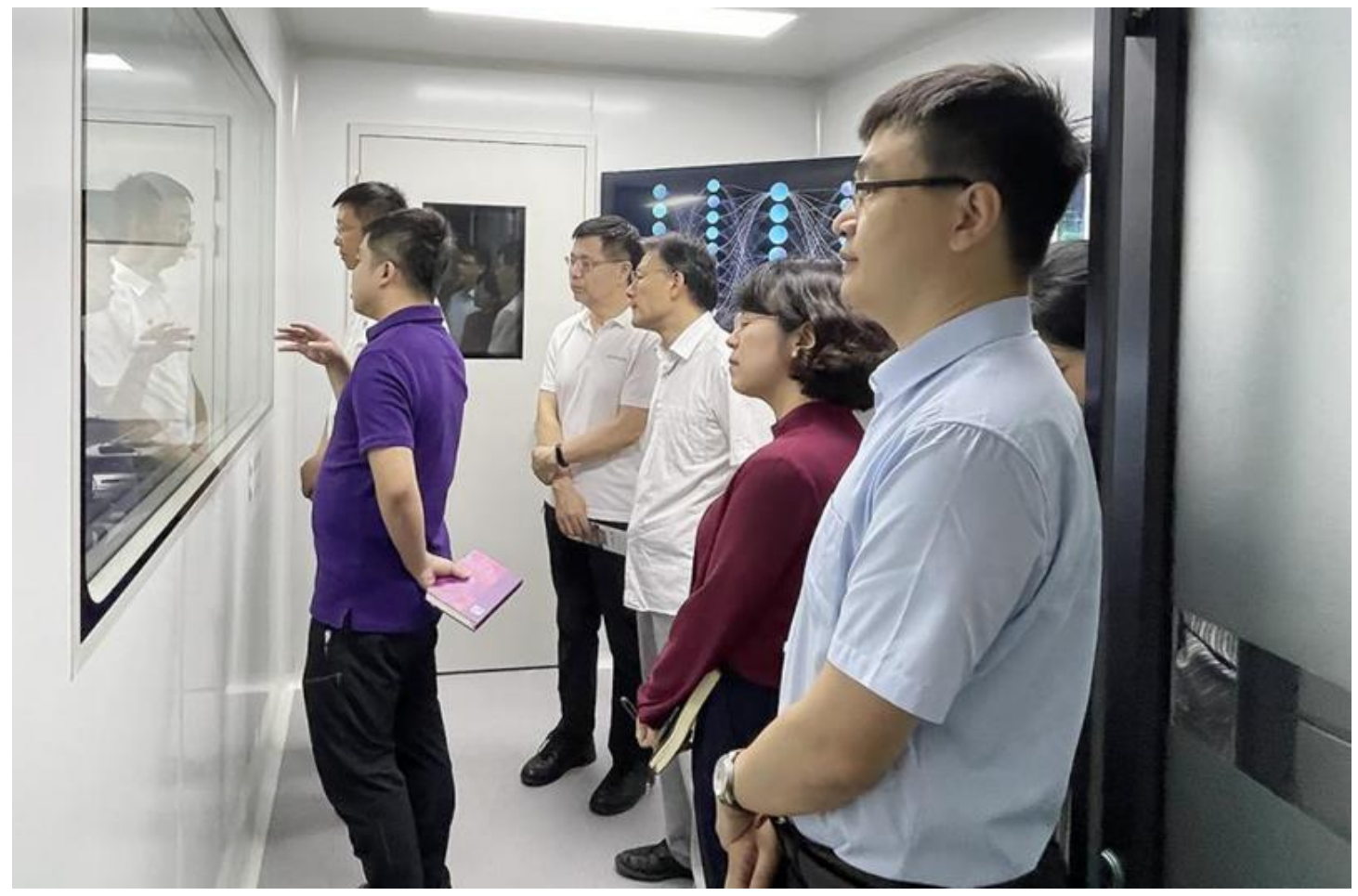

Figure 452: openness in China. You see the folks looking at the window of a lab. Go guess what they saw and understood!

I finally found out in August 2022 that they are working on some sort of coherent Ising machines using spiking neurons in an arXiv paper[1322].

**Duality Quantum Photonics** (2020, UK) is a Bristol-based startup created in February 2020. Its founder is Anthony Laing, from the Department of Physics at the University of Bristol where he developed a quantum simulator based on lithium niobate generated photons.

He targets drugs design for the pharmaceutical industry. They were supposed to create a prototype in 2021.

**TuringQ** (2021, China, $79M) creates lithium niobate on insulator (LNOI) optical quantum computer chips and femtosecond lasers. Not to be confused with Turing Quantum (USA) who is specialized in NV centers computing and Turing (USA) who develops quantum software.

**Quantum Computing Inc**. (*aka* QCI), also covered in the software vendors section, announced in February 2022 a "*business partnership and exclusive marketing agreement with QPhoton, Inc*".

In the current newspeak, it simply means an acquisition! QPhoton (USA) was a stealth quantum photonics computing and sensing company based in New Jersey. They hold a portfolio of patents on quantum hardware, authentication protocols, simulators, photonic Lidar, imaging and covert communications. They were mostly a contract research company working for DARPA, DoD, NASA and other US federal agencies who spent $30M on these projects. The company was created and headed by Yuping Huang, a professor from Northwestern University (Evanston, Illinois) and the Stevens Institute of Technology (Hoboken, New Jersey).

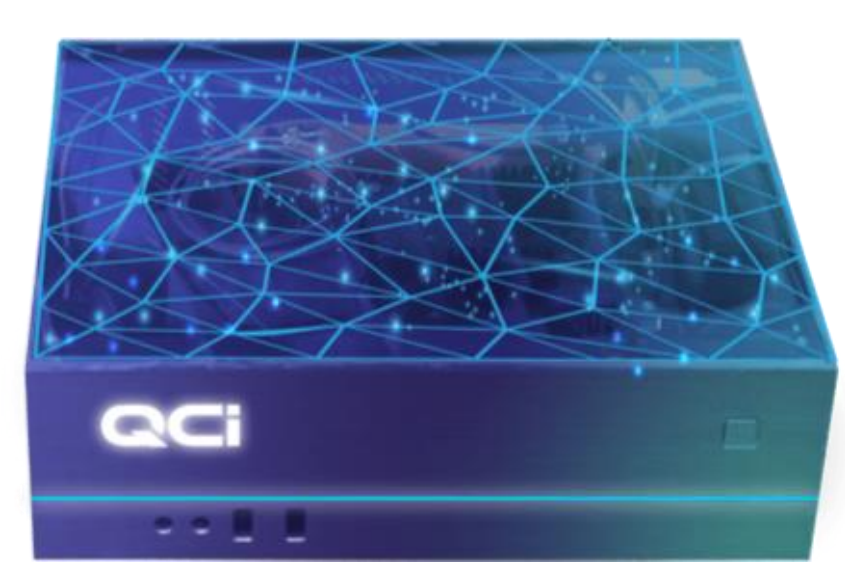

Figure 453: QCI photonic quantum computer package. Source: QCI.

[1321] Wen Kais thesis is Experimental study of tune-out wavelengths for spin-dependent optical lattice in $^{87}$Rb Bose-Einstein condensation by Kai Wen et al, September 2021 (9 pages). It relates to cold atoms qubits, not photons. But QBoson's communication is about photonic qubits controlled by lasers (source). All in all, one thing is sure: these guys don't want you to know what they are doing exactly.

[1322] See Combinatorial optimization solving by coherent Ising machines based on spiking neural networks by Bo Lu et al, August 2022 (6 pages).

How about their Quantum Photonic System (QPS)? It's a nonlinear system. That's all you need to know from them at this point!

QCI announced in July 2022 having solved a combinatorial problem with 3,854 variables with its "Entropy Quantum Computing" (EQC) hardware that was running 70x faster than QCI's 2021 hybrid DWave implementation[1323]. We can suspect it is based on some form of photonic coherent Ising model and on QPhoton QPS architecture. In September 2022, its Dirac 1 Entropy Quantum Computer (EQC) was launched as a cloud-based subscription.

**It's Q** (2022, Germany) is a photonic qubits startup created by Christine Silberhorn from the Institute for Photonic Quantum Systems (PhoQS) of the University of Paderborn.

The company is based on her research on frequency multiplexed qubits ("field-orthogonal temporal modes"), including the related pulsed photon pairs sources, and MBQC as well as quantum walks[1324], using $LiNbO_3$ on silicon oxide insulator circuits (lithium niobate). Christine Silberhorn is also investigating GBS (gaussian boson sampling) avenues as part of the German project PhoQuant led by Q.ANT. Quantonation is one of It's Q seed investors.

**Quantum Source Labs** (2022, Israel, $15M) is a photonic computer startup created by Oded Melamed (CEO), Gil Semo (R&D VP), Dan Charash (Chairman) and Barak Dayan (Chief Scientist, Associate Professor at the Weizmann Institute of Science, head of the Weizmann Quantum Optics group).

Likewise to It's Q, there's not much data available on what they are doing. They seem to run some contract research programs and among others, work with PsiQuantum. Given their founders background, you can infer that they have skills in designing photon guides in nanophotonic circuits, qubit conversions, nanofibers, MBQC and photon detection.

**LightOn** (France) announced a quantum photonic processor in 2021. It implements 8 input quantum states onto 19 distinct optical railings, performing 19x19 unitary linear operations with up to 8 entangled photons at minimal loss and a reconfigurability rate of 10Hz. This is based on using multimode fibers. It must be further documented to be fairly evaluated[1325].

HP conducted research in quantum computing at its Bristol laboratory, UK, covering quantum computing, cryptography and quantum communications. They invested in "The Machine", conceptually far from a universal quantum computer and uses an optical bus to link the different components of a supercomputer.

In partnership with HP, American and Japanese scientists proposed in 2008 the creation of an HPQC, High Performance Quantum Computer, with 3D qubit arrays realized in linear optics containing 7.5 billion physical qubits allowing to accumulate 2.5 million logical qubits[1326]. This project was left aside. HPE abandoned quantum computing entirely and explained it in 2019. They said they preferred to focus on neuromorphic processors and memristors[1327].

---

[1323] See QCI Solves 3,854-Variable Problem in Six Minutes in BMW Group, AWS Quantum Computing Challenge by Matt Swayne, The Quantum Insider, July 2022.

[1324] See for example Fabrication limits of waveguides in nonlinear crystals and their impact on quantum optics applications by Matteo Santandrea, Michael Stefszky, Vahid Ansari and Christine Silberhorn, March 2019 (16 pages).

[1325] See LightOn Qore, a novel Quantum Photonic Processor, June 2021 (2 pages).

[1326] See High performance quantum computing (7 pages).

[1327] See Why HPE abandoned quantum computing research by Nicole Hemsoth, April 2019.

Their photonics specialist is **Ray Beausoleil,** based in Silicon Valley. He was specialized in photonics and NV centers and abandoned this track, becoming a quantum computing skeptic. Somewhat along the lines of Gil Kalai, he believes that errors would increase faster than the growth in the number of qubits. Still, HPE invested in **IonQ** in October 2019 to show that it didn't entirely leave the quantum stage.

### Quantum computing hardware key takeaways

- Superconducting qubits are the most common nowadays, implemented by IBM, Google and Rigetti among others. But they are noisy and do not scale well. One solution may be cat-qubits which combine trapped microwave photons in cavities and superconducting qubits for their manipulation and readout (Alice&Bob and Amazon).
- Quantum dots spin qubits could scale well due to their small size, the reuse of classical CMOS semiconductors manufacturing known-how, their higher working temperature enabling the usage of control cryo-electronics. They have however been demonstrated at a relatively low scale at this stage.
- NV centers qubits have the benefit to be stable and to work potentially at ambient temperature but there are not many vendors involved there, besides Quantum Brilliance (Australia and Germany).
- Topological qubits could bring the benefit of being resilient to some quantum errors and to scale better than other solid-state qubits. It doesn't really exist yet, particularly the Majorana fermions species looked after by Microsoft.
- Trapped ions qubits have the best fidelities so far, but they are hard to scale beyond about 40 qubits, at least with their main vendor, IonQ, Quantinuum and AQT.
- Cold atoms qubits are mostly used in quantum simulation where it could scale up to a thousand qubit and it could potentially also be used in gate-based quantum computing although it's not really demonstrated at a large scale. Pasqal (France), Cold Quanta (USA), QuEra (USA) and Atoms Computing (USA) are the industry vendors in this field.
- Photon qubits are flying qubits, moving from a source to detectors and traversing optical devices implementing quantum gates. There are many investigated techniques, with the distinction between single/discrete variable photons and continuous variable photons. Scalability is also an issue, particularly with photon sources and the probabilistic nature of photons generation. Their limited quantum gates computing depth requires the implementation of specific computing techniques like MBQC and FBQC, this last one being used by PsiQuantum, the best funded quantum computing startup with IonQ as of 2022. Quandela (France), Xanadu (Canada) and Orca (UK) are other key players in that space. One key capability to implement MBQC is the generation of high volume cluster states of entangled photons.

# Quantum enabling technologies

Building a quantum computer and other second quantum revolution related products involve assembling a lot of various technologies, some being classical and others quantum-related themselves. This part of this book is dedicated to these various important enabling technologies. These are "enabling" in a sense that their characteristics and performances frequently have a direct impact on the performance and scalability, particularly with quantum computing. We'll see this with cryogenics, cabling, classical electronics, lasers and photonics.

We'll also look at the raw materials needed in quantum technologies, where it comes from, is it rare or not and how is it transformed. At last, we'll have a look at other unconventional computing technologies. They can both compete and, in some cases, complete quantum computers. Whatever happens, this coopetition is also enabling innovation.

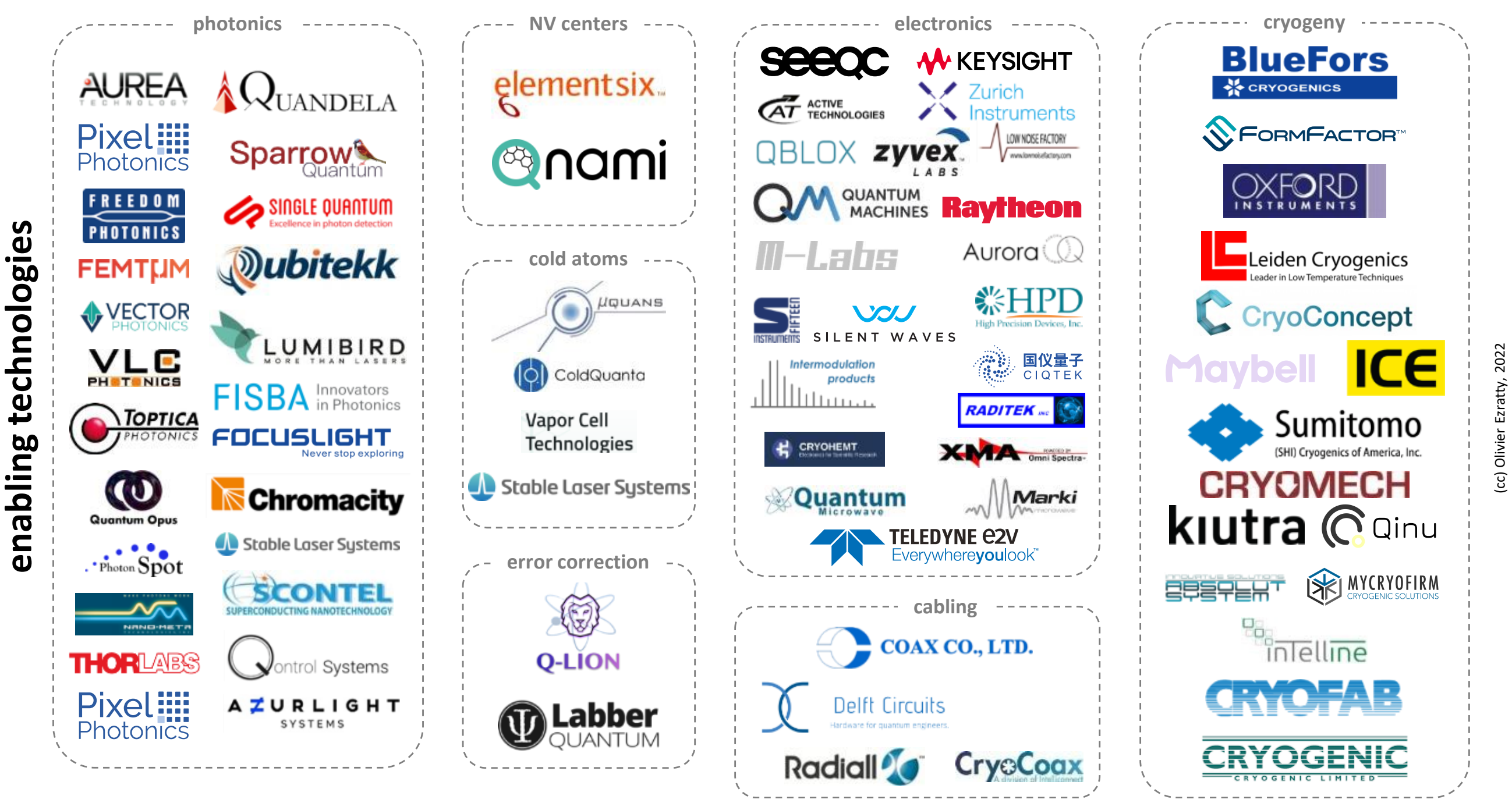

*Figure 454: a market map of key enabling technology vendors. (cc) Olivier Ezratty, 2022.*

## Cryogenics

Cryogenics is an important enabling technology used with most types of qubits, the most demanding being the very low operating temperatures of superconducting qubits, at 15 mK. Other technologies like photon qubits require lightweight cryogenics operating at 4K to 10K for their photon sources and detectors[1328]. Detectors must be cooled to avoid the photon dark count effect, when thermal noise originated photons are detected instead of useful photons.

In this part, we'll focus on the 15mK dry dilution refrigeration systems used by superconducting qubits.

[1328] By definition, cryogenics operates below 123K or -150°C. Bearing in mind that -153.15°C is the temperature below which permanent gases, i.e. gases in the ambient air, all condense into liquid at ambient pressure.

These qubits must be cooled to this low temperature to avoid noise sources from the environment, particularly when compared with the microwave pulses used to control qubits and handle their readout[1329].

These are the most complicated systems and also, those requiring some scalability in cooling power to accommodate the growth in number of physical qubits.

### Wet and dry dilution refrigeration

Superconducting quantum computer from IBM, Google and others are frequently presented with these mysterious gold chandeliers where the processor is housed, surrounded by an unlikely set of wires, devices and several layers of circular plates. This system is a mix of passive and active control electronics reaching the qubits processor and low temperature cooling system[1330]. The chipset must be as isolated as possible in terms of temperature, magnetism, vacuum and even mechanical vibrations.

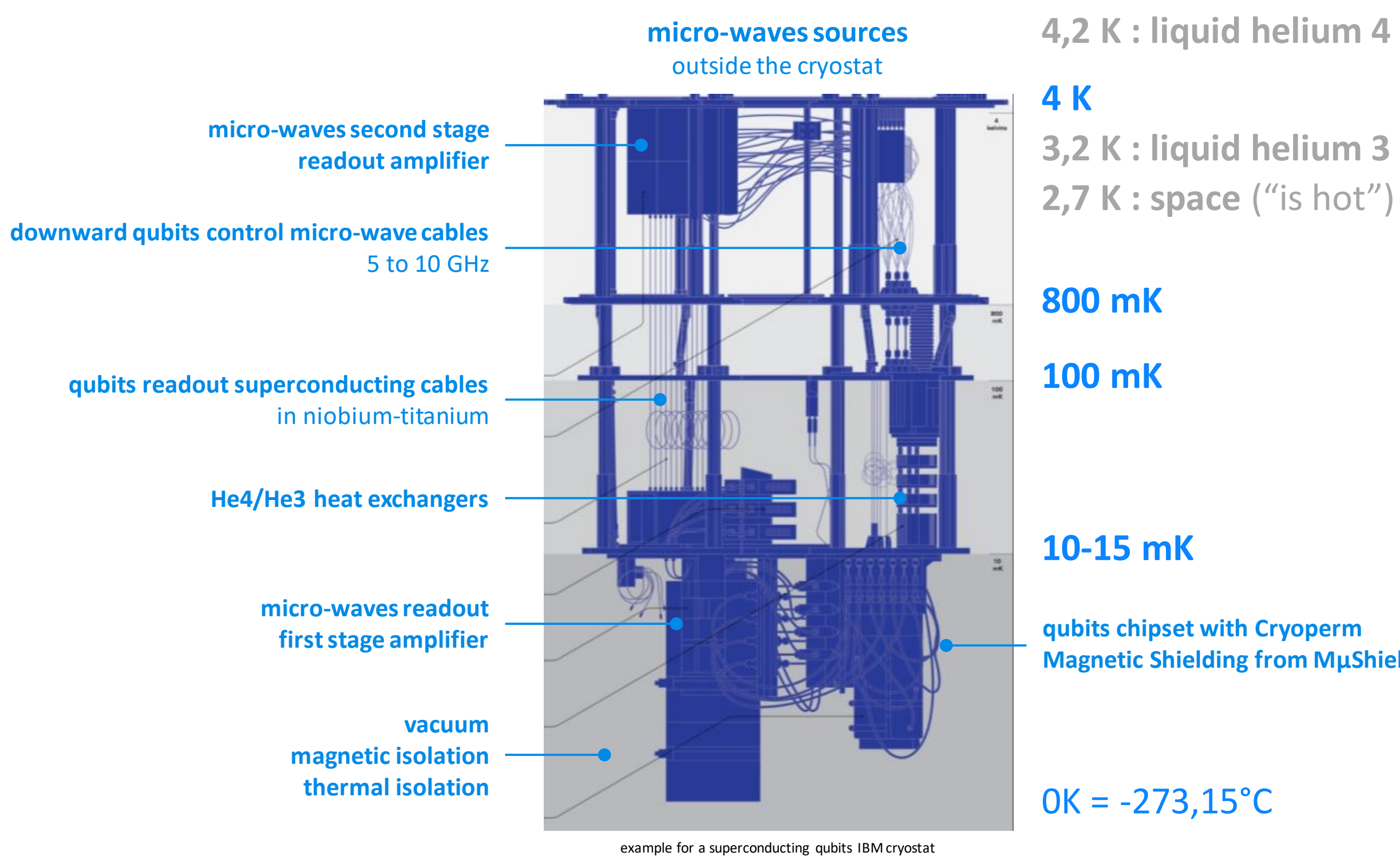

*Figure 455: a documented interior of an IBM superconducting qubit cryostat. Image source: Quantum Computers Strive to Break Out of the Lab, 2018. Legends by Olivier Ezratty.*

The refrigerated part of a quantum computer with superconducting qubits or silicon is generally organized in stages, knowing that the lower you go down in the stages, the colder it gets:

- On the upper level, a plate that is not usually seen in diagrams and picture is thermalized at 50K. This is where both the electronic cables for controlling and reading the qubits as well as the fluids used for refrigeration arrive in the cryostat.

[1329] It is governed by the equation $k_BT<\hbar\omega$. The Boltzmann constant multiplied by the temperature must be inferior to the product of the Dirac constant and the microwaves frequency in Hz. This leads us to adopt a temperature of about 15 mK for superconducting qubits.

[1330] A tour of the IBM Q Lab is available in the 2016 video A Tour of an IBM Q Lab.

- One level below, running at 4K, i.e. 4°C above absolute zero (273.15°C)[1331]. That's where sits the lower part of the so-called pulse tube.
- The below plate is at around 800 mK. Between these two floors is the lowest temperature in space, which is 2.7 K and also corresponds to the cosmic background radiation.
- Another plate is generally located at a temperature of 100 mK.
- The lowest stage plate is where the quantum processor sits, and is cooled between 10 and 25mK, usually around 15mK. It is also called the "mixing chamber cold plate". A cold plate is a one-stage copper plate and the mixing chamber is the last level at the bottom of the dilution refrigeration system that we will explore later[1332].

We will now study the detailed characteristics of the very low temperature cryogenics used in these superconducting quantum computers[1333].

It uses a **dilution refrigeration**, which is based on the association of two helium isotopes: helium 4 and helium 3, which have different and complementary physical properties[1334]. They have respectively a boiling temperature of 4.2K and 3.2K. Helium 4 is superfluid at 2.17K while helium 3 is superfluid at a much lower temperature of 2.5 mK, at ambient pressure. The cryostat exploits the combination of three phases: a gaseous $^3$He phase and two liquid phases, one with $^3$He and the other with a mixture of $^3$He and $^4$He, with evaporation of the $^3$He in a mixed chamber[1335]. Let's explain first why helium is so important for low temperature cryogenics. Hydrogen becomes liquid at 20.3K[1336], nitrogen at 77.4K and oxygen at 90.2K. These gases are useless for low temperature cryogenics.

On the other hand, $^4$He liquefies at 4.2K at room temperature and a $^4$He cryostat can reach 1K while $^3$He cryostats can go as low as 300 mK. The mix of $^4$He and $^3$He is used in so called dilution refrigerators reaching 15 mK[1337]. Note the low density of $^4$He which is 125g/L at 4.2K. There are two types of dilution refrigerators: "dry" and "wet".

---

1331 The Kelvin scale starts at absolute zero. This temperature where atoms literally no longer move is unreachable. If it were, Heisenberg indeterminacy would be broken! It is approached asymptotically. The lowest temperature record is 38 pK (pico-kelvin), reached in 2021. See Collective-Mode Enhanced Matter-Wave Optics by Christian Deppner, David Guéry-Odelin, Ernst M. Rasel et al, PRL, August 2021 (7 pages). See how fast these records have been broken over the last decades in Moore's Law for Low Temperature Physics by Pramodh Senarath Yapa, December 2021.

1332 In Top 5 Trends in Quantum Technologies to Look for in 2020 by QuantumXchange, January 2020, we find: "*Interestingly, IBM and Google are taking different approaches in the infrastructure of quantum computers. IBM's hardware resembles a chandelier with rings whereas the Google device looks like a chip*". Which shows that they did not understand at all that IBM and Google had both a candlestick and a chipset. So they did not explore the hardware architecture of a superconducting quantum computer!

1333 See Cryostats Design $^4$He and $^3$He cryostats by Guillaume Donnier-Valentin, CNRS Institut Néel, 2011 (91 slides), Some Fundamentals of Cryogenic and Module Engineering with regard to SRF Technology, Bend Petersen, ESY Cryogenic Group MKS (95 slides) and Development of Helium-3 Compressors and Integration Test of Closed-Cycle Dilution Refrigerator System, 2016 (5 pages).

1334 Helium was discovered indirectly in 1868 through the discovery of an unexplained spectral line in the light spectrum of the sun by astronomers Pierre Jules Janssen (1827-1907, France) and Joseph Norman Lockyer (1836-1920, United Kingdom). It was then isolated for the first time in 1895 by the Scottish chemist William Ramsay (1852-1916).

1335 See the video Quantum Cooling to (Near) Absolute Zero by Andrea Morello of UNSW which explains very well how dilutions work, 2013 (10 minutes). This illustration is inspired from a schema seen in inspired by Cryostat design below 1K par Viktor Tsepelin, October 2018. Bcc means body-centered cubic and hcp, hexagonal close-packed. These are two states of solid helium which are of no interest in dilution refrigerators. A phase diagram shows the phase of the element as a function of temperature (in X in logarithmic scale) and pressure conditions (in Y, 1 bar = atmospheric pressure). It shows that in the regime used below 1K, helium 3 is liquid and helium 4 is superfluid. This difference makes it possible to operate refrigeration at these low temperatures.

1336 Liquid hydrogen cryogenics uses spin variations of hydrogen, instead of isotopic ones. $H_2$ molecule exists in two forms, with both hydrogen atoms having the same spin (orthohydrogen) or an opposite spin (parahydrogen). At 300K, the ratio is 75% orthohydrogen and 25% parahydrogen. At low temperature, the ratio is different and the conversion between orthohydrogen and parahydrogen is exothermic, used in the refrigeration process.

1337 The first liquefaction of helium was achieved in 1908 in Leyden, Netherlands, by Kamerlingh Onnes. The dilution cryostat concept was proposed by Heinz London in 1951 and was tested in 1965 at the University of Leiden, when it reached 220 mK. The record temperature went down to 60 mK in 1972 and then to 1.75 mK in 1999.

In **wet dilution refrigerators**, a first system cools the enclosure to 4K with liquid $^4$He. A second so-called dilution system uses a mixture of liquid $^4$He and $^3$He with a flow circulating in ducts connecting the metal plates down to less than 15 mK in the bottom stage.

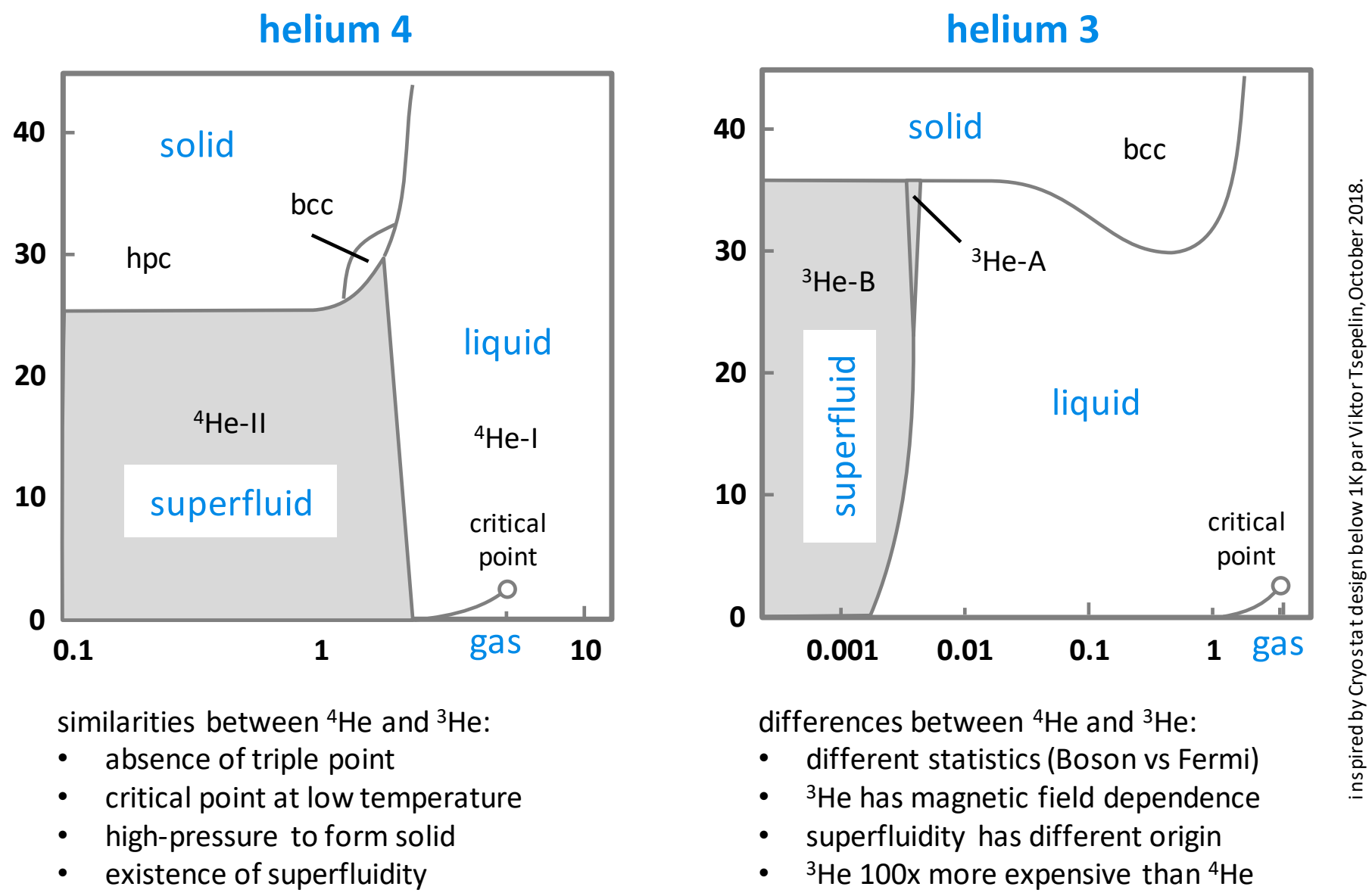

*Figure 456: phase differences between helium 3 and helium 4. Source: Cryostat design below 1K by Viktor Tsepelin, October 2018 (61 slides).*

Wet dilutions system was used until the early 2000s. It was then replaced by dry dilution systems that are simpler to operate, especially to create quantum computers that are easy to install at customer sites, thanks to avoiding liquid helium. However, wet dilution systems are still used for various physics experiments where the dry system is not appropriate, but usually not for quantum computing.

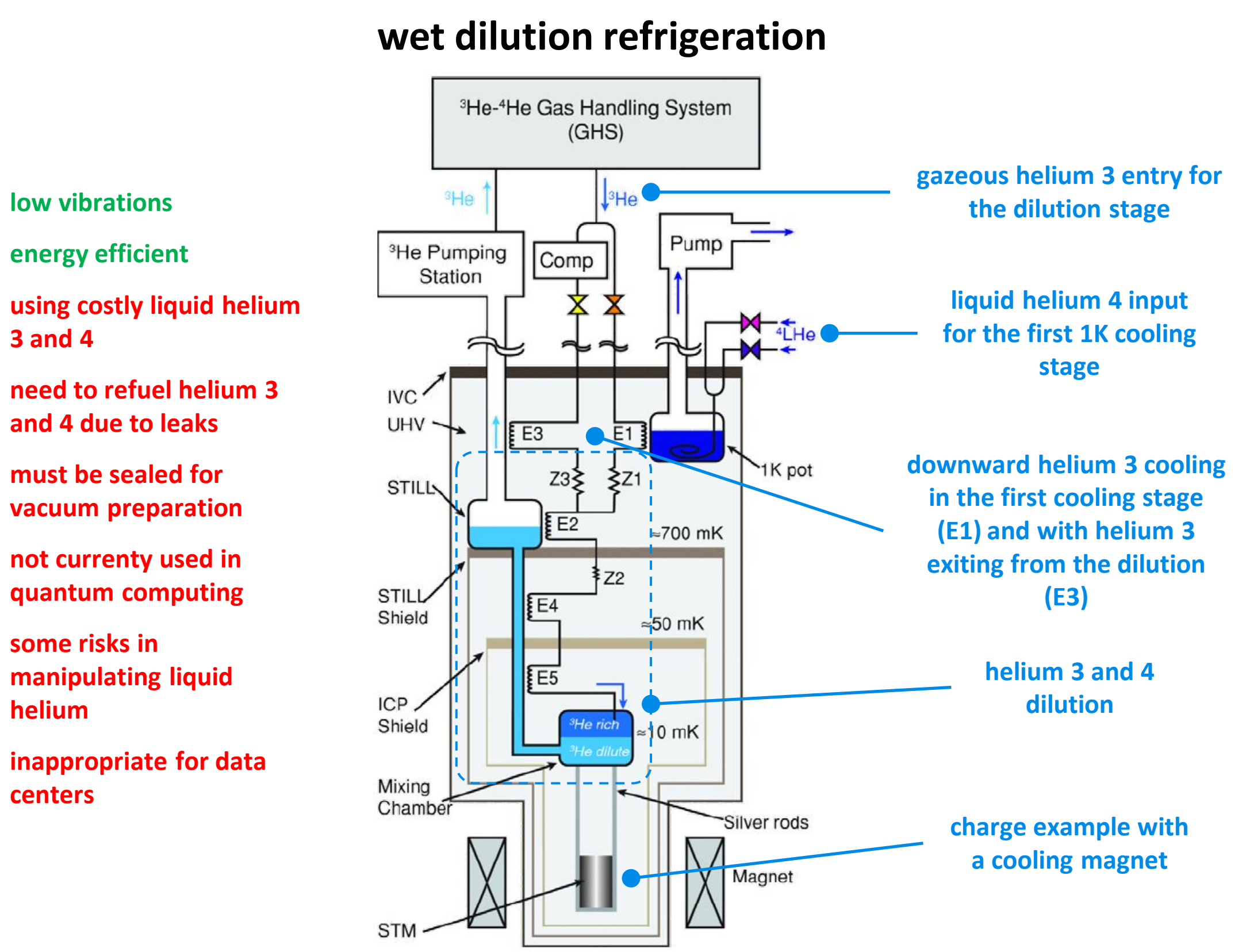

*Figure 457: wet dilution refrigerator operations. Schema from Source: Cryostat design below 1K by Viktor Tsepelin, October 2018 (61 slides) and legends from Olivier Ezratty, 2020.*

One interesting breed of wet dilution cryostats are inverted dilution refrigerators (IDR). I saw many of them at CNRS Institut Néel in Grenoble which has a dedicated cryogeny lab crafting custom cryostats for various use cases, including for astronomy. These inverted cryostats enable fast and easy experiment samples loading and fast cooling as well. Below is an example IDR from Nicolas Roch's laboratory at Institut Néel.

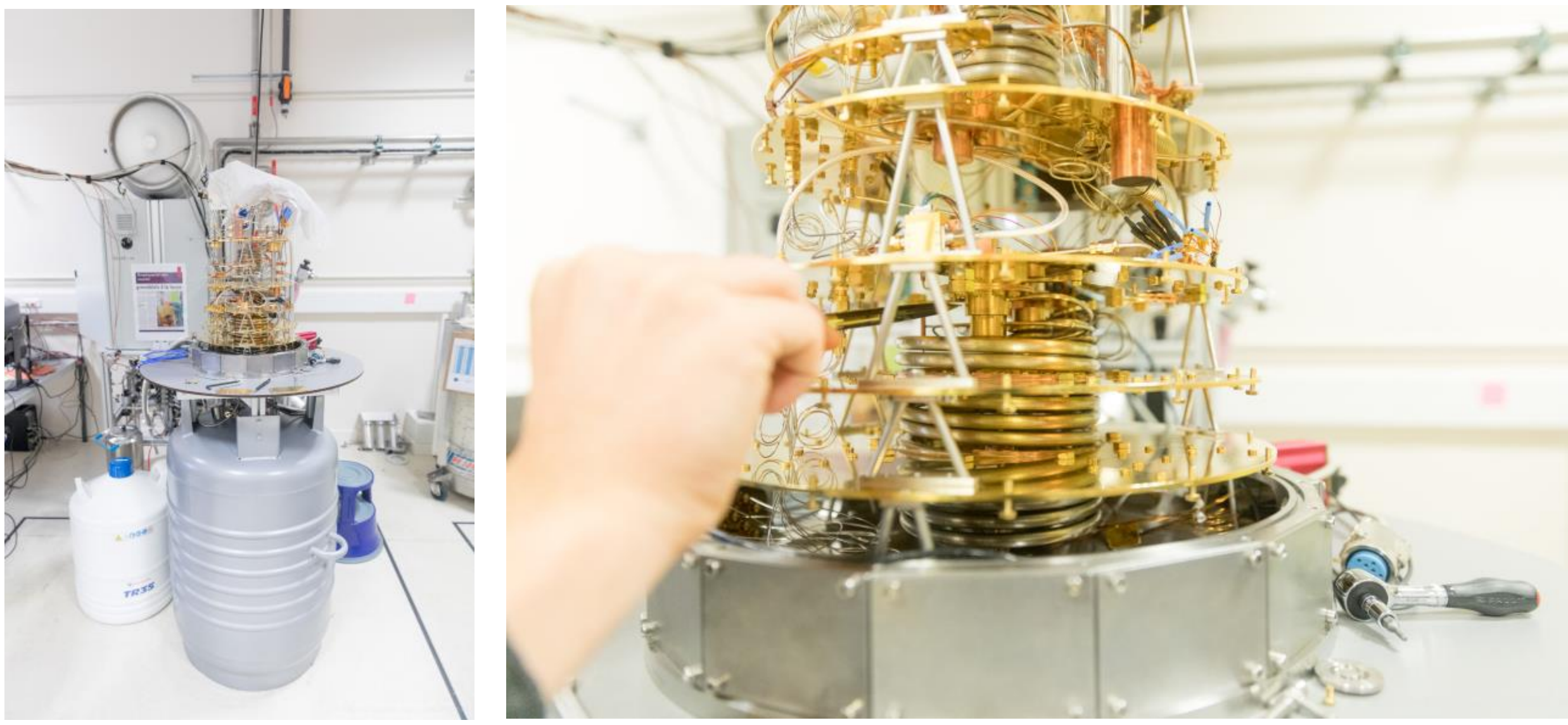

*Figure 458: custom made bottom-up cryostats made at CNRS Institut Néel in Grenoble. Pictures source: Olivier Ezratty.*

While most cryostats for quantum computers use dry dilutions, Fermilab is currently putting the final touch on an impressive wet dilution system developed to house a 3D supercomputing system, probably co-developed with Rigetti. Its numbers are impressive: it's got 625W of cooling power at 4.5 K. 50W at 2K and, above all, 300 µW at 20 mK for a load fitting in a 2m wide cylinder. And it's using 2000 liters of liquid helium[1338]. What is interesting here is that 625K cooling power at 4.5K could have another application: superconducting (non-quantum) computing.

**Dry dilution refrigerators** or so-called cryogen-free refrigerators do not use liquid helium. They are using only gaseous helium 3 and 4. Like wet systems, they have two stages: the lower dilution stage is about the same with controlled expansion of helium 3 which is bathed at the bottom in liquid helium 4 in a dilution chamber. This covers cooling to temperatures lower than 1K.

The upper stage relies on the pulsed tube technique that manages cryogenics down to about 2.8K with helium 4 gas and a large external water-cooled compressor. This technique has been mastered for about twenty years and has been progressing incrementally since then. Its arrival coincides with the first experiments with superconducting qubits. Dry dilution refrigerators are generally used for the cryogenics of qubits requiring to go below 1K. The schematic in Figure 459 explains how it works.

The pulsed tube is associated with a **Stirling** or **Gifford-McMahon** type compression and expansion system. The latter seems to be the most frequently used, particularly at **CryoMech**. It uses a piston. Stirling engines are used to cool infrared devices but not in dilution systems.

It can be seen in the curve on the right of Figure 461 that the available cooling power decreases rapidly with temperature. It is currently around 1W at 4K[1339]. There are no moving mechanical parts inside the cryostat, both in the pulse tube and in the dilution.

[1338] See A large millikelvin platform at Fermilab for quantum computing applications by Matthew Hollister, Ram Dhuley and Grzegorz Tatkowski, Fermilab, August 2021 (10 pages).

[1339] With larger liquid helium cryogenic installations like Helial SF from Air Liquide, a cooling power of 100W to 1kW can be generated at 4K.

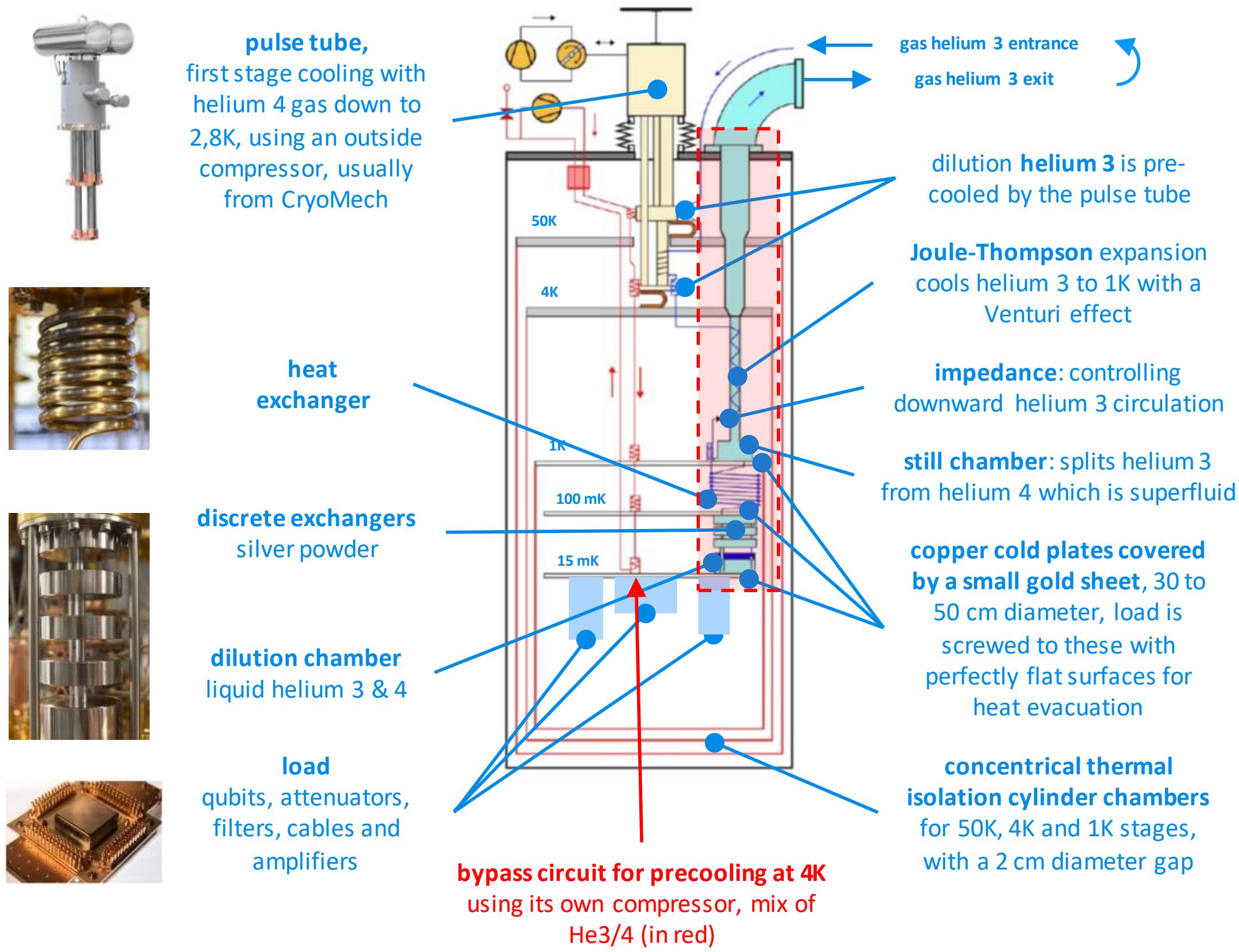

*Figure 459: dry dilution schematic inspired from Cryostat design below 1K by Viktor Tsepelin, October 2018 (61 slides), illustrations from CryoMech documentation, Janis, Dry dilution refrigerator with 4He-1K-loop by Kurt Uhlig, 2014 (16 pages) and IBM.*

This avoids the generation of unwanted vibrations that could disturb the wiring and the qubits which are very sensitive beasts. The flow of gases and liquids produces very little disturbance in the dilution process.

A refrigeration system is often evaluated in % of the Carnot cycle. This cycle describes a perfect thermodynamic cycle using four perfectly reversible thermodynamic processes involving work-heat exchange[1340]. The efficiency of a thermal machine is never perfect with 100% of this cycle.

For a pulsed tube, a perfect Carnot efficiency would be about 1.4%, i.e. it would take 70W of energy to extract 1W at 4.2K[1341]! In practice, it requires about 10 kW, i.e. 152 times more! We thus obtain a **Carnot efficiency** of less than 1%. That's <1% of 1.4%! Indeed, we spend more than 10 kW to get 1W of power at 4.2K. So... at 15 mK to get 10 µW? We do not evaluate the efficiency of the 15 mK stage of Carnot because it operates isobarically, i.e., at constant pressure, the thermal cycle being linked to a phase variation of helium 3. This stage is powered by heat exchanges between the pulsed tube and the helium 3 gas circuit.

[1340] See Cryogenic Systems by Pete Knudsen, 2018 (71 slides) which describes well the Carnot cycle principle.

[1341] See Lecture 5 Refrigeration & Liquefaction (Part 1) by J. G. Weisend II (17 slides).

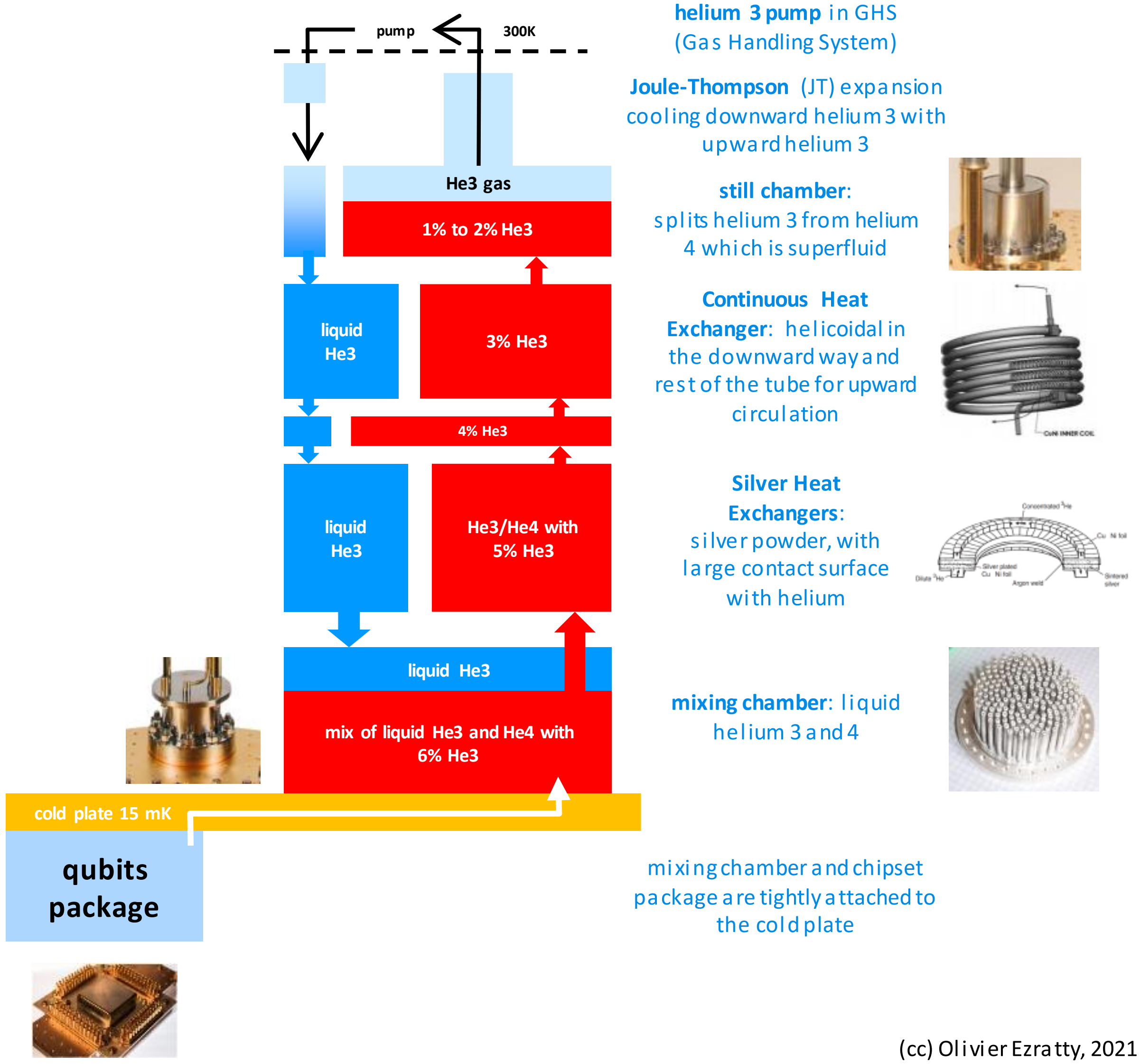

*Figure 460: details of the dilution inner working and the phases of helium 3 and 4 that are used. (cc) Olivier Ezratty, 2021.*

There is a circuit, shown in red in Figure 460, that is used to pre-cool the cryostat in the thermalization preparation. This is done in three steps: first, by starting the pulse tube which cools the 50K and 4K stages with helium 4 gas and the external compressor of about 12 kW[1342] (in yellow in the diagram). Then by using the pre-cooling circuit which will circulate a helium 3 and 4 mixture to the lower stages, and which will have been cooled by the pulse tube, in the circuit in red in the diagram.

Finally, the dilution system takes over from the second one and is launched to be able to go down to 15 mK in the lower cold plate (in light blue in the diagram).

By adopting a rocket analogy, the pulsed tube and its 7 to 12 kW compressor are the equivalent of the first stage of a Saturn V rocket. The pre-cooling system is the analogue of the rocket second stage and the dilution system is the equivalent of the third stage that sends the lunar module and the LEM to the moon, here, the chipset. Extracting the Earth's gravity over a large mass is equivalent to cooling a large metal mass inside the cryostat to 50K and 4K. While the dilution system is responsible for cooling a smaller mass from 4K to 15 mK, the lower cold plate and the payload attached to it.

[1342] At CryoMech, the compressors adapted to these dilution systems consume from 7.9 to 12kW; from PT410 to PT420. About 4kW must be added for the GHS (Gas Handling System) which manages the dilution circuits with their pumps and controls as well as for the computer and the assembly dashboard.

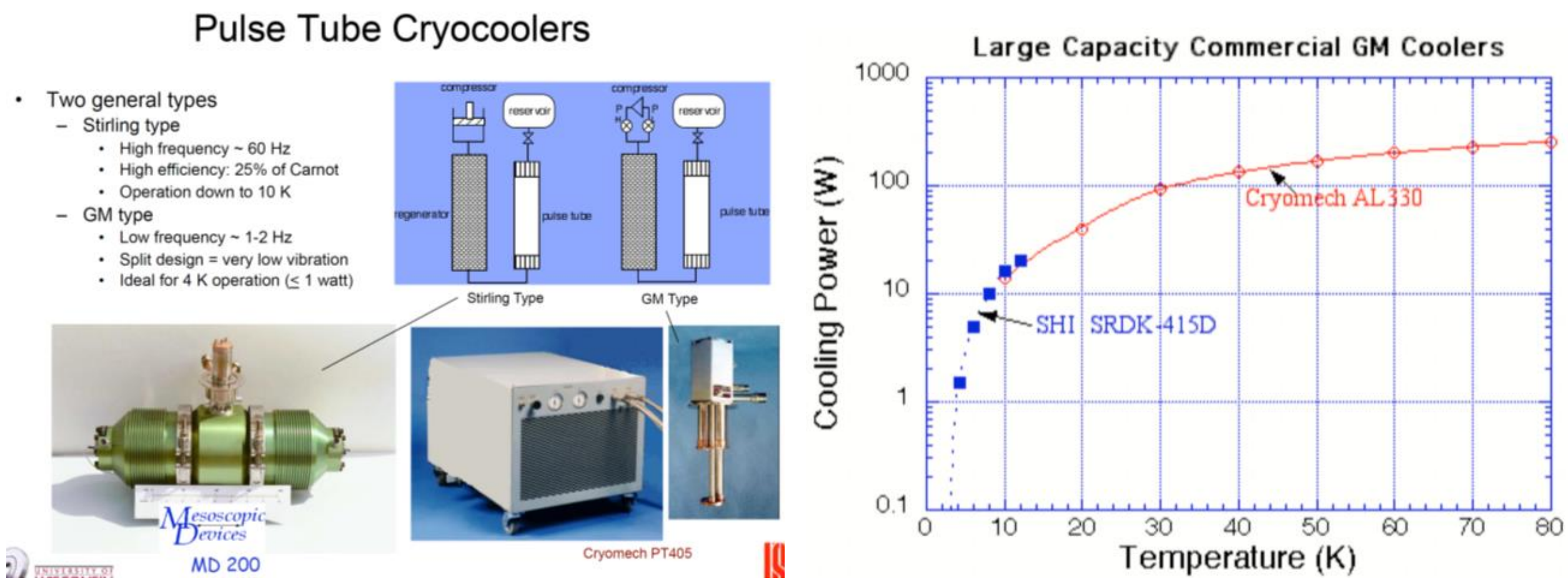

*Figure 461: pulse tubes models with Stirling and Gifford-McMahon types. And commercial capacities available. Source: [Lecture 2.2 Cryocoolers](), University of Wisconsin (25 slides).*

These systems require optimization with a large number of parameters. The modeling of a cryostat could one day benefit from quantum computation, especially since the fluids used are in a superfluid quantum state.

A good part of the power is used to lower the temperature to 1K, because the mass to be cooled is the most important. The cylinder that protects the part cooled at 4K receives the thermal radiation from the part at 50K. This makes a big thermal difference to absorb.

In a cryostat of about 16 kW, only about one third of this power is used in the dilution system, which is used to lower to 15 mK. It corresponds to the pumps in the GHS, the Gas Handling System, which contains all the pumps and gas circuits outside the cryostat, and to the share of the energy spent in the pulse tube to cool the dilution system.

The dilution system does not use a compressor. The helium 3 circulating outside is just driven by a pump located in the GHS. The reason is that the helium 3 that returns to the cryostat is cooled by the pulse tube. In practice, all the cryostat heat is evacuated by the compressor of the pulse head which is itself cooled by water.

The above diagram in Figure 460 details the operation of the dilution system as well as the phase (liquid or gaseous) and the concentration of helium 3 and 4 in each stage and component. It shows the descending circuit of helium 3 which becomes liquid from the condensation at the boiler.

In the circuit going up from the mixing chamber, a liquid mixture of helium 3 and 4 rises and the concentration of helium 3 goes down as the stages go up. It is only in the boiler that helium 3 becomes gaseous. Helium 4 remains liquid and is evacuated downwards. It has moreover a tendency to rise due to superfluidity. A trick is to cut this rising film and send helium 4 back down.

The helium 3 landing in the dilution chamber at the bottom must end up there at a temperature barely above 1mK of the chamber temperature. It is pre-cooled by the helium 3 that is moving upward. The only way to achieve this is to increase the contact surfaces, which is done in the discrete heat exchangers just below the cold plate at the 100 mK level.

These dry cryostats still use a cryogen, liquid nitrogen at 77K, to filter helium gas and remove impurities[1343]. This filtration is based on zeolite powder, made of microporous aluminosilicate crystals.

[1343] $LN_2$ for liquid nitrogen, gaseous nitrogen being a molecule of two nitrogen atoms.

The liquid nitrogen tank used for this pre-cooling is called a "cold trap"[1344]. This filtering is completed in the cryostat 4K stage by another filtering system based on activated carbon powder which works better at low temperatures and increases the contact surfaces with the gas to better filter it.

As a general rule, the complete thermalization of a quantum computing cryostat takes about 24 hours. The so-called "1K" stage was actually cooled at about 1.2K in wet cryogenics and is around 800 mK for dry cryogenics. The power consumption is identical between the thermalization phase and the temperature maintenance of the instruments once the thermalization is completed.

Cryogenics at 10-20 mK is specific to quantum computers whose qubits must be cooled at very low temperatures, mainly those based on electrons (superconductors, electron spin, Majorana fermions). Theoretically, silicon qubits should only be cooled down to 1K but for the moment, they are still cooled down to about 15mK. An Australian team created a proof of concept of silicon qubits running even at 1.5K and another one from Intel and Qutech at 1.1K[1345].

To reach **lower temperatures**, below 3 mK, a complementary technique is used, adiabatic nuclear demagnetization (ADR or Adiabatic Demagnetization Refrigeration)[1346]. It is not necessary for quantum computing. This type of refrigeration can be added to a wet or dry dilution cryostat. The principle consists in using a paramagnetic salt which is magnetized with a strong enough field, of 6 Tesla or more. This will heat the salt. The heat is evacuated via a 4K liquid helium bath. The suppression of the magnetic field cools the salt by expansion. The process complexity lies in the heating-cooling cycle which can disturb the cooled equipment. It is treated by combining several devices that take turns to smooth the temperature curve of the system. The process has been tried and tested for a long time, but the cooling power available is very low.

Even colder temperatures can be obtained with neutral atom cooling and Bose-Einstein condensates, below the nK threshold[1347]. It is based on using laser-based cooling, magneto-optical traps and the likes.

kiutra

**Kiutra** (2017, Germany) uses this technique to obtain more classical temperatures of a few hundred mK, one of its advantages being that it does not generate vibrations[1348]. These temperatures are interesting for cooling silicon qubits.

It is a startup from the TUM (Technical University of Munich) launched by Alexander Regnat. It was seed financed by APEX Ventures and German investors, but the amount is not known. Their cryostat range goes down to 100 mK (in pulsed mode) or 300 mK (in continuous mode), which is insufficient to cool superconducting Josephson effect quantum computers but could possibly be suitable for electron spin silicon chipsets that can theoretically be satisfied with a temperature of 1K.

---

[1344] Liquid nitrogen is also sometimes used to pre-cool the metallic structure of the cryostat during the warm-up. This is unrelated to the helium circuit. This can save up to five hours for the cryostat thermalization. But this process is not commonly used for quantum computer cryostats. It is used for precooling heavier payloads for physics experiments using equipment weighing up to several hundred kilograms, including superconducting magnets. This technique is not used for quantum computing.

[1345] See Hot qubits made in Sydney break one of the biggest constraints to practical quantum computers by UNSW, April 2020 related to Operation of a silicon quantum processor unit cell above one kelvin by Andrew Dzurak et al, April 2020 (in nature) and in February 2019 on arXiv. The test was performed on 2 qubits with a unit gate reliability rate of 98.6% quite average but in line with what is currently obtained with silicon qubits. See also Universal quantum logic in hot silicon qubits, 2019 (11 pages).

[1346] We owe the creation of the process to William Giauque (1895-1982, USA) in 1927. He was awarded the Nobel Prize in Physics in 1949.

[1347] See an interesting discussion on the limits of cooling in Landauer vs. Nernst: What is the True Cost of Cooling a Quantum System? by Philip Taranto, Marcus Huber et al, June 2021-September 2022 (61 pages) which makes a connection between Landauer's bound, the creation of quantum pure states and Nernst's unattainability principle, according to which infinite resources (time, energy) are required to cool a system to absolute zero temperature. They create a new Carnot-Landauer limit that generalizes Landauer's principle

[1348] See also Cryogenic Fluids by Henri Godfrin (now retired), 2011 (50 slides), from Institut Néel in Grenoble, which includes a leading research team on cryogenics. With a record of 100 µK obtained with the DN1 cryostat using nuclear demagnetization.

Their system uses the magnetocaloric effect which was discovered in stages in 1881, 1917 and demonstrated in 1933 to reach a temperature of 250 mK.

The Kiutra process is based first on this classical effect also called adiabatic demagnetization. It consists in magnetizing a solid material with magnetocaloric properties.

This makes it rise in temperature. This temperature increase is evacuated by a conventional heat transfer fluid, which is not specified. It may be helium 4 if it is a question of going down to a temperature of less than a few Kelvins. Then, the magnetization is stopped which leads the material to cool down.

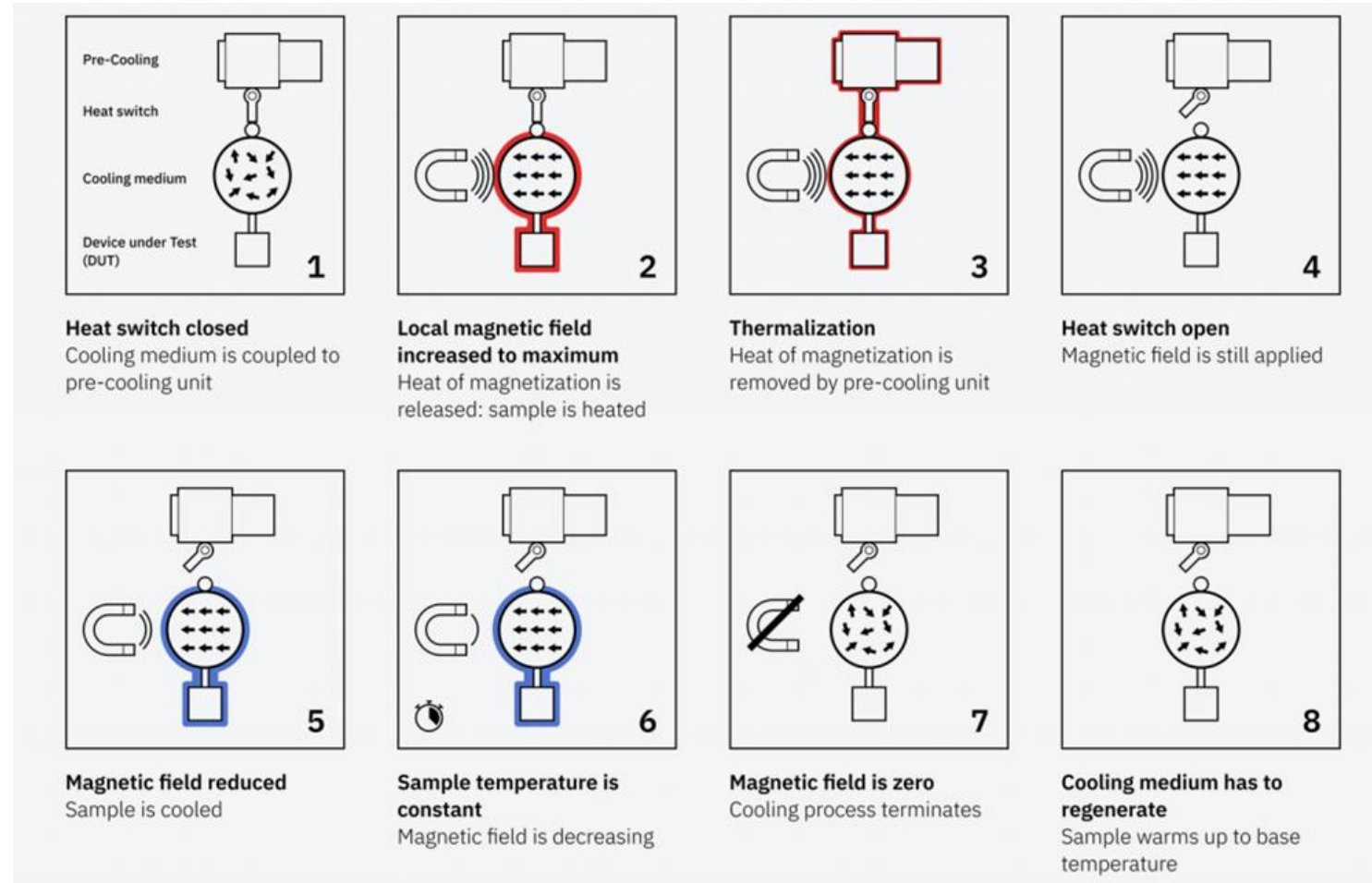

*Figure 462: the Kiutra magnetic refrigeration process. Source: Kiutra.*

To smooth in time and space this heating/cooling cycle, they combine several cooling units with what they call cADR (continuous Adiabatic Demagnetization Refrigeration)[1349] .

The apparatus proposed by Kiutra seems to be mainly designed to cool small samples and does not seem to be yet adapted to the usual architectures of quantum computers with their cooling stages stacked between 4K at the top and 15 mK at the bottom. On the other hand, some dry cryostats can reach temperatures situated between 5 and 10 mK.

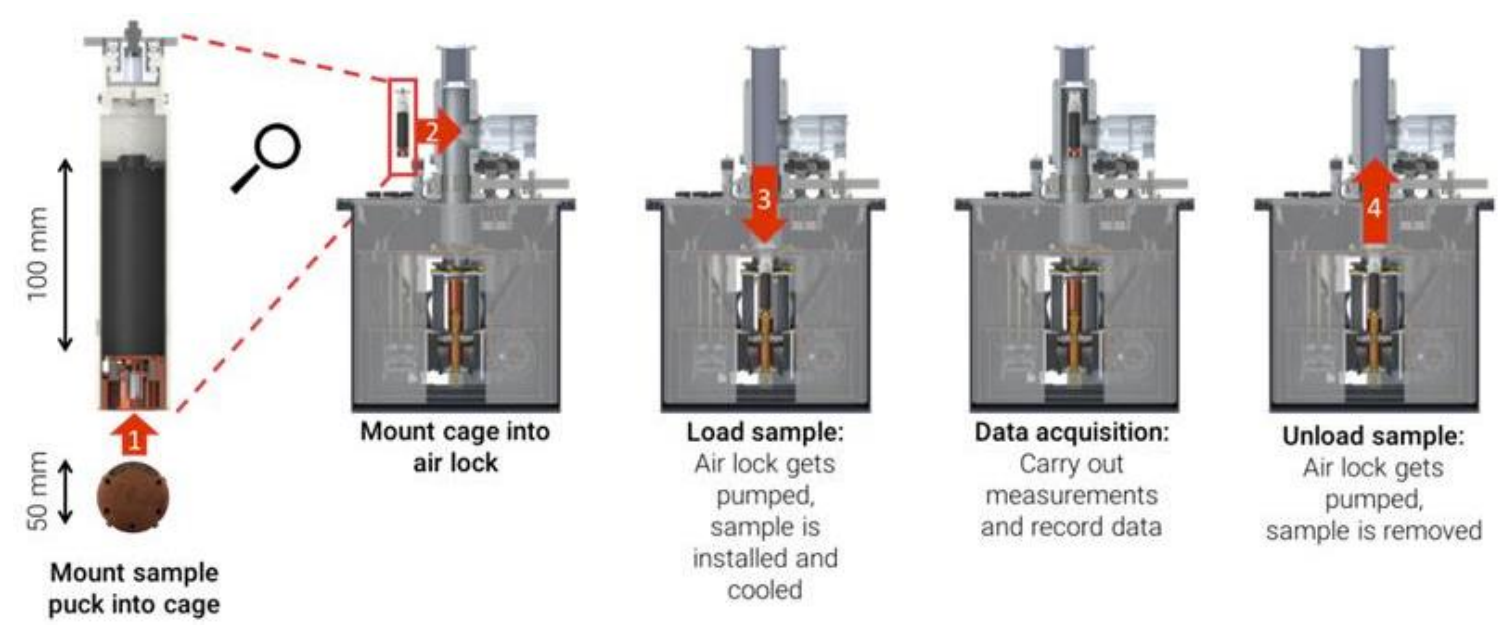

*Figure 463: Kiutra cooling process. Source: Kiutra.*

They are dedicated to physics experiments unrelated to quantum computing such as the search for dark matter (for the detection of WIMP, Weakly Interacting Massive Particles) and the analysis of cosmic radiation using calorimeters operating between 5 mK and 7 mK[1350].

### Dry dilution installation

A dry dilution refrigeration system is divided into two large parts with the compressor, pumps, liquid nitrogen and helium gas reservoirs positioned in one room, and the refrigerated enclosure in another room. This is quite logical since the compressor will generate heat that will have to be dissipated, via an incoming and outgoing water pipe[1351].

---

[1349] I discovered occasionally that this technique was also being explored at the Institut Polytechnique de Grenoble. See in particular the thesis Magnetic Refrigeration: Conceptualization, Characterization and Simulation by Morgan Almanza, 2015 (160 pages).

[1350] This is the case, for example, of the **CUORE** (Cryogenic Underground Observatory for Rare Events) bolometer installed in Italy. The cryostat comprises five pulse tubes and cools a 750 kg payload of tellurium dioxide to 10 mK. It was looking for signs of beta decay that could prove the existence of Majorana fermions. In the end, it did not find any.

[1351] See this well crafted detailed explanation of how a dry dilution cryostat works: Design and Analysis of a Compact Dilution Refrigerator by Jacob Higgins, 2017 (47 pages) and Dilution Refrigerators for Quantum Science by Matthew Hollister, 2021 (47 slides).

With dry dilution refrigerators, the safety constraints are quite light compared to wet versions. Wet dilution uses up to 80 liters of liquid helium which could drive a cryostat explosion if heated too abruptly because the expansion of the gas is important compared to its liquid state, with a ratio of 1 to 700. It was necessary to handle liquid helium canisters and fill tanks with protective equipment against splashes.

The oxygen level in the room could also dangerously decrease due to the accidental evaporation of nitrogen or liquid helium. Contact with cryogenic materials, particularly carbon steel, should also be avoided. Rooms must be large enough and care must be taken of in the higher zones in the room where helium can be concentrated since it is lighter than air.

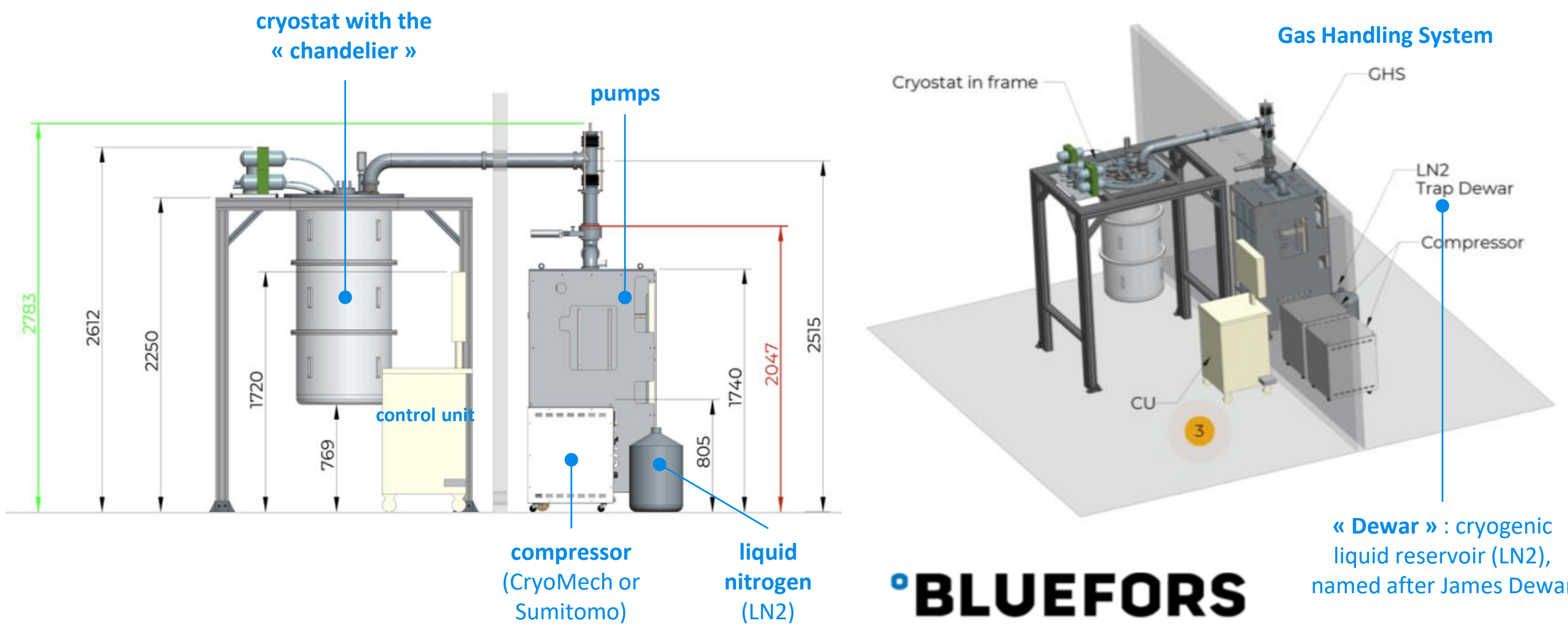

*Figure 464: Bluefors recommendations for setting up one of their dilution refrigerators. Source: Bluefors documentation.*

The wet dilution installation below is from CEA-IRIG in Grenoble, which deployed in June 2019 two systems from **Bluefors**. I visited it at the end of June 2019. These systems cost about €1 million each. The CEA teams installed a device that allows the tested sample to be changed in just 7 hours. Thermalization can thus be planning at night, and in the early morning, the experiments can be resumed.

The phenomena of materials **expansion and compression** are significant at very low temperatures. This has an impact on the design of the whole device and the choice of materials. The materials that can be used are special steels with nickel, chromium, aluminum, bronze, copper, composite materials, niobium-titanium for wiring, nickel-copper alloys, indium for joints, kapton and mylar for insulation.

The refrigerated system is usually placed in **vacuum**. The management of high vacuum and ultra-high vacuum (UHV) are industrial specialties. Knowing that cryostats of superconducting and silicon quantum computers only require high vacuum between 5 and 10 mBar. They use commercially available pumps from e.g. **Pfeiffer (**Germany). Pumping only takes place when the system starts up and is deactivated once the system is thermalized at low temperature. Cooling down to 15 mK does not require ultra-vacuum pumping because in practice, at this temperature, all the gases become solid and settle on the walls of the material, generating a very good vacuum.

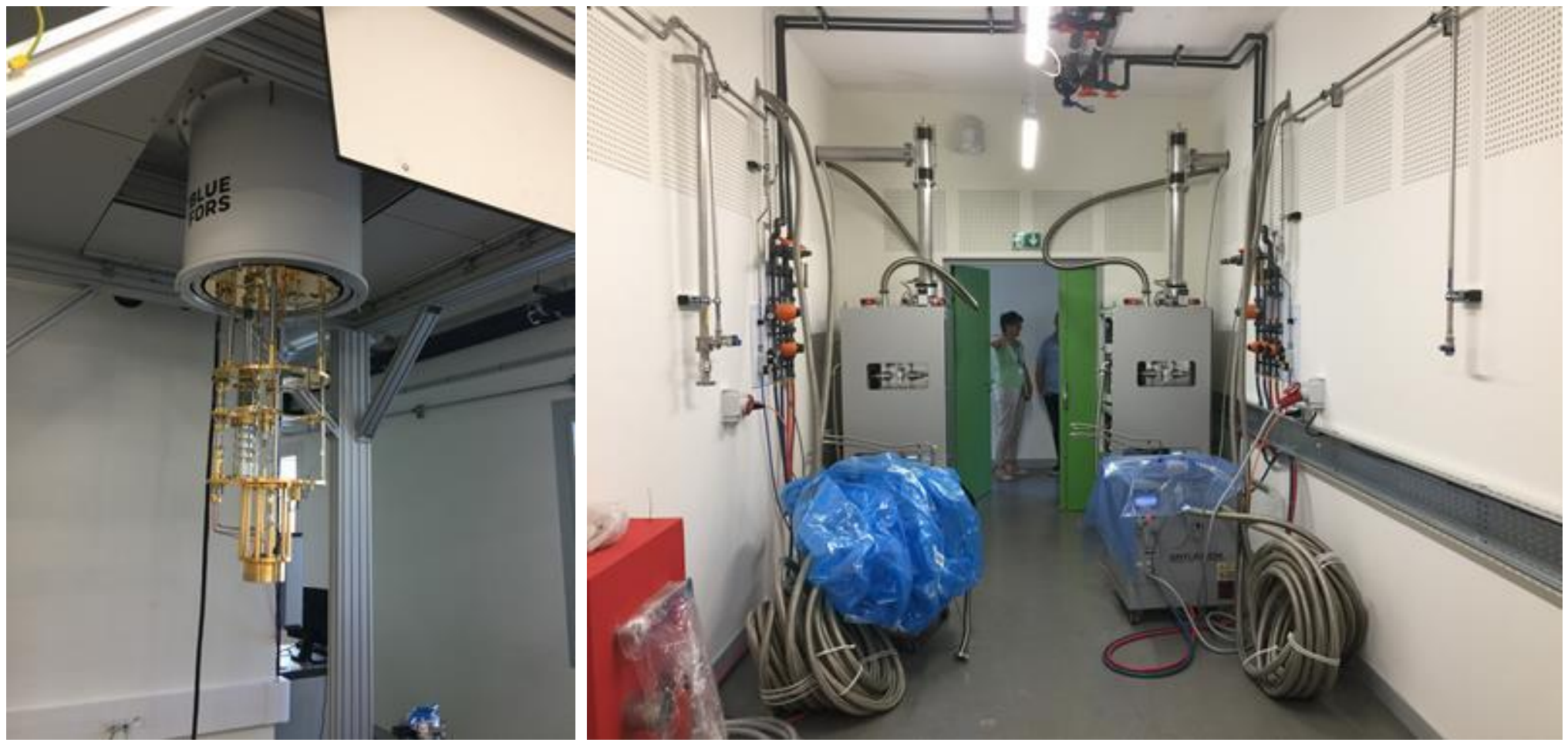

*Figure 465: Bluefors installation at CEA IRIG in Grenoble. Photos: Olivier Ezratty.*

Using too much pumping to generate ultra-high vacuum could propagate dust from these solidified gases, damaging the qubits or the rest of the equipment in the cryostat. Ultra-high vacuum is used for cold atom-based computers.

Thermal leaks are coming from cables entering and leaving the enclosure or radiation. Numerous layers of thermally insulating materials are integrated in the cryostat. They are cylinders stacked upside down like Russian dolls. It is made of aluminum, copper and steel. Each cylinder and plate acts as a thermal insulator vs the lower cylinder.

The quantum chipset is **magnetically isolated** from the outside. Magnetic isolation uses several Russian doll enclosures made of various alloys, including **Mu-metal,** an alloy of nickel, iron and molybdenum, aluminum alloys and other superconducting alloys. The quantum processor can also be magnetically shielded. IBM uses a Cryoperm Magnetic Shielding from **MµShield**.

Apart from this magnetic isolation, cryostats in research laboratories may be supplemented by **superconducting magnet** systems that occupy the lower part of the cryostat cylinder. They have a cylindrical shape that surrounds a measuring instrument. These magnets are also supplied with liquid helium to guarantee the superconducting effect that is used to generate intense magnetic fields of several Teslas.

These fields are used to set up various experiments, particularly in astronomy or fundamental physics. They are sometimes used in quantum computing, especially with silicon qubits for electrons spin control[1352]. At D-Wave, the magnetic field is reduced to one nano-Tesla (nT) in the computer enclosure, compared to the Earth's magnetic field, which can reach 65 micro-Teslas, giving us a ratio of 1 to 65,000. D-Wave communicates on a ratio of 1 for 50,000.

[1352] At CryoConcept, 8 or 14 Tesla magnets can be installed on the 4K stage next to the dilution unit.

The **cold plates** at each stage of the cryostat are generally made of 99.99% pure copper with very low oxygen content to maximize their thermal conductivity[1353]. It is covered with a thin a few microns thick layer of gold which serve as a protection against oxidation and radiation. It also has good thermal conductivity and is soft, which is very useful for solidly anchoring and cooling all the attached components.

On the right, an example of a BlueFors cryostat cold plate. These plates are custom perforated to allow all the cables to pass through, not including the cryostat components. All the holes must be used to avoid thermal leaks between the bottom and top of these cold plates.

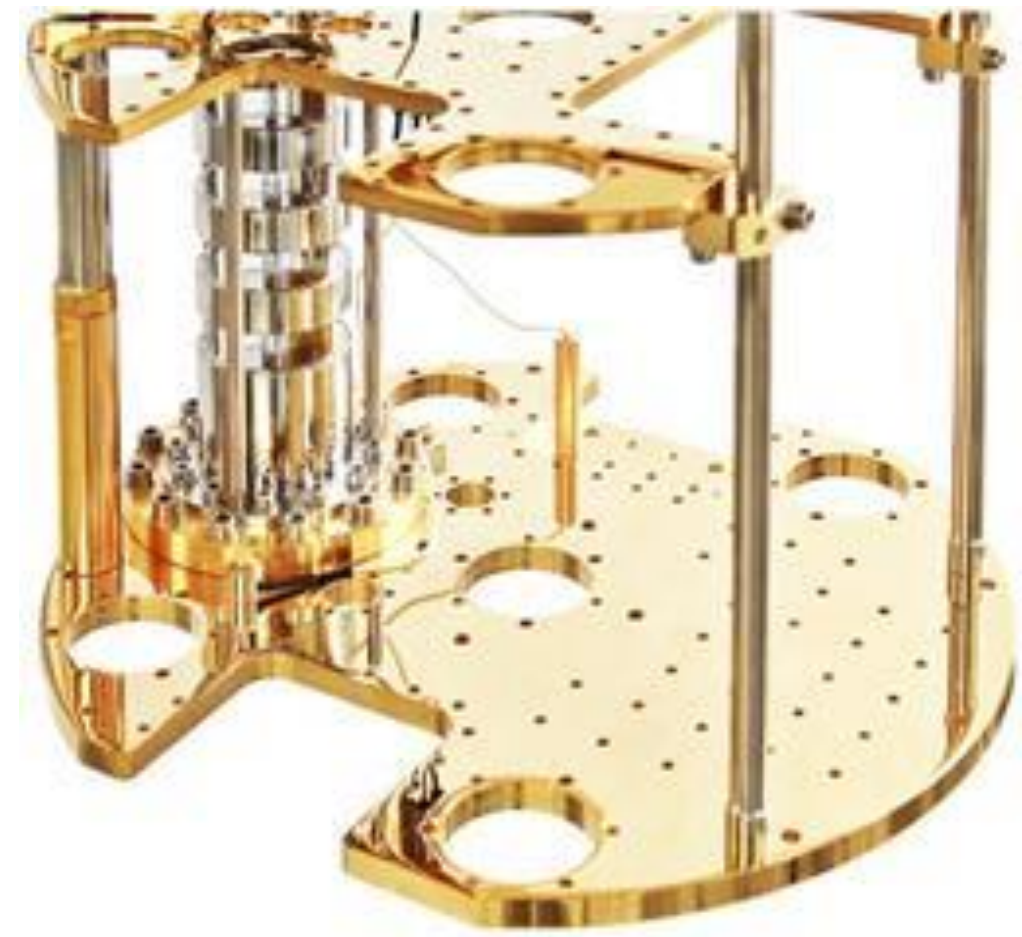

*Figure 466: cold plate with a gold finish which is used to facilitate the assembly with experimental devices and optimize thermal conductivity. Photo: Bluefors.*

Infrared rays must be prevented from passing from one level to another and generating downward heat leakage.

These plates must be optically totally watertight screens so as not to let a single photon pass through! The trend is to increase the size of the cold plates, with a diameter reaching 50 cm. Knowing that their size is slightly decreasing from the top to bottom cold plates because of the concentric cylinders shields surrounding them.

In cryostats for quantum computers, the current standard for the bottom plate is 30 cm to 40 cm for research and 50 cm in production, to accommodate more electronic components. It could soon reach 100 cm. Infrared photons are filtered with an **Eccosorb** resin that surrounds the superconducting cables in the lowest stage of the system. This resin is a mixture of epoxy and metal powder. It is injected into copper filters (OFHC) that surround the cables in the coldest stage of the cryostat as shown in Figure 467 [1354]. The resin is usually supplied by **Laird Performance Materials** (UK).

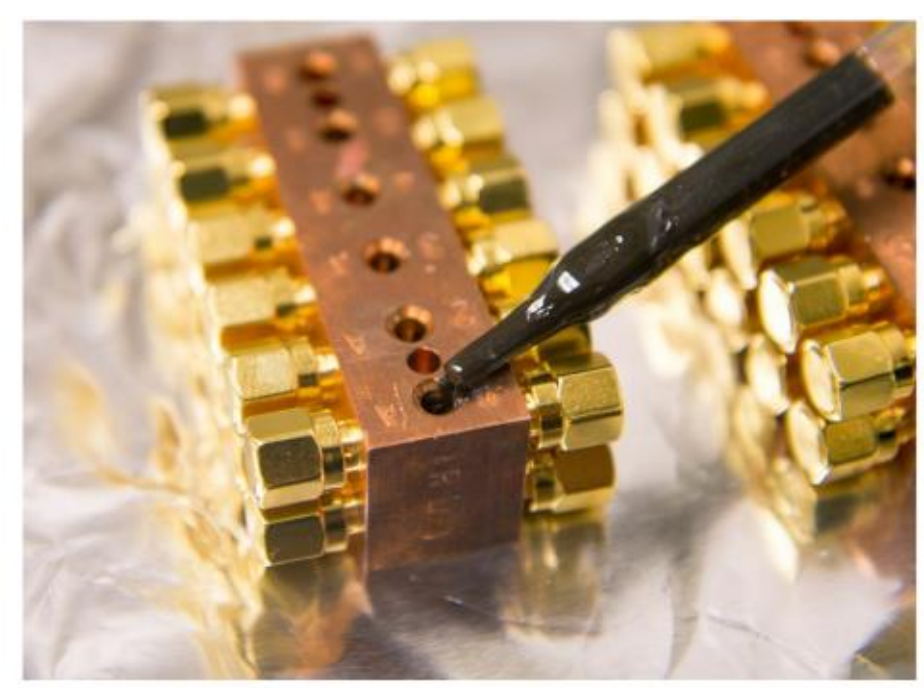

(A) ECCOSORB injection. Inject slowly and at a flat angle such that the liquid creeps onto the edge of the fill opening and into the cavity. Injecting too fast or steep will cause a planar bubble to form which blocks the opening (see Figure 3.8b).

*Figure 467: how the Eccosorb resin is injected in the filters.*

To reach ultra-low temperatures of 1 mK, the **Continuous Nuclear Demagnetization Refrigerator** technique can also be used, in complement with dry refrigeration[1355]. This temperature is required for some physics experiments but not with solid-state based quantum computers (superconducting or electron spin qubits). At such a low temperature, the cooling budget is equally super low, at just 20 nW.

---

[1353] It is OFHC for oxygen-free high thermal conductivity. Source of this information: Flying Qubit Operations in Superconducting Circuits by Anirudh Narla 2018 (219 pages).

[1354] See some explanations of the Eccosorb resin in Improving Infrared-Blocking Microwave Filters by Graham Norris, 2017 (114 pages) and Development of Hardware for Scaling Up Superconducting Qubits and Simulation of Quantum Chaos by Michael Fang, 2015 (56 pages). Eccosorb is a product from Laird, a subsidiary of Dupont. It came from the acquisition of Emerson and Cuming in 2012. Eccosorb is a laminated structure of polyurethane foam generating a controlled conductivity gradient.

[1355] See Development of a sub-mK Continuous Nuclear Demagnetization Refrigerator by David Schmoranzer, Sébastien Triqueneaux et al, Institut Néel, 2020 (7 pages).

### Cryostats vendors

The main suppliers of cryostats for quantum computers are **BlueFors Cryogenics** (Finland), which equips IBM, Rigetti and many others, **Oxford Instruments** (UK), which is used by D-Wave and Microsoft, **Form Factor** (USA), used by Google, **Maybell Quantum** (USA), **Leiden Cryogenics** (Netherlands)[1356]**,** which manufactures the most powerful cryostats on the market, used mainly for physics experiments, and **CryoConcept** (France), a branch of **Air Liquide** since 2020. The world market for cryogenic systems, all categories combined, was about $1.8B in 2020[1357]. Market share wise, BlueFors leads the pack with over 60% and >$100M revenue, followed by Oxford Instruments and the rest.

Let us recall that the science of low temperatures used in quantum computing has benefited from numerous advances from other fields: space and especially space telescopes where a large part of the instruments needs to be cooled such as infrared sensors or bolometers, particle accelerators with their superconducting magnets and finally, medical imaging, especially MRI, which also needs low temperatures to cool its superconducting magnets.

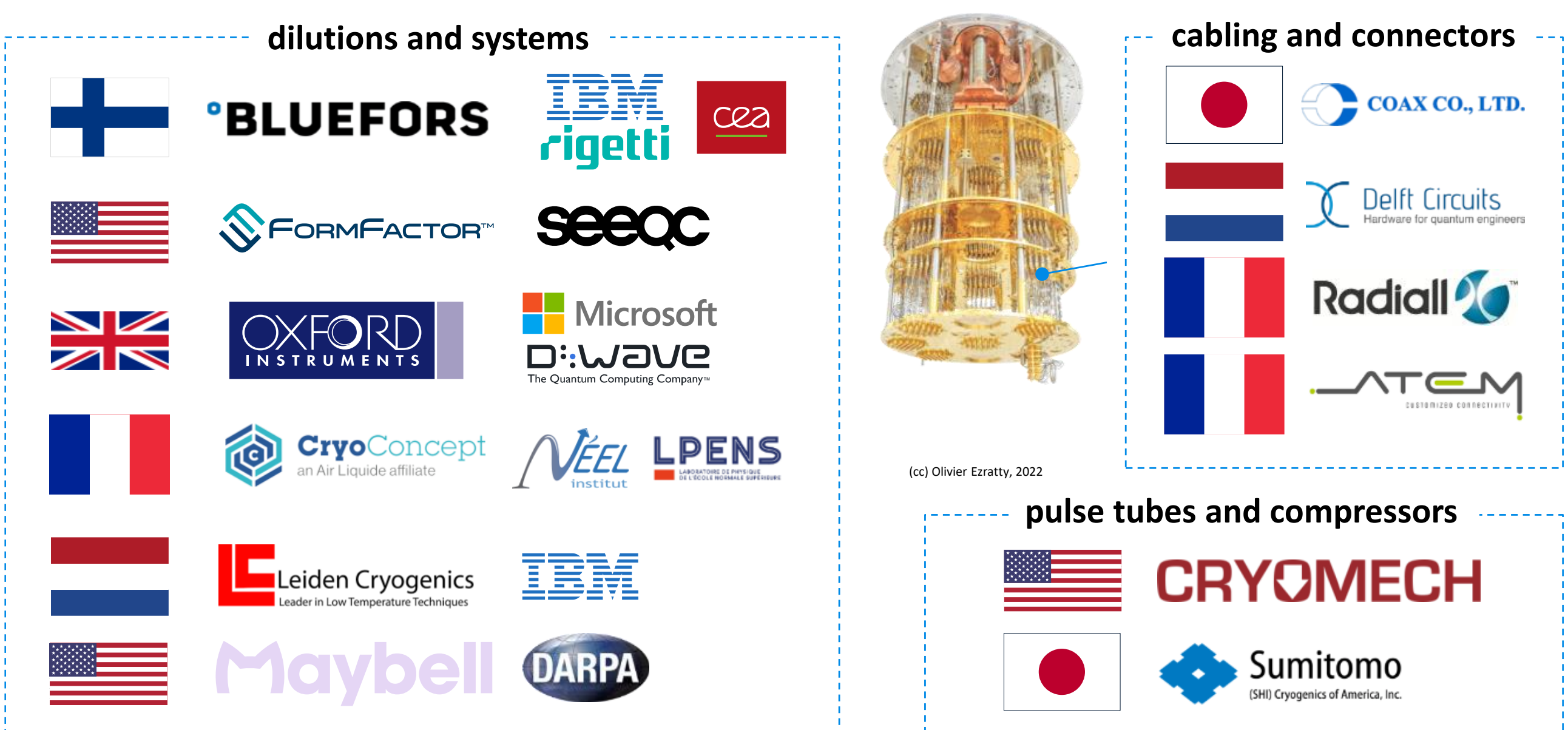

*Figure 468: the main vendors for quantum computer low temperature cryostats, their compressor, cabling and connectors. (cc) Olivier Ezratty, 2020-2022.*

**Bluefors** (2007, Finland) is the worldwide leader of low temperature cryogenic systems, using dry dilution. It's focused on the quantum computing market.

The spin-off from Aalto University delivered 600 systems with its 250 employees. It has a broad range of dry dilution systems, with some cabling (coaxial, ribbon, optical) and filters, QDevil X, codeveloped with **QDevil**.

[1356] See Leiden Cryogenics BV brochure (28 pages).

[1357] See Cryocooler Market by Type (GM, PT, JT, Stirling, and Brayton Cryocoolers), Services (Technical Support, Repair, Preventive Maintenance), Heat Exchanger Type (Recuperative and Regenerative), Application, and Geography - Global Forecast to 2022, December 2019. This market represented $1.4B in 2018 and is expected to grow 9.3% annually by 2027. But beware, the market for quantum computers cryostats is a rather small share of this market.

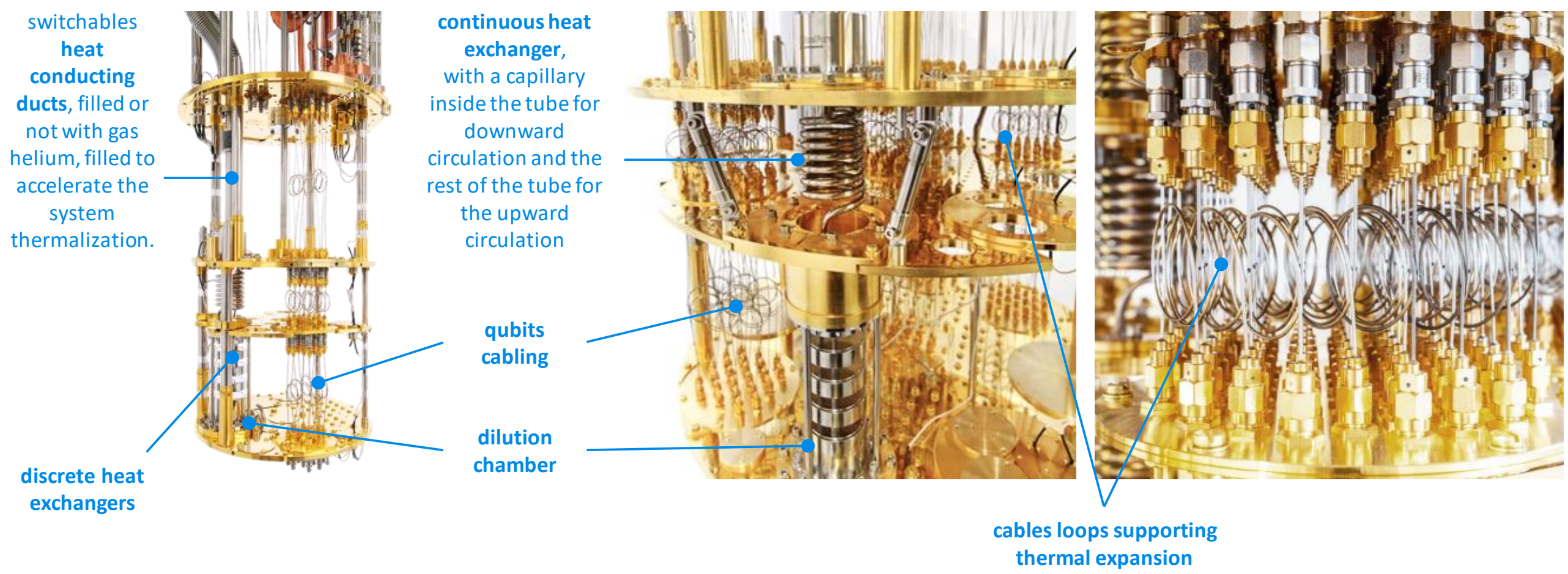

*Figure 469: details of a BlueFors cryostat with custom comments. Source: Bluefors.*

In March 2021, Bluefors announced a partnership with **Linde** (Germany) to create high-capacity cryogenic systems dedicated to scalable quantum computers. Linde is a gas producer competing with Air Liquide!

They also developed with **Afore** (Finland) a Cryogenic Wafer Prober, a system used for the characterization of 300 mm wafers at 4K. It was acquired by **CEA-Leti** in 2021 to test the quality of their silicon qubits wafers. **Intel** acquired a similar tool as well, for their own silicon qubits development efforts in their D1D fab in Hillsboro, Oregon, USA.

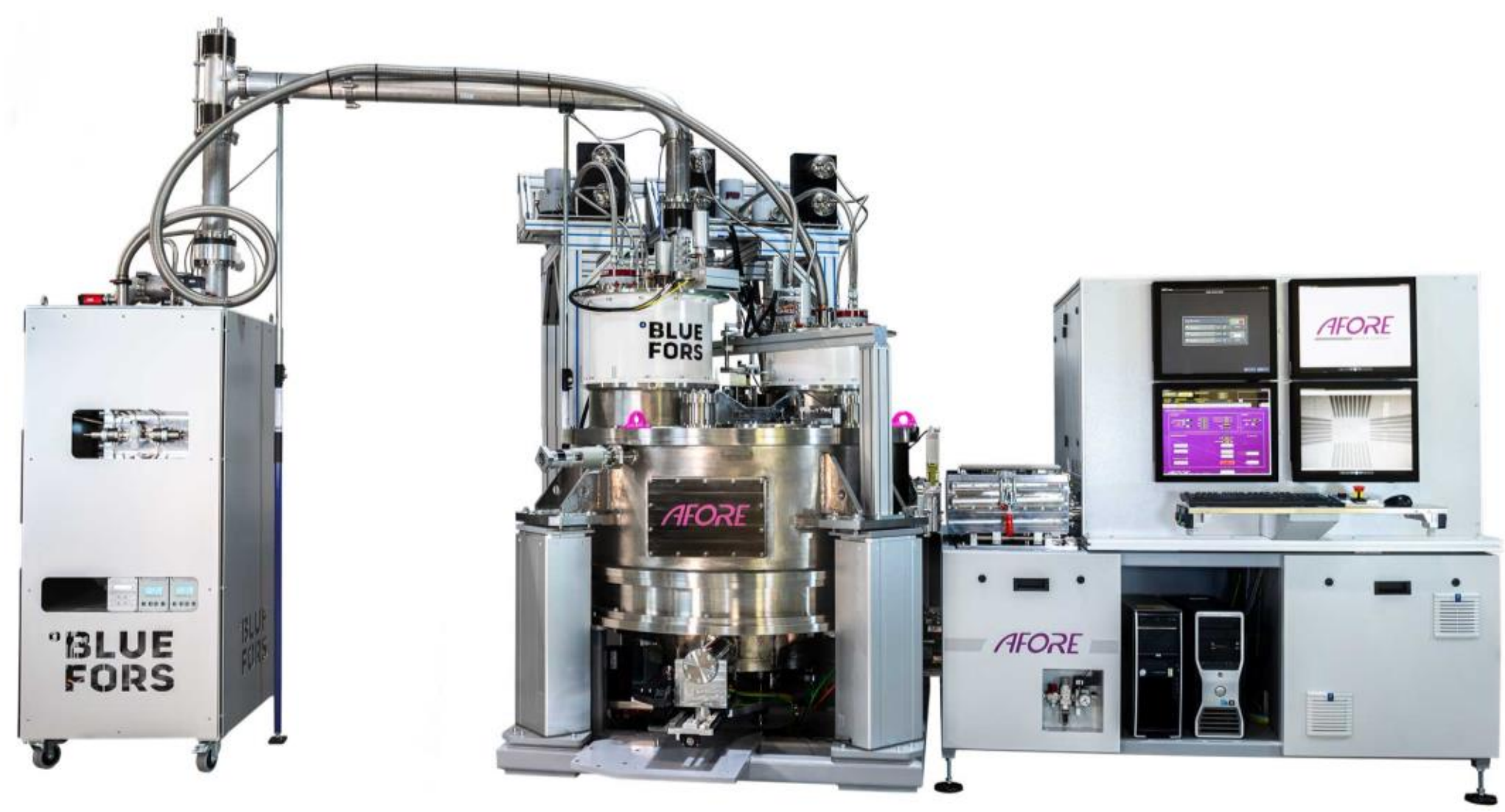

*Figure 470: the Bluefors/Afore cryoprober used by Intel and CEA-Leti.*

In November 2021, Bluefors announced KIDE, it new generation of cryostats, that has a hexagonal shape to make it easier to assemble them next to each other for distributed QPU setups and with three pulse tubes each. It will be produced in 2023 and used among others by IBM and Rigetti.

**JanisULT** (1961, USA) was initially Janis, a generalist cryostat manufacturer. In 2020, they sold their ‘classical’ laboratory cryostats business to Lake Shore.

They kept their ultra-low temperature cryostat business under the brand Janis ULT. They have an offering of wet and dry dilution refrigerators for various use cases, including quantum computing. Their high-end wet dilution refrigerator is the JDry-500-QPro with a 508 mm cold plate and >450 µW of cooling power at 100 mK achieved with a single pulse tube, coming from Sumitomo SHI.

**FormFactor** (1993, USA) is a provider of electronics test and measurement tools for the semiconductor industry with some products dedicated to quantum technologies.

They acquired in early 2022 the dry-refrigerator business line from Janis ULT. Their offering covers chipsets inspection and metrology, characterization, modeling, reliability, and design debug, to qualification and production test. They sell HPD IQ3000, a cryogenic probe stations for on-wafer and multi-chip measurements. It can embed IR-sensor test, radiometric test and DC and RF measurements. It supports 150 mm and 200 mm wafers at 4 K (while Bluefors' probe station also supports 300 mm wafers). They sell dry dilution refrigerators supporting 10 mK and below temperatures that are used for test and measurement (JDRY-250, JDRY-500, and JDRY-600, this last one offering a 630 µW cooling power at 100 mK and 17 µW at 20 mK), all coming from Janis ULT. They also sell one or two-stage ADR (Adiabatic Demagnetization Refrigerators) using a salt crystal to strong magnetic fields, complementing dilution refrigerators. Among others, FormFactor partners with SeeQC and Keysight.

**Oxford Instruments** (1959, UK) is an established British company, listed on the London Stock Exchange since 1999, specializing in scientific instrumentation including cryogenic systems capable of reaching 5 mK[1358].

They also provide CCD cameras to detect the state of trapped ion qubits, electron microscopes, vacuum deposition systems, X-ray sources and cameras, and nuclear magnetic resonance spectrographs. The company had acquired VeriCold Technologies (Germany) in 2007 to gain control of pulsed tubes used in the first stage refrigeration for dry dilution cryostats. Their last product is the Proteox, a high-end and flexible dry dilution system with removable cabling.

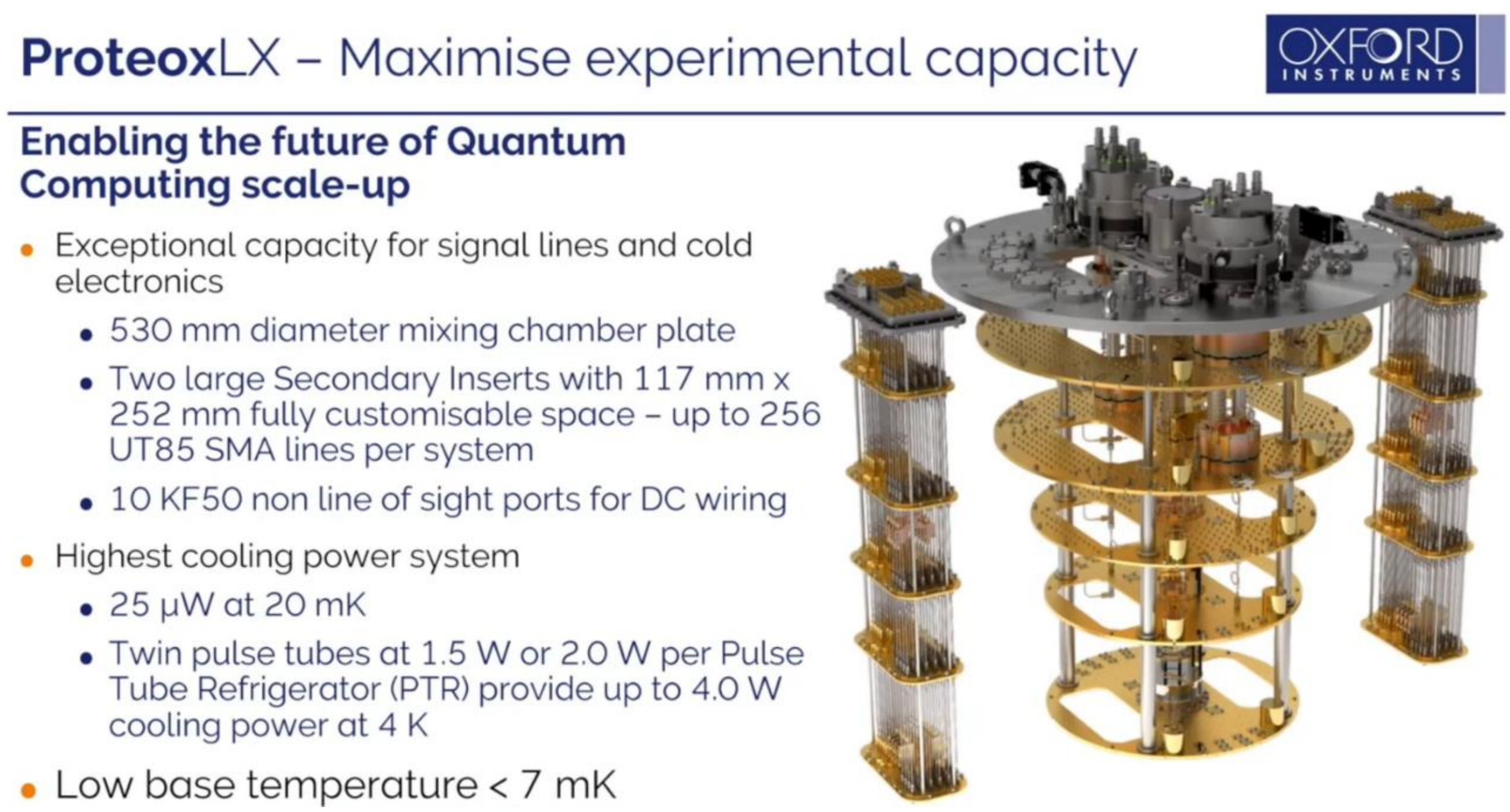

*Figure 471: Oxford Instruments ProteoxLX. Source: Oxford Instruments.*

In March 2021, they launched the ProteoxLX. It expands the qubits hosting capacity with a larger sample space and coaxial wiring capacity, low vibration and integration of cryo-electronics components. It offers a cooling power of 25 µW at 20 mK and 850 µW at 100 mK with twin pulse tubes providing up to 4 W cooling power at 4 K.

[1358] See Principles of dilution refrigeration by Oxford Instrument (20 pages) which also documents well the architecture of a cryostat.

They also designed a Q-LAN, a cryogenic link that could be used to connect two dilution fridges. The payload can reach 20 kg at 20 mK and 125 kg at 4K.

**CryoConcept** (2014[1359], France, acquired by Air Liquide in July 2020) stands out with cryostats ensuring a very low-level of vibration via their UltraQuiet technology.

They have deployed more than 120 cryostats in 13 countries for various players such as the CEA in Saclay and the ENS[1360]. Since 2018, CryoConcept has been collaborating with CEA-Leti to deliver two large cryostats to equip the QuCube project for silicon qubits. They sell worldwide including in the USA, Japan and South Korea in a market driven by dark matter research and bolometry. The unique low-level of vibrations of their cryostat is related to the absence of mechanical contact between the pulse tube and the cryostat.

By this mean, vibrations are reduced in the range from 1Hz to 1kHz. This absence of vibration is useful to preserve qubits coherence as for cryostats installations containing bolometers that are used to perform physics experiments such as in dark matter research. This experience in bolometry enabled CryoConcept to develop highly reliable dilution fridges, with systems running for more than one year without interruption. This reliability is a key attribute sought after to operate future quantum data centers.

Historically, CryoConcept started by manufacturing wet cryostats and kept an expertise in this field even though dry systems are now the most commonly manufactured dilutions. Now associated with Air Liquide, CryoConcept is working on coupling helium liquefiers with dilution refrigerators in order to overcome the current cooling power limitation at 4K, thanks to their refrigeration technology from 300 K down to 20 mK. This will ensure cooling power adapts as the number of qubits in the related quantum processors is growing.

**Leiden Cryogenics** (1992, Netherlands) was founded by Giorgio Frossati and Alex Kamper. The former had been working on dilution refrigeration since the 1970s. Among other things, he invented silver powder heat exchangers.

He started to work at the Centre de Recherche sur les Très Basses Températures in Grenoble, which became the research center on Condensed Matter and Low Temperatures (MCBT) of the Institut Néel of the CNRS. He then became a professor at the University of Leiden in the Netherlands. He designed there a dilution refrigerator reaching a record temperature of 1.85 mK with a cooling power of 25 µW at 10 mK. The heat exchanger technologies he developed were licensed to Oxford Instruments. At last, BlueFors was created by Georgio Frossati's post-docs! What a small world!

Leiden is behind what looks like the largest very low temperature cryostat ever build for a large load as shown in Figure 472. It was achieved between 2016 and 2018 for CUORE (Cryogenic Underground Observatory for Rare Events).

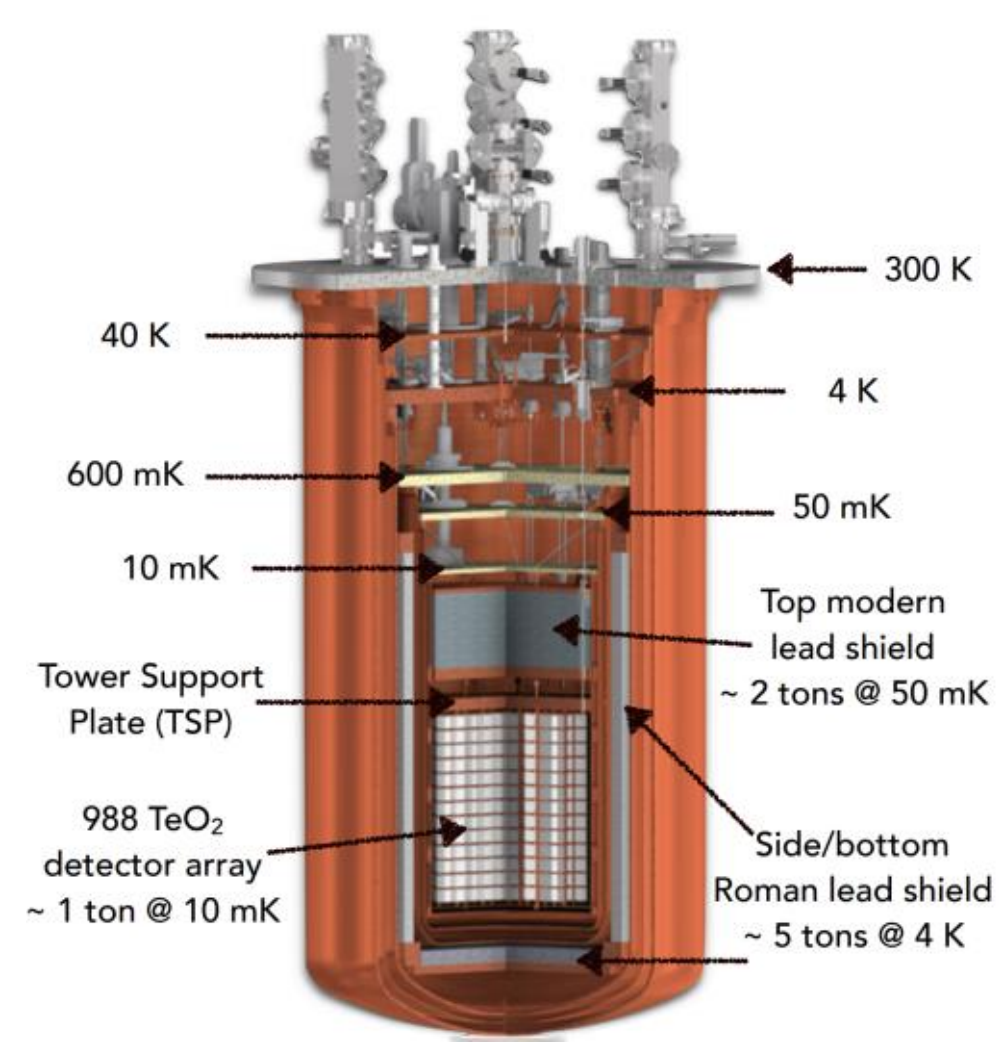

*Figure 472: the CUORE mega-cryostat cooling a load of one ton.*

[1359] CryoConcept was in fact created in 2001 by technology transfer from the CEA where Olivier Guia had worked. The company has had several different owners including French company Segula Technologies and American company CryoMagnetics. Olivier Guia took over the company in 2014. They then reintegrated the in-house R&D and in particular recovered the technological mastery that was at the CEA.

[1360] See the quantum equipment of the ENS (Ecole Normale Supérieure, in France) in their Labtour.

It is a bolometric experiment for neutrinoless double-beta decay detection in $TeO^2$ (tellurium dioxide) that is installed at the underground facility of Laboratori Nazionali del Gran Sasso (LNGS) in the Alps. The CUORE cryostat is cooling a one ton mass of metal with a one cubic meter size to 7mK. It is using a three stage cooling system with liquid Helium vapors for the first stage at 50K, 5 CryoMech PT415 pulse-tubes and compressors for the 4K stage and only a single Leiden modified dilution for the <10mK stage. Interestingly, the cooling power of the lowest stage is only of 3 $\mu$W at 12 mK, lower than the max cooling power at 15mK of most dilutions analyzed in this section. The difference is it took 26 days to cool down the experiment, including only 4 days for the last stage of one ton after 22 days to reach 3.4K[1361].

**ICE** (2004, UK) aka ICEoxford was created in 2004 by Chris Busby and Paul Kelly to design and manufacture custom wet and dry Ultra Low Temperature (ULT) cryostats (but not below 300 mK) and High Magnetic Field equipment for research applications.

**Maybell Quantum** (2022, USA, $500K) is a new company created in Colorado by Corban Tillemann-Dick (CEO, formerly at the BCG), Kyle Thompson (CTO, from the MIT Lincoln Labs and Janis ULT) and Brian Choo.

It develops a dilution refrigerator, the Icebox, that is intended to support three times more qubits in 10% of the usual space, all at a temperature below 10 mK. Above all, they announce a capacity of 4,500 cables to drive qubits with microwaves up to 12 GHz (meaning: superconducting qubits and electron spin qubits) thanks to their Flexlines, ultra-high-density RF ribbon cables and Super-Flex NbTi ribbons cables for the lower stages of the cryostat. Their Resito-Flex (CuNi-NbTi, BeCu-NbTi, CuNi-CuNi) and Atenu-Flex (SS-NbTi, SS-SS, SS-CuNi) ribbons are adapted to higher-temperature stages. It is optimized for 1000 qubits QPUs. They also provide classical coaxial cabling and fiber optics connectivity. The whole cryostat fits into two 19" server racks formats with extra space available for 9U of electronics and computing.

But it doesn't contain the compressor and the GHS (gas handling system) that controls the flow of helium in and out of the compressor and requires a space equivalent to their own system. The helium compressor can be cooled with air or water. All this looks like a game changer. On top of that, the Maybell Icebox experimental part is supposedly accessible with a simple door. It would probably require many Russian-doll doors given a cryostat have about five layers of isolations as portrayed in their own schematics below.

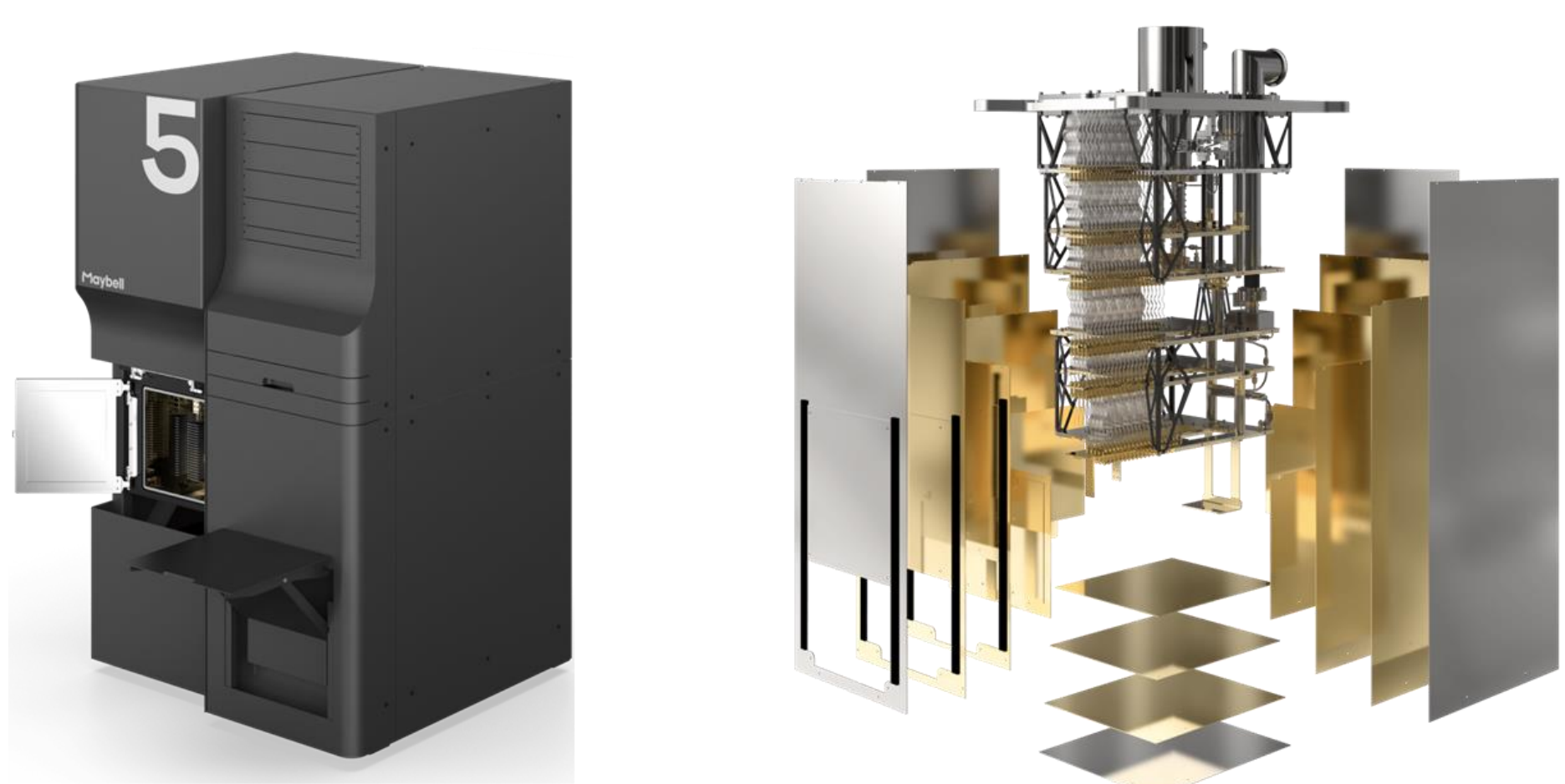

*Figure 473: the Maybell Quantum cryostat unveiled at the APS March meeting 2022 in Chicago. Source: Maybell Quantum.*

[1361] See The CUORE cryostat by A. D'Addabbo et al, August 2018 (8 pages).

The company initial funding came from Colorado's Advanced Industry Accelerator (AIA) venture fund and the US DoD National Security Innovation Capital fund (NSIC) which belongs to the Defense Innovation Unit. The company said it already has some contracts from DARPA.

**Absolut System** (2010, France) was created by Alain Ravex, former head of the low temperature department of the CEA in Grenoble in the 1980's and 1990's, then a consultant for Air Liquide. He sold the company to his partners and now consults for them and other players in the cryogeny field.

The company develops custom cryostats running at temperatures higher than 1.8K and targets a wide range of applications in research and in the industry, particularly for the production of liquid nitrogen. Their customers include CEA-Leti, Thales and Air Liquide. They are based near Grenoble.

They developed the ACE-Cube (Advanced Cryogenic Equipment), a cryogen free helium cryostat using a remote cooling technique. It is implemented for specific infrared detectors and semiconductors characterization and above 10K.

They also launched AFCryo (2017), a joint subsidiary in New Zealand, with Fabrum Solutions (2004) also based in New Zealand[1362].

**MyCryoFirm** (2013, France) produces dry cryostats, running at 3K with a cold plate of 250 mm diameter with a 300 mW cooling power at 4,2K. They rather target the field of research in quantum optics, quantum physics and quantum sensing.

They propose various experiment decks/plates adapted to creating magnetic fields, spectroscopy applications and the likes. In 2022, they did add a dilution option on their Optidry250 cryostat and made it operate at 50 mK at Météo France.

**Cryomech** (1963, USA) is a supplier of components for cryostats and in particular dry cooling systems comprising a pulsed tubes and a compressor which are integrated in the cryostats of most market players such as BlueFors and CryoConcept.

These pulse tubes and compressors are the first stage of dry dilution refrigeration systems. They use an expansion system of compressed gas outside the cryostat with no rotating parts in the cryostat [1363]. The compressor is water-cooled, with a flow rate of 5 to 12 liters per minute depending on the incoming temperature. But this water must also be cooled, and it can require up to an additional 10 kW of electric power unless the computer is located in a cool region.

Their pulse tubes range includes the PT415 and PT420 (*right*). Its main competitor is the SHI Cryogenics Group subsidiary of **Sumitomo** (Japan, *left*)[1364]. These compressors are sold combined with their related pulsed tubes.

[1362] See Commercial Cryocoolers for use in HTS applications by Christopher Boyle, Hugh Reynolds, Julien Tanchon and Thierry Trollier, 2017 (29 slides).

[1363] These pulsed tubes are used in particular in the semiconductor industry, in vacuum deposition machines (CVD, MOCVD) and plasma deposition machines. They are down to 10K, which is sufficient for semiconductor production.

[1364] There are other pulse head and compressor manufacturers such as Fabrum Solutions (New Zealand) but the latter only targets temperatures of 77K for liquid nitrogen production.

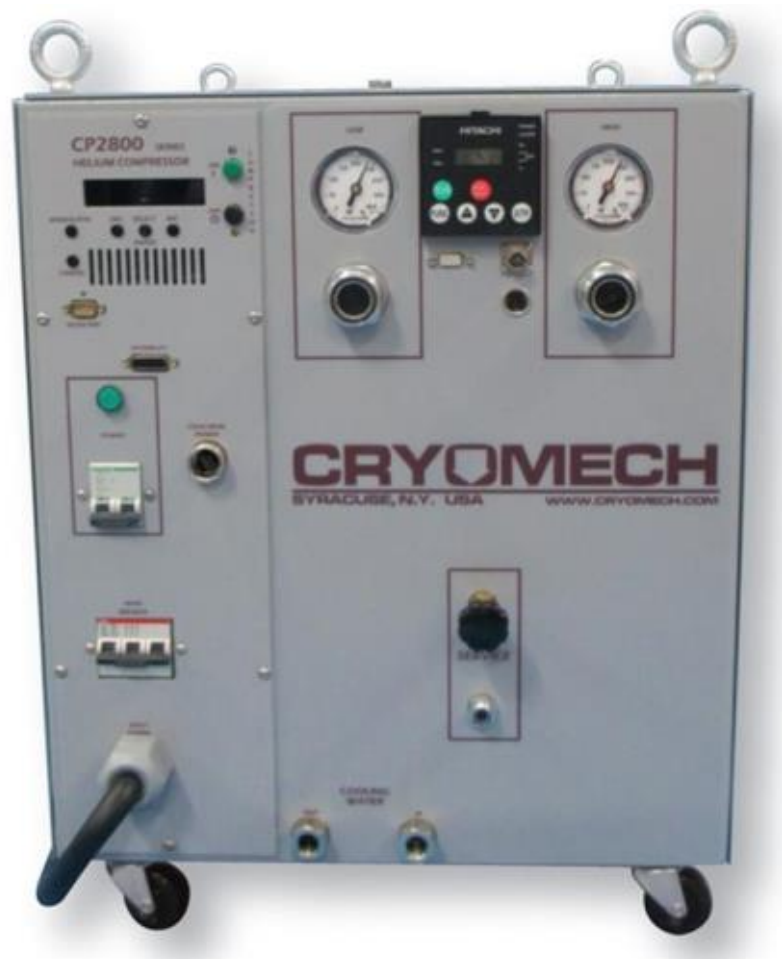

*Figure 474: a Cryomech compressor, that is connected to a pulse tube (on the right).*

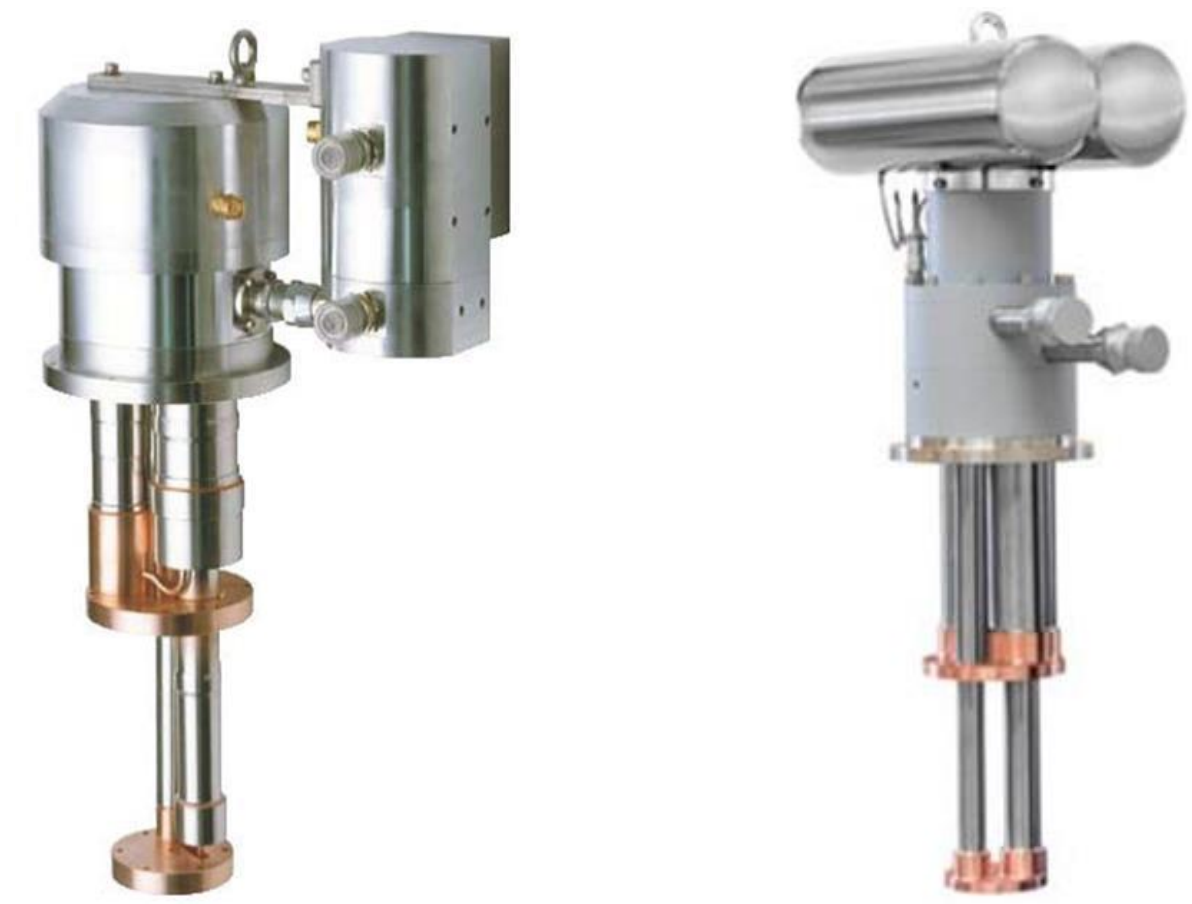

*Figure 475: Cryomech pulse tubes, that cool a cryostat down to 4K. It is also used to cool down the helium 3 and 4 mixture circulating in a dilution.*

**High Precision Devices** (1993, USA) develops cryogenic instruments adapted to superconducting quantum computers and in particular sensors. It is very specialized low-level instrumentation. They also develop ADR (Adiabatic Demagnetization Refrigeration) type cryogenic systems. It was acquired by **FormFactor** (1993, USA) in 2020, an advanced SoC and memory probe cards designer for the semiconductor industry.

**Intelline** (2018, Canada) produces customized cryogenic refrigeration systems that are expected to be more affordable than those of its competitors. But they seem to target markets other than quantum computer cryogenics, at least at temperatures below 1K.

**CryoFab** (1971, USA) provides liquid helium containers and related accessories.

**Cryogenic Limited** (1991, UK) provides a various set of cryogenic systems and superconducting magnets. It includes liquid helium systems and ultra-low temperature systems using their own magnet and an off-the-shelf cryostat from Leiden.

**Qinu** (Germany) is a new company selling mK and 4K cryostats. It was created by a former researcher from Institut Néel in Grenoble, which has its own cryostats design laboratory.

**Attocube Systems** (2001, Germany) has different line of businesses including cryostats mainly targeting the research community. It sells the attoDry series, closed-cycle dry cryostats (with cooling temperatures ranging from 1.65K to 4K), and the attoLIQUID series (300 mK), liquid helium cryostats.

**Montana Instruments** (2009, USA) develops cryostats and vacuum pumps, used in trapped ions (including IonQ) and NV centers computers as well as photon-sources (Sparrow Quantum).

One of their added value is to reduce the vibrations coming from the pulse tube. They cover temperatures ranging from 3.2K to 4.9K.

There are many other cryostats and cryogenic devices vendors but they are less specialized in serving the needs of quantum technologies providers[1365].

### Cooling budgets

The level of cooling power at ultra-low temperature is quite low. This limits the energy that can be released by the qubits themselves and by the microwave attenuation and amplification circuits used to read the state of the qubits. See in Figure 476 a comparison of these cooling power budgets by supplier.

| | cryostat | pulse tubes | minimum temperature | 20mK stage | 100mK stage | MC cold plate |
|---|---|---|---|---|---|---|
| BLUEFORS | LD250 | 1 | 10 mK | 12 µW | 250 µW | 30 à 50 cm |
| | XLD400 | 2 | 8 mK | 14 µW | 450 µW | 30 à 50 cm |
| | XLD1000 | 2 | 8 mK | 34 µW | 1000 µW | 30 à 50 cm |
| FORMFACTOR™ | JDry-500-QPro | 1 | 7 mK | 14 µW | 500 µW | 50 cm |
| OXFORD INSTRUMENTS | TritonXL | 2 | 5 mK | 25 µW | 1000 µW | 43 cm |
| | TritonXL-Q | 2 ou 4 | 7 mK | 25 µW | 850 µW | 50 cm |
| | Proteox | 1 | 10 mK | >25 µW | 500 µW | 36 cm |
| CryoConcept an Air Liquide affiliate | HD200 | 1 | 10 mK | 11 µW | 350 µW | 30 à 50 cm |
| | HD400 | 1 | 10 mK | 10 µW | 400 µW | 30 à 50 cm |
| Leiden Cryogenics Leader in Low Temperature Techniques | CF2400 Maglev | 2 | 4 mK | ? µW | 2000 µW | 49 cm |
| | CF1400 Maglev | 2 | 8 mK | ? µW | 1000 µW | 49 cm |

(cc) Olivier Ezratty, 2022

*Figure 476: cooling power per temperature and cryostat vendor. (cc) Olivier Ezratty, 2020-2022.*

The **BlueFors**' refrigeration thermal budget ranges from 12 µW (LD250) to 30 µW (XLD1000) at 20 mK, and from 250 µW (LD250) to 1000 µW (XLD1000) at 100 mK.

**Oxford Instruments**' TritonXL also has a thermal budget of 1000 µW at 100 mK but with two pulsed tubes, while the new Proteox reaches 500 µW ... with only one pulsed tube. It is completed by a removable system for qubit control cables supporting up to 140[1366].

The **Janis** JDry-500-QPro has a thermal budget of 14 µW at 20 mK and 450 µW at 100 mK (*above*, in-house compilation).

The current record can be found at **Leiden Cryogenics** with a recent cryostat with a thermal budget of 2000 µW on the 100 mK stage, but the budget at 20 mK is not indicated in their literature. On the 4K stage, the available thermal budget is around 1W. But beware, these extreme performances above 500 µW are often obtained with two pulsed tubes instead of one and thus, double the external compressor and power drain. All this with a double dilution refrigeration system to go below 1K. It is also possible to have systems with a single pulse tube and two dry dilution systems.

The thermal budget of the coldest stage is conditioned by the equation: $Q_m = 84\dot{n}_3T^2$ where $Q_m$ is the cooling power in W, $\dot{n}_3$ is the flow velocity in mol/s of helium 3 in the cryostat at this stage and T is the temperature of the stage in Kelvin. This law that can be simply called "Q=84NT2" explains that the thermal budget at 15 mK is very low compared to the cooling budget available at the upper stages (up to 25 µW at 15 mK, 1 mW at 100 mK and 1.5W at 4K).

There is another constraint related to the Kapitsa resistance. It limits heat exchanges between helium 3 and the heat exchanger. These exchanges are proportional to $T^4$. If we therefore want to multiply heat exchanges by 10, the exchange surfaces in the lower parts of the dilution system would have to be multiplied by 10,000! This is done with using silver powders integrated into the discrete heat exchangers above the dilution chamber. These powders are structured to maximize the heat exchange surface area with the helium gas flowing through them. Their deposition process must maximize the flat contact surface with the small tanks where they are located.

---

[1365] See 61 Ice Hot Companies Transforming The Cryogenics & Alternative Cooling Systems Industries, January 2021.

[1366] See the very interesting presentation 50 years of dilution refrigeration, by Graham Batey of Oxford Instruments, 2015 (26 slides).

It is possible to increase cryostats cooling power with adding more cooling stages, improving their Carnot efficiency and with multiplying the pulse tubes and dilutions (like what IBM has done with its Goldenye cryostat).

**Other cryogenics**

For the other types of qubits, the cooling requirements are different: trapped ion qubits are not theoretically refrigerated, but Honeywell's prototype ion trapped processors announced in early March 2020 are cooled to 12.6K, a temperature that can be obtained with helium 4 based cryostats.

In photon-based quantum processors, the optical components traversed by the photons (mirrors, prisms, interferometers, whether miniaturized in nanophotonics or not) are not refrigerated, but the photon sources and photon detectors are, at temperatures between 1K and 10K. The associated cryogenics is much lighter and consumes less energy compared to dilution cryostats.

Other techniques allow very localized cooling. This is the case of the **Doppler effect** which works on cold atoms suspended in a vacuum. Another solution developed by researchers from the VTT Technical Research Centre in Finland would cool silicon components with a phonon-based electronic cooling technique. It seems that this cooling's capacity is very low, very localized, and still requires pre-cooling the system to at least 244 mK. It is therefore still necessary to operate a helium 3 and 4 dilution cryostat [1367].

## Thales Cryogenics

**Thales Cryogenics** (France, The Netherlands) is a subsidiary of Thales Group which creates various specialized cryocoolers for military and commercial applications.

It includes rotary and linear Stirling coolers, mini-coolers, high-pressure gas compressors, miniature DC/AC rotary and linear converters, linear pulse tube cryocoolers etc.

Thales' NV centers-based quantum sensors use miniaturized cooling using liquid nitrogen and occupying only half a cubic decimeter[1368] . The required temperature is lower, around 70K which is quite hot compared to 15 mK! These cryostats are used for various breeds of quantum sensors.

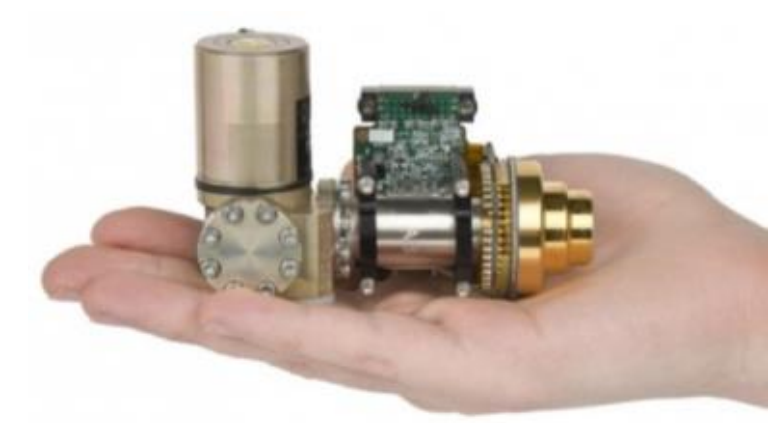

*Figure 477: a small Stirling cooler for embedded systems. Source: Closed Cycle Refrigerator by John Wilde, 2018 (11 slides).*

# Qubits control electronics

Most of the times, driving qubits with quantum gates and for their state readout requires sending them some sort of photons. For superconducting qubits and electron spin qubits, these photons are in the microwave spectrum. In a counterintuitive fashion, these microwaves are transmitted in coaxial cables and not over the air like radio waves. Their frequencies range between 4 and 8 GHz for superconducting qubits and between 12 and 26 GHz for electron spin qubits. These are in between higher-frequencies photons that can be transmitted in optical fiber and lower frequencies signals which are transmitted as classical electrical current in wires. These photons are generated as pulses of various shapes and duration (cosine signals shaped with an envelope, base pulses of diverse forms, cosine or other, and direct-current pulses which are the simplest to generate).

[1367] See Thermionic junction devices utilizing phonon blocking by Emma Mykkänen et al, 2020 (9 pages). It reads: "*The cooling power for this sample is about 2 pW/µm2 at 300 mK*". "*Our best-performing sample is S2 (subchip with 1-mm diameter and 0.4-mm height). Its maximal absolute and relative temperature reductions are 83 mK (at 244 mK) and 40% (at 170 mK), respectively*". Therefore, it is already necessary to reach 244 mK before starting, and it is therefore necessary to use a helium 3 and 4 cryostat.

[1368] These are usually systems using a Stirling engine. Thales Cryogenics produces such miniaturized refrigeration systems. The RM2 cools a payload to 77K for a mass of 275g and a thermal budget of 400 mW at this temperature. It is notably used for cooling infrared cameras in embedded systems. This type of small cryostats can also be found at SunPower (USA), capable of cooling down to 40K and with a larger mass of 1.2 kg. Ricor (1967, USA) is another manufacturer of this kind of mini-cryostats.

We'll look here into two sorts of micro-wave generation technologies: those coming from room temperature electronics and those generated within the cryostat at cryogenic temperature, including cryo-CMOS, SFQ superconducting electronics and other discrete electronic components working at these low temperatures like the TWPAs used for qubits microwave readout signals amplification.

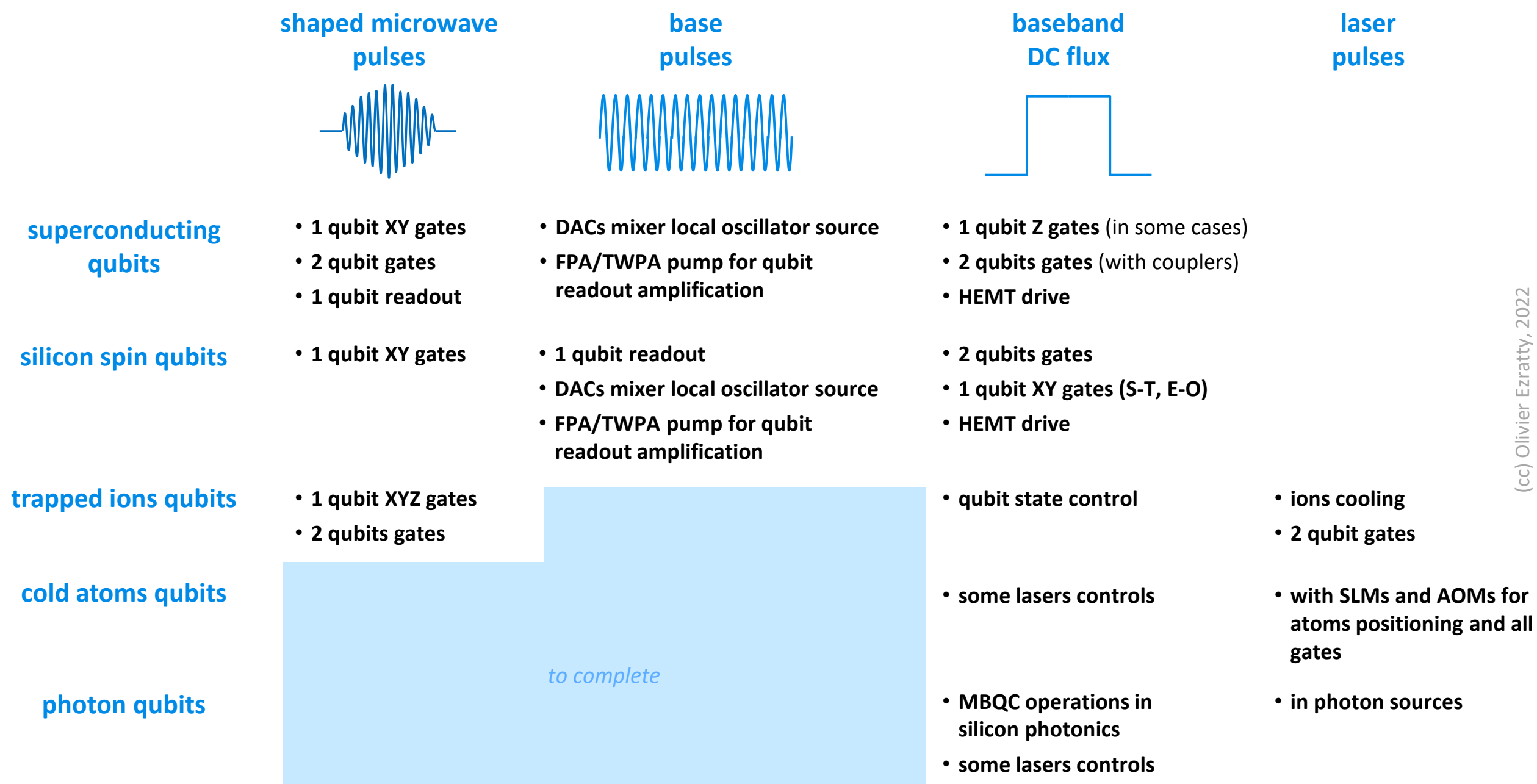

*Figure 478: compilation of the various electronic and photonic signals used to drive various types of qubits. This diagram will later be completed with more signals used to drive atoms and photon qubits. (cc) Olivier Ezratty, 2022.*

Direct current signals are also used to drive qubits, like with Z gates with some superconducting qubits and to drive some amplifiers[1369]. To be complete, we'll then also look at cabling and filtering components and their vendors.

With photon, cold atoms and trapped ions qubits, control techniques involve photons and lasers operating in the infrared and near visible spectrum which we don't cover here. There are missing parts in the schema above in Figure 478 with regards to the shaped and base pulses that may be used for the control of various electronic and photonic devices used in atom and photon based systems.

We'll try here to answer many questions: how are all these electronics affecting the quality of qubits? How is it scaling as you need to significantly increase the number of physical qubits to accommodate the requirements of fault-tolerant quantum computing? How do you optimize the existing cumbersome wiring? What are the solutions to run all or part of these electronic systems inside the cryostat? What is the power consumption of these various solutions? How can we scale with room temperature electronics and cryogenic electronics?

**Wiring**. How many wires are needed to control solid states qubits? It depends but there are usually half a dozen wires are needed to control a superconducting or quantum dot spin qubit. One or two microwave wires to drive qubit gates, one or two DC pulses wires to control other gates and then, two microwave wires for qubit readout (one in and one out of the qubit). One DC bias is sometimes used to change energy spacing for tunable superconducting qubits. When you scale the number of qubits, this creates a massive number of wires.

---

[1369] This excellent review paper Microwaves in Quantum Computing by Joseph Bardin et al, January 2021 (25 pages) provides an excellent overview of the challenges of microwave based qubit controls for superconducting, electron spin and trapped ion qubits. The table/chart in this page is inspired from this document. Z gates are driven by direct currents with Google's Sycamore qubits while IBM's are driven by microwaves, like the XY gates. See also Engineering cryogenic setups for 100-qubit scale superconducting circuit systems by S. Krinner et al, 2019 (29 pages) which makes a good inventory of energy consumption sources in the cryostat.

**Multiplexing**. Work is being done to multiplex these signals with different methods. One is frequency domain multiplexing that is already implemented for qubit readout with up to 8 microwaves frequencies in a single wire. A second would be frequency multiplexing with up and down convert to higher frequencies such as in the optical domain. Nanophotonic circuits sitting very close to qubit chipsets could be used for microwave up/down frequency conversion from microwave photons to optical photons and their multiplexing[1370]. One proposed conversion technique uses VCSEL (Vertical-cavity surface-emitting lasers) that are known to operate at low temperature of 2.6K, which is still high[1371]. A third method consists in using time domain multiplexing which probably has some limitations in scaling, making it impossible to run several gates simultaneously and therefore potentially slowing down computing and quantum error correction[1372]. To ensure scalability, these solutions must demux the signal as close as possible to the qubit chipset and have a very low power drain, compatible with the very low cryostat cooling budget at low cold-plate stages.

**Cryo-electronics options**. Many options exist that we'll cover in this part. Cryo-electronics can operate at the 4K stage (HorseRidge) down to the lower stages (15 mK to 100 mK). What are their limitations? Also, what are the pros and cons of cryo-CMOS vs SFQ electronics? All these cryo-electronics solution must be compared with detailed specifications. The generated microwave quality depends on the sampling rates used in their DAC and ADCs, on the number of points used in generating the wave envelope (16,384 points for HorseRidge 2, aka 14-bit sampling), their power consumption per qubit, their frequency range (targeting superconducting and/or electron spin qubits), their clock, their noise level and their real scalability potential with a large number of qubits. These electronic components must also be as isolated as possible from the qubit chipsets.

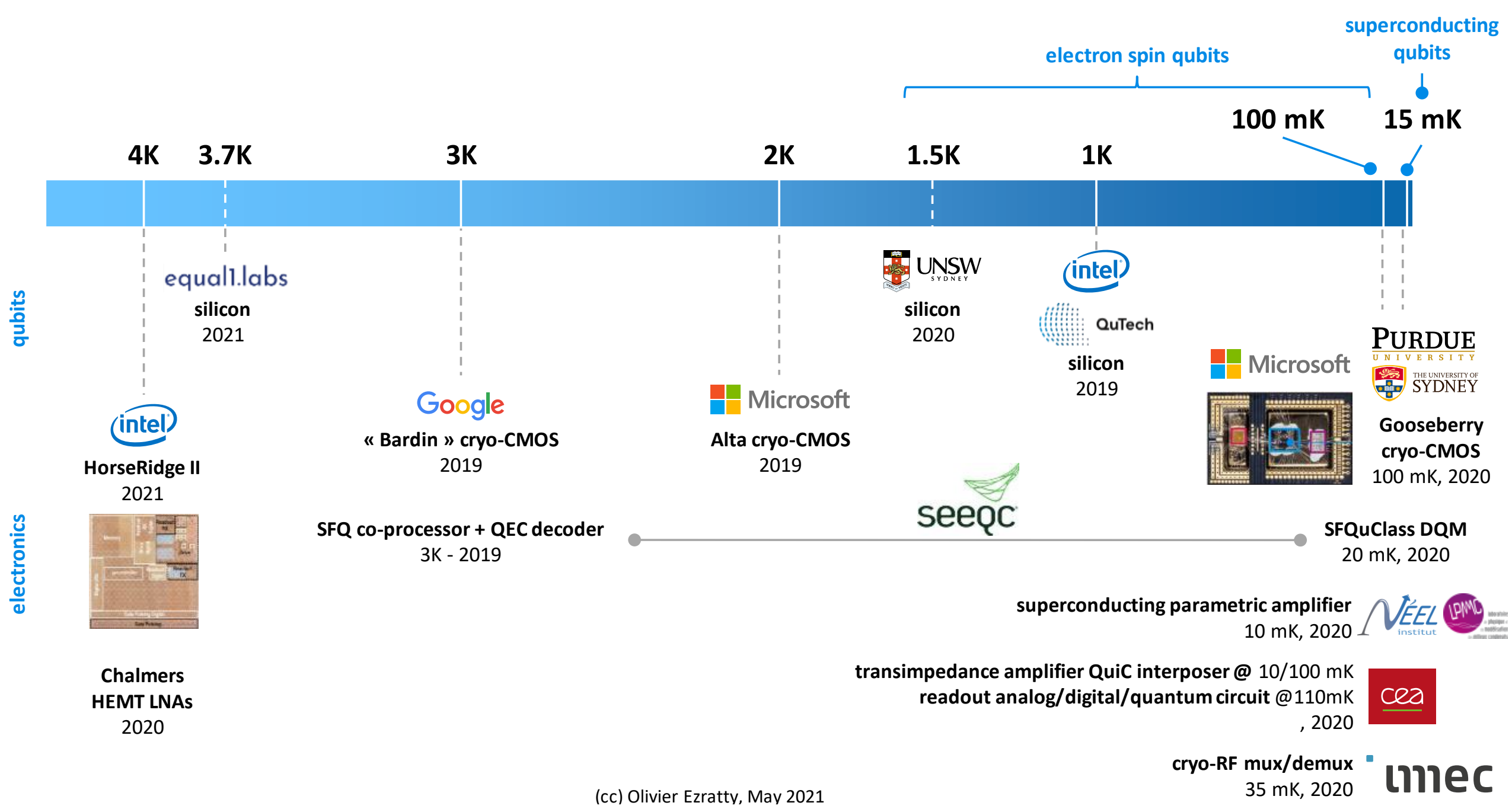

*Figure 479: comparison of the temperature and feature of various qubits and cryo-electronic chipsets. (cc) Olivier Ezratty, 2022.*

---

1370 See Supporting quantum technologies with a micron-scale silicon photonics platform by Matteo Cherchi et al, VTT, 2022 (17 pages) and Control and readout of a superconducting qubit using a photonic link by F. Lecocq et al, NIST, September 2020 (13 pages).

1371 See Recent Advances in 850 nm VCSELs for High-Speed Interconnects by Hao-Tien Cheng et al, February 2022 (27 pages) and Microwave-optical quantum frequency conversion by Xu Han et al, Optica, 2021 (15 pages).

1372 See Overcoming I/O bottleneck in superconducting quantum computing: multiplexed qubit control with ultra-low-power, base-temperature cryo-CMOS multiplexer by Rohith Acharya et al, IMEC, September 2022 (22 pages). The "1 to 4" drive multiplexer has a power consumption of around 0.24 µW of static power and 0.5 µW of dynamic power, with ~0.2 µW power consumption per qubit channel making it suitable to control 100 qubits and 1000 qubits with some optimizations, within the constraints of today's cryostats.

**Optimization**. Cryo-electronics are an interesting solution for many respects: it potentially reduces the wiring burden as seen above, it creates miniaturized electronics that can help constrain the total size and weight of a quantum computer, it has the potential to minimize the control/readout/processing cycle latency and its energy consumption can be much smaller than room temperature electronics.

If the cryo-electronics components are not at the same stage as the qubits, you'll still need some (expensive superconducting) wiring. And it has a significant indirect energetic cost. These cryo-electronics components generate heat that must be extracted from the cryostat. As you get in the lower cryostat stages, its cooling power gets drastically reduced proportionally to $T^4$, T being the cryostat stage temperature.

The current cooling budget at 15 mK is lower than 40 µW and goes up to 1W at the 4K stage. You can still bet on the creation of larger cooling power cryostats but their energetic cost will skyrocket. It explains why some cryo-electronics components are designed to stay at intermediate stages (2K to 4K). This balance depends also on the qubit operating temperature.

For superconducting qubits, the constraint is much bigger than with spin/silicon qubits which could operate at higher temperatures (100 mK to 1K). Recent modelling did show that for superconducting qubits, the control electronics energetic cost is surprisingly optimized with room temperature electronics[1373].

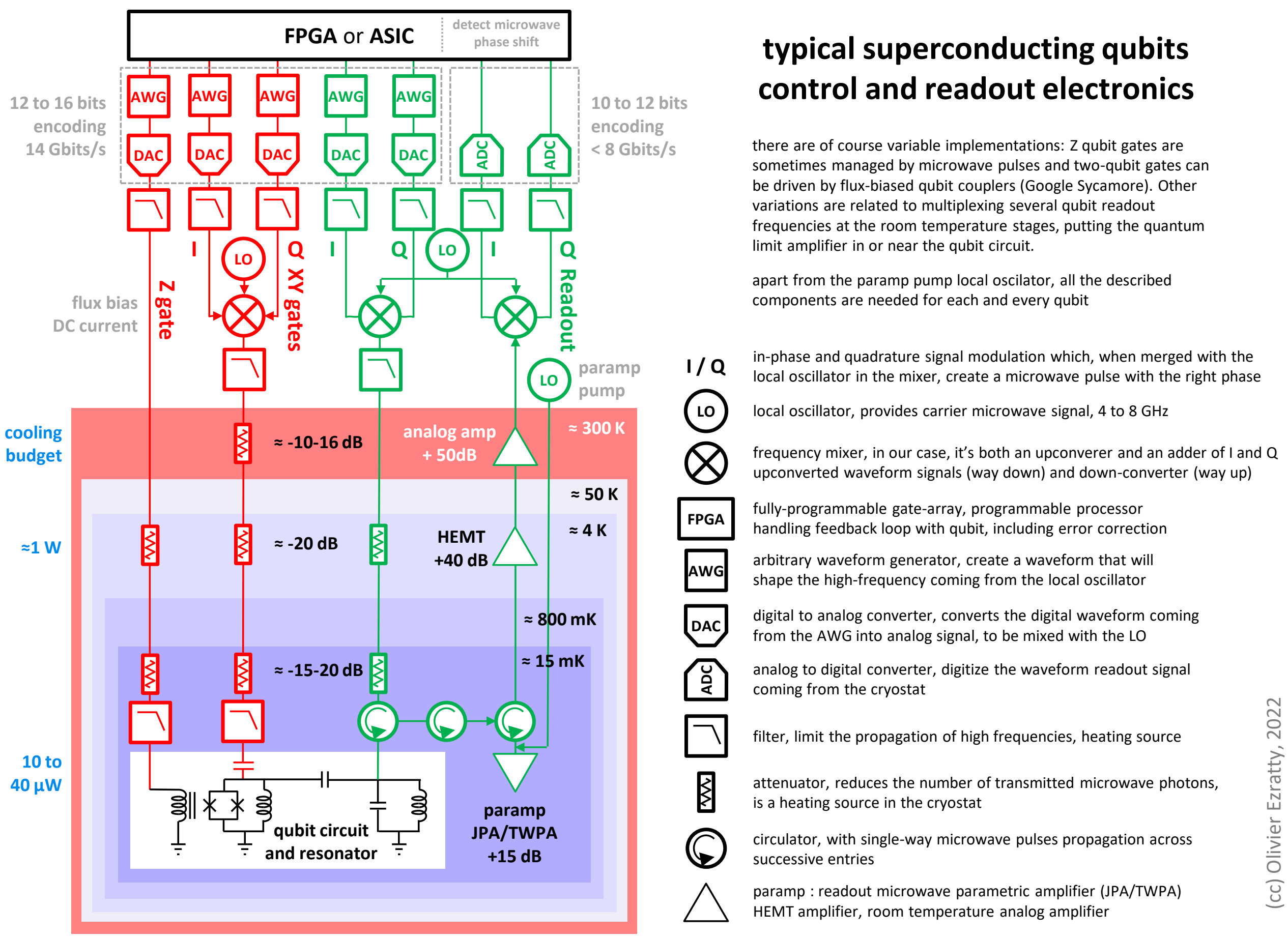

*Figure 480: description of the various electronic tools that control superconducting qubits. (cc) Olivier Ezratty, 2022.*

[1373] See Optimizing resource efficiencies for scalable full-stack quantum computers by Marco Fellous-Asiani, Jing Hao Chai, Yvain Thonnart, Hui Khoon Ng, Robert S. Whitney and Alexia Auffèves, arXiv , September 2022 (39 pages)

**Control electronics and qubit fidelities**. The relation between qubit fidelities (one and two qubit gates + idling + qubit readout) and control electronics precision has been widely studied. A paper from Intel and Dutch researchers from 2019 did show this correlation and created a model to reach 99,9% fidelities for all these operations. With solid state qubits, two sorts of control signals are generated: DC pulses and waveformed pulses in the microwave regime.

There's a clear link between qubit fidelities and the precision of microwave signal generation with regard to their duration, amplitude, frequency and phase. It demonstrated that it was a reachable goal for state of the art classical room temperature electronics[1374]. The noise affecting qubits and coming from control electronic comprises many aspects: phase noise coming from source clock jitter feeding the master and local oscillators, AWGs and DACs originated harmonics, leakage signals from mixers, various amplitude signal to noise ratios (SNR) and noise coming from reference voltage sources like the BVGs (bias voltage generators) that are used to generate DC pulses.

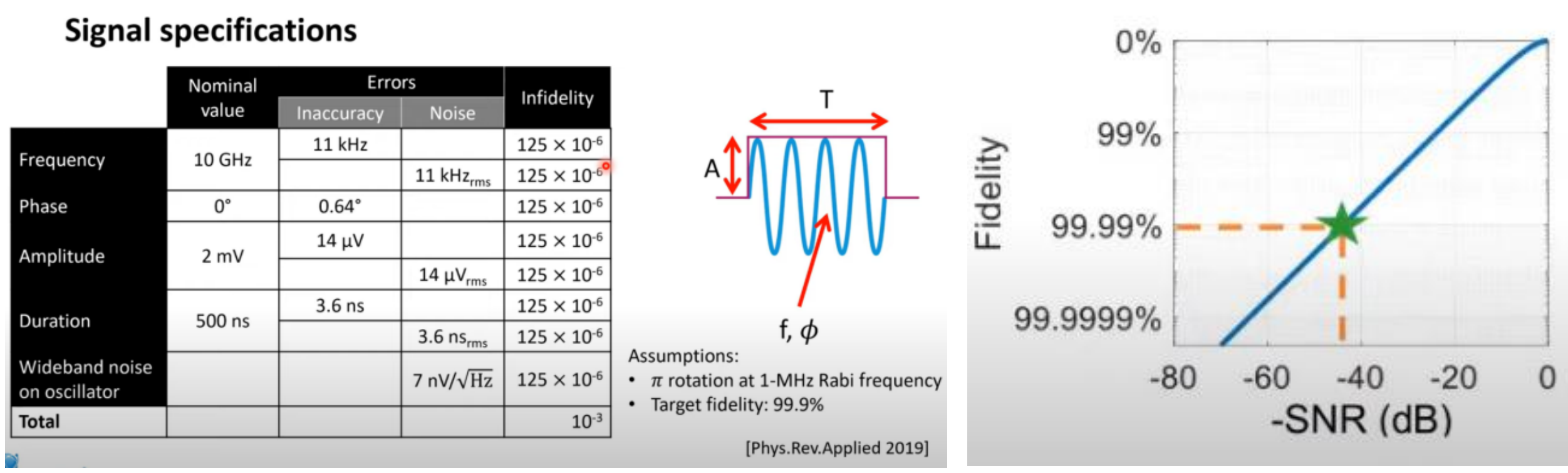

*Figure 481: specifications of a qubit control microwave pulse and of the infidelity sources. Data source: Impact of Classical Control Electronics on Qubit Fidelity by J.P.G. van Dijk, Menno Veldhorst, L.M.K. Vandersypen, E. Charbon, Fabio Sebastiano et al, PRA, 2019 (20 pages).*

Various improvements are thus sought in qubit control electronics:

**LO Phase Noise**. The local oscillators used in the AWG (arbitrary waves generators) must have a reasonable phase noise[1375]. Phase error noise becomes important as qubit fidelities are improved with better control of environmental sources of decoherence. Lab-grade oscillators may already limit the performance of qubits having microsecond scales gate times, like with trapped ions. Thus, the need to use low phase noise high precision local oscillators instead of traditional lab-grades LOs.

**Table 2.** Time until a qubit physical error rate *p* is reached solely due to phase fluctuations in the LO

| Time to Reach LO-Induced Error Rate *p* | | | | | |
|---|---|---|---|---|---|
| *p* | $10^{-3}$ | $10^{-4}$ | $10^{-5}$ | $10^{-6}$ | $10^{-7}$ |
| Lab-Grade LO | 4.0 µs | 900 ns | 200 ns | 30 ns | < 10 ns |
| Precision LO | > 100 ms | > 100 ms | > 100 ms | 80 ms | 600 µs |

These times may be viewed as an upper bound on the allowable QEC cycle period. Achievable cycle periods will be reduced because of other error sources. Error rates are derived from free-evolution calculations presented in Figure 2b. Driven operation error rates (see Figure 2c) yield similar results.

*Figure 482: The role of master clock stability in quantum information processing by Harrison Ball et al, NPJ Quantum Information, November 2016 (8 pages)*

There's a $10^{-4}$ difference in phase noise errors between lab-grade and precision LOs! Given superconducting gates time span 10 ns to 600 ns, a $10^{-5}$ error rate only due to LO phase noise could be reached during this time. We're not far from the required threshold for QEC!

---

[1374] See Impact of Classical Control Electronics on Qubit Fidelity by J.P.G. van Dijk, Menno Veldhorst, L.M.K. Vandersypen, E. Charbon, Fabio Sebastiano et al, PRA, 2019 (20 pages).

[1375] See The role of master clock stability in quantum information processing by Harrison Ball et al, NPJ Quantum Information, November 2016 (8 pages) and A 2–20-GHz Ultralow Phase Noise Signal Source Using a Microwave Oscillator Locked to a Mode-Locked Laser by Meysam Bahmanian and J. Christoph Scheytt, 2021 (11 pages).

**Low Latency control/readout cycles**. There's also a need to minimize the duration of the qubit control and readout cycle. It minimizes the impact of the errors that can happen during the cycle due to decoherence, enabling a higher quality QEC. The quantum feedback latency must be several orders of magnitude under superconducting qubits coherence times that are in the range of 50 µs to 100 µs, and could potentially exceed 1 ms. So, we're in for a maximum of a few 100 ns.

The control system must also be well synchronized across all qubits, which can be achieved with a distributed synchronous clock and trigger architecture[1376]. Labs also makes use of spectrum analyzers. Another feedback loop optimization comes with handling qubits gates AWGs, their DACs and readout data acquisition in the same FPGA[1377].

**From FPGA to ASIC**. At this point, the most advanced qubit control systems are FPGA based. Their advantage is sound economics for small scale use cases but at an energetic cost. Using CMOS ASICs could bring some energetic advantage on top of further reducing control cycle latency, but it has a significant cost few qubit developers can afford at least in research labs. Also, the ASIC design and manufacturing cycle is much longer than with an off-the-shelf FPGA board.

**Memristors DC control**. A Canadian-French team proposed to control spin qubits quantum-dot gate biases with DC sources with a cryogenic solution using $Al_2O_3$-$TiO_2$-based tunable memristors with a ±1V range and 100 µV resolution. Memristors are non-volatile systems with tunable resistance. It fits the need to tune these DC pulses due to qubits variability. It was demonstrated at 4.2K[1378]. It doesn't support the other needed type of electronic controls, microwave pulses, that are needed to drive single qubit gates. This solution simplifies the DC wiring between room temperature control electronics and the 4K stage in the cryostat, but it still requires DC lines between this stage and the qubits chipset at below 1K. These memristors are dissipating 1.77 mW. As a result, a current generation cryostat with 1.5W cooling power at 4.2K could accommodate about 800 such memristors. We're still far from the LSQ realm.

**Pulse control optimization**. At the software level, pulse control can benefit from some optimization techniques, requiring a verticalized approach crossing the usual layers between high-level gate-based code and pulse control. Such cross-layers optimizations are proposed by IBM and Q-CTRL[1379].

**Consolidated projects**. Several qubit control and electronics research and development projects were recently launched in Europe. In Germany alone, you have three related projects:

**QuMIC** (Qubits Control by Microwave Integrated Circuits, 6,3M€, 2021-2024) which involves four academic and two industry partners (Infineon, Supracon AG) and deals with miniaturization of RF electronics to control superconducting and trapped ions qubits.

**qBriqs** (2M€, 2021-2024) which also involves four academic and two industry partners (Rosenberger and Stahl Electronics) and deals with compact cryogenic connectors, qubit readouts TWPA and HEMT amplifiers, filters and attenuators, DACs and ADCs and DC flux current generators.

---

1376 In FPGA-based electronic system for the control and readout of superconducting quantum processors by Yuchen Yang et al, USTC China and Alibaba, February 2022 (12 pages), a Chinese team describes how it implemented such a system, to control 2 qubits for a starter. It's based on using FPGAs in 3U (3 units heights in electronics racks) PXIe modules, the instrumentation equivalent of the PCIe bus used in microcomputers and created by National Instruments in 1997. As an example, the 3U PXIe-1095 below has 18 slots. The team used a FS725 Rubidium Clock running at 10 MHz with ultralow phase noise, coming from Stanford Research Systems (1980, USA). They are also using the physical layer "Low Voltage Differential Signaling" system (LVDS) which has a low latency.

1377 See Hardware for multi-superconducting qubit control and readout by Zhan Wang et al, 2021 (11 pages). In this work, the feedback latency reached 178.4 ns. It's using a 28 nm Xilinx XC7K325T FPGA with 326K logic cells.

1378 See Memristor-based cryogenic programmable DC sources for scalable in-situ quantum-dot control by Pierre-Antoine Mouny et al, March 2022 (13 pages).

1379 See Summary: Chicago Quantum Exchange (CQE) Pulse-level Quantum Control Workshop by Kaitlin N. Smith, February 2022 (17 pages).

**HIQuP** (2021-2024, 2,2M€) with, again, four academic and two industry partners (Supracon AG and IQM Germany) which works on superconducting and cryogenic qubit control electronic circuits.

In France, the **QRYOlink** project combines CEA-Leti and Institut Néel from Grenoble, Radiall, ATEM, Air Liquide, C12 and Alice&Bob to develop a scalable architecture for cryogeny and cabling aimed at supporting solid-state qubits. Other related projects also cover scalable control electronics targeting large scale quantum computing architectures.

### Room temperature electronics

Room temperature electronics is the dominant solution used both in research labs and with most commercial vendors (IBM, Google, Rigetti, IQM, etc).

The key components are on the way in for each and every qubit:

- **AWGs** (arbitrary waveform generators) which create microwaves pulse forms and usually generate about 2 GigaSamples/s. These are used to create single qubit gates and also readout pulses. Alternative techniques are proposed which generate pulses width modulation (PWM) that would be less costly without jeopardizing qubit fidelities [1380].
- **DACs** (digital to analog converters) who use a 14-bit to 16-bit amplitude resolution to convert into analog format the output of the AWGs.
- A **mixer** of the waveform and a **LO** (local oscillator) signal in the used microwave range (around 5 GHz for superconducting qubits). The output is called a "heterodyne" signal.
- Some **direct current** sources to drive certain types of qubit gates *aka* bias drives.

Since we mentioned heterodyne measurement, let's make a pause with describing the three main different techniques used to measure an electromagnetic signal with homodyne measurement (one observable), heterodyne measurement (two orthogonal observables like in-phase and quadrature) and photon measurement or counting. These three techniques are used for optical frequencies photons and radio-frequency photon signals, with, of course, many differences and variations.

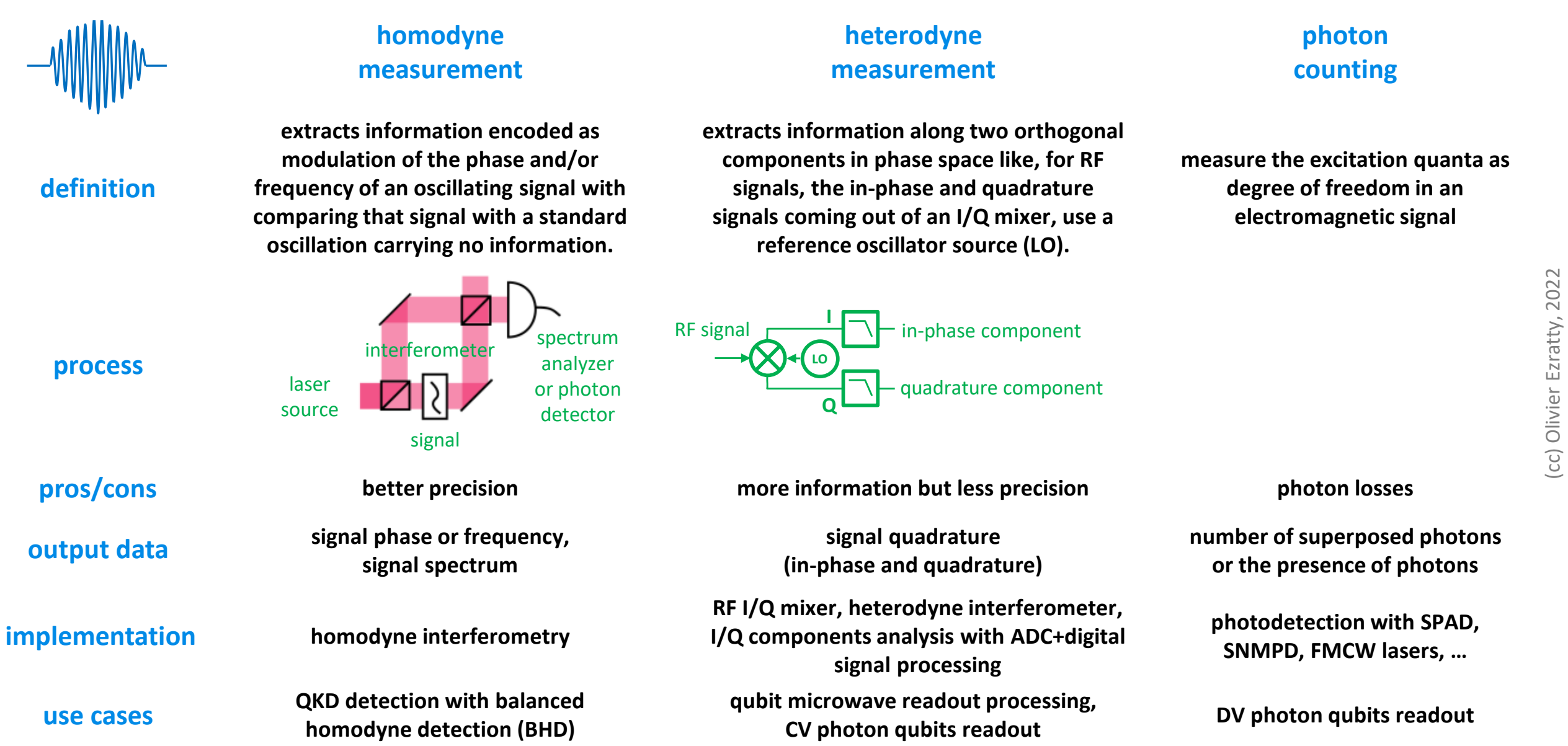

*Figure 483: explanation of the various ways to detect a photon or electronic signal with homodyne and heterodyne measurement and photon counting. (cc) Olivier Ezratty, 2022.*

[1380] See Quantum Optimal Control without Arbitrary Waveform Generators by Qi-Ming Chen et al, Aalto University, Princeton and Tsinghua University, September 2022 (14 pages). The paper however doesn't provide some indications of the associated power savings.

On the way out of the cryostat, we have:

- A last-stage **analog amplifier**, following the cryogenic amplification stages (TWPA and HEMT).
- **ADCs** (analog to digital converter) of the readout microwave signal, usually with a 1 GigaSamples/s sampling rate and a 8 to 12-bit encoding.
- **SoC** (systems on chip)**, FPGA** or **ASIC** circuits used to interpret the output of the ADCs to get the qubit state, and which may manage a closed-loop control of the whole cycle to implement error correction (QEC).

All these components are more or less integrated in one or several boxes, or boxes with interchangeable modules. The FPGAs are programmable, usually with Python and some extension (library or language extension).

This field is well covered by electronics industry vendors addressing research and commercial quantum computing markets. Beforehand, many quantum computing research laboratories were relying on generic micro-wave generator and readout systems coming from vendors like **Rohde & Schwarz, Tektronix** and **Keysight**[1381]. Over time, some of these vendors have developed specialized offerings for quantum computing, particularly through some acquisitions (for Rohde & Schwarz and Keysight). Specialized quantum computing electronics emerged like **Zurich Instruments**, **Qblox** and **Quantum Machines**. Large shops like **Google** also developed their own electronics.

Some hardware and software open source control systems have also been proposed like the **QICK** (Quantum Instrumentation Control Kit) from Fermilab[1382] and **QubiC** from Lawrence Berkeley National Lab, both from the DoE[1383]. They are cost efficient when compared to commercial solutions and adapted to the needs of research labs. These kits are all based on Xilinx FPGAs containing their own DACs and ADCs. Researchers from Chalmers and KTH in Sweden created **Presto**, a fully integrated room-temperature system on chip using a Zynq UltraScale+ RFSoC from Xilinx with full control operations for superconducting qubits[1384].

Zurich Instruments

**Zurich Instruments** (2008, Switzerland, $112K) is a manufacturer of electronic test and measurement equipment, including a range of microwave generation and analysis tools.

The company was acquired by Rohde & Schwarz in July 2021. Their offer is built around their Quantum Computing Control System, which bridges the gap between the quantum computer software control tools and the associated electronic instrumentation.

This system consists of several components.

---

1381 Tektronix provides an AWG that can be used to drive qubit signals, the 16-bit AWG5200 supporting up to 32 output channels with 2 Gbits/s sampling (and local oscillator frequency go up to 5 GHz) and the 6 Series Low Profile Digitizer for 4-channel qubits readout (up to 8 GHz).

1382 See The QICK (Quantum Instrumentation Control Kit): Readout and control for qubits and detectors by Leandro Stefanazzi et al, Fermi Lab, Princeton University, Seconda Università degli Studi di Napoli, GE Healthcare Institute, CNEA - Argentina, and University of Chicago, March 2022 (15 pages). It's based on Xilinx based RFSoC (Radio-Frequency System-on-chip) ZCU111 Evaluation Kit with a Xilinx XCZU28DR FPGA containing 8 14-bit DAC and 8 12-bit ADC. This is an "hybrid FPGA" with a programmable logic part and a more classical SoC part with a quad-core Arm Cortex A53 cores, various I/Os and memory. It supports microwaves output up to 6 GHz. The toolkit is programmed in Python. The QICK power consumption is 50 W and it seems able to drive 4 qubits. It could support up to 100 qubits with frequency domain multiplexing.

1383 See QubiC: An Open-Source FPGA-Based Control and Measurement System for Superconducting Quantum Information Processors by Yilun Xu, Irfan Siddiqi et al, Lawrence Berkeley National Laboratory and University of California at Berkeley, September 2021 (11 pages). QubiC uses a Xilinx Virtex-7 FPGA and separate DACs and ADCs from the Abaco Systems FMC120 and its four 16-bit ADC and four16-bit DACs. The conversion circuits come from Texas Instruments (ADS54J60 ADC and DAC39J84).

1384 See Measurement and control of a superconducting quantum processor with a fully-integrated radio-frequency system on a chip by Mats O. Tholén et al, Chalmers, June 2022 (14 pages). This type of chipset has a power drain of at least 70W (source). It seems it can handle about 8 qubits in total in a 2U 19-inch package.

First, the PQSC (Programmable Quantum System Controller, left of Figure 484 which is used to program and control all the devices. It is equipped with a Xilinx UltraScale+ FPGA that can be driven by the LabOne software using Python, C, MATLAB, LabVIEW and Microsoft's .NET framework. It controls up to 18 HDAWG (High-Density Arbitrary Waveform Generator, in Figure 484 *in blue*) microwave generators and manages up to a hundred qubits. LabOne became LabOne Q in October 2022 with some extensions easing the setting and optimization of Zurich Instruments tools.

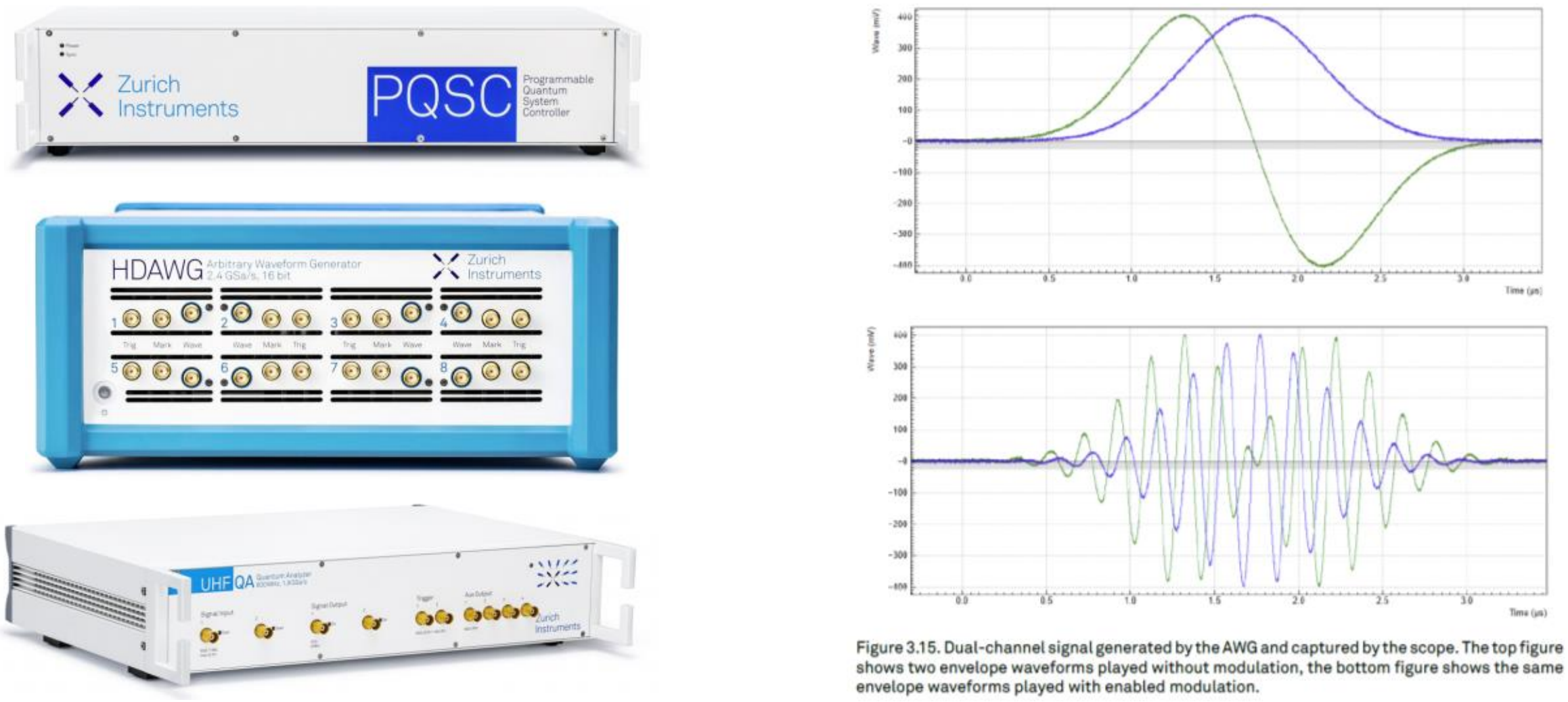

*Figure 484: Zurich Instruments PQSC and UHFQA for qubit control and readout. On the right, the types of microwave pulse signals generated. Source: Zurich Instrument product documentation.*

These are sold at 23K€. These generators create microwave pulses that combine a waveform (Gaussian or other, in Figure 484 *on the* right) modulated by a high-frequency signal, usually between 5 and 10 GHz, adapted to superconducting qubits drive and readout. It can control up to 8 channels. These microwaves are sent to the qubits to reset them to zero, activate quantum gates or handle state readout. The single-qubit quantum gates are generated by sending a modulated microwave that modifies the energy level of the qubits and change its state.

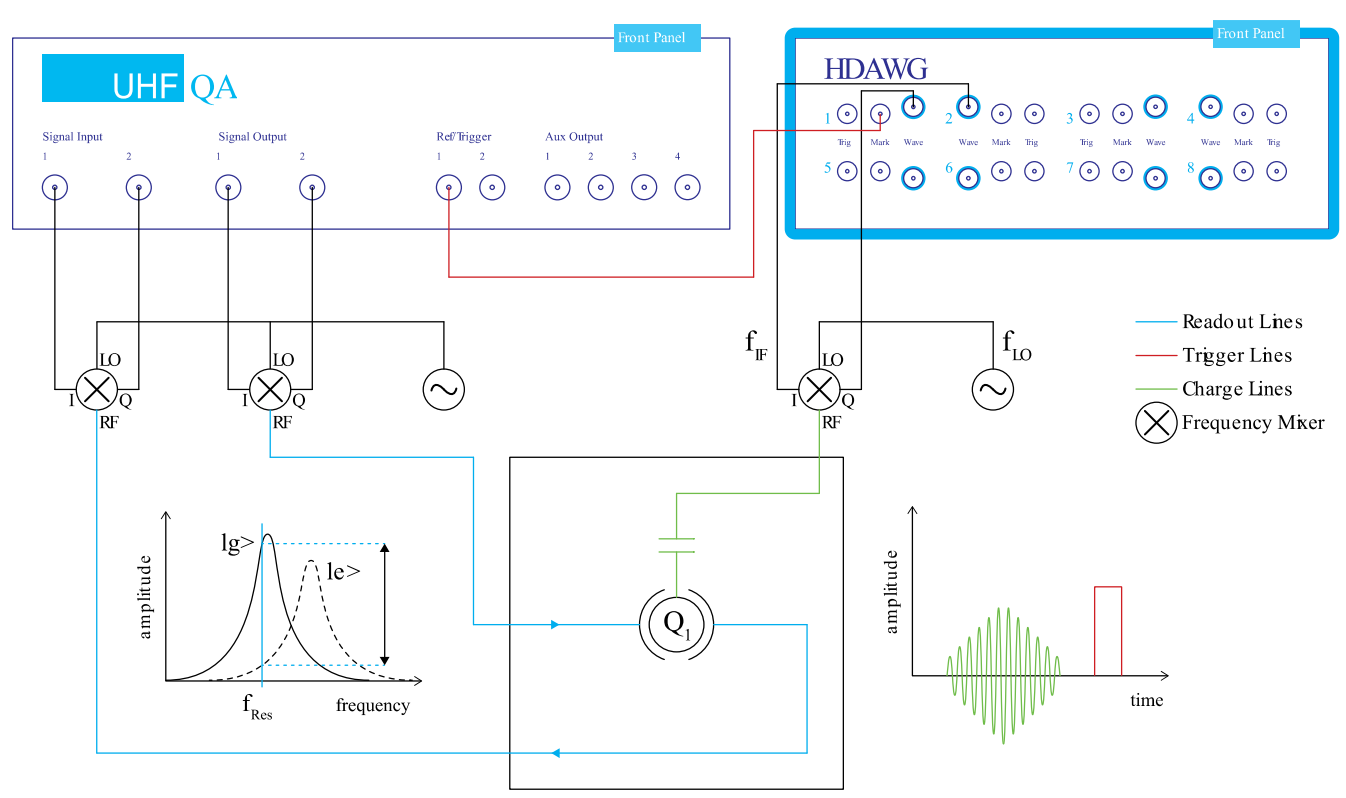

*Figure 485: UHFQA and HDAWG cabling. Source: Zurich Instruments.*

This is complemented by the UHFQA (Ultra-High Frequency Quantum Analyzer) which can analyze the readout state of 10 qubits. In the diagram, $F_{LO}$ is the frequency of the microwave signal to be modulated and $F_{IF}$ is the modulation waveform.

On the UHFQA side, the system detects the modulation or phase modification of the signal recovered through a resonator associated with the qubits, $l_g$ and $l_e$ respectively for ground states and excited states.

In April 2021, Zurich Instruments launched a new signal generator, the SHFSG with better microwaves signal spectral purity and stability. It can handle up to 144 qubit control channels and is accommodated with 4 or 8 channels, controlling up to 8 qubits. In August 2022, they introduced their SHFPPC (Super High Frequency Parametric Pump Controller), a room temperature tone pulse generator that feeds the parametric amplifiers like the TWPAs sitting at the lowest stage of the cryostat.

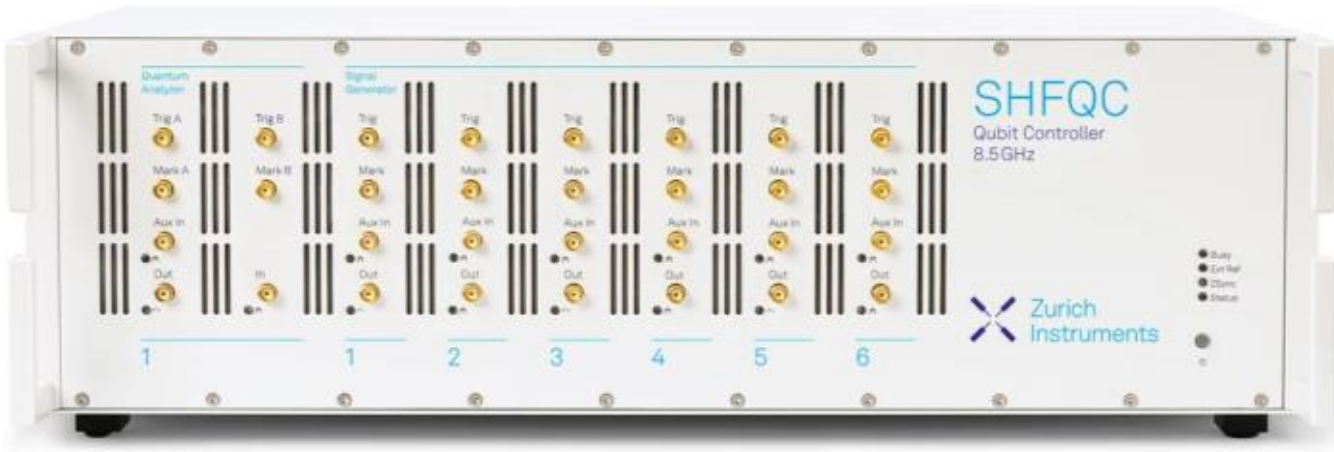

Figure 486: an SHFQC can control up to 16 qubits.

It was then completed by the software reconfigurable and programmable SHFQC launched in November 2021, which bundles 6 signal generator control channels and a readout channel analyzer supporting up to 16 qubits. Several SHFQC can be combined to support up to 100 qubits and "beyond".

QBLOX

**Qblox** (2018, Netherlands, $5M) is a spin-off from QuTech that develops scalable control electronics for superconducting qubits. Their latest generation controls up to 20 qubits with a 4U rackable system.

These clusters consume about 1 kW. The device contains both micro-wave generators for qubits gates (QCM module, blue) and qubits readout and electronics for qubit readout (QRM module, white). Each unit relies on small custom FPGAs. In a classical manner, it creates waveforms mixed with a microwave carrier signal after DAC conversion. Readout uses an ADC and a phase detection system.

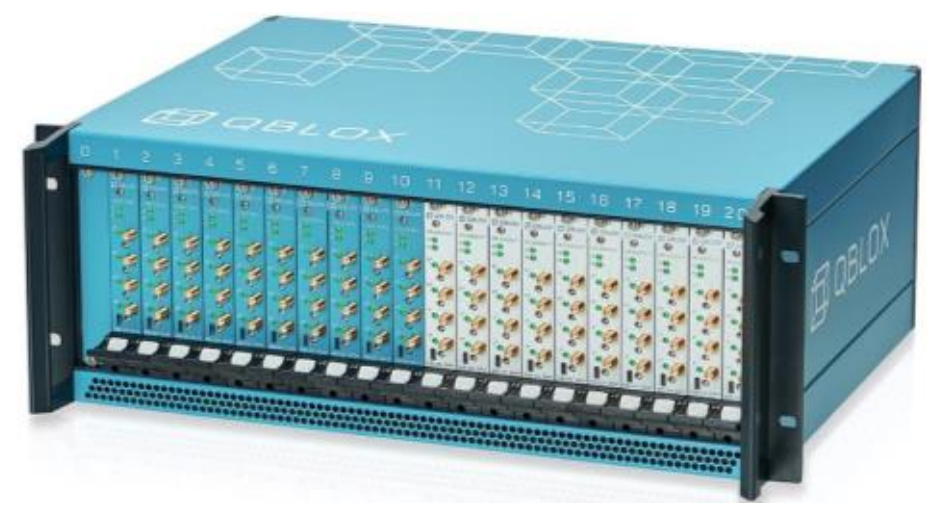

Figure 487: this Qblox system can control up to 20 qubits.

Their DACs/ADCs have a high sampling rate of 16 bits. A high sampling rate is important to create precise waveformed microwaves. This precision is a way to ensure a good fidelity for qubit gates generated by these generated microwave pulses.

Their architecture could scale up to controlling 1000 superconducting qubits. Calibration is done with the help from **Orange Quantum Systems** and cabling comes from **Delft Circuits**, two other spin-offs of Qutech in the Netherlands.

They also sell the desktop Pulsar QRM (quantum readout module) that handles a few qubits control in small factor format. As their Cluster modules, these can be coupled and synchronized together with using their homegrown protocols SYNQ (synchronized start within <<1 ns) and LINQ (distributing measurement outcomes in <200 ns).

QM QUANTUM MACHINES

**Quantum Machines** (2018, Israel, $73M) provides a qubit control layer for superconducting quantum computers that combines hardware and software[1385]. It is a spin-off from the Braun Center for Submicron Research Laboratory at the Weizmann Institute.

They developed their own classic qubit control processor, an FPGA operating at room temperature, which generates the pulses for controlling qubits and measuring their states either with microwaves and lasers [1386]. Packaged as their OPX/OPX+ systems, it supports superconducting, electron spin, NV centers, trapped ions and cold atoms qubits. In March 2022, they announced the availability of Octave.

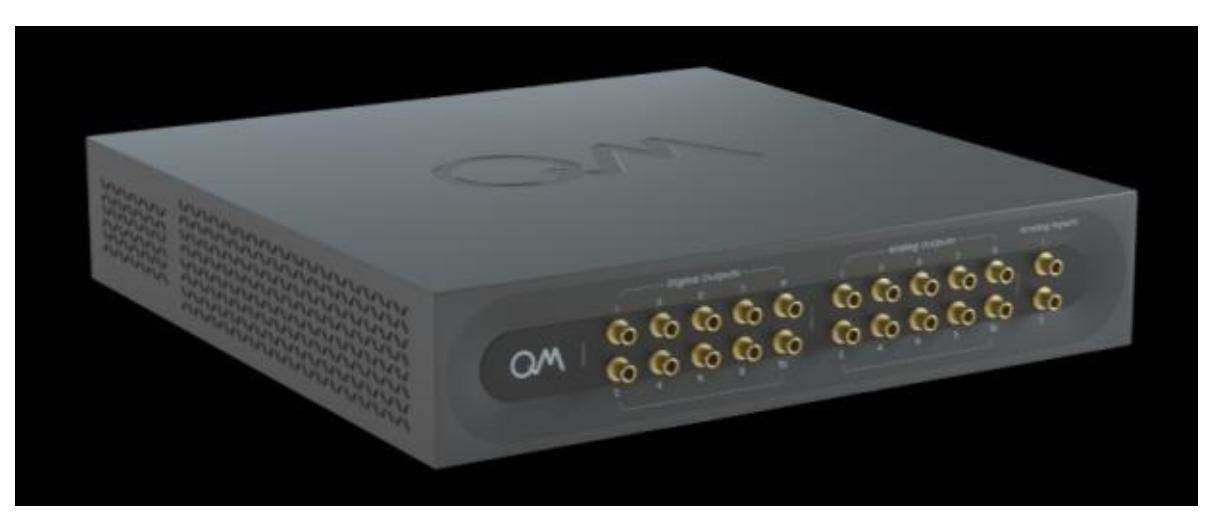
Figure 488: OPX+ is a full-stack solution for qubit control and measurement, enabling closed-loop error correction.

[1385] See The Story of the First Israeli Quantum Computing Startup by Eliran Rubin, December 2018.

[1386] See the video MLQ2021 Session Th2: Quantum Machines, March 2021 (46 mn) explaining their process.

It is a compact and rack-mountable all-in-one RF up/down-conversion module which completes their OPX systems. It contains its own built-in Local Oscillator (LO) sources and provides continuous self-calibration features.

They already have more than a dozen customers, including, in France, ENS Paris, ENS Lyon, Alice&Bob and Pasqal (just in France...). The company was created by Itamar Sivan (CEO, who did a Master's degree at the ENS Paris between 2009 and 2011), Yonathan Cohen (CTO, former Weizmann Institute managing director) and Nissim Ofek (Chief Engineer, who had a post-doc position in Rob Schoelkopf's lab in Yale University where he developed a FPGA based control and QEC code).

They also partner with Q-CTRL which develops qubits firmware level control software. Their processor is integrated into their "Quantum Orchestration Platform", which also combines a software layer[1387]. In June 2020, they announced the creation of the **QUA** language, positioned as a language for creating hybrid quantum and classical algorithms, such as VQE and QAOA, which need rapid feedback between classical and quantum processors. This programming language works with all types of qubits, superconductors, silicon, cold atoms and trapped ions. The compiler thus takes into account the differences in the implementation of qubits: their connectivity, the homogeneity or heterogeneity of their coupling, the coherence times, the error rates, etc.

In April 2022, Quantum Machines, together with their customers Alice&Bob, Benjamin Huard's team from ENS Lyon and Florian Marquardt of the Max Planck Institute for the Science of Light in Germany, announced the launch of Artemis, a 3-year EU funded project (900K€) as part of QUANTERA to use a real-time neural network to improve the accuracy of quantum controls and quantum error correction. It will lead to the creation of a full-stack QEC universal quantum controller. It will be complemented by an (unspecified) cloud-based quantum processor.

In March 2022, Quantum Machines made the acquisition of QDevil (Denmark) which gives them a foothold in the cryogenic electronics space with filters and a low noise DAC. The company had a total staff of 80 as of March 2022.

KEYSIGHT **Keysight Technologies** (USA) is an electronics measurement company spun-out of Agilent in 2014, which itself was coming from Hewlett Packard in 2000.

It then expanded its portfolio through several acquisitions: Signadyne (FPGA-based PXI digitizers and AWGs) in 2016, Ixia (software), Liberty Cal (calibration services) and ScienLab (test solutions in eMobility systems) in 2017, Labber Quantum in 2019 (MIT spun-out specialized in qubits control and software) and Quantum Benchmark in 2021 (quantum error diagnosis and suppression, and benchmarking software). The company has now a broad portfolio of measurement and control electronic systems widely used in the quantum real, in its three markets: computing, sensing and communications, including optical instruments. Most quantum research labs already have some Keysight test and measurement systems. In qubits control electronics, they provide pulse laser controls, basebands pulse controls and pulsed microwaves controls (AWGs like the M3202A PXIe Arbitrary Waveform Generator with 1 GSa/s, 14 bit sampling, DAC) as well as ADCs for qubit readout (like the M3102A 14-bit PXIe Digitizer). Interestingly, they address one source of electronics signals quality variability: the fluctuating room temperature. Thus, a solution to control rack temperature with air flow operating at 35°C.

[1387] See Quantum Machines raises $17.5M for its Quantum Orchestration Platform by Frederic Lardinois, March 2020, Israel gets ready to join global quantum computing race by Amitai Ziv, December 2019 and The quantum computer is about to change the world. Three Israelis are leading the revolution by Oded Carmeli, February 2020.

Their Quantum Control System which assembles various software and hardware components to drive single and multi-qubit lab experiments. The hardware part is the Quantum Engineering Toolkit (QET) and contains a PC workstation with a PXIe Interface Module, a PXIe chassis containing an AWG, a DAC, and In-Phase and Quadrature modulator and demodulator, a Vector Signal Generator and other optional electronics. Among other places, this toolkit is used since 2020 at the MIT EQuS (Engineering Quantum Systems Group) testbed.

Keysight is involved in several quantum computing related projects like the Boulder Cryogenic Quantum Testbed launched in 2019, a joint effort of Google, the NIST and the University of Colorado Boulder, housed in the JILA laboratory on the CU Boulder campus. It helps US researchers working on superconducting qubits at the characterization level. The lab is equipped with a 10 mK Janis JDry 250 mini dilution refrigerator. They also participate to MATQu, an EU funded German project which ambitions to produce superconducting qubits on 300 mm silicon-based process flows.

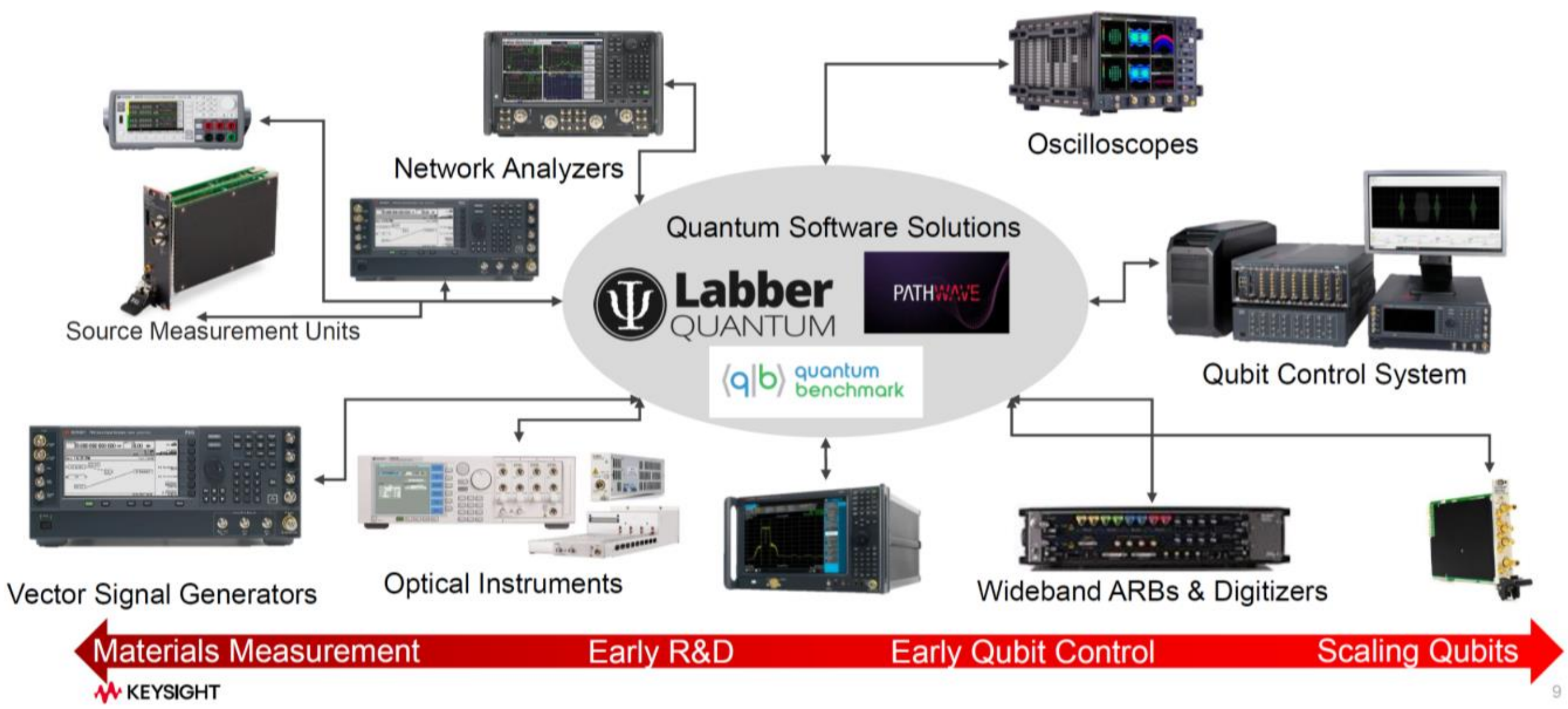

*Figure 489: the Keysight control electronics family, mostly used in research laboratories.*

Keysight hardware is also used in quantum sensing (AWGs, ADCs, DACs, analog signal generation, measurement of current-voltage ("I-V") and capacitance-voltage ("C-V"), oscilloscopes) and CV-QKD signal generation and detection (with waveform/pattern generation, oscilloscopes, lightware detectors and variable optical attenuators). In the vendors space, they partner with IBM on Qiskit Metal, an open source software solution to design your own superconducting qubits to understand qubits crosstalk effects.

In most of their solutions, Keysight uses Xilink FPGAs but it also has its own cleanroom facilities to design custom ASICs, like in photonics. They plan to implement control electronics with cryo-CMOS only in the long term, after 2030-2040. Until then, classical electronics will make the job.

Keysight announced in June 2022 its new generation of qubit control electronics, Quantum Control System (QCS), based on a proprietary ASIC integrating microwave signals AWGs, 14-bit DACs and 12-bit readout ADCs with the effect of reducing phase jitter in the generated signals and enabling fast closed-loop quantum error correction. It is packaged in PXIe cards format and a 4U box can control 20 qubits. Il is completed by a Python API. It seems to be the first offering of this type with some ASIC drive components. The readout lag is only 20 ns, from the reception of the readout microwave to outputting its result. The first announced customers of QCS are Alice&Bob and Rigetti.

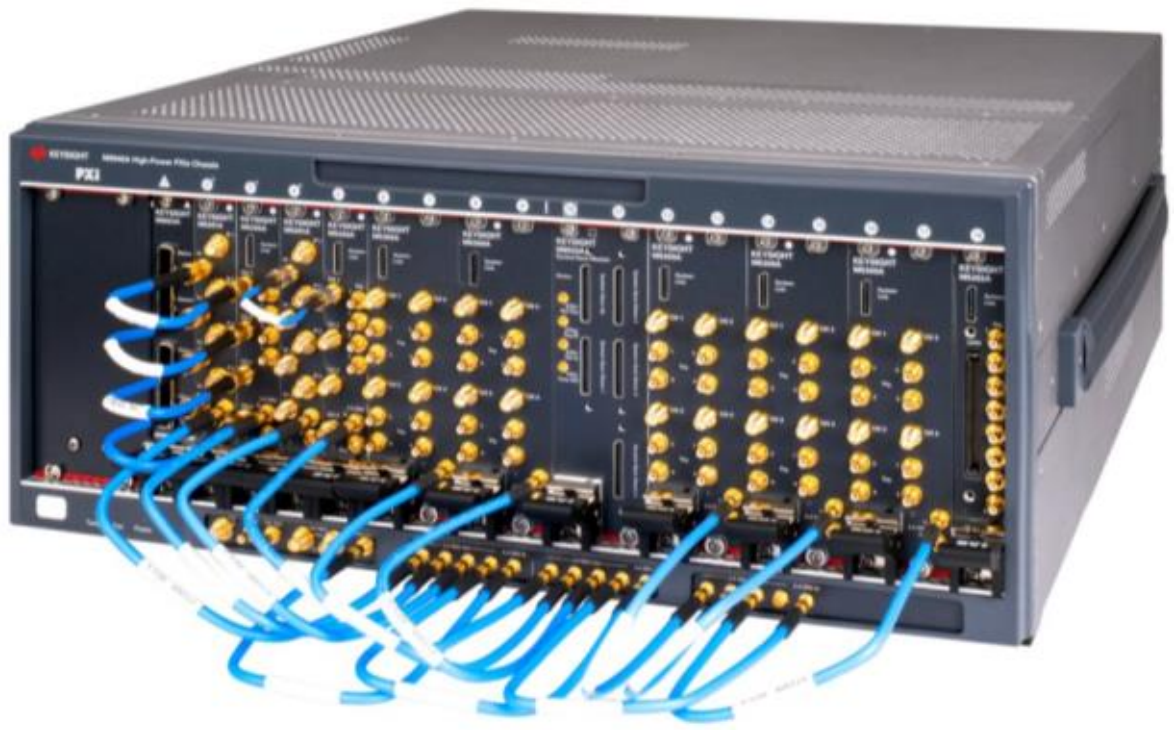

Figure 490: Keysight PXIe Quantum Control System.

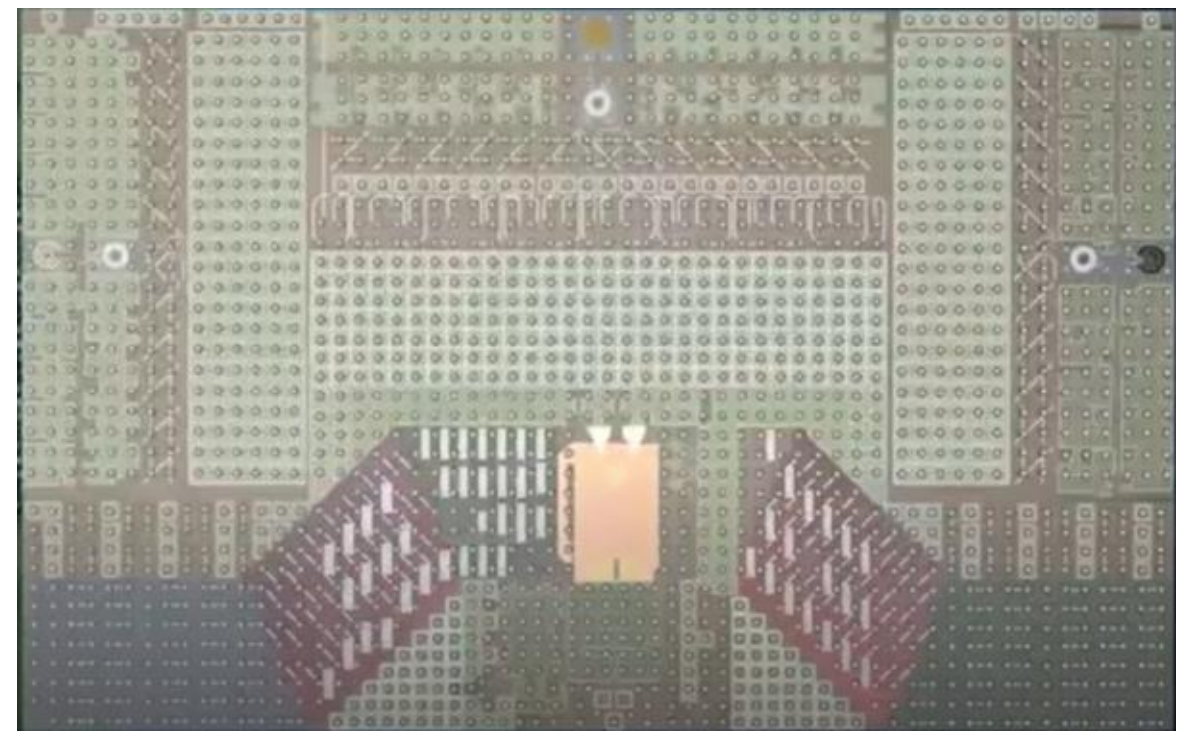

Figure 491: Keysight's first ASIC to control qubits.

**QuantrolOx** (2021, Finland-UK, $1.4M) is an enabling technology company creating a deep learning software solution to optimize the automatic tuning and optimization of qubits settings[1388].

Their solution that was developed with the help from DeepMind is using small training data sets and works in two steps: one with coarse tuning and a second with fine tuning. It efficiently adjusts the parameters of a large number of qubits and is applicable to all sorts of qubits but particularly with those who express the largest variability like superconducting and electron spins qubits. The company has already about a staff of 10 including Vishal Chatrath (CEO, UK), Andrew Briggs (Executive Chair, UK, Professor of Nanomaterials, University of Oxford), Natalia Ares (Chief Scientist, UK, with a strong background on quantum thermodynamics and machine learning, Professor at University of Oxford), Dominic Lennon (Head of Quantum Technologies, UK, also from Oxford University) and Juha Seppä (CTO, Finland). Mostly based in the UK, the startup is positioned as a Finish one, maybe to make it easier to get some EU funding! Their first investors are Nielsen Ventures, Hoxton Ventures, Voima Ventures, Remus Capital, Hermann Hauser (cofounder of Arm) and Laurent Caraffa.

We can also mention a Chinese project, a superconducting microwave generator for the control of superconducting qubits based on a Xilinx FPGA[1389]. In other similar and older projects, China's research teams showcased scalability claims that were not really sustained by a real scalable architecture[1390].

**Active Technologies** (2003, Italy) is a spin-off of University of Ferrara creating AWGs and Pulse/Pattern Generators. It can control experimental solid state qubits as well as electro-optical and electro-acoustic modulators used with cold atoms qubits. Its flagship product is the AWG-5000, a fast 16-bit AWG with 8 output channels.

**CIQTEK** (2016, China, $15M) aka Guoyi Quantum develops high-precision pulse generator (ASG) and arbitrary waveform generator (AWG) used in qubits control, electron parametric resonance spectroscopes and scanning electron microscopes. They also manufacture NV centers-based magnetometers and <Diamond I>, a 2-qubit computing system for educational purpose. It is based in Hefei and has 500 employees.

---

[1388] See Machine learning as an enabler of qubit scalability by Natalia Ares, Nature, 2021 (3 pages) and Learning Quantum Systems by Valentin Gebhart, Natalia Ares et al, July 2022 (26 pages) which provides a broad view on machine learning use-case for various types of qubits quantum error mitigation.

[1389] See Scalable and customizable arbitrary waveform generator for superconducting quantum computing by Jin Lin, 2019 (9 pages).

[1390] See High Performance and Scalable AWG for Superconducting Quantum Computer by Jin Lin et al, 2018 (5 pages).

**Intermodulation Products** (2018, Sweden) is a spin-off company of KTH, the Royal Swedish Institute of Technology. They market Vivace, a microwave generator in the 4 GHz band used to drive superconducting qubits.

**Quaxys** (2020, USA) provides hardware and software solutions for superconducting and spin qubits electronic control, including Quantuware 4840, a compact qubit control and measurement unit.

**Teledyne E2V** (USA/UK/France) is a designer, manufacturer and provider of DACs and ADCs circuits used for microwave processing with superconducting qubits, noticeably with IBM. These are designed and manufactured near Grenoble, France.

**VIQTHOR** (2022, France) is a stealth startup aiming to create scalable room temperature control electronics solutions for solid state qubits.

### Cryo-CMOS

Cryo-electronics sit inside the cryostat, control the qubits and manage their readout in place of the some of the external electronic devices we're just covered, totally or partially depending on the systems generation. Many research team and industry vendors are working on this strategic set of technologies which may help to unlock qubit scalability. We have among others the **University of Sidney**, **TU Delft** in the Netherlands[1391], **VTT** in Sweden, **CEA-Leti** and **CNRS Institut Néel** in France, **POSTECH** in South Korea and US vendors like **Intel** and **IBM**. Cryo-electronics could help save a lot of quite expensive and embarrassing cabling, filters, attenuators, amplifiers, and reduce thermal losses in the cryostat. It can also contribute to shorten the qubit gate to qubit readout cycle which can fasten the execution of quantum correction codes that will be required when operating large scale quantum processors.

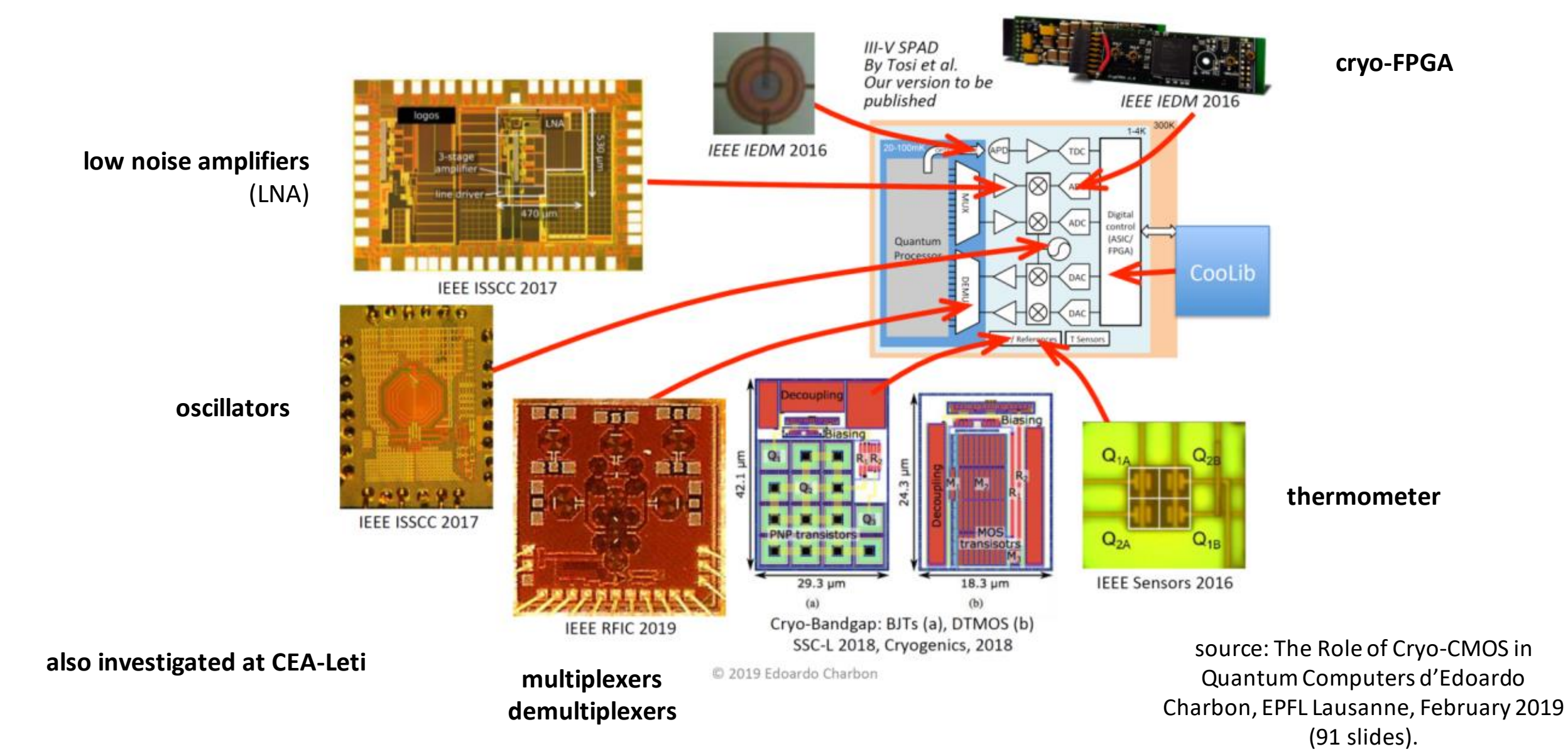

*Figure 492: initially, research labs tried to build specific cryo component chipsets for many qubit control functions. Then, players like Intel tried to consolidate these in fewer components. There are still many components around, even with integrated cryo-CMOS for qubit control and readout, like the parametric amplifiers and HEMT. Source: The Role of Cryo-CMOS in Quantum Computers by Edoardo Charbon, EPFL Lausanne, February 2019 (91 slides).*

[1391] See Large-Scale Quantum Computers: The need for Cryo-CMOS by Fabio Sebastiano, TU Delft, April 2021 (57 mn video).

It must meet rigorous specifications[1392]. Figure 492 describes the variety of component functions that can be integrated in the 1K-4K stages and even, when possible, at the qubit chipset stage at less than 20 mK[1393]. These components must be certified to operate at these temperatures. These are data multiplexers and demultiplexers, local oscillators, AWGs, DACs, ADCs, low-noise amplifiers, DC flux bias generators, thermometers and other various sensors.

The trend is to put within the cryostat a maximum of these electronic components. However, the heat they released is limited by the dilution refrigeration system cooling power[1394].It also conditions at which cold plate stage these components can operate. There's a complicated trade-off between the cryostat power overhead and what is saved by cabling and filters.

Starting in 2016, separate solid-state electronic components started to be designed and tested at cryogenic temperatures.

**IMEC** (Belgium) developed in 2020 a cryo-CMOS RF MUX multiplexing the in and out microwave signals used in qubit readouts and operating at 32 mK. Working at up to 10 GHz, it is suitable for superconducting qubits readouts and not yet for all electron spin qubits[1395]. It greatly simplifies the cabling between RF control and readout electronics.

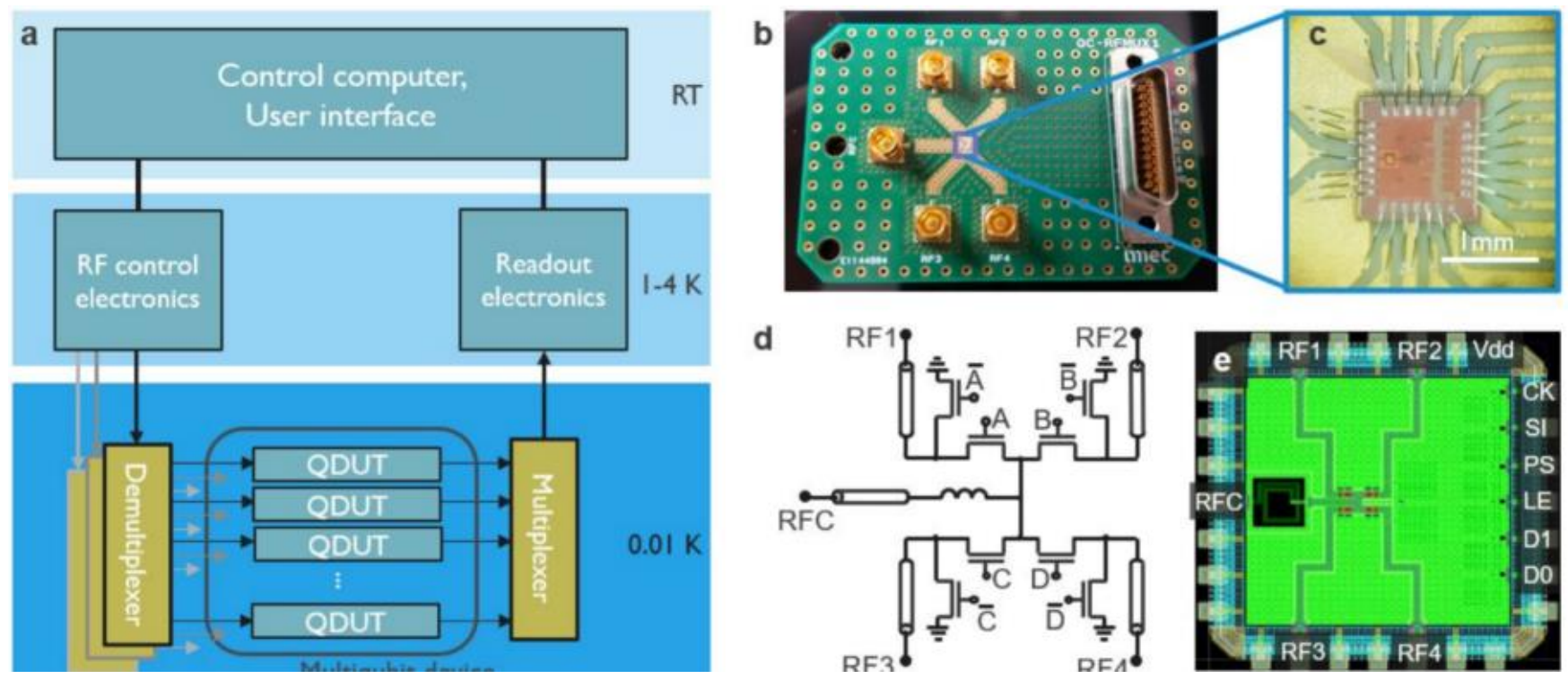

*Figure 493: a qubit control multiplexing solution developed by IMEC. Source: Millikelvin temperature cryo-CMOS multiplexer for scalable quantum device characterisation by Anton Potočnik et al, IMEC, November 2020 (35 pages).*

It sits at 4K and the qubit chipset sitting at below 20 mK. They created a similar solution operating at 15 mK in 2022 and suitable for time-domain microwaves multiplexing for the drive of superconducting qubits.

The trend is to integrate all these components in a minimum number of chipsets, preferably one, and working as close as possible to the qubits chipset. The best level of integration so far was reached with **Intel** HorseRidge 2 announced in 2021 and the coldest operation was achieved with the Gooseberry chipset from **Microsoft** and the **University of Sidney** as well as with a cryo-CMOS from **CEA-Leti**.

---

[1392] See Engineering cryogenic setups for 100-qubit scale superconducting circuit systems by S. Krinner et al, 2019 (29 pages) which describes the issues with superconducting qubit control. In 2018, they proposed an optimized approach of wiring and electronics allowing up to 150 superconducting qubits to be embedded in a cryostat.

[1393] Source of the diagram: The Role of Cryo-CMOS in Quantum Computers by Edoardo Charbon, EPFL Lausanne, February 2019 (91 slides). See also an earlier work from Purdue University and Australian colleagues: Cryogenic Control Architecture for Large-Scale Quantum Computing by J. M. Hornibrook, 2014 (8 pages) which describes well what should be done where in the cryostat.

[1394] See Cryogenic Control Beyond 100 Qubits by Ian Conway Lamb, 2017 (103 pages) which describes the technological challenges of components operating at cryogenic temperature, here for superconducting qubits. And the short version: Cryogenic Control Architecture for Large-Scale Quantum Computing by Ian Conway Lamb et al, 2017 (8 pages). See also Semiconductor devices for cryogenic amplification by Damien Prêle, 2013 (30 slides) and Cryo-CMOS Circuits and Systems for Quantum Computing Applications by Bishnu Patra et al, 2018 (14 pages).

[1395] See Millikelvin temperature cryo-CMOS multiplexer for scalable quantum device characterisation by Anton Potočnik et al, IMEC, November 2020 (35 pages).

The first approach was to miniaturize these circuits at the 4K stage of the cryostat. It was studied in 2019 at **TU Delft** for silicon qubits state readout with their QuRO, for Quantum Read-Out[1396]. The readout was using microwaves photon reflectometry. It sent an unmodulated RF frequency and evaluated the amplitude and phase of the reflected RF photon. The technique allows multiplexing qubits readout before sending the information out of the cryostat. This simplifies the output wiring. The prototype was based on a CMOS low noise amplifier (LNA) supplemented by a SiGe (silicon-germanium) transistor amplifier, followed by an analog-to-digital converter (ADC) implemented in a Xilinx Artix 7 FPGA. This FPGA made it possible to multiplex the readout state of several qubits. They use some copper cooling radiator in the 4K stage of the dilution refrigeration. They relied on standard market off-the-shelf passive and active components operating correctly at 4K. This prototyping did not deal with the waveform generation and DAC circuits driving qubit gates. The energy saving of this kind of system is related to the quantum error correction load on qubit measurement. Bringing readout electronics closer to qubits speeds up error correcting codes. It's also interesting for simplifying the connectivity and improving quantum computers scalability.

A similar approach was initially adopted by **Intel** in collaboration with **QuTech** for its 2020 HorseRidge superconducting and silicon qubits driver component capable of handling the microwave pulses of this frequency driver from 2 to 20 GHz. This component is sitting at the 4K stage of the cryostat[1397].

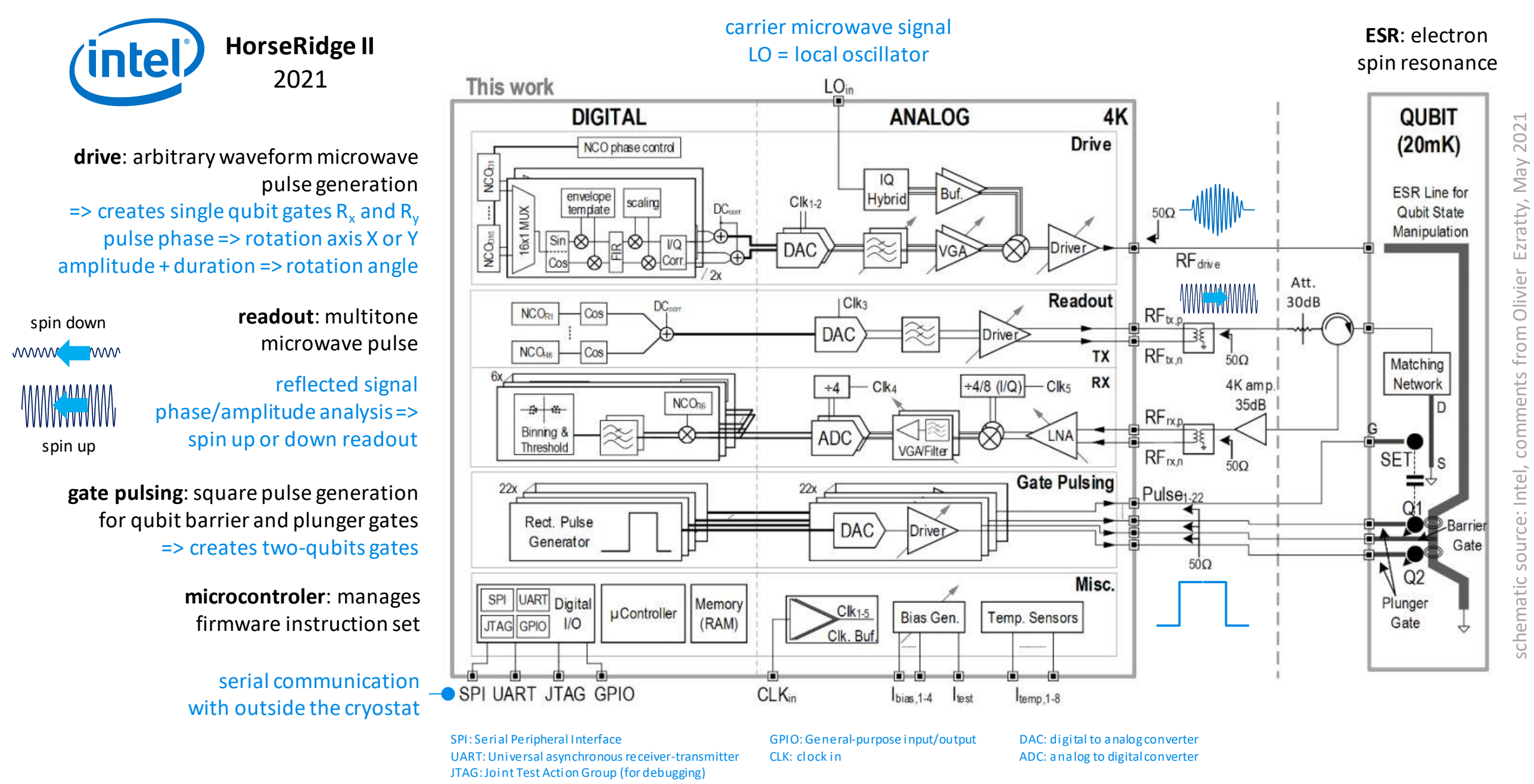

*Figure 494: Intel HorseRidge 2 presented in 2021 is probably the most integrated qubit control chipset being developed. Source: A Fully Integrated Cryo-CMOS SoC for Qubit Control in Quantum Computers Capable of State Manipulation, Readout and High-Speed Gate Pulsing of Spin Qubits in Intel 22nm FFL FinFET Technology by J-S. Park et al, February 2021 (3 pages).*

Introduced in 2021, **HorseRidge 2** improved cryo-electronics integration to an unprecedented level. It added multigate pulsing making it possible to control several qubits simultaneously, qubit readout and a programmable microcontroller. Gate pulsing create multi-qubit gates with square DC signals controlling the barrier and plunger gates of the quantum dots while single-qubit gates use modulated RF signals and qubit readout use regular RF signals. The chipset uses frequency multiplexing to reduce the number of RF cables for qubits drive and readout. It drives up to 16 spin qubits with frequency ranges between 11 and 17 GHz. It reads the state of up to 6 qubits simultaneously.

[1396] See Cryogenic electronics for the read-out of quantum processors by Harald Homulle, TUDelft, 2019 (185 pages).

[1397] See Cryo-chip overcomes obstacle to large-scale quantum computers by QuTech, February 2020.

The control chip contains 22 DACs to simultaneously control the gate potentials for many qubits. The chipset is manufactured in a 22nm low-power FinFET technology (22FFL), operates at 4K and contains 100 million transistors[1398]. In May 2021, Intel and Qutech demonstrated high-fidelity two-qubit control with this HorseRidge 2 control chipset.

In 2022, **POSTECH** from South Korea proposed a similar architecture to HorseRidge 2 with a CMOS SoC sitting at 3.5K. It adds local oscillators[1399]. Il was prototyped in 40 nm TSMC bulk CMOS and consumes about 15 mW per qubit.

In 2019, an American-Australian team from the **University of Sydney**, **Purdue University** and **Microsoft Research** designed Gooseberry, a CMOS circuit to control superconducting, electron spin or (yet to be seen) Majorana fermion qubits[1400]. Designed by David Reilly's team from the joint Microsoft Quantum Laboratories at the University of Sydney, it is operating at 100mK, just next to the qubit circuit on the same PCB support but without supposedly disturbing the qubits (in that case, only for silicon qubits since superconducting qubits would sit at the 15 mK cold plate stage). It seems to save power with a low sampling rate in AWG/DAC/DACs (4 bits).

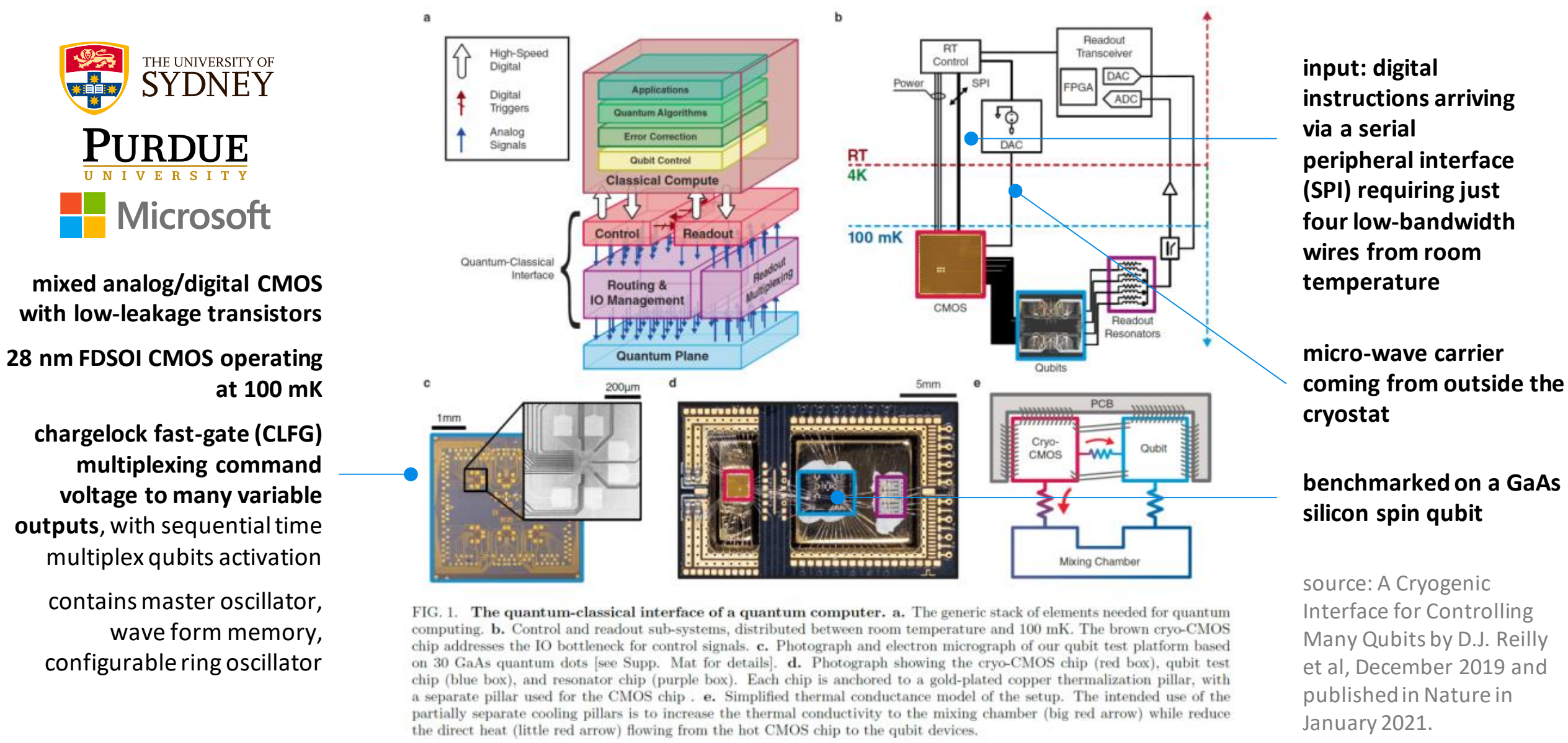

*Figure 495: Microsoft prototype another control chipset that support fewer functions than HorseRidge but it run next to the qubit chipset at lower temperature, suitable for silicon spin qubits. Source: A Cryogenic Interface for Controlling Many Qubits by D.J. Reilly et al, December 2019 (7 pages).*

The circuit is using a microwave carrier signal source (LO or local oscillator) sitting outside the cryostat. It is using a round-robbing scheme to distribute modulated micro-waves to each and every qubit in a sequential way. Qubit readout is done here with external circuits (ADC and FPGA). It is a bit the opposite of Harald Homulle's solution from TUDelft. The test CMOS is realized in FDSOI in 28nm. The chipset greatly simplifies the control circuitry coming from outside. This low-power chipset is generating control pulses of 100 mV at 18 nW per cell.

[1398] See A Fully Integrated Cryo-CMOS SoC for Qubit Control in Quantum Computers Capable of State Manipulation, Readout and High-Speed Gate Pulsing of Spin Qubits in Intel 22nm FFL FinFET Technology by J-S. Park et al, February 2021 (3 pages) and 41 slides (not free access).

[1399] See A Cryo-CMOS Controller IC for Superconducting Qubits by Kiseo Kang et al, August 2022 (14 pages). Computing the power per qubit was not obvious since the drain per function is not clearly presented in the paper (readout pulses vs readout pulses analysis).

[1400] See A Cryogenic Interface for Controlling Many Qubits by D.J. Reilly et al, December 2019 (7 pages). It was then published in Nature in January 2021.

The control of the qubits can also use superconducting microwave generation and reading circuits, their interest being a much lower thermal dissipation[1401].

In 2020, **CEA-Leti** in Grenoble created a mixed analog, digital and quantum cryo-CMOS circuit manufactured in 28 nm FDSOI and operating at 110 mK. It handles all the qubits driving and readout cycle with charge pumping, generating continuous tone GHz microwaves and measuring the induced current with a multiplexed transimpedance amplifier (TIA). At this experimental stage, it drives only a couple qubits but looks promising with regards to the ability to control quantum dots qubits at their operating temperature, at least for silicon qubits working between 100 mK and 1.5 K depending on their type and experimental settings. And the quantum dot qubits were in the circuit itself!

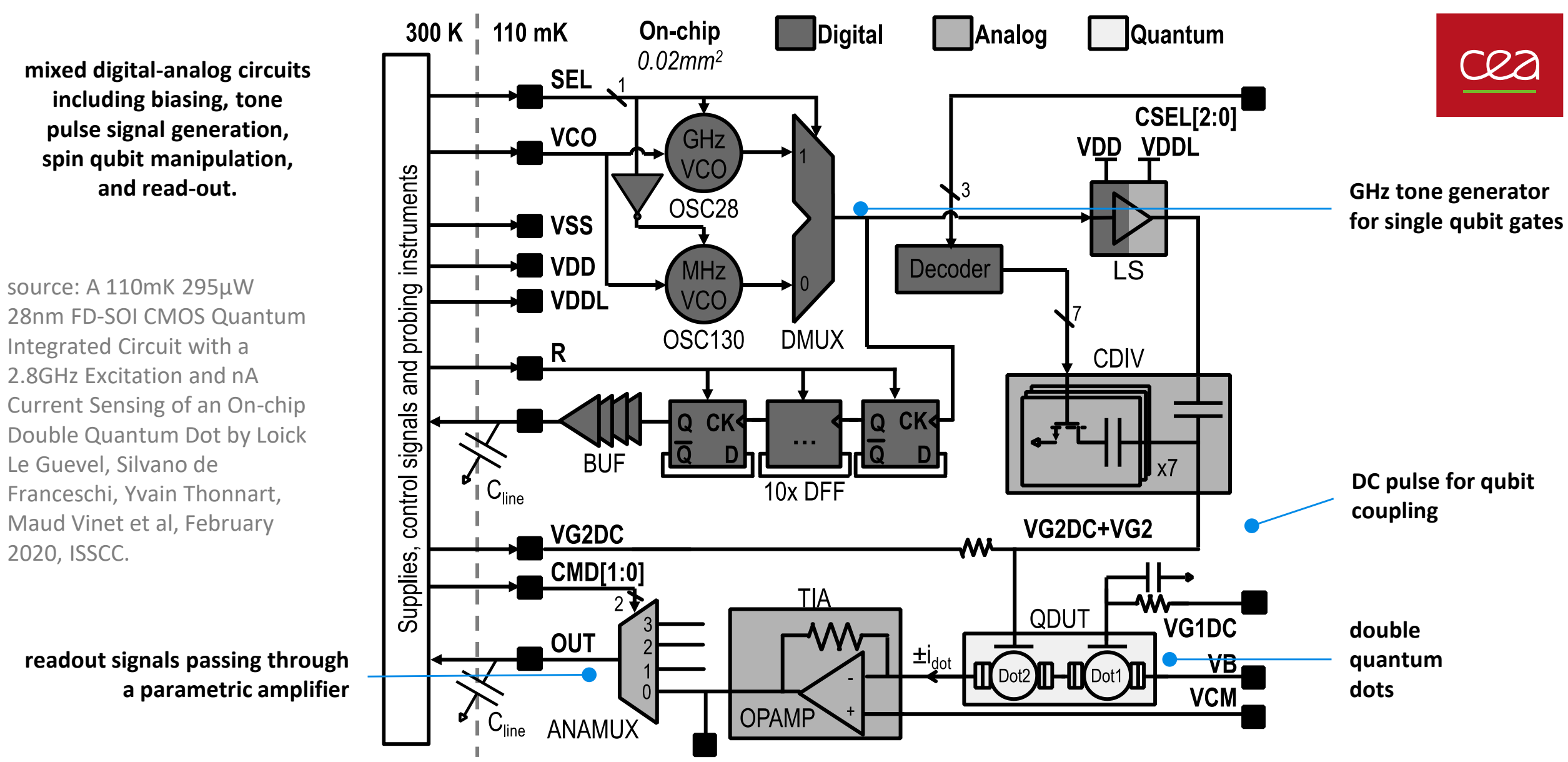

*Figure 496: this chipset from CEA-LIST runs at the same temperature as Microsoft's chipset seen before. It is tailored for silicon spin qubits control. Source: A 110mK 295µW 28nm FD-SOI CMOS Quantum Integrated Circuit with a 2.8GHz Excitation and nA Current Sensing of an On-chip Double Quantum Dot by Loick Le Guevel, Silvano de Franceschi, Yvain Thonnart, Maud Vinet et al, February 2020, ISSCC (12 pages).*

One key technology to master when assembling electronic components at the qubit level is packaging and connectivity. That's where a French team from **CEA-Leti**, **CEA LIST** and **CNRS-Institut Néel** made progress in February 2021 with building a prototype interposer enabling the integration of quantum and control chips fabricated from different materials, processes and sources. Named QuIC (Quantum integrated circuits with cryo-CMOS), the prototype demonstrator controls quantum chipsets with integrated control electronics and operating at below 1K. The integration uses a 3D flip-chip process. The control electronics are made on standard FDSOI 28nm by STMicroelectronics. Passive elements and filter devices will be integrated in future versions.

The integrated packaging increases the number of qubits that can be controlled with reducing the number of coaxial cables flowing through the cryostat from the upper stages. It also avoids chipset wire bonding since qubits and control electronics are coupled by routing lines on the interposer. The packaging allows thermal decoupling between the quantum chipset and the electronics control chipset. They also use a die-to-wafer process from CEA-Leti that are used to build interconnects working at under 1K.

[1401] See Quantum Computer Control using Novel, Hybrid Semiconductor-Superconductor Electronics by Erik P. DeBenedictis of Zettaflops, 2019 (15 pages), which describes an approach for controlling qubits mixing superconductors (JJ) and adiabatic circuits, Cryogenic Adiabatic Transistor Circuits (CATCs). The paper gives an overview of the energy efficiency of cryo-CMOS components and various known superconductors (RQL, AQFP, ...).

CEA is also replacing indium bumps with other materials that are compatible with existing CMOS manufacturing processes, like SnAg microbumps and directly bonded Cu pads from $Cu/SiO_2$[1402].

In another work published by an **EPFL** team in January 2021, a 40 nm CMOS chip operating at 50 mK hosts both 9 silicon quantum dots qubits organized in a 3x3 array and some digital electronics using analog LC resonators implementing time- and frequency-domain multiplexing for qubit readout, all operating at 50 mK[1403].

| | | DC pulses | LO | AWG | DAC | I/Q mixers | I/Q demod | readout parametric amplification | readout ADC | readout signal analysis | temp | power / qubit |
|---|---|---|---|---|---|---|---|---|---|---|---|---|
| cryo-CMOS | **Google - Bardin** 2019 | | | for X and Y gates | yes | yes | | | | | 3K | 2 mW |
| cryo-CMOS | **Microsoft/Sydney /Purdue**, FD-SOI 28 nm, 2019 | yes | yes | only external signal routing to qubit | | | | | | | 100 mK | N/A, low power |
| cryo-CMOS | **CEA List/Leti** FD-SOI 28 nm 2020 | Yes | yes | | | | | Yes TIA amplification | | | 110 mK | 295µW |
| cryo-CMOS | **QuTech& Intel** FinFET 22 nm 2020 | | | Yes, 8-10 bits | Yes | yes | | | | | | 12 mW |
| cryo-CMOS | **HorseRidge 2** FinFET 22 nm, Intel 2021 | yes | | yes, 14-bit | yes | Yes | yes | only LNA between paramp and HEMT | yes | | 4K | 27 mW (to check) |
| cryo-CMOS | **EPFL**, CMOS 40 nm 2021 | | | | | | yes | time and frequency domain readout signal multiplexing | | | 50 mK | |
| cryo-CMOS | **POSTECH** Korea CMOS 40 nm 2022 | | 2 | yes, 4 bits | yes, 4 bits | yes | yes | LNA | yes | | 3.5K | 15 mW |
| cryo-CMOS | **IBM**, FinFET 14 nm 2022 | | | Yes | | Yes | | | | | 3.5K | 23 mW |
| SFQ | **SeeQC**, SFQ 2019 **DigiQ**, USC, Chicago, Nvidia 2022 (*) | yes | | replaced by series of single amplitude pulses | | | | yes | | | 3K/600 mK (SFQ copro) & 20 mK (DQM) | 0,24 mW (*) |
| SFQ | **IBM QEC**, SFQ 2022 | | | | | | | | | partial surface code QEC | 4K | 10µW to 500µW per logical qubit |

(cc) Olivier Ezratty, 2022

*Figure 497: compilation of various cryo-chipsets developed so far. (cc) Olivier Ezratty, 2022. Sources: Google – Bardin: A 28nm Bulk-CMOS 4-to-8GHz <2mW Cryogenic Pulse Modulator for Scalable Quantum Computing, February 2019 (13 pages), Intel HorseRidge 2: A Fully Integrated Cryo-CMOS SoC for Qubit Control in Quantum Computers Capable of State Manipulation, Readout and High-Speed Gate Pulsing of Spin Qubits in Intel 22nm FFL FinFET Technology by J-S. Park et al, February 2021 (3 pages), Microsoft / Sydney / Purdue: A Cryogenic Interface for Controlling Many Qubits by D.J. Reilly et al, December 2019 (7 pages), CEA List/Leti: A 110mK 295µW 28nm FD-SOI CMOS Quantum Integrated Circuit with a 2.8GHz Excitation and nA Current Sensing of an On-chip Double Quantum Dot by Loïck Le Guevel et al, February 2020, ISSCC (12 pages). QuTech: A Scalable Cryo-CMOS Controller for the Wideband Frequency-Multiplexed Control of Spin Qubits and Transmons by Jeroen Petrus Gerardus Van Dijk, Menno Veldhorst, Lieven M. K. Vandersypen, Edoardo Charbon et al, November 2020 (17 pages). EPFL: Integrated multiplexed microwave readout of silicon quantum dots in a cryogenic CMOS chip by A. Ruffino et al, EPFL, January 2021 (14 pages), POSTECH: A Cryo-CMOS Controller IC for Superconducting Qubits by Kiseo Kang et al, August 2022 (14 pages). IBM: A Cryo-CMOS Low-Power Semi-Autonomous Qubit State Controller in 14nm FinFET Technology by David J Frank et al, IBM Research, ISSCC IEEE, February 2022 (no free access), SeeQC: Hardware-Efficient Qubit Control with Single-Flux-Quantum Pulse Sequences by Robert McDermott et al, 2019 (10 pages), DigiQ: A Scalable Digital Controller for Quantum Computers Using SFQ Logic by Mohammad Reza Jokar et al, February 2022 (15 pages). IBM QEC: Have your QEC and Bandwidth too!: A lightweight cryogenic decoder for common / trivial errors, and efficient bandwidth + execution management otherwise by Gokul Subramanian Ravi et al, August 2022 (14 pages).*

At last, **IBM** is also working on their own Cryo-CMOS component. They piloted the first one manufactured in a 14 nm process. It supports 4.5-to-5.5GHz RF AWG for pulse control generation, and doesn't rely on TDM or FDM (time or frequency multiplexing)[1404].

---

1402 See Die-to-Wafer 3D Interconnections Operating at Sub-Kelvin Temperatures for Quantum Computation, September 2020.

1403 See Integrated multiplexed microwave readout of silicon quantum dots in a cryogenic CMOS chip by A. Ruffino et al, EPFL, January 2021 (14 pages).

1404 See A Cryo-CMOS Low-Power Semi-Autonomous Qubit State Controller in 14nm FinFET Technology by David J Frank et al, IBM Research, ISSCC IEEE, February 2022 (no free access).

For a helicopter view, all these cryo-CMOS projects seem to make more sense to drive silicon spin qubits than superconducting qubits. One reason is the available cooling budget is much higher at the operating temperature of spin qubits that sits between 100 mK and 1 K while superconducting qubits operate at about 15 mK. Figure 497 contains a quick comparison of the various cryo-chipsets studied in the section and the next on superconducting logic. It shows a discrepancy of power consumption per qubit which is explained by several factor: the different electronic features supported by the chipsets, their mutualization across a given number of qubits and the manufacturing node technology.

It is completed by Figure 498 which lists which part in the table corresponds to which function in the pulse management sequences from qubit drive to qubit readout.

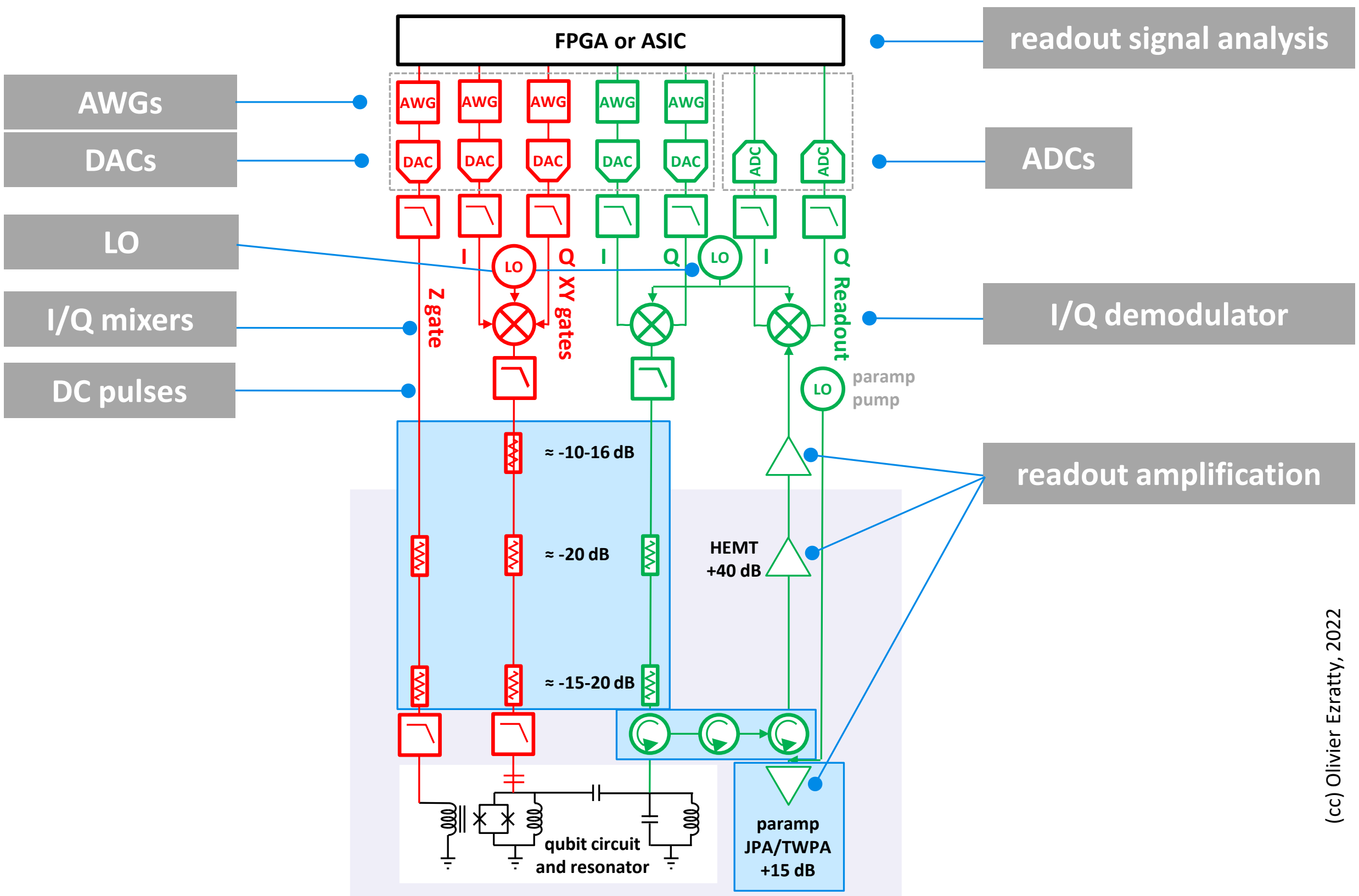

*Figure 498: feature list chosen for the table in Figure 497. (cc) Olivier Ezratty, 2022.*

## Superconducting electronics

The other option for qubit control and readout at low temperature is to rely on superconducting logic based on Josephson junctions[1405]. The most common one are SFQ, for "single flux quantum" and RSFQ for "rapid SFQ"[1406].Their potential benefit is a very low power consumption, up to 500 times less than CMOS logic[1407], the ability to operate at the same temperatures as superconducting qubits, and their enablement of a much simple cabling scheme within the cryostat[1408].

---

[1405] See Superconducting electronics at 4 K for control and readout of qubits by Adam Sirois et al, NIST, ASC 2020 (27 slides) and Flux Quantum Electronics by NIST which cover their broad research in the domain.

[1406] SFQ logic families are divided into two groups: ac-biased and dc-biased. Reciprocal Quantum Logic (RQL) and Adiabatic Quantum Flux Parametron (AQFP) are in the first group, and Rapid Single Flux Quantum (RSFQ), Energy-efficient RSFQ (ERSFQ) and energy-efficient SFQ (eSFQ) are in the second group. The dc-biased logic family with higher operation speed (as high as 770GHz for a T-Flip Flop (TFF)) and less bias supply issues are more popular than ac-biased logic family. Source: NISQ+: Boosting quantum computing power by approximating quantum error correction by Adam Holmes et al, Intel, University of Chicago and USC, April 2020 (13 pages).

[1407] Source: Superconducting Microelectronics for Next-Generation Computing by Leonard M. Johnson, February 2018 (27 slides).

[1408] With room temperature classical control electronics on a 1000 qubit system, you'd need between $5M and $10M of niobium-titanium cables for the 4K to 15 mK stages. Each such cable costs in excess of $2K.

The eSFQ variant can even potentially be $10^4$ more efficient than cryo-CMOS for some functions[1409].

However, this technology has some shortcomings: it may be a significant source of noise affecting qubits fidelities, there are some constraints on high-frequency power sources, classically generated AWG wave formed pulses are replaced by trains of single amplitude SFQ pulses of less than 2 ps duration which drives its own preparation overhead to create qubit gates[1410], limited Josephson junctions density and at last, SFQ logic can't be used to store data. Many solutions are investigated but are still in the making: cryogenic spintronics, magnetic tunnel junction (MTJ) [1411], RQL (reciprocal quantum logic)[1412], SQUID-based, hybrid Josephson–CMOS, JMRAM and OST-MRAM memories (I'll pass on the whereabouts of these).

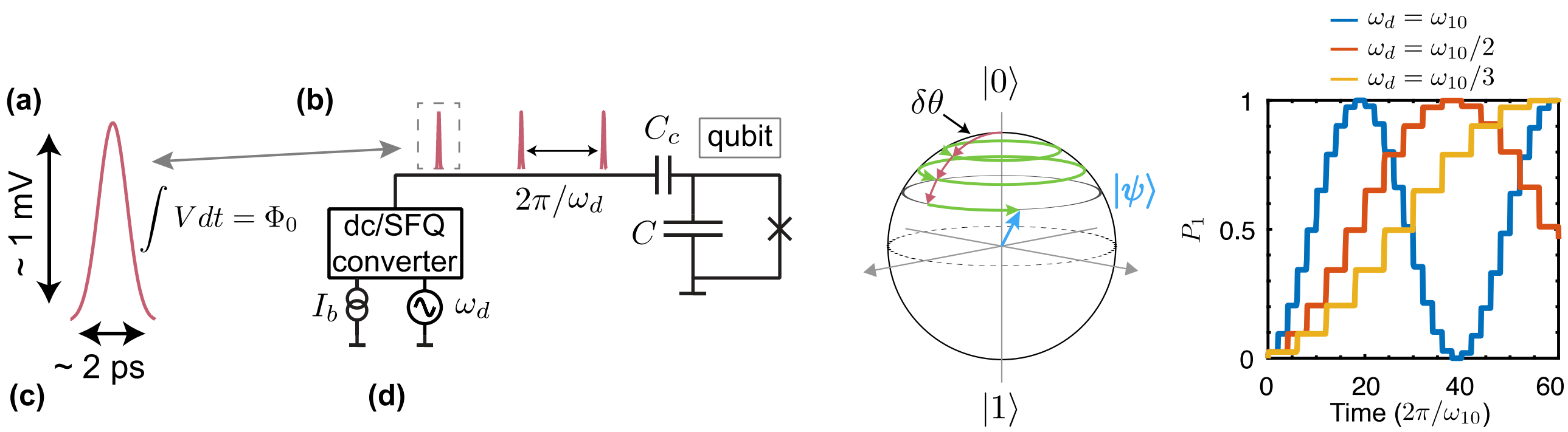

*Figure 499: SFQ based wave pulse generation process. Source: Digital coherent control of a superconducting qubit by Edward Leonard, Robert McDermott et al, 2018 (13 pages).*

There are many interconnected research fields here, and SFQ qubits drive logic is frequently of sub-product of more general research in superconducting electronics. Since the 1970s, there were many ups and downs with research in using superconducting electronics to override the apparent limitations of Moore's law with classical semiconductors. Also, superconducting electronics have other use cases like with single photon detection, magnetism sensing with SQUIDs and analog amplifiers working at the quantum limits (JPAs, SPMs, TWPAs, that we'll cover in the next part).

Qubit readout function can also be implemented with SFQ with readout signal generated through a JPM amplifier[1413] and converted with ADCs[1414]. Tone signal generation can also be implemented in SFQ logic[1415].

---

1409 See Quantum-Classical Interface Based on Single Flux Quantum Digital Logic by Roger McDermott, Oleg A. Mukhanov, Thomas A. Ohki et al, October 2017 (16 pages).

1410 At this point, in D-Wave annealers, SFQ circuits create DC signals and ramp currents with DACs (digital-to-analog converters) to configure the system  and drive the magnetometers used for qubits readouts.

1411 See Cryogenic Memory Architecture Integrating Spin Hall Effect based Magnetic Memory and Superconductive Cryotron Devices by Minh-Hai Nguyen et al, 2020 (11 pages).

1412 See Superconducting logic circuits operating with reciprocal magnetic flux quanta by O.T. Oberg, 2011 (337 pages).

1413 See Interfacing Superconducting Qubits With Cryogenic Logic: Readout by Caleb Howington, Alex Opremcak, Robert McDermott, Alex Kirichenko, Oleg A. Mukhanov and Britton L. T. Plourde, August 2019 (5 pages) with more details in the related thesis Digital Readout and Control of a Superconducting Qubit by Caleb Jordan Howington, December 2019 (127 pages).

1414 See History of Superconductor Analog-to-Digital Converters by Oleg Mukhanov, 2011 (19 pages) and Superconductor Analog-to-Digital Converters by Oleg A. Mukhanov et al, 2010 (21 pages).

1415 See A low-noise on-chip coherent microwave source by Chengyu Yan, Mikko Möttönen et al, November 2021 (14 pages) and Digital Control of a Superconducting Qubit Using a Josephson Pulse Generator at 3 K by L. Howe et al, PRX Quantum, 2022 (11 pages).

**IBM** studied classical electronics based on the Josephson junction from the 1960s to 1983, using lead and then lead/niobium. The technology was only supported by IBM and could not compete with CMOS processors, drive by Moore's law and the whole semiconductor industry, particularly with Intel. Japan's MITI had also launched a superconducting computing initiative throughout the 1980s leading to a 4-bit machine using 1 Kbits of RAM. The Bell labs also worked on niobium/aluminum oxide Josephson junctions.

**Hypres**. Then, the invention of the more efficient and energy efficient RSFQ in USSR in 1985 led to its transfer to the USA via its coinventor Oleg Mukhanov when he joined Hypres in 1991. It led to a short lived superconducting supercomputing project (1997-2001). Starting in the early 2000s, attention then turned to superconducting qubits with investments throughout the world (USA, France, Japan, ...) leading to major developments from IBM, Google and others. Hypres did use superconducting electronics for non-quantum use cases, particularly in the defense industry and with radars, and quantum sensing using SQUIDs. Interestingly, some of the interest in classical electronics made with SFQ and RSFQ came with quantum computing and the need for energy efficient control electronics.

**D-Wave** was probably the first to use SFQ electronics in its systems and since its inception. At their beginning, they hired skilled engineers coming from IBM, Stony Brooke University in New York and coming from Stellenbosch University in South Africa having a good experience in superconducting physics and electronics.

D-Wave's quantum annealers contain flux superconducting qubits and superconducting SFQ circuits handling signals control generation, control and qubit state readout, and for up to 5000 qubits. This is a little-known technological feat from D-Wave. It allows them to greatly simplify the wiring that leads to the quantum processor since all their SFQ electronics sits in the same chipset handling the qubits. Figure 500 shows how it looks like.

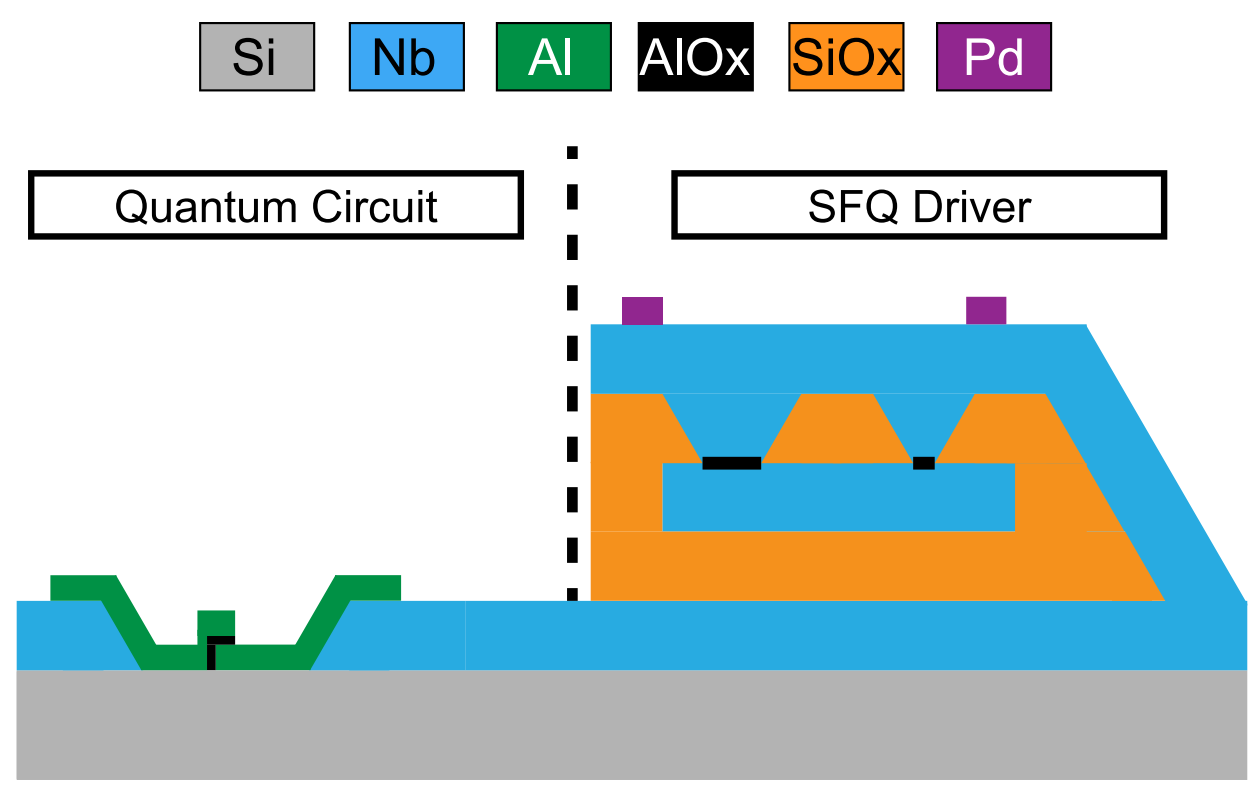

*Figure 500: a side-by-side comparison of the stacking of elements in a superconducting qubit (left) and with SFQ logic (right). Source: [Digital coherent control of a superconducting qubit](#) by Edward Leonard, Robert McDermott et al, 2018 (13 pages).*

Who else in working with SFQ electronics to control solid state qubits? Let's start with the USA who are the most active here.

- **SeeQC**, a spin-off / split-off from Hypres that we'll detail later and is specialized in superconducting electronics for qubits control.
- **Raytheon BBN** (USA) is investigating the usage of SFQ systems and a mix of SFQ and spintronics for controlling qubits[1416]. They have a wide-ranging partnership with IBM and some IBM researchers worked with BBN on SFQ back in 2018 but it doesn't tell whether IBM is keen to adopt SFQs to control their superconducting qubits.

---

[1416] See Quantum Engineering and Computing Group by Thomas Ohki, March 2021 (39 slides). The team has a staff of 20 and Digital coherent control of a superconducting qubit by Edward Leonard, Robert McDermott et al, 2018 (13 pages). It identifies a shortcoming of SFQ: quasi-particles poisoning that negatively impacts qubit fidelities. Back in 2018, they said it could be addressed with putting SFQ logic on a separate chip that would be bonded (with indium) to the qubit chipset. Nowadays, this is an available technology.

- **University of Wisconsin-Madison** has a Department of Physics run by Robert McDermott that investigates SFQ logic. He pioneered qubit control with trains of SFQ pulses[1417]. They authored the paper with Raytheon BBN on SFQ qubit control mentioned with Raytheon above.
- **MIT Lincoln Labs** has been working for a while on SFQ logic in the "beyond CMOS" roadmap funded by IARPA as part of the Quantum Enhanced Optimization (QEO) and Logical Qubit (LogiQ) programs [1418]. In 2017, they developed a 3D Integrated Superconducting Qubit Platform using three layers: the qubit chipset, an interposer with through-substate vias and a supporting chipset with a routing layer and a TWPA for qubit readout microwave amplification. They are also leveraging their own superconducting cleanroom. The Lincoln Lab is even providing many labs across the world with their own custom TWPA and for free.
- **University of Chicago** is also involved in the design of SFQ-based qubit control electronics. In 2022, they demonstrated low-error two-qubit operations using SFQ pulses drive working with fluxonium superconducting qubits [1419]. They also led the DigiQ project launched by the NSF.

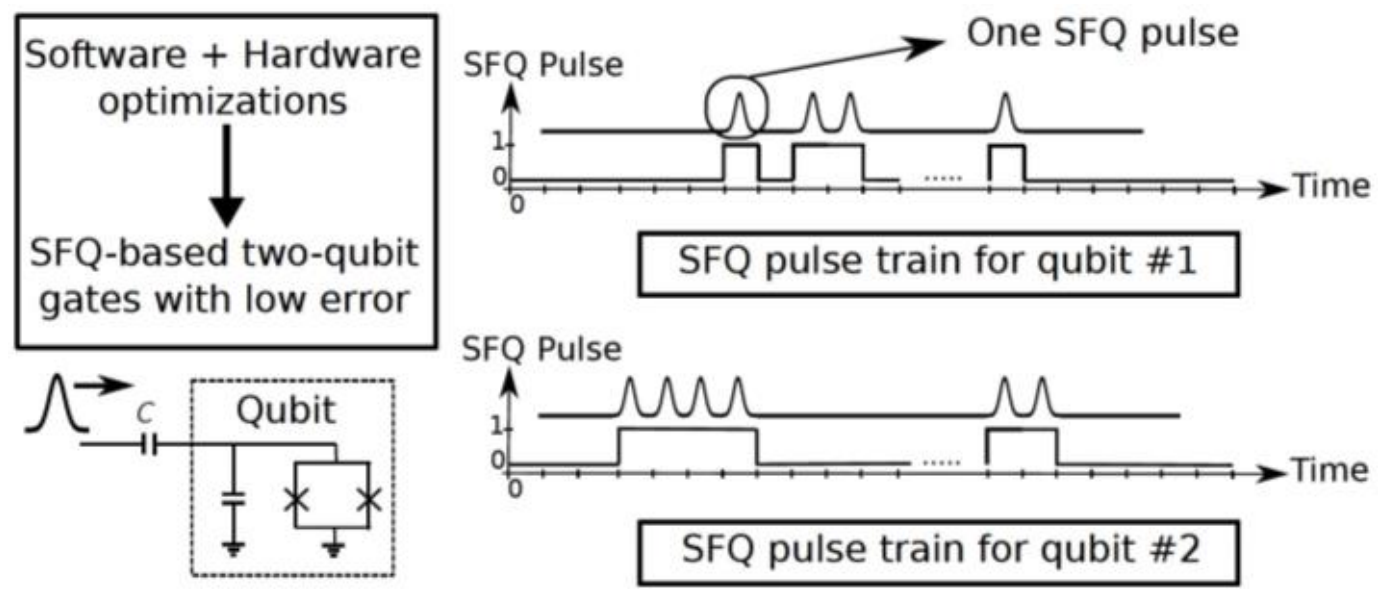

*Figure 501: SFQ wave packet optimization. Source: Practical implications of SFQ-based two-qubit gates by Mohammad Reza Jokar et al, February 2022 (11 pages).*

It was funded as part of the Enabling Practical-scale Quantum Computation (EPiQC)[1420].

- **Northrop Grumman** is also working on SFQ for qubit controls and even patented one related solution back in 2008[1421]. They also developed RQL techniques.
- **IBM** together with the Universities of Chicago and Southern California, and Super.tech (from ColdQuanta) presented in August 2022 the development of an SFQ-based cryogenic circuit to implement part of the logic of the most common errors in quantum error correction for surface codes[1422]. The decoder could support between 2000 and 100,000 logical qubits depending on the code distance.

And in the rest of the world:

- **Japan** has a very active group in SFQ, the group of Nobuyuki Yoshikawa from Yokohama National University. They are working in the field of superconducting electronics, SFQ and adiabatic circuits mostly as "beyond than Moore" solutions1423.

---

[1417] See Accurate Qubit Control with Single Flux Quantum Pulses by Robert McDermott and M.G. Vavilov, 2014 (10 pages), Quantum-Classical Interface Based on Single Flux Quantum Digital Logic by Robert McDermott et al, 2017 (16 pages) and Scalable Hardware-Efficient Qubit Control with Single Flux Quantum Pulse Sequences by Kangbo Li, Robert McDermott and Maxim G. Vavilov, 2019 (10 pages).

[1418] See Superconducting Microelectronics for Next-Generation Computing by Leonard M. Johnson, February 2018 (27 slides).

[1419] See Practical implications of SFQ-based two-qubit gates by Mohammad Reza Jokar et al, February 2022 (11 pages).

[1420] See DigiQ: A Scalable Digital Controller for Quantum Computers Using SFQ Logic by Mohammad Reza Jokar et al, February 2022 (15 pages). The project is run with Amazon, Nvidia, Super.tech and USC. The qubits are driven by series of small SFS pulses, not by arbitrary waveformed pulses. Theoretically, SFQ logic could still create these waveforms thanks to clock speed exceeding 100 GHz. It could create waveforms with basebands of 4 to 25 GHz. But this would require superconducting DACs and ADCs which happen to need resistances, thus being irreversible and dissipative, creating some thermal constraints.

[1421] See Method and apparatus for controlling qubits with single flux quantum logic patent.

[1422] See Have your QEC and Bandwidth too!: A lightweight cryogenic decoder for common / trivial errors, and efficient bandwidth + execution management otherwise by Gokul Subramanian Ravi et al, August 2022 (14 pages).

[1423] See a review paper he coauthored in 2004: Superconducting Digital Electronics by Hisao Kayakawa et al, 2004 (15 pages).

He participated to the development to AQFP (Adiabatic Quantum-Flux-Parametron), an energy-efficient superconductor logic element[1424]. Although they contribute significantly to the field of SFQ, they still do not seem to work on its implementation for qubits control.

- **Germany** has a couple labs and fabs looking at superconducting electronics and their potential usage in qubits control. You can count with Per J. Liebermann and Frank K. Wilhelm from Saarland University who work on improving qubit fidelities with varying the time distance between SFQ pulses in the train using control theory and (classical) genetic algorithms[1425]. The Leibniz-IPHT in Jena, Thuringia, has a cleanroom that works, among other things, on producing RSFQ circuits and SQUIDs for quantum sensing. Leibniz-IPHT is coordinating the German project HIQuP dealing specifically with superconducting qubit control electronics and partnering with IQM Germany and Supracon AG, itself a spin-out of the Leibnitz-IPHT that is specialized in SQUID based magnetometers. The PTB has also investigated SFQ circuits in the past[1426]. There was also the EU project RSFQubit from 2004 to 2007, involving many German players and coordinated by Chalmers University, Sweden, with a funding of 2,6M€.
- **Finland** also conducts some research in SFQ logic at VTT under the leadership of Matteo Cherchi[1427]. Their aCryComm project develops converters and input/outputs for simple SFQ processors. This work could also lead to some potential collaboration with IQM.
- **China** launched a 200M€ project on SFQ electronics. The Shanghai Institute of Microsystem and Information Technology's (SIMIT) Laboratory of Superconducting Electronics is studying SQUIDs (Superconducting quantum interference device used in sensing), SNSPD (superconducting nanowire single-photon detectors) and Superconducting large scale integrated circuits with a 50 persons team. They ambition to create a 64 bits SFQ-based microprocessor. They have their own cleanroom. As side-project of the later could well become SFQ-based qubits control chipsets.
- **Russia** has some researchers working on superconducting electronics and even on SFQ qubit drive electronics, particularly at Lomonosov Moscow State University and Lobachevsky State University of Nizhny Novgorod[1428].

**SeeQC** (2017, USA, $34,2M) was created as a subsidiary of Hypres, an American company specialized in the creation of superconducting electronics, by John Levy, Matthew Hutchings and Oleg Mukhanov[1429].

Its parent company Hypres (1983, USA, $50K) is a long-time specialist in superconducting electronic circuits. It was created by Sadeg Faris, a Libyan who invented the quiteron at IBM, a superconducting transistor. He created Hypres the same year IBM pulled the plug on superconducting electronics and used IBM patents under license.

---

1424 See Adiabatic Quantum-Flux-Parametron: A Tutorial Review by Naoki Takeuchi, Nobuyuki Yoshikawa et al, January 2022 (14 pages)

1425 See Optimal Qubit Control Using Single-Flux Quantum Pulses by Per J. Liebermann and Frank K. Wilhelm, Saarland University, 2016 (5 pages).

1426 See Low-noise RSFQ Circuits for a Josephson Qubit Control by M Khabipov, D Balashov, E Tolkacheva and A B Zorin, PTB, 2008 (7 pages).

1427 See Superconducting chips to scale up quantum computers and boost supercomputers by Matteo Cherchi, March 2021.

1428 See Beyond Moore's technologies: operation principles of a superconductor alternative by Igor I. Soloviev et al, 2017 (22 pages), Flux qubit interaction with rapid single-flux quantum logic circuits: Control and readout by N. V. Klenov, Low Temperature Physics, 2017 (11 pages), the excellent review presentation Superconducting digital electronics by Igor Soloviev, 2021 (115 slides) and Genetic algorithm for searching bipolar Single-Flux-Quantum pulse sequences for qubit control by M.V. Bastrakova et al, September 2022 (9 pages) that deals with the optimization of SFQ pulses to drive qubits, using a genetic machine learning algorithm.

1429 See Seeqc Cuts Its Own Path to the Quantum Era With Integrated Circuit Approach by Matt Swayne, The Quantum Daily, September 2020. By setting up offices in Milan and the UK, the startup found a way to secure European funding for its research. Otherwise they collaborate with Robert McDermott's team at the University of Wisconsin and the Syracuse team in upstate New York.

They've been the only superconducting electronics company for three decades and lived out of SBIR funding and some defense business, like with radars and spectrum analysis. They are a mix of nationalities with Indians, Russians, and a Lebanese. Hypres split in two in 2020. The RF business did stay at Hypres and SeeQC specialized in SFQ based qubit drive while keeping Hypres's cleanroom. The remained of Hypres then worked with other cleanrooms like with SkyWater and the MIT Lincoln Labs.

SeeQC stands for "Superconducting Energy Efficient Quantum Computing". It focuses on the creation of superconducting circuits completed with spintronic technology memories[1430]. The company was initially funded under IARPA's C3 project launched in 2016. Then SeeQC created a lab at Federico II University of Naples, Italy, and the UK, mostly to capture EU/UK public funding. It fared better with the UK than with the EU. They got grants from Innovate UK's Industrial Challenge Strategy Fund as part of four consortiums.

- The first, announced in April 2020, totaling £7M, is led by Oxford Quantum Circuits includes Oxford Instruments, Kelvin Nanotechnology, University of Glasgow (Martin Weides' team) and the Royal Holloway University of London, to create a superconducting qubit computer.
- The second, launched in September 2021, is NISQ.OS, totaling £5,363M, is focused on building an operating system and an hardware abstraction layer is led by Riverlane and includes, Hitachi Europe, Universal Quantum, Duality Quantum Photonics, Oxford Ionics, Oxford Quantum Circuits, arm and the UK National Physical Laboratory, another SeeQC partner in the UK. SeeQC and Riverlane announced in June 2021 that they had integrated Riverlane's operating system Deltaflow.OS with SeeQC's qubit driving components.
- The third consortium was launched in November 2021 in partnership with Merck who is also an investor in SeeQC as well as with Riverlane, Oxford Instruments, the University of Oxford and Medicines Discovery Catapult, with a total funding of £6.85M grant from Innovate UK's Industrial Strategy Challenge Fund (ISCF) to build a "*commercially scalable application-specific quantum computer designed to tackle prohibitively high costs within pharmaceutical drug development*" (aka QuPharma project). A bit like IQM's strategy, SeeQC is to create "*an application-specific quantum computer*" to simulate quantum chemistry, an idea that I found a bit questionable. The announcement didn't mention either a number of qubit or expected fidelities, but the project is due for completion "*in 18 months*".
- The fourth consortium is a project led by sureCore, with £6.5M to support the integration of SeeQC's technology with cryo-CMOS components for qubit controls. SeeQC's role is to "*determine what IP blocks the project will need to create for the Cryo-CMOS chips*". Other partners are Oxford Instruments, SemiWise, Synopsys, Universal Quantum and the University of Glasgow.

In December 2021, SeeQC also announced a partnership with QuantWare (The Netherlands) for the development of a QPU containing QuantWare's superconducting qubits and SeeQC's RSFQ cryogenic control electronics. It adds another superconducting vendor to SeeQC's partners, on top of OQC (UK). The difference here is that QuantWare will embed SeeQC's technology in its QPU.

SeeQC's architecture is based on two chipsets: a classical SFQ control chipset and the SFQuClass DQM, both using Josephson junctions in SFQ superconducting circuits. The first chipset runs at 3K or as low as 600 mK and uses an energy efficient ERSFQ or eSFQ variant of RSFQ logic.

[1430] See Single Flux Quantum Logic for Digital Applications by Oleg Mukhanov of SeeQC/Hypres, August 2019 (33 slides). Oleg Mukhanov also worked on a TWPA, in Symmetric Traveling Wave Parametric Amplifier by Alessandro Miano and Oleg Mukhanov, April 2019 (6 pages).

It controls the DQM and handles error corrections without requiring an external classical computer. The SFQuClass DQM (Digital Quantum Management) includes the microwave generators used to drive the qubits (with DACs, digital-to-analog signal converters) and for qubits readouts (with ADCs, analog to digital microwave signal converters).

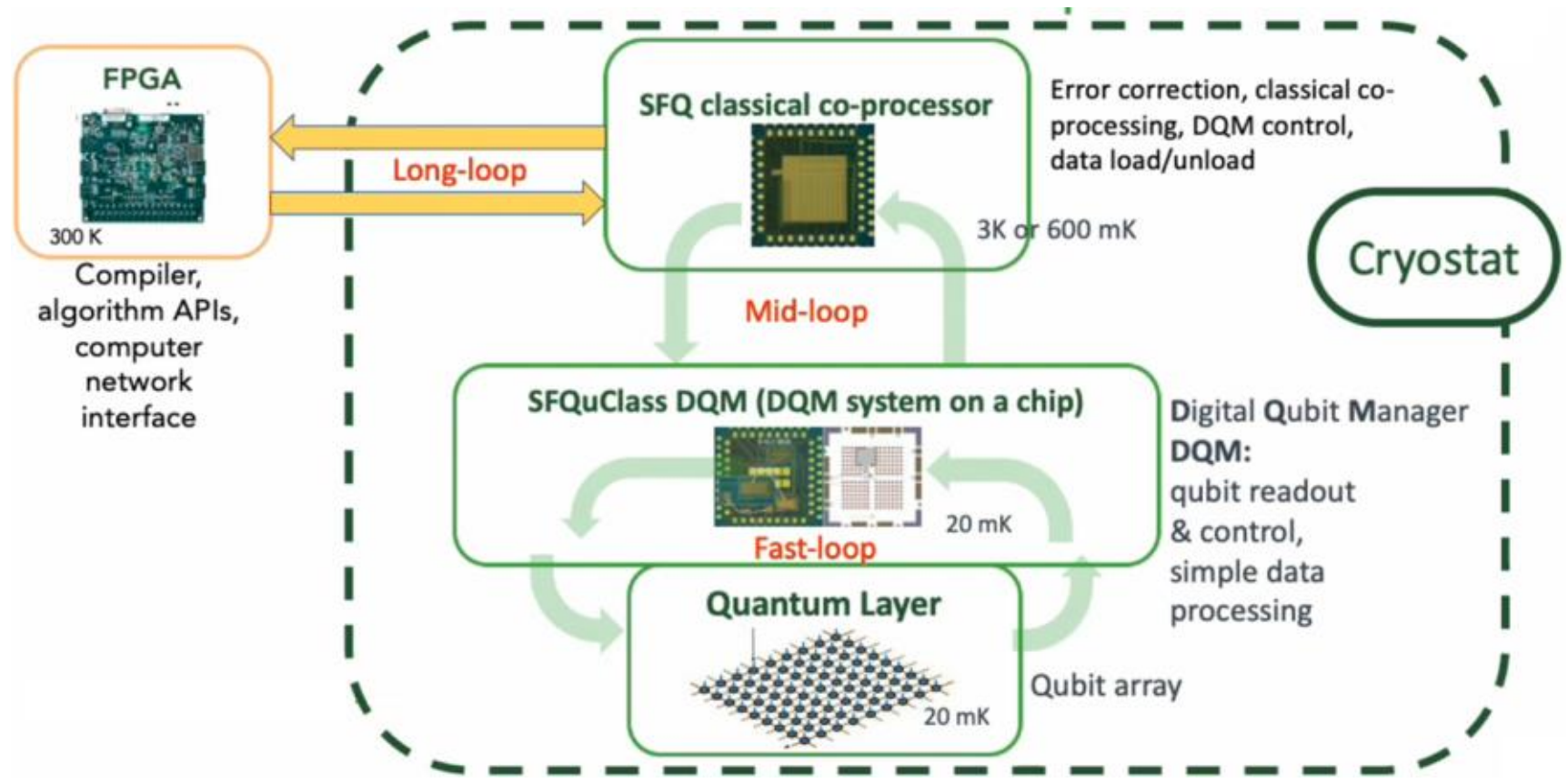

*Figure 502: SeeQC overall architecture with a classical coprocessor running at 3K/600mK and the DQM that site close to the qubit chipset at 20 mK. Source: SeeQC.*

Its power drain is only 0,0002 mW per qubit when it could reach over 20 mW per qubits with cryo-CMOS but we have to check these kinds of comparison making sure the same electronics functions are implemented or not implemented.

The mid-loop between the SFQ co-processor and the DQM also reduces the latency for qubits controls and is particularly interesting for implementing error correction codes[1431].

When we can generate qubit control signals inside the cryostat, it still must be exchanged both ways digitally with the outside of the cryostat. This can be done through signals multiplexed on copper, fiber optics, or even, it is under study, radio waves at very high frequencies (in THz). It also helps maximizing the thermal and vacuum insulation with the outside.

An optical fiber has the advantage of being made of glass, which does not generate thermal expansion and is a weak heat conductor. Still, SeeQC has scalability plans that will require using a growing number of wires with the number of qubits, with a better "Rent's rule" than with classical control[1432].

"Quantum" Rent's Rule

| Qubit Count | Qubit Control Wiring Overhead: Google | Qubit Control Wiring Overhead: seeqc | Application |
|---|---|---|---|
| **2** | 4 | **3** | Prototype |
| **10** | 20 | **3** | Prototype |
| **100** | 200 | **31** | Prototype |
| **1k** | 2,000 | **76** | Optimization / chemistry |
| **10k** | n/a | **330** | Optimization / chemistry |
| **100k** | n/a | **~1,600** | Big data / ML |
| **1m** | n/a | **~6,000** | Encryption |

*Figure 503: how many wires are necessary for controlling qubits comparing Google's Sycamore system and SeeQC's solution. Source: SeeQC.*

[1431] See Quantum-classical interface based on single flux quantum digital logic by Robert McDermott, 2018 (19 pages), Digital coherent control of a superconducting qubit by Edward Leonard Jr. et al, 2018 (13 pages). The diagram on page 10 suggests that microwave generation is always performed outside the cryostat. This is related to the fact that the experiment contains a double control of the qubits: by direct current to drive the microwave generation by the SFQ near the qubits, and in the traditional way outside the cryostat. This allows them to compare the fidelity of the two methods. And then Digital coherent control of a superconducting qubit, by Oleg Mukhanov (CTO and co-founder of SeeQC), Robert McDermott et al, September 2019 (39 slides) and Hardware-Efficient Qubit Control with Single-Flux-Quantum Pulse Sequences by Robert McDermott et al, 2019 (10 pages).

[1432] Rent's rule compute the maths of connections needed to control an electronic system and how it scales with size. See Microminiature packaging and integrated circuitry: The work of E. F. Rent, with an application to on-chip interconnection requirements, 2005 (28 pages) which describes the history of Rent's rule that dates from 1960 and Rent's rule and extensibility in quantum computing by D.P. Franke, James Clarke, L.M.K. Vandersypen and Menno Veldhorst, 2018 (8 pages) that describes how these rules could be applied in quantum computing.

## Circulators

Another key component are the circulators used in qubits readouts at the lowest level of a cryostat.

The generic role of a n-way circulator is to send the microwave from input i to input i+1 in a directional *aka* non-reciprocal way. The signal from i+1 can't be sent, or is sent with strong attenuation to input i. It is used to convey the readout microwave from their AWG/DAC source to the qubit resonator and its response microwave to the first amplification stage. The amplified microwave is sent upwards to the next amplification stage, without being sent back to the resonator. There are settings variations based on using between one and four circulators to improve the various components isolation in the readout food chain.

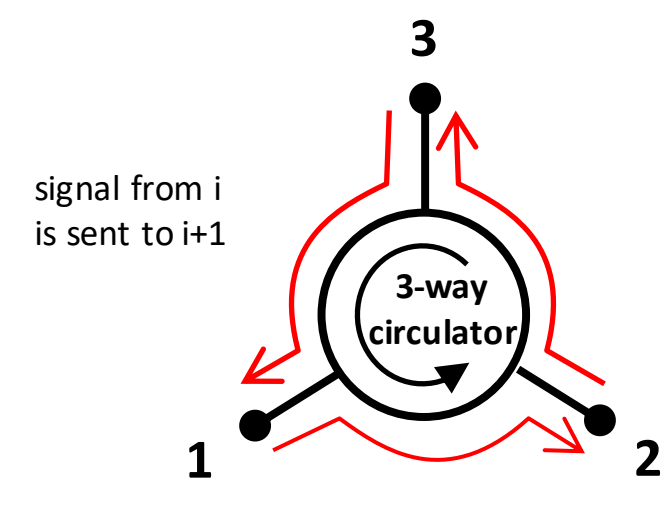

*Figure 504: principle of operation of a circulator which circulates microwaves in a one-way fashion.*

Indeed, the reciprocal protection is not perfect, as measured in decibels and is usually of about 17 to 18 dB. So, chains of circulators enable a protection of about 35 dB, if not over 50 dB. The circulator protection must exceed the first amplifier (or paramp) gain.

Circulators protect the qubits from unwanted noise coming from the output measurement chain. It avoids so-called back-action of the amplifier. They are key contributors to the qubit readout being "nondestructive" of the resulting quantum state (*aka* QND for quantum non-demolition measurement).

Isolators are similar symmetry breaking devices, but with only two connectors they enable a one-way microwave circulation. Conceptually, these are like "microwave diodes", letting microwaves be transmitted in only one direction.

Traditional circulators are passive devices using a ferrite magnet. A microwave entering the circulator through one port is subject to a Faraday rotation in the ferrite, changing its phase[1433]. It creates a constructive microwave interference in one direction of circulation and a destructive interference in the other direction. Such circulators can't be integrated in or near qubit circuits due to their ambient magnetism. Circulators can still be mutualized for the readout of several qubits with frequency-domain multiplexing used at the AWG/DAC and ADC/FPGA levels in the upper data processing stages[1434].

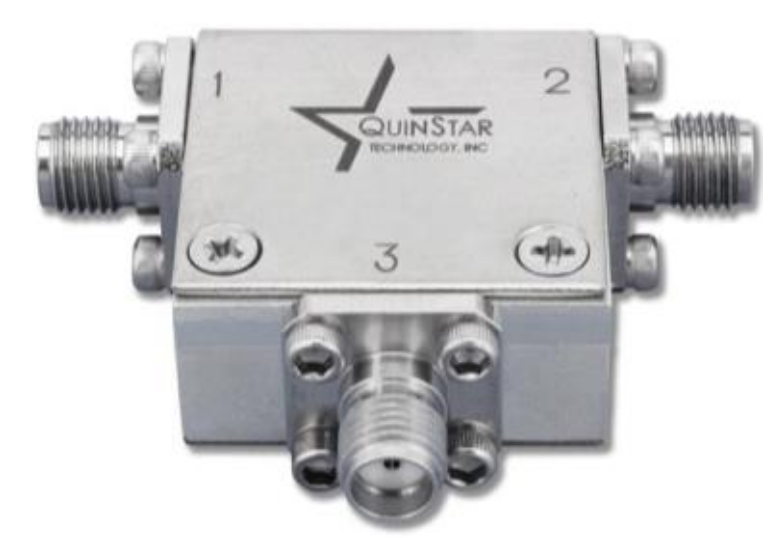

*Figure 505: a typical commercial bulky circulator.*

Nowadays, this multiplexing is not exceeding 8 qubits, but it could theoretically reach 100 qubits, such as with a 10 MHz bandwidth and equivalent 10 MHz spacing, spread in a 2 GHz bandwidth centered at 6.5 GHz, although 20-qubit multiplexing seems more reasonable given the specifications of existing parametric amplifiers (mostly TWPAs).

Existing Faraday circulators used with superconducting qubits are relatively large components of several centimeters wide. This length is conditioned by the microwave length, which is 5 cm for 6 GHz wavelengths. With cabling, filters and attenuators, circulators are the key components located in the cryostat that limit qubit scaling, thus the need for other solutions.

---

1433 This effect was described first in The Ferromagnetic Faraday Effect at Microwave Frequencies and its Applications by C. L. Hogan, 1952 (31 pages).

1434 See the interesting presentation Hall Effect Gyrators and Circulators by David DiVincenzo, Quantum Technology - Chalmers, 2016 (53 slides).

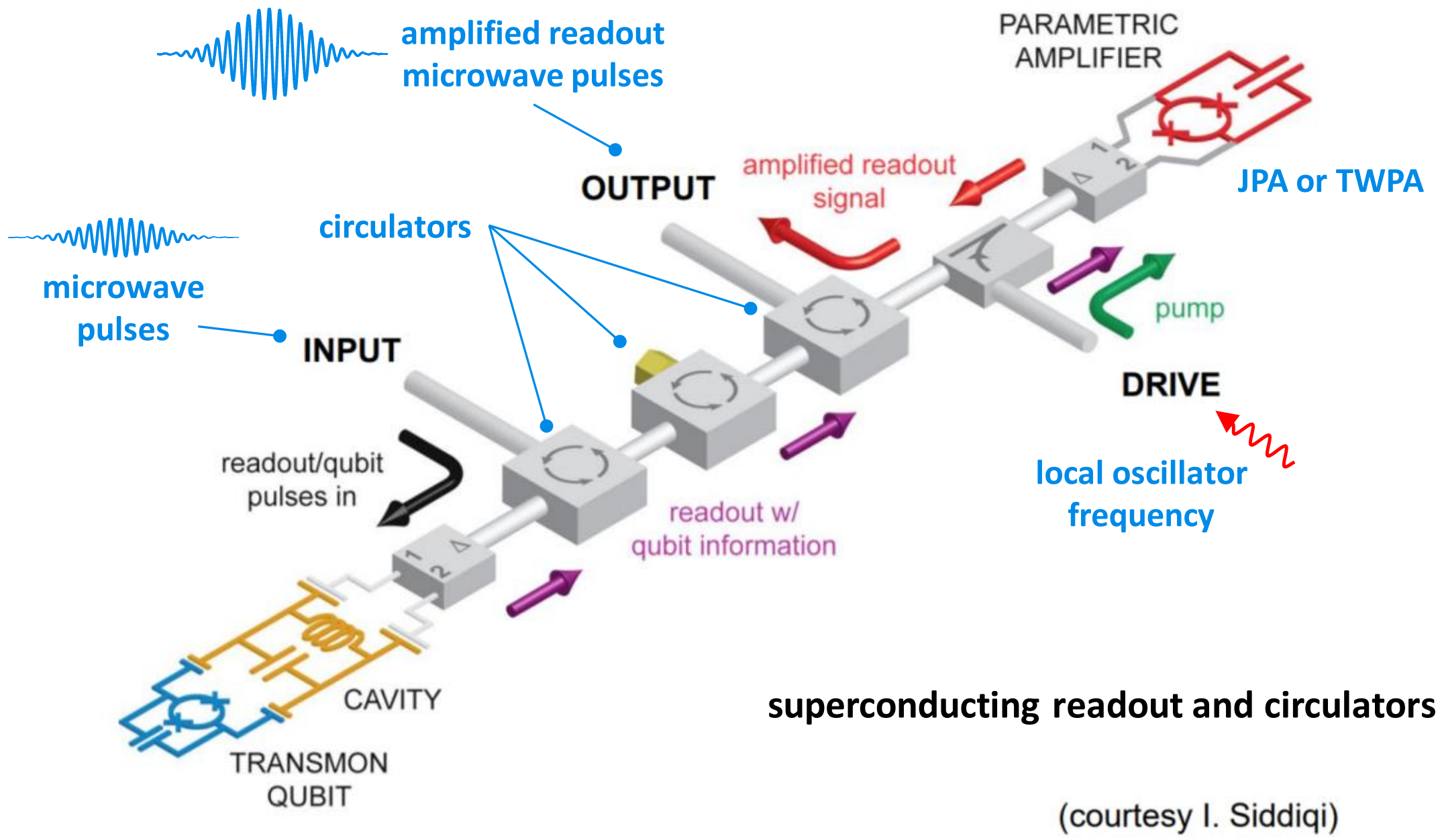

*Figure 506: several circulators are actually used for each set of qubits controlled through frequency multiplexing. Source: Irfan Siddiqi.*

Many alternatives are investigated to circumvent the shortcomings of existing ferrite-based circulators. Ideal circulators would fit into the qubit chipsets and use compatible (superconducting) circuits, have a high protection (in the 20 dB region), a large and controllable bandwidth (between 500 MHz and up to 2 GHz, to enable qubit readout multiplexing) and low power drain if they are active and depending on the cryostat stage where they are operating (15 mK or 4K). They'd also be combined with the first stage quantum-limit low-noise amplifier like a TWPA.

These new types of circulators can be segmented by their underlying physical process. I'll simplify this and have, first, classical electronics systems:

- **Hall effect circulators** as developed by David DiVincenzo in Germany and whose size is not constrained by the readout microwave wavelength[1435].
- **LC based cryo-CMOS circulators** as prototyped at TU Delft[1436].

And then, various superconducting circuits using Josephson junctions like:

- **Interferometric Josephson circulators**, developed at IBM[1437] and RIKEN[1438], with the advantage of being very low power hungry.

[1435] See Hall Effect Gyrators and Circulators by Giovanni Viola and David DiVincenzo, 2014 (18 pages) and On-Chip Microwave Quantum Hall Circulator by A. C. Mahoney et al, 2017 (9 pages) which is based on III/V GaAs and AlGaAs electronics.

[1436] See A Wideband Low-Power Cryogenic CMOS Circulator for Quantum Applications by Andrea Ruffino, Fabio Sebastiano and Edoardo Charbon, IEEE & EPFL, 2020 (15 pages). It uses resonant LC circuits level combining inductance (L) and capacitors (C). The circuit provides a 18 dB isolation and consumes a total of 10.5 mW. It runs at 4.2K and not at the lowest 15 mK. It was tested with a 40 nm node with an active area of 0.45 $mm^2$ in an experimental 1,5 mm wide square circuit.

[1437] See Active protection of a superconducting qubit with an interferometric Josephson isolator by Baleegh Abdo, Jerry M. Chow et al, IBM Research, 2018 (10 pages) and High-fidelity qubit readout using interferometric directional Josephson devices by Baleegh Abdo et al, IBM Research, 2021 (32 pages).

[1438] See Magnetic-Free Traveling-Wave Nonreciprocal Superconducting Microwave Components by Dengke Zhang and Jaw-Shen Tsai, RIKEN, 2021 (18 pages), with a bandwidth of 580 MHz around 6 GHz and an isolation of 20 dB.

- **Superconducting quantum tunnelling capacitors**, developed by Clemens Mueller at ETH Zurich and the University of Queensland[1439]. These are passive systems or can be controlled with just some DC (direct current) input, as shown in Figure 507[1440]. These could be potentially directly implemented in superconducting qubit chipsets.
- **On-chip microwave circulators** with a wide tunable frequency, developed at the University of Colorado in Boulder[1441], a variation with SQUIDs superconducting components.

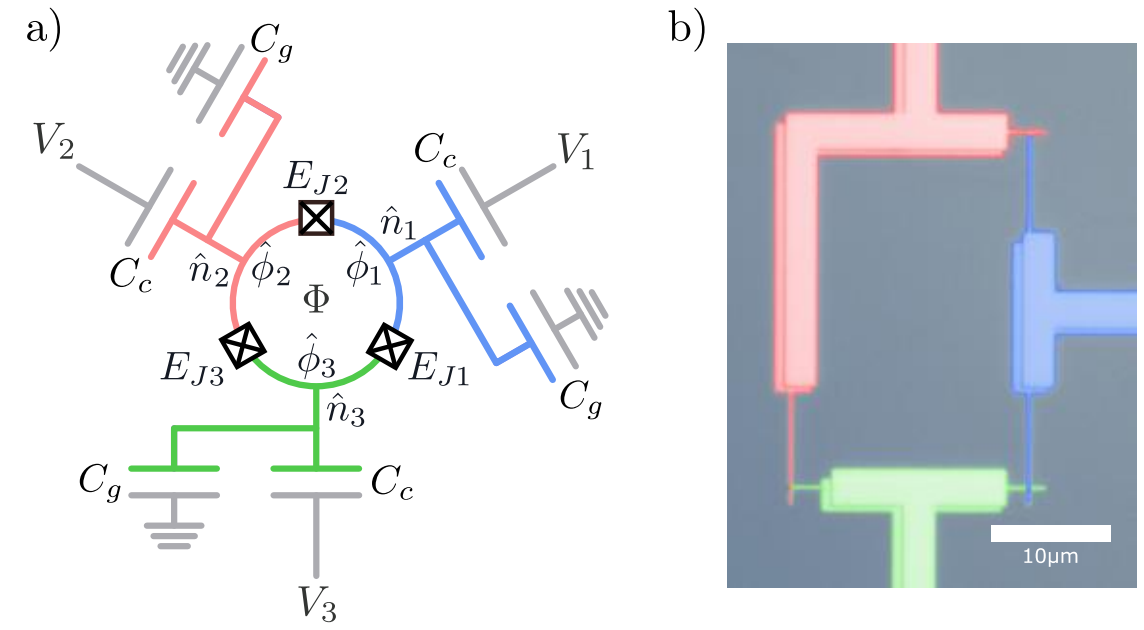

*Figure 507: a prototype passive superconducting circulator that could potentially be integrated in a superconducting qubit chipset. Source [Passive superconducting circulator on a chip] by Rohit Navarathna, Thomas M. Stace, Arkady Fedorov et al, August 2022 (11 pages).*

While there are many such prototype solutions, and we've not covered them all, they seem not having reached the commercial stage at this point, although some have use cases beyond qubits control like in the telecom and radar markets.

### Amplifiers

Analog qubit readout microwave signals have to be amplified several times before they are converted digitally with an ADC and then analyzed, usually with an FPGA programmable circuit. Superconducting readout microwaves are amplified at least three times: first, at the lowest cryogenic stage (15 mK) with a parametric amplifier ("paramp") operating at the quantum limit (JPA, TWPA), then at the 4K stage with an HEMT amplifier, then at ambient temperature with a classical RF analog amplifier that may also be an HEMT. The paramp serves as a low-noise signal preamplifier before the noisier HEMT. They all add a gain of respectively about 15 dB, 40 dB and 50 dB to readout microwaves.

**Paramps**. To make things short, two main generations of parametric amplifiers have been used for qubits readout. They are based on exciting and pumping a material with nonlinear polarization using an intense electromagnetic field. A weak microwave signal can then get amplified via the interaction with the medium. The first paramps used with qubit readouts were the JPAs (Josephson Parametric Amplifiers)[1442]. These are simple amplifiers, using one or two Josephson junctions, easy to manufacture, with a good gain of about 15 to 20 dB[1443]. Their main shortcoming is their narrow bandwidth which prevents their implementation with frequency-domain qubits readout multiplexing, where a single amplifier processes a readout signal coming from several chained qubits using different resonant frequencies. It is due to JPAs being based on cavities.

---

[1439] See Breaking time-reversal symmetry with a superconducting flux capacitor by Clemens Müller, Thomas M. Stace et al, 2018 (10 pages). This circulator uses a small superconducting capacitor, using the quantum tunnelling of the magnetic flux around it, allowing a flow of microwave energy in one direction.

[1440] See Passive superconducting circulator on a chip by Rohit Navarathna, Thomas M. Stace, Arkady Fedorov et al, August 2022 (11 pages).

[1441] See Widely Tunable On-Chip Microwave Circulator for Superconducting Quantum Circuits by Benjamin J. Chapman, Alexandre Blais et al, 2017 (16 pages) and the related thesis Widely tunable on-chip microwave circulator for superconducting quantum circuits by Benjamin J. Chapman, 2017 (144 pages). See also Design of an on-chip superconducting microwave circulator with octave bandwidth by Benjamin J. Chapman et al, 2018 (12 pages).

[1442] See Superconducting Parametric Amplifiers by Jose Aumentado, IEEE Microwave Magazine, August 2020 (15 pages) (*not open access*).

[1443] A Japanese team was able to reach a gain of 40 dB with a JPA in 2022, but with a narrow bandwidth. With a 20 dB gain, they obtain a bandwidth of only 400 kHz. See A three-dimensional Josephson parametric amplifier by I. Mahboob et al, May 2022 (5 pages).

This explains the interest for a relatively new generation of paramps, the TWPAs (travelling waves parametric amplifiers) which were pioneered by Bernard Yurke (USA) in 1996[1444] and successfully implemented in an array of SQUIDs in 2007 by Manuel A. Castellanos-Beltran from JILA (USA)[1445], and then, in 2008, with an intrinsic noise below the standard quantum limit (SQL) and over 20 dB of power gain[1446]. These amplifiers are used for qubit readout, in high-energy particle physics, radioastronomy and astrophysics for the detection of dark matter. TWPA (*aka* usually JTWPA) practically emerged in 2015[1447]. Recent TWPAs are based on long arrays of about 2000 series of Josephson junctions and as such are considered to be "meta-materials". They are more complicated to manufacture. Their broader bandwidth could potentially enable up to 10-20 qubits readout multiplexing[1448]. Their other figures of merit are their saturation and dynamic range (linked to minimum and maximum input signal power), and noise level (noise temperature, under 1K). New options arise with Floquet mode TWPAs from MIT, decreasing noise level and across a wide bandwidth of 6 GHz, further increasing qubits readout multiplexing capabilities. And it has a better directionality[1449].

Both JPAs and TWPAs are active components that are driven by a constant microwave pulse acting as a "pump". JPAs are fed with a pulse of -80 dB while TWPAs use a -70 dB or just a tiny 0,1nW. This is the power reaching the paramps. It is much larger from the outside local oscillator and has to be attenuated on the way down in the cryostat.

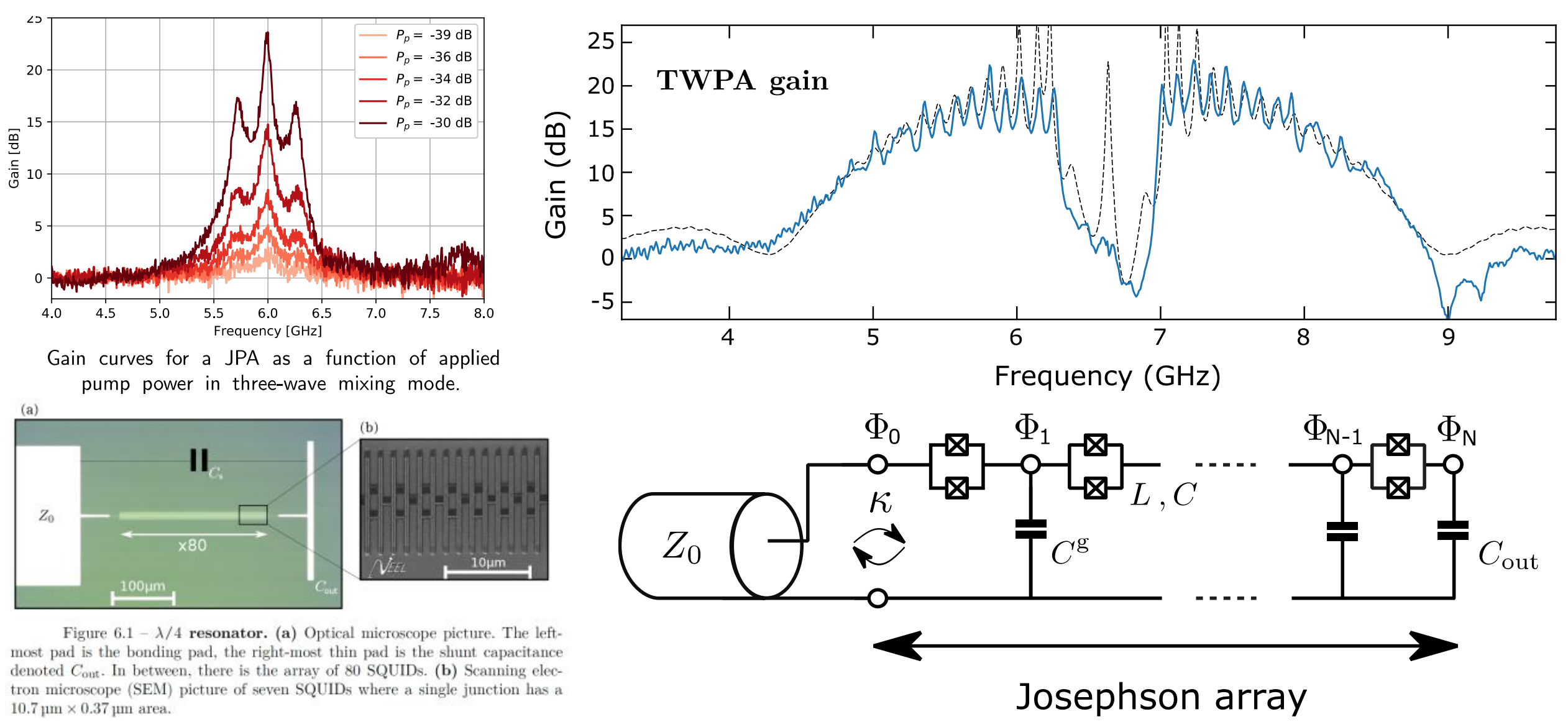

Gain curves for a JPA as a function of applied pump power in three-wave mixing mode.

Figure 6.1 – λ/4 **resonator.** **(a)** Optical microscope picture. The left-most pad is the bonding pad, the right-most thin pad is the shunt capacitance denoted $C_{out}$. In between, there is the array of 80 SQUIDs. **(b)** Scanning electron microscope (SEM) picture of seven SQUIDs where a single junction has a 10.7 µm × 0.37 µm area.

*Figure 508: top left, the typical narrow-band response curve of a JPA, and top right, the typical frequency response curve of a TWPA that has over 2 GHz available with a gain superior to 15 dB in two parts, below 6.2 GHz and above 7 GHz. Bottom left is a typical TWPA circuit, with 2000 series of Josephson junction bridges, as described on the right. Source: Resonant and traveling-wave parametric amplification near the quantum limit by Luca Planat, June 2020 (237 pages).*

1444 See A low-noise series-array Josephson junction parametric amplifier by Bernard Yurke et al, 1996 (4 pages).

1445 See Widely tunable parametric amplifier based on a superconducting quantum interference device array resonator by Manuel A. Castellanos-Beltran et al, 2007 (9 pages).

1446 See this good TWPA review paper Perspective on traveling wave microwave parametric amplifiers by Martina Esposito, Arpit Ranadive, Luca Planat and Nicolas Roch, September 2021 (8 pages).

1447 See Traveling wave parametric amplifier with Josephson junctions using minimal resonator phase matching by T.C. White, John Martinis et al, UCSB, 2015 (15 pages) and A near–quantum-limited Josephson traveling-wave parametric amplifier by C. Macklin, William D. Olivier, Irfan Siddiqi et al, 2015 (3 pages). Despite the first work involving John Martinis in 2015, Google Sycamore was using JPAs and not TWPAs. Google is still investigating, naturally, how to use TWPAs in their systems.

1448 TWPAs however generate undesired mixing processes between the different frequency multiplexed tones as described in Intermodulation Distortion in a Josephson Traveling Wave Parametric Amplifier by Ants Remm, Andreas Wallraff et al, October 2022 (11 pages).

1449 See Floquet Mode Traveling-Wave Parametric Amplifiers by Kaidong Peng et al, MIT, PRX Quantum, April 2022 (20 pages). This work from MIT was funded by Amazon AWS Center for Quantum Computing and by NEC.

With TWPAs, there are variations with three-wave mixing (3WM) using one pump photon yielding one signal photon and one idler residual photon and four-wave mixing (4WM) using two pump photons yielding one signal and one idler photon. The pump microwave usually comes from a local oscillator source outside the cryostat.

Other qubit readout options are investigated. One is based on a **JPM** (Josephson Photo Multiplier) that can be directly embedded in the qubit chipset. It doesn't require low-noise amplification with a JPA or a TWPA and provides a good readout fidelity in excess of 98% although it's a bit slow, lasting 500 ns[1450].

**CEA-Leti** prototyped in 2020 a low-noise cryo-CMOS amplifier that operates as low as 10 mK[1451]. All this is used to handle the first stages of qubits state readout within the cryostat. In another recent work involving the UK and CEA-Leti in France, quantum-dots based readout amplification is studied, and is adapted to silicon spin qubits.

With some improvements, it could reach gains of classical JPAs (15 dB) although on a small bandwidth[1452].

Before looking at cryogenic amplification commercial vendors, let's mention the **MIT Lincoln Lab** team from William D. Oliver who has been pioneering TWPAs for a while, and is providing many labs in the world with its own TWPAs since 2015, and for free (pictured in Figure 509). A gift for science development! On the industry vendors side, **IBM** and **Rigetti** developed their own TWPAs. **IQM** uses TWPAs from VTT.

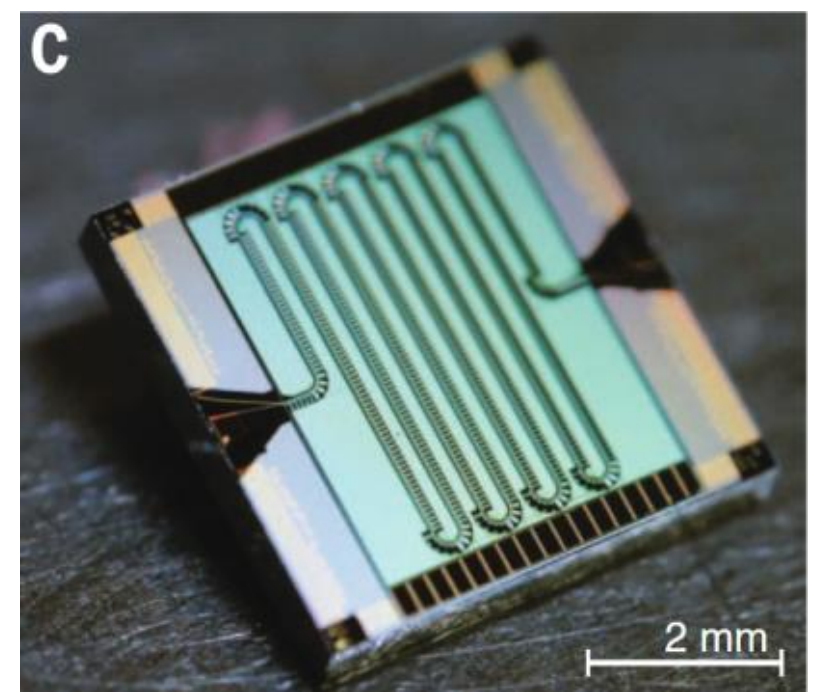

*Figure 509: an MIT Lincoln lab TWPA. Source: A near–quantum-limited Josephson traveling-wave parametric amplifier by C. Macklin, William D. Olivier, Irfan Siddiqi et al, 2015 (3 pages)*

In Finland, **VTT** is also manufacturing TWPAs using 1600 Josephson junctions. 40 nm CMOS. In Sweden, **Chalmers University** researchers are working on their own optimized 3WM TWPA[1453]. In Italy, various labs launched DARTWARS, for designing a TWPA with very large bandwidth covering 5 to 10 GHz with low noise temperature of 600 mK. It will use two techniques with Josephson junctions (JTWPAs) and kinetic inductances (KITWPAs)[1454]. There was also a 4-year EU TWPA project named **ParaWave**, run by German, Italian and British partners from 2018 to 2021[1455].

**Google** is not following the TWPA path but improving JPAs with their home-made SNIMPA or 'snake impedance matched parametric amplifier' which has a high dynamic range, large bandwidth, high saturation and where the active nonlinear element is implemented with an array of rf-SQUIDs[1456].

---

1450 See High-Fidelity Measurement of a Superconducting Qubit Using an On-Chip Microwave Photon Counter by A. Opremcak, Robert McDermott et al, February 2021 (13 pages). On top of give researchers from Wisconsin University, this work involves six researchers from Google and one from Syracuse University in New York.

1451 See Low-power transimpedance amplifier for cryogenic integration with quantum devices by L. Le Guevelet al, March 2020 (13 pages).

1452 See Quantum Dot-Based Parametric Amplifiers by Laurence Cochrane, Fernando Gonzalez-Zalba, Maud Vinet et al, PRL, May 2022 (7 pages).

1453 See Three-wave mixing traveling-wave parametric amplifier with periodic variation of the circuit parameters by Anita Fadavi Roudsari, Per Delsing et al, September 2022 (6 pages).

1454 See Ultra low noise readout with travelling wave parametric amplifiers: the DARTWARS project by A. Rettaroli et al, July 2022 (4 pages).

1455 See Josephson travelling wave parametric amplifier and its application for metrology, 2018 (7 pages).

1456 See Readout of a quantum processor with high dynamic range Josephson parametric amplifiers by T.C. White, Charles Neil, Frank Arute, Joseph C. Bardin et many al, September 2022 (9 pages). They tested it on a 54-qubit Sycamore processor.

**HEMT**. At last, at the 4K cryostat stage sit the second qubit microwaves readout amplifiers named HEMT (High-electron-mobility transistor). They provide a large gain amplification of about 40 dB with high dynamic range and a large-bandwidth >6GHz for the inbound microwave readout signal coming from the paramps from the first cryostat stage, which benefited from a first level 15 to 20 dB low-noise amplification at or near the quantum limit.

Many labs and vendors produce HEMTs for qubit readout, like Chalmers University of Technology in Sweden using indium phosphide (InP) transistors which are very efficient at 4K[1457]. The main vendor here is **Low Noise Factory** (described later, below) as well as **Cosmic Microwave Technology**, with HEMTs designed at Caltech.

Still, commercial HEMT amplifiers dissipate 10 mW of power, which can be a problem when the number of qubits scale, even though the cooling power at the 4K stage is quite larger than at 15 mK, with about 1W to 2W.

One solution would be to use a variety of weakly dissipative TWPAs sitting at the 4K stage, using superconducting materials operating at this temperature like NbTiN and put an HEMT at the 70K cryogenic stage[1458]. This would reduce the heat generated at 4K.

Then, we have room temperature analog amplifiers adding about 50 dB to the signal coming from the HEMT at 4K. These amplifiers are also usually HEMT-based. They consume about 250 mW, which is shared for the multiplexed readout signals of several qubits, usually between 5 and 10 with the potential to grow to 20 and even 100 qubits, depending on the readout speed. Indeed, the shorter the readout pulse, the broader the pulse frequency spectrum will be, limiting multiplexing over a bandwidth of about 2 GHz. The longer the pulse, the smaller the pulse spectrum will be, but it will be detrimental to the efficiency of error correction. That's another design trade-off to take into account in designing these systems.

Now let's look at the amplification vendors I have identified so far.

**Low Noise Factory** (2005, Sweden) designs and produces low-noise amplifiers (HEMT) operating at ambient or cryogenic temperatures as well as circulators and JPAs. They are part of the European **OpenSuperQ** project, led by the University of Saarland in Germany, VTT in Finland and Chalmers in Sweden, to create commercial TWPAs. The consortium demonstrated in 2019 a TWPA with a maximal gain of 10 dB over a 1.4 GHz bandwidth.

**Silent Waves** (2022, France) is a spun-out startup from CNRS Institut Néel in Grenoble. It was founded by Luca Planat, Nicolas Roch and Baptiste Planat.

Their offering is a very efficient TWPA, based on the research conducted by CNRS-Institut Néel-UGA and the LPMMC in Grenoble[1459]. Their commercial TWPA added noise is near the quantum limit of noise. It enables high-fidelity single-shot qubit readout on a wide frequency band and with a gain than can exceed 20 dB. Based on a patented fabrication process, Silent Waves' amplifiers are currently being manufactured in the clean room of the Institut Néel[1460].

---

[1457] See InAs/AlSb HEMTs for cryogenic LNAs at ultra-low power dissipation by Giuseppe Moschetti et al, 2020, Solid State Electronics (7 pages).

[1458] See Performance of a Kinetic-Inductance Traveling-Wave Parametric Amplifier at 4 Kelvin: Toward an Alternative to Semiconductor Amplifiers by M. Malnou et al, NIST and University of Colorado Boulder, October 2021 (11 pages). Their KI-TWPA dissipates only 1 µW.

[1459] See A photonic crystal Josephson traveling wave parametric amplifier by Luca Planat et al, October 2019 (17 pages).

[1460] TWPAs could have applications beyond superconducting qubits readout, in microwave photonics, quantum sensing and quantum information with continuous variables as described in Observation of two-mode squeezing in a traveling wave parametric amplifier by Martina Esposito, Olivier Buisson, Nicolas Roch, Luca Planat et al, first published in November 2021 and revised in April 2022 (16 pages).

They address both superconducting and silicon spin qubits readout. As a first stage, they will enable 5 and 10 qubits readout multiplexing.

**QuantWare** (The Netherlands) is not just providing custom superconducting qubits chipsets but also their Crescendo TWPA. It provides a gain of >18 dB gain on a bandwidth of over 1.5 GHz.

**Qubic Technologies** (2020, Canada) is developing JPAs and TWPAs, the aim being to produce more correlations and better filtered noise. They also use this technology to improve radars. The company was created by Jérôme Bourassa (CEO), a former researcher from the Institut Quantique at the Université de Sherbrooke. He also collaborates with the Institute for Quantum Computing from the University of Waterloo.

**Analog Quantum Circuits** (2021, Australia) is a TWPA manufacturer created in October 2021.

At last, let's mention again **Cosmic Microwave Technologies** (2016, USA) which produces cryogenic LNAs (low noise amplifiers) used for qubits readouts and is a spin-out of Caltech.

### Cabling, connectors and filters

In current quantum systems based on superconducting qubits, copper coaxial cables carry **microwave photon pulses** at frequencies between 5 and 10 GHz that act on the qubits for their reset, for implementing quantum gates and handling qubit readouts.

Microwaves are generated by devices generally located outside the refrigerated enclosure. Frequencies below 5 GHz and above 10 GHz are filtered out[1461]. These microwaves are also attenuated and filtered at the input on the 4K cold plate. An attenuation of 60db, carried out in three steps of 20 dB which each time divide by 100 the transmitted power. It is used to limit the thermal noise that is conveyed in the cables. It is reduced so as not to represent by more than one thousandth of the photons that end up in the qubits. Each filter absorbs energy that must be dissipated at the stages where they are placed.

The thermal conductivity of a cable Q is calculated as follows, using the product of the cable conductivity k, its cross-section A, the temperature gradient T2-T1 and L the length of the cable.

$$Q = kA\frac{T2-T1}{L}$$

**Coaxial superconducting cables** - having theoretically zero resistance at low temperature - connect the qubits to their reading system (thus, in the upward direction in the diagrams). They are made of niobium and titanium alloy (NbTi). They include loops to absorb the metal contraction that occurs during cryostat cooling and warm-up[1462]. With qubit readout, microwave signals are amplified at least twice before leaving the cryostat including once at superconducting temperature, below 1 K.

[1461] See Engineering the microwave to infrared noise photon flux for superconducting quantum systems by S. Danilin et al, 2022 (22 pages).

[1462] See Challenges in Scaling-up the Control Interface of a Quantum Computer by D. J. Reilly of Microsoft, December 2019 (6 pages) which states that superconducting cables have resistance and capacitance when microwaves are passed through them and therefore have a thermal release that must be taken into account.

These cables are used between the 4K and 15-100 mK cold plates in a cryostat. They come from various vendors including **CoaxCo** (Japan)**.** This company seems to be the only one in the world that produces NbTi cables[1463]. The 2 mm diameter cable consists of a conductive outer jacket and a central conductor, both made of niobium-titanium which are separated by a Teflon (PFTE) or Kapton insulation.

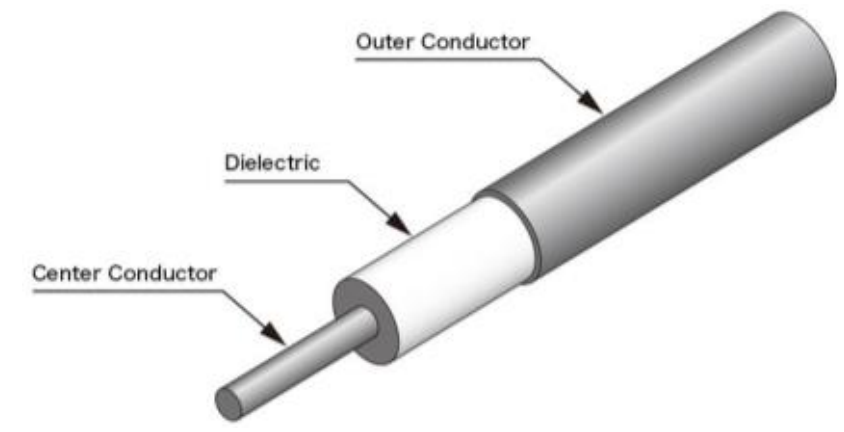

*Figure 510: a typical CoaxCo niobium-titanium cable. Source: CoaxCo.*

Other vendors like Delft Circuits are also proposing superconducting cables but they seem to rely on CoaxCo for the base cables they're then integrating in their own solutions.

Most vendors are now trying to miniaturize this cumbersome cabling, mostly with using flexible cables.

Other avenues are pursued, at the research level, using optical cables and frequency conversion from 5 GHz to 200 THz and the other way around, using phonon-based mechanical resonators cavity optomechanics, optomechanical crystal resonator[1464] or coplanar waveguide and optical cavity using the dark state protocol[1465]. You then have to look at the thermal cost of demultiplexing this signal in the cryostat and the quality of the microwave signal after its dual transduction to and from optical wavelengths. This is an option IBM is investigating for scaling its superconducting computer systems[1466]. Above 4K, the coax cables are using more regular materials like copper alloys.

Microwave qubit control downlink cables are made of various materials including copper-nickel, copper-beryllium or bronze alloys. After passing through the 4K stage, they are replaced by superconducting versions to limit their heat conduction. Between the two, 20 dB attenuators are inserted. In addition, conventional twisted pair cables carrying direct current are used to power the active electronic components integrated in the cryostat, in particular the qubit state readout amplifiers.

This creates significant **wiring clutter**. Figure 511 contains *on the left* a Google cryostat with its bunch of cables and wires connecting the different cold plates. This is the wiring for only 53 qubits (actually, you need to add the 88 coupling qubits). It seems that it is possible to miniaturize some of this, especially with flat ribbon cables. These various cables have another disadvantage: they are very expensive. The unit is several thousand dollars. For today's 53-qubit superconducting quantum computer, this cabling costs more than the entire cryostat, more than half a million dollars. This explains why cryostat manufacturers such as **Bluefors** also offer their own optimized cabling system, such as their 168-cable High-Density Wiring, which appears to be sized to support 56 qubits (*center*). The same is true with the removable cable system of the **Oxford Instruments** Proteox (*right*).

The most expensive cables are the niobium-titanium coax sitting between the 4K and 15 mK stages, up to $3K each. A 1000 qubit QPU could have $10M of cabling in its bill of materials!

[1463] See We'd have more quantum computers if it weren't so hard to find the damn cables, by Martin Giles, January 2019.

[1464] See Cavity optomechanics by Markus Aspelmeyer, Tobias J. Kippenberg, and Florian Marquardt, Review of Modern Physics, 2014 (65 pages), Two-dimensional optomechanical crystal cavity with high quantum cooperativity by Hengjiang Ren, Oskar Painter et al, 2020 (21 pages) and the review paper Mesoscopic physics of nanomechanical systems by Adrian Bachtold et al, February 2022 (87 pages).

[1465] See Proposal for transduction between microwave and optical photons using $^{167}$Er:YSO by Faezeh Kimiaee Asadi et al, University of Calgary, February 2022 (8 pages).

[1466] See Optomechanics with Gallium Phosphide for Quantum Transduction by Paul Seidler, May 2019.

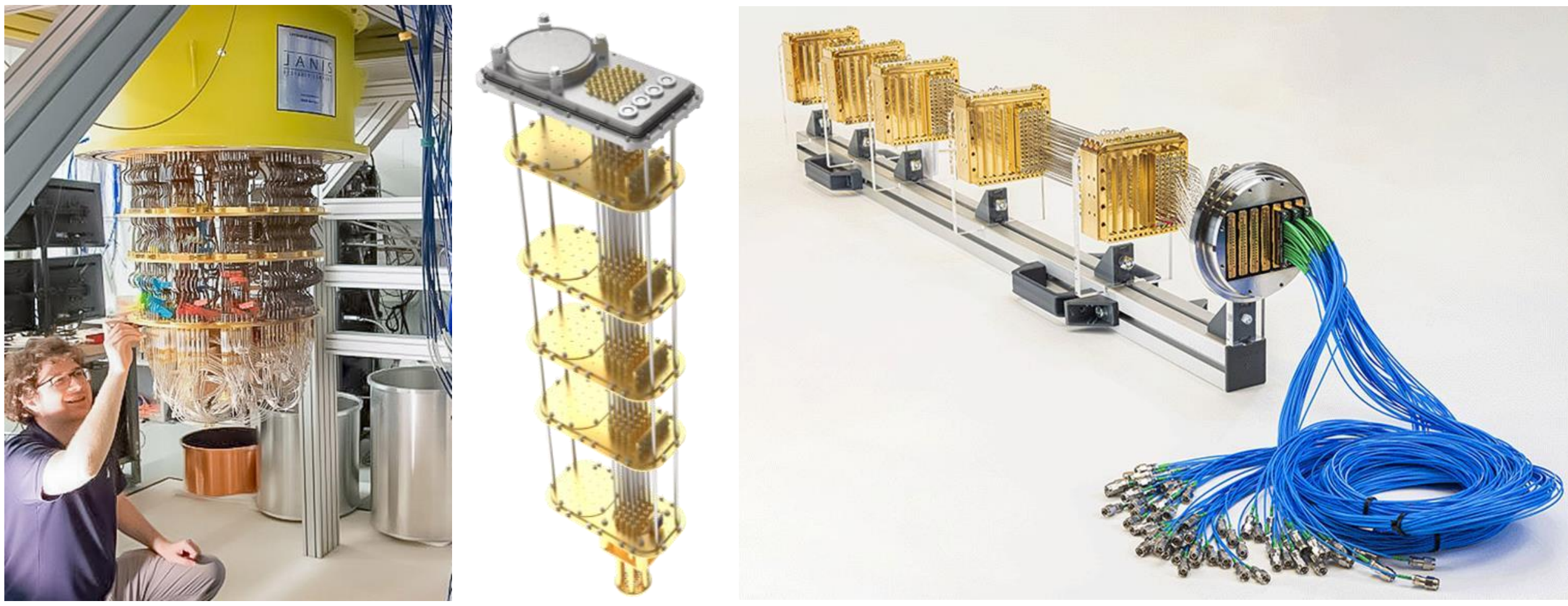

*Figure 511: from left to right, Google Sycamore cable clutter, BlueFors optimized cabling system and Oxford Instrument removable cabling system. Sources: Google, Bluefors, Oxford Instruments.*

Now, onto the vendors in this space…

**Delft Circuits** (2016, The Netherlands) was created by Sal Jua Bosman (CEO), Daan Kuitenbrouwer (COO) and Paulianne Brouwer (CFO), the first two coming from TU Delft.

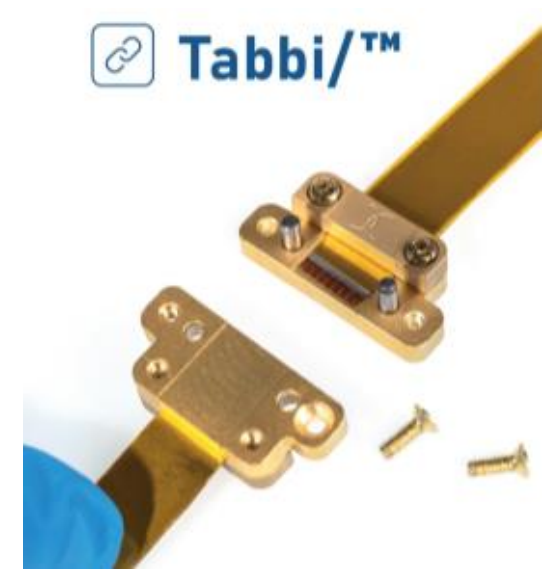

*Figure 512: a Delft Circuit Tabbi flat cable and connector.*

They offer cables and flexible mats used to carry the control microwaves of superconducting qubits such as CF3 (Cri/oFlex) and supporting frequencies ranging from 2 to 40 GHz with 8 embedded cables. Delft Circuit also introduced in March 2022 their Tabbi, an ultra-high density modular flexible microwave interconnect that consolidates the equivalent of 8 SMA cables in about 10 mm and embed its own filters. The startup had 24 people as of early 2022. They manufacture their products out of a 150 $m^2$ lab located at the Delft Quantum Campus. The company got financial support from many EU programs (AVaQus, MATQu and SPROUT).

**QDevil** (2016, Denmark, 1M€) sells filters used in cryostats including the QFilter, based on a collaboration between Harvard University and the University of Copenhagen. It's a cryogenic filter reducing electron temperatures below 100 mK.

They also sell the QDAC, a 24-channel low noise DAC, the QBoard, a PCB-based fast-exchange cryogenic chip carrier system, and the QBox, a 24-channel breakout box. They are partnering with Bluefors. The company was acquired by Quantum Machines (Israel) in March 2022.

**XMA Corporation** (2003, USA) provides the OmniSpectra product line comprising adapters, attenuators, couplers and other passive cryo-electronic components used in quantum computing.

**Radiall** (France) is an industry company specialized in connectors and cabling, very active in the aerospace vertical. They are now addressing quantum technology needs.

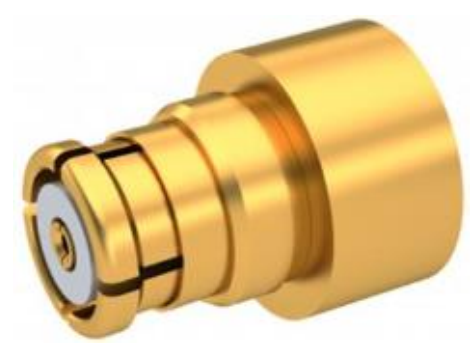

The company creates custom solutions for the quantum industry, including ultra-miniature microwave board-to-board connectors, 3D cabling and cryogenic switches. Radiall is currently expanding its product portfolio with microwave solutions supporting various quantum computing technology that require microwave components meeting strict electromagnetic compatibility/electromagnetic interference (EMC/EMI) constraints, cryogenic, non-magnetic and density specifications.

**CryoCoax** (UK/USA) is a division of **Intelliconnect** (UK/USA) that provides RF interconnect assemblies for various markets including quantum computing, based on niobium-titanium, cupro-nickel, beryllium-copper, stainless steel and brass.

They provide high-density multiway connectors using the SMPM interface (created by Carlisle Interconnect Technologies) that supports a 4.75mm pitch instead of a classical 15 mm pitch with SMA connectors. They also distribute the cabling solutions from Delft Circuits and passive components from OmniSpectra.

**Atem** (1990, France) is a coaxial cables designer and manufacturer. It wants to enter the quantum computers space with its Qryolink project to propose superconducting coaxial cables.

**Rosenberger Group** (Germany) is involved in the German qBriqs projects to build connectors, attenuators components, niobium-titanium and stainless cables assembly to be incorporated in 80 channel multiport connectors and flat cables (with a 1 mm pitch) developed by **Supracon AG** and **LPKF**. They plan to produce these flat cables with 3D printers.

At last, let's mention the IARPA funded **SuperCables** US program that aims to develop high-data rate and low-power transport solutions for cryogenic electronics.

It started in 2019. Its goal is to optical fiber connectivity between room temperature and cryogenic electronics to limit heating. It works on creating electro-optic modulators converting digital signals into and from optical data. It was a 2-year effort targeting a bandwidth of 50 Gbits/s. Which in itself is insufficient for qubits control!

### Other electronics vendors

Let's now look at other commercial vendors in the cryogenic electronics area, given they don't create any cryo-CMOS at this point.

**Atlantic Microwave** (1989, USA) produces and markets radio-frequency and microwave components operating at cryogenic temperatures.

They are used to control superconducting and silicon qubits in cryostats. This includes microwave attenuators, filters, microwave amplifiers and bias tees. It is a subsidiary of the British group ETL Systems, founded in 1984.

**Raditek** (1993, USA) is a designer and provider of RF signals processing systems, including the circulator magnet-based filters used between the first stage microwave amplifier (at 15 mK) and second stage amplifier (at 4K) used usually with superconducting qubits.

**QuinStar Technology** (1993, USA) is a vendor of cryogenic circulators, coming from the acquisition of **Pamtech** (USA) in 2010.

**RF-Lambda** (2003, USA) also provides circulators and low-noise amplifiers.

**CryoHEMT** (2019, France) is a company created by Quan Dong and Yong Jin in Orsay, France. It designs and manufactures low-noise HEMT microwave amplifiers which amplifies microwaves at the 4.2K cryostat stages.

Their technology is based on the PhD thesis from Quan Dong done under the supervising of Yong Jin in France in 2013. It seems not being used in quantum information systems.

**Diramics** (2016, Switzerland) is a spin-off from ETH Zurich creating ultra-low noise transistors in III-V materials (InP, indium-phosphorus) with a technology named pHEMT (pseudomorphic high-electron-mobility transistor, using junctions with two semiconductors with different band gaps).

It can be used in low temperature electronics. It is currently mostly used in astronomy applications.

**Marki Microwave** (1991, USA) is a supplier of microwave control components: amplifiers, bias tees, couplers, mixers and filters.

**Quantum Microwave** (2016, USA) creates microwave components operating at cryogenic temperatures for quantum computers, including JPA amplifiers, attenuators, frequency couplers, multiplexers, bias tees, diplexers, filters, image reject mixers and directional couplers.

One main applied research domain for **Raytheon** is related to superconducting qubits controls. They work on arbitrary pulse sequencers (APS) creating superconducting qubits control microwaves, an FPGA readout system using low noise parametric amplifiers and a custom made three-way mixing mode JPAs with 20 dB gain as show in Figure 513.

They are also exploring SFQ based control logic (Josephson gate-based logic) and spintronics based low-power memories. They are mainly found in superconducting qubits quantum computers (IBM, Google, Rigetti, D-Wave). However, they do not push forward the miniaturization of these components like what SeeQC is doing. In October 2021, they announce a technology partnership with IBM, with not many details[1467]. On top of that, they also develop Josephson junction based infrared photon detectors[1468].

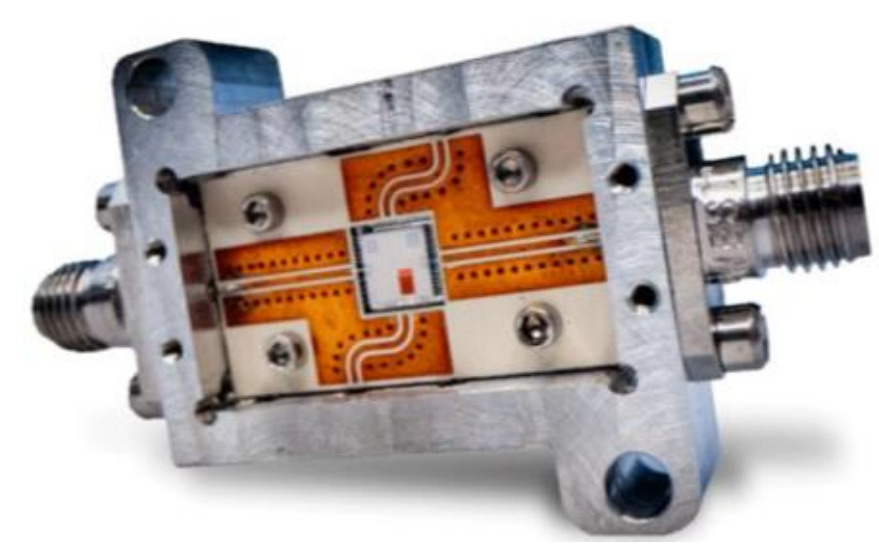

Figure 513: a Raytheon BBN JPA.

**Wenteq Microwave Corp** (2006, USA) provides low-noise amplifiers, attenuators, circulators and coaxial connectors, in the RF/microwave range.

**Holzworth Instrumentation** (2004, USA) is a provider of multi-channel RF sources and AWGs. They also provide phase noise analyzers. The company was acquired by Wireless Telecom in 2019.

**apitech** (1999, USA) is a provider of cryo-attenuators targeting the quantum computing market and covering signals from DC to 40 GHz, and with SMA, and SMPM connectors.

[1467] See Raytheon, IBM partner for quantum in defense, aerospace by Nicole Hemsoth, in TheNextPlatform, October 2021.

[1468] See Josephson junction infrared single-photon detector by Evan D. Walsh et al, April 2021 (12 pages).

**AnaPico** (2005, Switzerland) is a provider of low phase noise RF signals generators, for the local oscillators used in qubits control and readout, running up to 6,1 GHz.

**Scalinq** (2022, Sweden) was created by a team coming from Chalmers University who wants to help superconducting quantum computers companies design larger QPUs.

It designs a QPU chipset sample holder named LINQER, with currently 16, 36 and 80 connectors, and the potential to scale up to 300 connectors. It provides low crosstalk, an innovative magnetic shielding technology, and supports chipset sizes up to 20×20 mm$^2$. The company was created by Zaid Saeed (CEO), Lisa Rooth (VP) and Robert Rehammar (CTO).

## Thermometers

It is possible to measure **pressure** (ambient, gas), **temperature** (everywhere) and **flow** (of gas) at cryogenic temperature. Specific sensors are therefore installed for this purpose in the cryostat enclosure, attached to different places in the "candlestick".

Temperatures are measured with cryogenic thermometers! These are found in particular at **Lake Shore Cryotronics** with its Cernox thermometers which go down to 100 mK and resist well to the ambient magnetic field and its ruthenium oxide thermometers which go down to 10 mK. At less than 20 mK, noise thermometers using Josephson junctions are used (and the loop is closed...). Some thermometers are placed on the plates opposite the heat exchange tubes and the mixing box. Still, progresses need to be done even in this area, noticeably to measure precise temperatures in the 10 mK range and with no delay[1469].

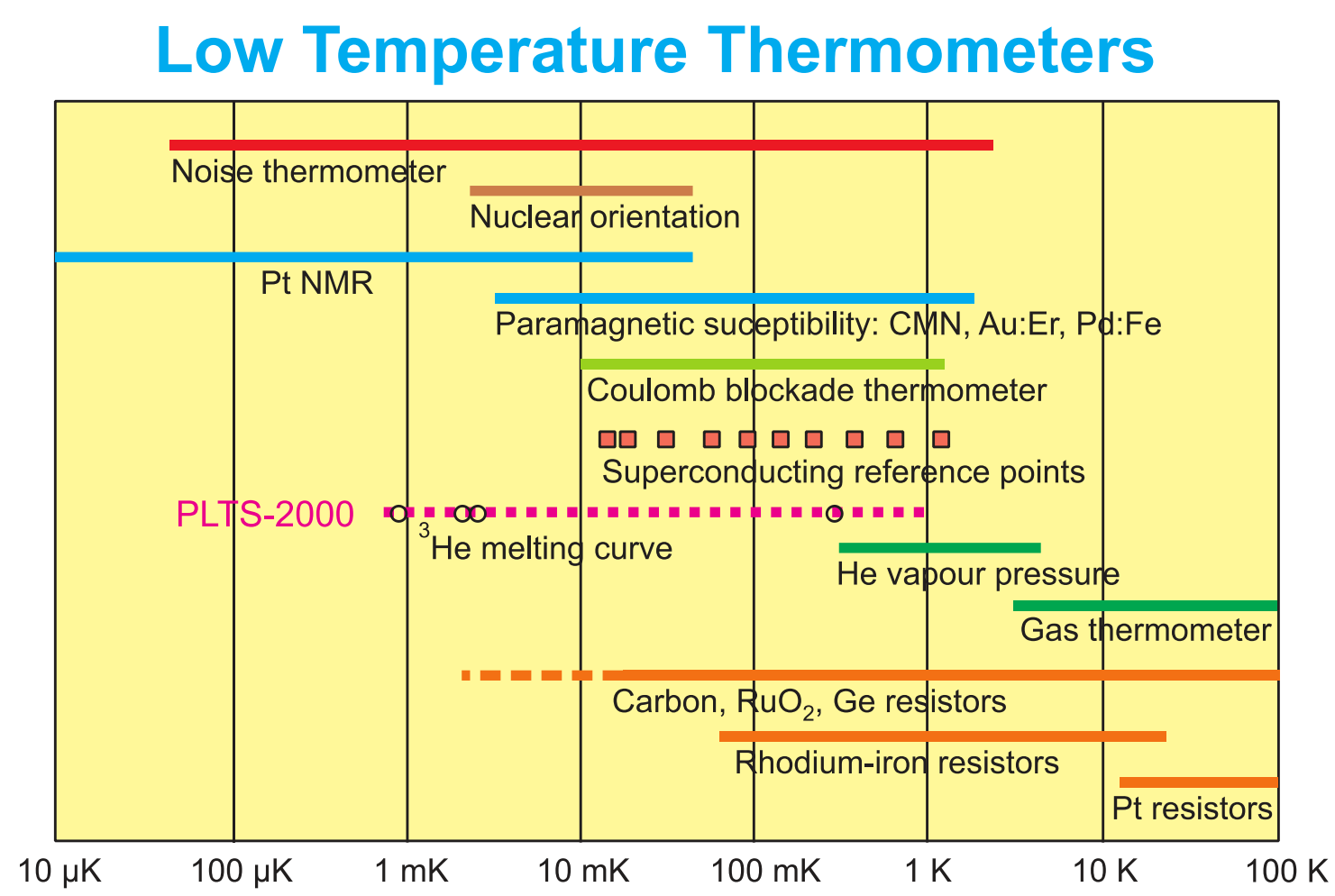

*Figure 514: categories of low-temperature thermometers. Source: [Thermometry at low temperature] by Alexander Kirste, 2014 (31 slides). We can see that there are about ten types of thermometers that go down to less than 1K. The most commonly used one exploits the Coulomb block based on tunnel junction. The electrical voltage of the junction varies linearly with the cryogenic temperature.*

## Vacuum

Besides photon qubits, most other qubit types require some form of vacuum to isolate the qubits from their environment. We usually make a distinction between different levels of vacuum. The most stringent ones are used with trapped ions and cold atoms which require ultra-high vacuum (UHV) conditions whereas solid-state qubits like superconducting and electron spins are less demanding. UHV starts at $10^{-9}$ mbar.

---

1469 That's what researchers from Chalmers in Sweden achieved in 2020. See Primary Thermometry of Propagating Microwaves in the Quantum Regime by Marco Scigliuzzo, Andreas Wallraff et al, December 2020 (14 pages).

One problem to avoid when creating vacuum is outgassing. It manifests with particles being ejected from the internal enclosure surfaces and materials, including residual water coming from the air. The phenomenon is avoided with carefully selecting the materials.

Cold atoms qubits require a pressure of $10^{-10}$ mbar while trapped ions goes down to $10^{-12}$ mbar. In both cases, low pressures and outgassing are obtained with heating the system enclosure above 200°C for several hours while the vacuum pumps are operating. This "bake-out" process removes water and other trace gases sitting on the chamber surface. Heating is done with heater stripes placed around the chamber. The chamber exterior can also be cooled with liquid nitrogen to contain any further gassing.

There are many vacuum and ultra-high-vacuum systems vendors. The most commonly seen in research labs come from **Pfeiffer** (Germany). Some pumps must be cooled at low temperature, like the 4K pump used by Pasqal to cool their atoms.

At last, measuring pressure in vacuum is also a challenge. Classical mechanical pressure measurement is of no use in the UHV to XUV (extreme ultra-vacuum) ranges covering $1 \times 10^{-6}$ to $1 \times 10^{-10}$ Pa. The NIST in the USA is proposing a solution applicable to cold atoms to cover these ranges of pressures[1470]. A dedicated part on quantum pressure measurement is located in page 892.

## Lasers

Masers and lasers are applications of three successive discoveries and inventions:

- **Fabry-Pérot resonant cavities**, named after Charles Fabry[1471] (1867-1945) and Alfred Pérot (1963-1925). Their system invented in 1898 was originally used to create an interferometer.
- **Stimulated emission**, formalized by Albert Einstein in 1917. It occurs when an excited atom receives a photon of energy equivalent to a transition between two energy levels. It then re-emits two photons identical to the received one and the energy level of the atom is reduced to its ground state.
- **Optical pumping**, invented by Alfred Kastler in 1949 at ENS in France, which earned him the 1966 Nobel Prize in Physics.

It generates a population inversion, creating a high proportion of atoms excited at level $E_2$ in the diagram below compared to level $E_1$. Optical pumping often excites atoms to energy levels higher than $E_2$ in Figure 515, with a non-radiative transition from these levels to the $E_2$ level and then from the $E_1$ level to the fundamental level of the $E_0$ atom. If pumping was performed only between levels $E_1$ and $E_2$, their proportion would balance and the laser effect could not be triggered. Three-level pumping is used with pulse lasers and four-level pumping with continuous lasers.

A laser is based on a resonant cavity filled with a gain or amplifier medium. The pumping of this gain medium is optical, electrical or chemical. Once at the high energy level ($E_2$ in Figure 515), the atom drops to the $E_1$ energy level either spontaneously or stimulated.

[1470] See Development of a new UHV/XHV pressure standard (Cold Atom Vacuum Standard) by Julia Scherschligt et al, 2018 (15 pages).

[1471] We owe to Charles Fabry the creation of the Institut d'Optique, of which he was the first director in 1926 of the engineering school that was originally called SupOptique or Ecole Supérieure d'Optique.

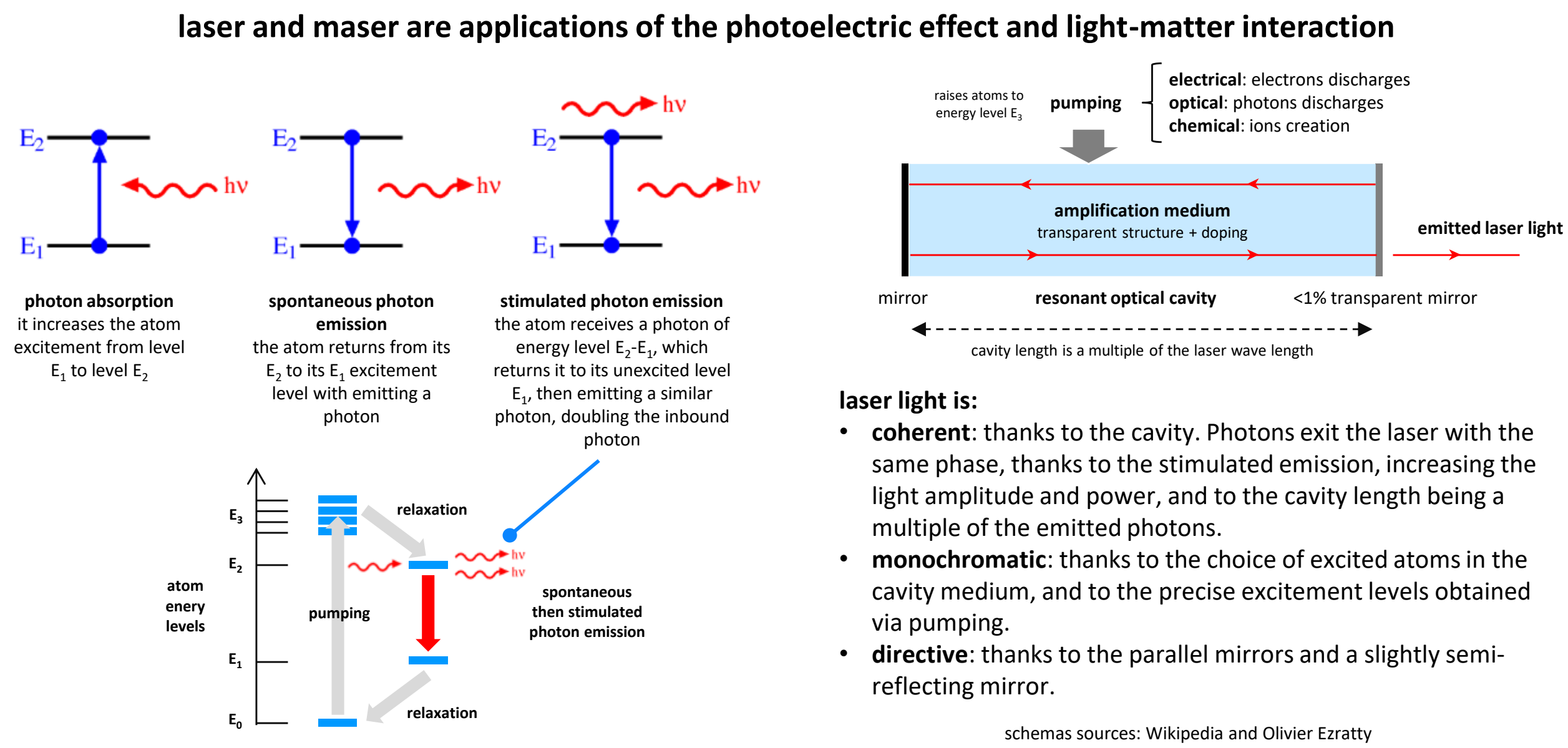

*Figure 515: how lasers work. (cc) Olivier Ezratty, 2021.*

The mechanism can be self-sustained since the spontaneously emitting photons then generate the stimulated emission of identical twin photons in frequency, phase and amplitude.

The stimulated emission is sustained by placing the atoms in a transparent cavity, filled with solid, liquid or gas, and parallel mirrors trapping the photons. One of the mirrors is slightly semi-reflective, allowing some of the amplified light to exit the laser.

This system of mirrors plays the role of a resonator. It reflects off-axis and thus undesirable photons out of the laser and the wanted on-axis photons back into the excited population where they can continue to be amplified thanks to the laser pumping.

The light resulting from this process is **directive** (thanks to the resonator and its parallel mirrors), **monochromatic** (thanks to the choice of excited atoms and the fineness of the cavity) and **coherent** (the photons are in phase and with the same wavelength/frequency thanks to the stimulated emission and the length of the cavity being a multiple of the laser wavelength). The laser photons frequency depends on the materials used in the cavity and the optical length of the cavity. As an order of magnitude, a 1mW red laser emits $3x10^{15}$ photons per second.

Lasers (light amplification by stimulated emission of radiation) appeared conceptually in 1958 in an article by Arthur Leonard Schawlow and Charles Hard Townes. The first **gas laser** was created in 1960 by Theodore Maiman, using helium-neon. **Excimer**-based gas lasers cover ultraviolet.

We then had successively **doped crystal lasers** (also named solid-state lasers, such as ruby which is $Al_2O_3$ doped with $Cr^{3+}$, or YAG, Yttrium garnet and Aluminum $Y^{33}$+$Al^{53}$+$O_{12}^{2}$-), **chemical** lasers (covering the infrared spectrum), **semiconductor diode lasers** (the most common today, usually based on gallium arsenide, or GaAs), **fiber lasers** (using rare earth elements like neodymium, erbium and thulium, mainly used in optical communications), and finally, **free electron lasers**, which we already briefly covered in relativistic quantum mechanics section.

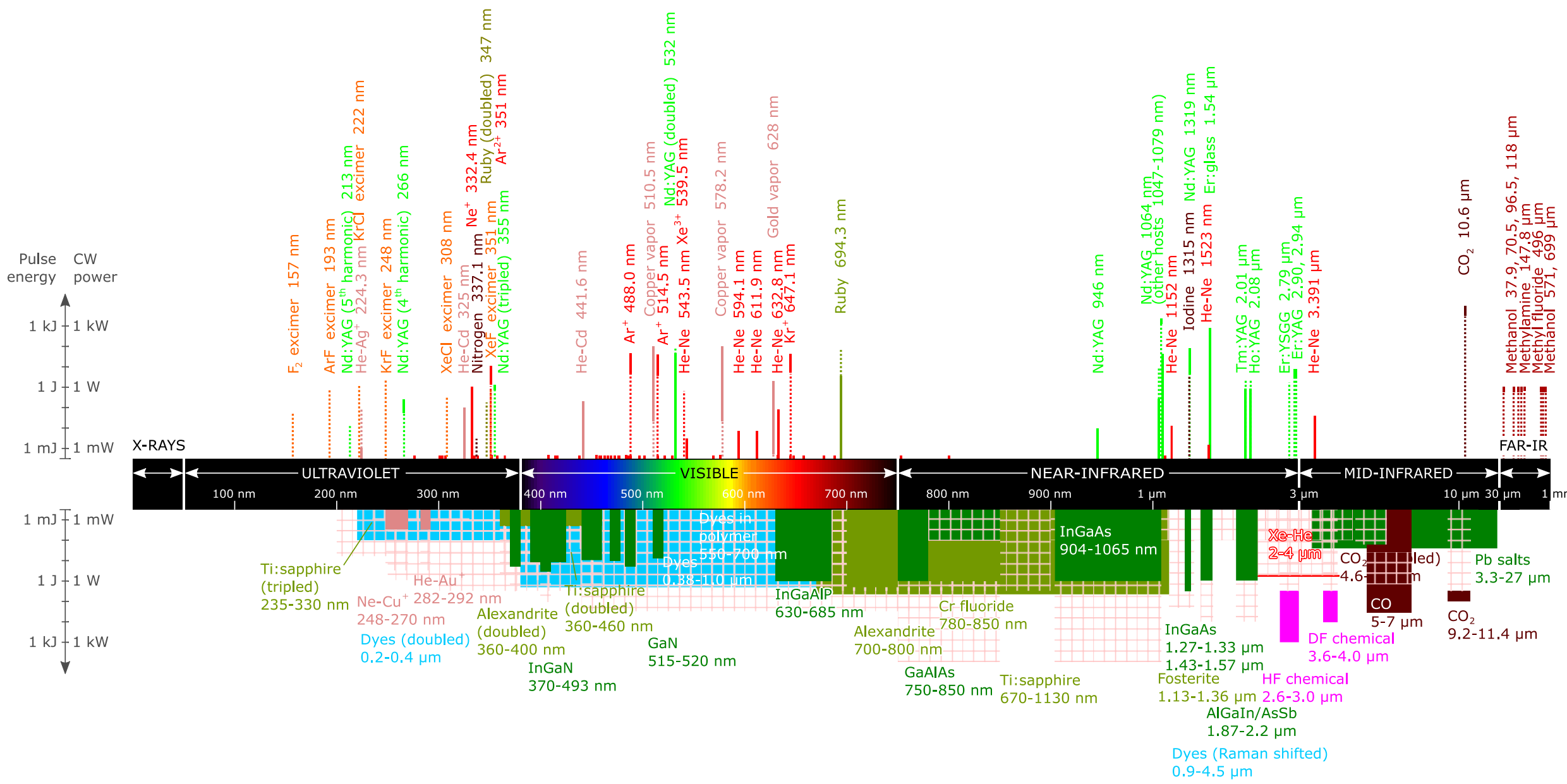

*Figure 516: the great variety of lasers covering the electromagnetic spectrum from ultraviolet to mid-infrared waves. Source: Wikipedia.*

Lasers operate either in pulses or continuously. The first mode is used to create very high-power levels. To create very powerful lasers, laser amplifiers are created which consists of chains of lasers with a primer laser that is connected to a series of lasers that successively amplify the light generated by the previous laser.

Lasers applications are quite various: industrial diamond drilling and cutting (1965), barcode readers (1974), laser printing (1981), office scanners, Laserdisc (1978), audio CDs (1982), DVDs (1995), surgery, particularly in ophthalmology (glaucoma, retinal detachment, refractive surgery), cosmetic surgery, in dermatology, for tattoo and hair removal, telecommunications, laser pointers, depth sensors, focus sensors for smartphones, iPhone FaceID sensor, measurement and alignment in construction, all sorts of LiDARs, stereolithography 3D printing, confocal microscopy (very shallow depth images), flow cytometry (cell counting), DNA chips analysis, video projector light sources, velocity measurement, the stripping of certain materials, various weapons, nuclear fusion, telescopes adaptive optics, atoms cooling, quantum telecommunications, quantum cryptography and finally, quantum computing, and on and on and on. In short, lasers are everywhere!

The frequency ranges covered by lasers range from infrared to ultraviolet. There are even types of lasers with adjustable frequency. Free electron lasers go as far as X-rays. Gamma-rays lasers - or grasers - do not yet exist.

In quantum technologies, the most commonly used laser wavelengths are 775 nm (beginning of the near infrared region next to red) and 1550 nm (middle of the near infrared region). The first one is used for quantum computing thanks to efficient photon generation and single photon detection (particularly with APD, avalanche photo diodes). The second is used in optical fiber for long distance communications, data transmission and QKD systems.

There are many solutions to up and down convert photons from/to these two wavelengths. For example, these conversions are mandatory when connecting several photon-based quantum computers through a fiber optic link. Solid-state qubits require another type of conversion, mostly from microwaves to 1550 nm infrared photons, given the conversion must convert the quantum information in the solid-state qubit to some encoding in the resulting photons, like their polarization.

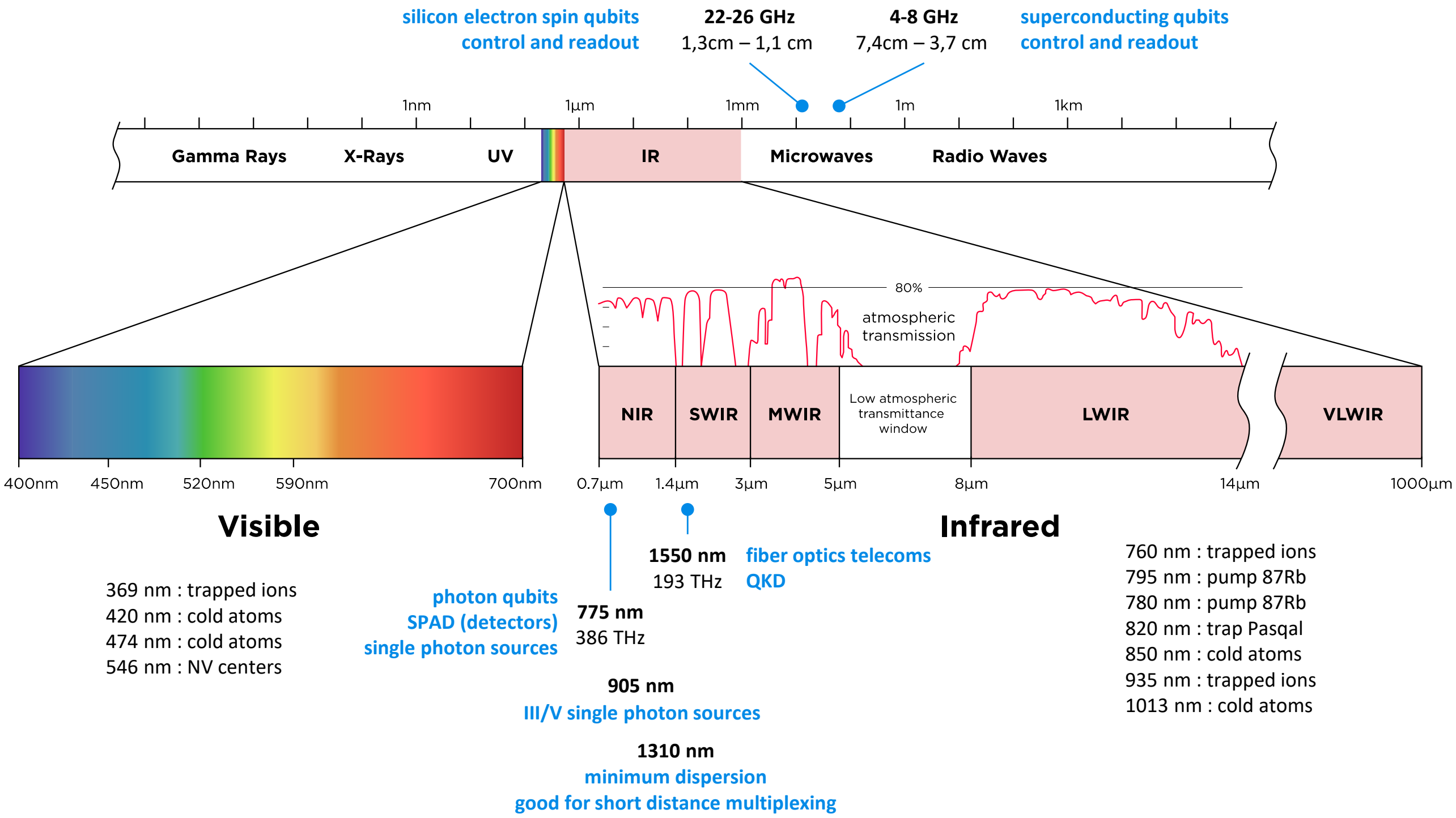

*Figure 517: lasers used in the visible and infrared spectrum. Source: http://www.infinitioptics.com/technology/multi-sensor/.*

Another breed worth mentioning are femtoseconds lasers, which create short pulses of coherent light in the range from the femtosecond ($10^{-15}$s) to the picosecond ($10^{-12}$s). They are use in micro-machining and various other tasks, including quantum sensing in relation with frequency combs[1472].

The **Maser** (1953) or "Microwave Amplification by Stimulated Emission of Radiation" was invented before the laser, in 1953, by Nikolay Basov, Alexander Prokhorov and Charles Hard Townes, who were awarded the Nobel Prize in Physics in 1964.

It is the equivalent of the laser, but emits microwaves instead of visible light. The first masers were made with ammonia and generated 24 GHz microwave photons. Hydrogen Masers followed in 1960.

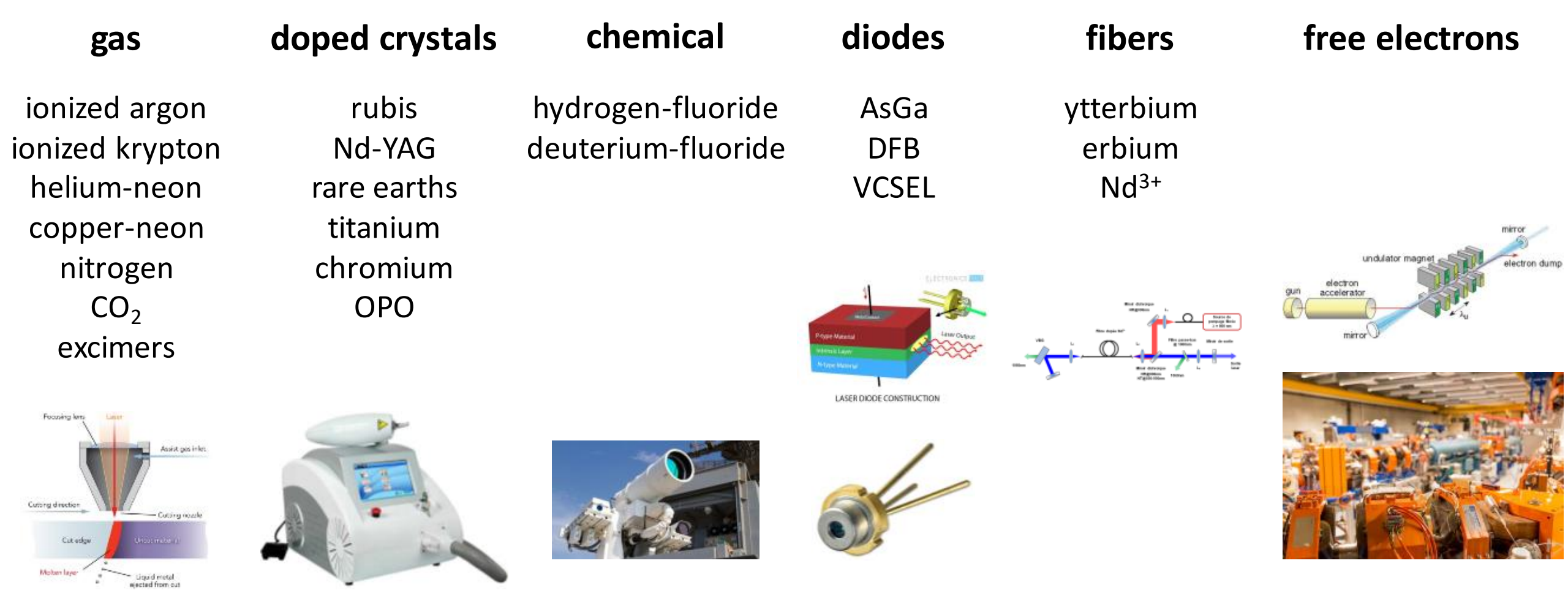

*Figure 518: the various types of lasers and their cavity materials. (cc) Olivier Ezratty, 2021.*

[1472] See 20 years of developments in optical frequency comb technology and applications by Tara Fortier and Esther Baumann, NIST, 2019 (16 pages).

There are many laser vendors who play a role in second revolution quantum technologies, both with photon qubits, quantum telecommunications, quantum cryptography and quantum sensing. Lasers are also used to control cold atom and trapped ions qubits.

**Toptica Photonics** (1998, Germany) is a photonics equipment manufacturer developing laser sources covering a wide range of frequencies from 190nm (UV) to Terahertz waves, including laser diodes and frequency combs. Their lasers can be used to control trapped ions and cold atoms. They employ over 320 people for a revenue of $82M and are a worldwide leader in their market.

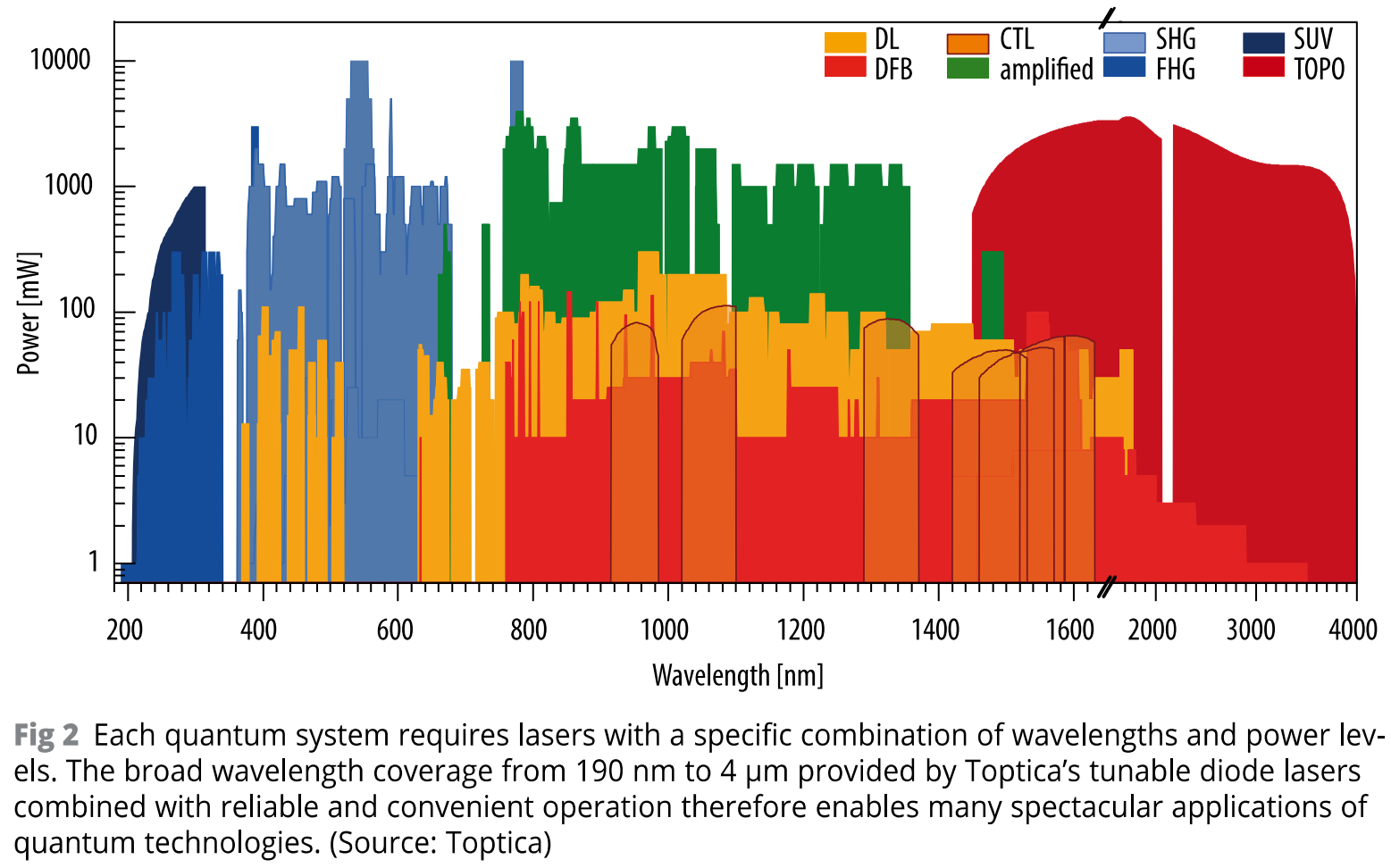

*Figure 519: wavelengths coverage of Toptica lasers. Source: The Control of Quantum States with Lasers in Photonics View, 2019 (3 pages).*

Their flagship product, the Chromacity OPO, has a tunable optical parametric oscillator that covers near-IR and mid-IR wavelengths. Some of their lasers can create entangled photons.

**Stable Laser Systems** (2009, USA) offers Fabry-Perot lasers and cavities that can be used for cold atom confinement.

The startup launched by Mark Notcutt is based in Boulder, Colorado, one of the nerve centers of quantum technologies in the USA, near NIST and the University of Colorado. His team also includes Jan Hall, winner of the 2005 Nobel Prize in Physics for the discovery of the effect that bears his name.

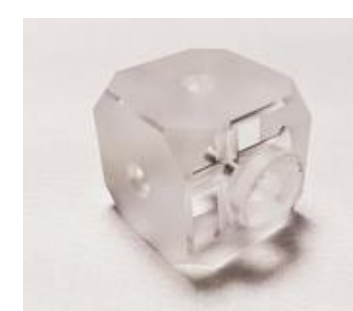

**Lumibird** (1970, France) is a supplier of lasers. Formerly Quantel and Keopsys, it is a large SME with more than 800 employees and a turnover of 110 M€, 80% of which is exported.

**Chromacity** (2013, UK) is a manufacturer of lasers targeting various industry and research needs, including quantum communications.

**DenseLight Semiconductors** (2000, Singapore) manufactures various laser products.

**FemTum** (2016, Canada) creates mid-infrared lasers which can be used in quantum optics and silicon photonics applications and quantum sensing using optical frequency combs. It is a spin-off from the Center of optics, photonics and laser (COPL) in Quebec City.

**FocusLight Technologies** (2007, China) produces laser diodes and laser optics components.

**Freedom Photonics** (2005, USA) manufactures lasers and photodiodes using InP and GaAs semiconductor, SiGe-based photonics and planar lightwave waveguides. It was acquired by **Luminar** in March 2022.

**GLOphotonics** (2011, France) sells hollow-core photonic crystal fiber (HC-PCF) and their functionalized form Photonic Microcells (PMC). These lasers use a proprietary fiber technology and gas photonics. They are partnering with CNRS XLIM lab in Limoges.

**iPronics** (2019, Spain) develops general-purpose integrated programmable photonic systems, where optical hardware complements software to perform multiple functions.

**Q.ANT** (2018, Germany) develops a green laser optimized for NV centers quantum sensors. It is a subsidiary of the TRUMPF Group. It also manufactures powerful lasers used in ASML lithography machines and light channels on silicon for qubits transport.

They are also working on photonic-based quantum computing and leader of a relate German consortium with 50M€ funding including 42M€ public funding from the German government.

**Silentsys** (2021, France) created a closed-loop voltage regulation electronic system that complements lasers and their servo controller to reduce the emitted laser noise and improve its frequency precision.

**UnikLasers** (2013, UK, £4.1M) sells ultra-narrow linewidth, high power lasers at the specific wavelengths related to the exact atomic transitions targeted for quantum sensing applications.

It can transform a MHz linewidth laser into an Hz linewidth laser. It is currently adapted to continuous lasers running in the 1550 nm and 1050 nm wavelengths and fits in a 2U rack system. Their OFD system (optical frequency discriminator) delivers a continuous voltage signal driving the laser diodes that is proportional to the frequency fluctuations of the input laser beam. The technology core is optical, using frequency combs. They also propose low power voltage power systems. One of their OFD can drive two lasers.

**Vexlum** (2017, Finland) produces high-power narrow-linewidth vertical external-cavity surface-emitting lasers (VECSELs) including blue and UV lasers, used among other things, to control trapped beryllium ions.

It is a spinoff from Tampere University of Technology Optoelectronics Research Centre (ORC).

And also: **Spectra Physics** (1961,USA), **Altitun** (1997, Sweden, $10M), **Calmar Laser** (1996, USA) and **Ampliconyx** (2016, Finland who manufactures short-pulses lasers), **Active Fiber Systems** (2009, Germany) which creates femtoseconds fiber lasers and is a spin-off from Fraunhofer IOF, **InnoLume** (2002, Germany, $26.8M, which sells laser diodes), **FISBA** (1957, USA) which develops multi-wavelengths lasers, **Intense Photonics** (1994, USA, $51M, which develops single and multi-mode monolithic laser array products, and high power laser diodes and was acquired by Orix Group), **Lytid** (2015, France) which manufactures terahertz cascade lasers, **Spark Lasers** (2015, France) and their picosecond and femtosecond lasers, **Amplitude Laser Group** (2001, France) and their femtosecond lasers, **neoLASE** (2007, Germany) a supplier of various laser products including laser amplifiers, **Alpes Lasers** (1994, Switzerland) which sell infrared quantum cascade lasers, **Luna Innovations** (1990, USA, $13.1M), **Vector Photonics** (2020, UK, £1.6M) is a spin-off from the University of Glasgow which develops semiconductor lasers based on PCSELs (Photonic Crystal Surface Emitting Lasers) and **OEwaves** (2000, USA, $15M) provides lasers, oscillators and optical/RF tests and measurement systems.

## Photonics

Let's now look at other photonics equipment manufacturers. They sometimes also manufacture lasers but even more.

**Accelink** (1976, China) sells optoelectronic components, including fiber optics modulation and demodulation systems, lasers and $SiO_2$/Si material plane optical waveguides. They probably play a role in the deployment of quantum telecommunication networks.

**Aurea Technology** (2010, France) is a photonics equipment manufacturer targeting various markets including quantum communications (QKD) and quantum sensing.

It sells twin photon sources (TPS), time correlated single photon detectors (Picoxea), ultra-low-noise NIR single-photon counting detection modules (SPD_A and SPD_OEM_NIR[1473]), time correlators (Chronoxea) and high-resolution fiber sensors (q-OTDR) and picosecond pulse lasers (Pixea). They also developed Fluoxea, a fluorescence lifetime imaging mapping system using time-correlated single photon counting that can be used to characterize semiconductors, qualify quantum dots or measure local magnetic fields (with the help of NV centers). The company has its own assembly plant in Besancon, France. It's a spin-off company from the optics department of FEMTO-ST, a public research lab based as well in Besancon.

**Azurlight Systems** (2010, France) develops high-power laser amplification systems in the visible and near-infrared spectra (from 488 nm to 1065 nm) using ytterbium-based fibers with low thermal dissipation. These can be used for atoms cooling and trapping.

**Qontrol Systems** (2016, UK) develops photonics components including photonics device status readout modules and backplanes (boards) on which several of these modules can be installed. These modules drive photonics devices via a 12V voltage and read signals with 18-bit accuracy. This is control electronics.

**Cailabs** (2013, France, €16.7M) is a company based in Rennes, France, which is a spin-off from the LKB of ENS Paris and markets photonics equipment and in particular spatial multimode multiplexing systems for optical fibers supporting up to 45 nodes.

This is what makes it possible to multiply the speed of the optical fibers of the telecom operators' networks. In particular, they have KDDI (Japan) as a customer. The startup is managed by Jean-François Morizur (CEO) and Guillaume Labroille (CTO) with Nicolas Treps from LKB being their scientific advisor.

**Quandela** (2017, France, €35M) is a startup specialized in the generation of indistinguishable photons with quantum dot fed by a laser. They target research, telecommunications and quantum computing use cases.

Quandela's team is composed of Valérian Giesz (CEO), an engineer from the Institut d'Optique with a PhD in photonics[1474], Niccolo Somaschi (CTO), PhD from the University of Southampton and Pascale Senellart (CSO), CNRS research director at C2N from CNRS and Université Paris-Saclay.

[1473] The SPD_OEM_NIR is using an InGaAs single photon avalanche diode (SPAD in the 900 to 1700 nm wavelength range with very low dark count rate noise (DCR<700 Hz). It is cooled with the Peltier effect. The supported wavelengths make is suitable to photon counting in telecom wavelengths based QKDs (1,550 nm). It also fits into a standard datacenter rack in a 2U package.

[1474] See his thesis in Cavity-enhanced Photon-Photon Interactions With Bright Quantum Dot Sources by Lavérian Giesz, 2016 (228 pages) where he describes his work in Pascale Senellart's team and the various evolutions of their quantum dot photon source that led in 2017 to the creation of Quandela.

It had a staff of about 50 people in Summer 2022 and several international customers, mainly in Europe, Russia and Asia. Their team also includes Shane Mansfield, who works on theory, algorithms and software.

With a trapped artificial atom comprised of a couple thousand atoms forced to emit periodically single photons in a given direction by laser-activated cavity quantum electrodynamics, they are able to generate photon streams that are well separated in time and with stable quantum characteristics, with wavelengths from 924 nm to 928 nm in the near infrared, this range being progressively extended[1475]. This creates a very bright photon source that can then be multiplied to create indistinguishable photons used in quantum photon processors and various other applications such as quantum telecommunications and quantum key distribution[1476].

They are developing single photon sources running at telecom wavelength as part of the project Paris-RegionQCI, a regional project, led by Orange, with an end goal to deploy a QDK-fibered link between Paris and the Paris-Saclay University. They also work on the generation of cluster states of entangled photons as part of the European Union FET Open Qluster project (2019-2023).

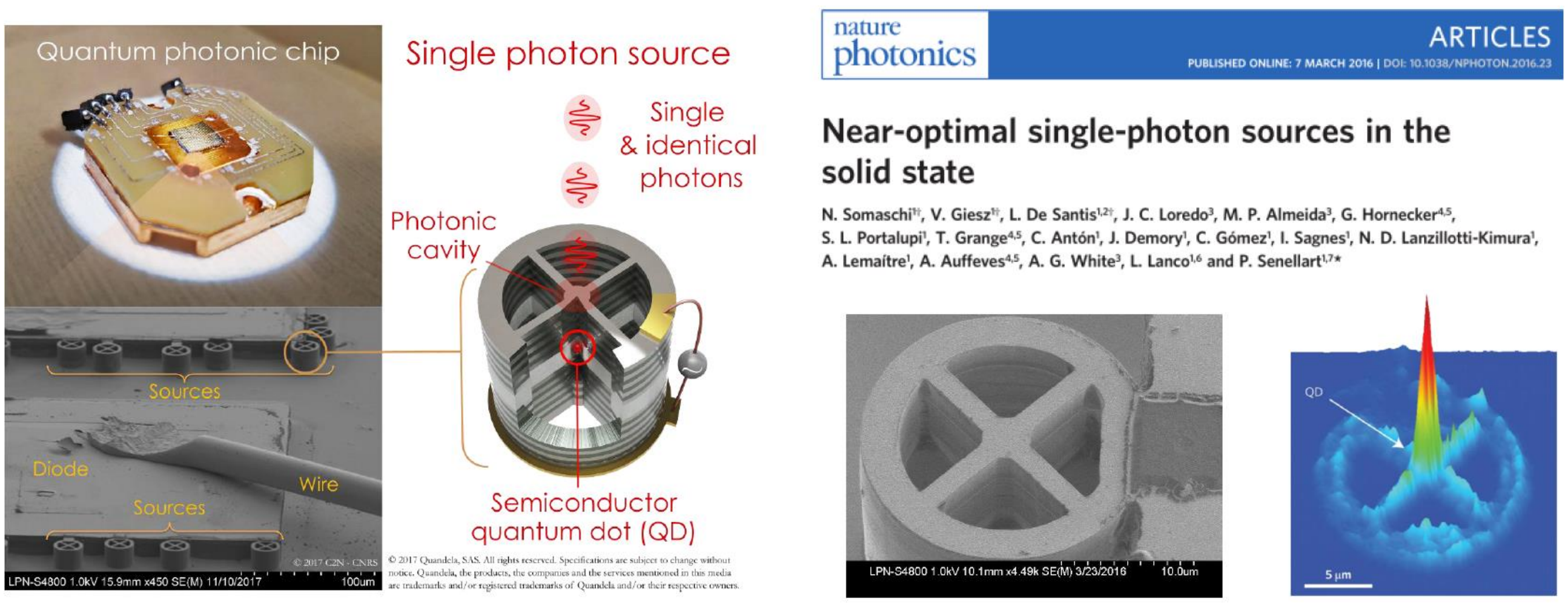

*Figure 520: Quandela's quantum dots single photon source. Source: Near-optimal single-photon sources in the solid-state by Niccolo Somaschi, Valerian Giesz, Pascale Senellart et al, 2015 (23 pages).*

The photon source must be cooled down to a temperature range of 5K-10K, which is achievable with compact cryostats costing only a few thousand Euros and using helium 4, such as the attoDRY800 from **Attocube** (Germany). These cryostats use a simple pulsed tube, reminiscent of the first cooling stage of the dry dilution cryostats used with superconducting and quantum dot spin qubits. These single photons are particularly indicated to allow the creation of quality quantum qubit gates and for measurement-based quantum computing (MBQC, which we cover starting page 450).

[1475] See Near optimal single-photon sources in the solid state, Niccolo Somaschi, Valerian Giesz, Pascale Senellart et al, 2016 (23 pages). The quantum dot is made with InGaAs (indium, gallium, arsenide) and is surrounded by stacked Bragg-reflectors made respectively with GaAs and $Al_{0.9}Ga_{0.1}As$ (aluminum, gallium, arsenide). Pascale Senellart describes in detail how Quandela's photon generators are made in her talk Quantum optics with artificial atoms in a Rochester Lecture in June 2018 (1h10mn). The prestigious Rochester Lectures are held once a year in Durham, UK. The 2017 edition welcomed Peter Knight and the 2012 edition Alain Aspect.

[1476] The process was improved in Pascale Senellart's laboratory in 2020 to generate even brighter and purer photon sources from a spectral and polarization point of view thanks to quantum dot excitation with phonons. See Efficient Source of Indistinguishable Single-Photons based on Phonon-Assisted Excitation by S. E. Thomas, Pascale Senellart et al, July 2020 (10 pages).

Quandela's photon generator was previously offered as a combination of two packaged products:

- The **Qubit Control Single Unit** which allows complete filtering of the single photons emitted by the sources in an Attocube cryostat of the laser used for quantum dot excitation. It is mainly composed of filters tuned to the energy of the optical transition of the emitter.

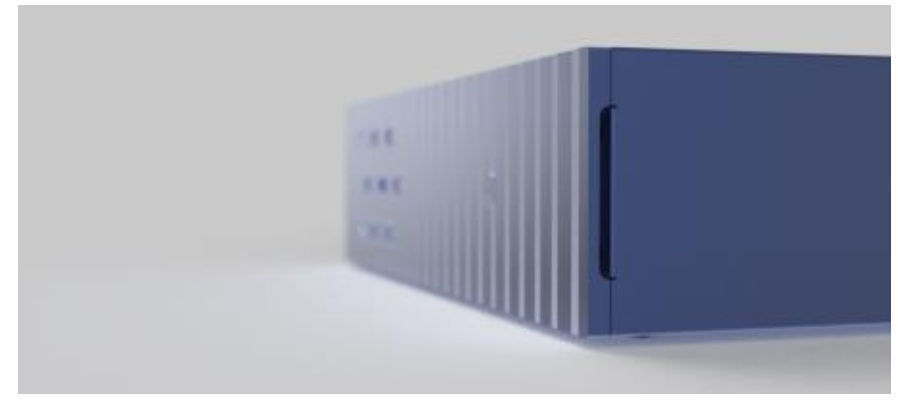

Figure 521: Quandela Control Single Unit.

- The **QShaper** is a more compact device that generates femto/pico-second laser pulses on an optical fiber that will then feed the QCSU quantum dot above. It is powered at the input by the customer's laser. It is used to prepare the laser beam with the right spatial and temporal shape. It is a device made up of various filters. It is calibrated to supply the semiconductor sources.

Quandela launched in 2020 a compact and integrated version of this whole set, fitting into a datacenter rack named Prometheus. The fiber is glued to the photon source, which will eliminate the mechanical part of the calibration. The pulsed head of the 4K cryostat is also integrated in this 3U rack, the compressor being outside and water-cooled at first. Eventually, it will be integrated in the rack and cooled by air.

The rack was designed by **Pentagram**, the same British designer that IBM used for the Q System One launched in January 2019. It is 1.75m high and 80cm wide. It stacks all the elements: the QShaper, the new QCF and a control computer with its keyboard. The whole thing consumes about 5 to 6 kW, the bulk of it coming from the cryostat.

Quandela and the C2N laboratory collaborate with research labs around the world to create advanced photonics platforms. In 2020, they published with a team from the **Hebrew University of Jerusalem** a paper on the creation of a photon cluster state for quantum computing (Figure 522). The idea is to use single photons and to entangle them with each other via a delay line, and inject them into a computing circuit based using cluster states and MBQC (measurement based quantum computing) method.

In Europe, they collaborate mainly with **Fabio Sciarrino**'s team in Italy, in the Netherlands with **QuiX**, in Spain with **INL** and other teams in Austria, the United Kingdom, Slovakia and Israel. They are part of the European FET project PHOQUSING for boson sampling led by Fabio Sciarrino's team.

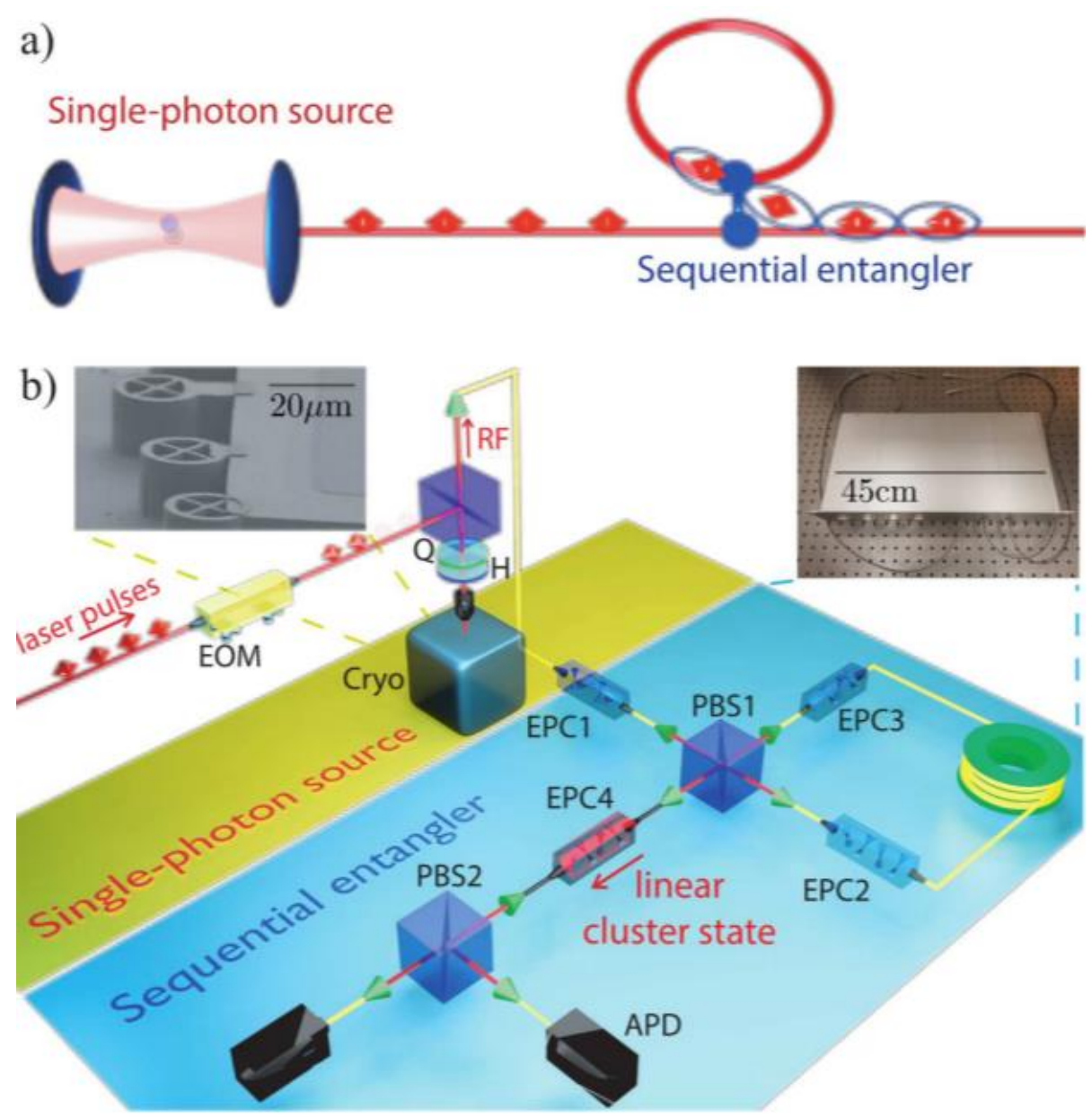

Figure 522: creation of cluster state photons with a serial entangler using a delay line. Source: [Sequential generation of linear cluster states from a single photon emitter](#) by D. Istrati et al, 2020 (14 pages).

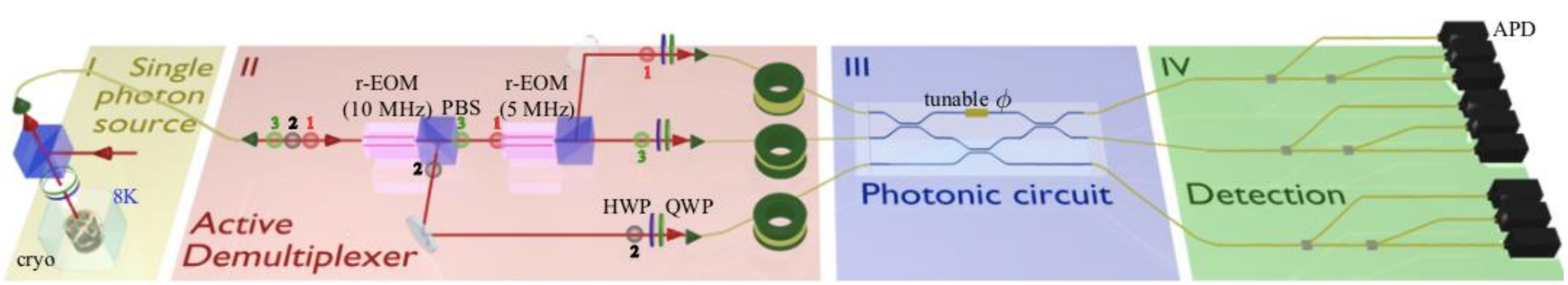

Figure 523: three entangled photon source using Quandela quantum dots. Source: [Interfacing scalable photonic platforms: solid-state based multi-photon interference in a reconfigurable glass chip](#) by Pascale Senellart et al, 2019 (7 pages).

In 2019, they experimented with Quandela's photon source to demultiplex it into three photons which were then injected into a photonic integrated circuit integrating a programmable quantum gate. The photonic circuit was precisely etched with a femtosecond laser (Figure 523).

A last, since 2020, Quandela has started working on creating a photon-based quantum computer using their own photon source. We cover this aspect of their business in the photonic qubits vendor section that starts page 455.

**Quantum Opus** (2013, USA) develops single photon detectors based on superconducting nanowires, the Opus One. The compact version Opus Two is an 8U data center rack-mount package, including cryostat[1477]. This company benefited from US federal funding, including $100K in 2015 and $1.5M in 2015 from DARPA and $125K from NASA in 2018. They are a provider of the Chinese team who did run the gaussian boson sampling experiment announced in December 2020.

**Qubitekk** (2012, USA, $5M) is a supplier of photon and entangled photon sources for use in the context of quantum cryptography (QKD). This technology can also be used to manage part of the communication between qubits in some types of quantum computers. It competes to some extent with Quandela.

**Scontel** (2004, Russia) offers single photon detectors in the visible and infrared (SSPD, for Superconducting Single Photon Detecting Systems). These detectors are cooled at 2.2K helium-4 using a Sumitomo SRDK 101 pulse head system with a water-cooled HC-4E compressor.

**Single Quantum** (2012, The Netherlands) offers Qos single photon detectors integrated in a 2.5K liquid helium cooled cryostat.

Their sensor uses the SNSPD (superconducting nanowire single photon detector) technique, made of a thin film of superconducting nanowires shaped into a flattened serpentine coil. This device captures a single photon from an optical fiber and have a detection efficiency of 85% to 90%, covering wavelengths from 800 nm to 1550 nm.

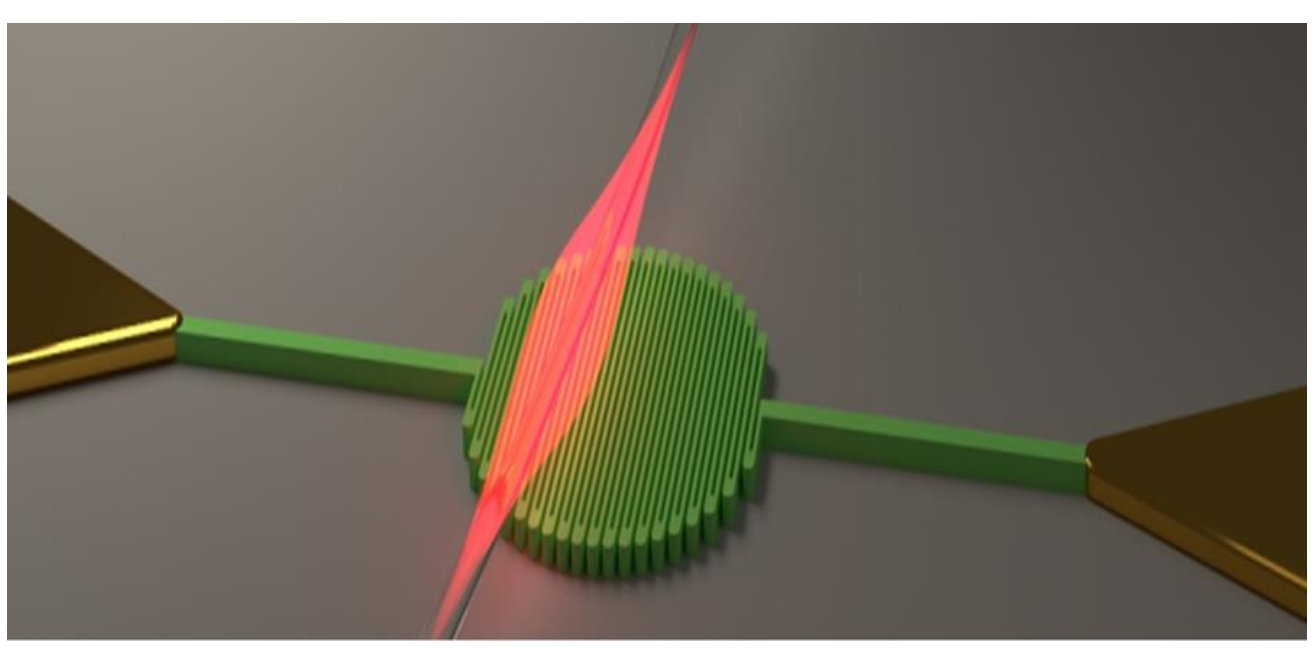

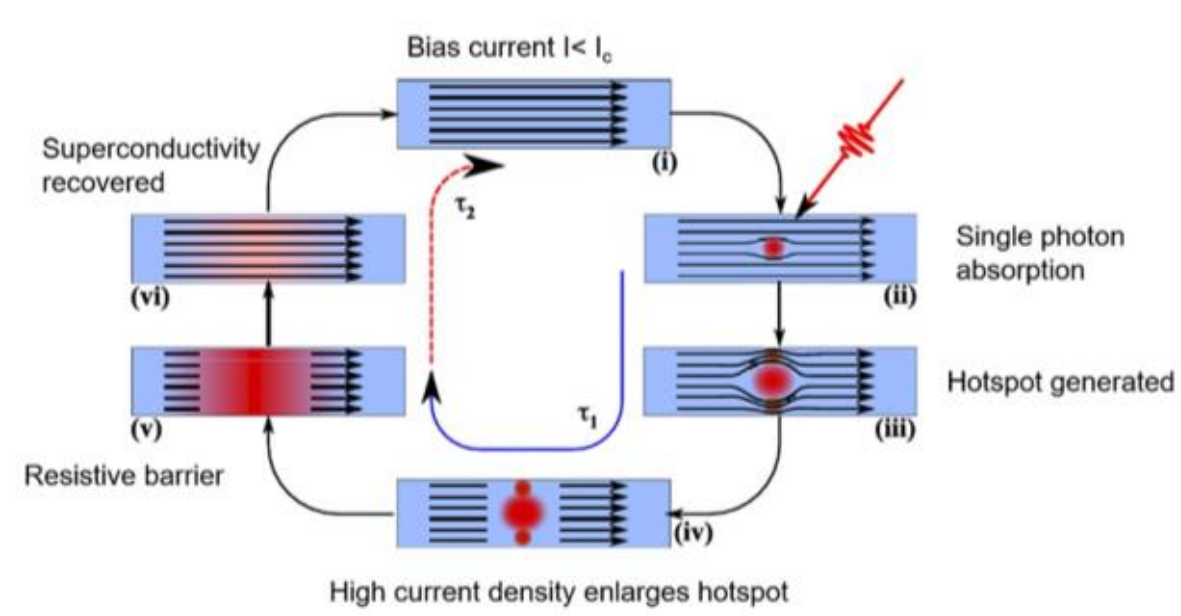

*Figure 524: SingleQuantum SNSPD photon source.*

**Sparrow Quantum** (2016, Denmark, $2.2M) is a spin-off from the Niels Bohr photonics research laboratory. Like Quandela and Qubitekk, they offer single photon sources.

Their solution is based on InAs quantum dots. Their engineering differentiation lies with the quantum dot efficient coupling with a slow-light photonic-crystal waveguide.

1477 See [Introduction to Quantum Opus and revolutionary superconducting detection systems]() (14 slides).

A laser is illuminating the quantum dots with using a confocal microscope. Their photon coherence indistinguishability is between 95% and 98% with their Sparrow Chip 2021 Resonant. They are generated in the 920-980 nm wave range. The photon generation system is cooled at 6K.

**VLC Photonics** (2011, Spain) produces photonics equipment and fabless design of photonic integrated circuits. The company is involved in European Flagship projects.

It is a spin-off of the University of Valencia. The company was founded by Iñigo Artundo, Pascual Muñoz, José Capmany and José David Domenech. They also market technical reports at prices ranging from 4K€ to 5.4K€ per piece.

**Excelitas** (2010, USA) sells various photonics devices including Single-Photon Counting Modules (SPCMs).

**Pixel Photonics** (2020, Germany, 1.45M€) develops single photon detectors (SNSPD) targeting quantum computing, QKD and imaging markets. With HTGF (Germany) and Quantonation (France) as seed investors.

**Hamamatsu Photonics** (1953, Japan) provides silicon photodiodes, electron multipliers for detecting electrons, ions, and charged particles, photon counters, LCoS based spatial Light Modulators (SLM) used for cold atoms controls, laser cooling systems, quantum imaging and image sensors for the detection of neutral atoms, trapped ions and NV centers fluorescence.

**Miraex** (2019, Switzerland) has two main quantum technologies in its portfolio: photonic based quantum sensors for vibration, acceleration, acoustic, pressure, electrical field and temperature measurement and a quantum system converting matter qubits into photon qubits and vice versa. It's a spin-off from EPFL.

**Micron-Photons-Devices** (2004, Italy) aka MPD creates Single Photon Counting Avalanche Diodes, "SPAD", fabricated using custom silicon, standard CMOS and InGaAs/InP technologies. It also sells photon counting based QRNGs.

**Qubitrium** (2020, Turkey) develops entangled photon sources, laser current drivers and single photons detectors.

**Teem photonics** (1998, France) creates lasers and integrated photonics components including erbium doped waveguide amplifiers and arrayed-waveguide gratings. Not the same AWGs than the arbitrary waveform generators used to control solid-state qubits, although these can be used to signals multiplexing/demultiplexing.

**Alcyon Photonics** (2018, Spain, $560K) is a spin off from IO (Instituto de Óptica) in Spain. Its expertise is on sub-wavelength grating (SWG) technology. They create complex photonic circuits like high-performance Application Specific Photonic Integrated Circuits (APICs). The company was cofounded by the researcher Aitor Villafranca.

**Scintil Photonics** (2018, France) develops mixed silicon and III/V photonic components, using their BackSide-on-BOX process, that mixes active and passive optical components.

Their technique bonds InP/III-V dies on the backside of processed Silicon-On-Insulators (SOI) wafers, only where it is needed. Their fabrication process is classical CMOS. Their components integrate lasers (WDM laser arrays and tunable lasers), modulators, waveguides, wavelength filters, surface fiber couplers, semiconductor optical amplifiers (SOA), and photodetectors.

**Muquans/ixBlue** (France) is also a provider of laser and intelligent frequency tunable lasers, laser frequency doublers and narrow-linewidth lasers used in various quantum technologies.

**Photon Force** (2015, UK) creates single-photons cameras of 32x32 pixels. It can be used in various photonic based quantum sensing applications. It enables time-tagging of incoming photons with a time resolution of 55 pico-seconds.

**Covesion** (2009, UK) develops laser frequency conversions devices that are used in many quantum optics applications. It's a spin-off from the University of Southampton.

We also have **Ibsen Photonics** (1991, Denmark) which provides spectrometers and various photonic equipment, **Lumiphase** (2020, Switzerland) which develops optical modulators, **Pixel Photonics** (2020, Germany, 1.8M€) which designs Single Photon Detectors (SNSPD), **Bay Photonics** (2007, UK) which provides photonic circuits assembly and packaging, **MenloSystems** (2001, Germany) and their optical frequency combs, terahertz systems and femtosecond lasers, **Qubig** (2008, Germany) which develops light modulators (amplitude and phase modulators, phase shifters, Pockels cells) that can be used in quantum computing or communications, etc.

# Fabs and manufacturing tools

Many quantum technologies components are nanofabrication based and must be manufactured somewhere. It's the case with superconducting qubits, superconducting electronics, quantum-dots based electron spin qubit circuits, quantum nanophotonic circuits, NV center based qubits and sensors, single-photon generating quantum dots, photon detectors, trapped ions supporting circuits, travelling wave parametric amplifiers and the likes. You could wonder how these circuits are manufactured and where. Like your regular smartphone chipset processor, is it coming from a giant $20B TSMC 5 nm fab in Taiwan? Well, most of the time, no!

### Foundries

We are in a very different technology and market realm. Quantum related components have some distinct characteristics compared to mass market semiconductors that you'll find in your TV, smartphone, laptop of tablet. They are very specialized and use sometimes special manufacturing processes and/or materials like III/V semiconductors or niobium/aluminum deposition for superconducting qubits and electronics. They are most of the time experimental with many try/error cycles. They are sometimes manufactured with specialized tools. And at last, they are produced in rather small quantities. Surprisingly, given the experimental nature of many components, the related fabs are usually less impressive in size and cost.

Fabs contain cleanrooms, where the concentration of airborne particles is controlled on top of temperature, humidity and sometimes, other parameters like ambient magnetism and vibrations. Cleanrooms are classified according to the number and sizes of particles suspended in the atmosphere. Cleanroom ISO classes range from 1 to 9, with an (exponential) increased number and size of particles per volume unit, 1 being the "cleaner". Most specialized quantum technologies fabs have a less stringent cleanroom class requirement than the most expensive and modern semiconductor fabs since they are not creating high-density chipsets and do not care so much about yield. They are rather class 100 to class 1000 cleanrooms.

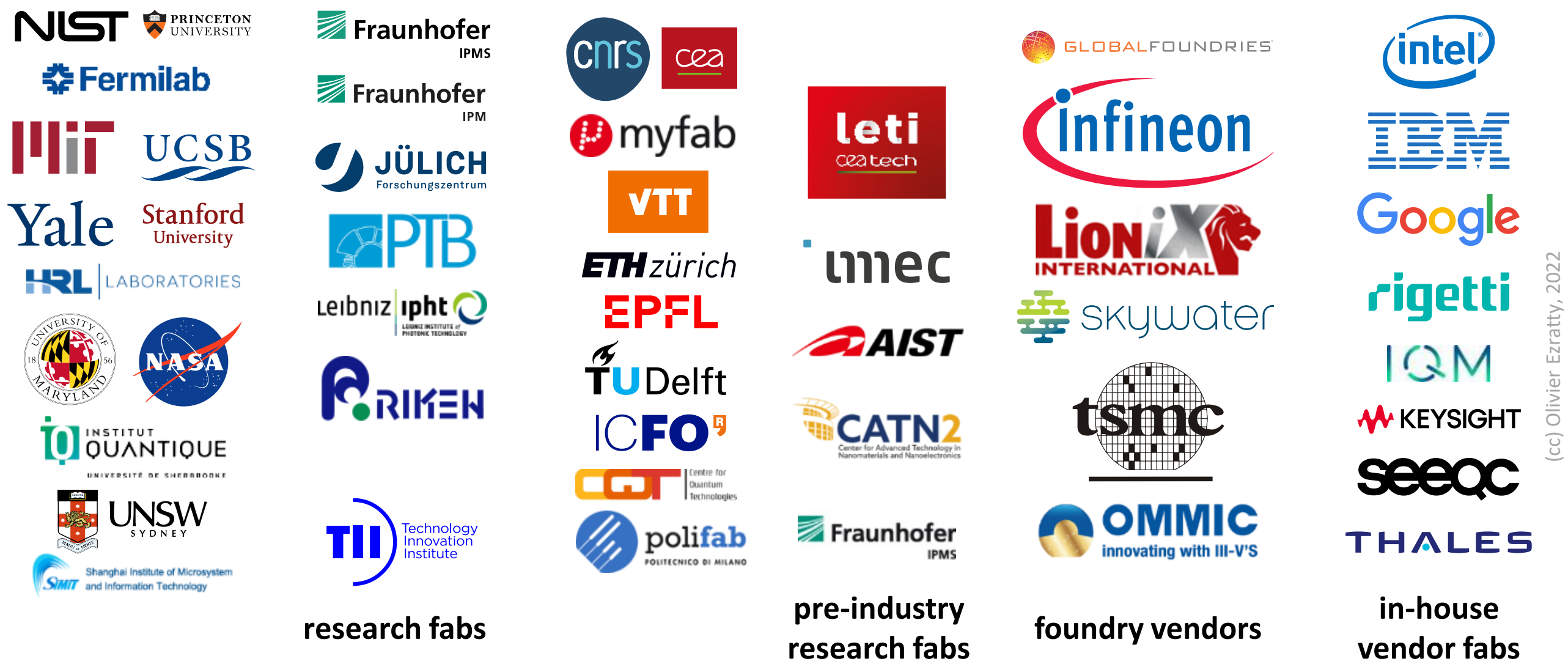

*Figure 525: research and industry cleanrooms fabricating semiconductors for quantum use cases. (cc) Olivier Ezratty, 2022.*

We can segment quantum technologies fabs in a couple categories:

**Research fabs**. These are most of the time fabs from national research organizations labs and universities. These fabs have cleanrooms with sizes ranging from 100 m$^2$ to 4,000 m$^2$. Their teams and the associated researchers are creating the "recipe" of new semiconductor technologies. These fabs usually produce 200 mm or smaller wafers. You have for example the famous MIT Lincoln Labs in the USA (superconducting electronics and qubits, trapped ions chipsets, with 1,629 m$^2$ of clean rooms), Princeton (1,350 m$^2$ cleanroom, silicon and III/V, up to 100 mm wafers), HRLabs in California (all sort of things in a 900 m$^2$ clean room), UCSB Nanofab (1,170 m$^2$, superconducting qubits, MEMS, photonics, imaging sensors), Harvard CNF (966 m$^2$), Yale University (108 m$^2$, superconducting qubits), Stanford (180 m$^2$, various quantum techs), NIST NanoFab (1,800 m$^2$ cleanroom), VTT in Finland (2,600 m$^2$ cleanroom for 150 and 200 mm wafers, superconducting qubits and electronics, photonics), CNRS C2N (2,900 m$^2$ clean room near Paris, photon quantum dots sources in GaAs, polaritons circuits, etc), CNRS Institut Néel (220 m$^2$, near Grenoble, superconducting electronics and qubits, graphene, diamonds growth[1478]), Van Leeuwenhoek Lab at TU Delft (3,500 m$^2$ cleanroom used by Qutech and TNO in The Netherlands), MyFab (Sweden), Fraunhofer IPM (400 m$^2$ clean room in Freiburg, various optical quantum sensors), Fraunhofer IPMS near Dresden (200 mm wafers 1,500 m$^2$ cleanroom) and also PTB, Leibniz IPHT and Jülich in Germany, EPFL and ETH Zurich in Switzerland and others like RIKEN and AIST (superconducting electronics, but also a strong CMOS 300 mm manufacturing capacity) in Japan or UNSW in Australia. PoliFAB from Politecnico di Milano created an interferometer used by Pascale Senellart's team to demonstrate the indistinguishability of photon clusters[1479]. The Shanghai Institute of Microsystem And Information Technology (SIMIT) has also its own fab.

**Pre-industry research fabs**. These are the likes of IMEC (Belgium) and CEA-Leti (France) who create new semiconductor components and design new manufacturing processes before they are volume-produced in commercial fabs. These are larger fabs than the aforementioned research fabs. CEA-Leti operates 11,000 m$^2$ of cleanroom in Grenoble. IMEC cleanrooms totals 12,000 m$^2$ in Leuven, Belgium. CEA-Leti produces silicon qubits wafers for its own usage as well as for vendors like Quantum Motion (UK). These fabs produce wafers up to 300 mm.

---

[1478] See Fabrication of superconducting qubits by Vladimir Milchakov (IQM), September 2020, who describes the Institut Néel superconducting manufacturing capability and process. This fab also produces the TWPA electronics from Silent Waves, a spin-off startup from Institut Néel created in 2022.

[1479] See Quantifying n-photon indistinguishability with a cyclic integrated interferometer by Mathias Pont, Pascale Senellart, Fabio Sciarino, Andrea Crespi et al, January 2022 (21 pages).

Likewise, the Center for Advanced Technology in Nanomaterials and Nanoelectronics (CATN2) from SUNY Polytechnic Institute in New York State has a cleanroom of 12,000 $m^2$ producing 200 mm and 300 mm wafers for AI, photonics, CMOS spin qubits, superconducting qubits and digital electronics.

**Foundry vendors**. These are independent foundries manufacturing semiconductors for third parties. GlobalFoundries manufactures nanophotonic chipsets for PsiQuantum and Xanadu in Malta, New-York State in their 41,400 $m^2$ clean room on top of classical CMOS chipsets like the IBM Power processors. Infineon's Villach fab in Austria manufactures trapped ions chipsets in its 23,000 m2 cleanroom for Oxford Ionics[1480]. In Germany, Infineon is an industry partner of many other projects with superconducting qubits, qubit control electronics and NV centers qubits and sensing[1481]. In the USA, SkyWater is the largest foundry for superconducting electronics, working among others for D-Wave, on top of working on various space applications and with DARPA. Formerly Cypress Semiconductor, Control Data and VTC, they consolidate a 7,360 $m^2$ clean room in Minnesota and another one of 3,300 $m^2$ in Florida and support 90 nm features geometries on 200 mm wafers. Lionix in The Netherlands manufactures nanophotonic circuits for its subsidiary QuiX. Larger foundries are usually needed for high-density chipsets, particularly with silicon qubits where patterns are relatively small, down to about 10 nm. OMMIC (2000, France) is a small foundry specialized in manufacturing III-V MMIC (monolithic microwave integrated circuit) which could comprise cryogenic amplifiers used in quantum computing.

**In-house vendor fabs**. These are the fabs from quantum technology vendors who are self-sufficient for this respect. Intel manufactures its own quantum dots spin qubits chipsets in one of its clean rooms at its Hillsboro facility in Oregon. IBM also has its own manufacturing capacity for superconducting qubits and high-density silicon chipsets with a cleanroom of 3,600 $m^2$ in Yorktown, New York State. Rigetti in the USA and IQM in Finland have their own small $20M fabs for their superconducting qubits chipsets. Google has also an in-house fab in Santa Barbara, California. SeeQC has a small 150 mm wafers 200 $m^2$ cleanroom dedicated to manufacturing superconducting electronics. Keysight also has its own III/V 1,200 $m^2$ cleanroom, the High Frequency Technology Center (HFTC) in Santa Rosa, California[1482]. Qilimanjaro relies on a fab that was put in place in 2021 at TII in Abu Dhabi. At last, Thales has an in-house 4000 $m^2$ III/V fab with CEA-leti as a partner. Having your own fab makes sense when you need to have a fast turn-around and test repetitively many generations of qubit chipsets. It is relatively affordable for producing superconducting qubits on small wafers.

The USA, European Union and China all want to increase their share, self-reliance and supply security with semiconductor manufacturing. In February 2022, the European Union launched the European Chips Act to "*foster development of capacities in advanced manufacturing, design and system integration as well as cutting-edge industrial production*", with a public/private funding of €43B until 2030. It includes international partnerships like when Intel is installing a new fab in Germany. The plan contains a provision for quantum technologies, to "*set up advanced technology and engineering capacities for quantum chips in the form of design libraries for quantum chips, pilot lines, and testing and experimentation facilities*". This may provide some additional funding for the extension of the many quantum-related fabs mentioned before.

In March 2022, the US Senate voted the CHIPS Act, with $52B funding. It was finally signed by POTUS in August 2022.

---

[1480] See Development of novel micro-fabricated ion traps by Gerald Stocker, November 2018 (96 pages) and Trapped ion quantum computing, Infineon.

[1481] See Infineon Participates in 6 Research Projects, Expands Commitment to Quantum Computing by Matt Swayne, The Quantum Insider, February 2022.

[1482] Their equipment is well documented in Keysight High Frequency Tech Center (HFTC) (15 pages).

It contains additional Federal budget of $152M per year for quantum technologies for the 2023-2027 period, although seemingly not specific to components manufacturing. Chipsets USA manufacturing market share is in the 12% mark, above EU's that sits around 9%. The rest is in Asia, mostly Taiwan, South Korea, China and Japan.

### Generic processes

We'll describe here the generic processes used to produce chipsets regardless of their use case, the most commonplace being bipolar, CMOS and BiCMOS chipsets[1483]. The story always begins with a wafer.

**Wafers** are usually made of monocrystalline silicon sliced with wired diamond saws out of ingots manufactured with the Czochralski crystal growth method. They are sometimes completed with a thin buried layer of $SiO_2$ (aka SOI, for silicon on insulator) and another thin layer of regular Si, using the SmartCut process invented by CEA-Leti and implemented by SOITEC and its licensing partners[1484]. SOI wafers have many interesting characteristics like reduced parasitic capacitances and low leakage currents. They are frequently used for nanophotonic circuits (like those from PsiQuantum manufactured by GlobalFoundries) or for silicon qubits chipsets (CEA-Leti, Qutech, ...).

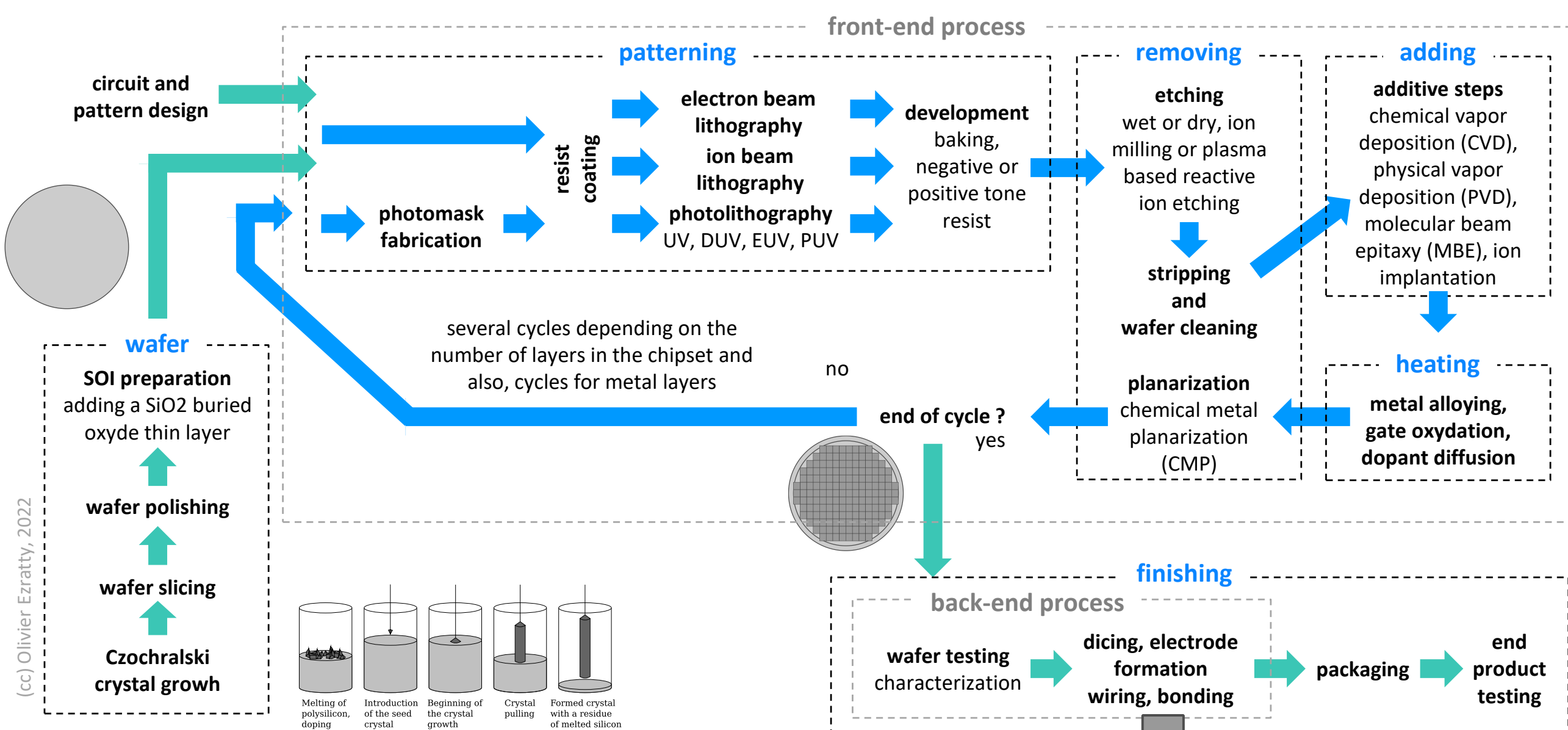

*Figure 526: a generic layout of a chipset manufacturing process. (cc) Olivier Ezratty, 2022.*

In other cases, wafers are made of III/V semiconducting materials (like GaAs or GaN, for manufacturing some nanophotonic circuits) or even sapphire (for some superconducting qubits and trapped ion circuits).

---

[1483] Bipolar transistors have a very high speed and are used in analog devices. CMOS transistors are slower and do not handle such high power as bipolar transistors but are aggressively scaled down in density and require far less power to operate. BiCMOS used both bipolar and CMOS logic that are co-integrated within the same chip, which requires additional process steps and incur higher costs.

[1484] The SmartCut process is not using a wire diamond saw like the ones used to slice wafers out of silicon ingots. It first creates a layer of $SiO_2$ on a Si wafer using Si thermal oxidation in wet atmosphere, PECVD or CVD. Then, an ionic implantation of H or He is made on another Si wafer creating a sort of "glue". The $SiO_2$ side of the first wafer is bounded with this "glued" Si wafer and a thin layer of Si of that wafer is deposed on the $SiO_2$ using a "layer splitting" process created by thermal annealing. The second Si wafer can be reused for another Si deposit cycle. Both the $SiO_2$ and the overlay Si layers can be as thin as 10 nm. A variant of the SmartCut process is also used to depose a thin Si layer on sapphire wafers and thermal annealing is created with laser beams in a so-called "LLO" process, for laser lift-off. I found interesting SmartCut process descriptions in The advanced developments of the SmartCut technology: fabrication of silicon thin wafers & silicon-on-something hetero-structures by Raphaël Meyer, 2018 (252 pages).

A wafer has a thickness ranging from 40 µm to 700 µm and its diameter ranges from a couple inches (for III/V and other small volume processes) to 300 mm and even 450 mm (for volume CMOS processes).

The most generic chipsets production processes then involve several cycles with the following successive steps with a repeat cycle ranging from resist coating to planarization including patterning, removing and adding matter.

When this cycle is over will all chipset layers added on top of the other, the process ends with various finishing steps up to a packaged chipset ready for integration.

***Patterning***

These are the process steps that define the places in the wafer where matter has to be removed with etching or added afterwards. It involves several steps that are defined during the design stages exploiting automated electronic design automation (EDA) software tools like those from ANSYS, Cadence, Keysight Technologies, Synopsis, Xilinx and Mentor Graphics (in Siemens group).

**Resist coating** is applied on the wafer with a photoresist liquid that will be later exposed during the lithography process and selectively removed during development. The coating is mechanically added with a dispenser nozzle positioned above the center of the rotating wafer attached to a chuck and spindle with vacuum pumping[1485].

EBR (edge bead removal) removes excess coating at the wafer edge with a solvent. Then a $N_2$ based soft bake evaporates most of the solvent.

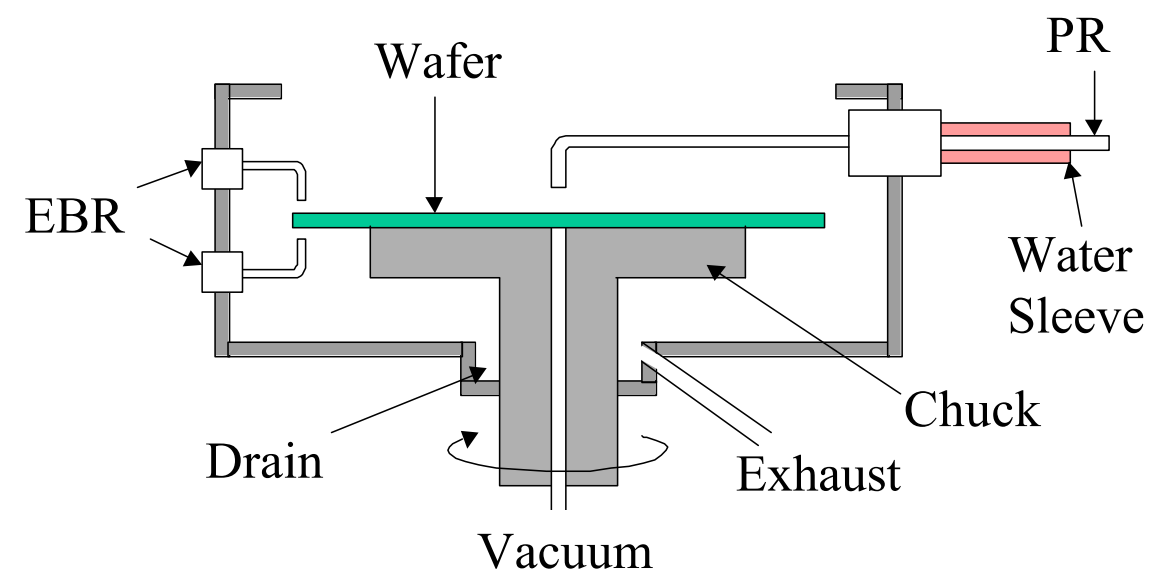

*Figure 527: resist spin coating. Source: Introduction to Semiconductor Manufacturing Technology by Hong Xiao (2148 slides).*

**Photolithography** is used to expose a special coating on selected areas. The photolithography technique makes use of a photomask and ultra-violet rays exposing a photoresist film or coating. It's being used to produce silicon qubits and nanophotonic chipsets. The usual photolithography process involves a stepper which moves the wafer under the camera to expose the wafer for each and every chipset to produce, and with a very high precision (<1 nm). It replaces mask aligners that are used when the photolithography process exposes the whole wafer in a single step. The size of the chipset is limited by the size of the photolithography reticle, which conditions the mask maximum size. Across time and density improvements, various ranges of ultraviolet wavelengths have been used. It started with UV, to DUV or deep ultra-violet under 248 nm, then to EUV (extreme ultra-violet) at 13 nm and PUV at 6,5 nm which is border line to soft XRays. Starting with EUV, there is only one provider of photolithography system, ASML.

**Electron beam lithography** (EBL) is another lithography technique, that is focusing a beam of electrons on an electron-sensitive resist film to remove matter in specified areas, without requiring a mask like with photolithography. EBL can reach precisions of 1 nm which is excellent and better then photolithography. It is used to create precision nanostructures like with photon-generating quantum dots and also superconducting qubit chipsets. It is a very slow process compared with photolithography, so adapted to low volume and custom productions. An EBL looks like an electron microscope. Existing electron microscopes can be converted to run EBL tasks.

Other less used varieties of lithography are ion-beam (using helium), laser lithography, the latter being used for resolutions above 500 nm, and STM (Scanning Tunnelling Microscopy) that can reach sub-nm resolutions.

---

[1485] See the incredibly rich Introduction to Semiconductor Manufacturing Technology by Hong Xiao (2148 slides) and its eponymous book published in 2012 (524 pages).

**Development** which removes the photoresist coating where it was or wasn't exposed during the photolithography or electron beam lithography step. It depends on the photoresist material used which is either a negative (insoluble after exposure) or positive resist material (soluble after exposure).

Positive photoresist enables better lithography resolution but is more expensive than negative photoresist. It uses so-called hard baking above 100°C to polymerize and stabilize the photoresist coating.

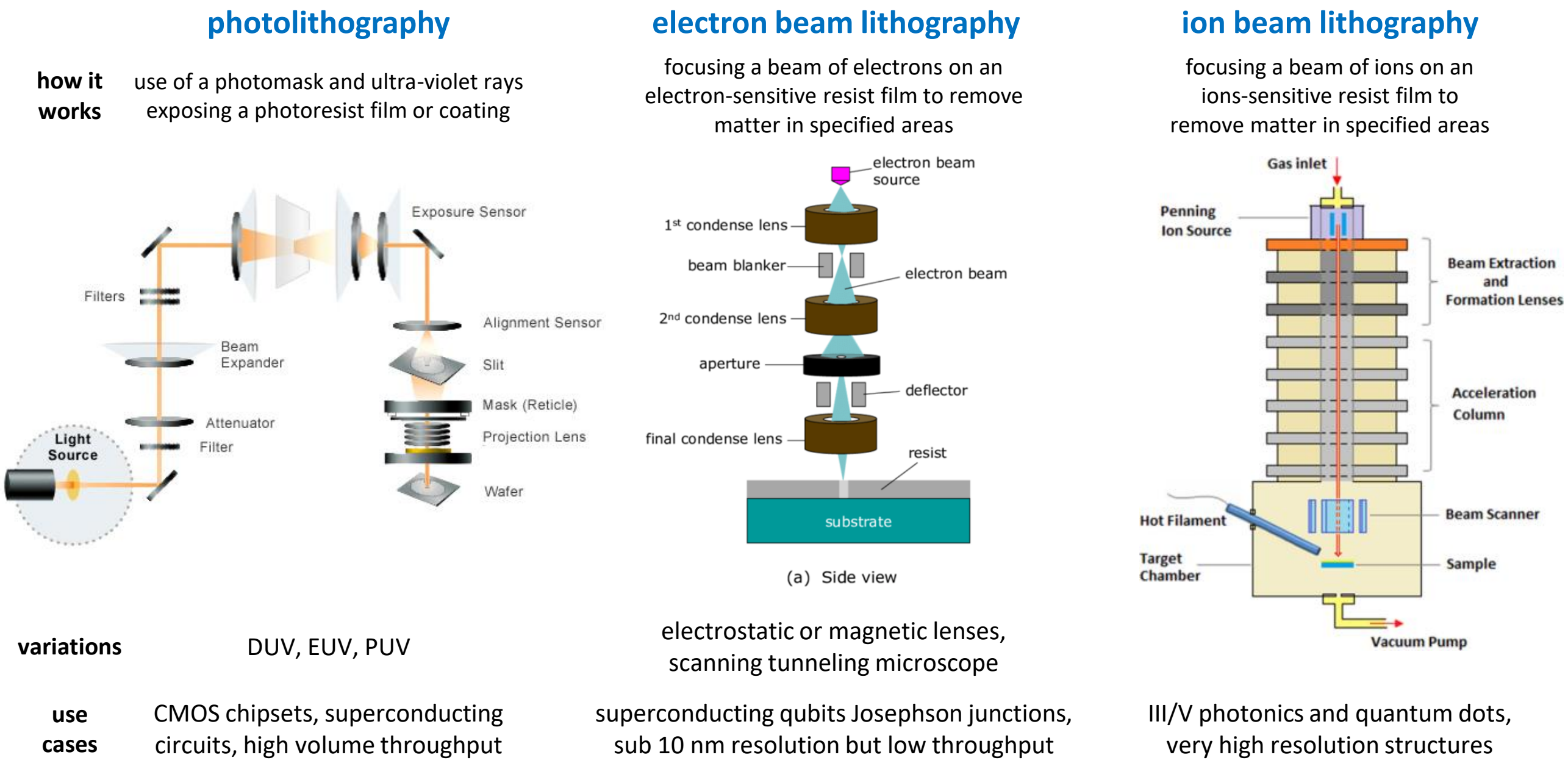

*Figure 528: the three main lithography techniques used for semiconductors manufacturing. Compilation (cc) Olivier Ezratty, 2022.*

There are many resist coatings depending on the process (photolithography or electron beam lithography) and whether we are using a negative or positive resist. With positive resist, coating can be made of long polymer chains with weak chain bonds. Exposure creates chain scission in the exposed areas. These are dissolved during the chemical development process while the longer chains do not dissolve. Another process consists in using resist creating hydrophilic product when exposed, which are then dissolved by water. Negative coating can be made of monomers that polymerize when exposed to light, and become non dissoluble by the solvent used in the development process.

The photolithography process contains in total about 10 stages (coating, soft-baking, exposure, cleaning, hard-baking, ...). In volume production, these are handled in cluster tool systems, using one or several robotized systems to move the wafer from one tool to the next in a controlled environment. This ensure both productivity and production quality. Photolithography clusters vendors include Dainippon Screen, who partners with ASML. Cluster tools are also in place for other parts of manufacturing seen later, like etching and additive steps. Multi-axis robots and roof conveyors like those from Kuka and Muratec move wafer cassettes (handling 25 wafers) from one cluster to the other.

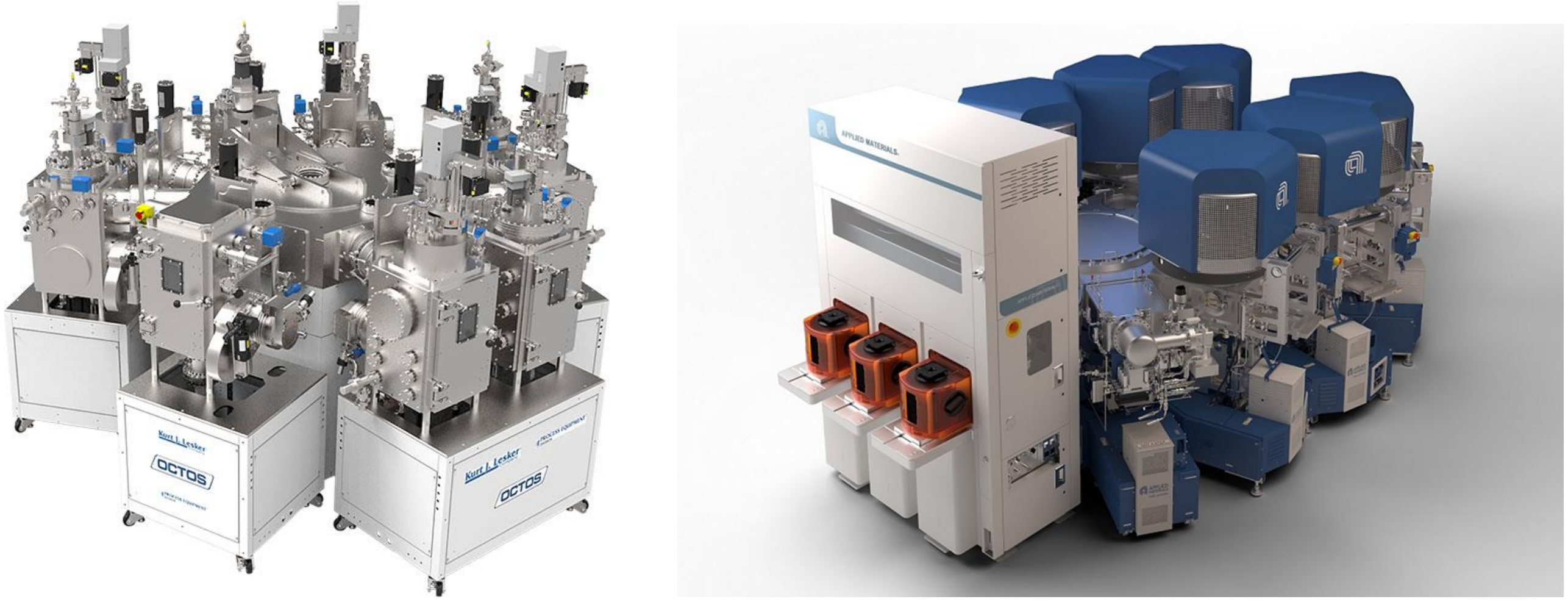

*Figure 529: two examples of such cluster tools, on the left with a Kurt Lesker OCTOS Automated Thin Film Deposition Cluster Tool ([source]) and on the right an Applied Materials Endura Clover MRAM PVD System ([source]).*

***Removing matter***

These steps correspond to the removal of matter on the wafer based on the zones defined by the lithography process.

**Etching** which removes matter in the uncovered areas, using wet chemical or dry physical methods, the dry methods being the most commonplace for high-density (VLSI) circuits. Various dry etching techniques include ion milling or sputter etching, and plasma based reactive ion etching. The first uses the projection of inert ionized noble gas while the second uses neutrally charged free radicals that react with the target surface. In general, a plasma is an ionized gas with the same proportion of positive and negative charges. There are also variants with anisotropic (orientation independent) or directional etching (orientation dependent).

| | ion milling or sputter etching | ion beam etching (IBE) | plasma based reactive ion etching |
|---|---|---|---|
| how it works | the ions of an inert noble gas like argon are accelerated from a wide beam ion source on the target in vacuum to mechanically remove material on the target. | the ions of active molecules like $CF_4$, $Cl_2$ or $CHF_3$ are accelerated from a wide beam ion source on the target in vacuum to mechanically remove material on the target. | energetic and neutrally charged free radicals reacting at the surface of the wafer, attacking the wafer from all angles, aka, being anisotropic. |

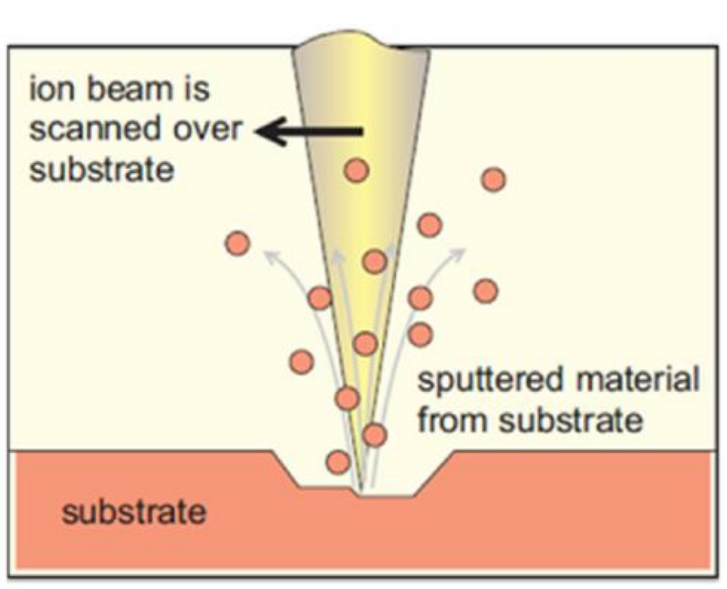

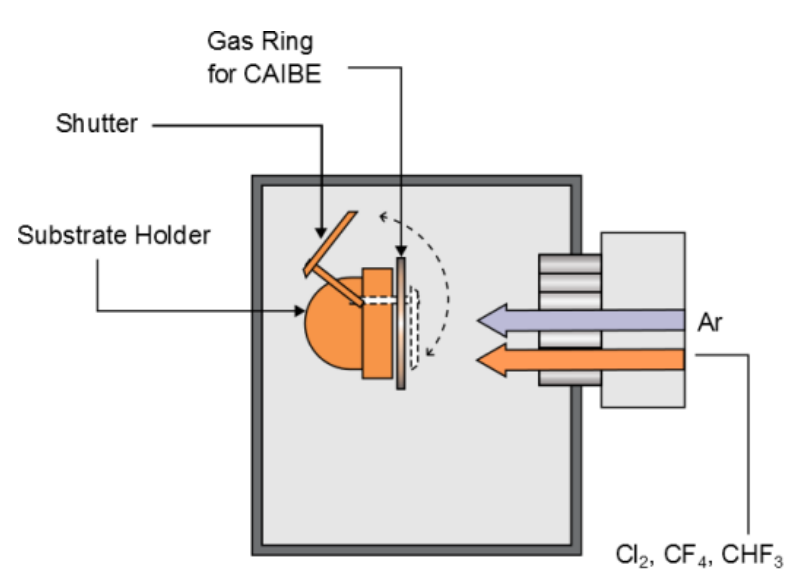

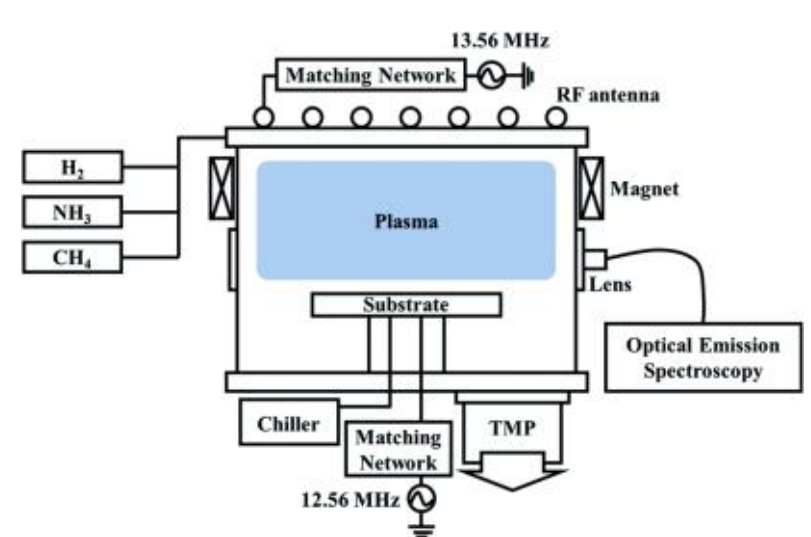

| | ion milling or sputter etching | ion beam etching (IBE) | plasma based reactive ion etching |
|---|---|---|---|
| variations | reactive ion etching<br>deep reactive ion etching | ion polishing, ion cutting,<br>reactive ion beam etching (RIBE, RIE),<br>chemically assisted ion beam etching (CAIBE). | MEMS, nanophotonics, niobium resonators in superconducting qubits |
| use cases | superconducting qubits | III/V chipsets, single photon quantum dots shapes, MEMS | |

*Figure 530: the various ways to remove matter in semiconductor manufacturing. Compilation (cc) Olivier Ezratty, 2022. Illustration sources: Wikipedia, others.*

**Stripping and cleaning** which removes the remainder of the photoresist material. In volume production clean rooms, etching and stripping is handled by cluster tools with robotized handlers moving wafers from the etcher to the stripper tool, with a loading and unloading station extracting wafers from its carrier box (usually, containing 25 wafers in volume production environments).

**Planarization** of the wafer uses physical polishing. It can be based on CMP (Chemical Metal Planarisation). It is generally implemented after additive steps.

### *Adding matter*

**Additive steps** consist in adding some materials in the visible areas, like silicon or doped silicon in classical CMOS transistors, using boron or indium (for p-doping) and arsenic (As), phosphorus (P), antimony (Sb) or aluminum (Al) (for n-doping). Some of these processes are implementing an epitaxy, creating a perfect crystalline structure with the added material, in the feature layers (doped silicon, gates). Other processes like PVD and sputtering are not epitaxial and are used for the production of superconducting qubits.

Additive steps can use various techniques like ion implantation, CVD (chemical vapor deposition, where the target surface is exposed to one or more volatile precursors, which chemically react and/or decompose on the target surface to leave a thin film deposit on the target, e.g. using silane $SiH_4$ to deposit Si on the wafer, generating 2 $H_2$ molecules), ALD (atomic layer deposition, a variation of CVD to create highly precise epitaxial atomic layers using repeat cycles), PVD (physical vapor deposition under low pressure, where the material to deposit with evaporation, sputtering or plasma and then condenses on the target surface), sputtering being one type of PVD (using ion projection to pull material from a source and deposit it on the target wafer or ionized gas like argon that, thanks to a high-voltage applied to the target, is projected on the target and creates a plasma with the target atoms that then condenses on the surface of the chipset), e-beam deposition (another variety of PVD using electron beams to evaporate the matter to deposit on the wafer), PLD (pulse laser deposition, using femtoseconds laser pulses to extract matter from a source and then sent to the target), MBE which is a variety of PVD (molecular beam epitaxy, for thin-film deposition of single orderly crystal structures). CVD can be plasma based.

In the last process cycles, these steps are related to the creation of several superposed metal layers connecting the various semiconducting circuits created in the earlier steps. With superconducting qubits, aluminum, and aluminum oxide (or niobium) sputtering is implemented in this step.

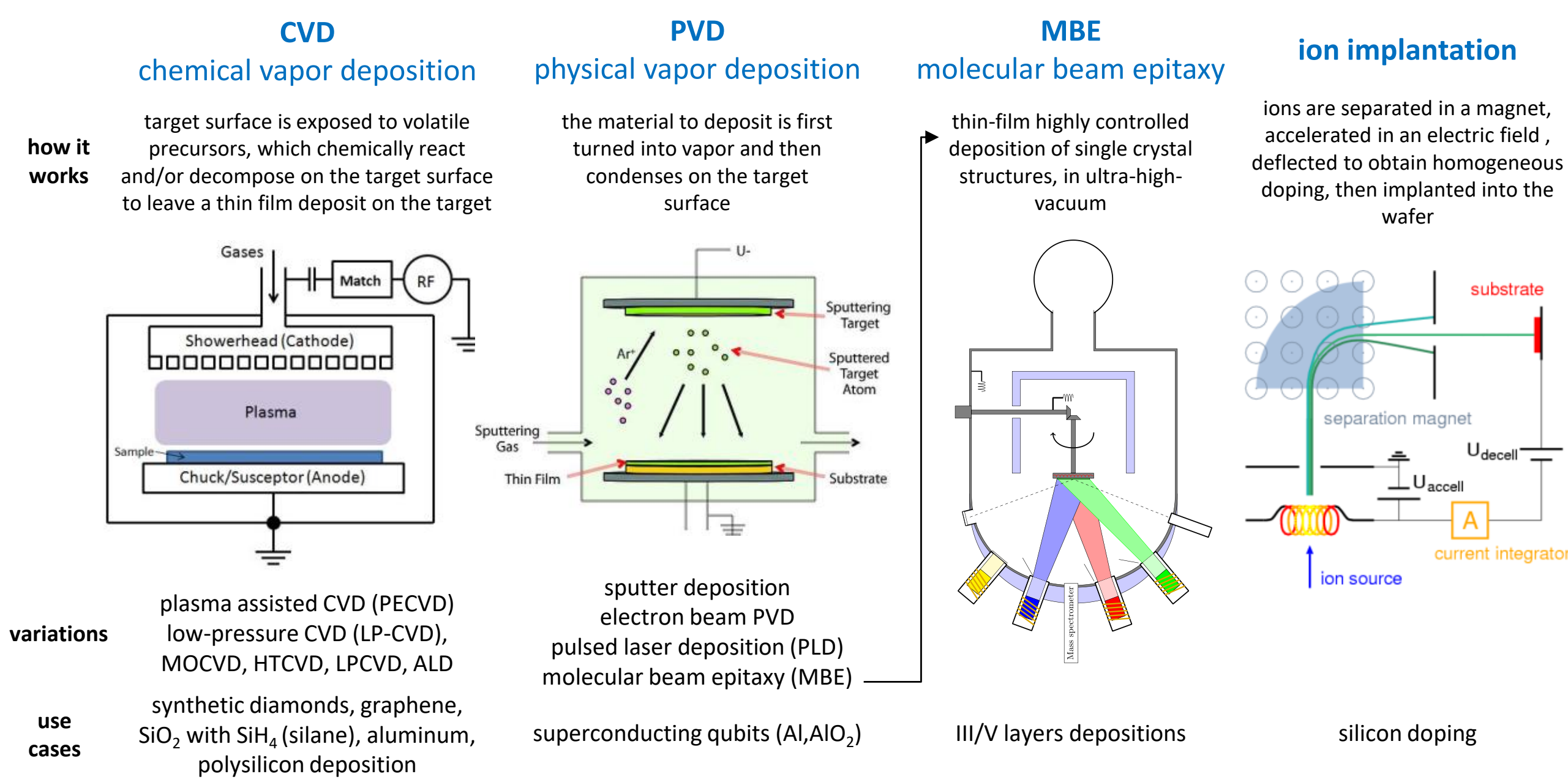

*Figure 531: the various ways to add matter in semiconductor manufacturing. Compilation (cc) Olivier Ezratty, 2022.*

**Metal layers**. When all cycles related to the functional parts of the circuits are finished, some electrodeposition of metal is made to connect the chipset to the outside world, usually copper or aluminum (in the case of superconducting qubits and electronics) and copper-aluminum alloys. Metal layers are created with a mix of lithography-etching and PVD/CVD. As metal layers are added, their density is decreasing. Then, some wiring may be added and bonding or bumps plus packaging. In CMOS designs, the "front-end-of-line" (FEOL) contains the individual active elements (transistors, capacitors, etc.) while the "back end of line" (BEOL) contains the metal layers.

So-called 3D chipsets like the superconducting qubits chipsets from IBM, OQC and others result from the assembly and perfect alignment of stacked chipsets. CEA-Leti (France) is collaborating with Intel in the design of such innovative 3D packaging technologies. 3D stacking makes use of TSV or through silicon vias, which establishes a metal connection from top to bottom of a chipset or from the active layer to the front plane through the wafer.

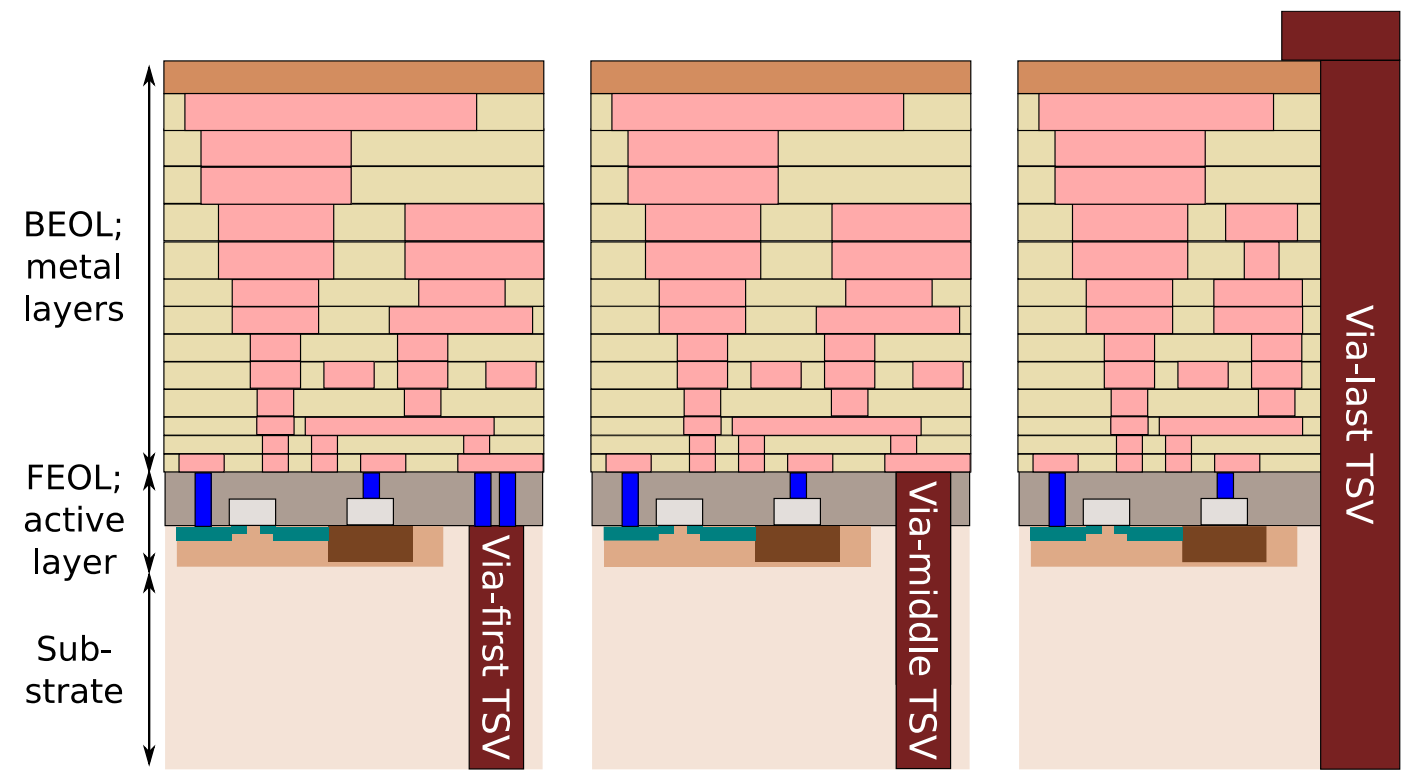

*Figure 532: typical metal layers of a semiconductor.*

A TSV hole is created with reactive ion etching, copper electrochemical deposition for creating a seed layer and electroplating to fill the hole[1486].

### *Heating*

**Thermal processes** are implemented for various purposes like dopant activation and diffusion, gate oxidation ($Si+2O \rightarrow SiO_2$), metal reflow which smooths its surface usually in an atmosphere of $N_2$ or $H_2O$, metal alloying and chemical vapor deposition. One used technique involves rapid thermal annealing, using a vertical or horizontal furnace.

### *Finishing*

These are the product finishing tasks undertaken when the patterning-removing-adding-heating cycle is completed (the front-end process). The aim here is to turn the chipset on its wafer into a functional component with its connectivity. It's also called the back-end process.

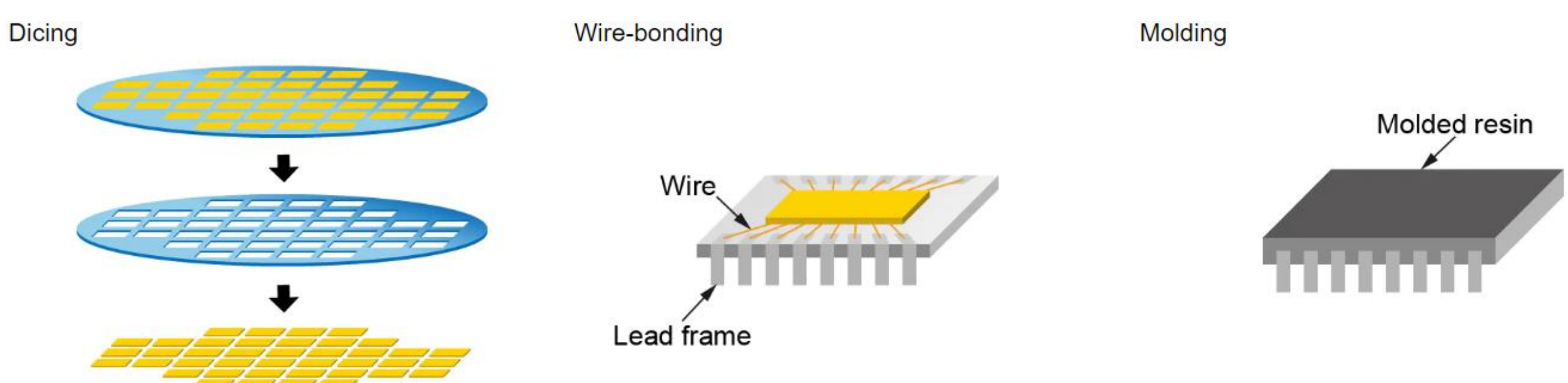

*Figure 533: the finishing steps of semiconductor manufacturing with dicing, wire bonding and molding. Compilation (cc) Olivier Ezratty, 2022 and [The semiconductor manufacturing process (back-end process)](), Matsusada, February 2022.*

[1486] See Tutorial on forming through-silicon vias by Susan L. Burkett et al, January 2020 (16 pages).

**Wafer testing**. Then, testing and characterization is done to make sure the manufactured components meet the required quality. One inspection tool used is electron microscopes, which can also be used to analyze the patterning quality between each patterning cycle. Wafers containing chipsets operating at cryotemperatures can be tested by a cryo-prober like the one provided by Bluefors/Afore and being used by Intel in Oregon and CEA-Leti in Grenoble.

**Electrode formation, wiring and bonding**. These are more traditional steps to add macro-elements to the circuit that will connect them to the outside world. This is done after the chipsets are extracted from the wafer with dicing.

**Packaging**. It is mostly about putting plastic and sometimes metal shielding for specific applications (space, military, quantum) around the chipset and its bonds/wires. The component can be then integrated in a system with its surrounding electronics.

**End-product testing**. The electronic circuit is fully functionally tested here before being used.

The manufacturing yield is the percentage of functional chipsets at the end of manufacturing. Each intermediate manufacturing step has its own yield, and the end yield is the result of the multiplication of each step yield.

### Quantum process specifics

Each and every chipset is manufactured with a specific recipe with many steps involving different tooling and dozens of parameters (tool, chemical compounds, temperature, pressure, angles, ...). Putting in place such processes is tedious and require very specialized skills. The whole manufacturing process for a chipset can last from a couple hours to a couple months depending on its complexity. All in all, a new chipset design, manufacturing and testing can last between a couple weeks to 2,5 years depending on the product and process.

#### *Superconducting qubits*

Manufacturing a superconducting qubit chipset is both rather specific and simple, at least, compared to classical silicon CMOS chipsets and their epitaxy processes, creating pure crystalline semiconducting structures. It explains why so many labs in the world have their own cleanroom able to prototype such chipsets. A superconducting chipset wafer can usually be produced in less than a week when it can last months for CMOS chipsets[1487]. There are of course many variations and as the superconducting chipsets become more complicated, assembling up to three stack chiplets, and with more lithography steps, the production cycle gets longer.

We'll describe here one of the methods to create a superconducting qubit chipset which is derived from the bridge-based Niemeyer-Dolan bridge technique[1488]. The superconducting qubit core feature is its Josephson junction made of three layers: a conducting metal like aluminum and its oxide insulator variant in between. It's surrounded by metal structures for creating capacitances and a resonator[1489].

---

[1487] See a couple examples of superconducting qubits manufacturing process descriptions in Manufacturing low dissipation superconducting quantum processors by Ani Nersisyan et al, Rigetti, January 2019 (9 pages), Simplified Josephson-junction fabrication process for reproducibly high-performance superconducting qubits by A. Osman et al, Chalmers, November 2020 (7 pages) and the thesis Micromachined Quantum Circuits by Teresa Brecht, Yale University, December 2017 (271 pages).

[1488] There are other Josephson junction techniques like the Manhattan bridge. See Improving Quantum Hardware: Building New Superconducting Qubits and Couplers by Thomas Michael Hazard, Princeton, 2019 (136 pages).

[1489] The schema below comes from Resonant and traveling-wave parametric amplification near the quantum limit by Luca Planat, June 2020 (237 pages). It describes the process for the creation of a Josephson junction in a TWPA, and is very similar to a superconducting qubit.

- The wafer substrate is made of either sapphire or intrinsic silicon (meaning monocrystalline and undoped). Silicon is commonplace but has its shortcomings: it must be deoxidized, since $SiO_2$ is damaging the qubit's quality. Sapphire can't be oxidized but is less commonplace. The wafer may be gold plated on its unpolished side to ensure good electric and thermal contacts between the chipset and the chip-carrier.
- Resist deposition is done using the spin-coating technique and with two layers of resist one on top of each other. The bottom one is more sensitive to the e-beam than the one above.
- E-beam lithography exposes some of the resist to an electron beam. This is a rather slow process. It uses a double insolation process with different strengths to attack the two resins layers.
- Development where the resist is removed from the hole exposed by the e-beam and etching which creates an undercut carved in the resist.
- First metal evaporation where a first layer of aluminum is deposited with an angle +θ. It creates the first layer of the Josephson junction.
- An aluminum oxide layer is grown during an oxidation step. The gate can be less than 1 nm thick.
- Second angled metal evaporation to create a new layer above the oxidized aluminum from the Josephson junction gate. It is done in the opposite angle -θ to cover a different area in the hole.
- The residual resist may in some situations be removed with a CMP process or more classically dissolved in solvent during the lift off step. For these, no additional layer or isolation layer is added on the Josephson junction.

All of this process was just about creating a single Josephson junction that is usually 200 nm wide. A classical superconducting qubit contains at least two other circuits: capacitances (about 100 µm to 600 µm wide), a resonator (*aka* superconducting coplanar waveguide) and a microwave network.

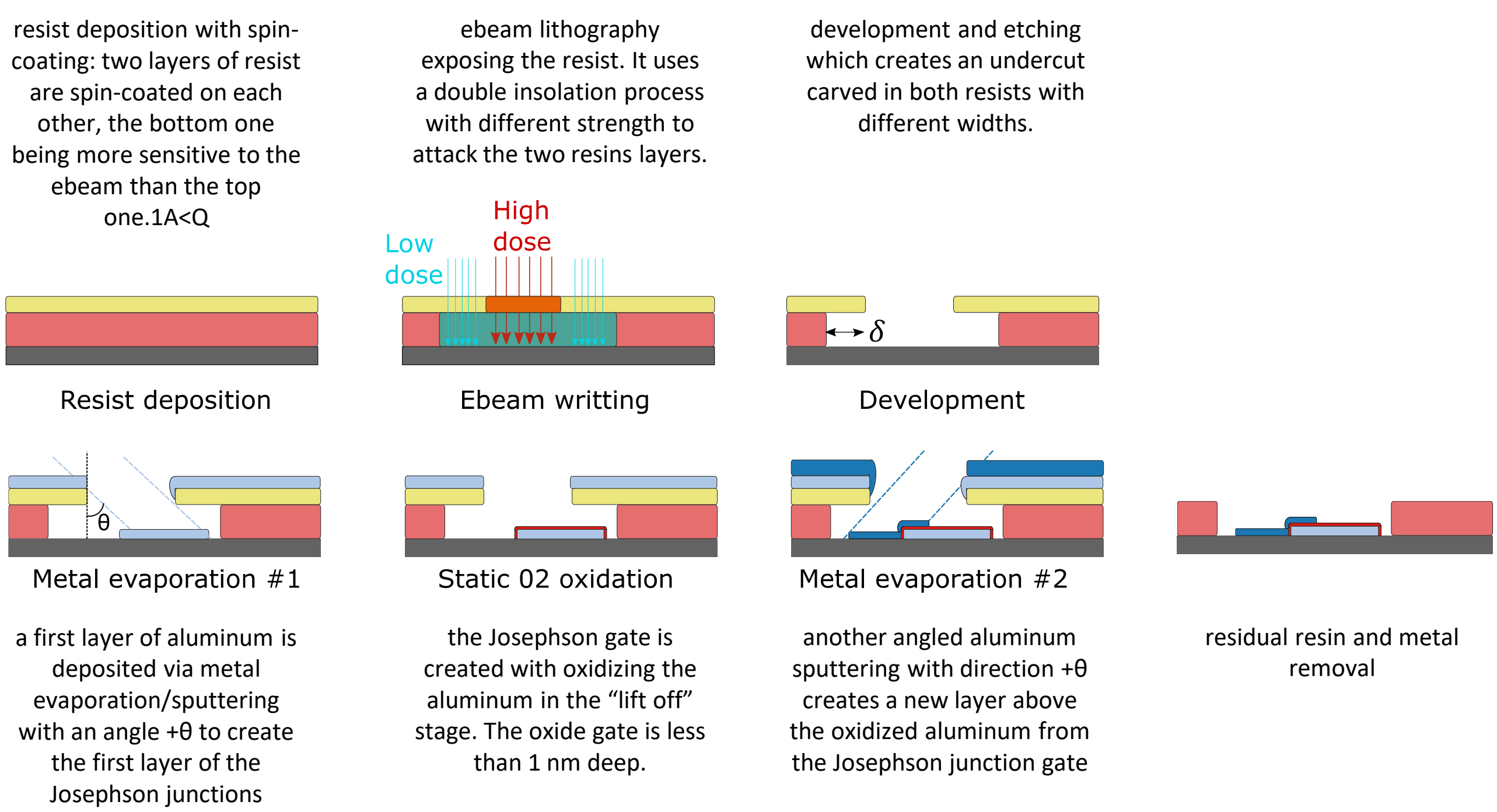

*Figure 534: the process of manufacturing a superconducting qubit or superconducting component like a TWPA. Source: Resonant and traveling-wave parametric amplification near the quantum limit by Luca Planat, June 2020 (237 pages). Comments added by Olivier Ezratty in 2022.*

The network and resonator can be created using 193 nm UV lithography or laser lithography and negative resist etching, meaning the metal is first deposited everywhere (except on the Josephson junction) and then, the resist coating is removed where it was not exposed. RIE (reactive ion etching) can also be used, particularly for creating resonators. This process can make use of aluminum, niobium, TiN (titanium nitride) as developed by John Martinis in 2013 and even indium[1490]. TiN is appreciated thanks to its ability to avoid oxidation. Titanium nitride (TiN) can be used as an isolation layer on top to a sapphire substrate to avoid dielectric losses between the various qubit's components[1491].

Superconducting qubits circuits are usually rather simple and only 2D with no additional metal layers and no insulator since there are no good insulators available. There's an empty space of a minimum 2 mm height above and below the chipset in its (copper) packaging. On the other hand, superconducting electronics dies can superpose up over 10 alternating layers of niobium and dielectric, usually $SiO_2$. The surrounding connections are themselves superconducting and the chipset edge is connected to gold or aluminum wires and bonds. Andreas Fuhrer Janett from IBM Zurich labs developed a process and apparatus to put the chipset in ultra-high vacuum (UHV) during assembly[1492]. According to IBM, the UHV can contribute to create less noisy qubits.

The connectivity constraints explain for example the limitations of the chimera structure in D-Wave superconducting qubits layout. The trend is to create 3D structures, assembling several chipsets, with one being dedicated to electronic signals controlling the qubits, like with Google, IBM and OQC.

The chipsets are assembled using wafer bonding, connecting metal layers using indium thanks to it ductility, even in low temperatures and at relatively low temperature (156°C).

The chipset density is not very high as compared with classical CMOS chipsets. The quality and fidelity of the superconducting qubits depends on several factors including materials purity[1493]. The manufacturing yield of superconducting chipsets can however be as low as 1% but is usually above 70%[1494]. One avenue to potentially improve the manufacturing quality of superconducting qubits would be to produce them with 300 mm CMOS fab technologies. That's what IMEC has been experimenting in 2022 with producing qubits of rather good quality (but not stellar), using argon milling and subtractive processes[1495].

We mentioned a lot aluminum so far. It is not the only superconducting metal used with Josephson junctions. While aluminum is used for superconducting qubits, niobium is used for superconducting electronics (SFQs, already covered) and SQUIDs sensors with the advantage of being a superconductor at 9K versus 1.2K for aluminum.

### *Superconducting nanowire single photon detectors*

Superconducting nanowire single photon detectors (SNSPDs) can be manufactured with NbTiN sputtering on sapphire[1496].

---

[1490] See Sputtered TiN films for superconducting coplanar waveguide resonators by S. Ohya, John Martinis et al, UCSB, 2013 (9 pages).

[1491] See Titanium Nitride Film on Sapphire Substrate with Low Dielectric Loss for Superconducting Qubits by Hao Deng et many als, Alibaba, May 2022 (10 pages). The use of TiN enables qubit lifetimes of up to 300 µs.

[1492] See Ultrahigh vacuum packaging and surface cleaning for quantum devices by M. Mergenthaler, Andreas Fuhrer et al, 2021 (6 pages).

[1493] See Material matters in superconducting qubits by Conal E. Murray, IBM Quantum, 2019 (98 pages).

[1494] It was the yield with IBM's 17 qubits chipsets in 2018 according to Towards Efficient Superconducting Quantum Processor Architecture Design by Gushu Li et al, 2019 (15 pages).

[1495] See Path toward manufacturable superconducting qubits with relaxation times exceeding 0.1 ms by J. Verjauw et al, npj, August 2022 (7 pages).

[1496] See NbTiN for improved superconducting detectors by Julien Zichi, KTH Sweden, 2019 (86 pages).

### Trapped ions circuits

These circuits implementing Paul or Penning traps are manufactured using a mix of techniques with e-beam metal evaporation-based deposition of titanium and gold, etching process, and femtosecond-laser machining for 3D surfaces shaping using tools like those from **FEMTOprint**[1497].

### Photon-generating quantum dots

Like the ones from CNRS C2N and Quandela are manufactured with adding about 100 layers alternating GaAs and GaAsAl compounds using molecular beams epitaxy. Adding these many layers can still be implemented in a couple hours. In the middle of the road, special techniques are used to deposit the planar λ cavity made of a couple hundred of InGaAs. The cylinder cut for the quantum dot enclosing is implemented with ion milling[1498].

### Silicon qubits

Their manufacturing is very close to traditional CMOS manufacturing techniques. It requires UV/EUV photolithography due to the relatively high features density in the chipsets (which can go as low as 10 nm). The etching processes are also rather similar. The materials purity is an important figure of merit to ensure the quality of the manufactured qubits as it is with superconducting qubits.

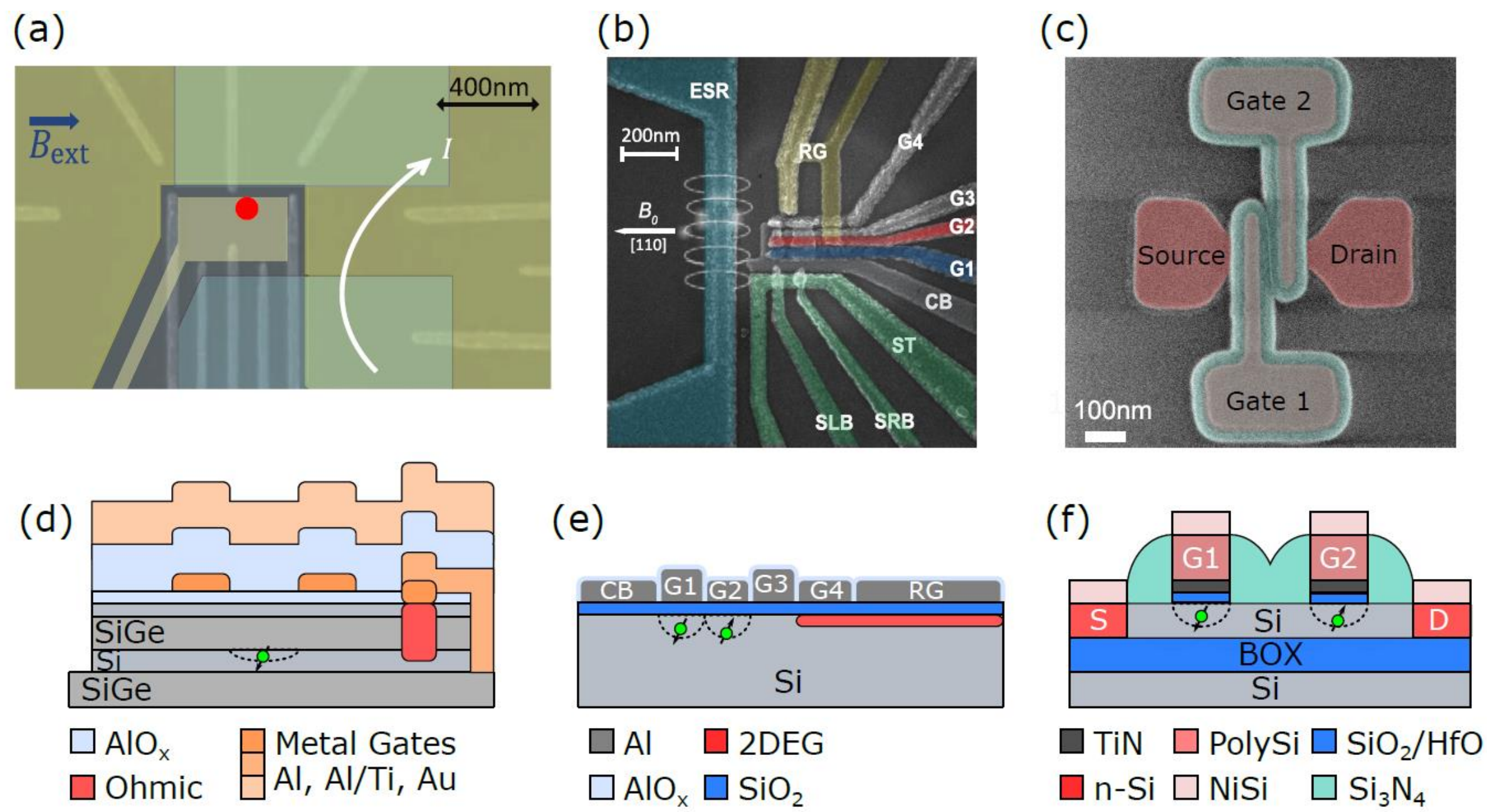

FIG. 1. Silicon quantum dot devices. (a) Scanning electron microscope (SEM) image of an accumulation mode Si/SiGe heterostructure. Two layers of gates, bottom (light grey) and top (dark green) are designed to form two quantum dots (centre of image) and a single-electron transistor for readout (right). The structure contains a micromagnet in an upper metal layer (light green) to produce a magnetic field gradient. The red dot indicates the position of a quantum dot used in ref. [16]. (b) SEM image of a metal-oxide-semiconductor multi quantum dot device with quantum dot gates (G1-4), confinement gate (CB), reservoir gate (RG), an integrated single-electron transistor (green) and microwave antenna for magnetic resonance spin control (blue) usied in ref. [17]. (c) SEM image of a CMOS p-type double quantum dot on an etched silicon-on-insulator nanowire used in ref. [18]. Cross-sectional view of the Si/SiGe quantum dot device (d) MOS double quantum dot device (e) and CMOS double quantum dot device (f) shown above in (a-c).

*Figure 535: various implementations of silicon spin qubits. Source: Scaling silicon-based quantum computing using CMOS technology: State-of-the-art, Challenges and Perspectives by M. F. Gonzalez-Zalba, Silvano de Franceschi, Edoardo Charbon, Maud Vinet, Tristan Meunier and Andrew Dzurak, November 2020 (16 pages).*

---

[1497] See the thesis Multi-wafer ion traps for scalable quantum information processing by Chiara Decaroli, ETH Zurich, 2021 (248 pages) which provides a lot of insights on trapped ion architectures and circuits manufacturing.

[1498] See the details in Near-optimal single-photon sources in the solid-state by Niccolo Somaschi, Valerian Giesz, Pascale Senellart et al, 2015 (23 pages).

The silicon wafers used to create spin qubits are covered by a layer of about 100 nm of $^{28}Si$ using a wafer scale CVD process.

All the other functional and isolation layers using silicon are also based on $^{28}Si$, mostly through silane ($SiH_4$). There are many variants with Si/SiGe heterostructures, Al/$AlO_x$ structures, etc.

The chipset vertical structure is relatively simple, with only a few metal layers, and some control electronics usually placed in a separate chipset that is bonded to the qubits in a 3D fashion. In its various research papers published at APS March meeting in 2022, **Intel** did showcase how manufacturing quality had an impact on the quality of quantum dots spin qubits. In October 2022, they added some information on the quality and yield of their silicon qubits wafers[1499].

### Tools

We'll cover here some specific manufacturing tools that are used for producing quantum technologies semiconductor components. The breadth of tools in semiconductor fabs is much broader with tools from vendors like **ASML** (UV and EUV photolithography) and **Applied Materials** (PVD, CVD, etching and stripping, ...).  Their tools are used in high-volume large fabs while many of the quantum-specific production tools are used more for research purpose and for small scale industrial production.

**Plassys Bestek** (1987, France) develops and manufactures vacuum and ultra-high-vacuum thin film deposition systems with a turnover of about 7M€. Most of their tools are based on physical vapor deposition (PVD) processes (vaporization of metal or compounds for deposition on a substrate, all under vacuum).

Positioned at the end of the 1990s as a key supplier of equipment for the fabrication of superconducting qubits, they have developed a wide range of electron beam deposition systems dedicated to controlled angle evaporation under the name "MEB" which makes Plassys the leader for this technology (Yale University, Rigetti Computing, QCI, NTT, Oxford, CEA Saclay, Qilimanjaro, TU Delft, VTT… rely on their tools). The "MEB" tools used an electron beam to melt and to evaporate materials that allows the deposition of aluminum films for forming the Josephson junctions or for resonators as well as of niobium as underlayer or as resonators. They also provide sputtering tools for depositing various kind of superconducting films (Al, Ti, Nb, Ta, MoSi, MoGe, nitrides…) and other elemental materials or more or less complex compounds. Sputtering tools integrates cathodes on which a bias voltage is applied under a controlled atmosphere of gas mixture including argon for generating a plasma around $10^{-3} – 10^{-2}$ mbar. Positive ions from the plasma are attracted by the cathodes on which a "target" made from the material source you want to deposit. The high energy of the ions sputters the target inducing then the generation of a vapor that condensates on to the substrate.

They also supply the SSDR150 chemical vapor deposition (CVD) reactor for the growth of ultra-pure diamonds which is the raw material for the development of NV center based technologies. This CVD process is using hydrogen and methane ($CH_4$) at a pressure around 100millibars[1500] with the assistance of a microwave source generating a high density plasma. They also handle diamond doping with nitrogen, boron, phosphorus….

Their R&D and production machines dedicated to quantum technologies are now grouped under the Qutek Series brand. In addition to the MEB systems, Qutek series includes MP systems (sputtering deposition for superconducting or photonic devices) and thermal evaporation system for indium bumps (used for connecting superconducting qubits).

[1499] See Intel Hits Key Milestone in Quantum Chip Production Research, Intel, October 2022.

[1500] The CVD diamond growth process is described in Diamond growth by chemical vapour deposition by J. J. Gracio et al, 2011 (75 pages).

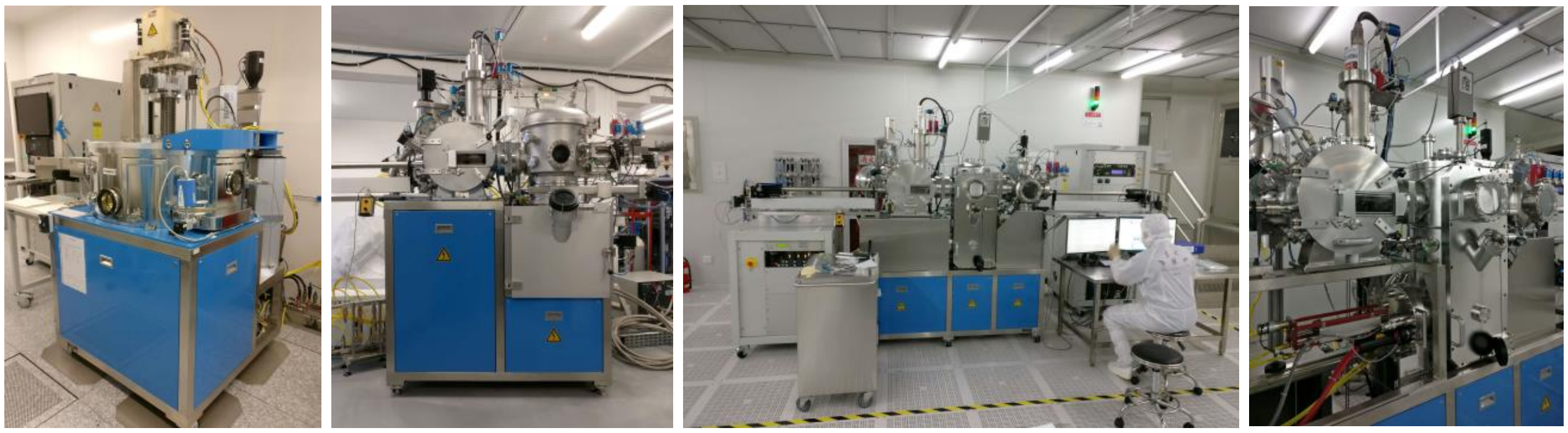

*Figure 536: various production machines from Plassys-Bestek. Source: Plassys-Bestek.*

**Angstrom Engineering** (1992, Canada) is a manufacturing tool vendor. Their Quantum Series line of physical vapor deposition (PVD) systems is adapted to the creation of Josephson Junctions, from using an electron beam source to deposit aluminum, magnetron sputtering for niobium and ion beam cleaning.

**Kelvin Nanotechnology** (2020, UK) is an electron beam lithography and nanofabrication tooling company. It manufactures various miniaturized MEMS and photonic components used in quantum technologies.

These include 3D ion traps, various photonic devices, MEMS gravimeters and lasers built on 200 mm wafers in features going as low as 20 nm. They are based at the James Watt Nanofabrication Centre (JWNC) in Glasgow, Scotland.

**Orsay Physics** (1989, France) is a subsidiary of **Tescan Orsay Holdings** (Czech Republic - France). It provides manufacturing tools for focused ion and electron beam processes.

Out of these, their nitrogen-FIB (i-FIB, for focused ion beam) is being used to create NV centers in nano- and micro-structures with high precision, like in NV center arrays with 2 µm separations between the centers[1501].

**Riber** (1987, France) is a manufacturer of MBE reactors, used mostly in III/V and II/VI multi-layers epitaxy processes.

They handle various MBE processes: solid sources MBE, Plasma-Assisted MBE (PAMBE), Metal-Organic MBE (MOMBE), Gas Source MBE (GSMBE) and full gaseous Chemical Beam epitaxy (CBE). A Riber MBE reactor is being used to manufacture the 100+ layers quantum dots based photon sources from Quandela.

**Raith** (1980, Germany) provides nanofabrication and electron beam lithography instruments. These tools are involved in the manufacturing of all sorts of qubits, trapped ions, superconducting, electron spin, topological qubits, NV centers and nanophotonics.

**Picosun Group** (2003, Finland, 17.4M€) is a manufacturer of atomic layer deposition (ALD) tooling used in the production of various electronic components (imaging sensors, LEDs and OLEDs, MEMS, etc).

Their technology can be used to create graphene structures among other things.

[1501] See i-FIB application note.

They are one of the Finish industry partners of **QuTI**, a 10M€ collaborative research project on quantum related components manufacturing and testing. They partner with VTT, Bluefors, Afore, IQM, Quantastica, Saab, Vexlum and the Finish offices from Rockley Photonics (USA) and CSC (USA).

**Encapsulix** (2011, France) develops ultra-short cycle time ALD systems. It's mostly used in the production of OLED in encapsulated quantum dots.

**Samco Inc** (Japan) is a provider of thin film deposition, microfabrication and surface cleaning, CVD and other treatment machines.

Their tooling includes PECVD systems (plasma-enhanced CVD), SiC CVD systems, ALD (atomic layer deposition) systems, reactive-ion etching (RIE) systems, Inductively Coupled Plasma (ICP) etching systems, Silicon Deep Reactive Ion Etching (DRIE) systems for MEMS device fabrication and TSV (through-silicon-vias) via-hole etching and plasma cleaners. These systems are used to produce various sorts of quantum components like niobium and tantalum based superconducting circuits, from qubits to surface acoustic waves filters[1502] and GaAs photonic components.

**ADNANOTEK** (Taiwan, 1999) is a provider of MBE, PLD (pulsed laser deposition which is a variation of PVD) and Laser MBE PLD, various sputtering systems, EBE (electron beam evaporators), Ion Beam Sputter Deposition (IBSD), ALD (atomic layer deposition), Plasma Enhanced Atomic Layer Deposition (PEALD) and various ultra-high vacuum equipment.

**CVD Equipment Corporation** (USA) is specialized in CVD and dry etching systems that can be used for various semiconductor production, including III/V and nanophotonic chipsets.

**Plasma-Therm** (USA) has a broad range of plasma and ion beam etching and deposition used among other things in GaAs components manufacturing.

The company made several acquisitions: Advanced Vacuum Europe of Lomma (1993, Sweden) in 2011, Nanoplas France in 2015, Nano Etch Systems (2009, USA) in 2016, Kobus and Corial (France) in 2018, JLS Designs Ltd (UK) in 2020. The company opened in 2018 its European Head Office in Grenoble, France, and in 2020, one process and technical support office in Singapore.

**Izovac Photonics** (Lithuania) provides the IZOVAC range of products vacuum coating equipment using vacuum sputtering (magnetron sputtering, Ion Beam Assisted Deposition, Ion beam sputtering, DLC (Diamond-Like Carbon) coating by PECVD (Plasma Enhanced Chemical Vapor Deposition). Their main market are the display and touch screen manufacturing. They also develop customized vacuum deposition equipment.

**Evatec** (Switzerland) provides a family of evaporation, sputtering and PECVD products covering various needs including in MEMS and photonics applications. They also develop and sell wafer cassette-to-cassette processing tools in their Clusterline family.

**Prevac** (Poland) has a breadth of semiconductors manufacturing tools including an UHV Magnetron Sputtering System working with 3-inches wafers and PLD systems. They also sell an UHV multichamber cluster tool to automate a process with a thin film layer growth deposition chamber, load-lock chambers and a transferring tunnel.

---

[1502] See Towards practical quantum computers: transmon qubit with a lifetime approaching 0.5 milliseconds by Chenlu Wang et al, NPJ, January 2022 (6 pages) and Niobium (Nb) Plasma Etching Process (RIE or ICP-RIE), Samco.

**Seki Diamond Systems** (Japan) is a subsidiary of **Cornes Technologies** (USA) that sells CVD diamond reactors producing synthetic diamonds and supporting Microwave Plasma CVD, Hot-Filament CVD and Low Temperature CVD. It covers broad industry use cases.

**Polyteknik** (2005, Denmark) provides PVD and coating systems, including their Flextura e-beam PVD system.

**Besi** (1995, The Netherlands) or BE Semiconductor Industries, is the worldwide leader in semiconducting assembly machines (die attach, packaging, plating). One key use case if 3D chiplets assemblies. The company participates to the EU project MATQu to create a manufacturing capacity of superconducting chipsets on 300 mm silicon wafers.

**SOITEC** (1992, France) is a company producing SOI wafer which contain an isolation layer of $SiO_2$. These wafers are commonplace in many quantum technologies semiconductor components.

SOITEC acquired EpiGaN (2010, Belgium, 4M€) in 2019. It adds GaN wafers production to their portfolio.

Some other vendors can be mentioned like **RECIF Technologies** (France) with its wafers handling and sorters, **Heidelberg** (Germany) and its mask writer, **Süss MicroTec** (Germany) and its photomask handling and mask aligners, Keysight Technologies (and its NX5402A silicon photonics hybrid wafer testing system), **Oxford Instruments** (UK) and its RIE plasma etchers from the Plasmalab family and ALD systems, **Vistec** (USA) and **STS Elionix** (USA) and their e-beam writers, **Transene Company** (USA) and their etching systems, **Pureon** with its diamond based Chemical Mechanical Planarization tools (CMP)[1503], **Polygon Physics** (2013, France) which provides ion, electron, plasma and atom sources based on ultracompact and ultralow power electron cyclotron resonance plasma technology (ECR) and Multi Beam Sputtering tools (MBS), **JEOL** (Japan) and **Multibeam** (USA) and their e-beam lithography systems, **Thermo Fisher** (USA) and its e-beam lithography and ion milling systems, **Veeco** (USA) and its lithography, MBE, CVD, PVD, ion beams, ALD and dicing systems, **Aixtron** (1983, Germany) and its CVD systems, **NuFlare** (Japan) and its mask writers and epitaxial growth reactors, **AJA International** (1989, USA) and its thin film deposition systems including magnetron sputtering, e-beam evaporation, thermal evaporation, and ion milling systems and **Denton Vacuum** (1964, USA) and its evaporation, sputtering, PE-CVD and ion beam deposition tools.

Let's add a couple software design tools:

**NanoAcademic Technologies** (2008, Canada) is a company created by Hong Guo rom McGill University and Yu Zhu and Leil Liu (all coming from China) which sells quantum materials software simulation tools like NanoDCAL. It is used to simulate the physics of quantum chipsets like superconducting qubits.

**QuantCAD** (2021, USA) is a company created by Michael Flatté and based in Iowa that develops and sells CADtronics and qNoise, a suite of simulation software that models noise and current in quantum devices. It is used to design various quantum components, including quantum sensors and optoelectronic devices.

[1503] In 2020, Microdiamant (Switzerland) acquired Eminess Technologies from Saint Gobain and was rebranded as Pureon.

## Other enabling technologies vendors

These companies are developing physical components and enabling technologies that can play a role in building quantum computers.

More often, as this market remains limited to research, these startups are more generalist and target broader markets than quantum computing, covering physics research in general and even various industrial applications.

**Aeponyx** (2011, Canada, $11,4M) is a fabless micro-optical switch semiconductor chips designer and manufacturer, specialized in Micro-Electro-Mechanical-Systems (MEMS) and Silicon Photonics.

**Alter Technology** (2006, Spain/Germany) is a subsidiary of the German group TÜV NORD specialized in micro and optoelectronics engineering for space and harsh environment applications.

It has labs in UK, France, Spain and Italy. They develop several quantum enabling technologies like frequency-stabilized lasers used to control cold atoms, an ion-trap chip carrier, entangled sources of photons for space based QKD, a squeezed light quantum MEMS gravimeter.

**AuroraQ** (2017, Canada) creates communication systems based on superconducting qubits, including quantum communication repeaters. It is complemented by the QSPICE Design software which allows the design of superconducting quantum circuits. In other words, this is an ultra-niche market[1504].

**DiamFab** (2019, France) is a spin-off of Institut Néel in Grenoble specialized in the growth of doped diamond layers on a diamond wafer substrate.

Among other markets, they also target NV center use cases in quantum technologies. Diamond is also used as a high-performance semiconductor for power applications for diodes and field-effect transistors.

**HiQuTe Diamond** (2022, France) is a company created by Riadh Issaoui, Ovidiu Brinza, Fabien Bénédic, Alexandre Tallaire and Jocelyn Achard, who are researchers from LSPM in Paris, France (Laboratoire des sciences des procédés et des matériaux). They produce high quality diamond crystals used in quantum technologies.

**Elementsix** (1946, Luxembourg) is a subsidiary of De Beers Group, the world's leading diamond producer, which, among other things, manufactures synthetic diamonds for use in NV centers based systems, mostly used in quantum sensing.

They hold a large number of patents in the related processes. In September 2021, they launched DNV-B14, a new chemical vapor deposition (CVD) made quantum-grade diamond with a uniform and x 10 higher density of NV spin centers.

**Hummink** (2020, France) developed a patented technology combining a nanometric "pen" with an oscillating macro-resonator to perform a capillary deposition of various liquids.

It can print conducting materials with an existing choice of 10 different materials. It can be used to add precision items on devices in 3D.

[1504] See The Geometry of a Quantum Circuit and its Impact on Electromagnetic Noise, 2018 (15 pages).

**Labber Quantum** (2016, USA) develops software solutions for controlling the qubits of experimental quantum computers with Python scripting handling electronics hardware control (AWGs, DACs, ADCs), data storage and visualization. They are used to calibrate qubits. The startup was acquired by **Keysight Technologies** in March 2020.

**LakeDiamond** (2015, Switzerland, €2M) produced synthetic diamonds used to create NV centers qubits in diamonds or with quantum sensing.

They use vacuum deposition with the CVD method (Chemical Vapor Deposition). The company closed in February 2020 after getting funding from an ICO in 2018 (Initial Coin Offering, using some crypto currency).

**Lucigem** (2016, Australia) manufactures fluorescent nano-diamonds that can be used in various quantum applications, particularly for medical imaging. The company is the result of work carried out at Macquarie University in Sydney.

**Diatope** (2021, Germany) creates diamonds with NV centers for quantum sensing and quantum computing applications. It is a spinoff from the Institute for Quantum Optics at Ulm University by Johannes Lang, Christoph Findler and Christian Osterkamp.

They produce NV centers using isotopically purified $^{12}C$ and do provide NV centers benchmarking services.

**Qzabre** (2018, Switzerland) creates NV center-based tips and probes to be used in scanning microscopes. They also sell a NV center microscope, the QSM. The startup was created by Christian Degen from ETH Zurich.

**Adamas Nano** (2010, USA) sells nanodiamond particles for various use cases including NV centers-based sensors. **Bikanta** (2013, USA, $1.7M), **Cymaris Labs** (2004, USA) and **FND Biotech** (2016, Taiwan) sell fluorescent nanodiamond targeting medical imaging applications. **Diamond Materials** (2017, Germany) is a manufacturer of various variations of diamonds including NV centers. **Quantum Diamant** (Germany) also produces NV centers diamonds, for quantum sensing.

It is a spin-off from TUM (the Technical University of Munich). **Photonanometa** (2011, Russia) is another producer of diamond with NV-center defects.

**M-Labs** (2007, Hong Kong), formerly known as Milkymist, is working on the ARTIQ (Advanced Real-Time Infrastructure for Quantum physics) project.

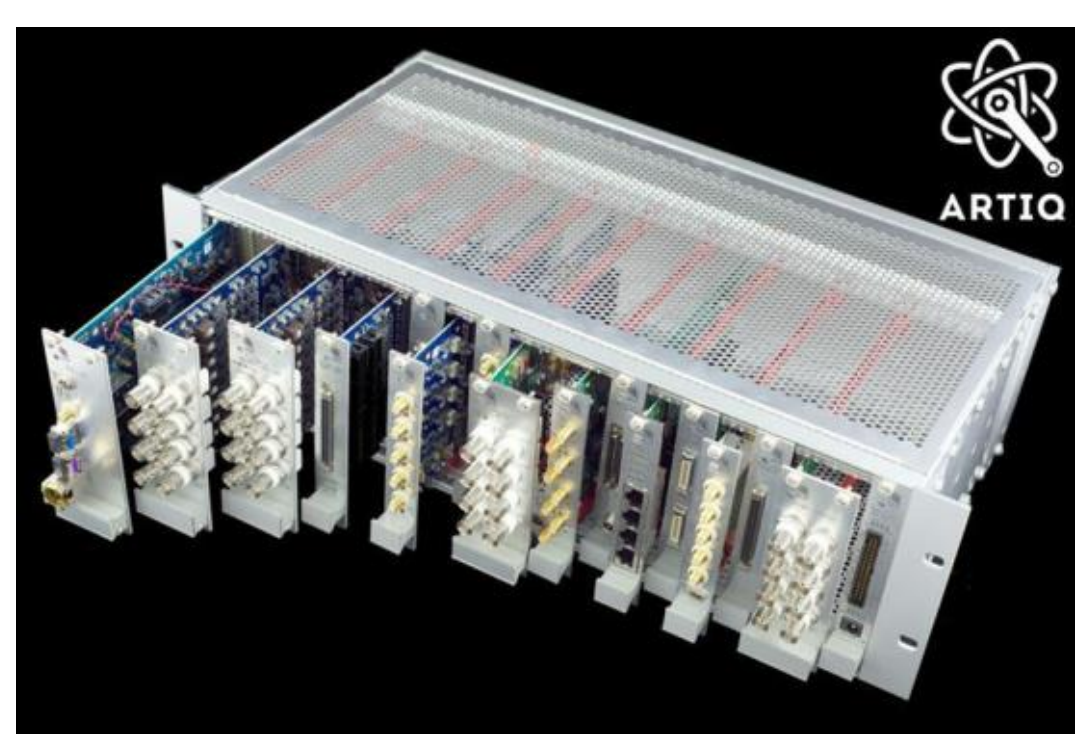

This system combines hardware and a real-time operating system to control quantum computer hardware based on trapped ions. It's a bit like the trapped counterpart of startups such as the Israeli Quantum Machines. They have developed their own FPGA circuit for ARTIQ, all programmed in Python. The solution has been developed with the Ion Storage Group team at NIST in the USA, working on ion trapped qubits.

The company was founded by a French engineer, Sébastien Bourdeauducq.

**Nano-Meta Technologies** (2010, USA) is a spin-off from the University of Perdue that aims to create a quantum information storage system. It is in fact a private contract research laboratory.

It commercializes intellectual property on technologies associating photonics and nanomaterials that could be used in quantum cryptography systems.

**Photon Spot** (2010, USA) develops nanowires based single photon detectors. They have received a DARPA funding of $100K in 2014 and $1.5M in 2015.

**QBee.eu** (2020, Belgium) is a sort of quantum accelerator or incubator created by Koen Bertels, who also leads the Quantum Computer Architectures Lab in TU Delft and also works at Qutech.

They run various research projects like defining a quantum micro-architecture for quantum accelerators using the OpenQL language from TU Delft, a quantum computing emulator, quantum genomics and quantum finance plus some services in education and consulting.

**Q-Lion** (2019, Spain) develops an error correction code solution for trapped ion qubits. The startup is a spin-off from the Bank of Santander's Explorer incubation program. It was created by Andrea Rodriguez Blanco, who was still working on a thesis in 2020.

**QuantTera** (2005, USA) is a contract R&D company created by Matt Kim that develops nano-engineered photonic devices targeting photonic telecommunications and wireless applications. It is mainly using silicon-germanium based photonics. It says it target quantum applications, with no details.

**QuTech** (2014, The Netherlands) is the quantum hardware spin off from TU Delft University. It collaborates with Intel in the development of superconducting qubits and with Microsoft in topological quantum.

The company is an applied contract research laboratory. It also develops software, such as the **Quantum Inspire** development platform, which enables quantum algorithms to be run on conventional computers in emulation mode. It provides a graphical programming interface in the QASM language. The code can then be executed in emulation mode in the cloud on a classic machine, the Dutch national supercomputer Cartesius, with 5, 26 and 32 qubits, depending on the chosen package.

Cartesius is equipped with thousands of Intel Xeon and Xeon Phi CPUs and a few dozen Nvidia Tesla K40m GPUs with 130 TB of memory delivering 1.84 PFLOPS. The equipment comes from Atos. Quantum Inspire also provides cloud access to QuTech 5 superconducting qubits and 2 electron spin qubits since April 2020.

**S-Fifteen Instruments** (2017, Singapore) is a spin-off from the renowned CQT laboratory and develops qubit control systems, entangled photon sources, single photons detectors and quantum cryptography solutions covering QKD and QRNGs.

**StarCryo Electronics** (1999, USA) creates SQUIDs sensors used mostly in quantum sensing and other cryo-electronics products (cables, connectors, ...).

**Vapor Cell Technologies** (2020, USA) provides alkaline atom capsules, mainly rubidium, for use in various miniaturized solutions using cold atoms[1505]. The company was founded by Doug Bopp, a former NIST researcher from Boulder, Colorado.

**Zyvex Labs** (1997, USA) develops atomic precise manufacturing (APM) solutions based on STM (Scanning Tunneling Microscopy) that can be used to produce components for use in quantum computing (such as the deposition of dopants for superconducting qubits and silicon) and quantum metrology.

They were funded by NIST, DARPA and the Department of Energy SBIR research programs. The company was founded by Jim Von Ehr. Zyvex announced in September 2022 a sub-nm version of its STM solution, the ZyvexLitho1.

## Raw materials

For any new hardware technology, it is now a common practice to wonder about its environmental friendliness. We've already been dealing with the energetic dimension of quantum computing. Another key aspect to investigate is the raw materials that are used. What are their sources of supply, their global reserves, their economic and environmental cost of extraction, consumable raw materials if any, and finally, the recycling processes of these materials?

In this exclusive content, I propose a first broad inventory of the different raw materials used in and around quantum technologies of all types, particularly in quantum computers. All these elements are positioned in an in-house Mendeleev periodic table of elements, *below*[1506].

We mainly have two types of materials to study: those used in qubits and the supplemental materials, particularly for cables and other supporting structures as well as the gases used in cryostats, mostly helium 3 and 4.

The materials used in qubits are sometimes quite rare (strontium, ytterbium, beryllium). Their selection is based on their energy transitions which correspond to laser or microwave wavelengths that can be used practically with market sources.

Other constraints explain their choice such as the stability of some of these energy levels. Some materials are very rare but their needs in quantum technologies remain marginal in proportion to their production and world consumption.

This is at least the case as long as millions of quantum computers using them are not manufactured. We are not yet at the stage where the consumption of certain elements would come mostly from quantum technologies, as may be the case for smartphones concerning certain rare earths and minerals such as the famous coltan[1507].

How about rare earth elements? Out of the 17 elements in that category who mainly sit in the lanthanum row in Mendeleev's elements table, about 6 of them are used in quantum technologies: yttrium, praseodymium, dysprosium, europium, erbium and ytterbium, the two later being commonplace in trapped ions computing.

---

[1505] See Chip-scale atomic devices by John Kitching, 2018 (39 pages) which makes a very interesting inventory of measurement components using this technology: magnetometers, gyroscopes, atomic clocks. You will say that this should go in the metrology section and you will be right.

[1506] See also this very nice illustrated poster: The Periodic Table of the Elements, in Pictures.

[1507] The coltan is the contraction of columbite-tantalite. It is used to recover tantalum and niobium. If it is an important source for tantalum, it is in fact secondary for niobium compared to other minerals. See USGS Mineral Commodity Summaries 2020, the equivalent of the French BRGM (204 pages) that helped me create this part.

In July 2022, Turkey announced the discovery of large reserves of rare earths minerals potentially exceeding China's reserves. But the announcement was probably overstated, preprocessed minerals getting out of mines and rare earth oxides produced after separation[1508].

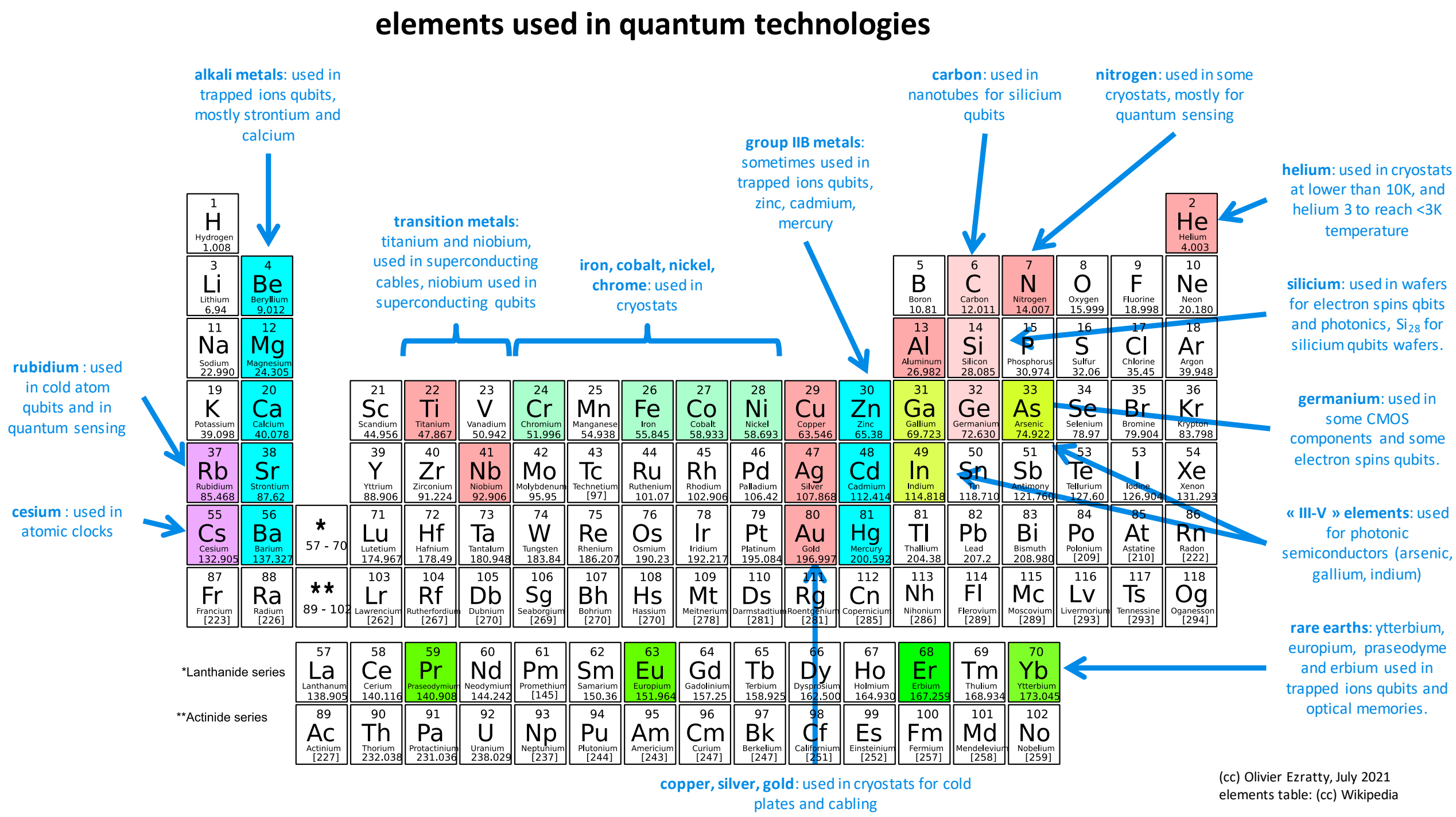

*Figure 537: table of elements and those who are used in quantum technologies. (cc) Olivier Ezratty, 2021.*

One differentiating aspect of quantum technologies relates to the isotopes used which are sometimes the rarest of their elements. This is the case for helium (3) used in cryogenics below 4K or for cesium (133) for atomic clocks or rubidium (87) in cold atoms. Silicon (28) is used in silicon qubits and, although it is the most abundant isotope, requires costly refining. Carbon (12) is also used in nanotubes like with the startup C12 Quantum Electronics, while Carbon 13 is used in some NV center structures. Some of these isotopes are purified with centrifugal separation, a technique well known in nuclear physics, both civil and military.

I do not mention in this inventory the materials used in the production of semiconductors, such as fluorine and other various solvents. And there are many of these!

We will also not deal with the recycling of quantum computers, an issue that has not yet arisen due to their current very limited number. However, it can be reduced to the more generic issue of recycling various electronic devices.

## Helium

Helium is a great paradox in the table of elements. It is the second most abundant element in the Universe after hydrogen. Nuclear fusion does the rest to create all the other elements in first- and second-generation stars. Yet, this element is quite rare on Earth and its reserves are dwindling. It is a noble, inert gas that does not interact chemically with any other element because its electron layer is complete with two electrons. Lighter than air, it tends to leave the atmosphere. As we have seen in detail in the cryostats section, page 465, helium is used for cooling superconductors and electron spins qubits systems.

---

[1508] See Turkey Discovers 694 million mt of Rare Earth Element Reserves, with Infrastructure Construction Starting This Year, July 2022 and Turkey Probably Hasn't Found the Rare Earth Metals It Says It Has by Chris Baraniuk, Wired, July 2022.

As soon as one needs to go below 1K, one must use a mixture of two helium isotopes, $^4He$ which is the most common and stable (with two neutrons) and $^3He$ which is much rarer (with only one neutron). For cryogenics above 1K, $^4He$ is sufficient.

For at least a decade, many specialists have been concerned about a shortage of $^4He$ supply. It is commonly used for cooling superconducting magnets in particle accelerators such as the CERN LHC and in MRI scanners or to inflate balloons. It is also used as a neutral gas for the production of semiconductors. Fortunately, new sources of natural gas from which $^4He$ can be extracted have emerged, notably in Tanzania and Qatar[1509] .

But a low annual growth in demand of just 1.6% is too high compared to production forecasts. Air Liquide is one of the major players in this global market, operating a large $^4He$ extraction and production unit in Qatar, linked to their gas operations. It seems however than the shortage is temporarily gone[1510].

The $^3He$ isotope is rather rare, therefore quite expensive! It was historically a by-product of the storage of tritium-based H-bombs. Tritium gradually disintegrated to produce $^3He$. It was therefore recovered from H-bomb stockpiles!

With the reductions in nuclear weapons stockpiles, the production of $^3He$ is now coming from specialized nuclear power plants. Tritium can be produced with irradiating lithium or with tritium-controlled decay in specialized nuclear facilities, such as those controlled by the US Department of Energy. Tritium is an isotope of hydrogen with one proton and two neutrons.

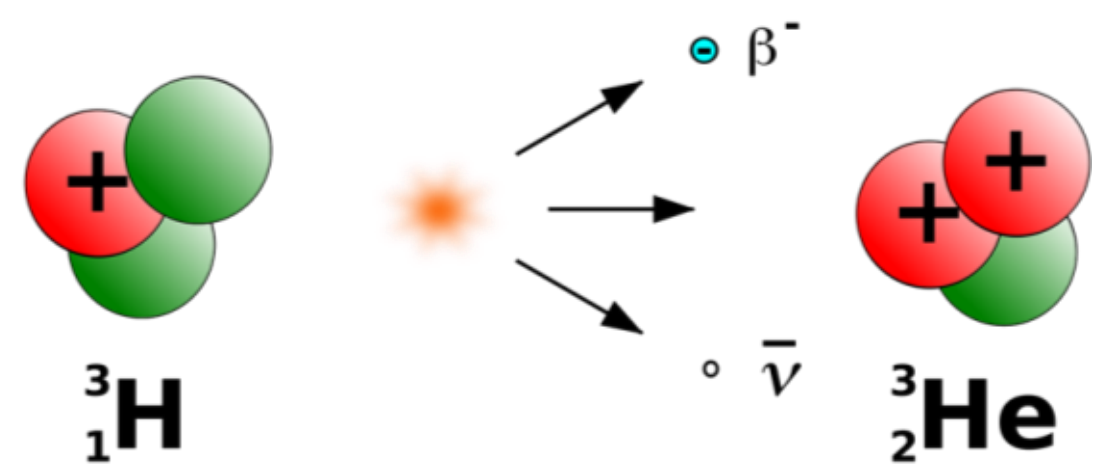

*Figure 538: Helium 3 is a by-product of tritium, an isotope of hydrogen with two neutrons.*

$^3He$ is produced at the U.S. Department of Energy's Savannah site in South Carolina and at the Canadian CANDU power plant[1511].

The price of $^4He$ gas is around €20 per liter while the price of $^3He$ gas is between €2K and €3K per gas liter.

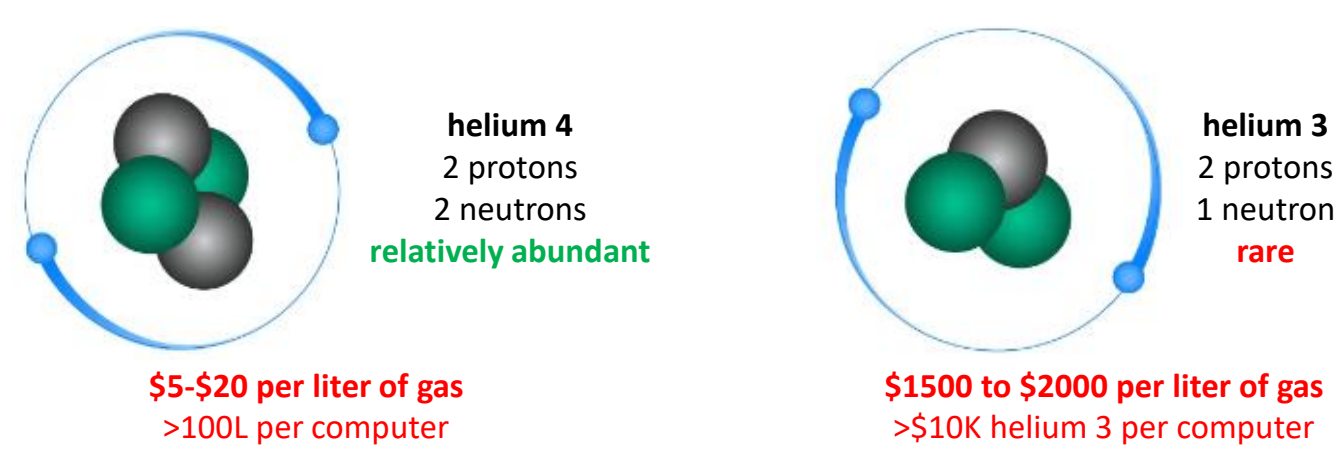

*Figure 539: price tags for helium 3 and 4… as gas!*

A typical dilution-based cryostat requires 15 to 18 liters of $^3He$ gas for a little over 100 liters of $^4He$ gas! The gases are purchased separately and mixed at the right dosage by the manufacturer of the dry cryostat.

At the end, it is therefore necessary to pay at least 30 to 40K€ of $^3He$ and $^4He$ per dry cryostat.

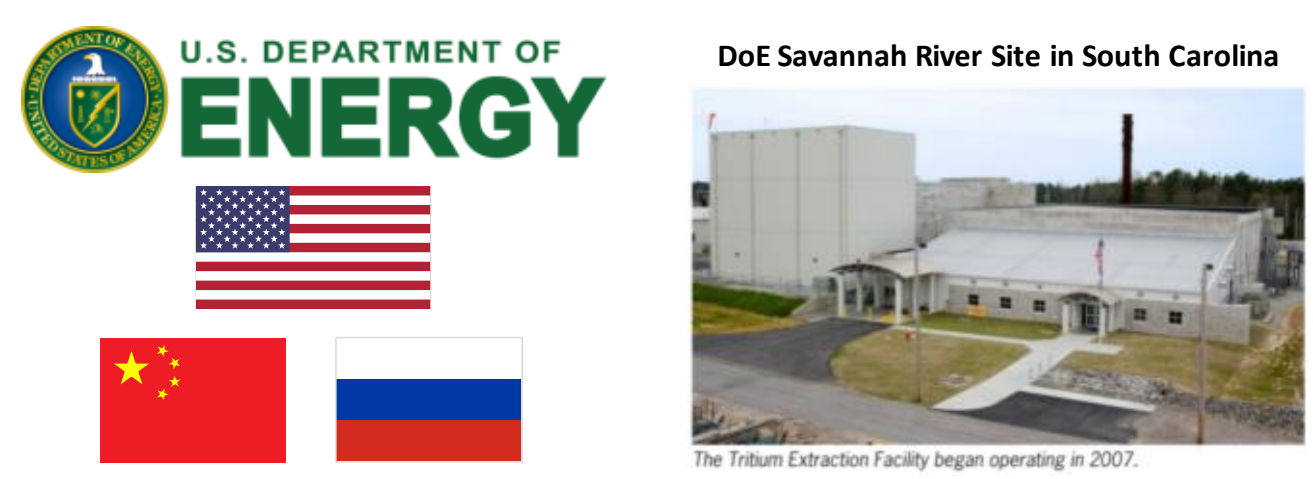

*Figure 540: Savanah River Site is one of the few places where helium 3 is produced in the world.*

The $^4He$ which feeds the pulsed head and passes through the large compressor must be highly purified.

---

[1509] See Helium - Macro View Update, Edison Investment Research, February 2019 (21 pages).

[1510] See Helium shortage has ended, at least for now, June 2020.

[1511] See Savannah River Tritium Enterprise (4 pages). Helium-3 is also exploited in various specialized applications: in neutron detectors used in security systems, in oil exploration, in medical imaging and in nuclear fusion research. Also see CANDU Reactor, Wikipedia.

France has some $^{3}He$ production capacities located in a CEA nuclear reactor in Grenoble. But it does not necessarily use them for quantum computers because this production is too expensive[1512].

We can also find $^{3}He$ on the surface of the Moon but it is not very practical to extract it and ship it back to Earth even if it is technologically possible[1513]! This isotope could be interesting to feed nuclear fusion reactions, pending its complicated technological development.

$^{3}He$ is therefore a real bottleneck in the production of superconducting and electron spin quantum computers! It cannot even be avoided for the latter, which requires a temperature of about 1K[1514].

**Silicon**

Silicon is the key element in many semiconductor components used in or around quantum processors. While being the second most abundant element in the Earth's crust after oxygen, the silicon used in semiconductors comes come from a few quartz mines. This is because quartz is composed of at least 97% silicon, which is easier to refine. After chemical-based refinement, silicon is turned into large cylindrical ingots which are then sliced into thin wafers. Wafers are then processed in semiconductor fabs with transistors that combine silicon oxide and different doping materials such as hafnium.

Silicon qubits require using $^{28}Si$, because the null spin of its nucleus does not interfere with the spin of the trapped electrons used as the qubit observable. The silicon wafers on which the qubits are etched are covered with a thin layer of $^{28}Si$. $^{28}Si$ is the most abundant variant of the element while $^{29}Si$ represents less than 4%.

$^{28}Si$ made headlines in 2010 when some German researchers created a perfect crystal ball made of $^{28}Si$ to accurately determine the Avogadro number, which determines the number of elements, here atoms, in a mole[1515]. The tests were carried out on a 5 kg sample at a cost of 1M€ In 2014, an American team improved the purity of $^{28}Si$ to 99.9998% with pumping silicon ions in a magnetic field, allowing it to be separated by mass[1516] .

This continued in 2017 with 99.999%[1517] $^{28}Si$ produced by a team of Russian and German researchers. The interest of $^{28}Si$ was to allow a precise counting of the number of silicon atoms in the mass considered, because of its perfect crystal structure, dimensioned by X-ray interferometry. The Avogadro number determined by the 2010 experiment was $N_A = 6.022\ 140\ 84(18) \times 10^{23}$. The ambition of these two projects was to create a new material standard of the kilogram, the 1889 material standard preserved in France that degrades by oxidation.

---

[1512] See Isotope Development & Production for Research and Applications (IDPRA), Supply and Demand of Helium-3, 2016, Responding to The U.S. Research Community's Liquid Helium Crisis, 2016 (29 pages) and How helium shortages will impact quantum computer research by James Sanders, April 2019.

[1513] See There's Helium in Them Thar Craters!. China is planning to harvest Helium 3 on the Moon.

[1514] Helium-4 is used to cool superconducting magnets in MRI systems. It is also used to cool the magnets of the LHC at CERN. The constraints are different: it is just a matter of obtaining superconductivity for the magnets that focus the particle beams. The required temperature is between 1.8K and 4.5K, much "hotter" than the 15 mK of electron-based quantum processors (superconductors, silicon, NV Centers, Majorana fermions). On the other hand, the volumes to be cryogenized are much larger. In some cases, however, the required temperature can fall below 1K, particularly for the search for dark matter. In CERN's LHC, 9 Tesla magnets are cooled to 1.8K with 18 kW cryostats that handle 120 tons of helium 4.

[1515] See An accurate determination of the Avogadro constant by counting the atoms in a 28Si crystal by B. Andreas, 2010 (4 pages). Silicon 28 was obtained by centrifuging silicon fluoride ($SiF_4$) gas, then transformed into $SiH_4$ which was then used to create the crystal by vacuum deposition of purified silicon. All this was carried out in different laboratories in Russia, in Nizhny-Novgorod and Saint Petersburg. The researchers involved also came from Italy, Australia, Japan, Switzerland and BIPM in France, from their respective weights and measures offices.

[1516] See Purer-than-pure silicon solves problem for quantum tech by Jonathan Webb, 2014 which refers to Enriching $^{28}Si$ beyond 99.9998% for semiconductor quantum computing by K J Dwyer et al, 2014 (7 pages).

[1517] See A new generation of 99.999% enriched $^{28}Si$ single crystals for the determination of Avogadro's constant by N V Abrosimov et al, 2017 (12 pages) which describes very well the process of purification of $^{28}Si$, the source of the illustration on this page.

Finally, in 2018, the Avogadro number was redefined in the international measurement system as a slightly different constant of 6.022 140 76 × $10^{23}$ $mol^{-1}$. Indirectly, however, these two experiments did advance the know-how of $^{28}Si$ purification, at a time when its interest in creating silicon qubits was barely in the radar. What a good illustration of serendipity in science!

The silicon purification process is complex. It involves the production of silicon tetrafluoride (SiF4) of all isotopes. Enrichment in $^{28}Si$ is carried out in a centrifuge, originally at the Central Design Bureau of Machine Building in St. Petersburg, in fact, a former plutonium enrichment plant reassigned for this use in 2004.

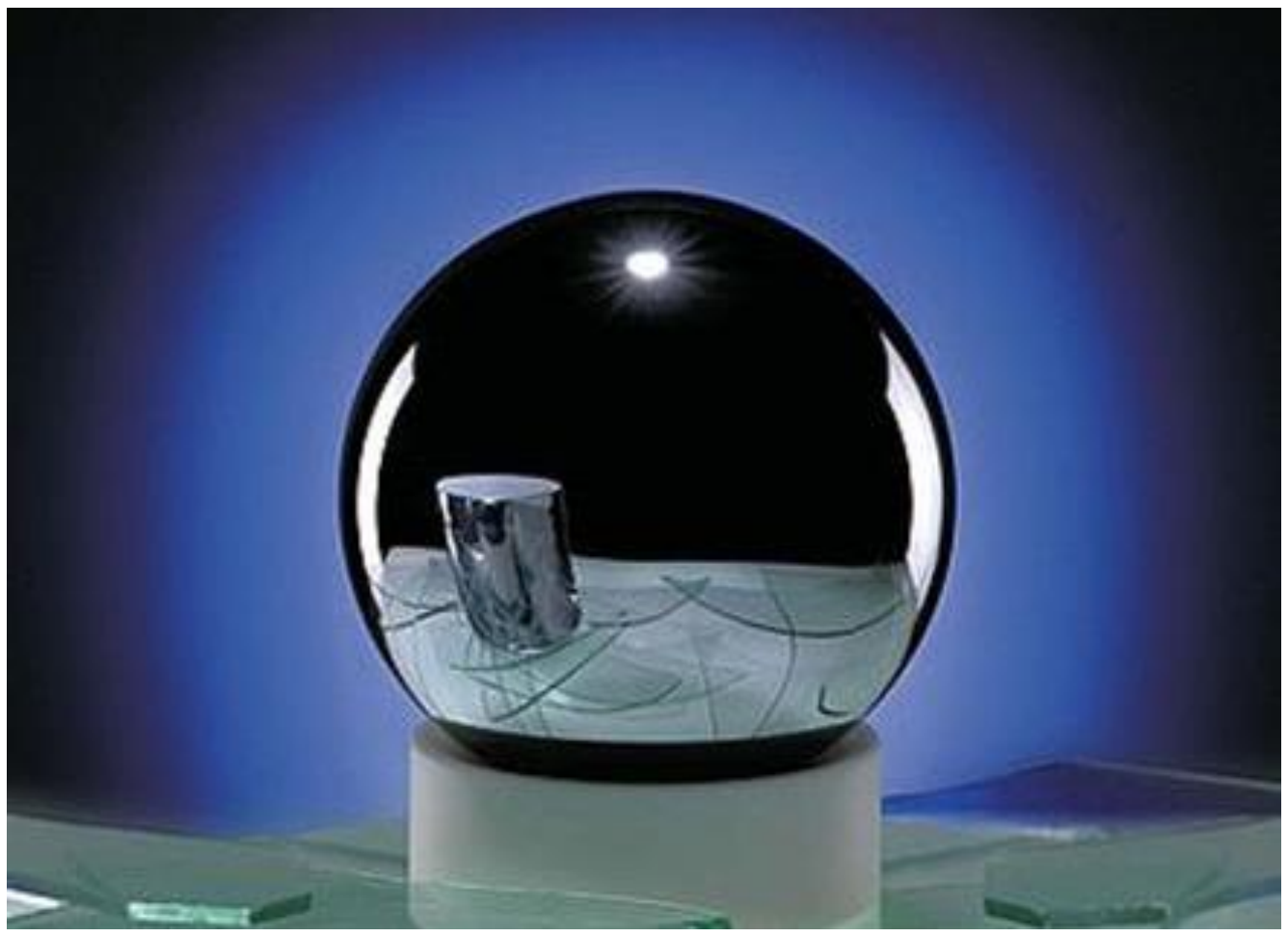

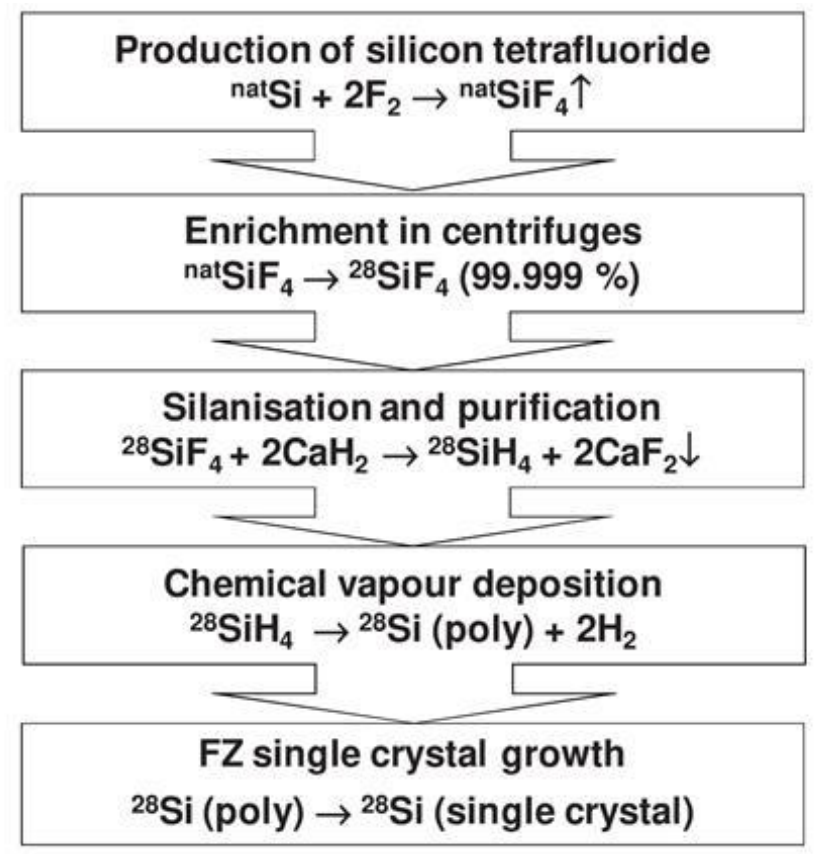

**Figure 4.** Main technological steps of the $^{28}Si$ crystal production ($^{nat}Si$: silicon of natural isotopic composition).

*Figure 541: silicon 28 was initially produced to create a replacement for the reference kilogram used in the international metric system, as a way to determine the Avogadro number. Purifying silicon 28 was a figure of merit of this quest that is now reused in the silicon spin realm.*

The gas is transformed into silane ($^{28}Si\,H_4$) at the **Institute of Chemistry of High-Purity Substances** of the Russian Academy of Sciences in Nizhny-Novgorod. It can then be deposited by vapor deposition (CVD) on silicon, releasing hydrogen. The resulting ingot can then be stretched to create a perfectly crystalline silicon ready to be sliced into wafers. CEA-Leti researchers are also working with Russian teams at Nizhny-Novgorod on the process for vacuum deposition of $^{28}Si$ on 300 mm wafers[1518]. In October 2021, **Orano** announced its ambition to produce $^{28}Si$ in France.

**Air Liquide** is also partnering with the Nizhny-Novgorod laboratory for this process of CVD (chemical vapor deposition) of $^{28}Si$ on a 30 to 60 nm thin film that is 99.992% pure[1519] above a conventional silicon wafer. Knowing that Air Liquide also masters the conversion of SiF4 into silane.

### Germanium

Germanium is a semiconductor metalloid that is part of the III-V family. It is used in many fields: in photonics, in SiGe heterojunction bipolar transistors which are used for the amplification of weak microwave signals as well as in electron spin qubits chipsets.

[1518] See 99.992% 28Si CVD-grown epilayer on 300 mm substrates for large scale integration of silicon spin qubits by V. Mazzocchi of CEA-Leti and colleagues from France and Russia, 2018 (7 pages).

[1519] See Quantum computing: progress toward silicon-28, April 2018.

With spin qubits, it must be isotopically purified to generate $^{73}$Ge which corresponds to 7.36% of its proportion (in purple in the chart *opposite*). It is a stable, natural and non-radioactive isotope. Germanium is generally extracted from zinc ores and also from zinc-copper ores. In 2019, 130 tons of germanium were produced, with China being the main supplier with 85 tons[1520]. Data on known reserves are variable and are estimated at approximately 9,000 tons, mainly located in China, Canada and the USA. Along with gallium and indium, which are also III-V materials, germanium is considered a critical resource.

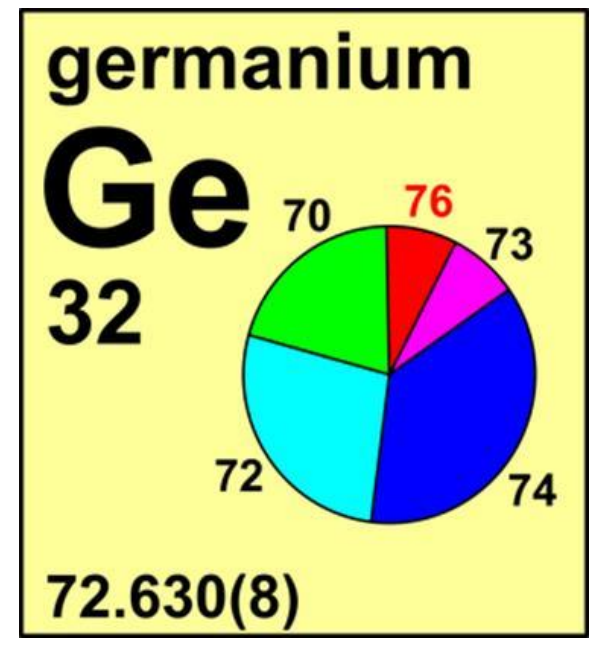

Isotopic purification of germanium is carried out by the same Russian teams at Nizhnii Novgorod as those producing $^{28}$Si. It uses a germanium tetrafluoride centrifugation process similar to the one used to produce germanium tetrafluoride and explained in Figure 541 [1521].

### Rubidium

Rubidium is an alkali metal used to create cold atom qubits that are excited into highly energetic Rydberg states. It is also used in quantum sensing, notably to create atomic clocks and micro-gravimeters. It is an alkaline, soft, silvery metal with a melting temperature of only 39.3°C (in Figure 542).

In a neutral atom computer, the metal is used very sparingly. It is supplied in ampoules of a few solid grams. It is heated in a small box to be sublimated into gas which then feeds the vacuum chamber where the lasers will trap individual atoms. The metal costs about $85 per gram and about $1600 per 100g. It is readily available from chemical companies.

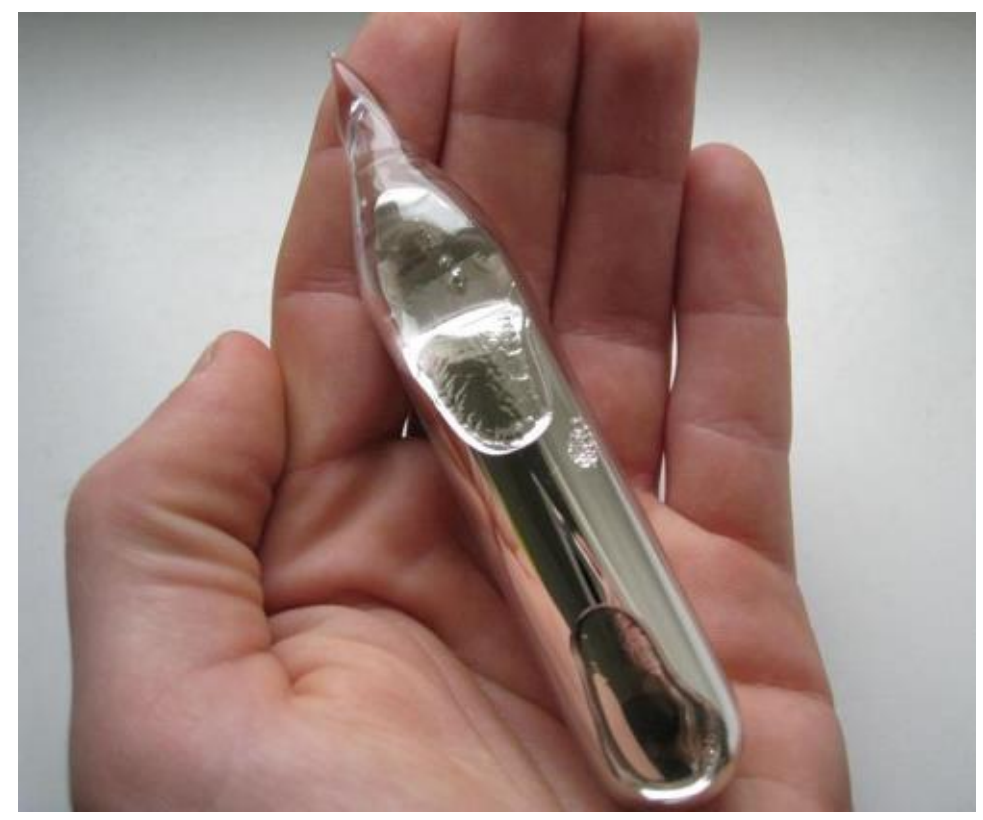

*Figure 542: rubidium in molten state. , in molten state. Source Wikipedia.*

Only 5 tons are produced annually worldwide, including China, Canada, Namibia and Zimbabwe[1522]. It is a by-product of the extraction of cesium and lithium. The isotope $^{87}$Ru is the most used and represents 27.8% of available rubidium. It is radioactive but with a half-life longer than the age of the Universe, so it is very stable. World reserves are estimated at 100,000 tons, which is enough to keep up with the current rate of production and consumption.

### Niobium

Niobium is a transition metal used in superconducting qubits as well as in microwave cables driving superconducting and electron spin qubits. **Coax Co** (Japan) has a monopoly in the manufacturing of these cable, which are very expensive, about $3K per half a meter segment. And three are needed per superconducting qubits, positioned between the 4K and 15mK cryostat cold plates.

In industry, it is used in the production of high-strength special steels, in superconducting magnets, in particle accelerators, in arc welding, in bone prostheses associated with titanium, in optics, as a catalyst for rubber synthesis, in aircraft engines and in gas turbines.

---

[1520] See Refinery production of germanium worldwide in 2021, by country, Statista.

[1521] See Production of germanium stable isotopes single crystals by Mihail Fedorovich Churbanov et al, April 2017 (6 pages).

[1522] Each human weighing 70 kg contains about 0.36g of it. However, we are not going to create a variant of Soylent Green to exploit it. Rubidium mining in Canada is carried out by Tantalum Mining Corporation, which belongs to the Chinese group Sinomine Resources since June 2019.

World production was estimated at 68,000 tons per year in 2018, with Brazil accounting for 88%, followed by Canada for just over 9%, generated by a single mine. It comes from the exploitation of pyrochlore, an ore combining calcium, sodium, oxygen and niobium.

It is not very expensive and is priced at $45 per kilogram, but in its ferroniobium form. The reserves are of 9 million tons, enough to last 130 years at the current usage rate. But in practice, niobium is considered a "risky" resource because its demand is growing rapidly even though it comes from relatively safe geopolitical places.

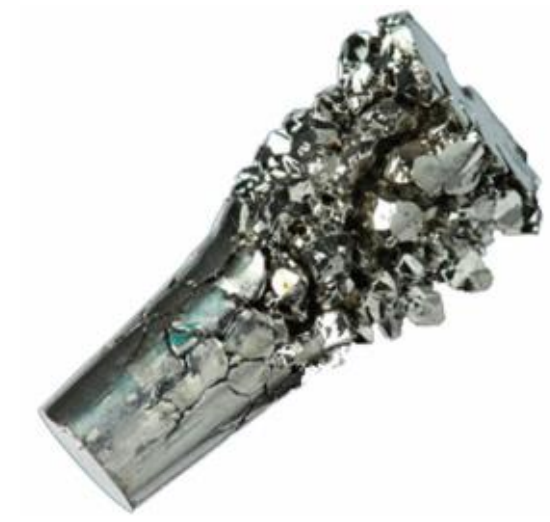

*Figure 543: niobium is a relatively cheap metal.*

**Ytterbium**

Ytterbium is a rare earth of the lanthanide series which is used in trapped ions qubits, quantum memories, atomic clocks, doping of certain lasers and, more rarely, in cold atom qubits.

Otherwise, it is used to reinforce certain specialized steels.

The metal is extracted from monazite, a tetrahedral crystalline rock structure of phosphorus oxide associated with various rare earths, which contains only 0.03% of it. Production follows a complex cycle using sulfuric acid and ions exchange. Quantum applications use isotope 171, one of the 7 non-radioactive isotopes of the element. It represents 14% of its proportion in the rocks from which it is extracted. This isotope is probably more expensive than the regular multi-isotope version which is sold between $500 and $1K per kilogram.

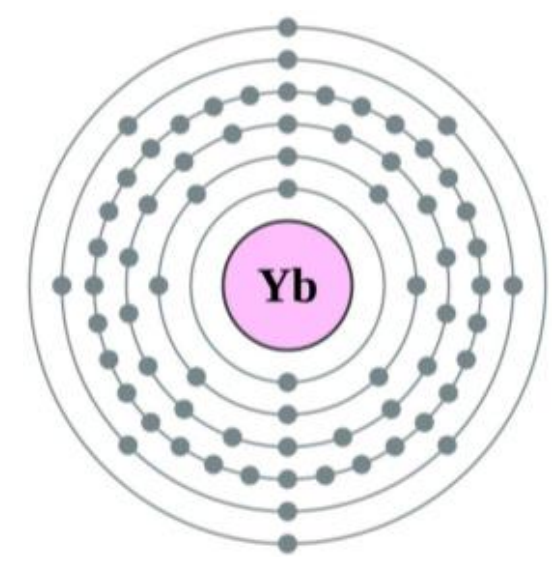

*Figure 544: ytterbium atomic structure.*

Approximately 50 tons are produced annually, mainly in China, the USA, Brazil, India and Australia, with reserves estimated at one million tons. Creating trapped ions computers require about one gram per quantum processor.

**Erbium**

This rare earth of the lanthanide family is used in quantum memories, in some fancy cold atom qubits and in certain lasers (Er:YLF type for yttrium lithium fluoride or Er:YAG type for yttrium aluminum oxide). It is found in some optical fibers used in optical amplifiers.

Finally, it can be used to create vanadium alloys found in cryostats thanks to its high thermal mass heat absorption capacity. China is the main producer, followed by the USA. It comes from extracting xenotime (phosphate ore) and euxenite (an ore also containing niobium, titanium and yttrium). The ore is processed with hydrochloric or sulfuric acid and then neutralized with soda ash. After a bunch of chemical treatments, erbium ions are extracted by ion exchange on polymer resins.

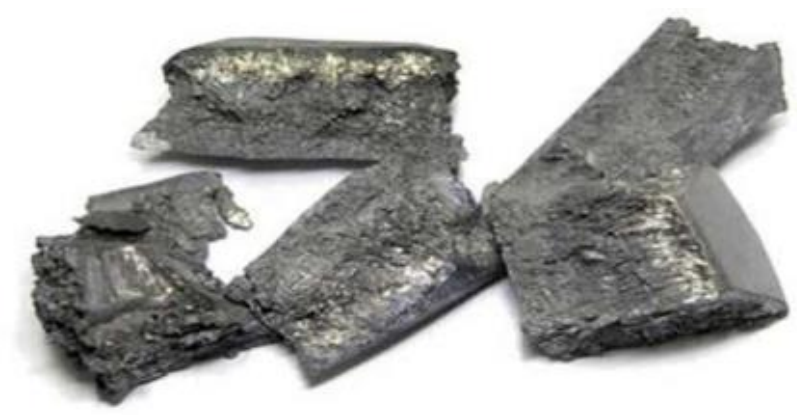

*Figure 545: erbium.*

Erbium is then obtained by heating its oxide with calcium at 1450°C in a neutral argon atmosphere. All this is a long and expensive chemical process, probably polluting of lot but carried out on small volumes. Erbium is produced at a rate of about 500 tons per year. Its price per gram is about $20, which is quite affordable to integrate it in memories or cold atoms qubits.

**Strontium**

Strontium is the most common alkali metal used to create trapped ions qubits, with its isotope 87, representing 7% of its five isotopes. It is used as a red dye in fireworks.

Mexico and Germany are the main producers, with an estimated world production of 220,000 tons per year and reserves of over one billion tons. It is notably used in bones anti-cancer radioactive chemotherapies.

Strontium is considered to be toxic. This is the case of all these rare metals which, being pure, oxidize quickly whatever happens. In particular, it explodes when being in contact with water.

### Gold

In quantum technologies, gold is mainly used as a thin layer covering the copper plates of the cold plates in cryostats. It prevents copper oxidation and adds good thermal conductivity. The volume used is quite small in relation to gold production and global reserves. About 3000T of gold are produced every year worldwide.

### Titanium

Titanium is mainly used in association with niobium in superconducting microwave cables.

In industry, it is used for its resistance to corrosion, particularly in the aerospace industry. Some submarines have an all-titanium hull. Titanium oxide is used as a painting white pigment. It is found in great quantities on Earth since it is the fifth most abundant metal. But only a few ores contain a high enough concentration of it to make its production profitable. The main producing countries are Australia, South Africa, Canada and Norway at a rate of 4.2 million tons per year. Reserves are in excess of 600 million tons.

### Nitrogen

Liquid nitrogen is used in cryostats to clean the gaseous helium that feeds them. It is also found in small quantities in NV Centers crystals. It is not a rare commodity. But its production in liquid form is quite energy consuming.

### Other materials

Many other relatively common materials are used in quantum technologies.

**Copper** is found in cryostat cold plates and with some of the various electrical connectors. It is purified at 99.99% to become free of impurities and oxygen (OFHC for oxygen-free high conductivity), in order to improve its thermal conductivity and electrical conductance. It is also widely used in trapped ions chambers. As far as its depletion is concerned, its consumption in quantum technologies is minor.

**Carbon** is exploited in a variety of places, including with carbon nanotubes from C12 Quantum Electronics. This carbon must be purified to keep only its isotope 12. $^{12}C$ is acquired in the form of methane in bottles acquired in the USA for $10K. It is 99.997% purified. The isotopic separation of $^{12}C$ uses a chemical process applied to $CO^2$. Carbon is also used in NV centers.

**Aluminum**[1523] is used in some superconducting qubits as well as for part of the connector technology in cryostats. It is abundant.

**Manganese** is used in very small quantities as a dopant in some superconducting qubits and can be used with trapped ions qubits.

**Silver** is mainly used in powder form in some heat exchangers in dilution refrigeration systems.

**Iron** is a commodity used in the form of steel in the structure of quantum computers.

---

[1523] The spelling is aluminum in American and Canadian English and aluminium elsewhere. This document is mostly in American English.

**Cesium** is mainly used in atomic clocks, in its isotope 133. Reserves are sufficiently abundant in relation to identified needs. They are mainly located in Canada.

In addition to germanium, **gallium** and **indium** play a key role in III-V components used mainly in photonics and **Neodymium** is used in lasers. This is one of the few areas of quantum technologies where there is a strong dependence on China as a source of supply. Finally, **beryllium**, **calcium**, **zinc**, **cadmium** and **mercury** can be used in trapped ion qubits, but the most commonly used are ytterbium, barium and calcium.

The below table in Figure 546 is a summary of this part with a list of materials, their main usage in quantum technologies, their main countries of production, rarity and production cleanliness.

| Element | Quantum computing | Quantum sensing and others | Main country sources | Rarity | Cleanliness |
|---|---|---|---|---|---|
| Helium 3 | Cryostats | | USA, Canada | | |
| Helium 4 | Cryostats | Cryostats | Qatar | | |
| Silicon 28 | Silicon Qubits | | Russia, France | | |
| Rubidium | Cold Atoms | Cold Atoms | China, Canada, Namibia and Zimbabwe | | |
| Niobium | Cables, supra qubits | | Brazil, Canada | | |
| Ytterbium | Trapped ions, memory | | China, USA, Brazil, India and Australia | | |
| Europium | Memories | Repeaters | Mongolia, China, Russia | | |
| Erbium | Cold atoms, memory | | China | | |
| Barium | Trapped ions | | UK, Romania, Russia | | |
| Strontium | Trapped ions | | Mexico, Germany | | |
| Neodymium | Lasers | Lasers | China, United States, Brazil, India, Sri Lanka, Australia | | |
| Gold | Cold plates | | Peru, Mexico, Indonesia | | |
| Titanium | Cables | | Australia, South Africa, Canada, Norway | | |
| Gallium | | Photonics | China, Germany, Kazakhstan, Ukraine | | |
| Germanium | | Photonics | China, Canada, Finland, Russia, USA | | |
| Indium | | Photonics | China, Belgium, Canada, Japan, Peru, South Korea. | | |
| Nitrogen | Cryostats | NV Centers | | | |
| Aluminum | Cryostats, supra qubits | | China, India, Russia, Canada | | |
| Silver | Cryostats | | Mexico, China, Peru, Chile, Poland | | |
| Caesium | | Clocks | Canada, Zimbabwe, Namibia | | |
| Carbon | NV Centers, nanotubes | NV Centers | | | |

*Figure 546: table with elements used in quantum technologies with their country or origin, rarity and environmental footprint. Consolidation (cc) Olivier Ezratty.*

## Quantum enabling technologies key takeaways

- Cryogeny is a key quantum computing enabling technology particularly for solid-state qubits which work at temperatures between 15 mK and 1K. These systems rely on a mix of helium 3 and 4 in so-called dry-dilution refrigeration systems. Other simpler cooling technologies target the 3K to 10K temperature ranges that are used with photon sources and detectors, as used with photon qubits systems.
- Cabling and filters play another key role, particularly with solid-state qubits. Superconducting cables are expensive with 3K€ per unit and come from a single vendor source from Japan. Signals multiplexing is on the way!
- Microwave generation and readout systems used with superconducting and quantum dots electron spin qubits are other key enabling technologies. The challenge is to miniaturize it and lower their power consumption and, if that makes sense, to put them as close as possible to the qubits, operate them at cryogenic temperatures and simplify system cabling. It's a key to physical qubits scalability. A lot of different technologies compete here, mostly around cryo-CMOS and superconducting electronics. Other components deserve attention like circulators and parametric amplifiers that we cover in detail in this new edition.
- Many lasers and photonics equipment are used with cold atoms, trapped ions and photon qubits and also quantum telecommunications, cryptography and sensing. It includes single indistinguishable photon sources as well as single photon detectors. The lasers field is also very diverse with product covering different ranges of wavelengths, power, continuous vs pulsed lasers, etc.
- Manufacturing electronic components for quantum technologies is a strategic topic covered extensively in this book for the first time with a description of generic fab techniques and some that are specific to quantum technologies like with the fabrication of superconducting qubits and quantum dots.
- Quantum technologies use a lot of various raw materials, some being rare but used in very small quantities. While some materials may have some incurred environmental costs, most of them do not seem to be scarce and they have multiple sources around the planet.

Page intentionally left blank.

# *Quantum algorithms*

It is now time to put aside quantum hardware that was the main topic of the previous parts of this book and to turn to quantum algorithms and software!

Gate-based quantum computers use quantum algorithms, some of which being theoretically much more efficient than their equivalents designed for classical computers. There are not that many algorithms and their relative performance compared to classical algorithms is not always obvious to prove. It is even sometimes contested. The assertion "*quantum computers are faster than classical computers*" is therefore debatable and must be discussed and analyzed on a case-by-case basis.

**Richard Feynman** described the idea of creating quantum simulators in 1982[1524]. His idea was to create devices using the effects of quantum mechanics to simulate them, which would be almost impossible with traditional computers. This corresponds today to so-called quantum simulators, a specific breed of analog quantum computers. But we're dealing here mostly with gate-based quantum computing, based on **Yuri Manin**'s idea from 1980 and then refined by **David Deutsch** between 1985 and 1992.

Mathematicians have been working since the mid- and late 1980s on creating algorithms for quantum computers and simulators, long before any hardware was available.

The first quantum algorithms were published in the early 1990s, while the first two-qubit quantum systems appeared around 2000/2002. Researchers have been regularly creating new algorithms for the past 25 years, regardless of the relatively slow progress with hardware. The Quantum Algorithm Zoo launched in 2011 identifies 62 classes in the scientific literature and 430 algorithms (as of April 2021), organized in 4 algorithms groups (algebraic and number theory, oracle-based problems, approximation and simulations, optimization - numerics and machine learning). The list is maintained by Stephen Jordan, a researcher at Microsoft Quantum. This is still a modest number compared to the thousands of non-quantum algorithms[1525]. Even though most classical computing developers don't know and use many algorithms in practice!

Quantum algorithms creation is thus a parallel research path with hardware progress, even though they might be sometimes closely related. This is not the first time in history. The emblematic **Ada Lovelace** did formalize the first algorithms and lines of code to run on **Charles Babbage**'s machine, which only saw the light of day in 2002 in London, 153 years after its conception (video) (see the sample program *below*). In 1842/1843, she had annotated a translation of her own of a paper by the Italian **Luigi Federico Menabrea** describing Babbage's machine. It took 102 years for the first electronic computers to see the light of day at the end of World War II! A beautiful game... of patience!

It is also reminiscent of **Leonardo da Vinci**'s helicopter designs dating from 1487-1490. A first human-powered helicopter created by the University of Toronto flew in 2013, AeroVelo (video) followed by another fairly close specimen from the University of Maryland flying in 2018 (video)! So, more than five centuries apart! And even taking into account the flight of the first motorized helicopter in 1907, the time lag is still over four centuries. This same University of Maryland is one of the most advanced in the world in quantum computers based on trapped ions!

[1524] See Simulating Physics with Computers, Richard Feynman, 1982 (22 pages).

[1525] For an extensive coverage of the key gate-based quantum algorithms, see Lecture Notes on Quantum Algorithms by Andrew M. Childs, April 2021 (181 pages) and Quantum Computing Lecture Notes by Ronald de Wolf, 2021 (184 pages).

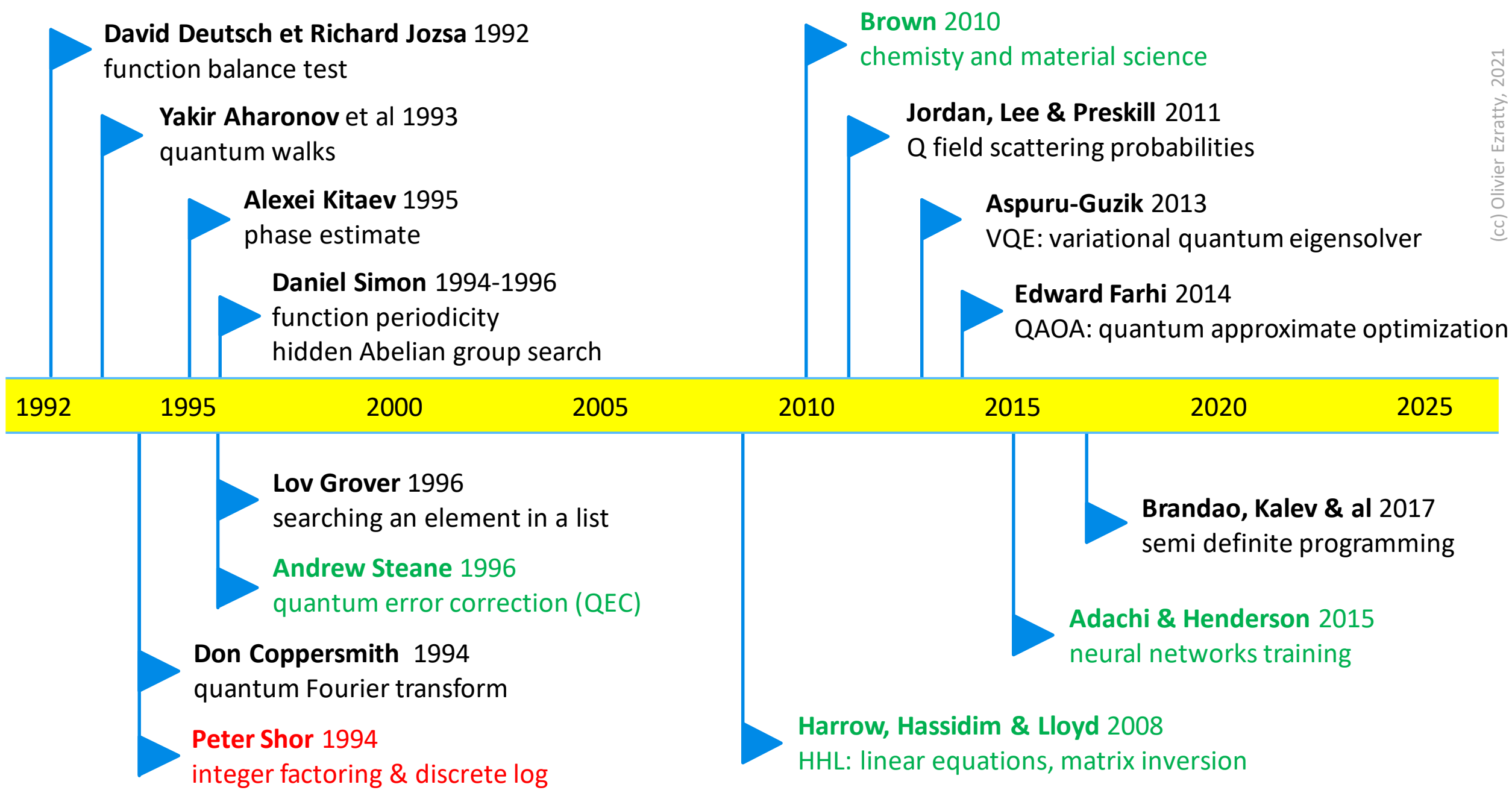

*Figure 547: a quantum computing algorithms creation timeline. It is a three-decade story. (cc) Olivier Ezratty, 2021.*

After the war, history repeated itself in part for much of the work in the vast field of artificial intelligence, where researchers were also working on algorithms, especially neural network-based algorithms, before any computers could execute them properly on a useful scale such as for objects recognition in images. The first computers running perceptrons, the ancestors of today's artificial neural networks, were rudimentary. The rise of deep learning since 2012 is partly linked to the power of machines and GPUs able to train such neural networks. Hardware has once again joined algorithms that were ahead of their time.

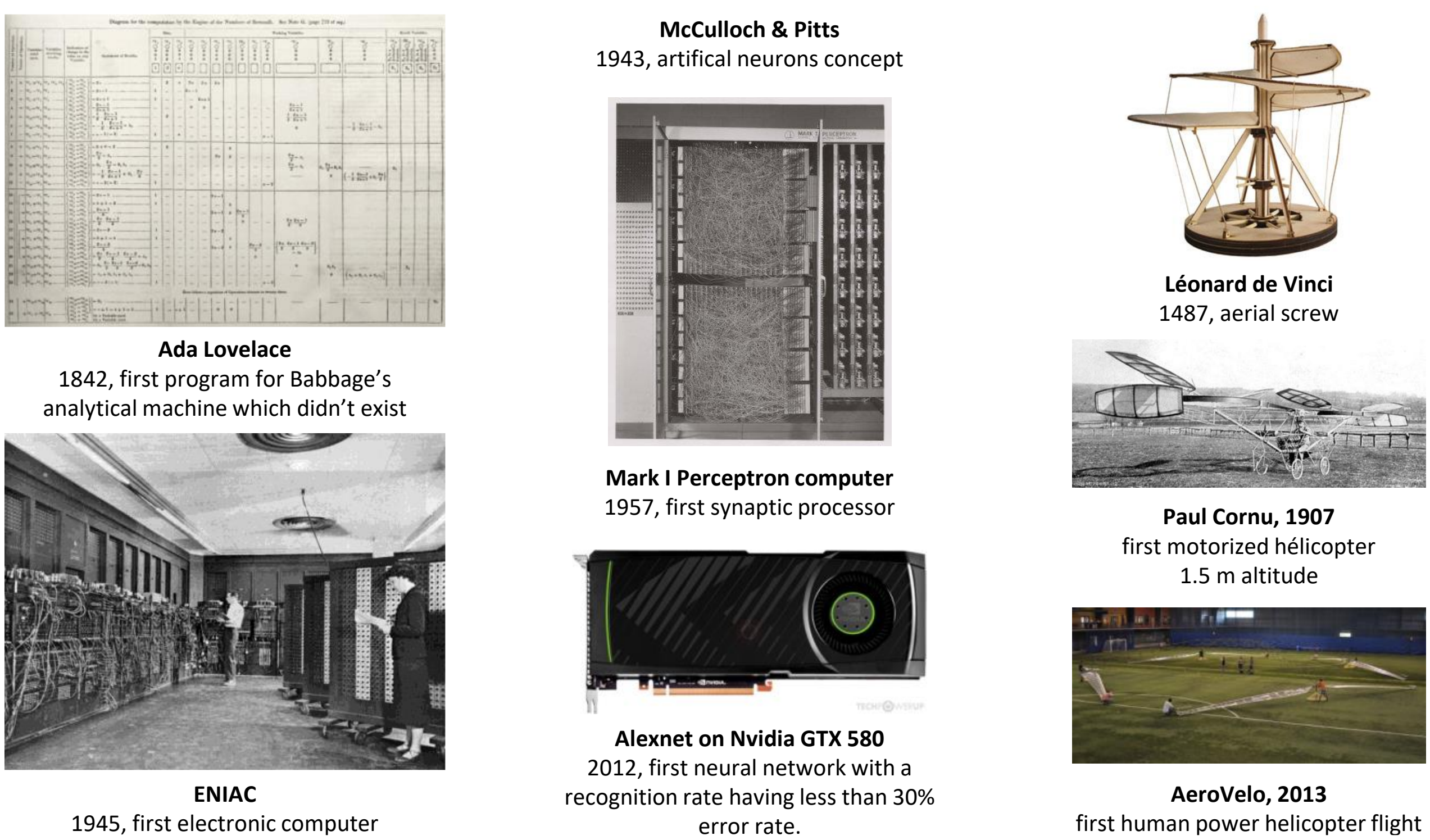

*Figure 548: a perspective on the time gap between algorithms creation and their underlying hardware. One century between Ada Lovelace's Bernoulli equations programming and the advent of the first electronic computer. And 6 decades to implement neural networks practically. Same for helicopters in another domain! (cc) Olivier Ezratty with various image sources. 2020.*

Even today, many of the quantum algorithms that are invented are not yet executable on a large problem scale on current quantum computers or on classical computing quantum emulators. There are not enough quality qubits available to be of any use and, more importantly, to be more powerful in any dimension than classical computers. Supercomputers can emulate about 50 qubits but no operational quantum computer can reach this number with error corrected qubits.

In another analogy with the History of Computer Science, we are still programming quantum computers with rather low layers of machine language, a bit like machine language or macro-assembler used 30 to 50 years ago, or more recently, for those who program low-level embedded systems or peripheral drivers. Today's quantum algorithms are mid- to low-level logical chunks of quantum code. Their assembly is even not yet done in practice.

**visual quantum circuits design**

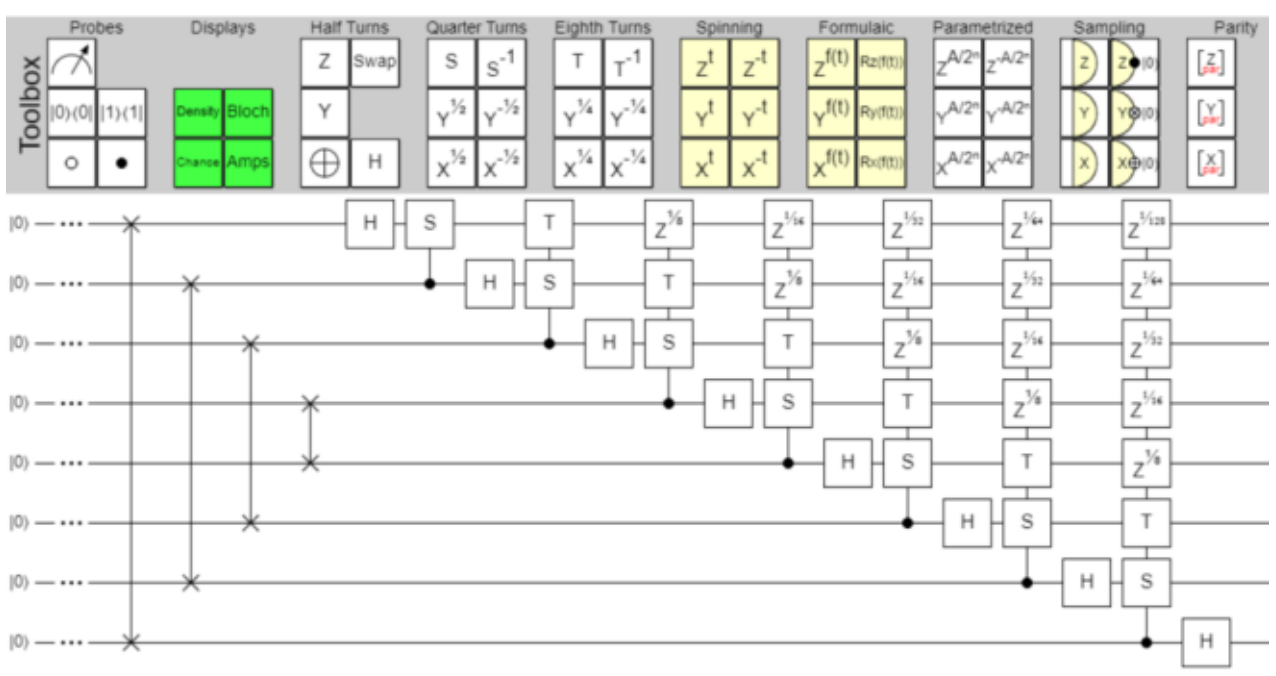

**https://algassert.com/quirk**

online open source tool to learn, program and emulate up to 16 qubits

**scripted Python code**

```
# Initialize counting qubits
# in state |+>
for q in range(n_count):
    qc.h(q)

# And auxiliary register in state |1>
qc.x(3+n_count)

# Do controlled-U operations
for q in range(n_count):
    qc.append(c_amod15(a, 2**q),
             [q] + [i+n_count for i in range(4)])

# Do inverse-QFT
qc.append(qft_dagger(n_count), range(n_count))

# Measure circuit
qc.measure(range(n_count), range(n_count))
qc.draw(fold=-1)  # -1 means 'do not fold'
```

**IBM Qiskit, Google Cirq, Atos myQLM**

*Figure 549: gate-based programming can be done graphically with tools like Quirk, mostly for learning purpose and also, to visualize interactively qubits values (Bloch sphere, vector state, density matrix) in emulation mode. Scripted code with Python is used for professional programming. (cc) Olivier Ezratty, 2022.*

The creation of quantum algorithms requires a capacity for abstraction that is beyond that of classical algorithms and programs, even taking into accounts object-based or events-based programming. We'll have to groom a new generation of mathematicians and developers capable of reasoning with the mathematical formalism of quantum programming as quantum computers mature. They will have to be able to conceptualize algorithms that are not easy to mentalize. Most of the times, though, quantum algorithms won't be simple language translation from classical programming language. They will solve problems that classical computers and classical programming languages can't solve efficiently.

One day, the abstraction level of quantum programming may rise to a point where it is no longer necessary to understand the low-level intricacies of quantum gates, Hilbert spaces, Hamiltonians and quantum interferences. But this is just a conjecture!

Today's classical quantum algorithms use quantum gates. But there are other variations:

- **Quantum annealing problem solving** such as those for D-Wave machines which are based on the initialization of relations between average quality qubits and on the search for a minimum energy based in particular on the tunnel effect. The basic algorithm there is about solving an Ising model. We will describe it when discussing about D-Wave.

- **Analog quantum simulators** are used to simulate quantum phenomena, for example to predict the organization of atoms in molecules. These include cold atom quantum simulators. An algorithm here is about preparing the state of the qubits in the system and their link weights. It's a process similar to D-Wave quantum annealing, with variations on the degrees of liberty handled in the system and qubits coherence.
- **Continuous variable quantum computers** that use quantum objects whose physical quantity can be measured as a continuous, not binary, quantity. This creates yet another programming model. They are mainly based on photons[1526].
- **Topological quantum computers**, which do not yet exist. This is the research path of Microsoft and some research laboratories, especially in China. We cover this on page 372. It should still be programmable with gate-based classical code.
- **Hybrid algorithms** combining traditional algorithms and quantum algorithms running on any of the above system[1527]. This is notably the case of the Variational Quantum Eigensolver (VQE) which allows the resolution of chemical simulation problems as well as neural network training.

We can also mention **Quantum inspired algorithms** which are algorithms running on classical computers that are inspired by quantum algorithms for solving complex problems. Their creation started long before the first experimental quantum computers were created.

In practice, the noisy intermediate scale quantum computers (NISQ) that are emerging now and will dominate the landscape for at least a good decade cannot run "deep" algorithms.

Namely, because of quantum gates and readout error rates is too high and limits the number of quantum gates that can be chained. We are thus limited to use algorithms that chain a rather small number of quantum gates.

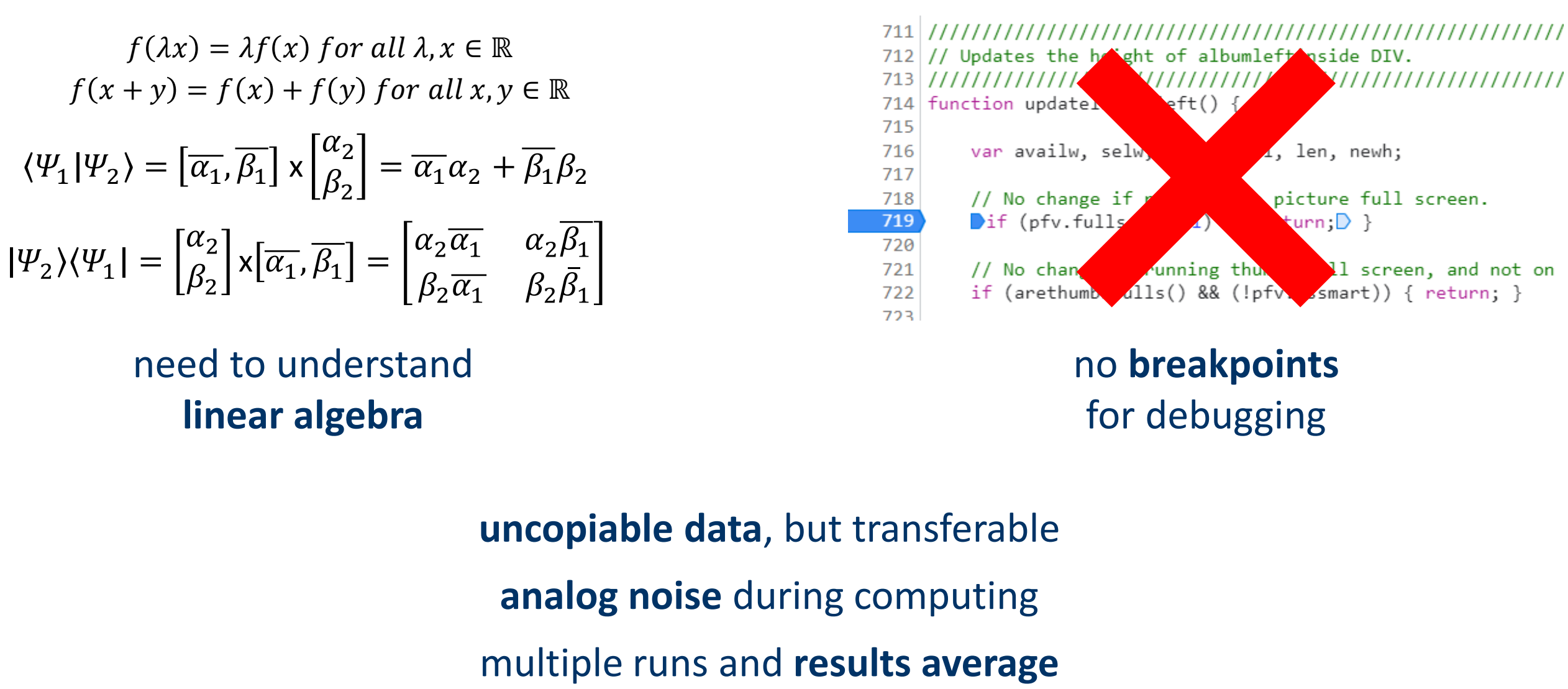

*Figure 550: the key differences with quantum programming. A need to understand linear algebra and do some maths, different debugging techniques and coping with the impossibility to copy data and playing with the probabilistic nature of quantum measurement. (cc) Olivier Ezratty, 2022.*

[1526] See for example Perspective: Toward large-scale fault-tolerant universal photonic quantum computing by S. Takeda et al, April 2019 (13 pages) and Continuous-variable quantum neural networks by Nathan Killoran et al, June 2018 (21 pages) which deals with the use of continuous variable qubits to create neural networks.

[1527] See Hybrid Quantum Computation by Arun, 2011 (155 pages).

This is the case for **VQE** (Variational Quantum Eigensolver), **QAOA** (Quantum Approximate Optimization Algorithm), **QAO-Ansatz** (Quantum Alternating Operator Ansatz, sometimes confusingly also named QAOA), Variational Quantum Factoring and some Quantum Machine Learning algorithms (Support Vector Machine, Principal Component Analysis and Quantum Variational Autoencoder). We will have the opportunity to study some of them later on.

# Algorithms classes

Before getting deep into quantum algorithms, let's take a detour with covering their practical usefulness known to date and for which category of quantum hardware they are designed. Then, how they are organized and what is the basic algorithms toolbox available to developers.

### Classes and use cases

Here's a simple classification of high-level algorithms by use cases[1528].

**Oracle function-based algorithms** can fasten the search of a needle in a haystack and find solution of some complex problems. Some are useful, some are not. The most famous oracle-based algorithms are Deutsch-Jozsa, Simon, Bernstein-Vazirani and Grover.

**Complex optimization problems**, particularly combinatorial problems like finding an optimal route for deliveries or automated drive, aka the traveling salesman problem. Such algorithms can also optimize the design of integrated circuits where one generally seeks to minimize the links between functional blocks, transistors and to minimize energy consumption, forms of sub-constrained optimization adapted to quantum processing. This category of algorithms can find solutions to combinatorial optimization problems (like discrete log, traveling salesperson problem and QUBO) and continuous optimization problems used in many other fields (linear programming, gradient descent, LPs, convex optimizations and semidefinite programming).

**Quantum machine learning** algorithms which under some circumstances could be more efficient than machine learning algorithms running on classical hardware, including GPGPUs and TPUs. This can impact both training machine and deep learning models and running inferences.

**Quantum physics simulations** is a broad field with applications in inorganic and organic chemical processes optimization and new material designs. This is based on simulating at the lowest level the interactions between atoms in molecules and crystal structures or magnetism, which themselves depend on the laws of quantum mechanics. This may help invent new solutions such as more efficient batteries that can be charged more quickly and with greater energy density, craft chemical processes for carbon capture or nitrogen fixation or create superconducting materials operating at room temperature.

**Biological molecule simulation** requires a much larger number of qubits, and therefore are positioned in the longer term. Quantum simulation may eventually help run simulations of biological molecules. This will start with the simulation of peptides, then polypeptides, and finally proteins folding and interactions. Biological molecules have the particularity of being overly complex, with structures that can reach tens of thousands of atoms. The top of the line would be the ability to simulate the assembly and then the operation of a ribosome, which is more than 100,000 atoms. It is the most magical molecular structure in living organisms, the one that assembles amino acids to build proteins from the messenger RNA code resulting from the transposition of gene DNA. This would be followed by the simulation of the functioning of a whole cell. But we are here bordering on with science fiction.

[1528] There are many such classifications around. I've used the most common one.

**Key factoring problems** relate to cryptography and breaking public encryption keys like RSA keys with Shor's integer factoring algorithm. These may be implemented over a very long-term, when highly scalable gate-based quantum computers are available.

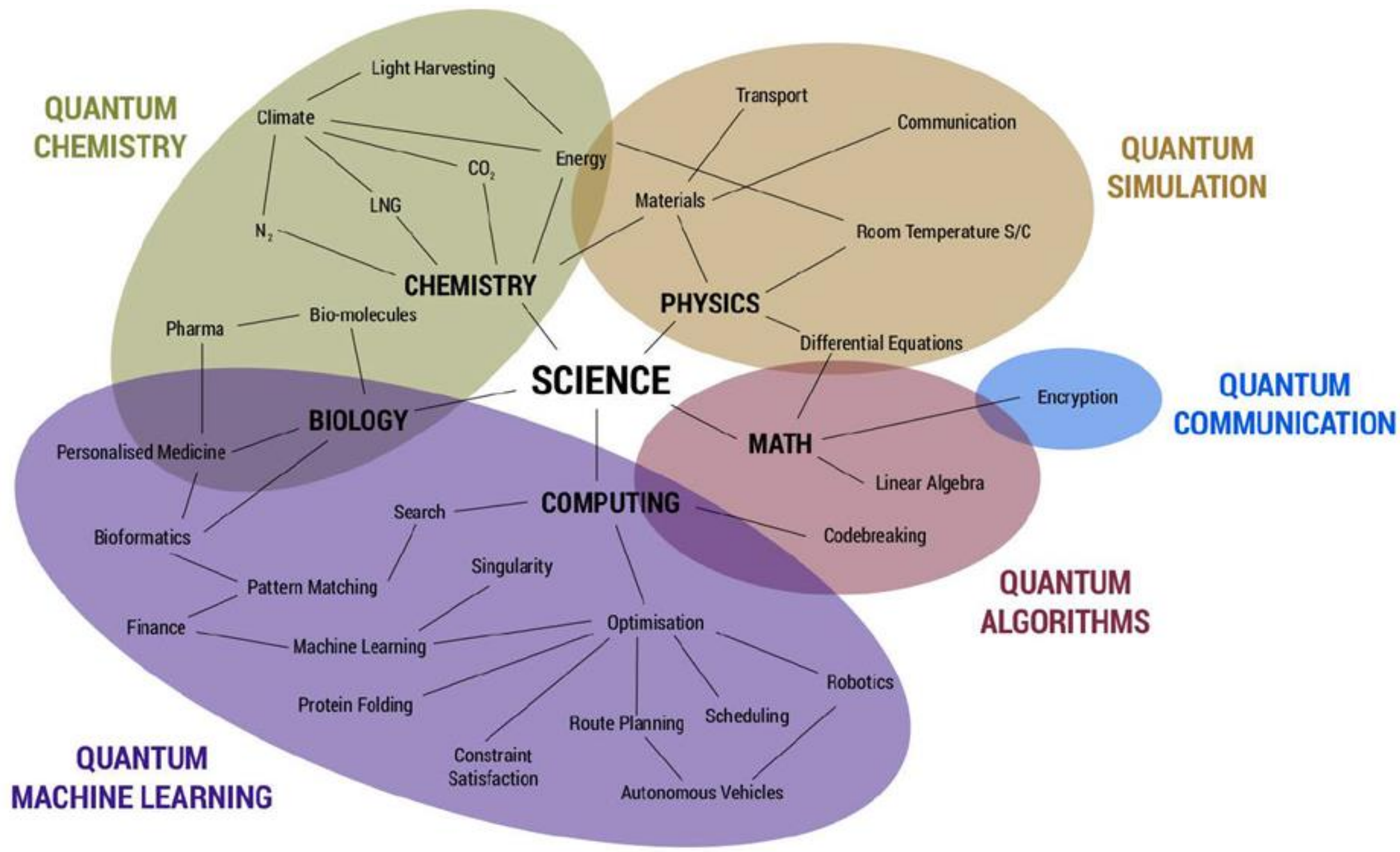

*Figure 551: the breadth of science domains covered by quantum algorithms. Source: Silicon Photonic Quantum Computing by Syrus Ziai, PsiQuantum, 2018 (72 slides).*

**Hybrid algorithms**, as already mentioned, use a mix of quantum and classical algorithms. These are mostly used for chemical simulations but also quantum machine learning. Most quantum algorithms are actually hybrid: QAOA, VQE, oracle-based algorithms when the oracle is retrieving classical data and even Shor's integer factoring algorithm.

**Quantum inspired algorithms** are classical algorithms inspired by quantum algorithms, particularly those that rely on interferences which are key characteristics of quantum algorithms.

See also this mapping of applications of quantum computing in Figure 551, which is however a bit fanciful, linking the learning machine to "singularity", which does not mean much.

### Algorithms and quantum computing paradigms

There's a relation between these broad classes of algorithms and the class of quantum computers they can run on. Reusing the quantum algorithms paradigm classification used earlier in this book, this gives an idea of what works where and when, given these computer classes span from available systems (quantum annealing), to NISQ systems in the short to mid-term, to very long-term availability (large scale quantum computing). It's still an open question to find relevant and useful algorithms running on NISQ systems, if not reaching a quantum advantage with these[1529].

[1529] See the review paper Noisy intermediate-scale quantum (NISQ) algorithms by Kishor Bharti, Alán Aspuru-Guzik et al, Review of Modern Physics, October 2021 (91 pages) and Simultaneous quantum circuits execution on current and near-future NISQ systems by Yasuhiro Ohkura et al, Dec2021 (10 pages) which addresses some limitations of NISQ systems with running smaller QPUs in parallel.

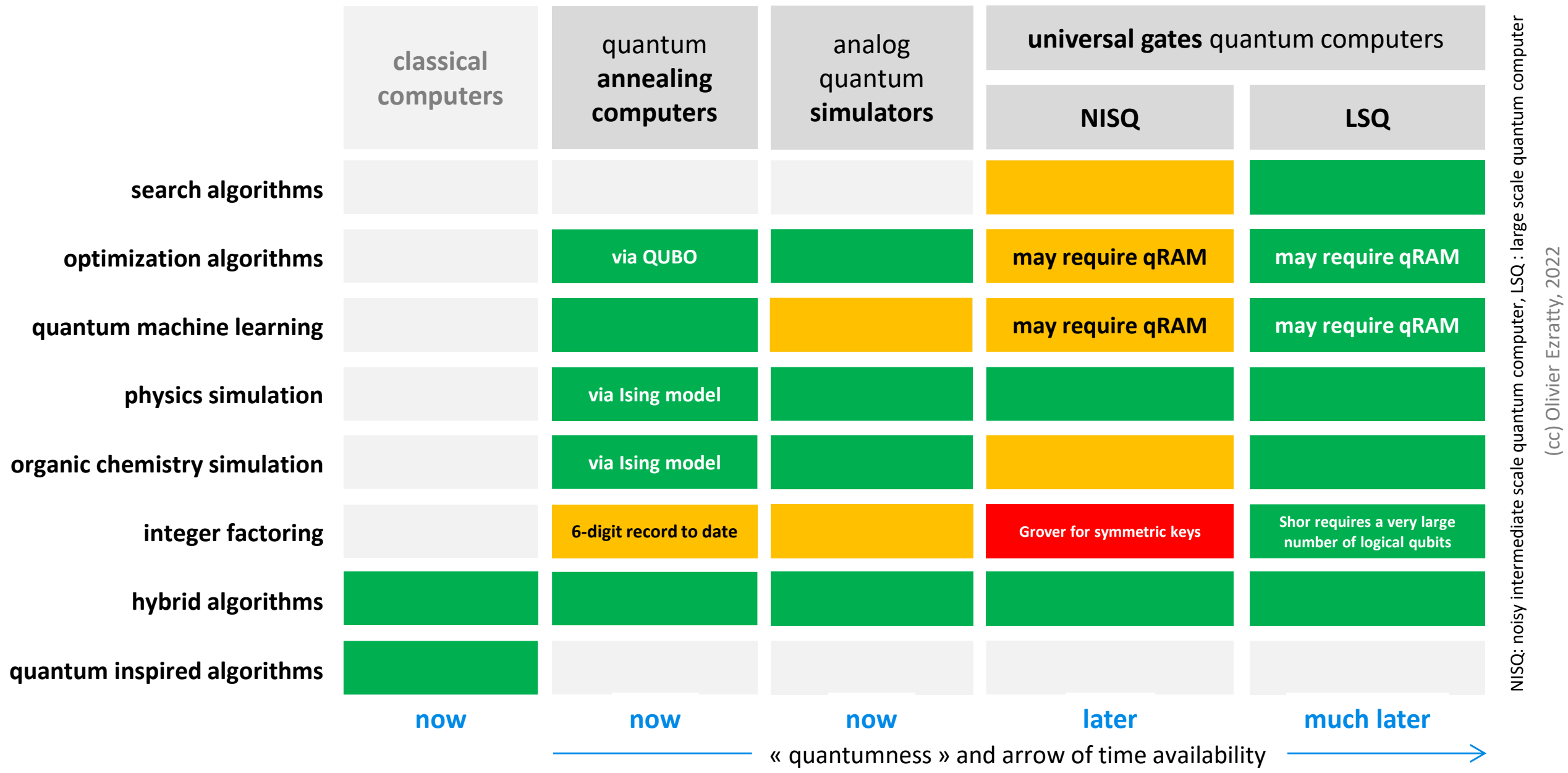

*Figure 552: classes of quantum algorithms, the quantum computing paradigm (gate-based, simulation annealing) they can run on and a time scale for their practical availability. Surprisingly, integer factoring algorithms are also available on quantum annealers and simulators, but it may not scale as well as future FTQC systems. (cc) Olivier Ezratty, 2021-2022.*

## Algorithms process and compilation

As we have seen in a previous section describing the structure of gate-based quantum computing, page 161, a quantum algorithm is built with three key parts: data initialization or preparation, computing and qubits readouts. This is always done on a data structure called a quantum register, made of N qubits. Data initialization, preparation and computing are implemented with quantum gates.

**Results**. The algorithm result comes from the classical measurement of some qubits giving out a mix of 0s and 1s bits. In general, it is necessary to run several times the algorithm entirely and compute an average of the generated results. How many times must it be done? It depends on the nature of the algorithm and the speed at which we'll move from a probabilistic output (one run) to a deterministic result (average of several runs).

**Time constraints**. The algorithm run must be compatible with the quantum computer characteristics. The main ones are qubits numbers, gates and readout fidelities and coherence time. These parameters will condition the usable depth of computing, *aka*, the number of series of gates that can be executed. This verification is generally performed by quantum code compilers. It will also have to consider the error correction codes that will be implemented in the hardware, either autonomously or through the control of the code compiler that will drive all logical qubits programming.

**Gates conversion**. Compilers play another key role: they translate the qubit gates used by the programmer into the set of physical qubit gates implemented at the hardware layer. Many quantum gates used by developers will be converted by the compiler into a set of universal quantum gates supported by the quantum computer. This will multiply the number of executed physical quantum gates compared to what shows up in the initial algorithm.

**Geometry**. They also take into account the physical geometry of qubits, i.e. how are they connected together. A simple two-qubit gate might require chaining a lot of SWAP gates because the two related qubits are far from each other in the quantum register physical layout.

**Efficiency**. An important consideration in creating quantum algorithms is to ensure that they are more efficient than their optimized counterparts for traditional computers or supercomputers. There are theories to verify this to evaluate the exponential, polynomial, logarithmic or linear rise in computing time as a function of the size of the problem to be solved, or a combination of all four. But nothing can replace experience!

**Everything is linear**. Quantum algorithms are practical applications of linear algebra, the branch of mathematics that handles vector spaces and matrix-based linear transformations. They are applied in large dimensional spaces, the vectors that define the states of qubit registers. Mathematically speaking, a qubit is a 2-dimensional vector space using complex number and N-qubit register manipulates a vector in a $2^N$ dimensional space of complex numbers. Their manipulation is based on matrix-based calculations that allow the qubit state to be modified without reading the content of the qubits. One way to look at a gate-based quantum algorithm would be the following: it's about finding the shortest path on a hypersphere of dimension $2^N$ from an initialized register to the problem solution[1530].

**Conditional programming**. Since quantum algorithms usually prohibits reading intermediate results, conditional programming is not obvious. Like running a given calculation depending on the value of some intermediate. However, multi qubits quantum gates (CNOT & co) are tools allowing conditional programming, but in another fashion than with classical computing. Conditional branching can be implemented in some situation and is implemented with hybrid algorithms using ancilla qubits for intermediate values measurements.

### Algorithms toolbox

All these algorithms are based on a small set of classical low-level algorithms that we'll describe in detail in the Basic algorithms toolbox section starting page 575:

- **Quantum Fourier Transforms** (QFT), which helps find periods in a signal. It's used in the famous Shor integer factoring, in many other algorithms (HHL, QML), discrete log search, solving the hidden subgroup problem (HSP) and even for simple reversible arithmetic[1531].
- **Quantum Phase Estimation** are relying on a QFT to find the eigenvalues or eigenvalues' phase of a unitary matrix or quantum subcircuit. It is used in HHL and many other quantum linear algebra algorithms.
- **Amplitude amplification** is used to amplify and select the desired state of a quantum superposition. It is used in the Grover algorithm and with combinatorial searches like the traveling salesperson problem search (TSP).
- **Quantum Amplitude Estimation** is a variation of Grover's amplitude amplification that is used to evaluate the average amplitude of an oracle function. It is used in many algorithms such as in the quantum variations of quantum Monte Carlo integrations.
- **Quantum phase kickback** is an interference trick used in most oracle-based search algorithms and quantum walks, and then in quantum machine learning.
- **Hamiltonian Simulation** are used to find a point of equilibrium of a complex system such as in quantum physics simulation, neural networks training, the search for optimal paths in networks or process optimization. It can be implemented in all quantum paradigms: annealing, simulation and gate-based computing.

[1530] It even seems to have a name: minimizing Wasserstein complexity as seen in Wasserstein Complexity of Quantum Circuits by Lu Li, Seth Lloyd et al, August 2022 (14 pages).

[1531] See A New Approach to Multiplication Opens the Door to Better Quantum Computers by Kevin Harnett, 2019.

- **Ising model** is the mechanism used to drive quantum annealing and quantum simulators. This is the underlying model used with D-Wave machines. Many physics simulation, combinatorial and optimization problems can be translated or converted into some Ising model problem.
- **Quantum teleportation** is also an algorithm basic, used mostly in cryptography and telecommunication. It will also play a key role in distributed quantum computing and also in some non-telecom related algorithms.

We'll add here several other key basic algorithms components:

- **Data preparation**: how is data loaded in an algorithm? This is particularly important for quantum machine learning and optimization algorithms.
- **Uncompute trick**: which consists in reversing some parts of an algorithm after it is run. It allows to get rid of garbage states and cleaning up ancilla qubits.
- **Oracle**: which are binary functions implemented as unitaries that can be used for parallelizing their operation on all computational state basis (all combinations of 0s and 1s in part of a qubits register).
- **Linear equations**: and the famous HHL algorithm.

Classifying quantum algorithms is a tedious task due to the many dependencies they have with each other. For example, a QFT is used in HSP and phase estimate algorithms which themselves are used in integer factoring and linear equations solving. I have found many different if not inconsistent algorithms classifications in the available literature (Wikipedia, John Preskill, Algorithm Zoo, etc). Some for example consider oracle-based algorithms as a separate algorithm class when other split these algorithms in various classes depending on the sub-algorithms they are using.

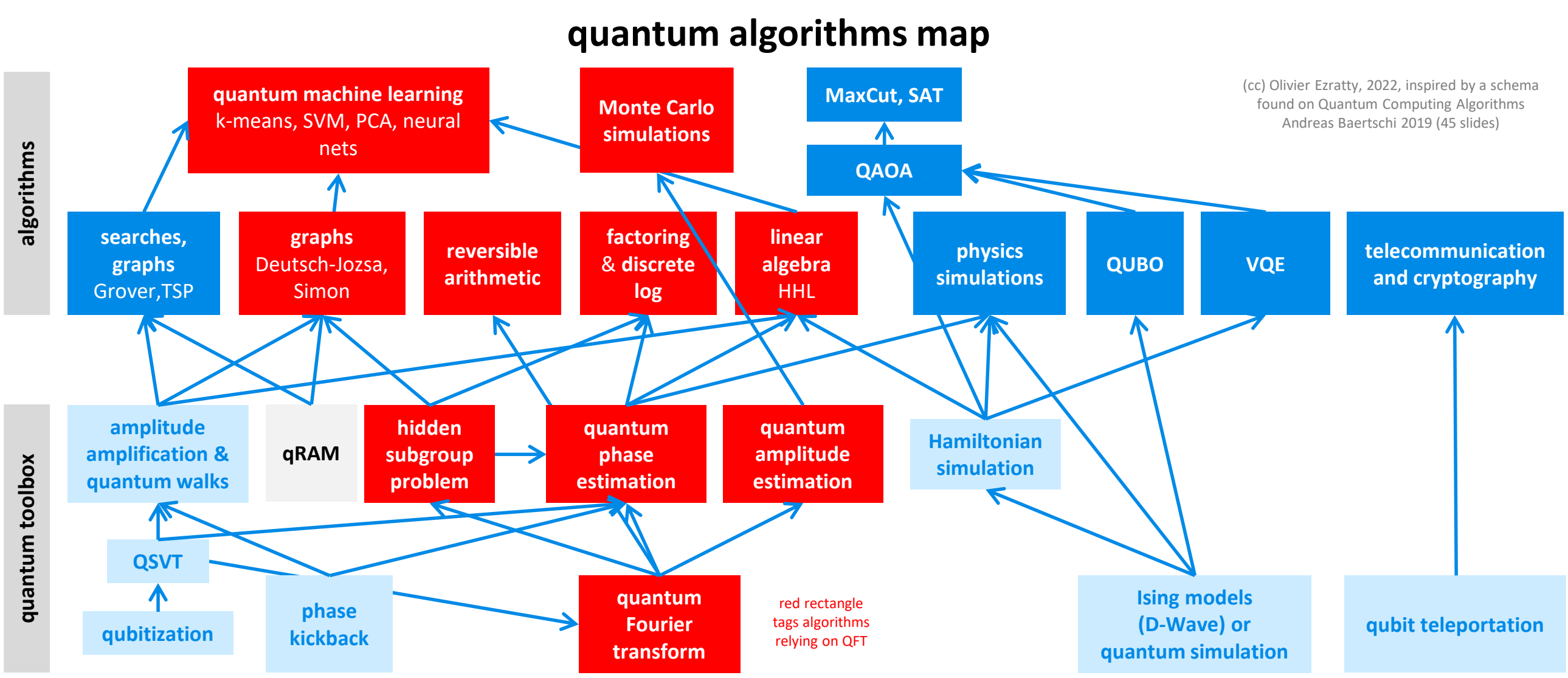

*Figure 553: a quantum algorithms map and their interdependencies. One interesting example comes with QSVT which can be used to generate search, phase estimation and Fourier transforms. (cc) Olivier Ezratty, 2022, inspired by a schema found on Quantum Computing Algorithms by Andreas Baertschi, 2019 (45 slides).*

The relationship between these low-level algorithms and higher-level ones is showcased in Figure 553. It shows qRAM for quantum RAM, which is not an algorithm per se, but a hardware tool that is indispensable to run the related algorithms, particularly Grover algorithm and a lot of quantum machine learning algorithms.

The chart in Figure 554 shows a more detailed connection between the QFT and the many algorithms that rely on it (those algorithms relying on a QFT are in red rectangles in Figure 553).

| Algorithm | Description | Reference |
|---|---|---|
| **Algorithms Based on QFT** | | |
| Shor's; $O\left(n^2 (\log N)^3\right)$ | Integer factorization (given integer N find its prime numbers); discrete logarithms, hidden subgroup problem, and order finding | Peter W. Shor, "Algorithms for Quantum Computation Discrete Log and Factoring," AT&T Bell Labs, shor@research.att.com |
| Simon's; *exponential* | Exponential quantum-classical separation. Searches for patterns in functions | Simon, D.R. (1995), *"On the power of quantum computation"*, Foundations of Computer Science, 1996 Proceedings., 35th Annual Symposium on: 116–123, retrieved 2011-06-06 |
| Deutsch's, Deutsch's – Jozsa, an extension Deutsch's algorithm | Depicts quantum parallelism and superposition. "Black Box" inside. Can evaluate the input function, but cannot see if the function is balanced or constant | David Deutsch (1985). "Quantum Theory, the Church-Turing Principle and the Universal Quantum Computer". Proceedings of the Royal Society of London A. 400: 97<br>David Deutsch and Richard Jozsa (1992). "Rapid solutions of problems by quantum computation". Proceedings of the Royal Society of London A. 439: 553 |
| Bernstein/Vazirani; *polynomial* | Superpolynomial quantum-classical separation | Ethan Bernstein and Umesh Vazirani. *Quantum complexity theory.* In Proc. 25th STOC, pages 11–20, 1993 |
| Kitaev | Abelian hidden subgroup problem | A. Yu. Kitaev. *Quantum measurements and the Abelian stabilizer problem*, arXiv:quant-ph/9511026, 1995 |
| van Dam/Hallgren | Quadratic character problems | Wim van Dam, Sean Hallgren, *Efficient Quantum Algorithms for Shifted Quadratic Character Problems.* CoRR quant-ph/0011067 (2000) |
| Watrous | Algorithms for solvable groups | John Watrous, Quantum algorithms for solvable groups, **arXiv:quant-ph/0011023**, (2001) |
| Hallgren | Pell's equation | Sean Hallgren. *Polynomial-time quantum algorithms for pell's equation and the principal ideal problem,* Proceedings of the thirty-fourth annual ACM symposium on the theory of computing, pages 653–658. ACM Press, 2002. |
| **Algorithms Based on Amplitude Amplification** | | |
| Grover's; $O\left(\sqrt{N}\right)$ | Search algorithm from an unordered list (database) for a marked element, and statistical analysis | Lov Grover, *A fast quantum mechanical algorithm for database search,* In Proceedings of 28th ACM Symposium on Theory of Computing, pages 212–219, 1996 |
| Traveling Salesman Problem; $O\left(\sqrt{N}\right)$ | Special case of Grover's algorithm | https://en.wikipedia.org/wiki/Travelling_salesman_problem |

*Figure 554: Source: Quantum computing (QC) Overview by Sunil Dixit from Northrop Grumman, September 2018 (94 slides).*

Quantum algorithms are classifiable and explainable at a high level, but their detailed understanding is not easy. You must develop some conceptual capacity in a rather analog world[1532].

One key thing developers have to learn is how to translate customer needs into existing quantum algorithms. How to assemble various quantum algorithms, frequently combined with classical algorithms, is another key skill.

### Algorithms figures of merit

It is rarely talked about, but what are the key figures or merits for quantum algorithms? In a generic way, they should showcase either a provable speedup compared with classical algorithms running on classical computers or some other quantitative advantage (precision, error rates for quantum machine learning, energy consumption advantage). They should generate a relative small-size data output since N qubits generate only N useful bits of information. At last, there should be some correctness guarantee on the results and of course, it should solve some useful problem. In the case of NISQ algorithms, they should add two requirements: a shallow depth circuit (not many quantum gate cycles) and a resilience to qubit noise[1533].

---

[1532] Here are a few sources of information to explore the topic: Quantum Computing Applications by Ashley Montanaro from the University of Bristol, 2013 (69 slides), an interesting course on the algorithmic part, An Introduction to Quantum Computing by Phillip Kaye, Raymond Laflamme and Michele Mosca, Oxford, 2017 (284 pages), Lecture Notes on Quantum Algorithms by Andrew M. Childs, University of Maryland, 2017 (174 pages), Quantum Computation and Quantum Information by Nielsen and Chuang, 2010 (10th edition, 704 pages) and A Course in Quantum Computing for the Community College by Michael Locef, 2016 (742 pages) which sets out in great detail the mathematical foundations of linear algebra with complex numbers, Euler formulas, vector and Hilbert spaces, matrix calculus, tensors, eigenvectors and eigenvalues, and quantum algorithms. It takes several weeks to be browsed and understood. It is a course for the second and third year of the Foothill Community College in Los Altos Hills, California (so Bac+1/+2 in French equivalent). In addition, here are some videos on this subject: Quantum Algorithms by Andrew Childs in 2011 (2h31), Language, Compiler, and Optimization Issues in Quantum Computing by Margaret Martonosi, 2015 (39 minutes and slides) and What Will We Do With Quantum Computing? by Aram Harrow, MIT, 2018 (32 minutes).

[1533] This inventory is inspired from the second page chart in Towards Quantum Advantage on Noisy Quantum Computers by Ismail Yunus Akhalwaya et al, September 2022 (32 pages) which presents a NISQ QML-based algorithm matching all these figures of merit. But … requiring 96 qubits with 99,99% 2-qubit gate and measurement fidelities which are yet to come, even with trapped ions.

# Basic algorithms toolbox

We'll describe here the overall structure of basic low-level gates-based quantum algorithms. We separate three stages: **data preparation**, **unitary transformations** (caveat: data preparation also relies on unitaries) and **measurement**.

We have already covered **error correction** in a previous chapter of this book, starting page 216.

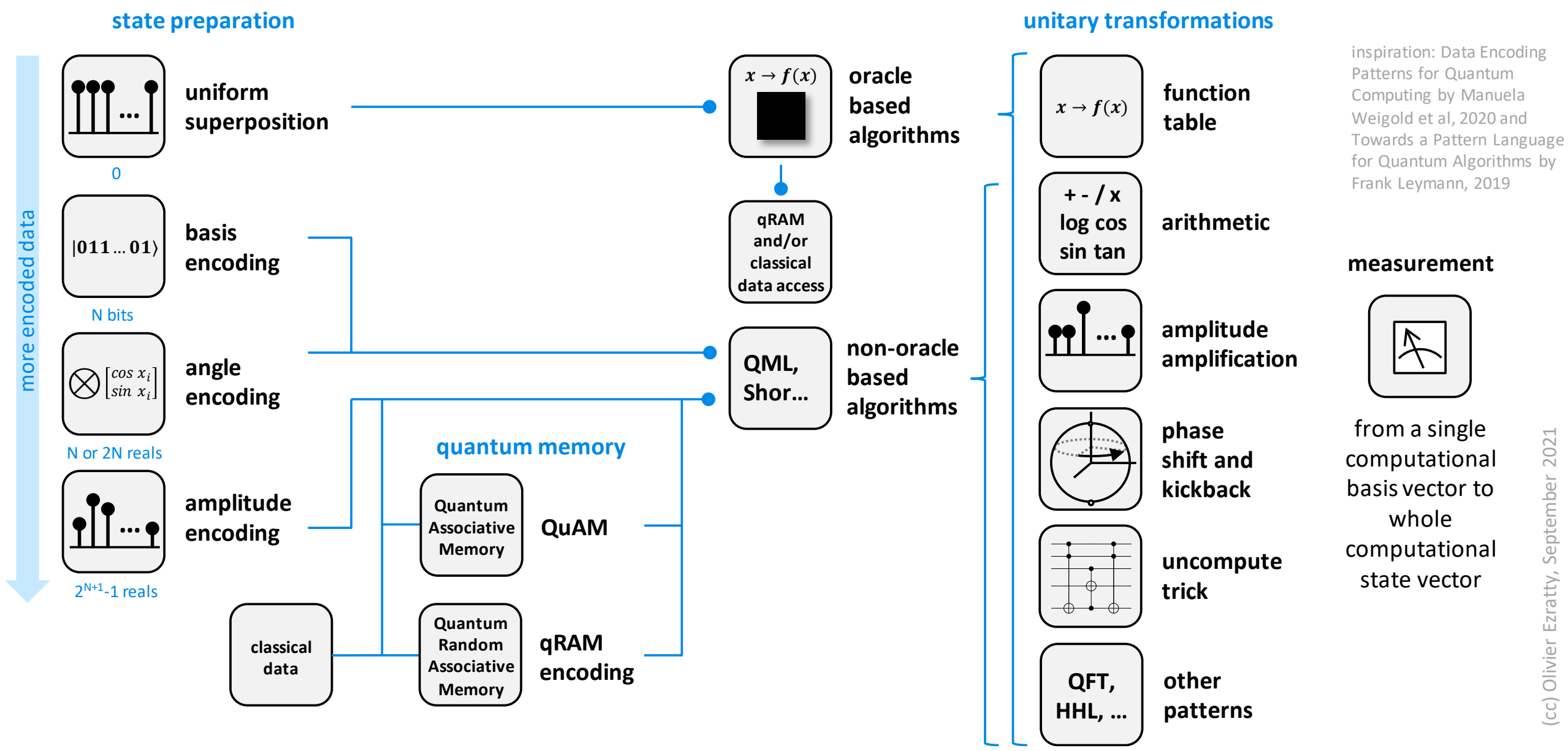

*Figure 555: how is data fed into a quantum algorithm depending on whether it uses or not an oracle. (cc) Olivier Ezratty, 2021.*

## Data preparation

The data preparation stage is also named data loading. Its complexity covers a large range from the simple process of uniform superpositions associated to oracle-based algorithms like Deutsch-Jozsa, Simon or Grover to the most complicated, linked to quantum machine learning algorithms requiring full computational basis state vector amplitude encoding.

Data loading is only implemented with non-oracle-based algorithms. It may be a long process for large sets of inputs and significant number of qubits. It thus may require using some form of quantum memory, a sort of qubits buffer used only for data preparation.

It can use addressable qubits like with qRAM where a program can ask to "*put this information in qubits at index i*". This memory can be a qubit register with a longer coherent lifespan than computing qubits or a classical data structure used along a quantum circuit to load a specific addressable quantum state. The data is not necessarily stored in some qubits. When the data is loaded in quantum memory, this one must be transferred to the computing qubits. This is a data transfer and not a data copy process due to the no-cloning theorem. All in all, quantum memory is just some sort of intermediate memory used before computing.

Let's first look at the various techniques used for data encoding[1534].

**Uniform superpositions** correspond to the simplest qubits register initialization with a register state where all the computational basis states have the exact same amplitude.

---

[1534] Here are the various sources I used to reconstruct this map: Loading Classical Data into a Quantum Computer by John Cortese and Timothy Braje, 2018 (38 pages), Circuit-centric quantum classifiers by Maria Schuld, Krysta Svore et al, 2018 (17 pages), Robust data encodings for quantum classifiers by Ryan LaRose and Brian Coyle, 2018 (24 pages), Towards a Pattern Language for Quantum Algorithms by Frank Leymann, 2019 (12 pages), Quantum linear systems algorithms: a primer by Danial Dervovic et al, 2018 (55 pages) and The Bitter Truth About Quantum Algorithms in the NISQ Era by Frank Leymann and Johanna Barzen, 2020 (42 pages).

It is used by an oracle-based algorithm where the "real data" sits in the oracle function f(x) that outputs such and such values depending on the entry (usually, a 1 for a single entry and 0 for all the others). The oracle can evaluate this function simultaneously for all superposed computational basis values in the prepared superposed register. This superposition is done with applying Hadamard gates on all computing qubits where some data must be prepared.

**Basis encoding** consists in directly transferring N classical bits in N qubits, using a set of X gates (Y gates would also make it), to change individual qubits from $|0\rangle$ to $|1\rangle$. It creates a simple encoding of a computational basis single state, combining 0s and 1s matching classical bits. The $2^N$ dimensions computational basis state vector thus contains only zeros and a single one related to this combination. Before this encoding, we select the method to encode the problem data which can be for example a floating-point number or an integer in a given number of classical bits before converting them on a computational basis state. Such a basis encoding is used in Shor's algorithm to provide the integer that must be factorized.

**Angle encoding** is about encoding a vector of real values of dimension N into N qubits. It's also named product encoding. Each qubit is individually encoded with single qubit gates $R_x$ (which themselves are usually decomposed in simple Pauli and T gates) to encode one of the (Bloch sphere) qubit angle. Since the register is the tensor product of each qubit, with no entanglement, we don't have any exponential gain in the encoding. The dense angle encoding variation uses two angles in the encoding for each qubit and can make use, additionally of $R_y$ and $R_z$ gates for the sake of adding some phase in each qubit state. We end up here with a maximum of 2N real numbers encoded in the N qubits register. And the qubit register is separable, its quantum state being separable into the quantum states of each of its qubits. It reminds us that without entanglement, you can't benefit from the exponential storage (or, better, data handling) capacity of quantum computing. In that case, data encoding requires a depth of $\log_2(N)$ gates.

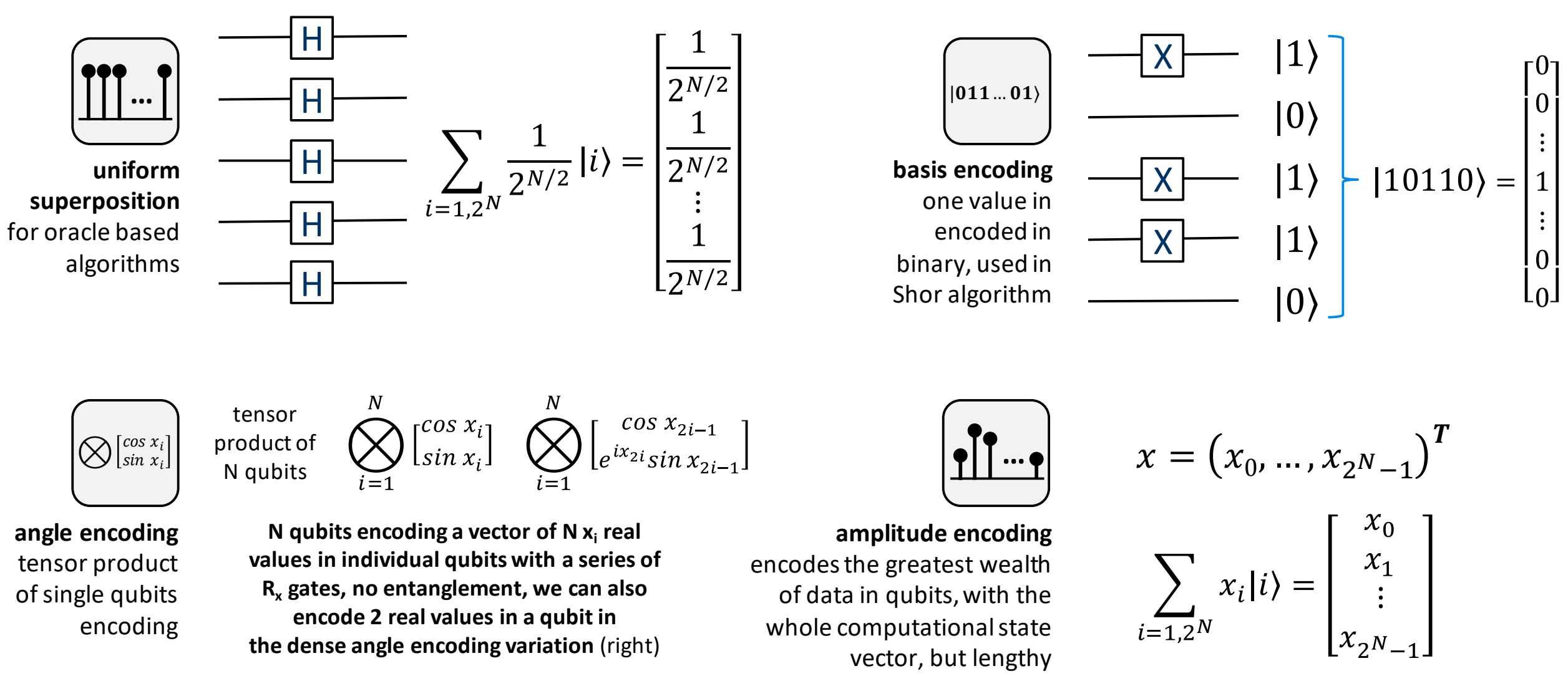

*Figure 556: details on the four ways to encode data in a qubit register, the most resource and time consuming being amplitude encoding. (cc) Olivier Ezratty, 2021.*

**Amplitude encoding** is about creating an arbitrary superposed state associating computational basis states with given real number amplitudes. It is also called an arbitrary state preparation, quantum embedding or wavefunction encoding. It creates a computational state vector with real numbers in several rows. To encode a vector of L real values, you need N=ceil($\log_2$(L+1)) qubits. Meaning you round up the $\log_2$ of the vector size and don't use the left-over values in the register vector. Why +1? Because of the normalization constraint, the sum of amplitude being equal to 1. So with 3 qubits, you have 7 available values, not $8=2^3$. Since the size of your encoded vector may be smaller than the $2^N$ states of your register, you'll pad the encoded vector with 0s.

To create an arbitrary amplitude set for N qubits, you need at least $\frac{1}{N}2^N$ gates operations combining single and two qubit gates since an arbitrary amplitude encoding will create an entangled state contrarily to a simple product encoding. The usual encoding algorithms use $2^N$ gates. So, unless the encoded data is sparse (with a lot of zeros), data preparation grows exponentially with the number of qubits, erasing any computing advantage we could get afterwards. It explains why quantum computing is not ideal for any big data computation task, and, at this point, for data intensive machine learning tasks[1535]!

**Encoding precision**. Angle and amplitude encoding theoretically deals with real numbers. But what is their precision, particularly on NISQ computers? It is at least bound by the cumulative error rates coming from the encoding qubit gates. It can easily reach a couple %, meaning the encoding precision is limited to a couple digits.

**Non linearities**. Quantum computing is based on using linear unitaries. This creates limitations on the kinds of computing that can be processed quantumly. But we can handle nonlinearities indirectly. One way lies with the way real numbers are turned into the raw data to be encoded with angle or amplitude encoding. Another way is to use angle encoding with repetition, creating powers of encoded values in the computational state vector of the input vector[1536]. Finally, we can apply non linearities on the classical data before it's quantumly encoded. It can also be implemented as a classical Boolean nonlinear circuit embedded in a quantum reversible circuit.

**How many registers?** In a classical microprocessor, the computing unit is handling data in multiple registers and the arithmetic logical unit can pull data from registers, make calculations and update registers with its results. In a quantum computer, there is usually only one register of N qubits. It can however be logically and dynamically partitioned by the algorithm. There are usually computing qubits and ancilla qubits. Computing qubits contain the input data related to the problem to be solved and that's on this data that most algorithm patterns will be executed, particularly with an oracle. The remaining qubits in the register will be used as accessory qubits and may also contain some part of the algorithm result. And this goes only with logical qubits. We've seen before that quantum error correction is using a lot of ancilla qubits.

### Black boxes and oracles

A black box based algorithm is a classical operation encoded with qubit gates that is applied simultaneously to various computational basis states. It's used in the famous Deutsch-Jozsa, Simon and Grover search algorithms. A black box contains reversible quantum equivalents of Boolean and arithmetic functions. It works on n entry qubits $x$ in superposed states and merges its result with m ancilla qubits $y$, that are usually initialized at 0. It leverages quantum parallelism with input initialized with Hadamard gates. If m = 1, the black box outputs a yes or no (1 or 0) and is branded as an "oracle"[1537].

There are many ways to implement an oracle. It can be entirely encoded with qubit gates or access some classical memory or functions, presumably through some qRAM addressing scheme. Presumably since the technology doesn't exist yet. Even the cost of implementing an entirely quantum oracle is unknown. For instance, just some complicated arithmetic functions can be highly costly in quantum gates since it may require implementing several QFTs (quantum Fourier transforms).

---

[1535] There are some ways to optimize amplitude encoding. See Quantum Resources Required to Block-Encode a Matrix of Classical Data by B. David Clader, William J. Zeng et al, Goldman Sachs, Caltech, AWS and Imperial College London, July 2022 (31 pages) also described in Goldman Sachs and AWS examine efficient ways to load data into quantum computers by Grant Salton et al.

[1536] See On nonlinear transformations in quantum computation by Zoë Holmes et al, December 2021 (10 pages) that describes another technique to introduce the support of nonlinearity in gate-based quantum computing.

[1537] See Inverse Problems, Constraint Satisfaction, Reversible Logic, Invertible Logic and Grover Quantum Oracles for Practical Problems by Marek Perkowski, May 2020 (62 slides).

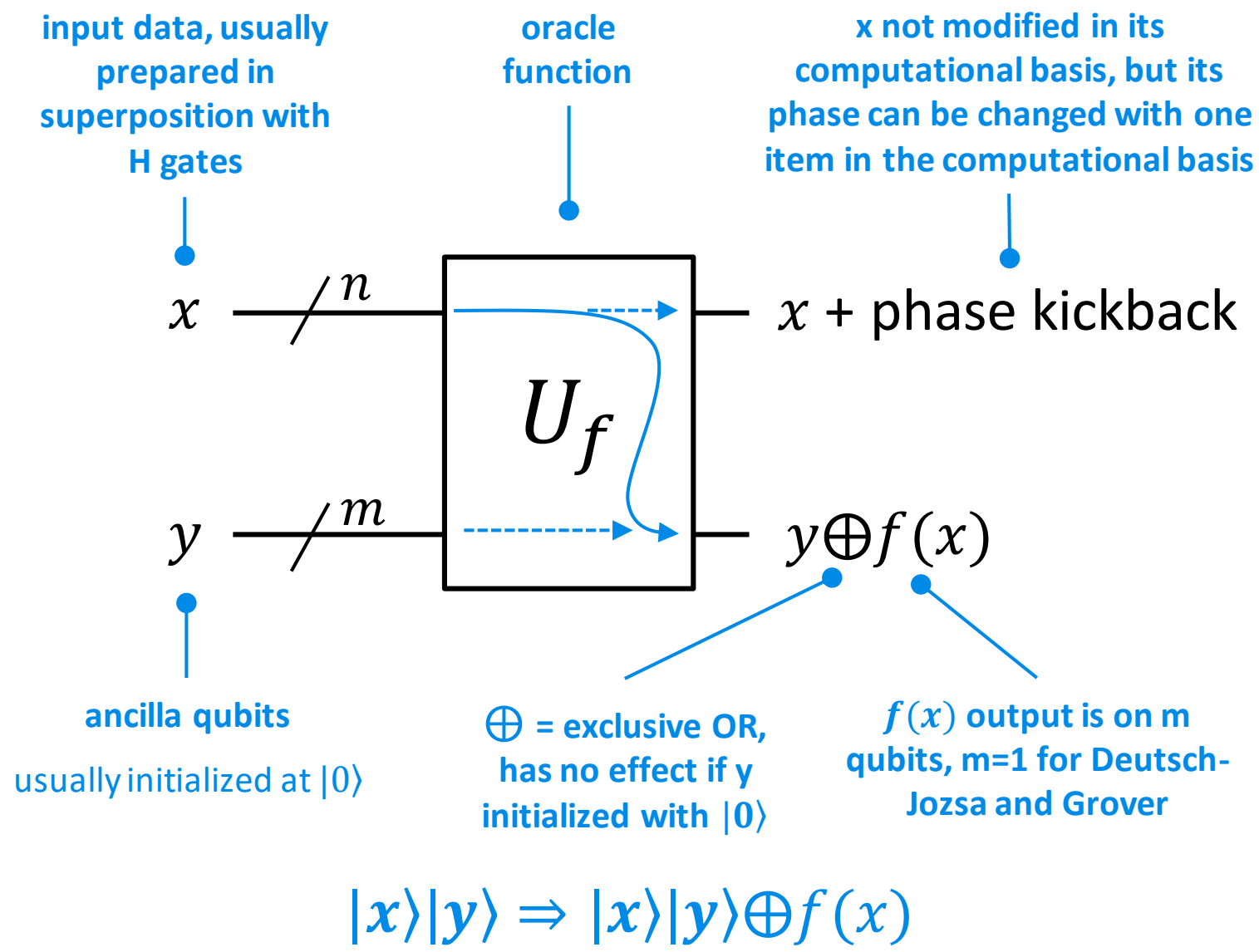

*Figure 557: how an oracle function is used in an algorithm, in complement of a phase kickback. (cc) Olivier Ezratty, 2021.*

Oracle-based algorithms speedup nearly never mention the potential computing overhead coming from the oracle itself. In an ideal world, the oracle implementation complexity should scale linearly and at worst polynomially with the number of handled qubits.

This overhead can be highly detrimental to any potential algorithm theoretical speedup. This is particularly concerning for Grover's algorithm that we'll look after later. This algorithm's speedup is only polynomial, before taking into account the oracle's cost, and of course, quantum error correction (although its cost is "only" polylogarithmic). In the end, Grover's algorithm may not bring any acceleration at all.

In other words, touting some oracle-based algorithm speedup is like saying that a car drives fast thanks to its aerodynamism, without mentioning anything about its engine specifications and power.

### Output encoding

The literature covering quantum algorithms rarely explains the format of the results they are generating. There are as many variations as in data encoding. This section echoes the one that was dedicated on the various sorts of qubits measurement, page 184. The simplest outcome of a quantum algorithm should be a computational basis vector with a series of $|0\rangle$ and $|1\rangle$, generating a classical bit string like with Shor's factoring algorithm. In this case a single run and measurement provides full characterization of this outcome (modulo the error rate of the system). This is the case for many classical quantum algorithms as described in the table below.

However, some algorithms like HHL (linear equations) may generate data encoded in amplitude. Exploiting directly an amplitude encoded vector state doesn't make much sense since you lose any exponential advantage coming from the algorithm. Decoding a full vector state indeed requires running the algorithm several orders of magnitudes of an exponential of the number of qubits. But such an algorithm may be an intermediate one feeding another algorithm. If we keep using quantum data from end to end, then it makes sense to use and create algorithms that output amplitude encoded data[1538].

[1538] This is what is proposed in Quantum advantage for differential equation analysis by Bobak T. Kiani, Dirk Englund, Seth Lloyd et al, April 2022 (21 pages). The authors use the output of a quantum differential equation solver as the input for a quantum machine learning algorithm.

In some cases, though, like with VQE algorithms, some generalized measurement of the state vector must be done[1539]. Other algorithms like in computational chemistry output will take the form of expectation values that are real numbers computed with averaging the results of many qubit readout values obtained with many computing runs. Its overhead requires significant optimizations[1540].

| algorithm | input | output |
|---|---|---|
| Deutsche-Jozzsa | oracle function | function is balanced if all output qubits are at ground state \|0⟩ |
| Bernstein-Vazirani | oracle function | (integer) secret string in basis encoding |
| Grover | oracle function | searched item index as integer in basis encoding |
| Simon | oracle function | parameters for a linear equation used to find a period, with average of basis encoding |
| Shor factoring | parametrized period finding function (quantum part) | integer in basis encoding |
| Shor dlog | integers in basis encoding | integer in basis encoding |
| QFT | series of complex amplitudes with amplitude encoding (any quantum input state) | Fourier coefficients in amplitude encoding, enabling the recovery of the main frequency |
| HHL | one vector and one matrix amplitude encoding | characteristics of inverted matrix x entry vector (= one vector) in amplitude encoding |
| VQE | cost function parameters encoded as an Hamiltonian with unitaries (quantum gates) | researched ground state in amplitude encoding |
| QML classification | object vector to classify encoded in amplitude | prediction result as an integer index in basis encoding |

*Figure 558: various algorithms and the format of their input and output data. (cc) Olivier Ezratty, 2021-2022*

### Quantum phase kickback

The role of an oracle is to change the phase of the found item in the computational basis state vector x. Instead of sending the phase to the ancilla qubit y, it is applied to the found result in the source x qubits thanks to the phase kickback mechanism. It is implemented for example in the Grover algorithm that we'll see later.

The Grover operator then amplifies the amplitude of the found item and attenuates the amplitude of the other items in the computational basis, leveraging this phase information injected in the x computational basis vector state. For this to work, the control qubits must be in a superposed state, created by Hadamard gates initialization, the target qubit $|\psi\rangle$ must be an eigenvector of the operator U applied to the target qubit $|\psi\rangle$ using the control qubits[1541].

This simple two qubit configuration explains what's happening. A control-phase gate ends up modifying the phase of the control qubit instead of the phase of the target qubit. It works in the example case since after the X gate being applied to the target qubit, the qubit state becomes an eigenvector of the control-S operation that is executed afterwards[1542].

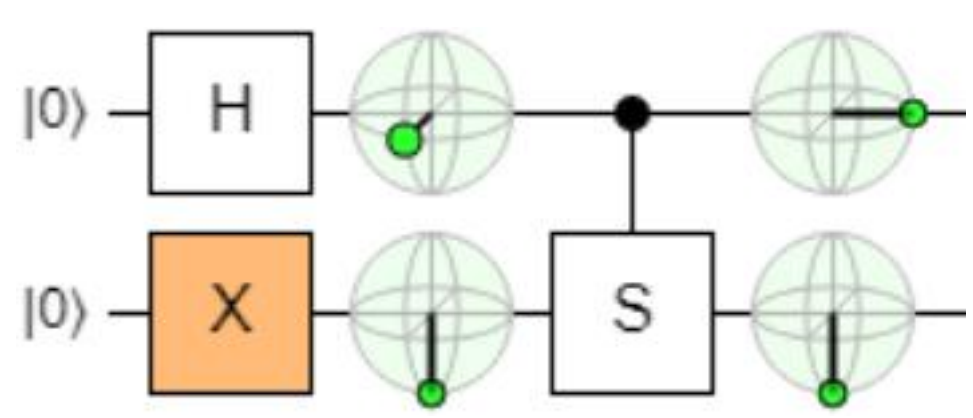

*Figure 559: a two-qubit phase kickback.*

[1539] See Learning to Measure: Adaptive Informationally Complete Generalized Measurements for Quantum Algorithms by Guillermo García-Pérez et al, PRX, November 2021 (18 pages) that proposes a POVM -based technique to undertake such measurement.

[1540] See Nearly Optimal Quantum Algorithm for Estimating Multiple Expectation Values by William J. Huggins, Ryan Babbush et al, Google AI, PNNL, Stanford and University of Toronto, November 2021-October 2022 (18 pages).

[1541] As explained in Phase Kickback by Eduard Smetanin, November 2019 (4 pages).

[1542] This is explained in A clever quantum trick by Emilio Peláez, January 2021. See also Quantum Phase Kickback - What I told you was true… from a certain point of view by Frank Zickert, March 2021.

It is not changed by a phase rotation. Since a control-phase changes the global phase of both qubits, the phase modification can only happen on the control qubit. Despite the entanglement created by the control-S gate, the qubits remain separable.

In a general case, when the target qubit $|\psi\rangle$ is an eigenvector of the unitary U (here, a S gate), the target qubit doesn't change after the control-U gate. Literally: $U|\psi\rangle = e^{i\phi}|\psi\rangle$. The control qubit changes and one of its computational state vector amplitudes gets multiplied by the eigenvalue of the eigenvector of U.

### Arithmetic

Arithmetic functions can be implemented in a quantum algorithm. It's mostly used in oracles. There are many quantum algorithms around that implement various arithmetic functions: adders, multipliers, dividers and even transcendental functions (exponential, logarithm, and trigonometric functions)[1543].

**arithmetic operations can be useful in many algorithms**

**quantum reversible adders/multipliers can be derived from their classical counterparts (ripple-carry adders) or use a QFT and IQFT (inverse QFT) to reduce the need for ancilla qubits**

| Adder type | Toffoli/T depth | Toffoli/T gates | Qubits required |
|---|---|---|---|
| Majority ripple [12] | $2n$ | $2n$ | $2n+1$ |
| Prefix-ripple [Section A.1] | $n$ | $3n$ | $2n+1$ |
| Carry look-ahead [13] | $4log_2(n)$ | $10n$ | $4n - log_2(n) - 1$ |
| Fourier transform basis [33] | $3log_2(1/\epsilon)$ | $3nlog_2(1/\epsilon)$ | $n$ |

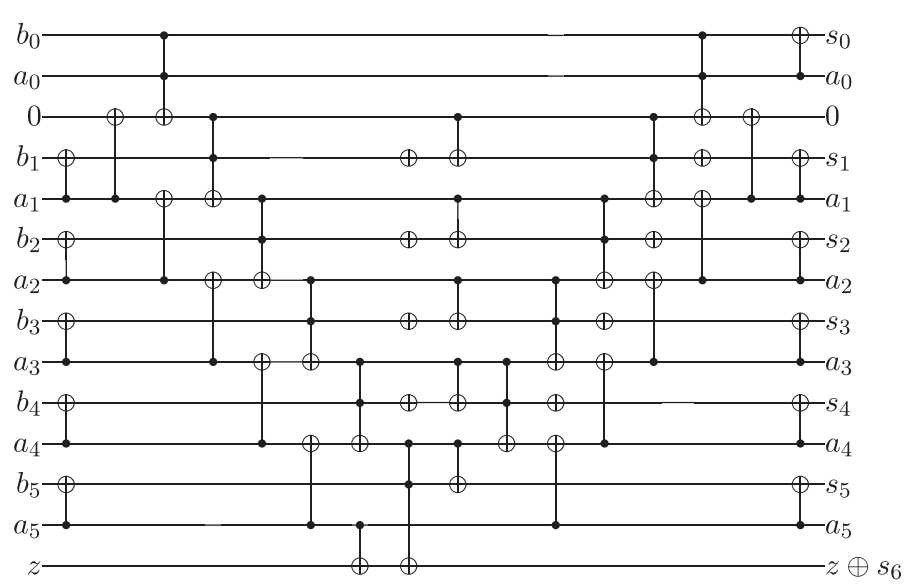

The ripple-carry adder for $n = 6$.

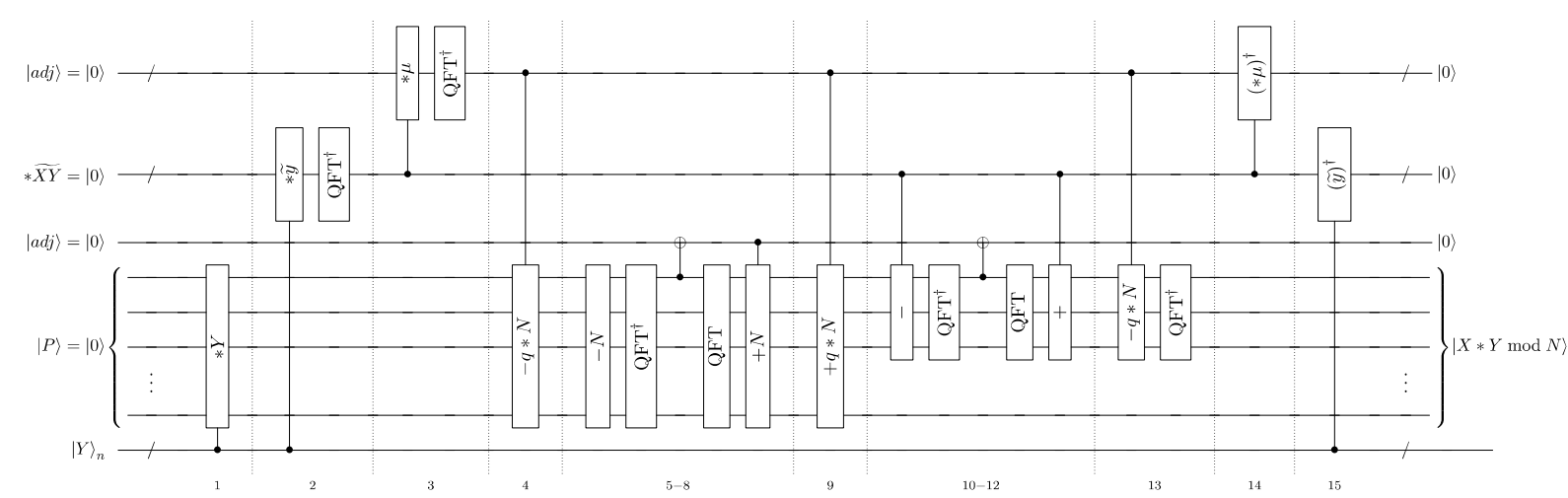

Figure 12: Barrett multiplication circuit using Fourier arithmetic. The numbers in the figure correspond to the steps of Algorithm 2.

**6-bit ripple-carry addition adder**

**Barrett multiplication circuit using QFT**

*Figure 560: various arithmetic computing can be implemented with quantum algorithms, mostly using a QFT. Sources: A new quantum ripple-carry addition circuit by Steven A. Cuccaro, Thomas G. Draper, Samuel A. Kutin and David Petrie Moulton, 2008 (9 pages) and High performance quantum modular multipliers, Rich Rinesy and Isaac Chuang, 2017 (48 pages).*

Quantum reversible adders and multipliers can be derived from their classical counterparts like with ripple-carry adders, or use a QFT and IQFT (inverse QFT) to reduce the need for ancilla qubits.

### Amplitude amplification

Amplitude amplification is a gate combination that is frequently connected to the phase kickback mechanism. It consists in amplifying one particular amplitude of the computational state vector of the control qubits that are submitted to an oracle function, at the expense of all the other amplitudes. It is used for example in the Grover operator from the Grover search algorithm as we'll see later.

In 2021, a researcher from the Fermi Lab at the DoE found a way to create an amplitude amplification working on a non-Boolean oracle[1544].

---

[1543] See A new quantum ripple-carry addition circuit by Steven A. Cuccaro, et al, 2008 (9 pages), High performance quantum modular multipliers, Rich Rinesyand and Isaac Chuang, 2017 (48 pages) and Quantum circuits for floating-point arithmetic by Thomas Haener, Mathias Soeken, Martin Roetteler and Krysta Svore, 2018 (13 pages) which were patented. See also Arithmetic on Quantum Computers: Addition, Faster by Sashwat Anagolum, October 2018, Arithmetic on Quantum Computers: Multiplication by Sashwat Anagolum, December 2018 and Everything You Always Wanted to Know About Quantum Circuits by Edgard Munoz-Coreas and Himanshu Thapliyal, August 2022 (18 pages) which provides a good overview of arithmetic algorithms among other topics.

[1544] See Non-Boolean Quantum Amplitude Amplification and Quantum Mean Estimation by Prasanth Shyamsundar, February 2021 (36 pages).

**Qubitization** was introduced by Low and Chuang in 2016 to facilitate the quantum computing of a time evolutive Hamiltonian $e^{iHt}$ to a given error $\epsilon$ for particular types of Hamiltonian resulting from the projection of an unitary oracle onto a state created by another oracle [1545].

It enables the creation of optimized amplitude amplification techniques. It led to the creation of the singular value transformation algorithm (QSVT) that brings an exponential speedup with applying polynomial transformations to the singular values of a block of a unitary[1546]. This is related to the notion of SVD (singular value decomposition) of matrices to find the values in their diagonal matrix[1547].

**Quantum Signal Processing** is a technique also created in 2016 to compute and analyze a Hamiltonian that depends on a value $\theta$ that can be viewed as an angle in some signal. The QSP is a simulation algorithm with potential exponential speedup. It is based on a linearization of the operator of a quantum walk using eigenvalue transformation using a constant number of queries[1548].

### Quantum Fourier Transform

Classical Fourier transforms are used to decompose a signal into its compound frequencies. In signal theory, this allows to identify the basic components of some sound by breaking it down into frequencies. In astrophysics, the atomic composition of stars is determined by a decomposition of the light spectrum, but this is done by an optical prism and not by Fourier transform. The same is true for Scio-type near-infrared sensors that determine the composition of food. A prism and the principle of diffraction therefore allow an optical Fourier transform to be performed.

The quantum Fourier transform was invented by **Don Coppersmith** (USA) in 1994.

A QFT is a quantum equivalent of a DFT or a FFT (Fast Fourier Transform). Its inverse operation, an inverse QFT is a QFT executed backwards, with its gates serialized in reverse order.

QFT is everywhere in the algorithm zoo as shown in red in Figure 553, page 573! Many known quantum algorithms are using it, including QPE (quantum phase estimation), HHL (linear equations), Shor's factoring algorithm and most QML algorithms. QFT helps find periodicity in a series of numbers, which is particularly helpful in Shor's algorithm.

A QFT is decomposing a series of qubits computational base states complex amplitudes in frequencies[1549]. The complex amplitude data encoding sits in the prepared register of qubits $|x_i\rangle$ with i=0 to n, as shown below.

These qubits contain a set of N=$2^n$ amplitudes $\alpha_j$ of the computational state basis orthogonal vectors $|j\rangle$, with j=0 to N-1.

The QFT implements a Discrete Fourier Transform (DFT) on these discrete amplitudes and converts it into a new computational state vector with amplitudes being the result of the QFT. The initial vector state can be written as in the formula on the right.

$$|\psi\rangle = \sum_{j=0}^{N-1} \alpha_j |j\rangle$$

---

[1545] See Hamiltonian Simulation by Qubitization by Guang Hao Low and Isaac L. Chuang, 2016-2019 (23 pages).

[1546] See Quantum singular value transformation and beyond: exponential improvements for quantum matrix arithmetics by András Gilyén, Yuan Su, Guang Hao Low and Nathan Wiebe, 2018 (67 pages).

[1547] See Everything about matrix factorizations by Tivadar Danka, 2022 which explains well what is SVD and visually.

[1548] See Optimal Hamiltonian Simulation by Quantum Signal Processing by Guang Hao Low and Isaac L. Chuang, 2016 (6 pages), Methodology of Resonant Equiangular Composite Quantum Gates by Guang Hao Low, Theodore J. Yoder and Isaac L. Chuang, PRX, 2016 (13 pages) and the tutorial A Grand Unification of Quantum Algorithms by John M. Martyn, Zane M. Rossi, Andrew K. Tan and Isaac L. Chuang, May 2021 (39 pages).

[1549] See Quantum circuit for the fast Fourier transform by Ryo Asaka et al, 2020 (20 pages) which describes a QFT variant using a faster basis encoding for the input register.

The QFT creates a new state vector $\text{QFT}_N(|\psi\rangle)$ with N $\beta_k$ amplitudes of the N computational basis vectors $|k\rangle$. It is a formula similar to the above starting point.

$$\text{QFT}_N(|\psi\rangle) = \sum_{k=0}^{N-1} \beta_k |k\rangle$$

The amplitudes $\beta_k$ are computed with a big sum using all the amplitudes $\alpha_j$ with the coefficient $\omega^{jk}$.

$$\beta_k = \frac{1}{\sqrt{N}} \sum_{j=0}^{N-1} \omega^{jk} \alpha_j$$

These coefficients $\omega^{jk} = e^{\frac{-2\pi i}{N} jk}$ explain the heavy use of $R_n$ phase rotation gates in the QFT algorithm as described on the right. You can remove the minus sign to obtain a reverse QFT.

$$\beta_k = \frac{1}{\sqrt{N}} \sum_{j=0}^{N-1} e^{-2\pi i \frac{jk}{N}} \alpha_j$$

In the end, the $\text{QFT}_\text{N}$ is a unitary matrix transformation $[\text{QFT}_\text{N}]$ with simple coefficients $[\text{QFT}_\text{N}]_{jk}$, as in a DFT.

$$[\text{QFT}_\text{N}]_{jk} = \frac{1}{\sqrt{N}} \omega^{jk}$$

When n=1 and N=2, the QFT becomes a Hadamard gate transform. The QFT is indeed presented as a generalization of the Hadamard operation, applied to dimensions N>2.

Since preparing such an arbitrary vector could take an exponential time with regards to the number of qubits, it is usually done through some faster preparation mechanism like in Shor's algorithm.

What are we really getting out of a QFT? Let's say we have 4 qubits and complex amplitudes with a rotating phase by 45° steps. It means we'll have a full phase periodic rotation for each 8 amplitudes and 2 full rotations for the whole state vector. The QFT will then output a register with the third qubit at $|1\rangle$ and all the others at $|0\rangle$.

This third qubit corresponds to the value 2, which is the frequency of the phase rotation. But we could have a more complex QFT with several added frequencies in the signal.

Getting all the $\beta_k$ coefficients and frequencies still wouldn't make much sense. Indeed, recovering a whole computational basis state would require running the QFT at least one or two orders of magnitudes of $2^N$. We'd lose any quantum speedup. What is usually done is to directly reuse this vector in the remainder of another quantum algorithm like Shor. Otherwise, after running the QFT a limited number of times, we can extract the computational basis state with the highest frequency. In other words, it means we'll have the main frequency extracted from the QFT, but not all of them.

The QFT relies on two types of logic gates: Hadamard gates to perform an overlay and two-qubit phase-controlled R gates whose phase is inversely proportional to 1 up to N. This creates a huge problem of accuracy in the calculation: the larger N is, the smaller the angle of rotation of the qubit in its Bloch sphere will be and the more impacting the phase errors will be. This requires a very precise control of the activation of the qubits.

In practice, phase-controlled R gates are generated by a combination of H, Z and T gates, plus a CNOT for the entanglement of the control qubit with the target qubit.

And it takes a lot! For example, for an $R_{15}$ gate, 127 H/Z/T gates must be used to obtain an accuracy of $10^{-5}$, which is enormous[1550]. This can be optimized with auxiliary qubits. And of course, we must integrate the associated error correction codes that add a good order of magnitude to the number of quantum gates in the depth of the calculation. This mainly impacts the calculation duration since the error correction codes are supposed to lengthen the duration of the qubit coherence.

[1550] See Efficient decomposition methods for controlled-R n using a single ancillary qubit by Taewan Kim et Byung-Soo Choi, 2018 (7 pages) and Approximate quantum Fourier transform with O(n log(n)) T gates by Yunseong Nam et al, 2020 (6 pages).

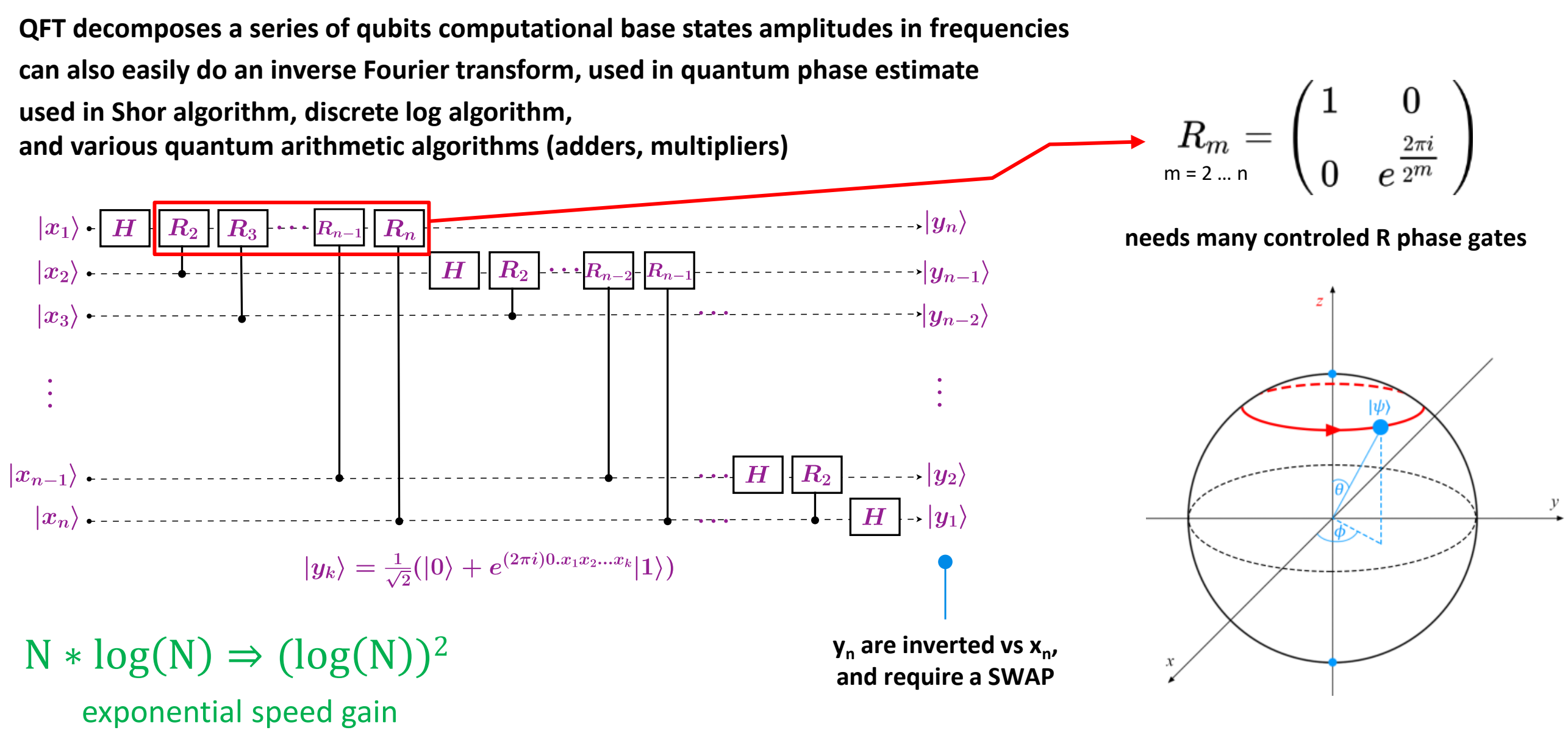

*Figure 561: the quantum gates resource constraint with a QFT are enormous as its size grows. It requires controlled R phase gates that are very costly to generate, using in many cases a long combination of tens of H and T gates. (cc) Olivier Ezratty and various sources.*

How about an $R_{2048}$ gate decomposition, the last of a long series of R phase gates to break a 2048-bit RSA key? That's about the same number of gates. This comes from the Solovay-Kitaev theorem according to which this decomposition depends only on the targeted error rate[1551]. In the case of superconducting qubits, the generation of variable phase gates is achieved by sending a shorter microwave pulse.

### Quantum phase estimation

Quantum phase estimation is an algorithm used to find the phase of an eigenvector of a unitary operator U. This operator can be implemented as an oracle function applied to a quantum state $|\psi\rangle$ that is decomposed in n controlled-unitaries operating on m qubits from $|\psi\rangle$. We are then looking for the phase angle $\theta$ according to $U|\psi\rangle = e^{2\pi i\theta}|\psi\rangle$.

This algorithm is based on an inverse QFT. Practically speaking, the QPE can estimate the angle $\theta$ with a precision $\epsilon$ with executing U for $O(1/\epsilon)$ times (meaning, with a high probability within an error $\epsilon$). The angle $\theta$ is encoded over n classical bits at the exit of the inverse QFT.

This algorithm was proposed by Alexei Kitaev in 1995[1552]. It is used among other domains in quantum chemistry and in Shor's factoring algorithm.

### Quantum amplitude estimation

Quantum amplitude estimation was proposed by Gilles Brassard et al in 2000. In simple terms, it is used to evaluate the average value of a quantum oracle. The original version is a combination of a quantum phase estimation and Grover's algorithm. Some newer versions avoid the quantum phase estimate step and are more suitable to NISQ architectures[1553].

---

1551 The main method of $R_n$ gate decomposition is documented in Optimal ancilla-free Clifford+T approximation of z-rotations by Neil J. Ross and Peter Selinger, 2016 (40 pages). It's cotton!

1552 New versions appear from time to time like Quantum Algorithm for the Direct Calculations of Vertical Ionization Energies by Kenji Sugisaki et al, University of Osaka, March 2021 (6 pages).

1553 See Quantum Amplitude Amplification and Estimation by Gilles Brassard, Peter Hoyer, Michele Mosca and Alain Tapp, 2000 (32 pages) and Amplitude estimation without phase estimation by Yohichi Suzuki et al, 2019-2022 (13 pages).

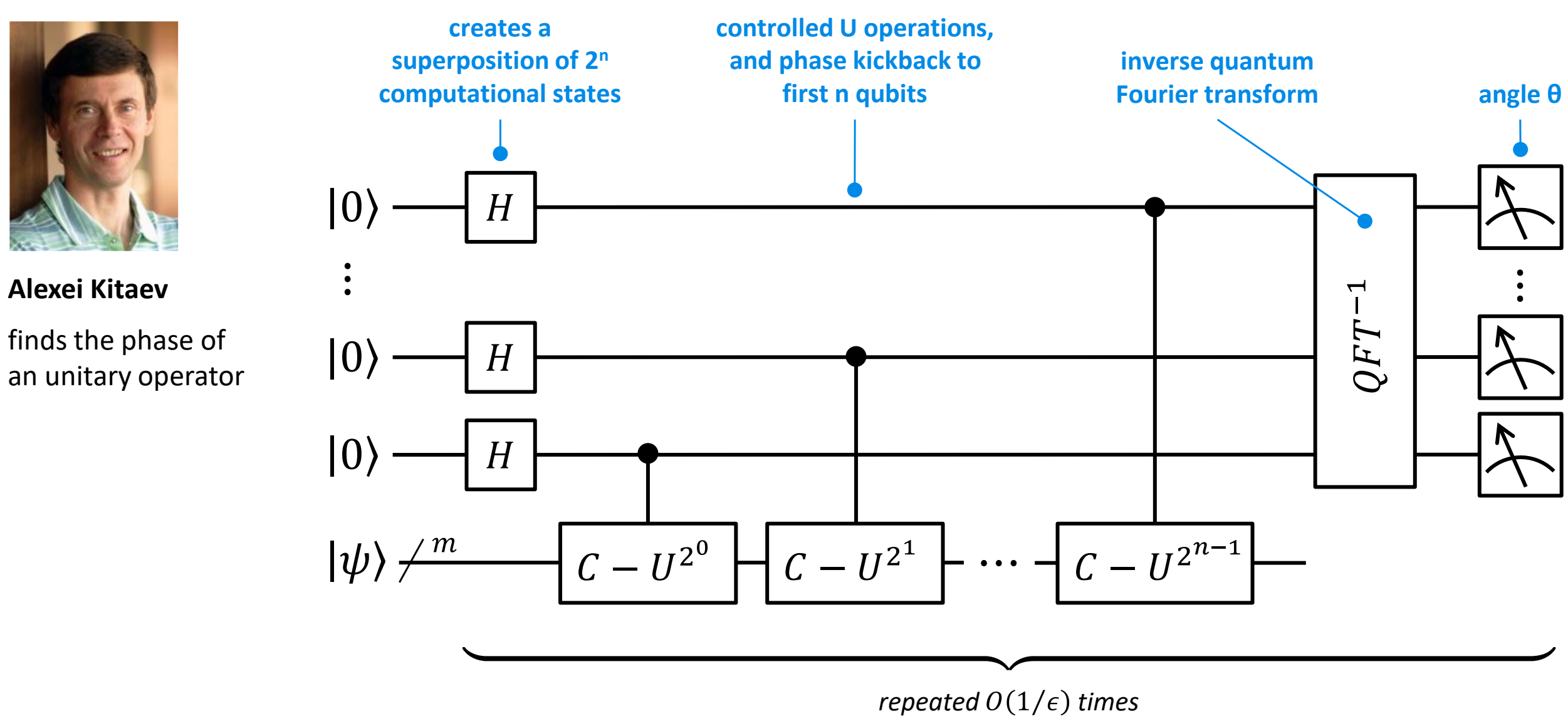

*Figure 562: the quantum phase estimate algorithm explained. The probed unitary U must be decomposed beforehand into components. (cc) Olivier Ezratty with various sources.*

## Uncompute trick

The uncompute trick was created by Charles Bennet in 1989. It is used to rewind some parts of an algorithm affecting ancilla or input qubits. It cleans up the state of a qubits register without requiring a qubit reset that may damage the stored values in the qubits with the algorithm results. It is also used to disentangle the ancilla qubits from the input qubits. It then makes it possible to go on using these ancilla qubits for the remainder of the algorithm. In a word, it cleans up the qubits register garbage at the end of some computing. The transformation works if the unitary $U_f$ is a reversible circuit which is the case for any combination of quantum gates (without any measurement done in between).

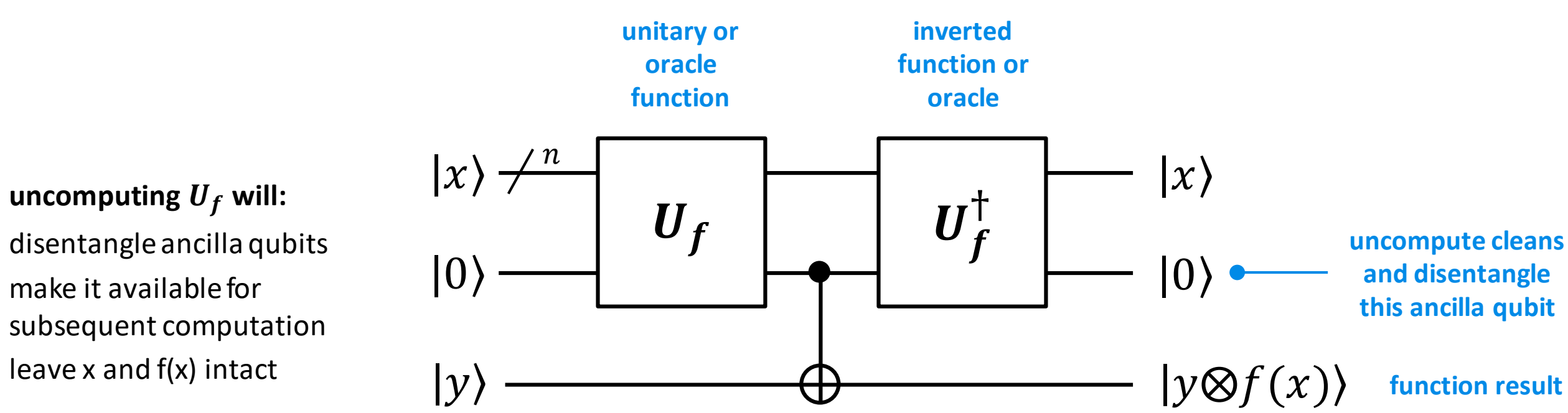

*Figure 563: the uncompute trick algorithm cleans up a register and its ancilla qubits with disentangling them from the data qubits while preserving the function result from the computed algorithm. (cc) Olivier Ezratty with various sources.*

The uncompute trick is often used when the algorithm is running an oracle function. But it presumes we are a working with "clean" qubits with no error. In a NISQ setting, an oracle inversion would generate so many errors that it would invert nothing.

## Linear equations

Many other quantum algorithms exist that allow complex mathematical operations such as solving differential equations, inverting matrices, or processing various linear algebra problems. They are then used elsewhere as in QML.

The best-known algorithm is the **HHL**, named after its creators Harrow, Hassidim and Lloyd, and created in 2009. It allows to solve linear equations, with an exponential performance gain.

The HHL algorithm input is a combining a $2^N x 2^N$ sparse Hermitian matrix A (or is prepared to be Hermitian, see the schema below in Figure 564) and an input state $|b\rangle$ with $2^N$ amplitudes. Its output is $|x\rangle = |A^{-1}b\rangle$. Namely, it inverts matrix A and multiplies it by $|b\rangle$. The processing is done in time O(N) with an exponential speed-up. But the state $|b\rangle$ must be prepared in some sort of qRAM that doesn't exist yet or be prepared quantumly. Also, input matrix A must follow a lot of constraints, with a having only a few nonzero values (sparsity).

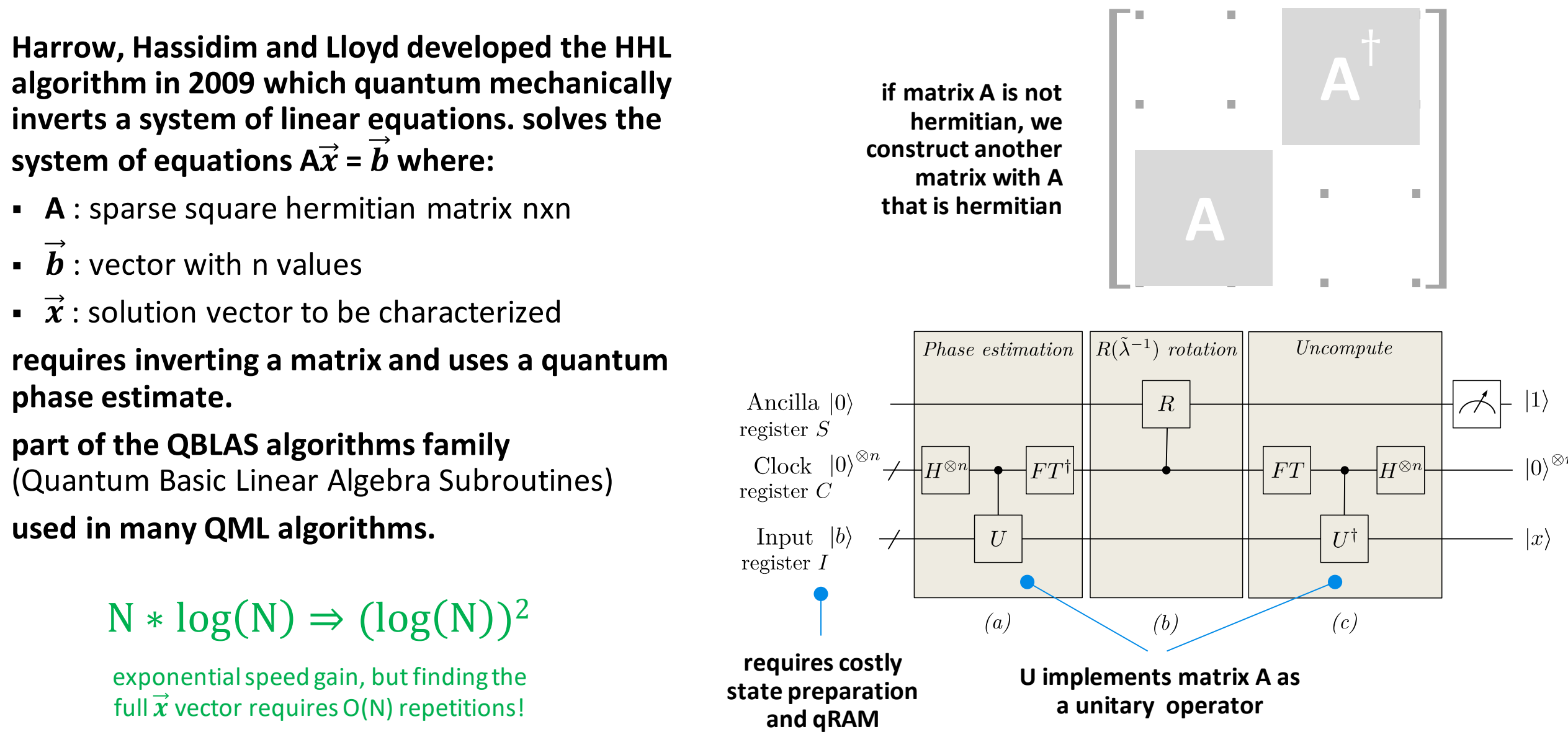

*Figure 564: the HHL linear equation solving algorithm. But its output is a quantum state that is costly to decode and should ideally be used with a subsequent quantum algorithm. The computed matrix must be Hhermitian. If it is not, it can be prepared to become Hermitian as explained on the top right. This preparation is named "block-encoding" or "standard-form" preparation.*

There's a caveat, that Scott Aaronson explained well in 2015[1554]. The HHL output is a quantum state $|x\rangle$ that can't be read right away. The vector can be read to get some statistical information about it or, among other stuff, an evaluation of a dot product between $|x\rangle$ and another vector $|z\rangle$. If you want to know everything about $|x\rangle$, you'll need to repeat the operation $2^N$ times and lose any exponential advantage gained in the first place. In the end, HHL is not really inverting the matrix A with a real exponential speedup.

Still, HHL is an algorithm that can be piggybacked by other algorithms to solve interesting problems. It is mostly used with quantum machine learning algorithms. It can also be used to solve physics and engineering problems involving Poisson's equation[1555].

### Hamiltonian simulation

Literally, as we've seen when describing Schrödinger's wave equation, a Hamiltonian of a quantum system is its description and evolution of its total energy, including kinetic and potential energy, over time. It's hard to evaluate for a given single quantum object and even harder for a multi-objects system. That's what Hamiltonian simulations are all about. One of their goals is to find the total energy of a system and its approximate ground state configuration which usually corresponds to its natural lowest-energy equilibrium state. It is particularly important in condensed matter physics and in organic chemistry.

---

[1554] See Read the fine print by Scott Aaronson, Nature Physics, 2015 (3 pages).

[1555] See Advanced Quantum Poisson Solver in the NISQ era by Walter Robson et al, September 2022 (4 pages) which shows however that such an algorithm doesn't work on an IBM 65-qubit system. A conclusion that shouldn't be generalized given future NISQ QPUs from various vendors including IBM could show up with better fidelities.

In that later case, it makes it possible to find the way molecules are naturally organized in three dimensions, from simple peptides to large proteins. It could also theoretically help simulate the interactions between different molecules.

Simulating a Hamiltonian was at the core of Richard Feynman's idea coined in 1981 when he wondered whether a quantum system could simulate another quantum system more efficiently than a classical computer, breaking down the fatal exponential growth of computing resources required to implement this kind of simulation on classical computers.

Such a simulation problem is described by a Hamiltonian which is a Hermitian matrix H of size $2^N x 2^N$, when working with N qubits or N two-states quantum objects like spin-1/2 particles. This is based on the hypothesis of a constant H or a slowly evolving one, as in the adiabatic theorem. The quantum system evolves over time according to (1), given $e^{itH}$ is the exponential of H (times i and t), is a unitary matrix. It is a solution to the Schrödinger equation (2):

$$(1) \quad |\psi(t)\rangle = e^{itH}|\psi(0)\rangle \qquad (2) \quad i\hbar\frac{\partial\psi(t)}{\partial t} = E\psi(t).$$

Simulating a Hamiltonian consists in finding matrix H or some characteristics of H. That simple. Or not.

The technique of the local Hamiltonian problem is a simplification of a Hamiltonian simulation. Thanks to special and general relativity, a Hamiltonian evolves according to local interactions. All Hamiltonian evolutions with only local interactions can be simplified as a combination of Hamiltonians acting on a limited space with at most $\ell$ of the total of N variables[1556].

Finding a system ground state of a local Hamiltonian is a QMA-complete class problem that can theoretically be efficiently solved on a quantum computer (QMA is defined later...)[1557]. Efficiently means with a polynomial instead of exponential growth in qubits. It mandates the usage of some quantum certificate or quantum proof, a validation technique used with QMA problems processing[1558].

There are of course many variations of Hamiltonian simulations depending on the type of quantum system to emulate and the characteristics we want to extract from H. It includes hybrid solutions associating classical and gates-based quantum computing including the quantum adiabatic algorithms.

If the Hamiltonian is of the family of an Ising model, it can be simulated using quantum annealers or quantum simulators. There are other families of Hamiltonians that can be simulated on quantum simulators or coherent quantum annealers with more than one degree of freedom (see Qilimanjaro).

### Quantum teleportation

One of the most intriguing quantum gate-based quantum algorithms is qubit teleportation. It was created by Charles H. Bennett (USA), Gilles Brassard (Canada), Claude Crépeau (Canada), Richard Jozsa (USA), Asher Peres (Israel) and William K. Wootters (USA) in 1993[1559] .

It allows to teleport the state of a qubit from one place to another. The principle of this algorithm consists in exploiting a pre-existing quantum entanglement channel to transmit the state of a qubit from one end of this channel to the other. Teleportation involves the transmission of two classical bits in the protocol that are used to reconstitute the qubit sent on arrival. As a result, the transmission of the latter cannot be faster than light.

---

1556 See Using Quantum Computers for Quantum Simulation by Katherine L. Brown et al, 2010 (43 pages).

1557 See QMA-completeness: the Local Hamiltonian Problem by Paul Fermé, based on lecture notes by Umesh Vazirani and lecture notes by Thomas Vidick, 2015 (6 pages).

1558 See Lecture 20: Local Hamiltonian ground state problems by Richard Kueng, on a course from John Preskill, December 2019 (17 pages).

1559 See Teleportation as a quantum computation by Gilles Brassard, 1996 (3 pages).

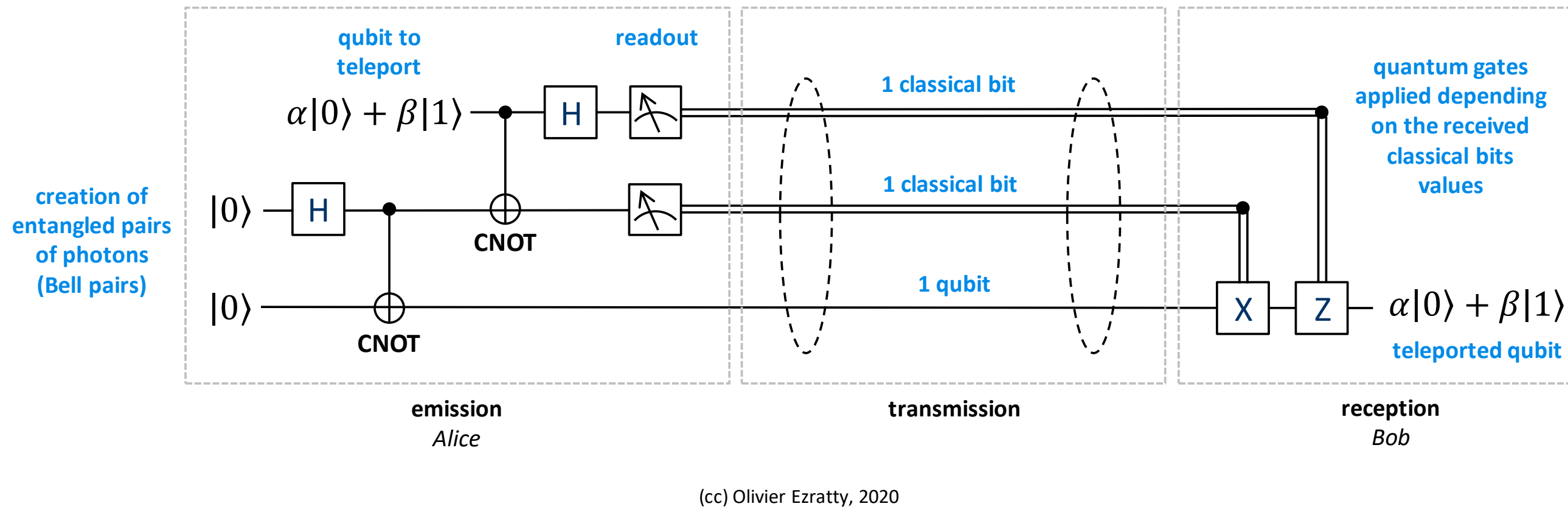

*Figure 565: the quantum teleportation algorithm and its two classical channels. (cc) Olivier Ezratty with various sources.*

Due to the quantum no-cloning theorem, this teleportation is a "move" and not a "copy" (or a "cut & paste" instead of a "copy & paste" to use an easy to understand analogy). The state of the transferred qubit is thus destroyed at its origin[1560]. The main use case of this algorithm and its many variants are in quantum cryptography and telecommunications systems that we will discover later.

It could also be used in distributed quantum computer architectures. Note that this algorithm can be tested locally in a quantum computer, as proposed by IBM in its Q Systems with Qiskit.

# Higher level algorithms

We'll now cover higher level algorithms which are based on the algorithm's toolbox described in the previous part[1561]. Quantum software engineering requires three main set of skills: having an understanding of both low and high-level algorithms, then some know-how about the way these algorithms can be assembled and also coupled with classical algorithms, and then, above all, how to find the ways to translate "business problems" into these algorithms.

### Oracle-based algorithms

One of the first quantum algorithms invented comes from David Deutsch, with its derivative called **Deutsch-Jozsa,** co-invented with Richard Jozsa and created in 1992. This algorithm makes it possible to characterize a function $f()$ called an "oracle" for which we know in advance that it will return for all its inputs, either always the same value, 0 or 1, or the values 0 and 1 in equal parts. The algorithm makes it easy to determine if the function $f()$ is balanced or not. It is working to a set of qubits n. Function $f()$ is making some classical computing on each $2^N$ values from the computational basis of n qubits.

The input qubits are all initialized to $|0\rangle$ except one which is initialized to $|1\rangle$. They are then all superposed between $|0\rangle$ and $|1\rangle$ with Hadamard gates. The qubits are thus said to have simultaneously all possible $2^{N+1}$ combinations of values.

It is easy to understand why this quantum algorithm is much more efficient than its traditional version: in traditional computation, more than half of the possible input values would have to be scanned sequentially, whereas in the quantum version, they are all analyzed at the same time by the oracle function working on all $2^N$ values of the first N qubits. The result is obtained with a few series of quantum gates, almost instantaneously, and it is perfectly deterministic.

[1560] See Quantum Teleportation in a Nutshell by Fabian Kössel, 2013 (35 slides).

[1561] See Quantum Algorithms by Ashley Montanaro, July 2016 (62 slides), Quantum algorithms: an overview by Ashley Montanaro, 2015 (16 pages), Quantum Algorithm Implementations for Beginners, 2020 (94 pages).

These superposed qubits are processed by the oracle which contains a set of gates implementing function $f()$ to be evaluated. The output is then measured to see if the function is balanced or not thanks to other Hadamard gates.

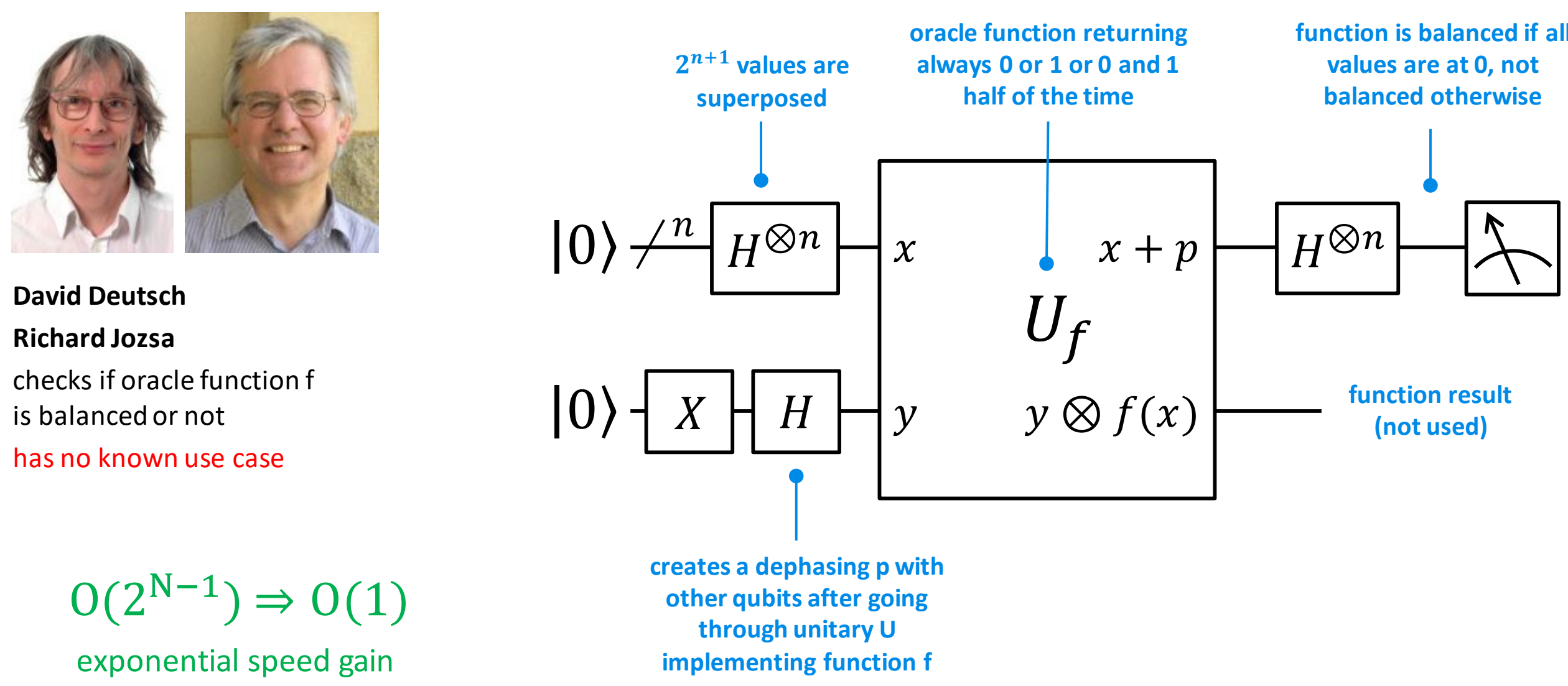

*Figure 566: the famous Deutsch-Jozsa algorithm which says if a function is balanced or not and doesn't have any known practical application as far as I know. (cc) Olivier Ezratty with various sources.*

The initialization of the last qubit to $|1\rangle$ is used to generate an interference with the other qubits that will impact the values leaving the H gates after passing through the oracle. The function $f()$ is constant if the final measurement gives $|000 \ldots 0000\rangle$ and unbalanced otherwise[1562].

What is the practical interest of such an algorithm given there are rather few functions $f()$ of this kind? This is an example of an ultra-powerful algorithm that has no known practical use to date. On top of that, there are very efficient classical probabilistic algorithms that are fast and cancel a good part of the quantum power gain coming from the Deutsch-Jozsa algorithm.

This is particularly the case with the Monte Carlo search algorithm which evaluates the oracle function on a limited number of randomly selected inputs.

The probability of errors depends on the number of evaluations and decreases very quickly[1563].

So, quantum computing is useless? Of course not. Other algorithms, less powerful but much more useful, have emerged since this patient zero of quantum algorithmics!

**Bernstein-Vazirani**'s algorithm is less talked-about in textbooks. This algorithm created by Ethan Bernstein and Umesh Vazirani in 1992 is a variant of the Deutsch–Jozsa algorithm. Instead of using two different classes of functions, it tries to learn a secret string encoded in an oracle function. The algorithm was designed to prove an oracle separation between complexity classes BQP and BPP. The speedup of this algorithm is polynomial but a derivative recursive version of the algorithm has an exponential speed gain[1564]. The algorithm has not much practical use cases although it could be used in some cryptography cases[1565].

---

1562 To find out how it works in detail, you can see the associated mathematical formulas as well as Eisuke Abe's Deutsch-Jozsa Algorithm presentation, 2005 (29 slides). But it is not that obvious!

1563 See on this subject the document Quantum Computation Models (30 pages).

1564 The algorithm is well explained in the Qiskit documentation.

1565 See Using Bernstein–Vazirani algorithm to attack block ciphers by Huiqin Xie et al, 2019 (22 pages).

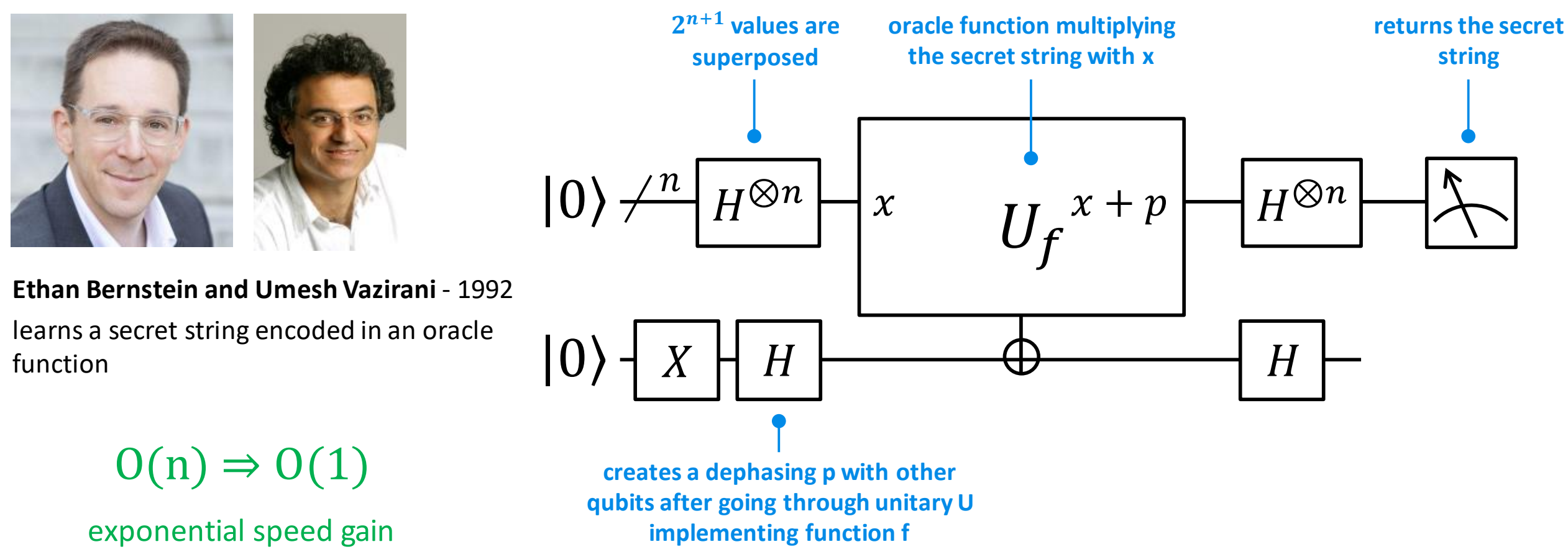

*Figure 567: Bernstein-Vazirani algorithm. (cc) Olivier Ezratty with various sources.*

**Simon**'s algorithm is a more sophisticated variant of the Deutsch-Jozsa algorithm[1566]. It consists in finding the combinations of values that verify a condition imposed by the oracle function. It solves the so-called hidden subgroup problem (HSP). Its performance gain is very interesting and, this time, the algorithm is useful, particularly to solve path problems in graphs like with quantum walks. The gain in performance is typical of what quantum computing can bring: we go from a classical calculation which is exponential time ($2^{N/2}$) to a linear time in N.

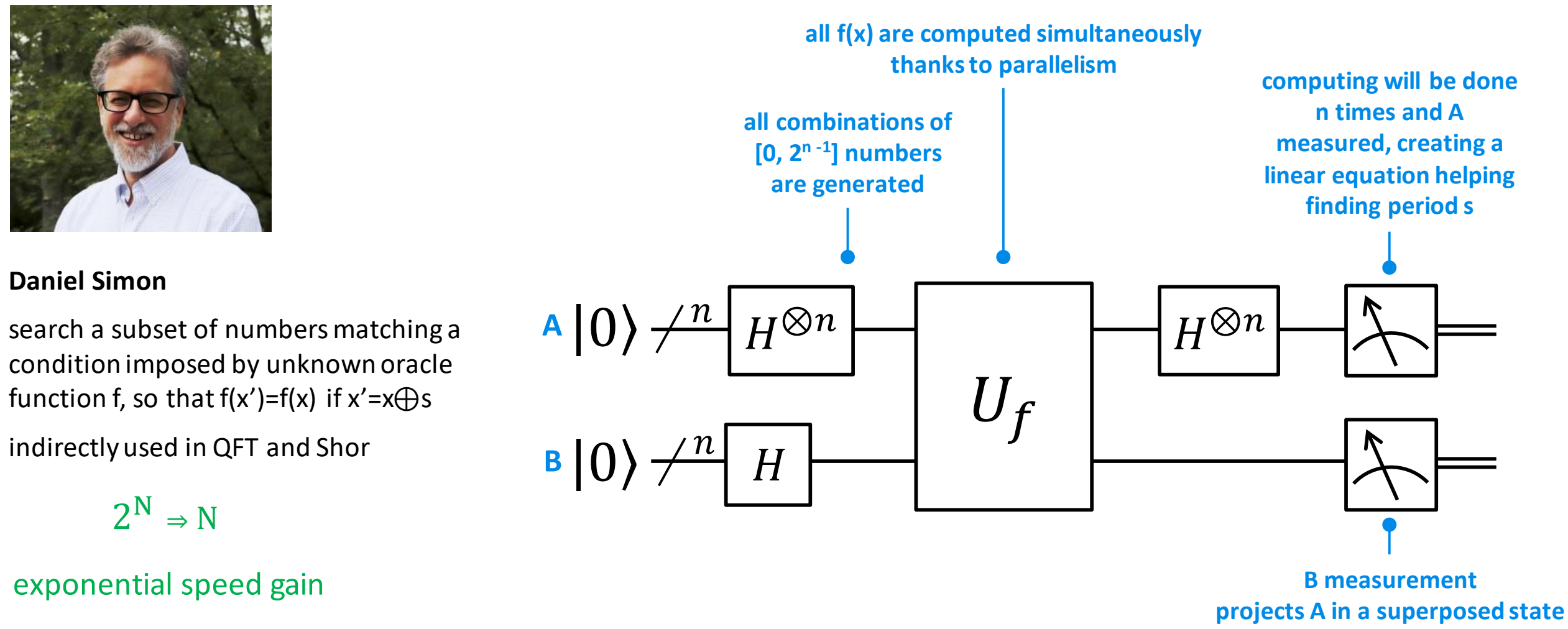

*Figure 568: Simon algorithm. (cc) Olivier Ezratty with various sources.*

The other best-known algorithm in this category is **Grover'**s algorithm, created in 1996 by Lov Grover. It allows to perform a fast quantum search in a database. It's however more generic: it can find an item in a long list that matches some specific criteria specified by an oracle function like finding the minimum item of an unsorted list of N integers, determining if a graph of N vertices is connected, or doing pattern matching searches, which can be useful in genomics.

The Grover oracle function is supposed to return 1 only for one combination of 0 and 1 with N bits. It also uses qubit state superposition to speed up processing compared to a traditional sequential search in an unsorted and non-indexed database. The performance improvement is significant compared to an unsorted database, except that in real life, we usually use indexed databases!

The question is to know if a 1 is yielded once and to which input combining 0 and 1s it corresponds. To do this, again with Hadamard gates, the algorithm will gradually amplify the combination of qubits

[1566] It is documented in [On the power of quantum computation]() by Daniel Simon, 1997 (10 pages).

of the result to an amplitude approaching 1 and make the other combinations of qubits converge to 0. This amplification operation is nicknamed the "global diffusion operator" and is repeated $\sqrt{N}$ times, N being the number of qubits. It explains why Grover's algorithm has only a quadratic speedup.

It will then be possible to measure the result and obtain the combination of qubits with the desired value (still, with repeating the algorithm several times and making an average of the results). This is well explained in Figure 569.

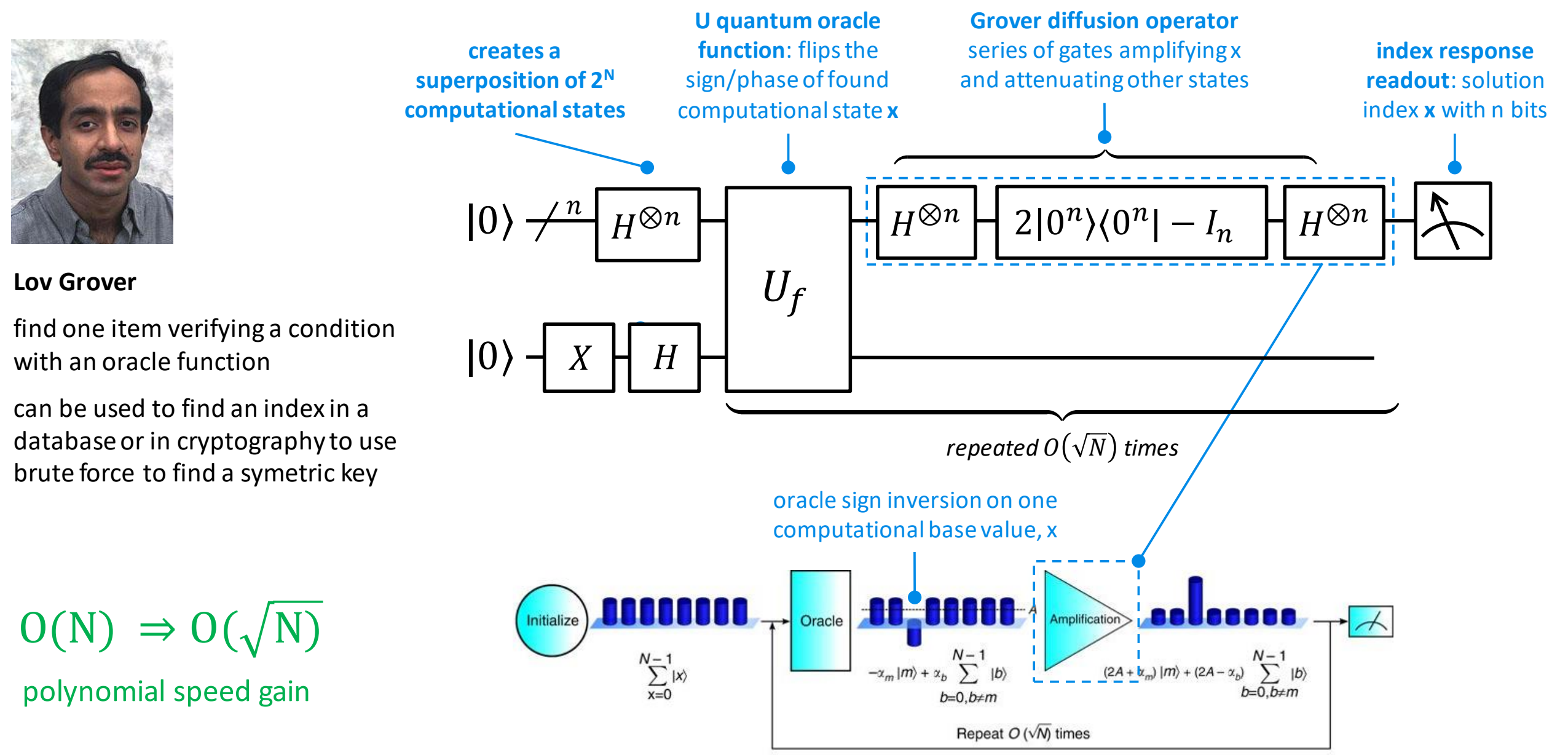

*Figure 569: Grover algorithm. (cc) Olivier Ezratty with various sources. And "Quantum Computing Explained for Classical Computing Engineers" by Doug Finke, 2017 (55 slides), broken link.*

The computing time is proportional to the square root of the base size and the storage space required is proportional to the logarithm of the base size. A classical algorithm has a computation time proportional to the size of the base. Going from a time N to √N is therefore an interesting gain, but it will not transform an exponential size problem into a polynomial size problem ($2^N$ to N power M).

On the other hand, this algorithm can be exploited to be integrated into other algorithms such as those that allow the discovery of the optimal path in a graph or the minimum or maximum number of a series of N numbers. Grover's algorithm is also used in some quantum machine learning algorithms like for measuring min/max/mean distances or other metrics between sets of data points and for automatic clustering. The settings are in the oracle, that encodes a constraint function or a function cost for which we are searching a minimum.

Note that Grover's search algorithm requires the use of quantum memory (qRAM) to "load" the related database in memory, in the oracle function[1567]! There are however some available optimization techniques available for quantum state preparation in the Grover oracle that could remove this qRAM requirement[1568].

In a 2002 lesson, **Serge Haroche** points out the known fact that these search algorithms have quantum optical interference equivalent implementations, as described in Figure 570 with 4 qubits. This has been described for a while, even trying to use only classical optical elements.

[1567] This is notably documented in Quantum algorithms for linear algebra by Anupam Prakash, 2015 (92 slides).

[1568] See Black-box quantum state preparation without arithmetic by Yuval R. Sanders et al, UNSW and Microsoft Research, 2018 (5 pages).

Various papers argue that, from a practical standpoint, these implementations don't scale well with a growing number of qubits, but they remind us that quantum algorithms are toying with waves and interferences and that optical analogies are well suited to understand their underlying processes[1569].

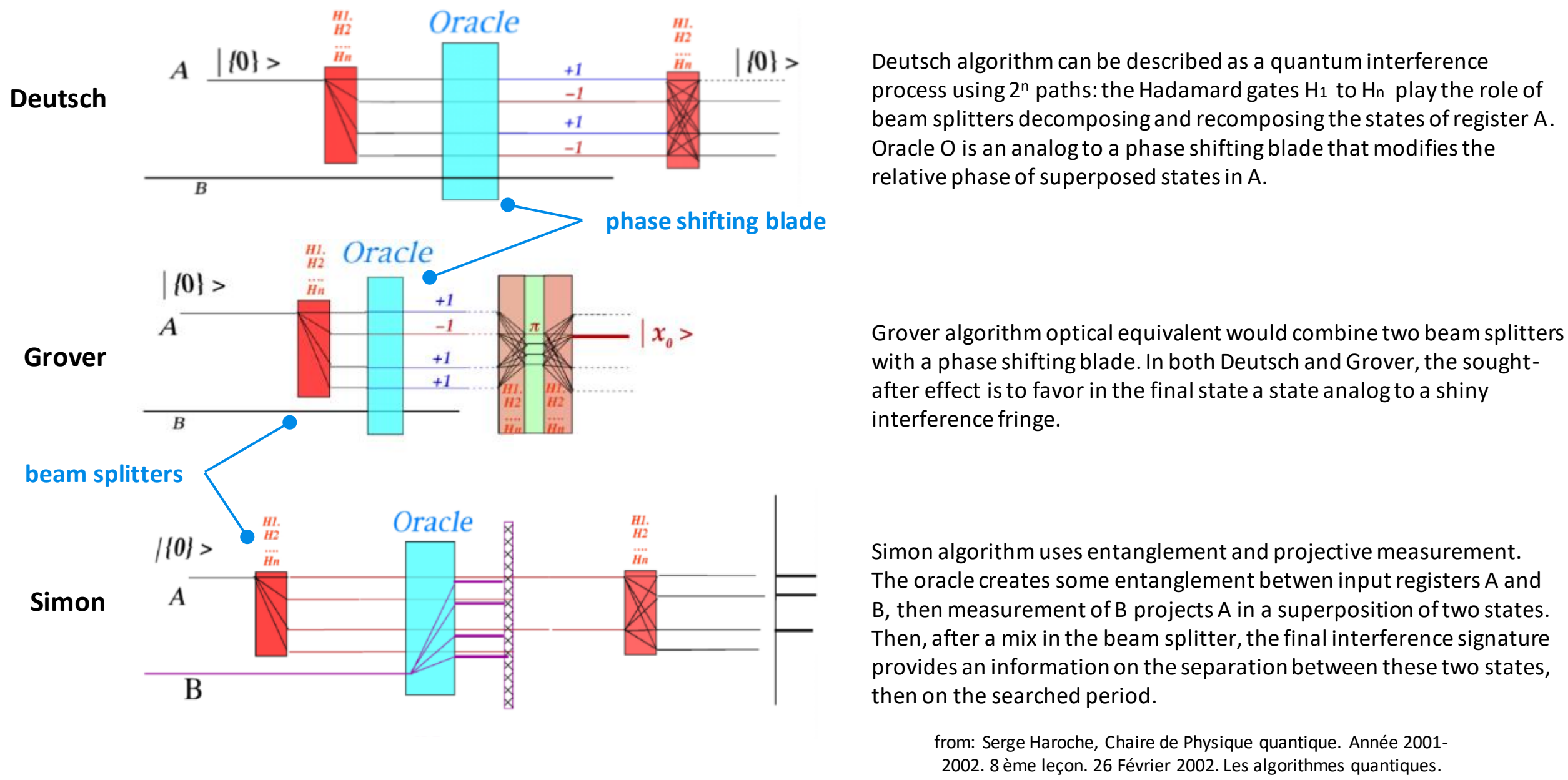

*Figure 570: quantum algorithms explained with interferences when implemented with quantum optics, by Serge Haroche.*

At last, let's remind again the reader that an oracle-based algorithm is efficient if the oracle itself is efficient, which depends on its implementation. If it is accessing some classical data or function, the algorithm's efficiency may be questionable in the end.

### Shor integer factoring

Shor's factoring allows you to decompose integers into prime numbers much faster than with a traditional computer. It works in two stages as described in the diagram below in Figure 571 [1570]:

- A **classical part** which reduces the factoring problem to and order-finding problem and produces a function f(x) implemented as a unitary in the quantum part of the algorithm.
- A **quantum part** itself made of three sub-stages with a set of Hadamard gates, repeated squaring for a modular exponentiation transformation (which could be replaced by a QFT), and an inverse QFT that extracts the solution using the periods found in the classical part. The QFT is responsible for the exponential speedup of Shor's factoring algorithm.

The gain in speed generated by Shor's algorithm compared to conventional calculation? The computation time goes from N*log(N) for the best simple Fourier transforms to $\log_2(N)$ for the QFT. We thus go from a linear order of magnitude to a logarithmic order of magnitude. But the state of the art of classical integers factoring is much better than the usual $O(\sqrt{N}/2)$ pointed out in textbooks, like:

$$\exp((1.923 + o(1))(\log N)^{1/3}(\log \log N)^{2/3})$$

[1569] See Grover's search algorithm: An optical approach by P. G. Kwiat et al, 1999 (6 pages), Implementation of quantum search algorithm using classical Fourier optics by N. Bhattacharya and al, 2002 (4 pages) and Classical wave-optics analogy of quantum information processing by Robert J. C. Spreeuw, 2001 (9 pages).

[1570] See On Shor's algorithms, the various derivatives, their implementation and their applications by Martin Ekera, 2019 (135 slides) which describes in detail how Shor's algorithm works.

One of the first implementations of Shor's algorithm took place in 2001 at IBM with an experimental quantum computer of 7 qubits, to factorize the number 15. Since then, we have just moved to a 5-digit number, 56153[1571], but with a different factoring algorithm than Shor's algorithm. It is in fact an optimization algorithm that was running on a D-Wave quantum annealer! A record was reached in 2016 with the factorization of 200,099 with 897 qubits on a D-Wave but with yet another algorithm than Peter Shor's[1572].

It is important to remember that Shor's algorithm theoretically allows to break the public keys of the RSA cryptography that is commonly used in Internet security. Public keys work by sending a very long integer number to a recipient who already has its divisor.

He just has to divide the large number received by his divisor to retrieve the other divisor and use it to decipher the encrypted message. Whoever does not have the divisor cannot exploit the complete key unless he has enormous traditional computing power to find his divisors.

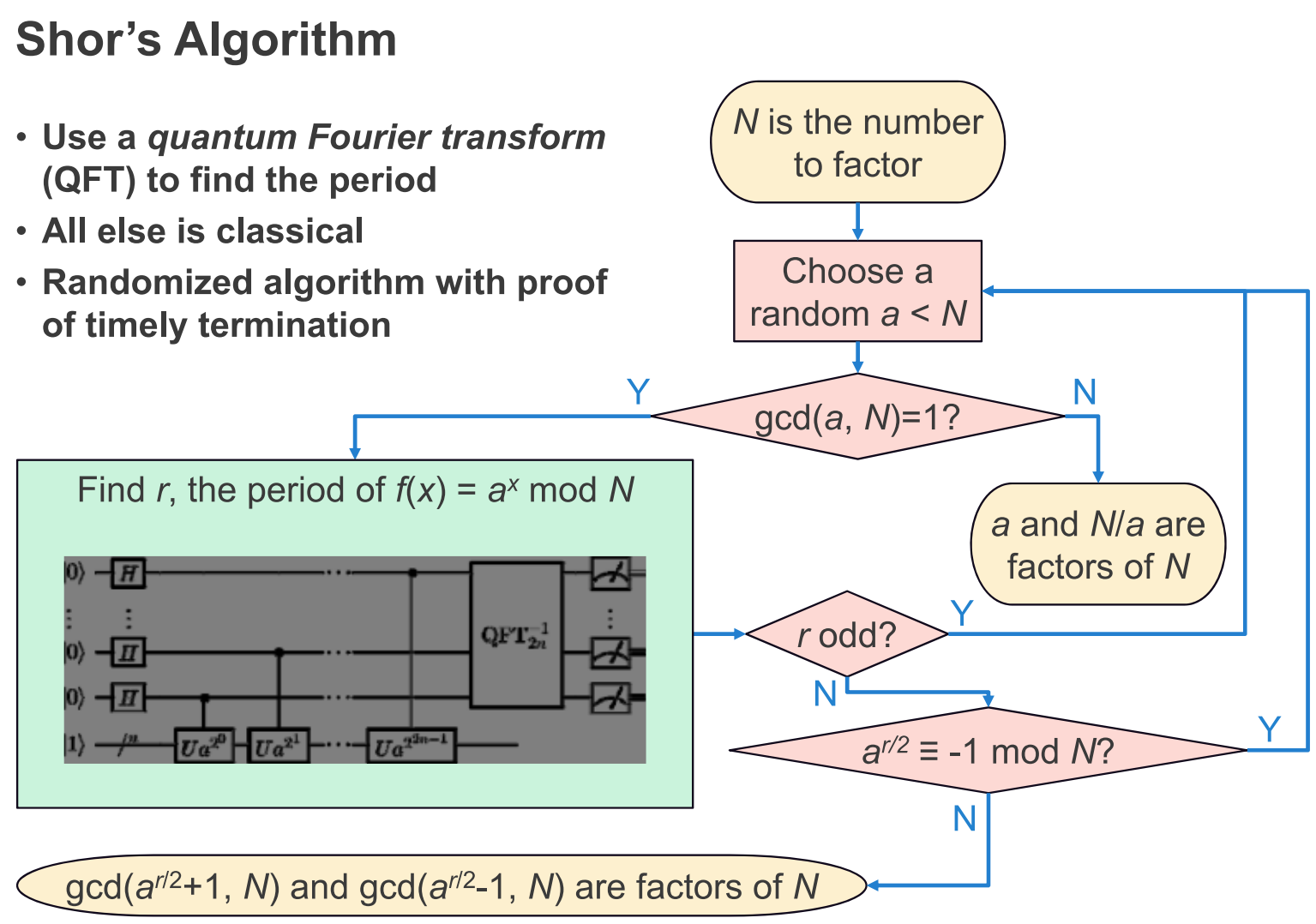

*Figure 571: Shor's algorithm high-level components. Source: Quantum Annealing by Scott Pakin, NSF/DOE Quantum Science Summer School June 2017 (59 slides).*

Until now, only NSA supercomputers have officially been able to break reasonably sized keys in the 256 to 800 bits range. But at 1024 bits and beyond, the task is inaccessible in a reasonable amount of time for these supercomputers. As far as we know!

In theory, this would become accessible someday to scalable quantum computers. To break a good 2048-bit RSA public key, one will still have to be patient because it requires to create quantum computers with a very large number of corrected qubits. It takes about twice as many logical qubits as there are bits in an RSA key. To factorize a 2048-bit RSA key, a minimum of 4098 logical qubits are required[1573]. Because of qubit noise, it is estimated that hundreds to tens of thousands of physical qubits per logical qubit would be needed.

Thus, such a RSA key break would require about 20 million qubits according to a famous Google algorithm from 2019. This algorithm based on a distance-27 surface code with about 10,000 logical qubits would require a bandwidth of 7.3 TBit/s per logical qubit for error processing[1574]. Another option would be to use some addressable quantum memory and reduce the qubits count to 13436[1575].

---

1571 This is documented in Quantum factorization of 56153 with only 4 qubits, 2014 (6 pages).

1572 The record was beaten in 2019, it was beaten by engineers from Zapata Computing and IBM with the factoring of 1,099,551,473,989 into 1,048,589 * 1,048,601, but using a variational hybrid algorithm on a few qubits, and with an undocumented speedup. See Analyzing the Performance of Variational Quantum Factoring on a Superconducting Quantum Processor by Amir H. Karamlou et al, 2019 (14 pages).

1573 The formula is 2xN+2 qubits. See Factoring using 2n + 2 qubits with Toffoli based modular multiplication by Thomas Haner et al, 2017 (12 pages) and Circuit for Shor's algorithm using 2n+3 qubits by Stephane Beauregard, 2013 (14 pages).

1574 See Hierarchical decoding to reduce hardware requirements for quantum computing by Nicolas Delfosse, January 2020 (8 pages).

1575 See Factoring 2048 RSA integers in 177 days with 13436 qubits and a multimode memory by Élie Gouzien and Nicolas Sangouard, March 2021 (18 pages). It requires some quantum memory of 2 hours storage time and qubits with a $10^{-3}$ error rate. The authors suggest realizing such an architecture with a microwave interface between a superconducting qubits processor and some multiplexed addressable quantum memory using the principle of photon echo in solids doped with rare-earth ions like Erbium or NV centers. Their physical qubits would use some 3D gauge color error correction codes.

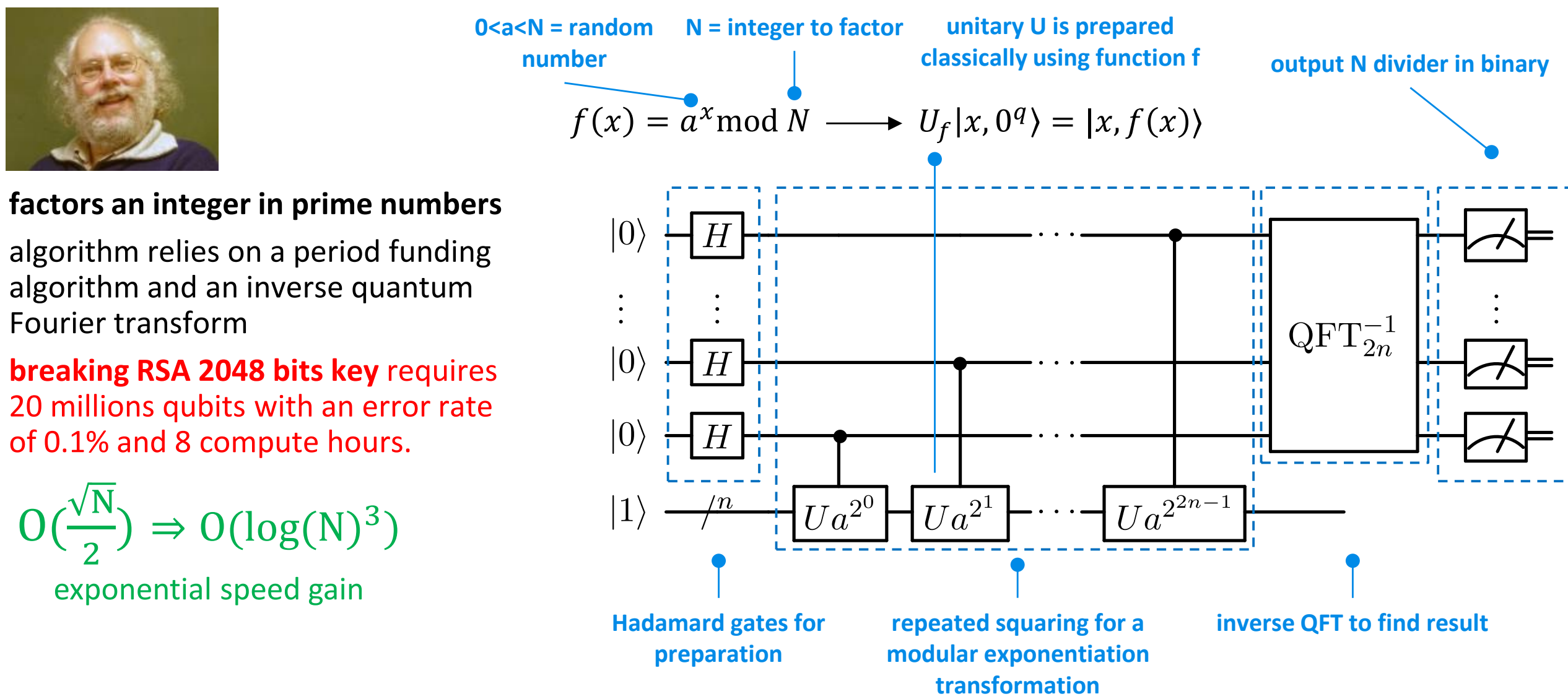

*Figure 572: Shor's algorithm with all its qubits. Source: Wikipedia description of Shor's factoring algorithm.*

Yet, another option would be to rely on qubits like cat-qubits for which the physical/logical qubits ratio would be much lower, in the 10-100 range.

Note that Shor's algorithm also allows to break cryptography using elliptic curves, which competes with RSA cryptography. By the way, some of the cryptography used in the Bitcoin protocol would also be broken by Shor integer factoring, which we will see later in this document, page 793.

In any case, Shor's algorithm has been terrorizing security specialists for a couple decades. This explains the interest in exploiting quantum keys distribution, which are supposed to be tamper-proof because their interception can be detected by their legitimate recipient, as well as post quantum cryptography, consisting in classical cryptographic algorithms and methods to make them (theoretically) tamper-proof by quantum computers using Shor's or any other algorithm.

But Shor's factoring is not the only quantum factoring algorithm created so far. Several other options are investigated like the **Variational Quantum Factoring** algorithm alternative that maps the factoring problem to the ground state of an Ising Hamiltonian that could be solved in a hybrid manner using the quantum approximate optimization algorithm (QAOA), running on a gates-based NISQ processor[1576]. But so far, the scalability of this algorithm with a large number of qubits and integer numbers is not proved.

### Shor dlog

Peter Shor did create his quantum dlog (*aka* discrete logarithm) algorithm simultaneously with his factoring algorithm in 1994 and solves another classically intractable problem. The discrete logarithm $k=\log_b(a)$ (or logarithm of a in base b) is an integer k such that $b^k = a$, where a and b are given integer numbers. You understand that this problem is intractable with digging in group isomorphisms logic, which I won't cover.

The dlog algorithm can help break Diffie-Hellman signatures, including those using elliptic curves.

Integer factoring and finding a dlog are both special cases of the hidden subgroup problem for finite Abelian groups (as seen just below).

---

[1576] See Variational Quantum Factoring by Eric R. Anschuetz, Alán Aspuru-Guzik et al, 2018 (18 pages).

## Hidden subgroup problems

The hidden subgroup problem is a generic problem which encompasses Shor's order finding, Simon's, the discrete log and the graph isomorphism problems. The definition of this problem is the following: let G be a group and H⊆G one of its subgroup. Let S be any set and f: G→S a function that distinguishes cosets of H, meaning that for all $g_1$ and $g_2$ in G, $f(g_1)=f(g_2)$ means $g_1H=g_2H$ (left cosets of H are equal). The hidden subgroup problem (HSP) is about determining the subgroup H using calls to function f with any combinations of g's in G.

Verstanden? Well, not really if you have no idea what's a group, a subgroup, a set and a coset. So let's define these:

- **Set**: arbitrary ensemble of elements.
- **Subset**: ensemble of some elements from a set.
- **Group**: a set coupled with an operation on the elements in the set, where any combination of two elements with this operation gives another item from the group. One example is the group $\mathbb{Z}$ of all integers associated with the addition. Any addition of integers yields an integer. A group also has an identity element (0 for integers) and all elements have an inverse element (inverse of integer a is -a). Operations are also associative: the order in which the operation is done is not important. For example, with integers: a+(b+c)=(a+b)+c.
- **Subgroup**: subset of the group G being also a group with regards to its associated operation. For example, with integers, the even set is a subgroup with addition since adding even numbers always give even numbers. It's however not true with uneven numbers given adding two uneven numbers gives an even number.
- **Coset**: set or ensemble of elements from G that contains all elements of H multiplied by a given item g from G. If you multiply all elements of H on the left by one element g of G, the set of products is a left coset. If multiplied by the right, it's a right coset (these operations may be non-commutative with some non-integer elements like matrices). A subgroup H of a group G may be used to decompose G into disjoint equal-size cosets. H cosets have the same number of elements as H.

Another definition of the hidden subgroup problem is: given a function f that is constant with all cosets of some subgroup H, find the subgroup H.

In its quantum version, the function f is usually implemented as an oracle. Solving HSP takes an exponential time classically with the size of log(|G|) whereas it can be solved efficiently for certain types of groups with quantum versions if done in a polynomial time of log(|G|), given log(|G|) is the logarithm of the number of elements in the group G[1577].

There are HSPs for Abelian and non-abelian groups given a group G is Abelian if xy = yx for all x, y in G. There is actually not a single quantum HSP algorithm but many of these that are applicable to different classes of groups and subgroups. It's a whole specialized field in itself.

One famous HSP problem is Pell's equation, a quadratic Diophantine equation of the form $x^2 - ny^2 = 1$ with n being a positive nonsquare integer, and x, y being integer solutions to the equation. A quantum algorithm to Pell's equation was created by **Sean Hallgren** at Princeton in 2002. It is based on a QFT[1578]. It has the particularity to be applied to an infinite group given we don't know in advance what are the bounds for x and y.

---

[1577] See a good overview of various HSP algorithms in The Hidden Subgroup Problem Master's Project by Frédéric Wang, 2010 (99 pages).

[1578] See Polynomial-Time Quantum Algorithms for Pell's Equation and the Principal Ideal Problem by Sean Hallgren, 2006 (21 pages).

Is solving that equation useful? It may be for some cryptographic purposes. The Hallgren algorithm finds one solution to the Pell equation, who has many. It has a (roughly) polynomial time vs an exponential time for its classical version, so we're in for some exponential speedup.

### Fluid mechanics

Fluid mechanics simulations are mostly based on solving Navier-Stokes equations. These are nonlinear partial differential equations, whose solution is essential to the aerospace industry, weather forecasting, plasma magneto-hydrodynamics and astrophysics. The problem with Navier-Stokes is nonlinear and quantum computing is implementing linear algebra. But there are some tricks available to turn nonlinear equations into linear ones.

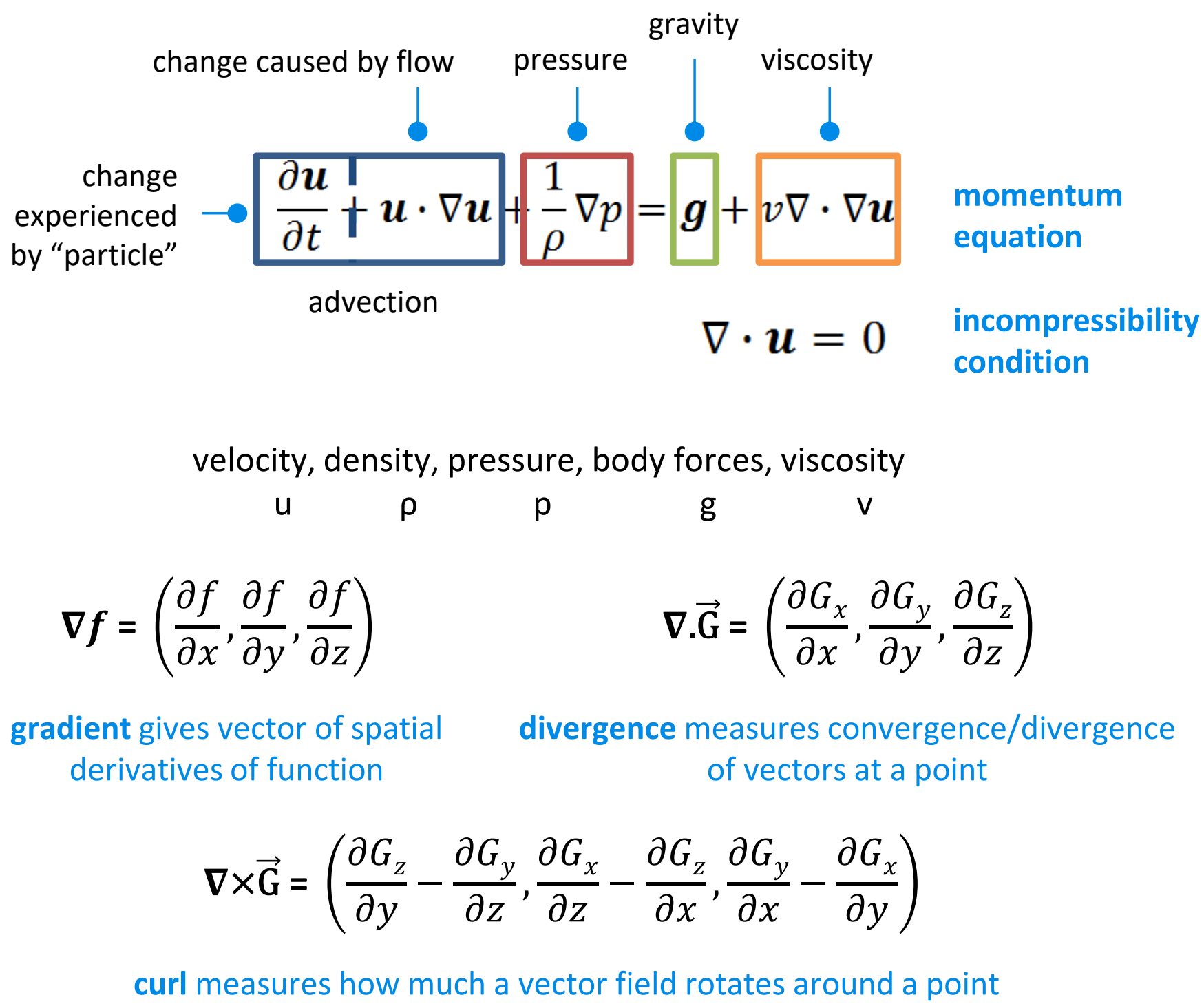

*Figure 573: Navier-Stoke equation explained. (cc) Olivier Ezratty with various sources.*

Various quantum algorithms have been designed to solve Navier-Stokes equations:

- **Hybrid computing** which makes use of a quantum nonlinear processing unit (QNPU) that is a unitary transformation implementing nonlinear operations[1579].
- **Continuous variable qubits** which implements nonlinearities with using multiple copies of the vector representing the state of the system to be investigated[1580]. Polynomial values are obtained with creating tensor products of all or part of these multiple copies of the vector state. It can simulate the dynamics of the nonlinear Schrödinger equation with quantum linear differential equation solvers.
- **Differentiable quantum circuits** to solve differential equations[1581].

---

[1579] See Variational quantum algorithms for nonlinear problems by Michael Lubasch et al, 2019 (15 pages).

[1580] See Quantum algorithm for nonlinear differential equations by Seth Lloyd et al, 2020 (17 pages).

[1581] See Solving nonlinear differential equations with differentiable quantum circuits by Oleksandr Kyriienko et al, 2020 (22 pages)

- **Converting a nonlinear system into a linear one** with transforming nonlinear problems into an array of linear equations[1582].
- **Quantum annealing for laminar plane channel flow problem** with a solution using a D-Wave quantum annealer[1583].

Worth mentioning, **Vorticity** (2001, USA) is developing custom (classical) DSA (domain specific accelerators) to solve Navier-Stokes equations with a $10^5$ speed gain over classical methods. They don't provide any technical information on their technology (FPGA, ASIC?).

### Quantum walks

Quantum walks are yet another weird beast of the quantum protocols zoo, based on sophisticated mathematical grounds. All in all, quantum walks are search algorithms in graphs. The concept was introduced in 1993 by Yakir Aharonov et al[1584]. They have many applications like searching a triangle in a graph or even Hamiltonian simulations[1585].

Andrew Childs demonstrated that quantum walks can be viewed as a universal quantum programming primitive, showing that an arbitrary set of qubit gates could be reduced to solving a quantum walk, which could be interesting with quantum systems implementing quantum walks at the hardware level one with photonic settings or even superconducting qubits[1586].

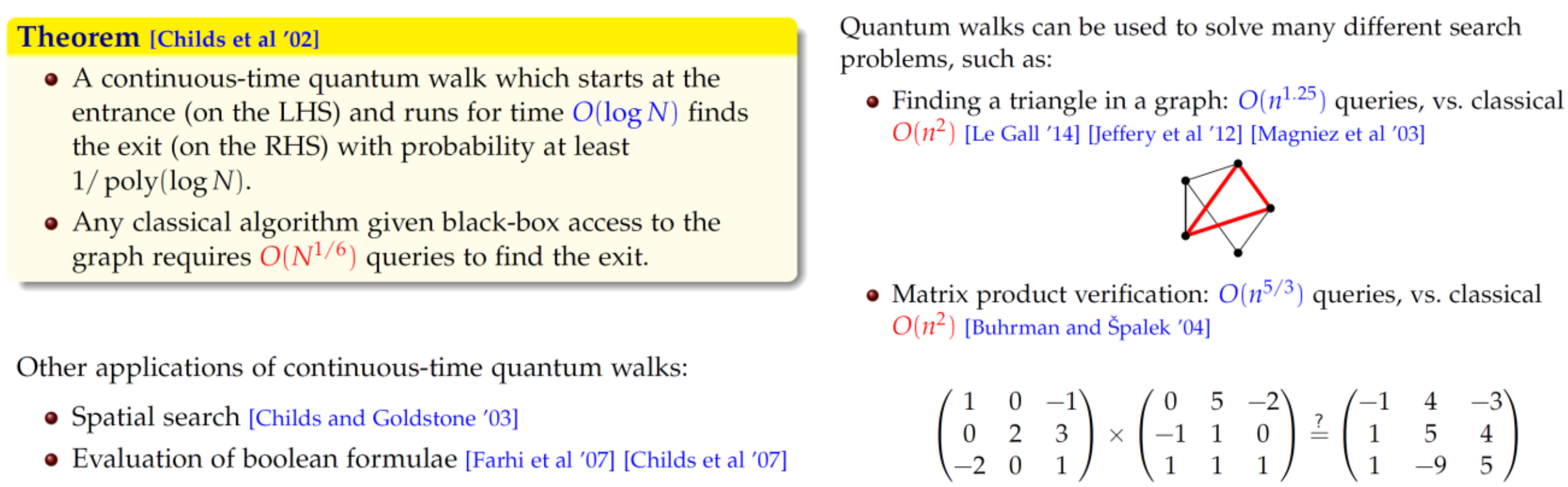

*Figure 574: quantum walks and their applications.*

That was the case with a 62 qubits system presented in China in 2021 that was dedicated to implementing a random quantum walk. The project was driven by Jian-Wei Pan. The processor is based on a 8x8 matrix of transmon superconducting qubits. It simulates a Mach-Zehnder interferometer. The matrix uses a nearest-neighbor connectivity like with Google Sycamore. Like Google's processor, two qubits were malfunctioning and disactivated, thus we have 62 instead of 64 qubits plus a non-functioning coupler. It runs at 10 mK and used 186 control lines, including 16 readout output lines with lines shared by 4 qubits[1587]. Quantum walks can also be implemented with photon qubits[1588].

---

[1582] See Efficient quantum algorithm for dissipative nonlinear differential equations by Jin-Peng Liu, Andrew M. Childs et al, March 2021 (36 pages) and New Quantum Algorithms Finally Crack Nonlinear Equations by Max G Levy, Quanta Magazine, January 2021.

[1583] See Towards Solving the Navier-Stokes Equation on Quantum Computers by N. Ray et al, April 2019 (16 pages).

[1584] See Quantum random walks by Yakir Aharonov (father of Dorit Aharonov), L. Davidovich and N. Zagury, 1993 (4 pages). See also this overview in Quantum walks by Martin Štefaňák, 2020 (44 slides).

[1585] See On the relationship between continuous- and discrete-time quantum walk by Andrew M. Childs, 2008 (22 pages).

[1586] See Universal computation by quantum walk by Andrew M. Childs, 2008 (9 pages).

[1587] See Quantum walks on a programmable two-dimensional 62-qubit superconducting processor by Ming Gong et al, 2021 (18 pages).

[1588] See Two-dimensional quantum walks of correlated photons by Zhi-Qiang Jiao et al, 2021 (22 pages).

In classical computer science, random walk or Markov chain are a algorithmic tools applied to search and sampling problems. Their quantum walks equivalent provide a framework for creating fast quantum algorithms. Quantum walks are based on the simulated coherent quantum evolution of a particle moving on a graph. Quantum walk algorithms use faster hitting (the time it takes to find a target vertex from a source vertex) and faster mixing (the time it takes to spread-out over all vertices after starting from one source vertex) [1589]. The quantum time gain can be exponential for hitting and quadratic for mixing. Since quantum walks are efficient ways to evaluate Boolean formulae, it can be used to solve satisfaction problems (MaxCut, SAT, 3-SAT).

In gate-based systems, quantum walks can be solved using a Grover search with an oracle function using an adjacency matrix for the searched walk. It can help find the shortest path in a graph (we're back at the traveling salesperson's problem), finding if a graph is bipartite (with all edges in one vertex connected to the edges in the other), finding subgraphs such as a triangle and solving maximal clique problems (used for example in social networks to find groups of people who know each other).

Then, you have quantum random walks that help reduce quantum walks query complexity to search and find graph properties, with the discrete time and continuous time variations[1590]. These are equivalent of the famous Galton's board.

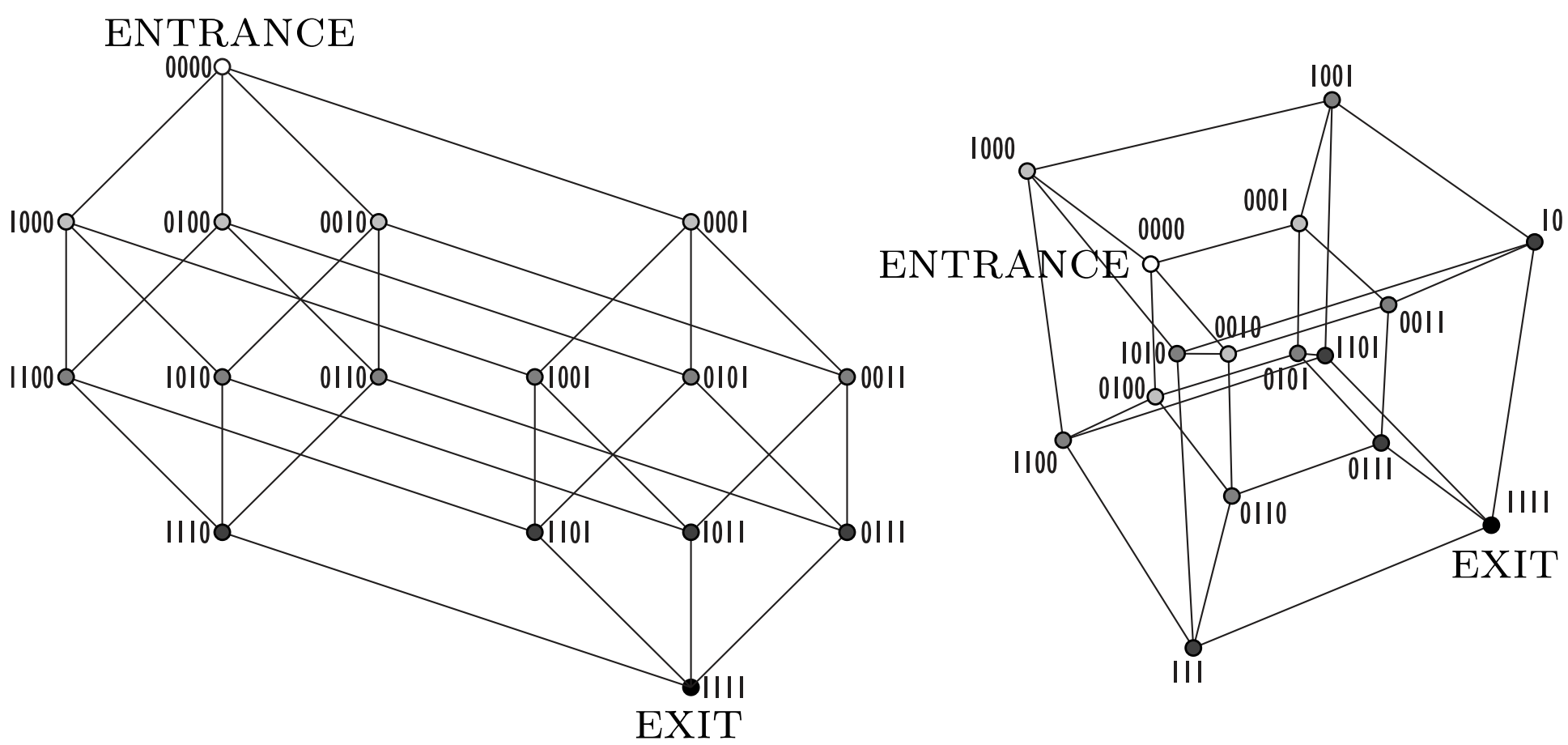

*Figure 575: Source: Quantum Walks by Daniel Reitzner, Daniel Nagaj and Vladimir Buzek, 2012 (124 pages), page 13*

On top of Andrew M. Childs, let's mention three great contributors to the quantum walk domain: **Stacey Jeffery**[1591], **Julia Kempe** and **Frédéric Magniez**.

### Quantum machine learning

What if quantum computing could accelerate machine learning and deep learning training and inferences? This is one of its potential domains of applications, but it is not that obvious. First, quantum computing does not seem to enable machine learning tasks that are impossible to implement with classical computing, including the many specialized hardware (tensors processing units, spiking neurons).

Second, their benefits are still hard to evaluate, particularly given all quantum machine learning are hybrid in nature.

---

[1589] I'm summarizing here the quantum walks description from Quantum algorithms an overview by Ashley Montanaro 2015 (16 pages).

[1590] See Quantum Algorithm Implementations for Beginners by Abhijith J. et al, 2020 (98 pages) provides many references related to quantum walks and quantum random walks algorithms.

[1591] See her thesis: Frameworks for Quantum Algorithms by Stacey Jeffery, 2014 (166 pages) and a lot of subsequent work in quantum walks algorithms.

In many cases, benchmarks tend to favor a comparison in the quality of the results like minimizing an error function and error rates more than proving a quantum speedup[1592].

Various quantum algorithms have been created in the last decades that cover the field of classical machine learning and with some variations in neural networks and deep learning[1593]. Many are based on linear algebra algorithms like the foundational HHL algorithm.

The literature on QML defines four models that connect how the data is fed into the model (classically, quantumly) and how the process is handled (classically or at least partially quantumly) as described in Figure 576 [1594]:

**CC** with classical data that are processed by classical algorithms. This is classical machine learning.

**CQ** with classical data that is encoded in quantum states and processed by quantum algorithms, which may need the use of a quantum RAM (qRAM). This is the most common method in NISQ systems.

**QC** with quantum data that is converted in a classical form and processed by classical algorithms, a solution implemented to analyze quantum physics and sensors measurement statistics, like for doing a qubit tomography.

**QQ** with quantum data that is processed by quantum algorithms which could be implemented with feeding a QML algorithm directly with quantum data coming from a quantum sensor.

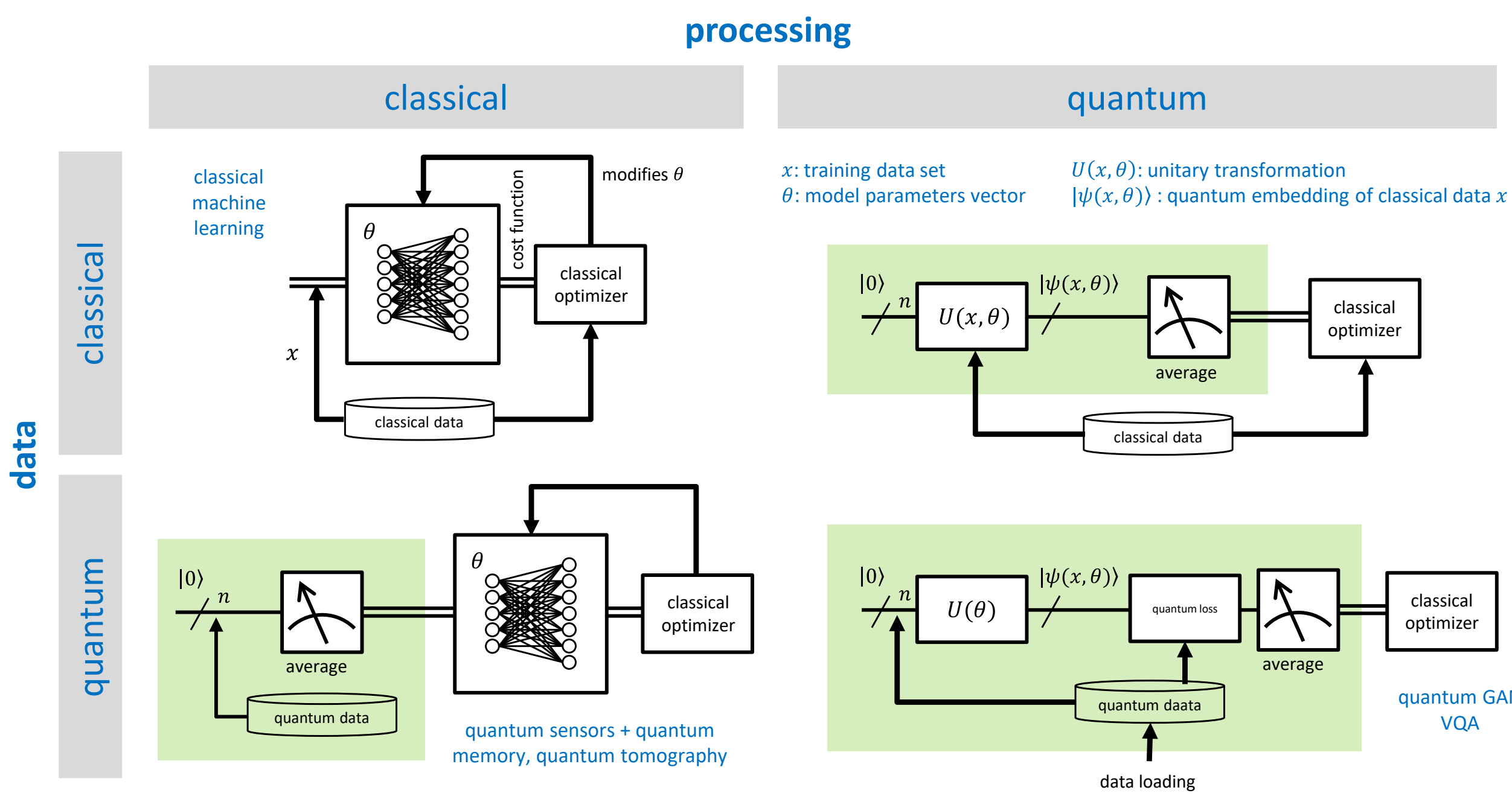

*Figure 576: the four main types of QML depending on whether data loading is classical or quantum and part of the processing is classical or quantum. Source: schema inspired by An Introduction to Quantum Machine Learning for Engineers by Osvaldo Simeone, July 2022 (229 pages).*

[1592] See Is quantum advantage the right goal for quantum machine learning? by Maria Schuld and Nathan Killoran, March 2022 (10 pages) and Why measuring performance is our biggest blind spot in quantum machine learning by Maria Schuld, Xanadu, March 2022, which provides an interesting perspective on why QML is not an obvious near-term candidate for some quantum advantage vs classical methods. And Quantum machine learning: a classical perspective by Ciliberto et al, 2020 (26 pages). It concludes with: "*Despite a number of promising results, the theoretical evidence presented in the current literature does not yet allow us to conclude that quantum techniques can obtain an exponential advantage in a realistic learning setting*".

[1593] See Machine Learning in the Quantum Era - Machine Learning unlocks the potential of emerging quantum computers by Loïc Henriet (Pasqal), Christophe Jurczak (Quantonation) and Leonard Wossnig (Rahko), November 2019. It highlights the potential of cold atom-based qubits for QML.

[1594] See An Introduction to Quantum Machine Learning for Engineers by Osvaldo Simeone, July 2022 (229 pages).

QML makes use of PQC, aka **parametrized quantum circuits** or parametric quantum circuits (not to be confused with post-quantum cryptography)[1595]. These are also labelled as "ansatz". These circuits contain rotation gates that contain the parameters of the problem to encode. A PQC model is usually defined by a classical optimizer. There are several types of data preparation ansatzes: with only single-qubit gates ("mean field ansatz") which require only N gates for N qubits and is simple and fast, then "hardware efficient ansatz" that also use a fixed number of entangling gates and then parametrized 2-qubit gates that are entirely tailored for the problem. These three models are presented below in Figure 577[1596].

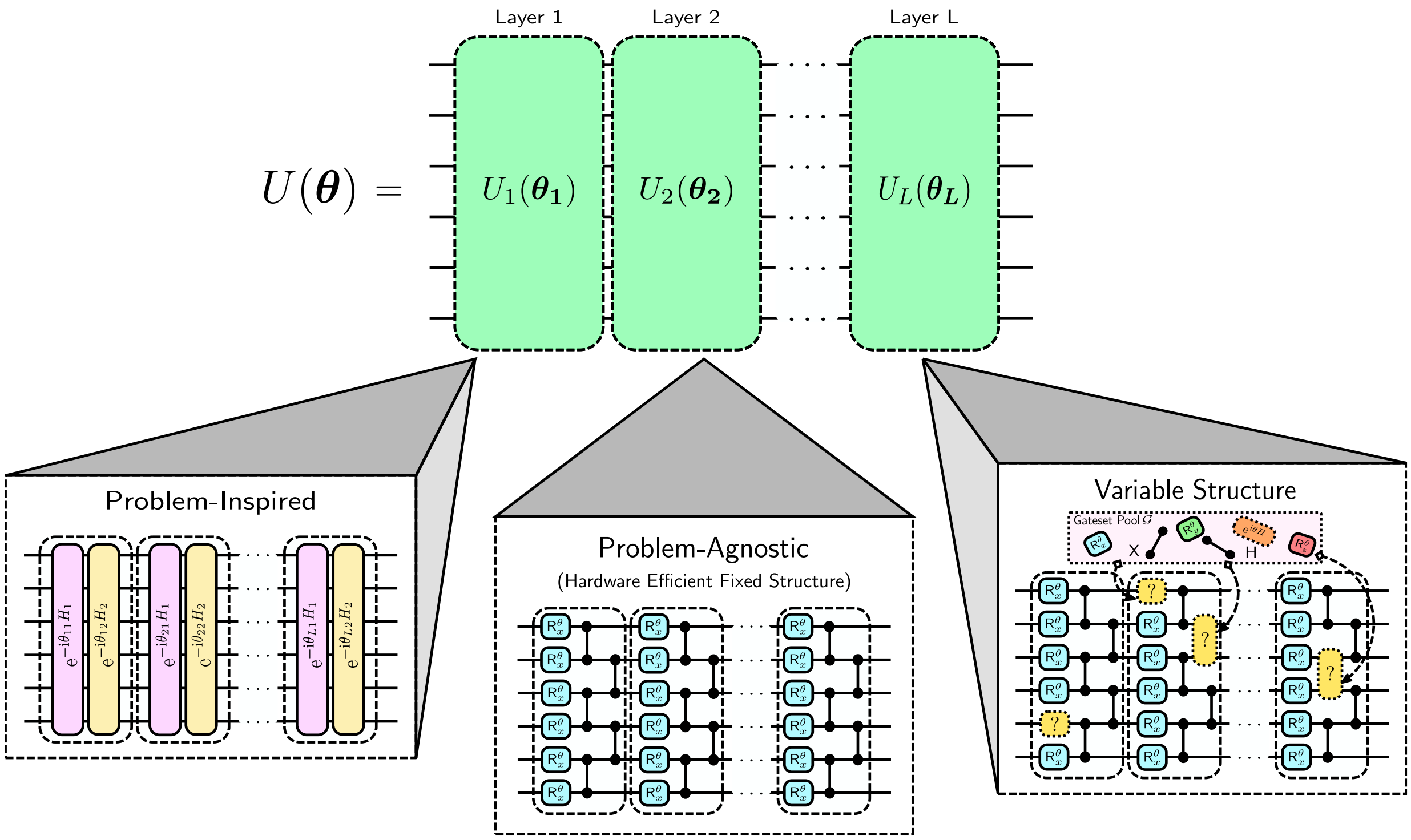

*Figure 577: various ways of preparing a QML ansatz or model, problem inspired, problem agnostic and with a variable structure. Source: [Machine learning applications for noisy intermediate-scale quantum computers](#) by Brian Coyle, University of Edinburgh, May 2022 (263 pages).*

Generically, QML algorithms from the CQ category rely on **variational circuits**, a family of hybrid algorithms that combine a quantum algorithm and a traditional algorithm that drives the latter[1597]. VQE and VQA are some of them[1598]. It allows finding global minimums. These algorithms are adapted to NISQ QPUs but their true quantum acceleration is not proven yet.

These algorithms are characterized by a cost function defining the problem, the quantum ansatz, how is the cost function optimized, the data encoded as the input to the VQA and the desired output.

---

[1595] See the review paper on PQCs Parameterized quantum circuits as machine learning by Parameterized quantum circuits as machine learning models by Marcello Benedetti, Erika Lloyd, Stefan Sack and Mattia Fiorentini, Quantum Science and Technology, 2019 (18 pages).

[1596] See the thesis Machine learning applications for noisy intermediate-scale quantum computers by Brian Coyle, University of Edinburgh, May 2022 (263 pages), done under the supervision of Elham Kashefi.

[1597] See Universal Variational Quantum Computation by Jacob Biamonte, 2019 (5 pages).

[1598] See Accelerated Variational Quantum Eigensolver by Daochen Wang, Oscar Higgott, and Stephen Brierley, 2019 (11 pages) which proposes a machine learning method to reduce the depth of the quantum circuits used (number of quantum gates to be executed). See also Quantum advantage with shallow circuits by Robert König et al, 2018 (97 slides). This list of quantum machine learning algorithms can be found in Quantum Machine Learning What Quantum Computing Means to Data Mining by Peter Wittek, 2014 (178 pages).

The breadth of QML algorithms is vast, covering all categories of classical machine learning and deep learning with supervised machine learning (to classify items or make predictions on time series), unsupervised machine learning (for automatic clustering), up to all sorts of neural networks (convolutional networks, recurrent networks, generative networks):

- Hybrid **quantum nonlinear regression** algorithm, one of the basic quantitative value prediction methods of the learning machine[1599].
- **SVM** (Support Vector Machine), a traditional method of segmentation that often relies on matrix inversions, based on its use of HHL[1600]. It can be used for text sentiment analysis.
- **PCA** (Principal Component Analysis) is used to determine the key variables in a data set[1601]. This is similar to searching for eigenvectors of a data set. Again, HHL is behind it.
- **Recommendation** systems useful in marketing or content[1602].
- **Gradient descent** used during training phase of neural network[1603].
- Unsupervised learning algorithm for **automatic data clustering**[1604].

Now onto deep learning algorithms:

- The implementation of **Perceptrons** which is the original neural network architecture designed by Franck Rosenblatt in 1957[1605].
- Quantum **convolutional neural networks**, still modest in size for the moment[1606]. These are now named **Quanvolutional Neural Network** algorithms[1607].

---

1599 See Nonlinear regression based on a hybrid quantum computer, 2018 (7 pages), from researchers in several laboratories in China.

1600 See Support Vector Machines on Noisy Intermediate-Scale Quantum Computers by Jiaying Yang, 2019 (79 pages) which discusses the use of SVM on NISQ computers, Quantum Machine Learning with Support Vector Machines by Anisha Musti, April 2020 and a practical example with Quantum support vector machines for aerodynamic classification by Xi-Jun Yuan et al, August 2022 (12 pages).

1601 See Quantum principal component analysis by Seth Lloyd, Masoud Mohseni and Patrick Rebentrost, from MIT and Google, July 2013 (9 pages) which lays the groundwork on the matter.

1602 See Quantum Recommendation Systems by Iordanis Kerenidis and Anupam Prakash, 2016 (22 pages, and video) is a proposed quantum machine learning algorithm for recommendation. The quantum algorithm of Iordanis Kerenidis had been challenged by a classical algorithm proposal by Ewin Tang in 2018. She "dequantized" Kerenidis's algorithm, meaning, she found a classical efficient equivalent. But Iordanis pointed out that with certain recommendation parameters, the quantum algorithm was still clearly superior. Always, as long as a machine is there to execute it. Both these algorithms are investigated in Exponential Advantages in Quantum Machine Learning through Feature Mapping by Andrew Nader et al, December 2020 (16 pages).

1603 See Quantum algorithms for feedforward neural networks by Jonathan Allcock, Iordanis Kerenedis et al, 2018 (18 pages) and Quantum Circuit Parameters Learning with Gradient Descent Using Backpropagation by M Watabe et al, 2020 (15 pages).

1604 See Quantum spectral clustering by Iordanis Kerenidis and Jonas Landman, April 2021 (20 pages). The method named spectral clustering consists in building a similarity graph with using distances between data vectors, extracting the eigenvectors from a matrix built with this graph and projecting the data onto this new orthogonal space and applying a classical k-means clustering method. But as seen frequently with QML algorithms, the best acceleration requires using some qRAM.

1605 See An Artificial Neuron Implemented on an Actual Quantum Processor by Francesco Tacchino et al, 2018 (8 pages).

1606 See Quantum Convolutional Neural Networks by Iris Cong et al, May 2019 (12 pages), Quantum Neurons: analyzing the building blocks of quantum deep learning algorithms by Zachary Cetinic et al, December 2019 (12 pages) and Quantum Algorithms for Deep Convolutional Neural Networks by Iordanis Kerenidis, Jonas Landman and Anupam Prakash, 2019 (31 pages). Also, Advances in Quantum Deep Learning: An Overview by Siddhant Garg and Goutham Ramakrishnan, May 2020 (17 pages) is focused on quantum neural networks including quantum convolutional neural networks and contains a good introduction to classical neural networks. And Realizing quantum convolutional neural networks on a superconducting quantum processor to recognize quantum phases by Johannes Herrmann et al, Nature Communications, 2022 (7 pages) which implements a QCNN algorithm on a 7-qubit QPU from IBM for a narrow problem (recognizing quantum phases) with some superiority in the results quality (and of course, not with a speedup given the low number of qubits).

1607 See Predict better with less training data using a QNN by Barry D. Reese, Marek Kowalik, Christian Metzl, Christian Bauckhage, and Eldar Sultanow, Capgemini, June 2022 (23 pages).

They seem to have the advantage of avoiding the ill-fated barren plateaus that make it difficult to converge a network[1608]. Variations of QCNN algorithms also exist for D-Wave quantum annealers[1609].

- Quantum **graph neural networks** have many applications, particularly in chemistry and biology[1610].
- **Feature mapping** in deep learning and convolutional neural networks, to detect patterns efficiently[1611].
- **Recurrent Neural Networks** used for MNIST handwriting recognition, an existing common task for classical OCR (optical character recognition)[1612].
- **Equivariant Neural Networks** (ENN) with better geometrical robustness and a better resistance to adversarial attacks[1613].
- **Generative Machine Learning** models, including the models based on so-called quantum circuit Born machines (QCBM). It can be used to create (quantumly generated) synthetic training data sets used in classical machine learning models[1614].
- **Quantum Generative Adversarial Networks** (qGAN) algorithms that generate synthetic content from existing content by checking its plausibility via a network of recognition neurons[1615].

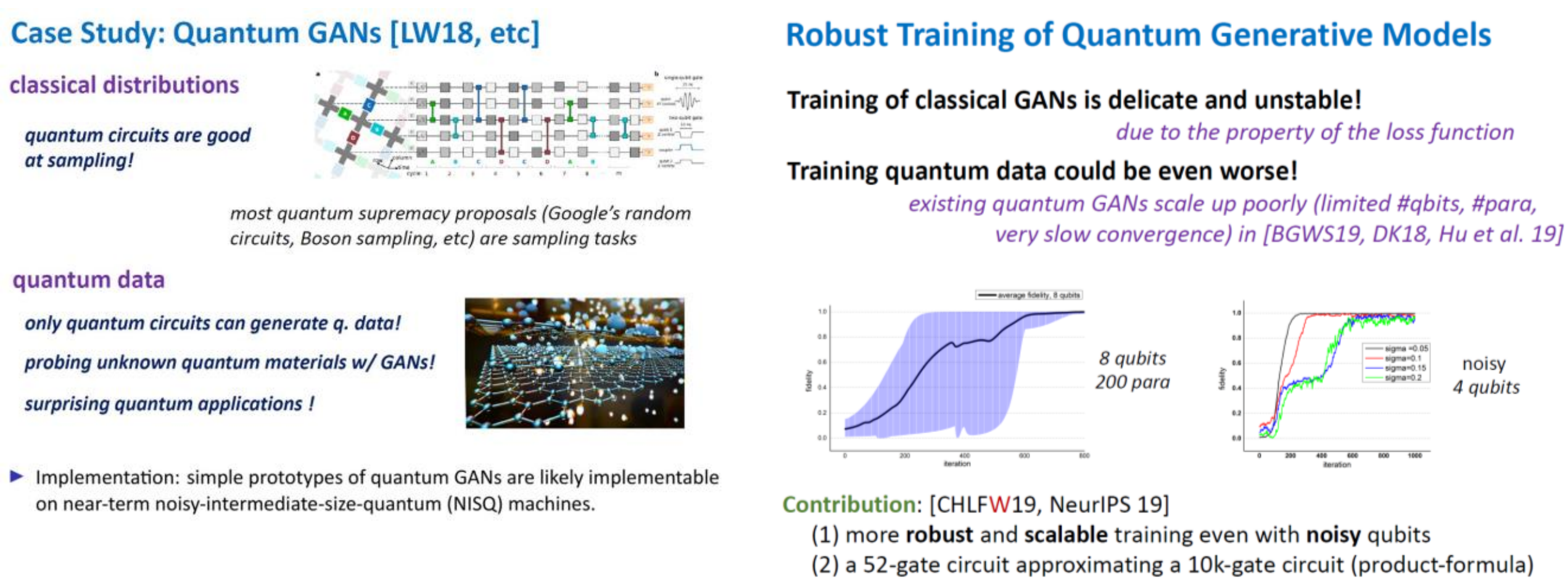

*Figure 578: quantum generative neural networks. Source TBD.*

---

[1608] See Absence of Barren Plateaus in Quantum Convolutional Neural Networks by Arthur Pesah et al, PRX, DoE Los Alamos Lab and UCL, November 2021 (26 pages).

[1609] See Adiabatic Quantum Computation Applied to Deep Learning Networks by Jeremy Liu et al, May 2018 (28 pages).

[1610] See Quantum Graph Neural Networks by Guillaume Verdon et al, 2019 (10 pages).

[1611] See Supervised learning with quantum enhanced feature spaces by Aram Harrow et al, 2018 (22 pages) which describes the use of quantum to detect complex shapes, far beyond what convolutional neural networks ("feature mapping") can do.

[1612] See Recurrent Quantum Neural Networks by Johannes Bausch (12 pages) and Quantum reservoir computing using arrays of Rydberg atoms by Rodrigo Araiza Bravo et al, Harvard and IBM Research, November 2021-July 2022 (10 pages).

[1613] See Introduction to Robust Machine Learning with Geometric Methods for Defense Applications by Pierre-Yves Lagrave and Frédéric Barbaresco, Thales, July 2021 (9 pages)

[1614] See Quantum versus Classical Generative Modelling in Finance by Brian Coyle, Elham Kashefi et al, August 2020 (17 pages) and The Born Supremacy: Quantum Advantage and Training of an Ising Born Machine by Brian Coyle, Daniel Mills, Vincent Danos and Elham Kashefi, April 2021 (47 pages).

[1615] This is well documented in Quantum generative adversarial learning by Seth Lloyd and Christian Weedbrook, 2018 (5 pages), Quantum generative adversarial learning in a superconducting quantum circuit, 2018 (5 pages) and Synthetic weather radar using hybrid quantum-classical machine learning by Graham R. Enos, Chad Rigetti et al, Rigetti, November 2021 (8 pages).

- **Reservoir Networks**[1616].
- **Graph Neural Networks**[1617].
- **Active Learning** algorithms which select the training data set to make learning more efficient and saves up to 85% of training time. It consists in labeling unlabeled data using an iterative supervised learning[1618].
- **Capsule Networks**, which can recognize features in images with taking into account their relative positioning[1619].
- **Various image analysis algorithms**[1620] with different use cases like improve satellite imaging interpretations[1621], improved edge detection[1622].
- Hybrid **transfer learning** which trains a quantum neural network using an already trained classical network[1623].
- **Federated machine learning** algorithms, which distributes training on several quantum computers to improve the training time while preserving privacy, with distributing and sharing the learned model instead of the training data[1624].
- **Spiking neurons** emulation although its speedup is not obvious[1625].
- And other fundamental advances like **group-invariant QML**[1626].

The table in Figure 579 positions the different quantum accelerations associated with various algorithms used in machine learning and deep learning[1627]. Accelerations in log(N) are more important than those expressed as the square root of N[1628].

---

1616 See the review paper Opportunities in Quantum Reservoir Computing and Extreme Learning Machines by Pere Mujal et al, February-July 2021 (14 pages) and Time Series Quantum Reservoir Computing with Weak and Projective Measurements by Pere Mujal et al, May 2022 (19 pages).

1617 See the thesis Quantum neural networks by Kerstin Beer, May 2022 (189 pages)

1618 See Active Learning on a Programmable Photonic Quantum Processor by Chen Ding et al, August 2022 (17 pages).

1619 See Quantum Capsule Networks by Zidu Liu and al, January 2022 (15 pages).

1620 See the review paper Quantum Image Processing by Alok Anand et al, Carnegie Mellon, March 2022 (10 pages) and Processing Images in Entangled Quantum Systems by S.E. Venegas-Andraca and J.L. Ball, 2010 (11 pages).

1621 See Towards Bundle Adjustment for Satellite Imaging via Quantum Machine Learning by Nico Piatkowski, Thore Gerlach, Romain Hugues, Rafet Sifa, Christian Bauckhage and Frederic Barbaresco, Thales and Fraunhofer IAIS, April 2022 (8 pages) which proposes keypoints extraction and objects alignment in pattern recognition applied to satellite imaging.

1622 See A hybrid quantum image edge detector for the NISQ era by Alexander Geng et al, Fraunhofer ITWM, March 2022 (19 pages).

1623 See Classical-to-quantum convolutional neural network transfer learning by Juhyeon Kim et al, August 2022 (15 pages).

1624 See Federated Quantum Machine Learning by Samuel Yen-Chi Chen and Shinjae Yoo, Brookhaven Lab, March 2021 (25 pages) and Federated Quantum Natural Gradient Descent for Quantum Federated Learning by Jun Qi, GeorgiaTech, August 2022 (9 pages).

1625 See An artificial spiking quantum neuron by Lasse Bjørn Kristensen, Alán Aspuru-Guzik et al, April 2011 (7 pages).

1626 See Group-Invariant Quantum Machine Learning by Martin Larocca et al, Google Brain, University of Waterloo, May 2022 (28 pages).

1627 The table is from The prospects of quantum computing in computational molecular biology by Carlos Outeiral, April 2020 (23 pages) which covers both QML algorithms and quantum simulation ones. It also mentions protein structures predictions. See also Quantum Machine Learning by Jacob Biamonte et al, May 2018 (24 pages).

1628 Also see Application of Quantum Annealing to Training of Deep Neural Networks (2015), Machine learning &... artificial intelligence in the quantum domain, 2017 (106 pages), On the Challenges of Physical Implementations of RBMs, 2014, with Yoshua Bengio and Ian Goodfellow among the authors, illustrating the interest of AI specialists for quantum and Quantum Deep Learning, 2014, all extracted from Near-Term Applications of Quantum Annealing, 2016, Lockheed Martin (34 slides). See also Quantum machine learning for data scientists, 2018 (46 pages).

**TABLE 1** Overview of the main quantum machine learning algorithms that have been reported in the literature, and complexities

| Algorithm | Classical | Quantum | QRAM |
|---|---|---|---|
| Linear regression | $\mathcal{O}(N)$ | $\mathcal{O}(\log N)^{*}$ | Yes |
| Gaussian process regression | $\mathcal{O}(N^3)$ | $\mathcal{O}(\log N)^{\dagger}$ | Yes |
| Decision trees | $\mathcal{O}(N \log N)$ | Unclear | No |
| Ensemble methods | $\mathcal{O}(N)$ | $\mathcal{O}(\sqrt{N})$ | No |
| Support vector machines | $\approx\mathcal{O}(N^2)$-$\mathcal{O}(N^3)$ | $\mathcal{O}(\log N)$ | Yes |
| Hidden Markov models | $\mathcal{O}(N)$ | Unclear | No |
| Bayesian networks | $\mathcal{O}(N)$ | $\mathcal{O}(\sqrt{N})$ | No |
| Graphical models | $\mathcal{O}(N)$ | Unclear | No |
| $k$-Means clustering | $\mathcal{O}(kN)$ | $\mathcal{O}(\log kN)$ | Yes |
| Principal component analysys | $\mathcal{O}(N)$ | $\mathcal{O}(\log N)$ | No |
| Persistent homology | $\mathcal{O}(\exp N)$ | $\mathcal{O}(N^5)$ | No |
| Gaussian mixture models | $\mathcal{O}(\log N)$ | $\mathcal{O}(\text{polylog } N)$ | Yes |
| Variational autoencoder | $\mathcal{O}(\exp N)$ | Unclear | No |
| Multilayer perceptrons | $\mathcal{O}(N)$ | Unclear | No |
| Convolutional neural networks | $\mathcal{O}(N)$ | $\mathcal{O}(\log N)$ | No |
| Bayesian deep learning | $\mathcal{O}(N)$ | $\mathcal{O}(\sqrt{N})$ | No |
| Generative adversarial networks | $\mathcal{O}(N)$ | $\mathcal{O}(\text{polylog } N)$ | No |
| Boltzmann machines | $\mathcal{O}(N)$ | $\mathcal{O}(\sqrt{N})$ | No |
| Reinforcement learning | $\mathcal{O}(N)$ | $\mathcal{O}(\sqrt{N})$ | No |

*Figure 579: main quantum machine learning algorithms. Source: The prospects of quantum computing in computational molecular biology by Carlos Outeiral, April 2020 (23 pages).*

Note the need for quantum memory for many of these algorithms, a type of memory that doesn't yet exists. None of these algorithms have been tested on a large scale, due to the absence of a quantum processor with more than fifty qubits.

QML is also one of the fields of application of D-Wave's quantum annealers. Annealers work well to find minimum energy of complex systems which is equivalent to searching for a minimum level of errors in the adjustment of the weight of neurons in a neural network[1629]. So far, they have tested an RBM (Restricted Boltzmann Machine) model[1630].

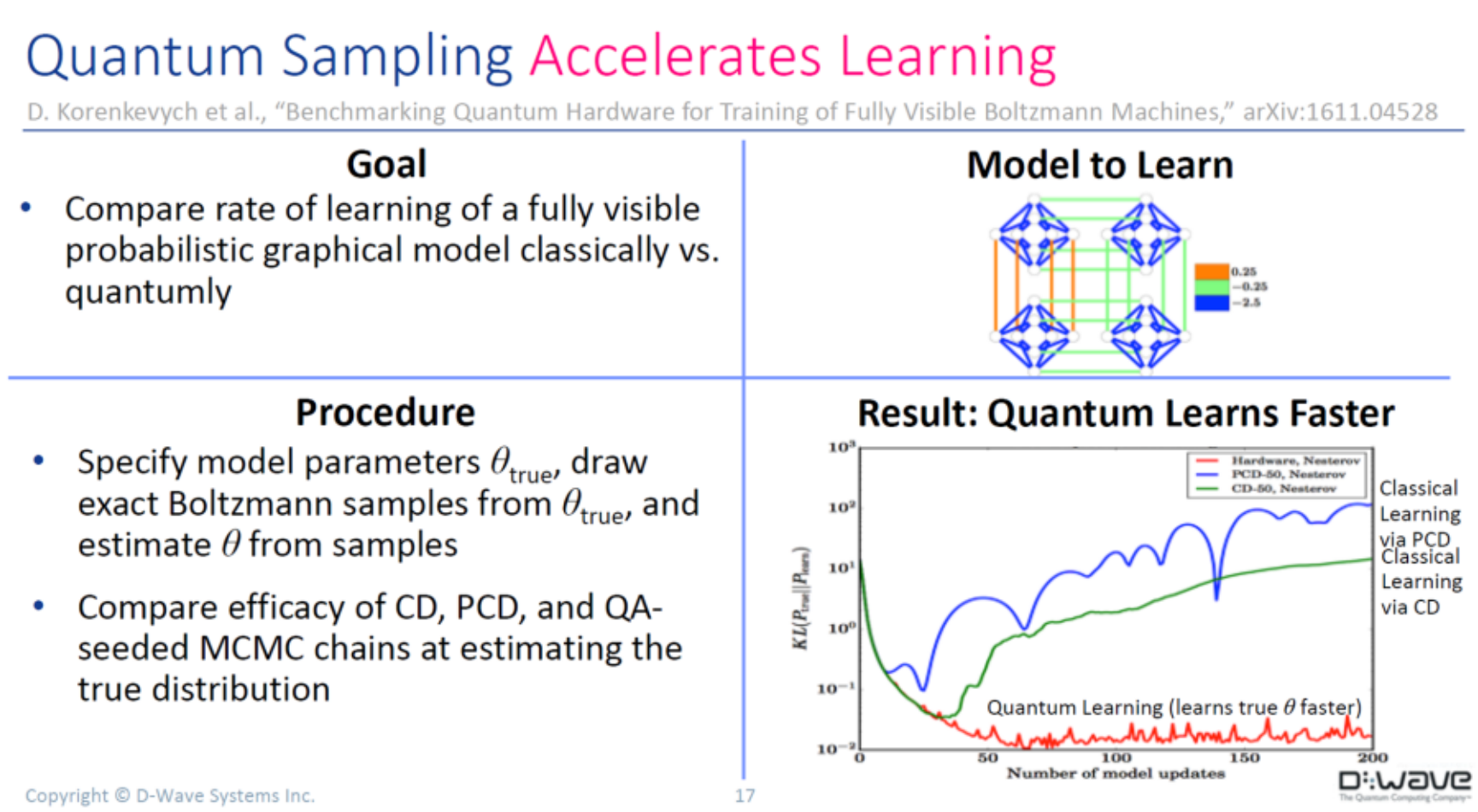

*Figure 580: Source: D-Wave Quantum Computing - Access & application via cloud deployment by Colin Williams, 2017 (43 slides).*

[1629] Examples source: D-Wave Quantum Computing - Access & application via cloud deployment by Colin Williams, 2017 (43 slides).

[1630] See Benchmarking Quantum Hardware for Training of Fully Visible Boltzmann Machines by Dmytro Korenkevych et al, Kindred AI et D-Wave, 2016 (22 pages).

They also did it with a hybrid algorithm for image recognition in a neural network, based on a variational circuit and a hybrid algorithm. But with very low-resolution images! D-Wave offers machine learning services in its Leap quantum cloud computing offering. But they are not the only ones. Many startups are specialized in Quantum Machine Learning, such as **QC Ware**.

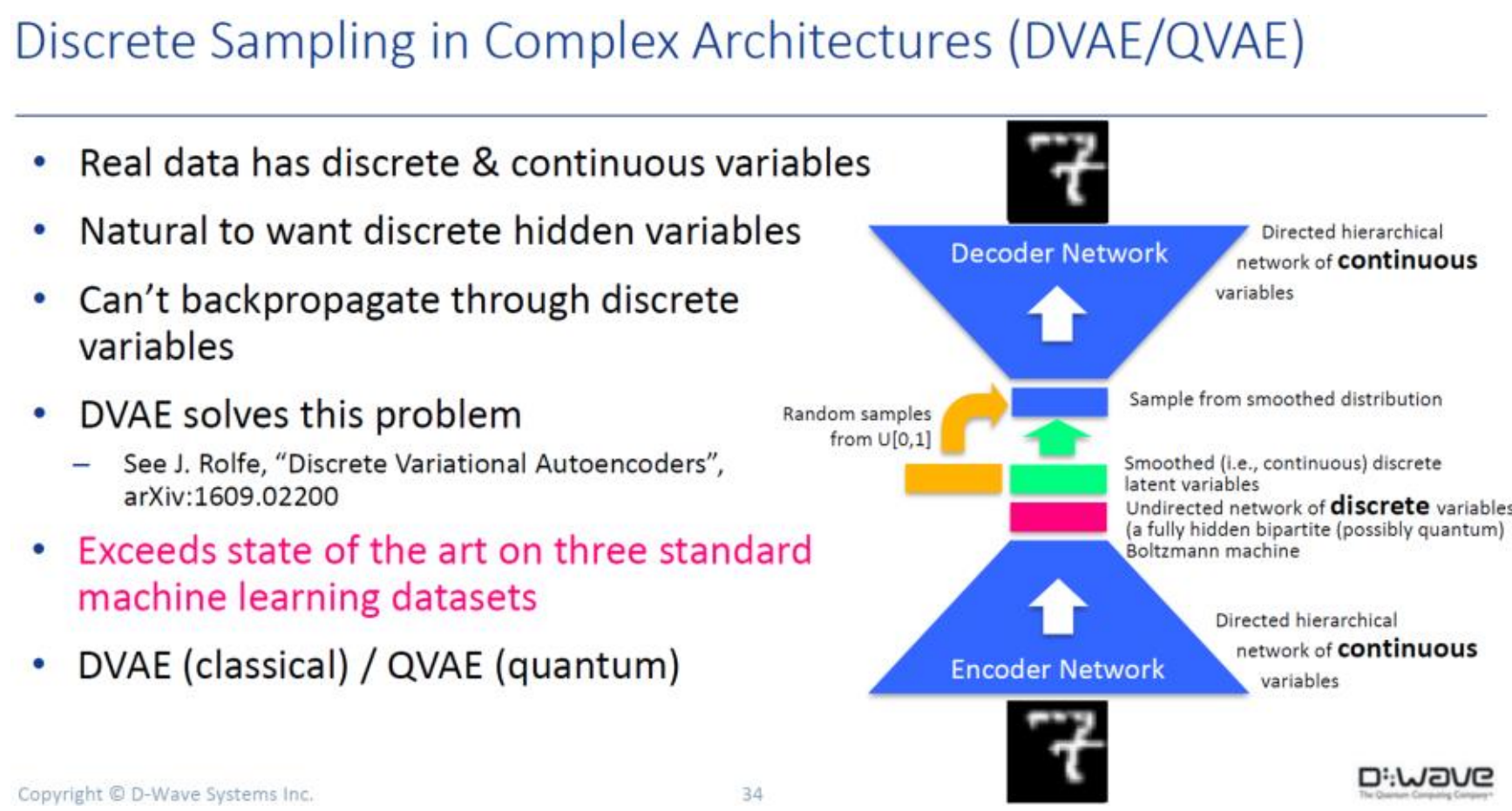

*Figure 581: Source: [D-Wave Quantum Computing - Access & application via cloud deployment](#) by Colin Williams, 2017 (43 slides).*

Many challenges remain to be addressed to operationalize QML beyond the emergence of sufficiently powerful and reliable quantum processors[1631]:

- **Loading training data** may take time and have a negative impact on the acceleration provided by QML. It also requires quantum random access memory (qRAM) which does not yet exist, even if some Quantum Data Loaders are proposed to circumvent this need[1632]. Some specific methods have to be created to load images efficiently[1633]. Other methods are developed that could require fewer training data than with classical machine learning[1634].
- **Avoiding the barren plateau** phenomenon which limits the efficiency of QML algorithms as the problem scales[1635].
- **Reading the results** of QML algorithms, particularly when they are classical data that take the form of real numbers.
- **Compressing** the neural networks to save quantum resources[1636].
- **Nonlinear activation functions** such as sigmoids used in classical neural networks are difficult to implement in quantum algorithms since quantum gates all apply linear transformations[1637].

[1631] See Quantum machine learning: a classical perspective by Ciliberto et al, 2020 (26 pages). It concludes with: "*Despite a number of promising results, the theoretical evidence presented in the current literature does not yet allow us to conclude that quantum techniques can obtain an exponential advantage in a realistic learning setting*".

[1632] See Quantum embeddings for machine learning by Seth Lloyd, January 2020 (11 pages). In Nearest Centroid Classification on a Trapped Ion Quantum Computer by Sonika Johri, Iordanis Kerenidis et al, December 2020 (15 pages), the QML algorithm "Nearest Centroid Classification" is implemented on a 11 trapped ions IonQ processor with an efficient amplitude data loader. See also Data compression for quantum machine learning by Rohit Dilip et al, April 2022 (8 pages) and the thesis Quantum Algorithms for Unsupervised Machine Learning and Neural Networks, by Jonas Landman, November 2021 (192 pages).

[1633] See A Novel Quantum Image Compression Method Based on JPEG by Jian Wang et al, 2017 (30 pages).

[1634] See Generalization in quantum machine learning from few training data by Matthias C. Caro et al, Nature Communications, 2022 (11 pages).

[1635] See Quark: A Gradient-Free Quantum Learning Framework for Classification Tasks by Zhihao Zhang et al, October 2022 (19 pages) which circumvents this problem with a quantum model and quantum optimizer.

[1636] See Quantum Neural Network Compression by Zhirui Hu et al, July 2022 (11 pages) and Experimental Quantum End-to-End Learning on a Superconducting Processor by Xiaoxuan Pan et al, March 2022 (10 pages).

[1637] The trick is explained in Quantum Neuron: an elementary building block for machine learning on quantum computers by Yudong Cao, Gian Giacomo Guerreschi and Alán Aspuru-Guzik in 2017 (30 pages).

There are workarounds, on which Iordanis Kerenidis has worked[1638] and some others[1639]. For example, quantum measurement can create the sough-after activation function nonlinearity in neural networks. There are also suggestions to use continuous variables qubits architectures to handle neural networks with nonlinearity provided by non-Gaussian qubit gates[1640], photonic quantum neural networks using Kerr nonlinearities[1641] and even other techniques not requiring measurement[1642].

- QML can take advantage of the **errors and noise** generated by quantum computation rather than be subjected to them. Work is going on in this direction.
- QML must prove that it brings a real **gain in computing time** compared to today's most advanced processors[1643].
- QML must also address the quest for **algorithms explicability**. The decomposition of the training and inference process of these quantum neural networks will probably be different from their implementation in more traditional processors[1644].

IBM published in 2021 a mathematical proof of a potential quantum advantage for a **quantum machine learning classification** task done with a quantum kernel method based on the Shor dlog algorithm. There was no actual experiment done due to the inexistence of sufficiently powerful quantum computers[1645]. We'll probably be stuck in this situation for several years from now.

On the other hand, QML's algorithm developments have served as a source of inspiration to improve algorithms that work with classical computation. As we will see in the section on quantum software vendors, page 726, there is no shortage of those who have specialized in QML. In general, they provide development tools and means to create QML proofs of concept.

In the AI domain, the European project H2020 **Quromorphic** launched in July 2019 aims to create a quantum processor dedicated to the execution of neural networks inspired by the brain[1646].

---

[1638] See Quantum Algorithms for Deep Convolutional Neural Network by Iordanis Kerenidis et al, 2020 (36 pages) which is discussed in Deep Convolutional Neural Networks for Quantum Computers by Jonas Landman, 2020.

[1639] See for example Continuous-variable quantum neural networks by Nathan Killoran et Al, June 2018 (21 pages).

[1640] See Continuous-variable quantum neural networks by Nathan Killoran, Maria Schuld, Seth Lloyd et al, 2018 (21 pages).

[1641] See Realistic quantum photonic neural networks by Jacob Ewaniuk et al, August 2022 (20 pages).

[1642] See Quantum activation functions for quantum neural networks by Marco Maronese et al, January 2022 (28 pages).

[1643] See Quantum Machine Learning: Algorithms and Practical Applications by Iordanis Kerenidis, QC Ware, Q2B Conference, December 2019 (34 slides) which makes an inventory of some potential gains with QML algorithms.

[1644] These techniques will be challenged by future memristor-based neuromorphic processors that will allow networks to converge more rapidly with backpropagation. Memristors will make it possible to place the neuron's computational functions and the associated memory in the same location in a semiconductor circuit, accelerating access to memory by several orders of magnitude during computations. This is another area of research, operated notably by Julie Grollier of the CNRS laboratory located at Thales TRT in Palaiseau.

[1645] See IBM shows quantum computers can solve these problems that classical computers find hard by Daphne Leprince-Ringuet, ZDNet, July 2021 that refers to Quantum kernels can solve machine learning problems that are hard for all classical methods, IBM Research, July 2021, itself referring A rigorous and robust quantum speed-up in supervised machine learning by Yunchao Liu, Srinivasan Arunachalam and Kristan Temme, Nature Physics, July 2021 (27 pages).

[1646] See Quantum computer: We're planning to create one that acts like a brain by Michael Hartmann and Heriot-Watt leads on next-gen computers, November 2018. The project is led by Michael Hartmann of IPaQS (Institute of Photonics and Quantum Sciences) at Heriot-Watt University in the UK, together with ETH Zurich, Delft University (Netherlands), Basque University (Spain), IBM Zurich and Volkswagen (Germany). 2.2M€ from the FET Open program were allocated to the project by the European Commission (details). My interpretation? The objective of the project has been adapted to the sauce of science fiction in order to recover community funding. The rest is about photonics.

It reminds us of the very controversial European flagship Human Brain Project led by Henri Markram from Switzerland. Quromorphic involves IBM Zurich, ETH Zurich, TU Delft, Volkswagen and Spanish and German Universities. Given the participants, we can guess that this will be based on superconducting qubits. The project got a funding of 2.9M€ in 2019 and is scheduled to end in 2022[1647]. This is quite reasonable.

We'll probably discover new fancy claims combining artificial intelligence and quantum computing and the devil will always be in complicated details[1648]. For instance, how about using these QML algorithms in robotics? Not so fast[1649]! It's still science fiction and *click-bait*.

On the other end, classical learning machine can be useful for quantum physics and quantum computing. We saw that Google used a deep learning algorithm to optimize the microwave frequency plan of the Sycamore processor's qubit control. Machine learning can also be used to model and simulate condensed matter, with an impact on the development of various qubits, especially superconducting qubits[1650].

Let us note finally the existence of an association promoting the field of AI and quantum computing, the **IAIQT** foundation based in Switzerland.

### Quantum physics simulation

Quantum simulation algorithms are used to reproduce matter at the quantum level in a computer. It can be used to simulate the interaction between atoms in molecules for the creation of new materials.

They can also simulate physical phenomena related to magnetism or the interaction between photons and matter. This amounts to solving "N-body problems", i.e. calculating the interaction between several particles according to the physical laws governing their interaction. Quantum simulation also helps studying how superconducting materials behave, particularly at (relatively) high temperature, superfluids at low temperature, the temperature-dependent magnetism of certain materials and the interactions between graphene and light[1651].

These algorithms run in qubit-based universal quantum computers as well as on quantum simulators and quantum annealers although we still lack data to compare their respective performance.

Starting with 50 electrons in a molecule, classical computers can no longer simulate their dynamics, which corresponds to just a few atoms. For simple molecules, the applications are in the field of materials physics: carbon or nitrogen capture, new batteries, discovery of superconducting mechanisms that can then be used in medical scanners, ideally operating at room temperature.

---

[1647] See Quantum computer: we're planning to create one that acts like a brain, January 2019.

[1648] Here is one good example with Using Pioneering Quantum Machine Learning Methods, CQC Scientists Offer Bright Forecast For Quantum Computers That Can Reason par Matt Swayne, 2021 referring to Variational inference with a quantum computer by Marcello Benedetti, April 2021 (17 pages) which is about apply some quantum version of MCMC (Markov-Chain Monte Carlo) algorithm using Born machines. These are described in The Born Supremacy: Quantum Advantage and Training of an Ising Born Machine by Brian Coyle, Elham Kashefi et al, April 2021 (10 pages).

[1649] As described in Daniel Manzano's The Rise of Quantum Robots, April 2018. And with Qubit or Qubot? Quantum Technology May Help Robots Learn Faster par Matt Swayne, 2021, Robots learn faster with quantum technology by University of Vienna, March 2021 pointing to Experimental quantum speed-up in reinforcement learning agents by V. Saggio et al, Nature, March 2021 (10 pages).

[1650] See some review papers like Machine learning & artificial intelligence in the quantum domain by Vedran Dunjko and Hans J. Briegel, 2017 (106 pages), Artificial Intelligence and Machine Learning for Quantum Technologies by Mario Krenn, Jonas Landgraf, Thomas Foesel and Florian Marquardt, August 2022 (23 pages) and Modern applications of machine learning in quantum sciences by Anna Dawid et al, April 2022 (283 pages).

[1651] See this interesting lecture by Jacqueline Bloch at the Academy of Sciences which makes an excellent overview: Quantum Simulators: Solving Difficult Problems, May 2018 (29 mn).

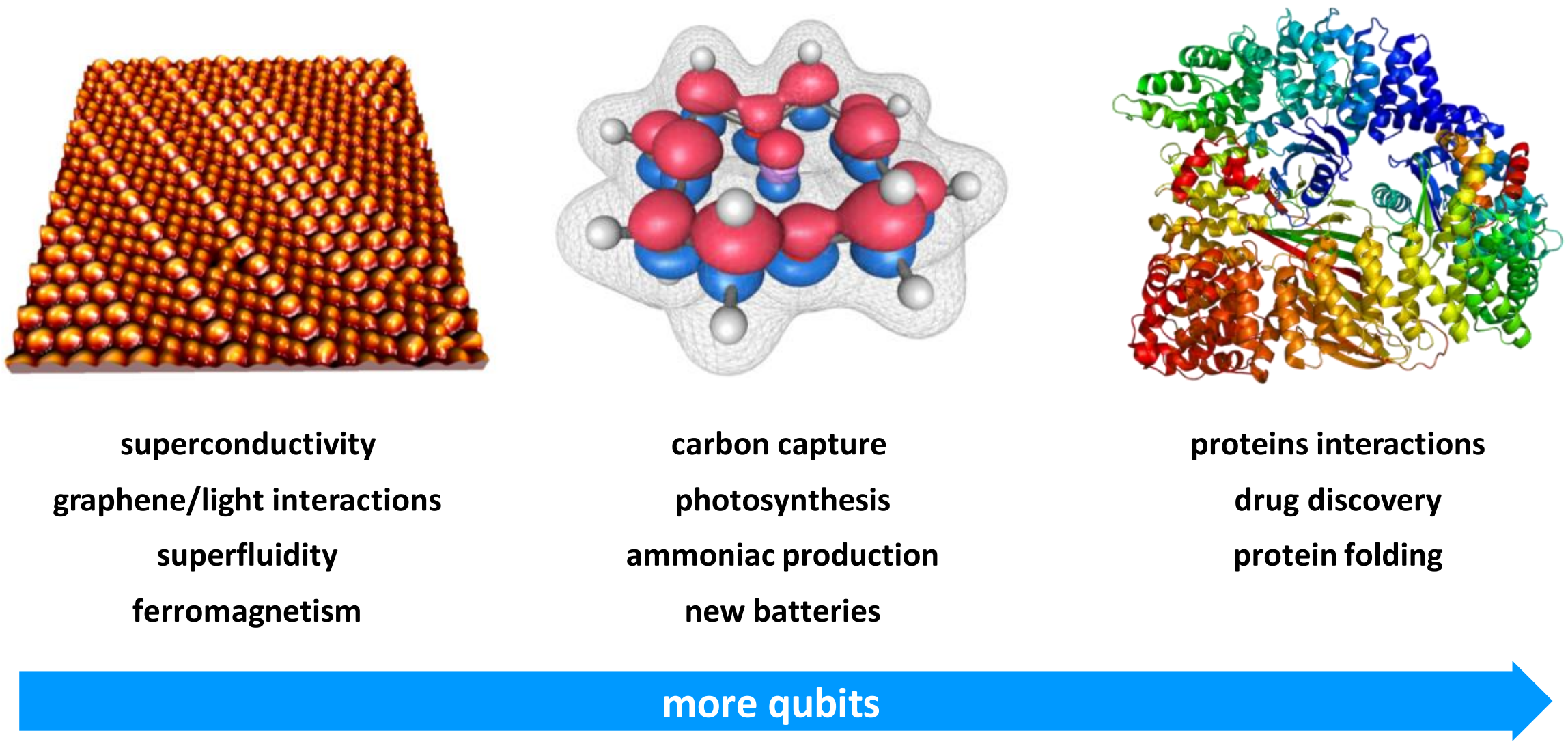

*Figure 582: quantum physics simulation applications and a grade of complexity. (cc) Olivier Ezratty, 2020.*

This should be accessible with universal quantum computers with 50 to a few hundred corrected logical qubits. For molecular biology simulations, it will probably take much longer before this is possible. We may need thousands or even hundreds of thousands of corrected qubits, which is far away in time. The diagram in Figure 583 positions the number of qubits needed to simulate the functioning of a mitochondrial protein, MRC2, in a fairly optimistic way.

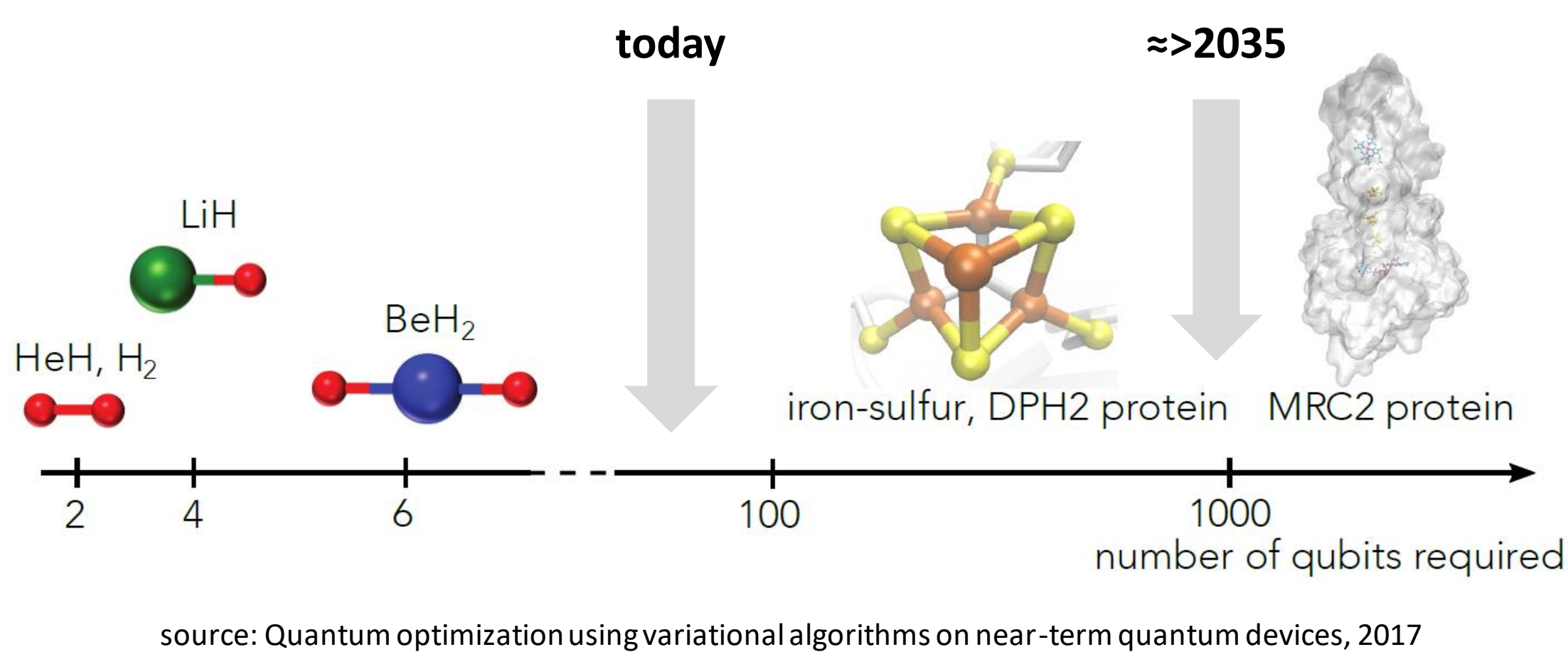

*Figure 583: another grade of complexity, for molecular simulations, in logical qubits. Source: from Quantum optimization using variational algorithms on near-term quantum devices by IBM researchers in 2017 (30 pages).*

Here are some examples of quantum simulation algorithms:

- Simulation of **quantum field theory** as reviewed by John Preskill[1652].
- Simulation of **hydrogen atoms** in superconducting qubits computers[1653].
- Simulation of **simple molecules** by Alán Aspuru-Guzik, one of the world's leading authorities in the field[1654]. It's centered around the $H_2$ molecule.

---

1652 Simulating a quantum field theory with a quantum computer by John Preskill, 2018 (22 pages).

1653 Computation of Molecular Spectra on a Quantum Processor with an Error-Resilient Algorithm, 2018 (7 pages).

1654 See Simulation of Electronic Structure Hamiltonians Using Quantum Computers by James Whitfield, Jacob Biamonte and Alán Aspuru-Guzik, 2010 (22 pages)

- Hybrid simulation of a **simple $CaH^+$ ion** also using two superconducting qubits[1655].
- Simulation of a **three atoms molecule**, beryllium hydride ($BeH_2$) with 6 qubits by IBM[1656].
- Determination of the **equilibrium state of simple molecules**[1657].
- Simulation of **water electrolysis** caused by light with use cases for the production of storable energy, particularly in fuel cells (hydrogen-based)[1658].
- Simulation of **semiconductor dynamics** using quantum annealing[1659].
- Simulation of the **cytochrome molecule**. With error rates of 0.1%, the run time would be 73 hours and require 4.6M physical qubits. With an error rate of 0.001% which is currently way out of reach, computing time could be down to 25 hours with 500K physical qubits. The equivalent classical computing is 4 years and requires 348 GB RAM and 2TB of storage[1660].
- Hybrid **molecular simulations** combining classical and quantum algorithms[1661].
- Simulation of **catalytic chemical processes** proposed by researchers from Microsoft and ETH Zurich with a requirement of about 4000 logical qubits[1662].
- Simulating organic molecules of medium complexity such as **cholesterol** would require about 1500 logical qubits and, above all, the ability to use billions of quantum gates[1663]. VQE algorithms can also be used there with a universal gate-based quantum computer using a reasonable depth of quantum gates (number of steps in the algorithm) [1664].
- Simulating **periodic solids**, which would require tens to hundred million of qubits[1665]!

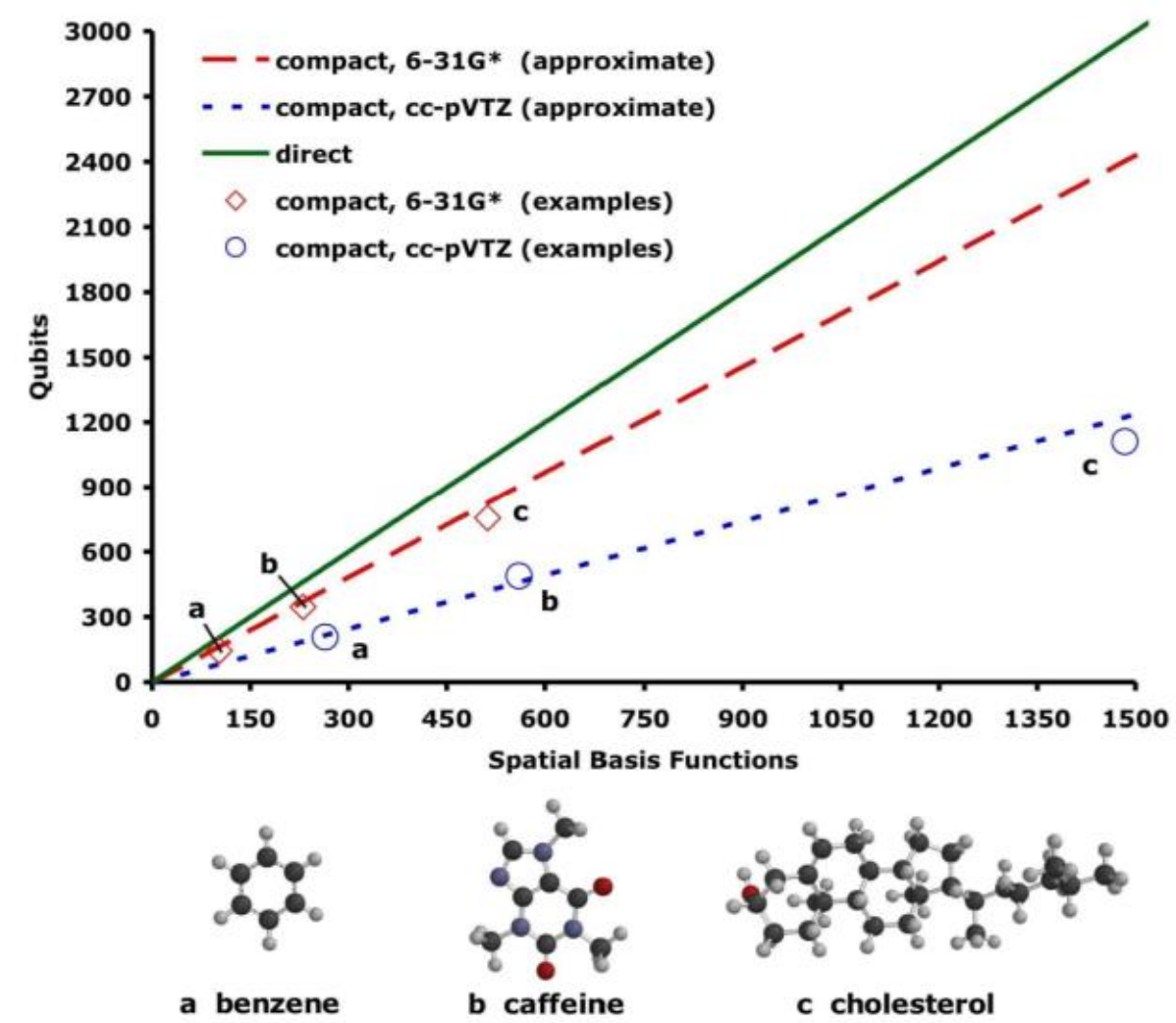

*Figure 584: how many logical qubits are actually necessary to simulate relatively simple molecules. Source: Simulated Quantum Computation of Molecular Energies by Wiebe, Wecker and Troyer, 2006 (21 pages).*

---

1655 See Researchers succeed in the quantum control of a molecule by Román Ikonicoff, May 2017 (38 pages), pointing to Preparation and coherent manipulation of pure quantum states of a single molecular ion, 2017 (38 pages).

1656 See Tiny Quantum Computer Simulates Complex Molecules by Katherine Bourzac, IEEE Spectrum, 201

1657 See Simulated Quantum Computation of Molecular Energies by Wiebe, Wecker and Troyer, 2006 (21 pages).

1658 This is one of the many examples from the presentation Enabling Scientific Discovery in Chemical Sciences on Quantum Computers, December 2017 (34 slides) by Ber De Jong from Berkeley

1659 See Solving strongly correlated electron models on a quantum computer by Wecker, Troyer, Hastings, Nayak and Clark, 2015 (27 pages)

1660 See Reliably assessing the electronic structure of cytochrome P450 on today's classical computers and tomorrow's quantum computers by Joshua J. Goings, Craig Gidney et al, Google, February 2022 (24 pages)

1661 See Quantum Machine Learning for Electronic Structure Calculations, October 2018 (16 pages).

1662 See Quantum computing enhanced computational catalysis by Vera von Burg, Matthias Troyer et al, July 2020 (104 pages).

1663 See Quantum Computation for Chemistry and Materials by Jarrod McClean, Google 2018 (36 slides).

1664 See An adaptive variational algorithm for exact molecular simulations on a quantum computer by Sophia Economou et al, 2019 (9 pages) which indicates in particular that "*VQE is much more suitable for NISQ devices, trading in the long circuit depths for shorter state preparation circuits, at the expense of a much higher number of measurements*".

1665 See Quantum Computation for Periodic Solids in Second Quantization by Aleksei V. Ivanov et al, Oct 2022 (29 pages).

One of the applications of molecular quantum simulation is to better understand how photosynthesis works in order to improve or imitate it, the involvement of different forms of ferredoxin, relatively simple iron and sulfur-based molecules that serve to transport electrons from the photoelectric effect used in photosynthesis in plants. Algorithmic research on this molecule simulation have downsized the duration of quantum theoretical simulation from 24 billion years to one hour in a few years! The simulation of photosynthesis can pave the way for better carbon capture, among others to produce synthetic fuel. Research is also advancing in this field, without quantum computation for the moment[1666].

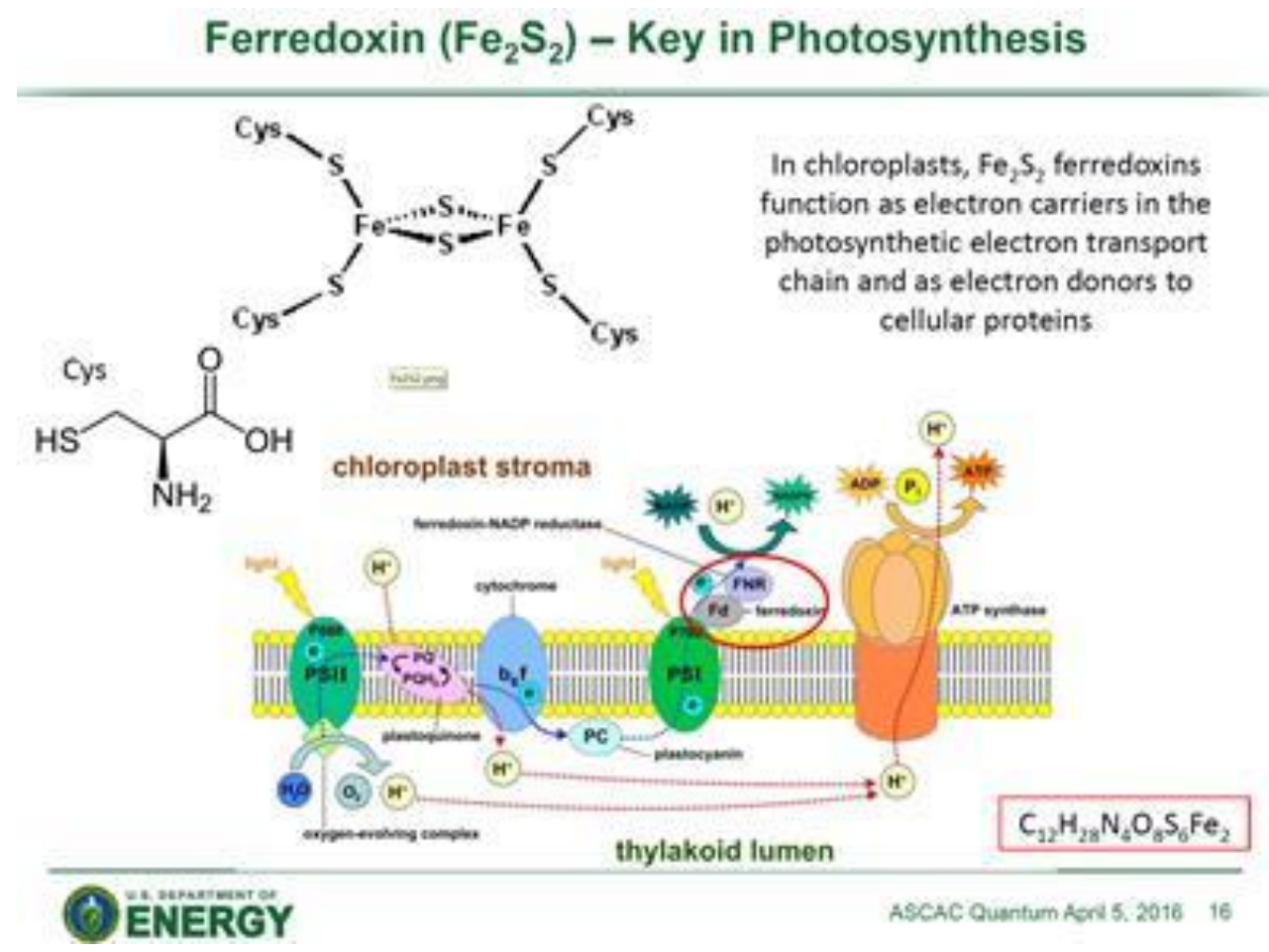

*Figure 585: Source: Quantum Computing (and Quantum Information Science) by Steve Binkley, US Department of Energy, 2016 (23 slides).*

Matthias Troyer explains how this algorithm has been optimized[1667].

At a higher abstraction level sits the simulation of atomic interactions in organic chemistry and molecular biology, going progressively from the smallest to the largest molecules: amino acids, peptides, polypeptides, proteins and perhaps much later, ultra-complex molecules such as ribosomes that fabricates proteins with amino acids using messenger RNA code.

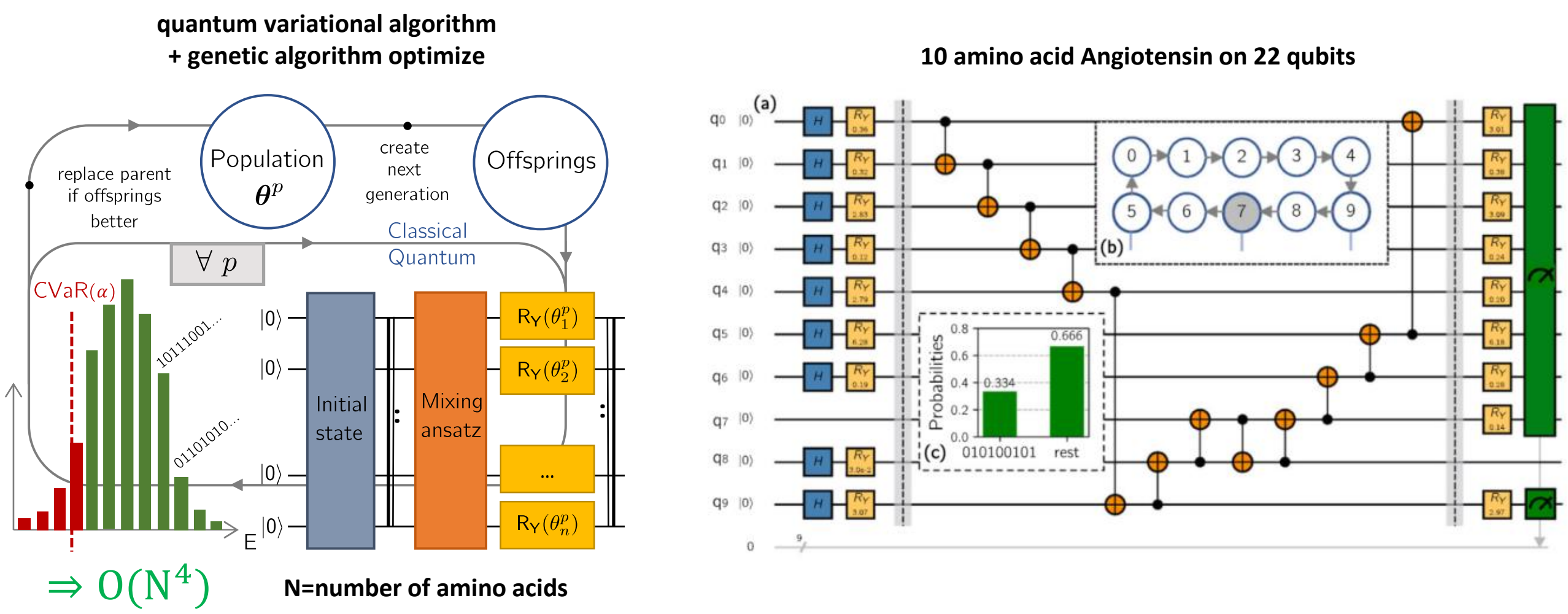

*Figure 586: a hybrid classical and quantum algorithm to fold proteins. Source: Resource-efficient quantum algorithm for protein folding, Anton Robert et al, 2020 (5 pages).*

Like classical algorithms, quantum simulation algorithms use approximation models based on known molecules, like what AlphaFold 3 from DeepMind does to predict the 3D structure of folded proteins. It works well for proteins which are close to those proteins used to train the model. For entirely new proteins (*aka* "de-novo proteins"), quantum simulation seems to be required[1668].

[1666] As seen in Semi-Artificial Photosynthesis Method Produces Fuel More Efficiently Than Nature, September 2018.

[1667] In What Can We Do with a Quantum Computer, Matthias Troyer, ETH Zurich, 2016 (41 slides), source for the illustration on the right.

[1668] See Evolution, energy landscapes and the paradoxes of protein folding by Peter Wolynes, 2015 (13 pages).

Various quantum algorithms have already been created for this purpose, including the Aspuru-Guzik algorithm in 2012, which was tested at a small scale on a D-Wave One[1669]. In 2020, researchers from IBM Zurich and from Institut Pasteur in France created an algorithm able to predict the 3D structure of a peptide, angiotensin, made of 10 amino acids and on just 22 qubits as shown above in Figure 586. Algorithms even exists to simulate RNA folding[1670].

The orders of magnitude of the quantum computers needed to solve these organic chemistry problems for large proteins have yet to be evaluated. It is not impossible that they are either impossible or extremely long-term even with the various optimizations that are proposed[1671]!

# Hybrid algorithms

Hybrid algorithms are another branch of quantum algorithms that has been steadily growing in recent years. These algorithms combine a classical and a quantum part. Any quantum algorithm requires the support of a classical computer for the control of the quantum computer and the activation of its quantum gates [1672]. Hybrid algorithms distribute actual computing on both sides and ensure that the quantum part of the algorithm only covers what cannot be executed efficiently as classical computing. Eventually, it is likely that a majority of quantum algorithms will be hybrid [1673].

These hybrid algorithms can be implemented in development tools and languages capable of controlling both the classical and the quantum part of a supercomputer or a distributed system.

This is particularly the case with the **XACC** (eXtreme-scale ACCelerator) programming model[1674]. It enables the development of hybrid code that takes into account the characteristics of the quantum computer, and in particular its error rate. It interfaces with IBM and Rigetti quantum accelerators programming models.

### Variational Quantum Eigensolver

One of the most pre-eminent hybrid algorithm class is the VQE (**Variational Quantum Eigensolver**), invented in 2013 by Alán Aspuru-Guzik[1675]. It allows the discovery of an energetic minimum of a complex equation[1676].

---

[1669] See D-Wave quantum computer solves protein folding problem by Geoffrey Brumfiel, 2012.

[1670] See A QUBO model of the RNA folding problem optimized by variational hybrid quantum annealing by Tristan Zaborniak et al, University of Victoria, August 2022 (12 pages).

[1671] See Quantum Information and Computation for Chemistry, 2016 (60 pages), which provides a good inventory of the various algorithmic works on quantum simulation of organic chemistry, A Comparison of the Bravyi–Kitaev and Jordan–Wigner Transformations for the Quantum Simulation of Quantum Chemistry by Andrew Tranter et al, 2018 (14 pages) that provides some solutions to reduce the gates count for quantum chemistry simulation with gate-based quantum computers and Creating and Manipulating a Laughlin-Type ν=1/3 Fractional Quantum Hall State on a Quantum Computer with Linear Depth Circuits by Armin Rahmani et al, November 2020 (7 pages).

[1672] See A Hybrid Quantum-Classical Approach to Solving Scheduling Problems, Tony T. Tran et al, (9 pages), Hybrid Quantum Computing Apocalypse 2018 (6 pages) according to which some Chinese team supposedly succeeded in running a Majorana fermion qubit, The theory of variational hybrid quantum-classical algorithms by Jarrod McClean et al (23 pages).

[1673] As such, the quantum algorithm patents filed by Accenture are worrying because they are at the limit of *troll patents*. See for example the Multi-state quantum optimization engine patent, USPTO 10,095,981B1, validated in October 2018 (20 pages). A second patent validated in April 2019 deals with a machine learning solution that helps an algorithm to decide which part to execute as classical and which part to execute as quantum. It is USPTO 10,275,721.

[1674] See Hybrid Programming for Near-term Quantum Computing Systems by A. J. McCaskey et al, Oak Ridge Laboratory, 2018 (9 pages).

[1675] VQE now belong to a broader category of hybrid algorithms, Variable Quantum Algorithms (VQA). See Variational Quantum Algorithms by M. Cerezo et al, Nature Reviews Physics, August 2021 (29 pages).

[1676] See an history timeline on Towards an experimentally viable variational quantum eigensolver with superconducting qubits, 2016 (18 slides). See also Variational Quantum Eigensolver explained, November 2019,

It is typically used to simulate the structures of molecules in inorganic and organic chemistry. It combines a classical part that determines an approximate starting point and a quantum part that refines the result. More precisely, the classical part prepares a so-called ansatz which is a set of parameters defining a quantum state, with using some nonlinear optimization techniques.

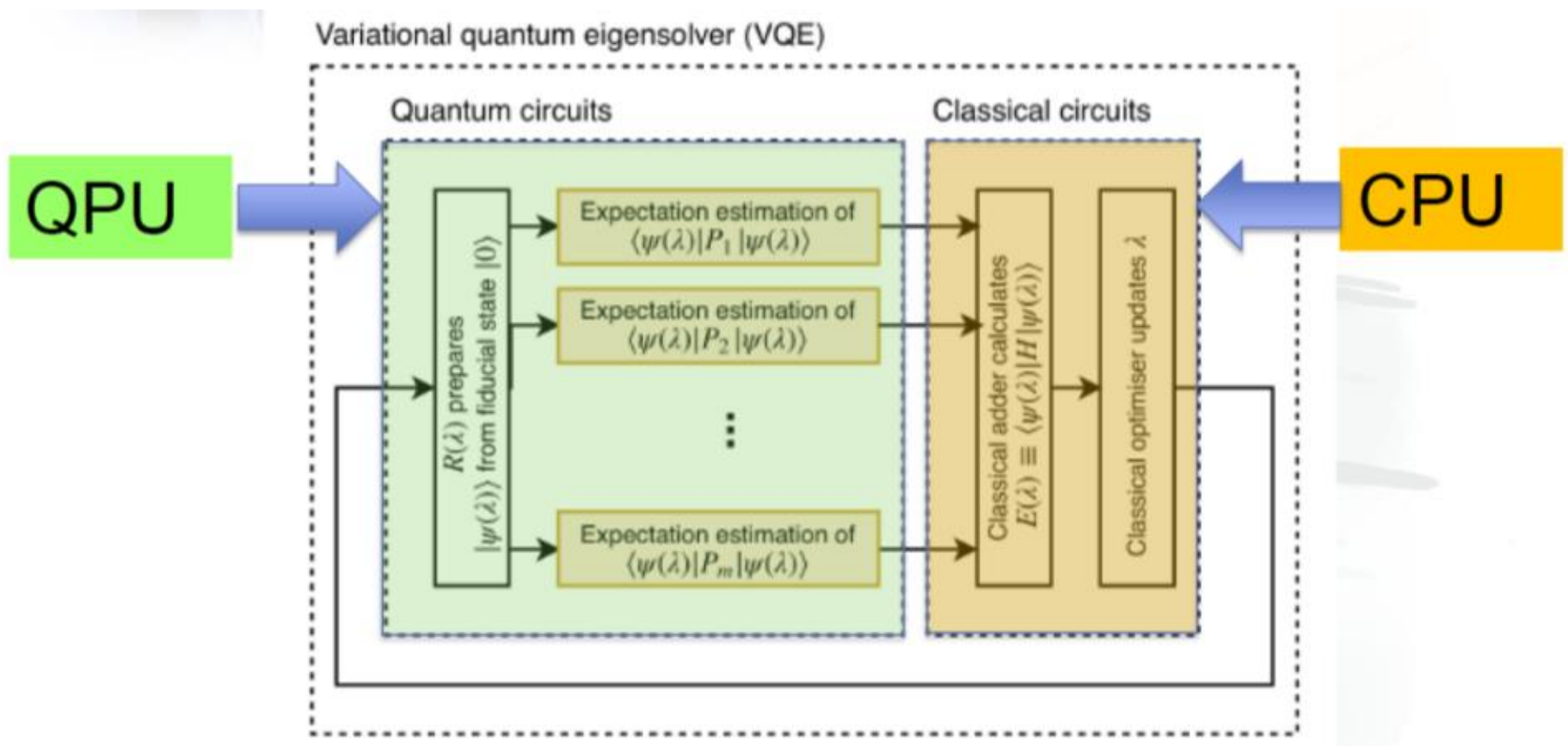

*Figure 587: Source: [Accelerated Variational Quantum Eigensolver]() by Daochen Wang, Oscar Higgott and Stephen Brierley, 2019 (11 pages).*

The quantum side of the algorithm is used to compute a cost function outputting a real number that we seek to minimize with varying the parameters of the ansatz in the classical part.

This type of algorithm has the advantage of being able to be processed in distributed architectures with several classical and quantum processors. The gain in VQE comes from the ability of quantum computing to explore the space of possibilities in parallel. The approach is iterative and the speed of convergence depends on factors related to the simulated physical system, the digital modeling and the desired quality of the result.

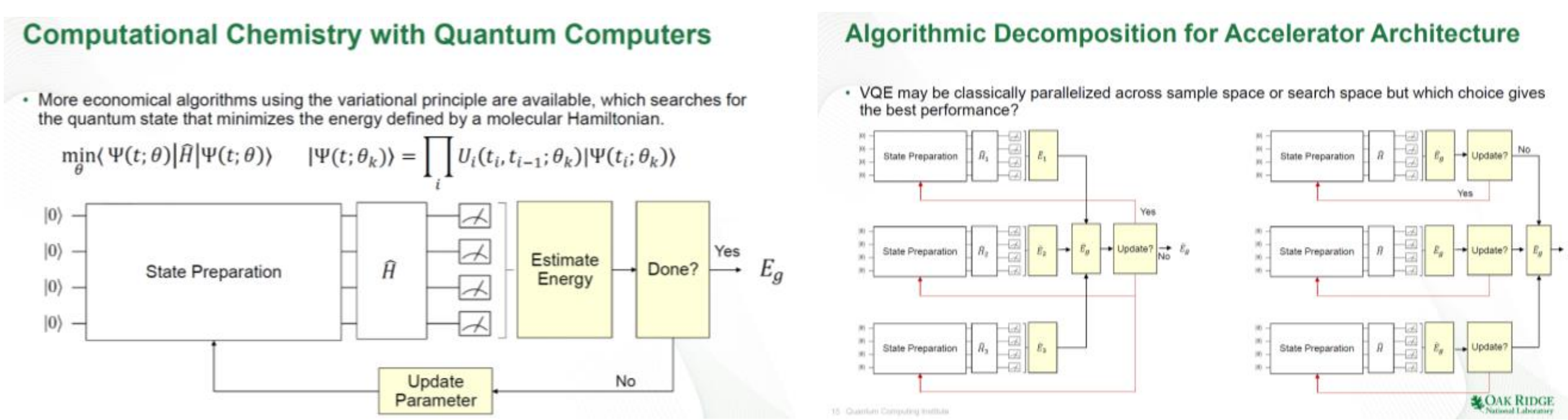

*Figure 588: Source: [Quantum Computing for Scientific Discovery: Methods, Interfaces, and Results]() by Travis Humble du Quantum Computing Institute, Oak Ridge National Laboratory, March 2018 (47 slides).*

VQE can also be used to train a machine learning model[1677].

### Quantum Approximate Optimization Algorithm

Another famous hybrid algorithm is the Quantum Approximate Optimization Algorithm **(QAOA)**, created by Edward Farhi in 2014. It is a combinatorial optimization algorithm used in particular in graph and slice management problems (MaxCut).

[1677] It is now possible to get rid of the classical part of the algorithm as explained in [An adaptive variational algorithm for exact molecular simulations on a quantum computer](), by Sophia Economou et al, 2019 (9 pages).

It has the advantage of requiring a low depth of quantum gates[1678].

### Hybrid quantum algorithms

I'd put in this category hybrid algorithms using different quantum computing paradigm. Given there are three main paradigms, it adds up to four combinations, by pairs and the 3 altogether. One of these is about mixing quantum annealing and gate-based quantum computing. It is proposed by a team of Taiwan researchers with their large-system sampling approximation (LSSA) algorithm that solves Ising problems more efficiently than quantum annealers alone[1679].

### Distribute quantum algorithms

In a section dedicated to distributed quantum computing using entanglement resources starting page 833, we have seen how quantum computer quantum interconnect could help create large QPUs. This is a rather long-term option and difficult to put in place.

Another sought option is to distribute some problem on several QPUs that are interconnected classically, using another form of hybrid algorithm. Such algorithms are designed for problems that can be split to run on several QPUs, with many constraints. This won't necessarily bring any exponential computing advantage but is an interesting path to work with NISQ QPUs[1680]. Among other scenarios, it is proposed to distribute some separable quantum neural network classification tasks to several QPUs[1681] and even more generic tasks[1682].

### Tensor networks

In our presentation of basic linear algebra, we have described the notion of tensor products to mathematically represent a qubit register (page 147). It's a multiplication of matrices and a qubit register is represented by an exponentially growing vector state of $2^N$ complex numbers. Tensor products are also heavily used with neural networks programming in classical deep learning, including with the famous TensorFlow SDK from Google.

Describing a many-body quantum system is making use of these exponentially large tensor products and they are hard to compute classically. So here comes the notion of tensor networks which helps factorize very large tensors into networks of smaller tensors. It can be viewed as techniques that "zips" the tensor representation of these many-body systems, providing up to an exponential gain in the number of computing parameters[1683]. One of the reason these systems can be compressed relies on the so-called area-law that says that entangled quantum systems are connected only to their

---

[1678] See Quantum Approximate Optimization Algorithm explained, May 2020, An Introduction to Quantum Optimization Approximation Algorithm by Qingfeng Wang and Tauqir Abdullah, December 2018 (16 pages), QAOA: Quantum Approximate Optimization Algorithm by Peter Shor (25 slides), Quantum Approximate Optimization Algorithm: Performance, Mechanism, and Implementation on Near-Term Devices, by Leo Zhou, Mikhail Lukin et al, 2019 (23 pages) and Quantum approximate optimization of non-planar graph problems on a planar superconducting processor by Matthew P. Harrigan et al, 2021 (19 pages) which uses a QAOA algorithm on Google's 53 qubits Sycamore.

[1679] See Hybrid Gate-Based and Annealing Quantum Computing for Large-Size Ising Problems by Chen-Yu Liu et al, August 2022 (14 pages).

[1680] See Quantum Divide and Conquer for Combinatorial Optimization and Distributed Computing by Zain H. Saleem et al, Argonne Lab, Princeton, University of Colorado Boulder and SuperTech/ColdQuanta, July 2021 (13 pages).

[1681] See Scalable Quantum Neural Networks for Classification by Jindi Wu, Zeyi Tao and Qun Li, Department of Computer Science William & Mary, Williamsburg, August 2022 (11 pages).

[1682] See Enabling multi-programming mechanism for quantum computing in the NISQ era by Siyuan Niu and Aida Todri-Sanial, LIRMM, March 2022 (23 pages).

[1683] I found the zip analogy in Tensor network states to compress the many body problem by Antoine Tilloy, Inria, November 2021 (15 slides).

neighborhood, thus enabling the split of many-body systems into separable smaller body systems[1684].Tensor networks have various use-cases in many-body quantum physics digital simulations, for classical simulations of quantum computers, in chemistry, as well as in machine learning and applied mathematics.

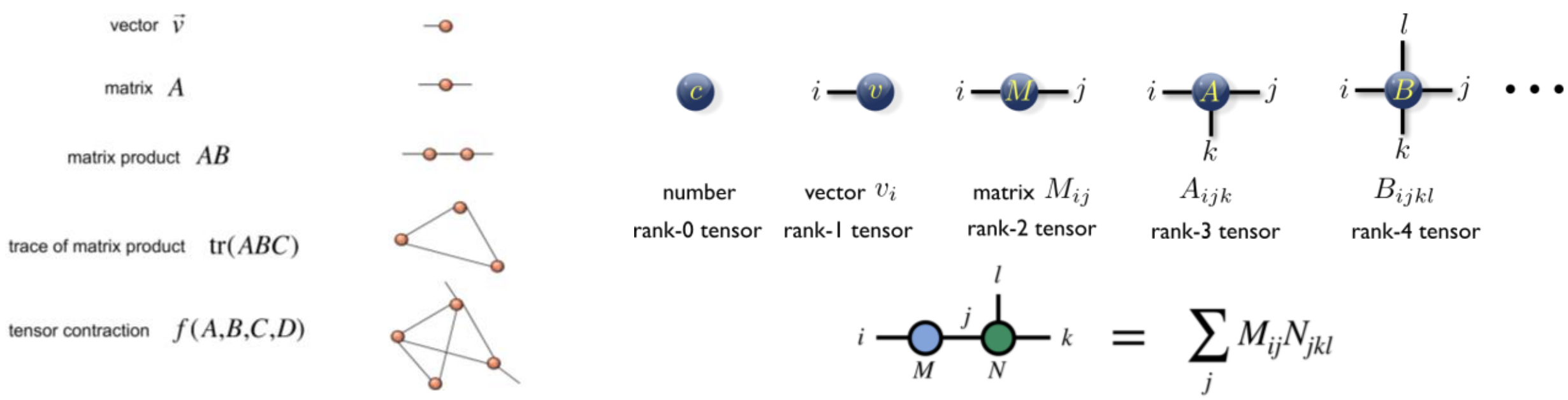

*Figure 589: Sources: [Introduction to Tensor Network States and Methods](#) by Román Orús, DIPC & Multiverse Computing, 2020 (229 slides) and [Lecture 1: tensor network states](#) by Philippe Corboz, Institute for Theoretical Physics, University of Amsterdam (56 slides).*

This is an entire new world of mathematics. How is it connected to quantum computing? The frontier is fuzzy.

Most of the tensor network literature deals with classical computing but tensor flow optimization techniques are also applicable to quantum computing. Some companies like Multiverse Computing make a lot of use of tensor networks techniques, sometimes embedded in the fuzzy concept of quantum inspired algorithms.

In typical graphical representations shown in Figure 589, tensor networks use graph notation connecting matrices (mid-points in graphs), vectors (line endpoints), traces of matrix products (triangles), and tensor contraction. Then, tensor network techniques are represented with these graphical views. As shown in Figure 590, the three main techniques are **MPS** (matrix product state represented in a 1D series of link with the preeminent DMRG variant [1685]), **PEPS** (Projected Entangled Pair States, used with graphs), **TPS** (Tensor Product States), and **MERA** (Multiscale Entanglement Renormalization Ansatz.

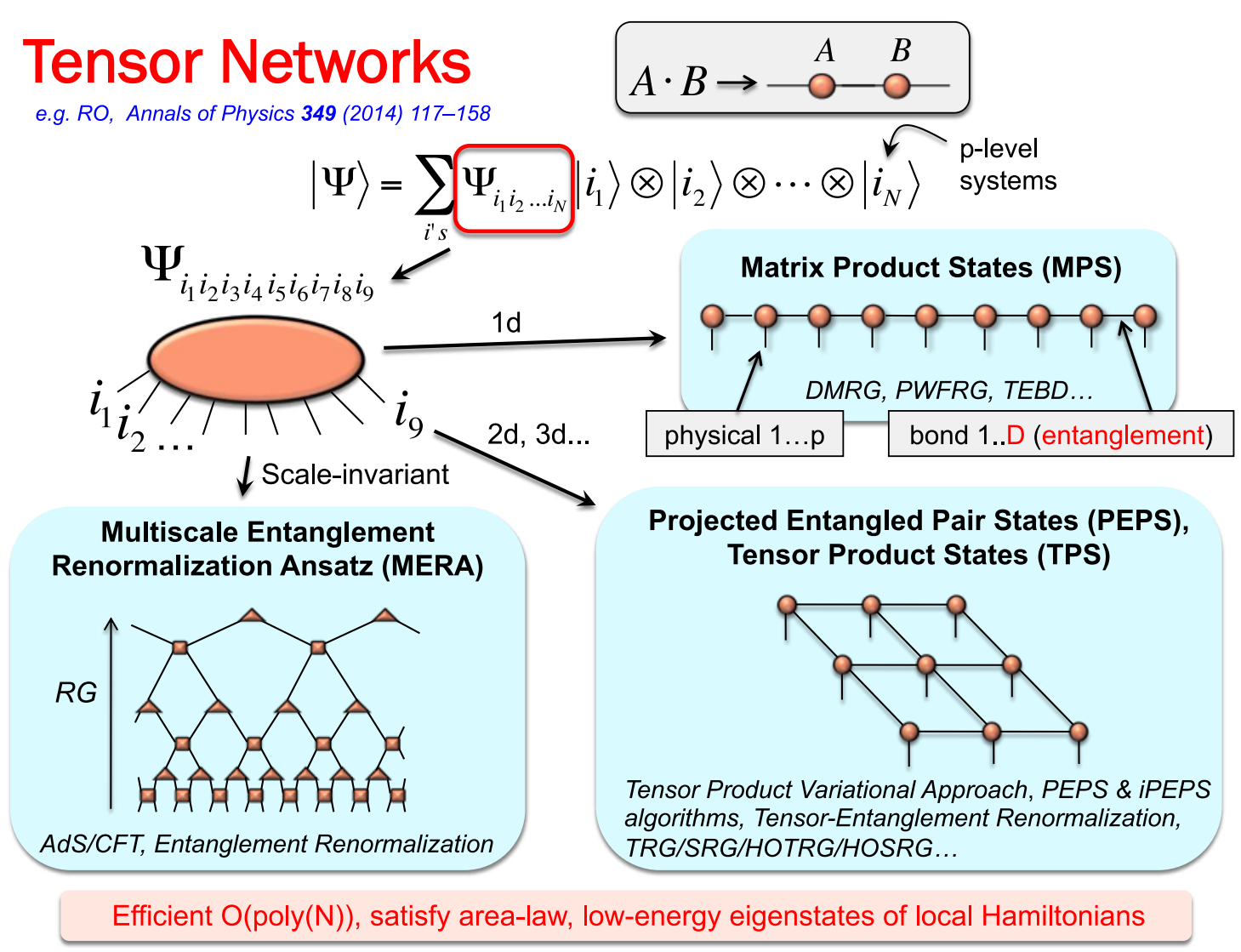

*Figure 590: how tensor networks are graphically represented using the above notation. Source: Same as above.*

[1684] For more on the area law, see [Colloquium: Area laws for the entanglement entropy](#) by J. Eisert, M. Cramer and M. B. Plenio, 2010 (28 pages) and [Area Laws for Entanglement](#) by Fernando G.S.L. Brandão and Michal Horodecki, Stanford University, 2014 (56 slides). To some extent, this law has indirect implications on the real scalability of quantum computers since on one hand, it seems to depend on the size of maximally entangled systems and in practice, these systems mays not exist due to the area law… !!!

[1685] Here are some key researchers behind tensor networks: Steven R. White, who created DMRG in 1992 and who's group at UCI in California maintains the ITensor software library for tensor network, Edwin Miles Stoudenmire who was his PhD and Ulrich Schollwöck from LMU München who also contributed to the development of DMRG.

The graphical representation of gate-based quantum circuits happens to be a special case of tensor networks[1686]! The ZX Calculus graphical language is also derived from the tensor network formalism[1687]. About 80 classical tools implement Tensor networks like Google TensorNetwork which relies on TensorFlow and was released in 2019[1688].

After you'll have spent some time understanding how tensor networks work[1689], you may wonder where quantum computing plays a role here. It seems mainly used as design tools to create quantum error correction codes, topological computing, solve condense matter physics problems[1690] and quantum machine learning algorithms[1691].

Tensor networks are also used in the creation of quantum code emulation software since these need various tensor contraction tools to save memory in storing the large quantum vector states if not density matrices of a number of qubits as large as possible[1692]. Xanadu's PennyLane framework contains tensor network circuit templates[1693].

Zapata Computing is also working on using tensor networks to prepare parametrized quantum circuits used in hybrid algorithms[1694].

## Quantum inspired algorithms

Quantum inspired algorithms are classical algorithms whose design is inspired by quantum algorithms and interference management, but not programmed as quantum algorithms run through a classical emulator. Quantum inspired algorithms can be helpful in solving linear algebra problems[1695], simulating quantum systems[1696], with portfolio optimization, recommendation systems (like with the famous solution from Ewin Tang), images classification[1697] and machine learning[1698].

---

1686 The Basics of Tensor Network - An overview of tensors and renormalization by Samuel Desrosiers, Glen B. Evenbly and Thomas E. Baker (5 pages) explain how we build tensor network algorithms using these representations. This graphical representation was created by Roger Penrose in 1971.

1687 See a use-case of ZX Calculus in the creation of a tensor network for the classical part of a variational circuit Barren plateaus in quantum tensor network optimization by Enrique Cervero Martín, September 2022 (26 pages).

1688 See The landscape of software for tensor computations by Christos Psarras et al, March 2021-June 2022 (16 pages) and Google TensorNetwork Library Dramatically Accelerates ML & Physics Tasks, 2019.

1689 See Tensor Networks in a Nutshell by Jacob Biamonte et al, 2017 (34 pages), Lectures on Quantum Tensor Networks by Jacob Biamonte, January 2020 (178 pages), Hand-waving and Interpretive Dance: An Introductory Course on Tensor Networks Lecture Notes by Jacob C. Bridgeman and Christopher T. Chubb, 2017 (62 pages), Tensor Network Contractions by Maciej Lewenstein et al, 2020 (160 pages), the review paper Tensor Network Algorithms: a Route Map by Mari Carmen Bañuls, May 2022 (14 pages) and Hand-waving and Interpretive Dance: An Introductory Course on Tensor Networks by Jacob C. Bridgeman and Christopher T. Chubb, 2017 (62 pages).

1690 See Applications of Tensor Networks: Machine Learning & Quantum Computing by Edwin Miles Stoudenmire, 2018 (142 slides).

1691 See Towards Quantum Machine Learning with Tensor Networks by William Huggins, Edwin Miles Stoudenmire et al, Berkeley and Flat Iron Institute, July 2018 (12 pages).

1692 See TensorCircuit: a Quantum Software Framework for the NISQ Era by Shi-Xin Zhang et al, May 2022 (43 pages).

1693 See Tensor-Network Quantum Circuits, June 2022.

1694 See Synergy Between Quantum Circuits and Tensor Networks: Short-cutting the Race to Practical Quantum Advantage by Manuel S. Rudolph et al, August 2022 (12 pages).

1695 See An improved quantum-inspired algorithm for linear regression by András Gilyén et al, January 2022 (23 pages).

1696 See one example in Classical algorithms for many-body quantum systems at finite energies by Yilun Yang, J. Ignacio Cirac and Mari Carmen Banuls, April 2022 (11 pages).

1697 See AutoQML: Automatic Generation and Training of Robust Quantum-Inspired Classifiers by Using Genetic Algorithms on Grayscale Images by Sergio Altares-López et al, August 2022 (13 pages) which improves medical imaging grey images classification using a quantum inspired machine learning algorithm.

1698 See the review paper Quantum inspired algorithms in practice by Juan Miguel Arrazola, Seth Lloyd et al, 2020 (24 pages).

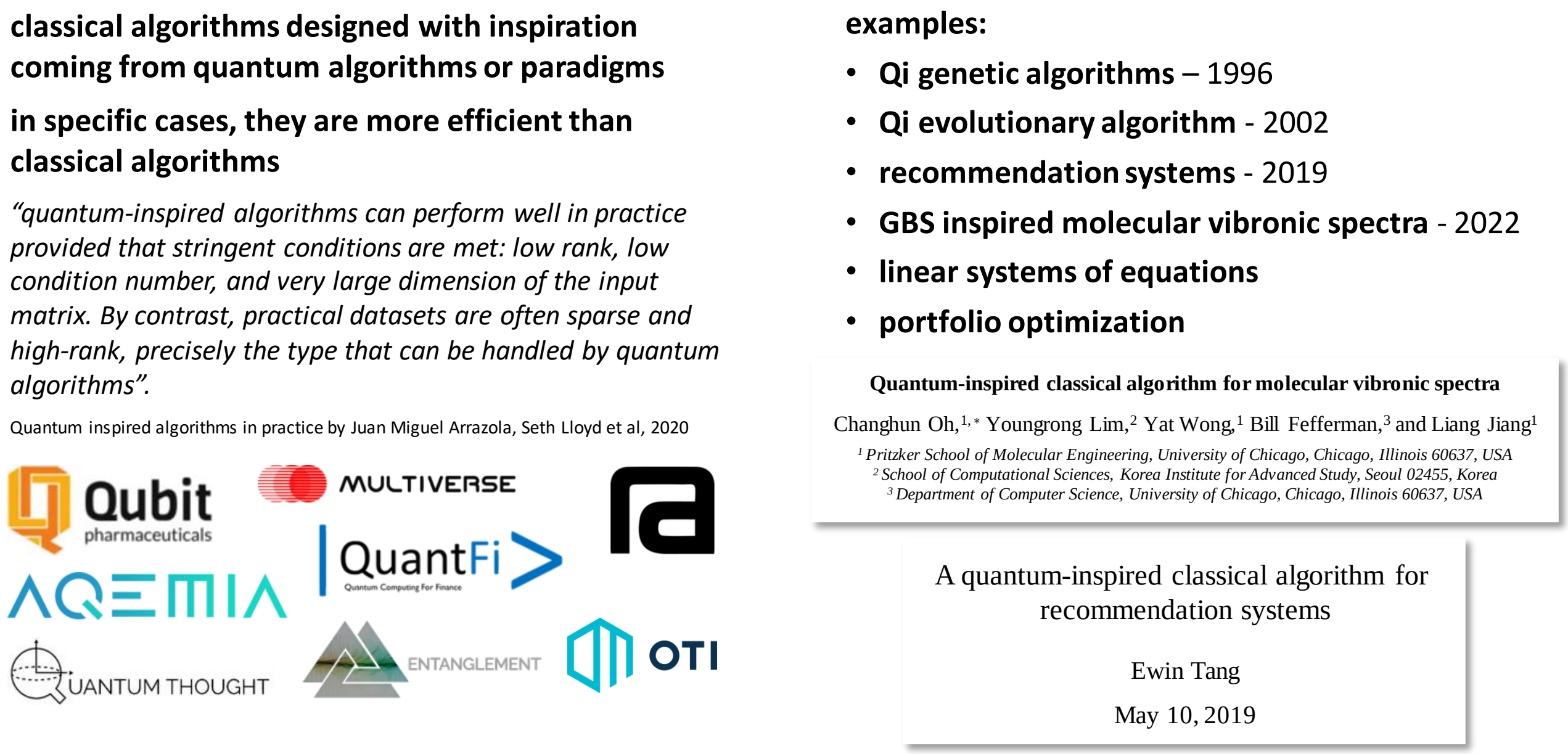

*Figure 591: quantum inspired algorithms examples. (cc) Olivier Ezratty and various sources.*

Creating a quantum inspired algorithm is sometimes said to rely on "dequantizing" or the "de-quantization" of a quantum algorithm[1699].

As Alastair Abbott and Cristian S. Calude wrote in 2010, "*de-quantization helps formulate conditions to determine if a quantum algorithm provides a real speed-up over classical algorithms. These conditions can be used to develop new quantum algorithms more effectively (by avoiding features that could allow the algorithm to be efficiently classically simulated), as well as providing the potential to create new classical algorithms (by using features which have proved valuable for quantum algorithms)*".In their paper, they also found that any algorithm in which entanglement is bounded is de-quantizable[1700].

Quantum inspired algorithms even exist that transposes the complicated gaussian boson sampling photonic based technique to simulate molecular vibronic spectra. Molecular vibronic spectra come from its light absorption that depends on the transitions between its different electronic states and changes in its vibrational energy[1701].

Quantum inspired algorithms are used in finance, healthcare and by many startups like Qubit Pharmaceuticals, Aqemia and Rahko in chemical simulations or QuantFi and Multiverse Computing in financial optimization. Quantum software startups find in quantum inspired algorithms a way to monetize their know-how while waiting for sufficiently powerful quantum computers.

## Complexity theories

So far, we have reviewed a lot of the most common quantum algorithms and their theoretical acceleration.

---

[1699] See one example in Sampling-based sublinear low-rank matrix arithmetic framework for dequantizing quantum machine learning by Nai-Hui Chia, Ewin Tang et al, October 2019 (79 pages).

[1700] See Understanding the Quantum Computational Speed-up via De-quantisation by Alastair Abbott and Cristian S. Calude, 2010 (12 pages).

[1701] See Quantum-inspired classical algorithm for molecular vibronic spectra by Changhun Oh, Liang Jiang et al, University of Chicago, February 2022 (19 pages).

Quantum computing is sometimes presented as a miracle solution to extend computing capacities beyond the limits of classical supercomputing. It allows solving so-called "intractable" problems on conventional computers.

But what is the nature of the problems that can be solved with a quantum computer and that cannot be solved with classical computers? And above all, what are the limits of quantum computers? Do we still have computational limits?

We will see that these limits are rather blurred and shifting over time. This deals with **complexity theories**, a rather cryptic field of science and mathematics. It is a very abstract world involving a cryptic semantic made of P, NP, BQP and other complexities classes. Mathematicians have been discussing for half a century whether **P = NP** or not. This is the science of problem complexity classes. Behind these mathematics of complexity lie key technical but also philosophical considerations that are fundamental to Man and his omnipotence and control desires.

Problem complexity classes are Russian dolls more or less nested with each other that describe the range of time it takes to solve a problem, to verify given solutions and also on the associated required memory space, with regards to the problem size. The size of a problem is often formulated as an integer N, giving the number of items in the problem.

As far as time scales are involved, there are many ways in which this problem solving time scale can grow with N. The key ones are: constant, logarithmic, linear, polynomial and exponential. In this time scale, a time is considered reasonable if it is polynomial or below polynomial in the growing scale. A polynomial time is proportional to a given power of N.

Quantum computing allows under certain conditions to solve certain exponential problems in a polynomial time. It must be translated in: a given problem that would require an exponential time to be solved on a classical computer would require a polynomial time to be solved on a quantum computer.

But what lies beyond exponential time? There are still various inaccessible problems with, for example, exponential of exponential time scales. And we have also exponential memory space which can add another complexity dimension. Quantum computers will not be able to solve all these problems, even when we will be able to align gazillions of logical qubits.

These limitations have an indirect impact on predictions about the creation of some omniscient artificial intelligences capable of transcending human reasoning and solving all problems. This hypothetical AGI (Artificial General Intelligence) will be limited by the data and concepts that feed it and by the impossibility of solving all complex problems.

Mankind will continue to confront impossible computing tasks. It will not be able to solve all the complex problems around! Quantum computing does not allow us to dominate nature, to put the whole Universe in equations and to predict how it will run with quantum precision. Chance and the unexpected will continue to play a role in a very indeterministic world, and for the better. It is a small lesson in humility for Mankind.

## Problem Complexity Classes

To dive into complexity classes, you need to define the main classes of problems by level of complexity. Here I am trying to simplify complexity, this time in the literal sense of the word.

Complexity classes often describe problems that are solved by using brute force with testing several combinations to find the ones matching some criteria (like with the so-called SAT problems) or with using mathematical equations defining complex systems (differential equations, Schrödinger's equation, ...).

Problem classes use the notion of Turing's deterministic and non-deterministic machines. Turing machines are conceptual models of computers created by Alan Turing before the Second World War.

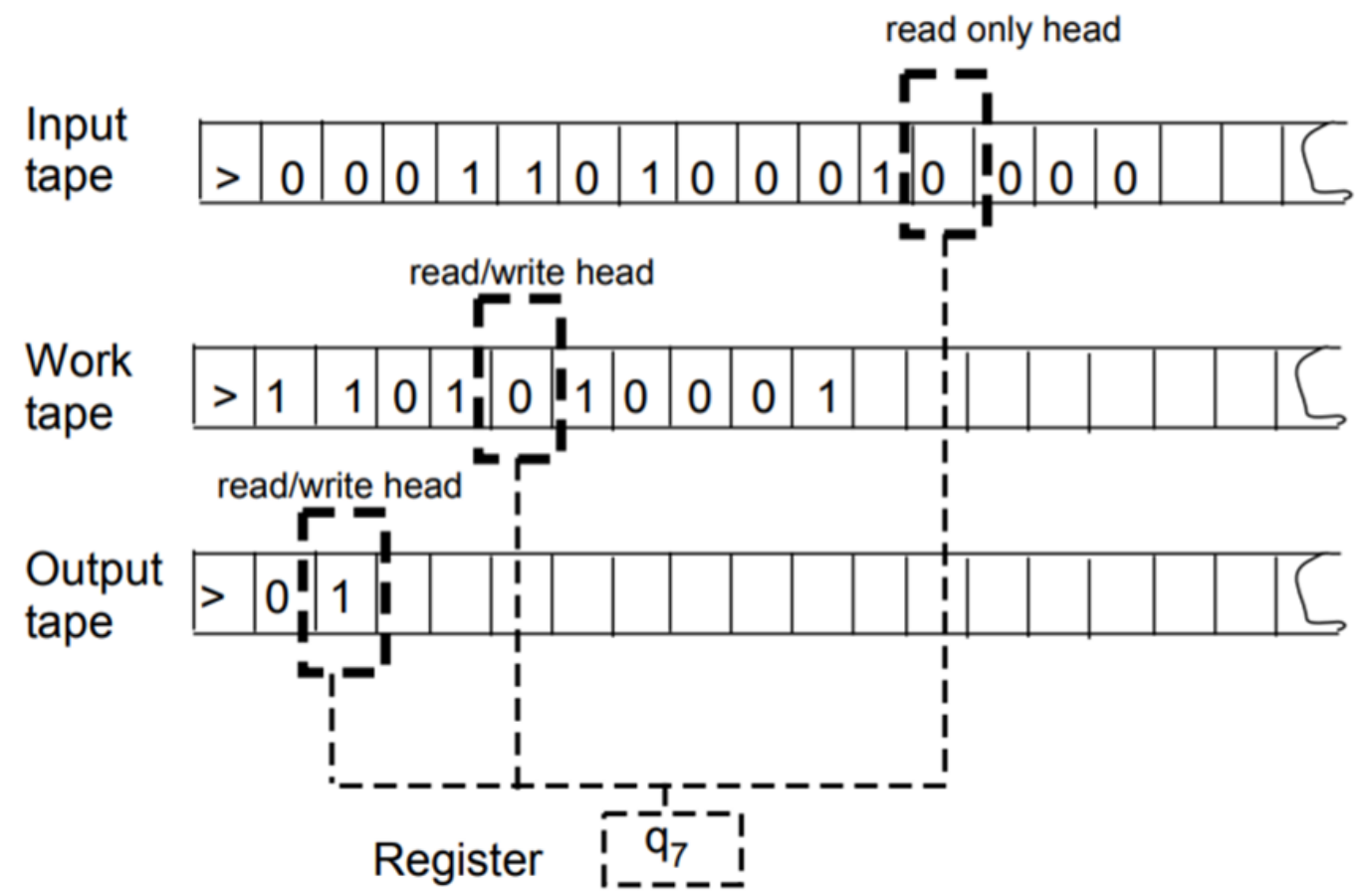

*Figure 592: the famous Turing machine. Source: Computational Complexity: A Modern Approach by Sanjeev Arora and Boaz Barak, 2007 (489 pages).*

They model computer processing based on the notion of programs and data, embodied by continuous and infinite rolls of paper, the first for the program, the second for the input data and the third for generating the results. Turing's theoretical model has long been used to define classes of problems that can or cannot be solved by a computer[1702].

Computers are all metaphorically Turing machines, reproducing this logic by reading program instructions and managing data in random access memory (RAM) or persistent storage (hard disk, SSD, ...). Associated with the notion of Turing machine is the notion of **Church-Turing's thesis**, named after Alonzo Church and Alan Turing, according to which there is an equivalence between computational problems that can be solved manually and with unlimited resources, those that can be handled with a Turing machine and those that can be solved with so-called recursive functions.

In a deterministic machine, the sequence of actions to be performed is predetermined and sequential. In the non-deterministic Turing conceptual machine model, computational rules can lead to execute several different operations for each evaluated situation. Basically, by exploring several paths in parallel and looking for a positive response to an algorithm component and closing parallel test loops once the sub-solutions are found.

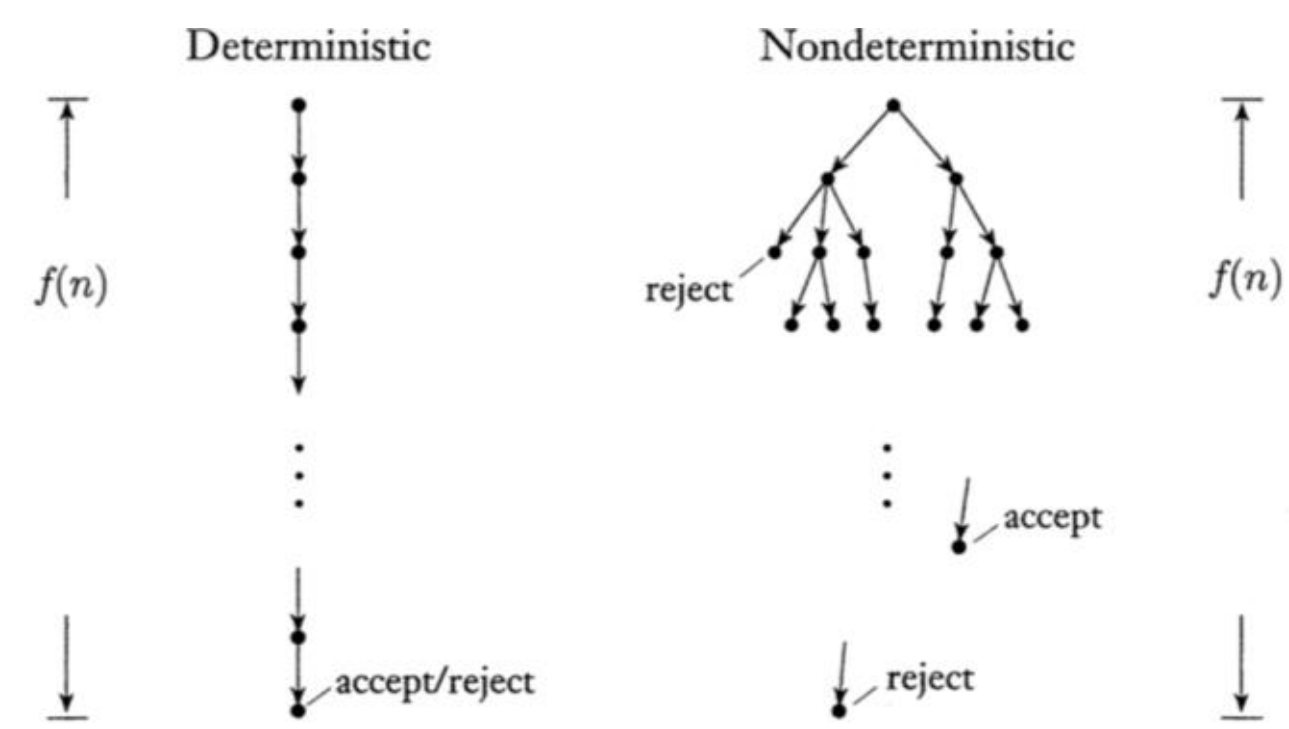

*Figure 593: deterministic and non-deterministic Turing machines.*

A non-deterministic machine increases the computational combinatorics compared to a deterministic machine. And this combinatorics usually jumps from polynomial to exponential.

### Generic complexity classes

The level of complexity refers to the computing time and memory space required for these calculations. We are usually limited by computing time long before being limited by the available memory. However, some problems such as scheduling reach memory limits before computation limits.

---

1702 See Computational Complexity: A Modern Approach by Sanjeev Arora and Boaz Barak, 2007 (489 pages) which is a good reference document on complexity theories. Students of a master's degree from ENS Lyon made a Turing machine in Lego in 2012 to celebrate the centenary of Alan Turing's birth (video) and it wasn't the only one of its kind (video)! Another one was made with wood in 2015 (video).

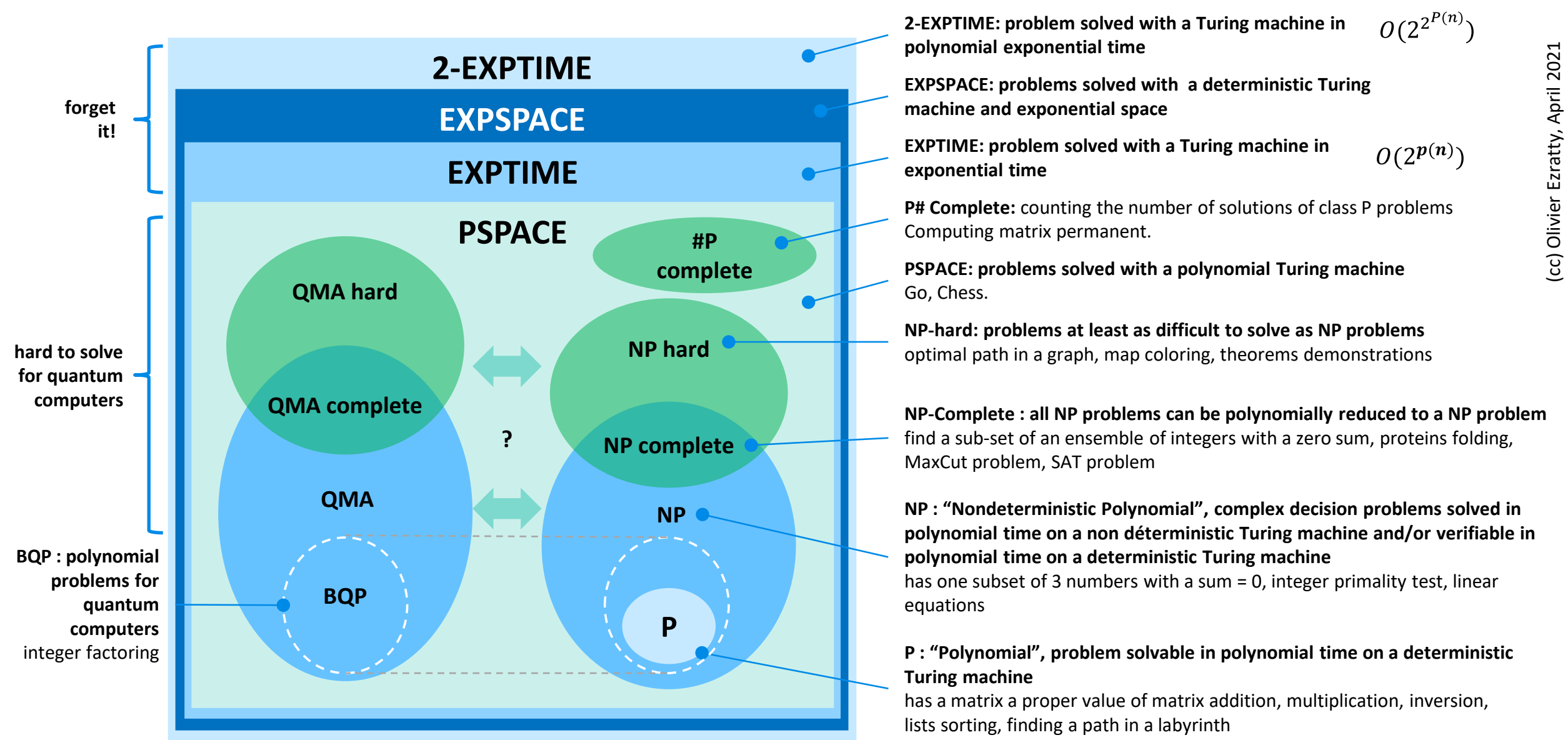

*Figure 594: quantum and classical complexity classes, compilation (cc) Olivier Ezratty, 2021.*

The association of a problem to a complexity class is related to the performance of the best-known algorithms to solve problems in that class.

Problem class levels in complexity theories are often based on black box or oracle models to which a system asks questions and gets answers based on the data provided. This is a logic of "brute force" and hypothesis scanning. The scale of the tested combinations depends on problem class.

So here are these classes by increasing level of complexity knowing that we will spend more time on NP related classes.

**L** or LSPACE, or DLOGSPACE, defines the class of problems that can be solved on a logarithmic scale of consumed memory and on a deterministic Turing machine, that is, on a traditional computer. Computational complexity increases slowly with the size of the problem. Unfortunately, very few complex problems sit in this class. These include queries in previously indexed relational databases, searches for DNA sequences, and generally speaking, search techniques that use pointers and optimize the use of computer memory.

**NL** is the class of problems solved on a logarithmic time scale on a non-deterministic machine. Complexity theory specialists are still trying to figure out whether L=NL or not! But they are less busy here than on determining whether P<>NP.

**P** covers problems that can be solved with a time growing polynomially with data to process and on a deterministic machine. If N is the size of the problem, the processing time is proportional to $N^M$, with M being an integer, preferably 2. It’s an easy problem to solve and said to be "tractable". This includes sorting lists, validating the existence of a path in a graph, searching for a minimum path in a graph, multiplying matrices or evaluating a number to see if it is prime.

**BPP** is a class of problems that can be solved by random approaches ("Bounded-Error Probabilistic Polynomial-Time"). It would seem that BPP=P but this has not yet been demonstrated.

**NP** describes the class of problems for which it is easy to check the validity of a solution, i.e., that it can be realized in polynomial time by a deterministic machine. The other definition of the class is that it contains problems whose solution time is polynomial on a non-deterministic machine. These more complex problems have a computing time that is at least exponential when the method used is said to be naive, testing all possible combinations. These are “intractable” problems. In practice, these are problems particularly suited to quantum computers because of their ability to evaluate in parallel $2^N$ combinations of some classical computation.

Some examples of NP problems are the Steiner tree to determine whether an electrical network can connect a number of houses at a certain price, checking that a DNA sequence is found in several genes and the distribution of tasks to different agents to minimize the time it takes to complete them.

These problems have very concrete equivalents in logistics, planning, production, transportation, telecom, utilities, finance and cryptography. Note that a "decidable" problem, i.e., one that requires exploring a finite space of options, is not necessarily feasible from a practical point of view. Even if it can be solved in a finite amount of time, its resolution may take too long. An exponential problem has an elegant solution if one can find one solution that has a polynomial or, at best, linear duration. Polynomial times scale better than exponential times!

A big debate has been going on since 1956 (Kurt Gödel) as to whether class P equals class NP. If P = NP, it would be as simple to find a result when one can also simply verify it. The general consensus is that P is not equal to NP[1703]. The demonstration of whether or not P<>NP is part of one of the seven Clay Mathematics Institute mathematical challenges launched in 2000, each with a prize of $1M (Figure 595)[1704].

Among these challenges are the demonstration of the Navier-Stokes fluid mechanics equations and of Riemann's hypothesis on the distribution of prime numbers.

On the P vs NP side, the wording of the challenge provides an example of such a problem: you have to allocate 50 rooms of two students to 400 candidates but some candidates do not need to live in the same room[1705].

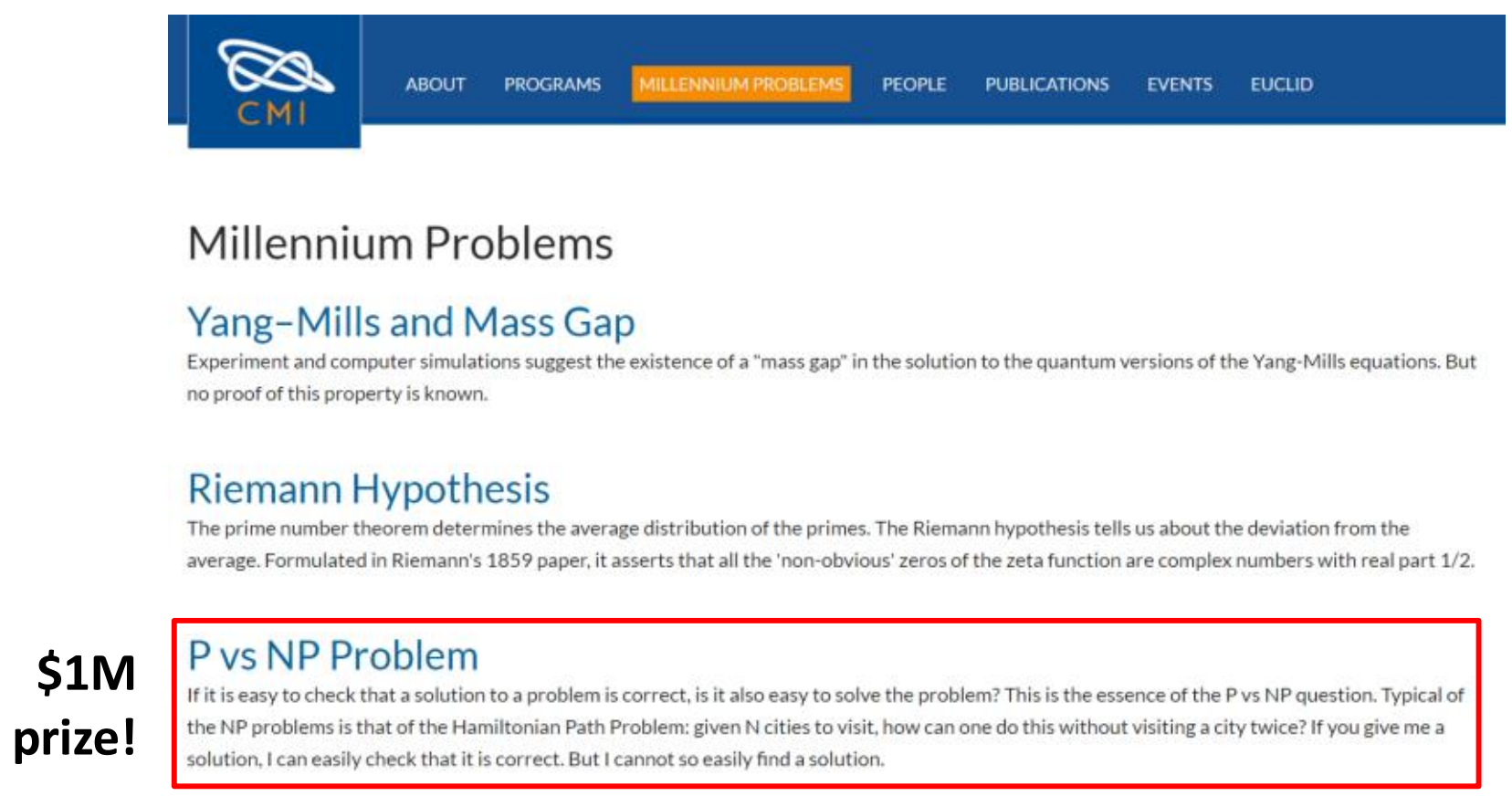

*Figure 595: the Millennium challenge and P vs NP problem. Source: Clay Mathematics Institute mathematical challenges.*

The combinatorial choice of 100 students among 400 is huge, so the problem is not easily handled with a supercomputer and brute force. It is indeed an NP problem because a given solution is easy to verify because it is simple to check that none of the rooms contains a forbidden pair of individuals. It is a bit like an all-or-nothing theory because if P = NP, all NP problems have an efficient polynomial solution. If P ≠ NP, none of the NP problems have a "pure" efficient solution[1706]. The definition of the problem classes NP and NP-Complete is relatively recent[1707].

[1703] See Fifty Years of P vs. NP and the Possibility of the Impossible by Lance Fortnow, 2021 (11 pages).

[1704] A Brazilian researcher, André Luiz Barbosa, published in 2010 his P ≠ NP Proof (25 pages) as well as a paper invalidating Cook's theorem that a Boolean SAT problem is NP-Complete, The Cook-Levin Theorem Is False, 2010 (11 pages). But this work seems ignored by specialists.

[1705] See The P versus NP problem by Stephen Cook (12 pages).

[1706] The classical method for solving these problems is to use heuristics allowing to obtain a satisfiable approximate solution, therefore not necessarily optimal, and in particular via probabilistic approaches.

[1707] It is derived from The complexity of theorem-proving procedures by Stephen Cook of the University of Toronto, 1971 (8 pages), best popularized in An overview of computational complexity (8 pages) and Reducibility among combinatorial problems by Richard Karp, 1972 (19 pages) and in Complexity and calculability by Anca Muscholl of the LaBRI, 2017 (128 slides).

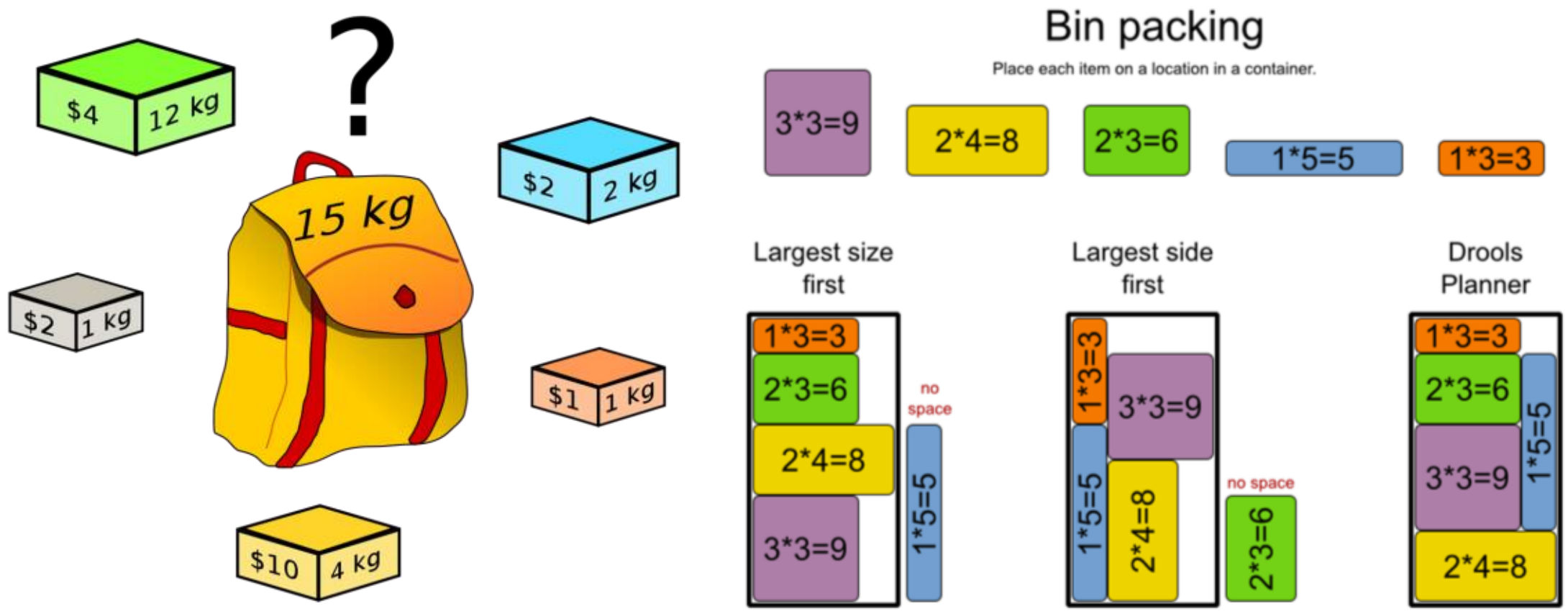

*Figure 596: the famous bin-packing problems. Ever filled your car's trunk when going to vacation? Sources: Wikipedia and Stackoverflow.*

**Complete NP** corresponds according to Richard Karp to problems in which other NP problems can be polynomially reduced. They have no known P (polynomial) solution. They are not accessible to quantum computers. It is in this class that we find the SAT or 3SAT type Boolean logic problems! More than 3000 NP-Complete problems are identified to date (list).

One of the typical problems is filling up the trunk of a car when you go on vacation or when you come back from Christmas with a bunch of presents for your family. And then the **bin packing** problem consisting in filling a backpack in an optimal way with a set of objects, to obtain the largest load and without exceeding a maximum weight (*aka* "bin packing" or "knapsack" problems).

It also includes the **subset sum problem of** finding a subset of a set of integers whose sum is equal to an arbitrary integer.

The **deminer's problem** consists in locating hidden mines in a field with only the number of mines in adjacent areas and the total number of mines in the field as an indication. All this without detonating them. It is a game well known by Windows users, launched in 1989!

The simulation of complex protein folding is also a NP-Complete problem[1708].

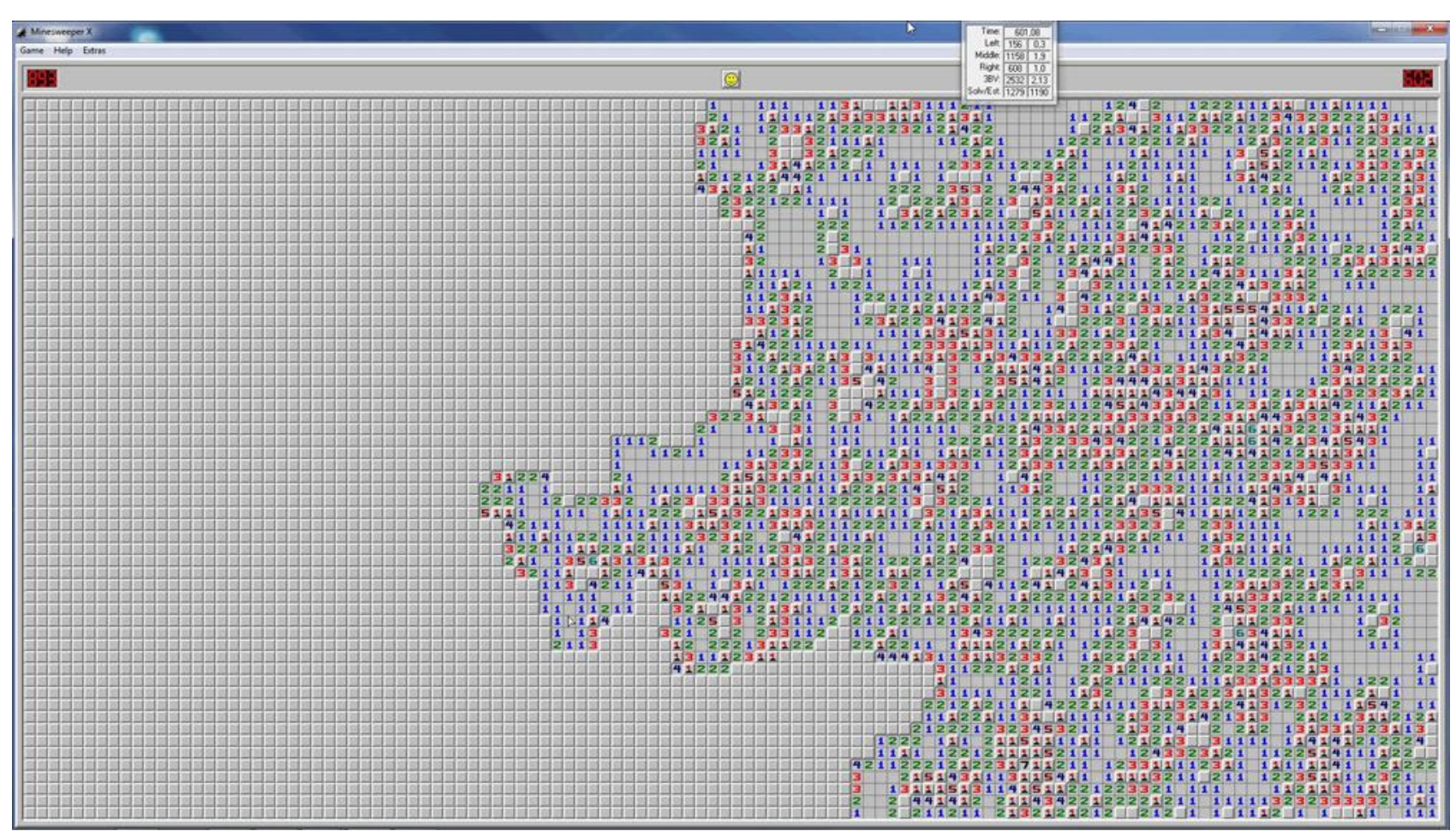

*Figure 597: the deminer's problem is also an NP complete problem. Source of the illustration.*

So, this would be a potentially very difficult problem to solve with a quantum computer with large proteins.

It is demonstrated that if an optimal solution to an NP-Complete problem could be found, all the solutions to problems in this class would be found. This is the important notion of problem reduction.

Graphs coloring with different colors for knots, branches or surfaces is part of the NP, NP-Complete and NP-Hard problems. The first two cases requiring a number of colors depending on the maximum number of connections between elements of the graph and the last case, relating to the coloring of

[1708] See Is protein folding problem really a NP-complete one? First investigations, 2013 (31 pages).

maps in different adjacent colors which requires a maximum of four, thanks to the computer demonstration of the four-color theorem done in 1976 by Kenneth Appel and Wolfgang Haken.

- **Graph nodes** coloring has applications in the placement of mobile antennas and in the allocation of memory registers for a compiler. The problem is NP-Complete for its resolution and NP-Difficult to find its optimal solution.
- **Branch** coloring has applications in the frequency allocation of multimode fiber optic networks. It also allows to optimize the placement of objects or persons according to their compatibility or incompatibility (*aka* the wedding tables problem). Optimum coloring is a NP-Hard problem.
- **Area map** coloring is used to define the coverage areas of mobile radio antennas or telecommunications satellites. It can even be used to allocate microwave frequencies for the activation of superconducting qubits. The coloring with three colors is an NP-Complete problem.

In general, many C problem classes have a subclass C-Complete and C-Hard. A problem is C-Hard if there is a type of reduction of problems from class C to this problem. If the problem C-Hard is part of class C, then it is said to be "C-Complete"[1709].

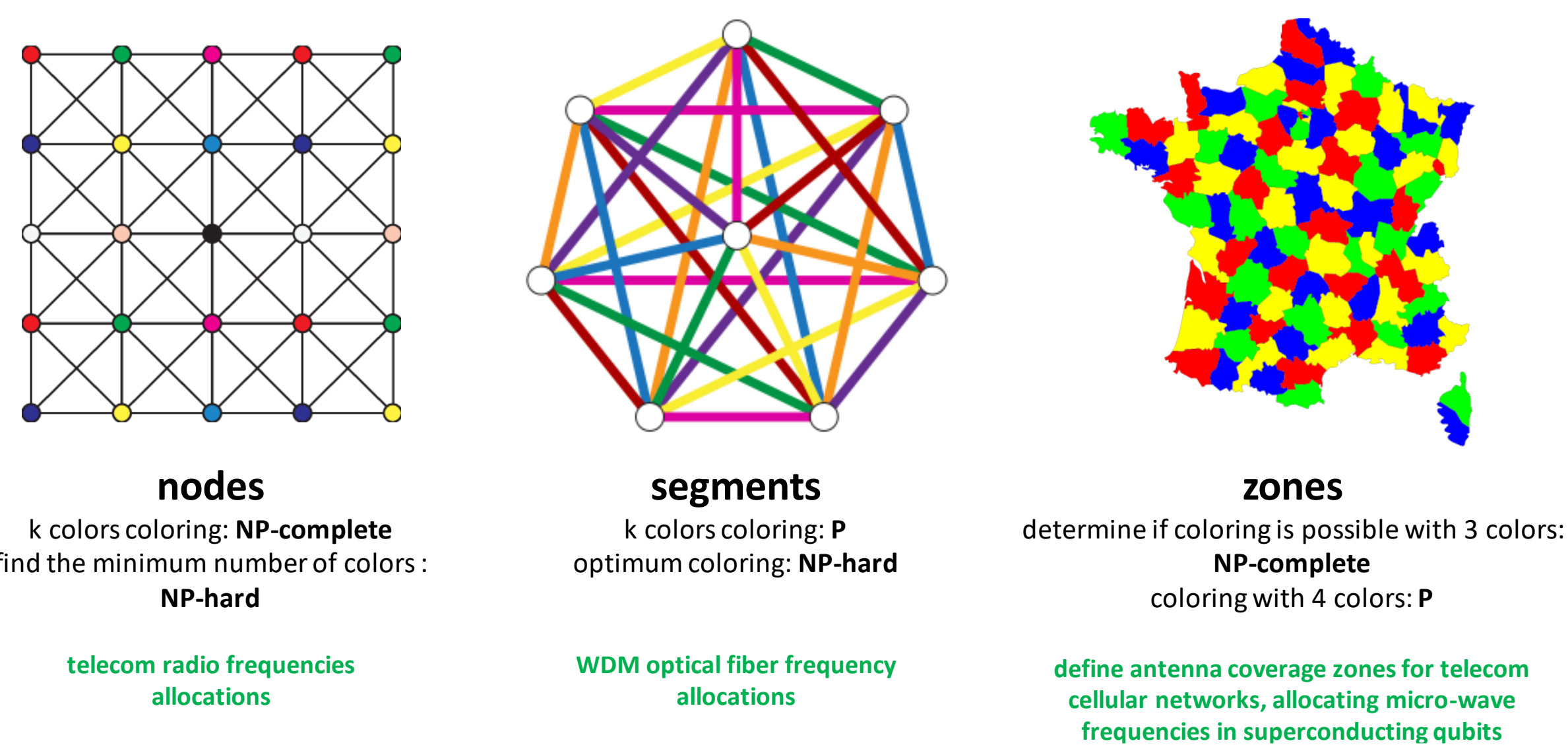

*Figure 598: graph problems with nodes, segments and zones coloring.*

**NP Hard** covers optimization problems where a minimum or a maximum is sought with a large combinatorics. A problem is NP-Hard if all NP-Complete problems can be reduced by polynomial simplification to this problem. It is the case of the solution of the **traveling salesperson problem** where one must test a large combination of routes to find the quickest one passing through a fixed number of cities. In this case, all solutions must be tested.

If a traveler has to go through 125 cities in less than 30 days, if there is a solution that works in that time frame, then the problem is NP.

[1709] For more information, see in particular Complexity Theory Part I (81 slides) and Part II (83 slides), which is part of a Stanford course on Complexity Theories, Calculability and Complexity- Some Results I Know by Etienne Grandjean of the University of Caen, 2017 (43 slides) as well as this video by Olivier Bailleux (2017, 20 minutes).

But nothing says that all the solutions have been found. Solving the problem below an arbitrary travel time with a return to the starting point is an NP-complete problem. This is called a Hamiltonian circuit: a path running a graph passing once and only once through each of the nodes and returning to its starting point. The determination of the shortest travel time is NP Hard.

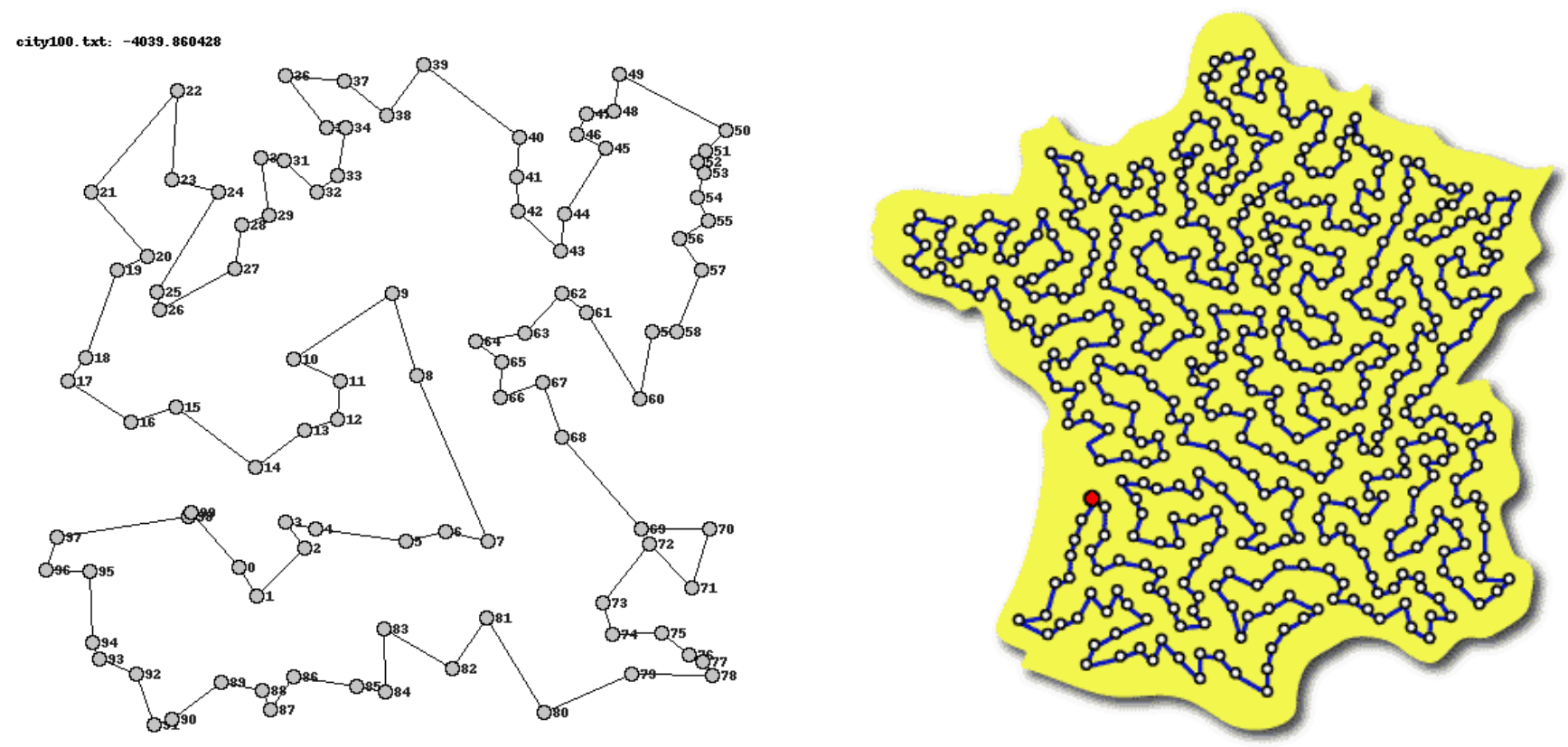

what is the shortest possible route that visits each place exactly once and returns to the origin place : **NP-hard**

*Figure 599: the TSP (traveling salesperson problem).*

The brute force algorithm to solve it has a time that depends on N! where N is the number of nodes in the network. The known optimum time is $N^2 2^N$. The problem is difficult to solve beyond 20 steps[1710]!

The NP-Hard problem class also contains a number of **Nintendo** games like Super Mario Bros, The Legend of Zelda and Pokemon[1711]. Quantum computing would not be able to solve the most complex NP-Hard problems.

**PSPACE** is the class of problems that can be solved in polynomial space on a deterministic machine. NPSPACE is the class of problems that can be solved in polynomial space on a non-deterministic machine. And NPSPACE = PSPACE according to Savitch's theorem.

**EXPTIME** is the class of problems decidable in exponential time by a deterministic machine. Precisely, the computation time of these problems is $2^{p(N)}$ where p(N) is a polynomial of N, N being the level of complexity of the problem. They are intractable with traditional machines. Some of these problems can be converted into problems that can be treated polynomially by quantum computers. Chess and Go games on arbitrarily sized grids belong to this category. In size-limited grids, the exponential effect has limits. These were exceeded for Deep Blue's chess game in 1996 and for DeepMind's AlphaGo game of Go in 2016 and 2017.

**NEXPTIME** is the class of problems decided in exponential time by a non-deterministic machine with unlimited memory space.

**EXPSPACE** is the class of problems that can be solved in exponential space. In other words, they are difficult to access on today's and even tomorrow's machines.

**2-EXPTIME** is a class including the previous ones that covers decision problems that can be solved by a deterministic Turing machine in double exponential time with an order of magnitude of $O(2^{2^{P(n)}})$where P(n) is a polynomial of n. In other words, it's an exponential of an exponential problem.

We should add the class **#P** of problems for counting the number of solutions of class P problems, which are solved in polynomial time. Proposed in 1979 by Leslie G. Valiant, it obviously has its associated classes **#P Hard** and **#P Complete**.

1710 See The Traveling Salesman Problem site which provides some examples of such problems such as the itinerary of all 49,687 English pubs or 49,603 tourist places in the USA.

1711 See Classic Nintendo Games are (Computationally) Hard, 2012 (36 pages).

The computation of the permanent of a square matrix filled with 0 and 1 is a complete #P problem according to Ben-Dor and Halevi's theorem demonstrated in 1993. In 2011, Scott Aaronson demonstrated that the calculation of the permanent of a matrix is a #P Difficult problem[1712]. All this is related to the numerical simulation of the boson sampling which is compared to its resolution by photon-based systems that we study in a section on photon qubits, page 434.

The classes PSPACE, EXPTIME, NEXPTIME, EXPSPACE and 2-EXPTIME do not correspond to practical problems that are easy to identify in everyday life. They cover the problems of predicting the behavior of ultra-complex systems with strong interactions. If it is possible that modeling the folding of a protein is an NP problem, what would be the class of problem to simulate the functioning of a whole living cell, or even a multicellular organism? There are so many interactions at the atomic, molecular and cellular level that the class of this kind of problem is probably well beyond NP-Hard level.

There are many other problem complexity classes that won't be described here: EXP, IP, MIP, BPP, RP, ZPP, SL, NC, AC0, TC0, MA, AM and SZK! You can find them in the Complexity Zoo site which inventories the zoo of problem complexity classes. There seems to be over a hundred of them[1713].

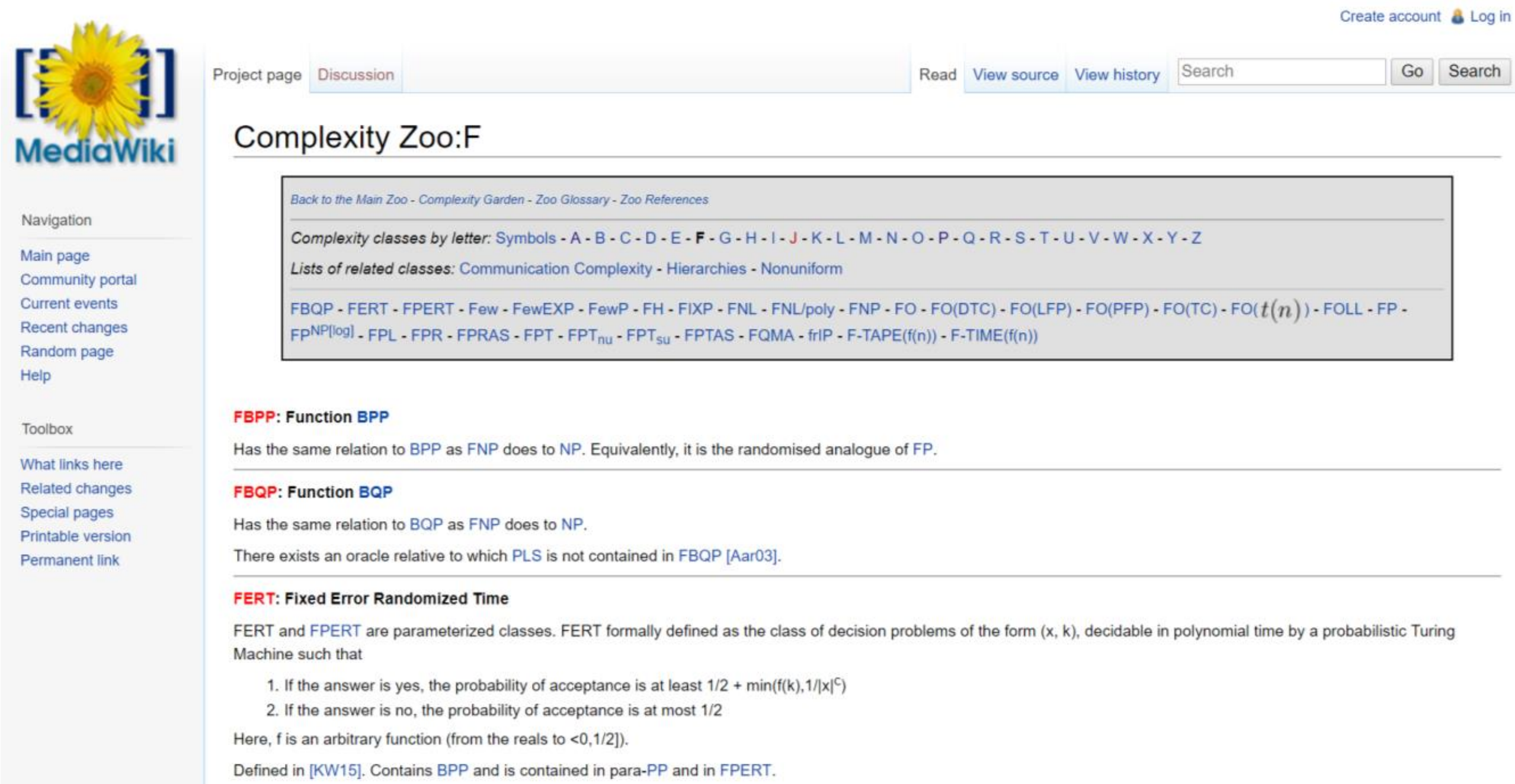

*Figure 600: there is even a zoo website for complexity classes! Source: the Complexity Zoo.*

### Quantum complexity classes

Let's now discuss the classification of problems that are theoretically addressable by quantum computers, the correspondence with the *above* classes being still a problem that is not entirely solved!

The classification is different because quantum computers can parallelize processing in an exponential way while classical computers like Turing machines cannot do it.

This is still very theorical since it doesn't take into account known constraints of quantum computers: their short coherence time and creates constraints on the number of quantum gates that can be chained together to solve a problem. This is a constraint that traditional computers do not have. But again, theoretically, this computation time constraint could be removed with using correction error codes.

---

[1712] In A Linear-Optical Proof that the Permanent is #P-Hard par Scott Aaronson, 2011 (11 pages).

[1713] To learn more about the subject of complexity theories, you can read the well-documented Computational Complexity A Modern Approach by Sanjeev Arora and Boaz Barak of Princeton University, 2007 (489 pages).

**PH** is a class of problems that generalizes the NP class. PH means "Polynomial Hierarchy".

**BQP** defines a class of problem that is solvable in polynomial time on a quantum computer with a constrained error rate of 1/3. This may in some cases correspond to P problems. The class was defined in 1993, when the first quantum algorithms appeared.

Whether the BQP class is really different from the P class is an ongoing debate. It has already been shown that the P class of polynomial problems is in BQP. But is NP in BQP? Seems not. It is however difficult to prove it in a generic way. The exact relationship between BQP and NP is still unknown.

The key point is to find algorithms that are part of BQP (processable in quantum) and that are not in PH (processable with any present and future classical architecture). This uncertainty has been removed only very recently[1714]. Oracle-based algorithms were found that are in BQP but not in PH.

**NISQ** is a class proposed by researchers from UC Berkeley, Harvard, Catltech and Microsoft in a paper published in 2022. It describes the class of problems that could be processed by a hybrid system using a NISQ quantum computer. It is showing that this class is in between BPP and BQP but it seems closer to BPP (problems accessible to classical computers) than BQP (problems accessible only to quantum computers) for Grover algorithm and farther for Bernstein-Vazirani algorithm[1715]. It doesn't include problems that are solvable by quantum annealers and simulators.

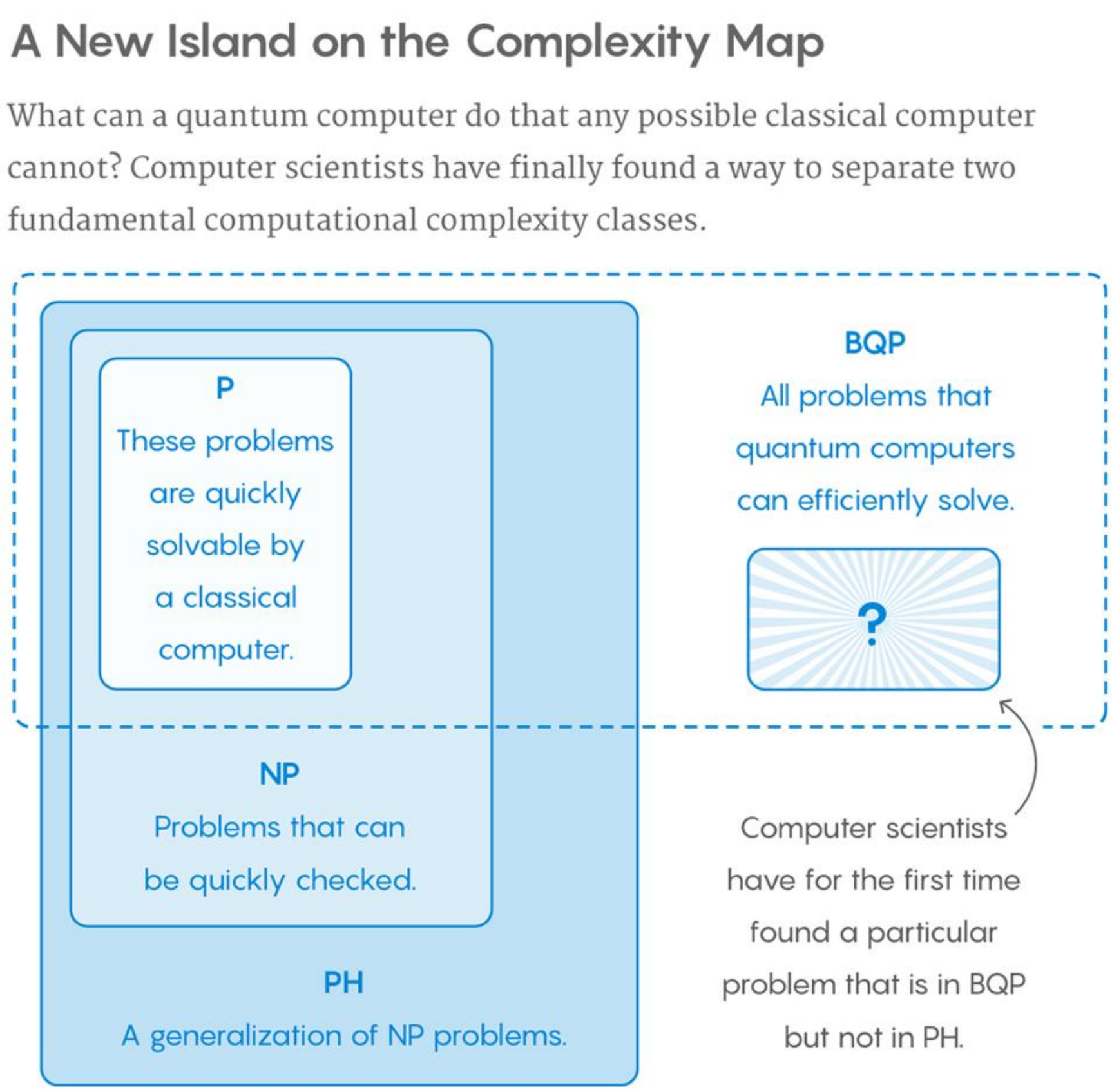

*Figure 601: how BQP relates to the P and NP complexity classes. Source: Finally, a Problem That Only Quantum Computers Will Ever Be Able to Solve by Kevin Hartnett, 2018.*

Therefore, these algorithms have a polynomial resolution time on quantum computers which remains exponential on their equivalent crafted for classical computers.

[1714] See Finally, a Problem That Only Quantum Computers Will Ever Be Able to Solve by Kevin Hartnett, 2018, referring to Oracle Separation of BQP and PH by Ran Raz and Avishay Tal, May 2018 (22 pages), presented in the Electronic Colloquium on Computational Complexity conference. This is the source of the illustration on this page.

[1715] See The Complexity of NISQ by Sitan Chen, Jordan Cotler, Hsin-Yuan Huang, Jerry Li, October 2022 (52 pages).

By the way, what about a possible complexity difference for problems manageable with universal gate quantum accelerators vs. quantum annealing accelerators? According to several researchers, there is no difference[1716]. Various theorems show that a problem that can be solved with universal quantum gates can also be solved with a D-Wave quantum annealing architecture and vice versa with only a polynomial time difference.

**QMA** (for Quantum Merlin Arthur) defines a class of problems that is verifiable in polynomial time on a quantum computer with a probability greater than 2/3. It is the quantum analog of the "traditional" NP complexity class. The QMA class contains the classes P, BQP and NP[1717]. Like the NP class, the QMA class has two derived subclasses, QMA Complete and QMA Hard. In practice, these are difficult problems to solve with quantum computers. Unfortunately, the literature on the subject does not describe its nature or provide examples. This is a pity for those who appreciate a practical sense of things!

**QCMA** is a hybrid problems class situated between QMA and NP. The proof is provided in classical polynomial time, but the resolution is at the QMA level and is performed in a quantum way.

Many publications point out the limitations of quantum algorithms and computers. A BQP problem that is not in PH gives the advantage to quantum computing. But exponential intractable problems for which the improvement brought by quantum computing is only a square root of classical time do not change their exponential nature. This is what Scott Aaronson points out[1718]. Complete NP problems and beyond remain inaccessible to quantum computers. Brute force has limits that even quantum computing cannot overcome in theory! This partly explains the difficulty of creating efficient quantum algorithms.

Finally, **NEEXP** is a class of problems that requires a double exponential computation time for its verification. Recent work shows that a result can be verified with several quantum computers with entangled qubits. This does not however enable us to solve this type of problems[1719]!

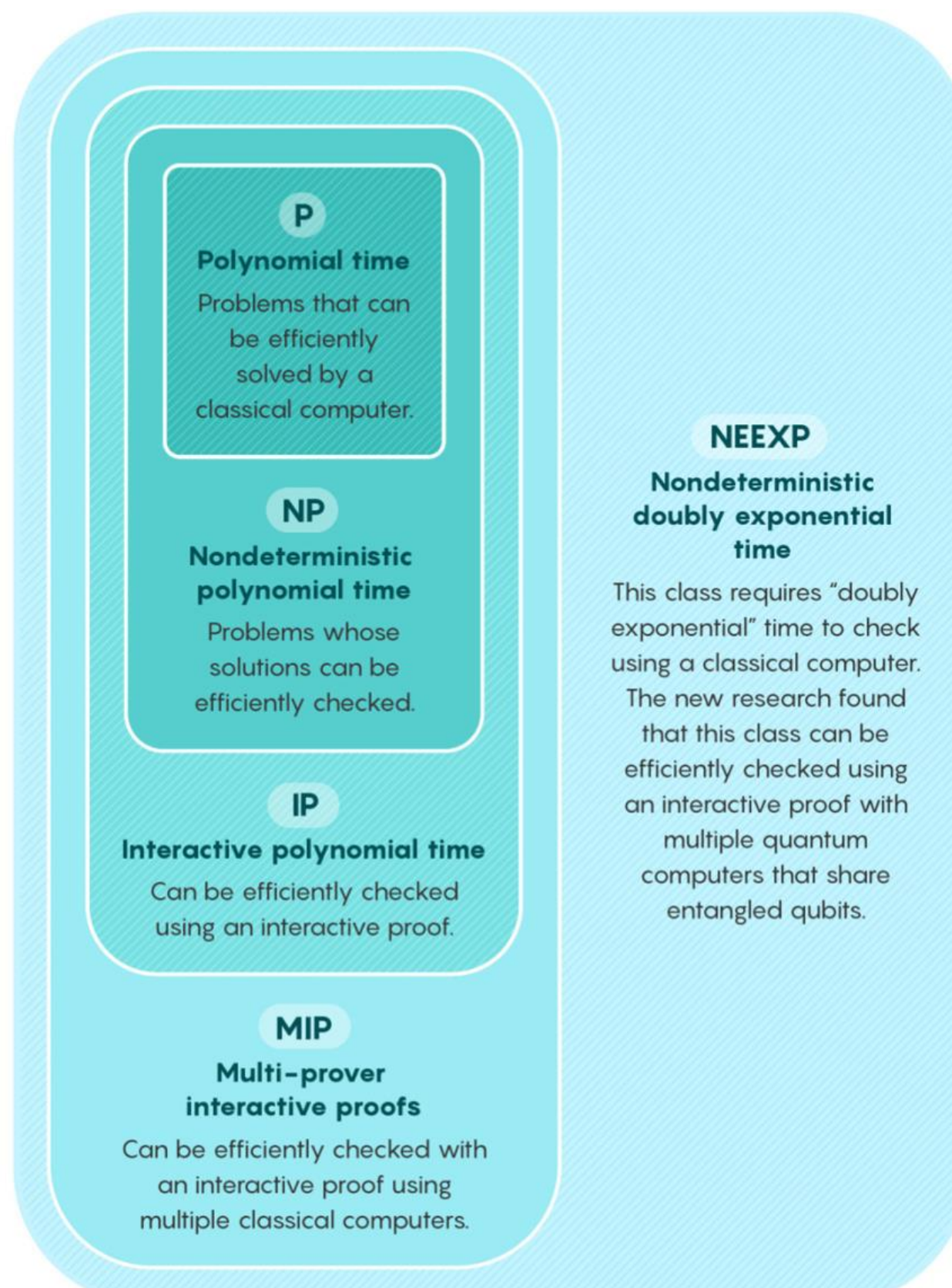

*Figure 602: the NEEXP complexity class. Source: [Computer Scientists Expand the Frontier of Verifiable Knowledge](), 2019.*

Some problems are undecidable, i.e., they cannot be solved by an algorithm, no matter how much time you have.

This is the case for the determination of the end of a Turing machine program.

---

[1716] See Adiabatic Quantum Computation is Equivalent to Standard Quantum Computation by Dorit Aharonov, Wim van Dam and Julia Kempe (from CNRS), 2008 (30 pages).

[1717] See QMA-complete problems by Adam Bookatz, 2013 (18 pages).

[1718] See The Limits of Quantum Computers (16 pages) and NP-complete Problems and Physical Reality (23 pages).

[1719] See NEEXP in MIP* by Anand Natarajan and John Wright, 2019 (122 pages) and Computer Scientists Expand the Frontier of Verifiable Knowledge, 2019.

In other words, there is no program that can determine whether any program written in a common programming language will stop or loop for an infinite amount of time.

However, in 2020, there was some progress with a demonstration that the classes MIP* and RE were identical [1720].

In the same order, **Rice's theorem** demonstrates that no non-trivial property of a program can be decided algorithmically. All this is to say that there is no automatic method to detect bugs in a program or to certify that it runs well. There are, however, formal methods of proof that can be used to certify the execution of specific programs. This involves the use of formal program specifications that serve as a reference for assessing how well a program is running. This is already widely used, without quantum, in industrial information technology and in critical systems such as aerospace.

## Quantum speedups

The chart in Figure 603 summarizes the theorical performance gains of some of the deterministic algorithms we have just seen. Complexity levels (exponential, polynomial, linear, ...) are generic.

The precise levels of complexity of each algorithm are roughly associated with these classes. Nlog(N) is the complexity of a classical Fourier transform and is nearly linear since N grows much faster than log(N) and $\log(N)^3$ is a log level complexity for the Shor algorithm and a QFT.

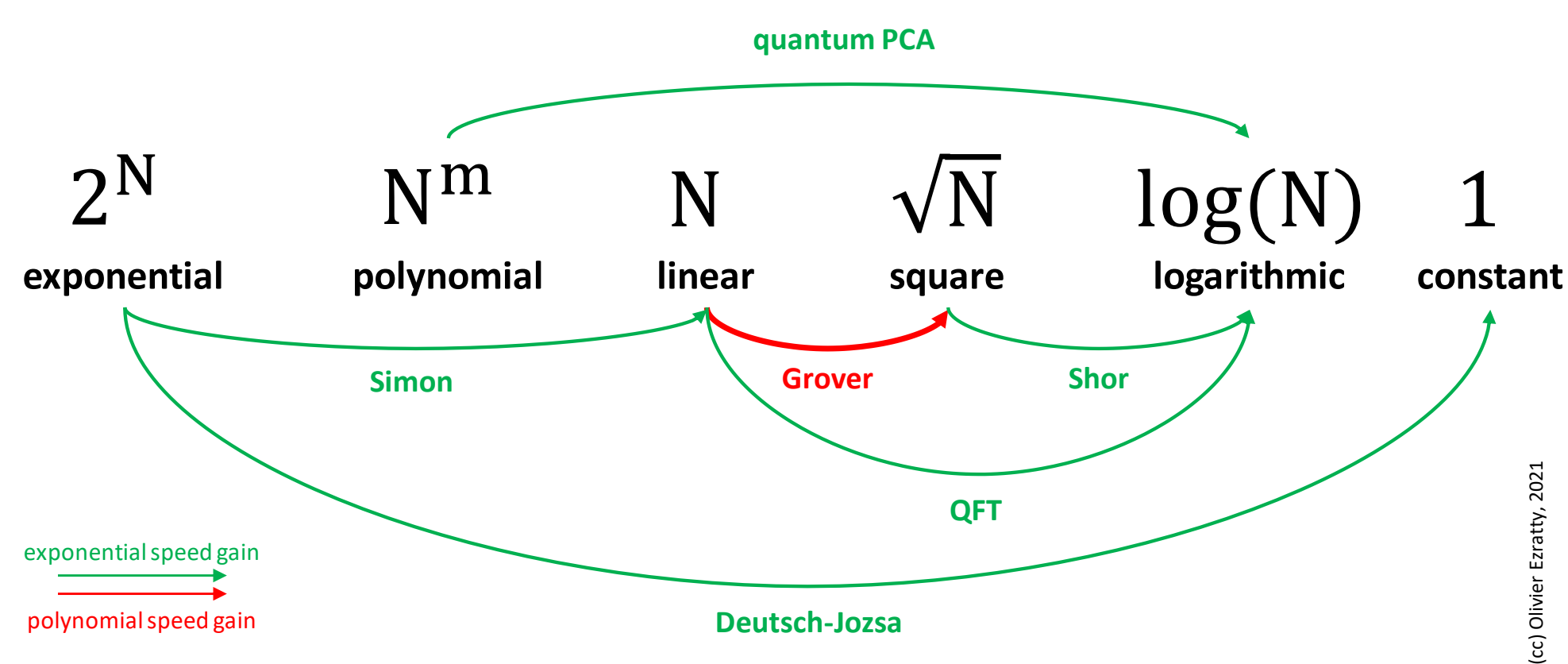

*Figure 603: how do O() compare for complexity classes in quantum computing. The arrows show how their classical and quantum solution compare. (cc) Olivier Ezratty, 2021.*

You can also visualize graphically the way computing time grows according to the big O() scale of problems below. A quantum algorithm's main goal is to move your problem's computation time from the red zone to the orange, or better, the yellow and green zones.

[1720] The MIP* class of problems that can be verified by quantum entanglement is equal to the RE class of problems that are no more difficult than the problems of program termination. See A quantum strategy could verify the solutions to unsolvable problems - in theory by Emily Conover, 2020 which refers to MIP*=RE by Zhengfeng Ji et al, January 2020 (165 pages) and seen in Mathematicians Are Studying Planet-Sized Computers With God-Like Powers by Mordechai Rorvig, 2020. See Landmark Computer Science Proof Cascades Through Physics and Math by Kevin Hartnett, March 2020.

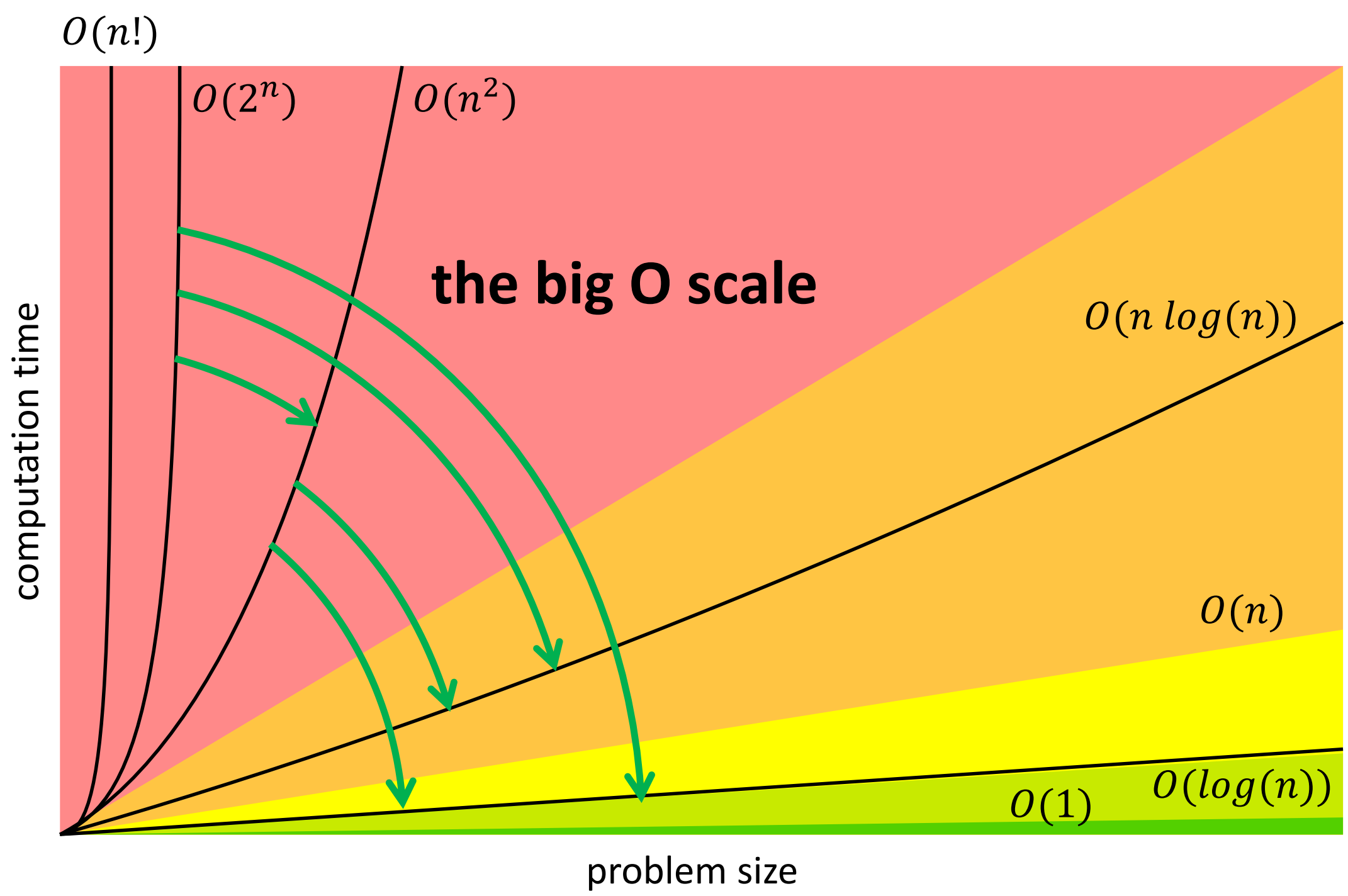

*Figure 604: another view of the big O() scale. Source: Wikipedia, reformatted.*

An exponential gain is also obtained when we move from N or √N to log(N). A QFT thus generates a theoretical exponential gain. The time scales are more meaningful in the table from Figure 605.

| Complexité | $n$ | $n \log_2 n$ | $n^2$ | $n^3$ | $1.5^n$ | $2^n$ | $n!$ |
|---|---|---|---|---|---|---|---|
| $n = 10$ | $< 1$ s | $< 1$ s | $< 1$ s | $< 1$ s | $< 1$ s | $< 1$ s | 4 s |
| $n = 30$ | $< 1$ s | $< 1$ s | $< 1$ s | $< 1$ s | $< 1$ s | 18 min | $10^{25}$ ans |
| $n = 50$ | $< 1$ s | $< 1$ s | $< 1$ s | $< 1$ s | 11 min | 36 ans | $\infty$ |
| $n = 100$ | $< 1$ s | $< 1$ s | $< 1$ s | 1$s$ | 12, 9 ans | $10^{17}$ ans | $\infty$ |
| $n = 1000$ | $< 1$ s | $< 1$ s | 1s | 18 min | $\infty$ | $\infty$ | $\infty$ |
| $n = 10000$ | $< 1$ s | $< 1$ s | 2 min | 12 jours | $\infty$ | $\infty$ | $\infty$ |
| $n = 100000$ | $< 1$ s | 2 s | 3 heures | 32 ans | $\infty$ | $\infty$ | $\infty$ |
| $n = 1000000$ | 1s | 20s | 12 jours | 31, 710 ans | $\infty$ | $\infty$ | $\infty$ |

*Figure 605: complexity classes and times scales. Heures = hours. Jours = days. Ans = years. Source: Complexity in time, Ecole Polytechnique (25 pages).*

The ideal performance gain consists in traversing several complexity scales, and particularly for an exponential problem. In practice, the main algorithms skip one or two complexity classes, but not necessarily from the exponential problem class. But my scheme is misleading. N can also grow exponentially depending on the size of a problem. The classic example is Shor's algorithm.

The starting point is an N which is actually a RSA key size which itself is evaluated in power of 2. A 1024-bit key is $2^{1024}$. If we move from $2^{256}$ to $2^{1024}$, the growth of the key size is exponential. With Shor's algorithm, we get an exponential performance gain by going from a square root of $2^{1024}$ to $\log(2^{1024})$, that is to say 1024 (in log base 2)! So, the time scale moves from $2^{512}$ to 1024, which is a perfectly exponential gain. Deutsch-Jozsa's algorithm has the particularity of traversing all levels of this scale, from an exponential time to a fixed time. We have unfortunately seen that it has no known practical application.

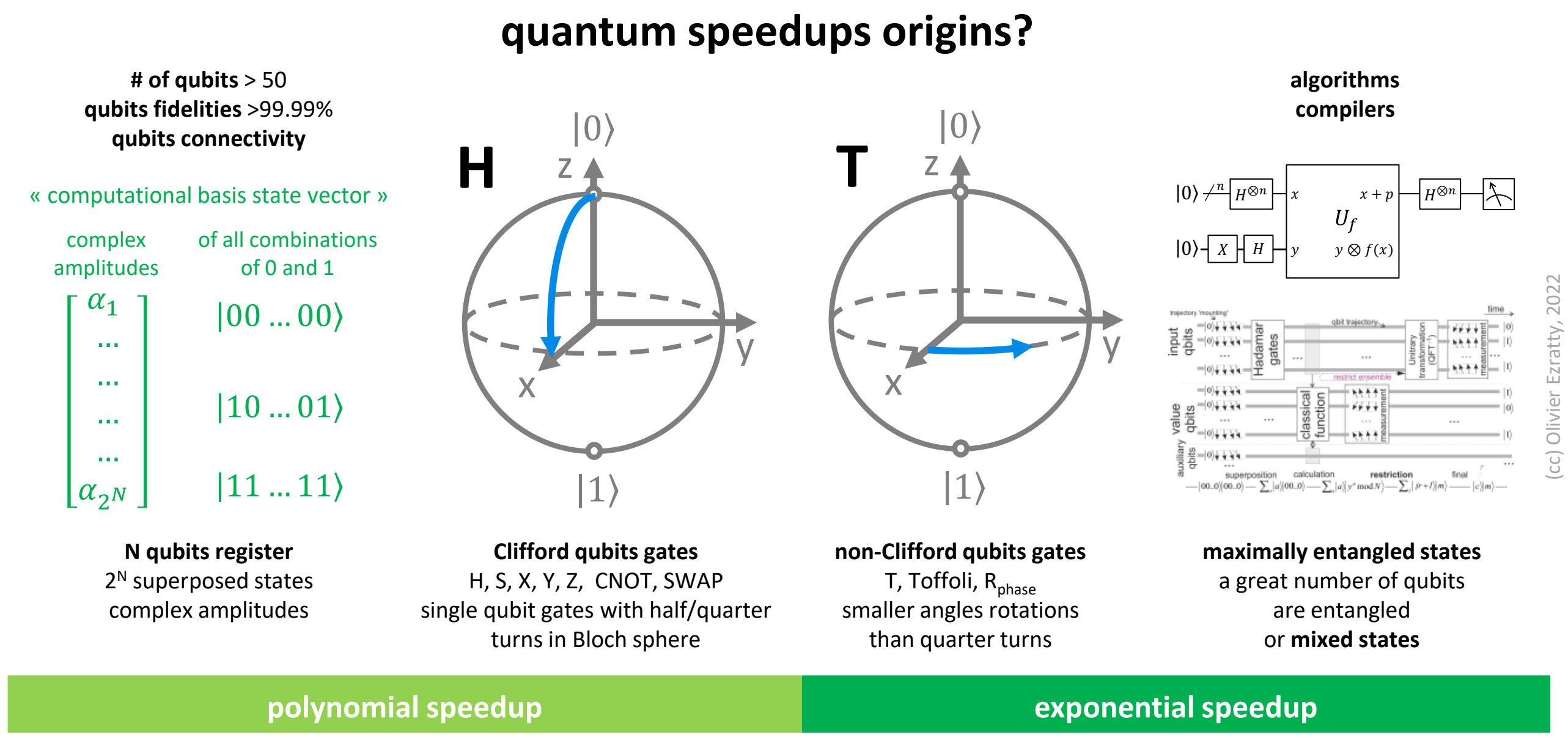

*Figure 606: the origins of quantum speedups are not obvious. It may be counter-intuitive but the exponential size of the computational vector space of N qubits doesn't explain any potential exponential gain in quantum computing. You need to have at least two other conditions: use non-Clifford qubit gates and have a N-qubit maximally entangled space. (cc) Olivier Ezratty, 2022.*

The speedup gain from an algorithm depends on the gates it's using. Shor's algorithm and any QFT based algorithm provide an exponential gain since it uses phase-controlled R gates. Grover's algorithm is providing a polynomial gain since it uses only Hadamard gates. But Deutsch-Jozsa's algorithm has an exponential gain although it is using only Clifford's group gates like the Hadamard gate.

Why so? Because it uses an Oracle function that may use non-Clifford's group gates[1721].

On top of using non-Clifford quantum gates, a quantum algorithm is also having an exponential speedup if it handles **maximally entangled states**[1722]. It means that there's a correlation of states between a maximum number of qubits in the register until the end of computing. If an algorithm handles islands of disconnected sets of qubits in the register, the speed-up will be constrained by the size of these islands.

The bigger the island, the bigger the Hilbert space managed by the algorithm. Let's explain this simply with a register of N qubits. If the algorithm is split into 4 different subspaces of N/4 qubits, you'll end-up managing a space with $4x2^{N/4}$ amplitudes, or $2^{2+N/4}$. That number is way smaller than $2^N$, starting with N=3. For N=50, the difference is between $2{,}3x10^5$ and $10^{15}$!

Studies on maximally entangled states and multipartite entanglement are numerous and accessible only to specialists[1723]. They deal with the theory, like with absolutely maximally entangled states (AME) that are multipartite quantum states and carry absolute maximum entanglement for all possible

[1721] See Focus beyond quadratic speedups for error-corrected quantum advantage, by Hartmut Neven et al , 2021 (11 pages) that also explain why quadratic speedups are not efficient due to the error correction overhead.

[1722] On maximally entangled states, see On the role of entanglement in quantum computational speed-up by Richard Jozsa and Noah Linden, 2003 (22 pages), Review on the study of entanglement in quantum computation speedup by ShengChao Ding and Zhi Jin, 2007 (6 pages) and Necessity of Superposition of Macroscopically Distinct States for Quantum Computational Speedup by Akira Shimizu et al, University of Tokyo and NTT, 2013 (16 pages). It refers to an indice p defined in Macroscopic entanglement of many-magnon states by Tomoyuki Morimae et al, 2005 (12 pages).

[1723] See A brief introduction to multipartite entanglement by Ingemar Bengtsson and Karol Zyczkowski, 2016 (38 pages).

subsystem partitions[1724], with how much entanglement is required for quantum algorithms[1725] and with how it can be checked with quantum measurement[1726].

As of 2022, the largest set of entangled quantum object was achieved with 3,000 atoms, but not for doing any computing and without a characterized quality of their entanglement[1727].

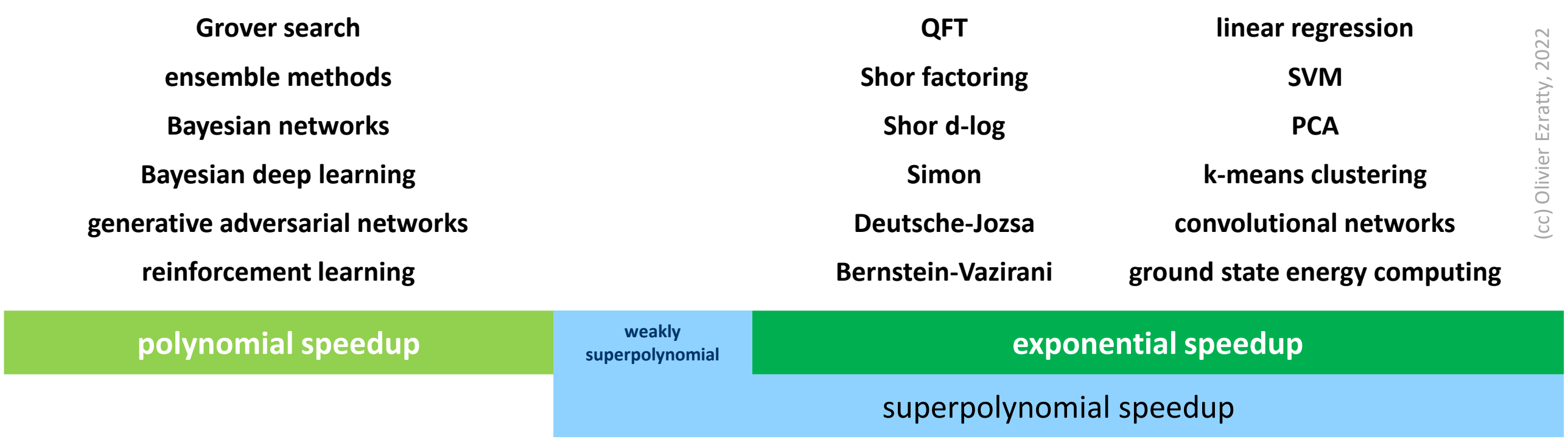

*Figure 607: polynomial, superpolynomial and exponential speedups and their corresponding most common quantum algorithms. (cc) Olivier Ezratty, 2022.*

We should here introduce the notion of superpolynomial speedup. A superpolynomial time is not bounded above by any polynomial. An exponential speedup is naturally superpolynomial but some superpolynomial speedups are not exponential, and said "weakly superpolynomial". Most common quantum algorithms showcase either a polynomial or an exponential speedup (*aka* strong superpolynomial speedup).

Other exponential speedups have been found for various algebraic algorithms like **estimating Gauss sums**[1728], **approximating Jones polynomial** that is not based on a QFT and still brings some exponential speedup, with some applications in topological quantum computing[1729] or counting solutions of **finite field equations**[1730].

It is also necessary to boil in the fact that the complexity of some problems can be addressed on conventional computers with probabilistic or heuristic approaches that also allow a significant reduction of the computing time of exponential problems. Practically, when moving from this kind of solution to a quantum algorithm, we replace one probabilistic approach with another since quantum computing is also highly probabilistic and prone to many computing errors.

All in all, quantum algorithms are interesting, but they are not always the only solution to cleverly solve a complex problem. This is amplified by the emulation going on between algorithms designers. Each and every new quantum performance challenge the classical supercomputers algorithms designers to improve the performance of their own tools.

---

[1724] See the thesis Symmetry and Classification of Multipartite Entangled States by Adam Burchardt, September 2021 (126 pages).

[1725] See How Much Entanglement Do Quantum Optimization Algorithms Require? by Yanzhu Chen, Linghua Zhu, Chenxu Liu, Nicholas J. Mayhall, Edwin Barnes and Sophia E. Economou, May 2022 (12 pages).

[1726] See Quantifying multiparticle entanglement with randomized measurements by Sophia Ohnemus, Heinz-Peter Breuer and Andreas Ketterer, July 2022 (17 pages) and Scalable estimation of pure multi-qubit states by Luciano Pereira, Leonardo Zambrano and Aldo Delgado, npj, May 2022 (12 pages).

[1727] See Entanglement with negative Wigner function of almost 3,000 atoms heralded by one photon by Robert McConnell, Hao Zhang, Jiazhong Hu, Senka Ćuk and Vladan Vuletić, Nature, March 2015 (11 pages).

[1728] See Efficient Quantum Algorithms for Estimating Gauss Sums by Wim van Dam and Gadiel Seroussi, 2008 (11 pages).

[1729] See A Polynomial Quantum Algorithm for Approximating the Jones Polynomial by Dorit Aharonov, Vaughan Jones and Zeph Landau, 2006 (19 pages).

[1730] See Quantum computing and polynomial equations over the finite field by Christopher M. Dawson et al, 2004 (7 pages).

This is what Toshiba did in 2019 with a classic optimization algorithm that was 10 times more powerful than the state of the art. That was fine even though a linear x10 gain is still not an exponential progress[1731].

This also explains why some are proposing to create a notion of "strong quantum speedup" using complexity classes instead of algorithms speeds[1732]. This being said, even if a polynomial gain is considered as a minor gain in complexity theories, it can still have a non-negligible practical value and make quantum computing attractive, without going through the Holy Grail of some fancy exponential acceleration.

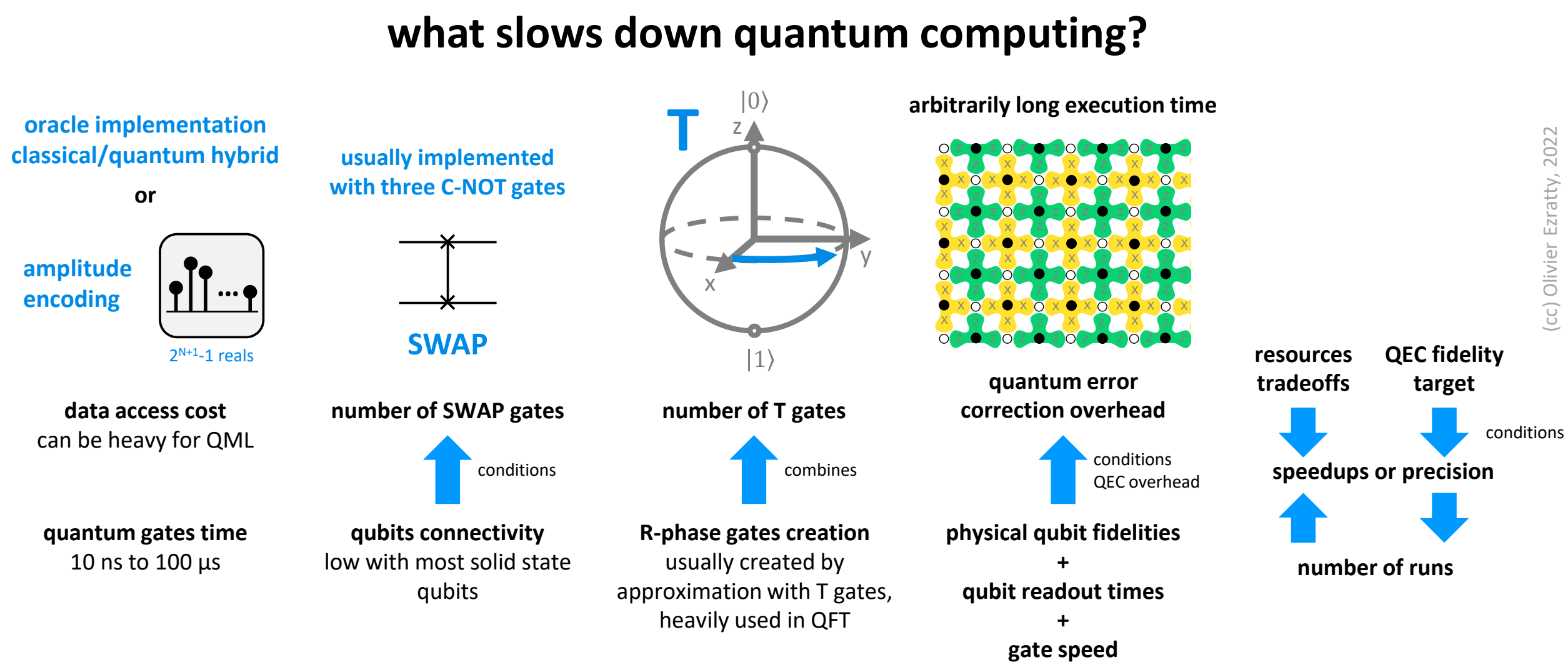

*Figure 608: quantum computing speedup must also include faces sources of slowdowns, which have to be known by algorithms developers. (cc) Olivier Ezratty, 2022.*

But the devils in the details must be cared about[1733] :

- First, the quantum speedup will be affected by the number of times the algorithm must be run. It's considered in some of the speedups like with Grover's algorithms but not always. These repetitions are also named shot count or shots. It depends on the problem, the number of qubits and the algorithm output (integers, real numbers). With Google's supremacy and its 53 qubits, it was 3 million shots. With IBM Quantum System One using from 5 to 28 qubits, the typical proposed shots number ranges from 1000 to 8000 shots.
- The quantum advantage is also depending on the number of qubits in your register that are put in superposition using Hadamard gates. Other qubits might be used elsewhere in the algorithm such as ancillas.
- Data preparation must also be handled, which is of particular importance for quantum machine learning algorithms[1734].
- The same should be said of oracle-based algorithms, particularly when they rely on some classical data access.

---

[1731] See Toshiba Promises Quantum-Like Advantage on Standard Hardware by Tiffany Trader, 2020, which references Combinatorial optimization by simulating adiabatic bifurcations in nonlinear Hamiltonian systems by Hayato Goto et al, April 2019 (9 pages).

[1732] See Measures of quantum computing speedup by Anargyros Papageorgiou and Joseph F. Traub, 2013 (4 pages). Proposing that "strong quantum speedup" is measured by a ratio between quantum complexity class and classical complexity class for a particular problem instead of comparing just cost of best-in-class classical algorithm vs cost of quantum algorithm "quantum speedup".

[1733] This is well explained in What Is the Quantum Advantage of Your Quantum Algorithm? by Jack Krupansky, February 2020.

[1734] See Information-theoretic bounds on quantum advantage in machine learning by Hsin-Yuan Huang, Richard Kueng and John Preskill, April 2021 (34 pages) which describes the conditions when QML algorithms provide a real speedup.

- Then, with large scale quantum computing, quantum error correction will add some additional burden. Using concatenation codes, you may lose polynomially on your speedup. Meaning that an initial polynomial speedup may be lost at this stage[1735].

So, as mentioned before, appreciating the real speedup of any quantum algorithm requires adopting an end-to-end approach considering all the parameters of the quantum algorithm execution time.

I would summarize the kinds of benefit coming from quantum computing over time like this:

**Better**. Some algorithms running on NISQ systems may bring a quality advantage in the solution they provide, like with quantum machine learning or various optimization tasks. This is due to the ability of quantum systems to better explore the problem computational space. For a QML problem, the better is expressed in precision %[1736].

**Faster**. We may then create solutions that are running faster than their best equivalent on classical computers. This may happen between the NISQ and the FTQC/LSQ era.

**Beyond**. In the longer term, and pending unlocking many showstoppers, large scale quantum computing with over 100 logical qubits may then be able to solve problems that are entirely not accessible by classical supercomputers, like in the quantum physics simulation realm.

### Quantum algorithms key takeaways

- Quantum algorithms have been created since the early 1990s, over ten years before any quantum computer was working out of a research laboratory.
- Quantum algorithms use very different concepts than classical programming, even including artificial intelligence development tools or object oriented programming. It's based on the manipulation of large matrices and using interferences.
- The main algorithms classes are oracle based and search algorithms, optimization algorithms, quantum physics simulation algorithms and quantum machine learning algorithms.
- A quantum algorithm is interesting if it provides some quantum speedup compared to their equivalent best-in-class classical version, including those that are heuristics based. These problems are said to be intractable on classical hardware. Most of the time, quantum speedups are theorical and do not incorporate the costs of quantum error corrections and of creating non-Clifford quantum gates. These gates are implementing small phase changes and are used in quantum Fourier transforms and implemented in many other algorithms. A quantum speedup that is not exponential is highly questionable. All of this requires some understanding of complexity classes like P, NP and BQP.
- Another key aspect of quantum algorithms is data loading and/or preparation. It is often overlooked and can have a significant time cost, on top of frequently requiring some form of not-yet-available quantum memory hardware. As a consequence, quantum computing is not adequate for big-data computations.
- Gate-based computers, quantum annealers and quantum simulators can exploit hybrid algorithms, combining a classical (preparation) part and a quantum part interacting with each other. This is particularly true with quantum machine learning.
- Quantum inspired algorithms are running on classical computers and are using some form of quantum mathematical models like interferences and signals decomposition (Fourier series/transforms).

[1735] See Focus beyond Quadratic Speedups for Error-Corrected Quantum Advantage by Ryan Babbush, Craig Gidney, Hartmut Neven et al, March 2021 (11 pages). Quadratic speedups requires more than 1M physical qubits!

[1736] See The Complexity of NISQ by Sitan Chen, Jordan Cotler, Hsin-Yuan Huang, Jerry Li, October 2022 (52 pages) which casts serious doubts on real quantum speedups that could be obtained with NISQ QPUs. In that case, we are left with better quality results as a potential quantum advantage.

# Quantum software development tools

We now need to explore quantum computing software to implement the various algorithms we've just uncovered. It is a completely new world with very different paradigms.

Quantum algorithms still require programming, programming languages and development environments.

As shown in Figure 609, quantum software is organized in layers with, starting from the bottom, the physical qubits followed by low-level machine language to drive them at the physical level (microwaves length and frequencies, readouts, etc).

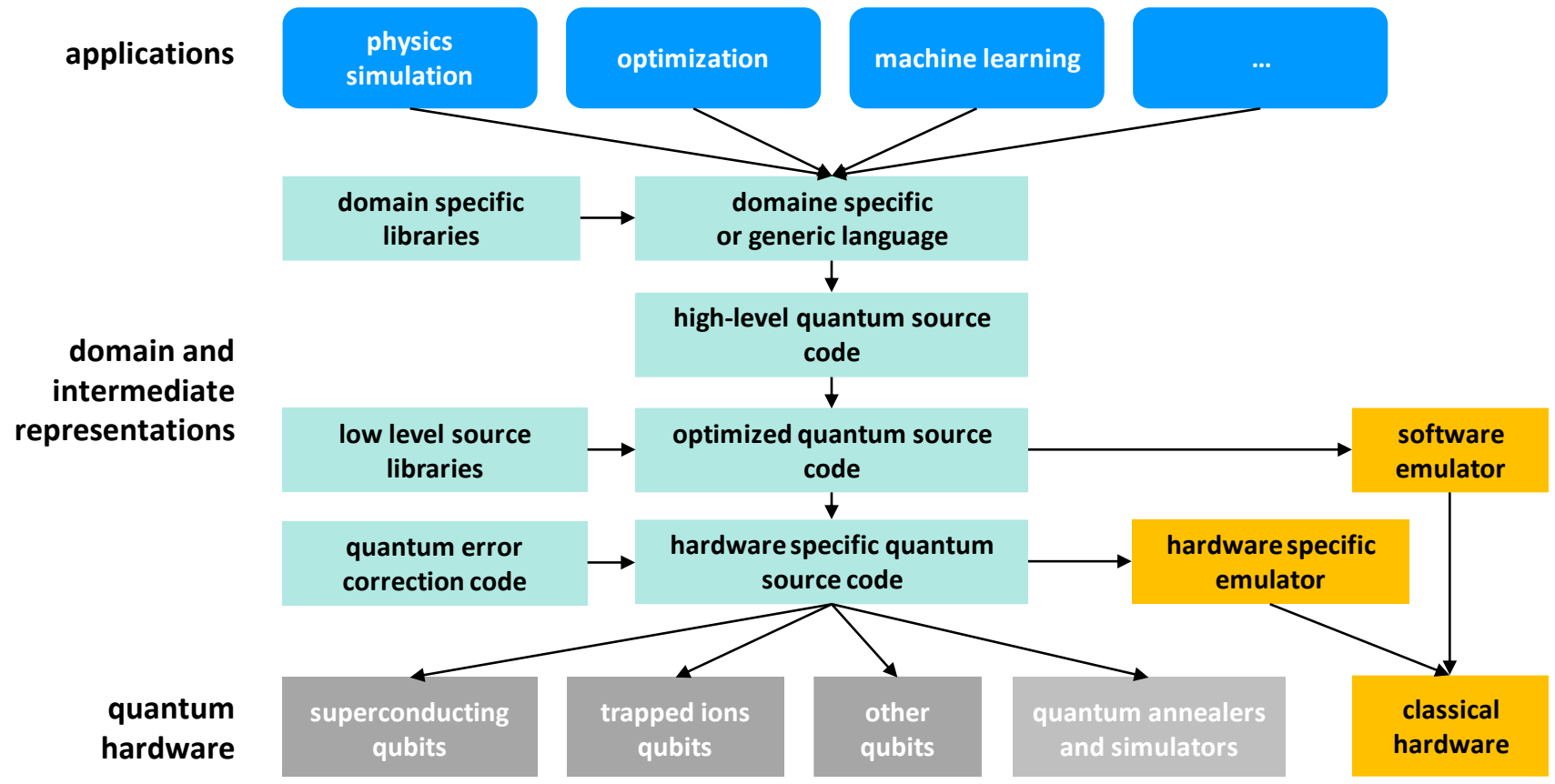

Figure 609: classification of quantum software engineering tools. (cc) Olivier Ezratty, 2021.

Next comes high level quantum source code which is in fact a kind of macro-assembler, able to take advantage of function libraries with ready-to-use algorithms (quantum Fourier transforms, phase estimates, etc.) and finally, high-level languages or libraries tailored to specific business needs.

In the lower layers between machine language and macro-assembler are functions for converting quantum gates into a set of universal quantum gates supported by the quantum hardware as well as error correction code systems that may require the execution of a large number of quantum gates.

A quantum compiler also implements many optimizations by for example removing quantum gate sequences that do not change the state of a qubit, such as two consecutive Hadamard or X (NOT) gates.

It also arranges these quantum gates to minimize the number of quantum gate steps in the solution. Quantum software architectures are generally hybrid. They manage side by side the execution of classical and quantum code, as shown in this diagram from Rigetti.

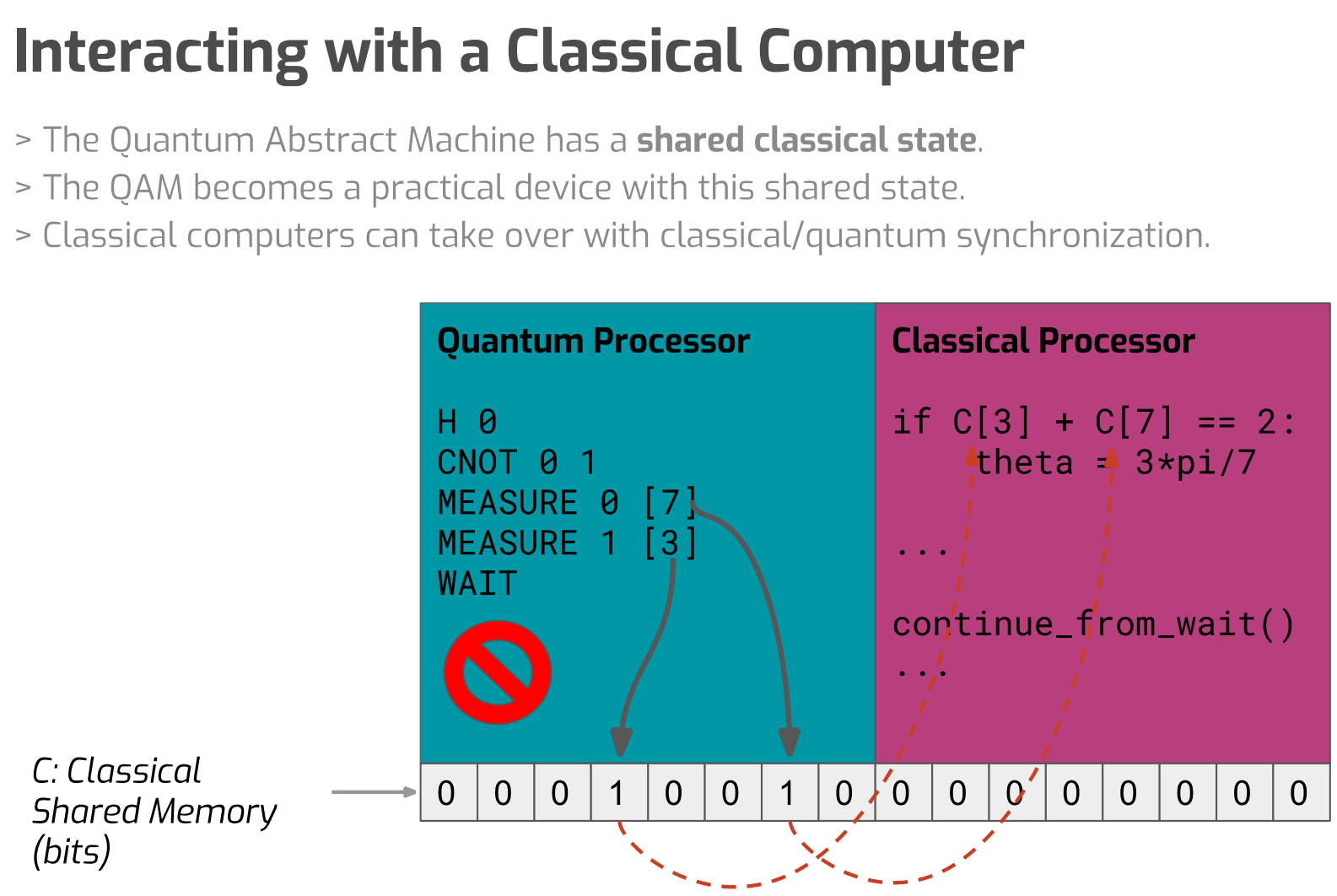

Figure 610: Source: [Quantum Cloud Computing](#) by Johannes Otterbach, January 2018 (105 slides).

At the very least, the classical computer is used to control the execution of quantum algorithms, if only to trigger the quantum gates at the right time, sequentially.

## Development tool classes

We can identify some major classes of tools for creating quantum software: graphic programming tools, scripting languages, intermediate languages, machine languages, compilers and application libraries.

### Graphical programming tools

They allow to visually define the sequence of quantum gates to create algorithms and execute it on quantum accelerators.

These tools can also emulate, run and visualize the status of qubits when their number is reasonable with various methods: Bloch sphere, register state and density matrix.

One example of such tools is the IBM Quantum Experience IBM Composer that is available online since 2016. Code can be executed on an IBM emulator or on one of the many IBM quantum systems available in the cloud, and for free up to 15 operational qubits.

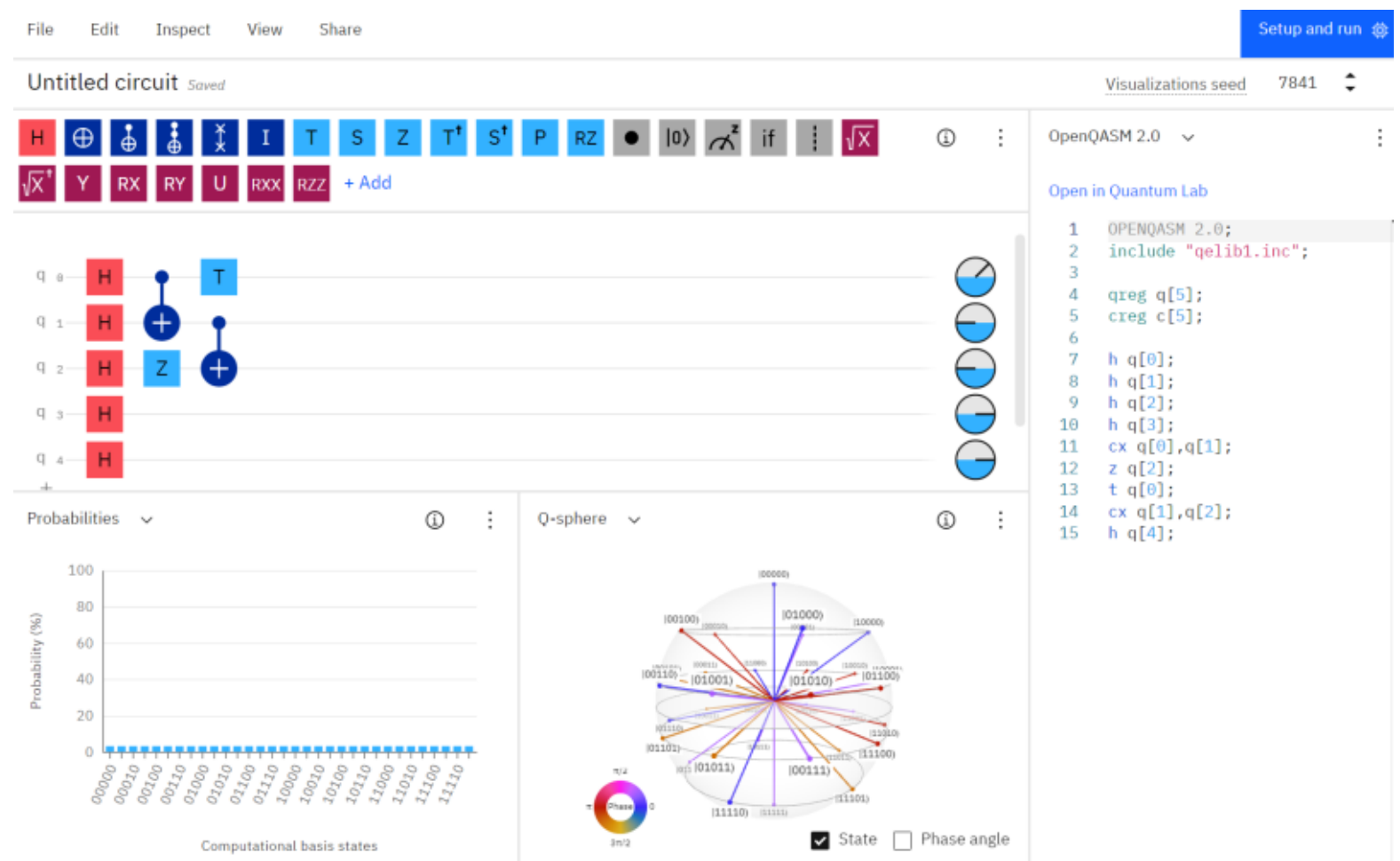

*Figure 611: IBM Quantum Experience visual interface. Source : IBM.*

There are also graphical simulators of qubits that can be used to understand how to chain quantum gates on a few qubits and visualize the result visually. The most open one is the open source tool Quirk which can simulate up to 16 qubits. It works online and can be downloaded to run on your own computer locally. Below is an example of a quantum Fourier transform performed in Quirk. It was developed by **Craig Gidney**, now a Google engineer specialized in algorithms and error correction codes.

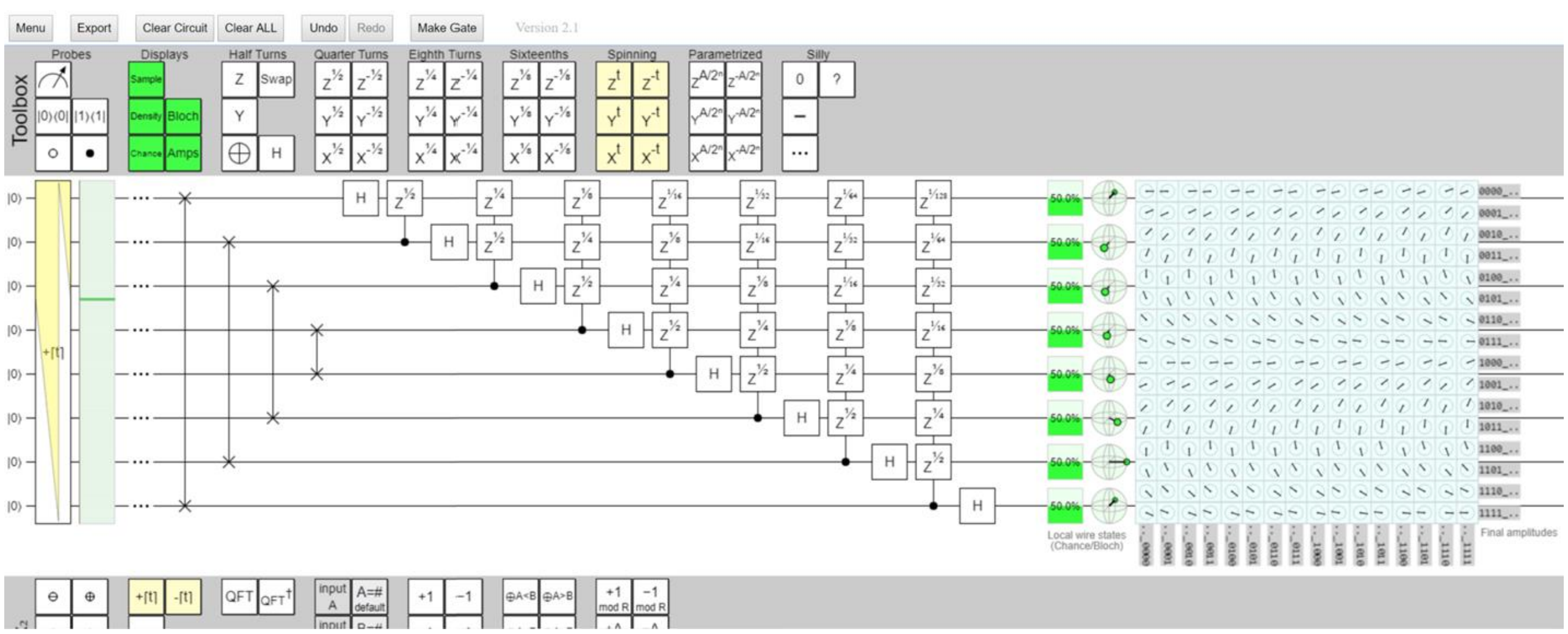

*Figure 612: Quirk's visual open sourced quantum programming tool, working in any browser. Source: Quirk Algassert.*

Finally, let's pay some attention to **ZX-calculus**, a graphical programming language that uses topological composition rules. It was created in 2008 by Bob Coecke and Ross Duncan[1737]. It visualizes the modifications made to a set of qubits and is based on transformations applicable to the geometric representation of quantum gates that simplify models.

[1737] See Interacting Quantum Observables: Categorical Algebra and Diagrammatics by Bob Coecke and Ross Duncan, 2009 (80 pages).

It is particularly useful for programming a quantum computer in MBQC (measurement base quantum computing) and for visually model error correction codes. It can also help optimize quantum code compiling[1738].

Contributors to the ZX-Calculus work include researchers from Loria under the responsibility of Simon Perdrix, a research laboratory located in Nancy and Dominic Horsman then at UGA LIG in Grenoble and now at Oxford University[1739].

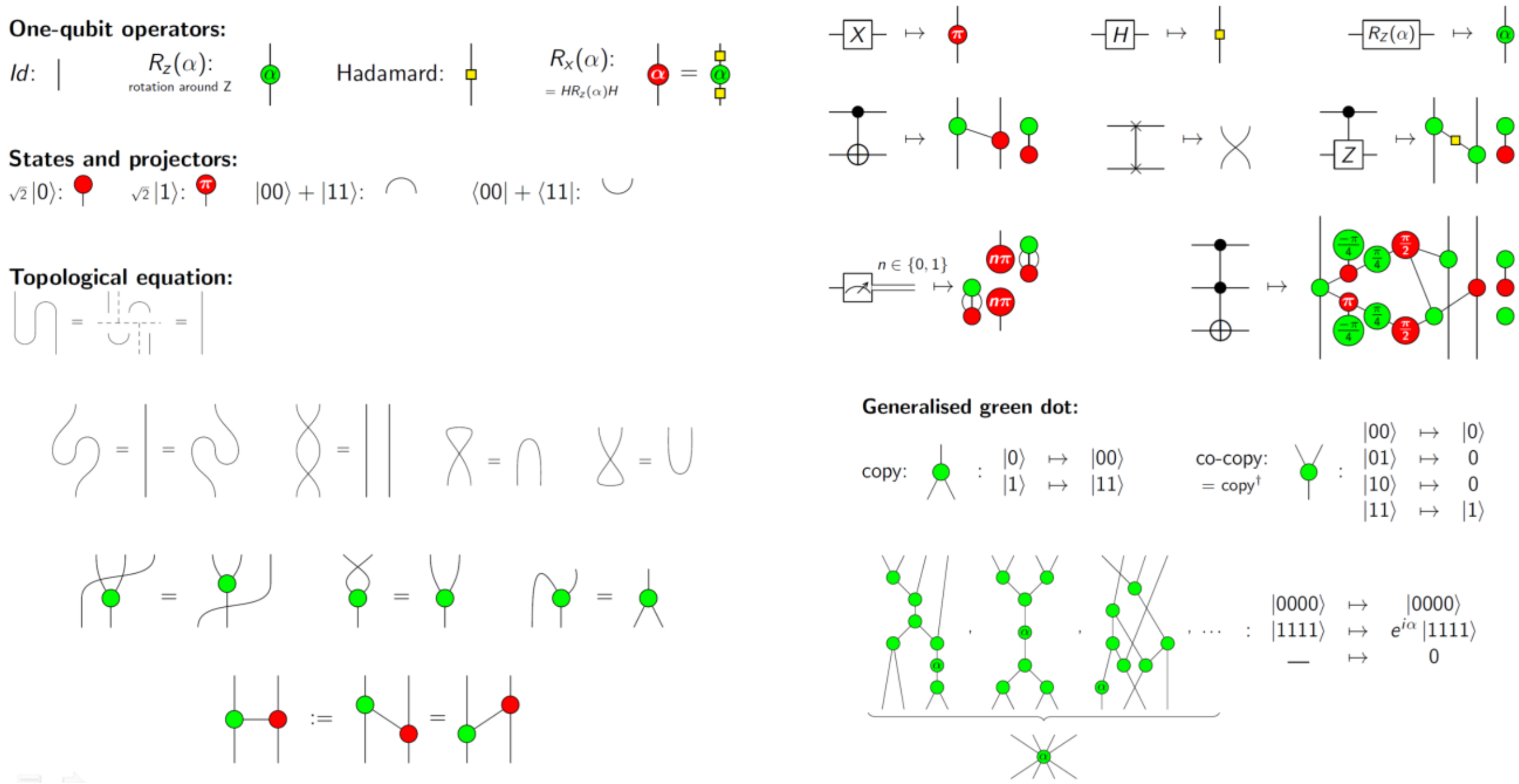

*Figure 613: ZX calculus graphical language key operations. Source: Completeness of the ZX-Calculus by Renaud Vilmart, 2018 (123 slides).*

ZX has its own zoo of extensions:

- **PyZX** is a Python tool created in 2019 that implements ZX-Calculus principles for the creation, visualization, and automated rewriting of large-scale quantum circuits.
- **SZC-calculus** (Scalable ZX-calculus) is a high-level extension of ZX-calculus for the design and verification of quantum computations with qubits registers. Among other things, it can be used to describe graph states used in MBQC and error correcting codes[1740].
- **ZXH-calculus** is a graphical language based on ZX-calculus helps modelize many-body states[1741].
- **ZW-calculus** is a variant that allows better entanglement modeling and **ZH-calculus** is used for the generalization of MBQC programming models thanks to the addition of a box implementing the Hadamard gate and an easier integration of Toffoli gates in its diagrams. It is associated with a development tool, **Quantomatic**, created by Aleks Kissinger and Vladimir Zamdzhiev then at of Oxford University and now at Inria in France [1742].

---

[1738] See Effective Compression of Quantum Braided Circuits Aided by ZX-Calculus by Michael Hank et al, November 2020 (13 pages).

[1739] See Completeness of the ZX-Calculus by Renaud Vilmart, 2018 (123 slides) and Completeness of the ZX-Calculus by Emmanuel Jeandel, Simon Perdrix and Renaud Vilmart (73 pages) which explains them.

[1740] See SZX-calculus: Scalable Graphical Quantum Reasoning by Titouan Carette, Dominic Horsman and Simon Perdrix, April 2019 (29 pages).

[1741] See AKLT-states as ZX-diagrams: diagrammatic reasoning for quantum states by Richard D. P. East et al, December 2020 (22 pages).

[1742] See Quantomatic: A Proof Assistant for Diagrammatic Reasoning, 2015 (11 pages).

They organize their own international conference, the **QPL** (Quantum Physics and Logic) but it covers broader topics than ZX calculus.[1743]

- **DisCoPy** (Distributional Compositional Python) is an open source toolbox for computing with string diagrams and functors[1744]. Its diagram data structure allows to encode various kinds of quantum processes, with functors for classical simulation and optimization, as well as compilation and evaluation on quantum hardware. It supports ZX calculus and its variants and linear optical quantum computing. Its photonics module that was under development as of May 2022 will allow to build linear optical circuits and interface with the quantum simulator **Perceval** from Quandela. DisCoPy encodes arbitrary string diagrams and interprets these for computation on various classical (NumPy, JAX, TensorFlow, PyTorch, Sympy) and quantum systems (Perceval, PyZX). DisCoPy is of particular use with quantum natural language processing (QNLP) and quantum machine translation applications[1745].

**Scripting languages**

They are used to program a quantum algorithm in text mode. These tools allow to associate classical programming with the chaining of quantum functions conditioned by the state of variables in classical memory.

There are two main types of quantum scripting languages: imperative and functional languages:

- **Imperative languages** are procedural programming languages where step-by-step algorithms are described. They include the usual languages such as C, C++, PHP or Java.
- **Functional languages** are used by defining various functions that are called on an ad-hoc basis by the program. The loops (for, while) are replaced by recursive functions and there are no modifiable variables. It uses high-level abstract data types that are manipulated by functions. It creates more concise code.

Many traditional programming languages can be used for imperative or functional programming, especially if they use function pointers or support event-driven logic.

To some extent, JavaScript and jQuery can be used as functional languages via their call-back functions. This is also the case with C++.

With quantum computer vendors such as IBM or Rigetti, two types of languages are sometimes offered: an intermediate language (Quil at Rigetti, openQASM with IBM) and a higher-level language in the form of extensions to the Python programming language (pyQuil at Rigetti, IBM Qiskit). A conversion tool converts the second one into the first one.

The most common programming language used with quantum libraries is Python. It provides language constructs, data types, for-loops code branching, modularity and classes. It can help build repetitive code structures, which would be harder than with graphical circuit design. It can also be used to create automatic testing tools.

---

[1743] See SZX-calculus: Scalable Graphical Quantum Reasoning by Titouan Carette, Dominic Horsman and Simon Perdrix, April 2019 (29 pages).

[1744] See DisCoPy for the quantum computer scientist by Alexis Toumi et al, CQC and University of Oxford, May 2022 (6 pages). A functor is "*a design pattern inspired by the definition from category theory, that allows for a generic type to apply a function inside without changing the structure of the generic type*" (Wikipedia).

[1745] See Towards Machine Translation with Quantum Computers by Irene Vicente Nieto, 2021 (48 pages).

Table 1: A selection of some quantum programming languages.

| Name | Style | Notes |
|---|---|---|
| QCL | Imperative | Has classical sublanguage, multiple high-level programming features. |
| qGCL | Imperative | Emphasis on algorithm derivation and verification. |
| LanQ | Imperative | Full operational semantics, proven type soundness. |
| Quipper | Functional | Focus on scalability, plans to include linear types for static checks (currently done at run-time). |
| QPL | Functional | Statically typed, denotational semantics in terms of CPOs of superoperators. |
| QML | Functional | Linearly typed, focused on weakening - not contraction. Quantum control and quantum data. |
| Qumin | Functional | Two sublanguages (untyped and linearly typed). Focus on ease of use and clean, functional style of programming. |

*Figure 614: imperative and functional quantum programming languages. Source: Qumin, a minimalist quantum programming language, 2017 (34 pages).*

### Machine languages

These are the lowest level programming languages of the quantum computer, which program the initialization of qubits and drive the physical signals sent to the qubits to implement universal gates and qubit readouts. They are generally specific to each type of quantum computer, or even to each quantum computer. Most quantum algorithms developers never use this type of low-level language.

### Compilers

Quantum compilers translate your code into the low-level sequences of qubit electronic controls for the target quantum accelerator expressed in a sort of machine language. It can also integrate quantum error correction codes (QEC). These compilers transform the program gates into universal physical gates operated by the quantum computer and then into control pulses of the qubits.

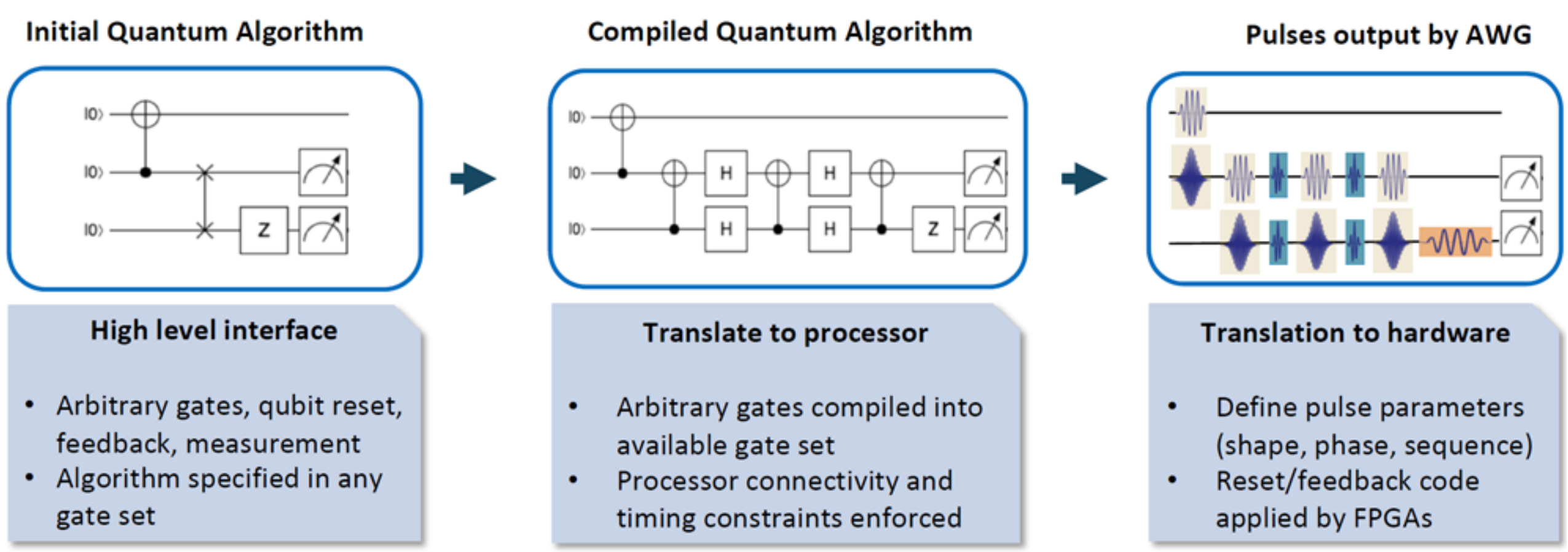

*Figure 615: the various roles of a quantum code compiler, first to translate high-level gate codes into primitive gates supported by the quantum processor, and then to turn these gates into low-level electronic controls driving qubit gates and readout. Source: How about quantum computing? by Bert de Jong, June 2019 (47 slides).*

It will also compute the gates activation times and verify that the accumulation of these activation times is within the range of the target accelerator's qubits coherence time.

Compilers can also carry out optimizations specific to certain types of algorithms and also hardware specifics[1746]. Many advances can reduce the number of gates used and create shallower circuits. It can be done with qubit mapping which optimizes SWAP insertion techniques that are used heavily when qubits connectivity is poor, like with converting SWAP gates into 3 CNOTs and removing redundant combinations of CNOTs[1747], with optimizing CNOT circuits generation[1748], optimizing the balance between CNOTs and T gates, particularly for fault-tolerant schemes that are very costly with T gates (it's about minimizing the T-count, the number of T gates used in the algorithm or its constituent parts like with quantum Fourier transforms)[1749] and also recycling unused qubits with automated uncompute set of gates[1750]. There are even strategies available to optimize the compilation of a VQE algorithm for NISQ QPUs[1751] and to distribute computing across multiple quantum computers[1752].

Like Atos' aQASM, these compilation tools can be cross-platform and support different gate-based quantum computer architectures. Quantum programming languages are generally able to combine classical procedural programming with quantum registers and gates programming. They allow parallel management of classical memory with quantum registers.

All these programming tools come from either research labs or from quantum computer vendors such as IBM, Rigetti and D-Wave[1753].

### Application specific frameworks

On top of scripting languages used to program quantum computers, annealers or simulators sit a wealth of application specific frameworks aimed at solving particular problems. The bulk of these frameworks are proposed by the many software vendors we inventory in a special part of this book, starting page 726. They usually support various underlying hardware architectures from classical digital emulators to QPUs running on-premises or in the cloud, the best ones being able to support heterogeneous classical and quantum platforms for the test and implementation of hybrid algorithms.

---

[1746] This is the case of Partial Compilation of Variational Algorithms for Noisy Intermediate-Scale Quantum Machines by Pranav Gokhale et al, 2019 (13 pages) which deals with a two-pass compiler optimized for Variational Algorithms (VQE). See also Spacetime tradeoffs when optimizing large quantum computations by Craig Gidney (Google AI Quantum, USA), IQFA 2020, December 2020 (60mn) and slides. He is turning serialized circuits into parallelized ones. With T state distillation, he gains three orders of magnitude in fidelities. At last, see Full-stack quantum computing systems in the NISQ era: algorithm-driven and hardware-aware compilation techniques by Medina Bandic et al, QuTech, April 2022 (6 pages) which summarizes well the full-stack architecture thinking influencing compilers designs.

[1747] See Architecture aware compilation of quantum circuits via lazy synthesis by Simon Martiel and Timothée Goubault de Brugière, Atos, December 2020 (31 pages), Not All SWAPs Have the Same Cost: A Case for Optimization-Aware Qubit Routing by Ji Liu et al, North Carolina State University, May 2022 (17 pages) and See Qubit Mapping Toward Quantum Advantage by Chin-Yi Cheng et al, October 2022 (14 pages).

[1748] And Decoding techniques applied to the compilation of CNOT circuits for NISQ architectures by Timothée Goubault de Brugière, Marc Baboulin, Benoît Valiron, Simon Martiel and Cyril Allouche, January 2022 (31 pages).

[1749] And Gaussian Elimination versus Greedy Methods for the Synthesis of Linear Reversible Circuits by Timothée Goubault de Brugière, Marc Baboulin, Benoît Valiron, Simon Martiel and Cyril Allouche, January 2022 (27 pages) and Reducing the Depth of Linear Reversible Quantum Circuits by Timothée Goubault de Brugière, Marc Baboulin, Benoît Valiron, Simon Martiel and Cyril Allouche, January 2022 (22 pages). On a way to reduce the T-count with a QFT: Halving the cost of quantum Fourier transform by Byeongyong Park et al, July 2022 (19 pages).

[1750] See Recycling qubits in near-term quantum computers by Galit Anikeeva et al, PRA, April2021 (11 pages).

[1751] See Policy Gradient Approach to Compilation of Variational Quantum Circuits by David A. Herrera-Martí, November 2021-September 2022 (17 pages).

[1752] See Qurzon: A Prototype for a Divide and Conquer Based Quantum Compiler by Turbasu Chatterjee et al, November 2021 (11 pages).

[1753] See this presentation which describes some of the tasks performed by quantum compilers: Opportunities and Challenges in Intermediate-Scale Quantum Computing by Fred Chong, 2018 (34 slides).

The most common proposed frameworks cover generic optimization problems (mostly using variational techniques mixing classical and quantum computing like QAOA and VQE/VQA), and others for chemical simulations and financial optimizations, which have been identified as the first go-to-markets for quantum computing so far.

Some application specific frameworks are also proposed by the large vertically integrated quantum computing vendors like IBM (with their Qiskit frameworks). Others can also be proposed by research labs like $Q^2$ Chemistry which comes from USTC in China[1754].

**Emulators**

Emulators are software and/or hardware tools that emulate the execution of quantum algorithms on classical computers. Their qubits emulation capacity is closely related to the amount of memory available and the emulation mode that is implemented[1755]. These emulators are also labelled "simulators" or "quantum classical simulators", which is creating some confusion with quantum simulators which are quantum systems simulating quantum physics systems.

Quantum code emulation serve multiple purposes: learning how to code, visualize a quantum algorithm internal data (which can't be done with a real quantum computer), develop quantum algorithms and test them at a low scale. It can also be used to simulate how quantum hardware behaves at a low level, with reproducing digitally their defects like noise and other imperfections. Some emulators are even able to simulate the inner physics of the underlying physical qubits, like Perceval from Quandela (for photon qubits).

The classical computing capacity grows very more or less exponentially with the number of supported qubits. On a laptop equipped with 16 GB of memory, you can simulate about 20 qubits. Specialized server appliances such as Atos' QLM fits in a datacenter rack, are designed to manage a very large amount of memory and can support the emulation of up to 40 qubits. More than 40 qubits can be emulated on massively parallel architectures and supercomputers or on a large number of clusters.

There are various methods for emulating a circuit of N qubits and a certain level of depth of quantum gate sequences which will resonate with what we've learned about registers computational basis and density matrices[1756]. We'll cover some of them from the hardest to the simplest with regards to computational resource requirements.

- **Density matrix** computing which requires $2^N x 2^N$ complex numbers, so $2^{2N+1}$ floating point numbers. It is the most memory-hungry method and is not used with a large number of qubits. It can be necessary if you need to emulate imperfect qubits with their noise and decoherence and their impact on quantum algorithm's execution.

- **Quantum state vector** which handles the complex amplitudes of all the Hilbert space managed by N qubits in memory. It requires $2^N$ complex numbers representing $2^{N+1}$ floating point numbers and $2^{N+5}$ bytes with double precision floating point numbers using 16 bytes. The action of quantum gates on this large vector consists in applying to it the quantum gates unitary matrices to one, two or three qubits which are respectively made of 2x2, 4x4 or 8x8 complex numbers. This method is implemented on supercomputers with huge memory capacities of the order of several Po. It is currently limited to about 50 qubits.

---

[1754] See $Q^2$ Chemistry: A quantum computation platform for quantum chemistry by Yi Fan et al, August 2022 (32 pages).

[1755] See the list of quantum algorithm simulation tools at https://quantiki.org/wiki/list-qc-simulators.

[1756] See Classical Simulation of Intermediate-Size Quantum Circuits, Alibaba, 2018 (12 pages). See also What limits the simulation of quantum computers? by Yiqing Zhou, Edwin Miles Stoudenmire and Xavier Waintal, March 2020 (14 pages) which provides a theoretical and practical framework for the optimization of quantum code emulation. Noteworthy is the work on the emulation of superconducting qubit modules with ... superconducting qubits. See Quantum computer-aided design: digital quantum simulation of quantum processors by Thi Ha Kyaw et al, 2020 (23 pages).

- **Tensor network** compression techniques are used to simplify emulation and ease its distribution across multiple classical computing nodes[1757]. It was used for example in September 2019 by Alibaba on a cluster of 10,000 servers with 96 CPUs. They simulated Google Bristlecone's 70 qubits (which never really run practically) over a depth of 34 quantum gates with 1449 instances of their Cloud Elastic Computing Service (ECS), each comprising 88 Intel Xeon chipsets with 160 GB of memory. So, a total of 127,512 processors[1758]! This method can be implemented with many compression techniques providing a lower accuracy[1759]. It was improved in October 2021, again by Alibaba, to classically simulate the Google Sycamore on the new Sunway supercomputer in 304 seconds[1760]. The chart below from this team shows the extent of tensor network compression capabilities as compared to state vector emulation.

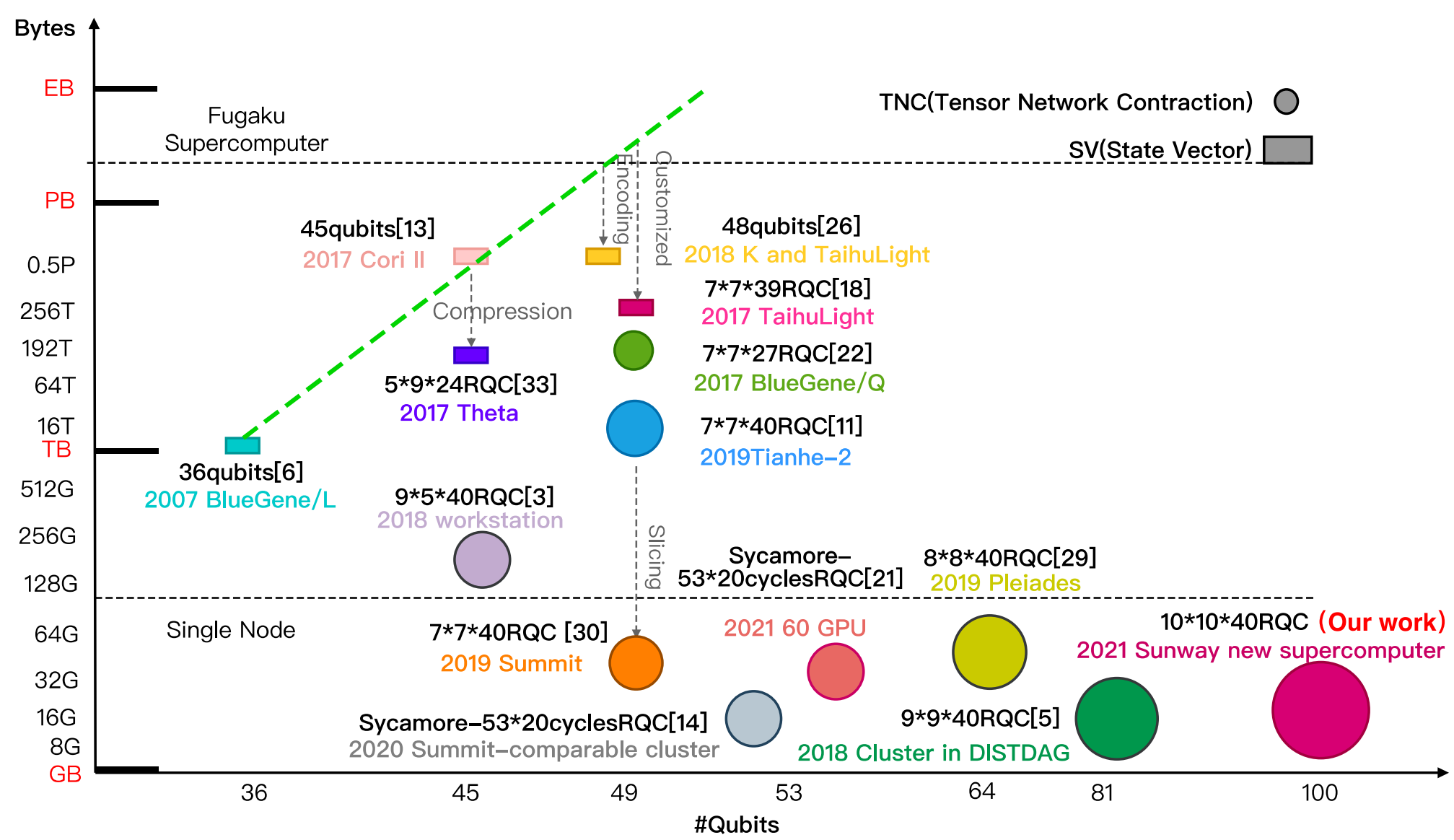

Figure 2: A summary of major classical RQC simulations. The x-axis denotes the number of qubits, while the y-axis shows the corresponding memory space required. The size of the circle/rectangular corresponds to the complexity (depth) of the circuit.

*Figure 616: the capacities of various quantum emulators in number of qubits (X) and memory capacity (Y). Source: Closing the "Quantum Supremacy" Gap: Achieving Real-Time Simulation of a Random Quantum Circuit Using a New Sunway Supercomputer by Yong (Alexander) Liu et al, October 2021 (18 pages).*

- **Weak simulation** which manages quantum state vector amplitudes without the phase, with $2^N$ floating point numbers and thus $2^{N+4}$ bytes, representing output measurement probabilities in the $|0\rangle$ and $|1\rangle$ computational basis[1761]. The method is easier to distribute over several servers. It is however not widely used.

---

1757 Partitioning methods for quantum simulation are well described in Distributed Memory Techniques for Classical Simulation of Quantum Circuits, Ryan LaRose of the University of Michigan, June 2018 (11 pages).

1758 See Alibaba Cloud Quantum Development Platform: Large-Scale Classical Simulation of Quantum Circuits, September 2019 (5 pages).

1759 See Full-State Quantum Circuit Simulation by Using Data Compression by Xin-Chuan Wu et al, 2020 (29 slides).

1760 See Closing the "Quantum Supremacy" Gap: Achieving Real-Time Simulation of a Random Quantum Circuit Using a New Sunway Supercomputer by Yong (Alexander) Liu et al, October 2021 (18 pages). The China team behind this was awarded the 2021 ACM Gordon Bell Prize. Their work was later contradicted by ORNL researchers who had developed Google's supremacy classical simulation in 2019. See China's exascale quantum simulation not all it appears by Nicole Hemsoth, NextPlatform, November 2021.

1761 See Just Like the Real Thing: Fast Weak Simulation of Quantum Computation by Stefan Hillmich et al, July 2020 (6 pages).

- **Hybrid quantum computing emulation** which emulates the quantum part, and executes the classical part as developed in China on a Sunway supercomputer[1762].

The main limitations of supercomputers for emulating quantum algorithms are more related to their memory (RAM) than to their processing capacity. It would take 16 Po of memory to fully simulate 50 qubits. How about 96 qubits? The memory requirement would be multiplied by $2^{46}$*2. Moore's law with memory cannot therefore keep pace with a linear increase in the number of used qubits in a quantum computer.

Nevertheless, the number of emulated qubits on supercomputers is still constantly increasing. China research teams have been the most active in this emulation race, particularly at Alibaba and Huawei, with several records set in 2018 up to 2021.

**Origin Quantum**, a Chinese multi-role (hardware, software) startup in partnership with the Guang-Can Guo team from the **University of Science and Technology of China**, emulated 64 qubits with a 22-depth algorithm on a cluster of 128 nodes in 2018[1763]. They used a method to transform combinations of CZ gates (conditional Pauli Z gates) and single-qubit gates into simpler sub-circuits that do not need to be interleaved. They also thought they could simulate 72 qubits over a depth of 23 gates on a supercomputer running for 16 hours. This work shows that two key parameters condition the emulation capabilities in classical computers: not only the number of qubits but also the number of quantum gate sequences. The larger the number of qubits emulated, the fewer quantum gate sequences we can simulate.

A second 2018 record coming from **Alibaba** was achieved with 81 qubits and 40 quantum gate sequences[1764]. Their Taizhang simulation exploited a method created by Igor Markov and Shi Yaoyun in 2005 that allows a quantum algorithm to be distributed over a farm of thousands of servers[1765]. The Alibaba Quantum Laboratory is managed by the same Shi Yaoyun, a professor at the University of Michigan. Their simulations included 100 qubits over 35 layers (10x10x35), 121 qubits over 31 layers (11x11x31) and 144 qubits over 27 layers (12x12x27). Another 2018 record came from **Huawei** and its "HiQ Cloud" service, capable of emulating 42 to 169 qubits[1766]. The method was similar to the one used by **Alibaba**. The 42 qubits were simulated in "full amplitude" mode. 81 qubits were simulated with "a single amplitude" and 169 qubits with a single amplitude and a small number of quantum gates. Other records have been broken in the USA in 2019, such as **Google'**s record with NASA, the University of Illinois and the Oak Ridge laboratory with 49 to 121 qubits on the IBM Summit[1767].

**IBM** broke a record of 56 qubits emulation in 2017 on a classic supercomputer of their own, the Vulcan BlueGene installed at the Lawrence Livermore National Laboratory in California. The same Oak Ridge laboratory is at the origin of **XAAC** (eXtreme-scale ACCelerator programming framework), a framework for Eclipse that manages hybrid calculations combining quantum computers and supercomputers such as the Titan equipped with Nvidia GPUs installed in Oak Ridge[1768] .

---

[1762] See Large-Scale Simulation of Quantum Computational Chemistry on a New Sunway Supercomputer by Honghui Shang et al, July 2022 (13 pages).

[1763] See Researchers successfully simulate a 64-qubit circuit, Science China Press, June 2018.

[1764] See Alibaba Says Its New "Tai Zhang" Is the World's Most Powerful Quantum Circuit Simulator, May 2018 et Alibaba announced that it has developed the world's strongest quantum circuit simulator "Taizhang", May 2018.

[1765] See Simulating quantum computation by contracting tensor networks by Igor Markov et Shi Yaoyun, 2005 (21 pages).

[1766] See Huawei Unveils Quantum Computing Simulation HiQ Cloud Service Platform, October 2018.

[1767] See Establishing the Quantum Supremacy Frontier with a 281 Pflop/s Simulation, May 2019 (11 pages). This Summit must have consumed a good part of the production of Nvidia V100! Here is also the list of qubit and qubit simulation records in https://quantumcomputingreport.com/scorecards/qubit-count/.

[1768] See Eclipse Science and Open Source Software for Quantum Computing, 2017 and the article describing XAAC: A Language and Hardware Independent Approach to Quantum-Classical Computing, July 2018 (15 pages).

It can transform quantum code for computers with quantum gates or quantum annealing models into executable code on any quantum architecture.

Many other IT players want to jump on the quantum emulation bandwagon. It was the case with **Dell** which announced in 2021 its hybrid solution combining classical computing and quantum emulation using its Dell EMC PowerEdge R740xd server appliance and IBM's Qiskit Runtime.

In March 2022, **Fujitsu** created "*the world's fastest quantum computer simulator*" supporting 36 qubits on a cluster system using Fujitsu PrimeHPC FX 700 and the in-house A64FX CPU that powers their supercomputer Fugaku. They run the quantum simulator software **Qulacs** from Kyoto University.

**Atos** (France) sells since 2017 an Intel-based quantum emulator appliance, the Atos Quantum Learning Machine (QLM).

It's been widely adopted worldwide, such as by the US DoE's ORNL, by the CEA, at the University of Reims, at the cybersecurity research department of the University of Applied Sciences of Upper Austria in Hagenberg, by the Hartree Science and Technology Facility Centre (STFC) in the UK, at the C-DAC (Centre for Development of Advanced Computing) in India, in its Quantum Computing Experience Center[1769], in Japan, in Finland at the CSC IT Center for Science Kvasi in collaboration with IQM, in the new Quantum Integration Centre (QIC) from LRZ (Leibniz Supercomputing Centre) of the Bavarian Academy of Sciences and in Spain (CESGA). A QLM was also sold to CERN in 2021.

Atos is also working with Total to develop quantum solutions to identify new materials and molecules in the energy transition.

In June 2020, Atos launched the QLM E, a new version of this emulator integrating from 2 to 32 Nvidia V100 GPUs, and multiplying computing power by 12 compared to the Intel-based initial version. This system was first delivered in December 2020 to the Irish HPC center (ICHEC). This was completed in early July 2020 with the support of a limited form of quantum annealing emulation.

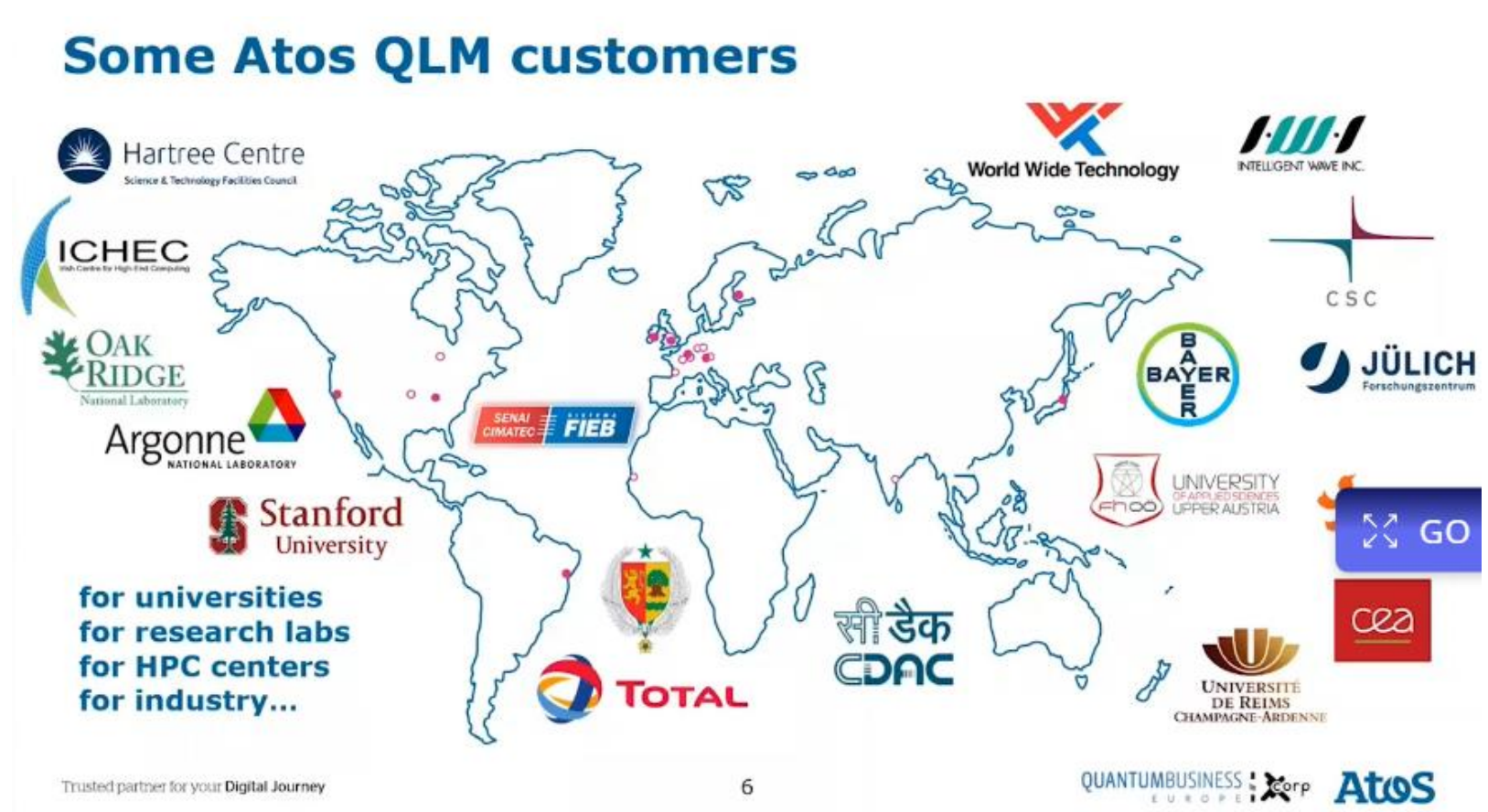

Figure 617: Atos QLM customers. Source: Atos.

In May 2019, Atos launched myQLM, a quantum programming tools for researchers, students and developers. It is a Python-based development environment that allows users to simulate quantum programs on their own computer. Programming is carried out in AQASM (Atos Quantum Assembly Language) and pyAQSM. To access a number of qubits that exceeds the current capacity of PCs, i.e. more than 20 qubits, developers can run their code on an Atos Quantum Learning Machine simulator in the cloud, but at a charge. Atos also enables the sharing of quantum practices, libraries and application codes. Atos offers one of the open source translators of myQLM code to other quantum programming environments. In September 2020, this software offer became free of charge to all audiences[1770].

---

[1769] See Atos and C-DAC sign a cooperation agreement to accelerate the development of quantum and exascale computing and Artificial Intelligence in India, August 2019.

[1770] See Atos roadmap in The Atos Quantum Program - Paving the way to quantum-accelerated HPC by Jean-Pierre Panziera, June 2021 (10 slides).

Atos announced in 2020 that they would launch a NISQ quantum accelerator by 2023. They are looking at several tracks such as superconductors (with IQM), trapped ions (with the University of Innsbruck and AQT), cold atoms (with Pasqal) and in the longer-term silicon qubits (with CEA-Leti). They participate in the European Flagship projects **AQTION** (quantum accelerator), **PASQuanS** (analog quantum simulator) and **NEASQC** (NExt ApplicationS of Quantum Computing, which they coordinate).

Atos is also heavily involved in the EuroHPC project, which includes the **European Processor Initiative**, an initiative to develop a processor adapted to the needs of supercomputers and on-board as well as autonomous vehicles[1771]. And, of course, in the $\langle \boldsymbol{HPC} | \boldsymbol{QS} \rangle$ project which will deploy in Finland, Germany and France three hybrid quantum solutions combining a supercomputer and a quantum computer.

**Nvidia** (USA) developed the cuQuantum SDK running on top of their GPGPUs. It implements gates-based programming emulation, announced in April 2021, beta in November 2021[1772] and released in March 2022[1773].

Thanks to the GPGPU tensors implementing matrix multiplications and fast HBM2E memory, the acceleration provided is clear, making it possible to emulate Google Sycamore processor with a depth of 20 gates in less than 10 minutes[1774]. It is to be supported by various cloud offerings from JUQCS-G (Julich), Qgate (NVAITC), Qiskit-AER (IBM), QuEST (Oxford), SV1 (Amazon Web Services) and Vulcan (QC Ware).

The cuQuantum SDK supports both state-vector emulation with 10s of qubits (**cuStateVec**) and a less resources-hungry tensor-network based emulation (**cuTensorNet**) that supports up to thousands of qubits. They integrated cuStateVec into qsim, Google Quantum AI's state vector simulator that can be used through Cirq. cuStateVec can also be used with Qiskit Aer, IBM's emulation framework.

Nvidia also proposes its quantum compiler **NVQ++** that targets the Quantum Intermediate Representation (QIR), a low-level machine language specification covering hybrid classical/quantum computing needs. It is supported by the Linux Foundation led **QIR Alliance** with contributions from ORNL, Rigetti, Quantinuum, Microsoft and Quantum Circuits Inc[1775].

Nvidia's software tools adopters include QC Ware (for quantum chemistry and QML using cuQuantum on the Nvidia A100-based Lawrence Berkeley National Laboratory Perlmutter supercomputer launched in 2021), ORNL (using cuQuantum in TNQVM, a framework for tensor network quantum circuit simulations), Xanadu (using cuQuantum in their PennyLane framework for QML and quantum chemistry), Classiq (in their Quantum Algorithm Design platform) and Zapata Computing (in Orquestra). Nvidia also works with Google Quantum AI, IBM, IonQ and Pasqal.

---

[1771] In July 2018, Atos also acquired Syntel for $3.4B in the USA, a $923M service provider specializing in the development and deployment of applications in the cloud with 22,500 employees, created in 1980 by Indo-Americans. This does not seem to have anything to do with quantum.

[1772] See NVIDIA Teams With Google Quantum AI, IBM and Other Leaders to Speed Research in Quantum Computing by Sam Stanwyck, Nvidia, November 2021.

[1773] See Nvidia Unveils Onramp to Hybrid Quantum Computing by Timothy Costa, March 2022.

[1774] See What Is Quantum Computing? by Dion Harris, April 2021 and Nvidia entangled in quantum simulators by Nicole Hemsoth, April 2021.

[1775] See https://github.com/qir-alliance.

In 2022, Pasqal (France) deployed an on-premises Nvidia DGX POD to run digital simulations of its quantum simulator using cuQuantum. In July 2022, Nvidia announced its quantum software emulation and hybrid computing, the Quantum Optimized Device Architecture (**QODA**) platform. It helps develop software that can run on both GPU-based classical emulation and on QPUs, including hybrid quantum/classical solutions.

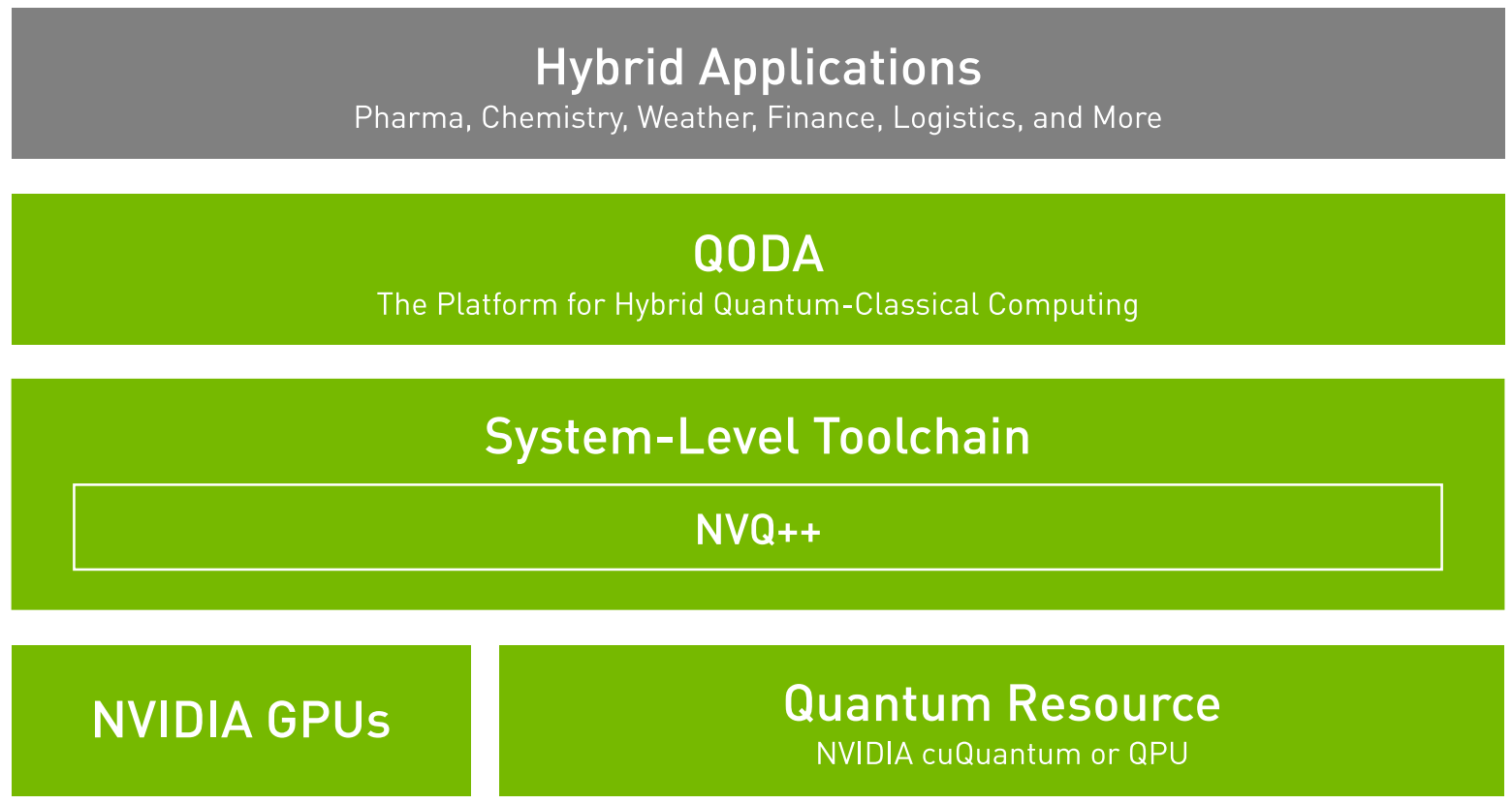

*Figure 618: Nvidia QODA architecture. Source: Nvidia.*

Beyond the solutions mentioned above, many software emulation tools are available and are mostly always open source [1776]:

**Quantum Circuit Simulator** runs under Android and was developed in 2013.

**Quirk** runs in your browser and even on your smartphone. It was developed in 2016 by Craig Gidney before he joined Google in 2017.

**Intel IQS** (formerly qHiPSTER) created by Intel in 2016[1777]. It supports up to 42 qubits pure states in state vector simulation mode.

**SimulaQron** from QuTech can run on their **Quantum Inspire** platform with two quantum processors using 2 spin qubits (Spin-2) and 5 superconducting qubits (Starmon-5) and two hardware emulators supporting 26, 31 and 34 qubits.

**Psitrum** is a software emulator developed in Saudi Arabia in Matlab. It computes the algorithm matrix, the density matrix of the simulated qubit register and the output state vector. It can simulate quantum noise and has a graphical circuit designer. It provides density matrices visualization tools[1778].

**QuEST** (Quantum Exact Simulation Toolkit) is a quantum emulator developed in C language and supporting QUDA APIs (not CUDA) and Nvidia's GPUs, created in 2017 by Simon Benjamin's Quantum Technology Theory Group from Oxford University and distributed in open source. The system can simulate 26 to 45 qubits depending on the available memory, respectively 2 GB and 256 GB.

**myQLM** was created by Atos in 2020. It supports the emulation of the three quantum computing paradigms: quantum annealing, quantum simulation and gate-based programming. myQLM is a freely available subset of the full QLM software simulation suite which can work in several modes: perfect state vectors, with using various compression methods (stabilizers, binary decision diagrams, tensor networks MPS), and with simulating qubit noise.

**Google**'s emulator qsim can simulate 30 qubits on a laptop and up to 40 qubits in Google Cloud.

**Qiskit Aer** is IBM's emulation software solution launched in December 2018. It supports simulations in state vector mode and density matrix mode as well as in the more exotic matrix product state (adapted to weakly entangled states) and stabilizer modes (supporting only Clifford group gates).

---

[1776] See this long list of emulators.

[1777] qHiPSTER stands for quantum High Performance Software Testing Environment. See qHiPSTER: The Quantum High Performance Software Testing Environment by Mikhail Smelyanskiy et al, 2016 (9 pages).

[1778] See Psitrum: An Open Source Simulator for Universal Quantum Computers by Mohammed Alghadeer et al, Saudi Arabia, March 2022 (27 pages).

**staq** : another emulator available as a GitHub repository, staq is a modern C++17 library for the synthesis, transformation, optimization and compilation of quantum circuits authored by softwareQ Inc. under the MIT License. It is usable either through the provided binary tools or as a header-only library that can be included to provide direct support for parsing & manipulating circuits written in the OpenQASM circuit description language. Inspired by Clang, staq is designed to manipulate OpenQASM syntax trees directly, rather than through an intermediate representation which makes retrieving the original source code impossible. In particular, OpenQASM circuits can be inspected and transformed (in most cases) without losing the original source structure. This makes staq ideally suited for source-to-source transformations, where only specific changes are desired. Likewise, this allows translations to other common circuit description languages and libraries to closely follow the OpenQASM source.

**Qrack** is a C++ quantum bit and gate simulator, with the ability to support arbitrary numbers of entangled qubits—up to system limitations. Suitable for embedding in other projects, the Qrack QInterface contains a full and performant collection of standard quantum gates, as well as variations suitable for register operations and arbitrary rotations. The developers of Qrack maintain a fork of the ProjectQ quantum computer compiler which can use Qrack as the simulator, generally. This stack is also compatible with the SimulaQron quantum network simulator. Further, it maintains a QrackProvider for Qiskit. Both ProjectQ and Qiskit integrations for Qrack support the PennyLane stack. For Qiskit, a fork of the Qiskit plugin provides support for a "QrackDevice".

**QX Simulator** is a universal quantum computer simulator developed at QuTech by Nader Khammassi. The QX allows quantum algorithm designers to simulate the execution of their quantum circuits on a quantum computer. The simulator defines a low-level quantum assembly language namely Quantum Code which allows the users to describe their circuits in a simple textual source code file. The source code file is then used as the input of the simulator which executes its content.

**QuIDDPro** is a fast, scalable, and easy-to-use computational interface for generic quantum circuit simulation. It supports state vectors, density matrices, and related operations using the Quantum Information Decision Diagram (QuIDD) data structure. Software packages including Matlab, Octave, QCSim, and libquantum, have also been used to simulate quantum circuits. However, unlike these packages, QuIDDPro does not always suffer from the exponential blow-up in size of the matrices required to simulate quantum circuits. As a result, we have found that QuIDDPro is significantly faster and uses significantly less memory as compared to other generic simulation methods for some useful circuits with many more than ten qubits.

**LIQUi|>** is a software architecture and tool suite for quantum computing. It includes a programming language, optimization and scheduling algorithms, and quantum simulators. LIQUi|> can be used to translate a quantum algorithm written in the form of a high-level program into the low-level machine instructions for a quantum device. LIQUi|> is being developed by the Quantum Architectures and Computation Group (QuArC) at Microsoft Research.

**Quantum Programming Studio** is a web-based graphical user interface designed to allow users to construct quantum algorithms and obtain results by simulating directly in the browser or by executing on real quantum computers. The circuit can be exported to multiple quantum programming languages/frameworks and can be executed on various simulators and quantum computers.

**Quantum Computer Emulator** (QCE) is a software tool that emulates various hardware designs of quantum computers. QCE simulates the physical processes that govern the operation of a hardware quantum processor, strictly according to the laws of quantum mechanics. QCE also provides an environment to debug and execute quantum algorithms under realistic experimental conditions. The software consists of a Graphical User Interface (GUI) and the simulator itself.

**SimQubit** is a GUI quantum circuit simulator, written on top of the Q++ (sourceforge.net/projects/qplusplus) quantum templates. It allows editing of quantum circuits and applying them to quantum states, with multiple ways to view the output probabilities.

**Qubit101** simulator is a user-friendly quantum circuit editor and simulator. The tool helps users to create, modify and save the quantum circuits. Along with this, users can simulate its effect over a predefined quantum state, watch the evolution of the state stage by stage, together with the possible measurements results, use other quantum circuits as gates, so complex circuits can be easily created and finally, simulate an almost arbitrary number of qubits. Supported platforms include Rigetti Forest, IBM Qiskit, Google Cirq and TensorFlow Quantum, Microsoft Quantum Development Kit, Amazon Braket and more.

**ScaffCC** is a compiler and scheduler adapted to the Scaffold programming language supporting the LLVM infrastructure. It supports QASM.

**Perceval** is a photon qubits emulator and simulator developed by Quandela.

**Pulser** is a cold atoms emulators developed by Pasqal.

# Research-originated quantum development tools

Here is an overview of the main quantum languages created to date, starting with languages that are independent of hardware architectures and that often originate from research laboratories.

They have the disadvantage that they are not generally linked to cloud quantum computer offerings. The related researchers are the equivalents of the Kernighan and Richie (creators of the C language) and Bjarne Stroustrup (creator of C++) in the quantum realm! A good number of these languages come from Europe.

- **QCL** or Quantum Computation Language has a syntax and data types close to those of the C language. This language is one of the first for quantum programming, created in 1998 by the Austrian researcher **Bernhard Ömer** from the Austrian Institute of Technology in Vienna. It is described in Structured Quantum Programming, 2009 (130 pages) which positions very well the conceptual differences between classical and quantum programming languages.

| Classical concept | Quantum analogue |
|---|---|
| classical machine model | hybrid quantum architecture |
| variables | quantum registers |
| variable assignments | elementary gates |
| classical input | quantum measurement |
| subroutines | operators |
| argument and return types | quantum data types |
| local variables | scratch registers |
| dynamic memory | scratch space management |
| boolean expressions | quantum conditions |
| conditional execution | conditional operators |
| selection | quantum if-statement |
| conditional loops | quantum forking |

Figure 619: classical and quantum programming concepts. Source: Structured Quantum Programming, 2009 (130 pages).

- **Q Language** is an extension of the C++ language that provides classes for programming quantum gates (Hadamard, CNOT, SWAP, QFT for quantum Fourier transform)[1779].
- **QFC** and **QPL** are two functional languages defined by Peter Selinger, from Canada, the first one using a graphical syntax and the second one using a textual syntax[1780].
- **QML** is a functional programming language created by Thorsten Altenkirch and Jonathan Grattage (UK)[1781].

[1779] It is documented in Toward an architecture for quantum programming, 2003 (23 pages), with as co-author, Stefano Bettelli from the Laboratory of Quantum Physics of the Paul Sabatier University of Toulouse.

[1780] They are described in Towards a Quantum Programming Language, 2003 (56 pages).

[1781] See A functional quantum programming language, 2004 (15 pages). The principles are well described in the presentation Functional Quantum Programming, (151 slides).

- **qGCL** or Quantum Guarded Command Language was created by Paolo Zuliani of the University of Newcastle[1782].
- **ProjectQ** is a scripting language from ETH Zurich that takes the form of an open source Python framework, released on GitHub since 2016. It includes a compiler that converts quantum code into C++ language for execution in a quantum simulator with a traditional processor[1783]. Launched in early 2017, it supports IBM's quantum computers via their OpenQASM language, which is normal since ETH Zurich is a partner of the latter, as well as simulation on a traditional computer via a C++ implementation that supports up to 28 qubits. ProjectQ is compatible with OpenFermion from Rigetti and Google.

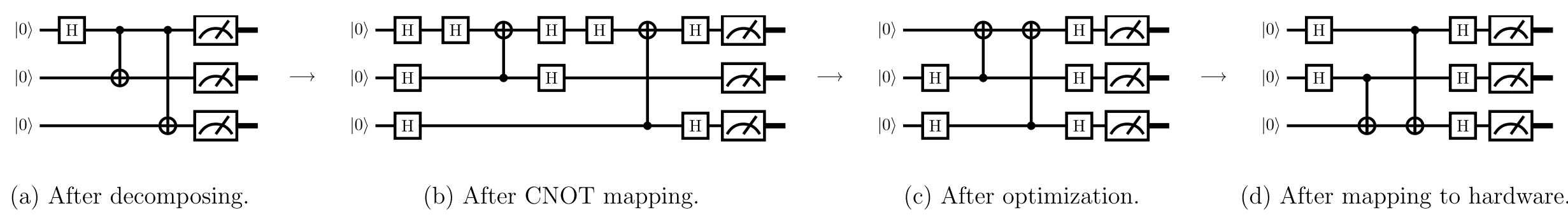

Figure 5: Individual stages of compiling an entangling operation for the IBM back-end. The high-level Entangle-gate is decomposed into its definition (Hadamard gate on the first qubit, followed by a sequence of controlled NOT gates on all other qubits). Then, the CNOT gates are remapped to satisfy the logical constraint that controlled NOT gates are allowed to act on one qubit only, followed by optimizing and mapping the circuit to the actual hardware.

*Figure 620: ProjectQ compiler entangling gates decomposition. Source: [ProjectQ: An Open Source Software Framework for Quantum Computing]() by Damian Steiger, Thomas Häner and Matthias Troyer, 2018 (13 pages).*

- **Quipper** is a language created in 2013 that builds on the classic [Haskell]() language, created in 1990, to which it provides extensions in the form of data types and function libraries[1784]. It manipulates a software version of qRAM, an addressable quantum memory register, that is essential for the execution of algorithms such as Grover and QMLs. The language does not seem to have evolved since 2016. One of its creators is Benoît Valiron who teaches quantum programming at CentraleSupelec in France[1785].
- **QWire** is another quantum programming language close to Quipper, launched in 2018, from the University of Pennsylvania [1786]. It is associated with a formal proof solution.

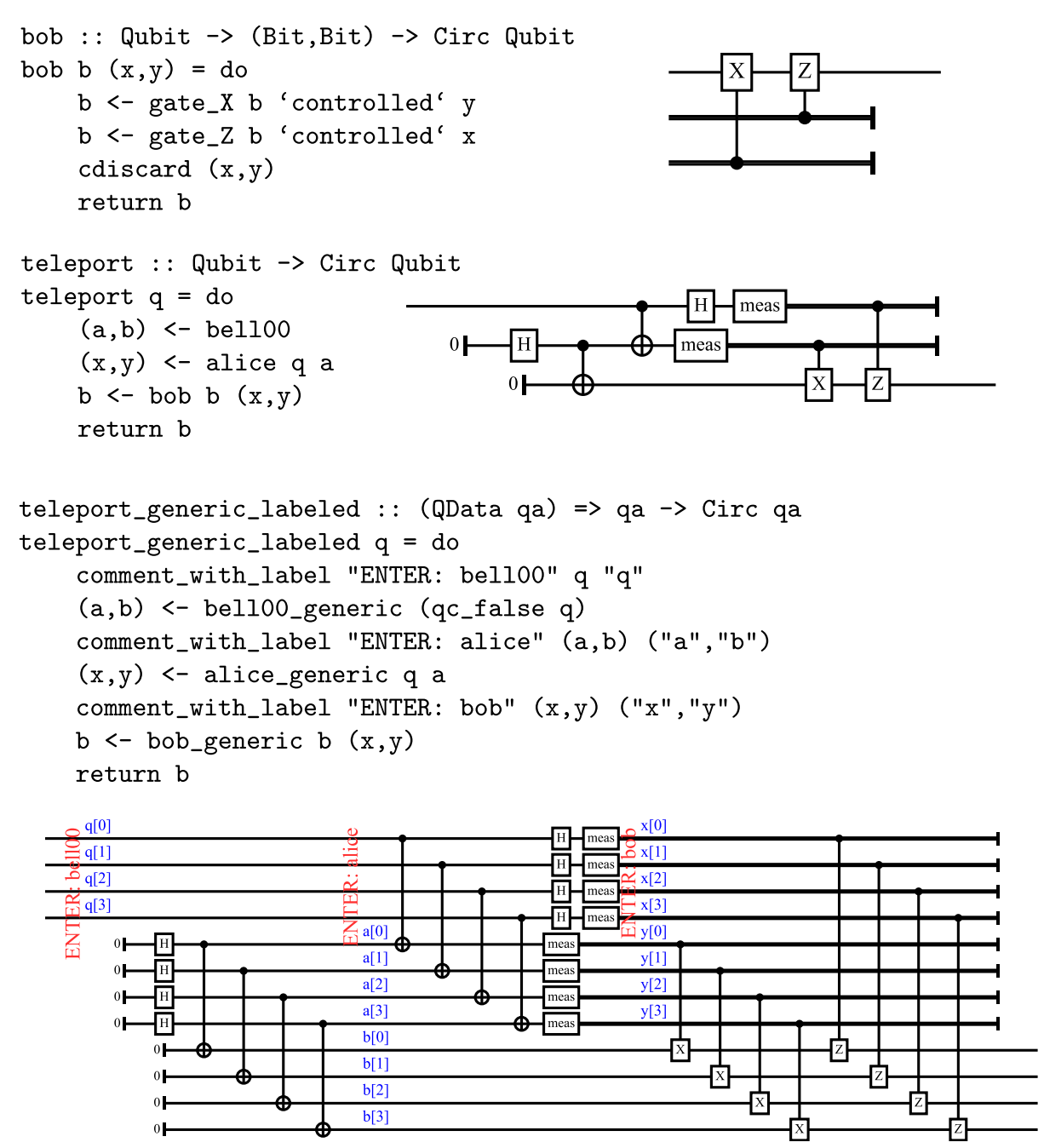

*Figure 621: example of Quipper code. Source: [An Introduction to Quantum Programming in Quipper](), 2013 (15 pages).*

---

[1782] See [Compiling quantum programs](), 2005 (39 pages).

[1783] See [ProjectQ: An Open Source Software Framework for Quantum Computing]() by Damian Steiger, Thomas Häner and Matthias Troyer, 2018 (13 pages) which explains how the compiler optimizes the code according to the gates available in the quantum computer.

[1784] It is documented in *[An Introduction to Quantum Programming in Quipper](), 2013 (15 pages)*. Its creation was funded by IARPA.

[1785] See his presentation [Programming a Quantum Computer](), 2017 (38 slides) and [Quantum Computation Model and Programming Paradigm](), 2018 (67 slides).

[1786] See [QWIRE: A Core Language for Quantum Circuits]() (13 pages) and [A core language for quantum circuits]() by Jennifer Paykin et al, 2017 (97 slides).

- **Qubiter** is an open source language developed in Python that can be used on top of IBM's OpenQASM and Google's OpenFermion. It was created in 2017.
- **Scaffold** is a language developed at Princeton University[1787]. It is used to program traditional code which is then automatically transformed into quantum gates via its C2QG (Classical code to Quantum Gates) function. In particular, Scaffold can generate QASM. It can be interesting to develop oracles for search algorithms. Figure 622 contains a sample Scaffold code, almost easy to understand! Its development was also funded by IARPA.
- **Qumin** is a minimalist open source quantum language designed in Greece in 2017[1788].

```
// Pauli X, Pauli Y, Pauli Z, Hadamard, S, and T gates
gate X(qreg input[1]);
gate Y(qreg input[1]);
gate Z(qreg input[1]);
gate H(qreg input[1]);
gate S(qreg input[1]);
gate T(qreg input[1]);

// Daggered gates
gate Tdag(qreg input[1]);
gate Sdag(qreg input[1]);

// CNOT gate defined on two 1-qubit registers
gate CNOT(qreg target[1], qreg control[1]);

// Toffoli (CCNOT) gate
gate Toffoli(qreg target[1], qreg control1[1], qreg control2[1]);

// Rotation gates
gate Rz(qreg target[1], float angle);        //Arbitrary Rotation

// Controlled rotation
gate controlledRz(qreg target[1], qubit control[1], float angle);

// One-qubit measurement gates
gate measZ(qreg input[1], bit data);
gate measX(qreg input[1], bit data);

//One-qubit prepare gates: initializes to 0
gate prepZ(qreg input[1]);
gate prepX(qreg input[1]);

//Fredkin (controlled swap) gate
gate fredkin(qreg targ[1], qreg control1[1], qreg control2[1])
```

*Figure 622: Scaffold code example. Source: Scaffold: Quantum Programming Language by Ali Javadi Abhari et al, 2012 (43 pages) page 15.*

- **Q.js** is a graphical quantum emulator launched in 2019, running in JavaScript and thus running in a browser[1789].
- **QuTiP** (Quantum Toolbox in Python) is another open source quantum code emulation tool developed by Paul Nation of IBM, Robert Johansson of Rakuten and Franco Nori of RIKEN (Japan) and the University of Michigan. The project started in 2011. It targets superconducting qubits.
- **QNET** is a language from Stanford University created in 2012, which allows to simulate the operation of quantum networks.
- Quantum implementation languages of **lambda calculus**, conceptualized by Alonzo Church and Stephen Cole Kleene during the 1930s, followed. This type of computation makes it possible to solve very complex and NP-complete problems, the class of problems that can be verified in polynomial time and whose resolution requires exponential time on classical computers and potentially polynomial time on quantum computers[1790]!
- **OpenQL** is an open source quantum programming language created by TU Delft in 2020. It includes a high-level quantum programming language, its associated quantum compiler and a low-level assembly language, cQASM[1791].
- **eQASM** is an intermediate quantum machine language from Delft University and its subsidiary QuTech. It sits in between high-level programming tools (QASM) and the quantum accelerator. It is a compiled language, hence the "e" for executable. The compiler manages the dependencies with hardware implementation specifics. Tests have been carried out with a 7-qubit superconducting chipset.

---

[1787] See Scaffold: Quantum Programming Language by Ali Javadi Abhari et al, 2012 (43 pages).

[1788] See Qumin, a minimalist quantum programming language, 2017 (34 pages).

[1789] See Quantum Programming: JavaScript (Q.js) - a drag and drop circuit editor by Stewart, 2020. And https://quantumjavascript.app/.

[1790] See A lambda calculus for quantum computation with classical control by Peter Selinger and Benoît Valiron, 2004 (15 pages).

[1791] See OpenQL : A Portable Quantum Programming Framework for Quantum Accelerators by N. Khammassi et al, 2020 (13 pages).

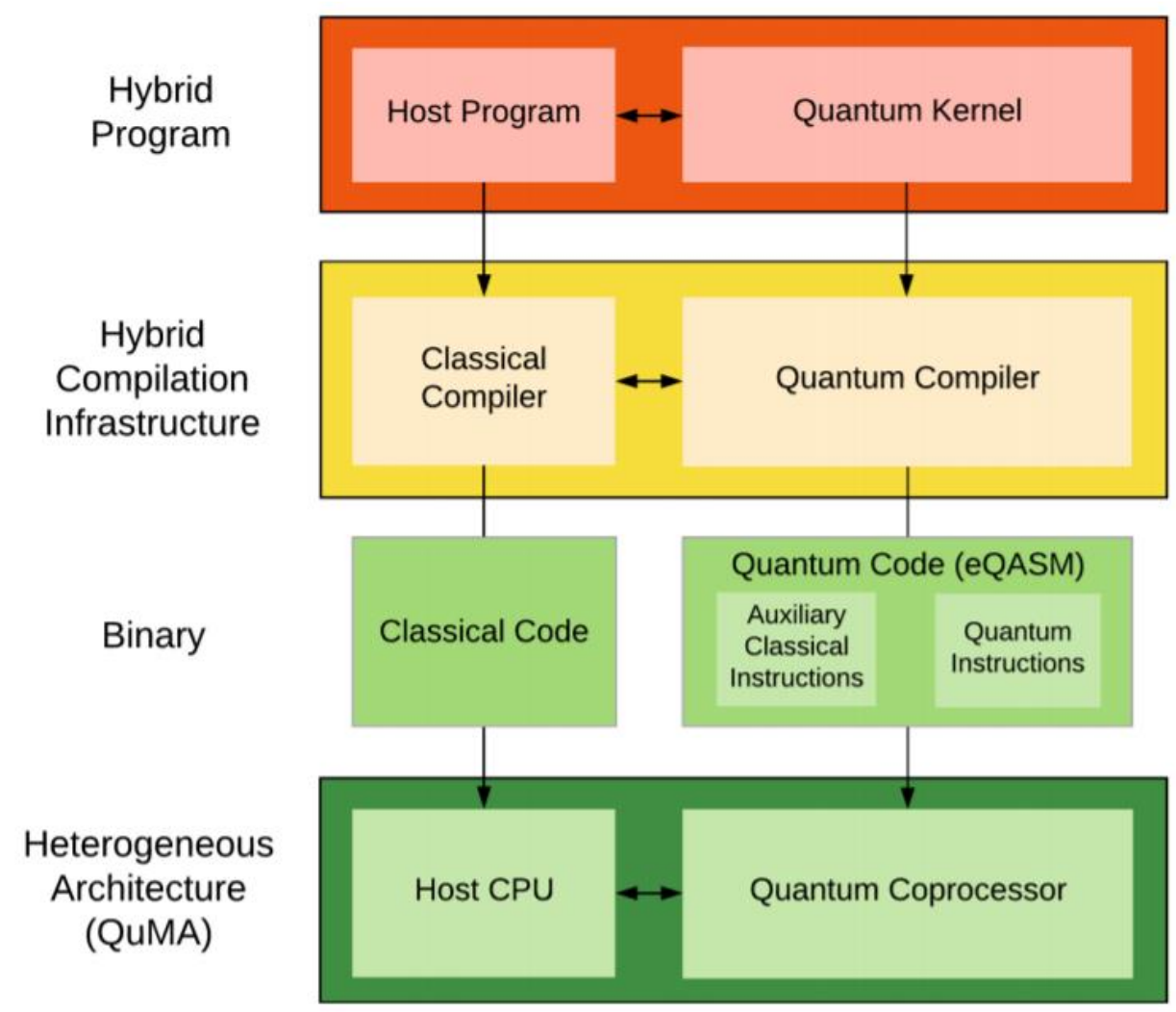

**Fig. 1. Heterogeneous quantum programming and compilation model.**

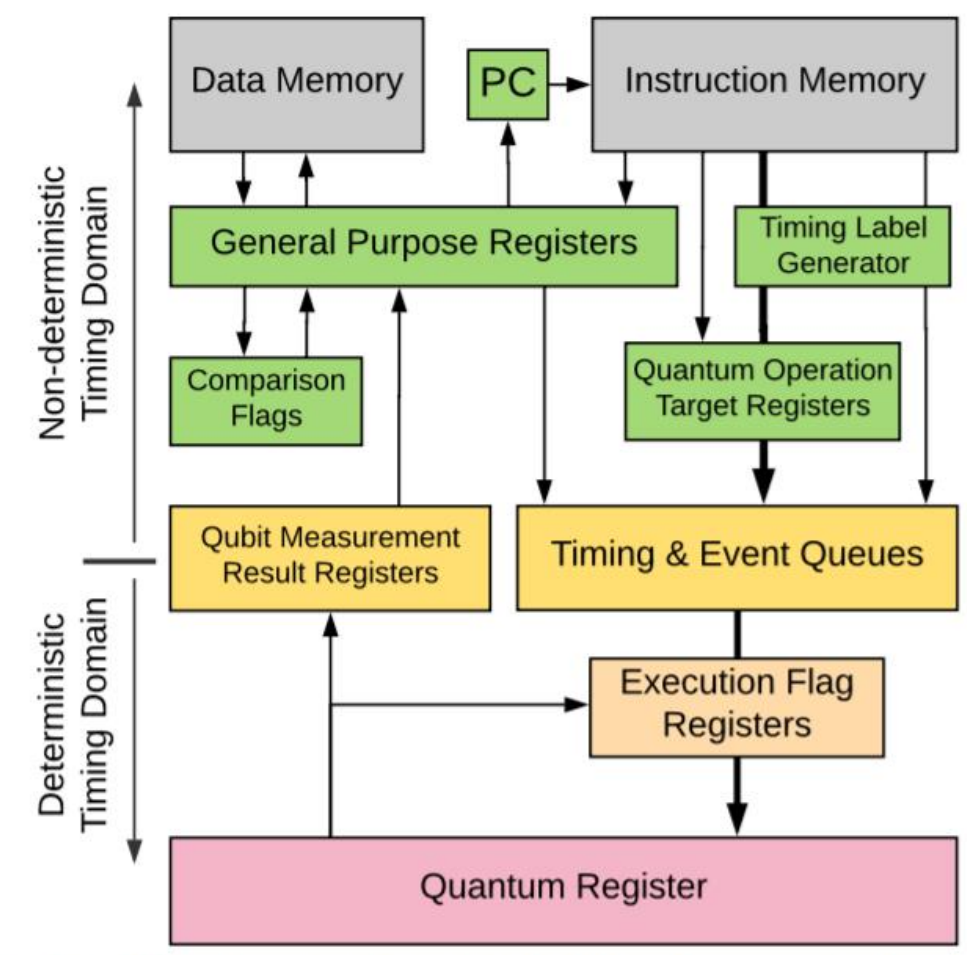

**Fig. 2. Architectural state of eQASM. Arrows indicates the possible information flow. The thick arrows represent quantum operations, which read information from the modules passed through.**

*Figure 623: eQASM architecture, from Delft University. Source: eQASM: An Executable Quantum Instruction Set Architecture, March 2019 (14 pages).*

- Researchers at the University of Chicago's Enabling Practical-scale Quantum Computation (**EPiQC**) laboratory proposed a compiler that can improve the speed and reliability of quantum computers by a factor of 10. Here again, the compiler has to adapt to the underlying hardware architecture[1792]. Their video explains the process. The team used Google's TensorFlow library to optimize the physical control parameters of the qubits.
- **Silq** is a concise and static quantum programming language proposed by a team from ETH Zurich[1793].
- **Yao.jl** is a package for the Julia language used for creating quantum circuits.
- **Qunity** is a language created in 2022 at the Universities of Maryland and Chicago, and at AWS[1794]. Its goal is to unify quantum and classical programming concepts in a single language. Its syntax uses familiar programming constructs that can have both quantum and classical effects like summing linear operators, using exception handling syntax with projective measurements and using aliasing to induce entanglement. It can also automatically construct reversible subroutines from irreversible quantum algorithms through the uncomputation of "garbage" outputs. It can for example create full quantum oracle functions for algorithms like Grover, Deutsch-Jozsa and Simon. Qunity is still being developed and will be compiled to generate OpenQasm lower level code.
- **MCBeth** is a language created at Yale University and Chicago tailored for MBQC (measurement based quantum computing) programming[1795]. It can represent, program and simulate measurement-based and cluster state computation. Its compiled code can be executed directly on MBQC hardware as well as on traditional gate-based QPUs. This language is based on the initial work by Vincent Danos, Elham Kashefi and Prakash Panangaden in 2007[1796].

[1792] See Research provides speed boost to quantum computers, April 2019.

[1793] See Swiss scientists launch high-level quantum computing language by ETH Zurich, June 2020.

[1794] See Qunity: A Unified Language for Quantum and Classical Computing by Finn Voichick et al, April 2022 (34 pages).

[1795] See MCBeth: A Measurement Based Quantum Programming Language by Aidan Evans et al, April 2022 (27 pages).

[1796] See The measurement calculus by Vincent Danos, Elham Kashefi and Prakash Panangaden, 2007 (46 pages).

- **TWIST** is a language created in 2021 by MIT's CSAIL lab that enforces how qubits are entangled or not, handles the notion of purity (a set of qubits not influenced by others) and enables the creation of safer programs. It introduces $\mu Q$, a small functional quantum language. Its creators plan to devise a higher-level abstraction language using TWIST[1797]. Although this project was jointly funded by IBM Research, it remains unclear whether IBM could reuse it in its quantum software toolbox.
- **QIRO** (Quantum Intermediate Representation for Optimization) is a two-dialect language proposal to enable quantum-classical co-optimizations[1798].
- **ScaleQC** is a framework developed by Princeton researchers for hybrid quantum and classical computing[1799].

Most quantum programming software tools are open sourced. Their differentiation is mainly concentrated on documentation and tutorials[1800]. However, in practice, few commercial application developers use the languages discussed in this section. Instead, they are hooked to the languages and toolkits provided by commercial quantum computer vendors listed afterwards. They are easily locked into "full stack" approaches that are proprietary in practice although also open sourced in principle.

The most interesting thing about all this is that many development tools allow us to get our hands on small-scale quantum algorithms before scalable and usable quantum computers are available. And most of them are also open sourced and free to install and use[1801].

Some optimization tools can also be mentioned here like **CutQC** which distributes in an optimized way a large quantum circuit onto several (non-connected) QPUs and classical platforms (CPU or GPU) for co-processing[1802]. It enables using NISQ QPUs at their optimum regime, when a small number of qubits have a sufficient fidelity. Obviously, it doesn't generate an equivalent system to the sum of the qubits of the used QPUs.

---

[1797] See Twist: Sound Reasoning for Purity and Entanglement in Quantum Programs by Charles Yuan, Christopher McNally and Michael Carbin, January 2022 (32 pages).

[1798] See Enabling Dataflow Optimization for Quantum Programs by David Ittah et al, Microsoft Research and ETH Zurich, January-August 2021 (15 pages).

[1799] See ScaleQC: A Scalable Framework for Hybrid Computation on Quantum and Classical Processors by Wei Tang and Margaret Martonosi, Princeton, July 2022 (12 pages).

[1800] As described in Open-source software in quantum computing, by Mark Fingerhuth, Thomas Babej and Peter Wittek, December 2018 (28 pages). It makes a detailed inventory of these different tools and gauges them against classical open source software features like source code documentation.

[1801] See on this subject the presentations of FOSDEM 2019 conference.

[1802] See Cutting Quantum Circuits to Run on Quantum and Classical Platforms by Wei Tang and Margaret Martonosi, Princeton University, May 2022 (11 pages).

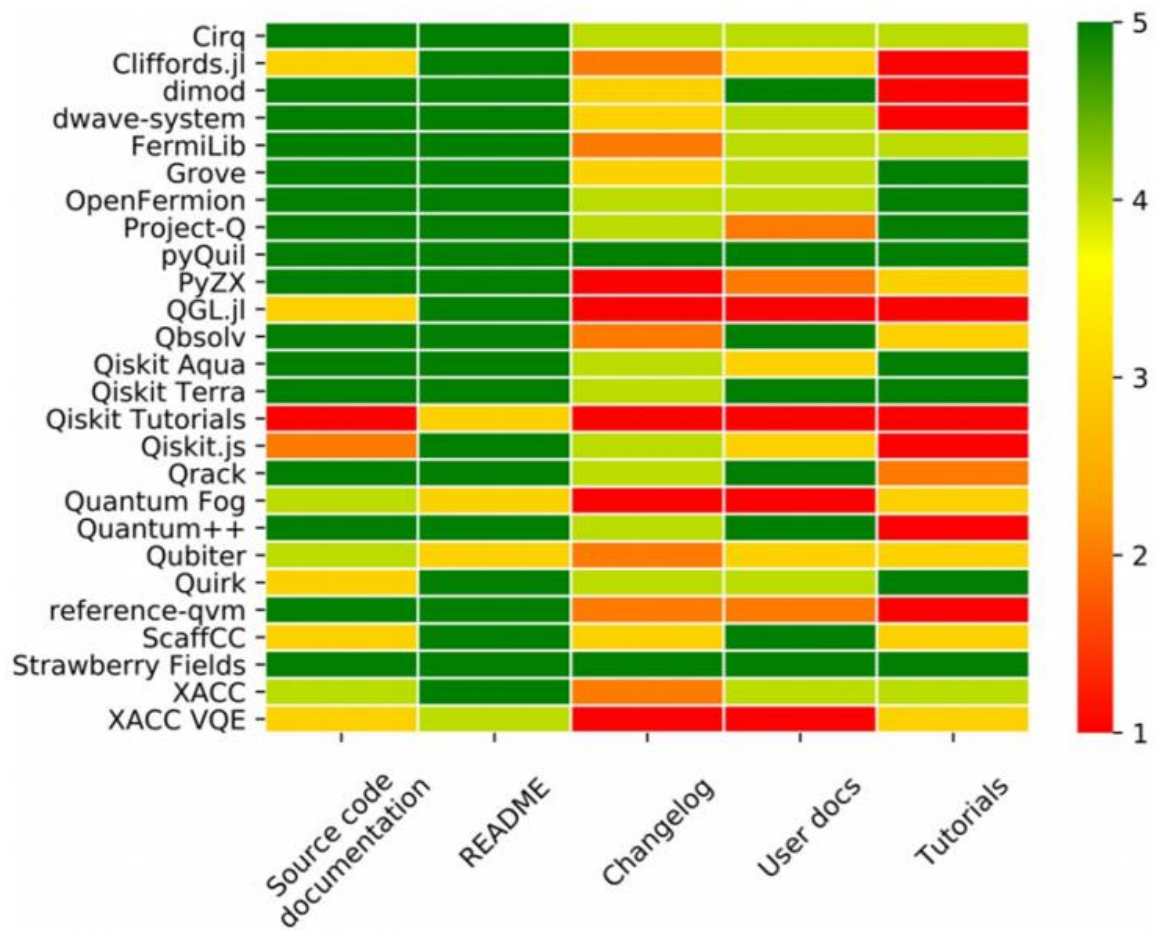

*Figure 624: heatmap of various quantum coding tools and their quality figures of merits. Source: [Open-source software in quantum computing](), by Mark Fingerhuth, Thomas Babej and Peter Wittek, December 2018 (28 pages).*

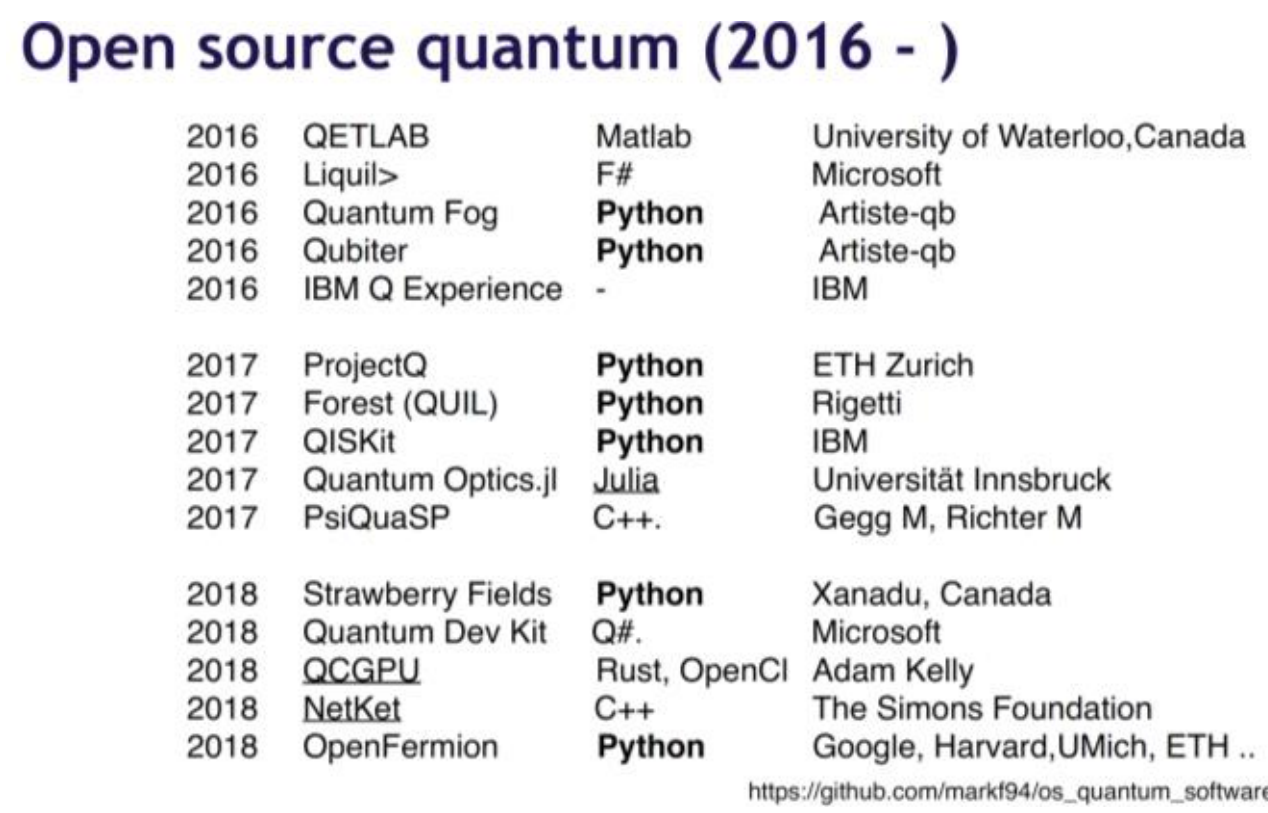

*Figure 625: a summary timeline of the appearance of various quantum development tools. Source: [Quantum Software Engineering Landscapes and Horizons]() by Jianjun Zhao, 2020 (31 pages) which provides an excellent overview of development tools covering the entire quantum software creation cycle, including the thorny issues of debugging and testing.*

| Year | Language | Reference(s) | Semantics | Host Language | Paradigm |
|---|---|---|---|---|---|
| 1996 | Quantum Lambda Calculi | [181] | Denotational | lambda Calculus | Functional |
| 1998 | QCL | [206–209] | | C | Imperative |
| 2000 | qGCL | [241, 312–314] | Operational | Pascal | Imperative |
| 2003 | $\lambda_q$ | [282, 283] | Operational | Lambda Calculus | Functional |
| 2003 | Q language | [32, 33] | | C++ | Imperative |
| 2004 | QFC (QPL) | [245–247] | Denotational | Flowchart syntax (Textual syntax) | Functional |
| 2005 | QPAlg | [141, 160] | | Process calculus | Other |
| 2005 | QML | [10, 11, 113] | Denotational | Syntax similar to Haskell | Functional |
| 2004 | CQP | [102–104] | Operational | Process calculus | Other |
| 2005 | cQPL | [180] | Denotational | | Functional |
| 2006 | LanQ | [188–191] | Operational | C | Imperative |
| 2008 | NDQJava | [298] | | Java | Imperative |
| 2009 | Cove | [227] | | C# | Imperative |
| 2011 | QuECT | [48] | | Java | Circuit |
| 2012 | Scaffold | [1, 138] | | C (C++) | Imperative |
| 2013 | QuaFL | [162] | | Haskell | Functional |
| 2013 | Quipper | [114, 115] | Operational | Haskell | Functional |
| 2013 | Chisel-Q | [175] | | Scala | Imperative, functional |
| 2014 | LIQUi\|⟩ | [292] | Denotational | F# | Functional |
| 2015 | Proto-Quipper | [234, 237] | | Haskell | Functional |
| 2016 | QASM | [212] | | Assembly language | Imperative |
| 2016 | FJQuantum | [82] | | Feather-weight Java | Imperative |
| 2016 | ProjectQ | [122, 266, 272] | | Python | Imperative, functional |
| 2016 | pyQuil (Quil) | [259] | | Python | Imperative |
| 2017 | Forest | [61, 259] | | Python | Declarative |
| 2017 | OpenQASM | [66] | | Assembly language | Imperative |
| 2017 | qPCF | [213, 215] | | Lambda calculus | Functional |
| 2017 | QWIRE | [217] | | Coq proof assistant | Circuit |
| 2017 | cQASM | [146] | | Assembly language | Imperative |
| 2017 | Qiskit | [4, 232] | | Python | Imperative, functional |
| 2018 | IQu | [214] | | Idealized Algol | Imperative |
| 2018 | Strawberry Fields | [147, 148] | | Python | Imperative, functional |
| 2018 | Blackbird | [147, 148] | | Python | Imperative, functional |
| 2018 | QuantumOptics.jl | [157] | | Julia | Imperative |
| 2018 | Cirq | [271] | | Python | Imperative, functional |
| 2018 | Q# | [269] | | C# | Imperative |
| 2018 | $Q\|SI\rangle$ | [174] | | .Net language | Imperative |
| 2020 | Silq | [35] | | Python | Imperative, functional |

*Figure 626: a timeline of quantum programming tools. Source: [Quantum Software Engineering Landscapes and Horizons]() by Jianjun Zhao, 2020 (31 pages).*

# Quantum vendors development tools

Even before general-purpose quantum computers are operational on an exploitable scale, the software platforms battle has already begun. The major quantum computing players have almost all adopted an end-to-end vertical integration approach from quantum processors to development tools. This is particularly the case at IBM, Microsoft, Rigetti and D-Wave. This is well illustrated in the chart in Figure 627, which also describes the main development environments for quantum applications from Rigetti and IBM.

The vertical offering of the above-mentioned vendors often integrates a low-level quantum language, then a higher-level language similar to the macro-assembler of traditional computers, then an open sourced framework that can be most often used in Python with ready-to-use functions, a development environment, possibly a quantum gates graphical coding tool, and often some access to their cloud based quantum accelerators and simulators.

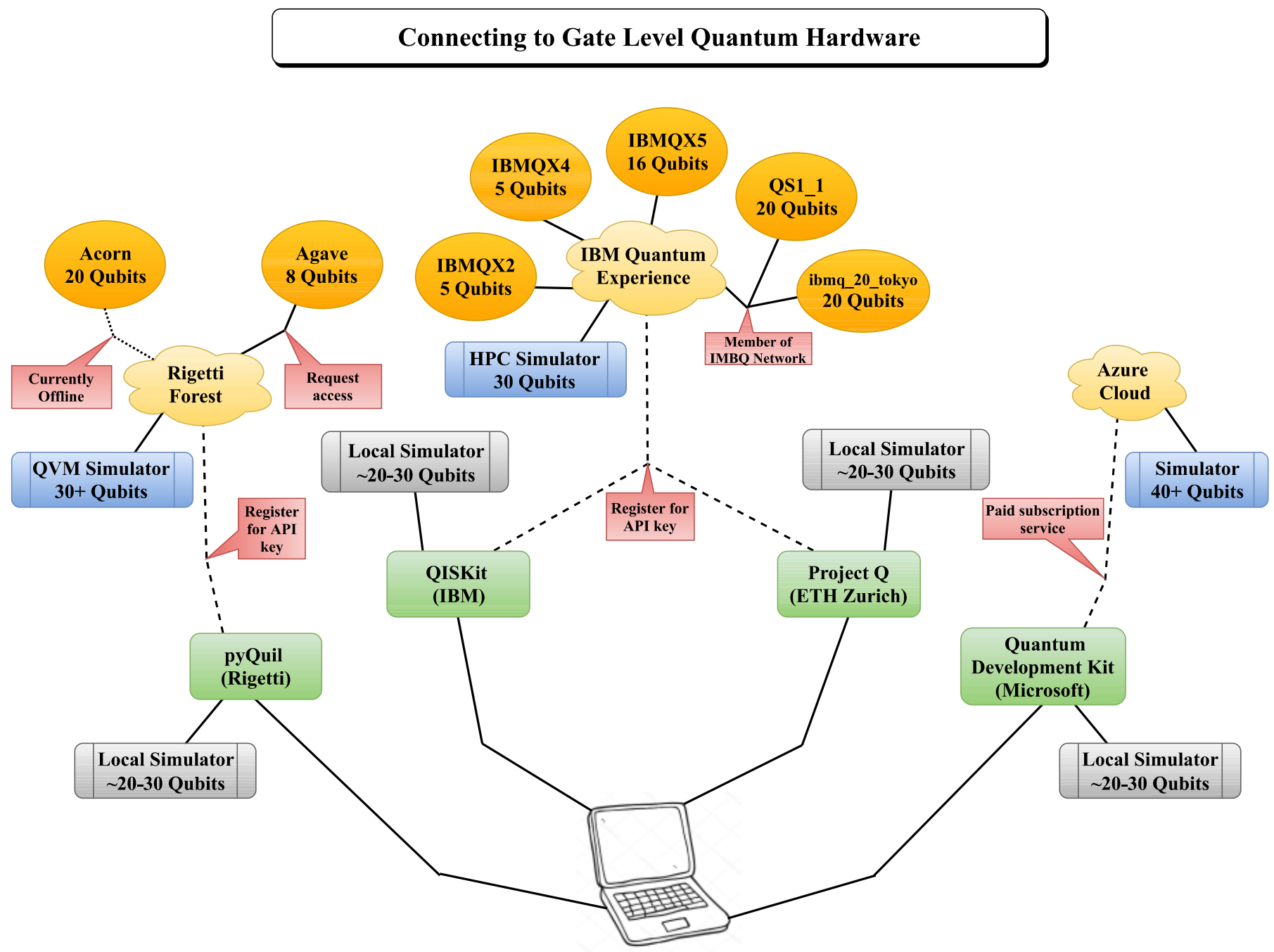

Figure 627: Source: Overview and Comparison of Gate Level Quantum Software Platforms by Ryan LaRose, March 2019 (24 pages).

One remaining tool to invent would be a higher level of abstraction tool to free developers from understanding the intricacies of quantum gates and interferences. Most recent supposed "higher level" languages are classical gate-based programming tools.

Another consolidation of these proprietary - although also open sourced - quantum software development platforms is shown in Figure 628.

| | IBM | rigetti | D-Wave | Xanadu | Google | Microsoft | aws | Atos |
|---|---|---|---|---|---|---|---|---|
| visual programming and integrated development environments | Q Experience | Forest | | | Quantum Playground | Visual Studio | | |
| thematic quantum libraries (chemisty, finance, machine learning, …) | QisKit Aqua | OpenFermion | Quadrant, Qsage, ToQ | PENNY LANE | OpenFermion | Quantum Chemisty PNNL | PENNY LANE | QLIB |
| generic quantum libraries / full-stack | QisKit | Grove QAOA | QUBO qbsolv | STRAWBERRY FIELDS | Cirq | Quantum Developer Kit | Braket SDK | pyAQASM |
| high level machine language (quantum circuits) | QisKit Terra | pyquil | QMASM | | | Q# | | AQASM |
| low level machine language | Open QASM | quil | QMI | Blackbird | many machine languages | | D-Wave, rigetti, IONQ, OQC | Cirq QPU |
| qubits and quantum gates | super-conducting | super-conducting | quantum annealing | photons | super-conducting | topologic, IonQ, Quantinuum | | any |

Figure 628: the various software stacks from large quantum vendors. (cc) Olivier Ezratty, 2022. Based on a schema found in Quantum Computing languages landscape by Alba Cervera-Lierta of the Quantum World Association, September 2018.

When all software tools are open source, it's not anymore a differentiating factor, or it is when you look at the fine prints. Is the open source software controlled by the vendor or by an independent third party?

Are all software tools really open sourced or just the lower layers with additional proprietary layers? Who are the main contributors to the open source tool? What is the exact open source license used?

### D-Wave

D-Wave proposes a complete range of software tools that have evolved a lot since its creation[1803]. The latest iteration of D-Wave's software platform is called **Ocean**. It includes low- and high-level building blocks for the development of quantum applications[1804]. The lowest level language is **QMI**, a kind of machine language for defining the links between the qubits to prepare the related Hamiltonian for an Ising model. QMI is usable from C, C++ Python and even Matlab, via the SAPI (Solver API) interface. Above QMI is a higher level of abstraction tool, **qbsolv**, an open source library launched in 2017.

It allows you to solve optimization problems by converting a QUBO (Quadratic Unconstrained Binary Optimization) problem into an Ising model ready to be processed by a D-Wave or even a classical computer.

Developers can also use the open source **QMASM** (Quantum Macro Assembler) language, which is a low-level language suitable for programming on a D-Wave annealer. It is a third-party tool coming from a D-Wave partner. Like qbsolv, QMASM is used to describe a Hamiltonian made of coupler-based qubit relationships. This method has a drawback: it is preferable to initialize the system in a state close to the search solution and this state can only be determined by classical calculations. It is in any case a very different programming model from the universal quantum gate model, even if there is a theoretical equivalence between quantum annealing and gate-based models as we saw in the previous section.

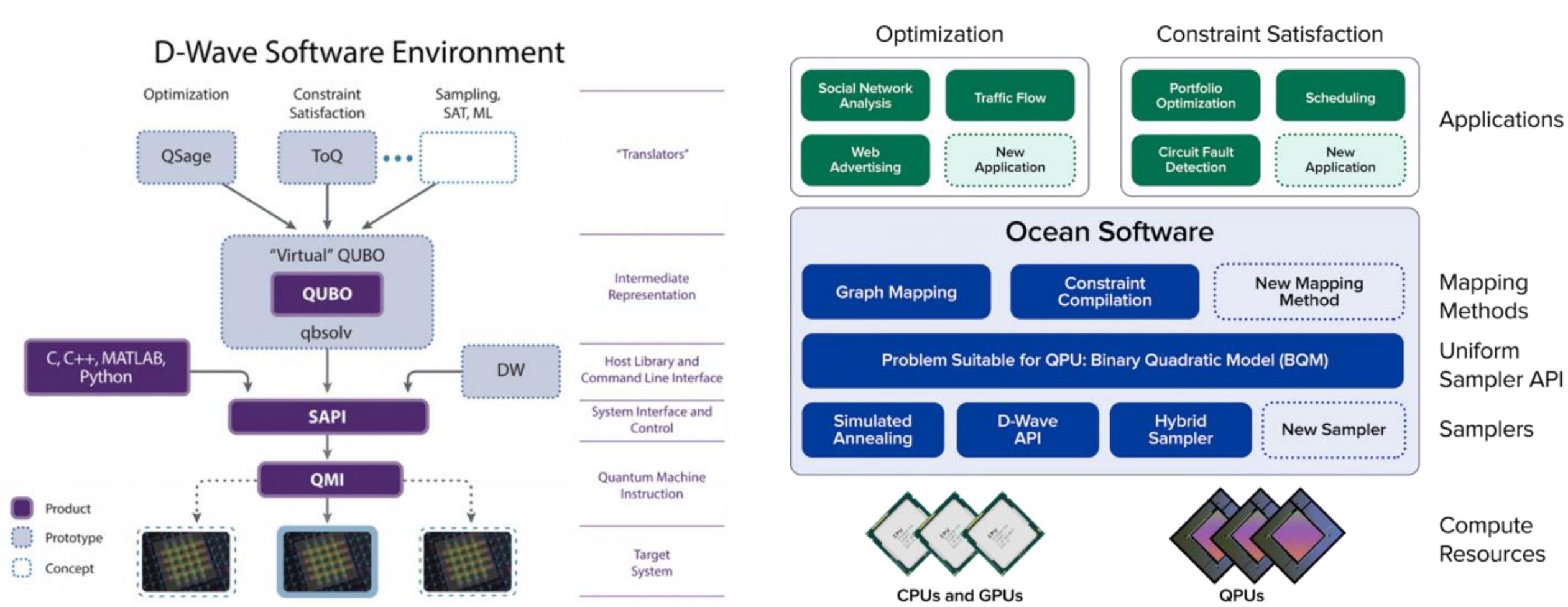

*Figure 629: D-Wave's software architecture components around the Ocean platform. Source: D-Wave.*

QMASM is also part of **Quadrant**, a comprehensive platform for the development of D-Wave's cloud-based solutions for machine learning launched by D-Wave in 2018[1805].

[1803] The source for the diagram on the left is D-Wave Initiates Open Quantum Software Environment, January 2017. And the one on the right is from: https://www.dwavesys.com/software.

[1804] D-Wave provides a very good document describing the problems that can be solved with their computers: D-Wave Problem - Solving Handbook, October 2018 (114 pages).

[1805] See D-Wave Announces Quadrant Machine Learning Business Unit, May 2018.

The D-Wave Ocean SDK also includes **Hybrid**, an open source framework for creating hybrid algorithms. We can add third party tools such as **QSage**, an optimization problems framework and **ToQ**, another framework for solving constraint satisfaction problems, as well as the SDK from **1Qbit**.

As of spring 2021, D-Wave, its partners and customers had prototyped over 250 algorithms and solutions. They have not necessarily generated any definite quantum advantage, but they do allow customers to learn quantum programming.

D-Wave's offering is mainly offered as a cloud-based resource, under the name **Leap**.

Leap V2 was launched in February 2020[1806].

It includes a new hybrid solver service that can handle optimization problems with up to 10,000 variables and a new interactive development environment using Python.

Prices range from $335 to $3000 per month for access to 10 to 90 minutes of quantum computing time.

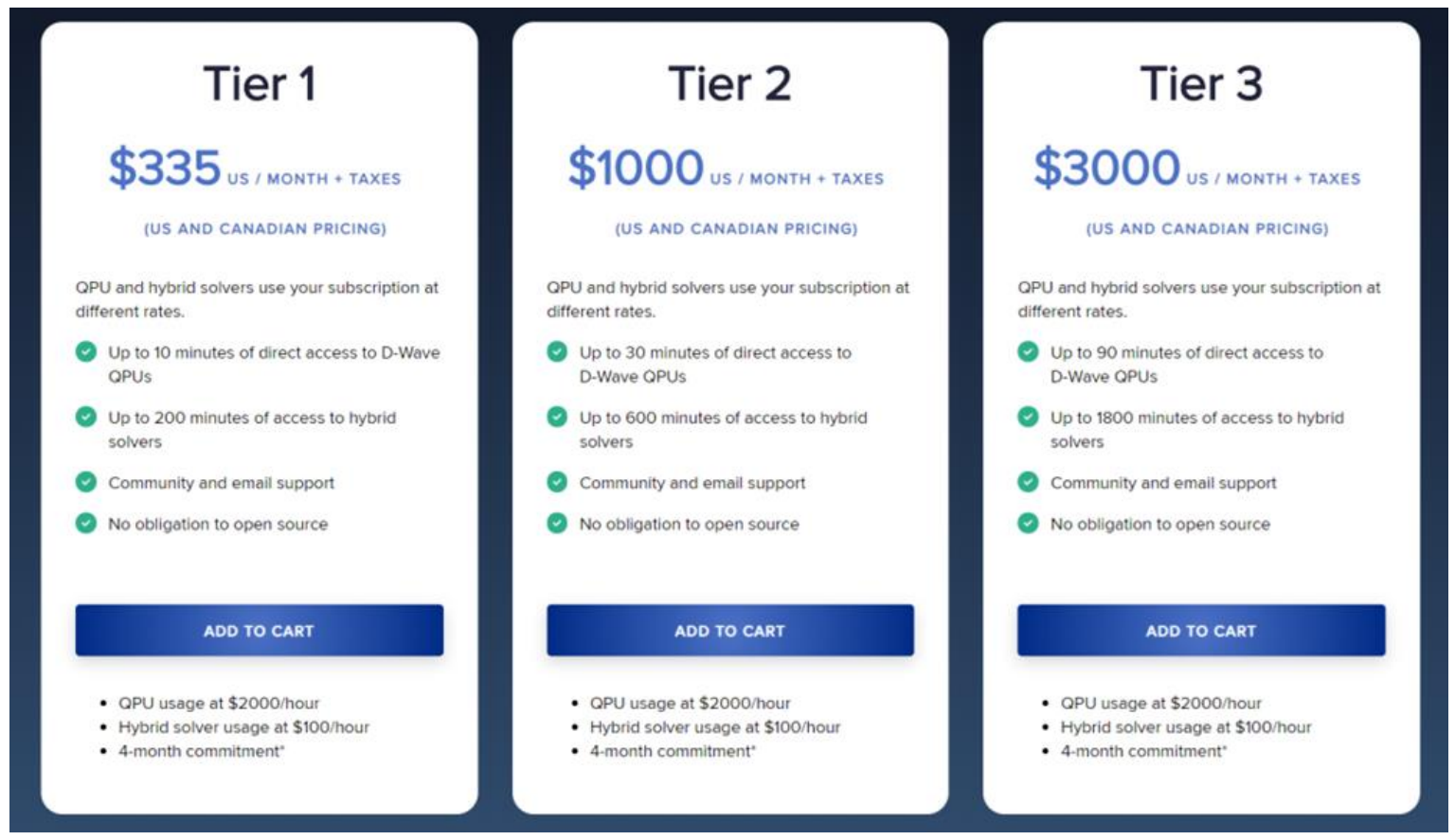

*Figure 630: D-Wave's Leap pricing as of 2021.*

### IBM

IBM's quantum software development platform is built around Qiskit and OpenQASM.

**OpenQASM** is an open sourced programming language introduced in 2017 that complements IBM's online graphical programming tool Q Experience Composer[1807]. The current version of OpenQASM is v3 and was codeveloped with AWS and the University of Sussex in the UK[1808].

It added support for arbitrary control flow, calling external classical functions, a description of quantum circuits at multiple levels of specificity, and extensions to drive gates timing, modifiers and even pulse control.

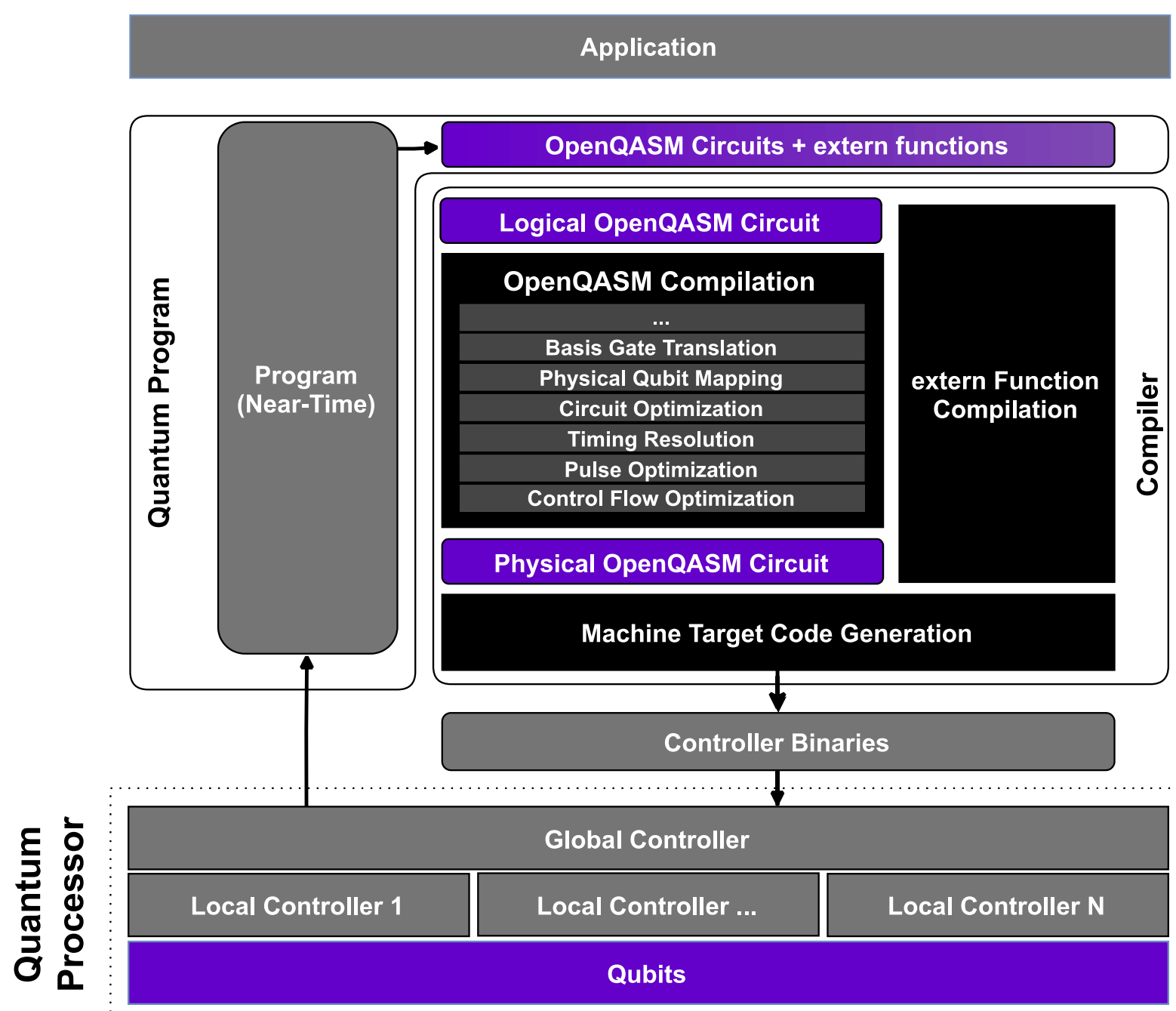

*Figure 631: IBM software architecture. Source: IBM.*

[1806] See D-Wave announces Leap2, its cloud service for quantum computing applications by Emil Protalinski, February 2020.

[1807] It is specified in Open Quantum Assembly Language, 2017 (24 pages), this document describing the many tasks performed by the associated compiler.

[1808] See OpenQASM 3: A broader and deeper quantum assembly language by Andrew W. Cross, Jay Gambetta et al, March 2022 (60 pages).

**Qiskit** is a high-level scripting library associated with OpenQASM. It can be used with Python, JavaScript and Swift (a general-purpose language from Apple) and on Windows, Linux and MacOS. It was launched in early 2017 and is also published in open source.

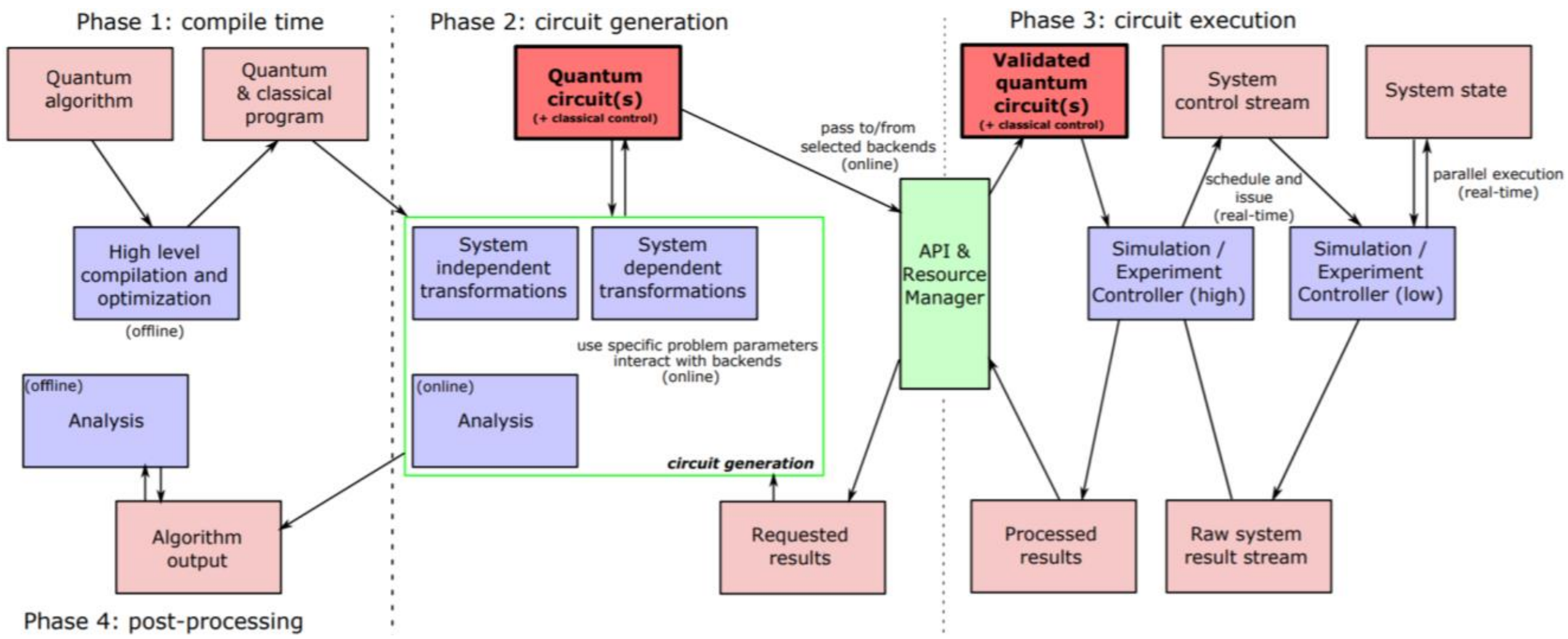

*Figure 632: Qiskit block-diagram of processes (blue) and abstractions (red) to transform and execute a quantum algorithm. Source: Open Quantum Assembly Language, 2017 (24 pages).*

Qiskit comes with numerous templates and sample codes to exploit a wide range of known quantum algorithms. These can be found in Qiskit documentation and Qiskit textbook on qiskit.org. It includes a graphical circuit-drawing function that generates a graphical visualization of quantum circuits using the open source document composition language LaTeX. Qiskit is or will be supported by other quantum computers vendors such as **IonQ** (USA) and **AQT** (Austria), both with trapped ions qubits, and **ColdQuanta** (USA) with cold atoms.

Qiskit is organized with modules around software building blocks:

- **Qiskit Terra** provides the circuit building and optimization functionalities and manages execution on the different backends like IBM's Qiskit Aer quantum simulator, and QPUs devices from various hardware providers including of course IBM.
- **Qiskit Aqua** is where use cases applications stood. It was deprecated to increase the library portfolio, with more specialized modules like Qiskit Finance, Qiskit Optimization, Qiskit Machine Learning and Qiskit Nature (for chemistry and materials science).
- **Qiskit Metal** is an open source EECAD (Electronic and Electrical Computer-Aided Design) used to design custom superconducting qubits chipsets and simulate their behavior and performance. It was launched in March 2021.
- **Qiskit Cold Atoms** supports quantum simulation models with fermionic modes and spins[1809].
- **Bosonic Qiskit** is a third-party extension which simulate bosonic qubits at the physical level, either purely photon based or in the quantum electrodynamic field, like with cat-qubits [1810].

Quantum code compilation takes place either on IBM's classic cloud-based HPC simulator or on a single quantum computer such as those from IBM that are available in the cloud with 5 and 7 qubits (free access), followed by 16, 27, 65 and 127 qubit versions (charged access) launched between 2019 and 2021.

---

[1809] See You Can Use Qiskit to Control Cold Atom Systems, May 2022.

[1810] See Bosonic Qiskit by Timothy J Stavenger, Eleanor Crane, Kevin Smith, Christopher T Kang, Steven M Girvin and Nathan Wiebe, DoE PNNL, NIST, Yale University, University of Toronto and University of Washington, September 2022 (8 pages).

The graphical **IBM Quantum Composer**, shown in Figure 634, is used to create quantum code graphically online and run it on a quantum emulator or on the various IBM quantum systems available online. It allows to interact indifferently with the text code on the right or with its graphical version in the middle. It shows vector states after running the code.

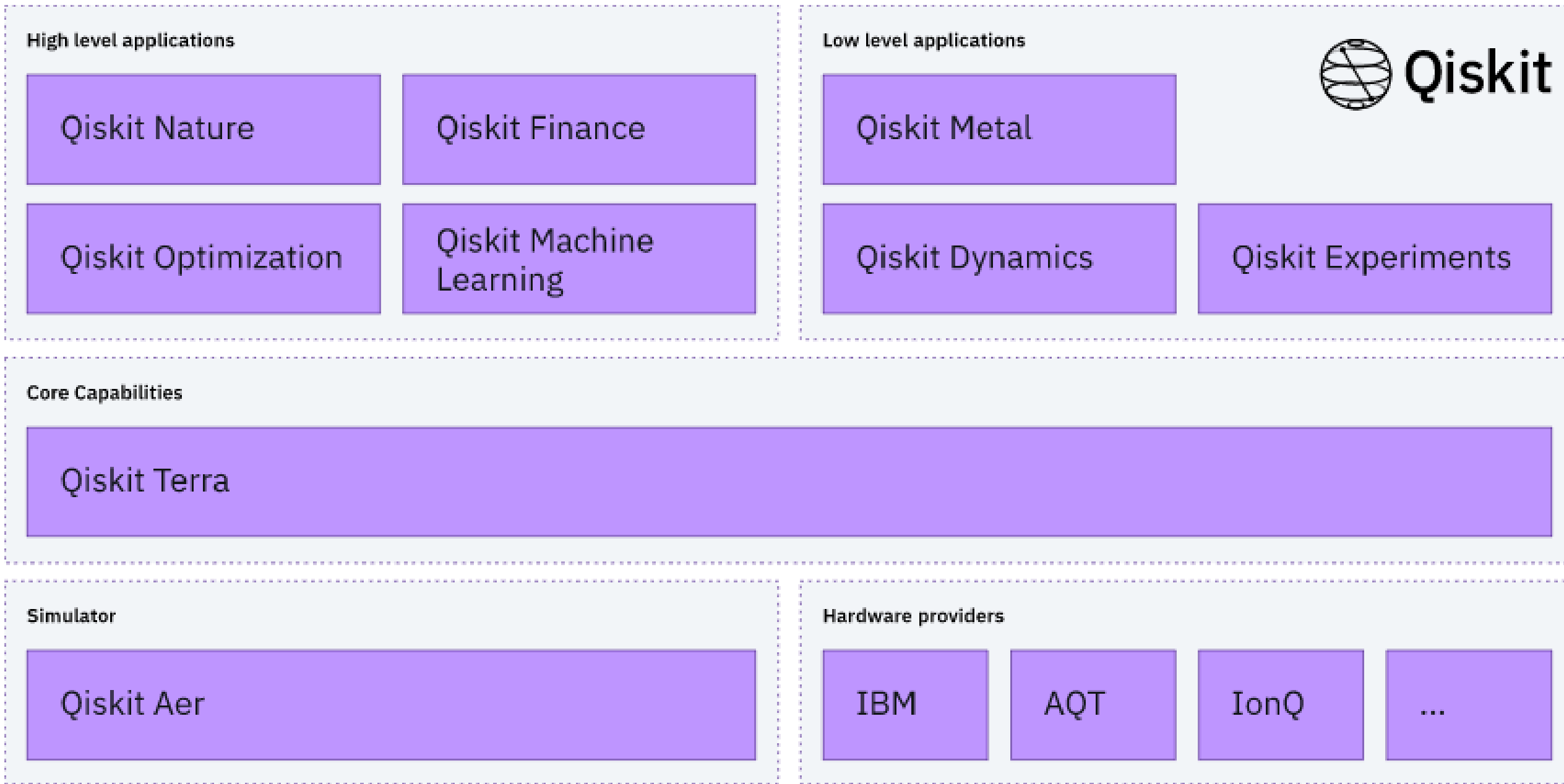

*Figure 633. Qiskit components, source qiskit.org.*

The 'ibm_qasm_simulator' emulates up to 32 qubits as if it was a real quantum device, including measurement pseudo-randomness, and also supports noise models injection. Circuits can be executed from 1 to 8192 shot(s). The local simulator included in Qiskit does not have this limitation but usually does not run efficiently above circa 15 qubits algorithms.

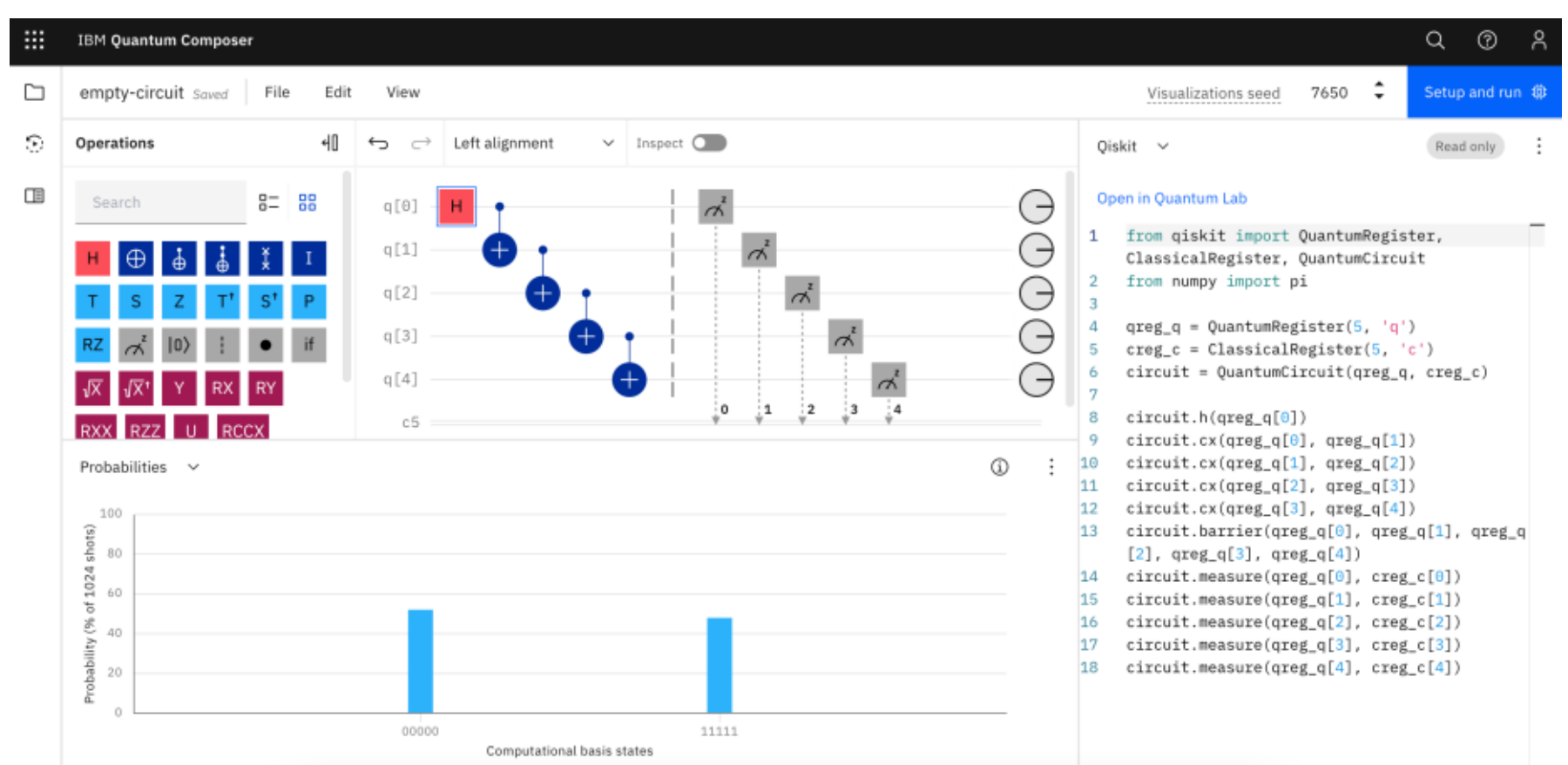

*Figure 634: IBM Quantum Composer, the graphical tool to design your quantum circuit, interacting with the language version on the left. Source: IBM Quantum Experience.*

Qiskit Aer also provides access to a 32 qubits state vector simulator, as well as 63, 100, 5000 qubit simulators, with some restrictions on the quantum gate sets. Since the batches are submitted one after the other, on the 9 open devices, and 400000+ registered users, one can wait up a long time, more than one hour for your code to be executed[1811].

Over the years since 2016, IBM has been building a user community, not only within the IBM Quantum Network with over 180 various partners, industry participants, startups and universities accessing premium devices and support, but also with a broad public, particularly students.

1811 There's a legal caveat in the tool terms of use: *"You may not use IBM Q in any application or situation where failure could lead to death or serious bodily injury of any person, or to severe physical or environmental damage, such as aircraft, motor vehicles or mass transport, nuclear or chemical facilities, life support or medical equipment, or weaponry systems"*. Given that with the few qubits offered, you can wonder how you could risk doing any of these nasty things.

Beyond free access to the quantum devices and simulators, IBM organizes Quantum Challenges, public and free Quantum Global Summer Schools in August since 2020 with two weeks of lectures and workshops, as an online and worldwide event, provides online Qiskit documentation in many languages including Japanese, Spanish, German and French, and online textbooks for learning quantum computing, a Qiskit channel on Youtube with learning content and weekly "seminar series" featuring scientists and technologists of this field. Setting the tone, a mobile application ("Hello Quantum") was launched for learning and playing with qubits and gates, as shown in Figure 635.

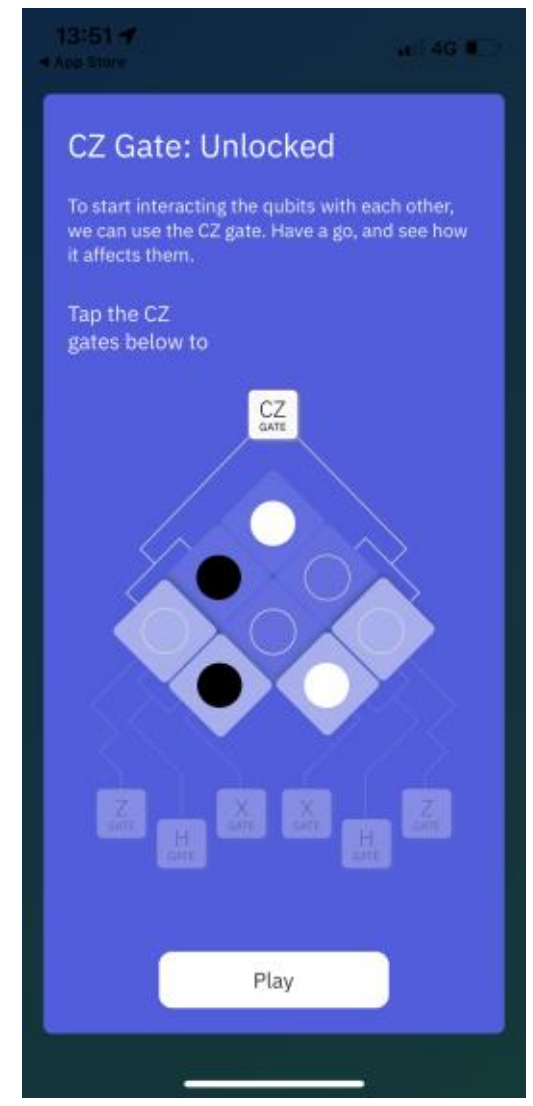

*Figure 635: Hello Quantum mobile app. Source: IBM.*

In February 2021, IBM complemented its hardware roadmap announced in September 2020 with a five-year software roadmap[1812]. Its main item was **Qiskit Runtime** bringing a 120x improvement on the time needed to run variational algorithms (such as VQE, which uses a classical optimizer iteratively with a call to the quantum processor until an exit condition is reached). With Qiskit Runtime both parts are run in the cloud, within the same job submission, avoiding returning to the queue for each iteration.

This x120 improvement can be broken down as follows: x4 from Qiskit itself, then x1.8 for the algorithm, x1.5 for system software, x4.2 with electronics control systems enabling faster readouts and, at last, x2.8 thanks to device fidelities. The rest of the announcement covered the willingness to address vertical markets with partners. They plan to package off-the-shelf libraries for natural science, optimization, machine learning, and finance with partners like **Strangeworks**.

IBM launched in March 2021 a certification program and test for Qiskit developers based on 60-question exam running on the Pearson VUE electronic testing solution[1813]. This is a typical tactic used to build technical communities. It was implemented a long time ago by Novell (Certified Novell Engineers) and Microsoft (Certified Professionals and Most Valuable Professionals programs).

Other software announcements were made in May 2022[1814]. They cover further improvements with Qiskit Runtime thanks to using "dynamic circuits" enabling a reduction of circuit depth and adding "threads", allowing the control of parallelized quantum processors. In 2024 and 2025, IBM will introduce error mitigation and suppression techniques. IBM will also improve Qiskit Runtime Service's primitives and process distribution across classical and quantum processors.

IBM will introduce "Quantum Serverless" in 2023 which transparently allocates classical and quantum processor resources to the developer. It will also enable "circuit knitting" and "entanglement forging", their techniques distributing large quantum circuits onto smaller circuits and reconstructing the results with consolidating their respected results. One of these techniques is "entanglement forging".

It is used to double the size of the quantum systems we could address with the same number of qubits, but it was used just in a particular case for the simulation of a single water molecule on 5 qubits[1815]. And it works only for weakly entangled states, those who do not bring a real exponential speedup!

---

[1812] See IBM's roadmap for building an open quantum software ecosystem by Karl Wehden, Ismael Faro and Jay Gambetta, February 2021.

[1813] See IBM offers quantum industry's first developer certification by Abe Asfaw, Kallie Ferguson, and James Weaver, IBM, March 2021.

[1814] See Expanding the IBM Quantum roadmap to anticipate the future of quantum-centric supercomputing by Jay Gambetta, May 2022.

[1815] See Scientists double the size of quantum simulations with entanglement forging by Robert Davis, IBM, January 2022 and Doubling the Size of Quantum Simulators by Entanglement Forging by Andrew Eddins, Sarah Sheldon et al, January 2022 (15 pages).

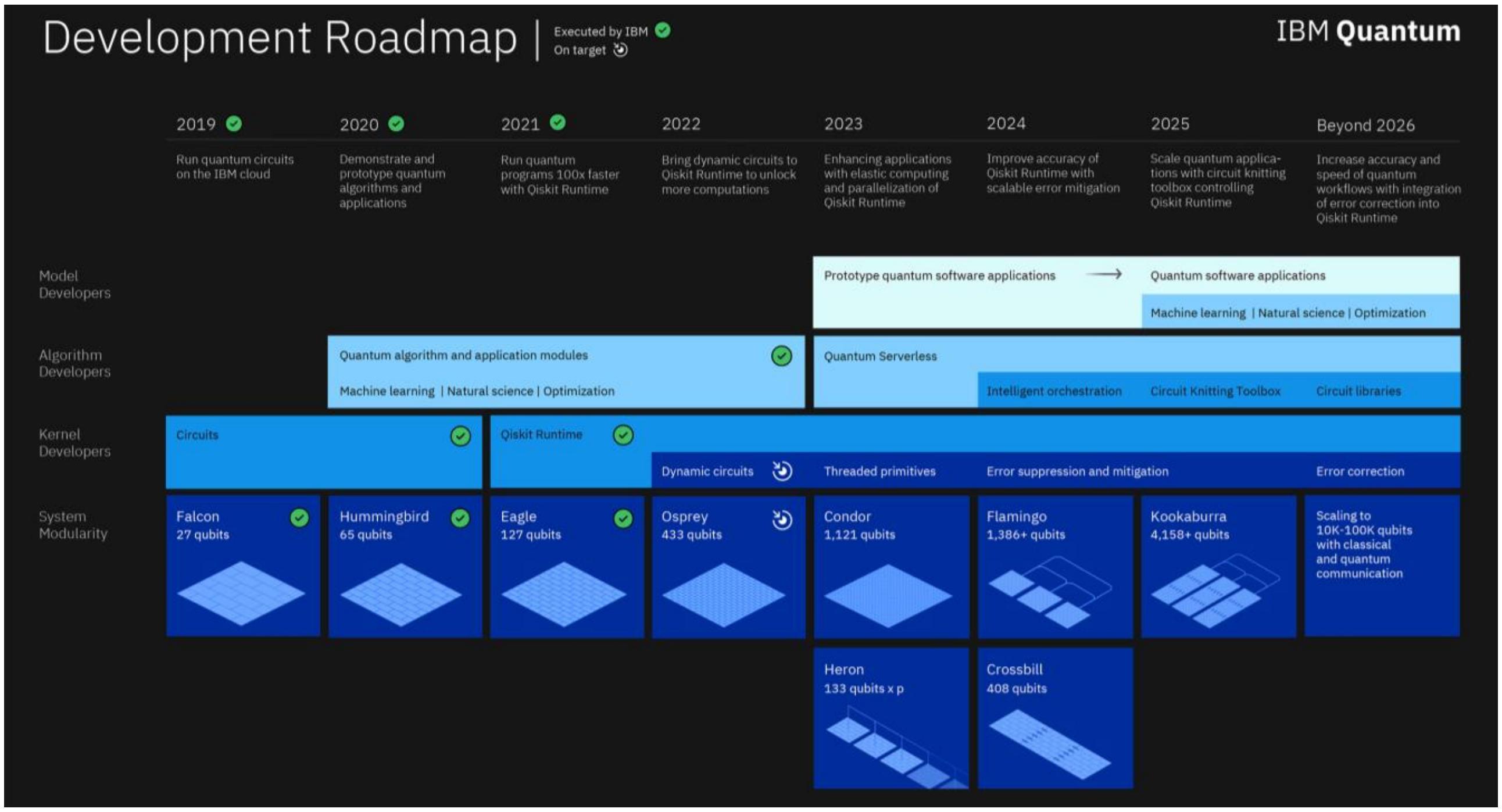

*Figure 636: IBM software and hardware roadmap as of May 2022. Source: [Expanding the IBM Quantum roadmap to anticipate the future of quantum-centric supercomputing](#) by Jay Gambetta, May 2022.*

## Rigetti

Rigetti proposes an integrated software development platform with the low-level language **Quil** that supports a mixed classical and quantum memory model[1816]. It runs on Windows, Linux and MacOS. The language uses the gates class to describe operations to be performed on qubits, indexed from 0 to n-1, for n qubits and with quantum gates.

The language allows you to create conditional programming based on the qubits state. It is completed by the open source library **pyQuil** launched in 2017 which includes the Grove library of basic quantum algorithms ([documentation](#)). It can be used with the Python programming language. The high level pyQuil (assembler) generates the low-level Quil language (machine code). Figure 638 contains a simple example with a single qubit activated by a Hadamard gate that creates a superposition of $|0\rangle$ and $|1\rangle$ to create a quantum random number generator.

pyQuil generates Quil
from pyquil.gates import X, CNOT, H, Z, RX, I
from pyquil.api import QVMConnection
from pyquil.quil import Program
import numpy as np
qvm = QVMConnection()
alice_register = 0
ancilla_register = 1
flip_correction_branch = Program(X(1))
phase_correction_branch = Program(Z(1))
prog = (Program()
.inst(H(0))
.inst(CNOT(0, 1))
.inst(RX(0.2 * np.pi, 2))
.inst(CNOT(2, 0))
.inst(H(2))
.measure(0, alice_register)
.measure(2, ancilla_register)
.if_then(alice_register, flip_correction_branch)
.if_then(ancilla_register, phase_correction_branch))
qvm.run_and_measure(prog, list(prog.get_qubits()), trials=10)
H 0
CNOT 0 1
RX(pi/5) 2
CNOT 2 0
H 2
MEASURE 0 [0]
MEASURE 2 [1]
JUMP-WHEN @THEN1 [0]
JUMP @END2
LABEL @THEN1
X 1
LABEL @END2
JUMP-WHEN @THEN5 [1]
JUMP @END6
LABEL @THEN5
Z 1
LABEL @END6

*Figure 637: pyQuil language example and the lower level Quil language generated on the right. Source: Rigetti.*

Used iteratively in a classical loop, the program can generate a random series of 0s and 1s with a 50% chance of having either one allowing to create a completely random single binary code.

[1816] It is documented in [A Practical Quantum Instruction Set Architecture](#), 2017 (15 pages).

```
# random number generator circuit in pyQuil
from pyquil.quil import Program
import pyquil.gates as gates
from pyquil import api

qprog = Program()
qprog += [gates.H(0),
          gates.MEASURE(0, 0)]

qvm = api.QVMConnection()
print(qvm.run(qprog))
```

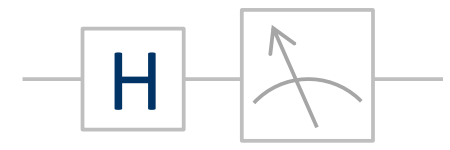

Hadamard gate on qubit 1

qubit readout on superposed state

readout output

**Listing 2:** pyQuil code for a random number generator.

*Figure 638: Hadamard gate programmed with pyQuil. Source: Rigetti.*

Rigetti offers the execution of quantum programs in its cloud systems and on conventional simulators via its QVMs, for **Quantum Virtual Machines**[1817]. Since 2020, it's also available on Amazon Braket cloud services. It is usable from the **Forest** development environment proposed by Rigetti. These tools are open source, but not cross-platform.

At the end of 2017, Google and Rigetti launched the open source initiative **OpenFermion**.

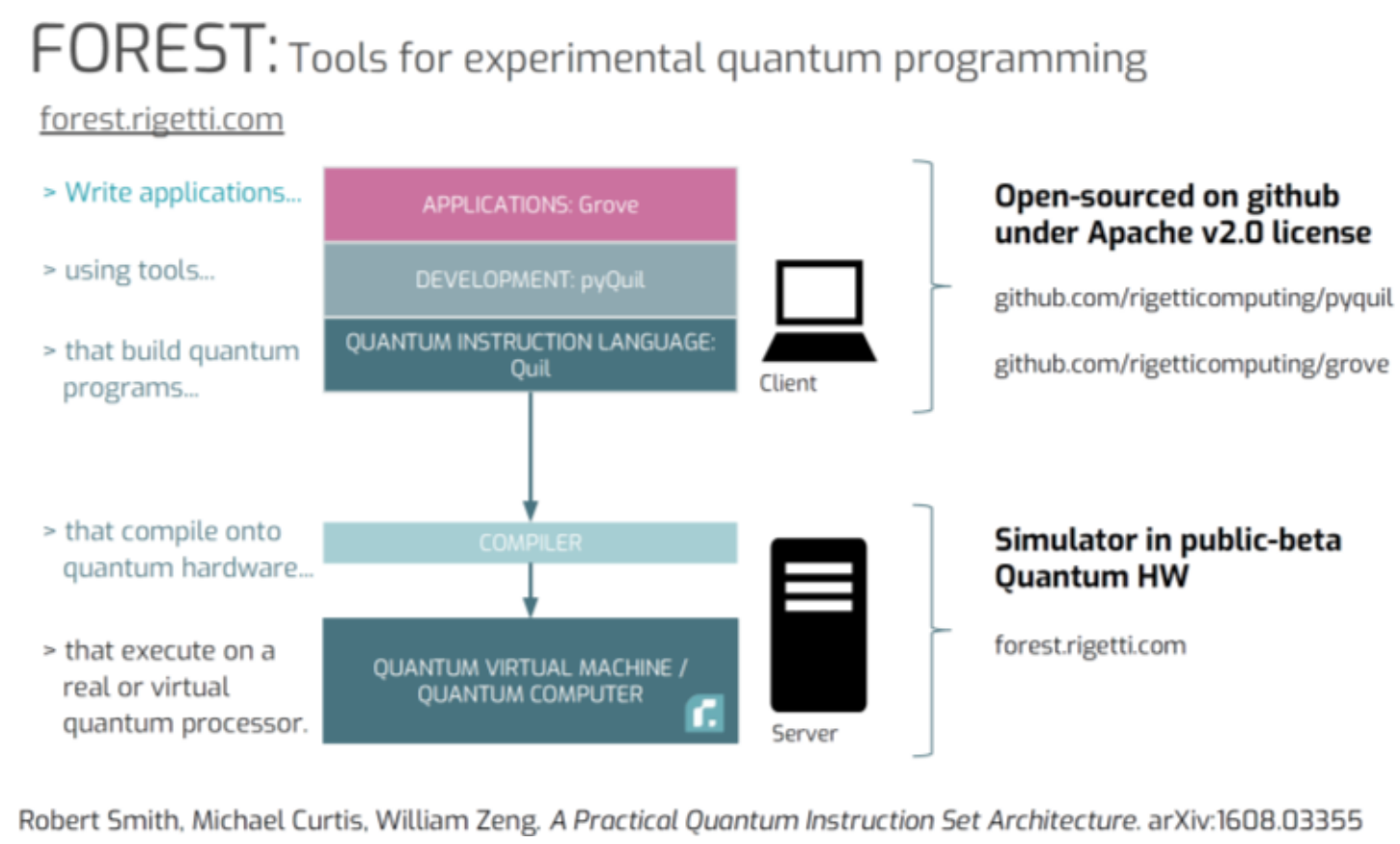

*Figure 639: Rigetti Forest software platform. Source: Rigetti.*

This framework developed in Python exploits research work from the Universities of Delft and Leiden in the Netherlands. It is a software solution for the creation of quantum algorithms for the simulation of chemical functions supporting any quantum computer, from Universal Quantum Computers to D-Wave annealers.

It complements Atos[1818]. In 2018, Rigetti finally launched a Quantum Algorithm Contest with a $1M prize, but with an interesting bias, comparing the creators of quantum algorithms with others seeking to create equivalents running on conventional computers.

The process could last 3 to 5 years and looks like the XPrize process[1819].

At last, Rigetti is also promoting **Quantum Programming Studio**, a web based interactive programming tool that can run your code on a Rigetti quantum computer in the cloud.

### Google

In addition to OpenFermion which is a high-level framework, Google launched on July 19, 2018, its own quantum framework **Cirq**, of course also in open source. It is a Python framework.

---

[1817] This is documented in pyQuil Documentation, June 2018 (120 pages) which contains many code examples like the one *above*.

[1818] See the announcement in October 2017, OpenFermion: The Electronic Structure Package for Quantum Computers, 2018 (19 pages) and Openfermion documentation.

[1819] See Can You Make A Quantum Computer Live Up To The Hype? Then Rigetti Computing Has $1 Million For You by Alex Knapp, Forbes, October 2018.

Since Google's superconducting Sycamore systems are not available on the cloud, Cirq is mainly used on a cloud simulator provided by Google[1820]. A tool for compiling OpenFermion code in Cirq is also proposed. It also supports IonQ trapped ions qubits that are supported in Google Cloud since 2021. It supports also Pasqal cold atoms systems and Rigetti superconducting qubits.

In March 2020, Google launched **TensorFlow Quantum**, an extension of the famous open source machine and deep learning framework. It provides hybrid classical/quantum computing functions for machine learning[1821]. Of course, the library supports Cirq.

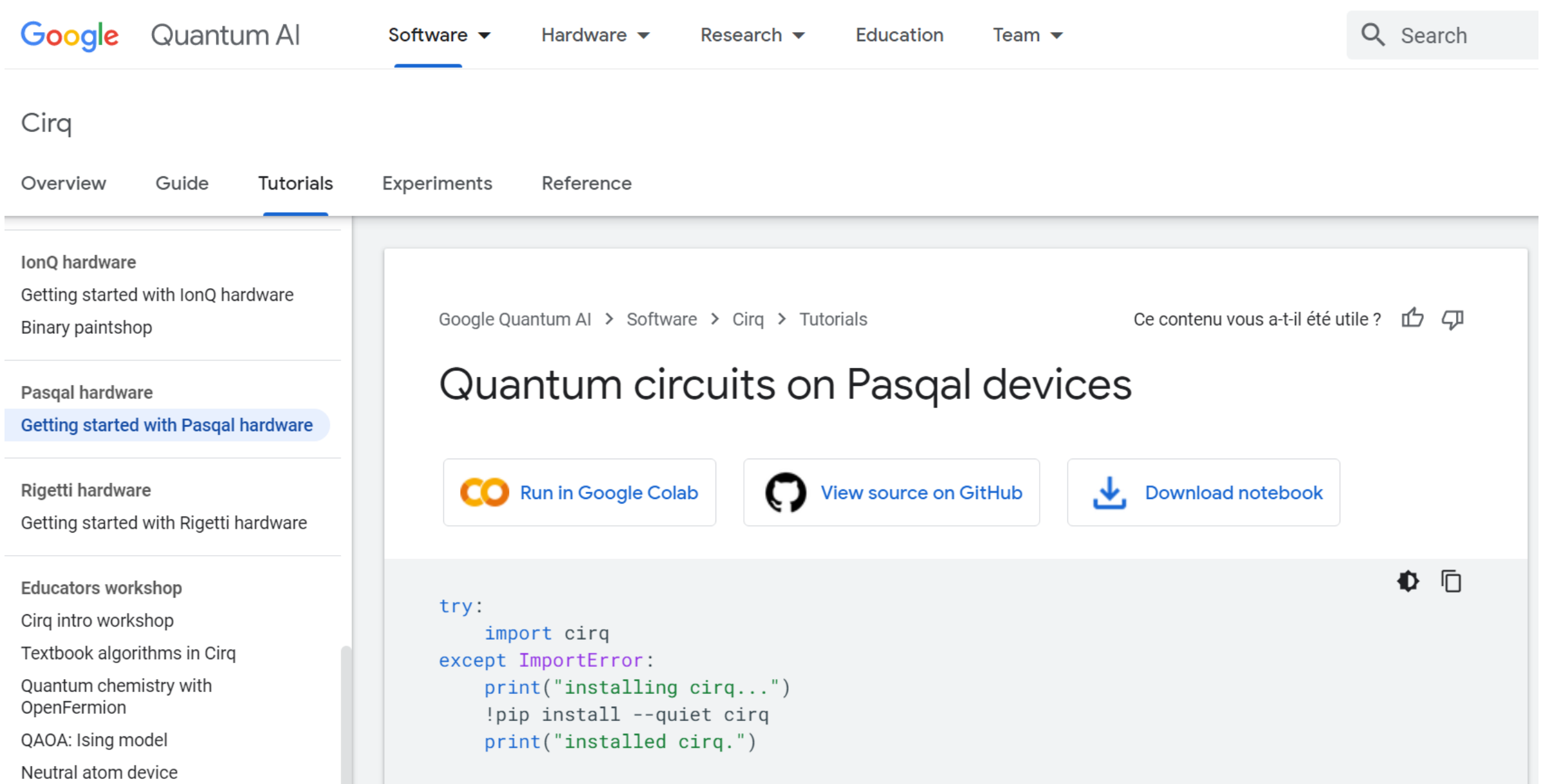

*Figure 640: Cirq support Pasqal cold atoms computer circuits. Source: Google Cirq tutorials.*

It is adapted to quantum simulators running on classical computers based on CPUs, GPUs and TPUs (Tensor Processing Unit, the specialized AI processors running in Google's data centers).

Eventually, QPUs (Quantum Processing Units) will be supported. Why doesn't Google use its 53-qubit quantum computer? Because it is a research object and not yet a production tool that could be integrated at this stage in a cloud offer. Meanwhile, you can access IonQ trapped ions in Google's cloud!

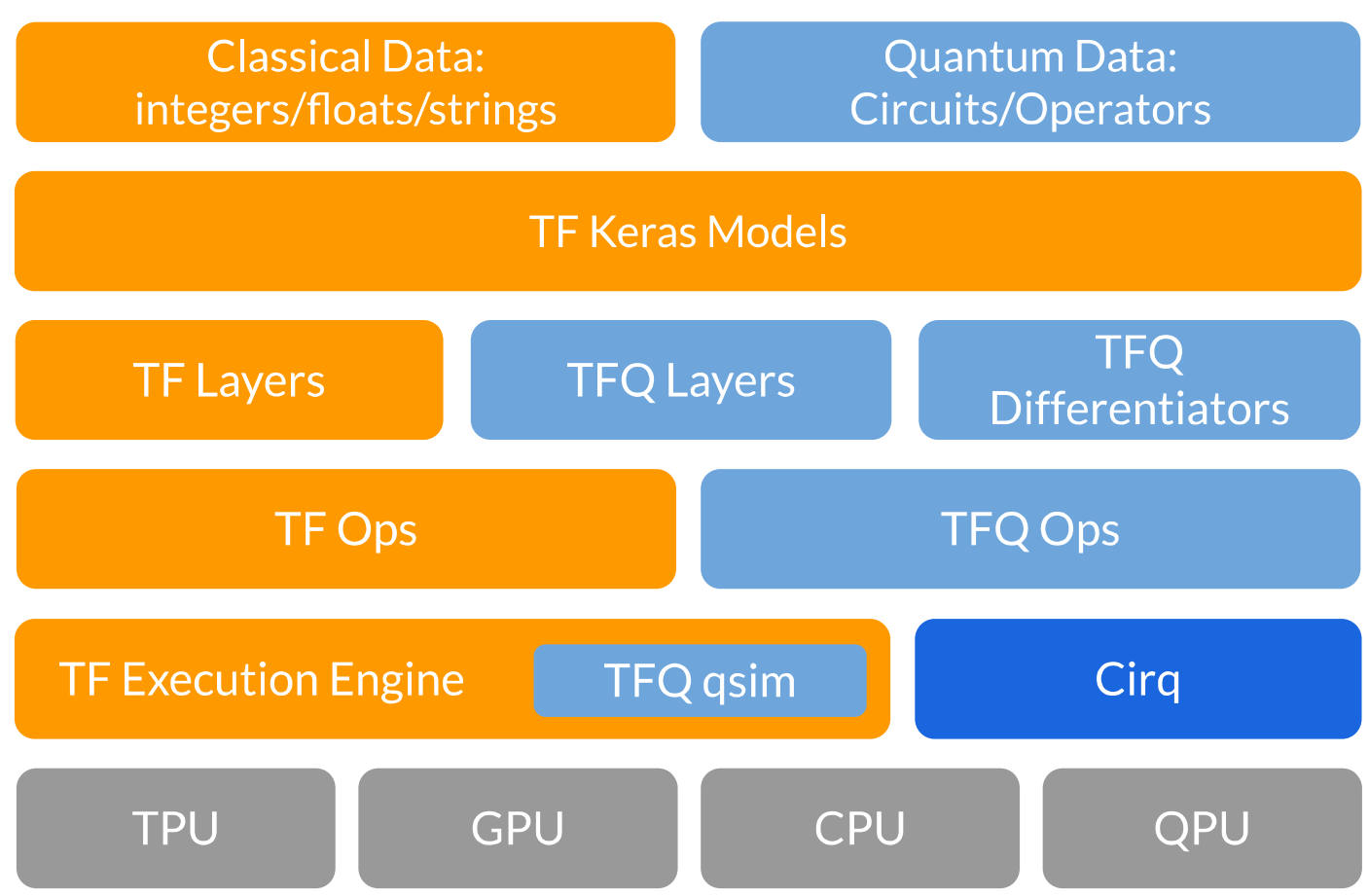

*Figure 641: Google's hybrid quantum classical software architecture. Source: [TensorFlow Quantum: A Software Framework for Quantum Machine Learning](#) by M Broughton et al, 2020 (39 pages).*

---

[1820] See explanations in Google Cirq and the New World of Quantum Programming by Jesus Rodriguez, July 2018.

[1821] See TensorFlow Quantum: A Software Framework for Quantum Machine Learning by M Broughton et al, 2020 (39 pages, and associated video).

## Microsoft

Microsoft's quantum efforts are proposed under the umbrella of Azure Quantum. It embeds various efforts ranging from fundamental hardware research to commercial offerings. Research covers Majorana fermions. Commercial offerings contain a wealth of software development solutions and a cloud offering hosting emulation software and third party quantum hardware vendors QPUs.

Let's try to decompose the Microsoft quantum platform.

**Q# language** is their cornerstone open source quantum language[1822] given Azure Quantum also supports Qiskit and Circ frameworks. Development is usually done with VScode, a lightweight and extensible integrated development environment. It's using Python as a scripting language which calls various modules developed in various languages including Python.

**Quantum Development Kit** is a software kit for Q# that support the development cycle and execute quantum code either on quantum hardware or on quantum simulators, all supporting Jupyter Notebooks. The QDK can run on your local computer and run with your preferred interactive development environment (Visual Studio or another). It contains several libraries supporting Hamiltonian (gate-based) simulations, amplitude amplification (for Grover search), phase estimation, arithmetic and quantum error correction codes.

**Quantum Katas** is an open source project launched in July 2018 that contains examples of quantum Q# code integrated into interactive tutorials[1823]. In December 2018, Microsoft introduced a chemical simulation library co-developed with Pacific Northwest National Labs, an equivalent of OpenFermion, which is co-developed by Rigetti and Google[1824]. The library complements PNNL's NWChem quantum chemistry simulation software package.

**QIR** (Quantum Intermediate Representation) is an intermediate representation for quantum programs launched in September 2020, serving as a layer between gate-based quantum programming languages like Q# and target quantum computers. It's based on **LLVM** open sourced intermediate language that was created in 2000 at the University of Illinois and is handled by the LLVM Foundation run by Tanya Lattner. It can also be used to run code on an emulator. The support is done with a compiler extension supporting that QIR with Q# but it can support other languages and frameworks like Qiskit. It is used by Azure Quantum to support the various hardware platforms it serves (IonQ, Honeywell, QCI). But it seems, not yet by any other vendor. It became the **QIR Alliance** in November 2021, to promote the adoption of QIR. It is now part of the Linux Foundation. The Alliance founding members are Honeywell, Microsoft, the DoE Oak Ridge National Laboratory, Quantum Circuits Inc. and Rigetti Computing[1825].

---

[1822] There's a long history behind Q#. Microsoft's first forays in quantum software developments started with the **LIQUi|>** extension of the **F#** scripting language which allowed to simulate quantum programs. In December 2017 was launched **Q#**, using a syntax derived from Microsoft's C# language. See Q#: Enabling scalable quantum computing and development with a high-level domain-specific language, 2018 (11 pages). It involves Alain Sarlette, Anthony Leverrier, Eric Fleury, Hélène Robak and Laurent Massoulie from Inria and Nicolas Delfosse from Microsoft Research.

[1823] See Learn at your own pace with Microsoft Quantum Katas, July 2018.

[1824] See Simulating nature with the new Microsoft Quantum Development Kit chemistry library, December 2018. PNNL is a research laboratory co-funded by the US Department of Energy and operated by the non-profit foundation Battelle Memorial Institute. Battelle operates numerous US laboratories such as Lawrence Livermore National Laboratory, Los Alamos National Laboratory and Oak Ridge National Laboratory.

[1825] See New Quantum Intermediate Representation Alliance Serves as Common Interface for Quantum Computing Development, November 2021.

**Quantum simulation** is proposed in several fashions, with full state simulation (limited to 30 qubits, operating the full quantum state vector), sparse simulation (sparse matrices and quantum states), a Toffoli simulator limited to X, CNOT and multicontrol X operations) and a noise simulator. They also have a resources estimator that computes the quantum resources necessary to run some quantum code (number of qubits, gates, CNOT, T and R gates, measurements, code depth).

**Quantum hardware** access is provided in Azure Quantum with an ever increasing breath of QPUs, starting with IonQ, Quantinuum, Rigetti and Pasqal. Access pricing is calculated at the quantum gate level with a different formula for each hardware. A single qubit gate costs \$0.00003 and two-qubit gates costing \$0.0003 on IonQ systems.

**Quantum Inspired** software models are also proposed in Azure Quantum. They are implemented classically. It contains various software libraries: Parallel Tempering (sort of digital annealing), Simulated Annealing (stochastic simulation method that mimics annealing), Population Annealing (walker simulation), Quantum Monte Carlo (a quantum-inspired optimization) and Tabu Search (an heuristic search algorithm used to solve QUBO optimization problems).

All that is summarized below in Figure 642 in an outdated slide from 2021. Microsoft has a lot to improve in it marketing of the Azure Quantum Platform.

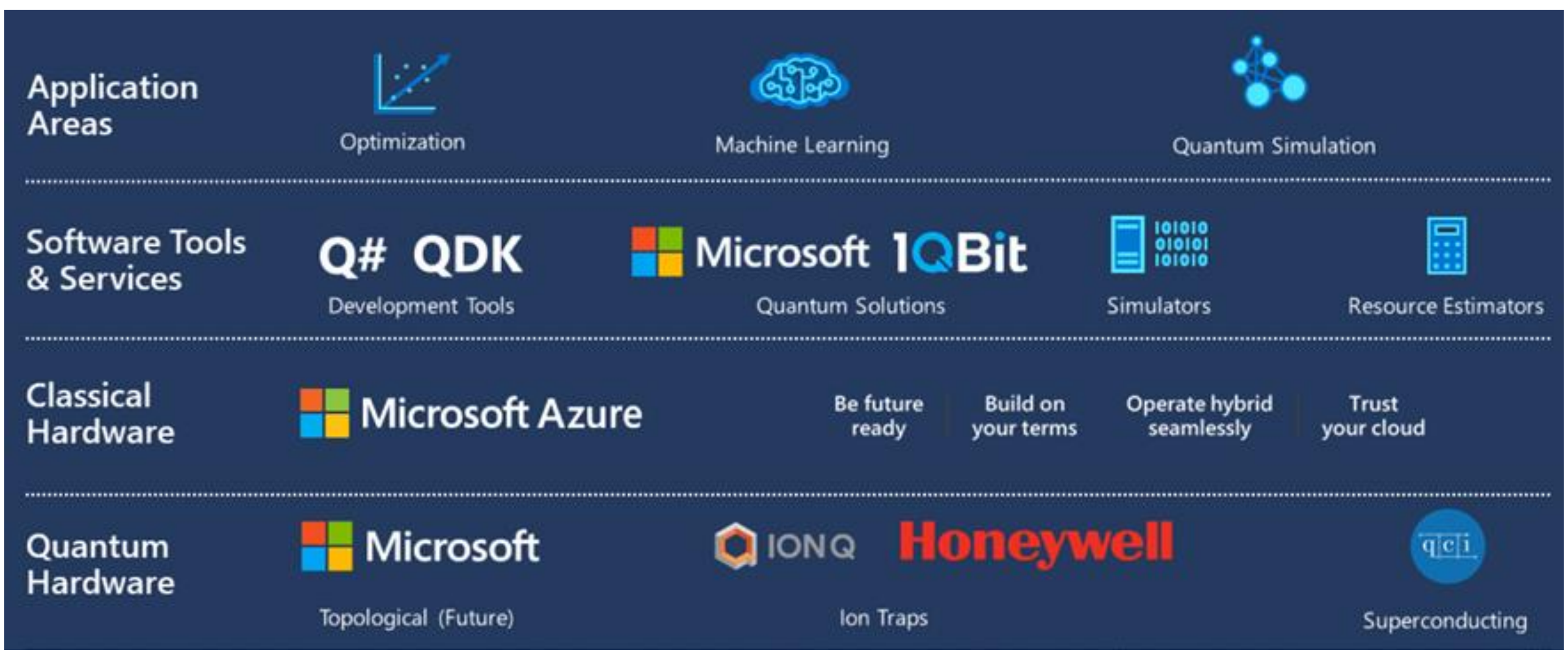

*Figure 642: Microsoft Azure Quantum overview. Since then, some new hardware vendors have been added or announced like Rigetti and Pasqal. QCI that is in this slide was announced in 2019 but never delivered a functional QPU. Source: Microsoft, 2021.*

### Amazon

Amazon's quantum software offering is organized in their Braket platform. It contains both a custom hardware independent development framework as well as the PennyLane framework from Xanadu, and all the tools to submit quantum code to the various AWS supported systems (IonQ, D-Wave, Rigetti and soon OQC) as well as their own classical computing emulators for testing and learning purpose.

### IonQ

Like Rigetti, IonQ also has its own "full stack" software offering adapted to their trapped ions quantum computer architecture and proposed in the cloud. It's also offered in Amazon and Microsoft's quantum cloud services.

### Intel

At this stage, Intel is not very advanced in the development of quantum software. They have created a quantum emulation software for classical computers, IQS, the first two authors working at Intel and the last one at Harvard. It can simulate up to forty qubits[1826].

### Huawei

At the end of 2018, Huawei launched its own quantum application development framework, compatible with ProjectQ, and including a graphical interface for algorithm creation. All this is integrated into their HiQ cloud-based quantum emulation service [1827]. It is provided free of charge for simulating up to 38 qubits. It can also simulate up to 81 qubits with a processing depth of 30 and 169 qubits with a computing depth of 20.

### Atos

Atos is not a manufacturer of quantum computers. Their partnerships with various players such as Finland's IQM and France's Pasqal suggest that at some point, they will integrate hybrid solutions mixing their classical server nodes and HPCs and quantum accelerators.

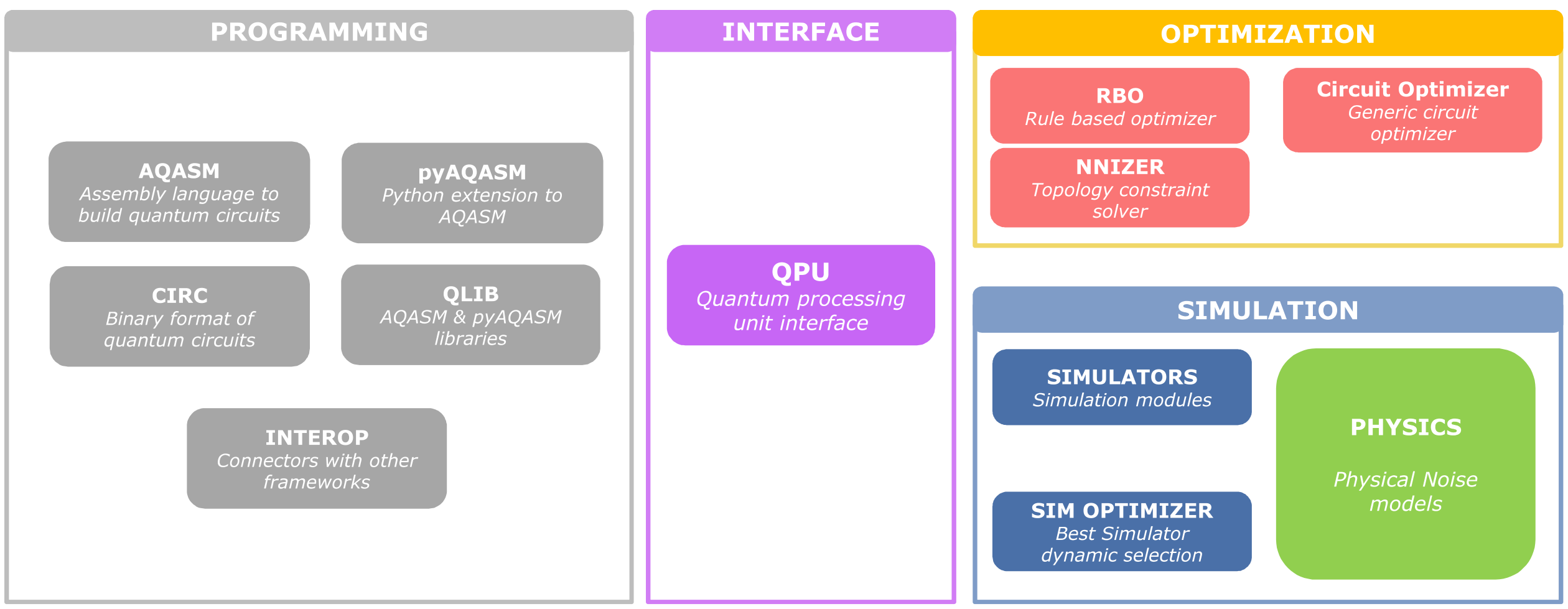

*Figure 643: Atos software platform around pyAQSM.*

For the time being, they are offering a quantum software emulation solution running on Intel processor machines and with their own optimized memory architecture, QLM.

They simulate 30 to 40 qubits depending on the QLM configuration and are agnostic regarding the physical qubit that are emulated. Since July 2020, they sell QLMe, a faster version of this emulator that runs Nvidia V100s GPUs. They can simulate a system's density matrix to model quantum noise up to 20 qubits.

aQASM (Atos Quantum Assembly Programming Language) is a programming language that complements Python to create quantum algorithms executable on QLM emulator or on any physical quantum computer architecture with universal gates. The language allows to define quantum gates using other quantum gates, equivalents of objects, functions or macros in traditional programming[1828].

---

[1826] Documented in qHiPSTER: The Quantum High Performance Software Testing Environment by Mikhail Smelyanskiy, Nicolas Sawaya, and Alán Aspuru-Guzik, 2016 (9 pages)

[1827] See Huawei Unveils Quantum Computing Simulation HiQ Cloud Service Platform, October 2018.

[1828] Source of the diagram: Atos QLM, a future-proof approach to quantum computing by Christelle Piechurski, Atos, March 2018 (26 slides).

Their compiler is also very efficient at optimizing code and removing useless combinations of quantum gates due to the addition of many SWAP and CNOT gates in relation to qubits connectivity limitations with their Lazy Synthesis feature[1829]. They can also interoperate with Qiskit, PyQuil and Cirq. A Qiskit code can be used to drive a QLM emulation backend and be used as a QPU by a QLM appliance.

aQASM is based on the OpenQASM standard language. It is completed by the PyAQASM Python library used to generate aQUASM files. The language helps programing the repetitive execution of looped gates and create reusable functions.

The aQASM code compiler generates CIRC binary code that is the low-level pivot language, which is then converted into the control language for specific universal quantum computers or for simulation supercomputers via the Quantum Processing Unit Interface (QPU). It is complemented by various optimization plugins that eliminates useless gates and tunes the generated low-level code for the targeted quantum accelerator hardware architecture.

Atos's QLM now support the three quantum computing paradigms: gate-based control, quantum simulations (with Pasqal) and quantum annealing (with D-Wave). They can classically simulate both noisy and noiseless hardware systems and at a low hardware level they call "below the gate".

At last, myQLM Power Access is a software tool that extends myQLM to submit quantum jobs to a QLM Appliance.

## Cloud quantum computing

A large share of quantum computers is intended to be offered through cloud services. It's even got a specific name: **QCaaS** (Quantum Computing as a Service). There are various estimates for the cloud quantum computing market including a very optimistic one of $26B by 2030 by The Quantum Daily[1830].

Let's mention a few of the many technical and economic reasons behind the ineluctable cloudification of quantum computing access. Most quantum computers are relatively expensive devices, costing at least a couple $M. D-Wave systems are sold at a price of about $14M. These are rapidly changing devices with one generation cancelling the previous one nearly each and every year. Also, many quantum algorithms are hybrid in nature, requiring a nearby classical computer, if not an HPC or supercomputer. All of this mandates some mutualization. Making a quantum computer accessible in the cloud requires putting in place a software infrastructure reminiscent of the old mainframe days. Indeed, QPUs are not "multitasking" machines. They are fed by classical computers through a queueing system in "batch mode". The batches first compile the code, execute it several times, usually a couple thousand times, then the result is sent in asynchronous mode to the user[1831]. Cloud quantum solutions usually also provide access to emulation solutions. This can be found at IBM, Microsoft, Google, Alibaba and, Huawei to name a few. If the cloud is to run a hybrid algorithm, it must also provision classical datacenters or HPC resources and synchronize their availability with the related QPU (and right now, we're using only one QPU at a time).

---

[1829] See Architecture aware compilation of quantum circuits via lazy synthesis by Simon Martiel and Timothée Goubault de Brugière, December 2020 (32 pages).

[1830] See Report: Quantum Computing as a Service Market to Hit $26 Billion by End of Decade by The Quantum Daily, August 2021 and Quantum Computing as a Service Market Sizing - How We Did It by The Quantum Daily, August 2021. These forecasts are fairly inconsistent with other quantum computing forecasts mentioned in this document, page 526, planning a total $2B in 2030.

[1831] See Quantum Computing in the Cloud: Analyzing job and machine characteristics by Gokul Subramanian Ravi et al, University of Chicago and Super.tech, March 2022 (13 pages).

When would customers prefer to purchase a QPU and run it on-premises in their own datacenters? It would make sense when economies of scale are large within the customer or if he handles very sensitive data and processes. The first on-premises installations will probably be deployed for military and intelligence use cases.

The first company to launch a quantum cloud offering was **IBM**, which started to make its first QPUs online in 2016. Is now proposes cloud access to over 24 QPUs with 1 to 127 superconducting qubits as of September 2022. Half of their QPUs with fewer than 15 qubits are accessible for free. **D-Wave** launched its Leap cloud solution in 2018 with its quantum annealers.

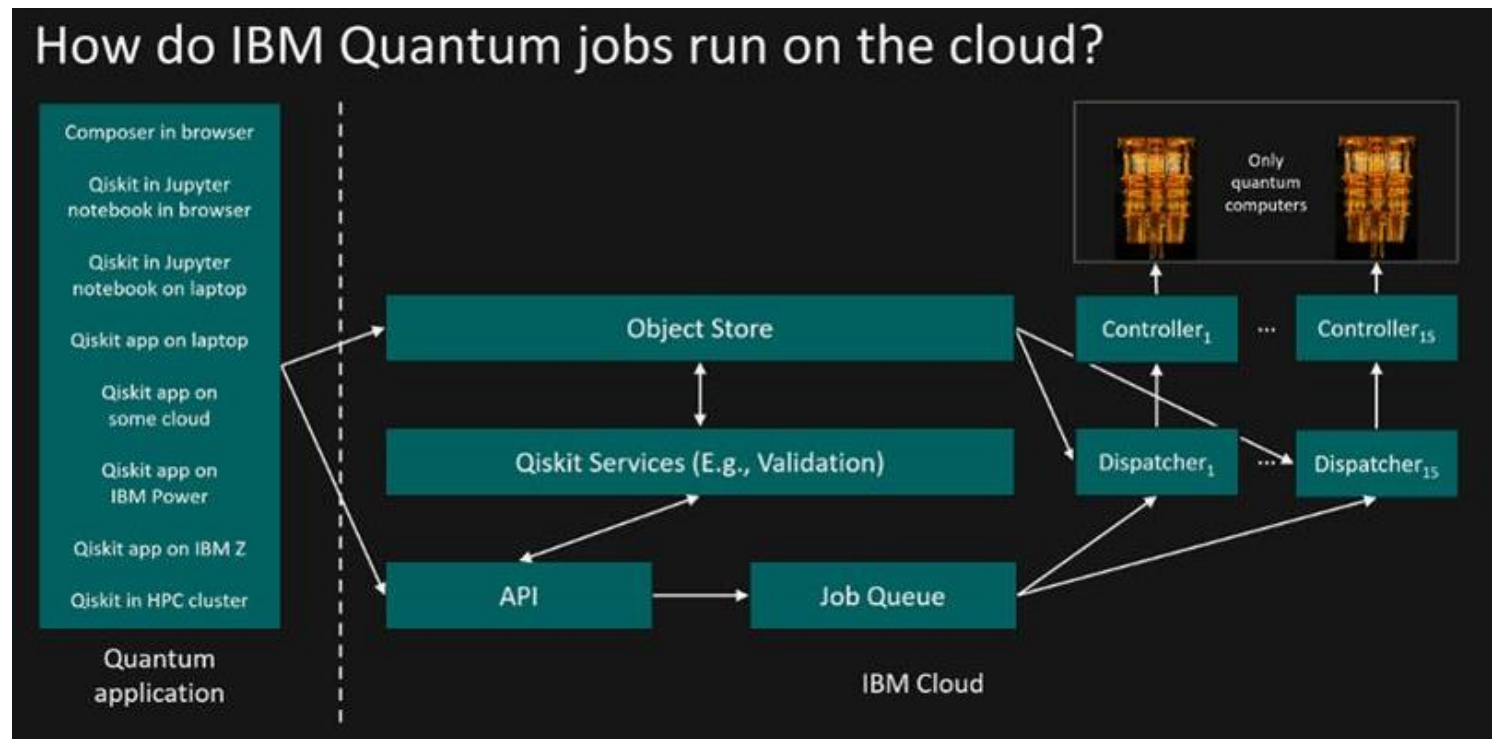

*Figure 644: how IBM is running a quantum job in the cloud. Source: IBM.*

**Rigetti** followed with its Cloud Services launched in 2019

**Alibaba** has a similar offer in China . We are here in the context of vertically integrated offers, the operator of the cloud service being the designer of the quantum computers.

Quantum cloud offering also started to be proposed in 2020 by cloud vendors selling access to third-party quantum computers, mixed with quantum emulation resources on conventional servers. This is what Amazon and Microsoft announced almost simultaneously at the end of 2019 and made available in 2020. Google followed-on in 2021.

**Microsoft** announced in November 2019 that it was integrating a quantum computing offering into the Azure cloud, and relying on **IonQ** and **Honeywell** (trapped ions qubits) as well as **QCI** (superconducting qubits). As of spring 2022, no QCI QPU was available online, seemingly since their qubits are not yet operational[1832]. **Rigetti** (superconducting qubits) was added to Azure Quantum in December 2021 (in “private previews” as of March 2022) and **Pasqal**’s quantum simulator in May 2022. Microsoft is associated among others with **1QBit** (Canada) to propose quantum software application layers[1833]. They promote quantum inspired algorithms that rely on traditional cloud resources, as in this case study of MRI scanner optimization at Case Western Reserve University[1834]. Microsoft and **KPMG** announced a partnership related to the development of quantum inspired algorithms in December 2022. Microsoft Azure Quantum also supports Qiskit and Cirq Python-based quantum code since October 2021. You can test Azure Quantum for free with a credit of $500 an even win a credit of $10K after submitting your applications. Microsoft is also selling the access to **Toshiba**’s Simulated Quantum Bifurcation Machine+ (SQBM+), a classical Ising model solver using quantum inspired models[1835].

---

[1832] Microsoft Azure Quantum was introduced step by step: announced in December 2019, released in limited preview in May 2020 and then in public preview in February 2021.

[1833] See Experience quantum impact with Azure Quantum, November 2019 and Microsoft Announces Azure Quantum with Partners IonQ, Honeywell, QCI, and 1QBit by Doug Finke, 2019. At the same time, Microsoft also announced that it has brought together many other quantum software partners: ProteinQure, Entropica Labs, Jij, Multiverse Computing, Qu&Co, QC Ware, OTI, Qubit Engineering, Qulab, QunaSys, Rahko, Riverlane, SolidStateAI, StrangeWorks, Xanadu, Zapata Computing. See the list here : Quantum Network- A community of pioneers by Microsoft, 2019.

[1834] See How the quest for a scalable quantum computer is helping fight cancer by Jennifer Langston, July 2019.

[1835] See Toshiba launches new SQBM+ quantum-inspired optimization provider on Azure Quantum, June 2022.

**Amazon** made its entrance in the quantum cloud market at the end of 2019 with the announcement of three components: Amazon Braket cloud services, AWS Center for Quantum Computing at Caltech University[1836], and Amazon Quantum Solutions Lab, a customer evangelization program reminiscent of IBM's Q initiative[1837]. Amazon also uses the brand Quantum Compute Cloud (QC2) for its offering. Braket is their whole quantum software infrastructure. It provides access to quantum computers from D-Wave (the 2000Q and Advantage annealers with respectively 2048 qubits and 5000 qubits), IonQ (it started with 11 qubits), Rigetti (16Q Aspen-4 with 16 superconducting qubits and their 31 qubits Aspen-9 version, and later, their 80 qubits system) and OQC (with 8 coaxmon qubits, in 2022). IonQ is thus proposed by both Microsoft and Amazon. Amazon announced in November 2021 that QuEra's cold atoms systems running in quantum simulation mode (*aka* "Hamiltonian simulation") would be added to AWS Braket in 2022. Late 2021, Amazon Braket Hybrid Jobs was added to their software portfolio with the capacity to run hybrid algorithms associating classical and quantum computation[1838]. Amazon also proposes software emulation of quantum algorithms on classical servers.

Amazon Braket is associated with an in-house SDK based on the classic Python language. Development is supported in the integrated open source environment Jupyter. It also includes support for the OCL (Object Constraint Language) constraint programming language. Like Microsoft, Amazon is also a partner of quantum software vendors. We find almost the same players with Xanadu, Zapata Computing , Rahko, QC Ware, 1Qbit and Qsimulate.

**Google** didn't offer any access to its own quantum computers in its cloud offering. Its Sycamore QPUs are only available to a handful of US Universities in a sort of "private cloud" access. Google still has a digital emulation offering supporting up to 40 qubits. In June 2021, it also announced the integration of IonQ's 11 qubits processor in its Cloud Marketplace. IonQ becomes de-facto the most distributed solution in the cloud with Amazon, Google and Microsoft.

At last, let's mention France's **OVHcloud** who entered the cloud quantum space in May 2022 with announcing that its datacenter was handling the classical part of **Pasqal**'s first entry in the cloud with its 100-qubits cold atoms-based quantum simulator. This is to be extended with the support of other European QPU solutions. They also host Perceval, **Quandela**'s photon qubits classical emulator.

Below is a summary of these cloud-based quantum computing offerings, distinguishing between emulating quantum code on classical computers and executing quantum code on quantum computers.

In China, we can add **Baidu** and its Quantum Leaf cloud offering. It's provided with **Paddle Quantum**, a quantum machine learning development toolkit based on the PaddlePaddle programming language, **QCompute**, a Python-based open source SDK and **Quanlse**, a machine-level programming tools controlling the pulse sent to emulated superconducting qubits.

Other cloud offerings can be mentioned like **Quantum Inspire** in The Netherlands and one from **Xanadu**.

[1836] The AWS Center for Quantum Computing is headed by Brazilian Fernando Brandao (1983), who is both a professor at Caltech and director of this Amazon AWS laboratory. He was previously a researcher at Microsoft Research. He is a good generalist, initially a physicist and now a specialist in quantum algorithms. In June 2020, John Preskill, also a professor at Caltech, announced that he would spend one day a week at the research center.

[1837] See Amazon Braket-Get Started with Quantum Computing by Jeff Barr, December 2019 and the presentation of the announcement Introducing Quantum Computing with AWS by Fernando Brandao and Eric Kessler (video and slides, featuring the Eiffel Tower of Rydberg atoms from French startup Pasqal in slide 15). I discovered in the hundreds of presentations at the Amazon Reinvent conference in December 2019 where this Braket announcement took place that Amazon was also presenting the QLDB, or Quantum Ledger Database, a blockchain management software brick. But that doesn't seem to have anything quantum at all.

[1838] See Introducing Amazon Braket Hybrid Jobs – Set Up, Monitor, and Efficiently Run Hybrid Quantum-Classical Workloads, Amazon, November 2021. See also PennyLane on Braket + Progress Toward Fault-Tolerant Quantum Computing + Tensor Network Simulator by Jeff Barr, December 2020.

Let's finish this with the newly launched hybrid quantum solutions launched by public organizations, mostly in the European Union. These are associating QPUs and supercomputers. The European project **HPC-QS** (for quantum simulation) will be deployed in three sites: in Finland (CSC LUMI), Germany (Munich at the Jülich Supercomputing Centre) and France (Bruyères-le-Châtel at CEA). It will start with attaching an Atos QLM appliance to these sites' supercomputers, themselves all running classical CPUs and GPGPUs from Nvidia.

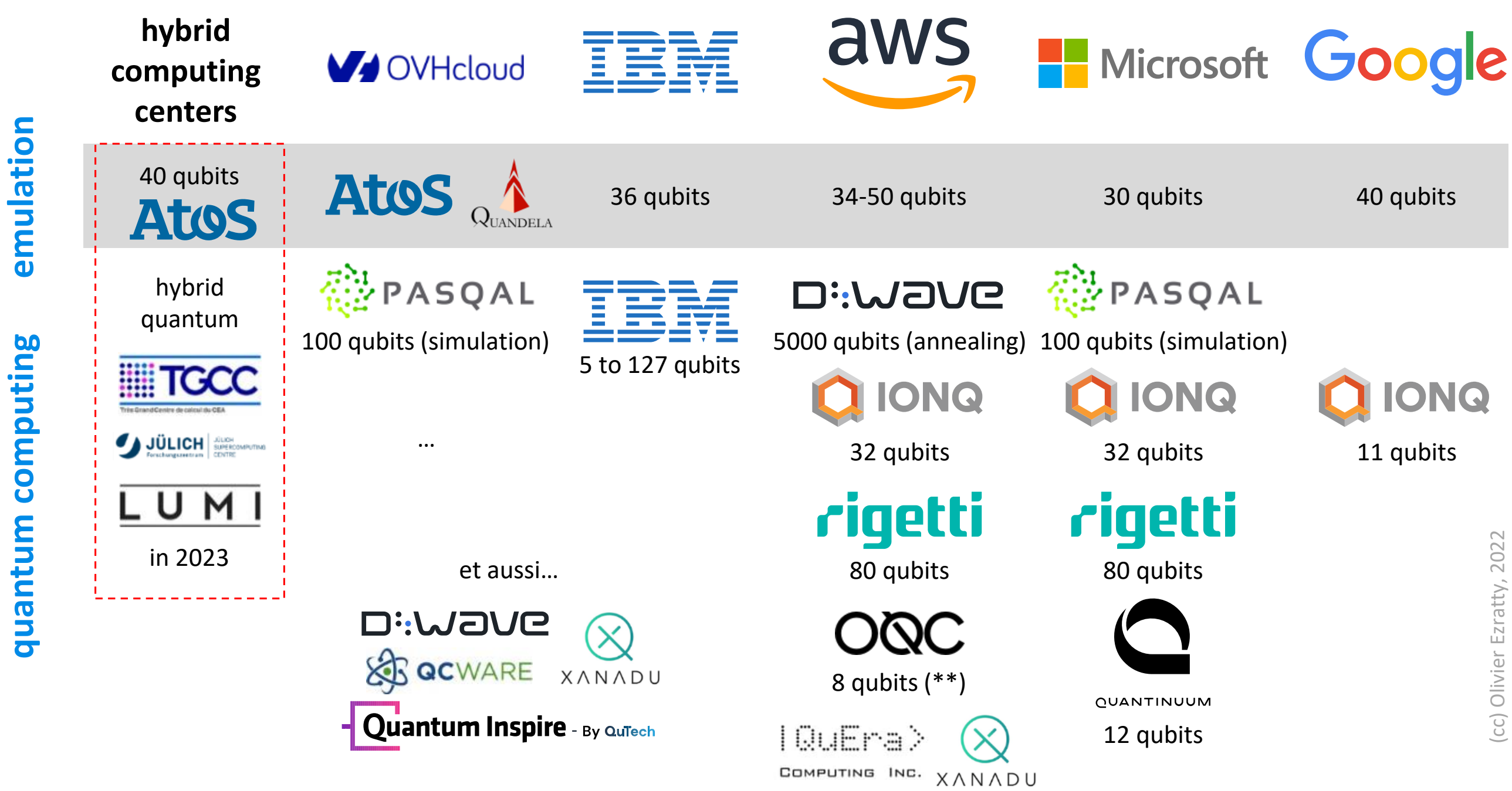

*Figure 645: main quantum cloud emulation and QPU offerings worldwide. (cc) Olivier Ezratty, 2022.*

Then, in 2023, they will be completed by a European QPU. Atos's QLM will both run quantum code emulation in the three paradigms (annealing, simulations, gates) and drive QPUs quantum code execution. In Germany and France, the first installed QPUs will be quantum simulators from Pasqal. In Finland, it will probably be superconducting qubits systems from IQM. In Germany, an IQM system will also be deployed with an Atos QLM at the Leibniz Supercomputing-LRZ center, as part of the project Q-EXA. These hybrid computing centers will mainly serve the needs of researchers and large companies.

# Quantum software engineering

### Certification and verification

The verification and certification of quantum algorithms and the results of their use is an important new topic. The factorization of integer numbers is obviously easy to verify. But when a quantum algorithm is used to simulate physical interactions such as those of atoms in molecules, it is less obvious.

Theoretical work shows that it is possible to prove polynomially that a result of a quantum algorithm is accurate[1839]. Unfortunately, we cannot explain in detail the origin of the result by breaking it down.

---

[1839] See How to Verify a Quantum Computation by Anne Broadbent, 2016 (37 pages) which demonstrates that all quantum algorithm results can be verified with classical polynomial algorithms by performing several tests and encrypting the input data. See also Verification of quantum computation: An overview of existing approaches by Alexandru Gheorghiu, Theodoros Kapourniotis and Elham Kashefi, 2018 (65 pages).

Nor can we prove that the result found, however valid it may be, is the best of all if there are several good ones[1840]. On top of that, we must make a distinction between error corrected hardware (LSQ) and noisy systems (NISQ). Surprisingly, while LSQ regimes will be inaccessible to classical hardware emulation and make verification difficult, verification is also complicated for noisy systems, particularly when they repeat some sequence of code iteratively.

The other key point is to make sure, in the case of the use of a remote quantum computer, that the recovered result corresponds to the submitted calculation and that an intruder did not interfere in the history nor was able to alter the calculation on the quantum computer side. Many research inroads have been made in the last 15 years for that respect.

One of the methods consists in relying on the concept of **blind computing** devised in 2009 by Anne Broadbent, Joseph. Fitzsimons and Elham Kashefi[1841].

**CEA LIST** announced in June 2020 that it had created **QBRICKS**, an environment for the specification, programming and formal verification of quantum algorithms. They used to do this for critical embedded systems where certification by formal proof is particularly important. They are now entering the field of quantum programming and have experimented their model with QPE, the quantum phase algorithm (QPE= that fits into Shor's model for integer factorization, the full Shor algorithm and Grover's algorithm. This work involves the joint LRI laboratory at the University of Paris-Saclay and CentraleSupelec[1842].

One of the major advances in the explicability of quantum algorithms comes from researcher **Urmila Mahadev,** whose work between 2012 and 2018 has led to the creation of a method for verifying quantum computer processing. She was a postdoc at Berkeley and supported by Scott Aaronson and Umesh Vazirani, two eminent researchers in quantum algorithmic research. Her work aims at proving that a quantum computer has indeed performed the treatments it has been asked to do. She shows that a classical computer coupled to a simple quantum computer can verify in a polynomial way the results of a quantum computer[1843] . The method exploits a technique of post-quantum cryptography that the verifier cannot break (LWE: Learning With Errors). LWEs are part of the Lattice-based cryptography (EN) or Euclidean networks (FR) class[1844]. The method was recently improved by an Austria team to work with an untrusted quantum computer[1845].

Other quantum programs verifiers[1846] from research laboratories include the **Path-sum** framework from the University of Waterloo[1847], **VOQC** (Verified Optimizer for Quantum Circuits) from the

[1840] See also Quantum cloud computing with self-check by Rainer Blatt et al, May 2019, which discusses quantum simulation calculations on 20 qubits of trapped ions with results controlled on the quantum computer as fast as on the PC.

[1841] See Universal blind quantum computation by Anne Broadbent, Joseph Fitzsimons and Elham Kashefi, 2008 (20 pages) and the associated presentation (25 slides) and A Framework for Verifiable Blind Quantum Computation by Theodoros Kapourniotis, Harold Ollivier, Elham Kashefi et al, June 2022 (33 pages) which shows a mathematical link between code verification and error correction.

[1842] See Toward certified quantum programming by Sébastien Bardin, François Bobot, Valentin Perelle, Christophe Chareton and Benoît Valiron, 2018 (4 pages) and An Automated Deductive Verification Framework for Circuit-building Quantum Programs by Christophe Chareton, Sébastien Bardin, François Bobot, Valentin Perrelle and Benoît Valiron, 2021 (30 pages).

[1843] See a description of the method in near-natural language in Graduate Student Solves Quantum Verification Problem, October 2018 and two reference publications: Classical Verification of Quantum Computations, September 2018 (53 pages) and Interactive Proofs For Quantum Computations, April 2017 (75 pages).

[1844] See this presentation describing the LWE protocol: An Introduction to the Learning with Errors Problem in 3 Hours (76 slides).

[1845] See Towards experimental classical verification of quantum computation by Roman Stricker et al, AQT and Innsbruck University, March 2022 (19 pages).

[1846] See the review paper Formal Methods for Quantum Programs: A Survey by Christophe Chareton, Sebastien Bardin, Dongho Lee, Benoit Valiron, Renaud Vilmart and Zhaowei Xu, September 2021 (66 pages).

[1847] See Towards Large-scale Functional Verification of Universal Quantum Circuits by Matthew Amy, University of Waterloo, 2018 (21 pages).

University of Maryland, itself based on **SQIR** (Small Quantum Intermediate Representation) supporting verification[1848] and QHL from Tsinghua University[1849].

### Debugging

Like any computer software, quantum software requires a set of quality control processes. Like most human-originated creations, they are prone to bugs and errors. Some classical computing methods can be reused for this respect but many require some adaptation to the specifics of quantum computing, whether done on gate-based systems or on analog quantum simulators.

A quantum circuit is not easy to debug! It will certainly require new debugging tools and approaches. And a majority of quantum code bugs are "quantum" in nature and not easy to spot with traditional methods[1850].

For the moment, simple circuits can be analyzed and debugged with a quantum emulator running on a classical computer, to understand how the qubit register vector state evolves step-by-step. But when quantum circuits in the advantage regime, beyond any classical emulation capacity, other means will have to be used.

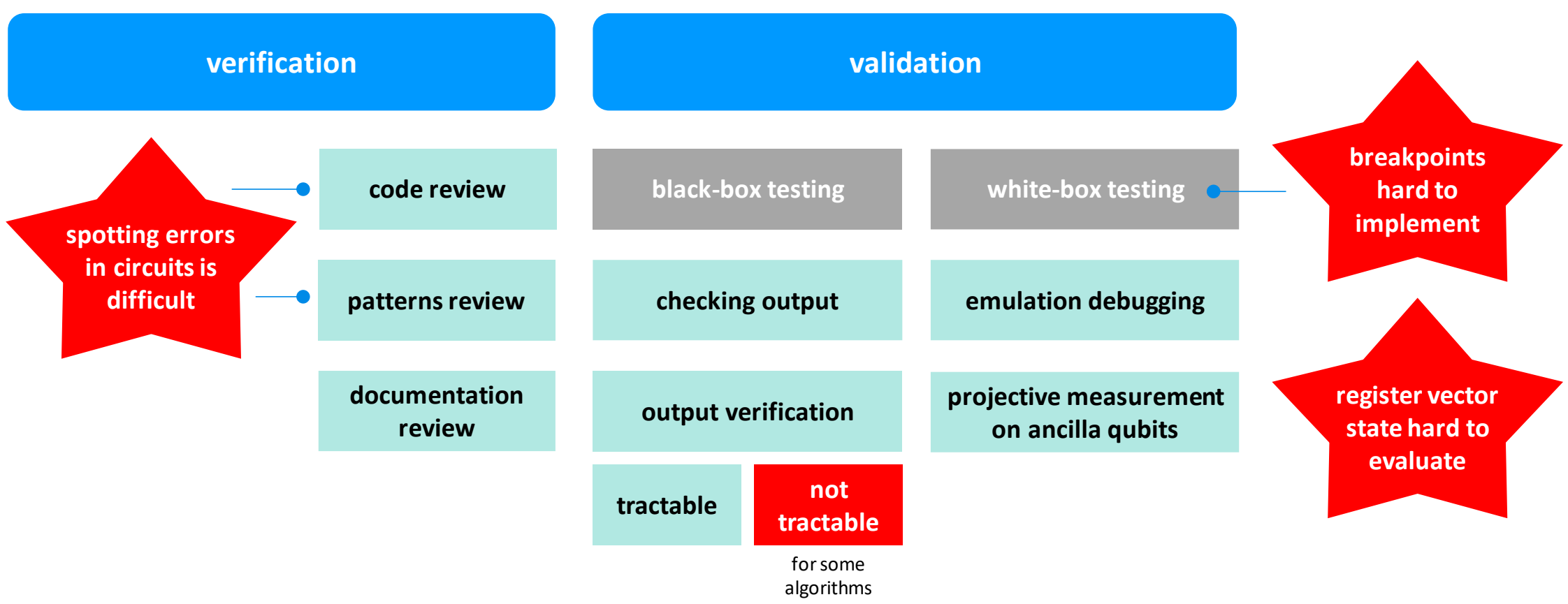

*Figure 646: some of the challenges with quantum software engineering. (cc) Olivier Ezratty, 2022.*

Software quality control usually goes through two main steps: verification and validation.

**Verification** deals with verifying that the code will run as expected. It includes checking code documentation, designs, circuits and the various software components or patterns that are used. Verification also deals with making sure that the specifications are correctly implemented by the system. It responds to the question: are we building the product right?

In classical programming, good programmers and code reviewers can spot an error with just looking at the code. Code inspection tools can also detect undeclared variables or variables used in the wrong context. These errors are way more difficult to detect visually on a quantum circuit, particularly with large ones. It may require the use of code decomposition in modules or patterns, like in object-oriented programming.

**Validation** concerns the program output and making sure it works as planned. It includes testing and validating the code against the user's needs. It responds to the question: are we building the right product?

---

1848 See A Verified Optimizer for Quantum Circuits by Kesha Hietala et al, University of Maryland (36 pages).

1849 See Quantum Hoare logic with classical variables by Yuan Feng et al, University of Technology Sydney, Australia, Chinese Academy of Sciences and Tsinghua University, China, April 2021 (44 pages).

1850 As studied in A Comprehensive Study of Bug Fixes in Quantum Programs by Junjie Luo et al, January 2022 (8 pages).

Quantum computing results validation is usually fast like with integer factoring (it requires a simple classical multiplication) or a Grover search (it requires checking the Oracle once in a classical way).

But some circuit validations may need to be done on a quantum computer, like with a boson sampling or with a QMA prover (Quantum-Merlin-Arthur). On top of that, contrarily to classical computers, quantum computations errors can also come from hardware imperfections and the fateful quantum noise[1851]. Compiler, code optimizers and even error correction codes can also generate software bugs and amplify some errors.

Quantum software bugs can have various sources: errors in the data preparation (which is itself based on quantum gates), incorrect operations and transformations, incorrect compositions and iterations and also incorrect qubits deallocations (or "uncomputations").

During validation, testing use the white-box and black-box approaches. **White box** testing tests internal data structures and program flow, and may include some interactive debugging.

One solution is to decompose manually or automatically the quantum code into accumulated slices of code with its incremental different parts like data preparation, oracle execution and amplitude amplifications in the case of a Grover algorithm, as pictured below. The debugging tool with run each accumulated slice followed by a measurement one after the other[1852]. It's not a classical pause-play like in classical interpreters like JavaScript but "pause" and "play again from the start", a probably with several shots being run and their results averaged to get a sampled output.

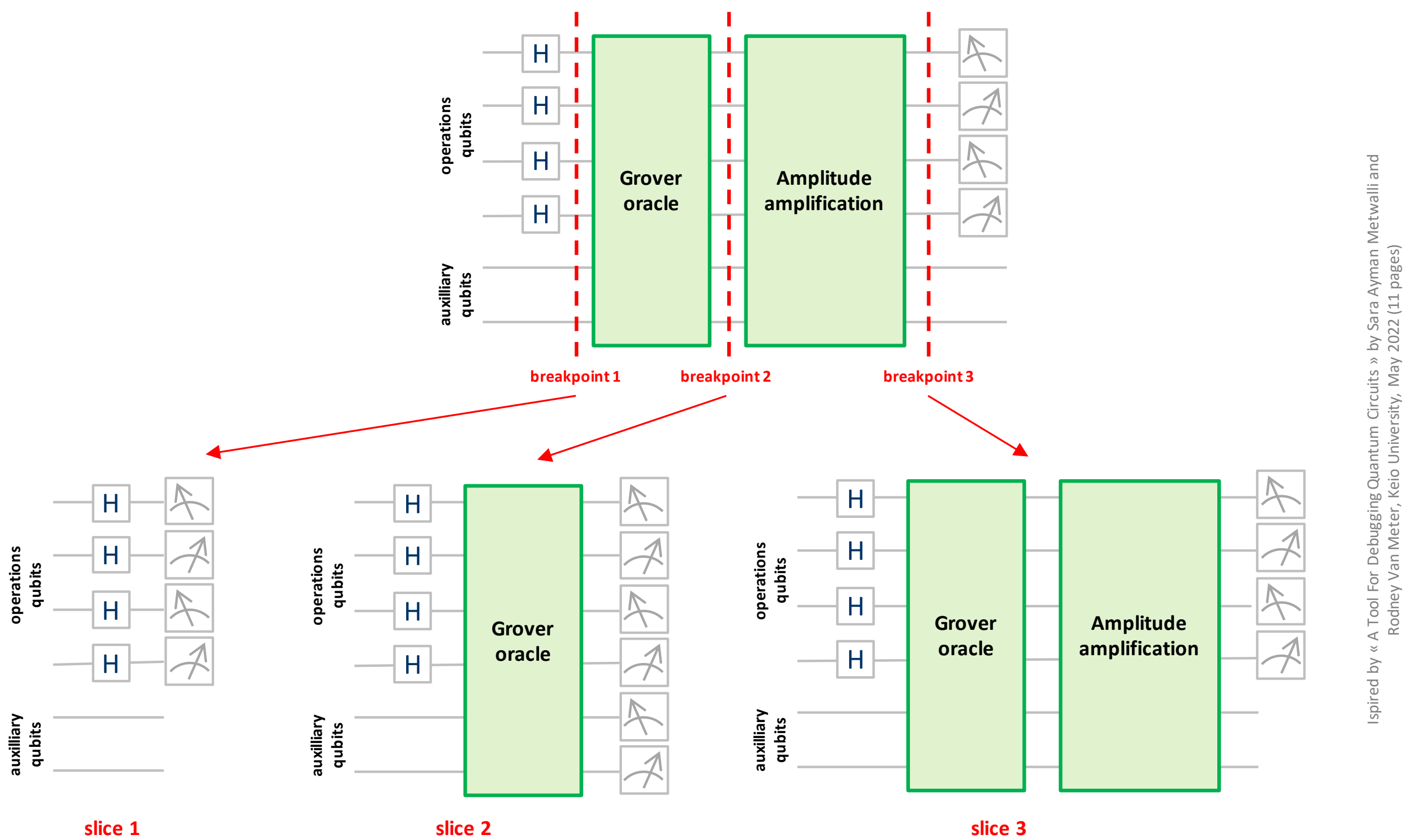

*Figure 647: a quantum code debugging approach with code slicing. Source: A Tool For Debugging Quantum Circuits by Sara Ayman Metwalli and Rodney Van Meter, Keio University, May 2022 (11 pages).*

**Black-box** testing looks at the functionality, ignoring the inner workings of the software, making sure the expected output is obtained with a given input?

[1851] See Formal Verification vs. Quantum Uncertainty by Robert Rand et al, University of Maryland, 2019 (12 pages) that pinpoints the role of hardware errors in quantum programs.

[1852] See A Tool For Debugging Quantum Circuits by Sara Ayman Metwalli and Rodney Van Meter, Keio University, May 2022 (11 pages).

How about using interactive debugging in the white-box approach? Right now, it can be done on quantum emulators but is limited by their computing/memory capacity. It can't exceed 40 qubits and practically 16 qubits. A state vector representation is quite difficult to visualize beyond 8 qubits.

On a real machine, implementing interactive breaking points in a quantum circuit is difficult due to the impact of measurement on the qubits register vector state and on the probabilistic nature of quantum computing. Let's say we'd like to implement a breaking point and line by line code execution. We'd need to run the quantum algorithm and stop it at the breaking point then make some measurement.

But good measurement, just to get a state in the computational basis would require running the code many times. And even way more times if we'd need to reconstitute the full vector state. Then, to move to the next series of gates, the circuit would have to be re-run the same number of times. And again and again and again. It would be worse if we were to check the entanglement within the register. Deciding if a register is separable is in itself an NP-hard problem. One way to proceed is to implement unit testing with splitting the code in trusted blocks and patterns. Other debugging tools can involve projective measurements on ancilla qubits or even gentle measurement techniques[1853]. And this deals just with classic gate-based quantum computing. Analog quantum computing and special techniques like MBQC or FBQC (from PsiQuantum) will mandate specific debugging techniques and tools.

At last, let's mention that many quantum algorithms are hybrid and aggregate classical and quantum algorithms, which requires another set of discipline and tools.

Research is going on in all these dimensions around the world. These are strategic components for quantum computing[1854].

## Benchmarking

Benchmarking quantum computing is becoming a strategic tool for both vendors and users. In the IT space, vendors have continuously relied on benchmarking to showcase new hardware advances and customers rely on it to assess the cost/benefit ratio of emerging technologies. Benchmarks participate to fostering innovation with driving the competition between vendors and technologies[1855].

In classical computing, raw computing power is measured in **FLOPs** (floating point operations per seconds). **Linpack** is used in the HPC TOP500 ranking and is based on a linear equation solving task. **MLPerf** is used in machine learning and for comparing GPGPUs like those from Nvidia and their competitors. Personal computers processors can be compared with **Passmark Software** benchmarks or alternative from other vendors.

Quantum computing benchmarking can have several purposes:

- Comparing quantum computing with classical computing, which involves coupling best-in-class algorithms and software in both cases, all being moving targets given the steady progress of classical computing, particularly with so-called "domain specific architectures", the most famous one being the tensor-based GPGPUs and ASICs. This sort of benchmarking can be exploited to assess a potential quantum advantage, where a quantum solution is either solving a problem in a shorter time than classical computers or able to solve a problem that can't be solved with existing classical

[1853] See Debugging Quantum Processes Using Monitoring Measurements by Yangjia Li and Mingsheng Ying, 2014 (7 pages) describes the process of interim measurement process within code and Projection-Based Runtime Assertions for Testing and Debugging Quantum Programs by Gushu Li et al, 2020 (29 pages) proposes to use some ancilla qubits to indirectly detect vector state characteristics. It uses projective measurements on a different basis than the computational basis of each qubit.

[1854] It includes the study Program Verification, Debugging, and QC Simulation — EPiQC, a IARPA funded project on quantum program verification and debugging.

[1855] See The Race to Quantum Advantage Depends on Benchmarking by Matt Langione, Jean-François Bobier, Lisa Krayer, Hanl Park and Amit Kumar, February 2022.

computers. The archetypal such benchmark is the cross-entropy benchmark (XEB) used by Google in 2019 to showcase its quantum supremacy with Sycamore[1856].

- Comparing different quantum computers competing with each other to solve particular tasks. It can be done with quantum computers using the same programming paradigm like gate-based systems (e.g. IBM's quantum volume), or even, quantum computers solving a given problem with different paradigms (annealing, simulation, gate-based; e.g. Atos Qscore). An objective comparison can be made with solving one or a broader set of specific problems and assessing the maximum problem size addressable by competing solutions (e.g., Atos Qscore, QED-C and IonQ's algorithmic qubits).
- Comparing different characteristics. The most common is computing time, but other metrics will become important as well such as precision, energy spent, weight, environmental footprint and total cost (e.g. DARPA's benchmarking project RFP won by Raytheon BBN and the Quantum Energy Initiative's Green 500 benchmark proposal).
- Comparing different algorithms solving a given problem on similar or different quantum hardware. These are not yet there.

Benchmarking tools already abound and are very diverse.

Following a bottom to top system approach, there are benchmarks for spare system level features (number of qubits, qubits gates and readout fidelities, connectivity, gates speed aka CLOPS with IBM, all of which are not benchmark results per se, entanglement quality). These are the quantum equivalent of number of cores, CPU clock, RAM, storage size and speed and network specs in classical computing.

| what and whom | what | pros | cons | timing / adoption |
|---|---|---|---|---|
| **IBM quantum volume** | breath/depth computing capacity, 2^#qubits | simple qualifier of qubits quality | doesn't work in advantage regime due to emulation needs requirements | published in 2019 IBM, Quantinuum |
| **Cisco MBQC quantum volume** | computing capacity for MBQC CV photon qubits | adapted to photon qubit using a different model than circuit based models | to be adapted to direct variable photons MBQC model | proposed in 2022 by Cisco |
| **IBM CLOPS** | circuit layers operations per seconds | complements QV for speed | N/A | announced in November 2021 |
| **cycle benchmarking** | qubits entanglement evaluation | useful to benchmark qubits quality | limited to one low-level feature | 2019, Canada, Denmark and Austria universities |
| **scalable benchmarks for gate-based QC** | six low-level structured circuits tests | tested 21 configurations from IBM, IonQ and Rigetti | low-level benchmark not usage based | published in 2021 QuSoft, Cambridge, Caltech |
| **PQF (photonic quality factor)** | assess performance of linear optics single photons multimode QPUs | covers many NISQ photon qubit implementations | limited to a specific photonic qubit configuration | published in 2022 by Quandela |
| **entanglement-based volumetric benchmark** | estimate size of maximum entangled qubit state | entanglement is a key feature of quantum acceleration | narrow and not usage oriented | proposed in 2022 par DoE Oak Ridge et al |

*Figure 648: low level benchmarking proposals. (cc) Olivier Ezratty, 2022.*

Then, some are benchmarking hardware capabilities at a higher level and regardless of the use case (quantum volume from IBM). At last, others are assessing the capability to solve one or several typical use cases, and usually, their maximum size (QED-C, Atos Q-score, IonQ algorithmic qubits, SupermarQ). At some point, benchmarking will be even more tricky when mixing quantum and classical computers.

---

[1856] This benchmarking technique like the quantum volume is limited by the capacities of classical emulation. One proposed way to overcome this problem is to run only Clifford gates, which reduces the resources requirements in the classical computer. See Linear Cross Entropy Benchmarking with Clifford Circuits by Jianxin Chen et al, June 2022 (30 pages).

| | what and whom | what | pros | cons | timing / adoption |
|---|---|---|---|---|---|
| multiple use cases | scalable benchmarks for gate-based QC | six low-level structured circuits tests | tested 21 configurations from IBM, IonQ and Rigetti | low-level benchmark not usage based | published in 2021 QuSoft, Cambridge, Caltech |
| | QED-C supported benchmark | set of low-level algorithms benchmarks | breadth of use cases | complicated visualization | published in 2021 QED-C, Princeton, HQS, QCI, IonQ, D-Wave, Sandia Labs |
| | IonQ Logarithmic Qubits | $\min(\#\text{qubits}, \sqrt{\#gates})$ | run on different use cases | a bit complicated | published in 2020 and refined in 2022, IonQ |
| | SupermarQ from Super.tech | suite of applications benchmark | also handles error correction benchmarking | | published in March 2022, Intel and Amazon |
| | Qpack by TU Delft | three sets of problems (Max-Cut, TSP, DSP) | measure différents metrics | Adoptions | proposed in April 2022 |
| single use cases | Atos Q-score | maximum size of solvable MAXCUT problem size | application need oriented works in advantage regime hardware independant | limited to MAXCUT problems marketing & adoption | published in 2020 Atos, be be expanded to other algos |
| | DoE ORNL | chemical simulation | works on existing superconducting hardware | limited to three 2-atom molecules simulations | published in 2020 DoE |
| | Zapata benchmark for fermionic quantum simulations | one-dimensional Fermi Hubbard model (FHM) VQE running on NISQ | tested on Google Sycamore with its tunable couplers | narrow use case | proposed in March 2020 |

*Figure 649: application level benchmarking proposals, either multiple or singe cases. (cc) Olivier Ezratty, 2022.*

These tools are created by different kinds of players, sometimes working together: research laboratories (e.g. DoE Sandia Labs), software and hardware vendors (IBM, IonQ, Atos, Zapata Computing, …), industry consortium usually working with a combination of these players (USA's QED-C) and then standards bodies (e.g. IEEE). The combination of these players is important.

| what and whom | what | pros | cons | timing / adoption |
|---|---|---|---|---|
| Unitary Fund Metriq | repository of benchmark results | N/A | N/A | announced in May 2022 |
| DARPA project | SWAP (size-weight-application-power) | hardware-agnostic and resource estimates | N/A at this point | awarded in 2022 to Raytheon BBN |
| IEEE QC Perf Metrics & Perf Benchmarking PAR | gate-based QC benchmarking | ongoing standardization project | | submission in Oct 2023 completion in Oct 2024 |
| Quantum Energy Initiative | QC energetics benchmarking consolidated approach, QGreen500 proposal could consolidate cryogeny benchmarks | methodology (MNR) to optimize QC energetics, first analysis done with superconducting qubits | research and industry must build coordination around this goal | joint research/industry Quantum Energy Initiative launched in 2022 |

*Figure 650: other benchmarks proposals. (cc) Olivier Ezratty, 2022.*

Industry vendors have some biasing interest. They don't want to favor the adoption of benchmarks that would be unfavorable to their offerings. This is amplified by the wide spectrum and diversity of quantum computing technologies and qubits including aspects like gate-times, fidelities and topologies. Some benchmarks can also be misleading if improperly extrapolated in the future, given all technologies don't have the same upscaling potential. An equivalent to Moore's law can't be applied in a simplistic way to all these technologies roadmaps. In all cases, quantum computing benchmarking will be difficult to interpret without the right technical background. You can be easily fooled by vendors given the complexity of the matter.

## Quantum Volume and CLOPs

Since late 2021, IBM uses three systems-level metrics and benchmarks to characterize its quantum computers: the scale with the number of qubits, the speed measured in CLOPS (circuit layer operations per second) and qubits quality measured with their homemade quantum volume metric.

The quantum volume was introduced in 2017 and was adopted by **Honeywell** in March 2020 and afterwards by **IonQ** in October 2020. Its use is also recommended by the **Gartner Group**.

Quantum volume is an integer that associates the quantity of qubits and the number of quantum gates that can be executed consecutively with a reasonable error rate. Indeed, having N qubits but being limited by the number of quantum gates that can be used can be detrimental to the execution of many quantum algorithms. Some are greedy for quantum gates, others are not[1857].

This quantum volume number is supposed to aggregate four performance factors:

- The **number of physical qubits of** the processor.
- The **number of quantum gates** that can be chained consecutively without the error rate being detrimental to the results.
- The **connectivity between these qubits**, which will impact the length of execution of an algorithm and potentially improve quantum volume for qubits with high connectivity such as with trapped ions. It reduces the number of SWAP gates that are required when connectivity is limited.
- The **number of quantum gates** that can be executed in parallel[1858].

The QV is evaluated using a random calculation benchmark consisting of chaining random quantum gates and which must give a correct result in two thirds of the cases. Why two thirds? Because quantum computing provides a probabilistic result. To obtain a deterministic result, the calculation is executed several times and the average of the results is evaluated, up to thousands of times as proposed by IBM in its cloud system. With an average of two-thirds good results, one can therefore statistically converge to a good result after a few measurements. The accuracy of the result will depend on this number, which is usually a few thousand.

In the first version in 2017, the quantum volume was the square of the maximum number of qubits on which the processor could perform this calculation[1859]. The definition then was changed in 2019 to become 2 to the power of this number of qubits[1860]. The following illustration explains how the 2017 and 2019 quantum volumes are evaluated.

$$d \simeq 1/(n\epsilon_{eff})$$

d = maximum computing depth
n = number of qubits
$\varepsilon_{eff}$ = % error rate of 2 qubits gates

$$V_Q = dn = 1/\epsilon_{eff}$$

base quantum volume = qubits # * computing depth

$$V_Q = \min(n, d)^2$$

quantum volume becomes min(n, depth)$^2$ to avoid tweaking the system with a low n=2 and a very high fidelity

$$V_Q = \max_{n' \leq n} \min \left[ n', \frac{1}{n' \epsilon_{eff}(n')} \right]^2$$

**2017 QV** : scans all combinations of n qubits below the available number of qubits, to run a random algorithm generating 2/3 good results.

$$\log_2 V_Q = \underset{m}{\operatorname{argmax}} \min[m, d(m)]$$

**2019 QV** : QV is a power of 2 of the number of qubits

*Figure 651: how is/was IBM's quantum volume calculated. (cc) Olivier Ezratty, 2021.*

[1857] Some algorithms can thus be satisfied with a limited number of quantum gates, such as Deutsch-Jozsa's and is satisfied with only four series of quantum gates. Peter Shor's integer factoring algorithm requires a depth of quantum gates equal to the cube of the number of qubits used.

[1858] Trapped ions quantum computers can't do that with two-qubit gates.

[1859] See Quantum Volume by Lev Bishop, Sergey Bravyi, Andrew Cross, Jay Gambetta and John Smolin, 2017 (5 pages).

[1860] See Validating quantum computers using randomized model circuits by Andrew W. Cross et al, 2019 (12 pages).

The diagram below from a paper by Robin Blume-Kohout and Kevin Young specifies how the m (number of qubits) and the d (computational depth) are evaluated[1861].

The number of qubits obtained to evaluate the quantum volume is much lower than the total number of qubits available: 8 for 16 in this case. The benchmark allows only 8 series of quantum gates in a row over 8 qubits, for 38 with only two qubits. In its 2017 version, the quantum volume was the grey square area containing the red lined squares.

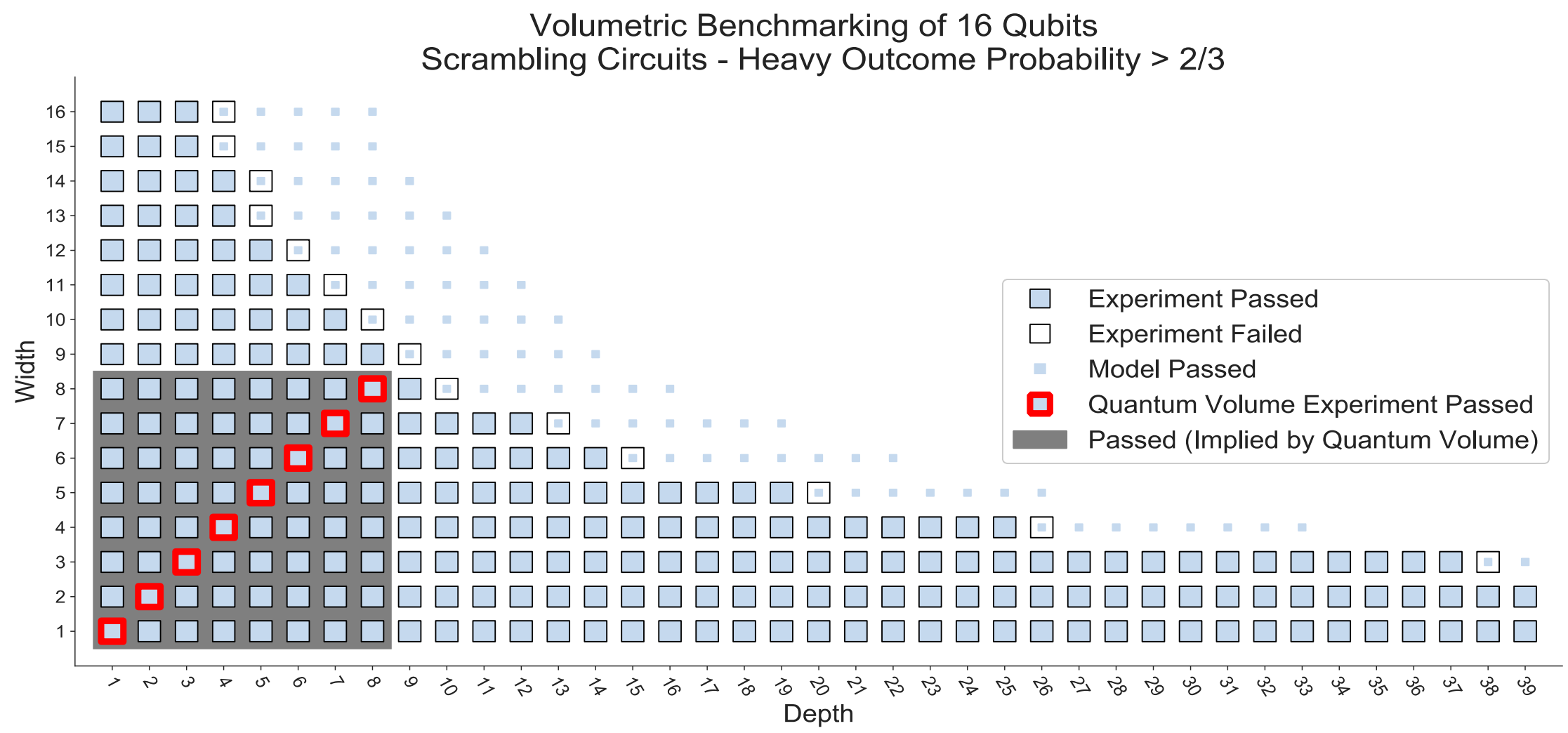

Figure 8(a). Volumetric benchmarking of a 16 qubit device using scrambling circuits. If at least 2/3 of the measurement results are heavy for a given width/depth pair, then the pair passes the test and is marked with a large, solid blue box. Using linear axes, the quantum volume experiments appear along the diagonal and are outlined with heavy, red lines. For this example, $\log_2(V_Q) = 8$. It is expected that scrambling circuits with both width and depth less than or equal to the quantum volume should succeed, and we highlight these with a gray background.

*Figure 652: a better visualization of how a quantum volume is evaluated. Source: A volumetric framework for quantum computer benchmarks by Robin Blume-Kohout and Kevin Young, February 2019 (24 pages).*

In its 2019 version, it became $2^{8,}$ or 256 instead of 64 ($8^2$). In the end, it is the dimension of Hilbert's vector space, i.e. the number of different superposed states that it is able to manage from a practical point of view with a depth of computation equal to the number of corresponding qubits.

When IBM states that their 27-qubit processor has a quantum volume equal to 128, it means that they only managed to validate their benchmark with 7 qubits among these 27 qubits.

IonQ announced in 2020 a quantum volume greater than four million corresponding to a QV of 4,194,304, representing $2^{22}$. So, with the ability to run 22 sets of quantum gates on 22 of their 32 qubits, with two-thirds correct results on the used random benchmark. This record seems to be related to the good connectivity of trapped ion qubits. These can all be directly entangled with each other, unlike superconducting qubits, which are at best entangled with their immediate neighbors. This allows the benchmark to be achieved in fewer series of quantum gates than on superconducting qubits, which require a lot of SWAP gates generating rapidly accumulating errors.

[1861] In A volumetric framework for quantum computer benchmarks, February 2019 (24 pages), Robin Blume-Kohout and Kevin Young propose volumetric benchmarks to evaluate the performance of quantum computers based on IBM's quantum volume. The latter also proposes its own quantum volume evaluation code.

When you look at the relative progress of QV between IBM, IonQ and Honeywell/Quantinuum systems, you see a clear difference: IBM has a tougher time to use all its qubits while trapped ions do it better, although with a rather limited number of qubits[1862]. But if and when IBM fixes some scalability issues with their qubits, like with reducing qubit crosstalk, they can potentially increase their QV to higher levels than those from trapped ions vendors.

| Year | Brand | Version | Hw Qubits | Log2(QV) | % |
|---|---|---|---|---|---|
| 2017 | IBM | Tenerife | 5 | 2 | 40% |
| 2018 | IBM | Tokyo | 20 | 3 | 15% |
| 2019 | IBM | Johannesburg | 20 | 4 | 20% |
| 2020 | Honeywell | | 4 | 4 | 100% |
| 2020 | IBM | Raleigh | 28 | 5 | 18% |
| 2020 | IBM | Montreal | 27 | 6 | 22% |
| 2020 | Honeywell | H0 | 6 | 6 | 100% |
| 2021 | IBM | Montreal | 27 | 7 | 26% |
| 2020 | Honeywell | H1-1 | 10 | 7 | 70% |
| 2021 | Honeywell | H1-1 | 10 | 9 | 90% |
| 2021 | Honeywell | H1-1 | 10 | 10 | 100% |
| 2022 | IBM | Manhattan | 127 | 6 | 5% |
| 2020 | IonQ | Aria | 32 | 22 | 69% |
| 2022 | Quantinuum | H1-2 | 12 | 12 | 100% |
| 2022 | Quantinuum | H1-1 | 22 | 13 | 59% |

*Figure 653: evolution of systems quantum volumes over time. (cc) Olivier Ezratty, 2022.*

QV is limited to about 50 operational qubits because it can only be evaluated with a benchmark comparing the qubits with their simulation on a conventional computer. This emulation is constrained by memory size, which reaches its limits between 50 and 55 qubits[1863].

Quantum computing scientists are circumspect about the interest of this indicator which is too simplistic[1864]. This use is contested by Scott Aaronson, a specialist in complexity theories and quantum algorithms [1865]. He reminds us that current QVs are more than easily emulable in a simple classical computer, if not on an Apple Watch! This does not make it particularly powerful. And when the QV will get significant, beyond the threshold of 50, we won't be able to measure it!

Scott Aaronson believed that this quantum volume indicator, which is a marketing simplification tool from IBM, should be avoided. He prefers a description of systems characteristics like the number of qubits, their connectivity, their coherence time (T1, T2), their reset, one and two-qubit gates and readout fidelities. With most vendors, these indicators are generally found in the scientific publications of researchers but not always in the vendors marketing literature. However, IBM publishes most of these data on their quantum systems available on Q Experience.

In August 2022, a team from **Cisco** proposed a way to evaluate a quantum volume for a MBQC-based photonic quantum processor. It is based on using a continuous-variable cluster state and the Gottesman-Kitaev-Preskill (GKP) encoding. It computes the system quantum volume based on the logical gate error channels of cluster state GKP squeezing and photon loss rate[1866].

In November 2021, **IBM** added the CLOPS speed metric (circuit layers operations per seconds), an equivalent to the clock of a classical CPU, given the numbers are different for resetting qubits, operating quantum gates and measuring qubits[1867]. As of 2021, IBM's systems CLOPS where between 1,5 and 2,4K, so about 2.000 layers of qubit gates per seconds. CLOPS are calculated as M * K * S * D / time taken, where M is the number of templates, K the number of parameter updates, S the number

[1862] See Quantum Volume in Practice: What Users Can Expect from NISQ Devices by Elijah Pelofske et al, March 2022 (19 pages).

[1863] See Why Is IBM's Notion of Quantum Volume Only Valid up to About 50 Qubits? by Jack Krupansky, October 2020.

[1864] Imagine an indicator of the power of your laptop aggregating the processor clock frequency, its number of cores, the power of its CPU, the RAM memory, the storage capacity, its type (hard disk, SSD) etc? And there, to ask yourself if you will be able to efficiently use your video editing, photo derush or video game software on augmented reality headphones!

[1865] In Turn down the quantum volume, Scott Aaronson, published just after Honeywell's February 2020 announcement.

[1866] See Quantum Volume for Photonic Quantum Processors by Yuxuan Zhang et al, August 2022 (22 pages).

[1867] See Scale, Quality, and Speed: three key attributes to measure the performance of near-term quantum computers by Andrew Wack, Hanhee Paik, Ali Javadi-Abhari, Petar Jurcevic, Ismael Faro, Jay M. Gambetta and Blake R. Johnson, October 2021 (8 pages). With their Falcon R5 processor, qubit reset takes 450ns while qubits readout takes 750 ns. It's much longer than gates.

of shots and D, the number of quantum volume layers, or $\log_2$(QV). IBM published a methodology where M=100, K=10, S=100 and D was dependent on the quantum volume of each of the benchmarked systems[1868]. IBM plans to reach 10K CLOPS in 2022 with its 433 qubits Osprey processor, thanks to using "dynamic circuits" that handle feedback and feedforward of quantum measurements and help fasten quantum error corrections.

### Algorithmic qubits

IonQ was initially supportive of IBM's quantum volume metric but now uses a QV variation denominated "algorithmic qubits". They wanted to create a more relevant single-number metric that is usage oriented and more telling for customers. They wanted to avoid using random circuits benchmarking as in IBM's QV.

It's not far from $\log_2$(IBM's QV) but not exactly. It was initially defined as the size of the largest circuit that could run with N qubits and $N^2$ two-qubit gates, but was then refined as $\min(\#\text{qubits}, \sqrt{\#gates})$ with algorithm success probability ≥50% (not the 66% from IBM's QV), assuming the algorithm requires $N^2$ two-qubit gates [1869].

Practically speaking, the #AQ benchmark is run on different algorithms, like the ones defined in the QED-C benchmark. You'll then get several numbers, one for each algorithm and for each tested machine. The #AQ must be represented in a 2D chart as shown in Figure 654, with these various algorithms' success probability represented as colored circles and two axis: on X, the 'depth' of the circuit represented by a log scale of number of two qubits entangling gates, and on Y, the number of qubits used.

IonQ touted in March 2022 having reached an algorithmic qubits record of 20 with its 32 bits Aria system. It was achieved with one of these algorithms, surprisingly, the most successful one, quantum phase estimation, which by the way, is very far from a practical "customer need". Other algorithms had #AQs of 4 to 16.

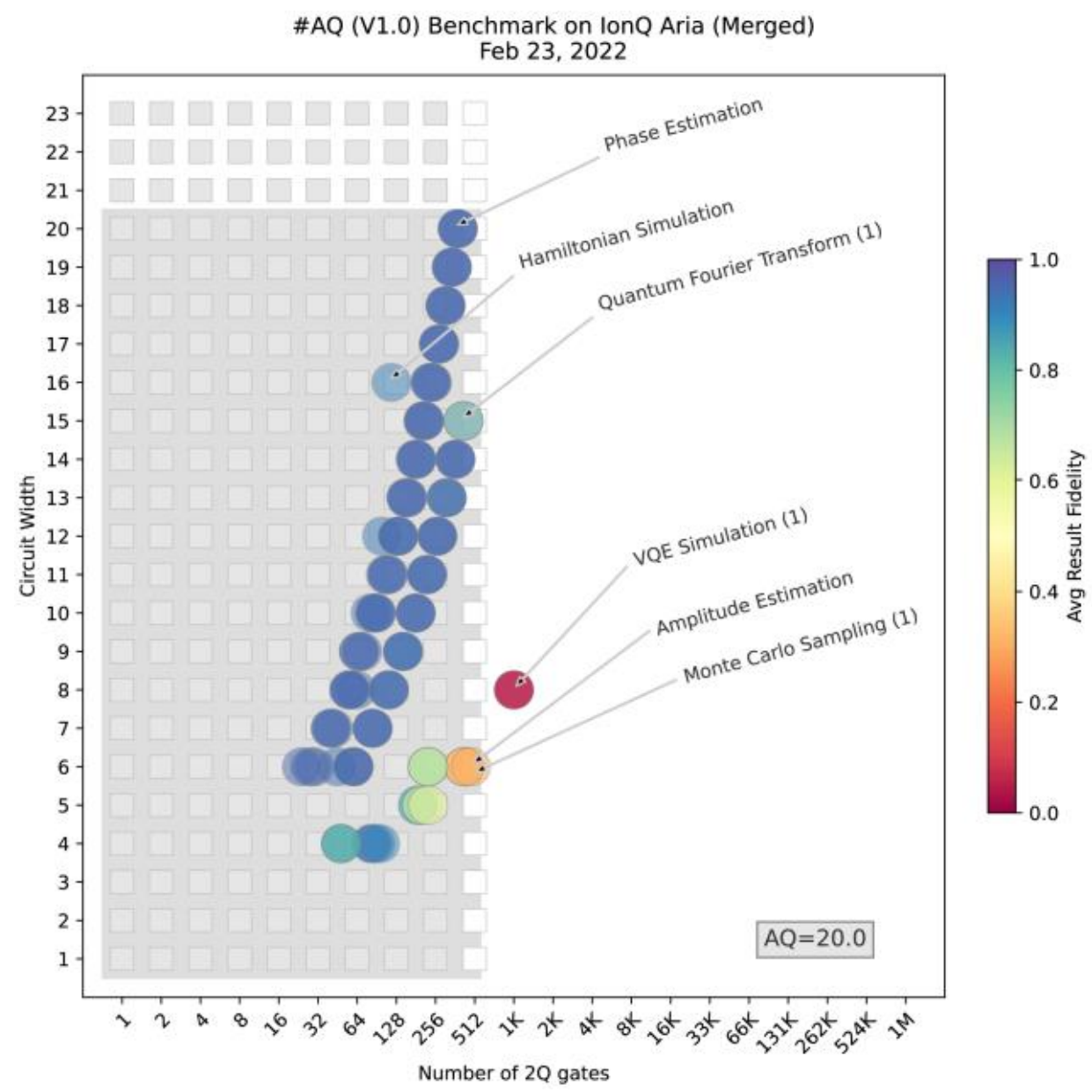

*Figure 654: Source: [Algorithmic Qubits: A Better Single-Number Metric](#) by IonQ, February 2022.*

With higher fidelities qubits, you could run an algorithm with a greater number of two-qubit gates and improve X. But in that case, your #AQ would still be constrained by the number of qubits used. However, with poor fidelities qubits, your #AQ could become much lower than this number of qubits, a bit like with IBM's quantum volume, but on a real algorithm and not with a randomized benchmark.

Since the suite of tested algorithms will change over time, IonQ will define release numbers for its #AQs.

### Other systems level benchmarks and metrics

Other various low-level systems benchmarks are worth mentioning:

---

[1868] See Driving quantum performance: more qubits, higher Quantum Volume, and now a proper measure of speed by Jay Gambetta, Ali Javadi-Abhari, Blake Johnson, Petar Jurcevic, Hanhee Paik and Andrew Wack, November 2021.

[1869] The algorithmic qubits benchmark is described in details in Algorithmic Qubits: A Better Single-Number Metric by IonQ, February 2022 and in QIP 2022 | The application oriented benchmarks for quantum computing (Luming Zhao, IonQ), (1h video).

**Cycle benchmarking** from a team involving Canada, Denmark and Austria which assesses the low-level quality of qubits entanglement, created in 2019[1870].

A proposal made by a team from QuSoft (The Netherlands), the University of Cambridge (UK) and Caltech (USA) in April 2021 is bound to measure the performance of universal quantum computers in a hardware-agnostic way with six structured circuits tests (Bell test, Schrödinger's microscope, Mandelbrot, line drawing, matrix inversion and platonic fractals). It's quite complex to interpret and reading out the graphical results is not straightforward, nor connected to an application need[1871].

In 2021, **DARPA** launched a research RFP for the creation of benchmarks in two categories: application-specific hardware-agnostic benchmarks (TA1, with $1.45M for 18 months) for quantum computing and hardware resource estimates for quantum computers (TA2, with a funding of $1.5M over 18 months). In the end, the project was awarded in March 2022 to Raytheon BBN and University of Southern California. Their benchmark targets all sorts of quantum technologies, both computing and sensing, and are summarized with the **SWAP** nickname corresponding to (size, weight, application and power)[1872]. In July 2022, as part of these benchmarking programs, DARPA awarded a three-year contract of $2.9M to Rigetti, the University of Technology Sydney, Aalto University and the University of Southern California to create benchmarks for large-scale quantum computers.

Another similar approach, but a narrower one, was proposed by researchers from **Brookhaven** and **Pacific Northwest** National Laboratories from the DoE which estimates hardware resources needed for key algorithms[1873]. Another DoE lab, the **Sandia Labs**, proposed a variation of randomized benchmarking that works in the quantum advantage regime[1874]. In 2022, DoE's **Oak Ridge** lab and several US universities proposed a volumetric benchmark qualifying the quality of qubit entanglement, that was tested first on IBM QPUs[1875].

And to be complete, another team from Berkeley, HRL Labs and University of Chicago devised a randomized benchmark measuring noise in non-Clifford quantum gates (those gates that are needed for a QFT and for generating an exponential speedup), extending the work from Google on their 2019 supremacy experiment[1876]. Other extensions are proposed by Alibaba USA with their Universal randomized benchmarking (URB) that scales better than the cross-entropy benchmark used by Google[1877] and with a randomized benchmark supporting a universal gate set[1878].

In a different realm, there is even a proposal of benchmark for **MBQC-based** computing systems, mostly used in photonic qubits[1879].

---

[1870] Presented in Characterizing large-scale quantum computers via cycle benchmarking par Alexander Erhard et al., 2019 (7 pages).

[1871] See Scalable Benchmarks for Gate-Based Quantum Computers by Arjan Cornelissen et al, April 2021 (54 pages).

[1872] See DARPA asks Raytheon BBN and USC researchers to test limits of quantum computing for military applications by John Keller, Military+Aerospace Electronics, March 2022.

[1873] See On the importance of scalability and resource estimation of quantum algorithms for domain sciences by Vincent R. Pascuzzi, Ning Bao and Ang Li, May 2022 (5 pages). They notices that for a QFT, the number of gates and CNOT gates scale exponentially with an increased number of qubits

[1874] See Measuring the Capabilities of Quantum Computers by Timothy Proctor et al, Sandia Labs, August 2020/January 2022 (4 pages) and Scalable randomized benchmarking of quantum computers using mirror circuits by Timothy Proctor et al, Sandia Labs, December 2021 (8 pages).

[1875] See An entanglement-based volumetric benchmark for near-term quantum hardware by Kathleen E. Hamilton, Sophia Economou et al, September 2022 (21 pages).

[1876] See Benchmarking near-term quantum computers via random circuit sampling by Yunchao Liu et al, April 2022 (43 pages)

[1877] See Randomized Benchmarking Beyond Groups by Jianxin Chen et al, Alibaba USA, March 2022 (35 pages).

[1878] See Demonstrating scalable randomized benchmarking of universal gate sets by Jordan Hines, Robin Blume-Kohout, Irfan Siddiqi, Birgitta Whaley, Timothy Proctor et al, August 2022 (31 pages).

[1879] See Measurement-based interleaved randomised benchmarking using IBM processors by Conrad Strydom et al, Stellenbosch University, March 2022 (17 pages).

## Q-score

Atos proposed its Q-score benchmark in December 2020. It's based on determining the maximum size of a standardized problem that can be solved on a given hardware on any quantum programming paradigm[1880]. The first selected problem is the classical combinatorial **MaxCut**. Its variations are used to solve the traveling salesperson problem or various graphs problems with applications in logistic, industry and finance. It can also be used to handle clustering in quantum machine learning. The Q-score benchmark is evaluated with using a hybrid classical+quantum algorithm like with QAOA (Quantum Approximate Optimization Algorithm).

This benchmark creates a simple metric (the number of variables that can be used in the optimization problem) and is independent from the computing paradigm used (gate-based or other) and it doesn't require a quantum computing emulation capacity like with the IBM Quantum Volume. And the algorithm solutions can be verified polynomially on a classical computer. The Q-Score software tools are also open source and published on Github.

Atos plans to publish the Q-scores of various QPU manufacturers. Right now, the record is at around 15Q. It could soon reach 20Q. And 60Q is needed to showcase a real quantum advantage. In August 2022, a Dutch team evaluated **D-Wave** 2000Q and Advantage annealers and obtained a record of 70Q and 140Q, which is much better than existing gate-based QPUs[1881]. But it's not yet sufficient to exceed the performance of classical computers. It seems however than hybrid solutions may under certain conditions outperform supercomputers. Another team, from **Pasqal**, implemented a Q-score benchmark on their neutral atoms simulator and obtained a Q-score of 80Q using a digital simulation of their system[1882]. They solved the benchmark MaxCut problem using a classical machine learning technique to reduce the number of runs on their QPU.

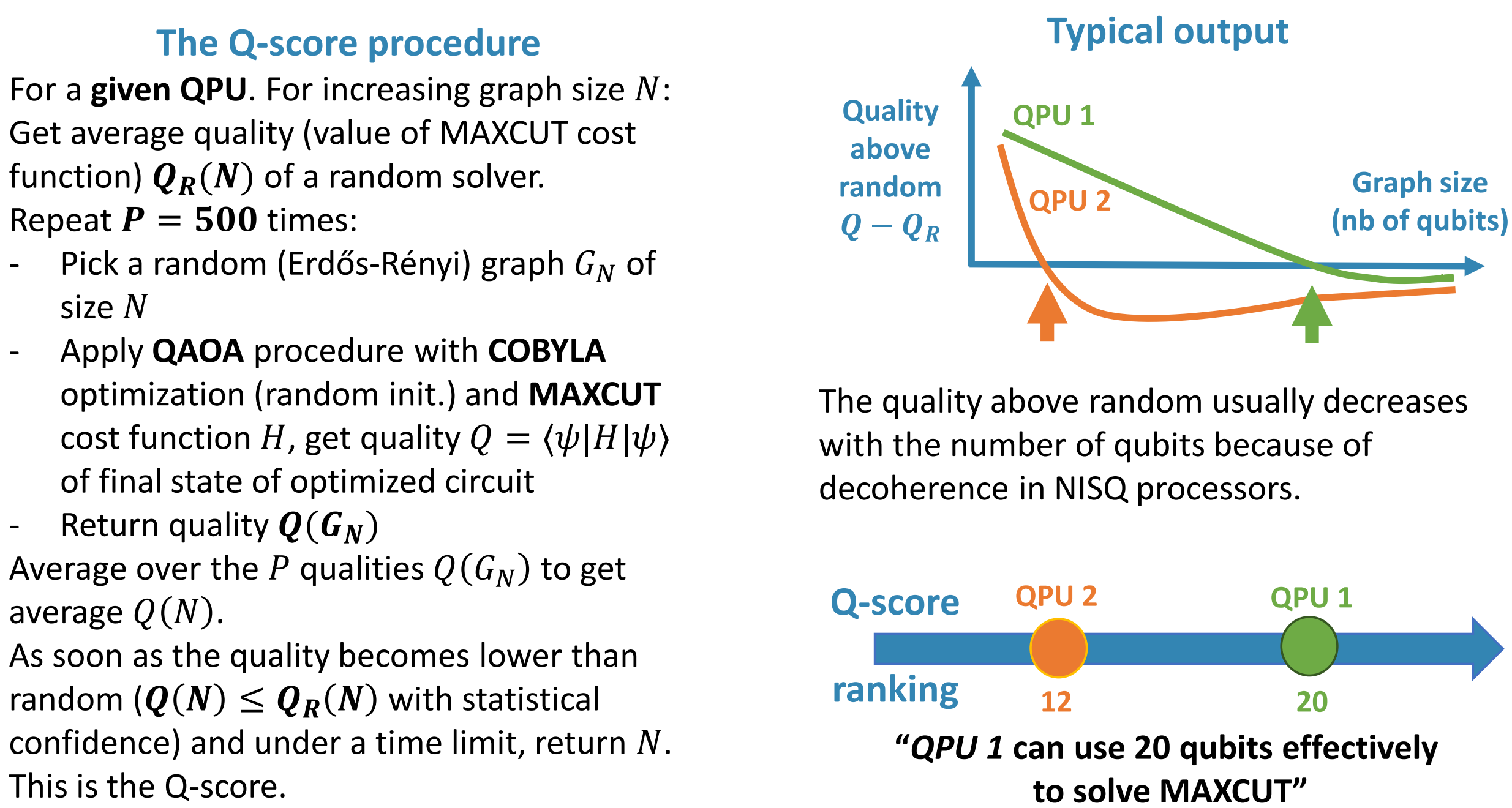

*Figure 655: Atos Qscore calculation method. Source: Atos.*

---

[1880] See Benchmarking quantum co-processors in an application-centric, hardware-agnostic and scalable way by Simon Martiel, Thomas Ayral and Cyril Allouche, IEEE Transactions on Quantum Engineering, February 2021 (11 pages).

[1881] See Evaluating the Q-score of Quantum Annealers by Ward van der Schoot et al, The Netherlands Organisation for Applied Scientific Research, August 2022 (8 pages).

[1882] See Efficient protocol for solving combinatorial graph problems on neutral-atom quantum processors by Wesley da Silva Coelho, Mauro D'Arcangelo and Louis-Paul Henry, August 2022 (23 pages).

This benchmark would need to be completed by other benchmarks such as one for quantum chemistry simulation (number of atoms in molecule) and another on the size of the maximum number that can be factorized (in power of 2).

### Other use case benchmarks

**QED-C** is supporting a series of application oriented benchmarks proposed by researchers from Princeton, HQS, QCI, IonQ, D-Wave and Sandia Labs in the USA[1883]. It mixes the volumetric benchmarking method from IBM and a comparison of performance with various standard algorithms. They did some comparison on actual quantum hardware from IBM, Rigetti, HQS and IonQ. It was launched in October 2021.

**SupermarQ** is a benchmark proposed in 2022 by Super.tech (acquired in May 2022 by ColdQuanta) and supported by Amazon and Intel (both of which have no functional quantum computer yet). It is a suite of application benchmarks that covers use cases in finance, pharmaceuticals, energy, chemistry and other verticals. SupermarQ also contains an error correction benchmark[1884].

**QPack** is a benchmark based on the Max-Cut problem (like Atos's Q-score), the dominating set problem (DSP) and the traveling salesperson problem (TSP), and using the QAOA algorithm as a benchmarking tool, proposed by TU Delft researchers. It measures the maximum problem size a quantum computer can solve, the required computing runtime and the achieved accuracy[1885].

The DoE **ORNL** (Oak Ridge National Laboratory) proposed a benchmark for chemical simulation[1886]. It deals with the simulation of three 2-atoms molecules (NaH, KH et RbH) which can be simulated on existing IBM and Rigetti 20 and 16 qubits superconducting systems. It's not generic and can't go beyond these molecule sizes.

Other single use-cases benchmarks have also been created: **Agnostiq** created a benchmark dedicated to optimizing financial portfolio using the Quantum Approximate Optimization Algorithm (QAOA)[1887], scientists from New-York created a benchmark related to Grover's search algorithm[1888] and Zapata Computing created one benchmarking on fermionic simulations[1889].

### IEEE and ISO

The **IEEE** has launched several benchmarking initiatives on its own with a standard to be submitted in 2024[1890].

---

[1883] See Application-Oriented Performance Benchmarks for Quantum Computing by Thomas Lubinski et al, October 2021 (33 pages).

[1884] See SupermarQ: A Scalable Quantum Benchmark Suite by Teague Tomesh et al, Princeton, University of Chicago, Super.tech and Intel, February 2022 (15 pages) and Applying classical benchmarking methodologies to create a principled quantum benchmark suite by Teague Tomesh et al, March 2022.

[1885] See QPack: Quantum Approximate Optimization Algorithms as universal benchmark for quantum computers by Koen Mesman et al, April 2022 (28 pages) and QPack Scores: Quantitative performance metrics for application-oriented quantum computer benchmarking by Huub Donkers et al, May 2022 (23 pages).

[1886] See ORNL researchers advance performance benchmark for quantum computers, January 2020.

[1887] See Wasserstein Solution Quality and the Quantum Approximate Optimization Algorithm: A Portfolio Optimization Case Study by Jack S. Baker et al, February 2022 (21 pages).

[1888] See Quantum search on noisy intermediate-scale quantum devices by Kun Zhang, Kwangmin Yu and Vladimir Korepin, February 2022 (12 pages).

[1889] See An application benchmark for fermionic quantum simulations by Pierre-Luc Dallaire-Demers et al, Zapata Computing, March 2020 (14 pages). See An application benchmark for fermionic quantum simulations by Pierre-Luc Dallaire-Demers et al, Zapata Computing, March 2020 (14 pages).

[1890] See P7131 - Standard for Quantum Computing Performance Metrics & Performance Benchmarking. It covers gate-based quantum computing. See also Metrics & Benchmarks for Digital Quantum Computing by Robin Blume-Kohout (18 slides) and Summary of the IEEE Workshop on Benchmarking Quantum Computational Devices and Systems, 2019. Also, see P2995 - Trial-Use Standard for a Quantum Algorithm Design and Development and P3120 - Standard for Quantum Computing Architecture.

They entertain several working groups with participating industry vendors and academic institutions.

- Trial-Use Standard for a Quantum Algorithm Design and Development (P2995, details).
- Standard for Quantum Computing Architecture (P3120, details).
- Software-Defined Quantum Communication (P1913).
- Standard for Quantum Computing Definitions (P7130).
- Standard for Quantum Computing Performance Metrics & Performance Benchmarking (P7131).

There is also an **ISO** working group working on quantum computing (ISO/IEC JTC 1/WG 14).

### Benchmarking tools

**Metriq** from **Unitary Fund** announced in May 2022 is a repository of benchmarking results.

**MQT Bench** aka the Munich Quantum Toolkit is a multiple-abstraction level suite of benchmarking tools covering both low-level abstraction building blocks (QFT, QPE, amplitude estimation) and higher-level ones (Grover, Shor, HHL)[1891].

**Arline** (Germany) is also proposing its own benchmarking tools suite that is used to compare compiler optimizers[1892].

**QUARK** (Germany) is another industry application-centric benchmarking proposal based on an opt-source framework[1893].

### Quantum supremacy and advantage

Quantum supremacy and advantage are two terms used to qualify the superiority of quantum computers as compared with the most powerful supercomputers.

**Quantum supremacy** was a term coined by John Preskill in a paper presented at the Solvay Congress in 2011[1894]. It is achieved when a problem, useful or not, is only solvable with a quantum algorithm running on a quantum computer, and there is no known classical algorithm for the most powerful supercomputers that could run in a reasonable human scale time[1895]. Quantum supremacy was a goal for researchers and vendors like Google and it became claims, with a peak in October 2019.

Quantum supremacy doesn't mean that a given quantum computer is supremely more powerful than all its contemporary supercomputers. The term is applicable to a combination of a specific problem and related quantum algorithm, a given quantum computer, and the best-in-class available classical algorithms adapted to the most powerful available supercomputers. The criteria are moving targets. Supercomputers have not said their last word and are classical algorithms are also improved[1896].

---

1891 See MQT Bench: Benchmarking Software and Design Automation Tools for Quantum Computing by Nils Quetschlich et al, TUM Germany, Johannes Kepler University Linz, Austria and Hagenberg GmbH (SCCH), Austria, April 2022 (7 pages).

1892 See Arline Benchmarks: Automated Benchmarking Platform for Quantum Compilers by Y. Kharkov et al, February 2022 (27 pages).

1893 See QUARK: A Framework for Quantum Computing Application Benchmarking by Jernej Rudi Finžgar, Philipp Ross, Leonhard Hölscher, Johannes Klepsch and Andre Luckow, August 2022 (12 pages).

1894 It is described in Quantum Computing and the Entanglement Frontier, 2011.

1895 See Quantum supremacy: Some fundamental concepts by Man-Hong Yung, January 2019 (2 pages) according to which there are three ways to demonstrate quantum supremacy: boson sampling, PQI and chaotic quantum circuits.

1896 See Quantum Algorithms Struggle Against Old Foe: Clever Computers by Ariel Bleicher, February 2018. This mentions the discovery of classical algorithms that are as powerful as their quantum equivalents, such as the one from Ewin Tang, already mentioned on page 61.

Robert König (Technical University of Munich), David Gosset (University of Waterloo, Canada) and Sergey Bravyi (IBM) demonstrated in October 2018 that quantum computers could perform operations inaccessible to conventional computers but based only on the case of a particular algorithm[1897].

Others are devising "*proofs of quantumness*", which are methods to demonstrate to a classical verifier that a quantum computer can perform some computational tasks that a classical computer with comparable resources cannot (meaning, the classical computer must achieve things in a tractable way, i.e. less than polynomial time)[1898].

Some D-Wave and Google benchmarks carried out in 2015 and showing the superiority of the quantum solution were then contradicted by the creation of algorithms optimized for supercomputers under certain conditions. In a few years' time, it will certainly come into play for a few algorithms that cannot have optimized supercomputer equivalents.

Google's quantum supremacy announced in October 2019 was touted as serious back then[1899]. It was later downgraded. It was based on a sort of *random numbers sampling* algorithm using 53 qubits. But there was only a 0,15% chance to get a good result, thus the need to run the algorithm 3 million times to compute an average. When Sycamore is used for useful algorithms, fewer than 20 qubits are used and we're far off any quantum supremacy or advantage.

Cristian and Elena Calude of the University of Auckland in New Zealand then argued that a high-performance limit, that of a precise quantum computer, is compared to a low limit which is the best performance in solving the same problem in a supercomputer[1900]. Quantum supremacy is thus a comparable between the existence of a quantum performance and the assumption of the non-existence of an equivalent performance in classical computing.

A 2020 paper from Yiqing Zhou (University of Illinois), Edwin Miles Stoudenmire (Flatiron Institute) and Xavier Waintal (CEA-IRIG) provided an interesting "reset" view on Google's quantum supremacy. It stated that emulating Sycamore's processes in a classical computing could use some compression technique to take into account the qubit's noise. With this compression, emulating Sycamore is much less costly and can be done on a simple microcomputer[1901]. But this corresponded to a 95% emulated gates fidelity. With a 99% fidelity matching Sycamore, it would still require a couple hundred cores and some TB of memory, fitting in a datacenter rack. There would be an energy advantage for Sycamore but going down from x500.000 vs the IBM Summit to about only x325 for a 500 core cluster server. A Chinese research team published an even better performance in 2021, using an improved tensor contraction technique and running for 15 hours on a cluster of 512 GPUs[1902].

In 2022, Xavier Waintal, Edwin Miles Stoudenmire and a team of Atos researchers including Thomas Ayral improved their classical simulation of Google's supremacy with a density-matrix tensor networks renormalization group (DMRG) algorithm. It did run on a few classical CPU cores from an Atos QLM and in a couple hours. The algorithm has a simulation cost scaling polynomially with the number of qubits and the depth of the circuit.

---

[1897] See First proof of quantum computer advantage, October 2018 and Quantum advantage with shallow circuits, April 2017 (23 pages).

[1898] See Simpler Proofs of Quantumness by Zvika Brakerski et al, May 2020 (12 pages).

[1899] See The power of random quantum circuits by Bill Fefferman, 2019 (25 slides) that explains the power behind the randomized benchmarking technique chose by Google.

[1900] In The road to quantum computing supremacy, 2017.

[1901] See What limits the simulation of quantum computers? by Yiqing Zhou, Edwin Miles Stoudenmire and Xavier Waintal, PRX, November 2020 (14 pages).

[1902] See Solving the sampling problem of the Sycamore quantum circuits by Feng Pan, Keyang Chen, and Pan Zhang, PRL, July 2022 (9 pages).

A consequence of this work is that to reach an exponential quantum computing advantage, it is more important to increase qubit fidelities than their number[1903].

In 2021, a Chinese team was even able to classically simulate the Google Sycamore cross-entropy benchmark with a single Nvidia A100 GPGPU running for 149 days with a fidelity of 73,9% while Sycamore's fidelity was only 0,2%[1904]. Another Chinese research team, from Alibaba, found a way to optimize Sycamore's emulation in September 2021 to reach only 20 days of computing on a system equivalent of the IBM Summit[1905]. And yet another one in October 2021 could simulate it on the new Sunway supercomputer in 304 seconds, using a tensor compression technique[1906]. At last, in October 2022, Gil Kalai et al produced an in-depth and well-crafted detailed analysis of Google's supremacy experiment and related protocoles[1907].

Likewise, and the other way around, a Google team did show that the 2021 Chinese gaussian boson sampling (GBS) supremacy experiment could be efficiently simulated classically (still, quadratically)[1908]. Another team, from the UK, achieved a similar feat in 2022, in which they did simulate a 100-mode and up to 60 click detection events GBS on a ~100,000-core supercomputer[1909].

In 2018, IBM researchers demonstrated that quantum supremacy was assured in the long run, even with quantum computers that can chain a finite and constrained number of quantum gates[1910]. In December 2020, they published a theorical model that could prove some quantum advantage, solving binary function problems, and tested on a low scale on a 27 qubits superconducting system[1911]. These various, sometimes convoluted, performances are very hard to compare and evaluate.

We cannot avoid mentioning the debates around the term supremacy with its bad social and political meaning. So far, I've seen only one replacement proposal, that hasn't been adopted yet, which consists in using the simpler term **primacy** with the same meaning[1912]. Another interesting fringe phenomenon is the usage of some quantum supremacy for things that are not at all related to quantum computing. It's borderline click baiting[1913]!

---

1903 See A density-matrix renormalization group algorithm for simulating quantum circuits with a finite fidelity by Thomas Ayral, Thibaud Louvet, Yiqing Zhou, Cyprien Lambert, E. Miles Stoudenmire and Xavier Waintal, August 2022 (25 pages).

1904 See Simulating the Sycamore quantum supremacy circuits by Feng Pan and Pan Zhang, March 2021 (9 pages). The authors improved their work in Solving the sampling problem of the Sycamore quantum supremacy circuits by Feng Pan et al, November 2021 (9 pages).

1905 See Efficient parallelization of tensor network contraction for simulating quantum computation by Cupjin Huang et al, Alibaba, September 2021 (10 pages).

1906 See Closing the "Quantum Supremacy" Gap: Achieving Real-Time Simulation of a Random Quantum Circuit Using a New Sunway Supercomputer by Yong (Alexander) Liu et al, October 2021 (18 pages). The China team behind this was awarded the 2021 ACM Gordon Bell Prize. Their work was later contradicted by ORNL researchers who had developed Google's supremacy classical simulation in 2019. See China's exascale quantum simulation not all it appears by Nicole Hemsoth, NextPlatform, November 2021.

1907 See Google's 2019 "Quantum Supremacy" Claims: Data, Documentation, and Discussion by Gil Kalai, Yosef Rinott and Tomer Shoham, October 2022 (32 pages).

1908 See Efficient approximation of experimental Gaussian boson sampling by Benjamin Villalonga, Hartmut Neven et al, September 2021 (15 pages).

1909 See The boundary for quantum advantage in Gaussian boson sampling by Jacob F.F. Bulmer et al, 2022 (8 pages).

1910 See Scientists Prove a Quantum Computing Advantage over Classical by Bob Sutor, October 2018, Quantum advantage with shallow circuits, Sergey Bravyi, David Gosset and Robert Koenig, 2017 (23 pages) and the video Quantum advantage with shallow circuits, IBM Research, December 2017 (44 minutes).

1911 See Quantum advantage for computations with limited space by Dmitri Maslov, Sarah Sheldon et al, IBM Research, December 2020 (12 pages). Also published in Nature Physics in June 2021.

1912 See Quantum Computing 2022 by James D. Whitfield et al, January 2022 (13 pages)

1913 See Quantum supremacy in mechanical tasks: projectiles, rockets and quantum backflow by David Trillo et al, IQOQI Vienna, September 2022 (18 pages). It deals with some relativistic physics phenomenon.

**Quantum advantage** is a different concept. It corresponds to a situation where a quantum computer executes a useful algorithm faster than on the most powerful supercomputers. It seems at first glance not as strong an argument as with quantum supremacy, but it happens that it's more difficult to reach a quantum advantage than a quantum supremacy.

But some are pushing various definitions for quantum advantage. Sometimes, it even has the same meaning than quantum supremacy but with a more politically correct terminology and for others, it's a stronger statement than quantum supremacy, meaning the same but for a useful algorithm.

One key aspect of all supremacy claims is that they implement some random benchmark that is difficult to simulate digitally. Like with boson samplings, these are physical processes that are difficult to simulate. Supremacies are obtained with algorithms using no input data. Thus, they don't solve any useful problem. An advantage is supposed to solve a useful problem with some input and output data.

There are much fewer advantage claims than supremacy claims. And sometimes, they are misnomers or applied to very specific cases, even beyond quantum computing[1914]. And they have no more input data than with quantum supremacy claims.

One first example comes from an interesting work from a team of researchers from France and Edinburgh announced in February 2021, including Eleni Diamanti and Iordanis Kerenidis[1915]. It involved a complicated photonics-based experiment that didn't do any real calculation. It was about putting in place a QMA (Quantum Merlin Arthur) verification protocol. The implemented protocol is an interactive test that requires, through a network, the verification of the solution of a complex NP-complete optimization problem without having to communicate the whole solution. The breakthrough that made this possible was the creation of a system encoding the solution result with partial information about the solution to be verified from one network node to another. The protocol compresses a large vector state describing the partial information on the solution, involving some entanglement and multi-mode photons quantum communications. This compression protocol would make it possible to verify the results in a much smaller time. No actual verification was done on the other end of the system.

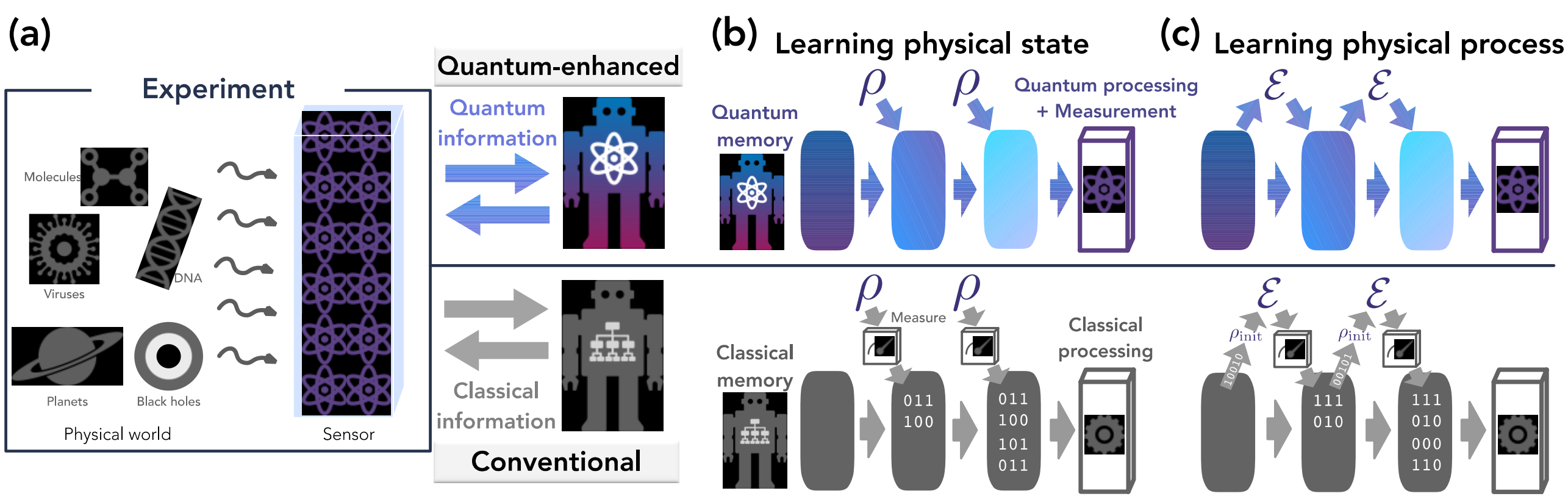

*Figure 656: a quantum advantage can come from connecting quantum sensors and quantum computers, avoid the tedious steps of quantum-to-classical and classical-to-quantum data conversions. Source: Quantum advantage in learning from experiments by Hsin-Yuan Huang, Hartmut Neven, John Preskill et al, December 2021 (52 pages) with 40 Sycamore qubits.*

[1914] See Quantum Advantage of Threshold Changeable Secret Sharing Scheme by Xiaogang Cheng et al, September 2022 (11 pages) which deals with secret key exchanges using quantum computing but not about solving an NPish problem.

[1915] See Experimental demonstration of quantum advantage for NP verification with limited information by Federico Centrone, Niraj Kumar, Eleni Diamanti, and Iordanis Kerenidis, published in Nature Communications, February 2021 (13 pages). This was a follow-up of Quantum superiority for verifying NP-complete problems with linear optics by Juan Miguel Arrazola, Eleni Diamanti & Iordanis Kerenidis, Nature, 2018 (8 pages).

We have here a quantum advantage coming from the way to connect a quantum computer solving an NP-complete SAT problem and another quantum computer verifying the solution with partial information. Both computers do not exist yet. Another view on this would be that it proposes an architecture to verify a solution to an NP-problem on an end-to-end solution.

Another example of quantum computing advantage could be reached with feeding a quantum computer with data coming from quantum sensors data and transmitted "quantumly" instead of classically. That's what demonstrated a team of Google and other researchers in 2021 and shown in Figure 656[1916]. This requires some form of quantum memory that is still to be created.

Many experts estimate that the threshold of 50-ish quality qubits, with a low error rate and a long coherence time, will be needed to achieve any real quantum advantage. These will probably be logical qubits, assembling physical qubits and some quantum error correction codes.

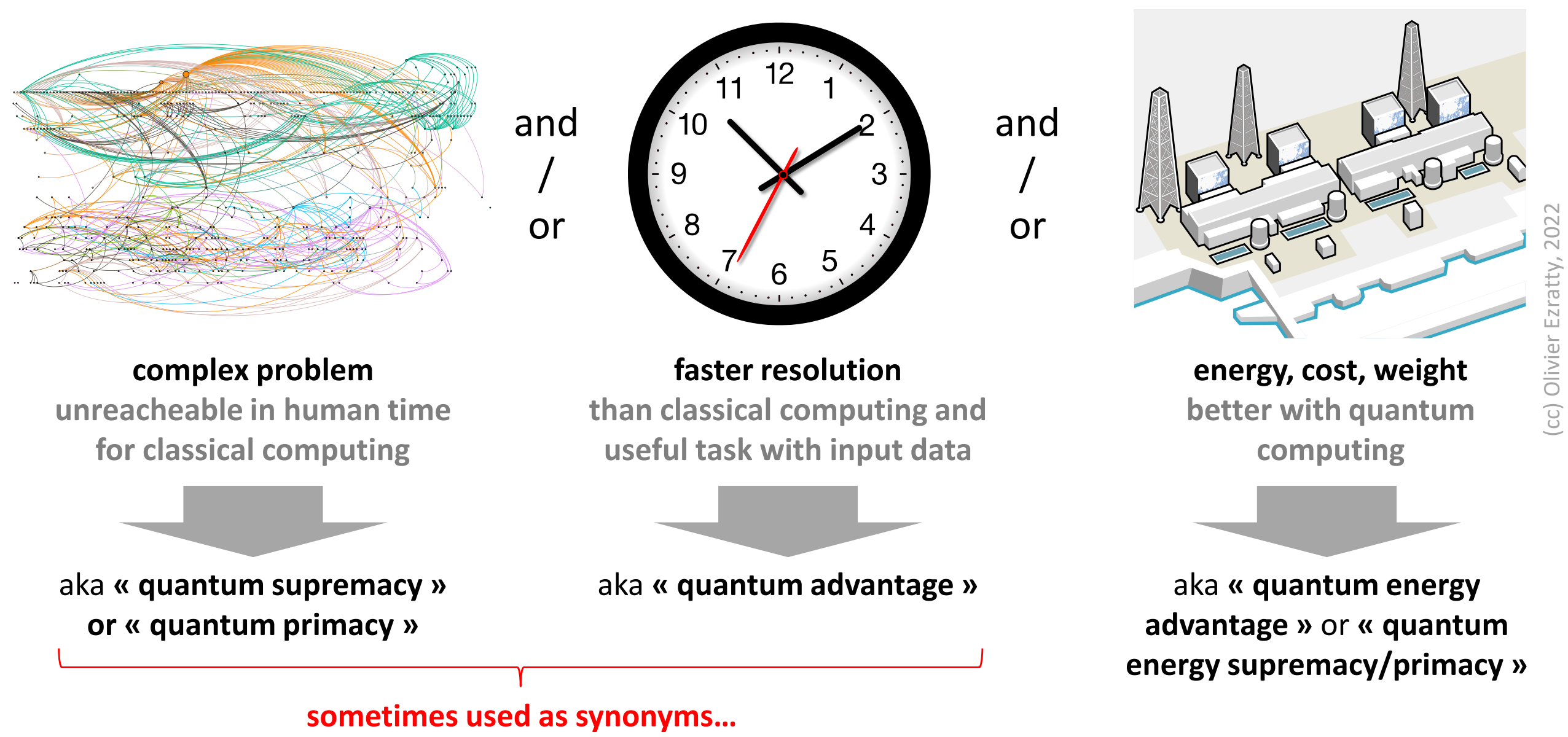

*Figure 657: trying to define quantum supremacy (or primacy) and quantum advantage. (cc) Olivier Ezratty, 2022.*

**Quantum energy advantage** is another threshold that may arise someday when on top of some computing time benefits, we could highlight the fact that quantum computers consume much less energy than supercomputers for solving similar problems. This is still a subject of research and dealt with as part of the QEI that is described page 251[1917].

Here's a tabulated consolidation of the various quantum supremacies and advantages announced since 2019[1918]. It shows that between 2019 and 2021, none of these achieved real useful computing with some application input data. And in 2022, we start to see appearing some real or potential interesting and narrow quantum advantages with input data.

The IBM et al example below was touting a quantum advantage, but it would require a NISQ QPU with 96-qubits and 99,99% 2-qubit gates and measurement fidelities which is extrapolated from a 12-qubit Quantinuum QPU having such fidelities. Problem is these trapped ion QPUs are quite hard to scale.

---

1916 See Quantum advantage in learning from experiments by Hsin-Yuan Huang, Hartmut Neven, John Preskill et al, December 2021 (52 pages) with 40 Sycamore qubits.

1917 See the Quantum Energy Initiative already discussed elsewhere in this book starting page 259.

1918 The Arizona performance is documented in Researchers demonstrate a quantum advantage by University of Arizona, June 2021, referring to Quantum-Enhanced Data Classification with a Variational Entangled Sensor Network by Yi Xia et al, June 2021 (17 pages). Their setting used variational quantum circuits for a classification of multidimensional radio-frequency signals using entangled sensors.

| who and when | architecture | algorithm | input data | comment |
|---|---|---|---|---|
| **Google**, Oct 2019 | Sycamore, 53 superconducting qubits | cross entropy benchmarking | **none** | running a random gates algorithm |
| **China**, December 2020 | 70 photons modes GBS (Gaussian Boson Sampling) | interferometer photons mixing | **none** | running a random physical process |
| **IBM Research**, December 2020 | IBM 27 superconducting qubits | symmetric Boolean functions | SLSB3 function parameters | **theoretical demonstration of quantum advantage** |
| **Kerenidis, Diamanti** et al, March 2021 | multi-mode photon dense encoding of verified solution | Quentin Merlin Arthur based verification | output from some quantum computation (not implemented) | **no actual computing done in the experiment** |
| **China**, April 2021 | Quantum walk on 62 superconducting qubits | simple quantum walk | simulating a 2-photons Mach-Zehnder interferometer | **no quantum advantage at all** |
| **University of Arizona**, May 2021 | supervised learning assisted by an entangled sensor network | variational algorithm, classical computing | data extracted from three entangled squeezed light photonic sensors | **not a quantum « computing » advantage per se** |
| **China**, June 2021 | 66 superconducting qubits and 110 couplers, Zuchongzhi 1, then 2.1 | cross entropy benchmarking | **none** | 56 used qubits |
| **China**, September 2021 | | | | 60 used qubits |
| **China**, June 2021 | 144 photons modes GBS and up to 113 detected events | interferometer photons mixing | **none** | **parametrizable photon phases could lead to a programmable system** |
| **Google, AWS, Harvard** et al, December 2021 | quantum sensors feeding a quantum computer | learning about the principal component of a noisy state | **quantum output from quantum sensors** | **requires some quantum memory** |
| **Xanadu** June 2022 | 216 squeezed photons modes GBS (Gaussian Boson Sampling) | time domain multiplexingand interferometer photons mixing | **programmable GBS with 1,296 parameters** | **first programmable GBS** |
| **IBM** et al, September 2022 | hybrid algorithm that could run on NISQ QPUs | QML-TDA unsupervised machine learning technique for extracting valuable shape-related data features | **small data sets related to cosmic microwave background** | **exponential speedup, resilient to noise, requires 96-qubit QPU with 2Q gate and measurement fidelity of 99.99%...** |

*Figure 658: an inventory of past quantum advantages/supremacies announcements and their underlying characteristics. (cc) Olivier Ezratty, 2022. Sources: Google 2019: Quantum supremacy using a programmable superconducting processor by Frank Arute, John Martinis et al, October 2019 (12 pages). China 2020: Quantum computational advantage using photons by Han-Sen Zhong et al, December 2020 (23 pages). IBM 2020 : Quantum advantage for computations with limited space by Dmitri Maslov et al, December 2020 (12 pages). Kerenidis / Diamanti 2021: Experimental demonstration of quantum advantage for NP verification with limited information by Federico Centrone, Niraj Kumar, Eleni Diamanti, and Iordanis Kerenidis, published in Nature Communications, February 2021 (13 pages). China April 2021: See Quantum walks on a programmable two-dimensional 62-qubit superconducting processor by Ming Gong, Science, May 2021 (34 pages). Arizona 2021: Quantum-Enhanced Data Classification with a Variational Entangled Sensor Network by Yi Xia et al, June 2021 (19 pages). China June 2021: Strong quantum computational advantage using a superconducting quantum processor by Yulin Wu, Jian-Wei Pan et al, June 2021 (22 pages). China September 2021: Quantum Computational Advantage via 60-Qubit 24-Cycle Random Circuit Sampling by Qingling Zhu, Jian-Wei Pan et al, September 2021 (15 pages). China June 2021: Phase-Programmable Gaussian Boson Sampling Using Stimulated Squeezed Light by Han-Sen Zhong, Chao-Yang Lu, Jian-Wei Pan et al, June 2021 (9 pages). Google, AWS, Harvard: Quantum advantage in learning from experiments by Hsin-Yuan Huang, Hartmut Neven, John Preskill et al, December 2021 (52 pages) with 40 Sycamore qubits. Xanadu: Quantum computational advantage with a programmable photonic processor by Lars S. Madsen et al, Xanadu, June 2022 (11 pages). IBM: Towards Quantum Advantage on Noisy Quantum Computers by Ismail Yunus Akhalwaya et al, September 2022 (32 pages) also discussed in Quantifying Quantum Advantage in Topological Data Analysis by Dominic W. Berry, Ryan Bab bush et al, September 2022 (41 pages) and contested in Complexity-Theoretic Limitations on Quantum Algorithms for Topological Data Analysis by Alexander Schmidhuber and Seth Lloyd, September 2022 (24 pages).*

## Quantum software development tools key takeaways

- Gate-based programming involves either graphical circuit design (mostly for training purpose) and (usually) Python based programming when qubit gates structures must be designed in an automated way.
- Python based programming is relying on libraries like IBM's Qiskit or Google's Cirq. There are however many development tools coming from universities and research labs like Quipper. Some tools like ZX Calculus are highly specialized and used to create quantum error correction codes or low-level systems.
- Quantum computing is based on running code multiple (thousands...) times and averaging the results. A single individual run yields a probabilistic outcome while many run averages will converge into deterministic ones.
- Most quantum computers are used in the cloud, through offerings coming from the computer vendors themselves like IBM or D-Wave or from cloud service providers like Amazon, Microsoft, Google and OVHcloud.
- Quantum emulators are very useful to learn programming, test it until the limits of classical emulation (about 40-50 qubits) and also help debug small-scale quantum algorithms. When these emulators include physical simulators of the underlying qubit physics like with Bosonic Qiskit and Quandela Perceval, they help create algorithms that are error-resilient and create quantum error correction codes. Quantum emulation is an indispensable part of any quantum cloud offering.
- Gate-based programs debugging is a significant challenge as it is difficult to implement equivalents of classical code breaking points. As a result, quantum code certification and verification is a new key discipline, particularly for distributed computing architectures such as the ones relying on the concept of blind quantum computing.
- Benchmarking quantum computers is an unsettled technique with many competing approaches. It includes the various techniques used to qualify so-called quantum supremacies and quantum advantages. Not a single of them, as of 2021, did show a real computing advantage compared to classical computing. The reasons were multiple, the main ones being that these experiments didn't implement any algorithm using some input data. But starting in 2022, we see appearing some relevant quantum advantage with actual data and useful algorithms running on NISQ hardware.

# Quantum computing business applications

Most algorithms mentioned before are generally very low-level. How about assembling them into business solutions, market by market? We are still far from having things settled for that respect. The quantum software industry is still very immature and for good reasons, since quantum computers are very limited at this stage. We are still in a stage equivalent where the computer industry was in the mid-1950s, when the software industry was in its infancy.

Still, you can discover here and there a lot of so-called case studies, mostly pushed by D-Wave and IBM and their customers or partners. These relate to proof-of-concepts and software prototypes. Most of these are not yet production grade nor bring any practical benefit compared to classical computing due to the limitations of existing hardware. Still, all this is very useful. This is an indispensable learning phase for research, startups and the industry. It's part of a readiness process that will speed things up when hardware will ramp-up. And this ramp-up will happen progressively.

## Market forecasts

Any new technology wave brings its market forecasts data born out of analysts and market survey companies. They have a very traditional closed-loop system in place: vendors want to get some ideas of customer demand or positive confirmation of their own biases, analysts poll large customers to get some understanding about their plans, and *voila*, you get your nice market predictions. It often looks like linear or simple nonlinear regressions. These predictions can become either self-fulfilling prophecies or total failures. The Gartner Group has turned its simplistic hype curve into a kind of Schrödinger's time and topic-independent wave equation of technology trends. But nobody really checked it, particularly when this curve was highly dependent on complicated scientific and technology challenges. It's more about probabilities than simplistic curves. Today, avoiding any cautious, the current analyst mantra is to claim that the uncertainty on the advent of quantum computing is not "if" but "when", based more on market assessments than on quantum physics and technologies roadmaps understanding.

So, how could you predict the size and shape of the quantum software market, vertical per vertical, when you have no idea of when actual useful quantum accelerators will show up? Will it follow an exponential market growth rate worthy of those of the microcomputer and smartphones industries? Let's look at what we have in store.

**BCG**'s quantum computing growth forecasts illustrate this strong uncertainty. They showcase predictions with an optimistic scenario, which starts seeing growth around 2030, and a very conservative one, which only takes off after 2040[1919]. In both cases, the quantum computing market grows linearly. They don't integrate a scenario of the emergence of the NISQ, or "Noisy Intermediate-Scale Quantum", or any advent of quantum simulators before NISQ become usable[1920].

BCG has however created in 2018 a good inventory of the current potential qualitative use cases of quantum computing per vertical market. This covers both case studies coming from D-Wave and prospective applications devises by research labs and with large industry companies including the usual suspects from the aerospace, chemistry, energy, pharmaceuticals and financial sectors.

Most of these big names worked with either D-Wave or IBM, and sometimes with some independent software vendors or large IT services firms like Accenture.

[1919] See The coming quantum leap in computing, BCG, May 2018 (19 pages).

[1920] See Quantum Computing in the NISQ era and beyond, John Preskill, 2018 (20 pages).

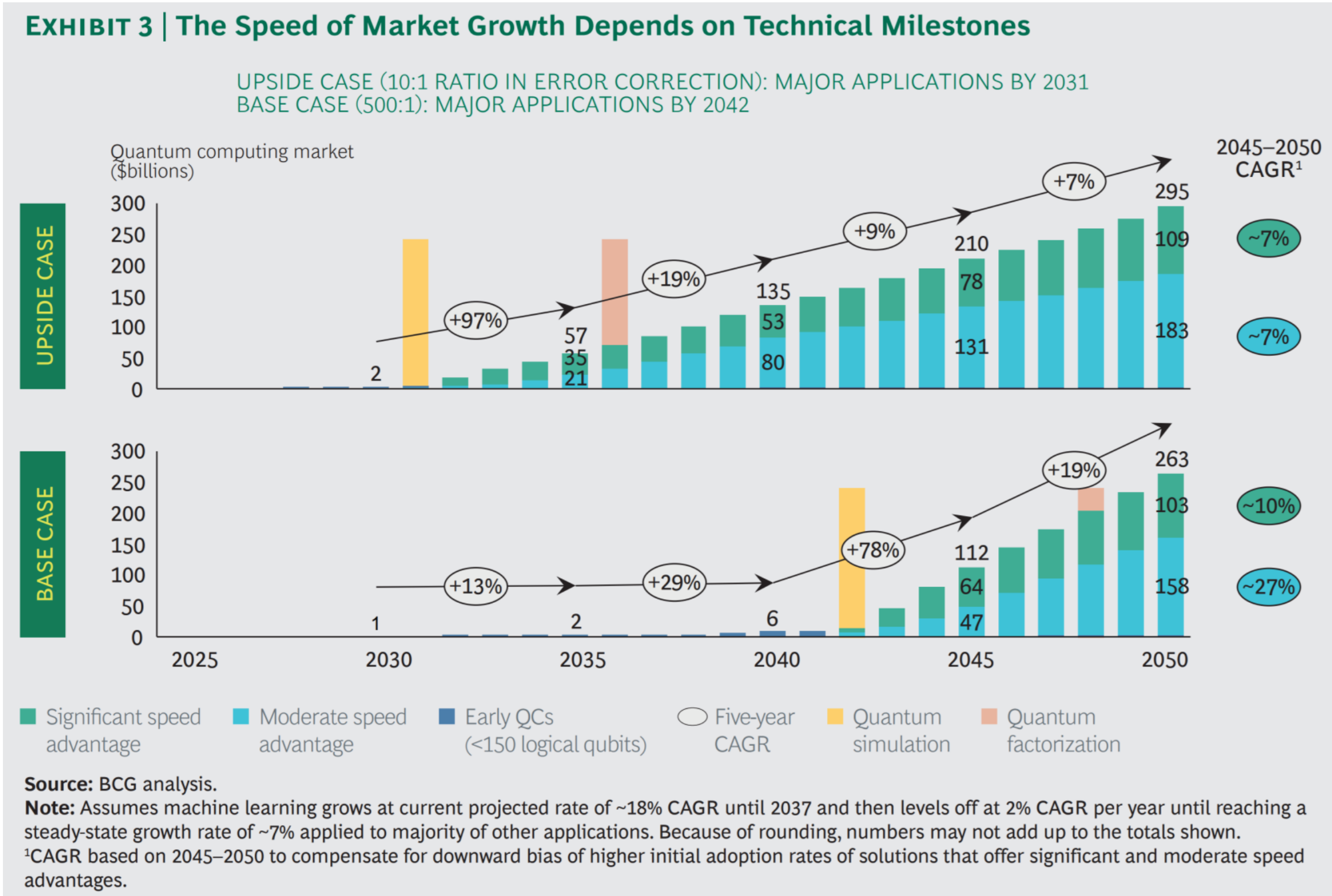

*Figure 659: BCG market forecast from 2018 with optimistic and pessimistic scenarios. Source: [The coming quantum leap in computing](), BCG, May 2018 (19 pages).*

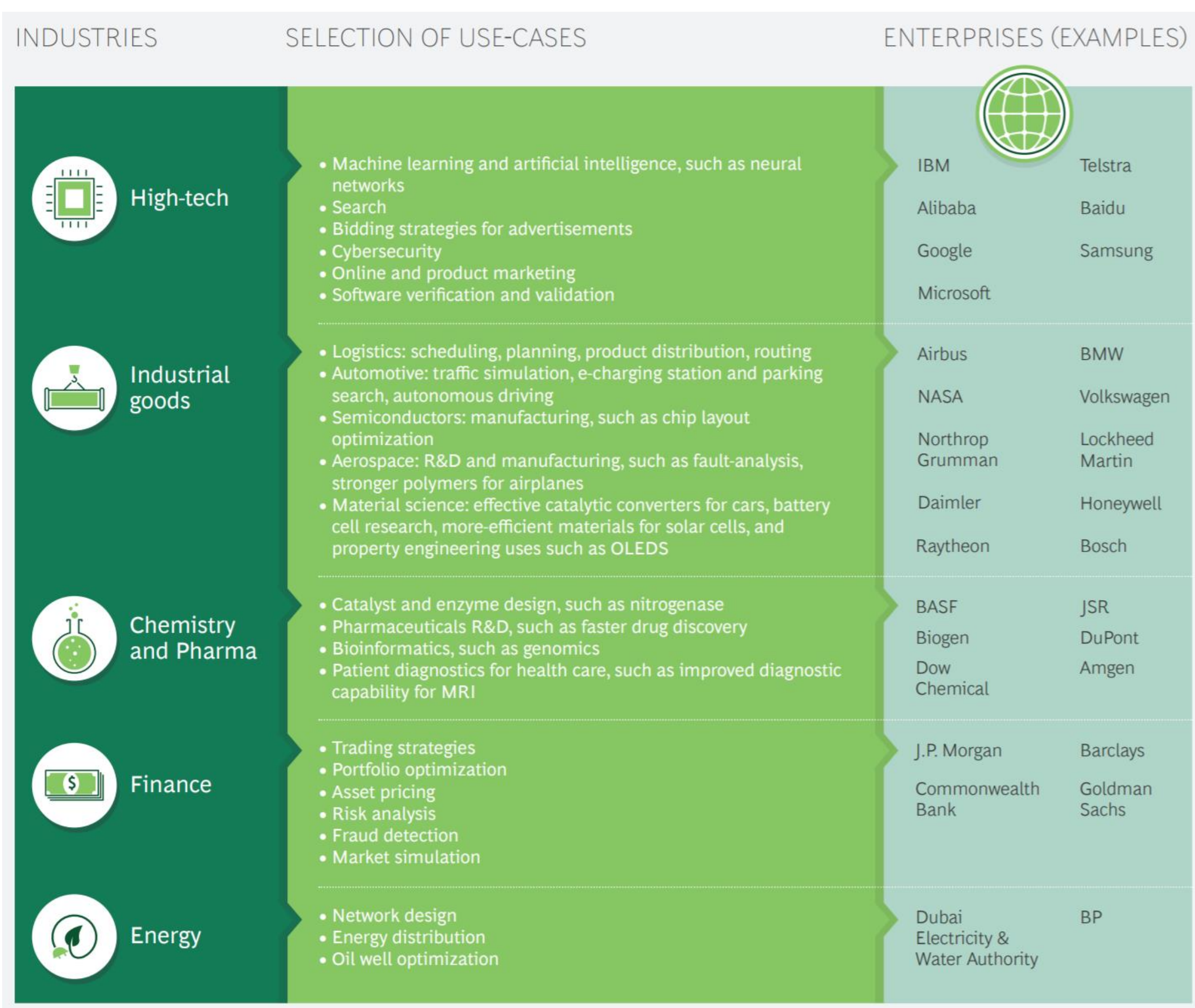

*Figure 660: quantum computing use-case scenarios per vertical. Source: [The Next Decade in Quantum Computing and How to Play]() by Philippe Gerbert and Frank Ruess, BCG, 2018 (30 pages).*

In another chart, BCG positions these vertical markets along two dimensions: business value and expected time of quantum advantage. This is more gut feeling than any real rationale thinking since there are too many variables to have any idea of where each of these industries sit in this fancy chart. We are at a too early stage of the quantum computing innovation cycle to make such predictions.

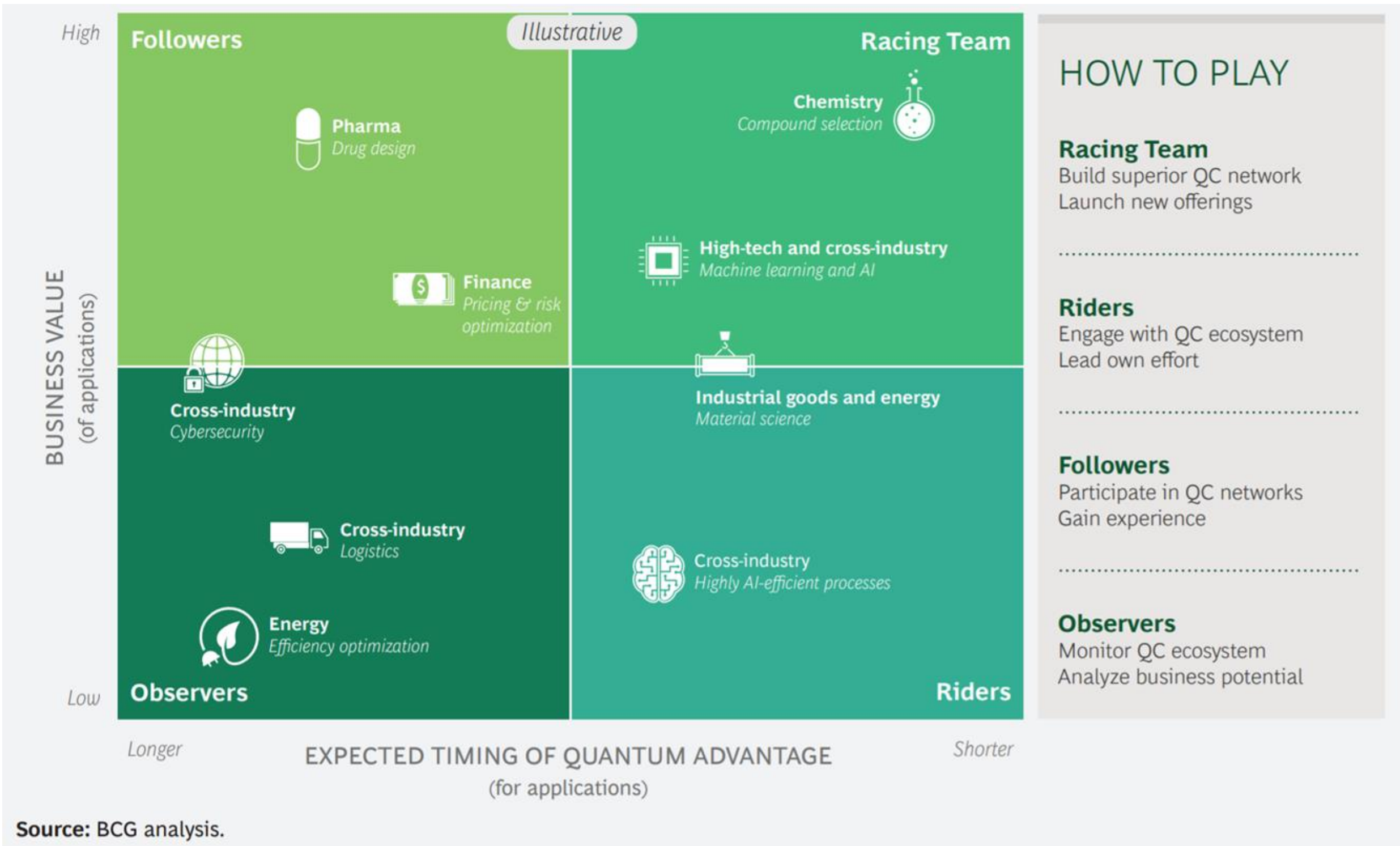

*Figure 661: correlation between use cases business value and expected timing for a quantum advantage. Four years later, this raw classification remains valid. Source: The Next Decade in Quantum Computing and How to Play by Philippe Gerbert and Frank Ruess, BCG, 2018 (30 pages).*

Predicting the size of the quantum computing market is indeed highly probabilistic. It's supposed to reach $553M in 2023 according to **Markets and Markets** (in 2017), $830M in 2024 for **Hyperion Research** (in 2021[1921]), $1,9B in 2023 for CIR and $2,64B in 2022 for **Market Research Future** (2018). Then we reached $8,45B in 2024 for **Homeland Security** (in 2018), $10B in 2028 for **Morgan Stanley** (as of 2017), $15B by 2028 for **ABI Research** (2018) and $64B by 2030 for **P&S Intelligence** (in 2020). **ResearchAndMarkets** predicted in May 2021 that the global quantum technology market would even reach $31.57B by 2026, including $14.25B for quantum computing[1922]. **IDC** planned for a 6-year compound annual growth rate (CAGR) of 50.9% over the 2021-2027 period with a market reaching $8.6 billion in 2027[1923]. At last, **The Quantum Insider** has its own predictions with a total quantum computing market between $300M and $1.3B in 2021 that could grow to between $3.5B and $10B by 2025 and between $18B and $65B by 2030 with a CAGR of between 70% and 80% from 2021 to 2025, to slow down between 39% to 45% between 2025 to 2030[1924].

[1921] See Quantum Computer Market Headed to $830M in 2024 by John Russell, HPC Wire, September 2021. Hyperion's forecasts have ups and downs. See New Study Estimates More Than 20 Percent Annual Growth of Global Quantum Computing Marketplace Through 2024, Hyperion Research, February 2022. The forecast is based on polling 112 quantum computing vendors from around the world and was funded by QED-C and QC Ware. A previous 2020 forecast did plan for a 27% annual increase and now, we are at 21,9%!

[1922] See The Worldwide Quantum Technology Industry will Reach $31.57 Billion by 2026 - North America to be the Biggest Region, May 2021.

[1923] See Quantum computing market landscape by TBR Research which is reviewing only 26 vendors, including two who have either nothing to sell (Intel) or doing nothing (Nokia) in quantum computing.

[1924] See Quantum Computing Market Size Expects Double-Digit Growth by Matt Swayne, December 2021.

Some forecasts can reach other crazy heights. For **Bank of America**, quantum technologies will be as important as smartphones. The main reason? Its potential applications in healthcare. To make sure, the point is made, Haim Israel from this bank also touted that quantum computing will be more important that the invention of fire which is a bit stretch[1925].

The only problem: many analyses behind these predictions gets confused between big data and quantum computing[1926]. Some are based on vendors expectations, other on customers fuzzy plans to adopt quantum computing, given nobody has a real clue of when and how it will work.

As of early 2020, **McKinsey** even predicted that quantum computing would be worth $1 trillion by 2035[1927]. It is easy to identify the forecast bias used here. It's based on a trick that was used a few years ago to evaluate the size of Internet of things and artificial intelligence markets. In 2022, they reused the same methodology to forecast the Metaverse would create $5T of value by 2030[1928]. It is not the market estimation for quantum technologies as such, but the incremental revenue it could generate for businesses, such as in pharmaceuticals, financial services or transportation. It is a bit like evaluating the software market (which was around $593B in 2021, including $237B in enterprise software, source Statista) by summing up the total revenue of the companies who use some software! This would be quite a large number and a significant share of worldwide GDP[1929]. Market predictions should focus on IT products, software and services and should be compared with existing reference markets. For example, the 2020 worldwide servers market size was $85.7B according to IDC[1930].

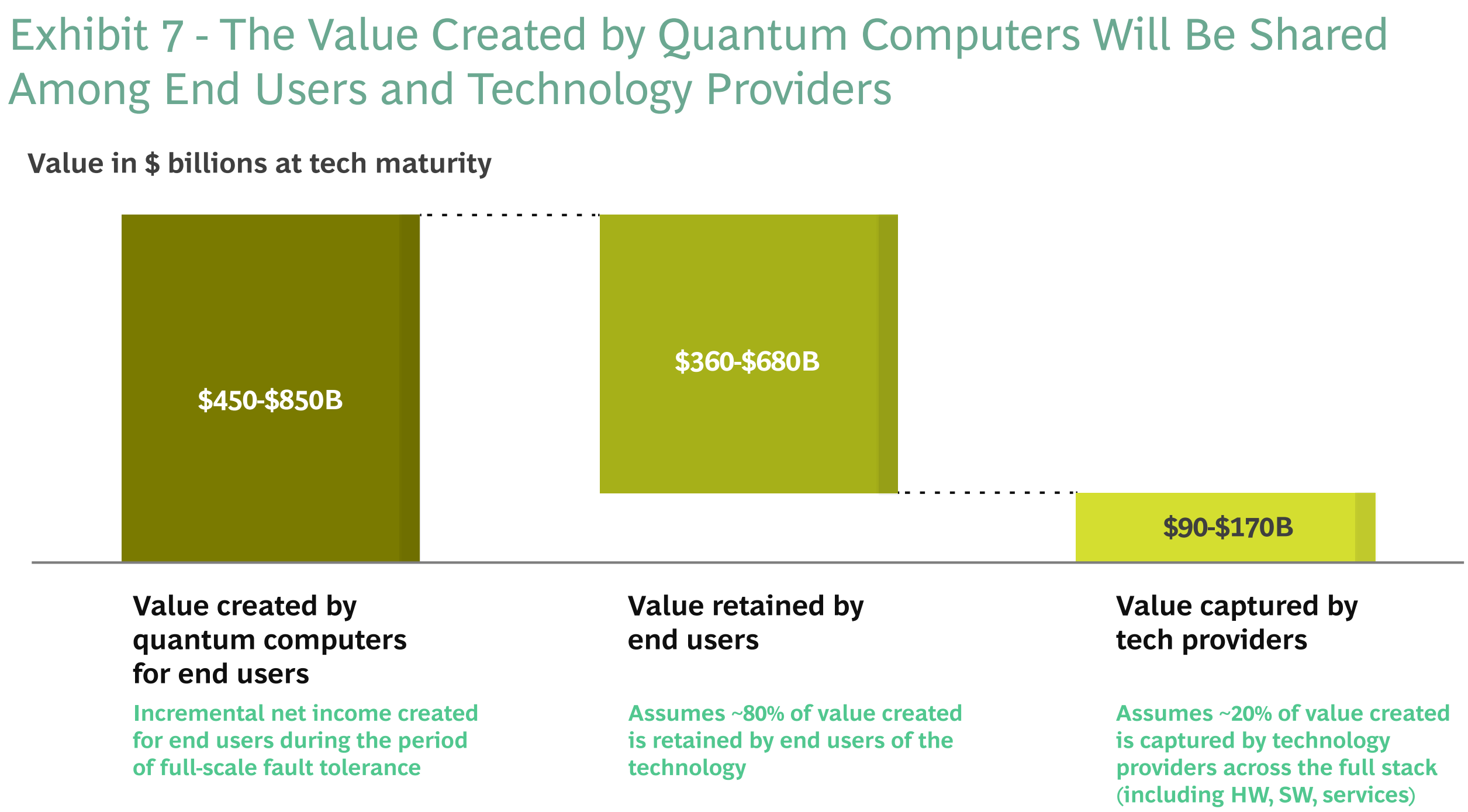

*Figure 662: BCG's 2021 estimation of the market value created by quantum computers and the share of this value that could be captured by the quantum industry, broadly estimated at 20%. By 2040! Source: What Happens When 'If' Turns to 'When' in Quantum Computing, BCG, July 2021 (20 pages).*

---

1925 See Quantum Computing Will Be Bigger Than the Discovery of Fire! By Luke Lango, August 2022.

1926 See Quantum computing will be the smartphone of the 2020s, says Bank of America strategist by Chris Matthews, December 2019.

1927 See Quantum computing will be worth $1 trillion by 2035, according to McKinsey, March 2020.

1928 See On the road to change Value creation in the metaverse, McKinsey, June 2022 (77 slides). Il mentions "$120B invested in 2022" but this includes M&As like the $69B acquisition of Activision by Microsoft and only $6B to $8B real investments in startups through venture capital.

1929 Analysis shared in McKinsey Forecasts Quantum Computing Market Could Reach $1 trillion by 2035, April 2020.

1930 IDC quarterly 2020 server market estimates: Q1, Q2, Q3 and Q4 with respectively $18,6B, $18,7B, $22,6B and $25,8B.

In a 2021 publication[1931], **BCG** estimated the size of the quantum computing market as 20% of its estimated generated value with customers of $850B, ending with a $90B to $170B market captured by technology providers, including software and services... some day after 2040, and a more reasonable $1B to $2B before 2030 and $15B to $30 after 2030.

So, we have here an uncertainty based on an unknown estimated with some fuzzy technology capability predictions. The problem is some vendors take these data for a market size in their investor pitch presentations, not as a generated value size[1932]. It is highly misleading.

On its end, The Quantum Daily forecasted in 2021 that the "Quantum Cloud as a Service" (QCaaS) market would reach $26B by 2030, and tried to document its methodology with reminding us that it was based on vendors questionable roadmaps[1933].

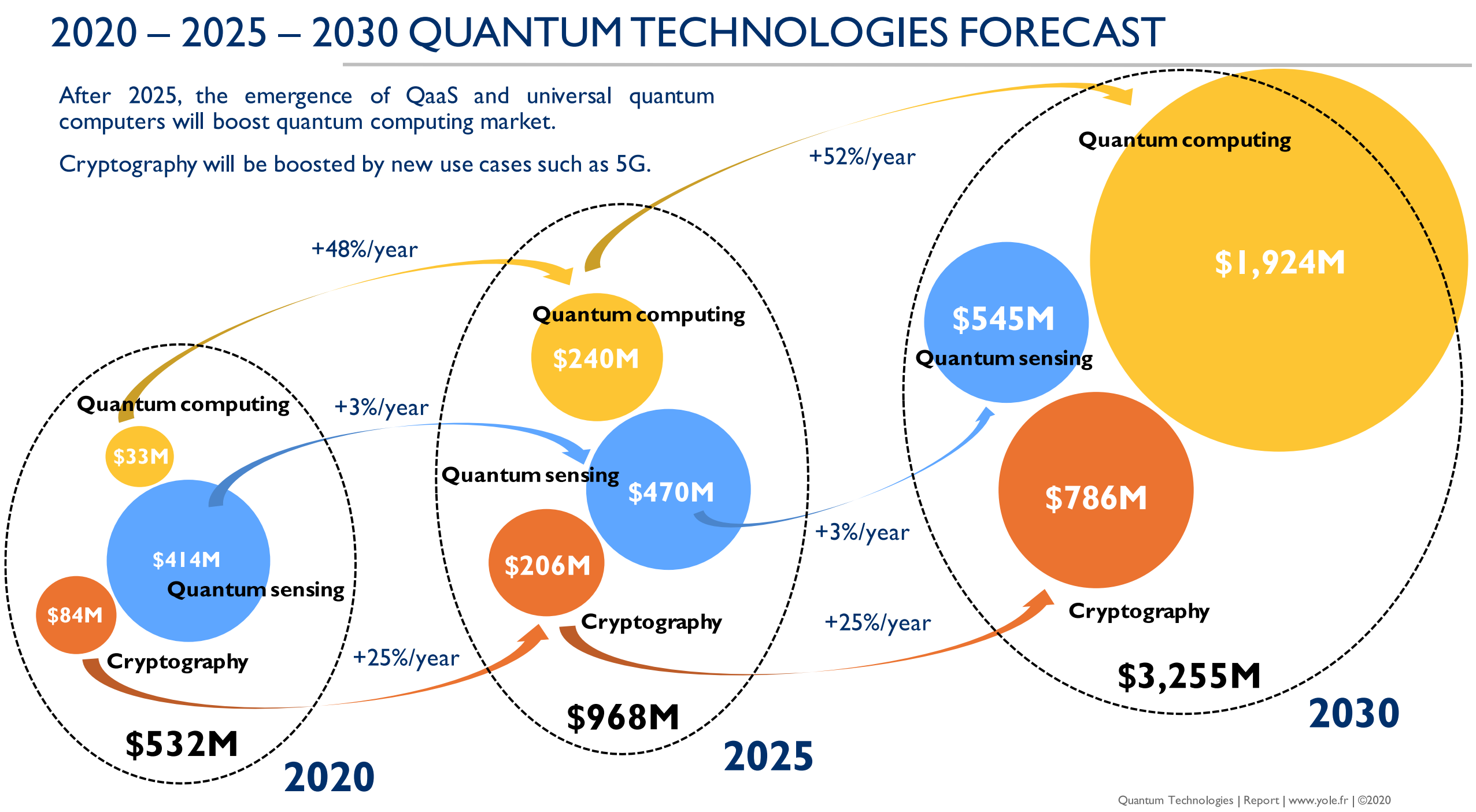

*Figure 663: Yole Development's sizing of the quantum technology market by 2030. Source: Quantum Technologies Market and Technology Report 2020 -Sample, Yole Development, 2020 (22 slides).*

A more detailed market size assessment was made in 2020 by **Yole Development** with seemingly more reasonable predictions for quantum technologies, with an increase to $3.2B per year by 2030, with 17% average annual growth, including $650M for hardware, $1.37B for cloud-based software and $785M for quantum cryptography (QKD)[1934].

The sensor market would grow from $400M in 2019 to $545M in 2030. This moderate growth seems a bit bearish since quantum sensors are the quantum objects with the lowest technological uncertainties and it is a market in its infancy.

---

[1931] See What Happens When 'If' Turns to 'When' in Quantum Computing, BCG, July 2021 (20 pages).

[1932] Two examples here with the Q2 Rigetti Quarterly report in August 2022 which describes these $850B as a highly ambiguous "*forecasted quantum computing generated operating income*" and Atlantic Quantum Emerges from MIT's Engineering Quantum Systems Lab, Raises $9M Seed Funding to Make Large-Scale Quantum Computing a Reality by James Dargan, from The Quantum Insider that is reusing a press release from Atlantic Quantum, July 2022, saying "*The enterprise quantum computing market could grow to a $450—$850 billion market in the next 15-to-30 years, according to Boston Consulting Group (BCG)*". It says it all about the confusion generated by these market value market data.

[1933] See Quantum Computing as a Service Market Sizing - How we dit it, The Quantum Daily, August 2021.

[1934] See Quantum technologies: a jump to a commercial state, Yole Development, 2020 and their sample Quantum Technologies Market and Technology Report 2020 -Sample, 2020 (22 slides).

The quantum team at **Total** constructed this interesting roadmap to provide an idea of the order in which practical quantum applications could emerge according to the number of available qubits.

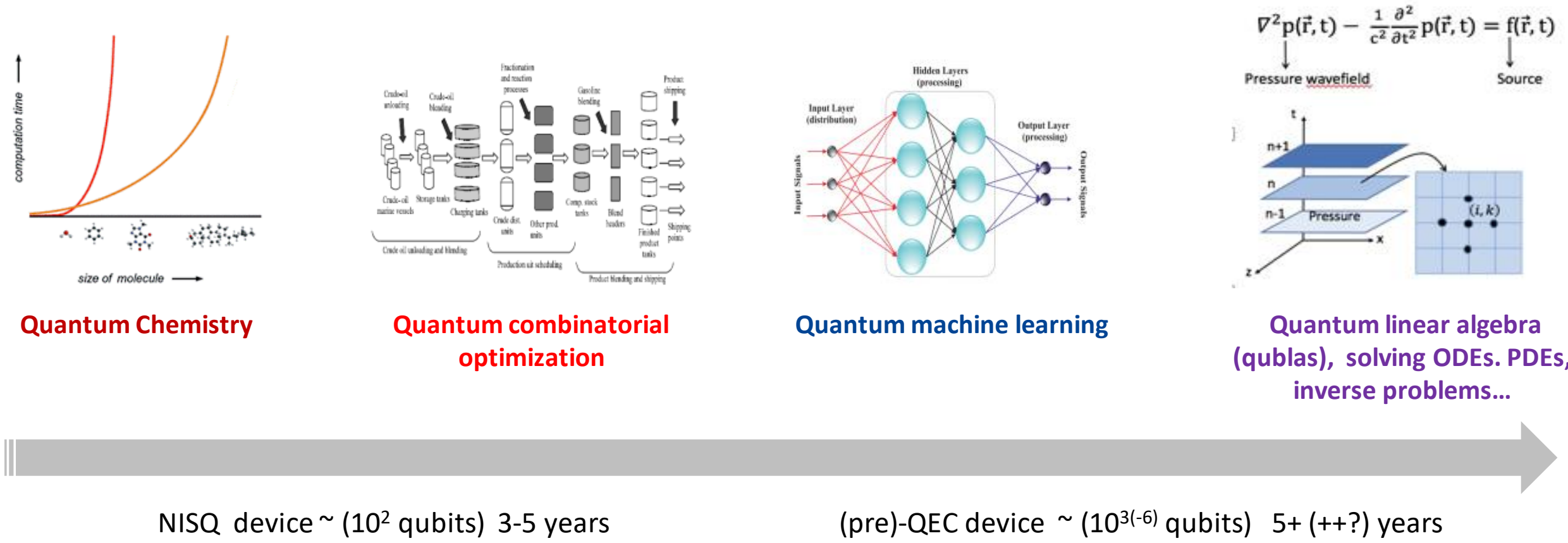

*Figure 664: practical quantum computing use cases emergence by domain. Source: Total QCB Conference, Paris, June 2019.*

Since we have no real idea of when scalable quantum computing will really work, let's try another exercise to determine the critical factors enabling some sort of technology commoditization for quantum computing:

**Technology**. The first factor is a mix of where and when we'll have first, useful quantum simulators, then useful NISQ, and then scalable **universal quantum computers** with more than a hundred logical qubits. Meanwhile, optimization solutions adapted to D-Wave's annealer will continue to be developed and may reach a point where they make a real difference with classical computing.

**Software Tools**. The second factor may be the consolidation of software development tools. These tools will continue to mature, raising their level of abstraction, and adapt to hardware evolutions. Libraries adapted to the needs of specific markets will undoubtedly consolidate, as in molecular simulation or finance. As the market matures, there will be some consolidation in this market.

**Skills**. One critical path to market growth as it's been the case for most previous major technology wave will be the availability of skilled workforce, particularly with developers. They will have, at least at the beginning, to handle abstractions levels that have nothing to do with the different forms of programming techniques that dominate today's computing, even in its event-driven programming variants that are common in the creation of websites and graphic applications. It's more an extension of the existing scientific computing community. A new generation of algorithm designers and developers will emerge. These will probably be young professionals who will have been able to digest new quantum computing concepts with a clean state mind.

**Startups**. The market will rely mostly on the fabric of startups, probably slightly ahead of traditional software publishers and IT services companies that may not necessarily venture first into this new world of quantum.

**Experience**. The first feedback from pilot projects, already underway, particularly with D-Wave, will be important. Most recent projects bring interesting learnings on the actual accelerations that quantum computing can provide. We will have to learn to make objective comparison between quantum algorithms, quantum hardware architectures and their equivalents running (or not) on supercomputers. It will also be necessary to sort out into "proof of concepts" and projects actually deployed.

**Mass impact**. At last, quantum computing commoditization will depend on the potential emergence of solutions that will have an impact on our daily lives. So, mostly consumer applications. It could come from healthcare and transportation. Who knows. Uses cases will move gradually from the research community, to the corporate world, and then to consumer applications.

## Healthcare

The healthcare market, and particularly pharmaceutical industries, is one of the most sought after by quantum computing players. It is the vertical markets with the largest number of dedicated startups. Most major pharmaceutical companies have been exploring and evaluating the potential of quantum computing for a few years, starting by conducting a few pilot projects with D-Wave[1935]. The dream is to extend the capabilities of today's supercomputers to simulate living organisms' molecules "in silico", mainly in order to create or discover new treatments. This is the field of "*in-silico drug discovery*".

This quest is linked to the pharmaceuticals industry worrisome situation, that is discovering fewer new treatments and seeing diminished portfolio of commercial patented drugs. The drugs development cycle from discovery to market is becoming increasingly expensive, particularly during clinical trials. It costs up to a $1B, if not more, and failure rates are numerous. 45% of cancer therapies clinical trials fail in phase III in the USA, and 97% of the new therapies tested are not approved by the FDA in the USA! If we could better digitally simulate the effects of new treatments before clinical trials, we might be able to increase these success rates. Also, quantum computing could be a critical tool to create digital twins of molecular complexes, used to find the right combinations and optimize their efficiency.

On top of news drugs discovery, pharmaceutical companies are also trying to leverage their existing portfolio with drugs re-targeting. It can speed up clinical trials since their adverse effects are already known. Even though this has not prevented the long controversy surrounding hydroxychloroquine in 2020! In any case, pharmaceutical players need simulation tools and in particular molecular simulation tools: to create molecules, from the simplest (peptides) to the most complicated (proteins, antibodies, vaccines), to model them in 3D, to analyze their interactions between their active sites and targets like cell surface proteins (transmembrane glycoproteins)[1936], and also to identify contraindications. Such treatments can be created ex-nihilo, but most often, they are derived from existing ones (known protein, enzyme, bio-inspiration, ...).

The pharmaceutical industry is organized in a couple consortia which help it share best practices, including in quantum computing adoption. The **Pistoia Alliance** was created in Pistoia, Italy, in 2007 by AstraZeneca, GSK, Novartis and Pfizer in 2007. It entertains a quantum computing community. **QuPharm** is a consortium of 17 pharmaceutical companies, including AbbVie, Bayer, GSK, Takeda, and Pfizer that is dedicated to quantum computing.

### Molecular simulations

Molecular simulations are based on the broad field of computational chemistry. It originated with the description of the nature of chemical bonds by **Linus Pauling** in 1928, which launched the vast field of quantum chemistry. Chemical bonds describe the way electrons of covalent liaisons are shared between atoms and the shape of their related orbitals.

Pauling's work came just after the creation of the **Born-Oppenheimer** approximation in 1927[1937] which simplified Schrödinger's equation for a molecule by separating the nuclei of the atoms from their electrons. The same year, **Llewellyn Thomas** (1903-1992, English) and **Enrico Fermi** (1901-1954, Italian American) created the later-called Thomas-Fermi model which describes the electronic structure of multi-atoms systems.

---

[1935] Like Abbvie, Amgen, AstraZeneca, Bayer, Biogen, Bristol-Myers Squibb, Johnson&Johnson, Merck, Roche, Sanofi and Taleda.

[1936] We could try to digitally simulate an entire cell with all its organelles. This would become quite complicated since a living cell comprises about 100 trillion atoms!

[1937] The Max Born from the probabilistic explanation of Schrödinger's equation and the Robert Oppenheimer from the atomic bomb.

The field of computational chemistry began much later, in 1964, with the creation of the two Hohenberg-Kohn theorems by **Walter Kohn** (1923-2016, Austrian then American) and **Pierre Hohenberg** (1934-2017, French American).

This was closely followed by **Kohn-Sham**'s equations from **Lu Jeu Sham** (1938, Chinese) in 1965. They are the basis of **DFT** (Density Functional Theory), a mathematical model that describes the structure of molecules at rest as a function of inter-atomic interactions and the structure of their electron clouds, and in a simpler way than with Schrödinger's equation which manipulates too many variables. Walter Kohn was awarded the Nobel Prize in Chemistry in 1998 for this work, along with **John Pople** (1925-2004, English) who had contributed to the modeling of electronic orbitals in molecules.

DFT was followed by the work of **Martin Karplus** (1930, American), **Michael Levitt (**1947, Israeli-American) and **Arieh Warshel** (1940, Israeli-American) who contributed to the digital modeling of chemical reactions in the 1970s. They were awarded the Nobel Prize in Chemistry in 2013 for their work. The DFT model was also simplified by **Axel Becke** (1953, Canadian) in 1993 with the hybrid DFT. Some quantum versions of DFT are developed for quantum computing and with some progress, even for NISQ QPUs[1938].

Molecular simulation faces quasi-quantum effects related to the continuous vibrations of molecules in their aqueous medium. Chemical bonds oscillate at a femto-second rate, atoms vibrate collectively at a one picosecond rate. On the other hand, more complex chemical processes such as the production and folding of proteins occur on scales ranging from micro-seconds to seconds.

The Holy Grail would be to understand how the assembly and then operations of ribosomes work. These molecular complexes are made of 73 proteins and 4 large RNA molecules. Ribosomes produce all proteins in our cells using messenger RNA code, which is itself synthesized from DNA through an also amazing biochemical process involving many complex molecules. Thousands of ribosomes operate in every living cell and each ribosome is made of about 250,000 atoms[1939].

One of the most recent protein folding algorithm was able to simulate the folding of the 10 amino acid Angiotensin on 22 qubits. The same method was also applied to folding a 7 amino acid neuropeptide using 9 qubits, all on an IBM quantum computer[1940]. Many other projects have been launched to improve quantum-based protein folding and there are still theoretical since no existing quantum computer is powerful enough to execute them[1941]. One common algorithm used here is QAOA but its applicability is questioned[1942].

Today, most molecular simulation calculations are carried out using algorithms running on classical supercomputers, increasingly using GPGPUs such as those from Nvidia or Google's TPUs.

---

[1938] See Toward Density Functional Theory on Quantum Computers? by Bruno Senjean, Saad Yalouz and Matthieu Saubanère, University of Montpellier and University of Strasbourg, April-October 2022 (19 pages).

[1939] The number of 2.5 or 3.5 million atoms is often mentioned, but this is not true. These are "Daltons" which are equivalent to one twelfth of the mass of carbon 12, or about the mass of a hydrogen atom. However, these organic molecules contain, in addition to hydrogen, a lot of carbon, nitrogen, phosphorus and oxygen. The latter contribute to a large part of the mass of the molecule, hence the fact that the number of daltons must be divided by 10 to obtain the number of atoms of an organic molecule.

[1940] See Resource-efficient quantum algorithm for protein folding by Anton Robert et al, 2021 (5 pages).

[1941] See QFold: Quantum Walks and Deep Learning to Solve Protein Folding by P A M Casares, Roberto Campos and M A Martin-Delgado, March 2022 (22 pages), Folding lattice proteins with quantum annealing by Anders Irbäck et al, Lund University and Forschungszentrum Jülich, May 2022 (21 pages) which runs on a D-Wave Advantage and Protein Folding Neural Networks Are Not Robust by Sumit Kumar Jha et al, September 2021 (8 pages).

[1942] See Peptide conformational sampling using the Quantum Approximate Optimization Algorithm by Sami Boulebnane et al, April 2022 (30 pages). Conclusion: "*these results cast serious doubt on the ability of QAOA to address the protein folding problem in the near term, even in an extremely simplified setting*".

Another test was published in 2021 by GSK[1943]. It was about solving a mRNA codon optimization problem. Each amino acid in a protein sequence can be encoded by as many as six different codons, these series of three DNA/RNA bases encoding one amino acid. The goal was to find the right combination of these codons. The codon selection in mRNA impacts protein folding and functions. The main task is to balance G and C bases in mRNA to optimize gene expression. This was one of the first published case studies using D-Wave Advantage annealer and its 5000 qubits and their Leap Hybrid Solver. It worked well with 30 amino acids and could scale up to 1000 amino acids.

The codon optimization problem is formulated as a Binary Quadratic Model that is itself close to an Ising model adapted to D-Wave annealers. It did fare well when compared to genetic and machine learning algorithms running on classical computers.

The first small-scale tests of simulation using quantum algorithms were done on D-Wave and superconducting qubits accelerators. However, the most common approach is based on hybrid algorithms associating HPCs and quantum accelerators.

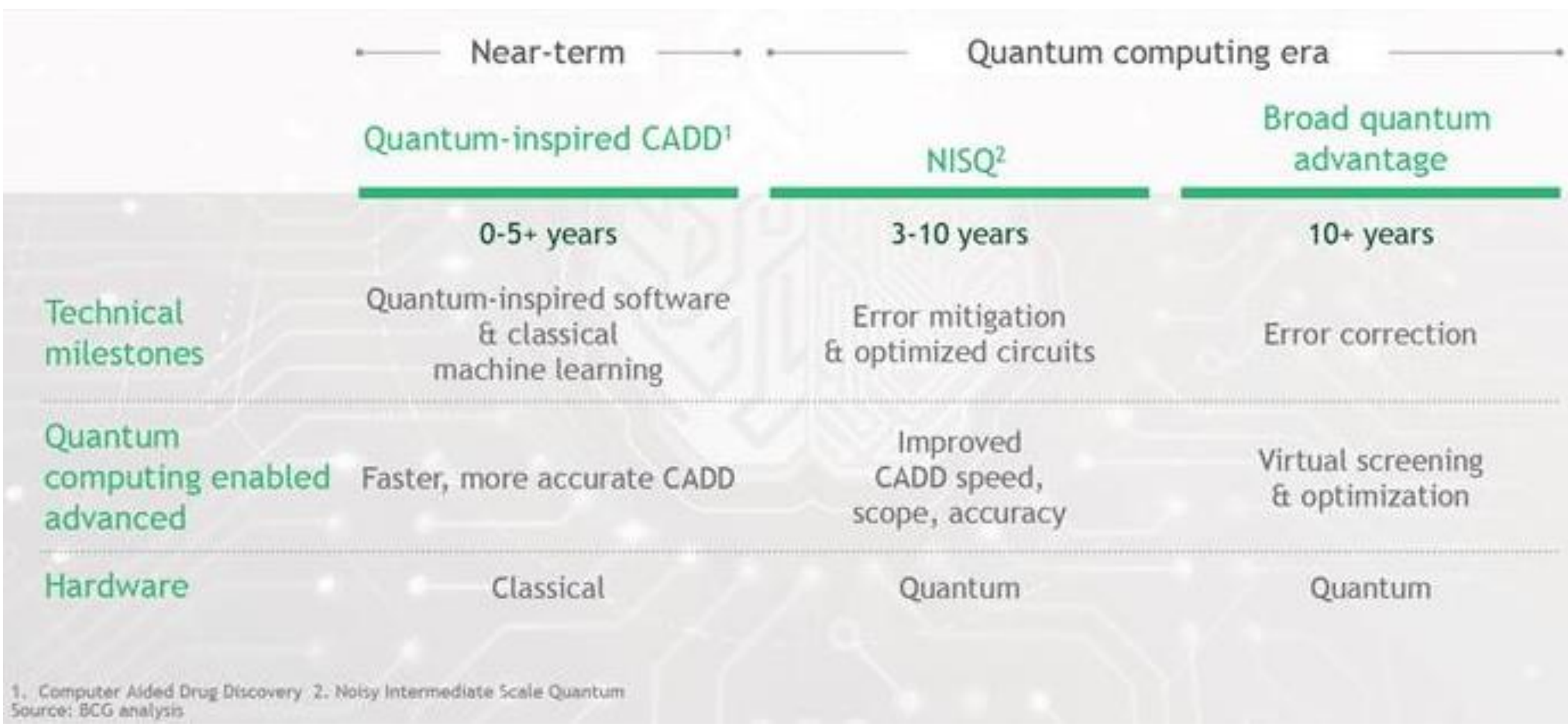

*Figure 665: Source: Will quantum Computing Transform Biopharma R&D? by Jean-Francois Bobier et al, December 2019.*

Quantum simulators are also machines suitable for simulating the interaction of atoms within molecules. BCG is thus presenting molecular simulation roadmaps spread out over time and following the rate of evolution of quantum computers between today's NISQ (intermediate-size noise computing) and LSQ (large scale quantum computing[1944].

One approach consists in relying on generic frameworks that can be distributed over classical computing in massively parallel architecture, and then progressively over quantum computing. This is the case of the **Tinker-HP** framework co-created by Jean-Philip Piquemal, co-founder of **Qubit Pharmaceuticals** and that the company plans to extend with hybrid quantum algorithms[1945].

[1943] See GlaxoSmithKline Marks Quantum Progress with D-Wave by Nicole Hemsoth, February 2021 pointing to mRNA codon optimization on quantum computers by Dillion M. Fox et al, February 2021 (35 pages).

[1944] See Will Quantum Computing Transform Pharma R&D by Jean-Francois Bobier, April 2020 (14 slides) and the written version Will quantum Computing Transform Biopharma R&D? by Jean-Francois Bobier et al, December 2019.

[1945] See Computational Drug Design & Molecular Dynamics by Jean-Philip Piquemal, April 2020 (28 slides) and Tinker-HP: a massively parallel molecular dynamics package for multiscale simulations of large complex systems with advanced point dipole polarizable force fields by Louis Lagardère, Jean-Philip Piquemal et al, 2018 (17 pages). The company plans to rely first on cold atoms simulators like those from Pasqal.

At this point in time, however, the priority is with quantum inspired algorithms[1946].

Most other quantum startups like **ApexQubit**, **HQS Quantum Simulations**, **MentenAI**, **ProteinQure** and **Qulab** are indeed adopting hybrid computing models, if only to have something practical to market[1947]. The most common hybrid method is the VQE (Variational Quantum Eigensolver)[1948].

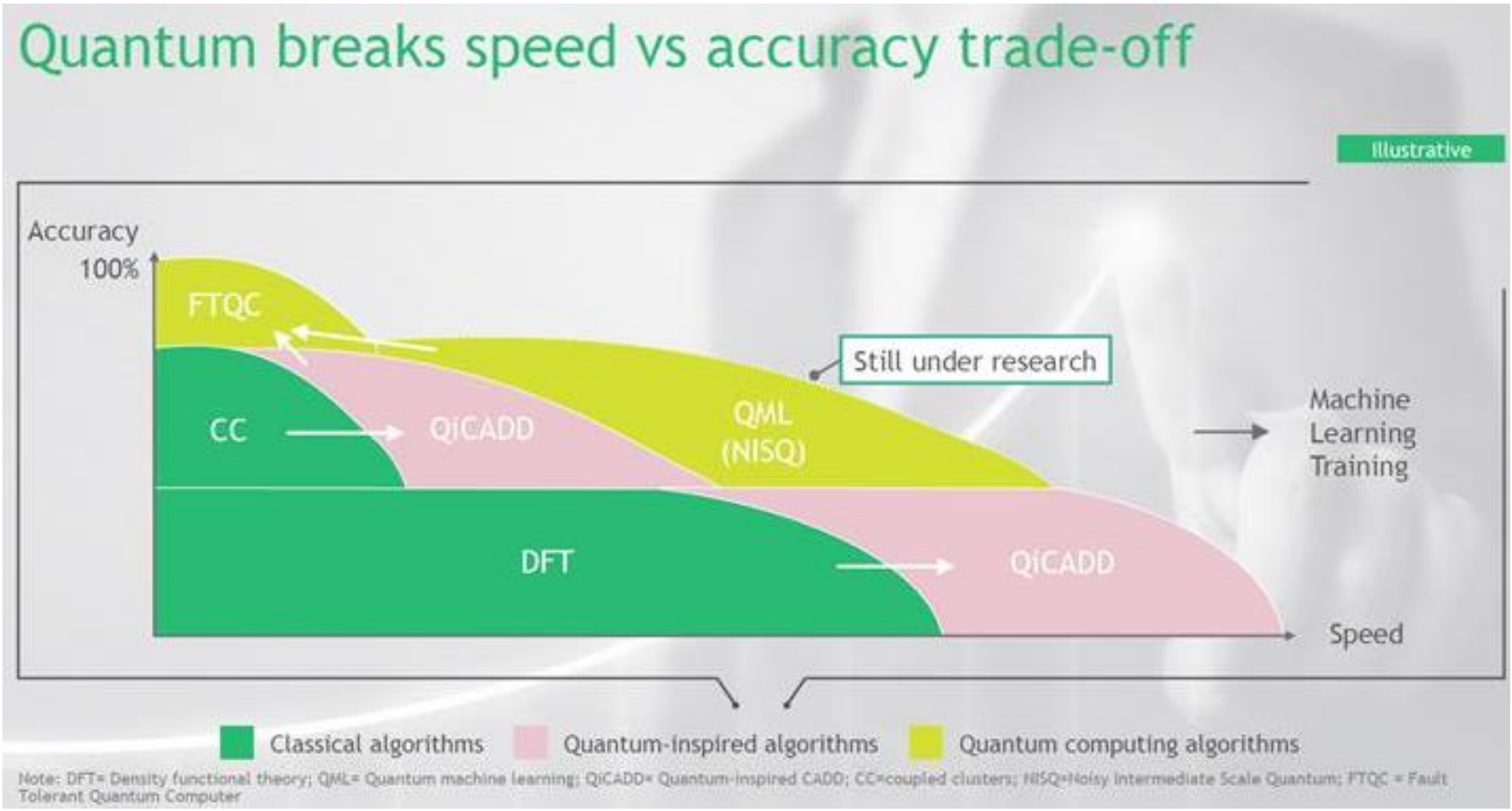

*Figure 666: Source: Will quantum Computing Transform Biopharma R&D? by Jean-Francois Bobier et al, December 2019.*

Another method is being developed to create quantum inspired algorithms, *aka* classical algorithms based on quantum algorithms[1949]. Quantum and quantum inspired computation complete the vast field of machine learning which is already very common in the discovery of therapeutic molecules[1950].

Molecules simulation can start with simple organic molecules like cholesterol up to protein folding which is many orders of magnitude more complex[1951]. This last feat is therefore bound to be a very long-term one.

Today, we can simulate peptides with about ten amino acids. The best algorithms require a number of qubits that evolves according to the power 4 of the number of amino acids[1952]. This simulation is also at the limit of feasibility in terms of complexity because it is in the class of NP-Complete problems as seen in the section dedicated to complexity theories, starting on page 615.

[1946] See Development of the Quantum Inspired SIBFA Many-Body Polarizable Force Field: I. Enabling Condensed Phase Molecular Dynamics Simulations by Sehr Naseem-Khan, Jean-Philip Piquemal et al, January 2022 (50 pages).

[1947] See Can Quantum Computing Play a Role in Drug Discovery? At least one Startup Thinks so by James Dargan, 2020, which mentions Menten AI.

[1948] See Quantum Chemistry and the Variational Quantum Eigensolver by S Kokkelmans et al, December 2019 (56 pages).

[1949] See Quantum and Quantum-inspired Methods for de novo Discovery of Altered Cancer Pathways by Hedayat Alghassi et al, 2019 (27 pages).

[1950] See Concepts of Artificial Intelligence for Computer-Assisted Drug Discovery by Xin Yang et al, 2019 (75 pages). A good review paper with 879 bibliographical references!

[1951] See Designing Peptides on a Quantum Computer by Vikram Khipple Mulligan, September 2019 (20 pages) which presents Rosetta, a protein quantum design tool running on D-Wave.

[1952] See Resource-Efficient Quantum Algorithm for Protein Folding by Anton Robert et al, August 2019.

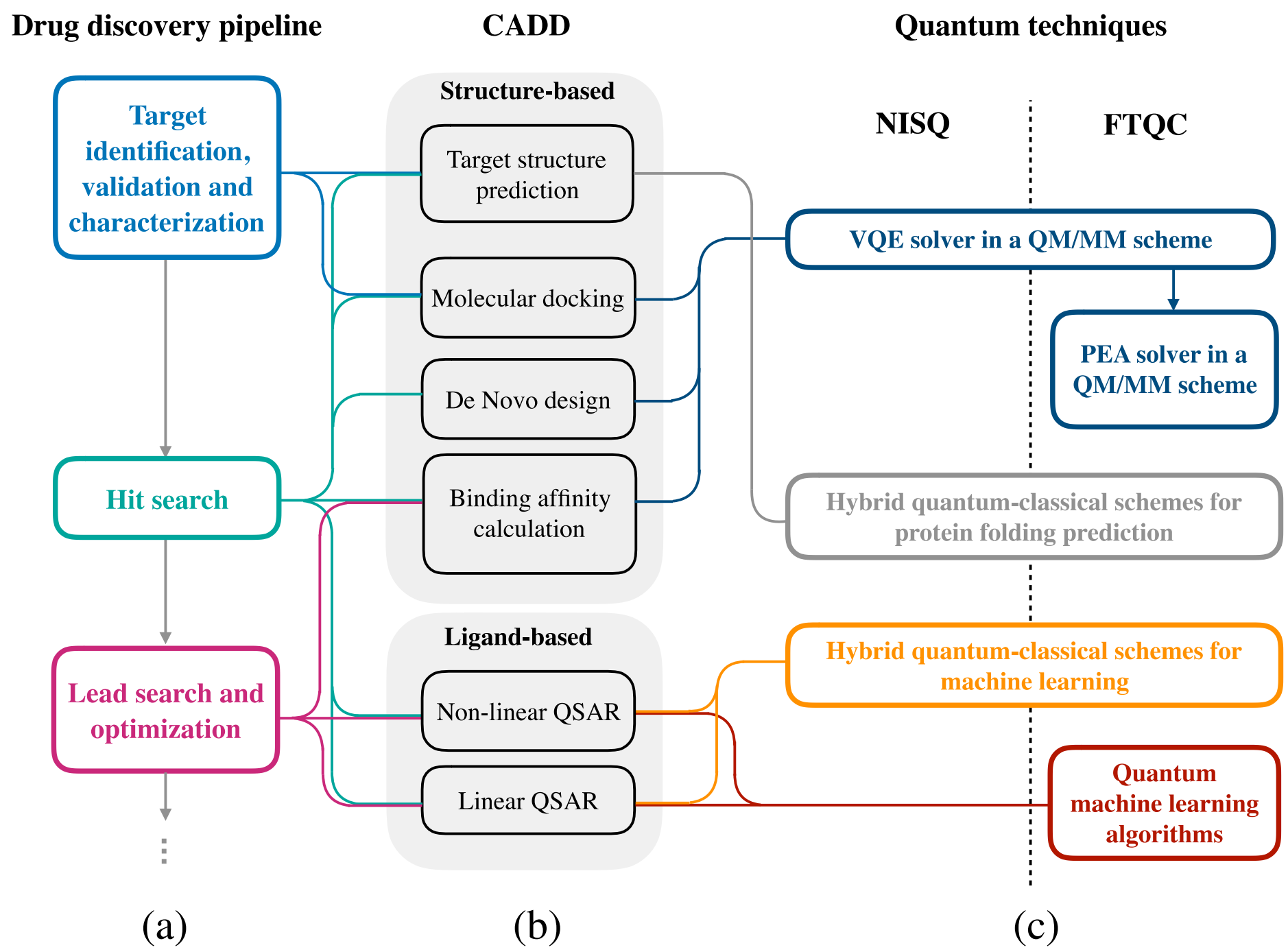

Fig. 1. (a) General workflow of drug discovery process. Here we focus on the early phase where computationally intensive quantum chemical analyses are involved. (b) Components of each stage of drug discovery that heavily involve quantum chemistry or machine learning techniques. (c) Quantum techniques that can be applied to the components listed in (b) and potentially yield an advantage over known classical methods. Here we make the separation between techniques for noisy intermediate scale quantum (NISQ) devices [21] and fault-tolerant quantum computing devices.

*Figure 667: a process flow for drug discovery. CADD = Computer Aided Drug Design. Source: Potential of quantum computing for drug discovery by Alán Aspuru-Guzik et al, 2018 (18 pages).*

78% of the search for therapies is focused on light molecules of less than 900 Daltons, i.e. about a hundred atoms. Its function is to associate itself with a target in the cells, often a specific protein that controls a metabolism that we want to attenuate or amplify[1953]. The discovery of small molecules of a few dozen atoms could be within the scope of the NISQ quantum computers within a few years. The first molecular simulation experiments were carried out on D-Wave. They work with the search for energy minima, which can be suitable in theory for the simulation of the organization of molecules.

A collaboration was launched in June 2017 between **Biogen,** the Canadian quantum software company **1QBit,** and **Accenture** for the creation of new molecules. **Biogen** (1978, USA) is a mid-size biotech company with 7300 employees specialized in the treatment of neurodegenerative diseases and leukemia.

Their use of quantum computing was aimed at retargeting therapeutic molecules, looking for matching between existing treatments and therapeutic targets in neurodegenerative or inflammatory diseases. **Amgen** is also active in the search for new therapies and is working since 2020 with **QSimulate** (2018, USA).

A similar project was launched in Spain with the consortium "QHealth: Quantum Pharmacogenomics applied to aging" launched in August 2020 with **aQuantum** (alhambraIT, Prologue Group) with **Gloin**, **Madrija** and various Spanish Universities. Its goals are to find correlations between physiological and genetic variables, drug usage history, side effects and/or potential lack of response of new drugs to fight aging. They plan to do simulations using quantum algorithms. The project totals 5.1M€ including a grant of 3.7M€ from CDTI awarded in November 2022 by the **Center for the Development of Industrial Technology** (CDTI) of the Ministry of Science and Innovation of Spain.

[1953] See Potential of quantum computing for drug discovery by Alán Aspuru-Guzik et al, 2018 (18 pages).

A hybrid quantum/classical machine learning algorithm was used for drug retargeting using Ligand Based Virtual Screening (LB-VS). It was developed in 2022 by **IBM** and **The Hartree Center** in the UK[1954].

In June 2019, **Merck** announced a three-year partnership with the Karlsruhe, Germany-based startup **HQS Quantum Simulations** for the development of quantum algorithms for chemical simulation. As already mentioned in the part related to SeeQC, **Merck** and SeeQC created together a consortium in 2021 to build a "*commercially scalable application-specific quantum computer designed to tackle prohibitively high costs within pharmaceutical drug development*". The project aw due for completion "*in 18 months*"! We are here in the pure overselling realm.

**cancers classification**
multi-omics: genomics + symptoms in QML
*source: D-Wave*

**de-novo proteins and polypeptides creation**
with hybrid computing, tests in research against the covid-19 virus.
*source: D-Wave*

**liver donor optimization**
NP-complet complete problem using QUBO
*source: Accenture, D-Wave*

**drug retargeting**
with Biogen, 1QBit and Accenture research
*source: D-Wave*

**radiotherapy optimization**
to minimized x-ray dose
*source: Roswell Park, D-Wave*

**cancer diagnosis**
quantum rule based to diagnose and treat invasive ductal carcinoma (breast cancer type)
*source: Atos*

*Figure 668: a couple quantum computing use cases in the healthcare industry. (cc) Olivier Ezratty, 2022.*

In September 2022, **Novo Nordisk** (Denmark) made an announcement that was unique in shape and form. It is launching the Novo Nordisk Foundation Quantum Computing Programme with the goal to build Denmark's generic quantum computer in 2034 with a funding of $200M, and obviously to become a platform for chemical simulations. It will fund the Niels Bohr Institute but also embed partners from the USA, the Netherlands and Canada. They will invest first in three different qubit platforms and select the best one after a first 7-year period.

### Genomics

The **DNA-Seq Alliance** combines the startup DNA-Seq and D-Wave, which also does molecular retargeting by combining genomics, protein kinase crystallography, quantum computing and the search for effective cancer treatments.

Let's also mention that quantum computing, both annealing and gate-based, could be used to accelerate **DNA sequencing**, particularly in de-novo mode, when no existing DNA mapping exists. It's about reconstructing a giant puzzle with small parts of DNA sequences which come out of sequencing[1955].

---

[1954] See Quantum Machine Learning Framework for Virtual Screening in Drug Discovery: a Prospective Quantum Advantage by Stefano Mensa et al, The Hartree Centre, STFC and IBM Research, April 2022 (16 pages).

[1955] See QuASeR - Quantum Accelerated De Novo DNA Sequence Reconstruction by Aritra Sarkar et al, TU Delft, April 2020 (24 pages).

As with any new technology, quantum computing specialists must learn to interact with bioinformatics specialists. Fortunately, bioinformaticians are already bridging the gap between molecular biology and computer science and are well positioned to learn quantum methods[1956].

### Radiotherapy optimizations

Also on D-Wave quantum annealing computers, an application of radiotherapy optimization was experimented as shown in Figure 669.

The principle consists in minimizing patients' exposure to X-rays while optimizing their efficiency. It is a complex problem of simulating the diffusion of electromagnetic waves in the human body.

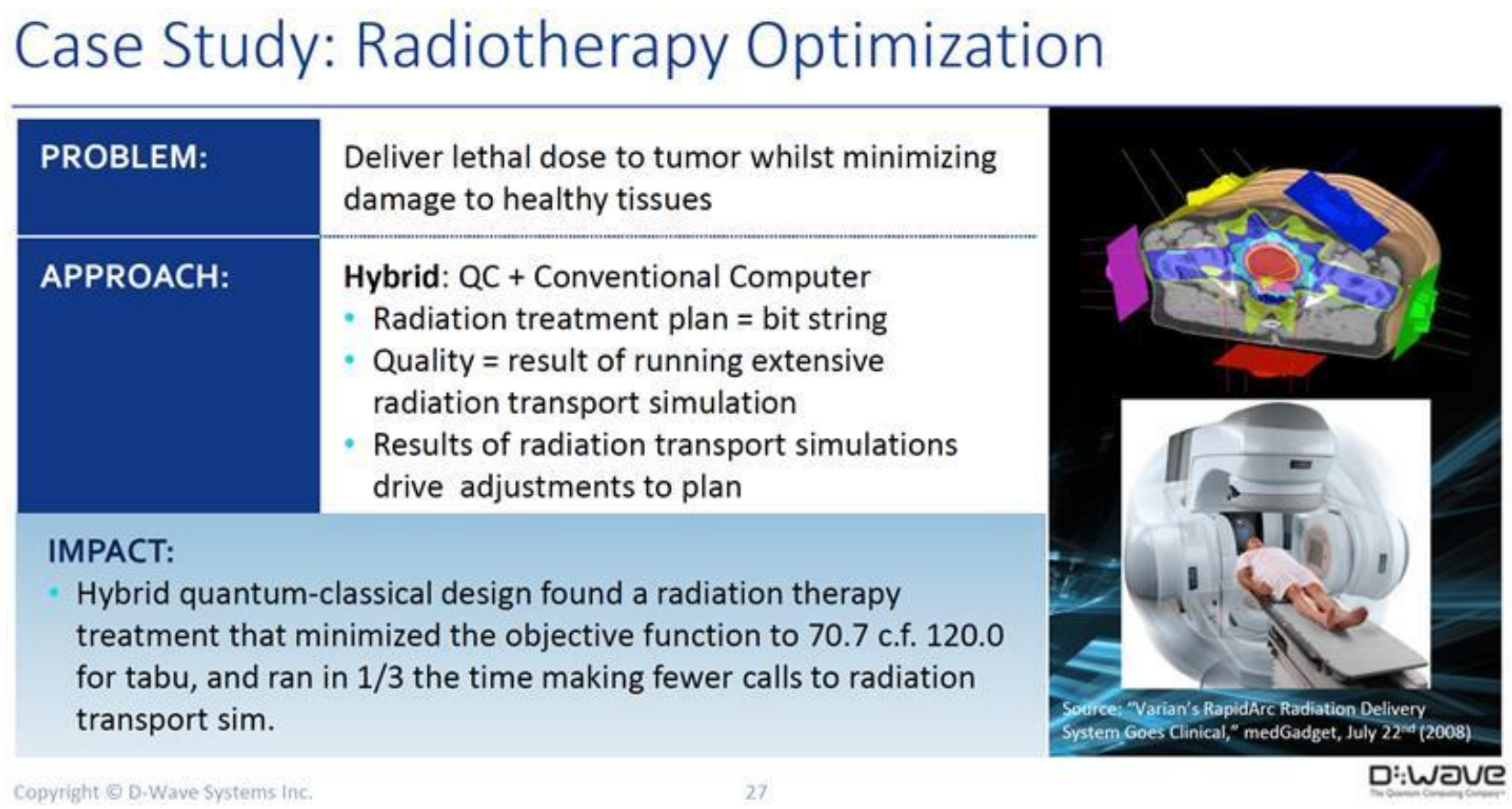

*Figure 669: quantum computing can be used to optimize cancer radiotherapy. Source: D-Wave.*

### Medical imaging

Using quantum machine learning algorithms to do image recognition is also investigated. It may help train these systems with fewer data. At this point in time, however, existing algorithms are just on par with their classical peers[1957].

### Precision medicine

David Sahner, who created his own consulting firm **Eigenmed**, is one of the promoters of precision medicine based on predictive machine learning techniques using D-Wave annealers[1958].

**Omnicom Healthcare** did not hesitate to promote in 2017 the use of quantum computing in healthcare in a white paper containing strictly no relevant information on the subject, especially since they seem to confuse machine learning applications analyzing data from connected objects with the ability of quantum computers to manage problems that are intractable by traditional computers[1959].

**Network medicine** analyzes disease pathogenesis, coupling information coming from Omics databases (protein-protein interaction, genomics data) and environmental factors with Bayesian networks and machine learning tools. The idea it to identify novel disease genes and pathways, their diagnostics and therapeutics. A 2022 paper from Algorithmiq and a Finland research team outlined how quantum computing could help there. Unfortunately, it didn't contain sizing recommendations for hardware and the problems it could solve[1960].

---

[1956] See Thirteen tips for engaging with physicists, as told by a biologist by Ken Kosik, January 2020 which describes how to bring physicists and biologists together.

[1957] See Quantum-classical convolutional neural networks in radiological image classification by Andrea Matic et al, April 2022 (12 pages).

[1958] See Predictive Health Analytics by David Sahner, 2018 (54 slides).

[1959] See Exponential Biometrics: How Quantum Computing Will Revolutionize Health Tracking, 2017 (7 pages).

[1960] See Quantum network medicine: rethinking medicine with network science and quantum algorithms by Sabrina Maniscalco et al, June 2022 (15 pages).

**Covid-19**

Finally, the covid-19 pandemic has led to a renewed interest in quantum computing. Several market players have put the cart before the horse in this respect. D-Wave has thus offered some of its machine time in the cloud for researchers in the field.

A couple tests were run in 2020 using D-Wave and IBM systems. In one case performed by Turkish scientists to classify CT scans images[1961]. Their quantum software generated 94% to 100% successful classifications while its classical counterpart did achieve 90% successful results. It was using only 4 qubits from 5-qubits systems (IBM Q-Rome and Q-London). It was a hybrid computing using the quantum transfer learning method. Quantum computing was used only for the classification part at the end of a convolutional network, not for the convolutions that remain classical, using 224x224 pixels versions of the CT scans, using a training set made of 2658 lung CT images with 1296 COVID-19 and 1,362 Normal CT images. They also tried emulators running with PennyLane, Qiskit and Cirq. The classification layer compressed 512 vectors into 4 vectors with a linear transformation.

Another test related to covid-19 research used a quantum assisted SVM classification using 16 qubits running on the IBM Q-Melbourne system and implementing feature mapping and hyperplane calculation with a variational quantum classifier[1962]. Classification was using time series of number of cases per counties in the USA. The classical part was done using the scikit-learn framework from Inria, France. Well, in the end, the research team observed that classical methods outperformed QML in accuracy, particularly with a high number of data points (>300).

Developed in China in 2022, **DeepQuantum** is a hybrid quantum deep learning algorithm was used to make predictions on mutant covid-19 variants[1963]. It was trained using a database of available mutated SARS-CoV-2 RNA sequences.

In practice, conventional HPCs helped molecules screening for therapies and to create 3D models of the covid virus and, in particular, of its glycoproteins that cling to the membranes of human cells in order to attack them and enable virus reproducing within cells[1964]. In 2022, a new SARS-CoV-2 main protease (Mpro) inhibitors was discovered using a classical GPU-based HPC using Atlas's platform from Qubit Pharmaceuticals[1965]. In the more or less distant future, quantum computing may have its say in similar pandemics[1966].

# Energy and chemistry

Energy and chemistry are gathered here for their tight relationship. Quantum computing could help save energy in chemical engineering like with ammonia production. It could help improve the chemistry of batteries or carbon capture processes. At last, it could also help energy utilities optimize power grids. Most of these applications have been prototyped at a low scale but are not yet production-grade. The associated research can help determine the characteristics of the quantum computers needed to solve these various problems. The associated research works are relying mostly on quantum annealing with D-Wave machines and/or hybrid quantum computing.

---

[1961] See COVID-19 detection on IBM quantum computer with classical-quantum transfer learning by Erdi Acar and Ihsan Yilma, November 2020 (16 pages).

[1962] Another one: Quantum-Enhanced Machine Learning for Covid-19 and Anderson Insulator Predictions by Paul-Aymeric McRae and Michael Hilke, December 2020 (25 pages).

[1963] See Quantum Deep Learning for Mutant COVID-19 Strain Prediction by Yu-Xin Jin et al, TuringQ, CAS and USTC, March 2022 (34 pages).

[1964] See an example in TACC Supercomputers Run Simulations Illuminating COVID-19, DNA Replication, March 2020.

[1965] See Computationally driven discovery of SARS-CoV-2 Mpro inhibitors: from design to experimental validation by Léa El Khoury, Jean-Philip Piquemal et al, Chemical Science, 2022 (14 pages).

[1966] See Covid-19: Quantum computing could someday find cures for coronaviruses and other diseases by Todd R. Weiss, April 2020.

When we move away from organic molecules and living organisms, everything suddenly becomes almost realistic in quantum computing even if we're still far away from having the appropriate quantum hardware to run it! The molecular structures to study and simulate here are generally simpler than with organic chemistry of living organisms, particularly proteins[1967].

The first plausible quantum computing applications deal with the creation of innovative materials. The energy and chemistry sectors are interested in solving complex analysis and optimization problems, in the in-silico simulation of crystalline molecules and structures, and in creating new materials[1968].

The first case studies were, not surprisingly, first carried out with D-Wave's annealers. These seem to be well suited for simulations of atomic interactions in materials even though the provided accelerations are not stellar[1969]. Simulations can also be done with air, water and other liquid flows, and in particular their turbulence. In particular, they can exploit the famous Navier-Stokes equations[1970]. Beyond aerospace applications, it could have some applications in energy production turbines optimizations.

### Chemistry

Most quantum chemical use case applications start with simulating molecules, and then, potentially, chemical reactions. Current quantum computers are able to simulate only molecules with a couple atoms[1971]. In September 2017, **IBM** simulated on a 16-qubit superconducting quantum computer a set of beryllium hydride molecules and their minimum energy balance[1972]. But this was done without reaching a quantum advantage, meaning, these simulations can well also be done on classical computers. It went up to 12 atoms from the benzene molecule, which was simulated with a quantum emulator supporting 35 qubits, by Total and Jean-Philip Piquemal (CNRS)[1973].

Many quantum computing hardware and software startups now propose development framework for quantum simulation of matter and chemical processes. They are mentioned in a later section with software vendors. We have for example **HQS** (Germany), **Quantinuum** who launched their InQuanto computational chemistry platform coming from CQC[1974], **Pasqal** (France, thanks to their 2021 M&A with Qu&Co), **Good Chemistry** (Canada), **Menten AI** (USA), **Q1t** (The Netherlands), **Qsimulate** (USA), **Qunasys** (Japan), **Riverlane** (UK) and **Zapata Computing** (USA).

---

[1967] See Enabling the quantum leap Quantum algorithms for chemistry and materials Report, January 2019 (115 pages) which provides a good overview of chemical simulation methods. It is a report of a workshop organized by the NSF.

[1968] See Quantum hardware calculations of periodic systems: hydrogen chain and iron crystals by Kentaro Yamamoto et al, September 2021 (13 pages) with some potential applications in steel manufacturing.

[1969] See Quantum Computing: Fundamentals, Trends and Perspectives for Chemical and Biochemical Engineers by Amirhossein Nourbakhsh et al, January 2022 (28 pages).

[1970] See Quantum Navier-Stokes equations by Pina Milišić from the University of Zagreb, 2012 (12 pages) and Navier-Stokes equations using Quantum Computing, July 2020.

[1971] See Is there evidence for exponential quantum advantage in quantum chemistry? by Seunghoon Lee, Ryan Babbush, John Preskill et al, August 2022 (81 pages). The study says that exponential speedups are not generically available.

[1972] See Hardware-efficient variational quantum eigensolver for small molecules and quantum magnets, October 2017 (22 pages).

[1973] See Calculating the ground state energy of benzene under spatial deformations with noisy quantum computing by Wassil Sennane, Jean-Philip Piquemal and Marko J. Rancic, March 2022 (11 pages). See also Open Source Variational Quantum Eigensolver Extension of the Quantum Learning Machine (QLM) for Quantum Chemistry by Mohammad Haidar, Jean-Philip Piquemal et al, June 2022 (39 pages).

[1974] See Quantinuum launches InQuanto, a state-of-the-art quantum computational chemistry software platform using quantum computers, 2022.

Some chemical industry corps are investigating quantum computing use cases. **Dow Chemical** has been a **1Qbit** partner since June 2017 for pilot projects in quantum chemical simulation. **Mitsubishi Chemical** and the **Materials Magic** subsidiary of **Hitachi Metals** are also testing quantum computing with IBM.

**Covestro** (Germany), a polymer chemical production company and QC Ware started a 5-year collaboration in 2022 to use quantum algorithms for the discovery of new materials and catalysts on NISQ hardware. They started to work in 2021 and already published some results on the way to reduce the quantum computing resources required for simulating and designing new materials and chemical processes and to compute energy gradients[1975].

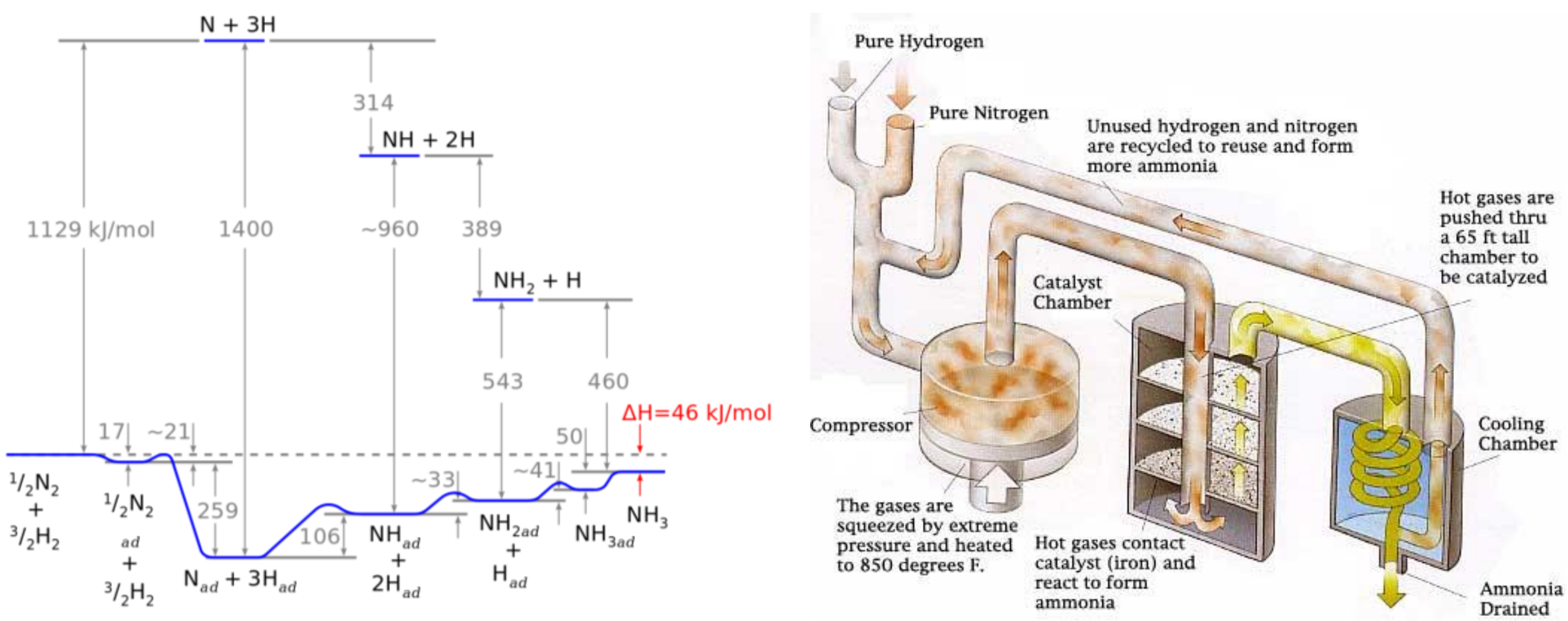

*Figure 670: the usual Haber-Bosch process. Source: Catalysis How Dirt and Sand Catalyze Some of the Most Important Transformations, by Justin J. Teesdale, Harvard Energy Journal Club, September 2017.*

However, one of the most famous chemical process that could benefit from quantum simulation would be to find a way to produce ammonia more efficiently. It's used in the production of fertilizers (88% of total) and also explosives. Right now, ammonia ($NH_3$) is produced using the famous Haber-Bosch chemical process that uses $N_2$ coming from the atmosphere and hydrogen usually coming from methane ($CH_4$). The Haber-Bosch process is currently responsible for 1% to 2% of global energy consumption and of 1.4% of $CO_2$ emissions.

**The result of quantum software optimization**

- Estimates for an example molecule: $Fe_2S_2$ with 118 spin-orbitals

| | |
|---|---|
| Gate count | $10^{18}$ |
| Parallel circuit depth | $10^{17}$ |
| Run time @ 10ns gate time | 30 years |
| Reduced gate count | $10^{11}$ |
| Parallel circuit depth | $10^{10}$ |
| Run time @ 10ns gate time | 2 minutes |

- Attempting to reduce the horrendous runtime estimates we achieved
  Wecker *et al.*, PRA (2014), Hastings *et al.*, QIC (2015), Poulin *et al.*, QIC (2015)
  - Reuse of computations: O(*N*) reduction in gates
  - Parallelization of terms: O(*N*) reduction in circuit depth
  - Optimizing circuits: 4x reduction in gates
  - Smart interleaving of terms: 10x reduction in time steps
  - Multi-resolution time evolution: 10x reduction in gates
  - Better phase estimation algorithms: 4x reduction in rotation gates

**From materials to models on quantum computers**

| | Material | Model |
|---|---|---|
| Orbitals per unit cell | ≈ 50 | 1 |
| Unit cells needed | 20x20 | 20x20 |
| Number of orbitals | $N$ ≈ 20'000 | $N$ ≈ 800 |
| Number of terms | $N^4$ | O($N$) |
| Scaling of algorithm | O($N^{5.5}$) | O($N^{0.5}$) |
| Estimated runtime | age of the universe | seconds |

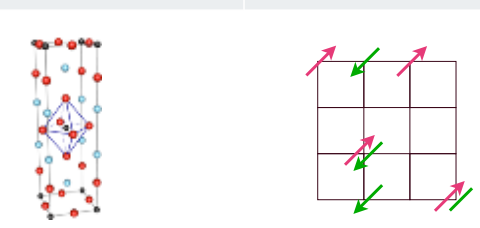

*Figure 671: resource estimates for simulating the spin-orbitals of $Fe_2S_2$ molecule. Source: TBD.*

The nitrogenase process uses a catalyst that is usually some iron doped with potassium and fixed on silica or alumina. Its natural equivalent is FeMoCo, a natural cofactor of nitrogenase. The Haber-Bosch process is highly energy-consuming particularly in the part producing pure $H_2$ and due to the heat and pressure needed in the main reactor (500°C and 100 bars).

---

1975 See Local, Expressive, Quantum-Number-Preserving VQE Ansatze for Fermionic Systems by Gian-Luca R. Anselmetti et al, May 2021 (26 pages) and Analytical Ground- and Excited-State Gradients for Molecular Electronic Structure Theory from Hybrid Quantum/Classical Methods by Robert M. Parrish et al, October 2021 (23 pages).

Two processes could be developed thanks to quantum simulation for improving the energy efficiency of ammonia production. The first would be to simulate the nitrogenase enzyme, FeMoCo, that converts nitrogen into ammonia in cyanobacteria at ambient temperature and pressure. The second would be to invent new catalysts serving to operate the Haber-Bosch process at lower temperature and pressure. One example is the design of Fe/K mixtures supported on carbon nanotubes. But the number of qubits and gates to implement for solving these problems seems quite mind boggling, even with corrected qubits[1976].

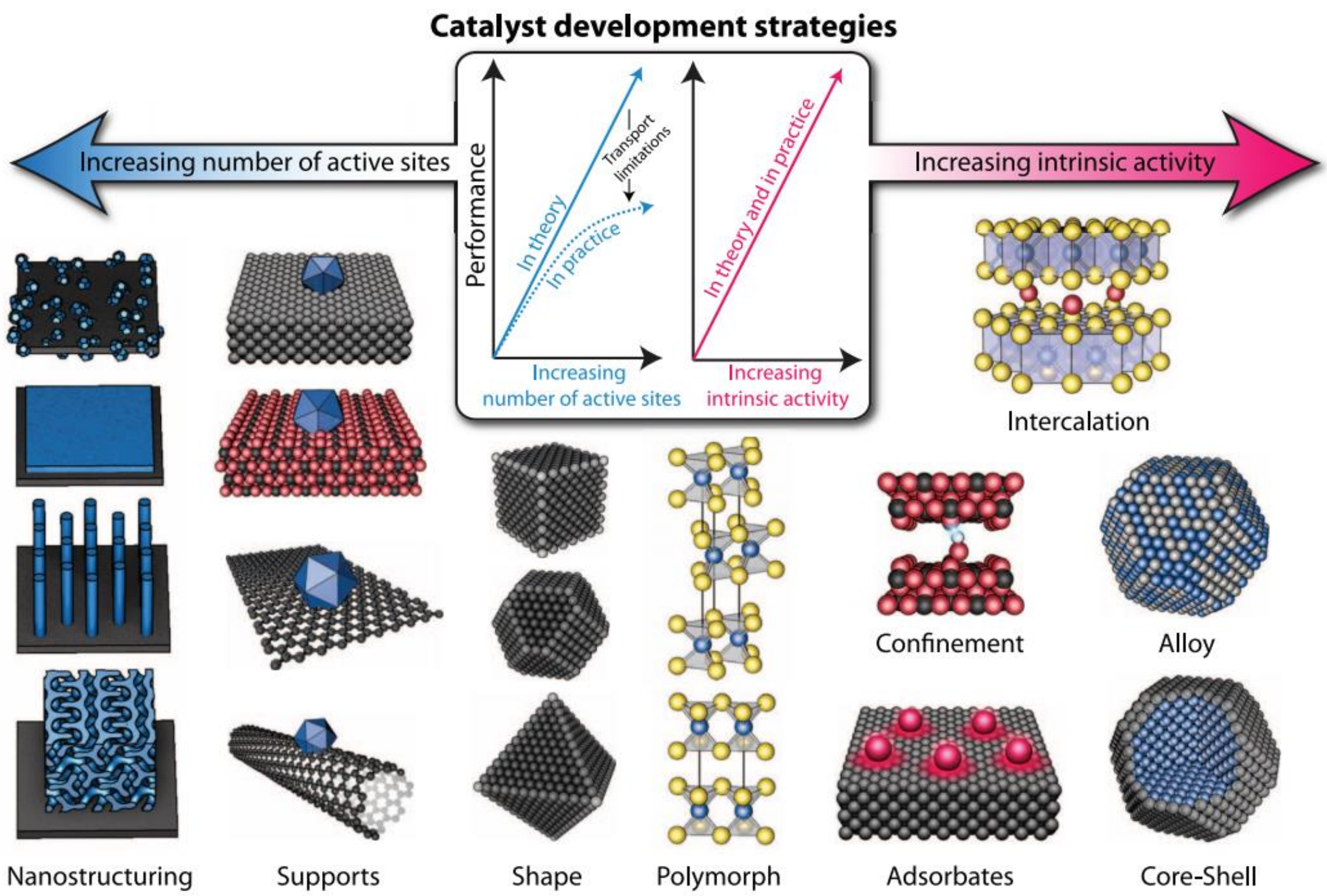

*Figure 672: Source: Combining theory and experiment in electrocatalysis: Insights into materials design by Jens Jaramillo et al, Science, 2017 (33 pages).*

You may wonder why cement production that is also responsible for significant $CO^2$ emissions is less talked about as a quantum case. Maybe is it because nature is not producing it and biomimetics harder to implement. Still, creating innovative cement production techniques is about testing many variants of clinkers who reduce $CO^2$ emissions during production. Quantum computing could help simulate these various material combinations to identify a more durable solution.

### Batteries

Research is well underway to create batteries that are more efficient in terms of energy density, charging speed and supported charging/discharging cycles[1977]. Simulation is most often used to understand the operations of the chemical reactions taking place on cathodes and anodes and in the crystalline intercalation structures and to find ways to improve energy density and avoid battery wear phenomena.

---

[1976] See Even More Efficient Quantum Computations of Chemistry Through Tensor Hypercontraction by Joonho Lee et al, July 2021 (62 pages). With this recent method, 4 million physical qubits with an error rate of 0,1% would still be necessary.

[1977] See The Promise and Challenges of Quantum Computing for Energy Storage (4 pages) and Can quantum computing fuel a leap in battery research? Not any time soon, but automakers and quantum companies are exploring it right now by Grace Donnelly, Emerging Tech Brew, April 2022.

**battery simulation**
lithium-oxygen
*source: IBM*

**battery simulation**
lithium-sulfur battery design
*source: IBM*

**battery simulation**
simulating magnetism and spins
*source: Samsung, Honeywell*

**battery simulation**
estimating the cost of electrolyte simulation on PsiQuantum's future QPU.
*source: PsiQuantum, Mercedes-Benz*

**battery simulation**
model lithium oxide to understand how batteries age over time
*source: Hyundai, IonQ*

**battery materials design**
simulating Mott insulator transitions in battery electrode materials and ceramic superconductors and discharge curve of $Li_xCoO_2$.
*source: Total, Pasqal*

*Figure 673: quantum computing use cases in the battery development domain. (cc) Olivier Ezratty, 2022.*

Several companies are pursuing this goal. IBM and **Mitsubishi Chemical** have simulated on a superconducting qubit computer the initial steps of the reaction between lithium and oxygen in lithium-air batteries[1978].

**Mercedes-Benz** worked with IBM to simulate lithium-sulfur batteries[1979] and later, with PsiQuantum to estimate the computing requirements for electrolyte simulation on PsiQuantum's system. For one of the simulation needs, they would need to have about 16K logical qubits. Given, PsiQuantum plans for requiring 10,000 physical qubits to create a single logical qubit. We end up with 160M physical qubits. PsiQuantum's plan is to reach 1M physical qubits by 2030[1980].

**Hyundai** and IonQ are developing a quantum application to model lithium oxide using IonQ's existing hardware. Lithium oxide is not used in EV batteries but its modeling could potentially help mitigate how batteries break down over time. But no IonQ computer has a computing advantage vs a classical computer! They use the Aria with 32 physical qubits and 20 "algorithmic qubits". The current project is focused on a relatively small molecules containing up to two lithium and one oxygen atom. In the end, it would require 2000 logical qubits that are far away beyond 2030 in IonQ's roadmap.

**Volkswagen** is also working on simulating lithium-ion batteries, teaming up with 1Qbit and Xanadu. They concluded that we are far off from being able to implement it[1981].

At last, quantum sensing could also help develop new batteries. That's what **AMTE Power** (UK) is doing with leading the 3 years Project Quantum to use quantum sensing to develop lithium-based battery manufacturing solutions, which got a funding of £5.4M from Innovate UK. This project is about using NV center-based magnetic quantum sensing to evaluate the battery aging process[1982].

[1978] See Computational Investigations of the Lithium Superoxide Dimer Rearrangement on Noisy Quantum Devices by Qi Gao et al, 2018 (6 pages).

[1979] See Quantum computation of dominant products in lithium–sulfur batteries by Julia E. Rice et al, Journal of Chemical Physics, 2021 (7 pages)

[1980] See PsiQuantum, Mercedes-Benz Study How Quantum Computers Can Accelerate EV Battery Design by Matt Swayne, The Quantum Insider, April 2022. Mentioning Fault-tolerant resource estimate for quantum chemical simulations: Case study on Li-ion battery electrolyte molecules by Isaac H. Kim et al, Physical Review Research, April 2022 (27 pages).

[1981] See How to simulate key properties of lithium-ion batteries with a fault-tolerant quantum computer by Alain Delgado et al, April-September 2022 (31 pages). The paper concludes that there is a need of 2,375 to 6,652 logical qubits to simulate $Li_2FeSiO_4$ with their 156 electrons.

[1982] See AMTE Power's Project Quantum Signals New Era of British Battery Manufacturing, September 2020.

### Carbon capture

Carbon capture is another issue and researchers are simulating its molecular modus-operandi by bio-mimetics.

**BASF** is trying to simulate synthetic polymers, first on HPCs, then eventually on quantum computers. They invested in both **HQS** (Germany) and **Zapata Computing** (USA).

Since 2017, **Dow Chemicals** has been collaborating with the Canadian software publisher **1Qbit** to create new molecules, using D-Wave annealers.

**TotalEnergies** is working with CQC from Quantinuum to capture carbon on Metal-Organic Frameworks using a VQE based algorithm. They did simulate the use of up to 16 qubits which is way below any quantum computing advantage threshold[1983].

### Oil exploration

Well, that stinks but oil companies are still trying to optimize oil exploration and extraction.

**BP** is working on the design of algorithms for optimizing oil exploration. This involves using data from various sensors, particularly seismic sensors, to consolidate models for simulating geological layers under the ground. They joined in 2020 the **IBM Quantum Network** as an industry partner to get access to IBM's 65-qubit quantum computer, software and experts.

It follows **ExxonMobil** who entered the program in 2019 as one of the major companies associated with IBM.

**TotalEnergies** also want to optimize oil prospection and reserves evaluation using seismic probes. They plan to address complex optimization problems such as **MINLP (**Mixed Integer NonLinear Programming[1984]) to optimize refining, planning, production and transportation[1985].

### Power grid

In 2019, the **Dubai Electricity and Water Authority** (DEWA) was working with Microsoft to solve complex power and water distribution problems. It was just a matter of testing a few algorithms on Intel simulators running on Azure. And for good reason, Microsoft did not yet have its own quantum computer[1986]. In 2020, DEWA announced that it would train their algorithms in D-Wave annealers[1987]!

**EDF** is another major French company that is studying the use cases of quantum computing very closely: quantum new materials simulations, material aging simulations particularly under radiations, safety statistics, combinatorial optimization for smart grids and battery management (teaming up with Pasqal and ParityQC[1988]) and also customer segmentation using Quantum Machine Learning.

---

[1983] See Modelling Carbon Capture on Metal-Organic Frameworks with Quantum Computing by Gabriel Greene-Diniz et al, March 2022 (18 pages), CQC and Total Announce Multi-Year Collaboration to Develop Quantum Algorithms for Carbon Capture, Utilization and Storage (CCUS), April 2020 and Total Exploring Quantum Algorithms to Improve CO2 Capture, May 2020.

[1984] A version of the MINLP problem solving algorithm exists for D-Wave via their QUBO framework. See Quantum Computing and Non-Linear Integer Optimization by Sridhar Tayur February 2019 (42 slides).

[1985] Total has partnered with private players (IBM, Atos, Rigetti QC Ware, Google) and various research laboratories around the world: PCQC (Paris), LIRMM Montpellier, CERFACS, Université ParisSud, Jülich Forschungszentrum (Germany) and the University of Leiden.

[1986] See Microsoft and DEWA bringing quantum computing to Dubai, June 2018.

[1987] See DEWA organises training sessions on quantum computing in partnership with D-Wave, February 2020.

[1988] See Qualifying quantum approaches for hard industrial optimization problems. A case study in the field of smart-charging of electric vehicles by Constantin Dalyac et al, June 2021 (29 pages).

**DTU** experimented in Denmark a grid optimization using an HHL algorithm with up to 7 IBM qubits QPUs. They found that they have scalability issues. Surprise[1989]!

## Transportation and logistics

Beyond energy matters mentioned above, transportation industries are mainly interested in algorithms for optimizing complex systems[1990]. Let's look at what can be done with road, rail, air and maritime transportation.

### Road

Quantum computing could be implemented in many domains to improve both vehicle production and their operations. In the product design phase, quantum computing could help optimize vehicle fluid dynamics thanks to solving complex partial differential problems (aka Navier Stokes equations), minimizing drag and improving battery/fuel efficiency. It could also simulate the physics of various materials to improve weight/strengths vehicle ratios and design new batteries (but in the very long term).

In the operations stage, QC could handle complex optimization problems like supply-chain optimizations, warehouse robots routing, improving the accuracy of demand forecasting both with end-users and for their suppliers and inventory optimization. You have also traffic flow optimization or multi-modal fleet operations[1991]. Machine learning based solutions could under some circumstances benefit from QC, for improving pattern recognition in images and various classification tasks used in manufacturing, predictive maintenance as well as in marketing[1992].

One less credible use case is autonomous vehicle control whether or not it's based on some machine learning technique. It doesn't make much sense for quantum computing due, at least, to the data-loading problem. Vehicles sensors are generating a huge stream of high-volume data that can't be ingested in a quantum computer, and particularly with one such computer sitting in the vehicle. So, let's forget this for a while.

Let's first look at the manufacturing of vehicles. And, surprise, they are mostly if not all… Germans! In 2022, **Volkswagen** Data:Lab published two papers coauthored with Terra Quantum AG related to hybrid computing associated with D-Wave quantum annealers in two areas: the optimization of assembly line work-flow scheduling[1993] and hybrid machine learning to enhance image recognition systems used for car classification[1994]. It is still not clear whether these applications really went into production.

---

[1989] See Quantum Computing for Power Flow Algorithms: Testing on real Quantum Computers by Brynjar Sævarsson et al, July 2022 (8 pages) and DTU first to use quantum computer for the power grid, 2022.

[1990] See this inventory of needs, but no solutions in Quantum Applications Transportation and Manufacturing by Yianni Gamvros, IBM, 2017 (20 slides).

[1991] See Quantum computing for transport optimization by Christopher D. B. Bentley et al, Q-CTRL, June 2022 (12 pages).

[1992] See Will quantum computing drive the automotive future? by McKinsey, September 2020. The listed use cases are interesting but the list contains some misleading case studies that are not relevant with quantum computing like "*processing vast amounts of data to accelerate learning in autonomous-vehicle-navigation algorithms*" (QC is not good at handling big data) and "*ensure car-to-car communications in almost real time*" (no quantum technology is bound to ensure real-time whatever communication). Most use cases related to autonomous vehicles operations are also quite unrealistic.

[1993] See Solving workflow scheduling problems with QUBO modeling by A. I. Pakhomchik et al, May 2022 (10 pages). The experiment is using QUBO on a D-Wave Advantage annealer. They found that the hybrid and classical algorithms were the most successful in solving the instances, although no solver was able to solve all QUBOs at all sizes. One caveat is that the paper doesn't mention computing time but only results quality.

[1994] See Hyperparameter optimization of hybrid quantum neural networks for car classification by Asel Sagingalieva et al, May 2022 (10 pages). It's about improving the accuracy of image recognition that is used for fault detection in manufacturing. They created an hybrid QML algorithm and deployed it on the QMware cloud from Terra Quantum running on D-Wave's annealers. The benchmark compared this new method over classical machine learning and quantified performance improvements in reduced expected run times and model fitness as the problem size scales. But it's very uncertain as of a real potential quantum advantage.

**Daimler AG** is one of the leading companies working on quantum technology with IBM, with applications for logistics and planning optimization and everything to do with autonomous vehicle routing at the forefront. In 2018, they also launched an initiative with IBM to develop lithium-sulfur batteries, which improve energy density and make it possible to do without metals such as cobalt and nickel. All of this will be achieved through quantum simulation.

BMW GROUP

**isolation PVC application in cars**
optimization
*source : BMW, D-Wave*

**compute metal forming applications modeling**
solve differential equations in order to better
*source : BMW, Pasqal*

VOLKSWAGEN

**vehicle recommandation**
hybrid computing
*source : D-Wave,*

**painting planning optimization**
*source : D-Wave*

**cars image recognition**
*source : Volkswagen, Terra Quantum*

**plant optimization**
*source : Volkswagen, Terra Quantum*

BOSCH

**factory digital twin**
Optimizing output, energy consumption and waste management
*source : Multiverse Computing.*

*Figure 674: a sampler of quantum computing use cases in the automotive industry. (cc) Olivier Ezratty, 2022.*

**BMW** is also willing to learn how to use quantum computing in various tasks, one being the optimization of its spare parts supply chain[1995]. They are partnering with Honeywell to do this as well as with Cambridge Quantum Computing, Zapata Computing and Entropica Labs. They started using the Honeywell trapped ions-based H0 and H1 with 10 qubits. They used a Recursive Quantum Approximate Optimization Algorithm (R-QAOA) to manage their combinatorial problem. The quality of these trapped ions qubits made the trial promising although of course not usable at all given the small number of available qubits[1996]. In May 2022, BMW started a partnership with Pasqal to solve differential equations in order to better compute metal forming applications modeling.

**Multiverse Computing** and a **Bosch** Automotive Electronics plant in Madrid announced in 2022 that they are exploring the creation of a digital twin of a factory with quantum computing. They plan to capture data to assess the performance of individual equipment and optimize quality control and overall efficiencies, including energy consumption and waste management. What they call digital twins are actually optimization systems.

In 2021, the German auto industry (BMW, Mercedes-Benz, Volkswagen, and Bosch) and research organizations (DFKI who runs contract research on AI and Forschungszentrum Jülich) even launched the **Q(AI)2** project to test AI applications for quantum computers[1997]. It particularly wants to study the usage of AI to solve various manufacturing optimization problems.

The deployment of autonomous vehicles fleets is a nice target application for quantum computers. The more autonomous the vehicles are, the greater the need for automation and route coordination. The problems to be solved will be to determine step-by-step the vehicle fleet routes in order to optimize each of these vehicles and passengers the travel time.

---

[1995] See this excellent use cases inventory: Quantum Computing: Towards Industry Reference Problems by Von Andre Luckow, BMW, April 2021.

[1996] See How BMW Can Maximize Its Supply Chain Efficiency with Quantum, Honeywell, January 2021.

[1997] See Quantum AI For The Auto Industry by DFKI, June 2021.

This was an experiment done in 2017 by **Volkswagen** on D-Wave annealers[1998]. Its goal was to optimize the routes of a (virtual) cab fleet in Beijing[1999]. The experiment was using the T-Drive data set published by Microsoft in 2008 which describes the routes of 10,357 cabs. The algorithm used was QUBO (Quadratic Unconstraint Binary Optimization). The diagram below shows the result of the optimization of the route of 418 cabs making the journey from the city center to the airport taking into account the route of 10,357 vehicles[2000].

There is a lack of hindsight in estimating the size of quantum computers needed to practically handle such large-scale problems. How many qubits would be needed to optimize a fleet of hundreds or even millions of autonomous vehicles? Volkswagen also experimented small-scale algorithms with D-Wave for optimizing vehicle recommendations and also optimizing cars painting planning. A fairly less ambitious version of this algorithm was used to optimize the individual travel routes of nine public-transit buses during the November 2019 Web Summit event in Lisbon.

So, we're off wondering what they really achieved or could achieve. Similar trucks routing algorithms were already explored by **Accenture** and **Denso** using D-Wave annealers. Annealers are so far in a better position to solve these problems than existing superconducting gate-based qubits systems from IBM. In an adjacent domain, **Toyota** worked with D-Wave to optimize traffic lights on a 50x50 road grid, using a D-Wave 2000Q[2001].

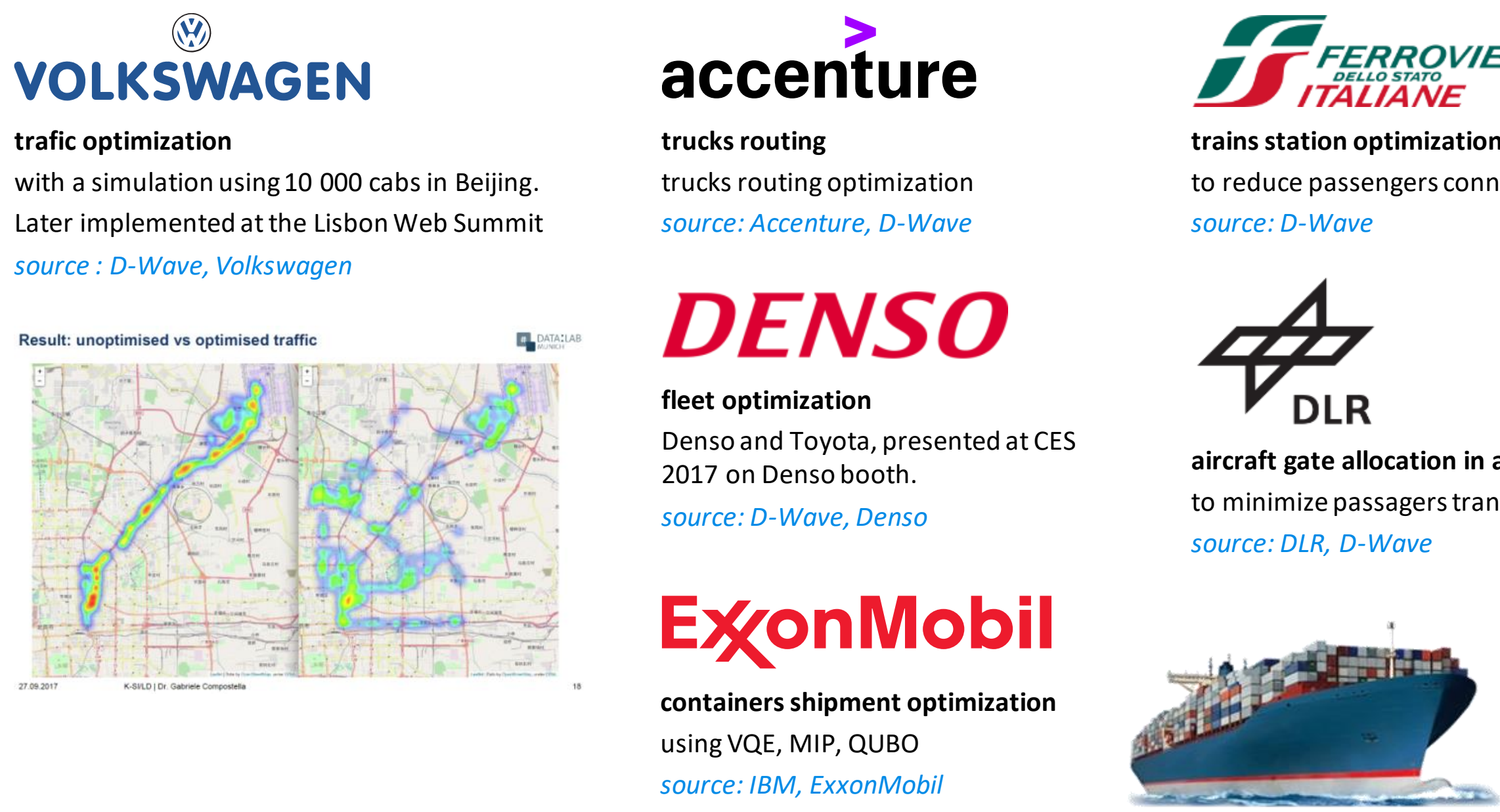

*Figure 675: a sampler of quantum computing use cases in the transportation industry. (cc) Olivier Ezratty, 2022.*

At last, another work conducted by **Hyundai** with IonQ is to improve 3D object detection with quantum computing. IonQ did demonstrate a "101" image recognition with 43 road signs (this is the basic test for a classical convolutional network running on a simple PC) and plans to expand it to recognize pedestrians and cyclists.

---

1998 See Volkswagen takes quantum computing from the lab to the factory by Volkswagen, August 2021.

1999 It is documented in Quantum Computing at Volkswagen Traffic Flow Optimization using the D-Wave Quantum Annealer, 2017 (23 slides).

2000 The results are published in Traffic flow optimization using a quantum annealer, August 2017 (12 pages). As with many case studies from D-Wave, this one is also contested by HPC specialists.

2001 See Toyota Central R&D Labs: A Quantum Approach to Transportation by D-Wave (2 pages) and Traffic Signal Optimization on a Square Lattice with Quantum Annealing by Daisuke Inoue et al, February 2021 (14 pages).

They believe they can do better with quantum computers than with classical NPUs (the neural processing units now embedded in most CPUs and mobile chipsets). They may forget the constraints of embedded systems in cars. An IonQ wouldn't fit in the trunk!

### Rail

An experiment was done by **Ferrovie dello stato Italia** to optimize train arrivals in railways stations, also to minimize passengers' connection times, again with D-Wave.

ALPHARAIL **AlphaRail** (2000, USA) is a railways software company using machine learning and quantum computing to improve railways operations.

They are relying on quantum and quantum-inspired approaches to solve routing and scheduling optimization problems.

### Air

With airlines, the current focus about optimizing aircraft fleets planning, to maximize the capacity to meet demand while optimizing the aircraft fill rate.

Also, quantum computing can enable optimizing airport and aircraft gates management, in order to minimize passenger waiting time, as tested by **DLR** in Germany[2002]. This is an NP-difficult problem that is difficult to deal with using conventional algorithms.

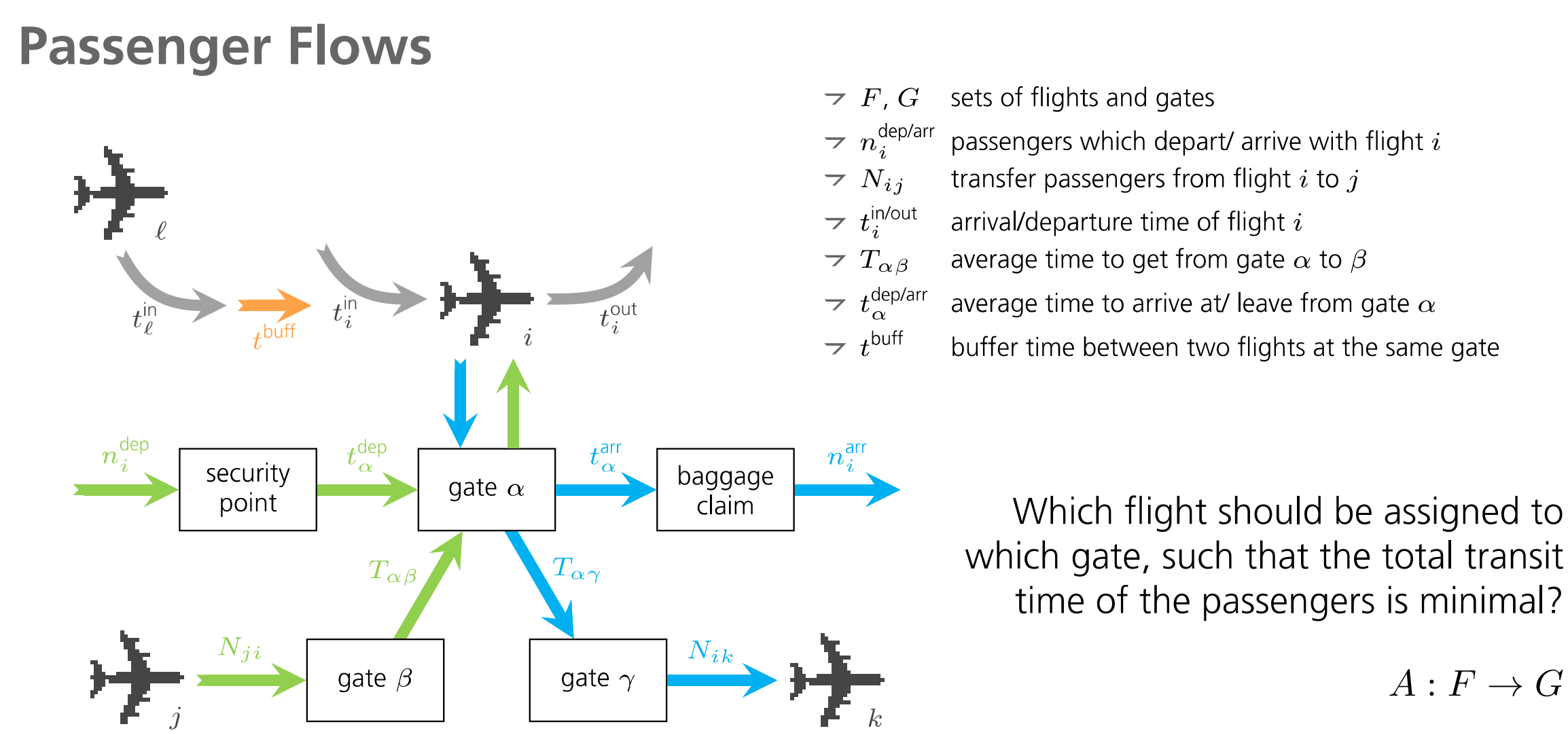

Figure 676: Flight Gate Assignment with a Quantum Annealer by Elisabeth Lobe and Tobias Stollenwerk, March 2019 (15 slides).

In Japan, **Sumitomo** launched in June 2021 a pilot experiment for optimizing flight routes for urban air mobile vehicles (air taxis and unmanned drones). They will rely on resources from Tohoku University. No precision on the problem sizing and on which quantum system they plan to pilot their solution even if D-Wave is a logic contender.

Researchers from **Chalmers University** in Sweden prototyped a promising QAOA hybrid algorithm solving the "tail Assignment problem", which is the task of assigning individual aircraft to a given set of flights, minimizing the overall cost for the airline. They said it worked with only 2 qubits, to optimize 2 flights and would scale well as flights are added[2003].

---

[2002] See Flight Gate Assignment with a Quantum Annealer by Elisabeth Lobe and Tobias Stollenwerk of the German Aerospace Center (or DLR for Deutsches Zentrum für Luft- und Raumfahrt e.V. ), March 2019 (15 slides). The case study uses a D-Wave. It shows that the solution is not obvious to develop.

[2003] See Two-Bit Quantum Computer Solves Real Optimization Problem by Matt Swayne, December 2020 pointing to Applying the Quantum Approximate Optimization Algorithm to the Tail-Assignment Problem by Pontus Vikstål et al, September 2020 (11 pages).

These are needs that can be addressed both by machine learning algorithms to take into account the past or with quantum optimization algorithms based on a description of the parameters of the problem. The former does prediction and the latter simulation. Simulations avoids the back-mirror bias that can be induced by prediction methods based on past data. A combination of the two methods is possible.

**Airbus** is also involved in quantum. In 2015, one of their teams based in Newport in the United Kingdom began working on the subject. In 2016, Airbus Ventures invested in the American startup QC Ware. They experimented with the use of a D-Wave for fault tree analysis (FTA), which is used to determine the origin of complex failures with a gain of a factor of 4 compared with traditional methods. This is a difficult NP-difficult combinatorial problem that is easier to solve in quantum programming. In January 2019, Airbus launched its "Quantum Computing Challenge", a way to outsource the development of quantum solutions to help them solve their business problems, in fluid mechanics, differential equations, flight optimization, wing design, cargo bay filling, etc.[2004]. As of May 2019, 475 teams from 57 countries had competed in this challenge. They came mainly from the USA and India, followed by Europe. They announced the challenge results in October 2020 with five finalists selected out of 36 contestants.

One team from **Capgemini** devised a new hybrid matrix inversion algorithm mixing the famous HHL algorithm and a QSVM (Quantum-enhanced Support Vector Machine (QSVM). Other teams worked on plane loading optimization problems, on quantum machine learning models and on fluids dynamics applied to aircraft design.

How about designing autonomous **quantum flying drones** using a quantum computer, a project from Indian researchers[2005]? It makes no sense from a practical standpoint, at least due to real-time constraints. Since the quantum computer can't sit in the drone, a fortiori, one from IBM, there's an inevitable lag in the communication between the drone and the quantum computer which, by the way, should be entirely dedicated to a single drone to work real-time. You already have autonomous drones and robots, thanks to their embedded computing systems, and they don't yet need a quantum computer to run their symbolic and connectionist artificial to fly.

### Maritime

**ExxonMobil** and **IBM** are working on finding algorithms to optimize maritime traffic routing. Existing solutions rely on heuristics and simplifications. They were willing to see whether quantum computing could transform these complex optimization problems and solve them more efficiently with quantum computing. Their vision is about container shipments volumes. They formatted their problem as a "vehicle routing problem with time windows" (VRPTW) which is a NP-hard problem. They compared various methods, using a QUBO algorithm that can be transformed on a lower-level VQE or QAOA hybrid algorithm and experimented it with Qiskit on the QasmSimulator IBM quantum emulator backend[2006]. As in many similar cases, the published paper does not provide any clear answers on the gain and on the quantum computer specification that would make it possible to solve a real-life problem. On top of that, it was not at all formulated as a container shipment optimization problem per se but as a simpler truck routing problem.

Maritime containers shipment is also a center of interest. **DP World**, the Dubai Port operator, is partnering with **D-Wave** to find ways to use quantum computing to optimize their port operations. At this stage, it's just exploratory work with no details at all about envisioned applications.

---

[2004] See Airbus gets aerodynamic with quantum computing by Michael Feldman, January 2019.

[2005] See Design and Simulation of an Autonomous Quantum Flying Robot Vehicle: An IBM Quantum Experience by Sudev Pradhan et al, June 2022 (7 pages).

[2006] See ExxonMobil & IBM Explore Quantum Algorithms to Solve Routing Formulations by Stuart Harwood et al, February 2021 which refers to Formulating and Solving Routing Problems on Quantum Computers by Stuart Harwood, January 2021 (17 pages).

The ones that are easy to guess are containers loading/offloading optimizations[2007]. Another such project was also undertaken with the **Port of Los Angeles** by SavantX, also using D-Wave annealers and quantum machine learning[2008].

## Retail

The retail sector could benefit from quantum computing generic use cases in transportation and logistics improvements as well as in some marketing optimization techniques. Robotized warehouse operations could also benefit from some quantum-driven optimizations like what **Ocado Group** (UK) is piloting with D-Wave[2009]. Beyond that, consultants and analysts have not much to say about it or they sometimes err in mention big data applications, which are out of scope of quantum computers[2010].

**PayPal** is also testing a D-Wave quantum annealer to solve some optimization problems and detect fraud using quantum machine learning[2011].

## Telecommunications

At this stage, the main quantum involvement of telecom companies is mainly related to quantum telecommunications and cryptography.

In the USA, **Verizon** launched a QKD trial on three sites in the Washington, D.C., area[2012]. In 2021, they also tested a VPN using a PQC (post-quantum cryptography) solution based on the Saber NIST competition finalist, for connection between two sites in the US and UK. **AT&T** is also investing resources in QKD networks, teaming up with Caltech.

In France, **Orange** participated to a QKD trial in the Nice region with partners including the InPhyNi research lab.

In the UK, **British Telecom** has also experimented a QKD setup to demonstrate some secured backup of datacenter resources. In October 2020, they deployed a pilot 6 km long QKD infrastructure in Bristol to connect several industry sites, in partnership with **Toshiba** as part of the **AQuaSec** (Agile Quantum Safe Communications) program, cofunded by UKRI[2013]. In October 2021, BT and Toshiba announced the deployment of a commercial QKD network in London, mixed with PQC for non photonic endpoints.

Quantum computing can still play a role to solve various optimization problems in the telecom industry. The most commonly presented are the placement, power and frequency assignment of overlapping cells in 4G/5G mobile networks, the configurations of paths and wavelengths on land line fiber optics networks and similar optimization problems for satellite communications[2014] and MIMO

---

[2007] See DP World explores quantum computing technology, April 2021.

[2008] See SavantX, D-Wave Collaborate on Quantum Algorithms to Tackle Supply Chain Problems at U.S.'s Busiest Port by Matt Swayne, The Quantum Insider, January 2022.

[2009] See Towards Real Time Multi-robot Routing using Quantum Computing Technologies by James Clark et al, SFTC Hartree and Ocado, 2019 (10 pages).

[2010] See Quantum computing in the consumer and retail sectors by Linda Ellett and Ian West, KPMG, July 2021. Extract: "*quantum computing could process data sets envisioned but not feasible at the current time to provide real-time consumer insights*".

[2011] See PayPal: Harnessing Quantum Computing in FinTech | D-Wave Qubits 2021, December 2021 (30 mn video).

[2012] See This Executive Director Is Leading Verizon Into the Future Through Quantum Computing by Joanna Goodrich, IEEE Spectrum, November 2020.

[2013] See BT and Toshiba install UK's first quantum-secure industrial network between key UK smart production facilities, October 2020.

[2014] See Heterogeneous Quantum Computing for Satellite Constellation Optimization: Solving the Weighted K-Clique Problem by Gideon Bass et al, Booz Allen Hamilton, 2017 (17 pages).

antennas optimization[2015]. It could also be used to enhance wireless networks energy efficiency with index modulation, a task that is a NP-hard problem[2016]. It uses a Grover Adaptive Search (GAS) providing a quadratic speedup (not an exponential one) and still requires a fault-tolerant large-scale quantum computers that doesn't exist yet.

In Italy, the telecom operator **TIM** used a D-Wave QUBO based algorithm to optimize the setup of 4G/5G radio cells frequencies.

In another realm, more in the home automation space, researchers from **Mitsubishi** developed a quantum human activity monitoring algorithm using Wi-Fi in the ultra-wideband range of 60 GHz and a variational quantum circuit (VQC) building a QNN (quantum neural network)[2017].

# Finance

Finance is another great playground for experimenting quantum technologies and particularly quantum computing[2018]. Both because this vertical is quite hungry with forecasting and optimization needs and also because it is a rather solvent market with many economic players having the critical mass to invest in new technologies.

Banks have a pressing need to transform themselves to adapt to constant technological and societal changes. They manipulate valuable data dumps. They need to optimize many facets of their business, starting with investment portfolios. It's all about optimizing portfolio returns, minimizing risks, manage regulations, mainly the Basel III framework rules with their liquid coverage radio constraints, and at last detect fraud risks as efficiently as possible. Also, assets are interdependent and transaction costs are variable depending on the type of assets. Their evolution responds to varying levels of uncertainty and risk.

| Question | Broad approach solution |
|---|---|
| *Which assets should be included in an optimum portfolio? How should the composition of the portfolio change according to what happens in the market?* | Optimization models |
| *How to detect opportunities in the different assets in the market, and take profit by trading with them?* | Machine learning methods, including neural networks and deep learning |
| *How to estimate the risk and return of a portfolio, or even a company?* | Monte Carlo-based methods |

Figure 677: examples of financial problems and approaches with quantum computing. Source: to Quantum computing for finance: overview and prospects by Roman Orus et al, 2018 (13 pages)

There are also some interesting mathematical relationships between some key equations in finance and in quantum physics.

This is the case of the Black-Scholes differential equation, which makes it possible to calculate the price of financial derivatives that are indexed on the price of underlying different financial instruments. It can indeed be considered as a variant of Schrödinger's wave function!

---

[2015] See Evaluation of quantum annealer performance via the massive mimo problem by Zsolt I. Tabi et al, August 2021 (14 pages). An Hungarian team assessed the use of D-Wave annealers to optimize MIMO antennas configurations. They found that some interesting results were obtained with the latest D-Wave Advantage annealer but that it will require at least another generation of better-connected qubits annealers to bring some form of quantum annealing advantage for this problem's solving.

[2016] See Quantum Speedup for Index Modulation by Naoki Ishikawa , IEEE Access, July 2021 (11 pages).

[2017] See Quantum Transfer Learning for Wi-Fi Sensing by Toshiaki Koike-Akino and Pu Wang, Ye Wang Toshiaki Koike-Akino, Mitsubishi, May 2022 (6 pages).

[2018] See overview in Quantum Computing and Finance from the Quantum World Association, August 2018, which refers to Quantum computing for finance: overview and prospects by Roman Orus et al, 2018 (13 pages) and the review paper A Survey of Quantum Computing for Finance by Dylan Herman et al, JP Morgan, Universities of Chicago, Delaware, DoE Argonne National Lab and Menten AI, January 2022 (56 pages).

A recent review paper from Isaiah Hull, Or Sattath, Eleni Diamanti and Goran Wendin from Sweden, Israel and France, describes the wealth of quantum algorithms that could be used in the economic and financial spheres[2019]:

- **Numerical differentiation** algorithms used in financial econometrics, structural microeconometrics, maximum likelihood estimation, dynamic stochastic general equilibrium (DSGE) modelling, and large-scale macroeconomic modelling conducted by central banks and government agencies.
- Solving **dynamic economic models** using interpolation algorithms.
- **Pricing of derivatives** using linear systems algorithms including matrices inversions, linear regressions, and matrix powers [2020].
- **Macro-economic models** using finite elements methods.
- **Portfolio optimization** using most of the time quantum annealers but that could run on a gates-based quantum computer [2021].
- **Macroeconomics**, **forecasting**, **modeling credit spread,** and **pricing financial derivatives** based on quantum machine learning algorithms including principal component analysis, including some work done by Goldman Sachs in partnership with IBM[2022].
- **Simulation of agent choices** over time using Monte Carlo simulations.
- **Investment portfolio optimization** including a model published in 2015 and running on a D-Wave quantum annealer and based on a QUBO model and graph modeling [2023].
- **Model market instability**, also on a quantum annealer [2024].
- **Random number generation** used beyond cryptography, for simulations and estimation, particularly with Monte Carlo simulations.

Most of the existing pilot algorithms have been tested with a small number of assets, in the 1-500 range, particularly on D-Wave quantum annealers, while real-world scenarios are based on thousands if not hundred thousand of them.

When you look at some of the referenced papers, you discover that gate-based solution hardware constraints are most of the time gigantic[2025].

---

[2019] See One Bit, Qubits, A Dollar: Researchers Say Economists Should Prepare for Quantum Money by Matt Swayne, January 2021, making reference to Quantum Technology for Economists by Isaiah Hull, Eleni Diamanti et al, December 2020 (120 pages). The report was published by Sveriges Riksbank, the employer of one of the contributors. They didn't fund any research related to this paper.

[2020] See Quantum computational finance: martingale asset pricing for incomplete markets by Patrick Rebentrost, Alessandro Luongo, Samuel Bosch and Seth Lloyd, September 2022 (31 pages).

[2021] See Quantum computational finance: quantum algorithm for portfolio optimization by Patrick Rebentrost and Seth Lloyd, 2018 (18 pages), Benchmarking the performance of portfolio optimization with QAOA by Sebastian Brandhofer et al, July 2022 (28 pages) and NEASQC financial application library v1.0 available, 2022, a quantum quantitative finance library released by the NEASQC European consortium.

[2022] See for example Study: Quantum Computers Can't Match Classical Computers in Derivative Pricing... Yet by Matt Swayne, December 2020 making a reference to A Threshold for Quantum Advantage in Derivative Pricing by Shouvanik Chakrabarti et al, , Goldman Sachs, IBM and University of Maryland, May 2021 (41 pages). This work was improved in Towards Quantum Advantage in Financial Market Risk using Quantum Gradient Algorithms by Nikitas Stamatopoulos, William Zeng et al, Goldman Sachs and IBM, November 2021 (20 pages) which reduced the QPU clock rate requirement to 30kHz, not far from the 2.5kHz record from existing IBM QPUs and their plans to reach 10kHz in the near future.

[2023] In Solving the Optimal Trading Trajectory Problem Using a Quantum Annealer, 2015 (13 pages).

[2024] See these slides in this D-Wave presentation. See also Applications of Quantum Annealing in Computational Finance, 2016 (29 slides) and the Quantum For Quants website they created.

[2025] See Credit Risk Analysis using Quantum Computers by Daniel J. Egger et al, 2019 (8 pages).

For example, some credit risks analysis and derivative pricing use cases mention a need for 7500 logical qubits (in Figure 678), which is a level above the 4098 minimum threshold for breaking a 2048 bits RSA key with Shor's algorithm[2026]. And it requires millions of physical qubits!

| | $(d, T)$ | | Error | | T-count | | T-depth | | # Logical Qubits | |
|---|---|---|---|---|---|---|---|---|---|---|
| Method | **Auto** | **TARF** | **Auto** | **TARF** | **Auto** | **TARF** | **Auto** | **TARF** | **Auto** | **TARF** |
| Riemann Sum | | | | | $\geq 10^{43}$ | $\geq 10^{18}$ | $\geq 10^{43}$ | $\geq 10^{18}$ | - | - |
| Riemann Sum (no-norm) | $(3, 20)$ | $(1, 26)$ | $2 \times 10^{-3}$ | | $1.4 \times 10^{11}$ | $5.5 \times 10^{10}$ | $1.9 \times 10^{8}$ | $1.7 \times 10^{8}$ | 24k | 15k |
| Re-parameterization | | | | | $4.2 \times 10^{9}$ | $3 \times 10^{9}$ | $4.6 \times 10^{7}$ | $6.2 \times 10^{7}$ | 7.5k | 9.5k |

TABLE I: Resources estimated in this work for pricing derivatives using different methods for a target error of $2 \times 10^{-3}$. We consider a basket autocallable (Auto) with 5 payment dates and a knock-in put option, and a TARF with one underlying and 26 payment dates. We find that Grover-Rudolph methods [10] are not applicable in practice (details in Appendix B) and that Riemann summation methods require normalization assumptions to avoid errors that grow exponentially in $T$. Even if those normalization issues were avoided, as detailed in the Riemann Sum (no-norm) row, the re-parameterization method still performs best. See Section IV A for a discussion of the Riemann summation normalization. The detailed resource estimation is discussed in Sections IV A 2 and IV B 3.

*Figure 678: Source: A threshold for quantum advantage in derivative pricing by Shouvanik Chakrabarti et al, Goldman Sachs, IBM and University of Maryland, May 2021 (41 pages).*

The Hull/Diamanti et al paper also reviews the broad topic of quantum money, and idea born circa 1969 and published in 1983 by **Stephen Wiesner**. The idea is to use quantum objects properties and the no-cloning theorem to avoid any counterfeiting and forging. Any bill has two unique identifying numbers: one classical serial number that is public and one secret random quantum number called a "random classical bill state". The central bank is the only one keeping the classical bill state. It's encoded using for example polarized photons on a 0° or 45° basis. Only the bank knows this sequence of encoding.

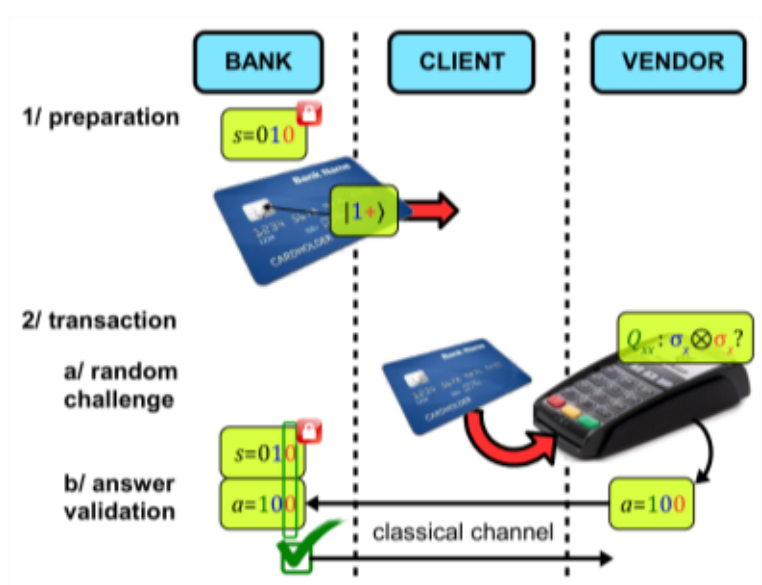

Figure 1. **Practical quantum money protocol.** The sequence of interactions between the credit card holder (client), the bank and the vendor involved in the transaction. In the preparation phase, the bank uses a secret key to prepare the quantum state loaded on the credit card, which is then given to the client. In the transaction phase, the vendor randomly selects one out of two challenge questions, measures the qubits and sends the outcome to the bank, who can then verify the validity of the credit card or detect a forgery attempt.

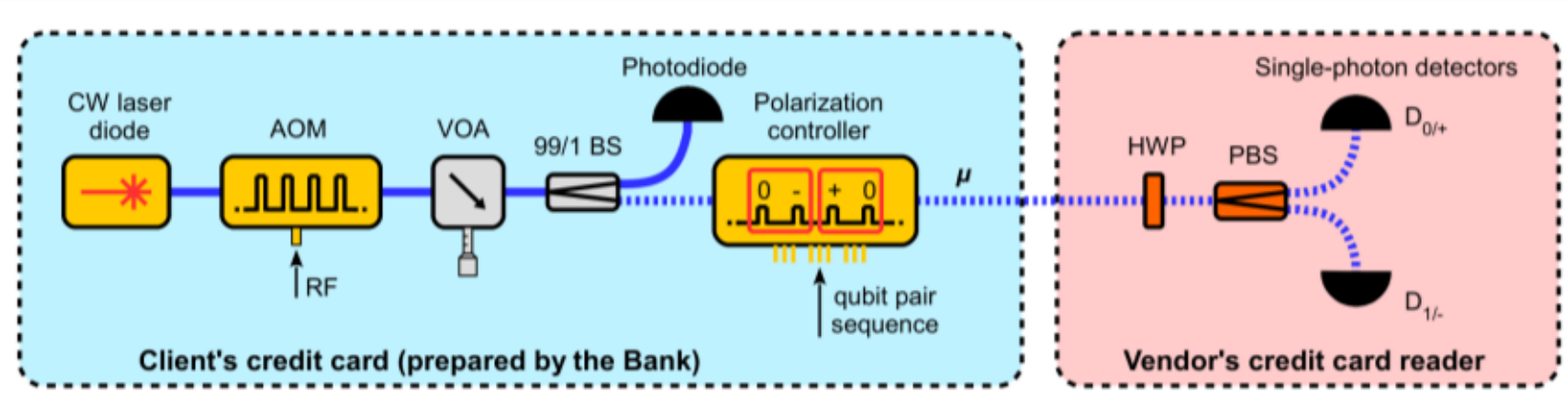

Figure 2. **Experimental setup of the quantum money system.** The credit card state preparation is performed using pulses carved from light emitted by a telecommunication wavelength laser diode using an acousto-optic modulator (AOM). A multi-stage polarization controller (EOSPACE) is then used to select the polarization states according to the protocol by applying suitable voltages. The average photon number of pulse $\mu$ is set by a variable optical attenuator (VOA) and is calibrated with a 99/1 beam splitter (BS) and a photodiode. The credit card reader is materialized by a standard polarization analysis setup including a half-wave plate (HWP), a polarization beam splitter (PBS) and two InGaAs single-photon avalanche photodiodes (ID201). The entire setup is synchronized using a multi-channel delay generator and is controlled by software incorporating the random state generation and data acquisition and processing.

*Figure 679: Source: Experimental investigation of practical unforgeable quantum money by Mathieu Bozzio, Iordanis Kerenidis, Eleni Diamanti et al, 2017 (10 pages).*

There are many variations of this concept of quantum money, with different degrees of anonymity and private and public schemes. Quantum money could be physical segmented into bills, coins and lightning schemes[2027]. An untraceable quantum coin proposal was made around 2010. But there are many shortcomings with these schemes which are just non implemented ideas at this stage.

[2026] See again A threshold for quantum advantage in derivative pricing by Shouvanik Chakrabarti et al, May 2021 (41 pages). 7.5K logical qubits and T-depth of 54 million are needed for pricing derivatives.

[2027] Quantum Lightning is a public key quantum money type.

One of these being that it requires quantum memory that doesn't exist yet, and which, by the way, is therefore not miniaturizable to be embedded in devices like a credit card[2028]. The Quantum Lightning variation prevents the bank to create multiple bills with the same serial number. You also have semi quantum money that requires no quantum communication infrastructure. All in all, it's quite difficult to assess the practicality of such quantum money ideas.

**BCG** published another review on quantum computing financial use cases including a roadmap against the estimated progress of quantum hardware in the next 5 to 20 years (in Figure 680)[2029]. For its part, McKinsey is more upbeat and is pushing banks to evaluate as fast as possible the potential use cases of quantum computing[2030]. D-Wave and IBM have been very active to push financial organizations to evaluate quantum computing benefits. So far, we have mainly proof of concepts and trials. Indeed, as the authors of another review paper published in 2020 point out: "*to date, quantum hardware is not advanced enough for solving any problem of practical relevance faster than classical computers*"[2031]. As a result, most banks are either using quantum inspired algorithms and small scale quantum annealing algorithms, or investigating highly demanding gate-based solutions that are positioned in the far-fetched future, in the 10-20 year timeframe at best.

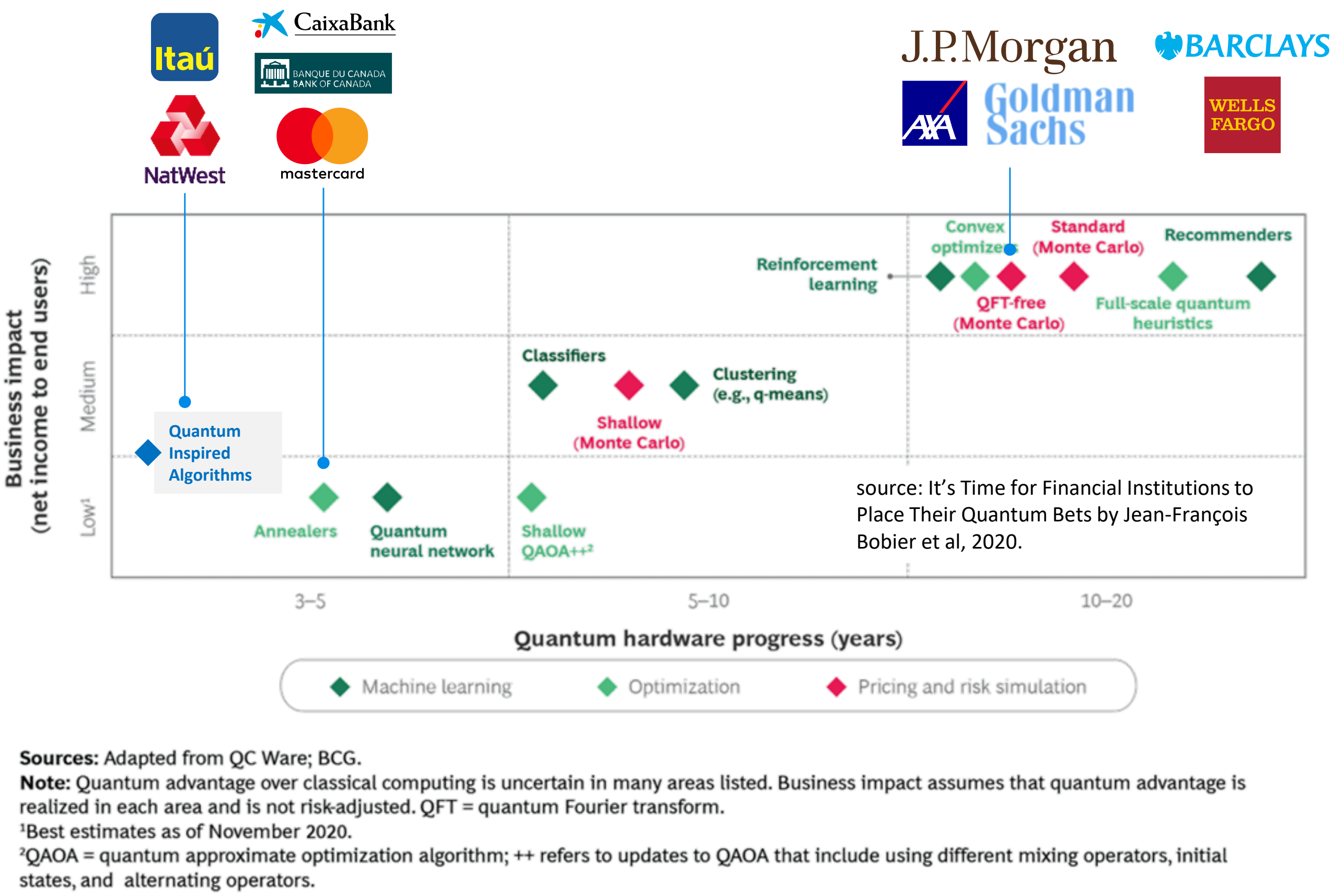

*Figure 680: Source: It's Time for Financial Institutions to Place Their Quantum Bets by Jean-François Bobier et al, 2020, and logos placed by Olivier Ezratty, 2022.*

[2028] See Experimental investigation of practical unforgeable quantum money by Mathieu Bozzio, Iordanis Kerenidis, Eleni Diamanti et al, 2017 (10 pages).

[2029] See It's Time for Financial Institutions to Place Their Quantum Bets by Jean-François Bobier et al, 2020.

[2030] See How quantum computing could change financial services by Miklos Dietz et al, McKinsey, December 2020.

[2031] See Quantum algorithms are coming to finance, slowly by Sarah Butcher, November 2020, mentioning the review paper Prospects and challenges of quantum finance by Adam Bouland, Iordanis Kerenidis et al, 2020 (49 pages). This paper documents the quantum speedups theoretically achievable with Monte Carlo simulation and portfolio quantum algorithms.

Since 2017, IBM has been highlighting partnerships with **JPMorgan Chase**[2032], **Barclays**[2033] and more recently with **HSBC**[2034] and **Wells Fargo**[2035] to study the uses of quantum in trading strategy optimization, investment portfolio optimization, pricing and risk analysis. Even if quantum algorithms that meet their business needs are conceivable, the capabilities of today's IBM quantum computers are insufficient to put anything into production.

In a recent review paper, IBM researchers describe some case studies of quantum computing in finance such as simulations (new customer identification, create new financial products, incorporate market volatility, improve customer retention), optimization and machine learning, options pricing, and quantum amplitude estimation with a quadratic speedup that can be used to estimate value at risk in an investment portfolio or in a credit[2036].

TABLE 5: Algorithms can improve computational efficiency, accuracy, and addressability for defined use case.

| | Quantum Algorithm | Description | Impact | Needs | Simulation | Optimization | ML |
|---|---|---|---|---|---|---|---|
| VQE | Variational Quantum Eigensolver | Use energy states to calculate the function of the variables to optimize | Procedure to assign compute-intensive functions to quantum and those of controls to classical | Qubit number increases significantly with problem size | | x | |
| QAOA | Quantum Approximate Optimization | Optimize combinatorial style problems to find solutions with complex constraints | Simplify analysis clauses for constraints and provide robust optimization in complex scenarios | Uncertain ability to expand to more optimization classes | | x | |
| AE | Quantum Amplitude Estimator | Create simulation scenarios by estimating an unknown property, Monte Carlo style | Handle random distributions directly, instead of only sampling, to solve dynamic problems quadratically speeding up simulations | High Quantum Volume required for good efficiency | x | x | x |
| QSVM | Quantum Support Vector Machines | Supervised machine learning for high dimensional problem sets | Map data to quantum-enhanced feature space to enable separation and better separate data points to achieve more accuracy | Runtime can be slowed by kernel computation and data structure | | | x |
| HHL | Harrow, Hassidim, and Lloyd | Estimate the resulting measurement of large linear systems | Solve high dimensional problems speeding up exponentially calculations | Hard to satisfy prerequisites and high measurement costs to recover solutions | | x | x |
| QSDP | Quantum Semidefinite Programming | Optimize a linear objective over a set of positive semidefinite matrices | Estimate quantum system states with less measurements to exponentially speedup in terms of dimension and constraints | High Quantum Volume required for good efficiency | | x | |

TABLE 6: Financial services focus areas and algorithms.

| Financial Services | Example Problems | Solution Approach | Quantum Algorithm |
|---|---|---|---|
| Asset Management | Option Pricing<br>Portfolio risk | Simulation | AE |
| Investment<br>Banking | Portfolio Optimization<br>Portfolio Diversification<br>Issuance: Auctions | Optimization | Combinatorial: VQE, QAOA<br>Continuous: QSDP<br>AE |
| Retail & Corporate Banking | Financial Forecasting<br>Credit Scoring (e.g. SME Banking)<br>Financial Crimes: Fraud + AML | Machine Learning | QSVM<br>HHL<br>AE |

*Figure 681: Source: Quantum Computing for Finance: State of the Art and Future Prospects by Daniel Egger et al, IBM Quantum, January 2021 (24 pages).*

**D-Wave** is at the origin with some of its customers such as **Deutsche Bank** of the creation of the Quantumforquants website, dedicated to the uses of quantum in finance. **Atos** also published a white paper on quantum applications in finance[2037] .

---

2032 J.P. Morgan recruited an IBM veteran in quantum computing, Marco Pistoia, who had contributed to the development of Qiskit Aqua. See JP Morgan Chase poaches an IBM 'Master Inventor' with 26 patents for quantum computing by Hugh Son, January 2020. This quantum activity is integrated in their "Quantitative Research Group". See also Option Pricing using Quantum Computers by Nikitas Stamatopoulos, Daniel J. Egger, Yue Sun, Christa Zoufal, Raban Iten, Ning Shen and Stefan Woerner, JPMorgan Chase, ETH Zurich and IBM, May 2019-July 2020 (20 pages).

2033 See Why banks like Barclays are testing quantum computing, de Penny Crossman, July 2018, Barclays demonstrates proof-of-concept quantum clearing algorithm by Cliff Saran, October 2019, Quantum Algorithms for Mixed Binary Optimization applied to Transaction Settlement by Lee Braine et al, October 2019 (8 pages) and Quantum Machine Learning in Finance: Time Series Forecasting by Dimitrios Emmanoulopoulos and Sofija Dimoska, Barclays, February 2022 (20 pages).

2034 See HSBC Working with IBM to Accelerate Quantum Computing Readiness, IBM, March 2022 and Unsupervised quantum machine learning for fraud detection by Oleksandr Kyriienko and Einar B. Magnusson, University of Exeter, HSBC and Canada Square, August 2022 (7 pages).

2035 See Wells Fargo prepares to take a quantum leap by Poornima Apte, CIO Magazine, June 2022.

2036 See Quantum Computing for Finance: State of the Art and Future Prospects by Daniel Egger et al, IBM Quantum, January 2021 (24 pages).

2037 See Quantum finance opportunities: security and computation, 2016 (20 pages). This is also the case of Everest Group with Quantum Computing in the Financial Services Industry-Infinite Possibilities or Extreme Chaos, 2018 (15 pages, $990... not really worth it).

In 2022, they announced an R&D partnership with **Mastercard** to create quantum-hybrid applications for optimizing consumer loyalty and rewards programs, cross-border settlement, and fraud management.

**NatWest** is experimenting quantum inspired algorithms running on traditional computers to optimize its investment portfolios (HQLA for High Quality Liquid Assets).

**Goldman Sachs** recruited Will Zeng, from Rigetti Computing, who had developed the Quil language. Will also works for the Unitary Fund which promotes open sourced quantum solutions, tools and benchmarks. With IBM, they work on different algorithms designs like the one on pricing derivatives already mentioned above. They are not yet operational but help define the hardware requirements to be able to run them. It's still at least in the mid-term.

**JP Morgan Chase** researchers created the NISQ-HHL approach, a hybrid version of the HHL algorithm that works on NISQ devices small-scale portfolio-optimization problems, using 6 and 14 assets from the S&P 500. It uses various techniques like mid-circuit measurement, Quantum Conditional Logic and qubit reset and reuse in the Quantum Phase Estimation routine used for eigenvalue estimation. The technique can reduce the number of ancillary qubits to just one and qubit connectivity requirements and (usually costly) SWAP gates. The test was implemented on a Quantinuum System Model H1 with 11 qubits. In other words, we are obviously far from getting any sort of quantum advantage in such a situation[2038].

**Microsoft** devised a way to make stock value predictions using topological computing, a farfetched idea given the state of the art of their Majorana fermion based qubits[2039].

**GE Research** with IonQ developed a risk management hybrid algorithm with 20 qubits from the IonQ Aria QPU[2040].

Also, noteworthy is the investment by the **Royal Bank of Scotland** (RBS) in 1Qbit along with Fujitsu and Allianz.

The **Bank of Canada** and **Multiverse Computing** created a prototype quantum simulation of the cryptocurrency market using a D-Wave quantum annealer. It simulated a financial network with 8-10 players[2041].

**Itaú Unibanco** (South America) and **QC Ware** (USA) undertook a four-month joint project in the customer retention domain with QML algorithms improving customer churn prediction models. The main outcome of the project is a new method to do this on classical computers and improved predictions by 2% and the model precision by 6,4% (from 71%). They used 180,000 data points[2042].

Of course, you must add to this quick review all the quantum software startups that are entering this market. They are either using quantum inspired algorithms or pilot projects using gate-based or annealing-based quantum computing. Among these are **1Qbit**, **Multiverse Computing**, **ApexQubit**, **JosQuantum** and **QuantFi**.

---

[2038] See NISQ-HHL: Portfolio Optimization for Near-Term Quantum Hardware by Romina Yalovetzky et al, January 2022 (14 pages). Presents a version of the HHL algorithms suitable for NISQ quantum computers.

[2039] As documented in Decoding Stock Market Behavior with the Topological Quantum Computer, 2014 (24 pages).

[2040] See IonQ and GE Research Demonstrate High Potential of Quantum Computing for Risk Aggregations, June 2022 and Copula-based Risk Aggregation with Trapped Ion Quantum Computers by Daiwei Zhu et al, June 2022 (10 pages). They say the results are better than full classical algorithms which always puzzles me given 20 qubits are very easily emulated on a classical computer.

[2041] See Bank of Canada and Multiverse Computing Complete Preliminary Quantum Simulation of Cryptocurrency Market by Multiverse, April 2022.

[2042] See QC Ware Applies Quantum Computing Principles to Increase Customer Retention at Itaú Unibanco by James Dargan, May 2022.

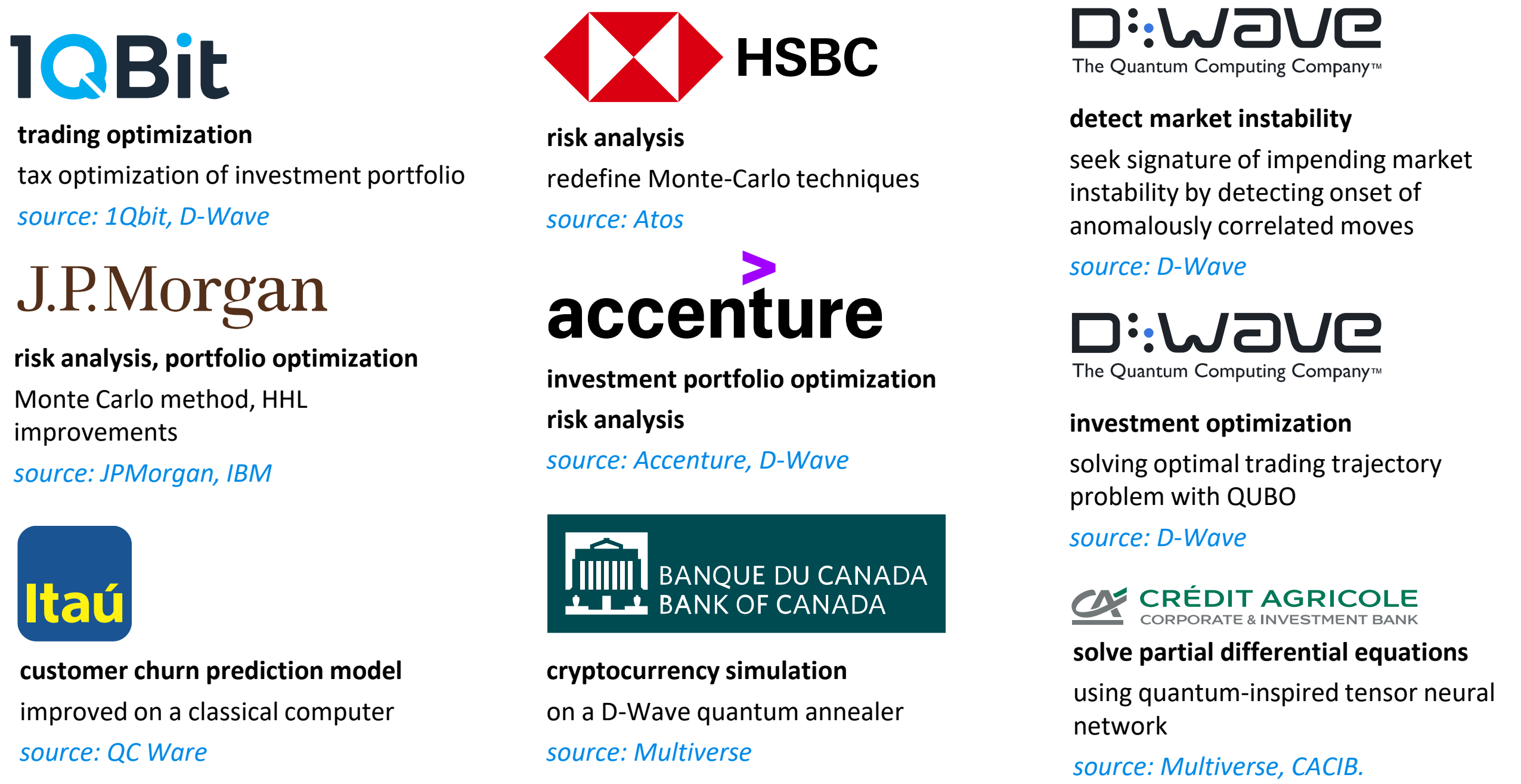

*Figure 682: a quantum computing use case sampler for financial services. (cc) Olivier Ezratty, 2022.*

## Insurance

The insurance market is a vertical that also must fix complex optimization problems, particularly related to risk modelling. The various related surveys and review papers I have found are not as rich as in the financial services vertical[2043].

Some analysts are using the usual Shor based codebreaking attacks cybersecurity red flag and explaining all the risks businesses may face in the future[2044]. The related reports are clearly misleading, stating for example that quantum-based communications could be "*quicker over long distances*" on top of being better secured[2045].

A 11 pages report was published late 2019 by **Novarica**, an US insurance consulting services company[2046]. Besides the usual generic description of the whereabouts of quantum computing, it contains only one and a half pages of insurance related use cases ideas. They are related to risk modelling and portfolio optimization. It also mentions quantum machine learning used to better detect and mitigate fraud, risks assessment with actuarial models for enhanced pricing and risk pooling precision, investments portfolio optimization and model life expectancy algorithms for large populations.

Another paper was published in 2022 by Michael Adam from **AXA Konzern AG** dealt with the potential applications of quantum computing in insurance. It is focused on the use case of valuation of insurance contracts based on quantum amplitude estimation, which would provide a quadratic speedup compared with classical Monte-Carlo based algorithms[2047]. However, it doesn't (yet) make an assessment of the quantum hardware resources needed to run this type of algorithm in production.

---

[2043] See The impacts of quantum computing on insurance - From theory to reality from Lloyds's, February 2021 (34 pages).

[2044] See Quantum computing a potential cyber risk for re/insurers by Charlie Wood, November 2019.

[2045] See Top 5 insurance quantum computing use cases by Danni Santana, January 2018. One speaker in Quantum Computing in Insurance - Interactive discussion, February 2021 (59 mn) estimates that the Shor threat can materialize in between 7 and 10 years, this being "conservative". Well that's kind optimistic.

[2046] See Quantum Computing to Affect Insurer Tech Strategies, December 2019.

[2047] See Potential Applications of Quantum Computing for the Insurance Industry by Michael Adam, AXA Konzern AG, October 2022 (43 pages).

One risk modelling algorithm created by **JoS QUANTUM** is indeed documented[2048]. Another use case was publicized by **Caixabank** with D-wave in 2022. They developed an investment portfolio hedging and portfolio optimization solution that generated a 90% decrease in time-to-solution when compared to their classical legacy system[2049].

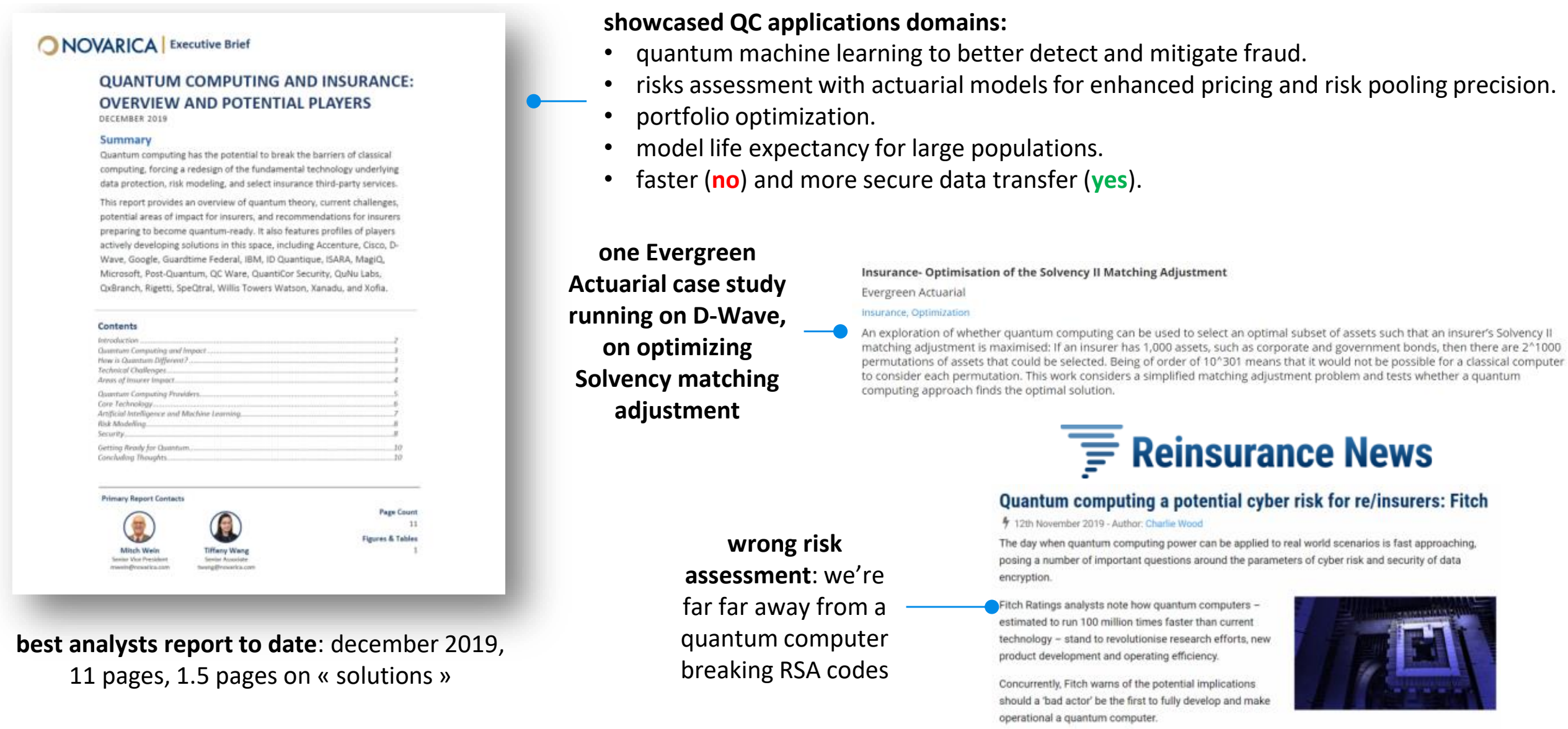

*Figure 683: some use cases and constraints for quantum computing in the insurance business. (cc) Olivier Ezratty, 2021.*

# Marketing

Marketing is also an area where optimization algorithms for complex systems based on quantum computers could be of interest. This concerns the optimization of the marketing mix, that of media plans, or the maximization of advertising revenues, various areas that are also invested by the AI field.

**Volkswagen** experimented a vehicle recommendation system in online sales sites, with a D-Wave.

Once again, predictive systems based on the exploitation of past data and simulation based on the knowledge of market operating rules are once again opposed to each other. However, these rules do not fall under the notion of AI expert systems, which manage logical predicates, but more complex causality models[2050].

# Content and media

Wondering how we could use quantum computing to create some content and art? That's the weird idea some have, at least with regards to music creation.

Computer have played a role in music creation for a while so why not quantum computers? Quantum mechanics is about waves, like music[2051]!

---

[2048] In A Quantum Algorithm for the Sensitivity Analysis of Business Risks by M. C. Braun et al, March 2021 (21 pages).

[2049] See CaixaBank Group, D-Wave Collaborate on Innovative New Quantum Applications for Finance Industry, March 2022.

[2050] See for example Display Advertising optimization by quantum annealing processor by Shinichi Takayanagi, Kotaro Tanahashi and Shu Tanaka of Waseda University and A quantum-inspired classical algorithm for recommendation systems by Ewin Tang, July 2018 (36 pages). The latter classical algorithm exceeds the performance of a quantum algorithm realized for D-Wave quantum computers.

[2051] See Quantum music Physics has long looked to harmony to explain the beauty of the Universe. But what if dissonance yields better insights? by Katie McCormick, May 2021.

The **Quantum Music** project[2052], was run by Volkmar Putz and Karl Svozil in Austria from 2015 to 2018. It led to the **QuTune** project[2053]. **Eduardo Miranda** from the Interdisciplinary Centre for Computer Music Research (ICCMR) at the University of Plymouth (UK) also works on using quantum computers to create music[2054]. He is currently preparing the release of the book "Quantum Computer Music", several of its chapters having already been published on arXiv[2055].

At this point, quantum music is about finding another source of randomness to create melodies (using quantum walks-based algorithms) and to generate credible synthetic voice. It led to the organization in November 2021 of a quantum music online event, organized, unsurprisingly, by the University of Plymouth (UK) with the sponsoring from IBM and Cambridge Quantum Computing[2056].

In 2019 **Quantum Sound** was the first music created and performed from measurements of superconducting qubits. It was not far from some form of random music generator. The project was run with the financial support of the Yale University Quantum Institute as well as from their top quantum scientists Michel Devoret and Robert Schoelkopf[2057].

At last, quantum music was generated in a more sophisticated way with a quantum annealer from D-Wave in 2021 by a very international team (India, Poland, Mexico, Estonia)[2058]. More recently, voice synthesis was studied using quantum algorithms[2059].

In 2022, a first “visual art” was created with the help of a quantum computer. The Quantum Prophet was created with the quantum artists trio Insig*ht* and **Kipu Quantum**. With the support of some AI, the author interfaced himself in real time with a quantum computer. Via motion capture, he manipulated qubits to modify aesthetically his animated 3D artwork which is sold with an NFT. All-in-all, the creative process was still largely in the hands of some humans!

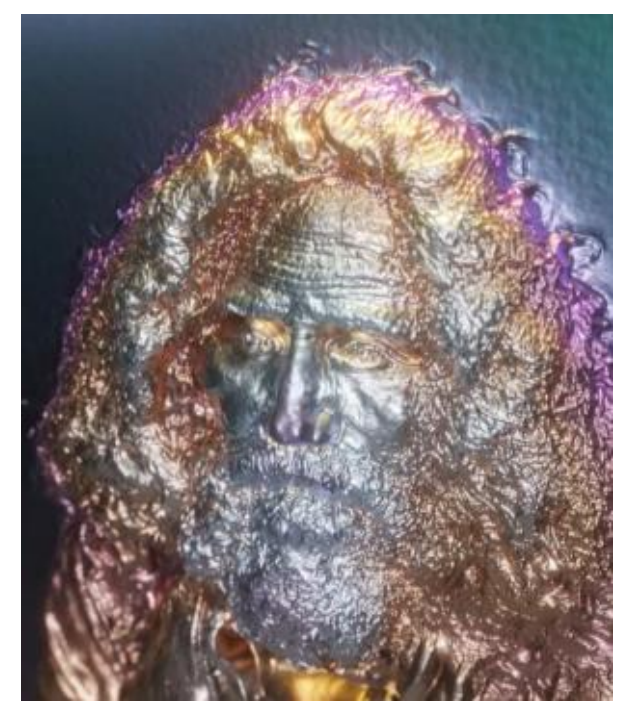

*Figure 684: The Quantum Prophet.*

A last content field explored are games. Proof of concepts of quantum computing based games were recently proposed for Mastermind[2060] and Go[2061] and of games using quantum principles[2062].

This is all early stuff. Don’t count yet to see quantum computers having an impact on classical video games and the Metaverse, its most recent incarnation.

---

[2052] See Quantum music by Volkmar Putz and Karl Svozil, 2015 (5 pages).

[2053] It was linked to QuTune Project Quantum Computer Music Resources, about making music with quantum computing, and making quantum computing with music. This project started in Spring 2021.

[2054] See Quantum Computer: Hello, Music! by Eduardo R. Miranda, June 2020 (32 pages), Creative Quantum Computing: Inverse FFT Sound Synthesis, Adaptive Sequencing and Musical Composition by Eduardo R. Miranda, 2021 (32 pages) and The Arrival of Quantum Computer Music by Eduardo R. Miranda, May 2020.

[2055] See Making Music Using Two Quantum Algorithms by Euan J. Allen, Jacob F. F. Bulmer and Simon D. Small, January 2022 (13 pages), New Directions in Quantum Music: concepts for a quantum keyboard and the sound of the Ising model by Giuseppe Clemente et al, April 2022 (14 pages), QuiKo: A Quantum Beat Generation Application by Scott Oshiro, April 2022 (23 pages) and A Quantum Natural Language Processing Approach to Musical Intelligence by Eduardo Reck Miranda et al, December 2021 (41 pages).

[2056] See 1st International Symposium on Quantum Computing and Musical Creativity.

[2057] See Superconducting qubits as musical synthesizers for live performance by Spencer Topel, Kyle Serniak, Luke Burkhart and Florian Carle, March 2022 (17 pages).

[2058] See Music Composition Using Quantum Annealing by Ashish Arya et al, January 2022 (29 pages).

[2059] See Teaching Qubits to Sing: Mission Impossible? by Eduardo Reck Miranda and Brian N. Siegelwax, July 2022 (31 pages).

[2060] See Winning Mastermind Overwhelmingly on Quantum Computers by Lvzhou Li et al, July 2022 (27 pages).

[2061] See Quantum Go: Designing a Proof-of-Concept on Quantum Computer by Shibashankar Sahu et al, June 2022 (7 pages).

[2062] See Defining Quantum Games by Laura Piispanen et al, May 2022 (19 pages) which deals with the games using some principles of quantum physics but not games handled by quantum computers.

Another form of quantum art without pretentions is to exhibit art to explain what quantum physics and technologies are about, like was done in Switzerland in 2020[2063].

## Defense and aerospace

The military-industrial complex has always been a big consumer of advanced IT. It is therefore not surprising that it is interested by quantum technologies. This is obviously the case in the USA but also in Europe, with Airbus being one of the first to take an interest in quantum applications, and also China and Russia to name a few others[2064].

The **US Air Force** has also identified various needs that can be covered by the four categories of quantum technologies with a special mention for quantum sensing in time measurement and navigation[2065]. They are also interested in quantum radars and, finally, in quantum computing applied to optimization problems.

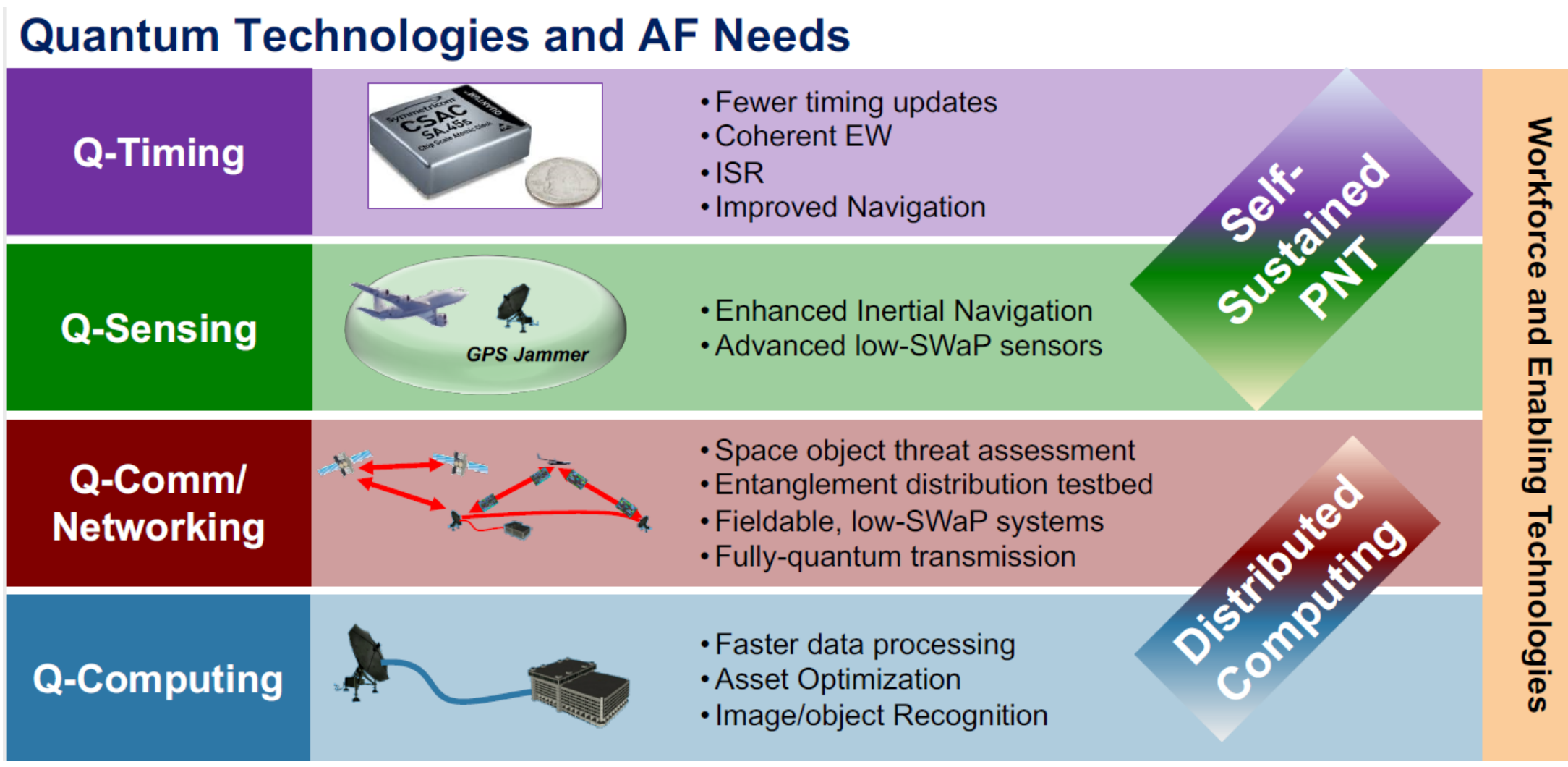

*Figure 685: quantum technology and US Air Force needs. Source: Quantum Information Science at AFRL by Michael Hayduk, December 2019 (21 slides).*

In France, the **DGA** has funded or co-funded since 2011 about twenty thesis on quantum and eight projects for €6.6M. The **Defense Innovation Agency** (reporting to the DGA) planned to launch a call for projects for quantum sensors in 2020 and in 2021 to fund a research project to support PQC in dedicated hardware. In July 2020, it became an ASTRID RFP on sensors, cryptography and quantum communications and on the creation of quantum computing algorithms[2066].

Quantum communications is also studies by the military around the world to secure communications[2067], particularly on the battlefield. **NATO** is experimenting it[2068]. Of course, **China** is working on it although with the visible part being only the civil use cases.

---

[2063] See "Do you speak quantum?" A quantum physics and art exhibition by Chiara Decaroli and Maciej Malinowski, June 2022.

[2064] See the review paper Quantum Technology for Military Applications by Michal Krelina, EPJ Technology, November 2021 (52 pages) that makes a pretty extensive inventory of defense use case for quantum sensing, communications and computing. On quantum radars, it showcase some adequate prudence. It still contains some misunderstanding like on page 27 on "*processing Big Data from surveillance and reconnaissance and identifying targets using quantum ML/AI*". Quantum computing is probably not bound to manage big data per se.

[2065] See Quantum Information Science at AFRL by Michael Hayduk, December 2019 (21 slides).

[2066] See Defense Research and Innovation: launch of a new ASTRID call for projects on quantum technologies, July 2020.

[2067] See Quantum Communication for Military Applications by Niels M. P. Neumann et al, TNO, November 2020 (11 pages).

[2068] See NATO cybersecurity center finishes tests of quantum-proof network by Jonathan Greig, March 2022.

Here are some various published case studies of quantum use in this vast sector.

It starts with **Lockheed Martin** partnering with **Google** and **NASA** to test D-Wave annealers staring in 2014. They developed with it a solution for formal proof of software operation. NASA co-founded the Quantum Artificial Intelligence Laboratory (QuAIL) with Google, operating a D-Wave Two. They test quantum optimization algorithms in different directions to optimize spacecraft filling, a variant of the bin-packing problem, on quantum versions of machine learning and deep learning algorithms, on problem decomposition and embedded computing[2069] .

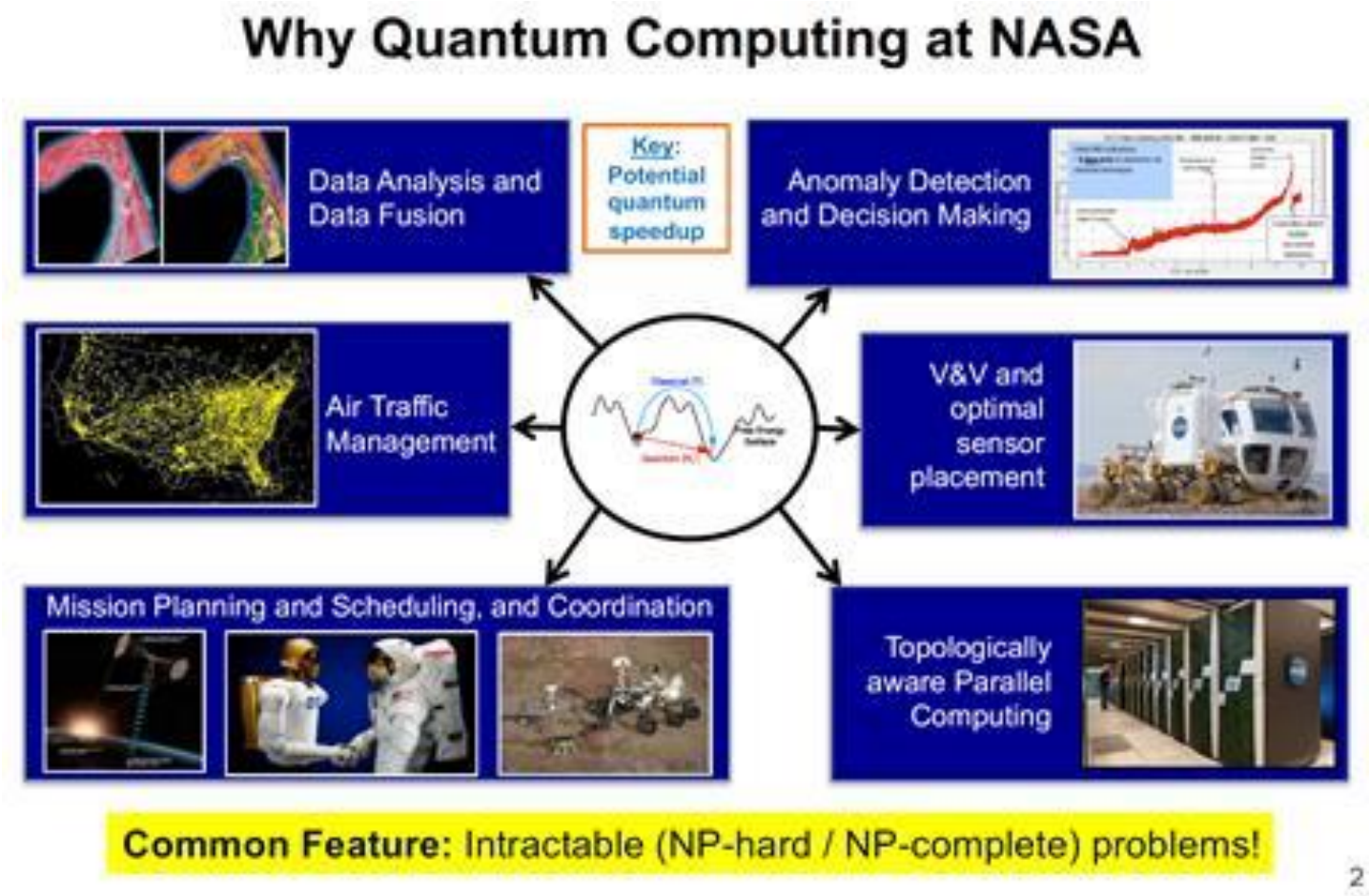

*Figure 686: Source: Quantum Computing at NASA: Current Status by Rupak Biswas, September 2017 (21 slides).*

In 2015, **Raytheon** and **IBM** demonstrated the efficiency of a quantum algorithm using a "black box" or "oracle" to reconstruct an unknown bit string, all running on an IBM 5 qubit general purpose quantum computer[2070]. This is obviously far from a real-world use case.

The **Airbus** group has created a team based at their Newport site in Wales, which is tackling the uses of quantum, particularly in the analysis of aerial imagery (not obvious... ) or for the design of new materials (more obvious). They also want to optimize the air flow on the wings, a problem that is nowadays dealt with by finite element simulation. They could try to optimize the air conditioning in airplanes, the biggest source of cabin noise, above the plane's engines!

In a different field, navies are interested in quantum sensing and particularly in micro-gravimetry measurement tools used to detect submarines.

**Thales** prototyped in 2021 a quantum annealing solution running on D-Wave Advantage (5000 qubits) to optimize radar configuration (the “Integrated Side Lobe Ratio” or ISLR NP-hard problem, which solution is “*the finding optimal sequences of phase shifts to minimize the mean squared cross-correlation sidelobes of a transmitted radar signal and a mismatched replica*”)[2071]. Yet, it didn’t reach any quantum computing advantage but the authors think future evolutions of D-Wave annealers may be promising for this respect.

In collaboration with **Fraunhofer IAIS**, Thales also prototyped a quantum image alignment solution for satellite images with quantum-based keypoints extraction and feature matching. It was tested with D-Wave quantum annealers and gate-based quantum computers.

The outcome is always the same: classical systems still deliver superior results but the proposed methods have the potential to outperform classical systems with future quantum computers[2072].

**Thales Alenia Space** is investing with CNES and DGA in quantum satellite telecommunications.

---

2069 This is well described in Quantum Computing at NASA: Current Status by Rupak Biswas, September 2017 (21 slides).

2070 This is documented in Demonstration of quantum advantage in machine learning (12 pages).

2071 See Phase-coded radar waveform AI-based augmented engineering and optimal design by Quantum Annealing by Timothé Presles, Cyrille Enderli, Rémi Bricout, Florence Aligne and Frédéric Barbaresco, Thales, 2021 (9 pages).

2072 See Towards Bundle Adjustment for Satellite Imaging via Quantum Machine Learning by Nico Piatkowski et al, April 2022 (8 pages).

The use of quantum technologies in the military field also gives rise to elucubrations that hybridize the plausible and the offbeat, such as those of the American political scientist **James Der Derian**, director of Project Q at the University of Sydney[2073].

Defense and quantum are also the playground of disinformation. For example, the battle is amplified between the USA and China, with many news reports touting an enormous China's quantum investment of $13B to $15B, which happens to be a misleading and exaggerated number[2074]. Very few sources, like a February 2022 report from **Rand Corporation** are pinpointing this and detailing China's real quantum investments which are probably either on par or slightly below similar investments in the USA and Europe, in the broad range of $2B to $4B across 10 years[2075]. The report estimated that only 1,700 students earned a PhD in quantum technology in China.

You also have weird plans like the UK military who seems interested to put quantum computers in tanks, which doesn't make much sense, at least for the next 20 years. The reason is they've been lured in this path by Orca which touts ambient temperature quantum computing[2076].

## Intelligence services

The world of intelligence and targeted eavesdropping is obviously on the lookout for the quantum. Shor's algorithm is the main application targeted by organizations managing electronic eavesdropping such as the **NSA** and all its colleagues. They are firefighters who are eager to decode information intercepted from various targets (embassy communications, economic intelligence, etc.) and to protect the sensitive communications of their own states against this type of decryption. They are therefore investing simultaneously in quantum computing (the "arsonist" dimension) and in quantum keys and post-quantum cryptography (the "firefighter" dimension).

On the other hand, these investments are not very public. The NSA has communicated well for almost ten years on the firefighter dimension but very little on the arson dimension.

They have surely acquired the various generations of D-Wave computers to get their hands on, in conjunction with **Lockheed Martin** which is one of their major suppliers. NSA also maintains a joint laboratory with NIST and the University of Maryland, **QuICS**, which will be launched in 2014.

One way to lift a veil on these activities is to detect laboratory and startup grants awarded by **IARPA**, the intelligence innovation agency led by the DNI (Director of National Intelligence), who oversees all American intelligence. It consolidates collaborative research funding for all intelligence agencies.

---

[2073] See Drones, radars, nuclear: how the quantum will change the war by Vic Castro, February 2020. Some remarks on this article: Rydberg atom-based qubits are only one of the types of qubits currently being studied. They are said to be "cold atom-based" and are moreover the specialty of Pasqal (France). There are many other types of qubits. The text also makes a big confusion in qubits and logical gates between qubits. These gates connect qubits together. They are often systems based on the diffusion of microwaves, photons emitted by lasers or magnetic couplers. Rydberg atoms are qubits and not qubit couplers.

[2074] See this example among others: The quantum tech arms race is on by Stuart Rollo, Asia Times, March 2022.

[2075] See An Assessment of the U.S. and Chinese Industrial Bases in Quantum Technology by Edward Parker, Rand Corporation, February 2022 (140 pages). Extracts on China: « *Even higher levels of announced investment are associated with the main Chinese quantum research facility, the Hefei National Laboratory for Physical Sciences at the Microscale (HFNL), which is led by Pan and a part of the University of Science and Technology of China in Hefei, Anhui Province. Chinese-language news media reported $1.06 billion in laboratory funding in 2017,6 and the Anhui Business Daily newspaper reported plans (though not confirmed funding) for $2.95 billion per year over the 2017–2022 period.* ***These announced spending goals are in stark conflict with the government-wide spending estimates*** *given in Table 4.11. The $1.06 billion start-up funding that Chinese news media announced in 2017 for Pan's quantum laboratory alone hugely exceeded Pan's own 2019 estimate for the PRC's total government spending over the same time period.* ***Our team's China experts assess that these conflicting reports of funding levels are not unusual in China; the PRC government often announces ambitious (and often highly politicized) spending goals, and it is not uncommon for these goals to go unmet****… […] In summary, official reports of the PRC's government investment in quantum R&D in recent years have varied widely, from a low of $84 million per year (Pan's estimate) to a high of at least $3 billion per year (the Anhui Business Daily's reported funding for Pan's laboratory). We are unable to assess from public information which figure is more accurate. By comparison, the U.S. government has spent $450–$710 million per year in recent years;* ***we cannot determine whether the PRC total is higher or lower than this amount****.”.

[2076] See UK Military Wants to Install Quantum Computers in Tanks for Some Reason by Lonnie Lee Hood, Future Society, June 2022.

It has already launched five programs around quantum technologies: in superconducting qubits (CSQ), logical qubits (LogiQ, with IBM), error correction (MQCO, also with IBM), the creation of development tools (QCS, with Raytheon and GeorgiaTech) and quantum annealing computation (QEO). But it is not clear that this has significantly advanced the state of the art.

Other Western intelligence services may also have acquired D-Wave, notably the British CGHQ. The NSA is also in contact with IBM and Google to explore the path of superconducting quantum general-purpose computers.

## Industry

Industry in the broadest sense of the term is another outlet for quantum computing. As soon as there is a complex optimization problem for scheduling, logistics or complex system design support, quantum will have its say.

The Japanese **JSR Corporation** is one of the companies working with IBM in the quantum field, mainly for the creation of new materials. **LG Electronics** announced a similar partnership with IBM in January 2022 to "*support big data, artificial intelligence, connected cars, digital transformation, IoT, and robotics applications*", you name it all.

It seems that quantum computing could be used within computer-aided design tools[2077]. But the document cited in the note comes back to the basics of quantum computing without being very elaborated on quantum computing uses in CAD, a very common phenomenon when quantum computing is pushed in various industries.

The routing of electronic circuits is also a NP-complete problem that could be partly handled by quantum algorithms, provided they have a sufficient number of qubits. This could be useful for designers of ASIC-type circuits and especially FPGAs, these circuits whose operating logic is dynamically programmable via two key parameters: the decision tables of the processing units and the links between these units.

## Science

Fundamental research is starting to test and use quantum computing, particularly in materials development and particle physics research[2078].

After having investigated quantum computing for a good number of years with a first workshop organized in November 2018, **CERN** launched a formal **Quantum Technology Initiative** (QTI) in September 2020[2079]. They want to use quantum computing to analyze the noisy data coming from their ultrasensitive particles detectors and to simulate the behavior of many-body quantum phenomena.

[2077] According to Computer-Aided Design for Quantum Computation by Robert Wille, Austin Fowler and Yehuda Naveh (Google and IBM), 2018 (6 pages).

[2078] See Applying quantum computing to a particle process by Glenn Roberts Jr., Lawrence Berkeley National Laboratory, February 2021, referring to Quantum Algorithm for High Energy Physics Simulations by Benjamin Nachman et al, February 2021 (6 pages). The algorithm used to detect particles using the 20 qubits IBM Q Johannesburg quantum system in the cloud is not providing any quantum advantage but would be promising with a larger number of qubits. See also the review paper Quantum Simulation for High Energy Physics by Christian W. Bauer et al, April 2022 (103 pages) and Simulating Collider Physics on Quantum Computers Using Effective Field Theories by Christian W. Baue et al, November 2021 (7 pages), Quantum computing hardware for HEP algorithms and sensing by M. Sohaib Alam et al, April 2022 (23 pages), Quantum Computing for Data Analysis in High-Energy Physics by Andrea Delgado et al, March 2022 (22 pages) and Snowmass White Paper: Quantum Computing Systems and Software for High-energy Physics Research by Travis S. Humble, March 2022 (17 pages).

[2079] It was later formalized in their CERN Quantum Technology Initiative Strategy and Roadmap by Di Meglio et al, October 2021 (46 pages).

CERN also participates to international quantum computing education and training with its online training resources. IBM worked with CERN to select/classify LHC events using QSVM[2080].

Back in 2017, **Caltech** used a D-Wave Two X quantum annealer with 1098 qubits to "rediscover" the Higgs boson using CERN LHC data and a QAML algorithm (quantum annealing machine learning)[2081]. Later in 2020, they improved it with their QAML-Z algorithm, quantum annealing machine learning model zooming in on a region of the analyzed energy surface[2082].

In astrophysics, superconducting qubits are also used to detect dark matter, in the form of axions, a dark matter candidate and hidden photons, that would interact with the photons[2083]. Other researchers are also using squeezed states to detect axions[2084]. Quantum computing is also tested to simulate exotic magnetic materials[2085].

Two scientific areas drive the attention with quantum computing, **weather forecast** and how to mitigate **climate change**. It's so politically trendy that many researchers and vendors are focusing their messaging on these application domains. They obviously tend to make significant overpromises since most of the related applications will require large-scale fault-tolerant quantum computers to be viable, like for the quantum simulations related to the improvements of the Haber-Bosch process or for designing new more efficient batteries. Overpromises on fixing climate change come for example from **McKinsey**[2086] or **PsiQuantum** who has no functional quantum computer yet[2087] and on weather modelling from **Pasqal** and **BASF**[2088]. **Quantinuum** is a little more credible with working on low-emission refrigerant production which deals with more simple problems although their existing 20 functional qubits are of no real use today[2089].

---

2080 See Application of quantum machine learning using the quantum kernel algorithm on high energy physics analysis at the LHC by Sau Lan Wu et al, Physical Review Research, 2021 (9 pages).

2081 See Solving a Higgs optimization problem with quantum annealing for machine learning by Alex Mott et al, 2017 (5 pages).

2082 See Quantum adiabatic machine learning with zooming by Alexander Zlokapa et al, Caltech, October 2020 (9 pages).

2083 See Searching for Dark Matter with a Superconducting Qubit by Akash V. Dixit et al, April 2021 (7 pages).

2084 See A quantum enhanced search for dark matter axions by K. M. Backes et al, 2021 (8 pages).

2085 See Quantum computing enables simulations to unravel mysteries of magnetic materials by Elizabeth Rosenthal, Oak Ridge National Laboratory, February 2021, using a 2000Q D-Wave annealer.

2086 See Quantum computing just might save the planet, McKinsey, May 2022.

2087 See PsiQuantum Announces Qlimate Initiative Developing Breakthrough Climate Technologies Enabled by Quantum Computing, May 2022.

2088 See Pasqal, BASF to collaborate on quantum compute-powered weather modeling, July 2022. This works seems serious from the pure mathematic standpoint, with modelizing weather models using differential equations and quantum neural network models labelled PINNs (physics-informed neural networks). What these stories don't tell is the size of the models that would be required in real use cases. Most of the time, these sizes are way beyond the capacities of the related quantum computers, even when taking into account their 5-year roadmap. On top of that, one unaddressed issue is how these quantum simulators are fed with training data. The larger the data set, the slower it is, and it requires a lot of classical pre-processing.

2089 See How Quantum Computing Can Help Keep Things Cool, 2022.

# Software and tools vendors

There are already many quantum software and development tools startups, particularly with regards to what suitable hardware is available. Initially, many of them were developing software running on D-Wave annealers. Then, as gate-based vendors like IBM started to put their hardware in the cloud, most software vendors adopted it. Many software vendors adopt hybrid approaches that combine business knowledge, associated algorithms and their execution on classical machines and quantum computers, hybrid classical-quantum algorithms, or so-called "quantum inspired" algorithms that run on classical computers. Only a few software vendors have adopted the quantum simulation paradigm which is a pity given these systems may be the most viable in the mid-term.

Even though quantum software won't solve all business and technical problems, it's time for legacy software vendors to give a look at the value it could provide[2090].

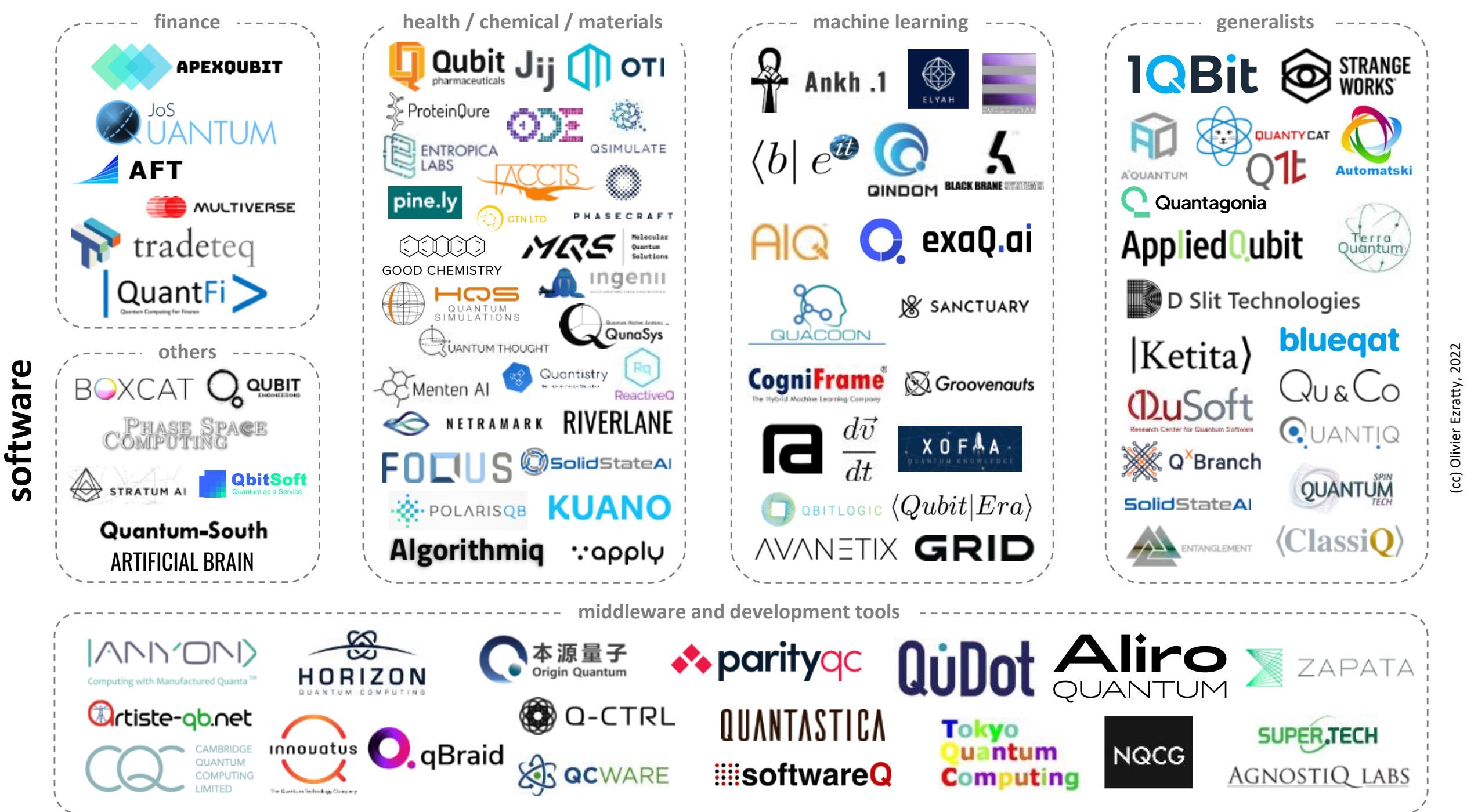

*Figure 687: a nice logo map of the quantum software industry. (cc) Olivier Ezratty, 2022.*

These approaches are essential for survival. Indeed, a startup cannot be exclusively dedicated to quantum computing at the risk of only being able to sell proofs of concepts on a very small scale that cannot generally be deployed industrially[2091].

There are real opportunities to position yourself in this emerging market! You will notice that this inventory does not include any Chinese startup. This is probably not by chance. This ecosystem is therefore still very young. It will evolve in parallel with the development of commercial quantum computers. You see already the startup scene maturing with many vendors adopting platforms approaches and developing partnerships models all over the place[2092]. China is not very well versed in software compared to hardware and seems to have put the quantum priority on cybersecurity more than on quantum computing.

[2090] See Why quantum software will be eating the world by Yuval Boger, June 2022.

[2091] This principle of reality is well described in The hard sell of quantum software by Jon Cartwright, 2019.

[2092] See a couple examples in Collaboration is Dominating Quantum Computing by Russ Fein, The quantum Leap, April 2022.

**1QBit** (2012, Canada, $35M) is a multi-sector quantum software company. It was funded among others by Fujitsu, as well as by Accenture and Allianz.

They have developed various low-level quantum algorithmic components that are hardware neutral. This includes, for example, the graphs processing that they apply in several markets, via their consulting activity. They cover financial markets, for the dynamic optimization of investment portfolios or to simplify the allocation of asset classes in a portfolio[2093].

They also developed QEMIST, a library for accelerating innovation in materials science and drug discovery. In addition to being a long-standing partner of D-Wave, they also work with IBM. The startup already has above 100 employees. Their customers include Dow Chemical (chemicals), Biogen (biotechs) and Allianz. In April 2020, they launched the "Quantum Insights Network", a network of around 100 experts and content in quantum computing.

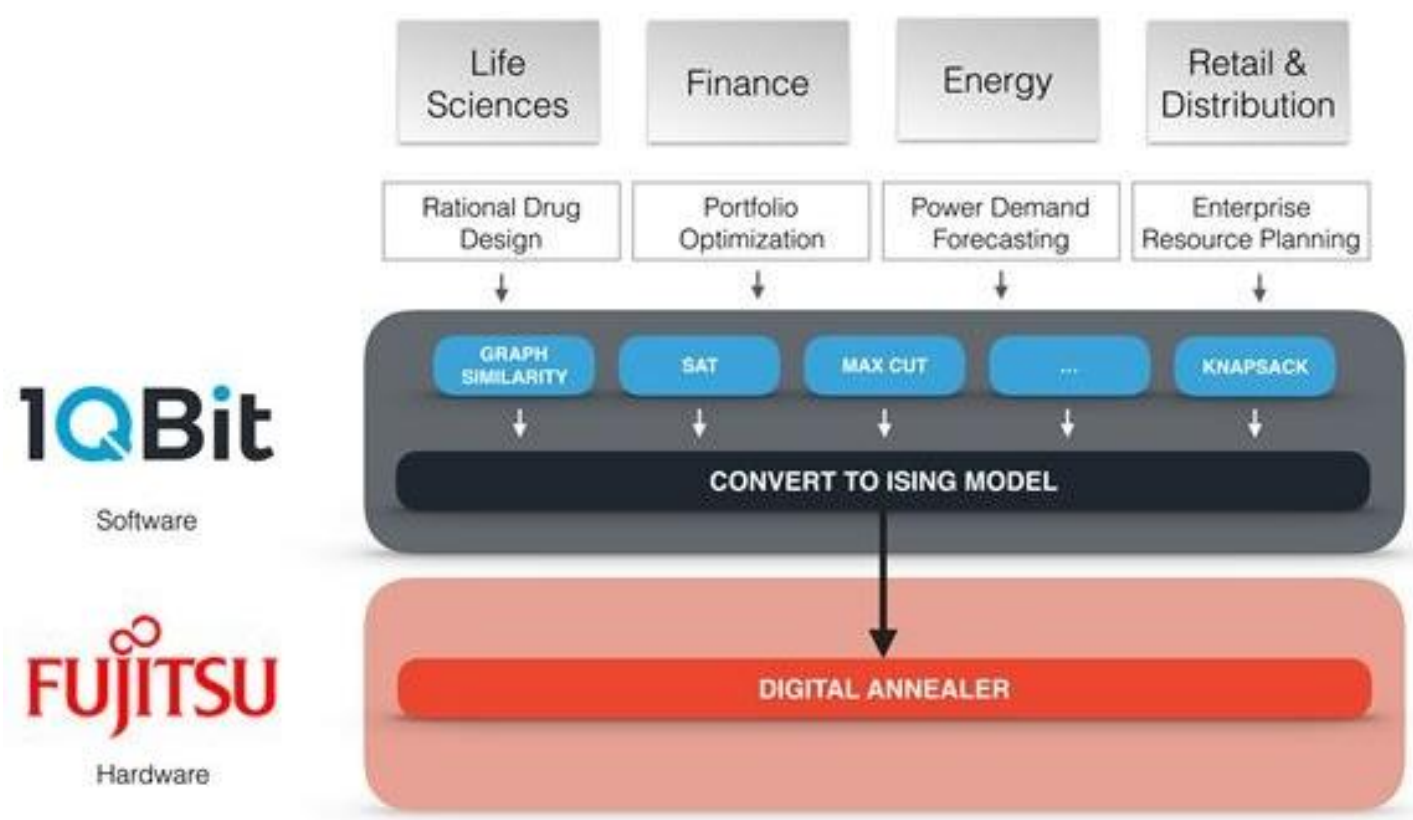

*Figure 688: 1QBit Software running on Fujitsu Hardware. Source: Fujitsu.*

**Adaptive Finance Technologies** (2020, Canada) came out of the Creative Destruction Lab. It was created by Roman Lutsiv, Vlad Anisimov and Edward Tang and develops investment and credit risk management software for the finance industry.

They used classical machine learning methods and are prototyping quantum machine learning solution running on D-Wave annealers.

**Algorithmiq** (2020, Finland) is a spin-off from the University of Turku which develop quantum software for life science and data science. They also created an online quantum science and technologies learning. Their CEO is Sabrina Maniscalco**.**

**AIQTECH Inc (**2018, Canada**)** is a machine learning specialist that explores the uses of QML. They are partners of the IBM Quantum Network. It is a two-person shop.

**Aliro Quantum** (2018, USA, $2.7M) is a startup that came out of the blue in September 2019. It started with developing software tools telling developers whether cloud quantum computing resources are available to perform calculations faster than on traditional processors, especially on GPUs.

They are partnering with HQAN (Hybrid Quantum Architectures and Networks, funded by the NSF, and belonging to the University of Illinois) to develop the seeds of a distributed quantum computing network in the USA. The startup founded by Prineha Narang (CTO, coming from Harvard) and Jim Ricotta (CEO) is positioned as a quantum Internet service company providing EaaS services, or entanglement as a service. In October 2022, they released AliroNet, a software solution providing emulation services of entanglement-based quantum networks, which can help design small scale pilots and universal entanglement-based quantum networks, including quantum networks connecting quantum sensors.

---

[2093] See Solving the Optimal Trading Trajectory Problem Using a Quantum Annealer, 2015 (13 pages).

**Apply Science** (2019, Italy) is an applied mathematics services company who is experimenting quantum computing in the IBM Quantum Network. They are testing QML algorithms for virtual drug development.

**ApexQubit** (2018, Belarus and USA) is a drugs discovery company that develops quantum software solutions for the pharmaceutical sector targeting rare diseases. They operate in project mode and publish some research papers on their web site.

**Aqemia** (2019, France, 31.6M€) is a software company developing drug discovery and retargeting algorithms using statistical physics, AI and quantum inspired algorithms. The startup is a spin-off from Ecole Normale Supérieure run by Maximilien Levesque and Emmanuelle Martiano. In December 2020, they announced a partnership with Sanofi to discover new treatments for covid-19. They raised 30M€ in October 2022 to fund their AI-enabled drug discovery pipeline, but not part of their future quantum-based algorithms efforts.

**aQuantum** (2018, Spain) is quantum software engineering service company doing contract research, development and consulting. They develop hybrid classical-quantum computing software and provide quantum software project management expertise, particularly in quantum machine learning.

They also developed |QuantumPath⟩ (*aka* Q|Path⟩), a quantum software development and lifecycle application platform. It contains all the tools to handle the whole software design and execution lifecycle covering both gate-based, quantum simulators and quantum annealing based computing, supporting Qiskit (IBM), Forest (Rigetti), Ocean (D Wave), ProjectQ and Quantinuum's |tKet⟩. They embed the open sourced Quirk graphical tool in their development environment. Since April 2022, QuantumPath is integrated with AWS and Amazon Braket making it possible for developers to access the quantum emulation and quantum hardware platforms supported by Amazon Braket.

**A*Quantum** (2018, Japan, $3M) specializes in the development of quantum software solutions for both annealers (including digital annealers from Fujitsu) and gate-based quantum computers (from IBM). Their ambition is to create high-level software libraries for users.

**Ankh.1** (2018, USA) has developed Anubis Cloud, a virtual machine in the cloud for data scientists that integrates with the open source solution Jupyter as well as with the TensorFlow and Keras learning machine frameworks.

**AppliedQubit** (2019, UK) presented itself as a publisher of quantum software for businesses.

In particular, they targeted the two main markets: finance and chemical simulation, in addition to generic optimization problems and predictive analysis.

They were developing both classical/quantum hybrid computing and quantum machine learning solutions. The company stopped operating in March 2021.

**Arclight Quantum** (2020, China, $3M) is a quantum software company spun out of the Institute of Software Research of the Chinese Academy of Sciences. With CAS, they developed isQ-Core, a quantum cloud software development platform that can execute gate-based quantum code on both classical emulators and quantum processors. They also created an EDA (electronic design automation) tool for the creation of quantum processors.

**Arline** (2020, Germany) is developing a compiler optimizing QML algorithms execution, reducing the number of quantum gates used and taking into account all qubits characteristics such as their connectivity.

They also propose Arline Benchmarks, an automated benchmarking platform for quantum compilers. It compares gate count, circuit depth and compiler runtime.

It can also be used to combine compiled circuits and optimization routines coming from different compilers in a custom pipeline to optimize algorithms performance.

**Artificial Brain** (2022, India/USA) is developing hybrid quantum-classical algorithms targeting energy, aviation, finance and climate change use cases.

Its first achievement was an algorithm identifying optimal locations for electric vehicle charging. It is running on D-Wave quantum annealers (probably from the Pegasus generation) using a mix of quantum annealing and genetic algorithm developed on QBSolv[2094]. It produces some result in 3 seconds for $8.5*10^{15}$ combinations. Its scalability depends on how D-Wave will expand the capacity of its annealers in the future. The company was founded by Jitesh Lalwani, who worked beforehand in the software industry.

**Artiste-qb.net** (2018, Canada) has a business model similar to that of 1Qbit: they develop algorithmic bricks of intermediate levels that they then assemble according to the needs of their customers.

They have even filed patents for certain methods. The startup was created by an international team including German researchers. They develop a Python based set of libraries in open source, available on Github.

**Automatski** (2014, USA) is a software company established in London, India in Bangalore and in California. They do applied contract research to develop quantum algorithms on any form of computer and quantum simulator. They have developed a software solution for emulating a large, unspecified number of qubits on conventional computers. They focus on creating biochemistry algorithms and claim to have solved protein folding and to cure diabetes, cancers and 4000 other diseases. Seems like some sort of overselling.

**Avanetix** (2019, Germany) develops hybrid algorithms dedicated to solving supply chain problems. They combine classical optimization methods, machine learning and quantum computing. They target the automotive and logistics markets. The startup is founded and managed by serial entrepreneur Naimah Schütter.

**Beit.tech** (2016, Poland, $1.4M) is specialized in quantum machine learning. It is mainly a research project funded by the European Union, covering the period 2017-2010. The founder Wojtek Burkot is a former Google employee who even tries to make D-Wave useless by creating algorithms for optimizing complex graphs that can run on traditional computers.

[2094] See Towards an Optimal Hybrid Algorithm for EV Charging Stations Placement using Quantum Annealing and Genetic Algorithms by Aman Chandra, Jitesh Lalwani Babita Jajodia, November 2021 (6 pages).

**Blueqat** (2008, Japan, $2.3M), formerly MDR for Machine learning and Dynamics Research, is creating algorithms integrating AI and chemistry, working among other with customers from the cosmetics industry like KOSÉ[2095].

They are working with D-Wave annealers. The startup was founded by Yuichiro Minato and various other alumni of the University of Tokyo.

**Boltz.ai** (2020, Canada) is specialized in the development of AI and quantum software for the agriculture business. They create crop field allocation optimization tools.

**BosonQ Psi** (2020, India) created the cloud-based BQPhy software suite, a computer-aided engineering (CAE) solution performing complex simulations with using hybrid quantum-classical algorithms. It covers structural mechanics, thermal analysis and design optimization. The startup was created by Abhishek Chopra, Rut Lineswala and Jash Minocha.

**Boxcat** (2017, Canada) is a startup created by Ystallonne Alves that develops image and video processing solutions based on hybrid quantum algorithms. They target the media and medical imaging markets. Their algorithms are based on currently available hardware architectures (D-Wave, IBM, Rigetti, Xanadu). The process they present on their site is an image realized on a D-Wave, which could have been realized with Nvidia's latest GPUs.

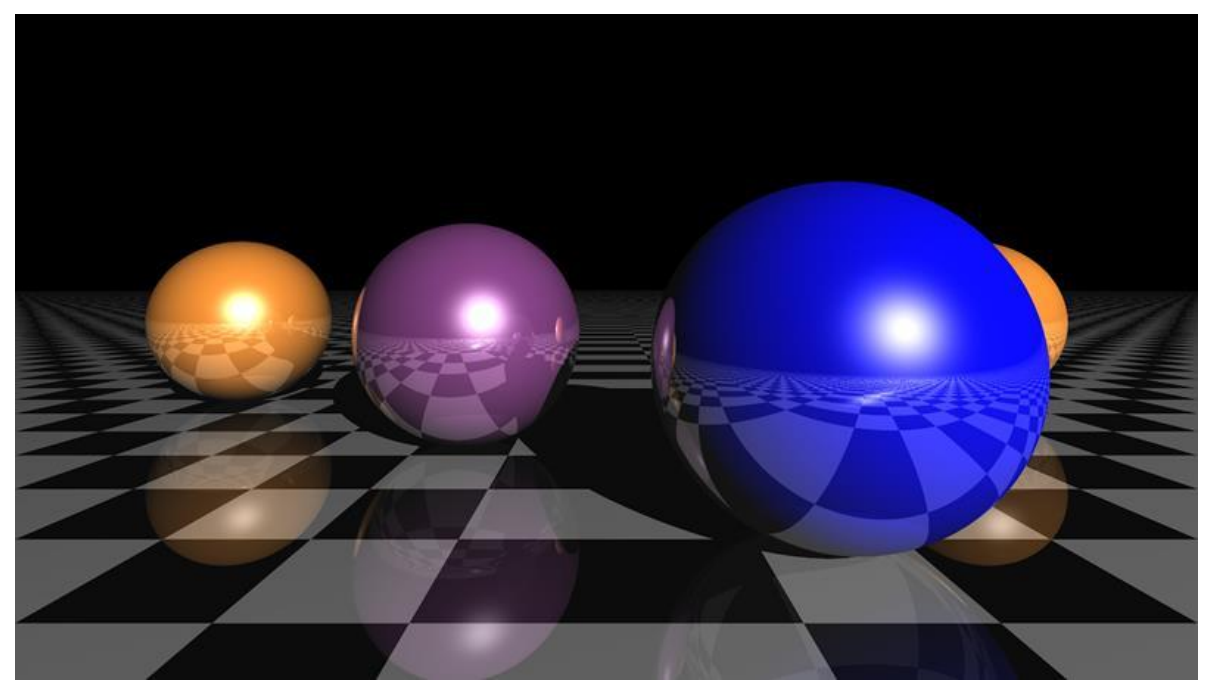

*Figure 689: an artificial image generated on a D-Wave 2000Q by Boxcat. Source: Boxcat.*

**Cambridge Quantum Computing Limited** (2015, UK, $72.8M) develops the $t|ket\rangle$ quantum operating system and various quantum algorithms including Arrow for machine learning[2096]. They are partnering with Oxford Quantum Circuits and with IBM which is one of their investors. CQC is also active in post-quantum cryptography.

$t|ket\rangle$ is available broadly and for free to everyone since February 2021 and also open sourced. It covers many quantum computing platforms (IonQ, Honeywell, AQT, IBM Qiskit, Rigetti, Amazon Braket and Azure Quantum) and incorporates circuit optimization and routing. It's interfaced with Python with Pytket. $t|ket\rangle$ is also used by EUMEN, CQC's quantum computational chemistry platform, and the company's QML framework. QQC is also partnering with Roche to use quantum algorithms for drug discovery targeting Alzheimer's Disease, as announced in January 2021.

In 2021, CQC launched a cloud software random quantum number generator. It is using a classical random generator, a quantum random number generator amplifying the randomness of the first and a Bell test used to check the resulting randomness, all running on an IBM quantum system[2097].

---

2095 Blueqat developed an algorithm that analyzes the distribution of cosmetics product features in a multidimensional space. It visualizes existing areas and reveals unknown product areas they were able to open up, to create possibilities for new cosmetic designs that humans never thought of. The solution was patented and thus, is not yet publicly documented.

2096 See t|ket> : A Retargetable Compiler for NISQ Devices, April 2020 (43 pages).

2097 See Quantum-Proof Cryptography with IronBridge, TKET and Amazon Braket by Duncan Jones, March 2021 and Practical randomness and privacy amplification by Cameron Foreman et al, 2020 (26 pages). It was branded as (origin)[cQ] in December 2021.

CQC has also been demonstrating how NLP (Natural Language Processing) could be implemented on current NISQ IBM quantum computers. Their researchers explain that the structure of natural language is natively quantum, which could lead to efficient translating, and better understanding of complete sentences and texts. The paper however doesn't provide much indication on the way data was actually encoded in the qubits. It lacks supporting data on actual system performance[2098]. In October 2021, this was packaged in an open source Quantum Natural Language Processing (QNLP) toolkit and library, lambeq.

The company announced a merger with Honeywell Quantum Systems in 2021. In December 2021, it became Quantinuum with a staff of over 350 employees.

**ChemAlive** (2014, Switzerland) is a quantum computational chemistry startup and contract research company. It provides simulation tools for getting molecular properties and synthetic reactions using basic 2D chemical syntax.

They deal with reaction mechanism elucidation and optimization, kinetic observed rate modeling, molecular design, virtual screening and drug discovery, molecular and spectroscopic property prediction, materials modeling and design, data and computation driven synthetic planning, experimental execution of chemical synthesis and experimental research on spectroscopic and electronic molecular properties. They developed ConstruQt, a software tool transforming molecular drawings into 3D structures and energies.

All of this is quantum... but seems to be computed classically. Quantum chemistry doesn't necessarily mean quantum-computed quantum chemistry. Once quantum computing hardware will scale, they'll naturally switch to it.

**ClassiQ Technologies** (2020, Israel, $51.8M) develops a quantum programming tool providing a higher level of abstraction than classical quantum gate programming[2099]. Their platform is on Amazon Bracket as announced in June 2022.

The company was created by Nir Minerbi (CEO), Amir Naveh (VP-R&D) and Yehuda Naveh (CTO, who spent 20 years at IBM Research in Haifa, including quantum, condensed matter specialist). Their CMO, Yuval Boger, also runs the Qubit Guy's podcast series

**CogniFrame** (2016, Canada) is a software publisher of data analysis platform software exploiting machine learning. They also develop hybrid algorithms for the financial sector based on D-Wave annealers.

One of their first customers is the Canadian investment bank Alterna Savings. The proposed applications are classic in the financial field: credit risk assessment and investment portfolio optimization. Late 2021, they launched **FirstQ Store**, an aggregation platform of quantum computing applications, provided as an application running on Windows, Mac and Linux desktops. It supports QUBO applications for Toshiba's Simulated Bifurcation Machine (a sort of digital annealer) and gate-based algorithms[2100].

**ColibrITD** (2019, France) is a quantum software R&D company created by Hacène Goudjil and Laurent Guiraud that develops vertical use case software components with a team of a dozen skilled PhDs/post-docs, focused on gate-based computing models.

---

[2098] See Foundations for Near-Term Quantum Natural Language Processing by Bob Coecke et al, December 2020 (43 pages). By the way, they are using ZX calculus in this work.

[2099] See A revolutionary approach to building quantum circuits by Amir Naveh and Yuval Boger, Classiq, November 2021 (32 minutes).

[2100] See Benchmark of quantum-inspired heuristic solvers for quadratic unconstrained binary optimization by Hiroki Oshiyama and Masayuki Ohzeki, December 2021 (11 pages).

They created a set of framework tools branded "QUICK" for "Quantum Innovative Computing Kit" for solving problems in the financial, pharmaceutical, material design and logistic fields. They also contribute to the advancement of quantum software engineering tools like in functional testing[2101].

**CreativeQuantum** (2010, Germany) is specialized in quantum physics-based R&D in the chemical and pharmaceutical industries.

They seem however to run these many algorithms with classical computers. Which makes sense given quantum computers are not yet powerful enough to run these physics simulation tasks efficiently.

**Culgi** (2004, Netherlands) is yet another computational chemistry company that will someday adopt quantum computing or simulation for its software. It was founded in 1999, changed its name in 2004 and was acquired by Siemens in 2020.

**Dirac** (2021, USA) is developing quantum software and algorithms for robotic applications. That New York City startup was created by Filip Aronshtein. I'd advise them to revisit their branding since the search engine optimization of Dirac is not obvious. On top of these nasty white logos on dark backgrounds!

**dividiti** (2014, UK) develops quantum algorithms, particularly in machine learning and using hybrid methods. Their solutions are open source. It is a service model, which is rather the standard in this market at the moment.

**D Slit Technologies** (2018, Japan) develops custom quantum software solutions for creating proofs of concept. Their website is not very talkative about their achievements.

**Elyah** (2018, Japan/Dubai) is developing quantum software to "*improve people's lives*". The company is made up of two people, a certain Salman Al Jimeely based in Dubai and an American, Sydney Andrew, based in Tokyo. I'm still looking for those developing software worsening people's lives, besides pirates.

**Entanglement** (2017, USA) is a quantum software development service company. One of their achievements was to create a quantum inspired software for vaccine distribution optimization in the USA in 2021.

**Entropica Labs** (2018, Singapore, $1.8M[2102]) is a startup dedicated to the creation of quantum (and non-quantum) algorithms for life sciences and in particular for genomics, based on quantum machine learning.

The result is faster development of therapies, in partnership with pharmaceutical companies. The company was founded by Tommaso Demarie (CEO), Ewan Munro (CTO), joined in 2018 by Joaquin Keller, a former Orange researcher based in France, who left them to later create **exaQ.ai**. It offers its Entropy Development Framework that manages the workflow of quantum software. They are working with Honeywell/Quantinuum and BMW to create proof of concepts for supply chain optimization.

**exaQ.ai** (2020, Singapore) is a quantum machine learning created by Joaquín Keller, who formerly co-founded Entropica Labs and was a teacher in France and a R&D lead at Orange.

[2101] See Principles of quantum functional testing by Nadia Milazzo, Olivier Giraud, Giovanni Gramegna and Daniel Braun, ColibrITD, Universität Tübingen and Université Paris Saclay CNRS LPTMS, September 2022 (13 pages).

[2102] See Singapore quantum computing startup Entropica Labs bags $1.8m in seed funding by Miguel Cordon, May 2020.

Their offering is based on the "polyadic QML Library[2103]" that does supervised quantum machine learning for multi-class classification on NISQ architectures.

It was tested on IBM Quantum hardware with accuracy levels similar to classical machine learning, doing a ternary classification of the Iris flower dataset that contains only 150 objects to classify in three classes. They also provide their ManyQ quantum computer emulator that is optimized for quantum machine learning and supports CPUs and GPUs.

**FAccTs** (2016, Germany) is a spin-off from Max Planck Institute for Chemical Energy Conversion that develops ORCA, a quantum-chemical software package.

**FAR Biotech** (2016, USA) does drug discovery based on quantum representation of molecular structures done, so far, on classical computing.

**Foqus** (2021, Canada) The company is a spin-out from University of Waterloo, led by Michele Mosca and Sadegh Raeisi which creates a software solution to improve the performance of MRI and NMR systems with machine learning and quantum computing. Quantonation is one of its investors.

**GenMat** (USA-Canada, $15M) aka Quantum Generative Materials is a stealth quantum computing software startup. They got a funding of $15M (plus $35M following development milestones) from **Comstock Mining** (USA), a gold and silver mining Company also invested in lithium-ion battery manufacturing.

**Good Chemistry Company** (2022, Canada) is a spin-off from 1Qbit which developed QEMIST Cloud, a cloud developers platform associating classical machine learning and quantum computing to undertake computational chemistry simulations. Accenture is an investor in the company.

**Grid** (2009, Japan) is a specialist in deep learning and learning by reinforcement with their ReNom platform.

They have adapted this library in a quantum version called ReNomQ. And they have been IBM Quantum partners since September 2019. On the other hand, their AI was probably not very efficient to help them find a company name easy to be found via search engines.

**Groovenauts** (2012, Japan, $4.5M) developed in 2016 a D-Wave based cloud service called Magellan Blocks to solve complex optimization problems.

They exploit hybrid algorithms combining machine learning and quantum algorithms. Their first customers include a Japanese retailer who optimizes its planning and Mitsubishi Estate who optimizes household waste collection[2104].

**Hafnium Labs** (2016, Denmark) develops software that provides physical property data for molecules and mixtures by combining quantum chemistry and AI. So, not yet a quantum software vendor.

---

2103 See Polyadic Quantum Classifier by William Cappelletti, Rebecca Erbanni and Joaquın Keller, Entropica Labs, July 2020 (8 pages) and https://dev.exaq.ai/.

2104 See Groovenauts and D-Wave collaborate on hybrid Quantum Computing, December 2019.

**Horizon Quantum Computing** (2018, Singapore, $15.23M[2105]) creates quantum development tools. Their ambition is to compile code from classical development tools such as Matlab and then run it on quantum computers, in order to make quantum computing accessible to traditional developers. In short, they want to democratize quantum software development.

They also work on quantum Internet software. As such, they announced in 2022 that they will establish a quantum communication on Singapore's National Quantum-Safe Network. The startup was launched by Joe Fitzsimons and Si-Hui Tan, both coming from Singapore's CQT research center. Their last funding of $12M in April 2022 came from Tencent Holdings.

**HQS Quantum Simulations** (2017, Germany, €14.3M) is a Karlsruhe Institute of Technology spin-out startup led by Michael Marthaler. They are developing quantum algorithms in the field of organic and inorganic molecular simulation of simple molecules (methane, light emission in OLEDs, diffusion of molecules in liquids). It mostly targets the chemistry industry.

They announced in July 2018 an open source porting tool for ProjectQ code (IBM platform) to Cirq (Google platform). They released their Quantum Assisted Design toolbox in 2020 and qoqo, a quantum circuit representation library in 2021. They already have BASF and Bosch as customers. In practice, they also develop classical versions of their algorithms, running on datacenters or supercomputers[2106] . Their latest 12M€ funding round in February 2022 was led by Quantonation.

**Ingenii** (2021, USA) is a startup created by Christine Johnson (CEO) and Marko Djukic (CTO, coming from Purdue University) that wants to simplify the work of scientists using quantum computers.

They started with creating a Python package to submit data science jobs like quantum chemical simulations to quantum hardware like IonQ through Microsoft Azure for a starter. These algorithms are of course limited by the current capacities of existing quantum computers.

**Innovatus Q** (2018, Singapore) is a spin-off from the Centre for Quantum Technologies in Singapore. They work on hybrid quantum algorithms based on trapped ions and superconductors.

**Jaynes Computing** (2019, Canada) is creating a cloud-based solution based on some quantum machine learning (QML) in the supply chain market. They are supposed to use some NISQ hardware, without any details. The startup was created by German Alfaro and was spun out of the Creative Destruction Labs.

**Jij** (2018, Japan, $1.9M) was created by researchers from the Tokyo Tech Institute of Technology. It develops software for quantum annealing, including OpenJij, an open source framework for implementing Ising models to model particle interactions, built on D-Wave's QUBO APIs. They are also partners of Microsoft Azure.

**JoS Quantum** (2018, Germany) develops quantum software solutions for the financial services industry, particularly in risk management and fraud detection. They also do contract research.

[2105] See the QSI Seminar presentation: Dr. Joe Fitzsimons, Horizon Quantum Computing, Abstracting Quantum Computation, April 2020 (1h26). Joe Fitzsimons create the blind quantum computing protocol with Anne Broadbent and Elham Kashefi in 2008.

[2106] See HQS Quantum Simulations: How to survive a Quantum winter by Richard Wordsworth, 2020.

**Ketita Labs** (2018, Estonia) develops unspecified quantum software for NISQ computers, and for good reason. It is a university spin-off.

**KiPu Quantum** (2021, Germany, 3M€) was created by Enrique Solano, a prolific and outspoken researcher working in Spain (Bilbao) and Germany. The startup's goals are to "*design and manufacture of modular and co-designed Quantum computers*" tailored to solve specific tasks with NISQ, without waiting for LSQ generations.

In 2021, they announced their NISQA paradigm merging digitized-counterdiabatic quantum computing (DCQC) and digital-analog QC (DAQC) for NISQ computers without error correction overhead. It uses digital and analog compression techniques to reduce the physical qubits required to solve specific industry problems (combinatorial and optimization problems, chemistry, QML, ...).

**Kuano** (2020, UK, $3.6M) creates quantum software solutions for the design of molecules and in particular for the inhibition of enzymes, which is used both in pharmaceuticals and to create protective agents in agriculture.

They use quantum emulation and quantum algorithms as well as machine learning. The company was founded by defectors of GTN, including their CEO Vid Stojevic who was the CTO of GTN.

**Menten.ai** (2018, USA, $4M) develops hybrid algorithms combining machine learning and quantum programming to simulate organic chemistry and design enzymes, peptides and proteins, working with D-Waves annealers.

**Molecular Quantum Solutions** (2019, Denmark) or MQS, provides computational tools for pharma, biotech and chemical industries. It's using HPC and quantum computers.

**Multiverse Computing** (2019, Spain, 24M€) develops quantum and quantum-inspired software for various markets, starting with financial services, with portfolio optimization, risk analysis, market simulation and fraud detection.

They announced in August 2021 their Singularity Spreadsheet app solution which gives access to some portfolio optimization algorithm running on D-Wave annealers in the cloud, directly from within Microsoft Excel (video). It is now part of their Singularity SDK for Portfolio Optimization. They have then expanded their reach to the mobility, energy, and manufacturing verticals. They also use more traditional techniques based on machine learning, digital annealing (with Fujitsu) and quantum inspired solutions using tensor networks.

They published several papers in 2022 on combinatorial problem solving using quantum annealing with ZF in Germany[2107], on artificial vision using both gate-based and quantum annealers with the research institute Ikerlan Technology Center in Basque country, Spain, showing qualitative improvements vs classical method[2108], on a quantum-inspired tensor neural network to solve partial differential equations, in partnership with CACIB (French investment bank)[2109], and on portfolio optimization with Ally and Protiviti on D-Wave annealers[2110]., as well as with Bankia[2111]. This is a good practice and track record for a quantum software vendor.

[2107] See Multi-disk clutch optimization using quantum annealing by John D. Malcolm et al, August 2022 (11 pages).

[2108] See Quantum artificial vision for defect detection in manufacturing by Daniel Guijo et al, August 2022 (11 pages).

[2109] See Quantum-Inspired Tensor Neural Networks for Partial Differential Equations by Raj Patel et al, August 2022 (14 pages).

[2110] See Financial Index Tracking via Quantum Computing with Cardinality Constraints by Samuel Palmer et al, August 2022 (8 pages).

[2111] See Hybrid Quantum Investment Optimization with Minimal Holding Period by Samuel Mugel et al, December 2021 (6 pages).

Their Fair Price quantum solution computes accurate fair prices for financial institutions and seems to run on IonQ quantum processors[2112]. The solution is leased as a cloud service for 100K€ to 200K€ a year.

The company is partnering with Xanadu, Microsoft, Fujitsu, IBM, Rigetti, DWave, NTT, Strangeworks, Pasqal (France), IQM (Finland), Objectivity IT (UK, an IT services company), among others. They have several offices outside Spain in Toronto (Canada, with one of their cofounders), Munich (Germany) and Paris (France) with about 55 people as of August 2022. The EIC (European Innovation Council) provided them with 12,5M€ comprising a mix of grant and venture funding. They have 22 patents pending in quantum computing algorithms.

**NetraMark** (2016, Canada) develops quantum software solutions for the pharmaceutical industry to define therapeutic targets. They are part of the Quantum Machine Learning program at the Creative Destruction Lab in Toronto. It was acquired by the brain biotech **Nurosene Health** (2019, Canada) in October 2021.

**Novarion** (2016, Austria) is a server storage and GPU server vendor who wants to create the first hybrid quantum computer by 2025, without mentioning what sort of quantum processor it will integrate.

In October 2020, they started a partnership with **Terra Quantum AG** (Switzerland) to create a Joint Venture to create '<Qa|aS> by QMware'. It supports machine learning and big data analytics. QMware's customers will be able to develop and effectively run completely Hybrid Quantum Applications. Applications built on QMware's Hybrid Quantum Cloud are supposed to run on upcoming native quantum processors when they show up. Meanwhile, it runs on some classical emulators, seemingly a QLM from Atos. And it's Gaia-X and GDPR compatible.

**Nordic Quantum Computing Group (**NQCG) (2000, Norway) does R&D in areas at the crossroads between AI and quantum computing. They are creating a platform agnostic quantum software using superconducting and photonic qubits.

**ODE L3C** (2018, USA) is an American NGO involved in the creation of chemical simulation algorithms. Its ambition is to solve "difficult NP" problems with quantum computation, which is far from obvious.

This sounds more like a service provider than a software publisher. The company was created by a certain Keeper Layne Sharkey.

**Opacity** (2020, Australia) offers Quiver, a quantum code optimization software compatible with IBM's Qiskit.

Their hardware-agnostic solution maps processor errors at the global and individual qubit level, including parasitic interactions between qubits. It then allows the code to be optimized to take into account this duly mapped noise. The tool seems to be dedicated to developers as well as to quantum computers designers.

**OTI Lumionics** (2011, Canada, $5.7M) is specialized in the design of new materials and in particular LEDs and OLEDs. They have developed quantum and quantum-inspired molecular simulation algorithms for this purpose.

[2112] See [Quantum portfolio value forecasting]() by Cristina Sanz-Fernandez et al, November 2021 (9 pages).

In particular, this allows them to predict the properties of the created materials like their color when being excited[2113], model chemical relationships and determine geometric structures. They are partners of Microsoft Azure (video).

**ParityQC** (2020, Austria) is a spin-off of the University of Innsbruck created by Wolfgang Lechner and Magdalena Hauser, the first being the scientist and the second, handling business aspects of the venture[2114].

By September 2022, the company had about 25 employees plus 15 researchers working at the University of Innsbruck. As architects, they connect the dots between hardware and software.

They develop software solutions to solve optimization problems (CADD, N-body problems, constraint problems) adapted to digital and analog quantum computers (qubits with universal gates or quantum simulators[2115]). Their ParityOS software suite optimizes the software parameters of the solution as well as those of the hardware control.

They support an architecture called LHZ, created by Wolfgang Lechner and Austrian colleagues Philipp Hauke and Peter Zoller, which is compatible with different hardware quantum platforms with 2D qubit connectivity architectures[2116].

Its principle consists in encoding a problem requiring n-to-n relations between qubits (all to all) to run it on a physical architecture where qubits are only connected to their closest neighbors as is the case in most quantum computers, except for some that rely on trapped ions. Their solution also includes an in-house error correction system[2117]. They are partnering with **Pasqal** whose cold atom-based architecture seems adapted to their model.

They announced in 2021 a partnership with **NEC** to help them with their superconducting qubits.

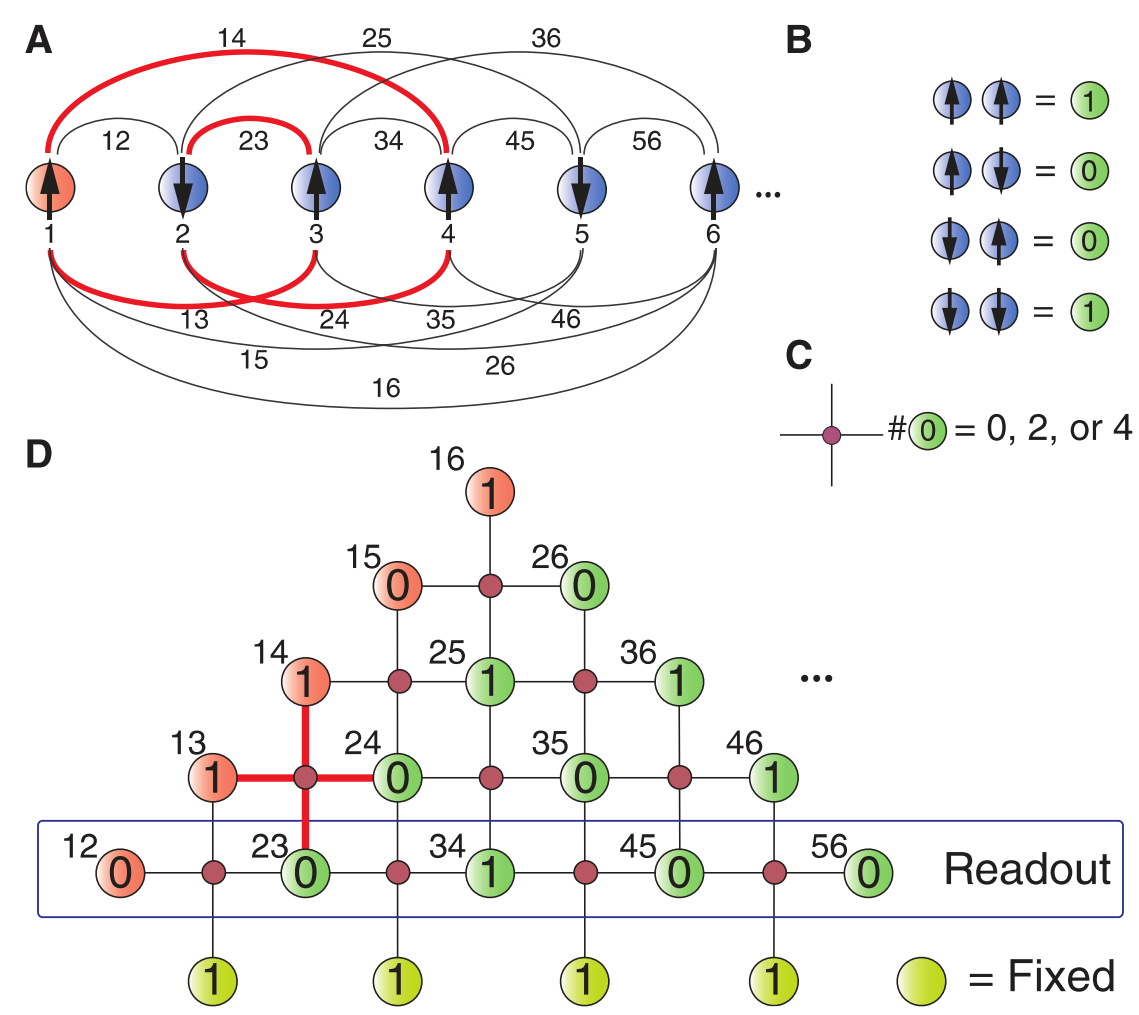

*Figure 690: LHZ architecture. Source: A quantum annealing architecture with all-to-all connectivity from local interactions by Wolfgang Lechner, Philipp Hauke and Peter Zoller, October 2015 (5 pages).*

**Phase Space Computing** (2017, Sweden) is a spin-off from the University of Linköping that develops training solutions on quantum computing for secondary and higher education.

[2113] See Estimating Phosphorescent Emission Energies in Ir(III) Complexes using Large-Scale Quantum Computing Simulations by Scott N. Genin et al, October 2021 (32 pages). It is based on using a 72 logical qubits classical emulator.

[2114] She is a family relative of Hermann Hauser, the co-founder of Arm, now a serial entrepreneur and investor in deep techs, including Graphcore (UK). He is also a member of the EIC Council.

[2115] Like in Rydberg blockade based parity quantum optimization by Martin Lanthaler, Clemens Dlaska, Kilian Ender, Wolfgang Lechner, October 2022 (10 pages) which describes a solution to quantum optimization problems using MWIS and UDG on quantum simulators using neutral atoms.

[2116] This LHZ architecture is documented in A quantum annealing architecture with all-to-all connectivity from local interactions by Wolfgang Lechner, Philipp Hauke and Peter Zoller, October 2015 (5 pages) for universal gated qubit platforms which they perfected in Universal Parity Quantum Computing by Michael Fellner, Anette Messinger, Kilian Ender and Wolfgang Lechner, May 2022 (6 pages). See also Rapid counter-diabatic sweeps in lattice gauge adiabatic quantum computing by Andreas Hartmann and Wolfgang Lechner, September 2019 (11 pages) for quantum annealing computing. See also Quantum Approximate Optimization with Parallelizable Gates by Wolfgang Lechner, 2018 (5 pages) which describes the implementation of a QAOA optimization algorithm with CNOT and unit gates.

[2117] See Error correction for encoded quantum annealing by Fernando Pastawski and John Preskill, 2015 (4 pages).

**PhaseCraft** (2018, UK, $1M) is a quantum software company spun out of University College London and the University of Bristol by Toby Cubitt, Ashley Montanaro and John Morton (also a cofounder of Quantum Motion). They plan to exploit quantum computing to create better energy collection and storage systems (batteries, solar PV, ...).

They launched in 2022 a partnership with Oxford PV, a perovskite photovoltaic panel company, to improve PV efficiencies thanks to quantum computer-based simulations.

Their researchers developed an optimized version of an algorithm solving the Fermi-Hubbard model with fewer resources which could be helpful to create high-temperature superconducting materials[2118]. In 2021, they also published a proposal for how to simulate a "kagome magnet". While it did not show any quantum advantage on existing quantum hardware and worked only with 20 qubits, it extrapolated that with 50 qubits, the problem could be solved by 200 two-qubit gates, something high-fidelities NISQ qubit system could achieve in the near future[2119].

**PiDust** (2019, Greece) is a startup launched by Vasilis Armaos, Paraskevas Deligiannis and Dimitris Badounas, who are alumni of University of Cambridge, Stanford and the University of Patras. They develop quantum algorithms in chemistry.

**Pine.ly** (2019, Canada) is positioned on software to assist in the creation of innovative materials with quantum computing. They aim in particular at the recycling of $CO^2$ emissions. The startup was created by three women, Nayer Hatefi, Shabnam Safaei and Rachelle Choueiri, all three scientists.

**POLARISqb** (2020, USA, $2M) is a startup that wants to use quantum computing to create new therapies. One more! It built the Tachyon platform.

But what kind of quantum computer and tools are used in Tachyon is a mystery. The startup was founded by Shahar Keinan (CEO) and Bill Shipman (CTO). They are partnering with Fujitsu, probably to use their conventional supercomputers and their digital annealing computer. Their idea is to use personalized medicine techniques to create ad-hoc therapies. Shahar Keinan received a PhD in chemistry from the Hebrew University of Jerusalem. She specializes in computational chemistry. In August 2022, they worked with **Auransa** to identify interesting target proteins for curing specific breast cancers using Tachyon and Auransa's SMarTR Engine based on classical AI.

**Allosteric Bioscience** (2021, USA, $920K) is a company integrating quantum computing and AI with biomedical sciences to create improved treatments for aging and longevity. In February 2022, the company announced it was teaming up with and investing in Polaris Quantum Biotech (Polarisdb). They first work on creating an inhibitor of a key protein involved in aging.

**ProteinQure** (2017, Canada) is a Toronto-based startup that uses various technologies including quantum computing to create and simulate new "*in silico*" therapies. They use quantum algorithms to simulate protein folding.

They are also developing hybrid algorithms also using GPUs. They support different hardware architectures including D-Wave computers.

---

2118 See Strategies for solving the Fermi-Hubbard model on near-term quantum computers by Chris Cade, November 2020 (27 pages) which turned into Observing ground-state properties of the Fermi-Hubbard model using a scalable algorithm on a quantum computer by Stasja Stanisic, Jan Lukas Bosse, Filippo Maria Gambetta, Raul A. Santos, Wojciech Mruczkiewicz, Thomas E. O'Brien, Eric Ostby and Ashley Montanaro, Nature Communications, October 2022 (11 pages).

2119 See Probing ground state properties of the kagome antiferromagnetic Heisenberg model using the Variational Quantum Eigensolver by Jan Lukas Bosse and Ashley Montanaro, October 2021 (12 pages).

In their experiments, they manage to simulate molecules with 6 atoms in universal quantum computers and reach 11 atoms with D-Wave. In practice, however, it would seem that they have put quantum computing on the backburner and are now focused on classical learning machines in the meantime.

**Q1t** (2018, Netherlands) creates mathematical models and software for classical and quantum computers in the field of quantum chemistry, quantum optics and financial analysis. They have tried these algorithms on quantum simulators and quantum optics.

They developed the q1tsim quantum simulation library published on Github which implements new quantum gates types for creating simpler circuits, the ability to simulate measurements without affecting qubit quantum states and the option to re-run a circuit starting with the previous quantum state for debugging purpose.

**QbitLogic** (2014, USA, $1.5M) is another startup that develops quantum machine learning applications, without more precision in their communication. They also develop an AI based system, CodeAI, to debug software.

**Qbraid** (2020, USA) provides a quantum online coding platform. Based in Chicago, the company was created by Kanav Setia who has a PhD from Dartmouth College in quantum simulation and worked with IBM research.

The platform supports the most popular programming frameworks (Qiskit, TensorFlow Quantum, Q#, Xanadu PennyLane, Braket, IonQ, Rigetti pyQuil, D-Wave...) and contains a transpiler that optimized the quantum code for the target hardware platforms. Access to IBM quantum computers seems embedded in the software solution which targets students. The company also developed quantum programming courseware.

**Quantum Computing Inc** aka QCi (2018, USA, $7,5M) is a quantum software, hardware consulting company that started with the creation of Qatalyst, a high-level software development and cloud provisioning tool.

It's mainly enabling developers to solve constrained optimization problems (QAOA, QUBO, graph optimization) with either gate-based accelerators, quantum annealers or classical computers. It's based on using six simple high-level APIs. It is commercially available. They support their home QikStart Program, a marketing initiative to accelerate real-world use cases with their customers. Qatalyst supports D-Wave, IonQ and Rigetti accelerators through Amazon's Braket cloud services. In March 2021, the company hired a "Chief Revenue Officer" (Dave Morris) and a Marketing VP (Rebel Brown) after having announced in December 2020 that they were filing for a Nasdaq IPO.

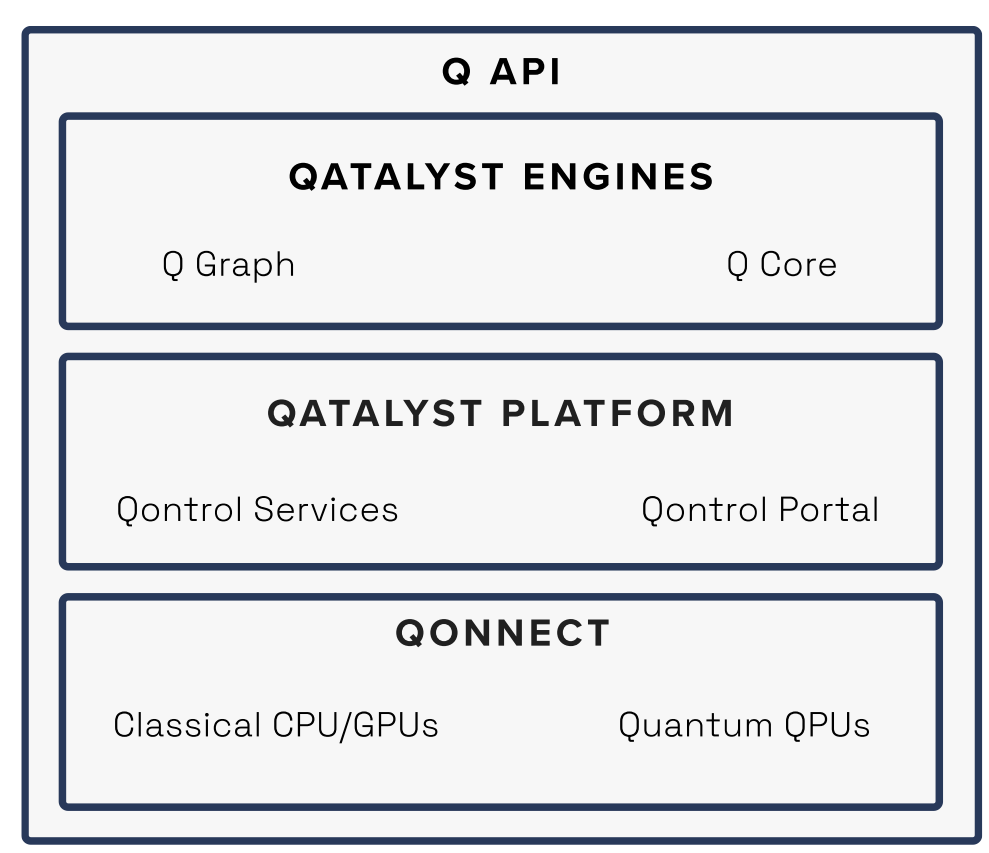

*Figure 691: QCi software platform.*

In July 2021, it became the first publicly traded pure player quantum computing company. QCi also created QUBT University in August 2021, an online initiative to train students on their Qatalyst tool. It's also a tactic to partner with academic institutions, like Notre Dame University in Indiana. They also have QCi Qonsulting practice, implementing their Path2Quantum (P2Q) methodology, a four-phase framework to planning quantum computing deployments.

In March 2022, QCi announced a partnership looking like an acquisition of a stealth company, QPhoton (2020, USA) from New Jersey and created by Yuping Huang (also assistant professor at the Center for Quantum Science and Engineering at Stevens Institute of Technology), and which develops some sort of photonic quantum computer of unspecified nature.

QCI with hook this system with its Qatalyst software. QPhoton was funded by various US federal agencies (DARPA, NSF, NASA, DoD). QCI launched in September 2022 its Dirac 1 Entropy Quantum Computer (EQC) cloud-based subscription based on Qphoton's technology.

In June 2022, QCi announced QAMplify, a solution supposedly increasing the processing power of quantum computers by x20. Of course, it deserves some scrutiny. Their patented process is supposed to increase the number of variables that can be handled in quantum hardware by a factor x5 for gate-based and x20 for quantum annealing. How they are doing this is not yet documented.

**Q-CTRL** (2017, Australia, $43.4M) is a startup created by Michael Biercuk of the University of Sydney. They develop a set of enabling software tools to improve the operations and programming of quantum computers.

Their Boulder Opal is quantum control infrastructure software working at the firmware level. It leverages machine learning to improve qubits control pulses and optimize quantum error correction codes. It is a Python toolkit used by quantum computers designers that works with IBM Qiskit, Rigetti and with Quantum Machines pulse generators. They implement error-correction techniques that increases the likelihood of quantum computing algorithm success between 1000x and 9000x on quantum hardware, as measured using the QED-C algorithmic benchmarks, but probably with a very low number of logical qubits. These impressive numbers relate to the impact of optimized quantum error correction.

They are relying on Google Cloud and TensorFlow to run the classical machine learning algorithms of their solution[2120]. Fire Opal is a set developer tools for quantum algorithm designers while Black Opal is an educational tool for students new to quantum programming.

In May 2022, Q-CTLR combined a 5-qubit QEC code (using 9 qubits, due to the additional helper qubits) with Fire Opal and could improve the code's ability to correctly identify errors by 70% on IBM QPUs (meaning: one qubit error detection rate was improved by 70%)[2121].

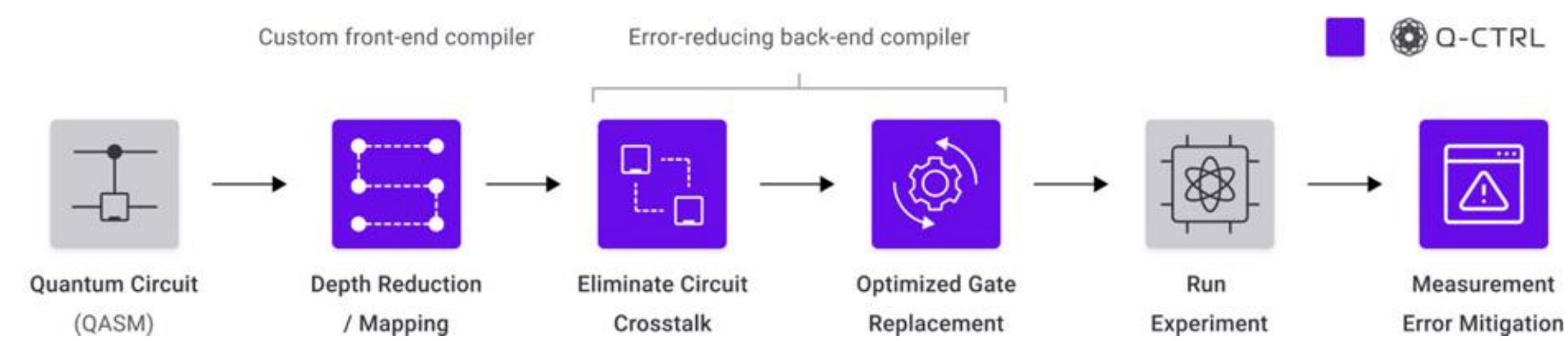

*Figure 692: Q-CTRL error reduction techniques. Source: Q-CTRL.*

They also work on optimizing quantum sensing solutions and are creating quantum sensors of their own for acceleration, gravity and magnetic fields measurement in space applications. They are partnering with the Australian company **Advanced Navigation**, which specializes in geopositioning. At last, they are also creating prototype algorithms for buses dynamic scheduling of buses in Sidney, Australia (*disclaimer: it can't be operational given the power of existing quantum computers*).

**QC Ware** (2014, USA, $39.7M) develops a platform for cloud-based quantum software development. They create quantum algorithms and software for large companies with two layers: their proprietary Forge platform and open source libraries for optimization, chemical simulation and machine learning.

---

[2120] See [Boosting quantum computer hardware performance with TensorFlow]() by Michael J. Biercuk, Harry Slatyer, and Michael Hush, October 2020.

[2121] Fire Opal's process is documented in [Experimental benchmarking of an automated deterministic error suppression workflow for quantum algorithms]() by Pranav S. Mundada, Michael J. Biercuk, Yuval Baum et al, September 2022 (16 pages).

They provide tools to load training data from learning machine models into memory more quickly. They have also developed an algorithm for calculating the distance between objects, which can be used to train both supervised (classification) and unsupervised (clustering) machine learning models. Their first customers include **Equinor** for oil exploration optimization, Japan's **AISIN** for certification testing of automatic gearbox software, **Airbus** for aircraft flight envelope optimization and **BMW** for autonomous vehicle route optimization. They are also targeting financial markets as well. It supports universal gate quantum computers (IBM, Rigetti), D-Wave quantum annealing computers and software emulators (IBM, Google, Microsoft, Rigetti).

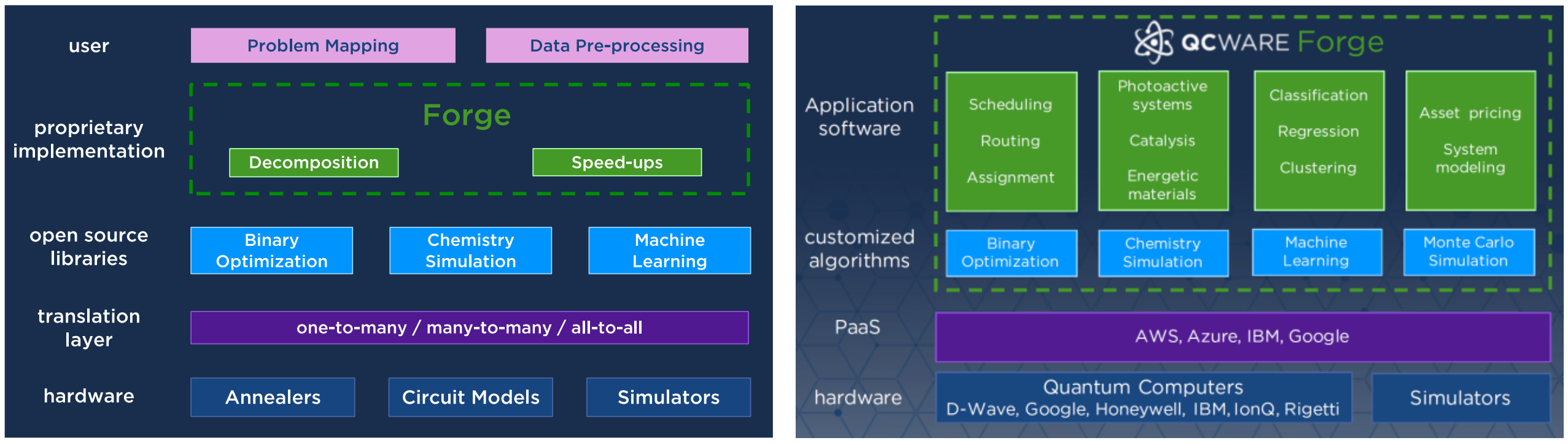

*Figure 693: QC-ware software platform. Source: Enterprise Solutions for Quantum Computing by Yianni Gamvros, December 2019 (25 slides).*

Airbus Ventures is one of their investors on top of the Koch group. They have received a $1M US public funding via the NSF in 2017. The startup, which already includes over 20 people, was created by Matt Johnson, who has a financial background, and Kin-Joe Sham and Randy Correll, who seem to have gotten late into quantum computing. The team also includes Iordanis Kerenidis, who is based in France and is a leading specialist in quantum machine learning. He oversees international algorithms development. Scott Aaronson is their Chief Scientific Advisor. Finally, the startup organizes an annual conference on quantum computing, the **Q2B**, the last edition of which was held in December 2020 online[2122].

**QEDma Quantum Computing** (2020, Israel) is a quantum software company created by Dorit Aharonov of the Hebrew University, Nathaniel Lindner of the Technion and Asif Sinai. It creates quantum algorithms and software tools.

**Qindom Inc.** (2018, Canada, $2M) is a startup developing quantum machine learning (QML) software running on D-Wave quantum annealers.

**Qoherent** (2019, Canada, $173K) is developing machine-learning based RF signals analysis tools that helps create adaptive RF communications and sensing systems in the Software-Defined Radio (SDR) realm. They analyze signals in the telecom ranges (4G, 5G, V2X, Satellite communications...). The company is investigating the usage of quantum machine learning techniques to improve its solutions.

**Qrithm** (2018, USA) develops quantum algorithms in diverse and rather disparate fields: machine learning, materials science, cryptography and finance.

**Qu&co** (2017, The Netherlands) was created by Benno Broer and Vincent Elfving, both quantum physicists who worked at TU Delft.

[2122] View presentation materials and videos from the 2019 Q2B conference, in December 2019.

They develop tailor-made quantum software solutions for large companies, accompanied by benchmark tools, particularly for chemical simulation applications based on DFT. They develop solutions for simulating fluid mechanics with nonlinear differential equations solved with VQE algorithms (hybrid variational quantum eigensolvers). They even solve Navier-Stokes 1D equations. They partner with IBM, Microsoft and Schrodinger (USA).

In March 2021, they launched a beta release of QUBEC, a chemistry and materials science toolkit. It comprises Q-time, a tool estimating when quantum advantage can be expected for solving chemistry or materials problems. QUBEC workflow manager supports quantum systems from IBM, IonQ and Rigetti. It is available through the IBM Quantum Experience and Amazon Braket platforms. LG Electronics is one of their customers.

In August 2021, they announced a new funding round with Quantonation, Runa Capital and SPInvest, with an undisclosed amount and in January 2022, the company merged with Pasqal (France). Benno Broer is now the CCO of Pasqal (Chief Customer Officer).

QUANSCIENT

**Quanscient** (2021, Finland) was created by Alexandre Halbach, Asser Lähdemäki, Valtteri Lahtinen and Juha Riippi.

It provides Quantum Simulation-as-a-Service (QSaaS) software. Its multiphysics simulation algorithms use finite element method and partial differential equations solvers.

They say hint about using hybrid classical/quantum computing but given the state of the art of quantum computers, it's probably wholly made of classical computing. They help simulate physics in the electromagnetism, mechanics and fluid dynamics fields.

QuantFi Quantum Computing For Finance

**QuantFi** (2019, France/USA) is a young startup specializing in the creation of quantum software solutions for finance.

It was created by Paul Hiriart (French, ex of Lehman Securities, who left the startup early in 2021), Kevin Callaghan (coming from the New York financial sector) and Gabrielle Celani (on sales and marketing). They are creating goals-based investment optimization algorithms, and also handle trends detection, derivatives pricing and risk management. In 2020, they joined the IBM Quantum network.

**Quanterro Labs** (2019, Abu Dhabi) is an association of researchers and entrepreneurs created by Kaisar Parvez and Ram Soorat and AiFi Technologies, an AI software startup, working in quantum information and security. They work on middleware and software development for D-Wave, Google, IBM and others. It's mostly a consulting services company.

QUANTICA COMPUTACAO

**Quantica Computacao** (2019, India) is the first Indian quantum startup and a software company working on creating a cloud development environment an QML algorithms. It's incubated in the Indian Institute of Technology Madras from Chennai.

QUANTOPTICON

**Quantopticon** (2017, UK) develops modeling and simulation software for the design of photonic components for quantum applications, particularly quantum communications and cryptography. It is a tool for the design of new materials. The company was created by a mother-daughter duo: Mirella Koleva (CEO) and Gaby Slavcheva (CSO).

⟨q|b⟩ quantum benchmark

**Quantum Benchmark Inc** (2017, Canada) provides an error-correcting code software solution for general-purpose quantum computers and error evaluation. It was thus apparently a competitor to Q-Ctrl.

They also offer a quantum computer performance validation system. The package is integrated into the True-Q suite, launched in 2018, with True-Q Design, which is used to evaluate the error rate of a quantum computer and to optimize its architecture, and True-Q OS, which helps optimize the accuracy of software solutions.

The target market is initially the manufacturers of quantum computers and those who evaluate them. Eventually, it will be that of user customers. Note that they have already tested Google's Cirq framework, having been part of Google's beta test program for this language, and that Google is using their solution, as it did in its 2019 supremacy experiment. They are also partnering with IBM (since 2018, on error characterization, mitigation, correction, and performance validation) as well as with Fujitsu Labs (since 2020, to develop quantum algorithms). The company was acquired by **Keysight Technologies** in May 2021, on top of Labber Quantum in 2020. Among its key people are Joseph Emerson (CEO), Joel Wallman (CTO) and Daniel Gottesman (Senior Scientist, specialized in error correction, MIT), Stefanie Beale and Kristine Boone (both researchers).

**Quantum Flytrap** (2020, Poland) is a quantum computing and cryptography software company. They created a real time browser-based emulator using photonic elements instead of quantum gates[2123]. They are behind the "Quantum Lab" that is being used at Stanford University and the University of Oxford.

**Quantum-South** (2019, Uruguay) is a spin-off from the University of Montevideo who is specialized in developing quantum optimization software, first targeting the cargo shipments in ships and airlines. The software was released in 2022.

They also target the financial sector which may be more dynamic although more crowded with many existing quantum software startups. In cargo shipments, they are partnering with Quantum Brilliance (Australia) with their prototype (5) NV centers qubits.

**QBaltic** (2019, Estonia) develops algorithms for quantum computing, quantum cryptography and artificial intelligence. QBaltic is a contract research spin-off from the University of Latvia, University of Tartu in Estonia and QuBalt, Germany and Latvia.

**Qsimulate** (2018, USA, $1.5M) develops quantum solutions for molecular simulation for healthcare and chemistry. They are partners of Amazon Braket and Google and are already working with Amgen. The company was co-founded by Toru Shiozaki and Garnet Chan, both specialized in chemistry.

**Quacoon** (2020, USA) is a small startup created by Tina Sebastian and Barbara Dunn that develops software solutions for the food supply chains combining AI and quantum annealing.

**Quantum Thought** (2019, USA) develops quantum or quantum inspired algorithms for chemical, AI and security markets. According to their website, it seems to be mainly a service company operating in project mode and doing consulting services. Their CEO is Rebecca Krauthamer.

**Quantumz.io** (2019, Poland) develops the Quantum Simulator Platform (QSP), a quantum program emulation solution running with GPUs. They are also developing a PQC (post-quantum cryptography) solution called banax, including some dedicated hardware to implement it.

**Quantagonia** (2021, Germany) is a quantum software company. Their Quantum Virtual Platform is a hybrid classical/quantum computing platform (gate-based and quantum-annealing based). It was created by Dirk Zechiel (CEO), Sabina Jeschke, Sebastian Pokutta and Philipp Hannemann.

[2123] See Visualizing quantum mechanics in an interactive simulation -- Virtual Lab by Quantum Flytrap by Piotr Migdał et al, May 2022 (29 pages).

**Quantastica** (2019, headquarter in Finland with offices in Estonia and Serbia) develops hybrid quantum algorithm software tools including Quantum Programming Studio, a graphical web development environment for creating quantum algorithms executable on quantum computers or simulators, including a classical simulator they have developed themselves.

**Quantistry** (2018, Germany) created a cloud-based simulation platform for material research and development, using quantum simulations, molecular dynamics and machine learning.

**Quantum Mads** (2020, Spain) was created in Bilbao by Eriz Zárate and Alain Mateo Armas and is positioned in the financial market. It offers four software tools with a mix of quantitative/classical/hybrid/quantum-inspired software.

With Q-MADS, an investment strategy analysis framework for traders, Q-RETAIL, a framework for retail banks, Q-ALLOCATE, for asset allocation optimization and Q-CRYPTO, a framework for optimal path finding in graphs. The whole is based on the HHL linear algebra algorithm.

**Quantopo LLC** (2017, USA) is a company specialized in machine learning algorithms. They focus on biotechs, supply chain and logistics. They are part of the Creative Destruction Lab in Canada. But as they don't have an active website, it is not certain that they still exist.

**Quantum Open Source Foundation** got a $4K grant from Unitary Fund. It publishes a list of quantum open source projects on Github, with various software development tools and libraries for gate-based quantum as well as and quantum annealing computing.

It's a repository of existing open source projects including IBM Qiskit and PennyLane from Xanadu. NISQ. Provides financial funding for quantum open source software projects. Organize events. More a community than a foundation like the Apache or Mozilla foundations.

**QuantyCat** (2020, USA) creates cloud-based APIs for quantum software development, supporting D-Wave, IonQ and Rigetti through Amazon Braket.

**Qubit Engineering** (2018, USA) was founded by University of Tennessee alumni. They develop classical and quantum optimization algorithms suitable for wind turbine design and location optimization. Another very niche market. They are partners of Microsoft Azure.

**Qubitera** (2018, USA) develops solutions combining AI and quantum.

**Qubitl Quantum Technologies** (2018, India) is a contract research laboratory developing quantum machine learning software. They were initially specialized in cybersecurity, having created a QRNG solution (Q-RandCon), a PQC protected healthcare solution (HealthCetra) and a Quantum Differential Phase Shift technology for QKD (Q-Shift).

**Qubit Pharmaceuticals** (2020, France/USA, $16.1M) is a startup co-founded by Jean-Philip Piquemal, a CNRS professor-researcher at Sorbonne University, a long-time specialist in molecular dynamics simulation that mathematically models the quantum mechanics of organic molecules.

The co-founders of the startup are based in Austin and at Washington University in Saint Louis. Its algorithms have been in use for a long time. Jean-Philip Piquemal is the co-author of the Tinker molecular simulation library and its Tinker-HP version adapted to supercomputers. It exploits massively parallel CPU-based systems and Nvidia GPU tensors, all with high-precision computation. In particular, it uses the Jean Zay computer from GENCI in Orsay, France, as well as those from the DoE in the USA. In this context, they have been involved in molecule screening for the search for covid-19 treatments by therapeutic retargeting. Re-targeting is easier to simulate than the 3D structure of the whole covid-19 which is more than 200,000 atoms or the simulation of protein folding.

What about quantum computing in all this? It could be used to define optimized parameters for classical simulation, in short, within the framework of hybrid algorithms. They will also be able to exploit quantum simulators in the future, such as those being developed with cold atoms at Pasqal[2124].

Jean-Philip Piquemal's laboratory at Sorbonne University has received a €9M ERC for the development of simulation solutions for organic systems of several million atoms[2125]. They finally welcomed the investment fund Quantonation in their capital in June 2020[2126].

**QbitSoft** (2022, France) is a new software and service vendor targeting in priority the retail and logistics market. It was created by Olivier Pegeon, a former business executive from IBM France.

**QuDot** (2018, USA) develops software for the simulation of quantum circuits on traditional computers, the QuDot Net. They use techniques based on Bayesian networks to optimize the in-memory representation of qubits.

**QunaSys** (2018, Japan, $12.8M) also develops quantum algorithms in chemistry and healthcare. From the universities of Tokyo, Osaka and Kyoto, they also maintain the Qulacs simulator developed at Kyoto University.

It was followed in 2021 by Qamuy, a quantum cloud platform. They also created in 2020 the Japanese consortium QPARC with 50 participating companies investigating various use cases of quantum computing. The company is also participating to the **Pistoia Alliance**.

**QuSoft** (2014, The Netherlands) is a spin-off from TU Delft University specializing in quantum algorithms and software. Like its sister company QuTech, it is more a private applied research laboratory than a startup.

**QxBranch** (2014, USA, $8.5M) was created by former Lockheed Martin employees. It offers solutions, probably tailor-made, for the financial, insurance, aerospace and cyber security markets.

Based in Washington DC, they already have offices in Hong Kong, London and Adelaide, Australia. They are partners of D-Wave and IBM. The startup was acquired by Rigetti (USA) in July 2019.

**Rahko** (2018, UK, £1.3M) is a quantum machine learning and chemical simulation software development company based in London. It was founded by Leonard Wossnig. They are among the first Amazon AWS partners for the use of quantum resources in the cloud, and the first in Europe.

In May 2020, they announced that they would work with Merck on "quantum inspired" algorithms, i.e. on conventional computers.

---

2124 See the presentation Computational Drug Design & Molecular Dynamics: an HPC perspective by Jean-Philip Piquemal, April 2020 (28 slides).

2125 See Extreme-scale Mathematically-based Computational Chemistry (EMC2), 2020.

2126 See Qubit Pharmaceuticals closes a pre-seed round with Quantonation, Quantonation, June 2020.

In 2021, they announced that they were collaborating with Honeywell and achieved excellent accuracy executing a Discriminative Variational Quantum Eigensolver algorithm on Honeywell's H0 6-qubit trapped ion system. The company was acquired by **Odyssey Therapeutics** in January 2022, a company specialized in drug discovery in the fields of inflammatory diseases and cancers.

**ReactiveQ** (2018, Canada) develops quantum simulation algorithms for the design of innovative materials such as high-temperature superconductors, all on a NISQ quantum computer.

**Riverlane** (2016, UK, $24.1M) is a spin-off from the University of Cambridge that provides services in quantum computing and develops new algorithms combining machine learning and quantum in chemistry.

They develop with dividiti Ltd, a one man shop created by a certain Grigori Fursin, the Quantum Collective Knowledge, a benchmark SDK for quantum hardware and software.

They have also developed what they call a quantum operating system, Deltaflow.OS dedicated to NISQ systems and which optimizes access to hardware resources for qubit control. It was deployed in mid-2020 at several sites in the UK and in partnership with SeeQC and, later, CQC.

In July 2021, Riverlane created a consortium with **Astex Pharmaceuticals** and **Rigetti UK** to develop quantum drug discovery algorithms running on Rigetti platforms in the cloud. This is part of a 18-month feasibility study funded by a grant from UKRI as part of the UK quantum plan[2127]. Of course, you may wonder what they will do with the 31 qubits from Rigetti.

**RQuanTech** (2018, Switzerland) develops RTranscender, a quantum machine learning tool for finance, healthcare, automotive, seismology and cyber security. It supports Fourier transforms, qubit-based arithmetic operations which can help craft oracles (additions, multiplications, divisions, exponentials), factorization, discrete logs, etc.

**Sanctuary** (2021, Canada) is a startup created by Geordie Rose, the cofounder and first CTO of D-Wave.

It wants to implement reinforcement learning algorithms with quantum computing. At this point, the company is hiring a bunch of scientists and developers, with no offering yet looming.

**SavantX** (2015, USA) was created by Ed Heinbockel and David Ostby who in the past worked for the FBI, the Department of Defense and the US Intelligence Community including the DIA where they developed OSINT collection tools (open source intelligence, using unstructured data).

They core skills are data search, discovery, analysis and visualization. They plan to leverage quantum computing to identify hidden patterns in data and help optimize complex systems. One of their first realization in that domain was implemented for the Port of Los Angeles in 2020, to optimize cargo flow and container handling using D-Wave quantum annealers and classical machine learning. Their target markets are the nuclear, healthcare, utilities and defense industries.

**Schrodinger** (1990, USA, $193) is a digital drug design company, mainly using molecules screening and doing drugs retargeting.

It is an established competitor of Qubit Pharma (France). They work with Sanofi. The company is listed on the NASDAQ. They inevitably became interested in quantum computing and have started a partnership with Qu&Co to ramp-up their skills in quantum computing.

[2127] See Riverlane and Astex Pharmaceuticals join forces with Rigetti Computing to drive drug discovery forward by Amy Flower, July 2021.

**Semicyber** (2018, USA) develops algorithms in various fields: data analysis (non-quantum), quantum and others for critical applications for the USA defense sector, particularly the US Air Force.

So they are probably closer to the service company than to the product-oriented startup. The startup is co-founded and managed by Kayla Farrow, an engineer specializing in algorithm creation and signal processing.

**Sigma-i Labs** (2019, Japan, $3.7M) is a consulting company and private laboratory that grew out of the Tohoku University Quantum Annealing Computing Research Laboratory, based in Sendai and led by their CEO Masayuki Ozeki. They started by doing consulting around the creation of software for D-Wave's annealers, using their cloud Leap platform since 2019[2128].

**SolidState.AI** (2017, Canada) develops machine learning solutions for the industry covering yield improvement, production calibration and predictive maintenance.

All of this is based on hybrid classical/quantum algorithms. They work with Bosch, Applied Materials, Mercedes-Benz as well as with D-Wave, Rigetti, Microsoft and IBM Q, among others.

**Spin Quantum Tech** (2018, Colombia) develops quantum algorithms in the field of cybersecurity that combine AI and quantum. They seem to create PQC (post-quantum cryptography) that exploits new encryption algorithms. They are also working on chemical simulation, which has nothing to do with it.

**SHYN** (2016, Bulgaria) develops solutions for the visualization of data coming from quantum calculations. So, some quantum dataviz! With a use case consisting in detecting quantitative fake news. It was co-funded by Google's Digital News Information Fund dedicated to the press. This €150M fund distributed funding of a few 100K€ to more than 400 projects in Europe.

**SoftwareQ** (2017, Canada) offers development software for quantum computing including a compiler, simulator and optimizer. The company was co-founded by Michele Mosca (on top of evolution) and Vlad Gheorghiu of the Canadian Institute of Quantum Computing.

It is a startup from the Quantum Machine Learning program at Creative Destruction Lab in Toronto.

**Strangeworks** (2018, USA, $4M) develops quantum software. Like many colleagues, they target the aerospace, energy, finance and healthcare markets. They are at the origin of the creation of a Q&A site on quantum computing, Quantum Computing Stack Exchange. The company was founded by William Hurley aka whurley, and is based in Austin, Texas, with a staff of about 15.

In 2019, they launched a beta of their multi-platform development environment for quantum applications supporting quantum computers or emulators from Rigetti Computing (Forest), DWave (Leap), Microsoft (Q#), Google (Cirq), IBM (Qiskit) and ColdQuanta Hilbert gate-based cold atom QPU (as of late 2021).

This environment seems to facilitate collaborative work and sharing of results. In February 2021, it turned into an initiative to “humanize quantum computing”.

[2128] See Sigma-i and D-Wave Announce Largest-Ever Quantum Cloud-Access Contract | D-Wave Systems, July 2019.

It's based on Strangeworks QS (Quantum Syndicate) which consolidates quantum hardware vendors solutions and software tools, Strangeworks QC (Quantum Computing), a free quantum computing ecosystem to learn quantum code using common quantum programming languages and Strangeworks EQ (Enterprise Quantum), an enterprise infrastructure solution consolidating QC and QS with better security, IP protection, quantum machine access, resource aggregation, custom integrations, private deployments, project management and the likes. In January 2022, Strangeworks announced a partnership with **Entangled Networks** (Canada) to support their future Multi-QPU Computers Entangled Networks and the associated MultiQopt code compiler. This scale-out approach to quantum computers is of course to be first validated at the hardware level and with physical (optical) connections between several QPUs before becoming operational.

**Stratum.ai** (2018, Canada) develops a quantum software dedicated to a very specialized market, the optimization of mineral prospecting, particularly in gold.

**Super.tech** (2020, USA, $150K) is a startup launched by Pranav Gokhale, Fred Chong and Teague Tomesh that develops a software stack dedicated to the control of quantum computing systems ranging from a hundred to a thousand qubits.

The solution is the result of the NSF-funded Practical-Scale Quantum Computation (EPiQC) research project involving five Chicago-area and MIT universities and stars such as Peter Shor and Aram Harrow. It is creating a software infrastructure targeting the development of NISQ solutions. The company was acquired by ColdQuanta in May 2022.

**Terra Quantum** (2019, Switzerland, $85.8M) was cofounded by Markus Pflitsch (CEO) with, as recent hires, Valerii Vinokur (co-CTO) and Dr Florian Neukart (CPO). Their recent $75M funding helps them expand their R&D effort and also, their international expansion, with the opening of new offices in Munich and in the Silicon Valley.

It develops quantum software solutions in all possible quantum fields: quantum computing, quantum cryptography and quantum sensing. They modestly position themselves as being in a position to build the "*European quantum ecosystem*"[2129] and working on the "*development of revolutionary quantum computing applications*", with supporting the development of more efficient fertilizers, batteries and power grids. Their "Quantum Algorithms as a Service" is an algorithms library with the usual optimizations and QML pieces. The company then customizes it according to the needs of their customers making them a mixed product/services company. Then, their "Quantum Computing as a Service" provides access to its "*high performing logical qubits to equip them with real quantum advantages already today*".

In January 2022, Terra Quantum launched an alpha release of its QMware hybrid quantum cloud data center (QMware), in partnership with Novarion Systems (Austria, covered earlier in this section). Their HPC and quantum emulation workloads are powered by Intel Xeon Platinum CPUs and Nvidia A100 GPGPUs. Their CPU-GPU-QPU interconnect is based on CXL (Compute Express Link), a unified in memory communication open standard protocol created in 2019 by Intel for high-speed CPU-to-device and CPU-to-memory connections, designed for HPCs and built on the PCI Express (PCIe) interface. Their hybrid quantum computing approach is described in a white paper where they claim it outperforms classical approaches[2130].

---

[2129] See Terra Quantum secures EUR10m to build the European Quantum Ecosystem by James Dargan, April 2020. On 1st April, their CTO declared: "*We plan to implement a useful quantum algorithm on the IBM machine with 20 qubits in order to test quantum supremacy*". Well, 20 qubits for a quantum supremacy? It's fine if it's an April's fool.

[2130] See Practical Application-Specific Advantage through Hybrid Quantum Computing by Michael Perelshtein et al, 2021 (14 pages).

Their architecture is based on a memory-centric compute architecture that supports processing with QPUs, CPUs and GPUs as well as hybrid quantum computing with 12 TB of NVRAM shared by all systems. They also developed a unified information theoretical model for classical and quantum information that allows for efficient QPU emulation via a hardware agnostic intermediate representation of the quantum circuits. In 2022, they launched QUANTON-HGX2, a new generation of servers with Nvidia GPGPU and AMD EPYC CPUs.

Their last offering is a "Quantum Security as a Service" using a QKD solution, given the physical architecture and related hardware are not specified.

**Tinubu Software** (France) is a credit insurance software company that is investigating the usage of quantum computing to extend its offering to improve prediction on time series. They had participated to the ClassiQ 2022 quantum computing coding challenge and were the bronze winners with Thomas Frossard, Ayoub El Qadi, Quoc Viet Nguyen and Marcelin Gallezot.

**Tokyo Quantum Computing** (2017, Tokyo) wants to develop quantum annealing computer simulation software like many of the software startups from Japan.

**Tradeteq** (2016, UK, $6.3M) is a financial trading platform that uses AI for risk assessment and portfolio optimization. Their ambition is to use quantum computing to develop their own quantum tools.

In April 2020, they announced that they would work in this direction with the Singapore Management University (SMU) and with quantum neural network algorithms.

**Turing** (2016, USA) is a company created by Seth Llyod and Michele Reilly that wants to solve key societal problems with hybrid classical-NISQ software solutions using AI and quantum machine learning techniques. Seth Lloyd is a famous prolific quantum scientist in quantum computing and Michele Reilly has been working on qRAM.

**Unitary Fund** (international) is a kind of equivalent of the Mozilla Foundation for quantum technologies. It's a non-profit organization creating open source quantum libraries, tools, hardware and content[2131].

They fund through a microgrant program ($4K) developers. They sponsored about 20 projects including the error mitigation framework mitiq[2132], Qrack (emulator accelerator on GPUs), OLSQ (Optimal Layout Synthesizer for Quantum Computing, a pre-compiler optimizer reducing the SWAP gates count), a quantum machine training textbook and Pulser, developed by Pasqal. They partner with Rigetti and IBM.

**Xofia** (2019, USA) develops software solutions based on quantum machine learning for classification. They want to distribute their software in open source. They exploit Atos' 40 qubit quantum emulator, a QLM server, sitting in the cloud.

**Zapata Computing** (2017, USA, $67.4M) is a quantum software and services company founded by Harvard researchers including Christopher Savoie and Alán Aspuru-Guzik from the University of Toronto who has developed many founding algorithms in chemistry quantum applications.

[2131] See their 2021 Annual Report, Unitary Fund (23 pages).

[2132] See Mitiq: A software package for error mitigation on noisy quantum computers by Ryan LaRose, William J. Zeng et al, September 2020-August 2022 (33 pages).

Their partners include Google, Rigetti and IBM. Also, Bosch (Germany) is one of their corporate investors. Honeywell invested in the company in March 2020. The company established an office in the UK in June 2021.

They initially developed a complete quantum operating system serving as a hub between application algorithms and quantum accelerators of all types. In April 2020, this took the form of Orquestra, a platform for managing quantum application workflows with:

**Building code** using their Orquestra Studio. It offers a set of code libraries supporting Cirq (Google), Qiskit (IBM), PennyLane (Xanadu), PyQuil (Rigetti), Q# (Microsoft) as well as pyAQASM (Atos).

**Orchestrate** its deployment across multiple QPU and classical emulation platforms using Zapata Computing Quantum Workflow Language (ZQWL), which is YAML-compatible and supports various quantum hardware architectures (NISQ, quantum annealing) and classical computing (quantum emulators such as those from Atos, supercomputers, cloud servers)[2133]. Orquestra includes tools for managing batch computing. The Orquestra Data Correlation Service (ODCS) collects treatment data in a MongoDB database which is then exported as Excel tables, a Jupyter notebook or for the Table software.

**Deploying** the Orquestra runtime locally or on the cloud.

Orquestra was in beta in April 2020 as part of an Early Access Program and is released since 2021.

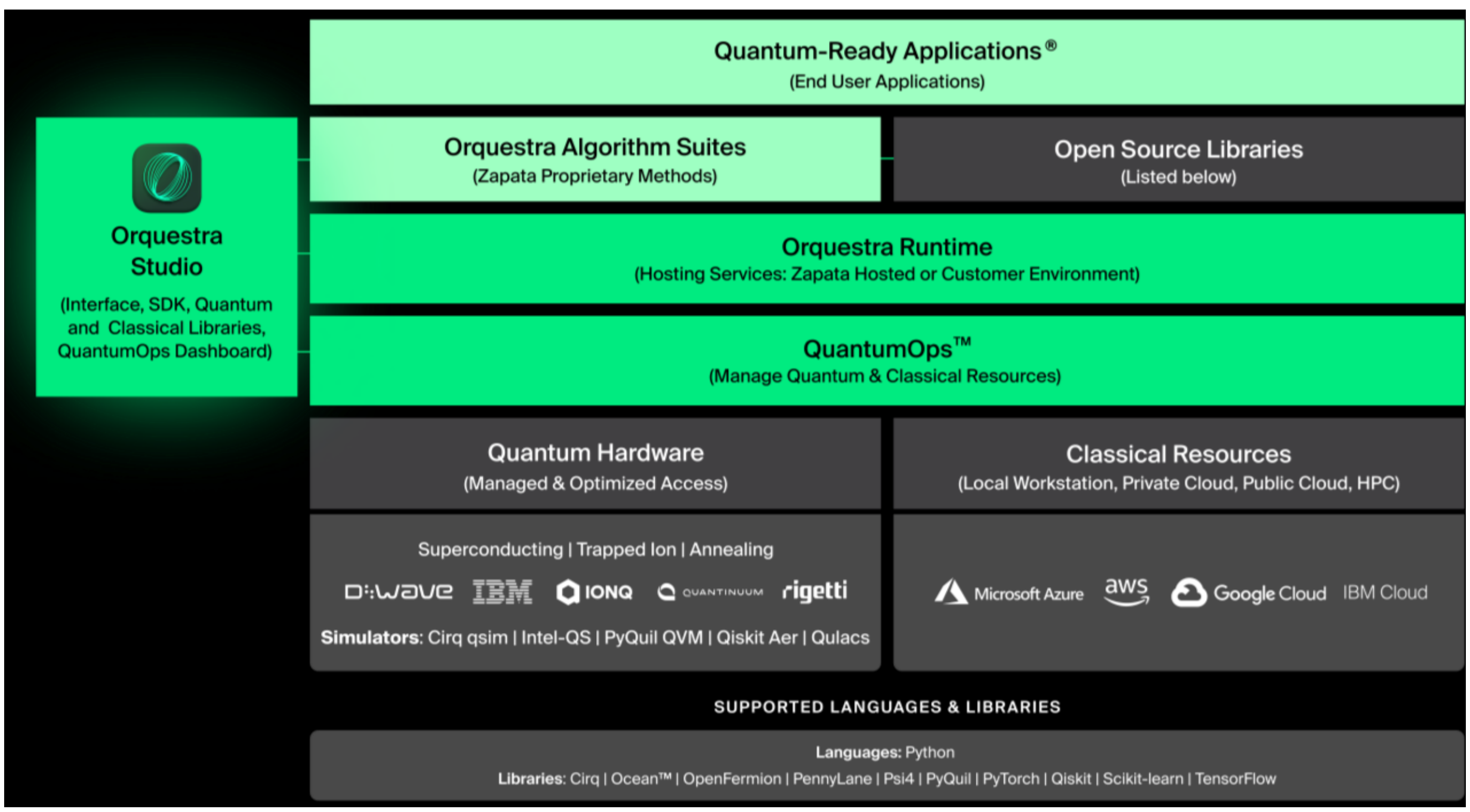

*Figure 694: Zapata Computing Orquestra platform. Source : Zapata Computing.*

# Service vendors

On top of software vendors, the service vendors industry is getting structure to work with end-user customers and help them adopt quantum technologies, mostly focused on quantum computing.

Large IT service vendors like **Accenture**, **Capgemini** and **KPMG** have launched quantum technology practices, often partnering with hardware and software vendors. Accenture teams up with IonQ, Quantinuum, D-Wave and 1Qbit. KMPG is with Microsoft, selling "quantum inspired" projects. Capgemini created its Quantum Lab (Q-Lab) early in 2022, partnering with IBM and positioned in

[2133] YAML is a language that dates back to 2001. It is used to create configuration files. It is used in conjunction with Python.

Germany, Portugal and India. It covers all branches of quantum technologies (computing, telecommunications and cryptography, sensing) and focuses on life sciences, financial services, automotive and aerospace verticals[2134].

Here are a bunch of more specialized quantum services and consulting companies throughout the world: **Safe Quantum Inc** (2020, USA) sells PQC services, **D-fine** (2002, Germany), **Q&I** (UK) also called Qandi, **QuantGates** (UK), **Kvantify** (2022, Denmark), **Quantum Phi** (2018, Czech Republic). **RayCal** (UK) which is an analyst firm in quantum technologies, **Inside Quantum Technology** (2018, UK) created by Lawrence Gasman with the help of 3DR Holdings, **Qureca (**Spain) whose name means Quantum Resources & Careers and which does training, recruitment, business development and events, **Max Kelsen (**Australia), **Quantum Quants** (Netherlands). **SoftServe** (1993, Ukraine), **StrategicQC** (USA) is also specialized in recruiting talent in quantum technologies. **Quantum Computing Engineering** (USA) or QCE, provides generic consulting services on quantum computing. **Aspen Quantum Consulting** (USA) does due diligence investigation of startups for investors. **AmberFlux** (2013, India) provides quantum computing consulting services on top of existing machine learning and data science services. **QRDLab** does about the same, also in India. **Quemix** (2019, Japan) is an IT company specialized in providing quantum computing solutions. **QuRISK** (2021, France) does consulting on quantum computing originated risks on cybersecurity and also develops quantum algorithms and **Q. BPO Consulting** (2020, France) seems to be a generalist quantum computing consulting shop. At last, **Reply Data IT** (Italy) is a quantum computing service company that deployed its custom MegaQUBO solver in the cloud, first on classical computers, and supposedly on quantum computers.

**Plantagenet Systems** (2014, UK) is a consulting service shop run by Roberto Desimone who runs projects for various UK organizations including their Ministry of Defense, InnovateUK, the UK National QT program, the National Quantum Computing Centre (NQCC) and the UK Quantum Hubs.

**Protiviti** (USA) is a large 7000 people consulting company with a quantum practice focused like many other on the potential applications from quantum computing and on securing IT infrastructures against the potential related threats (so, PQC et al). They are partnering with Multiverse Computing.

**Psi-Ontic** (USA) is a quantum consulting created in Florida by Alan Martin, a management consultant coming from the automotive industry and with a physicist PhD. background. He works with large end-user customers as well as with investment funds and startups. He also creates market studies and quantum strategy assessments for various government organizations.

**Quantum Computing Engineering Inc. (QCE)** (2019, USA) is a company created by Gonzalo Florez Giraldo that does consulting for putting in place quantum computing solutions in most addressable markets (optimization, chemistry, healthcare, finance, machine learning). It also addresses cybersecurity and cryptography. The company is based in Houston, Texas. The company has no visible web site.

**Q-iSIM** (2016, Germany) aka "Quantum Interdisciplinary Simulations" develops quantum annealing software in physics simulation and fluids dynamics.

**DN-Quantum Computing** (2019, India) aka DishaNitish Technologies | Quantum Computing is a service company helping corporations develop quantum computing solutions, noticeably in chemistry and with transmon qubits.

**QKrishi** (2021, India) develops quantum models, algorithms and kernels for applications in automotive, finance, agriculture, seismology, signal processing and other areas.

---

[2134] See [Capgemini, IBM Launch Quantum Lab to Promote Quantum Use Cases]() by Matt Swayne, The Quantum Insider, January 2022.

**Silicofeller** (2021, India) develops quantum solutions and advertises doing it in simulation, optimization and quantum machine learning which is fine, and then, crosses the line with mentioning the metaverse and Blockchain as application domains which are a bit farfetched.

**Quanta-ly** (2020, Libya) is preparing Libyan industries for the adoption of quantum products and services with training, consulting and advanced secure communications. No country, even in war, can escape quantum technologies!

**Quanvia** (2022, Spain) is a quantum consulting, service, research and training company targeting the usual suspect industries (retail, finance, logistics, energy, transport, automotive, pharmaceuticals, chemistry).

The company spun out of the University of the Basque Country, Bilbao, Spain. It is run by Enrique Solano (President, also cofounder of Kipu Quantum), Javier Mancilla Montero (Managing Partners), Dawoon Choi (Quantum Machine Learning Director) and Ariela Strimling (Quantum Computing Tech Lead). The team is spread in Spain, Chile and the USA. The Latin America expansion came from the acquisition of Nimoy Cognitive Computing, a data-science service shop.

**Qubitech** (2021, Greece) is a quantum consulting company cofounded by Alexis Askitopoulos (CSO). Not to be confused with QubitTech, a scam crypto trading company that "*uses quantum technology to generate a monthly ROI of 25% for its investors and a total of 250% in 10 months*".

**QuGeeks** (2022, Switzerland) is a jobs board and search company specialized in quantum technologies. Their outreach includes "The Quantum Hubs" newsletter and its related social network activity.

**Quant-X Security & Coding** (2019, Germany) is a quantum service company created by Xenia Bogomolec (CEO) and Peter Nonnenmann (Scientific Advisor and Analyst). They develop quantum algorithms for quantum annealing and gate based systems. They also integrate PQC for dynamic partner authentication in QKD networks.

**Unitary Zero Space** (2020, Finland) quantum consulting and training services company. Their services portfolio includes helping organization commercialize their own quantum technologies, educating customers on quantum technologies and cybersecurity expertise.

They packaged their training offering in a half-day quantum computing workshop format (that's quite short...). The company was created by Topias Uotila (quantum programming), Risto Hakala (cybersecurity) and Juha Muhonen (quantum physics).

## Quantum computing business applications key takeaways

- Most quantum computing market forecasts are highly optimistic and plan for an early advent of scalable quantum computers. They also sometimes tweak forecasts with pushing business value numbers instead of an actual market for quantum technologies.
- There are interesting potential use cases of quantum computing in nearly every vertical market, particularly in energy, chemistry, healthcare, transportation and then finance.
- Most of them are theoretical or have been evaluated at a very low scale given the capacity of existing quantum computers. Some may be useful with advanced noisy computers (NISQ) while most of them will require highly scalable fault-tolerant quantum computing systems (LSQ/FTQC). Others may find their way on quantum simulators.
- In some cases, the potential use cases are in the overpromising twilight zone like simulating very complex molecules, fixing global warming, curing cancers or optimizing large fleets of autonomous vehicles. All these are dubious long-term promises.
- The main purveyor of case studies is D-Wave with its quantum annealer although it has not demonstrated yet a real quantum advantage. IBM is second there, having evangelized a broad number of customers and developers since 2016.
- Beyond computing time, a quantum advantage can also come from the system energetic footprint and/or the precision of the outcome.
- There are already many software vendors in the quantum computing space. How do they strive as there are no real functional quantum computers around yet? They sell pilot projects, develop software frameworks, build quantum hybrid algorithms and create quantum inspired algorithms running on classical hardware. On top of being funded by venture capital! We also cover in this book the burgeoning IT and consulting services in quantum technologies.

# *Unconventional computing*

This part consolidates a set of technologies and companies that propose to significantly increase the computing power of computing machines while not relying on quantum technologies.

We will discuss supercomputing in general, and then cover a wealth of so-called **unconventional computing** paradigms like digital annealing, reversible and adiabatic classical computing, superconducting computing, probabilistic computing, photonics computing and chemical computing[2135].

Many of these avenues have been explored by major players such as IBM for superconducting components or by startups and with ups and downs. Some, such as MemComputing and InfinityQ, go so far as to tout exponential computing accelerations on more or less traditional architectures based on classic CMOS components. To the point of proving implicitly that P=NP in complexity theory, i.e. that the class of problems that can be solved in polynomial time with respect to their size is equal to the class of problems that can be verified in polynomial time. The consensus being that P<>NP, this is obviously questionable!

Unconventional computing paradigms differ in one way or another from Turing's machine-based technologies and Von Neumann's architecture based on control units, computing, registers and memory that are the basis of today's classic computers.

These do not all necessarily bring significant computing acceleration competing with quantum computing. This is particularly the case for the different domains that are part of **natural computing** that use physical elements from nature or are inspired by nature. This includes computers based on biological components like DNA, p-systems, chemical computers or membrane computers, spintronics and neuromorphic processors that are adapted to artificial intelligence processing[2136].

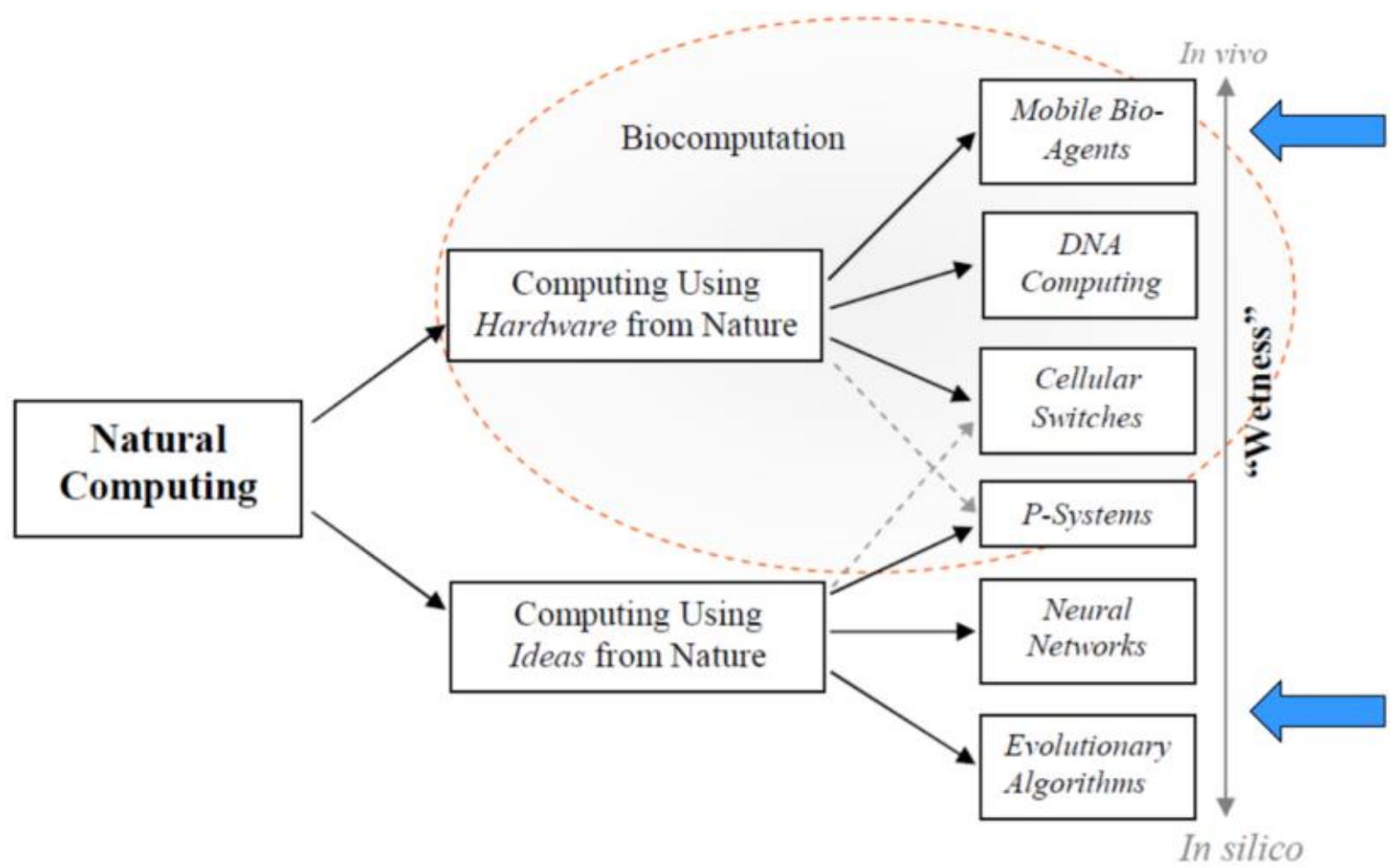

*Figure 695: how biomimetics is used in computing. Source: Unconventional Computing: computation with networks biosimulation, and biological algorithms by Dan Nicolau, McGill University, 2019 (52 slides).*

I won't cover here the special old case of analog computing made of analog electronics has been abandoned in favor of digital electronic systems due to their high flexibility and absence of noise. But for some respect, quantum computing is a return of the analog computing paradigms as we've seen.

Others, such as superconducting classical computing, could be useful to allow quantum computers to "scale". We can therefore also evaluate these different technologies from the point of view of their complementarity, rather than competition, with quantum computing. It explains why this broad topic is incorporated in the enabling technologies part.

[2135] I don't cover analog computing which seems entirely out of fashion. See Analog Computers by Francis Massen (80 slides).

[2136] See Unconventional Computation by Bruce MacLennan, University of Tennessee, who is a reference in the field, October 2019 (306 pages) and Unconventional Computing by Andrew Adamatzky et al, Springer, 2018 (698 pages).

# Supercomputing

The so-called "quantum supremacy" announced by Google in October 2019 systematically referred to power comparisons with supercomputers, in particular with the **IBM Summit** installed at the Oak Ridge laboratory of the US Department of Energy since 2018[2137].

This kind of supercomputer falls within the field of "High Performance Computing", which we will study briefly here to put it into perspective in relation to quantum computing.

The notion of HPC has not always been well defined, particularly since it is a moving target. The power of a supercomputer from the 1980s is now available in a simple recent server if not in your smartphone. However, it is possible to describe the category with its application requirements. HPC and supercomputers are essentially used for digital simulation and the analysis of complex data. These tools are provided to both researchers, public services and industry for their most advanced computational needs[2138].

HPC are used for weather forecasting[2139], organic and inorganic chemistry simulations, aerospace and automotive simulation, nuclear weapon simulation[2140], in finance, more recently in machine and deep learning and, we tend to forget, also to create computer graphics in movie and TV series productions. The mathematical models used in supercomputers are used in particular to solve partial differential equations and to carry out N-body simulations.

These systems are demanding in several ways: in computing capacity, often evaluated in floating-point operations per second, if possible, in double precision (FLOPS), in data storage capacity, and above all, in the ability to transfer data rapidly between storage, memory and processing units. It is in these areas that supercomputers are most distinct from commodity servers used in data centers.

| Function | Energy in Picojoules | |
|---|---|---|
| 8-bit add | 0.03 | computing |
| 32-bit add | 0.1 | |
| FP Multiply 16-bit | 1.1 | |
| FP Multiply 32-bit | 3.7 | |
| Register file access* | 6 | cache and memory access |
| Control (per instruction, superscalar) | 20-40 | |
| L1 cache access | 10 | |
| L2 cache access | 20 | |
| L3 cache access | 100 | |
| Off-chip DRAM access | 1,300-2,600 | |

*Figure 696: in classical computing, the biggest energy cost comes with moving the data and not computing. It may be the same with quantum computing. Source: The End of Moore's Law & Fater General Purpose Computing and a Road Forward, by John Hennessy 2019 (49 slides).*

[2137] We'll see in a later part that this comparison was non-sense and was mixing apples and oranges in an unfair way towards the IBM Summit and HPC overall. With factoring-in the noise generated by Sycamore, emulating it requires only the power of a simple PC.

[2138] See this good review paper on HPCs: Reinventing High Performance Computing: Challenges and Opportunities by Daniel Reed et al, Universities of Tennessee, Utah and ORNL, March 2022 (22 pages).

[2139] As for IBM Weather Channel and its GRAPH (Global Hi-Resolution Forecasting System) forecasting model which is accurate to within 3 km. It is based on an HPC, the Dyeus with 76 nodes of 4 V100 GPUs and 2 Power 9 CPUs. See High Performance Computing for Numerical Weather Prediction at The Weather Company, an IBM Business by Todd Hutchinson and John Wong, 2019 (18 slides).

[2140] This is the role of the supercomputer at CEA-DAM in Bruyères-le-Châtel in the Ile-de-France region, which is fed with data from the Megajoule laser located in Aquitaine.

This table describes the power consumption gap between computing and memory access, which are becoming increasingly expensive the further away memory is from computing. The ratio goes up to more than 1 to 1000! It explains the attempts to bring memory closer to computing units.

Historically, supercomputers such as **Cray's** relied on home-grown vector processors and various proprietary massively parallel systems[2141].

These systems have been swept away over the last decade by cluster-based architectures using standard market processors of the CPU type, complemented in recent years by GPGPUs and A100 (in 2020).

A cluster contains several nodes, each containing several CPUs and/or GPGPUs, themselves multicore, and fast interconnection between these nodes, between clusters, and fast access to data storage, increasingly based on SSDs, which are much faster than hard disks, and at last fast networking access.

Clusters based on standard microprocessors now account for 85% of the 500 largest supercomputers in the world[2142]. The CPUs most often come from Intel (Xeon), AMD (Opteron then EPYC), IBM (Power9) while Nvidia dominates the market with its GPGPUs (general purpose GPUs), including the famous V100 generation Volta launched in 2017 and its successor A100 Ampere announced in May 2020 and the H100 Hopper in March 2022 with 80 billion transistors and a data transfer speed of 3 TB/s with its associated HBM3 RAM. Two thirds of the top HPC now include such Nvidia GPGPUs. This trend accelerated as the OpenMP and OpenACC frameworks were ported to Nvidia GPUs, making these easier to use for a host of existing scientific applications.

This is a form of commoditization of supercomputing, even if these are heavyweight architectures to deploy in large clean rooms. The added value has shifted to the architecture of interconnection, memory and storage, and of course, software.

Interconnection in clusters uses technologies such as Nvidia's NVLink, which connects GPUs and CPUs at high speed. Clusters are interconnected by multiple 200 Gbits/s fiber optic links, often from **Mellanox,** Nvidia's subsidiary since 2019. On a larger scale, HPE is promoting the **Gen-Z** architecture optimized for data access in distributed "data-centric" systems.

Operations on supercomputers are programmed with different development tools. One example is **OpenFOAM**, an open source SDK used to simulate fluid mechanics, chemical reactions, heat transfer, solid mechanics, electromagnetism and also in finance. And besides **LS-DYNA** for structural simulation. Finally, the parallel application development library for Fortran, C and C++ **OpenMP** is very commonly used for scientific computing, as is **OpenACC**. Let's not forget also that there are many optimization algorithms based on practically acceptable approximations, such as for traveling salesman type problems.

Chinese and Japanese vendors are developing their own custom supercomputers microprocessors, in order to limit their dependence on USA companies. In Japan, the **Fujitsu** Fugaku supercomputer uses Fujitsu A64FX chipsets comprising 52 Arm cores and 32GB of HBM2 memory delivering a nominal power rating of 2.7 TFLOPS (processor layout *below*). The Fugaku, which is not a poisonous fish, has a total of 415 double precision PFLOPS with 396 racks and 152,064 processors. Its installation was completed in June 2020 and enabled Fujitsu to win first place on the podium of the world's most powerful supercomputers ahead of the USA with the IBM Summit[2143] .

---

2141 Cray was acquired by HPE in 2019.

2142 The Top 500 is based on a standardized benchmark, the HPL for High Performance Linpack. It is used to solve a set of linear equations using Gaussian elimination using dense matrices and floating number calculus. See the last published version as of the writing of this book: https://www.top500.org/lists/top500/2022/06/.

2143 See Fujitsu and RIKEN Take First Place Worldwide in TOP500, HPCG, and HPL-AI with Supercomputer Fugaku, June 2020 and Japanese Supercomputer Development and Hybrid Accelerated Supercomputing by Taisuke Boku, 2019 (59 slides), Supercomputer Fugaku, 2019 (13 slides) and The first "exascale" supercomputer Fugaku & beyond by Satoshi Matsuoka, August 2019 (80 slides).

China's largest supercomputer is the **Sunway TaihuLight** at the National Supercomputing Center in Wuxi. With a capacity of 93 PFLOPS, it uses 40,960 SW26010 256-core 64-bit RISC architecture home-built processors (with simplified instruction set).

As of mid-2022, China had deployed 35% of the world's Top 500 supercomputers, ahead of the USA with 26%, but only 4% of the TOP50 for 38% in the USA and 8% for Japan, France, Germany and Russia.

In September 2021, the DoE started the installation of its Frontier new generation supercomputer in its Oak Ridge lab, the Aurora system built by HPE with 9400+ Cray EX nodes, each equipped with one AMD Epyc CPU and four Radeon Instinct MI250X GPUs. Operational since 2022, it currently provides 1.1 exaflops of HPC and AI computing power and consume only 21 MW, reaching a record of 52 GLOPS/W[2144].

The European Union launched the **EPI** (European Processor Initiative), a project aiming to bring technology independence with supercomputers multicore microprocessors as well as in car embedded systems. It mainly involves German and French players, notably **Atos**.

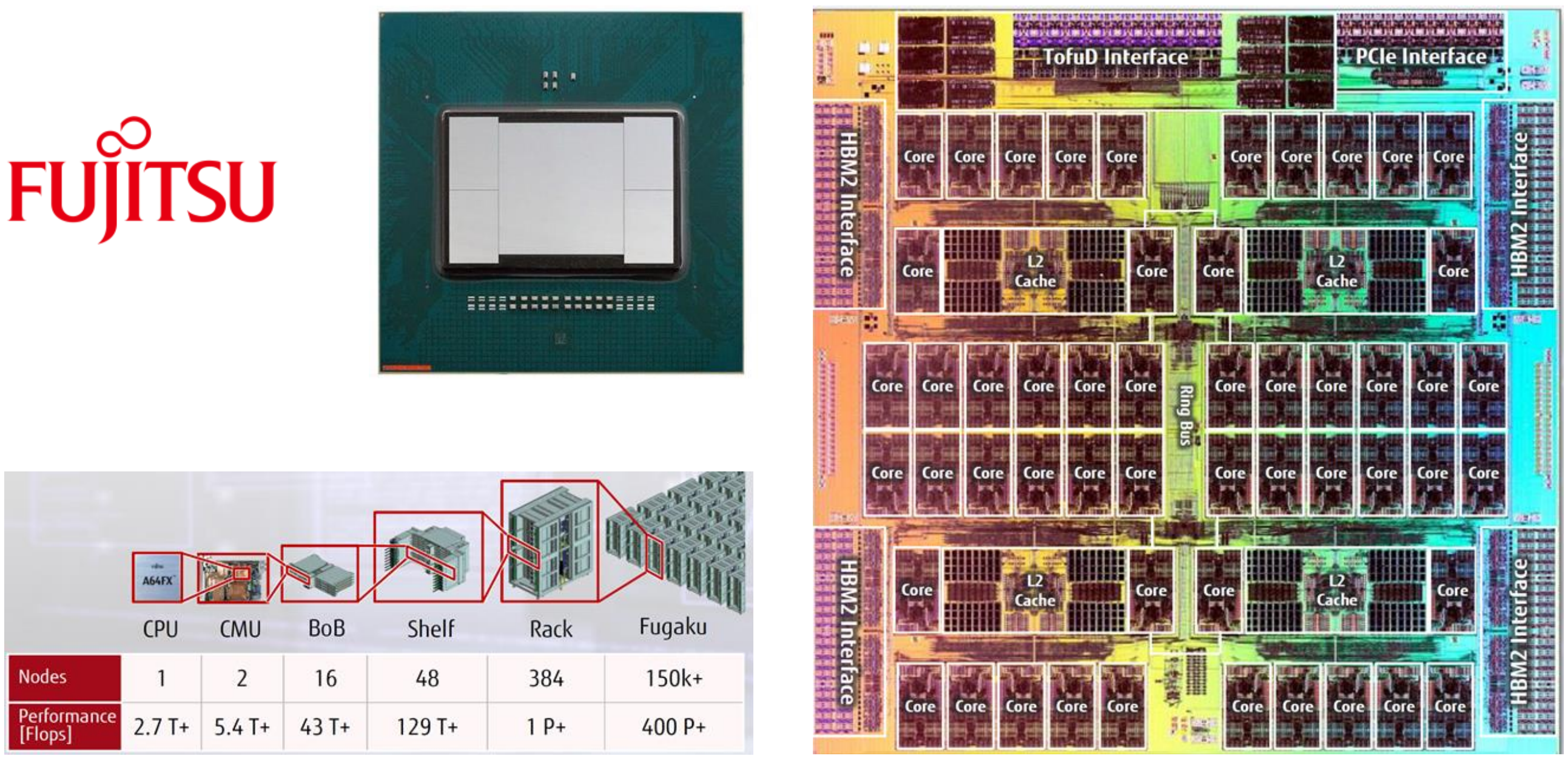

*Figure 697: Fujitsu's Fugaku supercomputer is one of the largest in the world. It uses Fujitsu A64FX chipsets containing each 52 Arm cores and 32GB of HBM2 memory. Source: Fujitsu.*

The effort is carried by the startup **SiPearl**, led by Philippe Notton. It is part of the **EuroHPC** project to create pre-exaflops and exaflops supercomputers, including one in Germany and one in France. The planned budget is 1B€, half of which will be allocated evenly between the European Union and its Member States.

[2144] See US Closes in on Exascale: Frontier Installation Is Underway by Tiffany Trader, HPCwire, September 2021.

In France, the **Jean Zay** supercomputer deployed at GENCI on behalf of the CEA, CNRS and Inria is equipped with 2696 Nvidia V100 GPUs and over 3462 Intel Xeon Cascade Lake CPUs. It is cooled by "hot water", from 30°C to 42°C. It was deployed as part of the French AI plan announced in 2018. Use cases are scientific simulation and machine learning.

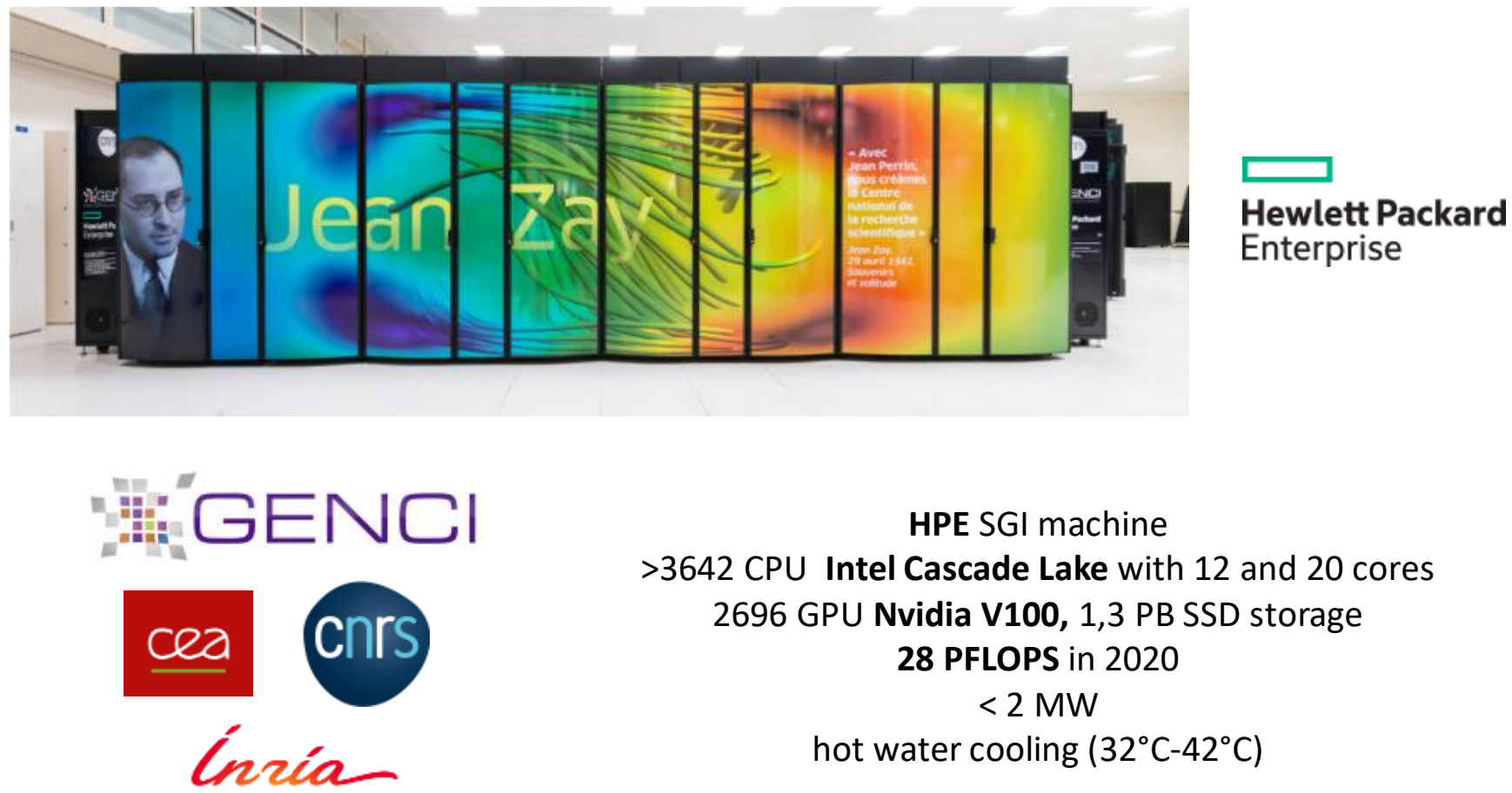

*Figure 698: the Jean Zay supercomputer in France is typical of the new generation of HPCs launched since 2018 with a mix of CPUs and GPGPUs from Nvidia. Source: GENCI.*

The GENCI computing center is due to house a quantum accelerator, probably from **Pasqal**, within a few years. It will be integrated into a hybrid computing architecture. Many other European HPC centers have similar plans, in Germany, Italy and the Netherlands, among others.

Since the advent of the cloud, HPC and supercomputing resources are now available on demand. Cloud data center do not necessarily provide HPC resources. This depends on the servers and clusters deployed architectures and on the packaging of the cloud vendor’s offering.

This notion can be associated with the notion of hyperscale, which covers the capacity of a cloud infrastructure to adapt to the increasing customer computing needs.

Machine learning and deep learning applications are the most recent applications implemented on supercomputers, particularly since they are using GPGPUs that run tensors enabling efficient matrix operations, which are very common in neural networks. In practice, however, most supercomputers continue to run scientific simulation applications.

The market for microprocessors dedicated to machine learning and deep learning acceleration has been booming for several years. A wide variety of approaches have been adopted by its vendors.

It can be represented as below on two axes: in the Y axis, the level of cores specialization, and on the X axis, the number of cores.

**Google**'s TPUs (Tensor Processing Units) are highly specialized for training neural networks, especially convolutional image recognition networks. Nvidia's GPUs contain thousands of classical arithmetic calculation cores as well as hundreds of tensors for matrix computing, which optimizes their versatility.

The most extreme processor is the **Cerebras V2** launched in 2021 with its 850,000 cores. It's a 21 cm square chipset containing a hefty 2.6 trillion 7 nm transistors and 40 GB of integrated SRAM ultra-fast cache memory. Their first version launched in 2019 is already deployed in two test servers at the DoE in the USA with one and two of these chipsets. The first benchmarks published late 2019 did show excellent performance in neural networks training. In 2022, TotalEnergies found 100x speedups vs conventional GPGPU with various material simulations applications[2145]. The company raised a total of $720M, a similar amount to PsiQuantum.

[2145] See Cerebras shows off scale up AI performance for big pharma and big oil by Timothy Prickett Morgan, TheNextPlatform, March 2022.

Finally, **FPGAs** are dynamically programmable circuits that allow the creation of custom circuits at rather low cost and high flexibility[2146]. These are used by some cloud vendors such as Microsoft (with its Brainwave chipsets) and Chinese cloud companies like **Alibaba** and **Baidu**.

Some of these cloud players are developing their own supercomputers. **Google** has created its TPU pods over several generations for its data centers. A TPU v3 board contains four TPU chips, each with two cores, with 16 GB of HBM memory for each TPU core. A TPU v3 Pod has up to 2048 TPU cores and 32 TB of memory.

**Graphcore** (2016, UK, $692M) is another contender in this crowder market. It is now using 3D chips bonding using a TSMC technology, the main chip containing processing units and the secondary underneath chip containing connectivity between the computing chipsets cores. Its current processors have 1,472 cores and 900MB on-chip SDRAM memory, which helps them outperform Nvidia chipsets on some tasks[2147].

**Nvidia** integrates its A100 GPUs in SuperPods totaling 140 DGX A100 and 1120 A100 servers and 4 Po of storage, for 700 PFLOPS. These FLOPS are however not necessarily the same as those used to evaluate the TOP 500 supercomputers. Vendor communication is sometimes misleading.

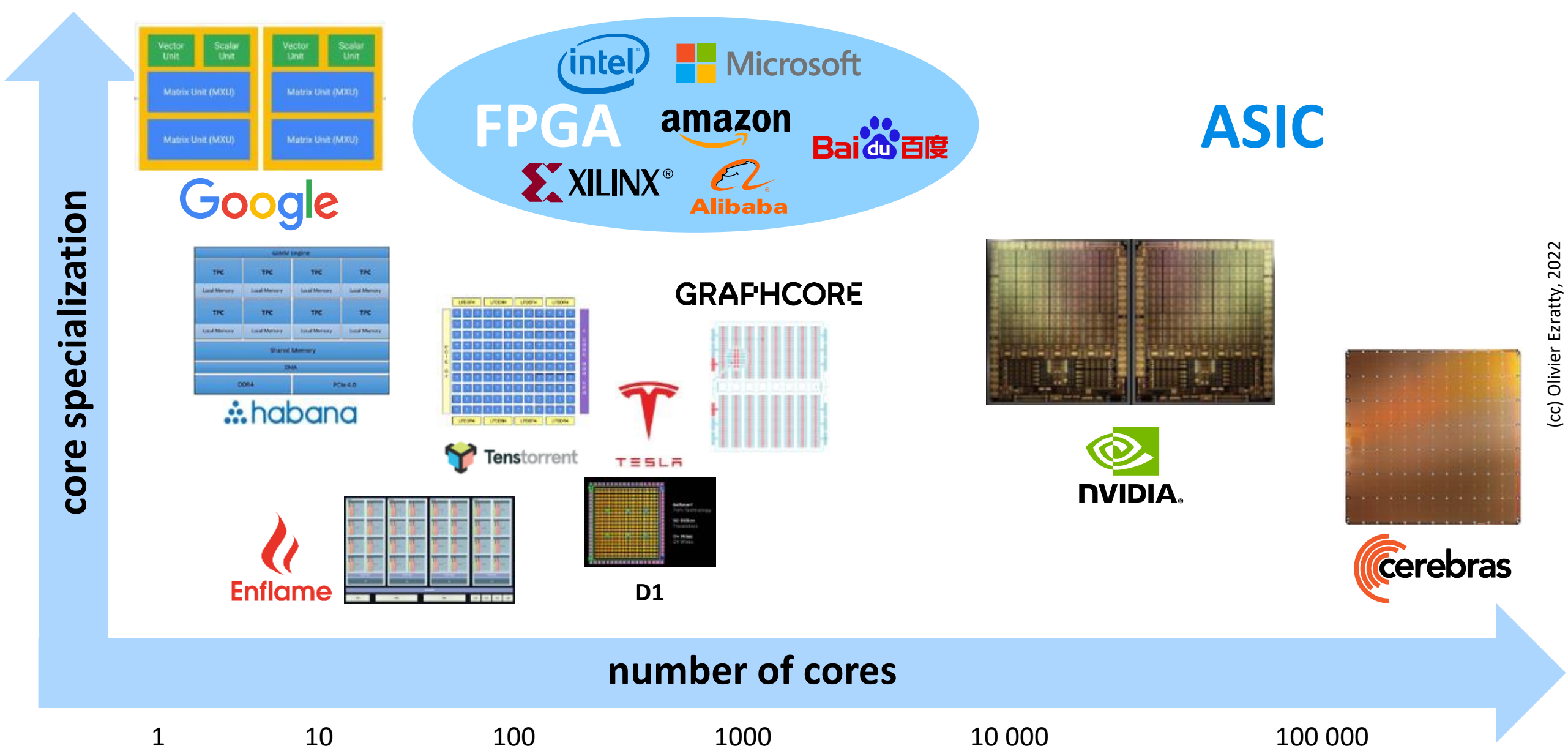

*Figure 699: a map of the tensor-based processors with two dimensions: the number of cores and their specialization. The more specialized cores are in Google's TPU with large tensor operations capacity (128x128 values) while Cerebras's wafer scale chipset has 845 000 relatively simple cores. (cc) Olivier Ezratty, 2020.*

Both to improve performance and to reduce energy consumption, there are a number of ways to make these calculations more efficient: **Approximate Computing**, which reduces precision in neural network training and/or inferences without affecting the results, **Quantization**, which switches from floating-point calculations to integer computing during or after training, **Binary Neuron Networks**, which is even simpler with 1 bit output neurons taking us back to the Perceptrons era of 1957 and **Sparse Computing**, which allows computations to occur on compressed representation of matrices, this being useful only for sparse matrices, without prior decompression.

---

[2146] The FPGA market is currently dominated by two vendors: Intel, after its acquisition of Altera in 2015, and AMD, after its acquisition of Xilinx in 2020. In 2019, Xilinx had a 52% revenue market share and Intel about 35%.

[2147] See Graphcore Uses TSMC 3D Chip Tech to Speed AI by 40% Unveils plan for $120-million "brain-scale" supercomputers in 2024 by Samuel K. Moore, IEEE Spectrum, March 2022.

Last but not least, there is the close integration between memory and computing capabilities. For example, the French startup **UpMem** offers DRAM memory modules integrating dozens of RISC-V cores to perform in-memory computing and speed-up certain processes by a factor of 10, particularly for big data applications. It is also possible to tune the cores clock frequency when they are waiting for data from memory.

Supercomputers are quite energy hungry. Their increasingly powerful microprocessors consume several hundred of Watts. A third of the electrical energy consumed by a data center is spent on cooling. Specialized server racks now easily consume more up to 30kW. It has now reached the point where liquid cooling is preferred for removing heat from components, usually with water. This provides greater efficiency.

Whatever happens with quantum computers, supercomputers will always be relevant. Applications using large amounts of data are not suitable for quantum computing, even with zillions of qubits. Indeed, data loading time in qubits is a huge bottleneck because it relies on very long series of quantum gates that are not as fast as classical data processing. Applications adapted to quantum computing should not rely on high-volume data feeds. This is the case with weather forecast which requires heavy data sets. It will rely on classical supercomputing for a long time despite some exaggerated claims[2148].

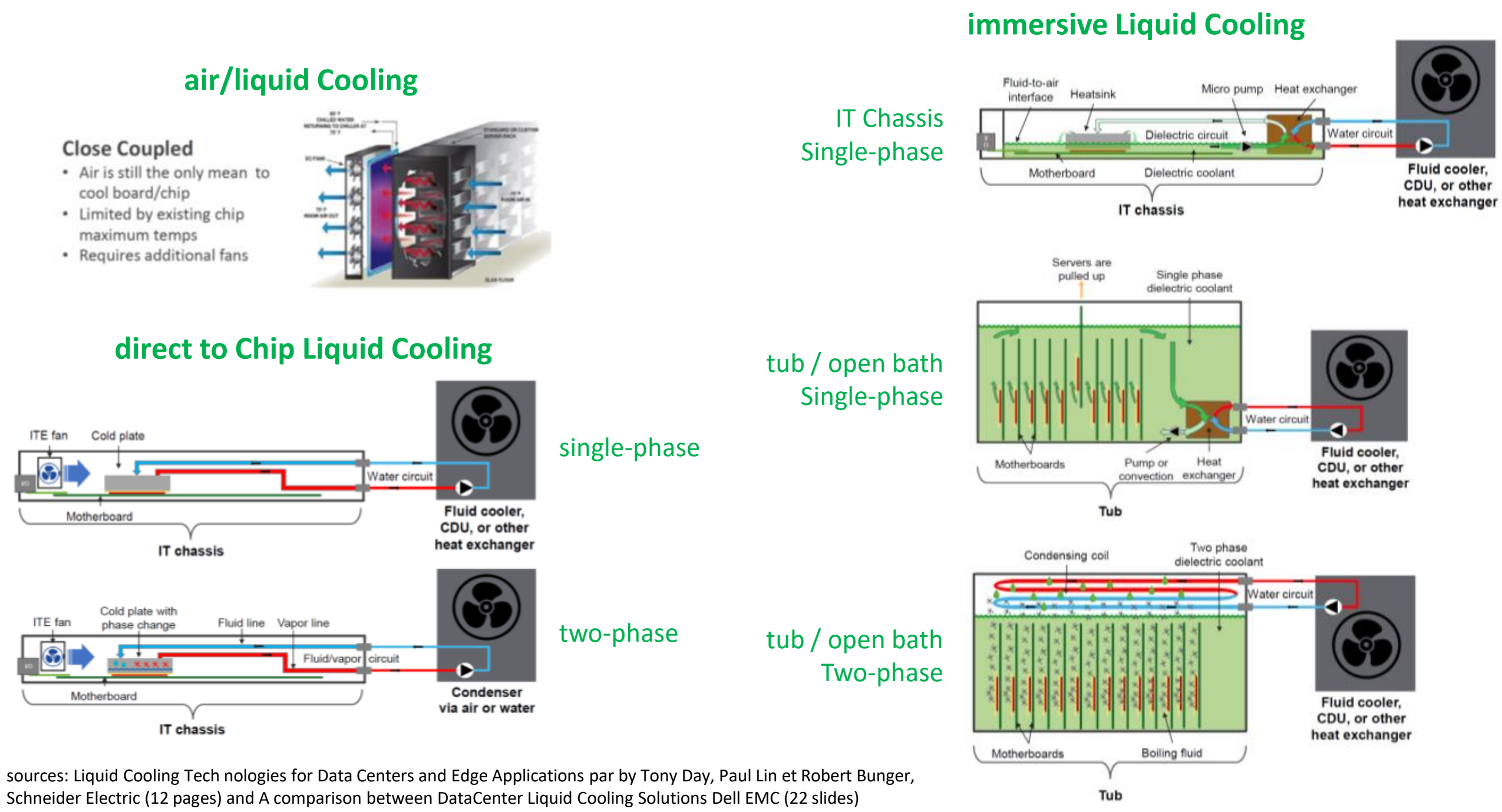

*Figure 700: the various ways to cool a server. Sources: Liquid Cooling Technologies for Data Centers and Edge Applications by Tony Day, Paul Lin and Robert Bunger, Schneider Electric (12 pages) and A comparison between DataCenter Liquid Cooling Solutions Dell EMC (22 slides).*

It is used to solve complex problems of combinatorial or minimum energy research. It is likely that for a long time to come we will have hybrid architectures combining classical computers or supercomputers and quantum accelerators. This is the approach adopted by supercomputer suppliers like **Atos**.

[2148] See Forecasting the Weather Using Quantum Computers by 1Qbit, 2017. The paper references CES 2019: IBM unveils weather forecasting system, commercial quantum computer by Abrar Al-Heeti, January 2019, which covers two entirely unrelated announcements from IBM, one on weather forecasting using classical computing and another, related to their Q System One, both introduced at CES 2019. See also Rigetti Enhances Predictive Weather Modeling with Quantum Machine Learning, December 2021 who does this with a mere 32 qubits!

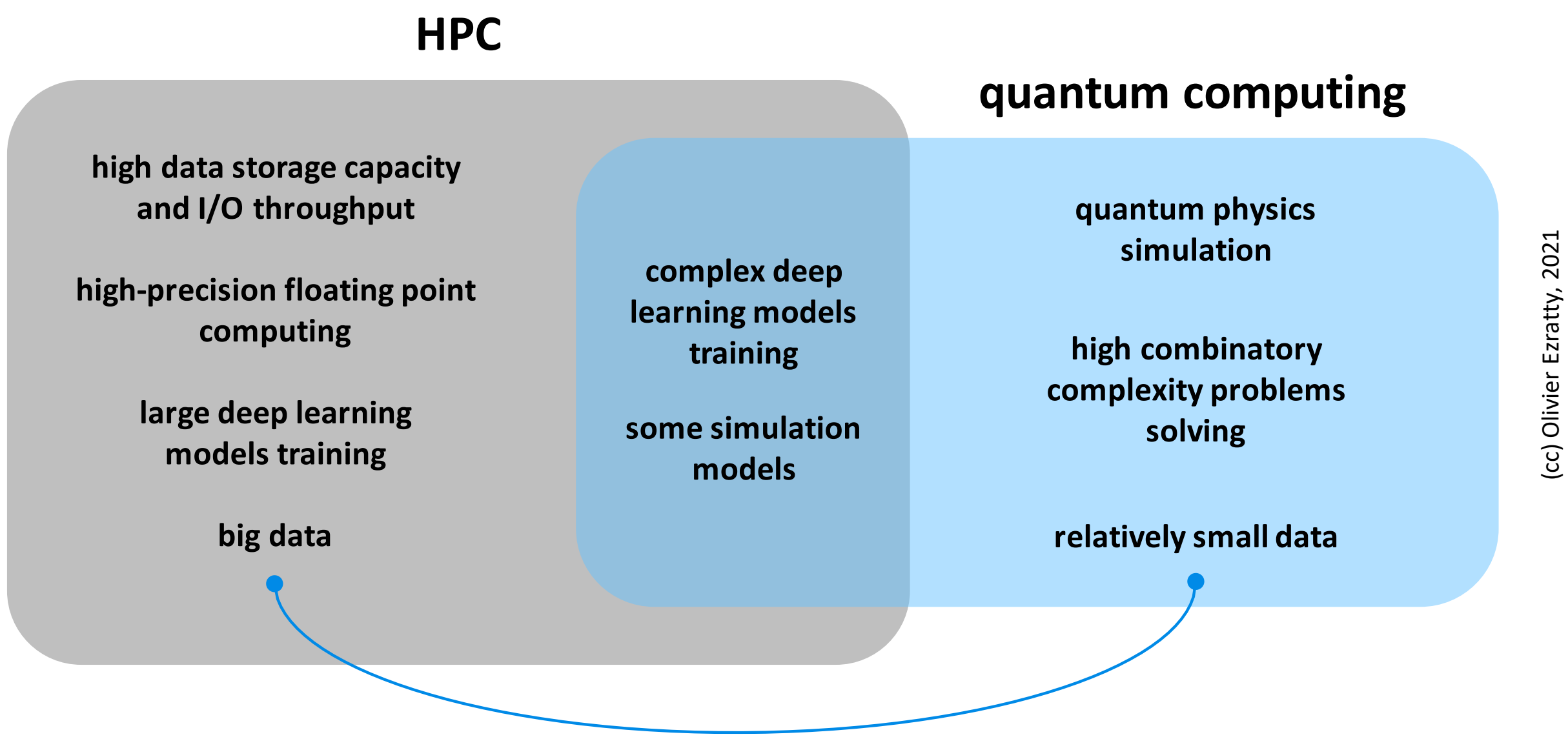

*Figure 701: how to position HPCs vs (scalable) quantum computers. HPCs are for big data and high-precision computing. Quantum computing will be adapted to high complexity problems but with relatively reasonable amounts of data. There's some cross-over between both systems and they will work in sync in many cases. (cc) Olivier Ezratty, 2021.*

## Digital annealing computing

Digital annealing is a non-quantum variant of quantum annealing used in D-Wave computers. It has the advantage of exploiting standard component production technologies in CMOS. The level of acceleration provided at the calculation level is not a priori exponential.

Several industry vendors are in this market with **Fujitsu** and **Hitachi**. This part also includes some related solutions coming from **MemComputing** and **InfinityQ**. Let's also mention the support of annealing simulation by the **Atos** QLM with a capacity to handle 50,000 variables and an optimized implementation of the SQA algorithm (simulated quantum annealing).

**Fujitsu** announced in early June 2018 a digital annealing computer operating at room temperature. Fujitsu is one of the world leaders in the supercomputer market with IBM, HPE and Atos. It was therefore logical that they explored ways to upscale their HPC offering.

It is supposed to scale much better than D-Wave quantum annealers[2149].

The technology is developed on CMOS classical components in partnership with the University of Toronto. It is already proposed as a cloud offering. It is used to solve optimization problems and to carry out molecule screening in biotechs. The dedicated chipset contains 1,024 bit update blocks incorporating memory to store their weights with a precision of 16 bits, logic blocks to perform value inversions and the associated control circuits. This is reminiscent of memristor-based neural networks. As with D-Wave, problems are loaded into the system in the form of matrices with biases in the links between elements and the system looks for a minimum energy state to solve the problem. It has some familiarity with the Ising model used in D-Wave.

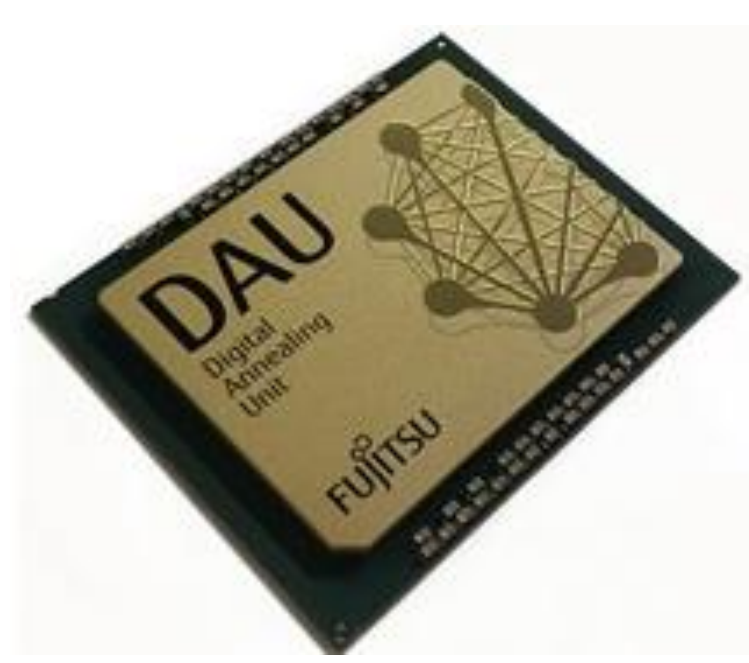

*Figure 702: Fujitsu's DAU processor for implementing optimized digital annealing. Source: Fujitsu.*

[2149] See Fujitsu's CMOS Digital Annealer Produces Quantum Computer Speeds, 2018.

Its designer, **Hidetoshi Nishimori** of the Tokyo Institute of Technology, believes that Fujitsu will be able to create solutions that outperform D-Wave. In 2019, Fujitsu announced its second generation of chips with 8,192 blocks. They expect to reach one million blocks thereafter.

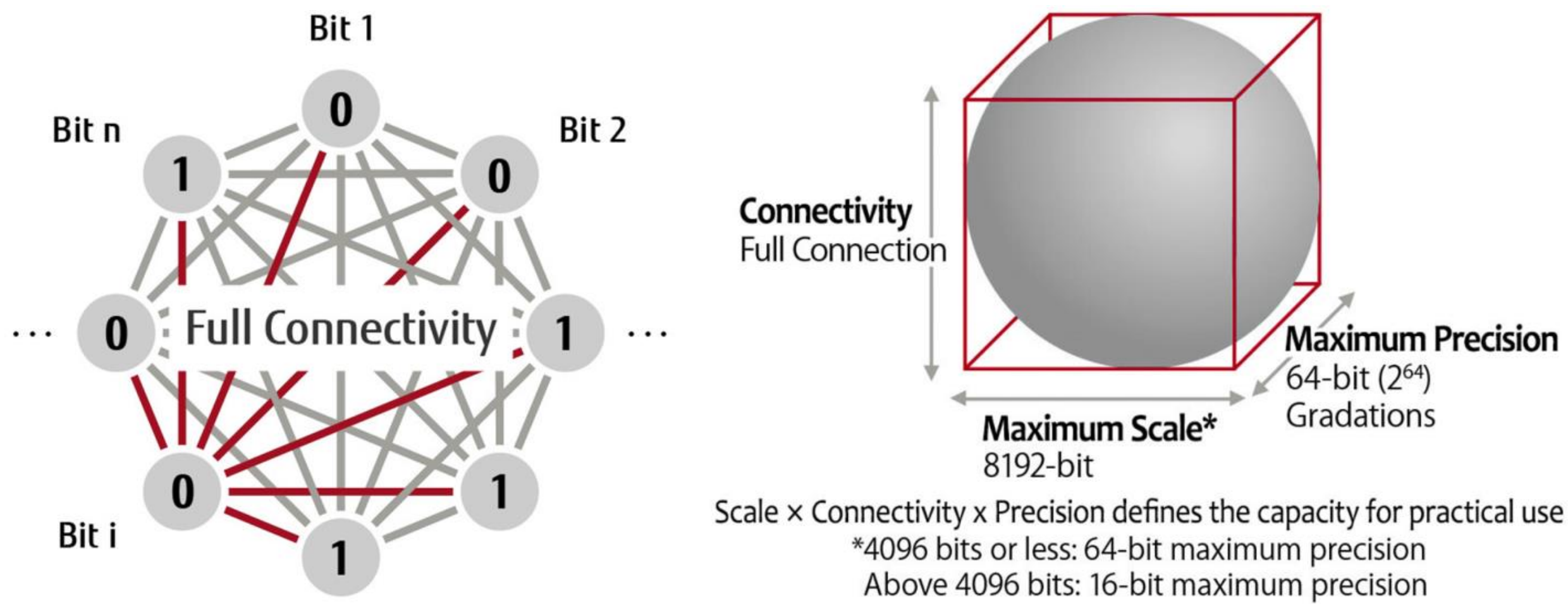

*Figure 703: Fujitsu's DAU high-level architecture. Source: Fujitsu.*

Development tools are provided by **1QBit,** in which Fujitsu has made an investment. Fujitsu has been collaborating since April 2020 with **Quantum Benchmarks (**Canada) on quantum algorithms and error suppression codes, based on an IA Fujitsu algorithm and their experience with their digital annealing[2150]. Fujitsu is also partnering with **Entanglement** (USA) which developed a Covid vaccine logistics optimization solution with its annealer.

HITACHI Fujitsu is not the only Japanese company exploring digital annealing. Hitachi also launched a related initiative although, contrarily to Fujitsu, it seems it did stay in research lab and didn't reach commercialization.

Their system is implementing a hardware solution to solve Ising models. It's mixing probabilistic and deterministic approaches running in a CMOS component to find some energy minimum of a combinatorial problem expressed as an Ising model[2151]. It doesn't find the absolute minimum to the problem but an acceptable solution. The CMOS uses SRAM to store the virtual "spin states" of the Ising model problem. Hitachi stated that it could help solve combinatorial optimization problems such as the travelling salesman problem efficiently. The architecture was first relying on FPGAs which makes sense given it didn't reach volume production.

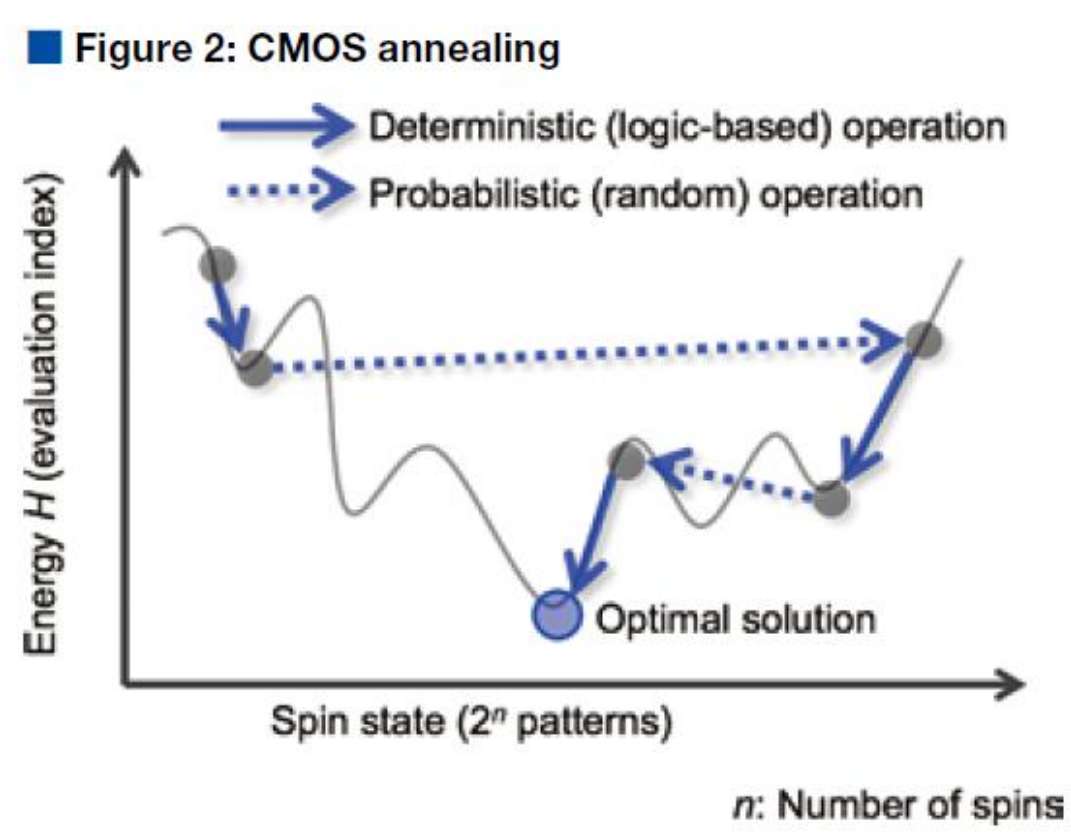

*Figure 704: CMOS annealing principle.*

[2150] See Fujitsu Laboratories and Quantum Benchmark Begin Joint Research on Algorithms with Error Suppression for Quantum Computing, April 2020 and Fujitsu Laboratories and Quantum Benchmark Begin Joint Research on Algorithms with Error Suppression for Quantum Computing by Fujitsu, March 2020.

[2151] See CMOS Annealing Machine – developed through multi-disciplinary cooperation by Hitachi, Ltd., November 2018, Overview of CMOS Annealing Machines by Masanao Yamaoka, January 2019 (4 pages) and CMOS Annealing Machine: an In-memory Computing Accelerator to Process Combinatorial Optimization Problems, April 2019.

The mysterious startup **MemComputing** (2016, USA) can be positioned in a category close to Fujitsu's offer. It is a solution inspired by quantum annealing computation. They use the principle of invertible computing units, able to circulate data both ways, from input to output and output to input.

It also uses oscillating Boolean gates implementing a non-Von-Neumann computing model and some sort of tunnel effect[2152]. Their hardware solution MemCPU Coprocessor is to place memory next to computing units in processing unit[2153]. These memristor-like computing cells have symmetrical inputs and outputs interconnected to neighboring cells. It computes exclusively with integer numbers. There is no floating-point computing at all. They would automatically find a complex balance of a parameterized system. This is the principle of SOLGs (Self Organizing Logic Gates) in Figure 706[2154].

The company positions its technology as competing with quantum computing with a market ready solution although its real existence and packaging is in question[2155].

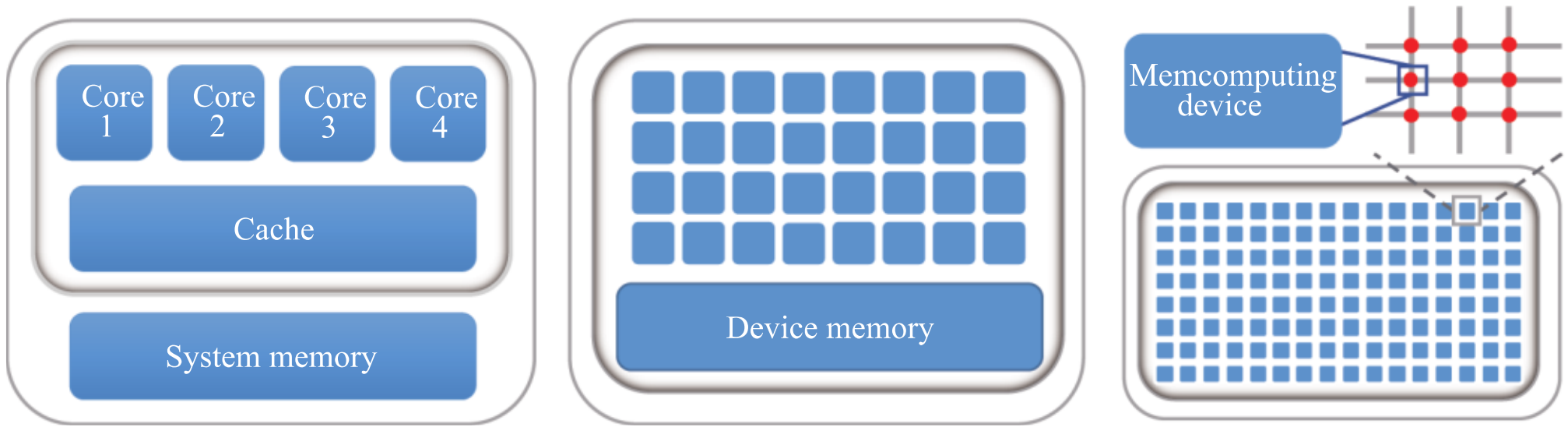

(Color online) Comparison of CPU, GPU, and MPU.

*Figure 705: MemComputing architecture compared with classical computing. It's basically a mix of in-memory processing with bidirectional computing. Source: MemComputing.*

The company was founded by **John Beane**, a serial entrepreneur, with two physics researchers, **Massimiliano Di Ventra** and **Fabio Traversa,** who have done extensive work on memory computing[2156].

Their architecture is supposed to solve various classes of NP-complete and NP-difficult problems in polynomial time such as 3-SAT problems[2157]. They tout significant performance gains such as four orders of magnitude for machine learning applications, i.e., performance multiplied by 10,000!

[2152] See Global minimization via classical tunneling assisted by collective force field formation by F. Caravelli, F. C. Sheldon and Fabio L. Traversa, February 2021 (15 pages).

[2153] It is described in Memcomputing: fusion of memory and computing by Yi Li et al, 2017 (3 pages) where this schema comes from.

[2154] SOLGs are described in the patent Self-Organizing Logic Gates and Circuits and Complex Problem Solving With Self-Organizing Circuits, March 2018 (37 pages). It is further detailed in Coupled oscillator networks for von Neumann and non von Neumann computing by Michele Bonnin, Fabio L. Traversa and Fabrizio Bonani, arXiv preprint, December 2020 (29 pages).

[2155] See MemComputing vs Quantum Computing by MemComputing, August 2022.

[2156] See Universal Memcomputing Machines by Fabio Traversa and Max Di Ventra, 2014 (14 pages) and Perspective: Memcomputing: Leveraging memory and physics to compute efficiently by Fabio Traversa and Massimilio Di Ventra, 2018 (16 pages).

[2157] See Memcomputing NP-complete problems in polynomial time using polynomial resources and collective states by Fabio Traversa, Massimilio Di Ventra et al, 2014 (10 pages) and Evidence of an exponential speed-up in the solution of hard optimization problems by Fabio Traversa et al, 2018. Then, See this Conference from Massimiliano Di Ventra at Berkeley in 2016 (26 minutes).

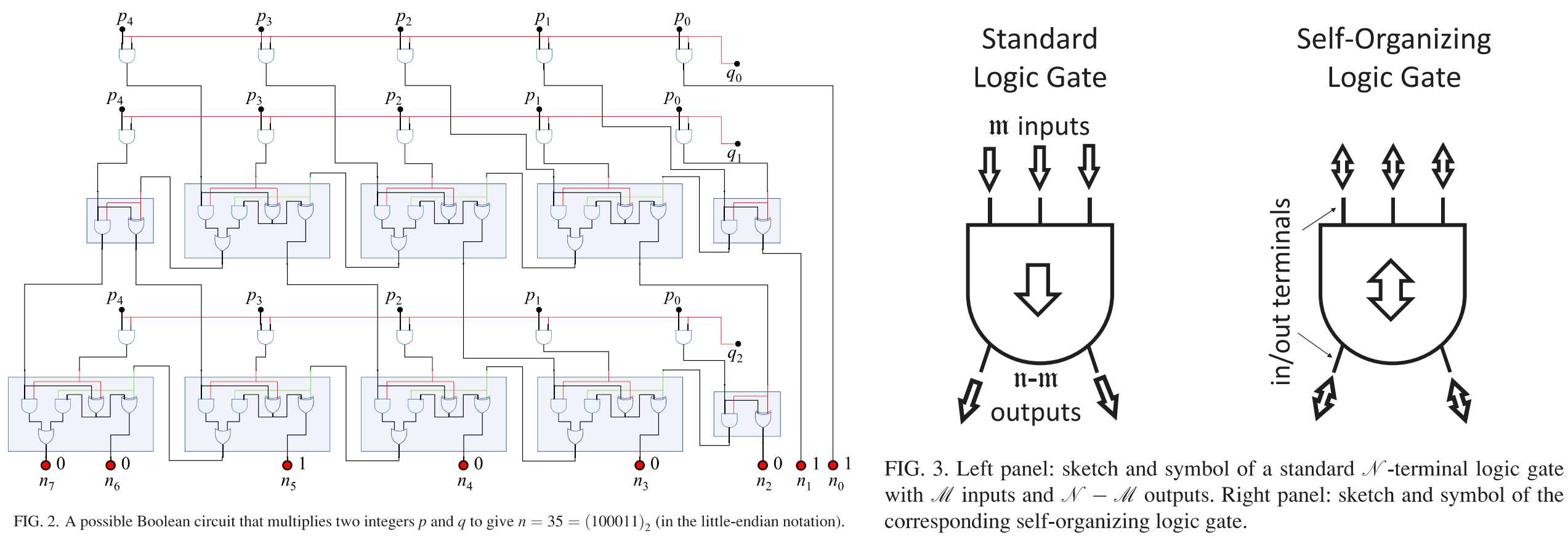

*Figure 706: SOLGs (Self Organizing Logic Gates) from MemComputing. Source: MemComputing.*

The application domains include the resolution of planning and optimization problems such as the traveling salesperson problem, combinatorial optimizations and searches[2158], bioinformatics, neural network training[2159] and even integer factoring[2160], each time with the promise of an exponential gain in computing time compared to traditional computing.

For the moment, their solution is only emulated in conventional computers like with the AMD EPYC microprocessor and provided as an SDK operated in the cloud they have designed in partnership with **Canvass Labs** (2017, USA). Their electronic component is not yet manufactured, even at the prototype stage, and it is not clear whether it is possible to manufacture it and would deliver the speedup promises. They announced that they would need to embed some memristor technology in their component, which can create significant delay in the manufacturing given this technology is not really mature.

They handle problems such as MIPLIB (Mixed Integer Programming Library), which are considered intractable with a 60-second response time on a server running Linux, and have even beaten a D-Wave. This is used to find a combination of given integers that can generate zero when added together (the "Subset Sum problem"). The startup manages to obtain a quantum scale advantage by emulating its process on traditional processors. This amounts to challenging all current theories of complexity. In short, it makes you dizzy.

In April 2020, MemComputing announced that it would make its XPC (Xtreme Performance Computing) software stack available in the cloud for researchers working on Covid-19[2161] .

So, is this technology simply revolutionary and could it nullify many efforts in quantum computing, or are there one or more shortcomings? There are plenty of them. How do you initialize the system so that it is close to a global minimum? What is their real capacity to create these SOLGs in current CMOS components? How is noise managed in their system[2162]?

---

[2158] See Stress-testing memcomputing on hard combinatorial optimization problems by Fabio Traversa, Massimiliano Di Ventra et al, 2018 (6 pages).

[2159] See Accelerating Deep Learning with Memcomputing by Haik Manukian, Fabio Traversa and Massimiliano Di Ventra, 2018 (8 pages).

[2160] See Polynomial-time solution of prime factorization and NP-hard problems with digital memcomputing machines by Fabio Traversa and Massimiliano Di Ventra, 2017 (22 pages).

[2161] See MemCPUXPC SaaS Platform available free for COVID-19 Research, 2020.

[2162] They provide an answer in Directed percolation and numerical stability of simulations of digital memcomputing machines by Yuan-Hang Zhang and Massimiliano Di Ventra, April 2021 preprint on arXiv (12 pages).

Is the system scaling well? Their approach would not be scalable according to several specialists including Scott Aaronson[2163].

In 2022, MemComputing published two papers in partnership with a scientist from the DoE Los Alamos lab on the potential effect of a tunneling effect with memristors to find global minimum, a task commonly used to solve various optimization and machine learning problems[2164]. All this remains very theoretical.

**InfinityQ** (2019, Canada, $6M) stated that they had built the "*first quantum computing CMOS microchip technology to work at room temperature*". It is supposed to be an analog quantum computer with between 1000 and 5000 qubits.

Their qubit architecture is "*based on an artificial atom in a lambda configuration*" which "*exploits superposition and entanglement to achieve quantum computing without the burden of maintaining very fragile quantum objects*". It is a cloud native platform and quantum analog-computing technology that can run any coding language, any problem up to 100,000x faster than an average laptop, with the same energy consumption as a lightbulb. Their first-generation machines is said to solve complex optimization problems, linear system and FFTs (fast Fourier transforms). That's quite a heavy stack of promises.

They created a proof of concept with 10 qubits late 2020 and announced a 100 qubits MVP in 2021 for and full commercialization after 2025. In April 2021, they mentioned that they had solved a traveling salesperson problem with 128 cities while other "*non-classical machines*" have solved such problems with a maximum of 22 cities (which is not really true). This system is currently programmed in C language.

The company was launched by Aurelie Hélouis (CEO) with Kristina Kapanova (CTO, who later left the company) and was supposedly backed by Philippe Dollfus, a CNRS Research Director in France who is specialized in computational nanoelectronics and John Mullen, former Assistant Director of the CIA. They also made business claims such as having projects starting with major unnamed UK and Canada banks, with a Swiss pharmaceutical and the Canadian government.

So, how can this feat be accomplished[2165]? How is it different than what MemComputing is doing? They share some similarities: use CMOS components and solve complex problems in a polynomial way. The difference seems that MemComputing is using an invertible digital logic while InfinityQ is based on some analog processing. It also looks like an analog annealing system. Their technology could be classified as some sort of reservoir computing running on CMOS doing rabi flops emulation[2166]. Their quantum, superposition and entanglement claims seem totally overexaggerated. In 2022, the company made a pivot, did reset its web site and claimed that its solution was "*A quantum-inspired technology to serve gaming & metaverse's intensive computation demands*". That smells quite fishy to move from the financial sector to the metaverse and gaming sectors.

---

[2163] See A Note on `Memcomputing NP-complete problems' and (Strong) Church's Thesis by Ken Steiglitz, 2015 (2 pages) which quickly demonstrates that this is not possible. The same goes for Memrefuting by Scott Aaronson in 2017 and for A review of`Memcomputing NP-complete problems in polynomial time using polynomial resources' by Igor Markov, 2015 (3 pages).

[2164] See MemComputing Announces Collaboration with Los Alamos National Lab, January 2022, Global minimization via classical tunneling assisted by collective force field formation by Francesco Caravelli et al, Science Advances, December 2021 (9 pages) and Projective Embedding of Dynamical Systems: uniform mean field equations by Francesco Caravelli et al, MemComputing and DoE Los Alamos lab, January 2022 (45 pages). PEDS, projective embedding of dynamical systems (PEDS).

[2165] See a video of their presentation, December 2020 (17 mn).

[2166] See Quantum reservoir computing with a single nonlinear oscillator by L. C. G. Govia et al, Raytheon, January 2021 (9 pages).

## Reversible and adiabatic calculation

Since the 1960s, researchers have been considering reducing computer power consumption by several orders of magnitude based on the principle of adiabatic reversible computing[2167].

The goal is primarily energy-based. It does not accelerate computing. In most cases, it is even contradictory with Moore's law, as the main techniques used result in a calculation speed decrease.

All this is due to our understanding, since the 1960s, of the link between computation and thermodynamic processes. **Rolf Landauer** created in 1961 the equation according to which the process of information processing that dissipates energy is related to memory erasure[2168]. The erased information is turned into heat sent outside the computer, increasing the environment entropy. Rolf Landauer estimated that the dissipated energy was always greater than **kTln(2)** per erased bit, k being Boltzmann's constant ($1.38\times10^{-23}$ J/K), T the temperature in Kelvin and ln(2) the logarithm of 2 (about 0.69315). At room temperature, this gives 0.017 eV. This is the famous Landauer limit[2169].

More generally, Landauer's limit illustrates the link between the notions of logical and physical reversibility of computation. The first is linked to the ability to determine the input values of a calculation according to the output values.

The second is that the unfolding of a physical process in reverse may not violate the laws of physics, including the inescapable second law of thermodynamics according to which the entropy of a thermodynamic system always increases unless the process is reversible.

Today, a CMOS component spends 5000 eV energy to erase one bit, almost 300,000 times more than the Landauer's limit. One could gain an order of magnitude and go down to 500 eV, but that would still be 30,000 times more than the Landauer limit. So, in order to reduce the computing energy consumption, why not avoid erasing information and, in the process, make all computing physically and logically reversible?

This would require a review of all current computational logic that relies at low-level on irreversible logic gates that destroy information, such as NAND or XOR gates that generate one bit from two bits.

In 1973, **Charles Bennett**, another IBM researcher and colleague of Rolf Landauer's, imagined a calculation method that would avoid this energy-dissipating erasure of information without requiring an infinite memory[2170].

---

[2167] To write this part, I used the many references from the excellent presentation Reversible Adiabatic Classical Computation - an Overview by David Frank, 2014, IBM (46 slides), as well as from The Future of Computing Depends on Making It Reversible by Michael P. Frank, 2017 and The Case for Reversible Computing by Michael P. Frank, 2018 (19 pages). See also Computers That Can Run Backwards by Peter Denning and Ted Lewis, 2017 and Theory of Reversible Computing by Kenichi Morita, 2017 (463 pages). See also the review paper Quantum Foundations of Classical Reversible Computing by Michael P. Frank and Karpur Shukla, April 2021 (70 pages).

[2168] See Irreversibility and heat generation in the computing process by Rolf Landauer, in IBM Journal of Research & Development, 1961 (9 pages).

[2169] Landauer's limit was experimentally verified fifty years later, in 2011, by a team from ENS Lyon in Sergio Ciliberto's group. See Experimental verification of Landauer's principle linking information and thermodynamics by Antoine Bérut et Al, 2011 (4 pages) and Information and thermodynamics: Experimental verification of Landauer's erasure principle by Antoine Bérut, Artyom Petrosyan and Sergio Ciliberto, ENS Lyon, 2015 (26 pages). Other experiments followed to validate this, with magnetic memories, such as Experimental test of Landauer's principle in single-bit operations on nanomagnetic memory bits by Jeongmin Hong et al, 2016 (6 pages). The principle consists in lowering the energy barrier of the bit state transition when an operation is required and then to raise it again to preserve the bit state. See also Delft's 2018 experiment in Quantum Landauer erasure with a molecular nanomagnet by R. Gaudenzi et al, 2018 (7 pages).

[2170] See Logical reversibility of computation by Charles Bennett, IBM Journal of Research and Development, 1973 (8 pages). Charles Bennett is also the creator of BB84 codes with Georges Brassard, which laid the foundations of quantum key distributions.

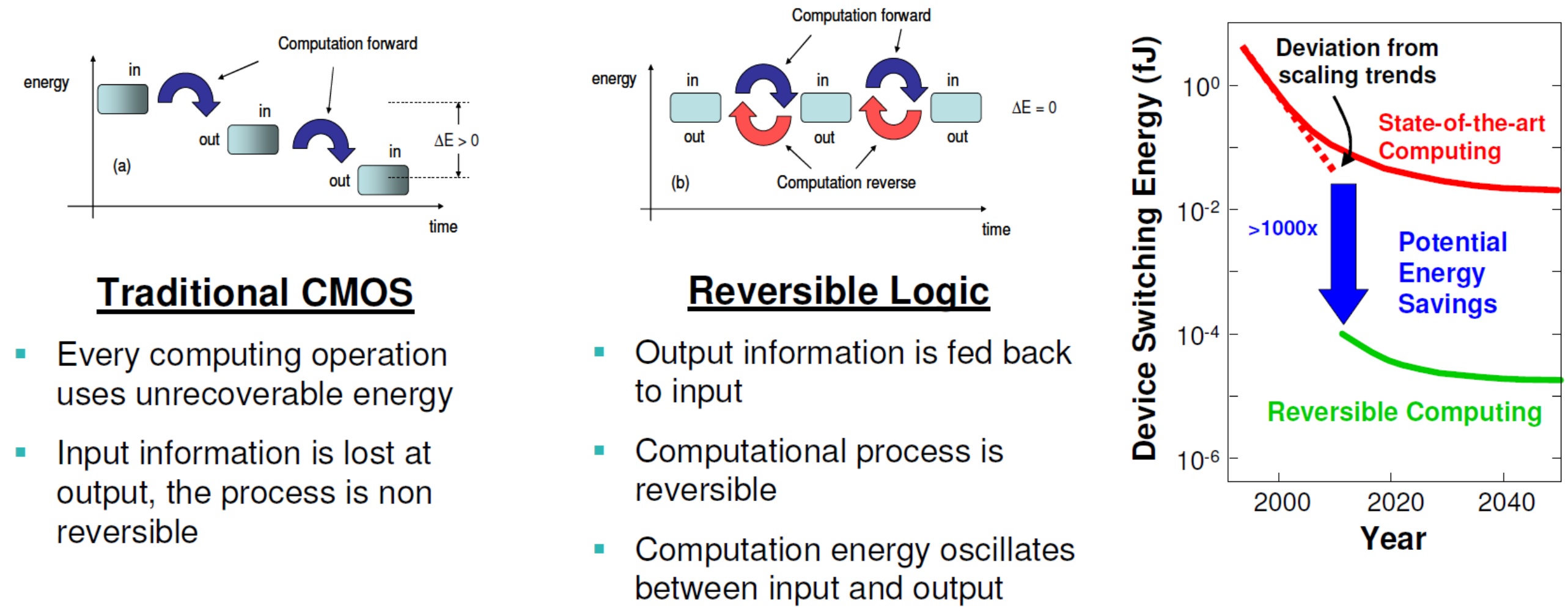

*Figure 707: why reversible logic could help save energy. But it won't make it faster. Source: Reversible Adiabatic Classical Computation - an Overview by David Frank, 2014, IBM (46 slides).*

He was followed by **Edward Fredkin** and **Tommaso Toffoli** who, in 1978 and 1982, imagined reversible logical gates inspired by a metaphorical physical model based on billiard balls, the BBM for billiard ball model[2171]. These logic gates have as many outputs as inputs and it is easy to understand why they become reversible. Although their model was not practically feasible with contemporary electronics, it was then applied to the quantum equivalent of these gates that we already covered.

**Konstantin Likharev** proposed in 1976, then in 1982, to implement this reversible computational logic by manipulating the energy levels of superconducting Josephson junctions, under the name of "parametric quantrons"[2172]. In 1991, this became the "quantron flux parametron" (QFP), capable of operating up to 10 GHz and developed by a Japanese team[2173].

This led then in 2003 to **Vasili Semenov'**s idea to use nSQUID circuits to realize these circuits, the n meaning "negative" because of a negative inductance that connects two SQUIDs of the device. As for any Josephson junction, it works under cryogenic temperature[2174].

Reversible computing is often associated with adiabatic computing but one could work without the other. The general principle of adiabatic computing is illustrated *below*: in a classical calculation, the energy barrier to switch the state of a system between a) and c) is high. In a quasi-adiabatic calculation, a physical system lowers the state energy transition barrier (in d) to trigger it (in e) and then in f, by raising the level of the barrier to its normal state.

The processing energy cost is thus lowered by approaching Landauer's limit. The high level of the non-calculation barrier guarantees the stability of the information managed outside of this operation. Lowering the barrier and raising it are often managed by trapezoidal voltage control of the transistors instead of looking like a square wave signal.

Between 1985 and 1993, reversible or partially reversible CMOS and CCD computing components were designed.

[2171] See Conservative Logic by Edward Fredkin et Tommaso Toffoli, International Journal of Theoretical Physics, 1982 (35 pages).

[2172] He tells this story in Josephson Digital Electronics in the Soviet Union by Konstantin Likharev, 2012 (18 slides).

[2173] See Quantum Flux Parametron: A Single Quantum Flux Device for Josephson Supercomputer by Mitsumi Hosoya et al, June 1991.

[2174] See an explanation of the process in Engineering and Measurement of nSQUID Circuits by Jie Ren, 2012 (26 slides). nSQUIDs are double SQUIDs connected by a negative inductance. SQUID = Superconducting Quantum Interference Device, a system used to accurately measure the magnetism of superconducting Josephson effect loops. These nSQUIDs were manufactured by Hypres.

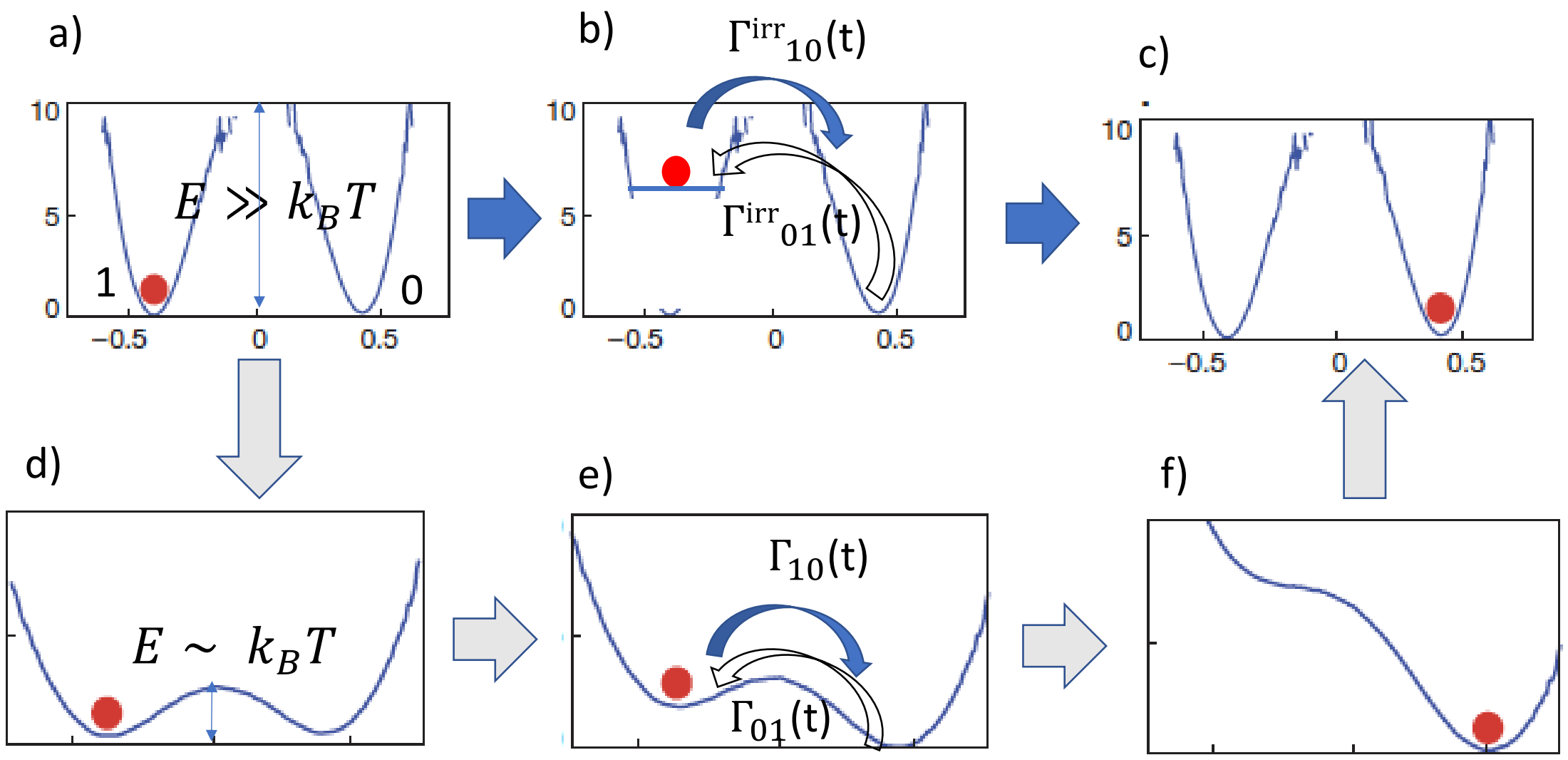

*Figure 708: the thermodynamic principle of reversible computing. Source: "Thermodynamics of computing, from classical to quantum" by Alexia Auffèves, May 2020 (11 pages), adapted from [Experimental verification of Landauer's principle linking information and thermodynamics](#) by Antoine Bérut et Al, 2011 (4 pages).*

**Craig Lent** then proposed in 1997 an adiabatic computation system based on quantum dots and cellular automata (QCA for Quantum dots Cellular Automata) to operate up to 100 GHz[2175] .

In the same manner, **Krishna Natarajan** suggested in 2004 to use MEMS (electro-micro-mechanical components) to drive the trapezoidal voltage control necessary to create adiabatic CMOS components with a very low energy dissipation of 1 eV[2176] .

The idea was pursued by a team from **CEA-Leti** and **Delft** in the Netherlands in 2017 and **Ralph Merkle** in 2019, with prototype circuits based on this kind of technology[2177].

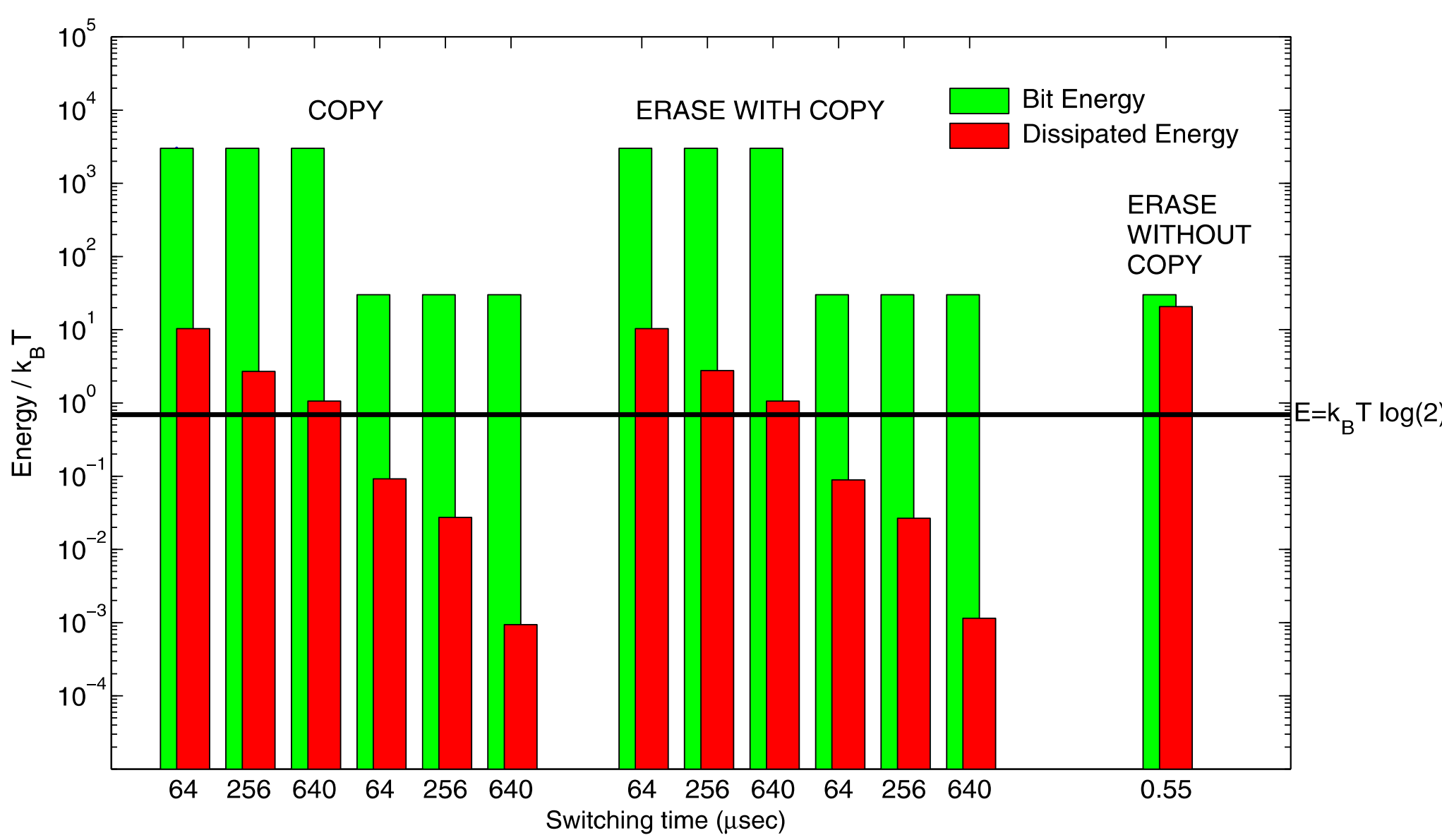

*Figure 709: Source: [Experimental Test of Landauer's Principle at the Sub-kBT Level](#) by Alexei Orlov, Craig Lent et al, 2012 (5 pages).*

---

[2175] See A Device Architecture for Computing with Quantum Dots by Craig Lent and Douglas Tougaw, 1997 (17 pages).

[2176] See Driving Fully-Adiabatic Logic Circuits Using Custom High-Q MEMS Resonators by Krishna Natarajan et al, 2004 (7 pages).

[2177] See Adiabatic capacitive logic: A paradigm for low-power logic by Gaël Pillonnet et al, CEA-Leti, 2017 (5 pages) and Mechanical Computing Systems Using Only Links and Rotary Joints by Ralph Merkle et al, 2019 (34 pages).

In 2012, **Alexei Orlov** et al. experimentally validated Landauer's limit and, above all, the possibility of overcoming it (from below) with reversible calculation, all with a few discrete classical electronic components, resistors and capacitors[2178]. Their experiment showed that a bit copy or erasing with copy could be done with an energy lower than Landauer's limit, at the price of slowing down the operation as shown in Figure 709.

Pure and simple erasing did consume more energy than Landauer's limit. The model was safe! And it all worked at room temperature. In 2019, Alexei Orlov's team from Notre Dame University in Indiana produced the equivalent of an 8-bit microcontroller using a subset of a RISC-type MIPS instruction set with 5766 transistors, 40% of which are adiabatic (Figure 710)[2179]. This seems to be, to date, the most successful realization of a reversible adiabatic processor. It remains however experimental and far from industry requirements. Its industrialization could be of interest to create microcontrollers for low power connected objects.

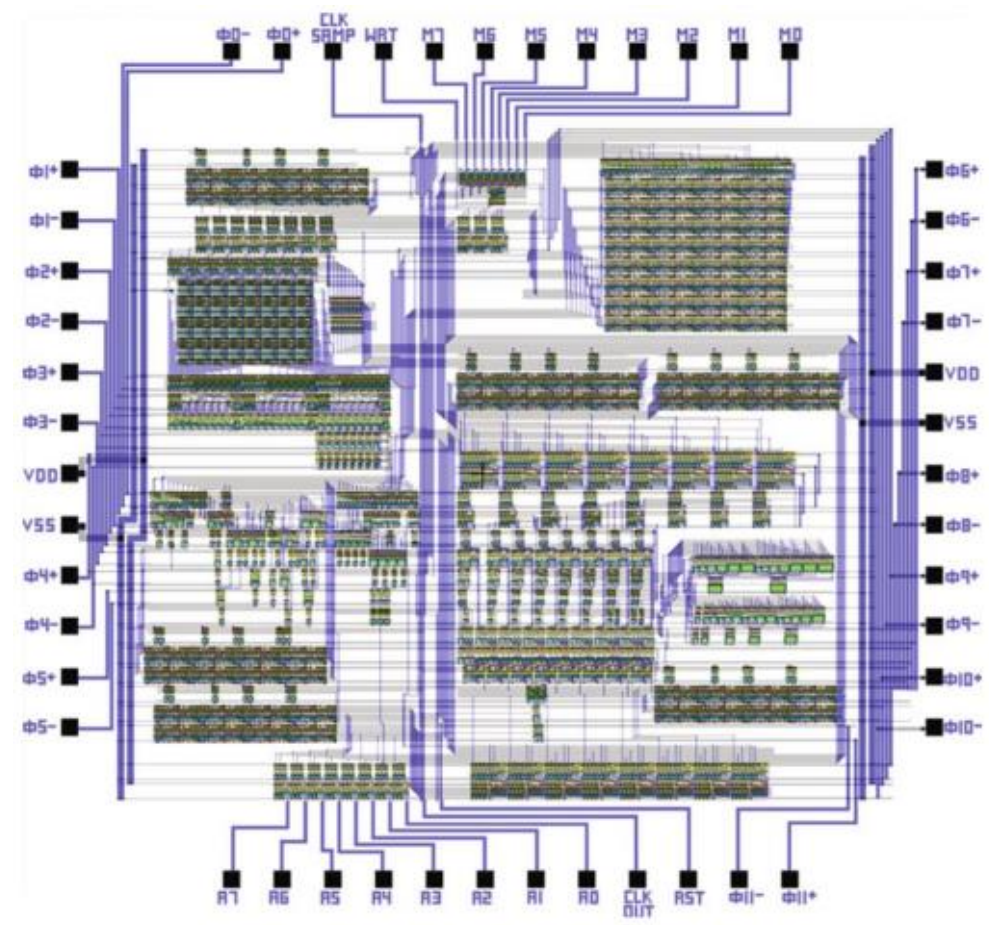

*Figure 710: Source: Experimental Tests of the Landauer Principle in Electron Circuits, and Quasi-Adiabatic Computing Systems by Alexei O. Orlov et al, 2012 (5 pages).*

However, this adiabatic CMOS technique requires a larger number of transistors. Therefore, emerges a new trade-off between a larger and more expensive to design and manufacture power saving component vs cheaper but more energy hungry conventional counterparts. Still, the environmental cause has recently revived interest in reversible and adiabatic computing. It is promoted by the Computer Community Consortium group of the American **Computing Research Association** with the lead from Michael P. Frank's team at **Sandia National Labs**[2180].

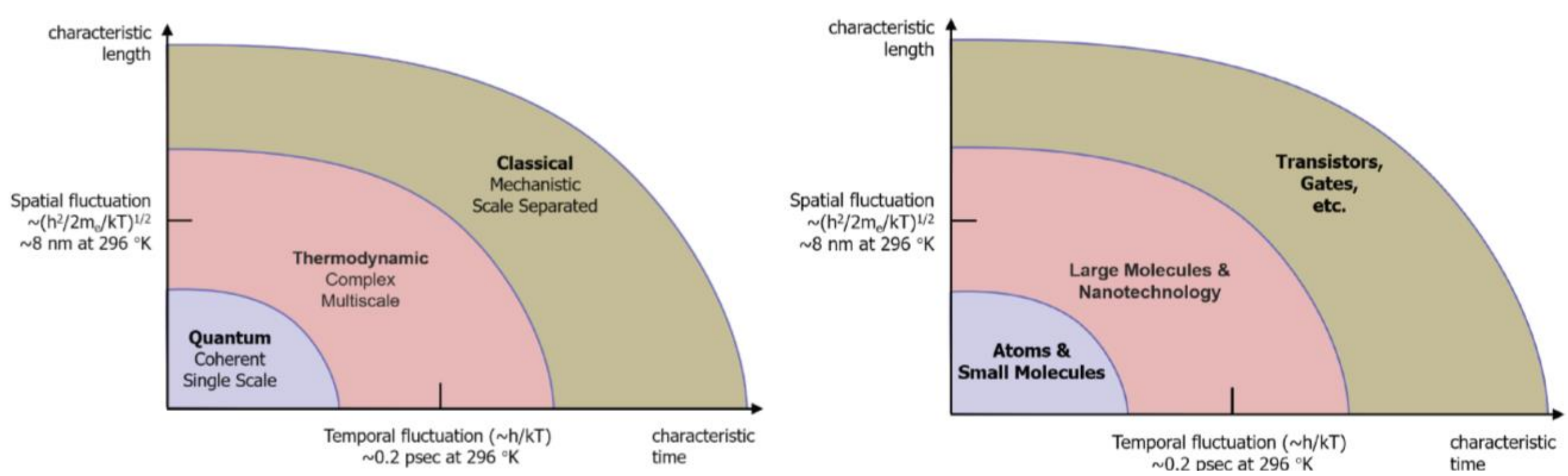

*Figure 711: Source: Thermodynamic Computing, Computer Community Consortium of the Computing Research Association, 2019 (36 pages).*

[2178] See Experimental Test of Landauer's Principle at the Sub-kBT Level by Alexei Orlov, Craig Lent et al, 2012 (5 pages).

[2179] See Experimental Tests of the Landauer Principle in Electron Circuits, and Quasi-Adiabatic Computing Systems by Alexei O. Orlov et al, 2012 (5 pages) which is integrated in Energy Limits in Computation by Craig Lent, Alexei Orlov et al, 2019 (245 pages).

[2180] See Thermodynamic Computing, Computer Community Consortium of the Computing Research Association, 2019 (36 pages). It is a manifesto to develop thermodynamically responsible computing inspired by biomimicry. The document is the result of a workshop with about 40 participants, almost all American except one researcher from London and another from Luxembourg, from universities and a few private actors such as Google, Rigetti, HPE, Knowm (which develops memristor-based circuits for AI applications, their kT-RAM technology) and Daptics (ex Protolife, created by Norman Harry Packard (1954) and which develops chemical simulation algorithms).

They position it in an intermediate architecture between classical and quantum computing, but on quantities that are not necessarily relevant (dimension of components and duration of state fluctuation, see Figure 711)[2181]. The purpose of the manifesto is to obtain US federal credits to finance this research. So, the story is not over!

## Superconducting computing

The idea of creating superconducting computers capable of taking advantage of the lack of resistance of low temperature electronic components dates back to the early 1960s. Its history evolves in parallel with reversible and adiabatic computing. It started with the discovery of the Josephson effect in 1962. This effect was later used to create two-states superconducting qubits with research starting in the early 1980s.

The expected benefits of superconducting transistors are an increase in clock frequency and a decrease in power consumption[2182]! The gain is more significant with the clock frequency than with energy consumption. For example, in a Japanese SFQ component realized in 2019, the clock was 32 GHz while the power drain was 2.5 TOPS per Watt, in the average of most deep learning CMOS chipsets[2183].

Several generations of superconducting components have been developed over time[2184]:

- **SFQ** (Single Flux Quantum) was a first-generation circuit, limited to a 1 GHz and 300 Mhz clock frequency. Work started at IBM in the 1960s. They had invested the equivalent of $100M today in a program that was partly funded by the NSA and which was abandoned in 1983 which are also based on the Josephson effect[2185]. D-Wave's superconducting qubit use SFQ-type components for generating and reading qubit control signals[2186].

---

[2181] See Quantum Foundations of Classical Reversible Computing by Michael P. Frank and Karpur Shukla, April 2021 (70 pages) and The Reversible Computing Scaling Path: Challenges and Opportunities by Michael P. Frank, February 2022 (40 slides).

[2182] See this very interesting presentation on superconducting electronics: Superconducting Microelectronics for Next-Generation Computing by Leonard Johnson, MIT Lincoln Labs, November 2018 (27 slides). The gain in power consumption would be between 10 and 1000. The integration level is currently low, of the order of 200 nm compared to 7 nm for the densest CMOS processors. But it is steadily increasing. There are even investigations to combine superconducting transistors, optoelectronics and neural networks. See Superconducting Optoelectronic Loop Neurons by Amir Jafari-Salim, 2018 (48 pages).

[2183] See 29.3 A 48GHz 5.6mW Gate-Level-Pipelined Multiplier Using Single-Flux Quantum Logic by Ikki Nagaoka et al, 2019.

[2184] See Single Flux Quantum (SFQ) Circuit Fabrication and Design: Status and Outlook by V. Bolkhovsky et al, Lincoln Laboratory at MIT, 2016 (34 slides) which provides an interesting view on their manufacturing process and metal layers, and IRDS Cryogenic Electronic and Quantum Information Processing, IEEE, 2021 (93 pages) which provides a good overview of the various SFQ technologies around.

[2185] See The Long Arm of Moore's Law: Microelectronics and American Science by Cyrus Mody, 2017 (299 pages), page 58.

[2186] See Architectural considerations in the design of a superconducting quantum annealing processor by P. I. Bunyk et al from D-Wave, 2014 (9 pages).

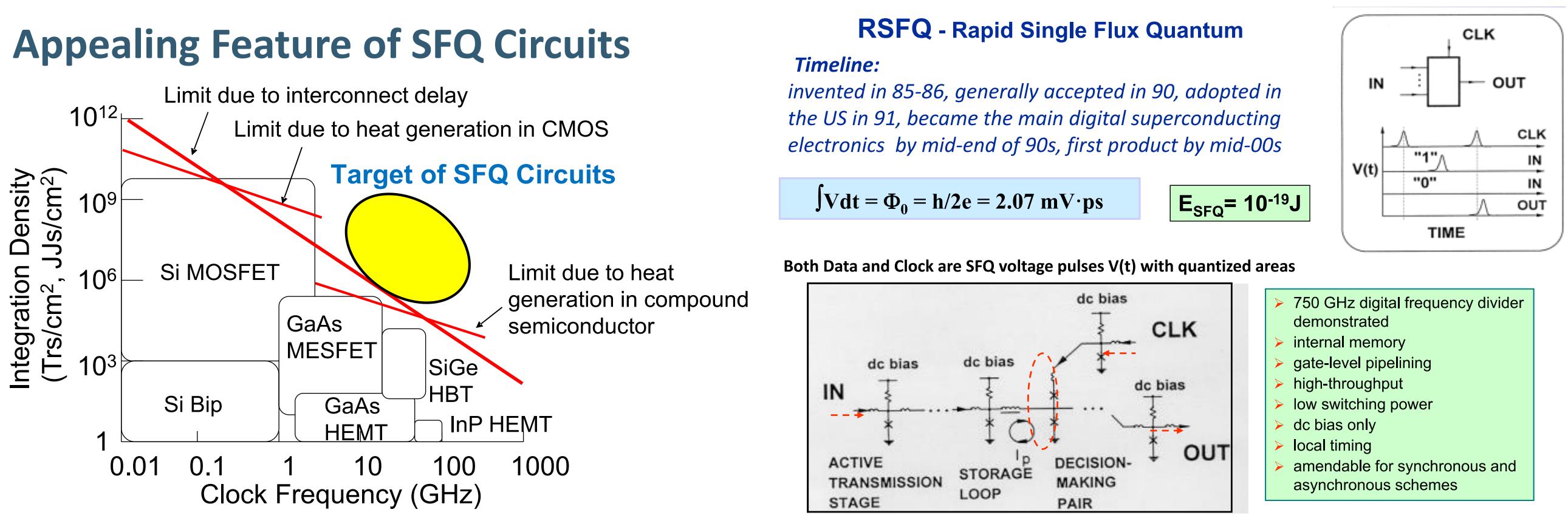

*Figure 712: schematic positioning SFQs in terms of clock speed and integration compared to traditional electronic components. Source: Impact of Recent Advancement in Cryogenic Circuit Technology by Akira Fujimaki and Masamitsu Tanaka, 2017 (37 slides).*

- **RSFQ** (Rapid Single Flux Quantum) was coinvented in Russia in the mid-1980s by Konstantin K. Likharev and his then-PhD student Oleg Mukhanov. Mukhanov then moved to the USA and joined Hypres in 1991 to launch the industry development of RSFQ. The first ones were produced in the mid-2000s[2187]. They have the advantage of being able to operate up to 750 GHz. They can be used to create ALUs (Arithmetic Logic Units[2188]) running at 20/30 GHz as well as ADCs (analog to digital converters) running up to 40 GHz[2189]. In RSFQ logic, binary information is managed in the form of quantum states of the Josephson junction flux, which is transferred as voltage pulses[2190]. However, the technology does not use state superposition and entanglement as in superconducting qubits. **Hypres** develops radio frequency reception systems using two superconducting components: **SQUID** (Superconducting Quantum Interference Device) based antennas that allow to capture magnetism with precision (invented in 1964) and an RSFQ chipset running at 30 GHz with 11K JJ (Josephson junctions)[2191]!
- **AQFP** (Adiabatic Quantum Flux Parametron) which comprises two superconducting Josephson loops connected together by an inductance, reminiscent of the nSQUID principle[2192]. The process is very energy efficient due to its ability to be reversible. A recent work from Japanese researchers prototypes a AQFP processor using 20,000 Josephson gates operating at 4.2K[2193].
- **RQL** (Reciprocal Quantum Logic)[2194], **eRSFQ** (Energy Efficient RSFQ) and **eSFQ** (Energy Efficient SFQ) are variants of the RSFQ that are more energy efficient due to the absence of bias resistance, replaced by an inductance. This is the variant chosen by Hypres and its subsidiary SeeQC. Their SFQs combine eRSFQs, of which they are the originators, and eSFQs. RQLs are studied to create superconducting memories.

---

[2187] Source: Single Flux Quantum Logic for Digital Applications by Oleg Mukhanov of SeeQC/Hypres, August 2019 (33 slides).

[2188] See for instance qBSA:Logic Design of a 32-bit Block-Skewed RSFQ Arithmetic Logic Unit, 2020 (3 pages).

[2189] This would be very useful to generate the microwaves to drive superconducting and silicon qubits.

[2190] Time management in logic programming must take this into account. On this topic, see A Computational Temporal Logic for Superconducting Accelerators by Georgios Tzimpragos et al, 2020 (14 pages).

[2191] See Superconducting Quantum Arrays for Wideband Antennas and Low Noise Amplifiers by Oleg Mukhanov et al, 2014 (36 slides).

[2192] See Adiabatic Quantum-Flux Parametron: Towards Building Extremely Energy-Efficient Circuits and Systems, by Olivia Chen et al, 2018 (10 pages) and Design and Implementation of a Bitonic Sorter-Based DNN Using Adiabatic Superconducting Logic also from Olivia Chen et al, 2019 (24 slides).

[2193] See MANA: A Monolithic Adiabatic iNtegration Architecture Microprocessor Using 1.4-zJ/op Unshunted Superconductor Josephson Junction Devices by Christopher L. Ayala et al, December 2020 (14 pages). They provide some impressive cryogenic needs for using this technology at a supercomputing scale.

[2194] See Ultra-Low-Power Superconductor Logic by Quentin P. Herr et al, 2011 (7 pages).

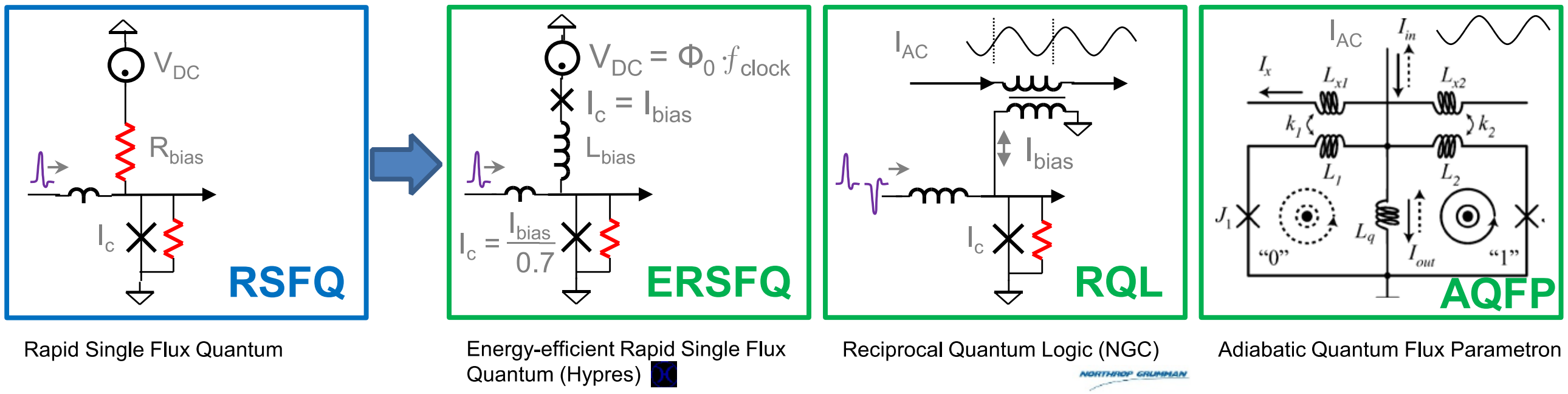

*Figure 713: RSFQ and its evolutions, ERSFQ, RQL and AQFP. Source: Single Flux Quantum Logic for Digital Applications by Oleg Mukhanov of SeeQC/Hypres, August 2019 (33 slides).*

- **SFETs** (Superconducting FETs, Field Effect Transistors) which implement a concept similar to adiabatic CMOS, but with a superconducting component. These components have been developed since the 1980s[2195].

There are a few other variants of superconducting components that I will just mention (SSV, SVJJ, STTJJ, S3JJ) because they do not seem to be common, on top of JMRAM that is investigated for implementing superconducting memory.

To date, the integration record for this type of component is only 144,000 Josephson junctions in a chipset, realized in a 248 nm integration[2196] if we don't account for D-Wave's Pegasus chipset with its million Josephson junctions including about 10,000 JJs used in their annealer qubits, the rest being used for qubits control logic and digital signals multiplexing/demultiplexing.

Another key component has been developed, superconducting diodes with zero resistance in one direction. They were experimented in 2022 with europium sulfide thin film separating a superconducting aluminum and a normal metal copper layer[2197] and, separately, with niobium bromide placed between layers of niobium diselenide[2198].

In the mid-2000s, the **NSA** invested $400M in the RSFQ over the 2005-2010 period. Its goal was to create a processor with one million logic gates running at 50 GHz. The NSA document describing the project is surprisingly detailed and highly informative. It reveals the scope of the related technological challenges[2199]. In particular, for creating cryogenic memories, superconducting or not: CMOS-Josephson junction hybrid, spintronic based[2200], SFQ or monolithic RSFQ-MRAM. Then the communication between the cryogenic electronics and room temperature components, with a 25 Gbits/s optical fiber that we would probably now reach 100 or 200 Gbits/s, leaving aside the question of the optical signal modulation and demodulation. Cryogenics must be sized to support a large number of components. For testing purpose, a simple pulsed tube would be sufficient but more imposing installations are planned as computing power would grow, as in the illustration in Figure 714: on the right.

---

[2195] See Josephson Junction Field-effect Transistors for Boolean Logic Cryogenic Applications by Feng Wen, 2019 (7 pages) and Superconducting silicon on insulator and silicide-based superconducting MOSFET for quantum technologies by Anaïs Francheteau, 2017 (153 pages).

[2196] See Advanced Fabrication Processes for Superconducting Very Large Scale Integrated Circuits by Sergey K. Tolpygo, 2015 (43 slides).

[2197] See Superconducting spintronic tunnel diode by E. Strambini et al, Nature Communications, May 2022 (7 pages).

[2198] And The field-free Josephson diode in a van der Waals heterostructure by Heng Wu et al, Nature, April 2022 (17 pages).

[2199] See NSA Superconducting Technology Assessment, 2005 (257 pages). The document is quite old but still very well crafted and interesting.

[2200] See Cryogenic Memory Architecture Integrating Spin Hall Effect based Magnetic Memory and Superconductive Cryotron Devices by Minh-Hai Nguyen et al, July 2019 (12 pages) and Modeling the computer memory based on the ferromagnet/superconductor multilayers by Serhii E. Shafraniuk, Ivan P. Nevirkovets and Oleg A. Mukhanov, July 2019 (27 pages).

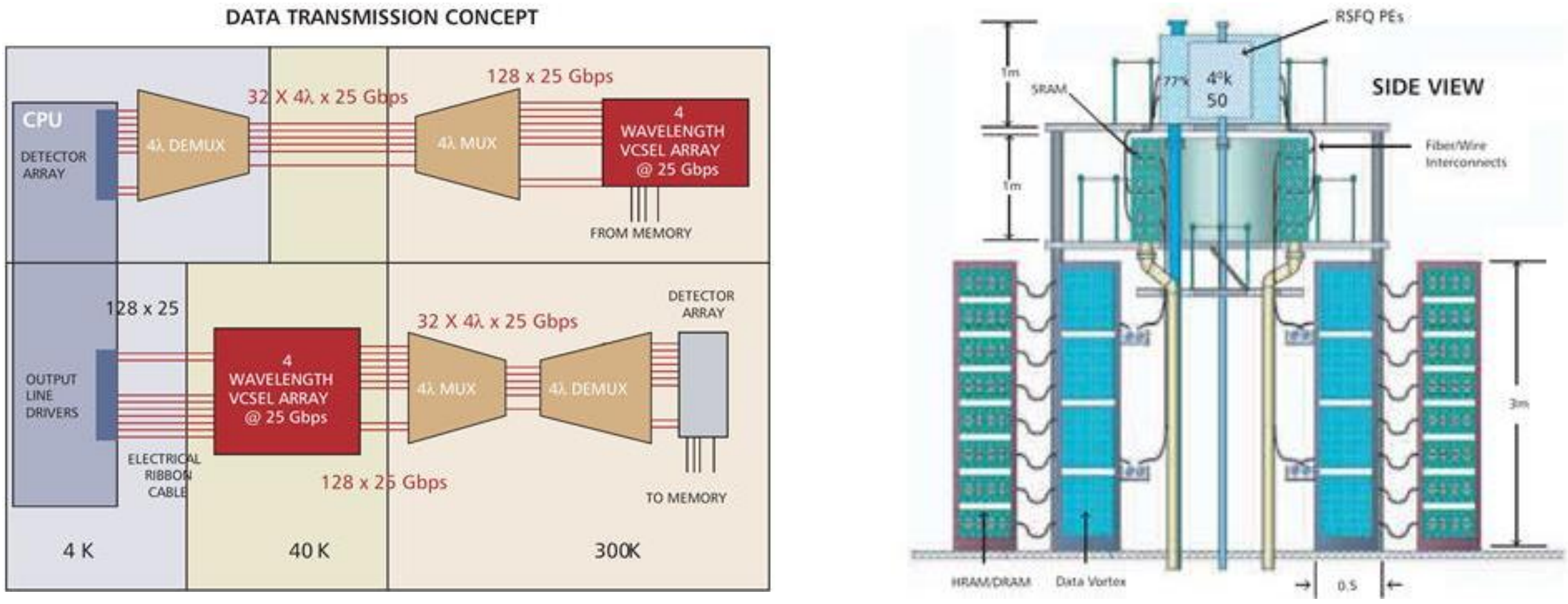

*Figure 714: Left: A 64-fiber system for bi-directional transmission totaling 6.4 Tbps between a superconducting processor operating at 4K and high speed mass memory at ambient temperature. Optical connections are shown in red and electrical in black. This technology was to be available by 2010. Right: concept for a large-scale system including cryogenic cooling unit for supercomputers. Source: NSA Superconducting Technology Assessment, 2005 (257 pages), pages 100 and 125.*

The project relied mainly on **Hypres,** the only American company entirely dedicated to the creation of superconducting components, who runs its own foundry since 1983, but then divested it to SeeQC and now seems to rely on SkyWater's foundry. They were supplying radio frequency components for military use cases. Out of this, they developed an 8-bit RSFQ processor and 28,000 Josephson junctions.

There is also **Northrop Grumman** with its foundry located in Linthicum, Maryland. Finally, **Chalmers University** in Sweden and various research laboratories in the USA (JPL, Berkeley, Stony Brook) as well as the **NIST** Boulder laboratory were also involved. **IMEC** has a lab in Florida that designs superconducting circuits with densities in the 10 nm to 20 nm range.

**TABLE 2-2. SUPERCONDUCTOR RSFQ MICROPROCESSOR DESIGN PROJECTS**

| Time Frame | Project | Target Clock | Target CPU Performance (peak) | Architecture | Design Status |
|---|---|---|---|---|---|
| 1997-1999 | SPELL processors for the HTMT petaflops system (US) | 50-60 GHz | ~250 GFLOPS/CPU (est.) | 64-bit RISC with dual-level multithreading (~120 instructions) | Feasibility study with no practical design |
| 2000-2002 | 8-bit FLUX-1 microprocessor prototype (US) | 20 GHz | 40 billion 8-bit integer operations per second | Ultrapipelined, multi-ALU, dual-operation synchronous long instruction word with bit-streaming (~ 25 instructions) | Designed, fabricated; operation not demonstrated |
| 2002-2005 | 8-bit serial CORE1 microprocessor prototypes (Japan) | 16-21 GHz local, 1 GHz system | 250 million 8-bit integer operations per second | Non-pipelined, one serial 1-bit ALU, two 8-bit registers, very small memory (7 instructions) | Designed, fabricated, and demonstrated |
| 2005-2015 (est.) | Vector processors for a petaflops system (Japan) | 100 GHz | 100 GFLOPS/CPU (target) | Traditional vector processor architecture | Proposal |

**TABLE 3.2-2. STATUS OF LOW-LATENCY CRYO-RAM CANDIDATES**

| Type/Lab | Access Time | Cycle Time | Power Dissipation | Density | Status |
|---|---|---|---|---|---|
| Hybrid JJ-CMOS (UC Berkeley) | 500 ps for 64 kb | 0.1 - 0.5 ns depending on architecture | 12.4 mW read 10.7 mW write (Single cell writing) | 64 kb in < 3x3 mm² | All parts simulated and tested at low speed |
| RSFQ decoder w/ latching drivers (ISTEC/SRL) | ? | 0.1 ns design goal | 107 mW for 16 kb (Estimate) | 16 kb in 2.5 cm² (Estimate*) | 256b project completed (Small margins) |
| RSFQ decoder w/ latching drivers (NG) | ? | 2 ns | ? | 16 kb/cm² * | Partial testing of 1 kb block |
| SFQ RAM (HYPRES) | 400 ps for 16 kb (Estimate) | 100 ps for 16 kb (Estimate) | 2 mW for 16 kb (Estimate) | 16 kb/cm² * | Components of 4 kb block tested at low speed |
| SFQ ballistic RAM (Stony Brook University) | ? | ? | ? | Potentially dense Requires refresh | Memory cell and decoder for 1 kb RAM designed |
| SFQ ballistic RAM (NG) | ? | ? | ? | Potentially dense Requires refresh | SFQ pulse readout simulated |
| MRAM (40K) | Comparable to hybrid CMOS | Comparable to hybrid CMOS | < 5mW at 20GHz (Estimate) | Comparable to DRAM (Estimate) | Room temperature MRAM in preproduction; Low temperature data sparse |

*Densities of JJ memories are given for the technologies in use at the time of the cited work. Greater densities can be expected when a 20 kA/cm² process is used. The symbol ? signifies insufficient data.

*Figure 715: various superconducting electronic projects launched about 20 years ago with processors on the left and cryo-RAM on the right, reminding us how long these projects can last or have their ups and downs. These projects became the C3 IARPA project that lasted until 2022. Source: NSA Superconducting Technology Assessment, 2005 (257 pages), pages 28 and 53.*

The **IARPA** agency has taken over with the **Cryogenic Computing Complexity** (C3) project launched in 2014. It involved IBM, Northrop Grumman (Quentin Herr, who works at IMEC in Belgium since May 2021), Raytheon and Hypres and was due to end in 2018[2201].

It actually ended in 2022. A total of $700M were spent in R&D there given the project was considered to be strategic. Some of the showstoppers there were the lack of EDA tools (electronic design automation) adapted to superconducting circuits and of scalable cryogenic cabling (leading to the Supercables project). There are solutions with signals frequency upconvert to optical wavelengths with VCSEL chipsets but these are consuming a lot of energy.

This project was part of the **National Strategic Computing Initiative** (NSCI) launched in 2015 by the White House, which focused on the development of supercomputers. It's difficult to find out what this project has achieved as of 2021.

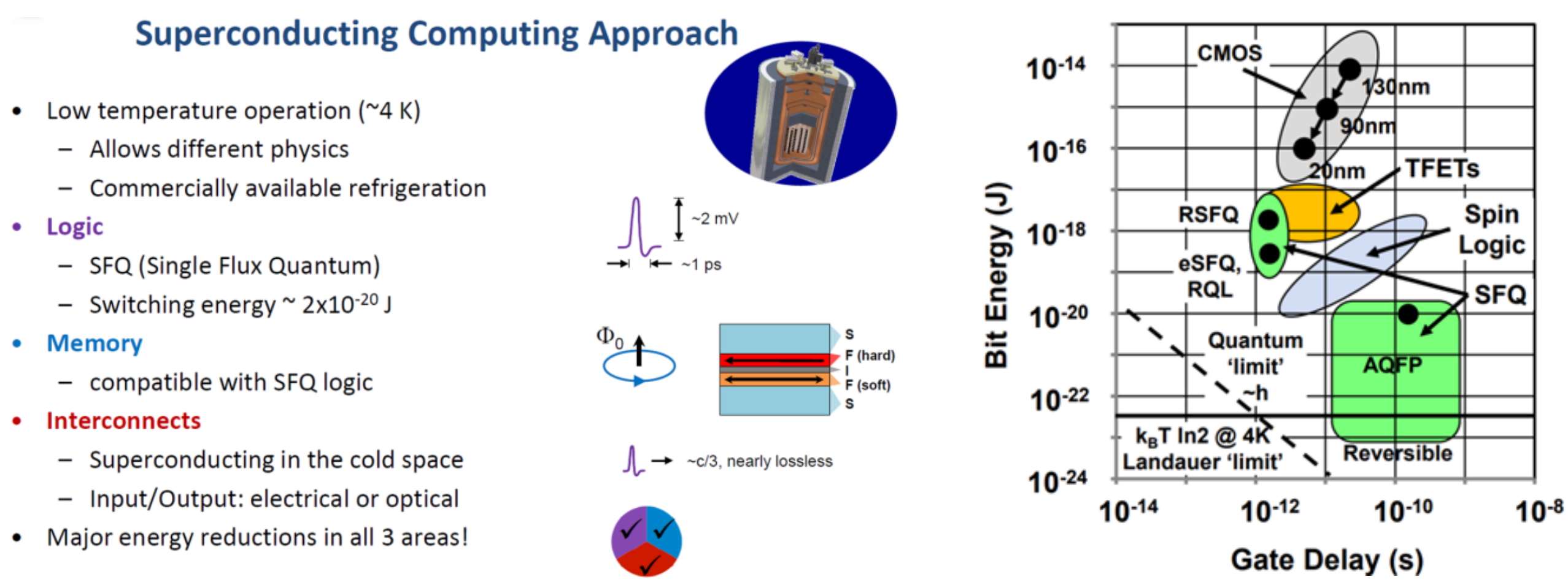

*Figure 716: superconducting approach and gate delay. Source: TBD.*

In the USA, the **DISCoVER** (Design & Integration of Superconductive Computation for Ventures beyond Exascale Realization) expedition is working on a "SuperSoCC" SFQ-based niobium chipset (superconductive system of cryogenic computing cores) operating at 4.2K. The project is led by Massoud Pedram, Timothy Pinkston and Murali Annavaram from USC. They are working on on-chip memory design and interface between room electronics and cryo-electronics. The team touts a potential energy gain of /100 vs classical computing or 10x speed improvement with the same energy consumption. With a relatively modest funding of $15M coming from the NSF, the project also involves Auburn University, Cornell University, Northeastern University, Northwestern University, University of Rochester and Yokohama National University in Japan.

Outside the USA, the **Japanese Superconducting Computing Program's** ambition in 2004 was to create a processor running at 100 GHz generating 100 GLOPS in SFQ, supplemented by 200 TB of DRAM running at 77K to generate a 1.6 PFLOPS system comprising 16,384 processors.

[2201] See Superconducting Computing and the IARPA C3 Program by Scott Holmes, 2016 (57 slides). All the presentations of the C3 conference are here.

| PROCESS | Current density [kA/cm²] | minimum area [µm²] | Maximum integration | Maximum frequency [GHz] |
|---|---|---|---|---|
| Hypres #03-10-45 | 0.03 1.0 4.5 | ~ 3.14 | 15,000 | 80 GHz RnIc=1.3mV @ 4.5 kA/cm² |
| Hypres #S45/100/200 | 0.1 1 4.5 10 20 30 | ~ 0.4 | 10,000 | 200 GHz @ 30 kA/cm² |
| MIT Lincoln Lab SFQx | 10 20 50 | ~ 0.06 | ~ 800,000 | 240 GHz RnIc=2.17 mV @50 kA/cm² |
| ADP2 | 10 | 1.0 | 1100 JJ/mm² | 80 GHz |
| STP2 | 2.5 - 20 | 0.25 - 4.0 | 100 JJ/mm² - > 2,000 JJ/mm² | 30 GHz - 150 GHz |
| HSTP | 10 | 1.0 | 70,000 | 80 GHz |
| Fluxonics standard | 1 | 12.5 | 100 JJ/mm² | 40 GHz RnIc =0.256 mV |
| INRiM SNIS | up to 100 | 25 | 1,000 JJ/mm² | 300 GHz RnIc =0.1mV - 0.7mV |
| NIST Nb/Nbx Si1-x/Nb | up to 110 | ? | 70,000 | 300 GHz |
| INRiM SNIS 3D FIB | up to 100 | 0.1 | 10,000 JJ/mm² | 300 GHz RnIc=0.1mV - 0.7mV |

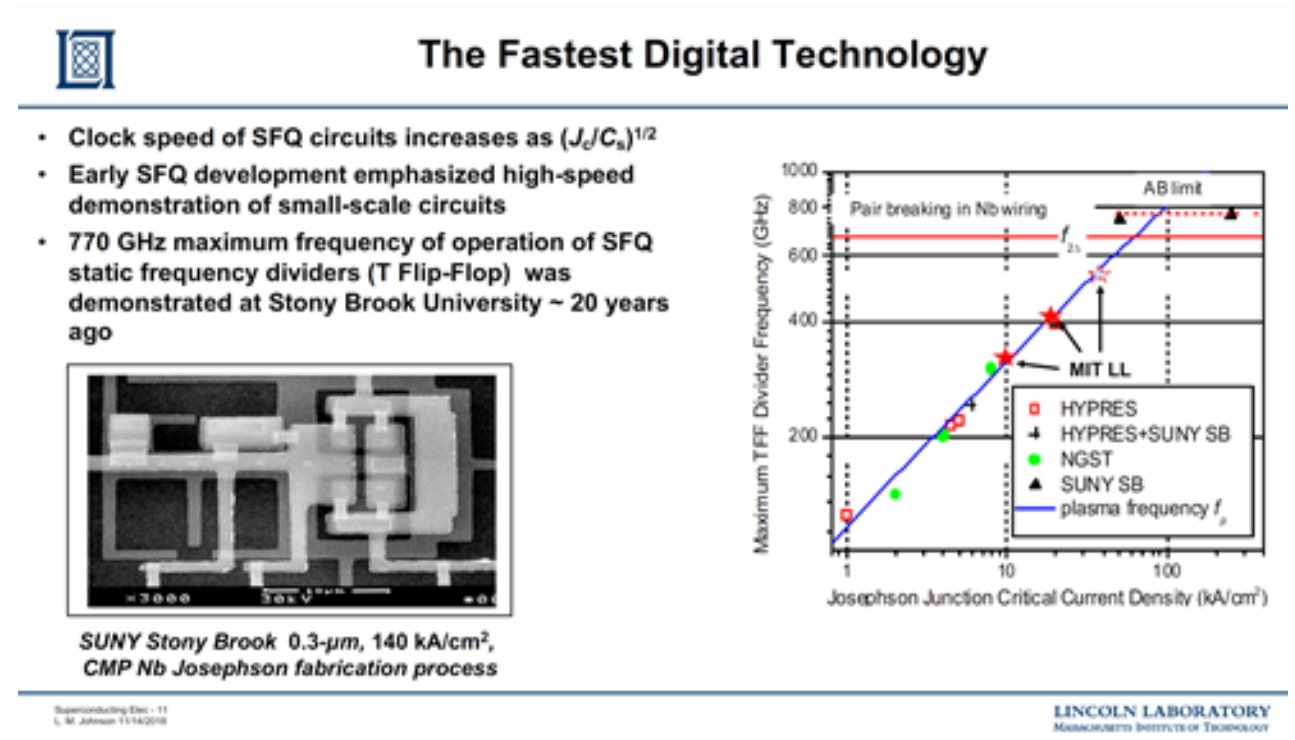

*Figure 717: the left table comparing different types of superconducting components comes from Superconducting Computing by Pascal Febvre, CNRS, 2018 (56 slides). The right slide comes from Superconducting Computing and the IARPA C3 Program by Scott Holmes, 2016 (57 slides).*

All this with a cryostat consuming 12 MW and generating a thermal power of 18 kW at 4.2K. It has not yet seen the light of day about 17 years later[2202]. Meanwhile, the IBM Summit supercomputer using traditional processors and GPUs generates 200 petaFLOPS consuming 13 MW, so why bother?

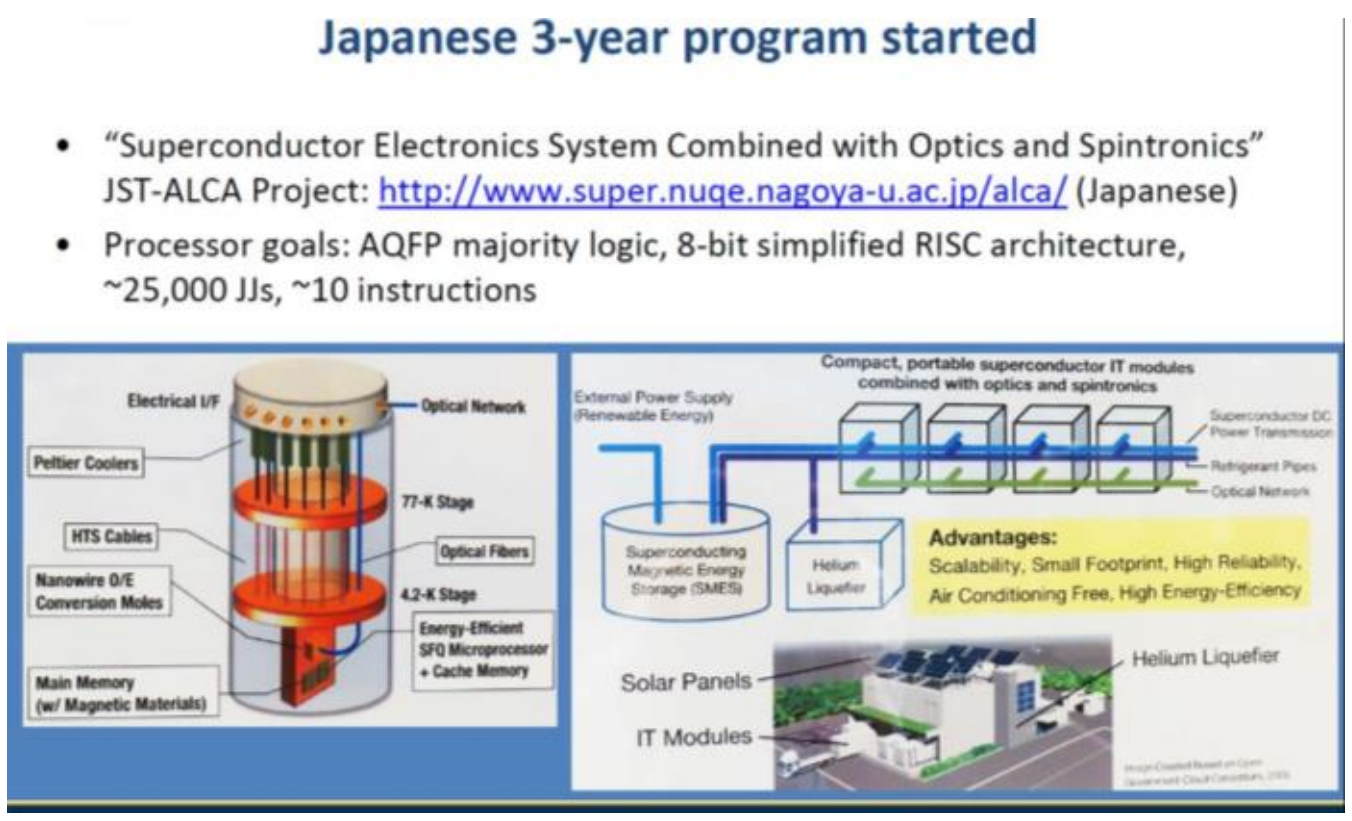

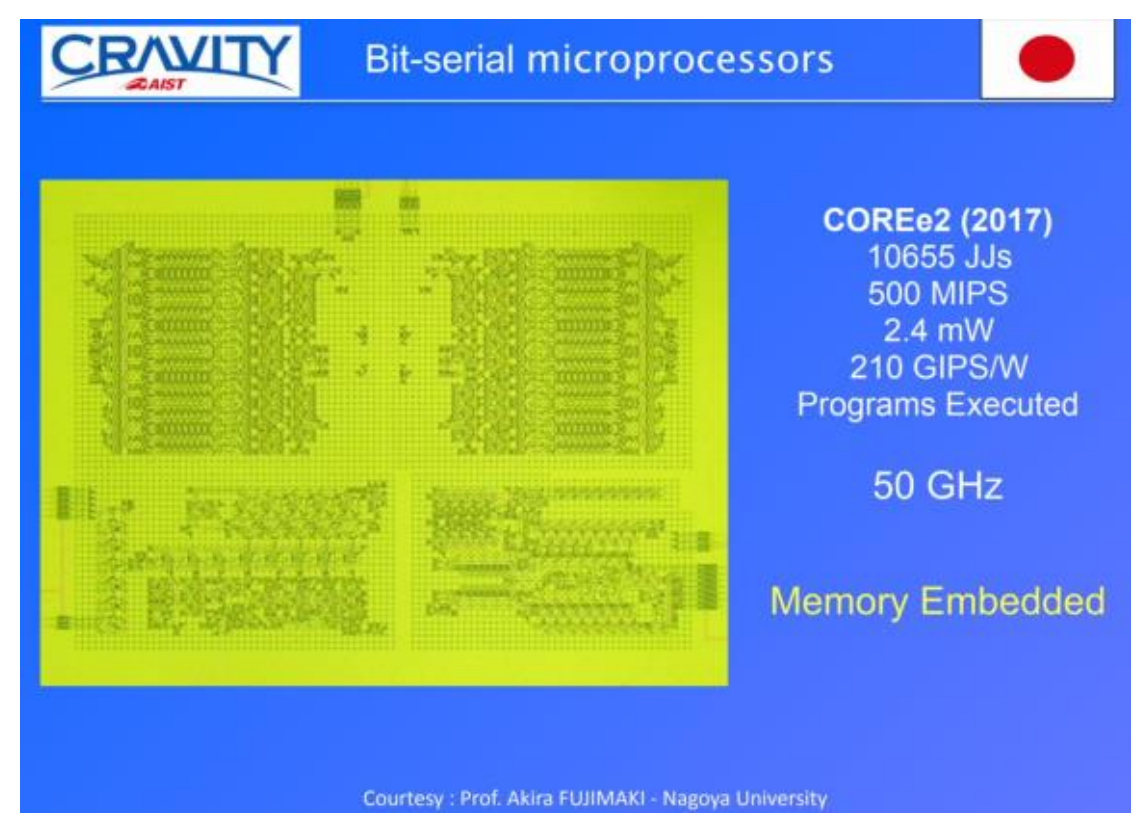

*Figure 718: the left slide comes from Superconducting Computing and the IARPA C3 Program by Scott Holmes, 2016 (57 slides) and the right schema comes from Superconducting Computing by Pascal Febvre, CNRS, 2018 (56 slides).*

**China** announced in 2018 a $145M plan to build a superconducting computer by 2022. They had then created a chip with 10,000 Josephson junctions. Not sure they are ready for this milestone. **Russia** also has ambitions in this field[2203].

In **France**, the laboratory CMNE (Micro Nano Electronic Components) of the IMEP-LaHC (Microelectronics, Electromagnetism, Photonics, Microwave) of the UGA (Grenoble) was working in this area, under the responsibility of Pascal Febvre who is based in Chambéry.

In Korea, researchers from **Seoul National University** have proposed another path, creating computers with CryoMOSFET chipsets operating at a relatively hot temperature of 77K, reducing cooling costs.

[2202] They managed to create the CORE1α in 2003 at 4999 JJ (Josephson junctions) running at 15 GHz, the CORE1β in 2006 at 10.955 JJ running at 25 GHz, the CORE1γ with 22,302 JJ also at 25 GHz, the CORE100 in 2015 at 3073 JJ and 100 GHz, the COREe2 in 2017 at 10,655 JJ and 50 GHz with an integrated memory. See Impact of Recent Advancement in Cryogenic Circuit Technology by Akira Fujimaki and Masamitsu Tanaka, 2017 (37 slides). This continued in 2019 with an 8-bit multiplier containing 20,251 JJ running at 48 GHz and consuming 5.6 mW. Source: 29.3 A 48GHz 5.6mW Gate-Level-Pipelined Multiplier Using Single-Flux Quantum Logic by Ikki Nagaoka et al, 2019.

[2203] See The Outlook for Superconducting Computers by R Colin Johnson, 2018.

Its benefits are less impressive than superconducting computing with an improvement of only 41% in single thread performance for the same power budget or a power reduction of 38% for the same performance, all of this including of course the cooling power budget[2204].

In the end, this branch of the superconducting computer industry is for the moment still immature. It has suffered from the uninterrupted advance of Moore's Law until the last few years and the difficulties of its practical implementation. It is not impossible that synergies will develop between quantum computing and this somewhat neglected branch. They can help each other, as can be seen with superconducting circuits for driving superconducting qubits or silicon. As we know, quantum computing will perhaps indirectly revive this sector[2205]!

## Probabilistic computing

Probabilistic processors are another variation of exotic processors. They use probabilistic p-bits that can fluctuate rapidly between 0 and 1 with a low transition energy level. They are supposed to allow the resolution of so-called "quantum" problems without relying on quantum mechanisms. p-bits can be realized with nanomagnets and also with regular transistors[2206].

Various applications are promoted such as the creation of neural networks called BSN (Binary Stochastic Neuron) and the resolution of optimization problems similar to those treated by quantum annealing and gate-based quantum computing. The accelerations obtained are not qualified as exponential. It may be just polynomial, which is still interesting.

Work in this direction is quite recent and comes from **Purdue University** in Indiana[2207] and from **Tohoku University** in Japan[2208]. **HawAI.tech** (France), a Grenoble-based startup, is positioned on the same niche and targets applications in the field of AI in embedded systems using data from various sensors[2209]. Their roadmap should lead to the creation of a complete probabilistic computer by 2024.

## Optical computing

Many research laboratories and startups are working on the creation of optical processors that are not based on photon qubits. Some are creating classical optical neural networks, others are adapted to convolutional neural networks or spiking neurons, the latter being closest to the way the human brain works. Others are even proposing to solve NP complete problems like Ising models, beyond the coherent Ising model machines I briefly discussed in the part dedicated to photon qubits.

---

2204 See CryoCore: A Fast and Dense Processor Architecture for Cryogenic Computing by Ilkwon Byun et al, 2020 (14 pages).

2205 At last, see this good review paper Beyond Moore's technologies: operation principles of a superconductor alternative by Igor I. Soloviev et al, Russia, 2017 (22 pages). It mentions the potential of a two orders of magnitude gain in energy efficiency with superconducting based supercomputers, cryogenics included. On top of the various variations of SFQ, RQL and SQUID superconducting circuits, the review also covers cryogenic memory. One of the limitations of superconducting circuits is their low potential miniaturization. Josephson junctions density seems limited to 2.5 million junctions per $cm^2$. To be compared with billions of CMOS transistors with 5nm/7nm nodes!

2206 See an explanation of p-bits in Waiting for Quantum Computing? Try Probabilistic Computing by Kerem Camsari and Supriyo Datta, IEEE Spectrum, March 2021, then Integer factorization using stochastic magnetic tunnel junctions by William A. Borders et al, 2019, p-Bits for Probabilistic Spin Logic by Kerem Y. Camsari, 2019 (11 pages), Stochastic for Invertible Logic by Brian Sutton et al, 2017 (19 pages) and Probabilistic computing with p-bits by Jan Kaiser and Supriyo Datta, October 2021 (8 pages).

2207 See From Charge to Spin and Spin to Charge: Stochastic Magnets for Probabilistic Switching by Kerem Y. Camsari et al, February 2020 and Hardware Design for Autonomous Bayesian Networks by Rafatul Faria et al, 2020 (10 pages).

2208 See Demonstrating the world's fastest spintronics p-bit by Tohoku University, March 2021 and Waiting for Quantum Computing? Try Probabilistic Computing by Kerem Camsari and Supriyo Datta, 2021.

2209 See Bayes from Cell to Chip by Pierre Bessière, 2018 (33 slides).

Let's mention **reservoir computing** which is a specific category of recurrent neural networks used to process time series (in language processing, finance, energy, robotics)[2210].

Their particularity is to use neuron weights and links between neurons randomly fixed in the reservoirs, all with nonlinear activation functions for these links. The hundreds of neurons in a reservoir are fed by input data stored in the reservoirs.

The activation functions nonlinearity makes this memory evanescent. The training parameters of these networks are located in the weights of the neurons that connect the reservoirs to the output data.

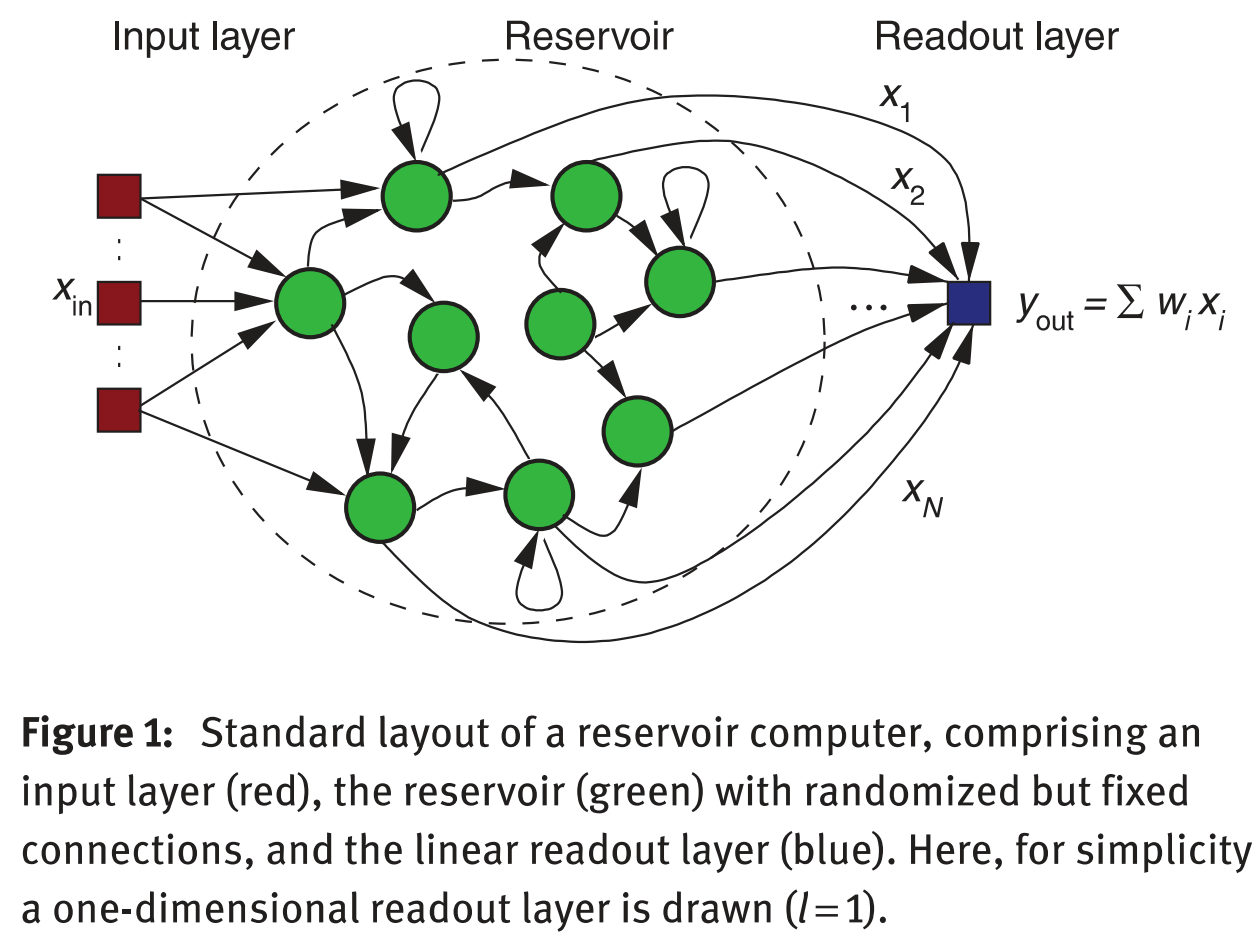

*Figure 719: the classical concept of reservoir computing in machine learning. Source: Advances in photonic reservoir computing by Guy Van der Sande et al, 2017 (16 pages) which provides an excellent focus on optronics based reservoir computing.*

There are classic reservoir computing projects, rather based on memristors[2211], five types of optical reservoir computing and even quantum versions, for quantum reservoir computing[2212]! A mapping of the different types of optical neural networks is proposed in Figure 720.

We will now focus on solutions based on optical processes using image diffraction from DMD or DLP chips illuminated by a laser and sent on various structures such as random matrices or various metamaterials. They are often based on the principle of the Optical Fourier Transform which allows to decompose a 2D image into spatial frequencies, itself represented in 2D. This transform is an image that contains key points representing shapes and repetitions in the analyzed images.

This can be leveraged to build convolution layers in convolutional neural networks. These serve to detect the presence of shapes in an image, the shapes being represented by filters. The Fourier transform helps to automatically identify these key shapes in the image. These systems capture the result with a CMOS sensor, usually with a very high resolution, much higher than that of the DLP or DMD chip used upstream. The diffraction thus carries out a projection in a space of larger dimension than the original image. All this is supported by serious mathematics[2213] .

[2210] The concept of reservoir computing dates back to 2007. See Toward optical signal processing using Photonic Reservoir Computing by Kristof Vandoorne et al, 2008 (11 pages). It is also described in Novel frontier of photonics for data processing - Photonic accelerator by Ken-ichi Kitayama, 2019 (25 pages) as well as in this beautiful presentation Introduction to Reservoir Computing by Helmut Hauser (282 slides). The notion is different from the one of reservoir engineering.

[2211] See Memristors and Beyond: Recent Advances in Analog Computing by Nick Skuda, 2019 (12 slides).

[2212] See Universal quantum reservoir computing by Sanjib Ghosh et al, from Singapore, 2020 (23 pages) as well as Integrated Nanophotonic Structures for Optical Computing by Laurent Larger et al, 2019 (50 slides). See Experimental photonic quantum memristor by Michele Spagnolo et al, Nature Photonics, March 2022 (7 pages) which describes a way to implement quantum memristors and quantum reservoir computing. As explained in Quantum memristor: A memory-dependent computational unit Quantum memristor bridges conflicting quantum requirements in single device by Chris Lee, ArsTechnica, April 2022, a quantum memristor is a memristor that will change its properties based on the qubit traversing it, in the case of photons. And at last, see Quantum reservoir neural network implementation on a Josephson mixer by Julien Dudas, Erwan Plouet, Alice Mizrahi, Julie Grollier and Danijela Marković, September 2022 (7 pages) which proposes to use a large number of densely connected neurons by using parametrically coupled quantum oscillators instead of physically coupled qubits.

[2213] See An optical Fourier transform coprocessor with direct phase determination by Alexander Macfaden et al, 2017 (8 pages) and Performing optical logic operations by a diffractive neural network by Chao Qian et al, 2020 (7 pages).

These different solutions are in their infancy. They can accelerate certain calculations for training complex neural networks. These accelerations seem to be rather polynomial and not exponential as quantum computing is supposed to generate. Except that they do not seem to be handicapped by noise issues as qubits are.

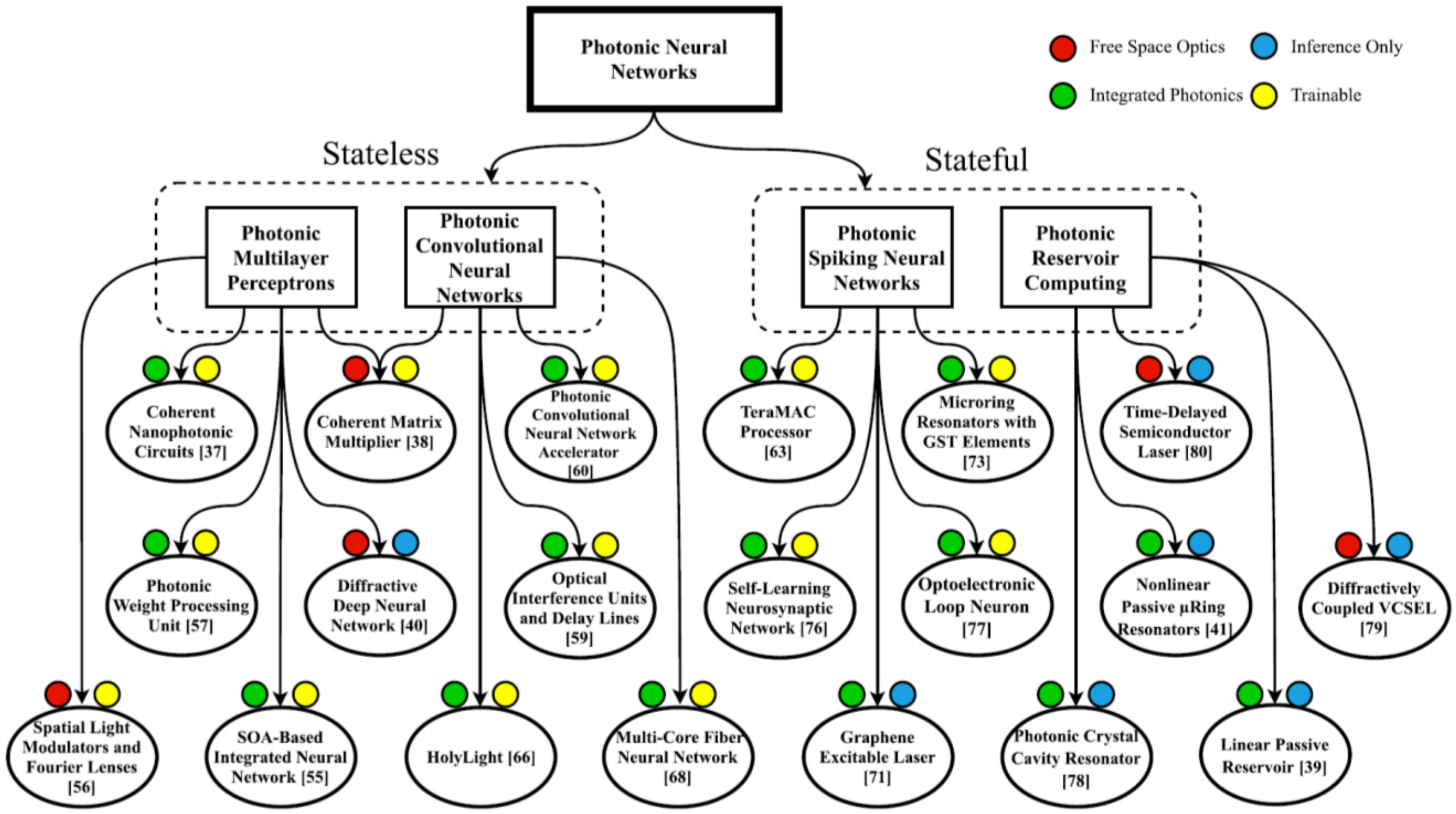

*Figure 720: source: Photonic Neural Networks: A Survey by Lorenzo de Marinis et al, 2019 (16 pages).*

Let's now have a look at various commercial vendors in this very specific market.

**Lighton.io** (2016, France, $3.7M) sells an optical coprocessor that accelerates neural networks training on large volumes of training data, such as with convolutional networks.

In their first system, a laser emits a beam that is magnified with some lenses to illuminate a parametrized DLP micromirror chip. The generated image then traverses a static random matrix playing the role of a scattering medium with focusing lenses L1 and L2 before and after the filter as illustrated in Figure 721. A monochrome CMOS imaging sensor then analyzes the resulting image[2214]. The sensor captures the interferences generated by the set and some mathematical computing interprets the result. This process enables a reduction of a complex data set dimensionality.

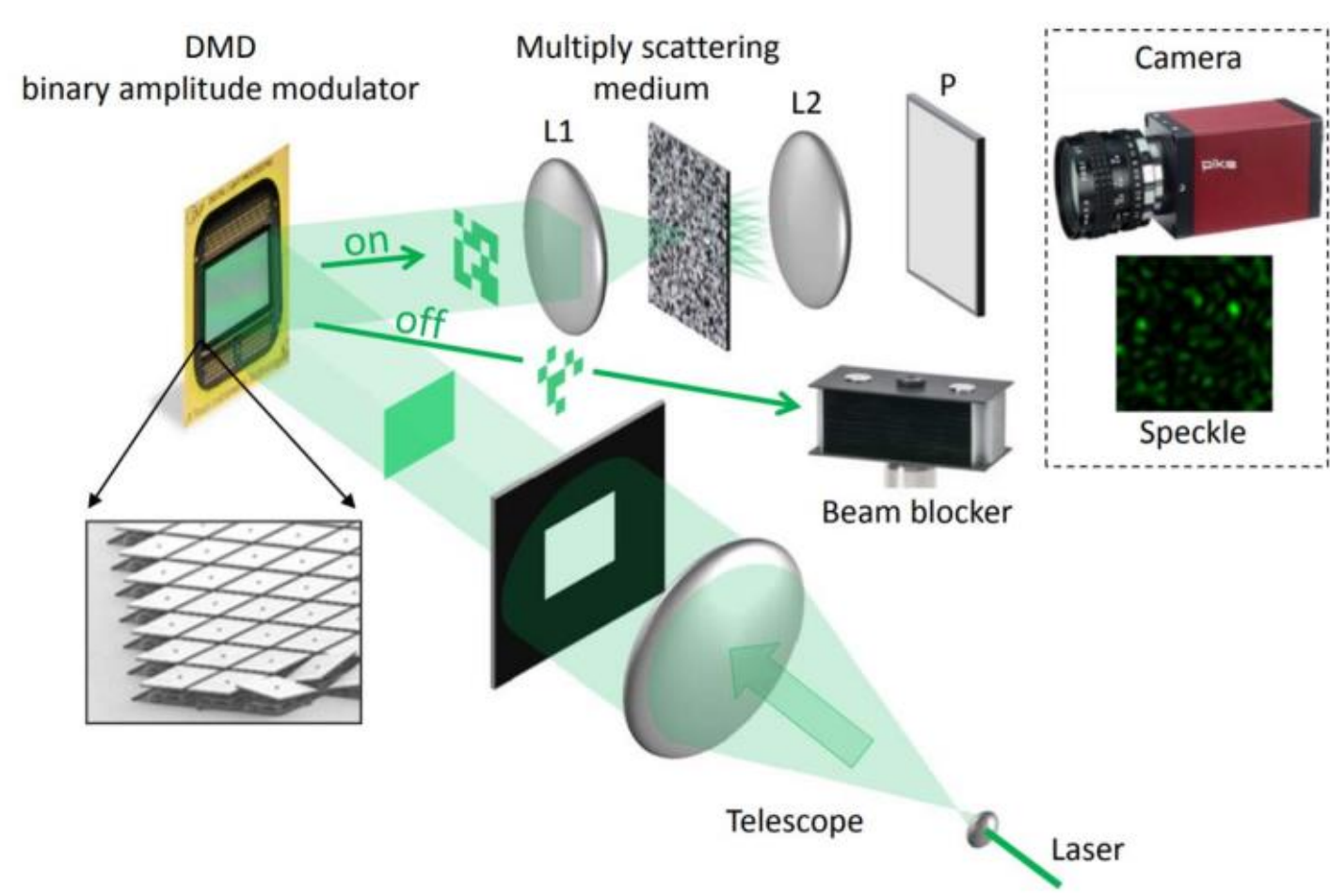

*Figure 721: the LightOn optical accelerator architecture. Source: Random Projections through multiple optical scattering: Approximating kernels at the speed of light, 2015 (6 pages).*

[2214] The process is described in Random Projections through multiple optical scattering: Approximating kernels at the speed of light, 2015 (6 pages).

The miniaturized device fits in a 2U server. The system computing power comes in particular from the DLP and the CMOS sensor resolutions, which is about several million pixels. Everything is driven by Python libraries developed with TensorFlow. The targeted applications are first and foremost genomics and Internet of Things solutions.

Since 2020, LightOn is also working on creating a photonic processor named QORE using reconfigurable linear optical networks to implement optical quantum information processing using multimode fiber with the programmable wavefront shaping of a SLM (spatial light modulator). It can have up to 8 laser inputs and 38 outputs, with fidelities >93% and losses <6.5dB. The device fits in a standard 19" server rack[2215]. The computing advantage brought by this platform is not obvious. However, its creators indicate that it could create high-dimensional entanglement from few single photon states, so these large photonic cluster states that are the Holy Grail of MBQC. The QORE project was funded as part of the H2020 OPTOlogic project.

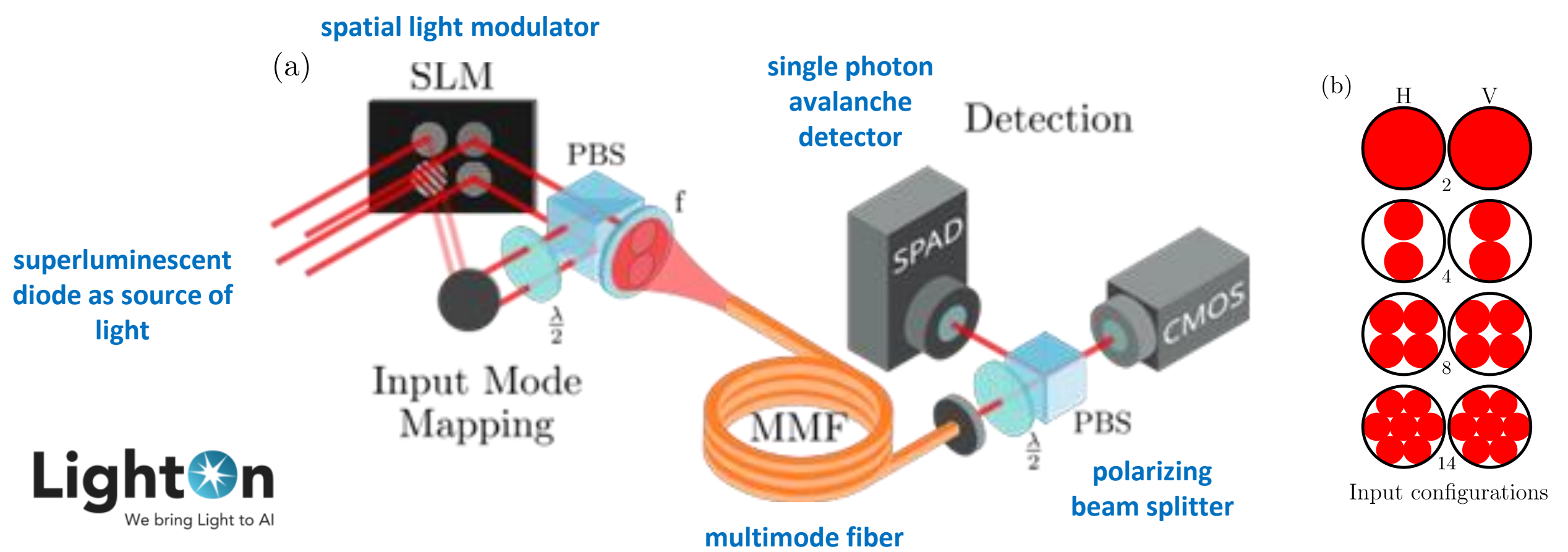

*Figure 722: schematic of the Light QORE quantum processor. Source: [A high-fidelity and large-scale reconfigurable photonic processor for NISQ applications]() by A. Cavaillès, Igor Caron, Sylvain Gigan et al, May 2022 (5 pages) with legends by Olivier Ezratty.*

**Optalysys** (2013, UK, $5.2M) is a spin-out of the University of Cambridge created by Nick New, Robert Todd and Ananta Palani. Their FT:X 2000 system is structured around the realization of optical fast Fourier transforms based on diffraction[2216].

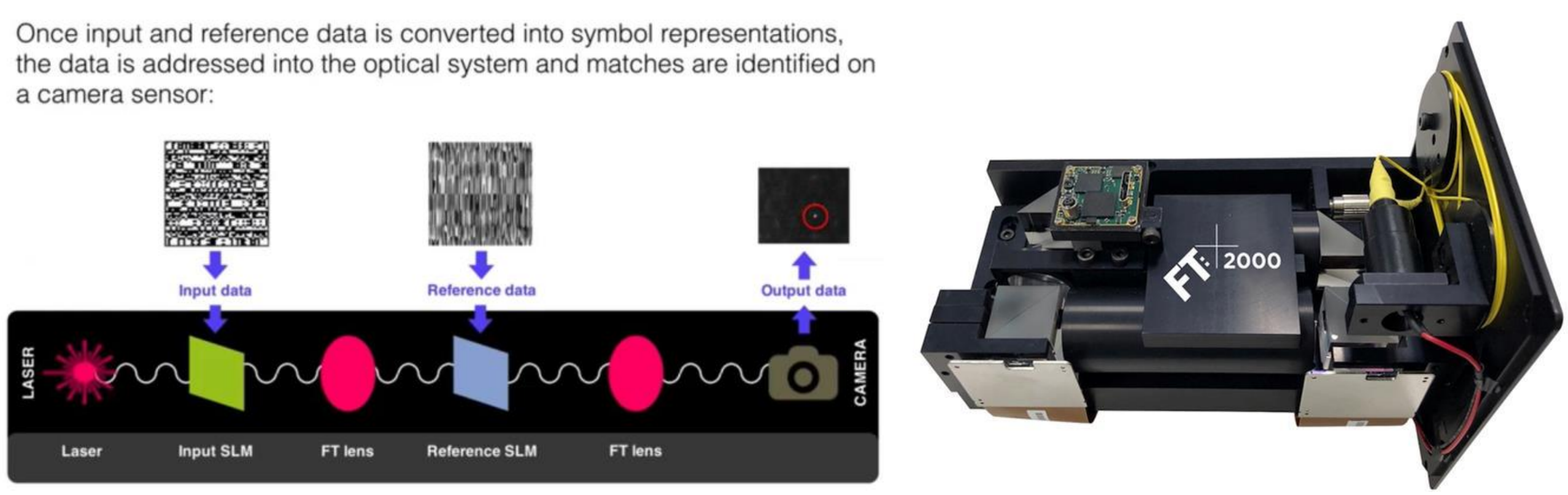

*Figure 723: Optalysys process and apparatus. Source: Optalysys.*

---

[2215] See [A high-fidelity and large-scale reconfigurable photonic processor for NISQ applications]() by A. Cavaillès, Igor Caron, Sylvain Gigan et al, May 2022 (5 pages).

[2216] See [Optalysys - Revolutionary Optical Processing for HPC](), September 2017 (23 mn).

They were involved in various projects, one in genetics to do genome sequence alignment, GENESYS, carried out with the **Earlham Institute**. The other project dealt with weather forecasting for the European center ECMWF and a third one for plasma and fluid dynamics simulation, for DARPA. They also did run a convolutional network in 2018 on a MNIST base with 60,000 letters for training and 10,000 for testing. Its success rate was of only 70%.

**Fathom Computing** (2014, USA, $2.3M) uses an "electro-optical" architecture capable of training memory and convolutional neural networks (LSTM). Their Light Processing Unit (LPU) would be able to read 90% of the tests from the MNIST handwriting database, using the same test as the one performed by Optalysys.

The system is adapted to linear algebra and matrix multiplication. They still have to miniaturize their device, which according to them should take at least two years, as of 2018. In 2021, it doesn't seem they achieved any result. The startup was launched by two brothers, William and Michael Andregg.

**Luminous Computing** (2018, USA, $115M) aims to create a high-performance optical component that would replace 3000 Google TPUs! It would exploit multicolor lasers and light guides. According to the publications of their CTO, Mitchell Nahmias, it seems that they use optical spike neurons[2217]. They can perform calculations very quickly, including in classical CMOS.

So, in optical computation, it should be faster. Wait and see, given they are still in a rather stealth mode.

These various startup solutions have so far demonstrated interesting small-scale computing capabilities for ad-hoc needs. What remains to be done is to scale them up and integrate them into computing solutions that are generally hybrid. Efforts are therefore focused both on solutions packaging so that they can be integrated into standard server racks, on scaling the architectures and on development tools.

Let's add to this the **Copac** project funded as part of a H2020 project. Its goal is to create an exotic quantum computing solution that does not exploit qubits. Its ambition is to enable the resolution of data analysis problems such as the simulation of complex systems or machine learning.

The processor would have the capacity to evaluate all the variables of a logical function in parallel. It is based on a quantum dots-based architecture that can be excited simultaneously on several frequencies by wideband lasers. The results are read by 2D spectroscopy of quantum dots. The machine would operate at room temperature. The process mixes classical computation (for the evaluation of functions) and quantum methods (to do it simultaneously on several sets of variables).

The project is conducted with the Universities of Liège (Françoise Remacle), Hebrew University of Jerusalem (Raphael Levine), University of Padua, the CNR Institute for Physical and Chemical Processes of Bari, the company **KiloLambda (**2001, Israel) which manufactures the quantum dots and **ProBayes** (France), a subsidiary of La Poste which produces the compiler of the solution (Emmanuel Mazer and David Herrera-Marti, who now works at CEA-LIST)[2218]. The great uncertainty on this project, as is often the case, concerns its scalability. It depends on the superposition of optical frequencies. The project documentation does not describe well the application domains of the possible in terms of complexity classes of addressable problems.

[2217] See Progress in neuromorphic photonics by Thomas Ferreira de Lima, Mitchell Nahmias et al, 2017 (23 pages).

[2218] See Coherent Optical Parallel Computing, 2017. The European project is funded until 2021. See more details in Coherent Optical Parallel Computing Project Summary.

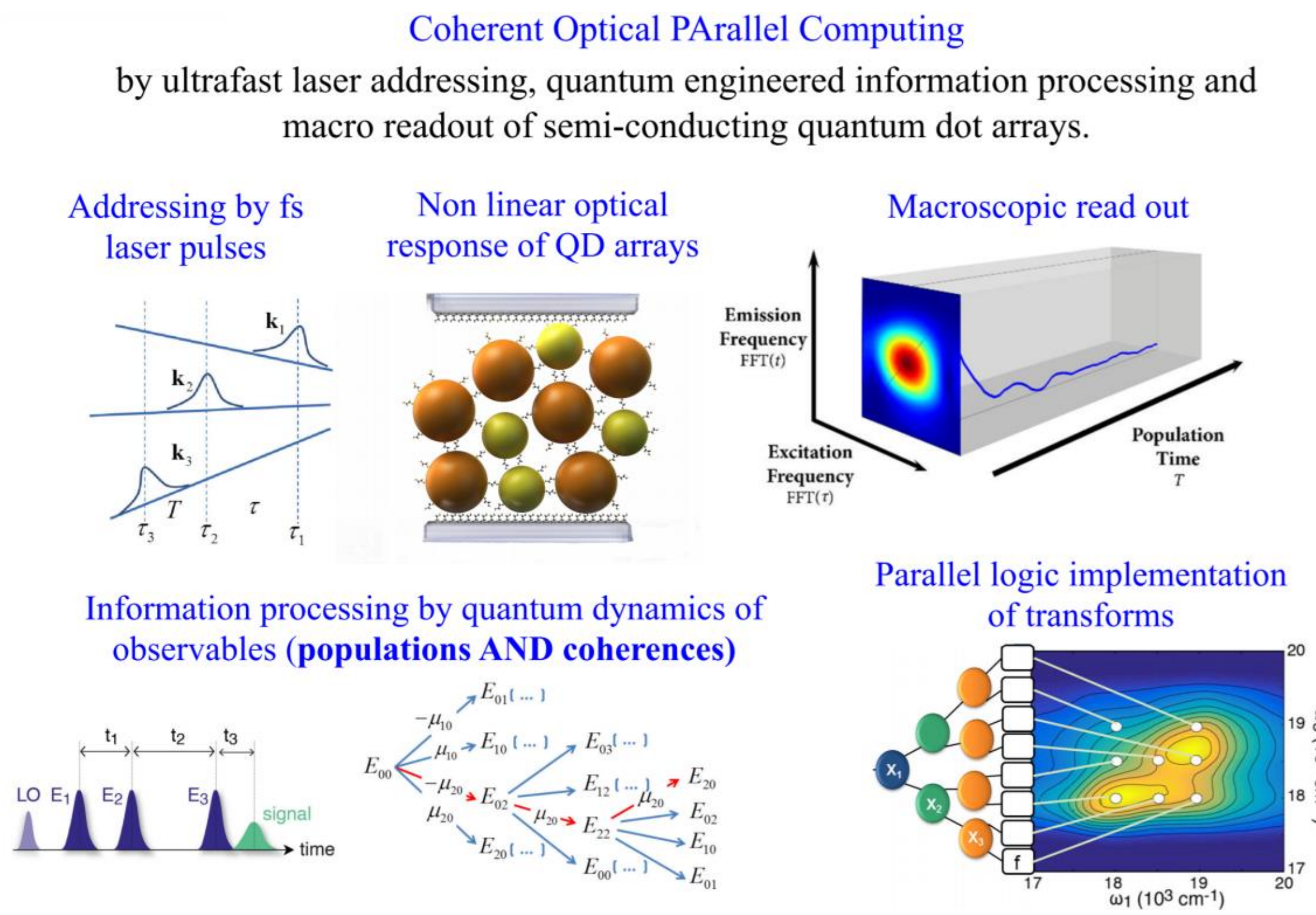

*Figure 724: the optical technology behind the European Copac project.*

Another collaborative effort in photonic chipsets development is with the **PhotonDelta Foundation** (The Netherlands, 1.1B€ including 470M€ from the Dutch government), a public and private European Integrated Photonics Ecosystem fund that supports photonic chip ventures, in collaboration with the MIT. They invested in 177 projects. The program launched in 2022 is supposed to last 6 years. IMEC, LioniX and Bright Photonics are among the industry partners of this fund which gathers 26 companies, 11 technology partners and 12 R&D partners.

**Salience Labs** (2021, UK, $11.5M) is a startup cofounded by Vaysh Kewada (CEO) and Johannes Feldmann (CTO) and spun out of Oxford and the University of Münster in Germany.

It develops a hybrid photonic-electronic silicon chip dedicated to AI applications. Other research participants are the Universities of Pittsburgh (USA) and Exeter (UK), EPFL and IBM Research Zurich. It implements fast tensor processing running at 100 GHz. It implements sort-of in-memory photonic computing implementing optical frequency combs coupled with a classical SRAM (Static Random Access Memory)[2219]. They could also use phase-change materials developed by Oxford University instead of SRAM[2220].

**Lightelligence** (2017, USA, $36M) develops PACE, another photonic processor that is supposed to be 100 times faster than a Nvidia RTX 3080 GPU. It is designed to solve Ising models problems like D-Wave quantum annealers and Fujitsu digital CMOS-based annealers.

[2219] It is documented in Parallel convolution processing using an integrated photonic tensor core by J. Feldmann et al, Nature, January 2021 (35 pages).

[2220] Oxford University researchers designed a light-based silicon photonic chip with hybridized-active-dielectric (HAD) nanowires enabling light polarization-selective tunability. They demonstrated the use of polarization as a parameter to selectively modulate the conductance of individual nanowires within a $Ge_2Sb_2Te_5$ (GST for germanium-antimony-tellurium, a phase changing material also used in RW optical discs and phase-change memories) and silicon multi-nanowire system. See Polarization-selective reconfigurability in hybridized-active-dielectric nanowires by June Sang Lee and al, Science Advances, June 2022 (8 pages).

Their chipset uses 12,000 thermal phase shifters and Mach-Zehnder interferometers (MZI) and runs at 1 GHz[2221]. It implements parallel photonic recurrent network[2222]. Now, how does it scale is yet to be shown.

So, imagine you didn't care about quantum computing because it's too complicated. Just with focusing on this optical classical and sometimes supposedly quantum photonic computing, you'd have to spend a significant amount of time to figure out which ones deserve some attention. It seems that you have many more projects run in parallel than specialists who could benchmark and compare them in a neutral fashion.

# Chemical computing

Chemical computing is yet another form of unconventional computing. The underlying physical phenomenon was discovered in 1951 by a Russian chemist, Boris Belousov, with a self-reversible chemical reactions alternating the color of solution mixing salts and acids. It would enable the creation of chemical logic gates. It was thought initially that it broke the second law of thermodynamics (a bit like time crystals…). It was found later that it was not the case and that some form of logic gate could be implemented using this phenomenon and a chemical oscillator that is akin to chemical concentration waves blocking or amplifying each other depending on their setting. One benefit is a low energy consumption[2223].

Research in the domain started to generate some experimental results in 1989 and onwards, particularly in the UK, at EMPA in Switzerland, Stanford, Harvard and the University of Washington. But the benefits of this architecture are not obvious, noticeably due to its relatively slow speed. The most recent work came in 2020 from the University of Glasgow who used 3D-printed programmable chemical processor with a 5 by 5 array of cells filled with a switchable oscillating chemical and magnetic stirrers to control their oscillations and compute binary logic gates to perform pattern recognition[2224]. It can control $2.9 \times 10^{17}$ chemical states.

The array is programmed with setting chemical relations between these chemical cells. It can implement a chemical autoencoder used for pattern recognition. Now, more comparisons have to be made with classical computing to see how it scales[2225]!

[2221] See Deep Learning with Coherent Nanophotonic Circuits by Yichen Shen et al, October 2016 (8 pages) and Accelerating recurrent Ising machines in photonic integrated circuits by Mihika Prabhu et al, Optica, April 2020 (8 pages). It seems the system can parametrize an Ising model only with coupling graph nodes with the equivalent of a longitudinal field, but not with setting longitudinal interactions at the graph node level like with D-Wave quantum annealers.

[2222] The system implements hybrid photonic/classical computing with the photonic part performing a unitary matrix product of a state-preparation rotation matrix R, the desired unitary (U or U†), and one of two homodyne detection matrices (h1 or h2). Phase-intensity reconstruction, a diagonal matrix multiplication, Gaussian noise addition, and a nonlinear threshold unit are performed classically.

[2223] See Chemical Computing, the Future of Artificial Intelligence by Javier Yanes, January 2019.

[2224] See A programmable chemical computer with memory and pattern recognition by Juan Manuel Parrilla-Gutierrez, Lee Cronin et al, Nature Communications, March 2022 (8 pages).

[2225] See the video The First Programmable Turing Complete Chemical Computer with Lee Cronin from the University of Glasgow (1h11mn).

### Quantum unconventional computing key takeaways

- We study various non-conventional computing technologies that may compete with quantum computing or even be of some help, like reversible and superconducting technologies that may be useful to create cryogenic electronics enabling the creation of scalable quantum computers. But these technologies are so diverse and with different underlying science that they would deserve a lot of time and energy to be properly evaluated and benchmarked with both physicists and computer science specialists.
- Digital annealing computing is mostly proposed by Japanese companies like Fujitsu and Hitachi, using classical CMOS chipsets. These solutions are supposed to solve untractable problems faster than classical-classical computers but their scalability remains questionable.
- Reversible and adiabatic computation has been researched for a long time and has not yet turned into commercial products. It could probably be more interesting to create energy saving solutions more than faster solutions with some potential use case in quantum computing enabling technologies.
- Superconducting computing is an interesting area of research to create more energy efficient supercomputers, despite the cooling cost that, hopefully is not as expensive as with superconducting qubits quantum computers. There are synergies between this research area and superconducting logic electronics that could be used to control superconducting and quantum dots spin qubits at low temperatures.
- Probabilistic and optical computing are interesting research areas. These solutions may be competitive to solve particular problems.
- Optical processors are mainly used in the deep learning space, to accelerate the training and inferences in some layers of convolutional networks.
- Chemical computing is one of the many other areas in unconventional computing that may be interesting but have probably various scaling limitations.

# *Quantum telecommunications and cryptography*

Quantum telecommunications cover a wide spectrum of use cases including quantum communications between quantum sensors, quantum computers and quantum key distribution used in cryptography. Some but not all of these technologies are based on using photon entanglement and quantum teleportation as resources. The field started to develop experimentally and industrially with quantum cryptography. It is already being deployed experimentally or at large scales like in China, while quantum telecommunications associated with quantum computers is still in its infancy.

Interest in quantum cryptography (using various quantum effects like measurement randomness and/or entanglement as resources to generate identical secured encryption keys between two communication endpoints) as well as in post-quantum cryptography (classical cryptography using techniques that are resilient to attacks by quantum computers) were triggered by the creation of Shor's algorithm in 1994. It theoretically enables integers factoring on a quantum computer in reasonable times, provided large scale quantum computers are available. This algorithm has been destabilizing computer security specialists since the advent of the first programmable quantum processors in the early 2000s and related scale-up promises and forecasts. It would make it possible to break the codes of many public key cryptography systems that are commonly used on the Internet. It is still highly hypothetical because quantum computers capable of executing Shor’s algorithms for RSA-2048 keys and requiring millions of high-fidelity qubits are really far in the distant future at best.

Once they are aware of the threat, however, governments, counterintelligence, intelligence and sensitive industries become seriously concerned or at least interested. The threat of quantum computing on cybersecurity even touches critical parts of the Bitcoin and Blockchain signatures and some proof of work protocols. Even though the threat is quite remote, the readiness inertia to counter this potential quantum threat means, for some, that it is necessary to launch it now.

Even before Shor's phantom menace materializes, the cyber security industry started to prepare countermeasures. The markets affected first will be the IT and telecommunications industry in general, which will have to update many software and hardware offerings, financial institutions, the energy sector, healthcare, and government utilities and the military.

In this part, we will describe:

- The basic principles of classical cryptography, in particular **public key cryptography,** with the example of RSA public keys.
- The **nature of the threat** coming from integer factoring and other quantum algorithms and the cryptographic solutions involved.
- **Quantum random number generators** which have become an indispensable complement to classical cryptographic solutions and also quantum cryptography as well.
- **Quantum Key Distribution** based systems that secure the physical part of communications for the safe distribution of random symmetric keys.
- **Post-Quantum Cryptography** that is used to distribute public encryption keys using classical computation that are resilient to codebreaking done by quantum computers.
- **Quantum telecommunications** applications, outside of those related to cryptography, and in particular for the creation of distributed quantum computing systems.

- **Quantum Physical Unclonable Functions** which are sitting in between QRNG, embedded systems and cryptography.
- **Vendors** in these sectors around the world, in a market that already includes many players and in particular many startups.

Encryption and cryptography involve mathematical concepts that are not always obvious. I'll share with you here what I have been able to understand about it and make is as accessible as possible.

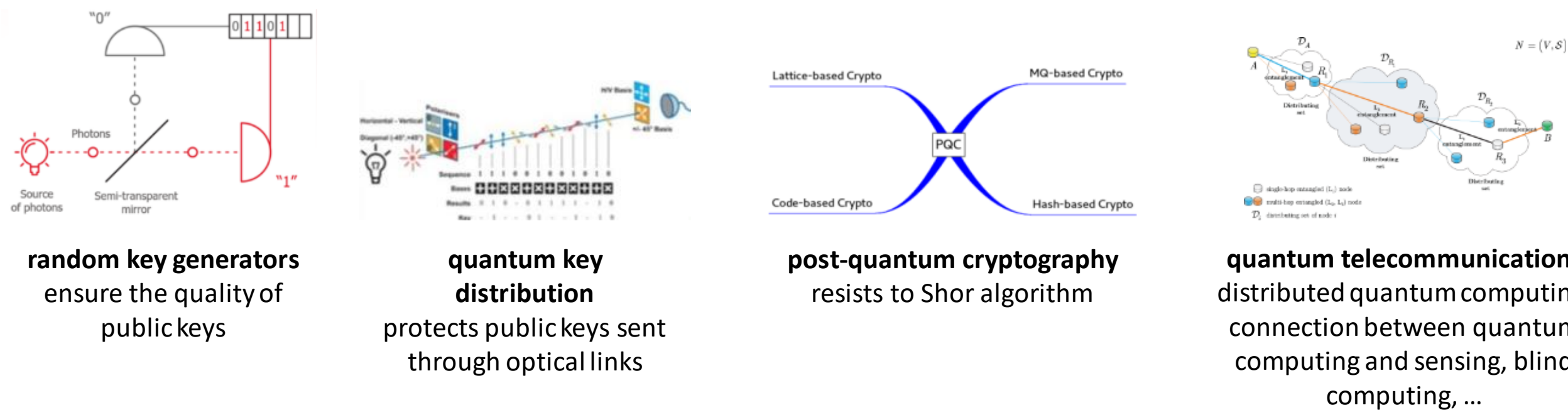

*Figure 725: the four classes of technologies covered in this part. (cc) Olivier Ezratty, 2020.*

# Public key cryptography

Cryptology is the science of secrets. It allows the transmission of sensitive information between a transmitter and a receiver in a secure manner. Cryptology includes cryptography, which secures transmitted and stored information, and cryptanalysis, which seeks to decrypt it by attack, or code breaking.

In the case of asymmetric public-key cryptography, encryption uses only public keys and decryption uses both public and private keys. Code-breaking exploits only the public keys, while trying to deduce the private keys through intensive computing.

Cryptography secures transmitted information in several ways:

- **Confidentiality**: only the recipient can retrieve the unencrypted version of the transmitted information.
- **Integrity**: the information has not been modified during its transmission.
- **Authentication**: the sender and receiver are who they claim to be.
- **Non-repudiation**: the issuer cannot deny having transmitted the encrypted information.
- **Access control**: only persons authorized by the issuer and the recipient can access unencrypted information.

Before computer telecommunications, confidentiality was ensured by the knowledge of a common secret between transmitters and receivers, the encryption and decryption codes, which could be the position of the wheels of a German Enigma machine during the World War II. This worked in closed environments such as for military communications or between embassies and their home countries.

With Internet communications, this modus operandi is inapplicable for consumer applications and for business relationships in general. Hence public key cryptography systems, such as RSA, which enabled most open Internet data exchanges. There are still highly protected systems using private and symmetrical keys, mainly used in government related applications (army, security, intelligence) as well as in various other cases (some file transfers, email encryption, server/client exchanges, in smart cards and associated payment terminals).

Asymmetric (public-key) cryptography is also exploited for pre-establishing common encryption keys between users of private-key systems, for managing the integrity of communications, and for authentication as in the TLS Internet protocol. Sensitive information is then encrypted with these keys and a symmetrical AES-type algorithm. AES is used to encrypt communications in WhatsApp, Messenger and Telegram. These applications often also use asymmetric cryptography for authentication, key exchange and communication integrity management. Asymmetric cryptography is very flexible while symmetric cryptography is more computationally efficient.In many cases, symmetric cryptography systems coexist with asymmetric (public key) cryptography systems. As a result, when you communicate securely over the Internet, different complimentary security protocols are activated.

With public key systems, different keys are used for encryption and decryption of the information transmitted, so that it is very difficult (if not sometimes impossible) to guess the private decryption key from the public encryption key. The message receiver sends its public key to the sender, who in turn uses it to encrypt the message. The receiver uses the private key that was kept to decrypt the received message. As explained in Figure 726, the private key is never transmitted. This is also called a PKI, for "Public Key Infrastructure".

The **RSA** cryptography system is the most widely used system for protecting public key information transmissions over the Internet. It was created in 1978 by **Ron Rivest** (1947, American), **Adi Shamir (**1952, Israeli) and **Leonard Adleman** (1945, American).

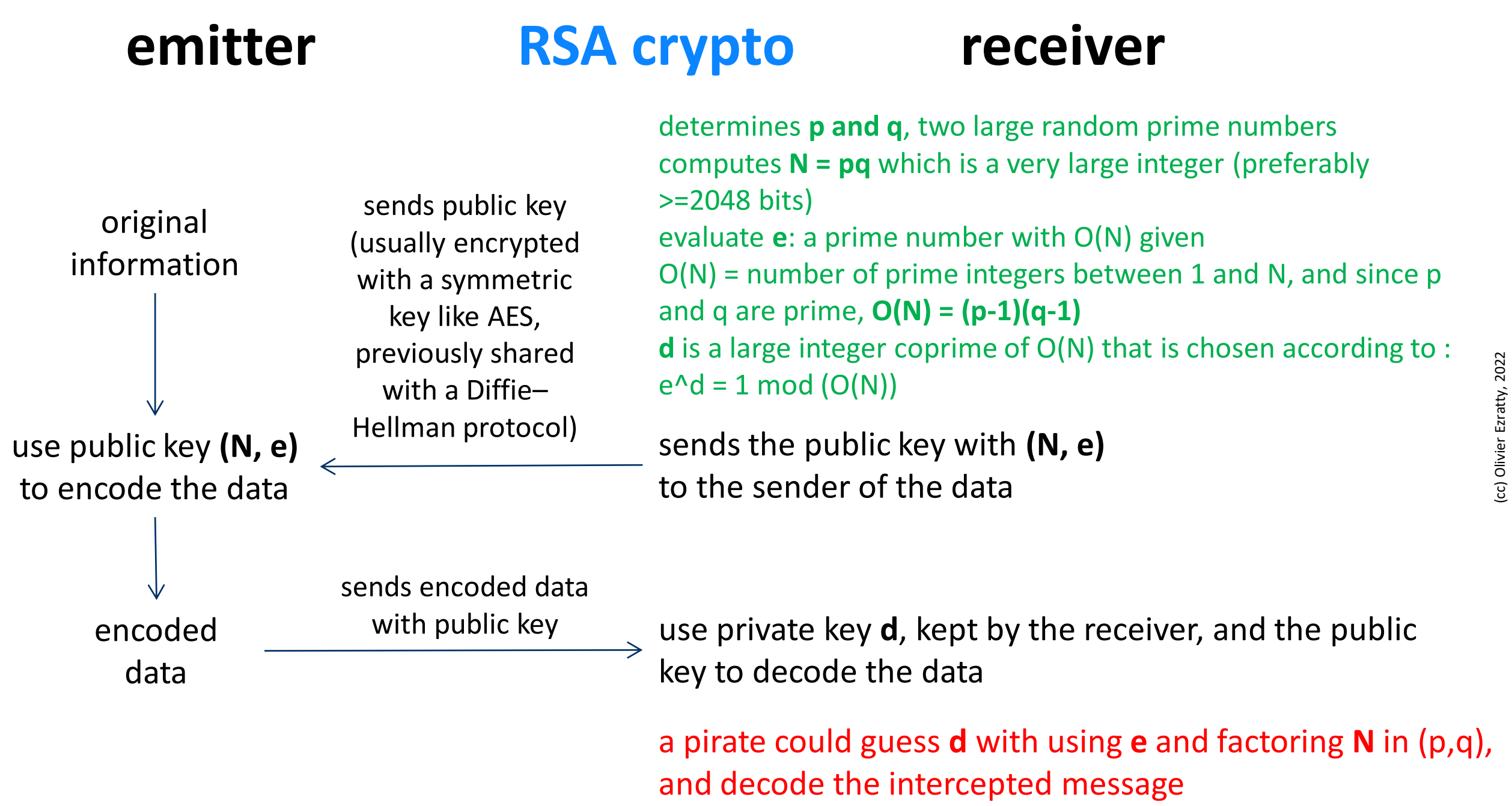

*Figure 726: description of the RSA key generation process. (cc) Olivier Ezratty, 2022.*

You don't necessarily need to understand the following that explains how keys are constructed. It starts by determining p and q, two large random prime numbers, with a "good" random number generator. We will see later that quantum physics can be used to create really random numbers. We calculate N = pq which is a very large integer. A good RSA key requires to have N stored on at least 2048 bits knowing that the NSA recommends 3072 bits keys for critical applications.

We then evaluate **e**, a prime number by exploiting O(N) which equals the number of prime integers between 1 and N relative to N, and which, as p and q are prime, equals (p-1)(q-1). d is a large integer which is co-prime of O(N) and is chosen according to: e^d = 1 mod (O(N)). At the end, we get a public key that includes the integers N and e, and a private key that includes d. All of this is based on the theory of numbers and uses in particular the Fermat's little theorem and Euler's theorem which make it possible to create two distinct keys that are the inverse of each other.

With that, anyone can encrypt a message using the public key and this message is being decipherable only by the person who has the private key that splits the public key into primitives.

A hacker could decrypt the information sent by intercepting e (the end of the public key) and factoring N, the other end of the public key, into the integers p and q, and then rebuilding the private key d from it.

To date, prime number factoring requires a traditional machine power that grows with the square root of the number to be factorized. The official RSA key factoring record was 768 bits in 2010, 795 bits in 2019 and 829 bits in February 2020[2226]. Even if this doesn't take into account undisclosed NSA records, it provides an idea of the problem scale. We're far from having a classical computer breaking an RSA 2048 bits code given the problem size is exponential with the size of the RSA key.

In the symmetric key realm, you'll find encryption systems using protocols like AES with usual keys of 256 bits, that is in use since 2001. These keys are shared by two parties in a channel and time that can be different from the encrypted data transmission. The best symmetric key system is called the "one time pad" with key size equaling the size of the encrypted data. It means that the key can be very large depending on the data size. But this is the best secured system. It however requires the ability to create a random and totally unpredictable key.

## Quantum cryptoanalysis threats

The diagram below in Figure 727 points out the main encryption algorithms vulnerable or not to known quantum algorithms[2227]. Broadly speaking, common public key encryption systems are vulnerable. Only post-quantum cryptography systems are supposed to be resilient. But things are moving fast and some SIDH, lattice and multivariate keys are now classically broken!

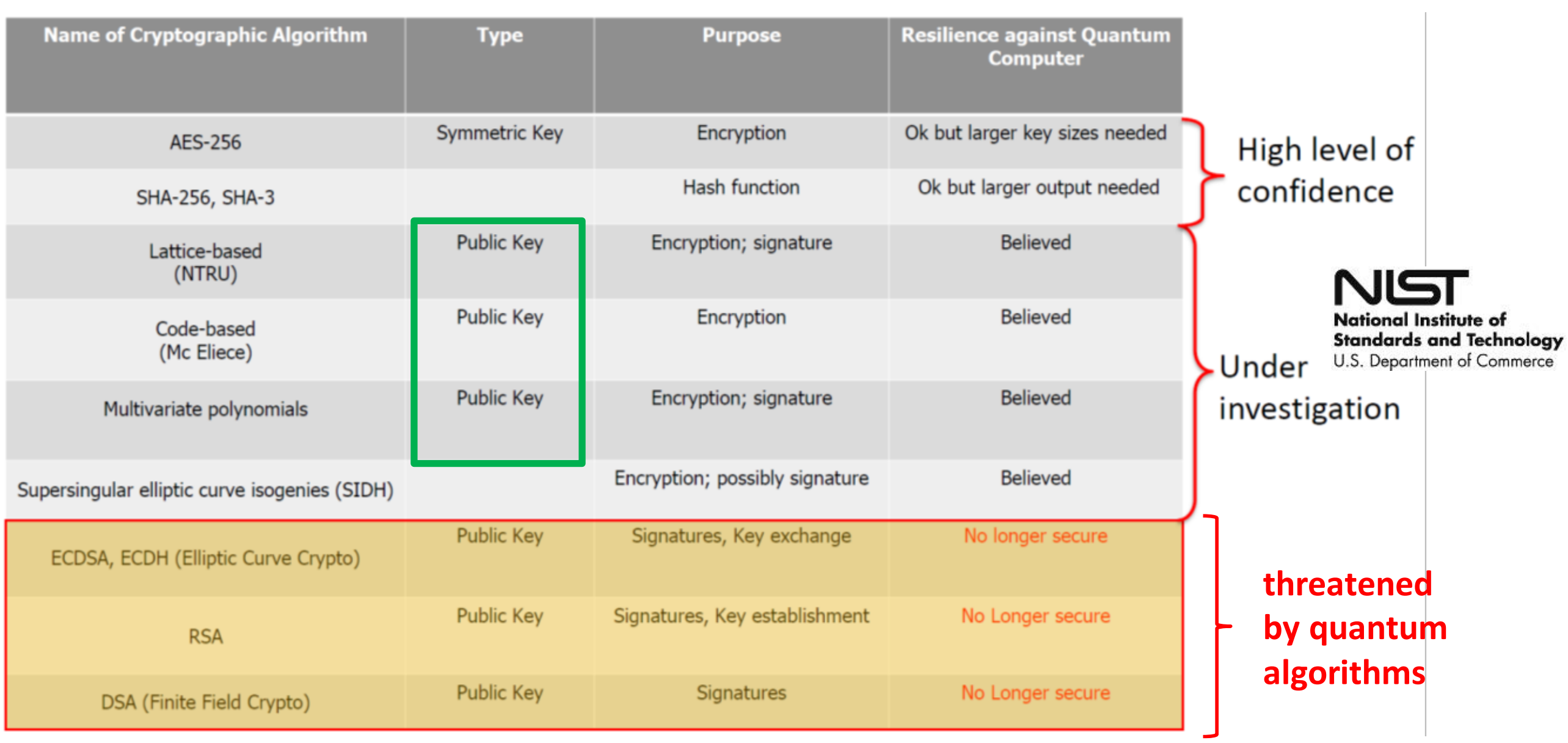

| Name of Cryptographic Algorithm | Type | Purpose | Resilience against Quantum Computer |
|---|---|---|---|
| AES-256 | Symmetric Key | Encryption | Ok but larger key sizes needed |
| SHA-256, SHA-3 | | Hash function | Ok but larger output needed |
| Lattice-based (NTRU) | Public Key | Encryption; signature | Believed |
| Code-based (Mc Eliece) | Public Key | Encryption | Believed |
| Multivariate polynomials | Public Key | Encryption; signature | Believed |
| Supersingular elliptic curve isogenies (SIDH) | | Encryption; possibly signature | Believed |
| ECDSA, ECDH (Elliptic Curve Crypto) | Public Key | Signatures, Key exchange | No longer secure |
| RSA | Public Key | Signatures, Key establishment | No Longer secure |
| DSA (Finite Field Crypto) | Public Key | Signatures | No Longer secure |

*Figure 727: what key generation services are quantum safe or not. Source: NIST.*

[2226] This factorization of an RSA-250 digits (829 bits) and the previous RSA-240 digits (795 bits) was achieved by an international team led by French researchers from Inria: Fabrice Boudot (Université de Limoges), Pierrick Gaudry (CNRS), Aurore Guillevic, Emmanuel Thomé and Paul Zimmermann (Inria) and Nadia Heninger (University of California). Computation used 2700 core-years, of Intel Xeon Gold 6130 CPUs running at 2.1 GHz. See The State of the Art in Integer Factoring and Breaking Public-Key Cryptography by Fabrice Boudot, Pierrick Gaudry, Aurore Guillevic, Nadia Heninger, Emmanuel Thomé and Paul Zimmermann, June 2022 (9 pages) and Factoring RSA-240 and computing discrete logarithms in a 240-digit prime field with the same software and hardware by Fabrice Boudot, Pierrick Gaudry, Aurore Guillevic, Nadia Heninger, Emmanuel Thomé and Paul Zimmermann, Inria, March 2021 (87 slides).

[2227] Seen in IDQ: Quantum-Safe Security relevance for Central Banks, 2018 (27 slides) and slightly supplemented by some captions.

## Shor's phantom menace

Peter Shor**'s** algorithm sparked interest in quantum computing when there was even no single working qubit around that was controllable by quantum gate! Shor’s algorithm enables theoretically to factor integers in a reasonable time that is proportional to their logarithm. It is thus a factorization in a linear time as a function of the number of key bits. This could be detrimental to all public key-based cryptography[2228].

But this will happen only in a relatively distant future! In 2019, Google researchers published an algorithm allowing to quickly break an RSA key (of 2048 bits) and with "only" 20 million qubits having an error rate of 0.1% and with a calculation lasting 8 hours. This is more "acceptable" than the billion qubits that were needed in previous instances of Shor’s algorithm. Today's quantum computers have a coherence time much shorter than a second, but the decoherence counter is reset after each error correction code that is used in the algorithm[2229].

Breaking a 2048-bit RSA key requires at least a number of logical qubits equal to twice the size of the key used +2, so 4098 qubits. Depending on the technologies used, this number should be multiplied by 30 to 20,000 for the number of physical qubits. This scalability is one of the greatest challenges for building viable quantum computers as we’ve seen in other parts of this book.

Moreover, as we have seen in the section on Shor's algorithm, the quantum Fourier transform underlying it uses phase controlled R-quantum gates whose implementation is far from obvious. Indeed, when the phase is an angle of 1/2048 times a 360° turn in the Bloch sphere of a qubit, the controlled rotation of the phase can be inferior to the error rate of a one or two-qubit quantum gate. We must therefore bet on the ability of error correction codes to handle this.

Is the threat assessment misplaced? At this point in time, the biggest number ever factored on a gate-based quantum computer with Shor’s algorithm is 35, with an IBM QPU[2230]. Using a hybrid Variational Quantum Factoring algorithm itself based on a QAOA algorithm, Zapata Computing factored the number 1,099,551,473,989 (=1,048,589 multiplied by 1,048,601) with 5 IBM superconducting qubits in 2021[2231]. Meanwhile, the largest number factored on a D-Wave 2000Q annealer was achieved in 2019 was 376,389 using a block multiplication table method [2232].

A 762-bit RSA key close to the 2010 record would require a **D-Wave** annealer computer with 5.5 billion qubits, far from the existing 5000[2233]. A D-Wave of 5893 qubits could do the job if all qubits could be arbitrarily coupled to the other, which is not possible due to the way these 2D chipsets are designed. And trapped ions and their any-to-any connectivity won’t save us since they don’t seem to scale. Also, we shouldn’t discount threats from quantum machine learning algorithms[2234].

---

[2228] See this presentation which describes in great detail how Shor's algorithm works: On Shor's algorithms, the various derivatives, their implementation and their applications by Martin Ekerå, 2019 (135 slides).

[2229] See How to factor 2048 bit RSA integers in 8 hours using 20 million noisy qubits by Craig Gidney and Martin Ekerå, 2019 (25 pages).

[2230] See An Experimental Study of Shor's Factoring Algorithm on IBM Q by Mirko Amico, Zain H. Saleem and Muir Kumph, 2019 (10 pages).

[2231] See Analyzing the performance of variational quantum factoring on a superconducting quantum processor by Amir H. Karamlou et al, npj, Zapata Computing October 2021 (6 pages).

[2232] See Breaking RSA Security With A Low Noise D-Wave 2000Q Quantum Annealer: Computational Times, Limitations And Prospects by Riccardo Mengoni et al, Cineca, 2019 (8 pages).

[2233] According to High-fidelity adiabatic quantum computation using the intrinsic Hamiltonian of a spin system: Application to the experimental factorization of 291311 by Nike Dattani, Xinhua Peng and Jiangfeng Du, June 2017 (6 pages).

[2234] See Capgemini and Fraunhofer IAIS lead study in quantum machine learning for it security commissioned by the German Federal Office for Information Security, November 2021.

Shor's menace was visualized in the diagram in Figure 728 from the European standardization organization ETSI[2235], based on very optimistic predictions on QPUs capabilities to exploit Shor's algorithm.

The orange part of the graph should be shifted into the future by at least 10 to 20 years, showing how the threat is usually exaggerated by cybersecurity specialists who are not aware of the various difficulties to create scalable quantum computers.

Shor's algorithm applied to RSA public key breaking could however have quite a negative impact on most Internet security since being integrated in the **TLS** and **SSL** protocols that protect websites and file transfers via **HTTPS** and **FTP**, in the **IPSEC** protocol that protects IP V4 in the IKE sub-protocol, in the **SSH** protocol for machines remote access and in the **PGP** protocol that is sometimes used to encrypt emails. RSA and derivatives are also used in many **HSM** (Hardware Security Modules) such as in cars ECUs (Electronic Central Units)[2236].

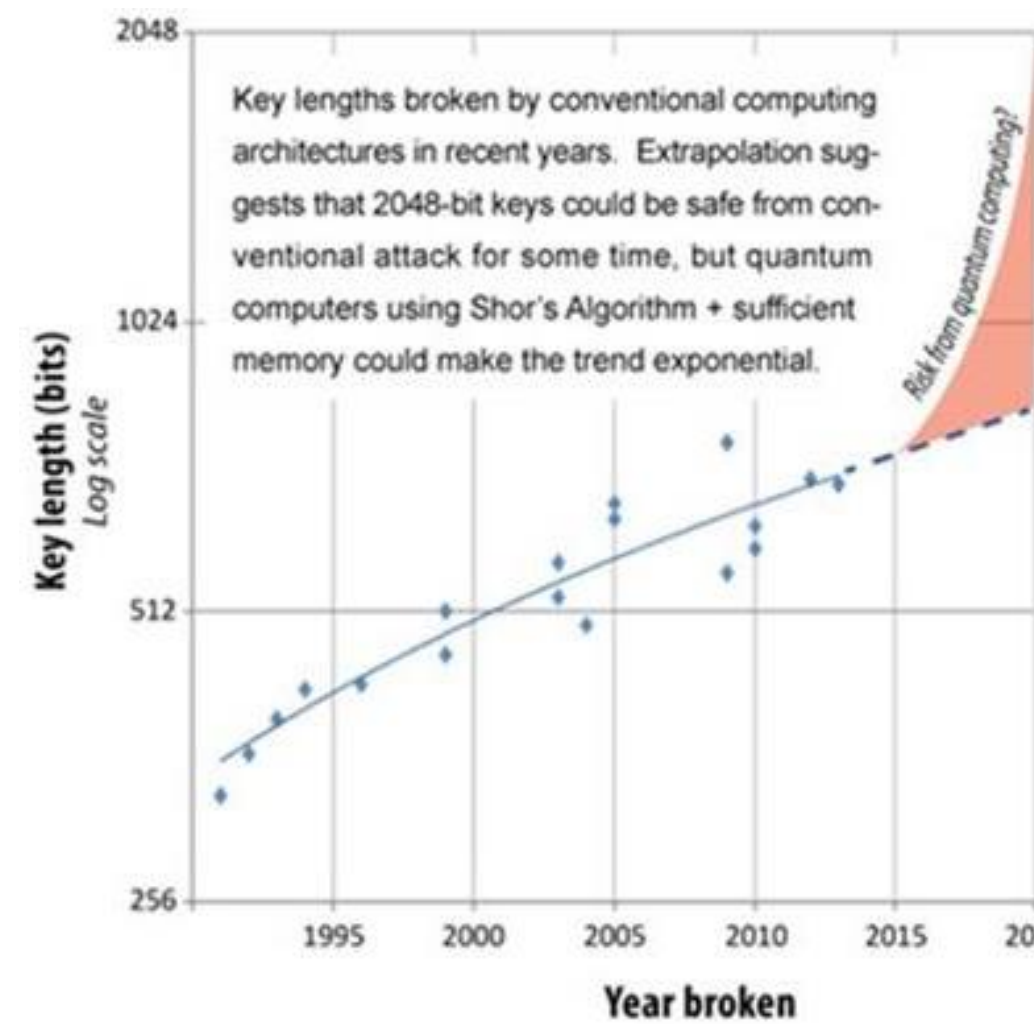

*Figure 728: how Shor's risk is usually overestimated with a past example. Source: Quantum Safe Cryptography and Security, 2015 (64 pages).*

The threat also deals with software **electronic signatures** and therefore their automatic updates, **VPNs** used for remote access to protect corporate networks, email security with **S/MIME,** various **online payment** systems, **DSA** (Digital Signature Algorithm, an electronic signature protocol), **Diffie-Hellman** codes (used for sending symmetrical keys) as well as **ECDH, ECDSA** and **3-DES** elliptic curve cryptography. The **Signal** protocol used in WhatsApp would also be in the spotlight. So a lot of Internet security is more or less in the line of sight.

**ECC** (Elliptic Curve Cryptography) is the first algorithm with elliptic curves, created in 1985 by Neal Koblitz and Victor Miller. The most common variants today are **ECDH** (Elliptic-curve Diffie-Hellman) and **ECDSA** (Elliptic Curve Digital Signature Algorithm, launched in 2005).

These variants were deployed from 2005 and more widely only from 2015, so 30 years after the creation of the first ECC! Incidentally, the elliptic curves allowed Andrew Wiles to demonstrate the Fermat's last theorem in 1992, which has nothing to do with cryptography[2237].

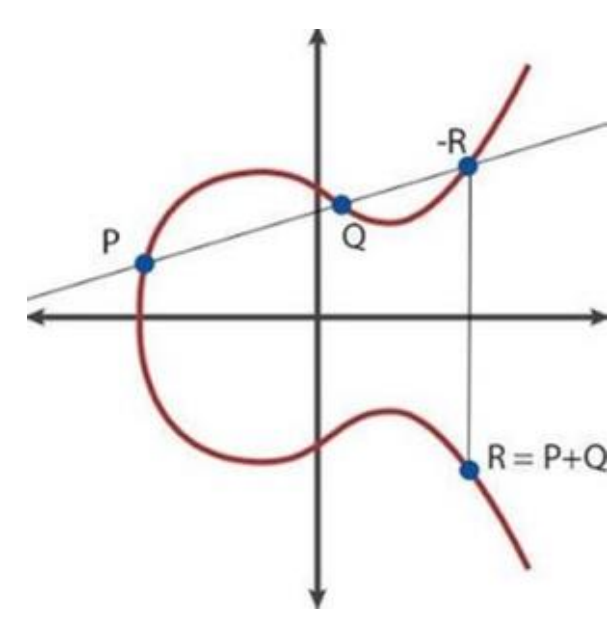

*Figure 729: an elliptic curve.*

One of the interests of elliptic curve-based codes is to use shorter public keys than with RSA encryption. But these elliptic curves are also theoretically breakable with quantum computing with a reasonable time because of our friend Peter Shor and the resolution of the discrete logarithm problem (DLP) [2238]. Moreover, an ECDSA backdoor was revealed by Edward Snowden in 2013, housed by the NSA in its Dual EC DRBG random number generator and not in the elliptic curve itself.

---

[2235] See Quantum Safe Cryptography and Security, 2015 (64 pages).

[2236] See Post-Quantum Secure Architectures for Automotive Hardware Secure Modules by Wen Wang and Marc Stöttinger, 2020 (7 pages).

[2237] As in Elliptic curves cryptography and factorization (86 slides).

[2238] As documented in Shor's discrete logarithm quantum algorithm for elliptic curves by John Proos and Christof Zalka, 2003 (34 pages). The discrete log problem consists in finding an integer k verifying $a^k = b$ modulo p, a, b and p being known integers. This allows to break the elliptic and Diffie-Hellman curve keys.

It was then recommended by NIST in 2014 and NSA in 2015 for the transmission of sensitive information[2239].

The other reason for alternative cryptography solutions is that today's sensitive communications can be stored for a long time by private or state hackers, and exploited much later, when quantum computers are up to the task. This category of attack is named **Store Now, Decrypt Later** (SNDL). Some present day information may have some value later, whether it is financial transactions, private communications, trade secrets or other state secrets. And forward secrecy, a feature used to protect the transport layer like SSL used with HTTPS with separate keys for each session, is not enough against quantum computing-based attacks. Quantum computing is thus a veritable Damocles sword whose fall is difficult to predict and rather distant in time by at least a very long decade. Beyond that time, it is almost impossible to make educated predictions.

Many predictions are entirely wrong whether they talk about a "quantum apocalypse[2240]", an equivalent of a "nuclear threat"[2241] or when some folks predicted in 2020 that the Bitcoin security would be broken by 2022[2242]. They were confusing physical and logical qubits requirements among other mistakes! The same mistake was reiterated a year and a half later with a cybersecurity poll[2243]. The main issue with these polls is that they are asking a question to cybersecurity specialists and other commentators who have no clue about the scalability challenges of quantum computing[2244]. And that's true as well for a former NSA director[2245]! The same could said about the interpretation of a 2021 paper by Nicolas Sangouard et Elie Gouzien which stated that theoretically, a RSA-2048 key could be broken by "only" 13,436 physical qubits in 177 days provided it could exploit a 40 million modes quantum memory that is even more long term than 13,436 physical qubits[2246].

Only a few analysts think that the quantum threat is overestimated[2247]. It is the result of a mix of an unbalanced knowledge of the actual quantum threat between quantum information and cybersecurity specialists. On top of that, cybersecurity vendors have a natural interest to increase the threat perception, in order to sell PQC-based upgraded security solutions[2248]. There's a nice potential market to address with $10B by 2030[2249]. So, if quantum computers were able to scale to a point to break today's asymmetric cryptography, it would indeed create a quantum apocalypse. The question is whether it is realistic or not. There is a bigger risk to kill all electronic devices with solar flares than asymmetric cryptography with a large scale quantum computer.

---

[2239] See Ben Schwennesen's Elliptic Curve Cryptography and Government Backdoors, 2016 (20 pages).

[2240] See What is the quantum apocalypse and should we be scared? by Frank Gardner, BBC, January 2022.

[2241] See 'Nuclear Threat to Cybersecurity' — Post-Quantum Cybersecurity Rapidly Gains Attention of U.S. Congress, Administration by Matt Swayne, The Quantum Insider, July 2022.

[2242] See Quantum computers could crack Bitcoin by 2022 by Robert Stevens, May 2020.

[2243] See Cybersecurity Experts Say Quantum, Advanced Technology Will Break Standard Encryption Within Two Years by Matt Swayne, The Quantum Insider, December 2021.

[2244] One good example is this paper Quantum Computing Threatens to Collapse the Grid by Alexander Boulden, December 2021 that forecasts doomsday for grid management due to quantum computing threat on cybersecurity. The chart from Statistica/CBinsights is both false and outdated, showing three qubit systems that never worked: Intel 49 superconducting qubits, Google's Bristlecone 72 qubits (it ended up being 53 functional qubits with Sycamore in 2019) and Rigetti's 128 qubits that were announced in 2018 and never released. They are up to 80 qubits as of 2022.

[2245] See Podcast with Adm. Mike Rogers - former NDA director by Yuval Boger, March 2022.

[2246] See Factoring 2048-bit RSA Integers in 177 Days with 13436 Qubits and a Multimode Memory by Nicolas Sangouard and Élie Gouzien, September 2021 (20 pages).

[2247] See Quantum Cryptanalysis: Hype and Reality by Chris Jay Hoofnagle and Simson Garfinkel, February 2022 think that the quantum cybersecurity threat is overestimated.

[2248] See The race to save the Internet from quantum hackers by Davide Castelvecchi, Nature, February 2022.

[2249] See The Quantum Insider Report Forecasts Quantum Security Market Worth $10 Billion by 2030, February 2022.

The fear is also fueled by governments' paranoia and the global fight between the USA and China for quantum technology dominance. On top of that, a lot of myths abound about the NSA and its capabilities, up to many folks thinking that this organization already owns an RSA-breaker QPU in its datacenters. While you can never efficiently prove that something doesn't exist, a good understanding on the scalability challenges with quantum computing make serious people think this thing is just a bad nightmare. But if the cost of getting protected against this potential threat it is reasonable, then it makes sense to implement PQC solutions.

### Mosca's inequality

Michele Mosca created an inequality that explains the time risk. It is expressed in the form $D+T>G_c$, where D is the length of time during which today's data circulating in encrypted form must be secured, T, the time needed to make my transition from its encryption systems to solutions resistant to quantum computing, and $G_c$, the time it will take to develop quantum computers capable of breaking the public keys of current encryption systems. You specify D. You can plan for T according to your information systems and available commercial solutions and standards. How about $G_c$? You must evaluate it with your gut feeling because current estimates span from 5-10 years to... never! For example, some researchers from the UK and USA tried in 2020 to predict when a FTQC would show up[2250]. With a sort of logistic regression, their model predicted that proof-of-concept fault-tolerant quantum computers will be developed between 2026 and 2033 with 90% confidence with the median in early 2030, and that RSA-2048 Shor attacks will become feasible between 2039 and 2058 with a 90% confidence and median in 2050. Making predictions with such a method for a 30 year timeframe seems preposterous.

Michele Mosca and Marco Piani from evolutionQ publish a report every year collecting the opinion of 46 respondent experts on the potential advent of a quantum threat to public-key cryptography. The 2021 edition seems to showcase similar results as in the 2020 edition[2251]. We see a broad spectrum ranging from 3 experts thinking that it would materialize in fewer than 5 years and most of them thinking it would do so before 30 years. My take is that the best prediction one should make is: "*we don't know*"! And when you are paranoid, it easily becomes "*who knows?*".

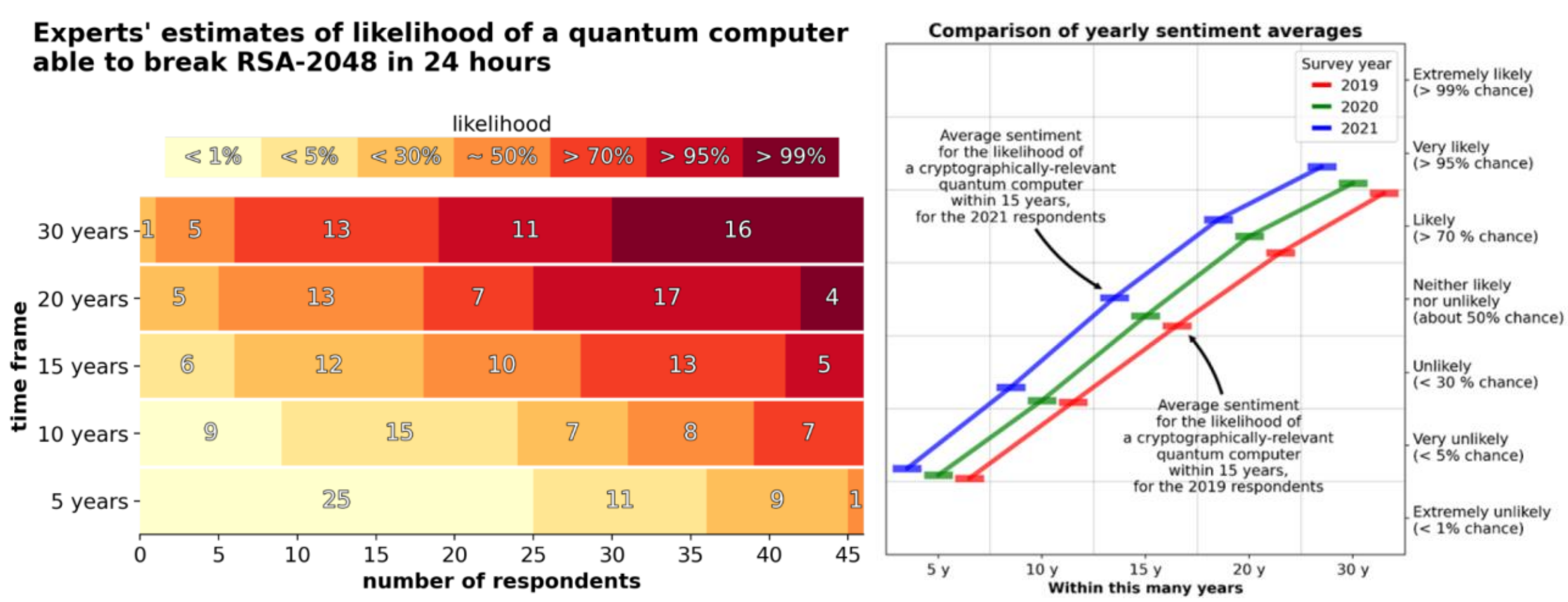

*Figure 730: evolutionQ's yearly report on the quantum cyber risk as estimated by specialists from various disciplines. Source: 2021 Quantum Threat Timeline Report by Michele Mosca and Marco Piani, 2021 (87 pages).*

[2250] See Forecasting timelines of quantum computing by Jaime Sevilla and C. Jess Riedel, December 2020 (23 pages).

[2251] See 2021 Quantum Threat Timeline Report by Michele Mosca and Marco Piani from evolutionQ for the Global Risk Institute, 2021.

As an illustration, a US report from 2004 prepared with the best scientists of this period planned by the year 2012 to see the implementation of a concatenated quantum error-correcting code and to obtain 50 physical qubits[2252]. These qubits came by 2019 and we don't have yet a sufficient number of qubits to implement concatenated codes. In 2012, the surface code was invented that is more efficient and has been experimented at very low scale.

### Grover, Dlog and Simon threats

Symmetric cryptography systems are not affected by Shor's algorithm. These include the **Data Encryption Standard** (DES) which uses keys of 64 bits or more and is outdated, replaced by the **Advanced Encryption Standard** (AES) which has been a US government standard since 2002, with private keys ranging from 128 to 256 bits.

To date, the best quantum breaking algorithms for symmetric **AES** keys would take more than the age of the Universe (13.8 billion years) to run on 128-bit keys. With AES-256 bits, we are therefore in peace, and it is the recommended key length to be quantum resistant! They are based on mechanisms that are quite different from the mathematical problem-solving of public key ciphers.

Keys are shared upstream of the exchanges and are generally themselves encrypted with the asymmetric **Diffie-Hellman** algorithm. But this Diffie-Hellman encryption is based on elliptic curves, which is breakable by Shor's algorithm. The problem lies then with the vulnerability of the majority of encryption systems using asymmetric keys which are used to share symmetric keys.

A hash function converts data of arbitrary size such as a file to a number of fixed size. This makes it possible to do quick searches to compare files. For example, it can be used to check that a file has not been altered during transmission.

SHA algorithms (Secure Hash Algorithms) are standard hash functions that consist in replacing data of arbitrary size by a unique key size. The **SHA-1** hash algorithm is resistant to Shor's algorithm, but it has been broken by other methods and is therefore considered outdated. It is the **SHA-3** which is the most up-to-date and since 2015. The SHA algorithm could be broken by Grover's search algorithm, but with a large number of logical qubits, at least 6,000 logical qubits for common keys[2253] .

| | | **SHA-256** | **SHA3-256** |
|---|---|---|---|
| Grover | $T$-count | $1.27 \times 10^{44}$ | $2.71 \times 10^{44}$ |
| | $T$-depth | $3.76 \times 10^{43}$ | $2.31 \times 10^{41}$ |
| | Logical qubits | 2402 | 3200 |
| | Surface code distance | 43 | 44 |
| | Physical qubits | $1.39 \times 10^{7}$ | $1.94 \times 10^{7}$ |
| Distilleries | Logical qubits per distillery | 3600 | 3600 |
| | Number of distilleries | 1 | 294 |
| | Surface code distances | $\{33, 13, 7\}$ | $\{33, 13, 7\}$ |
| | Physical qubits | $5.54 \times 10^{5}$ | $1.63 \times 10^{8}$ |
| Total | Logical qubits | $2^{12.6}$ | $2^{20}$ |
| | Surface code cycles | $2^{153.8}$ | $2^{146.5}$ |
| | Total cost | $2^{166.4}$ | $2^{166.5}$ |

**Table 3.** Fault-tolerant resource counts for Grover search of SHA-256 and SHA3-256.

*Figure 731: the staggering level of quantum resources required to beak SHA-256 symmetric keys with Grover's algorithm.*

This represents an order of magnitude close to the qubit requirements for breaking RSA keys with Shor's algorithm. For example, a hash key or fingerprint can be used to verify the integrity of content such as software or simply a password. The problem is to resist collisions, i.e., methods to find or create an object whose fingerprint would be the one you have, which is quite different from finding the original object (like an image) from its fingerprint, which is rather difficult.

The number of qubits needed to break keys depends on the size of the key. SHA-1 and SHA-2 have small key sizes that can be recovered in a reasonable time with **Grover**'s quantum search algorithm, but this is not the case for SHA-3 which exploits larger keys. This is the same logic as for AES.

---

[2252] See A Quantum Information Science and Technology Roadmap Part 1: Quantum Computation, 2004 (268 pages

[2253] Based on Estimating the cost of generic quantum pre-image attacks on SHA-2 and SHA-3, 2016 (21 pages), which is also the source of the table on this page.

At last, let's also mention **Shor's Dlog** (discrete log) algorithm that threatens Digital Signature Algorithms, Diffie-Hellman key exchanges and El-Gamal encryption. And **Simon**'s algorithm can also break Even-Mansour ciphers that are used in some disk encryption solutions, although with a lot of caveats, the attacker needing to encrypt a superposition of inputs all with the same keys[2254]. All in all, Shor is not the only quantum threat on cybersecurity!

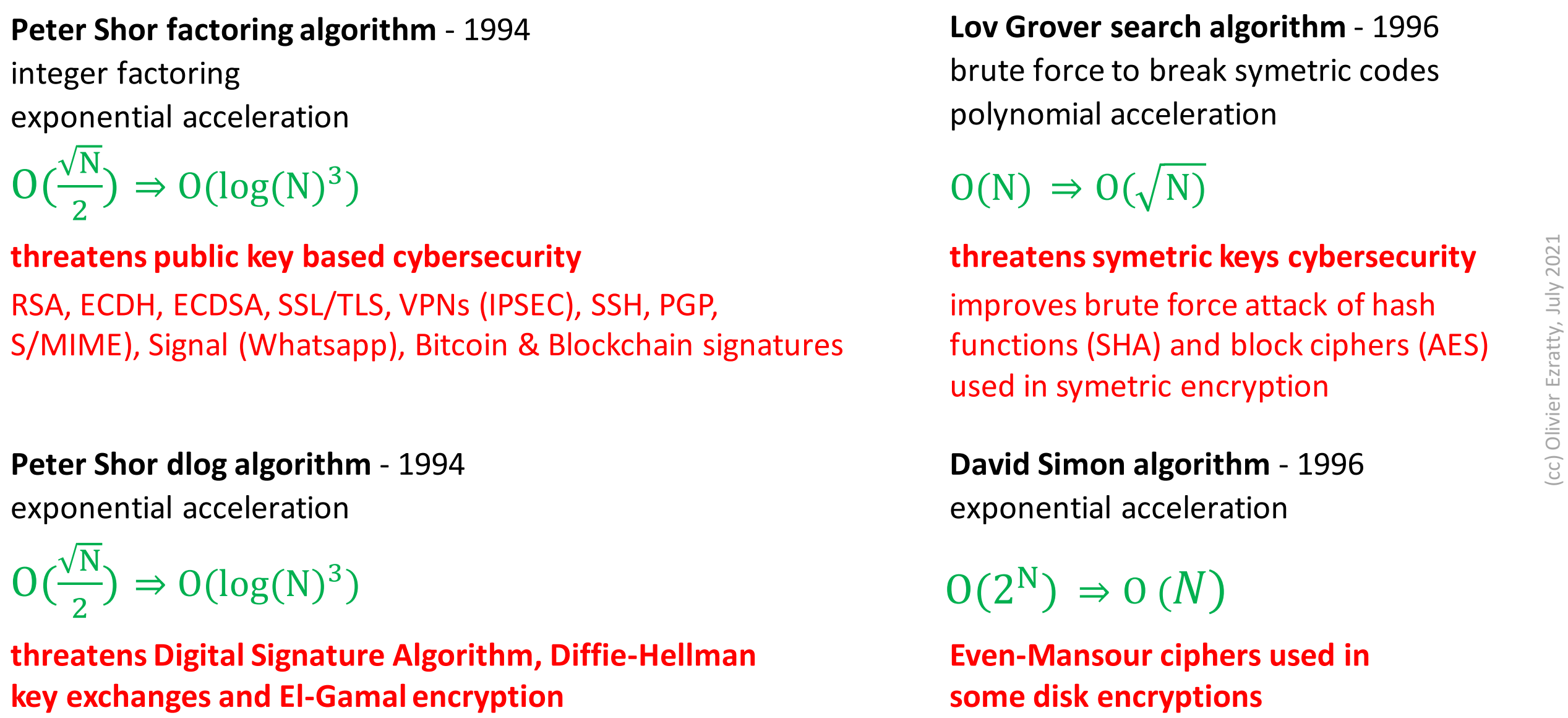

*Figure 732: Shor's algorithm is not the only quantum algorithm that could threaten existing cybersecurity. (cc) Olivier Ezratty, 2021.*

### Blockchain and Cryptocurrencies vulnerabilities

What about **Bitcoin**, other crypto-currencies and the **Blockchain**? The answer is summarized *below* with a good inventory of the cryptosystems used by use as a starting point[2255].

Otherwise, experts have opposite views on the quantum risk, from let's forget it[2256], to it will come sooner than expected[2257].

---

[2254] See Breaking Symmetric Cryptosystems Using Quantum Algorithms by Gaëtan Leurent with Marc Kaplan, Anthony Leverrier and María Naya-Plasencia, 2016 (58 slides).

[2255] The answer is well documented in The Quantum Countdown Quantum Computing and The Future of Smart Ledger Encryption by Long Finance, 2018 (60 pages).

[2256] See Here's Why Quantum Computing Will Not Break Cryptocurrencies by Roger Huang, December 2020.

[2257] See Q-Day Is Coming Sooner Than We Think by Arthur Herman, Forbes, June 2021. It mentions a crack of a RSA-2048 bit encryption key in 10 seconds with 4,099 stable qubits without mentioning the source of this performance. I have searched it and didn't found the source reference of these 10 seconds mentioned in various places. The only detail is it would require a perfect quantum computer executing one million operations per second.

Basically, the Blockchain is based on a patchwork of cryptographic algorithms including AES, RSA and SHA-3. It uses a hash algorithm to ensure the integrity of the chain of trust, and a digital signature to authenticate new transactions that are incrementally added to the Blockchain.

Bitcoin uses a SHA-256 crypto hash, which is quantum resistant, and a signature that exploits EC-DSA elliptic curves, which is not. In a similar manner, **Ethereum** uses a quantum resistant SHA-3 hash and a vulnerable ECDSA signature.

**Table 3. Main Algorithms Types Used for Cryptography, and Uses For Smart Ledgers**[19]

| Type of Algorithm | General Use | Example Algorithms of This Type | Example Uses for Smart Ledgers |
|---|---|---|---|
| Symmetric | Secret communications | AES, DES, 3DES, RC4 | Protection of resources stored on ledger |
| Public key | Secret communications (including key exchange) or digital signature | RSA, Diffie-Hellman, El Gamal, ECDSA | User authentication; signature of transactions, data or software |
| Hash | Generating fixed-length digest of arbitrary-length text | SHA-256, SHA-512, SHA-3 | Ensuring authenticity of blockchain |

*Figure 733: algorithms used in cryptos and smart ledgers. Source: The Quantum Countdown Quantum Computing and The Future of Smart Ledger Encryption by Long Finance, 2018 (60 pages).*

However, an Eth2 upgrade to Ethereum published in 2021 has replaced ECDSA based signatures by Lamport Q-R signatures which are quantum safe, with the inconvenient of being very large (over 200 times bigger than an ECDSA signature).

A recent review paper lists Bitcoin, Ethereum, Litecoin, Monero and ZCash as highly vulnerable to Shor's algorithm and all these, except Monero, to be moderately vulnerable to Grover search[2258].

All in all, quantum computing will not allow to alter the Blockchain, nor the proof of work used by Bitcoin which relies on the repeated use of quantum-resistant hash. The vulnerability of the Blockchain lies in the signature that relies on the ECDSA elliptic curve algorithm which can be broken with Shor's algorithm. This would make it possible to impersonate someone else in a transaction involving a Blockchain or Bitcoins. That's still whole lot of potential troubles! For example, if a Bitcoin transaction was intercepted to retrieve the sender's ECDSA signature, it could be exploited to transfer Bitcoins from that sender's wallet.

What is the "size" of the quantum computer that would break the ECDSA signature of the Bitcoin? The required number of qubits depends on the desired computing time. It would be 317 million physical qubits to break the key in one hour using a surface code, a code cycle time of 1 µs (IBM's cycle is right now at about 1 ms) and qubit fidelities 99,9% (IBM reached it with 27 qubits in 2021). In one day, the requirement is lower, at 13 million physical qubits. Whatever, as we've seen in studying the scalability of various quantum computing architecture and their enabling technologies, it's clear we are very far from seeing this threat to materialize[2259]. And like some like to asset, it doesn't matter since the Bitcoin may not outlast the current NISQ era!

Workarounds can obviously be created until a quantum threat to transaction integrity is confirmed. This can be done by encrypting the signatures used by the blockchains with a PQC system, as we'll see later[2260].

It is also possible to encrypt the data circulating in a Blockchain with a quantum computationally resistant algorithm such as AES-256, with the disadvantage that it is symmetrical and therefore requires keys to be exchanged beforehand. However, there are already some workarounds.

---

[2258] See Vulnerability of blockchain technologies to quantum attacks by Joseph J.Kearney et al, 2021 (10 pages).

[2259] See The impact of hardware specifications on reaching quantum advantage in the fault tolerant regime by Mark Webber et al, September 2021 (16 pages).

[2260] See Blockchained Post-Quantum Signatures by Chalkias Browny Hearnz, 2018 (8 pages).

A protocol using a longer validation time for Bitcoin transactions would allow to bypass the use of integer factoring to break the Bitcoin electronic signature algorithm, ECDSA[2261].But this would only amplify a key flaw of Bitcoin as a currency: a lengthening of transaction times that are already far from real time!

Bitcoin mining is potentially vulnerable by Grover's algorithm although its real practical speedup on a LSQC (large scale quantum computer) is questionable. Researchers are proposing some changes in the rules (and timing) applied by miners to mitigate this threat[2262].

We can also mention the open source Blockchain project resisting quantum attacks, Quantum Resistant Ledger. It is based on the XMSS (Extended Merkle Signature Scheme) electronic signature protocol[2263].

There is also a risk of attack at the mining level, with Grover's algorithm. But here again, there are solutions available[2264].The **Long Finance** document from which this table is extracted summarizes all these risks on Smart Ledgers by separating the transactions that are relatively protected and those that rely on vulnerable electronic signatures that are not vulnerable to hacking of the SSL and TLS protocols[2265].

**The Quantum Countdown**
**Quantum Computing And The Future Of Smart Ledger Encryption**

**Table 4. Risks to Blockchain Architectures from Quantum Computing**

| | Transactions | Data on Blockchain | Software on Blockchain |
|---|---|---|---|
| Read historical records without authorization | No (blockchains are intended to allow access to transaction information) | No, unless confidential and secured with vulnerable cryptography | No, unless confidential and secured with vulnerable cryptography |
| Alter historical records | No | No | May be able to run software without authorisation if signature used |
| Spoof ongoing records | Yes, possibly | Yes, possibly | Yes, possibly |

*Figure 734: Source: Source: The Quantum Countdown Quantum Computing and The Future of Smart Ledger Encryption by Long Finance, 2018 (60 pages).*

This section on threats would not be complete without mentioning the disagreements between cybersecurity specialists. Some are rather conservative and consider that one should not touch too much of what works well. They think Shor's threat is exaggerated. Others, such as the NIST in the US, are more alarmist and believe that the most critical cryptographic systems should be updated as soon as possible[2266]. And we also have arguments between the compared advantages of QKD and PQC, the two systems that can protect cybersecurity from quantum computing long-term threats[2267].

2261 It is documented in Committing to Quantum Resistance A Slow Defence for Bitcoin against a Fast Quantum Computing Attack, 2018 (18 pages).

2262 See On the insecurity of quantum Bitcoin mining by Or Sattath, February 2019 (22 pages).

2263 See also Blockchained Post-Quantum Signatures by Chalkias Browny Hearnz, 2018 (8 pages).

2264 See On the insecurity of quantum Bitcoin mining by Or Sattath, February 2019 (22 pages).

2265 For more information, see also The quantum threat to payment systems by Michele Mosca of the University of Waterloo, 2017 (52 minutes). Mosca is one of the world references in the quantum cryptography field. See also The Quantum Countdown Quantum Computing And The Future Of Smart Ledger Encryption, Long Finance, February 2018 (62 pages).

2266 Analysts are amplifying the fear, as in Executive's Guide to Quantum Computing and Quantum-secure Cybersecurity from Hudson Institute, a US conservative think tank, March 2019 (24 pages), Preparing Enterprises for the Quantum Computing Cybersecurity Threats by CSA, May 2019 or Global Risk Report 2020 from the World Economic Forum.

2267 See Quantum crypto-economics: Blockchain prediction markets for the evolution of quantum technology by Peter P. Rohde et al, February 2021 (12 pages) that modelize different scenarios depending on the reality of the quantum threat.

## Quantum Random Numbers Generators

Quantum, post-quantum and traditional cryptographic systems are all fed by random number generators. They have been around for ages. Random numbers are also used in a large set of applications beyond classical cryptographic protocols.

It includes gaming and casinos to draw lottery winning numbers, playing card shuffling, and various bets-related numbers, statistical analysis like the ones using Monte Carlo simulations in the finance sector, selecting random samples from large data sets like with machine learning, various scientific simulations and testings (like the Wheeler which-way or delay-choice experiment we already described), and smart networks simulations.

In all these use cases, the main concern is to create truly random numbers. Namely sequences of 0s and 1s without repetitions of any sequence and a balanced proportion of 0 and 1, as in the decimals of π. These numbers generation processes must also be non-deterministic, not reproducible and with no correlations, meaning that series of randomly generated numbers must be statistically independent. We could indeed generate good random numbers but if they were similar in time, it wouldn't be satisfactory at all.

Unfortunately, most commonly used random number generators are pseudo-random and happen to be deterministic. These are branded **PRNGs** (Pseudo-Random Number Generators).

Some mathematical formula deterministically produces a series of number and some randomness is introduced by using as seed parameters some highly variable elements such as time with a millisecond precision, GPS coordinates, thermal noise or other contextual information. It still generates deterministic sequences of numbers with some repeat period, although passing regular randomness tests successfully.

Most PRNG systems now use a randomness extractor merging the output of a random entropy source and a short random seed. Despite these initialization variables and various tricks of the trade, common random number generators still create some periods within their generated numbers[2268]. Still, it may be useful to use deterministic RNGs in some cases where reproducibility is mandatory. Also, PRNGs have the advantage to be fast[2269].

To avoid determinism, we must use a truly random physical process for the generation of numbers, aka **TRNGs** (True Random Number Generators), based on some chaotic physical phenomenon. One common technique consists in measuring the thermal noise of an electronic component or the atmospheric electromagnetic noise[2270].

Thermal noise TRNGs are implemented in most microprocessors like those from **Intel** since 2013 and from **AMD** since 2015 but with various identified weaknesses[2271]. It can for example rely on voltage randomness in resistive materials (Johnson's effect), Zener noise in diodes or, more commonly, on some amplified free-running oscillator.

---

[2268] However, there are still other solutions for generating non-quantum random numbers that need to be equally random, although this is still questionable. See for example Scientists Develop 'Absolutely Unbreakable' Encryption Chip Using Chaos Theory by Davey Winder, 2019.

[2269] See Quantum Random-Number Generators: Practical Considerations and Use Cases Report by Marco Piani, Michele Mosca and Brian Neill, evolutionQ, January 2021 (38 pages). This is the best document I found that explains the various subtleties of QRNGs, particularly about the device dependent and device-independent species.

[2270] Atmospheric noise is used by the service random.org operated by Randomness and Integrity Services Ltd (1998, Ireland).

[2271] Since 2013, Intel processors have been using the RDRAND function that is part of their 32 and 64 bits instruction set, returning a random number generated by an on-chip thermal noise based entropy source. AMD provides support for this instruction set since June 2015.

So here come **QRNGs** (Quantum Random Number Generators), a subclass of TRNGs. They rely on quantum physics laws and one that is particularly important: Born's probability rule, based on Schrödinger's wave equation. It replaces a generic chaotic system by a non-deterministic measurement of a physical property of some quantum objects, usually individual photons. In quantum physics, a quantum object's properties measurement is intrinsically random, at least, as far as we know[2272].

Quantum is the kingdom of randomness[2273]! But this randomness is not a guarantee to obtain truly nondeterministic random numbers. There are weaknesses in all these systems, particularly with their classical or semi-classical components like the beam splitters or photon detectors it is using, or with the software part handling the so-called randomness extraction. Its consequence is an intense competition between QRNG vendors. They all claim to generate "truly" random numbers contrarily to their QRNG competitors.

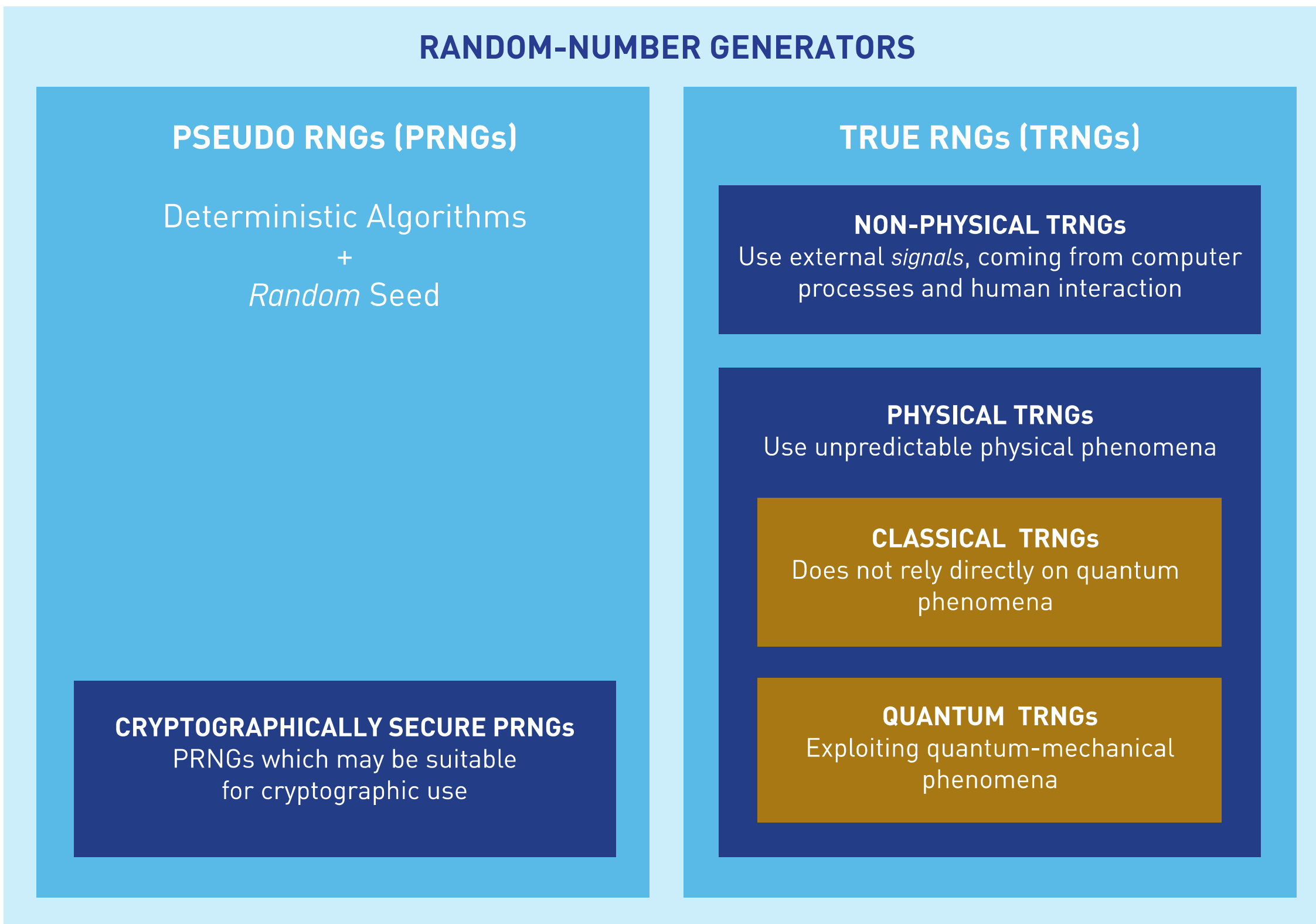

*Figure 735: taxonomy of random numbers generators. Source: Quantum Random-Number Generators: Practical Considerations and Use Cases Report by Marco Piani, Michele Mosca and Brian Neill, evolutionQ, January 2021 (38 pages).*

Many differentiation features are also important: the random numbers generation rate (in bits/seconds, some applications may be very demanding), the time it takes to warm up and stabilize the system (some QRNGs are slow to warm-up and may require hours to stabilize), is it device independent (impacts randomness quality but also RNG rates; but no such commercial systems are available yet), certifiability (some are black-boxes that are really difficult to audit, others are said to be self-certified) and other standard characteristics that may be important depending on the use case (weight, size, price and power drain).

[2272] See the review paper A Comprehensive Review of Quantum Random Number Generators: Concepts, Classification and the Origin of Randomness by Vaisakh Mannalath et al, March 2022 (38 pages) and Quantum Random Number Generators : Benchmarking and Challenges by David Cirauqui et al, June 2022 (15 pages).

[2273] See Can Free Will Emerge from Determinism in Quantum Theory? by Gilles Brassard and Paul Raymond-Robichaud, 2012 (22 pages). Even this is however arguable. See for example Quantum randomness is chimeric by Karl Svozil, April 2021 (16 pages) which comes back on the eternal debate of quantum measurement and its related randomness.

At last, vendor trust is a key criterion, particularly when you discover that some Switzerland cyber-security products contained backdoors created on behalf of the CIA[2274]. It also explains why, whatever the technology used, western countries may not and probably should not rely on Chinese or Russian TRNG/QRNG vendors.

It's now up to you to understand how these systems are benchmarkable and benchmarked to figure out whether such and such QRNG is safe or not. Many different QRNG techniques have been created to date. The most commonplace are those using photons, with components that are now easy to miniaturize, even in a smartphone.

**Photons counting** is the most common method, based on the measurement of single photons emitted individually in series, passing through a regular balanced beam splitter and analyzed by two detectors[2275]. The series of generated 0s and 1s are theoretically random. Quantum physics mathematical formalism and experiments say so!

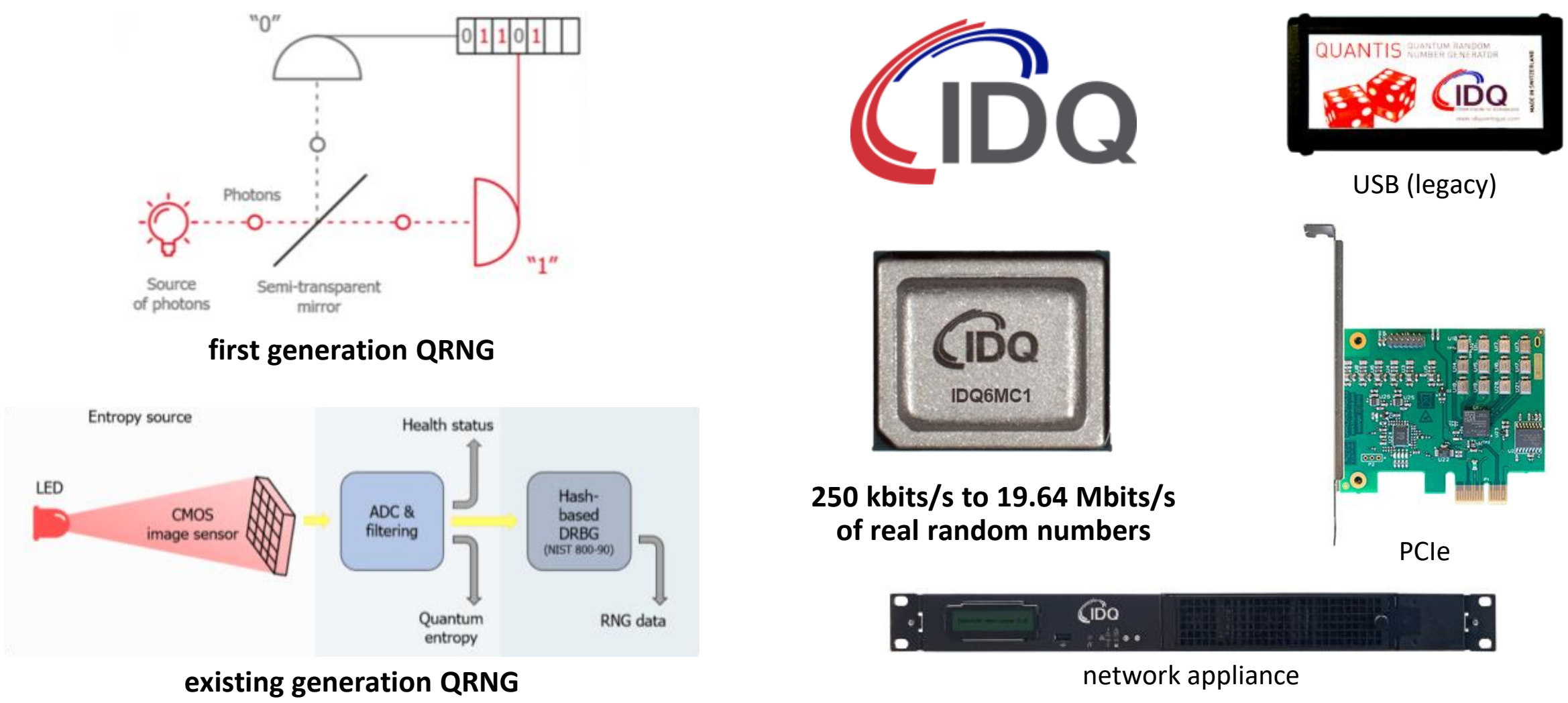

*Figure 736: IDQ quantum number generators. Source: IDQ.*

Each detected photon creates at most only one bit, but not all photons are detected. The detection speed is limited by the photon detectors bandwidth and their saturation level. This QRNG technique was pioneered by **IDQ** (Switzerland) in 2001. One of its shortcomings is the used light source that can't necessarily be certified. In some products, there's also some warm-up time before real random numbers can be generated. Some solutions can certify the numbers generation randomness in such situation, like using a first beam splitter and detector before photons are separated by a polarizing beam splitter[2276].

Nowadays, photon counting is (seemingly) done without a polarizing beam splitter. The light source is some LED diode, lighting a small CMOS image sensor, and generating "shot noise". That's what IDQ is now selling with its miniaturized Quantis chipset. It's adapted to mass market use cases, in smartphones, laptops and car. Such QRNG first appeared in a consumer product in 2020 in a version of the **Samsung** Galaxy A71 5G smartphone called Galaxy A Quantum, marketed by **SK Telecom** only in Korea. It probably won't change much in terms of user security, but it can make a lasting impression.

---

[2274] See The intelligence coup of the century' - For decades, the CIA read the encrypted communications of allies and adversaries by Greg Miller, Washington Post, February 2020. It deals with the Crypto AG company created in 1952 and dissolved in 2018.

[2275] See Quantum Random Number Generators by Miguel Herrero-Collantes, 2016 (54 pages). This setting is frequently referred to as a welcher-weg experiment, or "which way" experiment.

[2276] See Using the unpredictable nature of quantum mechanics to generate truly random numbers by Bob Yirka, 2021 that refers to Certified Quantum Random Numbers from Untrusted Light by David Drahi, December 2020 (32 pages). Its RNG output is 8.05 Gb/s.

In April 2021, Samsung announced a new version of this smartphone, the Galaxy Quantum2, adapted to 5G and with similar QRNG features using an IDQ Quantis chipset[2277]. The same chipset is found in a Vsmart Aris 5G smartphone, coming from Vietnam! **Q→NU**, **CryptaLabs** and **Qrypt** are also commercializing such type of QRNGs.

**Photon arrival time** aka "time bin qubits" is about evaluating the arrival time of successive single photons in a simpler setting coupling a photon source like a LED or a laser and a photon counter to a high-resolution counter, down to a couple nanoseconds[2278]. Practically, randomness comes from evaluating the variation of this arrival time compared with the decaying exponential waiting-time distribution. The system can also use a photon counting setting with a regular beam splitter and two photon detectors coupled each with a counter[2279]. It has low latency and is quickly up to speed. **Qnu Labs, PicoQuant** and **QuTools** are providers of such QRNGs. A variation of this technique recently developed uses a LED light illuminating a matrix of SPADs (single photon avalanche detectors) on a CMOS circuit, with a RNG capacity of 400 Mbit/s[2280].

**Quantum vacuum fluctuations** uses a balanced homodyne measurement of vacuum fluctuations of the electromagnetic field contained in the radio-frequency sidebands of a single-mode (usually 780 nm) laser diode[2281]. Two diodes compute the difference of the signals coming from the two exits of a polarizing beam splitter and the resulting signal is amplified and digitized, to be processed by a randomness extractor. Such a QRNG system is implemented in a web site run by ANU (Australian National University) with the qStream QRNG from **Quintessence Labs**, which generates keys at a >3.5Gbps rate[2282]. A record of 100 Gbps rate was obtained by a European team in 2022[2283].

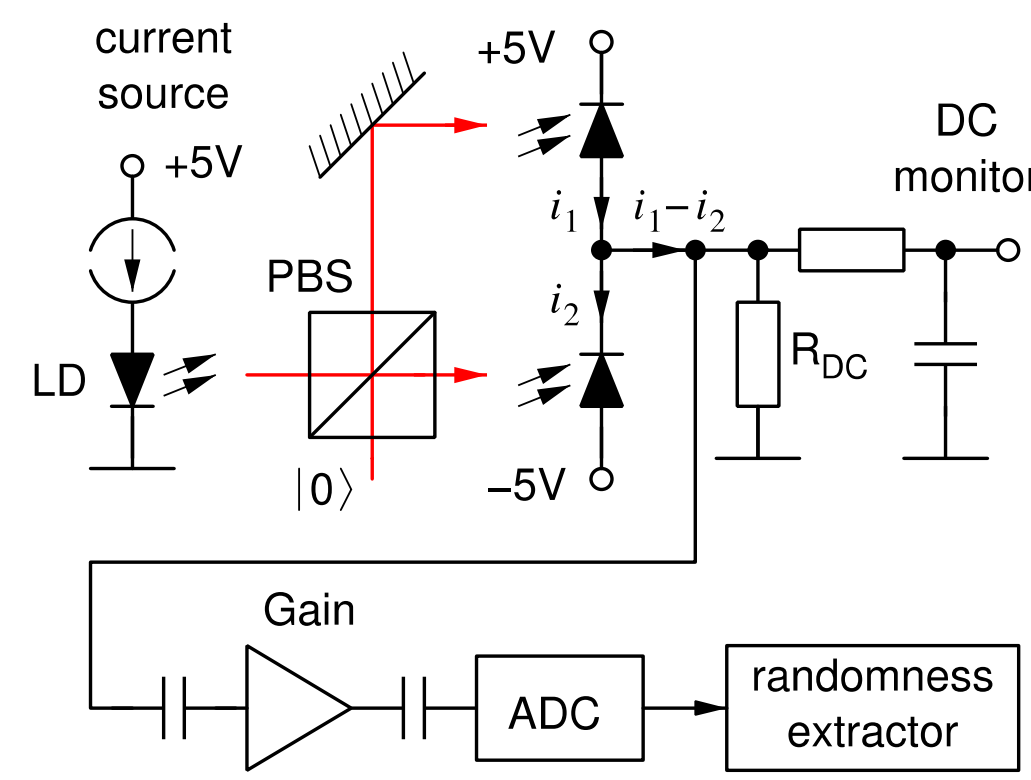

*Figure 737: quantum vacuum fluctuation QRNG. Source : Random numbers from vacuum fluctuations by Yicheng Shi et al, 2016 (5 pages)*

**Spontaneous emission** uses amplified spontaneous emission, detection and digitization of optically filtered amplified spontaneous emission noise from a light source such as superluminescent diode (SLD).

[2277] The QRNG chipset is a square of 2.5mm creating random codes by capturing noise from an LED and a CMOS sensor.

[2278] See Photon arrival time quantum random number generation by Michael A. Wayne et al, 2009 (7 pages) which describes the principles of this random numbers generation methods.

[2279] See First high-speed quantum-safe randomness generation with realistic devices, NTT, February 2021, which refers to A simple low-latency real-time certifiable quantum random number generator by Yanbao Zhang et al, 2021 (8 pages).

[2280] See A High Speed Integrated Quantum Random Number Generator with on-Chip Real-Time Randomness Extraction by Francesco Regazzoni et al, February 2021 (9 pages).

[2281] A homodyne measurement consists in extracting information encoded as modulation of the phase and/or frequency of an oscillating signal. In the mentioned case, it's the phase. See an example in Random numbers from vacuum fluctuations by Yicheng Shi et al, 2016 (5 pages) and in A homodyne detector integrated onto a photonic chip for measuring quantum states and generating random numbers by Francesco Raffaelli et al, University of Bristol, Quantum Science and Technology, February 2018 (10 pages).

[2282] See https://qrng.anu.edu.au/ and Real time demonstration of high bitrate quantum random number generation with coherent laser light, by T. Symul et al, 2021 (4 pages) and Maximization of Extractable Randomness in a Quantum Random-Number Generator by J. Y. Haw, 2015 (13 pages). The operations are linked to the offer of QuintessenceLabs.

[2283] See 100 Gbps Integrated Quantum Random Number Generator Based on Vacuum Fluctuations by Cedric Bruynsteen et al, September 2022 (10 pages).

**Phase noise** and **phase diffusion** (PD-QRNG[2284]) are variations of spontaneous emission QRNGs. It uses a photons counting method variation proposed in 2009[2285]. In one implementation, a VCSEL laser (single mode vertical cavity surface emitting laser) is associated with a phase noise measurement using a delay self-homodyne method.

The photons from the laser are traversing a beam splitter. Among its benefits are a very high-bit rate, of several tens of Gbits/s of random bits.

The technique is used by vendors like **Quside**, **QuantumeMotion** and **Kets**.

One way goes to the next beam splitter, and the other traverses a delay line, and is then merged back with the main line. An APD (avalanche photodetector), then counts the exiting photons and its signal is converted from analog to digital with an ADC[2286].

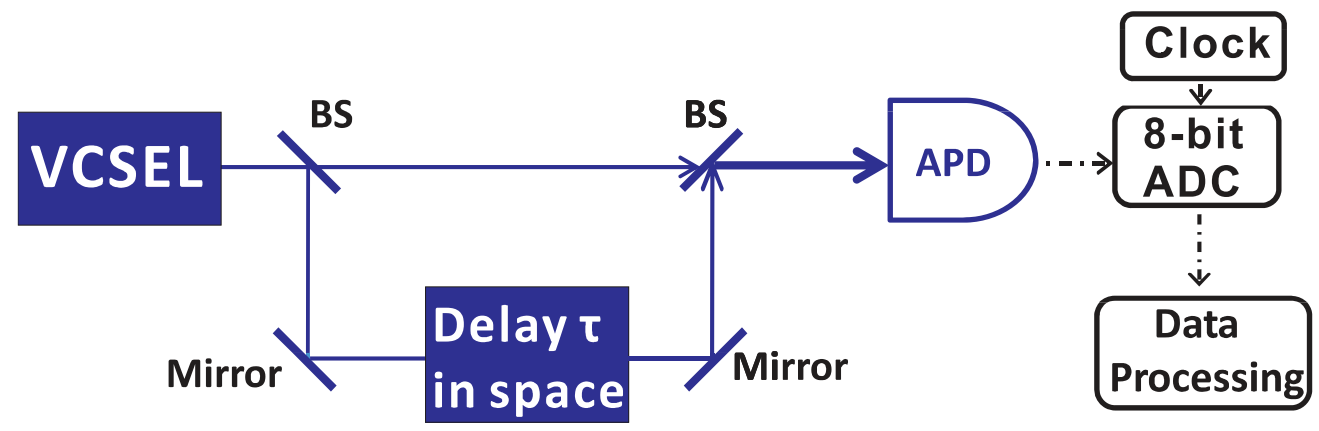

*Figure 738: Source: Truly Random Number Generation Based on Measurement of Phase Noise of Laser by Hong Guo et al, Peking University, January 2010 (4 pages).*

**Radioactive decay** was one of the first developed QRNG technologies, based on the random timing of decay of particular radioactive atoms, detected with a Geiger counter. It has limited bit rates and is not widely used, on top of being not very practical to implement given it's based on radioactive materials. There is however a vendor in that space, **EYL**.

Other various techniques are mentioned but seemingly not widely used: **laser chaos** that creates a time-delayed optical feedback via a reflector, **Raman scattering**, **attenuated pulse** and **Optical Parametric Oscillators (OPO)** and as proposed in 2022, **skyrmions**-based QRNGs[2287].

Another technique consists in merging in a single platform several QRNG methods. That's what a Chinese team released in July 2021 on an Alibaba Cloud server, mixing four types of QRNGs: single-photon detection, photon-counting detection, phase-fluctuations and vacuum-fluctuations[2288]. Three of these QRNG sources were off-the-shelf (Quantis-PCIe-16M from ID Quantique, QRG-100E from QuantumCTek and QRN-16 from MPD).

**Device Independence** deals with the difference between randomness and privacy. A SDI (source device independent) QRNGs ensures private randomness, where the created random numbers can't be known by any adversary. With SDI QRNGs, the randomness source is assumed to be untrusted but the measurement devices are trusted. It's more secured than a trusted device or device dependent QRNG where the device is well characterized and trusted. Device independence must also deal with detector attacks in the QRNG[2289].

---

2284 See Quantum entropy source on an InP photonic integrated circuit for random number generation by Carlos Abellan et al, Optica, 2016 (7 pages) and Real-time interferometric quantum random number generation on chip by Thomas Roger et al, Journal of Optical Society, 2019 (7 pages).

2285 In Experimental demonstration of a high speed quantum random number generation scheme based on measuring phase noise of a single mode laser by Bing Qi, Yue-Meng Chi, Hoi-Kwong Lo and Li Qian, 2009 (7 pages).

2286 See Truly Random Number Generation Based on Measurement of Phase Noise of Laser by Hong Guo et al, Peking University, January 2010 (4 pages).

2287 See Single skyrmion true random number generator using local dynamics and interaction between skyrmions by Kang Wang et al, Nature Communications, February 2022 (8 pages).

2288 See Quantum random number cloud platform by Leilei Huang, Hongyi Zhou, Kai Feng and Chongjin Xie, Nature, July 2021.

2289 See Source-independent quantum random number generator against detector blinding attacks by Wen-Bo Liu et al, April 2022 (14 pages).

These have a high bit rate in the Gbits/s range while SDI QRNGs have a much lower bit rate, in the kbits/s range due to a more complicated setup. SDI QRNG can rely on entanglement and non-locality or be based on quantum computation (this is a variation of the qubit measurement technique mentioned above). SDI QRNG enables real-time estimate of the output entropy which can quantify and certify the QRNG randomness without possessing a detailed knowledge of the entropy source device. The device independence certification comes with loophole free violation of Bell's inequalities.

A record rate of 17 GBits/s key generation with a SDI-QRNG was obtained in 2018 in an Italian lab[2290]. It was also experimented in a highly integrated photonic circuits, using a self-tested randomness expansion protocol with multi-dimensional encoding[2291].

There are also MDI-QRNGs, where the source is trusted and the measurement device is untrusted. It is for example used with time-bin QRNGs with a testing mode used to create a 4-quantum states ($|0\rangle$, $|1\rangle$, $|+\rangle$ and $|-\rangle$) tomography[2292].

**Qubits measurement** is a more generic way of generating quantum randomness than photon counting after traversing a polarizing beam splitter. It uses gates-based quantum processing units applied to one or several qubits and creating superposition. The simplest is a single Hadamard gate, but it's not sufficient to create enough randomness. The common practice is to create entangled states, or Bell states, which is done combining Hadamard and CNOT gates[2293]. The useful states are those where there is some correlation or anticorrelation between the qubits from a Bell pair, showing that they were random. **CQC** (UK, part of Quantinuum) is providing such a solution, named Quantum Origin, running on superconducting qubits in the cloud, in partnership with IBM and also running on Quantinuum trapped ions qubits. There are variations of QRNG exploiting computing qubits that are based on quantum walks[2294]. Still, I don't know how these solutions are certified and we're back at studying the real randomness, non-determinism and weaknesses of qubits preparation sources (microwaves, lasers) as well as qubit readout techniques (microwaves readouts, CCD/CMOS readouts for cold atoms, trapped ions and NV centers, and single photon detectors for photon qubits).

**Quality**. Quantum random numbers generators are not equal. The source may be a true RNG but other components may contain weaknesses and be hacked in some circumstances: the photon source[2295], the photon measurement system which could deviate or be defective, and at last randomness extractor various other weaknesses. Also, it can be difficult to distinguish classical hardware noise from the quantum randomness coming from the QRNG in evaluation tests.

**Evaluation**. Random series of numbers must be incompressible. This algorithmic randomness can be tested with Borel normality. An infinite sequence of binary numbers is random if every binary string of length n appearing in the sequence has a frequency of $2^{-n}$.

---

[2290] See Source-device-independent heterodyne-based quantum random number generator at 17 Gbps by Marco Avesani et al, 2018 (7 pages). It uses a POVM measurement of continuous variable observables.

[2291] See Multidimensional quantum entanglement with large-scale integrated optics by J. Wang et al, Science, 2018 (24 pages).

[2292] See Experimental measurement-device-independent quantum random number generation by You-Qi Nie et al, China, 2016 (16 pages).

[2293] See Quantum random number generators with entanglement for public randomness testing by Janusz E. Jacak et al, 2020 (9 pages) and Reference Standard RS-EITCI-QSG-EQRNG-PROTOCOLS-STD-VER-1.0, EITCI, 2019 (21 pages).

[2294] See Quantum Walk Random Number Generation: Memory-based Models by Minu J. Bae, University of Connecticut, July 2022 (12 pages).

[2295] See for example QRNG: Out-of-Band Electromagnetic Injection Attack on a Quantum Random Number Generator by P.R. Smith et al, January 2021 (12 pages).

There are various tests of algorithmic randomness like the NIST SP 800-22 1A Test[2296]. It contains 15 tests but other tests suites exist that complement the NIST set, totaling 40 tests[2297].

| Test | Defect detected | Property |
|---|---|---|
| Frequency (monobit) | Too many zeroes or ones | Equally likely (global) |
| Frequency (block) | Too many zeroes or ones | Equally likely (local) |
| Runs test | Oscillation of zeroes and ones too fast or too slow | Sequential dependence (locally) |
| Longest run of ones in a block | Oscillation of zeroes and ones too fast or too slow | Sequential dependence (globally) |
| Binary matrix rank | Deviation from expected rank distribution | Linear dependence |
| Discrete fourier transform (spectral) | Repetitive patterns | Periodic dependence |
| Non-overlapping template matching | Irregular occurences of a prespecified template | Periodic dependence and equally likely |
| Overlapping template matching | Irregular occurences of a prespecified template | Periodic dependence and equally likely |
| Maurer's universal statistical | Sequence is incompressible | Dependence and equally likely |
| Linear complexity | Linear feedback shift register (LFSR) too short | Dependence |
| Serial | Non-uniformity in the joint distribution for m-length sequences | Equally likely |
| Approximate entropy | Non-uniformity in the joint distribution for m-length sequences | Equally likely |
| Cumulative sums (cusum) | Too many zeroes or ones at either an early or late stage in the sequence | Sequential dependence |
| Random excursions | Deviation from the distribution of the number of visits of a random walk to a certain state | Sequential dependence |
| Random excursions variants | Deviation from the distribution of the number of visits (across many random walks) to a certain state | Sequential dependence |

*Figure 739: the NIST test suite for QRNG. Source: Random Number Generators: An Evaluation and Comparison of Random.org and Some Commonly Used Generators by Charmaine Kenny, April 2005 (107 pages).*

Recent TRNG/QRNG benchmarking tools use machine learning techniques and a convolutional network to detect patterns in the generated numbers[2298]. However, while these tests may detect weaknesses in QRNG randomness, it won't ensure real nondeterminism. Other tests are required, like loophole free Bell tests, already mentioned.

**QKD or PQC?** We'll describe these two cryptography solutions later on. Which one will make use of QRNGs? Post-quantum cryptography requires large classical random keys, so QRNGs will be very useful, particularly those who have a large throughput. Quantum Key Distribution (QKD) needs randomness to select its active basis choice for each and every detected pairs of photon like their polarization angle. So again, a good and fast QRNG will be mandatory. This QRNG functionality can however be embedded in some specific QKD systems with relying on the randomness of the time between photons detection in the SPCMs (Single Photon Counting Modules)[2299]. But QRNGs have a much broader addressable market: classical cryptography and all the businesses in need of random numbers like casinos, online gaming and lotteries.

---

2296 See A Statistical Test Suite for Random and Pseudorandom Number Generators for Cryptographic Applications by Andrew Rukhin et al, NIST, 2010 (131 pages). Then, Experimentally probing the algorithmic randomness and incomputability of quantum randomness by Alastair Abbott, Cristian Calude and al, UGA/Institut Néel France and University of Auckland, 2018 (17 pages) and Recommendations and illustrations for the evaluation of photonic random number generators by Joseph D. Hart et al, 2017 (29 pages).

2297 See Random Number Generators: An Evaluation and Comparison of Random.org and Some Commonly Used Generators by Charmaine Kenny, April 2005 (107 pages).

2298 See Machine Learning Cryptanalysis of a Quantum Random Number Generator by Nhan Duy Truong et al, 2019 (13 pages) and Benchmarking a Quantum Random Number Generator with Machine Learning, 2020 (26 slides).

2299 This is described in Practical random number generation protocol for entanglement-based quantum key distribution by G. B. Xavier et al, 2008 (10 pages).

Let's now look at the QRNG industry vendors landscape. As said before, this technique has been mastered for a long time by **ID Quantique** (IDQ), a company cofounded by Nicolas Gisin, which belongs to SK Telecom since 2018. Other players abound like **CryptoMathic**, **Crypta Labs**, **Quside**, **InfiniQuant**, **Kets**, **PicoQuant** and **Quantropi**[2300]. **Axion Technologies** (2017, Canada) also created a random number generator competing with the Swiss IDQ.

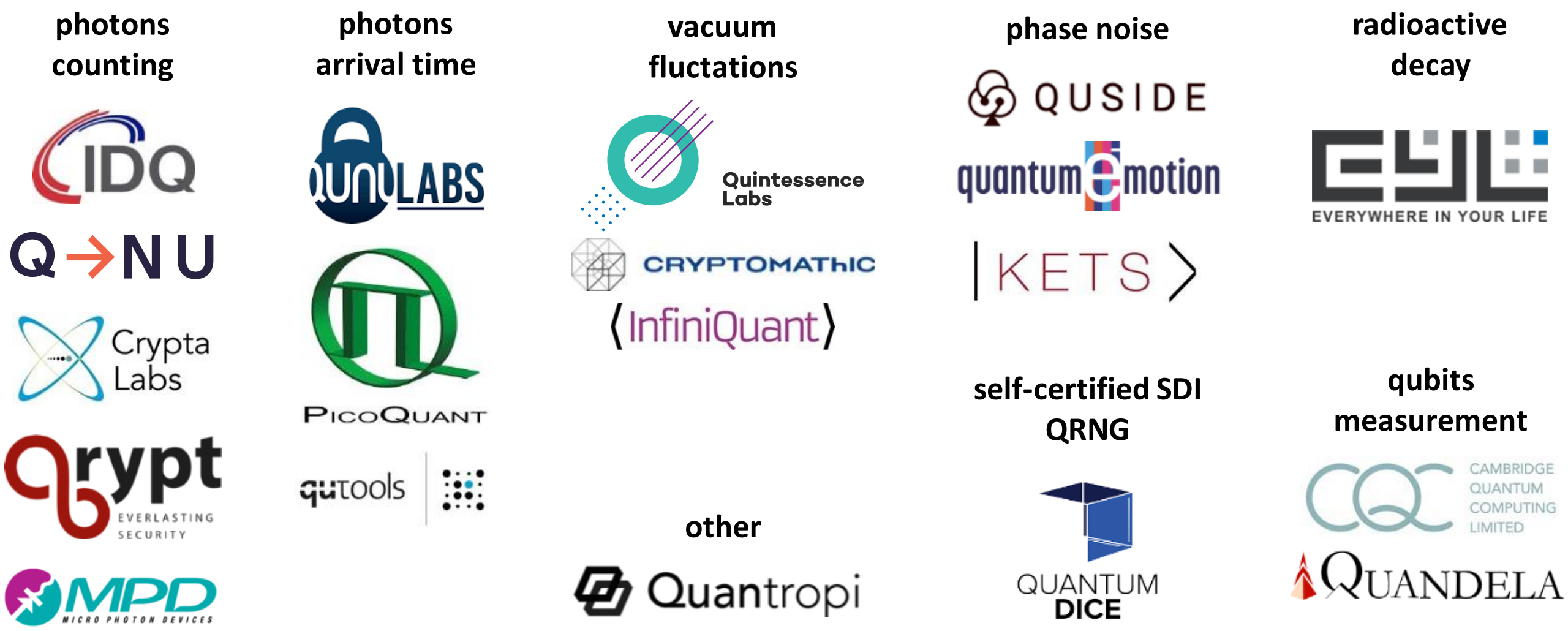

Figure 740: a map of QRNG vendors. (cc) Olivier Ezratty, 2022.

**CryptoMathic** (1986, Denmark) develops quantum random key generators and various keys generation systems.

**EYL** (2015, USA, $900K) sells radioactive isotopes-based entropy chips of 3mmx3mm and QRNG chips. They form factor is a USB key.

**PicoQuant** (1996, Germany) is a Berlin-based SME specialized in photonics and which markets photon counters and diode lasers. But they are here because they also offer a photon arrival time QRNG, the PQRNG 150, with a throughput of 150 Mbits/s. It is much less miniaturized than the random number generator component from IDQ that is integrated in Samsung's Galaxy 5G announced in May 2020[2301].

**Qrypt** (2018, USA) develops cryptographic solutions using a high speed QRNG powered by multiple entropy sources exclusively licensed from Oak Ridge National Lab and other labs. It will support NIST PQC standards when selected. Qrypt invested in Quside (Spain), which is developing high quality and high speed QRNGs.

**Qnu Labs** (2016, India) sells its Tropos Quantum Random Number Generation, which allows the creation of random numbers of any size and quite quickly, at a rate of up to 1.5 random Mbits/s, or even several tens of Gbits/s. It also sells a QKD solution, Armos and a quantum secure platform for key management, Hodos. They were helped by the Intel startup program.

---

[2300] These companies are described in the quantum telecommunication and cryptography vendors section since they provide some PQC or QKD solution on top of QRNGs.

[2301] They contribute to research projects in QKD infrastructures like in Ultrafast quantum key distribution using fully parallelized quantum channels by Robin Terhaar et al, July 2022 (13 pages)

**Quside** (2017, Spain) proposes a QRNG using phase diffusion, with a 400 Mbits/s key generation rate. It is a spin-off of ICFO, the Institute of Photonics of Barcelona. Quside QRNGs were used in many of the 2015 loophole-free Bell test experiments thanks to their high key rate.

**Quantum Dice** (2019, UK, £2M) is a spin-off from the University of Oxford selling a "true" "self-certified" and fast QRNG device. It uses a patented DISC protocol ensuring that randomness comes only from the quantum process and is protected from external influences.

Their October 2021 £2M investment round was led by French venture capital fund Elaia Partners. They also got a £1M non-dilutive grant from the Quantum Accelerator Group led by IP Group, in partnership with Innovate UK as part of the UK national quantum plan.

**Quantropi** (2018, Canada) is a company created in Ottawa by James Nguyen (CEO) and Randy Kuang (Chief Scientist).

It was initially created to distribute a software generator of "lightweight ultra-high-entropy" random encryption keys. It then evolved into selling a complicated cybersecurity offering mixing custom-made PQC, a specific scheme for quantum key distribution solution, all that packaged in a gibberish marketing lingua mixing classical and quantum crypto[2302].

Their QiSpace end-to-end quantum security SaaS platform contains a lot of stuff:

- **MASQ**, their asymmetric encryption offering, with a PQC for key exchange and digital signature containing a MPPK for Multivariate Polynomial Public Key, but they also have a Quantum Permutation Pad aka QPP. These things are supposed to support NIST PQC finalists and Quantropi's own PQC.
- **QEEP**, their symmetric encryption offering, providing "quantum-secure symmetric encryption that's up to 18 times faster than AES-256".
- **SEQUR**, their "quantum entropy as a service" (QEaaS) offering which contains some form of QRNG, branded QiSpace SEQUR NGen pseudo-QRNG (so, it's not a real QRNG) and their SEQUR SynQK, a sort of QKD that is supposed to deliver 5 simultaneous Quantum-key streams over distances ranging from 4,000 to 15,000 km at 130 to 190 megabits per second. What's quantum in-here? Their "coherent-based Two-Field Quantum Key Distribution (CTF-QKD)", a signal modulation scheme using coherent optical communications hardware and infrastructure[2303].

**ComScire** (1994, USA) is the developer of several random number generators including PureQuantum QRNG, which creates 4 to 128 million bits per second. Is it really quantum? Not really sure!

Its entropy source seems to be coming from CMOS shot noise, using an Altera FPGA[2304]. The company also markets QNGmeter 3.6, a software tool testing the randomness and nondeterministic nature of generated numbers.

**Terra Quantum** (Switzerland) also provides a photonic based QRNG generated 1.2 Mb/s of random bits.

---

[2302] This leads to some extreme marketing claims as seen on Startup: Only Quantum Cryptography Can Save The $100 Trillion Global Digital Economy by John Koetsier in Forbes, March 2021.

[2303] It seems documented in Quantum Public Key Distribution using Randomized Glauber States by Randy Kuang and Nicolas Bettenburg et al, 2020 (7 pages).

[2304] Its functioning is described in Entropy Analysis and System Design for Quantum Random Number Generators in CMOS Integrated Circuits by Scott A Wilber, 2013 (28 pages).

# Quantum Key Distribution

Quantum cryptography is based on "quantum key distribution" which consists in allowing the exchange of symmetrical encryption keys, by optical means (optical fiber, air link or satellite optical link) using a system to protect its transmission against intrusions [2305].

QKD protocols have the particularity of allowing the detection of any intrusion in the transmission chain and to indicate that someone has tried to read its contents or if, disturbances have occurred, "on the line"[2306].

## QKD principles

One early version was the **BB84** protocol invented by **Charles Bennett** and **Gilles Brassard** in 1979 and published in 1984 [2307]. They are even the creators in 1982 of the expression "quantum cryptography"[2308]. But it wasn't yet QKD per se. This protocol is about sending photon-based information with four types of rectilinear/diagonal polarizations, *aka* non-orthogonal states: 0°, 45°, 90° and 135°. Alice & Bob exchange through a classical channel their polarization basis, used for encoding by Alice and for measurement by Bob, after the photons have been sent to make sure Bob keeps only the relevant bits where his random measurement was done in the same as the polarization basis used by Alice.

The qubits read by an intruder would modify the key, by projecting their polarization at 0° or 90°, or 45°/135° depending on the case and randomly. Any reading intrusion would be detected by Alice and Bob during their classical communication because of the inevitable disturbances it would cause. If the protocol detects an intruder, it can take this into account and block the communication of sensitive information because the encoding key has been captured and, maybe, chose another quantum channel. And there are solutions to avoid a denial of service in such a case[2309].

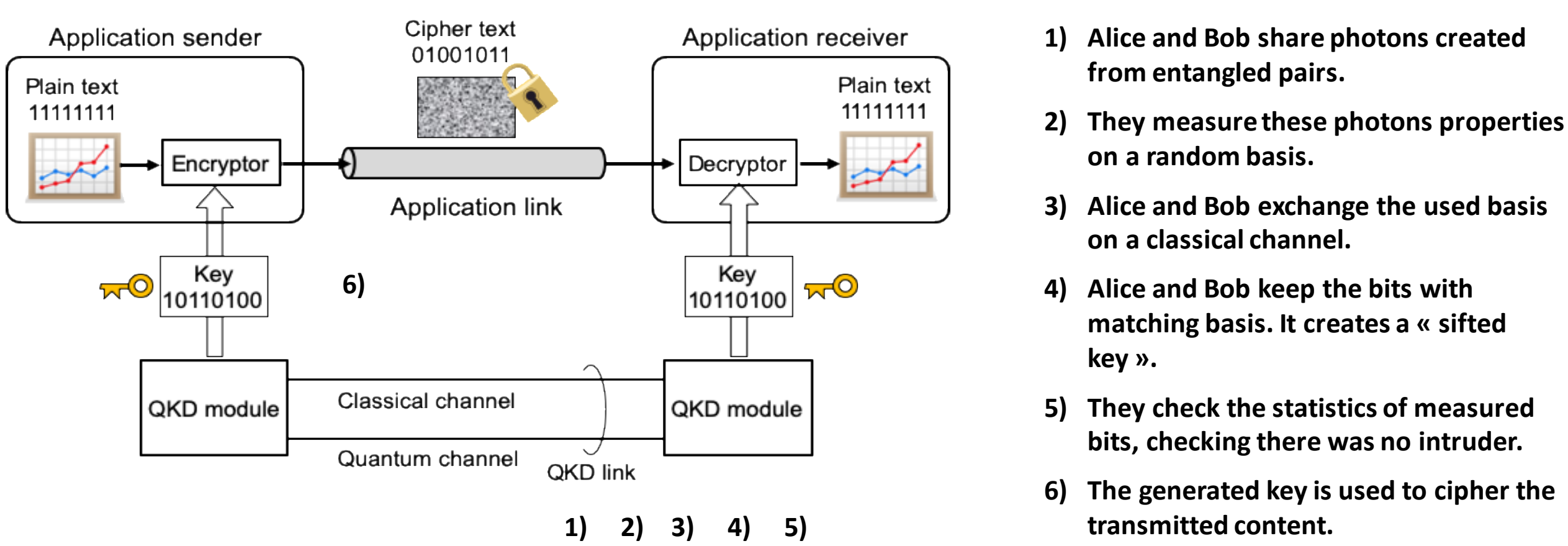

*Figure 741: general principle of quantum key distribution. Source: TBD.*

2305 See the review paper Quantum Key Distribution Secured Optical Networks: A Survey by Purva Sharma et al, September 2021 (35 pages) which contains a very good description of the various QKD technologies and protocols, their challenges and the way they are addressed and The Evolution of Quantum Key Distribution Networks: On the Road to the Qinternet by Yuan Cao, IEEE, 2021 (59 pages).

2306 See the review paper Advances in Quantum Cryptography by Stefano Pirandola et al, 2019 (118 pages).

2307 In Quantum cryptography : public key distribution and coin tossing, 1984 (5 pages).

2308 Here is a general overview of QKD and PQC: The Impact of Quantum Computing on Present Cryptography, March 2018 (10 pages).

2309 See for example A quantum key distribution protocol for rapid denial of service detection by Alasdair Price et al, from the University of Bristol, in EPJ Quantum Technology, 2020 (20 pages).

**Artur Ekert** then created the **E91** protocol in 1991 with using quantum entanglement and non-locality, avoiding the explicit transmission of photon information that is used in BB84 and that could be intercepted by an intruder [2310].

With E91, Alice and Bob share the photons created from entangled pairs. They can then share a randomly generated key with a sequential measurement of these photons. Like with BB84, this measurement must be done in a random orthogonal polarization basis that has to be shared afterwards between Alice and Bob. They will retain the randomly generated bits when their polarization was synchronized, creating a "sifted key". If an eavesdropper Eve intercepted the entangled photons, their projections would be different. To make sure there was no eavesdropper, Alice and Bob compute a Bell test statistic which must yield ideally a so-called Bell parameter $|S| = 2\sqrt{2}$, called Tsirelson's bound, otherwise, there was an eavesdropper.

One key difference between BB84 and E91 is the origin of randomness in the shared keys. With BB84, it must be generated by Alice with a random number generator, preferably a TRNG (true random numbers generator) and not a PRNG (pseudo-random numbers generator) as described in the earlier section on QRNGs, page 796. With E91, it comes directly from the randomness of the entangled photon pairs readouts. All in all, E91 consolidates a quantum communication protocol and a quantum random number generator.

QKD protocols have since made their way. They are at the origin of the creation of the whole field of quantum cryptography, which has now left the exclusive realm of research and experimentation to enter actual deployments like in China, even though there are still problems remaining to be fixed such as the creation of safe repeaters, to replace the commonplace unsafe trusted nodes (even in China) where the key bits are turned into classical data at each and every node station, given you need to have one of these about every 80 km.

QKD was expanded with CV-QKD (continuous variable) which modulates both the phase and the amplitude of the transmitted optical signal. It notably allows multiplexing several communications on the same optical fiber and to exploit the existing infrastructures of telecom operators. **Philippe Grangier** was one of its designers, along with **Frédéric Grosshans** from CNRS-LIP6, in 2002[2311]. CV-QKD complements discrete variables QKD (DV-QKD) as are called the previously mentioned QKD protocols, based on the properties of single photons, which requires some cooling on the single-photon detector side.

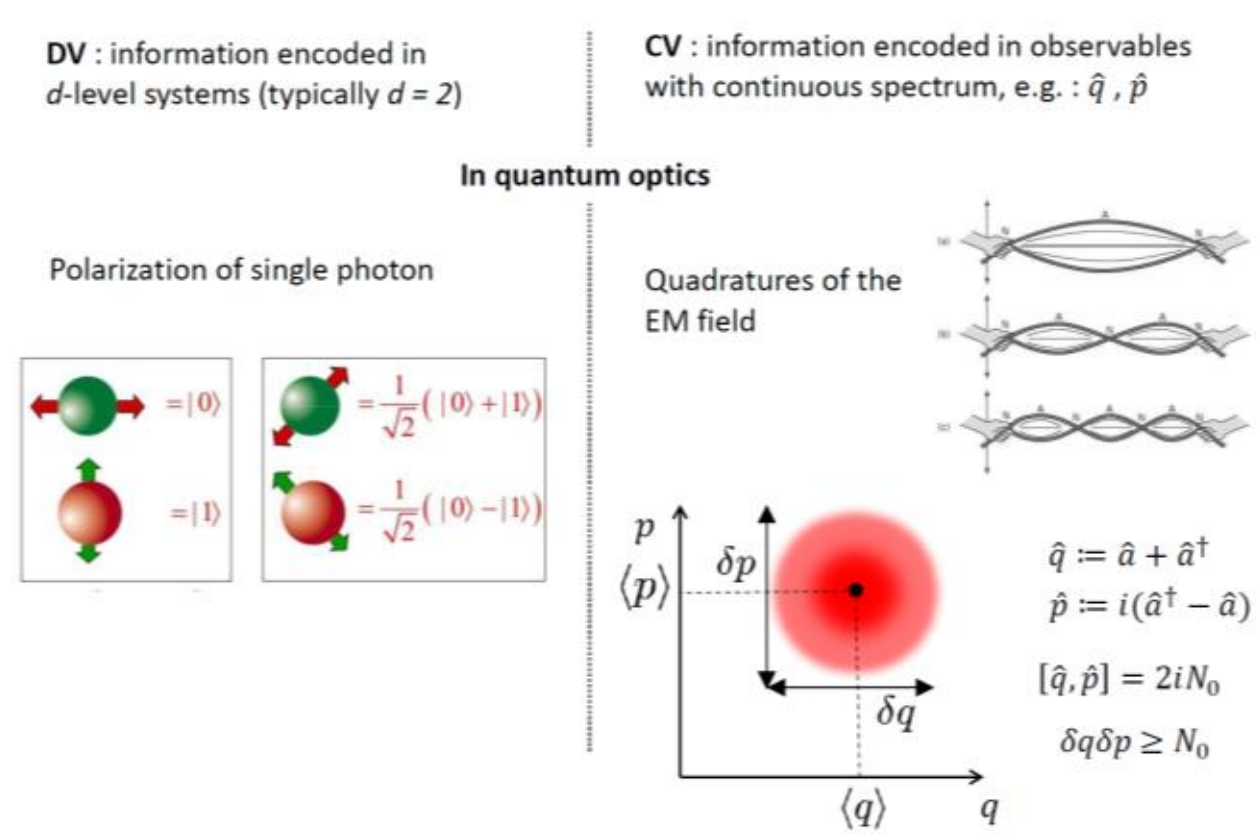

Figure 742: DV and CV in QKD. Source: TBD.

[2310] And published in the article Quantum Cryptography Based on Bell's Theorem (3 pages). Artur Ekert has been a member of Atos Scientific Council since 2016, along with Alain Aspect, Daniel Esteve, Serge Haroche, Cédric Villani and David DiVincenzo.

[2311] Their QKD protocol is baptized accordingly GG02.

Figure 743 makes a rough comparison between DV-QKD and CV-QKD. Nowadays, CV-QKD seems preferred for telecom fiber deployments.

Figure 744 describes a typical optical architecture for the implementation of a CV-QKD protocol based on BB84, without entanglement given CV-QKD can also be implemented with entanglement based protocols. It uses a simple photon source in the telecom wavelength band around 1550 nm followed by amplitude and phase modulators. On the Bob side, a homodyne detector will demodulate the signal.

| | DV-QKD | CV-QKD |
|---|---|---|
| Quantum state | Polarization, phase, or time bin of a single photon | Quadrature components of quantized electromagnetic field |
| Source | Single-photon source | Coherent-state or squeezed-state source |
| Detector | Single-photon detector | Homodyne or heterodyne detector |
| Channel model | Lossy qubit channel | Lossy bosonic channel |
| Distance limitation | Performance of single-photon detectors | Efficiency of post-processing techniques |

*Figure 743: comparison between DV-QKD and CV-QKD protocols. Source: The Evolution of Quantum Key Distribution Networks: On the Road to the Qinternet by Yuan Cao, IEEE, 2021 (59 pages).*

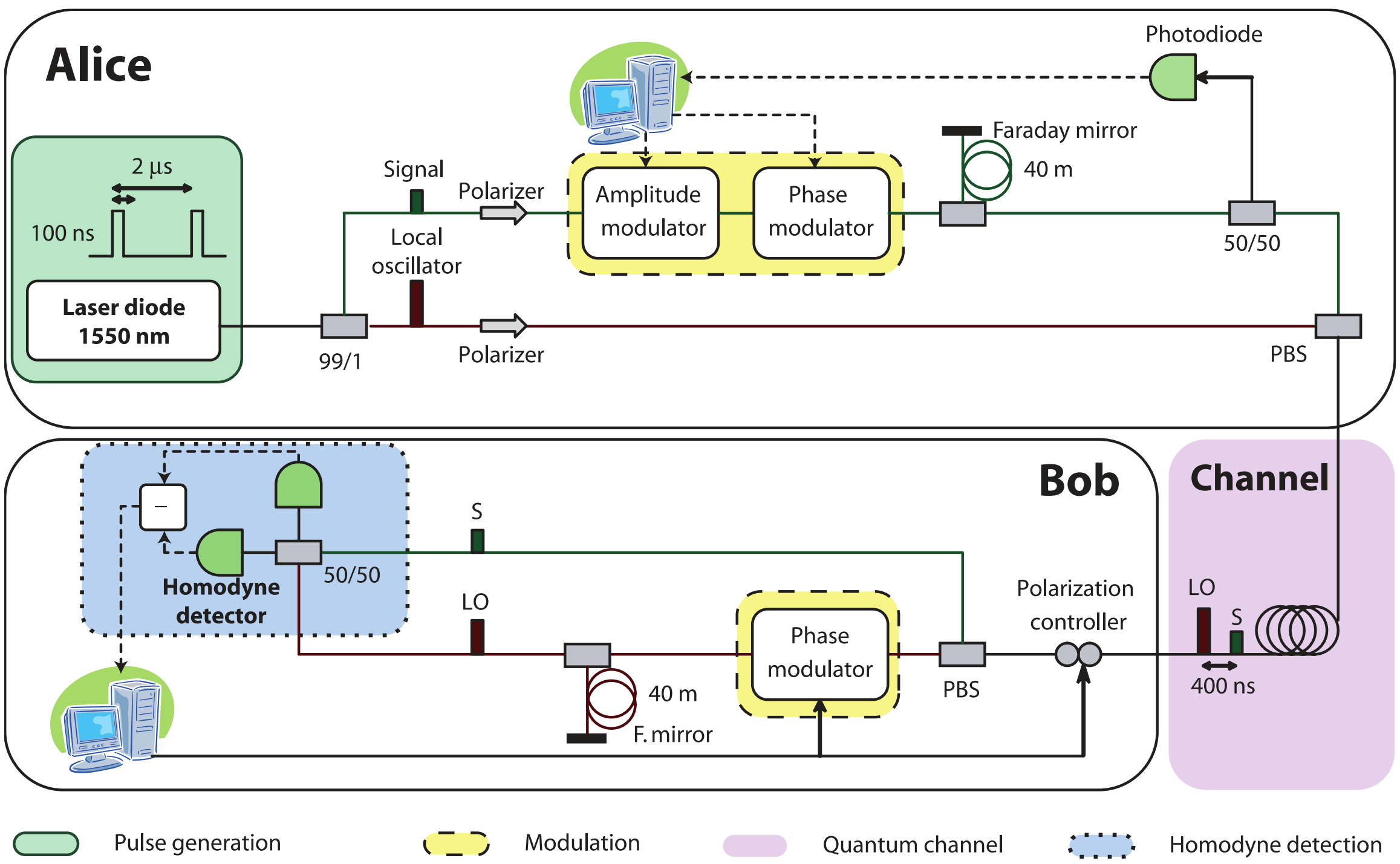

*Figure 744: an example of of CV-QKD implementation of the BB84 protocol. The SECOQC quantum key distribution network in Vienna by M. Peev, C. Pacher, Romain Alleaume, C. Barreiro, J. Bouda, W. Boxleitner, Thierry Debuisschert, Eleni Diamanti, et al, 2009 (39 pages).*

The integration of a QKD in conventional telecommunication optical fibers is typically done using three methods: by frequency multiplexing (WDM), for instance with a QKD signal on 1310 nm and data sent on 1550 nm, by time sharing (TDM) or by using a dedicated fiber embedded in a sheath (SDM). China has a very good experience in that domain[2312].

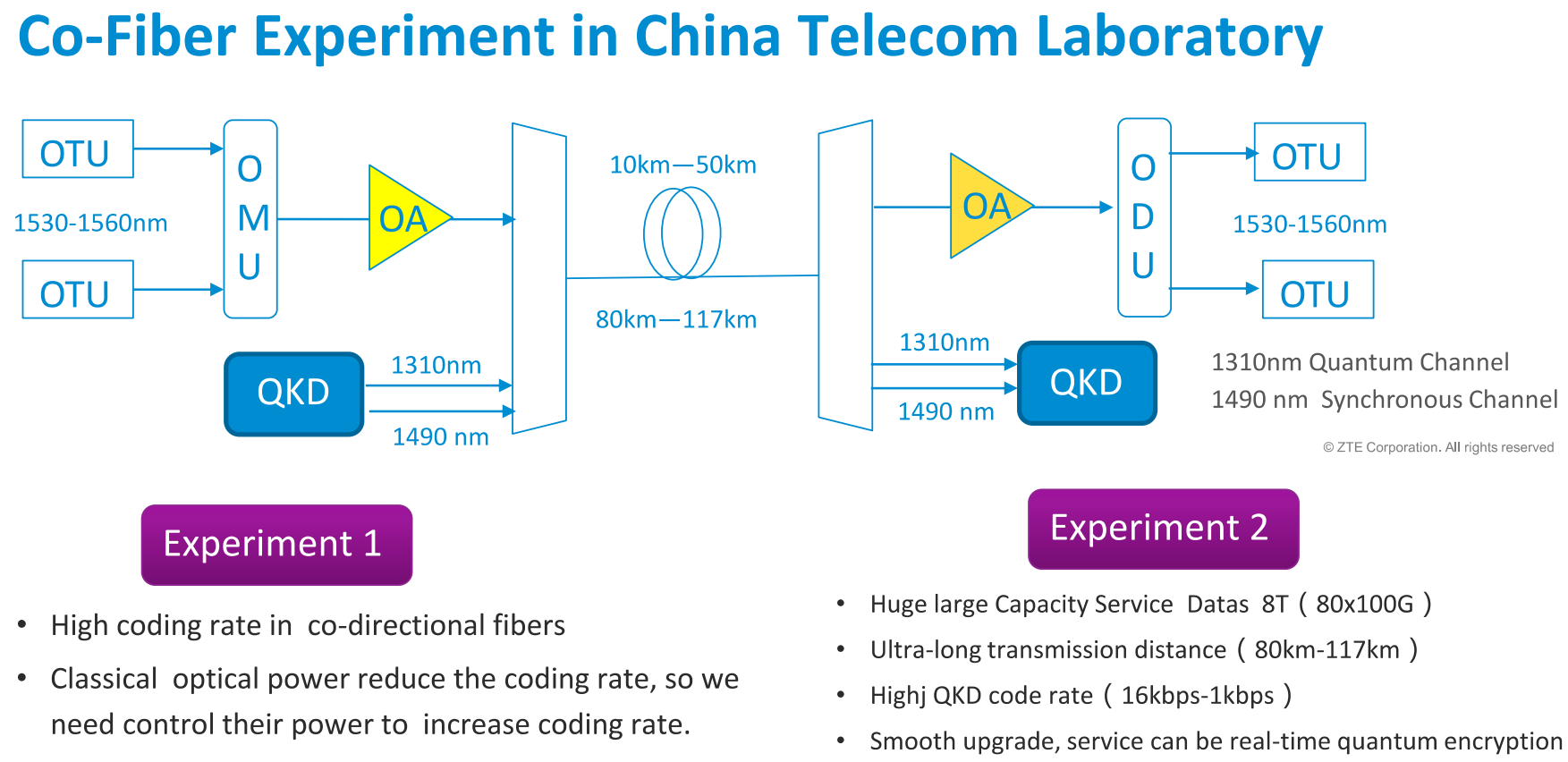

*Figure 745: co-fiber experiment in China Telecom laboratory distributing quantum keys over telecom fiber lines. Source: QKD Application: Coexistence QKD Network and Optical Networking the same optical fiber network by JiDong Xu, ZTE, June 2019 (15 slides).*

With entanglement-based CV-QKD protocols, the initial entanglement done before sending the two bits in the qubits avoids violating the Holevo theorem, already mentioned several times, according to which a set of qubits cannot carry more information than its equivalent number of classical bits. On the other hand, the information encrypted with the transmitted key is usually sent over a traditional channel[2313]. It is still often encrypted using SSL, which protects the relationship between your browser and the websites you visit and supports the secured https protocol.

In practice, keys transmission using a QKD is accompanied by a complex system of "key distillation" that manages the communication imperfections with classical error correction codes (which have nothing to do with the quantum error correction codes seen at the qubit level elsewhere in this document, page 216), an amplification of confidentiality and an authentication system using private keys already shared by the correspondents, making it possible to avoid so-called *man-in-the-middle* attacks by hackers pretending to be one of the interlocutors.

Error correction codes and the rest of the protocol generate on-line losses of about 80% of the quantum key communication[2314]. Implementing a QKD combines a quantum random key generator such as those from IDQ, an authenticated classical channel to exchange QKD basis information and a QKD channel to share random keys, which can generally be transported on a dark fiber from a B2B telecom operator.

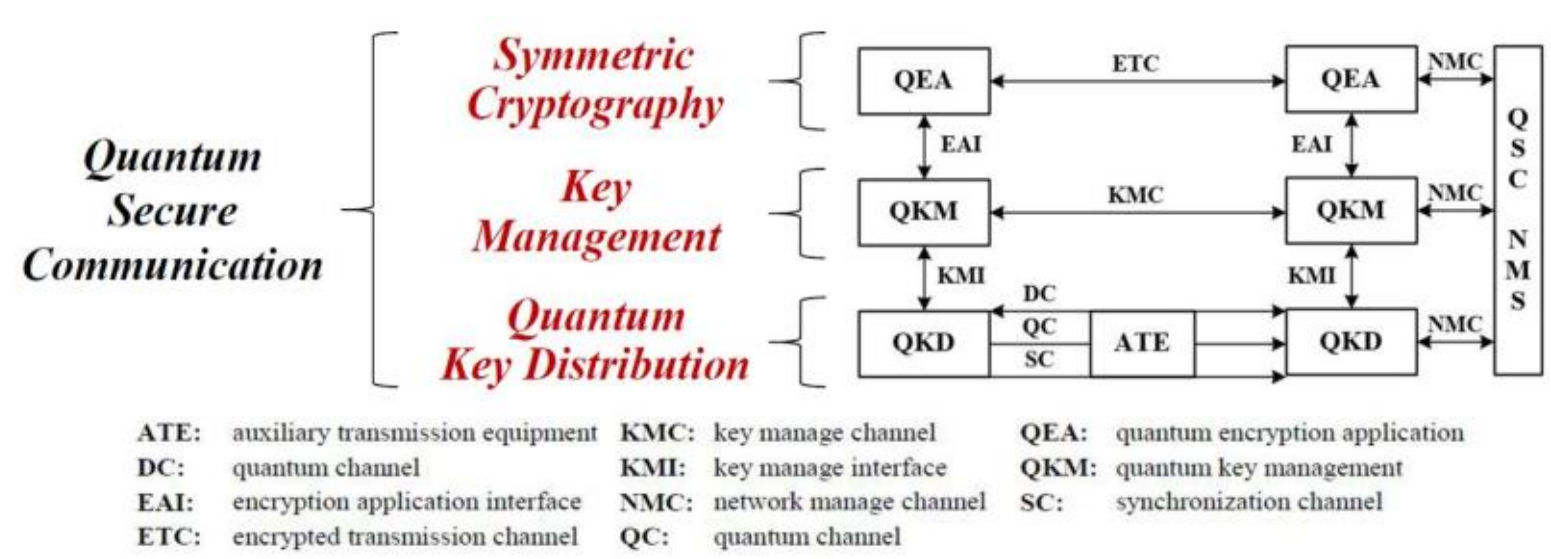

*Figure 746: the three components of quantum secure communication with a symmetric cryptography, key management and a QKD. Source: Development and evaluation of QKD-based secure communication in China by Wen-yu Zhao, June 2019 (15 slides).*

---

[2312] See QKD Application: Coexistence QKD Network and Optical Networking the same optical fiber network by JiDong Xu, June 2019 (15 slides). It's also described in Quantum Encrypted Signals on Multiuser Optical Fiber Networks Simulation Analysis of Next Generation Services and Technologies by Rameez Asif, 2017 (6 pages) and Quantum experiments explore power of light for communications, computing by Elizabeth Rosenthal, January 2020.

[2313] Classical information can take a very different path. For example, a quantum key can be transmitted by satellite and data can be transmitted terrestrially over fiber optics.

[2314] According to the excellent overview by Sheila Cobourne of the University of London Quantum Key Distribution Protocols and Applications, 2011 (95 pages).

The useful data is encrypted with the QKD generated key with classical encryption protocols like AES[2315] and transmitted over a traditional channel, which may also be a classical optical fiber or other physical communication media, even cellular communications.

This is well documented by ETSI[2316]. On arrival, a quantum key receiver and the system for decrypting the signal arriving via the traditional channel is used.

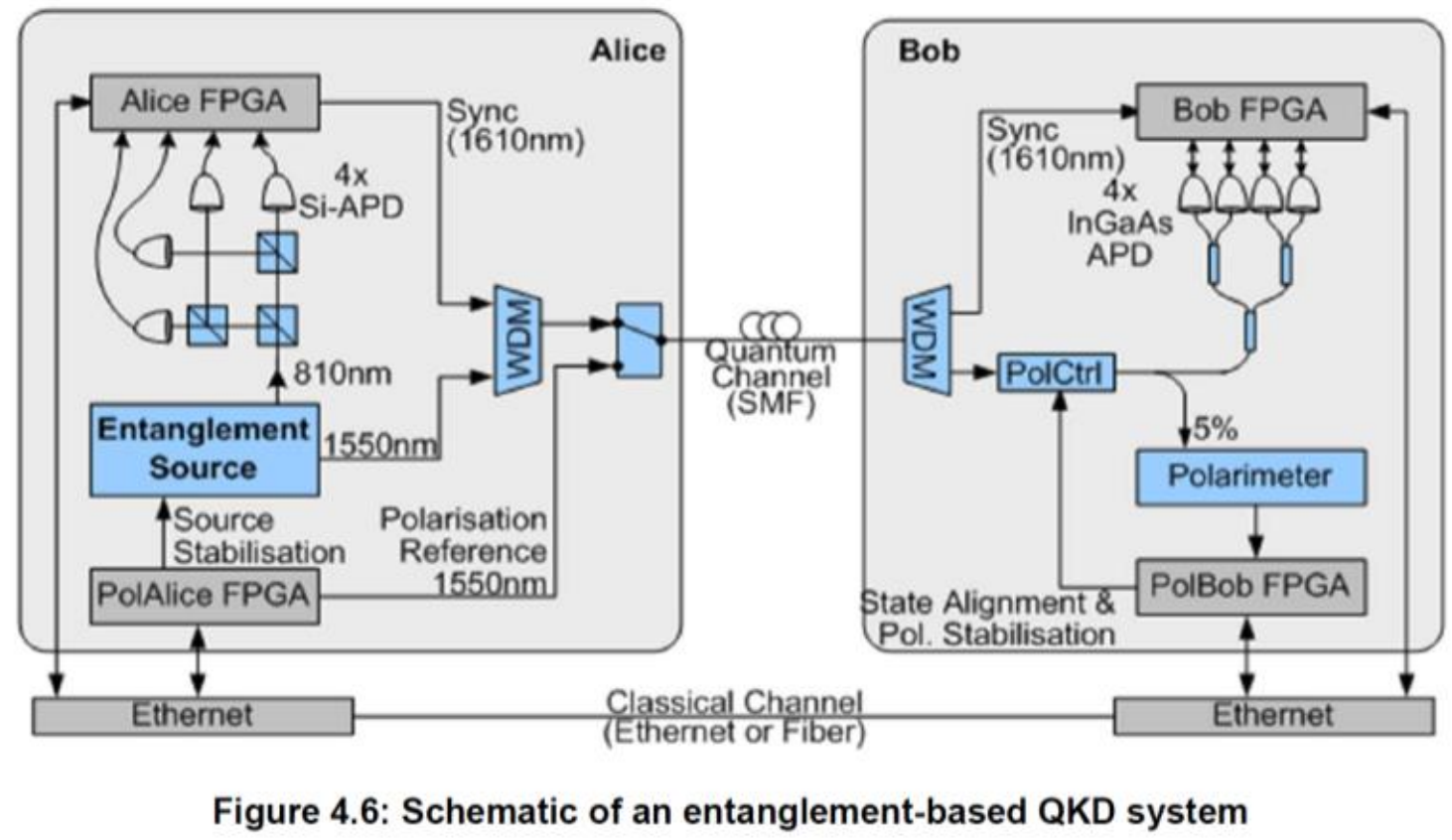

*Figure 747: Source: Quantum Key Distribution (QKD) Components and Internal Interfaces from ETSI, 2018 (47 pages).*

The secret keys throughput is an important issue and is currently at its maximum in Mbits/s vs. the Tbits/s of the operators' optical links[2317]. Many optimizations must be implemented to make QKD practical, first with closing the many discovered security loopholes. Other improvements can be achieved with implementing clock synchronization without an additional dedicated channel and develop protocols to increase raw key rates[2318].

The quantum keys transfer rate can be much lower than the physical data rate available. There are fundamental bounds on how much secret keys can be generated over a noisy quantum channel.

There are also other fundamental questions about the overall architecture and topology of QKD networks, most of the time working in "point to point" and relying on so-called trusted nodes, which can be considered as "classical" repeaters [2319].

Proposals abound to create large multi-point networks and routing protocols[2320] and on entanglement distribution optimization[2321].

[2315] One-time pads encryption techniques can also be used with QKD. It consists in creating a key that is as large as the content to be encrypted. This technique makes the content uncrackable with brute force.

[2316] In Quantum Key Distribution (QKD) Components and Internal Interfaces from ETSI, 2018 (47 pages) which describes the different QKD techniques available to date. It also describes very well the photon sources used in QKDs as well as the associated quantitative and qualitative parameters.

[2317] See Experimental Demonstration of High-Rate Discrete-Modulated Continuous-Variable Quantum Key Distribution System by Yan Pan et al, March 2022 (5 pages) which describes a record of high-speed CV-QKD distribution over distances of 5 to 50 km with keys generations ranging from 288 Mbits/s to 7.6 Mbits/s. A record key rate of 1 Gbits/s with a source of entangled photons in the 1550 nm telecom wavelengths was obtained in 2022 by an Austrian team but the practical key rate would be much lower in practical use cases due to fiber attenuation over long distances. See Experimental entanglement generation for quantum key distribution beyond 1 Gbit/s by Sebastian Philipp Neumann et al, September 2022 (11 pages).

[2318] See for example Operational real field entangled quantum key distribution over 50 km by Yoann Pelet, Sébastien Tanzilli et al, July 2022 (7 pages).

[2319] See an example of trusted node deployment with Trusted Node QKD at an Electrical Utility by Philip G. Evans et al, March 2021 (10 pages).

[2320] See A Quantum Internet Architecture by Rodney Van Meter et al, December 2021 (17 pages), Cost and Routing of Continuous Variable Quantum Networks by Federico Centrone, Frederic Grosshans and Valentina Parigi, April 2022 (20 pages), A quantum router architecture for high-fidelity entanglement flows in quantum networks by Yuan Lee, Eric Bersin, Axel Dahlberg, Stephanie Wehner and Dirk Englund, npj, June 2022 (8 pages) and An Efficient Routing Protocol for Quantum Key Distribution Networks by Jia-Meng Yao et al, April 2022 (27 pages).

[2321] See 40-user fully connected entanglement-based quantum key distribution network without trusted node by Xu Liu et al, January 2022 (15 pages) and Genuinely Multipartite Entanglement vias Quantum Communication by Ming-Xing Luo et al, April 2022 (9 pages).

There are a few other varieties of QKD protocols beyond the DV-QKD and CV-QKD as shown in Figure 748 with **DI-QKD**, a more secure Device-Independent QKD [2322] and **MDI-QKD** for "measurement-device-independent" QKD [2323].

| Protocol | Type | Approach | Year |
|---|---|---|---|
| BB84 | DV | Prepare-and-measure | 1984 |
| E91 | DV | Entanglement-based | 1991 |
| BBM92 | DV | Entanglement-based | 1992 |
| GG02 | CV | Prepare-and-measure | 2002 |
| DPS | DV | Prepare-and-measure | 2002 |
| Decoy-state | DV | Prepare-and-measure | 2003–2005 |
| SARG04 | DV | Prepare-and-measure | 2004 |
| COW | DV | Prepare-and-measure | 2005 |
| MDI | DV/CV | Prepare-and-measure | 2012 |
| TF | DV | Prepare-and-measure | 2018 |
| PM | DV | Prepare-and-measure | 2018 |

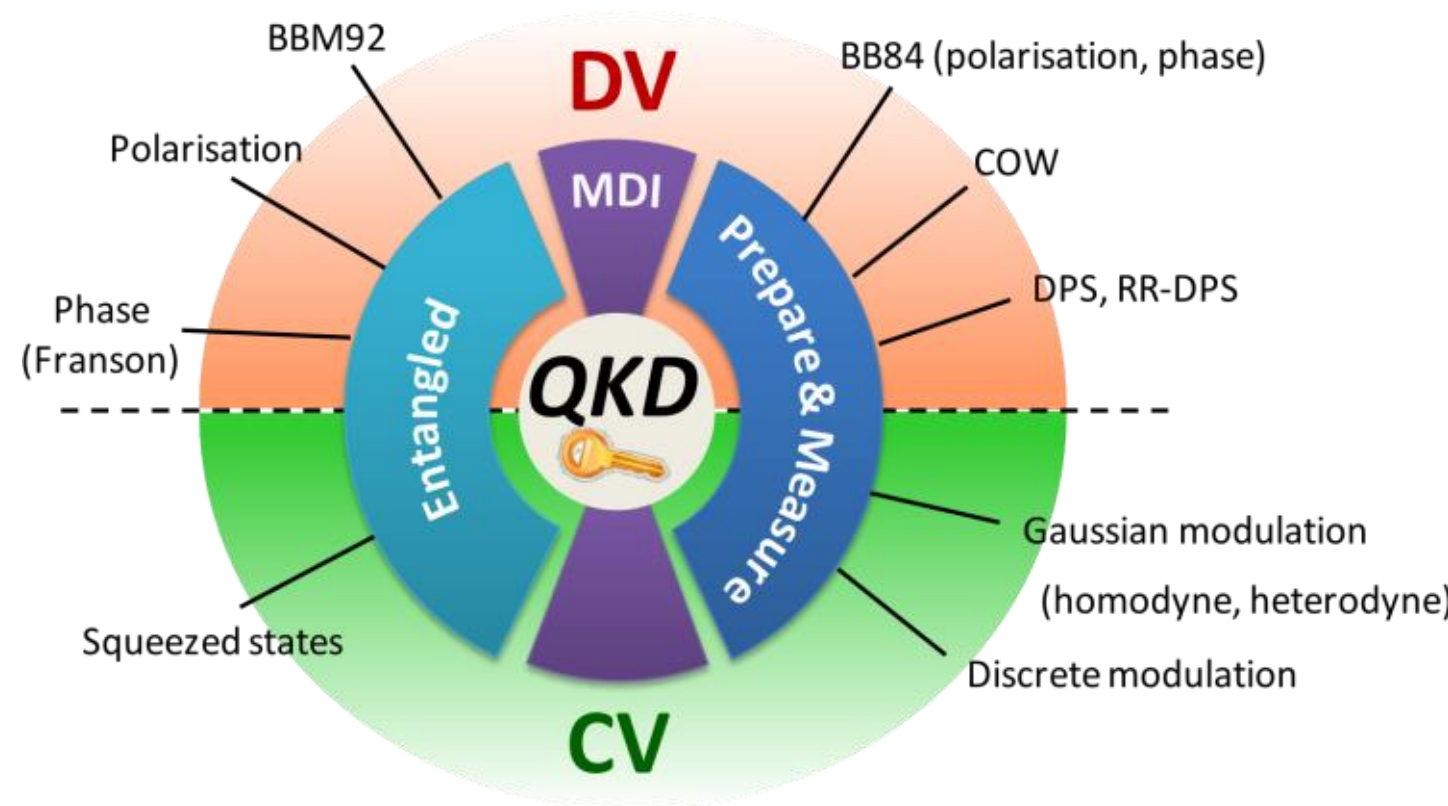

*Figure 748: a map of QKD protocols between DV and CV ones. I don't cover them all in this book. Sources: Source: The Evolution of Quantum Key Distribution Networks: On the Road to the Qinternet by Yuan Cao, IEEE, 2021 (59 pages) and VSCW19-QKD-part1. https://quantum-uniqorn.eu/wp-content/uploads/2019/06/VSCW19-QKD-part.pdf .*

## QKD experiments and deployments

Many breakthrough symbolic experimental deployments of QKD have been done both in the open air and with optical fibers.

Open-air QKD demonstrations started in 1996 in the USA on a 75m distance, then on 144 km to connect the islands of La Palma and Tenerife in the **Canary Islands** and conducted by Austrians in 2007 and 2010[2324], in 2019 in urban areas in Italy with a distance of 145m[2325] and in India in 2022 over 300 m[2326].

An experiment took place in **Vienna** in 2008 as part of the European **SECOQC** (*SEcure COmmunication* based on Quantum Cryptography) project launched in 2004, involving some 40 research laboratories and vendors, using a "mesh" architecture and a CV-QKD optical link [2327].

This went on in Switzerland with **IDQ** with local banks. In 2007 they also set up an election vote counting system based on a QKD.

---

[2322] See the international review paper Advances in device-independent quantum key distribution by Víctor Zapatero et al, August 2022 (15 pages).

[2323] See Measurement-device-independent quantum key distribution by Hoi-Kwong Lo, Marcos Curty and Bing Qi, 2011 (7 pages) and this good overview of QKD and its technical challenges in Practical challenges in quantum cryptography by Eleni Diamanti et al, 2016 (25 pages).

[2324] See Second Generation QKD System over Commercial Fibers, 2016 (5 pages) et Feasibility of 300 km Quantum Key Distribution with Entangled States, 2010 (14 pages).

[2325] See Full daylight quantum-key-distribution at 1550 nm enabled by integrated silicon photonics by Matteo Schiavon et al, July 2019 (7 pages). QKD key transmission over the 145m air link took place at 1550 nm in the infrared band that is commonly used in fiber optic transmissions and therefore compatible with many existing telecommunication equipment. They used a silicon chipset doing all the work with an error rate of only 0.5% and a data rate of 30 kbits/s. Their QCosOne ("Quantum Communication for Space-One") used telescopes with 120 and 315 mm optics for transmission and reception. It worked during daytime, but there were still problems in case of turbulence and depending on the time of day and the side effects coming from the sun. Performance was better in the evening. They used triple photon encoding: temporal, spatial and spectral.

[2326] See Indian Scientists Demonstrate Wireless Quantum Key Distribution Over 300 Meters by Matt Swayne, The Quantum Insider, February 2022.

[2327] See The SECOQC quantum key distribution network in Vienna by Romain Alléaume, Eleni Diamanti et al, 2016 (39 pages).

In 2018, the **UK** deployed its UK Quantum Communications hub between Bristol, London, Cambridge and Ipswich[2328].

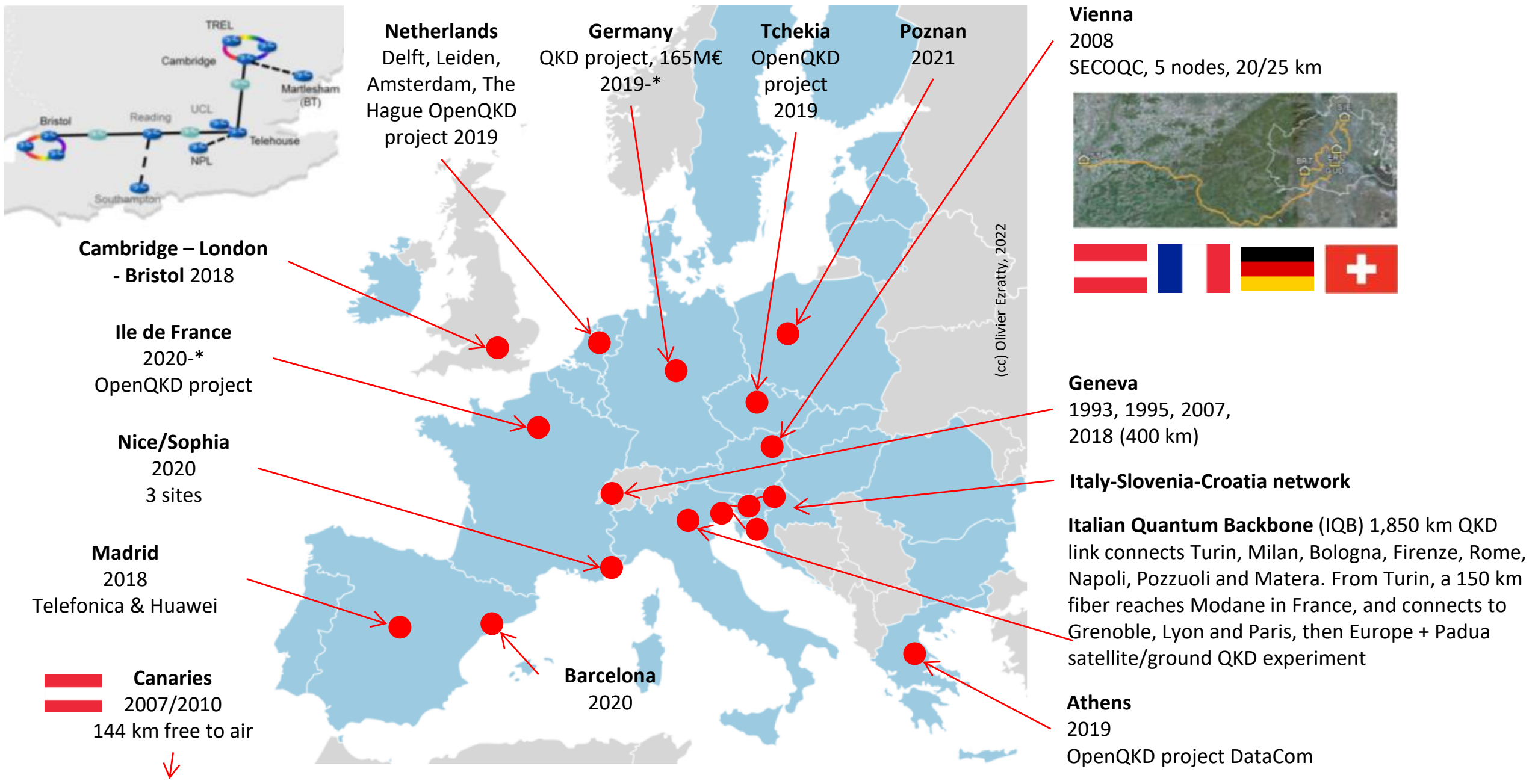

*Figure 749: a map of QKD test deployments throughout Europe. (cc) Olivier Ezratty, 2022.*

In France, **Orange** announced in May 2019 the launch of tests of a QKD protected communication with the University Côte d'Azur (UCA) which provides the solution via the InPhyNi laboratory. It connects the Valrose and Inria campuses in Sophia Antipolis with an access point on the Plaine du Var IMREDD campus in Nice, using dark fibers provided by the telecom operator[2329].

The test network was operational in September 2021. In particular, Orange is studying the reliability of trusted optical network nodes in such a configuration. The operator is also looking to combine QKD solutions to protect physical links and PQC solutions that could be used as a method of encrypting data transmitted in association with a QKD.

In **Denmark**, DTU launched a CV-QKD trial with Danske Bank[2330].

In **Greece**, Space Hellas and the University of Athens launched in 2022 a pilot QKD deployment after having started an OpenQKD project in 2019 with Datacom.

**Cyprus** announced in 2022 its deployment of a QKD network with a EuroQCI EU funding of 7.5M€ as part of the CYQCI project.

The European consortium **OpenQKD** is experimenting a terrestrial QKD network.

It prepares the ground for the launch of the **EuroQCI** network which would be an operational implementation of a terrestrial and satellite European QKD network[2331].

---

[2328] Seen in IDQ: Quantum-Safe Security relevance for Central Banks, 2018 (27 slides). The network was extended in Cambridge in 2019 as seen in Cambridge quantum network by J. F. Dynes et al, 2019 (8 pages).

[2329] See Experimenting quantum key exchange over the Côte d'Azur, Orange, December 2019. And for Orange's QKD projects in general: Orange and quantum technologies for secure data exchange, June 2020.

[2330] See First quantum-safe data transfer in the Nordic region by Anne Kirsten Frederiksen, February 2022.

[2331] See Nine more countries join initiative to explore quantum communication for Europe, December 2019 and Quantum communications infrastructure architecture: theoretical background, network structure and technologies. A review of recent studies from a European public infrastructure perspective by Adam M. Lewis and Petra F. Scudo, January 2022 (40 pages).

This involves in particular France, Germany, Austria, Italy, Spain, the Netherlands, Greece, Switzerland and Poland and vendors like Thales Alenia Space (satellite communication), Orange and Mellanox (a subsidiary of Nvidia). From a practical point of view, it's about deploying a large interoperable experimental QKD network on a European scale, exploited by applications in various fields (healthcare, energy grids, transportation, finance, government, education, and the likes). The consortium also intends to influence QKD standardization. And at last, as far as possible, it's also about contributing to the development of a European industry offering in QKD and associated technologies. There is at last a deployment of QKD across Italy, Slovenia and Croatia[2332]. In France, the test area will be the Paris region and its major research laboratories with the Institut d'Optique, Telecom Paris, LIP6 in Jussieu and Nokia labs in Villarceau. The project has received €15M in European funding from Horizon 2020, independently of Quantum Flagship. Thales announced in December 2021 that it was participating to the QSAFE consortium with Deutsche Telekom, Telefonica and the AIT (Austrian Institute for Technologies) to create a European quantum telecommunication infrastructure as part of EuroQCI, with both terrestrial and satellite links.

In **USA**, the first experiments were conducted in Boston by **DARPA** between 2004 and 2007. A QKD network piloted by **Battelle** was tested in Ohio in 2013[2333]. Tests were also conducted in 2015 at **MIT**, linking two sites 43 km apart. A commercial deployment of QKD on an unused 800 km fiber optic network connecting Boston to Washington DC is also being deployed by **Quantum Xchange** and **Zayo**, to connect Wall Street finance businesses with their back-offices in New Jersey. It uses some trusted node technology[2334]. An 85 km facility was also deployed in Chicago in 2019[2335]. In July 2020, the Department of Energy announced the expansion of the QKD network to link all of its research laboratory sites[2336].

In 2022, it connected the DoE Argonne and Fermi labs and enabled fast network synchronization of quantum and classical signals coexisting on the same optical fiber over a distance of 59 km[2337].

In **Canada** launched in November 2020 its Canada Quantum Network through a partnership between Xanadu, the Creative Destruction Lab and the startups service company MaRS around Toronto.

In **India**, the Defence Research and Development Organisation (DRDO) with IIT Delhi demonstrated a QKD for a distance of 100 km.

In **Japan**, Toshiba announced in September 2018 that a QKD solution co-developed with the Tohoku Medical Megabank Organization (ToMMo) at Tohoku University had achieved a QKD throughput of more than 10 Mbps during one month.

In **Singapore**, the Quantum Engineering Program (QEP, launched in 2018) from the National University of Singapore (NUS) launched the National Quantum-Safe Network (NQSN) which is starting classical and quantum network trials supporting both QKD and PQC.

---

2332 See Deploying an inter-European quantum network by Domenico Ribezzo et al, March 2022 (8 pages).

2333 See Battelle Installs First Commercial Quantum Key Distribution Protected Network in U.S., 2013.

2334 See New plans aim to deploy the first US quantum network from Boston to Washington, DC, October 2018. Schema source: From MIT : Semiconductor Quantum Technologies for Communications and Computing, 2017, 32 slides).

2335 See Argonne and UChicago scientists take important step in developing national quantum internet by Louise Lerner, February 2020.

2336 See Department of Energy (DOE) Unveils Blueprint for a U.S. Quantum Internet by Doug Finke, July 2020.

2337 See Quantum Network Between Two National Labs Achieves Record Synch by Matt Swayne, June 2022 and Picosecond Synchronization of Photon Pairs through a Fiber Link between Fermilab and Argonne National Laboratories by Keshav Kapoor et al, August 2022 (7 pages).

They plan to have 10 fiber nodes. This is done in partnership with Amazon Web Services and Thales. One of the challenges in deploying QKD is the miniaturization of its components. Whereas initially a complete rack of hardware was needed for quantum key transmitting/receiving stations, the goal is to fit everything in a photonics component a few mm long. This is what **NTU** researchers in Singapore did in 2019 to manage a CV-QKD supporting existing telecom operators' fiber infrastructures[2338]. But this miniaturization concerns here only the photonics part. These photonic circuits have to be completed by classical electronic components.

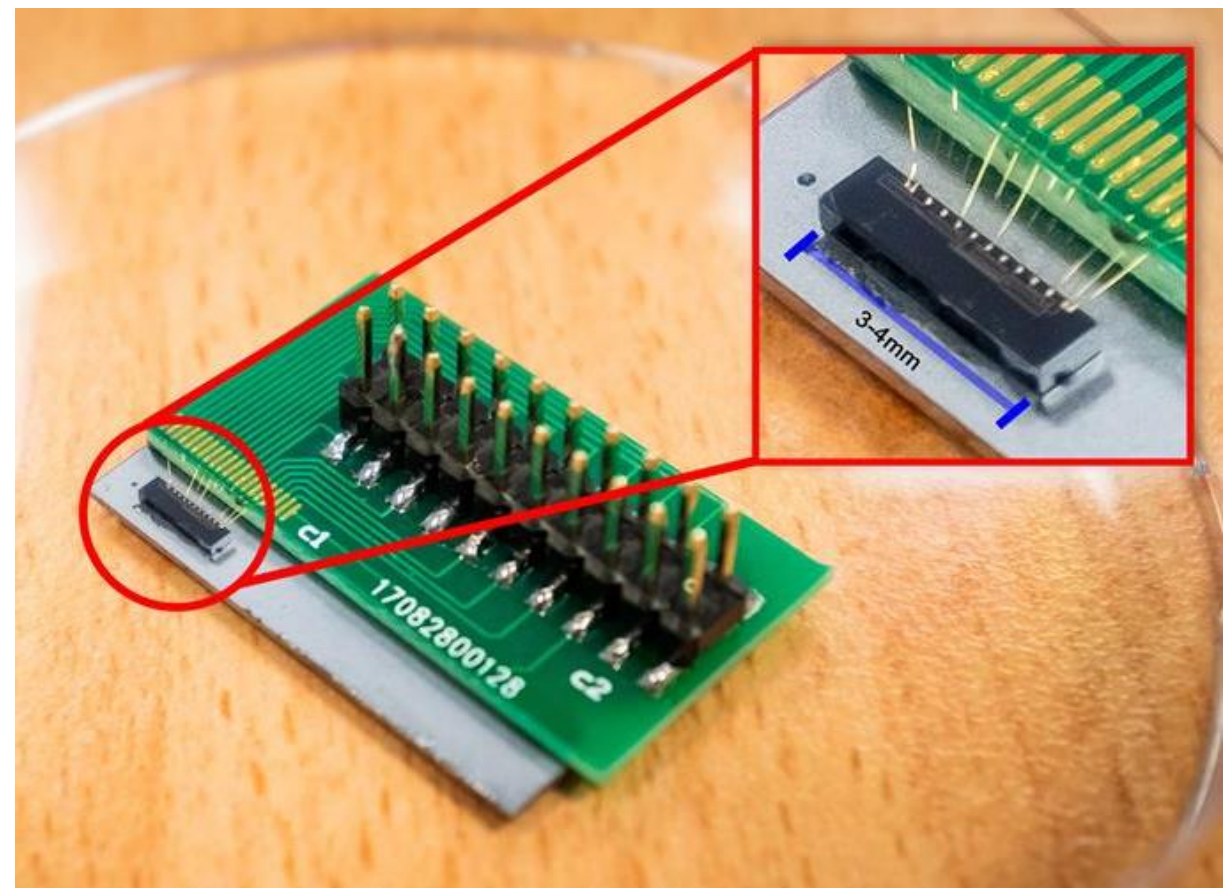

*Figure 750: Source: Researchers create quantum chip 1,000 times smaller than current setups, PhysOrg, October 2019*

The development of such integrated photonics components will also be supported in the framework of European Horizon Europe projects that follows Europe 2020 projects over the 2021-2027 period.

China stands out since 2016 with impressive QKD experiments and deployments. The country invests heavily in QKD with a now classical multi-pronged strategy: to protect its sensitive communications against any attacker and to develop an industry in a promising emerging technology field. A first deployment was carried out in 2012 in the Hefei area to link various Chinese government entities[2339].

It was then expanded with an installation of a QKD-secured fiber optic link between Shanghai and Beijing, covering 2,000 km. The line installed between 2013 and 2016 was deployed by a local startup, **QuantumCTek**.

The network relies on 32 transponders with secured physical access[2340]. Indeed, the signal attenuation was then too strong beyond about 50 km on one optical fiber.

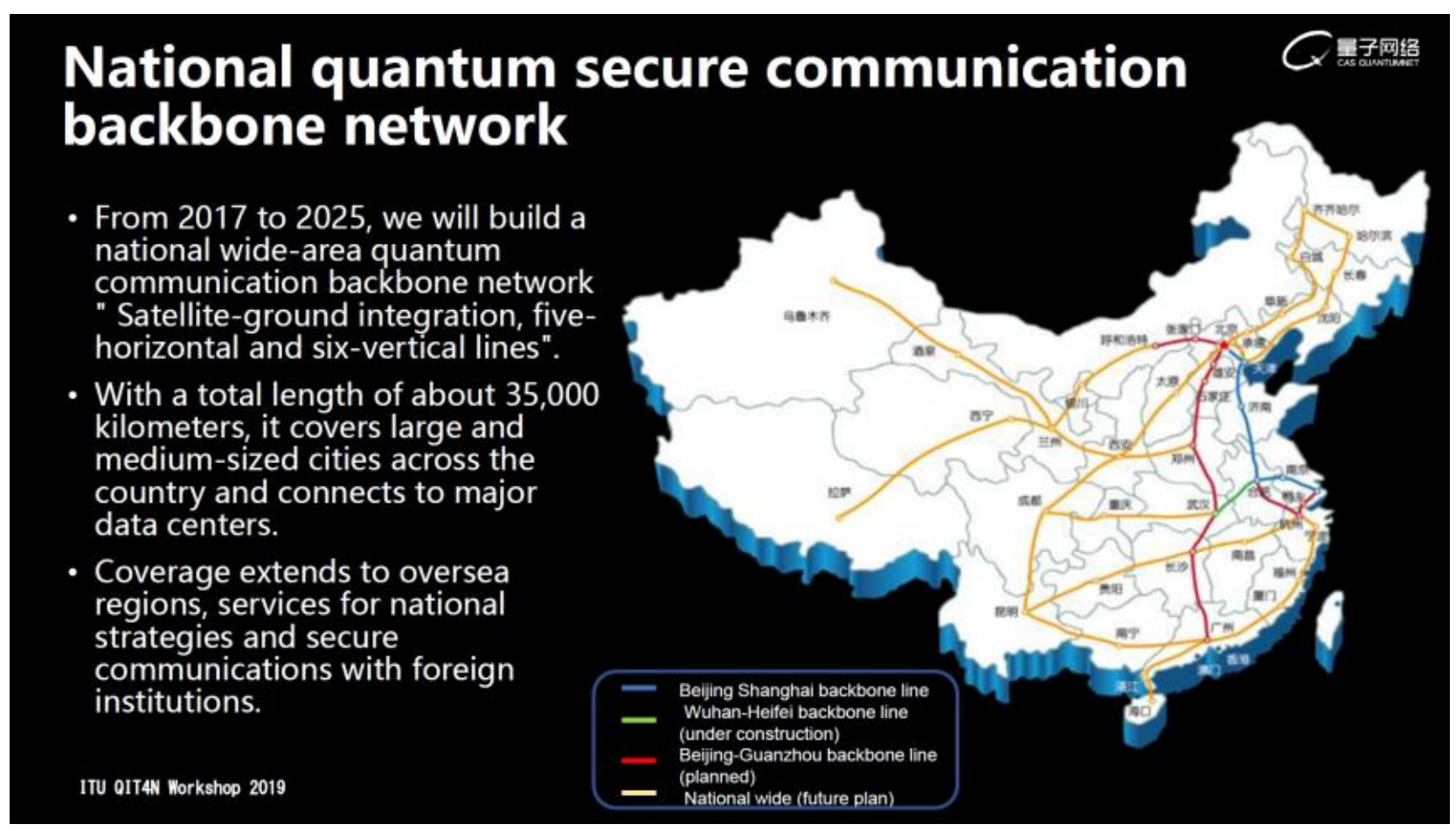

*Figure 751: China's QKD backbone. Source: 2019.*

The entities using this line are government agencies, including various financial sector regulatory agencies and banks. The country also plans to protect its energy grid infrastructure with this network[2341].

[2338] See Researchers create quantum chip 1,000 times smaller than current setups, PhysOrg, October 2019 which references An integrated silicon photonic chip platform for continuous-variable quantum key distribution by G. Zhang et al, December 2019 (5 pages).

[2339] See Unhackable Chinese Communication Network Launches Soon by Rechelle Ann Fuertes, 2017.

[2340] Source: Security assessment and key management in a quantum key distribution network by Xiongfeng Ma, June 2019 (21 slides).

[2341] See Application In Power Industry Promotes the Development of Quantum Cryptography Technology by Yonghe Guo, June 2019 (13 slides).

China then launched in 2017 a deployment of an additional 33,000 km of the **National Quantum Secure Communication Backbone Network**, to be completed by 2025. It had started with the creation of a Hefei-Wuhan link[2342]. Hefei is the city where Jian-Wei-Pan's main quantum technologies laboratory is located[2343].

### QKD by satellite, UAVs and underwater

Satellite quantum communication is another mean to distribute quantum keys that makes sense over long distance. These keys can be shared across two locations while the encrypted data may be transported in more classical terrestrial means. This would be theoretically simpler than using QKD distribution over large fiber networks full of (so far, non-existent) quantum repeaters.

One advantage is that the photon loss is much lower with satellites than with fibers. Indeed, with satellites, photon losses coming from the atmospheric absorption and scattering occurs only in the lower 10 km of the atmosphere, with about a 3 dB loss on a clear day. The rest of the distance is in near vacuum, with nearly no absorption and decoherence and the loss caused by beam diffraction is approximately proportional to the square of distance whereas the losses in fiber are mainly due to the absorption and scattering of the fiber medium, which is proportional to the exponent of the distance.

It means that for long communicating distances of several hundred km, satellite-ground channels have an advantage over fiber-based channels in terms of channel losses[2344]. China was a pilot country in that area, the use of the **Micius** satellite also named **Mozi** for a western/Chinese pronunciation and **QUESS** (Quantum Experiments at Space Scale).

It was used for running several different experiments: a satellite-to-ground decoy-state QKD with KHz keyrate over a distance of up to 1200 km and satellite-replayed intercontinental key exchange, a satellite-based entanglement distribution to two Earth locations separated by 1205 km and a ground-to-satellite qubit teleportation [2345]. The satellite weighs 640 kg and consumes 560W.

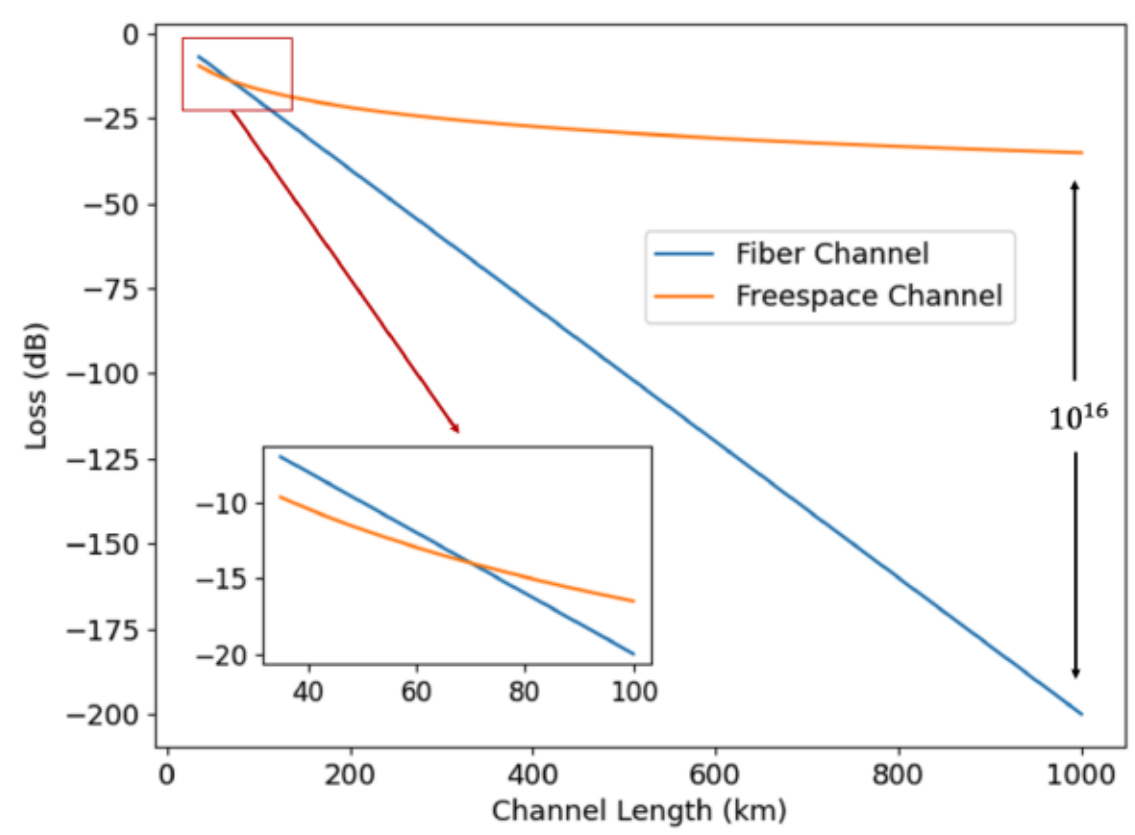

FIG. 10 Typical losses in fiber and free-space channels. The attenuation parameter of fiber is $\sim 0.2\ dB/km$. The parameters of free-space channel are based on the design of Micius satellite. The free-space channel shows advantage for a distance over $\sim$ 70 km

*Figure 752: how photon losses compare between fiber and freespace channel using satellite. Source: Micius quantum experiments in space by Chao-Yang Lu, Yuan Cao, Cheng-Zhi Peng and Jian-Wei Pan, August 2022 (53 pages).*

[2342] See Towards large-scale quantum key distribution network and its applications by Hao Qin, 2019 (17 slides).

[2343] The extended BB84 based QKD network is documented in Implementation of a 46-node quantum metropolitan area network by Teng-Yun Chen et al, September 2021 (14 pages). It describes the intra-metropolitan network infrastructure deployed in Hefei, with 46 nodes.

[2344] This paragraph is largely inspired from the excellent and well detailed review paper Micius quantum experiments in space by Chao-Yang Lu, Yuan Cao, Cheng-Zhi Peng and Jian-Wei Pan, August 2022 (53 pages).

[2345] See respectively Satellite-to-ground quantum key distribution by Sheng-Kai Liao, Jian-Wei Pan et al, Nature, August 2017 (18 pages), Satellite-Relayed Intercontinental Quantum Network by Sheng-Kai Liao, Jian-Wei Pan et al, PRL, 2018 (10 pages), Satellite-to-ground entanglement-based quantum key distribution by Juan Yin, Jian-Wei Pan et al, PRL, 2017 and Ground-to satellite quantum teleportation by Ji-Gang Ren, Jian-Wei Pan et al, Nature, 2017 (16 pages). The principle was first described in 1993 in Teleporting an Unknown Quantum State via Dual Classical and EPR Channels by Charles Bennett, Gilles Brassard (from Montreal), Claude Crépeau, Richard Jozsa, Asher Peres and William Wootters. See also Quantum Communication at 7,600km and Beyond by Chao-Yang Lu and Cheng-Zhi Peng, Jian-Wei Pan, November 2018.

A 2018 experiment was about creating a videoconference between China and Austria using a quantum key sent every minute[2346]. Why with Austria? Because Jian-Wei Pan did his PhD thesis in Austria under the supervision of Anton Zeilinger, who piloted the European part of the experiment.

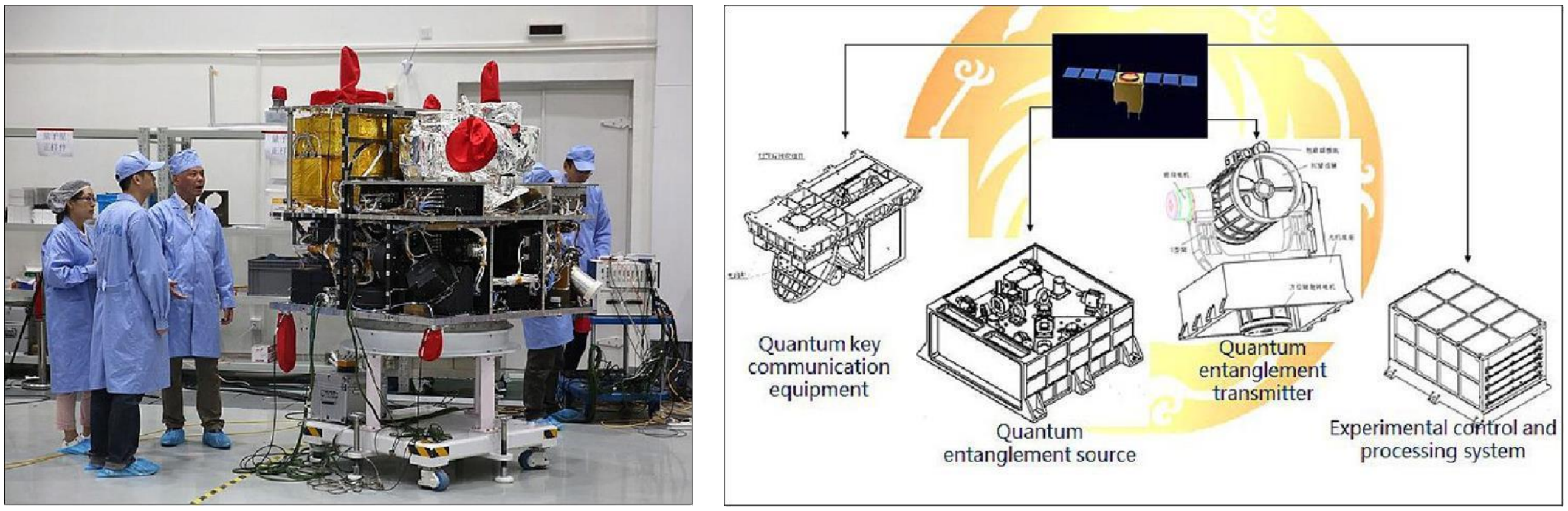

*Figure 753: a QKD satellite.*

China is planning to launch a cloud of satellites in low orbit by 2030 dedicated to sending quantum keys by repeating this process[2347]. If they are banking on QKD for symmetric key management, it seems that China is also investing in post-quantum cryptography but in a quieter way.

But this satellite experiment had limitations: it could handle 5.9 million pairs of entangled photons per second, but due to error corrections, only one useful photon pair was exploitable per second[2348]. As a result, Chinese scientists are studying scheduling approaches to load balance key distribution over time[2349]. Also, it works only by night!

At the beginning of 2020, China announced that it had miniaturized its ground receiving station for quantum key communication with the Micius satellite from 10 tons to 80 kg. The key bitrate was reduced, from 40Kbits/s to 4-10Kbits/s. The experiment took place on the Earth side in Jinan and Shanghai, so it seems at sea level[2350] . The Bank of China would already secure transactions by sending keys via the Micius/Mozi Beijing satellite and remote provinces.

In June 2019, Chinese researchers announced that they had demonstrated the use of optical QKD aerial links established within a network of 35 kg octocopter UAVs spaced 200 m apart during a 40-minute flight at an altitude of 100 m[2351].

---

[2346] See Real-world intercontinental quantum communications enabled by the Micius satellite, USTC, PhysOrg, January 2018. Experiments or equivalent experiences have been launched by European teams. See Quantum Photonics Technologies for Space, October 2018 (22 pages) and Nanobob CubeSat mission, 2018 (31 pages). This is also being done in the UK, where an experimental Cubesat micro-satellite project is planned to cover the country. See QUARC: Quantum Research Cubesat - A Constellation for Quantum Communication by Luca Mazzarella et al, 2020 (27 pages).

[2347] See some details on the satellite QKD deployment architecture in Approaches to scheduling satellite-based quantum key distribution for the quantum network by Xingyu Wang et al, 2021 (11 pages).

[2348] See A step closer to secure global communication by Eleni Diamanti, Nature, June 2020, which describes the practical conditions and limitations of these satellite key transmission experiments. And in particular the most recent one described in Entanglement-based secure quantum cryptography over 1,120 kilometers by Juan Yin et al, Nature, June 2020. The actual key bitrate was 0.12 bits per second!

[2349] See Approaches to scheduling satellite-based quantum key distribution for the quantum network by Xingyu Wang et al, 2021 (11 pages).

[2350] See China has developed the world's first mobile quantum satellite station by Donna Lu, January 2020.

[2351] See Drone-based all-weather entanglement distribution by Hua-Ying Liu et al, May 2019 (16 pages) and World's First "Quantum Drone" for Impenetrable Air-to-Ground Data Links Takes Off by Charles Q. Choi, IEEE Spectrum.

The payload handling quantum communication weighted 11.8 kg. It can still be miniaturized, since the Chinese are aiming to integrate it into mass-market UAVs.

China was also proud to announce in 2021 that they had created the world's first integrated quantum communication network, combining 700 terrestrial optical fibers with two ground-to-satellite links, achieving quantum key distribution over 4,600 km[2352]. China researchers also developed a portable ground satellite QKD station of only 100 kg[2353].

China announced yet another premiere in August 2022, with transmitting QKD keys from its **Tiangong-2** space lab to four ground stations who are also connected to Micius which is positioned in a higher orbit, thanks to using Tiangong-2 as a repeater[2354].

But China is not alone there.

**Singapore** and its universities and startups are working on CubeSat based tiny QKD distribution satellites[2355].

In **Europe**, there are also satellite QKD plans and architecture proposals[2356]. The EU is planning the launch of a 6B€ satellite program announced in February 2022 with QKD support, to be deployed by 2028. Eleni Diamanti's team in France is working on using adaptive optics to improve key sharing with satellite[2357]. Another project run by a 20-company consortium led by SES (Astra) and the European Space Agency with the support of the EU is planning the launch in 2024 of its EAGLE-1 QKD low-earth orbit satellite-based system.

**NASA**'s SEAQUE (Space Entanglement and Annealing QUantum Experiment) is a small experiment being launched in the ISS to test photon entanglement distribution in space, particularly to assess self-healing solutions against radiation damages. The project has contributions from the USA, Canada and Singapore. The experiment designed by **AdvR** (2019, USA) is to be docked outside the ISS in the "Bishop airlock" that is by Nanoracks, a private in-space services company[2358].

And beyond Earth, some are even devising about how some extraterrestrial civilization could communicate with us with entangled photons. I'm just wondering how these photon sources could be detected out of the noise coming from distant stars, even using all the technologies around to detect exoplanets (transit methods, spectrography, etc.)[2359].

After space, how about distributing QKD underwater? It's not a joke. It's being investigated in Turkey and in China[2360]! The first simulations deal with 10 to 40 m distances. It's far from being sufficient to enable communications with submarines, particularly of the nuclear breed.

---

2352 See Chinese Scientists Report World's First Integrated Quantum Communication Network by Matt Swayne, 2021.

2353 See Portable ground stations for space-to-ground quantum key distribution by Ji-Gang Ren, Jian-Wei Pan and al, May 2022 (9 pages).

2354 See Space–ground QKD network based on a compact payload and medium-inclination orbit by Yang Li et al, Optica, August 2022 (6 pages).

2355 See A CubeSat platform for space based quantum key distribution by Srihari Sivasankaran et al, April 2022 (6 pages).

2356 See Satellite-based Quantum Information Networks: Use cases, Architecture, and Roadmap by Laurent de Forges de Parny, Eleni Diamanti, Sébastien Tanzilli et al, February 2022 (21 pages).

2357 See Analysis of satellite-to-ground quantum key distribution with adaptive optics by Valentina Marulanda Acosta, Eleni Diamanti et al, November 2021 (17 pages).

2358 See NASA is launching a new quantum entanglement experiment in space by Charlotte hu, March 2022.

2359 See Viability of quantum communication across interstellar distances by Arjun Berera and Jaime Calderón-Figueroa, June 2022 (18 pages), a fancy topic that was covered everywhere like in Mathematical calculations show that quantum communication across interstellar space should be possible by Bob Yirka, Phys.org, July 2022.

2360 See On the Optimization of Underwater Quantum Key Distribution Systems with Time-Gated SPADs by Amir Hossein Fahim Raouf and Murat Uysal, Ozyegin University in Istanbul, June 2022 (6 pages) and Practical underwater quantum key distribution based on decoy-state BB84 protocol, by Shanchuan Dong et al, March 2022 (10 pages).

### QKD photon sources and detectors

We've already covered photon sources for quantum computing and seen some of their requirements like the creation of deterministic and indistinguishable photons, on top of the even harder challenge to create large clusters of entangled photons.

The challenges with photons generation for QKD are not the same. What is required are steady streams of individual photons, preferably generated in the telecom wavelengths between 1200 nm and 1550 nm. Entanglement based QKD also requires the generation of entangled pairs of photons. At last, these sources should be lightweight and easy to integrate in telecom infrastructures. Preferably, they shouldn't be power hungry and not require cryogeny, thus a preference for ambient temperature solid-state solutions.

The breath of technologies used or investigated for photon generation is amazing. Let's mention a few of these recent advances:

- **GaAs quantum-dot** single photon sources with high-source brightness ensuring high-speed quantum communication[2361]. A record of 175 km distance was broken in 2022 using GaAs/InGaAs quantum dots and a finite key rate of 13 kbps over 100 km[2362].
- **AlGaAs sources** with spontaneous parametric down-conversion (SPDC) enable the creation of polarization and/or frequency entangled sources of photons at telecom wavelengths[2363].
- **InAs** quantum dots embedded in GaAs with a conversion to telecom wavelength at 1550 nm generating indistinguishable photons[2364].
- **Silicon with carbon atoms defects** creating optical wavelengths photon at 1279 nm[2365].
- **Vanadium defects in silicon-carbide** operating between 100mK and 3K with the benefit from a relative long stability[2366].
- **Si-based SPDC** to create pairs of entangled photons[2367].

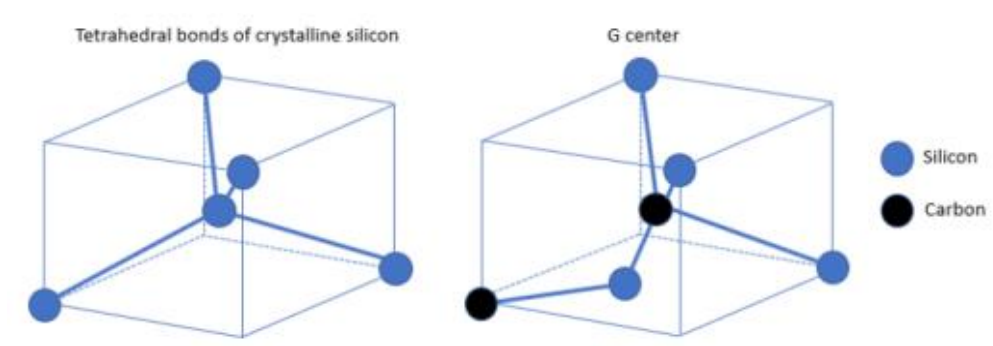

*Figure 754: a G center and its two carbon atoms.*

[2361] See Enhancing quantum cryptography with quantum dot single-photon sources by Mathieu Bozzio et al, April 2022 (37 pages).

[2362] See Single-emitter quantum key distribution over 175 km of fiber with optimised finite key rates by Christopher L. Morrison et al, September 2022 (9 pages).

[2363] See On-chip generation of hybrid polarization-frequency entangled biphoton states by S. Francesconi, Sara Ducci, Perola Milman et al, MPQ and C2N, July 2022 (10 pages) and Broadband biphoton generation and polarization splitting in a monolithic AlGaAs chip by Félicien Appas, Sara Ducci et al, August 2022 (16 pages).

[2364] See A Pure and indistinguishable single-photon source at telecommunication wavelength by Beatrice Da Lio et al, January 2022 (7 pages).

[2365] See Single G centers in silicon fabricated by co-implantation with carbon and proton by Yoann Baron, Anais Dréau et al, April 2022 (5 pages). The electronically active G center is a complex of two substitutional carbon atoms and an interstitial silicon atom. See also the review paper on silicon defects A bright future for silicon in quantum technologies by Mario Khoury and Marco Abbarchi, 2022 (12 pages). Another intrinsic silicon defect resulting from irradiation of Si crystals is also studied, aka the W-center. See Detection of single W-centers in silicon by Yoann Baron, Jean-Michel Gérard, Vincent Jacques, Isabelle Robert-Philip, Anais Dréau et al, April 2022 (10 pages). See also Wafer-scale nanofabrication of telecom single-photon emitters in silicon by M. Hollenbach et al, April 2022 (20 pages) that covers the controllable fabrication of single G and W centers in silicon wafers using focused ion beams (FIB) with a probability exceeding 50%.

[2366] See Vanadium in Silicon Carbide: Telecom-ready spin centres with long relaxation lifetimes and hyperfine-resolved optical transitions by T. Astner et al, June 2022 (9 pages).

[2367] See Near perfect two-photon interference out a down-converter on a silicon photonic chip by Romain Dalidet, Sébastien Tanzilli, Camille-Sophie Brès et al, InPhyNi at Nice in France, EPFL and Ligentec. February 2022 (8 pages).

- **Silicon-nitride MRR** (microring resonator) that are able to generate 8 pairs of heralded entangled photons directly in the telecom wavelengths around 1550 nm [2368]. Microring resonators are tiny optical waveguides looped back onto themselves in circle or spiral, usually implemented in silicon semiconductors. They enable interference phenomena, the creation of delay lines, and the likes[2369].
- **CV-QKD squeezed states** preparation with off-the-shelf telecom equipment[2370].

Then, we have photon detectors and counters. They use various techniques depending on the QKD protocol used (entangled-based or not, etc.).

**SPD** (single-photon detectors) are used after the photons traverse a polarization filter. The main variants are:

- **APD**, avalanche photodiodes.
- **SPAD,** single photon avalanche photodiode which can detect single photons.
- **SNSPD**, superconducting nanowire single-photon detector, which require <4K cooling[2371].
- Silicon-based **VLPC** (visible light photon counter).

### QKD nodes and repeaters

The range of QKD transmission over fiber has improved with only 30 cm in 1989 (IBM with Charles Bennett), 1100 m at the University of Geneva in 1993, then 23 km in 1995 with the BB84 protocol, all via fiber optics. China researchers created a 404 km QKD fiber connection without a repeater in 2016[2372], then extended this record in 2020 to 509 km of transmission without repeater[2373] and to 511 km using the TF-QKD protocol[2374]. The technique was improved in 2020 by a mix of British and American researchers to reach 600 km[2375].

The record was broken in 2022 with 833 km as shown in Figure 755 [2376]. In Austria, an entanglement distribution was achieved over 248 km in 2022[2377].At these large distances, the error rates are so high that it becomes impractical. The key rates are very low, getting under $10^{-7}$ for distances larger than 400 km. There is even a upper bound for these key rates, $-\log_2(1-\eta)$ with $\eta$ being the transmissivity of the lossy quantum channel, whatever the protocol[2378].

---

[2368] See High-quality multi-wavelength quantum light sources on silicon nitride micro-ring chip by Yun-Ru Fan et al, September 2022 (13 pages). Works with silicon chipsets and at ambient temperature. Creates 8 pairs of frequency multiplexed heralded entangled photons.

[2369] See the review paper Silicon microring resonators by Wim Bogaerts et al, Laser & Photonics Review, 2012 (27 pages).

[2370] See Plug-&-play generation of non-Gaussian states of light at a telecom wavelength by Mohamed Faouzi Melalkia, Sébastien Tanzilli, Virginia D'Auria et al, May 2022 (6 pages).

[2371] See an example with Heterogeneously integrated, superconducting silicon-photonic platform for measurement-device-independent quantum key distribution by Xiaodong Zheng et al, October 2021 (8 pages).

[2372] Documented in Measurement device independent quantum key distribution over 404 km optical fiber, 2016 (15 pages).

[2373] Seer Study achieves a new record fiber QKD transmission distance of over 509 km by Ingrid Fadelli, March 2020.

[2374] See Twin-Field Quantum Key Distribution over 511 km Optical Fiber Linking two Distant Metropolitans by Jiu-Peng Chen, Jian-Wei Pan et al, January 2021 (32 pages).

[2375] See 600-km repeater-like quantum communications with dual-band stabilization by Mirko Pittalugal et al, 2020 (14 pages).

[2376] See Twin-field quantum key distribution over 830-km fibre by Shuang Wang et al, Nature, January 2022 with a 140 dB loss!

[2377] See Continuous entanglement distribution over a transnational 248 km fibre link by Sebastian Philipp Neumann et al, March 2022 (23 pages).

[2378] See Fundamental Limits of Repeaterless Quantum Communications by Stefano Pirandola et al, 2017 (61 pages).

Quantum channels used for QKD are subject to noise and leaks. Transmitting a useful photon requires several trials and its number grows exponentially with distance. And when the photon arrives at destination, its state fidelity also decreases exponentially with distance. As a result, for large distances, we need nodes and/or repeaters. They must guarantee a good key rate and fidelity and be tolerant to errors. They are essential for distributing quantum keys over long distances, beyond 80 km[2379]. QKD networks also use classical optical switches using frequency multiplexing. It enables the routing of QKD signals from one emitter to different receivers, but once at a time. They also use SDN (software defined network) to dynamically configure the networks and their nodes, particularly trusted nodes.

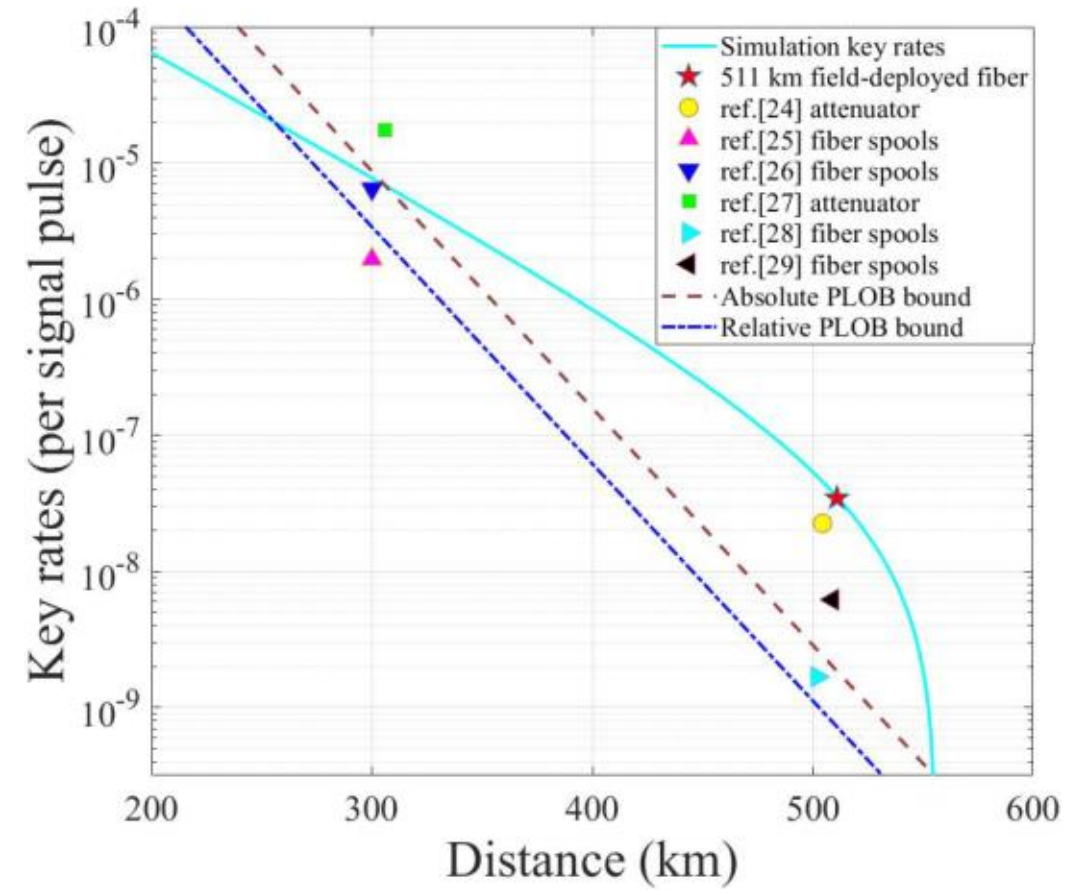

*Figure 755: relationship of QKD keyrates and distance without repeaters. Source: See [Twin-Field Quantum Key Distribution over 511 km Optical Fiber Linking two Distant Metropolitans](#) by Jiu-Peng Chen, Jian-Wei Pan et al, January 2021 (32 pages).*

There are three main kinds of nodes and repeaters:

**Trusted node relays** where keys must be revealed classically in the intermediate stations, thus the need for it to be in "trusted" locations. These nodes are mostly used with the BB84 protocol but can also work with entanglement-based QKDs. They can be implemented to create an arbitrary distance QKD connection. This is the dominant solution. It is used in the 2,000 km Beijing-Shanghai QKD line and in the various EU OpenQKD projects.

**Untrusted node relays** are safer than trusted node relays work but work only with the family of entanglement-based QKD protocols like MDI-QKD and for relatively short distance. It avoids security loopholes at the measurement side and allows the relay to be controlled by an eavesdropper without endangering the security of the shared keys. Complex networks can contain both trusted and untrusted nodes depending on their location and physical safety.

**Quantum repeaters** which are safer than trusted node repeaters. They use entanglement swapping techniques which keep the key sharing link quantum from end-to-end between Alice and Bob. They do not implement any measurement or cloning but, instead, quantum states purification which consists in keeping trusted entangled pairs selected out of many imperfect pairs. They must usually be equipped with some form of quantum memory to propagate the state of the photons to be transmitted through entanglement swapping[2380]. These quantum memories are still unproven and a field of fundamental research. These are different kinds of quantum memories than the ones that needed for quantum computing and that we covered in the quantum memory part starting page 244. These quantum memories for quantum repeaters need a single qubit per link.

[2379] Knowing that the record distance for quantum telecommunication without repeaters is 509 km as we have already seen. See also Viewpoint: Record Distance for Quantum Cryptography by Marco Lucamarini, Toshiba & Cambridge, November 2018 and Recent progress on Measurement-Device-Independent (MDI) Quantum Key Distribution (QKD) by Marco Lucamarini, 2018 (71 slides).

[2380] See Quantum Nodes for Quantum Repeaters by Hugues de Riedmatten, ICFO, January 2021 (60 slides).

It can be implemented with cold atoms[2381], trapped ions[2382], rare earth doped crystals[2383] and NV centers[2384], which can also be arranged in arrays[2385] and use all-photonic quantum processes (APQR)[2386].

These repeaters technologies are still at basic research stage and with some limitations[2387]. The DLCZ protocol created by Harvard, Austria and China scientists in 2001 made it possible to improve entanglement sharing on lossy communication channels and has been continuously improved since then[2388]. It is based on using clouds of identical atoms instead of individual atoms, beam splitters and single-photon detectors with moderate efficiencies with a communication efficiency that scales polynomially with distance.

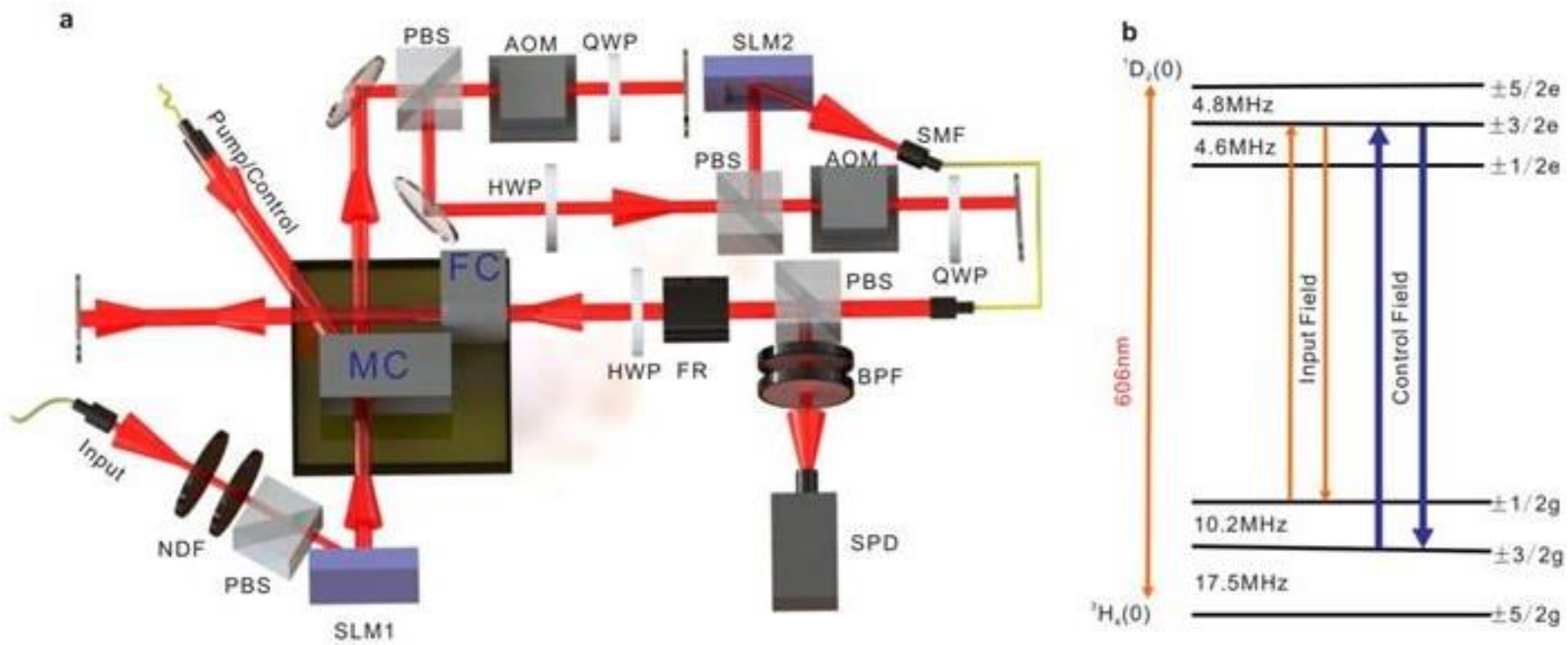

*Figure 756: schematic for an atomic based quantum memory for a quantum repeater. Source: Multiplexed storage and real-time manipulation based on a multiple-degree-of-freedom quantum memory, by Tian-Shu Yang et al, China CAS, 2018 (9 pages).*

---

[2381] See Entangling single atoms over 33 km telecom fibre by Tim van Leent et al, Nature, July 2022 (16 pages).

[2382] See the thesis A memory-based quantum network node with a trapped ion in an optical fibre cavity by Pascal Kobel, 2021 (222 pages). It uses ytterbium ions which have coherence time exceeding the second range and transitions in the microwave band. See also A telecom-wavelength quantum repeater node based on a trapped ion processor by Victor Krutyanskiy, Marco Canteri, Martin Meraner, James Bate, Vojtech Krcmarsky, Josef Schupp, Nicolas Sangouard and Ben P. Lanyon, October 2022 (31 pages),

[2383] See Storage of photonic time-bin qubits for up to 20 ms in a rare-earth doped crystal by Antonio Ortu, Adrian Holzäpfel, Jean Etesse and Mikael Afzelius, npj, March 2022 (7 pages) describes a 20 ms quantum memory using crystals doped with europium, operating near 0K, On-demand storage of photonic qubits at telecom wavelengths by Duan-Cheng Liu et al, January 2022 (11 pages) which uses erbium-yttrium crystals, Storage and analysis of light-matter entanglement in a fiber-integrated system by Jelena v. Rakonjac et al, Science Advances, July 2022 (6 pages) with praseodymium and yttrium, On-demand Integrated Quantum Memory for Polarization Qubits by Tian-Xiang Zhu et al, January 2022 (20 pages) with europium and yttrium, Remote distribution of non-classical correlations over 1250 modes between a telecom photon and a $^{171}Yb^{3+}$ $Y_2SiO_5$ crystal by Moritz Businger et al, Université de Genève, Chimie ParisTech and Sorbonne Université, May 2022 (8 pages) that uses ytterbium and yttrium and Multiplexed storage and real-time manipulation based on a multiple-degree-of-freedom quantum memory, by Tian-Shu Yang et al, China CAS, 2018 (9 pages).

[2384] See Proposal for room-temperature quantum repeaters with nitrogen-vacancy centers and optomechanics by Jia-Wei Ji et al, University of Calgary, March 2022 (20 pages).

[2385] See Electric-Field Programmable Spin Arrays for Scalable Quantum Repeaters by Hanfeng Wang et al, April 2022 (12 pages).

[2386] See Loss-tolerant all-photonic quantum repeater with generalized Shor code by Rui Zhang et al, March 2022 (8 pages).

[2387] See Tutorial on quantum repeaters by Rodney Van Meter and Tracy Northup, 2019 (178 slides), Overcoming the rate-distance limit of quantum key distribution without quantum repeaters, 2018 (5 pages), An Information-Theoretic Framework for Quantum Repeaters by Roberto Ferrara, 2018 (144 pages) and Quantum Internet Protocol Stack: a Comprehensive Survey by Jessica Illiano et al, February 2022 (25 pages) which explains how entanglement is used as a resource in quantum Internet and how it is handled in quantum repeaters. Quantum repeaters, namely, devices implementing the physical process called entanglement swapping and perform a BSM (Bell state measurement). It then covers higher level quantum Internet protocols proposals.

[2388] See Long-distance quantum communication with atomic ensembles and linear optics by Lu-Ming Duan, Mikhail Lukin, Juan Ignacio Cirac and Peter Zoller, Nature, May 2001 (11 pages).

In July 2019, Chinese researchers announced that they succeeded using a photonic repeater technology based on 12-photon interferometers without any quantum memory, encoding a qubit in a cluster state and using error correction in repeaters [2389]. In mid-2019, other Chinese researchers experimented the teleportation of qutrits, allowing a transmission of more information per photons[2390]. This could be used to increase the rate of QKD key transmission.

### Securing QKD

QKD is not the solution-that-fixes-all-problems. It can be subject to jamming, denial of services and various attacks which we cover later. Its safety also depends on the security of both ends of telecommunication like with any other solution.

Securing a chain depends on its weakest links and here it is the transmitters and receivers before they even exchange via a QKD. Furthermore, QKDs are not a panacea because they depend on a point-to-point link and not on a routing technique that allows several paths to be used.

| Attack | Target component | Tested system |
|---|---|---|
| **Distinguishability of decoy states**<br>A. Huang *et al.,* Phys. Rev. A **98**, 012330 (2018) | laser in Alice | 3 research systems |
| **Intersymbol interference**<br>K. Yoshino *et al.,* poster at QCrypt (2016) | intensity modulator in Alice | research system |
| **Laser damage**<br>V. Makarov *et al.,* Phys. Rev. A **94**, 030302 (2016); A. Huang *et al.,* poster at QCrypt (2018) | any | 5 commercial & 1 research systems |
| **Spatial efficiency mismatch**<br>M. Rau *et al.,* IEEE J. Sel. Top. Quantum Electron. **21**, 6600905 (2015); S. Sajeed *et al.,* Phys. Rev. A **91**, 062301 (2015) | receiver optics | 2 research systems |
| **Pulse energy calibration**<br>S. Sajeed *et al.,* Phys. Rev. A **91**, 032326 (2015) | classical watchdog detector | ID Quantique |
| **Trojan-horse**<br>I. Khan *et al.,* presentation at QCrypt (2014) | phase modulator in Alice | SeQureNet |
| **Trojan-horse**<br>N. Jain *et al.,* New J. Phys. **16**, 123030 (2014); S. Sajeed *et al.,* Sci. Rep. **7**, 8403 (2017) | phase modulator in Bob | ID Quantique |
| **Detector saturation**<br>H. Qin, R. Kumar, R. Alleaume, Proc. SPIE 88990N (2013) | homodyne detector | SeQureNet |
| **Shot-noise calibration**<br>P. Jouguet, S. Kunz-Jacques, E. Diamanti, Phys. Rev. A **87**, 062313 (2013) | classical sync detector | SeQureNet |
| **Wavelength-selected PNS**<br>M.-S. Jiang, S.-H. Sun, C.-Y. Li, L.-M. Liang, Phys. Rev. A **86**, 032310 (2012) | intensity modulator | (theory) |
| **Multi-wavelength**<br>H.-W. Li *et al.,* Phys. Rev. A **84**, 062308 (2011) | beamsplitter | research system |
| **Deadtime**<br>H. Weier *et al.,* New J. Phys. **13**, 073024 (2011) | single-photon detector | research system |
| **Channel calibration**<br>N. Jain *et al.,* Phys. Rev. Lett. **107**, 110501 (2011) | single-photon detector | ID Quantique |
| **Faraday-mirror**<br>S.-H. Sun, M.-S. Jiang, L.-M. Liang, Phys. Rev. A **83**, 062331 (2011) | Faraday mirror | (theory) |
| **Detector control**<br>I. Gerhardt *et al.,* Nat. Commun. **2**, 349 (2011); L. Lydersen *et al.,* Nat. Photonics **4**, 686 (2010) | single-photon detector | ID Quantique, MagiQ, research systems |

*Figure 757: QKD sources of vulnerabilities. Source: QKD Measurement Devices Independent by Joshua Slater, 2014 (83 slides).*

This could lead to a form of denial of service by blocking the used physical communication, but rerouting techniques are investigated[2391]. The table in Figure 757 lists a whole bunch of vulnerabilities in the QKD, many of which having since been removed [2392].

[2389] See Scientists Firstly Realize All-photonic Quantum Repeater, July 2019 and Experimental quantum repeater without quantum memory by Zheng-Da Li et al, 2019 (12 pages).

[2390] See Qutrits experiments are a first in quantum teleportation by Daniel Garisto in Scientific American, August 2019, which refers to Experimental multi-level quantum teleportation by Xiao-Min Hu et al, April 2019 (12 pages) and Quantum teleportation in high dimensions by Yi-Han Luo, June 2019 (23 pages).

[2391] On QKD vulnerabilities and methods to avoid them, see QKD Measurement Devices Independent by Joshua Slater, 2014 (83 slides).

[2392] See Certification of cryptographic tools by Vadim Makarov from Quantum Hacking Lab in Moscow, 2019 (15 slides).

All the side-channel attacks from the table in Figure 757 that can be typically fixed with counter-measures. But the more fundamental issue of device independence, linked to the need of loophole free Bell test will be very difficult to implement practically. Work is very active in the area[2393].

Cryptography is fascinating for the speed at which security devices can be broken by researchers before they are deployed en masse. Thus, QKDs would be vulnerable due to an implementation vulnerability associated with Bell's theorem that can be handled with better quality detectors[2394]. It's a never-ending race!

### QKD and Blockchain

Another example is this project to use QKD to secure a Blockchain. This is obviously delicate to deploy end-to-end on a large scale. Indeed, Blockchain users don't have a satellite link in the mountains or a secured fiber on hand, even when they are mobile.

But so be it. This is the proposal of Evgeny Kiktenko of the Russian Quantum Center in Moscow[2395] and Del Rajan and Matt Visser of Victoria University of Wellington in New Zealand[2396]. Why exactly is not all data transmitted protected in the same way as the QKD? It seems at least to be limited by the low bitrate of existing QKDs. Still, JPMorgan Chase, Toshiba and Ciena are piloting a QKD network of 100 km mixed fibers (handling both QKD and classical data distribution) to secure "mission-critical Blockchain application"[2397]. If it is so critical and requires a proprietary network, why does it need a Blockchain in the first place?

Other theoretical architectures have been proposed in India that would rely on quantum computing to improve its security[2398]. There are even proposals for a sort of quantum Bitcoin and smart contracts coming from Israel[2399].

When you mix a complicated system with a couple others that are as complicated, it doesn't make it simpler to grasp. So, I'll pass on.

### QKD over 5G

You may hear about plans to deploy QKD security in 5G networks. Of course, it doesn't deal with the radio portion of 5G and with your smartphone, but only about securing the backbone landline fiber networks of telecom operators[2400].

---

[2393] See Cryptographic Security Concerns on Timestamp Sharing via Public Channel in Quantum Key Distribution Systems by Melis Pahali et al, March 2022 (6 pages) and Improved Finite-Key Security Analysis of Quantum Key Distribution Against Trojan-Horse Attacks by Alvaro Navarrete and Marcos Curty, February 2022 (18 pages).

[2394] It is documented by Jonathan Jogenfors in Breaking the Unbreakable Exploiting Loopholes in Bell's Theorem to Hack Quantum Cryptography, 2017 (254 pages).

[2395] Documented in First Quantum-Secured Blockchain Technology Tested in Moscow, June 2017.

[2396] In Quantum Blockchain using entanglement in time, 2018 (5 pages).

[2397] See JPMorgan Chase, Toshiba and Ciena Build the First Quantum Key Distribution Network Used to Secure Mission-Critical Blockchain Application, February 2022. See also DV-QKD Coexistence With 1.6 Tbps Classical Channels Over Hollow Core Fibre by Obada Alia et al, March 2022 (7 pages) which documents the coexistence of QKD and classical communication on the same fiber.

[2398] See Quantum blockchain using weighted hypergraph states by Shreya Banerjee et al, Physical Review Research, 2020 (7 pages) and Quantum Blockchain based on Dimensional Lifting Generalized Gram-Schmidt Procedure by Kumar Nilesh and P. K. Panigrahi, January 2022 (16 pages). See The Next Generation of Blockchain: Quantum Blockchain Networks by Manan Narang from OneQuantum, March 2022, which confuses "quantum computing" and "quantum cryptography" and PQC to secure Blockchains.

[2399] See Quantum Prudent Contracts with Applications to Bitcoin by Or Sattath, April 2022 (49 pages).

[2400] See Quantum Key Distribution for 5G Networks: A Review, State of Art and Future Directions by Mohd Hirzi Adnan, Zuriati Ahmad Zukarnain and Nur Ziadah Harun, 2022 (28 pages).

**Market and standards**

What about the size of the QKD market? **Inside Quantum Technology** (a UK analyst company) made an estimate with a first $1B dollars reached in 2024, then an exponential growth leading to $7B in 2028[2401]. These are simplistic exponential growth curves, as usual. We'll see.

China is very active in defining a set of QKD standards[2402]. The ITU is also working on QKD standards[2403]. Europe is represented in the standardization work carried out at ISO, IEEE, ETSI and CEN-CENELEC, the European Committee for Standardization in Electronics and Electrotechnology.

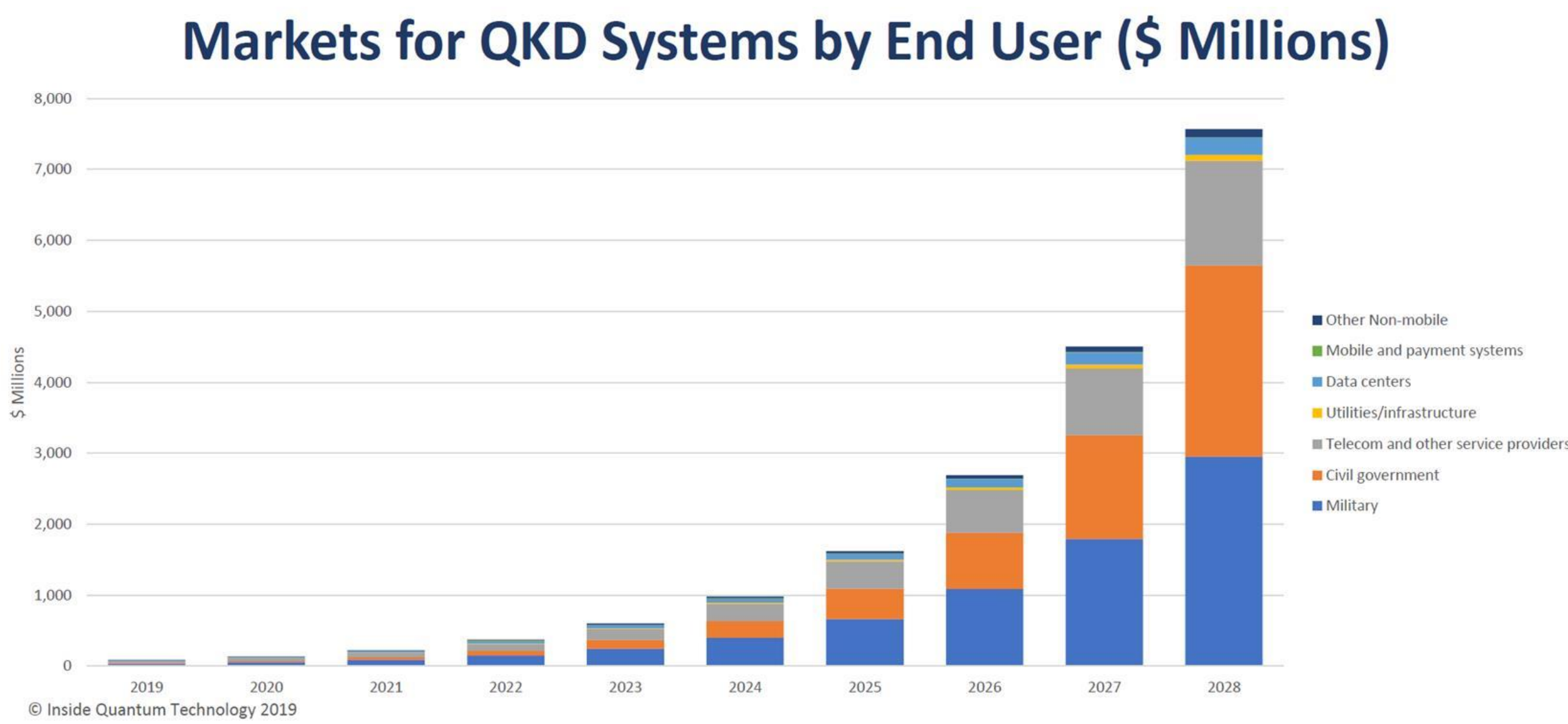

*Figure 758: Inside Quantum Technology's QKD market assessment as done in 2019. Source: The Future of the Quantum Internet A Commercialization Perspective by Lawrence Gasman, June 2019 (11 slides).*

# Post-quantum cryptography

A physical protection of symmetric key transmission is not easily applicable in a generalized way, if only because it requires some optical link (direct free to air or by optical fiber) between transmitters and receivers. This, for example, does not work with radio links like with smartphones.

So, cybersecurity also requires the creation of cryptography systems capable of resisting the onslaught of quantum computers whether coming from Shor's or Grover's algorithms. Breaking encrypted messages - without private keys - should be an NP-Complete or NP-Hard problem to withstand future quantum assaults.

Post-Quantum Cryptography (PQC) complements Quantum Key Distribution (QKD) for this respect. It is certainly easier to deploy on a large scale because it is independent from the telecommunications infrastructures.

[2401] See The Future of the Quantum Internet A Commercialization Perspective by Lawrence Gasman, Inside Quantum Technology, June 2019 (11 slides) and The Future of the Quantum Internet A Commercialization Perspective by Lawrence Gasman from Inside Quantum Technology, June 2019 (11 slides). Seen in ITU Workshop on Quantum Information Technology for Networks.

[2402] See Introduction of Quantum secured Communication Standardization in CCSA by Zhangchao Ma, June 2019 (16 slides) and An overview of current quantum information technology (QIT) standardization by Wei Qi, June 2019 (13 slides).

[2403] See ITU-T Focus Group on Quantum Information Technology for Networks (FG-QIT4N), 2019.

However, it can be combined by sending PQC public keys over physical QKD links or with using some PQC for authenticating the classical link used for the exchange of the photon measurement basis[2404]. Different PQC systems differ in many parameters and have different trade-offs between signature size, processing speed for encryption and decryption, and public key size.

Let's look at the PQC timeline[2405]:

- **1978**: the first algorithm resistant to quantum computers is created by **Robert McEliece** (details below) even before Richard Feynman even mentioned the idea of creating quantum computers and the creation of both Shor and Grover's algorithms!
- **2003**: the term "post quantum cryptography" (PQC) is created by **Daniel Bernstein**[2406].
- **2006**: the first international **PQCrypto** workshop is held in Belgium to study ways to circumvent quantum computer attacks at a time when you can barely assemble two qubits. The program consists in finding successors to the quantum-resistant public key cryptography algorithms RSA and ECC[2407]. The 12-person program committee includes among others Louis Goubin from the University of Versailles and Phong Nguyen and Christopher Wolf from the ENS. From this first edition, four of the five pillars of the PQC are established with the code-based crypto, lattice codes, hash Lamport signature and multivariate cryptography. The isogenies will arrive later. Two French researchers propose two of these four tracks: Nicolas Sendrier, from Inria, with "Post-quantum code-based cryptography" and Jacques Stern from ENS with "Post-quantum multivariate-quadratic public key schemes" [2408]. These workshops have since been held every one to two years around the world. The 2013 edition took place in Limoges, France.
- **2012**: the **NIST** (National Institute for Standards & Technologies) launches its first projects and a team on PQC.
- **2014**: the **European Union** launches a Horizon 2020 call for projects on PQC. At the same time, ETSI, the European Telecoms Standardization Body, also launches its working group on PQC.
- **2015**: **NIST** organizes its first PQC workshop. ETSI published a reference document on QC[2409]. The NSA woke up and declared that the transition to PQC would become a priority[2410]. The NSA is playing two roles each time: it wants to protect itself and the sensitive communications of the U.S. government with good encryption systems but at the same time maintain the ability to break the codes of standard commercial communications and those from other countries. This relies on the brute force of giant supercomputers and a highly asymmetrical technical resources. In 2015, the European project PQCrypto coordinated by Tanja Lange is launched[2411].
- **2016**: NIST publishes QCP Progress Report (15 pages) and an associated standardization roadmap.

---

2404 See Experimental authentication of quantum key distribution with post-quantum cryptography by Liu-Jun Wang et al, May 2021 (7 pages).

2405 I extracted a piece of it from Quantum cryptanalysis - the catastrophe we know and don't know by Tanja Lange, a researcher from the Netherlands, 2017 (33 slides).

2406 Daniel Bernstein is the author with Johannes Buchmann and Erik Dahmen of the impressive book Post-Quantum Cryptography, 2009 (254 pages) which describes well the challenges of PQC.

2407 The proceedings are in PQCrypto 2006 International Workshop on Post-Quantum Cryptography, May 2006 (254 pages).

2408 Source: Quantum Computing and Cryptography Today by Travis L. Swaim, University of Maryland University College (22 pages).

2409 See Quantum Safe Cryptography and Security (64 pages).

2410 See Commercial national security algorithm suite and quantum computing FAQ IAD (11 pages).

2411 It is documented in Post-Quantum Cryptography for Long-Term Security (10 pages).

- **2017**: marks the end of the PQC standardization proposal submissions to NIST. By the end of 2017, 69 applicants are accepted out of 82, mainly with Euclidean networks (lattice codes) and error correction codes (code based PQC). In the same year, the 8th PQCrypto workshop was held in Utrecht, The Netherlands. These candidates had to meet increasing security levels labelled SL1, SL3 and SL5 which corresponds to key size thresholds. These key sizes are between 30 bytes and 5 KB depending on the PKI/signature and the security level.
- **2019**: 26 candidates are selected by NIST in February to move to the second stage, including 17 candidates for public key encryption solutions and 9 for signatures[2412]. These include three projects involving Worldline, which until 2019 was part of the Atos Group. For its part, Inria (France) was involved in 7 of the 26 selected projects.

**2019 second round candidates,**
**2020 finalists and 2020 alternate candidates**

| | | |
|---|---|---|
| BIKE | LEDAcrypt | Rainbow |
| Classic McEliece | LUOV | ROLLO |
| CRYSTALS-DILITHIUM | MQDSS | Round5 |
| CRYSTALS-KYBER | NewHope | RQC |
| FALCON | NTRU | SABER |
| FrodoKEM | NTRU Prime | SIKE |
| GeMSS | NTS-KEM | SPHINCS+ |
| HQC | Picnic | Three Bears |
| LAC | qTESLA | |

*Figure 759: NIST PQC selection in 2019.*

- **2020**: results of the third round of NIST candidate selection in July, which kept 15 out of the 26 candidates from the previous round[2413]. This selection includes 7 teams that were finalists for this stage and 8 teams that propose lower quality solutions that need to be further evaluated (aka “alternate candidates”). See their list in the tables below[2414]. It must be noted that the NIST challenge embedded some constraints on intellectual property. Strictly said, NIST doesn’t object to the contestants having some patents related to their submitted protocols. But they favor royalty-free ones and IP licensing without compensation, under reasonable terms (RAND) and conditions that are demonstrably free of unfair discrimination. While this could certainly accelerate their adoptions, this may indirectly favor large cybersecurity vendors who already have an existing customer base.

  In Figure 760 and Figure 761 are the participants countries, research teams and vendor organizations per project. It shows in green the solutions that were later selected in July 2022. In red are the two solutions that were found to be defective in 2022.

[2412] See NIST Post-Quantum Cryptography - A Hardware Evaluation Study, 2019 (16 pages), Status Report on the First Round of the NIST Post-Quantum Cryptography Standardization Process, 2019 (27 pages) and Recent Developments in Post Quantum Cryptography by Tsuyoshi Takagi, November 2018 (38 slides).

[2413] See PQC Standardization Process: Third Round Candidate Announcement, July 2020.

[2414] The Bit-flipping Key Encapsulation (BIKE) was codeveloped by Intel. It’s a public key based encryption. Decoding can be done with 1.3 million operations at 110 MHz on an Intel Arria 10 FPGA in 12 ms.

**NIST finalists**

| | finalists | research teams | vendors teams |
|---|---|---|---|
| Public-Key Encryption/KEMs | Classic McEliece | **UK**: U. London, U. Plymouth.<br>**Switzerland**: ETH Zurich.<br>**USA**: U. Illinois & Chicago, U. Florida, Yale.<br>**Europe**: U.Ruhr Bochum, U. Eindhoven, U. Southern, Denmark, MPI, Inria (France).<br>**Taiwan**: Academia Sinica. | Google<br>PQ Solutions<br>PQShield |
| | CRYSTALS-KYBER | **USA**: SRI.<br>**Canada**: U. Waterloo.<br>**Europe**: Radboud U. Netherlands, Ruhr U. Bochum, ENS Lyon. | IBM Research Europe<br>Arm, PQShield<br>NXP Semiconductors |
| | NTRU | **Europe**: Radboud U Netherlands, Eindhoven U.<br>**USA**: Brown U.<br>**Canada**: U. Waterloo. | Qualcomm<br>NTT<br>Algorand, PQShield |
| | SABER | **Europe**: KU Leuven (Belgium).<br>**UK**: Birmingham U. | |
| Digital Signatures | CRYSTALS-DILITHIUM | **USA**: Florida Atlantic U.<br>**Switzerland**: ETH Zurich.<br>**Europe**: CWI Netherlands, Ruhr U. Bochum, MPI, ENS Lyon. | IBM Research Europe<br>Google, PQShield |
| | FALCON | **Europe**: ENS Paris, U. Rennes (France).<br>**USA**: Brown U. | IBM Research<br>PQShield, Qualcomm<br>Ethereum Foundation<br>Thales |
| | Rainbow | **Europe**: FAU Erlangen Nuremberg, U. Versailles.<br>**USA**: Cincinnati U.<br>**Taiwan**: Academia Sinica, National Taiwan U. | |

Grey: 2020 selection
Green: 2022 selection
Red: broken in 2022

(cc) compilation Olivier Ezratty, 2022

*Figure 760: NISQ finalists selection in 2020. In green, the 2022 selection. In red, broken PQC. (cc) Olivier Ezratty, 2022.*

And the alternate candidates:

**NIST alternate candidates**

| | finalists | research teams | vendors teams |
|---|---|---|---|
| Public-Key Encryption/KEMs | BIKE | **USA**: U.Washington, Florida U.<br>**Europe**: U. Limoges, ENAC & U. Toulouse, Inria, U. Bordeaux (France), U. Ruhr Bochum (Germany).<br>**IsraeL**: U. Haifa. | Intel<br>Google<br>IBM<br>**Worldline France** |
| | FrodoKEM | **USA**: U. Michigan. Stanford U.<br>**Netherlands**: CWI.<br>**Canada**: U. Waterloo.<br>**Middle-East**: Ege University (Turkey). | NXP<br>Microsoft Research<br>PQShield |
| | HQC | **France**: ISAE-Supaero, Limoges U., ENAC, U. Toulouse, Toulon U., Bordeaux U.<br>**USA**: Florida U. | Worldline France and Netherlands |
| | NTRU Prime | **Taiwan**: Academia Sinica, National Taiwan U.<br>**Australia**: U. Adelaide.<br>**Europe**: Eindhoven U (Netherlands), Hamburg U. (Germany), Tampere U. (Finland).<br>**USA**: Illinois U. | NXP |
| | SIKE | **USA**: Florida U.<br>**Canada**: Waterloo U., Toronto U.<br>**Europe**: Radboud U. Netherlands, U. Versailles (France). | evolutionQ<br>Amazon<br>Microsoft Research<br>Infosec Global<br>Texas Instruments |
| Digital Signatures | GeMSS | **France**: Inria, University of Versailles and Sorbonne Université. | CryptoNext<br>Orange |
| | Picnic | **USA**: Northwestern U., GeorgiaTech, U. Maryland., Princeton U.<br>**Europe**: Austrian Institute of Technology, TU Graz (Austria), Aarhus U. (Denmark), DTU (Denmark). | Microsoft Research<br>Dfinity |
| | SPINCS+ | **Europe**: U.Ruhr Bochum, KU Leuven, TU Graz, Eindhoven U, Radboud U. | Cisco, Infineon<br>Infosec Global<br>Genua, Taurus |

Grey: 2020 selection
Green: 2022 selection
Red: broken in 2022

(cc) compilation Olivier Ezratty, 2022

*Figure 761: NISQ alternate candidates selection in 2020. In green, the 2022 selection. In red, broken PQC. (cc) Olivier Ezratty, 2022.*

- **2022**: in January, the Biden administration published a Memorandum and Executive Order 14028 asking all Federal administration to prepare a PQC deployment plan in 2022[2415]. It even required the deployment of PQC solutions for stored and transiting data.

---

[2415] See Memorandum on Improving the Cybersecurity of National Security, Department of Defense, and Intelligence Community Systems, January 2022.

- **2022**: in July, NIST published a first final list of 4 validated PQC standards, 1 for a PKI and 3 for a digital signature[2416]. These are in green in the above tables. In the PQC standardized signatures, Falcon is recommended for those applications requiring smaller signatures than the ones generated by CRYSTALS-Dilithium. SPHINCS+ signatures are based on a different scheme, although more complex to implement. They still plan to standardize other PQC signatures. Just before this choice was published, after some delay, one of the 2020 finalist signatures PQC, Rainbow, was broken by IBM Zurich researchers at the first security level SL1 corresponding to small sized keys and, not surprisingly, not selected[2417]. In August 2022, the SIKE PKI was also seemingly broken by researchers from Belgium[2418].

  In July 2022, the NIST NCCoE (National Cybersecurity Center of Excellence), was tasked to work with Federal agencies and industry vendors to speed up the transition to PQC. The vendors are AWS, Cisco, Crypto4A Technologies, Cryptosense, InfoSec Global, ISARA Corporation, Microsoft, Samsung SDS, SandboxAQ, Thales and VMware. These players will provide deployment recommendations and their own software and services solutions. Simultaneously, the CISA (Cybersecurity and Infrastructure Security Agency), a US federal agency within the DHS (Department of Homeland Security) announced the creation of a PQC initiative to assist other federal agencies in the deployment of PQCs. Not surprisingly, many PQC and other security vendors are already providing NIST compliant solutions! They provide some encapsulation mechanisms for third-party and/or open source PQCs in their cybersecurity management tools.

- **2025**: NIST's target date for finalizing PQC standards. Deployments of these standards would begin with the rapid deployment of commercial solutions supporting these standards. Fast, for the simple reason that the candidates are often in the standardization consortia. Some of them are already testing their solutions.

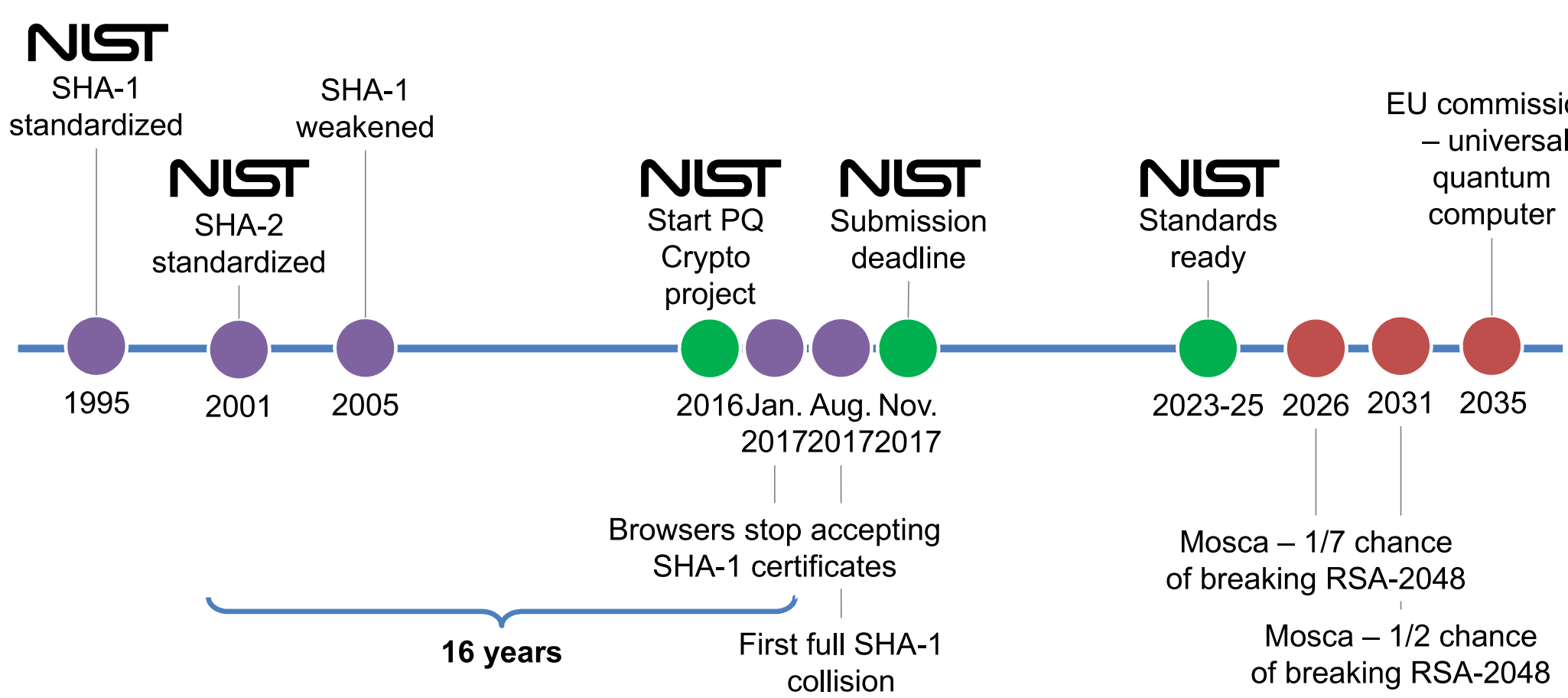

*Figure 762: NISQ PQC standardization planning as of 2019. Source: Introduction to post-quantum cryptography and learning with errors, Douglas Stebila, 2018 (106 slides).*

There are five distinct categories of PQC standards, as follows. I will not be able to technically describe them all except for the first category[2419]. In the last part of this section on cryptography, we will mention the case of some startups that are positioned in this market.

[2416] See NIST Announces First Four Quantum-Resistant Cryptographic Algorithms, NIST, July 2022.

[2417] See NIST PQC Finalists Update: It's Over For The Rainbow by Edlyn Teske. March 2022.

[2418] See Post-quantum encryption contender is taken out by single-core PC and 1 hour by Dan Goodin, ArsTechnica, August 2022, referring to An efficient key recovery attack on SIDH (preliminary version) by Wouter Castryck and Thomas Decru, August 2022 (15 pages). And some good explanations in "Quantum-Safe" Crypto Hacked by 10-Year-Old PC by Charles Q. Cho, August 2022.

[2419] See in particular A Guide to Post-Quantum Cryptography by Ben Perez, October 2018.

**Table 2 - Comparison on encryption schemes (RSA decryption = 1, size in bits, k security strength)**

| Algorithm | KeyGen (time compared to RSA decrypt) | Decryption (time compared to RSA decrypt) | Encryption (time compared to RSA decrypt) | PubKey (key size in bits to achieve 128 bits of security) | PrivateKey (key size in bits to achieve 128 bits of security | Cipher text (size of resulting cipher text) | Time Scaling | Key Scaling |
|---|---|---|---|---|---|---|---|---|
| NTRU | 5 | 0.05 | 0.05 | 4939 | 1398 | 4939 | $k^2$ | $k$ |
| McEliece | 2 | 0.5 | 0.01 | 1537536 | 64861 | 2860 | $k^2$ | $k^2$ |
| Quasi-Cyclic MDPC McEliece | 5 | 0.5 | 0.1 | 9857 | 19714 | 19714 | $k^2$ | $k$ |
| | | | | | | | | |
| RSA | 50 | 1 | 0.01 | 3072 | 24,576 | 3072 | $k^6$ | $k^3$ |
| DH | 0.2 | 0.2 | 0.2 | 3072 | 3238 | 3072 | $k^4$ | $k^3$ |
| ECDH | 0.05 | 0.05 | 0.05 | 256 | 256 | 512 | $k^2$ | $k$ |

**Note: in key scaling, the factor log k is omitted.**

*Figure 763: Comparison of key size of various encryption schemes. Source : Quantum Safe Cryptography and Security; An introduction, benefits, enablers and challenges, ETSI, 2015 (64 pages).*

Established companies are not left out. **IBM** announced in August 2019 a system for archiving information on magnetic bank that integrates post-quantum cryptography[2420]. They use encryption based on Euclidean networks. As it is usually long-term storage, it is necessary to keep the decryption software for the same length of time to avoid ending up with a pile of data that cannot be reused. IBM is also involved in the three consortia that responded to the NIST call for proposals. **Kudelski Security** (Switzerland) is also interested in PQC.

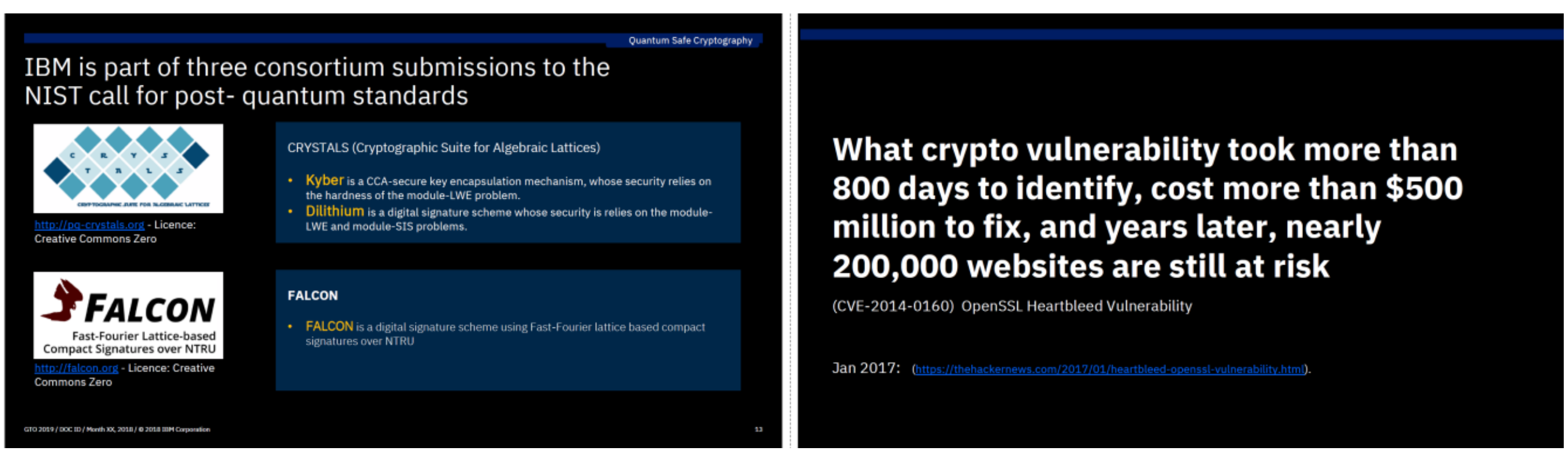

*Figure 764: IBM's stance on cybersecurity. They bet on the right horse given their slides in 2019 presented 3 of the 4 2022 NIST finalists!*

In France, the civil cybersecurity agency **ANSSI** published an information note in May 2020 in which it expressed certain reserves about the QKD[2421]. It highlighted the fact that it does not address a common problem, cannot guarantee perfect inviolability and requires dedicated optical infrastructures. Instead, it recommends focusing on PQC.

[2420] See IBM's quantum-resistant magnetic tape storage is not actually snake oil by Kevin Coldewey in TechCrunch, August 2019.

[2421] See L'avenir des communications sécurisées passe-t-il par la distribution quantique de clés? by ANSSI, May 2020 (6 pages).

This followed a memo of equivalent content from their British **NCSC** counterparts published in April 2020[2422]. A similar publication from the **NSA** was released in October 2020[2423]. It was renewed in 2021 and 2022.

### Code-based cryptography

This cryptographic system invented in 1978 by **Robert McEliece**, long before Shor's algorithm, has since resisted all cryptanalysis attacks, either classical or designed with quantum algorithms. It is the oldest of the PQC codes which was even a PQC before its time. The method consists in multiplying the data to be encrypted, represented as binary vectors (of length k), by a public and static matrix with more columns than rows (k x n), aka a "binary Goppa code".

This multiplication generates a vector larger than the original vector (with n bits). We then add a binary vector which adds random errors to the result but of constant value (vector z in schema with a given number of 1s). It is described as a "*uniformly random word of weight t*". It is a series of random bits containing a fixed number "t" of 1s called a Hamming weight. The public key sent by the receiver to the transmitter is the matrix Ĝ and this number of errors t.

The three matrices having created Ĝ constitute the private key. This matrix Ĝ is the multiplication of three matrices called SGP for "non **S**ingular", "generator matrix / **G**oppa code" and "**P**ermutation matrix". The message decoding uses inverses of matrix S, P and G. This is explained in this diagram. The G matrix is by designed crafted to remove the "t" errors introduced in the encryption phase.

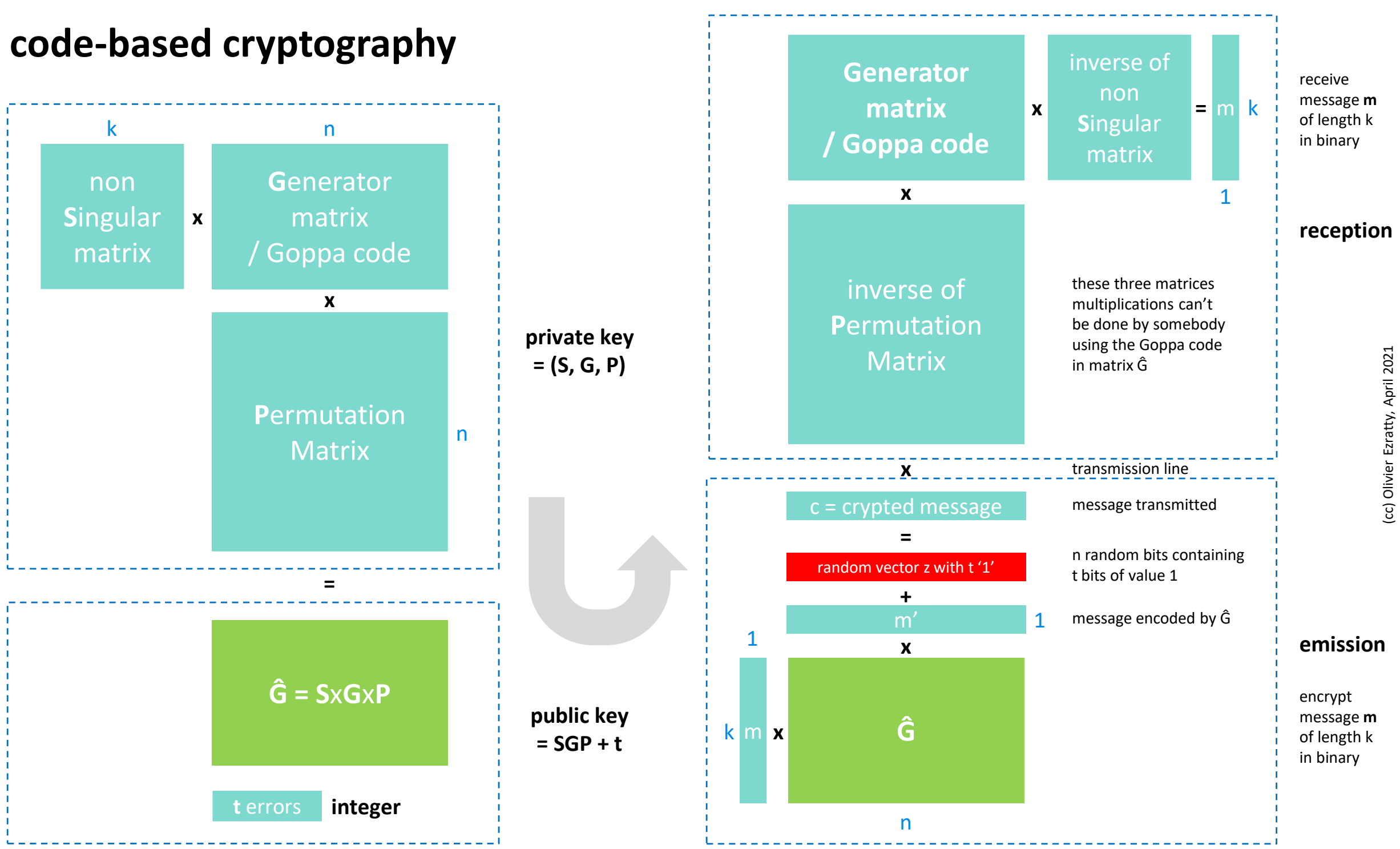

*Figure 765: how a code-based PQC key generation works. It's all about mixing and matching many non-square matrices. (cc) Olivier Ezratty, from various sources. 2021.*

This system generates public keys one hundred times larger than with RSA, of the order of 80 KB. It generates new vulnerabilities if you reduce their size.

[2422] See Quantum Security Technologies, NCSC, March 2020 (4 pages) and a detailed response in Quantum safe cryptography - the big picture - Fact Based Insight by David Shaw, 2020.

[2423] See NSA Cybersecurity Perspectives on Quantum Key Distribution and Quantum Cryptography, NSA, 2020.

The advantage of the PQC category is its good encryption and decryption speed. It can even be accelerated by using a dedicated FPGA chipset[2424].

Breaking this kind of encryption is an NP-Hard problem that is currently inaccessible to quantum computing, even though to resist quantum computing it would still require a fairly large key of at least 1 MB[2425].

### Lattice-based cryptography or Euclidean networks

The technique was proposed in 1996 by Miklos Ajtai, a researcher at IBM, and implemented in a public key system in 2005 by Oded Regev with its LWE (Learning With Errors) system and improved since then by many researchers.

The associated literature is inaccessible for non-specialists. It is not easy to understand how this encryption method works despite the elegance of the diagrams that present the notion of Euclidean network like the one in Figure 766[2426]. Basically, it is a matrix of dots that allows to locate points according to their coordinates according to a mark of different vectors between the public and private keys.

An error is added to the coordinates generated with the public key vector. Only the coordinate vectors of the private key can be used to retrieve the coordinate of the encrypted value. Initially, it suffered from performance problems, but effective solutions appeared such as NTRU, created in 1998 by Jeffrey Hoffstein, Jill Pipher and Joseph Silverman. The method advantage is to use small public keys. Its decryption is an NP-complete problem inaccessible to quantum computing. On the other hand, it is a method protected by many patents, so it is proprietary and potentially expensive[2427].

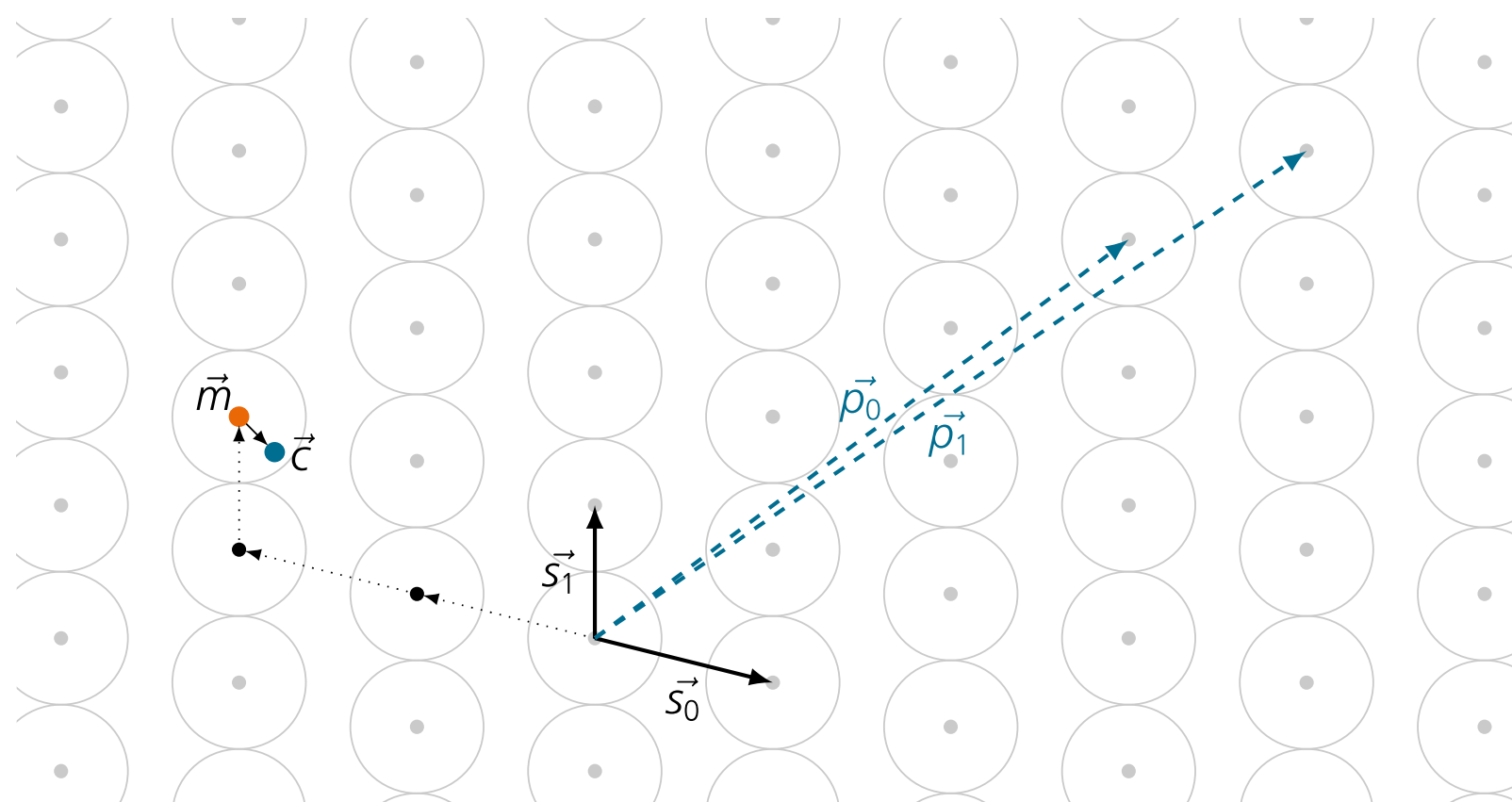

**Figure 3.2:** Example for lattice-based encryption in a two-dimensional lattice: The secret, well-formed base is $\{\vec{s_0}, \vec{s_1}\}$; the public, "scrambled" base is $\{\vec{p_0}, \vec{p_1}\}$. The sender uses $\{\vec{p_0}, \vec{p_1}\}$ to map the message to a lattice point $\vec{m}$ and adds an error vector to obtain the point $\vec{c}$. The point $\vec{c}$ is closer to $\vec{m}$ than to any other lattice point. Therefore, the receiver can use the well-formed secret base $\{\vec{s_0}, \vec{s_1}\}$ to easily recover $\vec{m}$ (dotted vectors); this is a hard computation for an attacker who only has the scrambled base $\{\vec{p_0}, \vec{p_1}\}$. For a secure scheme, the dimension of the lattice must be much higher than 2 as in this example.

*Figure 766: Euclidean network key generation. Source: Practical Post-Quantum Cryptography by Ruben Niederhagen and Michael Waidner, 2017 (31 pages).*

The PQC **New Hope** solution (CECPQ1) which was tested in 2016 for a few months by **Google** in Chrome and is based on Ring-LWE is in this class of methods. Since 2019, they have moved to CECPQ2 which includes a variant of the HRSS key exchange system that is among the bidders in the NIST competition and the selected in the last wave in the NTRU project[2428].

---

[2424] As seen in Code-Based Cryptography for FPGAs by Ruben Niederhagen, 2018 (73 slides).

[2425] The resistance of this method to attacks is documented in Code-Based Cryptography by Tanja Lange, 2016 (38 slides). For more information, see also Code Based Cryptography by Alain Couvreur, 2018 (122 slides) and Some Notes on Code-Based Cryptography, a thesis by Carl Löndahl, 2014 (192 pages).

[2426] See Practical Post-Quantum Cryptography by Ruben Niederhagen and Michael Waidner, 2017 (31 pages).

[2427] For more information, see the thesis Lattice-based cryptography: a practical implementation by Michael Rose, 2011 (103 pages), Lattice-based Cryptography by Daniele Micciancio and Oded Regev, 2008 (33 pages) and the slightly more pedagogical but still incomprehensible Overview of Lattice based Cryptography from Geometric by Leo Ducas, 2017 (53 slides).

[2428] See Experimenting with Post-Quantum Cryptography by Matt Braithwaite, July 2016 and then Google starts CECPQ2, a new postquantum key exchange for TLS, January 2019.

In France, a team from the IRISA-EMSEC laboratory is developing a cryptographic solution based on these Lattice base systems, also named Euclidean networks.

Damien Stehlé is another specialist of the domain, doing research at ENS Lyon. He participated to the creation of CRYSTALS - Kyber, a finalist in 2020 and 2022 of NIST's PQC competition.

### Isogeny-based cryptography

This variant of elliptic curves is even less easy to grasp than all of the above. It is a "*morphism of superimposed group and finite kernel between two elliptic curves*". Piece of cake! The system was proposed in 2006 by Alexander Rostovtsev and Anton Stolbunov and then broken by quantum cryptoanalysis by Andrew Childs, David Jao and Vladimir Soukharev. This led David Jao and Luca De Feo (Inria) to propose in 2011 the use of "super-singular" curves to correct this flaw[2429].

This cryptography is used in Supersingular isogeny Diffie–Hellman key exchange (SIDH).

Software publisher **Cloudflare** has released an open source security solution based on isogenies, CIRCL (Cloudflare Interoperable Reusable Cryptographic Library). It is published on GitHub. Their SIKE key encapsulation solution has been submitted to NIST.

In January 2019, they were among the 17 finalist candidates for public key encryption or key creation solutions[2430]. In 2022, it was broken using a single computer for one hour to 21 hours depending on the key size (from SIKEp434 to SIKEp751).

### Hash-based signatures

This post-quantum cryptography other method also predates the very notion of quantum computer imagined by Richard Feynman in 1982. It is based on the work of **Leslie Lamport** of the SRI in 1979 and her single-use hash-based "signatures". The method was then improved by using hash trees also called Merkle trees to sign several messages. It is based on public keys of reduced size, down to 1 kbits.

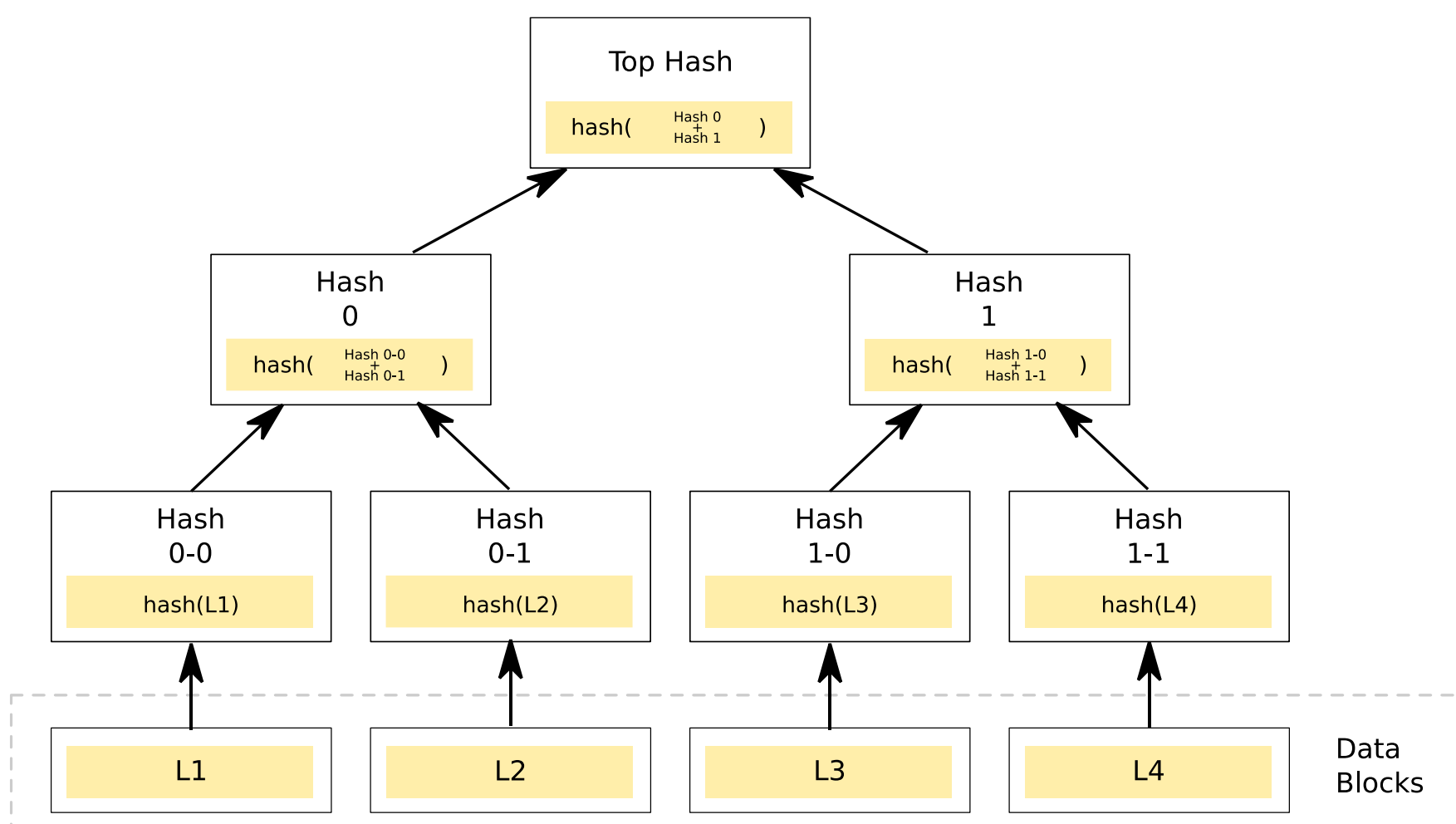

*Figure 767: source: Merkle Tree, Wikipedia.*

This method is mainly used for electronic signature[2431].

[2429] More on this with 20 years of isogeny-based cryptography by Luca De Feo, 2017 (84 slides), An introduction to supersingular isogeny-based cryptography by Craig Costello (Microsoft Research), 2017 (78 slides), Isogeny Graphs in Cryptography by Luca De Feo, 2018 (73 slides) and An introduction to isogeny-based crypto by Chloe Martindale, 2017 (78 slides).

[2430] See Cloudflare wants to protect the internet from quantum computing, June 2019 and Introducing CIRCL: An Advanced Cryptographic Library, June 2019.

[2431] If you are well versed in mathematics and cryptography, see Hash-based Signatures: An Outline for a New Standard (12 pages), Design and implementation of a post-quantum hash-based cryptographic signature scheme by Guillaume Endignoux, 2017 (102 pages) and SPHINCS: practical stateless hash-based signatures, 2015 (30 pages).

**Multivariate polynomial cryptography**

This last group of methods is reminiscent of error correction codes. The public key is a multiplication of several matrices, two of which are linear and one quadratic (with squared values), the three separate matrices constituting the private key used to reconstruct the encrypted message. As a result, the keys are extremely large.

Code breaking these keys is an NP-Hard problem, out of reach of quantum computing. The method dates from 2009 and was obviously then declined in several variants. The public keys are quite large, up to 130 KB (with the HFEBoost variant) [2432]. This encryption method is also rather used for electronic signatures.

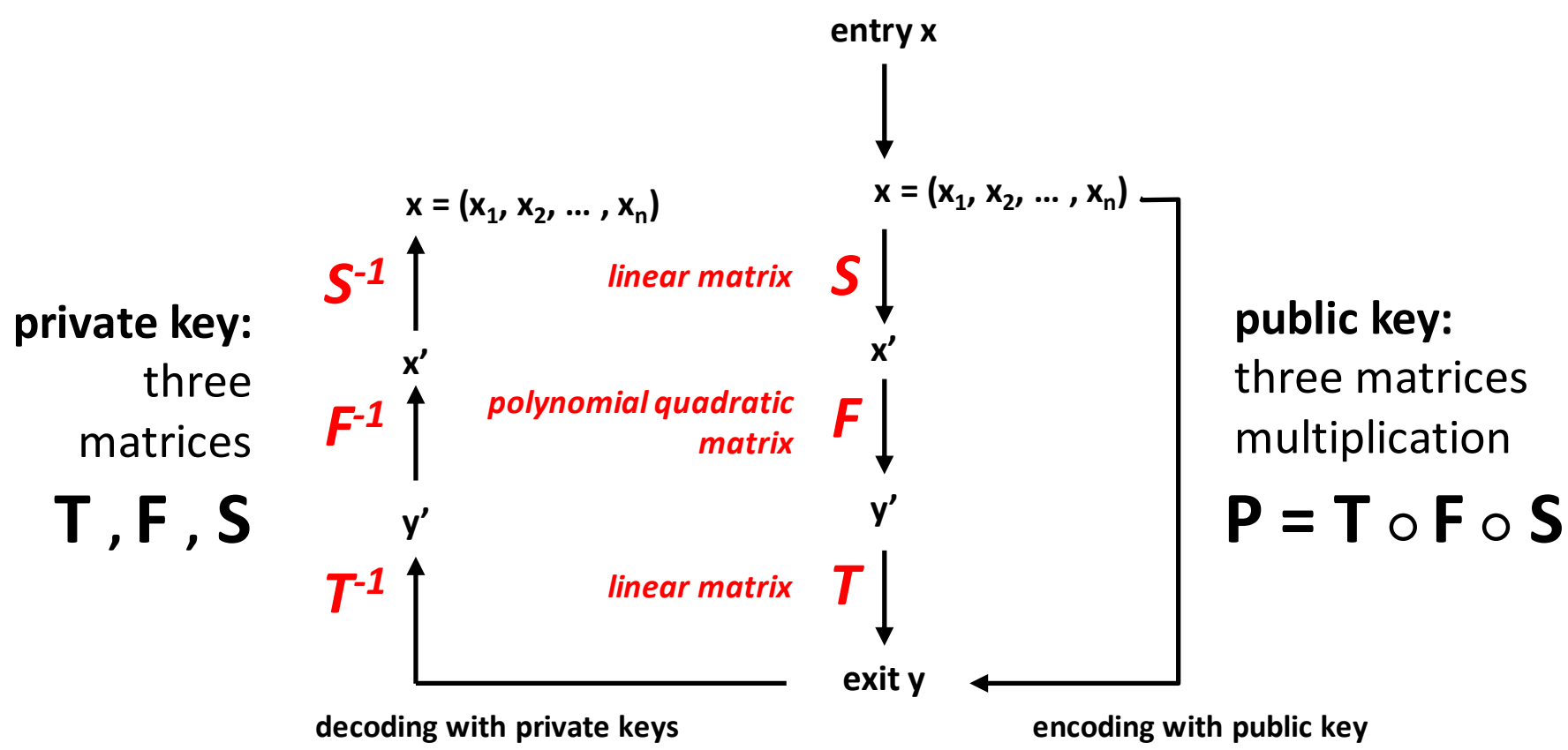

*Figure 768: multivariate polynomial cryptography. Source: (cc) Olivier Ezratty, reconstructed from other sources.*

The Rainbow PQC signature selected by NIST in 2020 as a finalist and broken in March 2022 was in that category.

We could imagine that QKD (physical protection of key distribution) could be combined with PQC (logical protection of encryption against quantum computer decryption). Actually, not really. QKD is rather dedicated to symmetric keys that assume protection of physical communication between correspondents, whereas PQC relies on public keys that do not need to be protected by QKD because their interception (without QKD) would already be useless to hackers.

However, QKD for key exchange can be combined with PQC for authentication and data encryption. QKD requires authentication, which can be provided upstream by PQC. On the other hand, QKD can be redundant with PQC used for key exchange[2433].

# Quantum homomorphic cryptography

Homomorphic cryptography consists in encrypting data that can then pass through a conventional processing in encrypted mode and give an encrypted result that will be decipherable at the end of the processing.

In machine learning and deep learning, this mode of encryption makes it possible to distribute training and inference processing of learning machine models in the cloud without the hacking of the transmitted data revealing the data content that feeds the model or inferences.

The disadvantage of this method is that it does not work with all learning machine models and is very expensive in terms of machine time for data encryption and decryption as well as computation.

Quantum homomorphic encryption is a similar approach for encoding data that will feed a quantum computer in the cloud and then decode the result of the processing.

[2432] Note the contribution of Jacques Stern from the ENS "Post-quantum multivariate-quadratic public key schemes" at PQCRYPTO 2006.

[2433] To learn more about PQC, see in particular Post-quantum cryptography - dealing with the fallout of physics success by Daniel Bernstein and Tanja Lange, 2017 (20 pages).

It is one of the tools for implementing so-called "blind computing" in the cloud, where servers cannot understand and interpret the data they process.

Various algorithms for encrypting quantum gate control programs have been proposed but are not yet commonly used[2434]. Some of the keys can be quantum-transmitted like a QKD. This is one of the conditions to be sure that the server part cannot interpret the processing it performs[2435].

# Quantum interconnect

As indicated at the beginning of this long section, QKD-based quantum cryptography is not the only application of quantum telecommunications[2436]. It is one of its applications. And, as we have already seen earlier in this book, quantum telecommunications are not about transmitting information faster than light[2437].

One application area is the creation of quantum networks linking quantum "endpoints" that are themselves quantum, which could be quantum computers and even quantum sensors[2438]. In the first case, links between quantum computers would make it possible to create distributed computing architectures, following the example of classical distributed computing architectures that exist on the Internet, in data centers and within supercomputers. This is a “scale-out” approach to grow the capacity of quantum computers, as opposed to “scale-in” approaches that consist in creating QPUs with a larger number of qubits. Since this number is limited, scale-out approaches are researched. We don’t know yet if scaling-out is going to be more or less difficult to achieve than scaling-in.

In the sensors case, networked quantum sensors can create a trusted system, improve the sensitivity of quantum sensing or even connecting quantum sensors directly with quantum computers, either for sending some form of quantum data without using conversion to classical data or to improve sensor/computer communication security[2439].

A direct link between quantum computers and quantum telecommunications would bring interesting benefits for implementing secured processing between several parties like a small quantum computer delegating some tasks to a larger one in a secured manner[2440]. This the concept of "blind computing" associated to the BFK protocol created in 2009 by Anne Broadbent, Joe Fitzsimons and Elham Kashefi[2441].

---

[2434] See Classical Homomorphic Encryption for Quantum Circuits by Urmila Mahadev, 2018 (7 pages), Quantum Fully Homomorphic Encryption With Verification, 2017 (30 pages and slides, 28 slides), Quantum Homomorphic Encryption: A Survey, 2017 (11 pages) et Quantum homomorphic encryption for circuits of low T-gate complexity by Anne Broadbent et Stacey Jeffery, 2015 (35 pages).

[2435] As indicated in On the implausibility of classical client blind quantum computing by Scott Aaronson, Elham Kashefi et al, 2017 (43 pages).

[2436] See the excellent Quantum internet: A vision for the road ahead by Stephanie Wehner et al, October 2018 (11 pages).

[2437] Let’s remind at least two key explanations: first, entanglement and non-locality is about having a correlation between two distant quantum objects values when measured sequentially, but this value is random by essence. You can’t set one quantum value at one end (Alice) and then measure it at the other location (Bob’s). You just decide to measure random values at both end that happen to be correlated. On a more practical reason, the teleportation algorithm that can send a qubit state to another location with using entanglement needs two classical communications links. So, we’re stuck with the speed of light. See No, We Still Can't Use Quantum Entanglement To Communicate Faster Than Light by Ethan Siegel, February 2020.

[2438] See Repeater-enhanced distributed quantum sensing based on continuous-variable multipartite entanglement by Yi Xia et al, 2018 (9 pages).

[2439] See one example of quantum sensors interconnect proposal in An elementary quantum network of entangled optical atomic clocks by B. C. Nichol et al, Nature, November 2021-September 2022.

[2440] See Equivalence in delegated quantum computing by Fabian Wiesner, Jens Eisert and Anna Pappa, June 2022 (43 pages).

[2441] See Universal blind quantum computation by Anne Broadbent, Joseph Fitzsimons and Elham Kashefi, 2008 (20 pages) and the associated presentation (25 slides), Blind quantum computing can always be made verifiable by Tomoyuki Morimae, 2018 (5 pages), Experimental Blind Quantum Computing for a Classical Client, 2017 (5 pages) and Blind Quantum Computation by Charles Herder (5 pages).

The principle consists in preparing computation in a quantum way at the starting point and sending it by a quantum link by teleportation to the remote quantum computer. It is a bit the quantum equivalent of the homomorphic encryption used in distributed machine learning.

However, connecting quantum computers is not an easy task. It's not about inputs/outputs or memory and storage sharing like with classical scale-out architectures. It should indeed be possible to convert the qubit state of these machines into quantum states of photons - usually in the infrared range at 1550 nm - for optical transmission. Apart from the photon-based systems case, qubits are most often electrons spins or atoms energy states. Hence the numerous efforts to make conversions between these qubits states and qubits encoded in transmissible photons. And setting up some quantum connectivity is not just about sending one photon from one computer to the other, but to connect them with entanglement resources and use the teleportation algorithm[2442].

So far, there are three known approaches to connect quantum computers: microwaves, photons and shuttling electrons or ions. In that space, there's a clear difference in photon-based solutions which could scale at a large level, leveraging fiber optics telecommunication infrastructures and other options (microwave, shuttling electrons of ions) which are by design "on premise" and won't rely on telecommunications infrastructures[2443].

### Microwaves interconnect

This type of QPUs interconnect is adapted to qubits that are driven by microwaves, so in order of priority, superconducting, silicon spin and to some extent trapped ions and cold atoms qubits.

But so far, it has been investigated mostly with superconducting qubits. This technology is adapted to relatively short range connectivity and could have a high efficiency.

Superconducting qubits belong to the field of circuit quantum electrodynamics (cQED) and are driven by microwave pulses. Microwaves are used for qubit readout so we know how to convert the state of a qubit into microwaves, which are in the 4-8 GHz range. These microwaves can be used for short distance communication between processing units as was first realized by an international team led by the University of Chicago in 2020.

They connected two nodes of three superconducting qubits, each arranged as an entangled GHZ state, and managed this entanglement transmission with microwaves on a distance of one meter on a niobium-titanium coax cable.

The entanglement transmission was done with a fidelity of 65% and 91% for a single qubit transmission. It's a first promising step[2444].

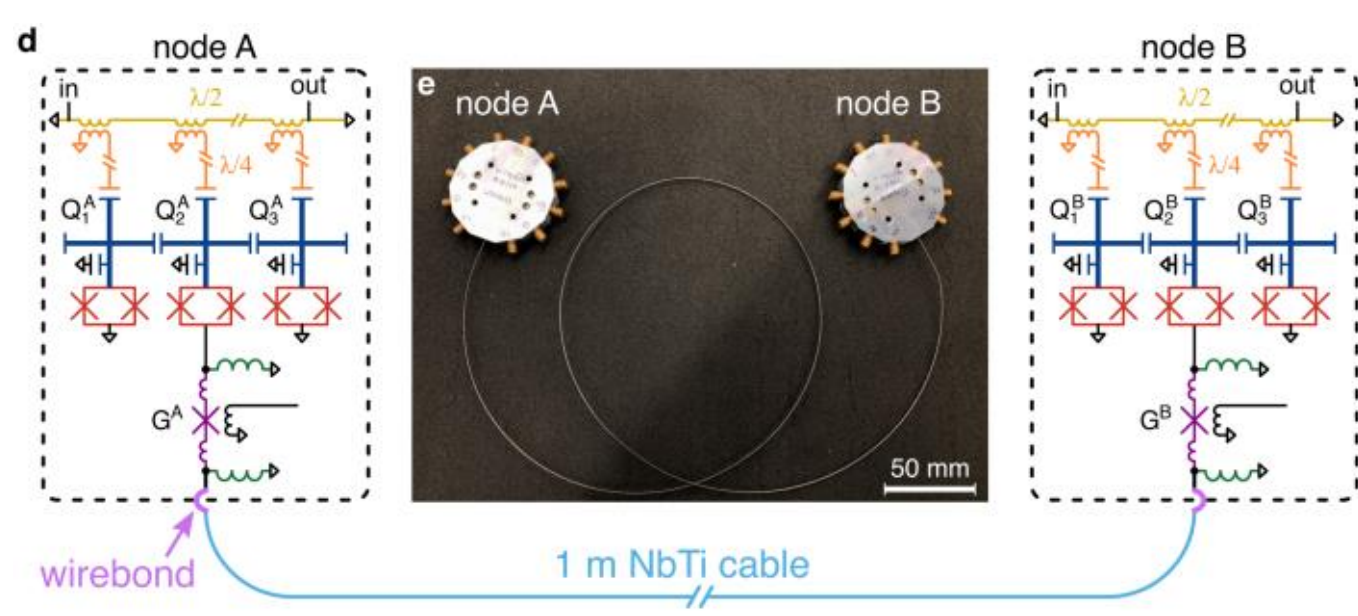

*Figure 769: Source: Deterministic multi-qubit entanglement in a quantum network by Youpeng Zhong, Audrey Bienfait (ENS Lyon), et al, November 2020 on arXiv and February 2021 in Nature (38 pages).*

---

[2442] See Distributed Quantum Computing: A path to large scale quantum computing by Stephen DiAdamo, August 2021.

[2443] See Development of Quantum InterConnects for Next-Generation Information Technologies by David Awschalom, Sophia E. Economou, Dirk Englund, Liang Jiang, Mikhail D. Lukin, Christopher Monroe, Jelena Vučković, Ronald Walsworth et al, 2019 (31 pages) which positions well different QPU interconnect technologies.

[2444] See Deterministic multi-qubit entanglement in a quantum network by Youpeng Zhong, Audrey Bienfait (ENS Lyon), et al, November 2020 on arXiv and February 2021 in Nature (38 pages).

Another team led by Andreas Wallraff from ETH Zurich with the University of Sherbrooke in Canada connected several superconducting units using connected cryogenic systems.

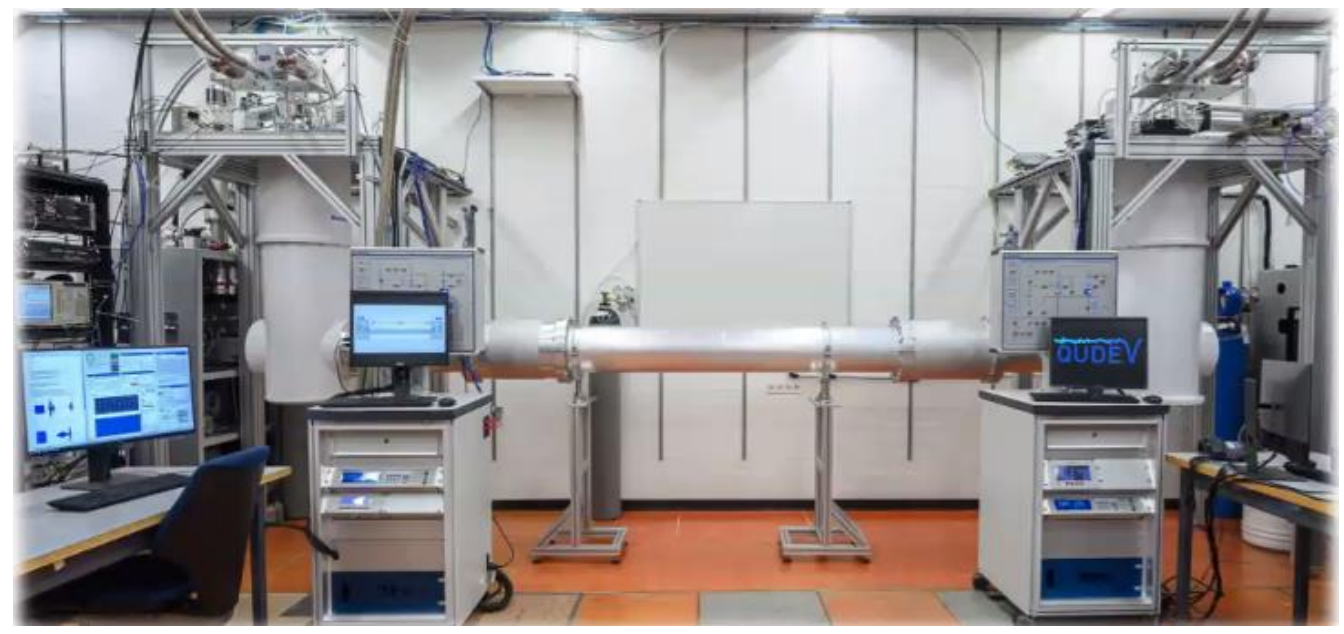

This enabled the connection between these units with microwave waveguides[2445]. The two processing units were separated by a 5 meters cryogenic link where the microwaves are transmitted. I don't know of any superconducting hardware vendor that plans to adopt this architecture.

*Figure 770 above and below, ETH Zurich 5 meter cryogenic microwave link. Source: [Microwave Quantum Link between Superconducting Circuits Housed in Spatially Separated Cryogenic Systems]() by Paul Magnard, Alexandre Blais, Andreas Wallraff et al, PRL, December 2020 (13 pages).*

In another experiment, a team from the Technical University of Munich with colleagues from RIKEN in Japan and Aalto University in Finland tested a microwave entangling connection of 45 cm between two superconducting qubits. They used two microwave parametric amplifiers (JPA) to create a pair of entangled and squeezed microwave photons to connect "Alice" & "Bob" qubits.

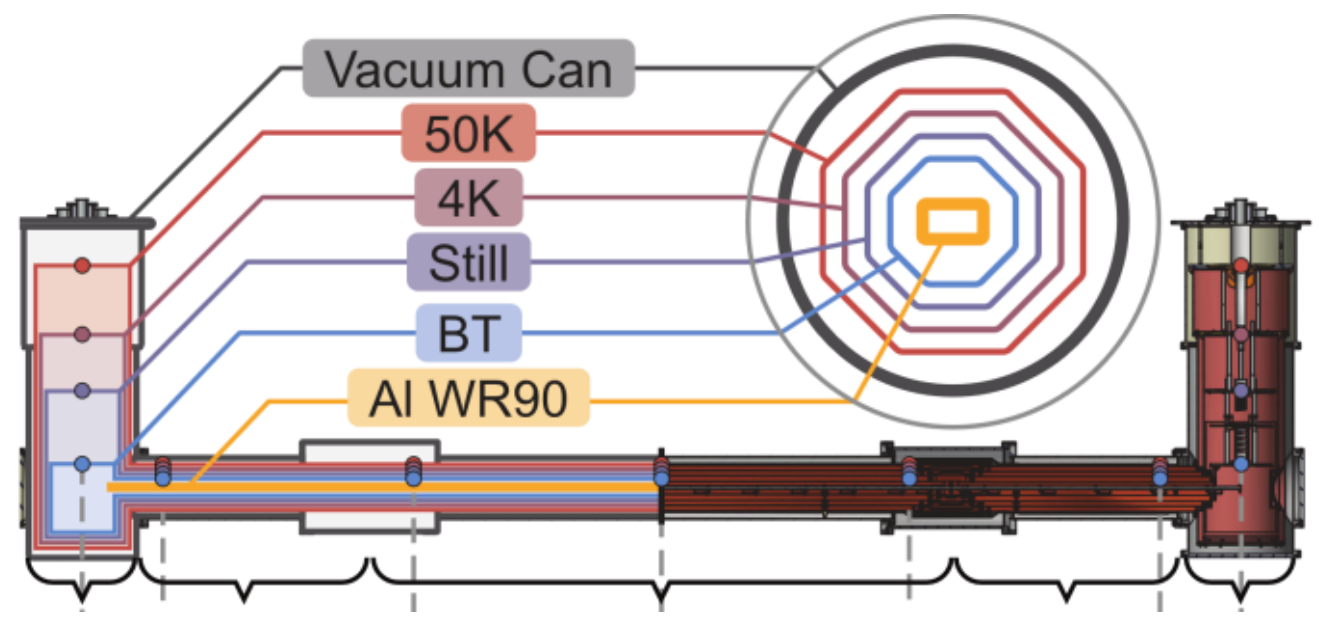

The resulting teleportation fidelity was of 69%. We'll see if that scales well with a growing number of qubits[2446].

Now, what is needed is way more sophisticated. Connecting different qubit chipsets would require to connect at least one qubit of nearby chipsets to each other. This would create a limited connectivity between the chipsets. The example using a simple hexagonal qubit topology typical of IBM's superconducting chipsets would require the connections in red. But this is for 27 qubit chipsets, not thousand qubit chipsets[2447]!

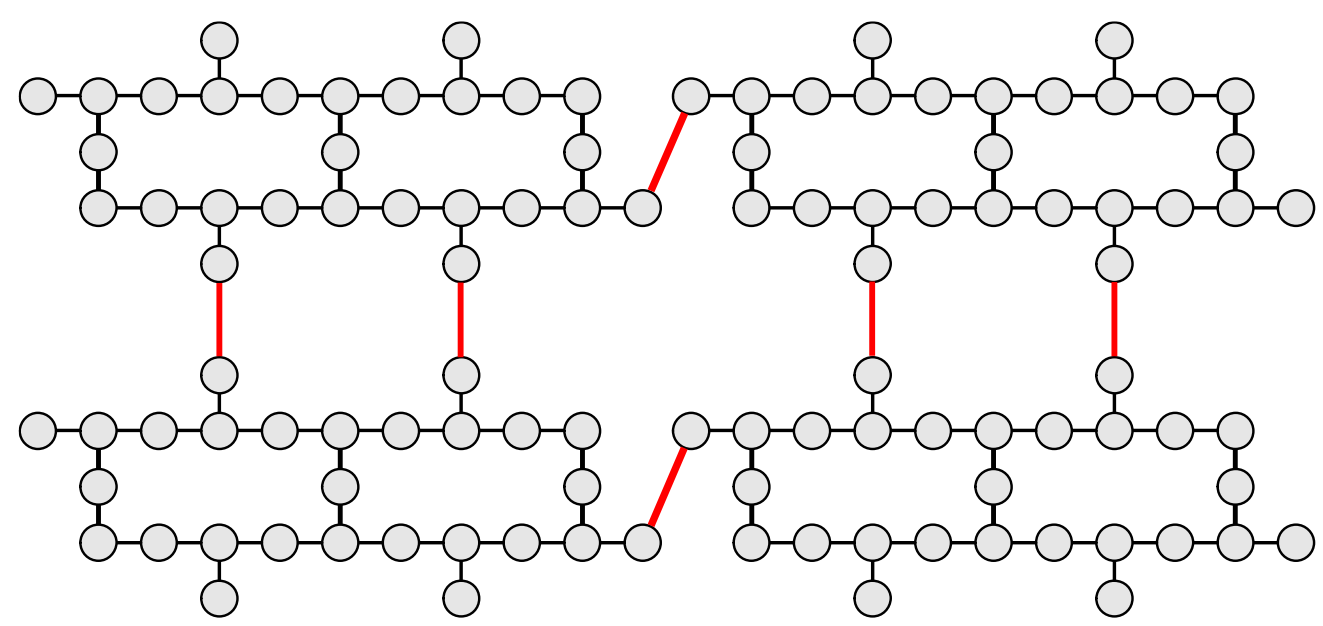

*Figure 771: Source: [Short-Range Microwave Networks to Scale Superconducting Quantum Computation]() by Nicholas LaRacuente et al, January 2022 (22 pages).*

That's also what a team from the Universities of Pittsburgh and Illinois with ENS Paris proposed in 2021, using a microwave router and a SWAP gate to connect several chipsets [2448].

---

2445 See Microwave Quantum Link between Superconducting Circuits Housed in Spatially Separated Cryogenic Systems by Paul Magnard, Alexandre Blais, Andreas Wallraff et al, PRL, December 2020 (13 pages). Paul Magnard now works for Alice&Bob.

2446 See Experimental quantum teleportation of propagating microwaves by K. G. Fedorov et al, December 2021 (7 pages).

2447 See Short-Range Microwave Networks to Scale Superconducting Quantum Computation by Nicholas LaRacuente et al, January 2022 (22 pages).

2448 See also A modular quantum computer based on a quantum state router by Chao Zhou, Matthieu Praquin et al, Universities of Pittsburgh and Illinois and ENS Paris, September 2021 (11 pages). With Praquin from ENS Paris. About linking transmon qubits with microwaves, starting with implementing SWAP gates between 4 single-qubit modules with relatively slow gates (750 ns). Quote: "*For atomic scale qubits communicating using optical frequency states, it is infeasible to couple photons into a communication channel with* ***very high efficiency****. This loss of information precludes light from simply being transferred from module to module, instead one must* ***herald instances*** *in which transmission is successful*". This doesn't bode well for photons interconnect.

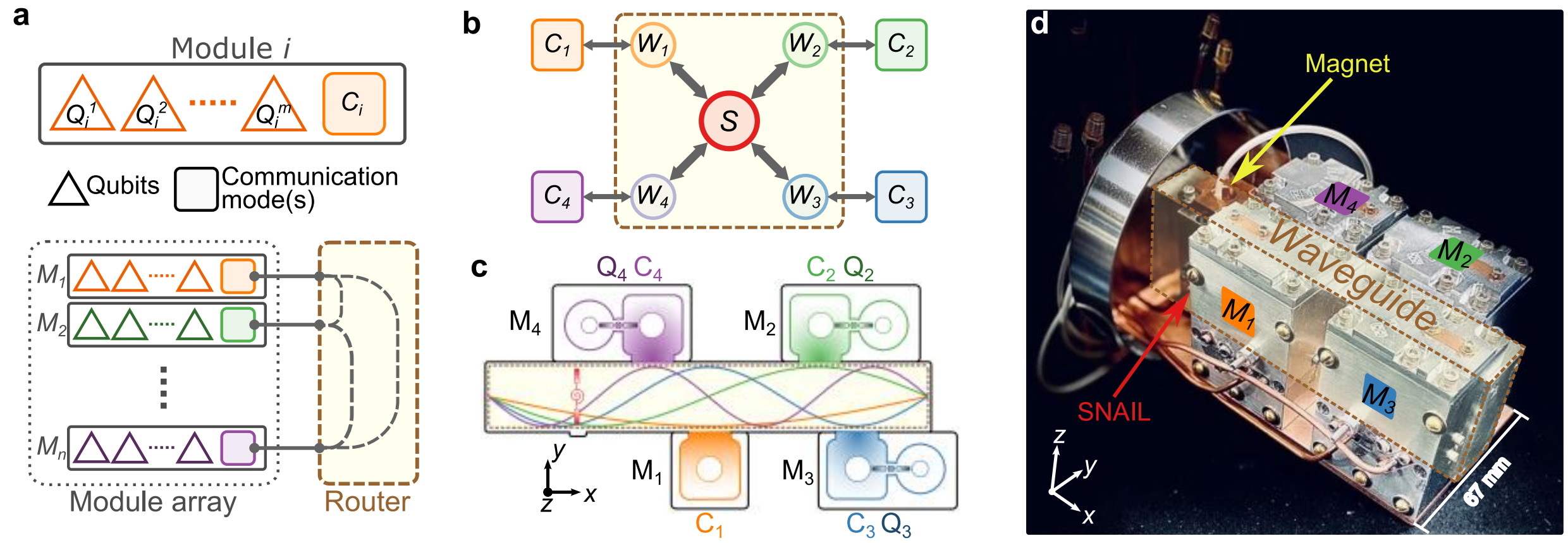

Figure 1. **Schematic representation and picture of the modular quantum computer device** (a) Basic structure of our modular quantum computer, in which a number of quantum modules are connected via their communication modes to a quantum state router. (b) Coupling scheme between the router and four communication modes. The brown dashed square represent the router with four waveguide modes ($W_1 - W_4$) and a SNAIL (S). Each waveguide mode is dispersively coupled to a single communication cavity mode ($C_1 - C_4$). (c) Schematic drawing of the full system consisting four modules and the central quantum state router. The colored curves inside the router represent the $E$ (electric) field distribution of the first four waveguide $TE_{10n}(n = 1, 2, 3, 4)$ eigenmodes. The SNAIL chip (represented in red) is placed at a location where it couples to all the waveguide modes being used. Each module (for $M_2$ to $M_4$) consists of a qubit ($Q_2 - Q_4$), a communication cavity ($C_2 - C_4$) and a readout cavity (for module $M_1$ the qubit has been omitted). (d) Photograph of the assembled device.

*Figure 772: Source: A modular quantum computer based on a quantum state router by Chao Zhou, Matthieu Praquin et al, Universities of Pittsburgh and Illinois and ENS Paris, September 2021 (11 pages).*

Another way would be the creation of more complicated many-to-many connectivity. If you have for example two superconducting chipsets ala Google Sycamore with, let's say, NxN qubits, you will first need to create N connections between the edge of the first chipset and the edge of the nearby second chipset.

Then, with entanglement sharing, you will have to make sure you can create two-qubit gates with these nearby connected qubits[2449]. What gate would be mandatory to enable a real scale-out? At least SWAP gates[2450]. You'd need to adopt a full stack approach, from the qubits, the QPUs interconnect and all related software concerns[2451].

Microwaves-based interconnect is also investigated to create quantum link between electron spin-based qubits[2452].

[2449] See Quantum transfer of interacting qubits by Tony J. G. Apollaro et al, May 2022 (24 pages) which addresses this point of N qubits connectivity between nearby chipsets.

[2450] See A modular quantum computer based on a quantum state router by Chao Zhou et al, April 2022 (21 pages) with all-to-all couplings among four independent quantum modules of superconducting qubits. They handle full-iSWAP time of 760 ns and average inter-module gate fidelity of 97% and Co-Designed Architectures for Modular Superconducting Quantum Computers by Evan McKinney et al, University of Pittsburgh, May 2022 (14 pages) which uses a superconducting nonlinear asymmetric inductive element (SNAIL) modulator and $\sqrt{iSWAP}$ gates. Hypercube 3D connections.

[2451] See Will Quantum Computers Scale Without Inter-Chip Comms? A Structured Design Exploration to the Monolithic vs Distributed Architectures Quest by Santiago Rodrigo et al, 2020 (6 pages) which makes some fully-stack architecture proposals. See also Towards a distributed quantum computing ecosystem by Daniele Cuomo et al, University of Naples, Italy, March 2020 (6 pages).

[2452] See Resonant microwave-mediated interactions between distant electron spins by F. Borjans, Jason Petta et al, Nature, December 2019 (6 pages) and Strong coupling between a photon and a hole spin in silicon by Cécile X. Yu, Simon Zihlmann, José C. Abadillo-Uriel, Vincent P. Michal, Nils Rambal, Heimanu Niebojewski, Thomas Bedecarrats, Maud Vinet, Etienne Dumur, Michele Filippone, Benoit Bertrand, Silvano De Franceschi, Yann-Michel Niquet and Romain Maurand, June 2022 (6 pages).

In August 2021, **AMD** published a patent designed to handle a local scale-out capacity for quantum computers, with a teleportation-based multi-SIMD architecture. SIMD stands for "Single Instruction Multiple Data" and is heavily used in parallel classical hardware architectures like vector processors or tensor processors and GPUs. Here, teleportation would be used to handle coordination between several quantum processing units and reduce both the number of qubits and quantum gates needed to run an algorithm. Unfortunately, this patent doesn't describe in any way a real quantum process, contains no physics, no maths, no compiling trick, no timing analysis and nothing about teleportation implementation and about any quantum algorithm parallelization. It also mentions a "global memory" like if creating qubits memory was some standard off-the-shelf technology. On top of that, none of the patent holders seem to have a quantum computing background and they never published any quantum-related paper visible on arXiv[2453].

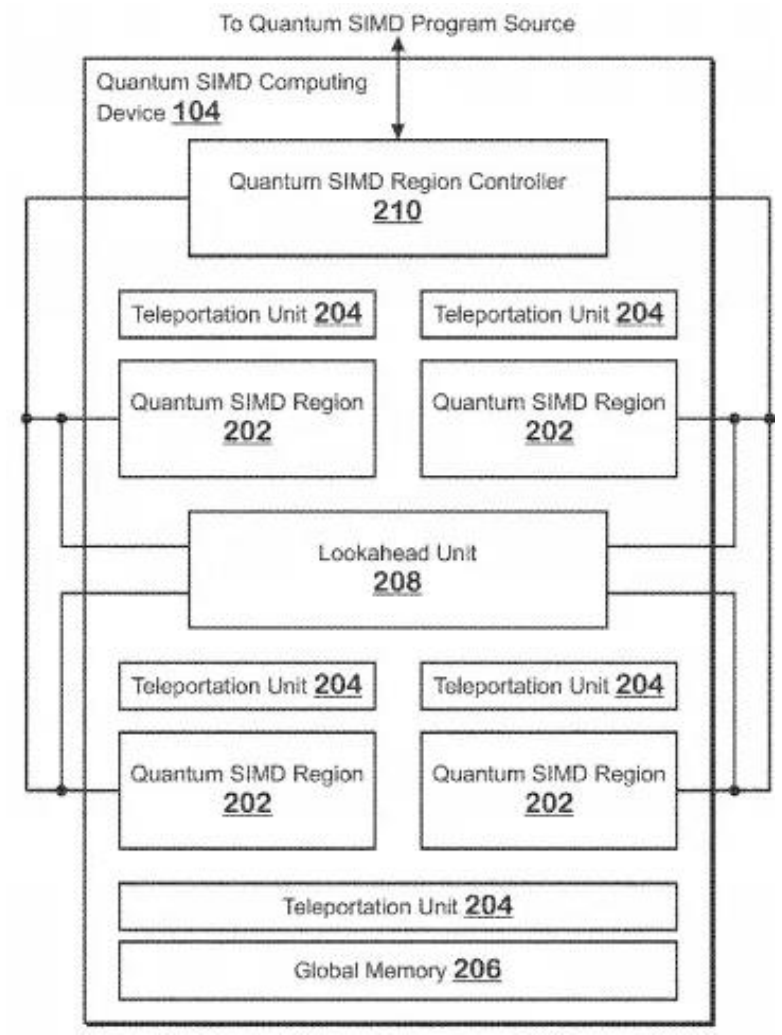

*Figure 773: AMD's weird patent.*

This has the flavor of a PR-driven approach that only a few scientists can fact-check, if not of a patent-troll. And it unfortunately worked[2454]!

In June 2022, **Huawei** also published a patent in China numbered CN114613758A related to the production of quantum computing chipsets and their scale-out. It describes M subchips of N qubits[2455]. In the provided schematics, you can see four qubits subchips (20), four coupling structures (30) and a central one with no connectivity with the others that is the cavity mode suppression structure of undetermined nature, shape and form (40). The asserted benefit from this architecture is resilience to manufacturing defects more than a scale-in architecture. These have undetermined internal structures and connections, nor any physical or experimental data attached. This wouldn't pass any scientific paper peer-reviewing process! Needless to say that like with AMD, this is borderline patent troll.

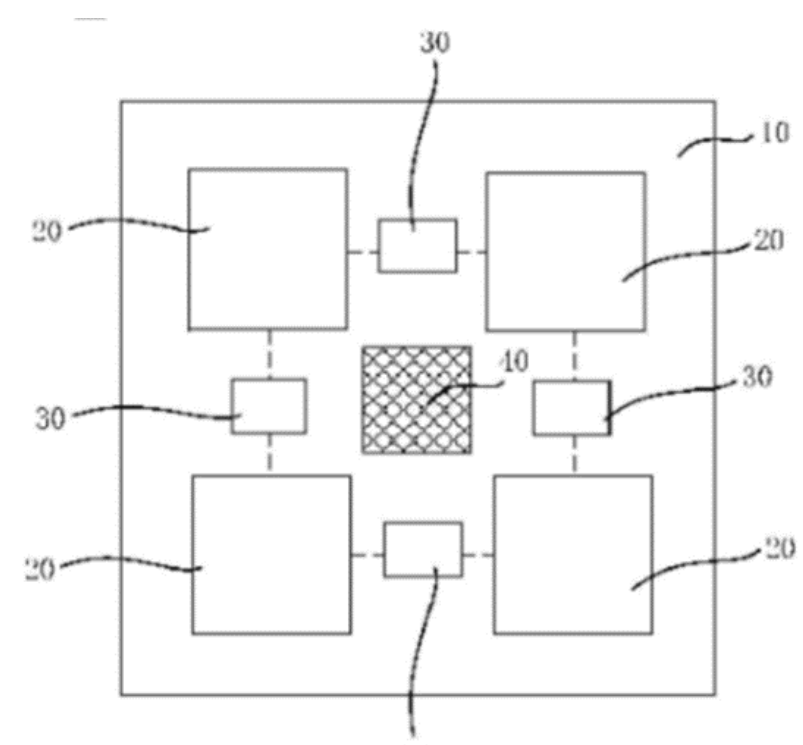

*Figure 774: Huawei also weird patent.*

### Photons interconnect

The most generic way to interconnect QPU, particularly over arbitrary distance, would be with photons and fiber optics. It requires a couple things like creating deterministic or heralded sources of entangled photons connecting qubits, qubits-to-photon or microwave-to-photon[2456] quantum state conversion, photons conversions to telecoms wavelengths[2457] and, most of the time, some form of quantum memory for synchronization purpose.

[2453] See New AMD Patent Proposes Teleportation to Make Quantum Computing More Efficient by Francisco Pires, August 2021.

[2454] See for example AMD patent reveals revolutionary teleportation-based quantum computer by Bogdan Solca, Notebook Check, August 2021.

[2455] See Huawei publishes patents related to Quantum chips and Quantum computers by Amit, Huawei, June 2022.

[2456] See Large-bandwidth Transduction Between an Optical Single Quantum Dot Molecule and a Superconducting Resonator by Yuta Tsuchimoto, Andreas Wallraff, Martin Kroner et al, PRX Quantum, 2022 (13 pages).

[2457] See Quantum frequency conversion of memory-compatible single photons from 606 nm to the telecom C-band by Nicolas Maring, Dario Lago-Rivera et al, IFCO, 2021 (7 pages).

It seems better adapted to either photon qubits (which don't need any conversion) and qubits that are controlled by photons in the visible or near visible spectrum like cold atoms, trapped ions and NV centers[2458]. Trapped ions and cold atoms are controlled by lasers, but converting their quantum state into a photon is no small matter. Silicon qubits use the spin of one or two electrons. Spin-to-charge and charge-to-photon conversions can then be performed.

At some point, as presented in a custom drawing below, interconnect architectures may someday mix various techniques, with short-range interconnect techniques using microwaves and longer range techniques based on photons entanglement.

Some photon-based interconnect experiments have also been done with silicon spin qubits[2459]. Another option to interconnect superconducting qubits is to entangle them first with NV centers spin qubits which themselves are then easier to interconnect with photons[2460]. Another option is to couple a SiV vacancies electron spin acting as a communication qubit to a $^{29}Si$ nuclear spin acting as a memory qubit[2461].

Other researchers are looking for ways to encode quantum information differently in transmitted photons.

Instead of using a classical polarization encoding, researchers from Caltech experimented quantum teleportation of time-bin qubits (with "time of arrival" encoding) using a standard telecommunication wavelength of 1536.5 nm with an average success superior to 90%[2462].

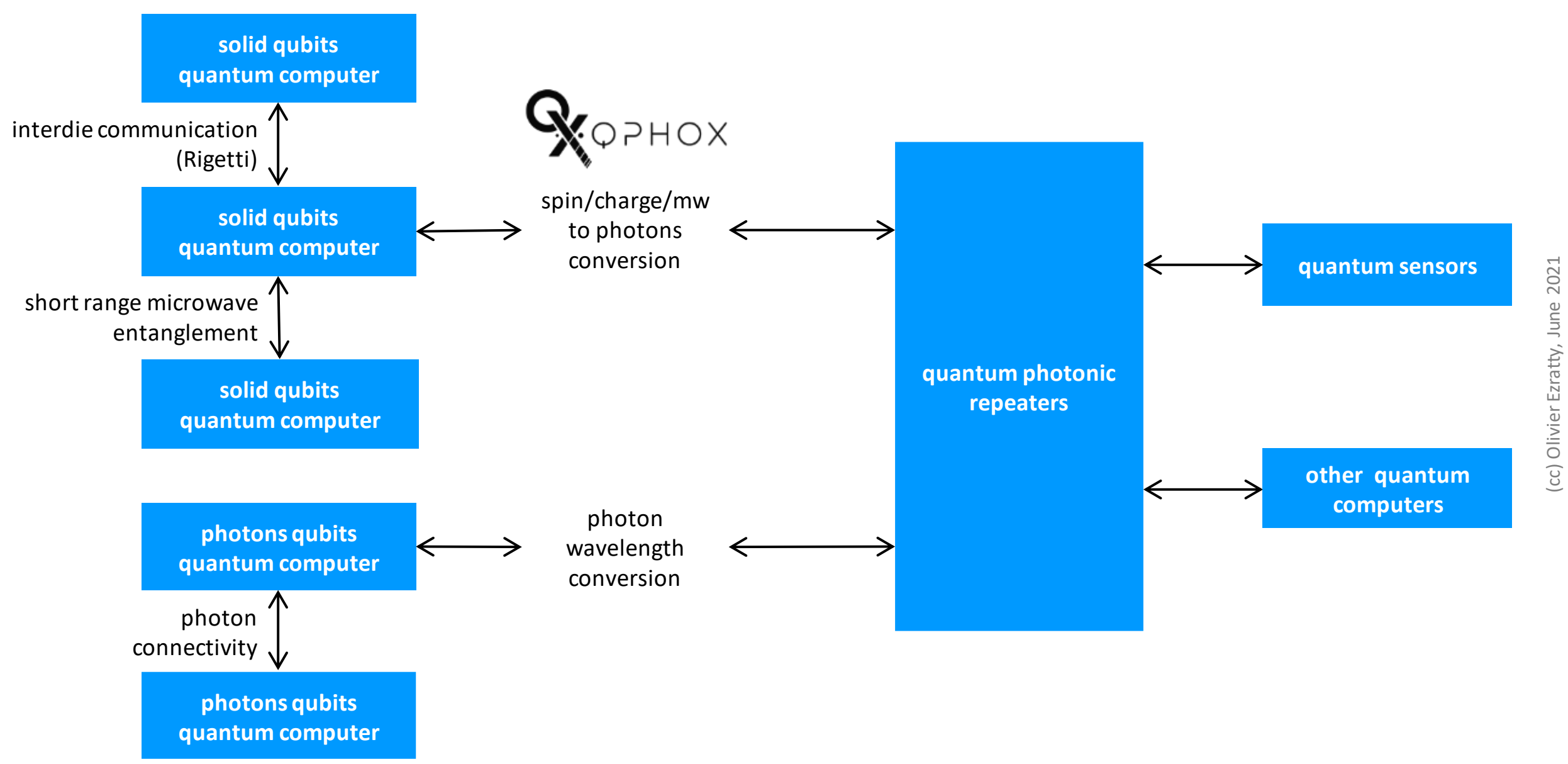

*Figure 775: various interconnect architectures. (cc) Olivier Ezratty, 2022.*

[2458] In 2021, a team from Qutech and TU Delft connected three NV centers qubits quantumly in an entangled GHZ state, beyond the traditional two nodes existing experiments. See Realization of a multi-node quantum network of remote solid-state qubits by Matteo Pompil, Sophie Hermans, Stephanie Wehner et al, February 2021 (28 pages).

[2459] See for example First chip-to-chip quantum teleportation harnessing silicon photonic chip fabrication by the University of Bristol, December 2019 which refers to Chip-to-chip quantum teleportation and multi-photon entanglement in silicon by Daniel Llewellyn et al, 2019 (48 pages). And the exaggerated version in The first "quantum teleportation" between two computer chips by Valentin Cimino, December 2019.

[2460] See Anisotropic rare-earth spin ensemble strongly coupled to a superconducting resonator by S. Probst et al, 2021 (7 pages).

[2461] See Robust multi-qubit quantum network node with integrated error detection by Pieter-Jan Stas, Mikhail D. Lukin et al, Harvard, July 2022 (24 pages).

[2462] See Teleportation Systems Toward a Quantum Internet by Raju Valivarthi et al, Caltech, 2020 (16 pages).

Trapped ions can also be interconnected with photons. It's the main scale-out plan for startups like IonQ[2463]. A European team succeeded in interconnecting trapped ions with photons over a distance of 230 m in 2022[2464].

Given it's still the dominant architecture, superconducting to photons connectivity is an intense field or research. In that case, microwaves are converted to another frequency range while keeping the quantum state.

This conversion can be done with opto-electromechanical systems[2465]. TU Delft researchers led by Simon Gröblacher experimentally achieved this in 2018, at 20 mK, close to superconducting qubits operating temperature[2466].

This led to the creation of **QPhoX** (2021, Netherlands, 8.9M€) by Simon Gröblacher, a startup seed funded by Quantonation. The research project turned into a "quantum modem for the quantum Internet"[2467]. Although QPhoX is a startup, it seems still operating in a field a fundamental and experimental research[2468]. In September 2022, the startup announced a partnership with IQM to scale-out superconducting qubit quantum computers.

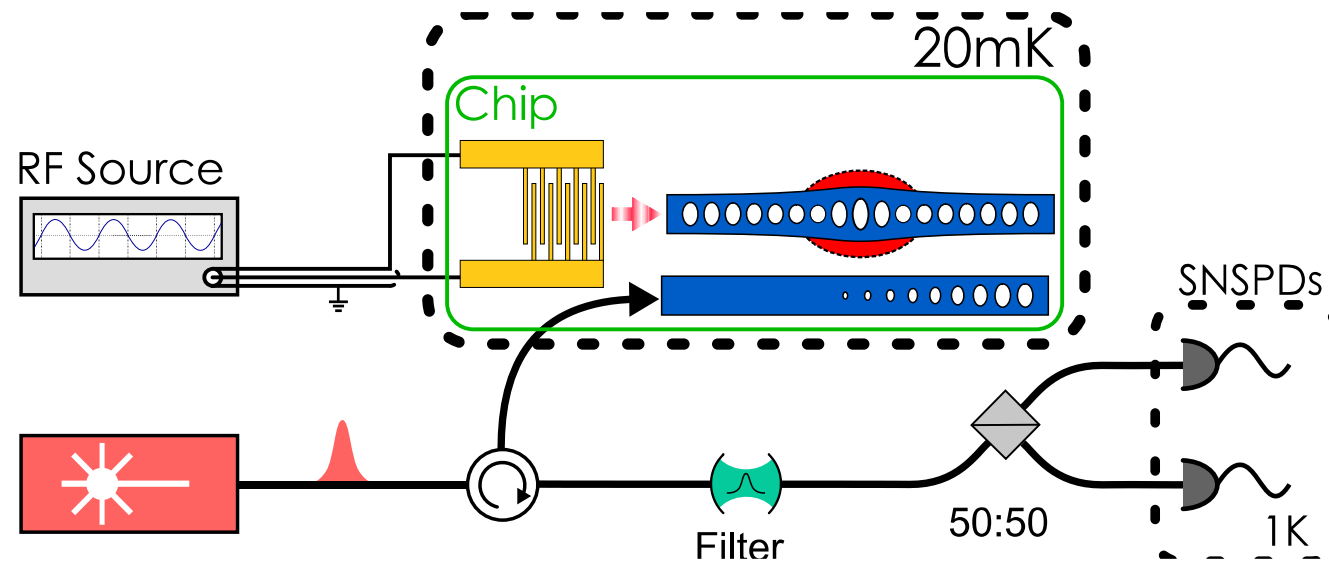

*Figure 776: Source: Microwave-to-optics conversion using a mechanical oscillator in its quantum ground state by Moritz Forsch et al, 2019 (11 pages).*

**Next Generation Quantum** (2019, USA) is CUNY university spin-off created by Shaina Raklyar (CEO) and German Kolmakov (CTO) developing commercial quantum computing applications and a hardware and software solution interconnecting multiple quantum computers to create quantum computer clusters.

They plan to manage this connection with photons and to use cavity polaritons, photons dressed with charges in a semiconductor optical microcavity that are sensitive to electric fields.

**WeLinQ** (2022, France) is a startup created by Tom Darras (CEO), Jean Lautier-Gaud (COO), Julien Laurat and Eleni Diamanti (both scientific advisors).

It is a spin-off from two laboratories from Sorbonne Université: the Laboratoire Kastler Brossel and the Laboratoire d'Informatique de Sorbonne Université.

---

2463 See Large Scale Modular Quantum Computer Architecture with Atomic Memory and Photonic Interconnects by Chris Monroe, July 2013 (16 pages), A high-fidelity quantum matter-link between ion-trap microchip modules by M. Akhtar, Igor Caron, Sylvain Gigan et al, March 2022 (8 pages).

2464 See Entanglement of trapped ion qubits separated by 230 meters by V. Krutyanskiy, Nicolas Sangouard, Tracy Northup, August 2022 (22 pages) with a fidelity of 88% but a very low success rate of $4\times10^{-5}$.

2465 See also A quantum microwave-to-optical transducer by Thibaut Jacqmin of LKB, 2019 (17 slides) which describes opto-electromechanical mechanisms for state conversion of superconducting qubits into transportable photons on optical fibers.

2466 See New horizons for connecting future quantum computers into a quantum network, October 2019 which references Microwave-to-optics conversion using a mechanical oscillator in its quantum ground state by Moritz Forsch et al, 2019 (11 pages).

2467 See The widely anticipated quantum internet breakthrough is finally here by Maija Palmer, May 2021 and A perspective on hybrid quantum opto- and electromechanical systems by Yiwen Chua and Simon Gröblacher, 2020 (7 pages). Simon Gröblacher also created Nenso Solutions, a quantum technology consulting company.

2468 See Optomechanical quantum teleportation by Niccolò Fiaschi, Simon Gröblacher et al, Nature, October 2021 (9 pages) and Coherent feedback in optomechanical systems in the sideband-unresolved regime by Jingkun Guo and Simon Gröblacher, June 2022 (12 pages).

They develop a unique full stack (hardware and software) quantum link solution to interconnect quantum processors to enable the scale up of quantum computing. Their quantum links are based on world record cold atom quantum memories. They enable the efficient interconnexion of quantum processors to increase their computational power and the implementation of world's first quantum repeaters to enable a secure access to quantum computing at a distance.

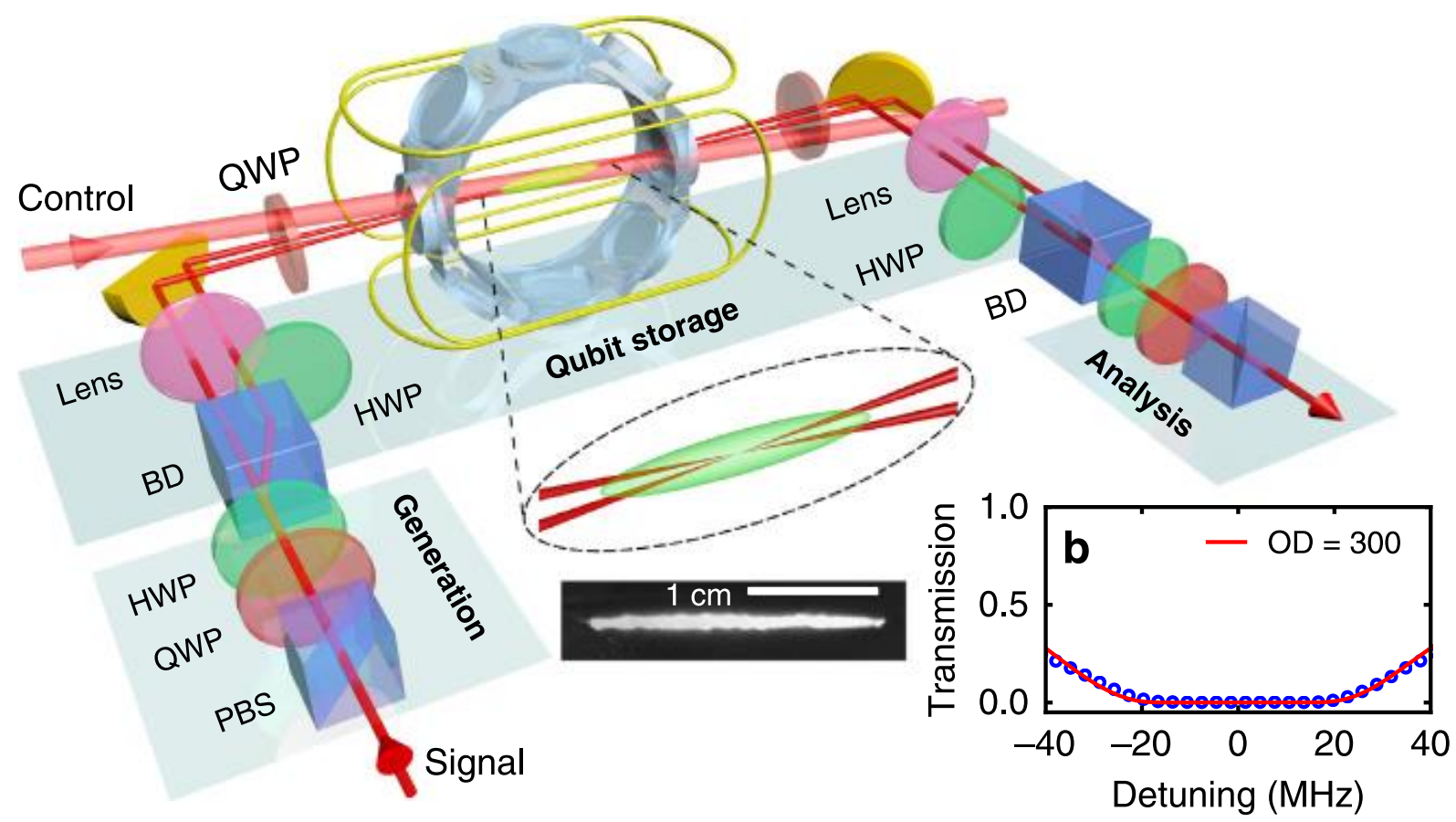

*Figure 777: Source: Highly-efficient quantum memory for polarization qubits in a spatially-multiplexed cold atomic ensemble by Pierre Vernaz-Gris, Julien Laurat et al, Nature Communications, 2018 (6 pages)*

Their first product, called *QDrive*, is a robust, compact, and transportable highly efficient quantum memory which will be deployed in quantum computing infrastructures to perform first proof of concepts with providers of quantum computing. Their memory technology is based on an elongated quasi-2D magneto-optical trap of alkali atoms cooled at a temperature close to 20 µK.

It enables the on-demand (i.e., the storage time is adjustable by the user) and efficient storage of photonic qubits and entangled states with a world record efficiency of 90%, qubit fidelity above 99% and a storage time of 15 µs[2469]. This is becoming a real commercial product.

In September 2022, WeLinQ announced a partnership with ixBlue for the procurement of lasers adapted to their needs.

In the quantum memory realm, Chinese researchers succeeded in 2019 to entangle two rubidium atom ensembles quantum memories via entangled photons at a distance of 50 km[2470]. But ensemble atoms are not perfect qubits for computing.

**PhotoniQ** (2021, Israel) is a startup developing quantum interconnection systems based on cold atoms chips. Its founder is Ron Folman, a quantum researcher from Ben Gurion Negev University. The system would be used both for quantum repeaters and for connecting several quantum processing units. This is still at the research stage and the startup is still stealth.

Other efforts are undertaken to connect heterogeneous quantum networks with hybrid entanglement swapping between DV and CV photonic systems[2471]. More classically, qubits can be distantly connected through a photonic link, as MPQ researchers in Germany did show in 2021[2472].

[2469] See Highly-efficient quantum memory for polarization qubits in a spatially-multiplexed cold atomic ensemble by Pierre Vernaz-Gris, Julien Laurat et al, Nature Communications, 2018 (6 pages) and Connecting heterogeneous quantum networks by hybrid entanglement swapping by G. Guccione, Tom Darras, Julien Laurat et al, April 2021 (15 pages) where they describe a higher-level interconnect architecture based on their technology.

[2470] See New Record: Researchers have entangled quantum memory over 50 kilometers by Stéphanie Schmidt, February 2020. The feat comes from Hefei's Jian-Wei Pan laboratory in China. This refers to the article published in Nature: Entanglement of two quantum memories via fibers over dozens of kilometers by Jian-Wei Pan et al, February 2020 and previously on arXiv in March 2019: Entanglement of two quantum memories via metropolitan-scale fibers (19 pages).

[2471] See Quantum Networking Demonstrated for First Time par Dhananjay Khadilkar, November 2018, referencing Connecting heterogeneous quantum networks by hybrid entanglement swapping by Giovanni Guccione, Tom Darras et al, May 2020 (7 pages).

[2472] See Quantum systems learn joint computing - MPQ researchers realize the first quantum-logic computer operation between two separate quantum modules in different laboratories, February 2021.

**Entangled Networks** (2020, Canada) is a startup developing a set of protocols and software for multi-QPU environment (multi-qubit gates implementation across distributed QPUs, performance optimization).

The hardware part requires optical quantum interconnect solutions including a high-resolution optical collection system and a low loss Bell state analyzer. The company was created by Aharon Brodutch (CEO) and Ilia Khait (CTO), two Israeli researchers established in Toronto.

**Bohr Quantum Technology** (2022, USA) is a stealth startup working on a quantum interconnect system based on some Caltech and Fermilab research. It is supposed to enable quantum computing scaley-out architectures with interconnecting QPUs and/or quantum memories.

The company was created by Paul Dabbar (CEO, a former Under Secretary for Science of the DoE who oversaw DoE research labs during the whole Trump administration) and Conner Prochaska (Chief of Strategic Partnerships, also coming from the DoE). No CTO or Chief Scientist? Bad omen. Their LinkedIn profiles tout that the company "*has developed and built the world's first commercial ready quantum networking system*".

### Electrons and ions shuttling

Electrons shuttling is a technique envisioned to enable interconnection of silicon spin qubits[2473]. It would have a very short range.

**QUASAR** is a semiconductor-based project using shuttling electrons with a QuBus (*pictured next*), a quantum bus to transport electrons and their quantum information over distances of 10 µm. The partners are Infineon, HQS, Fraunhofer (IAF, IPMS), Leibnitz Association (IHP, IKZ) and the Universities of Regensburg and Konstanz. The project will run until 2025 to create 25 coupled qubits. The resulting computer is to be deployed at JUNIQ.

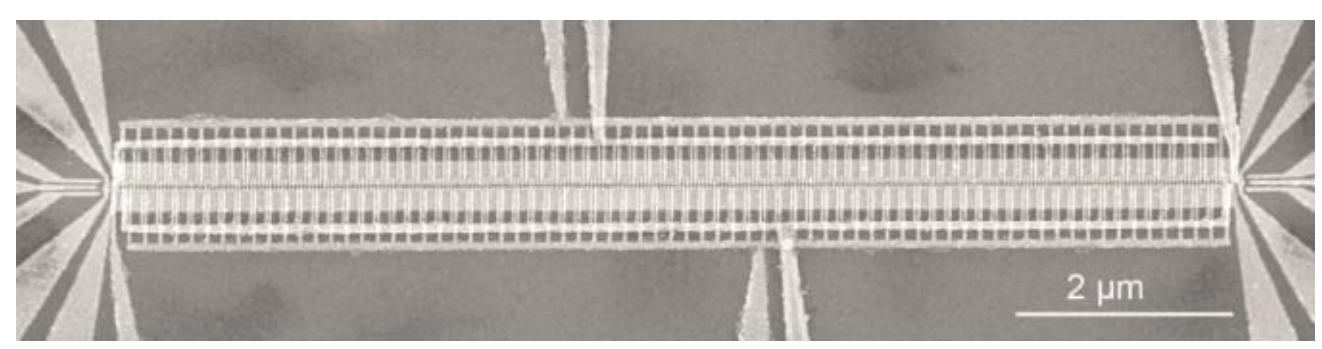

*Figure 778: an electron shuttling waveguide. Source: Quanten-Shuttle zum Quantenprozessor "Made in Germany" gestartet, Jülich, February 2021.*

Jülich is also participating to the European Flagship QLSI project driven by CEA-Leti in France. QUASAR got a 7.5M€ funding from BMBF. The resulting computer is to be deployed at JUNIQ. Jülich is also participating to the European Flagship QLSI project driven by CEA-Leti in France. QUASAR got a 7.5M€ funding from BMBF[2474].

Ion shuttling is an interconnect technique that we already quickly described with IonQ and Quantinuum.

At last, let's mention another interconnect technique, that ties superconducting qubits together with phononic communication using surface acoustic waves with the advantage that it fits into a solid-state circuit[2475].

---

[2473] See Multicore Quantum Computing by Hamza Jnane, Simon Benjamin et al, Quantum Motion, January 2022 (24 pages) which deals with interconnecting silicon based QPUs with microwaves or electrons shuttling.

[2474] See Quanten-Shuttle zum Quantenprozessor "Made in Germany" gestartet, Jülich, February 2021.

[2475] See Quantum Communication with itinerant surface acoustic wave phonons by E. Dumur, Audrey Bienfait, et al, University of Chicago and ENS Lyon, December 2021 (5 pages).

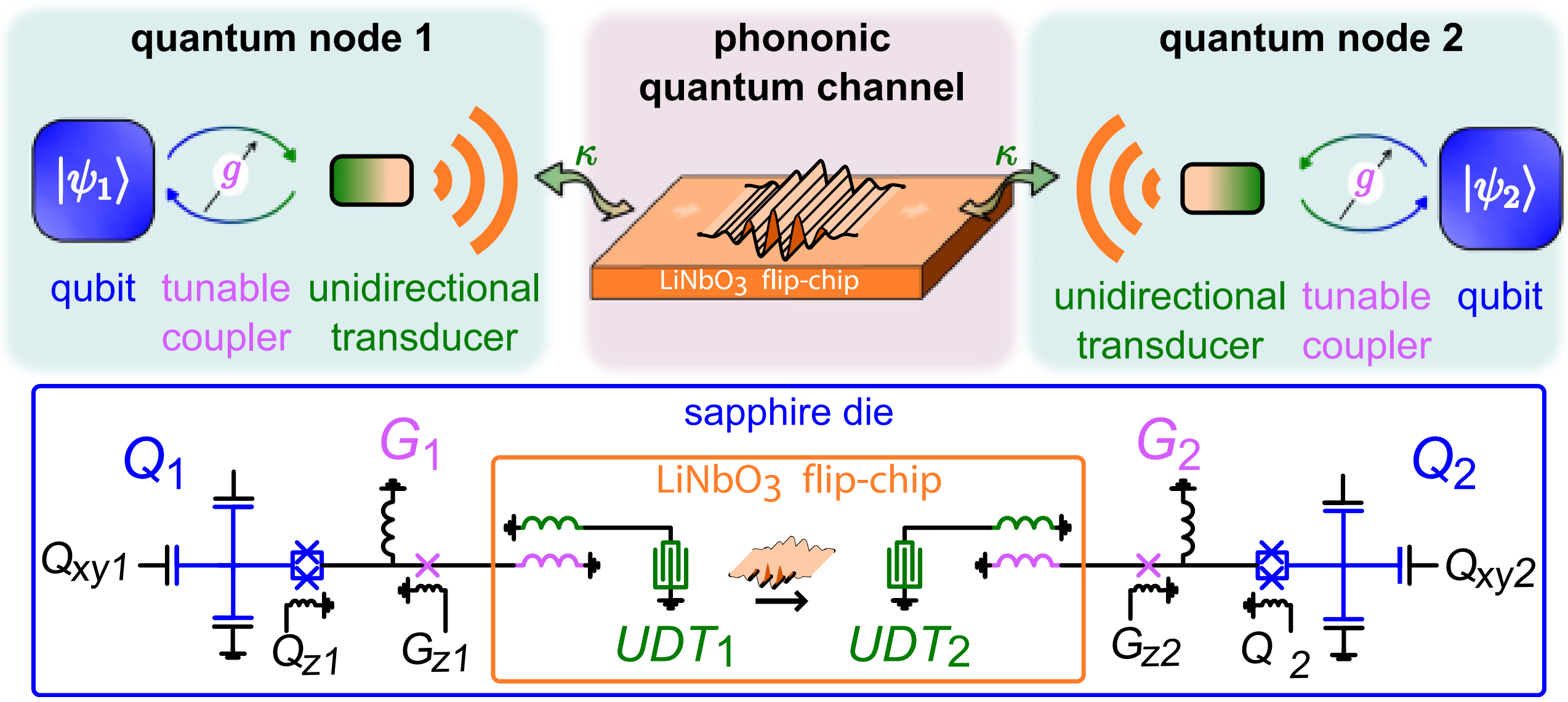

*Figure 779: Source: Quantum Communication with itinerant surface acoustic wave phonons by E. Dumur, Audrey Bienfait, et al, University of Chicago and ENS Lyon, December 2021 (5 pages).*

### Quantum telecommunications infrastructures

Three other areas must make progress to enable the deployment of quantum telecommunications networks: protocols[2476], repeaters[2477], switches[2478], quantum memories[2479], distributed quantum error correction schemes[2480] and also compiler and other various software tools to distribute processing over several QPUs[2481].

In the network layout domain, a team led by British and Austrian researchers established an any-to-any quantum communication link with 8 network nodes using dense wavelength division multiplexers (DWDM) and a single source of polarization-entangled photon pairs[2482]. Is this a "quantum Internet"? It's still a marketing buzzword since many-nodes technologies are not really available and quantum entanglement distribution is an additional feature and not a replacement for existing classical networks.

---

2476 Let's mention the Quantum Protocol Zoo initiative launched by LIP6 and VeriQloud which inventories about 56 quantum telecommunication and cryptography protocols. See Protocol Library. See also Benchmarking of Quantum Protocols by Chin-Te Liao, Elham Kashefi et al, July 2022 (16 pages).

2477 See Adaptive bandwidth management for entanglement distribution in quantum networks by Navin B. Lingaraju et al, 2021 (5 pages).

See Telecom-heralded entanglement between multimode solid-state quantum memories by Dario Lago-Rivera et al, Nature, IFCO, June 2021 (7 pages).

2478 See Development of Quantum InterConnects (QuICs) for Next-Generation Information Technologies by David Awschalom et al, 2019 (31 pages) and Quantum Switch for the Quantum Internet: Noiseless Communications Through Noisy Channels, 2020 (14 pages).

2479 See this general overview of quantum memories used in quantum networks: Optical Quantum Memory and its Applications in Quantum Communication Systems by Lijun Ma et al of NIST, 2020 (13 pages). The schema of this page on the Quantum Internet is derived from it. See also Towards real-world quantum networks: a review by Shi-Hai Wei et al, January 2022 (71 pages) that discusses the links between quantum computing resources and quantum memories.

2480 See Quantum teleportation of physical qubits into logical code spaces by Yi-Han Luo, William J. Munro, Anton Zeilinger, Jian-Wei Pan et al, July 2021 (5 pages).

2481 See Optimized compiler for Distributed Quantum Computing by Daniele Cuomo et al, December 2021 (15 pages) and Distributed Shor's algorithm by Ligang Xiao, July 2022 (15 pages).

2482 See A trusted node–free eight-user metropolitan quantum communication network by Siddarth Koduru Joshi et al, September 2020 (9 pages) and the subsequent work Flexible entanglement-distribution network with an AlGaAs chip for secure communications by Félicien Appas, Eleni Diamanti, Sara Ducci et al, NPJ Quantum Information, July 2021 (12 pages).

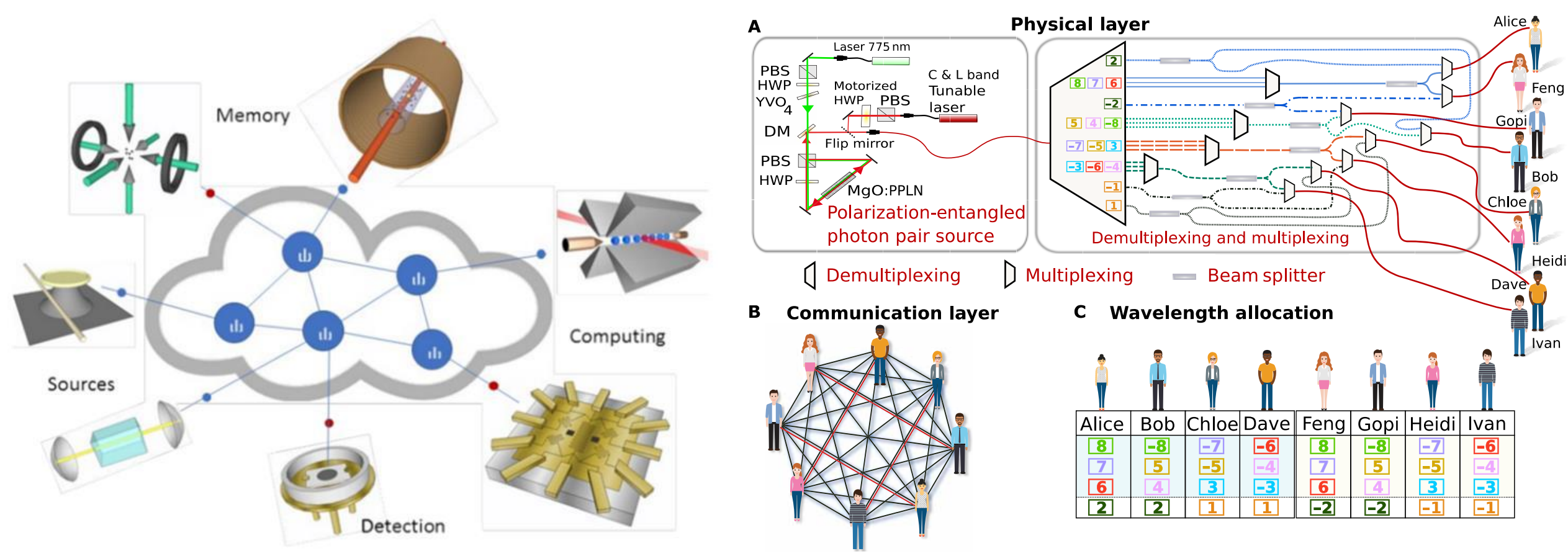

*Figure 780: Source: A trusted node–free eight-user metropolitan quantum communication network by Siddarth Koduru Joshi et al, September 2020 (9 pages).*

Other applications of quantum telecommunications should be mentioned in addition to distributed computing:

- **Quantum electronic signatures** that authenticate classic messages. They are transferable to third parties, non-repudiable and non-forgeable.
- **Connecting quantum sensors** which could be useful when it makes sense to consolidate quantum states from various quantum sensors, to improve their precision, like with telescopes interferometry, and also protect their content[2483].
- **Transmission of anonymous data**. It allows two nodes in a quantum network to communicate with each other without one node being able to identify the other node and also without the other nodes not involved in the protocol being able to identify the sender and the recipient. The communications leave no trace and are therefore not auditable. This replaces traditional anonymization proxies. It can be used as a basis for distributed processing, coupling this with classical or quantum data encryption. It is a means of ensuring the anonymization of the transmission of data such as survey or health data.
- **Quantum money** that applies a concept of Stephen Wiesner (1942-2021, Israeli) from 1970, and improved in 1983. It is based on tokens of verifiable integrity that can only be used once[2484].
- **Clock synchronization** which has been tested in the early 2000s and is useful for telecommunication networks and GPS operations[2485].
- **Detecting Extra-Terrestrial life**. Well, sort of. That's the plan from SETI who wants to analyze "quantum communications" coming from exoplanets, which would bear some differentiated signature. One can wonder how such communications could be sorted out from the planet's star random photon streams, but who knows[2486].

About the relationship with inter-QPU connectivity and data-transfer speed? Could these connections enable faster data-transfer than classical data links? Well, yes and no. If it's about transmitting classical data, an inter-QPU or quantum telecom link won't make things faster.

---

[2483] See one example with Distributed quantum sensing with optical lattices by Jose Carlos Pelayo, Karol Gietka and Thomas Busch, Okinawa Institute of Science and Technology Graduate University, August 2022 (7 pages).

[2484] A recent proposal of quantum money co-authored by Peter Shor is in Publicly verifiable quantum money from random lattices by Andrey Boris Khesin, Jonathan Z. Lu and Peter Shor, July 2022 (15 pages). It's still highly theoretical.

[2485] See Distant clock synchronization using entangled photon pairs by Alejandra Valencia et al, 2004 (10 pages).

[2486] See We could detect alien civilizations through their interstellar quantum communication by Matt Williams, April 2021 referring to Searching for interstellar quantum communications by Michael Hippke, April 2021 (14 pages).

Inter-QPU connectivity is mainly about creating a (much) larger QPU than the separate QPUs. The resource is entanglement. When you have two interconnected QPUs of N qubit each, the Hilbert space managed by the interconnected QPUs has a size of $2^{2N}$ complex numbers whereas the isolated QPUs handle only $2*2^N=2^{N+1}$ complex numbers. But in the end, when reading out the qubits in these isolated or consolidated QPUs, you don't capture more than 2N classical bits. But teleportation may be useful to teleport at a very fast rate an entangled state from one location to another. Let's say that while the data is quantum, interconnect and teleportation can bring some speed advantage. This could be the case when connecting quantum sensors and quantum computers.

## Quantum Physical Unclonable Functions

qPUF are not the most talked about quantum technologies. It is at the crossroads of quantum cryptography, QRNGs, embedded systems and sensors.

Generically, Physical Unclonable Functions (PUF) are systems containing a piece of classical hardware that contain some unique signature related to the random aspect of matter and physical disorder[2487]. This unique signature can't be reproduced. It can also be dynamically generated and always be different.

A PUF is implemented as a unique function and is hard to physically and logically clone. They operate on a challenge/response basis: a challenge is fed into the PUF, which generates a unique binary response coming from the hardware module, which is based on its unique material imperfections.

The physical imperfections used in PUF are manyfold: reflection of optical materials, random scattering of light, ring oscillators, coating materials capacitance up to some random characteristics of electronic components like SRAM and DRAM memories. The most common PUFs are using silicon IC and their various defects and variations that are even greater as their integration nodes become smaller, now down to 5 nm.

The generated binary strings are used in as keys or identifiers in cryptography systems and as anti-counterfeiting systems. These can also serve as generic true random number generators, either standalone or combined with other entropy sources. But these security systems are not perfect and are prone to various attacks[2488], done with various methods including machine learning, side attacks and even, potentially, quantum computing[2489].

qPUFs are quantum equivalents of PUF that are based on quantum states and their physical unclonability. There's even a variation with Quantum read-out of PUF (QR-PUF) which reads-out quantumly a non-quantum physical disorder in a physical device. These qPUFs are better than classical PUFs but their resistance to forgery depends on their detailed level of unforgeability, which can be "existential" and "selective"[2490].

---

[2487] See Physical One-Way Functions by Papu Srinivasa Ravikanth, 2001 (154 pages) and Physically Unclonable Functions - a Study on the State of the Art and Future Research Directions by Roel Maes and Ingrid Verbauwhede, 2010 (36 pages).

[2488] A side-channel attack collects information from a security system or influence its execution in an indirect manner, by collecting in a stealth way some data on hardware operations (quantity of data processed in a computer, heating, amount of gas in a car tank). Side-channels may be power dissipation, operations timing, system temperature, acoustic, radio or optical emissions or a mix of these.

[2489] Samsung Galaxy S10 launched in 2019 contains an Exynos 9820 chipset with PUF technology, for crypto wallets, associated with Samsung Knox, a built-in storage hardware for security keys used with Blockchain services and cryptocurrencies like Ethereum. According to Samsung, the Exynos "*PUF generates an unclonable key for data encryption by using the unique physical characteristics of each chip*" but this characteristic is not specified. Looks like this PUF was replaced in some subsequent Samsung smartphones by a QRNG coming from IDQ.

[2490] See Quantum Physical Unclonable Functions: Possibilities and Impossibilities by Myrto Arapinis, Elham Kashefi et al, June 2021 (32 pages) and A Unified Framework For Quantum Unforgeability by Mina Doosti, Mahshid Delavar, Elham Kashefi and Myrto Arapinis, March 2021 (47 pages), all from the University of Edinburgh and CNRS LIP6 Paris. See also On the Connection Between Quantum Pseudorandomness and Quantum Hardware Assumptions by Mina Doosti, Elham Kashefi et al, March 2022 (33 pages) which deals with the conditions of unforgeability in relation to quantum pseudorandomness.

What is the physical form of qPUFs? It can be based on photonic features and polarized prepared devices and multiple scattering medium.

**Quantum Base** (2014, UK, $1.1M) offers various quantum product authentication solutions such as the Q-ID Optical, an optically readable "atomic" fingerprint solution that uses Physically Unclonable Functions (PUFs) in the form of physical tags that cannot be copied at the atomic scale level[2491].

The tag exploits a thin 2D layer of graphene that contains unique irregularities at the atomic scale that cannot be cloned. These irregularities would be amplified by unspecified quantum phenomena. Moreover, these tags can be dynamically activated and deactivated. The project stems from the work of Lancaster University of Robert Young who is the co-founder of the startup. The whole is integrated in a home-made random number generator (Q-RAND) based on a semiconductor diode, which can be integrated in a chipset (video)[2492]. They also propose the Q-ID Electronic, a unique identifier generator.

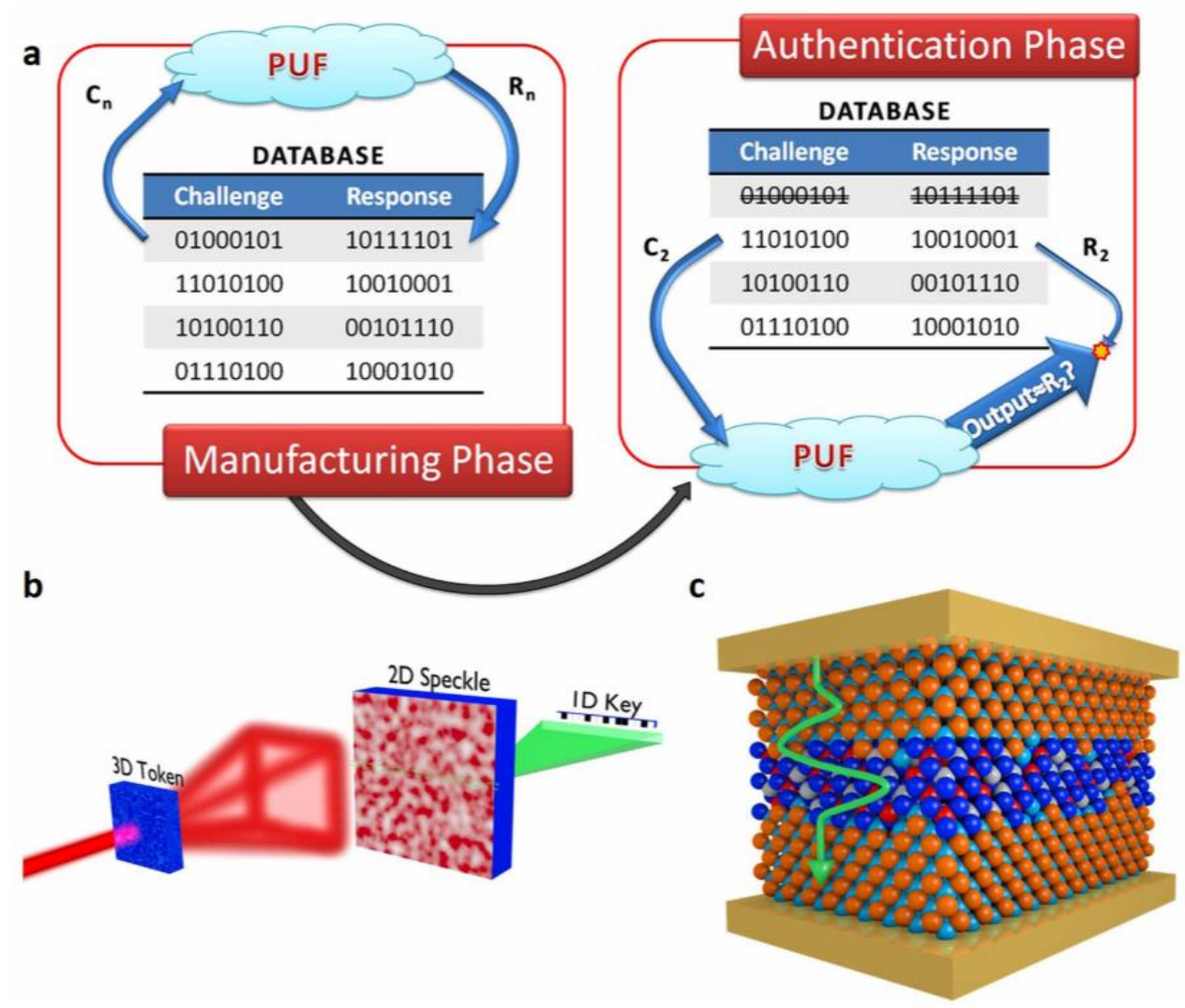

Figure 781: a qPUF made with photon and a scattering media.

## Vendors

Let's now review the startups in this vast sector of activity of quantum and post-quantum cryptography, trying to describe the nature of their offer and their differentiation when the information is publicly available! I have only kept here startups offering technology solutions and not integrated consulting and integration companies.

Note that on the market size side, the market for quantum and post-quantum cryptography is modest for the moment. A 2017 report estimated it at $2.5B by 2022[2493]. However, it is expected to gain momentum from this period onward, following the finalization of standardization by NIST and ETSI.

In addition to startups, commercial offers in quantum and post-quantum cryptography are also proposed or about to be proposed by various large IT players such as **Battelle**, **Infineon**, **Raytheon**, **IBM**,

[2491] This is documented in the USPTO patent US10148435B2 Quantum Physical Unclonable Function filed in 2015 (11 pages) and validated in 2018. It evokes a semiconductor component based on gallium arsenide, aluminum and antimony that generates a random spectral response that differs from one component to another. The process is described in Using Quantum Confinement to Uniquely Identify Devices by Robert Young et al, 2015 (8 pages).

[2492] The *resonant-tunnelling diode* (RTD) process is documented in Resonant-Tunnelling Diodes as PUF Building Blocks, by Ibrahim Ethem Bagci et al, 2018 (6 pages).

[2493] See New CIR Report States Quantum Encryption Market To Reach $2.5 Billion Revenues By 2022: Mobile Systems Will Ultimately Dominate, 2017.

**Cisco**[2494], **Atos**, **Gemalto** (part of Thales group), **Infineon**[2495], **Microsoft**[2496], **Intel**[2497], **NEC**, **Toshiba**, **Huawei**[2498], **KT** and **Samsung**. Even **Amazon** announced in 2022 the creation of its own quantum networking center in Boston and a partnership with Harvard University[2499].

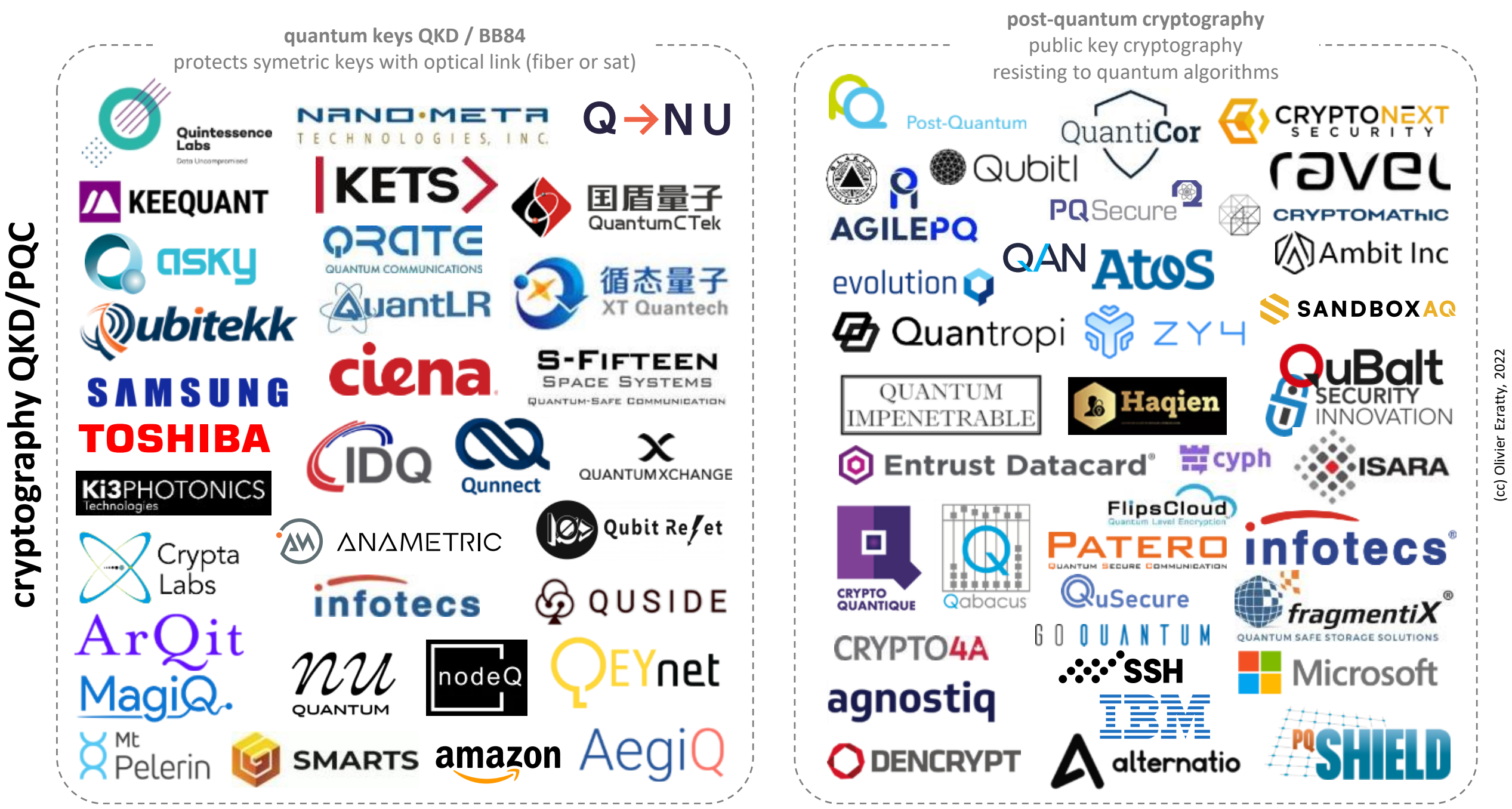

*Figure 782: the big map you were expecting on QKD (left) and PQC (right) vendors. (cc) Olivier Ezratty, 2022.*

**IBM** created a set of PQC protocols and participated to the NIST PQC competition. They were finalists of round 3 of the NIST selection in July 2020 for both lattice-based CRYSTALS-KYBER (secure key encapsulation mechanism) and CRYSTALS-DILITHIUM (secure digital signature). These were developed in partnership with **ENS Lyon** (France), **Ruhr-Universität Bochum** and **Radboud University** (Germany). These PQC solutions are embedded in an IBM TS1160 tape drive system with a modified firmware in combination with symmetric AES-256 encryption. It enables a secured long-term data storage. In November 2020, IBM also announced it would integrate these protocols in its cloud offering.

**Microsoft** also works on PQC algorithms. It includes FrodoKEM and SIKE which are PQC based key exchanges protocols. Then, qTesla and Picnic, PQC based signatures protocols.

---

[2494] Cisco's Quantum Research is led by Alireza Shabani with three full time researchers. The company is even organizing a yearly event, Cisco Quantum Summit.

[2495] With its Optiga TPM SLB 9672, a hardware trusted module supporting AES and SHA2-384 encryptions that are quantum resistant.

[2496] See the Microsoft site that describes their activity in the PQC.

[2497] Intel participated to the creation of the BIKE PQC PKI that was preselected in 2020 by NIST but is not so far a finalist, per the 2022 4 selected PQCs, including only one PKI. Intel PQC research is done at their Crypto Frontiers Research Center that was launched in 2021. This collaborative effort with Universities is plan to last only until 2024. See Intel Labs Establishes Crypto Frontiers Research Center, 2021.

[2498] See Continuous-Variable Quantum Key Distribution with Gaussian Modulation, the Theory of Practical Implementations, 2018 (71 pages). The Huawei team working on the QKD is partly located in their research center in Dusseldorf.

[2499] See Announcing the AWS Center for Quantum Networking by Denis Sukachev and Mihir Bhaskar, June 2022 and Announcing a research alliance between AWS and Harvard University by Antia Lamas-Linares, Mihir Bhaskar, and Denis Sukachev, September 2022.

Let's also mention **Toshiba**'s ambitions in QKD deployments, announced in October 2020. They are deploying a QKD based network for **NICT** (National Institute of Information and Communications Technology) in Japan, on top of a similar deployment undertaken with BT and their Openreach network in the UK in 2020 and demonstrations done with **Verizon** and **Quantum Xchange**.

Their hardware offer includes the "Multiplexed QKD system" which can transmit QKDs at a key rate of 40 kb/s over 70 km, and high-speed data on the same fiber. The "Long Distance QKD System" has similar features with a key rate of 300 kb/s and a range of up to 120 km but requiring two fibers. These solutions using the (quite old) BB84 protocol are manufactured in Cambridge, UK and fit in a 3U 19" rack format.

**ABCMintFoundation** (2017, Switzerland) created by Jin Liu and Jintai Ding is tasked with creating a quantum resistant Blockchain using a Rainbow Multivariable Polynomial Signature Scheme. It is a community driven open source project. It uses keys that can be as large as 1.7MB.

**Abelian** (2022, USA) is providing a Blockchain infrastructure enabling "gold 2.0" that is based on a lattice PQC. It secures financial transactions in smart contracts and so-called DeFI (decentralized finance based on a Blockchain). It even deals with the Metaverse and Web3 applications. The company is organized as a foundation and its founders are anonymous[2500].

**AegiQ** (2019, UK, $4,3M) is developing quantum cryptography systems based on III-V single-photon source semiconductors. It was created by Max Sich (CEO), Scott Dufferwiel (CTO) and Andrii Iamshanov (CFO) as a spin-off of the University of Sheffield. They plan to equip telecommunication data centers for their fiber optical infrastructures upgrades to QKD. They are also building a satellite-based QKD offering.

**AgilePQ** (2014, USA) provides a software platform for "post-quantum" security of communication between connected objects and the cloud, such as drones.

It includes AgilePQ C-code, a piece of software that runs on connected object microcontrollers and consumes little power, and AgilePQ DEFEND, an adaptive size-based key generation system. DEFEND generates codes that are harder to break than AES 256 and with 429 orders of magnitude difference. Specifically, we go from a key space of 10 to the power of 77 to 8*10 to the power of 506 (factor of 256)[2501] . The system that is patented seems to be a variant of linear random codes but with keys of reasonable size. It has been standardized at NIST and interfaces with SCADA (Supervisory control and data acquisition) control and supervision systems. The company is a Microsoft Azure partner.

**American Binary** or **Ambit** (2019, USA) sells PQC solutions to governments, enterprise customers and consumer products companies, including a VPN.

**Anametric** (2017, USA, $1.9M), formerly bra-ket science, is a startup that wants to create information storage systems in qubits operating at room temperature. It seems they are focused on quantum telecommunication systems.

[2500] See Abelian (ABEL) – A Quantum-Resistant Cryptocurrency Balancing Privacy and Accountability by Alice, Bob, Eve, and λ, February 2022 (140 pages).

[2501] See AgilePQ DEFEND Cryptographic Tests (11 pages).

**Agnostiq** (2018, Canada, $2.5M) is a startup coming out of the Creative Destruction Lab accelerator in Toronto. They develop workflow management tools to submit and scale jobs on hybrid quantum computers, cloud privacy and quantum obfuscation tools and pre-built applications in the finance sector for portfolio optimization and options pricing.

**Alternatio** (2016, Poland) develops a post-quantum cryptography IP core to be integrated in chipsets for the industry and connected objects.

**ArQit (**2016, UK, $70M) is developing QuantumCloud, a cloud-distributed symmetric keys mixing keys generation in local agents and key distribution via low earth orbit satellites and QKD.

They also claim, through a patent, that they can implement terrestrial QKD without trusted nodes or repeaters, which was debunked by an international team of researchers[2502]. In May 2021, the company announced a funding round of $400M using an IPO through a special purpose acquisition company like IonQ, this time with Centricus Acquisition Corp. As a result, the company published in May 2022, a revenue of $12.3M for H2 2021 with an operating loss of $14.3M, with $5,3M in H1 2022, compared to a loss of $5.5M in H2 2020. They also published a "myth busting" presentation[2503].

**BLAKFX** (2017, USA) develop quantum resistant software solutions around their Helix22 SDK that provides five layers of symmetric keys protections for end-user to end-user communications (AES1, TwoFish, AES2, ThreeFish, Snow3G). With sufficiently long keys, symmetric keys are quantum resistant.

**BraneCell Systems** (2015, Cambridge Massachusetts and Dusseldorf, $1.8M) is a startup launched by Wassim Estephan and Christopher Papile.

It started to develop a cold atom based QPU[2504]! They have filed a few patents, including USPTO patent 9607271, validated in March 2017[2505]. They however later pivot on creating an atom-based repeater.

**Ciena** (1992, USA) is an equipment manufacturer in the field of optical telecommunications. They integrate IDQ's offers in their solutions, and, their optical random key generators.

---

2502 See Long-range QKD without trusted nodes is not possible with current technology by Bruno Huttner, Romain Alléaume, Philippe Grangier, Hugo Zbinden et al, npj, September 2022 (5 pages).

2503 See Arqit Myth Busters by Arqit, 2022 (10 pages). I'd say OK with most myths, but not for their assessment of the quantum computing threat on classical security which, like all cybersecurity vendors, they tend to exaggerate.

2504 Their communication is cryptic to say the least, as BraneCell Systems Presents Distributed Quantum Information Processing for Future Cities, April 2018 and a partnership announced with the US government service provider, AST, in July 2018 in AST and BraneCell Announce Their Partnership to Improve Critical Government Functions Through the Power of Quantum Computing. They do not provide any technical or popularization information about their solution, qubits and error rates. They were also aiming for an ICO that would have been the first of its kind for a quantum computing startup. Their main goal was to create a secure communication system. They target the financial, energy, health, chemical and public sectors. A Quantum Theranos? At the very least, we have the right to doubt.

2505 Here is the patent description: "*The subject matter relates to multiple parallel ensembles of early stage spherical pulses radiated through engineered arrays forming the foundation for quantized computer processors taking advantage of integer thermodynamics. The materials, architecture and methods for constructing micro- and/or nano-scale three-dimensional cellular arrays, cellular series logic gates, and signature logic form the basis of small- and large-scale apparatuses used to execute logic, data bases, memory, mathematics, artificial intelligence, prime factorization, optical routing and artificial thought tasks not otherwise replicated in electron-based circuits*".

Although they do not yet have a structured QKD offering, they are very interested in standardizing it. In particular, they are participating in the Quantum Alliance Initiative launched in 2018 in the USA by the Hudson Institute, a conservative think tank, which is working towards this goal and creating proposed standards for QKD and QRNG (quantum random number generation).

**Crypta Labs** (2013, UK, $300K) develops post-quantum encryption solutions adapted to connected objects. In particular, they propose quantum random number generators that can be integrated into a cell phone (such as IDQ) and also work in space. The QRNG uses a LED or laser light source and a camera sensor. They are working with the University of Bristol.

**Crypto4A Technologies** (2016, Canada) offers an encryption solution based on a random number generation.

It includes a 19-inch QAOS format appliance server for generating entropy random numbers (without specifying the technology used) and another that generates quantum safe PQC encryption, the QxEdge Hardware Security Module (HSM). The PQC generation module is called QASM (Quantum Assured Security Module), which duplicates the quantum development language of the same name.

It is based on quantum safe hash-based signatures (HBS). These appliances are equipped with four Intel Core i7 chipsets, 16 GB of memory and 256 GB of SSD and run on a hardened version of Linux. They support algorithms certified by the NSA in the USA ("suite B") and future NIST PQC standards.

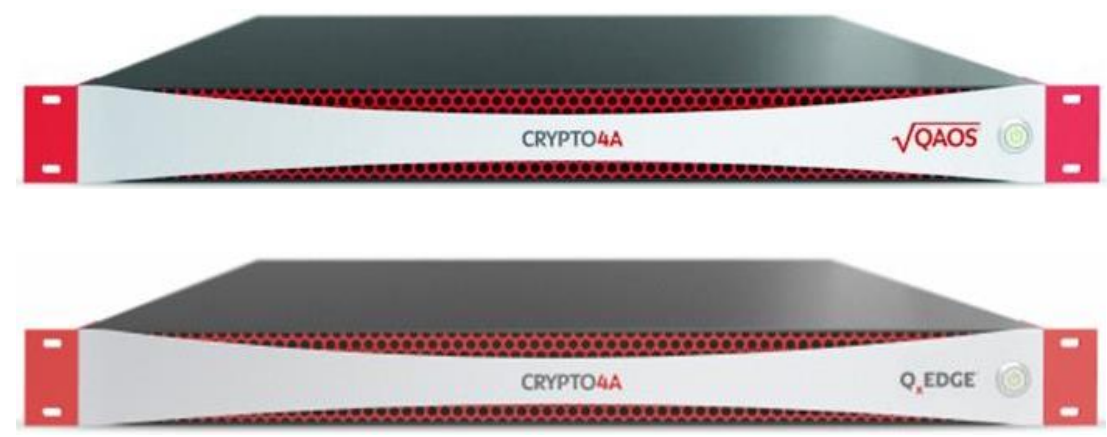

*Figure 783: CryptoAA QRNG based HSM products.*

**CryptoExperts** (2008, France) develops homomorphic encryption and post-quantum cryptography, and also offers services based on these technologies.

**CryptoNext Security** (2019, France) is a startup that develops a post-quantum cryptography solution. They were founded by Ludovic Perret (CPO, ex Inria, who left the company in 2022) and Jean-Charles Faugères (CTO, ex LIP6 Sorbonne) with Florent Grosmaitre as CEO.

Their software solution is developed in C language and assembler for performance reasons. It combines multivariate polynomials and hashing. Their solution can be integrated into RSA/ECC schemes by hybridization. CryptoNext is also one of the French teams who submitted a PQC proposal to the NIST which has been selected as an alternative candidate in 2020's round 2, GeMSS, which consolidates contributions from CryptoNext, Inria, Orange, University of Versailles and Sorbonne Université.

PQC standardization processors are used in practice by many organizations such as ISO, ITU (X509), IETF (TLS) and ETSI (algorithms). Their PQC should be integrated into R3's CORDA blockchain solution for banks. Note that China is also organizing a competition with a faster selection schedule than the NIST one. CryptoNext equips French special forces with their PQC, running on secure mobiles using Android.

**Crypto Quantique** (2016, UK, $8M) is a startup offering a cryptography solution to secure communication with connected objects targeting various markets ranging from automotive to finance. It uses a chipset that is installed in the object. It is a "quantum processor" in silicon technology that is used to generate a unique identification key for the object, which is tamper-proof and tamper-proof. It probably exploits photonics with a random number generator similar to the Swiss IDQ technologies.

Their technology is called Quantum Driven Physically Unclonable Function (QD-PUF) but they do not explain how it works or what encryption model is used[2506]. The founders are of Iranian, Italian and Greek origins, a beautiful patchwork. In July 2022, they announced that their platform was supporting the only selected PKI by NIST the same month, CRYSTALs-Kyber.

**Cyph** (2014, USA, $1M) sells PQC solutions. The company was created by Ryan Lester and Joshua Boehm, two engineers from SpaceX. Mars not interesting anymore?

**Dencrypt** (2013, Denmark) is a cybersecurity software provider protecting smartphone communications. They are working on creating PQC based solutions in partnership with the Technical University of Denmark (DTU).

**evolutionQ** (2015, Canada, $5.5M) is a startup that stands out especially for the pedigree of its creator, Michele Mosca, a specialist in post-quantum cryptography known for his Mosca scale to assess the risk of quantum computing on classical cybersecurity.

He is also the founder of the Institute for Quantum Computing at the University of Waterloo in Canada. The company provides what is known as "service equipped" to support companies in the adoption of post-quantum and quantum cryptography. It begins with a six-phase Quantum Risk Assessment product, documented in A Methodology for Quantum Risk Assessment, published in 2017.

Is it really a product? It looks more like a methodology to be implemented with consultants. The rest is of the same cream with integration and training services to evolve the company's cryptographic systems.

In July 2022, SandboxAQ announced a strategic partnership with evolutionQ including some funding a in series A round. It will help complement SandboxAQ's PQC offering with evolutionQ's QKD software platform, BasejumpQDN.

**Flipscloud** (2013, Taiwan) creates "quantum level" encryption software targeting IoT, cloud services and the big data market. It seems it is using some unspecified PQC and AES 256 bits keys. Software runs on embedded systems using Arm cores and Imagination CPUs.

**fragmentiX** (2018, Austria) offers a secure data storage management system that uses the technique of fragmentation and distribution of data on different physical media. All this is supplied in the form of appliances.

The distributed data is of course encrypted, but in a classical way. This is another way to create data protection that is resistant to Shor's algorithm. It is not the only company positioned in this niche.

They combined their appliances with QKD equipment from IDQ and Toshiba that was available at AIT. The theoretical proposal comes from researchers of TU Darmstadt in Germany and it has already been implemented a couple of years ago in Tokyo.

**GoQuantum** (2018, Chile) is willing to provide "*a post-quantum secure data transmission solutions through quantum-based hardware and radio link layer encryption*.". Translation: it's using PQC encryption algorithms with a photonic based QRNG for key generation.

**HaQien** (2019, India) designs post-quantum cryptography (PQC) solutions. But it is not quite clear because they also seem to use a random number generator to create classical keys.

[2506] See Physically Unclonable Functions: a Study on the State of the Art and Future Research Directions by Roel Maes and Ingrid Verbauwhede (36 pages) and Quantum readout of Physical Unclonable Functions by B. Skoric (21 pages).

**Hub Security** (2017, Israel, $55M) sells quantum secure FPGA based Hardware Security Module (HSM). These HSM modules contain QRNGs and support quantum-resistant algorithms acceleration in hardware (PQC).

**IDQ (**ID Quantique) (2001, Switzerland, $74.6M) is one of the oldest companies in the sector, cofounded by Swiss researcher Nicolas Gisin, a specialist in photonics and quantum entanglement.

The company offers a complete range of random number generators and QKD management systems as well as a high-efficiency (>95%) superconducting nanowire single photon detector (SNSPDs) as well as single-photon avalanche detectors (SPADs). These photon detectors are controlled by the ID1000 Time Controller Series introduced in January 2022.

Its Quantis random number generator, already described at the beginning of this section, is complemented by Cerberis, a QKD solution to protect the generation of encryption keys in a 6U rack and Centauris, a range of encryption servers supporting 100 GBits/s optical links. This FPGA-based server currently supports elliptic curve-based systems as well as AES-256, pending the standardization of PQC (post-quantum-crypto) protocols.

As of 2022, IDQ's Cerberis XG (enterprise market) and XGR (research market, supporting OpenQKD) Series was the fourth generation of this product line in a 1U form factor for both ends supporting various network topologies (ring, hub and spoke, meshed). It supports a 1.GHz key generation rate and includes a QRNG system for photon basis readout selection.

Since the beginning of 2018, the company belongs to the Korean group SKT Invest which is the Corporate Venture branch of **SK Telecom**. The fund invested $65M in what was modestly presented as a partnership while it's actually a full takeover.

IDQ's QKD offer is notably deployed in Korea to protect the 5G backbone of the operator SK Telecom. They are also partnering with Toshiba in Cambridge and in the OpenQKD project.

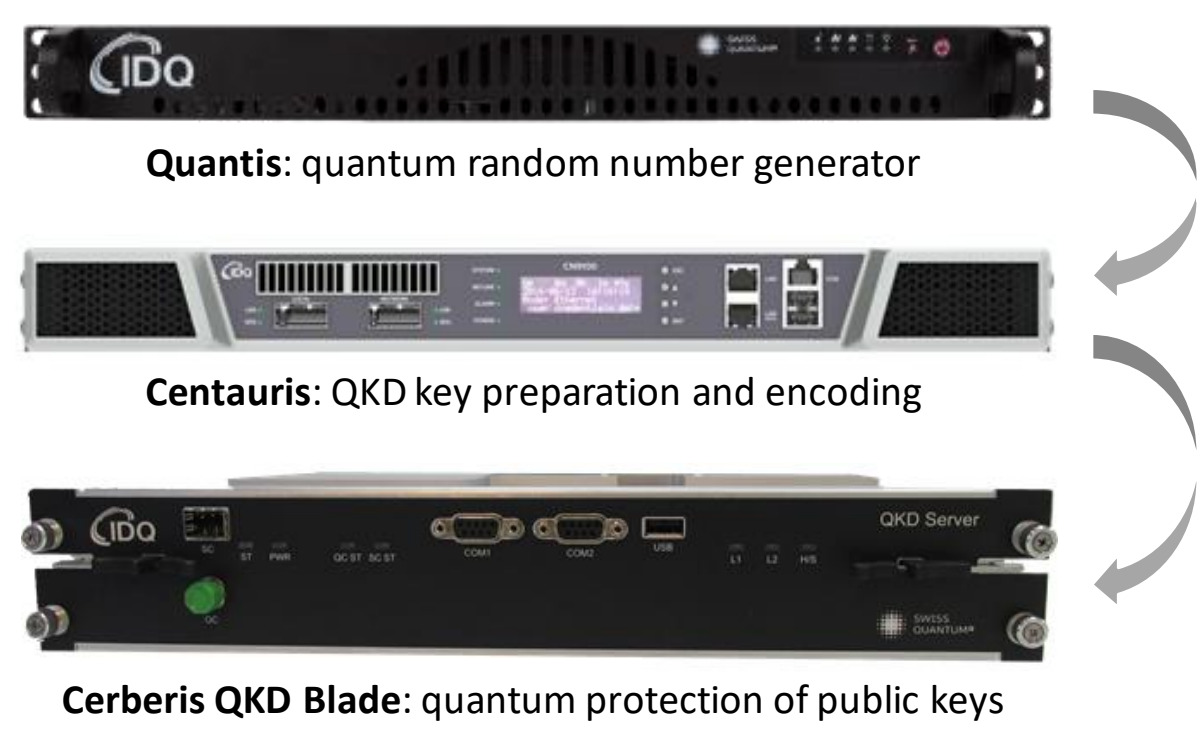

*Figure 784: IDQ's QKD offering.*

**Keequant** (2017, Germany, $1.7M), formerly InfiniQuant, is a spin-off from the Max Planck Institute for the Science of Light. They are developing a CV-QKD for use over fiber optics and satellite links.

This technique uses amplitude modulation in addition to phase modulation to transmit quantum keys. The startup is also working on a quantum random number generator.

**Infotecs** (1991, Russia) is a cybersecurity specialist historically specialized in VPN creation. It has developed a PQC solution in 2016, with an awkward communication that could make it look like QKD[2507].

But they develop many QKD solutions. In 2018, they launched their "ViPNet Quantum Phone", using their ViPNet VPN (ViPNet Client and ViPNet Connect) and a hardware QKD solution developed at Moscow University. What they call a "phone" is in fact a PC with an external box with a fiber optic link connecting it to a QKD key server[2508].

---

[2507] See Infotecs At The Forefront Of Quantum Cryptography, 2017.

[2508] See Infotecs has presented its ViPNet Quantum Phone, January 2018.

**ISARA** (2015, Canada, $27M) develops post-quantum encryption software solutions and PQC implementation consulting with "ISARA Radiate Security Solution Suite" which provides public keys and encryption algorithms.

They are visibly based on hash trees and combine PQC (post-quantum crypto) and traditional PKI (public-key infrastructure)[2509]. One of their investors is the Quantum Valley Investments fund, managed by Mike Lazaridis, co-founder of Blackberry RIM. He reinvested his Blackberry-related fortune in the development of the Canadian scientific and entrepreneurial ecosystem, particularly in quantum, where he has invested a total of $450M[2510].

**KETS Quantum Security** (2016, UK, £5.1M) develops a quantum random number generator (QRNG) and a QKD quantum key generator, all integrated in a single component miniaturized photonic and packaged in PCI cards.

All this is combined with a consulting activity for the deployment of the solutions. The company was founded by photonics researchers from the University of Bristol. They target the financial and public sector markets. They are prototyping UAVs with Airbus for QKD implementation in military or public security applications, with Airbus Defense. Their QKD chipset can also equip Cubesat-type microsatellites.

**Ki3 Photonics** (Canada) is a spin-off of the National Institute of Scientific Research on energy, materials and telecommunications from Montreal created by Yoann Jestin (CEO) and Piotr Rozgtocki (CTO)

It develops compact and energy efficient QKD hardware solutions to ease its deployment over standard telecommunications links with using signals multiplexing using frequency combs.

**Knot Communications** / **Artedys** (France) wants to launch a network of satellites to operate some sort of quantum blockchain satellite phone. They plan to launch a satellite between 2024 and 2027. This looks a little farfetched.

**MagiQ** (1999, USA, $7.5M) is a startup that initially started in 2003 with the creation of a QKD system. For about ten years, this company seems to have repositioned itself in the US service and defense industry. They have developed the Agile Interference Mitigation System (AIMS), a system for reducing electromagnetic communication interference.

**MtPellerin** (2018, Switzerland) is a startup specialized in the management of crypto assets via a dedicated mobile application ("Bridge Wallet"). They have created a quantum safe with IDQ, "The Quantum Vault", which is based on IDC's random number generator and QKD system.

**NodeQ** (2021, UK) is a software and service company dedicated to the deployment of quantum cryptography and telecommunications solutions. It was created by Stefano Pirandola and Samuel L. Braunstein. Stefano Pirandola pioneered continuous variable quantum cryptography, routing strategies on quantum networks using arbitrary topology, quantum repeaters, entanglement distillation and quantum computing optimization.

After working as a MIT researcher, he became a professor of quantum computing at the University of York (UK). Samuel L. Braunstein also worked on quantum teleportation with continuous variables and quantum teleportation networks.

---

[2509] This is documented in the white paper Enabling Quantum-Safe Migration with Crypto-Agile Certificates, 2018 (7 pages).

[2510] See The Co-Inventor of BlackBerry Is Building Canada's Quantum Brain Trust, Blomberg, 2018.

He also worked in quantum sensing and introduced the first bosonic model for universal quantum computing with Seth Lloyd in 1998[2511]. Both created the hybrid quantum Internet concept, association direct and continuous variables[2512].

**NuCrypt** (2003, USA) develops optical technologies for quantum communications and metrology, including entangled photon sources, optical pulse generators, single photon detectors, polarization analyzers and associated software.

**Nu Quantum** (2018, UK, £4.3M) is a spin-off from the University of Cambridge developing QKD optical links and satellite systems using proprietary single-photon components. They also created their own source of single photons and have some ambition to create a photon-based quantum computer of their own. The startup is co-founded and directed by Carmen Palacios-Berraquero.

**Origone** (2014, UK) develops cryptography solutions based on D-Wave computers. It targets the defense market as well as the railway industry. Their quantitative/post-quantum cryptography activity is an evolution of a traditional cybersecurity business.

**Patero** (2018, Germany) is a software PQC software and service vendor addressing the national security, critical infrastructures and supply chain markets.

It's bound to protect customers after "Q Day", the (elusive and very long term) moment when some quantum computer will break existing public-key based cryptography. It covers various parts of the IT stack: gateways, virtual cloud and edge computing. The company was cofounded by Henning Schiel who is their CTO.

**Post-Quantum** or **PQ Solutions** (2009, UK, $10.4M) is a startup initially created under the name SRD Wireless that created the secure PQ Chat messaging using the random linear codes invented by Robert McEliece.

The company was renamed as Post-Quantum or PQ Solutions Limited in 2014, or PQ Group. They offer a line of security products integrating post-quantum crypto algorithms. One of the co-founders, Martin Tomlinson, has developed the Tomlinson-Harashima pre-encoding, which corrects interference in telecommunication signals and various error correction codes. Their products also include PQ Guard, a post-quantum encryption system. Their CEO, Andersen Cheng is also the CEO of **Nomidio** (2020, UK), a provider of quantum-proof identity solutions, the Nomidio Private Identity Cloud.

**PQSecure Technologies** (2017, USA) is a provider of isogeny-based PQC solutions. It is a spin-off from the University of Florida Atlantic launched by Reza Azarderakhsh. Their SIKE algorithm is a finalist in the NIST PQC call for proposals.

**PQShield** (2018, UK, $26.9M) is Oxford University spin-off that develops PQC solutions, including HSM (hardware security modules).

They participate to several finalist teams in the PQC competition launched by NIST including the CRYSTALS-Kyber PKI and CRYSTALS-Dilithium's signature that were selected in July 2022. They have developed a licensed SoC (system on chip) that integrates their in-house Euclidian networks based PQC. Bosch is one of their first customers.

---

[2511] See Quantum computation over continuous variables by Seth Lloyd and Samuel L. Braunstein, 1998 (9 pages).

[2512] See Physics: Unite to build a quantum Internet by Stefano Pirandola and Samuel L. Braunstein, Nature, 2016 (3 pages).

**Qaisec** (2019, Bulgaria) develops cryptographic solutions targeting the AI, finance and telecommunications sectors. They seem to offer first a security audit service and then cryptography solutions that use quantum random number generators for keys. They are also creating a PQC-based blockchain.

**QANPlatform** (2019, USA) was cofounded by Johann Polecsak (CTO) and is developing a quantum-proof Blockchain relying on a Lattice-based PQC implemented in the RUST programming language.

**QuantLR** (2018, Israel) is a developer of an affordable QKD software solution that is supposed to reduce the cost of QKD deployment by 90% with "*no specific hardware*".

The startup was cofounded by Hagai Eisenberg, a professor at Hebrew University of Jerusalem. When looking at a recent paper he coauthored[2513], you get an idea of what they may be doing.

It consists in using a high-dimensional QKD architecture that can work on existing binary QKD hardware on distances of up to 40 km, thanks to using some time-bin encoding programmed on a FPGA component.

They also partner with a telecom equipment manufacturer, **PacketLight Networks** (2000, $18M), also based in Israel, which provides optical fiber WDM (wavelength-division multiplexing) equipment. They also tested their solution with **Medone**, an Israeli datacenter service provider. In a sense, the startup's go-to-market is original, relying on Israeli local players when many Israeli startups usually directly go and reach international players and partners, particularly in the USA.

**Qabacus** (2019, USA) is developing quantum computing and cryptography technologies and a complete cyber-security software stack.

**Qasky** (2016, China) commercializes research coming out of the Chinese Academy of Sciences. Funding comes from Wuhu Construction and Investment Ltd and the China University of Science and Technology.

They offer solutions for post-quantum crypto, QKD and photonics components. Their name is derived from CAS Key laboratory, CAS = China Academy of Sciences.

**QEYnet** (2016, Canada, $7M) is developing a QKD quantum cryptography satellite network. Funding for the startup comes from the Canadian government.

**QNu Labs** (2016, India, $5.3M) develops QKD-based solutions. They also offer their own quantum random number generator and are also working on the creation of a QKD solution operating on Li-Fi, W-Fi that uses the frequencies of visible light.

**Qrate Quantum Communications** (2015, Russia) sells QRate Key Distributor, a BB84 protocol-based QKD quantum key distribution solution in a 4U rack with a range of up to 100 km.

They also market a single-photon avalanche diode detector (SPAD) operating at 1550 nm and a quantum random number generator (QRNG).

[2513] See Fast and Simple One-Way High-Dimensional Quantum Key Distribution by Kfir Sulimany, Hagai Eisenberg, Michael Ben-Or et al, May 2021 (7 pages).

**QSpace Technologies** (2017, Russia, $1M) is a developer of QKD satellites that is building a CubeSat with a QKD system transmitter, to be launch in 2023. It is a spin-off from the Russian Quantum Center, itself coming out of QRate in 2021.

**QuantiCor Security** (2017, Germany) develops PQC solutions, particularly for Blockchain applications and connected objects, via offerings with Quantum-Multisign and Quantum IDEncrypt.

That they are supposed to be cheaper than traditional PKIs. They come from TU Darmstadt and target the healthcare market in particular.

**Quantum Blockchains** (2018, Poland) wants to create a quantum resilient blockchain based on QKD. That's an interesting long-term bet given the infrastructure required to make it happen at a large scale.

**QuDoor** (2016, China, $7,8M) aka **Qike Quantum**, aka **Quantum Door**, aka **Guokai Quantum Technology** designs various products for QKD distribution, a trapped ion quantum processor supposed to reach someday 100 qubits (Tiansuan 1) and a laser-based vibration quantum sensor. QuDoor's cofounder and R&D track record include a commercial QKD system (2003), a waveform generator (2007), ion trapping (2012), quantum computing sensing (2015) and ion-phonon-photon entanglement function (2018).

**Quantum Collective** (2021, Netherlands) is a European company created by Fabien Bouhier, Sebastien Le Goff and Floris Drupsteen that provides post-quantum security solutions. They name this "Quantum Security Solutions" but it's classical quantum-resilient security covering PKI, VPN and eMail security.

**QuantumCTek** (2009, China) is a provider of end-to-end quantum cryptography solutions: QKD, QKD repeaters, optical routers. The company is a spin-off of Hefei National Laboratory for Physical Science at Micro-scale (HFNL) and the University of Science and Technology of China (USTC).

They are behind the creation in 2014 of the "Quantum-Safe Security Working Group" with ID Quantique and Battelle, which promotes PQC. As we saw above, they have deployed the 2000 km QKD-protected link between Shanghai and Beijing. They ran an IPO (Initial Public Offering) in China in July 2020.

**Quantum Impenetrable** (2018, UK) is a Scottish startup that develops a security module (HSM) using a quantum random number generator and resistant to quantum key-breaking algorithms.

**Quantum Xchange** (2016, USA, $23.5M) distributes Phio Trusted Xchange, a key distribution system supporting both PQC and QKD.

They partner with the telecom infrastructure operator Zayo Group, from which they operate their dark fibers and use ID Quantique's QKD solutions. They began by deploying a 1,000-kilometer QKD network from Boston to Washington via New York and New Jersey[2514]. Since 2021, they also partner with **Cisco** for the support of the Cisco Secure Key Integration Protocol implemented in enterprise routers, for the non-quantum part of their hybrid key distribution architecture.

[2514] See Quantum Xchange Breaks Final Barriers to Make Quantum Key Distribution (QKD) Commercially Viable with the Launch of Phio TX, September 2019.

In 2022, they announced a partnership with **Thales** for the delivery of security keys with Thales High Speed Encryptors (HSEs) that supports QRNGs, QKDs and PQCs[2515].

**Quantum eMotion** (2007, Canada), formerly Quantum Numbers Corp**,** develops a cryptographic system based on a quantum random number generator (branded QNG2) and targets in particular mobile uses. It mainly communicates on the filing of an associated patent.

Let's hope it won't be a patent troller! The company seems to be licensing its technology to electronic component designers. It exploits research work from the Department of Physics at the University of Sherbrooke in Quebec. One of the patents relates to the generation of random numbers based on the random noise generated by some electron tunneling effect through a potential barrier[2516]. The QRNG speed reaches 1 Gbits/s and fits in USB key form factor. The company is listed on the Canadian Venture Exchange (CVE). The company expect to release a Blockchain using their QRNG in 2023.

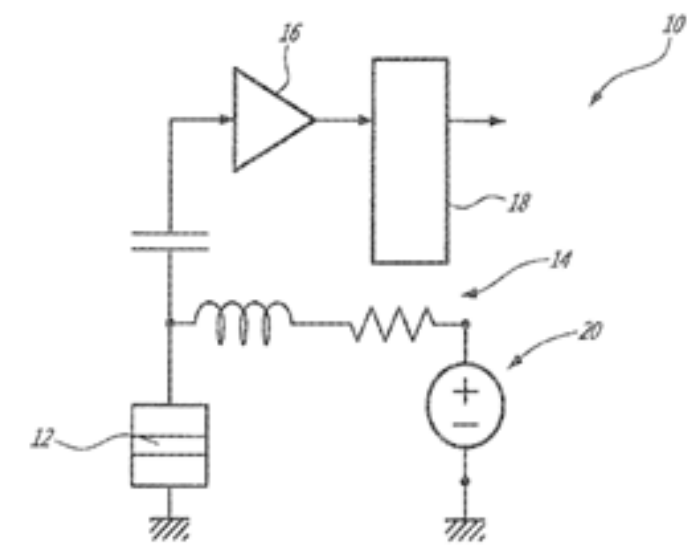

*Figure 785: a patented QNRG.*

**Quantum Trilogy** (2016, USA) created a secure communications solution based on applications (including for mail and voice communication) protected by an unspecified encryption, a QRNG to create real entropy in generated keys and servers that are protected at some level by a QKD.

**QuBalt** (2015, Germany) is a startup established between Germany and Latvia that develops solutions for post-quantum cryptography (PQC) and quantum algorithms.

**Qubit Reset** (2018, USA) develops quantum repeaters for QKD arrays. The company was founded by two Argentinians based in Miami. It is not listed in the Crunchbase and does not seem to have raised any funds, which seems to be a bad omen.

**Quintessence Labs** (2006, Australia, $49,3M) proposes a quantum random number generator and a QKD system. They use the CV-QKD technique which allows the use of existing fiber optic infrastructures of very high-speed telecom operators.

**Qunnect** (2017, USA, $12.4M) is a spin-off from Stony Brook University of Long Island created by Mehdi Namazi, Eden Figueroa, Noel Goddard and Mael Flament, Quantonation being one of their investors[2517]. Qunnect is developing a quantum product suite that enables long range quantum entanglement distribution protecting against photons loss.

It is based on a quantum memory operating at room temperature that can be used in quantum repeaters, using warm rubidium vapor and without any requirement for vacuum- and/or cryogenic- support[2518].

---

[2515] See Quantum Xchange Collaborates with Thales to Enable Quantum-Safe Key Delivery Across Any Distance, Over Any Network Media, November 2021.

[2516] I found the complete PDF of USPTO patent 10437559 on https://www.pat2pdf.org/.

[2517] They received $1.5 million of funding in April 2020 from the US Department of Energy under the Small Business Innovation Research Awards (SBIR) program. See Qunnect receives $1.5M award from the DoE - Swiss Quantum Hub, April 2020.

[2518] See Field-deployable Quantum Memory for Quantum Networking by Yang Wang, Alexander N. Craddock, Rourke Sekelsky, Mael Flament and Mehdi Namazi, May 2022 (16 pages). See also an explanation video, Quantum Memories for long distance quantum networking (3:40mn) and Quantum Repeaters which explains how it is deployed.

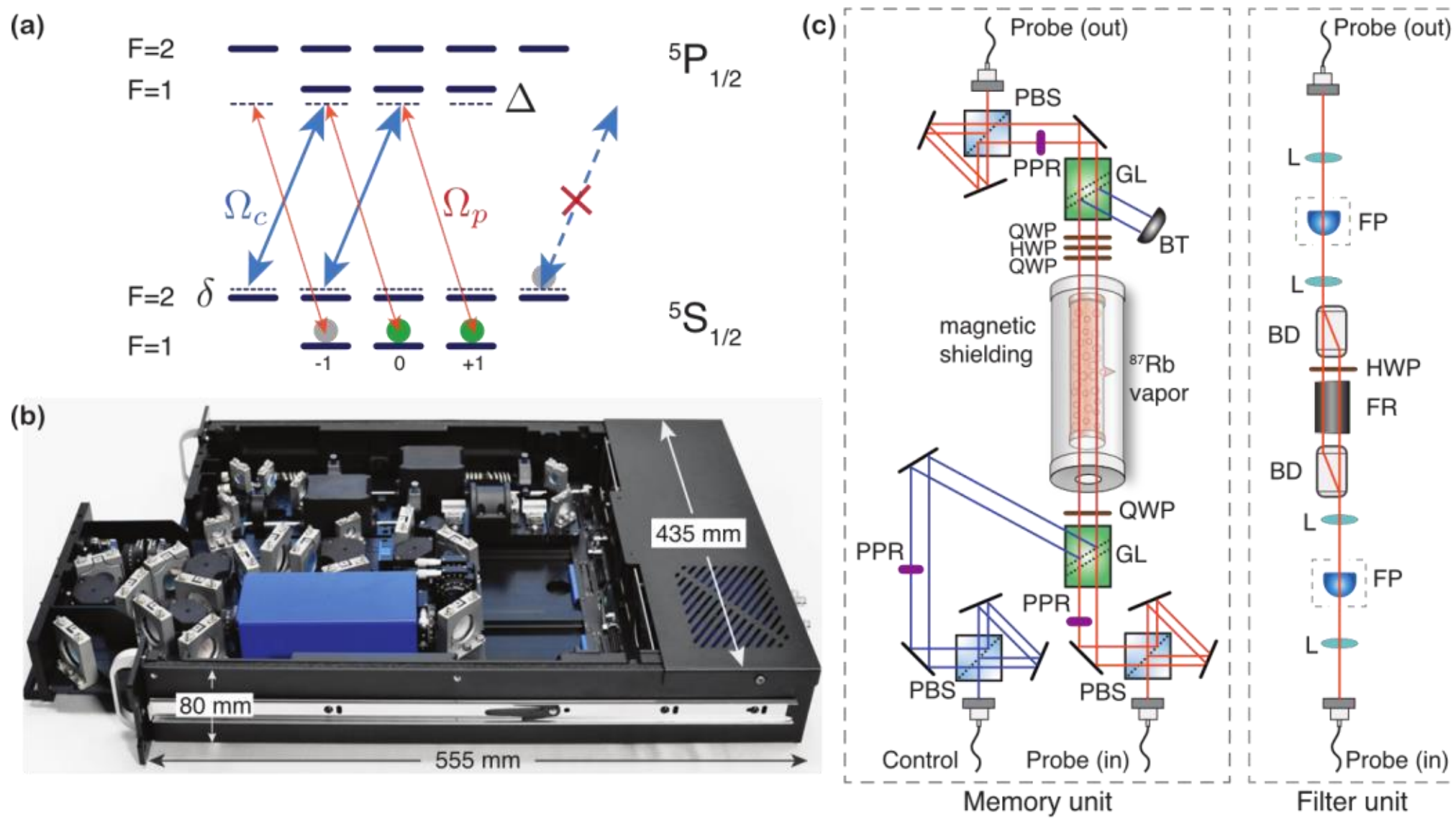

*Figure 786: Qunnect repeater architecture. Source: Field-deployable Quantum Memory for Quantum Networking by Yang Wang, Alexander N. Craddock, Rourke Sekelsky, Mael Flament and Mehdi Namazi, May 2022 (16 pages).*

In May 2022, the company got two SBIR awards from the US DoE for $1.85M to fund the development and commercialization of their quantum repeater product suite.

In October 2022, the startup got a series A funding of $8M led by Airbus Ventures Qunnect officially announces its Series A financing of over $8 million. The round was led by Airbus Ventures with Quantonation, SandboxAQ, NY Ventures, Impact Science Ventures and Motus Ventures as other investors.

**QuSecure** (2019, USA, $1.7M) develops a secure blockchain solution that is resilient to quantum code breaking.

It seems that they are also developing a Blockchain that would be secured via QKD, and Blockchain security testing protocols. The startup is also doing cybersecurity consulting and, in particular, audits for the deployment of PQC. It was founded by Rebecca Krauthamer, who also founded the **Quantum Thought** startup mentioned above. In 2022, they were awarded a SBIR Phase III contract to become, surprisingly, the sole-source provider of PQC for over a dozen US federal agencies with their QuProtect solution.

**Ravel Technologies** (2018, France) offers Ravel Homomorphic Encryption, a post-quantum and homomorphic encryption solution. The company was founded by Mehdi Sabeg.

**SandboxAQ** (2022, USA, >$300M) is a spin-out of Alphabet's Google AI that has a large breadth of activities, mostly around quantum related software applications and PQC software solutions.

Their first product offering is a multifunction Security Suite supporting PQC cryptography. They were selected in July 2022 as one of the 12 NIST partners to help the industry in the PQC migration path. The company also works on various quantum algorithms and on classical machine learning tools to exploit content from quantum medical image sensors The company's CEO is Jack Hidary, a serial entrepreneur and book writer ("Quantum Computing: An Applied Approach") and its chairman is Eric Schmidt. In 2022, they made a strategic investment in evolutionQ and acquired Cryptosense (2013, France, $5.7M).

**Secure-IC** (2010, France) is the leader of the RISQ project, which aims to create a French post-quantum crypto solution.

The company develops security hardware and software solutions that are used to evaluate the robustness of security solutions. The company is a spin-off from the Institut Mines-Télécom.

**SECQAI** (2021, UK) is a quantum cybersecurity company with a very broad goal to "*provide world leading technologies to organisations of all sizes*".

More precisely, they're in to "*bringing together their own quantum and AI tech to create disruptive data and security products and services for organisations looking to gain a commercial advantage*". The company was created by Rahul Tyagi, Dave Worrall and Martin Rudd. Rahul Tyagi developed a patented room temperature QRNG when he as the CEO of LyfGen. The patent USPTO 10606561 was filed in August 2018 and validated in March 2020. It's based on using slits-based diffraction and scattering entropy.

**Smarts Quanttelecom** (1991, Russia) proposes a quantum cryptography solution based on CV-QKD which exploits standard fibers of telecom operators.

Smarts was until 2015 a Russian mobile telecom operator. It has since become a telecom services operator with an offer of secure telecom links and data center and cloud services. Their QKD solution comes from Quanttelecom, a subsidiary of Smarts, developed jointly with the ITMO University of Saint Petersburg.

**Sonora Gold and Silver Corp** (2019, Canada) acquired in June 2021 a newly created startup, BTQ, that works on creating some quantum-safe Bitcoin solutions. BTQ was created in March 2021 in Lichtenstein.

**SpeQtral Quantum Technologies** (2017, Singapore), formerly S-Fifteen Space Systems, specializes in the distribution of QKD via satellite. They promote the work of the University of Singapore in the design of CubeSat-type pico-satellites for QKD key distribution. In September 2022, they started a partnership with Thales Alenia Space to test in 2025 a QKD distribution between their SpeQtral-1 satellites and Thales Alenia Space ground stations.

**SSH Communications Security** (1995, Finland, 14.1M€) is a cybersecurity company selling cybersecurity and SSH solutions. They sell NQX, a PQC-ready encryption solution among other cybersecurity solutions. In 2022, they released Tectia Quantum Safe Edition, a PQC protected SSH solution that protects remote access, file transfers and tunneling connections against the future quantum threat.

**Surrey Satellite Technology** or SSTL (1985, UK) wants to deploy a QKD-based quantum communications satellite built by Airbus Defense & Space (they are part of the Airbus group). The project is conducted in partnership with Eutelsat and ESA. The launch of the satellite was planned for 2020.

**Synergy Quantum** (2019, Switzerland) is a post-quantum encryption product and service startup that also supports data storage in a facility located in a former military bunker in the Swiss Alps.

They target governments, financial services, healthcare, smart Cities, energy and telecom sectors. The company is run by Jay Oberai (CEO), Arun K. Pati (Chief Scientific Officer), Mayank Sharma (VP Engineering), Vipin Kumar Rathi (VP Technology Architecture) and Manu Khullar (Chief Operating Officer). Any non-Indian there? Yes, with John Paukulis (CMO, USA) and Sasha Lazarevic (Head of Government Partnerships) and Jake Hwang (Head of Business Development and Investor Relations). It launched in 2022 a joint venture with the Quantum Technology Hub of the Indian Government. Not a coincidence. They seem to also offer some services in the QKD realm, including as part of this deal in India.

**Taqbit Labs** (2018, India) sell solutions for QKD. Their website and communication are so vague that it's impossible to understand whether they sell third party QKD products with some integration service of develop QKD hardware products of their own, and if so, in what category given most vendors don't reinvent the wheel to become full-stack vendors.

**ThinkQuantum** (2021, Italy) is a spin-off of the University of Padua which develops QKD and QRNG solutions for simple fiber optics deployments as well as for free space and satellite QKD.

**Ultimaco** (1983, Germany) sells lattice-based PQC embedded in Hardware Security Modules (HSMs). This solution is based on Picnic, one of the alternate candidate PQC digital signatures selected by the third round of the NIST PQC challenge in July 2020.

**VeriQloud** (2017, France) is a startup created by Elham Kashefi, and Marc Kaplan (France) and Joshua Nunn (UK). It is specialized in the creation of quantum software solutions adapted to quantum telecommunications for both QKD and distributed quantum processing.

Their offer is based on the Qline, a software solution that enables the deployment of a multi-point quantum network with a lower cost in hardware infrastructure. With this architecture, nodes that are very expensive can be replaced by simple modulators that are much more affordable. The operation is based on a kind of "time-sharing" of the line. This lowers the hardware addition by about two thirds on a typical installation below with two intermediate stations in the network. At both ends of the line there is on one side a laser and modulator-based photon generator and on the other side a single photon detector, the most expensive part of the equipment ranging from 20K€ to 100K€. The solution is deployable on networks totaling 100 to 200 km.

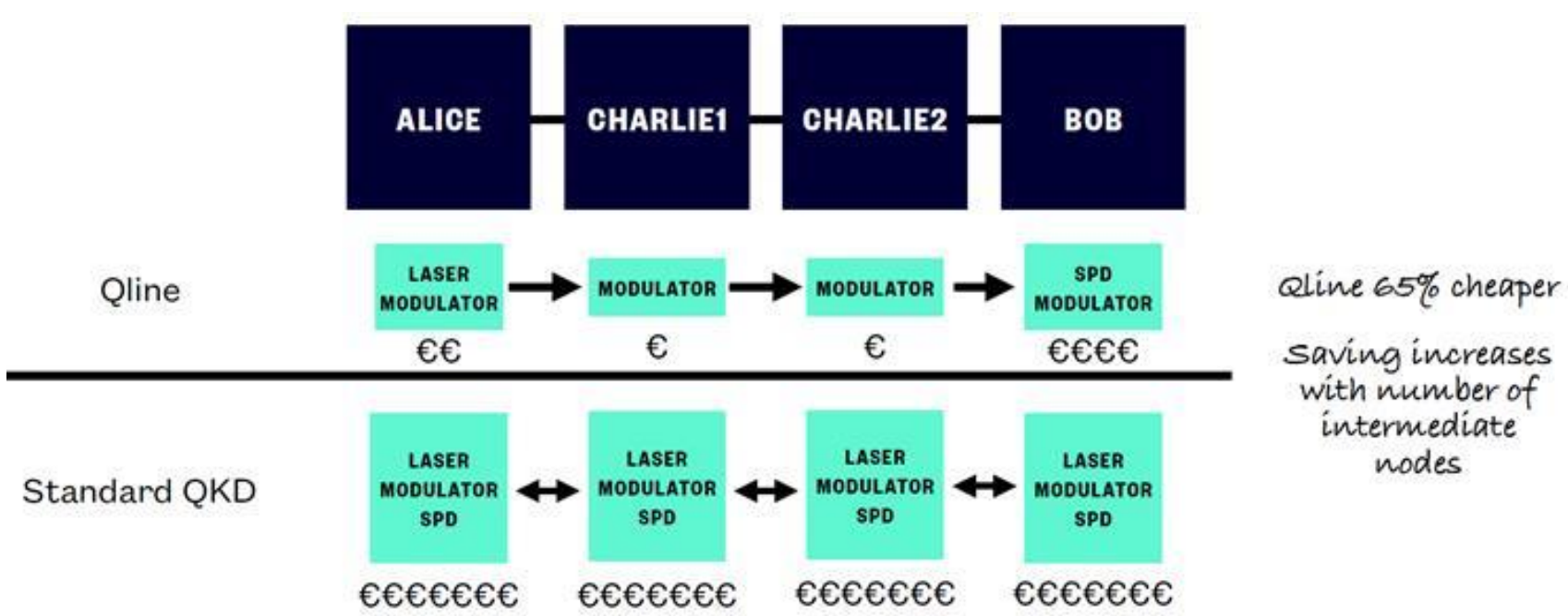

*Figure 787: VeriQloud qline architecture. Source: VeriQloud.*

The first application is QKD quantum key distribution. They can interoperate with classical QKD networks. Initially, secure file transfer and instant messaging are targeted as applications associated with QKD. The system can also be used to generate disposable masks[2519].

The VeriQloud solution is also relevant for securing data storage.

**XT Quantech** (2017, China) specializes in CV-QKD distribution equipment after initially focusing on DV-QKD solutions. The CV-QKD is essential because it can coexist in the fiber links of telecom operators.

They offer server appliances for QKD key encoding and decoding gateways. Its full name is Shanghai Xuntai Information Technology Co. The company also sells a QRNG, their XT-QRNG100 that is certified in … China.

**ZY4** (2014, Canada) develops post-quantum cryptography solutions based on their in-house concept of the Shannon Event Horizon which would be a new class of PKI and random number generation[2520].

[2519] See How to build quantum communication networks at a small scale by Marc Kaplan from VeriQloud, May 2020.

[2520] See their white paper Introducing the Shannon Event Horizon, 2019 (20 pages).

### Quantum telecommunications and cryptography key takeaways

- Quantum computing poses a theoretical threat to many existing cryptography systems, particularly those using public key distribution in asymmetric cryptography. This is due to Peter Shor's integer factoring algorithm that breaks RSA public encryption keys. But other algorithms than Shor are creating various threats, including for symmetric key distributions.
- As a result, two breeds of solutions have been elaborated. The first one is based on quantum key distribution, requiring a photonic transmission channel (airborne or fiber-based), and using sometimes quantum entanglement. Beware of a common misconception: these solutions are not based on quantum computers. Quantum computers are not (yet) making cryptography safer.
- The second option is to create classical cryptographic protocols generating public encryption keys that are not breakable by quantum computers. The USA NIST launched in 2016 an international competition to standardize a set of post-quantum cryptography (PQC) protocols. Four finalists were selected as NIST standard in July 2022, one for a public key interface (PKI) and three for signatures. Solutions deployments should happen next and be done way before the quantum computing menace will materialize, if it does some day. NIST may select other PQC solutions in the future.
- Quantum random numbers generation is/will be used for classical cryptography, and quantum cryptography. It provides sources of both random and non-deterministic numbers used in cryptography systems. It has other used cases when randomness is mandatory in classical computing like with lotteries and various simulation tools.
- Quantum telecommunications can also use quantum entanglement to enable communications between quantum computers and/or quantum sensors. Distributed quantum computing has two potential benefits: scale quantum computing beyond the capacity of individual quantum computers and enable safe communications between quantum computers. Distributed quantum sensing can enable better accuracy sensing.
- Quantum Physical Unclonable Functions are cryptographic solutions used to authenticate physical objects in an unfalsifiable quantum way. It is however still an unmature technology.
- There are already many startups in the QKD and PQC scene. Deployments have already started worldwide, particularly in China.

# Quantum sensing

Quantum sensing is about the various precision measurement solutions that rely on second generation quantum technologies and go beyond the limits of classical measurements systems. It also often allows non-invasive measurements to be carried out on various solid or organic materials. The main physical values measured are time, distances, gravity, magnetism, temperature and electromagnetic spectrum analysis.

## Quantum sensing use-cases and market

Many applications use these technologies like radars, sonars, very high sensitivity microphones or the field of imaging in general and particularly in medical imaging[2521].

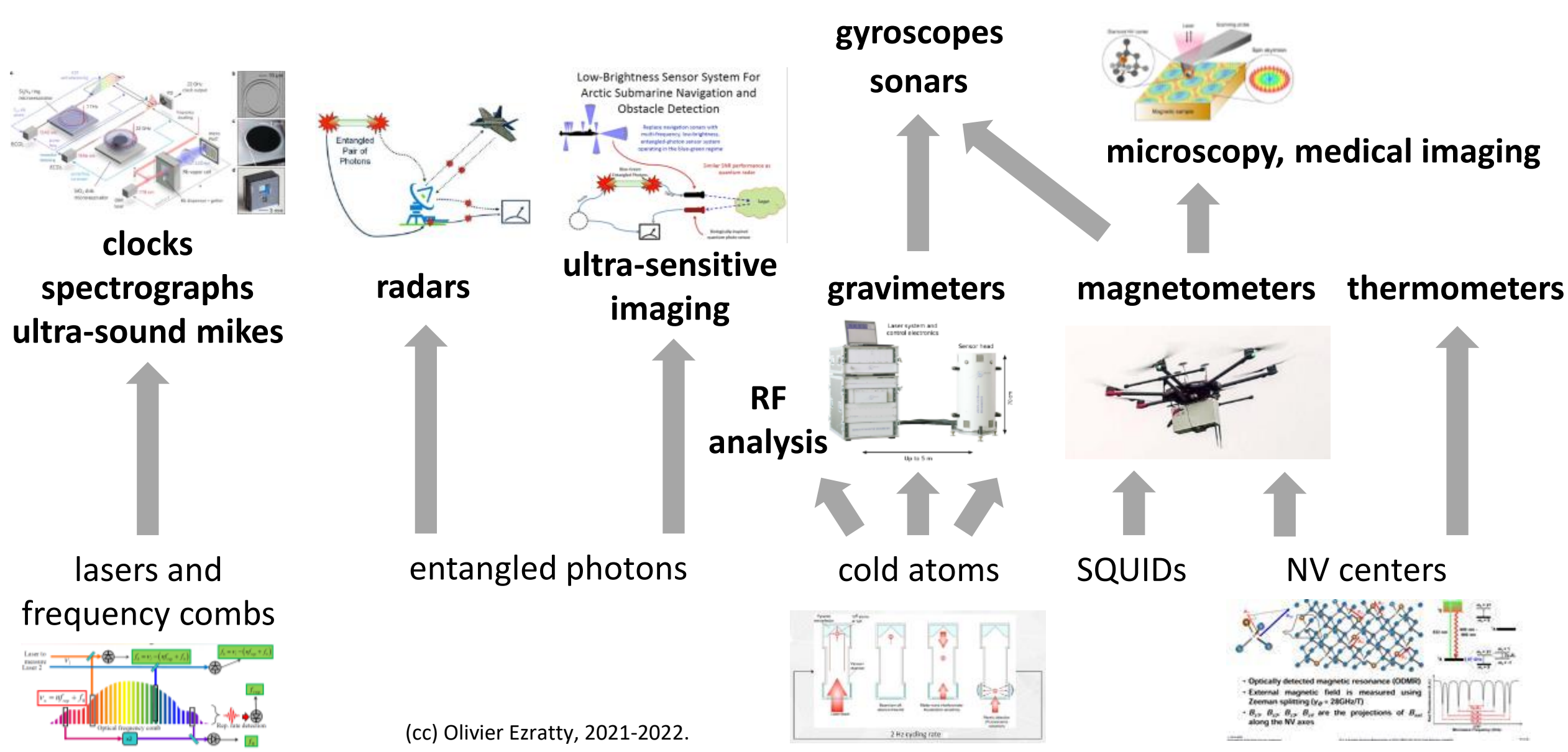

*Figure 788: a map of various quantum sensing basic technologies and use cases. (cc) Olivier Ezratty, 2022.*

Some of these technologies have commonalities with the various qubit types we have already explored in detail. This is particularly the case for cold atoms, NV centers and superconducting qubits. Precision magnetometers use NV centers as well as SQUIDs (Superconducting Quantum Interference Device), which also measure the direction of current in superconducting flux-type qubits and are used in particular by D-Wave in quantum annealers.

Many of these quantum measurement technologies make extensive use of photonic tools, either directly based on photons (lasers, frequency combs, entangled photons) or exploiting cold atoms and even NV centers, whose state is then evaluated by measuring their fluorescence.

[2521] See Quantum Sensing Use Cases 2022, SRI for QED-C, September 2022 (36 pages), a report on quantum sensors which makes an interesting inventory of use cases and market readiness, with an eye on national security issues. It is focused on four types of sensors, for time, inertial, magnetic field and electric field measurement. It doesn't cover other sensors categories reviewed in this part like frequency monitoring sensors, quantum radars (which may be mythical objects), quantum pressure sensors, quantum chemical sensors and quantum thermometers.

Some of these technologies are already commercially viable and continue to progress steadily. It is still a niche market made of many sub-niche markets, evaluated at around $1B and expected to double in a decade. The two largest applications markets are transportation and medical imaging.

But these forecasts may become wrong since some use cases might at some point become mainstream and drive more market growth. This is the case of GPS without satellite links using micro-magnetometers. It could for example someday equip many classical and autonomous vehicles.

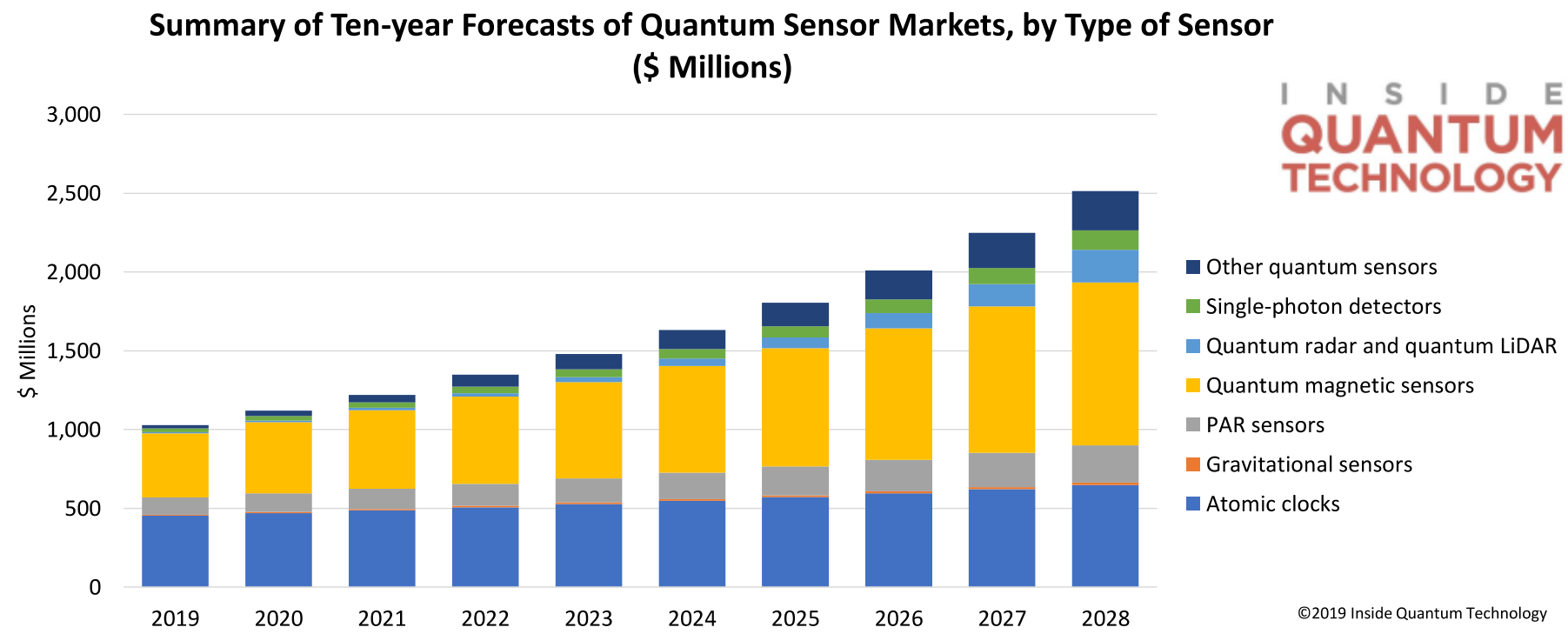

*Figure 789: Source: Quantum Sensors: Ten Year Market Projections by Lawrence Gasman, 2019 (7 slides).*

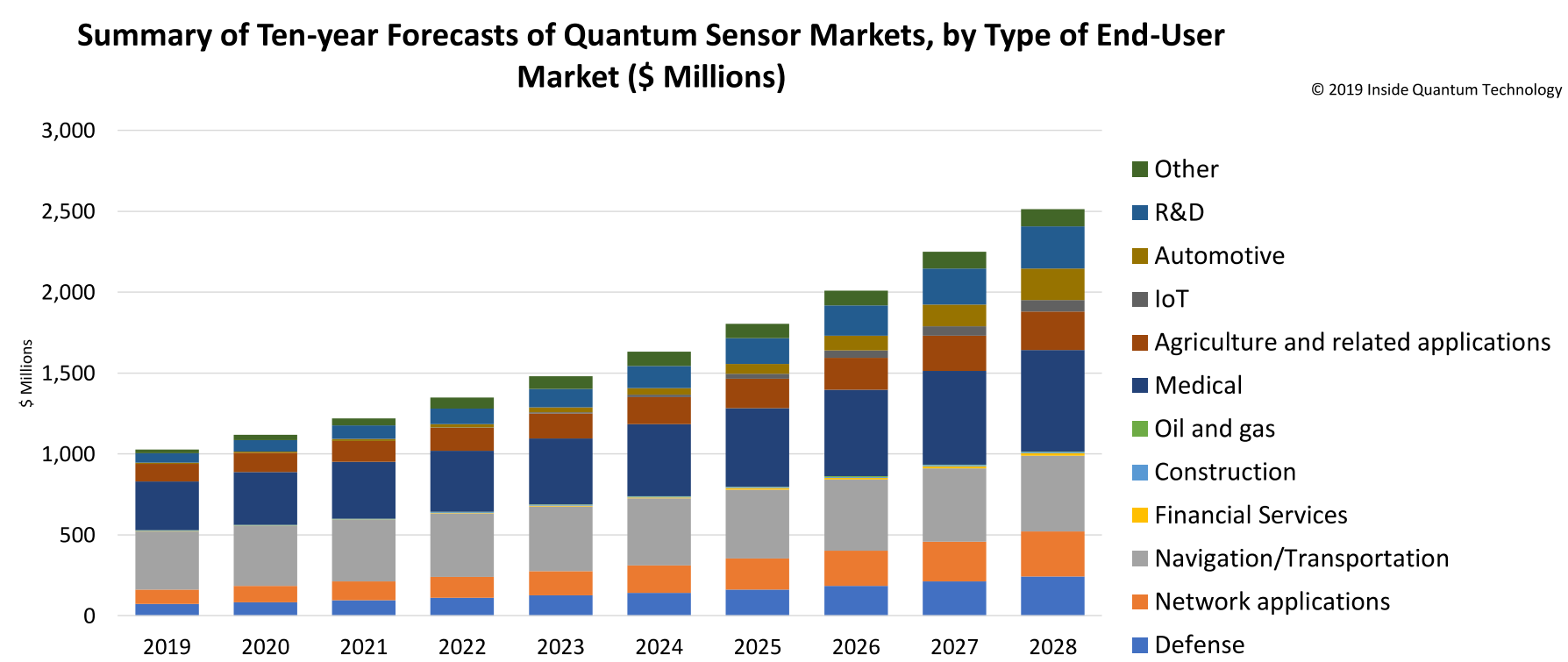

*Figure 790: Source: Quantum Sensors: Ten Year Market Projections by Lawrence Gasman, 2019 (7 slides).*

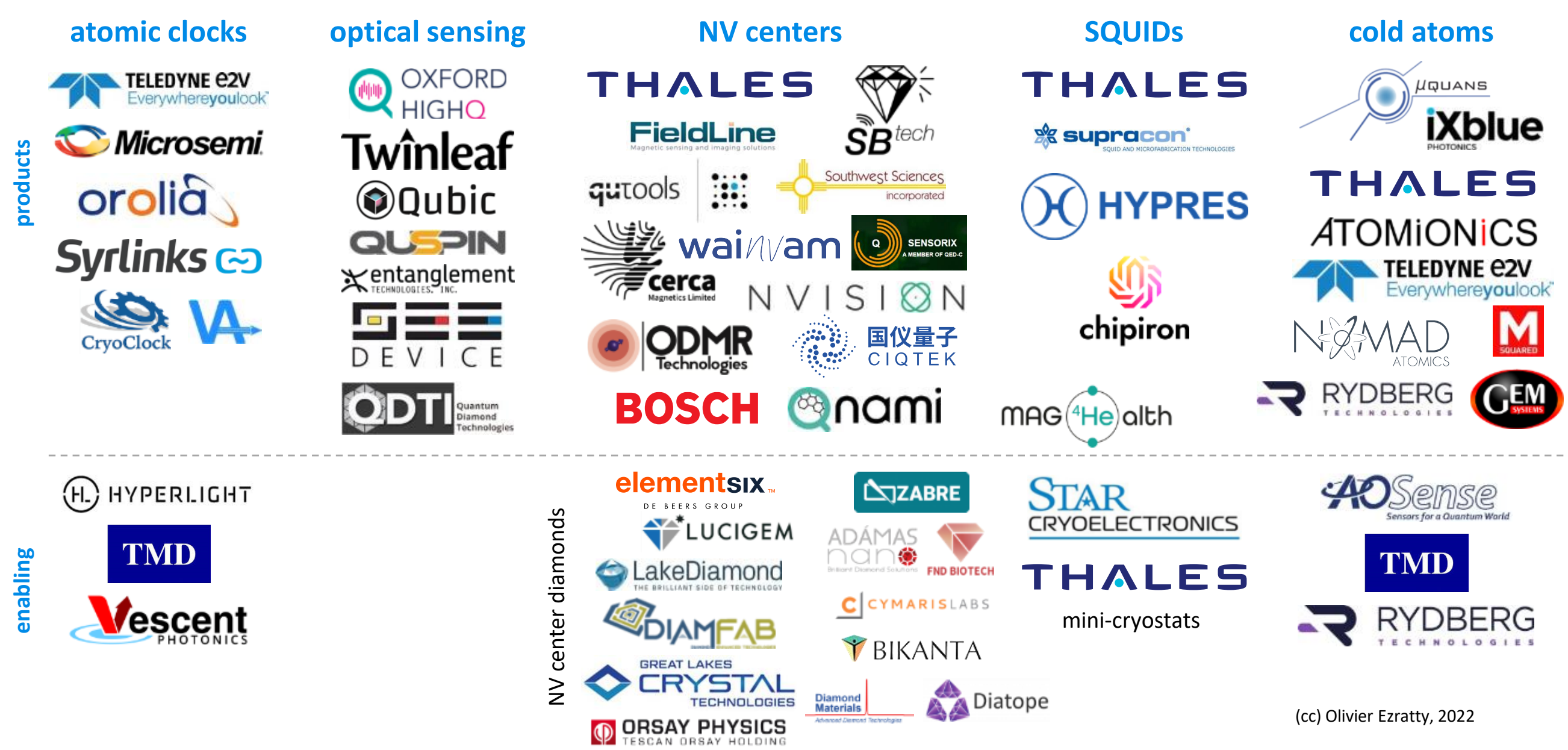

*Figure 791: and now ladies and gentlemen, here is the magnificent market map for quantum sensing, including some of their enabling technologies. (cc) Olivier Ezratty, 2022.*

Various other markets are relevant for quantum sensing: the energy and utilities sector[2522], telecoms, space and astrophysics research[2523], high-energy particles research[2524], civil engineering, manufacturing industries using many sensors like for quality control and non-destructive testing and control, and of course, the aerospace and defense.

## International System of Measurement

Sensing cannot be discussed without being connected to the **International System for Measurement** (or International System of Units, aka SI). It was recast after a unanimous vote on the Versailles Metre Treaty signed at the 26th General Conference on Weights and Measures (CGPM) in November 2018[2525]. Its implementation started on May 20, 2019.

The new 2019 SI updates the definition of the kilogram, ampere, kelvin and mole. It is built around seven fixed constants: a number of hyperfine transitions of cesium 133, the speed of light in vacuum[2526], the Planck constant, the elementary charge of an electron, the Boltzmann constant, the Avogadro number or constant and the luminous efficiency. From these constants are derived the seven basic units of the system: kilogram, meter, second[2527], ampere, kelvin, mole and candela.

It no longer relies on materials that are degrading over time, such as the standard kilogram kept at the BIPM in Saint-Cloud, or on the triple point of water (gel) which defined the kelvin, and which depended on its isotopic composition.

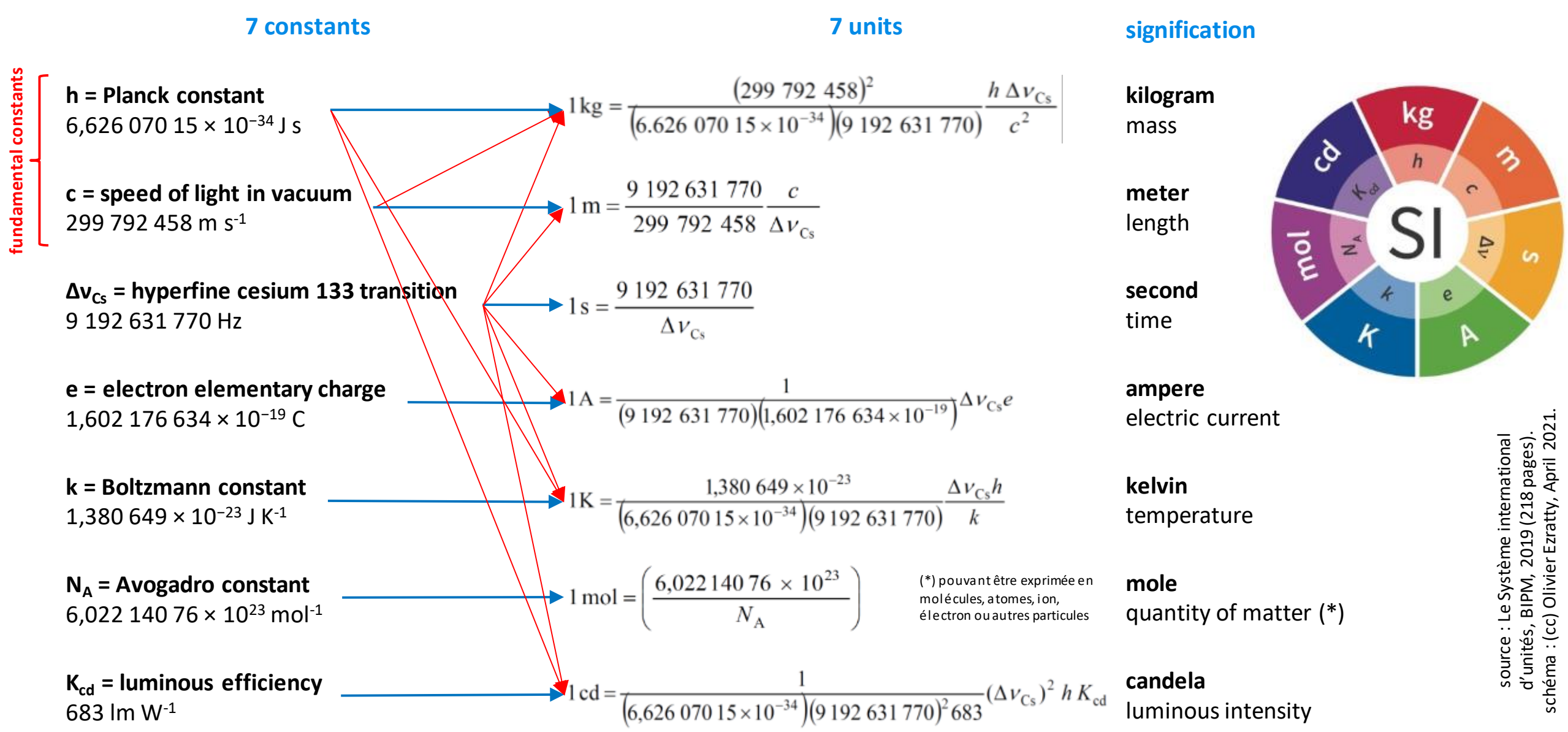

*Figure 792: reconstruction of the SI constants, units and their signification. Source: (cc) Olivier Ezratty, 2021.*

---

[2522] See Quantum Sensing for Energy Applications: Review and Perspective by Scott E Crawford, Roman Shugayev, Hari P. Paudel and Ping Lu, June 2021 (33 pages).

[2523] See Quantum Communication, Sensing and Measurement in Space by Baris I. Erkmen et al, 2012 (136 pages).

[2524] See High energy physics: Quantum Sensing for High Energy Physics by Zeeshan Ahmed, Yuri Alexeev, Giorgio Apollinari and Asimina Arvanitaki, March 2018 (39 pages) and Snowmass 2021: Quantum Sensors for HEP Science -- Interferometers, Mechanics, Traps, and Clocks by Daniel Carney et al, March 2022 (29 pages).

[2525] See The International System of Units (SI), NIST, 2019 (13 pages) and The International System of Units, BIPM, 2019 (218 pages).

[2526] The definition of the speed of light at 299 792 458 m $s^{-1}$ dates from 1983.

[2527] The second was defined with 9 192 631 770 periods of the hyperfine transition of cesium 133 at a temperature close to 0K since the 13th CGPM of 1967. Previously, it was a fraction of the solar day, which was not stable.

The mole was previously defined on the basis of 0.012 kg of $^{12}C$ [2528]. The standard meter, which is kept in the Archives in Paris, was no longer the reference since 1960. All other units of measurement such as hertz, joule, coulomb, lumen or watt are derived from constants and base units. This measurement system is branded as being "quantum" because it is based on the measurement of fundamental phenomena that bring us back to quanta, in particular for the definition of the second, which uses quantized energy transitions in the cesium atom, and that of the kilogram, which uses the Planck constant, itself a foundation of quantum physics.

Numerous quantum physics works related to these evolutions of the international measurement system, notably at NIST, have a link with the commercial devices discussed in this section.

# Quantum sensing taxonomy

Reusing a definition found on C.L. Degen et al's 2017 review paper[2529], quantum sensing describes the use of quantum systems, quantum properties and phenomena to measure physical quantities. The first quantum sensors were SQUID magnetometers, atomic vapors and atomic clocks. Then, the field was expanded to cover magnetic and electric fields, time and electromagnetic waves frequencies, mechanical rotations, temperature and pressure.

New "second generation" sensors work at the single-atom and spin level and may use quantum entanglement as a resource for increasing sensitivity.

Quantum sensing can use either or the following phenomenon corresponding to type I, II and III quantum sensors:

- **Type I**: using a quantum object to measure a physical quantity, characterized by quantized energy levels.
- **Type II**: using quantum coherence, wave-like spatial or temporal superposition states to measure a physical quantity.
- **Type III**: using quantum entanglement to improve the sensitivity or precision of measurement. It is sometimes labelled as "quantum metrology", "quantum-enhanced sensing" or "second generation quantum sensing".

By the way, the difference between quantum sensing and quantum metrology is not that clear in the scientific literature. For some people, it's the same. For others, quantum metrology is used for devices measuring units or fundamental constants such as an atomic clock or an atom interferometer measuring the fine structure constant, or constants that were used to create the last SI[2530]. Quantum metrology can have at least two other meanings: study of quantum-based precision measurements and quantum sensing using entanglement as seen in the above type I/II/III taxonomy.

The table in Figure 793 lists several types of known quantum sensors with their type, the physical nature of the measuring object ("qubit nature") and the measured physical quantities. For example, charged systems like trapped ions are sensitive to electrical fields while spin-based systems mainly respond to magnetic fields. Some quantum sensors may respond to several physical parameters or measure indirectly some physical quantity, which is the case for quantum thermometry with NV centers. Most of the time, quantum sensing exploits changes in the transition frequency or the transition rate in response to an external signal.

---

[2528] With the new SI, one gram of matter contains $N_A$ multiplied by the number of nucleons (protons and neutrons) of the element in question (atom, molecule). This comes from the fact that in an atom, the majority of the weight is in the atom nucleus. Electrons have a mass equivalent to 1/1836 times that of a single nucleon.

[2529] See Quantum sensing by C. L. Degen, F. Reinhard and P. Cappellaro, June 2017 (45 pages).

[2530] See What is the difference between quantum sensing and quantum metrology?, 2020.

| sensor type | | | qubit nature | type I | type II | type III | rotation | acceleration | force | pressure | displacement | time | frequency | refractive index | magnetic field | electric field | voltage | temperature | mass |
|---|---|---|---|---|---|---|---|---|---|---|---|---|---|---|---|---|---|---|---|
| neutral atoms | | atomic vapor | atomic spin | | X | X | X | | | | | X | X | | X | | | | |
| neutral atoms | | cold atom clouds | atomic spin | | X | X | | X | | | | X | X | | X | | | | |
| Rydberg atoms | | | Rydberg states | | X | X | | | | | | | | | | X | | | |
| trapped ions | | | electronic state | | X | X | X | | | X | | X | X | | | | | | |
| trapped ions | | | vibrational mode | | X | | | | X | | | | | | | X | | | |
| solid state | spin ensembles | NMR | nuclear spins | | X | | | | | | | | | | X | | | | |
| solid state | spin ensembles | NV/SiC center ensembles | electron spins | | X | | X | | | X | | | | | X | X | | X | |
| solid state | single spins | P donor in Si | electron spins | | X | | | | | | | | | | X | | | | |
| solid state | single spins | quantum dot | electron spins | X | X | | | | | | | | | | X | X | | | |
| solid state | single spins | single NV center | electron spins | | X | | X | | | X | | | | | X | X | | | |
| superconducting circuits | | SQUIDs | supercurrent | X | X | | | | | | | | | | X | | | | |
| superconducting circuits | | flux qubits | circulating current | | X | | | | | | | | | | X | | | | |
| superconducting circuits | | charge qubits | charge eigenstates | | X | | | | | | | | | | | X | | | |
| single electron transistor | | | charge eigenstates | X | | | | | | | | | | | | X | | | |
| optomechanics | | | phonons | X | | | | X | X | | | | | | X | | X | | |
| interferometer | | | photons, atoms | | X | X | X | | | | X | | | X | | | | | |

*Figure 793: quantum sensing taxonomy. Source: table reconstructed from Quantum sensing by C. L. Degen, F. Reinhard and P. Cappellaro, June 2017 (45 pages).*

The rest of this quantum sensing part is mostly organized along the first row and its green/blue measured values: gravity, time, magnetism, temperature, radiofrequencies and then to higher-level applications like imaging, radars and lidars and chemical sensors.

Quantum sensors are a bit paradoxical: after a measurement of a quantum sensing qubit, the only information you retrieve is a single classical bit (0 or 1). So how to you get a floating number with a large precision?

The answer is simple for the most basic cases: you make many measurements and average their results. But there are other subtleties. With NV centers sensors, you create a spin resonance spectrum with repeat measurements and the shape of the resulting curve enables you to determine the detected magnetic field with computing the distance between two curve peaks. It is an indirect measurement.

A quantum sensor measurement precision will depend on several parameters: the sampling rate (how many measurements are made with the same sensors or similar sensors in parallel), the sensor noise and the sensor sensitivity.

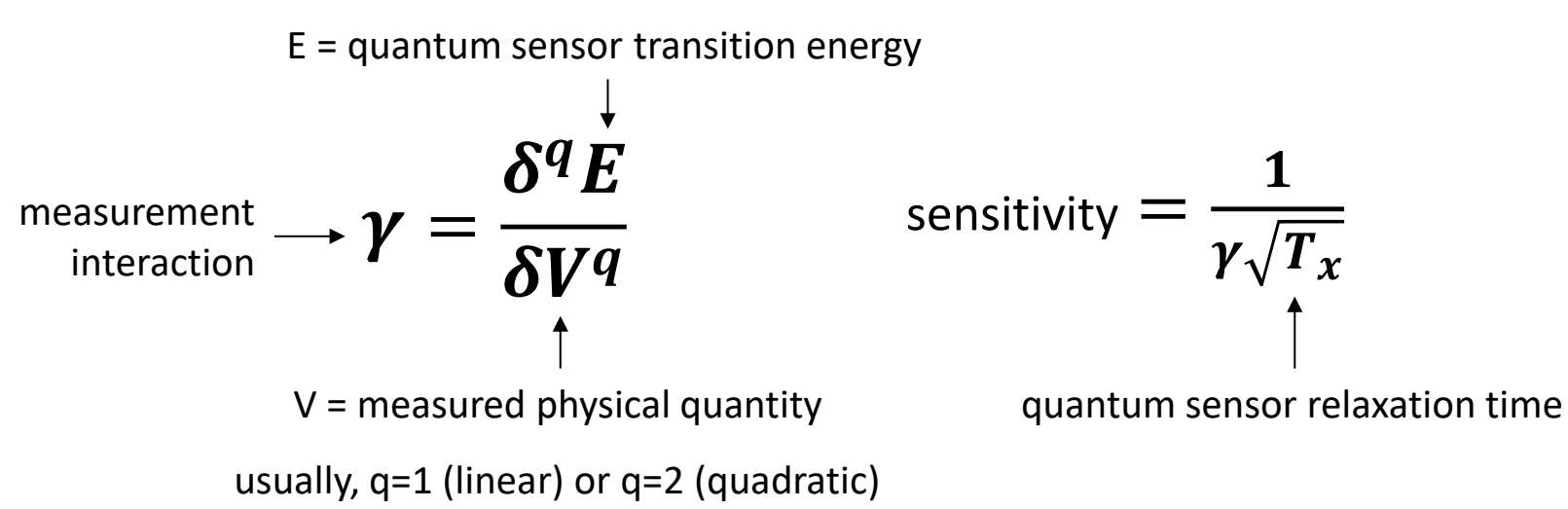

*Figure 794: calculating a quantum sensor precision.*

When using classical light as a measurement tool, the best precision possible with quantum measurement is the standard quantum limit (SQL) and it scales with $1/\sqrt{N}$, with N being the mean detected photon number. This threshold can be surpassed with using nonclassical states of light and nonstandard measurements, aka squeezed states, which can lower the precision down to the Heisenberg limit, with a better scaling of 1/N, generating a gain of $1/\sqrt{N}$ in precision[2531].

[2531] This is well explained in Basics of atomic clocks by Andrei Derevianko, University of Nevada, 2021 (77 slides).

At last, let's mention a developing field: quantumly networked sensors. This networking could have several benefits like connecting ensembles of quantum sensors to collectively improve their sensitivity[2532], to connect them to some quantum computers to help these directly capture some quantum data and accelerate the processing of sensed data, and also, to separate in a trusted way the state preparation, parameter encoding, measurement and data collection tasks[2533].

## Quantum gravimeters, gyroscopes and accelerometers

Quantum gravimeters measure gravity with a very high accuracy. These are useful in many scenarios: in seismic detectors and volcanoes monitoring, for precision autonomous navigation complementing GPS in airplanes, ships, submarines, and drones, for gravity field mapping, for detecting subterranean holes before undergoing constructions, for detecting groundwater, ice mass change, for oil and mineral exploration (well, to be restrained) and for the detection of gravitational waves in astronomy. One talked-about use case is the detection of nuclear submarines, which would destabilize or neutralize nuclear deterrence used by nuclear countries owning these submarines, but its practicality is still questioned[2534].

There are two categories of gravimeters. Absolute gravimeters measure the gravity per se measuring the free trajectory of a test mass in vacuum, and now, with cold atoms, their interferences. Relative gravimeters measure the variation of gravity over space and also time. They are usually calibrated using absolute gravimeters. The most precise relative gravimeters are superconducting based, but are not considered as quantum gravimeters.

The quantum measurement of gravity is generally performed with cold atom interferometers (CAI), taking advantage of the wave-particle of cold atoms[2535]. The technique has been developed since 1991 and perfected since then[2536]. The principle consists in creating a source of cold atoms free falling in suspension, generally rubidium, preparing their state with lasers, then passing them through a three-stages laser-controlled atom interferometer and then analyzing the results.

In the USA, NIST and NOAA in partnership with Institute for Applied Geodesy (IFAG), Germany, developed the FG5 absolute gravimeter that was tested starting in 1993. It was using a free falling rubidium atoms vacuum chamber and an interferometer. Stanford created a cold atom gyroscope in 2006 which led to the creation of AOSense. In France, SYRTE developed a six-axis inertial sensor using two magneto-optical-traps (MOTs), also in 2006.

A related field is gradiometry, which measures horizontal or vertical gravity gradients. It is used for the measurement of the variations or anomalies in the Earth's gravity field with creating gravity maps over the earth to either improve geopositioning or to detect changing underground features over time.

---

[2532] See Distributed quantum sensing with a mode-entangled network of spin-squeezed atomic states by Benjamin K. Malia, Mikhail Lukin, Ronald L. Walsworth et al, May 2022 (7 pages) and Quantum Logic Enhanced Sensing in Solid-State Spin Ensembles by Nithya Arunkumar, Ronald L. Walsworth et al, March 2022 (7 pages).

[2533] See Quantum Metrology with Delegated Tasks by Nathan Shettell and Damian Markham, December 2021 (20 pages).

[2534] An airborne array of SQUIDs magnetometry detection of submarine was supposedly prototyped in China in 2017. It requires cooling but that's not a big deal for such a use case. See China's quantum submarine detector could seal South China Sea by David Hambling, New Scientist, 2017. Found in Quantum Technology and Submarine Near-Invulnerability by Katarzyna Kubiak, European Leadership Network, December 2020 (18 pages).

[2535] See Experimental gravitation and geophysics with matter wave sensors by Philippe Bouyer, LP2N, 2018 (234 slides).

[2536] See Atomic Interferometry Using Stimulated Raman Transitions by Mark Kasevich and Steven Chu, PRL, July 1991 (4 pages) and Young double-slit experiment with atoms: A simple atom interferometer, from Olivier Carnal and Jürgen Mlynek, 1991 (6 pages) which describes a Young's double-slit interferometry experiment with helium atoms. See also Experimental gravitation and geophysics with matter wave sensors, LP2N, 2018 (234 slides).

It requires specific settings and techniques to reduce noise and vibrations within the gravimeter[2537].ONERA launched gravimetry experiments in 2009 (GIRAFE project) and in 2014-2016 (GIRAFE2)[2538]. It will be tested by the French Navy in a surface vessel in 2023 and deployed on 4 vessels by 2026/2027. The goal is to create a map of underneath the oceans.

Related sensing fields are rotation measurement and gyroscopes which are implementing it. Quantum gyroscopes can be implemented with optical interferometry[2539] and, surprisingly, with superfluid $^4$He[2540].

The figures of merits of quantum gravimeters are their **sensitivity** (smallest detectable change in gravity, measured in m/s$^2$ or cgs gals, for centimeter-gram-seconds, in reference to Galileo), **accuracy** (measurement uncertainty in reference to an absolute standard, is a %) and **stability**. Operational figure of merits are weight, size and warm-up time (which can last one hour). Best-in-class cold atom absolute gravimeters have a sensitivity of $10^{-9}$ m/s$^2$. Interesting miniaturization designs are also proposed, although they lead to a lower sensitivity measurement[2541].

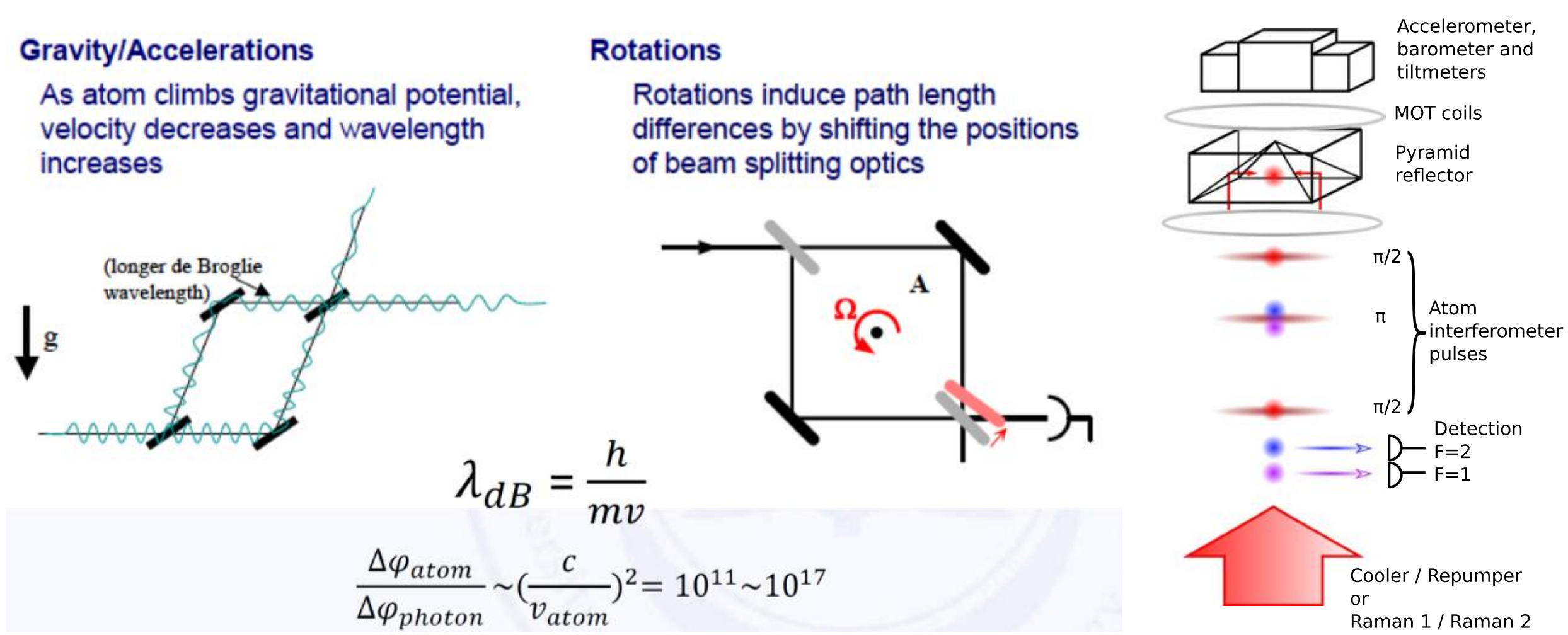

*Figure 795: how cold atom interferometry works to measure both gravity, accelerations and rotations. Source: Compact and Portable Atom Gravimeter by Shuai Chen, University of Science and Technology of China, June 2019 (22 slides) and Muquans.*

Other cold atom use case include magnetic field measurement in earth monitoring and temperature measurement. Hybrid sensors can associate electrostatic and cold atom acceleration sensors. Cold atom interferometry will even be used for gravitational waves detection in the MIGA experiment being setup in France[2542].

[2537] See recent advances with Quantum sensing for gravity cartography by Ben Stray, Kai Bongs et al, Nature, February 2022 (14 pages), Position fixing with cold atom gravity gradiometers by Alexander M. Phillips, April 2022 (10 pages) and Circulating pulse cavity enhancement as a method for extreme momentum transfer atom interferometry by Rustin Nourshargh, Kai Bongs et al, December 2021 (9 pages).

[2538] See ONERA invents with SHOM the "atomic" precision gravity mapping, February 2016. SHOM is the Hydrographic and Oceanographic Service of the French Navy.

[2539] See Quantum optical gyroscope by Lin Jiao and Jun-Hong An, January 2022 (7 pages).

[2540] See Superfluid helium-4 whistles just the right tune by Robert Sanders, UCBerkeley, 2005

[2541] See A Compact Cold-Atom Interferometer with a High Data-Rate Grating Magneto-Optical Trap and a Photonic-Integrated-Circuit-Compatible Laser System by Jongmin Lee, July 2021-September 2022 (21 pages). It uses a compact titanium vacuum package with a grating chip inside a tetrahedral grating magneto-optical trap (GMOT) using a single cooling beam. The sensitivity is 2x10-6.

[2542] See A gravity antenna based on quantum technologies: MIGA by B. Canuel, Philippe Bouyer et al, April 2022 (4 pages) and Exploring gravity with the MIGA large scale atom interferometer by B. Canuel, Philippe Bouyer et al, Nature Scientific Reports, 2018 (23 pages).

Here are the various vendors also positioned in this market, given many are still in the product development phase and don't have yet a commercial offering.

**AOSense** (2004, USA) creates and sells quantum gyroscopes, commercial optical clocks and develop a quantum gravimeter. It also provides instrumentation equipment with cold atom generators and laser frequency comb generators

They collaborate with IonQ for their quantum computers based on trapped ions. It is a spin-out from Stanford and a pioneer in the sector who is now highly diversified.

**M Squared** (2006, UK, $56M) has been developing a cold atom quantum gravimeter for a long time now[2543].

They work in partnership with the University of Birmingham and Imperial College London (UCL). The project was funded under the UK government's Quantum Initiative launched in 2013. Their last generation gravimeter has a precision of $10^{-8}$ m/s$^2$. It operates at 5 μK.

The original business of the Scotland startup is their range of SolsTiS lasers covering the spectrum from 200 nm to 4000 nm. These lasers are used in industry and in optical clocks.

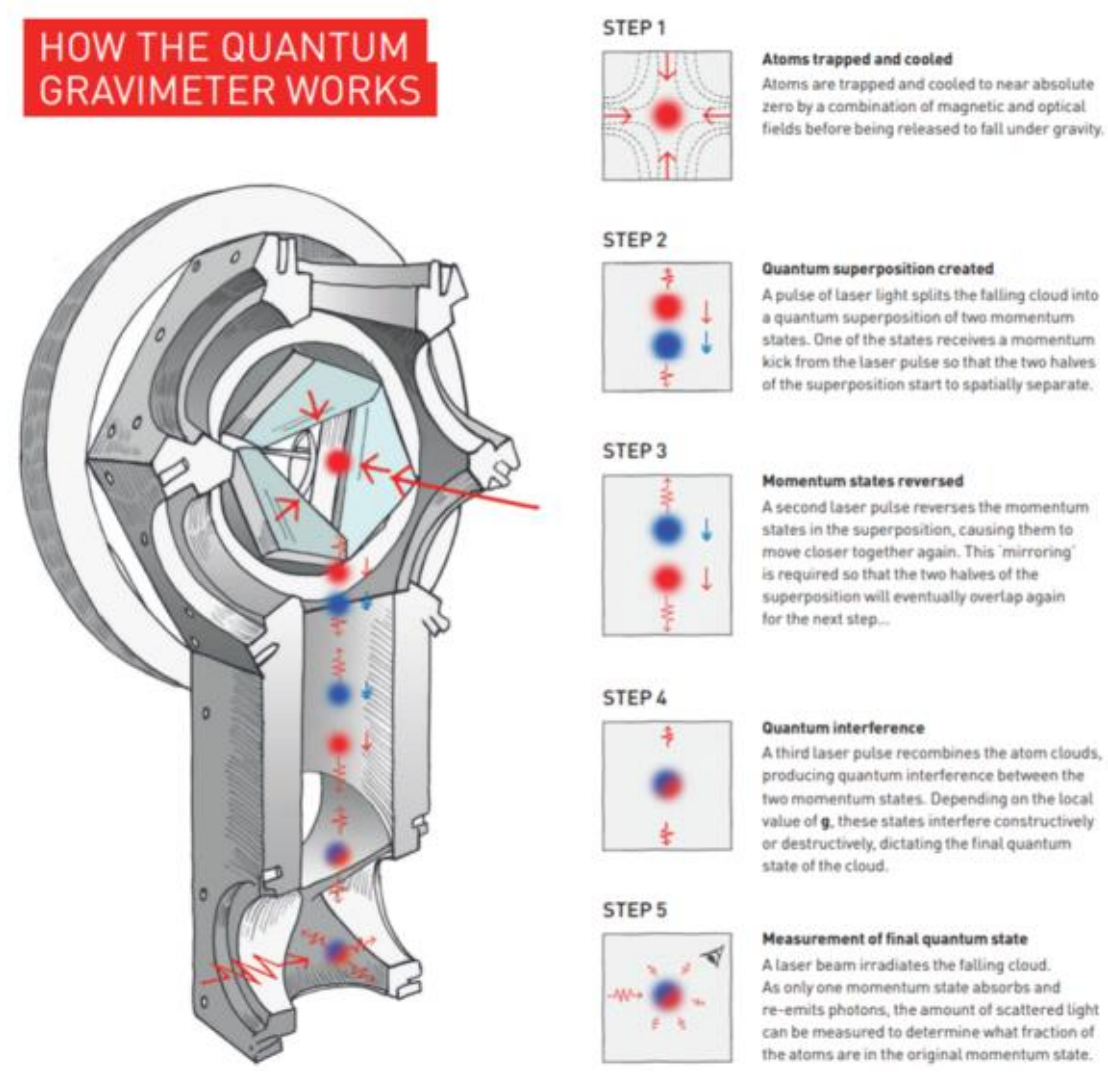

*Figure 796: M Square cold atom gravimeter. Source: M Squared.*

**iXblue Photonics** (2000, France) is the photonics branch of iXblue, which specializes in the design and manufacture of inertial and sonar power plants, with 700 employees.

It is specialized in the creation of lithium niobate optical modulators, microwave amplifiers and modulator bias controllers for the control of Mach-Zehnder interferometers. Their components are manufactured at their site in Lannion, Brittany and at Besancon. They are involved with the LP2N of Bordeaux in the creation of Ixatom, a quantum inertial sensor based on cold rubidium atoms[2544]. iXblue Photonics is the result of the acquisition of two companies: iXFiber in 2011, a specialist in passive optical components (FBG fiber grating filters, Fiber Bragg Gratings). Then Photline Technologies in 2013, a spin-off of the Femto-ST laboratory created in 2000 in Besancon. In May 2021, iXblue acquired **Muquans** and **Kylia** (a photonic equipment specialist, with its polarizers, delay line interferometers and multiplexers/multiplexers). In March 2022, Groupe Gorgé made the acquisition of iXblue to merge it with its subsidiary of **ECA Group**, a security, defense, and industry security technologies provider.

**Muquans** (created in 2011, France) is based at the Institut d'Optique in Bordeaux. They use joint research work done with the CNRS. Their quantum gravimeter is a commercial product that targets, for example, the detection of cavities in construction, oil exploration and the monitoring of volcanoes such as Etna in Italy[2545].

---

[2543] Illustration source: M Squared quantum gravimetry (4 pages).

[2544] See iXAtom - LP2N and iXblue Cold Atoms joint laboratory.

[2545] See Detecting Volcano-Related Underground Mass Changes With a Quantum Gravimeter by Laura Antoni-Micollier, Bruno Desruelle et al, June 2022 (9 pages).

Here is how their Absolute Quantum Gravimeter and other similar atom-based quantum gravimeters work. The usual description looks like the image in Figure 797 from Muquans. It's associated with a lot of schematics in the scientific literature that are quite hard to reconcile for the non-specialist.

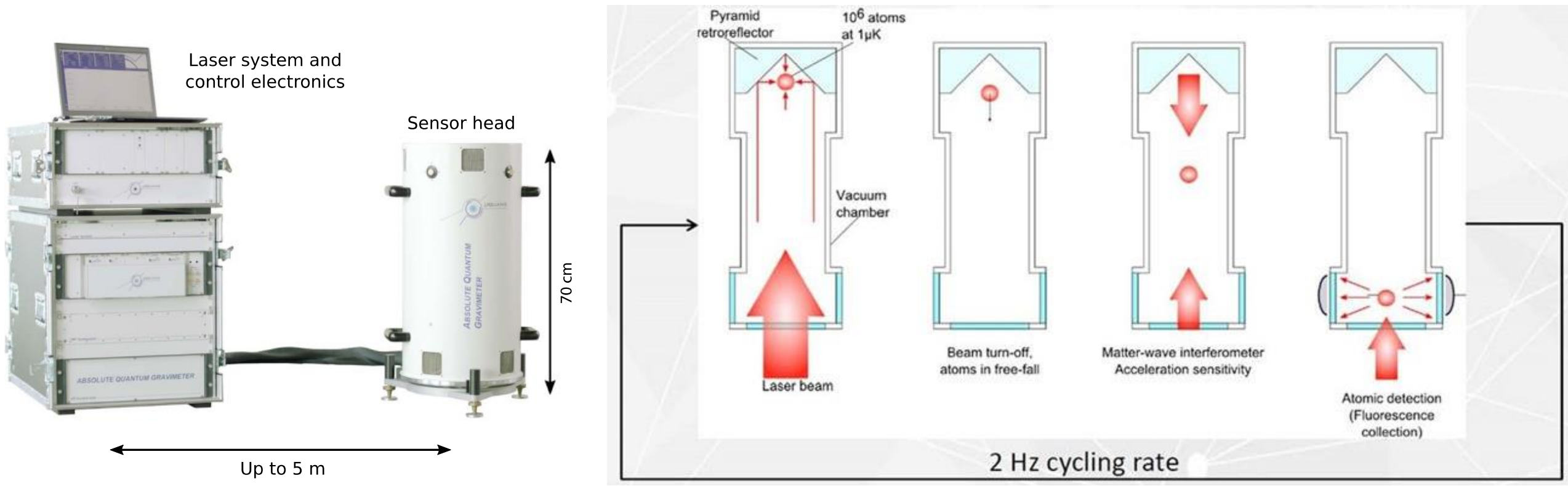

*Figure 797: how atom interferometry works, continued. Source: Muquans.*

I will decompose the various steps involved here:

① **Heating source**: a heated source at the top of the system prepares a small cloud of about $10^6$ atoms of rubidium $^{87}Rb$. The source of the atoms also contains an accelerometer that corrects the phase of the lasers in real time.

② **Magneto-optical trap**: a MOT as shown in Figure 798 is used to confine and cool the prepared source of atoms at 1µK using three pairs of counter-propagating laser beams in three orthogonal directions, applying the Doppler effect, and two anti-Helmholtz coils which magnetically trap the atoms[2546].

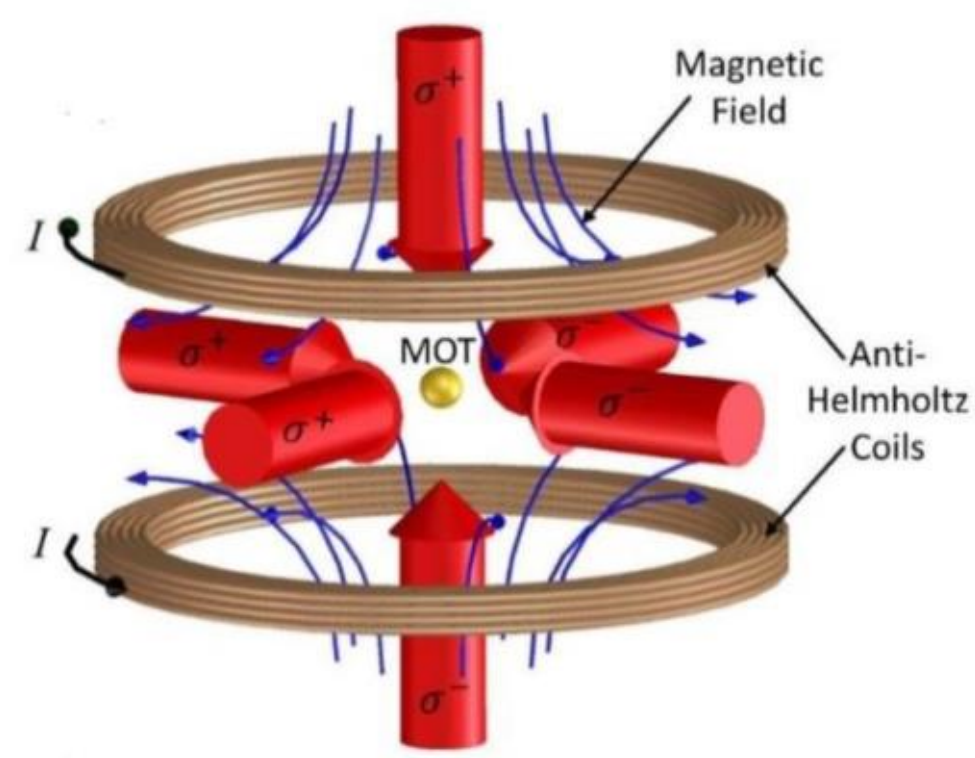

*Figure 798: a typical Magneto-Optical Trap used to cool and confine neutral atoms. Source: Cold atom interferometry sensors physics and technologies by Martino Travagnin, 2020 (47 pages).*

③ **Inverted pyramid retroreflection**: an inverted mirrors-based pyramid sits around the MOT and orient the prepared atom cloud downward in the instrument as shown in Figure 797.

④ **State preparation**: a laser beam coming from the bottom controls the atom cloud position. It is switched off and the cloud starts a freefall.

⑤ **Atom interferometer**: in their freefall, the atom cloud is exposed by two counter propagating vertical lasers pulses in three successive steps each generating a Raman transition with different durations and polarizations creating the equivalent of $\pi/2$, $-\pi$ and $-\pi/2$ gates (H, X and H)[2547]. The first step will have the effect of a coherent beam splitter and create two streams of atoms in ground and excited state whose proportion depends on the ambient gravity. The second step will have the effect of two mirrors like in an optical interferometer and inverse the population of ground and excited states. The third step will focus the two atom beams and create a coherent beam mixing. And each step changes the phase of the atom matter wave!

⑥ **Detection**: one or two other lasers excite the two streams of atoms exiting the interferometer, generating a fluorescence effect that is detected by two sets of several diodes.

---

[2546] The Muquans process is documented in Gravity measurements below $10^{-9}$g with a transportable absolute quantum gravimeter, 2018 (12 pages) and highlighted in Digging Into Quantum Sensors by Stewart Wills in Optics & Photonics, September 2019.

[2547] Double-photon Raman cooling uses two lasers. One excites the atoms to reach a high excited state and the other de-excites the atom to bring it down to a higher excited state than the initial state. This technique also contributes to cool atoms below a micro-Kelvin.

These diodes enable the measurement of the proportion of atoms in each interferometer output (excited/non excited). Their proportion will help compute the phase difference accumulated by the two atom streams, which itself will help calculate the ambient gravity. The process is repeated twice per second and a classical computer averages the results. ixBlue's gravimeters have a precision of $10^{-9}$ m/s$^2$.

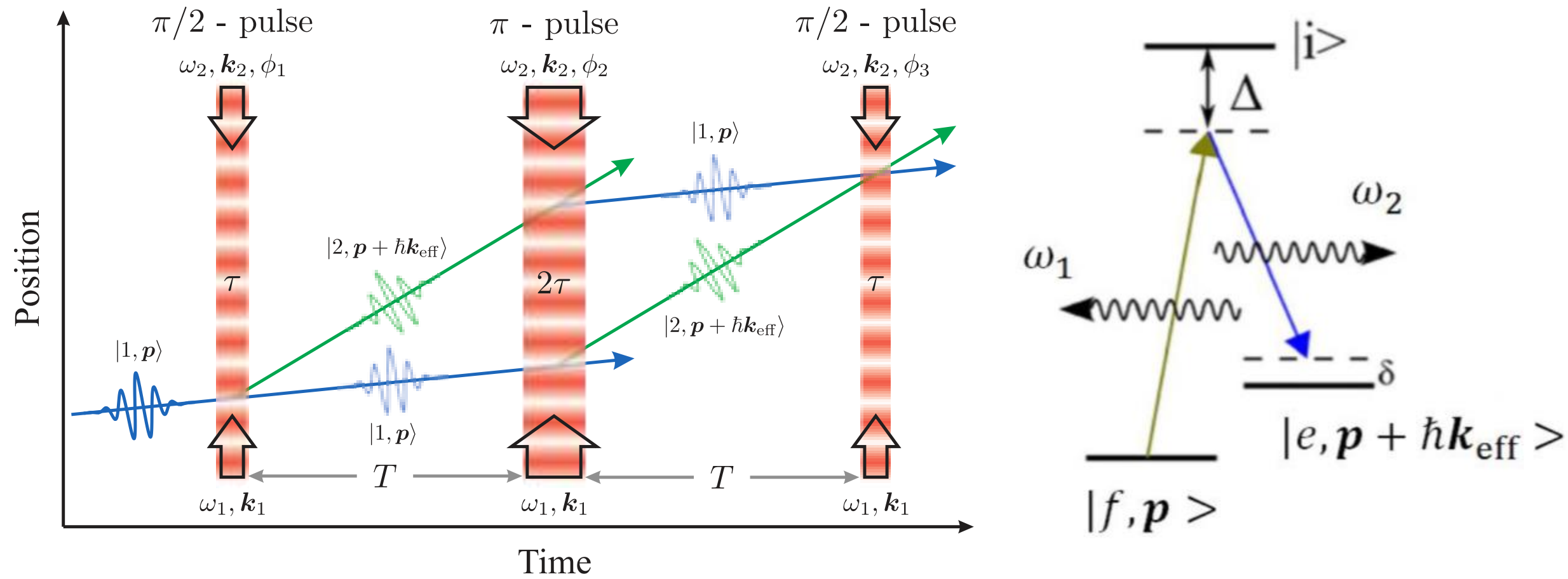

*Figure 799: the three steps of cold atoms interferometry used in a gravimeter. The figure is somewhat confusing since the position axis (Z) goes up as the atom falls. So, it's inverted. Otherwise, how would you measure atoms at the bottom of the gravimeter? Two counterpropagating lasers are used, one coming from the top at frequency $\omega_1$ and a wave vector $k_1$ and one from the bottom at $\omega_2$ and a wave vector $k_2$ to create a double-photon Raman transition that will modify displacement for a share of the atoms that depends on the gravity. The diagram on the right shows the Raman transitions created by lasers with pulses $\omega_1$ and $\omega_2$. f corresponds to the fundamental or ground state, e to the excited state. p is the atom momentum and its difference in the excited state is $\hbar k_{eff} = \hbar(k_1 - k_2)$. The width of the laser pulse $\tau$ (about 10 ms) corresponds to its duration which generates a superposition like a Hadamard gate in gate-based computing in step 1 and 3, and a population inversion in step 2 with a duration of $2\tau$. Meaning, the excited atoms (green lines) are turned in ground state atoms (blue line) and vice-versa, and also inverting their vertical velocity. If the lasers were used continuously, they would create a Rabi oscillation creating continuous change in the proposition of atoms in the ground and excited state over a couple ms. Sources: Mobile and remote inertial sensing with atom interferometers by B. Barrett et al, November 2013-August 2014 (63 pages) and Cold atom interferometry sensors physics and technologies by Martino Travagnin, 2020 (47 pages).*

iXblue is involved in the ESA's **NAVISP** program which plans to provide a supplementary navigation solution using gradiometry. The control of cold atoms has other applications[2548]. For example, Muquans participates in the European flagship project **Quantum Internet Alliance** to create hardware to extend the reach of QKD systems and with the French startup **Pasqal** which creates quantum processors based on cold atoms.

THALES

**Thales** Research & Technology (France) is developing miniaturized cold atom accelerometer, gyrometer and clock designed to be embedded. The whole with a "BEC on chip" component (for "Bose Einstein condensates on chip") in collaboration with the Charles Fabry laboratory of the Institut d'Optique (LCFIO). Atoms are vaporized in a glass cage glued to the chip, in which a good vacuum has been created. They are laser-cooled and trapped by a magnetic field and controlled by electromagnetic fields. This research project started around 2014.

*Photo credit: Ecliptique - Laurent Thion.*

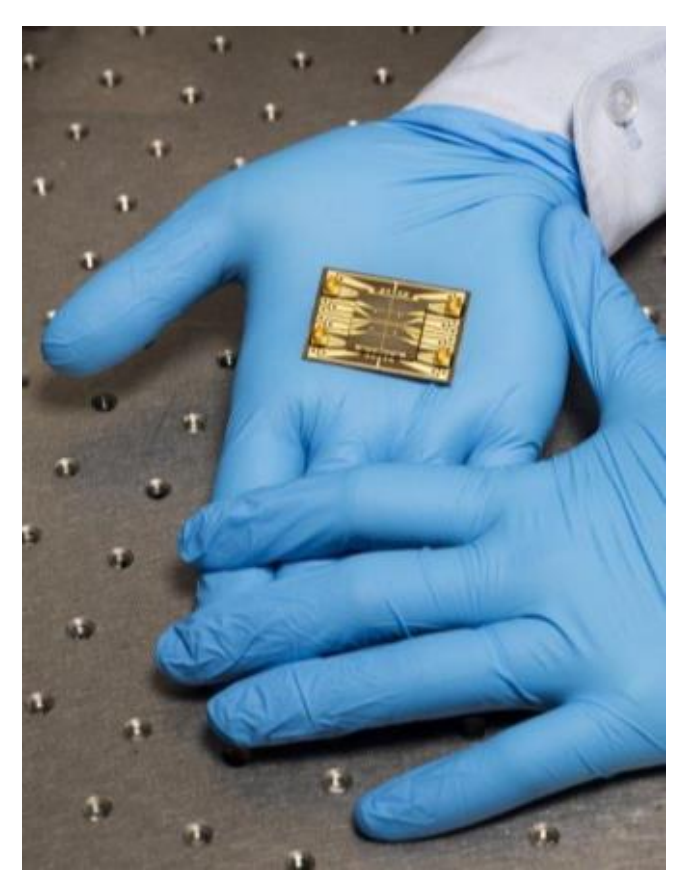

*Figure 800: a Thales BEC on chip.*

[2548] See Fifteen years of cold matter on the atom chip promise, realizations, and prospects by Mark Keil, Ron Folman et al, 2019 (46 pages) which makes a good inventory of scientific applications of cold atoms and Micro-fabricated components for cold atom sensors by J. P. McGilligan et al, Review of Scientific Instruments, September 2022 (28 pages).

**Atomionics** (2018, Singapore, $2.5M) is currently developing Gravio, a cold atoms interferometry based sensor measuring acceleration, rotations and gravity variations.

It can be used for navigation and resource exploration. It can also be used as an underground GPS.

Other laboratories are working on the same technology, such as the **Leibniz University** of Hannover[2549]. **Aquark Technologies** (2020, UK) is a spin-off from the University of Southampton which is in the same niche as Muquans. **AtomSensors** (2015, Italy) is a spin-off of the University of Florence which also develops cold atoms-based quantum sensors, including gravimeters. They also provide laser sources for spectroscopy and laser cooling of atoms. The Chinese are also in this field, but without having gone so far in miniaturization[2550].

**Teledyne e2v** (UK, a subsidiary of Teledyne US) is developing quantum gravimeters to maintains infrastructures with the detection of underground obstacles or cavities before construction works, and also do geothermal energy and groundwater reserves searches.

They are also involved in the creation of **CASPA** (Cold Atom Space PAyload), a small satellite weighing 14 kg and containing 6 CubeSat in a volume of 30x20 x10cm, including a cold atom gravimeter, which would be the first to operate in space. It was to be launched by ESA in 2020.

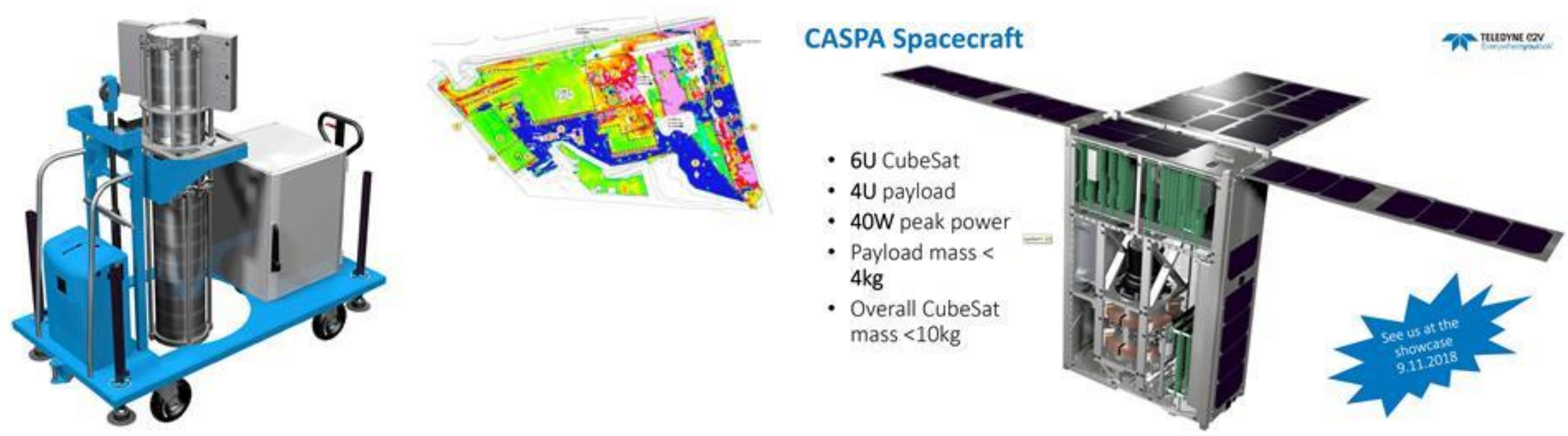

*Figure 801: Teledyne e2v cold atom sensors to be embedded in a satellite that was to be launched in 2020.*

**Nomad Atomics** (2018, Australia) develops compact cold atom-based quantum gravimeters and accelerometers. The company was launched by Kyle Hardman, Christian Freier and Paul Wigley, respectively researcher and post-docs at the Australian National University.

**Innoseis** (2020, Netherlands) was created by Mark Beker and Johannes van den Brand (who worked on gravitational waves detections instruments like those from LIGO), from Maastricht University. They develop MEMS based quantum gravimeters targeting seismic surveying.

**Zero Point Motion** (2020, UK, £2.58M) is a quantum optical inertial sensors company created by Ying Lia Li (Imperial College and University College London).

Their hybrid sensors use quantum photonics and cavity opto-mechanics to read the motion of MEMS masses. The product is being developed at the Quantum Technologies Innovation Centre at Bristol University with commercialization to start in 2024.

---

[2549] See Gravity measured using a Bose-Einstein condensate on a chip by Hamish Johnston, 2016 mentioning the work of Ernst Rasel of the Leibniz University of Hannover who refers to Atom interferometry and its applications by S. Abend et al, 2020 (48 pages).

[2550] See Compact and Portable Atom Gravimeter by Shuai Chen, 2019 (22 slides).

**OK Quantum** (2022, USA) or Oklahoma Quantum is a startup created by Saesun Kim (CEO) and Shan Zhong (Chief Scientist) who both come from the Center for Quantum Research and Technology (CQRT) of the University of Oklahoma.

They design and manufacture cold atom based quantum inertial sensors. Given their web site, they don't sell it yet.

**Rafael Advanced Defense Systems** (Israel) also has a department creating quantum sensing solutions, mainly gravitation sensors, with the Weizmann Institute of Science.

**Microg LaCoste** (1939, USA) develops absolute quantum gravimeters based on interferometry and free fall dropped mirrors, using a rubidium based atomic clock.

**Draper Labs** (1932, USA) also designs cold atoms sensors, mostly, gravimeters and accelerometers for navigation systems.

**Trumpf** (Germany) develops cold-atoms and VCSEL (lasers) based gyroscopes for satellites as part of the QYRO project funded by the German government and along with its subsidiary Q.ANT.

**Wideblue** (2006, UK) creates MEMS gravimeters. It's a consulting and engineering company.

Quantum accelerometers are also investigated to equip autonomous or assisted drive vehicles. German equipment manufacturer **Bosch** announced in February 2022 the creation its own quantum gyroscope for this purpose, aka IMU for inertial measurement unit as part of its new Bosch Quantum Sensing Unit, an internal startup led by Katrin Kobe and with a team of 15 people.

Finally, let's also mention a very special category of microgravimeters: the LIGO microgravimeters that are used to evaluate gravitational waves. They are based on optical interferometers of very high precision but of a size incompatible with all other imaginable uses[2551].

## Quantum clocks

Atomic clocks used in GPS provide resilience for navigation systems when standard satellite GPS signals are unavailable. They are also used in telecommunication infrastructure for the Internet and mobile communications, particularly with high-speed broadband landline and mobile infrastructures. Many industry sectors also rely on precision time measurement, like the financial sector and utilities power grids. Fundamental research and astrophysics also heavily rely on precision clocks.

Time measurement steadily progressed since the first mechanical clocks used between the 14th and 19th centuries. Quartz clocks appeared between the two World Wars. They were based on the piezoelectric effect demonstrated by Pierre and Jacques Curie in 1880. With a frequency of $2^{15}$ Hz, time is counted with using frequency dividers, with a drift of a few hundred microseconds per day.

The first cesium atomic clocks dates from the 1950s. They had a frequency of the order of 9 GHz and provide a time measurement accuracy ranging from $10^{-13}$ and recent generations can reach $10^{-16}$. The second is defined since 1967 as the duration of 9,192,631,770 periods of the radiation corresponding to the transition between the two "hyperfine" levels of the fundamental electronic state of cesium 133. The recent variants of these clocks are "fountain clocks". They operate at very low temperature, with laser cooling bringing the atoms at 1 µK, way much colder than superconducting qubits that are at 15 mK, but is however easier to obtain than with a dilution refrigerator. A frequency oscillator generates a transition between two levels of cesium energy. The frequency is locked with a servo loop.

---

[2551] See Advanced LIGO Just Got More Advanced Thanks To An All-New Quantum Enhancement by Ethan Siegel, December 2019. And a description of the quantum squeezing technique used in the latest version of LIGO: NIST Team Supersizes 'Quantum Squeezing' to Measure Ultrasmall Motion, 2019.

The precise measurement of frequencies has many applications: time measurement, synchronization of various devices on the Internet, even if only servers or scientific instruments, synchronization of moving objects to measure their position, astronomy (like with exoplanets and gravitational waves detection), absorption or emission precision spectroscopy, fiber optic transmissions and the generation of radio waves of arbitrary shape.

However, a better accuracy may be needed, thus the need for quantum clocks [2552]. For about 20 years, time measurement was implemented with optical measurement of frequencies and time.

The measurement of atomic vibrations can be replaced by the measurement of light waves generated by lasers and at $10^{15}$ Hz. It allows a gain in accuracy of five orders of magnitude ($10^{5}$). It was demonstrated in 2000 to generate an accuracy of a femtosecond. This earned the Nobel Prize in Physics in 2005 to **Theodor Hänsch** (1941, German) and **John Hall** (1934, American).

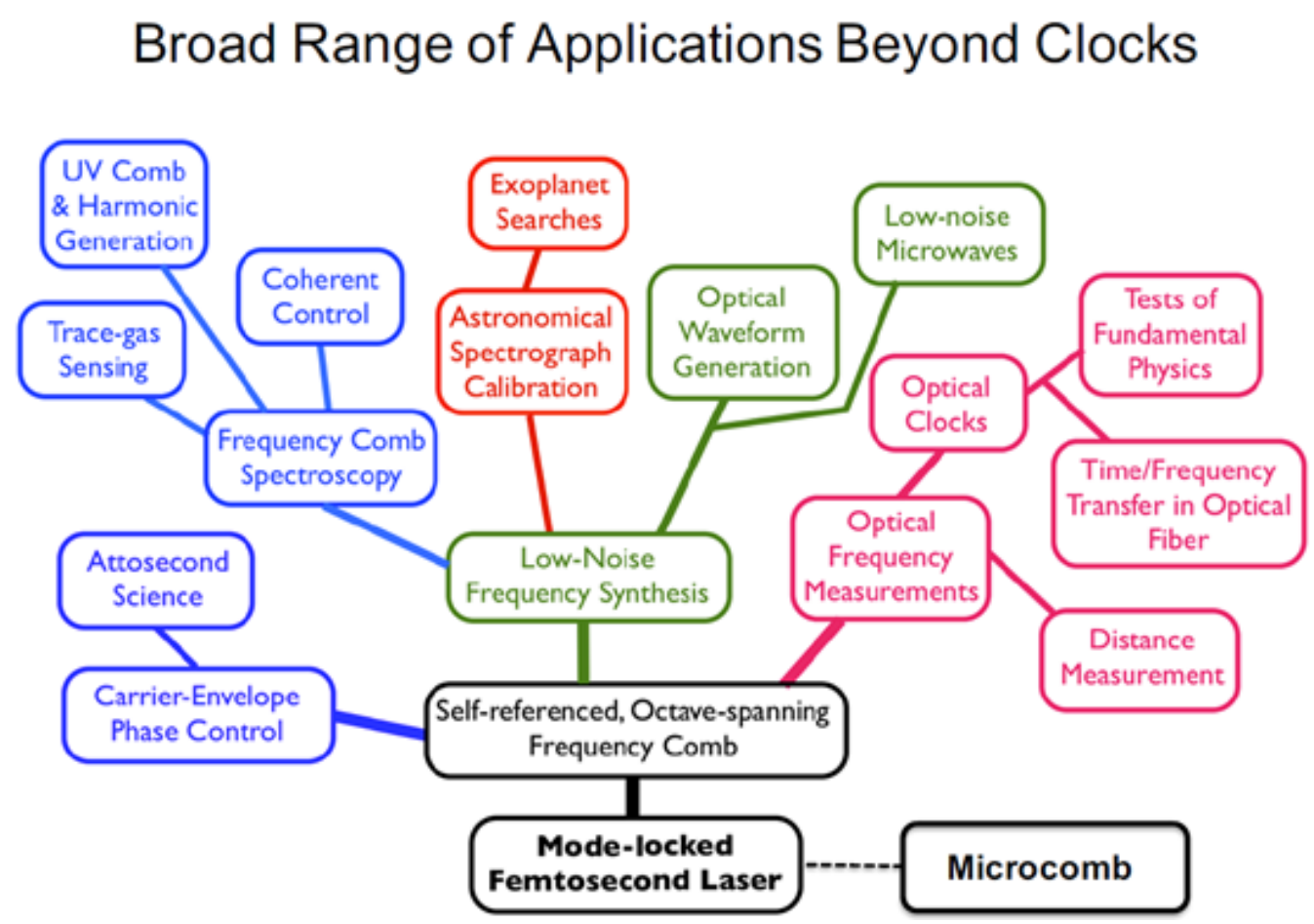

*Figure 802: femto lasers use cases in quantum sensing.*

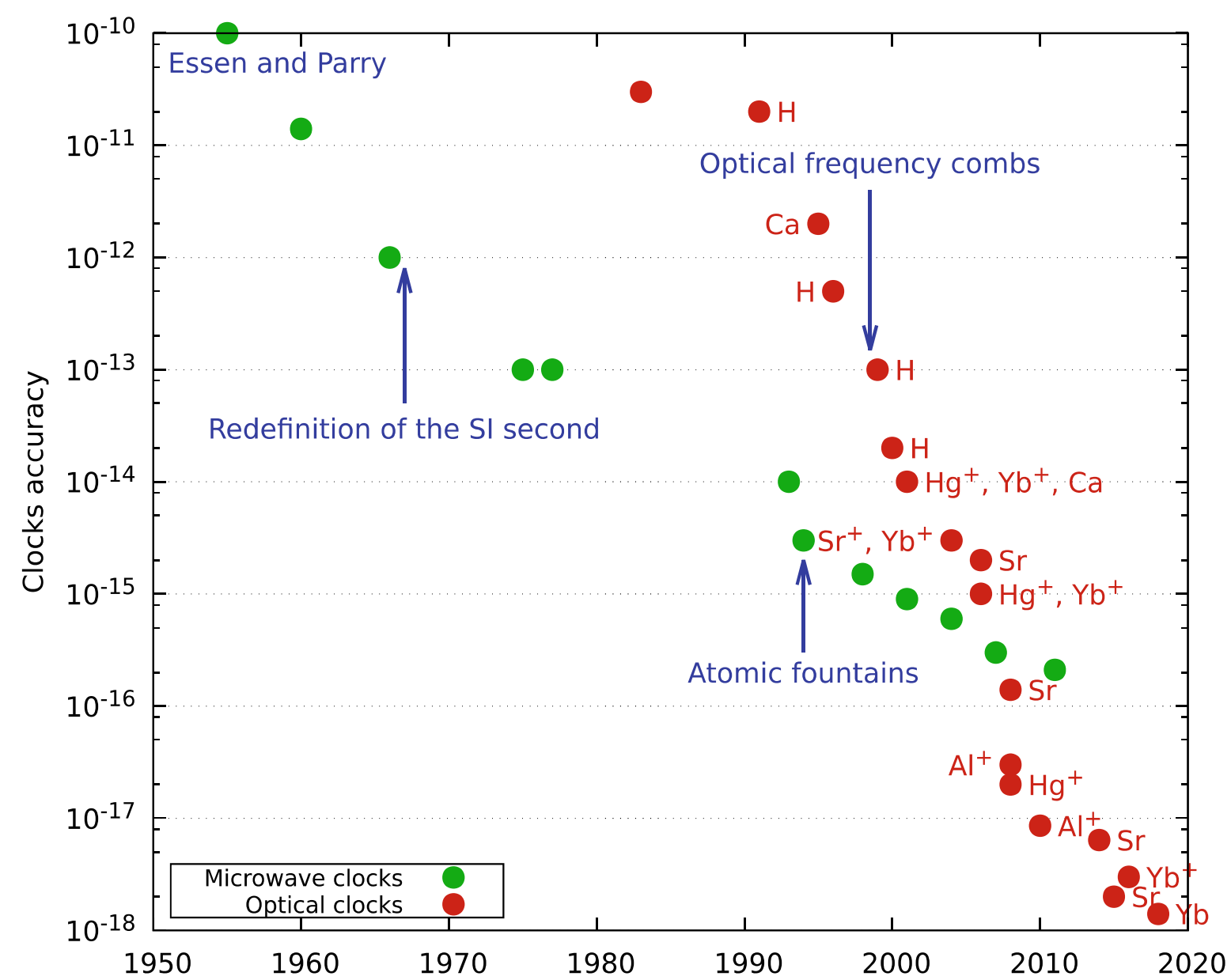

*Figure 803: how quantum clocks accuracy evolved over time. Source: Chronometric Geodesy: Methods and Applications by Pacome Delva, Heiner Denker and Guillaume Lion, 2019 (61 pages).*

In the USA, the **Chip Scale Atomic Clock** (CSAC) program funded by DARPA with NIST participation[2553] led to the creation of highly compact vapor-cells cesium based atomic clocks manufactured by Microchip (mentioned later, with a size of 1.6"x1.4"x0.45") after a decade of work starting in 2000 and \$100m spent. It generates pulses of 4596.3xMHz with a frequency tuning resolution of $1x10^{-12}$s [2554].

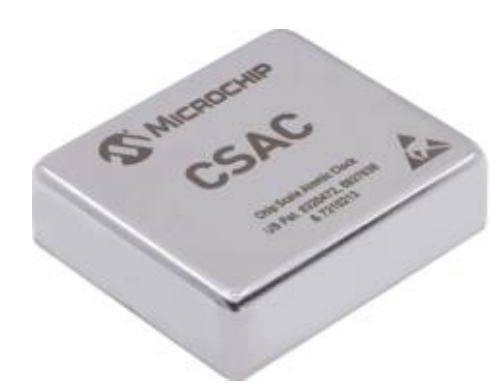

*Figure 804: a CSAC chipset.*

Atomic clocks use microwave frequency transitions while the new generation of quantum clocks are based on higher electromagnetic wave frequencies in the optical spectrum.

---

[2552] See Time and Quantum Clocks: a review of recent developments by M. Basil Altaie et al, 28 April 2022 (39 pages) and Quantum clocks are more precise than classical ones by Mischa P. Woods et al, February 2022 (75 pages).

[2553] See MEMS Atomic Clocks by Svenja Knappe, NIST, Comprehensive Microsystems, 2008 (45 pages).

[2554] CSAC are just one type of atoms chipsets, which have a broader spectrum of use cases in inertial sensing and electromagnetic field sensing. See Fifteen Years of Cold Matter on the Atom Chip: Promise, Realizations, and Prospects by Mark Keil, Ron Folman et al, 2016 (44 pages).

They use the technique of frequency combs. Since these frequencies are higher, these clocks can have a better precision, with a 3 orders of magnitude gain compared to classical atomic clocks.

**Frequency combs** were discovered with the first mode-locked lasers by Logan Hargrove in 1964[2555]. Before optical combs, light frequency harmonic generators were used with a combination of several lasers in complicated setups. To measure light high frequencies, these clocks use optical frequency combs, which subdivided optical (high) frequencies into microwave (lower) frequencies for frequency measurement and timekeeping. It uses blocked-mode lasers that generate very short pulses, which can be as short as a few femtoseconds.

The frequency decomposition of this kind of signal gives a Gaussian-shaped frequency comb with each tooth regularly spaced at a frequency equivalent to that of the laser pulse. This is related to the fact that the length of the laser cavity is a multiple of the length of the electromagnetic waves emitted by the laser. The greater the multiple, the greater the frequency generated.

The frequency spectrum resembles a Gaussian curve. Its envelope is equal to the envelope of the spectrum of an isolated pulse, which is continuous. The width of the frequency spectrum covered can be narrow, a few nm in wavelength, or cover the entire visible spectrum, thus a few hundred nm.

A calculation is used to determine the very high frequencies of the frequency comb ($f_n$). It uses several parameters: the reference frequency $f_{rep}$ of the laser pulses which is of the order of 250 MHz to 1GHz, n, the number of frequencies detected via spectroscopy (there can be hundreds of thousands) and the emission phase of the blocked mode laser which is added to each pulse and generates the frequency offset $f_0$, which is evaluated with a method described below and which is also of a lower order than GHz.

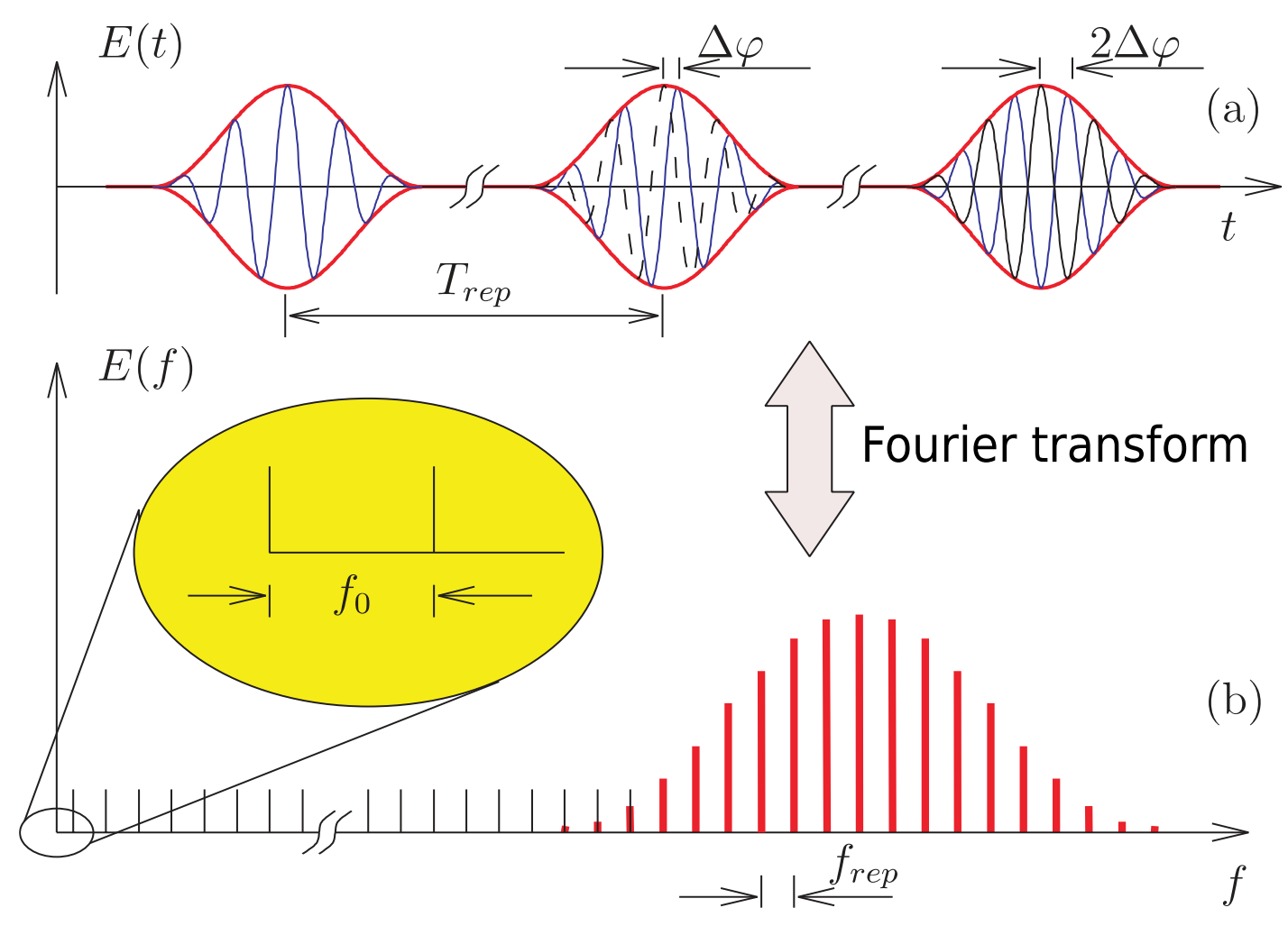

*Figure 805: how a frequency comb works. Source: Ultra-short light pulses for frequency metrology, CNRS (6 pages).*

The measurement of radio frequencies in the MHz/GHz wave range results in the measurement of frequencies in tens and hundreds of THz to the nearest Hz. The system thus acts as a frequency multiplier. The measurement of light frequencies is impossible with traditional electronics because of the frequencies used, which are several tens or hundreds of tera-Hertz.

These calibrated frequency combs are also used to measure a frequency difference with this standard[2556].

[2555] See Nobel Lecture: Defining and measuring optical frequencies by John Hall, 2006 (17 pages) and Light rules: frequency combs by Steven Cundiff, Jun Ye and John Hall in Pour la Science, 2008 (8 pages). John Hall describes frequency combs as the intersection of four initially independent fields of research: ultra-stable lasers, fast pulse lasers, nonlinear optical materials, and precision laser spectroscopy. This is a reflection of quantum computing and its many scientific and technological sources.

[2556] See Phase Coherent Vacuum-Ultraviolet to Radio Frequency Comparison with a Mode-Locked Laser by J. Reichert et al, 2005 (5 pages), Direct Link between Microwave and Optical Frequencies with a 300 THz Femtosecond Laser Comb by Scott Diddams et al, 2000 (4 pages), Fundamentals of frequency combs What they are and how they work by Scott Diddams (46 slides), Optical frequency combs and optical frequency measurements by Yann Le Coq, 2014 (38 slides) and Chip-scale Optical Atomic Clocks and Integrated Photonics by Matthew Hummon, NIST, 2018 (35 slides).

The frequency comb covers an octave, from one frequency (n) up to its double (2n). The evaluation of $f_0$ is done by extracting the frequency $f_n$, and doubling it with a crystal. By adding this frequency doubled with $f_{2n}$, we obtain a beat at the frequency of $f_0$.

$$2(f_0 + n \times f_{rep}) - (f_0 + 2n \times f_{rep}) = f_0$$

This is called **heterodyne detection**. The frequency comb becomes a kind of graduated ruler which is then used to position a frequency to be measured relative to the ruler. With that, you can build a new generation atomic clock[2557]!

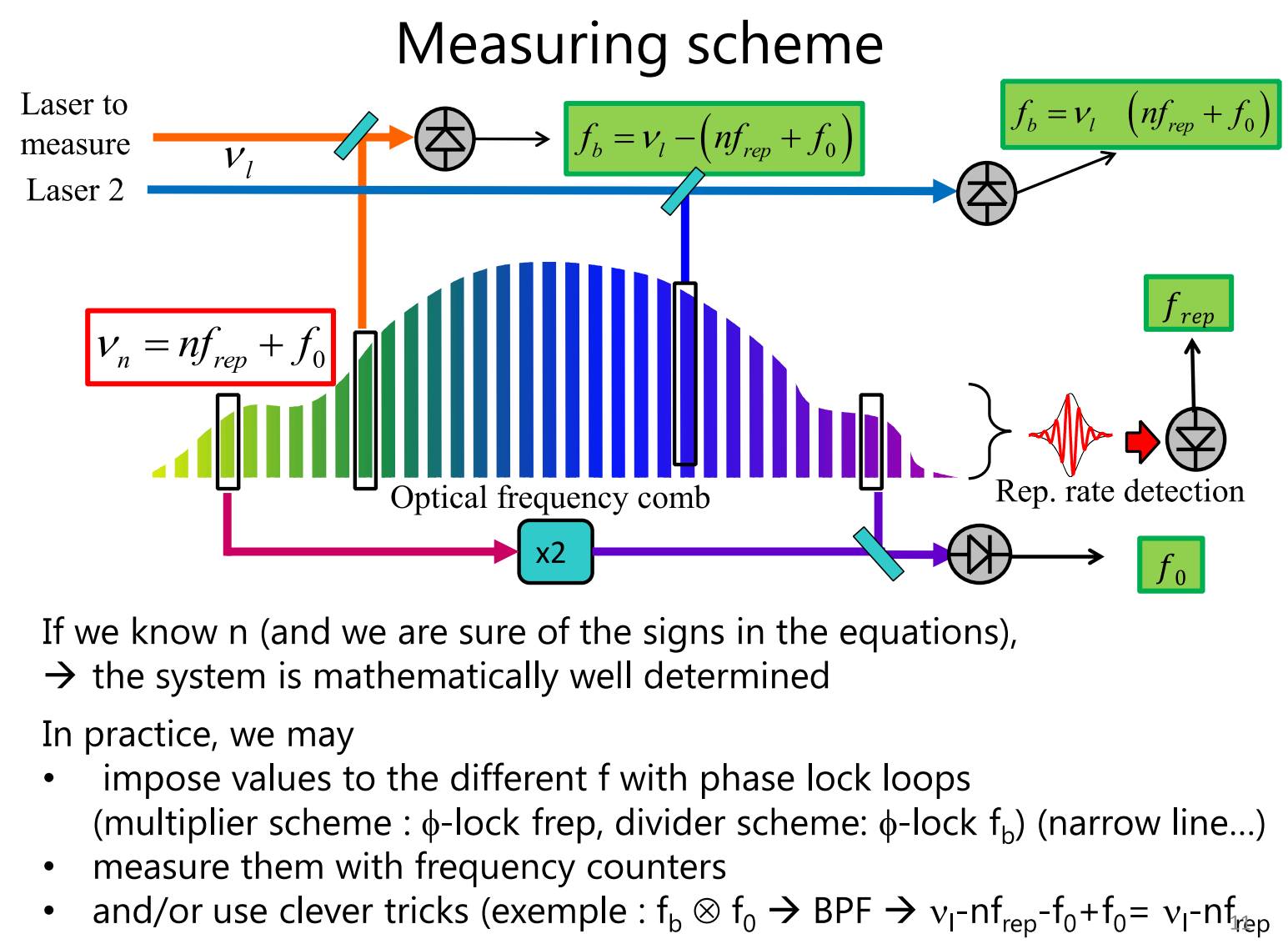

*Figure 806: frequency comb and heterodyne detection. Source: Optical frequency combs and optical frequency measurements by Yann Le Coq, 2014 (38 slides), slide 11.*

The readout of spectroscopy results using frequency combs can use CCD or CMOS cameras depending on the frequencies used in or around visible light[2558]. This measurement accuracy evolves with the use of lasers using a high pulse frequency.

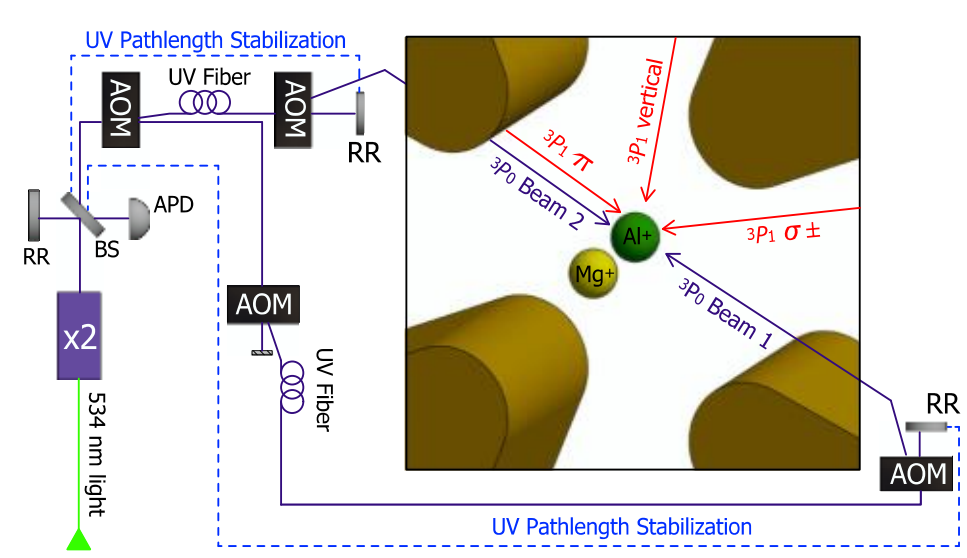

FIG. 1. Simplified schematic of the quantum-logic clock experimental setup. A frequency-quadrupled Yb-doped fiber laser is locked to the $^1S_0 \leftrightarrow {^3P_0}$ transition ($\lambda \simeq 267$ nm) by alternating the probe direction between two counterpropagating laser beams (shown in violet). An enlarged view of the trapping region is shown on the right. Three nominally orthogonal beams used for micromotion measurements are shown in red. Acousto-optic modulator (AOM), beam splitter (BS), retro-reflector (RR), frequency doubling stage (x2).

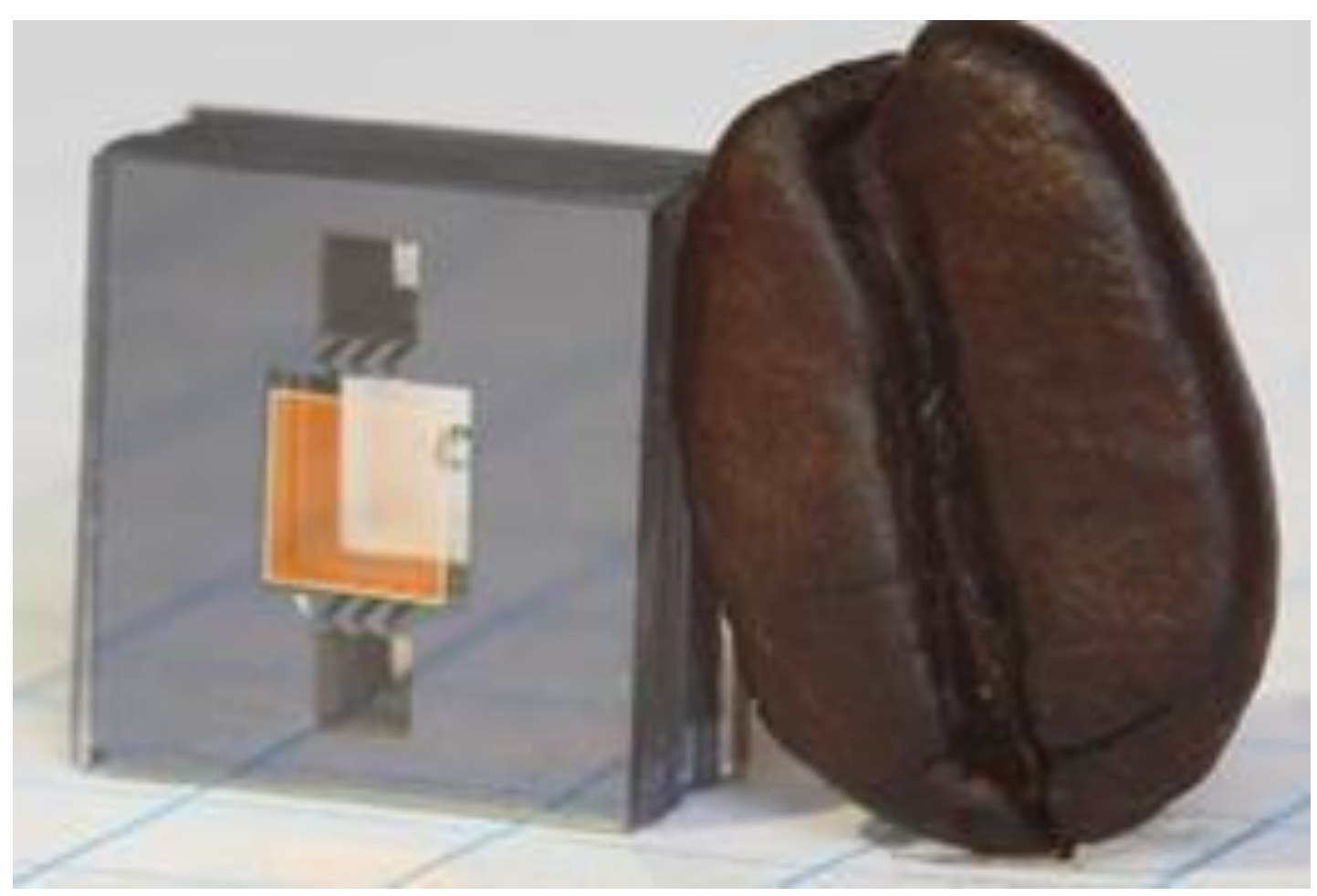

*Figure 807: Source: (right) Illustration source: $^{27}Al^+$Quantum-Logic Clock with a Systematic Uncertainty below $10^{-18}$, 2019 (6 pages).*

To date, the record for the accuracy of an atomic clock using spectroscopy is that of **NIST.** It is built with an aluminum ion associated with a magnesium anon. The aluminum ion is excited by two ytterbium lasers. Measurement is carried out using a quantum logic spectroscopy which is using the frequency combs seen in Figure 807, *left* [2559]. The clock reaches an accuracy of $10^{-18}$ seconds, a drift of one second per 33 billion years, 2.5 times the age of the Universe[2560].

---

[2557] This is explained in Optical Atomic Clocks by Andrew Ludlow, Martin Boyd, Jun Ye, E. Peil and P.O. Schmidt, 2015 (65 pages) and Optical atomic clocks by N. Poli et al, 2014 (70 pages). See also Photonic integration of an optical atomic clock by Z. L. Newman et al, November 2018 (12 pages).

[2558] See Frequency comb spectroscopy by Nathalie Picqué and Theodor Hänsch, 2019 (27 pages) which describes the many methods and use cases of frequency comb based spectroscopy.

[2559] See this explanation: Quantum Logic for Precision Spectroscopy by Piet Schmidt et al, 2009 (6 pages).

[2560] See $^{27}Al^+$Quantum-Logic Clock with a Systematic Uncertainty below $10^{-18}$, 2019 (6 pages).

In this market for optical quantum clocks, there are many research laboratories that produce their own equipment. These are usually based on titanium-sapphire with pulses of a few femtoseconds ($10^{-15}$ to $10^{-14}$ seconds).

NIST is also working on an atomic clock that would fit into a component the size of a coffee bean, using a double frequency comb and rubidium gas (in Figure 807, *right*). The whole thing consumes only 275 mW. This project was co-funded by DARPA[2561]. However, for the moment, the precision obtained is not yet satisfactory for industrialization.

One of the projects of the European Quantum Flagship, **iqClock** (Netherlands, €10M), also aims to create very high-precision, portable quantum clocks. The consortium brings together six universities and six private partners including Teledyne EV (USA), Toptica (Germany), NKT Photonics (Denmark), AckTar (Israel) and Chronos (UK).

Let's now look at some vendors in high-precision quantum time measurement.

In the private sector, **Teledyne** sells Minac (cesium atomic clock), T-CSAC (also cesium, integrated in a chip) and Synchronicity (ytterbium-based).

UK aircraft carrier HMS Prince of Wales has been fitted with the world's first atomic clock of its kind to help ensure pinpoint accuracy wherever she goes provided by BP and Teledyne e2v.

**MicroSemi** (1960, USA) sells its Quantum SA.45s, a miniaturized chip-scale atomic clock (CSAC). Among other use cases, it can be used in portable IED (improvised explosive devices) jammers. The company is a subsidiary of MicroChip Technology (1989, USA).

**HyperLight Corp** (2018, USA), based in Cambridge, near Boston, develops nanophotonic integrated circuits such as frequency combs or resonators that are used in quantum sensing.

**Cryoclock** (2016, Australia) develops sapphire-based cryogenic oscillators. The company was co-founded by John Hartnett. Applications include trapped ion quantum processors and atomic clocks.

**Orolia** (2005, France) creates atomic clocks with cesium, or based on rubidium oscillators. They mainly target the aerospace industry and provide the Galileo GNSS service.

**Syrlinks** (2011, France) develops miniature atomic clocks based on MEMS and cesium for embedded applications. Their MMAC is 40 x 35 x 22 mm and consumes less than 0.3 W.

With Thales, and with the help of CNRS (SYRTE lab) Syrlinks is developing Chronos, new quantum clocks for civil and military applications with error inferior to 1 second per tens of thousands of years. It will enable geolocalization when GNSS like GPS and Galileo are not available.

**TMD** (1969, UK) sells microwave amplifiers. They also develop atomic clocks and instrumentation for the manipulation of cold atoms.

**VectorAtomic** (2018, USA) markets rubidium atomic clocks for quantum inertial navigation systems that can then avoid using GPS.

[2561] The project is documented in Architecture for the photonic integration of an optical atomic clock, 2019 (6 pages).

**Vescent Photonics** (2002, USA) offers optical frequency comb generators for use in atomic clocks. They also master the laser-based technique used for controlling cold. They are based in Colorado.

In June 2022, **ColdQuanta** was announcing it was working for the US Navy with **LocatorX** (USA) to create an atomic clock using the Solid-state Miniature Atomic Clock (SMAC) technology created by LocatorX under license of Oxford University[2562]. The system is using fullerene molecules (C60) doped with nitrogen. It is not relying on cold atoms, the core technology mastered by ColdQuanta. It seems the two vendors are combining lightweight atomic clocks and more precise ones using cold atoms.

Finally, let us mention again **Muquans,** which also uses its cold atoms expertise to sell an atomic clock, the MuClock, designed in partnership with the LP2N laboratory in Bordeaux and the LNE-SYRTE. It is positioned as an alternative to cesium atomic clocks. The instrument weighs 135 kg and consumes 200W.

## Quantum magnetometers

Quantum magnetometers are used to detect small variations or levels of magnetism with high spatial accuracy. There are many uses cases: navigation, mineral exploration, current detection, magnetocardiography, magnetoencephalography, orientation of drones and autonomous vehicles in tunnels where GPS does not work[2563], sonar, detection of moving metal objects such as vehicles and cellular imaging[2564].

Different techniques are available for high-precision magnetometry including Bose condensates and **atomic spins in vapors**[2565], **SQUIDs** (superconducting effect with a Josephson junction as in superconducting qubits[2566]), **NV-centers** solid-state systems, particularly using optically detected magnetic resonance (ODMR), with a sensitivity (or noise) usually ranging from $10^{-9}$ to $10^{-15}$ $T/\sqrt{Hz}$ and decreasing as the signal frequency increases, and even **SiC-vacancies** (silicon vacancies in silicon carbide)[2567].

Measuring magnetism with NV centers exploit the variation of the spin resonance spectrum in the diamond cavity, which depends on the ambient magnetic field (see the chart above on the right on spin resonance spectrum). The distance between the two fluorescent light pulses (Y) generated is measured as a function of the electromagnetic excitation frequency used (X)[2568]. The spins preparation is performed with a laser and its modification with 3 GHz microwave pulses. NV-center based magnetometers can use large unordered ensembles of NV centers[2569] or individual NV centers[2570].

---

2562 See ColdQuanta and LocatorX Partner to Build Next Generation of Atomic Clocks by ColdQuanta, June 2022.

2563 A UAV solution using GPS in a tunnel is proposed by the startup Hovering Solutions (Spain).

2564 See Nitrogen-vacancy centers in diamond for nanoscale magnetic resonance imaging applications by Alberto Boretti et al, 2019 (24 pages).

2565 See the Rydberg atom technique described in Quantum sensing using circular Rydberg states by Rémi Richaud, LKB, November 2018 (41 slides). See also the thesis Rubidium vapors in high magnetic fields by Stefano Scotto, November 2017 (168 pages). See also Brief Review of Quantum Magnetometers by Ivan Hrvoic and Greg M. Hollyer, GEM Systems, not dated (15 pages).

2566 See this presentation of SQUID applications: SQUID Fundamentals and Applications by Robin Cantor, 2017 (48 slides).

2567 See Fiber-integrated silicon carbide silicon vacancy-based magnetometer by Wei-Ke Quan et al, CAS and Sichuan University, August 2022 (13 pages) which describes a proposal of silicon carbide vacancy-based room temperature ODMR magnetometer using a fiber for results measurements.

2568 After optical magnification, fluorescence can be analyzed by a CCD image sensor.

2569 See for example Picotesla magnetometry of microwave fields with diamond sensors by Zhecheng Wang et al, June 2022 (7 pages).

2570 See for example Scanning gradiometry with a single spin quantum magnetometer by W. S. Huxter et al, 2022 (9 pages).

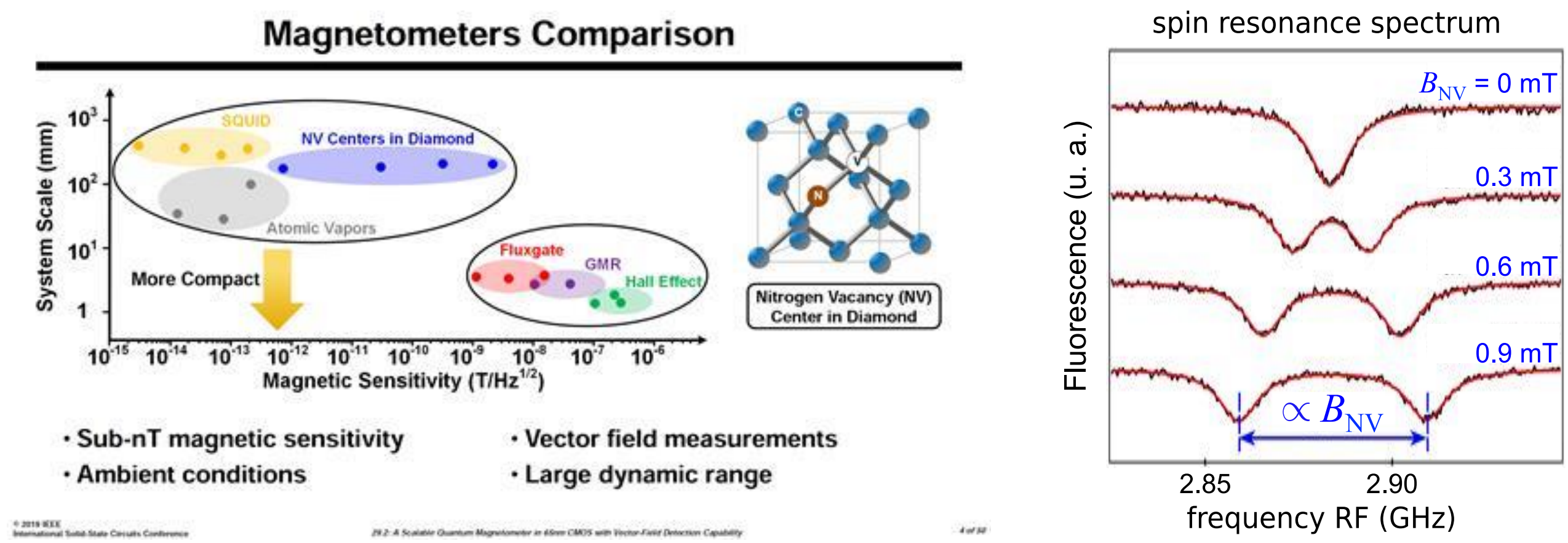

*Figure 808: NV center magnetometry principle using spin resonance spectrum analysis. The two energy gaps enable the evaluation of the current magnetic field. Sources: [A Scalable Quantum Magnetometer in 65nm CMOS with Vector-Field Detection Capability]() by Mohamed Ibrahim from MIT 2019 (51 slides) and [NV Diamond Centers: from material to applications]() by Jean-François Roch, Collège de France, 2015 (52 slides).*

The accuracy of magnetism measurement can reach a pico-Tesla, billions of times less than terrestrial magnetism[2571]. NV centers provide a lesser precision than cold atoms, but their use is more practical because the instrument is easier to miniaturize and most of them work at ambient temperature[2572]. Scanning probes magnetometers use a diamond nanocrystal containing a single cavity and a nitrogen atom, which ensures the accuracy of the measurement. The probe can be moved in space and used to analyze the magnetism of a material in 2D.

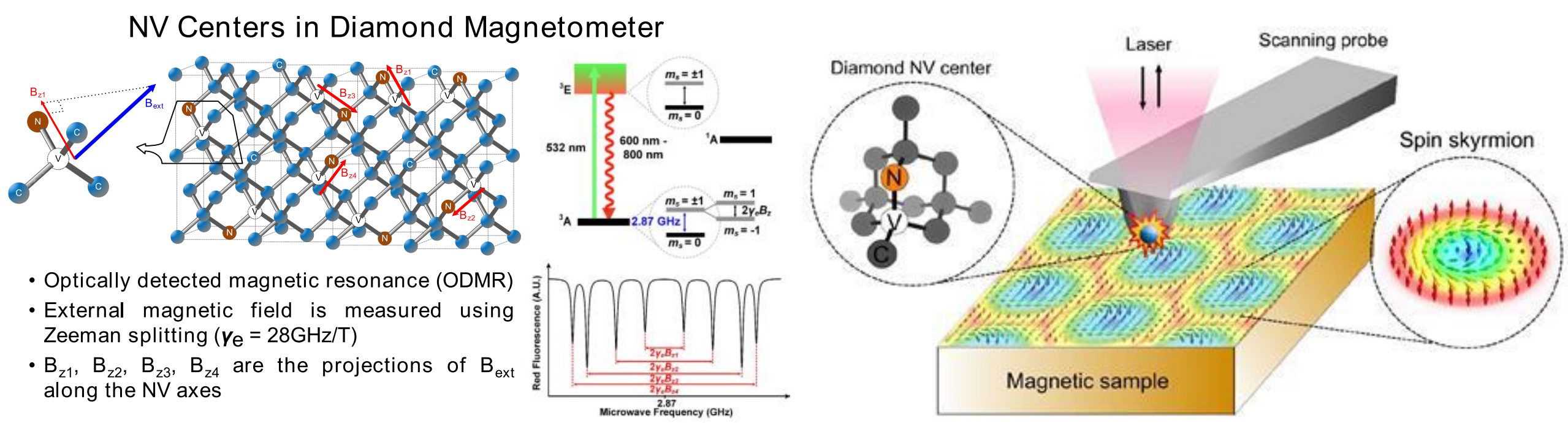

*Figure 809: NV centers used in ODMR for medical imaging of materials inspection. Sources: [A Scalable Quantum Magnetometer in 65nm CMOS with Vector-Field Detection Capability]() by Mohamed Ibrahim from MIT 2019 (51 slides) and [Probing and imaging nanoscale magnetism with scanning magnetometers based on diamond quantum defects](), 2016 (35 slides).*

The NV-centers technique appeared in 2009. It is notably developed in France by **Thales**[2573].

---

[2571] The accuracy of magnetometry with NV centers is evaluated with the formula $1\mu T/\sqrt{Hz}$. See [Picotesla magnetometry of microwave fields with diamond sensors]() by Zhecheng Wang et al, August 2022 (7 pages) which describe a heterodyne measurement technique enabling picotesla precision.

[2572] See [A Scalable Quantum Magnetometer in 65nm CMOS with Vector-Field Detection Capability]() by Mohamed Ibrahim from MIT 2019 (51 slides) which describes a miniaturization process of a quantum magnetometer combining a 65 nm CMOS circuit manufactured by TSMC and a diamond NV-center based system.

[2573] ASTERIQS (France, €9.7M) or "Advancing Science and Technology through diamond Quantum Sensing" is a European Quantum Flagship project launched in 2018 and led by Thales, which is expected to advance techniques for measuring magnetic, electric, temperature and pressure fields. There are many applications, such as sensors for vehicle battery monitoring, high-resolution sensors for nuclear medical imaging (NMR, nuclear magnetic resonance) or for creating radio frequency spectrum analyzers. The Swiss startup Qnami is involved in the project and provides artificial diamonds.

Laboratories in Bristol, the University of Ulm in Germany and Microsoft are working on the use of NV Centers techniques coupled with machine learning and Bayesian inference methods to correct the noise found at higher temperatures[2574].

Quantum magnetometry can also rely on mixed optomechanics-photonics systems, like in a 2022 proposal from a China research team. It couples a thin SiN mechanical membrane to Terfenol-D rods[2575] with a height that is sensitive to a static magnetic field. The membrane position modifies the phase a laser-originated photon that is reflected in a cavity containing the membrane. The magnetic field s converted in the photon phase which is measured with homodyne detection using a local oscillator. The sensitivity of this sensor could be excellent, reaching $10^{-15}$ to $10^{-17}$ $T/\sqrt{Hz}$ [2576].

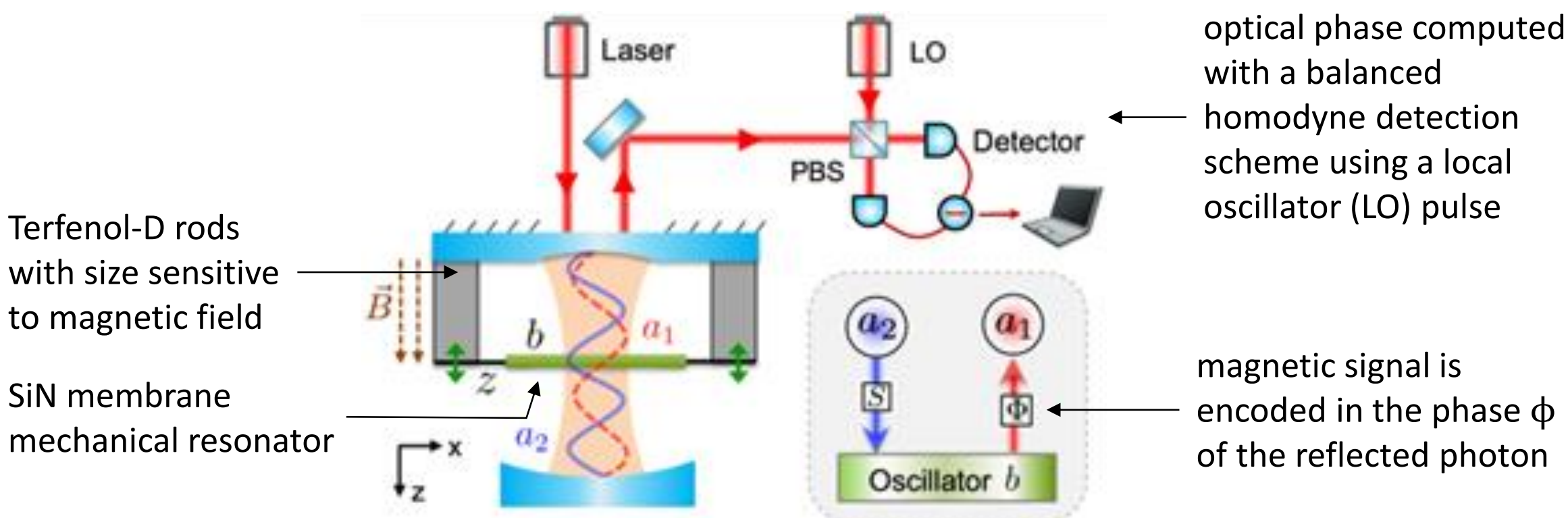

*Figure 810: Source: [Quantum Magnetometer with Dual-Coupling Optomechanics](#) by Gui-Lei Zhu et al, May 2022 (7 pages).*

Another amazing use case for NV center magnetometry is the precise measurement of battery charge and discharge[2577].

Let's look at some vendors in the quantum magnetometry space.

**qdm.io** (2021, USA) is a stealth company based in Maryland that creates NV centers sensors.

**Qubic** (2019, Canada) is a startup from the Institut Quantique from the University of Sherbrooke in Quebec that is working on microwaves-based quantum sensing tools for sensing, imaging and communications. It is led by Jérôme Bourassa.

**Qutools** (2005, Germany) offers its quNV quantum magnetometer kit, based on diamond NV-centers as its name suggests. It fits in a 3U rack. They also sell an interferometer for interferometric displacement measurement.

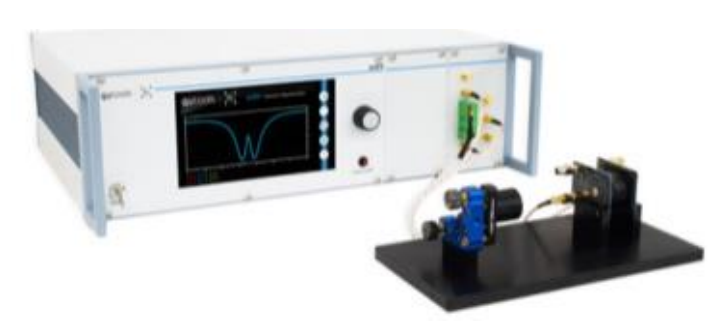

*Figure 811: a quantum magnetometer from qutools.*

---

[2574] See Magnetic-Field Learning Using a Single Electronic Spin in Diamond with One-Photon Readout at Room Temperature by Raffaele Santagati et al, 2018 (18 pages).

[2575] Terfenol-D is a magnetostrictive material made from an $Tb_xDy_{1-x}Fe_2$ ($x \approx 0.3$) alloy. Its name comes from terbium, iron (Fe), Naval Ordnance Laboratory (NOL, who developed the material in the 1970s), and D for dysprosium. The material is used to produce sonar systems.

[2576] See Quantum Magnetometer with Dual-Coupling Optomechanics by Gui-Lei Zhu et al, May 2022 (7 pages).

[2577] See High-precision robust monitoring of charge/discharge current over a wide dynamic range for electric vehicle batteries using diamond quantum sensors by Yuji Hatano et al, Nature Scientific Reports, September 2022 (10 pages)

Also in Germany, the University of Stuttgart is working with the Fraunhofer Institute to transfer NV-centered magnetometry technology as part of the **QMag**[2578] project.

**Supracon** (2001, Germany) manufactures magnetometers based on SQUIDs (Superconducting Quantum Interference Devices). It is a spin-off of the Leibnitz Institute of Photonics in Jena. It sells its sensors to astrophysics research laboratories and for geophysics prospection.

**Great Lakes Crystal Technologies** (2019, USA) is a supplier of diamonds for use in NV center applications, especially for quantum magnetometers. It is a spin-off from the University of Michigan and Fraunhofer USA.

**FieldLine Inc** (2020, USA) develops NV centers quantum sensing systems, particularly for medical brain imaging and non-destructive materials testing.

**Q-Sensorix** (2019, USA) develops NV centers magnetometer-based gyroscopes, launched by Alexey Akimov, Vladimir Shalaev and Yuri Lebedev. These are alumni of the University of Buffalo in New York State.

**Twinleaf** (2007, USA) develops precision magnetometers based on alkali metal lasers. The company is headed by Elisabeth Foley, a specialist in the field, and Thomas Kornack (CSO), both Princeton alumni.

**Cerca Magnetics** (2018, UK) is a spin-out from the UK Sensors and Timing Quantum Technology Hub at the University of Nottingham. It develops a wearable brain scanner magnetoencephalography (MEG) system, avoiding the use of cryogeny like heavy MRI scanners.

It uses an optically pumped magnetometers measuring weak magnetic fields. These are made with a laser illuminating a small glass cell containing a pressured gas of rubidium or cesium. A diode detects the transmitted light that depends on the local magnetic field perpendicular to the laser beam[2579]. The technology is competing with SQUIDs based MEGs.

**CIQTEK** (2016, China, $15M) manufactures quantum sensors targeting quantum computation, healthcare, food safety, chemistry and material science markets. These sensors are NV center magnetometers.

At last, the **Ivar Giæver Geomagnetic Laboratory** (IGGL) in Norway also uses SQUIDs to detect underground magnetism for paleomagnetic applications, to measure the magnetic remanence of ancient rocks. Using SQUIDs, their magnetometer must be cooled to 4K with a pulsed tube[2580].

**GEM Systems** (1980, Canada) is selling quantum magnetometers using optically pumped potassium (K-Mag).

**ODMR Technologies** (2015, USA) is a stealth spin-off from Berkeley which designs a magnetic resonance spectroscopic analysis system based on NV centers.

[2578] See Quantum Magnetometers for Industrial Applications, April 2019.

[2579] See Optically pumped magnetometers: From quantum origins to multi-channel magnetoencephalography by Tim M. Tierney et al, 2019 (12 pages).

[2580] See Instruments for Paleomagnetic Measurements WSGI (2G) Model 755 Superconducting Rock Magnetometer (SRM).

**Elta Systems** (1967, Israel) is a subsidiary from the IAI consortium. The company develops various electromagnetic sensors and radars for defense and intelligence.

They develop quantum magnetic sensors in collaboration with Israeli research groups and industry partners. These sensors can help detect IEDs (improvised explosive devices, unexploded ordinance, geophysical structures, vehicles and ships). They also work on quantum optical magnetometry and sub-pico-Tesla atomic magnetometers which can be used in medical imaging (MEG).

**Wainvam-E** (2020, France) is providing a set of magnetometry solutions based on NV centers targeting nondestructive measurement of various materials like steel and oxidation characterization in live cells.

## Quantum thermometers

NV centers have another use: temperature measurement with an accuracy of a few mK and with a very high spatial resolution, all with highly miniaturized sensors. It is currently the most powerful temperature measurement technology for these different dimensions. It allows, for example, to determine the temperature within living cells and organisms with a sub-mm precision [2581].

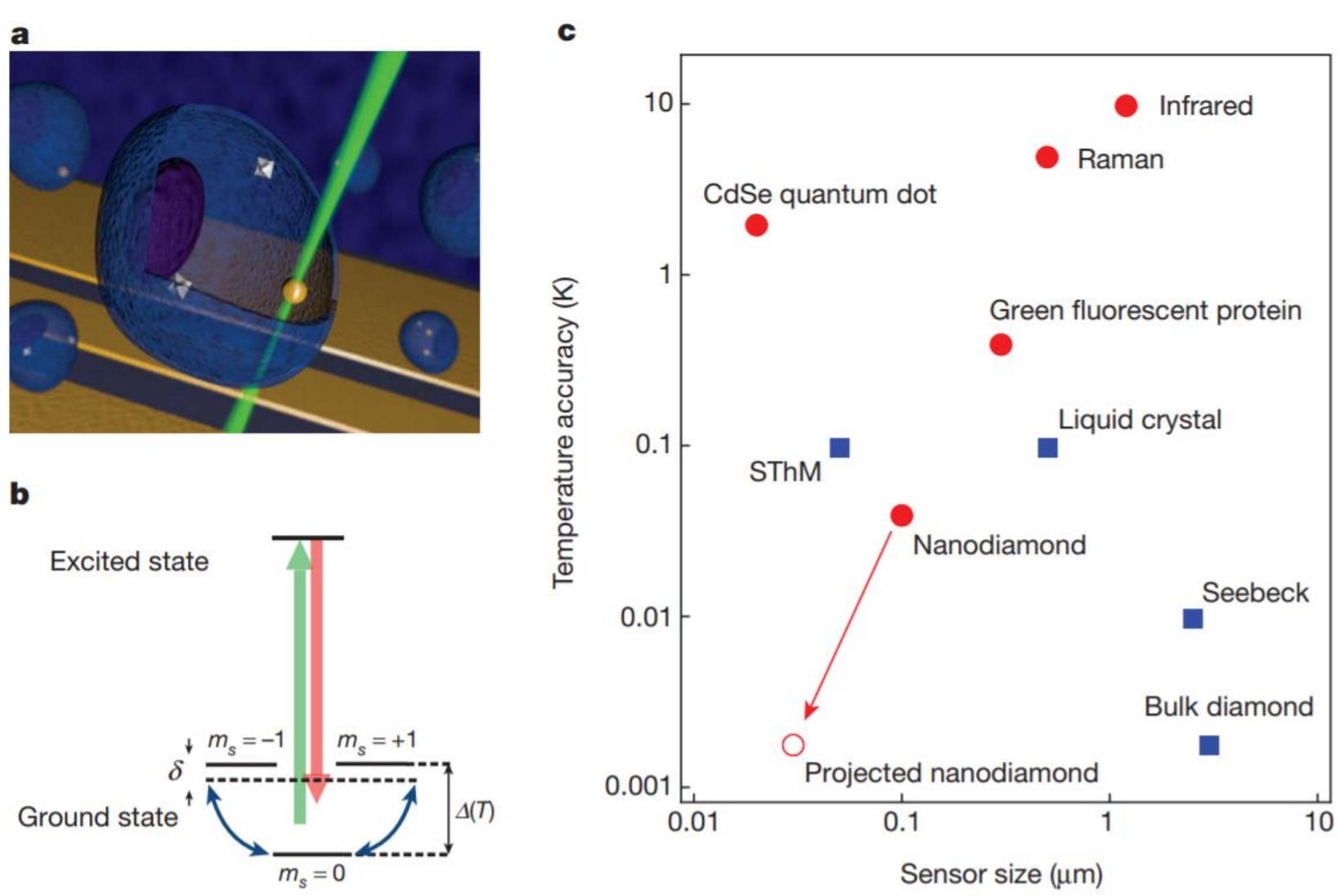

**Figure 1 | Nitrogen–vacancy-based nanoscale thermometry.** **a**, Schematic image depicting nanodiamonds (grey diamonds) and a gold nanoparticle (yellow sphere) within a living cell (central blue object; others are similar) with coplanar waveguide (yellow stripes) in the background. The controlled application of local heat is achieved by laser illumination of the gold nanoparticle, and nanoscale thermometry is achieved by precision spectroscopy of the nitrogen–vacancy spins in the nanodiamonds. **b**, Simplified nitrogen–vacancy level diagram showing a ground-state spin triplet and an excited state. At zero magnetic field, the $|\pm 1\rangle$ sublevels are split from the $|0\rangle$ state by a temperature-dependent zero field splitting $\Delta(T)$. Pulsed microwave radiation is applied (detuning, $\delta$) to perform Ramsey-type spectroscopy. **c**, Comparison of sensor sizes and temperature accuracies for the nitrogen–vacancy quantum thermometer and other reported techniques. Red circles indicate methods that are biologically compatible. The open red circle indicates the ultimate expected accuracy for our measurement technique in solution (Methods).

*Figure 812: Source: Nanometre-scale thermometry in a living cell, 2013 (6 pages).*

Quantum thermometry also has applications in very low temperatures measurement, such as within cryostats and physics experiments. Quantum thermometers sensitivity is assessed in $K/\sqrt{Hz}$ meaning it increased with the number and frequency of measurements.

[2581] See Nanometre-scale thermometry in a living cell, 2013 (6 pages) and Real-time nanodiamond thermometry probing in vivo thermogenic responses by Masazumi Fujiwara et al, September 2020 (10 pages).

How are these NV center quantum thermometers working? Several methods are used like spin-based thermometry with ODMR (optically detected magnetic resonance) spectrum measurement using microwave excitement. The diamond containing NV centers can be attached to a fiber[2582]. There are even hybrid thermometers associating NV centers and more classical magnetic nanoparticle thermometers[2583].

It uses the fact that the zero-field splitting frequencies are linearly dependent on the ambient temperature. All-optical methods use the correlation between the NV center ZPL (zero-phonon lines) and temperature[2584].

There are solutions for temperature measurement in biological matter by fluorescence that are based on quantum dots[2585].

In 2017, NIST produced a quantum photonics thermometer of very small size for optically measuring the surface temperature of metals. However, the picture does not show the control electronics associated with the sensor[2586].

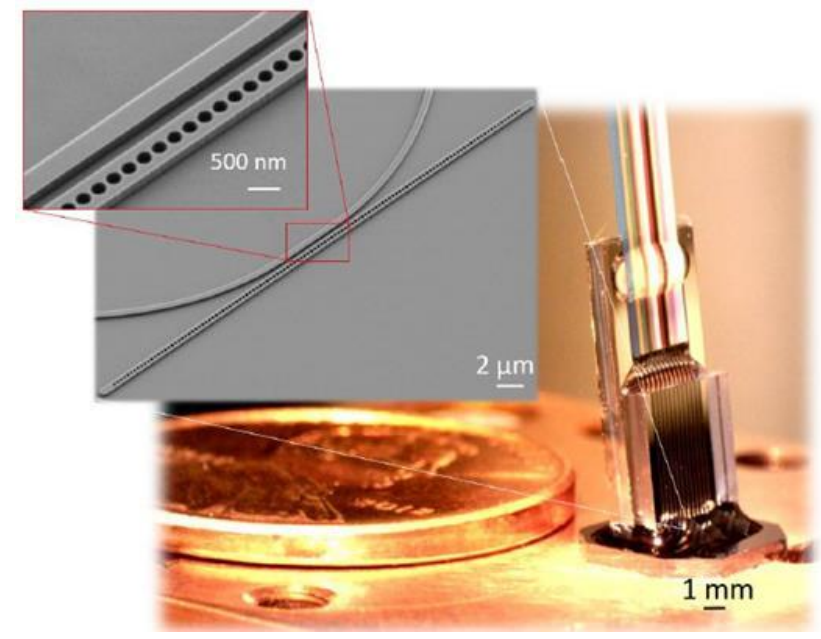

*Figure 813: quantum photonic thermometer from NIST.*

**Southwest Sciences** (1985, USA) develops optical temperature sensors based on NV centers for use in cryogenic systems. The company was founded by Alan C. Stanton and Joel A. Silver.

# Quantum frequencies sensing

Radio frequency analysis is an old matter, but it is also progressing thanks to quantum technologies often associating optronics with cold atoms. Wideband frequency quantum sensing can be implemented with neutral atoms and NV centers.

**Neutral atom**-based sensors paired with optical readout using coherent spectroscopy can analyze electromagnetic waves frequencies from direct current (0 Hz) to the THz range. This is due to their so-called high-energy Rydberg states which can help measure high-frequency electro-magnetic waves thanks to their strong dipole. It can help reduce the size of various antennas, improve radio frequency filtering and extend the range between mobile communications cellular towers. This sort of broadband spectrum analysis can have multiple use cases, particularly in the military and intelligence sectors. These systems don't necessarily showcase high-precision and can be used as "heads-up" spectrum analysis tools before more specialized tools are used for specific parts of the electromagnetic spectrum. But some progress is made to improve these neutral atom-based electromagnetic spectrography system, noticeably in the MHz bands[2587].

[2582] See Temperature dependence of nitrogen-vacancy center ensembles in diamond based on an optical fiber by Ke-Chen Ouyang et al, November 2021 (17 pages).

[2583] See Ultra-sensitive hybrid diamond nanothermometer by Chu-Feng Liu and al, 2019-2021 (9 pages

[2584] See the review paper Diamond quantum thermometry from foundations to applications by Masazumi Fujiwara and Yutaka Shikano, April-September 2021 (24 pages).

[2585] See Intracellular thermometry with fluorescent sensors for thermal biology by Kohki Okabe et al, 2018 (15 pages).

[2586] See Thermodynamic miniaturized sensors and standards and the quantum SI by Gregory F. Strouse, 2016 (39 slides).

[2587] See Highly sensitive measurement of a MHz RF electric field with a Rydberg atom sensor by Bang Liu et al, June 2022 (7 pages) and Quantum sensing of weak radio-frequency signals by pulsed Mollow absorption spectroscopy by T. Joas et al, 2017 (6 pages).

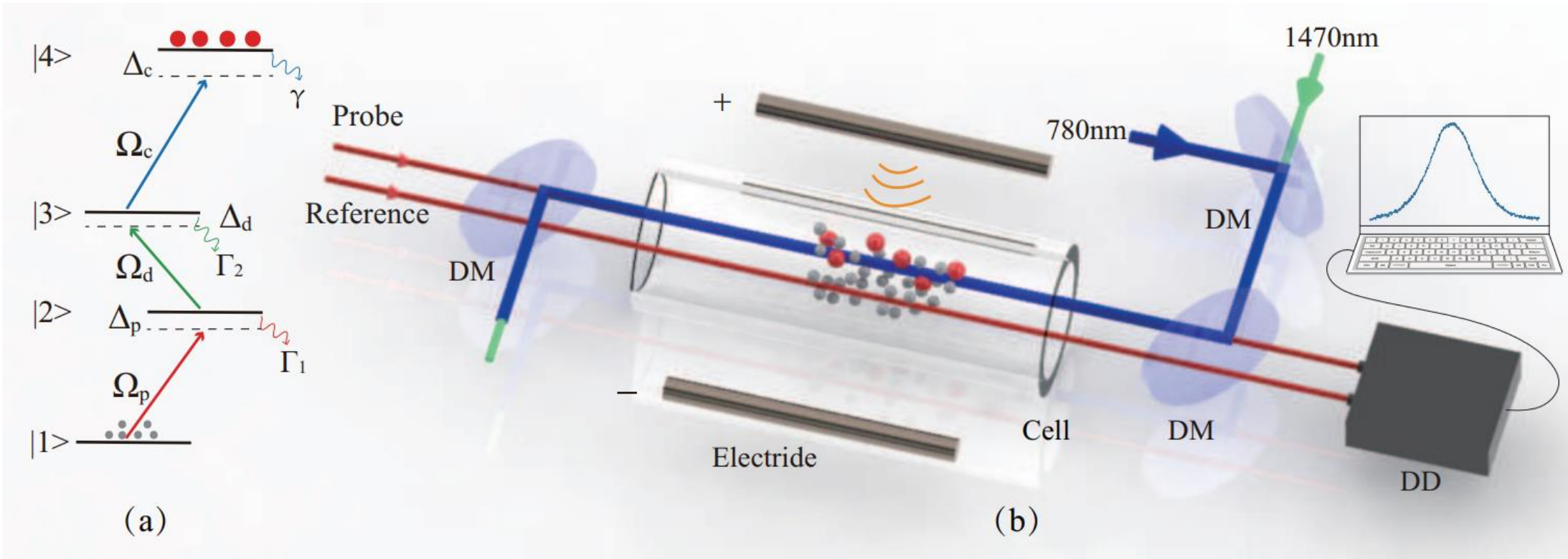

Figure 1. (a) Ladder-type four-level atomic energy diagram consisting of a ground state $|1\rangle$, two low-lying excited states $|2\rangle$ and $|3\rangle$, and a Rydberg state $|4\rangle$. An 852-nm probe light drives the transition $|1\rangle = |6S_{1/2}, F=4\rangle \rightarrow |2\rangle = |6P_{3/2}, F=5\rangle$, a 1470-nm dressing light couples the transition $|2\rangle = |6P_{3/2}, F=5\rangle \rightarrow |3\rangle = |7S_{1/2}, F=4\rangle$, and a 780-nm coupling light drives the transition $|3\rangle = |7S_{1/2}, F=4\rangle \rightarrow |4\rangle = |55P_{3/2}\rangle$ of cesium atoms. (b) Overview of the experimental setup. The probe light and the reference light propagate in parallel through a Cs vapor cell. The probe light (red) overlaps the counter-propagating coupling light (blue) and dressing light (green) to form an EIT configuration. The transmission difference between the probe and reference lights is detected by a differencing photodetector. Two electrode rods are placed parallel to each other on both sides of the vapor cell 4 cm apart. Labels: DM - dichroic mirror; DD - differencing photodetector.

*Figure 814: how cold atoms are used to measure electromagnetic waves frequencies spectrum in a highly sensitive solution developed in China, using a hot vapor cell of cesium atoms excited by lasers in their Rydberg states. The grey electrodes are connected to an RF antenna. Source: Highly sensitive measurement of a MHz RF electric field with a Rydberg atom sensor by Bang Liu et al, June 2022 (7 pages).*

A quantum sensor based on alkaline Rydberg atoms can analyze the radio spectrum from 1 kHz to 100 GHz [2588]. Other quantum sensors can analyze radio waves in the 1 THz band, that is intermediate between infrared and microwaves bands, with potential applications in the measurement of thickness of thin layers of heterogeneous materials[2589].

**NV centers** sensors can also help run RF signals spectral analysis, from 0 to 40 GHz with the advantage of being more lightweight than most neutral atoms sensors. NV centers sensors can even be arranged in sensor arrays enabling both a wideband spectrum coverage and a high-precision analysis of a target band[2590].

Some pure optics techniques can be used to implement frequency sensing in the THz range[2591].

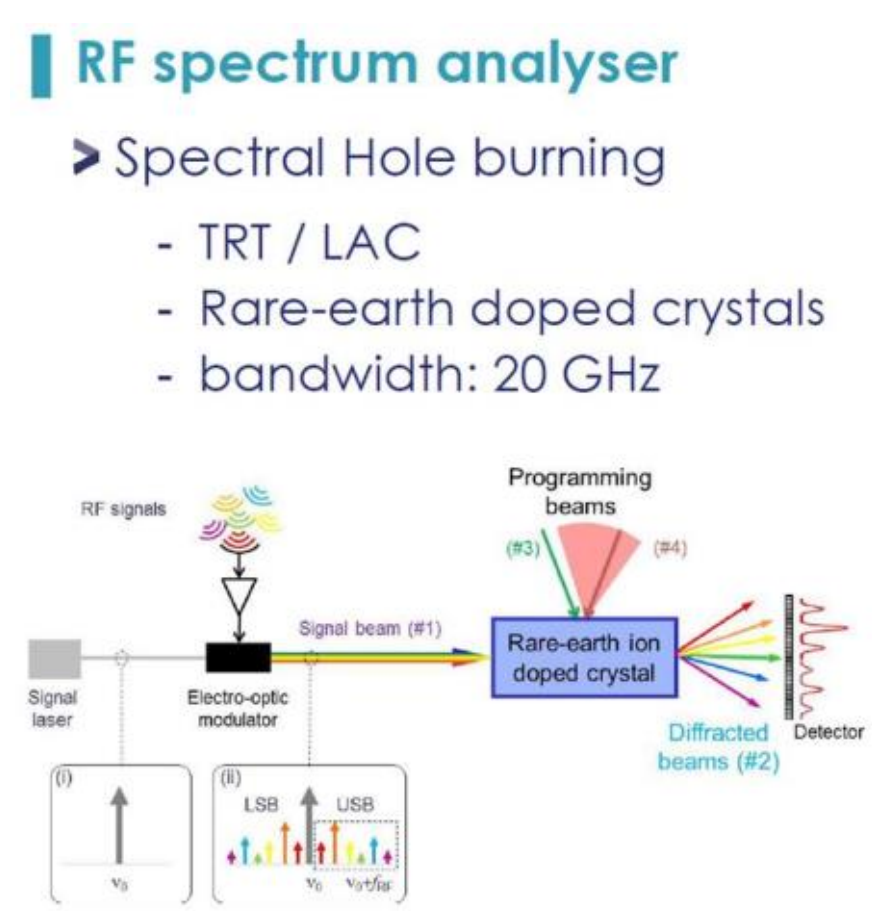

*Figure 815: RF spectrum analyzer with rare-earth doped crystals. Source TBD.*

And surprisingly, it can also work, in a different fashion, to create quantum optical microphones with the benefit to improve the quality of AI-based speech recognition[2592].

---

[2588] See Highly sensitive measurement of a MHz RF electric field with a Rydberg atom sensor by Bang Liu et al, June 2022 (7 pages) and Quantum sensing of weak radio-frequency signals by pulsed Mollow absorption spectroscopy by T. Joas et al, 2017 (6 pages). See also Assessment of Rydberg atoms for wideband electric field sensing by David H Meyer et al, January 2020 (16 pages). See also another less impressive performance from the same lab and published later: Waveguide-Coupled Rydberg Spectrum Analyzer from 0 to 20 GHz by David H. Meyer, 2021 (10 pages).

[2589] See Researchers demonstrate first terahertz quantum sensing, March 2020, which refers to Terahertz quantum sensing by Mirco Kutas et al, 2020 (9 pages).

[2590] See Sensing of Arbitrary-Frequency Fields Using a Quantum Mixer by Guoqing Wang et al, MIT, PRX, June 2022 (22 pages).

[2591] See Terahertz quantum sensing by Mirco Kutas et al, March 2020 (8 pages).

[2592] See A Quantum Optical Microphone in the Audio Band by Raphael Nold et al, April 2022 (7 pages).

**Rydberg Technologies** (2015, USA) provides cesium or rubidium specimens for cold atom-based metrology solutions. They also sell a Rydberg atoms-based radio-frequency probe, a RFLS (Rydberg Field Measurement System)[2593]. Their technology is also integrated in AM and FM radio-frequency receivers.

**Thales** is developing a real-time radiofrequency spectral analyzer *aka* a Quantum Diamond Signal Analyzer (Q-DiSA) with a bandwidth of 10 MHz to 25 GHz and a frequency resolution of 1 MHz, based on NV centers[2594]. The frequency adjustment is done with controlling the distance of the NV center with a small 1.3 cm magnetic sphere. The system is using a 532 nm laser and a CMOS sensor.

At last, we should mention **quantum antennas** where some non-classical phenomenon can occur like the creation of squeezed states[2595]. It is an emerging field that can address particular needs, like field shaping, quantum radars, quantum imaging and creating antennas for THz electromagnetic fields using quantum dots[2596]. Since 2022, **BAE Systems** (UK) is developing quantum antennas on behalf of DARPA.

## Quantum imaging

Quantum sensing is enlarging the scope of what is possible to do in imaging, at both the microscale and nanoscale levels, and in the biology and material inspection realms. These new tools are based on various techniques that we'll cover here: **SQUIDs**, aka superconducting qubits[2597], **NV-centers** based magnetometry to implement NMR spectroscopy which tend to dominate the market nowadays thanks to their space resolution and readout precision, and **OPMs**, aka optical-pumped magnetometers which are mainly used for neuron imaging (MEG, magnetoencephalography). We'll also have a look at some mysterious "ghost imaging" systems that take pictures of objects with a single pixel sensor and other special quantum sensors.

**SQUIDs** are classically used in MRI systems which require helium-4 based cryogeny. They are also implemented in various specific use case like in sub-mm astronomy and infrared (IR) imaging. One such imager with 10,000 pixels is embedded in the NIRSpec near-infrared telescope in the JWST (James-Webb Space Telescope)[2598].

**NV-centers** imagers can be used to analyze organic molecules with excellent spatial resolution. It makes use of electron spin resonance spectroscopy (ESR) at cryogenic temperature which allows to examine atoms and molecules at the level of their electron spin. This enables NMR (nuclear magnetic resonance) detection[2599]. The technique is usually integrated in scanning tunneling microscopes as well as in atomic force microscopes (AFM). The electron spin of the examined materials is excited by a magnetic field and microwaves. The NV-center technique allows the examination of a hard disk and semiconductors defects with a probe equipped with a single NV-center in Figure 816 and Figure 818.

---

[2593] See A vapor-cell atomic sensor for radio-frequency field detection using a polarization-selective field enhancement resonator by D. A. Anderson et al, 2018 (11 pages).

[2594] See A quantum radio frequency signal analyzer based on nitrogen vacancy centers in diamond by Simone Magaletti, Ludovic Mayer, Jean-François Roch and Thierry Debuisschert, Nature Communications, February-July 2022 (8 pages).

[2595] See the review paper Quantum Antennas by Gregory Ya. Slepyan, Svetlana Vlasenko and Dmitri Mogilevtsev, June 2022 (70 pages).

[2596] See Generating tunable terahertz radiation with a novel quantum dot photoconductive antenna by Andrei Gorodetsky, Ksenia A. Fedorova, Natalia Bazieva and Edik U. Rafailov, Aston University Birmingham, 2016.

[2597] See Magnetometry of neurons using a superconducting qubit by Hiraku Toida et al, NTT Research Labs, June 2022 (6 pages).

[2598] Seen in Micromachined Quantum Circuits by Teresa Brecht, 2017 (271 pages)

[2599] See Advances in nano- and microscale NMR spectroscopy using diamond quantum sensors by Robin D. Allert et al, May 2022 (42 pages).

It is also used for the characterization (quality control) of integrated circuits working with millimeter frequencies such as in 5G[2600]. Others are working on NV centers-based microscopy of living cells[2601].

There is even an application to qualify malaria patients by analyzing hemozoin nanocrystals appearing in red blood cells affected by the disease parasite[2602]. These techniques are used with confocal microscopy. This generates images with a very shallow depth of field of about 400 nm, creating optical sections of the sample to analyze. With modifying the position of the depth focal plane, a series of images are created which are then assembled to generate a 3D view of the analyzed sample. The light source is reflected or obtained by fluorescence in reaction to a laser beam. The result is a Confocal Laser Scanning Microscope (CLSM). NV-centers can also improve the accuracy of adaptive optics, which are used in astronomy[2603].

Imaging can also exploit an array of small NV centers that provide much better resolution than imaging systems based on SQUIDs magnetometers. The examples in Figure 818 show its architecture[2604].

The second was made to study bacteria that contain magnetic microelements. In other cases, magnetic markers can be used to attach themselves to the cells to be detected, typically in oncology.

NV centers could also be used in heart monitoring using magnetocardiography. It was tested on rats in Japan, as shown in Figure 817 [2605].

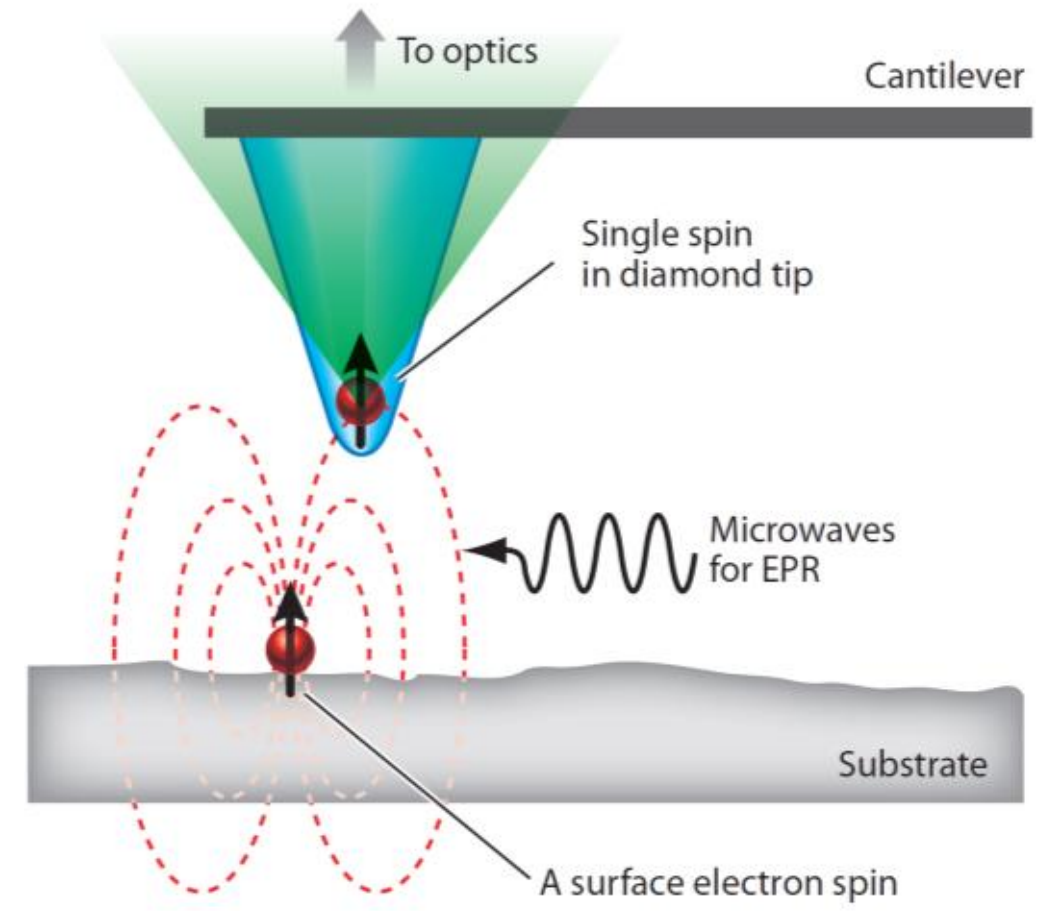

*Figure 816: a typical NV center in a diamond tip for various imaging applications. Source: [Nitrogen-Vacancy Centers in Diamond: Nanoscale Sensors for Physics and Biology]() by Romana Schirhagl, Kevin Chang, Michael Loretz and Christian L. Degen, ETH Zurich, 2014 (27 pages).*

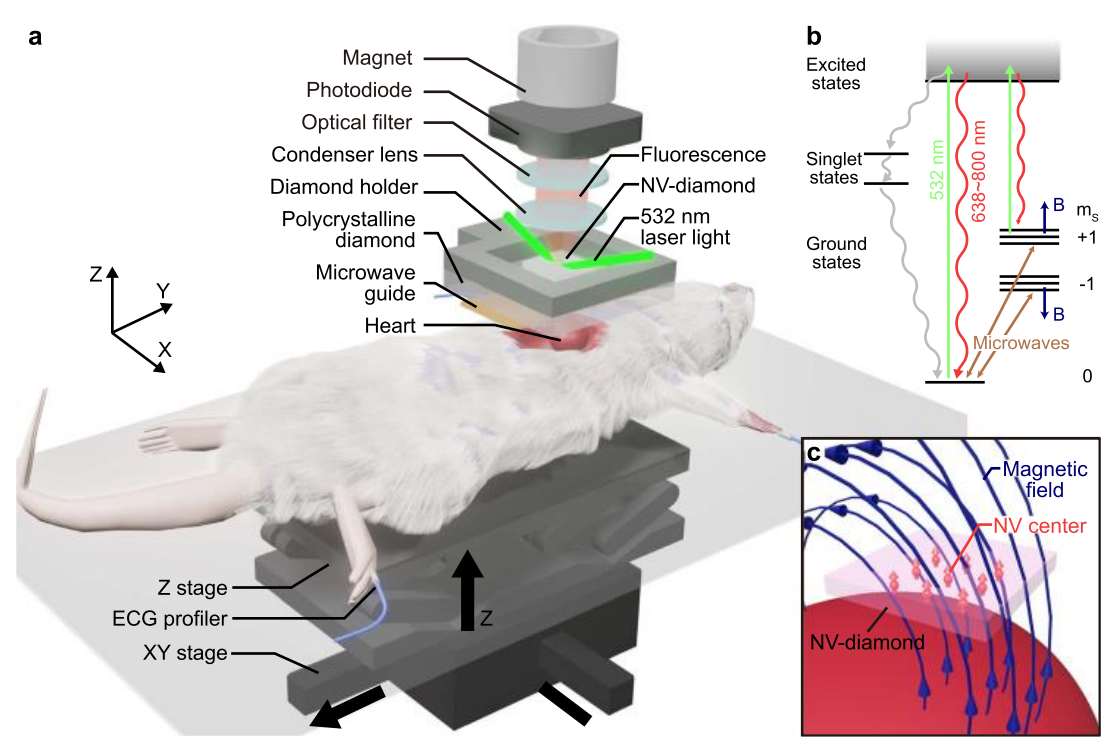

*Figure 817: NV center based magnetocardiography experiment on rats. Source: [Millimetre-scale magnetocardiography of living rats with thoracotomy]() by Keigo Arai et al, Nature Communications Physics, August 2022 (10 pages).*

---

[2600] See [Microwave Device Characterization Using a Widefield Diamond Microscope](), 2018 (10 pages) which involves in particular the LSPM of Paris.

[2601] See [A fluorescent nanodiamond foundation for quantum sensing in cells](), 2018 (147 pages) which discusses microscopy of living cells.

[2602] See [Diamond magnetic microscopy of malarial hemozoin nanocrystals]() by Ilja Fescenko et al, September 2018 (17 pages),

[2603] See [Nanodiamonds enable adaptive-optics enhanced, super-resolution, two-photon excitation microscopy](), 2019 (7 pages).

[2604] See [Enhanced widefield quantum sensing with nitrogen-vacancy ensembles using diamond nanopillar arrays]() by D. J. McCloskey, 2019 (7 pages). The NV centers matrices are 100 µm wide. The illustration comes from other work published in 2013, cited in the conference [Magnetic imaging using NV-diamond: techniques & applications]() by Ronald Walsworth, 2015 (51 min). Notably [Optical magnetic imaging of living cells](), Le Sage et al, Nature, 2013 (11 pages). See also [Principles and Techniques of the Quantum Diamond Microscope]() by Edlyn V. Levine et al, 2019 (47 pages) and [Atomic Scale Magnetic Sensing and Imaging Based on Diamond NV Centers]() by Myeongwon Lee et al, 2019 (17 pages).

[2605] See [Millimetre-scale magnetocardiography of living rats with thoracotomy]() by Keigo Arai et al, Nature Communications Physics, August 2022 (10 pages).

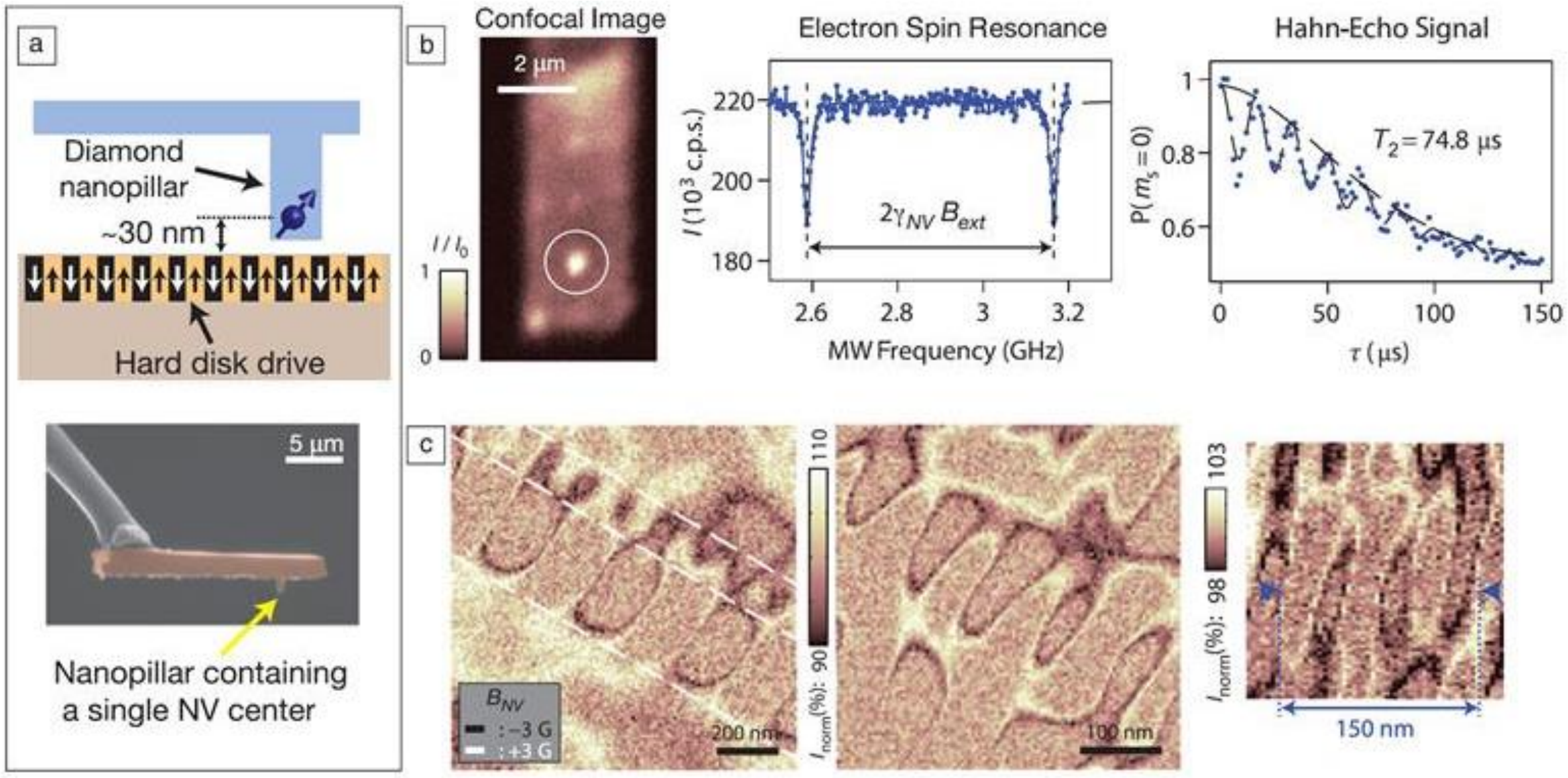

**Figure 4.** (a) Schematic of a monolithic diamond nanopillar probe (top) and representative SEM image of the nanopillar probe (bottom). (b) Characteristics of a nanopillar probe device. Confocal image of the device (left) clearly shows a localized fluorescence spot from a single NV center at the position of the nanopillar. Electron spin resonance (middle) was acquired with an enhanced fluorescence of 220,000 photons/sec. The coherence time of the measured Hahn-echo signal (right) is 74.8 µs, an order of magnitude longer than a typical Hahn-echo coherence time of commercial diamond nanocrystals (~5 µs). (c) Magnetic images of a hard disk drive acquired by the nanopillar probe. Alternating magnetic bits were imaged with varying sizes down to 25 nm (right), indicating the distance between a single NV center at the probe and the hard disk sample is roughly within 25 nm. Adapted with permission from Reference 19. © 2012 Nature Publishing Group.

*Figure 818: Source: Nitrogen-Vacancy Centers in Diamond: Nanoscale Sensors for Physics and Biology by Romana Schirhagl, Kevin Chang, Michael Loretz and Christian L. Degen, ETH Zurich, 2014 (27 pages).*

The European Quantum Flagship includes **MetaboliQs** (Germany, €6.7M), a diamond-based nuclear magnetic resonance cardiac medical imaging project. It also detects atrial fibrillation, a common cardiac pathology, with a rubidium-based atomic magnetometer[2606]. Another Flagship project, **PhoG** (United Kingdom, €2.6M) or Sub-Poissonian Photon Gun by Coherent Diffusive Photonics, is about creating stable light sources for various applications, particularly in quantum sensing. It involves researchers from Belarus, Germany and Switzerland.

**Optically-pumped magnetometers** (OPMs) are scalar-type quantum sensors enabling the measurement of very weak magnetic fields and based on the Zeeman effect[2607]. These sensors have emerged in the 1970s and expanded since the 2000s thanks to a better sensitivity, on par with SQUIDs.

One of their main use case is MEG (magneto-encephalography) to examine brain activity[2608] but it can also be used for materials inspections[2609]. OPMs are competing with MRI and NMR imaging systems. They are more lightweight, can be wearable, and don't require any cryogeny like with MRIs that are based on SQUIDs. OPMs can for example help track some neurodegenerative diseases like dementia, Alzheimer's and Parkinson's with tracking patients' brain waves and how their change over time[2610].

OPMs principle of operation is rather simple. A small laser beam which can be a vertical-cavity surface emitting lasers (VCSEL) illuminates a heated alkali atoms vapor (rubidium, cesium or potassium) trapped in a millimeter-scale glass cell whose size determines the sensor spatial resolution.

---

[2606] See New quantum technology could help diagnose and treat heart condition, March 2020.

[2607] See the review paper Optical magnetometry by Dimitry Budker and Michael Romalis, Berkeley, Nature Physics, 2007 (8 pages).

[2608] See Optically pumped magnetometers: From quantum origins to multi-channel magnetoencephalography by Tim M. Tierney et al, 2019 (11 pages).

[2609] See Optically Pumped Magnetometer Measuring Fatigue-Induced Damage in Steel by Peter A. Koss et al, 2022 (11 pages).

[2610] See Improved spatio-temporal measurements of visually evoked fields using optically-pumped magnetometers by Aikaterini Gialopsou et al, Nature Scientific Reports, November 2021 (11 pages).

The atoms nuclear and electron spin are influenced by the ambient magnetic field, which changes the vapor optical properties that will absorb more or less light depending on the probed magnetic field. Light is then measured by a photodetector after traversing a polarizing beam splitter[2611].

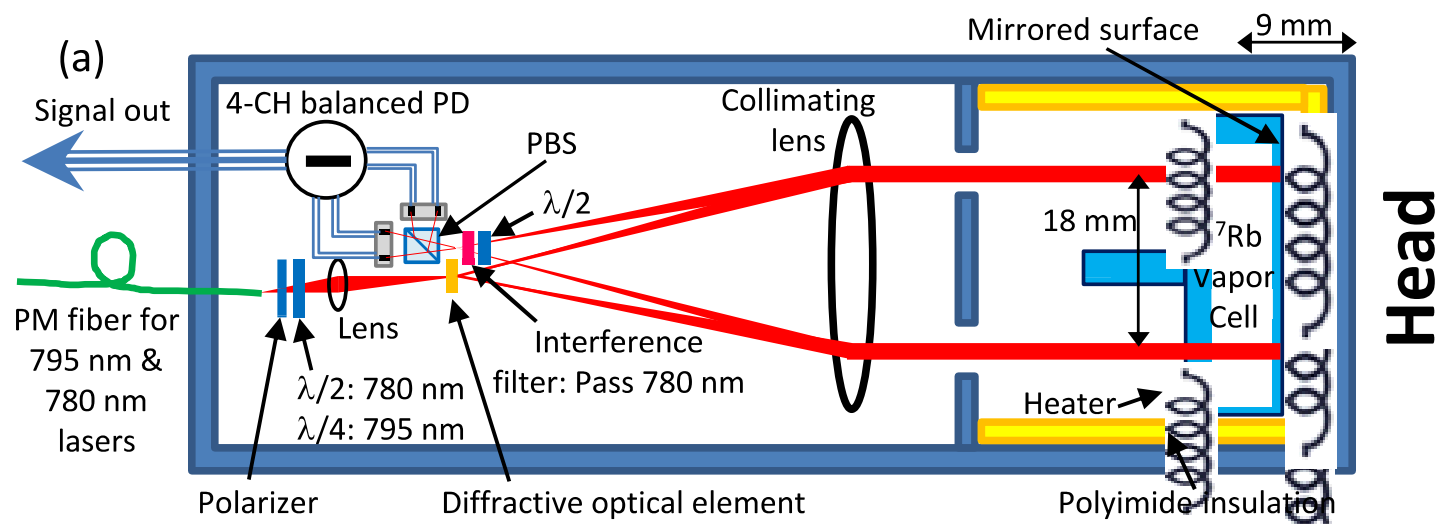

*Figure 819: Source: Four-channel optically pumped atomic magnetometer for magnetoencephalography by Anthony P. Colombo et al, 2016 (15 pages).*

Other quantum optical imaging techniques using **laser interferometry** make it possible to examine molecules at the atomic level in their environment and not in a vacuum and cryogenic cold[2612].

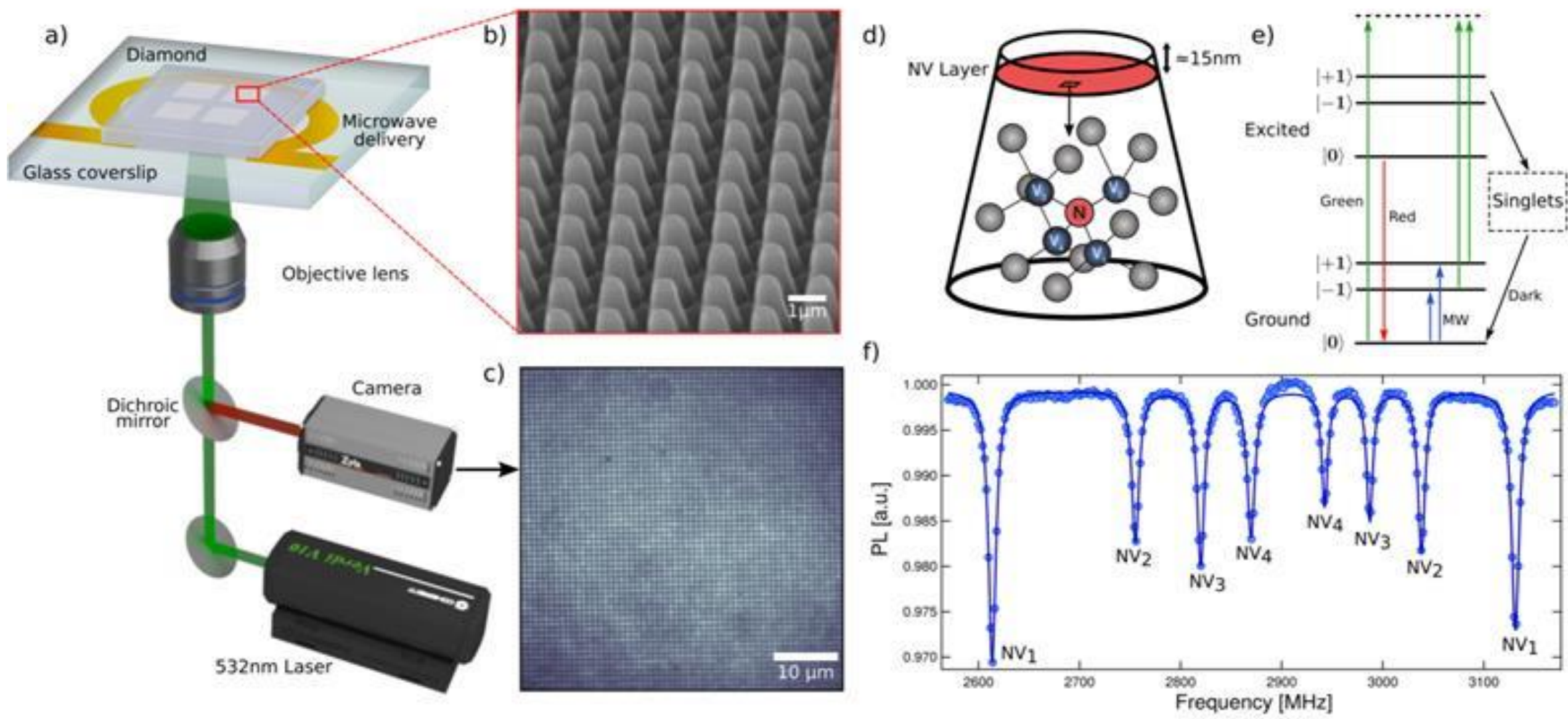

*Figure 820: a widefield array of NV centers to improve their sensitivity, developed in Australia. Source: Enhanced widefield quantum sensing with nitrogen-vacancy ensembles using diamond nanopillar arrays by D. J. McCloskey, 2019 (7 pages).*

The China laboratory of Jian-Wei Pan has developed a camera that analyzes the reflection of a single photon per pixel on the object to be observed. This is associated with algorithms filtering out the noise. Imaging is done in the infrared at 1550 nm and with polarized photons. This could be integrated in observation satellites[2613].

There is also a broad field of imaging that can be implemented with free electrons illumination and nanophotonics but it is at best in type I quantum sensors[2614].

[2611] See Four-channel optically pumped atomic magnetometer for magnetoencephalography by Anthony P. Colombo et al, 2016 (15 pages).

[2612] See An Entanglement-Enhanced Microscope by Takafumi Ono, Ryo Okamoto, Shigeki Takeuchi, 2014 (8 pages).

[2613] See A new camera can photograph you from 45 kilometers away, May 2019 which refers to Single-photon computational 3D imaging at 45 km by Zheng-Ping Li et al, April 2019 (22 pages). And the presentation Single Photon LiDAR by Feihu Xu, June 2019 (25 slides).

[2614] See the review paper Free-electron-light interactions in nanophotonics by Charles Roques-Carmes, August 2022 (34 pages). **Astrahl** (2022, USA) is a startup created out of the MIT which creates nanophotonics imaging systems based on free electrons scintillation. See A framework for scintillation in nanophotonics by Charles Roques-Carmes et al, Science, February 2022 (14 pages).

**QLM Technology** (UK, $14,64M) developed a "Quantum Gas Imaging Lidar", a quantum magnetometer solution that detects methane leaks in pipelines up to 100 meters away. The measuring system weighing a few kg can be embarked in a large drone flying at 50 km/h. They use a laser that illuminates a gaseous medium of variable opacity and a photodetector.

**IDQ** is involved in the creation of the solution at the LiDAR level. Their solution will be embedded in "Schlumberger End-to-end Emissions Solutions" (SEES), a product line that detects and eliminates oil and gas industry's polluting methane and flaring emissions. Their "Quantum Gas Imaging Lidar" will be embedded in "Schlumberger End-to-end Emissions Solutions" (SEES), a product line that detects and eliminates oil and gas industry's polluting methane and flaring emissions.

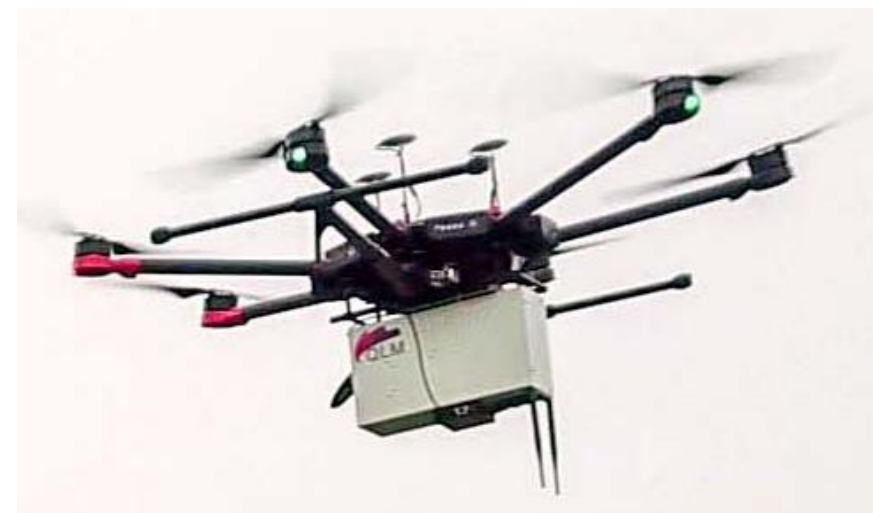

Figure 821: an airborne LIDAR to detect gas leaks.

**SeeDevices** (2017, USA) develops a quantum imaging system, the PAT-PD (Photon Assisted Tunneling Photo Detector), which exceeds the performance of traditional CMOS imagers.

This imager contains photosites using the tunneling effect that can detect single photons in a wide range of wavelengths from near infrared (1800 nm) to ultraviolet (up to UVA1, at 350 nm). This can be used for seeing in the dark and for medical imaging, like for detecting blood vessels in the infrared range.

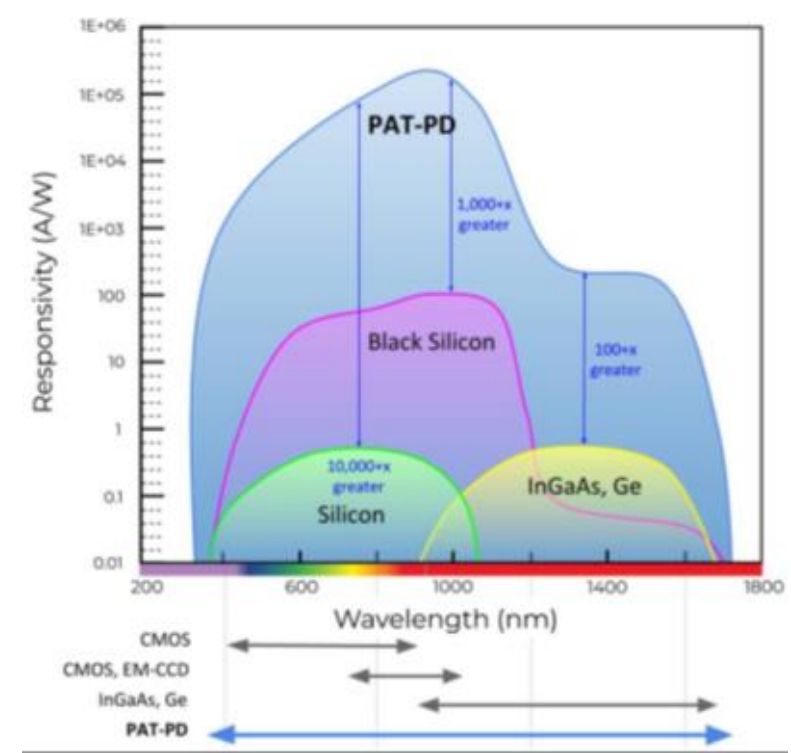

Figure 822: SeeDevice covered wavelengths.

**QDTI** (2012, USA) is the only known startup initially engaged in the development of a quantum computer based on NV Centers. Created by a team from Harvard University, it is logically based in Boston.

The startup is now focused mainly on medical imaging systems also using these NV centers, with the creation of precision magnetometers combined with MRI and immunological tests.

**Nvision Imaging** (2015, Germany) is developing an NV centers-based MRI medical imaging solution.

**SBQuantum** (2019, Canada) also known as SBTech and Shine Bright Technologies develops NV centers-based quantum magnetometers. They target the automotive market but are also working on integrating their technology into Cubesat-type satellites. It is also a startup coming from the Institut Quantique from the University of Sherbrooke.

They have developed a tri-axial version of their system that measures magnetism in the three X, Y and Z axis. Research on this product was funded by the NIH (National Health Institute). The product is mainly targeting magnetoencephalography (MEG) brain imaging.

**Chipiron** (2020, France) develops a portable and helium-free 2D and 3D MRI system using stacked SQUIDs quantum detectors (superconducting / Josephson effect based). The startup created by Dimitri Labat and Evan Kervella is first targeting osteoporosis and brain diagnosis. The product is to be sold by 2024[2615].

**Qnami** (2017, Switzerland, $7.3M) is a spin-off from the Quantum Metrology Research Laboratory at the University of Basel. Among other things, they produce artificial diamonds for various photonics applications.

Their targeted first the quantum sensing market with their Quantilever MX nano-diamonds probes.

They then launched in 2019 the ProteusQ range of NV center confocal microscopes, used to analyze ferromagnetic materials, based on the Quantilever probe.

Quantonation and Runa capital are among their investors. And one of the co-founders and the CEO, Mathieu Munsch, came from Grenoble Phelma and worked at the CEA in Grenoble.

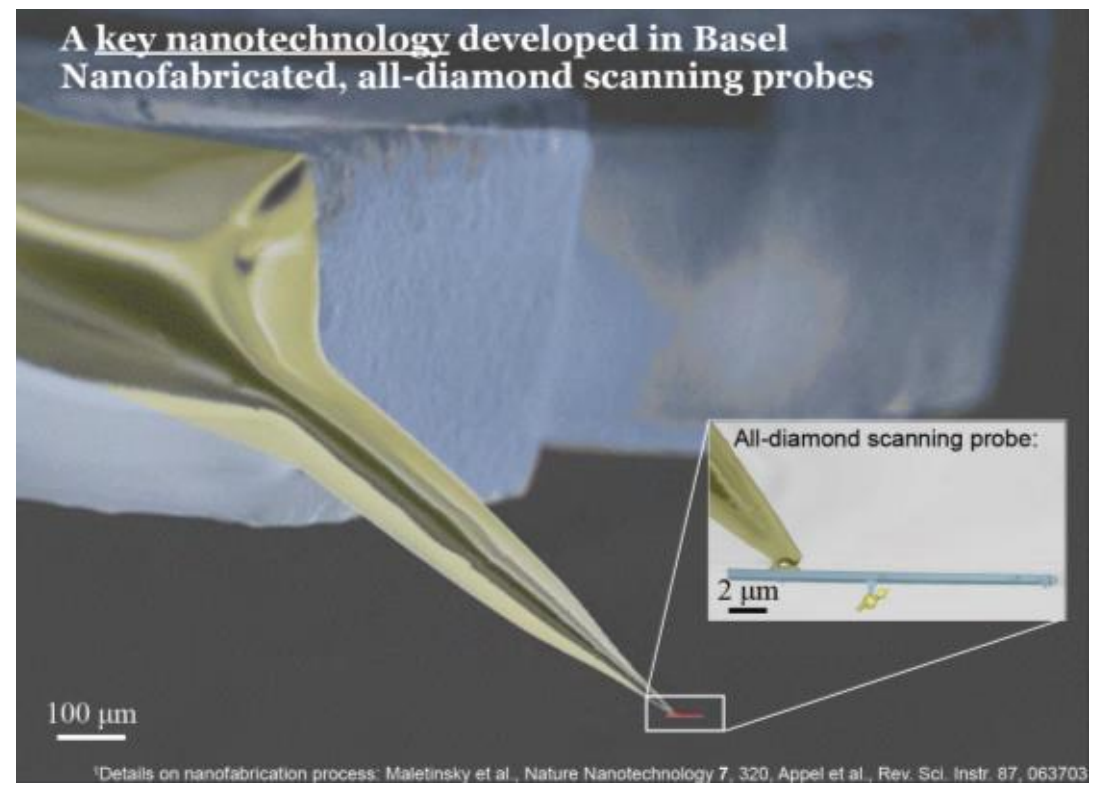

*Figure 823: Qnami NV center based imaging system.*

**Mag4Health** (2021, France) is a spin-off from CEA-Leti who develops an helium optically-pumped magnetometer based magnetoencephalography (MEG) system, working at room temperature[2616].

It is using quantum sensors that were developed at CEA-Leti. These record the brain's electromagnetic activity in real time, helping neuronal diseases diagnosis. Mag4Health MEGs are much lighter (about 150 kg) than classical systems using large superconducting magnets (5 tons).

**QuSpin** (2012, USA) is proposing an optical magnetometer (OPM) for neuroimagery.

**Siloton** (2020, UK, £470K) is developing optical coherence tomography (OCT) solutions for the assessment of age-related macular degeneration.

The company is the first investment from the Quantum Exponential Group launched by Steven Metcalfe. Their chip devices are designed by VLC Photonics in Spain and manufactured by Ligentec in Switzerland. OCT uses low-coherence light to capture micrometer-resolution, 2- and 3-dimensional images from within optical scattering media like the retina. It is also used in nondestructive testing (NDT) in the industry. It makes use of low-coherence interferometry with near-infrared light that easily penetrate the inspected medium. It probably belongs to Type I quantum sensing.

---

[2615] See Chipiron Ultra-low field MRI with SQUID detection by Dimitri Labat, February 2022 (13 pages).

[2616] See Performance Analysis of Optically Pumped 4He Magnetometers vs. Conventional SQUIDs: From Adult to Infant Head Models by Saeed Zahran et al, Mag4Health, Inserm and CEA, April 2022 (18 pages).

Quantum imaging could also use the curious technique of ghost imaging or quantum phantom imaging. It exists in many variations. The first one used a generator of infrared photons in 1995[2617]. One half of the photons illuminates the object and the other half a photo sensor, by crossing an optical delay line[2618]. The photons illuminating the object are entangled with those illuminating the camera, which has not seen the object at all! The obtained image is very noisy and requires some appropriate processing.

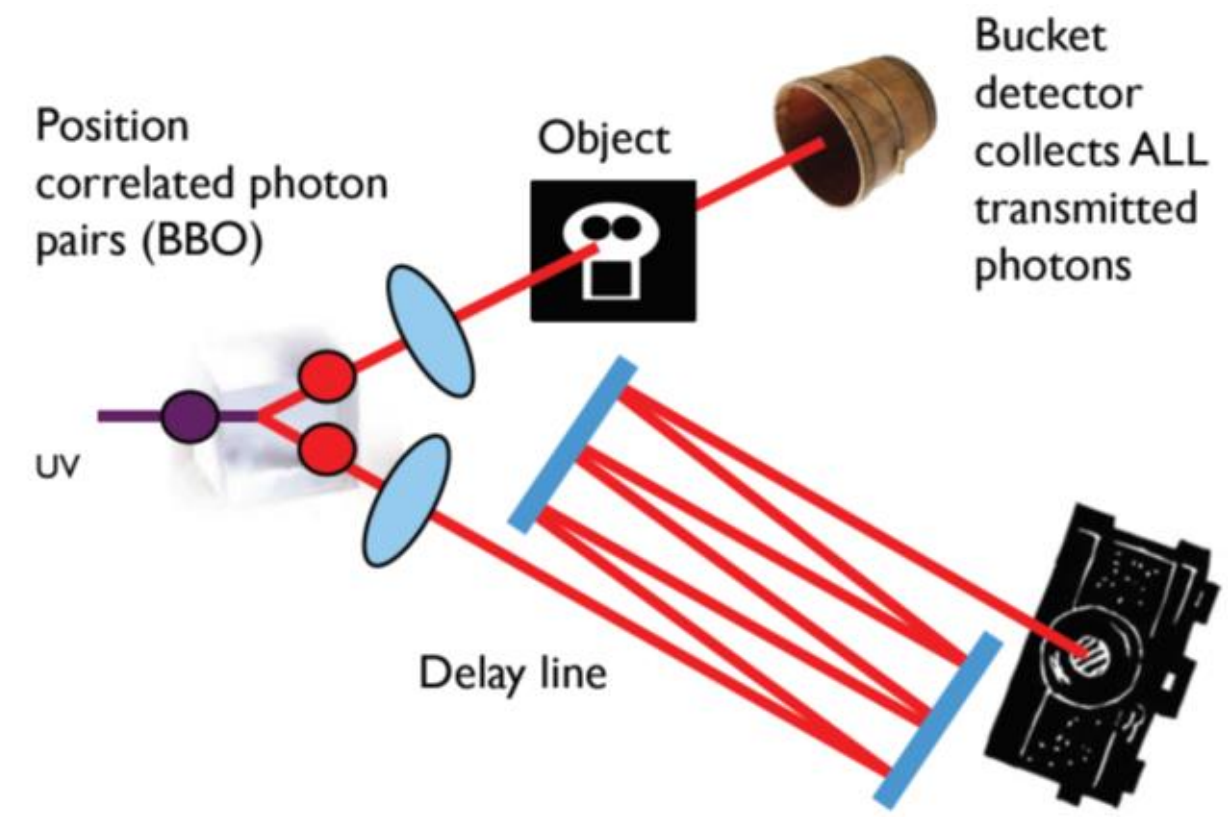

*Figure 824: ghost imaging principle. Source: An introduction to ghost imaging: quantum and classical by Miles Padgett and Robert Boyd, 2016 (10 pages)*

What is the purpose of this? Mainly to analyze objects with a very low photon number to avoid that they modify the object to be analyzed. This can be interesting in microbiology[2619]. The analyzed objects are seemingly always very small[2620].

Other non-quantum techniques use a single-pixel color imager that uses 1300 structured lights per second to illuminate the object for a few seconds, with up to one million iterations. The sensor prototyped at the University of Glasgow in 2013 contains four photodiodes positioned at different locations[2621]. This makes it possible to generate a 3D view of the object.

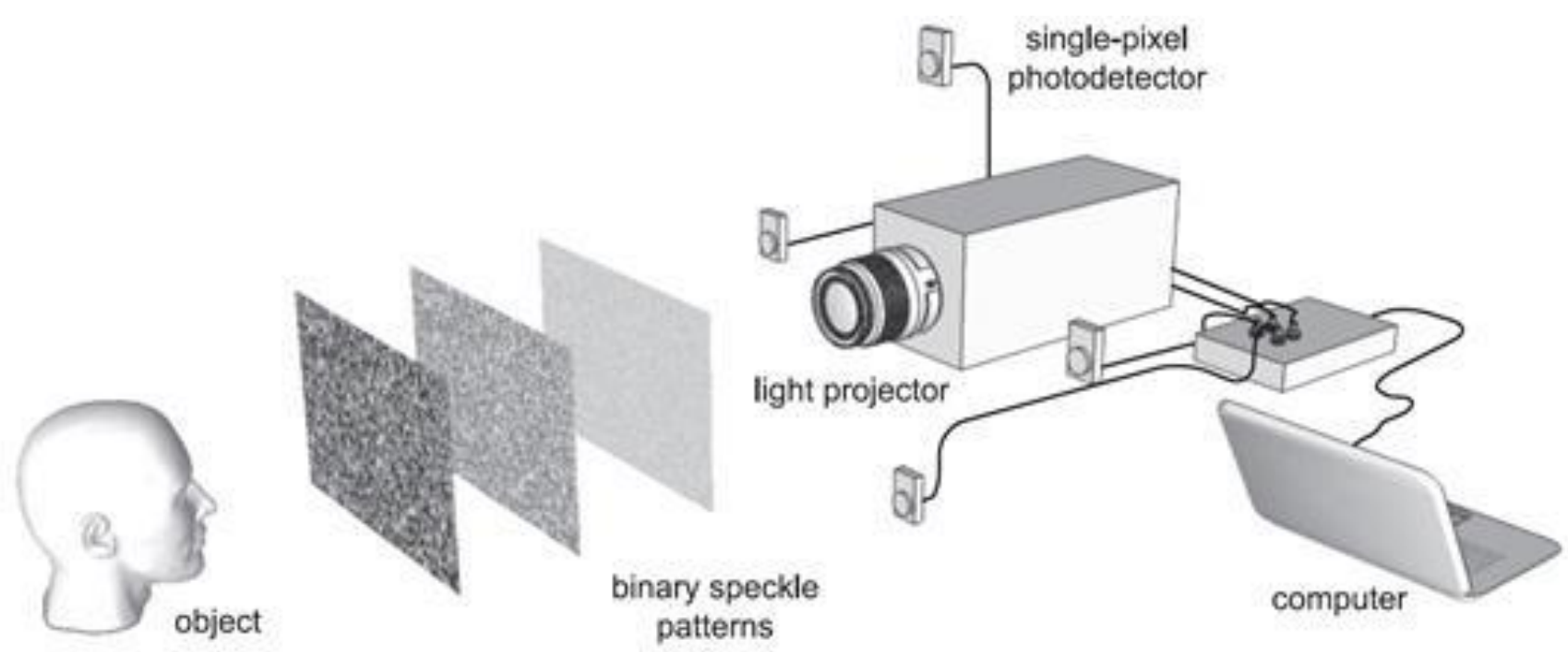

**Fig. 1. Experimental setup used for 3D surface reconstructions.** The light projector illuminates the object (head) with computer-generated random binary speckle patterns. The light reflected from the object is collected on four spatially separated single-pixel photodetectors. The signals from the photodetectors are measured and used to reconstruct a computational image for each photodetector.

*Figure 825: Source: 3D Computational Imaging with Single-Pixel Detectors, 2013 (4 pages).*

Finally, quantum imaging can also rely on the illumination of the object by entangled microwaves, using the principle of quantum radar that we will see in a following section. It is interesting for analyzing objects with low reflectivity, which would be useful in medical imaging as well as for creating short-range radars[2622]. Entangled photon can also be used to create holograms, as recently discovered by a team of physicists from the University of Glasgow[2623].

---

2617 See Optical imaging by means of two-photon quantum entanglement by Yanhua Shih et al, 1995 (4 pages), University of Maryland. And Observation of two-photon 'ghost' interference and diffraction by Yanhua Shih, 1995 (4 pages).

2618 See An introduction to ghost imaging: quantum and classical by Miles Padgett and Robert Boyd, 2016 (10 pages) provides a good overview of the subject. See also Quantum Ghost Image Identification with Correlated Photon Pairs, 2010 (4 pages).

2619 See The Dawn of Quantum Biophotonics by Dmitri Voronine et al, 2016 (30 pages).

2620 See this panorama of many ghost imaging methods: The promise of quantum imaging by Robert Boyd, 2016 (53 slides).

2621 See Fast full-color computational imaging with single-pixel detectors by Stephen Welsh et al, 2013 (7 pages). Also seen in 3D Computational Imaging with Single-Pixel Detectors, 2013 (4 pages) which extends this to the capture of 3D objects using four single-pixel sensors. The video projector creates patterns that illuminate the object and alternate with its negative. See finally Imaging with a small number of photons by Peter Morris et al, 2014 (9 pages) and Quantum-inspired computational imaging, 2019 (9 pages).

2622 See Experimental Microwave Quantum Illumination by S. Barzanjeh, S. Pirandola et al, August 2019.

2623 See Polarization entanglement-enabled quantum holography by Hugo Defienne et al, Nature, 2021 (31 pages).

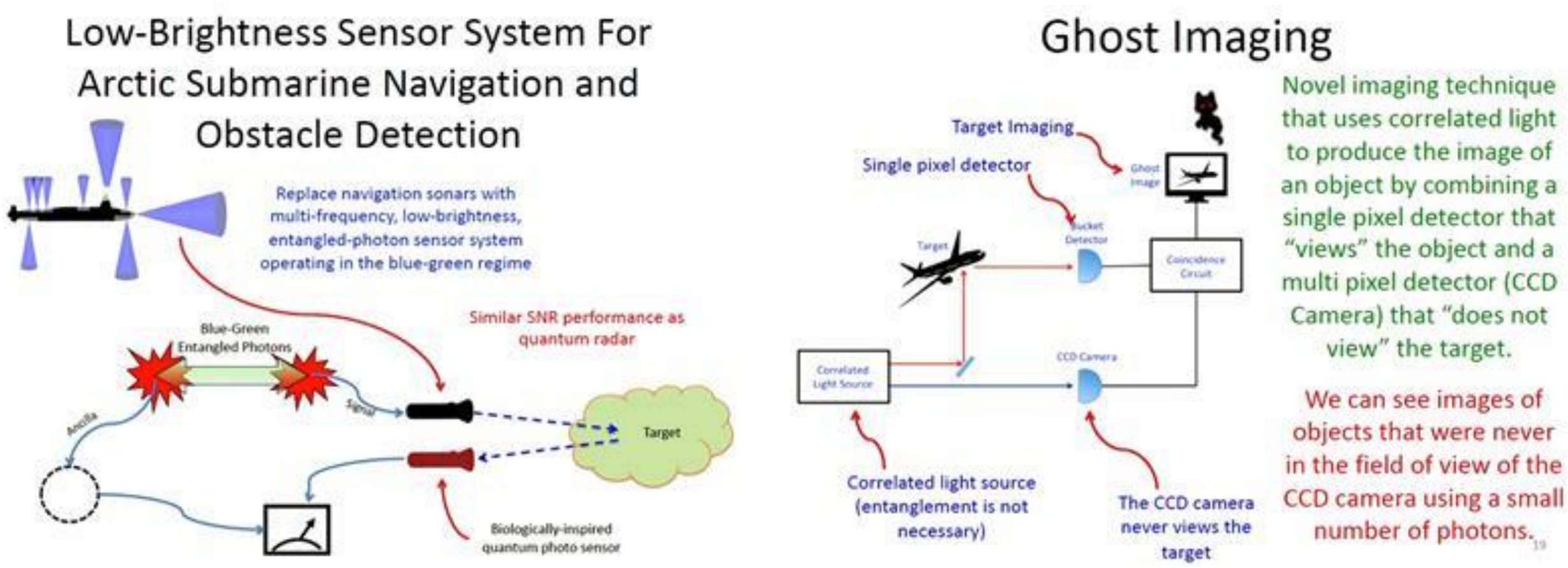

*Figure 826: ghost imaging. Source: [The Future of Quantum Sensing & Communications]() by Marco Lanzagorta of the US Naval Research Laboratory (USA), September 2018 (37 minutes).*

Another envisaged technique is the generation of **ghost images**, generated by a system coupling a camera that does not see the object to be captured and a single pixel sensor that sees the object. This kind of technique can be based on the entanglement of photons or of twister pairs of photons in the visible between the two sensors[2624].

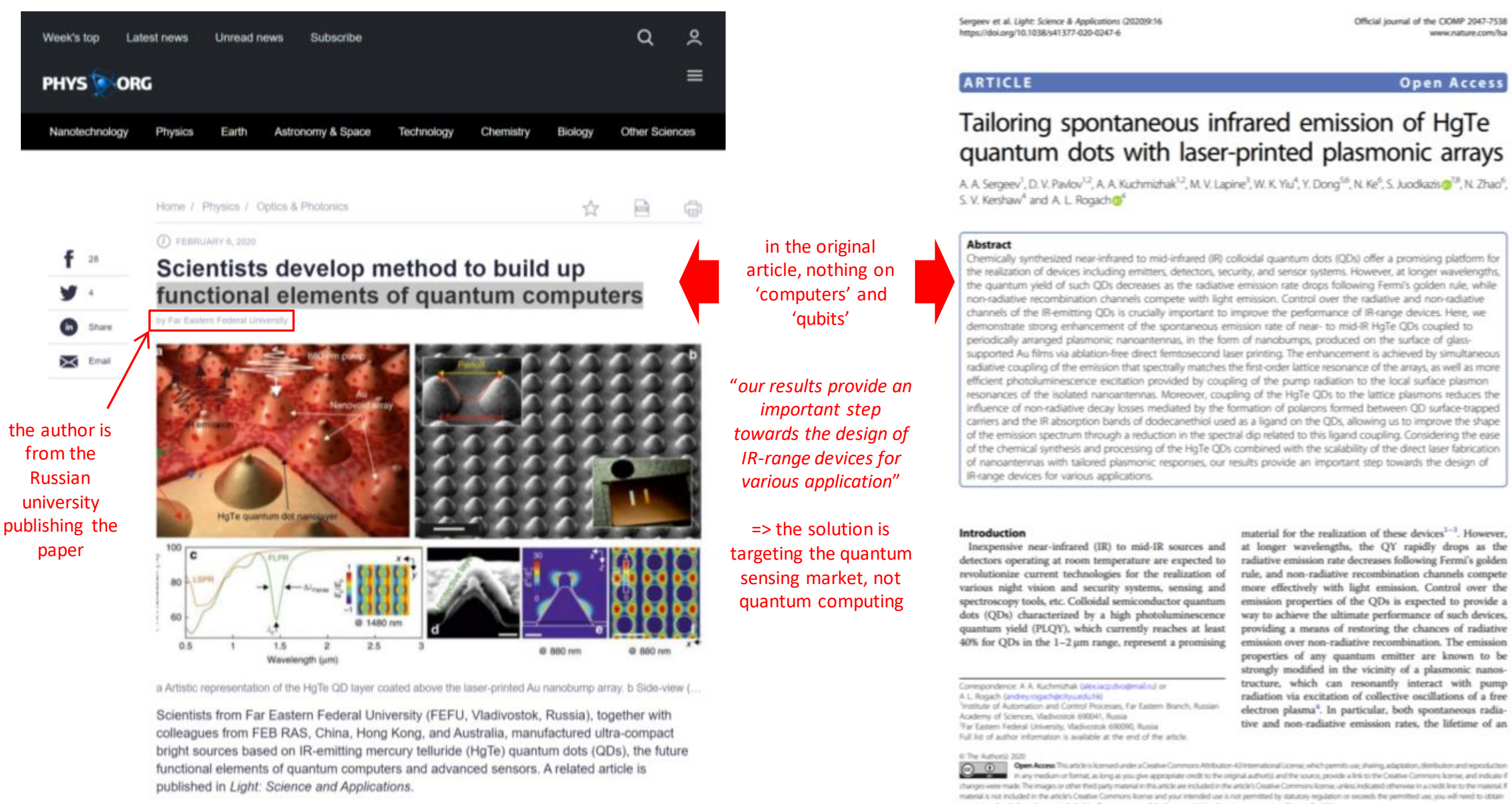

*Figure 827: another example of how some research works gets hype to an incredible extend. Source: [Scientists develop method to build up functional elements of quantum computers]() by Far Eastern Federal University, February 2020 and [Tailoring spontaneous infrared emission of HgTe quantum dots with laser-printed plasmonic arrays]() by A. A. Sergeev et al, 2020 (10 pages).*

Note that the topic of quantum imaging can sometimes be confusing. This is eloquent with this promotion of a particular type of quantum dot, confused with some fancy quantum computing technique (*above in* Figure 827).

2624 See [An introduction to ghost imaging]() by Miles John Padgett and Robert W. Boyd, 2017 (11 pages) and [Quantum imaging exploiting twisted photon pairs]() by Dianzhen Cui et al, June 2022 (6 pages).

And here, it was not a problem with some journalist since the "wrong" article comes from the laboratory promoting the research results[2625].

## Quantum pressure sensors

Quantum sensing also works with pressure measurement. Most pressure sensors are type I, not using superposition or entanglement features. They are used to measure pressure in specific conditions.

We have for example very thin ultra-high sensitivity pressure sensors used in medical monitoring[2626], for the measurement of low pressures using light interacting with helium gas and pressure dependent refractive index of helium[2627], high-pressures measurement up to 4 GPa using the shift of the optical spectra of quantum-wells made of III/V GaAs/AlGaAs materials[2628], with quantum dots used to measure pressure in liquids[2629], and with using SiV and GeV color centers to measure ultra-high pressure up to 180 GPa[2630].

## Quantum radars and lidars

Quantum radars are slowly emerging from research. The idea initially came from Seth Lloyd from the MIT in 2008 when he devised the concept of quantum illumination[2631]. They rely on photons in the visible spectrum, and in three different ways:

- The radar emits classical photons in the visible and receives the photon reflected by the target. This does not work very well because of clouds and light noise surrounding the object.
- Radar emits photons but uses quantum photo-sensitive sensors to improve its performance. It does not work better enough.
- The radar prepares pairs of entangled photons. One is sent to the target and reflected and the other remains in the radar. The reflected photon is compared with the one that remained in place. As they have a common past, it is possible to sort the photons received by the radar to keep only the photons reflected by the target.

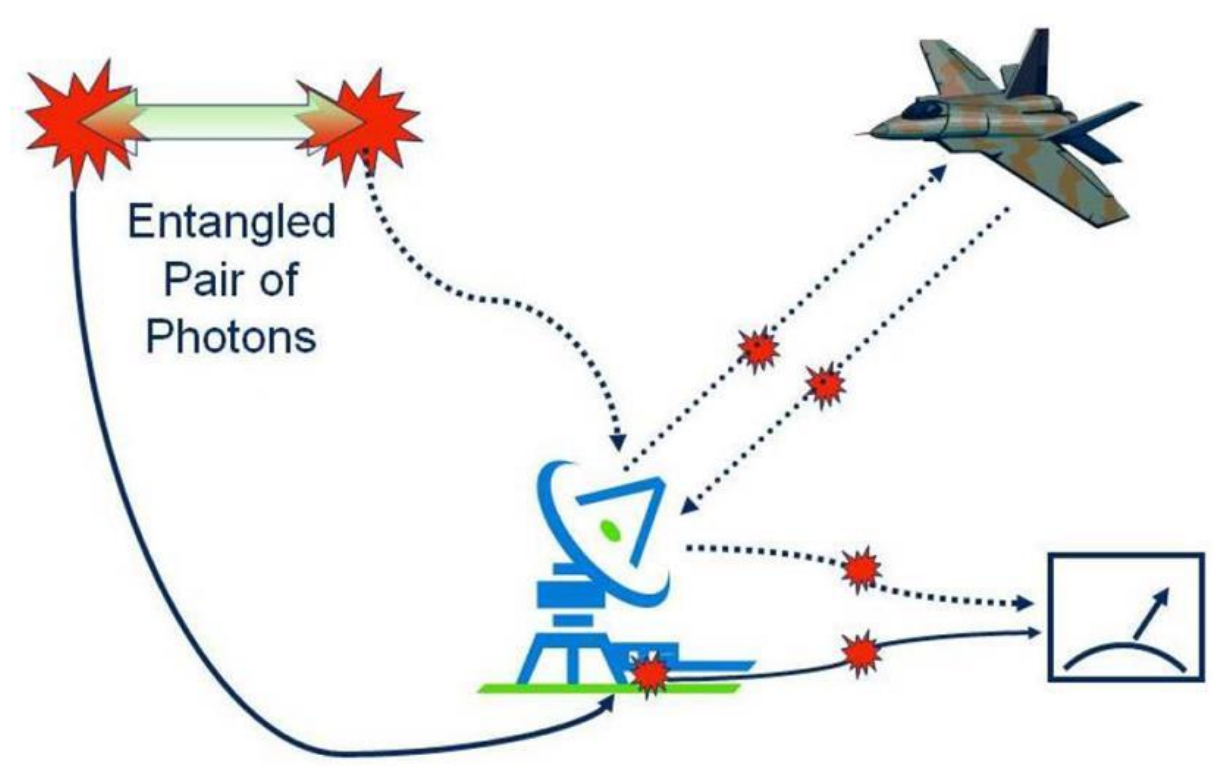

*Figure 828: the principle of quantum radar. Source: Quantum Radar by Marco Lanzagorta, 2012 (141 pages).*

[2625] See Scientists develop method to build up functional elements of quantum computers by Far Eastern Federal University, February 2020, which refers to Tailoring spontaneous infrared emission of HgTe quantum dots with laser-printed plasmonic arrays by A.A. Sergeev et al, 2020 (10 pages). Quantum dot seems to be more suited for night vision than for quantum computing. It is not a source of single photons. And the words "computer" and "qubit" are absent in the article.

[2626] See Quantum effect-based flexible and transparent pressure sensors with ultrahigh sensitivity and sensing density by Lan Shi et al, Nature Communications, 2020 (9 pages) that is based on thin films and spin-coating with carbon spheres dispersed in polydimethylsiloxane.

[2627] See Quantum-Based Photonic Sensors for Pressure, Vacuum, and Temperature Measurements: A Vison of the Future with NIST on a Chip by J Hendricks et al, NIST, 2021 (4 pages).

[2628] See Quantum-well pressure sensors by Witold Trzeciakowski, 1994.

[2629] See Quantum dots to probe temperature and pressure in highly confined liquids by Sayed M. B. Albahrani et al, RSC Advances, 2018 (12 pages).

[2630] See Optical properties of SiV and GeV color centers in nanodiamonds under hydrostatic pressures up to 180 GPa by Baptiste Vindolet, Jocelyn Achard, Alexandre Tallaire, Jean-François Roch et al, ENS, September 2022 (7 pages).

[2631] See Quantum Radars and Lidars - Concepts, realizations, and perspectives by Gregory Slepyan et al, June 2022 (21 pages) and A Study on Quantum Radar Technology Developments and Design Consideration for its integration by Manoj Mathews, Rowan University in New Jersey, May 2022 (8 pages) are relatively up-to-date theoretical papers on quantum radars.

It is in fact a variant of the third way which is studied[2632]. It consists in converting the photons sent to the target into a radio wave photon, while preserving their quantum state. A conversion of the same kind takes place for the photon remaining in the radar. This allows the radar waves to pass through bad weather, what photons in the visible cannot do.

This technique is expected to improve the accuracy of traditional radars and to improve its resistance to noise and interference. This kind of radar could theoretically detect stealth aircraft, modulo the fact that their flat reflective surfaces reduce their radar signature whatever the radar frequency[2633].

Entangled photons could also make it possible to effectively resist jamming systems. The first concepts saw the light of day in 2015[2634].

China is very interested in this technology and is working hard on it to be able to detect American stealth fighters or bombers like the F-22 and B-2. They announced a test of their first quantum radar in 2016, which was to become a prototype in 2018, produced by the government company **China Electronics Technology Group**[2635], with a range exceeding 100 km.

Other labs and companies are developing such radars, such as the **Institute for Quantum Computing** at the University of Waterloo in Canada[2636]. This project is funded by the Canadian Department of Defense for $2.7M.

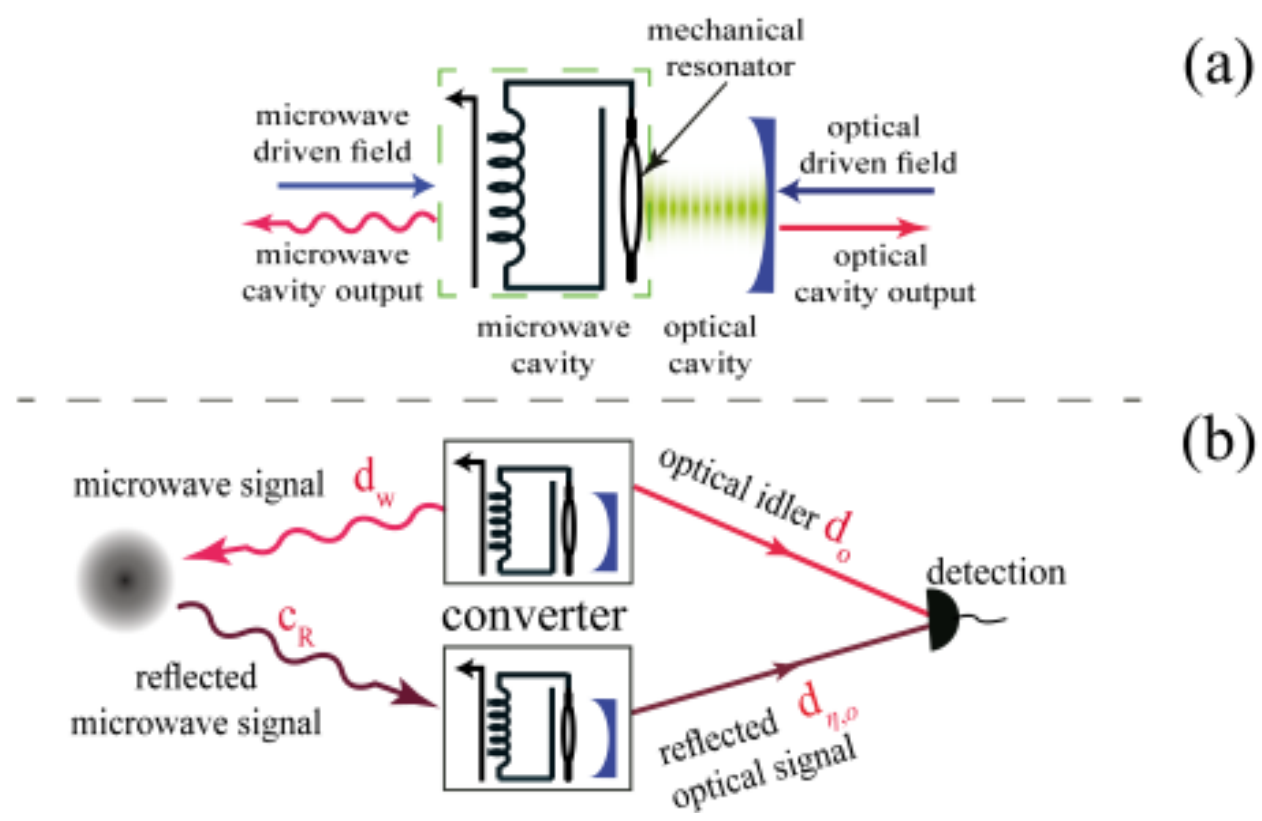

FIG. 1. (a) Schematic of the electro-opto-mechanical (EOM) converter in which driven microwave and optical cavities are coupled by a mechanical resonator. (b) Microwave-optical QI using EOM converters. The transmitter's EOM converter entangles microwave and optical fields. The receiver's EOM converter transforms the returning microwave field to the optical domain while performing a phase-conjugate operation.

*Figure 829: converting radar RF waves to/from photons. Source: Microwave Quantum Illumination by Shabir Barzanjeh et al, 2015 (5 pages).*

There are also some similar projects in Austria at the Institute of Science and Technology in Klosterneuburg. In the USA, **Lockheed Martin** is also invested in this emerging field. Specialists such as Marco Lanzagorta of the US Naval Research Laboratory believe that QKD satellites launched by the Chinese like Micius would have military applications of this type[2637]. In Europe, the Quantum Flagship **QMiCS** is dedicated to creating quantum microwave technologies operating at the single photon level that could help build quantum radars.

---

[2632] Another scenario proposes to use microwaves entanglement and a dual-receptor scheme to improve the angular detection of a target, in Entanglement-assisted multi-aperture pulse-compression radar for angle resolving detection by Bo-Han Wu et al, University of Arizona, Jul 2022 (18 pages).

[2633] See Can quantum mechanics improve radar technology? by Giacomo Sorelli and Nicolas Treps, November 2020. Which was maybe inspired by Quantum Flashlight Pierces the Darkness With a Few Percent as Many Photons by Adrian Cho, 2020.

[2634] See Focus: Quantum Mechanics Could Improve Radar, 2015, Microwave Quantum Illumination by Shabir Barzanjeh et al, 2015 (5 pages) and Enhanced Sensitivity of Photodetection via Quantum Illumination by Seth Lloyd, 2018 (4 pages).

[2635] See China's claim of developing "quantum radar" for detecting stealth planes: beyond skepticism by Ashish Gupta, 2016 (4 pages) and The US and China are in a quantum arms race that will transform warfare by Martin Giles, MIT Technology Review, January 2019. Some scientists, "under the radar", find that quantum radars features are overstated.

[2636] See Quantum radar will expose stealth aircraft, April 2018.

[2637] See The Future of Quantum Sensing & Communications by Marco Lanzagorta of the US Naval Research Laboratory (USA), September 2018 (37 minutes). He is the author of the book Quantum Radar by Marco Lanzagorta, 2012 (141 pages) which has been translated into Chinese by China, and officially bought the rights.

But quantum radars are still very theoretical devices[2638] that are currently plagued by photon losses, noise and detection issues[2639].

This technology could however be used in **LiDARs** to verify that the inbound photons correspond to those emitted by its own laser, avoiding unwanted optical interference from other LiDARs. Without malicious interference, this will be very useful when many autonomous vehicles equipped with LiDARs will have to coexist on the road[2640]. Quantum LiDARs can also have an improve resolution[2641]. Single-photon LiDARs could be used for remote wind detection at high resolution. It has been developed in China since 2014 and is used in transportable radars, including UAVs[2642].

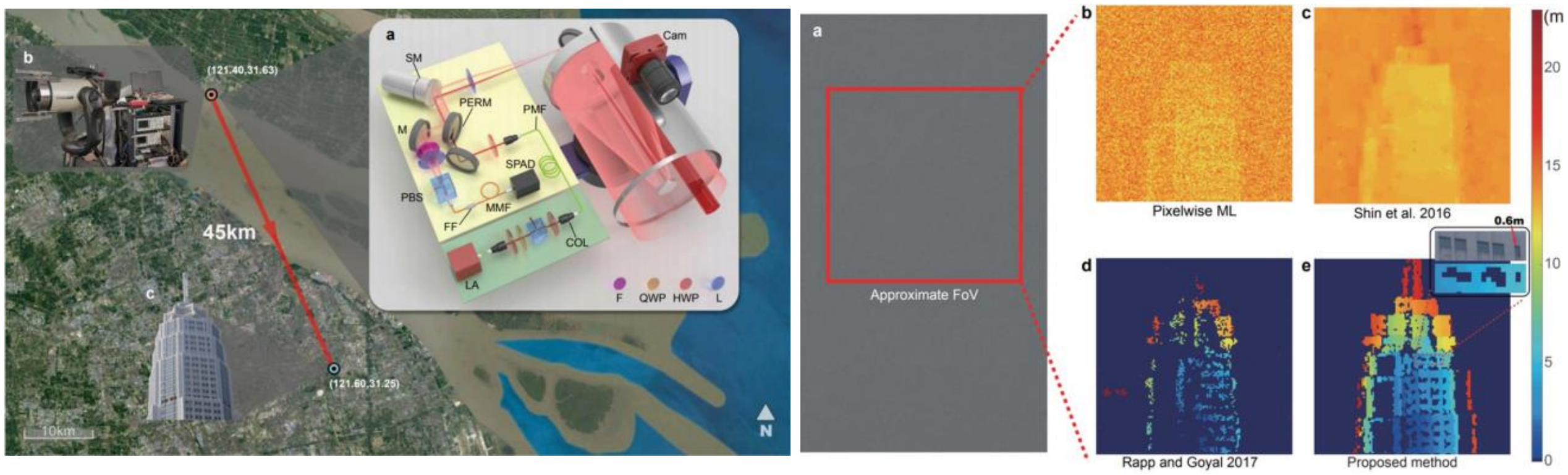

*Figure 830: Source: Single Photon LiDAR by Feihu Xu, June 2019 (25 slides).*

In a domain close to radar, **quantum sonars** could also emerge, so to speak. They use photons in the blue-green zone of the visible spectrum and would be usable for navigation in the Arctic Ocean. It would be a kind of quantum LiDAR. These systems could also implement optical communication with submarines via satellite, to replace radio waves that do not penetrate underwater well and are exploitable for very low-speed links.

On the other hand, one day, it may be necessary to find countermeasures against "quantum" coatings that allow the infrared signature of objects to be removed or reduced[2643].

## Quantum chemical sensors

Quantum sensing is also applicable with chemical sensors used to analyze the chemical composition of various materials and substances. It is commonly used with optical interferometers[2644].

Many such solutions could use NV centers, like:

- The detection of paramagnetic species in biological samples using a fiber equipped with an NV center detection probe[2645].

---

[2638] Like for example in Quantum-Enhanced Doppler Radar/Lidar by Maximilian Reichert et al, March 2022 (23 pages).

[2639] See Detecting a target with quantum entanglement by Giacomo Sorelli, Nicolas Treps, Frédéric Grosshans and Fabrice Boust, May 2020 (30 pages).

[2640] This approach has been studied since at least 2009. See Quantum Lidar - Remote Sensing at the Ultimate Limit, 2009 (97 pages).

[2641] See Two-photon interference LiDAR imaging by Robbie Murray and Ashley Lyons, 2022 (7 pages).

[2642] See Single-Photon Lidar for atmospheric detection by Haiyun Xia et al, June 2019 (22 slides).

[2643] See Camouflage made of quantum material could hide you from infrared cameras by Kayla Wiles, December 2019 which refers to Temperature-independent thermal radiation by Alireza Shahsaf et al, September 2019 (17 pages).

[2644] See Quantum Optical Technologies for Metrology, Sensing, and Imaging by Jonathan Dowling, 2014 (20 slides) and 12 pages, Advanced Micro- and Nano-Gas Sensor Technology: A Review by Haleh Nazemi et al, 2019 (23 pages).

[2645] See Nanodiamonds based optical-fiber quantum probe for magnetic field and biological sensing by Yaofei Chen et al, February 2022 (21 pages).

- A concentration sensors using NV center to detect electrochemical signals emerging from an electrolyte solution, and using the inhomogeneous dephasing rate of the electron spin of the NV center ($1/T_2^*$) as a signal[2646]. It is still theoretical work.
- Another theoretical NV center sensor, dedicated to the fast, cheap and low error (<1%) detection of the covid virus in biological samples. The NV sensor would be coated with cationic polymers such as polyethyleneimine (PEI), which can form reversible complexes with viral complementary DNA sequences. The detection is fairly indirect. A complicated biological reaction creates a RNA compound which diffuses in the solution, increasing the distance between magnetic molecules and the nanodiamond. The NV centers then senses less magnetic noise and has a longer $T_1$ time, which turns into a larger fluorescence intensity[2647].
- Various quantum sensing techniques, again, including NV centers, could help detect various breed of organic molecules[2648].

The University of Glasgow has developed and sells under license **IndiPIX**, an improved and simplified ambient temperature mid-wave infrared (MWIR) imager using indium antimonide (InSb) photodiodes on gallium arsenide (GaAs) transistors. It can detect outdoor gas leaks and plumes, mitigating explosion risks and reduce environmental impact[2649].

Ultracold chemical reactions could be used in cold atom sensors to detect very weak signals like in dark matter detection[2650].

OXFORD HIGHQ

**Oxford HighQ** (2017, UK) is a spin-off from the University of Oxford developing chemical and nanoparticles sensors using optical microcavities.

entanglement TECHNOLOGIES, INC.

**Entanglement Technologies** (2010, USA) is a spin-off of Stanford and Caltech.

It sells the AROMA (Autonomous Rugged Optical Multigas Analyzer) quantum gas detector that uses lasers and optical resonators similar to those used to detect gravitational waves in the LIGO, with a spectroscopy technique (CRDS: Cavity Ring-Down Spectroscopy). It allows the detection of dangerous gases in industry, especially in the extraction of fossil fuels. They were funded by EDF, via their Environmental Defense Fund.

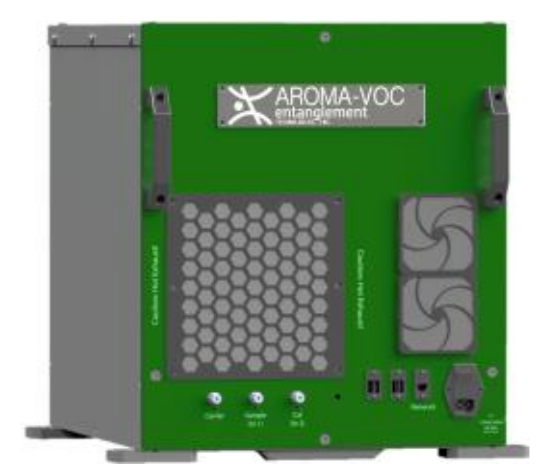

*Figure 831: Entanglement Technologies AROMA.*

## Quantum NEMS and MEMS

Nano or micro electromechanical structures are widely used in long-connected objects, such as accelerometers. They use many quantum phenomena, notably photonics-based, with mechanical resonators whose motion is analyzed by lasers and diodes.

[2646] See Sensing electrochemical signals using a nitrogen-vacancy center in diamond by Hossein T. Dinani et al, February 2021 (17 pages).

[2647] See SARS-CoV-2 Quantum Sensor Based on Nitrogen-Vacancy Centers in Diamond by Changhao L et al, University of Waterloo and MIT, December 2021 (14 pages).

[2648] See A molecular approach to quantum sensing by Chung-Jui Yu et al, April 2021 (12 pages).

[2649] See IndiPIX: Paving the way towards compact, portable, and cost-effective mid-wave infrared systems imaging systems, University of Glasgow (7 pages).

[2650] See Quantum metrology with ultracold chemical reactions by Seong-Ho Shinn et al, August 2022 (28 pages).

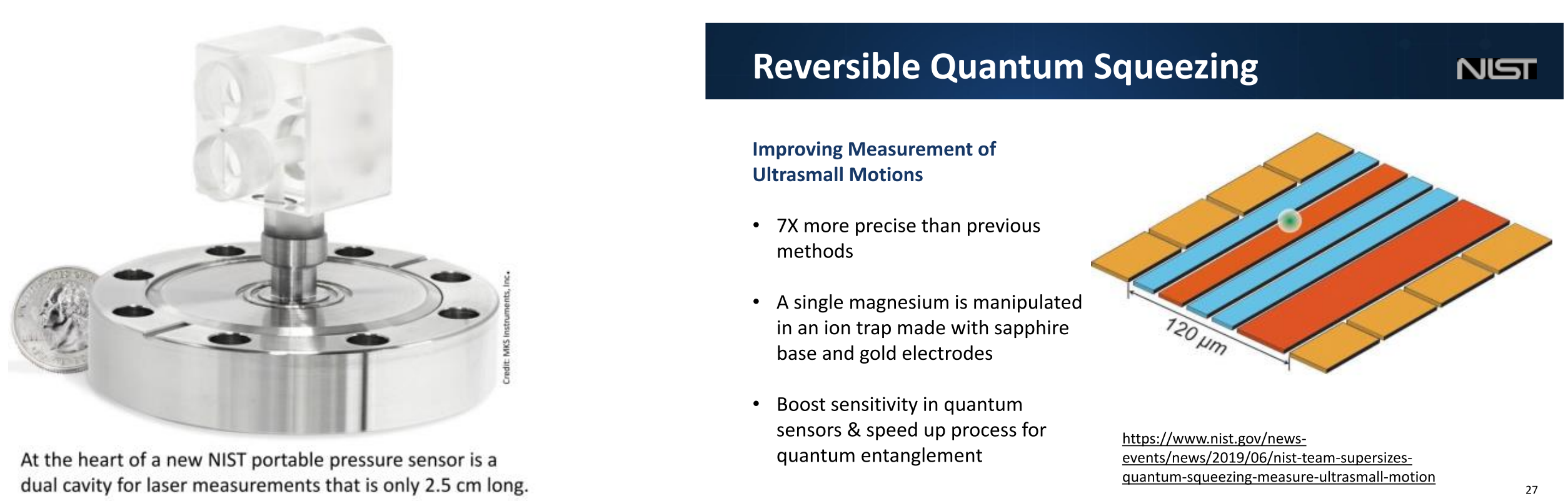

*Figure 832: quantum pressure sensors and quantum motion sensors. Sources: FLOC Takes Flight: First Portable Prototype of Photonic Pressure Sensor, February 2019 and Quantum Information Science & NIST - Advancing QIS Technologies for Economic Impact, 2019 (39 slides).*

They are found in quantum pressure sensors[2651] and motion detectors, both from NIST (*in* Figure 832[2652]). Other sensors are used to detect electrical resistance, temperature, mass and force, vacuum or voltage[2653] .

Finally, let's mention the European Quantum Flagship project **macQsimal** (Switzerland, €10.2M) or "Miniature Atomic vapor-Cells Quantum devices for SensIng and Metrology Applications", for the creation of quantum sensors aimed at the market of autonomous vehicle piloting and for medical imaging. This includes the creation of atomic clocks, gyroscopes, magnetometers, imaging systems using microwaves and electromagnetic fields in the tera-Hertz waves as well as gas detectors. In short, a fairly generalist approach. It is based on the use of cold atom vapor integrated in MEMS, a technique that Thales is also using.

## Quantum sensing key takeaways

- Quantum sensing is the most mature and underlooked market of quantum technologies.
- Quantum sensing enables better precision measurement of nearly any physical parameter: time, distance, temperature, movement, acceleration, pressure and gravity, magnetism, light frequency, radio spectrum and matter chemical composition.
- Quantum sensing has been extensively used to update the new metric system put in place in 2019.
- Lasers and the frequency combs technique is used to measure time with extreme precision, beyond atomic cesium clocks. It is based on blocked-mode lasers generating very short pulses, aka femtosecond-lasers.
- The most used quantum sensing technology is based on NV centers. It helps measure variations of magnetism and has applications in many domains like in medical imaging and non-destructive control. Indirectly, measuring magnetism can help measure many other physical parameters like temperature and pressure.
- Another one is cold atoms based interferometry that is implemented in micro-gravimeters, accelerometers and inertial sensors. It can also be used to analyze the radio frequency spectrum.
- China supposedly built some quantum radars using photons entanglement and up/down converts between visible photons and radar frequencies but the real performance of these devices is questionable and is driving a lot of skepticism in the Western world. But the related research is still going on.

---

[2651] See FLOC Takes Flight: First Portable Prototype of Photonic Pressure Sensor, February 2019.

[2652] See Quantum Information Science & NIST - Advancing QIS Technologies for Economic Impact, 2019 (39 slides).

[2653] See Quantum electro-mechanics: a new quantum technology by Konrad Lehnert from NIST JILA lab (47 slides), From micro to nano-optomechanical systems: light interacting with mechanical resonators by Ivan Favero (45 slides) and Progress of optomechanical micro-nano sensors a review, by Xinmiao Liu et al,  2021 (40 pages).

# Quantum technologies around the world

Quantum computing in the broadest sense is a strategic technology domain for various reasons. In cryptography, states sovereignty is at stake with the protection of sensitive communications. Quantum computing has critical applications that will extend the scope of digital computing beyond what is feasible today, particularly in the fields of healthcare, the environment and artificial intelligence.

In terms of maturity, quantum and post-quantum cryptography represent more established fields with economic players and commercial solutions, even if the standardization of post-quantum cryptography is not yet complete. However, it has fewer scientific and engineering unknowns compared to scalable quantum computing.

Quantum computing is much less mature. The feasibility of commercially and useful quantum computers remains an open question. There are significant technological challenges to overcome, including the thorny issue of qubits noise, quantum error correction, and how to scale the number of physical qubits by several orders of magnitude. So, quantum computing is full of scientific and technological uncertainties even before being economical and market ones. How countries deal with that is a good revelator of their innovation and forward-looking culture.

For the time being, fundamental research is mainly funded by governments in most countries, then from very large IT players who entertain many technology bets in parallel (IBM, Google, Microsoft, Intel, Honeywell, Alibaba), and a more or less well-funded startups, mainly in North America (D-Wave, IonQ, Rigetti, PsiQuantum, Xanadu) and, way behind, in Europe (IQM, OQC, Pasqal) and in other regions (SQC in Australia, Quantum Machines in Israel, etc).

The quantum computers software industry is in its infancy. The major players and startups creating quantum computers have all invested the software arena, starting with the low-level supporting tools and sometimes, for developing quantum applications.

Some systems are already available in the cloud, directly or through cloud services provides like Amazon, Microsoft and Google. Most of these and Atos also sell cloud access to classically run quantum emulators. One way of looking at things is coming from **Yuri Alexeev** of the Argonne National Laboratory in the USA, who draws a parallel between the history of quantum computing and that of artificial intelligence and anticipates the arrival of two winters, the first which occurred in the late 1990s, the second he expected in 2020 and the next around 2030[2654].

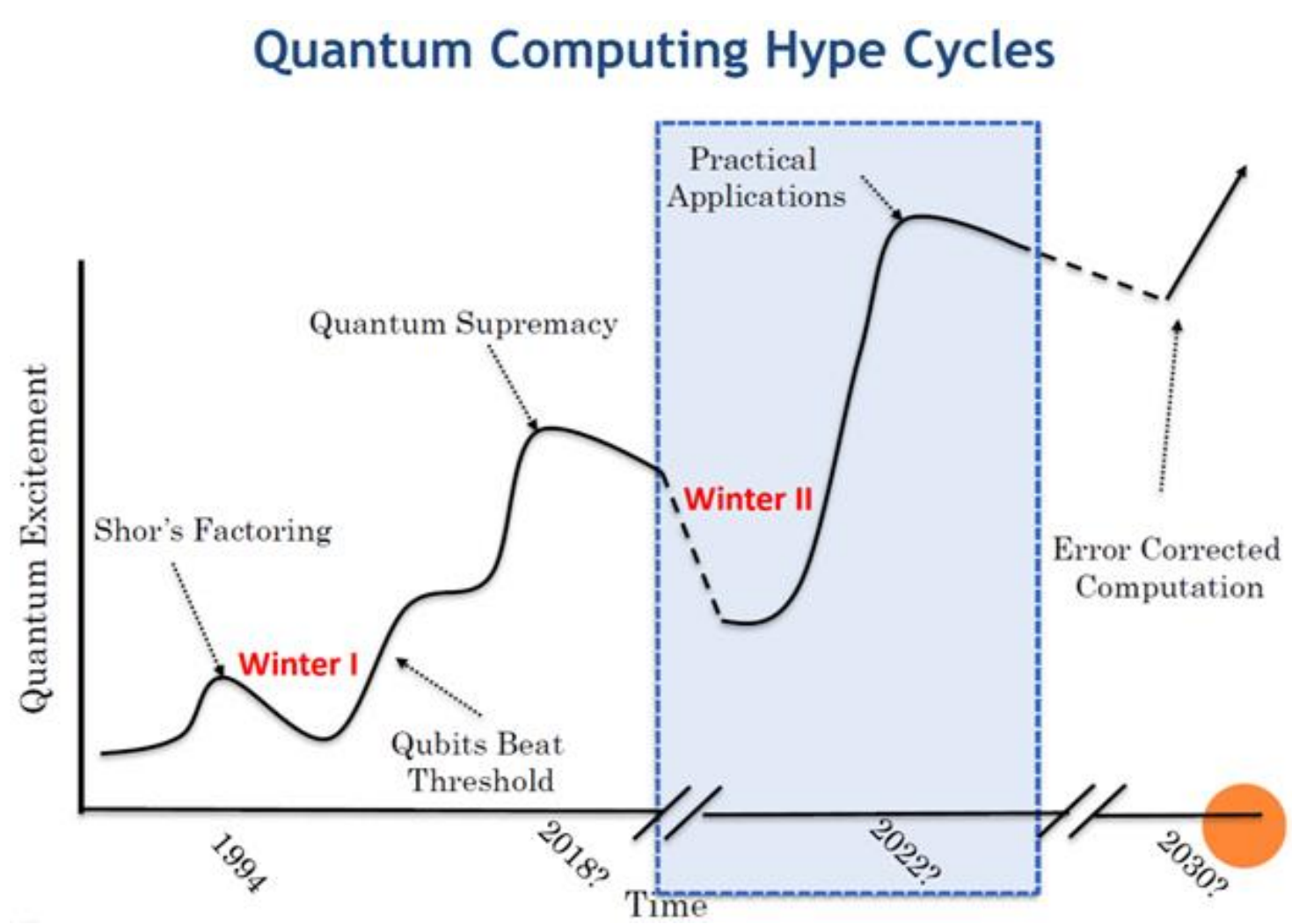

*Figure 833: one view of the quantum computing hype cycle. Source: Quantum Computing Trends by Yuri Alexeev, August 2019 (42 slides).*

Since this excitement is a somewhat fuzzy wave function and difficult to evaluate, it doesn't mean much. We can however anticipate at least a small winter with the startups of the sector and large customers engagements.

[2654] See Quantum Computing Trends by Yuri Alexeev, August 2019 (42 slides). You can also find an historical comparison with many other technology hypes in Mitigating the quantum hype by Olivier Ezratty, February 2022 (26 pages).

Hardware startups will have a hard time delivering useful machines, while software startups will not have a large enough addressable market due to the lack of hardware. But this will not prevent public research laboratories and large tech companies from doing fundamental and applied research.

## Quantum computing startups and SMEs

Mapping these vendors is a bit easier than in other deep techs like in artificial intelligence because there are not so many. There are many methods to inventory worldwide quantum startups and small businesses. I have accounted about 450 such companies, more than the 265 I had in store in September 2020 and their total funding is about $4.7B as of October 2021.

The increment comes mostly from existing quantum enabling technologies companies that I uncovered. I added them when and if I found that they were enabling "second quantum revolution" solutions. As a consequence, metal cutting lasers and classical telecommunications photonics are out!

This ecosystem began to take shape even before quantum computers were working on a small scale. It is fascinating to discover startups that make long-term bets, particularly with hardware. Software startups rely on a still limited hardware infrastructure but often reduce their risks by also supporting traditional computing architectures like Nvidia GPGPUs in machine learning. Their customers are large companies who test algorithms on a small scale to get their hands on quantum programming, often on D-Wave annealers and sometimes with IBM who is very pushy in its quantum evangelism efforts. To date, no application seems to have been deployed in production. We are therefore in the field of applied research and small scale prototyping within client companies. The software ecosystem is to be monitored closely. It will probably expand once hardware works on a larger scale, particularly with NISQ computers and quantum simulators[2655].

For their part, quantum cryptography systems are operational and correspond to a very separate market, just like the quantum sensing market that is more mature technologically but still in its infancy.

The stakes for many startups in this field are common with those of deep techs: how to develop real products with economies of scale, how to expand internationally rapidly, how to avoid falling into models that are too "service-oriented" and at last, how to resist what some people are already calling the quantum winter. Enabling technologies niche companies (photon sources, cryostats, ultra-vacuum, various sensors, electronics) can do well by reaching out diversified markets, notably targeting several different branches of research, telecommunications, military or aerospace applications.

### Investors

Quantum technologies investments may be impressive for the large rounds like those from PsiQuantum and IonQ and the associated "FOMO" factor ("*fear of missing out*"). But it's still small in volume in the technology sphere. The first investment funds more or less specialized in quantum technologies have already emerged with:

- **Quantonation**, a French seed fund created by Charles Beigbeder and managed by Christophe Jurczak, a physicist who got his PhD with Alain Aspect. They have already invested in over a dozen startups[2656].

---

[2655] See Some Teams Go For NISQ-y Business Some are NISQ-Averse by Doug Finke, February 2020.

[2656] Quantonation invested in LightOn (France), Spark Lasers (France), which offers laser sources not specifically dedicated to quantum computers, Pasqal (France, cold atoms), Quantum Benchmark (Canada, software), Kets Quantum Security (UK, QKD component), Orca Computing (UK, hardware, photonics-based computing), CryptoNext Security (France, PQC), Qunnect (USA, repeaters for QKD), Quandela (France, photon source), Qubit Pharma (molecular simulation), Qnami (Switzerland, NV center-based metrology), Orca Computing (photon qubits, UK), Foqus (Canada, quantum sensing software), Qu&Co (Netherlands, software), QPhoX (Netherlands, communications between quantum computers), HQS (Germany, software), Qubit Pharmaceuticals (France, software), evolutionQ (Canada, software), Inspek (France, chemical sensors), Multiverse (Spain, software), Pixel Photonics (Germany, single photons detector).

They organize quantum meetups and hackathons in Paris (with **QuantX**, the quantum alumni association from Ecole Polytechnique), the first edition having taken place in October 2018. It participated to the launch **Le Lab Quantique**, which was jointly created with **Bpifrance**. It is federating the country vendor and user quantum ecosystem.

- **Quantum Valley Investments** (QVI), a $100M Canadian investment fund, raised in 2013, dedicated to quantum technologies. Their founders had invested in 1984 in Blackberry / RIM. They do not disclose their investments, except in ISARA Corporation, part of which are spin-offs from the Canadian research laboratory Institute for Quantum Computing at the University of Waterloo in Ontario.
- **Black Quant** (Germany) is a quantum technologies dedicated investment fund created in 2022 by the QBN Network and CM-Equity, an investment company with teams in Germany, Slovenia and Croatia. They don't have any participation as of January 2022.
- **Quantum Exponential** (UK) invested £450K in May 2022 in Universal Quantum who plan to build a million qubit quantum computer. It also invested in Arqit, Siloton, Aegiq,
- **Quantum Ventures** is a quantum investment company launched in 2016. Its "Quantum Revolution Fund" is managed from London and Switzerland. It aims to raise €100M.
- **Quantum 1 Group** is an American investment fund specializing in quantum technologies since 2015.
- **Summer Capital** is a Dutch investment fund specialized in quantum technologies, data and finance. Their investments include Horizon Quantum Computing, Rigetti and Turing.
- **Parkwalk Advisors** is a British deep tech fund. As of 2021, they have invested in Phasecraft, Quantum Motion Technologies, Riverlane, nu quantum, nu nano and Oxford Quantum Circuits. This fund is part of IP Group Plc since December 2016.
- **Runa Capital**, created by Serguei Beloussov (Russian with a Singapore nationality, founder of Acronis) is an investor in many deep tech startups, including IDQ, Qnami, Qu&Co and Pasqal.
- **Phystech Ventures**, previously Quantum Wave Fund, created by Russians in the Silicon Valley, including Serguei Beloussov, and having already invested in the IDQ and Nano-Meta Technologies. Their fund is not 100% specialized in quantum. They also invest in robotics, drones, sensors and connected objects.
- **Machine Capital**, a UK fund focused on quantum and AI, which has so far invested in **QuantumX Incubator,** an incubator for quantum software projects launched jointly with **Cambridge** startup Quantum **Computing,** which specializes in the development of quantum software, with a 20-week incubation period.
- **SpeedInvest** is an Austrian investment fund specializing in deep tech start-ups, which invests among others in quantum technologies. They invest in seed stage with up to 1M€. They have invested in **QPhoX** and **Kets**.
- **Airbus Ventures** is an investment fund, headquartered in Silicon Valley, that operates independently from the Airbus Group which is one of its limited partners among several others. They invested in QCWare, IonQ, Q-CTRL, C12 and Qunnect.
- **BlackQuantFund** is an investment fund created in Germany by QBN and CM-Equity that is raising 100M€ and invested in seed in Aegiq[2657].

[2657] See QBN and CM-Equity Sets Up €100 Million Quantum Technologies Fund by Matt Swayne, January 2022.

- **2xN** (UK) launched in August 2022 a dedicated $120M fund focused on quantum computing startups with $3M to $5M pre-seed to Series A tickets. It was created by Lars Fjeldsoe-Nielsen and Niels Nielsen. Limited partners include the Danish Growth Fund and some family offices.

You then have generalist funds who invested in one or two startups, like the Canadian pension fund **PSP Investments** that invested in D-Wave with seats in the board, **UVC Partners** who invested in HQS in Germany, **Supernova Invest**, **Elaia Partners** and **Breega Capital** in Alice&Bob in France, **Omnes** in Quandela again in France. And of course, corporate venture funds like **TotalEnergies Ventures** and **Tencent Investments**.

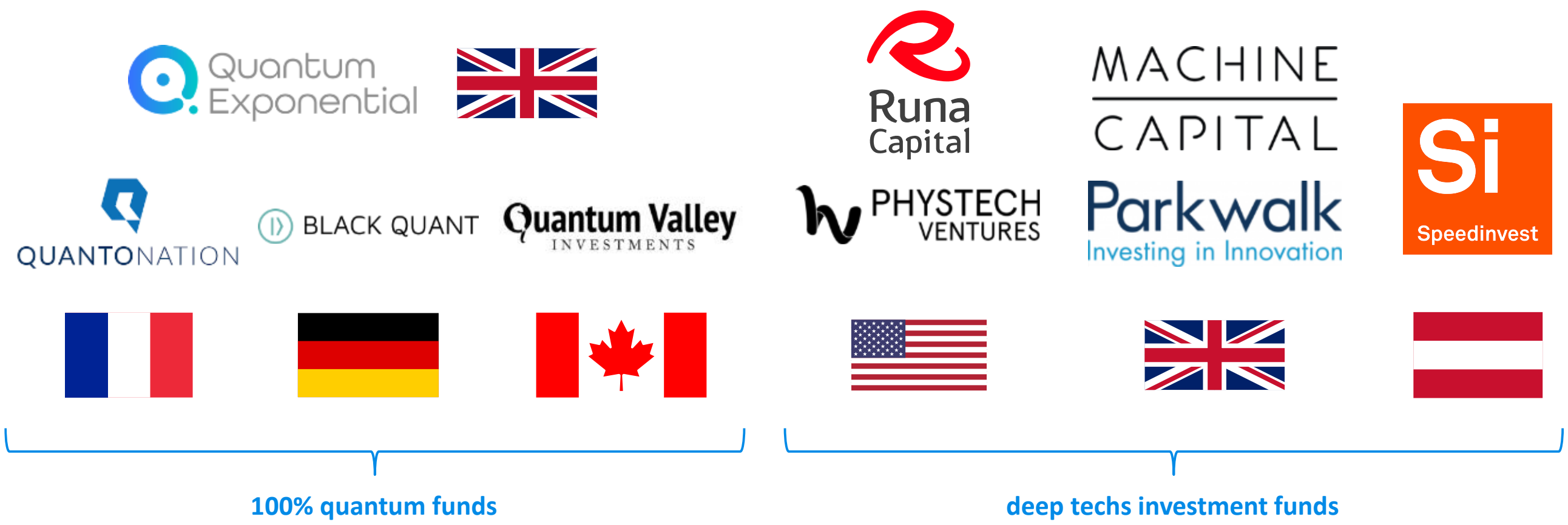

*Figure 834: a few of the key investors in quantum technologies. (cc) Olivier Ezratty, 2022.*

**The Quantum Insider** produced in early 2020 an [inventory of investors](#) in quantum technologies startups. Here is an excerpt with general and specialized VC investors. It contains a few mistakes such as Worldquant which is positioned as an investor specialized in quantum technologies whereas it is a generalist investment company created in 2007. The use of the word quantum or a piece of quantum is not a guarantee of specialization.

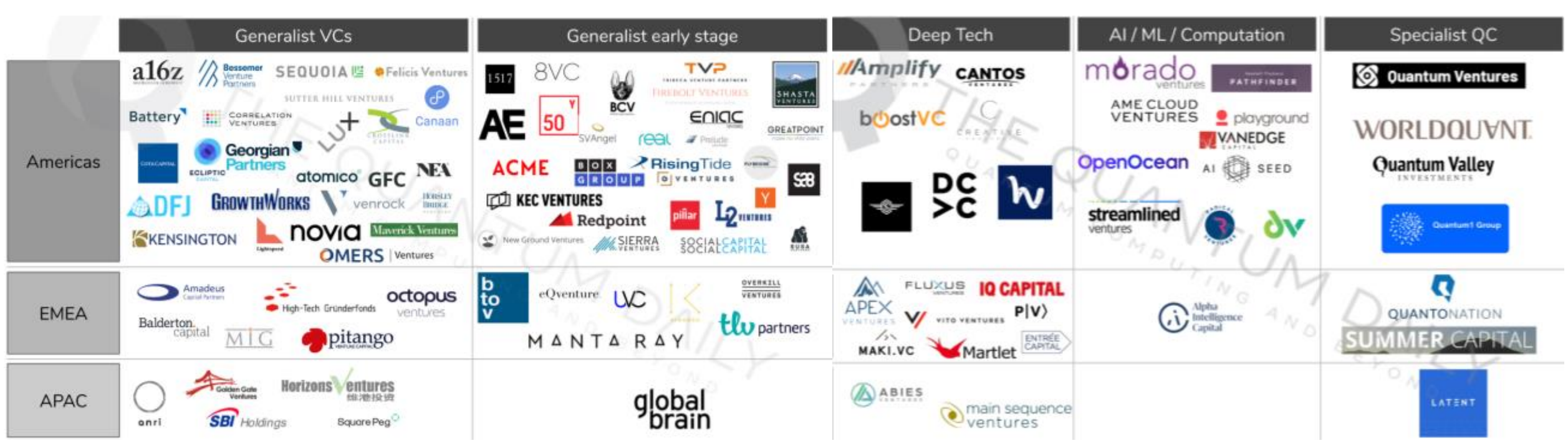

*Figure 835: a map of investors in quantum technologies by [The Quantum Investor](#).*

We must also describe the **SPAC** funding mechanism that was used by IonQ, Rigetti, D-Wave and Arqit. A SPAC is first an investment fund that is created before it finds where to invest the money[2658]. It plans to invest the money in one company. When the company is found, the SPAC buys shares of the company, usually with other limited partner investors like Corporate venture funds. At last, the SPAC puts the fund on the stock market and is traded with using the acquired company name. The process is a bit complicated. The SPAC business model is fairly unbalanced.

2658 See [What is a SPAC: the step by step process going public](#), May 2021.

The SPAC fund takes a significant cut of the deal and has a significant upside when the company IPO takes place, whatever the subsequent outcome in the stock market. There were hundreds of SPACs until 2022 but the phenomenon has dried up, particularly with assets reallocations in investors as inflation popped up after the start of the Ukraine war[2659].

IonQ and Rigetti's stock value trended down after their IPO as shown in the charts on the right, due to significant overpromises in their investor pitch decks. D-Wave's IPO in August went well but we'll have to wait and observe the stock value after a couple quarterly reports! But fears of a quantum investment winter are not limited to these ill-fated SPACs[2660].

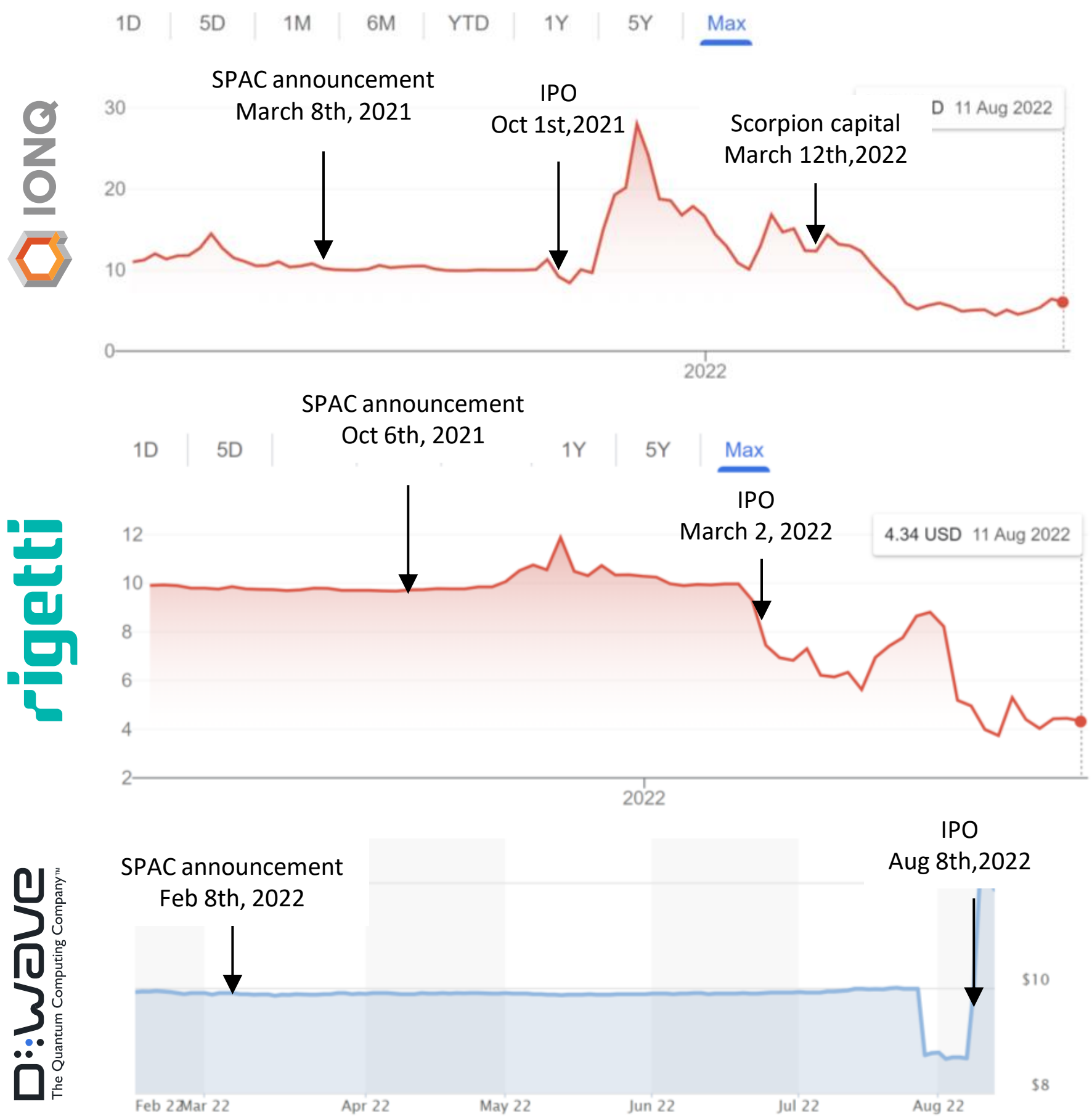

*Figure 836: evolution of the value of the SPACs of IonQ, Rigetti and D-Wave. (cc) Olivier Ezratty, August 2022. Source: Google Finance, MarketWatch.*

First, we had a sluggish start for 2022 with large fund rounds[2661]. Then, the current post-Covid/Ukraine war recession may drive assets reallocations unfavorable for long-term risky investments[2662]. This explains indirectly why governments and their aggressive national quantum plans find sideways to fund their local startups, like with creating an artificial market with public orders of non-existent or low performance experimental quantum hardware. And I don't account for the crooks in that market like this investment advisor in quantum technologies that selected 5 companies out of which only two happened to be in the quantum business (Rigetti, IonQ)[2663]. What a joke! What is going to happen? I'm not an oracle. Things will probably be tougher for some startups in their various rounds of investment, particularly in series B and C. The winning startups who'll escape the crisis will have good fundamentals: excellent teams, respected roadmap milestones, some IP and preferably some orders and/or revenue.

[2659] See SPACs out, the quantum community react by Karina Robinson, The Quantum Insider, June 2022.

[2660] See Is quantum computing headed for a financial reckoning? by Dan O'Shea, May 2022.

[2661] See Shifting Quantum Investment Dynamics by Russ Fein, The Quantum Leap, July 2022.

[2662] See How the recession will affect quantum tech vendors by André M. König, June 2022.

[2663] See 5 Quantum Computing Stocks to Watch towards 2022, December 2021.

Now, looking at the current revenue streams of these three companies, you find out that Rigetti, IonQ and D-Wave made respectively $2.1M, $2.6M and $1.3M in Q2 2022, probably with selling computing time in the cloud either directly or through cloud vendors like AWS, Microsoft and Google. Their cash burn rate is about $38M, $37M and $44M per year which corresponds to 4.8 and 10 years of "air supply" for Rigetti and IonQ. It is unclear for D-Wave when looking at their accounting[2664].

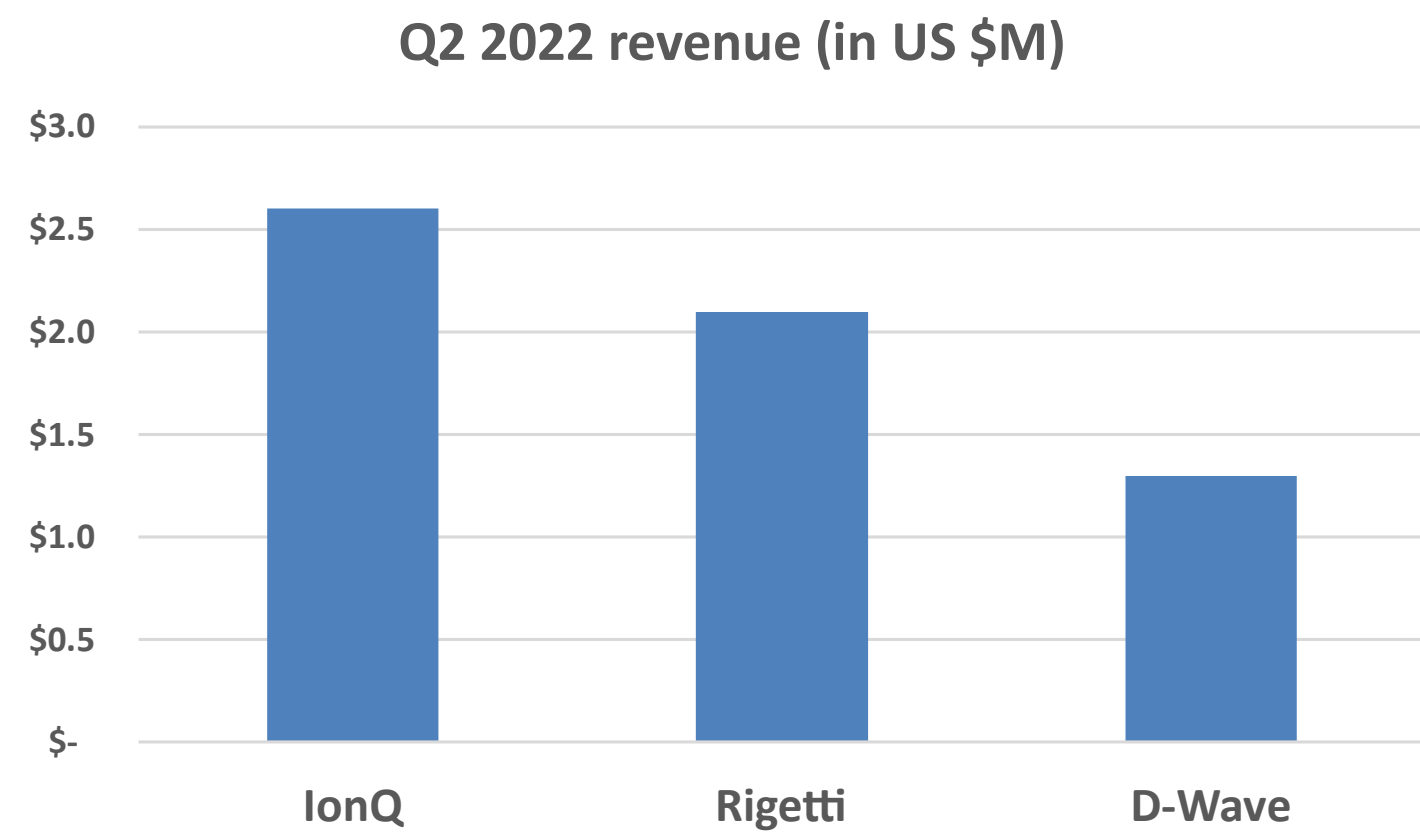

*Figure 837: IonQ, Rigetti and D-Wave Q2 2022 quarterly revenue. Source: their quarterly reports.*

## Startup maps

An inventory of quantum startups is available on the Quantum Computing Report website. It allowed me initially to identify a good number of the startups mentioned in this section. Some startups broadcast so little information about them that one may wonder if they are not scams. This lack of communication may simply be due to the fact that many creators may be uncommunicative researchers, that they are poorly funded, and that their projects have business prospects that are too remote and risky. Also, many times, they are so early-stage that they can't talk about anything that would drive interest, such as "*I have two functioning qubits*".

Many of the startups mentioned here are not yet in the "pure" form of the startupian model, i.e. they are far from having a scalable model or even, just a product. They are often either industrial small businesses targeting very low-volume niche markets, or startups where the scientific and technological risk is still very high before they can sell anything. And often, with a combination of both. They can then finance themselves with contract research and various consulting services for large companies or public institutions.

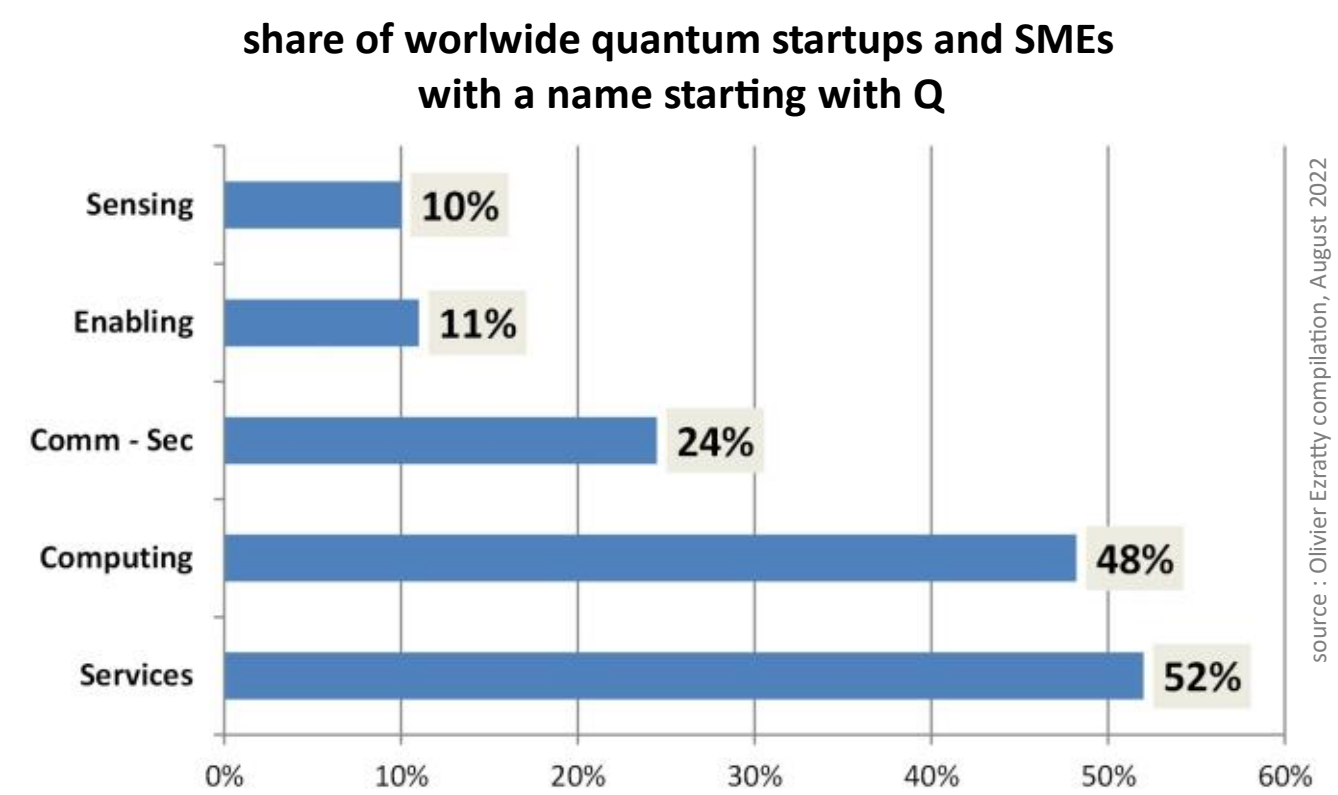

*Figure 838: fun fact: in some fields like telecommunications, cryptography and consulting services, quantum startups branding shows a lack of creativity with many names starting with a Q.*

In the vast majority of cases, I relied on public information available on the Internet to describe what these startups do in the various parts of this book. One way to find out what these startups do is to identify their founders, if they are researchers, and find their original laboratories, their past scientific publications, and their PhD thesis if it is available. Finally, search for possible patents filed by the startups. This is technology intelligence using open sourced information (OSINT). Of course, you can also meet with entrepreneurs, live or distantly.

This mapping also does not include companies that seem to offer only service and consulting in quantum computing, without having their own technology or products. We inventory some of them in an earlier section on Service vendors.

As far as data is concerned, here are some charts extracted from my database of startups and SMBs. The first one provides an indication of the number of startup creations per year. The second provides a breakdown by country.

[2664] Data source: Rigetti Q2 2022 quarterly report, IonQ Q2 2022 quarterly report and D-Wave Q2 2022 quarterly report.

Of course, other sources will publish different charts[2665]! As of October 2022, my database had 552 companies. The difference comes from both new startups and existing companies, mainly in the enabling technologies space, that happen to target "second quantum revolution" technologies use cases. The selection is sometimes not obvious, like in the very large photonics supplier sector. At least you can sort things out with these company creation date as shown in Figure 839.

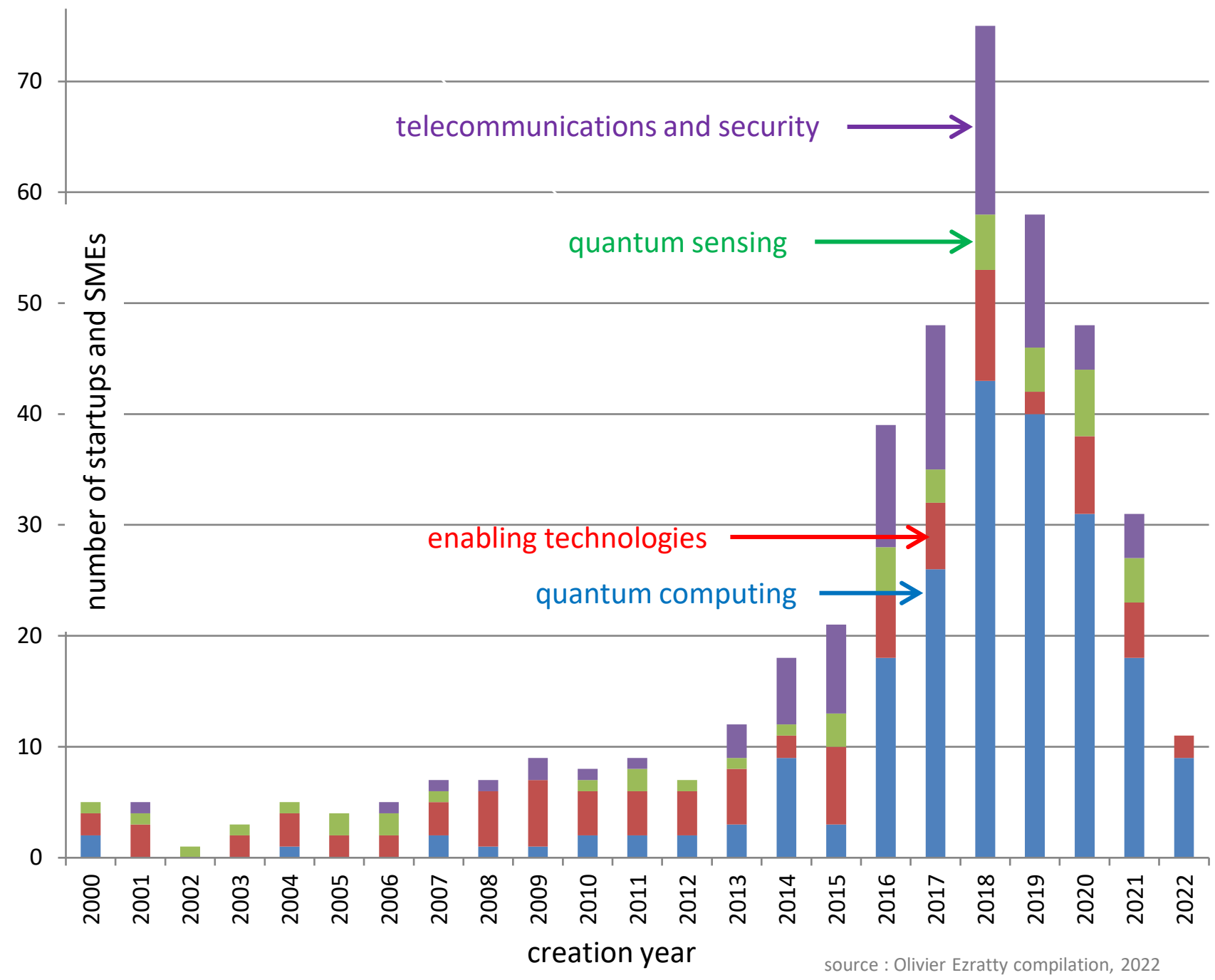

| Country | Computing | Enabling | Sensing | Services | Comm - Sec | Total |
|---|---|---|---|---|---|---|
| USA | 58 | 40 | 19 | 4 | 25 | 146 |
| UK | 20 | 11 | 8 | 5 | 18 | 62 |
| France | 10 | 27 | 10 | 4 | 5 | 56 |
| Canada | 31 | 6 | 4 | | 11 | 52 |
| Germany | 16 | 19 | 4 | 1 | 6 | 46 |
| India | 11 | | | 3 | 4 | 18 |
| Switzerland | 3 | 8 | 1 | | 4 | 16 |
| Netherlands | 7 | 4 | 1 | 1 | 2 | 15 |
| China | 6 | | 1 | | 6 | 13 |
| Japan | 10 | 2 | | 1 | | 13 |
| Spain | 7 | 3 | | | 1 | 11 |
| Finland | 4 | 3 | | 2 | 2 | 11 |
| Israel | 4 | 2 | 2 | | 3 | 11 |
| Australia | 6 | | 3 | 1 | 1 | 11 |
| Denmark | 3 | 6 | | | 1 | 10 |
| Singapore | 5 | 1 | 1 | 1 | 1 | 9 |
| Italy | 2 | 3 | 1 | 1 | | 7 |
| Sweden | 2 | 4 | | | | 6 |
| Russia | | 2 | | | 4 | 6 |
| Poland | 3 | | | | 2 | 5 |
| Austria | 3 | | | | 1 | 4 |
| Turkey | | 1 | 1 | | | 2 |
| Bulgaria | 1 | | | | 1 | 2 |
| Tchekia | | | | 1 | 1 | 2 |
| Estonia | 2 | | | | | 2 |
| UAE | 2 | | | | | 2 |
| Greece | 2 | | | | | 2 |
| Belgium | | 1 | | | | 1 |
| Bielorussia | 1 | | | | | 1 |
| South-Korea | 1 | | | | | 1 |
| Hong-Kong | 1 | | | | | 1 |
| Uruguay | 1 | | | | | 1 |
| Taiwan | | | | | 1 | 1 |
| Ukraine | 1 | | | | | 1 |
| Chile | | | | | 1 | 1 |
| Lybia | 1 | | | | | 1 |
| Luxembourg | | 1 | | | | 1 |
| Norway | 1 | | | | | 1 |
| Columbia | 1 | | | | | 1 |
| **Total** | **226** | **144** | **56** | **25** | **101** | **552** |

*Figure 839: chart of the creation year of small business and startups in "second revolution" quantum technologies. (cc) Olivier Ezratty, August 2022.*

Another chart in Figure 840 shows a different breakdown by country that highlights the largest funding. As usual, we see a significant financing gap between North America and Europe. One of the reasons is that European startups were created later or are more traditional small business which don't rely on venture capital for their development[2666]. The other is of course a different access to capital and markets. The size of the US market has always been an advantage for US-born companies although it doesn't prevent some worldwide leaders to emerge elsewhere in the world like ASML or SAP.

60% of worldwide startups funding went to the top seven startups: IonQ, PsiQuantum, Rigetti, D-Wave, Xanadu, Quantinuum all from the USA and Canada, plus Arqit from the UK[2667], while Europe has an equivalent number of companies compared to North America. If there are as many quantum startups and small businesses in mainland Europe than in North America, their visible equity funding represents only 10% of worldwide funding while North American companies got a hefty 70% as of August 2022, but it was 5% a year ago in September 2021. This doesn't include the investments from large IT corporations like IBM, Amazon, Microsoft, Google and Intel.

This funding discrepancy explains for example why PsiQuantum which was created out of UK became a US company. As a result, directly or indirectly, it could raise over $700M which a record to date in the quantum industry.

---

[2665] See The landscape of the quantum start-up ecosystem by Zeki Can Seskir et al, May and October 2022 (15 pages) that used this book as one source among others to create its own database of quantum companies with a total of 441 companies.

[2666] See New record looms in VC funding of quantum startups by Michel Kurek, September 2020 and The European Quantum Computing Startup Landscape by Alex Kiltz, October 2020.

[2667] Arqit's SPAC funding of $450M was not finalized as of January 2022.

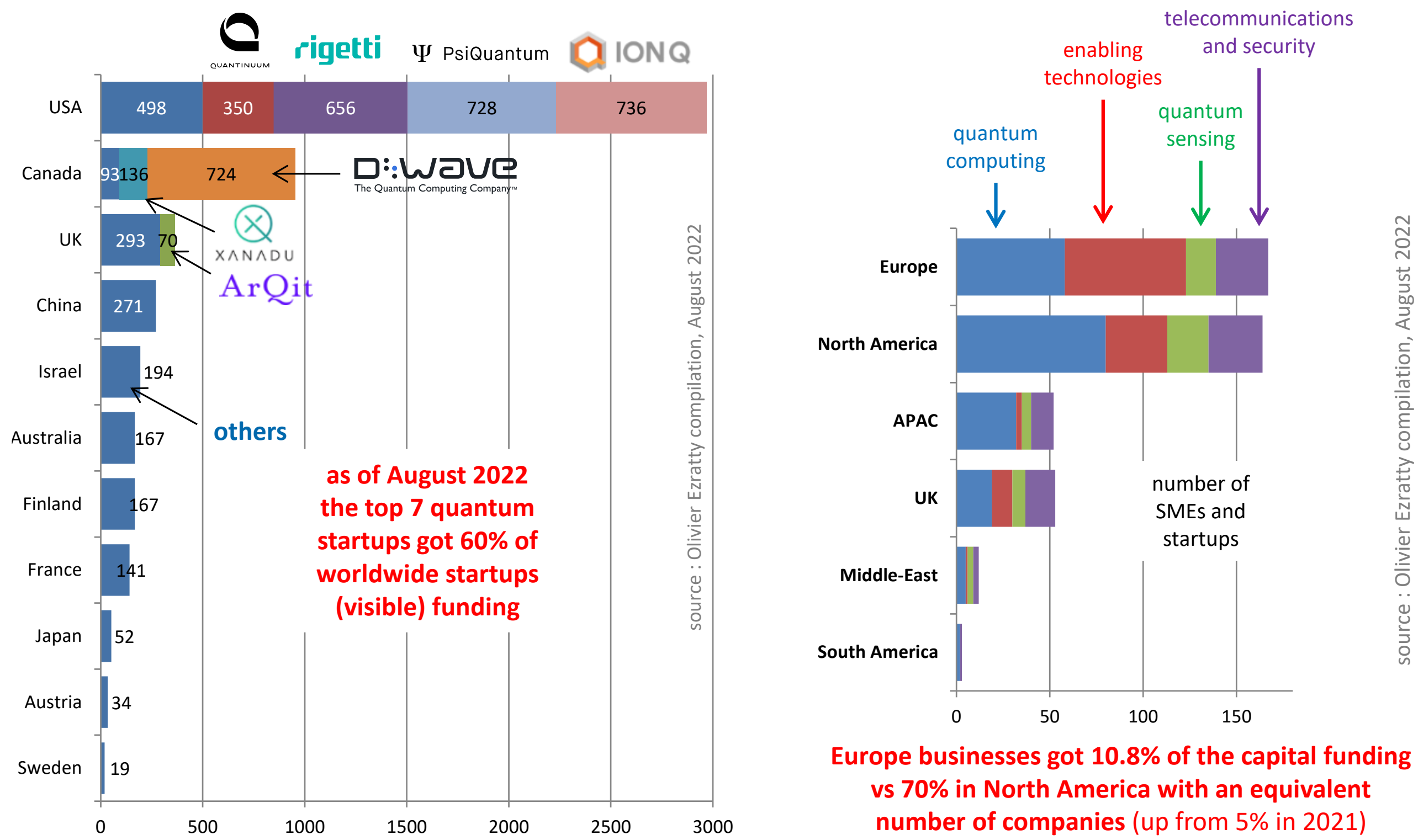

*Figure 840: chart showing where the investment money went by country and its concentration. (cc) Olivier Ezratty, 2022.*

**David Shaw** from Fact Based Insight (UK) created a very complete map of all quantum technologies in the global quantum ecosystem chart below[2668]. We cover most of these companies in different parts of this book, split between quantum computing (page 142), enabling technologies (page 464), software (page 726), telecommunication and cryptography (page 845) and sensing (page 861).

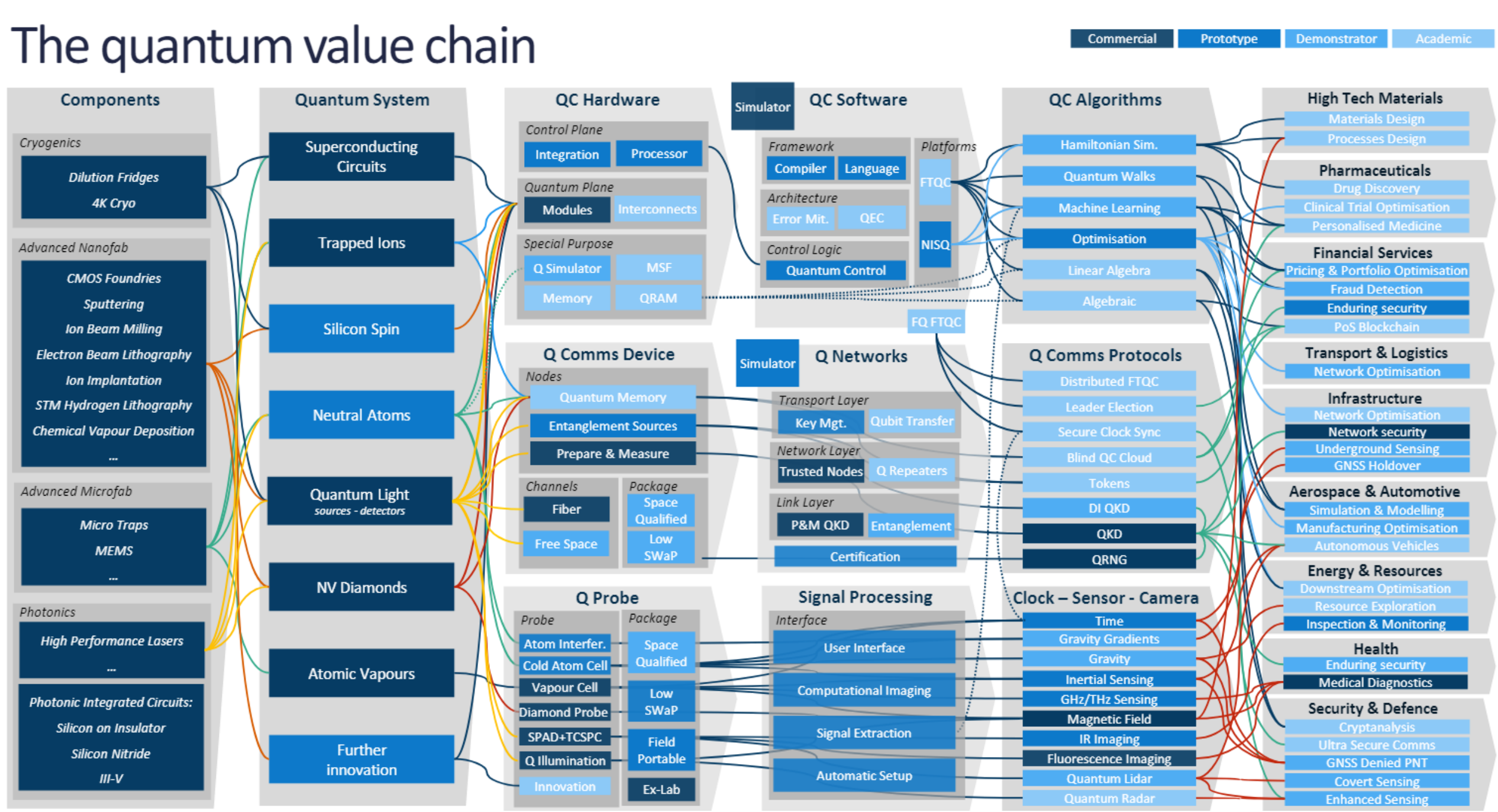

*Figure 841: David Shaw's quantum value chain. Source: [Quantum Value Chain Overview](), by David Shaw, Fact Based Insight, April 2021.*

---

[2668] See [Quantum Value Chain Overview](), by David Shaw, Fact Based Insight, April 2021.

### Quantum Startup incubation and acceleration

There are already a few incubation and acceleration programs for deep tech startups which host quantum startups in the world. One of the most famous is the **Creative Destruction Lab**, based in Toronto and other cities in Canada. **Xanadu** and **North** came out of it. Similarly, **Unit DX** is a deep tech incubator based in Bristol, UK, which started in the biotech industry and also helped some quantum startups[2669]. **Duality** is a quantum dedicated startup accelerator in the USA with a sponsorship from Amazon. **Quantum Startup Foundry** is the University of Maryland accelerator which was completed in 2022 by **Q-Cat** (Quantum Catalyzer), a startup studio.

In France, the **HEC Challenge+** program and the **Deeptech Founder** program created by the team behind the deep tech Hello Tomorrow event, accelerated a big share of France's quantum startups like Quandela, Pasqal and Alice&Bob. To create such acceleration programs, you need to be close to critical mass pool of talents, such as a dynamic academic and research zone.

### Disappeared startups

How about startups who disappeared? You would guess that there are many given the technology immaturity of the sector. Well, there aren't so many out of the 300 to 400 hundreds and so created so far.

**Quantum Factory** (Germany) was closed in January 2021. They were building trapped ions computers. **NextGenQ** (France) was pursuing the same goal and was also closed in 2021. The company had just a founder with not quantum physics skills nor any funding.

**SeQureNet** (2008-2017, France) was a spin-off of Telecom ParisTech specialized in the distribution of long distance CV-QKDs. It had been funded within the framework of the European research project SECOQC (secure communication based on quantum cryptography). The startup was based on work done by Philippe Grangier's team at the Institute of Optics and the Thales TRT laboratory in Palaiseau. The company closed down in 2017. It had been launched a little too early compared to the maturity of the market.

**Black Brane Systems** (2016, Canada) was a startup focused on the development of quantum machine learning solutions. They startup as a very "stealthy" company, got some undisclosed funding in 2018 and then closed their business.

We also have acquired startups like **CQC** (Quantinuum in 2021), **Labber Technology** and **Quantum Benchmark** (Keysight, 2020 and 2021), **Muquans** (ixBlue, 2021), **QxBranch** (Rigetti, 2019), **Qu&Co** (merged with Pasqal in 2022) and **Super.tech** (ColdQuanta, 2022). So far, no quantum startup was acquired by a large IT vendor like IBM, Google and Microsoft. But to some extent, the investment of SandboxAQ, a spin-off from Alphabet, in August 2022, is a first[2670].

## Global investments

What about global investments in quantum computing? A 2015 McKinsey study provided an overview of investments that likely compiled public research budgets. At that time there were 1500 researchers worldwide with a total budget of $1.5B. This number has since increased much. The USA and China were obviously leading that space. But the distribution of these investments, which probably include both quantum cryptography and quantum computers, is intriguing for other countries. As usual, Europe was fragmented with Germany, France, The Netherlands, Finland, Italy, Spain, the United Kingdom (then, in the European Union) and Switzerland (geographically in Europe). And we have strong quantum countries in other regions like Canada, Japan, Singapore and Australia.

---

[2669] See Incubators & Accelerators: Launchpads For Quantum Success? by James Dargan, 2020.

[2670] See SandboxAQ Announces a Partnership with evolutionQ as part of its New Strategic Investment Program, August 2022.

Quantum technologies have become a geopolitical issue, almost like nuclear deterrence[2671]. Governments are motivated to invest in quantum for strategic reasons: both in the idea of being able to decrypt existing or past telecommunications in the context of the activities of their intelligence services and to protect their own via quantum or post-quantum cryptography. More than almost any other digital technology, quantum is therefore a tool for the states strategic sovereignty[2672]. The public authorities in these different countries have mobilized in very different ways on quantum. Most developed countries governments coordinate efforts in the quantum field. Plans with up to $2B over 5 or 10 years periods have been announced here and there.

It's still quite difficult to compare these investments between countries and for a couple reasons:

- What is the **existing run-rate investment**? It's sometimes not easy to capture this data, particularly with highly decentralized research like in the USA and most European countries.
- What are the **undisclosed investments** in military and intelligence? It may be high in the USA and Russia. But lower in Europe, given these countries don't allocate a great share of their GDP to military expenses.
- Is the publicized funding **incremental** or includes existing investments? You can easily embellish things with the latter accounting method, or create misleading rankings of country investments like when McKinsey did showcase a chart with Germany investing more than the USA which was not true at all[2673].
- Are there any **double bookings** in the showcased investments? This can easily generate misleading information.
- Are some countries **overinflating** their investment? This is a hypothesis for China's investments which have been highly confusing. We provide as accurate data for this regard here.

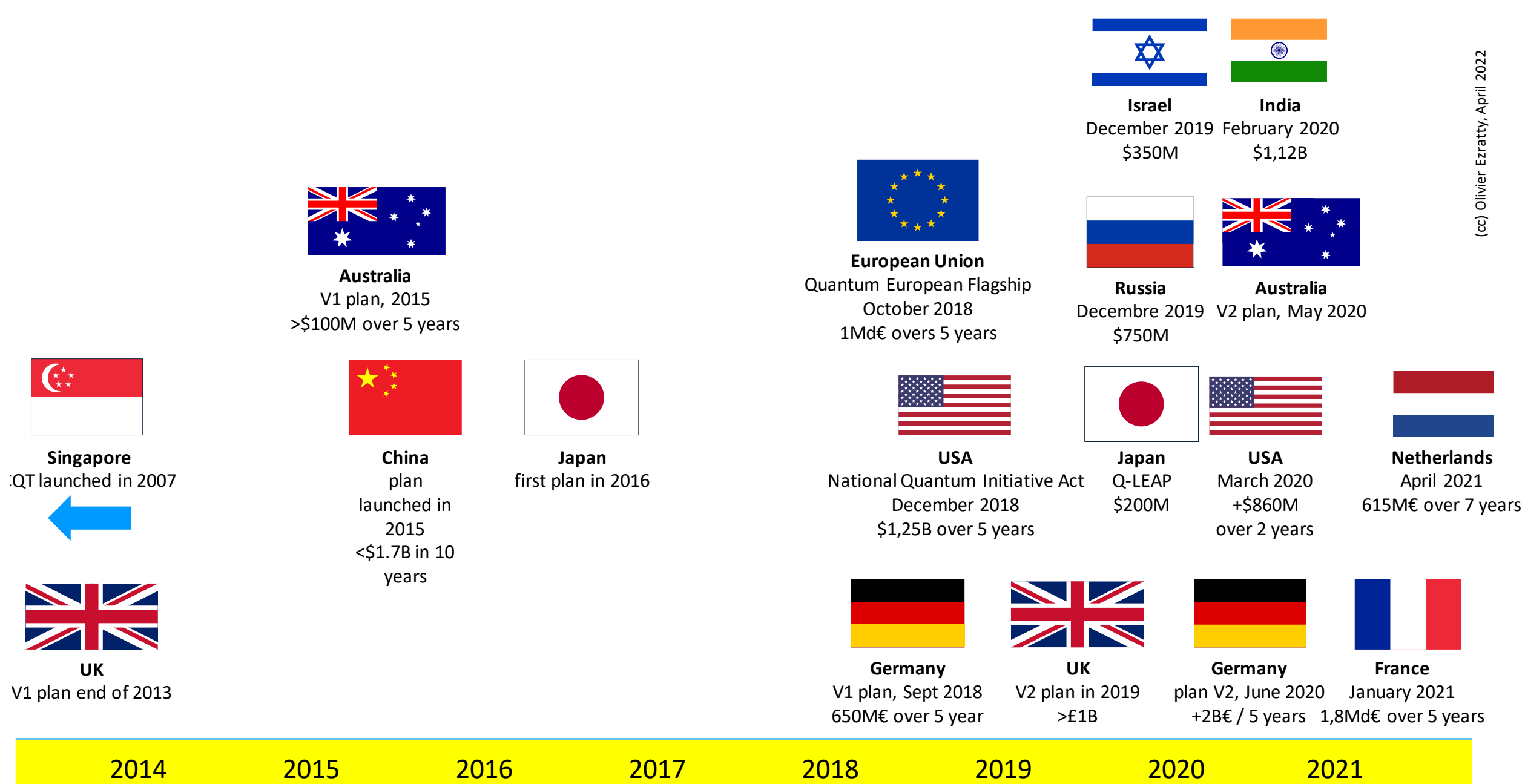

*Figure 842: national quantum initiative plans across the years. (cc) Olivier Ezratty, 2022.*

---

[2671] See Quantum, AI, and Geopolitics (3): Mapping The Race for Quantum Computing by Hélène Lavoix, December 2018.

[2672] See the forum Europe: Keys to Sovereignty by Thierry Breton, August 2020. He cites three pillars of this sovereignty: computing power, data control and secure connectivity. Quantum technologies have a key role to play in the first and third! However, the means cited to obtain this sovereignty are classic and relate to public funding for R&D. We know that this is clearly insufficient.

[2673] See A quantum wake-up call for European CEOs, McKinsey, December 2021.

Where's government money (your taxes) going? It usually funds the following in these various national quantum initiatives:

- Incremental **public research** funding efforts. This comes from both the country governments and sometimes, from local governments, like in Germany.
- Growing the **education effort** and addressing the skills shortage. This is particularly important as it deals with very highly skills roles, with a strong scientific background.
- Creating **quantum hubs** that consolidate quantum research and sometimes startup creation. These are handled within universities or public labs (like the DoE in the USA) and regions (like in Munich in Germany) or across several of them in a thematic way (UK). This may involve some real estate and building construction.
- Create **hybrid quantum-classical supercomputing centers**, based on existing supercomputing capacities, which is being done nearly everywhere, in the USA, Europe, etc.
- Create a **quantum technologies industry** with both encouraging startups creation and existing industries to adopt quantum technologies, like in the sensing domain or just, in software development. This goes with putting in place public/private funding mechanisms for long-term investments.
- Handle **procurement** with local startups with countries indirectly funding their local quantum computing startups champions through local procurement like in Finland with IQM, Australia with Quantum Brilliance, the USA with IonQ[2674], the UK with Orca and OQC or France with Pasqal. This is often related to the hybrid quantum-classical centers already mentioned.
- Foster **global partnerships** that are different in nature, between research laboratories within countries (as in the UK hubs), between particular labs across various countries, and between public and private research within the same country (CEA and Atos) or between different countries (Intel with Qutech). The raison d'être of all these partnerships can be identified: quantum computing is a complex scientific subject that cannot be mastered by a single laboratory or company. Collaboration is necessary to bring together talent from different specialties, between condensed matter physics, sensor and control technologies, optronics, cryogenics, semiconductor production, algorithmics and software development.
- Launch bilateral **countries partnerships** like the USA with Nordic countries, the UK and Australia, or France with The Netherlands and Singapore with Finland.

I tried here to consolidate a more global view with public and industry investments per country or region. This is based on some guestimates for large industry vendors in the USA (IBM, Microsoft, Google, Intel) but the public investment data is rather safe. What this shows is counter-intuitive: the first region in public investment is the European Union! It's investing more than the USA and even China. Still, the USA leads the international pack thanks to their industry investments, both from large IT vendors and from VCs in startups.

Europe's Achille's heel is not having these large IT vendors on one hand and investing less in startups in proportion to GDP. The result is some pressure put on European startups who have to find ways to get bigger fundings in Series B, C and beyond with investment sources outside the EU like the USA, Asia or even the Middle East oil countries.

I did this chart after seeing so many analysts providing wrong or outdated numbers on countries and region public investments in quantum technologies. The last one comes from a **BCG** report which tried in 2022 to compare of investments between the EU, the USA and China.

---

[2674] See IonQ Secures Contract to Provide Quantum Solutions to United States Air Force Research Lab, IonQ, September 2022.

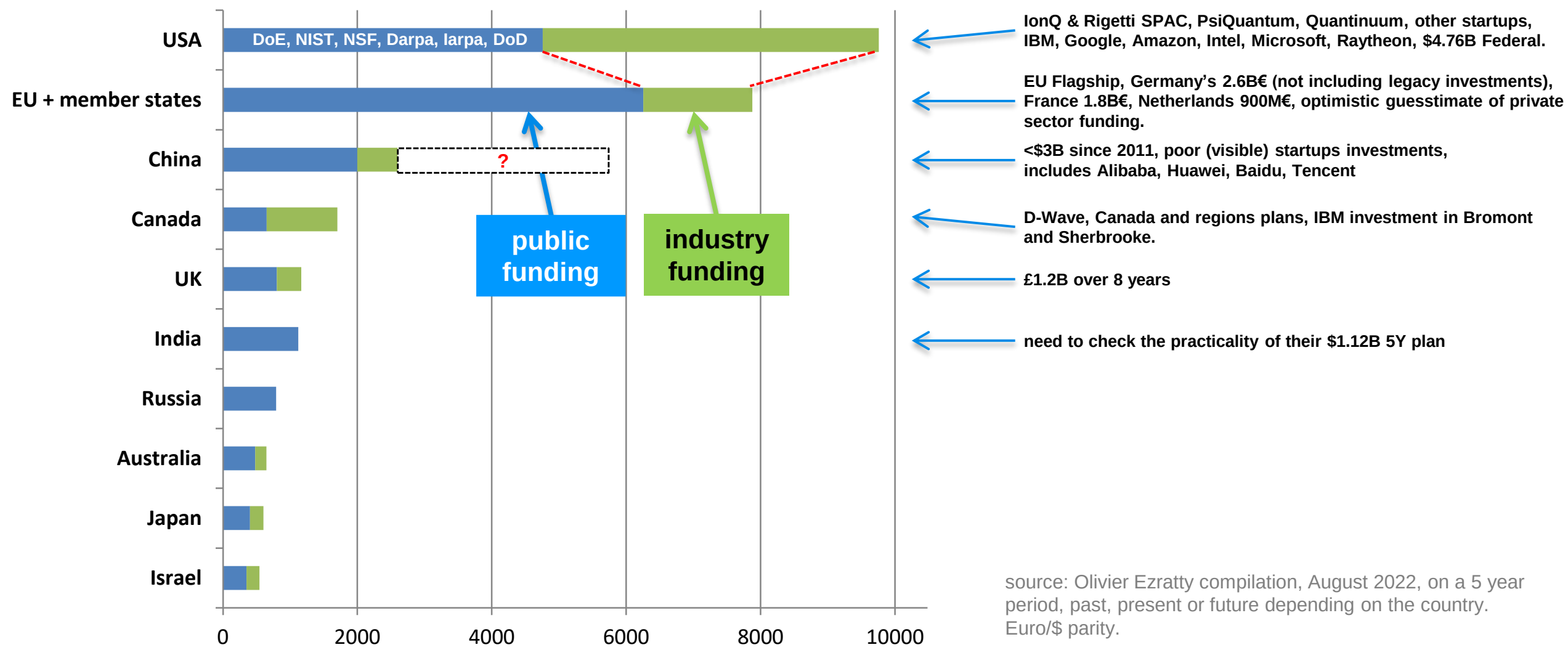

*Figure 843: a consolidation of quantum technologies public and private investments with some raw estimated for large IT vendors. It creates a very different picture than what is commonly thought about the place of China and Europe. (cc) Olivier Ezratty, 2022.*

Their data are wrong. Like many, in page 14, they overestimate China's investment (at $10B when it's probably way under $4B) and underestimate USA's which is about $4.7B as of August 2022 for 5 years instead of the mentioned $2.9B[2675]. They also showcase a $700M plan from France in 2019 that never existed. This is puzzling. But they showcase the harsh reality of a much smaller investment in startups in the EU compared with the USA, Canada and the UK. A similar report from the **World Economic Forum** contains the same wrong data on the USA, China, and also Japan[2676].

An evaluation of scientific publications in quantum computing done by **Insead students** in 2018 did show with no surprise to discover that the USA, Canada and China are the first countries to publish[2677].

A more detailed analysis was produced by **Michel Kurek** in September 2020 (sources of the illustrations in Figure 844) which did help relativize the influence of Chinese publications[2678]. Indeed, the Citations Per Publications is very low in China and also India, compared all Western countries.

The significant investments made by developed countries in quantum technologies raise fears that computing power could end up being concentrated in the hands of a few or even a single country or company. I don't believe this, at least not in the initial phase of development of these technologies. Knowledge on the subject is highly distributed, as are enabling technologies and strategic materials. I would rather situate the risk of concentration in a second phase of the maturation of this market, one that will see a market that was initially fragmented with many players concentrating through consolidation. It will probably do so for reasons that are more macro-economic than scientific or technological, through economies of scale and the platforming of offers. This explains why it is necessary to simultaneously keep an eye on the hardware, development tools and software applications of quantum computing.

Once the main scientific and technological uncertainties are lifted, the success of each company and country will depend on the classic key success factors of technology ecosystems: execution speed, team quality, funding levels, communication, marketing, sales, the ability to promote technology platforms to a maximum number of players and on a global scale.

---

[2675] See Can Europe Catch Up with the US (and China) in Quantum Computing? by François Candelon, Jean-François Bobier, Maxime Courtaux, and Gabriel Nahas, BCG, August 2022 (45 pages). You find the same China investment overstating in The Quantum Computing Arms Race is not Just About Breaking Encryption Keys by Adm. Mike Rogers and Nir Minerbi, Classiq, June 2022.

[2676] See State of Quantum Computing: Building a Quantum Economy, September 2022 (48 pages).

[2677] See VC investment analysis Quantum Computing, April 2018 (18 slides).

[2678] See Quantum Technologies : Patents, Publications & Investissements Landscape by Michel Kurek, September 2020 (52 pages).

This is where sovereigntist approaches combining protectionism of key players while ensuring maximum trade openness to the world to enable them to achieve economies of scale will have to be carefully adopted.

| | COUNTRY | TP | %TP | TC | %TC | CPP | RCI | %ICPEI |
|---|---|---|---|---|---|---|---|---|
| 1 | USA | 4,295 | 26.4% | 108,128 | 44.8% | 25.2 | 1.7 | 70% |
| 2 | China | 3,706 | 22.8% | 38,611 | 16.0% | 10.4 | 0.7 | 44% |
| 3 | UK | 1,428 | 8.8% | 32,435 | 13.4% | 22.7 | 1.5 | 120% |
| 4 | Germany | 1,400 | 8.6% | 38,339 | 15.9% | 27.4 | 1.9 | 123% |
| 5 | Japan | 1,106 | 6.8% | 20,996 | 8.7% | 19.0 | 1.3 | 99% |
| 6 | Canada | 1,056 | 6.5% | 23,104 | 9.6% | 21.9 | 1.5 | 124% |
| 7 | India | 991 | 6.1% | 5,847 | 2.4% | 5.9 | 0.4 | 33% |
| 8 | Australia | 777 | 4.8% | 20,777 | 8.6% | 26.7 | 1.8 | 130% |
| 9 | France | 699 | 4.3% | 14,016 | 5.8% | 20.1 | 1.4 | 117% |
| 10 | Italy | 635 | 3.9% | 10,522 | 4.4% | 16.6 | 1.1 | 116% |
| | Total 10 countries | 16,093 | 98.9% | 312,775 | 129,5% | 19.4 | 1.3 | 83.1% |
| | Total world | 16,279 | | 241,536 | | 14.8 | | |

*TP= Total Publication ; TC = Total Citation ; CPP = Citation par Publication = TC/TP ;
RCI = Relative Citation Index ; ICPEI = International Collaboration Publication Extended Index

*Figure 844: publications and patents on quantum tech per country. Source: Quantum Technologies : Patents, Publications & Investissements Landscape by Michel Kurek, September 2020 (52 pages).*

We'll go through the details, country by country, continent by continent. With one exception, Africa, which is little invested in the subject, at least as a producer of quantum technologies, maybe besides South Africa which seems to have started to get involved in the academic side[2679].

This summary shows which country best masters quantum computing technology per qubit type. All in all, we have a good balance between the USA and the European Union, although the USA have the benefit from having large IT vendors invested in the field in superconducting (Google, IBM), silicon (Intel), trapped ions (Honeywell, IonQ) and topological qubits (Microsoft).

What are the key success indicators of success for countries investing in the quantum race?

We'll probably have some analyst shops create their own quantum sort-of Shanghai ranking using composite metrics: public funding, scientific publications, patents and the likes, entrepreneurship spirit, number of startups, startups funding, large companies' investments, corporate adoption, skilled workforce and else. Guess what? US and China will probably rank first there. And smaller countries behind in some variable order. But what if Europe was consolidated?

[2679] See Will Africa miss the next computational revolution? by Amira Abbas, April 2020.

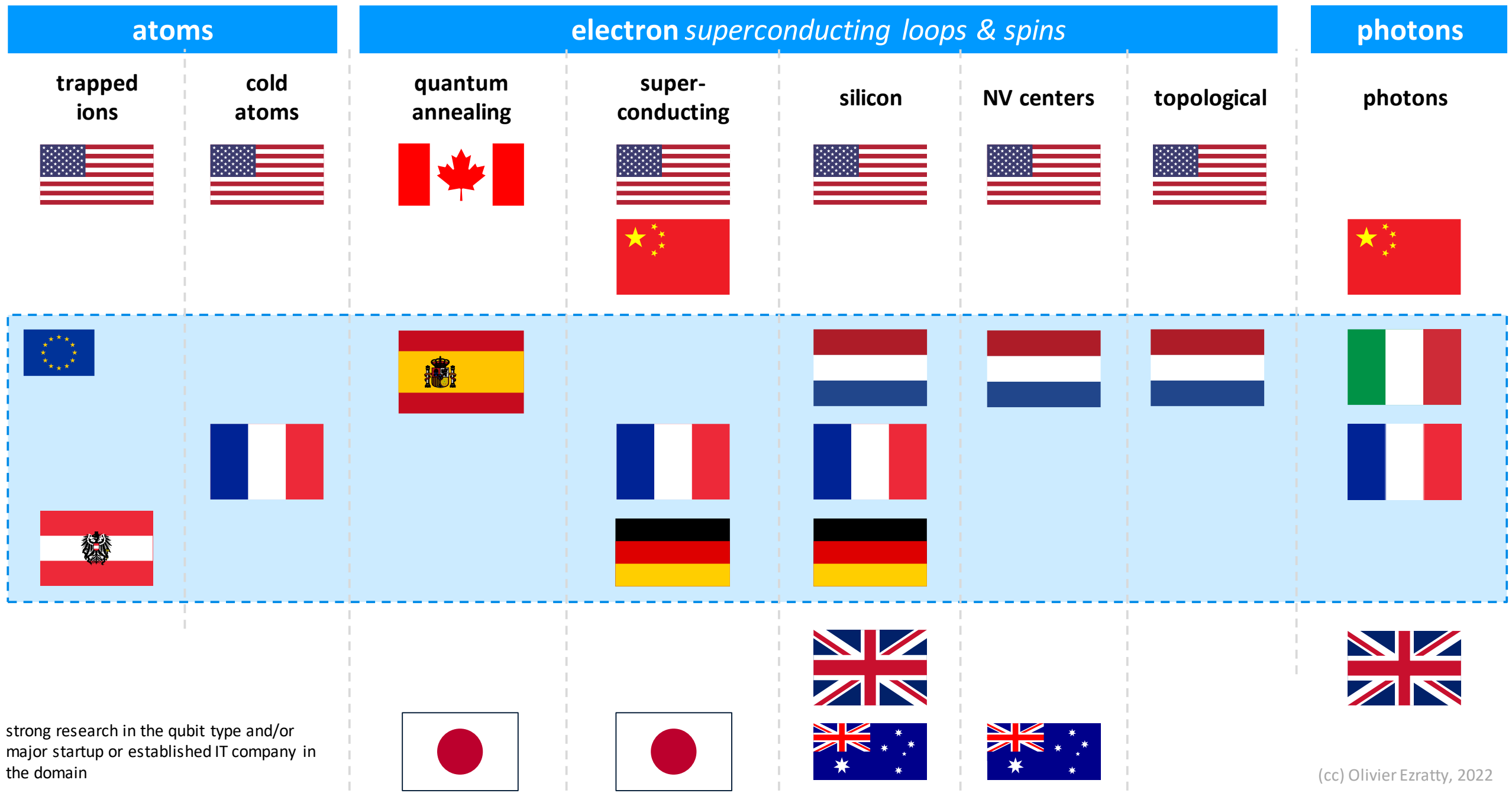

*Figure 845: key quantum computing technologies per qubit type and country of origin.*

# North America

## USA

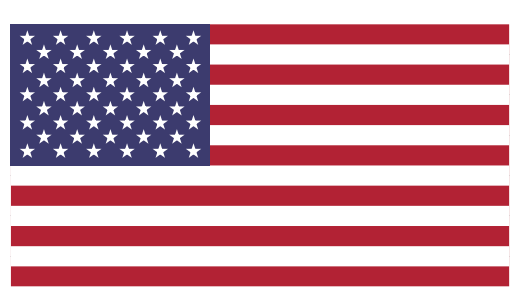

Whatever the metric you use, the USA is leading the world in quantum technologies. They mix three components no other country or region has: a powerful Federal government investing significant amounts in fundamental research, large IT companies investing a lot as well in research and industrialization and a healthy well-funded dense startup ecosystem.

The coordination of research in the different branches of quantum physics started in October 2014, the White House produced two reports[2680]. It was not a plan but rather an inventory of what existed. Like almost all countries, quantum technologies were segmented in four quantum communication, sensing, computing and simulation. In 2017, lobbying from the industry and research started to push the federal government to launch a national quantum plan. It started with U.S. House of Representatives organizing a hearing in October 2017 (video). For three hours, elected officials questioned a panel of scientists including James Kurose of the NSF and John Stephen Binkley of the Department of Energy, who explained the basics of qubits and the associated sovereignty issues.

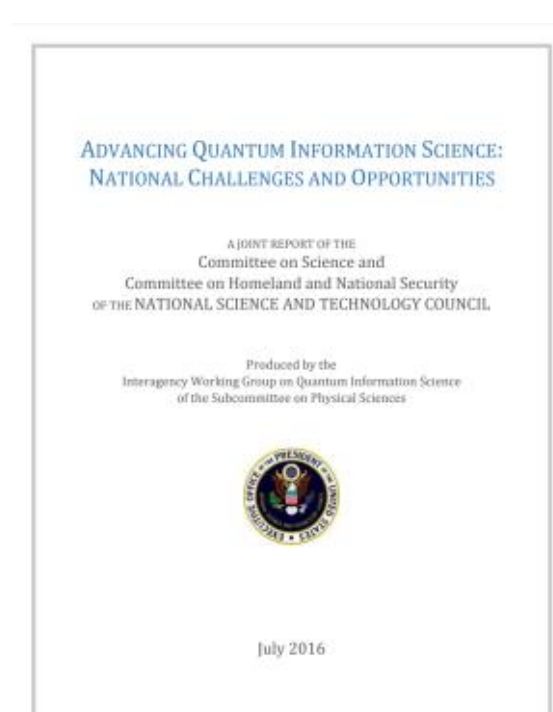
ADVANCING QUANTUM INFORMATION SCIENCE: NATIONAL CHALLENGES AND OPPORTUNITIES

A JOINT REPORT OF THE
Committee on Science and
Committee on Homeland and National Security
OF THE NATIONAL SCIENCE AND TECHNOLOGY COUNCIL

Produced by the
Interagency Working Group on Quantum Information Science
of the Subcommittee on Physical Sciences

July 2016

The Democrats were concerned about the Trump administration's proposed cuts in funding for civilian research in favor of defense budget increases and tax cuts. But the US Congress increased federal research budgets for fiscal year 2018 and beyond[2681], knowing that these budgets are then traditionally channeled mainly to American universities research laboratories. This is one of the few cases where the Republican-controlled Congress opposed the Trump administration.

[2680] The report Advancing Quantum Information Science: National Challenges and Opportunities, July 2016 (23 pages) was followed by a working meeting in October 2016.

[2681] See Trump, Congress approve largest U.S. research spending increase in a decade, Science, March 2018.

This happened consistently throughout all fiscal years of the Trump administration and will probably happen again with the Biden administration.

***National Quantum Initiative Act***

Then came the National Quantum Initiative Act that was first proposed on June 26, 2018 by the House of Representatives Science Committee (H.R. 6227, 25 pages). An equivalent proposal was done by the Senate on the same day. This project was the result of a proposal, the National Quantum Initiative-Action Plan, prepared by public and private research stakeholders (IBM, Google, Rigetti).

An intense lobbying campaign was carried out by several professional associations[2682], with the **National Photonics Initiative**, a professional association bringing together photonics physicists and industrialists in the sector, accompanied by the lobbying firm **BGR Group**.

This association, which wanted to make photonics a priority, was launched in 2012. It is sponsored by other entities: The Optical Society (OSA), SPIE (The International Society for Optics and Photonics), American Physical Society, IEEE Photonics Society, ALIA Laser Institute of America and a lot of other professional associations. The lobbying was also pushed by **Jonathan Dowling** (1955-2020, American), professor of physics at Louisiana State University and specialist in photonics[2683] .

The **Quantum Industry Coalition** brings together more generalist manufacturers such as Microsoft, Intel and Lockheed Martin as well as startups[2684]. This coalition is helped by the lobbying firm KL Gates. It is the 41st largest law firm in the world making $1B annually. The director of the Quantum Industry Coalition is Paul Stimers, a partner of KL Gate[2685]. To this should be added the **Quantum Alliance Initiative** launched in 2018 by the Hudson Institute, a conservative think tank, which creates proposed standards for QKD and QRNG and of course advocates for the development of this industrial sector in the USA.

**NIST** had also created with **SRI International** the **Quantum Economic Development Consortium** (QED-C) to develop the American quantum industry in the fields of communication and sensing[2686]. It is chaired by Joseph Broz, who is also vice-president of SRI, and by Celia Merzbacher, a semiconductor industry lobbyist who worked in the White House during the Bush 43 administration.

The Quantum National Initiative Act proposed allocating $1,275B over five years to fund civil quantum R&D, divided among the Department of Energy ($625M), NSF ($250M) and NIST which is focused on cryptography issues ($400M).

The Act also proposed the creation of a National Quantum Coordination Office within the White House Office of Science and Technology Policy. It asked the President of the United States to create a 10-year quantum plan, the first step being a five-year plan to be delivered one year after the passage of the law.

---

[2682] See Quantum computing finds its lobbying voice by Aaron Gregg, Washington Post, June 2018.

[2683] See Schrödinger's Killer App - Race to Build the World's First Quantum Computer by Jonathan P. Dowling, 2013 (445 pages) where the author was sending a warning about the risk to see China lead the quantum technology race. If the USA were not investing more: "*The future of the quantum Internet is in photons and the short circuiting of the development of optical quantum information processors in the United States means that the future quantum Internet will have 'Made in China' stamped all over it.* ", page 173.

[2684] See their Quantum Industry Coalition website.

[2685] See The US National Quantum Initiative by Paul Stimers, K&L Gates, October 2019 (6 pages).

[2686] See NIST Launches Consortium to Support Development of Quantum Industry, September 2018. And more details in U.S. Consortium Pulls Ecosystem Into Quantum by Susan Rambo, August 2019. As of July 2020, the association has 130 members from the private sector - large corporations and startups - and about 40 laboratories from universities and the U.S. public sector.

This bill was pushed by elected officials fearing that China will take over the quantum, especially in computer security[2687]. The USA likes to scare itself, even if, in the field of quantum technologies, it has nothing to be ashamed of with a density of public and private research laboratories and the major players having a large-scale industrialization capacity that almost no country can compete with. And their domestic market remains the largest in the world for enterprise computing applications.

The quantum bill was voted by the House in September and then by the Senate in December 2018[2688]. In September 2018, the White House published the National Strategic Overview for Quantum Information Science that included the terms of the congressional proposal. They emphasized research, training of scientists and international collaboration. At last, Donald Trump signed this law on December 21, 2018 just before the shutdown, but with no fanfare nor any scientists in the Oval Office[2689].

In December 2019, the **Quantum Information Edge** alliance was created, bringing together Lawrence Berkeley National Laboratory and Sandia Labs of the Department of Energy, the University of Maryland, Duke University (North Carolina), the University of Colorado at Boulder, Harvard, Caltech, MIT and the University of New Mexico[2690]. For the most part, the usual suspects of basic research in quantum computing, thus creating their "virtual hub" for coordinating research in this field. With a focus on error reduction at the qubit level, techniques for interconnecting qubits and the development of new quantum algorithms. For its part, the NPI embarked on a new lobbying campaign at the end of 2019 and early 2020 to increase once again the federal funds allocated to research in quantum technologies[2691]. In 2022, the campaign was still going on with pushes for more Federal investments in quantum technologies procurement[2692].

In February 2020, the White House published a memo from the National Quantum Coordination Office recommending the development of quantum networks[2693].

And in March 2020, the US executive proposed a new increase in quantum research budgets for the years 2020/2021[2694]. It included $450M for the Department of Energy, $330M for the NSF and $80M for the NIST. This was matched by a $1B increase for artificial intelligence research programs[2695]. In August 2020, the White House announced a 30% increase in the quantum and AI budgets for fiscal year 2021.

---

[2687] See How suspicions of spying threaten cross-border science by Patrick Howell O'Neill, December 2019, which discusses the direct and indirect methods used by China to plunder European and American quantum research and exploit it both civil and military, such as quantum radars, quantum sonars and QKD. Here is the link to retrieve the Quantum Dragon Strider study mentioned in the article, November 2019 (22 pages). You can indicate a bogus email to get it, the download does not go through an email. It evokes various partnerships in research that help the Chinese to exploit Western research. It is based on a few examples including the very detailed one from the University of Heidelberg in Germany. On the same subject, see also China's top quantum scientist has ties to the country's defense companies, December 2019, Quantum USA Vs.Quantum China: The World's Most Important Technology Race by Moor Insights and Strategy, October 2019 and New Warnings Over China's Efforts in Quantum Computing by Sintia Radu, January 2020.

[2688] See SIA Welcomes House Passage of Quantum Computing Legislation, September 2018.

[2689] See President Trump has signed a $1.2 billon law to boost US quantum tech by Martin Giles in the MIT Technology Review, December 2018. See Jeremy Tsu's The Race to Develop the World's Best Quantum Tech in IEEE Spectrum, December 2018, which discusses the CNAS report Quantum Hegemony-China's Ambitions and the Challenge to U.S. Innovation Leadership published in September 2018, which describes China's quantum strategy (52 pages). See also US intelligence community says quantum computing and AI poses an 'emerging threat' to national security by Zack Whittaker, December 2018.

[2690] See US alliance for quantum computing by David Manners, 2019.

[2691] See NPI Brings Quantum Experts to Capitol Hill to Advocate for Additional NQI Funding by Jo Maney, March 2020.

[2692] See The US government needs a commercialization strategy for quantum by Laura E. Thomas, senior director of National Security Solutions at ColdQuanta, December 2021. Pushing for federal procurement of US quantum computers through DARPA and the NSF.

[2693] See A Strategic Vision for America's Quantum Networks, White House, February 2020 (4 pages).

[2694] See Why is Trump funding quantum computing research but cutting other science budgets? The national security implications of this technology may be exaggerated by John Lindsay, March 2020.

[2695] See White House reportedly aims to double AI research budget to $2B by Devin Coldewey in TechCrunch, February 2020.

In December 2021, a memo was published by the NQI team showing for the first time the total Federal budget spent in quantum technologies by year with the legacy spendings and the NQI related spendings. It did show that the 5-year trend was a $4B plan[2696].

⟨quantum|gov⟩ The US NQI (National Quantum Initiative) is run by the **National Quantum Coordination Office** (NQCO), hosted by the White House Office of Science and Technology Policy (OSTP).

It has a web site since October 2020[2697]! Its director is Charles Tahan[2698]. And if you wonder about the bureaucracy in your own country, here you are also with several related committees: the **National Science and Technology Council** (NSTC) Subcommittee on Quantum Information Science (SCQIS) that coordinates Federal R&D in quantum technologies, the **National Science and Technology Council** (NSTC) Subcommittee on the Economic and Security Implications of Quantum Science (ESIX) that handles economic and security implications across federal agencies[2699] and the **National Quantum Initiative Advisory Committee** (NQIAC) that advises the President, the Secretary of Energy and the NSTC Subcommittee on QIS.

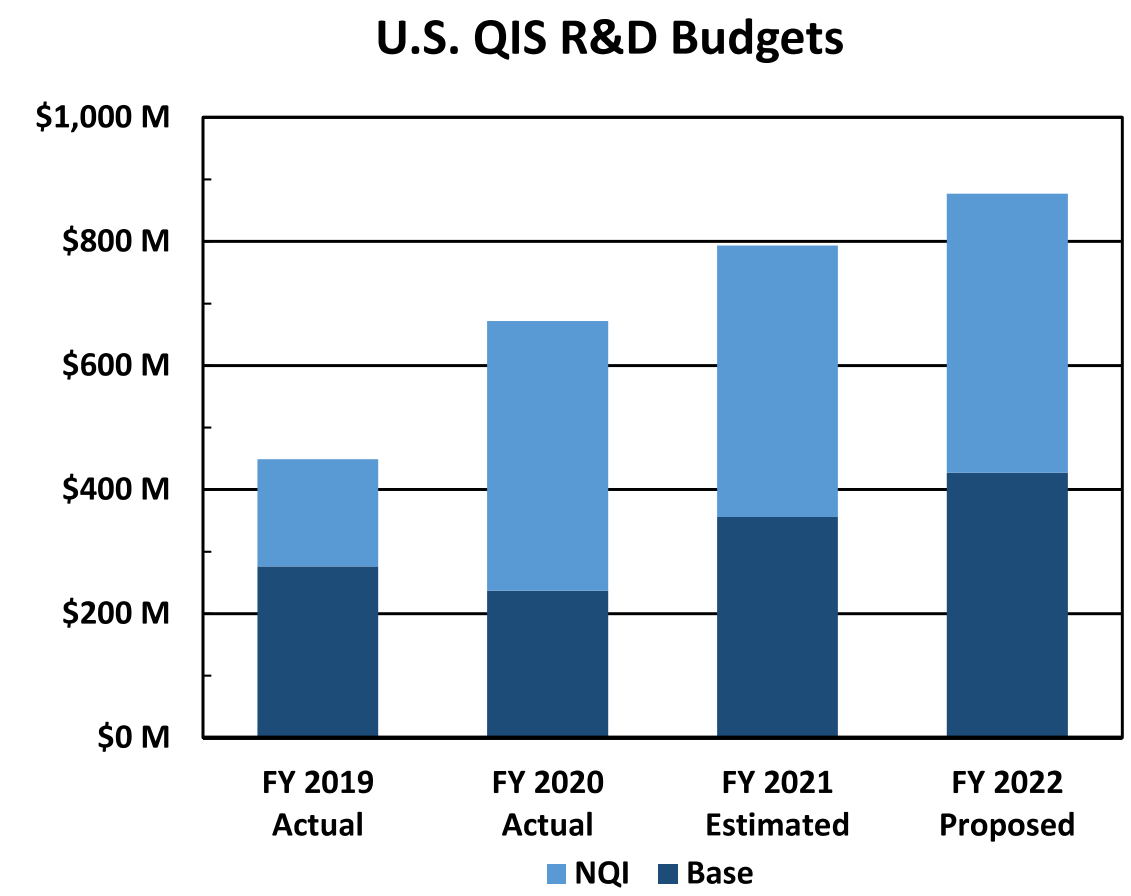

*Figure 846: the most accurate Federal investment report on quantum technologies. Source: [National Quantum Initiative supplement to the President's FY 2022 budget](), December 2021 (46 pages).*

In April 2021, the story went on with yet another Congress proposal, the **Quantum for Universal Advancement in Nationwide Technology Use and Modernization** (QUANTUM) for National Security Act of 2021[2700]. Two Senators introduced two bills to "*better position the United States to be globally competitive in quantum information science*". It's focused on developing Department of Defense quantum networking and telecommunications use cases and workforce developments.

This bill appears as a direct response to China's massive investments in quantum telecommunications infrastructures. It's kind of a military grade version of the National Quantum Initiative Act launched in 2018 that had mostly a civilian face despite the significant DoE funding it did incorporate[2701].

In February 2022, the NQI released a report on the development of quantum technologies workforce in the USA[2702]. It includes exposing the public to educational content about quantum technologies and to ensure quantum technologies are as inclusive as possible. The plan is mostly qualitative and doesn't provide means and objectives data.

---

[2696] See National Quantum Initiative supplement to the President's FY 2022 budget, December 2021 (46 pages).

[2697] The NQCO published the quick status report Quantum frontiers report on community input to the nation's strategy for quantum information science in October 2020 (32 pages).

[2698] Charles Tahan is a physicist specialized in condensed matter physics and quantum information science. He continues to publish some scientific papers from time to time, trying to not losing ground in his field.

[2699] In The role of international talent in quantum information science by National Science and Technology Council of the White House, October 2021 (20 pages), the NSTC worries about the quantum talent shortage in the USA and advocates a balanced approach between hiring international talent and protecting national security. The global hunt for talent is launched!

[2700] See Thune, Hassan Introduce Bills to Bolster the United States' Leadership in Quantum Information Science, April 2021.

[2701] And it's never enough! See for example America is Losing the Quantum Race with China by Theresa Payton, a former White House CIO, May 2022. Unfortunately, it contains only fainted documented anecdotal evidence and no significant hard data to prove that China is indeed ahead of the USA for all respects. Maybe the only real case is about QKD deployments but it's not even mentioned here.

[2702] See Quantum Information Science and Technology workforce development national strategic plan, February 2022 (34 pages.

The USA is concerned about managing its talent pool efficiently with, simultaneously caring about its national security. It is a concern fueled by a simple fact: a significant share of the quantum research talent pool in the USA are first generation immigrants.

Let's take this paper published by a research team from a University in Virginia and the DoE Los Alamos lab. All the names are Chinese! Are these Americans, first generation immigrants, or PhD students or post-docs who'll soon return to China?

**Quantum Neural Network Compression**

Zhirui Hu[1,2], Peiyan Dong[3], Zhepeng Wang[1,2], Youzuo Lin[4], Yanzhi Wang[3], Weiwen Jiang[1,2]

[1]Electrical and Computer Engineering Department, George Mason University, Fairfax, Virginia 22030, United States
[2]Quantum Science and Engineering Center, George Mason University, Fairfax, Virginia 22030, United States
[3]Department of Electrical and Computer Engineering, Northeastern University, Boston, MA 02115, United States
[4]Earth and Environmental Sciences Division, Los Alamos National Laboratory, NM, 87545, USA
(zhu2@gmu.edu; wjiang8@gmu.edu)

*Figure 847: an example of a paper published in the USA with authors all having a Chinese name. See Quantum Neural Network Compression by Zhirui Hu et al, July 2022 (11 pages).*

In May 2022, President Biden signed two Presidential directives[2703]. The first was an Executive order enhancing the governance of the NQI with the creation of an Advisory Committee. The second was a National Security Memorandum describing how the Federal government will prepare for the adoption of PQC cybersecurity to protect it from future quantum computing cryptology threats, in line with the first 4 NIST approved PQC standards announced in July 2022. The NIST is tasked with running a PQC migration project for the Federal government at the National Cybersecurity Center of Excellence and for the industry. The NSM also mandates new protections against IP theft and abuse.

In August 2022, the CHIPS Act was signed by POTUS with additional Federal funding of $152M per year for quantum technologies for the 2023-2027 period, going to NIST, the DoE and the NSF as usual. As a consequence, it adds about $760M to the 5-year $4B run-rate of quantum federal research expenditures[2704].

### *Military and intelligence federal agencies*

Public laboratories investing in quantum computing cut across much of the federal military-industrial complex with internal research or external research funded through calls for proposals or joint laboratories with universities:

**IARPA** (Intelligence Advanced Research Projects Agency) funds third-party projects on quantum computing and quantum algorithms. They run several quantum programs that happen to involve Universities outside the USA. The only that seems still in place is **LogiQ**. Its goal is to improve the quality of qubits. It involves TU Delft (Netherlands), the University of Innsbruck, Duke University and IBM. IARPA also funds programs conducted by third parties. It is a small agency that employs fewer than a hundred people.

**NSA** is investing heavily in quantum technologies, both in the race to implement Shor's algorithm for breaking RSA-based public-key protected communications and for protecting sensitive communications with quantum keys and cryptography. This work is obviously not public. The NSA subcontracts some of its research to private companies such as Lockheed-Martin. It is also part of a joint laboratory with NIST and the University of Maryland, QuICS, which was launched in 2014.

[2703] See Fact sheet: President Biden Announces Two Presidential Directives Advancing Quantum Technologies, White House, May 2022.

[2704] See Quantum in the CHIPS and Science Act of 2022, QuantumGov, August 2022.

**DARPA** funds three programs in quantum technologies: long-distance quantum communications, quantum metrology applied to imaging, and neurological trauma diagnosis and PTSD. Funding goes to projects led by universities, startups and established companies[2705]. In 2020, they launched a NISQ computation challenge which led to the selection of 7 research teams as part of the ONISQ program[2706] and QAFS, a program on quantum annealing involving among others the Lincoln Lab from the MIT.

**Army Research Office** also has its own quantum research program covering the entire spectrum from sensing to quantum computing, cryptography and quantum communications.

**US Air Force** and its Air Force Research Laboratory's Quantum Communications lab is focused on quantum cryptography (QKD). The AFRL announced in December 2020 that it planned to work with the Office of Naval Research to test quantum technologies with the "Five Eyes" countries (USA, Canada, Australia, New Zealand and UK) for a Naval exercise. Another lab, the Quantum Information and Sciences Laboratory, does applied research in superconducting qubits, photonic qubits, trapped ions qubits, quantum algorithms and quantum sensing. They even deployed their own superconducting qubits system prototype, created with the MIT.

**Office of Naval Research** (ONR) is working on the use cases of QKDs for the Navy and on using of quantum algorithms related to the Navy operational needs.

***Federal civil agencies***

In quantum science and technologies, the key Federal civil agencies are the Department of Energy, the NIST, the NSF and NASA.

**Department of Energy** (DoE) has many research laboratories that are big consumers of supercomputing capacities like in Oak Ridge and Argonne. It also operates the Los Alamos National Laboratory (LANL) and its Quantum Institute (QI) launched in 2002 that also invests in quantum computing and cryptography. In particular, they fund research at UNSW in Australia as well as in Maryland. The DoE also runs the Sandia National Laboratories**,** which also conducts applied research in all areas of quantum physics.

The DoE launched a call for proposals to award 158 grants totaling $32M to 118 SMEs through the SBIR program. The grants are delivered in two phases, a first phase of $200K followed by a second phase of $1.1M for the best projects, spread over a period of two and a half years.

The DoE also announced in August 2020 the funding of five research centers in quantum technologies, all led by DoE laboratories, with $300M coming from the DoE and the rest from relevant institutions and the industry (IBM, Microsoft, Intel, Lockheed Martin)[2707]. These new research centers are **Q-NEXT** (Next Generation Quantum Science and Engineering Center, David Awschalom) led by Argonne National Laboratory which focuses on the industrialization of quantum hardware, **C²QA** (Co-design Center for Quantum Advantage, Steve Girvin) led by the DoE Brookhaven National Laboratory which will focus on ways to achieve quantum advantage in scientific applications, the **SQMS** (Superconducting Quantum Materials and Systems Center, Anna Grassellino) led by the Fermi National Accelerator Laboratory which will focus on superconducting qubits, the **QSA** (Quantum

[2705] See The DARPA Model for Transformative Technologies, 2019 (511 pages) which tells the story of how the agency works. It has about one hundred program managers in total with a total budget of about $3.5B. It explains how it connects fundamental research to difficult technology challenges.

[2706] See DARPA Challenge May Boost Quantum Value of NISQ Devices by Matt Swayne, June 2020. One of the selected teams includes a certain Davide Venturelli who studied at the University of Grenoble.

[2707] See National Quantum Information Science Research Centers by Ceren Susut, December 2020 (17 slides). Unstable link.

Systems Accelerator Center, Irfan Siddiqi[2708]) led by the Lawrence Berkeley National Laboratory which works on quantum computing hardware and software and the **QSC** (Quantum Science Center, David Dean) led by the Oak Ridge National Laboratory which will focus on quantum computing scalability issues.

The DoE then launched a $30M program in March 2021 on nanoscale matter and their use case in energy applications. It will fund the five existing DoE Nanoscale Science Research Centers and their research partners over 3 years. The awards size is between $1M and $2.5M[2709]. It also launched a $25M program in April 2021 on Quantum Internet including quantum repeaters, quantum memory and quantum communication protocols, opened to the 17 DoE labs.

**NSF** funds various research projects[2710]. In 2019, it launched a call for **Quantum Leap Challenge Institutes**, to fund research institutes conducting interdisciplinary research projects advancing the state of the art in quantum technologies[2711].

Their format is reminiscent of the UK quantum program hubs. Three hubs were selected in July 2020 for a total of $75M spread over five years: a first dedicated to quantum sensing led by the University of Colorado, a second dedicated to quantum computing led by the University of Illinois - Urbana-Champaign and a third also on quantum computing and rather software side led by the University of Berkeley[2712]. These three hubs bring together 16 academic institutions, 8 national laboratories and 22 industrial partners. On top of that, the NSF is also funding the consolidation of other initiatives like the one around Purdue University in Indiana[2713]. The NSF also launched a **Quantum Algorithm Challenge** in March 2020[2714].

**NIST** is a federal research institute with the Department of Commerce. Its historical role is sensing and the definition of weights and measures. Its work on atomic clocks naturally led it to look after quantum technologies. It has an annual budget of $1.2B and employs 3,400 people on two campuses, one in Boulder, Colorado and another one in Maryland, next door to the University of Maryland and north of Washington DC. Several of its research groups are dedicated to quantum technologies with the Quantum Processing Group for quantum computing, another for spintronics, one for quantum sensing and another for superconducting electronics. On top of this, the Computer Security Division of the Information Technology Laboratory (ITL) manages the call for tenders on the standardization of PQC (Post-Quantum Cryptography) that we have already covered in a dedicated chapter after page 823[2715]. NIST's PQC standardization strategy has wide implications. It will sediment the market around a dozen standards that will be royalty-free. This may favor large cybersecurity vendors instead of enabling new players to disrupt the market.

---

[2708] The QSA was awarded a funding of $115M for 5 years in August 2020. See New $115 Million Quantum Systems Accelerator to Pioneer Quantum Technologies for Discovery Science by Dan Krotz, August 2020.

[2709] See DOE Announces $30 Million for Quantum Information Science to Tackle Emerging 21st Century Challenges, March 2021.

[2710] See for example NSF Awards $2M For Research on Quantum Machine Learning With Photonics, September 2019 for the University of Maryland.

[2711] See Quantum Leap Challenge Institutes (QLCI), NSF, 2019.

[2712] See NSF establishes 3 new institutes to address critical challenges in quantum information science, NSF, June 2021.

[2713] Purdue University launched in July 2021 a new Center for Quantum Technologies funded by the NSF with an established team of 50 quantum scientists and engineers coming from various research institutions in Indiana working on many quantum fields (atomic and molecular optics, solid state quantum systems, quantum nanophotonics, quantum information and communication).

[2714] See Dear Colleague Letter: Quantum Algorithm Challenge, Anne Kinney and Margaret Martonosi, NSF, March 2020.

[2715] See this overview of NIST's scientific activities: Quantum Information Science & NIST - Advancing QIS Technologies for Economic Impact, 2019 (39 slides).

NIST is also a stakeholder in and a co-founder of three joint laboratories with two major universities, each located near its own campuses in the states of Colorado and Maryland, the JQI, QuICS with the NSA and JILA with the University of Colorado.

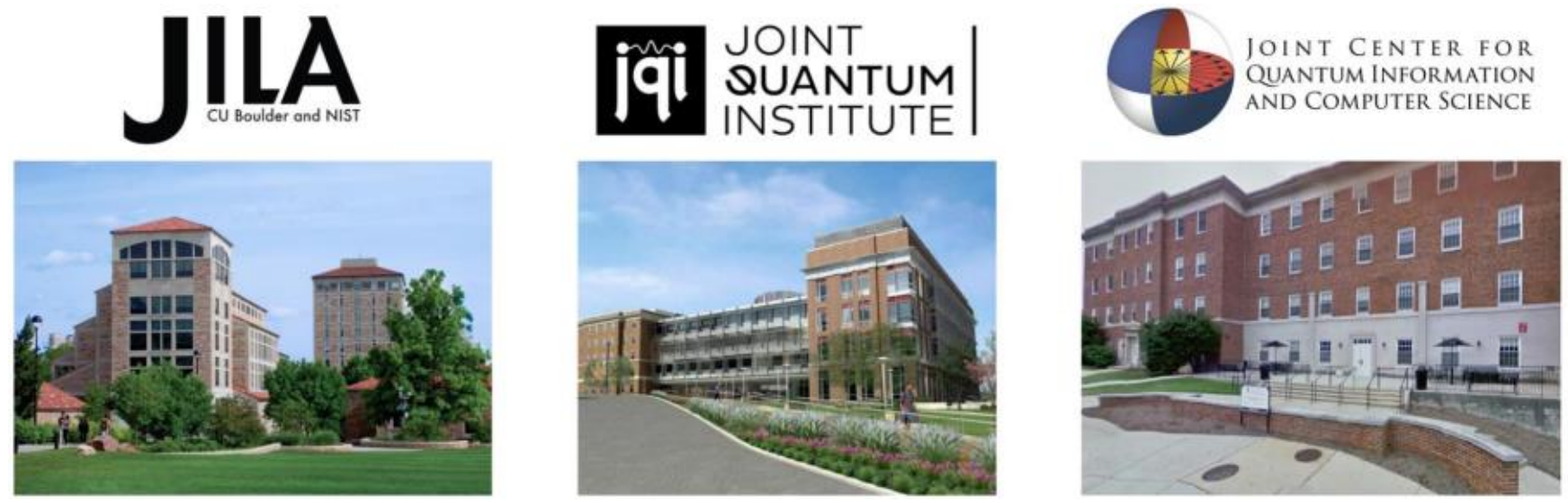

*Figure 848: the JV labs from NIST.*

The University of Maryland's **Joint Quantum Institute** (JQI**),** established in 2006 is a fundamental quantum physics laboratory. It is the home of David Wineland, a long-time specialist in ion control by laser cooling, who won the Nobel Prize in Physics in 2012 along with Serge Haroche. It is in this laboratory that the IonQ startup by Christopher Monroe was launched in 2015. Many alumni from this lab also joined Honeywell's quantum team in Denver, Colorado. This laboratory employs 35 permanent researchers, 55 post-docs and 85 PhD students with an annual budget of $6M supplemented by various external funding.

The **Joint Center for Quantum Information and Computer Science** (QuICS) at the University of Maryland (UMD) launched in 2014 in partnership with NSA's research directorate that focuses on quantum computing architectures, algorithms and complexity theories to complement the JQI.

The **JILA** at the University of Colorado at Boulder which is dedicated to sensing technologies[2716]. It is home to two Nobel Prize winners: Eric Cornell (in 2001, for his work on Bose-Einstein condensates) and John Hall (in 2005, for his work on laser frequency combs).

NIST employs a fourth Nobel Prize winner in physics, William D. Phillips for his work on atoms laser cooling using the Zeeman effect in 1997, shared with Claude Cohen-Tannoudji from France.

NASA created in 2013 the Quantum Artificial Intelligence Laboratory (QuAIL) jointly with Google, located at the Ames Research Center near the Google's headquarters in Mountain Views to explore the field of quantum algorithms, in particular on a D-Wave quantum annealer they acquired and installed there.

### *USA local quantum ecosystems*

The main geographical quantum hubs in the USA combine a mix of national labs like those from the DoE, Universities and commercial companies. So here we are:

- **California** with the Silicon Valley and in the Los Angeles area, with Stanford, UCLA[2717], Caltech[2718], UCSB, plus the labs from Google, Microsoft and Amazon, and also Rigetti. The Los Angeles area seems on part with, if not stronger than the Silicon Valley.
- **Massachusetts** with the MIT, Harvard and UMass Amherst.
- **Colorado** at Boulder, with Quantinuum, NIST and the University of Boulder Colorado[2719].
- **New Haven** with its very influential Yale University, particularly in the superconducting qubit domain, and Qci which is a spin-out startup from Yale.

---

[2716] JILA was created in 1962 as the Joint Institute for Laboratory Astrophysics but they now use only the acronym without this meaning given its extended activities beyond astrophysics.

[2717] UCLA got a $5M grant from Boeing to support the Center for Quantum Science and Engineering, as announced in May 2022.

[2718] In January 2022 was announced the creation of the "Dr. Allen and Charlotte Ginsburg Center for Quantum Precision Measurement" thanks to a donation from the couple.

[2719] The University of Boulder created the Qubit Quantum Initiative to foster interdisciplinary quantum research in a 4-floor building.

- **Illinois/Chicago** with two DoE labs (Fermi and Argonne), several universities and the Chicago Quantum Exchange ecosystem which regroups these labs, the University of Chicago, the University of Illinois, the University of Wisconsin and Northwestern University[2720]. The University of Chicago Polsky Center and the Chicago Quantum Exchange launched the first national quantum startups accelerator program in April 2021. In September 2022, the NRF (National Research Foundation of South Korea) awarded $1M (over 5 years) to David Awschalom and Liang Jiang from the University of Chicago to create a joint lab, "The Center for Quantum Error Correction".
- **New York State** with Princeton, the Flatiron Institute, IBM, GlobalFoundries and SeeQC.
- **Maryland** with the University of Maryland, IonQ, NIST, NSA and the Quantum Catalyzer quantum startups accelerator (Q-CAT) launched by Ronald Walsworth that creates quantum startups from scratch.
- **Tennessee** and **New Mexico** host three DoE labs and their quantum research centers.
- **Washington State** with the University of Washington and the Pacific NorthWestern DoE lab, given Microsoft and Amazon HQ are there but their quantum team mostly sits in part in Santa Barbara, California.
- Lesser developed ecosystems in **Indiana** (Purdue University, Midwest Quantum Collaboratory, a joint lab between Purdue University, Michigan State University and University of Michigan), **Virginia** (Virginia Tech), **Georgia** (GeorgiaTech) and **Florida** (Florida State University).

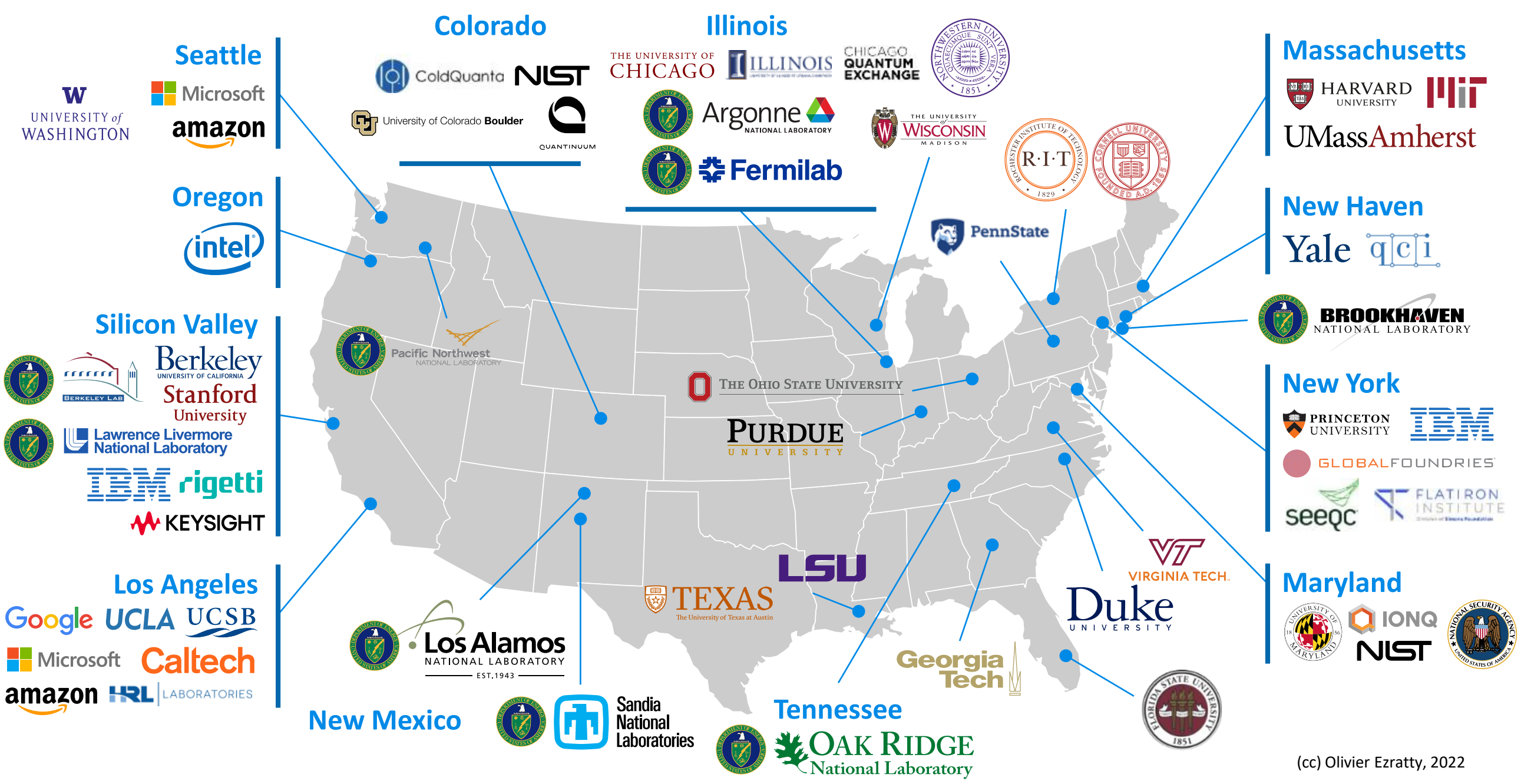

*Figure 849: another map! You want to know where Argonne and Sandia Labs are? Here it is with the DoE labs, large Universities and vendors. (cc) Olivier Ezratty, 2022.*

The above map in Figure 849 showcases this geographical distribution of USA's quantum technology R&D areas. The distribution is more even than in classical digital technologies, which are more concentrated on the country's West Coast and particularly in the Silicon Valley.

---

[2720] See Chicago Quantum Exchange welcomes new partners focused on manufacturing, computing, and the Chicago region, March 2022, and University of Chicago forges new bonds with European partners through Quantum and Sustainability conference in Paris, May 2022. It established a UChicago Center in Paris to foster collaboration between the Chicago and European quantum ecosystems.

Finally, let us recall a market reality that echoes economist **Maria Mazzucato'**s thesis on the public origin of technological innovations: the major American players are sourcing at different levels and throughout the USA and the world to advance their quantum technologies. Figure 850 is a good illustration of this phenomenon, showing how large IT players like IBM, Google, Intel, Microsoft, Amazon, Honeywell and even IonQ, surf on the work of publicly funded research labs not only in the USA but throughout the world. The last example being the creation of a Google AI lab in Australia in partnership with UNSW, the University of Sidney, Macquarie University and UTS, for the development of quantum applications.

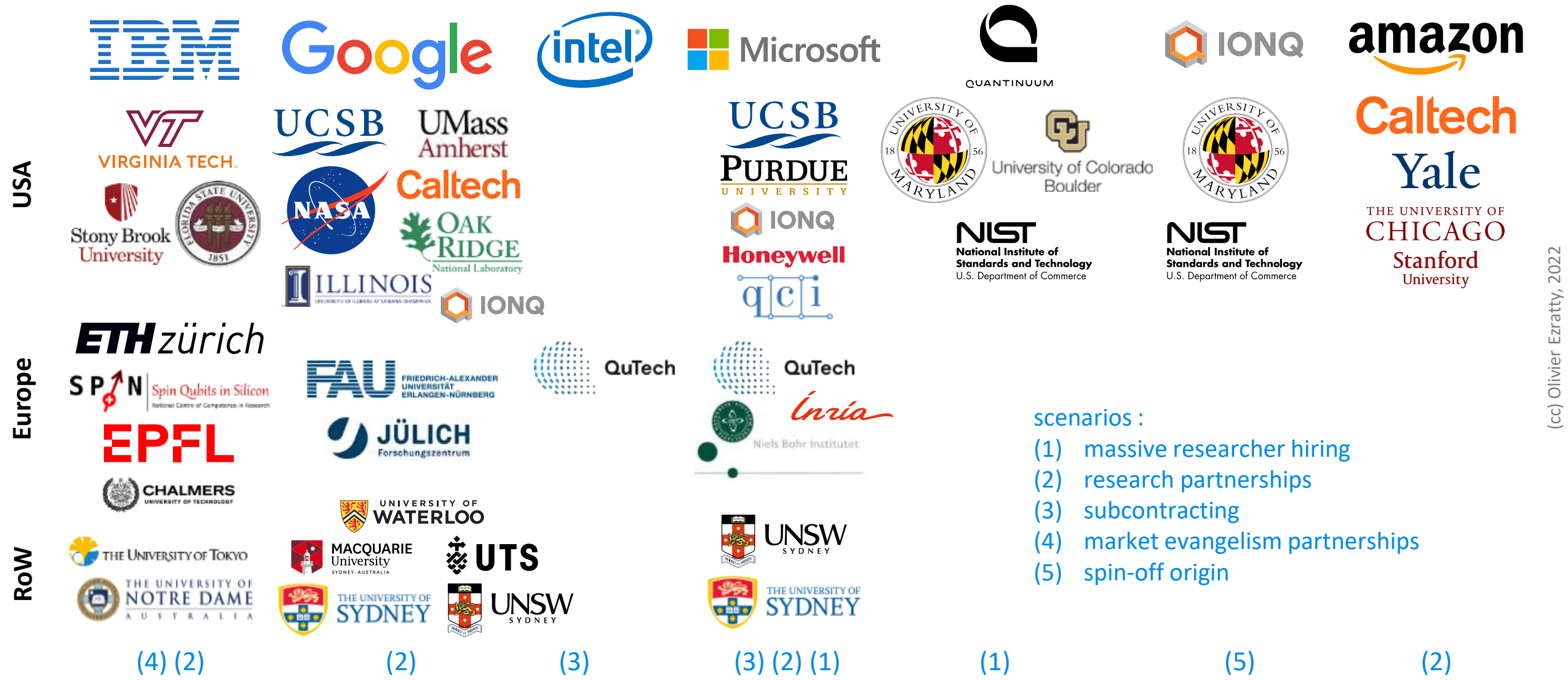

*Figure 850: large IT vendors are reusing a lot of research and talent from universities both in the USA and in the rest of the world. Here's a map of who works with whom. (cc) Olivier Ezratty, 2022.*

## Canada

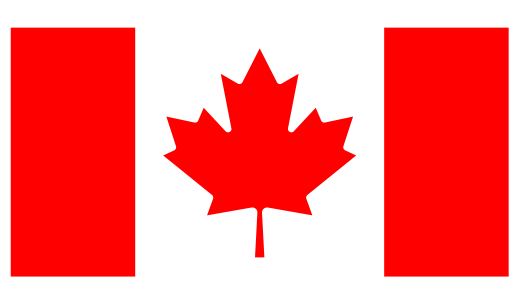

In Canada, a parallel can be drawn between artificial intelligence and quantum technologies. In both cases, the country's influence is far greater than its economic weight, at the basic research level, with a healthy startup ecosystem and best-in-class investment per capita.

This is due in particular to a constant and early-stage investments in research by government and universities and to a certain entrepreneurial dynamism.

Canada has two great quantum stars in research with **Gilles Brassard of the** University of Montreal who is with **Charles Bennett of** IBM Research the co-inventor of QKD's BB84 protocol.

### *Research*

Canada is distinguished by a strong investment in basic research in quantum computing, including more than $1B of public investment over a decade, mainly in three institutions[2721]. Let's look at the Canadian ecosystem from East to West.

[2721] See Quantum Canada by Ben Sussman, Paul Corkum, Alexandre Blais, David Cory and Andrea Damascelli, February 2019 (6 pages) for an overview on Canada's quantum investments.

In **Québec**, the University of Sherbrooke Quantum Institute**,** near Montreal, is home to Alexandre Blais, a recognized specialist in superconducting qubits. Their **QSciTech** training program organized with industry partners and the Q2 initiative encourage student entrepreneurship. Several startups came out of it such as SBTech (metrology), Nord Quantique (quantum computing) and Quantic (sensing). In February 2022, the Québec Region announced an impressive plan of public/private investment of CAN $435M (US $350M) in the Sherbrooke region. It includes CAN $131M to create "Sherbrooke Quantique" coming from the region and the rest from industry investors including IBM. The setup of an IBM quantum computer in Bromont near Sherbrooke corresponds to a region investment of CAN $68M and CAN $62M from IBM.

The rest of the region investments (about $62M) covers various research funding and real-estate investments, including the innovation platform PINQ² (that is not just related to quantum technologies). 1QBit, Pasqal and Eidos-Sherbrooke announced they would install an office in Sherbrooke, representing a forecasted investment of CAN $205M over 5 years. Eidos-Sherbrooke is a video games studio that is planning to use quantum computing. The plan also contains a region funding of CAN $3,6M out of a total of CAN $8,1M for setting up a collaborative platform on quantum software design with CMC Microsystems who also manufactures some electronic components (CMOS down to 22 nm density, III-V, Si Photonics, superconducting).

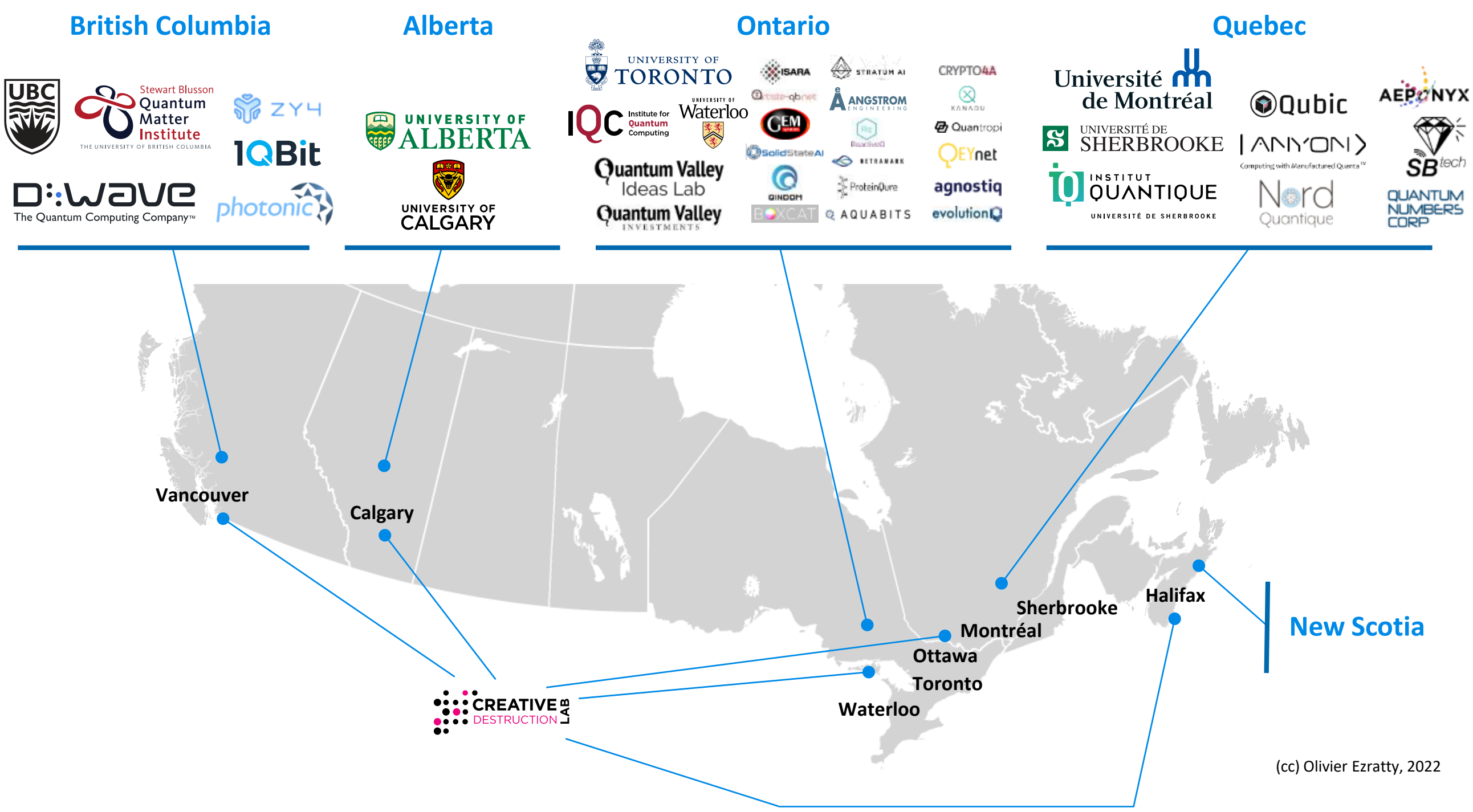

*Figure 851: a new map for Canada's quantum ecosystem from East to West where you see a startup concentration in Ontario. (cc) Olivier Ezratty, 2022.*

In **Ontario**, the quantum ecosystem is centered around Toronto and Waterloo. The University of Waterloo Institute for Quantum Computing, near Toronto obtained $120M in 2017 to fund its various quantum research institutes, complemented by $53M in Australian funding from UNSW, the operator Telstra and the Commonwealth Bank of Australia. The IQC does both research and teaching. They offer short courses of one to two weeks in the summer on quantum cryptography and quantum computing. The IQC is directed by Raymond Laflamme, one of the fathers of QEC. They cover all aspects of quantum technologies with about thirty teams of theorists and experimentalists, about fifty post-docs and 125 PhD students. A dozen startups were created since 2002. The IQC is leading the Transformative Quantum Technologies (TQT), a seven-year research commercialization program funded by the Canadian government and its First Research Excellence Funds to the tune of $76M.

In January 2020, TQT launched the Quantum Alliance, an umbrella for IQC and TQT to link them to their Canadian and international ecosystem, including the fabric of quantum startups. The University of Waterloo also runs its Quantum-Nano Fabrication and Characterization Facility (QNFCF) with cleanrooms located in the Lazaridis Quantum-Nano Centre. The Waterloo quantum ecosystem also hosts the Research Accelerator Center of Quantum Technologies.

In **Alberta**, the University of Calgary is working on quantum communications and has deployed an experimental QKD network of a few tens of kilometers. The University of Alberta in Edmonton, north of Calgary, is also involved in this work[2722]. In 2020, the government of Alberta dedicated $11.8M to the creation of an international hub for quantum computing, $3M of which will fund quantum research.

In **British Columbia**, the UBC (University of British Columbia) Quantum Matter Institute (QMI), located primarily in Vancouver.

### *Government funding*

As in most countries, the government is funding quantum research and the industry.

The Canadian government says it invested over CAN $1B in quantum research and education between 2009 and 2020. In April 2021, it announced its national quantum initiative with a CAN $360M (US $300M) plan spread over 7 years. Just before, in March, it announced a public funding of $40M for D-Wave[2723]. All this is packaged in highly optimistic business forecasts. The National Research Council of Canada (NRC) estimated that quantum technologies would generate CAN $139B turnaround and create 209,000 employments in Canada by 2045[2724]. It's quite optimistic given it's even larger than the most bullish worldwide forecasts! The forecast in 2022 was of 1,100 jobs creation by 2024.

### *Quantum industry*

In the industry side, you can't escape **D-Wave** and the quantum software specialist **1QBit**. With over 36 quantum startups and SMEs, they are the second largest ecosystem in the world in this respect after the USA and UK.

Private funding includes donations from Michael Lazaridis, one of the RIM BlackBerry co-founders, with $75M to the **Institute for Quantum Computing** at the University of Waterloo and $128M in 1999 to the **Perimeter Institute for Theoretical Physics** also located in Waterloo. Together with Doug Fregin, also co-founder of RIM, they also created the **Quantum Valley Investment Fund** with a total of $100M in funding and the **Quantum Valley Ideas Lab**.

Let us also note the existence of the **Creative Destruction Lab**, a deep techs startup acceleration structure with a specialty on quantum technologies. They are located in Canada (Toronto, Montreal, Vancouver, Calgary, Halifax), in the USA (Atlanta) as well as in Oxford and Paris.

In 2020, of group of industry vendors created **Quantum Industry Canada** (QIC), an association promoting the Canadian quantum industry. It includes D-Wave, 1Qbit, Xanadu, Zapata Computing Computing and ISARA.

---

[2722] See Quantum Communication Network Activities Across Canada by Barry Sanders and Daniel Oblak, June 2019 (10 slides).

[2723] See Government of Canada contribution strengthens Canada's position as a global leader in quantum computing, March 2021. This funding looks curious considering the company was created back in 1999. But it's probably not yet break even and has a strong need for cash to maintain its activity and leadership in a yet unmatured market.

[2724] Source: Economic impact of quantum technologies.

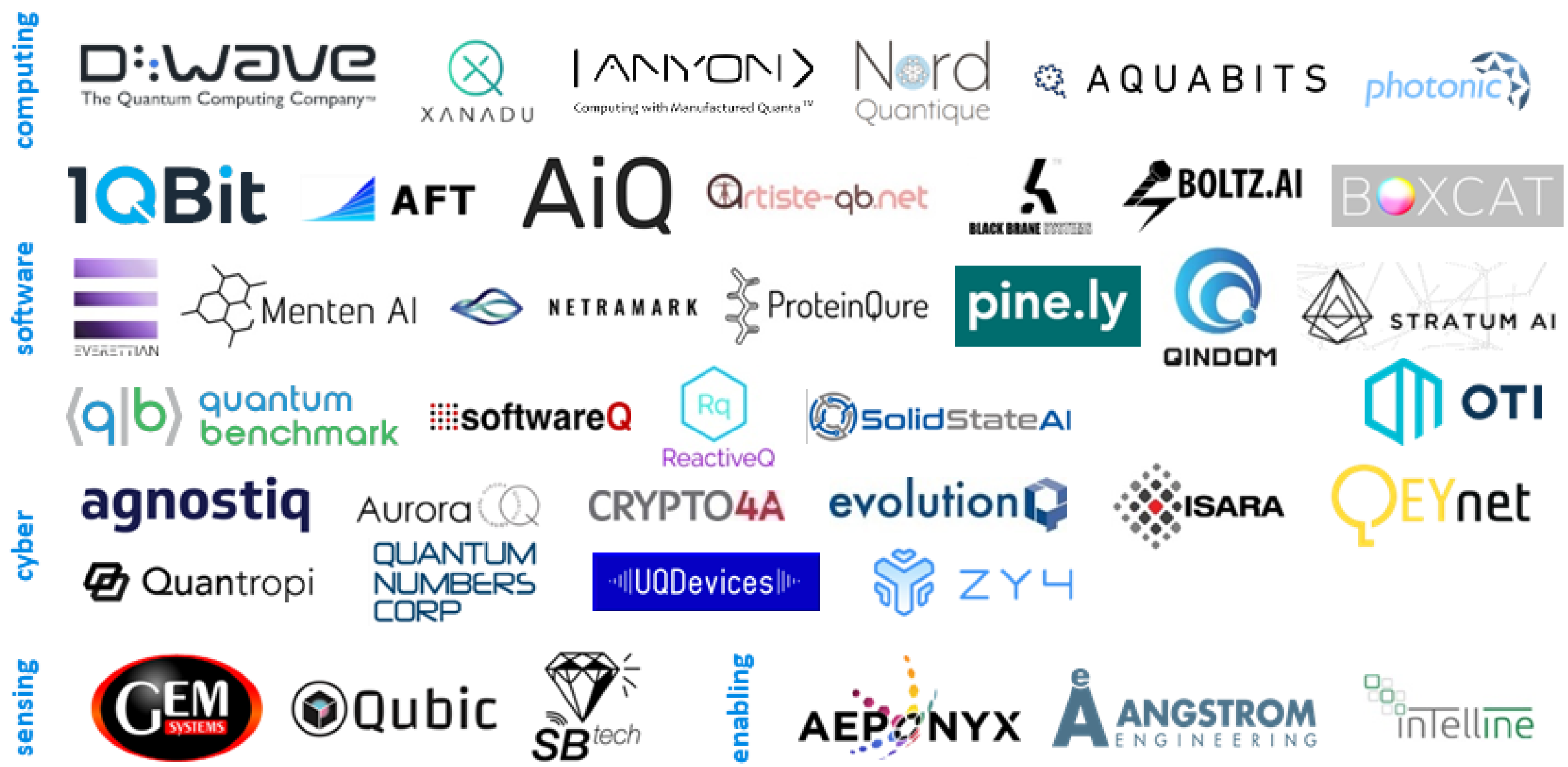

*Figure 852: the Canadian startup ecosystem by category. (cc) Olivier Ezratty, 2022.*

# Europe

Just making things clear, we're dealing here with geographical Europe, including European Union member states, the UK and Switzerland!

## United Kingdom

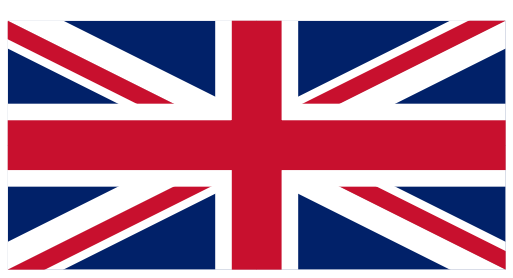

As with many continental European countries, the United Kingdom has contributed to many advances in quantum physics since the 18th century with precursors and founders, followed by a new generation of scientists in the second half of the 20th century.

Let's mention **Thomas Young** (1773-1829), **Ernest Rutherford** (1871-1937), **Joseph John Thomson** (1856-1940), **James Chadwick** (1891-1974), **Paul Dirac** (1902-1984), **Brian Josephson** (1940), **David Deutsch** (1953), **Andrew Steane** (1965) and even more recently the creators of the QML language, **Thorsten Altenkirch** and **Jonathan Grattage**.

### *Research*

In the UK, the main quantum research laboratories are located in the Universities throughout the country. All of these have one or several quantum physics laboratories. Their main specialties are found later in the UK quantum plan rollout with a lot of advanced photonics (Bristol, Oxford), electron spin (UCL), telecommunication and cryptography (nearly all of them), sensing (same) and the likes. See the list of UK Universities in the UK map in a forthcoming page.

### *Government funding*

At the instigation of the physicist Sir **Peter Knight** (1947), **UK** was the first large country to launch a quantum technology structured plan, the **UK National Quantum Technologies Programme**, announced in November 2013. It had an initial funding of £270M over five years[2725].

[2725] See The UK National Quantum Technologies Programme Current and Future Opportunities by Derek Gillespie, November 2014 (29 slides) and Delivering the National Strategy for Quantum Technologies (5 pages).

This represented a much larger amount of funding than for previous initiatives in innovative materials or robotics. The plan did not start from scratch. It was built on an existing ecosystem of university research laboratories in quantum physics.

The plan was and remains coordinated by the **EPSRC** (Engineering and Physical Sciences Research Council), a non-governmental organization funded and supervised by the government. The plan involves **Innovate UK** (basic research funding), the **Department for Business, Energy and Industrial Strategy**, the **NPL (**National Physical laboratory, where Peter Knight had been Chief Science Advisor, it is UK's metrology lab), the **CGHQ (**their NSA) and **dstl** (army research).

*Figure 853: the key public stakeholders of the UK quantum plan. (cc) Olivier Ezratty, 2021.*

In a fairly conventional way, the UK plan targeted all the usual quantum fields: computing, security, and sensing with a strong focus on medical imaging. Funding was based on thematic hubs bringing together universities and selected by call for projects (£124M), training, technology transfer and industrialization[2726].

From the outset, the plan showed a strong commitment to creating business and attracting private capital. The original plan was to move research into startups as quickly as possible.

Four quantum hubs cover the major fields of quantum technologies and bring together teams spread over the territory in some thirty universities. All the hubs managers are scientists, supplemented by a business development director and a board of 8 people including industry vendors CTOs.

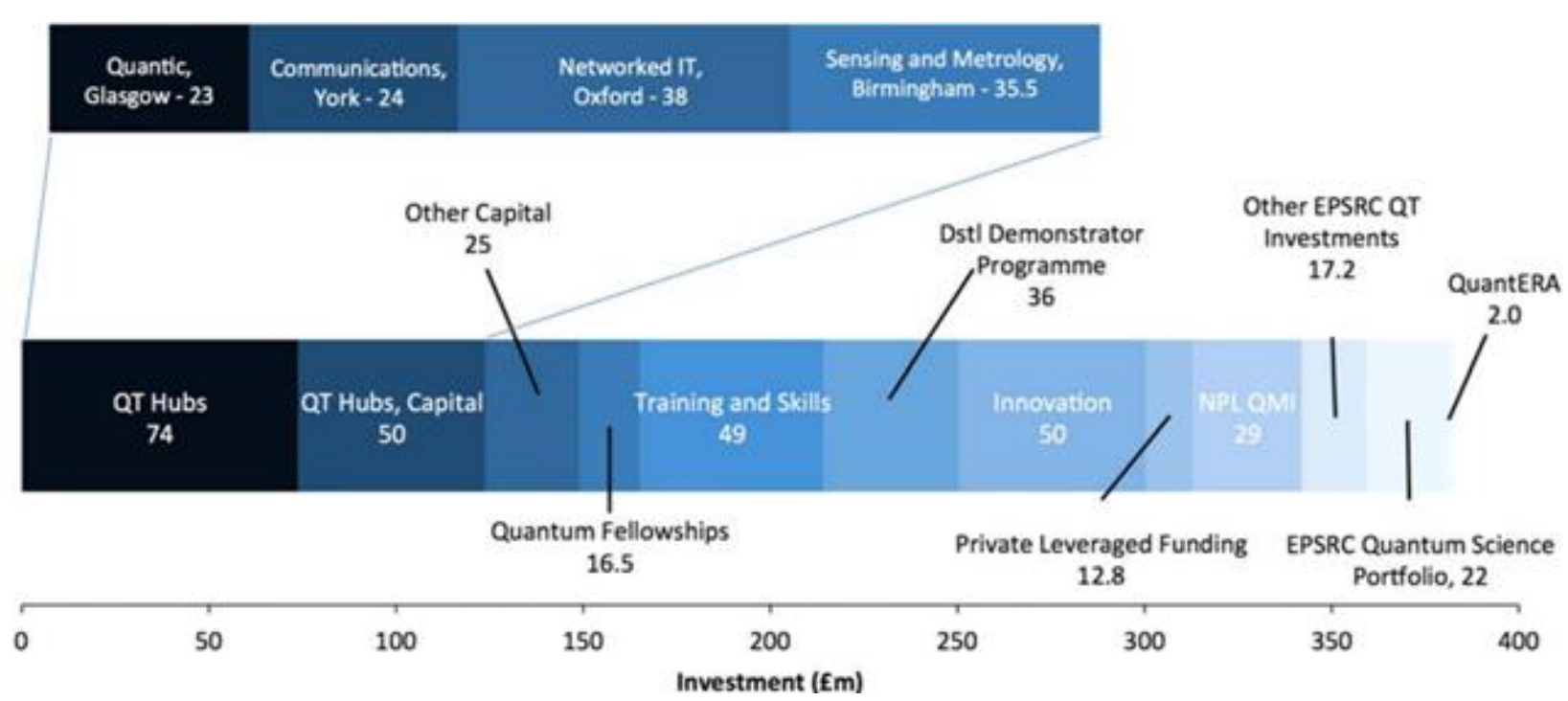

*Figure 854: UK's investments in quantum technologies in the first phase of their plan from 2014 to 2019.*

The **UK Quantum Technology Hub Sensors and Timing** covers sensing, including time measurement and involves the universities of Birmingham, Glasgow, Nottingham, Southampton, Strathclyde and Sussex.

The **Quantic** hub brings together the Universities of Glasgow, Bristol, Edinburgh, Heriot-Watt, Oxford and Strathclyde and focuses on quantum imaging. This gives us two hubs in the field of quantum sensing.

[2726] Diagram source: UK national quantum technology programme by Peter Knight and Ian Walmsley, October 2019 (10 pages).

The **Quantum Computing & Simulation Hub** brings together 17 universities and is led by Oxford University. It took over from the NQIT (Networked Quantum Information Technologies) hub in 2019. It focuses on computing and security issues[2727]. They are working on creating a network of trapped ions quantum computers.

The **Quantum Communications Hub** consolidates a dozen universities: Bristol, Cambridge, Glasgow, Heriot Watt, Kent, Oxford, Queen's Belfast, Sheffield, Strathclyde under the leadership of York University, companies such as Airbus, Toshiba, ID Quantique, Kets, and public agencies.

They are developing a quantum communication network between Bristol, Cambridge and Ipswich via the **UK National Dark Fibre Infrastructure Service** launched by the EPSRC (also linking Southampton and UCL in London)[2728].

This did not prevent the state security agency from expressing skepticism about the suitability of QKD in a four-page white paper published in April 2020[2729].

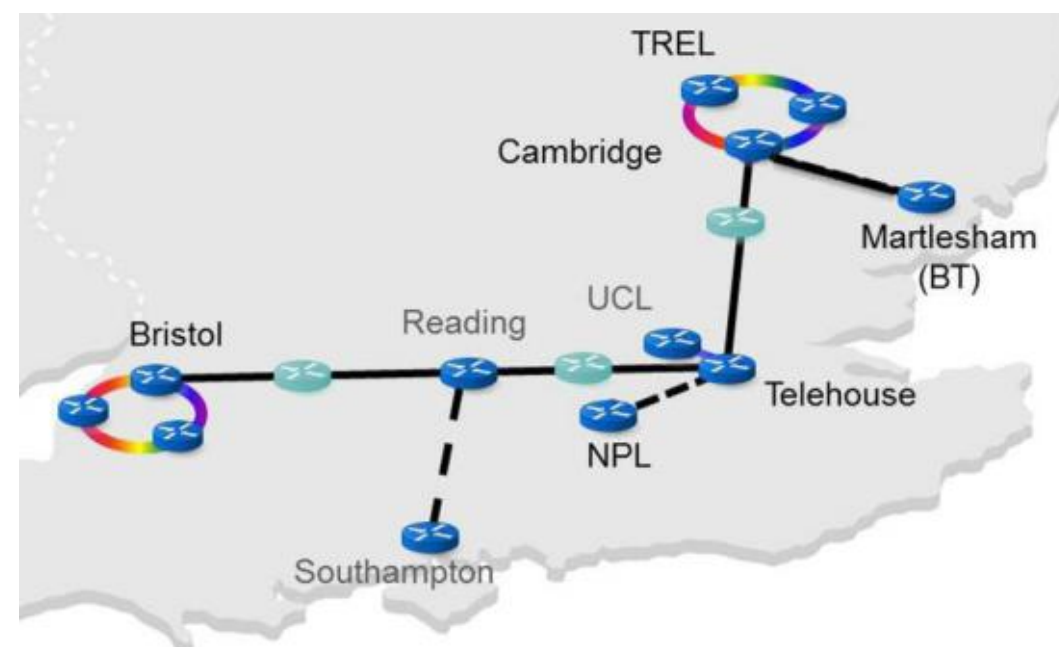

*Figure 855: the UK National Dark Fibre Infrastructure Service.*

These hubs are finally very multipolar, bringing together universities that are involved in several different hubs, according to the map on the next page[2730]. The United Kingdom has had a lot of ideas in managing this plan over the long-term.

In August 2022, as part of the NQTP, the EPSRC launched a new "Materials for Quantum Network" program led by Peter Haynes (Imperial College London) and Richard Curry (University of Manchester) to focus on quantum matter research. It seems that this topic will soon complement the usual computing/sensing/communications trio of all national quantum plans.

A progress report was published in 2015 by the EPSRC and Innovate UK followed by another interim report, the Quantum Age-Blackett review in 2016 investment launched in 2014 and extending the effort to the algorithmic and software part, in particular in liaison with the **Alan Turing Institute** and the **Heilbronn Institute for Mathematical Research** to propose case studies of computational problems to be solved[2731].

This was followed by a parliamentary report published in November 2018 which supported the continuation of the plan, the launch of a second phase of £350M over the period 2019-2024 and some fine tuning on the coordination between the different stakeholders (hubs, innovation centers, companies) [2732].

---

[2727] This includes the QuOpaL (Quantum Optimization and Machine Learning) initiative funded by Nokia and Lockheed Martin.

[2728] Diagram source: The Quantum Communications Hub, 2016 (11 slides).

[2729] See Quantum Security Technologies, NCSC, March 2020 (4 pages).

[2730] Map source: UK National Quantum Technologies Plan Strategic Intent, 2020 (38 pages). I added some of the large universities logos. See also UK national quantum technology programme by Peter Knight and Ian Walmsley, October 2019 (10 pages).

[2731] See The Quantum Age Technological Opportunities, 2016 (64 pages) and A roadmap for quantum technologies in the UK, 2015 (28 pages).

[2732] See Quantum technologies, House of Commons Science and Technology Committee, November 2018 (75 pages).

This led to the official announcement of Phase 2 in June 2019, following the recommendations of the House of Commons[2733]. With the expected private sector investments, the total of the two phases of the UK Quantum Plan was estimated at $1,227B.

Phase 2 funding renewed funding for hubs (£94M over 5 years), industrialization projects (£153M from the Industrial Strategy Challenge Fund, over 6 years[2734]), training (£25M over 5 years[2735]). It added the launch of the **NQCC (**National Quantum Computing Centre**)** for the development of quantum computing solutions, with £93M over 5 years[2736].

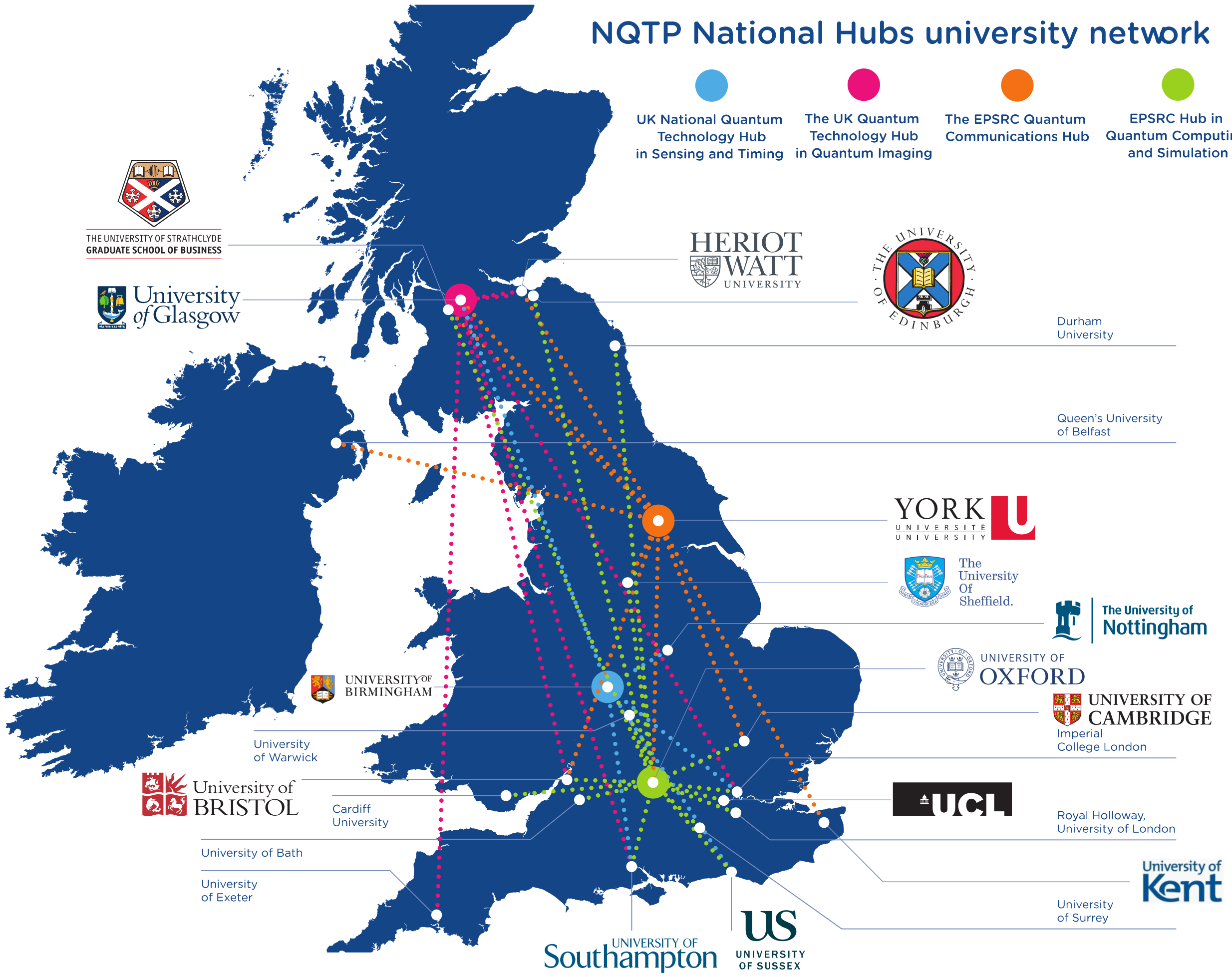

*Figure 856: the UK universities map. Source: UKRI and logos added by Olivier Ezratty, 2021.*

The first "UK" quantum computer was to be built by Rigetti (US). How can this be? It's linked to Rigetti having acquired a local startup, QxBranch and to its various connections with the local ecosystem and universities. But Oxford Quantum Circuits announced the launch of its cloud based superconducting qubits based computer in July 2021. All in all, the UK government has invested £100M per year in quantum technologies since 2014.

---

[2733] See UK government invests $194M to commercialize quantum computing by Frederic Lardinois.

[2734] The Industrial Strategy Challenge Fund (ISCF) was a multi-domains initiative of £2.6B backed by £3B of private investments, created to invest in challenges having a strong economical and societal impact. A dedicated Commercialising Quantum Technologies Challenge was then launched in two stages, first in 2018 with £20M and second with £153M completed by £205M from the private sector, in July 2019. To date, in 2021, over 40 such projects were funded. As of late 2020, over £200M were invested in UK startups.

[2735] Doctoral training in quantum technologies is not managed in the hubs but in doctoral centers such as the Quantum Engineering Centre for Doctoral Training in Bristol.

[2736] See Establishing the National Quantum Computing Centre (NQCC), August 2019 (64 slides). The construction started in September 2021.

The bulk of Phase 2 is the NQCC, which is led by **UKRI**, the **EPSRC** and the **STFC** (Science and Technologies Facilities Council), a government agency that conducts research in physics and astronomy and manages the country's major scientific instruments (particle accelerators, lasers, space engineering, etc.)[2737]. This center will produce NISQ and then LSQ computing demonstrators, develop quantum algorithms and software and their uses, and build a community of users around them. The center should open by the summer of 2021 and become fully operational in 2022. It will set up a NISQ machine that should be operational by 2025. In 2020, the preferred technologies were superconducting and ion-trapped qubits.

In May 2022, NQCC launched its SparQ Applications Discovery Programme and a collaboration with OQC. SparQ is a sort of directory aimed at UK-based companies and researchers who are looking for case studies of quantum computing[2738].

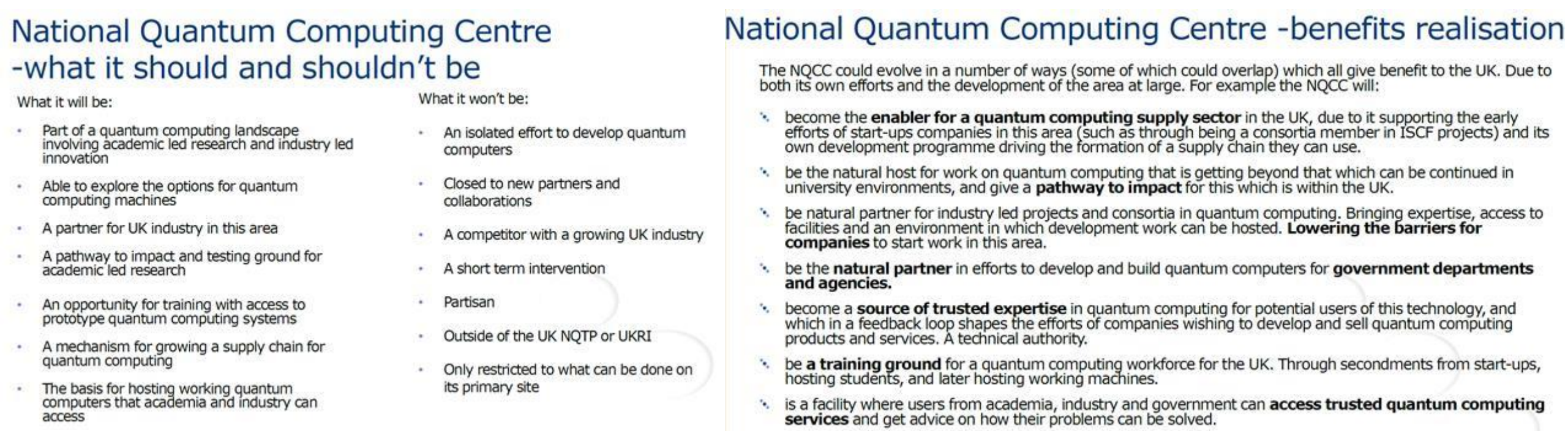

*Figure 857: NQCC positioning. Source: NQCC. 2021.*

The UK had recovered approximately 14% of the budgets for the first wave of European Quantum Flagship projects by October 2018. Despite the Brexit, the country will continue to benefit from it, as the collaboration with Europe on research survives the Brexit. For example, John Morton's UCL laboratory is part of the flagship project QLSI on silicon qubits driven by CEA-Leti and awarded in March 2020.

In November 2021, the UK signed a partnership agreement with the USA[2739], the first of a long list of USA bilateral partnerships later signed with Australia, all Nordic countries and Switzerland.

### *Quantum industry*

On the entrepreneurial side, around 40 quantum technologies startups were launched in the UK with a good balance by category. It is the third country in the world in terms of the number of startups, behind the USA and Canada. In October 2022 was created **UKQuantum**, a quantum industry association with Kets, Orca Computing, Arqit, Nu Quantum, Riverlane and Oxford Instruments among its funding members.

The intellectual property management company **IpGroup,** launched in August 2020 a £12M fund to fund startups, these being selected by the independent agency **UKRI**. Projects funding range from £125K to £2M. Let's also mention the **Quantum Technology Enterprise Centre** from the University of Bristol which was a sort of startups incubator and training program for quantum startup founders. The QTEC incubation program offered a 12-months salaried fellowship to quantum scientists during the build-up of their startup and business skills training.

---

[2737] UKRI (UK Research and Innovation) is an autonomous non-governmental organization created in April 2018 with an annual budget of £7B and consolidates seven former research councils including the EPSRC and STFC, Innovate UK and Research England.

[2738] See Industry engagement prepares UK for quantum transformation, PhysicsWorld, November 2021 and Early adopters position themselves for quantum advantage, June 2022.

[2739] See Cooperation in Quantum Information Sciences and Technologies Joint Statement, November 2021.

Since 2016, QTEC helped the creation of 31 startups including KETS, QLM, Nu Quantum, Quantum Dice and Vector Photonics. The program funding ended in 2021 and QTEC is looking for funding to launch a new “cohort” of quantum entrepreneurs.

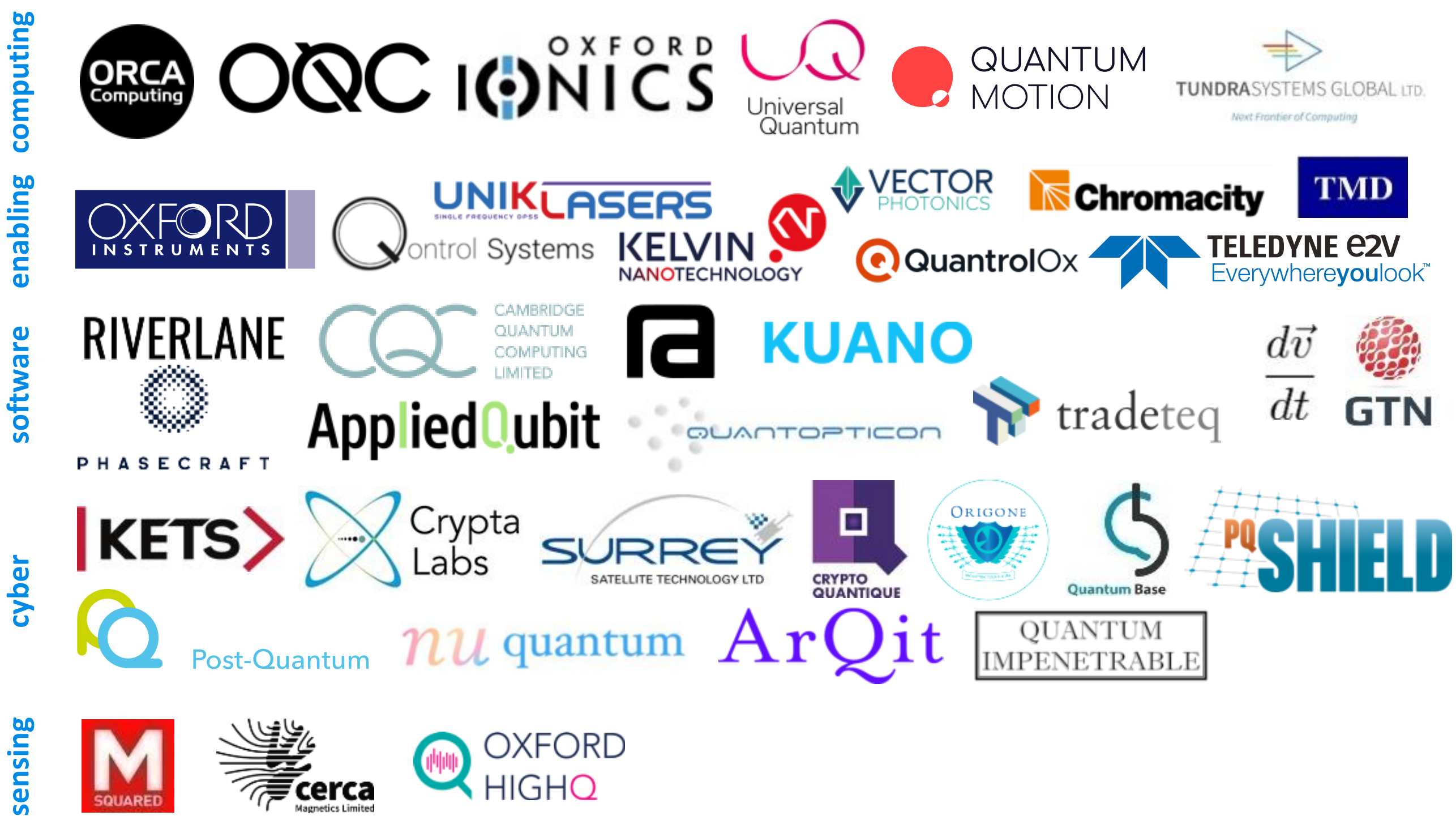

*Figure 858: the UK startup scene is the most active in Europe (the old Europe, with them, before Brexit...). (cc) Olivier Ezratty, 2022.*

Notable players are **Oxford Instruments** (cryogenics), **Oxford Quantum Circuits** (superconducting qubits[2740]), **Quantum Motion Technologies** (silicon qubits), **Cambridge Quantum Computing** (operating system, software, services, which merged with Honeywell Quantum Systems in 2021), **TundraSystems** (photonic qubits), **Orca Computing** (photon qubits) and **River Lane Research** (software). On the other hand, no major company in the country seems to be particularly invested in quantum computing, except perhaps in telecommunications.

### Germany

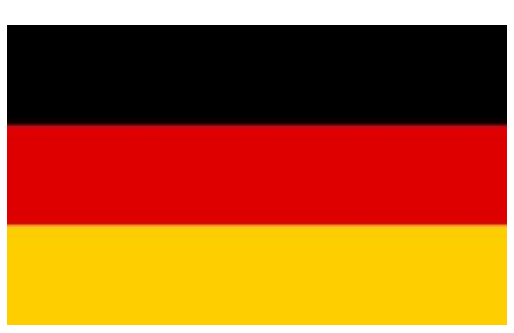

Germany is a land of dense basic research in quantum technologies. It builds on a strong history of the many German founders of quantum physics with **Max Planck**, **Albert Einstein**, **Werner Heisenberg** and many others[2741]. It also has a strong ecosystem of industry vendors, particularly in quantum enabling technologies.

#### *Research*

The main research organizations and laboratories involved in quantum technologies are:

**Max Planck Institute for Quantum Optics** (MPQ), based in Munich, is one of the 84 MPIs and their 24,000 employees. It was the home of Klaus von Klitzing who discovered the quantum Hall effect in 1980 and got the Nobel prize in physics in 1985.

---

[2740] Oxford Quantum Circuit obtained funding from Innovate UK in April 2020 in a consortium of four companies and two universities. See Oxford Quantum Circuits-led consortium wins Grant to Boost Quantum Technologies in the UK by Quantum Analyst, April 2020.

[2741] See The Innovation Potential of Second-generation Quantum Technologies by the National Academy of Science and Technology, July 2020 (96 pages) which contains a list of key people in German quantum research at the beginning.

They specialize in cold atom-based qubits in particular. This MPI is associated with the International Max Planck Research School also based in Munich. Two other MPIs are dedicated to information technology, but do not seem to be invested in quantum.

**Munich Center for Quantum Science and Technology** (MCQST) in Munich was launched in 2019 and brings together Munich's quantum research centers: the MPQ, the Walther-Meißner-Institute for Low Temperature Research (WMI) and the city's two leading scientific universities: Ludwig-Maximilians-Universität München and Technical University of Munich (TUM). It covers all quantum technologies (simulation, computing, communication, sensors). The whole with a budget of 31M€ over five years and about 55 permanent researchers. In February 2022, the Bavaria region announced an additional funding of 300M€ for the Munich quantum valley supplemented by 80M€ of federal funding. Among other things, the funding will help build the upcoming Center for Quantum Computing and Quantum Technologies (ZQQ), a center that will provide access to superconducting, ionic and atomic qubits quantum computers. Meanwhile, in January a D-Wave Advantage was installed at Julich in JUNIQ (Jülich UNified Infrastructure for Quantum computing) that was created in 2019.

**Fraunhofer Institutes** for Applied and Partnership Research with its 72 institutes and 26,600 people. They comprise several institutes specialized in quantum physics: the **IAF** (Institute for Applied Solid State Physics) in Freiburg, the **IOF** (Institute for Applied Optics and Precision Engineering) in Jena, the **ILT** (Institut for Laser Technology) in Aachen, **FOKUS** (Open Communication Systems) in Berlin, **SCAI** and **IAIS** (quantum machine learning) in Sankt Augustin and **ITWM** (quantum HPC) in Kaiserslautern. On top of that must be accounted cleanrooms from **IPM** in Freiburg and **IPMS** near Dresden.  In March 2022 was launched EIN Quantum NRW, a competence network for photonics-based quantum technologies in the North Rhine-Westphalia (NRW) lander, involving the Fraunhofer Institutes **FHR** (High Frequency Physics and Radar) and **IAIS** (Intelligent Analysis & Information).

**Helmholtz Association** groups 18 Research Centers that conduct basic research in response to major societal challenges, with a total of 40,000 people. It includes the Quantum Laboratory of the **Jülich Forschungszentrum** (*aka* Jülich FZ or Jülich Research Center) located between Aachen and Cologne and headed by Kristel Michielsen[2742], where Tommaso Calarco, who coordinates the European Quantum Flagship, also works. He is associated with the University of Aachen in the **JARA Institute Quantum Information** (IQI). The Helmholtz Network also includes the **Institute of Photonics and Quantum Electronics at the** Karlsruhe Institute of Technology (KIT).

**Leibniz Association** with its community of 96 centers conducting basic research includes the Institute for Solid State and Materials Research (IFW) in Dresden, Germany, which focuses on superconductivity and magnetism, the Institute of Photonic Technology (IPHT) in Jena, Germany, the Max-Born-Institute for Nonlinear Optics and Short Pulse Spectroscopy (MBI) in Berlin, Germany, and the Paul Drude Institute for Solid State Electronics (PDI) in Berlin, Germany.

**Institute for Complex Quantum Systems** at the University of Ulm between Stuttgart and Munich.

**PTB** is the federal office of sensing, which is obviously investing on quantum sensing like the NIST.

**BSI** is the federal office for information technology security[2743].

At last, let’s mention here the **Quantum Alliance** which regroups the German clusters of excellence and research centers working in quantum science and technology.

[2742] Jülich Forschungszentrum started in 1956 in nuclear research. It also houses a number of supercomputers, such as the CEA's DAM at Bruyères-le-Châtel in France or the various US DoE research centers across the USA.

[2743] In Germany, the federal agency that protects information systems, which is the counterpart of the French ANSSI, published in May 2018 the report Entwicklungsstand Quantencomputer (*State of the art of quantum computing),* which provided an update on quantum computing, focusing in particular on cybersecurity issues (231 pages, in English). This was a very good overview of global quantum computing research. It provided a surprisingly accurate inventory of efforts in the field, particularly in US public research. But things have changed a bit since 2018.

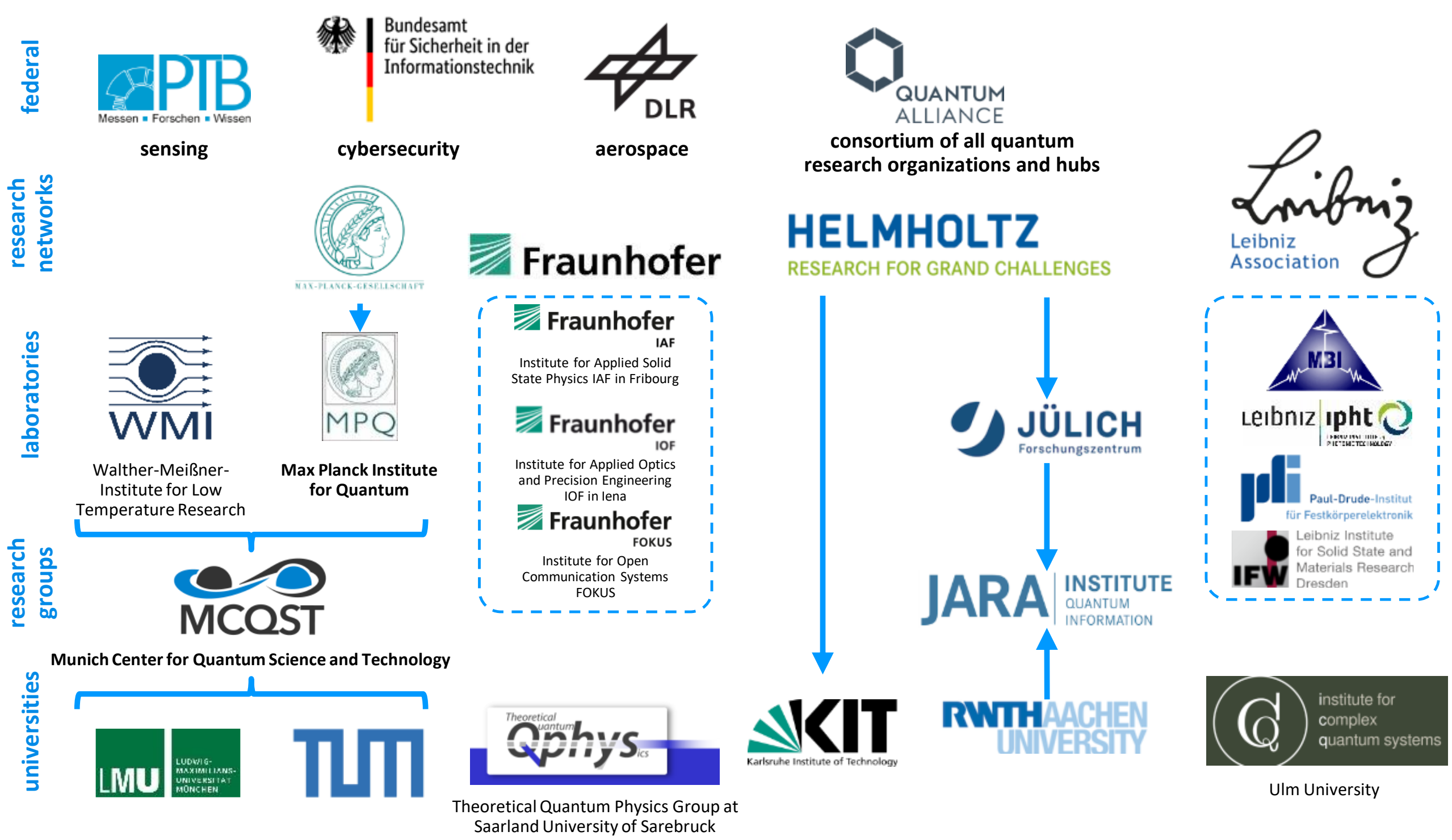

*Figure 859: understanding the research ecosystem in Germany. (cc) Olivier Ezratty, 2022.*

### *Government funding*

In September 2018, the German Federal Research Ministry announced €650M in funding for quantum technologies over four years (2018 to 2022)[2744]. Like all such plans, it funds projects in quantum computing, quantum communication and quantum metrology. In September 2019, IBM announced that it would join this plan and install a quantum computer in Germany addressing researchers and cloud usages. It is not certain that this is the best approach to develop a German and European quantum industry, at least on the hardware side. The computer was actually launched early in 2021 in the Stuttgart region in an IBM facility[2745].

In June 2020, the German government more than doubled its efforts by announcing a seemingly incremental €2B in funding for its quantum plan, including investment in two quantum computers[2746]. The $2B included the initial 650M€ of the 2018 plan. The German government put in place a scientific and industry experts board of 16 members to propose a roadmap and funding allocation with a Joint presidency of a scientist (Stefan Filipp from TU Munich) and an industry member (Peter Leibinger from Trumpf). It made some proposals in December 2020 including creating an independent coordination body, Deutschen Quantengemeinschaft (DQG). In January 2021, the BMWi disagreed with some of these proposals, estimating that too much funding was directed to fundamental research at the expense of startups.

In May 2021, the plan was split in two parts with 1,1B€ managed by the Federal Ministry of Education and Research (BMBF) and 878M€ by the Federal Ministry of Economic Affairs and Energy (BMWi), focused on applications developments.

---

[2744] See German Government Allocates €650M for quantum technologies, the German government's announcement (in German) and the plan itself (51 pages).

[2745] It lead to misleading information regarding a supposed investment of the German government in IBM of $717M corresponding to these €650M in German government to invest $717M in IBM quantum computing efforts, WRAL Techwire, September 2019. At best, only a small share of these 650M€ were dedicated to cofund the IBM initiative in Germany. Looking at what was done in 2022 in Canada, you can guess that the German government participation was around 40-60M€.

[2746] See Germany: 2 Billion euros for quantum technology, June 2020.

One key showcased goal for this plan is to build two national quantum computers with 24, then 100 and later 500 functional (physical) qubits. DLR (Germany's Aerospace Center) was to receive the bulk of this funding (740M€) to work with small, mid-sized and large companies and create two related consortia.

The Federal government is not the only public body funding quantum research. Landers added about 500M€ including the 300M€ from the Bavarian region (probably with half of it coming from the Lander and the rest from the industry). How is that possible? German Landers (regions) have a very large budget independent from the Federal government (in a 42%/58% ratio). Bavaria being a large Lander, they have the means to invest significantly on research[2747].

For example, the lander cofounded two quantum research projects, NeQuS (quantum networks between quantum computers) and IQ-Sense (quantum sensing) for a total of 3.5M€ in August 2022.

Like with each country, the German quantum plan covers quantum computing, communications and sensing. The 2B€ effort is planning many projects, particularly in quantum computing research. Let's list some identified projects.

**GeQCoS** (German Quantum Computer based on Superconducting Qubits) with 14,3M€ BMBF funding, involving Fraunhofer Fribourg and Infineon). It was launched in February 2021 and will seemingly use German originated quantum technologies. The ambition is modest with a goal of 50 qubits.

**QSolid** is another superconducting qubit computer that would be tied to a HPC at the Jülich Supercomputing Centre, including optimized firmware and software. It will start with a 10-qubit system. The project budget is 76.3M€ with 89.8% being funded by the BMBF over five years starting in January 2022. On top of Julich, it involves Fraunhofer IPMS and Fraunhofer IZM-ASSID, Karlsruhe Institute of Technology (KIT), Leibniz IPHT in Jena and the PTB.

**DAQC** is a project launched in February 2021 which got 12.4M€ from BMBF. It is coordinated by IQM Germany and involves Jülich, Infineon and ParityQC from Austria as well as the Leibniz Computing Center and Free University of Berlin. It will create a digital-analog superconducting qubits system using IQM's architecture. This project is related to the Quantum Flagship OpenSuperQ project.

**MANIQU** is a project running from 2021 to 2023 related to the usage of NISQ hardware to undertake quantum simulations to simulate new materials. The participants are HQS, Bosch, BASF, the Friedrich-Alexander University in Erlangen, and the Heinrich-Heine University in Düsseldorf.

**Q-Exa** is yet another superconducting qubits project announced in November 2021, involving IQM. Their systems will be integrated with HPCs. The project funding is of 45.3M€. It seems to be a follow-up project to DAQC.

**PhotonQ** is a photonic quantum computer project launched with a funding of 16M€ and led by the University of Stuttgart with participations from the Universities of Würzburg, Mainz and Ulm, TUM Munich, the Institute of Microelectronics Stuttgart and Vanguard Automation GmbH. The four-year project first goal is to create a demonstrator of 8 qubits using MBQC and deterministic photon sources, silicon photonics circuits and new single-photon detectors.

**PhoQuant** is another photonic quantum computer project, led by Q.ANT, a subsidiary of Trumpf, with the notable participation of Christine Silberhorn's Paderborn University lab and many others (Universities of Münster, Jena, Ulm, Humboldt Berlin, Fraunhofer IOF and IPM, HQS, Swabian Instruments, TEM Messtechnik, ficonTEC Service and MenloSystems).

[2747] A false number on German public investment circulated with a hefty 4,456M€ from Interference Advisors. This report is amazingly entirely wrong, making double bookings of the 2018 and 2020 announcements, using unsafe press reference ($717M allocated to IBM's effort...) and double-accounting subsequent quantum projects which are already part of the 2B€ 2020 announcement.

**ATIQ** is a trapped ions computing project. The German government is funding 81,1% of the 44,5M€ project that will last from 2021 to 2026. Participants include Toptica (lasers). The project is coordinated by the Leibniz University Hannover with participations from Johannes Gutenberg University Mainz, University of Siegen, TU Braunschweig, RWTH Aachen, Physikalisch-Technische Bundesanstalt, Fraunhofer-Gesellschaft, AMO, AKKA Industry Consulting, Black Semiconductor, eleQtron, FiberBridge Photonics, Infineon, JoS QUANTUM, LPKF Laser & Electronics, ParityQC, QUARTIQ, Qubig, AQT, Boehringer Ingelheim, Covestro, DLR-SI, Volkswagen and QUDORA Technologies.

**QUASAR** is a semiconductor-based project using shuttling electrons with a QuBus, a quantum bus to transport electrons and their quantum information over distances of 10 µm. The partners are Infineon, HQS, Fraunhofer (IAF, IPMS), Leibnitz Association (IHP, IKZ) and the Universities of Regensburg and Konstanz. The project will run until 2025 to create 25 coupled qubits.

The resulting computer is to be deployed at JUNIQ. Jülich is also participating to the European Flagship QLSI project driven by CEA-Leti in France. QUASAR got a 7.5M€ funding from BMBF[2748].

**QuaST** is an enabling technologies project (Quantum-enabling Services and Tools for Industrial Applications) that will develop high-level libraries automatically decomposing and optimizes a solution into classical and quantum parts. The project is run by the Fraunhofer Institute for Cognitive Systems IKS with other Fraunhofer Institutes (AISEC, IIS, IISB), the Leibniz Supercomputing Center, and the TUM, plus DATEV, Infineon, IQM and ParityQC. The project sponsor is German Aerospace Center (DLR). The project will last 4 years and got a funding if 7.7M€.

**QLindA** is a quantum machine learning project led by Siemens with participations from Fraunhofer IIS, IQM and others.

We also have three enabling technologies projects: **QuMIC** (Qubits Control by Microwave Integrated Circuits, 6,3M€, 2021-2024), **qBriqs** (2M€, 2021-2024, compact cryogenic connectors, qubit readouts TWPA and HEMT amplifiers, filters and attenuators, DACs and ADCs and DC flux current generators) and **HIQuP** (2021-2024, 2,2M€, superconducting and cryogenic qubit control electronic circuits).

Germany also launched the creation of two QKD-based telecommunications networks, both funded by BMBF:

**QuNET** (165M€) which uses a standard QKD associated with terrestrial and satellite links. The project involves several Fraunhofer Institutes including the Heinrich Hertz Institute (HHI), the Max-Planck Institute for the Physics of Light and the German Aerospace Center (DLR)[2749]. The project launched in November 2019 was scheduled to last seven years and aims to create a communications protection infrastructure for the German government. This should lead to the creation of a secure European network. The private sector is also involved with Deutsche Telekom, ADVA Optical Networking and Tesat-Spacecom. Test sites will be implanted in Bavaria, Saxony and Thuringia.

**Q.Link.X** (14.8M€) for the creation of a terrestrial network in optical fiber and QKD based on quantum repeaters, managed by the Fraunhofer HHI[2750].

Germany leads or participates to various European Flagship programs: **MetaboliQs** (NV center based medical imaging), **UNIQORN** (photon qubits chipsets), **S2QUIP** (hybrid photonic chipsets), **QRANGE** (QRNGs).

---

[2748] See Quanten-Shuttle zum Quantenprozessor "Made in Germany" gestartet, Jülich, February 2021.

[2749] See Germany's QuNET Receives €165 Million To Establish Quantum Communications Infrastructure, 2019, German ministry and research sector join forces to launch major quantum communications initiative, May 2019 and German Aerospace Center In QuNET Working On Satellite-Based Quantum Communication, November 2019.

[2750] See Germany splashes further €15m in quantum networks R&D project, October 2018.

At last, the German national plan is funding three other initiatives associating research labs and industry vendors: **BrainQSens** (medical imaging with NV centers, 2,8M€), **Opticlock** (compact optical clock to synchronize communication networks, 6M€) and **QUBE** (space QKD with Cube-Sat, 3,12M€).

***Quantum industry***

On the private sector side, Germany has a various set of quantum startups including **Avanetix** (hybrid algorithms), **InfiniQuant** (CV-QKD cryptography), **PicoQuant (**photon counters), **Kiutra** (magnetic cryogenics), **HQS Quantum Simulations (**algorithms), **JoS Quantum** (software in finance), **QuantiCor Security, QuBalt (**both in post-quantum cryptography) and **QuTools** (sensing).

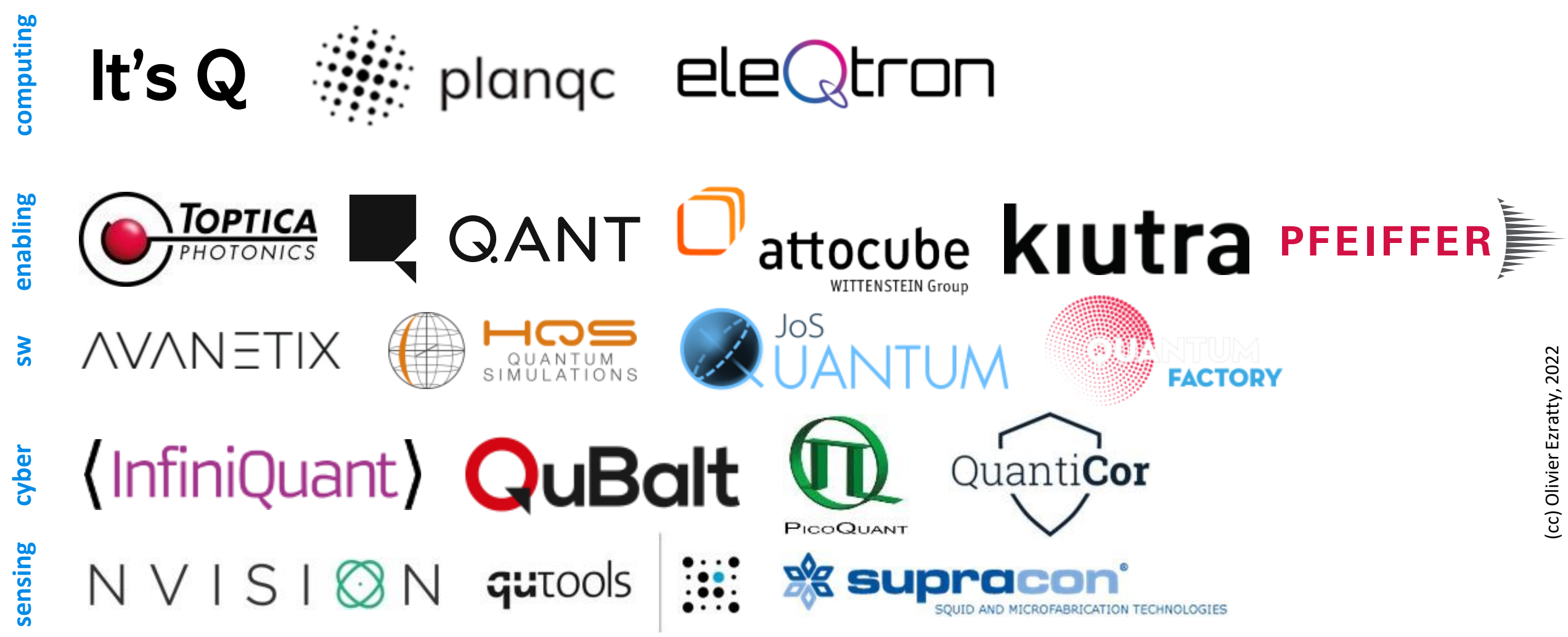

*Figure 860: the German quantum industry. (cc) Olivier Ezratty, 2022.*

Only a few quantum computing hardware startups have been created and recently like **It'sQ** (photonics) and **Planq** (cold atoms).

Many of the country's major industrial companies are also interested in quantum applications, particularly in chemistry (**BASF**), health (**Merck),** telecommunications (**Deutsche Telekom)**, components and automotive (**Bosch**, **Daimler**).

**PlanQK** (Platform and Ecosystem for Quantum-Assisted Artificial Intelligence) is a project to build a marketplace of quantum assisted artificial intelligence components, at first, quantum inspired algorithms. It gathers scientists from various universities (Stuttgart, Berlin, Munich) on top of Accenture, HQS, Deutsche Bundesbahn, Deutsche Telekom and other industries. It is supported by BMWi with a total funding of €19M.

In June 2021, ten German companies created **QUTAC** (Quantum Technology and Application Consortium) to develop quantum computing usable industrial applications in the technology, chemical and pharmaceutical, insurance and automotive industries.

The consortium was launched by BASF, BMW Group, Boehringer Ingelheim, Bosch, Infineon, Merck, Munich Re, SAP, Siemens, and Volkswagen. One of its goals is to create a cross-industry application portfolio.

There is another thing in Germany called **QuCUN** (Quantum Computing User Network) which also comprises SAP and BASF as partners. It is supported by the German Federal Ministry of Education and Research (BMBF).

The QuCUN platform launched in 2022 is "*designed to give potential quantum computing users – from SMEs to industrial giants – a central point of entry*". Does it mean that QUTAC is peripheral?

Let’s also mention **PushQuantum**, a student initiative born in Munich that organizes lectures, workshops and entrepreneurship labs for wannabee quantum entrepreneurs.

**Austria**

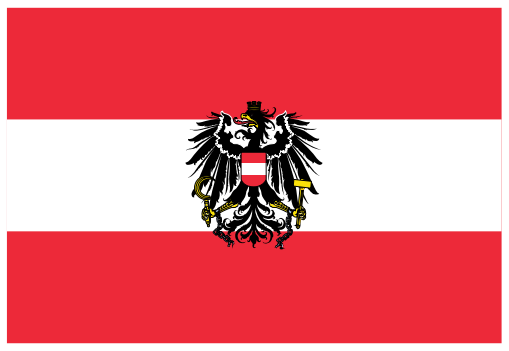

Austria**'s** investment in quantum computing is concentrated in the **IQOQI**, the Institute for Quantenoptik und Quanteninformation in Innsbruck and Vienna. It focuses in particular on the design of trapped ions qubits. This led to the start-up **Alpine Quantum Technologies**, founded by Rainer Blatt and Thomas Monz of the IQOQI.

It has received €12.3M in public funding and competes with the **IonQ** (USA), which is positioned in the same niche of ion-trapped qubits, as well as **Quantinuum**. One other notable Austrian startup is **ParityQC** which develops ParityOS and a related architecture to codevelop quantum hardware and software platforms.

The Vienna Center for Quantum Science and Technology (**VQC)** is a partnership between the University of Vienna, Vienna University of Technology and the Austrian Academy of Sciences. It brings together a critical mass of about 20 quantum physics research laboratories.

Austria is also invested in quantum cryptography and is associated with China, with whom it has conducted experiments in sending quantum keys via the Micius satellite to set up secure video communication. **IQOQI** is collaborating with the **Grenoble University Space Center** (CSUG) in the development of a CubeSat-type quantum key relay satellite, similar to the one in Singapore, in the **Nanobob** project (presentation, 13 slides).

The Austrian government announced a formal quantum plan in June 2021 with 107M€ over 2022-2026 covering research and quantum technology developments. Quantum Austria is part of the Austrian 10 years Research Technology and Innovation strategy 2030 launched in 2020, in a country that spends overs 3.18% of its GDP in R&D. Part of the funding comes from the European Resilience and Recovery Facility (NextGenerationEU). The plan is managed with RFPs from the Austrian Research Promotion Agency (FFG) and the Austrian Science Fund (FWF).

**France**

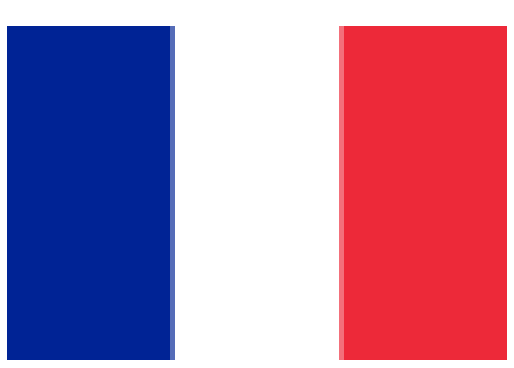

France has a good breadth of research and industry activities in quantum technologies. Let’s first mention its greatest scientists with **Henri Poincaré** (1854-1912), **Louis de Broglie** (1892-1987, Nobel prize in physics in 1929), **Alfred Kastler** (1902-1984, Nobel prize in physics in 1966), and **Claude Cohen-Tannoudji** (1934, Nobel prize in physics in 1997).

**Serge Haroche** (1944, Nobel Prize in Physics in 2012) is a pioneer in cavity quantum electrodynamics and on the interaction between photons in a superconducting cavity and Rydberg atoms passing through the cavity). Of course, we can add **Alain Aspect** (1946), who invalidated Bell's inequalities in 1982 and verified the principle of non-locality of entangled photons, a cornerstone of the second quantum revolution. Other domains worth mentioning are in quantum photonics with **Pascale Senellart**, neutral atoms physics with **Jean Dalibard** and **Antoine Browaeys**, silicon spin qubits with **Maud Vinet**, **Tristan Meunier** and **Silvano de Franceschi**, quantum cryptography and telecommunications with **Philippe Grangier**, **Frédéric Grosshans** and **Eleni Diamanti** and PQC with various cryptographers like **Damien Stehlé**.

***Research***

Public research is organized around three national research organizations: **CNRS**, **CEA** and **Inria**. The first is involved in fundamental research in physics, mathematics, and algorithms. The second also does fundamental research in physics, particularly on superconducting qubits, and applied research on electron spins qubits as well as on photonics. At last, Inria is doing research in computer science, and for quantum technologies, on quantum error correction, cryptography, and quantum algorithms. Many laboratories are joint research units ("Unités Mixtes de Recherche" in French) between Universities, these national organizations and sometimes industry vendors like Thales.

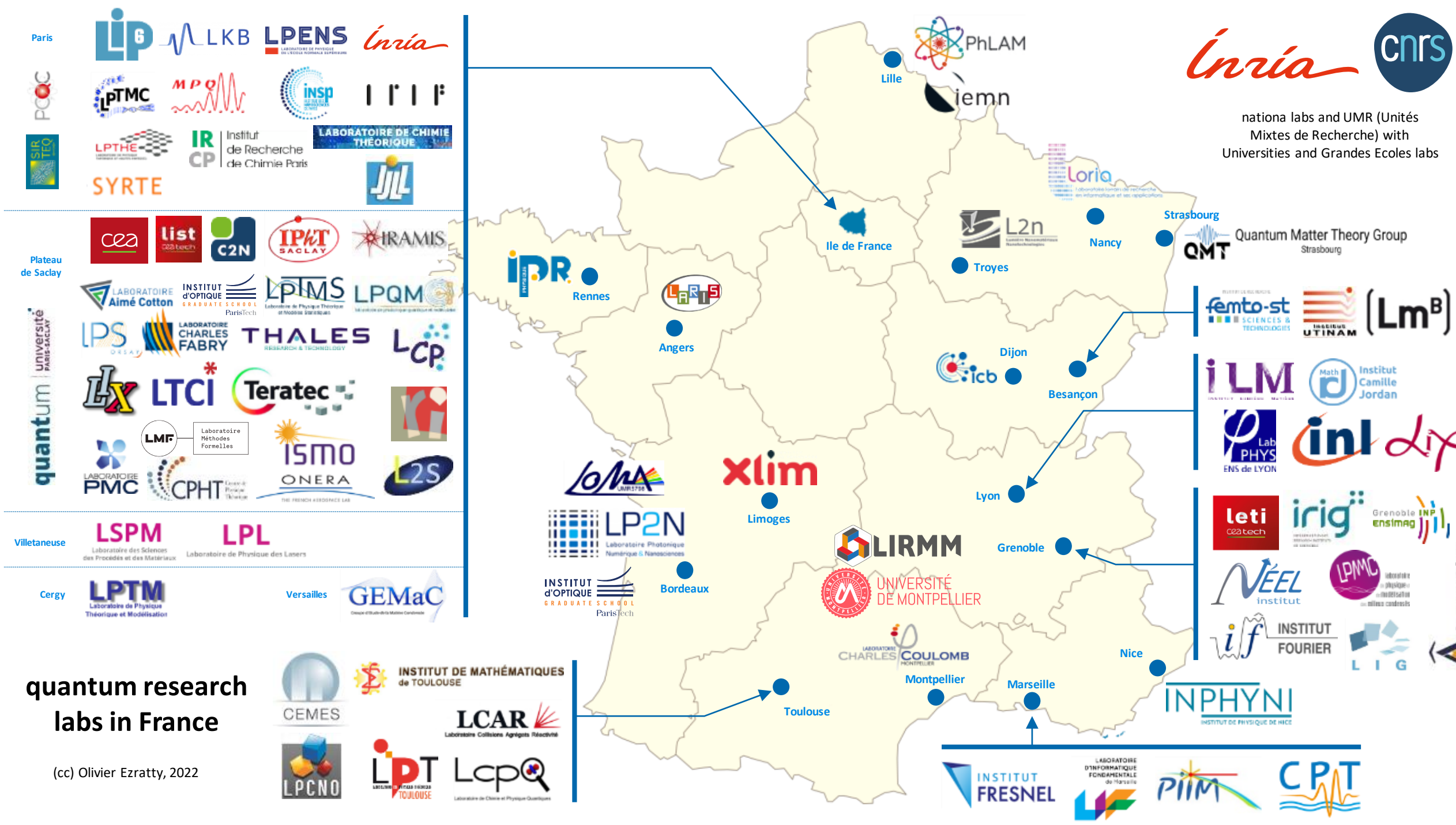

*Figure 861: a beautiful map of France's research labs. (cc) Olivier Ezratty, 2022.*

These research laboratories are mainly located in Ile de France and in Grenoble, but other regional locations are active such as Toulouse, Montpellier, Marseille, Lyon, Bordeaux, Besançon and Lille[2751]. Like many large countries, French laboratories are exploring many qubit tracks: superconducting, cold atoms, electron spins, photons and topological matter.

Public sector researchers get projects funding by answering various country and European RFPs[2752]. Of the more than 20 quantum startups in France 2021, 7 are from CNRS, two from Inria, two from ENS and one from CEA.

### *Ile de France*

Ile de France is home to a good half of the country's research laboratories devoted to quantum technologies. Let's start with the laboratories that are located within Paris.

[2751] For this purpose, I consulted the websites of these laboratories and the fields of research they present, plus, when they were easy to find, the scientific publications of the researchers of these laboratories.

[2752] Some obtain ERC Grants (European Research Council): Synergy Grants for a few handfuls of teams (up to €14M over 6 years), and more often Starting (young researchers, up to €1.5M), Consolidators (experienced researchers, up to €2M) and Advanced (emeritus researchers, up to €2.5M spread over 5 years). Then European FET funding, funding via the European Quantum Flagship, or finally through various calls for projects at the national level (ANR).

**Inria**'s efforts in the Paris region are concentrated in the Quantic (Quantum Information Circuits) team of Pierre Rouchon, Mazyar Mirrahimi, Zaki Leghtas and Alain Sarlette, which is a joint venture between the CNRS, ENS and the Ecole des Mines de Paris.

They work on mathematical models of superconducting qubits, on quantum error correction (including cat-qubits), on proof of superiority of quantum algorithms and on cryptographic issues[2753]. The Cosmiq team led by Anne Canteaut, works on cryptographic algorithms, and David Pointcheval's Cascade team, works in cryptography and PQC. Inria also jointly runs many other teams with various labs from CNRS.

Other Inria teams are dedicated to quantum science and technology: IQA (LTCI, Saclay) is working on networking aspects in quantum computing, cryptography and photonics and quantum machine learning, QI with LIP6, MOCQUA with LORIA, CAPP (LIG, Grenoble) on contextuality and quantum combinatorial games, AlgoComp with IRIF, MC2 with LIP Lyon and PACAP with IRISA and Inria Rennes working on mapping quantum circuits to particular architectures and the new QUACS team on quantum algorithms in Saclay.

**LIP6** (Laboratoire d'Informatique de la Sorbonne) hosts several recognized specialists in cryptography and quantum telecommunications (QKD): Eleni Diamanti was awarded a European Synergy Grand ERC for her work in the QUSCO (Quantum Superiority with Coherent State) project. Elham Kashefi is co-founder of the VeriQloud startup. She is also working on verified quantum computing, secure multiparty quantum computing, and features to achieve quantum advantages.

The **LPENS** (Laboratoire de Physique de l'Ecole Normale Supérieure) is the result of the merger in early 2019 of several physics research laboratories at ENS Paris, including the **LPA** (Laboratoire Pierre Aigrain), which specializes in nanotechnology and photonics.

They are working on numerous nanotechnologies used for the creation of qubits and the transport of quantum information: superconducting thin films, superconducting and microwave circuits for their control, two-dimensional electron gases with very high mobility, semiconductor quantum boxes, qubits based on carbon nanotubes.

Taki Kontos and Audrey Cottet's teams are at the origin of the creation of carbon nanotubes used as electron traps potentially usable in electron spin qubits, which led to the creation of the **C12** startup, already mentioned. The lab is also a participant on the work on cat-qubits related to the startup Alice&Bob.

The **LKB** (Laboratoire Kastler Brossel) from ENS Paris focuses on quantum information and photonics, interactions between light and matter (Nicolas Treps and Valentina Parigi), quantum simulation and precision sensing with cold atoms (Christophe Salomon).

Thibault Jacqmin is working on microwave photon generation with NEMS (nano MEMS).

The IRIF (Institut de Recherche en Informatique Fondamentale) from CNRS and the University Paris Diderot is led by Frédéric Magniez who also teaches at Collège de France and hosts Iordanis Kerenidis, Sophie Laplante and two Inria teams. It works in quantum computing, cryptography and communications.

[2753] This is specified in Inria strategic scientific plan 2018-2022, 2018 (93 pages), pages 47 and 48.

The **MPQ** laboratory (Materials and Quantum Physics) of the University Paris Diderot is particularly interested in the technique of ions trapped in the Quantum Physics and Devices (QUAD) and QITE (Quantum Information and Technologies) groups. But also, to the generation of entangled photon pairs (Sara Ducci).

The **LPTHE** (Laboratoire de Physique Théorique et Hautes Energies) of the University Paris Sorbonne works in condensed matter and statistical physics with applications in superconducting qubits.

The **INSP** (Institut des Nanosciences de Paris) of Paris-Sorbonne University is a generalist laboratory on nanosciences. They work in particular in different branches of photonics, on NV centers, on color centers qubits in silicon carbide, on spin and magnetism and on photonics components in III-V materials.

The **IRCP** (Institut de Recherche de Chimie Paris) associated with the Ecole Nationale Supérieure de Chimie ParisTech conducts research in innovative materials.

Philippe Goldner is working on the creation of qubits based on nanocrystals doped with rare earth ions such as europium or erbium, and is involved in the SQUARE project of the European Quantum Flagship, coordinated by the Karlsruhe Institute of Technology and also involving Thales. The laboratory is also involved in the European Quantum Flagship **ASTERIQS** project which is working on NV-based qubits in diamonds.

The **LPEM** (Laboratory of Physics and Study of Materials) of the ESPCI and the UMPC works in particular in superconductivity as well as on the fermions of Majorana.

The **LPTMC** (Laboratoire de Physique Théorique de la Matière Condensée) of the University Paris-Sorbonne has among other things several teams working on condensed matter physics like Jean-Noël Fuchs and Julien Vidal on topological insulators and Majorana fermions and Rémy Mosseri working on quantum information with topological qubits.

**SYRTE** (Laboratoire Systèmes de Référence Temps-Espace) located at Paris Observatory works in quantum sensing, in particular gravimetry, quantum gyroscopes and on time measurement with atomic and optical clocks. They are partnering with NIST. The quantum gravimeter and interferometry team is led by Franck Pereira dos Santos. SYRTE is led by Arnaud Landragin.

The **Laboratoire Jacques Louis Lions** (LJLL) is specialized in applied mathematics. It focuses on the analysis, modeling and high-performance scientific computation of phenomena represented by partial differential equations. Mario Sigalotti and Ugo Boscain, who specialize in the control of quantum systems and are also members of Inria, are among others.

The **Laboratory of Theoretical Chemistry** at Sorbonne University is directed by Jean-Philippe Piquemal (co-founder of Qubit Pharma) and is interested in computational chemistry, including quantum.

In September 2020, the **Quantum Innovation Center Sorbonne** (QICS) was inaugurated, a collaborative research structure associating LIP6, the LKB of the ENS and Inria.

The **Saclay plateau** has an even higher density of laboratories, located south-west of the Paris region. Most of these entities are consolidated in **Université Paris Saclay**.

At the **CEA**, Daniel Esteve's Quantronics team at the Iramis laboratory in Saclay has been working on superconducting qubits for nearly 20 years. Daniel Esteve's laboratory includes about fifteen people and is now managed by Hugues Pothier.

**IphT** (Institut de Physique Théorique de Saclay) associates CEA and CNRS. They work on the physics of condensed matter, including high-temperature superconductors, and on Majorana fermions. But their main focus seems to be mainly astrophysics.

The **LAC** (Laboratoire Aimé Cotton) is located at the ENS Saclay. It also works on cold atoms and interactions between atoms and light. In particular, they create qubits by combining an optically active erbium ion and a nuclear spin of yttrium.

The **C2N** (Centre des Nanosciences et des Nanotechnologies) of CNRS and Université Paris Saclay is a key quantum photonics laboratory. It is the home to Pascale Senellart and Jaqueline Bloch's labs. They work in particular on light-matter coupling in semiconductors. It also host quantum electronics teams (Frédéric Pierre).

The **LPS** (Laboratoire de Physique des Solides) works on magnetism, Josephson junction superconductors, thermodynamics, superconducting spintronics and quantum dynamics. They also develop codes for quantum and semi-classical dynamics and quantum control with applications in quantum information.

The **LPTMS** (Laboratoire de Physique Théorique et de Modèles Statistiques) has several strings to its bow in quantum physics without the link with quantum computing being immediately detectable.

The **LCP** (Laboratoire de Chimie Parisud) works on superconductors and on the dynamics and control of ions trapped by laser pulses. They develop hybrid computational models of quantum chemistry (quantum+traditional) using MCTDH (Multi-configuration time-dependent Hartree) which allows to solve the Schrödinger equation for the simulation of interactions between atoms in molecules.

On the program: condensed matter physics, modeling of classical and quantum systems via statistical physics, quantum chaos, number theory and quantum chaos, theoretical aspects of quantum information; cold atoms, quantum integrable systems, quantum groups, etc.

TelecomParistech's **LTCI** (Laboratoire Traitement et Communication de l'Information) is an industry laboratory operating with partnerships with the private sector and via chairs. Its "Quantum Information and Applications" (QIA) team specializes in the theoretical and experimental aspects of quantum communications.

They develop hybrid CV-QKD-based quantum cryptography protocols compatible with telecom operators' fiber networks and QKD repeaters. They are contributor, founding member and reporter to the ETSI QKD-ISG on the QKD standardization processor. The team is led by Isabelle Zaquine and includes Romain Alléaume.

**ISMO** (Institut des Sciences Moléculaires d'Orsay) works on quantum dynamics, interactions between heavy particles and electrons at low temperature, light/matter coupling and on software for the simulation of quantum physics.

The **CPht** (Centre de Physique Théorique de Polytechnique) is specialized among other things in the physics of condensed matter. But not to the point of creating superconducting qubits! We find there Karyn Le Hur's group, who is specialized in condensed matter physics.

The **Charles Fabry Laboratory** of the Institute of Optics Graduate school (IOGS) is specialized in lasers and quantum optics. It is home to Alain Aspect, Philippe Grangier as well as Antoine Browaeys, co-founder of the startup Pasqal and its laser-controlled cold atom qubits.

The **LIX** (Laboratoire d'Informatique de l'Ecole Polytechnique) is particularly active in post-quantum cryptography algorithms.

The **PMC** (Laboratoire de Physique de la Matière Condensée) is another laboratory of the Ecole Polytechnique. They work in particular on spin dynamics in semiconductors and magnetic thin films.

The **L2S** (Signals and Systems Laboratory) of CentraleSupelec is active in quantum systems research. In particular, the L2S is staffed by Zeno Toffano, who is focused on quantum states measurement.

The **LPQM** (Laboratory of Quantum and Molecular Photonics) associates the ENS Paris Saclay and the CentraleSupelec school. Their domains are coherence and quantum correlations.

The **LRI** (Laboratoire de Recherche en Informatique) located at Centrale-Supélec is managed by Benoît Valiron, who teaches and conducts research in quantum computing, a field that is still relatively under-taught in engineering schools.

The **LMF** (Formal Methods Laboratory) was created in 2021 as a joint research center of University Paris-Saclay, CNRS, ENS Paris-Saclay, Inria, and CentraleSupelec with a focus on formal methods, combining 100 members from the former Laboratoire Spécification et Vérification (LSV) and the VALS team of Laboratoire de Recherche en Informatique (LRI). They target computational paradigms ranging from classical to emerging ones such as biological and quantum computing.

Thales **RT** (Thales Research and Technology) carries out R&D to create industrialized quantum sensing solutions. In particular, they have developed expertise in diamond NV centers.

**Onera** studies quantum optics at its Palaiseau site. It is in this capacity that it coordinates the ASTERIQS project of the European Quantum Flagship, "Advancing Science and Technology through diamond Quantum Sensing".

They also have teams of researchers in photonics, in III-V semiconductor materials (gallium, ...) with a prototype manufacturing unit located in their premises in Palaiseau, in metrology (gravimeter, atomic clock, accelerometer) and in QKD.

Let's move on to other parts of the Ile de France: Cergy-Pontoise, Villetaneuse and Versailles.

The **LPTM** (Laboratoire de Physique Théorique et Modélisation) of the University of Cergy-Pontoise is interested in cold atoms, in liaison with the Institut Francilien de Recherche sur les Atomes Froids (IFRAF). They also study graphene, electronic quantum transport, topological phases and entanglement.

The **LSPM** (Laboratoire des Sciences des Procédés et des Matériaux) of the University of Paris 13 in Villetaneuse is working on the manufacturing processes of NV centers, carbon nanotubes and graphene centers and associated applications.

The **LPL** (Laboratoire de Physique des Lasers) of the University Paris 13 in Villetaneuse works in photonics and cold atoms, their traps and on quantum metrology. It is the laboratory of Hélène Perrin, already mentioned, who is its Deputy Director.

The **GEMaC** (Groupe d'Etude de la Matière Condensée) of Versailles also works in the field of diamonds and graphene, on spin electronics and magnetism. It also works on QKD and photonic quantum memory.

Launched in 2014, the **Paris Center for Quantum Computing** (PCQC) brings together several dozen researchers from various laboratories in the Paris region, including Philippe Grangier. The **CNRS** has informally grouped its efforts with the Quantum Computing working group which works more on the algorithmic dimension.

Finally, **QuanTiP** (Quantum Technologies in Paris Region) is a community that groups research laboratories in the Ile de France region that are focused on quantum communications technologies. According to them, there are 650 quantum researchers in the Ile-de-France region in all (physics, algorithms, telecommunications, cryptography) spread over 100 teams in 30 research laboratories. It followed-on SIRTEQ in 2022, that was created in 2017.

We can also mention the initiative of the high-performance computing cluster **Teratec** (based in Bruyères-le-Châtel, near the CEA's Military Affairs Department) around quantum physics[2754].

It aims to develop quantum algorithms, hybrid development methods, use cases, and to inform, train and animate a community. They benefit from an Atos QLM simulator installed at the CRTT (Centre de Calcul, Recherche et Technologie) of the CEA in Bruyères-le-Châtel.

### *Grenoble*

Grenoble's quantum ecosystem is dense, well-organized and very focused on the creation of qubits based on electron spins but also on superconductors, all with good skills in photonics. It is probably the place where coordination between research teams works best, particularly by integrating the key stages of industrialization.

Quantum research in Grenoble is led by different branches of the CEA (Leti in nanoelectronics and IRIG in fundamental physics), the CNRS with Institut Néel, LPMMC and two joint CNRS and CEA teams: NPSC (NanoPhysics and Semiconductors) focused on quantum sensing, quantum photonics,

---

[2754] Teratec brings together several private and public HPC players including Atos, CEA, CERFACS (European Center for Advanced Research and Training in Scientific Computing), Dassault-Aviation, EDF, IFPEN, PCQC (Paris Centre for Quantum Computing), Total and the University of Reims.

quantum thermodynamics and the quantum foundations, and Quanteca, created in 2019, which deals with all kinds of solid states qubits (electron spins, superconductors).

**CEA-Leti** (Electronics and Information Technology Laboratory) in Grenoble is the CEA's micro and nanoelectronics laboratory. It is notably at the origin of the SOI wafer technology that led to the creation of SOITEC. Leti is focused on CMOS electron spin qubit engineering. The project is coordinated by Maud Vinet and federates the efforts of several CEA, CNRS and UGA laboratories.

**CEA IRIG** (Grenoble Institute for Interdisciplinary Research) is the counterpart of Institut Néel in fundamental research at CEA. It includes the Laboratory PHotonique ELectronique et Ingénierie QuantiqueS (PHELIQS), which works on the physics of condensed matter.

**Institut Néel**[2755], launched in 2007, is a CNRS laboratory specialized in condensed matter physics with a critical mass of researchers in quantum physics. Its researchers are exploring the possibilities of electron spin qubits (Tristan Meunier), superconducting qubits (Nicolas Roch), topological matter (Adolfo Grushin) and photonics. It also works on thermodynamics and the energetics of quantum computing (Alexia Auffèves), cryogenics (Sébastien Triqueneaux) and quantum foundations (Cyril Branciard and also Alexia Auffèves).

The **LPMMC** (Physics of Condensed Matter) of the University Grenoble Alpes is a CNRS UMR focused on the theoretical physics of condensed matter and quantum physics, N-body quantum interactions, superconductivity and superfluidity, and on the temporal evolution of quantum systems under the effect of magnetic and electric fields.

The **IJF** (Institut Joseph Fourier) of the University of Grenoble is working on quantum dynamics and in particular on issues of decoherence and thermal quantum noise.

The **LIG** (Laboratoire d'Informatique de Grenoble) is interested in quantum algorithms in general. One of its members is the researcher Mehdi Mhalla, who works on the quantum resolution of graph problems.

Research in quantum computing in Grenoble is currently structured around three initiatives: **QuEnG**, **QuantECA** and **QuCube,** which are not on the same level.

[2755] The institute takes its name from Louis Néel (1904-2000, French), a physician of Lyon origin who was awarded the Nobel Prize in Physics in 1970 for his studies on magnetism and the discovery of antiferromagnetism. He is at the origin of the creation of the Polygone Scientifique de Grenoble, which brings together numerous research institutes and companies in the peninsula between the Isère and Drac rivers. The place hosted the first CEA site outside the Paris region in 1956, launched by Louis Néel. The CNRS established a foothold there in 1962, and in 1967 CEA-Leti was created. CEA-Leti is one of the world's largest civilian laboratories for applied research in nanoelectronics and nanotechnology. The Grenoble Science Park is also home to several international research organizations, the Institut Laue-Langevin, the European Synchrotron Radiation Facility and one of the branches of the European Molecular Biology Laboratory. In 2005 the CEA-Liten was created, a branch of the DRT specialized in new energies (photovoltaic solar, batteries, fuel cells, complete management of the carbon cycle, mixed energy management, innovative materials). In 2006, Minatec was launched, a nanotechnology commercial development center, later complemented by the Minalogic competitiveness cluster. In 2012, the Clinatec research center, founded by Alim-Louis Benabid, was launched, which is at the origin of the first complete exoskeleton for tetraplegics.

**QuantAlps** (formerly, between 2017 and 2021, QuEnG for Quantum Engineering Grenoble) is the Grenoble ecosystem ranging from philosopher to industrialist, a trans-laboratory, trans-disciplinary and trans-sectoral umbrella initiative.

The teams are working in physics on many other fields: in photonics, on superconducting qubits, electron spin qubits and qubits based on molecular magnets. Teams also make the link between quantum physics and philosophy. The initiative also includes training engineers in physics and quantum computing with various courses, including a project with Ensimag, Grenoble's leading computer science school. QuantAlps was launched by Alexia Auffèves and Anna Minguzzi. Anna is the Director of QuantAlps since Alexia Auffèves left Grenoble to work at Singapore's CNRS MajuLab in October 2022.

### *Lyon*

Research in Lyon is well balanced between the physical part and the mathematical and software part of quantum.

The **Physics Laboratory of ENS Lyon** studies condensed matter. The Quantum Circuit Group of Benjamin Huard is working on superconducting qubits and their error correction codes. He was notably joined by Audrey Bienfait in 2019, who works on electron spin resonance and its applications in quantum sensing. It was also there that Théau Peronnin finalized his thesis in 2020 while creating the startup Alice&Bob with Raphaël Lescanne.

The **INL** (Lyon Nanotechnology Institute) is located at Centrale Lyon (Ecully). They work on semiconductors and photonics. They have a technological platform for component prototyping, particularly in photonics.

The **iLM** (Institut Lumière Matière) of Lyon is specialized as its name says in photonics.

The **Camille Jordan Institute** in Lyon is a research laboratory in mathematics that works in particular on quantum probabilities. It is distributed on several sites: Villeurbanne, Saint-Étienne and on the Centrale Lyon campus in Écully.

The **LIP** (Laboratoire de l'Informatique du Parallélisme) of ENS Lyon associates CNRS, Inria and Claude Bernard Lyon 1 University. Its MC2 team works on theoretical computer science and complexity theory. It includes Omar Fawzi, CNRS 2019 bronze medalist and specialist in quantum information theory. He leads his work in the MC2 team at LIP.

### *Occitanie*

Quantum research in Toulouse is very focused on fundamental physics and quite far from quantum computing with the exception of **LPTT**. There are also two laboratories in Montpellier, one of which is associated with IBM. Let's mention the **QuantUM Hub** initiative launched by IBM Montpellier, the University of Montpellier, and the Occitanic Region.

The **Institute for Quantum Technologies in Occitanie** was created in January 2021 to consolidate all the Occitan research and industry organizations, including the research labs below from Toulouse and Montpellier.

The **CEMES** (Centre d'Élaboration de Matériaux et d'Etudes Structurales) in Toulouse is specialized in physics and optronics. It is interested in light-matter coupling at scale and the creation of sensors oriented more towards connected objects than quantum applications.

**LCAR** (Laboratoire Collisions-Agrégats-Réactivité) of the Paul Sabatier University of Toulouse works on Rydberg atoms. It is in the team of Juliette Billy and David Guéry-Odelin.

The **LPCNO** (Laboratory of Physics and Chemistry of Nano-objects) of INSA Toulouse is specialized in photonics and quantum electronics. They study electron and nucleus spins, quasi-particles and quantum dots. They aim at applications in quantum computing. Their research is looking at applications in the health sector.

The **ITM** (Institut de Mathématiques de Toulouse) of the University of Toulouse studies statistical and quantum physics. It is home to Clément Pellegrini who studies quantum information theory and quantum state measurement.

The **LPTT** (Laboratoire de Physique Théorique de Toulouse) works on superconductors and SQUID Josephson effect loops. They are involved in the Quantware project which has been co-funded among others by the NSA!

The **LCPQ** (Laboratory of Quantum Chemistry and Physics) of the Paul Sabatier University of Toulouse develops generalist quantum chemistry codes, contributing to molecular simulation efforts.

The **L2C** (Charles Coulomb Laboratory) of the University of Montpellier is working on quantum metrology, spin dynamics and graphene, with applications in magnetic microscopy.

**The University of Montpellier** is an IBM partner in the setting up of a joint laboratory on quantum which actually aims to evangelize customers on the general principles and tools of the IBM Quantum quantum platform.

The **LIRMM** (Montpellier Laboratory of Computer Science, Robotics and Microelectronics) focuses on the creation of quantum algorithms. It collaborates with IBM, Total and CERFACS.

Aida Todri-Sanial is one of their Research Director and works on quantum algorithms used for classical integrated circuits routing and on classical algorithms improving qubits gates mapping taking into account calibration data[2756].

### *Nouvelle Aquitaine*

The Nouvelle Aquitaine ecosystem is specialized in sensing and enabling technologies like lasers, with its industry ecosystem comprising **Muquans**, **Azurlight Systems** and **ixBlue**. Since March 2021, this ecosystem is federated under the umbrella **Naquidis**, as part of the AlphaLRH cluster and with the support of the Region.

Besides the local branch of IOGS (Institut d'Optique Graduate School), here are two quantum research labs in the region.

[2756] See A Hardware-Aware Heuristic for the Qubit Mapping Problem in the NISQ Era by Siyuan Niu, Aida Todri-Sanial et al, October 2020 (14 pages).

The **LP2N** (Laboratoire Photonique, Numérique et Nanosciences) of the Institut d'Optique de Bordeaux does research in photonics and metrology based on cold atoms (microgravitometry). This is where the startup Muquans started.

The **LOMA** (Laboratoire Ondes et Matières) from CNRS works on quantum matter and is investigating, among other things, nanomechanical qubits based on carbon nanotubes.

The **XLIM** (Limoges) does among other things photonics. They are notably partners with Thales TRT. They are working on applications in polariton metrology, in particular SPR (Surface Plasmon Resonance).

### *Sud*

There are also a few quantum physics laboratories in Marseille, three of which are directly related to the needs of quantum computing. And one laboratory in Nice.

The **Fresnel Institute of** Marseille is involved in photonics, so inevitably, it can contribute to advances in photon-based qubit management and QKD-based quantum cryptography.

The **CPT** of the Universities of Marseille and Toulon is working on quantum dynamics and wave diffusion in optical fibers and light guides. They are partners of various foreign universities: Aalborg University (Denmark), Pontificia Universidad Catolica de Chile, Karlsruhe Institute of Technology (Germany), Kyoto Institute of Technology and the Moscow Institute of Physics and Technology.

The **PIMM** (Physics of Ionic and Molecular Interactions) laboratory at the University of Marseille does research in plasmas, more related to the ITER nuclear fusion project than to quantum computing.

The **Laboratoire d'Informatique Fondamentale** de Marseille is particularly interested in quantum computing. Their Discrete Time Quantum Simulator project was launched in 2018. They are working on Quantum Walks and the Quantum Cellular Automata.

**INPHYNI** (Institut de Physique de Nice) of the Université Nice Côte d'Azur is interested in cold atoms, wave transport, interactions between light and atoms. It deploys a QKD test network between Nice in Sophia-Antipolis since 2019 in partnership with Orange. The quantum laboratory is directed by Sébastien Tanzilli.

### *Burgundy Franche-Comté*

Besançon is home to three quantum laboratories and Dijon to a fourth.

The **LmB** (Laboratoire de Mathématiques de Besançon) of the Université Bourgogne Franche-Comté studies quantum groups and probabilities.

The **UTINAM** Institute of the University of Besançon studies quantum decoherence, control, diagnosis, processing and transport of quantum information in the field of quantum sensing.

**Femto-St** is a research institute in Besançon focused on nanosciences, optics and optoelectronics. They work in particular on optical telecommunications, nonlinear optics, optics-based Ising machines and quantum imaging.

**Icb** (Interdisciplinary Carnot of Burgundy) of the University of Burgundy, based in Dijon, includes a team studying quantum and nonlinear dynamics (DQNL).

*Great East*

The region includes three quantum laboratories located in Strasbourg, Nancy and Troyes.

The **Quantum Matter Theory Group** from the University of Strasbourg is involved in condensed matter physics and also works on the interactions between light and matter, with Rydberg atoms. Run by Shannon Whitlock, the lab is developing cold atom-based quantum systems.

The **L2n** (Lumière Nanomatériaux Nanotechnologies) of the Technology University of Troyes is specialized in optoelectronics and photon sources.

The **Loria** (Lorraine Laboratory for Research in Computer Science and its Applications) is based in Nancy. Two teacher-researchers are interested in quantum computing and algorithms: Simon Perdrix and Emmanuel Jeandel. The first is one of the main contributors of ZX-Calculus. Since 2021, Simon Perdrix is a PI at Inria Nancy.

*Elsewhere in France*

And finally, here are a few quantum physics laboratories located in other regions, in Rennes, Lille, Bordeaux and Limoges, but with no apparent direct link to quantum computing.

The **IPR** (Institut de Physique de Rennes) is attached to the University of Rennes. They are interested in quantum dynamics, the evolution of quantum states over time.

The **LARIS** (Laboratoire Angevin de Recherche en Ingénierie des Systèmes) based in Angers deals with various IT subjects. Within it, François Chapeau-Blondeau and Etienne Belin are interested in the impact of noise on quantum algorithms.

The **PhLAM** (Laboratoire de Physique des Lasers Atomes et Molécules) in Lille is interested in photonics and cold atoms.

The **IEMN** (Institute of Electronics, Microelectronics and Nanotechnology) is a laboratory located on four sites in Lille, Villeneuve d'Ascq and Valenciennes. They specialize in the design of quantum nanostructures.

*International collaborations*

International partnerships are very common in research. Many of the works of French researchers are carried out with researchers from other countries, including the USA, the UK, Austria, the Netherlands and Germany, Japan and Singapore (notably with joint international units of the CNRS IFLI and MajuLab).

CEA-Leti is a partner of **IMEC,** its counterpart in Belgium, based in Leuven, covering AI and quantum computing[2757]. Like CEA-Leti in Grenoble, they have a clean room for etching up to 28 nm on 30-cm wafers and another on 20-cm wafers for MEMS.

Since 2017, the **Grenoble University Space Center** has been collaborating with the Austrian **IQOQI** on sending quantum keys via satellite in the Nanobob project.

And there is another international collaboration on quantum involving France, the Netherlands (QuSoft) and Latvia.

***Government funding***

After Atos launched in 2015/2016 its venture in quantum computing emulation, the French government started to look at the opportunity to launch a quantum plan. Back then, it was involved in the European Quantum Flagship which was announced in October 2018.

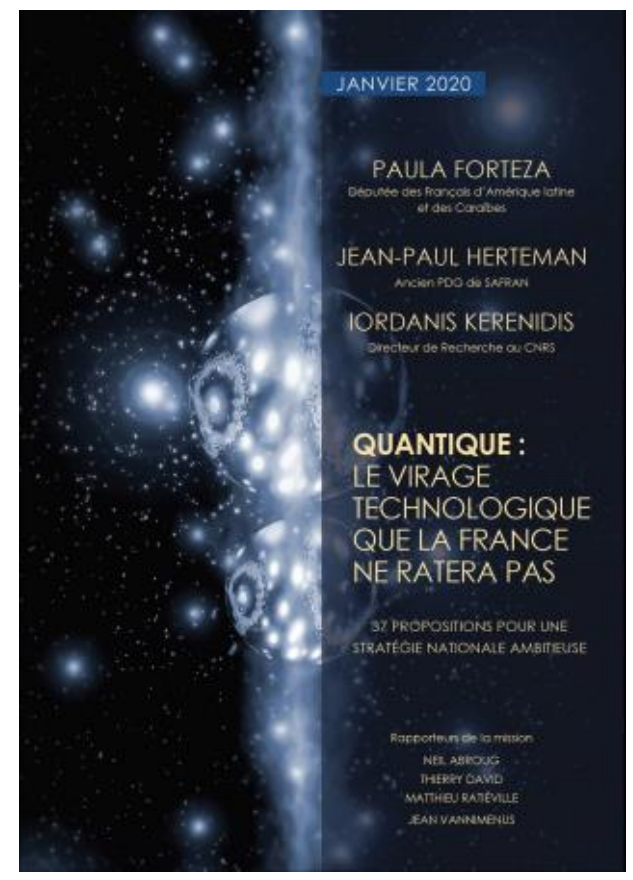

Things really started with the creation of a parliamentary investigation commissioned by the Prime Minister in March/April 2019 and led by MP **Paula Forteza**, accompanied by **Iordanis Kerenidis** (CNRS researcher specialized in quantum machine learning) and **Jean-Paul Herteman** (former CEO of Safran). The parliamentary mission submitted its report on January 9, 2020, titled "Quantum: the technology disruption that France will not miss". The report made fifty proposals, 37 of which were made public. The government then created a national quantum strategy that included some but not all of the parliamentary mission's proposals. All this during the early stages of the covid pandemic. It was finally announced a bit late, in January 2021, but by President Emmanuel Macron, a premiere in the western world.

The ambitions of the strategy and its roadmap revolve around rather classical themes: NISQ quantum computing, Fault Tolerant Quantum Computing, algorithms and software, quantum telecommunications overall (including quantum cryptography and distributed quantum computing), quantum sensing, and at last, enabling technologies. This includes cryogenics, cabling, control electronics, vacuum control, lasers and photon sources.

The plan is spread over 5 years from 2021 to 2025 with 1B€ public funding and an additional 850M€ funding expected from European funds and the private sector (industry R&D and startups funding)[2758].

In 2021 and 2022, many components of the French quantum strategy were launched. There was an incremental 150M€ research program handled by CNRS, CEA and Inria announced in September 2021 and launched in 2022. Then, a hybrid classical/quantum platform announced in January 2022, which will be located at the TGCC supercomputing center handled by CEA. It will use the Joliot-Curie supercomputer in association with a QLM classical server from Atos for emulation and QPU drive, and a Pasqal quantum simulator. This HQI (Hybrid HPC Quantum Infrastructure) project is partly funded by the EU HPC-QS program. An education program was launched with universities to expand training from license to PhDs. Other programs were launched for quantum startups accelerations, for enabling technologies and for the deployment of PQC cryptography systems.

[2757] See Partners Double-Team AI & Quantum Computing by Mathew Dirjish, November 2018.

[2758] See How France Is Becoming a Quantum Computing Power by Peter Suciu, The National Interest, January 2022 and What Europe can learn from France when it comes to quantum computing by Andersen Cheng, Sifted, November 2021.

*Quantum industry*

On the industry vendors scene, France has a handful ventures in quantum computing hardware front with **Alice&Bob** (cat-qubits), **C12** (carbon nanotubes electron spins qubits), **Pasqal** (cold atoms qubits), **Quandela (**single photon sources and photons qubits) and **Crystal Quantum Computing** (using trapped ions in Rydberg states).

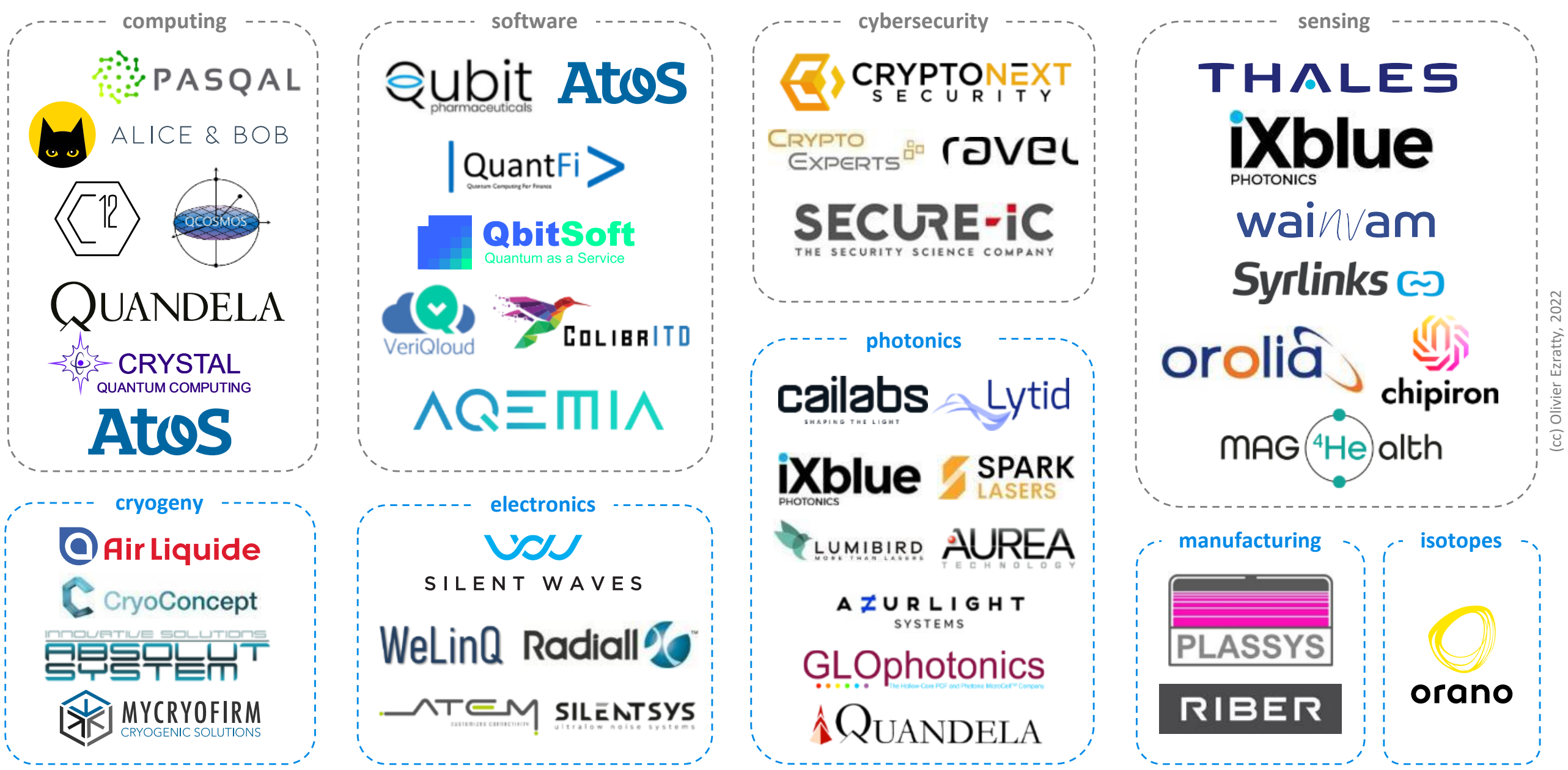

*Figure 862: France's quantum industry ecosystem. (cc) Olivier Ezratty, 2022.*

In the software side, we have **Qubit Pharmaceuticals** (healthcare), **QuantFi** (finance), **VeriQloud** (quantum telecommunications) and **Prevision.io** (quantum machine learning) plus a bunch of companies specialized in cryptography, mostly PQC with **CrytoNext**, **CryptoExperts**, **Ravel** and **Secure-IC**.

In quantum sensing, we have **Muquans** (microgravimeters, acquired by **iXblue** in 2021), **Chipiron** (NV centers imaging), **Wainvam** (NV centers imaging), **Mag4Health** (imaging) and **Thales** (NV centers, SQUIDs and cold atoms sensing, lightweight cryogeny).

In addition to **Bpifrance** and the investment fund **Quantonation,** the **Deep Tech Founders** trains entrepreneurs/researchers in deep techs. It is an international program created by the Hello Tomorrow team. All these are behind the creation of a structure to support the quantum ecosystem in partnership, the **Lab Quantique,** launched officially in April 2020.

The Lab Quantique is a think tank for the development of talent, particularly at the crossroads between science and entrepreneurship. From a practical point of view, Lab Quantique organizes regular meetings that bring together mainly quantum technology entrepreneurs from France and abroad. These meetings took place in the form of videoconferences on Zoom during the covid-19 pandemic period in 2020 and 2021.Its objectives are to connect industry players, startups and researchers, to build bridges with the international community, to launch a program to accelerate quantum startups and to organize a major annual high-level conference bringing together all the stakeholders in the ecosystem, as well as an International Prize (attracting talent).

In the end, it will also take the form of a trade association mixing the quantum industry (large organizations, small businesses and startups) and its users (mainly, large companies like EDF, Airbus and the likes).

One France specificity in Europe is its large corporations directly invested in quantum technologies and quantum enabling technologies : **Atos** (software, emulators, quantum accelerators), **Thales** (sensors), **Air Liquide** (cryogenics), **Orano** (isotopes production like silicon 28), **Radiall** (connectors, cabling, switches, attenuators, couplers, optical links), **ATEM** (cabling) and many in photonics (like **iXblue, Azurlight Systems, Aurea Technology, Lumibird** and **Cailabs**) and even semiconductor manufacturing machines with **Plassys Bestek** and **Riber**.

### The Netherlands

The Netherlands is one of the most active European countries in quantum technologies research and development, mainly around the University of Delft (**TU Delft**) and its **QuTech** branch.

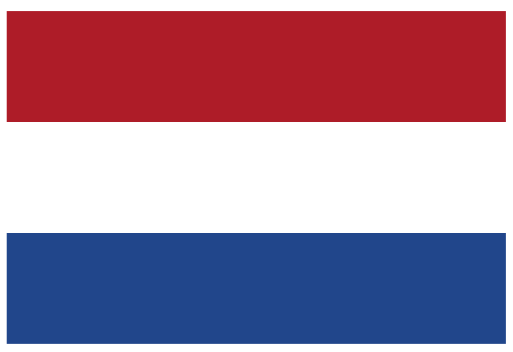

It has long been a historical melting pot of quantum physics research in Europe. We have thus cited many great names at the beginning of this book: **Hendrik Antoon Lorentz** (1853-1928), **Heike Kamerlingh Onnes** (1853-1926), **George Uhlenbeck** (1900-1988), **Hendrick Casimir** (1909-2000) and **Samuel Goudsmit** (1902-1978).

In 2015, the government launched a 10-year, 135M€ plan to create a quantum computer[2759]. The investment was made in **QuTech**, TU Delft's quantum research center launched in 2014 with a 10-year budget of 145M€, half of which comes from TU Delft University and the other half from the NWO, the national funding agency[2760]. Qutech employs more than 180 people in all, of which only 37% are Dutch, with 25 permanent researchers.

The Netherlands government then announced a 7 years 615M€ plan in April 2021, complemented by an addition of 228M€ in 2022[2761]. This makes the country probably the greater investor in quantum technology in proportion of its GDP.

This public funding should drive private sector investments of 3.6B€, a very ambitious goal in comparison with the similar 565M€ expected in France. It is managed by the non-profit foundation **Quantum DELTA NL** that was created in 2020[2762]. The Netherlands plans to create 30.000 high-tech jobs and create a cumulative economic impact of at least 5B€ with quantum technologies. The country plan is organized around the creation of three technology demonstrators, four generic action lines and shared cleanroom facilities.

- **Quantum Inspire**, their cloud superconducting computer service that is already available and got a funding of 90M€.
- **Quantum Network** project on quantum telecommunications and cryptography, connected to the related European projects, with a funding of 62M€. They expect to quantumly connect three quantum computers by 2023 and five by 2026.
- **LightSpeed** is a program connecting startups with investment funds. It’s overselling a bit its value touting access to 13.6B€ in investment capital, representing the totality of the various funds managed by these investors. The 2022 funding extension added 15M€ for a startup seed fund.

[2759] See the state of play of the Dutch National Quantum Plan in National Agenda for Quantum Technologies, Quantum Delta Netherlands, September 2019 (51 pages).

[2760] See QuTech's 2018 Activity Report (80 pages) as well as an independent valuation report published in 2019 and covering the period 2015-2018.

[2761] See Quantum Delta NL awarded 228 Million Euro for second phase of its programme to Accelerate Quantum Technology, April 2022.

[2762] See the plan details in Quantum Delta NL in a nutshell, 2021 (20 pages). Look also at the excellent Economic Impact of Quantum in the Netherlands, Quantum Delta NL, May 2020 (60 slides) which contains a lot of interesting market data. DELTA stands for Delft Eindhoven, Leiden, Twente and Amsterdam, completed by Nijmegen, Maastricht and Groningen.

- **House of Quantum** is a startup ecosystem facility to open in 2024 with a budget of 182M€. It would accelerate part of the 100 startups the country wants to consolidate by 2027. Within this house, the **Living Lab QT** that will focus on ethical, legal and societal aspects of quantum technology with research collaborations between universities, the public and private sector, with a funding of 20M€. They will open two related interdisciplinary university positions, create a desk and a toolkit for responsible innovation and entrepreneurship and create a covenant to be signed by private and public stakeholders promoting sustainable and safe use of quantum technologies.
- They also plan to invest 150M€ in the 5 cleanrooms from **NanoLabNL**, have a quantum sensors plan with 23M€ funding and a training program that should create 2000 PhDs and engineers by 2027 with a funding of 41M€.

QuTech is also associated with **Intel** and **Microsoft**. QuTech has received $50M in funding in 2015 from Intel as part of a partnership on their superconducting and electron spin qubits.

Microsoft has also been a partner of QuTech since 2010, which they have also depleted by hiring **Leo Kouwenhoven** in their Microsoft Research laboratory which is on site and working on topological quantum and fermion of Majorana in liaison with a team of QuTech dedicated to the same subject. The Netherlands looks like a brain reservoir for the American quantum industry.

Collaborative research approaches are making good progress, particularly with a view to recovering European funding. In October 2017, QuTech launched a partnership with the Institute of Photonic Sciences, the University of Innsbruck in Austria and the Paris Centre for Quantum Computer. QuTech is also a partner of the University of Aachen in the CMOS qubit. The University of Delft has also obtained for the European part of the QuNET project mentioned about Germany an ERC of 1,5M€ with a launch in November 2019 and an end planned for October 2024[2763].

Other initiatives with blurred contours have been launched such as **Quantum Helix,** funded under the European Quantum Flagship Program and Horizon 2020. The **Quantum Software Consortium** runs for 10 years from 2017 and has received €18.8M in public funding from the country's Gravitation Program. It brings together various Dutch laboratories: **TU Delft**, **QuTech** (part of the latter), **QuSoft** (a research laboratory dedicated to quantum software, launched by CWI, UvA and VU in 2015), **CWI** (Centrum Wiskunde & Informatica), the **University of Leiden, UvA (**University of Amsterdam) and **VU** (Free University of Amsterdam) to conduct research in quantum software and cryptography.

Other companies include **Delft Circuits** (superconducting cabling), **Leiden Cryogenics** (high-power dilution cryostats), **Qblox (**electronics for controlling superconducting qubits), **Single Quantum (**single photon detectors), **QuiX** (photonic processor, a subsidiary of Lionix, a foundry capable of producing photonics wafers in nitrates on SiO2), **Qu&Co** (quantum software), **QuSoft (**quantum software), **QPhoX** (quantum computer interconnection) and **ipCLock** (quantum clock).

In December 2020, the Dutch quantum industry created the **IMPAQT** consortium. The first members are Orange QS, Qblox, Delft Circuits, QuantWare BV (a new stealth spin-off from TU Delft creating superconducting QPUs) and Qu&Co. Their goal is to improve the coordination of how they are creating quantum computer enabling technologies.

At last, The Netherlands and France signed in September 2021 a Memorandum of Understanding to expand collaborations in quantum technologies, with Cédric O, the French Secretary of State for Digital and Electronic Communications and Mona Keijzer, the Dutch Secretary of State for Economic Affairs and Climate Policy. The bilateral collaboration includes research partnerships in silicon qubits as part of the European flagship project QLSI, research-industry collaboration involving companies like Atos and Qu&Co, the creation of a joint portal listing job opportunities in France and the

[2763] See A quantum network for distributed quantum computation, Cordis, 2019.

Netherlands (www.quantumjobs.fr and quantumjobs.nl) and collaboration to increase EU venture capital in the domain (involving Quantonation).

### Belgium

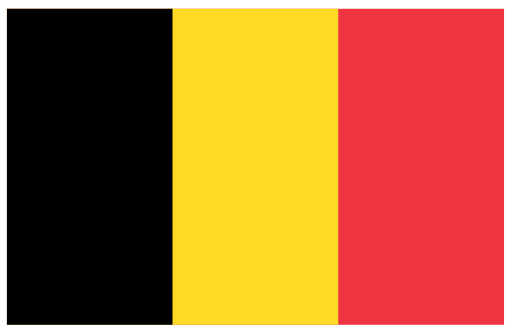

Belgium is the host of the famous Solvay conferences created in the early 20th by Ernest Solvay. Their presence in the quantum science and technology scene is exemplified by **IMEC**, the international semiconductor and nanotechnologies research center based in Leuven, an equivalent to CEA-Leti in France, with 4000 employees.

**IMEC**'s quantum technology activities are centered on producing superconducting and electron spin qubits on behalf of various laboratories and vendors as well as some cryoelectronics systems. Among other projects, they participate to the European Quantum Flagship QLSI project that is coordinated by CEA-Leti. They announced in August 2021 a partnership with **Xanadu**, for the development of fault-tolerant photonic qubits chipsets based on silicon-nitride.

Let's also mention the **Centre for Quantum Information and Communication** from the Free University of Brussels (Vrije Universiteit Brussel). It works on quantum measurement, quantum entanglement, quantum communication, quantum cryptography and quantum algorithms.

It has also worked on continuous-variable quantum cryptographic protocols, and developed quantum adiabatic algorithms.

In the vendor space, I have identified a company that was already mentioned, **QBee.eu**, a quantum accelerator and incubator created by Koen Bertels, who leads the Quantum Computer Architectures Lab in TU Delft and also works at Qutech.

### Finland

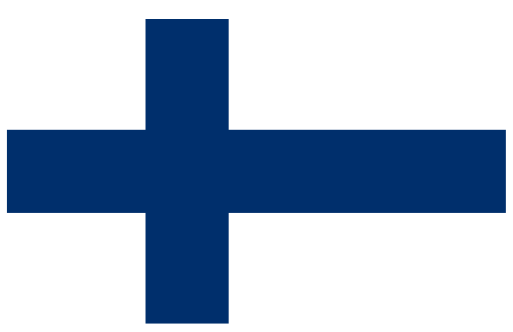

**Finland** has a couple very interesting assets in quantum technology. In research you can count with Aalto University, the University of Helsinki, Tampere University and VTT Technical Research Centre of Finland. The Finnish Quantum Institute federates the efforts of Aalto University, the University of Helsinki and VTT.

It is an organization fostering collaborative research (ResQ) and education particularly with public broad audiences (EduQ) and business adoption of quantum technologies within Finland (BusinessQ).

On the industry side, **Bluefors** is the worldwide leader in low temperature cryogeny used with quantum computers. **IQM** is by far the largest and best funded superconducting qubits quantum computing startup. Other Finish startups are **Algorithmiq**, **Ampliconyx**, **Kronus**, **Qplaylearn**, **Quanscient**, **Quantastica**, **SSH Communications Security**, **Unitary Zero Space** and **Vexlum**.

And you have the **CSC** computing center which hosts the LUMI supercomputer and will put in place an hybrid quantum/classical computing architecture with IQM as part of the EuroHPC HPCQS (High Performance Computer – Quantum Simulator hybrid) program.

In January 2022, Finland and VTT also launched **QuTI** with 10M€ to pool the expertise and resources of four research and eight industry partners over a three year period. It is centered around enabling technologies for quantum computing (materials, cryogeny, electronics, software, systems architecture). The program industry partners are Afore, Bluefors, IQM, Quantastica, Saab, Vexlum, and for companies outside Finland, Picosun and Rockley Photonics. At last, in April 2022, Finland and the USA signed a Joint Statement on Cooperation in Quantum Information Science and Technology very similar to the ones signed with Australia, Sweden, Denmark, Switzerland, Australia, and the UK in 2021/2022. A similar partnership was launched with Singapore in September 2022.

### Denmark

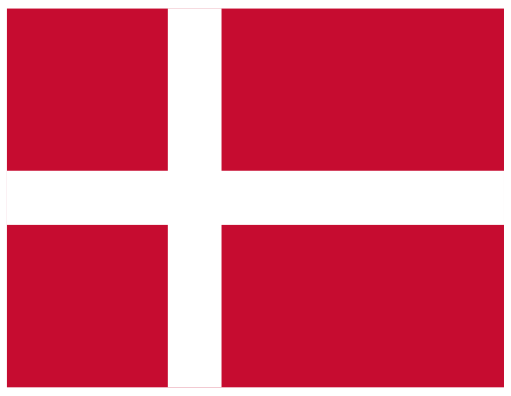

Quantum research in **Denmark** is organized around the Center for Quantum Devices (**QDev)** at the Niels Bohr Institute (NBI) at the University of Copenhagen[2764]. It is a quality laboratory focused in particular on topological qubits, with its director Charles M. Marcus who also worked for Microsoft Research in this field jointly with the MSR teams of Leo Kouwenhoven in the Netherlands (respectively until 2021 and 2022).

QDev is a laboratory of physicists focused on the study of condensed matter, i.e. the physical lower layers of qubits, as can be seen in their publications. The team seems to be only a dozen people. Unfortunately, they cannot then rely on Danish or European industry vendors to consider the transfer of their research into the production of quantum computers. DTU (Danish Technology University) also entertains its QuantumDTU Center for Quantum Technologies.

In April 2022, NATO announced the setup of a NATO Accelerator for Quantum Technologies in Denmark. This innovation accelerator is installed at NBI. This was completed in June 2022 by the announcement of a global quantum partnership with the USA.

The Danish government invested $12M on quantum research between 2017 and 2019. It was completed in September 2022 by the launch of a $200M initiative by **Novo Nordisk** to build a generic quantum computer in 2034, most of the funding going to the Niels Bohr Institute.

### Sweden

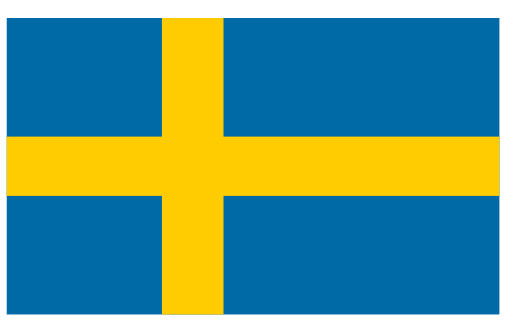

On the **Swedish** side, there is mainly the WACQT (Wallenberg Centre for Quantum Technology) which is part of the **Chalmers University** of Gothenburg and is co-financed by the Wallenberg Foundation. The WACQT has been funded under a 12-year plan with over $100M. As in all countries, the center targets all quantum technologies domains (computing, communications and sensing).

They are invested in superconducting qubits as well as in continuous variable qubits. They plan to create a 100 qubits superconducting computer. WACQT is also working on cold atom qubits from Rydberg... named after a Swedish physicist! Finally, it has launched a "Women in WACQT" initiative to develop gender diversity in quantum science. In March 2021, the Wallenberg budget nearly doubled to $9M per year, allowing the hiring of 40 more researchers.

In April 2022, Sweden and the USA announced a partnership on quantum technology similar to the ones with the UK, Finland, Denmark, Australia and Switzerland.

### Norway

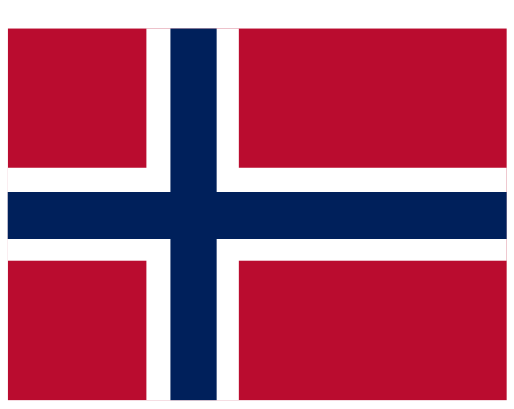

**Norway** feared in 2018-2019 to miss the second quantum revolution. In 2020, it created a Gemini-center on quantum computing with SINTEF, the University of Oslo, NTNU, and in 2021, Simula Research Laboratory, a fundamental research organization belonging to the Norwegian Ministry of Education and Research, where Shaukat Ali leads research in quantum software engineering (among other things)[2765].

[2764] See Quantum technology in Denmark by KPMG, November 2020 (34 slides). This report from a well-known consulting company is highly disappointing. It doesn't mention any research lab besides NBI, any scientist besides Niels Bohr or any startup from the Danish scene. It contains only generalities.

[2765] See QuSBT: Search-Based Testing of Quantum Programs by Xinyi Wang, Paolo Arcaini, Tao Yue and Shaukat Ali, April 2022 (5 pages).

The project was led by Franz Georg Fuchs from SINTEF, an independent research organization with the goal to make Norway "quantum ready".

Later in 2021, the Norwegian Quantum Computing Centre was created consolidating research from 13 scientists from three partner institutions (SINTEF, the Norwegian University of Science and Technology *aka* NTNU and the University of Oslo) including Jeroen Danon from the Center for Quantum Spintronics at NTNU[2766].

**Ireland**

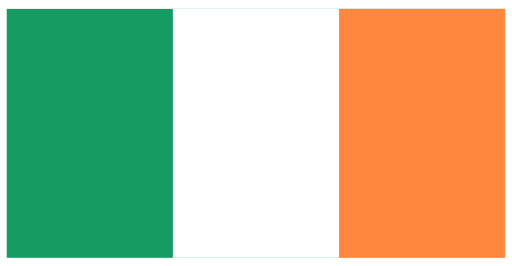

In Ireland, a first quantum computing initiative (QCoIr) was launched in 2020 with a funding of $11M. It included global companies like IBM and Mastercard, plus the Tyndall National Institute in Cork. It established a Quantum Center of Excellence.

**Italy**

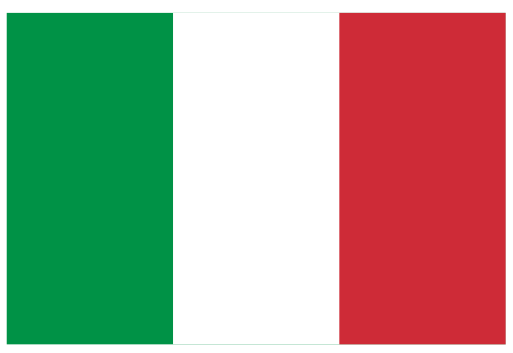

Italy has a very active research in place in various technologies in quantum computing. Photon qubits are explored by **Fabio Sciarrino** at Università La Sapienza in Rome. He is an European pioneer of boson sampling experiments and wants to make it programmable. And the **Università di Padova** launched its own Quantum Technologies Research Center that works on trapped ions computing.

**Francesco Tafuri** from the Università Federico II in Naples works on superconducting qubits. The Italian National Institute for Nuclear Physics (INFN) is also working with the US DoE on superconducting quantum materials at the FermiLab in Chicago, which happens to be run by Anna Grassellino, an Italian.

In the quantum communication realm, **Paolo Villoresi** from the Instituto Nazionale di Ricerca Metrologica in Turin pioneered photons polarization encoding with a satellite in 2015. Italy also deployed its **Italian Quantum Backbone** (IQB) with a total of 1,850 km fiber link based on commercial fibers. It connects INRIM's premises in Turin to Milan, Bologna, Firenze, Rome, Napoli, Pozzuoli and Matera. From Turin, a 150 km fiber reaches Modane in France, and connects to Grenoble, Lyon and Paris, then Europe.

The public supercomputing center **CINECA** entertains a quantum computing lab. It tests the capacities of various quantum computers, develops quantum algorithms and hybrid solution associating classical supercomputers and quantum accelerators.

The big shortcoming of Italy is its weak private sector with not many industry vendors and startups engaged in quantum technologies.

As part of its recovery plan announced in April 2021, the Italian government allocated a budget of 1.6B€ to fund 7 new research organizations, one of these being focused on quantum technologies[2767].

---

[2766] See Protected Solid-State Qubits by Jeroen Danon et al, October 2021 (6 pages).

[2767] See Italy's quantum scientists jostle for a superposition by Francesco Suman, April 2021.

### Spain

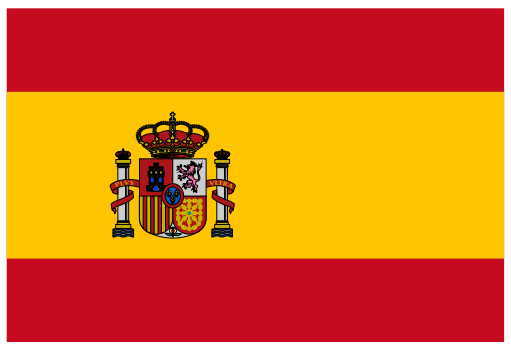

On the research side, most of Spain’s efforts are concentrated in the **ICFO** (Instituto de ciencias fotónicas) of Barcelona, which is mainly specialized in photonics. Other research in quantum is carried out at the Quantum Information and Computation Laboratory (GIC-UB) of the **University of Barcelona** as well as the **Autonomous University of Barcelona**[2768].

The **IFAE QCT** is the Quantum Technology Group from the IFAE (Institut de Fisica d’Altes Energies) from the Autonomous University of Barcelona opened its new lab and fab in October 2020.

On the startup scene, they have a startup, **Qilimanjaro** already mentioned, which develops mainly a cloud-based quantum software platform and a superconducting quantum annealer. Their chipset is manufactured at the IFA. And they have **Entanglement Partners,** a service provider that is clearly succeeding in selling quantum-related cybersecurity services.

They also animate the country's ecosystem, do evangelization and organize events. In 2017, the open innovation platform **Open Trends** launched **The Carrot Cake** to encourage projects in the quantum field. This complements the **Barcelona QBIT** think tank launched in 2015 and the **Quantum World Association** launched in 2017, which brings together Switzerland, Canada, Australia and Catalonia with startups such as ID Quantique, evolutionQ, h-bar and Entanglement Partners. Spain is networking, having realized that it could not go very far on its own.

In February 2022, UMIQ aQuantum-UEx was created, a Joint Quantum Software Research Unit between the Spanish quantum software and engineering company aQuantum and the University of Extremadura (UEx) consolidating a partnership that started in March 2020. They are collaborating on software research projects including the “QHealth: Quantum Pharmacogenetics Applied to Aging” project.

Seven Spanish quantum companies (Amatech, BBVA, DAS Photonics, GMV, Multiverse Computing, Qilimanjaro Quantum Tech and Repsol), five research centers (BSC, CSIC, DIPC, ICFO and Tecnalia) and the Universitat Politècnica de València launched in 2022 the CUCO Project to foster quantum computing research and development in Spain, particularly in the industry.

But regionalism works well in Spain as well, as another similar initiative was launched in the Basque country by the Provincial Council of Bizkaia. The goal is to foster the adoption of quantum computing in the industry. Participating members are Technalia (again), IBM, Telefónica, Accenture, the Bilbao City Council, Gaia, Silicon Europe, the UPV/EHU (the University of the Basque Country), the University of Deusto in San Sebastien and Mondragon Unibertsitatea (east of Bilbao).

### Portugal

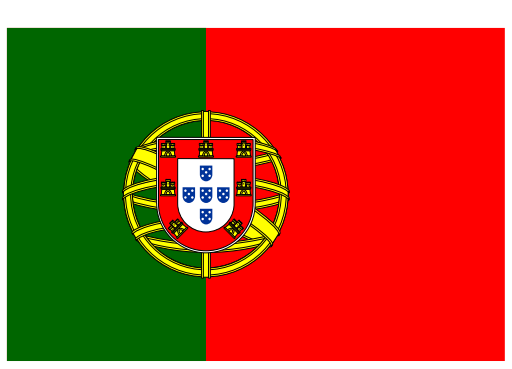

**Portugal**’s key investment in quantum technologies sits with **QuantaLab**, a collaborative research center launched by **International Iberian Nanotechnology Laboratory** (INL) and the **Universidade do Minho**, both in Braga, Portugal. It focuses on quantum materials and quantum technologies. Portugal is also participating to European quantum projects including the QCI (Quantum Communication Infrastructure).

[2768] See Quantum Technologies in Catalonia, July 2019 (43 slides) which describes very well the quantum ecosystem of this key region of Spain.

### Poland

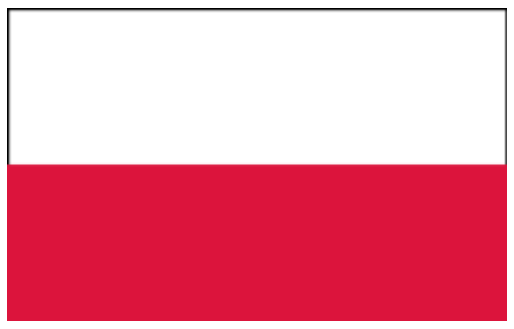

**Poland** launched its National Quantum Information Centre in Gdansk (KCIK) in 2007 with 9 research institutions. Other interesting labs are the Quantum Physics Research Center focused on quantum cryptography and the International Center of Theory of Quantum Technologies (ICTQT).

The University of Warsaw is also very involved in quantum research, particularly in photonics with the Center of Quantum Optical Technologies (QOT), led by Konrad Banaszek[2769].

The Polish National Science Centre also coordinates the international research network **QuantERA,** itself funded by the European Union's Europe 2020 budgets. It does this in coordination with the French ANR. The countries involved, in addition to those of the European Union, are Switzerland, Israel (Bar-Ilan University) and Turkey. About thirty research projects had been funded after a call for proposals in 2017, some of which were subsequently funded in the European Quantum Flagship, such as SQUARE. They are all quantum physics projects (photonics, cold atoms, ...).

Some other Polish quantum research groups worth mentioning are The Quantum Research Group from the Polish Academy of Sciences which works on quantum computing, qubit measurement, error mitigation and software engineering and the Quantum Resources Group from the Jagiellonian University in Kraków that was created in 2020 and works on quantum information science.

In the ecosystem, The Quantum AI Foundation was created in 2019 by Paweł Gora. It organizes meetings of the Warsaw Quantum Computing Group (WQCG) and hackathons.

### Hungary

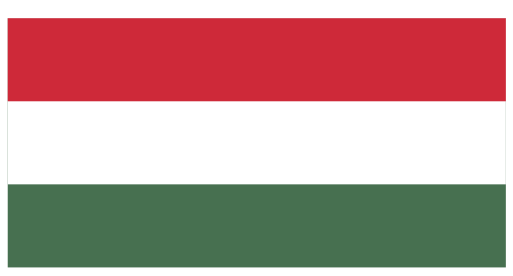

**Hungary** launched in 2018 its quantum plan with a funding of about $10M, consolidated in a new consortium named HunQuTech consolidating the countries quantum research groups from various institutions and industry vendors.

The participants are the Wigner Research Centre for Physics, the Institute of Physics the Faculty of Electrical Engineering and Informatics of the Budapest University of Technology and Economics, the Institute of Physics of the Eötvös Loránd University, completed by industry partners (Bonn Hungary Electronics, Ericsson Hungary, Nokia-Bell Labs, and Femtonics). The plan goals were to work on single photon sources for quantum telecommunications, pairs of entangled photons at telecom wavelength, free-space QKD systems, quantum memories, qubits and quantum gates and new quantum algorithms. Researchers also work on QRNG and PQC. In total, Hungarian researchers got 4 ERC from the EU in quantum physics and information science.

In 2022, the country launched the Quantum Information National Laboratory initiative that consolidates quantum research covering quantum computing and quantum telecommunications. Hungary also participates to the EuroQCI project since 2019. The country seems to have several intra-European partnerships like with Germany and the Netherlands.

In 2022, the Faculties of Science and of Informatics of the Eötvös Loránd University selected QuiX as a vendor for a photonic circuit to build their own research photonic quantum computer.

---

[2769] They however have not created the "first world's first quantum processor" as claimed in One-of-a-kind: Warsaw-based scientists build groundbreaking quantum processor by Jo Harper, The First News, February 2022. The related paper is Optical-domain spectral super-resolution via a quantum-memory-based time-frequency processor by Mateusz Mazelanik, Adam Leszczyński and Michał Parniak, Nature Communications, February 2022 (12 pages). It is about a (rather interesting) high-resolution spectrograph.

### Switzerland

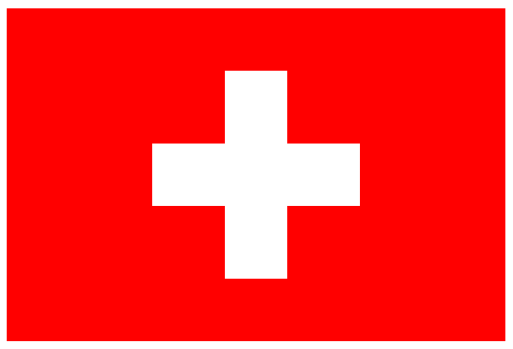

Switzerland is also mobilized on quantum technologies, particularly at **ETH Zurich**, which is collaborating with IBM and especially on quantum cryptography, notably with its startup **IDQ,** which is a leader in quantum random numbers generation used in quantum cryptography and elsewhere. And also with the Lausanne's **EPFL**.

The country has published a manifesto to promote its research and industrial efforts in quantum, Switzerland: At the Quantum Crossroads. The **Swiss Quantum Hub** brings together the Swiss quantum ecosystem.

The **Quantum Science and Technology Initiative** (QIST), a joint initiative of ETH Zurich and the University of Basel, which also involves the University of Geneva and EPFL Lausanne, has 34 faculty members and 300 students. It has been funded with $120M between 2010 and 2017.

It covers all the usual fields of quantum with, with a particular effort in quantum telecommunications. In August 2021, the EPFL launched though its own multidisciplinary Quantum Science and Engineering Center consolidating its research and academic efforts in all branches of quantum technologies.

The **Swiss Quantum Investor Club** was created in 2020 to link investors and quantum entrepreneurs and organize events in Geneva, Lausanne and Zurich, as well as the **Swiss Quantum Hub,** a think tank and accelerator for quantum startups, and the Quantum Computing Garage, a permanent hackathon.In November 2020, Martin Haefner, an alumn from ETH Zurich donated $44M to the ETH Foundation to have them build a quantum research facility. We could wish more wealthy people would make such long-term investments for their community! In another similar initiative, ETH Zurich and the Paul Scherrer Institute (PSI, which has its own proton accelerator and an electron synchrotron, the Swiss Light Source, that is equivalent to SOLEIL in France) created the Quantum Computing Hub in 2021, a joint quantum computing research center, focused on ion traps and superconducting qubits with the goal to host 30 researchers. ETH Zurich invested $36M there[2770].

In October 2022, Switzerland and the **USA** signed yet another bilateral quantum cooperation agreement[2771].

### European Union

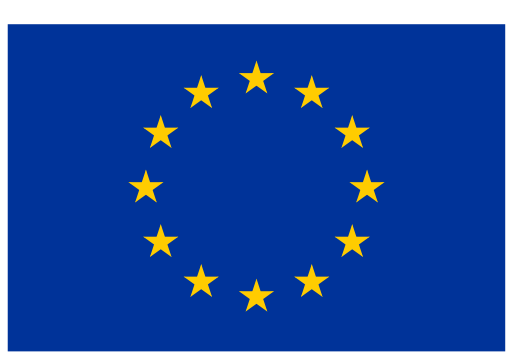

The **European Union** wants to consolidate its effort in quantum technologies. A "flagship project" germinated in 2016[2772] and was formally launched in October 2018 to fund collaborative research on all aspects of quantum information: sensing, communications, computing and simulation[2773].

---

[2770] See ETH Zurich and PSI found Quantum Computing Hub, May 2021.

[2771] See Joint Statement of the United States of America and Switzerland on Cooperation in Quantum Information Science and Technology, US State Department, October 2022.

[2772] It started with a Manifesto prepared by European quantum physics researchers. See Quantum hocus-pocus by Karl Svozil, 2016 (6 pages), a critic of the underlying scientific overpromises from the Quantum Manifesto published by European researchers, which led to the creation of the European Quantum Flagship program.

[2773] See the motivations behind the European Flagship: The Impact of quantum technologies on the EU's future policies: Part 1 Quantum Time, 2017 and Part 2.

It is theoretically endowed with 1B€ to be used for the development and diffusion programs of quantum technologies, spread over 10 years. In theory, because the budgets have not really been allocated at this level by the European Union. This Flagship is currently mainly focused on quantum computing fundamental physical. It has not yet looked at algorithms and software.

But the European Quantum Flagship is far from being the sole source of EU funding for quantum research and technologies. You also have to embed the ERC grants for individual researchers, other multi-partite projects (Europe Next, Qureca, H2020, Europe next) and startups funding via the EIC Accelerator.

This **Quantum Technologies Flagship** is one of the three European "flagships" that aim to place Europe at the forefront of major technological breakthroughs with strong community investment in research. The two other flagships are the "Human Brain Project" led by the Swiss Henri Markram and the Graphene project in nanotechnologies. The first phase of the Flagship included €132M spread onto 20 projects selected out of 140 applicants and for a period of three years. 130 additional projects will be later selected.

Launched by the European Commission on October 29, 2018 in Vienna (videos), the program covers the four usual quantum domains: computing, simulation, communication and sensing[2774].

Let's look at these projects. These projects involve an average of at least half a dozen countries, even partner countries like Switzerland and Israel.

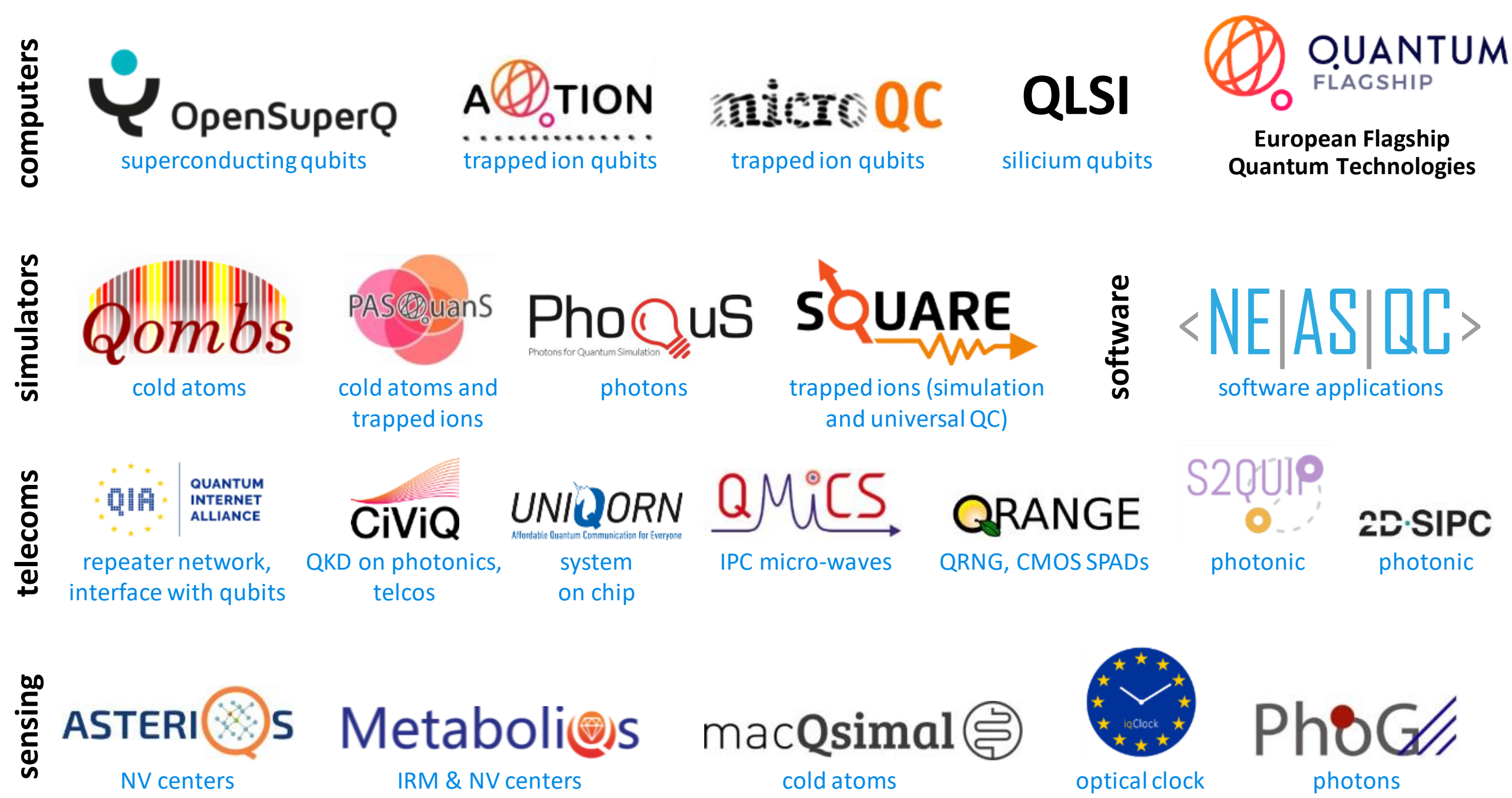

*Figure 863: the European Quantum flagship projects as of 2022. (cc) Olivier Ezratty, 2022.*

It starts with three side projects related to quantum computing:

- **AQTION** (Austria, €9.57M) is trapped ions qubit computer project, planning to reach 50 qubits. Austria has a long history here and is quite legitimate. Atos is participating in this project.

[2774] See the Press Kit (28 pages), the complete list of projects and Europe Accelerating the Industrialization of Quantum Technologies, October 31, 2018, the title of which is somewhat misleading in that the majority of projects funded are research projects and not industrialization projects. And then there is The quantum technologies roadmap: a European community view, October 2017 (25 pages), which takes stock of the state of the art in Europe and around the world. See also The EU Quantum Technology Flagship by Elisabeth Giacobino, 2018 (41 slides and video).

- **MicroQC** (Bulgaria, €2.36M) plans to create another trapped ion computer.
- **OpenSuperQ** (Germany, €10.33M) is a superconducting qubit computer project led by Saarland University that also involves Spain, Sweden, Switzerland and Finland and a total of 10 research laboratories. The ambition is to create a 100-qubit system. IQM is the probable vendor who will provide the quantum system for this project.

Then we have four quantum simulator projects:

- **PASQuanS** (Germany, €9.25M) is a quantum simulator project based on cold atoms and trapped ions up to 1000 qubits. It also involves the UK, Atos and Pasqal.
- **PhoQuS** (France, 3M€) is a photonic based quantum simulator. It is led by a team of PSL researchers. It involves the use of polaritons.
- **Qombs** (Italy, €9.3M) is another photonics-based quantum simulator project.
- **SQUARE** (Germany, €2.99M) a quantum simulator project using trapped ions. It is led by the University of Karlsruhe and involves laboratories from Denmark, Sweden, Spain and France, including Thales. It seems that they are also seeking to create a quantum processor with universal gates.

Let's continue with projects in quantum communication and telecom security.

- **Quantum Internet Alliance** (Netherlands, €10M) (QIA) aims at deploying an Internet network protected by quantum key distribution (QKD) in mesh network mode and not just point-to-point. The quantum nodes or relays will be made up of systems using cold atoms. They will start with a three or four-node network. The project is led by TU Delft University. The CNRS participates in it, notably Eleni Diamanti, Elham Kashefi and Iordanis Kerenidis. The Sorbonne University also participates. Other participants include Swiss, Germans, Danes and Austrians (complete list).
- **QRANGE** (Switzerland, €3.87M) is a project to improve quantum random number generation techniques.
- **CiViQ** (Spain, €9.9M), or Continuous Variable Quantum Communications, is another QKD-based fiber telecommunications security project. The project involves 21 stakeholders covering the academic and industrial world, including CNRS, Institut Mines-Telecoms, Nokia Bell Labs France, Inria, Orange, as well as Mellanox.
- **Uniqorn** (Austria, €9.9M) is in the same niche and is working on a random number generator and a QKD system. It associates 17 organizations from 9 countries (Austria, Netherlands, Italy). The Israeli Mellanox is also involved there.
- **S2QUIP** (Netherlands, 3M€), Scalable Two-Dimensional Quantum Integrated Photonics, is another QKD-based secure communication project.
- **2D-SIPC** (Spain, €2.9M) is a project for the development of photoelectronic components made for networks secured by QKDs.
- **QMICS** (Germany, 3M€) or "Quantum Microwave Communication and Sensing" is about creating a microwaves-based links and networks between superconducting network nodes with applications in distributed quantum computing and also in quantum sensing.
- **NEASQC** (NExt ApplicationS of Quantum Computing) is a collaborative project launched in September 2020, to develop practical applications of NISQ (noisy quantum computers, an intermediate step before scalable quantum computers). It is an H2020 project that brings together European players including Atos, Total, EDF, the Loria laboratory from the University of Lorraine, Astrazeneca, HQS Quantum Simulations, HSBC and the University of Leiden (Netherlands).

- **QLSI** (France) is a new Quantum Flagship awarded in March 2020 to fund four years of fundamental research in silicon qubits. It is being driven by the Grenoble team under the responsibility of Maud Vinet at CEA-Leti. The project funding is 14M€ spread over 19 organizations: Atos, STMicroelectronics, SOITEC, CNRS Institut Néel, TU Delft, University of Twente and TNO in the Netherlands, IMEC in Belgium, UCL and Quantum Motion in the UK, Infineon, RWTH Aachen, University of Konstanz, Fraunhofer and IHP Frankfurt in Germany, University of Copenhagen and University of Basel.
- **QTEdu** (Italy) is about creating the quantum education ecosystem in the European Union. It's funded by H2020[2775].
- **QFLAG** (Germany, €3.48M) was the project managing the coordination for European Quantum Flagship projects. It was followed by the project **QUCATS** that covers the 2022 to 2025 period and is coordinated by Philippe Grangier (CNRS, France).

Then we have five quantum sensing projects already seen.

We note the strong predominance of projects piloted by German research laboratories (5), followed by France (4), the Netherlands (3), Spain (2), Austria (2), followed by Italy, the UK and Switzerland, all driving a single project. Large countries are present in many of these projects. As an example, France is involved in many of these projects. CNRS (France) alone is involved in 14 of the 20 projects[2776]. These projects do not yet include efforts in software, to create algorithms, development tools and business software solutions adapted to quantum computers. Such projects will probably be funded in subsequent phases[2777].

But other European quantum projects are funded with other vehicles than the Flagship.

**QuantERA** I (2014) and II (2021) is an alliance of research funders from member states created to reinforce transnational collaborations in inspiring multidisciplinary quantum research. The QuantERA II Consortium assembles 38 Research Funding Organizations from 30 countries, some being extra-EU. It complements the Quantum Flagship in early stages, serving as an incubator of new ideas which then can get integrated in Quantum Flagship projects but comes from participating countries (45M€ for QuantERA I and 40M€ for QuantERA II) and the EU (11.5M€ for QuantERA I and 15M€ for QuantERA II). There were two calls for projects in QuantERA I and one in QuantERA II with a selection done late 2021 and projects funding starting in 2022.

**EQUIPE** (Enable Quantum Information Processing in Europe) project aims to advance the industrialization of the creation of quantum computing and telecommunications solutions for industry[2778].

**EuroHPC** projects include quantum computing deployments in hybrid datacenters with first deployments of quantum simulators planned in Germany and France as part of the HPCQS project. In October 2022, six other EU HPC sites were selected for these deployments in Czechia, Germany, Spain, France, Italy and Poland with a total investment of 100M€.

---

[2775] See Expanding the European Competence Framework for Quantum Technologies, January 2022.

[2776] See New Strategic Research Agenda on Quantum technologies, February 2020 (114 pages) which details the state of play of the European Quantum Flagship projects.

[2777] See The Quantum Technologies Flagship: the story so far, and the quantum future ahead by Thomas Skordas and Jürgen Mlynek, October 2020 which looks at the flagship progress two years after the program started.

[2778] See Simulation on / of various types of quantum computers by Kristel Michielsen (40 slides).

European research is federated under the umbrella of **QCN** (Quantum Community Network). Its industry counterpart is the **QuIC** (Quantum Industry Consortium) announced in June 2020 and formally launched in April 2021[2779].

Founding members are companies that were involved in at least two European Quantum Flagship projects. They include Bosch, SAP, Atos, Thales, Muquans, Airbus and many others.

The consortium has an extensive work plan covering market needs assessment, analysis of the quantum technology value chain, development of standards and regulations, sharing of best practices in intellectual property protection and market evangelization, access to infrastructure, linking startups and investors, skills development issues and coordination with public authorities.

But there is another association, **|QBN⟩**, the Quantum Business Network, another European Quantum Community, which connects the industry, users, vendors and research.

**LSQuanT** is an initiative funded by the EU that was launched in 2021 and is dedicated to promoting "large-scale quantum transport methodologies", which deals with the physics of quantum transport digital simulation, to invent new quantum materials and devices[2780].

**MATQu** (Materials for Quantum Computing) is developing an European value chain to manufacture superconducting qubits. This H2020 project running from 2021 to 2024 with a total cost of 21M€ with EU funding of 6,5M€ is led by Fraunhofer Mikroelectronics, IAF and IPMS with the participation of CEA-Leti, IMEC, Soitec, BE Semiconductor Industries (semiconductor assembly equipment), IQM, VTT, Keysight, Siltronic (silicon wafers production), Kiutra, Atos, Mellanox, Beneq (atomic layer deposition equipment), Orange Quantum Systems and Technic France (engineering).

**SPROUT** (Scalable Platform for Quantum Technology) is a project launched in November 2021 by Delft Circuits and kiutra to provide a scalable cryogeny platform funded by the EU Eurostars program. They develop a demonstrator for a <1K cryogent platform based on cryogen-free magnetic cooling (the Kiutra specialty) and a multi-channel electrical cabling.

## Russia

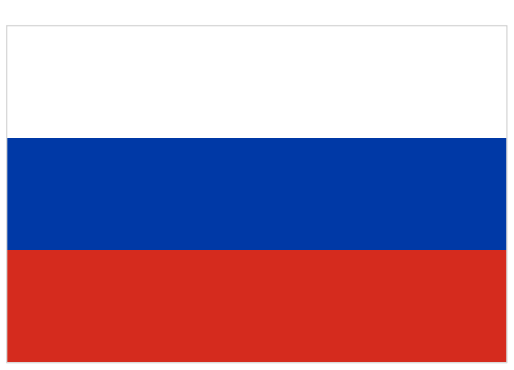

Russia is not very visible in the quantum scene, maybe because they have not built the same research and industry partnership that are seen in the western world. But like with AI, its government realized that quantum technologies were critical for sovereignty. In December 2019, Russia announced its own plan of attack on quantum technologies, which seemed very focused on military, intelligence and cryptanalysis applications[2781].

This plan got a five-year funding of $790M. In practice, it covers almost all fields of quantum technologies[2782]. In January 2022, Russia created it National Quantum Laboratory (NQL) run by Rosatom and with NRU HSE and MIPT, MISIS, P. N. Lebedev Physical Institute of RAS, Russian Quantum Center and the Skolkovo Foundation. The center will host a nano-fabrication center of 2,000 $m^2$ in Skolkovo. One might wonder how things will fare with exports restrictions to Russia after it started its war against Ukraine in February 2022. Many Western countries equipment vendors won't be in position to sell their hardware and Russia won't be able to count on China this time.

[2779] See Announcing the creation of the European Quantum Industry Consortium by Laure Le Bars (SAP), the first President of QuIC, April 2021.

[2780] See a related review paper: Linear scaling quantum transport methodologies by Zheyong Fan et al, December 2020 (61 pages).

[2781] See Russia joins race to make quantum dreams a reality by Quirin Schiermeier, December 2019.

[2782] Source: Quantum communication in Russia: status and perspective by Vladimir Egorov, 2019 (22 slides).

In January 2022, Rosatom announced a plan to build a trapped ions quantum computer that would be made available on the cloud by 2024. It is being developed by the Russian Quantum Center and the P.N. Lebedev Physics Institute of the Russian Academy of Sciences. They are starting with 4 qubits and, as such, are very late compared to state of the art quantum systems coming from AQT, IonQ and Quantinuum who have about 20 qubits in-store.

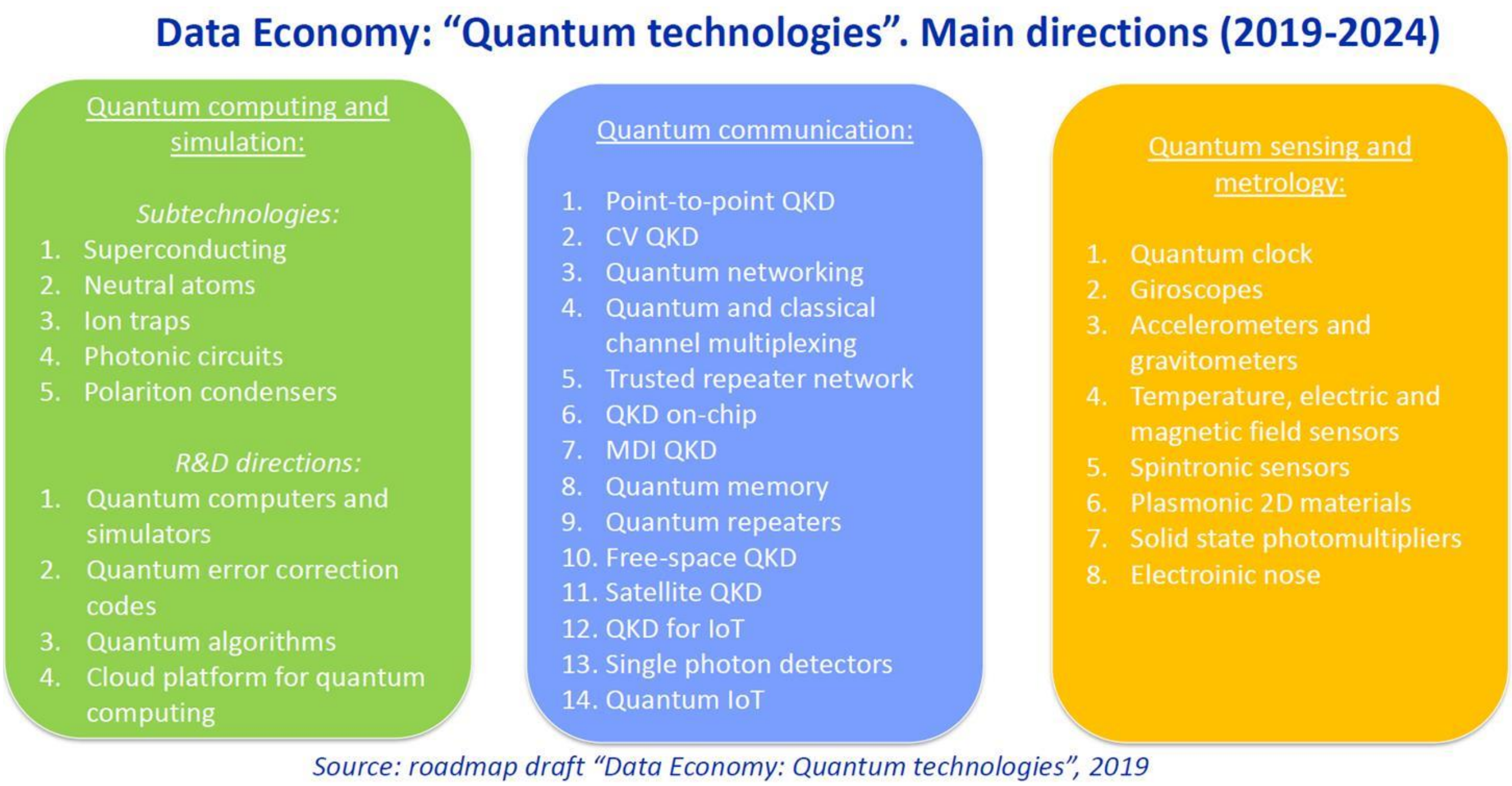

*Figure 864: Russia's quantum plan priorities as of 2019. Source: Quantum communication in Russia: status and perspective by Vladimir Egorov, 2019 (22 slides).*

Before all of that, the **Russian Quantum Center** was created in 2010, a private research center dedicated to the various application areas of quantum computing, including quantum cryptography. It employs over 200 researchers.

It covers many quantum computing branches: superconducting, trapped ions, photons and NV centers qubits, quantum sensing, QKD and a single photon detector. They collaborate with some international research organizations in the USA (MIT), Canada (University of Calgary), Germany (Max Planck Institute for Quantum Optics) and UK (University of Bath)[2783].

The St. Petersburg **ITMO University** has a QKD research laboratory as well as the **Kazan Quantum Center** which has deployed a QKD on a 160 km network in Kazan. The country also plans to launch a QKD quantum key communication satellite in 2023. A few other laboratories are invested in quantum technologies such as the **NTI Center for Quantum Communications** at MISiS University and the **NTI Technologies Centre** at Moscow University.

Industry wise, they are mostly in quantum cryptography with only one startup, **Qrate Quantum Communications,** the others being established companies, such as **Infotecs**, **Scontel** and **Smarts QuantTelecom**[2784].

[2783] This information comes from Evaluation Report of Russian Quantum Center, 2017 (7 pages). See also Quantum technologies in Russia, October 2019 (9 pages).

[2784] Here are some elements on this ecosystem: Quantum communication in Russia: status and perspective by Vladimir Egorov, 2019 (22 slides).

# Africa, Near and Middle East

### Israel

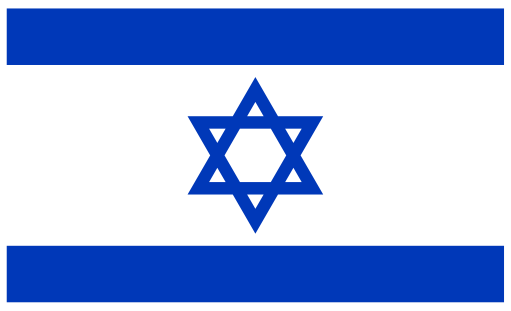

**Israel** was relatively quiet about quantum technologies until 2018 apart from Gil Kalai from the Hebrew University of Jerusalem who, since 2013, has shown a deep-rooted skepticism about the future of quantum computers. I was relatively surprised in 2018 to find out that the country was not very visible in the quantum research and entrepreneurship.

It was a stark contrast with other fields where this relatively small country has a significant impact worldwide: in software, artificial intelligence (where they have in excess of 400 startups), Internet, semiconductors, electronics, medtechs and biotechs to name a few.

#### *Government funding*

Then, things started to change[2785]. After a study carried out in 2017 by Uri Sivan from Technion to evaluate the country's quantum technologies efforts, a first initiative to better fund its research was launched in 2018 by the country's government and endowed with 75M€. It went mainly to Technion and the University of Haifa, which wants to design its own quantum computer and had also received a donation of $50M.

In December 2019, a panel of specialists commissioned by the government proposed a plan to invest $350M over 6 years in quantum technologies[2786]. In just a few months, the government approved this proposal which ended up with a funding of $400M, 60% of it supposed to fund academic research. The plan is fairly standard with an investment in human capital (faculty hiring and launching training courses), research and scholarships funding, attracting foreign researchers and the likes. The usual quantum technologies domains were picked with computing hardware and components, telecommunications, cryptography and sensing (with $40M), particularly with its military applications.

The Israel National Quantum Initiative (INQI) plan follows-up in naming the US plan announced in December 2018.

In March 2021, Israel announced it planned to create its own quantum computer, allocating a budget of $60M taken out of the national initiative. The goal is to create 30- to 40-qubit quantum systems. It was to take bids from both local players and international vendors, with a build or buy approach depending on the outcome. In July 2022, the government announced the creation of a quantum computing R&D center with a budget of $30M over a three-year. It selected their local star vendor, Quantum Machines to create the center that is supposed to work on superconducting qubits, cold atoms, trapped ions and quantum optics computers[2787]. The company will work with Classiq and Elbit Systems. Still, the ploy became a "buy" with Israel announcing the same month the procurement of an Orca Computing (UK) photonic quantum computer. They also work with ColdQuanta who already partners with Quantum Machines (like Pasqal in France) and Classiq[2788].

---

[2785] But not to the extent of this title; Israel has become a powerhouse in quantum technologies by David Kramer, Physics Today, December 2021 (4 pages).

[2786] See Israel joins the quantum club by Uri Berkovitz, December 2019 and Israel joins the race to become a quantum superpower by Anna Ahronheim and Maayan Hoffman, Jerusalem Post, December 2019.

[2787] See Quantum Machines to establish Israeli quantum computing center in $30 million deal by Meir Orbach, CTECH, July 2022.

[2788] See Inside ColdQuanta's Role in the Israel National Quantum Initiative (INQI) by Brian Siegelwax, The Quantum Insider, August 2022.

### *Research*

There are about 125 “principal investigators” in quantum technologies research in Israel spread in the following research institutions:

**Ariel University** with its Wireless Communication & Radars Lab, located in the Israeli settlement of Ariel in the middle of the Palestine West Bank. They work on millimeter wave and Terahertz sensors.

**Bar Ilan University** (Ramat Gan near Tel Aviv) with the Photonics and Optics group (Avi Peers, Eli Barkai and Dror Fixler) and the Institute of Nanotechnology and Advanced Materials which works on superconducting qubits (Michael Stern). These are integrated in QUEST (Quantum Entanglement in Science and Technology), a quantum research center launched in 2017.

**Ben-Gurion University of the Negev** (Beer Sheva) with the Sensing Technologies Lab (Asaf Gros), Quantum Magnetometry Group (Reuben Shuker) and the Atom Chip Group (Ron Folman). The startup AccuBeat (1993) which produces rubidium quantum atomic clocks, is a product of this university.

**Hebrew University of Jerusalem** with the Nano-Opto Group and Time Dissemination Group, and the Center for Nanoscience and Nanotechnology that works on superconducting qubits. The University also established its Quantum Information Science Center in 2013 to focus on secure quantum communication (QKD).

**Technion** (Haifa) with Quantum Information Processing lab (Tal Mor) which works on photonics, NMR and silicon qubits, the Russell Berrie Nanotechnology Institute created in 2005 (Gadi Eisenstein) and the Helen Diller Quantum Center which is working on photonics, quantum dots, superconducting qubits and cold atoms (Yosi Avron).

**Tel Aviv University** with its QuanTAU, their Quantum Science and Technology Center, working among other things on superconducting qubits.

**Weizmann Institute of Science** in Rehovot with the Center for Quantum Science and Technology (Adi Stern), the Quantum Circuits Group (Serge Rosenblum) working on superconducting qubits and the trapped ions group (Roee Ozeri) which created the first Israeli quantum computer in 2022 with 5 trapped ions qubits[2789].

### *Quantum industry*

The most visible startup in the field is **Quantum Machines** that sells qubit control hardware and software. Other Israeli startups and vendors in quantum technologies are Accubeat, ClassiQ (error control), Elta (sensing), Hub Security (PQC in HSMs), Mellanox Technologies (quantum communications, part of Nvidia), PhotoniQ (cold atoms interconnection), Qedma Quantum Computing (software), QuantLR (QKD) and Quantum Source Labs (photonic computing), Raicol Crystals (photonics), Tabor Electronics (RF electronics). Google's R&D lab in Tel Aviv also hosts researchers in quantum computing.

The **Quantum Technologies Consortium** created in 2019 assembles research institutions and industry vendors.

[2789] See A huge leap: Israeli researchers build country’s first quantum computer by Ricky Ben David, Times of Israel, March 2022 and Trapped Ion Quantum Computer with Robust Entangling Gates and Quantum Coherent Feedback by Tom Manovitz, Yotam Shapira, Lior Gazit, Nitzan Akerman and Roee Ozeri, PRX Quantum, March 2022 (14 pages). These qubits use strontium. Their fidelity if 99.64% for a single qubit gate and 97.3% for two-qubit gates, which is not stellar.

### Iran

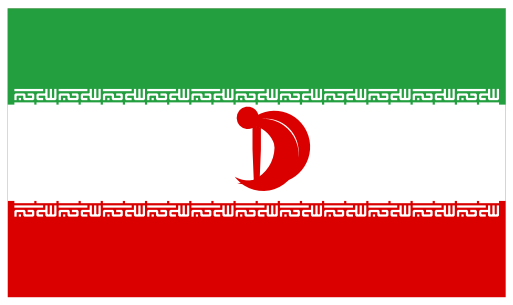

Israel is not the only country in the Near and Middle East that seems to be invested in quantum research. **Iran** is also involved with at least two research laboratories, **Sharif University** which is working on quantum physics in partnership with Canada and the **Quantronics Lab** of the Iranian Technological University which is dedicated to quantum communication (QKD)[2790].

The country even organizes its conference on quantum computing, the **IICQI,** since 2007[2791].

### United Arab Emirates

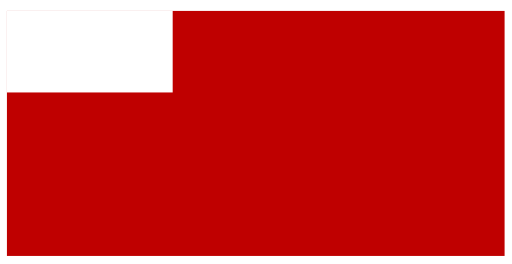

Each and every country wants « its » quantum computer. Even the emirate Abu-Dhabi got the quantum virus and decided to "build" its own quantum computer, even if "build" or "buy" are interchangeable in such a situation since it is a quantum annealing superconducting system coming from Qilimanjaro.

Still, it comes after the establishment of a Quantum Centre at the Technology Innovation Institute (TII) which hosts about 20 researchers coming from the Emirates and from various countries: Italy, Spain, Brazil, Greece, UK and Germany. This lab complements other labs in robotics, cybersecurity and energy. It even has some qubits manufacturing tooling.

Jose Ignacio Latorre is the chief scientist of this quantum research laboratory. He is a professor of theoretical physics at the University of Barcelona currently on leave, cofounder of Qilimanjaro and the director of the Singapore CQT since July 2020. Their key partners are Qilimanjaro, Universitat Catania in Sicilia and INFN, an Italian research network.

The QC-TII organized a webinar conference in June 2021, Atomtronics@Abudhabi with about 500 participants.

**Intqlabs** (2022, Dubai) is a contract research company created by Ankur Srivastava, from India, formerly an independent researcher, not affiliated with any lab. The company works on various fields: quantum computing, reversible classical computing, geo-magnetism and cybersecurity. They are talking about some form of "quantum and reverse computing" that would solve all current problems with quantum computing.

It's quite difficult to figure out whether it's some form of quantum computing or of classical reversible computing. All of this is patent pending, meaning that for at least a year and a half, you will have no idea what it's all about. Wait and see. They plan to license their technology to hardware manufacturers. They have otherwise created NGNSS (New Global Navigation Satellite System), an alternative to classical GPS solutions which as its name doesn't tell, is not requiring a satellite to function and uses magnetism detection and mapping.

### Qatar

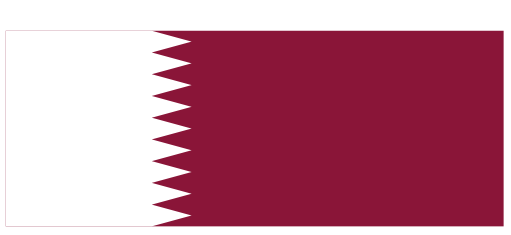

**Qatar** has also some ambitions in quantum technologies. In 2022, Hamad Bin Khalifa University (HBKU) announced the creation of the Qatar Center for Quantum Computing (QC2).

It received a $10M research grant from Barzan Holdings, a local defense industry vendor.

---

[2790] Source: [Iranian research in quantum information and computation](), June 2016.

[2791] See [http://iicqi.sharif.edu/](http://iicqi.sharif.edu/).

Let's also mention **QUANTUN**, the Quantum Tunisian Network launched in October 2021 to consolidate the work of quantum researchers in the country and **South Africa**'s Quantum Initiative (Sa QuTI) launched in September 2021.

# Asia-Pacific

### China

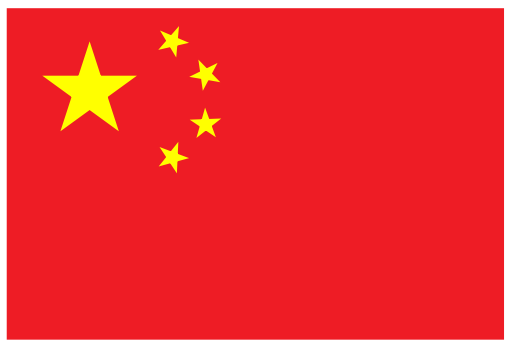

As in many technology sectors, **China** is loudly and clearly asserting its ambitions and power in quantum technologies[2792]. As in the UK, this investment was taken in hand early on by the executive and as early as 2013 with the involvement of Xi Jinping, the Chinese president, during a visit to the Anhui laboratory, focusing on quantum cryptography, combined with a training session.

#### *Government funding*

In 2015, Xi Jinping integrated quantum communication into the country's scientific priorities, in 13th plan covering 2016-2020. Maybe was it a benefit from having a government comprising a majority of politicians with some scientific background and also the result of Snowden's revelations on NSA's spying capabilities in 2013.

The amounts invested in quantum were respectively $160M in the 11th plan covering the period 2006-2010, $800M in the 12th plan covering 2011-2016 and $320M in the 13th plan starting in 2016, supplemented by $640M in funding from the regions[2793]. Later on, the Chinese government communicated an amount of $34B corresponding to several scientific priorities including quantum. This represented probably less than $1B between 2016 and 2020 and a total of $1.76B over 10 years. Other estimates are lower, in the $1.5B range for the 2006-2020 period[2794].

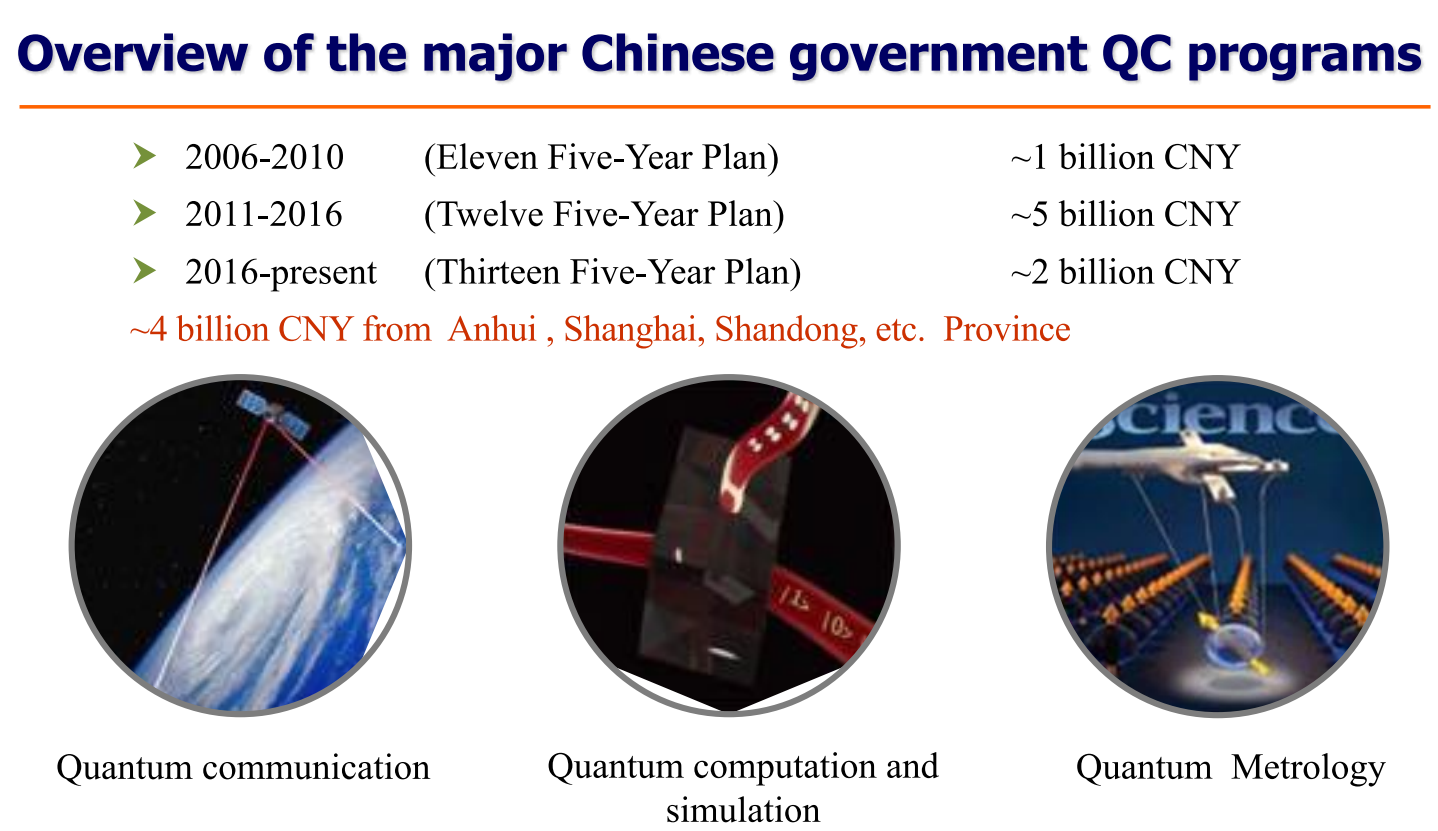

*Figure 865: China's quantum investments from 2006 to 2021 did not exceed $1.8B. This number is very different from the $10B to $15B investment showcased in various analyst publications. These >$10B numbers are false and based on fuzzy propaganda coming from China and amplified by various US interests. Source: Chinese QC Funding by Xiaobo Zhu, 2017 (35 slides). And… 1 CNY ≈ 0.14 US $.*

So, all the impressive $10B, $13B or $15B figures related to China's quantum technologies investments are probably completely off the mark. In 2021, China announced its new five year research plan, with a 11% global funding increase but with no details regarding quantum investments. At most, it would bring their 2006-2027 total quantum investments to about $3B, but definitively not $10B to $15B.

---

[2792] See Quantum Hegemony? China's Ambitions and the Challenge to U.S. Innovation Leadership, CNAS, 2018 (52 pages) and Quantum information technology development in China by Yuao Chen, June 2019 (25 slides).

[2793] The 2016 quantum roadmap is available in "Quantum Leap: The Strategic Implications of Quantum Technologies by Elsa Kania" and John Costello (part 1 and part 2). See also Chinese QC Funding by Xiaobo Zhu, 2017 (35 slides).

[2794] See FactBasedInsight's Quantum Landscape 2020: China, March 2020.

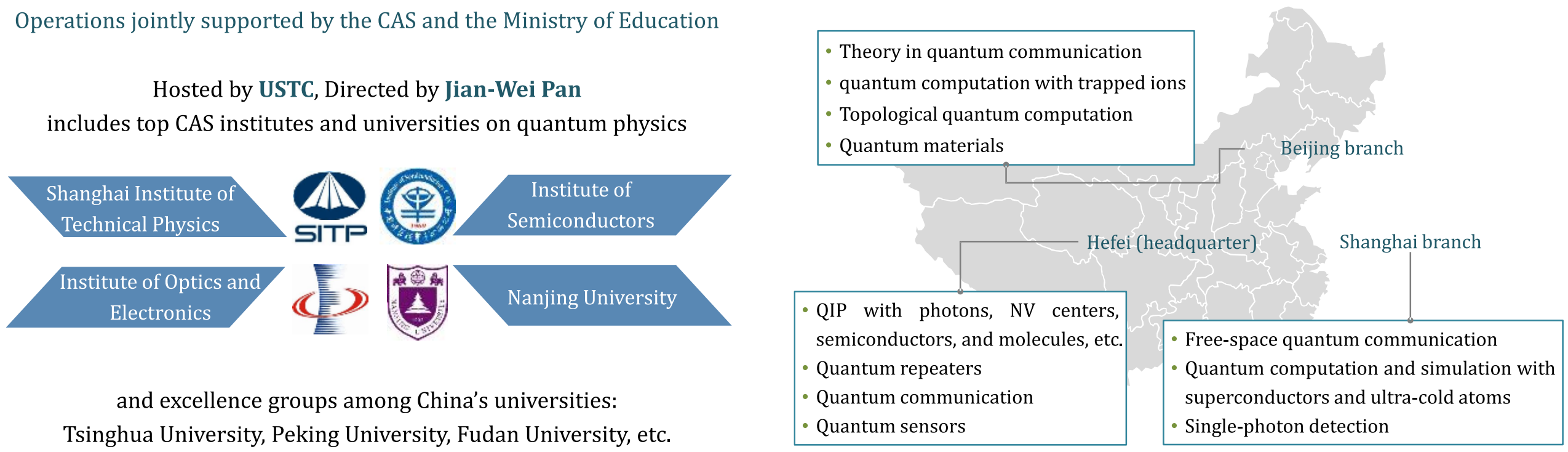

*Figure 866: China's quantum ecosystem. Source: Chinese QC Funding by Xiaobo Zhu, 2017 (35 slides).*

These investments are mainly spread over Beijing, Shanghai, and Hefei (500 km west of Shanghai). They specialize respectively in quantum communications, trapped ion computing, topological qubit computing and quantum materials for Beijing, silicon qubit computing, NV centers and photons, quantum communications and metrology in Hefei, and communication, superconducting and cold atom qubit computing and photon detection in Shanghai.

The Chinese plan is coordinated by the **USTC** (University of Science and Technology of China) of the Chinese Academy of Sciences (CAS) and under the leadership of Jian-Wei Pan[2795]. The most ambitious project is the "$10B" research center that partially opened in 2020, the **NLQIS (**National Laboratory for Quantum Information Sciences) of Hefei. This laboratory is focused on quantum computing and metrology for military and civilian applications.

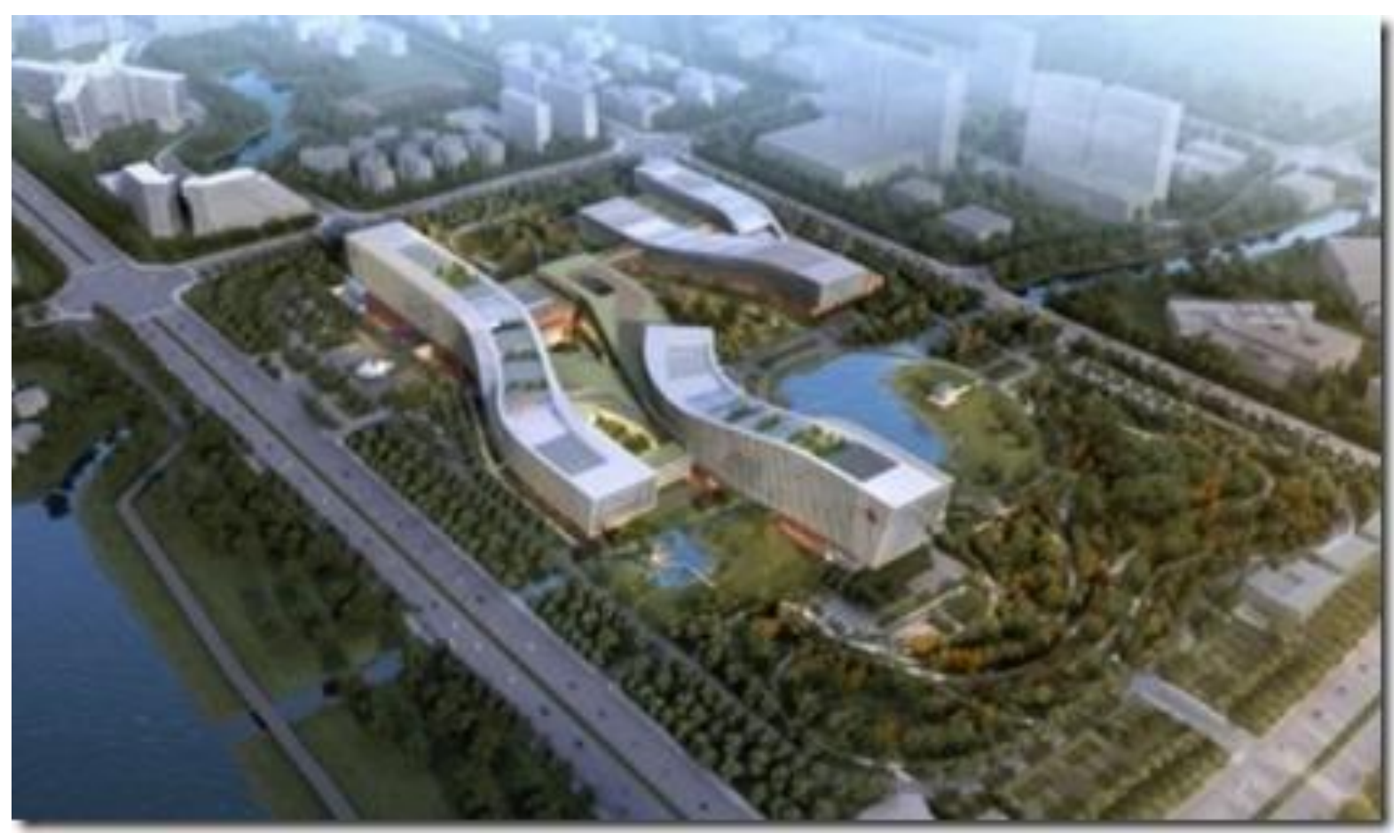

*Figure 867: Hefei's quantum lab.*

It will employ 1800 research people, including 560 full-time researchers spread across two labs, three universities and a fab[2796]. But it seems this research center is also dedicated to research in artificial intelligence, semiconductors and other domains, including the installation of a (costly) light synchrotron. So, again, the related investment amounts are certainly not entirely dedicated to quantum technologies.

In November 2021, the USA decided to restrict exportation of quantum technology goods to China related. The US Commerce Department added three entities in China: the Hefei National Laboratory for Physical Sciences at Microscale, QuantumCTek to the Entity List for acquiring and attempting to acquire US made items in support of military applications. The goal is to prevent China from developing counter-stealth technology like advanced radars and counter-submarine sensors. It also blocks US technology for breaking encryption (quantum computing) or develop unbreakable encryption (PQC)[2797]. Indirectly, this will impact exports from other countries related to the US, mainly from NATO.

---

[2795] See The man turning China into a quantum superpower by Martin Giles in MIT Technology Review, December 2018.

[2796] See Hefei's plan to create a national laboratory for quantum information science has been reported to the state council, EEworld, May 2018.

[2797] See Addition of Entities and Revision of Entries on the Entity List; and Addition of Entity to the Military End-User (MEU) List, November 2021.

***Research***

On the quantum computing side, Chinese laboratories are testing all imaginable qubit technologies and regularly announce technological progresses. They seem to be rather ahead in photon qubits as we have seen about their boson sampling experiments but not really with other qubit types.

In 2017, the Hefei laboratory announced the realization of a test system of 10 superconducting qubits in aluminum and sapphire[2798]. The two qubit gates error rate of 0.9% was not best in class.

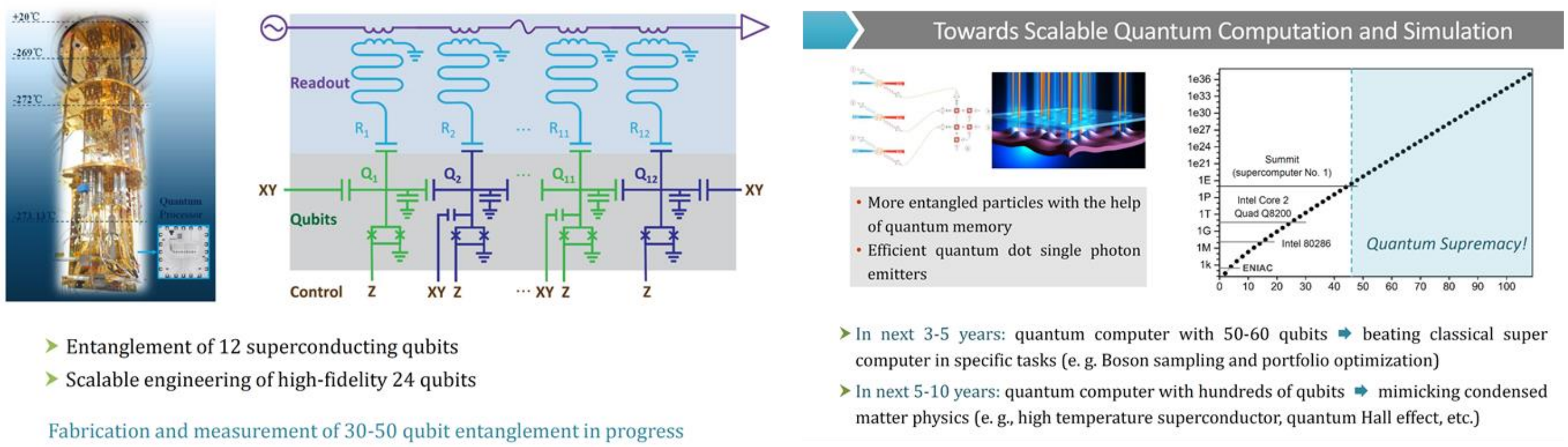

*Figure 868: Source: [10-qubit entanglement and parallel logic operations with a superconducting circuit]() by Chao Song et al, 2017 (16 pages).*

They were at 24 superconducting qubits in 2019. Their fidelity is 99.9% on single-qubit gates and 99.5% on two-qubit CZ gates, is much better[2799].

Their T1 duration, which defines the coherence time of the qubits is 40 µs, equivalent to what IBM obtains with its Q System One at 20 qubits. The Jian-Wei Pan team planned to reach 50 superconducting qubits by 2023.

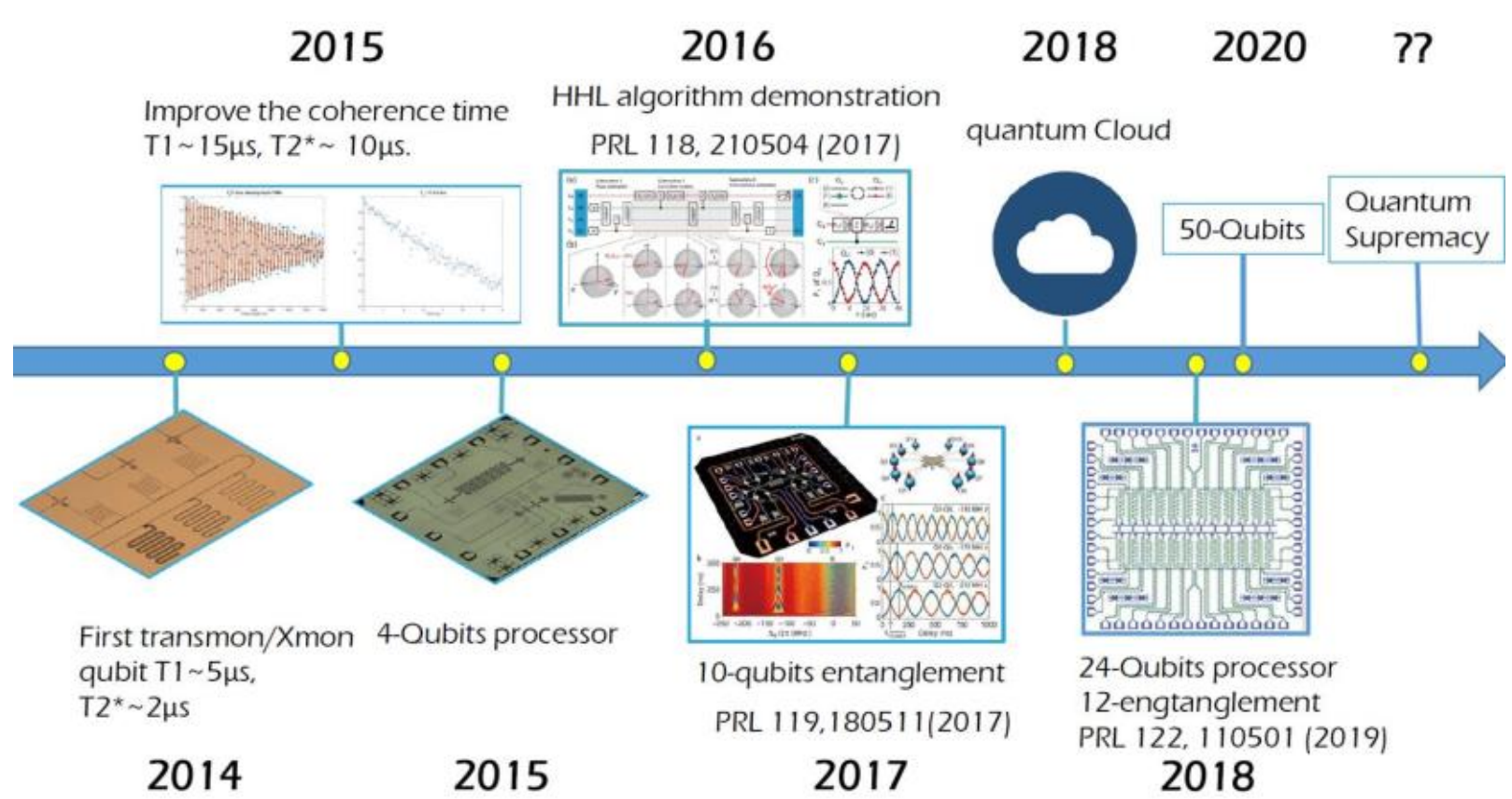

*Figure 869: Source: [Superconducting Quantum Computing]() by Xiaobo Zhu, June 2019 (53 slides).*

It delivered on this promise in May 2021 with 62 superconducting qubits, implementing quantum walks, which makes comparisons difficult, for example with IBM's 65 superconducting qubits system launched online in September 2020[2800]. They followed with the announcement of a 66 superconducting qubits system quantum advantage, being seemingly a copycat of Google's Sycamore processor architecture and benchmarking.

The Chinese scientific level is good but not yet stellar. They mostly improve technologies developed in Western countries and do not generate many new ideas. On the other hand, they create experiments like boson sampling or QKD deployments at a large scale.

---

[2798] See 10-qubit entanglement and parallel logic operations with a superconducting circuit by Chao Song et al, 2017 (16 pages).

[2799] Source: Superconducting Quantum Computing by Xiaobo Zhu, June 2019 (53 slides).

[2800] See Quantum walks on a programmable two-dimensional 62-qubit superconducting processor by Ming Gong, Science, May 2021 (34 pages).

China does not seem to have a significant influence in the academic world on quantum algorithms and programming tools. We must never forget the strategic role of software and platforms in the digital economic battles! It looks like History is repeating itself in China for this respect.

### *Quantum industry*

Public-private partnerships have been put in place, such as with **Alibaba,** who invested in the USTC to launch in 2015 the Shanghai Alibaba Quantum Computing Laboratory. It focuses on quantum cryptography and quantum computing. Quantum cryptography could be used to secure e-commerce transactions and data centers connections. In January 2018, Alibaba even launched a cloud-based 11 qubits system developed by USTC.

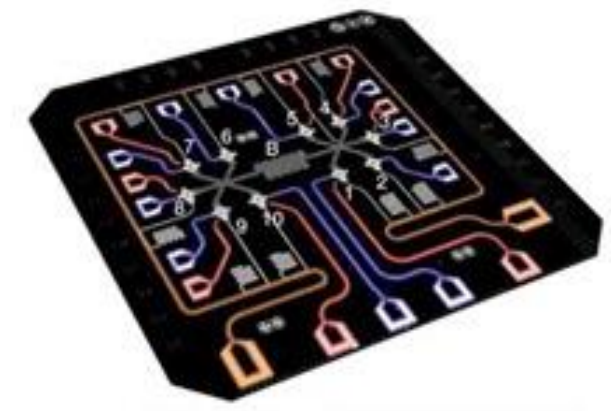

Figure 870: Alibaba's 11 qubit processor.

Alibaba is a serious contender in the superconducting qubit space with its fluxonium qubits variation.

**Baidu** launched in 2018 its **Institute for Quantum Computing,** which is being deployed in their Technology Park in Beijing with around ten people as of September 2019. It is led by Runyao Duan, a specialist in quantum information theory, with Artur Ekert as board member. They are mainly developing quantum software stacks in their QIAN full stack quantum software and hardware platform. They developed a quantum emulation solution in their cloud resources named Quantum Leaf[2801] and a bunch of other tools: Paddle Quantum (quantum machine learning), Quanlse (quantum pulse control for superconducting qubits), Qulearn (a quantum knowledge base), plus quantum error processing, measurement and control[2802], network architecture and quantum electronic design automation. They are also working also on a quantum Internet. In August 2022, they announced their first superconducting qubits quantum computer with 10 qubits named Qian Shi to be expanded later to 36 qubits.

**Tencent** also launched a Quantum Lab in 2018, led by Shengyu Zhang and based in Shenzhen. They plan to offer quantum computing resources in the cloud. The lab publishes work in quantum simulation and machine learning algorithms. It is developing QuAPE, a quantum control microarchitecture for superconducting qubits supporting multi-core QPUs parallelism that was prototyped with a FPGA circuit.

We can also mention the involvement of **ZTE** and many telecom operators and manufacturers in the deployment of secure fiber networks by QKD (**China Telecom, China Cable, China Comservice, China Unicom**) as well as various banks that use them.

Chinese startups are not very numerous at this stage, one of the reasons being that public research laboratories are well funded and have less incentive to create companies. One exception is **Origin Quantum Computing** which raised a record $163.4M to develop a full-stack superconducting qubits offering). Most other Chinese startups are in the quantum communication field, like **QuantumCTek** and **Qasky Science,** and have joined with ID Quantique and Battelle to form the **Quantum-Safe Security Working Group,** which federates the quantum cryptography industry.

## Japan

Let's move to the rest of Asia, starting with **Japan**. The country stands out for its very active and long-term oriented fundamental research initiation of two key technological waves in quantum computing.

---

[2801] See Introduction to Baidu Quantum Program by Shuming Cheng, June 2019 (9 slides). They notably propose the Paddle Quantum library, released on GitHub, which supports neural network QML, quantum chemistry and optimization tools. All this in quantum emulation on classical data centers.

[2802] See for example Efficient characterization of quantum nondemolition qubit readout by He Wang and Ya Cao, August 2022 (11 pages).

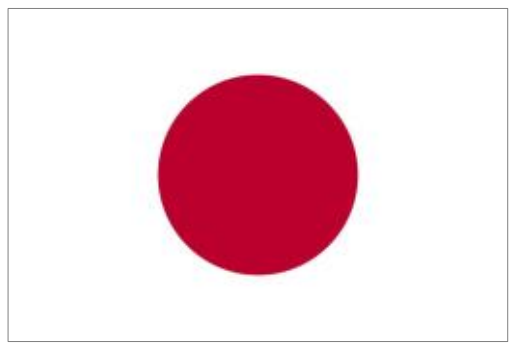

It started with the creation of the principle of quantum annealing by **Hidetoshi Nishimori** in 1998[2803]. Then, there was the creation of the first superconducting qubits in 1999 by **Yasunobu Nakamura**, **Jaw Shen Tsai** (both then at NEC) in liaison with **Yuri Pashkin** (Lancaster University, UK). Unfortunately, this was not turned into some industry lead.

***Research***

Japan's public research is conducted by several independent agencies attached to various ministries that fund public laboratories, university laboratories and research partnerships with companies[2804]:

- **JST** (Japan Science and Technology Agency) funded by the Ministry of Research and which funds deep techs research projects and also promotes science to the general public and international scientific collaboration. In 2016, JST launched a project by Yasunobu Nakamura of "Macroscopic Quantum Machines" to assemble 100 superconducting qubits.
- **RIKEN** (Institute of Physical and Chemical Research) also funded by the Ministry of Research (MEXT), with a total of about 3000 researchers. It includes a laboratory in theoretical quantum physics, headed by Franco Nori, and another in photonics, headed by Katsumi Midorikawa. They work in particular on silicon qubits. It collaborates with Fujitsu since 2020 to build a supercomputing qubits computer.
- **NICT** (National Institute of Information and Communication Technologies) includes the Quantum ICT Advanced Development Center, which specializes in quantum cryptography. In July 2017, the institute carried out a demonstration of quantum telecommunications using a microsatellite, reminiscent of the Chinese experiment with the Micius satellite carried out the same year.
- **NII** (National Institute of Informatics) includes a hundred or so researchers and focuses on research in theoretical quantum computing but also works on superconducting and silicon qubits.

  The Japanese-French Laboratory for Informatics (JFLI) created in 2009 is based in Tokyo and hosted at both the NII and the University of Tokyo. It brings together researchers from the Universities of Tokyo, Keio, NII, CNRS, Sorbonne University (LIP6), Inria and Université Paris-Sud. This multidisciplinary team ranges from fundamental physics to algorithms and studies the feasibility of large-scale quantum computing as well as quantum cryptography. The laboratory is co-directed by **Kae Nemoto**, from the NII, one of the few women in this whole panorama. Damian Markham from CNRS LIP6 has been working there since the beginning of 2020.
- **NEDO** (New Energy and Industrial Technology Development Organization) which is attached to the Ministry of Economy and Industry, METI. It is particularly invested in quantum annealing with a project running from 2018 to 2022 with $4.5M per year.
- **AIST** (National Institute of Advanced Industrial Science and Technology) also funded by METI. It employs about 2300 researchers in all. Several laboratories appear to be dedicated to nanomaterial sciences. There is also a research group on precision measurement.
- **QST (**National Institutes for Quantum and Radiological Science and Technology**)** was launched in April 2016 with an annual budget of $487M. This impressive amount is not exclusively

[2803] See Quantum annealing in the transverse Ising model by Tadashi Kadowaki and Hidetoshi Nishimori, 1998 (9 pages).

[2804] The most active quantum laboratories are located at the universities of Tokyo, Kyoto, Tohoku, Osaka, Nagoya, Keio, Tsukuka and Hokkaido. See Activities on Quantum Information Technology in Japan by Akihisa Tomita, June 2019 (19 slides).

allocated to quantum technologies. It mainly covers the vast field of quantum sensing and in particular medical imaging.

The Japanese government had launched various quantum initiatives such as **PRESTO** (since 2016) or the **CREST** cross-cutting program (also since 2016) as well as the **ERATO** projects in 1981 (Exploratory Research for Advanced Technology).

The country's quantum initiatives are currently part of its Fifth Science and Technology Plan, running from 2016 to 2022. In a typical Japanese way, this plan is linked to a societal goal "Society 5.0" to bring cyberspace and physical space closer together to solve society's social problems and create a human-centered society. All this with AI, quantum sensors and cybersecurity.

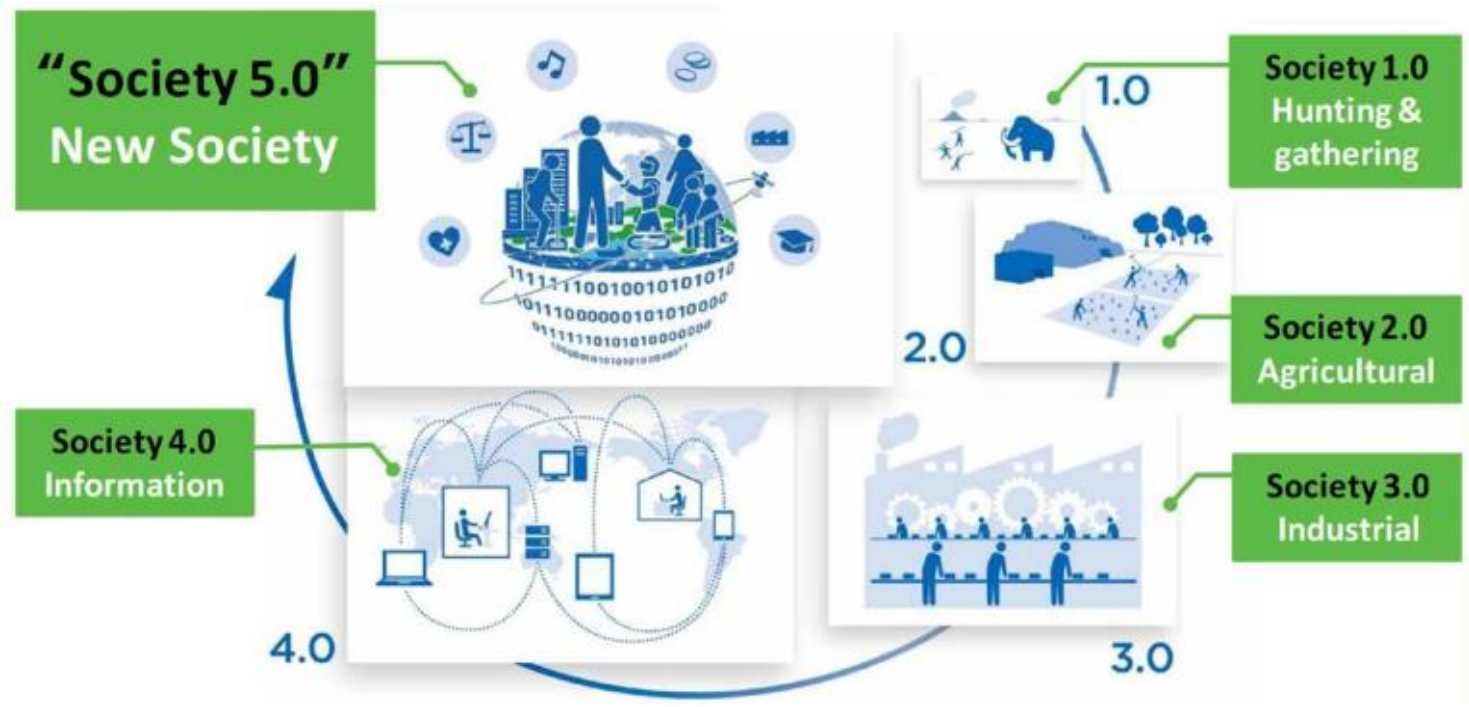

*Figure 871: Japan's classical societal angle to sell some new technology wave.*

Here are a few leading researchers in Japan in addition to those mentioned above[2805]:

**Akira Furusawa** of the University of Tokyo has the ambition to create a large-scale quantum computing solution with photon-based qubits. He's teaming up with NTT (see below) and plans to create a scalable computer by 2030[2806].

**Yoshihisa Yamamoto (**1950), a Stanford alumni and director of the NTT Physics and Computer Science Laboratory, who worked in photonics, QKD and quantum dots. He is very influential in Japan on the country's technological choices[2807]. He is the pilot of the Quantum Information Project (QIP), one of the national research program projects from FIRST selected in 2009 and which covered all branches of quantum applications[2808].

**Kohei Itoh** of Keio University has been managing the Q-LEAP project since 2018, which focuses on assembling different silicon isotopes into CMOS components and on NV center based quantum magnetometry (video). He is also a partner of IBM's Q Lab in Tokyo.

**Yasuhiko Arakawa** (1952) of the University of Tokyo specializes in semiconductor physics and optoelectronics, at the origin of new processes for the exploitation of quantum dots in sensing.

**Yasunobu Nakamura** (1968) who specializes in superconducting qubits and serves at the RCAST (Research Center for Advanced Science and Technology) of the University of Tokyo and at the CEMS (Center for Emergent Matter Science) of RIKEN[2809].

**François Le Gall** (1959) is a French researcher based at Kyoto University who specializes in quantum computing theory, mathematics, quantum algorithms and cryptography. He is also interested in distributed quantum computing (video). He has been living in Japan for more than 20 years.

---

[2805] Source: Q2B 2019 - International Government Panel, December 2019.

[2806] See Quantum computing: Japan takes step toward light-based technology - NTT, University of Tokyo and Riken aim for full-fledged system by 2030, Nikkei Asia, December 2021. The paper mentions a Japan $1.75B quantum plan. It is probably a mistake. See more reliable numbers in Concept of Quantum Technology Innovation hubs, 2021 (6 slides, broken link).

[2807] He is notably the co-author of the briefing note Quantum information science and technology in Japan, February 2019 (8 pages).

[2808] See First program overview.

[2809] See his presentation of the state of the art of quantum computing Development of quantum hardware towards fault fault-tolerant quantum computing by Yasunobu Nakamura (19 slides).

**Masahito Hayashi** of Nagoya University was originally a mathematician who then became a specialist in theoretical quantum computing. He coordinated the ERATO project on theoretical quantum computing.

**Masahiro Kitagawa** of Osaka University specializes in atomic nucleus spin-based quantum sensing in nuclear magnetic resonance with notable applications in medical imaging.

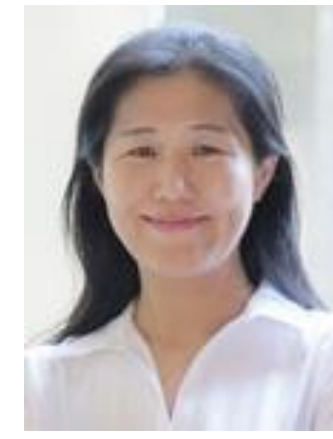

**Mio Murao** who created and manages the Quantum Information Group at the University of Tokyo that bears his name (Murao Group). This group specializes in distributed quantum computing, quantum systems simulation algorithms, quantum telecommunication protocols and quantum algorithms. She is very fluent in English, which has enabled her to serve as a connecting point between Japan and research teams in the USA (video).

**Nobuyuki Imoto** of Osaka University is leading research in quantum cryptography and telecommunications.

**Masahide Sasaki** of NICT leads much of Japan's quantum cryptography efforts. In particular, he has contributed to the SOTA project for quantum key communication using satellites[2810].

### *Government funding*

The *flagship* project **Q-LEAP** launched late 2019 by the Ministry of Research (MEXT) seems the most ambitious and aims to catch up with both China and the USA, even if an alliance with the USA also seems to be on the agenda[2811].

The roadmap extends to 2039 with $200M spread over 10 years. The program targets quantum computing, quantum sensing and next-generation lasers. Most qubits technologies are funded: superconducting, cold atoms, trapped ions and electron spin. This "Flagship" will run until 2027.

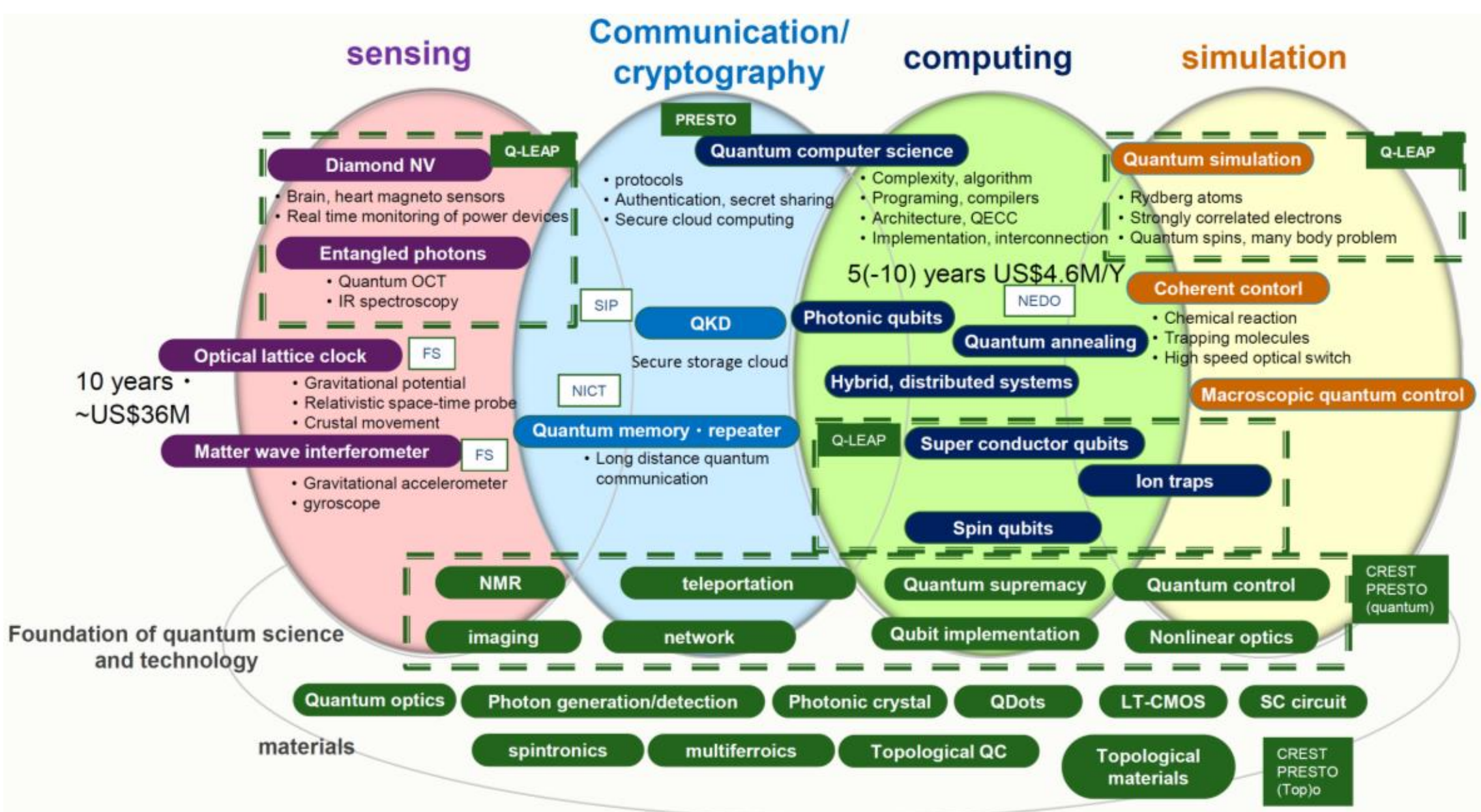

*Figure 872: Japan's quantum ecosystem and plans. Source: Activities on Quantum Information Technology in Japan by Akihisa Tomita, June 2019 (19 slides)*

[2810] See QKD from a microsatellite: the SOTA experience, October 2018 (10 pages).

[2811] See Japan plots 20-year race to quantum computers, chasing US and China by Noriaki Koshikawa, November 2019 and Land of the Rising Qubit: Japan's Quantum Computing Landscape by James Dargan, December 2019.

### *Quantum industry*

Japanese startups are rather specialized in software and in particular for quantum annealing computing running either on D-Wave quantum annealers or on Fujitsu digital annealers. We have **A*Quantum** (2018, QA software), **D Slit Technologies** (2018, software), **Groovenauts (**2012, QA software ), **Jij (2018,** QA software framework), **MDR** (2008, chemical simulation), **QunaSys (2018,** healthcare), **Sigma-I (**2019, QA software) and **Tokyo Quantum Computing** (2017, QA software).

**Softbank**'s investment fund abounded with Saud family's money up to $100B was also planning to invest in quantum technologies[2812].

However, four years after its announcement, the fund does not seem to have a single stake in quantum technologies.

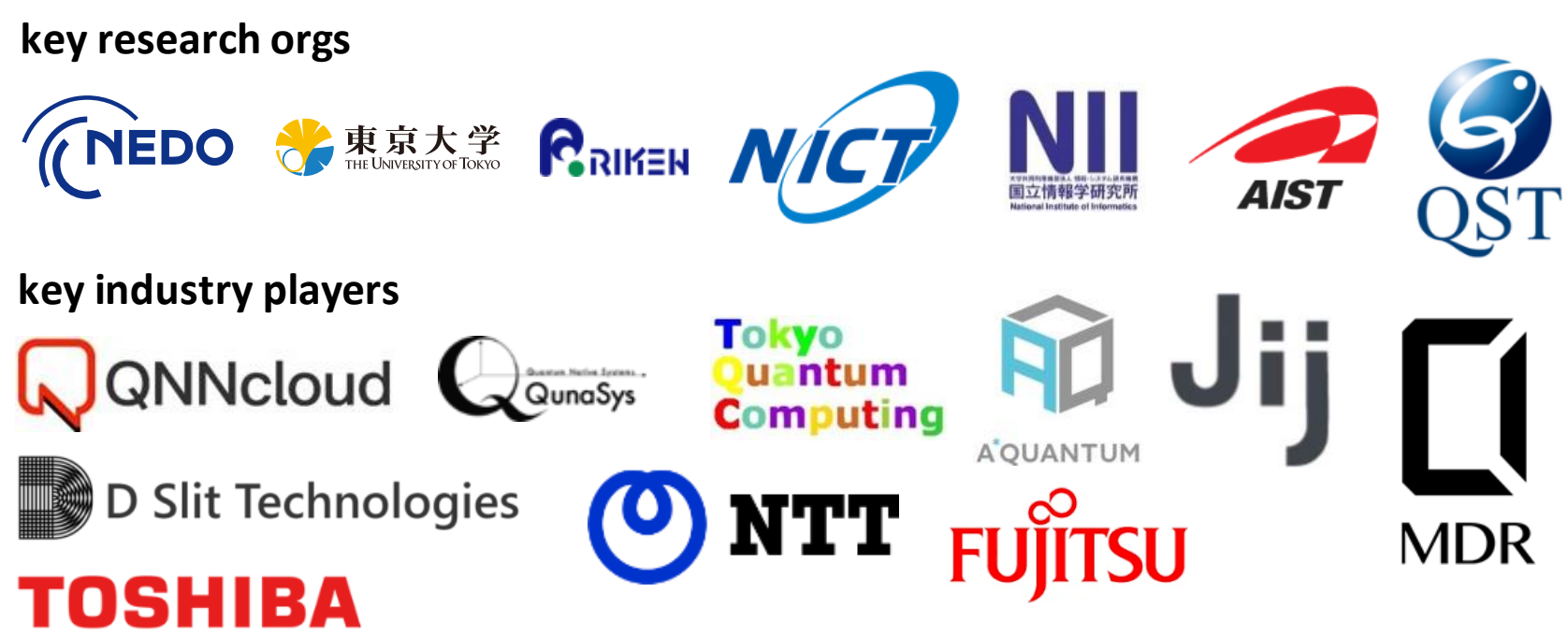

Figure 873: Japan's quantum industry vendors. (cc) Olivier Ezratty, 2022.

In the private sector, Japan's major industry groups are mainly focused on quantum telecommunications and cryptography, as well as on quantum and non-quantum annealing-based computing.

**Hitachi** also has a research laboratory located at the University of Cambridge (UK) that works on quantum key distribution, quantum computing and the creation of SQUID components for superconducting qubits. They are also working on silicon spin qubits quantum computing.

**Toshiba Corporation** has been involved in quantum cryptography since 2003. They are working on it with the Quantum Information Group (QIG) at the University of Cambridge, UK. They performed a first demonstration of quantum communication in 2014, sending 878 Gbits/s of secure data over a 45 km fiber between two areas in the Tokyo area over a cumulative period of 34 days, at a rate of 300 kbits/s. They were continuing the experiments in 2019 and beyond and with British Telecom in the UK[2813].

**NTT** maintains four applied quantum research laboratories, focused on quantum telecommunications and quantum cryptography, all with about 40 researchers[2814].

In 2017, telecom operator **NTT** launched a prototype photonics-based Quantum Neural Network (QNN) in collaboration with the **National Institute of Informatics** and the **University of Tokyo**. It was available on the cloud at qnncloud.com (video) but the service was discontinued in March 2019[2815]. This was done with Toshiba, NEC and the NICT in Tokyo with three nodes and 45 km apart[2816].They also work in the CMOS quantum dots qubits. NTT also developed LASOLV, a photonic based coherent Ising system with 2000 nodes[2817].

---

[2812] See SoftBank's Vision Fund Eyes Investment in Quantum Computing by Jeremy Kahn, Bloomberg, June 2017.

[2813] See Performance Limits for Quantum Key Distribution Networks by Andrew Shields, June 2019 (16 slides).

[2814] This leads to raising wages inflation for the most talented people, a bit like in Silicon Valley. See NTT offers researchers $1 million salaries in bid to lure top talent in cryptography, quantum computing, November 2019.

[2815] See Japan launches its first quantum computer by Walter Sim, November 2017.

[2816] See Tokyo QKD Network and its application to distributed storage network by Masahiro Takeoka, June 2019 (22 slides).

[2817] See LASOLV Computing System: Hybrid Platform for Efficient Combinatorial Optimization by Junya Arai et al, 2020 (6 pages),

Finally, several non-quantum annealing optimization computation projects on CMOS components have been launched. There is the **Fujitsu** offering, and also the NEDO project led by Masanao Yamaoka and Masato Hayashi at **Hitachi** in partnership with the AIST, RIKEN and NEDO (New Energy and Industrial Technology Development Organization, the equivalent of the energy branch of the CEA) laboratories[2818].

And then the **NEC** project in quantum annealing led by **Yuichi Nakamura** in liaison with Waseda University, those of Yokohama and Kyoto, AIST and Titech (Tokyo Institute of Technology). They are optimizing the classical part of annealing processing with NEC vector processors. The quantum part seems to be managed on D-Wave machines. NEC is also versed in quantum keys (QKD).

**IBM** announced at the end of 2019 the opening of a Q Lab in Tokyo in partnership with the University of Tokyo. IBM's investment in Japan follows a model already inaugurated in France in Montpellier in 2018, in Germany in September 2019 and in Canada with the Institut Quantique in June 2020: a partnership with a university, investments in training and above all, a technical and marketing investment to evangelize quantum among major customers[2819].

Finally, **Recruit Communications Ltd** (1960), a large $16B CDN company specializing in HR, communications and marketing, distinguished itself by launching a partnership with D-Wave in 2017 to develop quantum annealing-based solutions for the operational optimization of marketing, communications and advertising. In particular, they have developed the PyQUBO open source library, which simplifies the development of quantum annealing software applications[2820].

In September 2021 was launched Q-STAR (Quantum Strategic Industry Alliance for Revolution), an industry alliance to promote the usage of quantum technologies and particularly quantum computing and cryptography in various industries with the participation from Toshiba, Toyota, NEC, NTT, Hitachi, Fujitsu, Mitsubishi Chemical and Sumitomo among others.

### Singapore

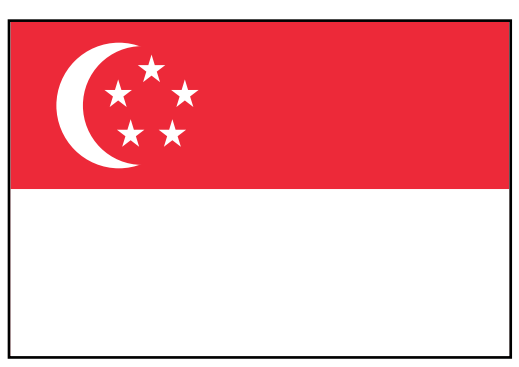

The small state of **Singapore** is known for its economic and entrepreneurial dynamism. Within the **National University of Singapore** (NUS), quantum research was consolidated in 2007 in the **Center for Quantum Technologies** (CQT) with an annual funding of about $15M. It was created thanks to some real political leadership, coming from the then Defense Minister of Singapore[2821].

---

[2818] See CMOS Annealing Machine - developed through multi-disciplinary cooperation, November 2018, Overview of CMOS Annealing Machines by Masanao Yamaoka, Hitachi, (4 pages) and A 2 x 30k-Spin Multi-Chip Scalable CMOS Annealing Processor Based on a Processing- In-Memory Approach for Solving Large-Scale Combinatorial Optimization Problems, November 2019.

[2819] See IBM Takes Its Quantum Computer to Japan to Launch Country-Wide Quantum Initiative by Anthony Annunziata, December 2019. In partnership with the University of Tokyo and IBM and the University of Tokyo Launch Quantum Computing Initiative for Japan by IBM, 2019. In August 2020, IBM embellished this partnership by announcing the creation of a consortium for the adoption of quantum technologies in Japan. See IBM Launches Global Consortium for Quantum Innovation by Chris Duckett, August 2020, which refers to an announcement that is really only about Japan: IBM and the University of Tokyo Unveil the Quantum Innovation Initiative Consortium to Accelerate Japan's Quantum Research and Development Leadership by IBM, August 2020.

[2820] See Recruit Communications and D-Wave Collaborate to Apply Quantum Computing to Marketing, Advertising, and Communications Optimization, May 2017.

[2821] Artur Ekert says he was persuaded to join Singapore in 2000 by Tony Tan, who was then the country's defense minister. He had met him at a conference where his visionary speech, for a politician, had impressed him. In 2005, Tony Tan took charge of the sovereign wealth fund Singapore Investment Corporation and then Singapore's National Research Foundation. He was at the origin of the strategy of targeted investment in cutting-edge research fields, which today we call deep tech. This Tony Tan then became the President of Singapore between 2011 and 2017. The CQT was launched in 2006. The story is told in the book 50 years of science in Singapore pages 362 to 387, February 2017. His personal credo: to be successful, you need to attract the right people, original ideas and then funding. Too often, this happens through funding.

It is vested in both quantum computing (cold atoms in Berge Englert's group, photons and superconductors in Dimitris Angelakis' group, trapped ions in Dzmitry Matsukevich's group), quantum cryptography (Kwek Leong Chuan's group) and quantum metrology (notably atomic clocks in Murray Barrett's group).

The CQT was led from its inception until July 2020 by Artur Ekert. Since then, it's run by José Ignacio Latorre. It brings together about twenty teams representing 22 permanent researchers, 60 research fellows and 60 PhD students, covering the four usual fields of quantum technologies. This represents a total of 300 people in all. Of the 22 research supervisors, about a quarter are Singaporeans who have usually done a thesis abroad. Singapore is doing well to attract talented foreigners and to ensure that they settle permanently in this country of five million people.

Six startups emerged from CQT with **Entropica Labs** (quantum algorithms), **Horizon Quantum Computing (**software), **Innovatus Q (**hybrid algorithms), **S-Fifteen Instruments (**quantum cryptography) and **SpeQtral** (satellite QKD).

Quantum communication is one specialty from Singapore. Since 2016, CQT has been associated with the telecom company **Singtel** and the NUS for the deployment of QKD on optical fibers with repeaters. At the end of 2019, a team from **Nanyang Technological University** (NTU) developed a 3 mm-sided chipset capable of integrating a CV-QKD, a continuous variable quantum key-based encryption system[2822].

In 2015, Singapore launched its Galassia-2U nanosatellite, created by CQT and used to experiment encrypted QKD based quantum communications. Galassia is integrated in a two-unit CubeSat format (two cubes on top of each other, see *opposite*). It weighs only 3.4 Kg in total. It was sent to space with 5 other satellites including the telecommunications satellite TeLEOS-1 (400 kg) at the end of 2015 by an Indian launcher.

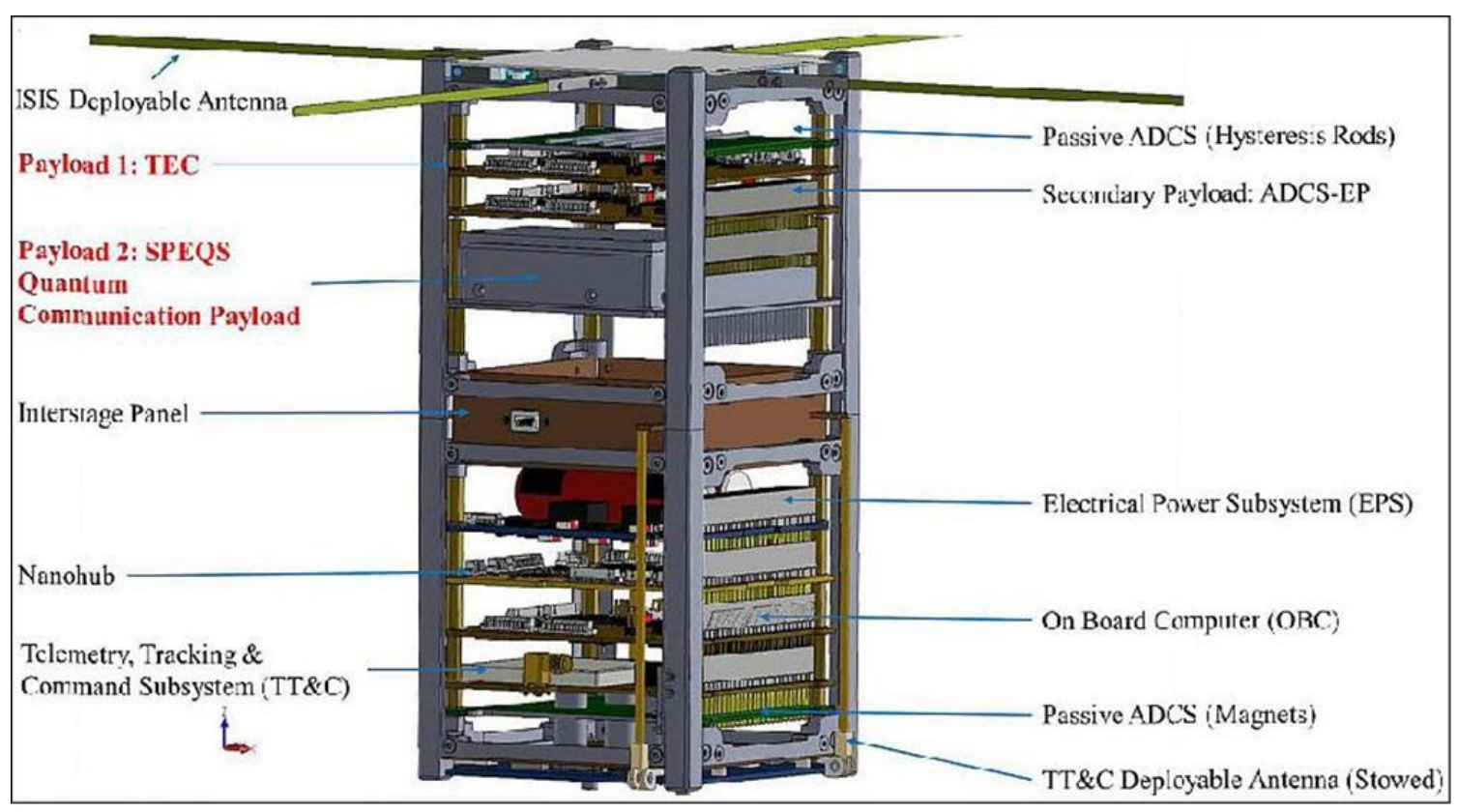

*Figure 874: Source: https://directory.eoportal.org/web/eoportal/satellite-missions/g/galassia*

In May 2022, as part of its Quantum Engineering Programme (QEP) started in 2018, Singapore launched three national platforms with a total funding of $23.5M for 3,5 years with a pooling of the resources and skills from CQT at NUS and NTU Singapore, A*STAR's Institute of High Performance Computing (IHPC) and the National Supercomputing Centre (NSCC).

- The **National Quantum Computing Hub** which will develop quantum computing capabilities and use cases for the industry.
- The **National Quantum-Safe Network** which will conduct trials of quantum-safe communication technologies for critical infrastructure.
- The **National Quantum Fabless Foundry** which will support microfabrication techniques for quantum devices and enabling technologies. Hosted at A*STAR's Institute of Materials Research and Engineering (IMRE), it will manage the micro and nanofabrication of quantum devices and related enabling technologies. This complements the Quantum Science and Engineering Centre (QSec) launched in December 2021 by NTU to design and manufacture various quantum chipsets using classical semiconductor manufacturing technologies.

[2822] See Quantum chip 1,000 times smaller than current setups, November 2019.

The lifetime of this type of satellite is six months[2823]. These experiments led to the creation of the S-Fifteen Space Systems. However, solutions have yet to be found to ensure that these satellites last longer in their low orbit and do not contribute even more to low orbit pollution.

Several partnerships and linking the French and Singaporean quantum ecosystems. The CQT welcomes several researchers from France, including **Steven Touzard**, **Miklos Santha** (CNRS) and **Christian Miniatura.** Christian is leading **MajuLab**, the joint CNRS-NTU research laboratory on quantum science. He is joined in October 2022 by **Alexia Auffèves** who was already partnering with a CQT research group in charge of studying noise and error correction codes led by **Hui Khoon Ng**. NUS is also partnering with the **Thales TRT** research lab based in Singapore, in security and sensing[2824]. **Sondra Lab** is another Franco-Singapore lab with CentraleSupelec, ONERA, NUS and DSO which is working in the fields of electromagnetism and signal processing applied to radar.

### South Korea

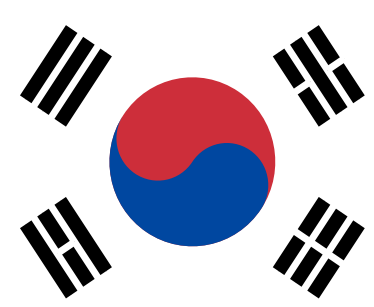

In South Korea, the telecom operator **SK Telecom is** investing in quantum telecommunications[2825]. They are partners with Florida Atlantic University. They have also invested in 2016 in the Swiss startup ID Quantique. SK Telecom is also partner since 2017 with Nokia in the QKD field as well as with Deutsche Telekom with whom they have established a "Quantum Alliance" to create secure telecommunications.

SK Telecom has deployed a QKD network in the backbone of its 4K network in the city of Sejong on two links of 38 and 50 km respectively[2826].

**Samsung** is also investing in QKD and cryptography. They integrated a quantum random number generator in a dedicated version of a Galaxy smartphone for the Korean market in April 2020, with a component from ID Quantique, the Swiss startup acquired by SK Telecom in 2018. A new version was launched in April 2021.

### Taiwan

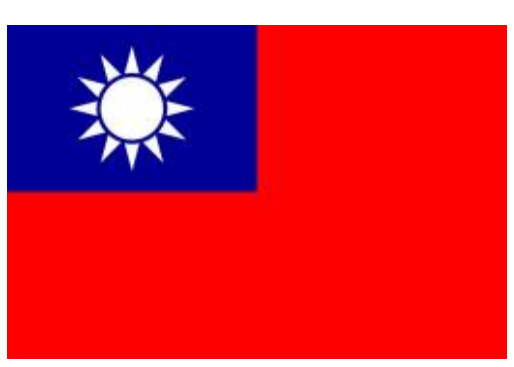

**Taiwan** is very advanced in semiconductors with TSMC, the leader in CMOS fab and the only one with Samsung that is able to go down to an integration level of 5 nm, soon 3 nm and with plans to reach 1 nm. It is also still very present in the PC components market. This is particularly the case with motherboards (MSI, Asus, Gigabyte) and PC manufacturing (Quanta, ...).

It was logical in these conditions that the country becomes interested in quantum computing.

We can identify initiatives in students training and with a conference organized in September 2019[2827]. **Quantum Design** provides measuring instruments but does not seem to exploit technologies of the second quantum revolution[2828]. Finally, IBM has established a foothold in the country to help it adopt quantum technologies.

---

[2823] See Quantum Tech demos on CubeSat nanosatellites (41 slides).

[2824] See Singapore's NUS and Thales developing quantum technologies for commercial applications by Jamilah Lim, October 2021.

[2825] See SK Telecom Continues to Protect its 5G Network with Quantum Cryptography Technologies, March 2019.

[2826] See Quantum Safe Communication - Preparing for the Next Era by Dong-Hi Sim, June 2019 (21 slides).

[2827] See Quantum Computer: Envision the New Era of Computing, a conference in September 2019 and Quantum tech summer program in Taiwan a success, Taiwan News, July 2019.

[2828] See Quantum Design Taiwan.

In December 2020, Taiwan launched a $282M quantum plan over 5 years. It will consolidate its investments in the Southern Campus of Academia Sinica, the national academy of Taiwan, in Tainan. They plan to create a Quantum Technology R&D center between 2022 and 2024. On top of that, Hon Hai (FoxConn) created a Quantum Computing Research Center in January 2021.

In April 2022, the Taiwanese government announced as part of this plan the selection of 17 research teams in universities and the hiring of 72 project directors and 24 IT companies. They are looking at ways to improve quantum hardware and software and to leverage, if possible, their semiconductor industry.

**Australia**

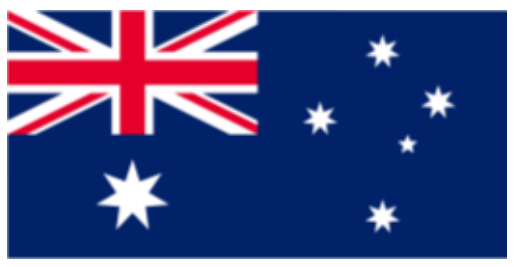

The Australian National Innovation and Science Agenda announced in 2015 included 24 initiatives and $820M in funding over 4 years, of which $19M were allocated to the Center for Quantum Computation and Communication Technology (CQCCT) over 5 years in quantum computing.

The country is prolific in public-private partnership projects associating Australia with other countries[2829].

In 2017, the University of New Wales (UNSW), the Commonwealth Bank of Australia and telecom operator Telstra provided $52M in funding for the creation of a silicon quantum bit processor. One could hope that Orange will do the same in France with the CEA and/or a startup!

An investment fund of the Ministry of Defence, the **Australian Next Generation Technologies Fund** allocated $730M to 9 areas including one on quantum technologies over 10 years[2830].

Assuming that these funds were distributed evenly among the 9 initiatives, this gives us $8M of additional funds per year on quantum technologies for military uses, including sensing.

In February 2019, **CQC2T** (Centre of Excellence in Quantum Computation and Communication Technology) was created at UNSW, headed by Michelle Simmons. The goal was to create an electron spin quantum computer. With federal funding of $33.7M, it brings together a community of 200 researchers[2831].

Australia also has **EQUS** (Arc Center of Excellence for Engineered Quantum Systems), a national quantum sensing research center. It partners with Microsoft, Moglabs and Lockheed Martin, among others.

On the entrepreneurial side, there are three startups in the field of quantum technologies with **QuintessenceLabs (**QKD optical keys), **QxBranch** (software and consulting, an American startup with an office in Australia, acquired by Rigetti in July 2019), **Silicon Quantum Computing (**silicon qubits), **Quantum Brilliance** (2019) on top of which should be added **Archer** and their carbon electron spins qubits.

[2829] See Charting the Australian quantum landscape, February 2019 (5 pages).

[2830] See Next Generation Technologies Fund, 2016.

[2831] In early 2019, UNSW's CQC secured an additional $33M in funding at its official launch. See Federal govt funnels $33.7 million towards UNSW's quantum research by Matt Johnston, February 2019.

In May 2020, Australia put its quantum strategy in order with the publication of a plan by CSIRO[2832]. Their (fairly optimistic) ambition is to turn it into a $4B industry creating 16,000 jobs by 2040 out of a projected global total revenue of $86B. The projected breakdown is $2.5B and 10,000 jobs for computing, $900M and 3000 jobs for sensing and $800M and 3000 jobs for telecommunications. The goals? To define a coordinated strategy, to finance research and business creation, to train talents and to create a coherent industrial value chain.

A relatively new point in such a plan, is to explore the ethical, social and environmental issues that could be raised by quantum technologies. The subject has been growing in importance since 2020. They also address the question of the supply chain of key components and materials for quantum technologies.

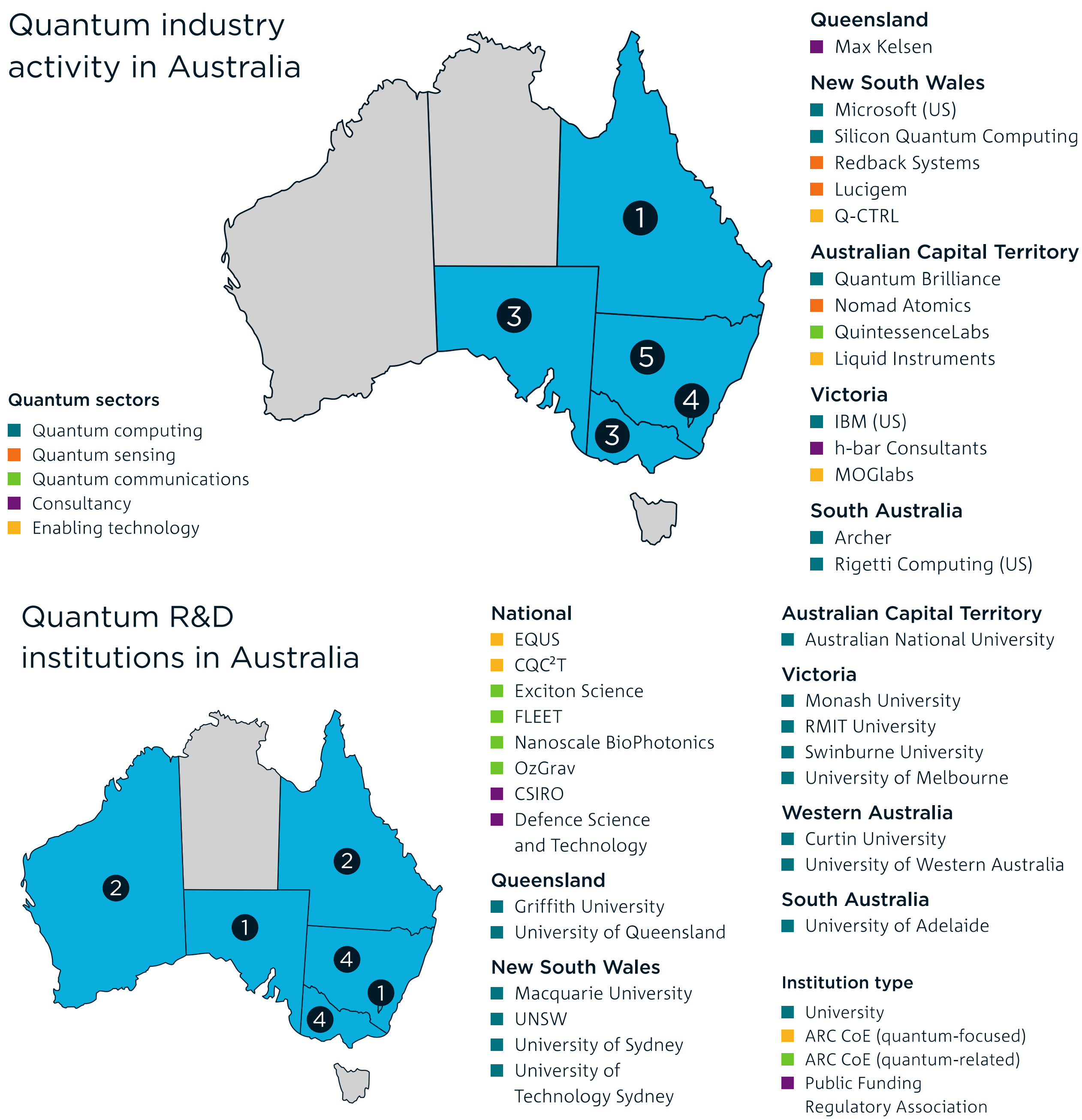

*Figure 875: Australia's ecosystem. Source: Growing Australia's Quantum Technology Industry by CSIRO, May 2020 (56 pages).*

[2832] See Growing Australia's Quantum Technology Industry by CSIRO, May 2020 (56 pages) and Australia could lose its quantum computing lead, CSIRO warns by John Davidson, May 2020.

As for companies and startups, they have some of them shown in this map. They highlight Microsoft and IBM. So be it. Rigetti because they have acquired the local startup QxBranch. And a few other startups, some of which are specialized around diamonds. In September 2022, the Tech Council formed the **Australian Quantum Alliance** to consolidate its quantum tech industry.

In terms of international partnerships, the country is associated with the University of Singapore for the creation of quantum telecommunication satellites.

In December 2020, Australia launched the **Sydney Quantum Academy,** a joint effort from Macquarie University, UNSW Sydney, the University of Sydney and UTS. It consolidates training offerings implemented by the partner Universities for undergraduates, PhDs plus some fellowships programs.

In November 2021, the Australian government allocated an additional US $80M to its quantum efforts, particularly for supporting the commercialization, adoption and use of quantum technologies and create new jobs.

It includes US $51M for the creation of a "quantum commercialization hub" with the task to build strategic partnerships with "like-minded countries" to sell Australia's quantum technologies, starting with the usual Commonwealth country partners and the USA. As a follow-up from the famous nuclear submarine deal with the USA (at the expense of an existing classical submarine deal with France) announced in September 2021, Australia cemented a global quantum partnership with the USA in November 2021, covering in a fuzzy way the exchange of quantum knowledge and skills. The University of Sydney is already part of an international consortium integrated in the US IARPA **LogiQ** program.

The Australian quantum ecosystem benefits from the country having Cathy Foley as its chief scientist given she led the development of a Quantum Technology Roadmap at CSIRO in 2020.

At last, in July 2022, Google announced new partnerships with Australian universities including UNSW and the University of Sydney. It extends what they already do with US universities, mostly for the development of quantum algorithms on Sycamore QPUs.

### India

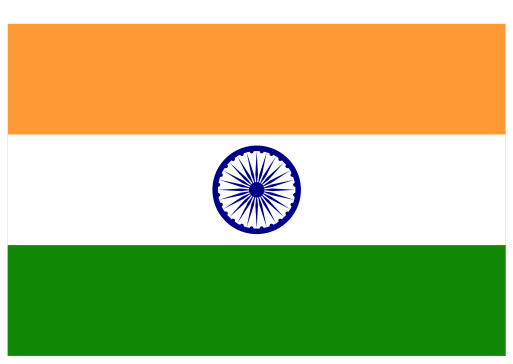

At the beginning of 2020, India launched an investment plan in quantum technologies, the **NMQTA** (National Mission on Quantum Technologies & Applications). This plan is well funded as a proportion of the country's GDP, with $1.12B over 5 years, at the same level as the American Quantum Initiative Act of 2018 or the European Flagship launched the same year[2833].

The plan covers the usual suspects: quantum computing, quantum telecommunications and quantum sensing. Ironically, the CEOs of IBM, Google and Microsoft who are strong investors in quantum computing are all of Indian origins (Arvind Krishna, Sundar Pichai and Satya Nadella)!

The Indian plan has an eye on China and wants to turn the country into a quantum leader, particularly in computing, telecommunications and cryptography.

In August 2021, **MeitY** (their Ministry of Electronics and Information Technology) launched the Quantum Computer Simulator (QSim) toolkit that was created by IISC Bangalore, IIT Roorkee and C-DAC. This software emulator of gate-based quantum code must not be confused with Qsim from Qsimulate and Google.

In January 2022, the **Indian Army** launched its Quantum Research Laboratory. So be it.

[2833] See India finally commits to quantum computing, promises $1.12B investment by Ivan Mehta, February 2020.

In February 2022, **Avasant** (an US consulting firm with a branch in India) and **NASSCOM** (the Indian software and IT service trade association) published a report on the opportunities of quantum technologies for India[2834]. It contained several nuggets like a quantum tech potential providing between $152B and $310B cumulative value to the Indian economy by 2030 with a maturity inflection point positioned in 2027.

They expect that 10K logical qubits will be available by 2030 (100 would be so nice…). But they expect a moderate workforce impact of 25K to 30K people in 2030. Also, India plans to develop a quantum computer with about 50 qubits by 2026. Also, they position the quantum Internet to be related to FTQC in their roadmap for 2027 and beyond. It does not really explain the connection between these two technologies. The document also shows the key role of the large Indian IT services companies in the adoption of quantum computing. As an example, in September 2021, Infosys Cobalt was partnering with AWS Braket to explore the business potential of quantum computing.

Among other things, the Indian plan has also accelerated the creation of startups in India, some being king in overselling their technology advances. This is the case of **QPI** and their projects of a one million silicon qubits and hybrid processor. Many of their new startups are multi-domains, such as:

- **QRLAB** (2020, India) who is a contract research, education and consulting company focused on quantum computing. They develop quantum inspired software, QML and also work on quantum Internet and cryptography.
- **Qulabs.ai** (2017, India) which builds quantum networks and has some expertise in QML in finance and for new drug discovery. That's quite broad in scope! Their QuAcademy facilitates students training.
- **Fractal Analytics** (2000, India, $685M) is an AI/data analytics company and a unicorn. It is creating an in-house quantum computing lab.

[2834] See The quantum revolution in India: betting big on quantum supremacy, Avasant, February 2022 (48 pages).

## Quantum technologies around the world key takeaways

- The quantum startup scene has seen its peak company creation in 2018. A small number of startups like D-Wave, IonQ, Rigetti, PsiQuantum and Xanadu collected about 70% of the worldwide quantum startups funding. The investors FOMO (fear of missing out) and the "winner takes all" syndrome explain this situation.
- Most developed countries now have their "national quantum plans" and want to lead that space, particularly with quantum computing. The first ones were Singapore in 2007 and the UK in 2013. Investment comparisons are not obvious since these plans accounting are not the same from country to country (incremental funding vs legacy+incremental, private sector included or not, European Union investments included or not). All these plans invest a lot in fundamental research and on developing a startup and industry ecosystem.
- China's quantum investments have been overestimated for a while, both because of the ambiguity of China's communication and since various lobbies in the USA were pushing for increased federal investments to counter China's perceived threat. This worked particularly well during the Trump administration and seems to persist with the Biden administration.
- Europe and the USA are the greatest investors in quantum science so far. The European Union as a whole is the largest region for public investments in quantum research. The USA has a larger industry investment than Europe due to its large IT vendors investments (IBM, Google, Microsoft, Honeywell) and a traditional lead in startups funding, and, certainly, with its domestic market size and dynamics.
- Many countries did put quantum technologies in the critical field of "digital sovereignty" like if it was some sort of nuclear weapon equivalent.
- Each country has its own strengths and specialty although most of them invest in all the fields of quantum technologies (computing, telecoms/cryptography and sensing).
- Some analysts are wondering whether we'll get soon into a quantum winter, like the ones that affected artificial intelligence in the 1970s and the 1990s. One way to avoid it is to limit overpromises.

# Corporate adoption

This book is intended for a wide audience interested in quantum technologies. It includes companies that may wonder what to do when facing such a deluge of hype, information, complexity and uncertainty. And this comes in addition to other technological waves to assimilate such as artificial intelligence, cryptos and other Blockchain, NFTs, the metaverse and 5G, not to mention cloud deployments and the classical business applications backlog.

The wave of quantum technologies is unique in that it is even more unpredictable and difficult to grasp than the other digital technologies waves. And yet, it is worth the attention, particularly in certain key verticals such as finance, healthcare, utilities and transportations. We are clearly in a technology push situation, meaning, here it is and it's up to us to imagine what to do with it. And quantum technologies, particularly computing, are not replacing legacy systems, but complementing them.

It's still a fairly green field and not only because scalable quantum computers are not yet available. It's also linked to having only a few people understanding how quantum computers are used and benchmarked. Innovation is still well ahead of us. And quantum technologies are not just about computing. It also deals with telecommunications, cryptography and sensing. This last domain might be under evaluated and could be strategic for many industries.

Large customers and their IT, digital transformation, innovation and R&D departments are exposed to a continuous stream of industry analyst and vendors pitches creating a sense of urgency for the adoption of quantum computing. It comes for example from **Capgemini**[2835], **McKinsey**[2836], **Deloitte**[2837], **Harvard Business Review**[2838], **Arthur D. Little**[2839], from **Pathstone**, a financial advisor company[2840]. It can take the form of a survey commissioned by a vendor, like **Zapata Computing**[2841]. Some even are completely off-the-mark on the role of quantum computing, such as in cybersecurity[2842].

I propose here a relatively simple and, all in all, fairly classic approach for corporations, which is laid out in a dozen points, some of which come from the experience of major large companies.

---

[2835] See Capgemini: Organizations need to get moving on quantum by Dan O'Shea, Fierce Electronics, April 2022.

[2836] See Quantum computing use cases are getting real—what you need to know, McKinsey, December 2021.

[2837] See Quantum computing in 2022: Newsful, but how useful? by Duncan Stewart et al, Deloitte, December 2021. Their assessment is the most honest of all, with cautious tales like: "*Many of the tasks that they currently do can be replicated on a standard laptop computer at a fraction of the cost. The problem with QCs' usefulness is not a lack of use cases, money, effort, or even progress. It's that current QCs are not yet powerful enough to tackle problems that can't be performed by traditional computers*".

[2838] See Quantum Computing for Business Leaders by Jonathan Ruane, Andrew McAfee, and William D. Oliver, HBR, January–February 2022. With an interesting statement: "*Quantum computing will enable businesses to better optimize investment strategies, **improve encryption**…*". How can you trust this sort of report with such misunderstanding of how quantum computers will impact cryptography (it potentially endangers it and you won't be saved by a quantum computer)?

[2839] See Quantum Computing - The state of play and what it means for business by Albert Meige, Rick Eagar, Lucas Könnecke and Olivier Ezratty. I indeed participated (pro-bono) to the fact-checking of this work.

[2840] See "Quantum Impact" - The Potential for Quantum Computing to Transform Everything by Pathstone, December 2021 (35 pages).

[2841] In Report: 74% of Executives Warn Either Adopt Quantum Soon, or Risk Falling Behind Forever by Matt Swayne, The Quantum Daily, January 2022, reporting on the first annual report on enterprise quantum computing adoption commissioned by Zapata Computing: The First Annual Report on Enterprise Quantum Computing Adoption, Zapata, January 2022 (42 pages).

[2842] See Quantum Computing: 5 Potential Applications, January 2022 where this nugget can be found: "*Quantum computing could also help in the development of new encryption techniques, known as quantum cryptography*". Somebody should tell them that quantum cryptography does not run on a quantum computer!

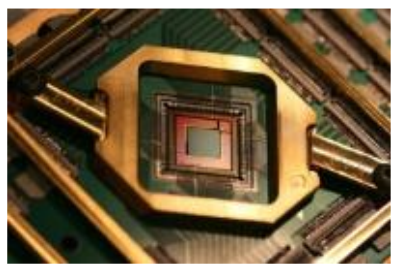

**technology screening**

- understand quantum technologies
- concepts and wording
- decipher vendor's messages and hype
- understand the news
- what can quantum algorithms do?
- case studies applicability and range

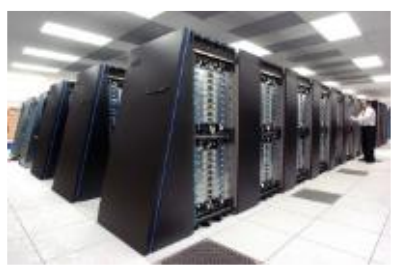

**needs analysis**

- existing unsolved problems or problems that are too lengthy or costly to solve?
- create an internal communauty
- involved security specialists
- security protocols mapping

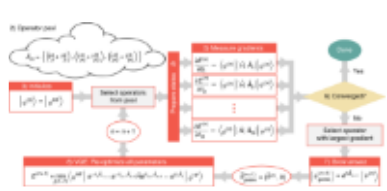

**evaluation**

- test some quantum algorithms at small scale
- on universal gates qubits as well as on quantum annealing or quantum simulators

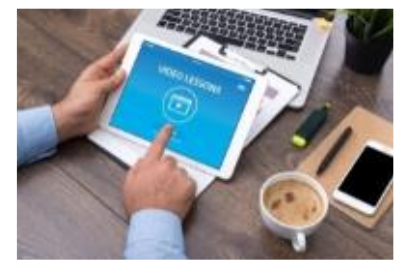

**education and training**

- some developers, IT architects and line of businesses R&D scientists.
- study the link between quantum computing and R&D unsolved problems.
- online training
- initial training

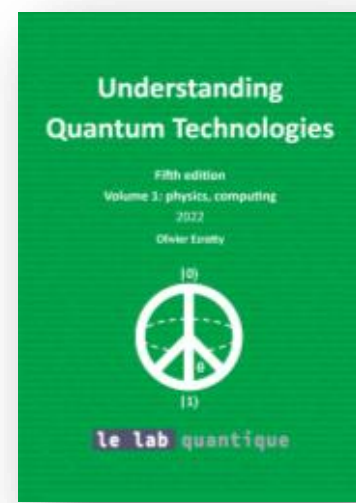

**resources**

- «Understanding Quantum Technologies» ebook (free, >1120 pages).
- ecosystem events (Q2B, QCB, Lab Quantique, …)
- vendors quantum offerings (IBM, Amazon, Microsoft, D-Wave, Pasqal, Quantinuum, IonQ, …)
- independant software vendors offerings (QC-Ware, Multiverse, …).

*Figure 876: a simple method to adopt quantum technologies. (cc) Olivier Ezratty, 2022.*

## Technology screening

Of course, you don't adopt any new technology after reading a news clip. But your management may push you to look at the trend and understand it. This top-bottom approach is amplified by the strategy consultants and analysts who are ringing the alarm bell in the direction of "business decision makers", up to, if they can, to CEO and executive boards.

The task is hardened with quantum computing because we are still in an intermediate exploratory phase where quantum computers are not yet really functional.

- Of course, you start with looking at the various **use cases** of quantum technologies in a "top-bottom" approach and what added value it could potentially bring to a business. This ebook can help you with its inventory of business applications, starting page 687. They are sorted by vertical market. If your market is not there, it doesn't necessarily mean that you shouldn't care. If you have a developer and/or mathematical background, you can have a look at what can be done at a lower level with quantum algorithms by looking at the algorithms part of this book, starting page **Error! Bookmark not defined.**.
- Then, you still need to understand the **technology dimensions** of quantum computing and related telecommunications and cryptographic matters. One thing is to understand what is the state of the art, how its changing over time, what are the scientific and technology challenges. This technology screening has to be done on a continuous basis. Things are changing fast in this domain.
- Don't miss the potential of **quantum sensing**. It may be enormous in various industries where precision is mandated. Quantum sensing helps measure with greater precision nearly any physical dimension: time, gravity/acceleration, magnetism, electro-magnetic waves and the likes.
- Learn how to **decode** analysts, research labs and vendors lingua, particularly in the field of over-promises. I provide a few examples in this book, about the fact that quantum computing is not a miracle solution that can speed up all computer processing. Learn some tricks to assess the real technology readiness level (TRL) of advertised quantum innovations. Also, understand that quantum computing is not adapted for big data applications. One important aspect here is the timing of innovations given the analysis timeframe is quite large, sometimes accounted in decades.

- Attend **ecosystem events** such as the QC Ware Q2B conference, Lab Quantique meet-ups, or quantum business conferences that are now organized all around the world. And real-life events are back after the long covid lockout period of 2020/2021.

## Needs analysis

Many vendors will push you to look at your needs even before describing what is really possible to do today. This is one reasonable approach but it should not be implemented independently from a real technology assessment.

- Identify **intractable problems** in the company's applications and business needs portfolio. This is a question that developers and data scientists can sometimes answer. For example, these are complex optimization problems involving the orchestration of many resources. You have also to look at your current existing or potential usage of high-performance classical computing. What if scenarios can also be built on the power quantum computing can bring. For example, what if you could solve such or such complex business problem that was never addressed, particularly related to some optimization process?
- Then, back to technology, look at the related case studies, existing algorithms that are supposed to solve these problems. Understand the **scope of existing case studies**: are they small scale pilot projects or deployable applications? Most of the time, they are in the first category. They interrogate vendors and independent specialists on the size and characteristics of the quantum computers and/or hybrid systems associating a quantum computer and classical computers that would solve the business problem.
- Create an **internal community** of engineers and business specialists interested by quantum technologies, as Goldman Sachs, Morgan Stanley, BMW, Volkswagen, Airbus, EDF and Total have done, for example. It can be fed with presentations from research labs and vendors and also sharing the understanding engineers have about quantum technologies, identify key questions to ask, brainstorm about business needs where quantum technologies could help.
- Launch a mapping of **security protocols** threatened by quantum computers and the infamous Shor integer factoring algorithm. What data in the present that could be intercepted now could have some value in the future for an attacker? If present data has some value more than 5 to 10 years from now, you may need to start worrying and looking at QKD and PQC solutions or even revisit the way you implement applications in the cloud.
- Look at what your **peer companies** and those from your own ecosystem are doing with quantum technologies. Some may be vocal, like in the financial sector, some less. But there's now no lack of industry events where this topic is discussed.

## Training

Training a core team of people will be necessary to launch the two previous steps.

- Train a **few developers** in quantum programming. This can be done by letting people interested in the matter spend time on it on their own. The information and tools are available online with IBM, Microsoft, Atos, D-Wave and many other places. Open-source cloud-based tools are already there. The youngest and most curious developers will probably be the ones who will best adapt to quantum computing programming paradigms, which are difficult to assimilate when being trained for classical programming. These must also have a stronger mathematical background than the average developer. Analog electronics engineers can also be interested with quantum programming giving the analog nature of the underlying processes like interferences between qubits.

- Understand the links between **quantum computing and artificial intelligence**. Quantum machine learning is a new sub-discipline of quantum algorithms that deserves to be explored and understood.
- The small hidden advertising in this book is here: I propose a **one or two days customized training** for corporate engineers, IT people, R&D and innovation specialists who are curious to discover the whereabouts of quantum computing and other quantum technologies.

## Evaluation

- Talk to the many quantum computing **independent software vendors**, particularly with those who are specialized in your vertical.
- Test **some algorithms** in the cloud with universal quantum computers (IBM, Amazon and Microsoft cloud, Xanadu, etc) or quantum annealer (D-Wave) or with emulators (Atos, IBM, Microsoft, Amazon, Google). The available case studies are discussed in this book in the section on algorithms and applications by market, starting page 693.
- Do not hesitate to test algorithms on D-Wave **quantum annealers** despite their relatively poor image among universal quantum computer purists. Quantum algorithms for these computers are suitable for solving complex optimization problems and represent a large part of what quantum computing can bring, whether in biology or finance, to take just two examples.
- Also keep an eye on **quantum simulations** which are useful for solving two main classes of problems: materials and chemical simulations, and complexity problems. Pasqal (France) and ColdQuanta (USA) are not far from delivering very interesting hardware here. The first Pasqal system with 100 cold atoms qubits is available in the cloud since mid-2022.
- Avoid the **do nothing approach**. Since quantum technologies adoption takes a while, you would be left behind against your competition. This may look contradictory with the need to avoid falling into the current quantum hype. Well no. Sort the hype and find what is useful! You’ll find stuff!

Congratulations, you have saved yourself an overpriced McKinsey or BCG study!

# Quantum technologies and society

We will leave quantum physics, hardware, mathematics and algorithms to focus on the links between quantum technologies and society. We are still at the very beginning of this technology revolution. What will follow is a mixture of observations and interpolations. Like with any digital technology wave, the quantum wave will affect society and industries at several levels, some of which can be anticipated, others less easily.

I am interested in connecting the potential impact of quantum computing with regards to mankind ambitions, the role of science fiction in the buildup of quantum imaginary, the philosophy of quantum physics, the way in which religions and spiritual movements may embed quantum whatever in their thinking, quantum technologies ethics, education and training in quantum computing, the role of gender balance in the sector and, at last, quantum vendors marketing side effects.

## Human ambition

Quantum computing is easily presented to the general public, or understood, as bringing a computational power defying imagination, going beyond anything that has been done so far. Quantum computing would thus be a way to circumvent the current sluggishness of Moore's law. It would make it possible to maintain some sort of eternal technology growth exponentiality. This may give the impression that, with quantum computing, mankind will have a tool providing him with infinite power and total control of information, in the line of many myths built around artificial intelligence and its ultimate mythical destiny, Artificial General Intelligence (AGI). In 2018, the futuristic American physicist and author **Michio Kaku** predicted that quantum computers will be the ultimate computers capable of surpassing human intelligence[2843]. Here we go again with the Singularity!

Artificial intelligence and quantum computing seem to have no boundaries. They illustrate mankind's desire for power and omniscience, to shape matter if not minds, and to have the capacity to predict the future, making it almost deterministic. So much that it would be the abandonment of free will[2844]. Of course, not!

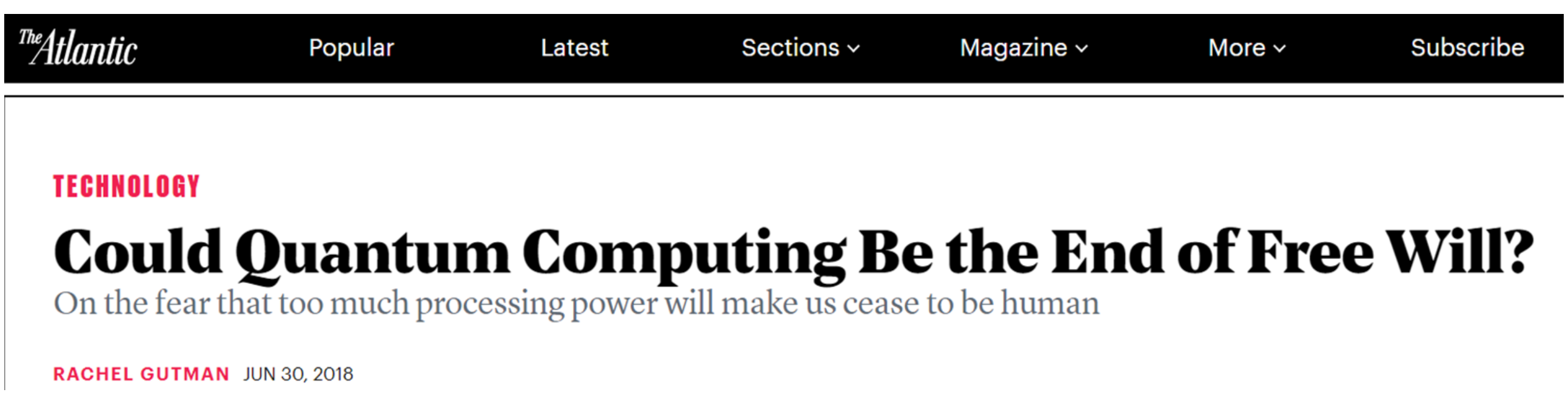
The Atlantic Popular Latest Sections Magazine More Subscribe

TECHNOLOGY

Could Quantum Computing Be the End of Free Will?

On the fear that too much processing power will make us cease to be human

RACHEL GUTMAN JUN 30, 2018

*Figure 877: no, quantum computers won't end free will!*

Quantum physics has generated its share of questions about the nature of the world. The indeterminism of quantum state measurement has become that of life. Quantum entanglement has given rise to pseudo-scientific explanations of telekinesis and the transmission of thought. We will see in the following section how quantum medicine mixes nano and macro worlds in a fancy way.

[2843] See The World's Most Disruptive Technology (That No One Is Talking About), Part II by Ian Connett, 2018.

[2844] As suggested by this article in The Atlantic of June 2018, the title of which has little to do with the content!

The mechanical nature or not of consciousness is at stake. For **epiphenomenalism** (definition), our consciousness is the result of physical phenomena in our body and brain but without direct external physical effects. Behavior is the result of the brain's action on the muscles.

For **mysterianism**, the understanding of consciousness is beyond the reach of Man. As consciousness depends at a low-level on quantum phenomena which govern a-minima the relations between atoms of the molecules of our brain, some people deduce a little quickly that quantum computing would allow AI to become general as in this debate! But these are at this stage fancy elucubrations.

Ambitious projects such as the European **Human Brain Project** led by Henri Markram aim to simulate the brain's behavior in a computer and thus to understand how it functions from start to finish, even if it is not possible to do so on even a molecular scale. In another fashion, the ability of quantum computers to simulate quantum phenomena has also sustained the idea that we are objects of a great simulation. An idea that ignores the constraints of dimensionality.

An exploration of the mysteries of quantum computing and complexity theories allows us to put our feet back on the ground. Complexity theories describe various limits to the nature of problems that can be solved with quantum computing. Computational omnipotence does not exist. We will always be obliged to use various forms of reductionism to simulate the world, i.e. we will only be able to do it correctly at "macro" scales and not at "micro" or "nano" scales for matters related to computational magnitude[2845]. A bit like predicting the weather thanks to the finite element method applicable to large portions of sky and not at the level of each water molecule.

The limits of the possible will be constantly pushed back, but they will remain. Just like those of understanding the world which are confronted with the temporal and spatial limits of the Universe. We will probably not be able to know what was happening before the big bang nor to evaluate the existence of multiverse. Being unverifiable, these interpretations of the world can only remain speculations and not become real science. In the same way, our physical means will probably never make it possible to simulate our world in-extenso.

Quantum physics also introduces a lot of chaos and randomness into biology that no computer will ever be able to fully simulate and control.

Finally, this quote from Scott Aaronson sums up the quest for quantum computing. This would be justified by the desire to counter those who say it is impossible. The rest is the icing on the cake[2846]. This is obviously some humor, not to be taken at face value!

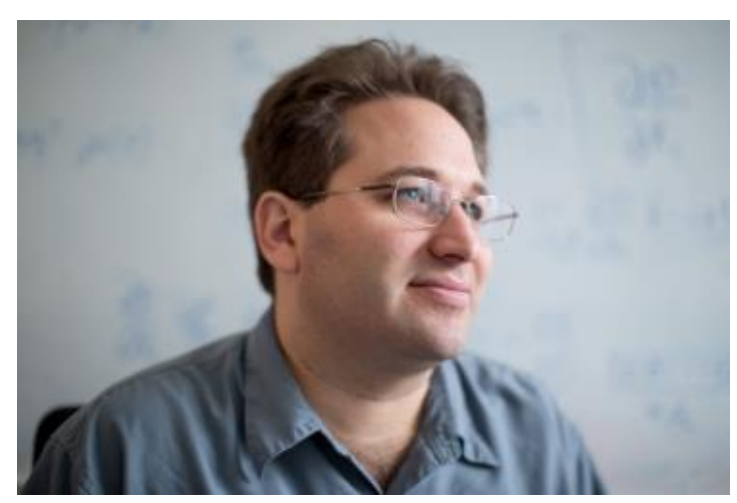

"*For me, the single most important application of a quantum computer is disproving the people who said it's impossible. The rest is just icing on the cake*"
**Scott Aaronson**

*Figure 878: Source: A tale of quantum computers by Alexandru Gheorghiu (131 slides).*

## Science fiction

Science fiction and particularly movies and TV series have been great sources of inspiration and also of delirium about the potential of quantum technologies.

---

[2845] See Three principles of quantum computing by Yuri I. Ozhigov, Moscow State University of M.V. Lomonosov, June 2022 (10 pages) which tries to address this topic.

[2846] The quote comes from A tale of quantum computers by Alexandru Gheorghiu (131 slides, slide 31). The Combination Problem for Panpsychism by David Chalmers (37 pages) and Why Philosophers Should Care About Computational Complexity by Scott Aaronson (59 pages) ?

They have created an imaginary world made of teleportation (**Star Trek)**, supraluminal speed transportation (**Star Trek**, **Star Wars)**, various entanglements and miniaturization (as in **Ant Man**[2847]), states superposition (**Coherence)**, parallel or multiverse worlds (**Fringe**, **Spiderman**, **Counterpart**, **Dark**, **Doctor Strange in the Multiverse of Madness**) or time travel (**Interstellar**, **Umbrella Academy**).

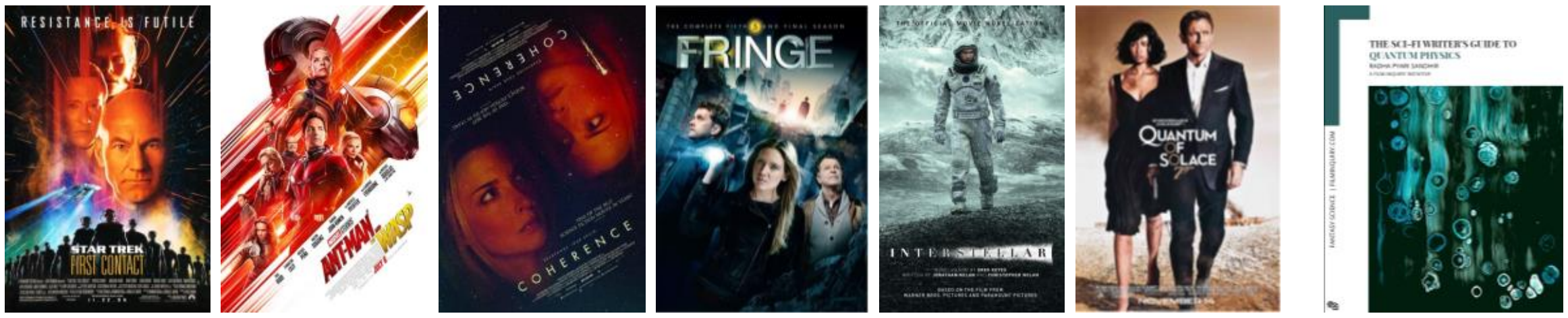

*Figure 879: quantum in science fiction movie and TV series.*

In some cases, the quantum term is used without any scientific connection to quantum physics, as in the 2013 James Bond **Quantum of Solace,** which means approximately "an ounce of consolation".

Or it plays the role of the "MacGuffin", popularized by **Alfred Hitchcock,** the gizmo that the protagonists will look after from the beginning to the end of a movie without us really understanding what's inside or all about. This is the case of the **Ronin** movie. We find another one in the movie **Hard Kill** with Bruce Willis released in February 2020. Bad guys are trying to get the "code" that will activate a "quantum AI", but its contours are quite blurred. All we know is that it could eventually do some "good" things just like hacking an airliner to crash it. In short, it's a banal "dual" civil-military solution. The bullets will rain down until the bad guy is dead without us really knowing what it's all about.

However, a small 11-page guide tries to explain quantum physics to screenwriters[2848]! It contains some language basics that can be used to create scripts. The usual scriptwriters do not hesitate to twist things, like **Christopher Nolan** with his elastic vision of time arrows in Interstellar or Tenet.

In March 2020, the eight-episode TV mini-series **Devs** was the first to be built around the prowess of a quantum computer capable of reconstructing the past up to Christ's crucifixion and predicting the future anywhere on Earth. With a video! Of course, this doesn't make any sense with today's technologies, but also with those of tomorrow[2849].

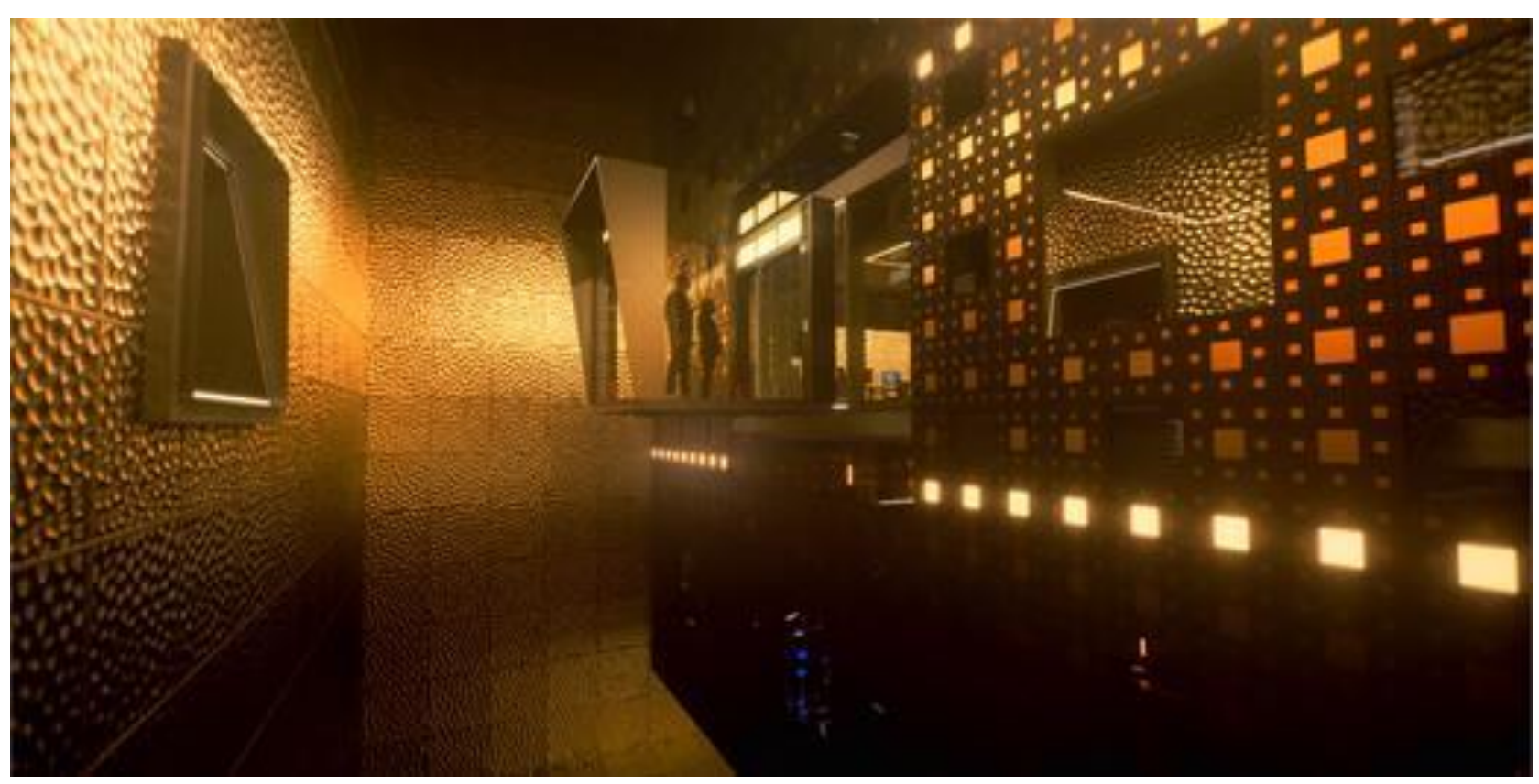

*Figure 880: Dev's series quantum computer is sitting in a suspended huge cage.*

[2847] See ['Ant-Man' science adviser explains the real-life physics behind the film]() by Denise Chow, July 2018, which explains the links between quantum physics and the scenario of the last [Ant Man.]() Well, knowing that there is none to enlarge or miniaturize a character.

[2848] See [The Sci-Fi Writer's Guide to Quantum Physics]() by Radha Pyari Sandhir, 2019 (11 pages).

[2849] The stylized quantum computer features an elongated candlestick that resembles those of IBM and Google quantum computers. It is not connected to anything at all from the top, but that's okay! The whole thing is enclosed in a huge cube that is suspended and magnetically isolated. See this beautiful analysis of the series: ["Devs" by Alex Garland: a quantum thriller in Silicon Valley]() by Romane Mugnier in Usbek&Rica, May 2020.

It's a level of complexity problem and also about getting the training data. Even assuming that the Universe is totally deterministic, it is impossible to capture the precise position of all particles in space to determine their past and future. And this comes up against one of the key principles of quantum physics, Heisenberg's indeterminacy.

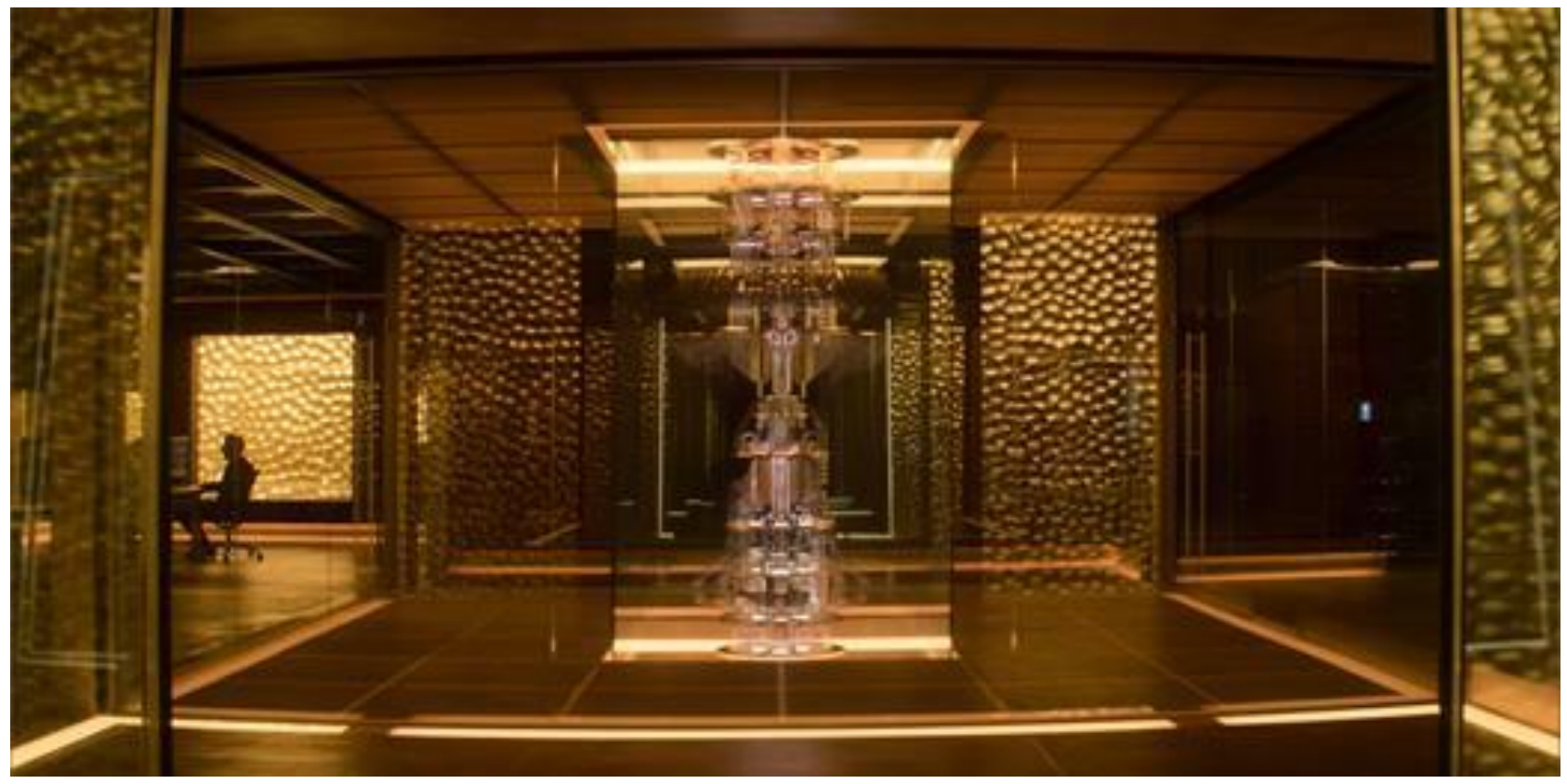

Figure 881: Dev's quantum computer is not well isolated…!

Its derivative states that one cannot accurately capture the position and velocity of an elementary particle. From this point on, everything falls into place to model and simulate the world with precision!

In 2015, episode 11 of season 2 of **Scorpions** featured a quantum computer made of lasers and a large plexiglass cube capable of injecting ransomware into the US Federal Bank with just 4 qubits! Quite a feat! The heroes hack the computer dressed as cosmonauts and by redirecting the beam of one of the lasers towards the luminous cube. We are far off from any realistic quantum computer here!

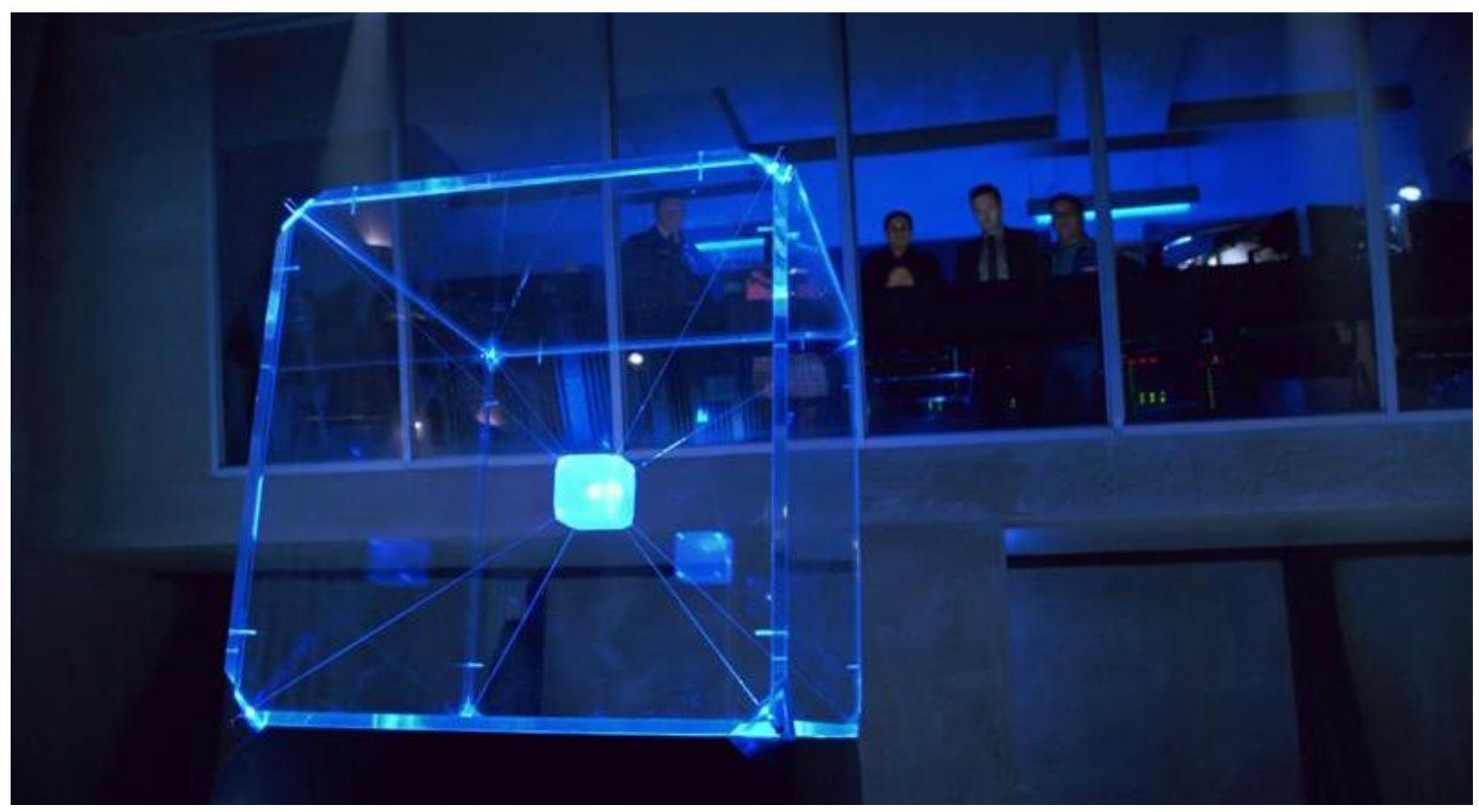

Figure 882: Scorpion's quantum computer could endanger banks with… 4 qubits!

Science-fiction is fine when it stays in the science-fiction realm. The problem starts when pseudo-researchers present science-fiction as if it was actual science instead of classifying it in a rough "hard science-fiction" category that is looking for some form of scientific credibility although being most of the time heavily farfetched. So, when some singularists tell you science and quantum physics could help resuscitate the dead using some fancy Dyson sphere and the likes, just forget it or just have some fun[2850].

These science-fiction dreams are far removed from the science of today and probably tomorrow. Their benefit is to create vocations. Dreaming drives innovation. Even when a young person discovers that science does not allow them to realize the scenarios of these fictions, they can discover the infinite field of applications of quantum physics and still be creative. If real-world quantum technologies are less impressive than Star Trek magic, it still can do wonders and bring new generation of researchers and innovators.

[2850] See A Dyson Sphere Could Bring Humans Back From the Dead, Researchers Say by Stav Dimitropoulos in Popular Mechanics, March 2021. Which refers to Classification of Approaches to Technological Resurrection by Alexey Turchin (Digital Immortality Now, Foundation Science for Life Extension) and Maxim Chernyakov (Russian Transhumanist Movement), not dated (39 pages). It suggests a Dyson sphere, some quantum algorithm based on a QRNG and some weird magic with an Everetian parallel universe could help resuscitate the dead. The science-fiction, not science at all. It also suggests quantum physics could help read data from the past, a bit a la Devs.

The use of quantum physics in Hollywood movies can also be used to convey other messages. As is often the case, they can agitate the potential of an external threat against which the USA should respond with strength. It would not be surprising to see fictions emerge in which the quantum threat comes from China. These movies often illustrate the myth of the hero who can get through adversity, also illustrating an alternative to the centralized powers of governments[2851].

In novels, fiction can also have pedagogical virtues. This is to some extent the case of **The Key of Solomon,** a novel by Portuguese author José Rodrigues Dos Santos published in 2015. In an affair mixing espionage and quantum computing, the hero spends his time teaching quantum physics to the other protagonists of the story. This gets the message across in a didactic way and without overly taxing science.

## Quantum foundations

Philosophy is a process of critical reflection and questioning about the world, knowledge and human existence. It creates a connection between all these dimensions. The discovery of quantum physics at the beginning of the 20th century created a real philosophical shock wave through the upheavals it brought to our understanding of the world at the microscopic level[2852]. It called into question key notions such as the links between reality and observations, or between ontology and epistemology. And the debates are still raging about it. If you meet a group of quantum physicists and want to have some fun, ask them the simple question: “what is a quantum state” or “what does the wavefunction mean?”! They may not agree on a (simple) answer.

### Quantum physics and its missing ontology

Science has always been closely linked to philosophy. It is not by chance that a doctorate is a "PhD", or Doctor of Philosophy, whether in humanities or in so-called hard or exact sciences.

The great physicists and mathematicians of the 19th and early 20th centuries were also philosophers, which is less common now, due to a process of accelerating specialization.

The creators of quantum physics were constantly questioning the impact and meaning of their discoveries. **Niels Bohr** was also both a physicist and a philosopher, influenced in particular by **Søren Kierkegaard** (1813-1855, Danish). **Erwin Schrödinger** was even more of a philosopher than a physicist. He had studied Western and Indian philosophy before creating the famous wave function that bears his name[2853]. An assistant to Niels Bohr, **Werner Heisenberg** had also invested a lot of his time in philosophy and it related well with his work around the mathematical modeling of quantum physics and the famous indeterminacy principle.

---

[2851] See Quantum Computing, Hollywood and geopolitics by Jean-Michel Valentin, March 2019. The author is a French specialist in strategic studies, sociology of defense and American strategy. The article relies heavily on the film Mortal Engines (2018), whose scenario only indirectly emphasizes quantum, with a past quantum war that ravaged the planet.

[2852] In practice, these upheavals occur mainly at the nanoscopic scale, that of atoms and their constituents, the nuclei and electrons. However, quantum effects can also be observed at the scale of large groups of particles that can be microscopic, as is the case with large molecules and their wave-particle duality, in Bose-Einstein condensates or superconducting currents. Knowing in all this that the frontier between quantum physics and classical physics has regularly evolved over the last century.

[2853] Michel Bitbol indicates that in the epilogue of "What is life? Mind and matter", Erwin Schrödinger wondered whether consciousness was singular or plural. If consciousness is only experienced in the singular, its extension to a global consciousness such as that of the Universe is only a risky extrapolation and difficult to prove experimentally. The thesis of a consciousness of the Universe is defended by some scientists. See for example Is the universe conscious? It seems impossible until you do the math by Michael Brooks, April 2020, which refers to the work of German mathematicians who try to define mathematically the notion of consciousness, allowing them to apply it then to the universe as a whole. Details are in The mathematical structure of integrated information theory by Johannes Kleiner and Sean Tull, 2020 (22 pages). It's cold and abstract!

The debates that agitated the physicists of quantum mechanics often took as much the form of philosophical jousting as of physical or mathematical debates, all the more so since the founders of quantum physics were not experimenters and were rather theoreticians[2854]. History has, moreover, forgot the names of the experimentalists[2855].

Quantum physics has generated endless debates since its beginnings because its formalism is difficult to associate with the principles of reality usually applicable in classical physics. Intuitive classical physics understanding has historically been associated with an ontology. In Newtonian physics, the notion of state with position and motion of an object and the laws of evolution of these properties allow the prediction of phenomena such as the motion of planets. These evolutions are perfectly observable and deterministic.

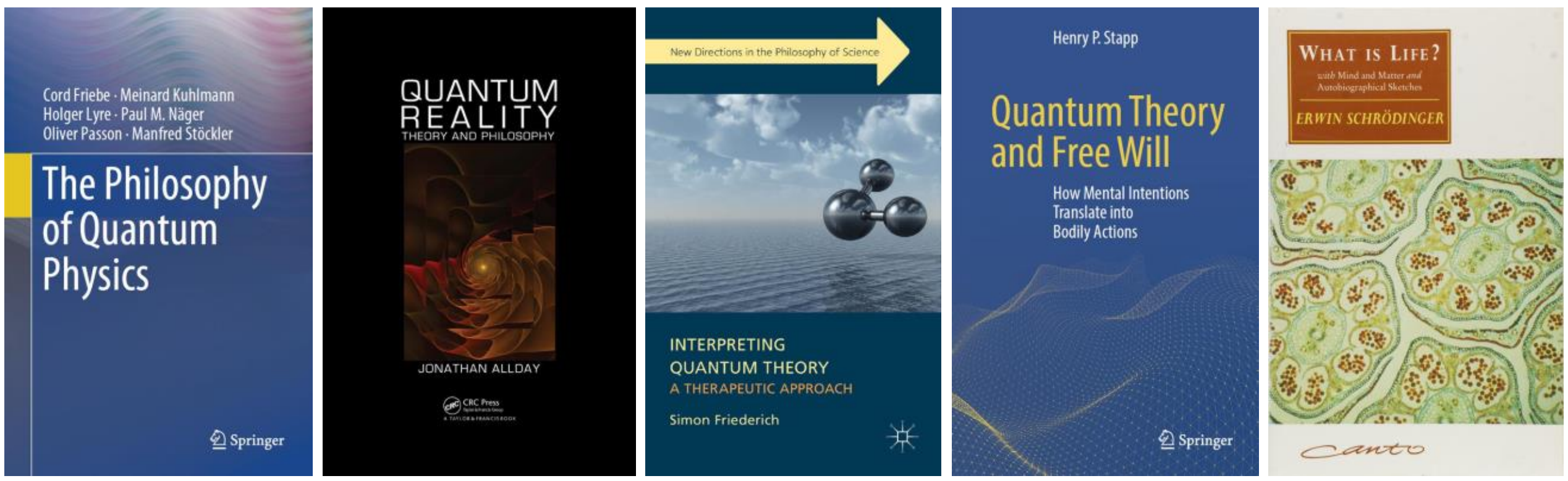

*Figure 883: some books on quantum physics and philosophy.*

Quantum physics was founded without such an ontology although it served to explain some known physical phenomenon like the blackbody radiation, the photoelectric effect or hydrogen's absorption and emission spectrum[2856].

It was created as a set mathematical postulates that could help predict experimental results. You have mainly the Schrödinger wavefunction for non-relativist massive particles and others like Dirac and Klein-Gordon equations for relativist particles. In quantum physics, the prediction instrument is a probabilistic wave function that is difficult to apprehend. It is coupled with a whole host of new notions that have no equivalent in the macroscopic and classical world: energy quantification, wave-particle duality which applies to matter (electrons, atoms) and photons (all have a momentum p related to a wavevector k using Planck's constant $\hbar$, as $p = \hbar k$), the influence of measurement on the quantities to be measured[2857], measurement indeterminacy and the notion of chance.

[2854] The book Fantaisies quantiques - dans les coulisses des grandes découvertes du xx<sup>e</sup> siècle by Catherine d'Oultremont and Marina Solvay, 464 pages (2020) tells the story of the famous 1927 Solvay conference. It is a very beautiful history of quantum physics that tells touching stories of its various protagonists in the first half of the 20th century. It seems the book is not yet available in English.

[2855] We mentioned many of them at the beginning of this book, such as Johann Balmer, Theodore Lyman, Friedrich Paschen, James Chadwick, Arthur Holly Compton, George Paget Thomson, Clinton Davisson and Lester Germer. The names of these experimental physicists generally do not ring a bell to the general public and scientists, whereas the general public has heard much more about Max Planck (with his constant more than for the black body radiation explanation), Albert Einstein (for the theory of relativity more than for the photoelectric effect explanation), Niels Bohr (for his model of the atom), Erwin Schrödinger (more for his cat analogy than for his wave equation) and Werner Heisenberg (for his indeterminacy principle, commonly called uncertainty, but not much for this huge work on quantum physics mathematical foundations). Among the founding fathers, Paul Dirac, Wolfgang Pauli and John Von Neumann were geniuses but are way behind in notoriety.

[2856] An ontology deals with what is, types and structures of objects, properties, events, processes and relationships in all areas of reality. It is usually opposed to epistemology, which covers how to obtain valid knowledge.

[2857] This is not valid only in quantum physics and the infinitely small. It works regularly at the macro scale, as in any survey with biased questions for example.

Quantum physics is a predictive, not descriptive theory. It doesn't describe physically the electrons and other particles when they behave quantumly. It doesn't physically explain entanglement nor wave-particle duality.

Einstein's position was that quantum physics was an incomplete theory when creating his famous EPR paradox in 1935. Werner Heisenberg asserted in 1927 that quantum physics established the final failure of causality! The knowledge of the present no longer made it possible to predict the future from the application of the laws of physics, all the more so as the knowledge of the present with precision is also impossible[2858].

Some like Niels Bohr concluded that it was useless and even counterproductive to create some quantum physics ontology. Many attempts were contradicting each other or weren't even realist per se. Some like Hugh Everett believed that reality was a universal $|\Psi\rangle$ function, David Bohm devised some pilot waves explanations, Christopher Fuchs et al are focused on the role of agents actions and experiences in quantum Bayesianism and its derivative QBism, CSM's ontology argues that states pertain to systems and contexts, and so on. We end up having competing postulated ontologies frequently enabling the same predictions. These are hard to sort out.

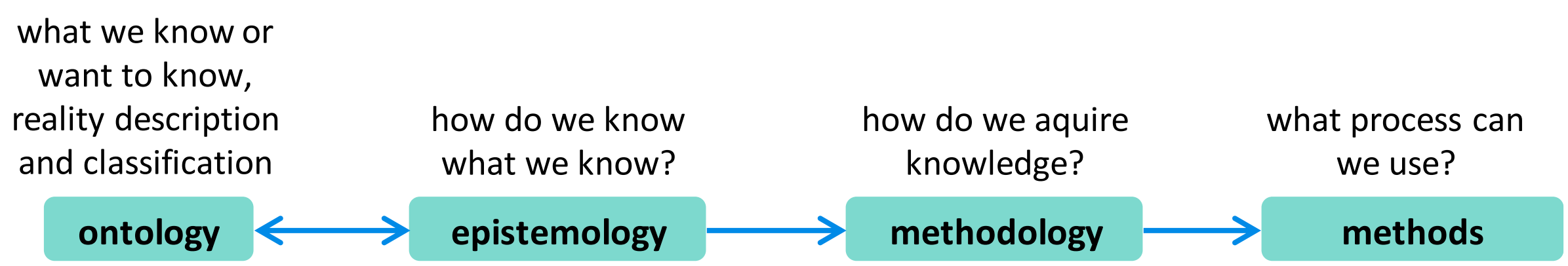

*Figure 884: ontology, epistemology, methodology and methods defined.*

The relationship between measured values, measurement and the observer is also debated. Would a true measurement be one that does not alter the quantity to be measured at all, a feat hard to attain in the infinitely small? In fact, quantum mechanics is contextual, the measurement depends on its context, which does not detract from its objectivity[2859].

The mathematical formalism of quantum physics from 1900 to 1935 was not at all disconnected from the observable physical world. It made it possible to explain experimentally studied phenomena such as black body radiation, interference from Young's slits with light and matter waves, or spectral excitation lines of atoms under a wide range of conditions. We have seen how important they are in photonics, with trapped ions, cold atoms or NV centers. Electron spin explained the hyperfine energy levels of atoms observed in 1922 in the Stern-Gerlach experiment[2860].

Relativistic quantum chemistry derived from Paul Dirac's equations explained spectral shifts of transitions involving low layer electrons of heavy atoms moving at relativistic velocities. The list is long.

If quantum physics explained experimental measurements, linking the observed reality and the theory, it was however insufficient to produce an unanimously accepted representation of reality. It is part of a history of science that described matter step by step, with nested Russian dolls. Atoms were initially abstract, theoretical entities before being embodied and accurately described and then directly observed as we do now with electron microscopes or cryogenic microscopy (Cryo-EM).

---

[2858] "*In the strong formulation of the causal law, 'If we know the present with exactitude, we can predict the future,' it is not the conclusion, but rather the premise that is false. We cannot know, as a matter of principle, the present in all its details.* " vu dans [One Thing Is Certain: Heisenberg's Uncertainty Principle Is Not Dead] by Ava Furuta in Scientific American, 2012.

[2859] This approach is challenged by the Bayesian quantum interpretation ([QBism] for Quantum Bayesianism) promoted from 2002 onwards by Carlton Caves, Christopher Fuchs, Rüdiger Schack and then David Mermin. See in particular QBism [The Future of Quantum Physics] by Hans Christian von Baeyer, 2016 (268 pages).

[2860] This one made a beam of heated silver atoms pass through a non-homogeneous magnetic field, which generated two distinct spots on a screen.

The very existence of atoms was debated at the end of the 19th century between Ludwig Boltzmann who believed in them and Wilhelm Ostwald and Ernst Mach who opposed them.

Protons and neutrons were then discovered. These were split into quarks and gluons with particle accelerators, turning the physical world into a maybe endless fractal. Obstacles to understanding it could simply be related to the enormous amount of energy that needs to be injected into particle accelerators, which is increasing the more elementary the particles are.

### Quantum Physics interpretations

Quantum physics philosophy belongs to the broad field of **quantum foundations**. It focuses essentially on the multiple possible interpretations of the same theory and their mathematical formalism. They all ask related questions such as: does reality exist independently of the observer? What is the physical meaning of the wave in the wave-particle duality? Is it a real wave of an indeterminate nature or is it a simple statistical and probabilistic mathematical model incomplete in its ability to describe physical reality[2861]? Other quantum foundation fields include work on the measurement problem and its indeterminism, the associated notions of causality, local realism and the completeness of quantum mechanics[2862].

Many of these theories deal with the complicated notion of **contextuality** in quantum physics according to which quantum measurements is not revealing pre-existing values but depends on the measurement context like the angle of a spin measurement in the famous Stern-Gerlach experiment or of a photon polarization angle[2863]. Contextuality is dealt with in the **Kochen-Specker** theorem *aka* the Bell–Kochen–Specker theorem demonstrated by John S. Bell in 1966 and by Simon B. Kochen and Ernst Specker in 1967. It's a "no go theorem" that creates constraints on the types of putative hidden-variable theories trying to explain the predictions of quantum mechanics in a context-independent way.

Several interpretations of quantum physics have thus emerged to try to provide answers to these many questions.

---

[2861] These different interpretations can be evaluated according to the criteria of scientificity of Karl Popper (1902-1994, Austrian/English), according to which a theory is scientific if it can be refuted by crucial experiments giving precise results. The theory cannot be shown to be irrefutable. A proven scientific theory is therefore always between two waters, in the state of a theory corroborated by facts, until proven otherwise. The history of physics has shown, however, that the "serious" theories of the past were mainly challenged by the broadening of their perspective and context: with large masses and high velocities (for relativity) and in the microscopic (for quantum physics). In their initial contexts, they remained perfectly valid. I like the very current example of the search for dark matter, which would represent 85% of that of the Universe. Its existence is not yet experimentally demonstrated but is assumed by the application of the laws of gravity and relativity applied to the cohesion of galaxies. It can be refuted or partially verified at present in at least three ways: by discovering elementary particles associated with dark matter (quantum detectors are built in this sense, and have so far given nothing), by modifying the laws of general relativity as the Israeli Morchedai Milgrom is trying to do, or by discovering hidden matter such as the dust of galaxies that could explain all or part of their cohesion without using dark matter. Belief in God and many areas of metaphysics are not part of science because they are neither demonstrable nor refutable. See on this subject the interesting debate between André Comte-Sponville and Jean Staune in André Comte-Sponville - Jean Staune: Will science refute atheism?, June 2007, where some allusions are made to quantum physics.

[2862] See Argument for the incompleteness of quantum mechanics based on macroscopic and contextual realism: EPR and GHZ paradoxes with cat states by Jesse Fulton et al, August 2022 (20 pages). The hidden-variables debate is not closed yet!

[2863] Works on contextuality frequently deal with discrete variable quantum systems. It can be extended to continuous variable systems. See The Interplay between Quantum Contextuality and Wigner Negativity by Pierre-Emmanuel Emeriau, April 2022 (221 pages), a thesis under the direction of Elham Kashefi and Shane Mansfield and the related Continuous-variable nonlocality and contextuality by Rui Soares Barbosa, Tom Douce, Pierre-Emmanuel Emeriau, Elham Kashefi and Shane Mansfield, May 2019 - April 2022 (44 pages).

**Copenhagen interpretation** is the canonical version of quantum foundations[2864]. It is essentially probabilistic. Quantum physics postulates and the wave function describe all that we can know about reality but not reality itself, which is neither accessible nor meaningful. It adopts the positivist approach according to which one sticks to observations, laws and phenomena, without seeking to know their intrinsic nature. It was the "Bohrian" side of the historical debate between **Niels Bohr** and Albert Einstein, mainly between 1927 (in the famous Solvay Conference in Brussels) and 1935 (the EPR paradox paper and subsequent discussions). Adopted by Werner Heisenberg, Max Born, Wolfgang Pauli, Paul Dirac, it is the classical and dominant interpretation of quantum physics that is still mostly taught like in the Cohen-Tannoudji/Laloe/Diu bible of quantum physics. It is satisfied with an essentially mathematical and probabilistic model that does not seek to physically describe the entire real world. There are, moreover, sub-branches in the Copenhagen interpretation, notably around the open and closed theories that had opposed Heisenberg and Dirac from 1929 onwards.

**Bohm interpretation** came from **David Bohm** (1917-1992, American then Brazilian and British). He proposed in 1952 a so-called deterministic version of quantum mechanics, called "De Broglie-Bohm theory". It was inspired by ideas initially promoted - but partly abandoned - by the French physicist and took up the idea of the existence of hidden variables, insinuated by Albert Einstein in the 1930s, and by Louis de Broglie, in the form of pilot waves[2865]. The existence of local hidden variables was disproved in 1982 with Alain Aspect's famous experiment. But the promoters of the therefore explicitly nonlocal Bohmian theory are still very active, including in France[2866].

**Many worlds interpretation** and its Universe wave function was proposed in 1957 by **Hugh Everett** (1930-1982, American) and after being almost forgotten, revived by **Bryce DeWitt** in 1970. It then became the multiple worlds or multiverse interpretation in an article published in Physics Today[2867]. It is said to be realistic in the sense that the Universe is a huge wave function with a (immensely) large number of parameters. It never collapses and the world is deterministic, but split in parallel branches. DeWitt's interpretation transforms quantum probabilities within this universal wave function into parallel worlds that exist simultaneously. Since it is impossible to verify that parallel worlds exist, the theory is not refutable. It's the case with all interpretations based on the same formalism. We are therefore far from an experimentally supportable interpretation[2868]. This theory has also been promoted by David Deutsch, also known for his quantum algorithms.

[2864] The "Copenhague interpretation" naming was created by Werner Heisenberg in 1955. It consolidated Heisenberg's and others contemporary views on the realism of quantum physics and concepts that were not promoted by Niels Bohr in the 1920s like the wave packet collapse and the views on the subjective observer in quantum measurement. It's a post-mortem consolidation on the quantum physics foundations. See Who Invented the "Copenhagen Interpretation"? A Study in Mythology by Don Howard, Department of Philosophy of University of Notre Dame, Indiana, 2004 (15 pages).

[2865] The Bohmian approach is well popularized in Quantum Physics Without Quantum Philosophy by Detlef Dürr, Sheldon Goldstein and Nino Zanghì, 2013 (304 pages). It is completed by Quantum solitodynamics: Non-linear wave mechanics and pilot-wave theory by Aurélien Drezet, July 2022 (29 pages).

[2866] With Aurélien Drezet from Institut Néel, Grenoble.

[2867] See Quantum mechanics and reality by Bryce S. DeWitt, 1970 (6 pages) as well as The Many-Worlds Interpretation of Quantum Mechanics by Bryce DeWitt and Neill Graham, 1973 (146 pages) which contains "The theory of the universal wave function" by Hugh Everett, 1957. DeWitt's interpretation is also called EWG for Everett-Wheeler-Graham. John Wheeler was supervisor of High Everett's thesis and Neill Graham, a student of DeWitt. Seen in Everett's pure wave mechanics and the notion of worlds by Jeffrey A. Barrett, 2011 (27 pages).

[2868] See Making Sense of the Many Worlds Interpretation by Stephen Boughn, 2018 (36 pages) which dismantles a bit the model of parallel universes, especially in terms of dimensioning. By calculating the number of bifurcations of the Universe since its birth, and taking Planck's time as a basis, we end up with a number of parallel worlds that is beyond comprehension and all imaginable analogies. As for Schrödinger's cat, the dead cat and the living cat cohabit in two parallel worlds and the matter is settled!

It feeds a many science fiction drams and mad mysticism, everything being linked to everything and vice versa, especially souls and consciences. Howard Wiseman extended this theory with embedding interactions between these many worlds[2869].

| | Copenhagen | Bohm | Everett / DeWitt |
|---|---|---|---|
| world entities | macroscopic quantum objects | wave function and particles position | wave function with quasi-classical world |
| determinism | indeterminism | determinist | deterministic |
| probabilities interpretation | objective | epistemic | objective |
| theories predictions goal | measurement results | particles position | agents bet |
| locality | non-locality | non-locality | locale |
| theory mathematical formalism | Schrödinger equation, projections, probabilities | Schrödinger equation and pilot waves | Schrödinger equation |

*Figure 885: the top three interpretations of quantum physics. Source: the excellent thesis The plurality of interpretations of a scientific theory: the case of quantum mechanics by Thomas Boyer-Kassem, 2011 (289 pages).*

**GRW theory** published in 1986 by the Ghirardi-Rimini-Weber trio proposes a different formulation of Schrödinger's equation with a spontaneous reduction of the wave function that is not simply related to the notion of measurement. This is a rare formalism that could experimentally validated.

**QBism** is a derivative from quantum Bayesianism, starting with some ideas by **Edwin Jaynes** (1922-1998, American, yes, the Jaynes from the Jaynes-Cummings Hamiltonian) and pushed by **Christopher Fuchs**, **David Mermin** et al, based on **Frank Ramsey**'s anti-realist interpretation of probability (1903-1930, British philosopher) and **Ludwig Wittgenstein**'s work (1889-1951, Austrian philosopher)[2870]. It interprets quantum physics through the eye of the observer agent actions and experiences.

**Relational Quantum Mechanics** (RQM) was crafted by **Carlo Rovelli** in 1994. This relational ontology considers that a quantum state is defined by the relation between any pairs of systems. One of them can be an observer. It is inspired by special relativity principles and its observer reference model.

**CSM ontology** was proposed in 2015 by **Alexia Auffèves** and **Philippe Grangier** in order to reconcile the Copenhagen interpretation with realistic models[2871]. CSM is a minimalist ontology designed to pacify somewhat these old debates.

---

[2869] See Quantum Phenomena Modeled by Interactions between Many Classical Worlds by Michael J. W. Hall, Dirk-André Deckert and Howard M. Wiseman, PRX, 2014 (17 pages).

[2870] See Quantum Wittgenstein - Metaphysical debates in quantum physics don't get at 'truth' – they're nothing but a form of ritual, activity and culture by Timothy Andersen, Aeon, May 2022.

[2871] CSM results from the creation with Nayla Farouki of the CEA of a group dedicated to the foundations of quantum mechanics in Grenoble. In 2013, they form a group with Philippe Grangier, who has long defended contextual objectivity. CSM is documented in several papers: Contexts, Systems and Modalities: a new ontology for quantum mechanics, January 2015 (9 pages) lays out the key principles of CSM ontology, tying physical properties to the system, and to the context in which it is embedded. Violation of Bell's inequalities in a quantum realistic framework, International Journal of Quantum Information, February 2016 (5 pages) reuses a lot of content from the first paper, commenting on observed "loophole free" violation of Bell's inequalities. Recovering the quantum formalism from physically realist axioms, Nature, December 2016 (8 pages) derives Born's probabilistic rule and unitary transforms from CSM. Then What is quantum in quantum randomness?, Philosophical Transactions of the Royal Society A, April 2018 ( 9 pages), Extracontextuality and Extravalence in Quantum Mechanics, Philosophical Transactions of the Royal Society A, May 2018 (7 pages), A generic model for quantum measurements, July 2019 (8 pages) and Deriving Born's rule from an Inference to the Best Explanation, October 2019 (6 pages). See one critic of CSM in Comments on New Ontology of Quantum Mechanics called CSM by Marian Kupczynski, 2016 (8 pages). And the more recent Contextual objectivity : a realistic interpretation of quantum mechanics by Philippe Grangier, 2001 (5 pages).

In this model, the properties that are measured, called **modalities,** are attributed to a **system** (studied system, as isolated as possible) within a **context** (completely specified measurement device like a photon polarizer or Stern-Gerlach experiment) Modalities are jointly associated to the system and its context, not just the system, building a contextual objectivity[2872]. In CSM, randomness doesn't just come from Heisenberg's indeterminacy principle but is a direct consequence of the quantization postulate and the contextual nature of reality.

CSM can also help explain the origin of probabilities, non-locality and quantum-classical boundary. Non-locality, aka the EPR paradox, has nothing to do with an action at a distance, but appears because a modality belongs to both a system and a context. It also solves the Wigner's friend thought experiment paradox based on a recursive observer of a measurement agent[2873].

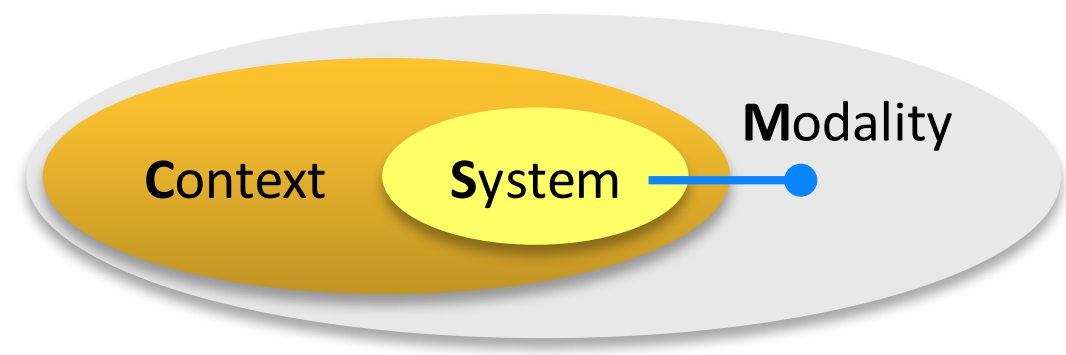

*Figure 886: CSM's simple view.*

There are many more interpretations of quantum physics than qubit types around! Like **superdeterminism** (which deterministically the observed violations of Bell's inequalities in entanglement experiments with some yet unknown hidden variable[2874]), c**onsistent histories** (which avoid the use of a wavefunction collapse to describe  physical processes and tries to get rid of the measurement theory[2875]), **modal interpretation** (class of interpretations created starting in 1972 by Bas van Fraassen which introduced a distinction between a dynamical state over time and a value state at a given time[2876]), **quantum darwinism** (which explains how the classical world emerges from the quantum world [2877]), **dynamic histories** (which reinterprets quantum mechanics as deterministically evolving dynamical world lines in a 5D universe not far from a many-worlds interpretation[2878]), **Quantum Coherentism** from Claudio Calosi and Matteo Morganti, **Foundationalism** (there must be a source of being), the **Geometrodynamic Model of Reality** from Shlomo Barak and the **Quantum Conceptual Turn** from Diederik Aerts and Massimiliano Sassoli de Bianchi[2879] and the **category theory** of Gennaro Auletta, a sort of generic unifying meta-logic theory applicable to quantum physics[2880].

---

[2872] Within the usual quantum formalism, a modality is a pure quantum state and a context is a complete set of commuting observables (CSCO). For a given context, CSM defines N distinguishable modalities that are mutually exclusive. If one modality is "true", or "realized", the others are "wrong" (or "false"), or "not realized". The value of N, called the dimension, is a characteristic property of a given quantum system, and is the same regardless of the context.

[2873] The Wigner's friend paradox is driving hot debates among physicists in quantum foundations. See for example A general framework for consistent logical reasoning in Wigner's friend scenarios: subjective perspectives of agents within a single quantum circuit by V. Vilasini and Misha P. Woods, ETH Zurich, September 2022 (47 pages).

[2874] See Rethinking Superdeterminism by Sabine Hossenfelder and Tim Palmer, May 2020 (13 pages) and What does it take to solve the measurement problem? by Jonte R. Hance and Sabine Hossenfelder, June 2022 (11 pages).

[2875] See Consistent histories and the interpretation of quantum mechanics by Robert B. Griffiths, Journal of Statistical Physics, 1984 (55 pages).

[2876] See The scientific image by Bas Van Fraassen, 1980 (248 pages) and Modal Interpretations of Quantum Mechanics, Stanford Encyclopedia of Philosophy.

[2877] Also see Quantum Darwinism, Decoherence, and the Randomness of Quantum Jumps by Wojciech Zurek, 2014 (8 pages).

[2878] See A Dynamic Histories Interpretation of Quantum Theory by Timothy D. Andersen, August 2020 (13 pages).

[2879] See Diederik Aerts and Massimiliano Sassoli de Bianchi - The quantum conceptual turn, May 2021 (48 mn) from the International Workshop on Quantum  Mechanics and Quantum Information, Quantum Ontology and Metaphysics, April 2021. See also Are Words the Quanta of Human Language? Extending the Domain of Quantum Cognition by Diederik Aerts and Lester Beltran, December 2021 (27 pages) which makes a symbolic projection from quantum physics phenomena to the way human language works.

[2880] The category theory is described at the end of The Quantum Mechanics Conundrum by Gennaro Auletta, 2019 (879 pages) which contains a good primer on quantum physics in its first 150 pages.

Other ontologies abound like **Structuralism**, **Perspectival Objectivity**[2881], **Pluralism** (atomism), **Monism** and **Infinitism**, but their scope goes beyond quantum physics.

Where are the schools of quantum foundations? In Europe, there is a quantum foundations epicenter in Italy but you find contributors in most countries. There are also two foundations who provide research grants on quantum foundations: the **Templeton Foundation** and **FQXi** (Foundational Questions Institute) which covers quantum foundations and cosmology and was created in 2005 by cosmologists Max Tegmark and Anthony Aguirre.

### Other interpretations

Among the physicists who have contributed to the field of quantum physics philosophy. **Pascual Jordan** (1902-1980, German) built a theory of free will according to which one is not freer by acting randomly or in a determined way, breaking the idea that quantum non-determinism would be a proof of human free will. **Henri Stapp** (1928, American) worked on consciousness and believes that it governs the world and reality and that it can only be explained by quantum physics[2882].

**Roger Penrose (**1931, English) considers that consciousness results from the reduction of the wave packet and **Elizabeth Rauscher** (1931-2019, American) was a physicist who first became interested in philosophy and then moved on to parapsychology[2883].

On the other hand, **Steven Weinberg** (1933-2021, American), Nobel Prize in Physics in 1979 for his work on the unification of the weak and electromagnetic forces, thought that philosophy is of little use in quantum physics other than to protect us against the errors of other philosophers[2884]. This view was shared by **Stephen Hawking** (1942-2018, English).

In France, in addition to the CSM ontology creators, **Michel Bitbol**, a biophysicist and philosopher of science, is interested in particular in the question of consciousness, **Etienne Klein,** originally an engineer and physicist, is specialized in the philosophy of science at the CEA, as well as **Alexei Grinbaum** and **Vincent Bontemps** who are both part of Etienne Klein's LARSIM laboratory.

Quantum physics raises other physico-philosophical questions such as does a total vacuum exist? Indeed, quantum physics describes the energy of the vacuum, which would always be crossed by various real and virtual particles. From a practical point of view, it is therefore difficult to create an empty space that is not crossed at all by electromagnetic waves or particles of all kinds. If therefore nothing exists, what was there before the big bang? And let's not talk about the nature of time, which is still a matter of debate.

---

[2881] See Perspectival Objectivity or: How I Learned to Stop Worrying and Love Observer-Dependent Reality by Peter W. Evans, University of Queensland, 2020 (16 pages).

[2882] See Mind, Matter and Quantum Mechanics by Henry P. Stapp, 2009 (303 pages). This is the kind of book that makes non-testable hypotheses that then become gospel for the quantum medicine quacks we are talking about in the dense section dedicated to quantum fumbling. And yet, the basic idea is nothing extraordinary: brain chemistry, like all chemistry, is based on many facets of quantum physics. This becomes complicated when the hypothesis is put forward of an implementation of entanglement in consciousness. Quantum medicine goes out of the scientific game when it claims that these mechanisms can be controlled from the simple will, without counting the action on the other organs (preferably sick ones) of the human body.

[2883] Elizabeth Rauscher was the cofounder of the Fundamental Fysiks Group in 1975 with George Weissmann to work on quantum foundations. She is co-author with Richard Amoroso of The Holographic Anthropic Universe, 2009 (510 pages). They discuss a model for creating a scalable quantum computer called "Universal Quantum Computing" that is difficult to grasp between real science and crackpot science and is based on a theory called "Unified Field Mechanics" that is difficult to evaluate. The subject is detailed in Brief Primer on the Fundaments of Quantum Computing by Richard L Amoroso, 2017 (140 pages). Richard Amoroso is Director of the Noetic Advanced Studies Institute in Oakland, California. Noetics is interested in the links between quantum states and consciousness. And this goes well beyond the realm of science with Pragmatic Proof of God (Part I & Part II, 2017 by Richard L. Amoroso (34 and 13 pages).

[2884] See the chapter "Against Philosophy" in "Dreams of a Final Theory", 1994, Steven Weinberg, which is contradicted in Physics Needs Philosophy / Philosophy Needs Physics by Carlo Rovelli, 2018. Carlo asserts that saying that science does not need philosophy is to be doing some sort of philosophy of science! See also The Trouble with Quantum Mechanics, 2016.

## Beyond Quantum Foundations

The current philosophical approach to quantum physics baffled me a bit. Most of the writings in this discipline are full of mathematics and physics. They must break records in this respect compared to any other subject covered by the field of philosophy. Above all, they do not deal much with human sciences per se.

What are the human consequences of these different interpretations of quantum physics? Are there philosophical questions other than these related to the interpretation of realism at low-level? There is much to be done in this area. The notions of uncertainty and indeterminism inevitably lead to the notion of free will and destiny (as seen by Pascual Jordan). The quantum philosophical focus on the microscopic and nanoscopic scale of physics could also be a form of reductionism preventing a wide-angle view of its societal impact.

Also, is the extension of the scientific field infinite? What are the limits of human knowledge that seeks to explain and interpret everything about the way the Universe works? What do we miss and why? What links can be made with our humility? What are the structural limits to our insatiable curiosity? I am only reformulating the very notion of Kantian metaphysics, the "*science of the limits of human reason*".

The philosophical question thus concerns the notion of the feasible and the unfeasible and its evolution over time, a perspective provided by the history and philosophy of science. What are the limits of human ingenuity? What is superhuman? Will we be able to create ultra-reliable and *scalable* quantum computers? The theories and classes of complexity, discussed in this book, should also serve as tools for this kind of thinking.

How to extend the interpretation of quantum physics to the metaphor of quantum computation: highly rich and complex inside but simple after measurement is done? Could it be used to simulate the living and create it in silico? This will then raise questions about man's power over nature and the associated responsibilities. We will also see the resurgence of debates on scientism, the "*science-led society*", as well as on technology solutionism, a concept promoted by **Evgeny Morozov**, which could provide answers to all problems, especially environmental and health problems, which cannot be treated properly with the required urgency.

These questions arise more and more at a time when precaution prevails over everything, when there are fears of technological blunders in almost every field (nuclear, GMOs, fertilizers, vaccines, artificial intelligence and 5G), when the very notion of scientific progress is no longer accepted and when cognitive relativism no longer allows us to distinguish between the serious and the farfetched, leading to a collective mistrust in science. In the following section, we will precisely study a question that belongs to the field of philosophy, the question of the ethics of quantum computing.

Are these questions really specific to physics and quantum computing? Aren't they recurrent as soon as a major new technology shows up? Perhaps, but these questions deserve to be asked, like those raised by the commoditization of artificial intelligence since 2012.

The interpretations of quantum physics are in any case there to remind us that in all matters, we must multiply the angles of view of problems to better analyze them. This is obviously full of lessons from a metaphorical point of view.

I wonder about all these questions by observing that, if they are not dealt with, they tend to become the field of esotericism and charlatanism as we will see in a following section dedicated to quantum fake sciences. It is a bit as if the philosophy of quantum physics had remained at the stage of fundamental research without moving on to the stage of applied research. In a way, it is in line with the level of market maturity of the technologies of the second quantum revolution. Let's bet that as quantum technologies will mature, the more this applied philosophy will develop and allow us to write a new chapter in this exciting history of science.

## Responsible quantum innovation

We'll cover here the broad topic of responsible innovation and ethics pertaining to quantum technologies. We'll drive some lessons from the current quantum hype and also from what happened with artificial intelligence. We'll look at the various initiatives around the world.

### Quantum hype side effects

The "quantum hype" has been perceived as being problematic for a while, mostly by many scientists who fell that the field was oversold, particularly by industry vendors of all sizes[2885]. It's hard to position the peak oil of the quantum hype ("Quipe" for some authors[2886]) but at least, with following the money invested in startups, 2021-2022 were defining years with the large funding rounds of startups like PsiQuantum and the initial public offerings (IPOs) through SPACs (special purpose acquisition companies) of IonQ, Arqit, Rigetti and D-Wave. Governments have been fueling this hype with their large national(istic) quantum initiatives and their aspirations for some technology sovereignty.

While technology hypes have always existed and contribute to drive emulation, innovations and a field attractiveness, it works well when scientists and vendors deliver progress and innovation on a continuous basis after a so-called peak of expectations. It fails with exaggerated overpromises and underdeliveries that last too long. In that case, it could cut short research and innovation funding, creating some sort of quantum winter[2887].

In the "Mitigating the quantum hype" paper published in January 2022[2888], I drove some lessons from past technology hypes and investigated the current quantum hype and its specifics. I laid out the structural changes happening like the vendors hype profound and disruptive impact on the organization of fundamental research. I then made some proposals to mitigate the potential negative effects of the current quantum hype including recommendations on scientific communication to strengthen the trust in quantum science, vendor behavior improvements, benchmarking methodologies, public education and putting in place a responsible research and innovation approach.

### Learnings from AI

Artificial intelligence ethical concerns became a real political issue in 2018. This was very apparent in France in the **Villani Mission Report on Artificial Intelligence published in** March 2018 as well as in a **Report of the House of Lords** published the same month and on the same subject in the UK[2889].

It highlighted the need to ensure, at least morally, but, if possible, practically, that AI-based solutions respect society and avoid generating or perpetuating training data-originated discriminations. Hence two salient topics such as the explicability of algorithms and the limits of the manipulation of our emotions, particularly via more or less humanoid robots and voice agents.

[2885] See Quantum computer researcher warns that the industry is full of ridiculous hype by John Christian, Futurism, March 2022, Quantum Computing Hype is Bad for Science by Victor Galitski, Maryland University, July 2021 and the quite exaggerated view from Nikita Gourianov from Oxford University, as described in Oxford scientist says greedy physicists have overhyped quantum computing by Tristan Greene, TheNextWeb, August 2022 and with a response in Separating quantum hype from quantum reality - Are the sceptics too sceptical? By Simon Benjamin, Financial Times, September 2022. See also Disentangling the Facts From the Hype of Quantum Computing by James Clarke, Intel, IEEE Spectrum, September 2022.

[2886] See Hope and Hype in Quantum Computing by Philip Nikolayev and Susmit Panda, Quantum Poet, June 2022.

[2887] See What if Quantum Computing Is a Bust? by Chris Jay Hoofnagle and Simson Garfinkel, January 2022.

[2888] See Mitigating the quantum hype by Olivier Ezratty, February 2022 (26 pages).

[2889] See AI in the UK: ready, willing and able?, March 2018 (183 pages).

The difficulty to explain how deep learning algorithms work has been exaggerated. If it is true that multilayer neural networks are somewhat abstract. But it is equally abstract for almost any software, with or without AI, that can affect our daily lives. But we've forgotten that a little. When a software from the Visa group rejects your credit card payment abroad, we almost never get an explanation of the whys and hows it was rejected and how to avoid it. Bayesian fraud and machine learning based detection techniques are not explained to consumers.

**Ethical quantum**

Quantum computation is likely to amplify this quest for explicability. It is even less obvious to satisfy with quantum algorithms, which follow a logic that few developers can grasp. Quantum algorithms are likely to be even more complicated and less understandable than those of today's AI. This is amplified since we cannot observe their inner working and intermediate quantum states. Only the "classical" result is measured at the end of the operations. Moreover, from about fifty qubits, it becomes impossible to emulate a quantum algorithm on a classical computer.

Their possible biases will not necessarily come from the data that feeds them because, for a certain period of time, quantum computers will probably not exploit large volumes of data. We can therefore speak literally of the term algorithm bias, whereas when we talk about AI, we are dealing more with training data bias rather than algorithms bias.

But this will be judged on a case-by-case basis. Depending on whether the applications of quantum computing optimize automobile traffic, manage energy distribution, optimize financial portfolios of the wealthy, create new molecules in chemistry or biology or help the NSA break the codes of private communications, the stakes will not be the same[2890].

Ethical questions related to quantum technologies will undoubtedly emerge. They will be associated with a whole range of applications of quantum computing: the simulation of the dynamics of organic molecules. It will probably be limited at the beginning to the simulation of relatively simple molecules. The simulation of complex proteins folding is a hypothesis that has not yet been validated. In a distant hypothetical future, we may be able to simulate larger biological ensembles.

When this is simulated and then altered, for example to create new therapies, the rejection of GMOs or vaccines will seem like distant memory. New fears will show up and scientists will have to get involved to prevent them from spreading. These irrational fears will emerge because of exaggerations about the capabilities of quantum computers. We already hear about "quantum robots", which means nothing, but can impress and drive wild thoughts.

The example below in Figure 887 is eloquent from this point of view with two titles in 2014 when quantum computing was nearly just about D-Wave and which, in practice, are only relaying a rather banal scientific publication, Quantum speedup for active learning agents (15 pages) describing quantum algorithms for the execution of agent networks used in robotics bringing a so-called "quadratic" performance gain, therefore... not exponential, therefore, not extraordinary[2891].

Each time, we will have to decode and take a step back. In 2022 emerged a proposal to quantumly link the brain with a quantum computer[2892].Not only is it really farfetch but it’s probably not a good idea to create trust with quantum computing to elaborate such crazy scenarios.

[2890] This is the point raised by Emma McKay in Should We Build Quantum Computers at All? A Q&A with Emma McKay, quantum physicist turned quantum skeptic by Sophia Chen, APS News, August 2022.

[2891] See Article 1 and Article 2.

[2892] See An approach to interfacing the brain with quantum computers: practical steps and caveats by Eduardo Reck Miranda, Enrique Solano et al, January 2022 (6 pages).

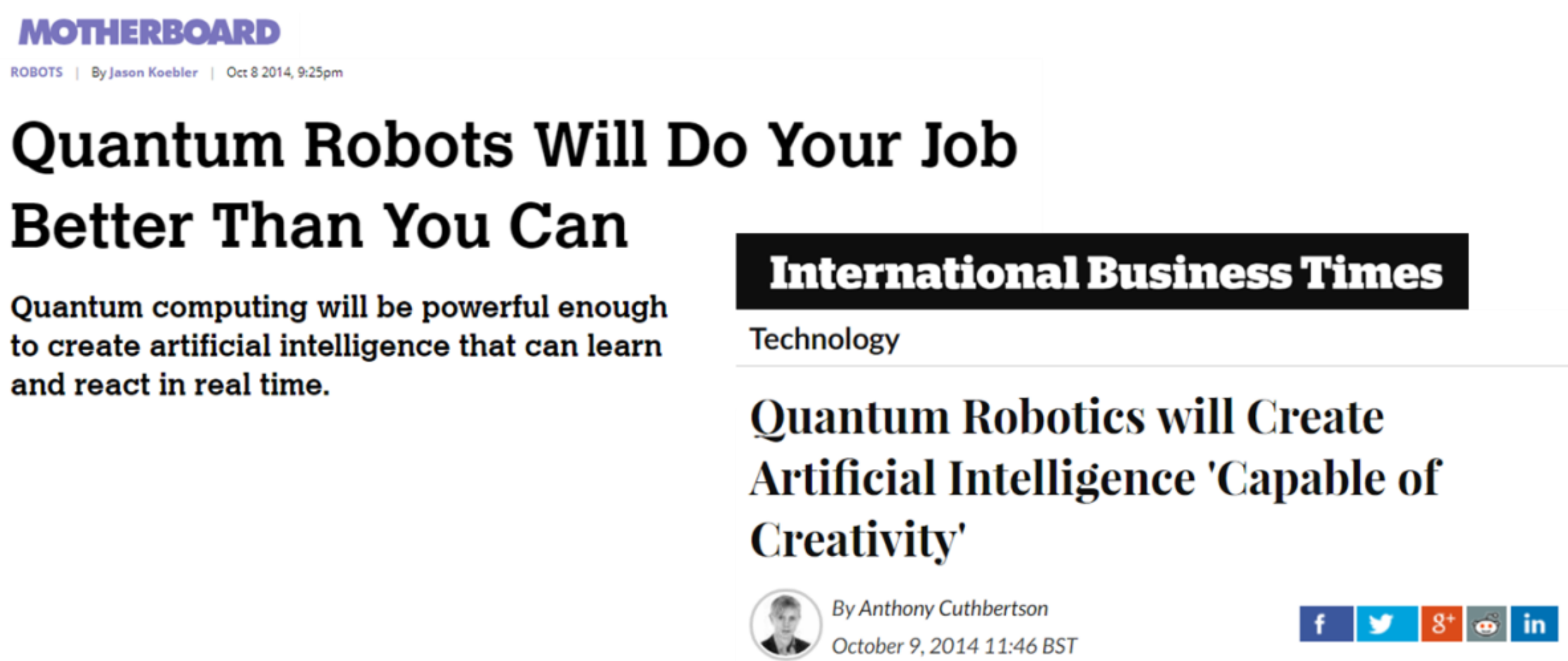
MOTHERBOARD

ROBOTS | By Jason Koebler | Oct 8 2014, 9:25pm

Quantum Robots Will Do Your Job Better Than You Can

Quantum computing will be powerful enough to create artificial intelligence that can learn and react in real time.

International Business Times

Technology

Quantum Robotics will Create Artificial Intelligence 'Capable of Creativity'

By Anthony Cuthbertson

October 9, 2014 11:46 BST

*Figure 887: the fuss about quantum robotics… in 2014! When science fiction is mixed with science, things get confusing.*

We have enough of it with Neuralink, the Elon Musk startup that works on neural implants that would supposedly be connected to some AI to augment the brain capacities, when it is more relevant for treating some neurodiseases like Parkinson's.

A good approach for the quantum scientific community would be to pre-empt these fears by analyzing them as early as possible and defusing them if possible, so as not to be in a situation that would block scientific progress and innovation useful to society because of these irrational fears[2893]. Also, good examples can be given with researchers who admit having published scientific erroneous papers[2894].

Various initiatives started to pop-up in 2021 around quantum ethics in Australia, the UK, the Netherlands, Canada and the USA. It's following a similar pattern than with artificial intelligence but earlier given the maturity of the sector. So far, contrarily to what's happening in the AI field, it has not yet been hijacked by large industry vendors or even regulators. Most initiatives were born out of the research community.

Still, there are already some similarities and overlaps between the AI and quantum ethics frameworks that are showing up.

On the AI side, many AI charters have been published since 2018. One of these comes from the **GPAI** (Global Partnership on AI[2895]) launched by 15 countries in June 2020 including France and Canada. Its goal is to foster the development of responsible and inclusive IAs based on human rights, favoring diversity, while driving innovation and economic growth. The GPAI did set up experts run working groups on responsible AI, data governance, the future of work and at last, on innovation.

OECD launched its **AI Policy Observatory** (OECD.AI) in February 2020, an online platform consolidating information to help states craft their public AI policies. OECD defined its own AI principles (OECD AI Principles) that were adopted by 42 countries in May 2019. Also in 2020, the **AI Rome Call for AI ethics** gathered the Vatican, Microsoft, IBM and others to whitewash about the same goals as GPAI.

---

[2893] See Ethics education in the quantum information science classroom: Exploring attitudes, barriers, and opportunities by Josephine Meyer, Noah Finkelstein and Bethany Wilcox, University of Boulder, Colorado, February 2022 (15 pages) where the authors argue that quantum ethics and social responsibility should be incorporated in quantum information science education from the beginning.

[2894] See one good example with On a gap in the proof of the generalised quantum Stein's lemma and its consequences for the reversibility of quantum resources by Mario Berta, Fernando G. S. L. Brandão et al, May 2022 (22 pages) which shows that an initial proof from one of the authors was incorrect. An author who works at Amazon!

[2895] With Canada, Germany, Australia, South Korea, USA, Italy, India, Japan, Mexico, New-Zealand, UK, Singapore, Slovenia and the European Union.

And these are just a couple initiatives among many others, frequently driven by industry vendors who are lobbying for self-regulation instead of tight government-based regulations.

In the quantum space, **Australia** was the first country that launched a quantum ethics initiative. Early on, in 2019, **CSIRO**, the Australian scientific research agency, mentioned the need to explore and address any unknown ethical, social or environmental risks that may arise with the next generation of quantum technologies[2896]. It was followed in 2021 by a white paper published by Elija Perrier from the Centre for Quantum Software and Information at the Sydney University of Technology[2897].

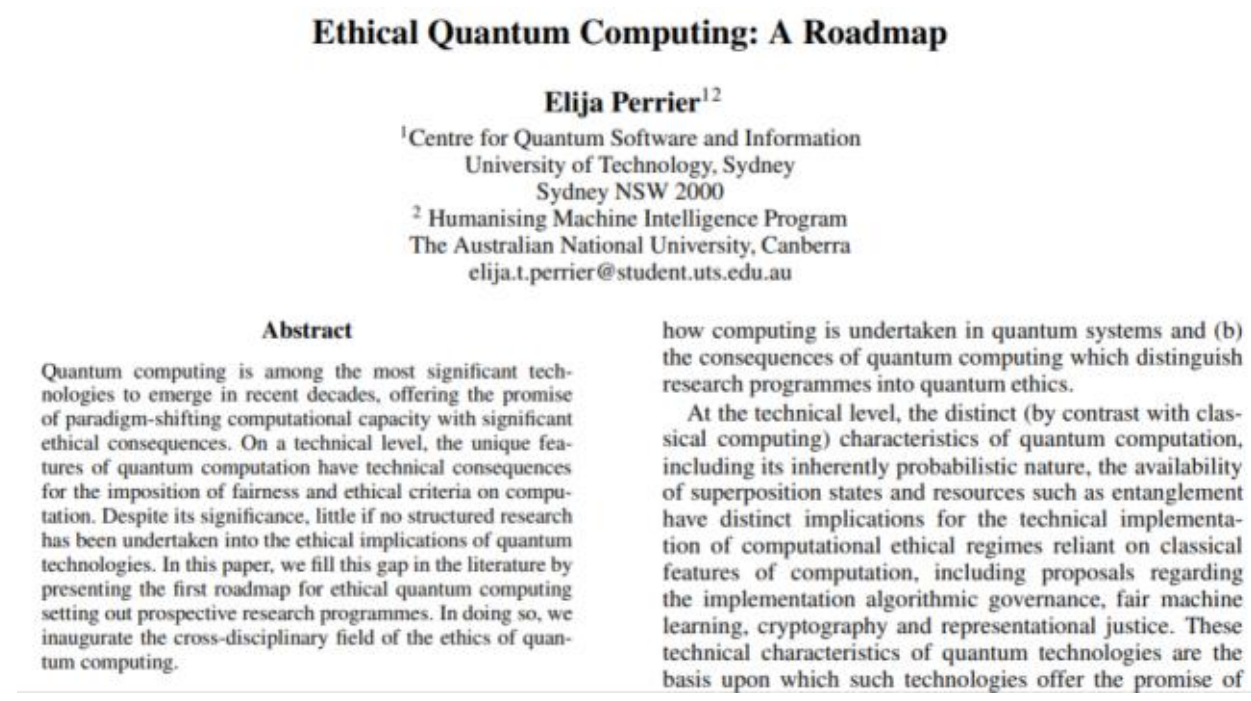
**Ethical Quantum Computing: A Roadmap**

**Elija Perrier**[1,2]

[1]Centre for Quantum Software and Information
University of Technology, Sydney
Sydney NSW 2000
[2] Humanising Machine Intelligence Program
The Australian National University, Canberra
elija.t.perrier@student.uts.edu.au

**Abstract**

Quantum computing is among the most significant technologies to emerge in recent decades, offering the promise of paradigm-shifting computational capacity with significant ethical consequences. On a technical level, the unique features of quantum computation have technical consequences for the imposition of fairness and ethical criteria on computation. Despite its significance, little if no structured research has been undertaken into the ethical implications of quantum technologies. In this paper, we fill this gap in the literature by presenting the first roadmap for ethical quantum computing setting out prospective research programmes. In doing so, we inaugurate the cross-disciplinary field of the ethics of quantum computing.

how computing is undertaken in quantum systems and (b) the consequences of quantum computing which distinguish research programmes into quantum ethics.

At the technical level, the distinct (by contrast with classical computing) characteristics of quantum computation, including its inherently probabilistic nature, the availability of superposition states and resources such as entanglement have distinct implications for the technical implementation of computational ethical regimes reliant on classical features of computation, including proposals regarding the implementation algorithmic governance, fair machine learning, cryptography and representational justice. These technical characteristics of quantum technologies are the basis upon which such technologies offer the promise of

*Figure 888: Source: Ethical Quantum Computing: A Roadmap by Elija Perrier, February 2021-April 2022 (40 pages).*

It was spun out of the Association for the Advancement of Artificial Intelligence (www.aaai.org). The paper starts with defining the quantum physics postulates[2898], then cover ethical quantum computation and asks many ethical related questions that could be asked for any kind of classical computing. They mention the complicated question of quantum algorithms auditing. Quantum algorithms indeed may become black box similarly to deep learning, leading to some explainability issues. So, on top of the various XAI (explainable AI) initiatives like the one launched by DARPA in the USA, will we see emerging the field of XQC for Explainable Quantum Computing?

It also mentions the need for some Quantum Fair Machine Learning (QFML). It may not be such of a problem since QML may not be used to process huge volumes of personal data due to quantum computing limitations in data loading techniques, which may last for a long time. They even go as far as asking whether quantum interferences implemented in quantum algorithms are ethical in nature. They also cover privacy and cryptography matters. Is Shor going to kill our private life? How could some differential privacy be implemented with quantum computing[2899]? Other topics involve distributional ethics and fair distribution which are classical economical questions arising with any new technology. At last, they wonder about the impact of quantum simulations and whether it could be implemented to simulate people's personal behavior.

The paper seems highly influenced by the works on ethical AI and sometimes mixes science-fiction with real state of the art understanding of what can and will be done with quantum computing. But it asks good questions. Another Australian paper focused more recently on the need to involve all stakeholders, beyond the classical market awareness creation[2900].

In the **UK**, ethical quantum computing became a topic promoted by the media The Quantum Insider starting in December 2020. They released a short video documentary trying to explain what quantum computing is and the related ethical issues involved with researchers like John Martinis and

[2896] See Growing Australia's Quantum Technology Industry, CSIRO, 2019 and The 'second quantum revolution' is almost here. We need to make sure it benefits the many, not the few by Tara Robertson, June 2021.

[2897] See Ethical Quantum Computing: A Roadmap by Elija Perrier, Centre for Quantum Software and Information, Sydney University of Technology, February 2021-April 2022 (40 pages).

[2898] They define only the first 4 quantum physics postulates and not the whole 6, and their fourth postulate doesn't correspond to the canonical Born rule related principle.

[2899] See Quantum Differential Privacy: An Information Theory Perspective by Christoph Hirche et al, February 2022 (26 pages).

[2900] See Talking about responsible quantum: Awareness is the absolute minimum... that we need to do by Tara Roberson, December 2021-September 2022 (15 pages).

entrepreneurs like Ilana Wisby[2901]. They are highlighting the need for democratizing quantum technology skills, mention the risks on privacy and security and the need to address quantum AI bias. They also pinpoint the "Hype-Fear-Disappointment Cycle" and recommend to set realistic expectations to avoid triggering fears and biases. Researchers from Oxford University are also studying responsible innovation in quantum technologies[2902].

Other ethical issues to be addressed cover the potential harmful manipulation of the human genome fears, the (positive) quantum use cases to find environmental solutions and the energetic footprint of quantum computing.

In **The Netherlands**, the government 615M€ initiative launched in April 2021 includes a 20M€ plan on quantum ethics and societal impact research run out of the Living Lab Quantum and Society spun out of Quantum Delta NL, the foundation established to run the Netherlands quantum program. They also create ethical, legal and societal standards for quantum technologies and their applications.

The **World Economic Forum** launched its Quantum Computing Governance initiative in February 2021[2903]. It wants to standardize an ethical framework enabling the responsible design and adoption of quantum computing. They ask the ever-lasting question: will the public trust technologies which they cannot understand and whose results they cannot verify as if they could do it with existing digital technologies. They advocate the use of preemptive involvement in technology design to make sure ethical issues are addressed as early as possible. With that, they are assembling a "*global multistakeholder community of experts from across public sector, private sector, academia and civil society to formulate principles and create a broader ethical framework for responsible and purpose-driven design and adoption of quantum computing technologies to drive positive outcomes for society*". They will frame the conversation, drive quantum ethical issues awareness, study quantum related risks, design quantum computing ethics principles and framework and test it with some case studies.

In **Canada, Q4Climate** is an initiative for using quantum technologies in climate research, an initiative coming from the Institut Quantique, the University of Waterloo and Zapata Computing. It looks like a small think tank. It explains how some quantum chemistry algorithms could potentially solve some environmental problems[2904].

In the **USA**, some spare initiatives are launched by academics like Chris Hoofnagle from Berkeley, or a while ago, by Scott Aaronson[2905].

Interestingly, none of these initiatives mention the field of quantum sensing, which could also have some underlying ethical issues to be addressed, particularly when used in the military. Quantum radars, quantum imaging, precision gravity measurement and its impact on underground resources exploitations are a couple examples.

---

[2901] See Quantum Ethics documentary, December 2020 (13 mn) published by TheQuantumInsider as part of a series of "conversations". It was followed by several posts like Quantum Ethics Series: Understanding the Issues and Expanding the Conversation by Matt Swayne, 2021.

[2902] See Asleep at the wheel Responsible Innovation in quantum computing by Philip Inglesant, Carolyn Ten Holter, Marina Jirotka & Robin Williams, October 2021 (14 pages).

[2903] See Quantum Computing Governance Principles, Insight Report, January 2022 (35 pages).

[2904] See Quantum technologies for climate change: Preliminary assessment by Casey Berger, Agustin Di Paolo, Tracey Forrest, Stuart Hadfield, Nicolas Sawaya, Michał Stęchły and Karl Thibault, June 2021 (14 pages).

[2905] See Law & policy for the quantum age : a presentation by Chris Hoofnagle, February 2021 (58 mn), Law and policy in the quantum age, by Chris Jay Hoofnagle and Simson L. Garfinkel, January 2020, free download (602 pages), and Why Philosophers Should Care About Computational Complexity by Scott Aaronson, 2011 (53 pages).

## Religions and mysticism

In recent millennia, the human race has developed the habit of devoting a cult to one or more higher divine powers of an imprecise nature, but explaining everything and everything else.

Mankind probably began to attribute this power to natural phenomena that he could not explain like the Sun or the stars. Mankind then went from multiple systems of gods to a single all-powerful God. In a way, the monotheistic religions realized before time the theory of unification so much sought after by physicists. This story is told with hindsight by **Yuval Harari** in Sapiens and with cynicism by **Richard Dawkins** in "The God Delusion".

For some scientists or believers in an afterlife, quantum physics renews the desire to explain the inner works of the Universe by some divine power. It gives the impression of providing an ultimate scientific explanation for everything, of God, and of his ability to control and supervise everything[2906]. The quantum function most often emphasized in these explanations is entanglement.

It makes it possible to envision a Supreme Being who, thanks to this physical phenomenon, can control all the particles of the Universe and at a distance. It would also explain strange synchronicity phenomena. The wave-particle duality also makes it possible to imagine or explain many magical scenarios such as remote healing, telekinesis or telepathy[2907].

Some of the protagonists of these theories are themselves quantum physics scientists. One of the best-known is **David Bohm** (1917-1992), already mentioned in the quantum foundations section, page 987, who came closer to Indian spiritualism in the 1960s, simultaneously with the Beatles! He was convinced that the laws of the Universe were governed by some spirit[2908]. He is one of the initiators of the theories of **quantum cognition**, a field of cognitive theories based on the mathematical formalism of quantum mechanics, and relying on analogies.

The literature on quantum derived spiritualism is sometimes mind blowing, such as Google's Quantum Computer May Point People to God, from 2013. According to the (anonymous) author, a perfect quantum computer could attempt to simulate the appearance of life on Earth and demonstrate through absurdity that it would not be possible without divine intervention. But who says that the result would not be the opposite? Quantum computing could invalidate classical theories of evolution.

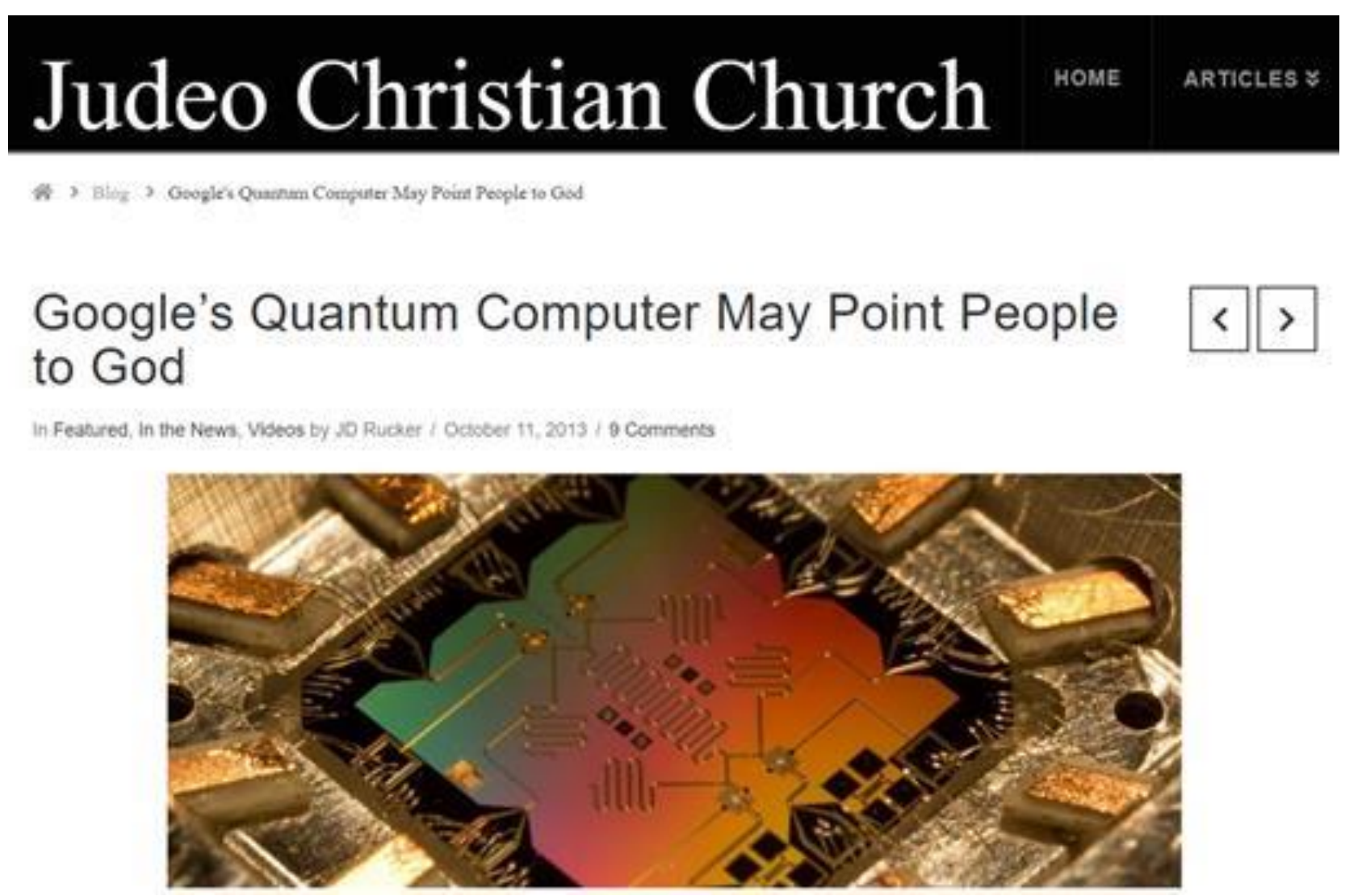

*Figure 889: quantum computer won't point you to God either! Source: Google's Quantum Computer May Point People to God, 2013.*

[2906] On this subject, see the Wikipedia fact sheet that briefly describes quantum mysticism.

[2907] A good inventory of these different debates can be found in The Quantum God An Investigation of the Image of God from Quantum Science, 2015 (81 pages) which evokes the notion of consciousness of the Universe. See also the almost parodic Nothing is solid "All is energy".

[2908] See Lifework of David Bohm - River of Truth by Will Keepin, 2016 (22 pages).

They do not specify the number of zillions of entangled qubits that would be required to support this. Of course, because they have no idea which algorithms to use. And who care about quantum error correction!

All of this is religion-science fiction and can generate heated debates with people who will never be on the same wavelength, some adopting a classical scientific approach and others a mystical and more emotional one.

# Public education

Quantum computing will amplify a situation observed with artificial intelligence: a huge gap between those who understand it and those who use it, coupled with a shortage of skills. It is right now definitely a world of specialists, and it is even harder to grasp than most other digital-related disciplines. Today, this world is balanced between specialists in condensed matter physics and quantum algorithms and software[2909].

By extrapolating a little and drawing inspiration from the history of computer science, we can anticipate that software will gradually take over when quantum computing becomes commonplace, especially if it leads to applications in all sectors of industry.

In today's digital economy, there are many more software specialists than there are with semiconductors. The economies of scale are actually much greater with the latter between producers and users. Quantum will probably not escape this, even if initially the market for quantum computers will not be a volume market.

In the short term, there is a great need to popularize the field and also avoid its technical jargon. You must proceed step by step, broadening the audience in a progressive way from the techie to the non techie[2910]. In parallel, training decision-makers in the industry and institutions must also be done. It is becoming even more important as the quantum technologies hype is peaking with the flurry of vendors and research labs announcements that are regularly showing up[2911].

Many initiatives around the world are launched to train the public on quantum physics and technologies. Let's mention the **National Q-12 Education Partnership** launched as part of the National Quantum Initiative and targets middle-school and high-school students. Industry participants include IBM, Google, Microsoft, AWS, Rigetti, Intel, Lockheed Martin, Boeing, Zapata Computing, APS Physics, Optica, IEEE USA and QubitbyQubit (a training organization). They organize the event QuanTime in spring 2022 with hundreds of quantum activity classrooms for K-12.

Other initiatives are gamifying the learning process like with **Quantum Odyssey** from Quarks Interactive (2020, Romania, 230K€) that were launched in 2020. They replace Dirac's bra-ket notation and linear algebra used by scientists by a visual puzzle-building approach. It's proposed to discover and learn gate-based quantum algorithms. **Quantum Chess** from Quantum Realm Games (2019, USA). The game embeds quantum phenomenon in the game play with pawns able to make multiple moves simultaneously in sort of superposition.

---

[2909] See Eleven risks of marrying a quantum information scientist by Nicole Yunger Halpern, 2020. A second degree but realistic inventory of the life of a quantum scientist in the USA.

[2910] See for example The Quantum Prisoner, a free scientific and technological video game is now available online, CEA, October 2020.

[2911] In Democratization of Quantum Technologies by Zeki C. Seskir, Steven Umbrello, Christopher Coenen and Pieter E. Vermaas, August 2022 (22 pages), the authors define the various aspects of a democracy and include educational efforts, like IBM's continuous evangelization efforts that started in 2016. They also pinpoint the asymmetry between the quantum ecosystem stakeholders and the general public which "has to be educated" but do not participate to a democratic decision making process. It also describes the counter-narrative on the democratization of quantum technologies like cybersecurity threats, its geopolitical dimension, and the habit to position quantum physics as impossible to understand, abusively quoting Richard Feynman. The authors do not describe the process followed by governments when launching their quantum plans, which embedded some participatory process with quantum ecosystem stakeholders but had their share of discretionary decision making, sometimes even escaping the minds of top policy makers.

Likewise, the very serious CEA in France launched **The Quantum Prisoner** in English in 2020, a free online adventure game inspired by quantum logic and targeting kids over 12 (meaning, adults are welcomed…). It has 10-12 hours of gameplay with a journey across the globe to find out what happened to a physician who mysteriously disappeared in the 1960s. Playing as Zoe, a young woman, gamers must solve over 30 technology, science and engineering-based puzzles.

Many other quantum games were created for educational purpose, like **Alice Challenge**, **Hello Quantum** and **Hello Qiskit** by IBM and the University of Basel, **Particle in a Box**, **Psi and Delta**, **QPlayLearn** (covered later), **Quantum Cats**, **Quantum Flytrap**, the **Virtual Lab** by Quantum Flytrap, **ScienceAtHome**[2912].

Some educational tools are specialized in quantum optics such **Quantum Games with Photons** from the MIT which is an open source puzzle game with 34 levels and a sandbox, and **The Virtual Quantum Optics Laboratory**, an optical lab running in a browser which enables you to build quantum optics experiments with all sorts of optical devices (lasers, PBS, depolarizer, etc, in Figure 890 *on the right*).

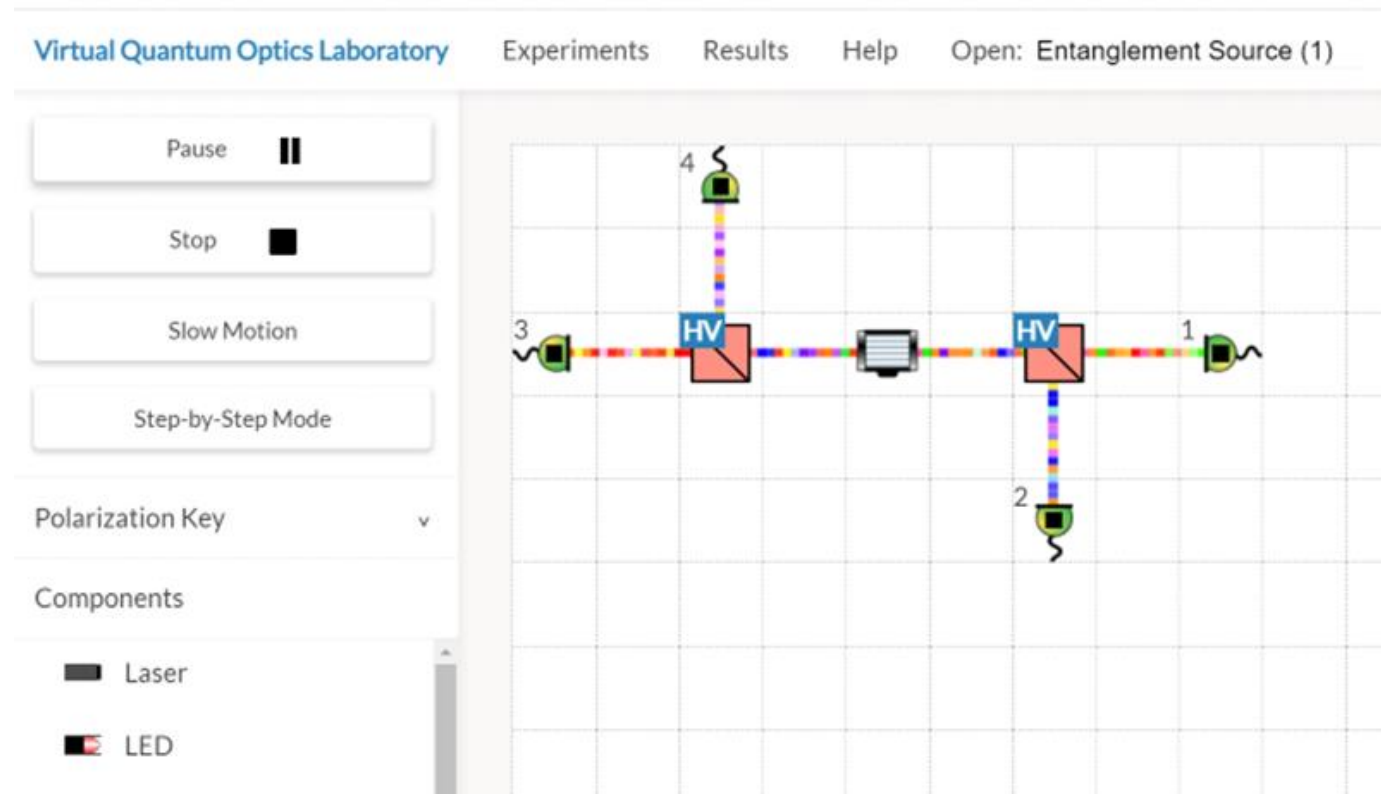

Figure 890: The Virtual Quantum Optics Laboratory.

**CoSpaces** is a similar simulation tool created in Italy and aimed at teaching quantum computing[2913].

## Professional education

All these countries launching well-funded quantum plans create a significant challenge with professional education along the whole cycle from bachelor to doctorate. Before being an industry competition, quantum technologies are a talents one. We can expect that there will be more money to spend than talent to hire with it for a while.

Existing quantum professional training is significant in quantum physics. Most universities in the world have special programs from bachelors/licences and masters to PhDs. Some groups are organizing summer schools for PhD level students like the famous **Ecole de Physique des Houches** doctoral and summer school in the French alps with various sessions on quantum physics. Look for example at the 2019 summer school agenda's speakers!

There are a couple new disciplines where more and more people will need to build knowledge and skills on. Quantum systems engineering create real machines that work from start to finish. This requires decompartmentalizing disciplines and bringing together physicists and engineers. The technologies involved are varied and include photonics and lasers, analog and digital electronics, thermodynamics, fluid mechanics, various components manufacturing techniques, and the design of complete systems. Quantum engineering involves many complementary disciplines[2914]. With AI, it is a new challenge for higher education that is being prepared.

---

[2912] See an inventory of quantum games in Quantum Games and Interactive Tools for Quantum Technologies Outreach and Education by Zeki C. Seskir get al, July 2022 (48 pages).

[2913] See Quantum computing teaching with CoSpaces by Francesco Sisini, Igor Ciminelli and Fabio Antonio Bovino, September 2022 (8 pages).

[2914] The schematic comes from the Introduction to Quantum Computing presentation by William Oliver from MIT at Q2B in December 2019.

In the purely mathematical and software fields, very important disciplines come into play for creating end-to-end quantum solutions: algorithms design, software tools design and applications software development. Added to this is the field of post-quantum cryptography.

The creation of business applications also requires skills at the crossroads between the above and vertical markets, which are often themselves scientific as in life sciences (organic chemistry, protein folding, photosynthesis, ...), materials sciences (battery chemistry, superconducting materials) or other branches such as portfolio management and risk assessment in finance or optimization problems in logistics, transportation and marketing.

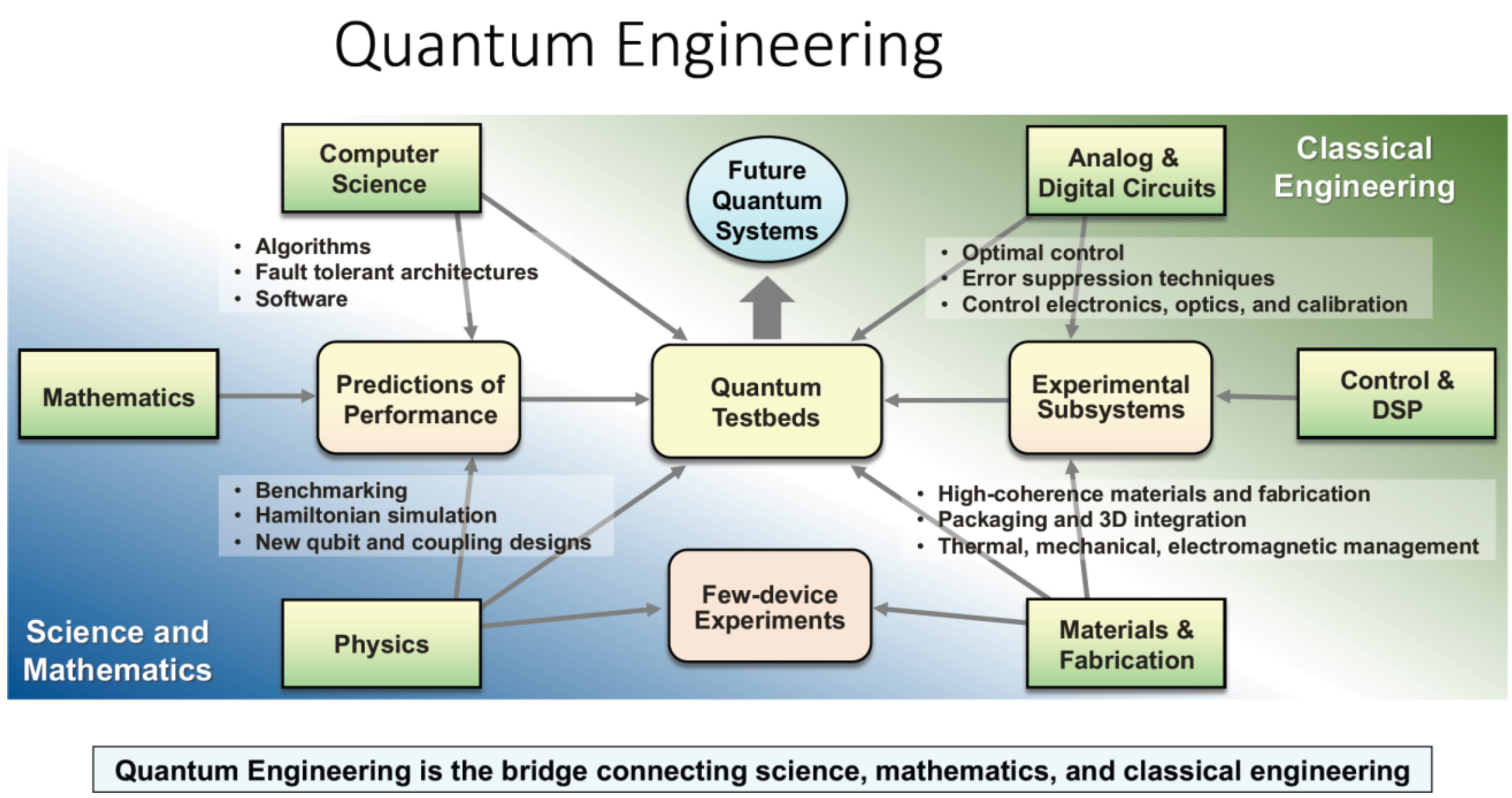

*Figure 891: quantum engineering defined. Source: Introduction to Quantum Computing by William D. Oliver, MIT, December 2019.*

Quantum technologies will be found with many different professions:

- **Fundamental physicists** (solid-state physics, condensed matter physics, light-matter interaction, quantum optics) who combine theoretical and experimental approaches to understand low-level phenomena.
- **Quantum technologies researchers** who turn fundamental discoveries into first proofs of concept in the laboratory. These research teams combine physics, technology and engineering researchers.
- **Design engineers** who create technical subsets of quantum computers to complete finished products. They essentially do the "D" of "R&D" by relying on the R of physicists.
- **Research engineers**, who participate in the development of new materials and new technologies in semiconductor fabs, or process engineers who design the manufacturing processes for these integrated circuit systems supporting qubits.
- **Technicians** for certain components manufacturing and/or for the deployment of technologies such as quantum cryptography in the telecom space. But only once this technology is deployed on an industrial scale, probably by generalist or specialized telecom operators.
- **Software tools developers** who must be associated with previous researchers and engineers. Indeed, for the time being, the design of these tools still has to take into account the physical characteristics of quantum calculators/accelerators.

- **Application developers**, whose numbers will increase as the computing power of quantum computers grows. Most of the time, they will need to have three key sets of technical skills: one is being able to program quantum computers, the second will be the ability to turn business problems into quantum programs (including the related mathematical/physics related know-how) and at last, they will have to know about classical programming since most quantum algorithms are hybrid.
- **Project managers** who manage projects and teams that combine these different professions.
- **Business strategists**. Brian Lenahan goes as far as defining the job of "quantum business strategist" which looks like an equivalent of the chief digital officer for quantum technologies related projects, creating the link between IT and business managers. This role is about crafting a quantum plan with mission, vision, goals, strategies, KPI's and tactics. In other words, it is an old-fashioned consultant[2915]!
- **Support activities** like in HR, marketing, business development (managing industry partners) and sales (to end-customers), communication, public relations, legal, finance, startup creation and acceleration.

As in many disciplines, researchers and engineers are increasingly required to be versatile. Teams must be structured around a strong interdisciplinarity and transversality. They need "technological polyglot" teams that link all these professions and skills. In particular, physicists will have to be increasingly interested in engineering and engineers in physics[2916].

Finally, when you turn to the business side with actual products that can be marketed and sold, you need the whole mix of skills usually found in technology marketing and sales: product marketing, operational marketing, business development and partnerships, creating ecosystems and, above all, pure and simple B2B sales for a starter. This is completed by the generic skills associated with deep techs startups creation (organization, business planning, recruitment, funding, etc.) and with intellectual property attorneys who must grasp the specificities of the quantum vocabulary.

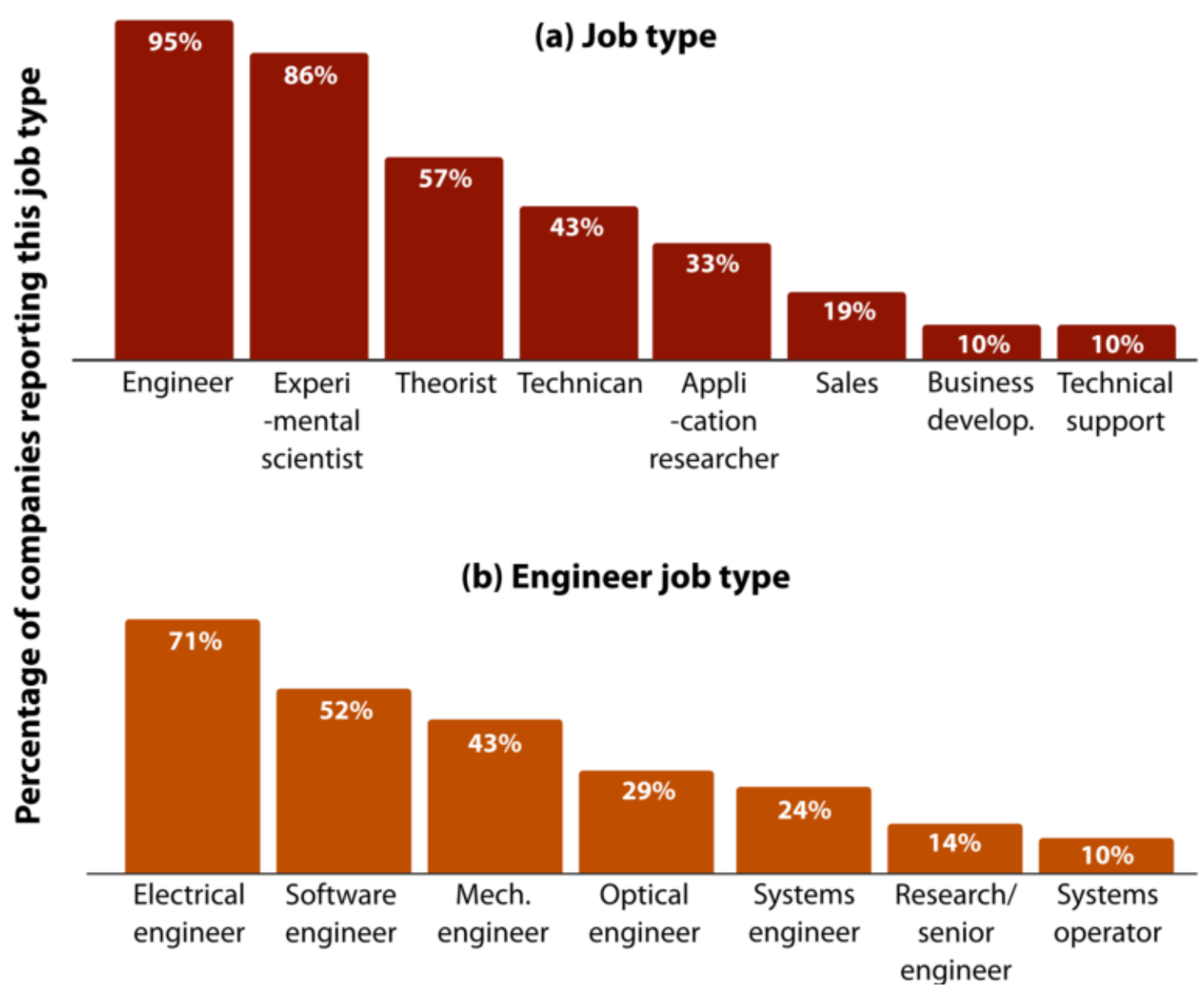

*Figure 892: an American inventory of engineering jobs and skills in quantum technologies. Source: Preparing for the quantum revolution -- what is the role of higher education? by Michael F. J. Fox, Benjamin M. Zwickl et H. J. Lewandowski, 2020 (23 pages).*

Quantum sensing products are beginning to be marketed, and in a market that is currently niche.

---

[2915] See What is a Quantum Business Strategist? by Brian Lenahan, April 2021 and his related book Quantum Boost: Using Quantum Computing to Supercharge Your Business by Brian Lenahan, May 2021. Brian Lenahan also created in September 2021 the Quantum Strategy Institute with various people from Spain, UK, France and the USA with the goal to bridge the gap between quantum science and businesses.

[2916] See Defining the quantum workforce landscape: a review of global quantum education initiatives by Maninder Kaur and Araceli Venegas-Gomez, Qureca, February 2022 (35 pages) which makes an inventory of the various quantum educational resources across the world and Building a Quantum Engineering Undergraduate Program by Abraham Asfaw, Alexandre Blais et al, 2021 (25 pages). See also the QTEdu, the European Quantum Flagship program on education which launched 11 pilot programs on education in quantum technologies and the associated report The Future Quantum Workforce: Competences, Requirements and Forecasts by Franziska Greinert et al, August 2022 (16 pages) which is based on a survey on quantum skills needs in Europe.

Quantum cryptography systems are in the experimental field phase and could be deployed on a larger scale in the coming decade.

Quantum communications with the objective of leading to quantum communications networks will develop in a second phase, combining fiber and satellite networks with quantum ground relays. This is a complementary field to the development of quantum computers.

Finally, quantum computing and simulation will progressively evolve and see their field of application widen as the qubits number and quality in quantum computers grows. It will be a process of continuous innovation.

As in the case of classical computing, the weight of software is bound to become dominant in skills requirements. This explains why many publications insist on the need for quantum application developers. This is what the major players such as IBM, Google and Microsoft, not to mention D-Wave, Rigetti and IonQ, are "evangelizing" about[2917].

Nevertheless, in parallel with the software market development, an intermediate phase will require a lot of skills in engineering and in the different branches of quantum technologies. In some cases, training can be shared between universities, particularly when teachers are scarce. That's what is implemented in the Université Paris-Saclay with the ARTEQ interdisciplinary year positioned before masters M1 and M2, to feed M2 masters in quantum physics and quantum information science.

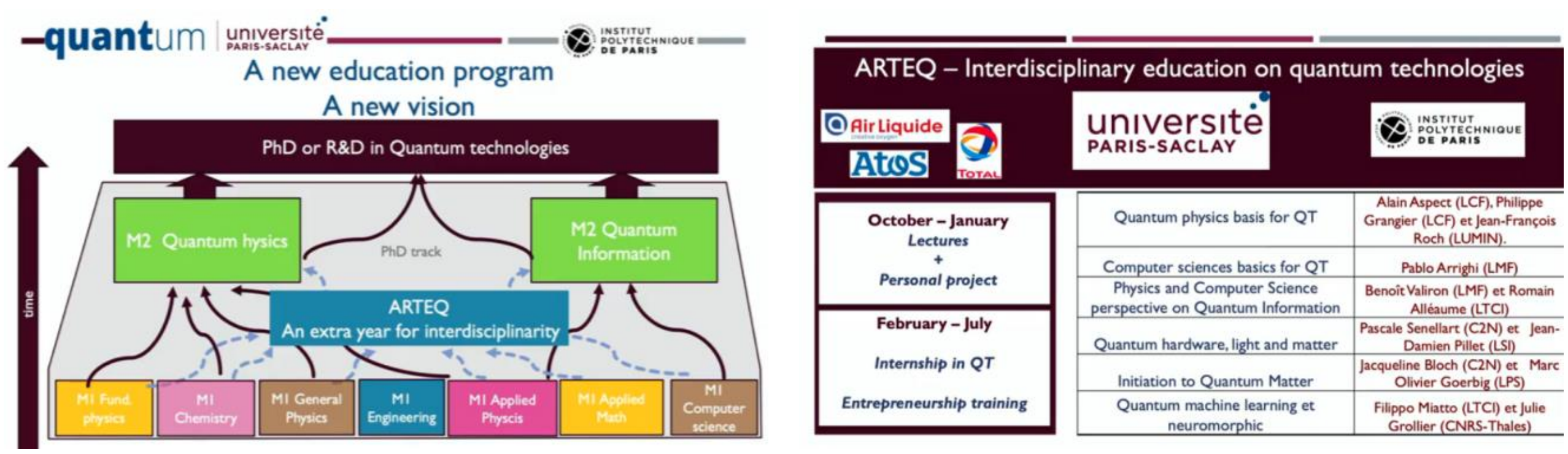

*Figure 893: ARTEQ training in Saclay.*

Training in public higher education should introduce quantum science and technology as early as possible in the bachelor's and master's degree programs. It will also be necessary to create master's degrees in quantum engineering, bringing the world of research and engineering closer together.

The training offer will depend on several parameters: funding for teacher-researchers or teaching positions, the creation of vocations, the ability to attract teachers and students from wherever they come.

Continuing education may include both scientific and technological courses (quantum physics, quantum communications, quantum algorithms and software) and strategic courses (understanding of the issues, knowledge of the players, economics of the sector, good practices). This is probably the less well address market need, so far. It can or could be delivered by private organizations, by higher education organizations as well as via online courses offered by Coursera and the likes.

Self-training allows enthusiasts to discover these sciences and technologies by themselves, but it is not self-sufficient as it is sometimes the case in artificial intelligence.

---

[2917] Look at it this way: Quantum Computing Demands a Whole New Kind of Programmer by Edd Gent, May 2017 (slightly ahead of schedule), The Hitchhiking Cat's Guide to Getting a Job in Quantum Computing by Jay Gambetta, October 2019, Building Quantum Skills With Tools For Developers, Researchers and Educators, IBM Research, September 2019 and Some useful skills for quantum computing by Chris Granada, January 2020, which also emphasizes mathematical and software skills.

It must be complemented by quality pedagogical support, if only to do and correct exercises. As far as the software part is concerned, this will perhaps change the day when development tools will be possible with higher levels of abstraction than today.

Scientific events organized by quantum hubs, research laboratories and companies serve to facilitate transdisciplinarity among researchers and engineers. They can be interdisciplinary symposia, thematic conferences or workshops.

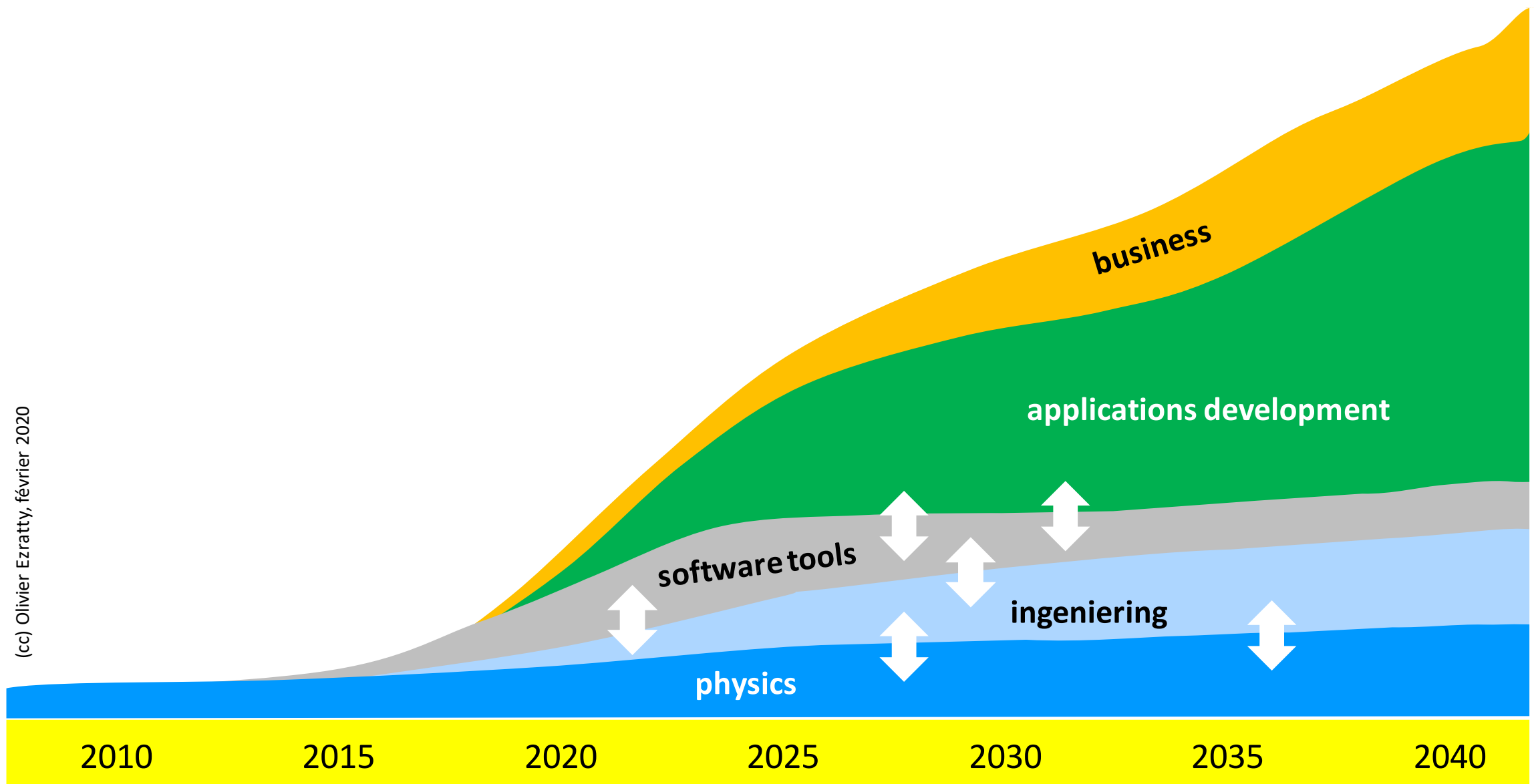

*Figure 894: how quantum tech skills need will evolve over time. More engineering and then more software and more business skills. (cc) Olivier Ezratty, 2020.*

It will also be necessary to attract as many women as men in these courses, otherwise there is a risk that a whole sector will develop, as in AI, which is far too masculine. Not to mention the increase in the diversity of students' social backgrounds, which remains a key means of republican promotion, despite its current decline.

Upstream of all these courses, the creation of vocations among young people is indispensable. Science fairs can also contribute to this. It is a long-term task, as is the creation of vocations in science in general and in the scientific and technical professions of the digital world in particular.

There are some pure players around in the quantum computing educational market, many of them offering open sourced eLearning contents:

**Q-munity** (2019) is a training organization and community connecting young individuals in quantum computing. With its 1000 members, it organizes summer camps (well, outside pandemics), conferences and workshops.

It was created by Anisha Musti, a quantum computer scientist who worked on Shor's algorithm, quantum teleportation and quantum machine learning.

**QubitbyQubit** (USA) is a quantum programming online learning initiative from The Coding School, created by a Brown University undergraduate in 2014. It was created by Kiera Peltz and is sponsored by IBM and Google.

**Qplaylearn** (2020, Finland) develops an online visual quantum programming training tool targeting a broad audience including high school students. They collaborate with various universities in Finland as well as with IBM.

**Quantum Country** (USA) is a tutorial web site on quantum programming created by Andy Matuschak and Michael Nielsen. It contains "mnemonic medium" that makes it easy to remember what you read. These are long reads including some good story telling and some exercising. It starts with the basics of quantum programming, then covers key algorithms like Grover search.

**QuTech Academy** is offering free online courseware on quantum technologies for engineers[2918].

**CERN** has a series of introductory conferences on quantum computing from **CERN** (7 x two hours tracks), broadcasted in November and December 2020, also targeting engineers[2919].

**Qureca** (2019, UK) sells online "Quantum for Everyone" courses for business people (at £400). These courses are delivered by Araceli Venegas-Gomez, the founder of Qureca, Bruno Fedrici, a French consultant and lecturer on quantum technologies and QuantFi, a French Startup specialized in Finance applications. The company also offers hiring services and other professional services for startups and businesses (community management, events organization, business development and strategy).

**EFEQT** (Empowering the Future Experts in Quantum Science and Technology for Europe) provides a learning experience between academic research and free interdisciplinary exploration for 25 students and young researchers. It takes the form a scientific hackathon, the first one was organized in October 2021 and ended with a graduation in September 2022. The best participants will receive a fast-track access to do a PhD or post-doc at EFEQT partner universities in Germany and Strasbourg in France. The program is supported by the Quantum Flagship's Quantum Technology Education Coordination and Support Action (QTEdu CSA).

## Jobs impact

Finally, what about the future of quantum-related employment, a question that Sophia Chen asked herself in Wired in June 2018? It's difficult to assess because we're thinking over several decades and about use cases that are still uncertain. There will be, as with AI, those who know and those who don't, those who code and those who use stuff, those who create wealth and those whose jobs are threatened.

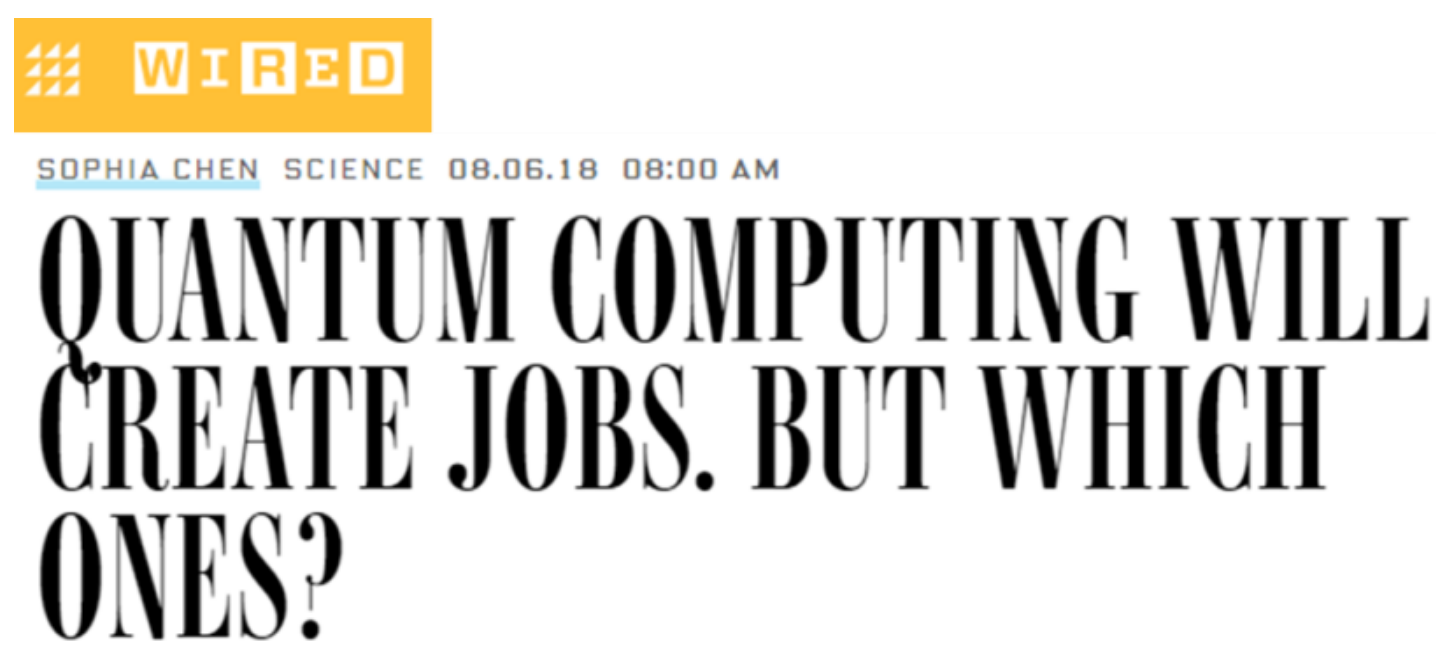

*Figure 895: who knows? There are so many uncertainties on the speed of how quantum tech will mature.*

For the moment, quantum computing does not generate any specific jobs threats, because it will enable us to do things that mankind can't do today. There is no logic of replacement, at most optimization as for applications based on graph optimization like those of the traveling salesman.

[2918] See QuTech Academy Online Learning.

[2919] See Online introductory lectures on quantum computing from 6 November, 2020.

# Gender balance

Gender unbalanced in all STEM jobs and particularly in computer science is a known fact and it has been so for a long time. You can look at all the statistics and they are not good. It started to go awry in computer science in the early 1980s when computing became mainstream. Many initiatives have been launched worldwide to rebalance gender in all these domains. They have mostly failed, or maybe did they just made things better than if nothing was done. Are quantum technologies different for gender balance?

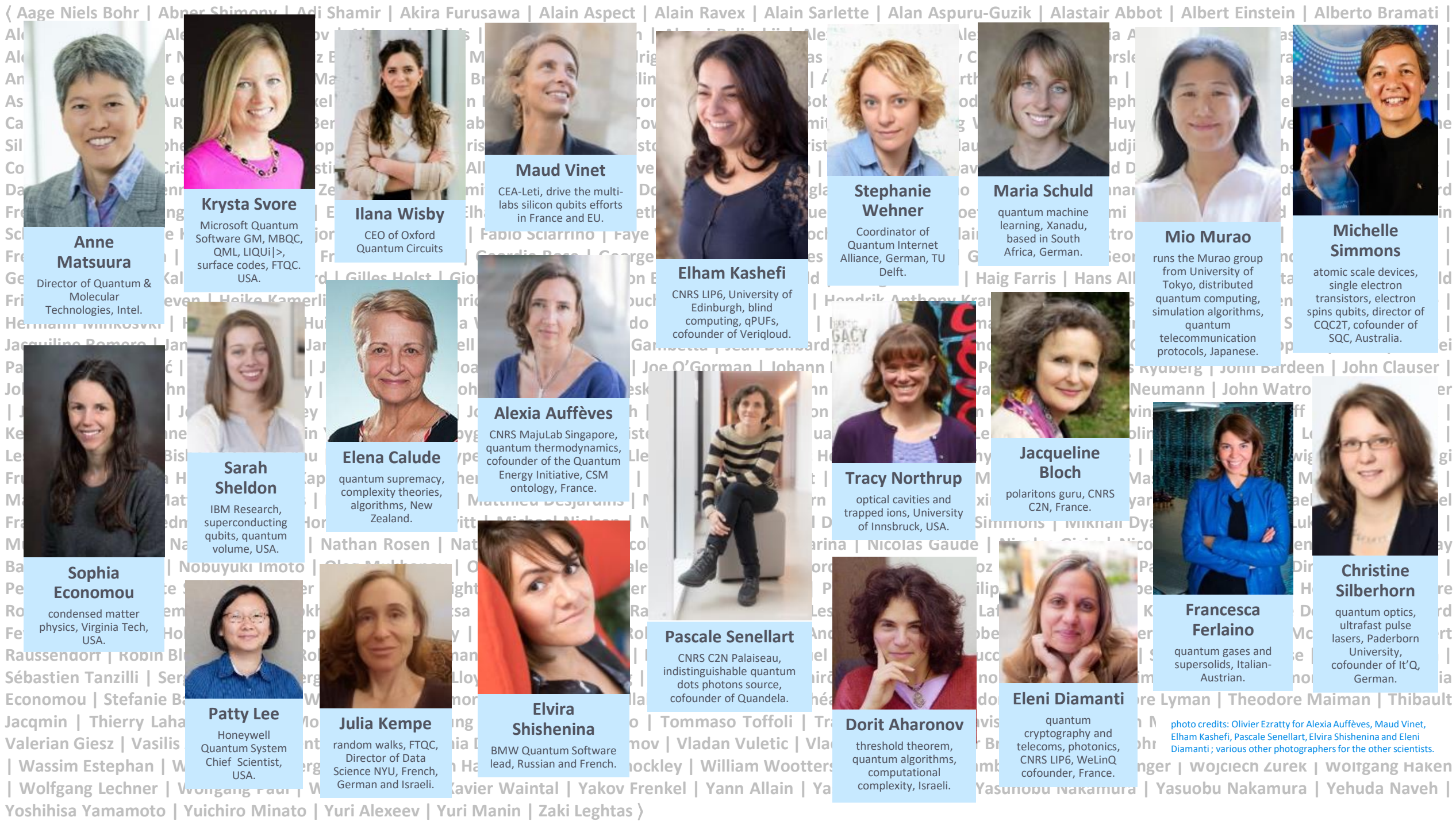

*Figure 896: some women role models around the world, from research to the industry. (cc) Olivier Ezratty, 2021-2022.*

### Problems

This domain is already highly male-dominated, in the lineage of computer science and artificial intelligence. The specialty is still too masculine as it stands. Quantum physics founder in History books are mostly men, particularly in the seven first quantum wave theoreticians narrow club with **Planck-Einstein-De Broglie-Schrödinger-Heisenberg-Dirac-Born**. You have to really dig into the History of science to recognize the role of **Emmy Noether** and **Chien-Shiung Wu**, the few female scientists of this era. Also, only three Nobel prizes were awarded to women with **Marie Curie** (1901), **Maria Goeppert** (1963, for her work on nuclear physics) and **Donna Strickland** (2018 for her work in pulse lasers). But besides Marie Curie, they don't yet have the recognition status of **Linda Lovelace**, **Grace Hopper** and **Margaret Hamilton** in computer science.

The statistics are depressing with only 20% women in STEM (in the USA) and it doesn't seem to be better in quantum science[2920]. Women's representations in culture, media and toys still play a leading role in crafting this unbalanced world. And this is just about gender balance. In the USA quantum scientific community, many groups are fighting against discriminations beyond gender balance issues.

[2920] See The Quantum Computer Revolution Must Include Women by Chandralekha Singh, Scientific American, January 2021 and The Upcoming Women In Quantum Summit III And Its Secret 70 Year-Old Legacy par Paul Smith-Goodson, December 2020. See also Women in Quantum Technologies - What are the challenges, February 2020.

Only a few countries are faring better, like in Asia. Is the condition of women in universities and research labs different than in business organizations? It probably depends on their values, leadership and culture. The scientific world seems as competitive and tough than the private sector even if its rules are different, based on h-indexes, conference talks and the likes. Still, in most places, research is a longer term activity which may create better conditions for women.

On top of that, the language used in quantum science is very masculine[2921]. It evokes the notions of superiority (supremacy) and auxiliaries (ancilla), the former echoing a higher authority, and the current "white supremacy" resulting from South African Apartheid and which is still stirring the US political scene. The second notion takes up the notion of "female servant" in Latin, slavery and racial segregation, whereas the technical term was coined in 1995. These are symbolic items but they deserve to be corrected. One solution is to talk about quantum advantage even if the meaning is slightly different from quantum supremacy (doing impossible things in classical computing vs doing things better). Some are advocating the usage of “useful advantage” in reference to use cases that provides some value with data inputs and outputs, but it doesn’t embed the distinction between supremacy and advantage. Another solution already mentioned consists in using the term **primacy**[2922].

A mix of other authors ask for creating a language void of competition connotations that should be comprehensible, specific and practical, open, accessible, responsible, culturally embedded and meaningful[2923].

### Hope

There’s still hope. It seems easier to identify dozens of women who are real inspiring role models and play a key role in quantum science and technologies and anywhere in the world. Many of these were at the origin of key scientific advancements in quantum technologies. You may know the famous threshold theorem co-demonstrated by **Dorit Aharonov**! There are a few startups created by women like **Silicon Quantum Computing** (SQC, Australia), created by **Michelle Simmons, Oxford Quantum Circuits that is led by Ilana Wisby, Quandela, co-founded by Pascale Senellart,** VeriQloud co-founded by **Elham Kashefi** and Qureca, created by **Araceli Venegas-Gomez**. In the Corporate world, **Krysta Svore, Patty Lee** and **Anne Matsuura** play leading roles at respectively Microsoft, Quantinuum and Intel. In Europe, **Laure Le Bars** leads SAP’s quantum research efforts on top of being the first President of the QuIC industry consortium.

The Quantum Insider launched in 2021 a series of interviews of key women working in the quantum industry. It included for example **Mercedes Gimeno-Segovia**, Systems Architecture VP for PsiQuantum, **Olivia Lanes** North-America lead for Qiskit at IBM, **Helena Liebelt** from Deggendorf Institute of Technology, Intel and SheQuantum, **Christine Johnson**, CEO of Ingenii, **Lior Gazit** quantum software engineering team lead at Classiq, **Ying Lia Li**, CEO from Zero Point Motion (UK), **Katerine Londergan**, CMO at Zapata Computing Computing, **Anindita Banerjee** Security VP at QNU Labs, and **Rojalin Mishra** senior hardware verification engineer at Riverlane.

Also, quantum tech is still a green field and it’s not too late to attract young women in this emerging and promising discipline. There are already women playing leading technical and business roles in quantum startups on top of the cofounders mentioned above[2924].

---

[2921] As Karoline Wiesner of the University of Bristol points out very well in her succinct The careless use of language in quantum information, 2017 (2 pages).

[2922] See Quantum Computing 2022 by James D. Whitfield et al, January 2022 (13 pages).

[2923] See Quantum Technologies and Society: Towards a Different Spin by Christopher Coenen, Alexei Grinbaum, Armin Grunwald, Colin Milburn and Pieter Vermaas, November 2021 (8 pages).

[2924] See 52 Wonder Women Working In Industry As Quantum Scientists & Engineers by James Dargan, The Quantum Daily, August 2021.

### Initiatives

Some initiatives have been launched around the world to promote and help women in quantum technologies. They are matching what has been done for a while in the computer science and information technology fields. Gender oriented actions are a mix of associations, events and media visibility initiatives. Too many of these are seasonal, and centered around the Woman's Rights day, on March 8th, each and every year.

Let's mention a few of these:

- **Women in Quantum** by **OneQuantum** is a think tank gathering quantum leaders worldwide in dedicated chapters, with the goal to influence government action, vendor relationships and the quantum ecosystem. It offers a resources, services and events platform for quantum startups to collaborate. Women in Quantum is one of the "chapters" of this organization, run by Denise Ruffner (also, Chief Business Officer of Atom Computing), organizing quarterly Women in Quantum events, the last one being held online in June 2021[2925].
- **Women in Quantum Development** is a professional network of quantum tech enthusiasts in the Netherlands, with events and mentoring programs. It belongs to a new trend, with national quantum plans containing specific initiatives around the ethics and social impact of quantum technologies. Netherlands is a good best practice for that respect. There's also a gender equality workgroup in the **EU Quantum Flagship**.
- The **University of Bristol** organized a two-day Women in a Quantum Engineering event in December 2019.
- Some research labs and organizations showcasing their women quantum scientists and engineers like at the **Lawrence Berkeley National Lab** from the DoE in the USA[2926], at the **Harvard** Center for Integrated Quantum Materials[2927], at **Yale University**[2928], with **IBM**[2929] and **Microsoft**[2930].
- **SheQuantum** (2020, India) is an eLearning provider offering quantum computing education content targeting women.
- In France, the association **Quelques Femmes du Numérique!** promotes women in tech, particularly engineers and scientists using quality photography portraits with over 750 women in various fields (artificial intelligence, Blockchain, cybersecurity, IT, etc) and over a dozen in quantum techs[2931]. It launched many initiatives including promoting quantum science to female teenagers.

### Solutions

Like in any domain, particularly in social science, there's not yet a common agreement on what should be done to create a better gender balance in STEMs and in quantum science.

---

2925 See the casting of the Fall 2020 edition.

2926 See Women of Quantum Computing Go Tiny in Big Ways by Elizabeth Ball, June 2021.

2927 See Ask a Scientist: Women in Quantum Science and Technology, November 2020.

2928 See WIQI (Women in Quantum Information) Group.

2929 See Encouraging more women in quantum: four insights from four women, IBM UK, March 2021.

2930 See Women of Microsoft Quantum Part 1 and Part 2, March 2020.

2931 Disclaimer: I'm the cofounder and photographer of this association. Whenever I can in media and events speaking opportunities, I propose to create a duo with one the quantum scientist women I know well.

Should we encourage some affirmative actions or not? Some are worth the effort like the European Union ERC Grant program which extends since 2010 the age limit by 18 months per child plus other anti-bias measures. Paternity leaves are also taken into account[2932].

In the way women scientists and entrepreneurs are promoted, I believe we should be more engaged but with subtlety. For example, it's more efficient to value scientists and entrepreneurs for their achievement and who happen to be women instead of doing this explicitly because they are women. An implicit communication is sometimes more efficient than an explicit one. Finding women talents should be a sort of backstage work. It requires some discipline. When organizing training and events, and with any media speaking opportunity, make sure gender balance is respected. It involves having some knowledge of the field ecosystem and of its female leaders. Don't say "there are only a few of them", but "where are they?" and look for them. Also, let them talk about their science.

We should also promote a broad range of role models in different fields and jobs to inspire young talents.

It's also about building inclusive and welcoming work environments in universities, research labs and commercial vendors.

Of course, in a broader scale, media and fiction play a key role. The geek in TV series and movies is too frequently an introverted male. We need more Felicity Smoak, the geek from the TV Series Arrow! At last, do that all year long and not just on March 8$^{th}$.

## Quantum technologies marketing

The last point to be mentioned here is the role of marketing and propaganda. Quantum technologies are the perfect spot to broadcast extraordinary and impressive claims that few specialists can fact-check. It's a world of superlatives and exaggerations. It started in 2019 with Google's supremacy claim.

We are going to be drowned in innovation propaganda that will blur things. Scientists in the field will no longer recognize their creations. Popular news related to quantum computing will continue to start explaining qubits with their superposed states 0 and 1 and... stop there!

Consulting firm will also strive in simplifications. This **BCG** set promoting quantum computing in the pharmaceutical industries is quite amazing although, hopefully a bit dated (see Figure 897). It mentioned the ability of a quantum computer to solve an "*infinite number of problems simultaneously*", confusing, infinity and exponentiality, and then also, superposition and problems. They did estimate the quantum computing market in the pharmaceutical industries in the USA to sit between $15B and $30B with no precise date. A market forecast from 2018 expected that global IT spending dedicated to drug discovery would have reached $5.3B by 2020[2933]! That is not really consistent.

Marketing and communication are all about making fancy claims and simplifying facts with wild exaggerations. One can wonder, how is the bullshit created in marketing when there's so much science behind most projects? It starts with the businessification of quantum technologies. The rules of the game for a staring looking for some VC funding is to talk about customer use cases and market size, creating an echo chamber to the crazy numbers published by industry analyst firms. You will therefore have plenty of quantum computing hardware and software companies web site presenting the same story about the beauties of quantum computing in pharma, financial services, transportation and the industry, if not to fix climate change, but nearly nothing on their actual technologies and products.

---

[2932] See ERC Gender Actions, 2021 (14 slides). It provides some data on the share of women applicants vs men who get ERCs and H2020 grants based on the discipline. Across the board, women have about 20% less chance to get a funding.

[2933] Source: Growth Insights Report: Global Pharmaceutical Drug Discovery IT Solutions Market 2017-2020 - Key Initiatives by Big Pharmaceutical Companies, January 2018.

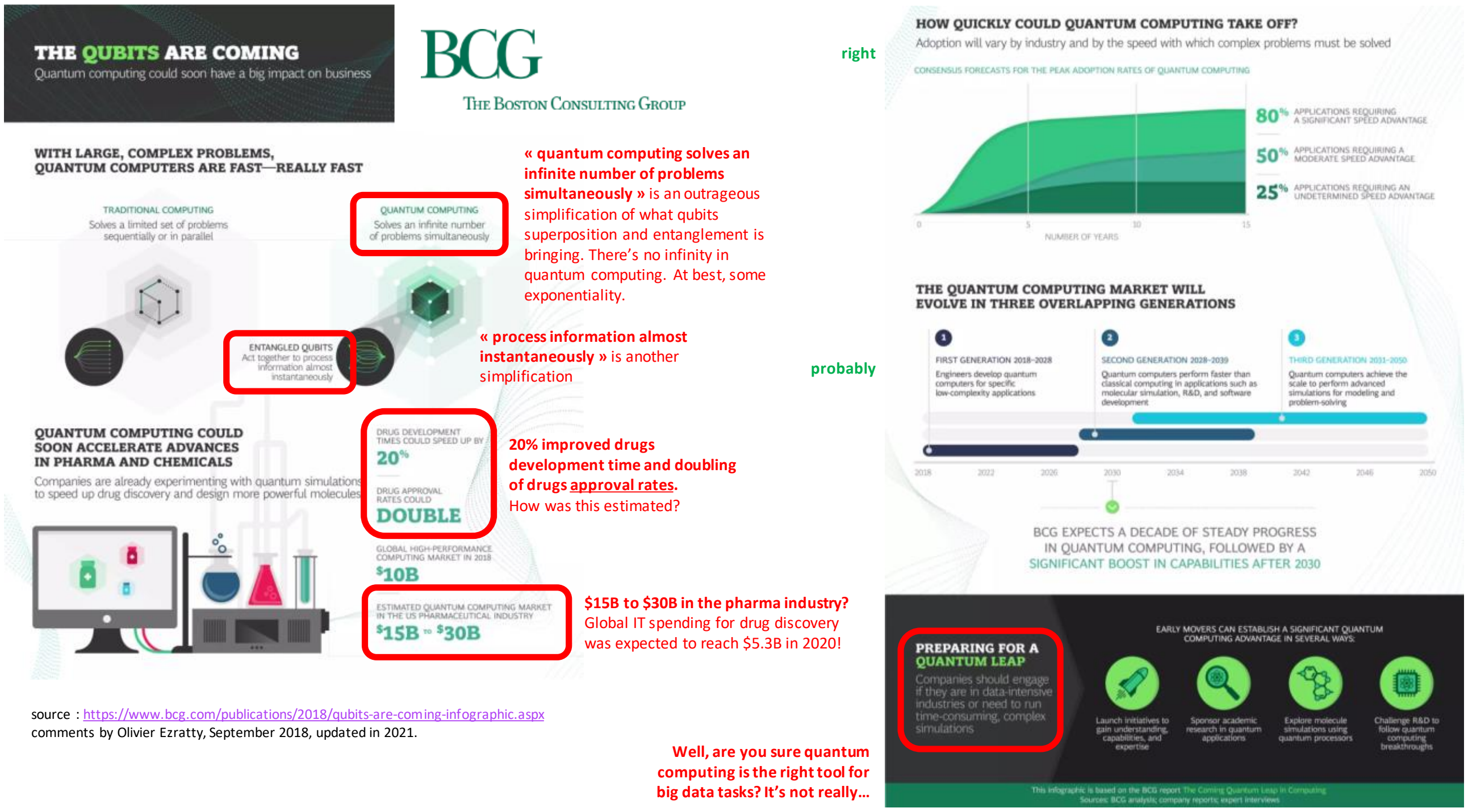

*Figure 897: an example of fact-checking on a BCG forecast related to the healthcare industry. Source: The Qubits are coming, BCG Henderson Institute, June 2018, extracted from the report The Coming Quantum Leap in Computing. Comments by Olivier Ezratty, September 2018, updated in 2021.*

A key form of bs shows-up when quantum hardware startups are hiding simple information like the number of qubits of their QPU (a practice from Anyon Systems in Canada as of 2022 and OQC in the UK in 2021). It usually means that they are too shy to say that they have fewer than 5 operating qubits and therefore are not competitive against companies like IBM (127 qubits as of August 2022) and Rigetti (80 qubits as of the same date). How about qubit fidelities? They are usually hidden as well. Or like with Anyon, they are published without a number of qubits which tells a lot (it's hard to have good fidelities over a large number of qubits, and if their fidelities are bad with an undisclosed number of qubits that must be small, it smells fishy).

Another doubtful practice from quantum hardware vendors connects the dots between customer orientation and unmature hardware platforms is to say: we create application-specific quantum hardware. While it may make sense in some cases, it's an economic and technology non-sense. Successful hardware companies create economies of scale, like IBM in the 1960s with its family of IBM 360 mainframes. Hardware must be generic for a large range of applications. Like Nvidia GPGPUs that are used for both machine learning and scientific applications thanks to a broad software support. On top of that, if one hardware platform has so many limitations that it's bound to be used for only one category of application, you as a customer will be locked-in. Then, you can listen to the technical rationale behind custom-hardware platforms. But specialists will tell you it doesn't make much sense in general.

At last, some quantum hardware vendors will sell you fancy customer-oriented application benefits, without presenting any real quantum advantage whether in results precision, execution time, solution price or spent energy for the environmentally conscious as compared to best-in-class classical solutions.

And yet, I am the first to be convinced of the benefits of quantum computing in pharmaceutical applications, and in particular for simulating the behavior of organic molecules. These die-cut exaggerations are delirious and remind me of those that were made about the Internet of things a few years ago.

In the same vein, the **quantum transistors** evoked in this presentation by Movimento Group for the autonomous vehicles of 2030, stem from a lack of knowledge of the state of the art of quantum computing, its speed of progression and the physical nature of qubits[2934]. Bearing in mind that transistors have been using quantum phenomena since their creation!

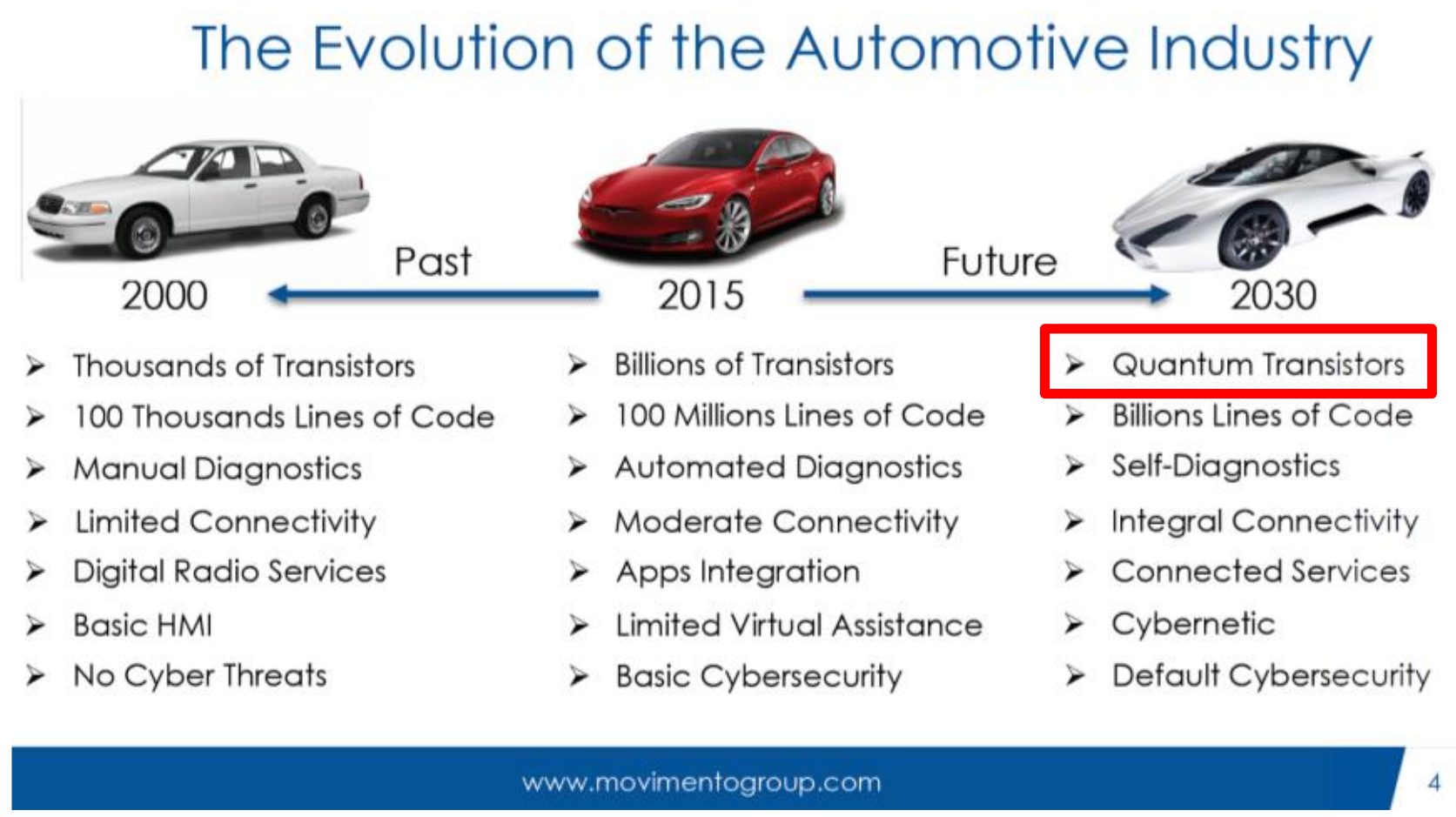

*Figure 898: quantum transistors for the automotive industry? Well, maybe not!*

## Quantum technologies and society key takeaways

- Quantum technologies can become one of the artefacts of Mankind's technology ambitions, pushing the limits of what can be achieved in the line of some works done in artificial intelligence. It may give the impression that mankind's power has no limit. A sound scientific mind will however understand that quantum computing has its own limits. The world can't be simulated, the future can't be predicted, and apparent free will can persist.
- Science fiction has built an imaginary of what quantum technologies could achieve, with teleportation, supraluminal traveling speeds, various entanglement and miniaturization feats, parallel or multiverse worlds and time travel. While none of these things are possible given our current scientific knowledge, it can create scientific vocations and drive new generations to solve actual problems.
- Quantum foundations is the branch of science philosophy that aims to build some understanding of the real world. Quantum physics' formalism is difficult to associate with the principles of reality usually applicable in classical physics. While classical physics understanding has historically been associated with an ontology with objects position and motion enabling the prediction of phenomena such as the motion of planets. Quantum physics lacks such an ontology describing the physical world. Beyond the canonical Copenhagen interpretation (psi and the wave equation), many scientists tried to create such ontologies and the debate is still raging.
- The quantum scientific community is starting to investigate the ethics of quantum technologies. Like with artificial intelligence, it will be questioned on algorithms explainability and auditability, on what it will do to simulate if not tweak matter and life and on how to handle public education. Some related initiatives have already been launched by scientists in Australia, The Netherlands, Canada and the UK.
- The education challenge around quantum sciences and technologies is enormous, both for the general public and with specialists. There's a need for better pedagogy, accessible educational content and also for sound fact-checking information.
- Gender balance is already an issue in quantum technologies with a low share of women in the field, particularly with vendors. Hopefully, there are many top women scientists and entrepreneur role models around who can inspire a new generation of women teenagers. Many initiatives around the world have been launched for that respect.
- At last, quantum technologies vendors marketing must be watched carefully. It is and will be full of exaggerations and approximations. The worse will happen with vendors outside the quantum technology sphere.

[2934] See Protecting Autonomous Vehicles and Connected Services with Software Defined Perimeter, 2017 (21 slides).

# *Quantum fake sciences*

One of the most fascinating topics in the mainstream impact of quantum physics is the way some people integrating it into alternative dubious scientific approaches. The vast framework of "quantum medicine" is a fairly coherent stream of thought and practice from this point of view. It has given rise to the proliferation of gurus of all kinds and to voluntary or involuntary scams based on miracle machines for detecting electromagnetic waves or vague energies, and restoring your body balance. It is at best a subset of the vast placebo effect industry!

Other fields took over quantum physics and long before quantum computing became a visible subject: management and marketing, not to mention politics[2935]. Quantum physics is essentially used there as a source of inspiration by analogy. But the “gurutisation” of these sectors is also quite common, linking together currents of thought that revolve a lot around magical thinking.

## Quantum biology

The starting point of quantum medicine is, however, scientifically relevant and interesting. Some low-level biological phenomena can be well explained at a low-level by quantum physics. Of course, since everything is quantum at this scale!

To mention just a few examples, this is obviously the case of **photosynthesis** in plants, which uses the photoelectric effect transforming a photon into electron displacement, leading after the Calvin cycle to the production of glucose that is used to store energy. The same applies to **retina cones and rods** which capture light. **UV-B rays** participate in the synthesis of Vitamin D3 precursors in the skin again using the photoelectric effect but with a different wavelength[2936]. Quantum physics also explains the **capture of terrestrial magnetism** in the brains of many birds via a special protein called cryptochrome. This mechanism relies on the protein's ability to detect magnetic variations through some electron quantum entanglement[2937]. Quantum biology is a serious scientific domain and it deserves its proper attention[2938].

So far so good.

---

[2935] The concept of quantum politics is still in its infancy. Here is some literature from economic and social researchers on the subject. For example, Quantum like modelling of the non-separability of voters' preferences in the US political system by Polina Khrennikova, University of Leicester, 2014 (13 pages) seeks to model the choices of US voters and the entanglement or not of the choice of presidential candidate and congressional candidates showing that it can decouple under certain conditions. And Quantum Politics: New Methodological Perspective by Asghar Kazemi, 2011 (15 pages) creates a link with chaos theory and the butterfly effect. The paper was written just after the 2011 Arab revolutions. See also Schrodinger's Cat and World History: The Many Worlds Interpretation of Alternative Facts by Tom Banks, who uses Bryce DeWitt's Multiple Worlds Thesis to explain the election of Donald Trump in 2016 by a giant tunnel effect. That maaaayyyy be a little exaggerated! In 2022, some writer tried to explain the Russia invasion of Ukraine with quantum physics. See Quantizing the Invasion of Ukraine by Nicholas Harrington, 2022 (not precisely dated…) and another paper described a quantum parliament, in A two-party quantum parliament by Theodore Andronikos and Michael Stefanidakis, January 2022 (23 pages).

[2936] See The Relationship between Ultraviolet Radiation Exposure and Vitamin D Status by Ola Engelsen, 2010.

[2937] See Resonance effects indicate a radical-pair mechanism for avian magnetic compass by Thorsten Ritz et al, 2004 (4 pages), Cellular autofluorescence is magnetic field sensitive by Noboru Ikeya and Jonathan R. Woodward, January 2021 (6 pages) and Magnetic sensitivity of cryptochrome 4 from a migratory songbird by Jingjing Xu et al, June 2021.

[2938] See The Future of Biology is Quantum - A proposal for a new scientific research organization by Arye and Clarice D. Aiello, May 2022 which calls for the creation of a dedicated research lab on quantum biology, that would be directed by one of its authors.

Then, some renowned scientists want to explain the origin of consciousness with quantum physics. Several major schools of thought are related to each other like the **Orch-OR theory**, the **holographic** dimension of DNA and **biophotons**. And then there are all the works around **structure of water** and **water memory**.

None of these works obtained the agreement of a majority of scientists, but it still deserves a little review. If only to understand how they are quickly being misused by the quantum medicine charlatans over the world.

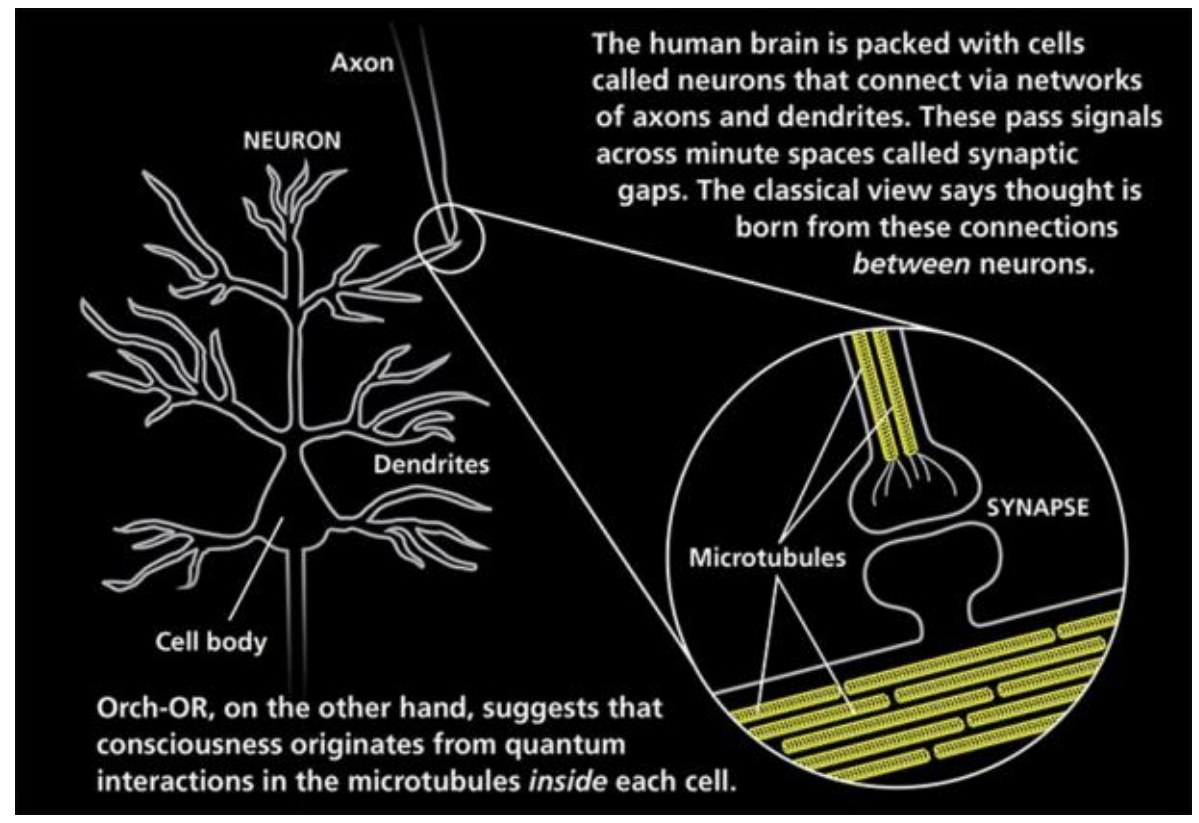

*Figure 899: Orch-OR top-level view.*

## Orch-OR Theory

According to **Roger Penrose** (English, 1931[2939]) and **Stuart Hameroff** (American, 1947), consciousness is housed and managed by microtubules, the complex fibrous structures that, together with actin filaments and intermediate filaments, constitute the structure of neuron cells, called the cytoskeleton, and in the case of neurons, the dendrites, synapses and axons[2940].

In 1996, they proposed the Orch-OR (Orchestrated Objective Reduction) model according to which these microtubules were coherent quantum systems explaining consciousness.

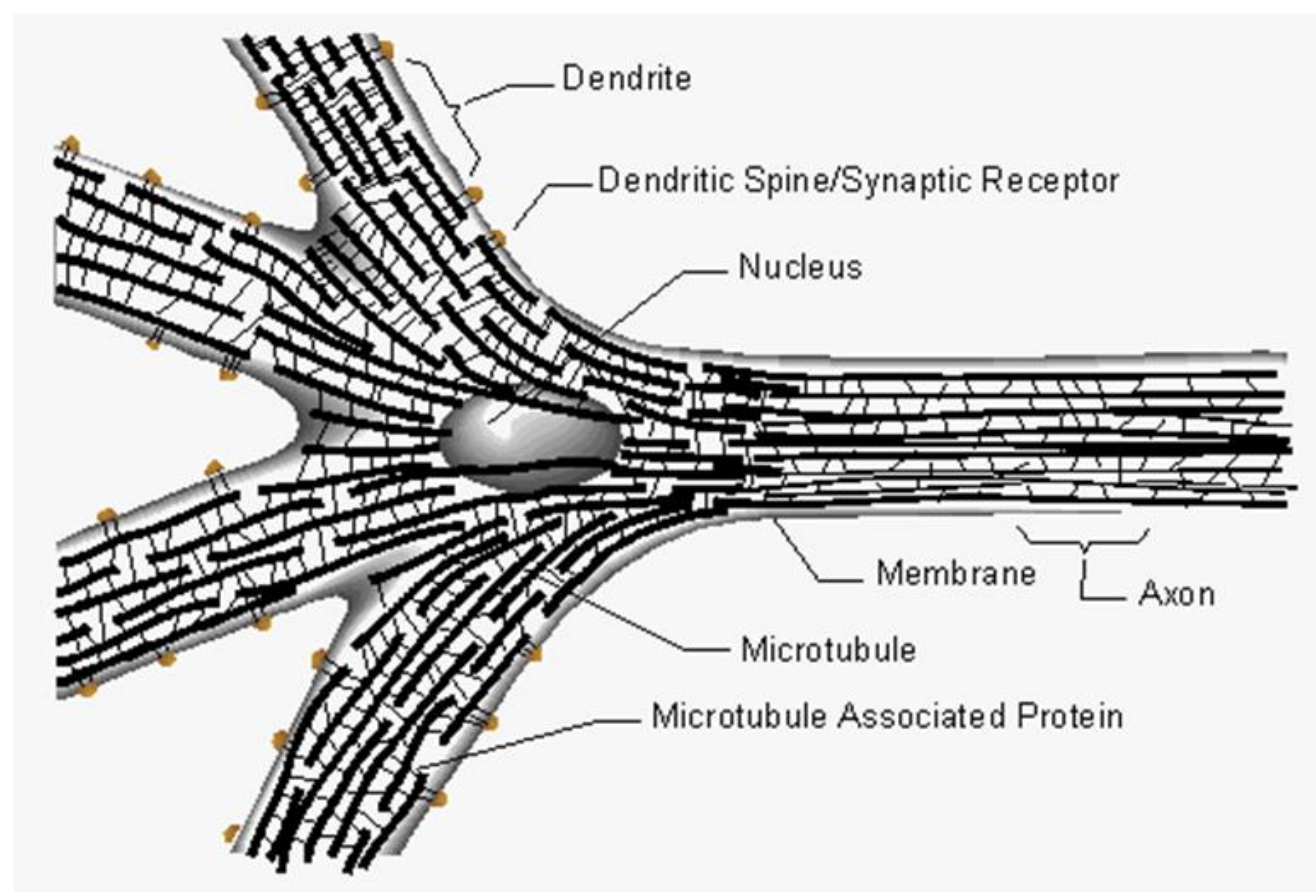

*Figure 900: Orch-OR low level view with neurons and their microtubules.*

For them, consciousness is managed in the neurons within these microtubules and not by their interconnections via dendrites/synapses pairs. In 2011, Roger Penrose and Stuart Hameroff even suggested that these microtubules would be quantum nanocomputers capable of managing qubits and associated calculations[2941]. If this were true, the power of this computer in number of qubits would be immeasurable because a single neuron comprises about 100 million tubules, the brain 86 billion neurons and more than 600 trillion connections between neurons! These theories obviously do not specify how the entanglement between these qubits would work on this scale.

[2939] He was awarded the Nobel prize in physics in 2020 for his seminal work on black holes.

[2940] Illustration source: [Is our brain a quantum computer?]() by Laurent Sacco, April 2018.

[2941] Other theories think that quantum entanglement also works elsewhere in the brain, at the level of phosphorus atoms associated with calcium. This would allow the creation of quantum bonds between neurons. See [Quantum Cognition: The possibility of processing with nuclear spins in the brain]() by Matthew Fisher, 2015 (8 pages). As the article indicates, this raises questions but does not provide answers! Therefore, any rather rapid interpretation of the "quantum brain" is to be taken with a grain of salt.

Ironically, the indirect impact of this gargantuan sizing would be to push back even further in time a possible singularity, the moment when a computer would reach the computing capacity of a human brain in raw computing power[2942]. We are dealing here with another current of thought, promoted in particular by **Ray Kurzweil**.

The Orch-OR theory was revived in 2014 with the discovery of quantum vibrations in microtubules by **Anirban Bandyopadhyay** from the National Institute for Materials Science in Japan[2943]. But that doesn't explain anything. Roger Penrose and Stuart Hameroff also asserted that this behavior is influenced by some type of gravity-related wavefunction collapse[2944]. Consciousness is a "macro" phenomenon. Trying to explain a "macro" phenomenon by a single "nanoscopic" process is meaningless because it completely gets rid of the entire biological hierarchy between the two and the other nanoscopic mechanisms at stake in the nervous system: neurons themselves, neurotransmitters, synapses and dendrites, neurons nucleus, brain regulatory glial cells, and on a larger scale, senses and brain macro-organization[2945].

For example, we can explain a good part of living things via the weak hydrogen-hydrogen bonds - (which are of quantum nature, of course) that are linking together the two DNA strands, or with the oxygen and phosphorus bonds, in DNA and RNA, which are strong and can thus explain the cohesion of these fundamental molecules of living things.

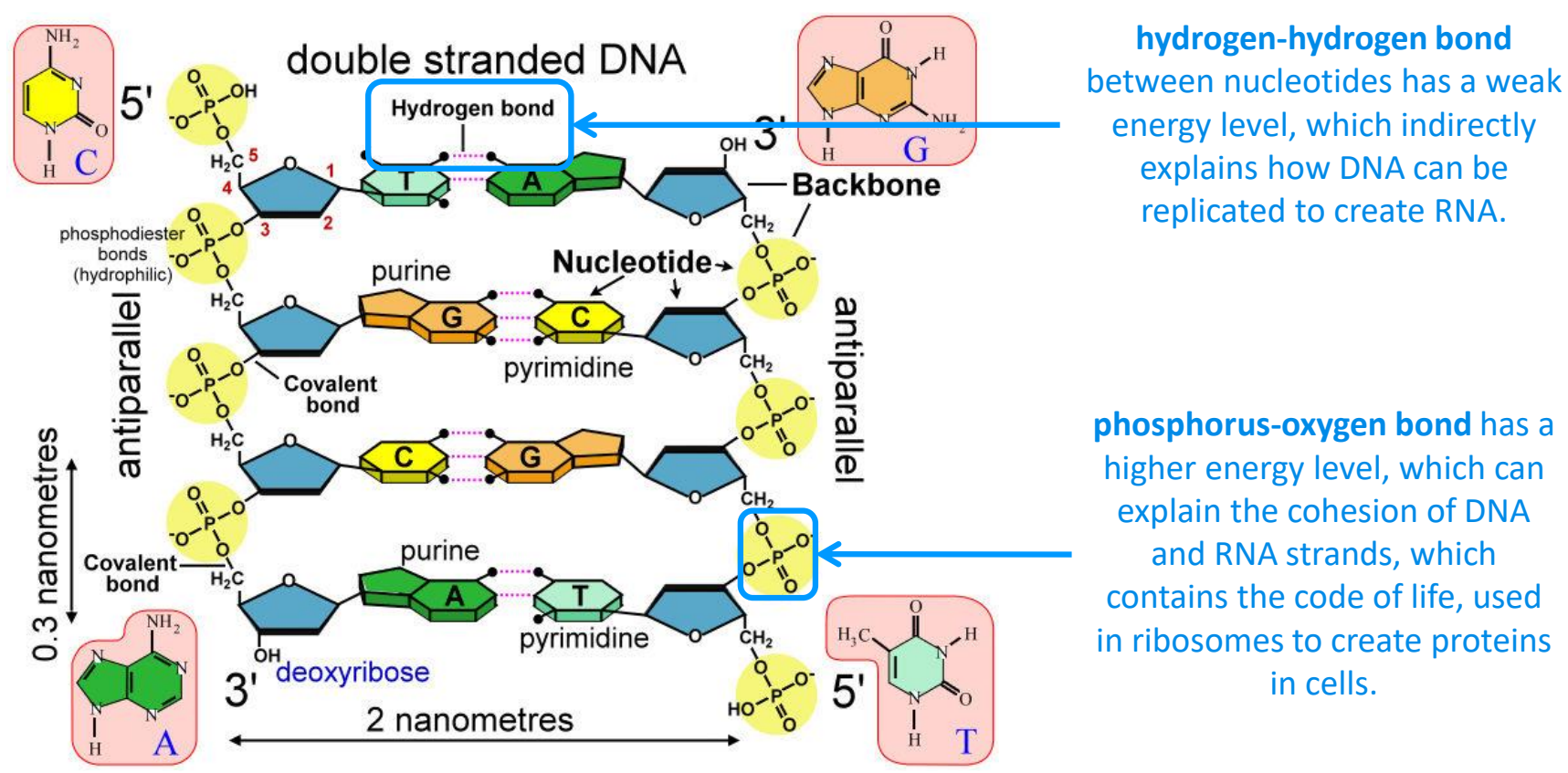

*Figure 901: you can build a whole explanatory theory on life with just two chemical liaisons (hydrogen-hydrogen and oxygen-phosphorus). Source: http://universe-review.ca/F11-monocell08.htm.*

However, this is obviously not enough to explain consciousness or how your heart, eyes and kidneys work. One could also easily build a bozo theory associating consciousness with electrons if not with quarks and gluons. Indeed, without electrons, there's no chemistry and no consciousness! It explains the chemical bonds between atoms.

---

2942 See Consciousness in the Universe Neuroscience, Quantum Space-Time Geometry and Orch OR Theory by Roger Penrose, 2011, 50 pages). All this is documented in Orchestrated Objective Reduction of Quantum Coherence in Brain Microtubules: The "Orch OR" Model for Consciousness, 1996 (28 pages) as well as in Consciousness, Microtubules, & 'Orch OR' A 'Space-time Odyssey' by Stuart Hameroff, 2013 (28 pages), Are Microtubules the Brain of the Neuron by Jon Lieff, 2015 and popularized in The strange link between the human mind and quantum by Philipp Ball, 2017. Roger Penrose has collaborated with Stephen Hawking on gravitational singularities and radiation emission from black holes. Hawking had developed a cosmological theory combining the theory of relativity and quantum physics.

2943 The discovery is disputed by Matti Pitkanen in New Results about Microtubules as Quantum Systems, 2014 (18 pages).

2944 An physics experiment did invalidate most of this theory. See Quantum theory of consciousness put in doubt by underground experiment, PhysicsWorld, July 2022 referring to At the crossroad of the search for spontaneous radiation and the Orch OR consciousness theory by Maaneli Derakhshani et al, Science Direct, September 2022.

2945 See this interesting discussions on Orch-OR in Why is Orch-OR ignored by the mainstream scientific community?, Quora, and also Falsifications of Hameroff-Penrose Orch OR Model of Consciousness and Novel Avenues for Development of Quantum Mind Theory by Danko Dimchev Georgiev, 2006 (32 pages) which debunks many of Stuart Hameroff and Roger Penrose assertions in the Orch-OR model with an in-depth neurobiology analysis.

Fortunately, nobody has yet ventured into this kind of explanation. In short, explaining consciousness by the possibly quantum nature of a particular structure of neurons is the most simplistic reductionism possible, ignoring all the other knowledge available... or yet unavailable[2946].

DNA would also have a quantum function. A curious paper of Russian, German and English origin describes quantum and non-localized phenomena in DNA, verified in a famous experiment based on laser light diffraction, in Figure 902[2947].

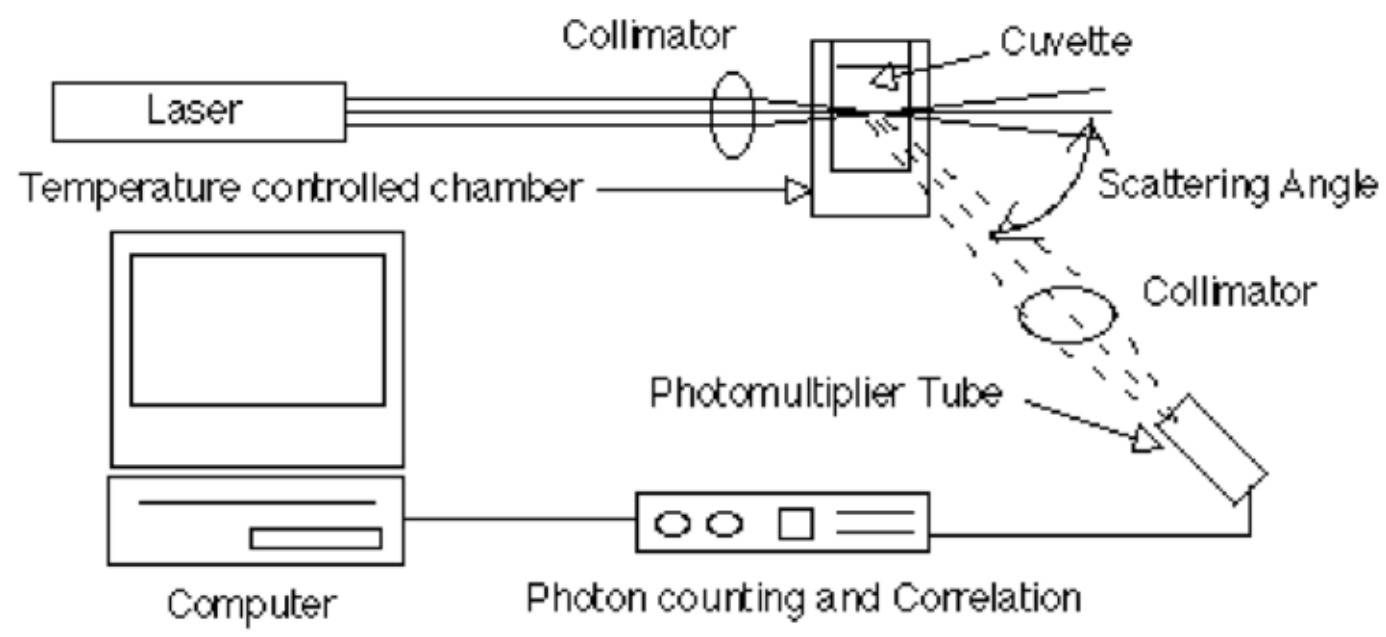

*Figure 902: Source: DNA as Basis for Quantum Biocomputer, 2011 (22 pages),*

The bio-digital DNA wave (20 pages) explains that DNA is in fact a hologram, which interacts with its environment with laser radiation.

Through quantum entanglement, the chromosomes of several cells would interact with each other via these radiations. The Russian of history and leader of this work is a certain **Peter Gariaev,** creator of the concept of BioHolograms within his **Wave Genetics Institute** in Moscow[2948].

Other attempts to explain consciousness by quantum physics have been created. **Matthew Fisher** from UCSB wanted to investigate the brain's potential for quantum computation, based on phosphorus ions spin entanglement[2949]. He launched his **Quantum Brain Project** (QuBrain) with a 1M€ funding in 2018 from the Heising-Simons foundation. Since then, the Project was discontinued. Others like **Johnjoe McFadden** from the University of Surrey in the UK try to explain consciousness with electromagnetic waves circulating in the brain[2950]. At last, some journalists invent some quantum influence where it doesn't exist[2951].

### Biophotons

Another alternative school of thought is related to **biophotons**. These are the low light emissions in the visible generated by living beings. They were discovered in 1922 by the **Alexander Gurwitsch** (Russia). The theory of biophotons was perfected by the **Fritz Albert Popp** (Germany). It complements at a low-level the hologram DNA thesis.

It describes the emission of photons from molecules such as DNA, but also the emission of photons related to the energy metabolism of cells such as the transformation of ADP molecules into ATP in the mitochondria of cells.

---

[2946] A similar reductionism process shows up in Scientists think quantum tunneling in space led to life on Earth by Tristan Greene, TheNextWeb, March 2022, that refers to A pathway to peptides in space through the condensation of atomic carbon by S. A. Krasnokutski et al, Nature Astronomy, February 2022 (13 pages). They explain the appearance of life on Earth for some low-level chemical reaction that could happen on an asteroid. But this reaction is even more likely on Earth given the conditions on the planet!

[2947] See DNA as Basis for Quantum Biocomputer, 2011 (22 pages).

[2948] The history of the theme is explored in Quantum BioHolography A Review of the Field from 1973-2002 by Richard Alan Miller, Iona Miller and Burt Webb (23 pages), but these texts do not give any idea of its scientific validity.

[2949] See Quantum Cognition: The possibility of processing with nuclear spins in the brain by Matthew P. A. Fisher, 2015 (8 pages).

[2950] See Integrating information in the brain's EM field: the cemi field theory of consciousness by Johnjoe McFadden, September 2020 (13 pages) covered in New research claims that consciousness itself is an energy field - a professor says this could be the key to building conscious machines by Victor Tangermann, in Futurism, October 2020.

[2951] See Your brain might be a quantum computer that hallucinates math by Tristan Greene February 2022 referring to Neuronal codes for arithmetic rule processing in the human brain by Esther F. Kutter et al, Science Direct, March 2022 (15 pages).The words quantum and entanglement do not appear in the scientific paper. Ergo the first title is pure clickbait.

The biophotons are ultraviolet and visible light emissions, at levels that are much lower than the mid-infrared emission occurring at around 12 microns wavelength. Up to a few hundred photons per square centimeter of organ analyzed could be detected, often at the skin level.

These biophotons are also made of coherent light - photons with the same frequency. They would constitute a form of inter-cellular communication[2952].

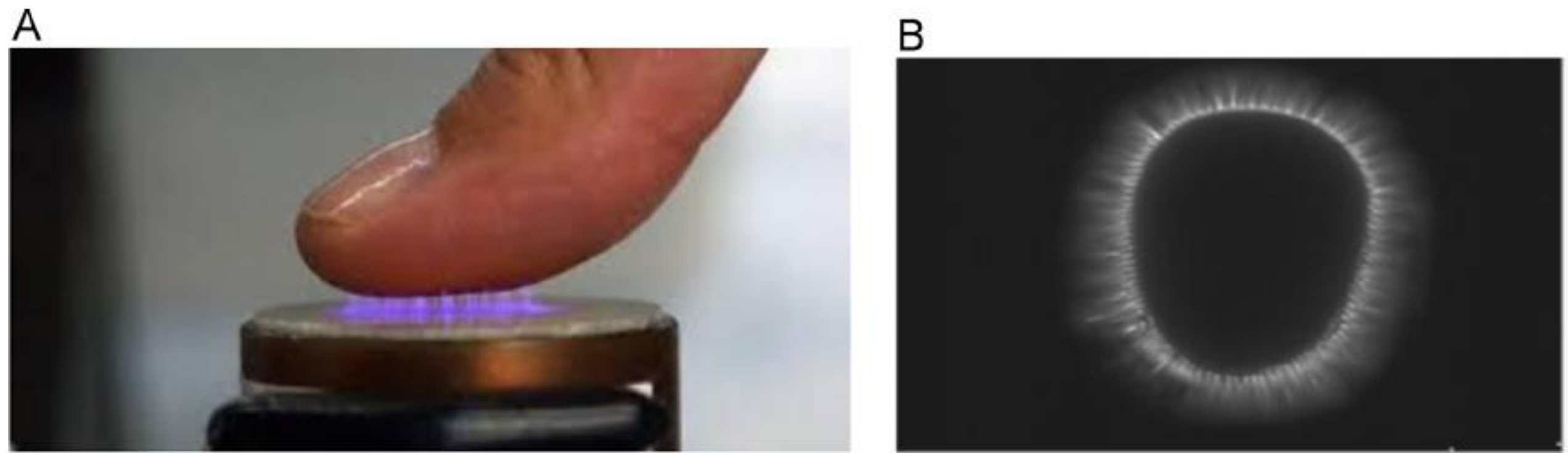

*Figure 903: Biophotons. Source TBD.*

I wonder how this communication works: at what range, due to the obvious attenuation of photons scattering, and with what precision targeting (direction, orientation).

According to Fritz Albert Popp, raw foods emit more biophotons than cooked foods, and organic raw plants emit five times more biophotons than traditionally grown plants. Conclusion: eat raw and organic! This is also a reason to have prehistoric men regret having discovered fire!

In any case, the detection of biophotons on the 10 fingers of the hand would make it possible to detect cardiac pathologies[2953]. The **ClearView** scanner used exploits a curious process: it sends a high-voltage pulse that creates an electromagnetic field around the finger that amplifies the biophotons that are emitted. This excites molecules in the air, creating a plasma between the sensor and the finger (*above left*) that ionizes the air, generating the emission of UV and visible light. This is the **Kirlian effect**, discovered by the Russian Semyon Kirlian in 1939.

The ionization that is captured by the camera (*above right*). The software analyzes the generated shape and compares it to a pathology database. I have a hard time figuring out the exact link between bioluminescence and this process! And what about the receptors of these biophotons?

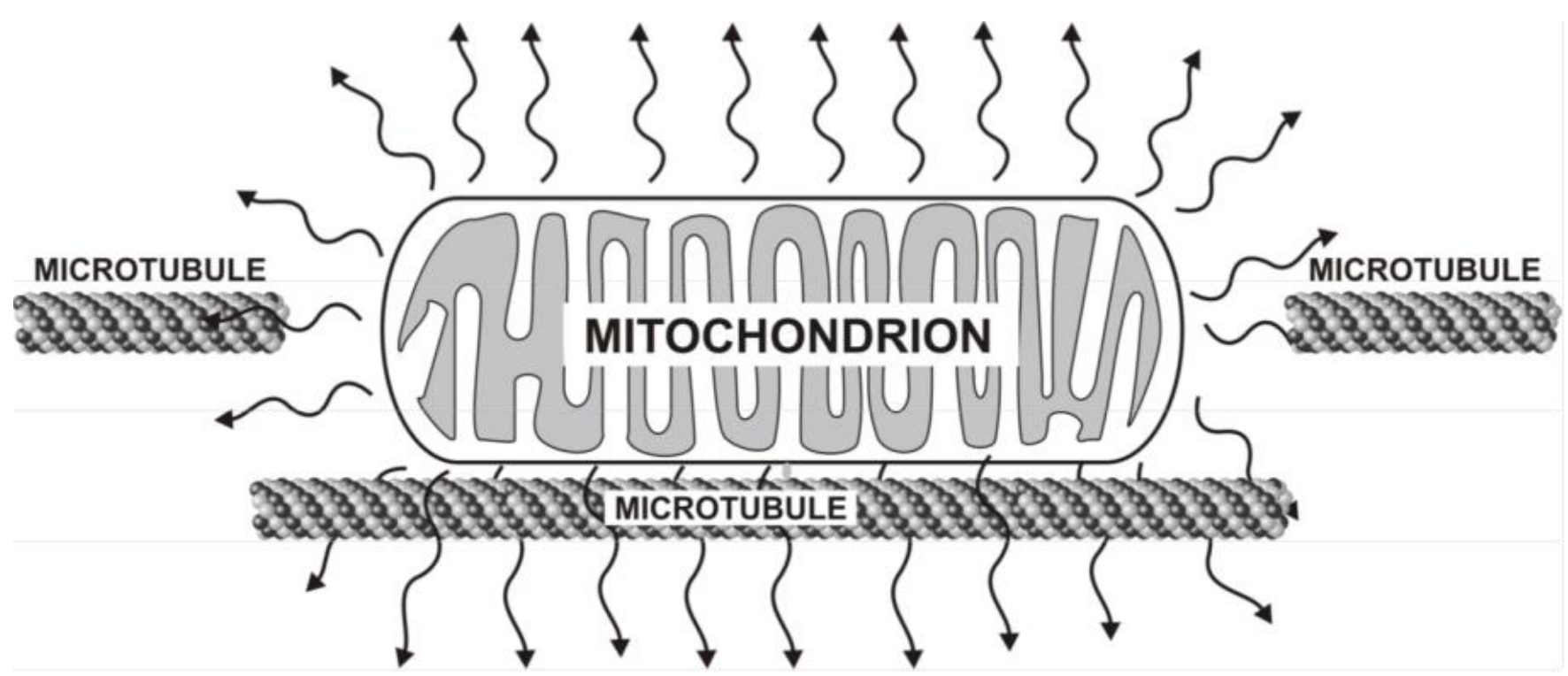

*Figure 904: Source: [Emission of Mitochondrial Biophotons and their Effect on Electrical Activity of Membrane via Microtubules](), 2010 (22 pages).*

[2952] As described in Photonic Communications and Information Encoding in Biological Systems by S.N. Mayburov, 2012 (10 pages) and popularized in Biophoton Communication: Can Cells Talk Using Light?, 2012 in the MIT Technology Review.

[2953] According to Detecting presence of cardiovascular disease through mitochondria respiration as depicted through biophotonic emission by Nancy Rizzo, 2015 (11 pages).

Well, it comes from the neuron microtubules, of course, closing the loop[2954]! According to Popp: "*matter would only be condensed light*"[2955]. By the way, biophotons would be a way to explain chi.

David Muehsam mentions many biological effects of biophotons, which would be involved in the regulation of neurotransmitters secretion (for rats) but without the distinction between correlation and causality being visibly made in the associated publications[2956].

If all that was just science and research! But hell no. It helps snake oil vendors to sell miracle healings through the control of the body by conscience. Practitioners of quantum medicine are very often psychosomaticians exploiting mysticism and autosuggestion to generate, in the best of cases, a good placebo effect that can work with certain mild pathologies. Even so, they justify their methods on the contested work of researchers such as Roger Penrose and Stuart Hameroff, already mentioned, but also Karl Pribram and Henry Stapp, who want to explain human consciousness by quantum phenomena intervening at a low-level in the brain that would also explain a so-called immortality.

Wikipedia's Quantum Mind fact sheet reports on the evolution of this branch and the associated criticisms. It underlines the fact that there is no way to apply possible quantum phenomena such as entanglement at the scale of macroscopic brain molecular or cellular structures.

Entanglement is even less justifiable to connect the brain at long distance to the "*holographic global consciousness of the Universe*" promoted by **Karl Pribram** and **Paola Zizzi**[2957]. In the same way, it does not necessarily make sense to link mind and matter as waves and particles and their famous duality. This leads otherwise to absurdities that explain psychic phenomena of synchronicity by the collapse of the wave function of consciousness, an explanation as absurd as Schrödinger's cat thought experiment. Even if the theories of Penrose and Hameroff were verified, the shortcut would be a little hasty, moving quite too fast from a nano-phenomenon to a macro-phenomenon!

The other commonly proposed method involves the use of various electromagnetic waves, including the famous and smokey **scalar waves**. The idea is to exploit them to restore the balance of unbalanced organs, exploiting the wave-particle duality and the ability to restore the basic energy level of... we don't know. Particularly given the proposed waves are not really targeted.

It is notable, however, that few scientific specialists in quantum medicine mention the capabilities of future quantum computers to simulate the operations of organic molecules and create new therapies. Maybe because known applications of quantum computing in health care are part of traditional allopathic medicine, that they usually avoid or at least complement.

However, I found a vague trace of with **Matti Pitkanen** (Finland) who, in the framework of his work on TGD (Topological Geometrodynamics), proposes a unified theory of physics, and puts forward the idea of creating DNA-based quantum computers[2958]. He believes that DNA communicates "with the Universe". It is also based on Luc Montagnier's experiments on DNA. Matti Pitkanen provides the basis for highly speculative theories on the supposed consciousness of the Universe[2959]. His theories of the unification of physics are so complex that they are impossible to understand, and eventually to validate by experience or to refute.

---

[2954] This is what comes out of Emission of Mitochondrial Biophotons and their Effect on Electrical Activity of Membrane via Microtubules, 2010 (22 pages).

[2955] See Introduction of Consciousness in Matter from Quantum Physics to Biology (18 pages) by Jacqueline Bousquet, a former CNRS researcher who died in 2013.

[2956] See The Energy That Heals Part II: Biophoton Emissions and The Body of Light by David Muehsam, April 2018.

[2957] See Consciousness and Logic in a Quantum-Computing Universe, 2006 (25 pages).

[2958] See Quantum Mind, Magnetic Body, and Biological Body by Matti Pitkanen, August 2018 (186 pages).

[2959] See TGD Universe as a conscious hologram, February 2018 (612 pages).

In the field of light-based therapy, one puzzling solution being sold comes from **Bioptron AG** (1988, Switzerland), part of **Zepter Group** (1986, Switzerland), since 1996. Its "Bioptron Quantum Hyperlight" uses "hyperpolarized light" generated with fullerene ($C_{60}$), a molecule also used by Archer for trap its electron spin qubits. Among other benefits, it treats injuries pain, avoiding pain killer drugs. So far, so good. The system generates some vertically linearly polarized light which passes through a filter containing these fullerene molecules which happen to rotate at a $1.8x10^{10}$ frequency per second. It creates "*perfectly ordered hyperpolarized light*" that is supposed to have some quantum properties similar to those of the biomolecules inside our bodies. Practically, this light is made of both vertically and horizontally polarized photons that "*without exaggeration, [...] reestablish the balance and harmony of energetic processes in biostructures and to harmonize cells, bringing them to back to their initial state of natural equilibrium*". Contrarily to many of the pseudo-quantum scams we'll cover later in this section, this offer is fairly well documented, even scientifically[2960]. You're flooded by tons of scientific information, historical references, links to Nobel prize inventions and scientific publications. But many indices generate serious doubts[2961]. Among others, it mentions these dubious Emoto's research on water structure and the way it can be changed with music and good mood.

### Water memory

The last area on the fine line between science and charlatanism is that of water. It features a model of thought close to Roger Penrose's **Orch-OR** theory, which consists in explaining everything about life based on a few isolated physical phenomena at the microscopic level. The phenomenon of the **memory of water**, its explanation by **electromagnetism**, and parallel theories on the **water structure** are all mixed together.

One of the starting points around the role of water is **Jacques Benveniste**'s work on water memory. This immunology and allergies specialist was a director of an Inserm research laboratory in Clamart, France. He conducted experiments that led to the conclusion that "*water could preserve a memory, a print, of substances that have passed through it*". With Israeli, Italian and Canadian researchers, he published a landmark article in Nature in 1988, which was soon contested[2962]. He described a series of experiments that showed the effectiveness of anti-IgE (anti-immunoglobin E) causing the loss of histamine-containing granules by a type of white blood cell, basophilic cells, even when this anti-IgE is repeatedly diluted to the point where no anti-IgE molecule can be found in solution. For this to work, solutions must be shaken vigorously after each dilution, using the "dynamization" principle!

---

[2960] See the Bioptron Quantum Hyperlight brochure (60 pages) and Hyperpolarized light 2018 (318 pages) by Djuro Koruga.

[2961] Some are well documented in an extensive analysis, although a bit dated, in Cancer and the magic lamp, February 2009. It shows that most scientific surveys were of small scale and non audited and with no control group trials. It was done only on wounds healing. But the vendor web site touts many medical indications that their device is supposed to treat, without any scientific evidence, beyond wounds healing: osteoarthritis, arthroses, lowered motivation and the inability to feel happy. All are good indications, in the best case, of some placebo effect. On top of that, the Zepter also sells blue and red LED light therapy devices, for 500€. The Bioptron is priced at about 1000€.

[2962] See Human basophil degranulation triggered by very dilute antiserum against IgE, Jacques Benveniste et al, June 1988 (3 pages) and Ma vérité sur la mémoire de l'eau by Jacques Benveniste, 2005 (122 pages). The book contains a preface by the Nobel Prize winner Brian Josephson. In this book, published after his death in 2004, Jacques Benveniste recounts his experiences, his tumultuous relations with the medical mandarins over several decades, the story of the publication of his famous article in Nature in 1988 and other experiments conducted during the 1990s and early 2000s.

In the article, Benveniste hypothesized that the phenomenon could be explained by the creation of structured networks in water or by persistent electric or magnetic fields. They would constitute some sort of "water memory" which would "record" the allergen characteristics and reproduce its effects on basophilic cells. This was supposed to explain high dilutions used in homeopathy!

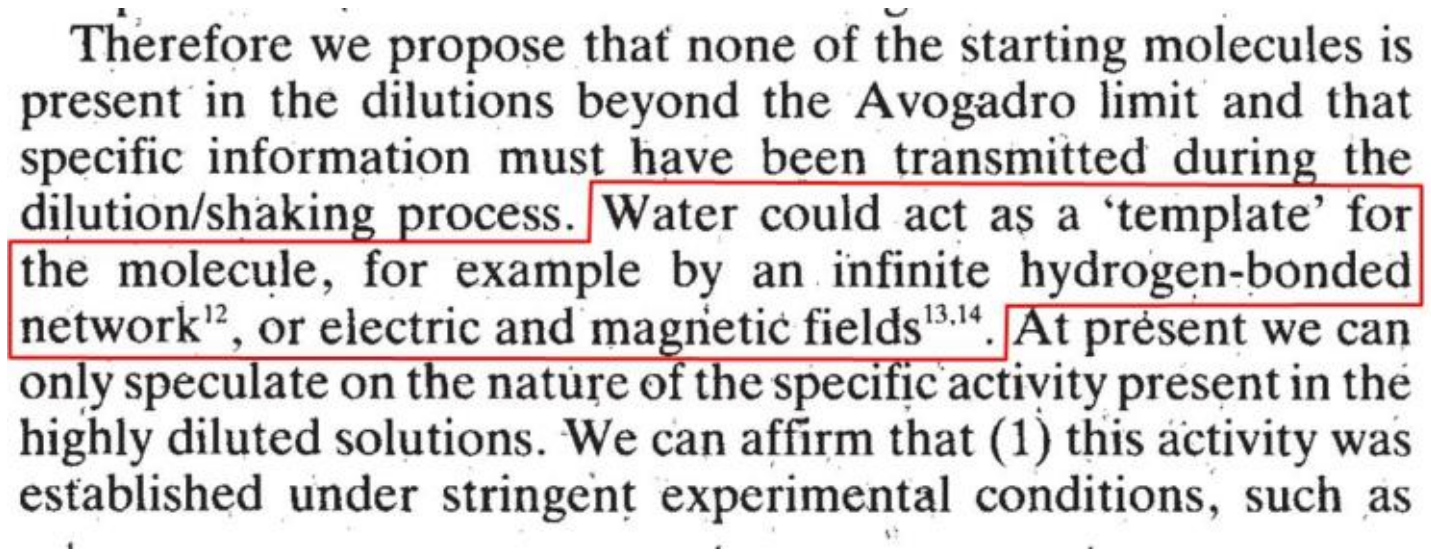
Therefore we propose that none of the starting molecules is present in the dilutions beyond the Avogadro limit and that specific information must have been transmitted during the dilution/shaking process. Water could act as a 'template' for the molecule, for example by an infinite hydrogen-bonded network[12], or electric and magnetic fields[13,14]. At present we can only speculate on the nature of the specific activity present in the highly diluted solutions. We can affirm that (1) this activity was established under stringent experimental conditions, such as

*Figure 905: water memory key description in Benveniste's Nature paper. Source: Human basophil degranulation triggered by very dilute antiserum against IgE, Jacques Benveniste et al, June 1988 (3 pages).*

The promoters of this empirical medicine devised by **Samuel Hahnemann** around 1810 and explained in the book "The Organon " thought they had finally found their scientific support.

Testing and evaluation protocols were flawed in many ways. Solutions were not analyzed by spectrographic analysis to deduce their molecular composition[2963]. Only electrophoresis was used to detect the presence of ions[2964]. The presence of histamine resulting from the release of granules from the basophiles had not been assessed.

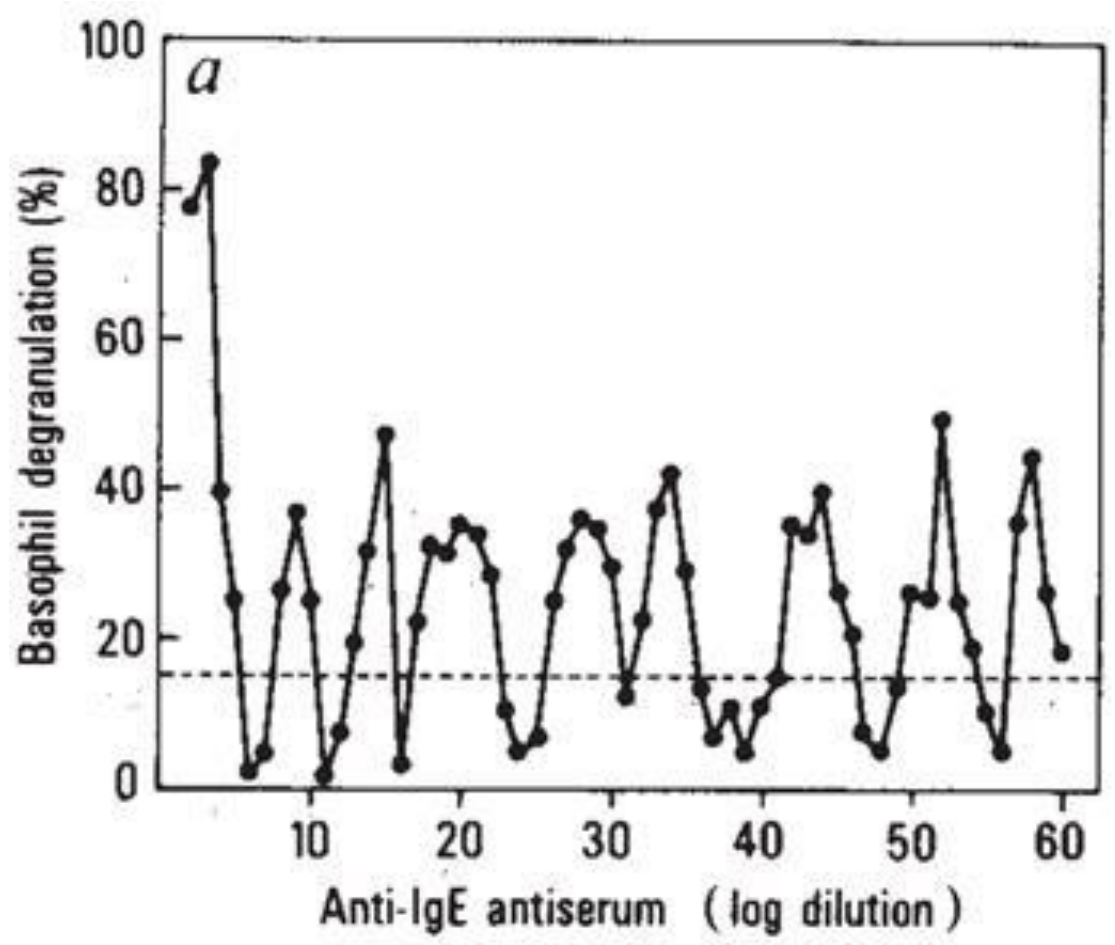

*Figure 906: Source: Human basophil degranulation triggered by very dilute antiserum against IgE, Jacques Benveniste et al, June 1988 (3 pages).*

It was realized in other experiments that there was none! Moreover, the phenomenon presented a cyclic character of a period of 8 dilutions (in Figure 906), according to the successive dilutions, but being out of phase by four dilutions from one experiment to another. No explanation is given for this cyclic phenomenon[2965]. The electromagnetic theory that would explain the phenomenon is his other Achilles' heel.

It is weakly substantiated. These waves are not characterized, measured nor their source explained. The story of Jacques Benveniste is the story of a curious experimenter who lacks, however, the bases in adjacent disciplines around electromagnetism.

However, he did investigate long-range electromagnetic fields, inspired by the work of Italian physicists specialized in quantum electrodynamics, **Giuliano Preparata** (1942-2000) and **Emilio Del Giudice** (1940-2014). In 1990, he set up an experiment with the CNRS Central Laboratory of Magnetism in Meudon, France, which showed that the activity of the diluted solution is modified by prolonged exposure to a magnetic field. The experiment used animal hearts with an electrical apparatus invented by **Oskar Langendorff** (1853-1908). In another experiment carried out over several years, he also uses an amplifier using a sound card from a microcomputer to transmit the properties of a solution to another neutral liquid. This leads to the concept of "digital biology"[2966].

---

[2963] Raman spectrometry will be used in other experiments, much later from 2007, on various homeopathic strains.

[2964] At high dilutions, electrophoresis showed that there was no anti-IgG molecule left in the active ingredient.

[2965] Ironically, the process used does not prevent allergic reactions as is expected in homeopathy, which wants to treat evil with evil, but in low doses. Here the anti-IgE causes the production of histamine and does not prevent it. Some debunking came with "Memory of Water" Experiments Explained with No Role Assigned to Water: Pattern Expectation after Classical Conditioning of the Experimenter by Francis Beauvais, 2018 (20 pages).

[2966] This story is well told in L'âme des molécules, une histoire de la mémoire de l'eau by Francis Beauvais, 2007 (626 pages). The author was one of Jacques Benveniste's experimenters.

After the death of Jacques Benveniste in 2004, his work was taken over by **Luc Montagnier** (1932-2022, French), who created the first AIDS treatment and got the Nobel Prize in medicine in 2008. He described low frequency waves (7 Hz) that would be emitted by DNA strands. He set up an experiment in which the waves of DNA molecules are transmitted through a coil fed at 7 Hz to pure water in another test tube. A PCR is then used to regenerate the DNA in this test tube (DNA multiplication process, "polymerase chain reaction").

And gel electrophoresis is used to decode the replicated DNA! In the experiment, this DNA corresponds exactly to the original DNA. His code would have been transmitted by electromagnetic wave[2967]. But the documentation does not specify which DNA was used as a primer for PCR! Indeed, a PCR does not start from zero and a bunch of nucleotides, but uses DNA strands to replicate them[2968]. The work of Luc Montagnier is related to that of the Italian **Emilio Del Giudice**, again, on the structure of liquid water[2969]. It will not surprise you to learn that this kind of discovery is rather controversial among specialists[2970].

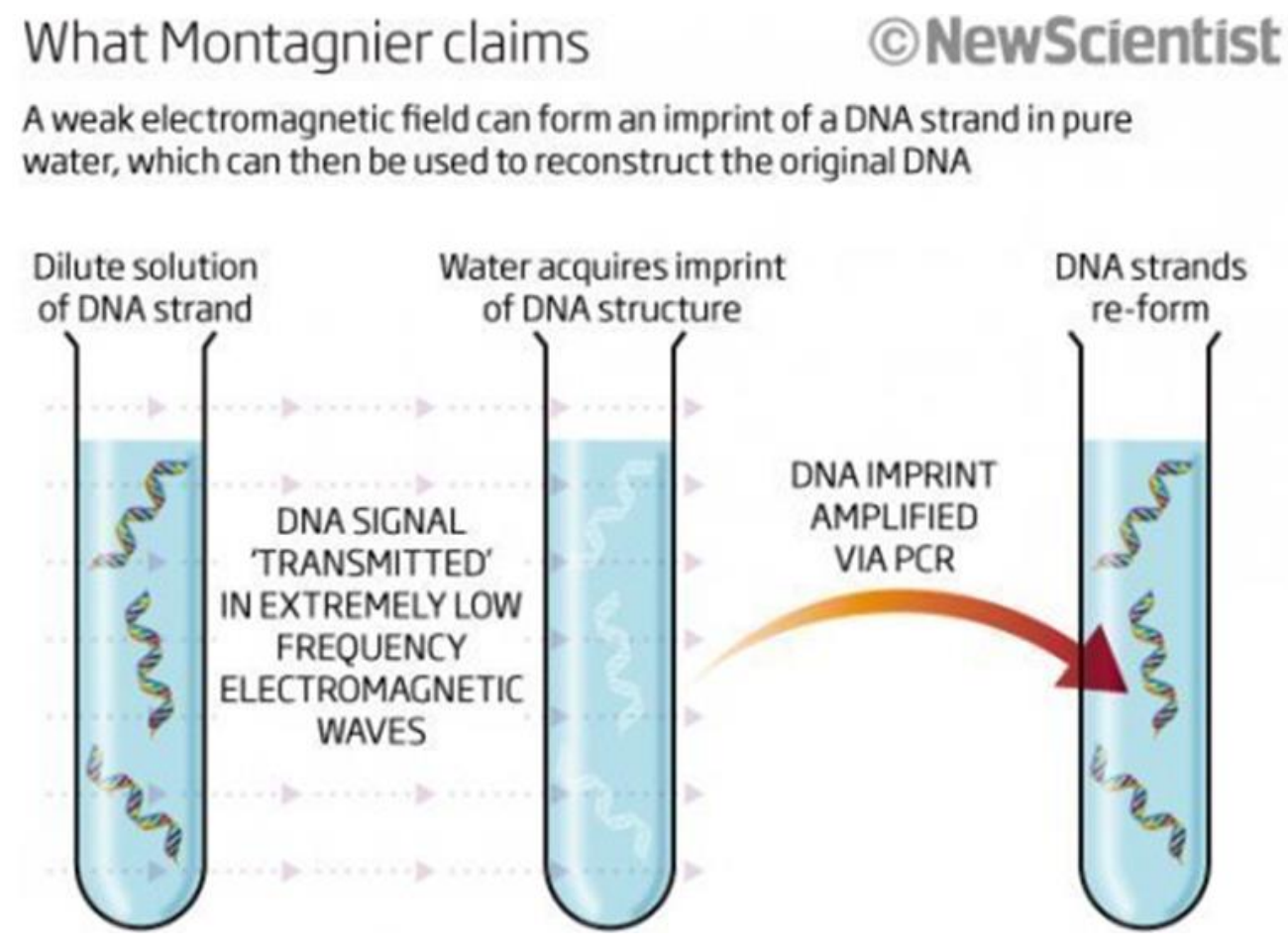

*Figure 907: Montagnier claimed DNA could be created in water containing just… water molecules. How did carbon, phosphorus and nitrogen atoms appear? Source: NewScientist.*

And Luc Montagnier's publication was not made in a peer-reviewed journal. But he continues to publish, with international teams, on interesting research explaining by the quantum field theory how DNA polymerase works[2971]. The relationship between water and quantum physics is being emulated by others and drove the creation of many scams selling structured water and the likes[2972].

**Konstantin Korotkov** (Russia) did some experiments supposed to show that projecting negative emotions on water reduced its energy level and vice versa[2973]. This guy created IUMAB (International Union of Medical and Applied Bioelectrography), an organization that promotes the use of bioelectrography devices[2974]. He is promoting DGV Bio Well cameras, aura detection systems around patients that would materialize the chakras, via the analysis of the "gas discharge".

---

[2967] See explanations in Luc Montagnier's article DNA waves and water, January 2010 (10 pages). Montagnier and the quantum teleportation of DNA by Vincent Verschoore, January 2011, is the source of the illustration.

[2968] This PCR problem is noted in The Nobel disease meets DNA teleportation and homeopathy, January 2011.

[2969] See Mae-Wan Ho's Illuminating Water and Life, 2014 (18 pages) which describes the theories of Emilio Del Giudice, who died that same year.

[2970] See Luc Montagnier and the Nobel Disease by David Gorski, June 2012.

[2971] See Water Bridging Dynamics of Polymerase Chain Reaction in the Gauge Theory Paradigm of Quantum Fields by Luc Montagnier et al, 2018 (18 pages).

[2972] See Hypotheses quantum of mechanism of action of high homeopathic dilutions, is a doctoral thesis by Mathieu Palluel, 2017 (252 pages). Its first part is a fairly well-supplied history of homeopathy. It also covers the experiences of Jacques Benveniste and Luc Montagnier. The quantum part starts on page 181 and is quite weak. This PhD student was definitely not a physicist. He makes a countersense on Schrödinger's equation on page 189. He uses quantum field theory and quantum electrodynamics in a weird context, water at room temperature. On page 201, the paper states that water molecules have a diameter of approximately 3 nm while it is 0.27 nm. It also talks on page 221 about the Nobel Prize of "Serge Laroche" instead of Serge Haroche. In short, this thesis document was poorly reviewed by the people who validated it, and who were not at all up to date in quantum physics.

[2973] See The First Korotkov Intention Experiment by Konstantin Korotkov, January 2018 as well as The Intention Experiment on H2O, 2007 (18 pages) which reproduced his experiments in the USA.

[2974] He is also the author of The Emerging Science of Water: Water Science in the XXIst Century by Vladimir Voeikov and Konstantin Korotkov, 2018 (253 pages), a work or current of thought that certainly influenced Marc Henry's work, unless the opposite is true.

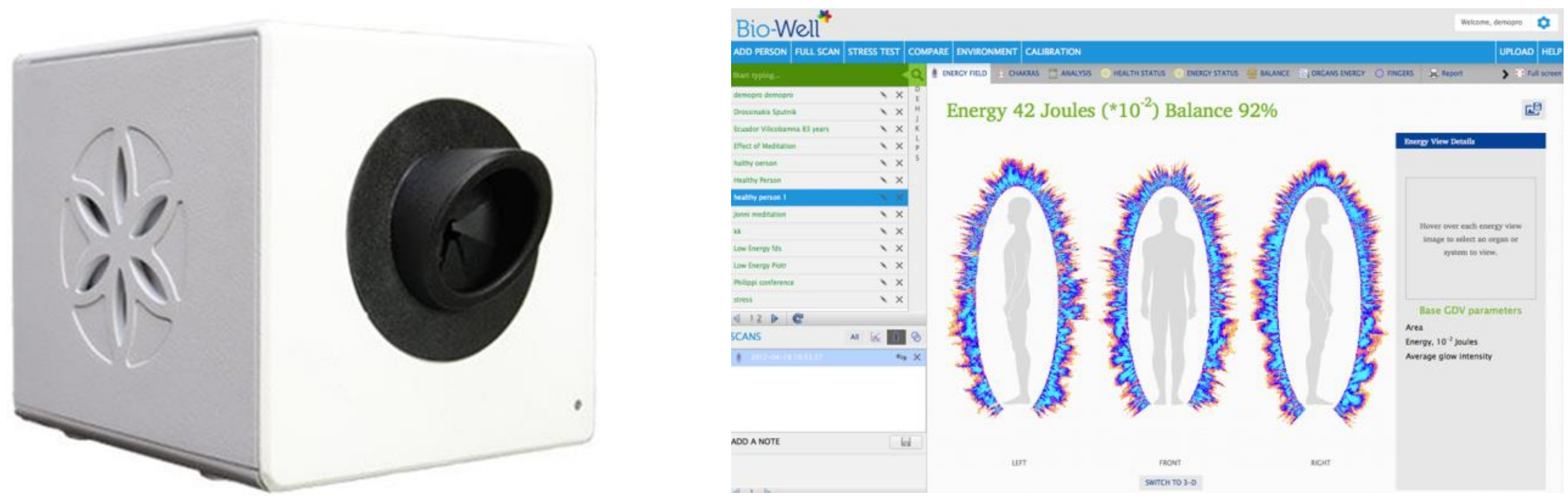

*Figure 908: Bio-Well measure human energy using bio-electrography, inspired by Konstantin Korotkov. These are clearly scams.*

We then have **Mazaru Emoto**'s MRA (Magnetic Resonance Analyzer) (1943-2014). He conducted experiments analyzing the impact of emotions on the structure of water. Experiments that were never reproduced independently[2975]. You probably guessed it!

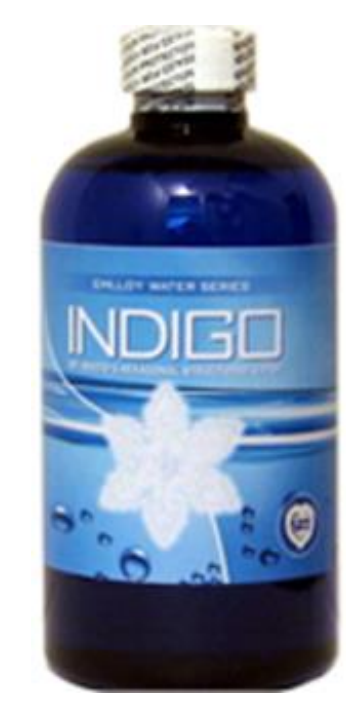

OK, emotions can generate infrared waves and gases that can be exhaled, producing in turn a minute reaction on exposed water[2976]. This makes it possible to sell a concentrated structured water that can be used to prepare distilled water, **Indigo Water** (*opposite*, source). Here is the description: "*A geometrically perfect water with the "Message" your body is waiting to receive. Dr. Emoto's Indigo Water contains eight ounces of highly charged hexagonally structured concentrate. By mixing one ounce of concentrate with one gallon of distilled water, you are creating eight gallons of structured water from this 8 ounce Indigo water. This is about a one month supply of structured water*". For $35. By the way, it doesn't mention if it's drinking water or shower water!

The delirium continues with the structured water of **Rustum Roy** (American). Structured water is said to be an antibiotic: "*One molecule of structured water in 100 million molecules of drinking water can destroy all germs present in a wound. The American army has used this water in Iraq and Afghanistan. Obama uses structured water to wash his hands*". Verification made, the only example that can be found is the healing of a foot wound and it's water associated with money[2977]. And how can we restructure water, so to speak? Simple: by heating it, with vortexes, magnetic fields, music, the force of thought, "frequencies" or minerals!

The concept of wormholes comes from the astronomer **Nicolaï Kosyrev** (1905-1983) who discovered lunar volcanism and the biologist **Rupert Sheldrake** (1942), who became an expert in telepathy. This led to **Vodaflor**'s Voda vortexors which generate vortexes in water to structure it with models ranging from 936€ to 3300€ depending on the desired water structuring rate.

More recently, the discourse around the benefits of water in homeopathy was renewed with the integration of quantum electrodynamics as an explanatory feature. Why not, since almost nobody can understand anything about it, except the few physicists in this domain[2978]. Not to mention the lack of

---

2975 This is well explained in The pseudoscience of creating beautiful (or ugly) water by William Reville, 2011. See also the site Structure-altered water nonsense which makes a good inventory of commercial offers of structure water in the USA. The 1995 style layout serves the site but the inventory of solutions is edifying. Masaru Emoto also certified an effect of exposing zam zam water that is produced at Mecca to Quran. The water is supposed to have similar miraculous effects, a bit like Lourdes' water in France.

2976 See The experiments of Masaru Emoto with emotional imprinting of water, 2018 (11 pages).

2977 In Ultradilute Ag-Aquasols with extraordinary bactericidal properties : the role of the system Ag-O-H2O, 2006 (13 pages). Rustum Roy is also the author of The Structure Of Liquid Water; Novel Insights From Materials Research; Potential Relevance To Homeopathy by Rustum Roy, 2009 (33 pages).

2978 See Explaining Homeopathy With Quantum Electrodynamics by Antonio Manzalini and Bruno Galeazzi, 2018.

experimental protocols to verify anything. Again, we are confronting a fake science because it cannot be refuted[2979]!

The structured water business has evolved a little. Instead of selling structured water, some companies are now selling bottles that create this structured water with regular water. It is just a bottle, or sometimes contains a blender. Gullibles can buy it for about $60 on Amazon.

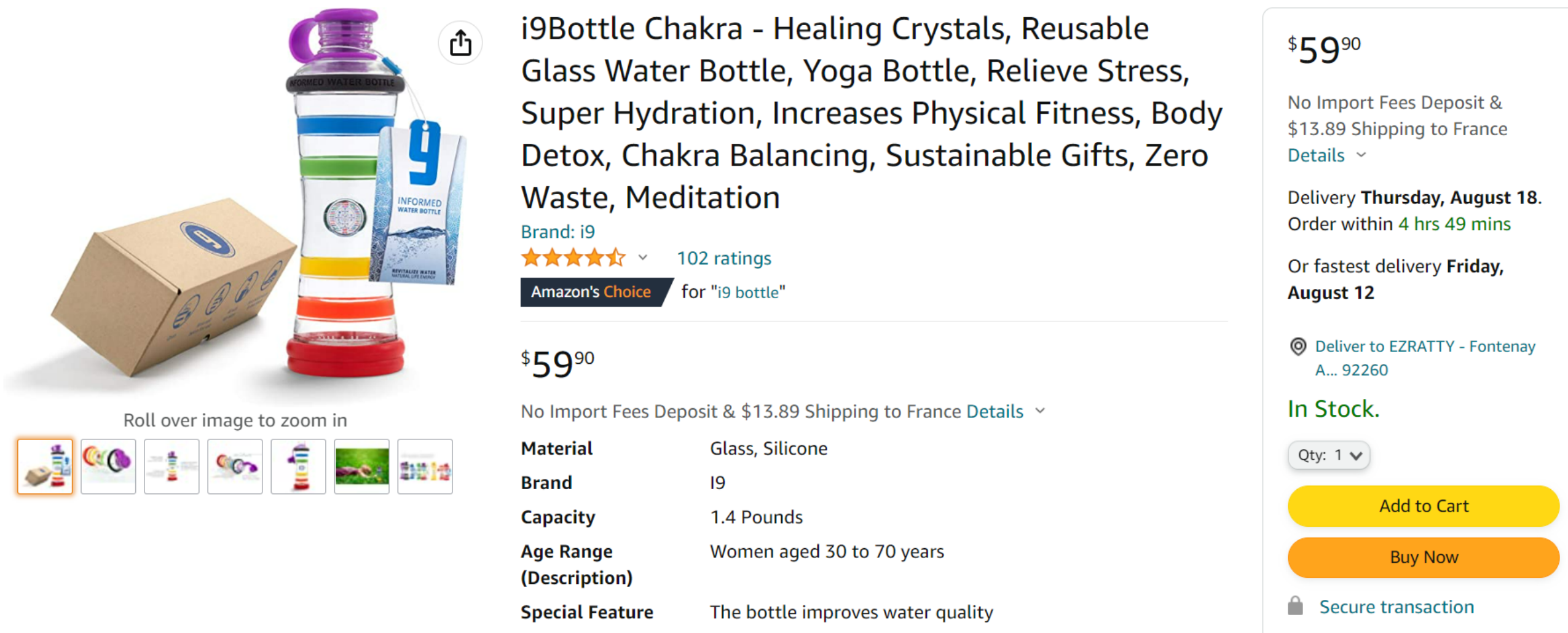

*Figure 909: you can buy plastic bottles that will structure your drinking water. It's even not a thermos!*

To conclude this part before moving on to the most beautiful scams of pseudo-quantum medicine, let us recall that there is a fine line between low-level science and its high-level interpretation, especially when it is then exploited by unscrupulous entrepreneurs.

And we're not done finding more of the same such as some weird quantum behavior of water in carbon nanotubes[2980], superconductivity in the brain[2981] or other elucubrations on quantum cognition[2982]. This will undoubtedly fuel new waves of quantum mysticism!

# Quantum medicine

As Wikipedia's quantum healing page on quantum medicine points out, this discipline misuses the jargon of quantum physics to make people believe in magical cures for certain pathologies that traditional medicine, well or badly practiced, cannot treat properly[2983]. The phenomenon is already over a decade old.

### Method for detecting false science

The methods used to promote false quantum science in health (and in general for that matter) are easily detectable to an educated person, or just with some common sense:

- It starts with some **scientific statement** associating very quickly humanities and biology and making approximate shortcuts on quantum physics.

[2979] Fortunately, some scientists address this nonsense, such as L'homéopathie confrontée à la physique by Alain Bonnier, 2014 (34 pages), which dismantles homeopathy in a very didactic way, relying in particular on Planck's constant.

[2980] See Evidence of a new quantum state of nano-confined water by G. F. Reiter et al, 2011 (5 pages).

[2981] See Possible superconductivity in brain by P. Mikheenko, 2018 (10 pages).

[2982] See What is quantum cognition? Physics theory could predict human behavior by Nicoletta Lanese, January 2020.

[2983] These methods are also well described in Richard Monvoisin's Quantox - Ideological Misuses of Quantum Mechanics, published in 2013 (in French).

- The solutions are being promoted with some **esoteric jargon** using unprecise terms like wave, matter, vibration, vortex and energy[2984].
- When they exist, **tests are performed with small samples** that are not statistically representative. The arguments are often based on non-verifiable anecdotes. The miraculous healings observed in Lourdes, France, are even better documented and, moreover, as probable as those occurring in the hospital environment[2985], i.e., between 1/350,000 and 1/100,000 cases.
- Many specialists sell various, rather expensive, **healing materials or devices**, not considered as medical devices, and whose effectiveness is clearly related to the placebo effect.
- These solution's marketing target **vulnerable people** (sick, elderly, etc.). It can be seen in the media used for advertising it.
- The **vague side of the pathologies covered**. Some are related to pain management or to what can be treated by placebo effect, such as psychonomy [2986]. Others target all the major pathologies of the moment: chronic diseases, cancers and in some cases even neurodegenerative diseases.
- **Extended resumes** with impressive diplomas and scientific guarantees to be taken with a grain of salt for many quantum medicine specialists. There are even "diploma mills" in the USA, where you can buy a doctorate in medicine or another junk discipline at a reasonable price. A bit like in the late Trump University.
- Rare **scientific publications** and when they exist, rarely published in peer-reviewed journals, knowing that this validation is already not enough to be a guarantee of seriousness. These therefore become "private" publications. Or it can't be falsified, like with this paper on quantum immortality that is based on a mathematical approach the many-worlds interpretation[2987].
- Some **conspiracy theories** about the pharmaceutical companies lobbying and other healthcare professionals who will do anything to prevent alternative solutions from emerging.

Nonetheless, there are positive comments from readers of these books that show that the market for gogos is a thriving one. It takes place in a context of loss of confidence in politics, media and science and the development of many conspiracy theories, fueled by the fluidity of the Internet and social networks.

### Quantum medicine marketing

Let's review some of the reference books that promote this curious quantum medicine.

**Quantum Healing** by Deepak Chopra (1988) seems to be foundational. It comes from a former endocrinologist. He became an Ayurvedic practitioner, coming from traditional Indian medicine. According to him, quantum thinking explains some cases of psychosomatic healings that resemble self-healing. The author is a star in the field, especially in India and the USA, with a total book sales of over 10 million copies and a personal fortune estimated at over $80M[2988]. The content of his works

[2984] You find a marvelous example with the Quantum Field Medicine web site that consolidates all these fancy alternative quantum medicines, mostly all based on placebo effect. You have consciousness awareness techniques, acupuncture, homeopathy, electro-magnetic resonance, Timewaver (another electrical product scam), color and light therapy and sound/music therapy.

[2985] See Miracles de Lourdes, Charlatans.info, March 2022.

[2986] Which is yet another false science associating mind and body.

[2987] See Theoretical Quantum Immortality and its Mathematical Authority by Ce Han, February 2021 (8 pages).

[2988] See Alternative medicine is not medicine by Joel Gottsegen, Stanford Daily, October 2014.

is of course quite weak scientific speaking, especially when he deals with quantum physics, mostly in metaphorical terms[2989]. Of course, all of this is plain bullshit and has been debunked[2990].

Amit Goswami's **The Quantum Doctor** (2004) is along the lines of Deepak Chopra's theories. The author is an Indo-American physics teacher who practiced in Oregon between 1968 and 1997, but in nuclear physics. He defines himself as a quantum activist who even has his own Quantum University which seems to be to healthcare what Trump University was to business schools. According to him, quantum activism through consciousness can save civilization. He also demonstrates scientifically (!) the existence of God by building upon Deepak Chopra's consciousness of the Universe thesis. In his work, he explains the therapeutic effectiveness of "integral medicine" which combines allopathic medicine and more or less soft, alternative and traditional medicines, particularly Indian and Chinese. But god's existence can also be proven with some laser beams[2991]!

The scientific content of the book fits on a tiny postage stamp. It looks even like a giant quantum joke. The idea is the following: your organs are born in good health. A time passes, like a qubit would become after a Hadamard gate, it becomes superposed in good and bad health. Then, with the strength of your consciousness, you could provoke a quantum wave function collapse of your organs into the health version. That simple! It's a scam version of this poor Schrödinger's cat.

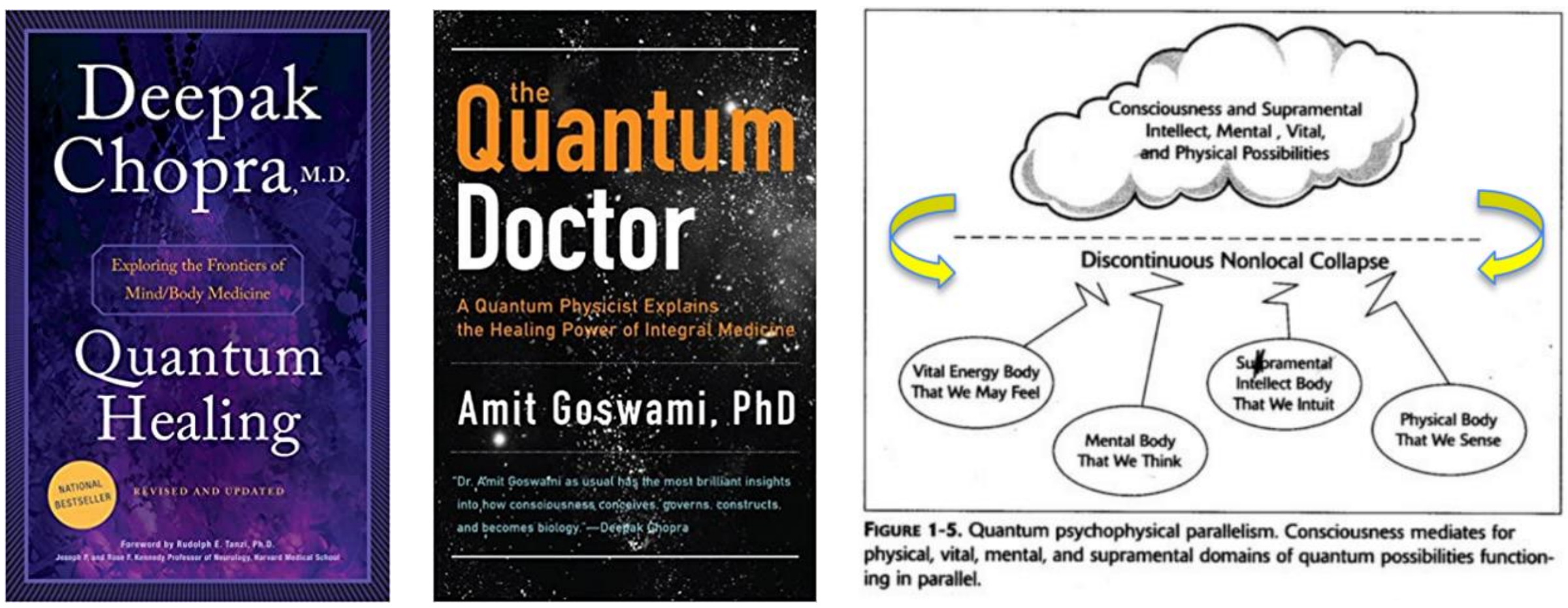

*Figure 910: Deepak Chopra and Amit Goswami are promoting a quantum medicine with no scientific content. At best, it's placebo.*

[2989] On this subject, I watched the enlightening debate between Deepak Chopra and Richard Dawkins (Mexico, 2013, 1h13) which highlights the difficulty of reconciling Chopra's emotional and symbolic approach with Dawkins' rationalist and scientific approach. At one point, the debate focuses on the supposed Universe intelligence that exists according to Chopra and at all levels, from elementary particles to the entire Universe. While this makes no sense to Dawkins beyond biological beings with brains, or computers imitating them. It is a homothetic debate with the link between consciousness and the pathologies that consciousness would or would not necessarily control. The other interesting part of this debate concerns the notion of quantum leap on the appearance of language or certain biological evolutions that are a view of the mind for Richard Dawkins. The latter even denounces Chopra's "deliberate obscurantism". For Richard Dawkins, consciousness is explained or will be explained by neuroscience and certainly not by Deepak Chopra's meta-consciousness galimatias.

[2990] Chopra's discourse has been thoroughly debunked in Problems of Deepak Chopra's discourse: a metalinguistic analysis of "Quantum Healing" by Caderno Brasileiro de Ensino de Física, December 2021 (29 pages). He was also awarded the "Ig Nobel prize" in 1998 "*for his unique interpretation of quantum physics as it applies to life, liberty, and the pursuit of economic happiness*".

[2991] See Interference of Two Independent Laser Beams – Scientific Evidence of God by Henok Tadesse, December 2021 (5 pages). Published on viXra!

The work also seeks to explain the effects and precepts of oriental medicines (chakras, reincarnation, ayurvedic medicine, acupuncture)[2992]. Here are a few selected excepts with the "*morphogenetic fields of the vital body*", "*when the mind creates the disease, sometimes healing is impossible to achieve on the mind level. One must then make a quantum leap to the supramental to heal*" or "*quantum collapse is also fundamentally non-local. Therefore, the non-locality of healing, as in healing through prayer, finds a clear explanation within the framework of quantum thinking*". With quantum entanglement, one can relate everything to everything and explain everything.

Amit Goswami mentions distant healings through prayer by referring to an experiment by physicist **Randolph Byrd** in 1988. The statistical representation was very weak with 6 healings out of 26 patients of not well specified cardiac pathologies. It may not surprise you to find out that it was demonstrated that prayers did not have any large-scale effects[2993].

He also quotes the telepathy experiment of **Jacobo Grinberg-Zylberbaum** (Mexico)[2994]. It involved measuring EEG waves on a participant to assess the impact on him of a flash of light arriving on one of the participants, both of whom were in Faraday cages. The experiment was repeated later between 2000 and 2004 using MRI[2995].

A small technical detail: there cannot be any radio waves transmission between the participants who are in Faraday cages, no photon either, nor particles with a common history in the brain of the participants.

Others have a slightly more scientific view of the quantum nature of consciousness, such as **Ervin Laszlo**, even if the latter relies a bit too much on quantum entanglement in his explanations[2996].

Other pseudo-scientists promote fancy theories related to the so-called quantum medicine.

**James Oschman** (USA) promotes a concept of life energy, based on electric currents and water related quantum phenomenon. He invented the concept of perineural brain cells, which are obviously only the glial cells that surround neurons, but with a different name and which generate energy that goes to the hands[2997].

**Kiran Schmidt** is a German who does "information medicine". He also promotes strange machines that are supposed to cure everything, especially under the brand **Inergetix CoRe**.

**Nassim Haramein** deals with the energy of creation and also water memory. He is selling fancy products through his Resonance Science Foundation. The starting point? Some work on his unified field theory, an old Holy Grail of fundamental physics[2998]. This scientist thinks he has discovered an infinite source of energy. Of course, none of the work of this "scientist" was validated by his peers.

---

2992 Illustration source: Messengers and Messages-then, now, and yet to come (15 pages).

2993 See Studies on intercessory prayer, Wikipedia.

2994 Documented in The Einstein-Podolsky-Rosen Paradox in the Brain: The Transferred Potential, 1994 (7 pages).

2995 See details and results.

2996 In Why Your Brain Is A Quantum Computer, 2010. This thesis is partly deconstructed in The Myth of Quantum Consciousness, 2002 (19 pages), although it is an earlier work.

2997 He is the author of Energy Medicine, James L. Oschman, 2000.

2998 His list of scientific publications deals with neutrons and protons. A part of the articles have been published in the journal Neuro Quantology which is not considered as being serious and whose review committee does not include any scientist in quantum physics or neuroscience. This publication process is known and exists in other fields such as medicine.

This guru markets ARK crystals, which are magical crystals that heal or improve the performance of athletes. They even publish a study on how to improve athlete performances. It used a double-blind method with a placebo effect for half of the test subjects. Given the study involved only 10 athletes, 5 men and 5 women, with progress of about 10%, thus within the margin of error of the sample. The study was done by the Energy Medicine Research Institute laboratory, versed in studies of fancy products such as LifeWave placebos marketed in a Tupperware-style pyramidal model.

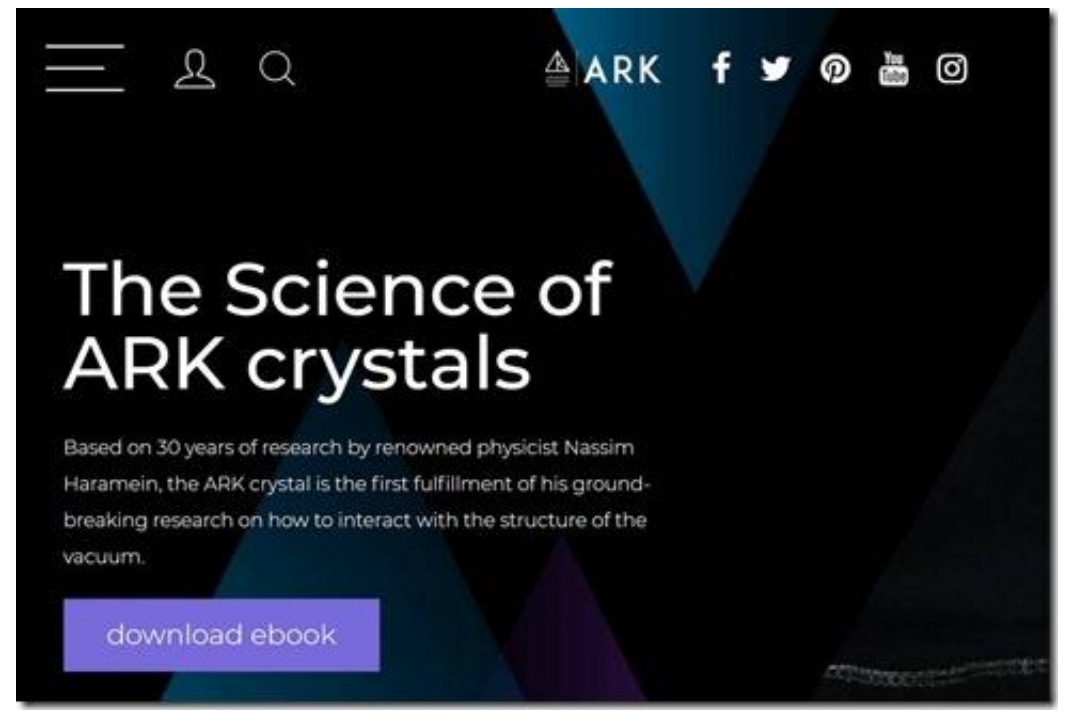

*Figure 911: there's no science in ARK crystals, it's a scam.*

These crystals would also help accelerate plant growth! Prices range from 277€ to 1850€. This is part of a trend in the sale of magic crystals that dates from a few years ago and where the offer is plethoric[2999].

Frank J. Kinslow's **The Secret of Quantum Healing** (2011) introduces the notion of "Quantum Training", a "*scientific, fast and effective* method that *reduces pain and promotes healing*". In a few words, it is about having your consciousness send vibratory waves to your organs to heal them. By the play of interferences, they will cancel the evil.

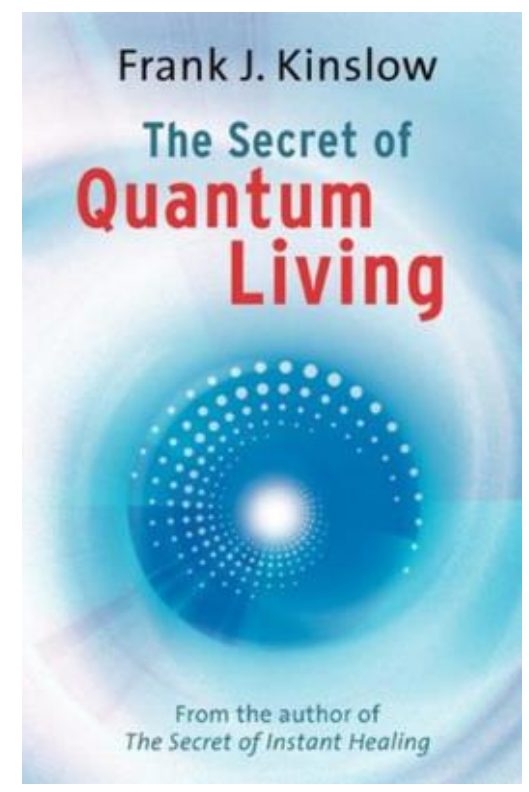

Another Schrödinger's cat trick with the application of the quantum mechanics of the pico (elementary particles) to the macro (the organs). It is mainly aimed at physical and emotional pain. It is a variant of meditation. It should be avoided for the treatment of hypothyroidism or about anything else by the way!

This kind of work has the particularity of always being very vague on the notion of pathology treated, especially if a pseudo-medical apparatus is involved, as is the case here. Even if the "Quantum Training" is supposed to work remotely.

**Quantum Intronic Medicine** from Christian Daniel Assoun deals with quantum biology. It is a form of epigenetic treatise describing DNA memory by quantum mechanics. According to him, "*WATER is the first quantum liquid: its current state is liquid* whereas *its state should be gaseous*". This book describes the presence of a third DNA catenary in the form of physical plasma (hydrogen)[3000]. He states that his "*work is currently focused on the INTRONIC parts which represent 95% of our DNA and which are unfairly classified as silent or even useless*". Intronic is used in the sense of DNA "introns", the part of DNA genes that is transcribed into RNA when the genes are expressed.

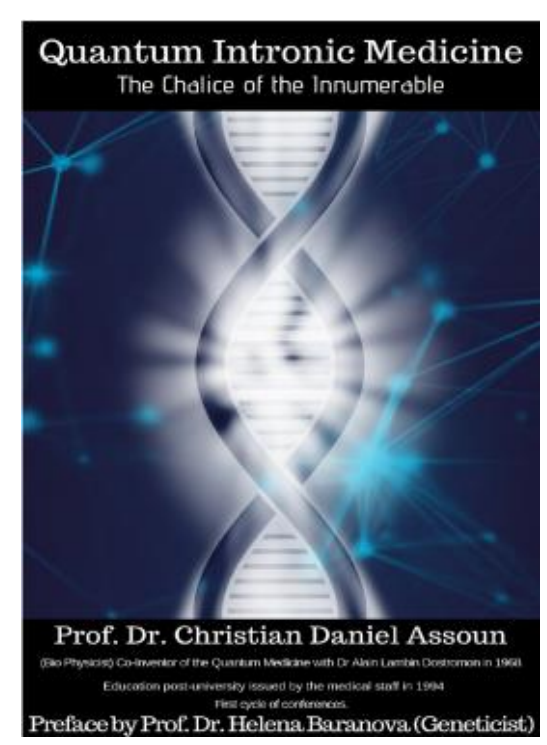

[2999] See Dark crystals: the brutal reality behind a booming wellness craze by Tess McClure dans The Guardian, September 2019, A Cynic's Search for the Truth About Healing Crystals by Katherine Gillespie in Vice, September 2017 and The Sickening Business of Wellness by Yvette d'Entremont, December 2016. Those guys also promote the water memory theory. See Does Water Have Memory?, ArkCrystals, July 2021.

[3000] Ebook downloadable here. It is also documented in The 3rd Strand (or 3rd Catenary) of DNA by the same author and which dates from 2011/2012.

These are eliminated during splicing which generates mature RNA that will then be used in the ribosomes to make proteins. In fact, introns represent only 25% of human DNA. The rest, about 73%, corresponds to sequences that are effectively non-coding in the DNA of our chromosomes, but whose role in the regulation of genes is progressively revealed with research. Exons, the coding part of genes, represent 1.5% of human DNA (source).

Christian Daniel Assoun believes that DNA could be strengthened with "the *help of new tetravalent elements such as Germanium or Silicon (reverse optoquantic properties)*". Why germanium and silicon? Because they are in the same column of Mendeleev's table as carbon with four free electrons. This is a good idea for creating extraterrestrial life. So why didn't life on Earth use silicon, which is as abundant as carbon? One of the reasons is that silicon oxide ($SiO_2$) is inert and solid while carbon oxides ($CO$, $CO_2$) are gaseous and therefore more easily recombinable with other atoms and molecules. Also, carbon is more abundant than silicon on the surface of the Earth.

Christian Daniel Assoun is also the founder of the **Glycan Group**, in 1996, a company selling organic silicon for various uses and notably as a food supplement. Their subsidiary Glycan Pharma was struck off the commercial register in 2012. The company is in competition with Silicium Espana, a company linked to Loic Le Ribault, who died in 2007, who was also passionate about organic silicon. The two companies had a legal dispute in 2011 over the use of the G5 trademark.

At last, you also can count on the many books on Transurfing by **Vadim Zeland** who introduces himself as a physicist. This quantum model of personal development is based on the idea that "*When the parameters of mental energy change, the organism moves to another lifeline. When the parameters of mental energy change, the organism moves towards another life line*". So be it!

### Scalar wave generators

The best of the quantum medicine scams are the **scalar wave generators**. These are electromagnetic waves associating a supposedly horizontally polarized electro-magnetic wave and another vertically polarized wave of the same frequency but 90° or a quarter wavelength out of phase.

Scalar waves were initially promoted by a certain **Thomas Bearden** in the USA as well as by the Russian **Sergei Koltsov** with his Functional State Correctors (CEF[3001]). Bearden explains this in a 1991 interview. He had also invented a **MEG** (Motionless Electromagnetic Generator) capable of extracting free energy from the vacuum and thus, of generating more energy than it consumed. A product that has of course never been commercialized.

Scalar waves were also promoted by a German scientist, **Konstantin Meyl**, with a paper that had to later be retracted[3002]. The general public propaganda on scalar waves is a big fantasy and always linked to alternative medicine literature. These waves would come from the Sun with neutrinos and have no energy loss over distance. The brain is supposed to produce and sense scalar waves with its own interferometer. It would explain telepathy and other paranormal effects. Well well.

[3001] Watch this video Functional State Correctors (FCS) - Koltsov Plates, 2014 (55 minutes) which is a good digest of any scientific theory.

[3002] See "Way out there" paper claiming to merge physics and biology retracted, RetractedWatch 2013 and Scalar Wave Transponder device by Konstantin Meyl, 2005 (70 pages).

Scalar waves would also make it possible to treat diabetes (I or II? Who cares...), kidney stones, Parkinson's disease, heart attacks, osteoarthritis, cancer and also aging. As for type I diabetes, which is linked to the autoimmune destruction of beta cells of the islets of Langherans in the pancreas, it is not clear how waves of any kind would bring dead cells back to life. The proposed solution?

Scalar wave generators such as the **INDEL** at 8820€. Given its price, it targets professionals in a kind of Ponzi model. This generator produces a scalar wave field with a voltage of 2V. It also includes a music modulation accessory for therapy practices and wellness centers. It is also available at **QuWave**.

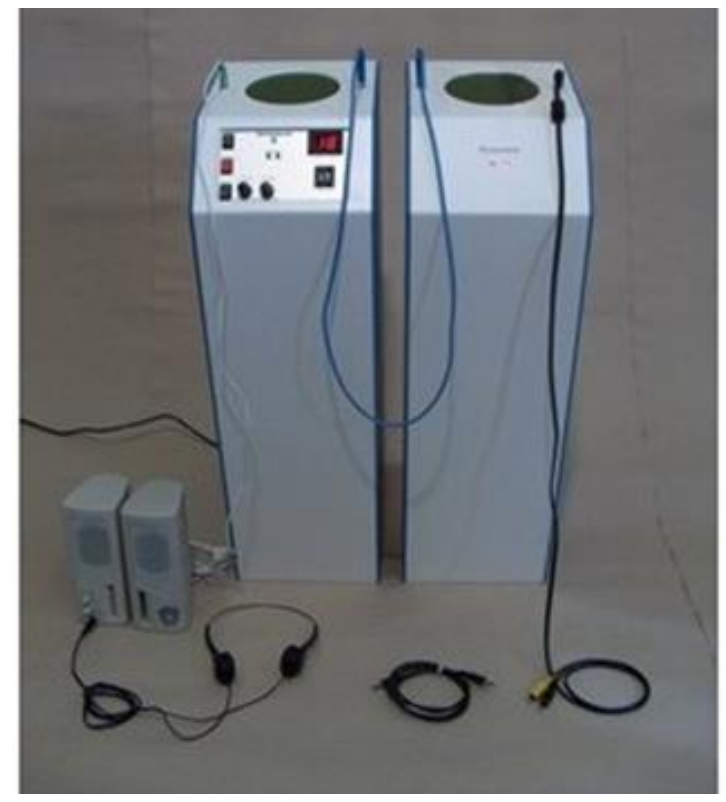

*Figure 912: scalar waves cost a lot and do nothing.*

You can also (not safely) rely on the **ETHX-SCIO Biofeedback** from William Nelson, which combines global therapies and advanced quantum physics (in Figure 912). The device scans the body on 10,400 different frequencies to detect many pathologies. It then rebalances the body's energy with quantum biofeedback. The toy also runs 200 biofeedback therapies with the world's largest health software that integrates Western and Eastern philosophies[3003]. The EPFX-SCIO includes a wave diffuser box, connected to the patient with sensors attached to his ankles, wrists and skull. One could almost do both an EEG and an ECG with it! All this for getting some placebo!

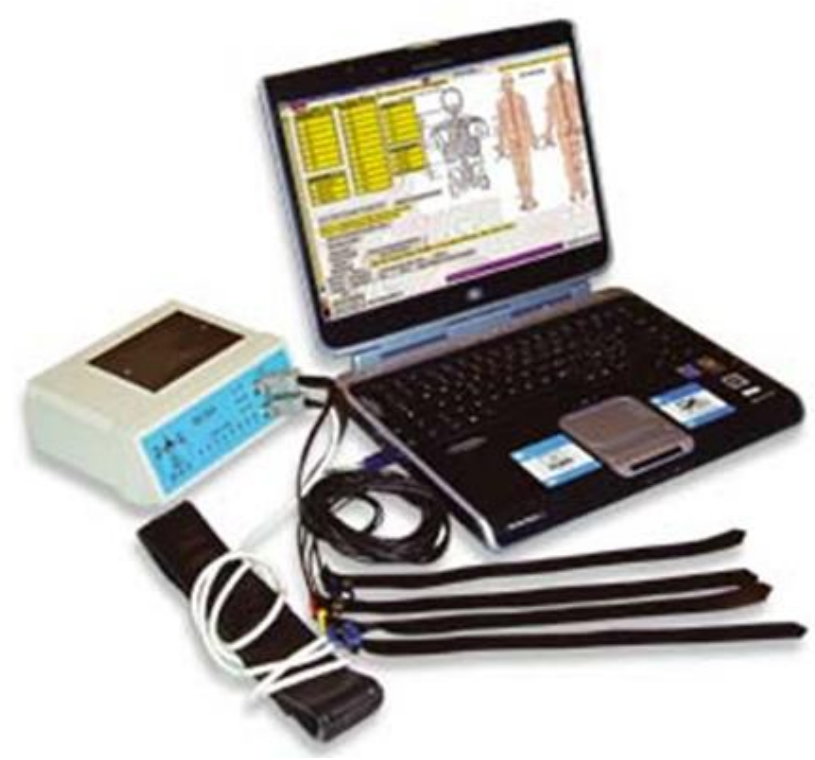

*Figure 913: SCIO Biofeedback is not better.*

In the scam devices category, you also find the **Healy** and its bioresonance features using some electrodes and supposed to cure many illnesses[3004]. At best, it can be a temporary pain killer. Another device, the **TimeWaver**, is based on "quantum field theory" from a certain Burkhard Heim (1925 - 2001, German) on the 12-dimensional composition of the universe where "*the light quantum effect communicates mainly with the Global Information field (GIF) i.e. at a nonenergetic, non-phenomenal and therefore more causative level*". It looks like a biofeedback device similar to the one above. Burkhard Heim did try to unify all quantum theories but he was neither a Dirac or a Feynman[3005]! The TimeWaver site also mentions of **Kozyrev mirrors** using cylindrical aluminum sheets that were used for extrasensory perception experiments in Russia. But it doesn't seem to be involved in the TimeWaver device.

Other various Russian 'quantum' scientists, dead or alive, are frequently used in support of these scam devices like Nikolai Kozyrev, Vlail Kaznacheev or Alexander Trofimov. When you look at their biographies online, you quickly find that they were not at all mainstream quantum scientists. This is all full of esoterism, not science.

---

[3003] See How one man's invention is part of a growing worldwide scam that snares the desperate ill.

[3004] See A Skeptical Look at the Healy "Bioresonance" Device by Stephen Barrett, July 2020.

[3005] See TimeWaver System website. They hopefully have a: « *Disclaimer: Science and conventional medicine does not acknowledge the existence of information fields their medical and other important TimeWaver systems and their applications due to lack of scientific evidence. The said application is based on, treatment options, experiences and anecdotal reports from the practice*".

### Quantum medallions

Quantum medallions for smartphones have become commonplace for several years and target another phobia, electromagnetic waves and 5G. This is the case of **Quantum Science**'s Quantum Shield medallions (on Amazon and Alibaba). One also finds some in the form of USB keys **5G BioShield** which contain a "*quantum holographic catalyst*".

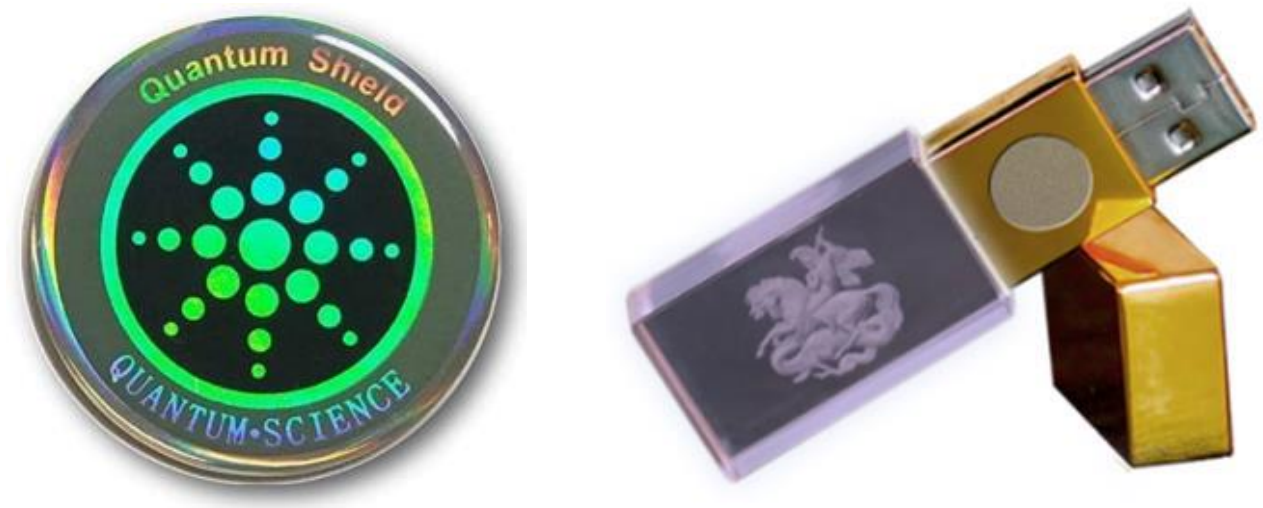

*Figure 914: quantum medallions and 5G quantum keys are fancy gadgets for the gullible. That's a huge market!*

It is obviously a huge quantum bullshit of the first kind. It is accompanied by a scientific justification that is not worth a lot of money[3006]. The American FTC has flagged these products as vulgar scams[3007]. It was even later discovered that some of these medallions were radioactive due to their component metals. And worn on a long period of time, they could actually be dangerous[3008].

In the field of wacky quantum devices, let's finish with the **Quantum 5 Ozone Generator** using Neos Technology from the **Longevity Resources** (sources). It uses a quartz electrode. It's supposed to help purify indoor air. There is one major drawback: ozone can also be toxic to the human body and cause respiratory problems. It can also affect plants health. In short, quantum medicine may one day emerge in the wake of scientific discoveries, but the ones proposed today is for the time being full blown charlatanism.

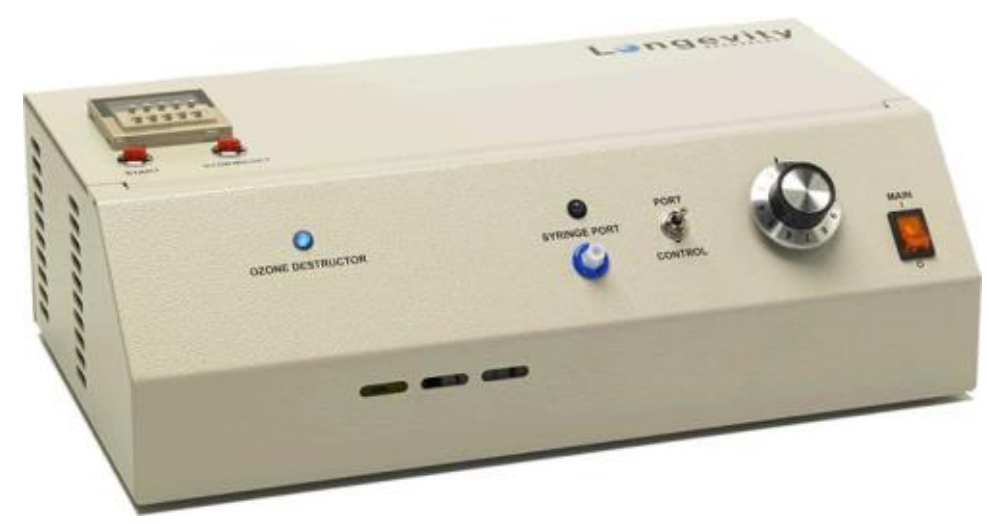

*Figure 915: a quantum ozone generator to purify indoor air. It may work but it is not quantum.*

They have the advantage of generating at least a placebo effect for users and filling the wallets of their promoters. Except that this can be dangerous if the placebo effect is used instead of a traditional treatment that is essential to stay alive.

I will not, however, trash all the techniques and approaches mentioned here. Some may make sense, even though there is still a lack of both a scientific corpus and more solid evidence to support them. But most are fake sciences and are quite easily detectable.

### Quantum skin care

I discovered the scam category of **quantum cosmetics products**, coming mainly from China that was mentioned in a 2022 Rand Corporation on China and the USA investments in quantum technologies. In it, Jian-Wei Pan is said to have criticized companies claiming to sell "quantum skin care" products in China[3009].

QUANTUM
FACIAL AESTHETICS

Let's make a roundup of these scams. We have for example a Quantum Health Super Lysine+ Cold-Stick sold on Amazon, Energecia Quantum Beauty sells quantacosmetics bullshit stuff, BioEqua sells energized nanospray skin care, probably with stirred magic water, Age Well Fundamentals sells Phyto5 quantum energy facial skin care that balances the energy vitals and is made in Switzerland

---

[3006] See "Aton" True Cell, Atom and Particle Concept by Ilija Lakicevic, 2019 (8 pages).

[3007] See Cell Phone Radiation Scams, 2011.

[3008] See Anti-5G "quantum pendants" are radioactive by Jennifer Ouellette, ArsTechnica, December 2021.

[3009] See An Assessment of the U.S. and Chinese Industrial Bases in Quantum Technology by Edward Parker, Rand Corporation, February 2022 (140 pages).

(but not at ETH Zurich), Quantum Botanika, Halcyon Skin Care, Ratzilla Cosme, Quantum Aestetics, Quantum Life and Igesto. In the same vein, Boiron is selling X-ray homeopathy products for curing skin rash. You read it well!

## Quantum management

Quantum management is a new and fashionable practice that seeks to draw inspiration from the general principles of quantum mechanics. Its practitioners are frequently followers of more or less occult sciences who have converted to target corporate markets that are more financially attractive than consumer markets. The vulnerability of educated executives and managers to the most outlandish proposals is always amazing.

However, we can indeed identify many analogies between quantum physics and management in the broadest sense of the term. For this purpose, I have pushed the envelope and reused advanced quantum physics fundamentals and applied it to your usual business life. Any resemblance with a real-life situation would be totally fortuitous or entirely intentional, as you will guess[3010]. As a warning, I must precise that all of this is not serious at all. It is a way to make fun of many things, both the various gurus using quantum physics in scams and, also, life in the enterprise.

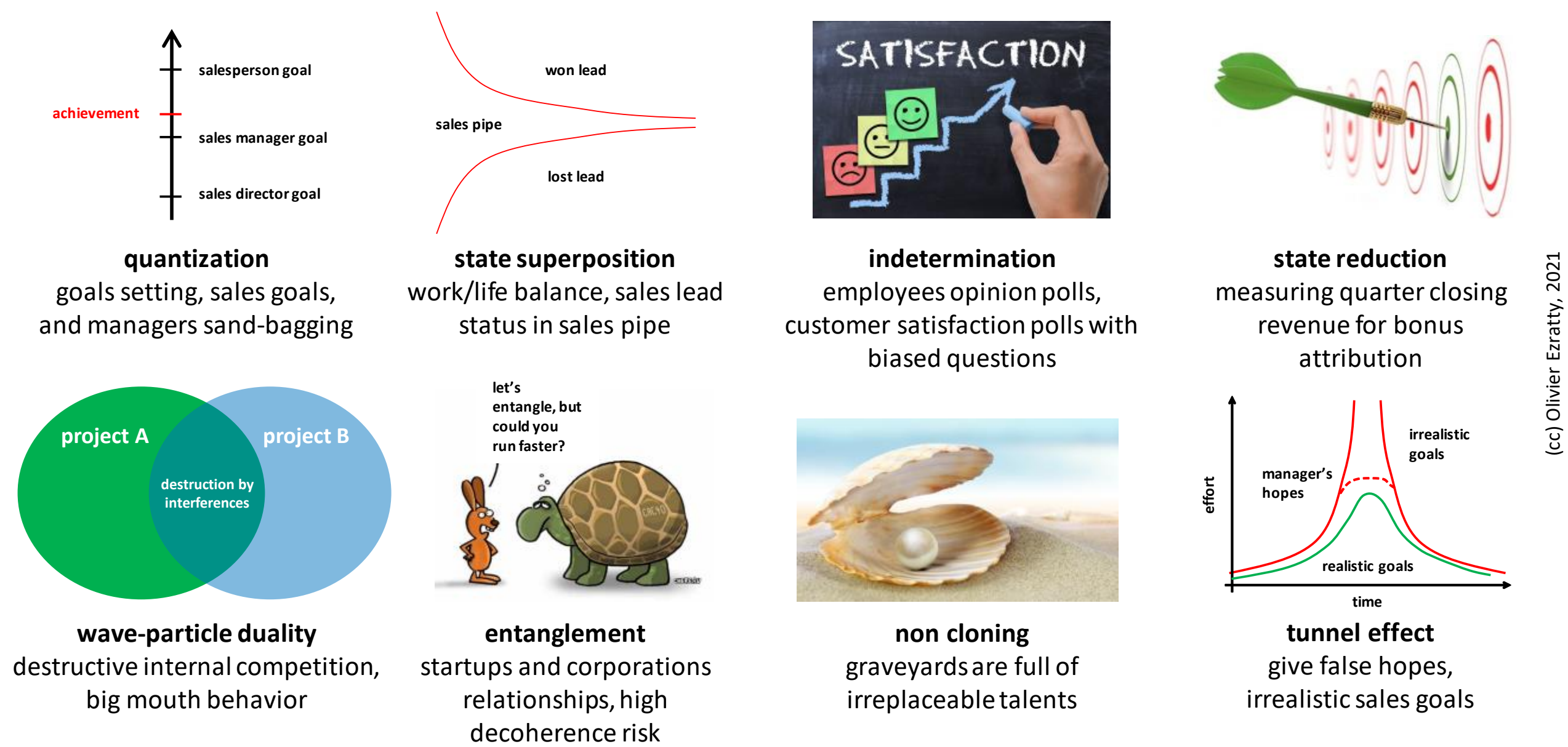

*Figure 916: my useless framework for quantum management. (cc) Olivier Ezratty, 2022.*

**Quantization** means that certain physical values can only be very precise, discontinuous and not arbitrary, like the energy levels of a hydrogen atom. After all, an employee is just a cell in a spreadsheet. He's there one day and gone the next. Workforce management is indeed quantum. A company's workforce at a given time is a discrete integer number.

But if we average it over a period of time, taking into account departures during the period, part-time employees, fixed-term contracts, apprenticeships, subcontractors and people whose real activity we are not sure of, it is no longer an integer but a number of FTEs (full time employees) or FTEs (full time equivalents) that is at least a sum of fractions. Fortunately, it is never a complex number and one escapes Hilbert's spaces to represent them.

[3010] I didn't rely on the proposal in Toward a Quantum Theory of Humor by Liane Gabora and Kirsty Kitto, January 2017 (10 pages) that is quite poor in its scientific and mathematical content.

Quantization also manifests itself with sand bagging, when a sales manager is distributing his own sales goals to his team by adding a quantum margin of safety. The last link in the chain, the unfortunate poor salesperson, will be assigned a goal that is greater than that of all the layers of management above. Only certain layers of sales management have this flexibility. The end result is that salespersons become Rydberg state atoms: they are excited with a very high level of energy and they sometimes burn out. This system is designed so that the base salesperson does not reach his or her objective and is penalized on the bonus side, unlike managers above him or her. Particularly if you want to fire him or her. Judgment about individuals is also subject to quantization. A person is often smart or nice, or a total moron. Personal judgments are rarely nuanced in grey. Yet, in a purist application of quantum physics, this kind of judgment should be a more vague and subtle wave function, until you measure it during a stressful experience.

The top of quantization? Those nasty Internet popups where the given choice is "OK" or "Later". For the quantum measurement guru, it's a real-life example of an exaggerated POVM (see Glossary).

**Superposition** is very common in business. For example, thanks to smartphones and other laptops, employees are kept both at work and in their personal lives all day long. It can also manifest itself in regulatory compliance, which is variable geometry in many companies. And then, of course, in the application of the company values defined in Powerpoint slides and rehashed by managers or the HR department. States superposition also manifests itself in the evaluation of leads that are closed or not in a sales pipeline. They are usually assigned a closing rate which is an amplitude and phase $|\psi\rangle$ until it is known whether the deal is lost or won, which is like the wave-packet collapse happening with quantum measurement, on a basis state $|0\rangle$ or $|1\rangle$. This collapse also occurs if an external event creates a lead quantum state decoherence. For example, a competitor who wins the deal under the nose of the salesman. This quantum analogy, however, will not help you improve your sales pipe closing rate.

**Indeterminacy** works with the measurement of employee satisfaction, where the measuring tool always influences the quantity to be measured. This is true as well in the questions asked in opinion polls, which are often oriented. More generally, the measurement of any parameter in a company by a consulting firm like McKinsey, particularly during an audit, will probably lead to changes in the measured quantities (e.g., downsizing, management change, reorganization and the likes). You just hope that your enterprise won't become a planar wave afterwards.

One variation of Heisenberg's principle of indeterminacy is that one cannot accurately measure both the position and velocity of a particle. The analogy in business would be the observation of a growing startup: by the time one understands where it is at a given moment, it has already changed its situation (headquarters, staff, CEO, turnover, M&A, company name, product, done a pivot). This is why it takes an infinite amount of energy to create an up-to-date startup base in one country or worldwide, even with only quantum technologies startups. So, thank you Crunchbase for the effort!

**Measurement** is in line with the history of quantization when measuring revenue at the end of a fiscal quarter. In this case, one is obliged to provide numbers and not to rely on some closing rates fuzzy logic. If only to determine the bonuses of sales representatives. Otherwise, Bill Gates said loud and clear in 1997 that "*bad news should travel fast in efficient companies*". But not too fast my dear, otherwise you'll get fired. That's what is called a non self-destructive measurement.

**Wave-particle duality** manifests itself with real people in companies who work on competing projects and happily annihilate each other. It is the phenomenon of interference linked to the waveform aspect of each and every projects! You also have the loudmouthed managers facing their teams (thus, in the state of a solid particles) who turn into wipes in front of their own management (thus, in the state of very low-energy waves).

This behavioral duality is also often observed with irascible managers who become docile sheep once at home, or who fail to properly educate their children. Can a trendy startup Chief Happiness Officer be quantum? In any case, this person must fight on a daily basis against a universal phenomenon: a good number of passions quickly fade with time, such as the amplitude of a Rabi oscillation, which is commonly observed in quantum physics and is related to superconducting qubit decoherence (Figure 917).

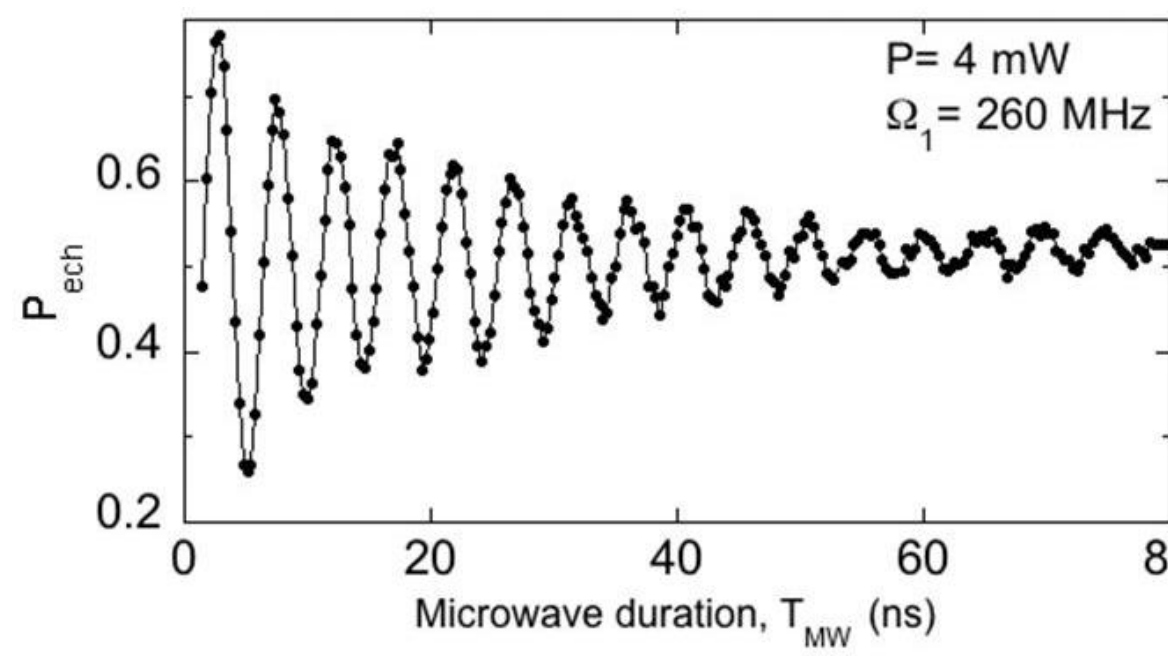

*Figure 917: the Rabi oscillation of your motivation over time.*

The Doppler effect also allows to indirectly put an end to a messed-up project with light, for example, via a well-managed leak in the media. Remember Theranos?

**Entanglement** applies to startups that are integrated into the open innovation programs of large companies. Everything goes well until the appearance of decoherence between the startup and the corporation! I create a product and you want a customized solution, I need speed and you're too slow, etc!

Entanglement also occurs with the teleportation of rumors faster than light. It is also known that the coherence time of qubits is linked to their good physical, magnetic and vacuum insulation, often at very low temperature, in order to avoid any external disturbance. This is the opposite of the open spaces in companies where employees are crammed together! It has long been claimed that open spaces improve teamwork, whereas their main purpose is to compress real estate costs!

**No-Cloning Theorem** says that it is impossible to identically clone the state of a qubit or quantum, has an application in business life with all those people who are believed to be irreplaceable until the day they leave or die. The theorem applies in particular when the departing manager is not replaced and whose role is then distributed among several existing managers, a bit like a quantum error correction with ancilla qubits and projective measurements. The theorem also works with successful entrepreneurs who find it difficult to replicate a success from one area to another.

**Tunnel effect** makes it possible to implement change management. It consists in presenting a wonderful future situation and making people forget the difficulties to get there. The principle could also be adopted by the Gartner Group with its famous innovation adoption cycle curves ("hype cycle"), as some technologies do not necessarily pass through the valley of death, as was the case for smartphones. It had benefited from the reality distortion field of a certain Steve Jobs, a great adept of quantum management principles. By the way, the trajectory Apple-Next-Apple was a great application of the tunnel effect, Next being a relative failure while both Apple were successes.

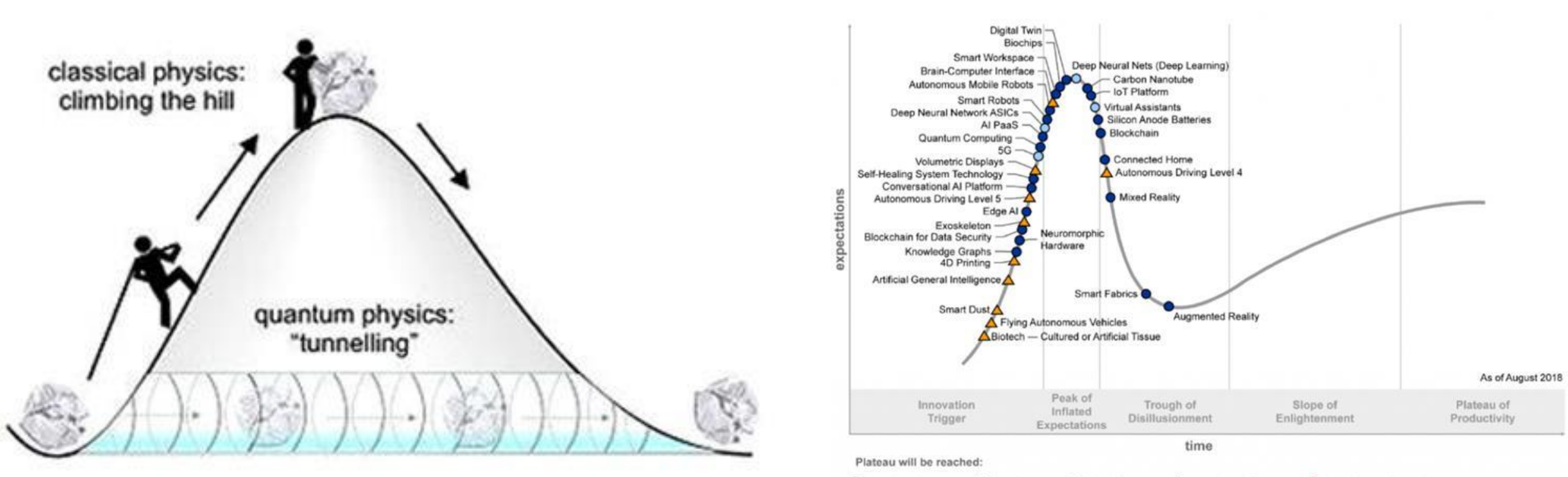

*Figure 918: the quantum tunnel effect and hype effects. Sometimes, hype is so strong that it creates a pass-through from hype to success without a valley of death.*

**Superconductivity** is linked to meeting rooms. Employees and managers are conditioned to be even-spin bozos, who can be assembled in meeting rooms or covid-zooms. Organizational superconductivity also avoids resistance to change. You freeze employees and their resistance to change disappears. Which is a bit paradoxical because once frozen, you are as solid as a rock, and defrosting is not obvious. If we take the principles of Deepak Chopra's quantum pseudo-medicine, a company is in a superposed state between a healthy leader and a declining star. The strength of leadership should theoretically allow the wave packet function of the company's quantum state to collapse in the healthy leader state. In real life, this collapse is tricky to achieve and companies simply collapse. The processes that lead the company to find itself in a declining situation are most often irreversible and linked to a slow decoherence with the environment, competitors and customers who have not waited to adapt. Corporate life is not a reversible quantum gate nor any sort of linear algebra. It's mostly nonlinear. Try, for example, to turn Nokia into the leader of Android smartphones!

**Universal Gates Quantum Computing** has a beautiful analogy in the life of companies with the management of calls for tenders such as those for communication agencies. The responses of candidate agencies are superposed states of a quantum register.

They undergo a simultaneous evaluation process, as in an oracle-based quantum algorithm. In the end, only one offer emerges: the winner. But during this process, there may be some quantum entanglement affecting the winner final proposal. Translation: the elements of certain answers will magically appear in the winner's answer. Again, perhaps via the enterprise quantum tunnel effect.

Finally, let us mention this other universal principle, the very famous **quantum teleportation** of human stupidity to large sections of the company or in the population. It uses superdense falsehoods encoding. And contrarily to actual teleportation, it travels faster than light. It is so fast that it is the only plausible explanation.

All of these analogies are amusing but not very useful for improving management. Even if its scientific dimension is more than questionable, parody is finally an interesting form of pedagogy!

## Other exaggerations

There are many startups or ventures surfing on the quantum technology wave with various intentions. Some are just quantum startups with fluffy claims and others have only quantum in their name but nothing else.

One classical exaggeration comes with making cross-predictions between one digital trend and quantum computing. This one predicting that we'll have **quantum digital twins** is based on the usual wrong premise that quantum computing is made for compiling huge swaths of data[3011].

**Dark Star Quantum Lab** (2020, USA) introduces itself as a contract defense and space research company covering applied quantum physics and quantum information science (QIS). They develop tons of quantum stuff: quantum software and quantum emulation solutions, unspecified hardware, a 'Sentinel' mobile phone including a QRNG.

They also claim to have developed a Qloud high-frequency trading, a Qoin (quantum-secured cryptocurrency), a BloQchain (quantum-secured blockchain working with Qoin), a use-case of Nash embedding to create error-free qubits and, also, some Star Trek Tricorder fancy stuff.

[3011] See Why we need Quantum Digital Twins by Ian Gordon, Head of Data at Houses of Parliament Restoration & Renewal, January 2021.

This laundry list of things is not credible. And they don't seem to have any real defense customer. Looks like it's not really serious[3012].

Also quite weird is this **Quantum_AI Group Of Companies** with its 15 branches dealing with aerospace, artificial intelligence, naval, finance, energy, automotive, electronics and... quantum computing. They develop, take it or leave it: Nano-Flux, a range of flux-qubits superconducting computers, Q-optic, the most powerful optical qubits quantum computer, BEC, the fastest Bose-Einstein condensate based trapped ions computer (seems they mixed some things here) and also SSL, a solid-state quantum computer and Infinity-Q a high-performance heavy load quantum computer. Interestingly, these 4 ranges of systems have respectively 40, 200, a 1.6 and 128 billion qubits and they look the same in their 3D rendered pictures.

They are supposed to be based in Stanford, Boston, India, Abu Dhabi, Dublin and Tel Aviv. They still have a CEO, a certain Ranobijoy Bhattacharya. If it's not an April's fool, what is it? Some new form of mythomania?

Quantum physics abuse can be found in various other product categories. In China, for example, a so-called **quantum satellite camera** was used to produce high-resolution panoramas. The view presented was that of Shanghai with 195 billion pixels. Practically, the pictures were captured from the top of a skyscraper - there is no shortage of them in Shanghai - and not by any satellite. It used conventional high-resolution cameras that have nothing more quantum than the very classic photoelectric effect used in CMOS sensors to transform photons into electric current. The information is totally bogus and was only used to generate buzz.

Unfortunately, many media outlets around the world have taken the bait without any doubt [3013].

For its part, a French SME **What-Innove** from the East of France, specialized in renewable energies, claims it is creating an engine that captures energy from vacuum. How does it work? An unlikely mix combining a quantum field generator, the creation of photons from vacuum energy exploiting the Casimir effect, the combination of magnetodynamics and space-time, ambient temperature and pressure superconductors (which would win them a Nobel Prize if it worked), and negative entropy. They just need €2.7M of funding to move ahead!

You are also entitled to a beautiful **quantum cooler** from **Chillout Systems** that has only quantum in its name. It uses a compact classic compressor[3014].

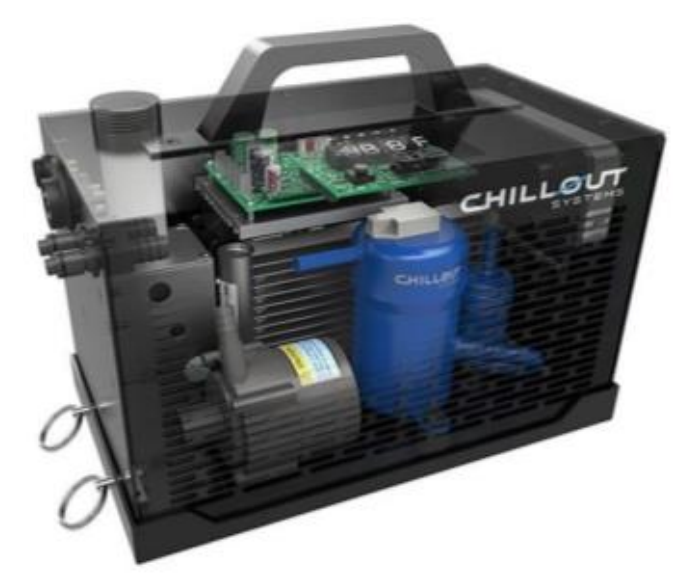

*Figure 919: the non-quantum cooler from Chillout.*

Other cases extrapolate to the macro scale of quantum phenomena observed at a nano scale. This is the case of **time inversion** with quantum computing, a view of the mind that is linked to the reversible nature of quantum gates but does not mean that one can go back in time scale in macroscopic practice[3015].

---

3012 See one scientific publication of their own, which is quite short: How many physical qubits are needed exactly for fault-tolerant quantum computing? by Faisal Shah Khan, Dark Star Quantum Lab, December 2021 (4 pages).

3013 See 60 seconds over sinoland: quantum satellite camera used to do movable, panoramic photos of Shanghai, December 2018 (video) and Truth Behind Viral 24.9 Billion Pixel Image Taken By Chinese "Quantum Satellite" by Anmol Sachdeva, December 2018 and the Bigpixel website to view the view.

3014 See Chillout Systems Quantum Cooler. It is sold for $2199.

3015 See Arrow of time and its reversal on the IBM quantum computer by G. B. Lesovik et al, 2018 (14 pages) and Does the IBM quantum computer violate the second principle of thermodynamics?, 2019.

We also have equally wild theories willing to **predict the future** with quantum computing. If it is true that quantum computing allows us to evaluate all the solutions of a complex problem, it is reduced to simple problems in view of the complexity of macroscopic life, even if it could be deterministic[3016].

The next step is to consider that we are actually living in a **simulation**.

This is the theory presented in Rizwan Virh's The Simulation Hypothesis. The author presents himself as an MIT Computer Scientist, whereas he is more of an entrepreneur in video games, more accustomed to books on entrepreneurship than on science. This kind of simulation scenario is roughly equivalent to believing in a kind of omnipotent God who controls everything or who created the simulation tool. The question can moreover be recursively implemented: if this creator has developed a simulation tool, who created his universe and isn't this one also a simulation?

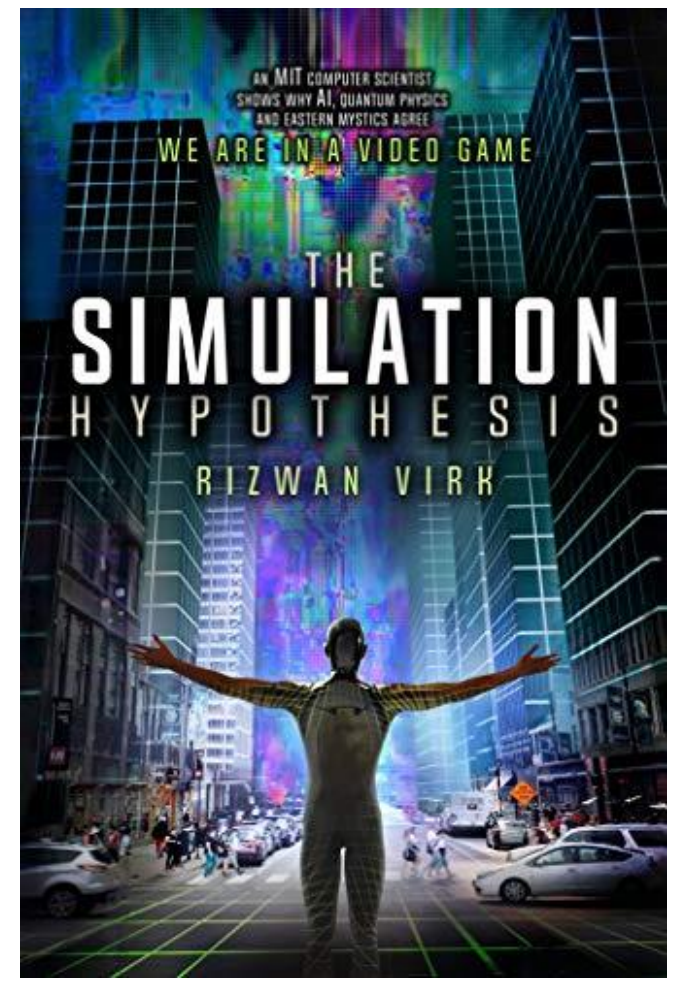

Another case that should inspire the utmost caution is that of this curious company **Precog Technologies**, which claims to offer solutions for teleportation, time travel and anti-gravity systems. Miracles one stop shopping!

The company was created to valorize the intellectual property of a certain Anisse Zerouta that is covered in a dubious scientific paper[3017].

Another guy, from **Quanta QB** (South Africa) thinks he has also found a qubit architecture that showcases a miraculous 0% error rate[3018]. Good luck with that!

We also saw the first quantum financial scam appear in 2018 with this fake article in the Guardian reporting a quantum computer project for Elon Musk's finance[3019]. The trained eye quickly detects that it is a montage, like this series of **QuantumAI quantum** computers that are nothing more than D-Waves annealers whose logo has just been photoshopped.

Second, the article is supposed to come from the Guardian, but not the url! It quotes a number of scientists from research laboratories around the world, all with Russian names. The article points to **QuantumAI**'s online service which would be able to go around robbing the rich and redistributing the money to the poor. And the site indicates that the startup has Jeff Bezos and Bill Gates as advisors and IBM, Microsoft and OpenAI as partners.

There is another, called **Quantum Code**. Obviously, run away! It is in fact a scam designed to rob users of their savings, but in an indirect way. The site offers to create an account by providing its coordinates. These are then resold to unscrupulous companies of shady financial products that exploit leads of prospects easy to fool.

[3016] See Interfering trajectories in experimental quantum-enhanced stochastic simulation by Farzad Ghafari et al, 2019 (7 pages).

[3017] See The Seed Theory: Unifying and replacing quantum physics and general relativity with "state physics" by Anisse Zerouta, 2017 (28 pages) which develops a theory of parallel worlds that does not seem to meet the criteria of a scientific publication worthy of the name. Anisse Zerouta is a company manager in Paris born in 1973, first with Elysee Communication (2008-2011) then with Avenir Optique, an optician (2011-*), companies with only one employee, its founder (source). Created in September 2018, Precogtec would have a CTO, a certain François Bissege, who has a PhD in sociology (source) and another employee, Julien Darivel, who has a DUT and worked at PSA. It's bizarre!

[3018] See I made the Quantum Breakthrough, June 2019.

[3019] See Elon Musk to Step Back From Tesla And SpaceX, Jumps on Quantum Computing Financial Tech (not dated).

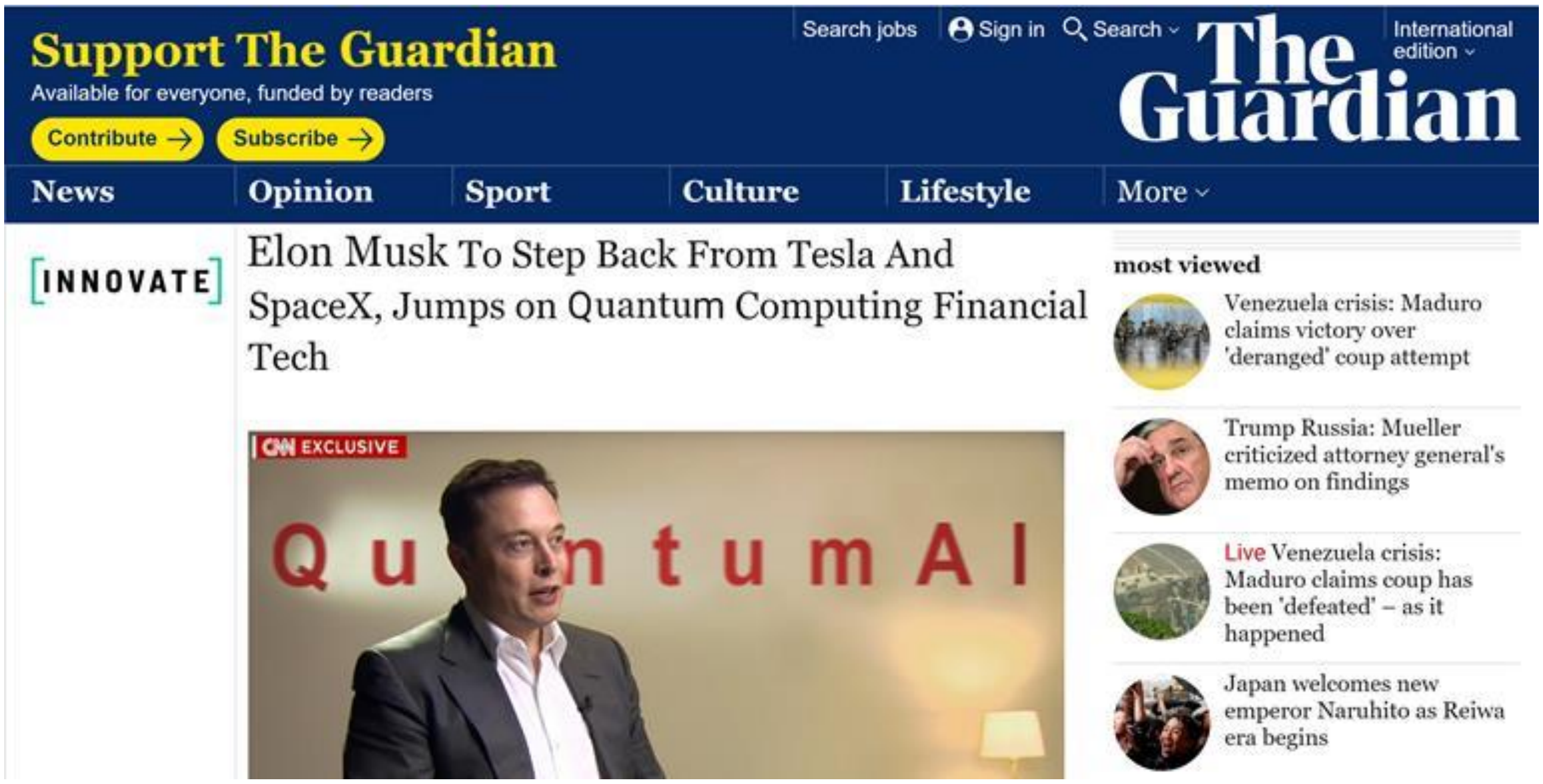

*Figure 920: do you spot the scam?*

Other financial scams promising skyrocketing return on investment on mysterious quantum investments in the stock market are now current[3020].

We can also quote **Qubole** which launched its Quantum SQL server, which has nothing quantum[3021]. The **Samsung** Quantum 8K processor launched in 2018 was not particularly quantum either, except via its classical CMOS transistors.

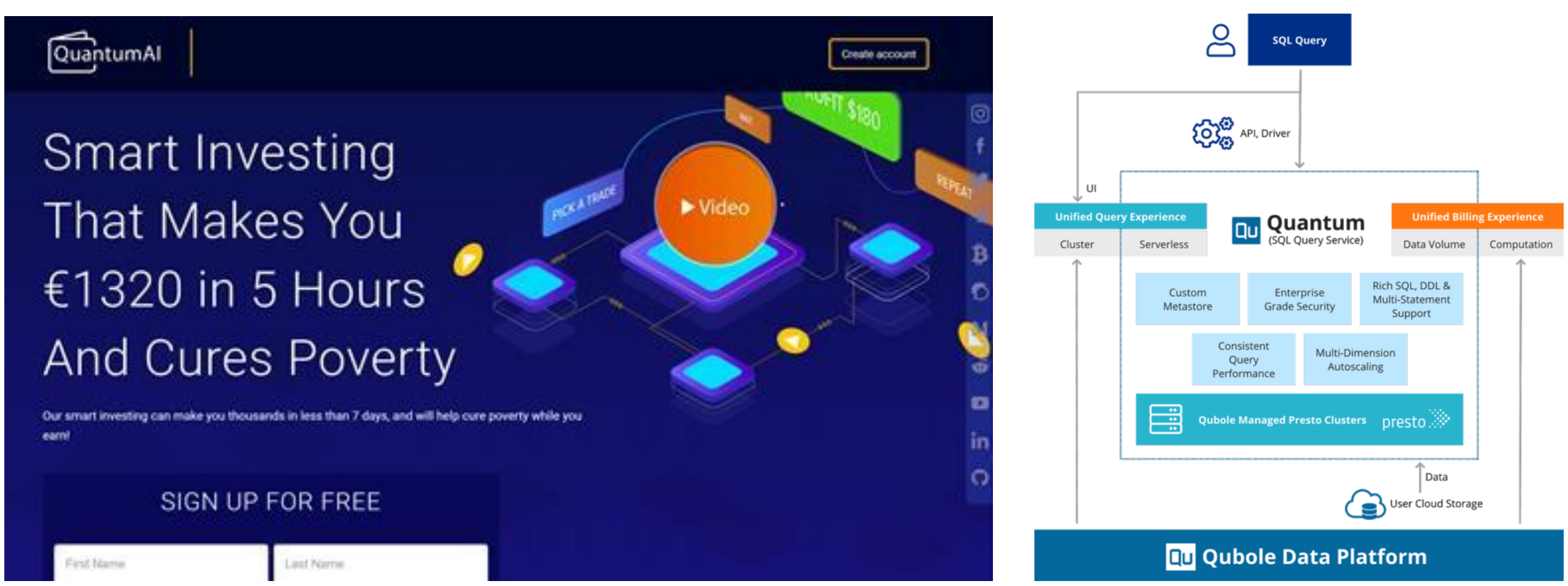

*Figure 921: QuantumAI and its financial scam.*

In consumer products goods, we have this washing powder **Quantum Max** of the brand **Finish** from the Reckitt Benckiser group. And also **Quantum American** PQ rolls and Quantum red wine from **Beringer,** a brand from Napa Valley, California.

[3020] That one is a concentrated feat of bullshit: https://secure2.wealthdaily.com/o/web/362894 using the one line sentences tactic to strengthen its messaging. Each and every line would deserve some debunking.

[3021] See Qubole launches Quantum, its serverless database engine by Frederic Lardinois, June 2019.

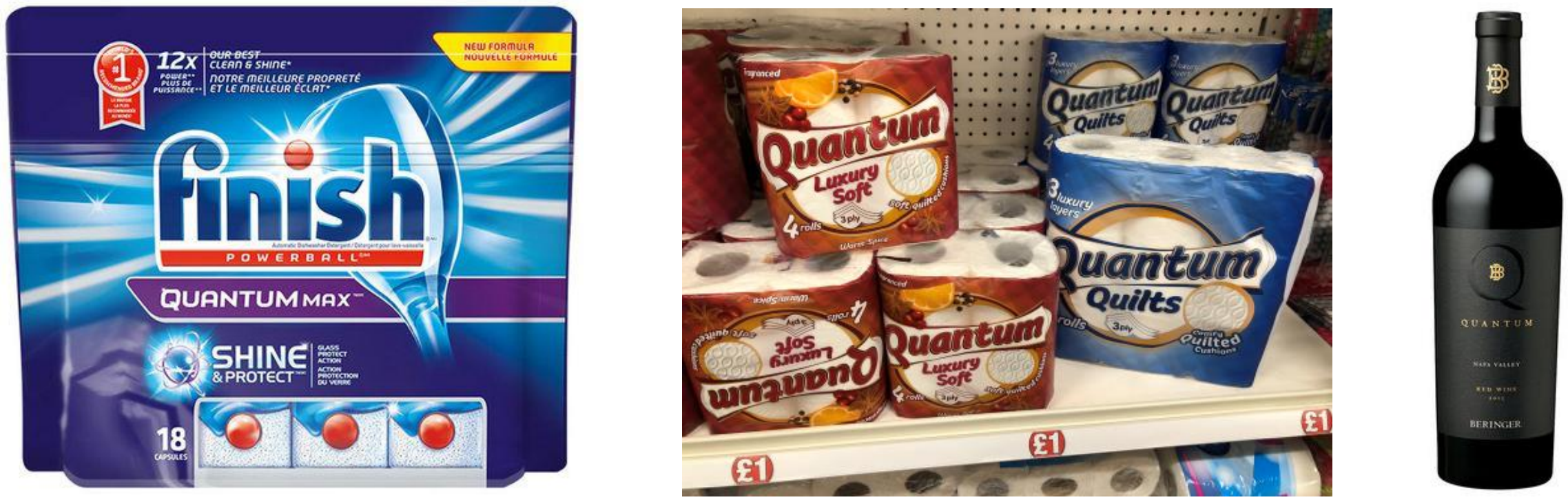

*Figure 922: anything can be quantum, your washing machine powder, your toilet paper and your wine or beer.*

Otherwise, **Quantum Corp** (USA) does nothing quantum and just manages tape storage. The same goes for **Quantum Entanglement Entertainment** (Canada) which, as its name indicates, is in the content market. **Quantum Surgical** (France) makes surgery robots for liver cancer, which have nothing quantum. **QuantumScape** is a solid-state lithium battery manufacturer in the USA. **QuantumSi** created a silicon-based DNA advanced sequencing machine that doesn't seem to use any second-generation quantum technology, **Quantic** Executive MBA has only quantum in name. **Quantum Metric** is providing cloud based digital content design software tools. There is even a Chinese company created in 2016 named **Quantum Technologies** (officially "Guandong Technology") that sells some augmented reality product with nothing quantum at all. **Quantum Switch** is not a quantum switch from the physical standpoint. It states that it creates "*A New Era In Quantum Defined Data Centres*" but does just service classical colocation data centers.

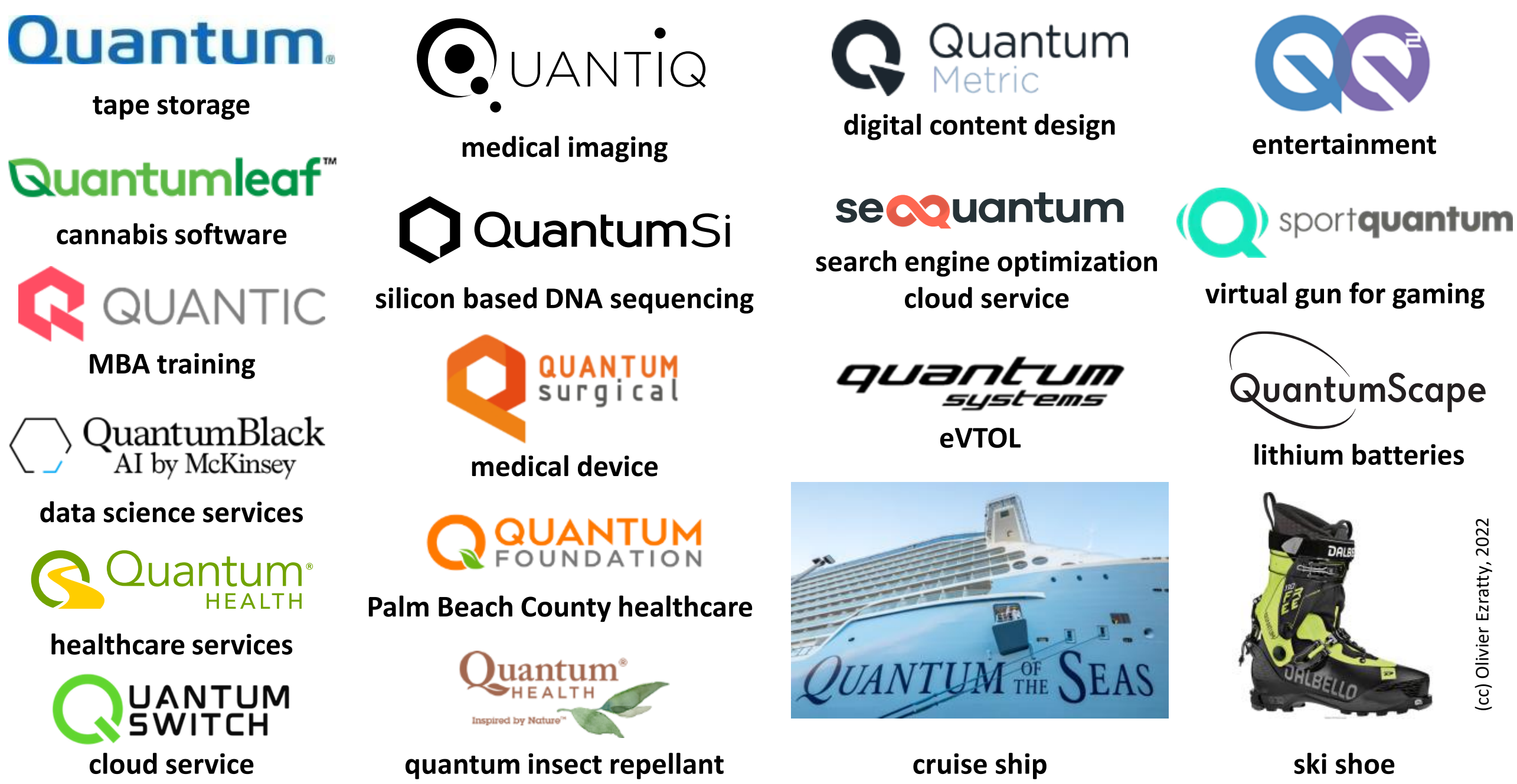

*Figure 923: all those companies chose to use quantum in their name, but they have nothing quantum.*

At last, **QuantumLeaf** is a cannabis software company servicing the cannabis industry in the USA. **Quantum of the Seas** is a famous cruise ship covering all oceans. Even **McKinsey** got into this game, labelling its data-science consulting and service offering QuantumBlack! And don't be confused between **Quantum Health** which provides healthcare services to company's employees and **Quantum Health** which sells among other things an insect repellant. Hopefully, despite the company name, they don't claim the repellant use a specific quantum effect. It's just branding washing, not more. We're safe!

## Quantum fake sciences key takeaways

- Quantum physics has been for a while integrated in highly dubious offerings, particularly in the healthcare and energy domains.
- There is a proliferation of gurus and scams-based miracle cures machines for detecting electromagnetic waves or vague energies, and restoring your body balance. It is at best a subset of the lucrative placebo effect industry targeting the gullible!
- The shift from some low-level physics studies on water and matter led some scientists to explain consciousness with quantum physics. This form of reductionism is unproven. It's the same with scalar-waves detectors or generators, miraculous healing crystals, structured water and other quantum medallions.
- This part proposes a simple methodology to detect these healthcare related scams, with using some common sense.
- We uncover some other scams in the free energy generation category. These systems are supposed to extract some energy from vacuum when their only actual effect is to pump money out of your wallet.
- Quantum physics is sometimes used in management and marketing. This book offers you a nice in-depth parody of these methodologies.
- At last, we showcase a few companies using quantum in their branding when they have nothing quantum at all to offer.

# *Conclusion*

Quantum technologies perfectly symbolize the world of innovation and extreme entrepreneurship: it is full of uncertainties, risks and failures. There is "test & learn", the crossroads of many sciences, the need to invest well in advance of economic success, with a critical role for government investments, the only ones able to invest in the long term, more than 10 years ahead. Numerous parallel paths of exploration have been launched by researchers and entrepreneurs. Only a few will succeed. A new industry is emerging from all of this.

After reading part or the totality of this book, you may find out that its premise was misleading. “Understanding quantum technologies” is indeed a quest and a journey, but you never reach the destination. There’s always something you didn’t understand well and need to review again and again. For a starter, you may spend some time on the Bloch sphere trying to grasp what it represents and its various angles, then try to understand either how quantum algorithms work or how circuit quantum electrodynamics operate in a superconducting qubit, or, say, how quantum photonics really work up to the mysteries of MBQC.

To write this book over the course of the last four years, I downloaded and compiled more than 4500 documents freely available on the Internet, viewed dozens of hours of conferences and courses on YouTube, and met dozens of researchers and entrepreneurs.

Like in a thesis, I have here to thank many people and friends who helped me craft this book over my 5-year journey in quantum science and technologies. First of all, **Fanny Bouton**, with whom I started this quantum adventure in 2018 and who now runs quantum operations at OVHcloud.

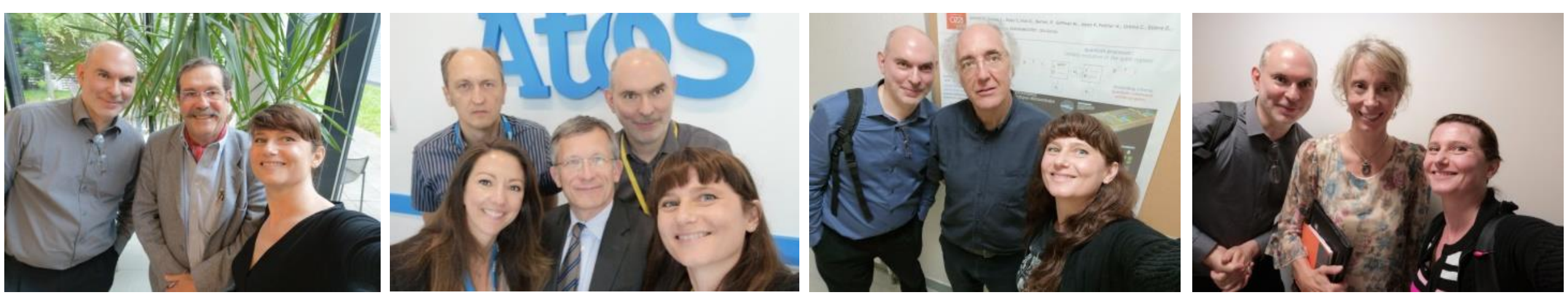

*Figure 924: the first quantum scientists we met in 2018, Alain Aspect, Cyril Allouche, Philippe Duluc, Daniel Esteve and Maud Vinet.*

In chronological order, here are the first scientists and other people we met back in 2018: **Alain Aspect** (IOGS), **Daniel Esteve** (CEA Quantronics), **Christian Gamrat** (CEA LIST), **Maud Vinet** (CEA Leti in Grenoble), **Tristan Meunier** (CNRS Grenoble), **Alexei Tchelnokov** (CEA Grenoble), **Laurent Fulbert** (CEA-Leti Grenoble), **Cyril Allouche** and **Philippe Duluc** (Atos), **Bernard Ourghanlian** and **David Rousset** (Microsoft), **Pat Gumann** (IBM), **Etienne Klein** (CEA), **Christophe Jurczak** and **Zoé Amblard** (Quantonation), **Nicolas Gaude** (Prevision.io) and **Françoise Gruson** (Société Générale).

Then, in 2019, with **Philippe Grangier** (IOGS), **Elham Kashefi** (LIP6 and VeriQloud), **Marc Kaplan** (VeriQloud), **Pascale Senellart** (C2N and Quandela), **Franck Balestro** (UGA, Institut Néel) and **Alexia Auffèves** (CNRS, then at Institut Néel in Grenoble and now MajuLab in Singapore, we are now teaming up as cofounders of the Quantum Energy Initiative), **Matthieu Desjardins** (C12), **Jacqueline Bloch** (C2N), **Iordanis Kerenidis** (CNRS), **Heike Riel** (IBM Zurich) and **Vern Brownell** (then D-Wave CEO).

In 2020, **Artur Ekert** (CQT Singapore), **Patrice Bertet (**CEA SPEC), **Xavier Waintal (**CEA IRIG), **Yvain Thonnart** (CEA LIST), **Rob Whitney** (LPMMC Grenoble), **Damian Markham** (CNRS LIP6 and JFLI in Tokyo), **Robert Whitney** (CNRS LPMMC, also part of the QEI), **Bruno Desruelle (**Muquans), **Georges-Olivier Raymond** and **Antoine Browaeys (**Pasqal), **Théau Peronnin** and **Raphaël Lescanne (**Alice&Bob) as well as **Jeremy O'Brien (**PsiQuantum), **Magdalena Hauser** and **Wolfgang Lechner (**ParityQC), **Roger McKinley** and **Peter Knight** (UK) and the IBM Zurich research teams. I also had discussions with the teams from **Qblox**, **Qilimanjaro**, **Quantum Motion**, **Strangeworks** and **IQM**.

There were these countless discussions with **Jean-Christophe Gougeon** of Bpifrance, **Neil Abroug**, who is since 2021 the coordinator of the quantum strategy in France, as well as **Charles Beigbeder** and **Christophe Jurczak** from Quantonation and Le Lab Quantique, who wrote the foreword of this book, page vii. I should also mention the numerous exchanges related to quantum investments with **Cédric O** and his team, in the French government. He was onboard early on and became its driving force within the government.

The fourth edition in 2021 benefited from the contributions of **Alexia Auffèves** (measurement, energetics of quantum computing, quantum foundations, photon qubits), **Antoine Browaeys** (IOGS and Pasqal, cold atoms), **Christophe Chareton** (CEA LIST, linear algebra, quantum algorithms and development tools), **Cyril Allouche** (Atos, supercomputing, emulators, European projects), **Daniel Esteve** (CEA DRF, superconducting qubits), **Eleni Diamanti** (CNRS LIP6**,** quantum telecommunications and cryptography), **Elvira Shishenina** (BMW, proof-read all the document), **Frédéric Nguyen Van Dau** (Thales, quantum sensing), **Georges Uzbelger** (IBM, quantum algorithms and software tools), **Jonas Landman** (CNRS IRIF, quantum algorithms), **Léa Bresque** (CNRS Institut Néel, quantum physics 101, quantum postulates and measurement), **Marc Kaplan** (Veriqloud, quantum telecommunications and cryptography), **Michel Kurek** (who patiently proof-read several times all the book and checked all hyperlinks), **Peter Eid** (Arm, classical and unconventional computing, telecommunications/cryptography), **Philippe Grangier** (Institut d'Optique, quantum foundations), **Pol Forn-Díaz** (Qilimanjaro, superconducting qubits), **Théau Peronnin** and **Jérémie Guillaud** (Alice&Bob, cat- qubits) and **Valérian Giesz** (Quandela, photon qubits and photonics).

In 2022, I had a chance to discuss about the quantum ecosystem with **Rainer Blatt** (AQT and Munich ecosystem), **Jonathan Home** (ETZ Zurich), **Tommaso Calarco** (Jülich) and **Jay Gambetta** (IBM). I also met with countless other quantum physicists and entrepreneurs.

Reviewers for this fifth and 2022 edition were **Jean-Philippe Fauvarque** (Plassys-Bestek, for the fabs part), **Antoine Gras** (Alice&Bob, also for the fabs part), **Frédéric Wyczisk** (for the quantum matter part), **Michel Kurek** (the proof-reading master who does an incredible work spotting all the details), **Antoine Browaeys** (Pasqal/IOGQ, neutral atoms computing), **Bruno Desruelle** (ixBlue, quantum sensing), **Christophe Jurczak** (updated foreword), **Marco Fellous-Asiani** (energetics, control electronics, superconducting qubits), **Clément Barraud** (MPQ, quantum matter), **Georges-Olivier Raymond** and **Nicolas Proust** (Pasqal, cold atoms qubits), **Léa Bresque** (Institut Néel, Quantum technologies energetics), **Luc Gaffet** (Air Liquide, cryogeny), **Olivier Hess** (Atos, software), **Jérémie Guillaud**, **Blaise Vignon** (Alice&Bob, superconducting qubits), **Thomas Ayral** (Atos, various pats), **Loïc Chauvet** (CACIB, software tools), **Daniel Vert** (CEA, quantum annealing), **Xavier Vasques** and **Jean-Michel Torres** (IBM, on IBM quantum software stacks), **Stéphane Louise** and **Christophe Chareton** (CEA LIST, on algorithms), **Marc Kaplan** (VeriQloud, quantum telecommunication and cryptography), **Maud Vinet** (history, silicon qubits) and **André M. Konig** (Global Quantum Intelligence, various places).

And maybe you, next time :)!

Cheers,

Olivier Ezratty, September-October 2022

# Bibliography

Here are a few books and other sources of information on quantum technologies that I consulted or discovered to prepare and update this book.

## Events

There are numerous conferences on the different scientific branches of quantum technologies and a growing number of quantum “business” conferences associating some scientific content, industry vendors talks (and sponsoring) and customer use cases testimonials.

I found online various inventories of quantum related scientific events on Waset.org. quantum.info (which also inventories some quantum physics predatory journals), Conference service, Conference Index and Quantum Computing Report.

Many of these events are fee based for both participants and speakers. It costs up to $1,000 to participate as an attendee, plus extra-fees for speakers and poster sessions. It’s a business!

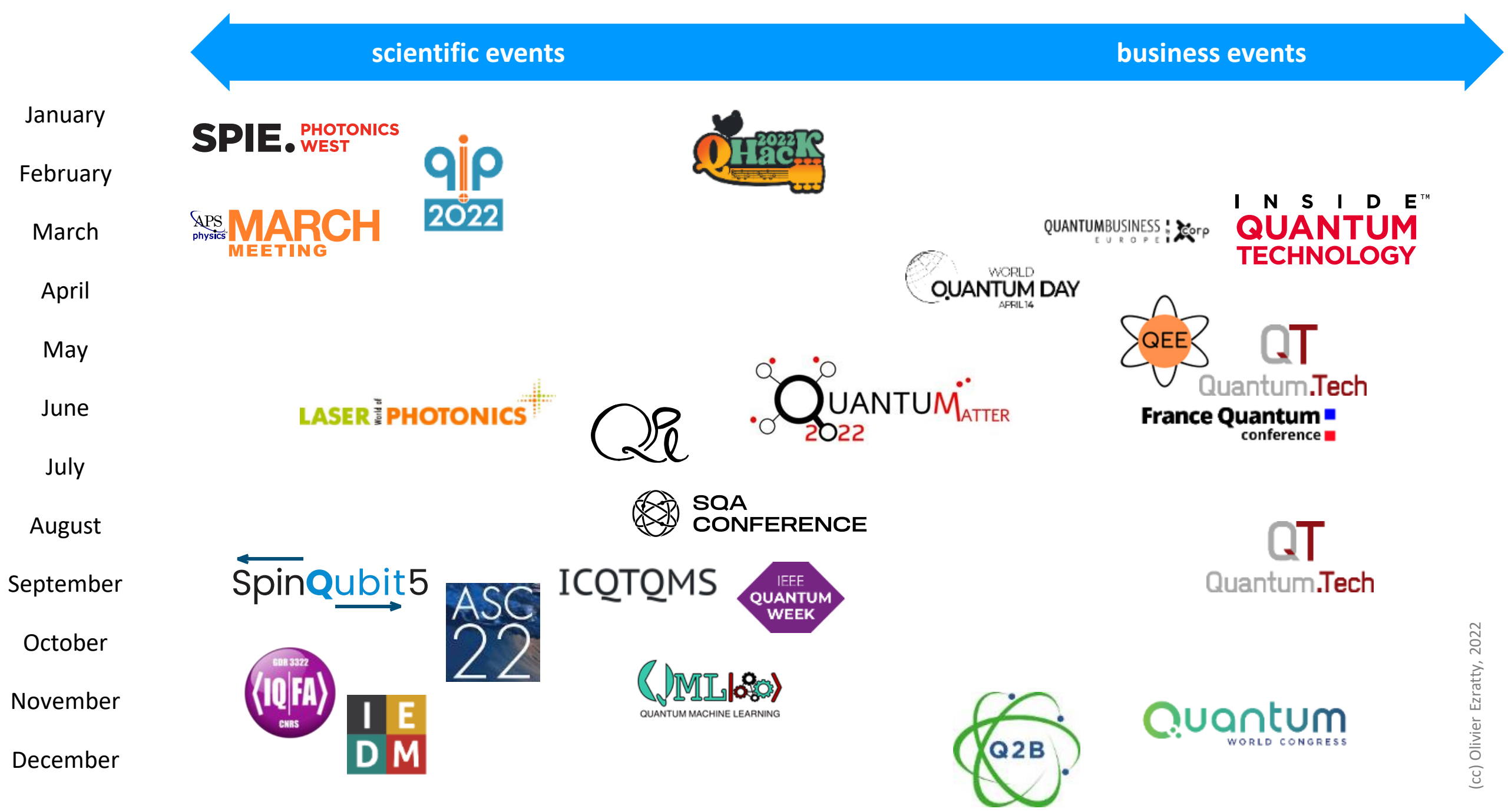

*Figure 925: a yearly timeline of some notable quantum events, from science to business. (cc) Olivier Ezratty, 2022.*

### Quantum Scientific events

Let’s quickly cover the main quantum related events where the audience is mainly made of scientists and the content is likewise highly scientific[3022].

**APS March Meeting** is the largest physicists conference in the world with over 12K attendees, hundreds of sessions, thousands of presented papers, and a significant part of them are in the quantum physics disciplines, including enabling technologies like cryogeny. Other talks cover different parts of physics likes high-energy particles physics or astronomy. They have over 100 exhibitors. Their industry tracks showcase scientific advancements from the main quantum computing players like

[3022] See Q-Turn: Changing Paradigms In Quantum Science by Ana Belén Sainz, February 2022 (9 pages) about how to organize scientific quantum events.

IBM, Google, IonQ and others. The 2022 edition was in Chicago and will be in Las Vegas in 2023. This four-day event is organized by the American Physical Society which also publish the reference journals PRX Quantum, PRX, PRL (Physical Review Letters) and PRA (Physical Review A), that are frequently mentioned in this book bibliographical references.

**Photonics West** in January/February is the largest photonics related conference and vendors exhibition, including quantum photonics, their related enabling technologies (lasers, photon counters, …) and also covering medtechs, organized at the Moscone Center in San Francisco. It also hosts industry related events.

**Laser** World of **Photonics** and the **World of Photonics Congress** is another photonic major congress, happening in Germany. The 2023 edition takes place in June in Munich with over 30K visitors and 6000 congress participants.

**QIP** (Quantum Information Processing) is an important quantum information conference with prestigious scientific speakers. The 2022 edition was organized in Caltech, California (attendees picture below) and the 2023 edition is planned in Ghent, Belgium in February.

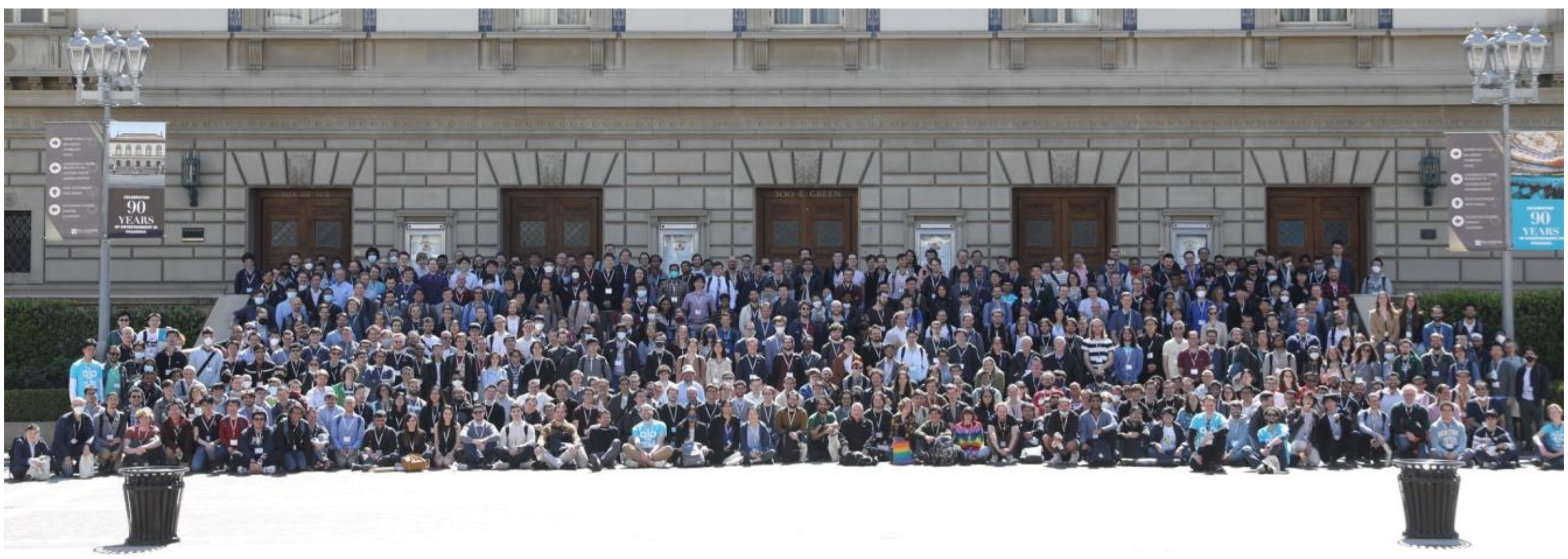

*Figure 926: QIP2022 group photo at Caltech.*

**QUANTUMatter** gathers various communities in quantum information and quantum matter involved in all branches of quantum technologies (computing, sensing, telecommunications). Its second edition was organized in Barcelona in June 2022.

**QPL** (Quantum Physics and Logic) is about the mathematical foundations of quantum computation, quantum physics and related areas with a focus on the use of mathematics, formal languages, semantic methods and other mathematical and computer scientific techniques to the study of physical systems and processes. The 2022 edition was organized in Oxford, in June/July.

**IEEE** organizes many quantum related scientific conferences including QSW (International Conference on Quantum Software), in Barcelona, Spain in July 2022, the IEEE Quantum Week, in September, 2022 in Broomfield, Colorado and IEDM (International Electron Devices Meeting) in September in San Francisco which covers silicon spin qubits and qubit control electronics among other topics.

**ASC** (Applied Superconductivity Conference) is a superconducting related event. The 2022 edition happens in Honolulu, Hawaii in October. It covers quantum systems, computation, sensing and networking, control and readout electronics, fabrication, packaging, and scalable infrastructure, and hybrid or novel quantum systems.

**SQA** (Superconducting Qubits and Algorithms Conference) covers science, technology, and algorithms related to superconducting quantum computing organized by IQM. The 2022 edition takes place, not surprisingly in Helsinki, in August. With prestigious speakers like William D. Oliver (MIT), Andreas Wallraff (ETH Zurich) and Vladimir Manucharyan (Maryland University).

**Spin Qubit 5** is a conference on spin qubits organized in Pontresina, Switzerland in September 2022. It covers NV and SiC centers qubits, quantum dots spin qubits and the likes. The conference chair is none other than Daniel Loss.

**ICDCM** (International Conference on Diamond and Carbon Materials) is the conference were NV center and SiC cavities are talked about. The 2022 edition happened in Lisbon, Portugal, in September.

**ICQTQMS** is the International Conference on Quantum Technologies, Quantum Metrology and Sensors. The 2022 edition is planned in September in Rome, Italy.

**QTML** (Quantum Techniques in Machine Learning) focuses not surprisingly on quantum machine learning, gathering researchers and industry players. The 2022 edition takes place in Naples, Italy in November.

**QRE** is a yearly workshop on Quantum Resource Estimation also dealing with benchmarking and performance analytics. The 4$^{th}$ edition took place in June 2022 in New York.

**QTD2022** (Quantum Thermodynamics Conference) took place virtually on Zoom in June/July 2022.

**FQMD** (Frontiers in Quantum Materials and Devices).

**GDR IFQA** gathers international quantum scientists covering all fields, with tutorial sessions, docs/post-docs sessions and poster sessions. It has been organized by the research group on quantum technologies from CNRS in France every November since 2011. The 2022 edition takes place in Palaiseau in the Paris region in November.

### Quantum Business conferences

And now, onto the quantum business events which usually provide a mix of scientific and business related content.

**Q2B** is a conference organized by QC-Ware since 2017. Initially happening in California each December, it will also happen in Tokyo in July 2022 and later in Europe. It is a good opportunity to learn about the scientific progress from major industry players in quantum computing.

**Inside Quantum Technology** is a series of quantum business conferences organized in several like The Hague (February 2022), San Diego (May 2022), New York (October 2022) and San Jose, California (April 2023).

**Quantum.Tech** is another conference focusing on industry use cases of quantum technologies. The last edition was in Boston in June 2022. A London edition takes place in September 2022.

**Commercialising Quantum** was organized in May 2022 in London by the Economist, London with a mix of in-person and virtual events.

**QHack** is a mix of a fan expo, hackathon and scientific conference for quantum software developers. The third edition was organized online in February 2022.

**Quantum Business Europe** is an international conference launched in 2021 that had a second edition in March 2022 with a mix of vendors, scientists and user talks.

**World Quantum Day** is a virtual global event, consolidating events happening one day in April all over the world, targeting broad audiences.

**Quantum World Congress** gathers experts from the industry with plenary sessions and breakout tracks on market acceleration, science and engineering, government and security, complemented by an exhibition. The 2022 edition takes place in Washington DC in November/ December.

**France Quantum** is a conference that was launched by Startup Inside and OVHcloud in June 2022 at the Eiffel Tower to promote the French quantum ecosystem. The 2023 edition is planned in Paris in June.

**Quantum Eastern Europe** is a 2-day online event gathering quantum stakeholders from Eastern Europe. The 2022 edition happened in May.

**Investment Summit for Quantum Startups** was organized in October 2021 at Maryland University as a gathering of investors and startups.

**Metaverse Quantum Computing Summit** is not a joke. Or still, yes, it's really a gigantic joke. You even have the opportunity to "*Learn best practices, strategies and ideas you can implement today*". This is an extreme case of mixing two trendy B.S. into one compound B.S (source). By the way, none of the speakers talk about the metaverse, but only about crypto-financial stuff.

## Websites and content sources

The Quantum Insider (USA, formerly The Quantum Daily) is a quantum news site. Many of the news are broadcasting press releases from vendors, government and research labs and some others are original posts. It is completed by a job board and professional services, exploiting a proprietary database of quantum industry vendors and stakeholders. The company is run by Alex Chahans out of the USA. In October 2021, The Quantum Insider integrated Entangle, a global quantum technology community. The Entangle team is now focused on a crusade around using quantum tech to solve the climate crisis.

Quantum Zeitgeist (USA) is another quantum news media that also maintains an online database of startups in the field and a quantum jobs board. Their about talks about "we" but they provide no name of who's behind the site. The site belongs to a company named Hadamard LLC based in Wyoming, and created in July 2021.

Fact Based Insight (UK) is/was an analyst shop run by David Shaw in the UK. It publishes very interesting charts and analysis on the quantum ecosystem, including some insights on quantum hardware.

Global Quantum Intelligence is an analyst shop created by André M. Konig, David Shaw and Doug Finke that consolidated in 2022 the activities from The Quantum Report (Doug Finke), Fact Based Insight (David Shaw) and André M. Konig's consulting activities. They provide data and insights on an annual subscription basis.

The Quantum Leap is a blog of news on the quantum ecosystem published by Russ Fein, a venture investor passionate of quantum computing and based in the USA.

AzoQuantum (UK/Australia) an information site on quantum science news with a supplier's directory, focused on manufacturing equipment and some interviews, mostly written by young researchers.

Quantum Journal - the open journal for quantum science, a site of scientific news on quantum physics, mostly showcasing preprints published on arXiv.

Quantiki is an information site on quantum computing. It looks like they became mainly a jobs board.

The Qubit Report (USA) is a news media focused on quantum vendors launched in 2017.

Qosf, a site that inventories guides and training for developers of quantum applications.

The Quantum Hubs newsletter (Switzerland) is broadcasting in his website and email newsletter the latest news from quantum research and vendors.

# Podcasts

Quantum Tech Pod are podcasts by Christopher Bishop on quantum technology news, published on Inside Quantum Technology's web site. These are mostly half-an-hour interviews of quantum startup founders. It started mid-2021.

Consulting firm Protivity also launched its own series of podcasts, in May 2021. Like Chris Bishop's podcasts, it's about interviewing quantum startup founders. They also last half an hour.

Quantum Computing Now by Ethan Hansen covers quantum computing news, basic concepts, and what people in the field are doing. The first episode was aired in July 2019. They are biweekly and cover news as well as science and learning tutorials.

The Quantum Analysts Roundtable is a podcast launched in January 2022 and run by Doug Finke, David Shaw, Shahin Khan, James Sanders and André M. König.

The Qubit Guy's podcast series is run by Yuval Boger, Classiq's CMO with short-format interviews.

QViews is a news podcast from Anastasia Marchenkova launched in May 2022. It's a pity since it just reads the titles and headlines from press releases.

The Quantum podcasts that I have been recording regularly (in French) since September 2019 with Fanny Bouton (OVHcloud). They are available on all audio platforms (Spotify, iTunes, Deezer, ...) as well as on YouTube in video version. It covers quantum news including what's happening in the ecosystem, with startups and in research. We decipher many lead scientific announcements. Our first episode was on Google's supremacy!

They are complemented by the Decode Quantum interviews that we have been publishing since March 2020 with a great variety of personalities (lead researchers, startup founders, investors, user companies, public servants, etc.) in partnership with Frenchweb. The first episodes featured Pascale Senellart, Alexia Auffèves, Maud Vinet, Eleni Diamanti, Elham Kashefi, Théau Peronnin and Raphaël Lescanne from Alice&Bob, Alain Aspect, Philippe Grangier, Michel Devoret, Daniel Esteve and many others since we had about 50 episodes in-store as of September 2022! They last about one hour.

insideQuantum is an equivalent series of podcasts in English which started in 2022 out of Spain with slightly shorter formats.

# Books and ebooks

If you wander in Amazon or your other preferred real-life of virtual scientific bookstore or University library, you'll find an abundant literature on quantum physics and quantum information. Many people willing to learn in these domains have a hard time finding the "right" book that is adapted to their existing knowledge and particularly, to their fluency in mathematics. Here's a not-too long list of books for this purpose. It's mostly adapted to students since, if you work in the industry, you probably won't have much time to read many of these thick books. Many of these ebooks are open source and/or free to download[3023].

### Quantum physics

Quantum Mechanics, Volume 1: Basic Concepts, Tools, and Applications, Second Edition, 2017, by Claude Cohen-Tannoudji, Bernard Diu and Franck Laloë (879 pages) is an undergraduate reference series of books to learn quantum physics. It is considered to be the reference or the bible by many students and teachers of quantum physics.

[3023] See Publicly available quantum computing books (WIP) for a list of open access book on quantum computing, a bit lazily done with urls without titles.

Quantum Mechanics, Volume 2: Angular Momentum, Spin, and Approximation Methods, Second Edition, 2017, by Claude Cohen-Tannoudji, Bernard Diu and Franck Laloë (688 pages).

Quantum Mechanics, Volume 3: Fermions, Bosons, Photons, Correlations, and Entanglement, Second Edition, 2017, by Claude Cohen-Tannoudji, Bernard Diu and Franck Laloë (747 pages).

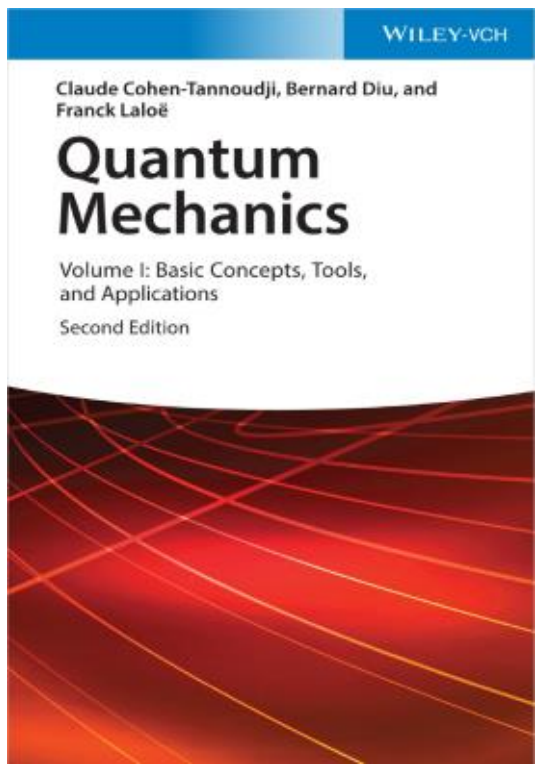

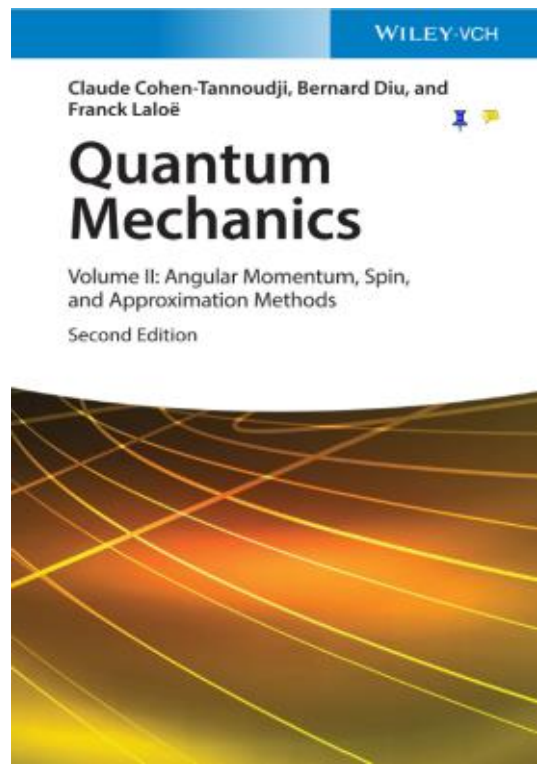

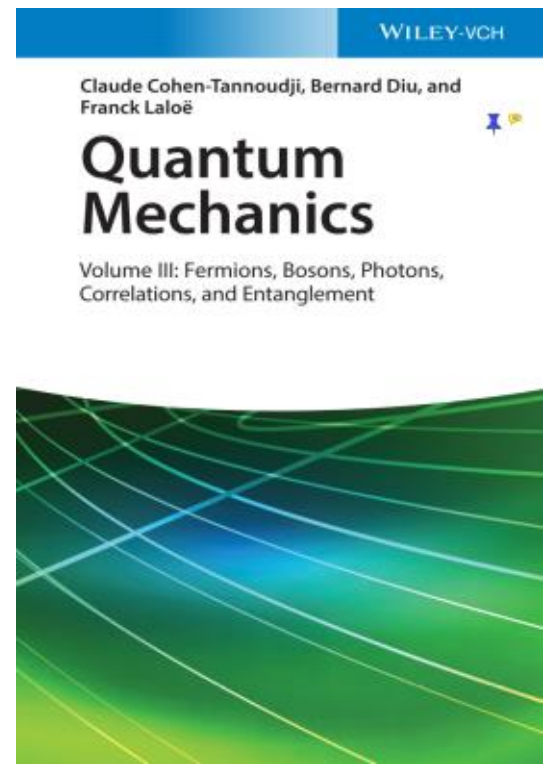

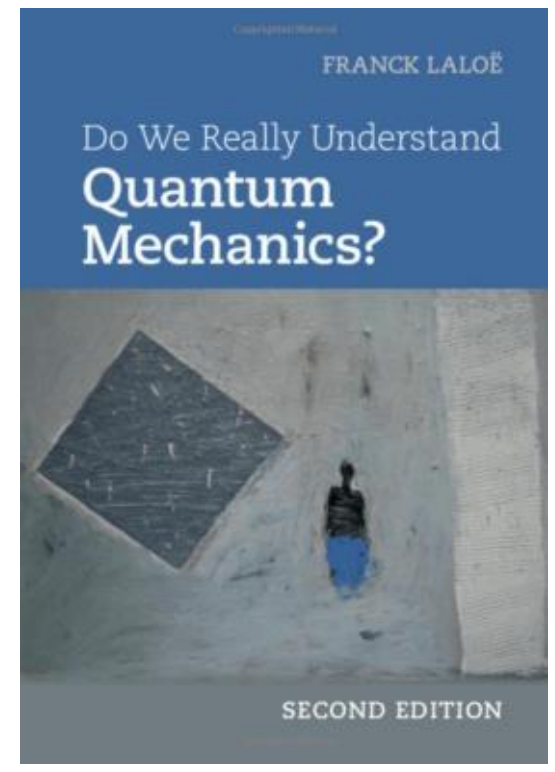

Do we understand quantum mechanics? Second Edition by Franck Laloë, 2019 (550 pages) is an interesting piece that documents the debates on quantum foundations and how to interpret quantum physics. Do we really understand quantum mechanics? by Franck Laloë, 2004 (118 pages) is a shorter and older version of this book, in public access.

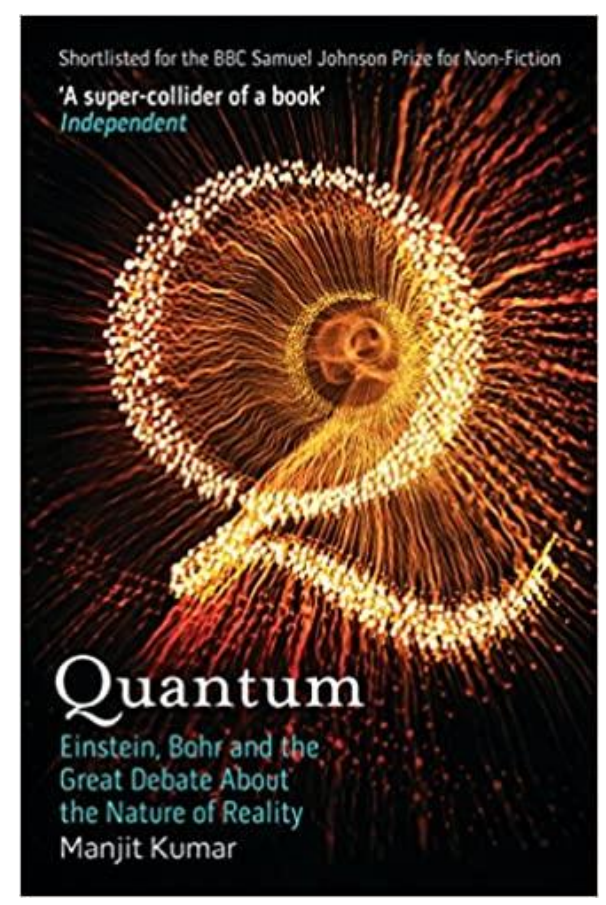

Quantum: Einstein, Bohr, and the Great Debate about the Nature of Reality by Manjit Kumar, 2009 (480 pages) is an excellent history book about the creation of quantum mechanics. It centers a lot on the works from Max Planck, Niels Bohr, Albert Einstein, Max Born, Werner Heisenberg and Erwin Schrodinger. It's a good account of the history of ideas and how quantum physics saw the day of light. It also showcases a lot of lesser known scientists who played key roles around the most famous ones and the balance between theoreticians and experimentalists. On top of that, the book scientific content is quite good and easy to understand, without any mathematics! Other history books and papers, mostly available in open access, are also mentioned throughout this book, particularly in the History and Scientists section, starting page 21.

Lecture notes on Quantum Mechanics by Frédéric Faure, 2015 (397 pages) which provided me with some leads to link quantum mechanics to its mathematical formalism and notably to explain the Born equation.

The Feynman Lectures on Physics - Volume III on Quantum Mechanics by Richard Feynman, Robert Leighton and Matthew Sands, (688 pages). It contains lecture notes of Feynman legendary courses from the early 60s. These are treasures of pedagogy with a content that is still up to date to grasp the fundamental of quantum physics. One advantage is it doesn't make any abuse of mathematics.

Quantum Optics An Introduction by Mark Fox, 2015 (397 pages) an excellent coverage of the broad field of quantum optics and the second quantization.

The Quantum Theory of Light by Rodney Loudon, 1973-2001 (450 pages) is a classic book on quantum light, that is useful to later better understand the physics of photon qubits used in quantum computing, telecommunications and cryptography. It classically starts with Planck's radiation law, then covers lasers, light-matter interactions, Mach-Zehnder interferometry, light quantization, single mode, multi-mode and continuous-mode optics and nonlinear optics.

Fundamentals of Photonics by Bahaa Saleh and Malvin Teich, 2019 (1401 pages) is a comprehensive quantum optics books that also covers instrumentation, which means it’s good stuff for experimentalists.

Introduction to Optical Quantum Information Processing by Pieter Kok and Brendon W. Lovett, 2010 (506 pages) is another classic quantum photonics books covering quantum information systems.

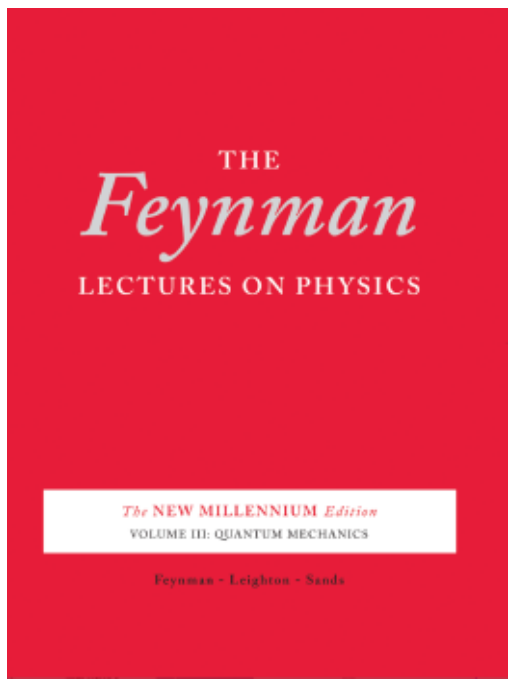

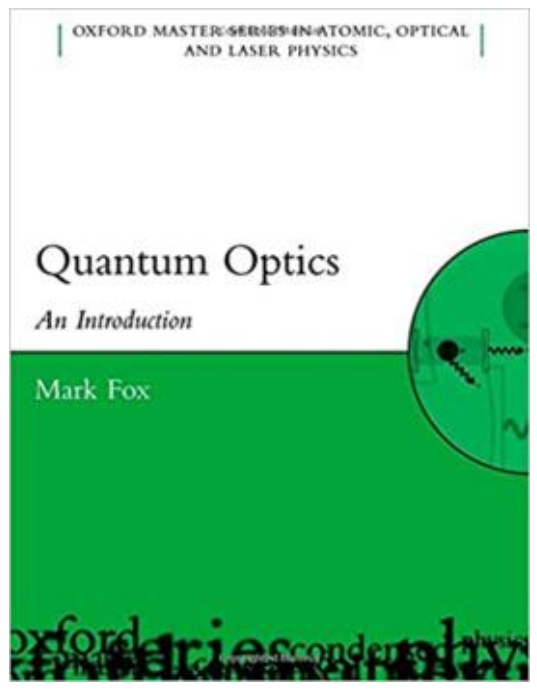

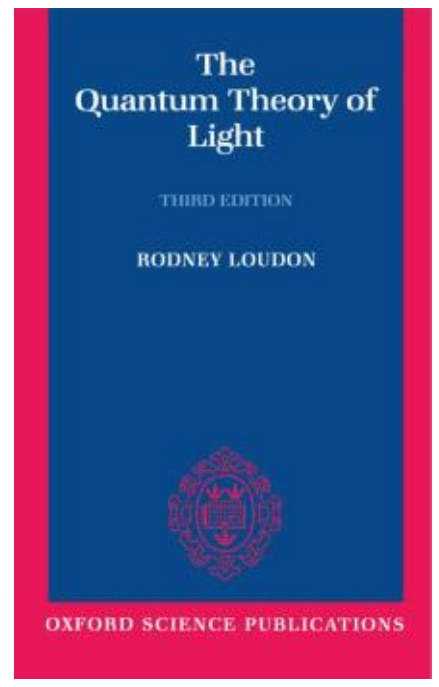

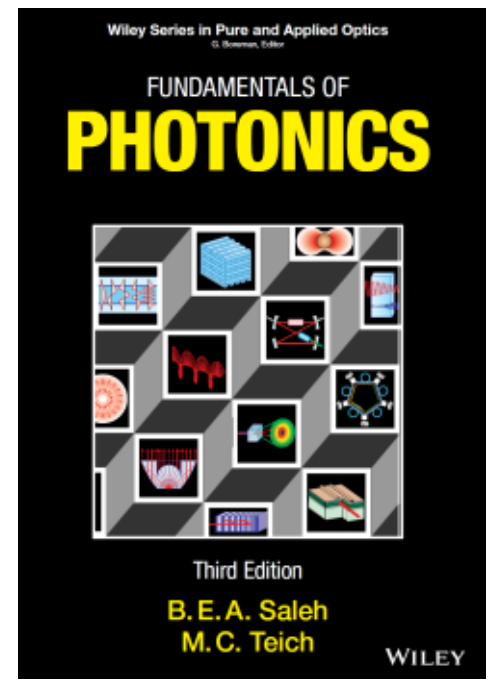

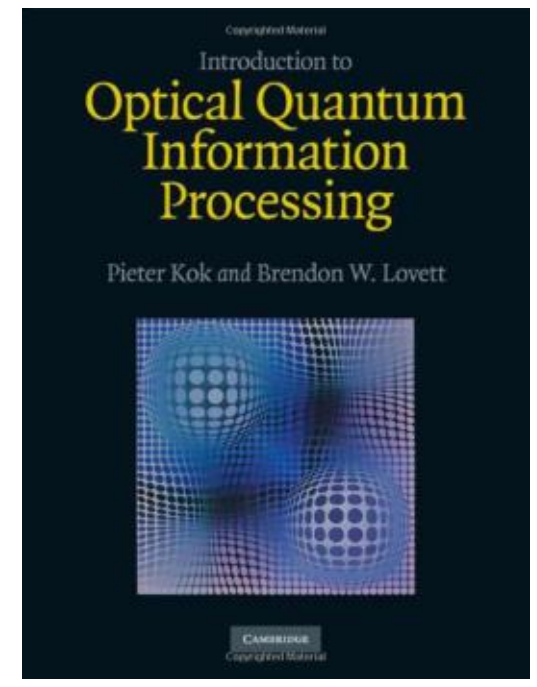

I Don't Understand Quantum Physics by Douglas Ross, 2018 (104 pages) is a nice primer to conceptualize and visualize many quantum phenomena. It describes the founding experiments of quantum physics (blackbody radiation, photoelectric effect, Compton scattering, etc), the wave-particle duality, matter wave, indeterminacy, Schrödinger’s equation, superposition and the EPR paradox.

**Quantum information**

Quantum Computation and Quantum Information by Michael Nielsen and Isaac Chuang, 2010 (10th edition, 704 pages, public access) is the definitive reference on the basics of quantum computing. It answers many key questions, in particular on the mathematical models of linear algebra used in quantum computing. It also covers the basics of quantum physics, quantum postulates, problems complexity classes, quantum measurement, quantum algorithms, how qubits are realized (harmonic oscillators, trapped ions, photons, NMR), the impact of quantum noise and decoherence, how quantum error corrections work, what is fault-tolerant quantum computing, how about Shannon and Von Neumann information entropy and the likes. It also covers quantum key distribution and cryptography.

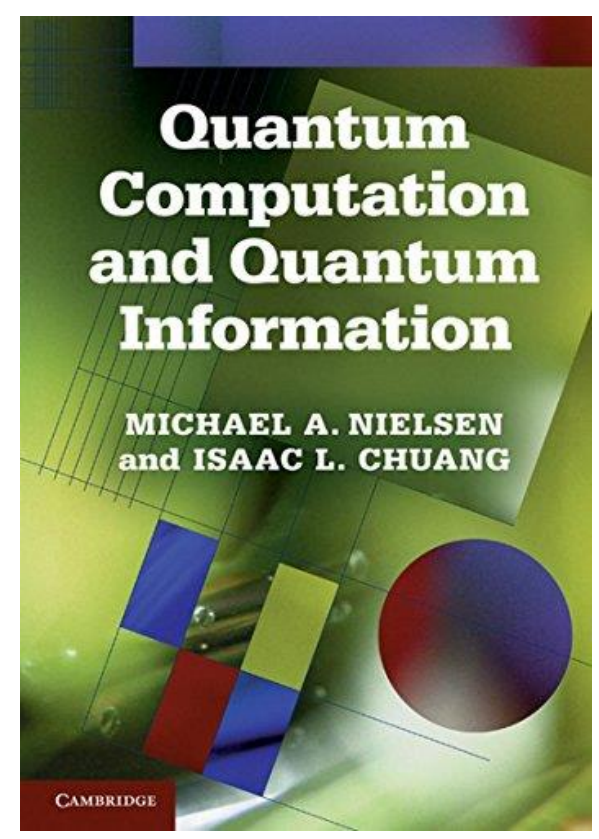

Programming Quantum Computers - Essential Algorithms and Code Samples by Eric R. Johnston, Mercedes Gimeno-Segovia and Nic Harrigan (2019, 336 pages) is an excellent and detailed description of key quantum algorithms like the QFT, phase estimation and the likes.

Introduction to Classical and Quantum Computing by Thomas Wong (2022, 400 pages), a book that is free to download in PDF and available in paperback on Amazon ( 😊 ). It makes a good comparison between classical and quantum computing. Like most quantum books for developers, it covers only classical gate-based algorithms with nothing on quantum annealing and quantum simulation. And it says nothing about the hardware and what it can do. It also makes some confusions between phase flip errors and decoherence related errors, which create mixed state when a simple phase error preserves a pure state.

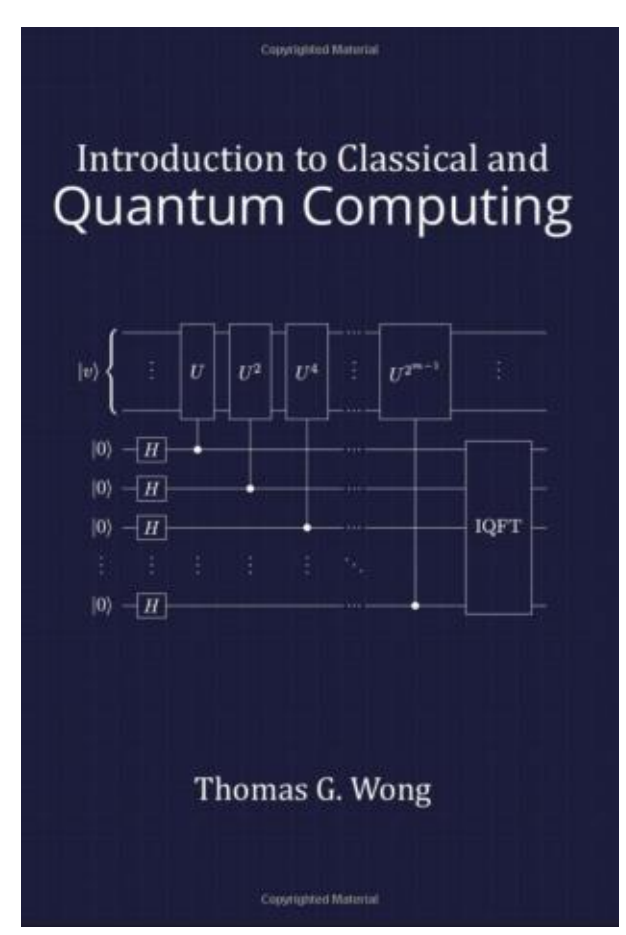

Elements of Quantum Computing by Seiki Akama (133 pages), which is at the same time concise, precise and quite complete on the nuts and bolts of quantum physics and quantum computing, with a good historical overview.

Quantum Computing: An Applied Approach (2021, second edition, 445 pages) by Jack Hidary, a fairly comprehensive book covering quantum algorithms and their mathematical foundations. It briefly describes the different architectures of quantum computers.

Quantum Machine Learning - What quantum computing means to data mining, by Peter Wittek, 2014 (176 pages) is a good introduction to machine learning and quantum machine learning although many progresses were made since this book's publication.

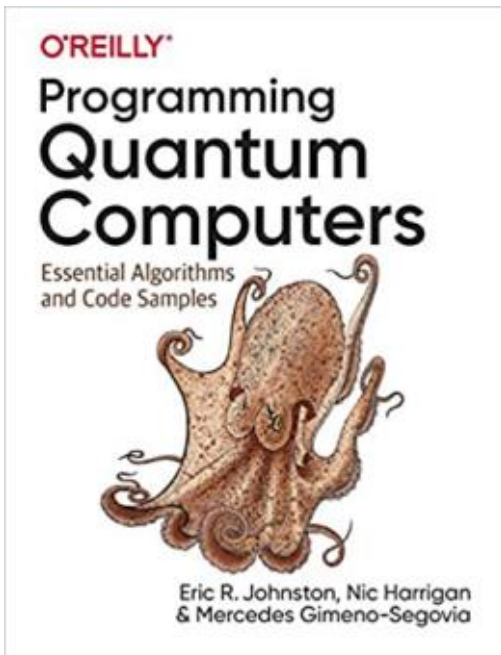

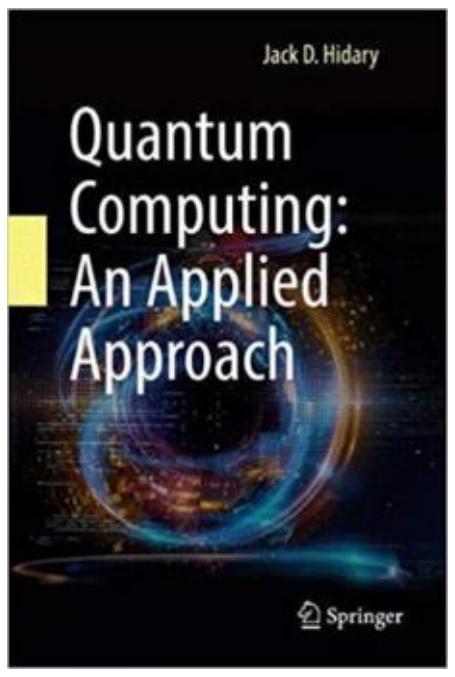

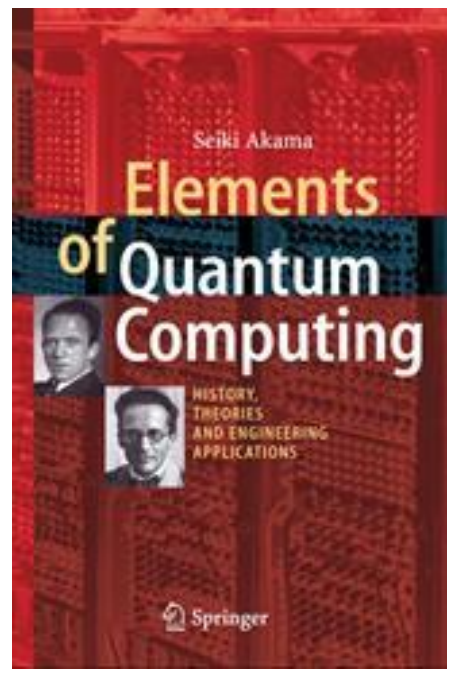

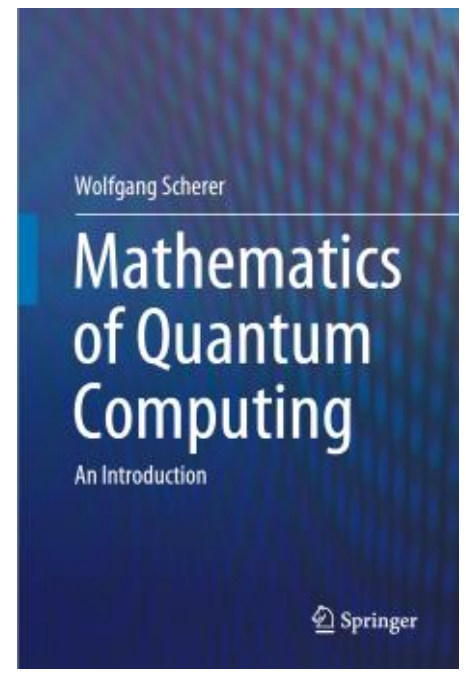

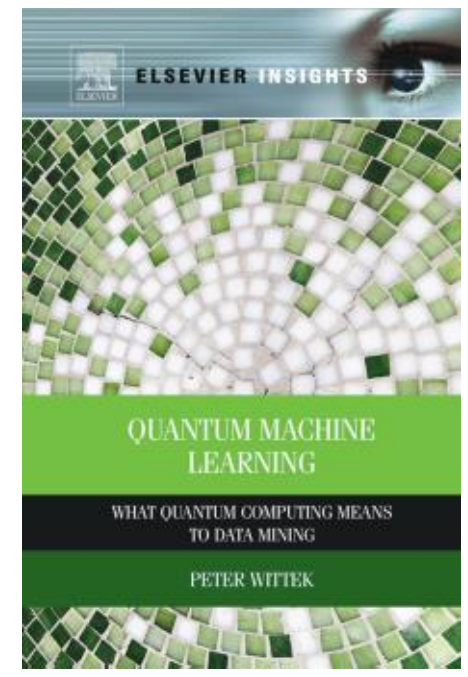

Quantum Information Meets Quantum Matter by Bei Zeng, Xie Chen, Duan-Lu Zhou and Xiao-Gang Wen. It is available in a February 2018 version on arXiv as a free download (373 pages).

Quantum computing- from quantum physics to quantum programming in Q# by Benoit Prieur, 2019 (244 pages). It starts with the general principles of quantum physics. The section on quantum computers themselves is rather thin and explores only a few technologies (superconductors and NMR, which is little used). The rest is dedicated to learning programming in Q#, Microsoft's quantum programming language.

Principles of Quantum Computation and Information, A Comprehensive Textbook by Giuliano Benenti, Giulio Casati, Davide Rossini and Giuliano Strini, December 2018 (598 pages).

Quantum Computing for Everyone by Chris Bernhardt, 2019 (216 pages) which describes the basics of quantum computing starting with the inevitable qubit, quantum gates, accelerations brought by quantum algorithms and the main components of a quantum computer.

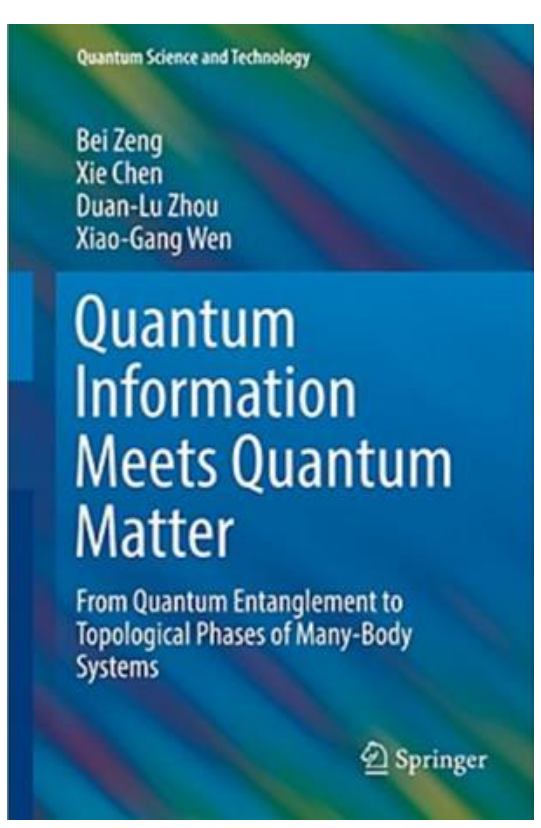

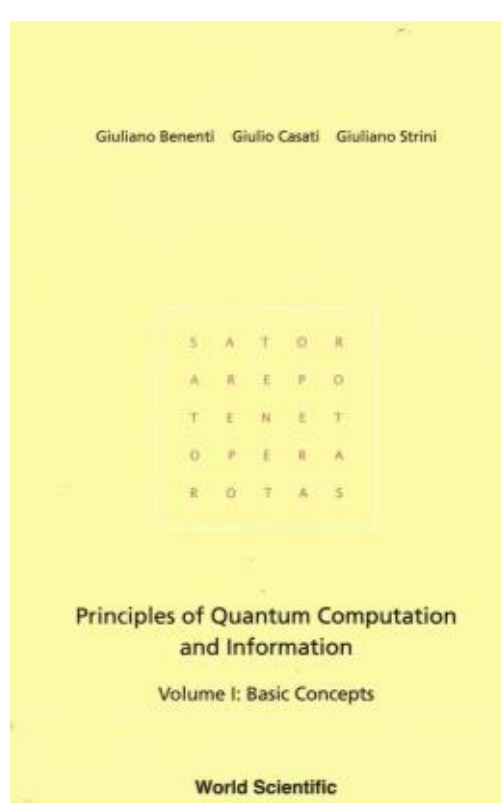

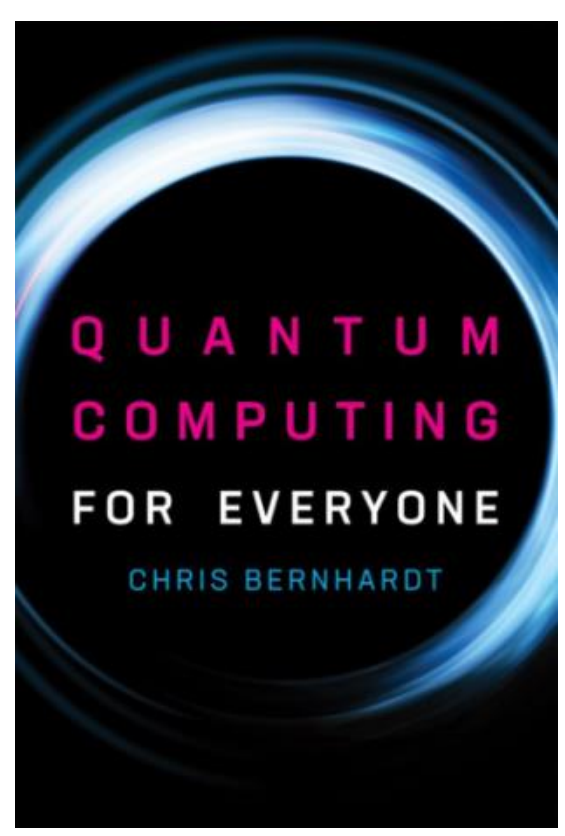

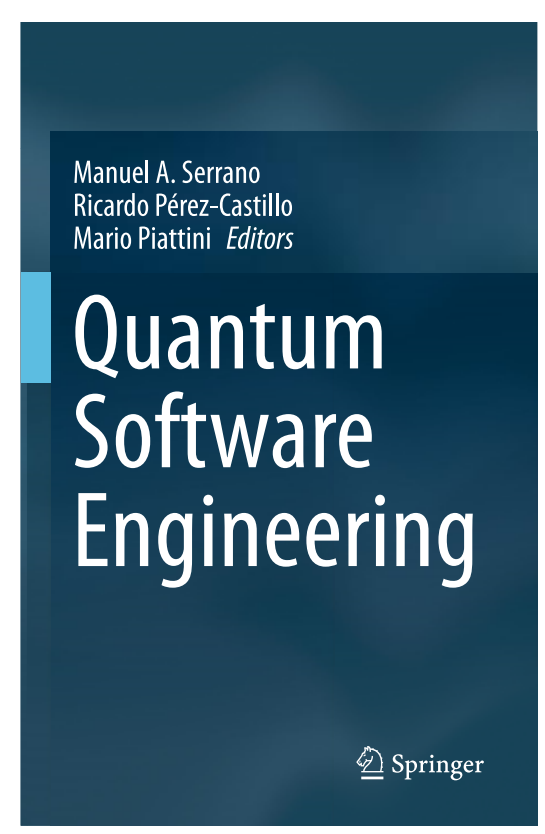

Quantum computing by Joseph Gruska (1999, 390 pages), another fairly comprehensive base covering all aspects of quantum computing and communication.

An Introduction to Quantum Computing by Phillip Kaye, Raymond Laflamme and Michele Mosca, 2007 (284 pages) which starts with some mathematical foundations of quantum physics and quantum computing. By reference authors such as Raymond Laflamme (Canada) who is one of the fathers of error correction codes.

Introduction to quantum computing algorithms by Arthur Pittenberger, 2001 (152 pages) which describes classical quantum algorithms with a good part dedicated to error correction codes.

Quantum Software Engineering is a book edited by Manuel A. Serrano, Ricardo Pérez-Castillo and MarioPiattini from University of Castilla-La Mancha in Spain, 2022 (321 pages). It provides an overview of this emerging discipline, describing key concepts, the related vocabulary and formal methods. It also presents Q-UML, a quantum modeling language.

Quantum Internet, a 60-page magazine presenting the different sides of quantum computing, published by TU Delft (2019).

Quantum computing for dummies by William Hurley, April 2023, has yet to deliver its nuggets!

Cryogenic Engineering and Technologies by Zuyu Zhao and Chao Wang, October 2019 (386 pages) is a reference book on cryogenic issues with a very extensive and well-documented history. There is an excellent chapter on dry dilution cryostats used in quantum computers. This helped me to prepare the part of this book on cryogenics (starting page 464) in addition to an interview with the team of the French startup CryoConcept and with researchers from CNRS Institut Néel in Grenoble.

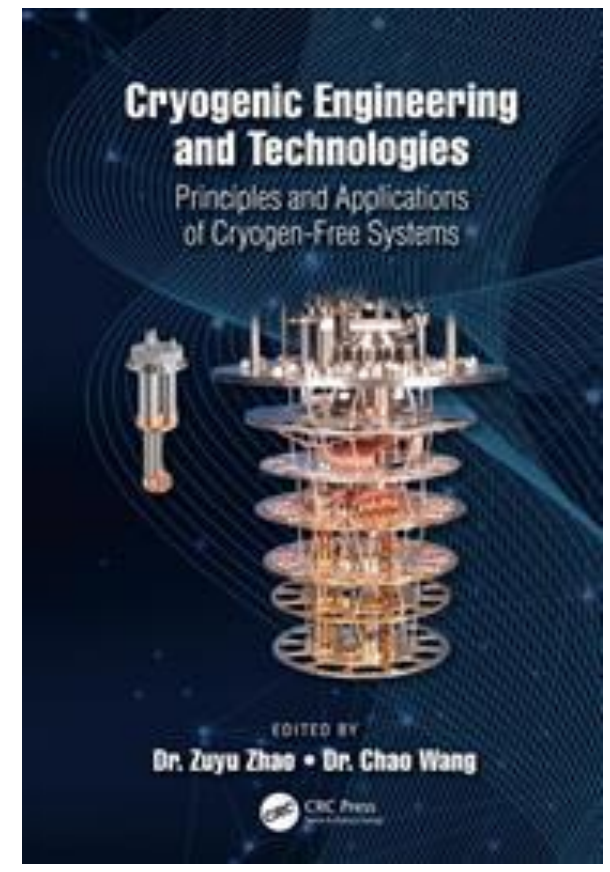

Unconventional Computation by Bruce MacLennan, University of Tennessee, October 2019 (304 pages) which discusses the energy issues of computation (reversible, non-reversible) and various alternative methods of computation including quantum computing and molecular computing.

## Comics

Quantum Computing for babies by Chris Ferrie and William Hurley, April 2018, is aimed more at children or even older children. The book popularizes the major concepts of physics and quantum computing in a very colorful way. It comes from a professor of quantum computing at Sydney University of Technology and the founder of the American startup Strangeworks. William Hurley is the founder of **Strangeworks**. Two other books in the same vein from Chris Ferrie were also released: Quantum Physics for babies and Quantum Information for babies, all for less than $10. I "read" the last one (24 pages) and I'm not sure even an adult would really understand how quantum computing after looking at it. This is the danger of oversimplification and information dilution.

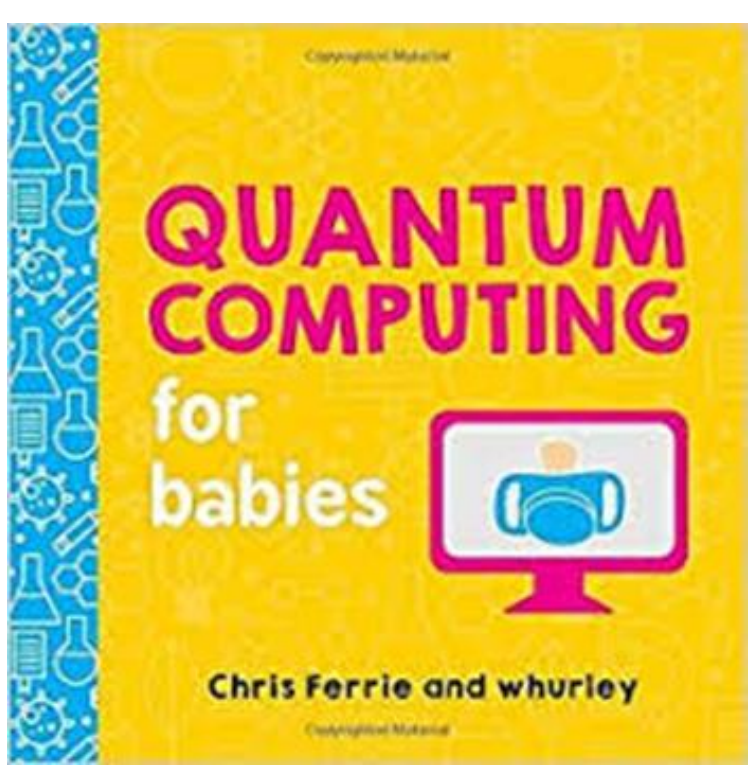

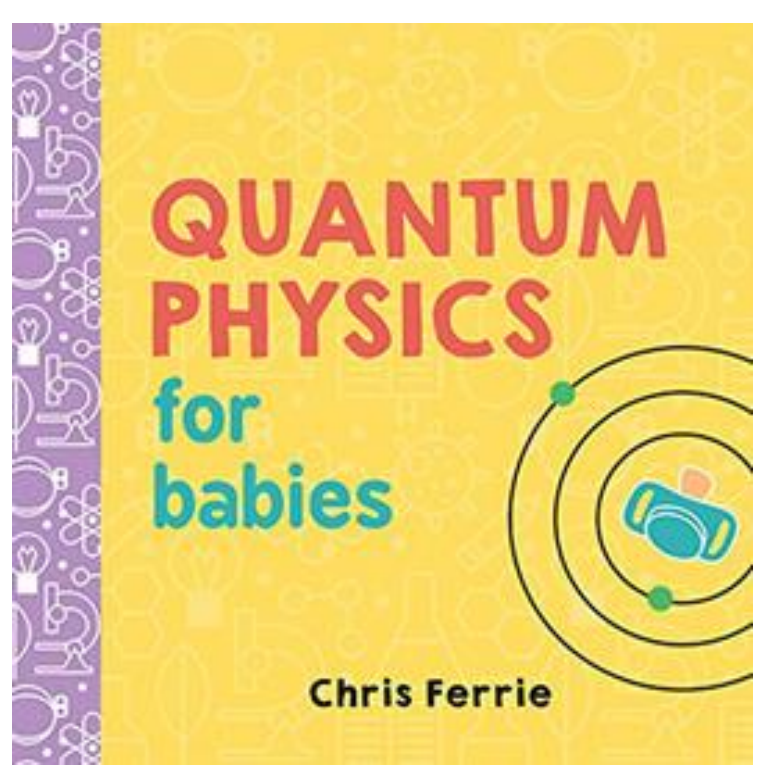

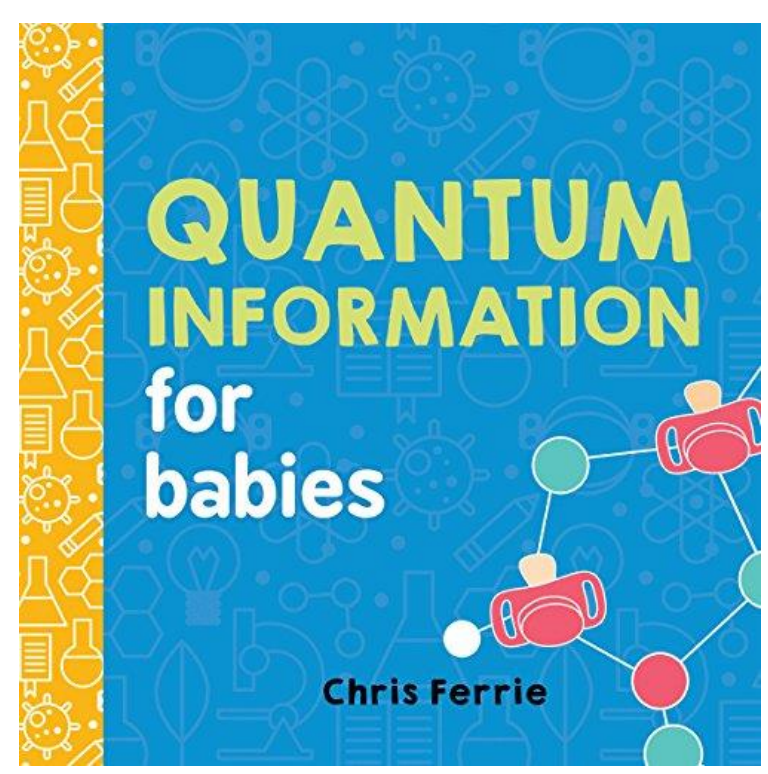

## Presentations

Here are a few conferences and presentation materials rather well done to popularize quantum computing.

Quantum computing for the determined by Michael Nielsen, a series of 22 videos on quantum computing, 2011, accompanied by a long text of explanations.

CERN's Quantum Computing for High Energy Physics workshop in November 2018, with presentation materials and videos and interesting talks by various players in quantum computing, including Intel, who are not often seen. The specific content may be overtaken by the basic principles remain valid.

Quantum computing Overview by Sunil Dixit, September 2018 (94 slides) is a presentation by Northrop Grumman that takes a fairly broad look at quantum computing and the underlying mathematical models.

A Practical Introduction to Quantum Computing From Qubits to Quantum Machine Learning and Beyond, by Elias Combaro, CERN course, 2020 (251 slides) is a good course on quantum algorithms, debugging, validation, verification and benchmarking.

Quantum computing, a four-part course by Hélène Perrin at Université Paris 13, February 2020 (lecture 1 of 77 slides on trapped ions, lecture 2 of 36 slides on superconducting qubits, lecture 3 of 39 slides on silicon, molecular and NV Centers qubits, lecture 4 of 75 slides on cold atoms). This requires a good background in physics to be understood from start to finish. The references provided allow to deepen the topics covered.

## Training

Berkeley courses for 2013: Quantum Mechanics and Quantum Computation on YouTube.

Videos from the Stanford Quantum Computing course.

The Quantum Computing Fundamentals course offered by MIT.

An online training course on quantum programming offered by Brillant in partnership with Microsoft.

The QSIT course (FS 2016) at ETH Zurich with its slides and lecture notes.

Quantum Computing as a High School Module, a curriculum with exercises on the basics of quantum physics for students at the BAC level.

## Reports

Inside Quantum Technology, an analyst company dedicated to quantum technologies, which sells industry reports.

You can also find analysts reports on quantum technologies with McKinsey, BCG, Yole Development and other analyst companies.

## Miscellaneous

Designing and Presenting a Science Poster, Jonathan Carter, Berkeley (20 slides) which is intended to help researchers design a good research project presentation poster.

# Glossary

*What is the purpose of a glossary? It allows you to find your way around in a new terminology and to step back to understand new concepts. For the author, it was also a good checkpoint of his own understanding and ability to popularize scientific and technological concepts. Some of these descriptions are simplified versions derived from Wikipedia definitions. Welcome to the lingua franca of quantum sciences and technologies!*

**137**: constant used to compare different equivalent quantities in quantum physics. It turns out that 1/137 is a value that corresponds approximately to the fine-structure constant, a ratio that is found in several places in quantum physics and compares data of the same dimension. It is for example the ratio between the speed of an electron in the lower layer of a hydrogen atom and the speed of light or the probability of emission on the absorption of a photon for an electron. 137 is a bit like 42 in quantum physics. Wolfgang Pauli died after an operation for pancreatic cancer, while his hospital room was number 137.

**ADC**: analog-digital converter. Converts an analog signal into a digitized signal. In quantum technologies, it is used to convert the reflected microwave signals used in superconducting and electron spin qubits readout.

**Adiabatic**: quantum computation method used in particular with D-Wave quantum annealing computers. A complex Hamiltonian describing a complex system is first determined whose fundamental state describes a solution to the problem under study. A system with a simpler Hamiltonian is then prepared and initialized in its fundamental state. This Hamiltonian then adiabatically (meaning, with no energy or mass exchange with the outside environment) evolves into the complex Hamiltonian. According to the adiabatic theorem, the system remains in its fundamental state, and its final state describes a solution to the problem under consideration.

**Adiabatic theorem**: quantum mechanics concept created by Max Born and Vladimir Fock in 1928. It states that a quantum mechanical system subjected to gradually changing external conditions adapts its form, but when subjected to rapidly varying conditions there is insufficient time for the functional form to adapt, so the spatial probability density remains unchanged. This can be used to find Hamiltonian energy minimums with quantum algorithms running on various architectures: gate-based, annealing and quantum simulation.

**Advantage**: see quantum advantage.

**Aka**: shortened "also known as".

**Algorithm**: a method of problem solving that is made up of a finite sequence of operations or instructions. The word comes from the name of the 9th century Persian mathematician, Al-Khwârizmî.

**Algorithmic qubit**: benchmark metric proposed by IonQ which corresponds to the number of qubits usable with an equivalent computing depth with a randomized benchmark producing a good result in 2/3rd of the runs. It's actually $\log_2$ of IBM's quantum volume.

**Amplitude**: this term has various meanings depending on the context. It can be the classical amplitude of a wave, i.e. half of its maximal variation, as opposed to its phase. For a quantum object, it can be the complex amplitudes of its basis states or eigenvectors. With a qubit in its Bloch sphere representation, the amplitude is related to the projection of the qubit vector on the z axis. But the $\boldsymbol{\alpha}$ and $\boldsymbol{\beta}$ describing the qubit vector are also amplitudes, although, precisely, complex amplitudes. These complex amplitudes define the qubit amplitude (1-cos $(\theta/2)$) and its relative phase (angle $\varphi$).

**Anharmonic oscillator**: contrarily to harmonic quantum oscillator that have the same energy difference between ach each consecutive energy levels, an anharmonic quantum oscillator has different energy differences between consecutive energy levels. This is the case of superconducting qubits, in order to create two manageable energy levels that are controlled with microwaves with the highest energy transition level of the oscillator.

**Angular momentum**: generally speaking, speed of rotation of a rotating object. In quantum physics, angular momentum is quantized and can have only discrete values.

**Ansatz**: another name for the parametrized quantum circuits or parametric quantum circuits that contain rotation gates and the parameters of the problem to encode in a variational quantum optimizer defined by a classical optimizer. Typically used is most QML algorithms.

**Anyons**: type of elementary particle found in two-dimensional systems. It is a generalization of the concept of bosons and fermions. Anyons have intermediate statistical behaviors between the two types of elementary particles. They are in fact virtual particles that live in two spatial dimensions and are generally based on electrons or electron gaps moving in superconducting metallic 2D structures. Anyons are a particular type of quasi-particles. They are used in topological quantum computers and would be used in particular in computers based on the hypothetical Majorana fermions studied at Microsoft.

**arXiv**: Cornell University's site that allows researchers to publish scientific papers prior to publication in peer-reviewed journals such as Nature, Science or Physical Review. It can take up to 9 months between publication of an article on arXiv and publication in a peer-reviewed journal. In the latter case, the article will have eventually evolved. The interest of arXiv in literature search is that publications are open and free of charge whereas most of the peer-reviewed journals are not free. The disadvantage is that the articles are not necessarily validated and that one has to make his own evaluation. It should be noted that in a researcher's publication, there are often several authors, up to several dozen. The first author is generally the PhD student who has carried out a large part of the work, particularly its experimental part. Others are contributors who helped him/her. The last author is the thesis director or principal investigator (PI), the group leader or the laboratory director who has closely or remotely supervised the project. He/she probably contributed significantly to the article writing and cleanup.

**Atoms**: the smallest constituent element of matter that manifests chemical properties. It consists of a nucleus, with one or more positively charged protons and zero or more neutrally charged neutrons, around which negatively charged electrons gravitate. In a neutral atom, the number of electrons is equal to the number of protons. Otherwise, the atom is negatively or positively charged and forms an ion. The number of protons determines the nature of the atom in Mendeleev's Elements Periodic Table. An atom with one proton is hydrogen, with two protons it is helium, etc. Uranium has 92 protons. The nucleus represents the bulk of the atom mass. The isotopes of an element correspond to variations in the number of neutrons. In general, the number of neutrons of an element is equivalent to that of protons. Electrons are distributed in layers whose number depends on the atomic number. They are numbered from 1 to 7. Each layer can contain a maximum of $2n^2$ electrons, n being the number of the layer (thus 2, 8, 18, 32, 50, 72 and 98). This model was developed by Niels Bohr between 1909 and 1913. The chemical properties of the element depend on the number of electrons of the last layer which is called the valence layer. If this number is $2n^2$, the atom will be inert and will not combine chemically with other atoms. Carbon has three layers of electrons, the last one having 4 which allows it to combine with other atoms such as hydrogen (1 layer, 1 electron) or oxygen (6 electrons in the last layer).

**Autonomous quantum error correction** (AQEC): quantum error correction codes and architectures that doesn't require error syndrome measurement. It replaces real-time feedback by analog feedback circuits using engineered dissipation with the reservoir engineering technique. It couples the system with a dissipative reservoir to transfer the entropy created by errors to an ancillary system, the reservoir. This entropy is then evacuated via the strong dissipation of the ancilla. This technique is used in cat-qubits.

**Back action**: in quantum measurement, this is the physical impact of the measurement device on measured quantum objects. Quantum measurement usually modifies the state of the measured quantum object unless it is already in a basis state (mathematically, one of the eigenvectors of the measurement observable operator...). After measuring a quantum object state, performing the same measurement on the already measured system will not provide any additional information. In order to increase our knowledge on the final state of some quantum computation, the only solution is to start again the computation from the beginning and measure again the final state. The subtlety being that this new final state has not yet been measured and thus corresponds exactly to the one we are trying to infer. Then, we compute an average of the obtained results across several experiments. Some measurement techniques like gentle measurement or weak measurement are designed to minimize this back action and are sometimes used in quantum error correction codes.

**Balmer series**: set of four spectral emission or absorption lines with the hydrogen atom, generated by electron transitions between the second and higher energy levels of the atom.

**Beam splitter**: optical device that splits a beam of light in two. It's usually made with two glued triangular glass prisms. Polarizing beam splitters are a particular class of beam splitters that use birefringent materials to split light into two beams of orthogonal polarization states.

**Balanced beam splitter**: beam splitter where the light is equally divided in two streams.

**Baryon**: class of elementary particles of the first level of the nuclei of atoms. It contains protons and neutrons.

**Bell inequalities**: Bell's 1964 theorem proves that no hidden variable theory - imagined by Einstein in 1935 - can reproduce the phenomena of quantum mechanics. Bell's inequalities are the relations that measurements on quantum entangled states must respect under the hypothesis of a local deterministic hidden variable theory. Experiments shows that Bell's inequalities and related statistics are systematically violated, forcing scientists to give up one of the three following hypotheses on which Bell's inequalities are based. The first is the locality principle according to which two distant objects cannot have an instantaneous influence on each other, which means that a signal cannot propagate at a speed greater than the speed of light in a vacuum. The second is causality, according to which the state of quantum particles is determined solely by their experience, i.e. their initial state and all influences received in the past. The third is realism, which means that individual particles are entities that have properties of their own, carried with them (source).

**Bell states**: or EPR states are maximally entangled states of two qubits. There are four of them: $(|00\rangle + |11\rangle)/\sqrt{2}$, $(|01\rangle + |01\rangle)/\sqrt{2}$, $(|00\rangle - 11\rangle)/\sqrt{2}$ and $(|01\rangle - |01\rangle)/\sqrt{2}$. The first of the Bell state is generalized in GHZ states with n qubits and the second is generalized as W states.

**Bell test statistic**: it is a test of correlation of quantum state detection with two entangled quantum objects which can have values A and A', and B and B'. Quantum entanglement showing a correlation of the values of these two objects will yield and average value of: $|S| = \langle AB\rangle_{lim} + \langle AB'\rangle_{lim} + \langle A'B\rangle_{lim} - \langle A'B'\rangle_{lim} = 2\sqrt{2}$ (about 2.828). In this formula, $\langle A'B\rangle_{lim}$ is the probability to have outcome A with the first quantum object and outcome B' for the second. It's usually a photon polarization. This test is also a way to evaluate the entanglement of two qubits in quantum computers.

**Black body**: a body that is in thermal equilibrium with the radiation it emits. It can be the inside of a furnace or a star. It is by studying the radiation of the black body and its frequency distribution as a function of the body temperature that Max Planck uncovered the existence of energy quanta in 1900. Also written blackbody or black-body depending on the source.

**Blind Quantum Computing**: technique for distributing quantum processing in remote quantum processors and securing the confidentiality of the processing.

**Bloch sphere**: geometric representation of a qubit state with a vector in a sphere of radius 1. The qubit ground state is an upwardly directed vector $|\mathbf{0}\rangle$ and the excited state is a downwardly directed vector $|\mathbf{1}\rangle$. An intermediate state vector is defined by its amplitude and phase, in line with the wave-particle duality of the qubits. It models of the state of a qubit using polar coordinates with two angles, one indicating the amplitude of the quantum and the other its relative phase.

**Born rule**: postulate of quantum mechanics created by Max Born in 1926 giving the probability that a measurement of a quantum system will yield a given result. It states that the probability density of finding a particle at a given point, when measured, is proportional to the square of the magnitude of the particle's wavefunction at that point.

**Bose-Einstein Condensate**, or BEC, state of very low density boson gas cooled to a temperature close to absolute zero (-273.15°C) where a large part of the bosons are in the lowest possible quantum energy state and exhibit particular properties such as interferences. A special case of BEC is superfluid helium, discovered in 1938, which, at very low temperatures, has no viscosity, i.e. it can move without dissipating energy. These condensates were imagined and theorized by Satyendra Nath Bose and then Albert Einstein in 1924. Their existence was demonstrated experimentally in 1995 by Wolfgang Ketterle, Eric Cornell and Carl Wieman who were awarded the Nobel Prize in Physics in 2001. In quantum computing, this field is related to the field cold atom-based qubits.

**Boson sampling**: typical experiment with photons qubits that mixes photons in an interferometer. It's hard to emulate on a classical computer and is used to show a specific quantum advantage. The only caveat is these experiments are not programmable and are therefore entirely useless and irrelevant to compare any calculation capacity between systems.

**Boson**: particles with gregarious behavior, which can accumulate in arbitrarily large numbers and in the same state. Bosons comprise photons and composite objects with whole integer spin such as hydrogen, lithium-7, rubidium-87, carbon and silicon atoms in crystalline structures. These particles escape Pauli's exclusion principle. They have a symmetrical wave function.

**Bosonic codes**: hardware system that implement quantum error corrections with bosonic modes, using a quantum harmonic oscillator with a continuous energy levels. It includes cat-qubits, GKP codes (Gottesman-Kitaev-Preskill) and binomial codes.

**BQP** (problem class): complexity class of problems that can be handled by quantum algorithms. Means a bounded-error quantum polynomial time. It is the class of problems that can be solved in polynomial time relative to the size of the problem with a probability of obtaining an error not exceeding one third of the results. This class is positioned between the class P (problems that can be solved in polynomial time on a classical machine) and NP (problems for which a solution can be verified in polynomial time on a classical machine).

**Bra-ket** (notation): A notation model describing the state of a quantum and a qubit in the form $|\boldsymbol{\psi}\rangle$ and $\langle\boldsymbol{\psi}|$. It was created by Paul Dirac in 1939. A bra psi vector is a quantum state described as a column vector. A ket is its transpose, a row vector. It facilitates the writing of operations with quantum states, like inner products $\langle\boldsymbol{\phi}|\boldsymbol{\psi}\rangle$, outer product $|\boldsymbol{\phi}\rangle\langle\boldsymbol{\psi}|$ and projection $\langle\boldsymbol{\psi}|\boldsymbol{A}|\boldsymbol{\psi}\rangle$.

**Cavity quantum electrodynamics** (CQED): field of quantum physics coupling trapped atoms in physical cavities and microwaves. It is about the interactions between photons and electrons and atoms.

**Chandelier**: nickname of the quantum computing system located inside the cryostat of a superconducting or electron spins quantum computer. It contains several stages made of copper disks covered with gold. These disks are crossed by numerous coaxial cables that are used to drive the qubits and read their state with microwaves. It is completed by filters, attenuators and amplifiers for the microwaves that circulate in these wires, various sensors, and heat exchangers that cool the copper disks, which in turn cool the elements that are placed on them.

**Chi**: Greek letter used to define the level of nonlinearity of an optical medium. A $\chi^{(i)}$ medium had a nonlinearity of level i. When i=2, the medium has a second order nonlinearity. It is used for example for the frequency doubling of laser light. With i=3, it is a third order nonlinearity, which is used for example in four-wave mixing. Chi is a coefficient of the polynomial relation between the phase P change of a light beam traversing the medium and the beam energy E, with the formula $P = \epsilon_0(\chi^{(i)}E^i)$.

**Circuit quantum electrodynamics** (cQED): architecture used in solid state qubits systems using superconducting qubits and microwave photons. Science behind the interactions between microwaves and electromagnetic circuits.

**Clifford group**: group of unitary quantum gates that can be easily simulated in polynomial time on classical computers according to Gottesman-Knill's theorem. A Clifford gate is a quantum gate that can be decomposed into gates of the Clifford group. It is sufficient to have one unitary gate rotating on the X axis and another on the Z axis to create a complete set of Clifford gates. They must be completed with at least one two-qubit gate as a CNOT. These gates make quarter turns or half turns in the Bloch sphere. They are not sufficient to create a universal gates set. You need non-Clifford gates like the T gate.

**Circuit**: describes a set of quantum gates applied in an orderly fashion to a register of qubits. In other word, it is a quantum program for a gate-based quantum computing system. A graphical representation of a quantum circuit shows qubits in stacked horizontal lines, quantum gates as boxes labelled X, Y, Z, H and others applied to single qubits and two or three quantum gates with vertical lines connecting qubits. The X axis represents time and gates are executed from left to right.

**Cluster state**: the starting point for an MBQC (Measurement Based Quantum Computing) calculation with a grid of embedded qubits that are usually initialized in an entangled state. Used mostly with photon qubits.

**CMOS**: a common semiconductor fabrication technique used to produce processors and memory, and which is reused to create qubits that manipulate electron spins.

**Code distance**: notion used in quantum error correction codes and with the stabilizer codes formalism which is linked to the smallest number of simultaneous qubit errors that can be fixed with a given quantum error code. A code distance d means that the error correction code can correct errors for up to (d − 1)/2 qubits. These are usually even numbers (3, 5, 7, …). So a code distance of 7 can correct at most 3 qubits.

**Coherence**: quantum coherence is the ability of a quantum system to demonstrate interference. The coherence between different parts of a wave function allows for the famous double-slit interference and the formation of short quantum wave packets propagating in space. Two wave sources are coherent when their frequency and waveform (or phase for an electromagnetic signal) are identical. There are temporal coherence (same waveforms with some time delay), spatial coherence (in 2D or 3D such as with plane waves) or spectral coherence (waves of different wavelengths but with a fixed relative phase form a wave packet). In quantum physics, coherence comes with linear superposition of various states of a quantum system containing one or several quantum objects (represented by a wave due to the wave-particle duality). Quantum coherence progressively degrades naturally due to the interactions with the environment and ends after a certain time for qubits (the coherence time) and also when measuring the state of a qubit.

**Cold atoms**: atoms cooled at very low temperatures, generally with techniques using lasers and the Doppler effect. They are used in certain types of quantum computers called cold atom quantum computers. The atoms used are neutral atoms (not ionized) and quite often rubidium, an alkali metal.

**Compatible properties**: physical properties of a quantum system that can be measured in any order or simultaneously.

**Commutativity**: mathematically, two variables A and B commute when A×B=B×A. They do with integers but not with non square matrices. Even square matrices don't necessary commute. They are then "noncommutative".

**Commutator or commutation operator**: Characterize the level of non-commutativity between two variables, usually matrices. For two matrices A and B, their commutator is [A,B]=AB-BA.

**Complementarity**: principle of quantum physics introduced by Niels Bohr in 1927 according to which quantum objects have certain pairs of complementary properties which cannot all be observed or measured simultaneously. These are incompatible properties. Another version of this principle is that it's not possible to simultaneously observe a quantum object as a particle and as a wave, like in the Young slit experiments.

**Complementary variables**: pairs or complementary variables or properties according to the Bohr complementary principle.

**Complex number**: set of complex numbers created as an extension of the set of real numbers, containing in particular an imaginary number noted i such that $\boldsymbol{i^2 = -1}$. Any complex number can be written in the form a + ib where a and b are real numbers. These numbers are used in particular to describe the state of a qubit and to represent the phase of a quantum object with its complex component.

**Complexity** (theory): branch of theoretical computer science and mathematics that plays an important role in quantum computing to evaluate its performance compared to traditional Turing/Neumann machine computing. It defines classes of problems by levels of complexity, in terms of computing time or even the memory space required, with, in particular, problems that are solved in polynomial time in relation to their complexity (class P) and whose results are verifiable in polynomial time (class NP). The methods used to solve these problems are most often based on the brute force of navigating through an increasingly large space of combinations to be evaluated according to the size of the problem to be solved.

**Compton effect**: effect which demonstrates that photons can have some momentum and behave as particles, that was demonstrated by Arthur Holly Compton in 1922 with scattering of X rays and gamma rays photons by atomic electrons.

**Computational basis**: naming of the basic states of a qubits register. For a single qubit, this corresponds to the $|\mathbf{0}\rangle$ and $|\mathbf{1}\rangle$ states. For a register of N qubits, the computational basis is made of the $2^N$ combinations of series of N 0s and 1s, named in Dirac's notation $|\mathbf{000 \dots 000}\rangle$ to $|\mathbf{111 \dots 111}\rangle$. All these states are mathematically orthogonal with each other. A N qubits register in a pure state mode is a linear superposition of all these states using complex amplitudes.

**Concatenated codes**: describes the recursive application of error correction codes where in an error correction code, a physical qubit is replaced by a logical qubit, and so on.

**Condensed matter physics**: branch of physics that studies the macroscopic properties of matter (solids, liquids, glasses, polymers) and in systems where the number of constituents is large and the interactions between them are strong. Condensed matter physicists seek to understand the behavior of these phases using the laws of physics (quantum mechanics, electromagnetism and statistical physics). In practice, it mainly covers low temperature superconducting, ferromagnetic, antiferromagnetic and ferrimagnetic phases of spins in crystalline lattices of atoms, spin glasses, spin liquid, and Bose-Einstein condensates. Physicists working on superconducting qubits are part of this discipline.

**Conjugate variables**: pairs of dynamic variables describing the state of a quantum object, like position and momentum, that are related to the other with the Heisenberg indeterminacy principle which prevents a precise measurement of both variables.

**Continuous variables quantum computing** (CV): a type of quantum computer that uses qubits whose values are continuous and not binary. Used in two types of quantum computers: analog quantum simulators (particularly based on cold atoms) and CV photon-based systems.

**Cooper pair**: pairs of tightly coupled electrons creating electric current flow in superconducting materials, usually at very low temperatures and without resistance. Cooper pairs have an integer spin because they accumulate two electrons with a spin of ½. They become bosons and can accumulate and form macro quantum objects.

**Copenhagen interpretation**: interpretation of quantum physics elaborated by Niels Bohr in Copenhagen and by Werner Heisenberg, although it was never clearly formalized. Applied to individual quantum objects, it is mostly based on Bohr's correspondence and complementarity principles, Heisenberg's indeterminacy principle, Born's probability interpretation of the Schrodinger wave function and on the wave function collapse and its fundamental indeterminism. It avoids describing any reality beyond what can be measured like an exact position of an electron. The completeness of this theory was challenged by Albert Einstein. Physicists are still debating about this interpretation, as part of the quantum foundation field.

**Correspondence principle**: principle formulated by Niels Bohr in 1920 which states that the behavior of systems described by quantum physics matches classical physics in the limit of large quantum numbers (large orbits and large energies or electron quantum numbers).

**Coulomb blockade**: decrease of electrical conductance at small bias voltages of small electronic devices containing at least a low-capacitance tunnel junction. As a result, the conductance of the devices may not be constant at low bias voltages, but disappears for biases under a certain threshold, i.e. there is no current flows.

**Coulomb force**: electrostatic force between electrically charged particles like electrons and protons. Its strength is inversely proportional to the square of their distance and proportional to the product of their respective charge.

**CPTP map**: completely positive and trace preserving map or operator also referenced as a quantum channel or superoperator. It is a linear operator that turns a density matrix describing a mixed state system into another density matrix. Its size is then the square of the density matrix size, so $2^{4N}$ for a system of N qubits. It can describe any operation on a mixed state system: some quantum gates, any sort of measurement, quantum filters, as well as feedback networks in quantum control theory.

**Crosstalk**: in the qubit field, is a phenomenon where an action on a given qubit or set of qubits has a side effect on other qubits. Of course, various techniques are employed to minimize it like, in the case of superconducting qubits, with using tunable couplers between qubits. Then, there are several crosstalk types like IX, IY, IZ, ZX, ZY, ZZ which depend on the type of interactions between qubits.

**Cryogenics**: cooling technology. Very low temperature cryogeny is used with superconducting and electron spin qubits computers. The temperatures required to stabilize qubits and reduce their error rate are very close to absolute zero: around 15 mK. The most commonly used systems are dilution refrigerators that use helium-3 and helium-4. Cryogenics is also used for photon generators and photon detection systems, but at a higher temperature situated between 2K and 10K.

**CSCO**: complete set of commuting observables, the most complete measurement of a quantum system comprised of compatible properties that can be measured in any order.

**CVD**: chemical vapor deposition, an additive manufacturing technique for semiconductors, where the target surface is exposed to one or more volatile precursors, which chemically react and/or decompose on the target surface to leave a thin film deposit on the target, e.g. using silane $SiH_4$ to deposit Si on the wafer, generating 2 $H_2$ molecules.

**DAC**: digital-analog converter. Classical electronic device converting a digital signal into an analog signal. Is used in the microwave generation systems implemented to control superconducting and electron spin qubits.

**Dark count**: photons detected by photon detectors that come from the environment and thermal or tunneling effect. This explains why most single photon detectors must be cooled at a temperature usually below 10K.

**De Broglie wavelength**: wavelength of a particle calculated with its momentum p with h/p, with h being Planck's constant.

**Decoherence**: marks the end of the coherence of a quantum object or a qubit. It is notably caused by the interactions between the quantum objects and their environment. One often uses indifferently the expression coherence time (time during which the qubits are in a state of superposition and entanglement with other qubits) or decoherence (time at the end of which this superposition and entanglement end), which is the same.

**Degenerate**: a quantum system energy level is degenerate if it corresponds to two or more different measurable states with different quantum numbers. Mathematically, a quantum state is degenerate when several linearly independent eigenvectors may have the same eigenvalue. A normalized linear combination of these eigenvectors is also an eigenvector with the same eigenvalue. The number of linearly independent eigenvectors having the same eigenvalue corresponds to the degree of degeneracy of the quantum system. The number of different eigenstates corresponding to a particular energy level is the degree of degeneracy of the level. This happens for example when the energy level alone is not sufficient to characterize the state of a quantum system. That's where we need other quantum numbers to characterize the state. This is the case with the hydrogen atom electron. Its energy level depends only on its principal quantum number n (the electron layer), and not the three other electron quantum numbers (orbital angular momentum, magnetic moment and spin, although this degeneracy can be broken with using relativistic quantum mechanics and hyperfine structure splitting of electron energy levels). But you also have degenerate quantum error correction codes, which are supposed to correct more errors than they actually detect, particularly with noisy quantum channels (meaning practically qubits gates). Another example is an atom's nucleus energy level that is only dependent on its orbital angular momentum in the absence of magnetic field. Different energy levels arise with a magnetic field due to the nucleus magnetic quantum number.

**Density matrix**: matrix of complex numbers used to describe the statistical state of a physical system that is more precise than the computational state vector used in quantum computing. Density matrices are useful to describe so-called mixed states versus pure states that are sufficiently described with state vectors. They are used to describe what happen to subsystems of entangled systems, when decoherence happens and also, during measurement.

**Dequantization**: said about some quantum algorithm where an efficient classical equivalent is found. Term initiated with Ewin Tang's work on recommendation systems in 2018, when she found a dequantized classical equivalent to a quantum recommendation algorithm devised in 2016 by Iordanis Kerenidis and Anupam Prakash. Interestingly, quantization is a term used in artificial intelligence and deep learning when the numbers used in these models are integers (or even binary numbers) instead of floating point numbers.

**Determinism**: situation when events are determined completely by previously existing causes and parameters. Applied to classical mechanics to predict objects position and momentum based on initial conditions. Contrarily, in quantum physics, it is not possible to determine simultaneously the position and momentum of any particle at any instant. This indeterminism is observed with quantum measurement when the quantum object is in a superposed state.

**Deutsch-Jozsa** (algorithm): quantum algorithm created in 1992 by David Deutsch and Richard Jozsa. It can check whether a given function is balanced or not, i.e. whether it always returns 0 or 1, or 0 and 1 in equal proportions. The alternative between equilibrium (as many 0's as 1's) or not (as many 0's or 1's in output) is a starting postulate. The gain in performance compared to classical algorithms is exponential. In the case of N qubits, the function should be classically evaluated on at least half of the possible input values, i.e. $2^{N-1}+1$. Unfortunately, this algorithm is not very useful.

**DFT** (Density Functional Theory): mathematical model used to describe the structure of molecules at rest as a function of inter-atomic interactions. Used in high-performance computing as well as in quantum computing for chemical simulation.

**Diffraction**: phenomenon created when a wave encounters an obstacle or opening, like a small hole or slit. It is generated by the bending of photon waves around the corners of the obstacle. It creates interferences between the passing waves as they are detected in a plane further down the waves path. The phenomenon can be described classically with the Huygens–Fresnel principle that considers points in the hole or slit as a collection of individual spherical wavelets. The interference pattern shows up with laser light and can also be explained by the photon wave functions and their probability distribution. All-in-all, you can consider that a Young single-slit experiment also create quantum interferences!

**Dilution refrigerator**: name given to the very low temperature cryostats used to cool quantum computers below 1K. They cool superconducting or electron spin chipsets to respectively 15 mK and 100 mK. Dilution is related to the use a mixture of two helium isotopes (3 and 4), which are diluted in a mixing chamber, the two isotopes having slightly different physical properties. A helium 4 cryostat only goes down to about 2.8K, a helium 3 cryostat goes down to 300mK while a cryostat using both goes down to 10mK. The most common variant is the "dry" as opposed to "wet" dilution refrigerator. This version uses less helium and leaves more space in the chandelier to house electronic and quantum devices.

**Dirac's notation**: see bra-ket.

**Dirac constant**: Planck constant divided by $2\pi$, also called reduced Planck constant and denoted $\hbar$ (h-bar). Some physicists called sometimes, abusively, this constant "Planck constant".

**Discrete log problem**: mathematical problem consisting in finding a log of a number that happens to be an integer. It is used in finding the solution of cryptographic problems with quantum algorithms. Shor's dlog algorithm is a quantum algorithm solving discrete log problems.

**Distillation**: technique used in quantum error correction codes based on magic states. It consists in combining several magic state qubits to feed others with a lower error rate. Distillation has the effect of purifying the state of qubits, meaning turning mixed states into pure states.

**Doppler effect**: shift in the electromagnetic spectrum due to the speed at which the source moves away from or closer to the observer. If the source moves away from the observer, the light wavelength is shifted towards the red (redshift), otherwise towards the blue. This effect is used in particular in the technique of atoms laser cooling. It consists in illuminating atoms that are in thermal motion with a wavelength that is just below the absorption level of the atoms. Those atoms moving towards the laser beam will absorb the photon, which will reduce their kinetic energy and movement. Those atoms moving in the other direction will not absorb it because the apparent frequency of the photon will be too low to change the energy state of the atoms. As atoms get cooled, the cooling laser wavelength has to be adjusted. This technique allows atoms to be cooled to below mK (milli-Kelvin).

**D-Wave**: Canadian company designing quantum annealing computers. They do not have the same power as universal gate quantum computers with equal numbers of qubits. The current generation of D-Wave "Advantage" using Pegasus chipsets includes 5000 qubits.

**EBL**: electron beam lithography, a lithography technique that is focusing a beam of electrons on an electron-sensitive resist film to remove matter in specified areas, without requiring a mask like with photolithography. It is used to create 1nm precision nanostructures like with photon-generating quantum dots and also superconducting qubit chipsets. It's a rather slow process compared with photolithography that is adapted to low volume and custom productions.

**Eigenstate**: for a quantum object, these are the elementary wave functions in which it is possible to decompose it. They are represented by eigenvectors.

**Eigenvalue**: see eigenvector.

**Eigenvector**: for a square matrix A, an eigenvector x of A is a vector that verifies the equation Ax = λx, λ being a real number called eigenvalue. Their direction do not change once multiplied with matrix A.

**Electromagnetic spectrum**: all electromagnetic radiation from the largest radio waves to X-rays and gamma rays. Visible light is only a very small part in the middle of this spectrum. An EM wave is decomposable in a number of photons, the smallest elementary unit of an EM wave.

**Electron**: elementary particle found in atoms, orbiting the nucleus, but also in freeform traveling between atoms and creating something we know as being electric current. According to Bohr's model developed in 1913, there is a finite number of electron orbits around the nucleus of atoms. The movement of electrons from one orbit to another corresponds to the absorption or emission of a photon. Electrons are elementary particles in the standard model because it is not composed of sub-particles, unlike neutrons and protons which are composed of quarks. According to quantum physics, the electron as many other particles behaves as a particle and as a wave. Electrons are often used in qubits, in the form of electrons circulating in semiconductors loops or who are trapped in quantum dots or electromagnetic cavities and whose spin is controlled.

**Electron gas**: describes the behavior of free valence electrons in metals and semiconductors when they move around free of the atom nucleus. Their behavior is governed by the Pauli exclusion principle (1925), Dirac-Fermi statistics (1926) and Arnold Sommerfeld's quantum theory of metal (1927). Electron gas enable the modeling of electric conductivity, electron heat capacity and electric thermal conductivity as well as the Hall and quantum Hall effects. There are 1D, 2D and 3D electron gases. 1D electron gas are observed in semiconductor nanowires and carbon nanotubes. 2D electron gas show up in semiconductor quantum wells and in graphene sheets.

**Elliptic Curves Cryptography** (ECC): a type of public key cryptography that is potentially broken by Shor's quantum algorithm. One of its advantages is that it requires small keys, about three times smaller in number of bits than RSA public keys.

**Energetics of quantum technology**: cross-disciplinary research field and sector studying the energetics of quantum computing but also quantum telecommunications, cryptography and sensing. It's about making sure quantum technologies are not power hungry and also dealing with the energetic constraints related to quantum computing scalability. It's about balancing the act between cooling requirements, cabling, control electronics to ensure quantum computers can scale in number of physical and logical qubits.

**Entanglement**: quantum phenomenon where two quantum objects are related with each other in a way that a measurement done on these two objects generates a correlated (but random) value. Mathematically speaking, two quantum objects are entangled when their quantum state (psi, vector state) cannot be expressed as the tensor product of individual quantum states. This process is used to link qubits together through two or three qubit quantum gates in quantum computers. It is also used in quantum cryptography and telecommunication systems based on entangled photons in QKDs.

**Entropy**: measures the degree of disorder and randomness of a physical system. Key concept related to the second law of thermodynamics that states that the entropy of an isolated system cannot decrease spontaneously. In quantum mechanics, the (Von Neumann) entropy of a system is $-tr(\rho log \rho)$ where $\rho$ is the density matrix describing the system state.

**ERC Grants** : European Research Council grants. Funding of European research projects with several levels, the top of which is the Synergy Grant which funds "moonshots" in European research associating at least two principal investigators (PIs) from public or private research laboratories. 14M€ is the maximum funding for such a project with 10M€ of core funding and 4M€ which can notably finance heavy investments or access to large infrastructures. Other levels include the Starting grants with up to 1.5M€ for 5 years and the Consolidator grants with 2M€ also for 5 years.

**Ergodicity**: capacity of a moving system to explore all parts of the space in which it can move in, in a uniform and random manner. The phenomenon occurs with many physical systems like with electrons. Quantum ergodicity states that in the high-energy limit, quantum objects tend to a uniformly distribute in the classical phase space.

**Ergotropy**: maximum amount of work that can be obtained from a quantum system.

**Error Correction Codes**: describes both logical methods and physical architectures to correct physical errors happening in both classical and quantum computing and telecommunication technologies.

**Errors**: a major concern in the operation of quantum computers. Operations on qubits: one and two qubit gates and qubit readouts generate errors that must be minimized. Error rates are in 2021 between 0.1% and 2% for quantum gates. When several quantum gates are chained together, the rates of correct results (1 - error rate) multiplies quickly to the point of distorting everything. This is avoided either by reducing the physical level error rate like with cat-qubits, using shallow algorithms (low number of gates) or with error correction code systems.

**Eta letter**: η, used to describe error rates or efficiency, Carnot engine efficiency, and also Landau symmetry breaking.

**Exclusion principle**: see Pauli exclusion principle.

**Expectation value**: average or mean value of an observable. With an observable operator A on a quantum state $\psi$, the expectation value is $\langle a \rangle = \langle \psi | A | \psi \rangle$. In other words, it's a scalar product of the $\psi$ vector and the vector resulting from the projective measurement of $\psi$ using the observable A. In layman's term for a qubit, it is either the average value that would be obtained when doing an experiment, a large number of times and measure the value of the qubit yielding 0 or 1, and making an average, or the result of its mathematical evaluation if you have a clear idea of the qubit quantum state description. For example, after an Hadamard gate is applied to a $|0\rangle$ qubit, the expectation value of its measurement in in the typical z basis will be 0,5. Some algorithms output like in chemical simulations output real number values that are obtains through expectation value assessment using a large number of computing runs.

**Euclidean networks**: class of encryption algorithms used in post-quantum cryptography (PQC).

**Fabry-Pérot cavities**: equipment used in lasers that combines two parallel mirrors, one of which is semi-reflective. This contributes to the creation of the laser effect in the cavity. The length of the cavity is generally a multiple of the laser light wavelength, at least if we want to emit coherent light with photons having all the same phase. The name of the cavity comes from the French scientists Charles Fabry (1867-1945) and Alfred Pérot (1963-1925).

**FBQC**: fusion-based quantum computation, a variant of MBQC crafted by PsiQuantum that is based on micro-clusters states with groups of 4 qubits connected together and using Resource State Generators (RSGs). It's replacing measurement of entangled states in MBQC with double measurement of non-connected adjacent qubits to create entanglements between them.

**Fermi sea**: electrons filling the lowest atom orbits or degenerate low-energy states within a solid and at very low temperature near 0 K. It corresponds to low-energy states that do not participate in materials thermal activity.

**Fermions**: particles with individualistic behavior. Two particles of this type cannot be in the same state at the same place. This includes electrons, quarks, half-integer spin composite objects. For example, deuterium, lithium-6, potassium-40 atoms (source: Jean Dalibard). In contrast, integer spin bosons such as photons and some atoms can accumulate in the same state. In a word, bosons are communists and fermions are ultra-liberals.

**Fine structure**: splitting of an energy level or spectral line into several distinct components that take into account electron spins and relativistic corrections to Schrodinger's wave equation.

**Floquet code** or Planar Honeycomb Code is a family of quantum error corrections codes created by Matthew B. Hastings and Jeongwan Haah from Microsoft in 2021. It simplifies toric codes with fewer qubits and stabilizers and is adapted to qubits architectures implementing pair-wise qubit measurements like with the elusive Majorana fermions.

**Fluxonium**: variation of flux superconducting qubit. It has a better coherence time than transmon, above 100 µs but two-qubit gates are more difficult to implement, and this architecture seemingly has not yet been tested beyond 10 functional qubits.

**Flying qubits**: qubits that can move, as opposed to stationary qubits that do not move. They are usually photons but there's a small branch of flying qubits studying flying electrons.

**Fock space**: mathematical object of algebra used to describe the quantum state of a set of identical particles whose number is variable or unknown. It is a Hilbert space made up of the sum of the tensor products of Hilbert spaces for the particles that make up the set.

**Fock state**: defines a group of quantum objects, like photons, who have the same quantum numbers and are indistinguishable. They are defined by their number, a photon number in the case of photons, and their common quantum numbers describing the quantum objects state.

**Flux biasing**: technique used to control with some direct current the resonant frequency of a frequency-tunable superconducting.

**Fourier Transform**: mathematical decomposition of a time domain signal into elementary single frequency signals with their frequency, amplitude and phase. It is a complex value function of time with, for each frequency, a magnitude (real part) and a phase offset (complex part) of the sinusoid of this elementary frequency. The inverse Fourier transforms that frequency decomposition function back into its original compound signal.

**FPGA**: Field Programmable Gate Arrays. Integrated circuit where some or all functions can be dynamically defined and program on-demand. It can have analog and digital features. Modern FPGAs also embed full-fledge processing units (Arm cores, GPUs, neural processing units, networking units). FPGAs are used in qubit control electronics for reading out the signals coming from the resonators attached to superconducting and electron spin qubits. It measures the phase and amplitude of the reflected microwaves after they are converted from analog to digital with an ADC (analog-digital converter) that can be embedded in the FPGA.

**Fredkin gate**: quantum gate operating on three qubits that inverts the state of the second and third qubit if the first qubit is 1. Also called CSWAP gate (conditional SWAP).

**FTQC**: Fault-Tolerant Quantum Computer. Error-resistant quantum computer that is based on logical qubits made of many physical qubits and implementing quantum error correction. Fault-tolerance is based on the error correction making sure errors don't propagate to many qubits.

**FTDQC**: Fault Tolerant Distributed Quantum Computation, and extension of the FTQC concept to distribution quantum computing.

**Gate-based quantum computing**: the broader category of quantum computing system based on qubits and quantum circuits implementing quantum gates on 1, 2 and 3 qubits at a time.

**Gaussian Boson Sampling (GBS)**: variation of a boson sampling experiment that uses Gaussian states photons as input. It's a physical model that is even more difficult to digitally simulate than a boson sampling since the underlying mathematical object is a Hafnian instead of a permanent, that is even more complicated to compute.

**Gaussian state**: describe particular photon states that are classical. The gaussian curve is the form in three dimensions of Wigner's function which describe the phase and amplitude distribution of the photon. It is opposed to non-Gaussian states which are non-classical, with some negative Wigner function values and a non-Gaussian form for the 3D function.

**GHZ**: means something other than giga Hertz in quantum computing! It is a three-qubit Greenberger-Horne-Zeilinger superposed state that allows to demonstrate the inexistence of hidden variables in the quantum entanglement of at least three particles and with a finite number of measurements. The concept dates back to 1989 and has been experimentally validated in 1999.

**GKP qubits**: error corrected qubits according to a method proposed by Gottesman, Kitaev and Preskill that encodes a qubit in a harmonic oscillator. It is works with photon qubits using linear elements for Clifford gates.

**Gleason's theorem**: according to Andrew M. Gleason's theorem proved in 1957, the functions assigning probabilities to measurement outcomes are projection operators that must be expressible as density operator and follow the Born rule. This determines the way to calculate probabilities and the set of possible quantum states.

**GPGPU**: General Purpose Graphical Processing Unit, used for simulation, scientific computing and machine learning, like the Nvidia V100, A100 and H100. These are coprocessors which are mostly not anymore used for graphics software or gaming but more for machine and deep learning and scientific computing, thus the "general purpose" nickname addition.

**Grotrian diagram**: diagram used to show the various electronic energy transitions for a given atom, introduced in 1928 by the German physicist Walter Grotrian. The indicated frequencies of transitions to higher energy levels provide an indication of their source like lasers (in the hundreds of nm wavelengths) or microwaves (in the 4-20 GHz frequency regimes).

**Ground state**: lowest energy state of an atom, other states being excited states. The hydrogen atom ground state happens when its electron occupies the lowest energy level (with main quantum number = 1). More generally, is said of a qubit that is in its ground basis state $|\mathbf{0}\rangle$.

**Grover** (algorithm): quantum algorithm for finding an element in a non-indexed array or a unique element for which an oracle function returns 1.

**H-bar**: see Dirac constant.

**Hadamard** (gate) : quantum gate creating a superposed state between $|\mathbf{0}\rangle$ and $|\mathbf{1}\rangle$ in a qubit when starting with $|\mathbf{0}\rangle$ or $|\mathbf{1}\rangle$.

**Hall effect**: production of a voltage difference across an electrical conductor that is transverse to an electric current in the conductor and to an applied magnetic field perpendicular to the current. The effect was discovered by Edwin Hall in 1879.

**Hamiltonian**: equation describing the total and potential energy of a system of quantum objects. It is the global operator of the right part of Schrödinger's equation. This notion is used in D-Wave quantum annealing computers and with quantum simulators. "Preparing a Hamiltonian" in this kind of computer is equivalent to setting up a matrix of qubits linked together by potentials and which will seek a minimum energy resulting in a balanced Hamiltonian corresponding to the solution of the problem to be solved. The solution is about finding the right combination of qubits states (up/down for quantum annealing) that minimizes the energy of the whole system.

**Hamming distance**: metric used to compare two binary data strings of equal lengths. It is the number of bit positions in which the two bits are different. For two strings a and b, it is denoted as d(a,b).

**Harmonic oscillator**: in classical mechanics, system that, when displaced from its equilibrium position, experiences a restoring force proportional to its displacement x with a frequency that does not depend on the amplitude. Quantum physics formalize the whereabouts of many harmonic oscillators including photons in cavities, superconducting qubits, phonons, diatomic molecules, etc.

**Hartree-Fock**: method to compute atomic structures using the time dependent Schrödinger's wave equation.

**Heisenberg** (principle of indeterminacy): fundamental principle of quantum mechanics which postulates that there is a lower limit to the precision with which one can measure two independent parameters relating to the same object such as its speed and position or the energy emitted and the duration of emission.

**Heisenberg limit**: in quantum sensing, like with interferometry, the optimal rate at which the accuracy of a measurement can scale with the energy used in the measurement. More precisely, not every quantity is a quantum observable that can be measured directly. The estimation of such quantity, however, can be performed by measuring a state whose probability distribution depends on it. To evaluate the accuracy of this estimation, one often considers the variance of the estimated quantity. When using, for instance, an ensemble of photons as the meter probing our parameter of interest, if these photons are allowed to be initially entangled, then this variance is lower bounded by the fundamental Heisenberg limit. As for the standard quantum limit, it implies that the more resources, the more accurate the measurement. However, only quantum probing resources can reach the Heisenberg limit which states that our estimation's standard deviation is at best inversely proportional to the size of the meter, hence here the number of photons. The Heisenberg limit can be reached in quantum sensing with using entangled objects with a precision that scales better at a rate of 1/N instead of $1/\sqrt{N}$ with the standard quantum limit. The Heisenberg limit is reached with using so-called NOON states superposing N (bosonic) objects (like photons) in a state with all the objects being in one or the other of two modes.

**Helium 3**: a rare isotope of helium that is used in cryogenic quantum computer systems to generate temperatures below 1K as part of dilution refrigeration systems. It is usually produced from tritium in specialized nuclear power plants, including the US Department of Energy's Savannah River nuclear power plant.

**Helium 4**: a common helium isotope that is also used in cryogenic systems.

**Heralded single-photons**: pairs of single photons can be generated in highly correlated states from using a single high-energy photon to create two lower-energy ones. One photon from the resulting pair is detected to "herald" (or "signal") the other so its state is pretty well known prior to its own detection or whereabouts. The two photons need not be of the same wavelength, but the total energy and resulting polarization are defined by the generation process. Two commonly types of heralded single-photon sources are SPDC (spontaneous parametric down-conversion with line width in the THz range) and SFWM (spontaneous four-wave mixing with line width in the MHz range or even narrower). It's used with QKD.

**Heterodyne measurement**: method used for extracting information from an oscillating signal along two orthogonal components in phase space like the in-phase and quadrature signals coming out of an I/Q mixer. In this type of measurement, two conjugate operators are measured simultaneously, which create added noise.

**Hilbert** (space): vector space of real or complex numbers with a Euclidean or Hermitian scalar product, which is used to measure distances and angles and to define orthogonality. It is an n-dimensional extension of the concept of three-dimensional Euclidean space. In quantum mechanics, the state of a quantum is represented by a vector in a Hilbert space with as many dimensions as the number of basic (or observable) states of this quantum. These are geometrical spaces which are used in particular to measure lengths and angles, to make projections on dimensions and to define the orthogonality between vectors.

**Hidden variables**: interpretation proposals of quantum physics based on the use of (yet) unobservable hypothetical entities what would explain phenomena like entanglement and describe reality. Bell's theorem implies that local hidden variables of certain types cannot exist. Based on the assumption, promoted by Albert Einstein in the famous 1935 EPR paper, that quantum physics is an incomplete theory.

**Homodyne measurement**: method used for extracting information encoded as modulation of the phase and/or frequency of an oscillating signal. It compares that signal with a standard oscillation carrying no information. Homodyne detection uses a single frequency where heterodyne detection uses dual frequencies. Since we measure only one characteristic of the signal, it can yield a better precision than with heterodyne measurement which captures two characteristics using two conjugate operators.

**HPQC**: High Performance Quantum Computing, a quantum analogue of HPC (High Performance Computing). These are currently theoretical models of quantum mainframes comprising giant matrices of qubits that can be partitioned for shared use by several users. See High Performance Quantum Computing, 2011 (7 pages).

**Hubbard model**: physics simulation model of mixed conducting and isolated systems based on a simple Hamiltonian. Mentioned in the sizing benchmark used by Amazon for its cat-qubits fault-tolerant quantum computing system being currently designed.

**Hybrid quantum algorithm**: an algorithm that combines classical processing running on classical computers and some processing performed on quantum computers, where needed.

**Hyperfine structure**: small splitting of atomic energy levels or spectral lines with electrons with the same quantum numbers into several distinct components that are explained by the interactions between the nucleus and electron clouds.

**Indistinguishability**: relates to bosons quantum objects that have the same quantum state in a given location and are impossible to separate with any measurement tool.

**Indistinguishable photons**: see Indistinguishability, photons being a common type of boson.

**Integrated Quantum Photonics** (IQP): technologies exploiting photons as quantum information carriers and implemented on chipsets using wafer-scale fabrication, mostly in silicon-based CMOS or with III/V materials like gallium arsenide (GaAs) and indium phosphide (InP). IQP is used in quantum telecommunications and computing. It is using optical waveguides to guide and route single-photons, provides miniaturized split and phase control circuitry, entangled state generation, overall manipulation and sometimes even photons generation and photons detection.

**Interference**: fundamental phenomenon of quantum physics used with the wave aspect of quantum objects, when several waves can add or annihilate with constructive and destructive interferences. Is the basis of gate-based quantum algorithms!

**Invertible computation**: involves computations that run both forwards and backwards so that the forward/backward semantics form a bijection. In classical computing, it can correspond to some symmetric logical circuits that can process data forward and backwards with both ends used as inputs and outputs. It's used for example in MemComputing classical processors. The principle was created by Supriyo Datta from Purdue University in Indiana, USA.

**Irreversible**: said in computing of a calculation that makes it impossible to compute the initial values with using the result of the calculation. This is the case with all two-classical bits gates (NOR, OR, AND). Contrarily, quantum computing gates are mathematically reversible since relying on unitary transforms that, multiplied with their transconjugate, generate an identity operator. In plain language, if you apply an unitary (set of quantum gates) to a set of qubits, you can reverse this computing with the transconjugate of this unitary. Practically, it means playing in reverse order the gates initially applied. This technique is used in the uncompute trick that we describe elsewhere.

**Ion**: non-neutral atom, which has a positive or negative electric charge. It is negative if its number of electrons exceeds the number of protons (anions) and positive in the opposite case (cations).

**IonQ**: an American startup from the University of Maryland that pioneered the first commercial quantum computers using ion traps. Their operational record as of 2021 was 11 qubits with 32 qubits to be made readily available.

**I/Q mixer**: in phase and quadrature mixer, which adds two pulse signals of same frequency but with different amplitude and phase. The In-phase signal is a sinusoid and the quadrature signal is a cosine. They have a delay of π/2 or 90°. When added up in the mixer, the sum of both signals create an arbitrary phase and amplitude signal of the same frequency.

**Ising** (model): a statistical physics problem that can be simulated and solved using quantum algorithms, especially on quantum annealers like those from D-Wave. It models the interactions between two-levels particles (spin, ferromagnetism). All algorithms for D-Wave annealers are reduced to solving an Ising model.

**Isotopes**: variations of atoms where the number of protons and electron is the same, sharing the same atomic number, but when the number of neutrons is different. For example, helium can exist in the for $He^3$ and $He^4$ with one and two neutrons. Many materials involved in quantum technologies are used with particular isotopes, like $Si^{28}$ in silicon wafer used with electron spin qubits, the reason being the number of neutrons has an influence on atom nucleus spins, that can interfere with their electron spins.

**IT**: information technologies.

**Jaynes-Cummings Hamiltonian**: Hamiltonian used to describe the total energy of a system linking a resonator usually implemented as a coplanar waveguide (CPW) resonator with a superconducting circuit.

**JJ**: "jay-jay", nickname for Josephson junctions.

**Josephson** (effect): physical phenomenon happening in a superconducting current loop traversing a thin insulating barrier known as a Josephson junction (JJ) like some non-superconducting metal thanks to the tunneling effect. It enables the creation of a multiple level energy or phase state for the superconducting current. This technique is used in superconducting qubits from quantum systems such as those of IBM and Google. It is also used in quantum sensing with SQUIDs (superconducting quantum interference devices) that are used as very sensitive magnetometers.

**JPA** (Josephson Parametric Amplifiers): simple amplifiers, using one or two Josephson junctions that are used for the first stage amplification of readout microwaves in superconducting or silicon spin qubits. Their narrow bandwidth prevents their implementation with frequency-domain qubits readout multiplexing.

**Kerr effect**: when some materials refractive index is modified in a nonlinear (quadratic, second-order nonlinear) manner as a function of the electric field applied to them. Is a variant of Pockels effect.

**Ket**: vertical vector describing in Dirac's notation the state of a quantum object, with the symbols | and ⟩ forming a psi vector noted $|\psi\rangle$. It contains complex number amplitudes defining the relative weights in the computational basis. For a qubit, it's a 2 complex numbers vector. For a register of N qubits, it's a $2^N$ size vector of complex numbers defining the amplitudes of each combination of N 0s and 1s, which are orthogonal states in the $2^N$ state vector Hilbert space.

**Kochen-Specker theorem**: no-go theorem that states that it is impossible to assign simultaneously values with certainty to all observables in all possible contexts. This simple observation contradicts classical physics, where such an assignment is quite possible. It is the formal proof of quantum contextuality.

**Larmor frequency**: frequency of the Larmor precession (magnetic moment rotation). It is frequently mentioned in papers related to electron spins qubits.

**Larmor precession**: rotation of the magnetic moment of an object like an electron when it is exposed to an external magnetic field. This rotation happens along the axis of the magnetic field.

**Laser**: coherent light source invented in 1960 and used in many fields such as CD and DVD players, fiber optic communications, surgery, ophthalmology and dentistry, LiDARs. They are also often found in quantum computing to control cold atoms or manage photon-based qubits as well as in quantum cryptography and telecommunications (QKD & co). Laser means Light Amplification by Stimulated Emission of Radiation. It is a source of coherent light, i.e. it consists of photons of the same polarization, phase and wavelength, and emitted in the same direction in a narrow beam. Light amplification uses a process of stimulated emission in an amplifying active medium made of solid, fiber, liquid, gas or semiconductor which is placed in the center of a resonant optical cavity with a reflecting mirror on one side and a semi-reflecting mirror on the other side, which allows the light beam to exit. The wavelength and power of the light radiation depends on many parameters. The energy comes from an excitation or pumping system: primary laser, laser diode, flash lamp or electric discharge.

**Leggett-Garg inequality**: mathematical inequality fulfilled by macroscopic physical theories and systems. It says that a macroscopic object which has two or more distinct states, is at any given time in of those states. And it is possible in principle to determine which of these states the system is in without any effect on the state itself, or on the subsequent system evolution. This inequality is violated by quantum systems when superposition and entanglement are put in play like in interference processes.

**Lindblad equation**: equation describing the time evolution of the density matrix ρ of a quantum system that preserves the laws of quantum mechanics, meaning it preserves the trace and positiveness of the matrix. But the transformation is usually not a unitary due to decoherence. Also named a Lindbladian, a quantum Liouvillian, and in the long form, a Gorini–Kossakowski–Sudarshan–Lindblad equation (GKSL equation, for Vittorio Gorini, Andrzej Kossakowski, George Sudarshan and Göran Lindblad).

**Linear algebra**: branch of mathematics that is used in quantum physics and quantum computing. It is based on the manipulation of vectors and matrices within Hilbert spaces. In particular, the state of a sets of qubits is represented by vectors in a Hilbert space of size $2^N$ when N is the number of qubits. Computing with qubits consists in applying linear transformations.

**Linear optics**: field of quantum mechanics that manipulates photons based on their classical properties: polarization, phase or frequency.

**Locality** (principle): in classical physics, principle according to which distant objects cannot have a direct influence on each other. An object can only be influenced by its immediate environment. This principle derived from Albert Einstein's restricted relativity is questioned by quantum mechanics, non-locality and quantum entanglement observed experimentally since at least 1982 with photons, in Alain Aspect's famous experiment (with Philippe Grangier and Jean Dalibard). But there are various interpretations of quantum physics which explain entanglement without resorting to non-locality.

**Logical Qubit**: an assembly of physical qubits implementing hardware and software quantum error correction. Seen from the software developer's point of view, it creates a virtual logical qubit with a very low error rate. The fidelity of logical qubits depends in particular on the number of physical qubits they contain, the quality of the error correction codes and the qubits fidelity stability with the increase in the number of physical qubits.

**LSQC**: Large Scale Quantum Computing also frequently called FTQC for fault tolerant quantum computing. Category of future fault tolerant quantum computers. These will be based on the use of numerous physical qubits assembled into logical qubits with a very low error rate as seen from the software. Precisely, an LSQC implementing fault-tolerance has error corrections codes with at least two characteristics: it must not propagate errors broadly in the physical qubits and it must be able to implement non-Clifford group qubit gates like the single qubit gate T or the three qubits gate Toffoli. But LSQC definition is not clear yet. It could pertain to a large number of physical qubits (not necessarily arranged in logical qubits) or a large number of logical qubits, way beyond the first generations of FTQC. The jury's out to settle dusts with this terminology.

**Magic states distillation**: process that converts a set of noisy qubits into a smaller number of qubits with a lower noise. It is particularly useful for non-Clifford group quantum gates that bring universal computing power and exponential speedup. It is one of the ways to create fault-tolerant quantum computers but it has a high overhead cost with physical qubits. It was proposed in 2004 by Emanuel Knill, Sergey Bravyi and Alexei Kitaev.

**Magneto-Optical Trap** (MOT): device used to cool down and trap a cloud of neutral atoms. It uses a combination of magnetic trapping using two coils and Doppler effect in three orthogonal directions for cooling. The technique is used in cold atom interferometry (cold atoms gravimeters) and cold atom computers.

**Majorana fermion**: an electron-based quasiparticle in superconducting materials that could be used to manage reliable qubits in so-called topological computing. This virtual particle was imagined by Ettore Majorana in 1937. Microsoft intends to build a quantum computer based on these quasiparticles. But their very existence has not yet been really demonstrated.

**Manifold**: corresponds to the discrete controllable states of a quantum object.

**Matrix**: mathematical object made of rows and columns of values.

**Matter wave**: principle of quantum physics enacted by Louis De Broglie in 1924 according to which massive objects can also behave as waves. The De Broglie wavelength of a massive particle is the Planck constant divided by its momentum.

**MBE**: molecular beam epitaxy, a variety of PVD process (thin-film deposition), used to create a single orderly crystal structure in semi-conductor manufacturing. Is notably used to produce semi-conductors quantum dots in III/V materials.

**MBQC**: Measurement Based Quantum Computing, a quantum computing method invented in 2001 by Robert Raussendorf and Hans Briegel that uses a high number of groups of pre-entangled qubits, called cluster states, embedded in two-dimensional grids in which qubit state readouts modify the grid structure and help create quantum gates. The last measured qubit gives the result of the algorithm. This technique is particularly useful with flying qubits like photons because it can be implemented in a highly parallel way and support the finite depth of quantum gates that these qubits enable.

**Mesoscopic**: subdiscipline of condensed matter physics that deals with materials of an intermediate size. The size ranges from a couple atoms to a µm.

**Metal layers**: in semiconductor chipsets, correspond to the layers containing wires connecting the various transistors and other electronic elements. These layers are surrounded by some insulator (silicon oxide or other). In typical CMOS processors, you have over 12 metal layers of decreasing density as you move father from the logic layer, In superconducting qubits, you have no metal layers on top of the Josephson junctions since it must be as isolated as possible.

**Microring resonator**: tiny optical waveguides looped back onto themselves in circle or spiral which enable interference phenomena, the creation of delay lines, and various other optical devices used for example in entangled photon generation.

**Mode-locked laser**: pulse laser generating streams of very short pulses of light formed of wave-packets in the picosecond to femtosecond range. These pulses are generated thanks to the emitted photons being synchronized in phase. A synonym of mode-locked is phase-locked!

**Mott insulator**: material that are expected to conduct electricity but are actually insulators, particularly at low temperatures, and under certain conditions which can be controlled, leading to so-called Mott transitions.

**Mott transition**: change in a condensed matter material's behavior from insulating to metallic generated by multiple factors like and ambient electric field changing the band structure of the material like with some metal oxides.

**MVP**: Minimum Viable Product. Concept used mostly in startups consisting in creating the simplest form of a product before starting to sell it. Opposite to full-fledged product with tons of R&D and an ever-lasting perfectionist approach.

**MINLP**: Mixed Integer Non Linear programming, a class of complex problems that can potentially be solved with quantum algorithms. It is about finding the minimum(s) of nonlinear functions and under constraints that aim to respect nonlinear functions. The variables in the equation are a combination of integers and floating-point numbers. The applications are numerous in all cases where one seeks to optimize a constrained function (energy distribution, optimum take-off of an aircraft, optimization of financial portfolio, minimizing risk in insurance or credit, etc.).

**Mixed state**: quantum objects state that is a classical statistical combination of several pure states. They can be prepared with physically associating several sources of pure states, like with merging two laser beams in one beam. A subsystem of an entangled quantum objects system is also a mixed state. A mixed state is mathematically represented by a density matrix operator, providing all the information that can be obtained about the related quantum system.

**Momentum**: physical property of an object or particle that for a massive particle is equivalent to its mass multiplied by its velocity. Usually denoted p. A (massless) photon has a momentum equal to their wavelength multiplied by Planck's constant.

**Multimode**: said of an optical fiber with a larger core (about 50 to 62 µm) where several light beams can be transported, usually with different wavelengths. Light propagation use bouncing inside the fiber walls. These fibers are used for short distances communications of less than a kilometer and with bit rates reaching 200 GBit/s. The contrary of multimode fibers are monomode fibers. Also said of multimode photons, with an entirely different meaning and a way more complicated one, never explained in plain language by quantum photonicians. Its contrary is single mode photons. A single mode photon has one complex amplitude while a multimode photon is a mixed state of single mode photons with several independent complex amplitudes. If you want to know more, you get to use a complicated mathematical formalism.

**Mutually unbiased bases**: it is a concept mostly used in quantum key distribution. A bases is a set of orthogonal vectors in a Hilbert space. Mutually unbiased bases are two bases where the measurement of one of the vectors of one basis will yield a random result on the other base.

**NISQ**: Noisy Intermediate-Scale Quantum, a name for current and near future gate-based quantum computers, which are intermediate in terms of number of qubits (a few tens to hundreds) and subject to quantum noise that limits their capabilities. This acronym was created by John Preskill.

**No-go theorem**: theorem that demonstrates that a physical phenomenon is not possible. In quantum physics, famous no-go theorems are Bell's theorem and the Kochen–Specker theorem which con-strain hidden variable theories trying to explain non-locality and entanglement with an underlying deterministic model featuring hidden states and variables. You also have the no-cloning and no-deleting theorems which prevents the cloning and deletion of a quantum object state.

**Non-Clifford gates**: said of quantum gates that are outside the Clifford group itself based on combining Pauli gates (half-turns in Bloch's sphere), Hadamard gates (quarter turns) and CNOTs for entanglement. To make things simple, non-Clifford gates enable the creation or arbitrary rotations in Bloch's sphere and their multi-qubits gates derivatives. The single qubit T gate (one eighth turn in Bloch's sphere) is the minimum additional gate, that, combined with the others, enable by approximation the creation of any arbitrary gate and unitary transformation.

**Non cloning theorem**: prohibits the identical copy of the state of a quantum. Therefore, it is impossible to copy the state of a qubit to exploit it independently of its original. Any copy destroys the original!

**Nonlinear optics**: field of optics where the optical properties of materials depend on the light amplitude and lead to the creation of new frequencies. Nonlinearity qualifies the response of a medium to an excitation that is generally quite energetic from intense fields, mainly from lasers, especially femtosecond pulsed lasers. In this case, the response of a material to the sum of two electromagnetic fields is not equal to the sum of the response to each individual field. Nonlinear optics can be used to create two-photon quantum gates with continuous variables photons. See also Chi.

**Non-locality**: principle allowing a (quantum) object to influence the state of another (quantum) object at a distance, which can be very large. Contradicts the principle of locality, which means that an object can only influence another object at close range. Photons quantum entanglement at great distances verifies the non-locality. However, the initial quantum state of both objects is always random. So it doesn't transmit a predetermined information per se from one place to the other.

**NMR**: Nuclear Magnetic Resonance, a type of qubit that was investigated in the 1990s and early 2000s and was then nearly abandoned. The reason is it didn't scale well at all and these were very noisy qubits and difficult to entangle. It was based on exploiting quantized states of atoms nuclei spins. However, the Chinese startup SpinQ is offering a desktop NMR-based quantum computer with 2 to 5 qubits. It's useful only for educational tasks.

**Non-classical light**: forms of light and electromagnetic fields treated as quantum systems. It contains single photon wave packets, pairs of entangled photons and squeezed states of light.

**Non-demolition measurement**: see QND.

**NOON state**: many-body entangled state superposing N quantum objects, usually bosons, in two modes. Namely, it superposes all the objects in one mode and all the objects in the other mode. This kind of superposition is used in quantum metrology to obtain a precision reaching the Heisenberg limit, which is better than a standard quantum limit based measurement.

**Normalization**: in quantum physics, normalization is used in many situations like with scaling wave functions so that the sum of probabilities equal one. This 1 is considered as a normalization constant or constraint.

**NP** (problem class): class of problems whose solution is verifiable in a polynomial term relative to the size of the problem. It includes the so-called exponential or intractable problems, whose solution time is exponential with respect to their size. A quantum computer is supposed to solve some NP problems in a tractable way, meaning, not exponential time.

**NP-complete** (problem class): decision problem for which it is possible to verify a solution in polynomial time and for which all problems of the NP class are reduced to it via a polynomial reduction. This means that the problem is at least as difficult as all other problems of the NP class. The problems of the traveling salesperson and the knapsack problem are Complete NP problems. The concept dates from 1971 and comes from Stephen Cook.

**NP-difficult** (problem class): problem to which any problem of the NP class can be reduced by a polynomial reduction. If it is also in the NP class, it is said to be an NP-complete problem. If $P \neq NP$, then NP-difficult problems cannot be solved in polynomial time.

**Observable**: equivalent in quantum mechanics of a physical quantity in classical mechanics, such as position, momentum, spin or energy. In quantum physics, an observable is a mathematical operator used for the measurement of one property.

**ODMR**: optically detected magnetic resonance is a quantum sensing testing is a double resonance technique where the electron spin of a crystal defect like a NV center is optically pumped for initialization and readout with a green laser light. It radiates some red light or nothing depending on the cavity electrons spin. It uses the Zeeman effect in unpaired electrons. With $NV^-$ centers, it is used for high-precision magnetometry and medical imaging with a sensitivity ranging from $10^{-9}$ to $10^{-15}$ $T/\sqrt{Hz}$, the unit of magnetometry precision.

**On-premises**: said of hardware that sits in a customer site or datacenter. It is the opposite of sitting in a datacenter from a cloud vendor.

**Optical molasses**: gas of cold neutral atoms whose cohesive strength is of the viscous type. It is cooled with lasers using the Doppler effect, usually with three pairs of lasers in three orthogonal directions.

**Optical pumping**: technique used to modify the states of atoms by increasing their energy level using polarized photons. Alfred Kastler, invented it in 1950 and was awarded the Nobel Prize in Physics in 1966. The technique is used in lasers and quantum sensing. Optical pumping passes through three to four energy levels of atoms (E0, E1, E2, E3). Pumping moves an atom from its fundamental level E0 to E3. A (mechanical) relaxation brings the atom back from the E3 state to E2. In lasers, this generates a population inversion between the E1 and E2 states, so that there are more atoms in the E2 state than in the E1 state. The spontaneous and stimulated emission of photons of E2-E1 energy can then take place. The atom in the E1 state then returns to the E0 state by relaxation.

**Orbital angular momentum** (OAM): is one of the two angular momenta of photons with spin angular momentum. Discovered in 1992 by Les Allen et al from Leiden University, this phenomenon is more difficult to visualize than spin angular momentum. With OAM, the photon itself is rotating along its propagation axis or vector. One analogy with the Earth is its own rotation (spin angular momentum, defining days) and its rotation around the Sun (orbital angular momentum, defining years). This orbital angular moment is quantified with integers times the reduced Planck constant. It can be any integer! One record OAM number of 10.100. Being quantified, it can lead to superposition and entanglement. It can also be used to encode information on fibers.

**P** (problem class): problem that can be solved in polynomial time with respect to its size, on a deterministic Turing machine.

**Paramp**: parametric amplifier using a parametric nonlinearity and a pump wave. Paramps exist for photos in the visible spectrum (these are OPA for optical parametric amplifiers) as well as for microwaves. In this last case, they are used to amplify readout microwaves from superconducting or silicon spin qubits. The most recent breed of paramps are the TWPAs.

**Pauli** (exclusion principle): postulates that two fermion particles cannot be in the same quantum state. Two electrons or two neutrons cannot be in the same place with the same energy level. If an external force such as gravitation forces them to be in the same place, they cannot have the same energy, i.e. the same speed. If a set of fermions has to be in the same place, they will have to have different velocities. Fermions have half-integer spins.

**Permanent**: real number resulting from n! additions of multiplications of n values of a square matrix n*n. They are used to evaluate the complexity of matrices representing graphs.

**Phase Estimation Algorithm**: algorithm created by Alexei Kitaev in 1995 and used to find the phase of an eigenvector of a unitary operator U. This algorithm is based on an inverse QFT. It is used as part of period finding in Shor's factoring algorithm and in quantum chemistry algorithms.

**Phase**: an important physical properties of quantum objects given they all can behave as waves. It explains interferences between all sorts of quantum objects, like electrons on top of photons.

**Phasor diagram**: two-dimensional diagram describing electromagnetic field quadratures positioning the statistic characteristics of a photon source, with X1 and X2 orthogonal axis corresponding to two oscillating electric fields that are out of phase by 90°.

**Phonon**: collective excitation in a periodic, elastic arrangement of atoms or molecules in condensed matter, specifically in solids and some liquids. In quantum information technologies, it is mostly used with trapped ions to provide a n-to-n connectivity between qubits.

**Photoelectric effect**: emission of electrons from a material like a metal when electromagnetic radiation above a certain minimum frequency strikes it, independently of its intensity. Formalized by Albert Einstein in 1905.

**Photolithography**: patterning process in semiconductor manufacturing used to define in which zones matter must be removed or added in subsequent steps. It uses ultraviolet rays illuminating a photomask that exposes a photoresist film or coating. For very high densities, the exposure is done with extreme ultra-violet waves. The related manufacturing tools are now produced by a single company in the world, ASML (The Netherlands).

**Photon**: quantum of energy associated with electromagnetic waves ranging from radio waves (long waves, low frequencies) to gamma rays (very short waves, very high frequencies) through visible light. Its mass is zero. Its spin is 1 and it is therefore part of the bosons. Photons are absorbed an emitted by atoms during energetic levels changes.

**Photon measurement**: measurement of a photon where the degree of freedom is the excitation quanta. It can yield a number of superposed photons that we try to detect, but not their characteristics like their phase or frequency, which requires homodyne or heterodyne measurement.

**PKI**: Public Key Infrastructure, set of roles, policies, hardware, software and procedures used to create, manage, distribute, use, store and revoke digital certificates and manage public-key encryption.

**Planck constant**: fundamental constant of quantum physics ($h=6.626x10^{-34}$ Js). Created in 1900 with Max Planck's explanation of black body radiation spectrum and then used in most other quantum physics equations, including Schrodinger's wave equation.

**Pockels effect**: effect used in optical modulators where a medium refraction index changes in a linear manner as a function of the electric field applied to it.

**Polarized Beam Splitter** (PBS): class of beam splitters that use birefringent materials to split light into two beams of orthogonal polarization states.

**Polaritons**: quantum quasi-particles with strong interactions between light and matter in semiconductors. It results from the coupling between photons and an electrical polarization wave which occurs in particular in plasmons (oscillations of free electrons in metals), phonons (oscillations of atoms, especially in crystalline structures) and excitons (pairs of electron holes generated by photons in semiconductors).

**POVM**: Positive Operator-Valued Measure, quantum measure generalizing Projection-Valued Measures (PVMs) which is useful when the measurement basis is not made of orthogonal states in their Hilbert space. POVMs that are not PVMs are called non-projective measurements. They have many use cases like enhancing quantum states tomography, help detect entanglement and allow unambiguous state dis-crimination of non-orthogonal states, with applications in quantum cryptography and randomness generation

**PQC**: Post Quantum Cryptography, cryptography resistant to quantum computers-based codebreaking algorithms. It is based on the use of public keys that are not decomposable with conventional or quantum computers.

**PQS**: Programmable Quantum Simulator, or analog quantum computers.

**Private Key**: key used in private key encryption systems. Keys are exchanged beforehand by the parties using an encryption algorithm, often hash or Diffie-Hellmann algorithms.

**Property**: physical characteristic of a physical object. In quantum physics, observables are the mathematical operator used to compute properties values using the quantum object state vector. For a photon, it can be for example its phase, polarization and wavelength. In quantum physics, it's not possible to evaluate the values of all properties of quantum systems to describe it, due to Bohr's complementarity principle.

**Public key**: an encryption system that involves sending a public key to an interlocutor who will use it to encrypt a message sent in the other direction. The elements used to create this public key are used to decrypt the message sent. It is normally impossible or very difficult to decompose the public key to find the elements that were used to create it. PQCs are based on public keys.

**Purcell effect**: relaxation or loss of energy of a superconducting qubit through its readout resonator. More generally, it's the enhancement of a quantum system's spontaneous emission rate by its environment, discovered in the 1940s by Edward Mills Purcell with the spontaneous emission rates of atoms when incorporated into a resonant cavity.

**Purcell filter**: high and low-band filter that reduces the Purcell effect between a superconducting qubit and its readout resonator.

**Pure state**: quantum state of an isolated quantum system of one or several objects constructed as a linear superposition of the states from its computational basis.

**Purification**: process applied to a mixed which integrates it in a larger system to create or recon-struct a pure state. It can be applied to a set of entangled qubits as well. It is used in some error-correcting codes both in quantum telecommunications and quantum computing.

**PVD**: physical vapor deposition, is a material deposition technique in semiconductor manufacturing where the material to deposit is first turned into vapor and then condenses on the target surface. There are various PVD methods like sputtering that is using ion projection to pull material from a source and deposit it on the target,

**PVM**: Projective Value Measurement, used in quantum computing, consists in doing a geometrical vector projection of your qubit pure state on any axis in the Bloch sphere.

**Q factor**: quality factor, dimensionless value defined as the ratio between the energy stored in a resonator and the energy dissipated per oscillation cycle times $2\pi$. With the frequency of the oscillator, it provides an indication on the oscillator lifetime. In a superconducting qubit, it characterizes the stability of its oscillation and determines its $T_1$ or relaxation time. The greater the Q factor is, the longer the $T_1$ is. The higher the better, this factor can exceed $10^7$. The dissipation comes from cavity losses and depends on the materials and structure of the electromagnetic cavity. Another definition for Q factor for an oscillator is the ratio between the main resonance frequency and its bandwidth.

**QCaaS**: quantum computing as a service, a fancy acronym for quantum computing running in the cloud.

**QFHE**: Quantum Fully Homomorphic Encryption. A method of quantum information encryption allowing to perform processing on encrypted data.

**QFT**: Quantum Fourier Transform. Quantum variation of the Fourier transform. The classical Fourier transform allows to decompose a signal (as in audio) into frequencies (or frequency spectrum). The QFT does this on a sequence of integers and determines its largest observable frequency.

**QIP**: Quantum Information Processing, a name sometimes used to information tools based on second-generation quantum technologies. It contains quantum computing, quantum simulation, quantum cryptography and quantum telecommunications.

**QHO**: quantum harmonic oscillator.

**QKD**: Quantum Key Distribution, a secure protocol for sending symmetrical keys via an optical link based on quantum entanglement (fiber or satellite). These keys are tamper-proof, or at least an interception of the key is detectable.

**QLM**: Quantum Learning Machine, name of the Atos quantum emulator appliances using classical hardware (Intel and/or Nvidia).

**QMA**: Quentin Merlin Arthur, a class of problems that is verifiable in polynomial time on a quantum computer with a probability greater than 2/3. It is the quantum analogue of the "traditional" NP complexity class. QML: Quantum Machine Learning. Branch of quantum algorithms used in machine learning.

**QML**: Quantum Machine Learning. Class of quantum algorithms implementing machine learning or deep learning techniques.

**QND**: quantum non-demolition measurement, a sequence of measurements where results are completely predictable using the result of the first measurement. Practically, it stays the same. Let's say we measure the state of a qubit that yields a $|0\rangle$ or a $|1\rangle$. The measurement is a QND one if a new measurement will yield the same $|0\rangle$ or a $|1\rangle$ coming out of the first measurement and so on. In mathematical parlance, it means the measurement observable commutes with itself at different times. How about a destructive measurement of a qubit? It can happen for example with a photon detector which absorbs it. The photon is then entirely destroyed (and converted to some current in the detector) and cannot be measured a second time. Usually, a QND signal is quantum and extremely weak and is obtained with a quantum probe.

**QRNG**: Quantum Random Number Generator, the optical random number generators used in quantum cryptography, like those of the Swiss IDQ.

**QSVT** (Quantum Singular Value Transformation): quantum algorithm that performs a polynomial transformation of the singular values of a linear operator embedded in a unitary matrix. Was created by András Gilyén, Yuan Su, Guang Hao Low and Nathan Wiebe in 2018.

**Quantization**: in quantum physics, happens with quantum objects having some physical properties that are discontinuous and not continuous, like electron energy levels and electron spins.

**Quantum accelerator**: quantum computer used as a complement to a supercomputer or HPC, usually to run hybrid algorithms like VQE (Variational Quantum Eigensolvers) combining a classical part that prepares the data structure that feeds a quantum accelerator.

**Quantum advantage**: occurs when a quantum computer executes some processing faster than its optimum equivalent adapted to a supercomputer, with a useful algorithm. This advantage can be declined on another aspect than the duration of the calculation. For example, a quantum energy advantage relates to energy consumption instead of computing time.

**Quantum annealing**: technique used to find the global minimum of a given objective function over a given set of candidate solutions, based on using quantum fluctuations. It is used to solve combinatorial optimization problems with a discrete search space. This computational process is in D-Wave quantum computers.

**Quantum channel**: transformation of a quantum state resulting from any kind of interaction with a quantum environment. It is modelized with a density matrix super-operator. It is useful to modelized subsystems, decoherence, quantum error correction and qubits noise.

**Quantum Chaos**: branch of physics studying how chaotic classical dynamical systems can be described with quantum theory. It deals with the relationship between quantum mechanics and classical chaos and with the boundaries between classical and quantum physics in modelling chaos.

**Quantum chromodynamics**: describes the strong interaction, one of the four fundamental forces, that governs the interactions between quarks and gluons and the cohesion of atomic nuclei. Why "chromo"? Because we describe the states of elementary particles with color codes: blue, green and red for particles, then anti-blue, anti-green and anti-red for anti-particles. This theory is based on the quantum field theory. This part of quantum physics is not used in the creation of qubits. It is used for the physics of elementary particles and is verified in large particle accelerators such as the CERN LHC in Geneva.

**Quantum circuit**: see circuit.

**Quantum cognition**: descriptive model of the functioning of human knowledge (language, decision-making, memory,

conceptualization, judgment, perception) based on the mathematical formalism of quantum mechanics, proceeding mainly by analogy, without going through physical explanations or quantification of the neurosciences, which themselves fall within the "quantum mind" field resulting from the work of Roger Penrose.

**Quantum dots**: we can mention at least three different types of quantum dots: the powders used in LCD screens that convert the blue backlighting LED light into green or red light based on their grain size. Then we have the quantum dots used to generate single photons. At last, we have quantum dots used to trap electron spins in spins qubits.

**Quantum Electro-Dynamic** (QED): branch of quantum physics, or QED, which is "*a physical theory that aims to reconcile electromagnetism with quantum mechanics using a relativistic Lagrangian formalism. According to this theory, electric charges interact by photon exchange*" (Wikipedia). This is the basis of the quantum field theory which applies to all elementary particles.

**Quantum emulator**: a software and/or hardware system using a conventional computer to run and test some software programmed for a quantum computer. This makes it possible to test quantum programs without a quantum computer. The execution speed is not as good as on a quantum computer as soon as you exceed a few tens of qubits. And beyond about fifty qubits, the capacity of classical machines is insufficient to perform it properly. Emulation should not be confused with quantum simulation, which simulates quantum physics phenomena with an analog quantum processor like those using cold atoms.

**Quantum engineering**: is about developing quantum technologies in computing, telecommunications, cryptography and/or sensing with a pluridisciplinary approach merging quantum physics and other related sciences and technologies like thermodynamics, cryogeny, electronics, semiconductors, cabling, mathematics, information theory, programming and the likes.

**Quantum foundations**: branch of science philosophy that aims to build some understanding and description of the real world in quantum physics and, as such, associate it to some ontology.

**Quantum gates**: operations modifying the state of one or several qubits. Multi-qubit gates (Toffoli, Fredkin, ...) exploit the principle of quantum entanglement. The operations of quantum gates are generated by physical actions on the qubits which depend on their nature. For superconducting qubits, this involves sending microwaves between 5 and 10 GHz via electrical conductors. For trapped ions, these are laser-controlled operations. For electron spins qubits, these are a mix of electrical voltages and microwave pulses. For qubits based on mass particles (electrons, ions, cold atoms), quantum gates act on the qubits but these do not move in space. For flying qubits based on photons or electrons, these circulate and cross quantum gates which modify their state (phase, frequency, or other).

**Quantum Hall effect** (QHE): or integer quantum Hall effect is a quantized version of the Hall effect which is observed in two-dimensional electron systems at low temperatures and under a strong magnetic field, in which the Hall resistance Rxy has quantized values. It is related to the field of quantum matter. The effect was discovered by Klaus von Klitzing from the MPI in Germany in 1980 and was awarded the Nobel prize in physics in 1985.

**Quantum hydrodynamics**: studies the hydrodynamic effects of quantum systems such as superfluid elements (helium at very low temperature) or polaritons and associated light fluids.

**Quantum Internet**: marketing term describing a quantum network enabling the quantum telecommunications based on entanglement, particularly to connect quantum systems like quantum computers and quantum sensors. By extension, it also includes quantum key distribution infrastructures that are used to secure information exchange with encryption keys that are shared quantumly between senders and receivers.

**Quantum medicine**: in general, false science and charlatanism based on a totally fanciful interpretations of quantum mechanics.

**Quantum Non Demolition measurement** (QND): type of measurement in which the uncertainty of the measured observable does not increase from its measured value during the subsequent normal evolution of the system. QND measurements are the least disturbing type of measurement in quantum mechanics. In other words, for a qubit, it would mean that after a $|\mathbf{0}\rangle$ or $|\mathbf{1}\rangle$ is measured, subsequent measurements will always yield the same $|\mathbf{0}\rangle$ or $|\mathbf{1}\rangle$ that was obtained in the first place.

**Quantum number**: variables describing quantum objects physical quantities or variables that are discrete. Electrons have four quantum numbers: principal quantum number (energy level or electron shell), angular momentum also named azimuthal or orbital quantum number describing electron subshell, magnetic quantum number describing the electron energy level within its subshell and spin projection quantum number, being either +1/2 or -1/2, in a given spatial direction.

**Quantum postulates**: basis of quantum physics formalism. These are postulates and not laws because it describes a mathematical formalism that cannot be proved per se. There are many different presentations of quantum postulates in reference sources (Nielsen & Chuang, Preskill, Cohen-Tannoudji, Wikipedia, ...). Depending on the sources, you'll find 3, 4, 5, 6 or even 7 of them.

**Quantum Physical Unclonable Functions** (qPUF): quantum based physical identifiers that can be used to create unique and unclonable security keys.

**Quantum reservoir computing**: specific category of recurrent neural networks used to process time series. It uses a set of neuron weights and links between neurons randomly fixed in the reservoirs, all with nonlinear activation functions. The hundreds of neurons in a reservoir are fed by input data stored in the reservoirs. The activation functions nonlinearity makes this memory evanescent. The training parameters of these networks are located in the weights of the neurons that connect the reservoirs to the output data

**Quantum reservoir engineering**: set of techniques for managing qubits through their interaction with a "quantum thermal bath" (quantum bath) to reduce energy consumption, reduce the duration of the measurement of the state of the qubit and allow a non-destructive and reversible measurement of this state ("Quantum non-demolition" or QND). Reservoir engineering is used in cat-qubits.

**Quantum simulator**: name given to analog quantum computers that are capable of simulating quantum objects and solving related problems, particularly in materials physics. By abuse of language, the name is used for supercomputers capable of executing quantum algorithms by numerical simulation. In this case, it is preferable to use quantum emulator.

**Quantum state**: mathematical object used to compute at a given time the probabilities of a quantum object or set of object property values that would be obtained when measuring it and to predict their evolution over time. It is usually represented by a vector in a Hilbert space (linear, metric and complete). This is however only the case for a pure state. A mixed state is represented by a density matrix. The notion of quantum state is usually the first quantum postulate.

**Quantum state tomography**: technique used to characterize the quality of qubits and qubits gates or any quantum channel. It is used to experimentally reconstruct a density matrix of a set of qubits. It also requires a lot of classical computing to process the experimental data obtained with repeated state preparation and measurements.

**Quantum Steering**: quantum measurement phenomenon when one subsystem can influence the wave function of another subsystem by performing specific measurements.

**Quantum supremacy**: describes a situation where a quantum computer can perform some computation that is inaccessible to the best current supercomputers with the best classical algorithm and in a humanly reasonable time. The computing time differential between quantum computing and classical computing must be several orders of magnitude. It can deal with a useful calculation or not. Thus, the quantum supremacy claimed by Google in October 2019 dealt with a random algorithm that had no practical interest. The term was coined by John Preskill in 2011. Nowadays, the trend is to use the quantum advantage denomination.

**Quantum switch**: consists in creating a series of qubit transformations that can be implemented simultaneously in different orders, creating an indefinite causal order computing flow.

**Quantum teleportation**: technique used to transfer the state of one qubit to another location. It is usually performed with three communication links: a pair of previously entangled photons and two classical bit links. It has many uses such as in quantum cryptography (QKD). The no-cloning theorem also says that the state of a teleported quantum disappears from the source after teleportation. It can be used to transmit a rich quantum state of several qubits and can enable distributed quantum computing.

**Quantum variational circuits**: type of quantum algorithm used to implement machine learning.

**Quasi-particles**: physical concept which treats elementary excitations in solids like spin waves, as particles. As the particles do not consist of actual matter, they are called quasi particles. Majorana fermions and polaritons are examples of quasi-particles.

**Qubit** or physical qubit: the elementary unit of information in quantum computing in quantum computers and quantum telecommunication. It stores a quantum state associating two distinct states of a particle or of a quantum system (electron spin, energy level of a superconducting loop, energy level of a trapped atom or ion, polarization or other property of a photon). Its mathematical representation is a vector comprising two complex numbers in a Bloch sphere.

**QUBO**: Quadratic Unconstrained Binary Optimization problem. It is a generic NP hard combinatorial optimization problem and its related algorithm which can help solve many applications in finance, logistics and other domains. Many classical combinatorial problems like maximum cut, graph coloring and the partition problem, can be turned into a QUBO problem. QUBO problems can be solved on all three quantum computer paradigms (gate-based, annealing, simulation).

**Qudits**: generic form of qubit that has d possible quantum states instead of two. The approach is rarely used, at least in quantum computers outside research laboratories.

**Qunat**: another name for qubits based on continuous variables.

**Qutrit**: it is a form of qubit which instead of having two possible quantum states, has three. It is a special case of qudits.

**Rabi** (oscillation): oscillations between states of a two-level system excited at a frequency close to its resonance. This phenomenon is observed between two spin states in nuclear magnetic resonance as well as when an electric field acts on the transitions from one electronic state of a system to another for an atom or molecule. The curve describing the oscillation resembles a sinusoidal curve that attenuates over time. Isidor Isaac Rabi is an American physicist of Hungarian origin (1898-1988) who was awarded the Nobel Prize in Physics in 1944. Rabi's oscillations can be found almost everywhere, especially in the operation of superconducting qubits with microwave pulses.

**Raman cooling**: variant of the Doppler effect using the Raman effect used to cool atoms below the limit of Doppler-based cooling, under 1μK. It uses two counter-propagating laser beams. This effect is used in cold atom based interferometry in absolute gravimeters. Also known as Raman sideband cooling.

**Raman effect or Raman scattering**: shift in wavelength of an inelastically scattered radiation where an incident monochromatic photon energy and momentum are both changed. Discovered by Chandrasekhara Venkata Raman (1888-1970, India), Nobel Prize in physics in 1930. This is a small effect that accompanies the predominant Rayleigh scattering of light (unchanged wavelength). The incident polarized light is scattered at its original frequency (Rayleigh elastic scattering) and with higher and lower frequencies (Raman stokes and anti-stokes anti-elastic scattering).

**Raman spectroscopy**: determines vibrational and rotational level spacings from the energy (wavenumber) shifts of inelastic scattered light (*aka* Raman scattering). It is used to analyze multi-atoms molecules through their vibrational modes, particularly in organic chemistry.

**Raman transition**: couples two atomic levels by the absorption of a photon in one Raman beam (pump beam) and by stimulated emission of another one in the other beam (Stokes beam). It is used in cold atom interferometry to split a cold atom cloud into two superposed matter waves of different energy levels and vertical velocity.

**Ramsey experiment**: technique used to measure the $T_2^*$ of a superconducting qubit, with applying one Hadamard gate, waiting a time t, then applying another Hadamard gate, and measuring the output. The sinusoid curve amplitude slowly decreases around a probability 0.5. $T_2^*$ is obtained when the probability reaches 1/e.

**Rayleigh scattering**: predominantly elastic scattering of electromagnetic radiation by particles that are much smaller than the radiation wavelength. Elastic scattering happens with incident photons whose direction is changed but not their energy (color or wavelength). It explains why the sky is blue, linked to blue light being more scattered than green and red light, and also polarized.

**Realism**: in science, philosophical view according to which there exists a reality independently of an observer. The Copenhagen interpretation of quantum physics is non-realist since it believes reality is only what can be observed and measured.

**Reduced Planck constant**: see Dirac constant.

**Reflectometry**: technology used with superconducting and electron spin qubits readout. It consists in sending a microwave to the qubit and to analyze the reflected microwave, which can have different phase and amplitude depending on the measured qubit state.

**Register**: set of bits or qubits. In the case of qubits, it provides an exponentially growing computational base space with the number of qubits.

**Relaxation**: corresponds to the $T_1$ or lifetime of a qubit, which defines when the qubit loses its amplitude.

**Renyi entropy**: generalized version of entropy that can be used as a measure of entanglement. Shannon entropy is a special case of Renyi entropy.

**RIE**: reactive ion etching, a process used to remove some material on a semiconductor target using accelerated molecular or atomic ions in vacuum.

**RSA**: a public key encryption system based on the difficulty of factoring a public key formed by multiplying two very large prime numbers. This factorization is theoretically possible with Peter Shor's quantum algorithm. However, it requires a very large number of qubits to break the most common RSA keys at 1024 or 2048 bits. For 2048-bit keys, 20 physical million qubits with a 99,9%+ fidelity are required, which is very long-term in quantum computer roadmaps.

**Rydberg** (atoms): excited state of an atom having one or more electrons and whose principal quantum number n (index of the electron layer in the atom which is an integer between 1 and the number of electron layers in the atom) is very high. These atoms are generally of large size, proportional to $n^2$, and with very strong inter-atomic interactions. These interactions are used to build entanglement between atoms. These atoms have been used by Serge Haroche's team to detect non-destructively the presence of a photon in a cavity, and thus study quantum decoherence. Hydrogen can also be a Rydberg atom if it is excited with high energy levels.

**Sapphire**: aluminum oxide crystals ($Al_2O_3$) that is sometimes used as a substrate instead of silicon for the manufacturing of superconducting qubits chipsets. In that case, wafers are made of synthetic sapphire.

**SAT**: class of logic problem or Boolean satisfiability problem, of 0-order logic. It is a decision problem, which, given a propositional logic formula, determines whether there is an assignment of propositional variables that makes the formula true. As when looking for Boolean variables x, y and z that satisfy the equation $(x \vee y \vee z) \wedge (\bar{\mathbf{x}} \vee \bar{\mathbf{y}}) \wedge (\bar{\mathbf{x}} \vee y \vee z)$, $\wedge$ meaning "and", and $\vee$ "or" or "and". $\bar{\mathbf{x}}$ being the negation of x. The problem becomes very complex if the number N of variables becomes very high because to test their combinatorics with brute force, we will have to test $2^N$ combinations. This problem has been highlighted by Cook's theorem according to which the SAT problem is NP-complete. The SAT problem also has many applications, notably in constraint satisfaction, classical planning, model verification, diagnostics, up to the configurator of a PC or its operating system: we go back to propositional formulas and use a SAT solver.

**Scale-out**: generic information technology term describing the capacity to expand computing power with several processors connected to the other. This is done in classical server clusters and datacenters, using both hardware (multiple processors on same board, high-speed connectivity between boards and servers, high-speed data storage, ...). Such techniques are envisioned with quantum computing, consisting in connecting different processing units, usually with using photons and entanglement resources.

**Scattering**: deflection of moving particles by some physical medium or radiations.

**Schrödinger** (equation, wave function): describes the evolution in time and space of the wave state of a quantum object with a mass like an electron, i.e. the probabilities of finding the object at a given place and time in time.

**Schrödinger wave function collapse**: in the case of a qubit, happens at the end of the coherence (superposed state) which is generated by its state readout, bringing it back to one of its basis states ($|\mathbf{0}\rangle$ or $|\mathbf{1}\rangle$). This collapse is also caused by the interaction between the qubit and its environment and after qubit measurement.

**Second quantization**: field of quantum physics that deals with many-body quantum systems. It was introduced by Paul Dirac in 1927 and developed afterwards by Vladimir Fock and Pascual Jordan.

**Second quantum revolution**: covers advances in quantum physics since the 1980s, when we began to control the properties of individual quanta, at the level of photons (polarization, ...), electrons (spin) and atoms and also use superposition and entanglement. It covers in particular the uses of these properties in cryptography and telecommunications, quantum computing and quantum sensing. The term was created simultaneously in 2003 by Alain Aspect, Jonathan Dowling and Gerard Milburn.

**Semi-classical light**: describes interactions between quantized matter (atoms, electrons) and classical light fields. Laser light belongs to this category.

**Shor** (algorithm): integer quantum factorization algorithm invented by Peter Shor in 1994. It would theoretically allow to break RSA public keys by decomposing them into prime numbers.

**Silicon 28**: Silicon isotope allowing the creation of silicon wafers suitable for the creation of silicon qubits. Silicon 28 has a zero spin that does not affect the spin of the trapped electrons used to manage the qubits. It is purified in Russia and can then be deposited in a thin layer in the gas phase on conventional silicon.

**Single mode**: said of an optical fiber using a small core (around 9 µm) and transporting a single light beam that doesn't bounce off the inside walls of the fiber. It has low loss and is adapted to long distance transport, usually in the 1310 nm or 1550 nm wavelengths. These cables still use multiple wavelengths, with WDM (wave-division multiplexing). Also said of a single mode photon, see Multimode.

**SPAC**: special purpose acquisition company. A funding mechanism used by IonQ and HQS (Honeywell Quantum Systems) consisting in getting acquired by an investment fund creating a dedicated fund for the company and raising money on both limited partners (individual corporate ventures and the likes) and on the stock market like the NASDAQ.

**SPAM**: State Preparation And Measurement, a sequence of operations after which the fidelity of qubits is measured. This fidelity reflects that of an initialization sequence, the application of single qubit gates and the measurement of the qubit state.

**Spectral lines**: lines obtained graphically after decomposing an electromagnetic radiation into frequency components, usually with some spectrography apparatus. You have absorption and emission spectral lines depending on the source of light (indirect, direct). Each line corresponds to the emission or absorption of photons in atoms at particular energy levels, then wavelength and frequency.

**Spectral decomposition**: mathematically, spectral decomposition of a pure state vector in a Hilbert space is its eigenstates $|i\rangle$ and eigenvalues $\lambda_i$. It can be related to the wave-duality aspect of all quantum objects. A quantum object pure state is indeed decomposable in a coherent superposition of elementary waves, the eigenstates.

**Spin**: quantized angular momentum of elementary (like electrons or photons) or composite particles (like atoms) that cannot be described or explained in classical physical terms. The spin of composite particles is the addition of its components spin. A proton and a neutron have a spin of 1/2. An electron has a spin of +1/2 or -1/2. A photon also has a spin, which relates to its circular polarization. Spin help distinguish fermions who have half integer spins from bosons who have integer spins.

**Spintronics**: a set of technologies based on the manipulation of electron spin. It is found in memristors as well as in hard disks using giant magnetoresistance (GMR). The latter was discovered by Albert Fert (France) and Peter Grünberg (Germany) independently and the same year, in 1988. This got them the Nobel Prize in Physics in 2007.

**Spontaneous emission**: when an atom emits a photon resulting from the transition of an electron from an excited to a lower energy state.

**Spontaneous Four-Wave Mixing** (SFWM): photons pairs source category based on pumping nonlinear optical waveguides or cavities.

**Spontaneous Parametric Down-Conversion** (SPDC): system converting high-energy photons into pairs of photons of lower energy, based on pumping nonlinear optical waveguides (crystals) or cavities. It can be used to create pairs of entangled photons as well as single photons sources.

**Squeezed states of light**: correspond, in a quadrature or phasor diagram representation, to wave functions which have an uncertainty in one of the quadrature amplitudes (phase or photon number) smaller than for the groundstate corresponding to the vacuum state. It can be generated by different means like a parametric down conversion. In other words, it's a way to increase the measurement precision of one of the photons characteristics at the expense of another characteristic.

**SQUID**: Superconducting Quantum Interference Device, a magnetometer that measures the direction of current in a superconducting qubit. It is notably used by D-Wave and in some quantum sensors.

**Stabilizer gates**: quantum gates that are used in error correction systems: CNOT, H (Hadamard) and P (phase).

**Standard quantum limit**: to estimate a system's parameter, one usually uses light as the meter by making it interact with the system and thereby extracts some information. The standard quantum limit, also known as shot noise, states that the variance of this estimation is larger than the inverse of the square root of the number of times the measurement is made. It limits the precision of quantum sensing using non-entangled quantum states. See Heisenberg limit.

**Stark shift or Stark effect**: shifting and splitting of spectral lines of atoms and molecules due to the presence of an external electric field. This is the electric-field analogue of the Zeeman effect which is linked to the effect of the magnetic field,

**State reduction**: consequence of the measurement of the state of a quantum or a qubit, which modifies its (superposed) state into a stable state (not superposed). For a qubit, it is one of the two basic states: excited or non-excited, horizontal or vertical polarization for a photon, spin orientation for an electron, excited state for an ion or a cold atom, etc.

**State vector**: Hilbert space vector representing a pure state of a quantum object.

**Stationary qubits: stationary** (or static) qubits, which do not move in a circuit. This is the case of superconducting qubits, trapped ions and cold atoms qubits as well as electron spin qubits. They are opposed to flying qubits that move, like photons.

**Stern-Gerlach experiment**: deviation of projected atoms placed under an intense magnetic field, which is explained by electron intrinsic angular momentum or spin.

**Stimulated emission**: when an incident photon is not absorbed by an excited atom but stimulates the atom to emit a second photon with the same wavelength. This principle is used in lasers to amplify light in their cavity.

**Sturm–Liouville problem**: mathematical problem consisting in solving some second-order differential equations where the unknown is a density function and finding eigenvalues and eigenvectors and satisfying bound limits. Solving Schrodinger's wave equation is a particular case of such a problem.

**Superconductivity**: the ability of some materials to conduct electricity without resistance. It generally occurs at low temperatures. It is linked to the behavior of electrons in some crystalline structures who happen to gather in pairs, Cooper pairs, who become bosons, and have a collective behavior enabling them to move around within the structure. Superconducting and electron spins qubits use this effect. The first with superconducting loops traversing a Josephson barrier and all of them with cabling and some surrounding electronics.

**Superdense coding**: technique used to send two bits on a single (optically transmitted) qubit between two points when they are already connected by a pair of entangled photons. It is a communication protocol imagined by Charles Bennett and Stephen Wiesner in 1992 and experimented in 1996 by Klaus Mattle, Harald Weinfurter, Paul Kwiat and Anton Zeilinger. The initial entanglement preceding the transmission of the two bits in the qubits avoids violating Holevo's theorem that a set of qubits cannot carry more information than the equivalent number of classical bits.

**Superoperator**: linear operator that transforms a linear operator like a density matrix. It must be a CPTP map, completely positive and trace preserving map (see CPTP definition).

**Superposition**: property of quantum objects and qubits to be able to be in several states at the same time. This can be explained by the wave-like nature of quantum objects. A superposition is a linear combination of quantum eigenstates (the $|0\rangle$ and $|1\rangle$ in the case of qubits).

**Surface codes**: type of quantum error correction code that is tolerant to high qubits error rates and require a larger number of physical qubits per logical qubits and have a design constraint for physical qubits that must be connected to their immediate neighbors in a 2D structure. This is the QEC architecture chosen by Google.

**SVD**: singular value decomposition, a mathematical process to factorize any mxn complex values matrix into three matrices: a unitary mxm matrix, a rectangular diagonal mxn matrix (which has no zero values only in its diagonal) and a unitary nxn matrix. SVD is used in QSVT algorithms.

**SWAP**: quantum gate that inverts the state of two qubits. It is very useful since most qubit geometries don't allow an any-to-any qubit connection. The SWAP gate enables this kind of connection that is mandatory for many quantum algorithms.

**Symmetry**: of Schrödinger's wave function, with bosons. Fermions have an antisymmetric wave function. It is the mathematical consequence of Pauli's exclusion principle which states that two fermions with the same quantum numbers cannot cohabit while two similar bosons can.

**T gate**: single qubit gate implementing a "quarter turn" phase rotation around the Z axis in the Bloch sphere. It is a very important gate enabling (on top of the three-qubit Toffoli gate) the creation of a universal gate set.

**T-count**: number of T gates required in a quantum algorithm. It is an important metric since T gates are the most expensive to correct with FTQC.

**T-depth**: number of circuit layers implementing one or several T gates in a quantum algorithm.

**$T_1$**: qubit amplitude coherence time, which indicates the end of coherence of the qubits linked to a loss of amplitude ("energy relaxation"). Aka qubit lifetime.

**$T_2$**: phase related coherence or time when some phase shift occurs, i.e. a rotation around the z axis in the Bloch sphere of the qubit state.

**Tensor**: in multilinear algebra and differential geometry, a tensor designates a very general object whose value is expressed in a vector space. In quantum physics and computing, tensors are used to describe the state of a compound quantum object with several quanta or qubits. A qubit is represented by a vector of 2 complex numbers. A register of N qubits is represented by a vector with $2^N$ complex numbers resulting from the tensor product of N vectors of 2 complex numbers. In a way, the tensor product represents the combinatorial space of the values that a combination of qubits can take. Before entanglement comes into play to mix things up and create non separable vector states, i.e., which cannot be expressed as tensor products of individual quantum states.

**Thermodynamics first law**: the internal energy of an isolated system is a constant, applying the principle of the conservation of energy. Inside the system, the form of energy can however be transformed.

**Thermodynamics second law**: the entropy of a closed system cannot decrease. In other words, heat does not flow spontaneously from cold to hot objects. Was formalized by Rudolph Clausius in 1854.

**Time crystal**: also, DTC for discrete time crystal is a topological state of condensed matter at very low temperature where atoms in their ground state are periodically arranged in both space and time with in a permanently oscillating structure with a given period (for discrete time crystals).

**Time domain**: deals with the evolution of some value and signal over time. It's frequently opposed to frequency domain where a signal is analyzed with decomposing it into frequencies (mathematically, with a Fourier transform).

**Time reversal**: clearly a misnomer. It describes situations when from a given point in time, for some physical property like energy, a system presents a symmetry when you evaluate it with look forward or backward at time. It's a mathematical symmetry. You don't change the arrow of time backwards. Time reversal is not a time machine!

**Toffoli** (gate): also called CCNOT is quantum gate operating on three qubits which modifies the value of the third qubit if the value of the first two is 1.

**Topological**: topological quantum computing is based on the notion of anyons which are "quasi-particles" integrated in two-dimensional systems. The anyons are asymmetric and two-dimensional physical structures whose symmetry can be modified. This allows the application of topology principles with sets of successive permutations applied to pairs of anyons that are in proximity in circuits. The associated algorithms are based on the concepts of topological organizations of braids or nodes ("braids"). There is an algorithmic equivalence between computation with universal gated qubits and topological qubits.

**Transmon**: transmission-line shunted plasma oscillation qubit, variation of superconducting qubit with superconducting current oscillating at two different frequencies across a Josephson junction. The difference between these two frequencies corresponds to the energy of the microwave pulses sent to the qubit to drive single qubit gates.

**Transpiler**: source-to-source compilers used in classical computing and quantum computing to optimize the source code based on some hardware constraints like to fit with the universal gate set available in the quantum processor. It can help reduce the number of gates to execute and as a result, reduce errors (particularly with NISQs) and reduce the algorithm execution time.

**Transpilation**: code conversion and optimization achieved by transpilers.

**Transversal gates**: relates to quantum error correction and fault tolerance. These are gates implemented with QEC where there is a 1-1 correspondence and link between all qubits from a given corrected qubit with a similar corrected qubit, when they are assembled through concatenation. This mechanism limits the propagation of errors between logical and physical qubits.

**Trapped ions**: these are ions used in certain types of quantum computers. They are usually trapped magnetically or electrically, and their state is controlled with lasers. Their readout uses a laser excitation and an imager readout of the resulting ions fluorescence.

**Tunnel effect**: property of a quantum object to cross a potential (or energy) barrier even if its energy is less than the minimum energy required to cross this barrier. This effect is used in D-Wave's quantum annealers to quickly determine the minimum energy of a complex system ("Hamiltonian" implemented as an Ising model).

**Two-Level Systems** (TLS): other descriptor of quantum systems used to implement qubits. A qutrit is a three-level system.

**TWPA** (Travelling Waves Parametric Amplifiers) microwave readout amplifiers implemented in a long array of SQUIDs. Their broad bandwidth that can reach 2 GHz with a 15 dB amplification enables up to 20 qubits readout multiplexing. But it depends on the gate speed since the faster the gate, the smaller the microwave pulse bandwidth will be large.

**UHV**: Ultra High Vacuum, the ultra-high vacuum required to operate certain types of qubits. It is mainly used for cold atoms and trapped ions. Superconducting qubits are integrated in a vacuum cryostat that does not require ultra-high vacuum.

**Ultraviolet catastrophe**: expression of Paul Ehrenfest, linked to the Rayleigh-Jeans law proposed in 1900 to explain the black body radiation spectrum, which was diverging to infinite values as the temperature was growing, when reaching ultraviolet wavelengths. Planck's law based on quanta solved the problem and got rid of the ultraviolet catastrophe.

**Unary gates**: single qubit gates. Not to be confused with unitary operations that are the result of the combination of all qubit gates on a given set of qubits. A unitary transformation of the computational state vector is a matrix operator that is equal to its transconjugate. It is a mathematically reversible operation.

**Unconventional Computing**: computing methods that do not fall under the classical computing principles of Turing and Von Neuman machines. Covers non-traditional tools and methods that include, but are not limited to, quantum computers. It also includes molecular computers and neuromorphic processors.

**Unitary operation**: linear operation on a vector that preserves its length. In the case of qubits whose vector always has a length of 1, the unitary quantum gates apply on it a transformation that preserves this length and is also reversible. In the representation of qubits in the Bloch sphere, the operation rotates the vector representing the state of the qubit in this sphere.

**Universal quantum computer**: most generic form of a quantum computer exploiting a universal quantum gate set, and which can both simulate quantum physics and implement any operations of a classical computer.

**Universal quantum gates**: sets of quantum gates from which all other quantum gates can be reproduced to create any unitary transformation on any number of qubits (by approximation and under a given error rate). It requires a non-Clifford group gate, like a T gate or a Toffoli gate.

**Unruh effect**: relativistic thermodynamic effect with a black body radiation showing up in vacuum at relativistic speed. Il has some connection with quantum noise.

**VQA** (Variable Quantum Algorithm) more generic quantum hybrid algorithm than VQE. It combines a classical optimizer that is used to train a parametrized quantum circuit. It could lead to obtain some quantum advantage with NISQ quantum computers. VQA has a broad set of applications: finding ground and excite states, quantum simulations, machine learning and optimizations.

**VQE** (Variational Quantum Eigensolver): hybrid quantum algorithm used in chemical simulation created in 2013. Its main contributor is Alán Aspuru-Guzik, a researcher at the Zapata Computing startup. It is also used in machine learning tasks. VQE was the first proposed VQA.

**Wave packets**: is a burst of electromagnetic wave that travels as a unit. It is formed by the addition of an infinite number of sinusoidal waves of different frequencies, phases and amplitudes creating constructive and destructive interferences on a small region in space, and destructively elsewhere. Wave packets are used in many quantum technologies such as with microwaves sent to superconducting and electron spin qubits or by femto- and picoseconds lasers. In these cases, their decompositions in frequencies lead to so-called frequency combs.

**Wave-particle duality**: the property of elementary particles such as electrons, neutrons, atoms and photons to behave as both particles with momentum and waves that can generate interference. It is verified with the famous Young's slits experiment which shows these interferences with both photons and electrons.

**Wien's displacement law**: describes the relationship between peak wavelength and temperature in black body energy spectrum. Discovered by Wilhelm Wien in 1893.

**Wigner function**: representation of a quantum state used to measure the level of quantumness of a light pulse. It has the particularity of having negative values for entangled and non-gaussian states. It is usually visualized in a 3D chart with peaks and lows. Also called Wigner quasiprobability distribution or Wigner-Ville distribution. It was created by Eugene Wigner in 1932.

**X**: quantum gate at a qubit that inverts its amplitude, goes from $|\mathbf{0}\rangle$ to $|\mathbf{1}\rangle$ or from $|\mathbf{1}\rangle$ to $|\mathbf{0}\rangle$ for the basis states.

**XY gates**: two qubit entangling gate.

**Y**: single-qubit quantum gate that performs a 180° rotation around the Y axis in the Bloch sphere.

**Z**: quantum gate to a qubit that applies a sign change to the β component of the qubit vector, i.e. a phase inversion and a 180° rotation with respect to the Z axis. More generally, Z gates is also a denomination for phase change gates.

**Zeeman effect**: splitting of spectral lines when atoms are placed in a static magnetic field. Explained by the different electron's magnetic moment and/or by the atom nucleus spin for the nuclear Zeeman effect. There are two Zeeman effect, the normal and abnormal effect, differentiated with the off/even spectral rays generated and generated by different orbital angular momentum (normal) and spin (abnormal).

**Zeeman cooling or slower**: use of the Zeeman effect to cool atoms at a lower temperature than with a simpler Doppler effect. Invented by William D. Phillips (Nobel prize in physics in 1997), it consists in adjusting the atoms resonant frequency with a magnetic field as the atoms are slowing down when implementing the Doppler effect.

**ZX calculus**: graphical language and formalism used to visualize in quantum programming the notions of entanglement, complementarity, causality and their interactions. It can be used for Measurement Based Quantum Computing (MBQC), the creation of error correction codes and compiler optimization techniques.

# Index

# Table of figures

# Revisions history

This book improved over time with successive revisions. The first editions from September 2018, 2019 and 2020 were published in French and this fifth one is the second published in English. It is freely downloadable on:

https://www.oezratty.net/wordpress/2022/understanding-quantum-technologies-2022/

Different PDF formats are available: **A4 in full resolution in a single volume** (87 Mb), **A4 in reduced resolution in two volumes** (to fit under the 32 Mb threshold for some ebook readers) and the same in **Letter format** for USA and Canada readers who would like to print it on their own.

After a post-publication proof-reading and correction couple weeks period, I will also publish it during October 2022 on **arXiv** and in printed paperback editions on **Amazon**.

| Version and date | Modifications |
|---|---|
| 1.0 (332 pages)<br>September 29th, 2018 | First version of this document published in French and consolidating 18 posts published between May and September 2018 on www.oezratty.net. |
| 2.0 (504 pages)<br>September 20th, 2019 | Second edition, also in French. New content on superconductors, superfluidity, quantum sensing, quantum supremacy, quantum computing emulation, cryogeny, hybrid algorithms, algorithms certification, quantum teleportation, blind computing. Addition of a glossary and bibliography. |
| 3.0 (684 pages)<br>September 7th, 2020 | Third edition, also in French. New content on Maxwell, Schrödinger and Dirac's equations, relativistic quantum chemistry, how research works, lasers and masers, polaritons, extreme quantum, linear algebra, quantum gates classes, quantum error correction codes, cryo-electronics, MBQC, quantum cloud, qubits technologies, unconventional classical computing, quantum hype cycle, quantum foundations and on the influence of science fiction. |
| 4.0 (838 pages)<br><br>September 27th, 2021 | Fourth edition, the first one in English. Main new features vs the 3.0. on top of updates nearly everywhere:<br><br>New section on quantum physics postulates, page 86.<br><br>More on wave-particle duality and on photon qubits physics.<br><br>Improvements and extensions on linear algebra, on quantum measurement and quantum memory. New part on vacuum in enabling technologies.<br><br>Much improved section on algorithms, including data preparation and debugging.<br><br>Expanded part on QRNG.<br><br>Went from 265 to over 450 vendors covered in the various sections of the book.<br><br>More on ethical issues and gender balance.<br><br>Additional covered countries: Belgium, Portugal, Italy and Abu Dhabi and nice ecosystems maps for the USA and the UK.<br><br>Added an index with company names, people and some scientific terms and many terms in the glossary (from 179 to >280). |
| From October 2021 to January 2022 | The two-volumes printed version of the book was made available to purchase at an affordable price on most **Amazon** sites.<br>Added a new graph explaining Dirac $\langle A\|B\|C\rangle$ notations in the measurement section starting page 185.<br>Added a roadmap for managing the energetic footprint of quantum computing.<br>Added American Binary (Ambit) in the quantum cryptography vendors section that starts page 845.<br>Small update related to D-Wave gate-based quantum computing announcement in the quantum annealing section that starts page 277.<br>Added Energetics of quantum technology, Quantum postulates, QML, QLM and scale-out in the glossary. Corrected wrong naming for Atos QLM (instead of Atos QML or Atos aQML).<br>Update on Origin Quantum and their cloud quantum emulation offering.<br>Added StarX Electronics in the inventory of enabling technology vendors and Chipiron in quantum sensing imaging, starting page 882.<br>Added Two-Level Systems in glossary.<br>Added a chart from Joseph Bardin describing the various microwave and other signals used to drive superconducting, electron spin and trapped ions qubits.<br>Added Algorithmic qubit, anharmonic oscillator, cQED, CQED, fluxonium, Q-factor, Dirac and Reduced Planck constants, homodyne detection and Universal quantum computer in the glossary.<br>Updates on IBM and its 127 qubits processor announced in November 2021, the related qubits fidelities chart with the best-in-class IBM 27, 65 and 127 QPUs.<br>Some updates on IQM and OQC (on Amazon Braket).<br>Updates on Kipu Quantum and on Quantinuum (new name for HQS/CQC merger).<br>Added a mention on Alexia Auffèves quantum energy initiative.<br>Updates on Quandela, Qu&Co (acquired by Pasqal) and Rahko (acquired by Odyssey Therapeutics).<br>Added Mikhail Lukin in the quantum computing physicists section.<br>Added Black Quant and Runa Capital in the quantum investors section. |
| 5.0 (1128 pages in single volume version)<br>September 2022 | Many updates everywhere, including…<br>**Why**: new intro, rearranged and updated the part on Moore's law. Removed the long abstract from the previous version.<br>**History and scientists**: updated scientists with more stuff on Marc Benioff, David Deutsch, David Hilbert, Daniel R. Simon, Erwin Schrodinger and new bios for Andrew G. White, Gaston Floquet, Bruce Kane, Daniel Kleppner, Daniel Loss, Frank Wilczek, Gerhard Rempe, Herbert Walther, James Clarke, Jeff Kimble, Jian-Wei Pan, Leo Kouwenhoven, Menno Veldhorst, Steven Girvin, Rob Schoelkopf, Roy J. Glauber, Jay Gambetta, Alexandre Blais, Robert Raussendorf, Maciej Lewenstein, Ronald Walsworth and Philip W. Anderson. More details on how research works, papers published and evaluated.<br>**Quantum physics 101**: added a table listing the various quantum physics postulates versions, added nuclear quantum numbers in quantization, added a part on quantum matter, including on quantum batteries, time crystals and skyrmions.<br>**Gate-based quantum computing**: various updates on quantum gates, added illustrations on quantum computing dimensionality, better differentiation between quantum emulation and quantum inspired software.<br>**Quantum computing engineering**: simplified qubit type descriptions, update content on exotic qubits, better explained how error rates are measured, many updates on quantum error |
| | corrections, definitions of FTQC, universal QC and LSQC. Described in detail the whereabouts of the Quantum Energy Initiative. Added a definition of quantum switch.<br><br>**Quantum computing hardware**: added vendor investment comparisons per type of qubit, types of use cases per qubit type, inventory of scalability challenges per type of qubit, rearranged and standardized the presentation of each qubit type with history, science, qubit operations, research and vendors. More schematics. More science on quantum annealing. Genealogy of superconducting qubits. Updated content on IBM, Google, Rigetti, IQM, OQC and added Atlantic Quantum, Baidu and Toshiba in superconducting computers vendors. Comparisons of different types of quantum dots spin qubits. Added Diraq in quantum dots spin qubits vendors, XeedQ in NV centers qubits vendors, It's Q, Quantum Source Labs and TuringQ in photonic qubits vendors and Crystal Quantum Computing and Planqc in neutral atoms qubit vendors, and Hon Hai / Foxconn in the trapped ions vendors list. Added coherent Ising machine in quantum photonics systems. Archer Materials, Atom Computing, Bleximo, BosonQ Psi, Nord Quantique and QBoson.<br><br>**Quantum enabling technologies**: rearranged the section on control electronics. Added ICE, Maybell Quantum and FormFactor, and updated myCryoFirm in cryostats vendors, Active Technologies, Keysight, QuantrolOx and Scalinq in control electronics vendors, Qubic Technologies, Raditek, QuinStar Technology, RF-Lambda, Wenteq Microwave Corp, Holzworth Instrumentation, apitech, Analog Quantum Circuits, CryoHEMT and Silent Waves in cryoelectronics vendors, CryoCoax, XMA and Rosenberger Group in cable and filtering vendors and Alcyon photonics, Teem photonics and Scintil Photonics in photonic enabling technologies vendors, AnaPico, Diatope, HiQeTe Diamond, Orsay Physics, QuantTera and Quantum Diamant in other enabling technologies vendors. Added a new part on fabs, processes and manufacturing tools. Added here many manufacturing tools vendors like BESI, PlasmaTherm, Picosun Group, NanoAcademic Technologies and QuantCAD. Updates on raw materials.<br><br>**Quantum algorithms**: added a part on tensor networks. Significant updates on quantum machine learning.<br><br>**Quantum software development tools**: updates on emulation software, restructured and updated the part on benchmarking.<br><br>**Quantum business applications**: updated all vertical case studies lists. Added Arclight Quantum, Allosteric Bioscience, Artificial Brain, ColibrITD, Dirac, Foqus, GenMat, Good Chemistry Company, Ingenii, Qbraid, QEDma, Qoherent, Quanscient, Quantagonia, Sanctuary, SavantX, Tinubu Software and Turing in the software and tools vendors inventory. Created a section on IT service vendors working on quantum technologies. Added DN-Quantum Computing, Kvantify, Plantagenet Systems, Protiviti, Psi-Ontic, Quantum Computing Engineering, Quanvia, quGeeks, Quant-X Security & Coding, Qubitech and Unitary Zero Space. Updated information on AegiQ, Azurlight Systems, Cogniframe, exaQ.ai, Horizon Quantum Computing, HQS Quantum Simulations, Multiverse Computing, and Nomidio, Phasecraft, OTI Lumionics, Q.ant, Quantum Computing Inc, QunaSys, Q-Ctrl, Strangeworks and Terra Quantum.<br><br>**Quantum enabling technologies**: I singled out this part and moved it to the second volume. Content has been sporadically updated.<br><br>**Quantum telecommunications and cryptography**: added a part on quantum photon sources and detectors in the QKD section, improved description of trusted nodes and repeaters, rearranged and enriched the quantum interconnect and telecommunication part. Update on PQC with NIST 2022 selection results. Added Abelian, ComScire in QRNG vendors, Bohr Quantum Technology, Photoniq and Entanglement Networks in quantum telecommunications, NodeQ, Patero, QANplatform, Quantum Collective, SandboxAQ, Synergy Quantum and ThinkQuantum in quantum telecommunications and cryptography vendors. Updated information on Post-Quantum, IDQ, Qnu Labs, QphoX and, Qunnect.<br><br>**Quantum sensing**: added new parts on quantum sensing taxonomy and on quantum pressure sensing. More details on quantum thermometers. Added OK Quantum and Zero Point Motion in quantum gravimeters and accelerometers, Improved the technical description of a quantum gravimeter. QuSpin, Siloton, QLM Technology and Mag4Health in imaging sensors and qdm.io and Elta Systems in quantum magnetometers. Updated information on Chipiron. |
| | **Quantum technologies around the world**: updates in investments data, new part disappeared startups, SPACs, government spending and quantum national initiatives rationales. Added Finland, Norway, Hungary, Ireland, and Qatar in countries overview and Intqlabs startup in UAE. Updated nearly all other countries quantum activities. Added a map of Canada's quantum ecosystem.<br><br>**Quantum technologies and society**: added EFEQT in scientific education and on quantum technologies marketing.<br><br>**Bibliography**: reshuffled the quantum events section, added an events timeline.<br><br>**Glossary**: added Ansatz, Bell state, Chi (nonlinearity order), Circuit, Coulomb blockade, crosstalk, CVD, Fermi sea, Floquet code, EBL, electron gas, FTDQC, Flux biasing, Gaussian Boson Sampling (GBS), Hall effect, Heterodyne and Homodyne measurements, I/Q mixer, Jaynes-Cummings Hamiltonian, JPA, Leggett-Garg inequality, MBE, Magneto-Optical Trap, mesoscopic, metal layers, microring resonator, Mott insulator and Mott transition, Mutually unbiased bases, no-go theorem, normalization, on-premises, ODMR, paramp, photolithography, Purcell effect, Purcell filter, purification, PVD, Quantum Hall effect, QHO, QND, QSVT, QUBO, quantum steering, quantum switch, Ramsey experiment, relaxation, Renyi entropy, RIE, Sapphire, surface code, SVD, time crystals, time reversal, TWPA, Stark shift, Unruh effect and Zeeman cooling.<br><br>Added over 900 figure captions with sources and sorting out how figures are referenced in the text (mostly). |
| 5.1 (1128 pages in single volume version)<br><br>October 12th, 2022 | Added Nobel prize mentions for John Clauser, Alain Aspect and Anton Zeilinger.<br><br>Integrated some corrections suggested by André Konig.<br><br>Modifications in the FTQC/LSQ nomenclature proposed in Figure 246 with suggestions from Alastair Abbott and Tristan Meunier and thoughts about the progressive growth of logical qubits fidelities.<br><br>Presentation improvements in quantum physics simulation algorithms.<br><br>Various improvements in the part on quantum error correction page 235.<br><br>Various edits in satellite QKD, China funding, Google Sycamore, NV centers frequency analyzers in sensing. |
| 5.2 (1130 pages in single volume version)<br><br>October 16th, 2022 | Updates on Alain Aspect's experiment and their explanations.<br><br>Updates in insurance use cases.<br><br>Added the class NISQ in quantum complexity classes.<br><br>Added quantum amplitude estimation in the algorithms toolbox.<br><br>Precisions on transpilers and transpilation in the gates section and with the glossary.<br><br>Added T gate, T-count and T-depth in the glossary.<br><br>Added IonQ, Rigetti and D-Wave quarterly revenue in the investor section.<br><br>Added a reference to the book "Quantum Software Engineering" in the bibliography.<br><br>Index cleanup and various edits elsewhere. |
| 5.3 (1132 pages)<br>October 23th, 2022 | Changed page formatting at the beginning of volume 1 to accommodate Amazon's stringent rule for paperback edition support on KDP.<br><br>Added a new partnership between Switzerland and the USA. |
| 5.4 (1132 pages)<br><br>October 25th, 2022 | Updated the record "non-Shor" integer factoring algorithms in the quantum cryptoanalysis threat section. |

I update the book on a regular basis as I find editorial issues, mistakes, misspellings, and the likes. You can submit me any comment, correction, suggestion or even request for digging into some untapped topic (olivier@oezratty.net).

page intentionally left blank.

page intentionally left blank.

page intentionally left blank.

back-cover flip page.

# le lab quantique

⟨ Aage Niels Bohr | Abner Shimony | Adi Shamir | Akira Furusawa | Alain Aspect | Alain Ravex | Alain Sarlette | Alán Aspuru-Guzik | Alastair Abbott | Albert Einstein | Alberto Bramati | Alexander Andreev | Alexander Holevo | Alexander Prokhorov | Alexandre Blais | Alexandre Zagoskin | Alexei Bylinskii | Alexei Grinbaum | Alexei Kitaev | Alexia Auffèves | Alfred Kastler | Alfred Pérot | Alonzo Church | Amir Naveh | André Luiz Barbosa | Andrea Morello | Andrea Rodriguez Blanco | Andreas Wallraff | Andrew Childs | Andrew Cross | Andrew Gle[illegible] | Andrew White | Andy Matuschak | Anne Broadbent | Anne Canteaut | [illegible]nger | Aram Harrow | Arieh Warshel | Arthur Holly Compton | Arthur Leonard Schawlow | Artur Ekert | Astrid Lambrecht | Audrey Bienfait | Axel Becke | Benjamin Huard | Benoît Valiron | Bettina Heim | Bob Wiens | Boris Podolsky | Brian Josephson | Bruce Kane | Bruno Desruelle | Bryce DeWitt | Carl Wieman | Carlo Rovelli | Chad Rigetti | Charles Bennett | Charles Fabry | Charles Hard Townes | Charles Hermite | Chien Shiung Wu | Christiaan Huygens | Christian Weedbrook | Christine Silberhorn | Christophe Jurczak | Christophe Solomon | Christophe Vuillot | Christopher Monroe | Christopher Papile | Claude Cohen Tannoudji | Claude Weisbuch | Clauss Jönsson | Clinton Davisson | Cornelis Dorsman | Cristian Calude | Cristina Escoda | Cyril Allouche | Damien Stehlé | Daniel Esteve | Daniel Gottesman | Dave Wecker | David Bohm | David Deutsch | David DiVincenzo | David Gosset | David Hilbert | David J. Thouless | David Wineland | Dennis Dieks | Dieter Zeh | Dirk Bouwmeester | Don Coppersmith | Don Misener | Donna Strickland | Dorit Aharonov | Douglas James Scalapino | Earle Hesse Kennard | Ed Sperling | Edward Farhi | Edward Fredkin | Edward Tang | Elena Calude | Eleni Diamanti | Elham Kashefi | Elisabeth Giacobino | Elvira Shishenina | Emanuel Knill | Emmy Noether | Enrico Fermi | Eric Cornell | Ernest Rutherford | Ernst Ising | Erwin Schrödinger | Etienne Klein | Ettore Majorana | Ewin Tang | Fabio Sciarrino | Faye Wattleton | Felix Bloch | Francesca Ferlaino | Franck Balestro | François Le Gall | Frédéric Grosshans | Frédéric Magniez | Freeman John Dyson | Friedrich Hund | Friedrich Paschen | Geordie Rose | George Uhlenbeck | Georges Paget Thomson | Georges Zweig | Georges-Olivier Reymond | Gerard Milburn | Gerrit Jan Flim | Gil Kalai | Gilles Brassard | Gilles Holst | Giorgio Frossati | Gordon Baym | Gordon Gould | Guang-Can Guo | Haig Farris | Hans Albrecht Bethe | Hans Jürgen Briegel | Hantaro Nagaoka | Harald Fritzsch | Hartmut Neven | Heike Kamerlingh Onnes | Heinrich Hertz | Hélène Bouchiat | Hélène Perrin | Hendrik Anthony Kramers | Hendrik Casimir | Hendrik Lorentz | Henri Poincaré | Hermann Minkoswki | Hermann Weyl | Hugh Everett | Hui Khoon Ng | Ilana Wisby | Ilya Mikhailovich Lifshitz | Immanuel Bloch | Iñigo Artundo | Iordanis Kerenidis | Irfan Siddiqi | Isaac Chuang | Isaac Newton | Itamar Sivan | Jacqueline Bloch | Jacques Salomon Hadamard | Jacq[illegible]mero | James Chadwick | James Clarke| James Clerk Maxwell | James Park | Jared Cole | Jason Alicea | Jay Gambetta | Jea[illegible] | Jean Michel Raimond | Jean-François Roch | Jean-Michel Gérard | Jean-Philip Piquemal | Jian-Wei Pan | Jelena Vučko[illegible]ć | J[illegible]lena Vucokic | Jeremy O'Brien | Joannes van der Waals | Joe O'Gorman | Johann Balmer | Johannes Pollanen | Johan[illegible]n Bardeen | John Clauser | John Frank Allen | John Hall | John Levy | John Martinis | John Morton | John Pr[illegible] | John Rowell | John Smolin | John Stewart Bell | John Von Neumann | John Watrous | John Wh[illegible] | [illegible] Koomey | Jonathan P. Home | José Capmany | José David Domenech | Jose Ignacio Latorre | [illegible] | Josh Nunn | Juan Ignacio Cirac | Julia Kempe | Julian Schwinger | Julien Bobroff | [illegible] | Kevin Young | Kirill Tolpygo | Kohei Itoh | Kristel Michielsen | Krysta Svore | [illegible] | Le Si Dang | [illegible] | Léon Brillouin | Leon Neil Cooper | Leonard Adleman | Lester Germer | [illegible] | Llewellyn Thomas | Loïc Henriet | Louis Cauchy | Louis de Broglie | [illegible] | Lu Jeu Sham | [illegible] | Luigi Frunzio | Magdalena Hauser | Marcus Doherty | Marcus Huber | [illegible] | Martin Karplus | Masahide Sasaki | Masahiro Kitagawa | [illegible] | Mathieu Munsch | [illegible] | Matthieu Desjardins | Maud Vinet | Max Born | Max Planck | [illegible] | Michael Ben-Or | Michael Frank | Michael Freedman | Michael Horne | [illegible] | Michelle Simmons | Mikhail Dyakonov | Mikhail Lukin | [illegible] | Nathanaël Cottet | Niccolo Somaschi | Nick Farina | Nicolas Gaude | [illegible] | Nikolay Basov | Nir Minerbi | Nobuyuki Imoto | Oleg Mukhanov | Olivier Carnal | [illegible] | Pascual Muñoz | Patrice Bertet | Patty Lee | Paul Benioff | Paul Dirac | [illegible] | Peter Knight | Peter Leek | Peter Shor | Peter Zoller | Philip W. Anderson | [illegible] | Pierre Hohenberg | Pierre Rouchon | Pieter Zeeman | Pranav Gokhale | Pyotr Kapitsa | [illegible] | Raymond Laflamme | Rebecca Krauthamer | René Descartes | Richard Feynman | [illegible] | Richard Murray | Rob Schoelkopf | Rob Whitney | Robert Andrews Millikan | Robert Dennard | Robert König | Robert McDermott | Roy J. Glauber | Robert Raussendorf | Robert William Boyd | Robin Blume-Kohout | Roger Penrose | Roland [illegible] | Roman Lutsiv | Ron Rivest | Roy Glauber | Ronald Walsworth | Rudolf Grimm | Ryan Babbush | Samuel Goudsmit | Sandu [illegible] | Sara Ducci | Sarah Sheldon | Satyendranath Bose | Scott Aaronson | Sébastien Balibar | Sébastien Tanzilli | Serge Haroche | Sergey Bravyi | Seth Lloyd | Shengtao Wang | Shi Yaoyun | Shin'ichirō Tomonaga | Silvano de Franceschi | Simon Benjamin | Simon Perdrix | Simone Severini | Sophia Economou | Stefanie Barz | Stephanie Wehner | Steve Girvin | Steve Lamoreaux | Taki Kontos | Tauqir Abdullah | Terry Rudolph | Théau Peronnin | Theodor Hänsch | Theodore Lyman | Theodore Maiman | Thibault Jacqmin | Thierry Lahaye | Thomas Monz | Thomas Young | Tommaso Calarco | Tommaso Toffoli | Tracy Northup | Travis Humble | Tristan Meunier | Umesh Virkumar Vazirani | Urmila Mahadev | Valerian Giesz | Vasilis Armaos | Vincent Bouchiat | Virginia D'Auria | Vlad Anisimov | Vladan Vuletic | Vladimir Fock | Walter Brattain | Walter Kohn | Walter Houser Brattain | Walther Meissner | Walther Nernst | Wassim Estephan | Werner Heisenberg | Wilhelm Wien | William Rowan Hamilton | William Shockley | William Wootters | Willis Eugene Lamb | Winfried Hensinger | Wojciech Zurek | Wolfgang Haken | Wolfgang Lechner | Wolfgang Paul | Wolfgang Pauli | Wolfgang Ketterle | Xavier

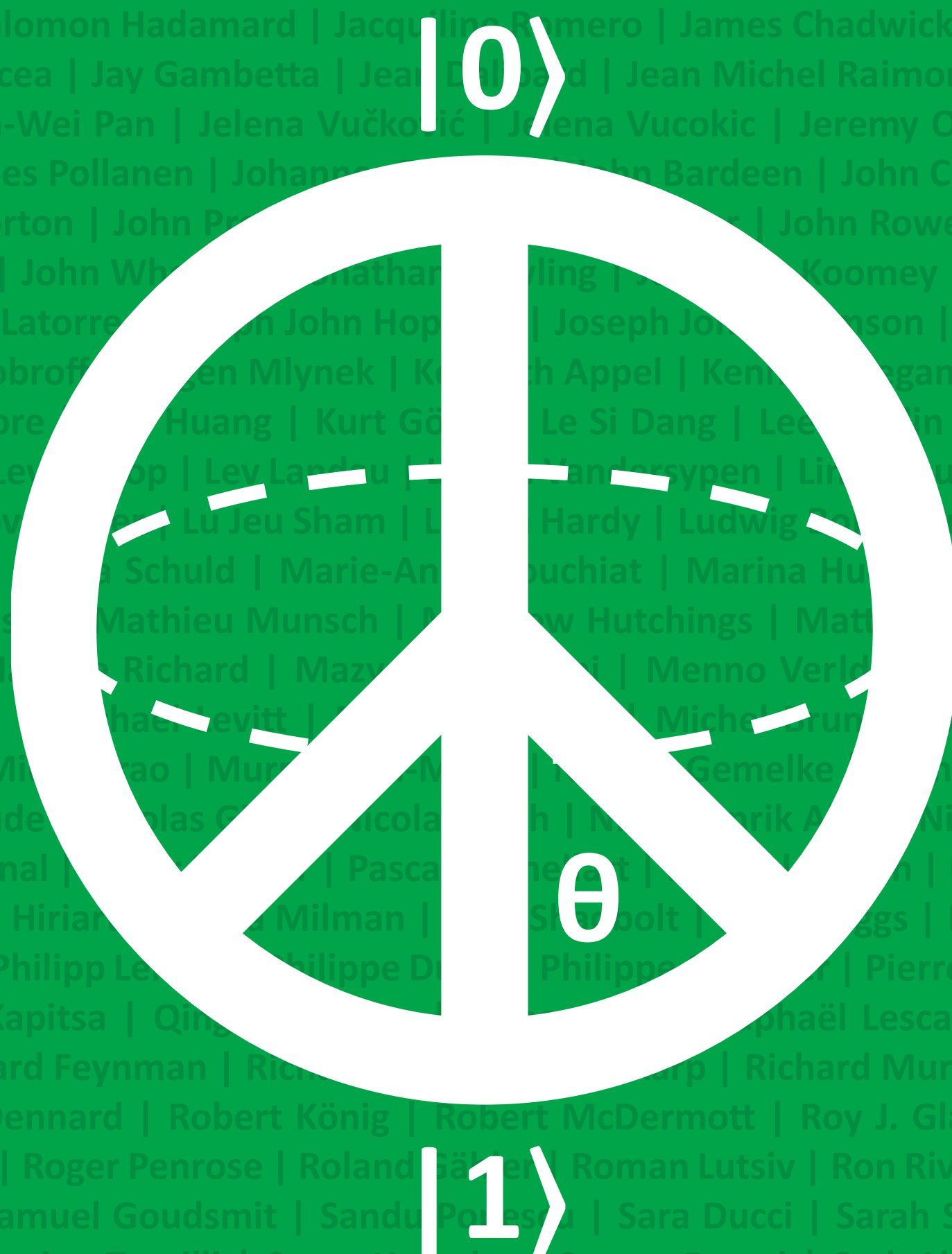